# The Physics of the $B$ Factories

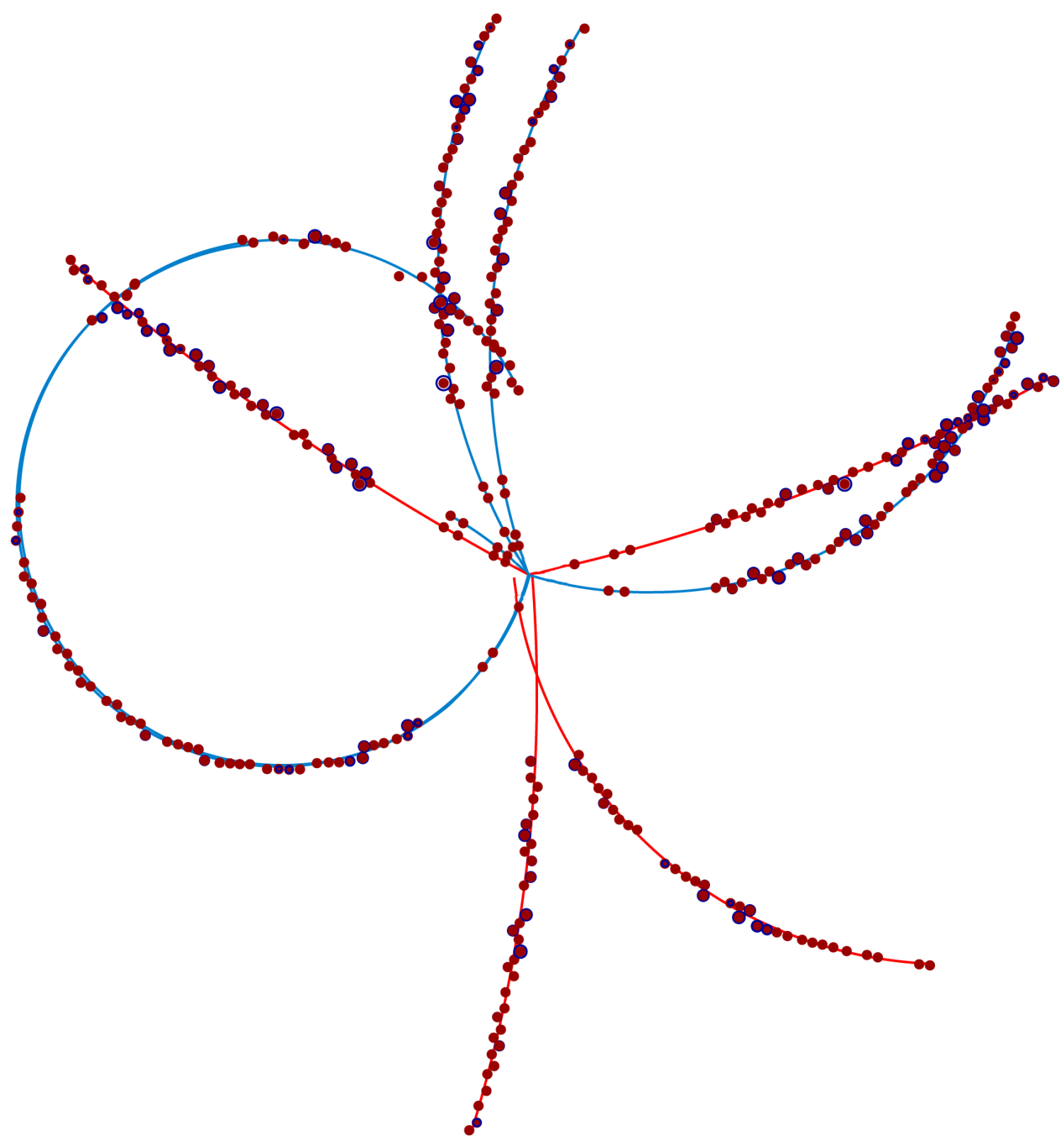

# Foreword

"The Physics of the *B* Factories" describes a decade long effort of physicists in the quest for the precise determination of asymmetry — broken symmetry — between particles and anti-particles. We now recognize that the matter we see around us is the residue — one part in a billion — of the matter and antimatter that existed in the early universe, most of which annihilated into the cosmic background radiation that bathes us. But the question remains: how did the baryonic matter-antimatter asymmetry arise? This book describes the work done by some 1000 physicists and engineers from around the globe on two experimental facilities built to test our understanding of this phenomenon, one at the SLAC National Accelerator Laboratory in California, USA, and a second at the KEK Laboratory, Tsukuba, Japan, and what we have learned from them in broadening our understanding of nature.

Why is our universe dominated by the matter of which we are made rather than equal parts of matter and antimatter? This question has puzzled physicists for decades. However, this was not the question we addressed when we wrote the paper on *CP* violation in 1972. Our question was whether we can explain the *CP* violation observed in the *K* meson decay within the framework of the renormalizable gauge theory. At that time, Sakharov's seminal paper was already published, but it did not attract our attention. If we were aware of the paper, we would have been misled into seeking a model satisfying Sakharov's conditions and our paper might not have appeared.

In our paper, we discussed that we need new particles in order to accommodate *CP* violation into the renormalizable electroweak theory, and proposed the six-quark scheme as one of the possible ways introducing new particles. We thought that the six-quark scheme is very interesting, but it was just a possibility. The situation changed when the tau-lepton was found and it was followed by the discovery of the Upsilon particle. The existence of the third generation became reality. However, it was still uncertain whether the mixing of the six quarks is a real origin of the observed *CP* violation. Theoretical calculation of *CP* asymmetries in the neutral *K* meson system contains uncertainty from strong interaction effects. What settled this problem were the *B* Factories built at SLAC and KEK.

These *B* Factories are extraordinary in many ways. In order to fulfill the requirements of special experiments, the beam energies of the colliding electron and positron are asymmetric, and the luminosity is unprecedentedly high. It is also remarkable that severe competition between the two laboratories boosted their performance. One of us (M. Kobayashi) has been watching the development at KEK very closely as the director of the Institute of Particle and Nuclear Studies of KEK for a period of time. As witnesses, we appreciate the amazing achievement of those who participated in these projects at both laboratories.

The *B* Factories have contributed a great deal to our understanding of particle physics, as documented in this book. In particular, thanks to the high luminosity far exceeding the design value, experimental groups measured mixing angles precisely and verified that the dominant source of *CP* violation observed in the laboratory experiments is flavor mixing among the three generations of quarks. Obviously we owe our Nobel Prize to this result.

Now we are awaiting the operation of the next-generation Super *B* Factories. In spite of its great success, the Standard Model is not an ultimate theory. For example, it is not thought to be possible for the matter dominance of the universe to be explained by the Standard Model. This means that there will still be unknown particles and unknown interactions. We have a lot of theoretical speculations but experimental means are rather limited. There are great expectations for the Super *B* Factories to reveal a clue to the world beyond the Standard Model.

Makoto Kobayashi
Honorary Professor Emeritus
KEK

Toshihide Maskawa
Director General
Kobayashi-Maskawa Institute for the Origin of Particles and the Universe
Nagoya University

# Preface

The inspiration for this book came from François le Diberder. During his term as spokesperson for *BABAR* he laid down a vision for the two $B$ Factory detector collaborations, *BABAR* and Belle, to work together on a book that would describe the methodologies used and physics results obtained by those experiments. A key ideal emphasized from the outset was that this book should be written from a pedagogical perspective; it should be of interest to the student and expert alike. This vision was presented during a *BABAR* collaboration meeting on the island of Elba in May 2008 and a follow up Belle collaboration meeting at KEK, with visiting colleagues from the *BABAR* collaboration, and was embraced by the community. A number of workshops involving people from the theoretical community as well as the two collaborations were held on four continents over the following years. The resulting book, "The Physics of the $B$ Factories", is a testament to the way that this concept captured the zeitgeist on both sides of the Pacific Ocean.

This book is divided into three parts, the first of which provides a brief description of the $B$ Factories, including a short (though not exhaustive) historical perspective, as well as descriptions of the detectors, ancillary data acquisition systems and data (re)processing systems that were built by the two detector collaborations in the late 1990's. The second part of the book discusses tools and methods that are frequently used when analyzing the data collected. These range from details of low level reconstruction algorithms and abstract summaries of statistical methods to high level prescriptions used when evaluating systematic uncertainties on measurements of observables. The third part of the book is devoted to physics results. This includes sufficient theoretical discussion in order for the reader to understand the context of the work being described. We are indebted to our colleagues from the theoretical community who have helped us achieve our goal of explaining the physics of the $B$ Factories in a broader context.

It should be noted that both $B$ Factory experiments are still actively publishing results and as a result the work presented here is a snapshot of the output of the $B$ Factories up to some point in time. Where appropriate, measurements from other experiments have been mentioned, however the focus of this book is on the output of the $B$ Factories. As a result, any brief description of important work by others should be interpreted as a suggestion for further reading on a given topic.

Just as there are two $B$ Factories, many of the observables studied or used in this book have a dual notation in the literature. While preparing this book we have placed the emphasis on the physics rather than trivialities such as convention. The most notable instance of this issue found here is that of the nomenclature used for the angles of the Unitarity Triangle. In order to retain a pedagogical approach we chose a method for selecting between the two notations that is symbolic of their equivalence from the perspective of physics. This choice was decided on the outcome of a coin flip.

It has been a privilege for us to work with our colleagues from the experimental and theoretical communities while compiling this book. The journey of preparing this tome has been as rewarding as being a part of the individual collaborations. This book has come into existence because of the efforts of the many people who have devoted their time and effort writing contributions found herein, and it belongs to the community who helped create it.

Adrian Bevan
Queen Mary University of London

Boštjan Golob
University of Ljubljana
Jožef Stefan Institute

Thomas Mannel
University of Siegen

Soeren Prell
Iowa State University

Bruce Yabsley
University of Sydney

# How to cite this work:

Please note that this is the official version of *The Physics of the B Factories.* An auxiliary version of this book will be made available online, both on `arXiv` and the INSPIRE database, under the same entry as the official version of the book. The official version of the book uses the notation $\phi_1$, $\phi_2$, $\phi_3$ for the angles of the Unitarity Triangle, and the auxiliary version uses the notation $\beta$, $\alpha$, $\gamma$.

# A note on conventions:

This book follows common practice in particle physics by using a relaxed system of natural units. The reduced Planck constant $\hbar$ is set to unity, and electromagnetic expressions include the fine structure constant $\alpha$ rather than dimensionful constants. Nevertheless, the units of energy (GeV, MeV, etc.) are distinguished from those of momentum ($\mathrm{GeV}/c, \mathrm{MeV}/c$) and mass ($\mathrm{GeV}/c^2, \mathrm{MeV}/c^2$); when length and time are explicitly mentioned, and especially in detector-related discussions, meters and seconds are used rather than the reciprocal of energy.

The treatment of charge conjugation depends on the context. Many analyses are motivated by possible differences between the behaviour of $B^0$ and $\overline{B}^0$: in such cases, samples of the two states are distinguished. When describing the method, however, if the text specifies reconstruction of $B^0 \to \pi^+ D^-$ with $D^- \to K^+\pi^-\pi^-$, it is usually implied that the equivalent procedure is followed for the charge conjugate mode $\overline{B}^0 \to \pi^- D^+$ with $D^+ \to K^-\pi^+\pi^+$. From time to time, explicit statements are made to resolve potential ambiguities.

Citations follow the author-year format, used in a flexible way. The most common form is surrounded by parentheses (Kobayashi and Maskawa, 1973). However, about 20% of cases incorporate the names of the authors into the grammar of the sentence, as when referring to the classic paper of Kobayashi and Maskawa (1973). Variant forms are used within the text of a parenthesis; all should be clear from the context.

The only unusual feature is the use of three bibliographies: one for *BABAR* papers (page 806), one for Belle papers (page 822), and one for other references (page 835). To avoid tedium, the "et al." is omitted for $B$ Factory papers, citing only the first author of full *BABAR* Collaboration authorlists (Aubert, 2001e), and either the first member (Choi, 2011) or the whole of the first-authorship group (Mizuk, Danilov, 2006) for full Belle Collaboration authorlists. Long authorlists for "other" references are treated normally. The great majority of *BABAR* papers have either Aubert, del Amo Sanchez, or Lees as first author; most early Belle papers have Abe, but from 2002 onwards show great variety. Results are described as being from *BABAR* or Belle if the responsible experiment is not already apparent from the context. Occasionally, a *BABAR* paper and a Belle paper will be cited together, for example in a quoted average or in the body of a table. It should always be clear which bibliography is meant.

In such a long work, there is inevitably some variation in style and usage. As editors, we have endeavoured to keep this to a minimum.

A. J. Bevan[*1], B. Golob[*2,3], Th. Mannel[*4], S. Prell[*5], B. D. Yabsley[*6],

K. Abe[§7], H. Aihara[§8], F. Anulli[§9,10], N. Arnaud[§11], T. Aushev[§12], M. Beneke[§13,14], J. Beringer[§15], F. Bianchi[§16,17], I. I. Bigi[§18], M. Bona[§16,17], N. Brambilla[§13], J. Brodzicka[§19], P. Chang[§20], M. J. Charles[§21], C. H. Cheng[§22], H.-Y. Cheng[§23], R. Chistov[§12], P. Colangelo[§24], J. P. Coleman[§25], A. Drutskoy[§12,26], V. P. Druzhinin[§27,28], S. Eidelman[§27,28], G. Eigen[§29], A. M. Eisner[§30], R. Faccini[§10,31], K. T. Flood[§22], P. Gambino[§16,17], A. Gaz[§32], W. Gradl[§33], H. Hayashii[§34], T. Higuchi[§35], W. D. Hulsbergen[§36], T. Hurth[§33], T. Iijima[§37,38], R. Itoh[§7], P. D. Jackson[§10,31], R. Kass[§39], Yu. G. Kolomensky[§15], E. Kou[§11], P. Križan[§2,3], A. Kronfeld[§40], S. Kumano[§41,42], Y. J. Kwon[§43], T. E. Latham[§44], D. W. G. S. Leith[§45], V. Lüth[§45], F. Martinez-Vidal[§46], B. T. Meadows[§47], R. Mussa[§16,17], M. Nakao[§7], S. Nishida[§7], J. Ocariz[§48], S. L. Olsen[§49], P. Pakhlov[§12,50], G. Pakhlova[§12], A. Palano[§24,51], A. Pich[§52], S. Playfer[§53], A. Poluektov[§27,28], F. C. Porter[§22], S. H. Robertson[§54], J. M. Roney[§55], A. Roodman[§45], Y. Sakai[§7], C. Schwanda[§56], A. J. Schwartz[§47], R. Seidl[§57], S. J. Sekula[§58], M. Steinhauser[§59], K. Sumisawa[§7], E. S. Swanson[§60], F. Tackmann[§61], K. Trabelsi[§7], S. Uehara[§7], S. Uno[§7], R. van der Water[§40], G. Vasseur[§62], W. Verkerke[§63], R. Waldi[§64], M. Z. Wang[§20], F. F. Wilson[§65], J. Zupan[§3,47], A. Zupanc[§3],

I. Adachi[¶7], J. Albert[¶55], Sw. Banerjee[¶55], M. Bellis[¶66], E. Ben-Haim[¶48], P. Biassoni[¶67,68], R. N. Cahn[¶15], C. Cartaro[¶45], J. Chauveau[¶48], C. Chen[¶5], C. C. Chiang[¶20], R. Cowan[¶69], J. Dalseno[¶70], M. Davier[¶11], C. Davies[¶71], J. C. Dingfelder[¶45,72], B. Echenard[¶22], D. Epifanov[¶8], B. G. Fulsom[¶45], A. M. Gabareen[¶45], J. W. Gary[¶73], R. Godang[¶74], M. T. Graham[¶45], A. Hafner[¶33], B. Hamilton[¶36], T. Hartmann[¶64], K. Hayasaka[¶37,38], C. Hearty[¶75], Y. Iwasaki[¶7], A. Khodjamirian[¶4], A. Kusaka[¶8], A. Kuzmin[¶27,28], G. D. Lafferty[¶76], A. Lazzaro[¶67,68], J. Li[¶49], D. Lindemann[¶45], O. Long[¶73], A. Lusiani[¶77,78], G. Marchiori[¶48], M. Martinelli[¶24,51], K. Miyabayashi[¶34], R. Mizuk[¶12,50], G. B. Mohanty[¶79], D. R. Muller[¶45], H. Nakazawa[¶80], P. Ongmongkolkul[¶22], S. Pacetti[¶81,82], F. Palombo[¶67,68], T. K. Pedlar[¶83], L. E. Piilonen[¶84], A. Pilloni[¶10,31], V. Poireau[¶85], K. Prothmann[¶70,86], T. Pulliam[¶45], M. Rama[¶9], B. N. Ratcliff[¶45], P. Roudeau[¶11], S. Schrenk[¶47], T. Schroeder[¶87], K. R. Schubert[¶88], C. P. Shen[¶89], B. Shwartz[¶27,28], A. Soffer[¶90], E. P. Solodov[¶27,28], A. Somov[¶47], M. Starič[¶3], S. Stracka[¶67,68], A. V. Telnov[¶91], K. Yu. Todyshev[¶27,28], T. Tsuboyama[¶7], T. Uglov[¶12,26], A. Vinokurova[¶27,28], J. J. Walsh[¶77,92], Y. Watanabe[¶93], E. Won[¶94], G. Wormser[¶11], D. H. Wright[¶45], S. Ye[¶95], C. C. Zhang[¶96],

S. Abachi[97], A. Abashian[†84], K. Abe[98], K. Abe[75], N. Abe[99], R. Abe[100], T. Abe[7], T. Abe[32], G. S. Abrams[15], I. Adam[45], K. Adamczyk[19], A. Adametz[101], T. Adye[65], A. Agarwal[55], H. Ahmed[55], M. Ahmed[102], S. Ahmed[102], B. S. Ahn[94], H. S. Ahn[49], I. J. R. Aitchison[45], K. Akai[7], S. Akar[48], M. Akatsu[38], M. Akemoto[7], R. Akhmetshin[27], R. Akre[†45], M. S. Alam[102], J. N. Albert[11], R. Aleksan[62], J. P. Alexander[7], G. Alimonti[103], M. T. Allen[45], J. Allison[76], T. Allmendinger[39], J. R. G. Alsmiller[104], D. Altenburg[105], K. E. Alwyn[76], Q. An[106], J. Anderson[36], R. Andreassen[47], D. Andreotti[107], M. Andreotti[107,108], J. C. Andress[109], C. Angelini[77,92], D. Anipko[27], A. Anjomshoaa[53], P. L. Anthony[45], E. A. Antillon[32], E. Antonioli[110], K. Aoki[7], J. F. Arguin[111], K. Arinstein[27,28], K. Arisaka[97], K. Asai[34], M. Asai[112], Y. Asano[113], D. J. Asgeirsson[75], D. M. Asner[114], T. Aso[115], M. L. Aspinwall[116], D. Aston[45], H. Atmacan[73], B. Aubert[85], V. Aulchenko[27,28], R. Ayad[117], T. Azemoon[45], T. Aziz[79], V. Azzolini[46], D. E. Azzopardi[1], M. A. Baak[118], J. J. Back[44], S. Bagnasco[119,120], S. Bahinipati[121], D. S. Bailey[76], S. Bailey[122], P. Bailly[48], N. van Bakel[45], A. M. Bakich[6], A. Bala[123], V. Balagura[12], R. Baldini-Ferroli[9], Y. Ban[124], E. Banas[19], H. R. Band[125], S. Banerjee[79], E. Baracchini[10,31], R. Barate[85], E. Barberio[126], M. Barbero[103], D. J. Bard[45], T. Barillari[32], N. R. Barlow[76], R. J. Barlow[76], M. Barrett[103,127], W. Bartel[61], J. Bartelt[45], R. Bartoldus[45], G. Batignani[77,92], M. Battaglia[15], J. M. Bauer[128], A. Bay[129], M. Beaulieu[111], P. Bechtle[45], T. W. Beck[30], J. Becker[32], J. Becla[45], I. Bedny[27,28], S. Behari[7], P. K. Behera[21,130], E. Behn[36], L. Behr[131], C. Beigbeder[11], D. Beiline[27], R. Bell[†45], F. Bellini[10,31], G. Bellodi[1], K. Belous[132], M. Benayoun[48], G. Benelli[39], J. F. Benitez[45], M. Benkebil[11], N. Berger[45], J. Bernabeu[46], D. Bernard[131], R. Bernet[53], F. U. Bernlochner[55], J. W. Berryhill[63], K. Bertsche[45], P. Besson[†62], D. S. Best[133], S. Bettarini[77,92], D. Bettoni[107], V. Bhardwaj[34], W. Bhimji[116], B. Bhuyan[134], B. Bhuyan[135], M. E. Biagini[9], M. Biasini[81,82], K. van Bibber[136], J. Biesiada[91], I. Bingham[25], R. M. Bionta[136], M. Bischofberger[34], U. Bitenc[3], I. Bizjak[3], F. Blanc[32], G. Blaylock[137], V. E. Blinov[27,28,138], E. Bloom[45], P. C. Bloom[32], N. L. Blount[139], J. Blouw[117], M. Bly[65], S. Blyth[140], C. T. Boeheim[45], M. Bomben[48], A. Bondar[27,28], M. Bondioli[133], G.R. Bonneaud[48], G. Bonvicini[141], M. Booke[†128], J. Booth[133], C. Borean[142,143], A. W. Borgland[15], E. Borsato[144,145], F. Bosi[77], L. Bosisio[142,143], A. A. Botov[27], J. Bougher[146], K. Bouldin[45], P. Bourgeois[62], D. Boutigny[85], D. A. Bowerman[116], A. M. Boyarski[45], R. F. Boyce[45], J. T. Boyd[109], A. Bozek[19], C. Bozzi[107], M. Bračko[3,147], G. Brandenburg[†122], T. Brandt[88], B. Brau[39], J. Brau[139], A. B. Breon[15], D. Breton[11], C. Brew[65], H. Briand[48], P. G. Bright-Thomas[148], V. Brigljević[136], D. I. Britton[54], F. Brochard[131], B. Broomer[32], J. Brose[88], T. E. Browder[103], C. L. Brown[149], C. M. Brown[55], D. N. Brown[15], D. N. Brown[146], M. Browne[45], M. Bruinsma[133], S. Brunet[111], F. Bucci[77,92], C. Buchanan[97], O. L. Buchmueller[45], C. Bünger[64], W. Bugg[150], A. D. Bukin[†27,28], R. Bula[102], H. Bulten[118], P. R. Burchat[66], W. Burgess[45], J. P. Burke[25], J. Button-Shafer[15], A. R. Buzykaev[27], A. Buzzo[119], Y. Cai[45], R. Calabrese[107,108], A. Calcaterra[9], G. Calderini[48], B. Camanzi[127], E. Campagna[77,92], C. Campagnari[63], R. Capra[119,120], V. Carassiti[107], M. Carpinelli[77,92], M. Carroll[25], G. Casarosa[77,92], B. C. K. Casey[103], N. M. Cason[18], G. Castelli[144], N. Cavallo[151], G. Cavoto[10], A. Cecchi[107],

R. Cenci[77,92], G. Cerizza[67,68], A. Cervelli[77,92], A. Ceseracciu[45], X. Chai[21], K. S. Chaisanguanthum[122], M. C. Chang[152], Y. H. Chang[80], Y. W. Chang[20], D. S. Chao[22], M. Chao[133], Y. Chao[20], E. Charles[15], C. A. Chavez[25], R. Cheaib[54], V. Chekelian[70], A. Chen[80], A. Chen[117], E. Chen[22], G. P. Chen[96], H. F. Chen[106], J. -H. Chen[20], J. C. Chen[96], K. F. Chen[20], P. Chen[20], S. Chen[32], W. T. Chen[80], X. Chen[125], X. R. Chen[153], Y. Q. Chen[20], B. Cheng[125], B. G. Cheon[154], N. Chevalier[109], Y. M. Chia[76], S. Chidzik[91], K. Chilikin[12], M. V. Chistiakova[15], R. Cizeron[11], I. S. Cho[43], K. Cho[155], V. Chobanova[70], H. H. F. Choi[55], K. S. Choi[43], S. K. Choi[156], Y. Choi[157], Y. K. Choi[157], S. Christ[64], P. H. Chu[20], S. Chun[97], A. Chuvikov[91], G. Cibinetto[107], D. Cinabro[141], A. R. Clark[15], P. J. Clark[53], C. K. Clarke[1], R. Claus[45], B. Claxton[65], Z. C. Clifton[32], J. Cochran[5], J. Cohen-Tanugi[131], H. Cohn[150], T. Colberg[88], S. Cole[6], F. Colecchia[144,145], C. Condurache[65], R. Contri[119,120], P. Convert[62], M. R. Convery[45], P. Cooke[25], N. Copty[153], C. M. Cormack[1], F. Dal Corso[144], L. A. Corwin[39], F. Cossutti[142], D. Cote[111], A. Cotta Ramusino[107], W. N. Cottingham[109], F. Couderc[85], D. P. Coupal[45], R. Covarelli[81,82], G. Cowan[149], W. W. Craddock[45], G. Crane[45], H. B. Crawley[5], L. Cremaldi[128], A. Crescente[144], M. Cristinziani[45], J. Crnkovic[158], G. Crosetti[119,120], T. Cuhadar-Donszelmann[75], A. Cunha[63], S. Curry[133], A. D'Orazio[10,31], S. Dû[11], G. Dahlinger[88], B. Dahmes[63], C. Dallapiccola[137], N. Danielson[91], M. Danilov[12,26], A. Das[79], M. Dash[84], S. Dasu[125], M. Datta[125], F. Daudo[16], P. D. Dauncey[116], P. David[48], C. L. Davis[146], C. T. Day[15], F. De Mori[16,17], G. De Domenico[62], N. De Groot[65], C. De la Vaissière[48], Ch. de la Vaissière[48], A. de Lesquen[62], G. De Nardo[151,159], R. de Sangro[9], A. De Silva[160], S. DeBarger[45], F. J. Decker[45], P. del Amo Sanchez[85], L. Del Buono[48], V. Del Gamba[77,92], D. del Re[10,31], G. Della Ricca[142,143], A. G. Denig[33,161], D. Derkach[11], I. M. Derrington[32], H. DeStaebler[†45], J. Destree[32], S. Devmal[47], B. Dey[73], B. Di Girolamo[16], E. Di Marco[10,31], M. Dickopp[88], M. O. Dima[32], S. Dittrich[64], S. Dittongo[142,143], P. Dixon[1], L. Dneprovsky[†27], F. Dohou[131], Y. Doi[7], Z. Doležal[162], D. A. Doll[22], M. Donald[45], L. Dong[5], L. Y. Dong[96], J. Dorfan[45], A. Dorigo[144], M. P. Dorsten[22], R. Dowd[126], J. Dowdell[65], Z. Drásal[162], J. Dragic[7], B. W. Drummond[95], R. S. Dubitzky[101], G. P. Dubois-Felsmann[45], M. S. Dubrovin[47], Y. C. Duh[152], Y. T. Duh[20], D. Dujmic[69], W. Dungel[56], W. Dunwoodie[45], D. Dutta[134], A. Dvoretskii[22], N. Dyce[109], M. Ebert[45], E. A. Eckhart[117], S. Ecklund[45], R. Eckmann[163], P. Eckstein[88], C. L. Edgar[76], A. J. Edwards[164], U. Egede[116], A. M. Eichenbaum[125], P. Elmer[91], S. Emery[62], Y. Enari[38], R. Enomoto[7], E. Erdos[32], R. Erickson[45], J. A. Ernst[102], R. J. Erwin[22], M. Escalier[62], V. Eschenburg[128], I. Eschrich[133], S. Esen[47], L. Esteve[62], F. Evangelisti[107], C. W. Everton[126], V. Eyges[5], C. Fabby[47], F. Fabozzi[151], S. Fahey[32], M. Falbo[165], S. Fan[45], F. Fang[103], F. Fang[22], C. Fanin[144], A. Farbin[36], H. Farhat[141], J. E. Fast[114], M. Feindt[161], A. Fella[110], E. Feltresi[144,145], T. Ferber[61], R. E. Fernholz[91], S. Ferrag[131], F. Ferrarotto[10], F. Ferroni[10,31], R. C. Field[45], A. Filippi[16,17], G. Finocchiaro[9], E. Fioravanti[107], J. Firmino da Costa[11], P.-A. Fischer[5], A. Fisher[45], P. H. Fisher[69], C. J. Flacco[30], R. L. Flack[116], H. U. Flaecher[149], J. Flanagan[7], J. M. Flanigan[63], K. E. Ford[148], W. T. Ford[32], I. J. Forster[25], A. C. Forti[76], F. Forti[77,92], D. Fortin[55], B. Foster[109], S. D. Foulkes[73], G. Fouque[131], J. Fox[45], P. Franchini[107], M. Franco Sevilla[63], B. Franek[65], E. D. Frank[166], K. B. Fransham[55], S. Fratina[3], K. Fratini[10], A. Frey[167], R. Frey[139], M. Friedl[56], M. Fritsch[33], J. R. Fry[25], H. Fujii[7], M. Fujikawa[34], Y. Fujita[7], Y. Fujiyama[99], C. Fukunaga[168], M. Fukushima[7], J. Fullwood[76], Y. Funahashi[7], Y. Funakoshi[7], F. Furano[144], M. Furman[15], K. Furukawa[7], H. Futterschneider[88], E. Gabathuler[25], T. A. Gabriel[104], N. Gabyshev[27,28], F. Gaede[32], N. Gagliardi[144,145], A. Gaidot[62], J.-M. Gaillard[85], J. R. Gaillard[116], S. Galagedera[65], F. Galeazzi[144,145], F. Gallo[16,17], D. Gamba[16,17], R. Gamet[25], K. K. Gan[39], P. Gandini[67,68], S. Ganguly[141], S. F. Ganzhur[62], Y. Y. Gao[169], I. Gaponenko[45], A. Garmash[27,28], J. Garra Tico[170], I. Garzia[107], M. Gaspero[10,31], F. Gastaldi[131], C. Gatto[151], V. Gaur[79], N. I. Geddes[65], T. L. Geld[47], J.-F. Genat[48], K. A. George[1], M. George[25], S. George[149], Z. Georgette[62], T. J. Gershon[7,44], M. S. Gill[15], R. Gillard[141], J. D. Gilman[32], F. Giordano[158], M. A. Giorgi[77,92], P.-F. Giraud[62], L. Gladney[166], T. Glanzman[45], R. Glattauer[56], A. Go[80], K. Goetzen[87], Y. M. Goh[154], G. Gokhroo[79], P. Goldenzweig[47], V. B. Golubev[27,28], G. P. Gopal[65], A. Gordon[126], A. Gorišek[3], V. I. Goriletsky[171], R. Gorodeisky[90], L. Gosset[62], K. Gotow[84], S. J. Gowdy[45], P. Graffin[62], S. Grancagnolo[142,143], E. Grauges[170], G. Graziani[62], M. G. Green[149], M. G. Greene[172], G. J. Grenier[21], P. Grenier[45], K. Griessinger[33], A. A. Grillo[30], B.V. Grinyov[171], A. V. Gritsan[169], G. Grosdidier[11], M. Grosse Perdekamp[57,158], P. Grosso[16], M. Grothe[30], Y. Groysman[15], O. Grünberg[64], E. Guido[119,120], H. Guler[103], N. J. W. Gunawardane[116], Q. H. Guo[166], R. S. Guo[173], Z. J. Guo[169], N. Guttman[90], H. Ha[94], H. C. Ha[94], T. Haas[45], J. Haba[7], J. Hachtel[32], H. K. Hadavand[174], T. Hadig[45], C. Hagner[84], M. Haire[175], F. Haitani[98], T. Haji[8], G. Haller[45], V. Halyo[45], K. Hamano[55], H. Hamasaki[7], G. Hamel de Monchenault[62], J. Hamilton[45], R. Hamilton[21], O. Hamon[48], B. Y. Han[94], Y. L. Han[96], H. Hanada[176], K. Hanagaki[91], F. Handa[176], J. E. Hanson[22], A. Hanushevsky[45], K. Hara[7], T. Hara[7], Y. Harada[100], P. F. Harrison[44], T. J. Harrison[148], B. Harrop[91], A. J. Hart[148], P. A. Hart[76], B. L. Hartfiel[97], J. L. Harton[117], T. Haruyama[7], A. Hasan[45], Y. Hasegawa[177], C. Hast[45], N. C. Hastings[8], K. Hasuko[57], A. Hauke[105], C. M. Hawkes[148], K. Hayashi[7], M. Hazumi[7], C. Hee[45], E. M. Heenan[126], D. Heffernan[178], T. Held[87], R. Henderson[160], S. W. Henderson[69], S. S. Hertzbach[137], S. Hervé[62], M. Heß[64], C. A. Heusch[30], A. Hicheur[85], Y. Higashi[7], Y. Higasino[38], I. Higuchi[176], S. Hikita[179], E. J. Hill[174], T. Himel[45], L. Hinz[129], T. Hirai[99], H. Hirano[179], J. F. Hirschauer[32], D. G. Hitlin[22], N. Hitomi[7], M. C. Hodgkinson[76], A. Höcker[11], C. T. Hoi[20], T. Hojo[178], T. Hokuue[38], J. J. Hollar[125], T. M. Hong[63], K. Honscheid[39], B. Hooberman[15], D. A. Hopkins[149], Y. Horii[37,38], Y. Hoshi[98], K. Hoshina[179], S. Hou[20,80], W. S. Hou[20], T. Hryn'ova[45], Y. B. Hsiung[20], C. L. Hsu[20], S. C. Hsu[20], H. Hu[125], T. Hu[117], H. C. Huang[20], T. J. Huang[20], Y. C. Huang[173], Z. Huard[47], M. E. Huffer[45], D. Hufnagel[39],

T. Hung[45], D. E. Hutchcroft[25], H. J. Hyun[180], S. Ichizawa[99], T. Igaki[38], A. Igarashi[113], S. Igarashi[7], Y. Igarashi[7], O. Igonkina[139], K. Ikado[38], H. Ikeda[7], H. Ikeda[7], K. Ikeda[34], J. Ilic[44], K. Inami[38], W. R. Innes[45], Y. Inoue[181], A. Ishikawa[7], A. Ishikawa[176], H. Ishino[99], K. Itagaki[176], S. Itami[38], K. Itoh[8], V. N. Ivanchenko[27], R. Iverson[45], M. Iwabuchi[43], G. Iwai[100], M. Iwai[7], S. Iwaida[113], M. Iwamoto[182], H. Iwasaki[7], M. Iwasaki[8], M. Iwasaki[139], T. Iwashita[34], J. M. Izen[95], D. J. Jackson[178], F. Jackson[76], G. Jackson[76], P. S. Jackson[149], R. G. Jacobsen[15], C. Jacoby[129], I. Jaegle[103], V. Jain[102], P. Jalocha[19], H. K. Jang[49], H. Jasper[105], A. Jawahery[36], S. Jayatilleke[47], C. M. Jen[20], F. Jensen[15], C. P. Jessop[18], X. B. Ji[96], M. J. J. John[48], D. R. Johnson[32], J. R. Johnson[125], S. Jolly[127], M. Jones[103], K. K. Joo[7], N. Joshi[79], N. J. Joshi[79], D. Judd[175], T. Julius[126], R. W. Kadel[15], J. A. Kadyk[15], H. Kagan[39], R. Kagan[12], D. H. Kah[180], S. Kaiser[88], H. Kaji[38], S. Kajiwara[178], H. Kakuno[168], T. Kameshima[113], J. Kaminski[45], T. Kamitani[7], J. Kaneko[99], J. H. Kang[43], J. S. Kang[94], T. Kani[38], P. Kapusta[19], T.M. Karbach[105], M. Karolak[62], Y. Karyotakis[85], K. Kasami[7], G. Katano[7], S. U. Kataoka[34], N. Katayama[7], E. Kato[176], Y. Kato[38], H. Kawai[182], H. Kawai[8], M. Kawai[7], N. Kawamura[183], T. Kawasaki[100], J. Kay[65], M. Kay[25], M. P. Kelly[76], M. H. Kelsey[45], N. Kent[103], L. T. Kerth[15], A. Khan[127], H. R. Khan[99], D. Kharakh[45], A. Kibayashi[7], H. Kichimi[7], C. Kiesling[70], M. Kikuchi[7], E. Kikutani[7], B. H. Kim[49], C. H. Kim[49], D. W. Kim[157], H. Kim[45], H. J. Kim[180], H. J. Kim[43], H. O. Kim[180], H. W. Kim[94], J. B. Kim[94], J. H. Kim[155], K. T. Kim[94], M. J. Kim[180], P. Kim[45], S. K. Kim[49], S. M. Kim[157], T. H. Kim[43], Y. I. Kim[180], Y. J. Kim[155], G. J. King[55], K. Kinoshita[47], A. Kirk[148], D. Kirkby[133], I. Kitayama[95], M. Klemetti[54], V. Klose[184], J. Klucar[3], N. S. Knecht[75], K. J. Knoepfel[18], D. J. Knowles[148], B. R. Ko[94], N. Kobayashi[99], S. Kobayashi[185], T. Kobayashi[7], M. J. Kobel[88], S. Koblitz[70], H. Koch[87], M. L. Kocian[45], P. Kodyš[162], K. Koeneke[69], R. Kofler[137], S. Koike[7], S. Koishi[99], H. Koiso[7], J. A. Kolb[139], S. D. Kolya[76], Y. Kondo[7], H. Konishi[179], P. Koppenburg[7], V. B. Koptchev[137], T. M. B. Kordich[172], A. A. Korol[27,28], K. Korotushenko[91], S. Korpar[3,147], R. T. Kouzes[114], D. Kovalskyi[63], R. Kowalewski[55], Y. Kozakai[38], W. Kozanecki[62], J. F. Kral[15], A. Krasnykh[45], R. Krause[88], E. A. Kravchenko[27,28], J. Krebs[45], A. Kreisel[32], M. Kreps[161], M. Krishnamurthy[150], R. Kroeger[128], W. Kroeger[45], P. Krokovny[27,28], B. Kronenbitter[161], J. Kroseberg[30], T. Kubo[7], T. Kuhr[161], G. Kukartsev[15], R. Kulasiri[47], A. Kulikov[45], R. Kumar[186], S. Kumar[123], T. Kumita[168], T. Kuniya[185], M. Kunze[87], C. C. Kuo[80], T. -L. Kuo[20], H. Kurashiro[99], E. Kurihara[182], N. Kurita[45], Y. Kuroki[178], A. Kurup[149], P. E. Kutter[125], N. Kuznetsova[63], P. Kvasnička[162], P. Kyberd[127], S. H. Kyeong[43], H. M. Lacker[184], C. K. Lae[169], E. Lamanna[10,31], J. Lamsa[5], L. Lanceri[142,143], L. Landi[107,108], M. I. Lang[69], D. J. Lange[136], J. S. Lange[187], U. Langenegger[101], M. Langer[62], A. J. Lankford[133], F. Lanni[67,68], S. Laplace[11], E. Latour[131], Y. P. Lau[91], D. R. Lavin[53], J. Layter[73], H. Lebbolo[48], C. LeClerc[15], T. Leddig[64], G. Leder[56], F. Le Diberder[11], C. L. Lee[122], J. Lee[49], J. S. Lee[157], M. C. Lee[20], M. H. Lee[7], M. J. Lee[49], M. J. Lee[15], S.-J. Lee[21], S. E. Lee[49], S. H. Lee[49], Y. J. Lee[20], J. P. Lees[85], M. Legendre[62], M. Leitgab[158], R. Leitner[162], E. Leonardi[10], C. Leonidopoulos[91], V. Lepeltier[†11], Ph. Leruste[48], T. Lesiak[19,188], M. E. Levi[15], S. L. Levy[63], B. Lewandowski[†87], M. J. Lewczuk[55], P. Lewis[45], H. Li[125], H. B. Li[96], S. Li[45], X. Li[49], X. Li[137], Y. Li[84], Y. Li[146], L. Li Gioi[10,31], J. Libby[45,189], J. Lidbury[65], V. Lillard[36], C. L. Lim[43], A. Limosani[126], C. S. Lin[137], J. Y. Lin[152], S. W. Lin[20], Y. S. Lin[20], B. Lindquist[45], C. Lindsay[55], L. Lista[151], C. Liu[106], F. Liu[73], H. Liu[153], H. M. Liu[96], J. Liu[102], R. Liu[125], T. Liu[91], Y. Liu[47], Z. Q. Liu[96], D. Liventsev[7,12], M. Lo Vetere[119,120], C. B. Locke[55], W. S. Lockman[30], F. Di Lodovico[1], V. Lombardo[67,68], G. W. London[62], D. Lopes Pegna[91], L. Lopez[24,51], N. Lopez-March[46], J. Lory[48], J. M. LoSecco[18], X. C. Lou[95], R. Louvot[129], A. Lu[63], C. Lu[91], M. Lu[139], R. S. Lu[20], T. Lueck[55], S. Luitz[45], P. Lukin[27,28], P. Lund[150], E. Luppi[107,108], A. M. Lutz[11], O. Lutz[161], G. Lynch[15], H. L. Lynch[45], A. J. Lyon[76], V. R. Lyubinsky[171], D. B. MacFarlane[45], C. Mackay[109], J. MacNaughton[7], M. M. Macri[119], S. Madani[65], W. F. Mader[88], S. A. Majewski[66], G. Majumder[79], Y. Makida[7], B. Malaescu[11], R. Malaguti[107], J. Malclès[48], U. Mallik[21], E. Maly[88], H. Mamada[179], A. Manabe[7], G. Mancinelli[47], M. Mandelkern[133], F. Mandl[56], P. F. Manfredi[190], D. J. J. Mangeol[54], E. Manoni[81], Z. P. Mao[96], M. Margoni[144,145], C. E. Marker[149], G. Markey[65], J. Marks[101], D. Marlow[91], V. Marques[62], H. Marsiske[45], S. Martellotti[9], E. C. Martin[133], J. P. Martin[111], L. Martin[48], A. J. Martinez[30], M. Marzolla[144], A. Mass[109], M. Masuzawa[7], A. Mathieu[131], P. Matricon[131], T. Matsubara[8], T. Matsuda[191], T. Matsuda[7], H. Matsumoto[100], S. Matsumoto[192], T. Matsumoto[168], H. Matsuo[†193], T. S. Mattison[75], D. Matvienko[27,28], A. Matyja[19], B. Mayer[62], M. A. Mazur[63], M. A. Mazzoni[10], M. McCulloch[45], J. McDonald[45], J. D. McFall[109], P. McGrath[149], A. K. McKemey[127], J. A. McKenna[75], S. E. Mclachlin[†54], S. McMahon[25], T. R. McMahon[149], S. McOnie[6], T. Medvedeva[12], R. Melen[45], B. Mellado[125], W. Menges[1], S. Menke[45], A. M. Merchant[15], J. Merkel[105], R. Messner[†45], S. Metcalfe[45], S. Metzler[22], N. T. Meyer[21], T. I. Meyer[66], W. T. Meyer[5], A. K. Michael[32], G. Michelon[144,145], S. Michizono[7], P. Micout[62], V. Miftakov[91], A. Mihalyi[125], Y. Mikami[176], D. A. Milanes[46], M. Milek[54], T. Mimashi[7], J. S. Minamora[22], C. Mindas[91], S. Minutoli[119], L. M. Mir[15], K. Mishra[47], W. Mitaroff[56], H. Miyake[178], T. S. Miyashita[66], H. Miyata[100], Y. Miyazaki[38], L. C. Moffitt[126], G. B. Mohanty[44], A. Mohapatra[130], A. K. Mohapatra[125], D. Mohapatra[114], A. Moll[70], G. R. Moloney[126], J. P. Mols[62], R. K. Mommsen[133], M. R. Monge[119,120], D. Monorchio[151,159], T. B. Moore[137], G. F. Moorhead[126], P. Mora de Freitas[131], M. Morandin[144], N. Morgan[84], S. E. Morgan[148], M. Morganti[77,92], S. Morganti[10], S. Mori[113], T. Mori[38], M. Morii[122], J. P. Morris[39], F. Morsani[77], G. W. Morton[116], L. J. Moss[45], J. P. Mouly[62], R. Mount[45], J. Mueller[60], R. Müller-Pfefferkorn[88], M. Mugge[136], F. Muheim[53], A. Muir[25], E. Mullin[73], M. Munerato[107,108], A. Murakami[185], T. Murakami[7], N. Muramatsu[194], P. Musico[119], I. Nagai[38], T. Nagamine[176], Y. Nagasaka[112], Y. Nagashima[178], S. Nagayama[7], M. Nagel[32], M. T. Naisbit[76], T. Nakadaira[8],

Y. Nakahama[8], M. Nakajima[176], T. Nakajima[176], I. Nakamura[7], T. Nakamura[99], T. T. Nakamura[7], E. Nakano[181], H. Nakayama[7], J. W. Nam[157], S. Narita[176], I. Narsky[22], J .A. Nash[116], Z. Natkaniec[19], U. Nauenberg[32], M. Nayak[189], H. Neal[45], E. Nedelkovska[70], M. Negrini[107], K. Neichi[98], D. Nelson[45], S. Nelson[45], N. Neri[67], G. Nesom[30], S. Neubauer[161], D. Newman-Coburn[†1], C. Ng[8], X. Nguyen[111], H. Nicholson[195], C. Niebuhr[61], J. Y. Nief[11], M. Niiyama[193], M. B. Nikolich[116], N. K. Nisar[79], K. Nishimura[103], Y. Nishio[38], O. Nitoh[179], R. Nogowski[88], S. Noguchi[34], T. Nomura[193], M. Nordby[45], Y. Nosochkov[45], A. Novokhatski[45], S. Nozaki[176], T. Nozaki[7], I. M. Nugent[55], C. P. O'Grady[45], S. W. O'Neale[†148], F. G. O'Neill[45], B. Oberhof[77,92], P. J. Oddone[15], I. Ofte[45], A. Ogawa[57], K. Ogawa[7], S. Ogawa[196], Y. Ogawa[7], R. Ohkubo[7], K. Ohmi[7], Y. Ohnishi[7], F. Ohno[99], T. Ohshima[38], Y. Ohshima[99], N. Ohuchi[7], K. Oide[7], N. Oishi[38], T. Okabe[38], N. Okazaki[179], T. Okazaki[34], S. Okuno[93], E. O. Olaiya[65], A. Olivas[32], P. Olley[65], J. Olsen[91], S. Ono[99], G. Onorato[151,159], A. P. Onuchin[27,28,138], Y. Onuki[8], T. Ooba[182], T. J. Orimoto[15], T. Oshima[38], I. L. Osipenkov[15], W. Ostrowicz[19], C. Oswald[72], S. Otto[88], J. Oyang[22], A. Oyanguren[46], H. Ozaki[7], V. E. Ozcan[45], H. P. Paar[174], C. Padoan[107,108], K. Paick[175], H. Palka[†19], B. Pan[102], Y. Pan[125], W. Panduro Vazquez[116], J. Panetta[166], A. I. Panova[171], R. S. Panvini[†197], E. Panzenböck[34,167], E. Paoloni[77,92], P. Paolucci[151], M. Pappagallo[24,51], S. Paramesvaran[149], C. S. Park[49], C. W. Park[157], H. Park[180], H. Park[32], H. K. Park[180], K. S. Park[157], W. Park[153], R. J. Parry[25], N. Parslow[6], S. Passaggio[119], F. C. Pastore[119,120], P. M. Patel[†54], C. Patrignani[119,120], P. Patteri[9], T. Pavel[45], J. Pavlovich[146], D. J. Payne[25], L. S. Peak[6], D. R. Peimer[90], M. Pelizaeus[87], R. Pellegrini[67,68], M. Pelliccioni[16,17], C. C. Peng[20], J. C. Peng[20], K. C. Peng[20], T. Peng[106], Y. Penichot[62], S. Pennazzi[81,82], M. R. Pennington[44], R. C. Penny[148], A. Penzkofer[32], A. Perazzo[45], A. Perez[77], M. Perl[45], M. Pernicka[†56], J.-P. Perroud[129], I. M. Peruzzi[9,82], R. Pestotnik[3], K. Peters[87], M. Peters[103], B. A. Petersen[66], T. C. Petersen[11], E. Petigura[15], S. Petrak[45], A. Petrella[107], M. Petrič[3], A. Petzold[105], M. G. Pia[119], T. Piatenko[22], D. Piccolo[151,159], M. Piccolo[9], L. Piemontese[107], M. Piemontese[45], M. Pierini[125], S. Pierson[45], M. Pioppi[81,82], G. Piredda[10], M. Pivk[48], S. Plaszczynski[11], F. Polci[10,11,31], A. Pompili[24,51], P. Poropat[†142,143], M. Posocco[144], C. T. Potter[139], R. J. L. Potter[1], V. Prasad[135], E. Prebys[91], E. Prencipe[33], J. Prendki[48], R. Prepost[125], M. Prest[142], M. Prim[161], M. Pripstein[15], X. Prudent[85], S. Pruvot[11], E. M. T. Puccio[66], M. V. Purohit[153], N. D. Qi[96], H. Quinn[45], J. Raaf[47], R. Rabberman[91], F. Raffaelli[77], G. Ragghianti[150], S. Rahatlou[174], A. M. Rahimi[39], R. Rahmat[139], A. Y. Rakitin[22], A. Randle-Conde[58], P. Rankin[32], I. Rashevskaya[142], S. Ratkovsky[45], G. Raven[118], V. Re[190], M. Reep[128], J. J. Regensburger[39], J. Reidy[128], R. Reif[45], B. Reisert[70], C. Renard[131], F. Renga[10,31], S. Ricciardi[65], J. D. Richman[63], J. L. Ritchie[163], M. Ritter[70], C. Rivetta[45], G. Rizzo[77,92], C. Roat[66], P. Robbe[85], D. A. Roberts[36], A. I. Robertson[53], E. Robutti[119], S. Rodier[11], D. M. Rodriguez[32], J. L. Rodriguez[103], R. Rodriguez[45], N. A. Roe[15], M. Röhrken[161], W. Roethel[65], J. Rolquin[62], L. Romanov[27], A. Romosan[15], M. T. Ronan[†15], G. Rong[96], F. J. Ronga[7], L. Roos[48], N. Root[27], M. Rosen[103], E. I. Rosenberg[5], A. Rossi[81], A. Rostomyan[61], M. Rotondo[144], E. Roussot[131], J. Roy[32], M. Rozanska[19], Y. Rozen[63], Y. Rozen[198], A. E. Rubin[5], W. O. Ruddick[32], A. M. Ruland[163], K. Rybicki[19], A. Ryd[22], S. Ryu[49], J. Ryuko[178], S. Sabik[111], R. Sacco[1], M. A. Saeed[102], F. Safai Tehrani[10], H. Sagawa[7], H. Sahoo[103], S. Sahu[20], M. Saigo[176], T. Saito[176], S. Saitoh[199], K. Sakai[7], H. Sakamoto[193], H. Sakaue[181], M. Saleem[127], A. A. Salnikov[45], E. Salvati[137], F. Salvatore[149], A. Samuel[22], D. A. Sanders[128], P. Sanders[116], S. Sandilya[79], F. Sandrelli[77,92], W. Sands[91], W. R. Sands[91], M. Sanpei[98], D. Santel[47], L. Santelj[3], V. Santoro[107], A. Santroni[119,120], T. Sanuki[176], T. R. Sarangi[199], S. Saremi[137], A. Sarti[107,108], T. Sasaki[7], N. Sasao[193], M. Satapathy[130], Nobuhiko Sato[7], Noriaki Sato[38], Y. Sato[176], N. Satoyama[177], A. Satpathy[47,163], V. Savinov[60], N. Savvas[76], O. H. Saxton[45], K. Sayeed[47], S. F. Schaffner[91], T. Schalk[30], S. Schenk[101], J. R. Schieck[36], T. Schietinger[45,129], C. J. Schilling[163], R. H. Schindler[45], S. Schmid[56], R. E. Schmitz[30], H. Schmuecker[87], O. Schneider[129], G. Schnell[200,201], P. Schönmeier[176], K. C. Schofield[25], G. Schott[161], H. Schröder[†64], M. Schram[54], J. Schubert[88], J. Schümann[7], J. Schultz[133], B. A. Schumm[30], M. H. Schune[11], U. Schwanke[174], H. Schwarz[45], J. Schwiening[45], R. Schwierz[88], R. F. Schwitters[163], C. Sciacca[151,159], G. Sciolla[69], I. J. Scott[125], J. Seeman[45], A. Seiden[30], R. Seitz[111], T. Seki[168], A.I. Sekiya[34], S. Semenov[12], D. Semmler[187], S. Sen[32], K. Senyo[202], O. Seon[38], V. V. Serbo[45], S. I. Serednyakov[27,28], B. Serfass[62], M. Serra[10], J. Serrano[11], Y. Settai[192], R. Seuster[103], M. E. Sevior[126], K. V. Shakhova[171], L. Shang[96], M. Shapkin[132], V. Sharma[174], V. Shebalin[27,28], V. G. Shelkov[15], B. C. Shen[†73], D. Z. Shen[203], Y. T. Shen[20], D. J. Sherwood[127], T. Shibata[100], T. A. Shibata[99], H. Shibuya[196], T. Shidara[7], K. Shimada[100], M. Shimoyama[34], S. Shinomiya[178], J. G. Shiu[20], H. W. Shorthouse[1], L. I. Shpilinskaya[171], A. Sibidanov[6], E. Sicard[111], A. Sidorov[27], V. Sidorov[†27], V. Siegle[57], M. Sigamani[1], M. C. Simani[136], M. Simard[111], G. Simi[144], F. Simon[70,86], F. Simonetto[144,145], N. B. Sinev[139], H. Singh[153], J. B. Singh[123], R. Sinha[204], S. Sitt[48], Yu. I. Skovpen[27,28], R. J. Sloane[25], P. Smerkol[3], A. J. S. Smith[91], D. Smith[148], D. Smith[116], D. Smith[45], D. S. Smith[39], J. G. Smith[32], A. Smol[16], H. L. Snoek[118], A. Snyder[45], R. Y. So[75], R. J. Sobie[55], E. Soderstrom[45], A. Soha[45], Y. S. Sohn[43], M. D. Sokoloff[47], A. Sokolov[132], P. Solagna[144], E. Solovieva[12], N. Soni[123,148], P. Sonnek[128], V. Sordini[11,10,31], B. Spaan[105], S. M. Spanier[150], E. Spencer[30], V. Speziali[190], M. Spitznagel[69], P. Spradlin[30], H. Staengle[137], R. Stamen[7], M. Stanek[45], S. Stanič[205], J. Stark[48], M. Steder[61], H. Steininger[56], M. Steinke[87], J. Stelzer[45], E. Stevanato[144], A. Stocchi[11], R. Stock[206], H. Stoeck[6], D. P. Stoker[133], R. Stroili[144,145], D. Strom[139], P. Strother[1], J. Strube[139], B. Stugu[29], J. Stypula[19], D. Su[45], R. Suda[168], R. Sugahara[7], A. Sugi[38], T. Sugimura[7], A. Sugiyama[185], S. Suitoh[38], M. K. Sullivan[45], M. Sumihama[207], T. Sumiyoshi[168], D. J. Summers[128], L. Sun[29], L. Sun[47], S. Sun[45], J. E. Sundermann[88], H. F. Sung[20], Y. Susaki[38], P. Sutcliffe[25], A. Suzuki[15], J. Suzuki[7],

J. I. Suzuki[7], K. Suzuki[38,45], S. Suzuki[185], S. Y. Suzuki[7], J. E. Swain[53], S. K. Swain[45,103], S. T'Jampens[131], M. Tabata[182], K. Tackmann[15], H. Tajima[8], O. Tajima[7], K. Takahashi[99], S. Takahashi[100], T. Takahashi[181], F. Takasaki[7], T. Takayama[176], M. Takita[178], K. Tamai[7], U. Tamponi[16,17], N. Tamura[100], N. Tan[208], P. Tan[125], K. Tanabe[8], T. Tanabe[15], H. A. Tanaka[45], J. Tanaka[8], M. Tanaka[7], S. Tanaka[7], Y. Tanaka[209], K. Tanida[49], N. Taniguchi[7], P. Taras[111], N. Tasneem[55], G. Tatishvili[114], T. Tatomi[7], M. Tawada[7], F. Taylor[69], G. N. Taylor[126], G. P. Taylor[116], V. I. Telnov[27,28], L. Teodorescu[127], R. Ter-Antonyan[39], Y. Teramoto[181], D. Teytelman[45], G. Thérin[48], Ch. Thiebaux[131], D. Thiessen[75], E. W. Thomas[32], J. M. Thompson[45], F. Thorne[56], X. C. Tian[124], M. Tibbetts[116], I. Tikhomirov[12], J. S. Tinslay[45], G. Tiozzo[144], V. Tisserand[85], V. Tocut[11], W. H. Toki[117], E. W. Tomassini[32], M. Tomoto[7], T. Tomura[8], E. Torassa[144], E. Torrence[139], S. Tosi[119,120], C. Touramanis[25], J. C. Toussaint[62], S. N. Tovey[126], P. P. Trapani[16], E. Treadwell[210], G. Triggiani[77,92], S. Trincaz-Duvoid[11], W. Trischuk[91], D. Troost[15], A. Trunov[45], K. L. Tsai[20], Y. T. Tsai[20], Y. Tsujita[113], K. Tsukada[7], T. Tsukamoto[7], J. M. Tuggle[36], A. Tumanov[91], Y. W. Tung[20], L. Turnbull[175], J. Turner[45], M. Turri[30], K. Uchida[103], M. Uchida[99], Y. Uchida[199], M. Ueki[176], K. Ueno[7], K. Ueno[20], N. Ujiie[7], K. A. Ulmer[32], Y. Unno[154], P. Urquijo[126], Y. Ushiroda[7], Y. Usov[27,28], M. Usseglio[62], Y. Usuki[38], U. Uwer[101], J. Va'vra[45], S. E. Vahsen[103], G. Vaitsas[149], A. Valassi[11], E. Vallazza[142], A. Vallereau[48], P. Vanhoefer[70], W. C. van Hoek[32], C. Van Hulse[200], D. van Winkle[45], G. Varner[103], E. W. Varnes[91], K. E. Varvell[6], G. Vasileiadis[131], Y. S. Velikzhanin[20], M. Verderi[131], S. Versillé[48], K. Vervink[129], B. Viaud[111], P. B. Vidal[1], S. Villa[129], P. Villanueva-Perez[46], E. L. Vinograd[171], L. Vitale[142,143], G. M. Vitug[73], C. Voß[64], C. Voci[144,145], C. Voena[10], A. Volk[88], J. H. von Wimmersperg-Toeller[125], V. Vorobyev[27,28], A. Vossen[211], G. Vuagnin[142,143], C. O. Vuosalo[125], K. Wacker[105], A. P. Wagner[45], D. L. Wagner[32], G. Wagner[64], M. N. Wagner[187], S. R. Wagner[32], D. E. Wagoner[175], D. Walker[109], W. Walkowiak[30], D. Wallom[109], C. C. Wang[20], C. H. Wang[140], J. Wang[124], J. G. Wang[84], K. Wang[73], L. Wang[30], L. L. Wang[11], P. Wang[96], P. Wang[96], T. J. Wang[96], W. F. Wang[45], X. L. Wang[84], Y. F. Wang[106], F. R. Wappler[102], M. Watanabe[100], A. T. Watson[148], J. E. Watson[53], N. K. Watson[148], M. Watt[65], J. H. Weatherall[76], M. Weaver[45], T. Weber[45], R. Wedd[126], J. T. Wei[20], A. W. Weidemann[153], A. J. R. Weinstein[45], W. A. Wenzel[15], C. A. West[63], C. G. West[32], T. J. West[76], E. White[47], R. M. White[153], J. Wicht[7], L. Widhalm[†56], J. Wiechczynski[19], U. Wienands[45], L. Wilden[88], M. Wilder[30], D. C. Williams[30], G. Williams[95], J. C. Williams[76], K. M. Williams[84], M. I. Williams[1], S. Y. Willocq[137], J. R. Wilson[153], M. G. Wilson[30], R. J. Wilson[117], F. Winklmeier[117], L. O. Winstrom[30], M. A. Winter[149], W. J. Wisniewski[45], M. Wittgen[45], J. Wittlin[137], W. Wittmer[45], R. Wixted[91], A. Woch[111], B. J. Wogsland[150], E. Won[122], Q. K. Wong[39], B. C. Wray[163], A. C. Wren[149], D. M. Wright[136], C. H. Wu[20], J. Wu[122], S. L. Wu[125], H. W. Wulsin[45], S. M. Xella[65], Q. L. Xie[96], Y. Xie[53], Y. Xie[9], Z. Z. Xu[106], Ch. Yèche[62], Y. Yamada[7], M. Yamaga[176], A. Yamaguchi[176], H. Yamaguchi[7], T. Yamaki[212], H. Yamamoto[176], N. Yamamoto[7], R. K. Yamamoto[†69], S. Yamamoto[168], T. Yamanaka[178], H. Yamaoka[7], J. Yamaoka[103], Y. Yamaoka[7], Y. Yamashita[213], M. Yamauchi[7], D. S. Yan[203], Y. Yan[45], H. Yanai[100], S. Yanaka[99], H. Yang[49], R. Yang[91], S. Yang[22], A. K. Yarritu[45], S. Yashchenko[61], J. Yashima[7], Z. Yasin[73], Y. Yasu[7], S. W. Ye[106], P. Yeh[20], J. I. Yi[76], K. Yi[45], M. Yi[69], Z. W. Yin[203], J. Ying[124], G. Yocky[45], K. Yokoyama[7], M. Yokoyama[8], T. Yokoyama[179], K. Yoshida[38], M. Yoshida[7], Y. Yoshimura[7], C. C. Young[45], C. X. Yu[96], Z. Yu[125], C. Z. Yuan[96], Y. Yuan[96], F. X. Yumiceva[153], Y. Yusa[100], A. N. Yushkov[27], H. Yuta[183], V. Zacek[111], S. B. Zain[102], A. Zallo[9], S. Zambito[16,17], D. Zander[161], S. L. Zang[96], D. Zanin[16], B. G. Zaslavsky[171], Q. L. Zeng[117], A. Zghiche[85], B. Zhang[48], J. Zhang[7], J. Zhang[32], L. Zhang[73], L. M. Zhang[106], S. Q. Zhang[96], Z. P. Zhang[106], H. W. Zhao[96], H. W. Zhao[128], M. Zhao[69], Z. G. Zhao[106], Y. Zheng[69], Y. H. Zheng[103], Z. P. Zheng[96], V. Zhilich[27,28], P. Zhou[141], R. Y. Zhu[22], Y. S. Zhu[96], Z. M. Zhu[124], V. Zhulanov[27,28], T. Ziegler[91], V. Ziegler[45], G. Zioulas[133], M. Zisman[15], M. Zito[62], D. Zürcher[129], N. Zwahlen[129], O. Zyukova[27,28], T. Živko[3], and D. Žontar[3]

[*] General Editor
[§] Section Editor
[¶] Additional Section Writer
[†] Deceased

[1] Queen Mary, University of London, London, E1 4NS, United Kingdom
[2] Faculty of Mathematics and Physics, University of Ljubljana, 1000 Ljubljana, Slovenia
[3] J. Stefan Institute, 1000 Ljubljana, Slovenia
[4] Theoretische Physik 1, Naturwissenschaftlich-Technische Fakultät, Universität Siegen, Walter-Flex-Straße 3, D-57068 Siegen, Germany
[5] Iowa State University, Ames, Iowa 50011-3160, USA
[6] School of Physics, University of Sydney, NSW 2006, Australia
[7] High Energy Accelerator Research Organization (KEK), Tsukuba 305-0801, Japan
[8] Department of Physics, University of Tokyo, Tokyo 113-0033, Japan
[9] INFN Laboratori Nazionali di Frascati, I-00044 Frascati, Italy
[10] INFN Sezione di Roma, I-00185 Roma, Italy
[11] Laboratoire de l'Accélérateur Linéaire, IN2P3/CNRS et Université Paris-Sud 11, Centre Scientifique d'Orsay, F-91898 Orsay Cedex, France

[12] Institute for Theoretical and Experimental Physics, Moscow 117218, Russia
[13] Physik Department, James-Franck-Straße 1, Technische Universität München, D-85748 Garching, Germany
[14] Institut für Theoretische Teilchenphysik und Kosmologie, RWTH Aachen, D-52056 Aachen, Germany
[15] Lawrence Berkeley National Laboratory and University of California, Berkeley, California 94720, USA
[16] INFN Sezione di Torino, I-10125 Torino, Italy
[17] Dipartimento di Fisica, Università di Torino, I-10125 Torino, Italy
[18] University of Notre Dame, Notre Dame, Indiana 46556, USA
[19] H. Niewodniczanski Institute of Nuclear Physics, Krakow 31-342, Poland
[20] Department of Physics, National Taiwan University, Taipei 10617, Taiwan
[21] University of Iowa, Iowa City, Iowa 52242, USA
[22] California Institute of Technology, Pasadena, California 91125, USA
[23] Institute of Physics, Academia Sinica, Taipei, Taiwan 115, Republic of China
[24] INFN, Sezione de Bari, via Orabona 4, I-70126 Bari, Italy
[25] University of Liverpool, Liverpool L69 7ZE, United Kingdom
[26] Moscow Institute of Physics and Technology, Moscow Region 141700, Russia
[27] Budker Institute of Nuclear Physics SB RAS, Novosibirsk 630090, Russia
[28] Novosibirsk State University, Novosibirsk 630090, Russia
[29] University of Bergen, Institute of Physics, N-5007 Bergen, Norway
[30] University of California at Santa Cruz, Institute for Particle Physics, Santa Cruz, California 95064, USA
[31] Dipartimento di Fisica, Università di Roma La Sapienza, I-00185 Roma, Italy
[32] University of Colorado, Boulder, Colorado 80309, USA
[33] Johannes Gutenberg-Universität Mainz, Institut für Kernphysik, D-55099 Mainz, Germany
[34] Nara Women's University, Nara 630-8506, Japan
[35] Kavli Institute for the Physics and Mathematics of the Universe (WPI), University of Tokyo, Kashiwa 277-8583, Japan
[36] University of Maryland, College Park, Maryland 20742, USA
[37] Kobayashi-Maskawa Institute, Nagoya University, Nagoya 464-8602, Japan
[38] Graduate School of Science, Nagoya University, Nagoya 464-8602, Japan
[39] Ohio State University, Columbus, Ohio 43210, USA
[40] Fermi National Accelerator Laboratory, Batavia, IL 60510, USA
[41] KEK Theory Center, Institute of Particle and Nuclear Studies, KEK 1-1, OHO, Tsukuba, Ibaraki, 305-0801, Japan
[42] Particle and Nuclear Physics Division, J-PARC Center 201-1, Shirakata, Tokai, Ibaraki, 309-11-6, Japan
[43] Yonsei University, Seoul 120-749, South Korea
[44] Department of Physics, University of Warwick, Coventry CV4 7AL, United Kingdom
[45] SLAC National Accelerator Laboratory, Stanford University, Menlo Park, California 94025, USA
[46] IFIC, Universitat de Valencia-CSIC, E-46071 Valencia, Spain
[47] University of Cincinnati, Cincinnati, Ohio 45221, USA
[48] Laboratoire de Physique Nucléaire et de Hautes Energies, IN2P3/CNRS, Université Pierre et Marie Curie-Paris6, Université Denis Diderot-Paris7, F-75252 Paris, France
[49] Seoul National University, Seoul 151-742, South Korea
[50] Moscow Physical Engineering Institute, Moscow 115409, Russia
[51] Dipartmento di Fisica, Università di Bari, I-70126 Bari, Italy
[52] Departament de Física Teòrica, IFIC, Universitat de València – CSIC
Apt. Correus 22085, E-46071 València, Spain
[53] University of Edinburgh, Edinburgh EH9 3JZ, United Kingdom
[54] McGill University, Montréal, Québec, Canada H3A 2T8
[55] University of Victoria, Victoria, British Columbia, Canada V8W 3P6
[56] Institute of High Energy Physics, 1050 Vienna, Austria
[57] RIKEN BNL Research Center, Brookhaven, NY 11973, USA
[58] Southern Methodist University, Dallas, Texas 75275, USA
[59] Institut für Theoretische Teilchenphysik, Karlsruher Institut für Technologie, D-76131 Karlsruhe, Germany
[60] University of Pittsburgh, Pittsburgh, PA 15260, USA
[61] Deutsches Elektronen-Synchrotron, 22607 Hamburg, Germany
[62] CEA, Irfu, SPP, Centre de Saclay, F-91191 Gif-sur-Yvette, France
[63] University of California at Santa Barbara, Santa Barbara, California 93106, USA
[64] Universität Rostock, D-18051 Rostock, Germany
[65] Rutherford Appleton Laboratory, Chilton, Didcot, Oxon, OX11 0QX, United Kingdom
[66] Stanford University, Stanford, California 94305-4060, USA
[67] INFN Sezione di Milano, I-20133 Milano, Italy
[68] Dipartimento di Fisica, Università di Milano, I-20133 Milano, Italy
[69] Massachusetts Institute of Technology, Laboratory for Nuclear Science, Cambridge, Massachusetts 02139, USA
[70] Max-Planck-Institut für Physik, 80805 München, Germany
[71] SUPA, School of Physics and Astronomy, University of Glasgow, Glasgow, G12 8QQ, UK

[72] University of Bonn, 53115 Bonn, Germany
[73] University of California at Riverside, Riverside, California 92521, USA
[74] University of South Alabama, Mobile, Alabama 36688, USA
[75] University of British Columbia, Vancouver, British Columbia, Canada V6T 1Z1
[76] University of Manchester, Manchester M13 9PL, United Kingdom
[77] INFN Sezione di Pisa, I-56127 Pisa, Italy
[78] Scuola Normale Superiore di Pisa, I-56127 Pisa, Italy
[79] Tata Institute of Fundamental Research, Mumbai 400005, India
[80] National Central University, Chung-li 32054, Taiwan
[81] INFN Sezione di Perugia I-06123 Perugia, Italy
[82] Dipartimento di Fisica, Università di Perugia, I-06123 Perugia, Italy
[83] Luther College, Decorah, IA 52101, USA
[84] Virginia Polytechnic Institute and State University, Blacksburg, VA 24061, USA
[85] Laboratoire d'Annecy-le-Vieux de Physique des Particules (LAPP), Université de Savoie, CNRS/IN2P3, F-74941 Annecy-le-Vieux, France
[86] Excellence Cluster Universe, Technische Universität München, 85748 Garching, Germany
[87] Ruhr Universität Bochum, Institut für Experimentalphysik 1, D-44780 Bochum, Germany
[88] Technische Universität Dresden, Institut für Kern- und Teilchenphysik, D-01062 Dresden, Germany
[89] Beihang University, Beijing 100191
[90] Tel Aviv University, Tel Aviv, 69978, Israel
[91] Princeton University, Princeton, New Jersey 08544, USA
[92] Dipartimento di Fisica, Università di Pisa, I-56127 Pisa, Italy
[93] Kanagawa University, Yokohama 221-8686, Japan
[94] Korea University, Seoul 136-713, South Korea
[95] University of Texas at Dallas, Richardson, Texas 75083, USA
[96] Institute of High Energy Physics, Beijing 100039, China
[97] University of California at Los Angeles, Los Angeles, California 90024, USA
[98] Tohoku Gakuin University, Tagajo 985-8537, Japan
[99] Tokyo Institute of Technology, Tokyo 152-8550, Japan
[100] Niigata University, Niigata 950-2181, Japan
[101] Universität Heidelberg, Physikalisches Institut, D-69120 Heidelberg, Germany
[102] State University of New York, Albany, New York 12222, USA
[103] University of Hawaii, Honolulu, HI 96822, USA
[104] Oak Ridge National Laboratory, Oak Ridge, Tennessee 37831, USA
[105] Technische Universität Dortmund, Fakultät Physik, D-44221 Dortmund, Germany
[106] University of Science and Technology of China, Hefei 230026, PR China
[107] INFN Sezione di Ferrara, I-44100 Ferrara, Italy
[108] Dipartimento di Fisica e Scienze della Terra, Università di Ferrara, I-44100 Ferrara, Italy
[109] University of Bristol, Bristol BS8 1TL, United Kingdom
[110] INFN CNAF I-40127 Bologna, Italy
[111] Université de Montréal, Physique des Particules, Montréal, Québec, Canada H3C 3J7
[112] Hiroshima Institute of Technology, Hiroshima 731-5193, Japan
[113] University of Tsukuba, Tsukuba 305-0801, Japan
[114] Pacific Northwest National Laboratory, Richland, WA 99352, USA
[115] Toyama National College of Maritime Technology, Toyama 933-0293, Japan
[116] Imperial College London, London, SW7 2AZ, United Kingdom
[117] Colorado State University, Fort Collins, Colorado 80523, USA
[118] NIKHEF, National Institute for Nuclear Physics and High Energy Physics, NL-1009 DB Amsterdam, The Netherlands
[119] INFN Sezione di Genova, I-16146 Genova, Italy
[120] Dipartimento di Fisica, Università di Genova, I-16146 Genova, Italy
[121] Indian Institute of Technology Bhubaneswar, SatyaNagar, 751007, India
[122] Harvard University, Cambridge, Massachusetts 02138, USA
[123] Panjab University, Chandigarh 160014, India
[124] Peking University, Beijing 100871, PR China
[125] University of Wisconsin, Madison, Wisconsin 53706, USA
[126] School of Physics, University of Melbourne, Victoria 3010, Australia
[127] Brunel University, Uxbridge, Middlesex UB8 3PH, United Kingdom
[128] University of Mississippi, University, Mississippi 38677, USA
[129] École Polytechnique Fédérale de Lausanne (EPFL), 1015 Lausanne, Switzerland
[130] Utkal University, Bhubaneswar, India
[131] Laboratoire Leprince-Ringuet, CNRS/IN2P3, Ecole Polytechnique, F-91128 Palaiseau, France
[132] Institute for High Energy Physics, Protvino 142281, Russia

[133] University of California at Irvine, Irvine, California 92697, USA
[134] Indian Institute of Technology Guwahati, Assam 781039, India
[135] Indian Institute of Technology Guwahati, Guwahati, Assam, 781 039, India
[136] Lawrence Livermore National Laboratory, Livermore, California 94550, USA
[137] University of Massachusetts, Amherst, Massachusetts 01003, USA
[138] Novosibirsk State Technical University, Novosibirsk 630092, Russia
[139] University of Oregon, Eugene, Oregon 97403, USA
[140] National United University, Miao Li 36003, Taiwan
[141] Wayne State University, Detroit, MI 48202, USA
[142] INFN Sezione di Trieste, I-34127 Trieste, Italy
[143] Dipartimento di Fisica, Università di Trieste, I-34127 Trieste, Italy
[144] INFN Sezione di Padova, I-35131 Padova, Italy
[145] Dipartimento di Fisica, Università di Padova, I-35131 Padova, Italy
[146] University of Louisville, Louisville, Kentucky 40292, USA
[147] University of Maribor, 2000 Maribor, Slovenia
[148] University of Birmingham, Birmingham, B15 2TT, United Kingdom
[149] University of London, Royal Holloway and Bedford New College, Egham, Surrey TW20 0EX, United Kingdom
[150] University of Tennessee, Knoxville, Tennessee 37996, USA
[151] INFN Sezione di Napoli, I-80126 Napoli, Italy
[152] Department of Physics, Fu Jen Catholic University, Taipei 24205, Taiwan
[153] University of South Carolina, Columbia, South Carolina 29208, USA
[154] Hanyang University, Seoul 133-791, South Korea
[155] Korea Institute of Science and Technology Information, Daejeon 305-806, South Korea
[156] Gyeongsang National University, Chinju 660-701, South Korea
[157] Sungkyunkwan University, Suwon 440-746, South Korea
[158] University of Illinois at Urbana-Champaign, Urbana, IL 61801, USA
[159] Dipartimento di Scienze Fisiche, Università di Napoli Federico II, I-80126 Napoli, Italy
[160] TRIUMF, Vancouver, BC, Canada V6T 2A3
[161] Universität Karlsruhe, Institut für Experimentelle Kernphysik, D-76021 Karlsruhe, Germany
[162] Faculty of Mathematics and Physics, Charles University, 121 16 Prague, The Czech Republic
[163] University of Texas at Austin, Austin, Texas 78712, USA
[164] Harvey Mudd College, Claremont, California 91711, USA
[165] Elon University, Elon University, North Carolina 27244-2010, USA
[166] University of Pennsylvania, Philadelphia, Pennsylvania 19104, USA
[167] II. Physikalisches Institut, Georg-August-Universität Göttingen, 37073 Göttingen, Germany
[168] Tokyo Metropolitan University, Tokyo 192-0397, Japan
[169] Johns Hopkins University, Baltimore, Maryland 21218, USA
[170] Universitat de Barcelona, Facultat de Fisica, Departament ECM, E-08028 Barcelona, Spain
[171] Institute for Single Crystals, National Academy of Sciences of Ukraine, Kharkov 61001, Ukraine
[172] Yale University, New Haven, Connecticut 06511, USA
[173] National Kaohsiung Normal University, Kaohsiung 80201, Taiwan
[174] University of California at San Diego, La Jolla, California 92093, USA
[175] Prairie View A&M University, Prairie View, Texas 77446, USA
[176] Tohoku University, Sendai 980-8578, Japan
[177] Shinshu University, Nagano 390-8621, Japan
[178] Osaka University, Osaka 565-0871, Japan
[179] Tokyo University of Agriculture and Technology, Tokyo 184-8588, Japan
[180] Kyungpook National University, Daegu 702-701, South Korea
[181] Osaka City University, Osaka 558-8585, Japan
[182] Chiba University, Chiba 263-8522, Japan
[183] Aomori University, Aomori 030-0943, Japan
[184] Humboldt-Universität zu Berlin, Institut für Physik, D-12489 Berlin, Germany
[185] Saga University, Saga 840-8502, Japan
[186] Punjab Agricultural University, Ludhiana 141004, India
[187] Justus-Liebig-Universität Gießen, 35392 Gießen, Germany
[188] T. Kościuszko Cracow University of Technology, Krakow 31-342, Poland
[189] Indian Institute of Technology Madras, Chennai 600036, India
[190] Università di Pavia, Dipartimento di Elettronica and INFN, I-27100 Pavia, Italy
[191] University of Miyazaki, Miyazaki 889-2192, Japan
[192] Chuo University, Tokyo 192-0393, Japan
[193] Kyoto University, Kyoto 606-8502, Japan
[194] Research Center for Electron Photon Science, Tohoku University, Sendai 980-8578, Japan

[195] Mount Holyoke College, South Hadley, Massachusetts 01075, USA
[196] Toho University, Funabashi 274-8510, Japan
[197] Vanderbilt University, Nashville, Tennessee 37235, USA
[198] Technion, Haifa, Israel
[199] The Graduate University for Advanced Studies, Hayama 240-0193, Japan
[200] University of the Basque Country UPV/EHU, 48080 Bilbao, Spain
[201] Ikerbasque, 48011 Bilbao, Spain
[202] Yamagata University, Yamagata 990-8560, Japan
[203] Chinese Academy of Science, Beijing 100864, PR China
[204] Institute of Mathematical Sciences, Chennai 600113, India
[205] University of Nova Gorica, 5000 Nova Gorica, Slovenia
[206] University of Frankfurt, 60318 Frankfurt am Main, Germany
[207] Gifu University, Gifu 501-1193, Japan
[208] Tokyo University of Science, Chiba 278-8510, Japan
[209] Nagasaki Institute of Applied Science, Nagasaki 851-0123, Japan
[210] Florida A&M University, Tallahassee, Florida 32307, USA
[211] Indiana University, Bloomington, IN 47408, USA
[212] Sugiyama Jogakuen University, Aichi 470-0131, Japan

# Contents

## C The results and their interpretation 178

# Part A
# The facilities

## Chapter 1
## The $B$ Factories

***Editors:***
*David Leith (BABAR)*
*Kazuo Abe, Stephen L. Olsen (Belle)*

**Additional section writers:**
*Peter Križan, Leo Piilonen, Blair Ratcliff, Guy Wormser*

### 1.1 Introduction

In their classic paper, Kobayashi and Maskawa (1973, "KM") pointed out that *CP* violation could be naturally incorporated into the Standard Model (SM) as an irreducible complex phase in the weak interaction quark-flavor-mixing matrix if the number of quark flavors was six. This was remarkable because at that time only the three quarks of the original Gell-Mann (1964) and Zweig (1964a) quark model — *i.e.*, the $u$-, $d$- and $s$-quarks — were experimentally established. The situation changed dramatically in late 1974 with the discovery of the $c$-quark at Brookhaven (Aubert et al., 1974) and SLAC (Augustin et al., 1974) and the 1977 discovery of the $b$-quark at Fermilab (Herb et al., 1977). By 1980, the KM idea, by then embodied in the Cabibbo-Kobayashi-Maskawa (CKM) quark flavor mixing matrix (Cabibbo, 1963; Kobayashi and Maskawa, 1973), was accepted as an integral component of the Standard Model, even though its *raison d'etre*, the *CP*-violating complex phase, had not been measured (Kelly et al., 1980).

#### 1.1.1 Testing the KM idea

In the early 1980's, when the experimental state-of-the-art in $B$ meson physics was defined by the CLEO experiment, where the measurements were based on data samples of a few tens of events (Bebek et al., 1981; Chadwick et al., 1981), Bigi, Carter, and Sanda published papers exploring the possibilities of using $B$ meson decays to test the validity of the KM six-quark mechanism for *CP* violation (Bigi and Sanda, 1981, 1984; Carter and Sanda, 1980, 1981). They concluded that for a relatively small range of the CKM-matrix parameter-space that was allowed at that time — a range that corresponds to a substantial probability for $B^0 - \overline{B}^0$ mixing and a long $B$ meson lifetime — large *CP* violation might be observable in neutral $B$ meson decays to *CP* eigenstates, such as $B^0 \to J/\psi K^0_S$. However, in the early 1980's, no decays of this type had been seen; we now know that their branching fractions are $\sim 0.1\%$ or less. A reasonable conclusion that could be derived from these papers at that time was that definitive tests of the KM idea were hopelessly impractical.

#### 1.1.2 Three miracles

Subsequently, three remarkable developments occurred that completely turned the tables. These included the first emergence of evidence of a long $B$ meson lifetime from experiments at SLAC (Fernandez et al., 1983; Lockyer et al., 1983), and the unexpected discovery by the ARGUS experiment at DESY in 1987 of a substantial rate for $B^0 - \overline{B}^0$ mixing (Albrecht et al., 1987b). These measurements indicated that the CKM-matrix parameters are, in fact, in the range that is accessible to tests of the KM idea. This was helped along by many well-attended international workshops[1] developing each of the different technical approaches and refining the requirements and specifications for each. It became clear that *CP* violation, at the level manifest in the Standard Model, could be experimentally observable somewhere other than in neutral kaons: namely, in the $B^0 - \overline{B}^0$ system. Moreover large *CP* violation was expected, rather than the one-in-a-thousand effect seen in $K$ decay. In addition, Bigi and Sanda (1981) had shown that a measurement of *CP* violation in neutral $B$ meson decays to *CP* eigenstates could be clearly interpreted without theoretical uncertainties. However, an experiment to observe *CP* violation in $B$ decays would require about a thousand-fold larger data samples of $B$ mesons than had been gathered heretofore.

The two fortuitous circumstances mentioned above were accompanied by a third "miracle": extraordinary improvements in the performance of $e^+e^-$ storage rings, with order-of-magnitude luminosity improvements occurring approximately every seven years. In 1980, the original CESR collider typically produced $\sim 30$ $B\overline{B}$ meson pairs per day; thirty years later, the two $B$ Factories, KEKB and PEP-II, routinely produced more than one million $B\overline{B}$ meson pairs per day, a nearly five orders-of-magnitude improvement! The $B$ Factories built on the success of CESR at Cornell and DORIS-II at DESY to achieve these production rates. These developments were accompanied by less miraculous, but still impressive, advances in the capabilities of large solid-angle detectors, especially in the ability of data acquisition systems to handle the huge event rates associated with the available luminosities, precision tracking and vertexing devices, and the software and storage technologies required to deal with these large data samples.

[1] The main workshops include: Heidelberg (Schubert and Waldi, 1986), Stanford (Bloom, Friedsam, and Fridman, 1988; Hitlin, 1990), Courmayeur (De Sanctis, Greco, Piccolo, and Tazzari, 1988), Zuoz (Locher, 1988), Los Angeles (Cline and Fridman, 1988; Cline and Stork, 1987), Blois (Cline and Fridman, 1991), Syracuse (Goldberg and Stone, 1989), Tsukuba (Kikutani and Matsuda, 1993; Ozaki and Sato, 1991; Yoshimura, 1989), Vancouver (MacFarlane and Ng, 1991), and Hamburg (Aleksan and Ali, 1993).

The remainder of this chapter discusses the historical route taken to develop the ideas necessary to build a $B$ Factory (Section 1.2), followed by an overview of the two storage rings that were built to provide a source of $B^0\overline{B}^0$ meson pairs to explore the $B$ Factory scientific program (Section 1.3). A review of general issues concerning the detector requirements for a $B$ Factory is presented in Section 1.4; a more detailed discussion of the two detectors realized can be found in Chapter 2. We conclude with a brief look at the early physics discoveries of the $B$ Factories (Section 1.5).

## 1.2 The path to the $B$ Factories

### 1.2.1 Requirements for a $B$ Factory

The time-dependent method for testing the KM idea is based on the fact that there are decays with interfering amplitudes (see Fig. 1.2.1) where the interference term contains $V_{cd}^* V_{cb} V_{td} V_{tb}^*$. The phase of this quartet of CKM matrix elements is $\phi_1 = \beta$. Note that the *BABAR* experiment uses $\beta$ to denote this angle, whereas the Belle experiment reports results in terms of $\phi_1$; further notational differences are discussed in Chapter 16. In the following we will use the $\phi_1$ notation for this phase. The "golden observable" for its determination is the *CP* asymmetry between $B^0 \to J/\psi K_S^0$ and $\overline{B}^0 \to J/\psi K_S^0$. At the $B$ Factories, neutral $B$ mesons are created in pairs at a center-of-mass energy corresponding to the $\Upsilon(4S)$. As a result the wave function of the $B^0\overline{B}^0$ pairs is in a $P$-wave entangled state, until one of the mesons decays. A further complication arises as neutral $B$ mesons mix with a characteristic frequency $\Delta m_d$, so one computes the asymmetry as a function of the proper time difference between the decays of two mesons in an event, and uses knowledge of $B^0\overline{B}^0$ mixing to infer the flavor of one of the $B$ mesons (decaying into a *CP* eigenstate) relative to that of the other $B$ decaying into a flavor specific final state. This initial state preparation at the $\Upsilon(4S)$ enables one to determine the flavor of the $b$ quark for the flavor specific final states with a high efficiency.

The amplitude for the direct decay $B^0 \to J/\psi K_S^0$, shown in the upper right panel of Fig 1.2.1, is proportional to the $V_{cb}$ CKM matrix element. The decay can also proceed via the two-step process $B^0 \to \overline{B}^0 \to J/\psi K_S^0$, shown in the bottom-right panel of the figure. The phase difference between these two amplitudes is $2\phi_1$.

The technique for performing the interference measurement is illustrated in Fig. 1.2.2. A $B^0\overline{B}^0$ pair produced via $\Upsilon(4S) \to B^0\overline{B}^0$ decay is entangled in a coherent quantum state until one of the mesons decays. Most $B^0$ meson decays produce flavor-specific final states, *i.e.*, the final-state particles can be used to determine whether the decaying meson was a $B^0$ or a $\overline{B}^0$. For example, a $K^+$ meson in the final state signals a high likelihood for the $B \to \overline{D} \to K^+$ decay chain and, thus, a higher probability that the parent meson was a $B^0$ rather than a $\overline{B}^0$. Such a decay is called a "flavor-tag" decay. At the time this $B$ meson decays ($t_1$ in the figure), the accompanying $B$ meson's flavor is specified as being the opposite.

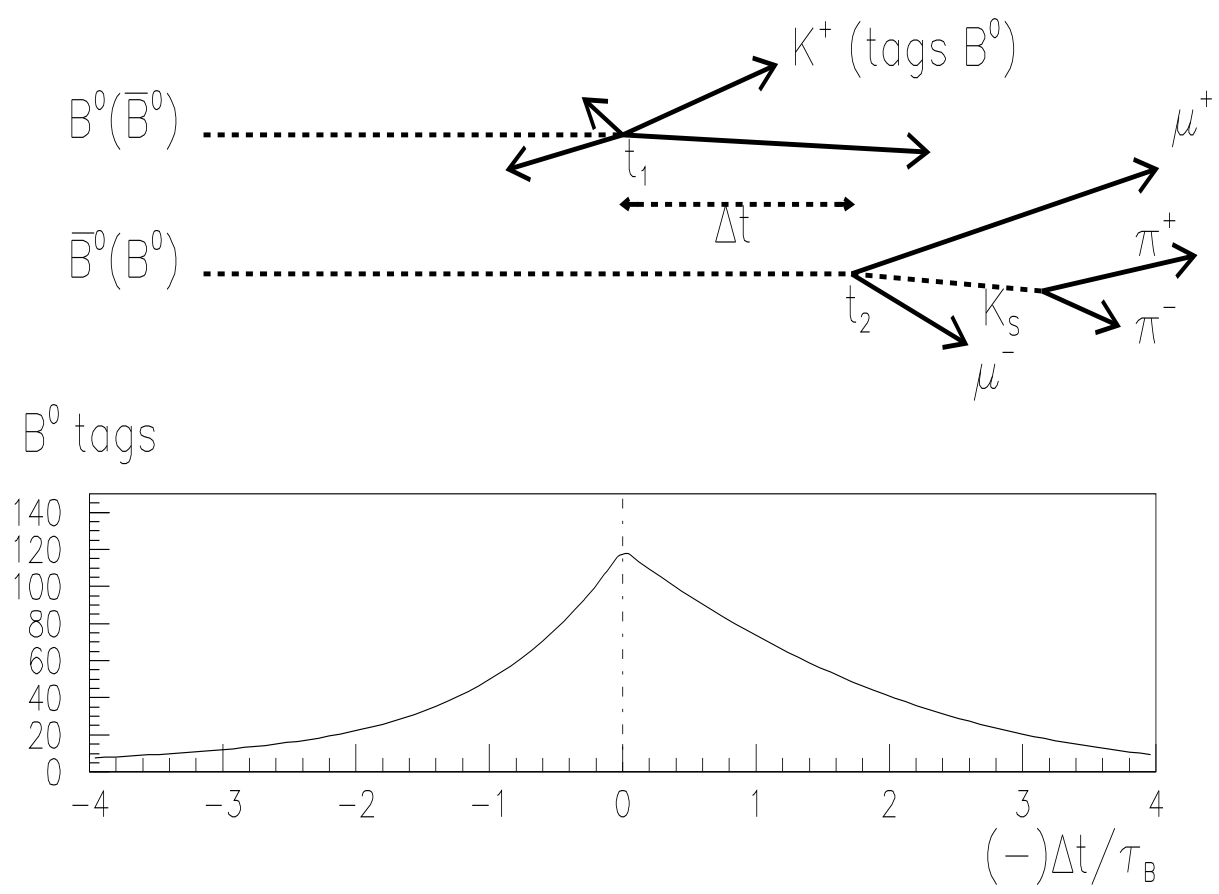

**Figure 1.2.2.** An illustration of the $B$ Factory flagship measurement of $\sin 2\phi_1 = \sin 2\beta$.

This accompanying meson then propagates in time and the quark flavor content can oscillate from an unmixed state into a mixed one, until it decays (at time $t_2$). If it decays into a *CP* eigenstate such as $J/\psi K_S^0$, the unmixed and mixed flavor components interfere, producing different decay rates for $B^0$-tagged and $\overline{B}^0$-tagged mesons. A similar pattern occurs for those cases where the *CP* eigenstate decay occurs before the flavor tag decay (*i.e.* $t_2 \leq t_1$) except that in this case the common phase from the mixing diagram has opposite sign. Thus, for $B^0$-tagged events, the interference is destructive for negative values of $\Delta t = t_2 - t_1$ and constructive for positive $\Delta t$ values, as indicated in the graph in the lower part of the figure, where the $\Delta t$ dependence for $B^0$-tagged events is shown in units of $\tau_B$, the $B^0$ lifetime ($\approx 1.5$ ps). The time-integrated asymmetry is zero; asymmetries only show up in the decay-time-dependence of the flavor-tagged distributions. The interference in $\overline{B}^0$-tagged events has the opposite pattern, *i.e.*, constructive interference for negative $\Delta t$ and destructive interference for positive $\Delta t$. Detailed discussions of flavor tagging and time-dependent *CP* asymmetry measurement techniques used by the $B$ Factories can be found in Chapters 8 and 10, respectively.

These considerations set the base-line requirements for an experiment to measure the *CP*-violating phases using the time-dependent *CP* asymmetry technique at the $\Upsilon(4S)$:

**High luminosity:** The branching fraction for the $B^0 \to J/\psi K_S^0$ decay, the most prominent mode that is useful for these measurements, is $\sim 0.04\%$ and that for $J/\psi \to \ell^+\ell^-$ (where $\ell = e, \mu$) is $\sim 12\%$. Thus, tens of millions of $B^0\overline{B}^0$ pairs are needed. For an $e^+e^-$ collider operating at the $\Upsilon(4S)$, this requires integrated luminosities of $\sim 30$ fb$^{-1}$ or more.

**Boosted $B^0\overline{B}^0$ pairs:** The $B^0$ and $\overline{B}^0$ mesons must have decay lengths in the laboratory that are sufficiently long so that the time sequence of their decays can be measured. Also it should be noted that $\Upsilon(4S)$ mesons produced in symmetric colliders are almost at rest in the laboratory frame, and as a consequence one can only measure functions of $t_1 + t_2$, for which any *CP* asymmetry vanishes. Both of these reasons impose the requirement of an asymmetric energy $e^+e^-$ collision in the laboratory frame of reference (see Section 1.2.3).

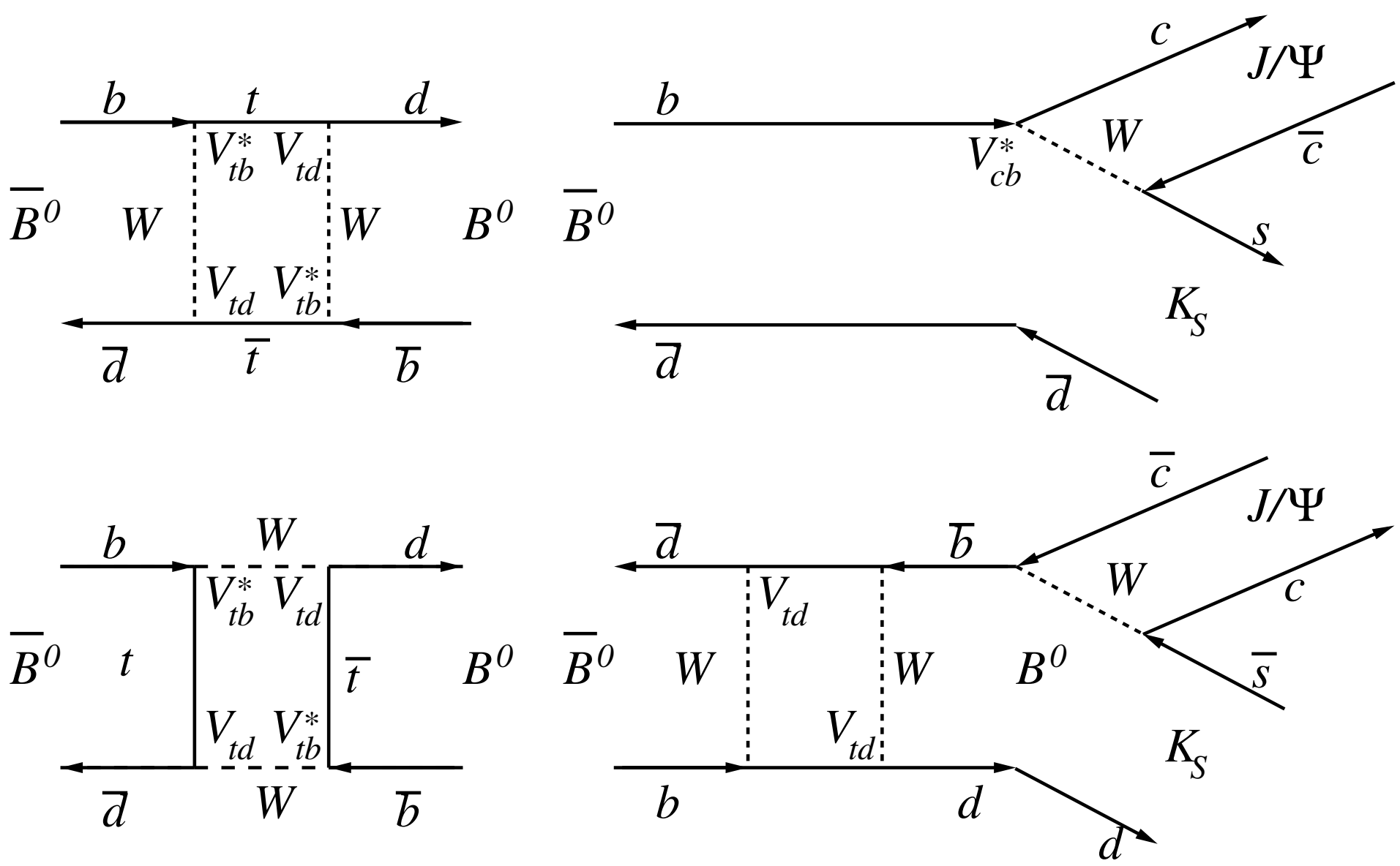

**Figure 1.2.1.** (left) The dominant quark-line diagrams for $B^0 - \overline{B}^0$ mixing. (right) The interfering diagrams used for the $\phi_1$ measurement. As the direct $\overline{B}^0$ decay produces $\overline{K}^0$, and the $B^0$ decay produces $K^0$, the relative phase between $\overline{B}^0 \to B^0 \to J/\psi K^0_S$ and $\overline{B}^0 \to J/\psi K^0_S$ contains an additional term due to $K^0 - \overline{K}^0$ mixing (not shown).

**High-resolution and large-coverage detector** with excellent particle identification: The measured amplitude of the *CP*-violating asymmetry is directly proportional to the detector's ability to reconstruct and flavor-tag the accompanying $B$ meson.

### 1.2.2 Early proposals

At the time, the most successful studies of $B$ mesons were being performed at the CESR and DORIS II $e^+e^-$ colliders operating at the center-of-mass (CM) energy corresponding to the $\Upsilon(4S)$ resonance, which, because it decays into $B\overline{B}$ (and nothing else) nearly 100% of the time, is a copious source of $B$ mesons in a clean, low-background environment. Also, luminosities of $\sim 10^{32}$ cm$^{-2}$ s$^{-1}$, while a significant advance over previous machines, are two orders-of-magnitude too low to provide samples of $B$ meson decays that are adequate for the *CP* violation measurements.

During the late 1980's a very large number of concepts (twenty-two in all) emerged on the international scene to test *CP* violation in $B$ mesons. Both Hitlin (2005) and Schubert (2007) have presented detailed reviews of these proposals, and how they synergistically evolved to the two $B$ Factories that were eventually built.

### 1.2.3 Asymmetric colliders

In the late 1980s, as the TRISTAN program at KEK (High Energy Accelerator Research Organization, Tsukuba, Japan) and the SLC program at SLAC (SLAC National Accelerator Laboratory, Stanford, USA) were winding down, workshops and task forces were formed at both labs to investigate possible facilities to attack the *CP* violation problem. In 1987, at a specialized workshop at UCLA that was focused on possibilities for using linear $e^+e^-$ colliders for $B$ physics, Pier Oddone proposed a novel concept of an asymmetric-energy, circular $e^+e^-$ collider. This would operate at the $\Upsilon(4S)$ and produce $B$ mesons with a lab-frame boost sufficient to enable decay-time-dependent measurements (Oddone, 1987), as discussed in Section 1.2.1. The experimental and analysis details on how one might effectively detect *CP* violation in such asymmetric decays are described in Aleksan, Bartelt, Burchat, and Seiden (1989).

Within the US, the 1990 HEPAP Panel on "The HEP Research Program for the 1990's" (Sciulli et al., 1990), recommended that the US should study the science opportunities and technical requirements of a $B$ Factory as a possible component of the future US accelerator program, and vigorously support the necessary R&D funding. Two years later, the next HEPAP Panel (Witherell et al., 1992) recommended that a $B$ Factory be constructed in the US under all budget scenarios under consideration. In the fall of 1992 the Office of Management and Budget (OMB) and the White House were assembling the budget proposal for fiscal year 1994, and included possible initial funding for a $B$ Factory. Both California and New York congressional delegates were working towards the interests of their constituencies. In April 1993 the OMB asked the DOE and NSF to convene a joint review of the two projects, both having already done careful reviews of their respec-tive proposals — SLAC by DOE and Cornell by NSF. This review (Kowalski et al., 1993) was charged to look at both projects separately and non-competitively, and assess their suitability for the task ahead and the risks that each project posed with respect to achieving the goals, the schedule, and the cost. That fall, Congress recommended incremental growth for HEP funding, including \$36 million to start the construction of a $B$ Factory, with the choice of site awaiting the decision from this review. In October 1993 on the basis of this review Secretary of Energy Hazel O'Leary made the decision to go ahead with the construction of the SLAC facility (O'Leary, 1993), and that same month President Clinton announced the construction of a $B$ Factory at SLAC, as a Presidential Initiative, with a four year financial profile (Clinton, 1993). A management team was immediately formed to design and build the PEP-II collider under the leadership of Jonathan Dorfan (SLAC), together with Tom Eliof (LBL) and Robert Yamamoto (LLNL). Complementing this team, an Interim International Advisory Committee was formed by the lab management to advise on the formation of the *BABAR* collaboration's first committees. The detector evolution from this point onward is discussed in more detail in Section 1.4. In the shadow of the cancellation of the Superconductiong Super Collider (SSC) project in Texas in October, 1993, the HEPAP Panel on "The Vision for the Future of HEP" (Drell et al., 1994) was quickly assembled and charged; it met through the short period December 1993 and March 1994. They presented HEPAP, DOE, and Congress with a strong vision of how to pull the US HEP program back from the brink caused by the SSC cancellation decision, and set a path to a healthy, competitive international research program. This plan strongly recommended continuing forward with both the main Injector project at FNAL and the $B$ Factory at SLAC. The three-lab (SLAC, LBL, LLNL) $B$ Factory team worked well together, smoothly solving the problems that arise in all high-tech construction projects, and bringing the project in "on-time" and "on-budget". The high energy ring was completed and beam stored by mid 1997, and the low energy ring was completed, with beam stored, a year later. First collisions were observed that same month, and first collisions with the *BABAR* detector in place were observed in May 1999. Design luminosity was achieved in the fall of 2000.

In Japan, the first official presentation for a $B$ Factory construction took place at the TRISTAN Program Advisory Committee (TPAC) in March 1991. The committee members heard the progress report on the feasibility studies for the machine design and detector configuration that were accumulated from the past several year's work. The committee was convinced that constructing a $B$ Factory at KEK was sufficiently feasible and the project should nicely fit in as a third stage of the TRISTAN project. The committee recommended that KEK should proceed with its construction and, due to the highly competitive situation worldwide, aim for the earliest possible completion of the project. With this official TPAC recommendation, and expression of support from the international community in the form of letters from prominent figures and presence at well-attended meetings, the KEK management began to talk to the funding agency of the Japanese Government and to rearrange the laboratory resources toward the new project.

Of the original leading $B$ Factory proposals mentioned in Section 1.2.2 above, only these two $B$ Factory projects, both based on the Oddone concept of asymmetric energy electron-positron storage rings, PEP-II (PEP-II, 1993) and KEKB (Abe et al., 1993), were to survive. *BABAR* at PEP-II was approved in 1993, and Belle at KEKB was approved the following year, in 1994.

### 1.2.4 A different approach

Meanwhile a different approach, aimed at using $B$ mesons produced in hadron collisions, was pursued by HERA-B (Hartouni et al., 1995; Padilla, 2000). Here, the plan was to place thin metal targets inside the halo of the proton beam in the HERA electron-proton collider and run parasitically with other HERA experiments. A drawback was that the cross section for producing $B$ mesons in proton-nuclear collisions at the available CM energy is a tiny fraction ($\sim 10^{-6}$) of the total hadronic cross section. Although serious difficulties were anticipated with this approach, the project was approved in 1995 with an expected data-taking start in 1998, one year ahead of the expected start-up of PEP-II and KEKB. Ultimately, however, the huge non-$B$ meson background turned out to be too difficult to contend with and this approach proved not to be competitive with the asymmetric $e^+e^-$ collider approach.

In 1994, the year that the SLAC and KEK $B$ Factories were approved, three sets of proponents for a dedicated $B$ physics experiment at the LHC were encouraged to "join together to prepare a letter of intent for a new collider mode $b$ experiment to be submitted to the LHCC" (Kirsebom et al., 1995). The three projects were called COBEX, GAJET, and LHB, and the merger resulted in the LHCb experiment. The experimental design for LHCb is similar to that of HERA-B, in that it is a single-arm spectrometer. Unlike HERA-B, which relied on a target to create $B$ mesons, LHCb relies on production of $B$ mesons from $pp$ collisions at the LHC. A dedicated spectrometer in the forward region is chosen to take advantage of the large cross section in the forward-backward direction. The LHCb experiment started taking data in 2008, when the LHC started collisions. Another proposed experiment to study $CP$ violation in a hadronic environment was put forward, with the aim of using the Tevatron at Fermilab. This was called the BTeV experiment and it was to have been a two-arm spectrometer, each arm being similar in design to LHCb (Santoro et al., 1999). Only the HERA-B and LHCb experiments were constructed and took data.

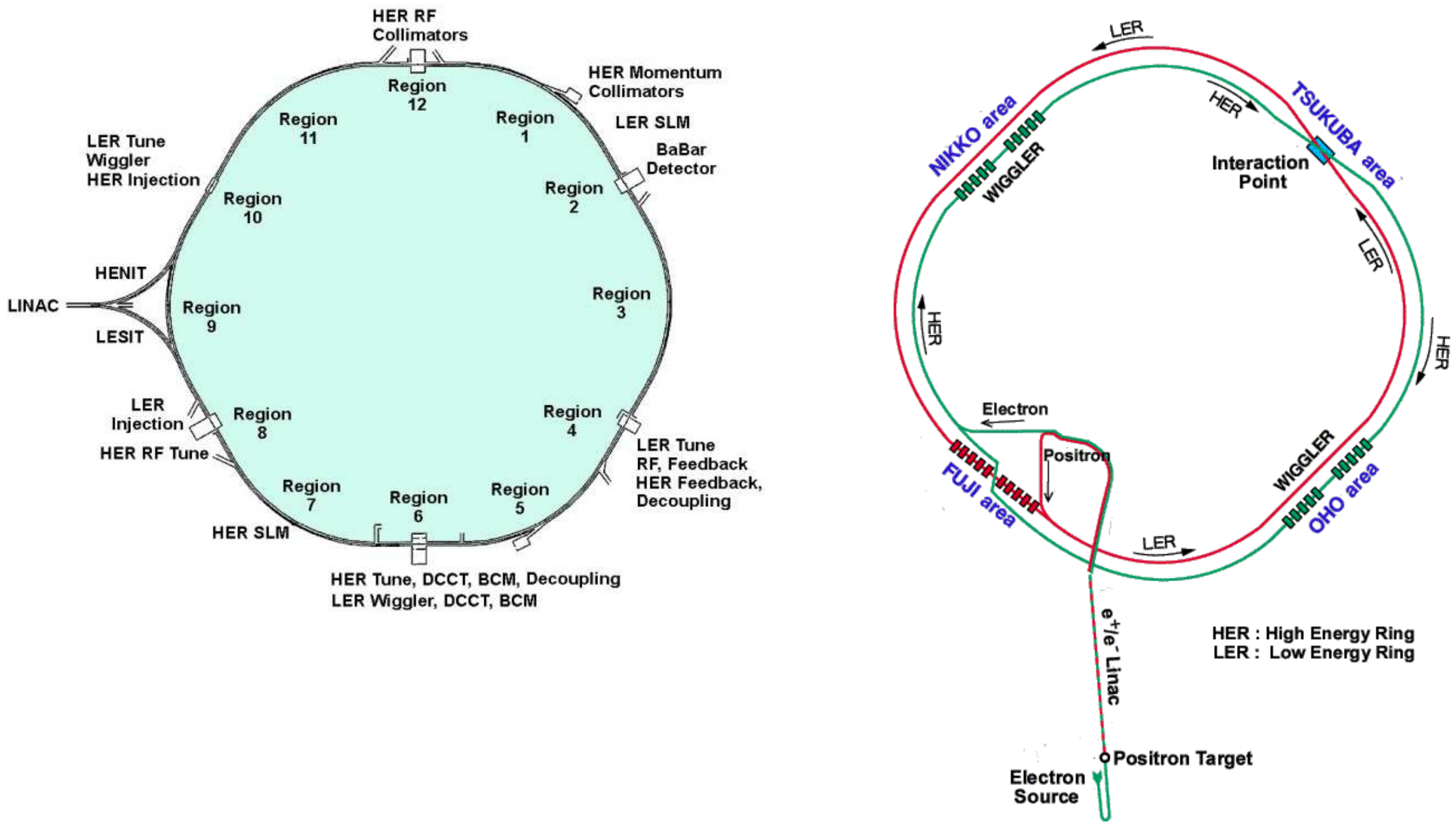

**Figure 1.3.1.** Schematic view of the PEP-II (left) and KEKB (right) rings. At PEP-II, the two beams are stacked one on top of the other; the *BABAR* experiment is located in an experimental hall at the single interaction region, within region 2 of the PEP-II complex. At KEKB, the two beams are side-by-side, and intersect in the Tsukuba area experimental hall where the Belle detector was placed.

## 1.3 PEP-II and KEKB

PEP-II was located in the tunnel that had housed the 32 GeV center-of-mass energy PEP $e^+e^-$ storage ring,[2] while the KEKB ring was in the 64 GeV center-of-mass energy $e^+e^-$ TRISTAN storage accelerator tunnel. Figure 1.3.1 shows a schematic overview of the PEP-II and KEKB rings.

Both projects included conversions to meet the $B$ Factory requirements, namely an instantaneous luminosity in excess of $10^{33}$ cm$^{-2}$ s$^{-1}$ and a boost factor (of the CM frame relative to the laboratory) sufficient for observing the time evolution of $B$ decays. To achieve these requirements, however, some considerable challenges had to be addressed.

Asymmetric energies mean a dedicated ring for each beam. In order to reach a high integrated luminosity one requires an intense positron source and on-energy injection for both rings. For KEKB, this meant that the injection linear accelerator (Linac) energy had to be raised from 2.5 GeV to 8 GeV in order to provide for on-energy injection of 8 GeV electrons and sufficient production of 3.5 GeV positrons. PEP-II had the advantage of the existing powerful SLAC Linac, which could provide the required electron and positron beams with minimal modifications. Both facilities used high-energy electron beams and low-energy positron beams in order to avoid beam-instability problems due to ion trapping, which are most serious at lower energies. Both facilities had only one interaction region (IR) for the detector in order to optimize the luminosity. The luminosity of an $e^+e^-$ storage ring is given by

$$\mathcal{L} = \frac{N_b n_{e^-} n_{e^+} f}{A_{\rm eff}} \qquad (1.3.1)$$

where the numbers of electrons and positrons in each bunch are given by $n_{e^-}$ and $n_{e^+}$, $N_b$ is the number of bunches, $f$ is the circulation frequency, and $A_{\rm eff}$ is the effective cross-sectional overlapping transverse area of the beams at the interaction point (IP). While the five parameters are independent at lower beam currents, at high beam currents $A_{\rm eff}$ becomes strongly beam-current dependent. As the product $N_b n_{e^-} n_{e^+}$ is increased, $A_{\rm eff}$ increases, thereby limiting the luminosity.

Particles inside a beam bunch are deflected when they pass through the collective electromagnetic fields of the oncoming beam bunch at the IP; as a result, the oncoming bunch collectively acts as a focusing lens. However, these beam-beam effects are highly non-linear and produce spreads in the operating point in the betatron-oscillation tune plane, causing considerable complications in the machine operation. These beam-beam interactions, which become larger as the bunch charges are increased, also limit the luminosity by enlarging $A_{\rm eff}$.

Attempts to raise the luminosity by raising $N_b$, the number of bunches in each ring, face a different prob-lem. When a beam bunch circulates with small separation intervals from other bunches, it feels some effects of the other bunches caused by residual oscillating electromagnetic fields produced in the beam chambers and other ring components by the preceding bunches. These effects can drive coupled-bunch instabilities throughout the entire ring that grow as the beam currents increase. Coupled-bunch instabilities in the electron ring are also caused by the presence of residual-gas ions and, for the positron ring, clouds of photoelectrons generated by synchrotron X-rays hitting the beam chamber walls and by photoelectrons reaccelerated by the beam striking the walls to make secondary yields.

[2] A maximum center-of-mass energy of 29 GeV was achieved during the lifetime of PEP.

In addition to driving coupled-bunch instabilities, the presence of ions and electron clouds enlarges the beam sizes, sometimes leading to beam losses throughout the ring. In fact, this effect turned out to be the most serious problem for both projects, especially "blow-up" of the positron beam caused by the photoelectron clouds.

Large beam currents also imposed serious challenges for the hardware components along the rings. A high-quality vacuum had to be kept in the beam chambers to ensure reasonably long beam lifetimes in an environment where the chamber walls were constantly bombarded by huge fluxes of synchrotron X-rays. Heat energy accumulated in the ring components had to be removed efficiently. Tireless efforts were made throughout the entire period of operation to keep improving the performance of critical hardware components and for finding optimum operating conditions, which were often far from those carefully developed during the design stage. Movable masks used to scrape away unwanted beam-halo particles turned out to be a particularly difficult challenge.

A background simulation effort started in *BABAR* immediately to focus on the ingredients that should be integrated in the PEP-II machine design, namely collimators and synchrotron radiation masks.

The conclusions of these early simulations were clear:

- The background would be severe.
- The uncertainties in the simulation were very large due to many reasons (incomplete knowledge of the physical sources, incomplete description of the machine, crude assumptions on the machine vacuum, etc.).
- An experimental approach to try to control all these approaches was mandatory. This led to the creation of a commissioning detector which started in 1996 (see Section 1.4.3.1).
- The detector design, which was proceeding, had to adopt a safety factor of 10 relative to all background predictions. This "administrative" rule turned out to be extremely difficult to meet initially, and led to changes in the technical implementation of several detector components, but turned out to be very wise and had many pay-offs in the long term.

Many collimators were proposed, with fixed or movable jaws, for inclusion at key locations. It turned out that it was difficult and very costly to implement them all, so only a select few were installed. At Belle, several versions of movable masks were used, each version being a gradual improvement on the previous one.

In a two-ring machine with small bunch spacings, a beam-separation scheme is needed to divert the beams as they leave the IP in order to avoid parasitic interactions. PEP-II used a head-on collision scheme with near-IP bending magnets to steer the $e^+$ and $e^-$ beam bunches away from each other as soon as possible after the collision. KEKB, on the other hand, used a scheme in which the two beams collide with a small ($\pm 11$ mrad) crossing angle. While this scheme had the considerable merit of allowing for shorter bunch spacing and more available space for the detector components near the IP, it was not without risk. A previous attempt to use a small but finite-angle crossing scheme in the DORIS ring at DESY (Piwinski, 1977) had problems that were attributed to beam instabilities from unwanted couplings between betatron and synchrotron motions caused by the crossing angle, and it was generally believed that this effect would get worse at larger crossing angles. However, a theoretical study (Hirata, 1995) concluded that a large horizontal crossing angle in KEKB would, in fact, not be very harmful; based on this, a finite crossing angle was incorporated at an early stage of the design process. Ultimately, crossing-angle-induced transverse-longitudinal couplings were canceled by the use of the world's first operational set of superconducting crab cavities that realign the directions of the beam bunches so they pass through each other head-on (Hosoyama et al., 2008). These were installed in January 2007; with the cavities, and with chromatically corrected IP beta functions, KEKB eventually reached a peak luminosity of $2.1 \times 10^{34}$ cm$^{-2}$ s$^{-1}$, more than twice the original design goal.

As a result of due care and attention in the design of the machines, the excellent performance of the KEKB and PEP-II colliders was comfortably sufficient to allow *BABAR* and Belle to verify the Kobayashi-Maskawa theory of *CP* violation, and, in addition, provide opportunities for a number of other measurements and discoveries, many of which were well beyond the scope of the original physics goals listed in the 1994 Belle Letter of Intent (Cheng et al., 1994) and the *BABAR* Physics Book (Harrison and Quinn, 1998). The machine parameters for the two $B$ Factories during the final stages of their operation are given in Table 1.3.1.

## 1.4 Detectors for the $B$ Factories

The $B$ Factories have a common set of design requirements which are driven by the physics goals laid down in Section 1.2. The resulting detector designs for *BABAR* and Belle are, broadly speaking, quite similar, with similar operational performance. Any differences resulted from conditions expected from the PEP-II and KEKB accelerator complexes and the technical competences and available resources of the groups who built the various sub-systems. The main requirements are as follows

**Light material (*i.e.* high $X_0$) for the inner detector:** The beam pipe, for the length corresponding to the solid angle subtended by the active region of the $B$ Factory detectors, was made of beryllium with a cooled channel between inner and outer walls. Beryllium was chosen to minimize the amount of material in terms of radiation length, to reduce multiple scattering and energy loss of particles crossing the beam pipe.

**Table 1.3.1.** Machine parameters of PEP-II and KEKB during the last stage of their operation.

| Parameters | | | PEP-II | KEKB |
|---|---|---|---|---|
| Beam energy | | (GeV) | 9.0 ($e^-$), 3.1 ($e^+$) | 8.0 ($e^-$), 3.5 ($e^+$) |
| Beam current | | (A) | 1.8 ($e^-$), 2.7 ($e^+$) | 1.2 ($e^-$), 1.6 ($e^+$) |
| Beam size at IP | $x$ | ($\mu$m) | 140 | 80 |
| | $y$ | ($\mu$m) | 3 | 1 |
| | $z$ | (mm) | 8.5 | 5 |
| Luminosity | | (cm$^{-2}$ s$^{-1}$) | $1.2 \times 10^{34}$ | $2.1 \times 10^{34}$ |
| Number of beam bunches | | | 1732 | 1584 |
| Bunch spacing | | (m) | 1.25 | 1.84 |
| Beam crossing angle | | (mrad) | 0 (head-on) | $\pm 11$ (crab-crossing) |

**Vertexing capability:** The key to measuring *CP* violating asymmetries is the precise determination of the decay vertex of each $B$ meson in an event. The only viable technology to use at the time the $B$ Factories were being constructed was a silicon-strip-based vertex detector.

**Particle identification:** In order to classify particles in the final states of interest, over a broad range of momentum, it is not possible to rely on a single particle identification technology. Both experiments constructed drift chambers with sufficiently good specific energy loss ($dE/dx$) measurement capability to perform charged particle identification for low momentum tracks. This was supplemented at Belle by a Time-Of-Flight system, and an aerogel-based Cherenkov detector for characterizing high momentum particles. At *BABAR*, high momentum track identification was achieved via the Detector of Internally Reflected Cherenkov light (DIRC), which was proposed by Blair Ratcliff (Ratcliff, 1993; Schwiening et al., 2001).

**Electromagnetic calorimetry:** Many final states of interest, including $B^0 \to J/\psi K^0_S$ where $J/\psi \to e^+e^-$, require that one is able to measure the energy of both electrons and neutral particles. The technology adopted by the $B$ Factories was inspired by the CLEO electromagnetic calorimeter (Kubota et al., 1992): both experiments used CsI(Tl) crystal calorimeters.

**$K^0_L$ and muon identification:** The expected *CP* asymmetries in $B^0 \to J/\psi K^0_S$ and $B^0 \to J/\psi K^0_L$ are equal in magnitude and opposite in sign: it was realized that to verify any observation of *CP* violation in $B$ decays, it would be important to measure both of these modes. Given the lifetime difference between $K^0_S$ and $K^0_L$ mesons, the $K^0_S$ mesons would be expected to decay in the beam pipe or silicon detector, whereas most $K^0_L$ mesons would pass through the inner part of the detector without decaying. Detection requirements for $K^0_L$ mesons were similar to those required for efficient muon identification, which was important in order to detect the $J/\psi \to \mu^+\mu^-$ contributions for *CP* asymmetry measurements. As a result, the outer parts of the two $B$ Factory detectors were instrumented with layers of active detector sandwiched between absorber material. Belle adopted float-glass based Resistive Plate Chambers (RPCs) operating in limited-streamer mode. *BABAR* initially adopted a Bakelite-based RPC solution for its $K^0_L$ and muon identification. However, soon after operation started it was clear that this needed to be replaced, and a system of Limited Streamer Tubes (LST's) was successfully installed to replace the RPCs for the remainder of *BABAR*'s operational lifetime (see Sections 1.4.3.6 and 2.2.5).

**Data handling capability:** The design goals of the $B$ Factories were ambitious. If these were to be met, then a significant amount of data would have to be transferred from the detector system front-end, classified by a trigger system, and stored for subsequent processing. As the $B$ Factory design luminosity was surpassed, the data flow and offline computing systems had to be adapted in order to keep up with the output of the machine, and allow members of the Collaborations to produce the physics results that appear in this book.

A more detailed discussion on the $B$ Factory detectors and readout can be found in Chapter 2, and an overview of data taking and Monte Carlo production required for physics analysis can be found in Chapter 3.

### 1.4.1 The BABAR detector collaboration

The SLAC management decided that with the approval of the $B$ Factory as a new element of the national HEP accelerator program, it should explore how CERN had managed the growing of the large, international collaborations which had designed, built and operated the large detectors at that laboratory. CERN Research Directors Pierre Dariullat and Lorenzo Foa were very generous in providing access to the lab archives, and engaged in full discussions on the CERN procedures and processes, identifying both the strengths and weaknesses. These visits were very helpful in guiding the initial planning at SLAC. Several other visits to Europe allowed gathering a "temporary international advisory committee" (see below) to listen to their collective wisdom, and advice on moving forward with the formation of national core groups for the detector communities within Italy, France, Germany, UK and the US.

The CERN discussions emphasized the central importance of gathering representatives of all the international agencies involved, to oversee their investments in the scientific collaboration. It was the first time that SLAC, or indeed any DOE Office of High Energy Physics (OHEP) lab, organized an external group of representatives of funding agencies from around the world to regularly review one of its experiments, and the first time that major construction and operational funding from non-DOE sources came to a SLAC experiment. All of this was done through the International Finance Committee (IFC), which will be described later. This committee was a major player in the story of the construction of the *BABAR* experiment, but also in continuing operational support, and indeed was a central figure in solving the serious computing problem in 2001 that was caused by the accelerator team outperforming the PEP-II design luminosity (Section 1.4.3.5).

The international community working on the detector design for the SLAC-hosted asymmetric $B$ Factory held its inaugural gathering at the end of 1993, as the culmination of a two year period of many workshops and detector meetings preparing for a $B$ Factory, hopefully to be built at SLAC. Over the next year there were seven more collaboration meetings preparing the Letter of Intent and the Technical Design Report, and working through the final choices of technology and performance specifications for each detector sub-system. SLAC management recruited a short-lived, yet very important, Interim International Advisory Committee in 1993, to advise the lab on formation of the *BABAR* collaboration's first committees and identify and recruit those top level scientists. The target committee was an Interim International Steering Committee formed in early 1994 with a very important charge. It was to advise the laboratory on creating a detector R&D program (which was funded originally by SLAC, but later substantially supplemented by DOE/OHEP); to select an initial Executive Board of the collaboration; to write the original governance document and socialize it within the collaboration; and to choose the first Collaboration Council. This they did in short order and, having completed their job, the group just as quickly dissolved, with the thanks of the laboratory management.

The first Collaboration Council, in May 1994, quickly gave formal blessing to the collaboration's Governance document, and chose a Nominating Committee to search for the first spokesperson of the detector collaboration, following the search process defined in the newly passed governance rules. The Council ratified the Executive Board selection, and voted on the name for the collaboration, establishing the little French Elephant *BABAR* on "his" way to having an impressive citation count.[3] It was a productive first Council meeting, and a great kick-off for the *BABAR* collaboration. Just seven weeks later, at the July 1994 Collaboration Meeting, the Council formally ratified the nomination of David Hitlin as the first *BABAR* Spokesperson. Indeed, he had been filling the role of interim spokesman of this proto-*BABAR* community since the late 1980's, and had coordinated and led the first five formal meetings of the collaboration.

The detector collaboration had a single spokesperson through the entire construction and commissioning periods, and through the first years of data taking. From that point forward a new spokesperson was chosen from the collaboration every two years.[4] This group of seven individuals were able stewards of the scientific life of the *BABAR* collaboration. Their distinct visions on how to guide the experiment forward, their use of the associated strong management teams and their scientific judgment was no small part of the scientific success of *BABAR*. Within the *BABAR* collaboration the spokesperson is the chief officer of the collaboration, responsible for all scientific, technical, organizational, and financial affairs of the collaboration, and represents the collaboration to the SLAC laboratory, to the DOE/OHEP, and to the international funding agencies, represented by the IFC. The spokesperson is assisted in this heavy responsibility by a Senior Management Team for day-to-day decisions, and by an Executive Board which the spokesperson chairs. The Senior Management Team is chosen by the Spokesperson and ratified by the Executive Board and the Council.[5] The Executive Board is representative of the regional composition of the collaboration, and consists of members distinguished by their scientific judgment, their technical expertise, and their commitment to the experiment, and is chosen by the Council through an election process. The technical life of the collaboration was managed by the Technical Coordinator, who chaired the Technical Board. This was normally a twenty member group comprised of the detector system managers, the lead engineering staff, the computing leadership, and representatives from the accelerator collider team. For an important period of the life of *BABAR*, starting in 1999 for about two years, this group was expanded to include a much broader membership and called the Augmented Technical Board, which included all of the old Technical Board but also all of the leaders from electronics, online and off-line monitoring, computing, physics planning, and analysis machinery — a cadre of about 50 staff. For these two years this body worked hard and was a very important part of the *BABAR* story; they can take a lot of the credit for bringing the detector operations and the physics production activity into a true "factory mode," alongside the operations of the PEP-II accelerator complex. The collaboration has been well served by the five strong scientists who served as the *BABAR* Technical Coordinator[6] providing sound technical judgment and strong commitment to top level detector performance and to high efficiency uptime.

[3] The name *BABAR* is derived from $B$ and $B$-bar. The *BABAR* elephant and the many distinctive likenesses of that character, are used with permission of Laurent de Brunhoff, negotiated by David Hitlin. All copyrights were reserved to the owner, which changed to Nelvans after the late 1990's.

[4] *BABAR* Detector Spokespersons: David Hitlin (1993–2000), A. J. Stewart (Stew) Smith (2000–2002), Marcello Giorgi (2002–2004), David MacFarlane (2004–2006), Hassan Jawahery (2006–2008), François Le Diberder (2008–2010), J. Michael Roney (2010–).

[5] As part of the transition from detector construction to operation and data taking and physics analysis, a Senior Management team was formed in 2000, which included the Spokesperson, the Technical Coordinator, a senior technical advisor and lab contact if not covered by the Technical Coordinator, the Physics Analysis Coordinator, the Computing Coordinator and deputy, the past Spokesperson, and the Spokesperson-elect.

The collaboration is represented by a Council[7] with an elected chair and deputy, and made up of representatives from each institution participating in the detector collaboration. The Council is the principal governing body of the collaboration. The Council selects the Spokesperson Nominating Committee, ratifies the Spokesperson nomination, and the selection of the Executive Board. The Council appoints the operating committees of the collaboration — Membership, Speakers Bureau, and Publications Board. The Council has the unusual power to request a full review from the Spokesperson of any decision or action for which it deems such accountability was necessary, and could remove the Executive Board, or even the Spokesperson, under very strict conditions, if this unlikely situation should occur. This served as a balance to the strong and independent authority given to the *BABAR* Spokesperson under the collaboration's governance (see above).

The experiment began in 1993 and by 1995 had 483 members from 77 institutions, drawn from 10 countries — Canada, China, France, Germany, Italy, Norway, Russia, Taiwan, the UK, and the US. By 2005, the collaboration had grown to 625 members, from 80 institutions and 12 countries — with Israel, India, Netherlands and Spain having joined in the meantime, and China and Taiwan leaving. By January 2013 the active membership was still 325, of whom 51 were postdoctoral researchers and 56 graduate students. The experiment has produced 505 PhD theses, a number which is still growing, and is a remarkable testament to the intellectual life of the experiment and the breadth of its academic reach. The collaboration has produced more than one paper each week during a six year period (2004 through 2009) in the world's leading peer-reviewed journals, and a total by fall 2012 of 507 papers. We can celebrate that not only have both the *BABAR* and Belle experiments been "factories" of physics, producing new results over a broad spectrum of topics, but they have been veritable factories in producing candidates for new academic appointments for universities around the world from the pool of graduate students and post doctoral researchers who received their training on the *BABAR* and Belle experiments. They have outstanding training with both technical and operational experience with large detectors and running accelerators, and computing and data production on a factory scale, and hands-on development of creative data analyses in a small group environment.

In order for collaborators to be considered as authors on *BABAR*, they first must perform a substantial service to the experiment, either through the construction or operation of hardware, or by taking on some technical or administrative role required to maintain the quality of physics output from the experiment. Having qualified for authorship, a *BABAR* collaborator automatically signs papers. The authors appear in the author-list in alphabetical order by institute. As a result there is, in general, no direct correlation between the lead authors of a given analysis and the initial authors of a given *BABAR* paper. On occasion, where non-*BABAR* collaborators (mainly students) have made significant contributions to an analysis, requests have been made for those people to be added to the author list on the paper describing that analysis in detail. Such requests, while never a foregone conclusion, were generally granted.

### 1.4.2 Formation of the Belle collaboration

The Belle collaboration was officially formed at a one-day meeting held at Osaka University on October 7, 1993, where it was formally decided that the results of the previously held workshops (Abe et al., 1993) were encouraging enough to merit proceeding towards the development of a Letter of Intent during the next year (Cheng et al., 1994). This was followed by a series of meetings at which details of the detector design and issues of collaboration governance were discussed.

The collaboration organization was discussed at a second meeting at KEK on November 19–20, 1993. Here, it was decided that there would be three co-spokespersons, one representing each of the major constituencies of the collaboration: the KEK group, non-KEK Japanese groups, and groups from outside of Japan. All three spokespersons were elected by the full collaboration. In the beginning they served for a three year term that could be renewed. This rule was later changed to a two year term and limiting renewals to a single term. In addition, it was decided to have an Institutional Board (IB) comprised of the spokespersons and one representative from each of the collaborating institutions,[8] to deal with organizational and personnel issues, and an Executive Board (EB) consisting of about ten members selected by the spokespersons to advise them on technical and scientific issues.[9] Important matters are discussed in the IB or EB and then proposed to a general meeting of the collaboration. The general organizational principle has been that, insofar as possible, decisions are made at general group meetings, either by consensus or by a vote of those present. Urgent decisions are made by the spokespersons in consultation with the EB. This organization proved to be reasonably successful; when the experiment switched from the construction to the operating phase in 1999, a task force was formed to re-examine the organizational structure, but eventually only minor changes in the basic structure were adopted.

[6] *BABAR* Technical Coordinators: Vera Lüth (1994–1997), Jonathan Dorfan (1997–1999), A. J. Stewart (Stew) Smith (1999–2000), Yannis Karyotakis (2000–2003), Bill Wisniewski (2003–2011).

[7] The *BABAR* Collaboration Council was formed under action of the Steering Committee (chaired by Pier Oddone), in May 1994 with the first chair being Livio Piemontese (1994), followed by Bob Wilson (1996), Erwin Gabathuler (1998), Patricia Rankin (2000), Klaus Schubert (2002), Frank Porter (2004), Gerard Bonneaud (2006), David Leith (2008), George Lafferty (2010), Brian Meadows (2012), and Fabrizio Bianchi (2014).

[8] Belle Institutional Board chairs: Yasushi Watanabe (1994–2000), Seishi Noguchi (1994–2000), Leo Piilonen (2000–2012), Christoph Schwanda (2012–).

The name "Belle" (proposed by A. Abashian, Virginia Tech) was adopted by a group vote at the third group meeting held in January 1994 at Nara Women's University.[10] The Belle logo (proposed by T. Matsumoto, Tohoku) was selected by a vote at the sixth group meeting at Tohoku University in February 1995.

The experiment began in late 1993 with 136 members from 39 institutions from 7 countries — Japan, China, India, Korea, Russia, Taiwan and the US. The first spokespersons[11] were F. Takasaki, S. Suzuki, and S. Olsen. By 2008, the Collaboration had grown to 275 members, from 60 institutions and 15 countries — with Australia, Austria, Czech Republic, Germany, Italy, Poland, Slovenia, and Switzerland having joined in the meantime. Up to fall 2012, the Collaboration published 370 papers in scientific journals.

Two unique features of the Belle publication policy, developed after considerable discussion and finalized at a meeting at KEK in November 2001, are worth noting:

**Authorship confirmation:** In Belle, there is no default author list and authorship on a Belle paper is not automatic. An important rule is that after a paper draft has received approval from its internal referees and the relevant physics conveners, it is posted for general review by all eligible authors.[12] During the review period, a collaborator is required to confirm his/her authorship by submitting the statement: "I have read this paper and agree with its conclusions. Please include me as an author." Only then is he/she included in the author list.

**Author-list name order:** In principle, the order of the names in the author list is alphabetic. However, the persons responsible for preparing a paper can propose to the spokespersons that a single person or a small group of people be listed as first authors. In general, the spokespersons have approved such requests, the exceptions being for important papers central to the main goals of the Belle program (*e.g.*, precision measurements of $\sin 2\phi_1$) or cases where the proponents cannot agree on the specific name order. In these cases, the author list is strictly alphabetic.

When this policy was adopted, it was with the explicit proviso that it could be re-examined and modified at any time. However, it has proven to be quite popular among Belle collaboration members and has never been modified. Almost all Belle papers since 2002 have had a first-author group, with up to seven collaborators appearing out of alphabetical order at the start of the list; the number of confirming authors has been, on average, about half of the total number of eligible authors.

### 1.4.3 Building the *BABAR* detector

The *BABAR* collaboration faced a set of design challenges as they prepared their Letter of Intent (LOI) during the period spring 1993 through summer 1994. These included a long list of issues demanding detailed analysis to arrive at conclusions — inheriting an Experimental Hall which was smaller, and had too low a beam height, for an optimal "start-from-scratch" design; determining how to meet the stringent specifications for the silicon vertex detector and drift chamber tracker to manage both the spatial resolution to measure the separated $B$ decay vertices and at the same time handle measuring with adequate precision the broad momentum spectrum of the produced tracks; meeting the strong specifications for the charged particle identification along with good photon detection for both position and energy measurement, and for reliable muon and $K^0_L$ detection. The actual LOI document was produced over a few months, was completed in June 1994, and quickly approved by the SLAC Experimental Program Advisory Committee (EPAC) in July, only one month later.

As with all high tech projects, the detector design, construction, and commissioning came along with its problems. Fitting the collaboration's ambitions to the available budget was a stringent constraint at the outset. A great deal of hard work went into defining the technical details for the final sub-systems in the short nine month period between the submission of the *BABAR* Letter of Intent and the submission of the Technical Design Report, in February 1995. The TDR had essentially the final vertex detector geometry and technical description, a new Drift

[9] Belle Executive Board chairs: Kazuo Abe (1994–2000), Dan Marlow (2000–2002), Alex Bondar (2002–2010), Simon Eidelman (2010–2012), Tom Browder (2012–2013), and Toru Iijima (2013– ).

[10] The name Belle is a pun on beauty, the quark of primary interest for the $B$ Factories, which led to a natural choice for the name of the commissioning detector discussed later in this chapter: BEAST. The name can also be decomposed as B-el-le implying electrons (el) and their opposite — positrons (le) — colliding to produce $B$ mesons.

[11] Belle Detector Spokespersons: Fumihiko Takasaki (1994–2003), Shiro Suzuki (1994–2000), Steve Olsen (1994–2006), Hiroaki Aihara (2000–2006), Masanori Yamauchi (2003–2009), Tom Browder (2006–2012), Toru Iijima (2006–2012), Yoshihide Sakai (2009–), Leo Piilonen (2012–), Hisaki Hayashii (2012 –).

[12] Eligible authors are those members of the collaboration that actively contributed to Belle for at least six months in form of construction, maintenance or operation of the detector, software development, contributing to ongoing analyses, etc. They are also required to take a certain number of experimental shifts.

Chamber design with flat aluminum end plates instead of a cleverly shaped carbon fiber construction, the choice of the internally reflected Cherenkov detector, DIRC, and its quartz bar radiators for the particle identification system, and finalizing the choice of the muon detector technology as Resistive Plate Chambers, RPC's. Later on there were other surprises that emerged and had to be dealt with promptly; the flux return iron for the magnet had production schedule problems from the Japanese supplier as did the superconducting magnet coil from Italy, but the IFC came through with an added incentive clause to the magnet steel contract, and the lab management's connections to the US Air Force helped bring the delayed superconducting coil to SLAC on time, via "air mail" on a C5A, as part of a crew training flight. Learning how to grow the cesium iodide crystals and managing the salt delivery schedule for the large electromagnetic calorimeter, and how to successfully polish the quartz bars for the DIRC particle identification system to the exacting dimensional optical specifications, were time-consuming problems that emerged during construction, looked as though they might cause serious schedule problems, required creativity and focused commitment, but were finally solved in time for detector turn-on.

#### 1.4.3.1 The PEP-II commissioning run

Immediately after PEP-II approval in June 1993, it was realized that, because of the existence of the PEP tunnel and the significant reuse of PEP machine components, that PEP-II machine would be ready one or two years before the *BABAR* detector would be. This was considered as a good opportunity to be able to tune the machine without the complications of detector protection and to provide a fast start for *BABAR*. The machine had to reach a luminosity 100 times higher than previously achieved and was doing so with much higher currents. The potential threat posed by backgrounds induced by such currents was considerable. A few years previously at SLAC, muons from the SLC tunnel had been compromising the Mark-II/SLC detector performance, and therefore there was a high degree of consciousness of these issues among members of the PEP-II machine group.

In 1996 there was a call proposing the instrumentation, at minimal costs, of the PEP-II IR in the absence of *BABAR* during two running campaigns: a short one, in 1997, where only the HER ring would be available, and another one in 1998 with both rings. The goal of this instrumentation was manifold:

- understand and quantify the various background sources in both rings,
- provide to the machine reliable background sensors, so background could be reduced while tuning the machine,
- test prototypes of final *BABAR* elements to understand their sensitivity to background,
- test the radiation protection and abort mechanism system.

It was of course not possible to cover all these issues with a very small number of detectors since some of the requirements were potentially conflicting with each other. *BABAR* therefore adopted a "wideband" approach where a variety of detectors were assembled for the commissioning detector. PIN-diodes, silicon strip detector modules, similar to the final *BABAR* ones, a newly built mini-TPC, and reused straw tubes were used to understand the background resulting in charged particles, whereas a newly built movable ring of thallium-doped CsI (or CsI(Tl)) crystals, similar to the *BABAR* ones, were used to monitor neutral background. DIRC and IFR prototypes complemented this equipment.

This set-up and the 1997 and 1998 campaigns turned out to be successful. The large backgrounds observed were mostly due to the not-yet-scrubbed state of the rings; their various sources were understood, and their variation with current properly measured. After the required tuning, simulations were able to reproduce the observed background to within 50%. The correct strategy for a fast start to the *BABAR* experiment in 1999 was established, together with a flexible and reliable abort system.

#### 1.4.3.2 The *BABAR* background remediation effort and detector commissioning

Since *BABAR*'s high potential vulnerability to PEP-II background had been demonstrated both from simulations and from the 1997–1998 background measurement campaign described above, in 1998 a background remediation effort was set up to precisely quantify the adverse effects engendered by high background on the *BABAR* detector and physics analysis. Four areas were identified:

1. long term degradation due to integrated dose,
2. immediate damage due to a radiation burst,
3. high occupancy in the detectors leading to ghosts or to inefficiency,
4. large dead-time in electronics read out leading to dead time and/or inefficiency.

This remediation group took many important decisions to protect *BABAR* in both the short and long term, based on background extrapolations taking account of future running conditions: a very comprehensive set of dosimeters were installed throughout the detector, and an abort strategy was put in place to avoid item (2). The weakest points in the data acquisition (DAQ) chain were identified as bottlenecks two years before they needed upgrading. As a result the DIRC and drift chamber electronics were partially upgraded in good time and without limiting data taking. Good running conditions were defined in order that *BABAR* did not accumulate data that would prove not to be useful.

A strict policy to use up allowed radiation exposure as a function of the integrated luminosity was defined. A 10% occupancy limit in the drift chamber and the vertex detector were thus defined so as to guarantee good physics output, and were correlated to real time background sensors incorporated in the machine diagnostics system to prevent running in worse conditions.

Another crucial aspect of this task force was to prepare a set of 25 machine-detector interface experts that provided 24-7 support in the PEP-II control room, during the first four years of *BABAR* data taking. These background shifts proved invaluable to further the understanding and control of the background issues and to disseminate background related issues to the PEP-II operations crew.

The first short run took place in May 1999, to be followed by a short shut-down to install the full DIRC system, and then operations began again in late October. Physics running began in late 1999 and continued through 2008, when the experiment was turned off with the PEP-II collider having achieved design luminosity ($3 \times 10^{33}$ cm$^{-2}$s$^{-1}$) within one year of operation. During its final year the PEP-II collider ran regularly at a daily integrated luminosity of over seven times the design value, with record high circulating currents of both electrons and positrons, and accumulating 557 fb$^{-1}$ of data in the *BABAR* detector. Background issues were always present during the lifetime of *BABAR*, but these were successfully managed to prevent them from seriously damaging the experiment. Once routine operation of PEP-II and *BABAR* had been achieved, the background remediation effort underwent a transition to the Machine-Detector-Interface (MDI) working group that was responsible for maintaining a watchful eye on the background conditions expected within the detector, and over time learned (with the help of accelerator physicists from PEP-II) to use data from both the machine and the detector to measure beam parameters such as emittances, the betatron oscillation amplitude at the IP, and estimates of the beam sizes for bunches of electrons and positrons (Kozanecki et al., 2009). This background remediation and MDI effort was key to *BABAR*'s high luminosity running and was the result of the hard work of many people from all parts of the PEP-II and *BABAR* teams.

#### 1.4.3.3 Other beam-related backgrounds encountered

In addition to the expected background effects dominated by beam-gas terms, some unexpected sources came along the way:

- A luminosity term was readily observed in addition to single beam backgrounds and to backgrounds induced by beam-beam effects. This luminosity term was traced to the presence of off-momentum electrons or positrons after radiative Bhabha scattering. The unfortunate presence of a dipole magnetic field at the IP made *BABAR* very sensitive to these luminosity terms that became relatively more and more important as the machine was getting scrubbed and its peak luminosity increased.
- Electron cloud effects were analyzed in early studies in 1993-1994: they cause bunch-to-bunch instabilities believed to be damped by the proposed feedback systems. In 1999 electron cloud effects were experimentally observed by huge pressure increases in the LER above thresholds and by intra-bunch size enlargement unaffected by bunch-by-bunch feedbacks. The machine was immediately equipped wherever possible with cable coils around the beam pipe providing a 50 Gauss protecting field that pushed the current thresholds far away. Nevertheless, the electron cloud effect was responsible for a significant increase of the positron beam size with current that would finally limit the maximum achievable luminosity.
- Neutron induced background, where neutrons are produced by few-MeV gamma photonuclear reactions, were found to be quite significant in some sub-detectors and even dominant in the case of the IFR.

#### 1.4.3.4 *BABAR* reviews and oversight committees

The detector design and construction were formally overseen by two committees that were standard to the normal SLAC way of doing things — a DOE Lehman Review process for agency oversight of construction readiness and budget soundness, and the usual laboratory Experimental Program Advisory Committee, which had stewardship over the SLAC experimental program. There were two other new, and very important, very helpful, international committees as partners in the detector building story — a Technical Review Committee (the Gilchriese Committee), and the International Finance Committee, the IFC. The Technical Review Committee worked closely with the Detector collaboration, met twice per year through the construction period, and provided advice to both the Spokesperson and the laboratory. The committee worked in sub-committees on specific aspects of the detector construction, or as requested by either the Spokesperson or the Research Director. In practice, the collaboration used this committee in its preparation for the formal technical reviews by DOE — the Lehman Reviews. The IFC met twice per year to review progress of the construction, discuss with the lab management and the Spokesperson progress and concerns, and to set homework for lab and collaboration. Members of the group were very used to working together from many years doing just this same exercise at CERN, trusted each other and the agencies they represented, and took a strong, stewarding responsibility for their new charge — the fledgling North American-hosted *BABAR* experiment. They met by phone in between regular face-to-face sessions when serious, time-urgent problems came up, and were very effective in finding solutions to the unexpected problems when they arose. The IFC were able to ensure that *BABAR* could draw together a critical mass of manpower and institutional support from each of the regions working on the experiment, to ensure success on the central areas of the experiment construction. They, as a group, appreciated that SLAC and the US would carry the largest share of the expenses for building and operating the experiment, but participated in solving all of the many problems that arose as "our joint problem". Largely because of their long history on other experiments at CERN, and the mutual trust they had built up, they were a very important component in guiding and enabling an extraordinary experiment. Both committees continued their important stewardship roles beyond the end of the construction.

The Technical Review Committee was called back when the lab and the experiment ran into computing problems because the machine performance surpassed the design luminosity, causing a computing load that could not be handled by the laboratory alone, without severe financial hardship. They were also called to help as the detector proposed hardware upgrades to several sub-systems. They performed spectacularly, once again.

The IFC, by the constitution, continued the twice-a-year oversight of the detector collaboration through the operational phase of the *BABAR* experiment. Again, this was a familiar role, as they worked in a similar way at CERN.

The IFC determined the Common Fund component of the construction budget and the operating budget, and negotiated with the lab and the Spokesperson on both of these important, but thorny issues. The financial needs of the collaboration were presented by the Spokesperson after discussion with the laboratory management, while the decision making on what financial support would actually be provided was the IFC's job. They also defined how each region would meet their share of these costs. Typically this was by a negotiated mix of head-count and system responsibility determining the cost sharing in the construction phase, and essentially it was by participating head-count for the operational phase. During construction this Common Fund was around \$4 M per year, totaling \$15.4 M over the construction period, and about \$2.7 M per year during the operations period, until the computing crisis (see the following two sections).

The DOE Lehmann Committee formally base-lined the detector budget in late 1995. Each member of the Technical Review Committee had an individual system assignment, and through the full construction and commissioning schedule these connections were maintained and provided timely advice to the construction team and up-to-date information to the review panel as a whole, and to the lab management. The IFC was a very helpful resource for both the laboratory and for the experiment. They brought a different kind of management layer into the lab — a technically savvy group, and a small enough group to have strong working relationships between each other, in command of substantial financial resources, and very committed to the success of the *BABAR* project. The Technical Review Committee was rather stable in its membership throughout the period of construction, with only a few people stepping down and requiring replacement. However the IFC was rather different, in that the heads of each of the international partner agency offices rotated quite frequently.

#### 1.4.3.5 Computing

From the beginning SLAC had proposed that the lab would provide the computing hardware resources, both processing and data storage, for the *BABAR* experiment. The collaboration, on their part, was to provide the required trained manpower needed to create the software tools and handle the data analysis. Early on, the IFC agreed to support a model for computing where computer professionals were hired to work alongside computer-savvy collaboration physicists. This was a very important early investment that strategically enabled the rest of the *BABAR* computing story and bolstered the scientific output of the experiment. The cost of this manpower was borne by the Common Fund.

Computing became a serious problem around the year 2000 as the PEP-II collider luminosity climbed past the design luminosity and eventually grew to three times that. The cost of upgrading the *BABAR* computing center to handle the increased data analysis and data processing was more than the lab budget could handle. In addition, the existing *BABAR* computing model did not scale to the large number of machines that would be required to keep up with the data taking. The IFC was sympathetic, but requested that the Technical Review Committee examine the problem, and carefully review the technical details of the collaboration's proposal along with the proposed cost model. The new costs were much too large for the non-US countries to support directly with cash. This turned out to be a blessing in disguise because the European IFC members proposed an alternative in which the computing load would be distributed among several "Tier A" computing centers in Europe, in addition to SLAC. Europe had built up a large computing capacity in anticipation of the coming LHC experiments, most of which was lying fallow as the LHC turn-on was delayed. The proposed *BABAR* computing model successfully passed the technical review by the Technical Review Committee, and at a special meeting in Paris in January 2001 the IFC formally agreed that the costs of computing for the *BABAR* experiment, beyond those to support the original PEP-II design luminosity, should be shared by the whole collaboration. In retrospect this spark of creativity not only saved the *BABAR* experiment, but helped set the stage for international grid computing in HEP.

As part of the examination of the computing crisis, the collaboration rethought the needed changes to the existing computing model, and a small, passionate, very focused group worked to implement an entirely new computing model. The largest change was moving from the *Objectivity* data base system to a `Root`-based system, which was done in 2003-2004, but beyond that there were continued optimizations over the following years. The implementation of the new arrangement for handling computing at the distributed agency computing centers was put in place in 2003, with the international Tier A site system set up with SLAC, CCIN2P3 Lyon (France), INFN Padova and CNAF (Italy), GridKa (Germany), RAL (UK) and latterly U. of Victoria (Canada) making up the nodes. The core computing (CPU and disk) came two thirds from SLAC and one third from the other sites. Two years later, this sharing was fifty-fifty through the intense analysis period. This high volume, distributed computing environment was the first successful example of large scale production distributed computing (also known as Grid Com-puting) in HEP in an actual data-taking experiment. The *BABAR* collaboration set up a Computing Steering Committee, which twice a year examined the foreseen needs for processor power and storage, and reported to the IFC. This ranks among the great achievements of the collaboration and of the funding agencies within the IFC. It built on the large international investment in Grid computing, and on very good international networking. The new computing model, including the change to the `Root` data analysis framework, was in operation by mid 2003 (well ahead of schedule), and allowed the experiment to keep processing the data, even at the higher luminosities.

#### 1.4.3.6 Sub-system upgrades

*BABAR*, the lab, and the Technical Review Committee engaged in a review of each of the detector sub-systems in the 2003, with the outcome that all of the systems were expected to manage the increases in luminosity promised by the accelerator team, with just nominal improvements (even the expected increased backgrounds), with one exception — the muon system's IFR chambers. The IFR sub-system had become a serious problem around 2001, with dropping efficiency as the accumulated radiation dose increased. New muon chambers had to be designed and built, and then the installation of the new technology successfully implemented without an undue hit to data taking. This was another multi-lab and multi-nation effort to execute this detector upgrade rapidly while still taking data. The collaboration made a heroic effort and made very good progress in production of replacement detectors — this time LST's, which were essentially completed by the end of 2004. Installation in the detector was not completed until 2006, due to a chain of unfortunate accidents unrelated to *BABAR*. The new chambers worked very well, and for the remaining running *BABAR* had high efficiency muon tagging.

### 1.4.4 Building the Belle detector

As with *BABAR*, the construction phase of the Belle detector had to resolve a number of technical challenges in order to provide a design that would work sufficiently well to deliver the physics goals of the $B$ Factory. As is typical with particle physics experiments, some of the sub-systems under consideration for Belle had proposed variants that had to be studied in detail (Section 1.4.4.1). Along the way the Belle detector team were also presented with several unexpected problems that required timely resolution (Section 1.4.4.2). The commissioning period and the first years of full Belle operation are discussed in Sections 1.4.4.3 and 1.4.4.4 respectively.

#### 1.4.4.1 Design choices and related issues

**Beam pipe:** The beryllium beam pipe section is made of two concentric cylinders and an intermediate cooling channel. The only supplier for beryllium in such a configuration was Electrofusion in California. Because of its toxicity, it was only with considerable difficulty that the import of beryllium was allowed by Japan Customs officers.

**Silicon:** The silicon detector — a key component for the success of a $B$ Factory experiment — was originally planned to use a custom designed Application Specific Integrated Circuit (ASIC), however as discussed in Section 1.4.4.2, the then-standard Honeywell technology for radiation-hard ASIC design could not be used for a project in Japan. As a result, the choice of which ASIC to use had to be changed to allow a working vertex detector to be assembled and installed in time for data taking, while a suitable radiation-hard design was developed for a subsequent detector.

**Drift chamber:** The Central Drift Chamber (CDC) design originally envisaged two chambers: an inner "precision chamber" with two wire layers and three cathode-strip readout surfaces that focused on high spatial resolution and the provision of $z$-direction information for triggering and an outer 48-layer closed-cell drift chamber for momentum and $dE/dx$ measurements.

Since most of the particles produced in $B$ meson decays have relatively low momentum, multiple scattering is a major contributor to momentum measurement precision. Because of this, and in order to maximize the chamber's transparency to synchrotron X-rays, considerable effort was made to increase the effective radiation length of the chamber. This included the use of a helium-based chamber gas and aluminum field wires with no gold plating, both unique features at that time (Uno et al., 1993). Eventually, the inner precision chamber and the outer tracker were both incorporated into a single, common gas vessel and their intervening gas barrier was eliminated.

**Particle identification:** A number of technologies were investigated for an efficient charged particle identification system for higher momentum tracks. These included a Time-Of-Flight (TOF) system, an array of aerogel radiators (ACC) developed in collaboration with Matsushita Electric (Enomoto et al., 1993), and DIRC for the barrel region following the design concept developed for *BABAR* (Ratcliff, 1993) and a focusing DIRC for the forward end-cap region (Kamae et al., 1996; Lu et al., 1996). The choice of aerogel for both the barrel and end-cap was finally made by an ad-hoc task force appointed by the spokespersons. Their main reason for the selection of the aerogel option was its overall simplicity and minimal impact on the design of the accelerator and other detector components. The aerogel system served the Belle experiment well.

In addition to a cylindrical array of 128 4-cm-thick scintillators as a TOF system, for additional charged particle identification capability, it was also decided to include a second layer of 64 4-mm-thick counters (the TSC) to form a track-trigger. The TSC-TOF was initially considered to be a unnecessary redundancy. The subsequent issues with regard to de-scoping the SVX trigger capability (Section 1.4.4.2) meant that the provision of this redundancy proved to be a wise choice. The fast L0 triggers generated from TSC-TOF coincidences are an essential part of the Belle DAQ system.

**$K_L^0$-muon detector:** For the detection technology of the "KLM" (Belle's instrumented return yoke) LST's and RPC's were considered; RPCs were finally selected because of their robustness and simplicity. The key components in an RPC are the highly resistive planar electrodes that require very smooth surfaces in order to avoid non-particle induced electromagnetic discharges. Various electrode materials were studied including oil-covered Bakelite, dry Bakelite, ABS and PVC plastic, and float glass. It was found that ABS plastic and float glass had acceptable efficiency and lifetime properties and glass electrodes were selected because of their availability and low price (Morgan, 1995). This was the first use of glass electrodes in a large-scale RPC system. These worked well as long as care was taken to avoid any moisture contamination in the operating gas.

#### 1.4.4.2 Belle construction: two major crises

The Belle and KEKB Letters of Intent, submitted in April 1994, resulted in the approval of the project by the Japanese government, and construction started soon thereafter. The detector construction had two major crises: the failure of the initially planned technology for the silicon vertex detector and the collapse of the support structure for the CsI crystals of the barrel electromagnetic calorimeter.

*SVX failure*

A complication arose in the design of the ASIC chip, called SMAASH (with both analog and digital pipelines, on-board data sparsification, and trigger signals derived from 32-bit digital OR circuits; Yokoyama et al., 1997), intended for front-end readout of the SVX detector. US export restrictions meant that the chip design program also had to incorporate the development of the required radiation-hard techniques for the SMAASH ASIC chip. In early 1997, technical problems with the chip development caused the SVX subsystem project to fall well behind the schedule needed to be ready in time for the August 1998 installation date. Following a June 1997 recommendation of a review panel of international experts chaired by P. Weilhammer of CERN, Belle abandoned the SVX and redesigned the entire system, settling for a more modest arrangement, SVD1, based on commercially available, non-radiation-hard components, that met the angular acceptance and signal-to-noise requirements, but with no triggering capability and a marginally acceptable data acquisition rate. SVD1 (Alimonti, 2000) was a three-layer array of double-sided silicon detectors (DSSD) that were fabricated by Hamamatsu Photonics using a design that was originally developed for the DELPHI experiment's microvertex detector. The readout was based on the VA1 front-end chip that was commercially available from the IDE AS company in Oslo, Norway. In a crash program involving a close collaboration among thirteen different groups in Belle and the KEK mechanical shop, SVD1 was designed and constructed and ready to be installed in Belle by the beginning of October 1998. By that time, the Belle roll-in date had been shifted to February 1999. To compensate for SVD1's lack of internal trigger capabilities, a fast L0 trigger derived from TSC-TOF coincidences was used to latch the SVD response for potentially interesting beam-crossings while the slower L1 trigger decision was being made.

Because it was a relatively primitive system, enough spare parts and a prototype frame were available to permit the assembly of a spare device, SVD1.1. During all of the data-taking prior to the installation of SVD2 in 2003 (Natkaniec, 2006), Belle maintained a spare, replacement vertex detector that was ready to be installed. The original version was eventually replaced after radiation damage in summer 1999, and was replaced again by a more radiation-hard version a year later.

*Collapse of the CsI crystal support frame*

In the Belle calorimeter design, the crystals are supported by a honeycomb cell structure formed by 0.5-mm-thick aluminum fins stretched between a 1.6-mm-thick aluminum inner cylinder and an 8-mm-thick stainless steel outer cylinder. The fins and the inner cylinder were originally welded together and bolted to the outer supporting cylinder.

In May 1998, when the loading of the crystals into the structure and the associated cabling was nearly complete, and just weeks before the scheduled date for installation of the ECL into the Belle structure, severe deformations to the structure were evident and loud ominous sounds were heard when the partially filled support structure was rotated. These were caused by failures of many of the welds between the thin aluminum vanes and the inner cylinder. After removing all of the crystals and cables, a major renovation of the structure was undertaken that stiffened the outer support cylinder and used bolts and washers to connect the aluminum vanes to the inner cylinder. This required a delay of the Belle roll-in date from August 1998 until February 1999. The modifications to the support structure were completed by mid-August and crystal re-installation and re-cabling were completed in September.

#### 1.4.4.3 KEKB/Belle commissioning and early running

The original schedule, in which Belle and KEKB were commissioned at the same time, was changed. The initial KEKB commissioning occurred without Belle in place. Instead, a modest commissioning detector, called BEAST, was installed to provide feedback to KEKB on background conditions during the machine study and tuning period. During the initial KEKB beam commissioning period, the fully assembled Belle was commissioned in the rolled-out position using cosmic rays.

*KEKB commissioning run*

The initial KEKB commissioning run started in December 1998 and was reasonably successful, but not without mishap. The injection system, including the positron source, worked well, although sometimes positron injection produced large radiation doses in BEAST. The closed-orbit deviations in both rings were corrected to less than 1 mm, which indicated that the magnets were well aligned. In the high-energy ring (HER), a 250 mA electron beam ($\sim$ 0.25 times the design value) was stored with a respectable 60 minute lifetime. In the low-energy ring (LER), a 370 mA positron beam was stored ($\sim$ 0.15 times the design value).

In February, during high-current operation of the HER, the intense synchrotron radiation fan generated in the downstream superconducting IR quadrupoles — through which the exiting electron beam passes off-axis and, thus, in a region of high field — burned a hole through a downstream section of the aluminium beam-pipe, causing a catastrophic vacuum system failure. A replacement pipe section, made from aluminum, was quickly fabricated and installed. Subsequent simultaneous running of both the LER and HER produced collisions with a luminosity that was estimated to be $\sim 10^{30}\ \mathrm{cm}^2\,\mathrm{s}^{-1}$. BEAST measurements indicated that the SVD occupancy rates would probably be tolerable, but the large radiation doses that sometimes occurred during positron injection posed some danger. In addition, BEAST results indicated that the CDC occupancy levels and CsI pedestal widths would be very high during high-current operation of the HER.

*Belle commissioning run*

The commissioning of the fully assembled Belle detector and solenoid with cosmic rays in the rolled-out position also started in December 1998. This allowed for a complete relative alignment in space and time of all the detector subsystems and exposed some problems with the detector and the data acquisition system. The SVD1 and CDC spatial resolutions and the overall $p_T$ resolution of the CDC were measured to be near the design value. The other subsystems, including the trigger and the DAQ software, also performed well. One major problem was an efficiency drop in the resistive-plate chambers of the $K_L^0$-muon detector, which was caused by minute levels of water vapor contamination in the chamber gas. This was cured by replacing all 5 km of polyolefin tubing in the gas distribution system with copper.

#### 1.4.4.4 Early operation

Belle rolled into place on May 1, 1999 and saw first collisions (25 mA positron beam on a 9 mA electron beam) on June 1. Early running was plagued by high occupancy in the CDC caused by synchrotron radiation produced by the electron beam. The origin of this problem was traced to back-scattered X-rays from the aluminum section of the down-stream beam-pipe that was installed during the KEKB commissioning run. In addition, in July, there was an abrupt deterioration in the performance of the inner-most layer of SVD1.0. This was found to be due to low-energy synchrotron X-rays produced in one of the upstream correction magnets in the HER.

The first run managed to map out the $\Upsilon(4S)$ peak, and was then terminated in August. In the ensuing two-month shutdown, the downstream aluminum pipe was replaced with a copper version, SVD1 was replaced by the SVD1.1 spare, the CDC grounding was improved and additional beam halo masks were incorporated inside the HER to reduce backgrounds from spent electrons. Software current limits were established on the upstream correction magnets to prevent a repetition of the conditions that destroyed SVD1.0. Although the front-end electronics for SVD1.1 were not radiation hard, subsequent versions of the VA1 chip were fabricated with smaller feature sizes, and these were found to be quite radiation hard (Taylor, 2003).

*Electron cloud instability*

These fixes were effective and in the next run—Belle's first physics run—Belle collected a $28\,\mathrm{pb}^{-1}$ data sample at the $\Upsilon(4S)$ peak containing 76k hadronic events with all detector sub-systems operating at near-design performance levels. The peak machine luminosity was $3.1\times10^{32}\ \mathrm{cm}^2\mathrm{s}^{-1}$ but attempts to go above this level were stymied by a blow-up of the positron beam size. This was traced to the electron cloud instability, in which photo-electrons from the vacuum chamber wall produced by synchrotron X-rays from one positron bunch experience a Coulomb attraction to the following positron bunch. The cure for this was the establishment of a weak magnetic field near the vacuum chamber wall that bends the photo-electrons back into the wall. The first attempt at doing this in the LER involved attaching a large number of small permanent magnets to the beam pipe, which was only modestly successful. The real cure to the problem was achieved by the painstaking wrapping of solenoidal coils around all exposed sections of the LER beam pipe, as was the case with PEP-II.

## 1.5 Physics at last

The KEK and SLAC $B$ Factories were under constant examination to improve the respective accelerator teams' understanding of beam optics, accelerator controls, and all aspects of collider operations; the instantaneous luminosity increased gradually and steadily with the passage of time. Both machines quickly passed their design luminosities. The PEP II luminosity passed $1\times10^{33}\ \mathrm{cm}^2\,\mathrm{s}^{-1}$ in 1999, and reached $2\times10^{33}\ \mathrm{cm}^2\,\mathrm{s}^{-1}$ early in 2000. The KEKB peak luminosity passed $1 \times 10^{33}$ cm$^2$ s$^{-1}$ in February 2000 and reached $2 \times 10^{33}$ cm$^2$ s$^{-1}$ by the summer, later reaching $3.4 \times 10^{33}$ cm$^2$ s$^{-1}$ in April 2001, the largest luminosity then achieved in colliders. Over the life of the $B$ Factories there was a further improvement by a factor of six at both facilities: see Table 1.3.1.

There were two very different kind of collaborations going on between the two $B$ Factory communities: the collaboration between the accelerator groups, and that between the detector and physics analysis groups. The accelerator collaboration was both close and collegial. On each occasion when one or the other group were faced with a new phenomenon on their suite of accelerators or control systems or simulation systems, they would be in touch, and most often a small crew of experts from the "other team" would appear in their control room trying to help diagnose the new behavior. This joint facing of each new problem was certainly a component of the increased performance of both machines. The competition between the two teams was also very important in motivating careful attention to up-time, and to optimum performance. The detector/physics teams, by contrast, were relatively separate. One explanation of this comes from the desire not to share "too much" the details of data analysis, so that any discovery made would be independently verified by the other experiment. A secondary concern was to ensure that knowledge of analysis techniques and systematic uncertainties from one experiment did not unintentionally lead to a bias on the results, and "too good" agreement between Belle and *BABAR*. In any case, while there were occasional requests for help or advice on problems, details of on-going analyses were treated as confidential.

The initial aim of both experiments was to present first results at the ICHEP 2000 meeting in Osaka, Japan. Belle submitted 17 papers to this conference, most of these using 5.6 fb$^{-1}$ of data, whereas *BABAR* submitted 15 papers based on a data sample of 9.8 fb$^{-1}$. Belle's first journal paper, a measurement of the $B^0 - \overline{B}^0$ mixing parameter $\Delta m_d$, was submitted to Physical Review Letters in November 2000 (Abe, 2001b) and the first *BABAR* paper accepted for publication was measurement of time-dependent $CP$ asymmetries in $B^0$ meson decay and was submitted to Physical Review Letters in February 2001 (Aubert, 2001a). These first publications were a taste of things to come.

### 1.5.1 Establishing $CP$ violation in $B$ meson decay

Following the initial results shown in Osaka, the two $B$ Factories continued to work in competition with one another toward the goal of determining the level of $CP$ violation manifest in $B$ meson decay. The two experiments had similar strategies: to accumulate as much data as possible in time for the next summer conference season. By the time of the 2001 summer conference season *BABAR* and Belle had accumulated, and processed for physics analysis, approximately 29 fb$^{-1}$ of data each at the $\Upsilon(4S)$ peak.

At the 2001 Europhysics Conference on HEP *BABAR* announced the result $\sin 2\beta = 0.59 \pm 0.14(\text{stat}) \pm 0.05(\text{syst})$, a $4.1\sigma$ deviation from the $CP$ conserving solution of $\sin 2\beta = \sin 2\phi_1 = 0$. At the same time this result was submitted for publication. A few weeks later at the 2001 Lepton-Photon conference, Belle announced their result $\sin 2\phi_1 = 0.99 \pm 0.14(\text{stat}) \pm 0.06(\text{syst})$, a $6\sigma$ deviation from the $CP$ conserving solution. These *BABAR* (Aubert, 2001e) and Belle (Abe, 2001g) results were published as back-to-back articles in the August 27, 2001 issue of Physical Review Letters. The Belle and *BABAR* central values straddled predictions based on the KM model — and they were consistent with each other. Together the $B$ Factory results clearly established the existence of $CP$ violation in the $B$ meson system. More details of these and subsequent measurements of $\phi_1 = \beta$ can be found in Chapter 17.6.

### 1.5.2 The premature end of BABAR data taking

As a result of budgetary decisions within the US, data taking with *BABAR* was curtailed and the experiment stopped running in 2008. However, the *BABAR* management, supported by SLAC, was able to work with the funding agency representatives in order to ensure that a series of planned special runs at center-of-mass energies away from the $\Upsilon(4S)$ would be allowed to go ahead before the shut-down. As a result *BABAR* accumulated data at the $\Upsilon(3S)$ and $\Upsilon(2S)$, and performed an energy scan above the $\Upsilon(4S)$. The most significant result from these runs was the discovery of the $\eta_b$, the long-sought-after ground state of the $b\bar{b}$ system (Section 18.4). The measurement of the ratio of hadrons to di-lepton pairs can be used to obtain a precision determination of the $b$ quark mass as discussed in the same section.

### 1.5.3 The final Belle data taking runs

The final beam abort ceremony of KEKB/Belle took place at KEK on June 30, 2010. The last data taking period was devoted mainly to an energy scan around the $\Upsilon(5S)$, collecting more than 21 fb$^{-1}$ of data (see Section 3.2 for details on data taking).

The end of Belle data taking was triggered by two considerations. First, Belle accumulated data in excess of 1 ab$^{-1}$ in accordance with the plan put forward before the start of operation. Second, it was time to start work on the upgrade of the facility, both the accelerator (to SuperKEKB) and the detector (to Belle II).

# Chapter 2
# The collaborations and detectors

***Editors:***
*Nicolas Arnaud (BABAR)*
*Hiroaki Aihara, Simon Eidelman (Belle)*

**Additional section writers:**
*I. Adachi, D. Epifanov, R. Itoh, Y. Iwasaki, A. Kuzmin, L. Piilonen, S. Uno, T. Tsuboyama*

## 2.1 Introduction

The *BABAR* and Belle detectors have been primarily designed to study *CP* violation in the *B* meson sector. In addition, they aimed to precisely measure decays of bottom mesons, charm mesons and $\tau$ leptons. They also searched for rare or forbidden processes in the Standard Model. As described in detail in this book, all these original goals have been reached and in many cases exceeded, thanks to the very high integrated luminosity delivered by the two *B* Factories (PEP-II and KEKB, see Chapter 1), to the quality of the physics analysis stimulated by the fruitful competition between the two experiments, and, last but not least, to the excellent performance of the two detectors, maintained over almost a full decade-long operation period. In the following, the main characteristics of *BABAR* and Belle are reviewed and compared, while the main information about the evolution of these detectors during the data taking period can be found in Section 3.2. These two chapters, however, only provide an introduction to the two *B* Factory detectors and to their years of operation. For more details, the reader should consult specific detector papers from *BABAR* (Aubert, 2002j, 2013) and Belle (Abashian, 2002b; Brodzicka, 2012), as well as the references therein. A summary of the two detector main characteristics can be found in Table 2.2.1 located at the end of this chapter.

Both $e^+e^-$ colliders operated mainly at the center-of-mass energy of 10.58 GeV which corresponds to the mass of the $\Upsilon(4S)$ resonance which decays almost exclusively (with branching fraction greater than 96%) to charged or neutral *B* meson pairs (Beringer et al., 2012).

In a $\Upsilon(4S)$ decay, neutral *B* mesons are produced in a coherent quantum state $|B^0,\overline{B}^0\rangle = (|B^0\rangle|\overline{B}^0\rangle - |\overline{B}^0\rangle|B^0\rangle)/\sqrt{2}$, which means that, until one meson decays, there is always one $B^0$ and one $\overline{B}^0$ in spite of $B^0 - \overline{B}^0$ mixing. Studying their decays often requires one to reconstruct *B* decay vertices and to measure the flight times of these mesons – in particular for time-dependent *CP* violation analysis. As they are produced almost at rest in the $\Upsilon(4S)$ rest frame – the mass of the resonance is just above the $B\overline{B}$ production threshold – the only way to have *B* vertices displaced from the $e^+e^-$ collision point is to boost these particles. This is achieved by choosing different energies for the two beams – see Table 2.1.1.

Neglecting a very small beam crossing angle (in KEKB), the kinematic parameters of $\Upsilon(4S)$ in the laboratory frame (*i.e.* detector rest frame) are:

$$\beta = \frac{p_{\Upsilon(4S)} \times c}{E_{\Upsilon(4S)}} = \frac{E_- - E_+}{E_- + E_+} \quad (2.1.1)$$

$$\gamma = \frac{1}{\sqrt{1-\beta^2}} = \frac{E_- + E_+}{2\sqrt{E_-E_+}} \quad (2.1.2)$$

$$\beta\gamma = \frac{E_- - E_+}{2\sqrt{E_-E_+}} \quad (2.1.3)$$

Asymmetric colliders require asymmetric detectors, designed to maximize their acceptance. By convention, their 'forward' and 'backward' sides are defined relative to the high energy beam. With the large boost, more particles are produced on average in the forward direction, as shown on the *BABAR* and Belle protractors displayed in Figure 2.1.1. Therefore, both detectors have more instrumentation on the forward side (extended polar angle coverage including a forward electromagnetic calorimeter) and they are offset relative to the interaction point (IP) by a few tens of centimeters in the direction of the low energy beam.

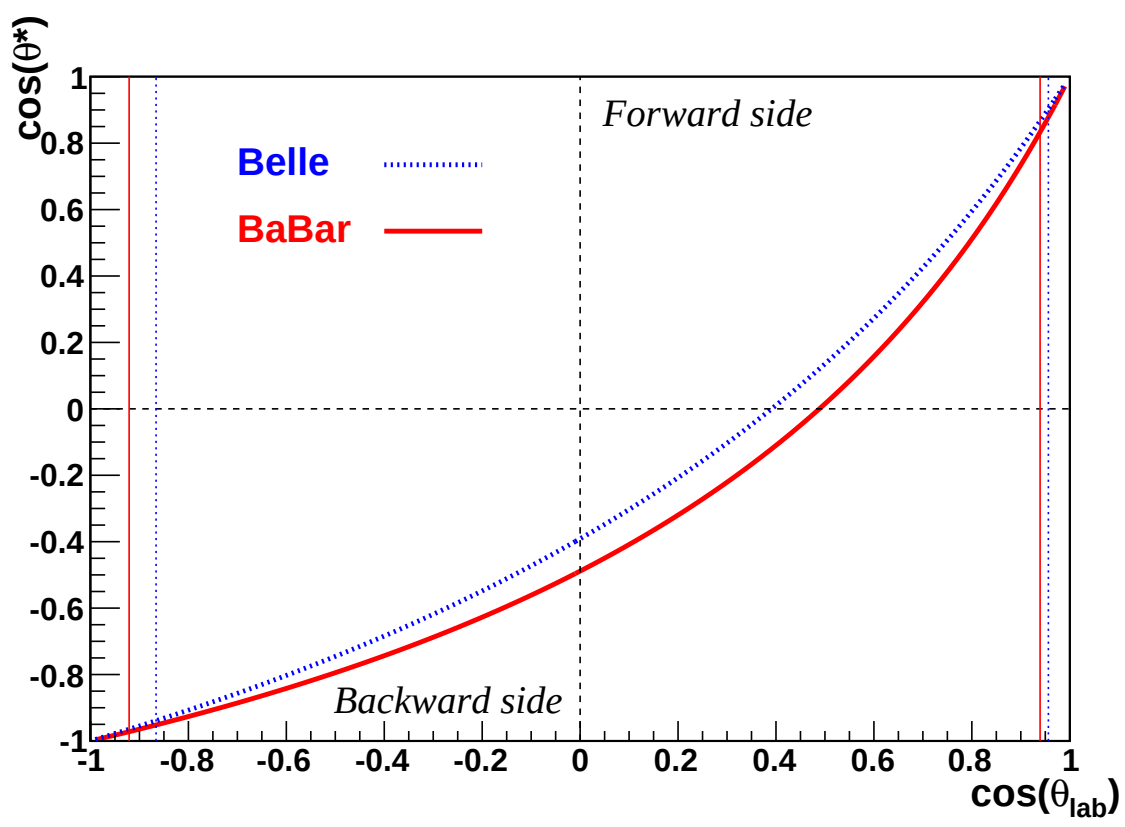

**Figure 2.1.1.** This plot shows the relationship between polar angles in the center-of-mass and laboratory frames for *BABAR* (red curve, solid line) and Belle (blue curve, dotted line). The corresponding vertical lines define the angular acceptance of the two detectors.

The Belle and *BABAR* detectors must fulfill stringent requirements imposed by the physics goals of the two experiments.

- An acceptance close to $4\pi$ and extended in the forward region, as explained above.
- An excellent vertex resolution ($\sim$100 $\mu$m), both along the beam direction and in the transverse plane.
- Very high reconstruction efficiencies for charged particles and photons, down to momenta of a few tens of MeV/$c$.
- Very good momentum resolution for a wide range of momenta, to help separating signal from background.

**Table 2.1.1.** Beam energies, corresponding Lorentz factor, and beam crossing angle of the $B$ Factories for the nominal $\Upsilon(4S)$ running.

| $B$ Factory | $e^-$ beam energy $E_-$ (GeV) | $e^+$ beam energy $E_+$ (GeV) | Lorentz factor $\beta\gamma$ | crossing angle $\varphi$ (mrad) |
|---|---|---|---|---|
| PEP-II | 9.0 | 3.1 | 0.56 | 0 |
| KEKB | 8.0 | 3.5 | 0.425 | 22 |

– Precise measurements of photon energy and position, from 20 MeV to 8 GeV in order to reconstruct $\pi^0$ mesons or radiative decays.
– Highly efficient particle identification for electrons and muons, as well as a $\pi/K$ separation over a wide range of momenta – from $\sim$0.6 GeV/$c$ to $\sim$4 GeV/$c$.
– A fast and reliable trigger, and online data acquisition system able to acquire good quality data, to process the data live, and finally to store it pending offline reconstruction
– A high radiation tolerance and the capability to operate efficiently in the presence of high-background levels.

Both detectors have the same structure with a cylindrical symmetry around the beam axis. They are of compact design with their size being a trade-off between the need for a large tracking system and the need to minimize the volume of the calorimeter, by far the most expensive single component of the detector. The forward and backward acceptances are constrained by the beamline geometry. Although the *BABAR* and Belle collaborations made different technological choices for their detector components, they have similar subdetectors, each with well-defined functions. Going from the inside to the outside of the *BABAR* and Belle detectors, one finds successively:

– A charged particle tracking system, made of two components.
  – A silicon detector, known as the SVT ('Silicon Vertex Tracker') in *BABAR*, and the SVD ('Silicon Vertex Detector') in Belle, made of double-sided strip layers to measure charged particle tracks just outside the beam pipe. This detector is used to reconstruct vertices (both primary and secondary), measures the momentum of low-energy charged particles which do not reach the outer detectors due to the strong longitudinal magnetic field and provide inputs (angles and positions) to the second tracking detector, a drift chamber, which lies just beyond its outer radius – see below for details.
  – A drift chamber, known in *BABAR* as DCH ('Drift CHamber') and in Belle as the CDC ('Central Drift Chamber'), which measures the momentum and the energy loss ($\mathrm{d}E/\mathrm{d}x$) of the charged particles which cross its sensitive volume. The latter information is useful for particle identification (PID).
– A solenoid cryostat located between the electromagnetic calorimeter and the instrumented flux return – these two detectors are described below. The cryostat is needed by the superconducting solenoid that provides a 1.5 T longitudinal magnetic field in which both tracking devices are embedded.
– PID detectors designed to distinguish the numerous pions from the rarer kaons from a momentum of about 500 MeV/$c$ to the kinematic limit of 4.5 GeV/$c$.
  – *BABAR* is using a novel device called DIRC (Adam, 2005) – 'Detector of Internally Reflected Cherenkov light' – which covers the barrel region.
  – Belle has two types of PID detectors: Aerogel Cherenkov Counters ('ACC') covering both the barrel and the forward regions; additional Time-Of-Flight ('TOF') counters in the barrel region with a $\sim$100 ps resolution which makes them efficient in separating charged particles up to 1.2 GeV/$c$, as the particle flight path from the IP to the TOF counters is about 1.2 m.
– The *BABAR* (EMC) and Belle (ECL) calorimeters; these are highly-segmented arrays of thallium-doped cesium iodide – in short CsI(Tl) – crystals assembled in a projective geometry. The *BABAR* EMC consists of a barrel and a forward end cap while the Belle ECL includes a barrel, a forward end cap and a backward end cap. Both calorimeters cover about 90% of the total solid angle. In addition to the ECL, Belle developed a special extreme forward calorimeter (the EFC), made of radiation-hard BGO (Bismuth Germanate Oxide or $Bi_4Ge_3O_{12}$) crystals. Mounted on the final quadrupoles close to the beam pipe, it provided information on the instantaneous luminosity and the machine background which helped optimize KEKB operation.
– An instrumented flux return, designed to identify muons and to detect neutral hadrons (primarily $K^0_L$ and neutrons), and divided into three regions: central barrel, forward and backward end caps. The *BABAR* IFR ('Instrumented Flux Return') consists of alternative layers of glass-electrode-resistive plate chambers (RPC's) and steel of the magnet flux return. Originally, there were 19 RPC layers in the barrel and 18 in the end caps. Second-generation RPCs were installed in the forward end cap in 2002 while RPCs were replaced by Limited Streamer Tubes (LSTs) in the barrel in the period 2004-2006. Belle $K^0_L$ and Muon detection system (KLM) was designed designed similarly and employed alternating layers of RPC's (15 in the barrel and 14 in the end caps) and 4.7 cm-thick iron plates.
– A two-level trigger with a hardware Level-1 (L1) followed by a software Level-3 (L3). The L1 trigger com-

bines track and energy triggers with information from the muon detectors and the decision to accept/reject an event is taken by a central trigger system called GLT ('GLobal Trigger') by *BABAR* and GDL ('Global Decision Logic') by Belle. The L3 trigger level runs on the online computer farm. The two trigger systems have similar design characteristics: a L1-accepted rate of O(kHz) and L3-accepted rate of O(100 Hz), for a few percent dead time and an event size of about 30 kB. Obviously these parameters have evolved during the data taking as luminosity and backgrounds increased. Both the *BABAR* and Belle triggers have been found to be robust, reliable and efficient in a wide range of data taking conditions, including runs at lighter $\Upsilon$ resonances or at $\Upsilon(5S)$ and above.

### 2.1.1 The BABAR and Belle collaborations

*BABAR*

The size of the *BABAR* collaboration reached a maximum in 2004-2005 with more than 600 collaborators. At the end of 2012, there were still 325 *BABAR* collaborators belonging to 73 institutions.

The *BABAR* collaboration is led by a spokesperson whose term is three years. He/she is selected by an *ad hoc* search committee whose choice is then validated by the *BABAR* Council. The Council is the main body of the collaboration and gathers representatives from all *BABAR* institutions. All important decisions (changes in the *BABAR* management, turnovers in the various *BABAR* committees, application of a new institution wishing to join *BABAR*, etc.) are subject to ratification by the Council. During the first year following his/her election, the spokesperson-elect works in the senior management team with the current spokesperson who is ending his/her term. The other members of the senior management are the technical coordinator, the physics analysis coordinator (PAC) and the computing coordinator. The PAC and computing coordinator are usually aided by a deputy who is expected to become the head of the corresponding office later. The two other *BABAR* boards are the Executive Board which includes representatives from the different countries involved in *BABAR* and the Technical Board (TB). The TB focuses on the detector running; each *BABAR* system (the various sub-detectors, the online and trigger groups, the machine detector interface, etc.) is represented there by two system managers, at least one of whom is based at SLAC.

The physics analysis organization is led by a PAC and a deputy-PAC (DPAC). The PAC term is two years: one as DPAC, the other as PAC on charge. Analysis Working Groups (AWGs), led by up to three people depending on the workload, gather together analysis topics which belong to the same field, *e.g.* 'charmonium' or 'charmless $B$-decays'. Analysts regularly report the progress of their work at AWG meetings during which group discussions help the analysis to move forward. Analysis developments and details are described in *BABAR* Analysis Documents (a.k.a. 'BADs') stored in the *BABAR* CVS repository. Usually, an analysis has one or more 'supporting BADs' (which are private *BABAR* documents) and one journal draft BAD which will ultimately be submitted for publication. Readers from within the AWG are chosen to read in detail the supporting BAD(s) of an analysis once it is in an advanced stage. When this part is completed, a Review Committee (RC) made up of three people (not all from the AWG) is formed. The RC and the analysts then work in close contact (phone or in-person meetings, exchanges on internal forums, etc.) to finalize the analysis, validate its results and complete the journal draft.

The *BABAR* collaboration as a whole has two main ways to get involved with the review of an analysis which is close to completion. One is the 'Collaboration Wide Talk' (CWT) which is held during either a physics meeting or a plenary session of a *BABAR* quarterly collaboration meeting. The CWT describes the whole analysis, usually including systematic uncertainties and the unblinded results – the permission for unblinding is given by the RC (see Chapter 14 about blind analysis). The last global step is the 'Collaboration Wide Review' (CWR), a two week-period during which *BABAR* collaborators proof read the draft of the written document which summarizes the whole analysis – either a journal paper or a physics note if the result is initially only to be shown at conferences. Finally, a journal draft is examined by two 'Final Readers' (FR) prior to being submitted. The PAC and the DPAC follow all the on going analyses in parallel and can step in at any time to request more information, clarify a potential issue, remind about the coming deadlines, etc. The CWR and FR steps are managed by the 'Publication Board' which also follows the correspondence between analysts and journal referees. Finally, the assignment of *BABAR* talks (obtained by the PAC who is in direct contact with conference organizers) is the responsibility of the 'Speakers Bureau'.

The analysis review process described above has been continued since the completion of the data taking so as to maintain the high quality of the *BABAR* scientific production. An internal forum system and various databases provide permanent documentation of the on-going analyses and of their review process, to the whole collaboration. The Authorship of each paper is automatically granted to all current members of the *BABAR* collaboration; people who contributed significantly to this paper without being official *BABAR* members are added to that particular author list. People usually start signing *BABAR* papers one year after becoming a *BABAR* member, and remain author one year after leaving the collaboration.

*Belle*

The size of the Belle collaboration grew with time and reached a maximum in 2012, two years after data taking ended, with about 470 collaborators from 72 institutions in 16 countries.

The Belle collaboration is led by three spokespersons whose term is two years with a maximum of three con-secutive terms. One spokesperson is from KEK, one from Japanese Universities and one from the non-Japanese institutions. The spokespersons are elected by the staff members of the whole collaboration. Spokespersons are responsible for running the collaboration, representing its interests in the institutions and with national funding agencies, and for allocating the available resources among the different subgroups.

The main body of the collaboration assembles three times a year at the Belle General Meeting (BGM), and between BGMs, decisions are enacted by the spokespersons and the Executive Board (EB). The role of the Executive Board, which is made up of the three spokespersons, three members from KEK, three members from Japanese institutions, and three members from institutions outside Japan, is to advise the spokespersons on scientific and technical matters, and to ratify all important decisions. The EB usually meets monthly.

Each collaborating institution selects a representative to sit on the The Institutional Board (IB), which meets at each BGM. The IB deals with organizational, management, and personnel issues, including admitting new collaborators, modifications of the group's organization, initiating the spokespersons' selection process, etc. The IB also makes recommendations concerning potential new members during a general meeting. The resignation of members or institutions is treated similarly. The IB also functions as a "KEKB users' organization". It gathers complaints and/or suggestions regarding KEK and asks KEK for improvements. Various institutional matters are also discussed by the IB, *i.e.* items concerning each institution's interest, such as students' thesis topics, etc. The Belle management also includes two physics analysis coordinators and the computing coordinator.

The organization of the physics analysis is similar to *BABAR*. Working Groups (WG) led by one or two persons gather together analyses that belong to the same field, *e.g.*, charmonium or charmless $B$ decays. Analysts report regularly the progress of their work at WG meetings during which group discussions help the analysis to move forward. Analysis developments and details are described in written documents - so called Belle Notes. Usually, an analysis has one or more supporting Belle Note resulting in a journal draft to be submitted for publication. When an analysis is judged to be mature enough, a refereeing committee (RC) of three collaboration members is formed. The RC and the analysts then work in close contact (phone or in-person meetings, E-mail exchanges, videoconferences etc.) to finalize the analysis, validate its results and complete the journal draft.

In addition to BGMs the results of analyses close to completion are discussed at Belle Analysis Meetings (BAM) usually held three times a year. When the RC and the analysts decide that the analysis is complete, a collaboration-wide review starts, a two week-period during which Belle colleagues proof read the final document, a draft of a journal publication. These steps are managed by the Publication Council which follows up on the correspondence between analysts and journal referees and has the general task of maintaining high quality of the Belle papers. Finally, a so called authorship confirmation procedure is started by the general consent of the referees. Authorship of each paper is not automatic in Belle. Those eligible for authorship are supposed to read the final draft and choose one of the three possibilities: agreement with the paper conclusions and willingness to become an author, non-authorship because of disagreement with the conclusions or because of insufficient contribution.

The assignment of Belle talks is the responsibility of the spokespersons who are in direct contact with conference organizers and inform the collaboration about the forthcoming scientific meetings.

### 2.1.2 The BABAR detector

Figure 2.1.2 (Aubert, 2002j) shows longitudinal and end views of the *BABAR* detector. The end view shows the forward side of *BABAR*; on the backward side one would see the toroidal water tank (also called 'StandOff Box', in short SOB) which contains the 10,752 DIRC photomultipliers (PMTs) detecting the Cherenkov photons created in the quartz bars. The right-handed *BABAR* coordinate system is shown on both pictures: the $z$-axis coincides with the axis of the DCH, which is offset by about 20 mrad relative to the beam axis in the horizontal plane – this rotation helps to minimize the perturbation of the beams by the *BABAR* solenoidal field which is parallel to the axis of the DCH. The $y$-axis is vertical and points upward while the $x$-axis points away from the center of the PEP-II rings. One commonly uses another coordinate system as well, with $z$ unchanged, $\theta$ the polar angle defined with respect to this axis ($\theta = 0$ corresponds to the most forward direction), and $\phi$ the azimuthal angle – unless otherwise stated, the *BABAR* detector is assumed to have a cylindrical symmetry. Figure 2.1.3 shows photographs of the *BABAR* detector seen from the backward end (left picture) and of the SVT (right picture).

### 2.1.3 The Belle detector

The schematic longitudinal cross section of the Belle detector is shown in Figure 2.1.4. Individual subdetectors as listed in Section 2.1 are denoted in the figure. The full detector is composed of the barrel part and of the forward (in the direction of the incoming $e^-$ beam) and the backward (in the direction of the incoming $e^+$ beam) endcaps. The coordinate system used is similar to that of *BABAR*; the $z$-axis is in the opposite direction of the $e^+$ beam (note that this is not exactly the same as the direction of the $e^-$ beam due to a finite crossing-angle of the beams), the $y$-axis is vertical and the $x$-axis horizontal away from the center of the KEKB ring.

Photographs of the Belle detector are shown in Figure 2.1.5.

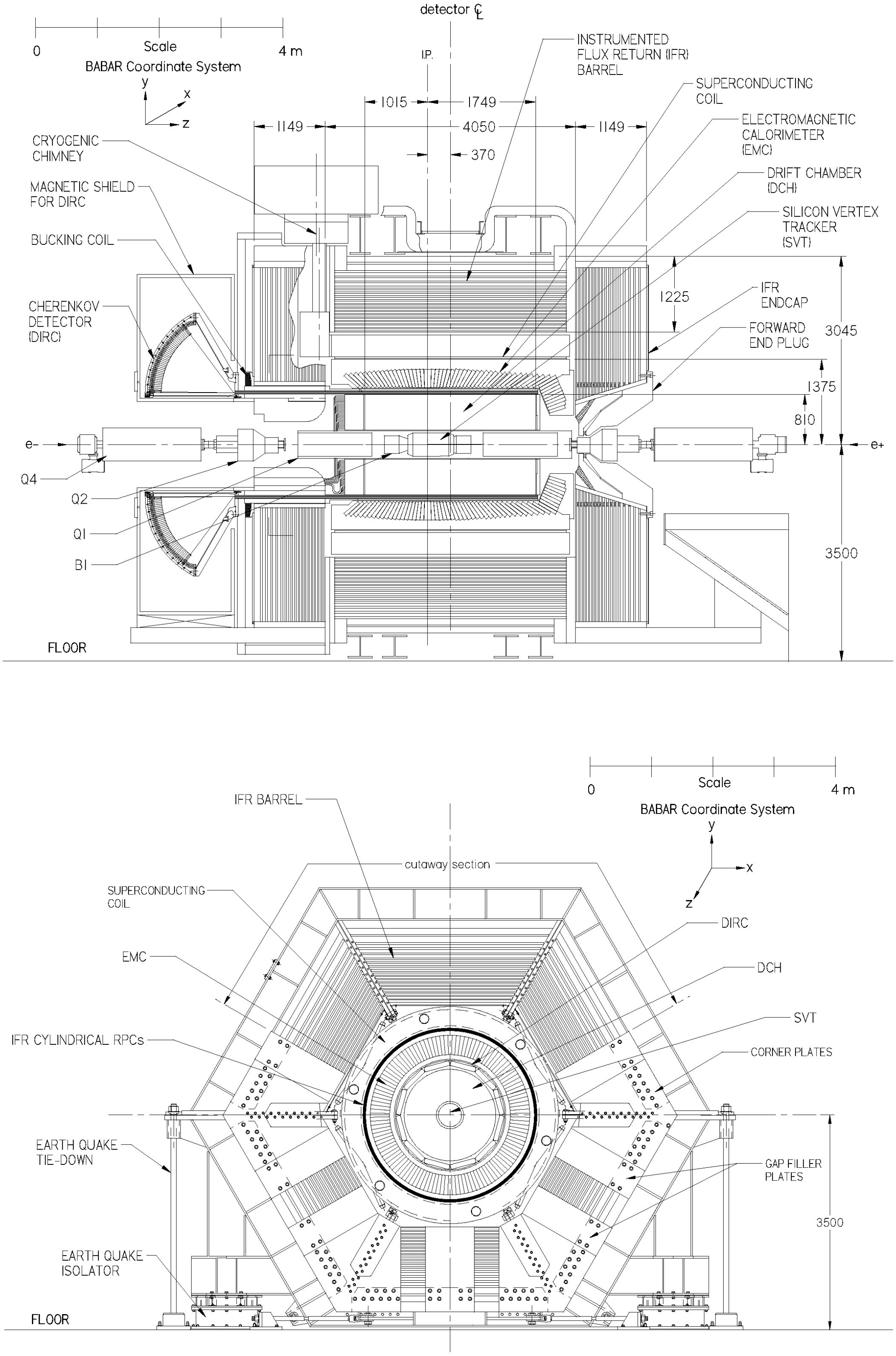

**Figure 2.1.2.** (top) Longitudinal and (bottom) end view of the *BABAR* detector (Aubert, 2002j).

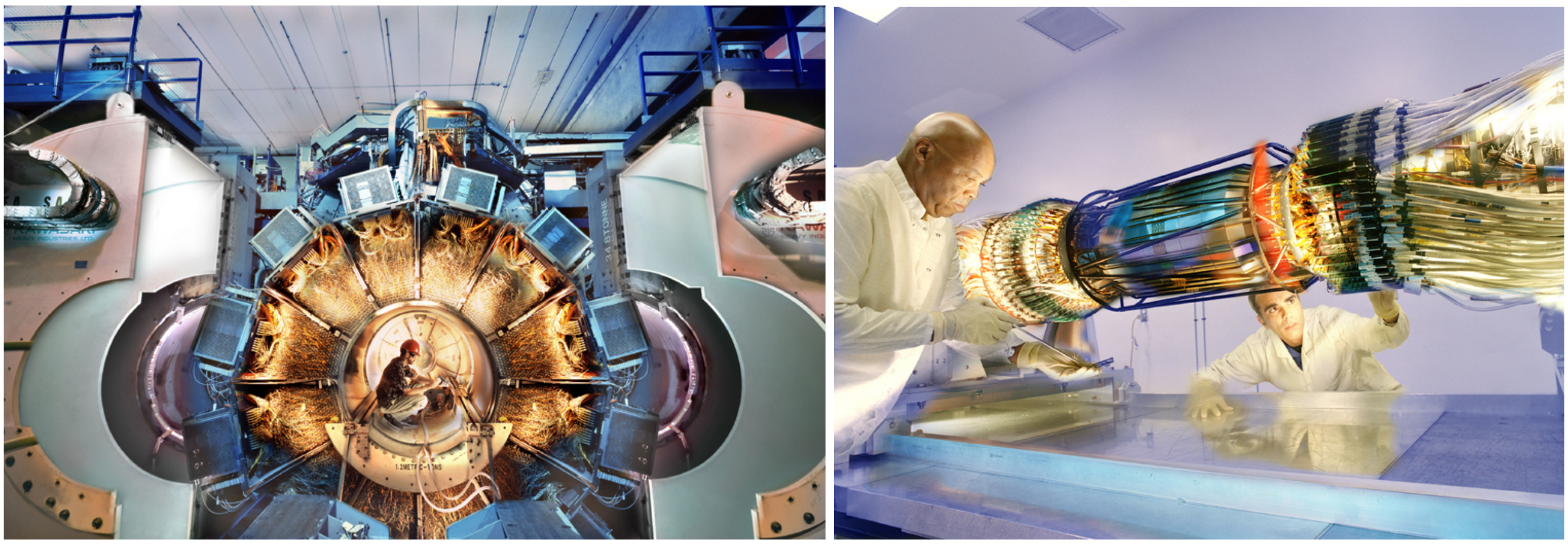

**Figure 2.1.3.** (left) View of the *BaBar* detector from the backward end, with the magnetic shield rolled out of the way to reveal the PMTs of the DIRC. The central support tube, with the SVT as well as the B1 and Q1 (dipole and quadrupole) magnets of the interaction region beam delivery system (right) was removed from the detector for maintenance at the time this photograph was taken.

## 2.2 BaBar and Belle comparative descriptions

This section provides a comparison of the different *BABAR* and Belle components, classified by function: first the subdetectors, then the trigger, the online and Data AcQuisition (DAQ) systems and finally the background protection system. As previously mentioned, the detector journal publications from each collaboration should be consulted for more detailed explanations of the detectors discussed below. Information about the PEP-II trickle injection system can be found in Section 3.2.2. Also, a casual reader not interested in the technical details of the detector setup and performances can move directly to Section 2.2.9 in which a summary of the comparison between the two detectors is provided.

### 2.2.1 Silicon detector

*BABAR*

As shown on Figure 2.2.1, the *BABAR* SVT is made of five layers: three close to the beryllium beam pipe to perform impact parameter measurements and two at a larger radius to help pattern recognition in the tracking system (SVT and DCH) and to perform stand-alone low-$p_T$ tracking: only tracks with momentum greater than 120 MeV/$c$ can be reliably measured in the DCH. The inner three layers are primarily used for vertex measurements while the outer two, located much further away, help the track extrapolation to the DCH. The end view in Fig. 2.2.1 shows the number of SVT modules: 6, 6, 6, 16 and 18 for layers 1 to 5 respectively. It also shows that the two outer layers are divided into two sub-layers each, located at slightly different radii to ensure a small azimuthal overlap between modules. A similar overlap exists for the inner 3 layers which are tilted by 5°. The three inner layers are straight while the outer two are arch-shaped to minimize the amount of silicon required to cover the solid angle and hence the amount of silicon that a track would have to pass through in the forward or backward regions of the SVT: only about 4% $X_0$.[13] The angular coverage is from 20 degrees to 150 degrees in the laboratory frame: 90% of the solid angle is covered in the center-of-mass frame. The total active area of silicon is close to 1 m$^2$ for about 150,000 channels. Each SVT module is divided electrically in two half-modules which are readout at the ends. All sensors are double-sided: on one side, the strips are parallel to the beam and measure the azimuthal angle $\phi$ and the radius of the hit $r$; on the other side the strips are transverse and measure the $z$ coordinate. The SVT consists of 340 sensors which are aligned *in situ* relative one-another using dimuon and cosmic ray events. This local alignment is quite stable over time: it only needs to be updated when something 'significant' occurs in the *BABAR* detector hall: a detector access or a quench of the superconducting coil for instance. Once this is done, the SVT is considered as a rigid single body and one can check its alignment with respect to the DCH. This global alignment is updated after every run (about once an hour): the newly computed alignment constants are then used to reconstruct tracks during the following run, data from which a new set of constants is extracted and so on. This procedure, called rolling calibration, is used by most of the *BABAR* systems and allows one to monitor changes in detector calibration which occur for the whole detector about once a day, between two successive periods of data taking.

[13] The quantity $X_0$ is called the radiation length.

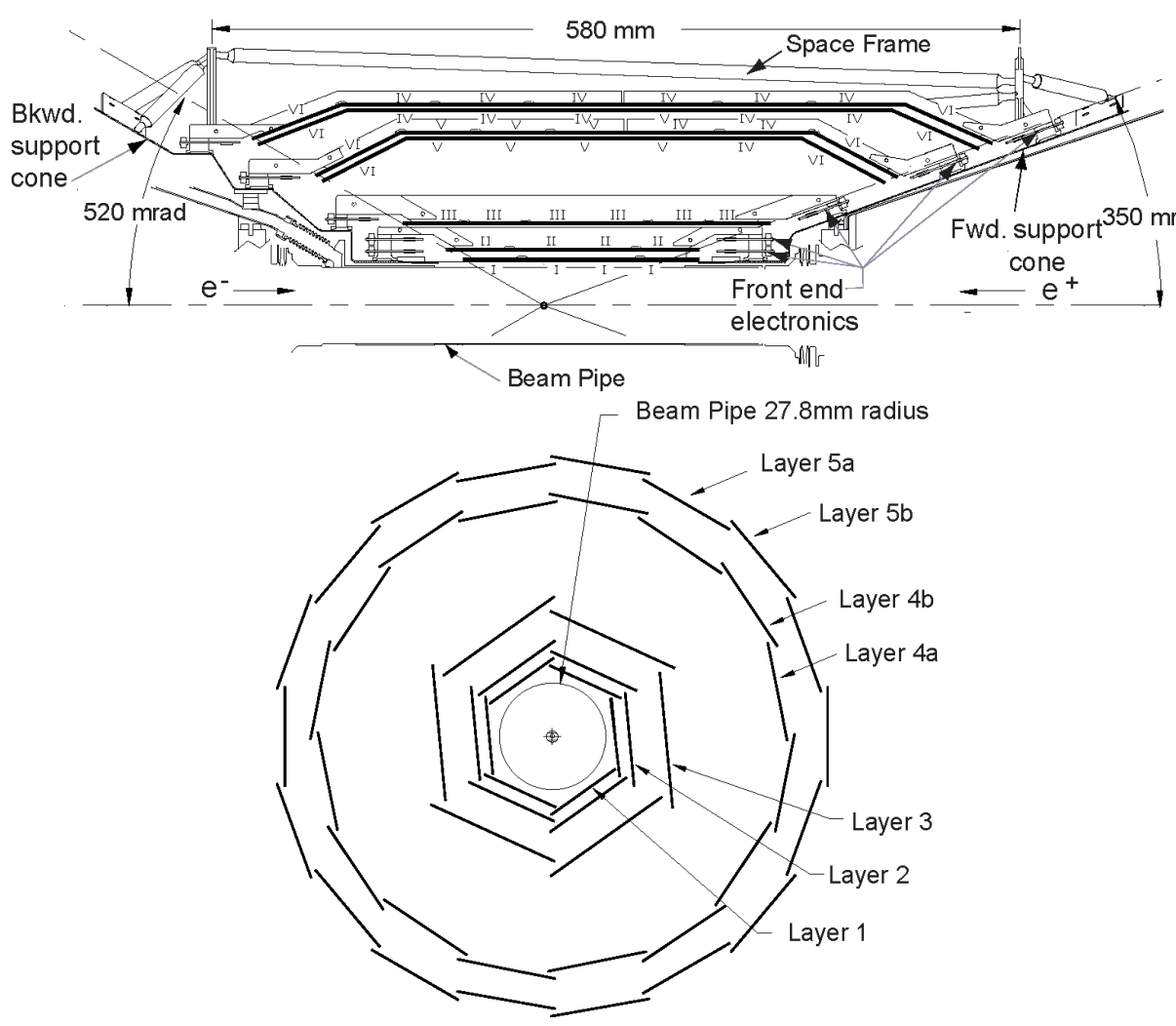

**Figure 2.2.1.** Longitudinal – unless otherwise mentioned, all subdetectors are axially symmetric around the detector principle axis – and transverse sections of the 5-layer *BaBar* SVT (Aubert, 2002j). The 27.9 mm diameter beampipe visible in the center of the SVT is composed of two beryllium layers with a water channel between them for cooling purpose.

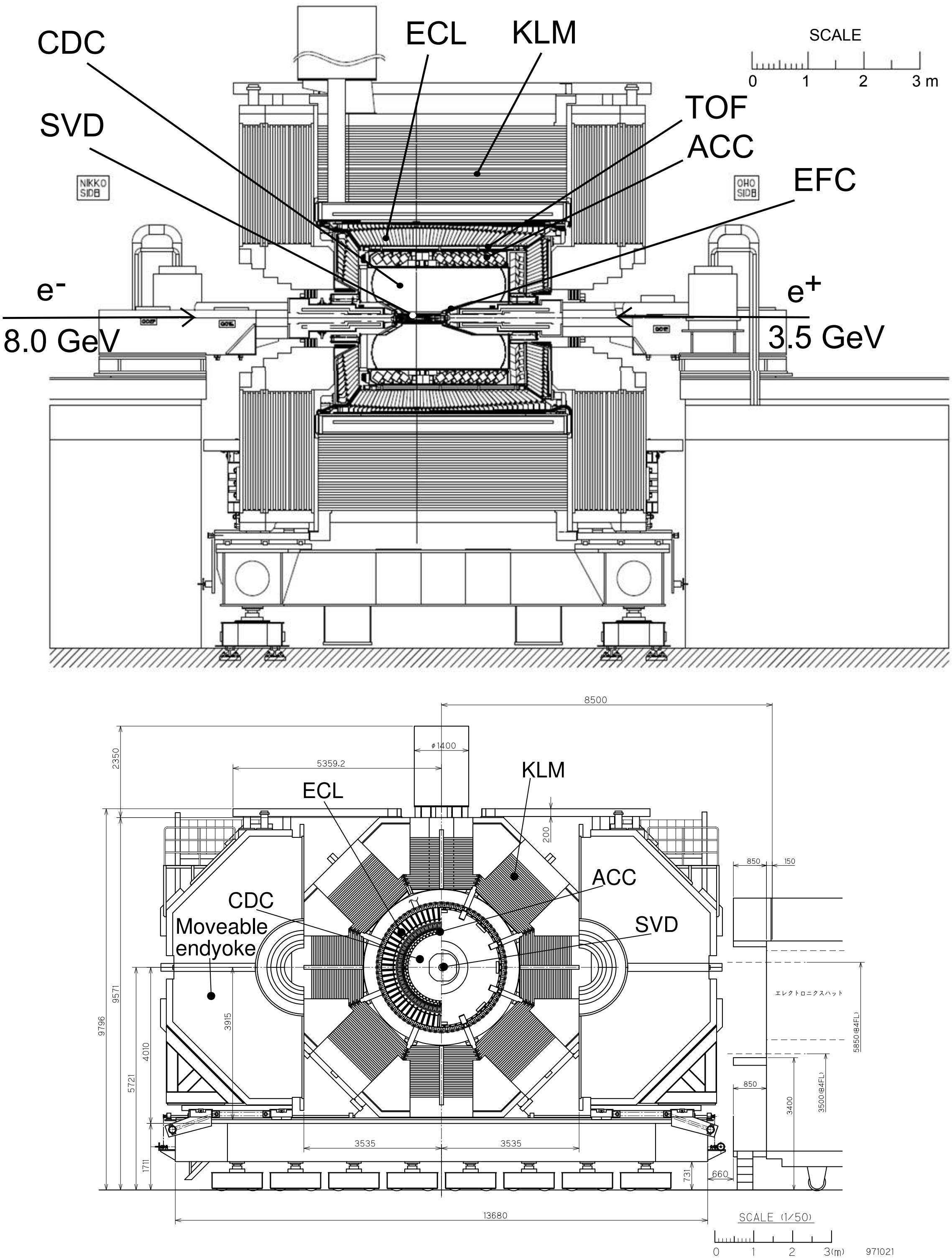

**Figure 2.1.4.** Longitudinal (top), adapted from (Abashian, 2002b), and transverse (bottom) cross sections of the Belle detector.

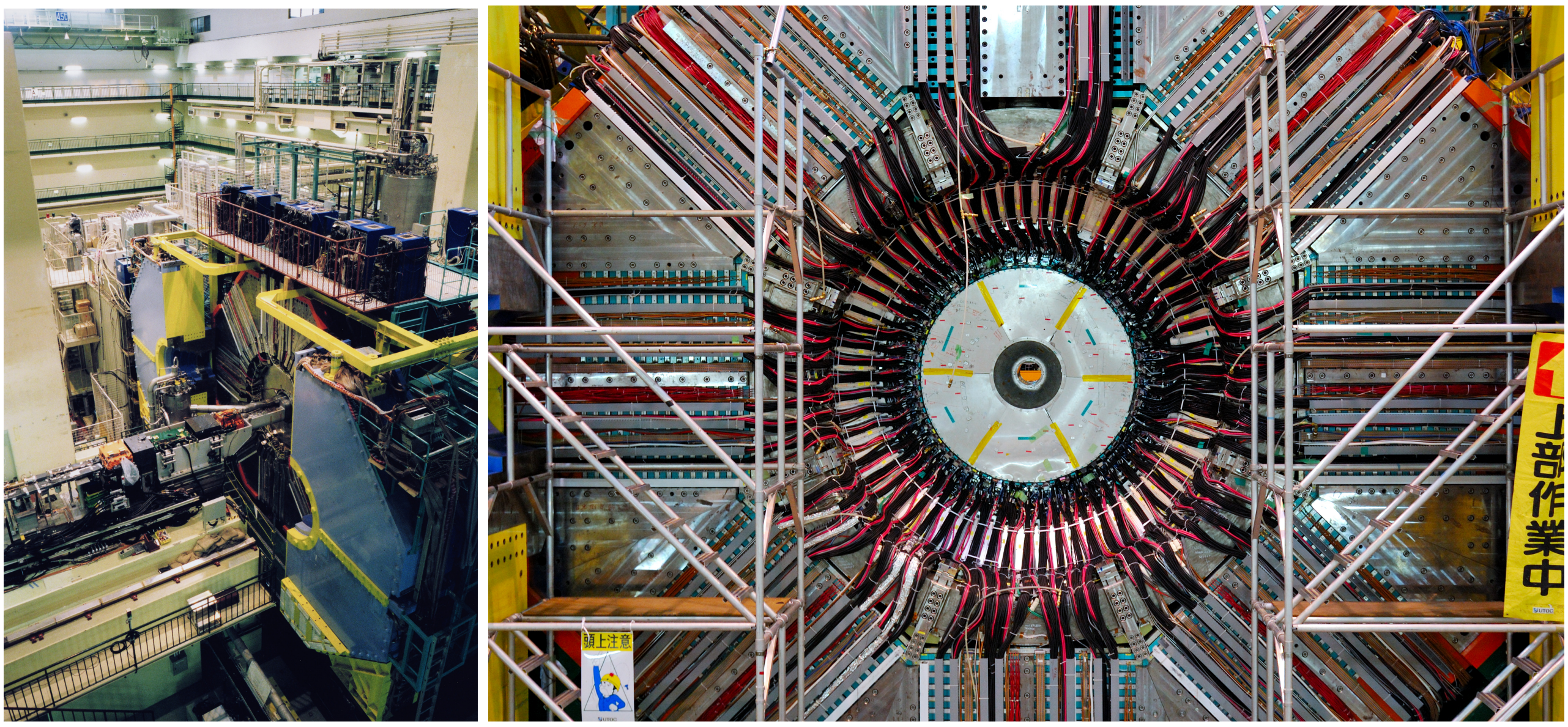

**Figure 2.1.5.** Left: View of the Tsukuba detector hall with the Belle detector. The beamline enters from the bottom left through the detector end cap. Right: Beamline view of the detector. From the outer to the inner part the KLM modules, ECL modules, ACC PMT's and the CDC end flange can be seen (see the text for description of subdetectors).

Obviously the SVT is a very sensitive device which could be damaged by radiation as it is very close to the IP. Damage could come from two effects: either a huge burst of radiation destroying instantaneously some channels, or the integrated dose exceeding the SVT radiation budget and leading to permanent damage. To mitigate such problems, a dedicated system called SVTRAD has been developed: this continuously monitors the radiation levels in the SVT and can either temporarily inhibit the injection or even force a beam abort if the instantaneous dose is deemed to be too high. More information about the SVTRAD system can be found in Section 2.2.8 below.

During the whole data taking period, the SVT performance was constantly monitored while studies were done regularly to predict future performance based on the expected increase of the beam currents and of the luminosity. The main effects of the evolving running conditions to the SVT were twofold: occupancy-induced damage and radiation damage. While the former is an instantaneous effect which can be mitigated by limiting the occupancy in the most affected layers, the latter gets integrated over time. Both the modules and the front-end electronics suffer from this degradation. There is no way to recover the lost performance, except by replacing any damaged components – which was not attempted on the SVT. The consequences of these effects are the reduction of the collected charge and the increase of the noise. Both effects limit the SVT performance and have been taken into account to define the operating mode of this sub-system.

Over the nine years of operation, the average efficiency of the SVT modules (computed for each half-module by dividing the number of hits associated to tracks with the number of tracks crossing that particular module) was above 95%, excluding a few percent of defective half-modules. Some half-modules had issues with individual channels; however, these had no significant impact on the overall efficiency as usually two or more strips are used to detect charge in a given layer crossed by a charged particle. The $z$ and $r\phi$ resolutions range from $\sim$15 to $\sim$40 $\mu$m depending on the layer and on the measured quantity. The best results are obtained for tracks with a polar angle close to 90° while resolution degrades slowly in the forward and backward directions. Measurements of $\mathrm{d}E/\mathrm{d}x$ allow the SVT to achieve a $2\sigma$ separation between kaons and pions up to a momentum of $500\,\mathrm{MeV}/c$.

*Belle*

The Belle SVD has been improved step by step after the commissioning of the Belle detector in 1999. In the first 3 years, the first system, called SVD1, which consisted of 3 layers of AC coupled double-sided silicon-strip detectors (DSSD) read out with VA1 readout chip (Gamma-Medica, 1999), was used. As SVD1 was the first silicon vertex detector built at KEK, a conservative design was chosen. Its coverage was $23° < \theta < 140°$ while the full acceptance of the Belle detector was $17° < \theta < 150°$. The limited radiation hardness of the VA1 chip AMS 1.2 $\mu$m (200 krad) and its long shaping time (2.8 $\mu$sec) discouraged aggressive operation of the KEKB collider. In addition, since the Belle readout electronics were set to the ground level, and the bias voltage was applied across the dielectric in the coupling capacitor of the DSSD, a few pinholes appeared in the dielectric each year.

Because of these problems, the Belle collaboration started the upgrade of the SVD before the start of KEKB operation. In 2000, all SVD ladders were replaced utilizing an upgraded VA1 AMS 0.8 $\mu$m chip (Aihara, 2000b) whose radiation tolerance improved to 1 Mrad.

A major upgrade was done in summer 2003. The second generation silicon vertex detector, SVD2 (Natkaniec, 2006), consisting of 4 layers of DSSD and covering the full angular acceptance ($17^\circ < \theta < 150^\circ$), was installed (Fig. 2.2.2). The inner radius of the beam pipe was reduced from 20 mm to 15 mm (Abe, 2004i). The radii of the SVD2 layers are 20 mm, 44 mm, 70 mm and 88 mm. As the KEKB luminosity increased after SVD2 was installed, 85 % of Belle data were taken with SVD2.

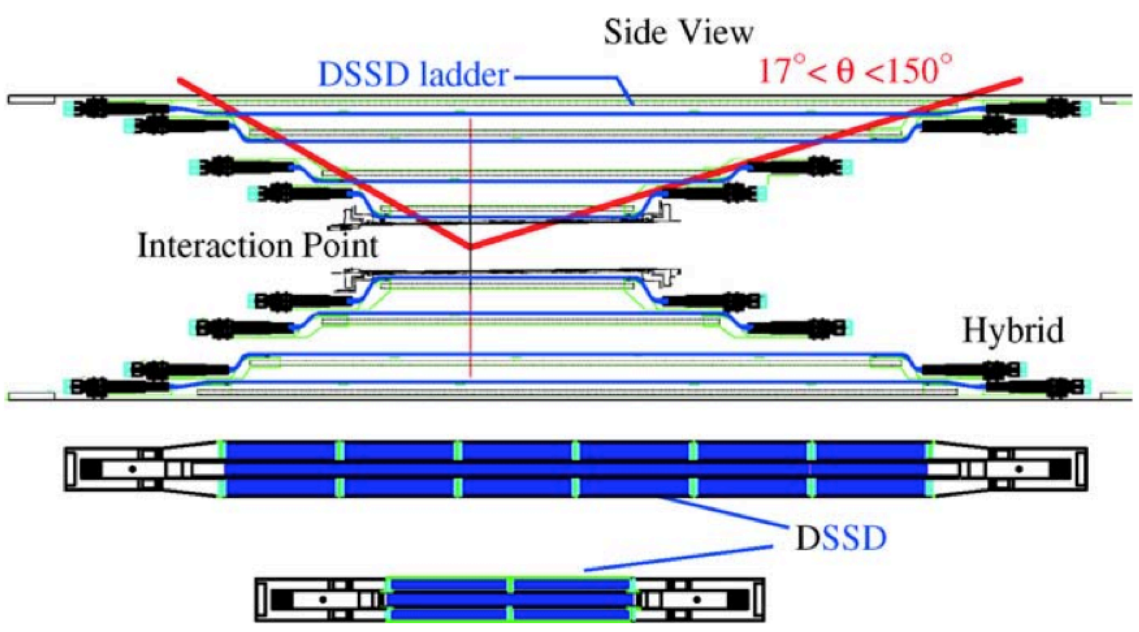

**Figure 2.2.2.** The longitudinal cross section of Belle's SVD2 (Natkaniec, 2006). The layer 1 and layer 4 ladders are also depicted. The radii of layers 1 to 4 are 20, 44, 70 and 88 mm, respectively. SVD2 covers the whole Belle acceptance ($17^\circ < \theta < 150^\circ$) shown by dashed lines.

SVD2 also utilized a newly-developed chip, VA1TA, which had a 0.8 $\mu$sec peaking time and a radiation tolerance of 20 Mrad (AMS 0.35 $\mu$m technology) (Yokoyama, 2001). The control register was made of triple-module-redundancy logic to avoid and detect single-event upsets (SEUs). Thanks to the short shaping time, the contribution of the dark current to the overall noise was not substantial. The voltage from the low-voltage power supply was increased to be above the bias voltage and the rate of pinhole appearance was reduced dramatically. SVD2 was operated for eight years without major problems.

The material in front of the CDC innermost layer is the beam pipe (0.62% $X_0$), four layers of strip sensors (1.71% $X_0$), the SVD CFRP (carbon fiber reinforced polymer) cover (0.23% $X_0$) and the CDC inner CFRP cylinder (0.17% $X_0$) totaling 2.73% $X_0$. The SVD sensor alignment is done among DSSDs (internal) and with respect to the CDC (global). Both internal and global alignment parameters are determined for every KEKB run period. No significant change in alignment parameters was observed throughout the experiment.

The impact parameter resolution in $r$-$\phi$ and $r$-$Z$ was measured to be $\sigma_r = 21.9 \oplus 35.5/p$ $\mu$m and $\sigma_Z = 27.8 \oplus 31.9/p$ $\mu$m, respectively, where $p$ represents the track momentum in GeV/$c$ and the $\oplus$ sign denotes summation in quadrature (Abe, 2004h).

The hit occupancy in the inner most layer remained in the range 5-7% at the highest luminosity of $2{\times}10^{34}/\,\text{s}/\,\text{cm}^2$ without degradation of the detector performance.

There is an important difference in the positioning of the silicon detector and hence its role as a part of the tracking system between *BABAR* and Belle. In the case of *BABAR* the SVT is installed inside a support tube. As a result, the innermost radius of DCH is 236 mm and the radius of the outermost layer of the SVT is 140 mm. Therefore, efficient low-momentum track-reconstruction capability of the SVT was required and the 5-layer design was a natural choice. In the case of Belle, the SVD is supported by the CDC, with the radii of the outermost SVD layer and the innermost CDC layer being 90mm and 110mm, respectively. The reconstruction of low $p_t$ tracks can be done by the CDC. Thus, the main purpose of the Belle SVD is to extrapolate the tracks reconstructed in the CDC to the decay vertices inside the beam pipe. The reconstruction of low $p_T$ tracks with the CDC is efficient down to 70 MeV/$c$ (Dungel, 2007).

### 2.2.2 Drift chamber

*BABAR*

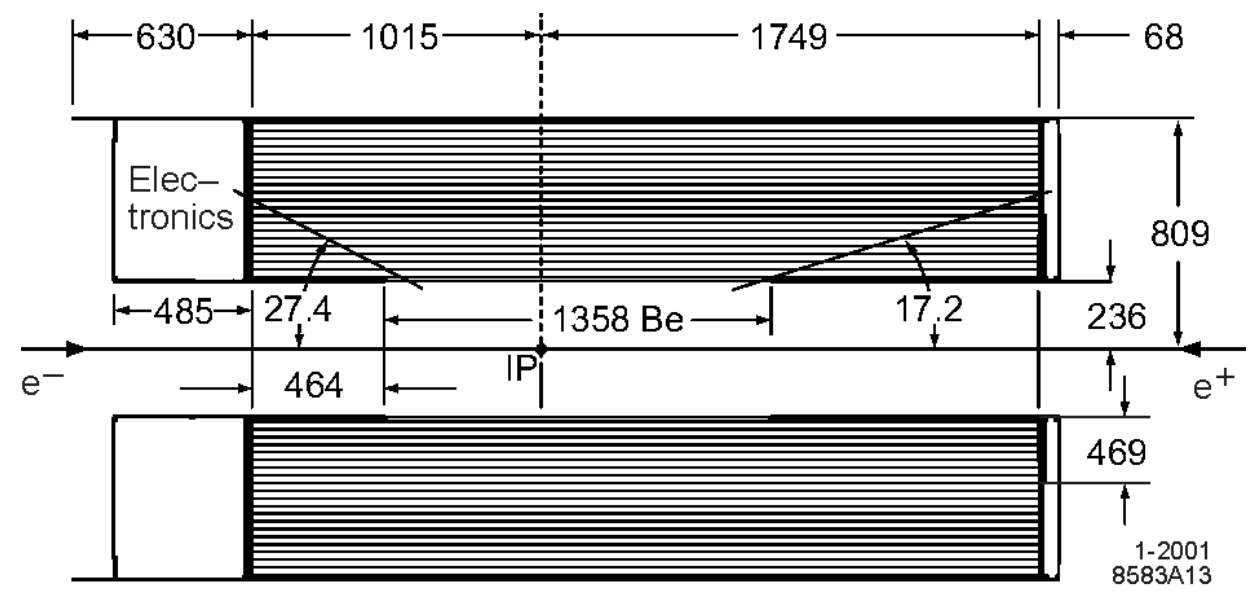

**Figure 2.2.3.** Longitudinal section of the *BABAR* DCH (Aubert, 2002j) with the principal dimensions given in millimeters. Like the whole *BABAR* detector, the 40-layer drift chamber is offset by 370 mm from the IP. The electronics are located behind the backward end plate. The DCH coverage, defined by requiring that at least half of the layers are traversed by the tracks, extends from 17.2° to 152.6° in polar angle.

Figure 2.2.3 shows a longitudinal section of the *BABAR* DCH which performs both the tracking and part of the PID for charged particles – the latter is possible thanks to measurements of track ionization losses (d$E$/d$x$). Indeed, low momentum tracks do not reach the DIRC and so only the tracking system can help identify them. Moreover, the DIRC only covers the *BABAR* barrel section which means that the DCH is the only detector available to perform PID on the forward side of *BABAR*. The DCH is also a key component of the L1-trigger level. The DCH readout electronics, mounted on the backward end plate of the chamber, were upgraded in 2004-2005 to cope with the trigger rate increase associated with the increase of the PEP-II luminosity and with the corresponding increase of background. In particular, the new readout boards included FPGAs responsible for performing the feature extraction step (extraction of physical signals from the raw data; gain and pedestal corrections; data sparsification and data formatting) prior to transferring the data from the front-end boards to the DAQ modules. Previously, feature extraction was performed in the DAQ modules. These new chips were sensitive to SEUs occurring at a rate of a few per day in the whole DCH electronics. Therefore, a dedicated system was set up to monitor the behavior of the new DCH front-end boards and to reload in a few seconds the chip firmware, should errors be detected.

The DCH counts 40 layers of small hexagonal cells of which 24 are placed at small stereo angles (about 50-70 mrad) to provide $z$ information. The field wires are made of aluminum and the gas mixture is 80:20 Helium:Isobutane in order to minimize multiple scattering inside the DCH (the material inside the chamber only counts for 0.2% $X_0$). The 40 layers are gathered in 10 'superlayers' in which all layers have the same orientation. Labeling 'A' an axial DCH superlayer (which stereo angle is null), 'U' a superlayer with positive stereo angle and 'V' a superlayer with negative stereo angle, the pattern of the *BABAR* DCH can be written: 'AUVAUVAUVA'. This particular alternation optimizes the performance and the reliability of the DCH.

Like the SVT, the DCH is a delicate system which must be monitored continuously and carefully to detect any unsafe condition and mitigate it in the appropriate way. Particular examples of monitoring (with hardware and software systems) included the DCH gas mixture composition and potential gas leakage, and the high-voltage (HV) settings of each group of wires. The monitoring systems were continuously improved over the years to minimize the dead time of the DCH without bypassing safety requirements. In the final implementation, if the current of a given channel was found to be too high, the corresponding voltage was reduced until the current fell below a safe threshold, at which point the HV would be ramped up again. During this process, all the other HV settings were unchanged, allowing data taking to proceed. In addition, a real time software process was able to predict the DCH current during running, using several monitoring variables that were independent (beam currents, various background levels readout by sensors, etc.). In this way, the DCH would only switch from the injectable voltage level to the running one if the beam conditions were good enough to ensure a safe operation of the chamber when it would reach its working point. Apart from a small number of wires which were damaged by a HV incident during the *BABAR* commissioning phase, the whole DCH worked well during the whole data taking period. The DCH nominal HV was regularly raised during the data taking to correct for gain losses due to aging: while the nominal HV level was 1960 V, the initial setting was 1900 V; by the end of data taking, it had been raised to 1945 V – one volt corresponds to about 1% on the gain. Loss of gain due to wire aging was 11% over the life of the chamber. The DCH performed as expected during all the *BABAR* data taking, both as the main component of the tracking system and as an important contributor to *BABAR* PID, with a measured $\mathrm{d}E/\mathrm{d}x$ resolution of about 8%, close to the design value of 7%.

### *Belle*

The Belle Central Drift Chamber (CDC) plays several important roles. First, it reconstructs charged particle tracks, precisely measures their hit coordinates in the detector volume, and enables reconstruction of their momenta. Second, it provides particle identification information using measurements of $dE/dx$ within its gas volume. Low-momentum tracks, which do not reach the particle identification system, can be identified using the CDC alone. Finally, it provides efficient and reliable trigger signals for charged particles.

Since the majority of the particles in $B$ meson decays have momenta lower than 1 GeV/$c$, minimization of multiple scattering is important for improving the momentum resolution. Therefore, a gas mixture of 50% He and 50% $C_2H_6$ was chosen, which, because of the low Z nature of the gases, provided optimal momentum resolution while retaining good energy loss resolution.

The structure of the CDC is shown in Fig. 2.2.4. It is asymmetric in the $z$ direction with an angular coverage of $17^\circ \leq \theta \leq 150^\circ$ and has a maximum wire length of 2400 mm. The inner radius of the CDC lies at 80mm, and the detector has no inner wall in order to minimize multiple scattering in the material that lies within the radius of the first wire layer and to ensure good tracking efficiency for low-$p_t$ tracks. The outer radius is 880 mm. In the forward and backward directions at small $r$, the CDC has the shape of a truncated cone. This allows for the necessary space to accommodate the accelerator components while keeping the maximum available acceptance. The chamber has 50 cylindrical layers, each containing between three and six either axial or small-angle stereo layers, and three cathode strip layers. The CDC has total of 8400 drift cells. The two innermost super-layers are composed of three layers each and the three outer stereo super-layers are composed of four layers each. When combined with the cathode strips, this provides a high-efficiency fast $z$-trigger. For each stereo super-layer, the stereo angle was determined by maximizing the $z$-measurement capability while keeping the gain variations along the wire below 10%. The sense wires are made of gold-plated tungsten and have the diameter of 30 $\mu$m, while the aluminum field shaping wires have the diameter of 126 $\mu$m.

In all layers, except the three innermost, the maximum drift distance is between 8 mm and 10 mm. In the radial direction the thickness of drift cells ranges from 15.5 mm to 17 mm. In the innermost layers the cells are smaller and signals are read out by cathode strips. Staggering of the neighboring radial layers within a super-layer in the $\phi$ direction by half cell helps in resolving left-right ambiguities.

The CDC read-out electronics consists of Radeka-type pre-amplifiers which amplify the signal and send it to modules performing shaping, discrimination and charge(Q)-to-time(T) conversion. These modules are placed in the electronics hut and are connected to pre-amplifiers via ~30 m long twisted pair cables. The technique used is a simple extension of the ordinary TDC/ADC readout scheme, but allows Belle to measure both, timing and charge of the signals, using multi-hit TDC's only.

In summer 2003, the cathode part, which corresponds to the inner most three layers, was replaced with a new chamber in order to provide space for SVD2. The new chamber consists of two layers with smaller cells about 5 mm × 5 mm due to limited space and reducing the occupancy. The maximum drift time becomes shorter; less than 100 nsec in the 1.5 T magnetic field.

The high voltage applied to the sense wires was kept for 11-years of operation without serious radiation damage. After detailed alignment and calibration, the overall spatial resolution is around 130 $\mu$m, as expected. The tracking system consisting of the SVD and CDC provides rather good momentum resolution, especially for low-momentum tracks thanks to the minimization of material inside the inner radius of the CDC:

$\sigma p_T/p_T = 0.0019 p_T \oplus 0.0030/\beta$ [$p_T$ : GeV/$c$].

The resolution on $dE/dx$, which is important for PID, was 7% for minimum-ionizing particles. The $r-\phi$ trigger of the CDC provides a highly efficient and reliable trigger signal. The $z$ trigger that uses the cathode strips works well in reducing the rate of the charged trigger by a factor of three without sacrificing any physics events.

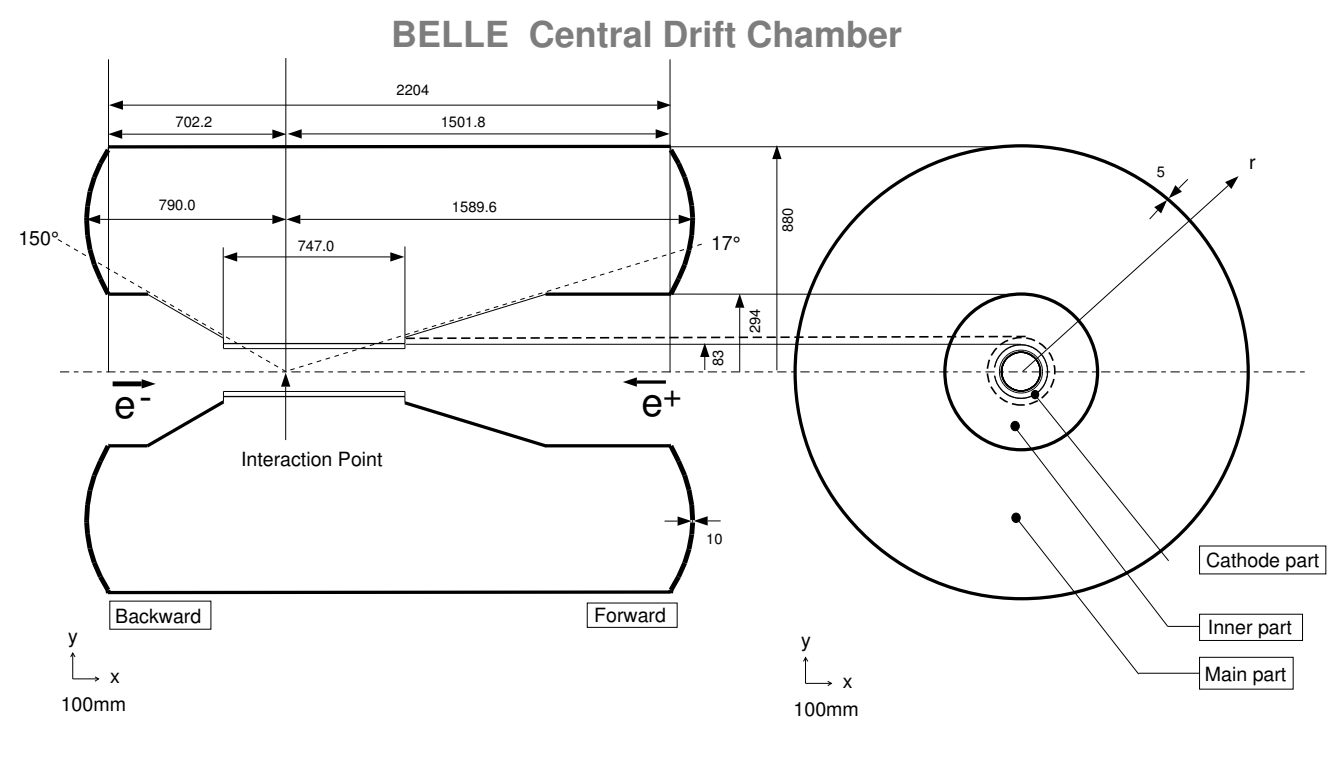

**Figure 2.2.4.** Belle CDC structure.

## 2.2.3 Charged particle identification

Principles of the charged particle identification and their technological realization used in both detectors are described in the following. Readers interested mainly in the methods and performance of the PID systems may obtain more details from a separate chapter on charged particle identification, Chapter 5.

### 2.2.3.1 *BABAR*

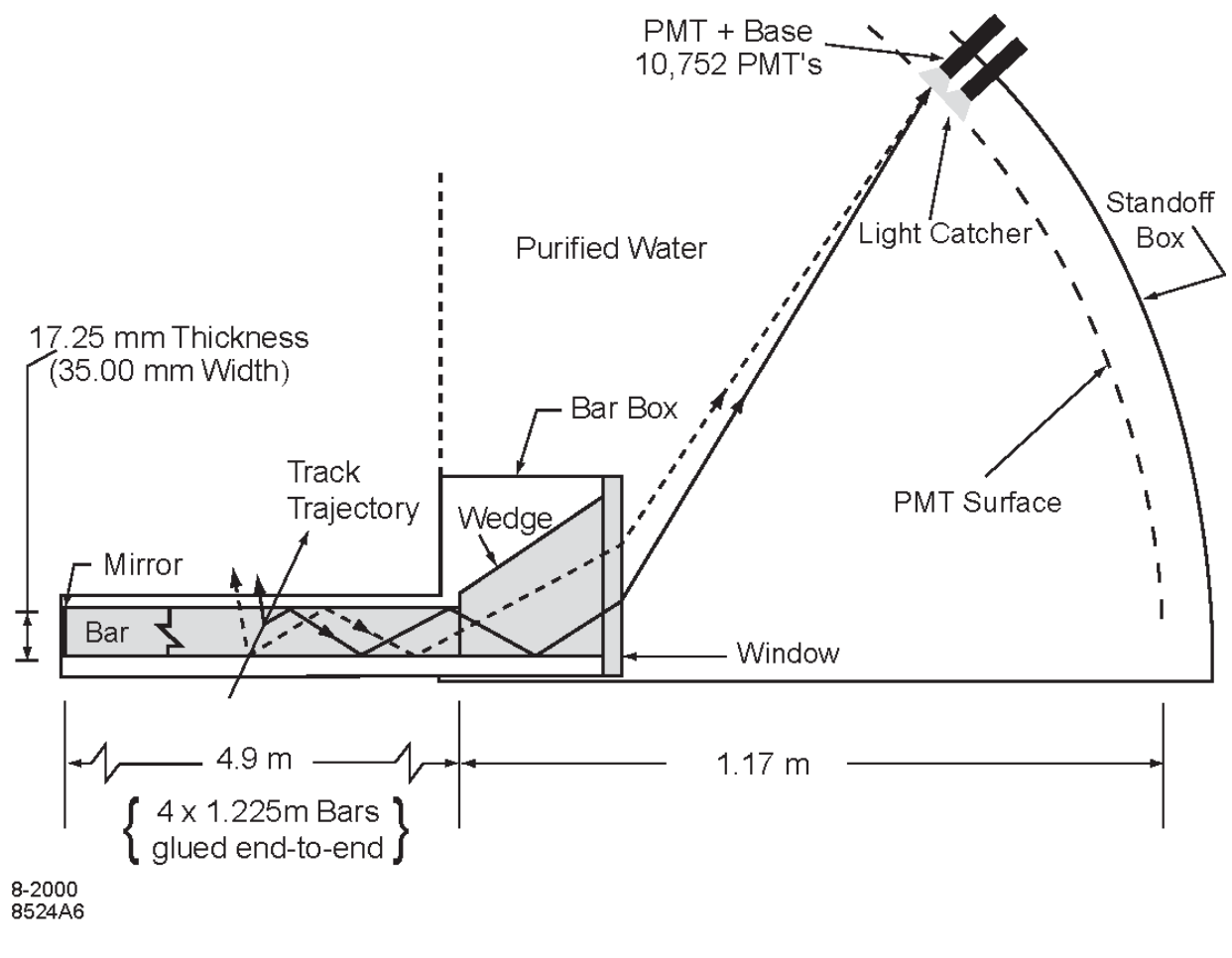

**Figure 2.2.5.** Principle of the *BABAR* DIRC (Aubert, 2002j) – note that this schematic is inverted with respect to the other pictures showing longitudinal sections of *BABAR* or of one of its components: the forward (backward) side of the detector is on the left (right) side of the picture.

Many detectors contribute to the *BABAR* PID system: the SVT and DCH via measurements of the specific energy loss d$E$/d$x$ for charged particles crossing their active area; the EMC for electron identification and the IFR for the muons. But its main component is the DIRC which dominates the $\pi/K$ separation power at high momentum by measuring the emission angle $\theta_C$ of the Cherenkov light produced by a charged particle crossing a quartz bar radiator (see Fig. 2.2.5). The dimension of each quartz bar is 4.9 m × 6 cm$^2$.

Charged tracks crossing a quartz bar at a velocity greater than the speed of light in that medium produce light through the Cherenkov effect. A fraction of these photons propagate to the backward bar end through total internal reflection – the forward bar end is instrumented with a mirror to reflect forward photons backward. Then, they exit the quartz bar through the quartz wedge which reflects them at a large angle with respect to the bar axis. Traveling through the ultra-pure water contained in the SOB, they are finally detected by one of the 10,752 PMTs located about 1.2 m away from the bar end (located beyond the backward end of the magnet). Not only the positions of the detected photons but also their arrival times are used to reconstruct the Cherenkov angle at which they were emitted.

The large water tank of the *BABAR* DIRC was sensitive to backgrounds resulting mainly from neutrons interacting with the $H_2O$ molecules. Moreover, it was a permanent concern as water could leak in the boxes containing the DIRC quartz bars (called 'barboxes') and from there reach other parts of the *BABAR* detector, causing serious and permanent damage to the apparatus. Therefore, the DIRC group had to design a sophisticated system monitoring in real time the humidity outside the SOB and triggering a quick water dump, should a leak be detected. In addition, $N_2$ was continuously flowing in the DIRC barboxes to keep the quartz bars dry – drops of water would have spoiled the quartz optical properties. Lastly, the SOB was full of ultra-pure water (a potential environmental hazard) running in closed circuit and which had to be continuously purified by a dedicated water plant.

The DIRC reconstruction associates PMT hits with charged tracks crossing the quartz bars with a momentum above the Cherenkov threshold. In addition to background hits which can potentially 'hide' the image of the Cherenkov ring on the PMT array, a complication arises from the fact that the actual path of a given photon between its origin, somewhere along the charged particle track in the quartz, and its detection is unknown. For each detected photon, there are 16 ambiguities coming from our ignorance in the number and of the nature of the reflections undergone by the photon in the quartz. Fortunately, most of them can be rejected as un-physical or leading to an inconsistent timing for the hit – the DIRC is truly a 3D-imaging device, which uses both the position and timing information to reconstruct its data. The ambiguities reduce typically to three and are used in the reconstruction algorithms based on an unbinned maximum likelihood formalism – see Chapter 11. Their outputs are usually a likelihood value for each of the five 'stable' charged particle types (e, $\mu$, $\pi$, $K$ and $p$) plus an estimation of the Cherenkov angle $\theta_C$ and of the number of signal and background photons, if enough photons have been found for that particular track. The angle resolutions achieved are typically 10 mrad per photon and 2.5 mrad per track, a level only 10% larger than the DIRC design goal. This is sufficient to separate kaons from pions by more than 4 $\sigma$ at 3 GeV/$c$.

### 2.2.3.2 Belle

Particle identification at Belle, in particular for kaons and pions, is performed by combined information from three detector elements; the time-of-flight detector (TOF), aerogel Cherenkov counter (ACC) and d$E$/d$x$ in the CDC. In this section, brief specifications of two of these detectors (TOF and ACC) are summarized. A description of the CDC is given in Section 2.2.2.

#### *Time-of-flight system*

The time-of-flight (TOF) system consists of a barrel of 128 plastic scintillator counters and can distinguish between kaons and pions for tracks with momenta below 1.2 GeV/$c$. The system is designed to have time resolution of 100 ps for muon tracks (Kichimi, 2000).

One TOF module (the entire system comprises 64 modules) is shown in Figure 2.2.6. Each module consists of two TOF counters and one thin trigger scintillation counter (TSC). Fine-mesh PMTs are attached to the both ends of the TOF counter and the backward end of the TSC counter. The acceptance is 33° - 121° in the laboratory polar angle, and the minimum transverse momentum to reach a TOF counter is 0.28 GeV/$c$. The two-layer configuration of TSC and TOF counters with 1.5 cm air gap removes photon-conversion triggers due to a huge photon background caused by spent particle hits on the beam pipe near the interaction region.

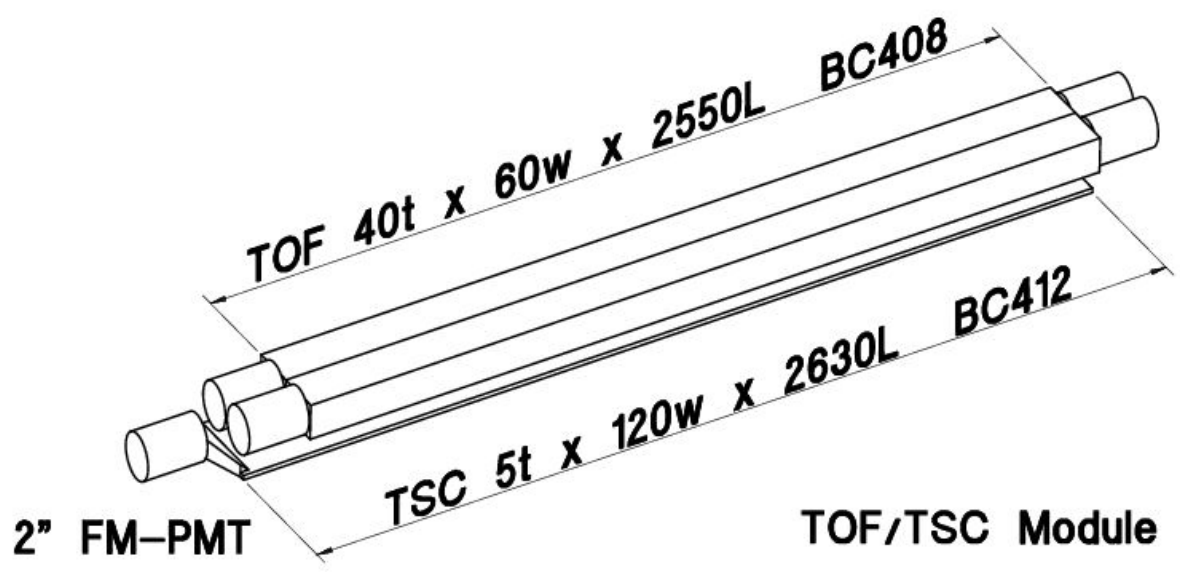

**Figure 2.2.6.** One TOF module consisting of two TOF counters and one TSC counter. The scales are in mm.

The TOF readout system records a set of charges $Q_i$ and timings $T_i$ from the rising edges of discriminator outputs for each PMT signal from the TOF detector. Figure 2.2.7 shows the block diagram of the timing measurement utilizing the Time Stretcher (TS) circuit. The circuit finds the first rising edge $T_2$ of the TS reference clock (reduced radio-frequency - RF - signal of the KEKB accelerator with a frequency of 508.9 MHz) following the rising edge $T_1$ of the TOF signal, and expands the time interval $(T_2 - T_1)$ by a factor of 20, for the timing of the following pulse $(T_3 - T_2)$. These measured times are read out with Belle standard FASTBUS TDCs with a 0.5 ns least significant bit (LSB), providing a 25 ps LSB as a result. A further time-walk correction is applied for timing variation due to a pulse charge, $\Delta T_i \sim 1/\sqrt{Q_i}$.

The TOF system measures time of flight for charged tracks reconstructed by the CDC and requires additionally the beam collision time for each event, $t_{\rm IP}$. It is determined by the RF clock signal used as a reference, and the time offset is calibrated offline on a run-by-run basis using a large sample of $\mu$-pair events ($\gamma\gamma \to \mu^+\mu^-$) with a purity better than 98%. The expected TOF for each muon track is calculated, taking into account its flight length measured by the CDC, and the offset is tuned to give a zero deviation on average between the calculation and the TOF measurement for each PMT.

Determination of the collision timing for TOF measurement has an ambiguity of an integer multiple of 1.96 ns in each event corresponding to the period of the RF clock. This ambiguity can be solved in almost all cases, assigning the velocity of light to high momentum tracks in an event (or, equivalently, assigning the pion mass to the tracks). When the pion-mass assumption fails, the kaon or proton mass is tried.

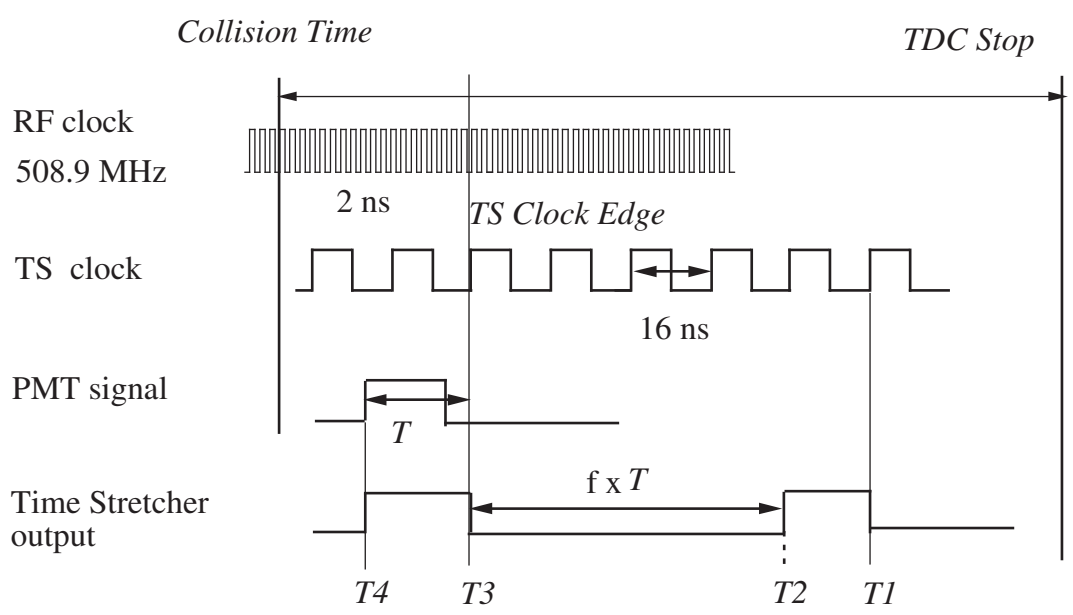

**Figure 2.2.7.** Time Stretcher TDC scheme for Belle's TOF sub-system. The TS reference clock of approximately 8 ns is generated from the KEKB RF signal of 508.9 MHz (Abashian, 2002b).

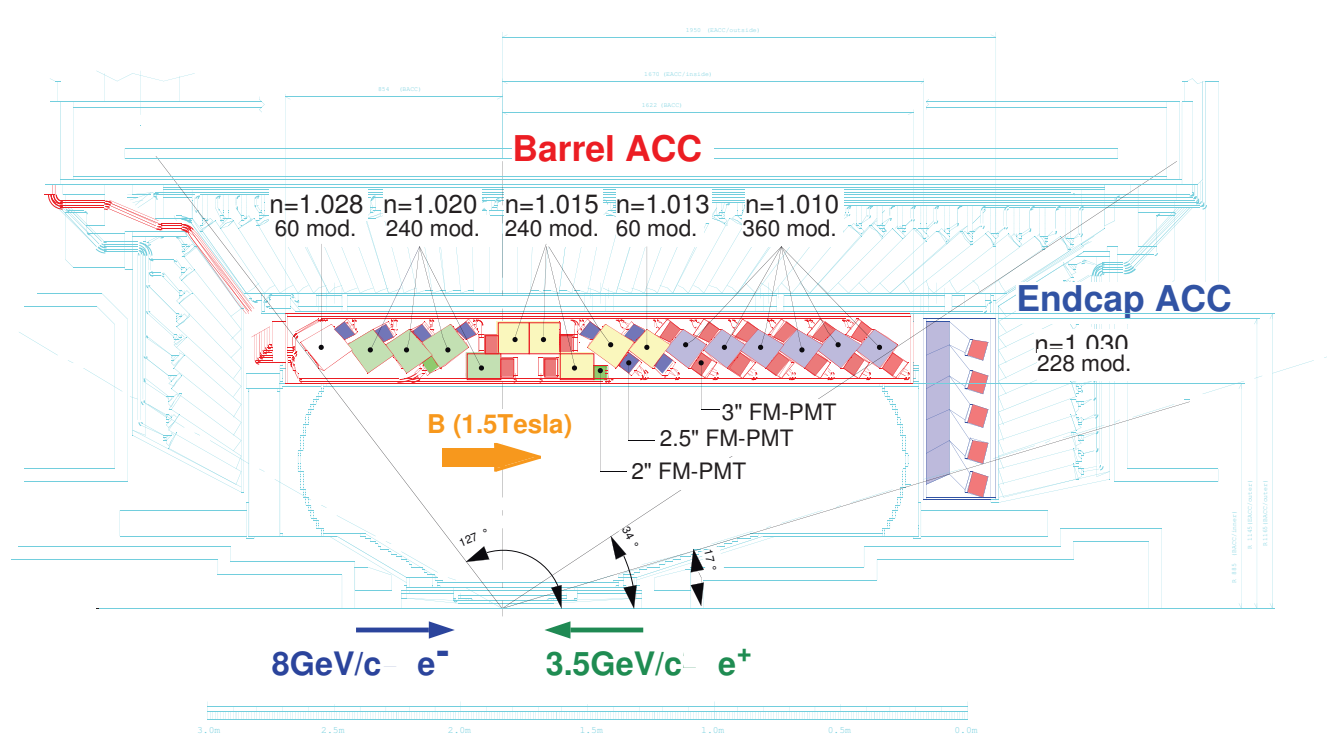

**Figure 2.2.8.** From (Abashian, 2002b). Layout of the ACC system consisting of 16-module lineup for the barrel and 5-layer modules for the end cap regions of the Belle detector.

Long-term variation of the time resolution of the TOF system was monitored using the $\mu$-pair samples. The resolution of 110 ps measured in 2008 (Kichimi, 2010) was degraded from the initial resolution of 96 ps obtained in 1999. The 110 ps resolution includes a systematic error of 40 ps in total from timing jitters in the detector and accelerator electronics, calculation from $\mu$-track information, and the collision position spread due to a beam bunch length. The degradation in timing performance is mainly due to aging, a reduction of the attenuation length and light yield in the TOF scintillation counters over the ten year running period. Pion tracks have a slightly worse average time resolution, typically by 10 ps, due to a nuclear scattering effect.

*Aerogel Cherenkov counters*

Figure 2.2.8 shows the configuration of the Belle Aerogel Cherenkov counter (ACC; Iijima, 2000). The polar angle coverage is $33.3° < \theta < 127.9°$ in the barrel, and $13.6° < \theta < 33.4°$ in the forward endcap. The detector is built from aerogel modules of ten distinct types, varying in refractive index ($n = 1.010, 1.013, 1.015, 1.020, 1.028$, or $1.030$), and in the number (one or two) and size (2-, 2.5-, or 3-inch diameter) of photomultiplier tubes used to detect photons, according to their position in polar angle.

The barrel device consists of 60 identical sectors in the $\phi$ direction, and 16 modules are arranged in each sector. The typical size of one module is approximately $120 \times 120 \times 120$ mm$^3$, occupied with a silica aerogel radiator. The aerogel radiator volume is covered with a white reflector with high reflectivity (larger than 93%); it is supported by a 0.1 mm thick aluminum wall.

Each counter is viewed by one or two fine-mesh PMT(s) to detect Cherenkov light in an axial magnetic field of 1.5 T. The PMT diameters were chosen to be either 2", 2.5", or 3", depending on refractive indices since larger index aerogel generates more photons and the acceptance of a PMT can be smaller as a result.

The end cap device is divided into 12 identical sectors in $\phi$, and each sector contains 19 modules, which are configured to have 5-layer structure in the radial direction. Each counter module contains a $\sim 100 \times 100 \times 100$ mm$^3$ radiator volume followed by an air light-guide, and then one 3" PMT is attached. This module is made of 0.5 mm-thick CFRP to reduce material while remaining rigid. The CFRP inner wall is covered with the same white reflector as used for the barrel. As there is no TOF coverage in the endcap regions, in order to achieve the required $K$–$\pi$ separation for tracks with momenta $< 1.5\,\mathrm{GeV}/c$, the ACC endcap aerogel system has a refractive index of 1.03.

Output signals are amplified by front-end electronics attached to the PMT backplane and are sent to a charge-to-time conversion circuit and subsequently digitized using a TDC.

The calibration constants for all PMTs are obtained by $\mu$-pair events collected in the beam collisions and daily PMT responses during experiments are monitored by the illuminating LED system, which is installed on all counter modules. The effective number of photoelectrons extracted from LED data as a function of the integrated luminosity for a typical PMT is plotted in Figure 2.2.9. The luminosity range plotted (up to 300 fb$^{-1}$) corresponds to almost 6 years from the beginning of operation. The variation is less than 5% over this period and this stability is found to be sufficient.

### 2.2.4 Electromagnetic calorimeter

*BABAR*

Figure 2.2.10 shows the longitudinal cross-section of the *BABAR* EMC. Its polar angle coverage ranges from 15.8° to 141.8° which corresponds to around 90% of the solid angle in the center-of-mass system. The cylindrical barrel is divided into 48 rings of 120 CsI(Tl) crystals each while the end cap holds 820 crystals assembled in eight rings. These add up to a total of 6,580 crystals among which only three had their readout chain permanently broken by the end of the data taking period. The penetrating particles – in particular electrons and photons – initiate showers in crystals and cause the CsI to scintillate; the amount of light depends on the energy deposited in the calorimeter by each particle. The crystals are supported at the outer radius to avoid pre-showers (*i.e.* particles producing showers in the material in front of the calorimeter). It is worth noting that the crystals are organized in a quasi-projective geometry: they all point to a position near the IP, offset just enough to avoid the possibility of having particles going completely through non-instrumented gaps of the EMC. The amount of material between the IP and the EMC ranges between 0.3 and 0.6 $X_0$ except for the 3 most forward rings of the forward end cap, which see elements of the beamline and of the SVT readout system. These rings are shadowed by up to 3 $X_0$ and have been mainly included to ensure shower containment close to the end of the calorimeter acceptance.

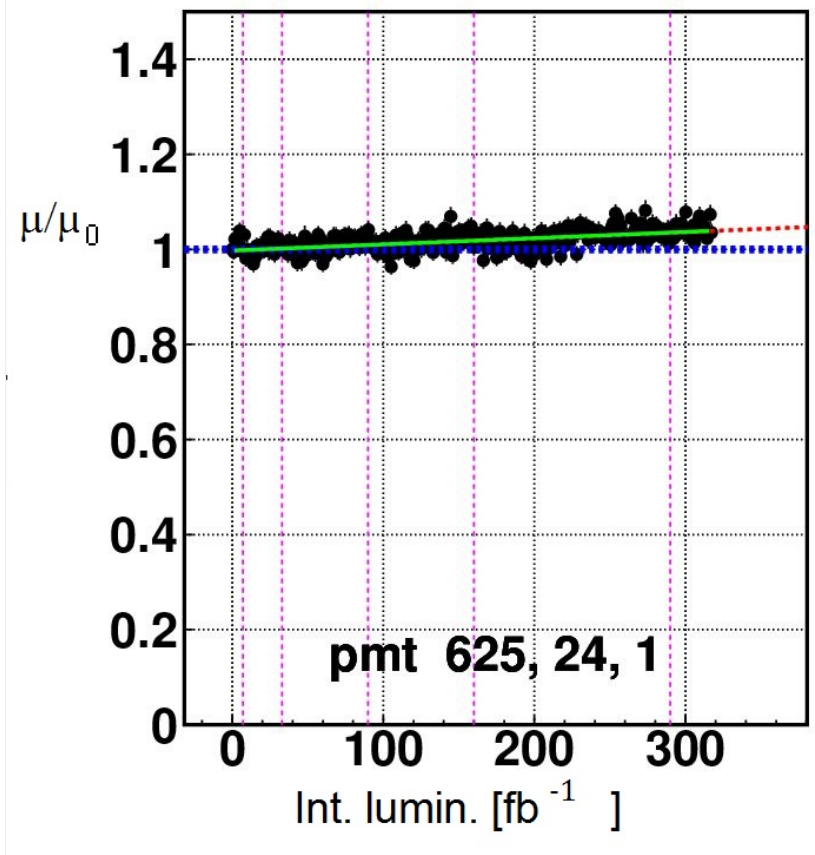

**Figure 2.2.9.** The relative pulse height as a function of integrated luminosity for a typical PMT of ACC.

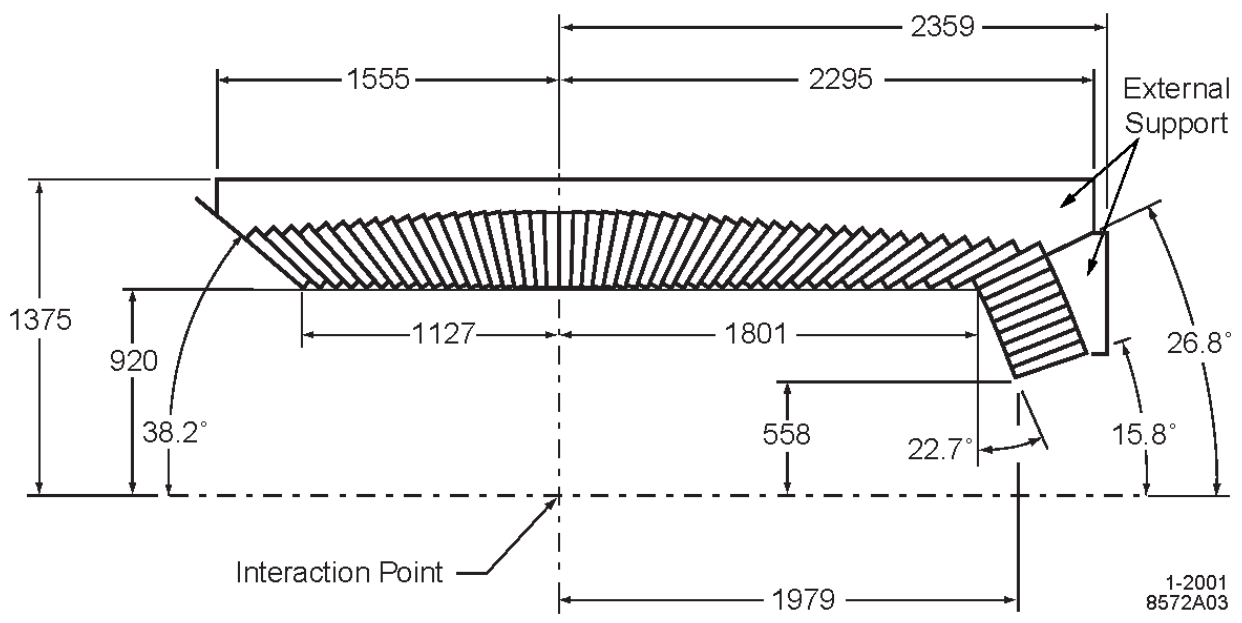

**Figure 2.2.10.** Longitudinal section of the *BABAR* EMC (Aubert, 2002j) showing the arrangement of the 56 CsI(Tl) crystal rings: 48 for the barrel and 8 for the forward end cap. All dimensions quoted on the drawing are in mm.

There are two kinds of calibration for the EMC: a low-energy calibration using a 6.13 MeV radioactive photon source (fluorinert irradiated by neutrons) and a high-energy calibration using reconstructed Bhabha events. The source (Bhabha) calibration was performed about once every 1-2 weeks (a few times a year). In addition, a light pulser was used to monitor the light response of each individual crystal on a daily basis in order to identify potential problematic areas. The radiation dose received by the EMC over the years of data taking had no significant impact on its performance.

The EMC energy resolution $\sigma_E/E$ varies from 5% at 6.13 MeV to about 2% at 7.5 GeV, an energy probed using Bhabha events. The angular resolution is 12 mrad (3 mrad) at low (high) energy. The $\pi^0$ measured mass is in agreement with the PDG value and has a resolution of about 7 MeV/$c^2$. Finally, the EMC provides the main discrimination variable to identify electrons: the ratio $E/p$ of the shower energy to the track momentum – other PID inputs are the DCH $\mathrm{d}E/\mathrm{d}x$ and the $\theta_C$ value measured by the DIRC. The electron identification probability is around 90% on average with a pion contamination of $15-30$%, depending on the track momentum and polar angle.

### *Belle*

The overall configuration of the Belle calorimeter, ECL, is shown in Figure 2.2.11.

The ECL consists of a barrel section and two end caps of segmented arrays of CsI(Tl) crystals. The former part is 3.0 m long and has an inner radius of 1.25 m. The end caps are located at $z = +2.0$ m and $z = -1.0$ m. The ECL is composed of 8736 CsI(Tl) crystals in total. The scintillation light produced by particles in the crystals is detected with silicon photodiodes.

Each crystal has a tower-like shape and points almost to the interaction point. The crystals are tilted by a small angle in the $\theta$ and $\phi$ directions to prevent photons escaping through the gaps between the crystals. The angular coverage of the ECL is $17.0° < \theta < 150.0°$ (total solid-angle coverage of 91% of $4\pi$). Small gaps are left intentionally between the barrel and end cap crystals providing the necessary space for cables and supporting parts of the inner detector (these gaps result in a loss of acceptance at the level of 3%).

The amount of material in front of the ECL ranges between 0.3 to 0.8 $X_0$.

The calorimeter is calibrated using Bhabha scattering and $e^+e^- \to \gamma\gamma$ events. For the two innermost layers of crystals in the forward and backward end caps, cosmic ray interactions are used for calibration. The Bhabha calibration is performed once every 1-2 months. The electronic channel transition coefficients are monitored every day with a test pulse generator.

The radiation dose received by the ECL varies from 100 rads for barrel crystals to about 700 rads for forward end cap crystals. The degradation of the light output due to the overall dose was less than 5% and had no significant impact on ECL performance.

The ECL energy resolution varies from 4% at 100 MeV to about 1.6% at 8 GeV. The angular resolution is about 13 mrad (3 mrad) at low (high) energies. Such an energy and angular resolution provides a $\pi^0$ mass resolution of about 4.5 MeV/$c^2$. The ECL provides the main parameter for electron/hadron separation: the ratio $E/p$ of the shower energy to the track momentum.

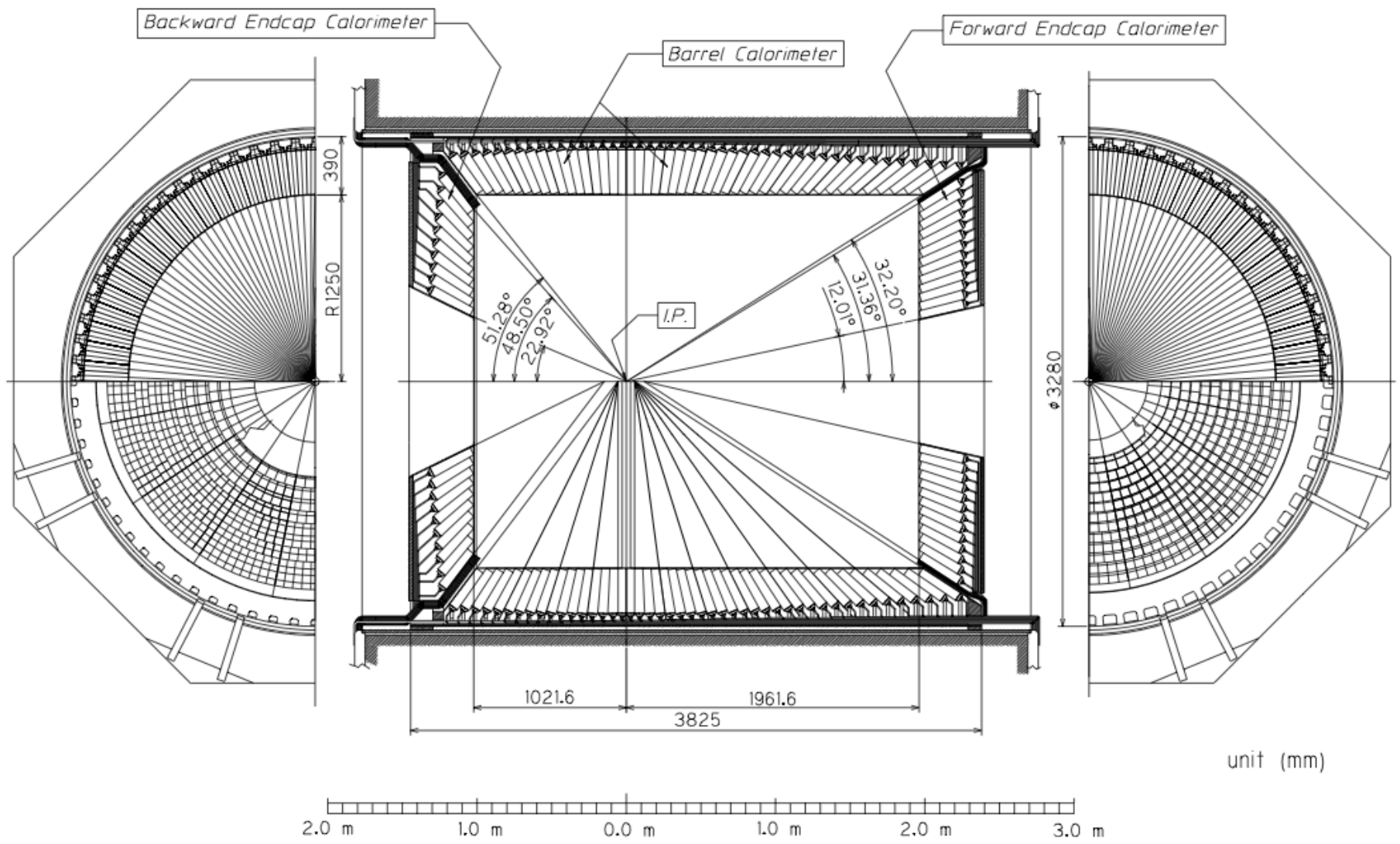

**Figure 2.2.11.** From (Abashian, 2002b). Overall configuration of the Belle ECL.

In addition the ECL is used to provide the Belle online luminosity monitoring system. The rate of Bhabha events is measured using geometrical coincidences of high energy deposits in the forward and backward ECL. This system provides a stable accurate luminosity measurement during an experimental run as well as during injection periods.

### 2.2.5 Muon detector

*BABAR*

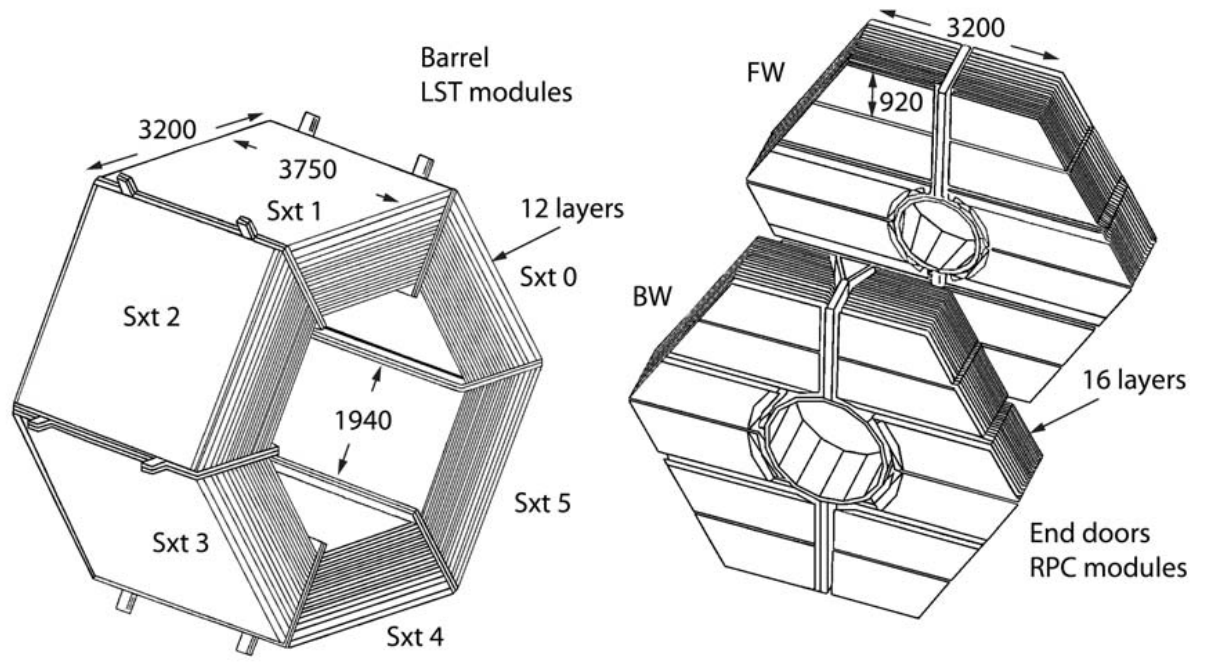

**Figure 2.2.12.** Overview of the *BaBar* IFR at the end of the data taking period (Aubert, 2013): the barrel sextants made of 12 LST layers are visible in the left picture while the forward and backward end doors appear on the right. The forward RPCs (16 layers) have all been changed whereas the backward ones are still the original detectors.

The final layout of the *BABAR* IFR – with, in particular, LST modules in all sextants of the barrel region – is shown in Figure 2.2.12.

The steel of the magnet flux return is finely segmented into 18 plates of increasing thickness: from 2 cm for the nine inner plates to 10 cm for the outermost ones. When data taking started, the *BABAR* IFR was instrumented with more than 800 RPCs, organized in 19 layers in the barrel region (divided itself into six sextants) and 18 in the end doors. These detectors quickly showed serious aging problems (Anulli, 2002, 2003; Piccolo, 2002, 2003) and the deterioration of their performance lead directly to a reduction of the *BABAR* muon identification capability. Overall, 6-17% of the muons were lost due to problems in the IFR. Although several attempts were made to fix the RPCs and to limit the rate of degradation, it was finally decided to replace most of these detectors. This was by far the largest *BABAR* upgrade and it was successfully completed in a 4-year period in various steps.

The RPCs in the backward end cap were never replaced. Due to the boost, they had low rates and covered a small solid angle. In 2002, more than 200 new RPCs were installed in the forward end cap (Anulli, 2005a). Their performance was significantly improved with respect to the original RPCs (Anulli, 2005b). These detectors nevertheless required constant maintenance and upgrades (Band, 2006; Ferroni, 2009) until the end of the data taking, in order to maintain their efficiency and their reliability while the luminosity was increasing. In particular, the chambers with the highest rates were operated in avalanche mode from 2006.

The first two barrel sextants were replaced during the summer 2004 shutdown, only one and a half years after the decision to proceed with this upgrade had been taken. An extensive review process lead to the choice of the Limited Streamer Tube (LST) (Andreotti, 2003) technology to replace the existing RPCs. The procedure consisted of replacing 12 RPC layers by LSTs and to fill the remaining gaps with brass – the outermost layer (#19) could not be instrumented due to a geometrical interference. Increasing the total absorber thickness allowed the improvement of the pion rejection of the muon PID algorithms. The last four barrel sextants were replaced during the fall 2006 shutdown.

The LST efficiency was measured using di-muon events. On average, it was 88% at the end of the data taking, slightly below the geometrical acceptance of 92%. The difference was mainly due to a few misfunctioning or broken channels.

*Belle*

The muon and $K_L$ detector subsystem of Belle identifies $K_L$ mesons and muons above 600 MeV/$c$ with high efficiency. The barrel-shaped region around the interaction point covers a polar angular range of 45° to 125° while the forward and backward end caps extend this range to between 20° and 155°.

This system consists of alternating layers of double-gap resistive plate counters and 4.7 cm thick iron plates. There are 15 detector layers and 14 iron layers in the octagonal barrel region and 14 detector layers and 14 iron layers in each end cap. The iron plates provide a total of 3.9 interaction lengths of material (in addition to the 0.8 interaction lengths in the ECL) for a hadron traveling normal to the detector planes. The hadronic shower from a $K_L$ interaction determines its direction (assuming an origin at the $e^+e^-$ interaction point) but not its energy. The range and transverse deflection of a non-showering charged particle discriminates between muons and hadrons ($\pi^\pm$ or $K^\pm$).

The active elements are double-gap glass-electrode RPCs operating in limited streamer mode. Each 2 mm gas gap is sandwiched between float-glass electrodes with a bulk resistivity of $10^{12-13}\,\Omega\cdot$ cm (Figure 2.2.13). The non flammable gas mixture consists of 62% HFC-134a, 30% argon, and 8% butane-silver.[14] An ionizing particle traversing the gap initiates a streamer in the gas that results in a local discharge of the electrodes. This discharge is limited by the high resistivity of the glass and the quenching characteristics of the gas. A discharge in either gas gap induces signals on both of the orthogonal external copper-strip planes. Each ~5 cm wide strip forms a ~50 Ω transmission line with an adjacent ground plane. In the barrel (but not the end caps), a 100 Ω resistor connects the pickup strip to ground at the readout end to minimize reflections; it also reduces the pulse height into the front-end electronics by a factor of two.

The barrel RPCs, made in the US, use 2.4 mm thick float glass (73% $SiO_2$, 14% $Na_2O$, 9% CaO, and 4% trace elements). The end cap RPCs, made in Japan, use 2.0 mm thick float glass (70–74% $SiO_2$, 12–16% $Na_2O$, 6–12% CaO, 0–2% $Al_2O_3$, and 0–4% MgO).

[14] Butane-silver is a mixture of approximately 70% n-butane and 30% iso-butane.

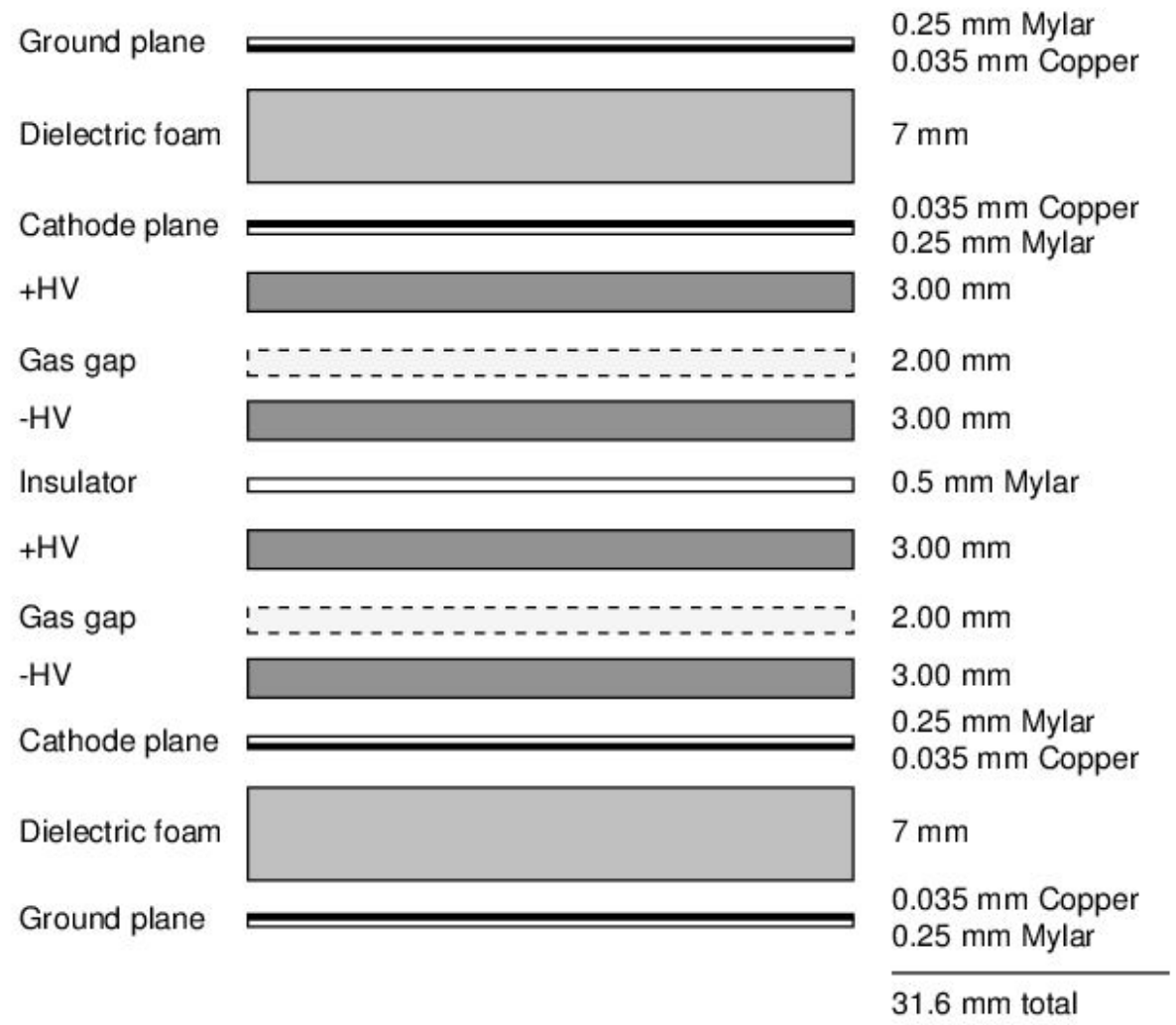

**Figure 2.2.13.** Exploded cross section of a Belle superlayer double-gap RPC module.

The VISyN system by LeCroy (now Universal Voltronics) is used to distribute high voltage, with Model 1458 mainframes and 1468P and 1469N modules. For each RPC, a positive voltage of +4.7 kV (+4.5 kV) is applied to the barrel (end cap) anode plates and −3.5 kV to the cathode plates. Eight (five) anode plates in the barrel (end cap) are driven by a common HV channel while each cathode plane is driven by its own HV channel. The dark current is approximately $\sim 1\,\mu$A/m$^2$ or 5 mA total; most of this flows through the noryl spacers.

Pulses travel from the 38,000 RPC cathode strips along twisted-pair cables, between 3 and 6 meters long, to front-end electronics on the magnet yoke periphery. The typical 100 mV pulse has a FWHM of under 50 ns and a rise time of under 5 ns. The dark rate in a typical detector module is under 0.03 Hz/cm$^2$ with few spurious discharges or after pulses. The signal threshold for discriminating these pulses is 40 mV (70 mV) in the barrel (end caps). The double-gap design results in a superlayer efficiency of over 98% despite the lower (90% to 95%) efficiency of a single RPC layer. Robustness against several failure modes is achieved by having independent gas and high voltage supplies for each RPC layer within a module. Hit position is resolved to about 1.1 cm when either one or two adjacent strips fire, resulting in an angular resolution of under 10 mrad from the interaction point.

The Belle RPCs have performed reliably and without evidence of failures or physical deterioration throughout Belle's lifespan. However, the RPCs are rate-limited by the glass-electrode resistivity, so the efficiency of the modestly shielded end cap RPCs suffered during high-luminosity operation from soft neutrons produced in beamline structures. This was mitigated by the addition of external polyethylene shielding outside the end caps in Belle's later years, but more such shielding would have been needed to eliminate the efficiency drop.

### 2.2.6 Trigger

*BABAR*

As already discussed above, the *BABAR* trigger is implemented as a two-level hierarchy, with the L1 (hardware) followed by the L3 (software). Its combined efficiency at the $\Upsilon(4S)$ resonance energy matches its requirements: more than 99% for $B\overline{B}$ decays, more than 95% for continuum decays ($u\overline{u}$, $d\overline{d}$, $s\overline{s}$ $c\overline{c}$) and still around 92% for $\tau\tau$ events. This trigger was very flexible, as illustrated by the quick and complex modifications of the L3 trigger lines implemented during the last few months of the *BABAR* running, when data were taken at the $\Upsilon(2S)$ and $\Upsilon(3S)$ resonances and a final energy scan above the $\Upsilon(4S)$ was performed. It was also robust against background: trigger rates much higher than the design values for both L1 and L3 were achieved as luminosity was increasing, while the dead-time remained relatively constant, around the 1% design value.

The *BABAR* L1-trigger uses information coming from the DCH for charged tracks, from showers in EMC and from the IFR. The corresponding first two triggers – Drift Chamber Trigger (DCT) and ElectroMagnetic Trigger (EMT) – fulfill all trigger requirements independently and are highly redundant, which boosts the global L1 efficiency and allows one to measure the efficiency of these components using data. Originally, the DCT only provided $r$ and $\phi$ information; in 2005, 3D-tracking was implemented in L1 to add $z$-information which allowed one to reject background events (scattered beam-gas particles hitting the beam pipe) where tracks were produced tens of centimeters away from the IP. This upgrade gave the system more headroom to follow the increases of luminosity and background without generating a significant dead time, especially during the final period of data taking. The third L1 input trigger, the IFR Trigger (IFT), is mostly used for tests: IFR plateau measurements, cosmics trigger, etc. Some work was required after the IFR barrel upgrade to align in time the RPC and LST signals, the latter coming in about $0.6\,\mu$s later.

Information coming from the three components described above are received by the GLT which processes all these primitives and sends out some triggers to the central *BABAR* DAQ system. At this stage, a trigger can be masked (for instance if it corresponds to a known temporarily noisy EMC crystal) or prescaled (meaning that not all selected events are registered; in particular, events identified as Bhabha at the trigger level are prescaled). If a valid trigger remains at this stage, the DAQ system issues a *L1 Accept* signal and the entire event is readout.

The *BABAR* L3-trigger refines and augments the L1 selection methods. It has been implemented in such a way that a wide range of algorithms can be used to select events independently of one another. Their logic and their parameters are set in software and these filters have access to the full event to make their decision. First, L3 input lines are defined by using a logical OR of any number of L1 output lines. Then, one or more scripts are executed for each firing L3 input line and return a yes/no flag depending on whether the event passes this step. Finally, L3 output lines are the logical OR of selected L3 script flags; these flags can also be used as vetoes, for instance to reject Bhabha events which would have been accepted otherwise. Thanks to the spare capacity planned for at the time the L3 system was designed, it could log data at a much higher rate than anticipated: close to 800 Hz at the end of the $\Upsilon(3S)$ data taking, to be compared with the initial expectation of 120 Hz.

Moving from the regular $\Upsilon(4S)$ data taking to the $\Upsilon(3S)$ run during which new physics (NP) decays were sought after, the trigger had to identify completely different topologies of events. Indeed, part of the signal decays were containing particles invisible to *BABAR* which would take away a significant fraction of the energy-momentum available for the collision. Whereas $B\overline{B}$ events exhibit large visible energy, high multiplicity or high transverse activity, the decays of interest of the $\Upsilon(3S)$ are characterized by low visible energy and low multiplicity. This new approach was implemented in three successive steps which required the design of new L1 and L3 trigger lines, such as new L3 filters. These updates were done carefully, checking at each step that the trigger rates would not exceed the capabilities of the system. They were successful, allowing the *BABAR* collaboration to collect large datasets at the $\Upsilon(2S)$ and $\Upsilon(3S)$ resonances.

*Belle*

The trigger system of the Belle detector consists of sub-triggers and the global decision logic (GDL) - constituents of the Level 1 (L1) hardware trigger - and of Level 3 (L3) software trigger. The sub-triggers are formed by signals from the CDC, ECL, TOF, and KLM sub-detectors. The GDL receives summary information from each sub-trigger, then makes a logical combination of sub-trigger information to trigger on hadronic ($B\overline{B}$ and continuum) events, Bhabha and $\mu^+\mu^-$ pair events, etc. Three independent triggers are prepared for the hadronic events; they require either three or more charged track candidates, high levels of deposited energy in the ECL (with a veto on the ECL trigger for Bhabha events) or four isolated neutral clusters in the ECL. The L3 software trigger ran on the online computer farm (see Section 2.2.7). Events triggered by L1 as Bhabha, $\mu^+\mu^-$ pairs, two-photon events, cosmic rays or events with high deposited energy in the ECL, bypass the L3 trigger decision. The events triggered by the presence of charged track candidates are passed to the L3 trigger to determine the presence of actual good charged particle tracks, thus reducing the size of the raw data being recorded.

The efficiencies of the L1 triggers for hadronic events can be measured using the redundancy of the three selection requirements mentioned above because they are almost independent. The overall efficiency for hadronic events is estimated to be more than 99%.

At the beginning of the experiment Belle experienced a high trigger rate caused by the beam background. Signals arising from this background caused the trigger rate to be nearly the DAQ upper limit of 200 Hz even while running at very low luminosity. The rate of the two-charged-track trigger was especially high because of the low $p_t$ tracks originating from the beam-nucleus interactions. To reduce such a high trigger rate, the requirement of coincidence with outer sub-triggers, such as a TOF hit and/or an ECL isolated cluster, was added. Figure 2.2.14 shows the average trigger rate as a function of the experiment number.[15] The green curve shows the average total current of KEKB. The highest total current was 3000 mA around experiment 50. The sudden drop of the total current at experiment 57 was due to the crab cavity installation at KEKB. The red curve shows the average trigger rate. It was as high as 500 Hz around experiment 50, which corresponds to the highest total current and luminosity. In early experiments, high background was indicated by the normalized trigger rate, the blue curve in Figure 2.2.14, defined as the the average trigger rate divided by the average luminosity (called the effective cross section). This rate is normalized to the trigger rate with the luminosity $1\times10^{34}$ cm$^{-2}$ s$^{-1}$. It was higher than 1200 Hz at the beginning of operation, and dropped dramatically as the total current increased (and hence the luminosity increased). After experiment 33 the rate was stable below 400 Hz, which was interpreted as an amelioration of the vacuum around the IP with the higher beam current. In a special run in experiment 47, the luminosity components in the trigger rate were measured to be about 190 Hz in the normalized trigger rate. The noise-to-signal ratio (N/S) was calculated to be about 5.6 in experiment 7, and about 1 or smaller after experiment 37, an indication of the cleaner environment of KEKB operation.

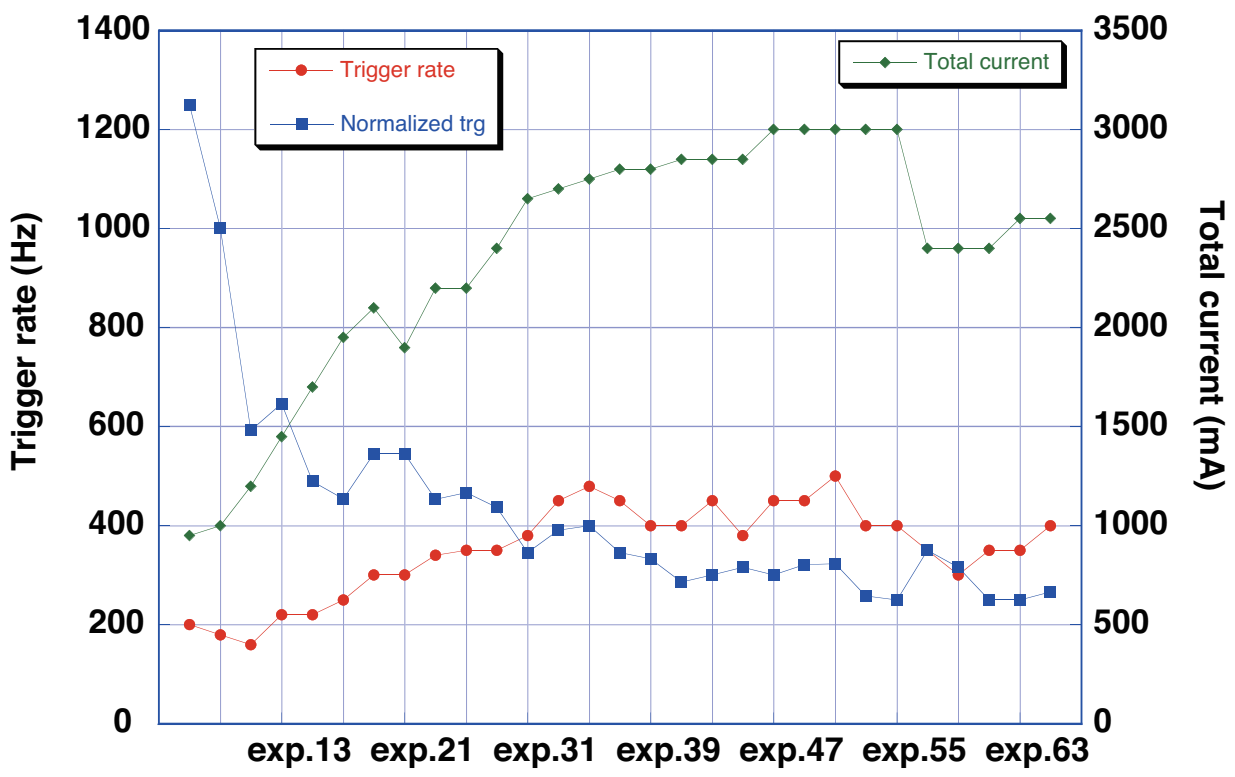

**Figure 2.2.14.** Average trigger rate as a function of the experiment number for Belle.

[15] An extended period of operation is referred to as an experiment within Belle, see Chapter 3. The corresponding nomenclature on *BaBar* is a Run.

### 2.2.7 Online and DAQ

*BABAR*

The high-level design of the *BABAR* online system (Aubert, 2002j) remained unchanged during the whole data taking period. The DAQ chain starts from the common front-end electronics, includes the embedded processors in the readout modules (which start processing the data fragments coming from the detector after a Level 1 accept), the network event builder, the Level 3 trigger and the event logging system. While the design remained constant, the system itself evolved significantly over the time to follow the progress in hardware technology, and to cope with the changes in data taking conditions: higher luminosity, larger backgrounds, longer periods of data taking thanks to the trickle injection mode (see Section 3.2.2 for details), and so on. Several other developments were made with the intent of making the overall system more robust, better performing, and easier to use. For all upgrades, the philosophy was first to maximize the performance of the existing hardware, and only then to plan a hardware upgrade.

With PEP-II operated in trickle injection mode, data taking could occur continuously during one day or more. Therefore special emphasis was put on the data taking efficiency. The aim was to minimize the time spent by the detector in any non-data-taking state (calibration, error recovery, transition from ‘injectable’ mode to ‘runnable’ mode, procedure to begin a new run, etc.). Maximizing the *BABAR* duty cycle required a continuous monitoring of the whole system and attention to detail. While the online system had already been designed to minimize the DAQ dead time, new features were introduced, parts of the system were improved, and procedures modified to increase the detector uptime despite the more challenging environment. One concrete example of this evolution was the reduction of staffing for the detector operation, as the online control and monitoring system was simplified and automated.

Moreover, as explained in Chapter 3, the PEP-II operation in trickle injection mode required developments in the trigger and the DAQ, in order to make the detector insensitive to the background bursts associated with the continuous injection. Dedicated monitoring was added to allow detailed data quality analysis in real time.

The CPUs and the operating system used by the *BABAR* online system evolved over the years, switching from vendor-specific products to commodity systems. This allowed control of the cost of the upgrades of the online system and to provide enough headroom to anticipate the increase of luminosity and background. Most of the online software was written in C++; various scripting languages were used as well, such as Java for graphical tools.

An important evolution of the online system was the replacement of Objectivity-based databases by `Root`-based ones. Several reasons explain this migration, which culminated in 2006 with the decision to stop using Objectivity in *BABAR*. Indeed, there were many concerns regarding the support and the maintenance cost of this software, plus some technical issues. All these changes were carefully planned to make sure they would have no impact on the data taking.

### *Belle*

The original requirement for the Belle Data Acquisition System (Belle DAQ) was to read out event fragments from 8 detector subsystems with a total data size of 40 kbytes at a maximum rate of 500 Hz, and to record the data after event building and data reduction by real time processing.

Figure 2.2.15 shows the configuration of the DAQ system at the beginning of the experiment. The readout system is designed to utilize the unified technology based on the Q-to-T conversion combined with the common FASTBUS multi-hit TDC (LeCroy 1877S), except for the SVD readout. The data are read by the VME processor and collected by the specially-designed event builder, and then processed by the online computer farm equipped with a large number of VME processor modules where high level software triggering is performed. The data are finally sent to the KEK Computer Center via ∼2 km optical fiber links and recorded on digital video tapes.[16]

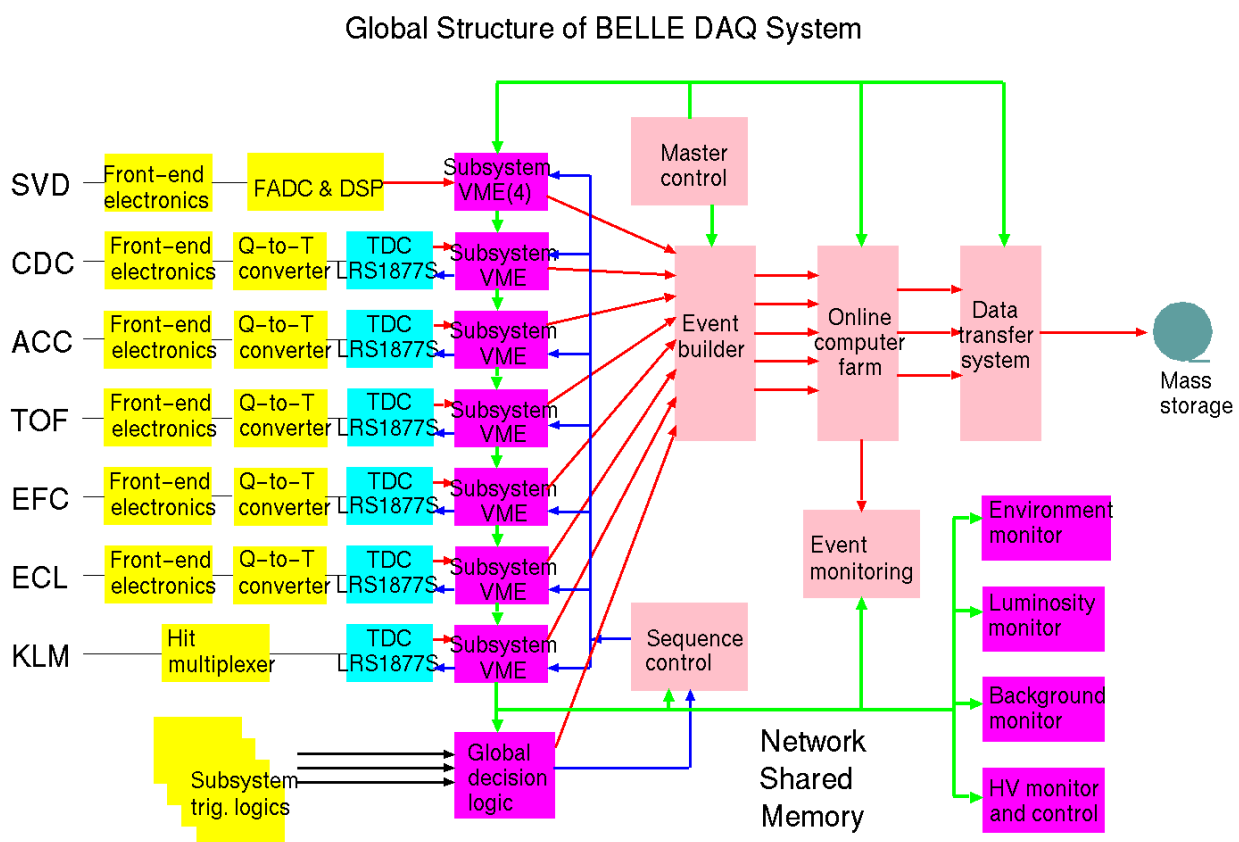

**Figure 2.2.15.** The configuration of the Belle DAQ system at the beginning of the experiment.

However, since the system was implemented using 1990's-era information technology, maintenance of the system was difficult in the long run. In addition, the FASTBUS based readout system is not pipelined and it has a readout dead time of more than 10% at the design maximum trigger rate of 500 Hz. The trigger rate at the beginning of data taking was  200 Hz and the dead time was manageable, but the rate increase was foreseen as the luminosity improves.

Belle started the 'continuous' upgrade of the system to keep up with the luminosity increase. The first step was made in 2001 to replace the event builder and VME based online computer farm with a set of Linux PC servers (EFARM) connected via Fast Ethernet fibers. The level 3 data reduction which was performed in VME processors was ported to the EFARM. The system became more maintainable for a longer term operation as a result of this upgrade.

In 2003, the real time reconstruction farm (RFARM) was introduced. The system is a large scale PC farm directly fed by the event builder, and real time full event reconstruction is performed utilizing parallel processing of events. The processing results such as the reconstructed IP position were also fed back to the accelerator control, which greatly contributed to the improvement of luminosity. In the same year, the improvement of the FASTBUS readout was also made so as to reduce the readout dead time by a factor of four.

An improvement to the back-end system was made in 2005, when a second EFARM and RFARM were added in order to have sufficient bandwidth and processing power to cope with the expected increase in luminosity.

For further reduction of the readout dead time, an upgrade of the FASTBUS readout system, to a pipelined version, was started. A new TDC was developed based on COPPER, a common pipeline readout module developed at KEK (Figure 2.2.16). The TDC is designed to be plug-compatible with LeCroy 1877S, allowing the use of the same detector front-end electronics without any modifications. The upgrade was performed detector by detector starting from the CDC in 2007 utilizing the short shutdown time during summer and winter. By 2009, five detector subsystems were upgraded resulting in a reduction in dead time to less than 1%. Figure 2.2.17 shows the Belle DAQ configuration at the end of data taking.

## 2.2.8 Background and mitigation

### *BABAR*

Predicting accurately the background level using dedicated simulations is not an easy task, whether the detector plans to run at the intensity or at the energy frontier. Yet, background is a major concern for any HEP experiment as it can severely impact the data taking: first, by slowing down the acquisition system and creating dead time; then, by decreasing the quality of the logged data when signal signatures get lost in a mass of random hits; finally, by degrading or even destroying detector components. Therefore, special care is given to design detectors able to handle background levels corresponding to the predictions (with significant safety margins added), while numerous probes monitor the background during the data taking. When the conditions become unsafe for the detector, automated systems make its HV ramp down to safer levels and can even dump the beams.

Figure 2.2.18 shows an overview of the *BABAR* background monitoring system: several probes monitor quantities sensitive to background (radiation doses, rates recorded by scaler boards, channel currents, etc.) in real time and compare the measured values with pre-defined alarm levels. The status of each variable (in alarm or not) is indi-cated by the color of the display. New alarms produce visual and audio alerts in the control room while automated systems can modify the detector state or even abort the beams if the background becomes worrisome.

[16] These are the same tape format as previously used by some TV broadcasting companies.

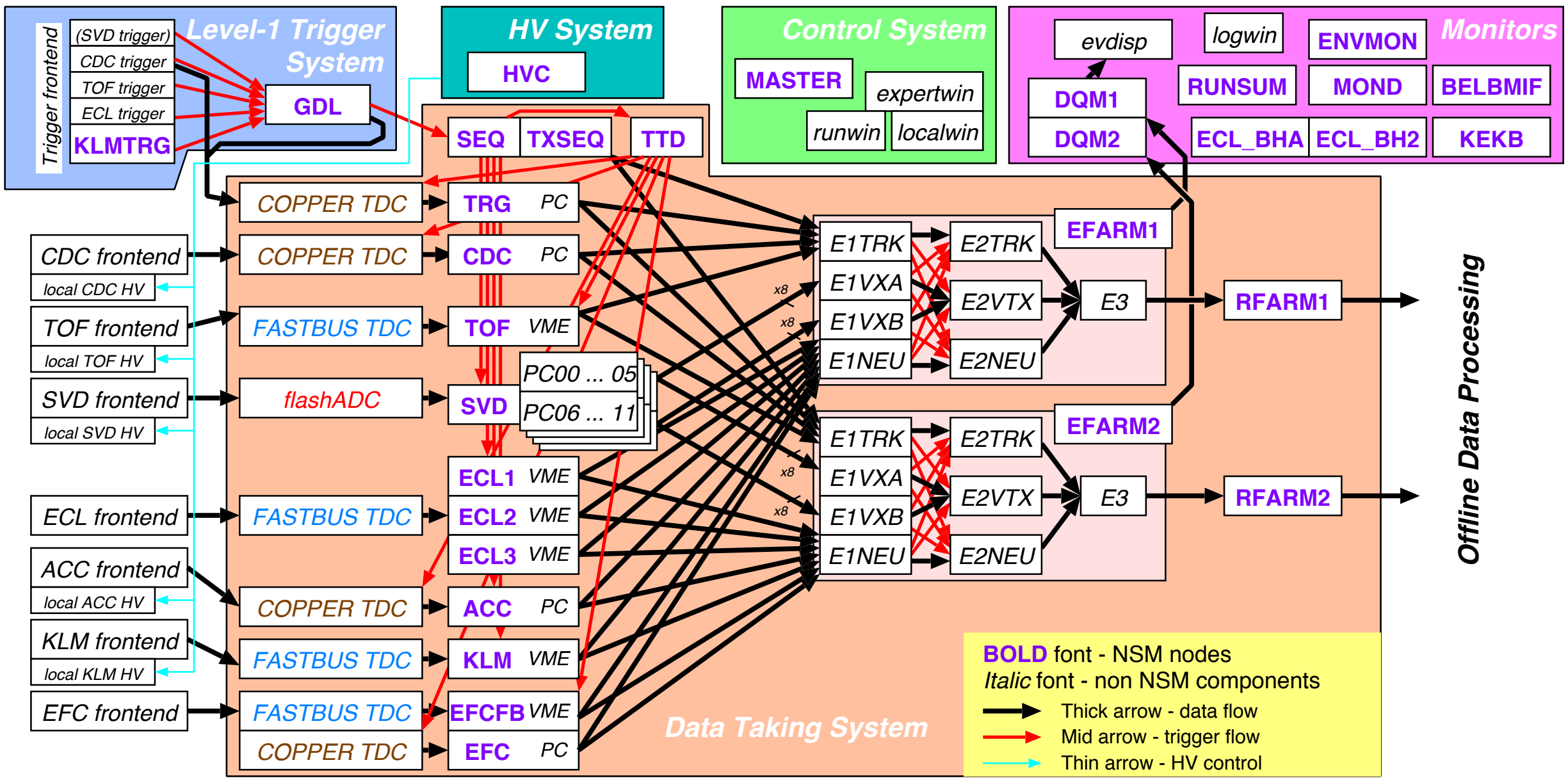

**Figure 2.2.17.** The configuration of the Belle DAQ system at the end of data taking.

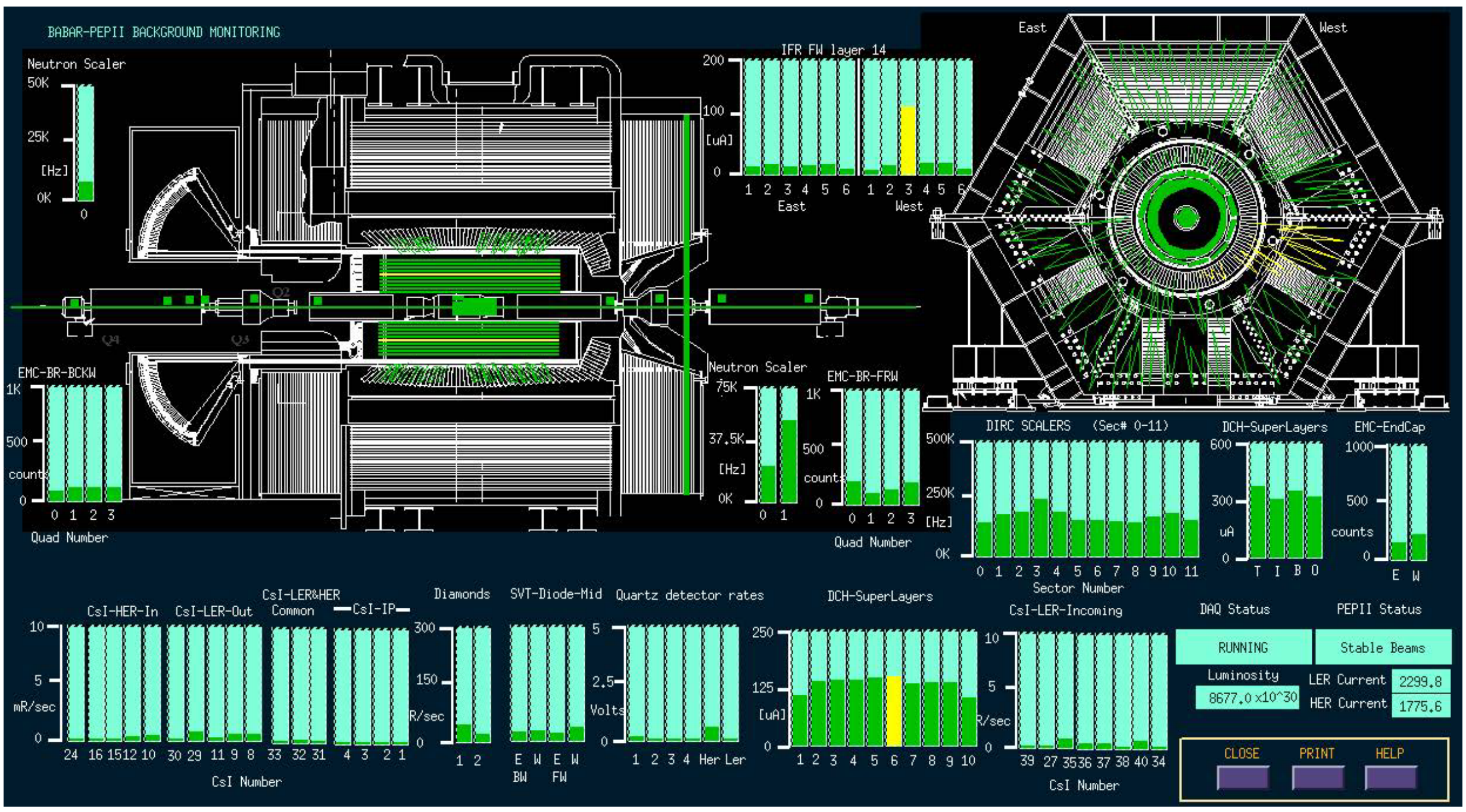

**Figure 2.2.18.** Snapshot of the global *BABAR* background display (Aubert, 2013) taken at a time when the background was low: all but a couple of probes are green which, in the *BABAR* framework, means 'safe level' – alarm states are indicated by yellow (warning level reached) and red (concern) colors. This display was available 24/7 in the control room to help shifters get a real time overview of the background levels around the *BABAR* detector. The longitudinal and end cross-sections show the locations of the background probes which survey all systems: SVT radiation monitors, current levels in the DCH superlayers, rates in the DIRC, EMC and IFR or neutron rates on both ends of the beampipe.

There were two main active detector protection systems in *BABAR* to ensure a safe operation of the sensitive tracking system. First, the SVTRAD which monitored both the instantaneous and the integrated radiation doses received by the SVT. Originally, rates were measured by 12 PIN diodes located on both ends of the SVT in three horizontal planes (one at the beam level, the other two 3 cm above/below it) and on the inside and outside of the PEP-II rings. As expected, the middle-plane diodes accumulated the highest radiation doses and started to become less reliable due to damage. Therefore, in 2002 two diamond sensors were added to the SVTRAD system – this was the first time such sensors were used in a HEP experiment – and they worked well until the end of the data taking. Another advantage of these detectors with respect to the PIN diodes is that they are insensitive to temperature fluctuations. The maximum total dose after nine years of operations was measured to be around 4 MRad, *i.e.* less than the SVT radiation budget, set to 5 MRad. The SVTRAD was also able to abort the beams, either when instantaneous doses were too high or because the integrated dose was consistently above some threshold during 10 consecutive minutes. Beam aborts induced by the SVTRAD protection system occurred a few times a day on average. When PEP-II started to deliver beams in trickle injection mode (particles are injected in existing bunches at a few Hz frequency, see Section 3.2.2 for details), the SVTRAD was modified to monitor in addition the dose associated with each injection of particles in the collider rings. This provided a complementary feedback on the trickle injection quality. The second active protection system was based on the monitoring of the DCH currents and was used to prevent damage to the drift chamber wires and the associated front-end electronics; it is described above in Section 2.2.2.

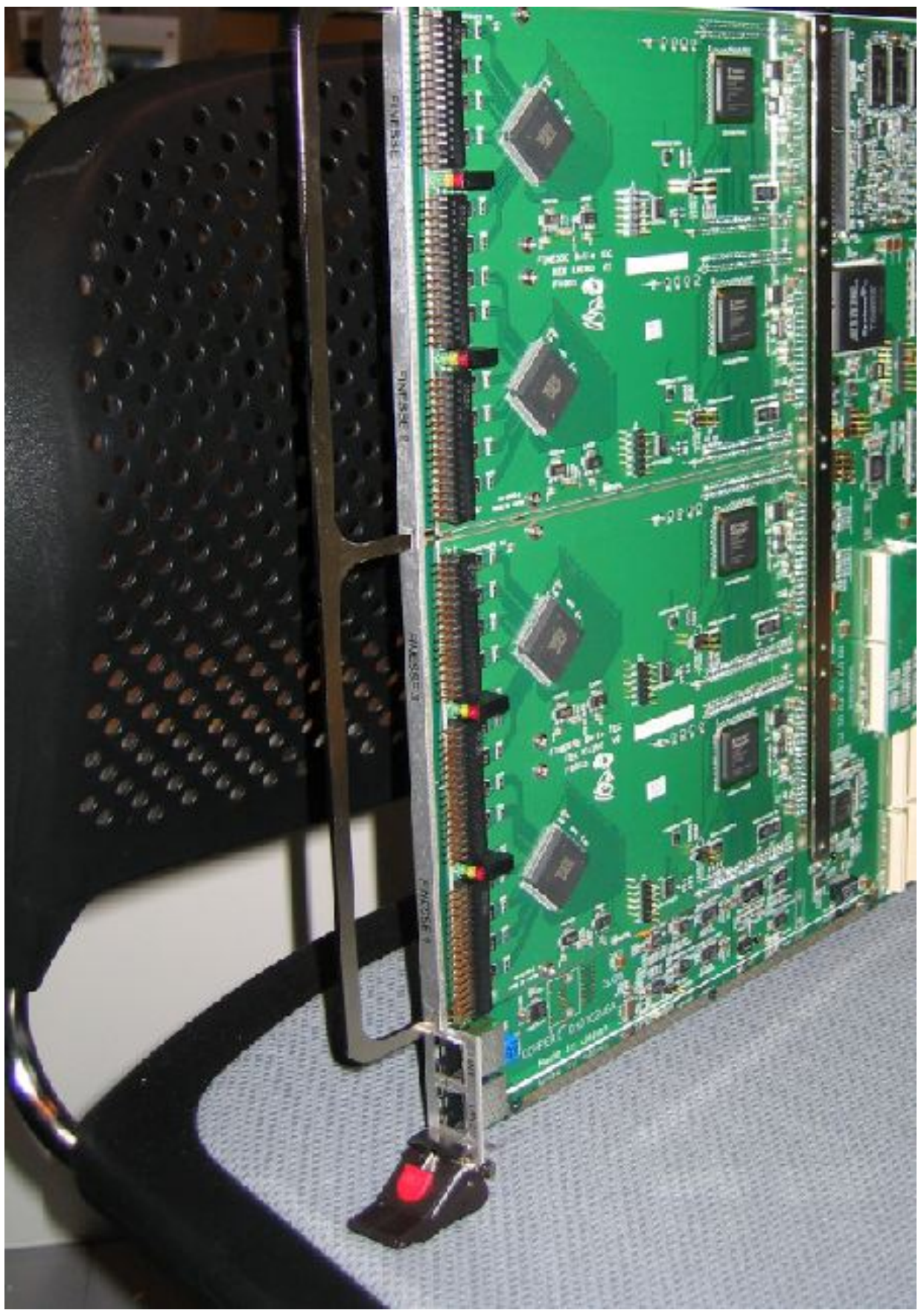

**Figure 2.2.16.** A pipeline TDC module based on COPPER.

The main *BABAR* background probes were also displayed in the accelerator control room, providing valuable information about the beam status and helping operators reduce the background levels. For instance, the accelerator crew was notified when the SVTRAD 10-minute counter was enabled; this signal would tell them that the beams were to be tuned and that they also had some time to try and fix the problem before a beam abort would be issued.

In addition to the real-time monitoring and protection system, various shieldings around *BABAR* have been built and improved over the years. The main additions with respect to the original detector design have been a DIRC shielding around the beamline components at the backward end and shielding walls on the forward side of *BABAR* to protect the outer IFR layers.

### 2.2.9 Conclusion: main common points, main differences

Table 2.2.1 summarizes in a single page the typical performances of the *BABAR* and Belle detectors. Of course the signals detected by the individual subdetectors need to be combined and converted into data used for physics measurements. Various methods and tools are used for this data reconstruction which are beyond the scope of this book. Typical performances of combined tracking, charged particle identification and neutral particle reconstruction are also given in Table 2.2.1. More information can be found in the detector articles published by the two collaborations and in this book, in particular for PID – see Chapter 5 – and for tracking and vertexing – see Chapter 6.

Both detectors reached their design performance and were robust enough to keep them almost constant while the luminosity delivered by the colliders was increasing. Both data taking periods were about a decade long, allowing *BABAR* and Belle to collect huge datasets which made possible the impressive harvest of physics results achieved by the two collaborations. The detector upgrades described in Section 3.2 were mainly driven by the luminosity increase although both experiments had a subdetector weaker than the others: the silicon tracker for Belle and the muon detector for *BABAR*. Several technological and conceptual breakthroughs were made by the $B$ Factories, among which the *BABAR* DIRC (a new concept of ring-imaging Cherenkov PID detector), the use of the object-oriented language C++ for the experiment software, or the development of distributing computing. Now, they all are well-established in the HEP community.

**Table 2.2.1.** Summary of the *BaBar* and Belle detector main characteristics. The *BaBar* numbers provided in this table are representative of the detector performances; they vary with the type of events reconstructed. Moreover, the PID selectors can be tuned depending on the analysis requirements – looser or tighter cuts. [a]Until summer 2003 Belle used a 3 layer SVD. [b]Number of photo-electrons. [c]For Bhabha events. [d]L3 trigger was operated partially from 2004 to 2007. [e]The maximal trigger rate is determined at the end of the DAQ chain. [f]For $B\overline{B}$ events. [g]For momenta above 0.8 GeV/$c$. [h]For $\pi^0$'s reconstructed from photons in hadronic events.

| Detector type | Belle abbreviation | Belle Type | Belle $\theta$ Coverage | Belle Illustrative Performance | *BaBar* abbreviation | *BaBar* Type | *BaBar* $\theta$ Coverage | *BaBar* Illustrative Performance |
|---|---|---|---|---|---|---|---|---|
| Tracking | SVD | Silicon<br>3/4 layers[a]<br>Two-sided | $[17°; 150°]$ | Single hit resolution:<br>$12\,\mu$m ($R\phi$)<br>$19\,\mu$m ($z$) | SVT | Silicon<br>5 layers<br>Two-sided | $[20.1°; 150.2°]$ | Single hit resolution:<br>$\sim$10-15 $\mu$m (inner)<br>$\sim$40 $\mu$m (outer) |
| | CDC | Drift<br>chamber | $[17°; 150°]$ | Single hit resolution:<br>130 $\mu$m ($R\phi$)<br>200-1400 $\mu$m ($z$)<br>$\sigma(\mathrm{d}E/\mathrm{d}x)\sim7\%$ | DCH | Drift<br>chamber | $[17.2°; 152.6°]$ | Single-cell hit<br>resolution: $\sim100\,\mu$m<br>(center of the cell)<br>$\sigma(\mathrm{d}E/\mathrm{d}x)\sim8\%$ |
| Particle ID | TOF | Time of flight<br>scintillator | $[34°; 130°]$ | $\sigma_t = 100$ ps | DIRC | Cherenkov | $[25.5°; 141.4°]$ | $\sigma_{\theta_C}\sim2.4$ mrad |
| | ACC | Threshold Cherenkov<br>with aerogel | $[17°; 127°]$ | $N_{\mathrm{p.e.}} \geq 6$[b] | –<br>– | –<br>– | –<br>– | –<br>– |
| Calorimetry | ECL | CsI(Tl) | $[12.4°; 31.4°]$<br>$[32.2°; 128.7°]$<br>$[130.7°; 155.1°]$ | $\sigma_E/E\sim1.7\%$[c] | EMC | CsI(Tl) | $[15.8°; 140.8°]$ | $\sigma_E/E\sim3\%$ |
| Muon and $K_L^0$ detector | KLM | RPC | $[20°; 155°]$ | $\sigma_\theta = \sigma_\phi = 30$ mrad<br>for $K_L^0$ | IFR | RPC, LST | $[20°; 154°]$ | LST layer eff. $\sim$88% |
| Trigger | L1<br>L3[d]<br>L1+L3 | Hardware<br>Software | Full<br>Belle<br>acceptance | Max. rate $\sim$0.5 kHz[e]<br>Physics mode eff. $> 99\%$[f] | L1<br>L3<br>L1+L3 | Hardware<br>Software | Full<br>*BaBar*<br>Acceptance | Max. rate $\sim$5 kHz<br>Max. rate $\sim$1 kHz<br>Physics mode eff. $\sim$99% |
| PID Algorithms | $\mu^\pm$<br>(KLM)<br>$K/\pi$<br>(TOF,ACC,CDC)<br>$e^\pm$<br>(CDC,ECL) | | | $\langle\mu$ eff$\rangle = 90\%$[g]<br>$\langle\pi$ misID$\rangle = 2\%$<br>$\langle K$ eff$\rangle \geq 85\%$<br>$\langle\pi$ misID$\rangle \leq 10\%$<br>$\langle e$ eff$\rangle = 90\%$<br>$\langle\pi$ misID$\rangle \sim 0.3\%$ | $\mu^\pm$<br><br>$K/\pi$<br><br>$e^\pm$ | | | $\langle\mu$ eff$\rangle = 59-65\%$<br>$\langle\pi$ misID$\rangle = 1.4-0.8\%$<br>$\langle K$ eff$\rangle = 84\%$<br>$\langle\pi$ misID$\rangle = 1.1\%$<br>$\langle e$ eff$\rangle = 90-95\%$<br>$\langle\pi$ misID$\rangle < 0.2\%$ |
| Tracking | (CDC,SVD) | | | $\sigma_{p_T}/p_T = 0.0019 p_t$ [GeV/$c$]<br>$\oplus\, 0.0030/\beta$ | SVT + DCH | | | $\sigma_{p_T}/p_T\sim0.5\%$ |
| Neutrals | (ECL) | | | $\sigma(m_{\pi^0}) = 4.8$ MeV/$c^2$ [h] | EMC | | | |

# Chapter 3
# Data processing and Monte Carlo production

***Editors:***
*Fabrizio Bianchi and Nicolas Arnaud (BABAR)*
*Shoji Uno (Belle)*

**Additional section writers:**
*Concetta Cartaro, Christopher Hearty, Ryosuke Itoh, Leo Piilonen, Teela Pulliam, Dennis Wright*

## 3.1 Introduction: general organization of the data taking, data reconstruction and MC production

The *BABAR* and Belle experiments have collected around one Petabyte of raw data each. These data have been calibrated, the events reconstructed, and collections of selected events produced. Monte Carlo events (MC) have been generated and reconstructed with the same code used for the detector data. The total amount of data produced by *BABAR* and Belle were over six Petabytes and over three Petabytes respectively. Over the years, both collaborations have developed computing models that have proven to be highly successful in handling the amount of data produced, and in supporting the physics analysis activities. The main elements of the two computing models are outlined in this introductory section and will be described in more detail in the remainder of this chapter.

The 'raw data' coming from the detectors have been permanently stored on tape, calibrated, and reconstructed usually within 48 hours of the actual data taking. Reconstructed data have been permanently stored in a format suitable for subsequent physics analysis.

Many samples of Monte Carlo events, corresponding to different sets of physics channels, have been generated and reconstructed in the same way. In addition to the physics triggers, the data acquisition also recorded random triggers that have been used to create 'background frames' that have been superimposed on the generated Monte Carlo events to account for the effects of the machine background and of electronic noise, before the reconstruction step.

Detector and Monte Carlo data have been centrally 'skimmed' to produce subsets of selected events, the 'skims', designed for a specific area of analysis. Skims are very convenient for physics analysis, but they increase the storage requirements because the same event can be present in more than one skim.

The quality of the detector data and of the simulated events has been monitored through all the steps of processing.

From time to time, as improvements in detector calibration constants and/or in the code were implemented, the detector data have been reprocessed and new samples of simulated data generated. When sets of new skims become available, an additional skim cycle has been run on all the events.

*BABAR* has been one of the first experiments to adopt the C++ programming language to write offline and online software. In the mid-nineties, when this decision was taken, the dominant language in the High Energy Physics (HEP) community was `Fortran 77`. However, problems and limitations associated with this language were becoming very clear and *BABAR* chose early to commit to the C++ technology because there was the perception that the HEP computing model was a very good match to an object-oriented design. At first, the C++ expertise was limited to few collaborators, who started offering tutorials. Starting in 1996, formal training courses were offered to the collaboration members and rapidly produced a shared vocabulary and set of concepts that were immensely helpful in the actual software development. The final outcome of this effort was the over 3 million lines of code that today constitutes the *BABAR* offline software.

Belle data processing and analysis code (called Belle AnalysiS Framework - `basf`) was developed in C++ with an extensive use of adjoined tools (*e.g.* the CLHEP library (CLHEP, 2008) for which some of the Belle members were the initial developers). The simulation tool, `GEANT3` (Brun, Bruyant, Maire, McPherson, and Zanarini, 1987), on the other hand, was written in `Fortran`.

Belle data were stored using the `PANTHER` banks event store based on the entity-relationship model (Putzer, 1989) and developed specifically for this experiment. `PANTHER` banks (Adachi, 2004) offered a satisfactory storage throughout the data taking and reliable usage in the data analysis process. Due to the large volume of recorded data centralized skimming was used (see Section 3.5) in order to facilitate subsequent analysis of events. Furthermore, at the level of specific analysis, additional skimming was performed, resulting in the so called index files, providing unique event identifiers that enable processing of selected events only.

Similarly large data volumes produced by *BABAR* were anticipated to make it impossible to routinely run on all the data. At first, *BABAR* decided to use an event store based on the object-oriented database technology that was expected to solve the problem of an efficient and scalable access to the data. The end result of this work was what, at the time, was the world's largest object-oriented database. Unfortunately, it soon became clear that data volumes and usage patterns were exceeding the capabilities of the technologies that were available at that time. A lot of effort went into mitigating these problems. Finally, the working solution identified was to handle data persistency using `Root` I/O which offers the advantages of its lightweight interface and built-in data compression. In this context, client/server data access was a very important issue and the bundled data server, `rootd`, was insufficient for *BABAR*'s need. A better performing solution was developed starting from `rootd` and taking advantage of the experience made with the object-oriented database.

The result of this effort was a data server named `XRootD` (Furano and Hanushevsky, 2010).

*BABAR* was the first HEP experiment to effectively use geographically distributed resources, because the amount of computing needed to satisfy the production and analysis requirements exceeded what was possible at SLAC. Grid computing tools became available too late for the $B$ Factories and *BABAR* solved the problem by assigning specific production tasks and datasets to different computing centers. Only 20-30% of Monte Carlo data where produced using Grid resources with the aid of specific software tools.

Belle (re-)processed the recorded data centrally at KEK while the production of simulation was dispersed among the collaborating institutions. As with *BABAR*, a significant part of MC simulation was produced at remote sites.

## 3.2 Data taking

*BABAR* started taking physics data in October 1999 after an extensive period of commissioning of both the collider and the detector. The data taking ended on April $7^{th}$ 2008, about six months earlier than planned, due to budget constraints at the US Department Of Energy (DOE) level. The *BABAR* data taking can be divided into seven main periods, called 'Runs',[17] for which details are given below. The equivalent of the *BABAR* Run is called 'Experiment' at Belle.

Two consecutive *BABAR* Runs are separated by a shutdown period usually lasting a few months and during which various operations are performed by the PEP-II and *BABAR* teams: repairs, fixes and maintenance, both at the hardware and software levels. The longest *BABAR* shutdown took place between Runs 4 and 5 (from August 2004 to April 2005) as the start of the new data taking period was delayed due to an electrical accident at SLAC: all work procedures had to be reviewed and improved in order to reinforce the site-wide safety best practice.

*BABAR* Runs 1 to 6 data were taken at (or near) the energy of the $\Upsilon(4S)$ resonance (10.58 GeV). About 90% of these data were taken at the peak of the resonance ('on-resonance' data) to maximize the number of produced $B\overline{B}$ pairs. The remaining $\sim$10% were taken about 40 MeV below ('off-resonance' data) to study non-$B$ backgrounds, in particular the production of light quark and $\tau$ pairs called 'continuum'. Taking advantage of years of continuous improvements and upgrades, both on the machine and detector sides, Run 7 was expected to increase the size of the *BABAR* dataset by 50% in about a year. Once this goal would have been achieved, it was planned to end the data taking by running at other energies, below and above the $\Upsilon(4S)$ resonance. When it became clear shortly before Christmas 2007 that Run 7 was going to be much shorter than anticipated due to the lack of funding, the *BABAR* management reacted quickly and decided to stop the $\Upsilon(4S)$ resonance data taking – which had just restarted a week earlier. Instead, data were taken at the $\Upsilon(3S)$ resonance during two months; then, the collision energy was moved to the $\Upsilon(2S)$ resonance for about a month. In both cases, on- and off-resonance data were recorded. Finally, the energy region above the $\Upsilon(4S)$ resonance up to 11.2 GeV was scanned during the last 10 days of data taking.

Although originally designed to be a fixed-energy machine, PEP-II performed remarkably well during Run 7 and all of the CM energy changes were done by moving the energy of the HER beam, keeping the LER one fixed. At the $\Upsilon(2S)$ energy (10.02 GeV), the HER orbit was quite close to the vacuum beam pipe in the interaction region (IR), leading to a trade-off between luminosity and background. At 11 GeV and above, synchrotron radiation became the dominant issue and the HER current had to be decreased, which had a direct impact on the delivered luminosity. On the *BABAR* side, the trigger was the main system impacted by the changes of the running energy as the data taking goal moved from selecting $B\overline{B}$ events with large visible energy, high multiplicity and/or high transverse energy to looking for decays with low visible energy and low multiplicity. These changes had to be made while the data taking was ongoing and occurred thanks to the flexibility of the *BABAR* trigger design.

Belle started taking data on June $1^{st}$ 1999. After that, data taking has been continuous for 6-9 months every year until the final shutdown on June $30^{th}$ 2010. After each major shutdown a new "Experiment" started. Hence the Belle data are grouped into experiments 7 to 73, where only odd numbers are used.[18] Experiments 7 - 27 are recorded using the first Silicon Vertex Detector (SVD1) and the rest with the second (SVD2) detector (see Section 2.2.1). There were two scheduled shutdowns every year, in summer and winter. The summer shutdown took about three months or more for maintenance and hardware replacements within the Belle detector as well as in the KEKB accelerator. The winter shutdowns were shorter, typically one month long. In the last three years of operation, the winter shutdowns were slightly extended due to budget constraints. Beside these shutdowns one day every two weeks was devoted to maintenance of the accelerator and detector. Typically after each experiment cosmic ray data was taken with the Belle solenoid turned off for the purpose of detector alignments. Belle took data mostly at the energy of the $\Upsilon(4S)$ resonance in order to study $B$ meson decays. For the purpose of the background estimation arising from the non-$B$ meson events the off-resonance data was collected 60 MeV below the resonance peak energy, for around 10% of the running time, approximately every two months. Similar off-resonance data taking was performed also for the data taken at other $\Upsilon$ resonances. Note that the *BABAR* off-resonance data taking was performed 40 MeV below the $\Upsilon(4S)$ mass, a difference which has no impact on the usage of this data.

[17] In the following the word "run" is used to identify a small data acquisition batch up to a few hours long, *i.e.* the basic unit of the *BABAR* and Belle data taking system, not to be confused with the "Run" defined here.

[18] For various reasons some experiment numbers are not used: experiment 29, 57 and 59.

The first Belle non-$\Upsilon(4S)$ data was taken at the energy of $\Upsilon(5S)$ resonance for 3 days in 2005. During the following year in the last week of February, $\Upsilon(3S)$ resonance data was taken to enable the search for invisible particles from decays of the $\Upsilon(1S)$ resonance. The last $\Upsilon(4S)$ resonance data was taken in June 2008. After that, $\Upsilon(1S)$ (second half of June 2008), $\Upsilon(2S)$ (December 2008 and November 2009) and $\Upsilon(5S)$ resonance data were taken, and energy scans between the $\Upsilon(4S)$ and $\Upsilon(6S)$ were carried out in the last two years of operation. The $\Upsilon(1S)$ The CM energy change was rather smoothly performed, keeping the same ratio of the beam energies in the KEKB rings. During that time, the magnetic fields of the Belle solenoid and super-conducting final focusing magnet were kept at the same values. The luminosity decreased at lower CM energies for reasons which have not been well understood. The beam background did not change by a large amount when running at different energies. The same was true for the trigger rates, where the increase of the cross-section at lower energy resonances was canceled by a lower luminosity. Looser trigger requirements were adopted for two charged track events in the case of the $\Upsilon(3S)$ data taking to achieve the physics goals of the $\Upsilon(3S)$ programme.

### 3.2.1 Integrated luminosity vs. time; luminosity counting

The integrated luminosity collected by Belle for each CM energy is listed in Table 3.2.1 and is calculated using Bhabha events, where the final state electrons are detected in the barrel part of the detector, and after removing runs deemed to be unusable for physics studies (so-called bad runs) because of detector-related issues. The Belle integrated luminosity as a function of time is shown in Fig 3.2.1. As well as the luminosity measurement, the counting of recorded $\Upsilon(nS)$ events is done using the method described in Section 3.6.2. The yields obtained are presented in Table 3.2.2.

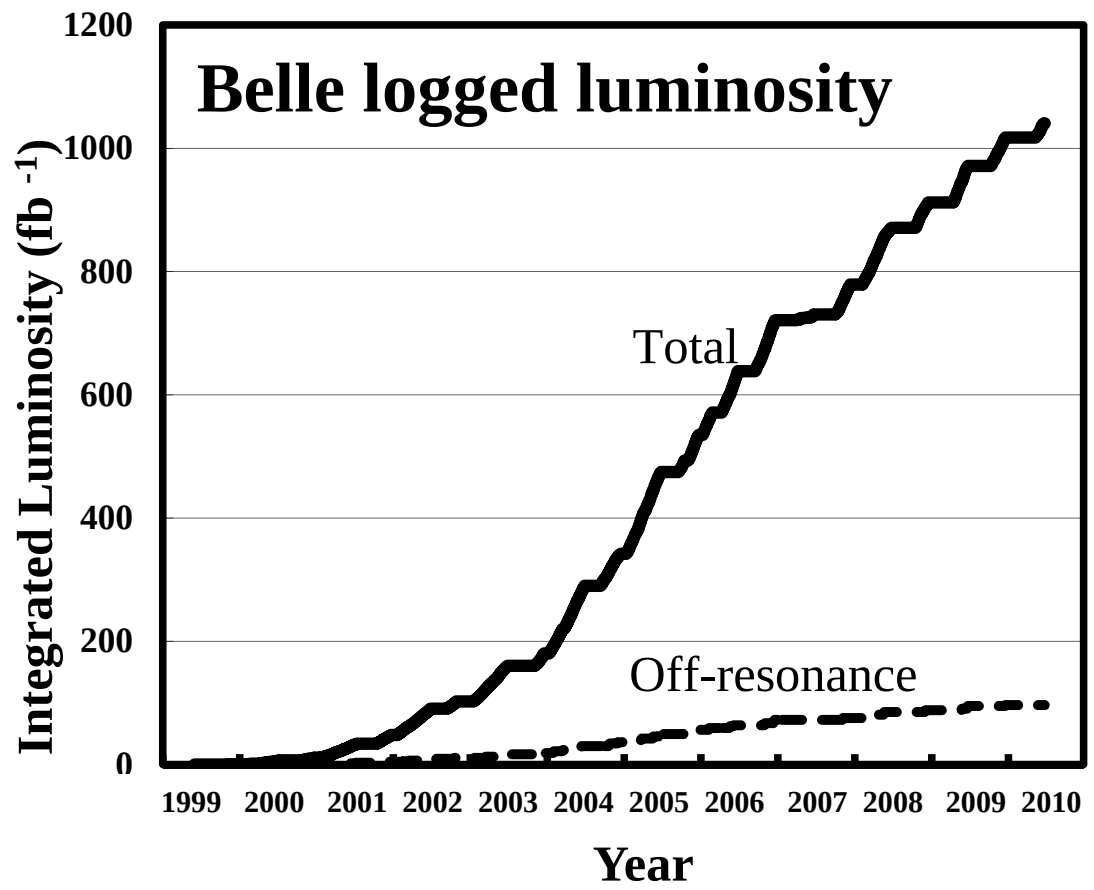

**Figure 3.2.1.** Evolution of the Belle integrated luminosity. A detailed breakdown of datasets is given in Table 3.2.1.

The systematic error on the luminosity measurement is about 1.4% and the statistical error is usually small compared to the systematic error. The latter is dominated by the uncertainty of the Monte Carlo generator used to calculate the cross-section for Bhabha events. The $\Upsilon(4S)$ dataset is split into two periods, named SVD1 and SVD2, which correspond to different configurations of the Silicon Vertex Detector, as explained in the following section. All other resonance and scan data were taken in the SVD2 configuration.

Lees (2013i) describes the methods used to measure the *BABAR* time-integrated luminosities at the $\Upsilon(2S)$, $\Upsilon(3S)$, and $\Upsilon(4S)$ resonances, as well as in the continuum regions below each of these resonances. For each running period at fixed energy, the luminosity was computed offline, using Bhabha ($e^+e^- \rightarrow e^+e^-$) and dimuon ( $e^+e^- \rightarrow \mu^+\mu^-$) events for Runs 1-6 and only Bhabha events for Run 7 – due to uncertainties in the large $\Upsilon \rightarrow \mu^+\mu^-$ background. No detailed analysis could be performed for the final scan data because of the short duration of the running at each scan point (only about 5 $\text{pb}^{-1}$). Therefore, the corresponding luminosity is only an estimation taken from (Aubert, 2009x). The systematic error on the luminosity measurement is about 0.5% for the data collected at the $\Upsilon(4S)$ and 0.6% (0.7%) for data collected at the $\Upsilon(3S)$ ($\Upsilon(2S)$). Table 3.2.1 and Fig. 3.2.2 show the luminosity integrated by *BABAR*, broken down by CM energy.

In addition to measuring the luminosity, the number of $\Upsilon$ particles in the different datasets is also computed using a common method referred to as '$B$-counting' for the $\Upsilon(4S)$ running. This number is found by counting the hadronic events in the on-resonance dataset and subtracting the contribution coming from the continuum, estimated using off-resonance data and properly scaled to the peak energy – see Section 3.6.2 for details. The final results are shown in Table 3.2.2.

**Table 3.2.2.** Number of $\Upsilon$ particles in the different *BABAR* and Belle datasets

| Experiment | Resonance | $\Upsilon$ number |
|---|---|---|
| *BABAR* | $\Upsilon(4S)$ | $(471.0 \pm 2.8) \times 10^6$ |
| | $\Upsilon(3S)$ | $(121.3 \pm 1.2) \times 10^6$ |
| | $\Upsilon(2S)$ | $(98.3 \pm 0.9) \times 10^6$ |
| Belle | $\Upsilon(5S)$ | $(7.1 \pm 1.3) \times 10^6$ |
| | $\Upsilon(4S)$ - SVD1 | $(152 \pm 1) \times 10^6$ |
| | $\Upsilon(4S)$ - SVD2 | $(620 \pm 9) \times 10^6$ |
| | $\Upsilon(3S)$ | $(11 \pm 0.3) \times 10^6$ |
| | $\Upsilon(2S)$ | $(158 \pm 4) \times 10^6$ |
| | $\Upsilon(1S)$ | $(102 \pm 2) \times 10^6$ |

**Table 3.2.1.** Summary of the luminosity integrated by *BABAR* and Belle, broken down by CM energy.

| Experiment | Resonance | On-resonance Luminosity ($fb^{-1}$) | Off-resonance Luminosity ($fb^{-1}$) |
|---|---|---|---|
| *BABAR* | $\Upsilon(4S)$ | 424.2 | 43.9 |
| | $\Upsilon(3S)$ | 28.0 | 2.6 |
| | $\Upsilon(2S)$ | 13.6 | 1.4 |
| | Scan $> \Upsilon(4S)$ | n/a | $\sim$4 |
| Belle | $\Upsilon(5S)$ | 121.4 | 1.7 |
| | $\Upsilon(4S)$ - SVD1 | 140.0 | 15.6 |
| | $\Upsilon(4S)$ - SVD2 | 571.0 | 73.8 |
| | $\Upsilon(3S)$ | 2.9 | 0.2 |
| | $\Upsilon(2S)$ | 24.9 | 1.7 |
| | $\Upsilon(1S)$ | 5.7 | 1.8 |
| | Scan $> \Upsilon(4S)$ | n/a | 27.6 |

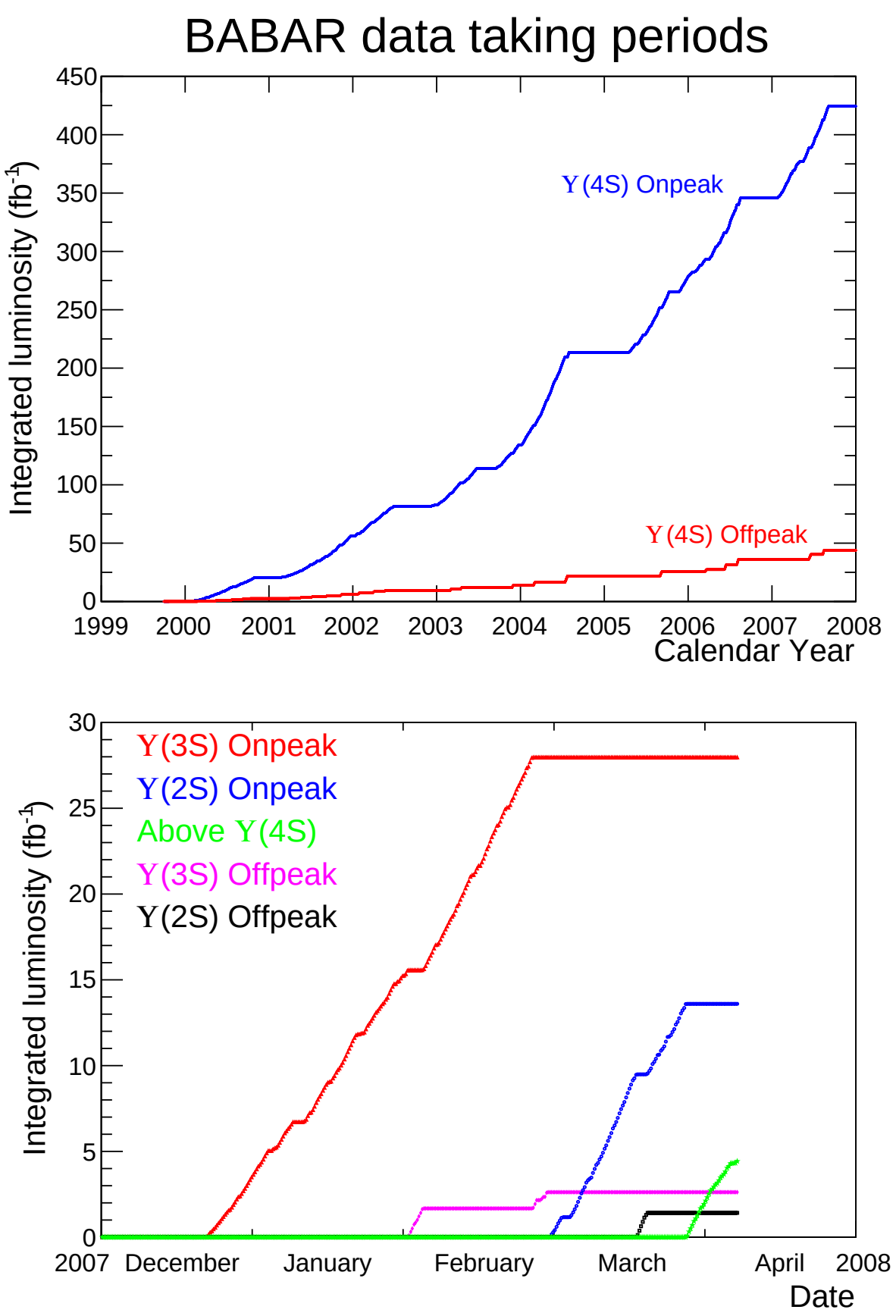

**Figure 3.2.2.** Evolution of the different *BABAR* datasets with time. The top plot shows the luminosity integrated during the $\Upsilon(4S)$ running periods, on-resonance (blue curve) and off-resonance (red curve) operation. The bottom plot focuses on the last *BABAR* running period (Run 7) which lasted less than four months and during which three different data taking phases occurred: $\Upsilon(3S)$ (red curve shows the on-resonance dataset; the purple one the off-resonance), $\Upsilon(2S)$ (blue curve for the on-resonance data, black for the off-resonance) and finally a scan between the $\Upsilon(4S)$ energy and 11.2 GeV (green curve).

### 3.2.2 Major hardware/online upgrades which modified the quality of BABAR data

This section summarizes the upgrades to the *BABAR* detector and also describes the 'trickle injection' mode which allowed PEP-II to keep the luminosity (and hence the detector data taking conditions) stable during most of the run. In the following, the 'forward' and 'backward' sides of the detector are defined relative to the high energy beam.

#### *Detector upgrades*

Over the years, the main *BABAR* activities during the shutdowns between Runs were related to the Instrumented Flux Return (IFR). Indeed, from the very beginning of the data taking, the resistive plate chambers (RPCs) showed severe aging all around the detector. Attempts were made to slow down the performance degradation but it became clear soon enough that the whole system needed an upgrade involving the replacement of most of the muon chambers. This project was completed with the following sequence:

- 12 forward RPCs were replaced between Run 1 and Run 2.
- The remaining forward RPCs were replaced between Runs 2 and 3. Brass was installed in the forward IFR to increase the total absorber thickness.
- The first two Limited Streamer Tube (LST) sextants were installed between Runs 4 and 5 and the last four between Runs 5 and 6.

Another important area of detector-related work was background mitigation and system upgrades, to cope with the instantaneous luminosity increase over the years – PEP-II exceeded its design luminosity goal by a factor four. These issues were addressed in various ways:

- Addition of shielding in various places around the detector (inside the PEP-II tunnel entrance on the backward side, in front of the IFR end-cap, etc.).
- Replacement of the Detector of Internally Reflected Cherenkov light (DIRC) and Drift Chamber (DCH)

front-end electronics to deal with the increase of the instantaneous luminosity over time.

– Online software developments (mainly for the Silicon Vertex Tracker (SVT), the DIRC and the EMC) to speed up the readout of the detector after a L1-accept and hence to be able to run at higher trigger rate while keeping the DAQ dead time low.

The trigger system also underwent upgrades, primarily the inclusion of 3D-tracking information in the L1 DCT trigger to remove background events in which scattered beam-gas particles would hit the beam pipe about 20 cm away from the IP. This new system was tested in parallel to the old one at the end of Run 4 and was used from Run 5 onwards. The IFR component of the trigger also had to be updated when RPCs were replaced by LSTs which had a different latency. Finally, as explained above, several changes were made to the trigger system (both to L1 and L3) in early 2008 during Run 7, as the characteristics of the events needed for the physics analysis during this period were completely different from those recorded at the $\Upsilon(4S)$ resonance.

#### *Trickle injection*

When *BABAR* started taking data, PEP-II was operating in fill-and-coast mode during which the injection and data taking periods were clearly separated. No attempt was made to inject the beams during a data taking run. Therefore, both currents (and consequently the instantaneous luminosity) were slowly decreasing over time. When they had dropped by about 30-50%, data taking was ended and the detector HV ramped down. Once *BABAR* was in a safe mode insensitive to the potentially-high injection backgrounds, the beams were replenished. Then, the HV were raised again and the DAQ restarted when they had reached their nominal values. The whole procedure (end of the actual run; *BABAR* transition from runnable to injectable states; beam injection; *BABAR* transition from injectable to runnable; beginning of a new run) would take around 5 minutes. The duration of each fill was adjusted depending on the machine conditions, in order to maximize the amount of integrated data. But the average luminosity delivered by PEP-II was only ∼70% of the peak luminosity.

A major improvement took place in 2004 when a new mode of operation called 'trickle injection' was introduced. The beam currents were kept constant thanks to a continuous injection of particles into the least filled bunches, without interrupting the data taking. The average luminosity immediately grew up by about 40% and the increase of the integrated luminosity was even larger than the gain directly provided by improving the duty cycle of the machine. Indeed, operating the accelerator near the peak luminosity at all times and with constant currents, allowed the PEP-II crew to improve the tuning of the beams and to reach new standards of performance and stability from which *BABAR* benefited as well.

The main challenge of this new running mode (first established with one beam, some months later with both) was to inject enough particles into the rings, while keeping the background levels low for *BABAR*. Quickly, it became clear that the newly-injected bunches were causing background bursts: events with many hits in all detector components were saturating the DAQ and causing high dead time. This background in phase with the injected bunch lasted up to a few thousands revolutions after the injection, until the excitations induced by the particles added to the bunch got damped. As it was possible to know exactly which bunches had been recently refilled and where they were located in the ring at any time (technically speaking, the *BABAR* clock was locked to the PEP-II timing system and markers were recorded for each injected bunch), the solution to this problem was to inhibit the trigger when one such bunch was close enough to the detector. These online vetoes were extended offline when the data were reconstructed, to make sure that the trickle injection background would not impact the physics. Indeed, no significant difference was ever found between events recorded just outside the trickle injection inhibit windows and those selected far away from any injected bunch. The trickle injection frequency was 5 Hz for the HER and 10 Hz for the LER, resulting in a dead time of 1.4% for the HER and 1.9% for the LER, 3.3% in total.

At the end of the commissioning phase which lasted a few months in total, the trickle injection mode became the default configuration for PEP-II. A constant and detailed monitoring, both on the detector and machine sides, allowed to operate the $B$ Factory safely in these conditions until the end of the data taking, not only at the $\Upsilon(4S)$ resonance, but also from the $\Upsilon(2S)$ energy up to 11.2 GeV. The veto regions did not change over time and induced a dead time of ∼1% (∼0.5%) for the LER (HER) beam. As the HER and LER inhibit windows did not overlap due to the low injection frequencies, the total dead time was the sum of the two contributions, a small price to pay for the significant increase in integrated luminosity described above. As described in Section 2.2.7, the *BABAR* online system had to be significantly modified to follow this significant change of the machine operations: not only had the detector control system to allow injection during data taking, but the DAQ system also had to accommodate much longer periods of continuous data taking.

#### *Summary*

Both the detector improvements and the PEP-II trickle injection mode allowed *BABAR* to accumulate good data at a rate which increased over the years. More information about these different types of upgrades can be found in (Aubert, 2013).

### 3.2.3 Major hardware/online upgrades which modified the quality of Belle data

#### *Detector upgrades*

Belle encountered serious beam background in the beginning of the experiment. The radiation damage on the read-out electronics chips of the silicon vertex detector (SVD1) was serious and the detector was replaced several times. Finally, the second type of silicon vertex detector (SVD2), which used so-called radiation hard electronics, was installed during the summer of 2003. At the same time, the inner part of the central drift chamber was also replaced with a compact small cell type drift chamber in order to make space for four instead of only three SVD layers. The diameter of the beam pipe was changed from 40 mm to 30 mm enabling the radius of the innermost SVD layer to be reduced to 20 mm in order to achieve a better vertex resolution; also the angular coverage of the silicon vertex detector was matched to that of other detectors ($17° \leq \theta \leq 150°$). This was the only major hardware change in the whole running period of the Belle detector and more information can be found in Chapter 2.

Other detector modules have been used without any major replacement. Unfortunately, the outermost two layers of the 14 resistive plate chambers used in the muon and $K_L$ detector could not be operated due to the neutron background created by the radiative Bhabha events. However, the muon identification capability was not significantly affected. After the summer of 2003 the beam background was not so serious despite an increase of the luminosity to twice the design value.

Apart from the silicon vertex detector, the Belle data acquisition system used one type of multi-hit TDC module. The module did not have a pipe-line readout scheme. Therefore, the readout dead time was larger than at *BABAR*. Several efforts have been made in order to reduce the dead time. Finally, the readout modules were replaced gradually with a pipe-line TDC for most of the sub-detectors rather late in the running period.

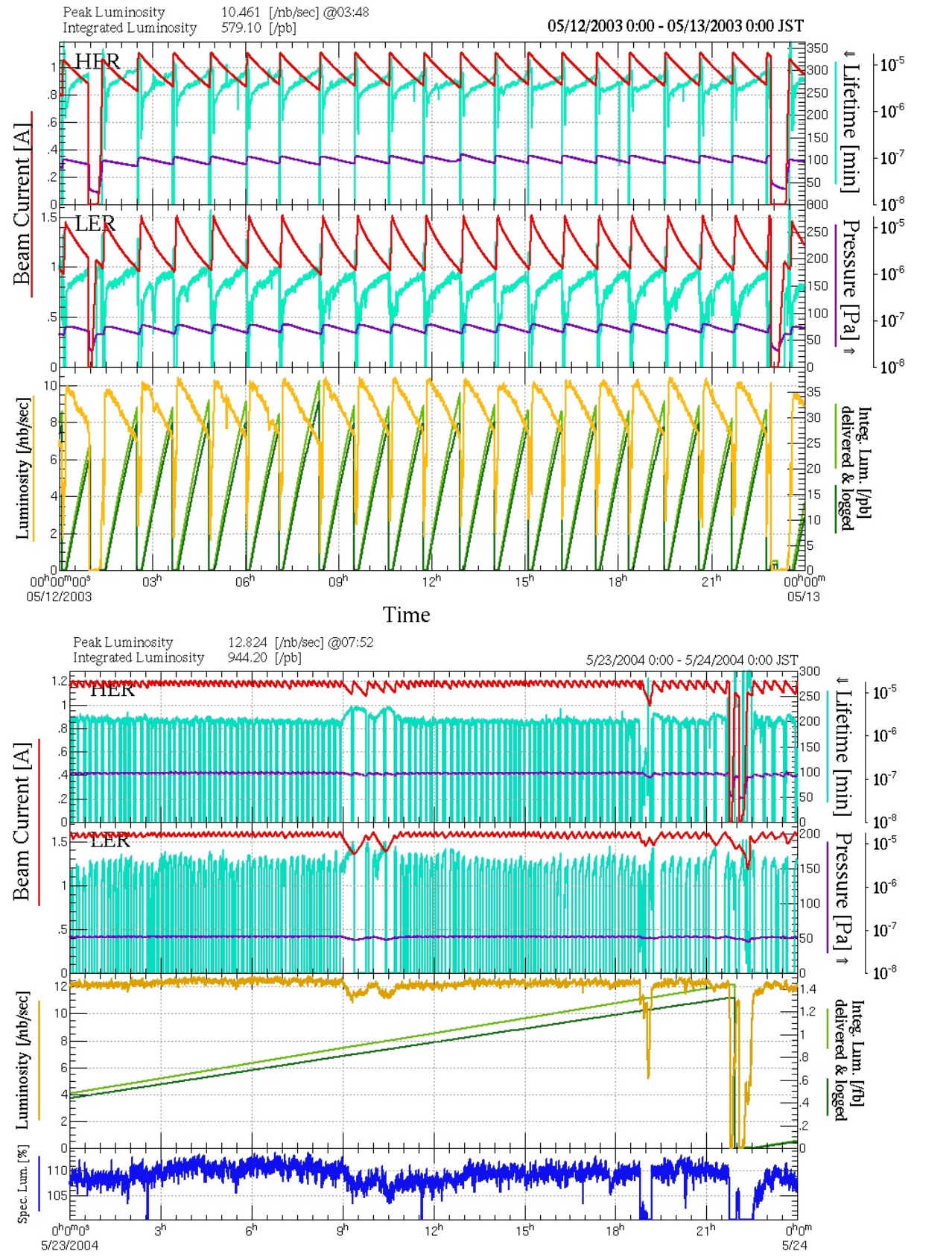

**Figure 3.2.3.** Comparison of beam currents and luminosity of KEKB before (top) and after (bottom) adoption of the continuous injection scheme. The top two panels of each plot show the electron and positron beam currents (red) and the third panel shows the luminosity (yellow). From (Abe et al., 2013).

### *Continuous injection and Crab cavities*

Belle turned off the detector high voltage during beam injection as commonly done at other experiments. The injection time took slightly longer than at PEP-II causing a slightly lower average luminosity. In order to reduce such a time loss, a continuous injection scheme was adopted from January of 2004. The detector high voltage was kept on and the trigger signals were vetoed for 3.5 ms just after each beam injection. The scheme caused 3.5% dead time only in the case of a 10 Hz injection rate. After adopting continuous injection, the KEKB machine beams became stable and the peak luminosity was improved due to the constant beam currents. The obvious difference in the beam currents and luminosity before and after adoption of the continuous injection scheme is shown in Fig. 3.2.3 (Abe et al., 2013). The effect of the scheme can also be seen in Fig. 3.2.1 as an increased slope of the integrated luminosity after the beginning of 2004.

Another important upgrade of the beam optics took place in February 2007. At that time Crab cavities (Yamamoto et al., 2010) were introduced. These are RF deflectors providing the electron and positron bunches inside the KEKB accelerator rings, which at the interaction point have a crossing angle of 22 mrad, with a rotational kick in order to undergo a head-on collision. The schematic principle of the Crab cavities operation is shown in Fig. 3.2.4. The installation of the cavities into the KEKB was not without problems, as can be also observed by a short-lasting plateau at the beginning of 2007 in the integrated luminosity curve (Fig. 3.2.1). While the increase in the luminosity after the installation was modest, the beam induced backgrounds were reduced.

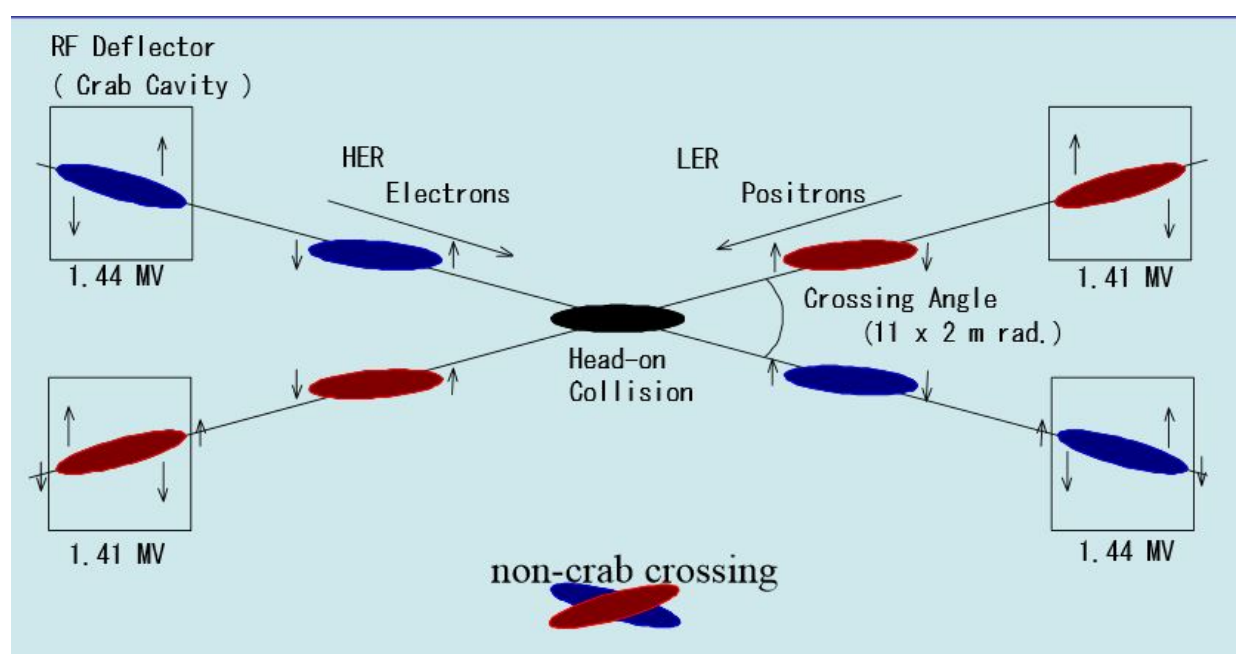

**Figure 3.2.4.** Schematic principle of Crab cavities operation leading to head-on collisions in KEKB despite the finite crossing-angle of electron and positron bunches.

## 3.3 Data Reconstruction

### 3.3.1 Introduction

Both *BABAR* and Belle developed tools to process raw data in a timely way. The reconstruction also provides another layer of data quality checks besides those performed in the control room by looking at strip charts and histograms filled during data collection.

### 3.3.2 The *BABAR* prompt reconstruction

#### 3.3.2.1 Data processing

The *BABAR* data are processed in a two pass Prompt Reconstruction (PR) system. The raw data (XTC files) are read in each pass, once to compute time-dependent or detector specific calibration constants, and then again to fully reconstruct the data. The system is named Prompt since the calibration pass is done within a few hours of collecting data and the reconstruction pass is completed within 12 hours.

The first pass, the Prompt Calibration (PC) fully reconstructs a representative subset of the raw data. The actual percentage of data used depends on the number of events in the XTC file. The PC pass computes various calibration constants which are recorded in the Conditions Database (CDB). The CDB tracks information related to the detector systems and the beam conditions as a function of data-collecting time. Most calibrations are calculated for each run, but a subset of these calibrations needs information collected over multiple runs and were then called Rolling Calibrations. A separate database was used to collect inputs from each run for the Rolling Calibrations. When enough statistics were collected, or some other criteria met, a Rolling Calibration was performed. An example of this procedure is the determination of the beamspot rolling calibrations described in Section 6.4. The output of both single run and rolling calibrations are written to the CDB with a validity period corresponding to the span of runs used. A copy of the updated conditions database is made available for the full event reconstruction (the second PR pass), for more specific physics event selection (skimming) and for general data analysis.

The second pass, the Event Reconstruction (ER), reads the raw data from the XTC files, the conditions and calibrations from the CDB, and performs the full physics event reconstruction; track finding, vertexing, PID, etc. Interleaved with this processing are two stages of event-filtering. The first uses only L3 output-line information (Section 2.2.6) to reduce the contribution of events collected solely for diagnostic or detector-calibration purposes (*e.g.*, Bhabha events, used for EMC calibration, etc., are reduced by a factor of 15 beyond the factors already applied in L3). The second, which follows DCH-track and EMC-cluster reconstruction, tests events against about a dozen physics-motivated filters. One filter is highly efficient for $B\overline{B}$ final states, but much less efficient for some other processes. Hence additional filters address particular low-multiplicity states relevant to tau physics, two-photon physics, and so on. If an event satisfies any of these filters, or the earlier L3-based filter stage, it is saved. Reasons for saving it are recorded with the event.

The output of the ER pass, the reconstructed events, is written to data collections which are archived and then made accessible to the skimming system and to the analysts. *BABAR* originally used an object-oriented database technology (Objectivity/DB) to store both the conditions and the reconstructed events, but later switched to a file-based `Root` I/O system (`XRootD`), first migrating the data storage (2003) and then also the conditions database (2007).

#### 3.3.2.2 Reprocessing

During the life of *BABAR*, as in any active experiment, the data reconstruction algorithms and the detector calibrations are constantly being improved. In order for the physics analysis to benefit from these improvements, it is necessary to reprocess the accumulated dataset, starting from the raw data. In *BABAR*, this reprocessing was done about once a year, in parallel with the prompt processing of the incoming data. The total throughput and resources needed for the reprocessing often exceeded the corresponding need for the current data. The allocation of resources needed to perform a reprocessing of the *BABAR* dataset was driven by several facts: first, the moment when a stable and improved reconstruction framework was available; then, the deadline by which to make the reprocessed data available for physics analysis, in order to prepare results for the next round of conferences; finally, the size of the particular dataset to reprocess.

The optimization of resources for the reprocessing is accomplished by breaking the conditions time-line into intervals and running separate instance of the two-pass processing system for each interval. The calibrations are computed within each separate interval and data run ranges corresponding to each interval can be processed in parallel. The reprocessed condition intervals are then merged into the Master CDB covering the whole time-line. The Master CDB is then used for accessing the current and reprocessed data.

A comprehensive bookkeeping system, based on a relational SQL database (Oracle or MYSQL), keeps track of all processing and reprocessing jobs indexed by run number. It records the date, time, software release and calibration used for that (re)processing of the data run, as well as status of the job (completed, failed, etc.) and other statistical quantities.

### 3.3.3 The Belle data reconstruction

#### 3.3.3.1 Data processing

There are three major periods in the Belle data processing scheme, designed to cope with increasing event rate as well as to monitor data quality more reliably.

In the $1^{st}$ period from 1999 to 2003, raw data acquired in the Belle DAQ system are recorded to tape. Then, once a tape becomes full and is released from the drive, offline processing starts reading raw data to perform event reconstruction (Adachi, 2004). This method only allows one to monitor data quality with a delay of several hours since one has to wait for a tape release to trigger the processing. In this first processing step, detector calibration constants are not updated and are usually taken from the previous experimental period with some necessary extrapolations applied. If one needs to process a run immediately after it has finished, that is possible, but only by forcing a change of tape. The delay in having processed data is reduced for that run at a cost of adding an overall delay, corresponding to the tape change, for processing all data. The reconstructed data are written to tape as a data summary tape (DST). Then the next step called "skimming" is done by reading DST (see Section 3.5.3), where one creates datasets containing physics events such as Bhabha events, $\mu$-pair events, and hadronic events on disk which can be accessed by users. Those physics datasets are used for checking detector response and producing calibration constants.

To improve the reconstruction chain, a computing cluster (PC farm) for a real-time reconstruction (RFARM) was introduced in 2003. Data sent by the DAQ system are received by the PC farm and reconstruction is done in parallel to the data acquisition (Itoh, 2005a). Output data are written in a hierarchy mass storage system (HSM) consisting of disks with a tape library as backend (Katayama, 2005). This upgrade enables Belle to obtain reconstructed events shortly after online data-taking, and precise data monitoring becomes much more reliable. The data quality assurance is one of the duties of persons on shift during the data taking. The skimming to select physics events is also carried out in the same way as before to provide calibration data for detector experts.

Following the initial success of the first RFARM system, the computing power in the RFARM doubled in order to be able to keep up with increasing luminosities in 2005. This configuration can process events at the highest KEKB luminosity without delay.

The Belle experiment employs a unique software framework `basf` (Belle AnalysiS Framework) and traditional data manipulation system with a `zlib` compression capability (`PANTHER` ) throughout for all phases in event processing and this simple management was scalable using the processing scheme mentioned above (Adachi, 2004). The software has been widely used not only for event reconstruction, but for all physics analyses without any serious issues.

#### 3.3.3.2 Reprocessing

Belle reprocesses all of the raw data once the detector calibration constants are obtained (Ronga, Adachi, and Katayama, 2004). Usually the first half of the annual data recorded from spring to summer is reprocessed to produce analysis datasets used to obtain new results to be presented in the summer conferences and the rest of raw data from autumn to winter is reprocessed for the winter conferences. The calibration constants used for reprocessing are computed by the detector experts using the physics events described above, once the experimental period (a couple of months) is completed, and another set of constants computed directly from data. Once constants for all detector elements are updated in the database (based on PostgreSQL) the reprocessing is carried out. In this step, output data are recorded in a compact form effectively used for physics analysis (mini-DST, MDST) on disk. Major physics analysis skims such as events containing $J/\psi$ candidates from $B$ decays are produced in an organized fashion to speed up individual analysis. More background-tolerant tracking algorithms (combination of Hough and conformal transformation) and improved calibration constants (polar angle dependent threshold for shower clusters in the ECL, new SVD alignment resulting in smaller $\Delta z$ bias for several experiments - see Section 6) are developed using a large amount of data, making detailed studies of detector response possible. These new features are applied in a consistent way by reprocessing the raw data sample of $\sim 560\,\text{fb}^{-1}$ (experiment 31 to 55) taken with the SVD2 vertex detector (see Section 2.2.1), in the so called "grand reprocessing", and the data processing of later experiments. The "grand reprocessing" was started in July 2009 and completed (including the calibration part) by February 2010. Due to lack of time and manpower available, many shorter runs of the earlier part of the Belle data sample, taken with the SVD1 vertex detector, were not included in this effort. At the same time new sets of Monte Carlo events are simulated with up-to-date decay information to improve the understanding of the nature of background. All Belle final physics results are in principle obtained from datasets produced in the grand reprocessing.

## 3.4 Monte Carlo simulation production

### 3.4.1 Introduction

Several Monte Carlo event generators are used to simulate the final states of $e^+e^-$ collisions. A final state is represented by a set of four-vectors originating from a common vertex near the $e^+e^-$ interaction point or from the source of a particular background. Once produced, the four-vectors are passed by the software framework to the detector simulation where they are tracked in the detector, taking into account the interaction between the particles and the different materials, and the electronic signals which mimic the detector response are computed.

In the following sections the Monte Carlo simulation production at *BABAR* is described, followed by Section 3.4.5 detailing some differences in the approach taken by Belle.

### 3.4.2 Event generators

The generators depend on theoretical models of interactions to calculate the four-vectors. A combination of events from both signal and background generators is required in order to produce a simulated event stream realistic enough to be essentially indistinguishable from real data. A variety of generators makes this possible, as well as allowing individual sources of signal or background to be studied independently.

#### 3.4.2.1 Signal generators

The production of hadronic events from the $e^+e^-$ collision through the decay of the Upsilon resonances and the direct production of $u\overline{u}$, $d\overline{d}$, $s\overline{s}$ and $c\overline{c}$ pairs, is handled by the `EvtGen` (Lange, 2001) package and the `Jetset` generator, otherwise known as `Pythia` (Sjöstrand, 1995). Collision vertices are sampled from beam parameters in the PEP conditions database or ASCII files. These parameters include beam energies, boosts and spot sizes.

$B$ decays, including *CP*-violating and other complex sequential decays are simulated using `EvtGen`. `EvtGen` is a framework in which new decay simulations can be added as modules. It uses decay amplitudes instead of probabilities for each node in the decay tree in order to simulate the entire decay chain, including all angular correlations. It also has detailed models for semileptonic decays and an interface to `Jetset` for the generation of continuum events ($u\overline{u}$, $d\overline{d}$, $s\overline{s}$ and $c\overline{c}$ production), and for generic hadronic decays including those of $B$ mesons.

Lepton pair events were simulated with `KK2F` (Jadach, Ward, and Was, 2000), which is a high precision electroweak Standard Model generator for $e^+e^- \rightarrow \tau^+\tau^-$ and $e^+e^- \rightarrow \mu^+\mu^-$ events, amongst others. It takes into account QED radiative corrections (up to second order), including hard bremsstrahlung. When $\tau$ pair events are produced, the $\tau$ decays are handled by the `TAUOLA` generator (Davidson, Nanava, Przedzinski, Richter-Was, and Was, 2012).

`AfkQed` (Czyz and Kühn, 2001) was used to generate hard photons from initial and final state radiation using lowest-order QED calculations. Other generators used included `Gamgam`, which produces exclusive 2-photon decays of $B^0$'s, `Diag36` (Berends, Daverveldt, and Kleiss, 1986), which generates 4-lepton final states, and `SingleParticle`, which generates one particle per event, using user-specified parameters.

To compute the PEP luminosity and the Bhabha scattering cross section `BHWIDE` (Jadach, Placzek, and Ward, 1997), a wide-angle Bhabha generator which has a theoretical accuracy of 0.5%, and `BHLUMI` (Jadach, Placzek, Richter-Was, Ward, and Was, 1997), a small-angle Bhabha generator, were used.

#### 3.4.2.2 Background generators

In real data, several background processes contribute to events and mimic (or hide) real signals. Some of these backgrounds may be removed during the data analysis, while others may not. In either case, it is necessary to simulate them in order to aid background subtraction or to mix them with the simulated signal. These backgrounds include Bhabha scattering, bremsstrahlung, QED background, initial state radiation, machine background, and cosmic rays.

Luminosity backgrounds from electrons or positrons striking the beamline or other machine elements outside the nominal detector acceptance, were simulated using `BHWIDE` and `BHLUMI` .

Lepton pair and two-photon events from QED background were generated by `Bkqed` (Berends and Kleiss, 1981) which also includes effects from radiative photons.

Machine backgrounds due to electrons and positrons striking apertures and photons from Compton scattering and bremsstrahlung from beam gas are simulated by `TurtleRead` (Barlow et al., 2005) which reads ASCII files written by the `Decay Turtle` ray-tracing program.

Cosmic ray muons were another source of background triggers for *BABAR*. To estimate this, the `HemiCosm` code shot muons inward from the upper hemisphere surrounding the volume of the simulated detector. The muons were sampled from the usual zenith angle distribution and one of three available momentum spectra.

All these background generators where mostly used during the design and construction phase of the *BABAR* detector. After the start of the data-taking the effect of the different backgrounds processes, including machine background and background hits from the detector electronic noise, was simulated by superimposing recorded random triggers to the signal events.

### 3.4.3 Detector Simulation

The purpose of the *BABAR* detector simulation is to take four-vectors from the generator stage and transport them through the detector geometry, where energy loss, production of secondaries, multiple scattering and decays can occur. As these particles pass through sensitive regions of the detector, their energy, charge and angle information is collected in order to generate raw, idealized hits, which consist of positions and energy deposits in the detector. These quantities are stored in persistent containers in the database for later use in the simulation of the detector response where idealized information is converted to realistic detector hits, blended with background data, and digitized. The resulting realistic hits are then passed to the reconstruction code where the full simulated event is built for later comparison with real events.

#### 3.4.3.1 `Bogus`, `SimApp`, and `GEANT4`

The software package which handled the generation of the raw hits on *BABAR* is called `Bogus`. It was an application layer built on top of the `GEANT4` simulation toolkit (Agostinelli et al., 2003) and was designed to model the *BABAR* detector geometry and materials, propagate particles through a varying magnetic field, perform particle interactions and decays, and provide scoring of detector hits.

`Bogus` was integrated into the *BABAR* software framework and designed to be fully compatible with its event scheme, allowing Monte-Carlo truth information to be added to the simulated *BABAR* event. The code which accomplished this, `BfmModule`, initialized the `GEANT4` kernel, extracted event generator tracks from the framework event, invoked `GEANT4` to propagate these tracks through the detector and wrote the propagated tracks and produced secondaries into the event framework.

This event was then passed to `SimApp`, the package responsible for simulating the detector response. Beginning with hits from `Bogus`, it converted them to digitizations which mimicked the real electronic output of the detector, that is, the ADC and TDC words. These were then mixed with corresponding digitizations from background frames obtained from random triggers recorded by the data acquisition.

Trigger conditions corresponding to a particular month of data-taking (see Section 3.4.4) were finally applied to the full event which was then sent to the reconstruction.

#### 3.4.3.2 Physics and transport processes

The physics of the initial $e^+e^-$ collisions and the decays of short-lived hadrons were handled by the event generators discussed above. All other physics processes, including $K_s$ and $\Lambda$ decays, and $\pi$ and K decays in flight, were supplied by `GEANT4`. In terms of shower development in the detector, by far the most important are the standard electromagnetic processes of multiple scattering, ionization, bremsstrahlung, pair production, Compton scattering and photoelectric effect. These processes are sufficient to describe accurately the energy distribution in the EMC. Hadronic processes, though less frequent, are important for the propagation of hadrons produced in the initial interaction and the hadronic secondaries they in turn produce. The processes used included elastic scattering and capture, as implemented by the `GEANT4` version of the `Gheisha` hadronic code (Fesefeldt, 1985), and inelastic scattering as implemented by the `GEANT4` version of the Bertini cascade (Bertini and Guthrie, 1971). The latter was especially useful for a reasonable propagation of kaons from $B$ decays.

The decay of long-lived particles was also handled by `GEANT4`, which used PDG (Beringer et al., 2012) branching ratios to determine the final state of the decays.

The default particle transport code in `GEANT4` is a Runge-Kutta stepper, but for *BABAR* this was deemed too slow. It was replaced by a specialized helical stepper which took advantage of the near-uniform *BABAR* magnetic field by taking large steps and using exact calculations of the intersection of helical tracks and volume boundaries.

### 3.4.4 MC production systems

Quite early in the history of the *BABAR* experiment, the simulation production used computing resources coming from over 17 production sites across the globe. Such distributed production was possible because the only data that needed to be available at the production sites were the background event collections and the conditions. Moreover, a missing production due to failed jobs was simply replaced with a new production of the same decay mode, but with different random number generation seeds. All this resulted in simple production management tools that were easy to install at production sites.

In *BABAR*, simulation production is done on a 'per month' basis, using background frames and conditions and calibrations corresponding to a specific month of data taking. Conditions and calibrations are read from the MySQL conditions database and were previously computed during the prompt calibration pass of the reconstruction of raw data or with a special offline analysis of the raw data for those conditions that require data samples larger than a single run.

The production is carried out in cycles corresponding to major updates in the simulation or reconstruction code. In all cycles, the number of Monte Carlo events was much larger than the number of events collected by *BABAR*. In the final cycle, the number of $b\bar{b}$ and $c\bar{c}$ events corresponded to a luminosity ten times higher than the luminosity of the detector data and to a luminosity three times higher for continuum events.

Unlike the detector data, the simulated data are automatically marked 'good' in the bookkeeping database.

Before simulation production at a site can start, a test production must be run and compared to the exact same production performed at SLAC. This tests the release installation, the accuracy of the conditions exported to the site, and the availability of the background collections. Recently, most of the major simulation productions have been done off-site while specialized productions were mostly done at SLAC (for maximum control). However, having multiple sites has been very useful when several varieties of production needed to be done at the same time.

All the Monte Carlo event collections are imported at SLAC and stored in a High Performance Storage System (HPSS, a large tape storage robot). From SLAC, they are exported to the remote sites according to the requests of the Analysis Working Groups (AWGs) that are doing their analysis at that site.

Currently, simulation production remains distributed although an eventual collapse back onto SLAC is foreseen.

### 3.4.5 Differences between BABAR and Belle simulations

Rather than implement stand-alone programs for event generation and simulation in Belle, these codes were integrated into the `basf` as user modules. In this way, a user could run an entire Monte Carlo production sequence — generation, simulation, reconstruction, skimming, and

analysis — in one `basf` job and therefore is able to take advantage of the parallel-processing of events built into `basf` if desired.

In practice, event generation in Belle was done in single-processing mode to avoid inadvertent repetition or overlap of random number sequences. The output files from this generation step were fed to the subsequent parallel-processing job for simulation and analysis (generated events were processed by several processors, one event at a time by each processor).

#### 3.4.5.1 Generators

In addition to `EvtGen` (Lange, 2001), Belle used the `qq98` (CLEO, 1996) event generator in the early years for $B$ decays. Other generators used by Belle included `CTOY` (written for Belle based on the `HemiCosm` code) for cosmic ray muons, `SG` for single tracks (including cosmic rays), `BHLUMI` (Jadach, Placzek, Richter-Was, Ward, and Was, 1997) for lepton pairs (with `TAUOLA` (Davidson, Nanava, Przedzinski, Richter-Was, and Was, 2012) for subsequent $\tau$ decays), `KK` (Jadach, Ward, and Was, 2000) for fermion pairs, and `AAFH` (Berends, Daverveldt, and Kleiss, 1986) for two-photon production of fermion pairs.

#### 3.4.5.2 Detector simulation

Belle used the `Fortran`-based `GEANT3` (Brun, Bruyant, Maire, McPherson, and Zanarini, 1987) toolkit for detector simulation (this was the dominant motivation for Belle's continued support of `Fortran`, alongside C++, in its software library). C++ wrappers were incorporated around the `GEANT3` toolkit to embed it within the `basf`. `GEANT3` was supplemented with a Cherenkov-light simulation (written in C++) to model light propagation within the Aerogel Cherenkov Counters (ACC). Four-vectors of the generated particles in an event were passed to `GEANT3`, which then pass them through the model of the Belle geometry and generate hits in the sensitive elements. Decays of long-lived particles such as $K_S^0$ mesons were handled by `GEANT3`. The simulation accounted for the evolution of the real detector's behavior (dead or hot channels, efficiency changes, geometry changes, and trigger-parameter tuning) via information tabulated in the master database by experiment and run number. Through user hooks provided in `GEANT3`, these hits were digitized (simulated ADC, TDC and latch responses) tailored to the detector element so that the output data stream would mimic the appearance of the real data, supplemented with the additional "truth" information from the simulation. At the conclusion of the simulation of each event, additional hits from real background events (recorded with a random trigger and filtered to avoid any events with reconstructed tracks or clusters) were superimposed on the event to mimic the background activity in each detector element. The method developed consists of overlaying a random-triggered real beam background event to a simulated signal event. The random-triggered event is taken during a beam run with a typical rate of 1-2 Hz. The beam background file, the collection of the random-triggered events, is created for each run. The beam background overlay procedure is applied to the output after the detector simulation. Thanks to this method, the run-dependent beam background effects can be reproduced in the simulation. However, because this overlay process is done after the digitization step, it is not possible to consider a pile-up effect of electric charge before the digitization. A data file containing these background events was recorded for each run. Background events were selected at random from the files for a given Experiment when simulating Monte Carlo data early on within Belle. Later in the life of the experiment background events were selected sequentially from the corresponding background file for a given run.

#### 3.4.5.3 Geometry

The detailed Belle detector geometry was modeled for `GEANT3` in a manner similar to that of *BABAR* for `GEANT4`. The magnetic field in Belle's interior was obtained from a tabulated map of the field's radial and axial components that extended from the beamlines to the yoke's exterior surface; this field was used by `GEANT3` for charged particle propagation. No uniform-field approximations were made in the Belle simulation.

#### 3.4.5.4 Physics and transport processes

Propagation, decay and interactions of all particles except the Cherenkov photons in the ACC were handled by the `GEANT3` toolkit. Also for the most demanding part, the Belle electromagnetic-calorimeter (ECL) simulation, no fast (*i.e.*, parametric) simulations were used. The `Fluka` (Fasso, Ferrari, Ranft, and Sala, 1993) code embedded in `GEANT3` was used to simulate hadronic interactions.

#### 3.4.5.5 Post-simulation track extrapolation

In the analysis phase of each event, whether simulated or real, Belle utilized the GEANT track-extrapolation package distributed with `GEANT3` to extrapolate each reconstructed charged track from the outer surface of the Central Drift Chamber (CDC) through the outer detectors; ACC, Time-of-Flight (TOF), ECL, and $K_L^0$ and $\mu$ detector (KLM). This proved quite useful in matching tracks to hits in these outer detectors.

#### 3.4.5.6 MC production systems on Belle

Generation, simulation, and reconstruction of $e^+e^- \to \tau^+\tau^-(\gamma)$ was done for the most part at Nagoya University and the output data files were stored there.

Monte Carlo production of generic $B\overline{B}$ decays, continuum ($e^+e^- \to q\overline{q}$) processes, and other specific signal processes were handled by KEK and the other institutions with significant computing resources. Grid computing became available for Belle's use fairly late in its lifetime and therefore did not play a significant role in Monte Carlo production. In Belle, the Monte Carlo Production Manager utilized a web-based production scheme that harnessed the CPU and storage capabilities at the remote institutions; the grid was treated as one of these 22 remote sites.

Each production cycle was defined by a set of experiments (and all of the real-data runs within each experiment) and the Belle software library that had been used to process the real data therein. Ten times the real integrated luminosity in $b\bar{b}$ events and six times that in continuum events (with $c\bar{c}$ handled separately from the lighter quarks) were produced in each MC production cycle. For data samples taken at energies other than $\Upsilon(4S)$ six times the accumulated luminosity in the data were simulated.

The Production Manager would first coordinate with each of the Site Managers to ensure that the remote site had the proper Belle software library installed and operating properly; this was done by exercising the remote library via several test jobs and then comparing several thousand output histograms with the reference histograms at KEK. The Site Manager at each validated site was then permitted to request the simulation of a sequence of experiments and runs via the web interface, upon which the KEK-generated event files and the corresponding background-event files were delivered to the remote site for MC production. Each job's simulated, reconstructed, analyzed and filtered outputs were delivered to KEK and tracked by the Site Manager, who was responsible for restarting any failed jobs. Each output file was read back in entirety upon delivery to KEK to verify its integrity. Once all jobs in the requested sequence were completed and delivered successfully, the Site Manager would record this via the web interface. On rare occasions when a site fell behind significantly in its commitment to deliver the requested sequence, the Production Manager would consult with the other Site Managers and then transfer the sequence to another site with spare capacity. KEK produced about half of Belle's generic-MC events; the other institutions produced the remainder (see Fig. 3.4.1).

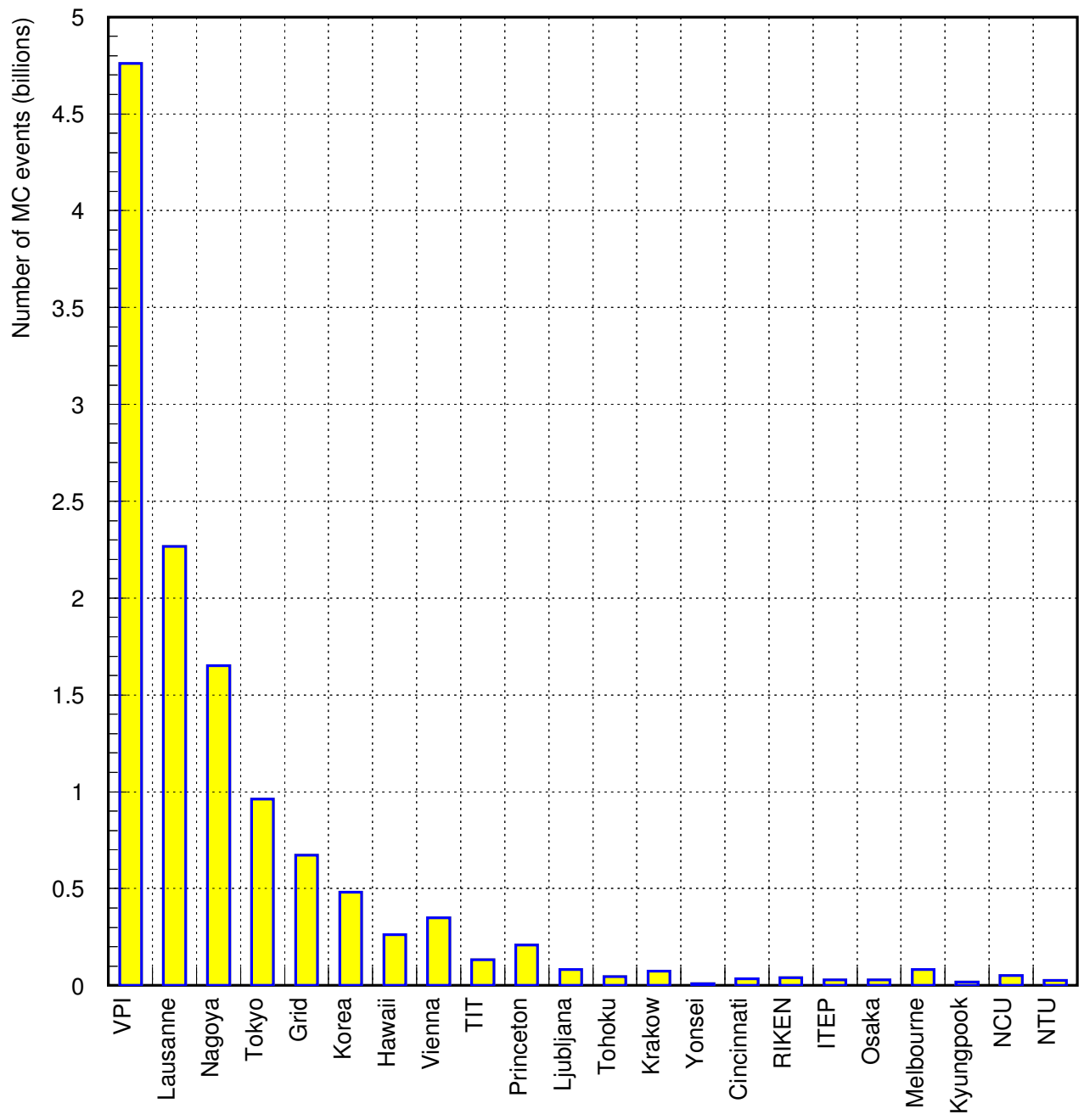

**Figure 3.4.1.** Generic MC production in Belle at remote sites (circa 2008).

## 3.5 Event skimming

### 3.5.1 Introduction: purpose of event skimming

The amount of detector and Monte Carlo data is such that it would be highly inefficient to have all analysts reading the full data sample. The identified solution was to centrally run an extra production step, the skimming, where events passing different sets of physics-motivated criteria were written to separate streams, the skims.

Each skim was optimized for a group of analyses using common sets of selected events as input. The fact that some analyses reached completion and new analyses started, resulted in skim definitions that were changing with time, new skims being added to production while others became obsolete and were removed. The two experiments adopted different skimming philosophies, *BABAR* introduced a large number of skims specific to analysis topics, whereas Belle had a limited number of skims strongly related to the selection of events produced in a type of process. The *BABAR* methodology is described below, and is followed by a more detailed description of the Belle skims as an illustration of how one can identify events of a given type.

### 3.5.2 Skimming in BABAR

*BABAR* analysis effort is organized into AWGs and each AWG is assigned to a particular site for the bulk of their analysis work. The skims relevant to a specific AWG are exported to the site of the AWG.

Events are organized into lists referred to as 'collections'. Events from the full reconstruction steps go into the 'AllEvent' collections. The outputs of the skimming step consist of the 'AllEventSkim' collections (with all the events that passed the skimming step) and of a set of collections for each skim. Skims can either be a full copy of the selected events (deep copy skims) or pointers to the events in the 'AllEventSkim' collections (pointer skims). The choice of the type of skims used depends on the fraction of selected events, on the need for detailed detector data, and on the availability of the 'AllEventSkim' collections at the AWG site.

Skim production was done in Skim Cycles and a couple of cycles had more than 200 output streams. Each 'AllEvents' collection, corresponding to a single run, was bro-ken into pieces and each piece was skimmed. The output streams coming from the pieces of the same 'AllEvents' collection are then merged. Finally, in order to create skimmed collections with a reasonable number of events, streams coming from different AllEvents collections were merged.

Only the 'AllEvents' collections declared 'good' by the 'Data Quality Group' (see Section 3.6) were used as input for the skimming procedure. All the skim jobs must have been completed and the output streams merged successfully to declare the skimming of several AllEvents collections which are part of the same skimming job as good. To have an efficient skimming production, monitoring, job crash recovery, disk clean up, and efficient data distribution are all critical elements. A set of software tools was developed to make this production possible.

As mentioned above, the level of analysis pre-selection that is available in skimming depends on the AWG requirements. To illustrate this one can consider the example a number of different Charmless $B$ decays to four-particle final states (where each particle is one of the following: $\pi^{\pm}$, $K^{\pm}$, $\pi^0$, $K^0_S$) which are studied within the so-called Quasi-Two-Body AWG within *BABAR*. A set of skims associated with these final states was developed by members of that working group to isolate $B$ decays of particular interest. While each of the possible final states is topologically similar, and in turn the analysis strategies for these decays are similar, there are different requirements placed on different channels. Hence analyses would use dedicated skims for a given combination of topology and final state. The decay $B^0 \to \rho^+\rho^-$ has two charged and two neutral pions in the final state. This used the 'BFourHHPP' skim variant, where H denotes a charged hadron (without any PID constraints imposed), while P denotes a neutral pion decaying into two photons. Similarly the decays to the four charged track final states $B^0 \to \rho^0\rho^0$ and $B^0 \to K^*K^*$ (with subsequent $K^* \to K\pi$ decay) used the 'BFourHHHH' skim. In this way each of these skims can be used to study a number of similar final states minimizing the time required by the data analyst to process the data. The ensemble of similar four body skims was also made available as the 'BFourBody' skim. This skim methodology is applied across the *BABAR* AWG system, where some skims are specific to the analysis of a given decay, while others are usable for a set of similar decays.

### 3.5.3 Skimming in Belle

After data processing, events taken by Belle are classified into several categories. Some of the categories such as Bhabha events, muon pair events and $\gamma$ pair events are used for detector calibration, while the following three categories are used for physics analyses:

1. a skim for hadronic events, called *HadronBJ*, which is mainly used for analyses of $B$ and charm mesons,
2. a skim for $\tau$-pair events, called *TauSkim*, which is mainly used for analyses of $\tau$ leptons, and
3. a skim for low multiplicity events, called *LowMult*, which is mainly used for two photon analyses

Further skims that contain smaller categories of physics events are made from these three basic skims and provided to individual analyses, so that users usually do not need to run over a huge number of events in the basic skim. Details of the second stage skim are described in the section of each analysis. Classification conditions for the three basic skims are described in the rest of this subsection.

*Hadronic event skim:*

*HadronBJ* events are selected primarily based on track multiplicity and visible energy: the event must have at least three charged tracks with a transverse momentum greater than 0.1 GeV/$c$ that originate from the vicinity of interaction point ($|\Delta r| < 2\,\text{cm}$ and $|\Delta z| < 4\,\text{cm}$), and the sum of the energy of charged tracks and reconstructed photons ($E^*_{vis}$) must be greater than 20% of $\sqrt{s}$. Note that all observables denoted by an asterix are measured in the CM frame.

These two selection criteria remove the majority of beam gas background and two-photon events. Beam gas background is further reduced by requiring the primary vertex position of the event, when the vertex is well-reconstructed, to be $|\Delta r| \leq 3.5\,\text{cm}$ and $|\Delta z| \leq 1.5\,\text{cm}$. Background events from radiative Bhabha and higher multiplicity QED processes are suppressed by requiring that two or more ECL clusters are detected at large angle ($-0.7 < \cos\theta^* < 0.9$), the average ECL cluster energy below 1 GeV, and the total ECL cluster energy ($E^*_{sum}$) to be below 80% of $\sqrt{s}$. $E^*_{sum}$ is also required to be greater than 18% of $\sqrt{s}$ since there are $\tau$-pair, beam gas and two photon events that have low energy sum. However, this condition is rather tight for light quark pair production events ($e^+e^- \to q\,\overline{q}$ with $q = u, d, s, c$), and hence a conditional selection is applied: $E^*_{sum} > 0.18\sqrt{s}$ or $HJM > 1.8$ GeV/$c^2$, where $HJM$ stands for heavy jet mass, which is the invariant mass of particles found in hemispheres perpendicular to the event thrust axis. The $HJM$ is the most effective variable to remove $\tau$-pair events, and it is required to exceed 25% of $E^*_{vis}$. However, in order to regain $q\overline{q}$ events, a conditional selection is required: $HJM/E^*_{vis} > 0.25$ or $HJM > 1.8$ GeV/$c^2$. These general conditions to select hadronic events turned out not to be very efficient for inclusive $\psi$ events. Therefore, the events with $J/\psi$ and $\psi(2S)$ candidates are explicitly added to *HadronBJ*.

*Tau pair events: TauSkim*

Signatures of the $\tau$-pair production, $e^+e^- \to \tau^+\tau^-(\gamma)$, are low-multiplicity and missing-momentum. Since at least two-neutrinos are missing in $\tau$-pair events, tight kinematic constraints can not be applied. So *TauSkim* is designed to reduce well defined Bhabha, $q\overline{q}/B\overline{B}$, two-photon and beam-gas background.

*TauSkim* events are selected primarily based on track multiplicity and the position of the event vertex: the number of charged tracks in an event must be at least two and less than 8, where each track must have a transverse momentum greater than 0.1 GeV/$c$ and originate from the vicinity of the interaction point ($|\Delta r| < 2$ cm and $|\Delta z| < 5$ cm). The net charge of the event $Q$ must be $|Q| \leq 2$. Beam gas background is reduced by requiring the primary vertex position of the event to be $|\Delta r_{\rm v}| \leq 1.0$ cm and $|\Delta z_{\rm v}| \leq 3.0$ cm.

Background from (radiative) Bhabha events is suppressed by requiring the sum of ECL clusters in CM ($E^*_{sum}$) to be below 11 GeV, and the polar angle of the missing momentum in the CM frame to be between 5° and 175° for two track events.

Background from two-photon events is reduced by requiring the maximum of the transverse momentum of the charged tracks ($P_t^{\rm max}$) to be greater than 0.5 GeV/$c$ and the sum of the visible energy $E^*_{vis}$ greater than 3 GeV, where $E^*_{vis}$ is the sum of the absolute momentum of charged tracks multiplied by $c$ and the photon-cluster energies in the CM: the photon cluster is the ECL cluster to which no charged tracks are associated. Even if $E^*_{vis}$ is less than 3 GeV, the events are accepted if $P_t^{\rm max} > 1.0$ GeV/$c$.

In order to further reduce the (radiative) Bhabha events, events with 2-4 charged tracks are rejected if the total energy $E^*_{tot}$ is greater than 9 GeV and the number of clusters in the barrel region ($30° < \theta^* < 130°$) is less than two, where $E^*_{\rm tot}$ is the sum of of the visible energy and the absolute value of the missing momentum ($E^*_{tot} = E^*_{vis} + c|p^*_{miss}|$). This condition reduces (radiative) Bhabha events where one electron or positron is detected in the Barrel calorimeter, but the energies of the other electron or photons are not measured correctly either by starting to shower in the tracking volume or missing energy from the shower in the gap between the barrel and end cap of the calorimeter.

With these selection criteria, about 80% of tau-pair events are kept while Bhabha and two-photon events are reduced to an acceptable level. If the events are passed by both the *TauSkim* and *HadronBJ* conditions, the events are kept in *HadronBJ*, while the remaining ones are kept in *TauSkim*. As a result both *HadronBJ* and *TauSkim* events are processed in physics analyses using the *TauSkim* sample.

*The low-multiplicity skim*

The low-multiplicity (LowMult) skimming of Belle data processing provides event-data collections mainly for analyses of zero-tag two-photon processes with an exclusive final-state system, $\gamma\gamma \to X$, including charged tracks in the final state (see Chapter 22 for the description of two-photon processes). The charged multiplicity of the target events is required to be two or four because of charge conservation, and the total visible energy is expected to be much smaller than the energy of the $e^+e^-$ collision.

The minimum requirement of the transverse momentum $p_t$ for charged tracks in two track events is chosen to be 0.3 GeV/$c$. Tracks must originate from the vicinity of the interaction point, which is $|\Delta r| < 1$ cm and $|\Delta z| < 5$ cm. For the four track events the additional two tracks are required to satisfy looser selection criteria, $p_t > 0.1$ GeV/$c$, $|\Delta r| < 5$ cm and $|\Delta z| < 5$ cm. For the four-track events, a looser constraint for the impact parameter of tracks is adopted to collect the $K^0_S K^0_S$ final-state events.

Only events with smaller visible energy, with the sum of absolute momentum of tracks $\Sigma|p| < 6$ GeV/$c$ and the sum of calorimeter cluster energies $E^*_{sum} < 6$ GeV, are collected, thus rejecting QED backgrounds with the full energy of beam collision deposited in the detector.

A further requirement on the missing-mass squared $MM^2 > 2\,{\rm GeV}^2/c^4$ is imposed to reject radiative events such as $\mu\mu\gamma$ where the photon travels in the forward direction and remains undetected. Any constraints originating from the trigger or particle-identification are not included in the requirements, in order to avoid introducing systematic uncertainties on the skimming efficiency from these sources.

In two photon events an approximate transverse-momentum ($p_t$) balance is expected. This was used in skimming of events with two charged tracks, applying loose selection on $p_t$ balance (where in the calculation of $p_t$ one also takes into account the calorimeter energy deposits for any number of $\gamma$ or $\pi^0$ candidates).

In addition, to salvage physics events where a track is mis-reconstructed or originates from noise (or from secondary interactions), a sub-category of events is skimmed using a condition on the visible energy $E^*_{vis} < 4$ GeV, when the event has at least two tracks. Processes with six tracks, such as $D^+D^-$ production, can be explored in this sub-category, although the skimmed data must be used together with the TauSkim and/or HadronBJ skims to recuperate events with the visible energy exceeding the above condition.

## 3.6 Data quality and $B$ counting

### 3.6.1 The control of data quality

Data quality control is crucial at each step of the data acquisition, from the initial readout of the detector following a positive trigger, to the final physics analysis. Therefore, *BABAR* and Belle have defined detailed procedures to validate each step of the data processing and to identify as quickly as possible any new hardware or software problem. These prescriptions have evolved over the years while the experiments were gaining experience. In the following, we will mainly focus on the final versions of the data quality procedures which were in use at the end of the data taking.

#### 3.6.1.1 Online data quality control in *BABAR*

The first level of data quality control is done in the control room. The shift crew relies on information from the slow control monitoring and DAQ systems to make sure that the detector is taking good data in a smooth way. Should an unexpected event occur, the diagnostics of the situation and the following actions are guided by well-established recovery procedures. If needed, the shift crew can also seek help by contacting a team of on-call experts – at least one per critical system of the experiment.

In *BABAR*, the standard shift crew was made of two people: the 'pilot', in charge of controlling the flow of the main data acquisition elements, and the 'Data Quality Manager' (DQM), whose main task was to check monitoring plots continuously. These histograms, classified by subsystem (SVT, DCH, etc.), accumulated data in real time during a run (usually about an hour long, unless a beam abort or some hardware problem ended it prematurely). About 15-20% of the events accepted by the L1 (hardware) trigger level were used for this fast monitoring. Most histograms could be directly compared with reference ones, automatically selected by the control system depending on the data taking conditions (colliding beams, single beam or cosmic events). Detailed guidance was also provided by each *BABAR* system to help the shift crew assess the quality of the runs. Therefore, it was very easy to spot a change in the behavior of a given hardware component (readout section with an occupancy unusually low or high, noisy channels, etc.) and to react appropriately. This information, combined with the detector status provided by the slow monitoring system (high voltage, low voltage, gas flow, temperature, etc.), allowed the shift crew to flag each run after it had ended. Flags assigned at the subsystem level included 'good', 'bad', 'unknown', and 'flawed'. The first three have obvious meanings while the fourth one was used to mark data in which the quality was not perfect, but would be worth processing for offline checks by experts. The global run flag was the worst among the subsystem flags: for instance, one subsystem flagged 'flawed' while the other ones got the mark 'good' would result with the run being assigned 'flawed' as global flag. Shift crews had two hours to flag a run after its end. This delay gave shifters the opportunity to get expert advice when needed. To avoid PC processing delays, it was crucial to give the right flag to each run in a timely manner as only colliding beam runs with 'good' or 'flawed' flag were automatically processed. Runs initially marked 'bad' and re-qualified as 'good' later could only be processed during the next round of reprocessing; in the meantime, their data were unavailable.

Most of the raw data that was marked 'bad' suffered from hardware failure. Although such a failure may have occurred in the final part of a run, all its data were potentially lost as the entire run would not be processed. In the worst case, up to an hour of *BABAR* data would be declared unusable, even if the failure occurred only in the last few seconds of data taking. Therefore, a software tool was developed during the final reprocessing to truncate these problematic runs and recover some good data. This procedure was conceptually simple, but involved significant bookkeeping subtleties. Ultimately, this tool added about $1\,\mathrm{fb}^{-1}$ to the final $\Upsilon(4S)$ dataset.

#### 3.6.1.2 Control of the data processing quality in *BABAR*

Data processing procedures could be subject to various problems, even when working with raw data designated as 'good'. To handle such complexities, this stage required dedicated quality assurance (QA) procedures which had to be (re)done for a given run each time it was (re)processed. Only runs that were declared good after data processing were included in the datasets used for physics analysis.

The two steps of the *BABAR* processing (PC and ER) generated a large number of `Root` histograms. The Data Quality Group (DQG), led by an experienced *BABAR* member, analyzed the primary histograms produced by the processing, and was responsible for the quality control of data produced by the experiment. This group also checked the consistency of the skimmed data, and validated software releases used to generate Monte-Carlo events. The DQG met weekly at SLAC – to facilitate face-to-face collaboration between the online and offline teams – to assess the quality of the runs processed in the past week. Experts (one per subsystem) used logbook entries and QA histograms to flag each processed run. They could also look at stripcharts showing the run-by-run evolution of key QA quantities (both at the detector level and after the event reconstruction) versus time. These were very useful to help identify trends which could indicate a developing problem. The processing classification was similar to the one used for online data: a run could be declared 'good', 'flawed' (meaning worth reprocessing, either immediately or after some further data correction) or 'bad'. This global flag, with optional related comments, was added to a database which kept track of all these checks and ensured that at most a single processing of a given run was used by analysts. Selecting good runs was of course a key task for the DQG group; but experts were also working hard to distinguish runs which were bad for well-identified and permanent reasons from those which might be later reprocessed successfully. To give an idea of the amount of work performed by the DQG, one can note that the whole *BABAR* dataset ($\Upsilon(4S)$, $\Upsilon(2S)$, $\Upsilon(3S)$ and the final energy scan) contains more than 35,000 physics runs in total. Only the common and constant efforts of both the operations and computing teams allowed *BABAR* to log 95% of the luminosity delivered by PEP-II and to give the analysts 99% of this dataset for physics. Indeed, a few $\mathrm{fb}^{-1}$ of data were recovered during the final reprocessing of the $\Upsilon(4S)$ dataset in 2008.

#### 3.6.1.3 Data quality monitoring in Belle

The monitoring of data quality was done in two levels at Belle. The first was the real time monitoring of detector signals based on sampled level 1 triggered events, which is called the Data Quality Monitor (DQM). The data of 10-20% of triggered events were sampled at the event builder and sent to the monitor PCs. The data were analyzed to examine the detailed operating status of each detector, and histograms were accumulated including the detector hit-map, the gain variation, etc. The histograms were placed in a shared memory so that the contents could be referred to without interrupting the data taking and are transferred to the browsing PC on request over the network. The task of monitoring the data was performed every 15 minutes by one member of the Belle shift crew, the so called "non-expert" shifter. Of course the title is misleading since the physicist on shift needed to be well acquainted with the detector in order to observe any deviation of the monitored distributions of recorded events from the expected ones. However it is true that the second shift member, the "expert" shifter, was usually a more senior member of the collaboration responsible for the data acquisition and the slow control monitors. In case of deviations evident in the DQM which the "expert" shifter was unable to resolve the corresponding detector experts were called in order to resolve any issues.

The second level of data quality check is the monitoring of data quality of the full event reconstruction and event classification. During the DST production, various higher level quantities were accumulated in histograms to facilitate maintaining a high data quality for physics analysis. This system is called the Quality Assurance Monitor (QAM), and is maintained by the QAM group. The histograms are checked whenever the DST for one run was made. At the beginning of the experiment, the DST production was performed offline and it took a few days to to obtain the result from the QAM. Therefore, timely feedback to the team responsible for data taking was difficult. After the introduction of RFARM in 2003, the DST production was fully integrated as a real time processing step, and the QAM was merged with the DQM. The RFARM was capable of full event reconstruction together with the event type classification, and the versatile monitoring of specific physics quantities became possible.

A mechanism to collect histograms from nodes processing data in parallel was implemented in RFARM and the histograms were collected and merged every 3 minutes during data taking. The resulting histograms were sent to the monitor PC of the DQM over the network so that they could be treated as a part of DQM histograms. The shifters checked both of DQM and QAM histograms in real time to verify and ensure the high quality of data being recorded.

The real time monitoring of QAM provided by RFARM was a powerful tool for the special runs such as the energy scan. For example, the distribution of the Fox-Wolfram moment ratio ($R_2$, see Chapter 9) could be obtained for hadronic events during data taking, giving the fraction of $B\overline{B}$ events in the sample in real time, and it was possible to know the beam energy of the current scan point precisely. It enabled "on-the-fly" determination of next scanning point so that the energy scan could be performed efficiently.

### 3.6.2 $B$-counting techniques

Knowing with the best possible precision and with well understood errors the number of $B$ meson pairs in the used data sample is of paramount importance for many of the analyses performed at the $B$ Factories. The techniques developed by *BABAR* and Belle to compute this number for a given set of data were made part of the central production activities to enforce quality control and consistency of the results.

#### 3.6.2.1 $B$-counting in *BABAR*

For the $\Upsilon(4S)$ running periods, the number of $B\overline{B}$ events in *BABAR* was computed by subtracting the number of hadronic events due to continuum interactions from the total number of the events in the on-resonance data set:

$$N_{B\overline{B}} = (N_H - N_\mu \cdot R_{off} \cdot \kappa)/\epsilon_{B\overline{B}} \tag{3.6.1}$$

where

- $N_H$ is the number of events satisfying the hadronic event selection in the on-resonance data;
- $N_\mu$ is the number of events satisfying muon pair selection criteria in the on-resonance data;
- $R_{off}$ is the ratio of selected hadronic events to selected muon pair events in the off-resonance (continuum) data;
- $\kappa \equiv \frac{\epsilon'_\mu \cdot \sigma'_\mu}{\epsilon_\mu \cdot \sigma_\mu} \cdot \frac{\sum_i \epsilon_i \cdot \sigma_i}{\sum_i \epsilon'_i \cdot \sigma'_i}$ corrects for the changes in continuum production cross section ($\sigma$) and efficiency for satisfying the selection criteria ($\epsilon$) between on and off-resonance center-of-mass energies. Off-resonance quantities are denoted by a prime. The subscript $\mu$ refers to muon pair events; the various contributions to the continuum hadronic cross section, primarily $e^+e^- \to q\overline{q}$, are denoted by the subscript $i$. Since the muon pair and $q\overline{q}$ cross sections vary similarly with $\sqrt{s}$ (0.7% difference between on- and off-resonance), $\kappa$ has a value close to 1. The quantity $N_\mu \cdot R_{off} \cdot \kappa$ is then the number of continuum hadronic events in the on-resonance dataset.
- $\epsilon_{B\overline{B}} = 0.940$ is the efficiency for produced $B\overline{B}$ events to satisfy the hadronic event selection, calculated under the assumption that

$$\mathcal{B}(\Upsilon(4S) \to B^+B^-) = \mathcal{B}(\Upsilon(4S) \to B^0\overline{B}^0) = 0.5. \tag{3.6.2}$$

Variations in the amount of non-$B\overline{B}$ decays of the $\Upsilon(4S)$, and in the branching ratios of $B^+B^-$ and $B^0\overline{B}^0$, are included in the systematic error, but are not significant.

The numbers of hadronic events and muon pairs for each run was found as part of the skimming process (see Section 3.5 above). The hadronic event selection was based on the number of charged tracks ($\geq 3$), the total measured energy, the event shape, the location of the event vertex, and the momentum of the highest momentum track. Muon pair events were selected using the invariant mass of the two tracks, the angle between them, and the energy associated with each track in the calorimeter. When no energy was associated with either track, at least one of the tracks was required to be identified as a muon in the IFR. This happened in roughly the 0.5% of the events, when backgrounds in the calorimeter (such as out-of-time Bhabha events) would cause a timing mismatch between the calorimeter and the tracking system.

The selection criteria were tuned to maximize efficiency for $B\overline{B}$ and $\mu^+\mu^-$ events while minimizing sensitivity to beam backgrounds. In particular, the analysis minimized the time variation of the efficiency for simulated $q\bar{q}$ and $\mu^+\mu^-$ events.

The residual non-statistical time variations of the efficiencies result in an uncertainty in $\kappa$ and a corresponding 0.27% systematic error on $N_{B\overline{B}}$. The other significant contributions to the overall 0.6% uncertainty on $N_{B\overline{B}}$ include 0.36% from the uncertainty in the fraction of events that fail the selection criteria, mostly low multiplicity $B\overline{B}$ decays that fail the requirement on the number of charged tracks, and 0.40% from the uncertainty in the modeling of the total energy distribution that translates into an uncertainty on the fraction of the events that fail the energy cut.

The total number of $B\overline{B}$ events (McGregor, 2008) in the nominal full dataset is $N_{B\overline{B}} = (471.0 \pm 2.8) \times 10^6$. In addition to the overall number quoted above, $N_{B\overline{B}}$ was tabulated for each run so that analysts could obtain $B$-counting and luminosity values for any subset of the full $\Upsilon(4S)$ dataset.

The numbers of $\Upsilon(3S)$ and $\Upsilon(2S)$ mesons produced in data sets collected at these resonances have been found using a similar analysis. In this case, the off-resonance continuum scaling was performed using $e^+e^- \to \gamma\gamma$ events, due to the non-negligible $\Upsilon \to \mu^+\mu^-$ branching fraction. The hadronic selection criteria were also modified for these analyses.

The $\Upsilon(3S)$ and $\Upsilon(2S)$ datasets contain $(121.3 \pm 1.2) \times 10^6$ and $(98.3 \pm 0.9) \times 10^6$ Upsilons, respectively. These numbers are calculated using hadronic events, with a correction for the fraction of leptonic decays that fail the hadronic selection.

The primary contributions to the systematic errors are uncertainties on the efficiency of the total energy selection (0.6%), the requirement on the number of tracks (0.4%), and the uncertainty on the $\Upsilon \to \ell^+\ell^-$ branching fractions (0.5%).

#### 3.6.2.2 $B$-counting in Belle

The final Belle $\Upsilon(4S)$ dataset contains $(771.6 \pm 10.6) \times 10^6$ $B\overline{B}$ events. As in the *BABAR* $B$-counting scheme, this number is obtained by a subtraction of off-resonance hadronic contributions, as measured by the number of events in the previously described $HadronBJ$ skim, from the total number of on-resonance hadronic events. In the Belle case, this is calculated as:

$$N_{B\overline{B}} = \frac{N_{on} - r(\epsilon_{q\bar{q}})\alpha N_{q\bar{q}}^{off}}{\epsilon_{B\overline{B}}} \qquad (3.6.3)$$

where

- $N_{on}$ is the number of events satisfying the hadronic event selection in the on-resonance data;
- $r(\epsilon_{q\bar{q}})$ is the ratio of efficiency for $q\bar{q}$ events off-resonance to the efficiency for those on-resonance;
- $\alpha$ is the ratio of the number of Bhabha ($e^+e^-$) events or $\mu$-pair events observed on-resonance to those observed off-resonance. This is described in more detail below;
- $N_{q\bar{q}}^{off}$ is the number of events satisfying the hadronic event selection in the off-resonance data;
- $\epsilon_{B\overline{B}}$ is the efficiency of the $\Upsilon(4S) \to B\overline{B}$ event selection criteria for on-resonance data.

The values of $\epsilon_{B\overline{B}}$ remained relatively stable throughout the lifetime of Belle. Although it was evaluated on an experiment-by-experiment basis, typical values were around 99% and differed by less than 0.5% over all experiments. The efficiency for $q\bar{q}$ events showed no strong dependence on energy, so $r(\epsilon_{q\bar{q}})$ was determined to be very near to 1, with variations of less than 0.3% over all data taking periods.

Aside from differences in these numerical constants, there is a notable difference from the *BABAR* approach. For most data taking periods, the off-resonance contributions are scaled using Bhabha events, rather than $\mu$-pair events. Originally, the average of $\alpha$ as calculated with Bhabha events and $\mu$-pair events was used for the final calculation. However, for data taken after spring of 2003, the $\mu$-pair efficiency became significantly less stable. This is attributed multiple effects, including changes to the trigger masks used in the dimuon event identification, as well as some inherent instability due to intrinsic timing variations in a subset of these trigger masks. For data following this period, only Bhabha events are used to calculate the value of $\alpha$.

Since the rate of fermion pair production is identical regardless of the type of fermion produced, the approach is effectively equivalent, regardless of whether $e^+e^-$ or $\mu^+\mu^-$ events are considered. However, the periods when both methods can be used to calculate $\alpha$ allow an estimate of systematic uncertainty on this value. This was determined to be a 0.5% uncertainty. This value is considered representative of the uncertainty on $\alpha$, even during data taking periods when $\mu$-pair events were not used for this calculation.

Systematic uncertainties are also assigned on the value of $r(\epsilon_{q\bar{q}})$, but these are a minor contribution to the overall error, less than 0.2% for all experiments. This uncertainty is consistent with the level of variation seen in $q\bar{q}$ efficiency as a function of run range during a single experiment, as evaluated by Monte Carlo events generated with conditions matched to those of the corresponding running period. A sideband in the $z$-position of the measured event vertex is used to study systematic uncertainties due to the inclusion of beam gas events, but such uncertainties are below 0.1%.

Ultimately, the uncertainty on $N_{B\overline{B}}$ is dominated by the systematic uncertainty from $\alpha$, and is approximately 1.5% for most of the Belle data.

The $B$-counting and $\bar{b}$ cross section measurement methodology used by Belle in the context of $B_s^0$ mesons collected at the $\Upsilon(5S)$ is discussed in detail Chapter 23.

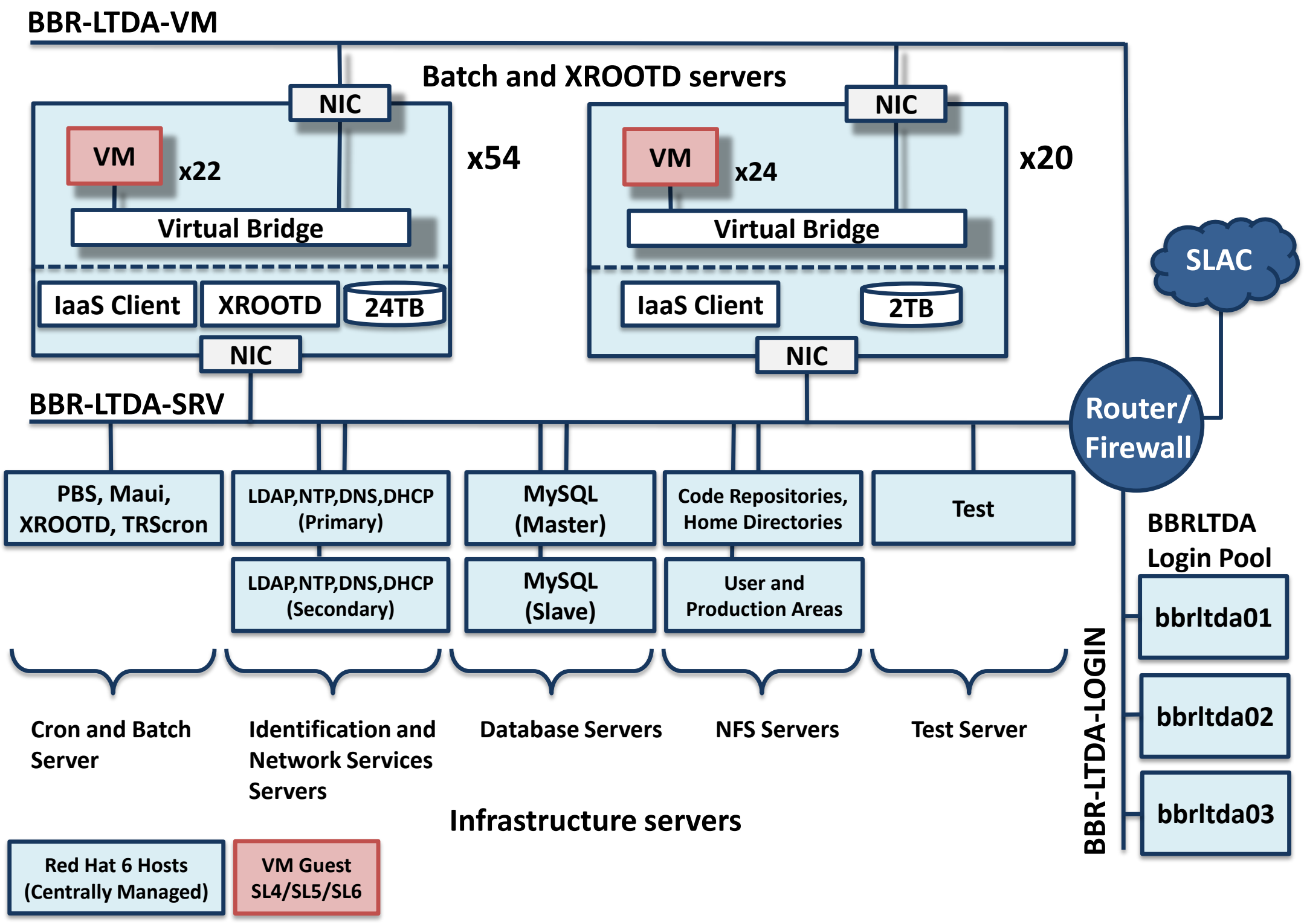

**Figure 3.7.1.** The LTDA cluster provides both storage and CPU resources in order to support analysis of *BABAR* data in the future. It includes database servers, code repositories, user home directories, working areas, production areas, and XRootD disk space. The isolation of back versioned components running on the batch system is implemented with firewall rules: virtual machines (VM's) are not allowed to connect to either the SLAC network or the world, and only well defined services are allowed between the VM network and the service network – see text for details.

## 3.7 Long Term Data Access system

### 3.7.1 The BABAR approach

The Long Term Data Access (LTDA) system is designed to preserve the capability of analyzing the *BABAR* data until at least the end of 2018. This requires the support of code, repositories, data, databases, storage, and CPU capacity. Special attention has to be devoted to the documentation. The system maintenance effort has to be minimized, including hardware maintenance, operating systems (OS) upgrades, tool upgrades, code validation, etc. The use of a contained system offers a controlled environment and simplifies documentation and user support. The *BABAR* analysis environment is supported with a frozen operating system infrastructure rather than actively migrating to future software environment as needed. The *BABAR* framework preserves its full capability of expansion and development, and is able to support future new analyses.

A long-lived frozen *BABAR* environment has to be maintained despite the evolving nature of hardware and OS. Also the support of back versioned OS is difficult, because future security exploits will require unknown patches. Hardware virtualization solves the hardware support problem for the foreseeable future and the use of OS images on virtual machines (VM's) solves the system administration problem, replacing it with the easier management of a small number of OS images.

The design of the LTDA cluster architecture takes into account the possibility that systems can be compromised from the security point of view and, in order to reduce risk to an acceptable level, a risk-based approach has been taken:

- Assume that systems that can be compromised, are actually compromised.
- Compromised components of the LTDA will be detectable by logging and monitoring.
- The LTDA will prevent accidental modification or deletion of the data.
- The dynamic creation of VM's from read-only images adds a small layer of security, by avoiding the compromised elements from being persisted beyond the destruction of the VM.

A representation of the cluster together with the layout of the network is shown in Figure 3.7.1. All sessions requiring back versioned platforms, including interactive sessions for debugging, run in VM's on the batch system. The isolation of the back versioned components is realized through firewall rules that are implemented in the LTDA switch. The LTDA network is composed of three subnets to which different elements of the cluster are attached. All the back versioned components (VM's) are connected to a VM subnet (BBR-LTDA-VM) and connection rules are enforced with the service network, (BBR-LTDA-SRV) including the VM's physical hosts and other infrastructure servers (always patched and up to date), and the login network (BBR-LTDA-LOGIN, always patched and up to date). The login pool is the only point of access for the users.

The LTDA batch resources are managed by PBS Torque (Torque, 2012) and Maui Scheduler (Adaptive Computing, 2012) is used as the job scheduler. The virtualization layer is implemented using QEMU (Qemu, 2012) and KVM (KVM, 2012). The data to which the user jobs need to access are managed by `XRootD` and staged on the disks of the batch servers on demand. Each batch and `XRootD` server has 12 disks of 2 TB, 11 of which are dedicated to `XRootD`. The last 2 TB disk of each server is used as a scratch area by the VM's running on the node. Each batch server has 12 physical cores of which one is dedicated to the host itself and the `XRootD` service. The other 11 cores are used to run virtual machines. With hyper-threading on, each node can run up to 22 VM's. The cluster also includes 20 servers used uniquely as a batch resource.

The LTDA cluster has been running in production mode since March 2012. All the active *BABAR* users have an account on the cluster with a 1GB NFS home directory. So far about 50 users have run jobs on the system while about 15 of them have made heavy use of the system. About 2 million jobs have been completed in the last year.

In almost one year of active use some fine tuning has been necessary. NFS connection parameters have been adapted to handle the high number of NFS accesses occurring when the queues are filled to their maximum capacity. On two occasions an upgrade of the host kernels has disrupted the system network. We have now established a validation procedure which allows us to test all the upgrades on a test machine, configured exactly like a batch server, before they are deployed to the entire cluster.

Monitoring of the servers, the services and the batch queues is also implemented. So far the cluster has met and exceeded the expectations.

### 3.7.2 The Belle approach

The Belle group recently discussed their policy on data preservation (Akopov et al., 2012). It was decided that the Belle data will not be released to the public domain until the time the statistics of Belle II supersedes the Belle data and all Belle members (and Belle II members) lose interest in Belle data. This situation will likely occur around 2017-2018, a couple of years after the commissioning of the SuperKEKB accelerator. Two approaches are considered to provide an environment to access Belle data even in the Belle II experiment period. One is porting the Belle software to the new computing system for the Belle II experiment. The other is converting the Belle data to the data format adopted in Belle II so that it can be read in the Belle II software framework. The former approach does not require significant modifications of the current software. However, every time the computing system is replaced with a new one (which typically takes place every three or four years at KEK) the portability of the data has to be confirmed. For the latter case, one needs to prepare conversion software from the Belle data format to the Belle II one. Furthermore, the Belle data conversion has to be done in a systematic manner considering the available hardware and human resource. But once it is converted, Belle users can keep using it in the Belle II software framework. In both cases, the current Belle data has to be migrated to a new format.

# Part B
# Tools and methods

## Chapter 4
## Multivariate methods and analysis optimization

***Editors:***
*Frank Porter (BABAR)*

**Additional section writers:**
*Piti Ongmonkolkul*

Multivariate analysis (MVA) is widely used to extract discriminating information from data. This chapter provides a general discussion of the most relevant MVA tools used by *BABAR* and Belle, their mathematical properties, and optimization methods. Specific multivariate algorithms used for charged particle identification (PID), $B$-flavor tagging and discrimination against background are described in Chapters 5, 8, and 9, respectively.

### 4.1 Introduction

The goal of analysis optimization is to make optimal use of the available data to perform a measurement of physical interest. Depending on the circumstance, the exact meaning of "optimal" may differ. However, the essential notions are those of efficiency (minimizing variance) and robustness. The goal of efficiency must be interpreted in the context of being unbiased, or negligibly biased. Robustness is used here in the broader sense, including both sensitivity to model errors and sensitivity to statistical outliers. An analysis that minimizes statistical uncertainties may not be optimal if the systematic uncertainties are large.

With the large, complex event samples from present experiments, plus the improvements in computing technology, analysis methods have evolved. This evolution has been aided by advances in the available statistical methodologies.

The optimization problem may be viewed as a problem in classification: For example, we wish to classify a set of events according to "signal" or "background". Thus, we have the problem of optimizing a binary decision process. This may be generalized to more than two classes, but the binary decision covers much of what we do. Another possible approach is to define some weight, or probability for each event to belong to the various classes. The technique of ${}_s\mathcal{P}lots$, discussed in Chapter 11, provides such an example.

It should be remarked that there are many variations on the methods presented. The discussion here is introductory rather than comprehensive. The reader is referred to the text by Hastie, Tibshirani, and Friedman (2009) for a more complete treatment of most of this material.

### 4.2 Notation

As is common in physics, we adopt an informal notation eschewing a notational distinction between a random variable and an instance. Our variables may be discrete or continuous, but for convenience the treatment here is in terms of continuous variables. The particle physics notion of an "event" maps easily onto the statistical concept of "event".

We suppose that each event corresponds to an independent identical random sampling in an $\ell$-dimensional sampling space. An event is described by the vector $\boldsymbol{x} = (x_1, \ldots, x_\ell)$. The variables used to optimize the selection of events are called "selection variables". We'll denote these with the symbol $\boldsymbol{s} = (s_1, \ldots, s_k)$. These are functions of the sampling vector, $\boldsymbol{s} = \boldsymbol{s}(\boldsymbol{x})$. In some cases, $\boldsymbol{s}$ is simply a subset of the $\boldsymbol{x}$ variables. The dimension, $k$, of $\boldsymbol{s}$ may itself be varied during the optimization process. The term multivariate is used to describe situations where we analyse a multi-dimensional hyperspace $\boldsymbol{s}$, using some well defined methodology.

The means of the selection variables are denoted $\boldsymbol{\xi} = (\xi_1, \ldots, \xi_k)$. The covariance matrix is

$$\Sigma = E\left[(\boldsymbol{s}-\boldsymbol{\xi})(\boldsymbol{s}-\boldsymbol{\xi})^T\right], \tag{4.2.1}$$

where the "$E$" denotes expectation value. Uncertain parameters of the distribution of the selection variables are denoted with $\boldsymbol{\theta}$. If there are $r$ such parameters, we denote them as $\boldsymbol{\theta} = (\theta_1, \ldots, \theta_r)$. The quantities $\boldsymbol{\xi}$ and $\Sigma$ may be functions of $\boldsymbol{\theta}$. Estimators for $\boldsymbol{\theta}$ are denoted $\widehat{\boldsymbol{\theta}}$. If the sampling distribution for the selection variables is multivariate normal, the corresponding density is

$$N(\boldsymbol{s};\boldsymbol{\xi},\Sigma) \equiv \frac{1}{\sqrt{(2\pi)^k \det \Sigma}} \exp\left[-\frac{1}{2}(\boldsymbol{s}-\boldsymbol{\xi})^T \Sigma^{-1}(\boldsymbol{s}-\boldsymbol{\xi})\right]. \tag{4.2.2}$$

### 4.3 Figures of merit

We often reduce the optimization of an analysis to the problem of maximizing or minimizing the expected value of a figure of merit (FOM). "Loss functions", typically making some estimate of error rate, are often used for this, and are discussed, for example, in Hastie, Tibshirani, and Friedman (2009). Here, we mention some of the more common FOMs used specifically in particle physics.

If we are looking for some yet unobserved new effect, we might optimize on the expected significance of that new effect. Suppose $S$ is the expected number of signal events after selection (depending on the analysis), and $B$ is the expected number of background events, which we assume we can estimate from known processes. The total number of events observed is $N$, including both signal and background. The size of a possible signal is estimated according to $\widehat{S} = N - B$. An estimate for the size of fluctuations in background is $\sqrt{B}$. Thus, $S/\sqrt{B}$ is related to the significance of a possible signal. In such a measurement, this provides a figure of merit to be maximized. The left side of Fig. 4.3.1 shows an example of this (with detection efficiency substituting for $S$, that is, the efficiency is $S$ divided by expected number of produced signal events in the dataset) in the Belle analysis searching for $\tau \to \ell hh'$ lepton flavor violating decays (Miyazaki, 2013). Another example can be found in Section 18.4.4.2, where the analyses that resulted in the observation of $\eta_b(1S)$ and $\eta_b(2S)$ mesons used the test statistic $S/\sqrt{B}$ to optimize event selection criteria.

Another approach to a figure of merit for the case of a search for a new effect has been suggested by Punzi (2003b). This approach defines a "sensitivity region" for the possible parameters, $m$, of the new effect. This definition is based on the confidence level of the region for $m$ that will be quoted if evidence for a new effect is not claimed. The figure of merit then corresponds to maximizing the size of the sensitivity region. A simple form of this figure of merit is

$$\frac{\epsilon}{n_\sigma/2 + \sqrt{B}}, \tag{4.3.1}$$

where $\epsilon$ is the efficiency to observe a signal event, $B$ is the expected background, and $n_\sigma$ is the desired one-tailed significance (in order to claim a discovery) of an observation expressed in standard deviations of a Gaussian probability distribution. This FOM has been used in some analyses, for example, in *BABAR*'s search for for $B^+ \to \ell^+\nu_\ell$ recoiling against $B^- \to D^0\ell^-\bar{\nu}X$ (Aubert, 2010a).

On the other hand, we may wish to get the most precise measurement of some known process. In this case, the signal is proportional to $S$, and the estimated error on the signal is $\sqrt{S+B}$ (*i.e.*, the expected fluctuation on the total number of events). Thus, to optimize on precision (of signal yield), we wish to maximize the expected value of $S/\sqrt{S+B}$. This can be viewed in an equivalent form: suppose that there are a total of $N_S$ signal events in the dataset before event selection. Selection involves some efficiency, $\epsilon$, to select signal events, so that we expect $S = \epsilon N_S$. Then this FOM may be expressed as $\sqrt{N_S}\sqrt{\epsilon \cdot S/(S+B)}$, where the factor $S/(S+B)$ is the signal purity in the selected sample. This makes explicit the trade-off between efficiency and purity in the optimization.

Of course, the idea of optimizing precision applies more generally than measurements of signal strength, for example in the measurement of $CP$ asymmetries. An example of optimizing on expected precision is shown in the right side of Fig. 4.3.1, for the Belle analysis measuring $y_{CP}$ in $D^0 - \overline{D}^0$ mixing (Staric, 2007).

In practice, in a complicated analysis, the optimization process is usually broken into more-or-less disjoint aspects, such as topological background suppression (*e.g.*, Chapter 9) or particle identification (Chapter 5). For these situations we often optimize on the signal purity, $S/(S+B)$, or equivalently, the "signal-to-noise": $S/B$. For example, in the optimization of PID, the goal is to get the best efficiency for the desired particle type for a given contamination probability, or variations on this idea. A useful graphical tool is known (from its engineering origins) as the "receiver operating characteristic", a plot showing the trade-off between efficiency and purity, or variants. Fig. 4.4.1 provides an example in the context of PID, discussed later in this chapter. The idea is used as well in $B$ meson reconstruction, for example in Fig. 7.4.3. Depending on the application, it may be acceptable to have a greater or lesser contamination. That is, we may not optimize strictly on the particle identification purity in the context of a given analysis. This leads to the provision of several PID selectors. In principle, the particle identification could be optimized along with the subsequent analysis, but this is unwieldy, and the provision of a choice of selectors approximates this. Providing pre-defined selectors also facilitates re-use of work done to estimate systematic uncertainties.

There are still other figures of merit that may be used in classification problems. The misclassification error, equal to the fraction of the sample that is incorrectly classified may be used. In building decision trees, two variants of this idea are commonly adopted, the "Gini index" and the "cross-entropy". These FOMs are available in most multivariate classification packages in use in HEP and are defined below in the discussion on decision trees, although their application is not limited to decision trees.

## 4.4 Methods

Statistical methods and tools of increasing sophistication used to optimize analyzes are described in the remainder of this chapter. Beforehand, it is important to stress that for many methods to be successful, two mandatory steps are required : training and validation. There are a few exceptions to this rule, where one can analytically compute the parameters required to perform an optimization.

It is dangerous to optimize a selection with the actual data that is to be used in the measurement. Such an approach is prone to tuning on fluctuations and the production of biases. For a simple example, suppose we are tuning an analysis for a particular signal, using the actual data. If we try to optimize $S/B$, say, we will find selection criteria that tend to favor signal-like events, tuning on any upward fluctuations. This will tend to bias our measurement of the signal strength toward high values. Nevertheless, this has been done extensively in particle physics, sometimes successfully, but sometimes with disastrous results. With an awareness of the issues, *BABAR* and Belle have gone to some length to avoid relying on the measurement data for the optimization. Note that these issues are discussed in a somewhat different context in Chapter 14.

Thus, *BABAR* and Belle take the approach of using a training dataset for the optimization. This could be simulated data, sidebands to the data that will not be used in the measurement, or a dataset that has similarities with the measurement data. A feature of the training dataset is that it is known (or known well enough) which class each event belongs to, so that the FOM may be computed. The selection criteria are optimized using the training dataset, then applied to perform the desired measurement.

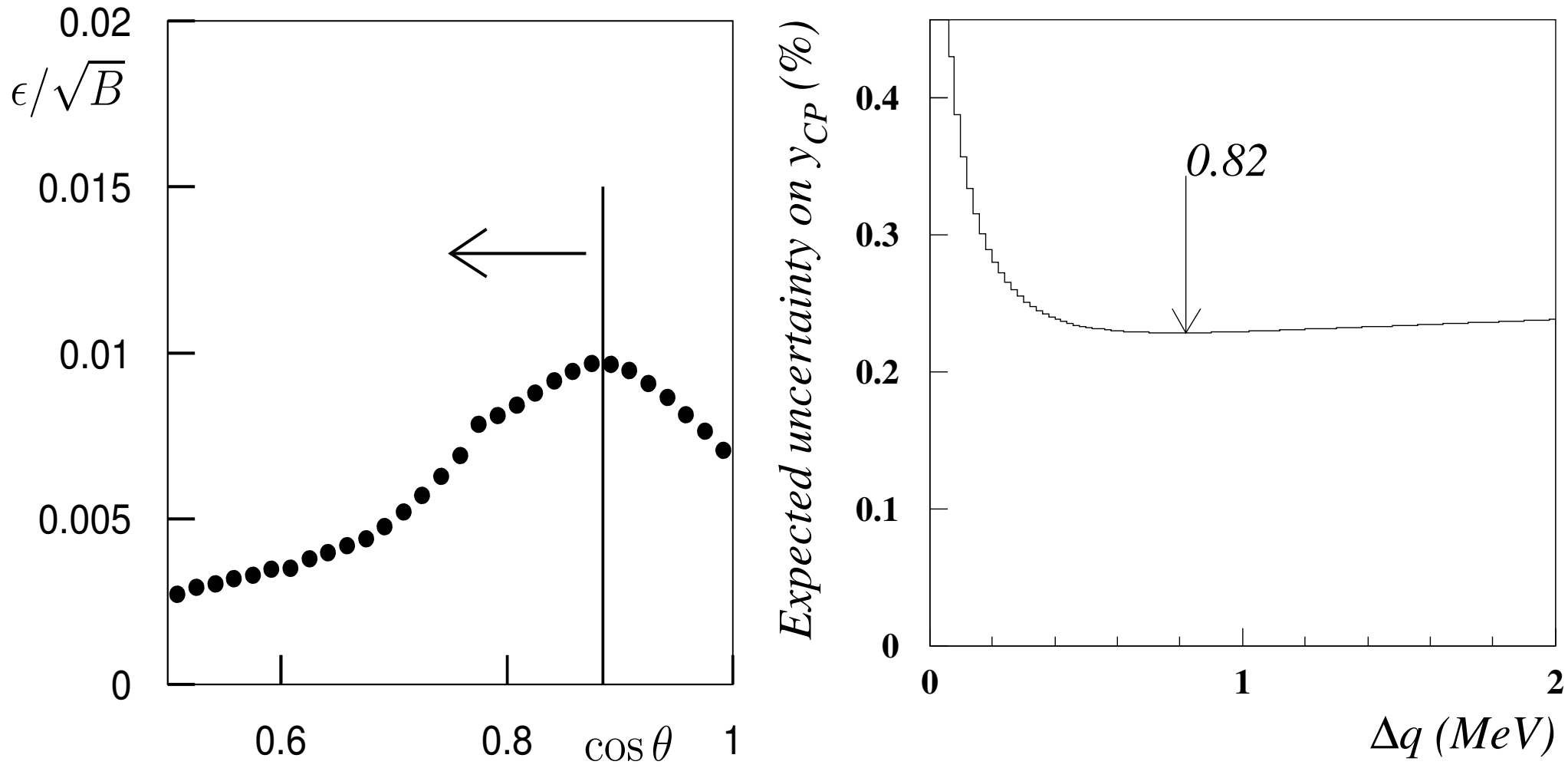

**Figure 4.3.1.** Examples of figures of merit used in optimization of Belle analyses. Left: Optimization on $\epsilon/\sqrt{B}$ in the search for the lepton flavor violating decay $\tau \to \mu\pi\pi$ (see Chapter 20). The horizontal axis is the cosine of the angle between the missing momentum vector and the direction of the tagging charged particle, in the CM frame. Belle internal, from the (Miyazaki, 2013) analysis. Right: Optimization on expected uncertainty in the measurement of the $y_{CP}$ parameter in $D^0 - \overline{D}^0$, see Section 19.2.3. The horizontal axis is the measured kinetic energy released in the candidate $D^*$ decay. Belle internal, from the (Staric, 2007) analysis.

A further refinement in method is the notion of validation. It is possible that the training dataset contains fluctuations that result in criteria that are not broadly optimal. This is related to the problem of "over-training", in which the training provides a model exquisitely tuned to the training sample, but with no real advantage on an independent sample. Effectively, the model is made very complicated when the underlying distribution is simpler. Since the training must be useful on an independent sample (it has to "generalize"), this erratic tendency has to be regularized in some way. For example, another dataset may be used to "validate" the selection and stop the optimization procedure (training) when no further improvement is obtained. This helps to avoid the phenomenon of over-training. A variant on this is "cross-validation", in which the training dataset is split into multiple equal subsets, and each of the subsets is used to validate the training on the remaining (aggregated) subsets.

The estimate of the efficiency obtained using the training/validation datasets may be biased too high. This is because the final selection criteria actually depend on both the training and validation datasets, and fluctuations in either dataset may affect the tuning in the optimization. To avoid this, a further independent "test" dataset, not used in the optimization process, may be used to obtain an unbiased efficiency estimate.

Some classification methods lend themselves more easily than others to interpretation, for example, in deciding how important the various inputs are. However, for a complicated problem a dedicated procedure may be required to understand which variables are most important, and perhaps eliminate ones that are not useful. A simple approach is to remove one or more variables at a time to see the effect of this on the classifier performance.

### 4.4.1 Rectangular cuts

When variables are uncorrelated, a selection may be optimized by looking at the effect of each variable in turn. This gives a selection region that is a hyper-rectangle in the space of selection variables, with sides aligned with the coordinate axes of the selection variables. Such selection criteria are known as rectangular cuts. They have the merits of ease of application, optimization, and interpretation. They are widely used, especially in "pre-selection" (*e.g.*, skim production) where the selection is still relatively inclusive, and more sophisticated optimization is not essential.

This simple approach may be used even if variables are correlated, however the result may no longer be optimal. In this case it may be possible to do considerably better with more sophisticated methods. For example, a refinement is possible, in which arbitrary regions of sample space may be approximated by sequences of rectangular cuts. A form of this approach is the technique of the decision tree, described further below.

When there are correlations among variables, we may also look for transformations that produce a set of uncorrelated variables, and then apply rectangular cuts in the transformed space.

### 4.4.2 Likelihood method

The likelihood function provides a mapping of the observations with often beneficial properties. This is employed,

for example, in the "likelihood method" for particle identification (Chapter 5). In this approach, detector measurements such as $dE/dx$, time-of-flight, calorimeter response, and muon detector response are combined by multiplying their likelihoods for a given particle type interpretation. Then rectangular cuts are applied to ratios of these likelihoods for different particle hypotheses. This approach to combining the available information has the merits of ease of application and interpretation. It also has some motivation from the fact that the likelihood ratio provides a uniformly most powerful test in the case of simple hypotheses. Table 5.2.1 shows a comparison of "cut-based" (that is, making rectangular cuts on the basic detector quantities) and "likelihood based" muon selection: for an efficiency loss of less than 10%, the likelihood method decreases the pion contamination by approximately 30%.

The likelihood function is constructed from the sampling *p.d.f.*, so the form of the distribution must be known including any correlations among variables. This can be a difficulty with this approach if this information is not readily available. The "supervised learning" methods (neural networks and decision trees) described below have an advantage in this respect, because subtle features, including correlations, are usually included automatically in the training samples. Maximum likelihood fits have been used widely at the $B$ Factories and are discussed in Chapter 11.

### 4.4.3 Linear discriminants

A linear discriminant is some linear function of the sample event variables:

$$L = A + \boldsymbol{B} \cdot \boldsymbol{s}, \tag{4.4.1}$$

where $A$ and $\boldsymbol{B}$ are independent of $\boldsymbol{s}$. The idea here is that $L$ may be such that it tends to take on different values for different classes (*i.e.*, signal or background) of event. Thus, $L$ may be useful for event classification. The optimization process here is to select those values of $A$ and $\boldsymbol{B}$ that produce the best FOM.

The most commonly used linear discriminant is the "Fisher discriminant" (Fisher, 1936), motivated in the case of multivariate normal sampling. If signal is described by $f_S(\boldsymbol{s}) = N(\boldsymbol{s}; \boldsymbol{\xi_S}, \Sigma_S)$ and background by $f_B(\boldsymbol{s}) = N(\boldsymbol{s}; \boldsymbol{\xi_B}, \Sigma_B)$, we may form the logarithm of the likelihood ratio for an event to be signal or background:

$$\begin{aligned}\ln\lambda &= \ln\frac{w_S f_S(\boldsymbol{s})}{w_B f_B(\boldsymbol{s})}\\ &= \ln\frac{w_S}{w_B} - \frac{1}{2}\ln\frac{\det\Sigma_S}{\det\Sigma_B} - \frac{1}{2}\left(\boldsymbol{\xi}_S^T\Sigma_S^{-1}\boldsymbol{\xi}_S - \boldsymbol{\xi}_B^T\Sigma_B^{-1}\boldsymbol{\xi}_B\right)\\ &+\boldsymbol{s}^T\left(\Sigma_S^{-1}\boldsymbol{\xi}_S - \Sigma_B^{-1}\boldsymbol{\xi}_B\right) - \frac{1}{2}\boldsymbol{s}^T\left(\Sigma_S^{-1} - \Sigma_B^{-1}\right)\boldsymbol{s},\end{aligned} \tag{4.4.2}$$

where $w_S$ and $w_B$ are the probabilities (weights) for an event to be signal or background, respectively. If the covariance matrices for signal and background are the same, $\Sigma_S = \Sigma_B = \Sigma$, then

$$\begin{aligned}\ln\lambda = \ln\frac{w_S}{w_B} &- \frac{1}{2}\left(\boldsymbol{\xi}_S^T\Sigma^{-1}\boldsymbol{\xi}_S - \boldsymbol{\xi}_B^T\Sigma^{-1}\boldsymbol{\xi}_B\right)\\ &+\left(\boldsymbol{\xi}_S - \boldsymbol{\xi}_B\right)^T\Sigma^{-1}\boldsymbol{s}.\end{aligned} \tag{4.4.3}$$

This is now a linear expression in $\boldsymbol{s}$, referred to as the "Fisher discriminant".

If any of $\boldsymbol{\xi}_{B,S}$ or $\Sigma_{B,S}$ are unknown, they must be estimated, for example with a least-squares or maximum likelihood fit to the entire dataset. It is important to remember the assumption that $\Sigma_B = \Sigma_S$. There is no general reason why this should be true. If not equal, improvement (possibly substantial) in the analysis may sometimes be obtained with the full "quadratic discriminant" of Eq. (4.4.2). This is discussed and demonstrated with a simple example in Narsky (2005b, Section 2.1). Linear discriminants have been used widely at the $B$ Factories, for example see Section 9.5 which contains a detailed description of the Belle strategy for continuum background suppression for $B$ meson decay analyses.

### 4.4.4 Neural nets

The basis of the neural net (see, for example, Haykin (2009); MacKay (2003) for thorough developments) is a model for biological neurons, in which the firing of a neuron occurs once the summed "inputs" cross some threshold. In practice, this discontinuous behavior is smoothed out to a continuous function such as the sigmoid:

$$\sigma(X) = \frac{1}{1+e^{-X}}, \tag{4.4.4}$$

where $X$ is a parameterized function of the inputs (for example, Eq. (4.4.5) below). As with other classification methods, the neural net is trained, validated, and tested on datasets with known outcomes. The training involves optimizing the values of parameters in the net to, for example, minimize classification error.

The simplest neural net consists of one "neuron". Suppose the function $X$ is of the linear form $X = \sum_{i=1}^k w_i s_i + b$ (which is the same form as a Fisher discriminant). To use this net as a binary classifier, we choose a threshold $X_c$ such that if $X > X_c$, the net returns a one, otherwise it returns a zero. Such a basic element is called a "perceptron", which represents a decision boundary in the problem space. Complex networks may be built out of these. Note that the function of the parameters $\boldsymbol{w}$ is to assign weights to the different inputs, and the parameter $b$ acts as a "bias", changing the location of the decision threshold but not the relative weightings.

A feed-forward neural net (or "multilayer perceptron") consists of layers – an input layer, an output layer, and any number of "hidden" layers in between. Each layer has a number of nodes that take inputs from the next lower layer and provide outputs to the next higher layer. The input layer consists simply of the $k$ selection variables $s_i, i = 1, \ldots, k$, each variable represented by a node. Let us suppose for this discussion that our network has a single hidden layer. Each node in the hidden layer represents a numeric value obtained by a non-linear transformation on a linear combination of the input nodes. For example, using the sigmoid, the hidden nodes $h_1, \ldots, h_p$ compute the values:

$$h_i = \sigma\left(\sum_{j=1}^{k} w_{ij} s_j + b_i\right), \quad i = 1, \ldots, p. \qquad (4.4.5)$$

The inclusion of a constant bias term, $b_i$, may be thought of as including a linear term corresponding to an additional input equal to the constant one.

The output layer may consist of multiple nodes for multiple classes; often we have two output nodes, which logically may be taken as a single output, as appropriate for the two-class "signal" vs "background" selection. We will assume this case here. The output is computed from the hidden layer by taking linear combinations of the hidden layer results,

$$y_j = a_j + \sum_{i=1}^{p} c_{ji} h_i, \quad j = 1, 2. \qquad (4.4.6)$$

We may then obtain a number between 0 and 1 expressing the output of the neural net, for example, by

$$t = e^{y_1} / \left(e^{y_1} + e^{y_2}\right), \qquad (4.4.7)$$

where $y_1$ is the "signal" class output. In the two class problem, a single output is often taken using the sigmoid where $t \equiv \sigma(y_2 - y_1)$; Eq. (4.4.7) is a generalization that may be extended to an arbitrary number of classes. Once the neural net is trained, large values of $t$ indicate signal; an analysis can make an event selection based on $t$. It may be remarked that the difference between the neural net and a linear model is the use of non-linear "activation functions"; in the present example, the sigmoid.

Training of the neural net consists in searching for optimal values of the net parameters, where optimal is defined in terms of minimizing a measure of the classification error rate. For the example net, this training corresponds to finding values for the $p \times k$ parameters $\boldsymbol{w}$, the $p$ parameters $\boldsymbol{b}$, the two parameters $\boldsymbol{a}$, and the $2 \times p$ parameters $\boldsymbol{c}$. The optimal values are often found by a gradient descent method, referred to as "back propagation" in this context.

A popular methodology is the Bayesian neural network (for example see the discussion on hadronic tag reconstruction for Belle in Section 7.4.1). In this case, the output of the net is interpreted as a posterior probability to be, *e.g.*, signal. Regularization of the network may be achieved with the help of prior distributions (often Gaussian) in the parameters.

### 4.4.5 Binary decision trees

The idea of a binary decision tree [see for example Hastie, Tibshirani, and Friedman (2009, Chapter 9)] is a recursive search for the best binary selection over the set of variables. Given a (training) dataset, we search for the variable and a selection (or "cut") value which provides the best FOM. This split results in two "nodes", one classified as "signal", the other as "background". A new search is applied to each of these nodes, resulting in two further splits. The process is repeated until further splits do not improve the FOM or fall below a specified minimum number of events. Trees that are grown by the latter criteria may be "pruned" to eliminate splits that fail some worthiness criterion. The result is a set of rectangular regions in our selection variable space, each classified as either signal or background. In the tree analogy, the set of final nodes at the end of the chain are called "leaves"'.

In binary decision trees, a commonly used FOM, besides simply computing the average error (misclassification error), is the "Gini index", $G(p) = -2p(1-p)$, where $p$ is the fraction of correctly classified events at the given node. For example this FOM has been used in a number of inclusive $B \to X\ell^+\ell^-$ analyses described in Section 17.9. A similar alternative is the "cross-entropy", $Q(p) = p \log p + (1-p) \log(1-p)$. At each split, the values of $Q$ of the two daughter nodes are added, weighted by the numbers of events (or other weights). The split that maximizes this sum is chosen. However, these FOMs are not necessarily the ones we really wish to optimize on, and some available tools permit user-defined FOMs.

An individual decision tree is a "weak" classifier (or "weak learner") in general. That is, it has a probability greater than random of making a correct classification, but possibly not much greater. It has been trained with a particular set of assumptions, such as the relative importance of training events. Better predictive power may be obtained with methods that combine decision trees trained in different ways. We introduce some of these techniques below.

A feature of decision trees is that they are intuitive. We can follow the progress along the tree and see how decisions are being made as well as see the relative importance of the different inputs as discriminators. By studying the trees produced in a given problem, we may eliminate variables that have little separation power, or are redundant with other variables.

### 4.4.6 Boosting

The idea of boosting [see for example Hastie, Tibshirani, and Friedman (2009, Chapter 10)] is to take a set of weak learners and combine them in such a way as to obtain a "strong learner": roughly, a classifier whose output error can be made arbitrarily small in a computationally efficient manner. Here, we introduce the technique in the context of boosting decision trees, although it can be used as well with other classifiers, such as neural nets.

In boosting trees, we take the results of training a tree and increase the weight ("boost") of misclassified events in forming a new tree. This process is repeated, and the outputs of the trees combined. For example, we consider the popular adaptive "AdAboost" methodology (Freund and Schapire, 1997; Hastie, Tibshirani, and Friedman, 2009):

– Start by assigning an equal weight to each event.
– Train a tree with these weights.
– Compute the weighted average error $\epsilon$ over all events.
– Compute $\alpha = \log[(1-\epsilon)/\epsilon]$.
– Increase the weight of misclassified events by a factor of $e^{\alpha}$.
– Repeat the training with these weights, using the same classification algorithm.

After some desired number of iterations, the classification of an event is computed as an average over all of the trees, weighted by their values of $\alpha$. The AdAboost is set as a default option within Toolkit for Multivariate Analysis (TMVA) and is used for the final *BABAR* PID algorithm discussed in Chapter 5.

### 4.4.7 Bagging and random forest

In "bagging" [Bootstrap AGGregatING; see, *e.g.*, Hastie, Tibshirani, and Friedman (2009, Chapter 8)] decision trees (or other classifiers in general) are constructed many times on bootstrap replicas of the training data. A bootstrap replica is a sampling, with replacement (that is, the datum is "returned" to the sample before the next sampling), of events from the training dataset. An event may appear multiple times in the replica. The point of the bootstrap is that the dataset itself is used as an empirical estimator for the underlying sampling distribution. Hence, multiple occurrences of an event are simply a consequence of identically distributed, independent samplings from this density estimator. The bootstrap replication results in another training dataset of the same size as the original. The final classifier is obtained by taking the majority vote of the individual classifiers.

If each bagging replica is passed through the same training algorithm, there will generally be significant correlations among the resulting decision trees. This tendency can be mitigated by the "random forest". In a random forest, each decision begins with choosing a random subset of the selection variables to be used in determining the split for that node. The sum of exclusive $b \to s\gamma$ analysis from *BABAR* described in Section 17.9.2.4 uses two random forest classifiers, one to perform best candidate selection and a second to provide background suppression.

### 4.4.8 Error correcting output code

We may consider the situation with multiple output classes, but where one is still interested in the binary question of determining whether the event belongs to a particular class or not. For example, suppose we have the classes $e$, $\pi$, $K$, $p$. There may be discriminants among all of these, and we may train classifiers to distinguish among binary partitions of this set of classes. That is, we might have a classifier that preferentially returns a 1 for classes $e$ or $\pi$ and a $-1$ for $K$, or $p$. We could train different classifiers for every such partition of the classes, resulting in an "exhaustive matrix". The aggregate of these classifiers is used in classifying an event. The technology of digital error correction may be used for this, in a method referred to as "error-correcting output codes" (ECOC) (Dietterich and Bakiri, 1995).

An event is classified by evaluating each of the classifiers to give a vector consisting of the numbers $-1$ and 1 for the event. The soft Hamming distances (Hamming, 1950) between this vector and the expected vectors for each class are calculated, where the soft Hamming distance between two binary strings of equal length is the sum of squares of the differences at each position of the vector. This yields a vector of numbers with length equal to the number of classes. In the simplest case we can take the class with minimum soft Hamming distance to be the resulting class. The idea is that an individual classifier might make an error, but this error may be corrected by the redundancy in the combination of the classifiers. For instance in *BABAR* many analyses have different PID requirements on the efficiency and mis-identification rate implying different levels of tightness in the selection. Instead of assigning the class with the minimum soft Hamming distance, a cut is applied based on the soft Hamming for the particular class and the ratios of soft Hamming distance of the particular class to those of the other classes. For example, for electron selection, we cut on $S_e$ and $S_e/S_K$, $S_e/S_\pi$, $S_e/S_p$ where $S_x$ is the soft Hamming distance for class $x$. The disadvantage of the ECOC approach is in the need to build the classifiers for the exhaustive matrix, which becomes daunting if the number of classes becomes large.

*BABAR* eventually applied the ECOC approach in the evolution of its particle identification algorithm (Chapter 5), where the results of several bagged decision tree classifiers are combined. We may get an idea of the impact from Fig. 4.4.1, which compares three methods for particle identification: a likelihood-based selector (Section 4.4.2); a selector using bagged decision trees (Section 4.4.7) with a non-exhaustive error correction matrix; and a selector using bagged decision trees with an exhaustive error correction matrix. In the case of the non-exhaustive matrix, the classifiers used are one-vs-one classifiers, comparing the pion with kaon hypothesis, pion with electron, etc.

In Fig. 4.4.1, top (for $\pi-K$ separation), we see that the non-exhaustive ECOC performs similarly with the likelihood selector. When we go to an exhaustive ECOC selection we find a notable improvement in mis-identification for the same efficiency. In the bottom plot (for $e-\pi$ separation) the non-exhaustive ECOC is tuned to somewhat higher efficiency, but yields much poorer mis-identification than the likelihood selector. Note that this is in contrast with the situation for the $\pi-K$ separation: relative classifier performance can depend substantially on the problem. Finding the optimal approach may require extensive study, including consideration of systematics as well as performance. However, in this case tuning an exhaustive ECOC to the same efficiency as the likelihood selector again provides a lower misidentification for the same efficiency.

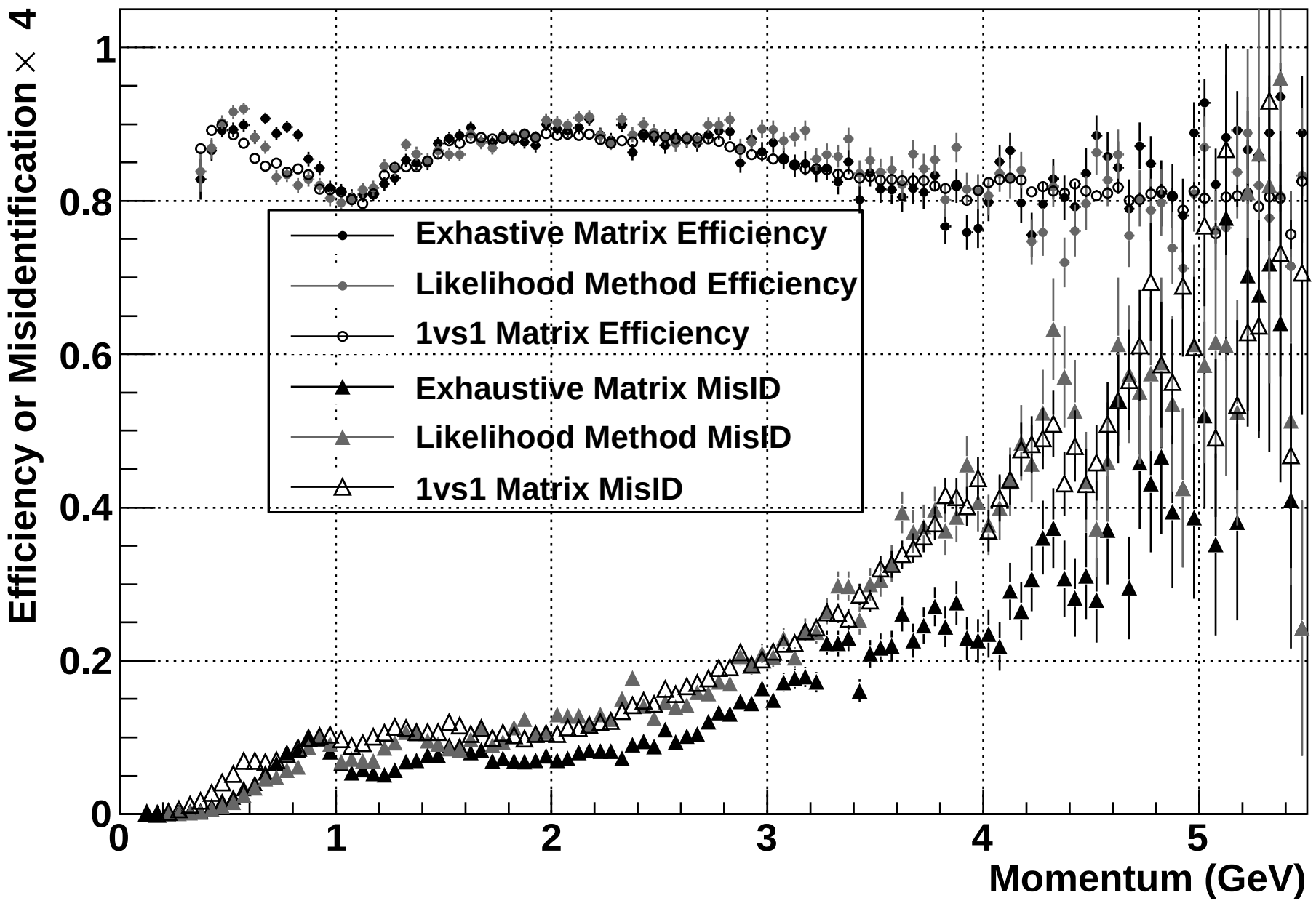

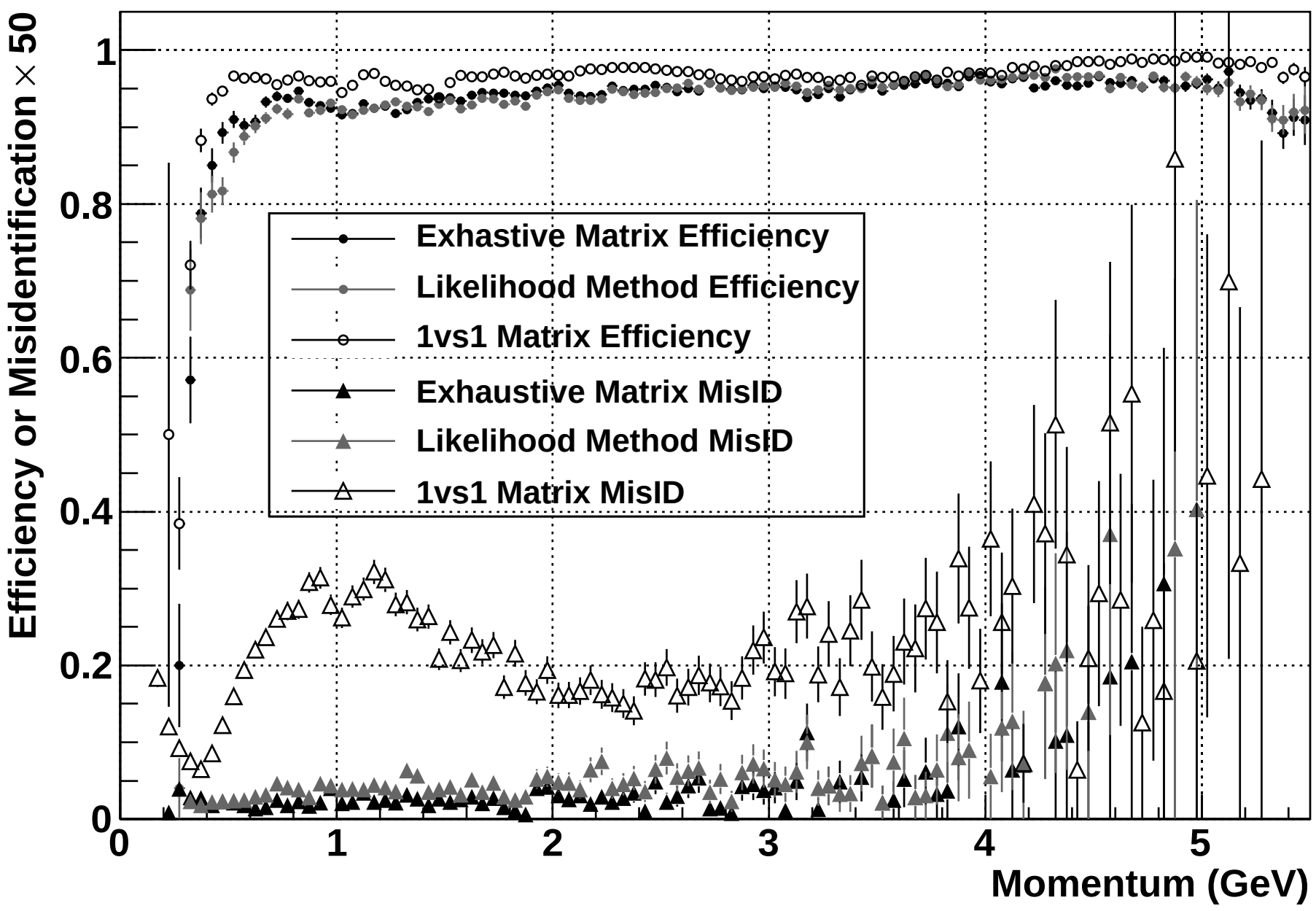

**Figure 4.4.1.** Performance of various particle identification selections in *BaBar*. The horizontal axis is momentum, and the vertical axis is either efficiency (circles) or a factor (for visibility) times the mis-identification probability (triangles). Gray symbols indicate a selector based on a likelihoods; open symbols indicate a selector based on bagged decision trees with a non-exhaustive error correction matrix (see text); black symbols indicate a selector based on bagged decision trees with an exhaustive error correction matrix. Top: Performance of kaon selection. The pion mis-identification probabilities are multiplied by four. Bottom: Performance of electron selection. The pion mis-identification probabilities are multiplied by fifty.

## 4.5 Available tools

There are two general purpose toolkits implementing many of these algorithms that have become the most widely used in our analyses:

– StatPatternRecognition (Narsky, 2005b)
– TMVA ("Toolkit for Multivariate Analysis"; Hoecker et al., 2007)

For neural nets, popular packages are:

- Stuttgart Neural Network Simulator (SNNS; Zell et al., 1995)
- NeuroBayes (Feindt and Kerzel, 2006; Phi-T, 2008)

Implementations of various classifiers may be found as well in the broader toolkits:

- The R project (R Project Contributors, 1997)
- S-PLUS (TIBCO, 2008) (a commercial alternative to R)
- MATLAB (MathWorks, 1984)

These should not be taken as exhaustive lists, only providing those packages most commonly seen in the present context.

# Chapter 5
# Charged particle identification

***Editors:***
*Alessandro Gaz (BABAR)*
*Shohei Nishida (Belle)*

## 5.1 Introduction

In this chapter we present the implementation and performance of charged particle identification (PID) at Belle and *BABAR*.

After a brief introduction, the algorithms and statistical tools used by the two experiments are discussed (Section 5.2). The PID algorithms that give the ultimate performance are based on multivariate techniques, described in detail in Chapter 4. Some examples of the typical performance of the particle identification algorithms (PID selectors) are then given, along with a discussion on PID-related error sources, for both *BABAR* (Section 5.3) and Belle (Section 5.4).

The identification of charged particles stable enough to be detected (electrons, muons, pions, kaons, and protons) plays a central role in the physics program of the *BABAR* and Belle experiments. Not only are very good PID capabilities required for separating hadronic final states of $B$ decays such as $\pi^+\pi^-$, $K^\pm\pi^\mp$, $K^+K^-$, and many others, but the PID performance is crucial for the flavor-tagging of the $B$ mesons (see Chapter 8). $B^0$ candidates are distinguished from $\overline{B}^0$ candidates based on the identification of their decay products such as high-momentum charged leptons ($e$ or $\mu$) or charged kaons. More generally PID very often provides powerful tools to reduce the backgrounds arising from final states which differ from that under study by swapping one of its particles with one of different flavor.

### 5.1.1 Definitions

The performance of a PID selector dedicated to the identification of charged particles of type $\alpha$ ($\alpha = e$, $\mu$, $\pi$, $K$, $p$) is characterized by an efficiency and a set of mis-identification probabilities.

The PID efficiency of particle type $\alpha$ is computed as the fraction of successfully identified $\alpha$ tracks among all the $\alpha$ tracks reconstructed and selected for a particular analysis, while the mis-identification probabilities are the probabilities that particles of type $\beta, \gamma, \ldots$, are incorrectly identified as $\alpha$.

In many cases the quantities defined above depend on the momentum and on the polar and azimuthal angles of the tracks. Therefore the performance of PID selectors is studied and determined in bins of $(p, \theta, \phi)$.

### 5.1.2 Subdetectors providing PID information

*BABAR* uses the information from all of its subdetectors as inputs for the PID selectors. Measurements of the energy loss $\mathrm{d}E/\mathrm{d}x$ of a charged track are provided by the SVT and the DCH. The number of Cherenkov photons and the measurement of their angle with respect to the incident track are provided by the DIRC, while the EMC is responsible for the measurement of the deposited energy and of quantities describing the shape of the shower associated with a track (such as the lateral and the Zernike moments (Zernike, 1934)), which can be used to distinguish leptonic and hadronic tracks. Finally most information (such as the number of iron layers traversed by the candidate track, and variables related to the shape of the cluster) relevant to the identification of muons is provided by the IFR.

Belle uses similar input information. Measurements of the $\mathrm{d}E/\mathrm{d}x$ of a charged track are provided by the CDC. A TOF counter measures the time of flight of a charged particle from the interaction point to the counter, from which the velocity of the particle can be measured (Kichimi, 2000). The number of Cherenkov photons at the ACC provides separation for higher momenta (Iijima, 2000). Information from the ECL, together with that from the CDC and ACC, is used for electron identification (Hanagaki, Kakuno, Ikeda, Iijima, and Tsukamoto, 2002). The KLM is responsible for muon identification (Abashian, 2002a).

## 5.2 PID algorithms and multivariate methods

In the most simple method, PID selectors are based on cuts applied to the most relevant variables for every particle type (*e.g.* $E/p$ for electrons, the distance traveled in the return yoke for muons, the Cherenkov angle for $K/\pi$ separation, ...). Better performance is obtained with the use of likelihood based selectors, in which the information from the various subdetectors is used to compute a set of likelihoods $L_k$ that the measured properties of the charged track in question would be produced by a true $k$-particle. For an example of implementation of a selector based on likelihood ratios, see Eq. (5.2.1). Belle has always used selectors based on likelihood ratios throughout the whole life of the experiment.

Cut and likelihood based selectors are very stable over the data-taking periods and do not need re-tuning to compensate for the aging of the detectors and the changes introduced by the reprocessing of the data. However, significant improvements can be achieved by considering a larger set of variables, even some with very mild discrimination power, in the implementation of PID selectors. *BABAR* uses more sophisticated statistical tools such as *Neural Networks* (NN), *Bagged Decision Trees* (BDT), and *Error Correcting Output Code* (ECOC) algorithms, to accommodate a large number of input variables (up to 36) and the significant correlations among them.

Due to their higher sensitivity to variations in the performance of the detector, the selectors based on multivariate methods need to be re-trained on data control samples (see Section 5.3) after every major change in the reconstruction algorithms. Particularly important for *BABAR*, which was affected by large variations in the performance of the IFR, is the inclusion of the data taking period as one of the input variables, in order to take into account the loss of efficiency in specific regions of the detector.

In the following sections the more refined algorithms implemented at Belle and *BABAR* will be described.

### 5.2.1 Belle algorithms

The PID at Belle is based on likelihood ratios. For hadron identification, likelihoods for a candidate particle $\alpha$ are calculated based on $\mathrm{d}E/\mathrm{d}x$ information from the CDC ($L_\alpha^{\mathrm{CDC}}$), time of flight from the TOF ($L_\alpha^{\mathrm{TOF}}$) and the number of photons from the ACC ($L_\alpha^{\mathrm{ACC}}$), respectively. Then, the likelihood ratios

$$L(\alpha:\beta) = \frac{L_\alpha^{\mathrm{CDC}} L_\alpha^{\mathrm{TOF}} L_\alpha^{\mathrm{ACC}}}{L_\alpha^{\mathrm{CDC}} L_\alpha^{\mathrm{TOF}} L_\alpha^{\mathrm{ACC}} + L_\beta^{\mathrm{CDC}} L_\beta^{\mathrm{TOF}} L_\beta^{\mathrm{ACC}}} \quad (5.2.1)$$

are calculated and used for identification. For example, pions (kaons) can be selected by requiring a low (high) value of $L(K:\pi)$, and protons are typically identified with requirements on both $L(p:K)$ and $L(p:\pi)$. The cut value applied to the likelihood ratios can be optimized depending on the analysis.

For electron identification, in addition to $L_\alpha^{\mathrm{CDC}}$ and $L_\alpha^{\mathrm{ACC}}$, information from the ECL (matching of the positions of the track and the energy cluster, $E/p$, and transverse shower shape) is used to form likelihood ratios. There is a small region around $\theta \sim 125°$ with low electron identification performance because of a small gap between the barrel ECL and backward endcap ECL. For muon identification, reconstructed hits in the KLM are compared to the extrapolation of the CDC track, using the difference $\Delta R$ between the measured and expected range of the track, and the statistic $\chi_r^2$ constructed from the transverse deviations of all hits associated to the track, normalized by the number of hits. Likelihoods for the muon, pion, and kaon hypotheses are formed based on *p.d.f.*s in $\Delta R$ and $\chi_r^2$. The likelihood ratio $L_\mu/(L_\mu + L_\pi + L_K)$ is then used as a discriminating variable.

### 5.2.2 BABAR algorithms

In *BABAR*, the ultimate performance in the selection of muons is achieved with an algorithm based on *Bagged Decision Trees* (Narsky, 2005a; also discussed in Section 4.4.7 of this Book). The algorithm takes as input 30 variables: in addition to variables related to the length and the shape of the IFR cluster associated to the candidate track and the measurement of the energy deposited in the EMC, the variables related to the shape of the cluster in the calorimeter, the number of Cherenkov photons, the opening angle of the Cherenkov cone, and the number of DCH hits and the $\mathrm{d}E/\mathrm{d}x$ measured in the DCH are also used.

The training of the selectors is performed on high purity data samples of muons and pions, subdivided in 720 bins of $p$, $\theta$, and charge. Candidate tracks are randomly discarded in order to have the same number of muons and pions in the same bin. This allows the use of the $p$, $\theta$, and charge variables in the tree without introducing any bias due to the different $(p,\theta)$ spectrum of the source sample. The source sample is then randomly split into a training and a testing sample. Four different levels of tightness are designed for the muon selector (VeryLoose, Loose, Tight, and VeryTight); the cuts on the output of the classifier are designed such that either the muon selection efficiency or the pion mis-identification probability are kept constant. The target efficiencies (besides the very low-momentum part of the spectrum, where few muons can be identified) are 90%, 80%, 70%, and 60% and the target pion mis-identification probabilities are 5%, 3%, 2%, and 1.2%. Two additional selectors, optimized for muons in the momentum range [0.3, 0.7] GeV/$c$, with a target efficiency of 70% and 60% have been developed. With roughly the same efficiency, the BDT based muon selectors are significantly more effective in rejecting the pion contamination with respect to the selectors based on Neural Networks, as can be seen from Table 5.2.1.

For the other charged particles (electrons, pions, kaons, and protons), a class of selectors based on the *Error Correcting Output Code* algorithms (Dietterich and Bakiri, 1995) is used. The discrimination is based on 36 variables from the four inner subdetectors: SVT, DCH, DIRC, and EMC. Candidate $e$, $\pi$, $K$, and $p$ are separated by means of several binary classifiers (in our case BDT's) combined through an *exhaustive matrix* (see Chapter 4). The use of the exhaustive matrix ensures the robustness of this type of selector against potential mis-classifications of some of the binary classifiers. The selectors are trained on high purity data samples (see Section 5.3) and the cuts on the outputs of the binary classifiers are tuned in such a way that the selection efficiency matches that of the analogous likelihood based selectors. Six levels of tightness are provided (SuperLoose, VeryLoose, Loose, Tight, VeryTight, and SuperTight). At the same level of efficiency, the mis-identification rate for the ECOC algorithms is significantly lower than that of the likelihood based selectors (see Table 5.2.1).

**Table 5.2.1.** Efficiencies and mis-identification rates (averaged over the momentum and polar angle spectra) for different kinds of muon and kaon *BABAR* PID selectors, all using Tight requirements. The quoted uncertainty represents the typical statistical uncertainty in each bin of the tables that measures the performance of the supported selectors. No systematic uncertainty has been included.

| Muon selector | efficiency (%) | $\pi$ mis-id rate (%) |
|---|---|---|
| Cut based | 65.0 ± 0.5 | 1.43 ± 0.05 |
| NN | 60.5 ± 0.5 | 0.97 ± 0.05 |
| BDT | 59.4 ± 0.5 | 0.76 ± 0.05 |
| Kaon selector | efficiency (%) | $\pi$ mis-id rate (%) |
| Cut based | 80.2 ± 0.2 | 1.39 ± 0.07 |
| Likelihood based | 83.0 ± 0.2 | 1.47 ± 0.07 |
| ECOC | 84.2 ± 0.2 | 1.10 ± 0.07 |

## 5.3 BaBar PID performance and systematics

The tuning of the PID selectors and the assessment of their performance takes advantage of high purity samples of tracks selected from the data. A large number of electron and muon tracks is selected from $e^+e^- \to e^+e^-(\gamma)$, $\mu^+\mu^-(\gamma)$ processes, with minimal cuts on the kinematics of the event, on the quality of both the candidate track and of the other track in the event, and on the basic PID properties (to distinguish electrons from muons). For some low-statistics cross-checks, a sample of electrons (muons) from the decays $B \to J/\psi K^{(*)}$, $J/\psi \to e^+e^-$ $(\mu^+\mu^-)$ has also been used.

$K$ and $\pi$ candidates are selected from $D^{*+} \to D^0\pi^+$, $D^0 \to K^-\pi^+$. The $K/\pi$ assignment is done based on the charge of the soft pion from the $D^{*+}$ decay. The purity of the sample is increased by applying quality cuts on the reconstructed tracks, and rejecting fake $D^0$'s using cuts on the invariant mass of the reconstructed $D^0$ candidate and on the likelihood that the $K$ and $\pi$ tracks originate from a common vertex. Additional $\pi$ samples, especially important for measuring the mistagging of pions as muons at high momentum (where the population of $D^0 \to K^-\pi^+$ is low) are obtained from $K^0_S \to \pi^+\pi^-$ decays and from $e^+e^- \to \tau^+\tau^-$ events where one $\tau$ (tag) has one charged particle among its decay products and the other decays to a final state with three charged particles. Finally a high-purity sample of protons is obtained from $\Lambda^0 \to p\pi^-$ decays, by taking advantage of the long lifetime of the $\Lambda^0$ baryon. The purity of the sample is enhanced by applying cuts on the quality of the candidate tracks and on the probability that the proton and pion tracks are consistent with originating from the same displaced vertex. Some examples of performance of the *BABAR* selectors are displayed in Table 5.2.1 and in Figure 5.3.1.

These high purity samples are utilized in the training of the more advanced PID algorithms and in establishing the performance of all the selectors. Depending on the available statistics, the control samples are divided into several bins with different $(p,\theta)$. In the case of the muon selectors at *BABAR*, the samples are also subdivided in 6 bins of $\phi$, to better characterize the degradation of the RPC chambers and the staged upgrade of the barrel section with LST detectors (see Chapter 2). Each of the selectors is applied to every bin of the control samples and the efficiencies for both the data ($\varepsilon_{\rm data}$) and the simulation ($\varepsilon_{\rm MC}$) are computed. The tables of efficiencies thus built are then used to correct the simulation so that its PID performance matches that of the data. One of the most widely used algorithms to apply this correction in *BABAR* is the so-called *PID-tweaking*. In the case where $\varepsilon_{\rm data} = \varepsilon_{\rm MC}$, no correction is applied, whereas if $\varepsilon_{\rm data} < \varepsilon_{\rm MC}$ a MC track that passes the selector is randomly discarded with probability

$$\frac{\varepsilon_{\rm data}}{\varepsilon_{\rm MC}}\,. \qquad (5.3.1)$$

In the case $\varepsilon_{\rm data} > \varepsilon_{\rm MC}$, a MC track that does not pass the selection is accepted with probability

$$(\varepsilon_{\rm data} - \varepsilon_{\rm MC})\frac{1}{\varepsilon_{\rm MC}}\,. \qquad (5.3.2)$$

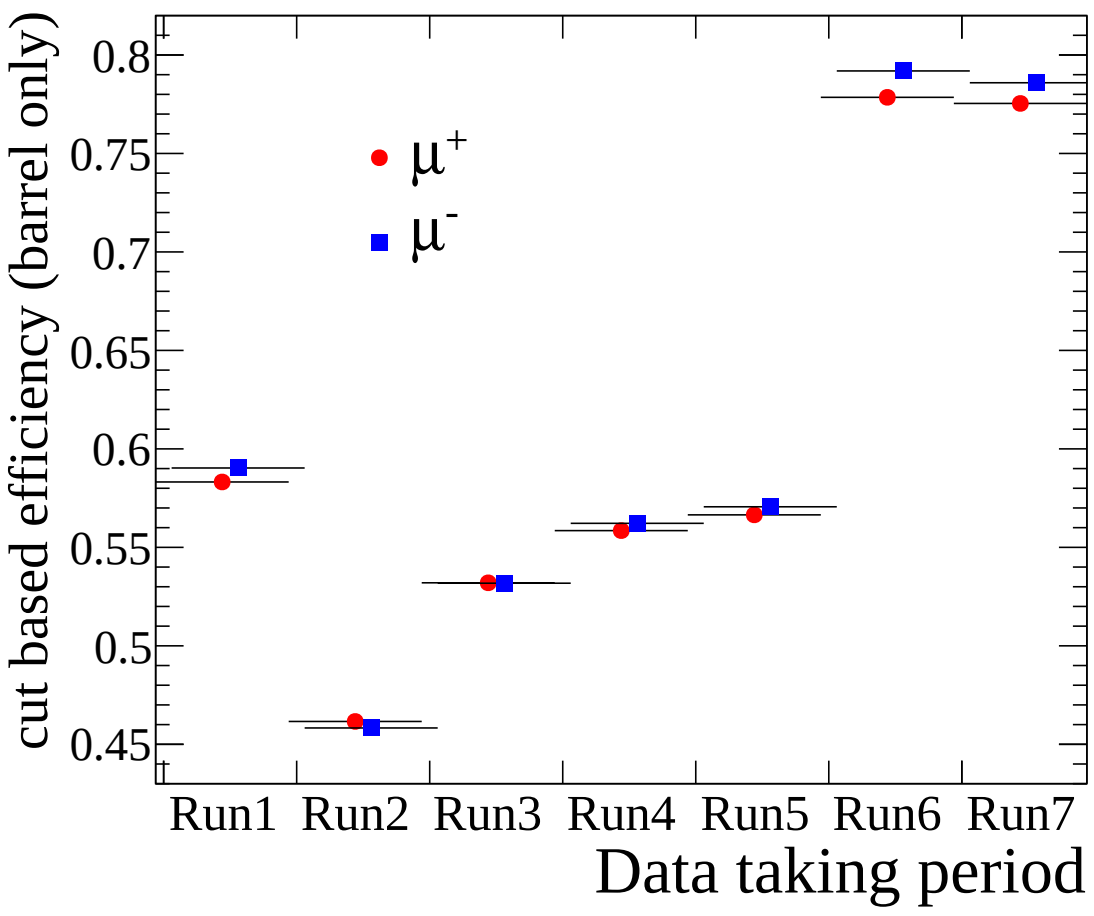

**Figure 5.3.2.** Muon selection efficiencies for a typical *BABAR* cut-based muon selector as a function of the data taking period. The efficiencies are computed only for the barrel region. The loss in performance due to the degradation of the RPC detector during the early phases of the data taking is evident, as is the full recovery with the installation of the LST's, completed after the end of Run5.

At the end of the *BABAR* experiment, the size of the typical correction applied by the PID-tweaking algorithm was about one percent.

### 5.3.1 History of PID performance in BaBar

For the *BABAR* experiment, the most important issue affecting the stability of PID performance was the degradation of the efficiency of the RPC chambers (see Chapter 2). This is visible from Fig. 5.3.2, which shows the efficiency of one of the cut-based muon selectors as a function of the data-taking period. This loss of performance was also one of the main motivations to develop muon selectors relying on variables in addition to those measured by the IFR.

### 5.3.2 Systematic effects

Both experiments rely on high-purity data samples to assess the performance of PID selectors and correct the simulation so that it matches the data as much as possible. Several ways exist to estimate the systematic uncertainty in a measurement related to PID requirements. It is not possible to establish a recommended way to proceed for all analyses, since in general the performance of each selector can be sensitive to the charged and neutral multiplicity of the events studied. For example, the performance of electron and muon selectors is studied in low multiplicity events, thus some care must be taken when applying these selectors to $B$-decays, where the multiplicity of the final states is substantially higher.

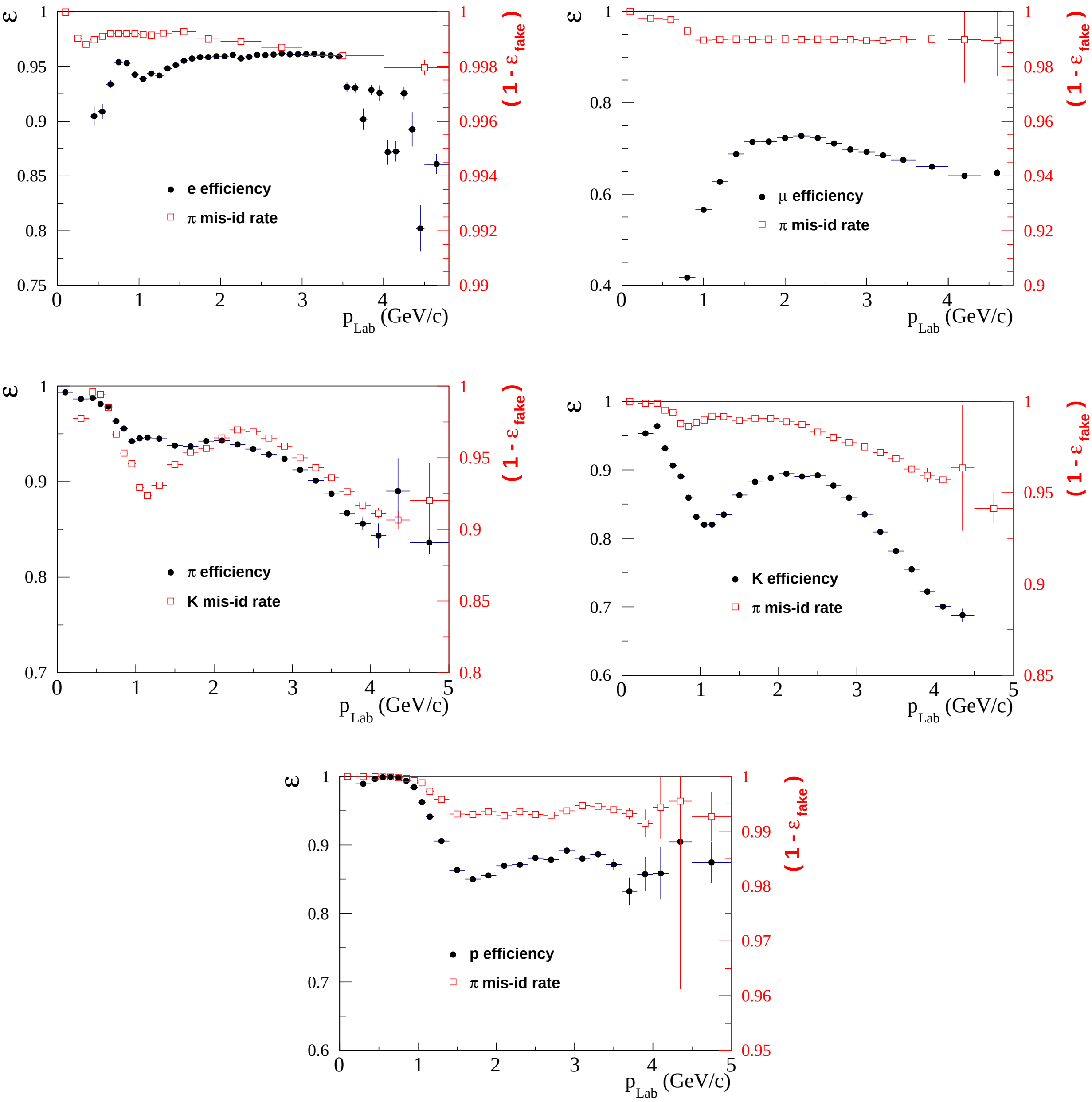

**Figure 5.3.1.** Performance of some typical *BABAR* PID selectors for electrons (top left plot), muons (top right), pions (middle left), kaons (middle right), and protons (bottom) as a function of the momentum of the candidate charged track. The solid (black) dots represent the efficiency, which can be read off the left axis, of the particular selectors, while the empty (red) squares show the complement (*e.g.* kaon for the pion selector, and pion for all other selectors) mis-identification probability (right axis). Note that the vertical scale differs from plot to plot.

In *BABAR*, many of the analyses estimate the systematic uncertainty on the PID performance by taking the difference of the signal reconstruction efficiency in the simulation obtained by applying or not applying the correction (usually the PID-tweaking) based on the efficiency tables described above. For some analyses where the relative contribution of the PID to the total systematic uncertainty is large, or there is a sizable dependence on the multiplicities and the topologies of the events, alternative strategies have been applied, and where possible the performance of the chosen selector(s) has been checked in control samples with similar track multiplicities and topologies of the channel under study.

## 5.4 Belle PID performance and systematics

In Belle, the PID performance of the kaon and pion identification algorithm is estimated using the decay $D^{*+} \to$

$D^0\pi^+$ followed by $D^0 \to K^-\pi^+$, similar to *BABAR*. Figure 5.4.1 (a) and (b) shows typical curves of the efficiencies and mis-identification rates for the kaon and pion identification in the barrel region, studied with this control sample. Discrepancies between data and MC can be seen, especially in the mis-identification.

In the study of the kaon and pion identification, the control sample is divided into 384 bins, *i.e.* 32 momentum ($p$) bins and 12 polar angle ($\theta$) bins. The momentum range is divided into 100 (200) MeV/$c$ bins below (above) 3 GeV/$c$. The polar angle subdivision is based on the structure of the ACC: one $\theta$ bin for the backward endcap (with no ACC), and one bin for each of the ten types of aerogel counter module in the barrel and forward endcap, except for the large polar angle range covered by the $n = 1.010$ modules, which is divided in two (see Figure 2.2.8, and the accompanying text in Section 2.2.3).

For each bin, the efficiency and mis-identification rate for $K$ and $\pi$ are estimated both for the data and the MC for different PID selections. The relevant value for general analyses is the ratio of the efficiency or mis-identification rate between the data and the MC: $R_l = \epsilon_l^{\rm data}/\epsilon_l^{\rm MC}$ and its uncertainty, where $l$ is the bin index. These quantities are provided as a look-up table for general use in Belle analyses. The efficiency (mis-identification rate) ratio and its uncertainty for a given analysis, which is quoted as the systematic uncertainty from PID, can then be calculated by

$$R = \frac{1}{N}\sum_l n_l R_l, \tag{5.4.1}$$

and

$$\delta R = \frac{1}{N}\left(\sqrt{\sum_l \left(n_l \delta R_l^{\rm stat}\right)^2} + \sum_l n_l \delta R_l^{\rm syst}\right) + \delta R_{\rm const}, \tag{5.4.2}$$

where $R_l$ is the efficiency ratio in bin $l$, $n_l$ is the number of tracks in that bin (analysis dependent), and $N = \sum n_l$. The parameters $\delta R_l^{\rm stat}$ and $\delta R_l^{\rm syst}$ are respectively the statistical and systematic uncertainties in bin $l$ obtained from the control sample study; $\delta R_{\rm const}$ is an additional systematic uncertainty, independent of $(p, \theta)$, based on variations in efficiency between different data taking periods ("experiments" in Belle nomenclature: see Section 3.2). In this way, the correction factor and the systematic error can be automatically calculated. The typical systematic uncertainty $\delta R$ for kaon and pion identification at Belle is 0.8%. In physics analyses that measure a direct *CP* asymmetry, the systematic error due to an asymmetry in the PID efficiency between positive and negative charged tracks needs to be estimated. This error can be calculated by using the tables for $R_l$, $\delta R_l^{\rm stat}$, and $\delta R_l^{\rm syst}$, which are provided separately for positive and negative particles.

The study of the proton identification is performed with $\Lambda \to p\pi^-$, using the same binning for $\theta$ as above, but with only 12 bins for momentum. The typical proton efficiency is shown in Fig. 5.4.1 (c).

For the study of the lepton identification, the two-photon process $e^+e^- \to e^+e^-\ell^+\ell^-$ ($\ell = e, \mu$) is used to obtain high statistics electron and muon samples. The control sample is divided into 70 bins (10 momentum bins in 500 MeV/$c$ steps and 7 polar angle bins). The efficiencies of the lepton identifications estimated using this process are shown in Fig. 5.4.1 (d) and (e). Since the above process leads to low track-multiplicity events, inclusive $J/\psi$ events ($J/\psi \to \ell^+\ell^-$) are also used as a control sample, mainly for the estimation of a possible performance difference between low-multiplicity events and hadronic events. The mis-identification rates of the lepton identification for pions and kaons, are studied using a control sample of $K_S^0 \to \pi\pi$ and $D^{*+} \to D^0\pi^+ \to K^-\pi^+\pi^+$.

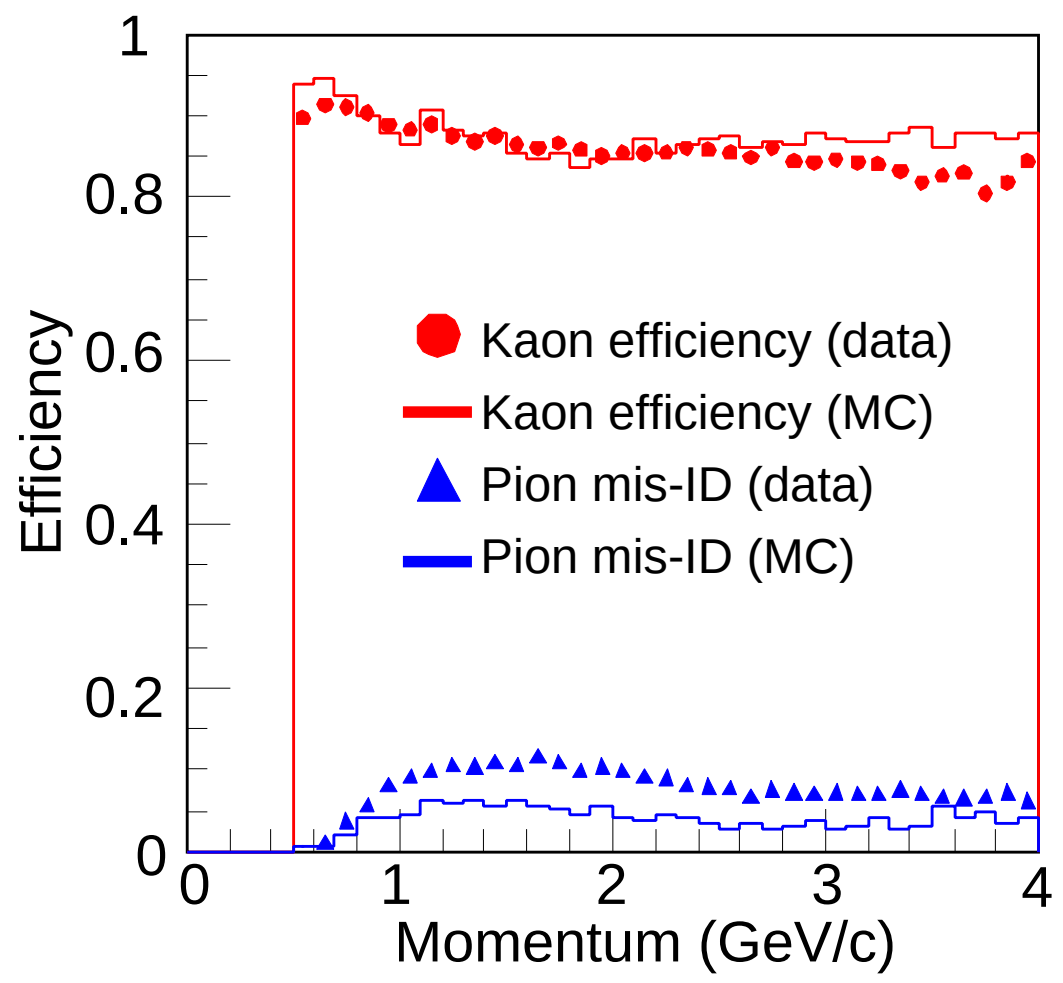

(a) Kaon identification.

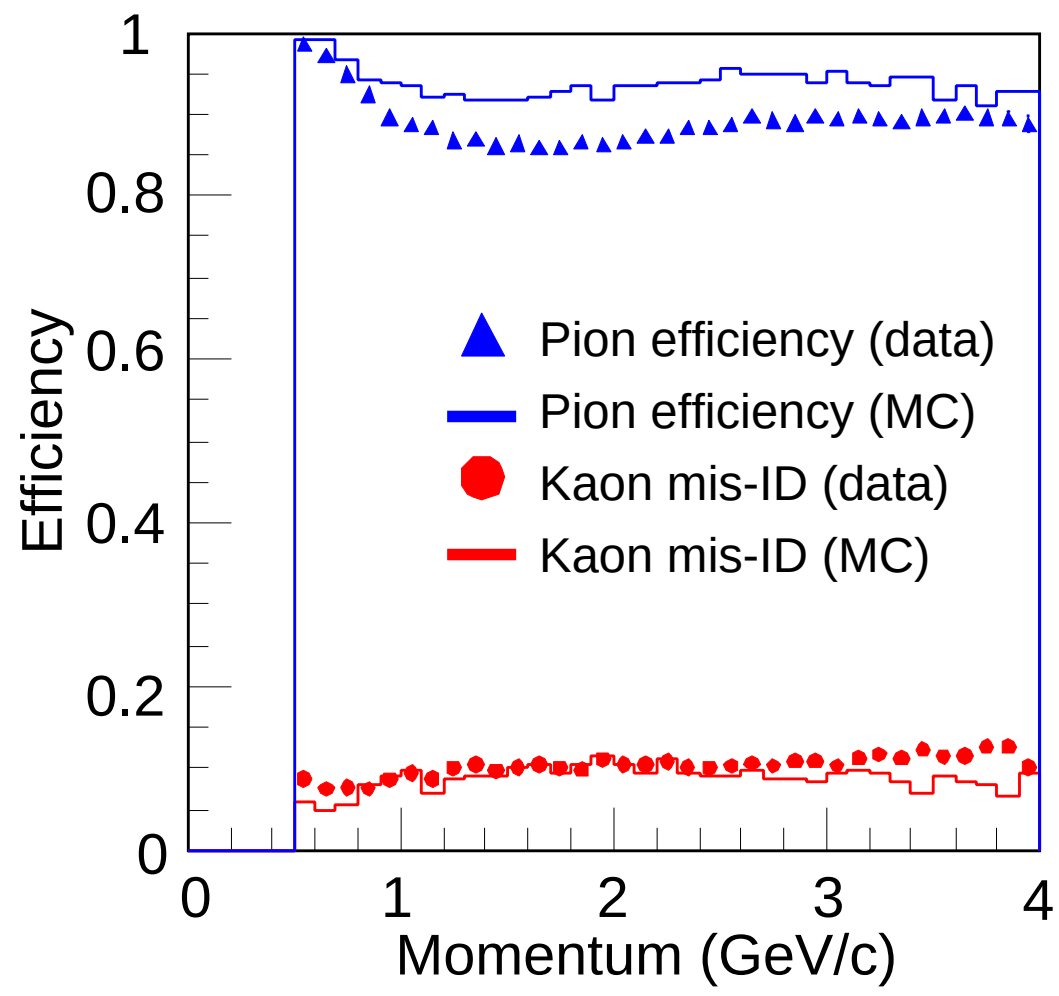

(b) Pion identification.

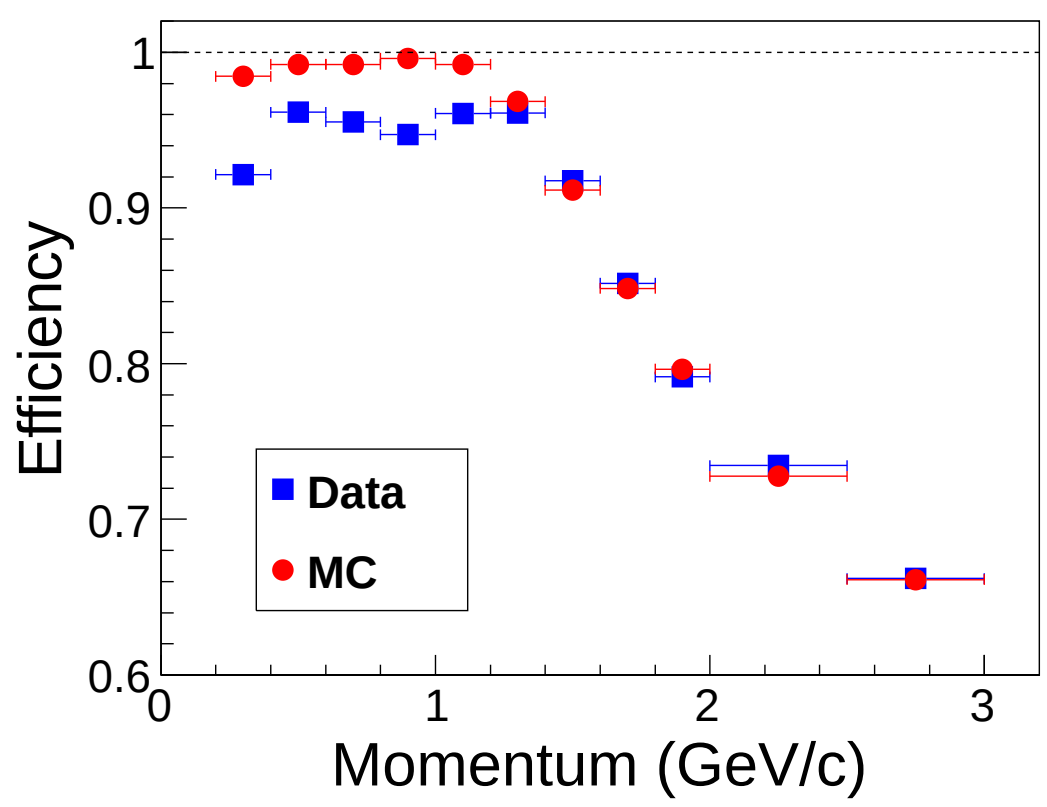

(c) Proton identification.

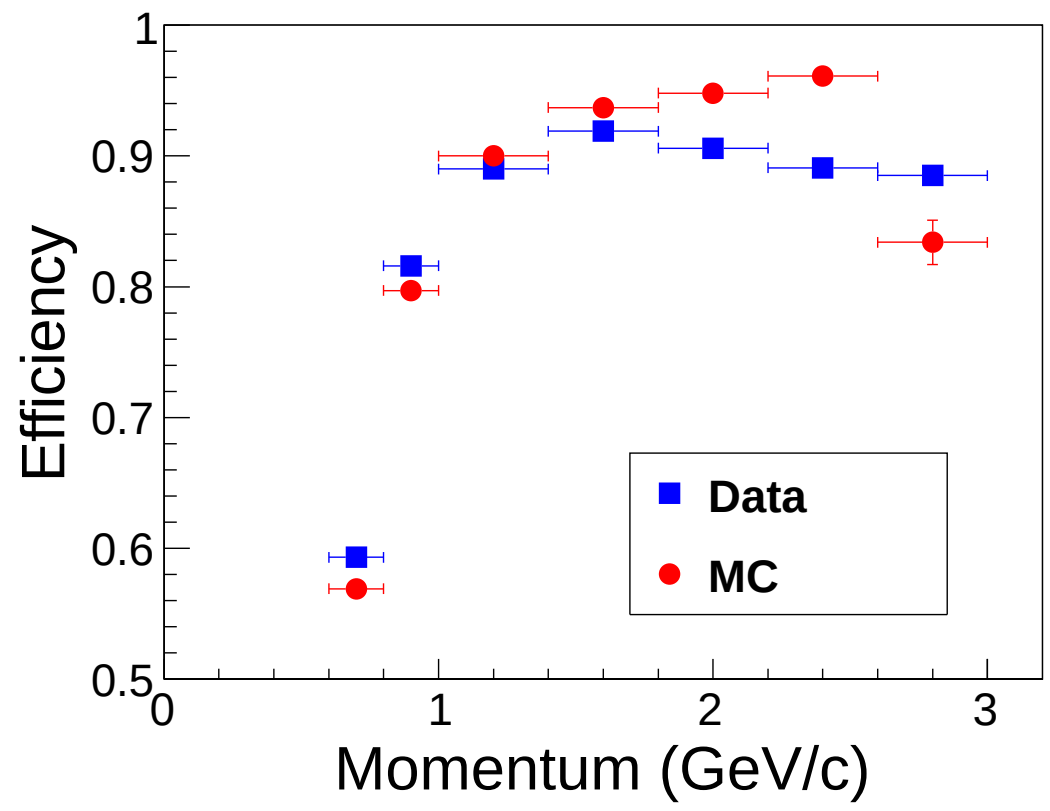

(d) Electron identification.

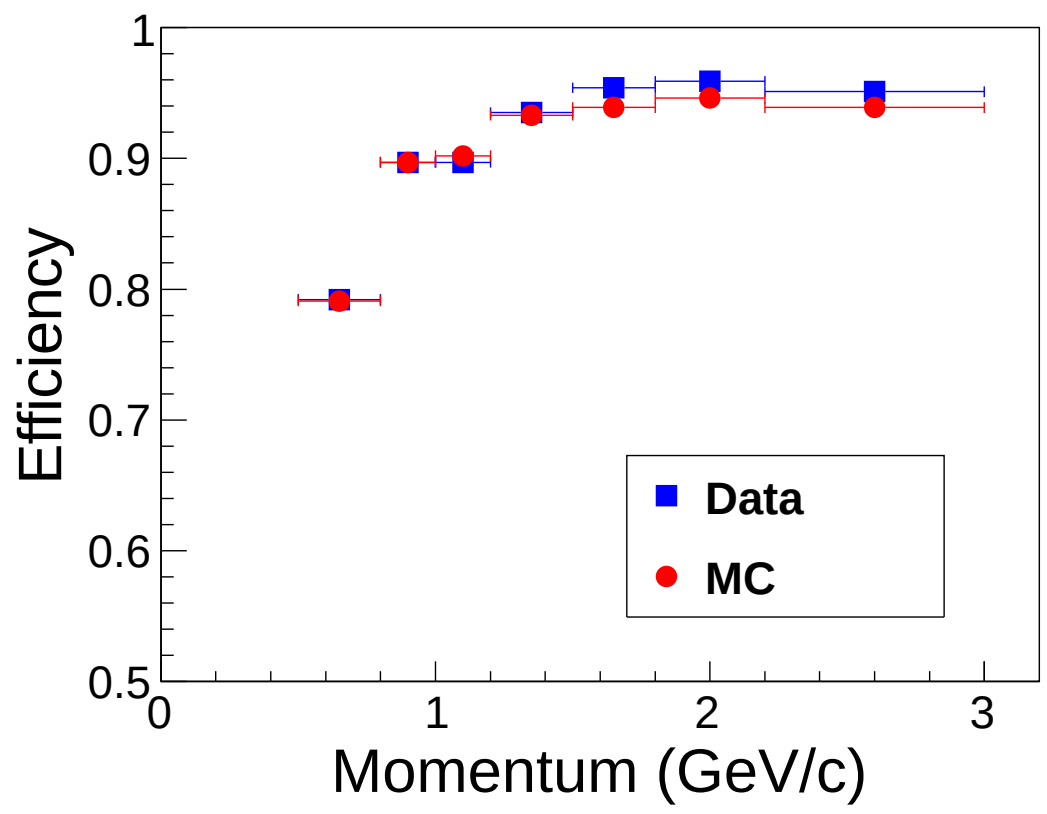

(e) Muon identification

**Figure 5.4.1.** Performance of the PID at Belle as a function of the momentum of the candidate charged track for the data and MC-simulated events. (a) Performance of kaon identification: kaon efficiency and pion mis-identification rate. (b) Performance of pion identification: pion efficiency and kaon mis-identification rate. (c) Performance of proton identification. (d) Performance of electron identification. (e) Performance of muon identification. In (c), (d) and (e), only efficiencies respectively for protons, electrons and muons are shown for the data and MC simulated events.

# Chapter 6
# Vertexing

***Editors:***
*Wouter Hulsbergen (BABAR)*
*Takeo Higuchi (Belle)*

**Additional section writers:**
*Maurizio Martinelli*

## 6.1 The role of vertexing in the $B$ Factories

A vertex algorithm is a procedure by which the parameters of a decay vertex or interaction vertex are determined from the reconstructed parameters of the outgoing particles. In the simplest case the outgoing particles are charged particles that are either stable or have a large $c\tau$ (where $\tau$ is the particle lifetime) compared to the dimensions of the detector, namely electrons, muons, protons and charged pions and kaons. These particles are reconstructed as charged particle trajectories (or 'tracks') in the tracking detectors and their reconstructed parameters are the track parameters. More complicated vertex algorithms involve final states that include not only tracks, but also photons or other decaying particles.

The role of vertexing algorithms in the $B$ Factory experiments can roughly be divided in three parts. First, vertex fits are used to obtain the parameters of reconstructed 'composite' particles from their decay products, *i.e.* charged particle trajectories and photon calorimeter clusters. These parameters are usually the vertex position, momentum and invariant mass of the decaying particle. However, also the decay length of an unstable particle inside a decay chain (such as the $D$ meson in a $B \to D\pi$ decay), or the decay time difference $\Delta t$ of the two $B$ mesons from an $\Upsilon(4S)$ decay, can be computed with a vertex fit.

Second, the $\chi^2$ of a vertex fit is used to suppress combinatorial background in the selection of composite particles. Apart from a few cases of decays in flight (pions and long lived strange hadrons), the decay products from most composite particles all originate from a small region around the interaction point. The track parameter resolution of $B$ Factories is just sufficient to separate the decay vertices of bottom and charm mesons. When searching for exclusive decays a requirement on the vertex $\chi^2$ provides an efficient way to reject wrong combinations from the composite particle candidates. The $\chi^2$ plays a similar role in the reconstruction of the primary interaction vertex or in the reconstruction of the 'second' $B$ vertex for the determination of $B$ meson decay time difference. In that case the contribution of individual tracks to the vertex $\chi^2$ is used to select the subset of tracks that best determines the vertex position.

Finally, vertexing is used in the calibration and monitoring of the position and size of the interaction region. As we shall see, information on the average position of the primary vertex can be used as a constraint in vertex fits. In the *BABAR* and Belle experiments the beam parameters are also fed back in real time to the accelerator for diagnostics.

This chapter is organized as follows. The parameterization of reconstructed tracks, which defines the input to the vertex algorithms, is described in Section 6.2. Vertex fitting algorithms are discussed in Section 6.3. The calibration of the interaction region for use in vertex fits is described in Section 6.4. An important application of vertex fits is the determination of decay times, in particular the $B$ meson decay time difference $\Delta t$. The demands on vertex resolution in the $B$ Factory experiments are primarily determined by the requirement that $\Delta t$ be measured with sufficient precision to probe $B^0\overline{B}^0$ oscillations. The procedures by which the decay time difference is estimated and its resolution calibrated are discussed in Sections 6.5.

## 6.2 Track parameterization and resolution

If stochastic processes like energy loss and multiple scattering in detector material are ignored, the trajectory of a charged particle in a magnetic field can be described by five parameters. In a uniform magnetic field the trajectory follows a helix. The helix axis is parallel to the magnetic field, which in the $B$ Factory solenoids is almost parallel to the $e^+$-$e^-$ beam axis.

Even in the case that the field is not uniform or material effects cannot be ignored, the track can locally be parameterized as a helix. With respect to a conveniently chosen pivot point, the parameters can be defined as (see Chapter 2 for the definition of the coordinate system)

| | |
|---|---|
| $d_\rho$ or $d_0$ | signed distance in the $x$-$y$ plane from the pivotal point to the helix, |
| $\phi_0$ | azimuthal angle from the pivotal point to the helix center, |
| $\kappa$ or $\omega$ | inverse of the track transverse momentum times charge of the track, $\kappa = e/p_t$ |
| $d_z$ or $z_0$ | signed distance along the $z$ axis from the pivotal point to the helix, |
| $\tan\lambda$ | tangent of the dip angle. |

The two experiments follow a slightly different notation and definition. When two names are shown in the first column of the table above, the first is for Belle and the second for *BABAR* . The sign of the inverse transverse momentum $\kappa$ coincides with the sign of the charge of the particle. If the pivot point is the origin, then $d_\rho$ is the (signed)[19] minimum distance to the $z$-axis and $d_z$ is the $z$-coordinate of the point-of-closest approach to the origin. The azimuthal coordinate $\phi$ is the angle of the transverse momentum vector with the $x$ axis in *BABAR* while the coordinate $\phi + \phi_0$ is the angle of the transverse momentum vector with the $y$ axis in Belle. In the following we use the Belle definition, illustrated in Fig. 6.2.1. The *BABAR* definition can be found in (Hulsbergen, 2005).

[19] Sign of $d_\rho$: for $e > 0$ and the pivot point lying outside the helix projection to the $(x, y)$ plane then $d_\rho > 0$; for the pivot point inside the helix projection $d_\rho < 0$. For $e < 0$ this definition is reversed.

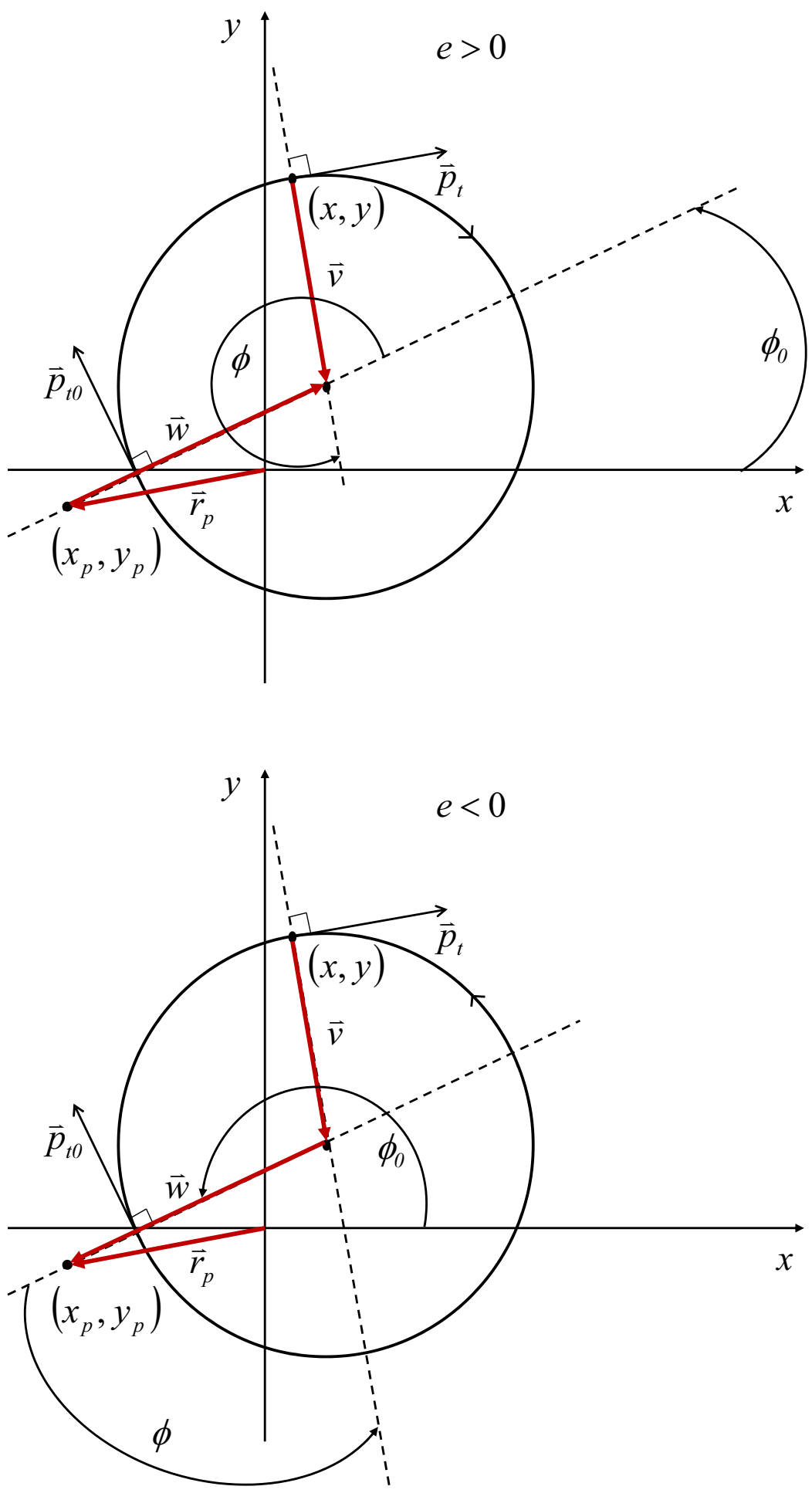

**Figure 6.2.1.** Schematic representations of the helix parameterization for a positively (top) and negatively (bottom) charged track in the $(x, y)$ plane used at Belle. Magnetic field is in the direction of the $z$-axis. Vector $\boldsymbol{r}_p$ determines position of the pivot point. Other vectors in the figure are defined as $\boldsymbol{r} = \boldsymbol{r}_p + \mathrm{sgn}(e)\boldsymbol{w} - \boldsymbol{v}$, where $\boldsymbol{w} = \mathrm{sgn}(e)(d_\rho + \rho)(\cos\phi_0, \sin\phi_0)$, $\boldsymbol{v} = \rho(\cos(\phi_0 + \phi), \sin(\phi_0 + \phi))$.

The charged particle position along the track trajectory can be represented using a running parameter $\phi$ as

$$\begin{aligned} x(\phi) &= x_p + d_\rho \cos\phi_0 + \rho\{\cos\phi_0 - \cos(\phi_0 + \phi)\}, \\ y(\phi) &= y_p + d_\rho \sin\phi_0 + \rho\{\sin\phi_0 - \sin(\phi_0 + \phi)\}, \\ z(\phi) &= z_p + d_z - r\phi\tan\lambda, \end{aligned} \tag{6.2.1}$$

where $(x_p, y_p, z_p)$ is the pivot point and $\rho = 1/B_z\kappa$ is the (signed) curvature radius with $B_z$ representing the strength of the magnetic field. Using $p_\mathrm{t} = e/\kappa$ the momentum vector along the trajectory is given by

$$\begin{aligned} p_x(\phi) &= -p_\mathrm{t}\sin(\phi_0 + \phi), \\ p_y(\phi) &= \ \ p_\mathrm{t}\cos(\phi_0 + \phi), \\ p_z(\phi) &= \ \ p_\mathrm{t}\tan\lambda. \end{aligned} \tag{6.2.2}$$

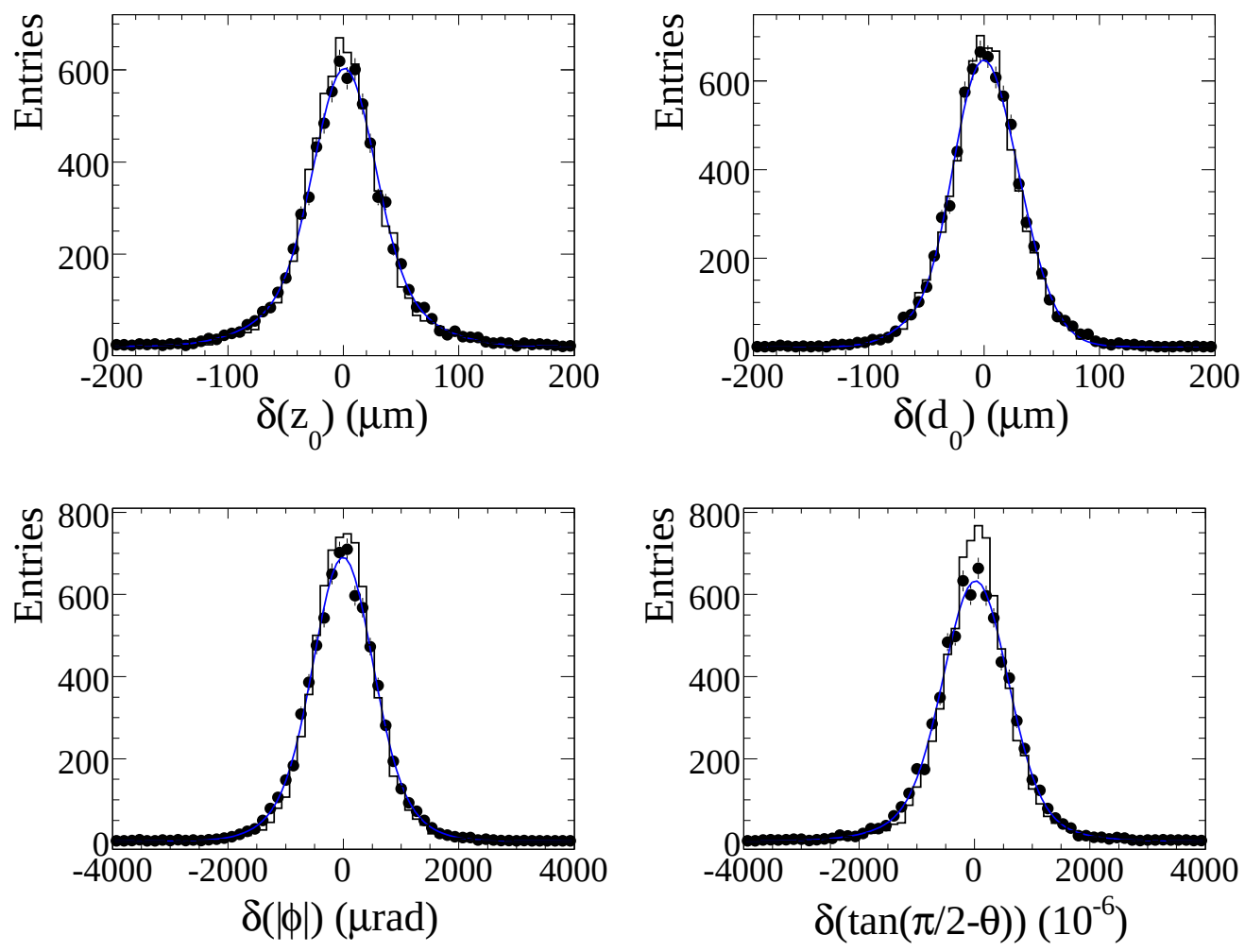

**Figure 6.2.2.** Measurements of the differences between the fitted track parameters of the top and bottom stubs of cosmic ray muons with a momentum above 2 GeV/$c$ in *BABAR*. The data are shown as points, Monte Carlo simulation as histograms. The (blue) smooth curves are the results of a Gaussian fit to the data. From (Brown, Gritsan, Guo, and Roberts, 2009).

Expressions for the inverse transformation — from position and momentum vector to helix parameters — and the corresponding Jacobian can be found in (Hulsbergen, 2005).

The helix track parameters are determined by a fit to the measured hit coordinates along the track. Both Belle and *BABAR* use a track fit based on a Kalman filter (Fruhwirth, 1987). The *BABAR* track fit is described in (Brown, 1997). The track parameter resolution is determined by the number of hits and the hit resolution, and by multiple scattering and energy loss. For the resolution on the direction and position of the track the first two layers in the vertex detector are most important. However, for the extrapolation to the interaction point the curvature resolution is relevant as well. Both $B$ Factory experiments feature a multi-layer vertex detector (Section 2.2.1) with a hit resolution in the range $10 - 50$ $\mu$m to precisely measure impact parameters. A precise curvature resolution is facilitated by a large drift chamber.

An estimate of the track resolution in data can be obtained from cosmic ray events. The muon trajectory is reconstructed as two separate segments in the top and bottom halves of the tracking detector. The difference or 'residual' between the reconstructed parameter of the segments at their point of closest approach is representative for the actual parameter resolution, after a correction with a factor $\sqrt{2}$. The distribution of the residuals is shown in Figure 6.2.2 for muons with momenta above 2 GeV/$c$ in *BABAR* data and Monte Carlo. From a fit with a single Gaussian to these distributions the single-track resolution in data is estimated as 29 $\mu$m for $z_0$, 24 $\mu$m for $d_0$, 0.45 mrad for $\phi_0$ and $0.53 \cdot 10^{-3}$ for $\tan\lambda$ (Brown, Gritsan, Guo, and Roberts, 2009) (see Chapter 2 for a discussion of the $p_T$ resolution). The parameter resolution in Belle is similar. It should be noted, however, that due to the contribution from multiple scattering the resolutions are a rather strong function of momentum. For example, in *BABAR* the $d_0$ resolution at $p_T \approx 0.1$ GeV/$c^2$ is over a factor 5 worse than at $p_T \approx 3$ GeV/$c^2$ (Aubert, 2002j).

Besides the track parameters the track fit also computes a track parameter covariance matrix, which can be used in vertex fits. The covariance matrix is among others a function of the estimated uncertainty in the hit coordinates and the estimated RMS of the scattering angle distribution. Due to pattern recognition mistakes and simplifications in the track model, the estimated track parameter uncertainty may not perfectly reflect the RMS of the error distribution. In Belle these imperfections are compensated by scale factors that depend on track $p_T$ and $\tan\lambda$. The scale factors are calibrated with cosmic ray events and simulations. In Babar such scale factors are not used.

## 6.3 Vertex fitting by $\chi^2$ minimization

The $B$ Factory experiments have deployed several implementations of vertex fits. A complete description of these algorithms is outside the scope of this book. In the following we sketch the formalism of a generic minimum $\chi^2$ vertex algorithm. A pedagogical introduction to vertex fitting can be found in the lectures by P. Avery (Avery, 1991, 1998).

To start, we consider a collection of $N$ charged tracks and use a $\chi^2$ minimization algorithm to determine the best vertex out of which they emerge. Once that is done, the vertex can be improved by adding neutral particles, enforcing mass constraints to the in-going or some of the outgoing composite particles, and requiring consistency of the vertex location with the collider luminous region. The goodness of a fit is measured by testing the compatibility of the minimum $\chi^2$ with the expected probability distribution of a $\chi^2$ with the relevant number of degrees of freedom.

Following the notation in (Fruhwirth, 1987) we denote the reconstructed helix parameters of track $i$ by $\boldsymbol{p}_i$ and the corresponding covariance matrix by $V_i$. Given a set of $N$ outgoing tracks each labeled with an index $i$, the $\chi^2$ of the vertex can be generically written as

$$\chi^2 \; = \; \sum_{i=1}^{N} \left[\boldsymbol{p}_i - \boldsymbol{h}_i(\boldsymbol{x}, \boldsymbol{q}_i)\right]^T V_i^{-1} \left[\boldsymbol{p}_i - \boldsymbol{h}_i(\boldsymbol{x}, \boldsymbol{q}_i)\right] \quad (6.3.1)$$

where $\boldsymbol{x}$ is a 3D vector representing the fitted vertex position, $\boldsymbol{q}_i$ is the fitted momentum vector of the outgoing track and $\boldsymbol{h}_i$, the measurement model, is a function of $\boldsymbol{x}$ and $\boldsymbol{q}_i$ that expresses the parameters of the helical trajectory of the charged particle emerging from the vertex with momentum $\boldsymbol{q}_i$.

The solution to the vertex fit is the set of parameters $\widehat{\boldsymbol{\xi}} \equiv (\boldsymbol{x}, \boldsymbol{q}_1 \ldots \boldsymbol{q}_N)$ that minimizes the $\chi^2$. In case the function $\boldsymbol{h}_i$ is linear in the parameters $\boldsymbol{\xi}$, the solution can be expressed generically as

$$\widehat{\boldsymbol{\xi}} \; = \; \boldsymbol{\xi}_0 - \left[\frac{d^2\chi^2}{d\boldsymbol{\xi}^2}(\boldsymbol{\xi}_0)\right]^{-1} \frac{d\chi^2}{d\boldsymbol{\xi}}(\boldsymbol{\xi}_0) \quad (6.3.2)$$

where $\boldsymbol{\xi}_0$ is an arbitrary starting point for $\boldsymbol{\xi}$. The inverse of the second derivative matrix on the right hand side is also half the covariance matrix for $\widehat{\boldsymbol{\xi}}$. If the derivative of $\boldsymbol{h}_i$ is denoted by $\boldsymbol{H}_i$, this leads to the well known expression for the linear least squares estimator,

$$\widehat{\boldsymbol{\xi}} \; = \; \boldsymbol{\xi}_0 - C \sum_i \boldsymbol{H}_i^T V_i^{-1} \left[\boldsymbol{p}_i - \boldsymbol{h}_i(\boldsymbol{x}, \boldsymbol{q}_i)\right] \quad (6.3.3)$$

with the covariance matrix

$$C \; = \; \left(\sum_i \boldsymbol{H}_i^T V_i^{-1} \boldsymbol{H}_i\right)^{-1}. \quad (6.3.4)$$

For vertex fits to helix trajectories the function $\boldsymbol{h}_i$ is not linear and hence its derivative $\boldsymbol{H}_i$ not constant. In that case the minimum is obtained by starting from a suitable expansion point $\boldsymbol{\xi}_0$ and iteratively applying Eq. (6.3.2) until a certain convergence criterion is met, usually a minimum change in the $\chi^2$.

There are two flavors of measurement models for tracks in vertex fits: If the parameters $\boldsymbol{p}_i$ are helix parameters, the measurement model is given by the inverse of Eq. (6.2.1) and Eq. (6.2.2) above. Alternatively, the track parameters can also be translated into position and momentum space using Eq. (6.2.1) and Eq. (6.2.2). In this case the measurement model is trivial, but has one dimension more than the original five parameter helix. Furthermore, since the transformation only applies to a particular point on the helix, it needs to be repeated if the vertex position estimate changes between iterations.

The number of degrees of freedom of the computed $\chi^2$ is $N_{DOF} \equiv 2N - 3$, *i.e.* the difference between the number of measurements, $5N$ (5 helix parameters per track) and the number of fitted parameters $3(N+1)$ (3 vertex coordinates and 3 momentum components per track). Assuming that the uncertainties on the track parameters are correctly estimated *i.e.* that they are representative of the RMS of the error distribution, the minimum $\chi^2$ follows the probability distribution of a $\chi^2$ variate with $N_{DOF}$ degrees of freedom whose expectation value equals $N_{DOF}$. A goodness of fit requirement is usually derived from $\chi^2$ and $N_{DOF}$ to retain the acceptable $N$-prong vertices *e.g.* in the selection of event data samples.

The vertex fitting formalism can be extended with additional constraints, such as prior knowledge of the vertex position (for example from knowledge of the interaction point, IP) or the known mass of the decaying particle. Such constraints always take the form of a constraint equation

$$f(\boldsymbol{\xi}) \; = \; 0. \quad (6.3.5)$$

A distinction can be made between exact constraints and constraints that have an associated uncertainty. The latter are sometimes called '$\chi^2$ constraints'. Mass constraints are usually (but not always) implemented as exact constraints while IP constraints are an example of a $\chi^2$ constraint. Exact constraints can be implemented by using a Lagrange multiplier. They add a term to the $\chi^2$

$$\Delta\chi^2 = \lambda f(\boldsymbol{\xi}) \tag{6.3.6}$$

where the Lagrange multiplier $\lambda$ is treated as an additional parameter in the vertex fit. An alternative (more efficient) method to deal with exact constraints is discussed in (Hulsbergen, 2005). For one-dimensional constraints with an uncertainty $\sigma$ the $\chi^2$ contribution is

$$\Delta\chi^2 = \frac{f(\boldsymbol{\xi})^2}{\sigma^2}. \tag{6.3.7}$$

This expression can be generalized to more than one dimension by writing it in a matrix notation. Note that each independent constraint adds one degree of freedom to the $\chi^2$.

The vertex fit can also be extended to include reconstructed neutral particles. Photons reconstructed as calorimeter clusters do not add position information to the vertex, but they contribute to the momentum, and affect the $\chi^2$ minimization once mass constraints are applied.

Several vertex fits are implemented in sequence to reconstruct decay trees that involve more than one decay vertex, *e.g.* $B \to DX$ transitions. Such decay trees are usually reconstructed by starting from the most downstream vertex and working towards the mother of the decay trees: first fit the $D$ vertex, then use the result to fit the $B$ (this approach is sometimes called leaf-by-leaf fitting). Other more global associations of constraints are implemented for decay trees with leaves or branches with many neutral particles (Hulsbergen, 2005).

The vertex fits applied in the $B$ Factory experiments are essentially extensions of the scheme above – see in particular (Tanaka, 2001) for Belle and (Hulsbergen, 2005) for *BABAR*. Implementations of the vertex fitting algorithm differ both in the parameterization of the problem and in the way the $\chi^2$ is minimized. As outlined above, tracks can be parameterized in terms of helix coordinates or (locally) in terms of Cartesian coordinates. The latter leads to simpler expressions for derivatives, but may lead to slower convergence because derivatives vary more rapidly along the track.

For the minimization both the global $\chi^2$ fit technique described above and the Kalman filter are used. Even for algorithms that seemingly use the same minimization scheme, the implementations may differ. To our knowledge, the most efficient method to fit tracks to a common vertex is the algorithm developed by Billoir, Fruhwirth, and Regler (1985), presented in slightly different from in (Fruhwirth, 1987). This algorithm was extended with a mass constraint in (Amoraal et al., 2013).

Not all algorithms are applicable to all vertexing problems. The general leaf-by-leaf approach for decay tree fitting cannot easily be applied to the reconstruction of *e.g.* $K_S^0 \to \pi^0\pi^0$ or $B^0 \to K_S^0\pi^0$. For these types of decay trees a 'global' decay tree fit can be used (Hulsbergen, 2005). The latter also has the advantage that one has access to the vertex-constrained parameters of all particles in the decay tree. However, this algorithm computes a single covariance matrix for all of the parameters in the decay tree, making it noticeably slower than a leaf-by-leaf approach. The CPU consumption of vertex algorithms is often a concern because of the combinatoric background in the reconstruction and selection of composite particles.

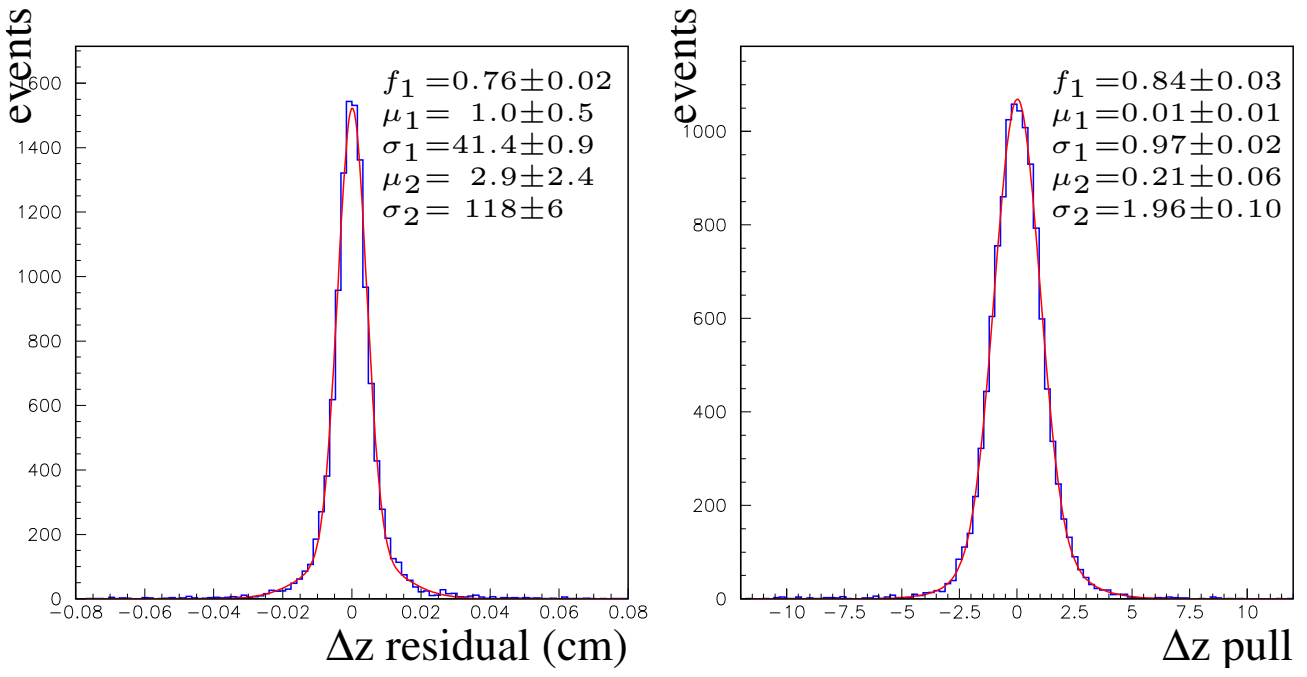

**Figure 6.3.1.** Residual (left) and pull (right) of the decay vertex $z$ position of reconstructed $B^0 \to J/\psi K_S^0$ candidates in a *BABAR* simulated data sample. Fits to a double Gaussian are superimposed.

A strict control on the accuracy of the vertex reconstruction is mandatory for the $B$ Factory experiments where the primary goal is to determine time-dependent *CP* asymmetries from the distance between two vertices. This is illustrated in Figure 6.3.1 which shows the residuals and pull[20] for the decay vertex $z$ position of reconstructed $B^0 \to J/\psi K_S^0$ ($J/\psi \to \mu^+\mu^-$) candidates from a sample of simulated data taken from *BABAR*. The vertex resolution depends on the topology of the decay and the direction and momenta of the final state particles and especially on whether the $K_S^0$ particles decays inside or outside the vertex detector volume. These effects are accounted for in the per-event reconstruction uncertainty, the estimate of which is computed by the vertex fit algorithm. Due to spread in the estimated uncertainty, the vertex resolution is not a Gaussian distribution. However, the pull distribution is reasonably Gaussian with an RMS value close to unity, indicating that the uncertainties are correctly estimated.

For this decay the $z$ residual distribution has an RMS of about 70 $\mu$m. A double Gaussian fit returns a core component, which corresponds to about three quarters of the distribution, with a standard deviation equal to 40 $\mu$m. The resolution in the transverse coordinates is comparable to that in $z$: about 50 $\mu$m.

Figure 6.3.2 shows the reconstructed mass of $B^\pm \to J/\psi K^\pm$ decays in data, from *BABAR*, fitted both with and without a mass constraint on the $J/\psi \to \mu^+\mu^-$ decay. The mass constraint improves the accuracy of the derived $J/\psi$ momentum and this leads to a large improvement in the $B^\pm$ invariant mass resolution. The improvement in mass resolution is comparable to what one would obtain by considering the $B^{\pm}$-$J/\psi$ mass difference instead of the $B^{\pm}$ mass. However, the advantage of applying the mass-constrained vertex fit is that the resolution on both the vertex position and on the $B$ momentum are improved.

[20] A 'pull' is a residual divided by its estimated uncertainty. See also Section 11.5.2.

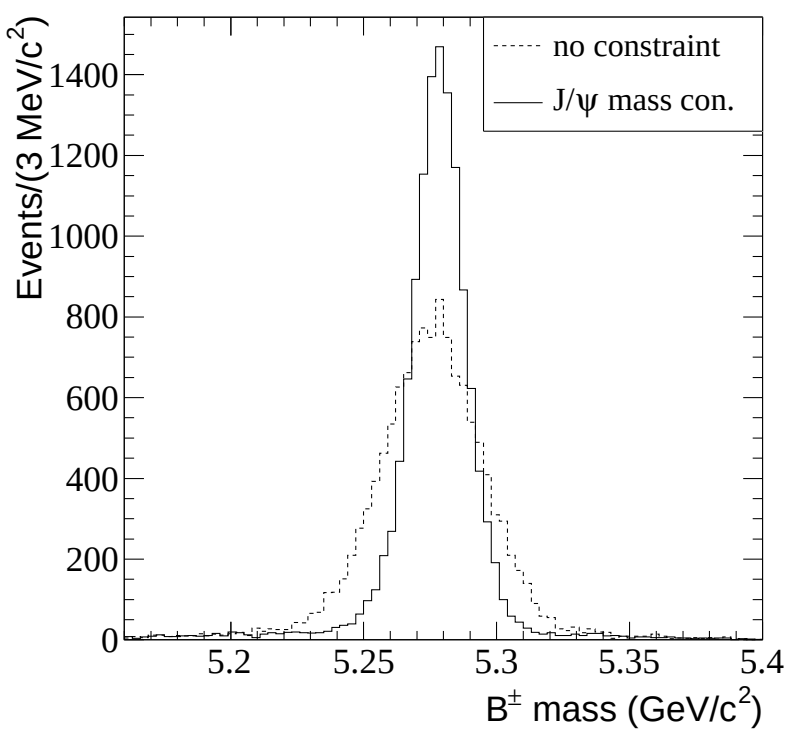

**Figure 6.3.2.** Distribution of the reconstructed invariant mass of $B^{\pm} \to J/\psi K^{\pm}$ decays in *BABAR* data with and without enforcement of a mass constraint on the $J/\psi \to \mu^+\mu^-$ decay vertex leaf.

## 6.4 Primary vertex reconstruction and beamspot calibration

The majority of beam-beam collisions occur in a tiny region in the center of the detectors, the interaction region or *beamspot*. The size of the interaction region is determined by beam optics and has varied through the $B$ Factory runs. It is typically 1 mm along the beam ($z$), 100 $\mu$m in the horizontal direction ($x$) and a few $\mu$m in the vertical direction ($y$).

The position and size of the beamspot are used as a constraint in the reconstruction of the $B^0\overline{B}^0$ decay time difference $\Delta t$. Since the beamspot is smallest in the vertical plane, the vertical coordinate is the most constraining. In the directions along $x$ and $z$ the beamspot is not smaller than a typical $B$ decay length, which is about 25 $\mu$m in the transverse plane and about 200 $\mu$m along the $z$-axis, and its constraint plays a marginal role.

The position and shape of the interaction region vary with time and needs to be carefully calibrated and monitored. The calibration is based on the spatial distribution of reconstructed primary vertices (PVs). In the production of a $B^0\overline{B}^0$ or $B^+B^-$ pair at the $\Upsilon(4S)$ resonance there are no particles originating from the primary collision point other than the $B$ mesons themselves. Consequently, the primary vertex cannot be directly reconstructed in these decays and the beamspot calibration instead relies on continuum events. Bhabha and di-muon events have the advantage that there are only two tracks in the event, that have both relatively high momentum and are guaranteed to originate from the PV. Hadronic events have more tracks and consequently a smaller statistical per-event uncertainty on the vertex position, but they are polluted by a $b\bar{b}$ contribution. The calibration in *BABAR* relies both on two-prong events and on multi-hadron events with at least 5 tracks. The calibration procedure in Belle uses only multi-hadron events (Tomura, 2002a).

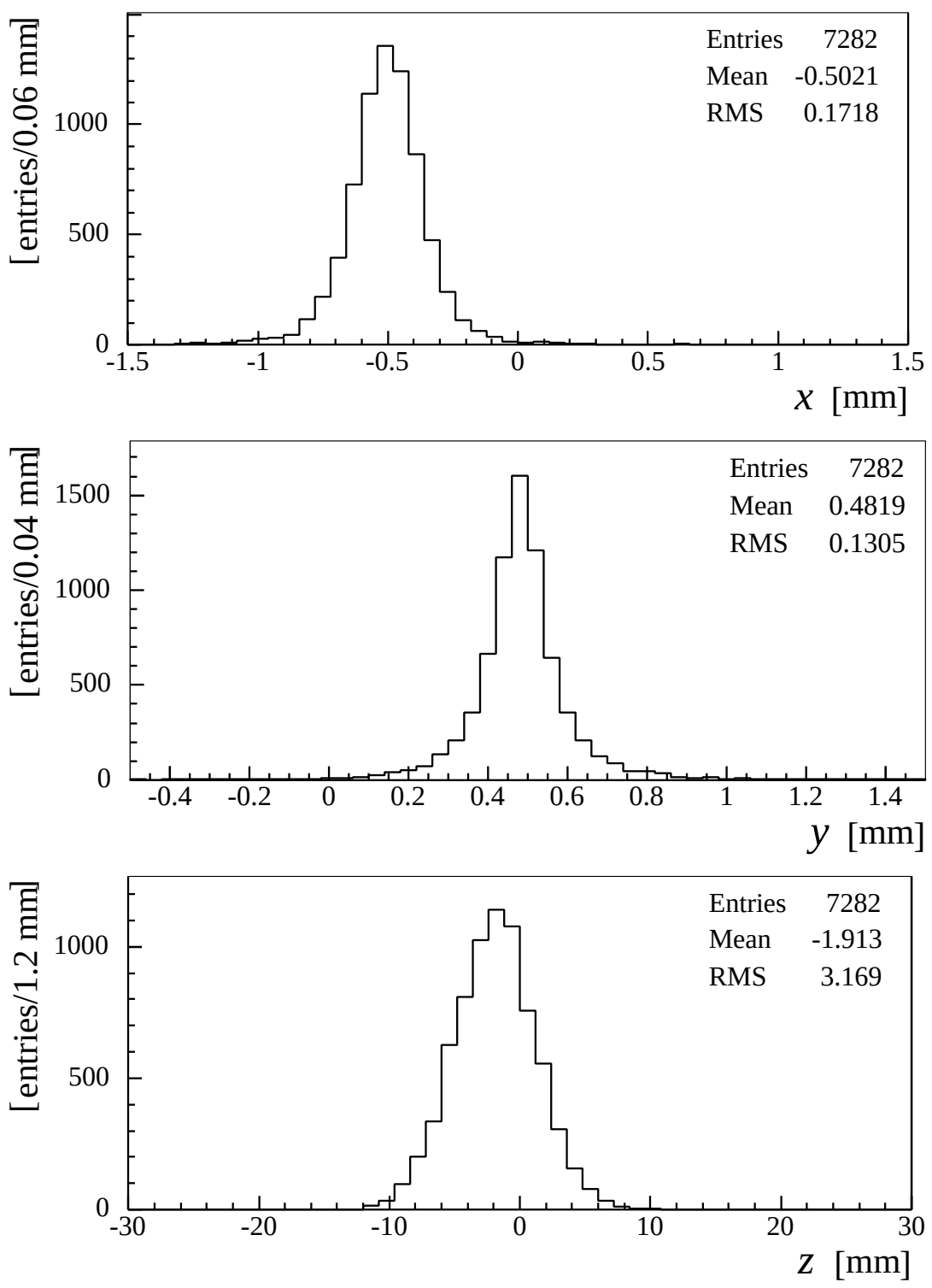

**Figure 6.4.1.** Distribution of the $x$ (top), $y$ (middle), and $z$ (bottom) position of reconstructed primary vertices in a typical Belle run (Exp. 5, run 333). From (Tomura, 2002a).

An example of the distribution of the position of reconstructed primary vertices in hadronic events in a typical Belle run is shown in Figure 6.4.1. In the $y$ direction the RMS of the distribution is dominated by the vertex resolution. In the $z$ direction it is dominated by the beamspot size, while in the $x$ direction it is a combination of both.

The distribution of PV positions is characterized by an average position, the direction of its three principal axes (which are close, but not identical to the $x$, $y$ and $z$ axis; see Chapter 2) and the RMS along each axis. The calibrated position, rotation and sizes are determined from moments of (*BABAR*) or fits to (Belle) the $(x, y, z)$ distribution of PVs.

To determine the size of the beamspot the vertex resolution must be 'subtracted'. In the vertical direction since the resolution is so much wider than the beam size, the beam spread must be estimated by other means. In *BABAR* the size in $y$ is computed from the luminosity reported by the accelerator (Chapter 1). In Belle it is obtained from measurements of the size of the HER and LER beams by the accelerator (Tomura, 2002a). When the beamspot is used as a constraint in vertex fits, its size always appears in quadrature with the actual vertex resolution. Hence, it is important to know the size in the vertical direction precisely.

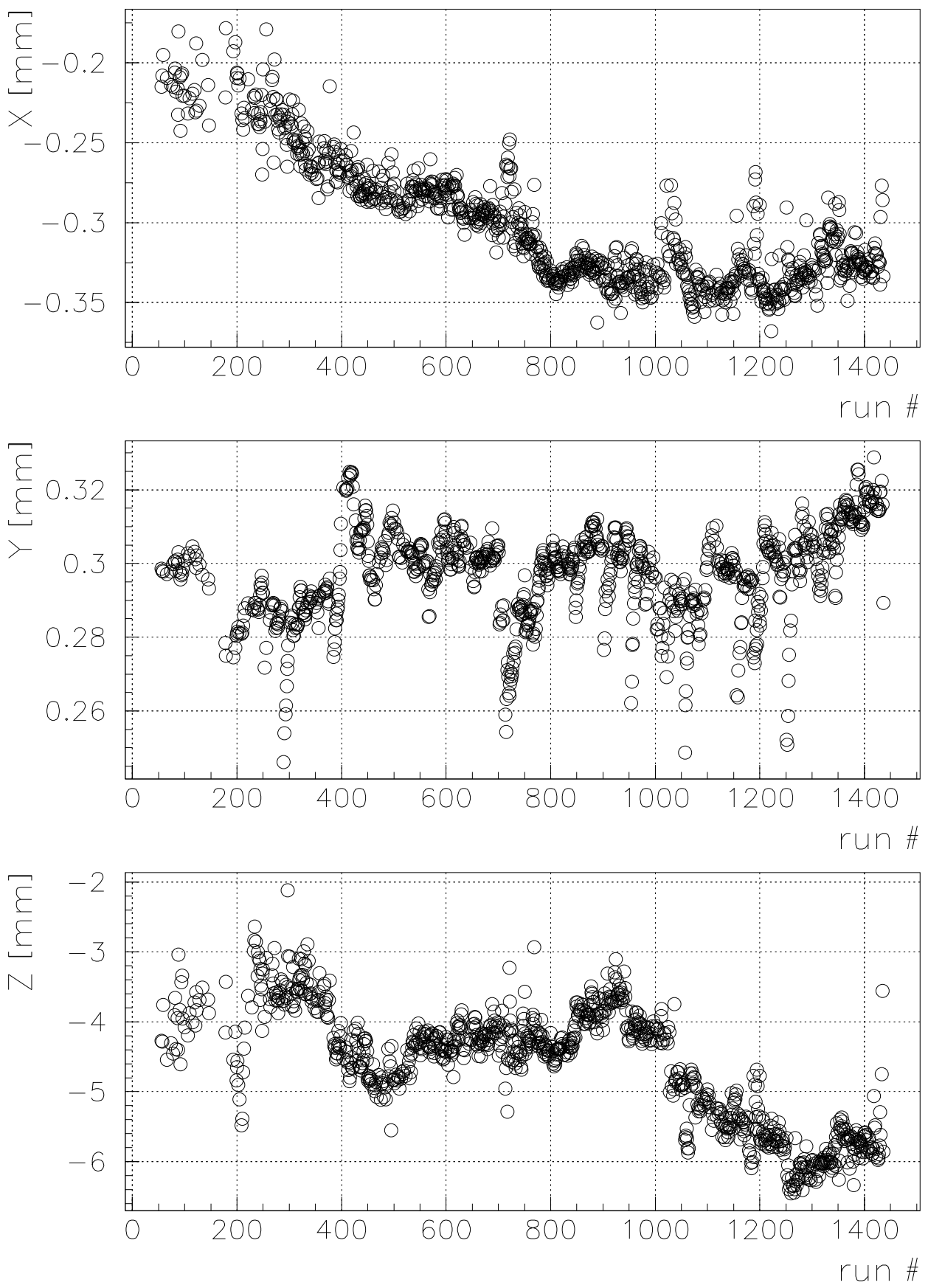

**Figure 6.4.2.** Average primary vertex position in $x$ (top), $y$ (center) and $z$ (bottom) as a function of run number in Belle data. From (Tomura, 2002a).

To accommodate variations over time the calibration procedure is performed in time slices. Belle fits the mean position with the other parameters (the widths and the rotation angles) fixed for every $\mathcal{O}(10^4)$ events. *BABAR* updates all parameters every $\sim$ 10 minute interval, corresponding to approximately the same number of selected events. Figure 6.4.2 shows the average primary vertex position as a function of run number in the early days of Belle. In this period the typical duration of a run was about 2 hours. Under stable conditions, the variation of the position within a run is much smaller, typically of the order of 10 $\mu$m in $x$, 1 $\mu$m in $y$ and 100 $\mu$m in $z$ in both experiments.

In vertex reconstruction the average beamspot can be used as a constraint on the *production* vertex of the $B$ (or $D$, or $\tau$) particle. The $\chi^2$ contribution takes the form, *cf.* Eq. (6.3.7),

$$\Delta\chi^2 = \begin{pmatrix} x_{\rm p} - x_{\rm IP} \\ y_{\rm p} - y_{\rm IP} \\ z_{\rm p} - z_{\rm IP} \end{pmatrix}^T V_{\rm IP}^{-1} \begin{pmatrix} x_{\rm p} - x_{\rm IP} \\ y_{\rm p} - y_{\rm IP} \\ z_{\rm p} - z_{\rm IP} \end{pmatrix} \qquad (6.4.1)$$

where $\boldsymbol{x}_{\rm p}$ are the parameters of the production vertex in the vertex fit, $\boldsymbol{x}_{\rm IP}$ is the position of the center of the beamspot and $V_{\rm IP}$ is a $3 \times 3$ covariance matrix, representative of the size of the beamspot. In Belle the constraint is only applied to the coordinates in the transverse plane; in *BABAR* both the 2D and 3D constraint are used, depending on the vertex algorithm. Figure 6.4.3 shows the $D^{*+} - D^0$ mass difference in $e^+e^- \to D^{*+}X$ continuum events where we have selected $D^{*+} \to D^0\pi^+$ decays with $D^0 \to K^-\pi^+$ with and without the constraint that the $D^{*+}$ originates from the beamspot. Due to its low momentum the direction of the soft pion is very sensitive to multiple scattering. Requiring it to originate from the interaction region substantially improves the mass resolution.

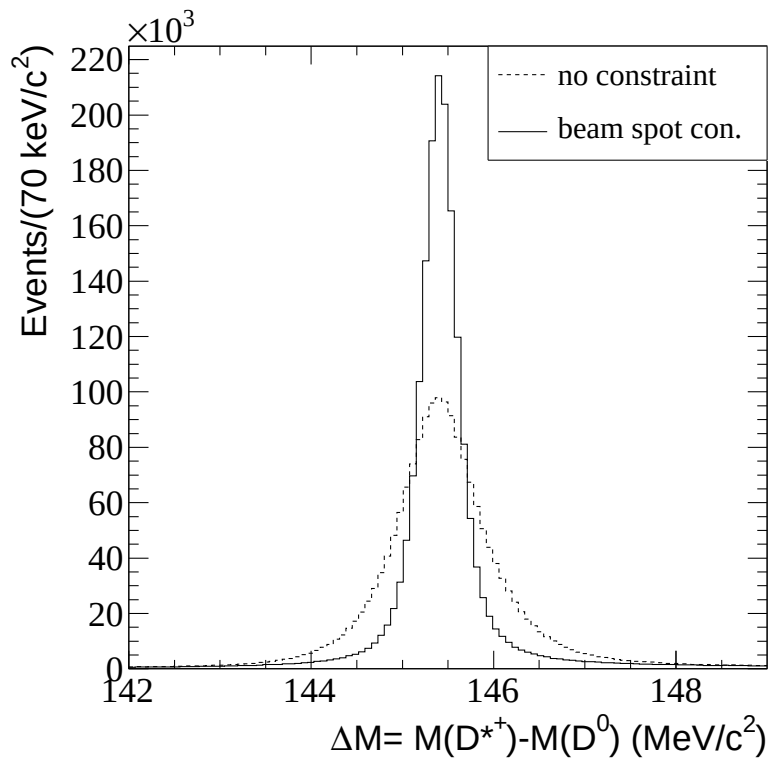

**Figure 6.4.3.** Distribution of the reconstructed $D^{*+} - D^0$ mass difference in $D^{*+} \to D^0\pi^+$ decays with $D^0 \to K^-\pi^+$ from *BABAR* continuum data with and without a primary vertex constraint.

In some applications, such as for $D^*$ from $B$ decays or the reconstruction of the associated $B$ vertex for $\Delta t$ reconstruction in Belle, the beamspot is used as a constraint on a *decay* vertex. In this case the size of the beamspot must be increased with the effective width of the decay length distribution of the (mother) particle, schematically,

$$V_{\rm IP,tot} = V_{\rm IP} + V_{\rm flight} \, . \qquad (6.4.2)$$

Both experiments add the RMS of the $B$ decay length distribution in the transverse plane (about 25 $\mu$m, see Figure 6.4.4) in quadrature with the calibrated beamspot size to obtain an effective size appropriate for $B$ decay products. This mostly affects the size in $y$.

Finally, although these quantities do not directly pertain to the vertex algorithms, it is convenient in the characterization of the beamspot, to mention the calibration of the beam kinematics. The beam energies are used in the computation of *e.g.* the beam-energy-constrained mass (Chapter 9) and the proper decay time. In principle, there are six unknown parameters related to the incident beams, namely the 3-momenta of the electron and positron beam. In practice, the beam-directions are close enough to their nominal direction that only the relative direction matters, reducing the number of degrees of freedom to four. These are parameterized by the center-of-momentum energy $\sqrt{s}$ and by the boost vector.

Both experiments calibrate $\sqrt{s}$ with the kinematics of fully reconstructed hadronic $B$ decays. In particular, a deviation of $\sqrt{s}$ from nominal can be directly inferred from a shift of the beam-energy-constrained mass relative to the nominal $B$ mass. The uncertainty is dominated by the uncertainty on the nominal $B$ mass.

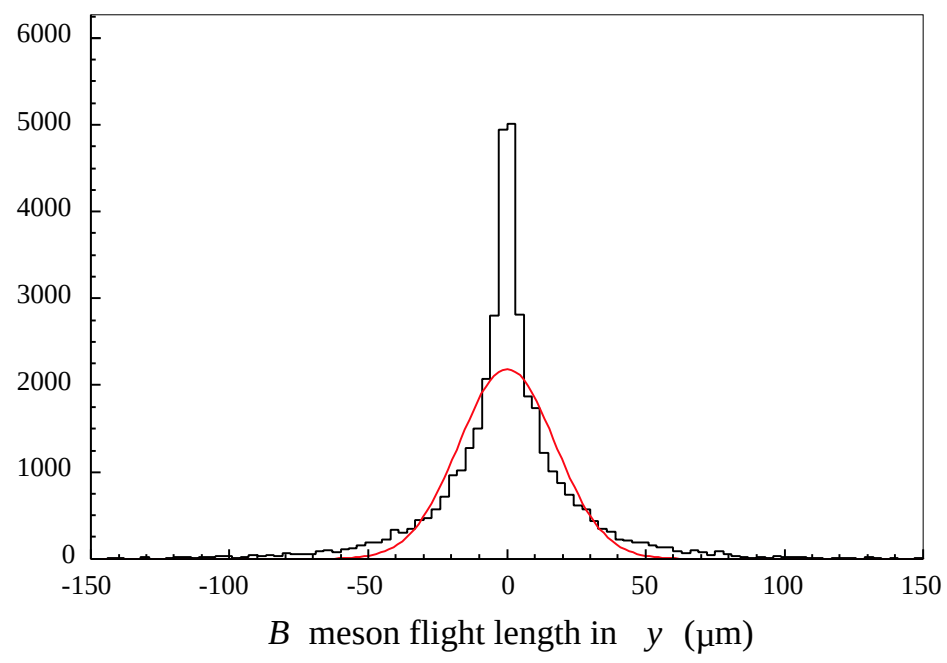

**Figure 6.4.4.** Distribution of the $B$ meson flight length in the $y$ direction in Belle simulated data. A fit to a single Gaussian (red), with a width of 25 $\mu$m, is superimposed.

In Belle the boost vector is sufficiently constant that it has been fixed to its nominal value for the entire period of data taking. In *BABAR* the boost vector is calibrated on a run-by-run basis using the four-momentum sum in dimuon events. Note that due to effects of initial and final state radiation, the latter is not a very sensitive probe of $\sqrt{s}$.

## 6.5 $\Delta t$ determination

The analysis of time-dependent *CP* violation in decays of neutral $B$ mesons at the $e^+e^-$ $B$ Factories requires measurement of the decay time difference $\Delta t$ of the two $B$ mesons in the event (see Chapter 10). The procedures to reconstruct the vertex of the 'tagging' $B$ and extract the decay time difference are described in (Tajima, 2004) for Belle and (Aubert, 2001c, 2002a) for *BABAR*, to which the reader is referred for details.

We denote the reconstructed $B$ meson that decays to the final state of interest as $B_{\rm rec}$. We label the other $B$ meson by $B_{\rm tag}$, because its decay products are used to determine the flavor of $B_{\rm rec}$ at $\Delta t = 0$. In an asymmetric $e^+e^-$ $B$ Factory the determination of $\Delta t$ is derived from the measurement of the difference in the decay vertex positions of $B_{\rm rec}$ and $B_{\rm tag}$ along the boost axis, which is approximately the $z$ axis. Consequently, we talk about the $\Delta z$ *measurement* and the $\Delta z$ *to* $\Delta t$ *conversion*.

By far the dominant contribution to the resolution on $\Delta t$ is the $\Delta z$ resolution. For most analyses the latter is in turn dominated by the $B_{\rm tag}$ vertex resolution. The determination of the $B_{\rm rec}$ vertex position is performed with a standard vertex fit, as described above. The reconstruction of the $B_{\rm tag}$ vertex position is more complicated since it requires the selection of the subset of tracks that directly originate from the $B_{\rm tag}$ vertex.

### 6.5.1 Reconstruction of the $B_{\rm tag}$ vertex

Figure 6.5.1 shows schematically the topology of an event with the $B_{\rm rec}$ and $B_{\rm tag}$ decays. Since there are no other particles in the event beside the two $B$ mesons, all tracks that are not associated to $B_{\rm rec}$, *i.e.* tracks from the *rest of the event* (ROE), necessarily originate from the $B_{\rm tag}$ decay. However, a couple of experimental complications make the reconstruction of the $B_{\rm tag}$ vertex position non-trivial. First, in only a small fraction of events, are all the decay products of the $B_{\rm tag}$ inside the acceptance of the detector, hence a strategy based on a full reconstruction is excluded.[21]

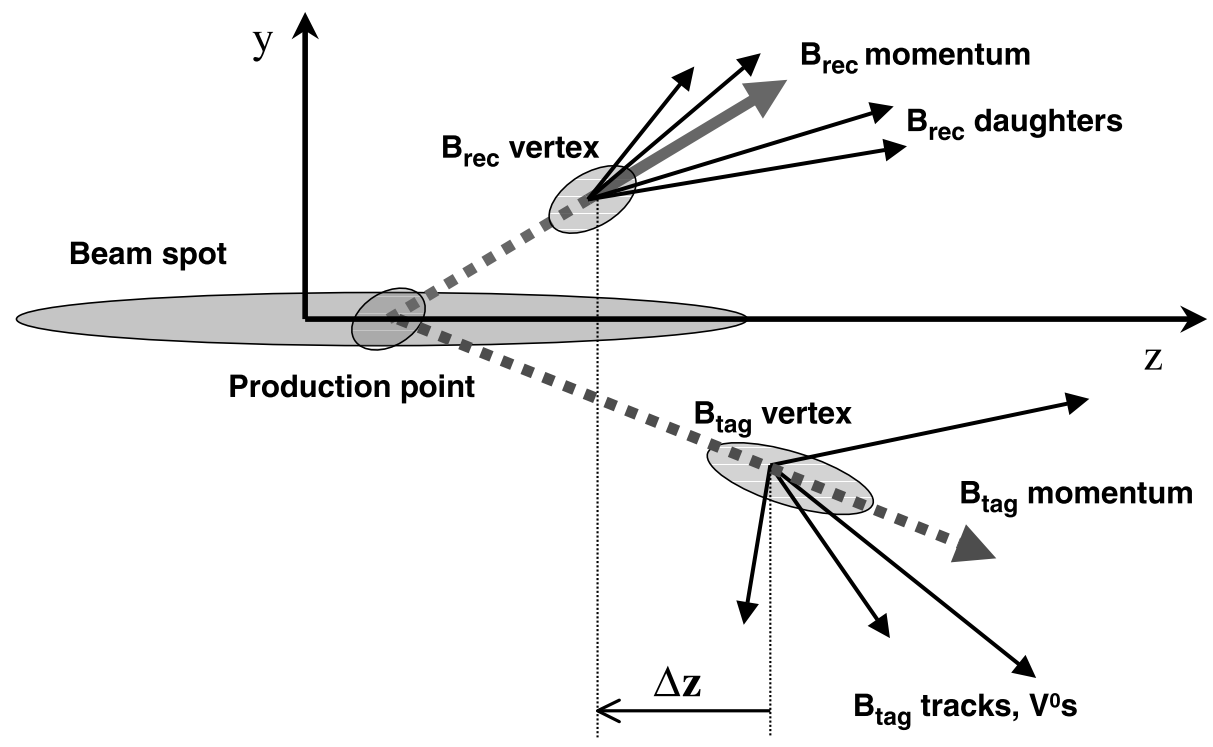

**Figure 6.5.1.** Schematic view of the geometry in the $yz$ plane for a $\Upsilon(4S) \to B\overline{B}$ decay. For fully reconstructed decay modes, the line of flight of the $B_{\rm tag}$ can be estimated from the (reverse) momentum vector and the vertex position of $B_{\rm rec}$, and from the beamspot position in the $xy$ plane and the $\Upsilon(4S)$ average boost. Note that the scale in the $y$ direction is substantially magnified compared to that in the $z$ direction. From (Aubert, 2002a).

Second, most $B_{\rm tag}$ mesons decay to an open-charmed particle with at least one additional vertex after a flight length comparable to the decay length of a $B$ meson. The confusion in the assignment of the tracks between these vertices biases the measurement of the $B_{\rm tag}$ position and degrades the $B_{\rm tag}$ vertex resolution.

The strategy to select the optimal set of tracks is similar in both experiments. First, from the tracks in the ROE a subset is selected that satisfies requirements like a minimum number of vertex detector hits and a maximum transverse distance to the interaction region. Tracks from reconstructed photon conversions and $V^0$ decays (a neutral particle decaying into two charged tracks, for example $K^0_S \to \pi^+\pi^-$) are either removed or replaced with the mother particle. Subsequently, all tracks are combined in a single vertex using the interaction region as a constraint. If the $\chi^2$ of the vertex is larger than a certain criterion, the worst track is removed and the vertex refitted. This procedure is repeated until the criterion is satisfied or no tracks are left. In *BABAR* the criterion is a maximum contribution to the $\chi^2$ of 6 for each track, while in Belle the criterion is a maximum vertex $\chi^2$ of 20 per degree of freedom (since a track contributes two degrees of freedom, the *BABAR* criterion is substantially tighter than the Belle criterion). In Belle tracks that have been identified as high $p_T$ leptons by the flavor tagging algorithm are always kept since those have a large probability to originate from the $B_{\rm tag}$ vertex.

[21] Also the sum of branching fractions of decays used in typical full reconstruction, see Chapter 7, is small.

If the beamspot is used as a constraint in the $B_{\rm tag}$ vertex reconstruction, even vertices with a single track can be reconstructed. The experiments exploit the beamspot differently. In Belle the constraint is an ellipsoid in the $xy$ plane, increased in size to account for the $B_{\rm tag}$ transverse motion, as explained in Section 6.4. This use of the beamspot leads to a small bias that is proportional to the $B_{\rm tag}$ decay time and is treated as a systematic uncertainty. In *BABAR* the $B_{\rm tag}$ direction and origin are reconstructed with a vertex fit using the $B_{\rm rec}$ vertex and momentum and the calibrated beamspot position and $\Upsilon(4S)$ momentum. This $B_{\rm tag}$ 'pseudo-particle' is subsequently used as any other track in the $B_{\rm tag}$ vertex reconstruction. The advantage of this approach is that there is no bias due to the beamspot constraint. However, it can only be applied to analyses with a fully reconstructed $B_{\rm rec}$.

Since the $B_{\rm tag}$ vertex has in general fewer tracks than the $B_{\rm rec}$ vertex and may be contaminated by $D$ daughter tracks, the $\Delta z$ resolution is dominated by the $B_{\rm tag}$ $z$ position resolution. The latter is in the range $100-200$ $\mu$m, which has to be compared to a typical resolution of the $B_{\rm rec}$ vertex of 50 $\mu$m. As the total resolution is of the order of the $B$ mixing period, accurate knowledge of the resolution is essential when $\Delta t$ is used in maximum likelihood fits to extract the parameters for time-dependent *CP* violation. The calibration of the so-called *resolution function* is discussed below.

### 6.5.2 From vertex positions to $\Delta t$

To be sensitive to time-dependent *CP* violating effects the vertex resolution must be sufficient to resolve the oscillations due to $B^0\overline{B}^0$ mixing in the decay time distribution. Given a proper decay time $t$ and a momentum vector $\boldsymbol{p}$, the difference between the production and decay vertex positions of a $B$ meson is given by

$$\boldsymbol{x}_{\rm decay} - \boldsymbol{x}_{\rm prod} \;=\; \frac{\boldsymbol{p}c}{mc^2}\,ct \qquad (6.5.1)$$

where we have explicitly included factors $c$ to express momentum and mass in units of energy. At the $\Upsilon(4S)$ resonance the $B$ momentum in the $\Upsilon(4S)$ rest frame is $p^*_B \approx 340\,{\rm MeV}/c$. With a lifetime of 1.5 ps, the $B^0$ decay length in the $\Upsilon(4S)$ frame is only $\sim 30$ $\mu$m, small compared to the typical resolution of vertex detectors. This is the main motivation for constructing an asymmetric $B$ Factory: the boost of the $\Upsilon(4S)$ system increases the decay length, making the measurement of the decay time possible.

If the $z$-axis is chosen along the boost direction, the experimental resolution on the $B$ meson decay time difference is dominated by the resolution on the decay vertex $z$ position. The displacement in $z$ of one of the $B$ mesons is related to its proper decay time $t$ by

$$z_{\rm decay} - z_{\rm prod} \;=\; \gamma\left(\alpha\cos\theta + \beta\sqrt{1+\alpha^2}\right)ct \quad (6.5.2)$$

where $\gamma$ and $\beta$ are the boost parameters from the $\Upsilon(4S)$ frame to the lab frame and $\theta$ and $\alpha = p^*_B c/m_B c^2$ are the polar angle and boost factor of the $B$ in the $\Upsilon(4S)$ frame.

Since no tracks originate from the production vertex, the sensitivity to the decay time difference of the $B$ mesons comes mainly through the difference in the $z$ positions of the decay vertices. As the polar angles of the two $B$ mesons are exactly opposite, the difference in the $z$ positions can be expressed as

$$z_1 - z_2 \;=\; \gamma\beta\sqrt{1+\alpha^2}c(t_1-t_2) + \gamma\alpha\cos\theta c(t_1+t_2). \quad (6.5.3)$$

If the small parameter $\alpha \approx 0.06$ is ignored, one obtains the well known approximation

$$\Delta t \;=\; \Delta z/\gamma\beta c. \qquad (6.5.4)$$

This expression is used for all time-dependent analyses in Belle and for those without a fully reconstructed $B_{\rm rec}$ in *BABAR*. The average value for the boost factor is $\beta\gamma = 0.55$ in *BABAR* and $\beta\gamma = 0.42$ in Belle. It is calculated directly from the beam energies and has a typical uncertainty of 0.1%. For a typical $\Delta z$ resolution of 100 $\mu$m, the $\Delta t$ resolution is 0.6 ps, a bit less than half the $B$ lifetime and small compared to the $B^0$ oscillation period of $\sim 12.5$ ps.

Ignoring the second term in Eq. (6.5.3) leads to a $\cos\theta$ and decay time dependent bias. If the detection efficiency is symmetric in $\cos\theta$, the expectation value of the bias is zero.[22] Ignoring the acceptance and taking $P(\cos\theta) \propto 1-\cos^2\theta$, the RMS of this term is $2\gamma\alpha c\tau_{B^0}/\sqrt{5}$, or about 30 $\mu$m (taking $\langle t_1+t_2\rangle \sim 2\tau_{B^0}$). Consequently, its contribution to the resolution is small but not negligible.

In the case of a fully reconstructed $B_{\rm rec}$ the momentum direction is measured with sufficient precision to correct for the $B$ momentum in the $\Upsilon(4S)$ frame. However, as can be seen in Eq. (6.5.3) the correction depends on the sum of the decay times, $t_1+t_2$, which can only be determined with very poor resolution. *BABAR* has used the estimate $t_1 + t_2 = \tau_B + |\Delta t|$ to correct for the measured $B_{\rm rec}$ momentum direction and extract $\Delta t$ from Eq. (6.5.3), giving

$$\Delta t \;=\; \frac{\Delta z/c \,-\, \gamma\alpha\cos\theta\,\tau_B}{\gamma\beta + s\gamma\alpha\cos\theta} \qquad (6.5.5)$$

where $s$ is the sign of $\Delta z$ and terms quadratic in $\alpha$ have been ignored. The distribution of the event-by-event difference between $\Delta t$ computed with Eq. (6.5.4) and Eq. (6.5.5) has an RMS of 0.20 ps. Therefore, for a typical resolution of 0.6 ps, the $\cos\theta$ correction improves the $\Delta t$ resolution by about 5% (Aubert, 2002a).

Equation (6.5.5) is used for most $B$ decays to hadronic final states in *BABAR*, while Eq (6.5.4) is used for semileptonic modes. In Belle the correction is not applied, but included in the resolution model. The contribution to the resolution is computed on a per-event basis for fully reconstructed final states and empirically parameterized from simulated events for the semi-leptonic modes.

[22] Assuming that also the distribution of events is symmetric in $\cos\theta$, which is valid in the case of $B\overline{B}$ events.

The time-dependent analysis of decays $B^0 \to K_S^0\pi^0$ and $B^0 \to K_S^0\pi^0\gamma$ is particularly challenging because there are no tracks directly originating from the $B_{\rm rec}$ vertex. In early analyses in *BABAR* (Aubert, 2004q), the $B_{\rm rec}$ vertex position was estimated from the intersection of the trajectories of one or both $K_S^0$ daughters with the beamspot. The implementation was similar to the reconstruction of $B_{\rm tag}$ vertices with a single track and the standard $\Delta z$ to $\Delta t$ conversion (see above) was used. This method suffers from a bias, small compared to the resolution, but irreducible.

Eventually *BABAR* developed a third method that makes use of a decay tree fit (Hulsbergen, 2005) which was applied to a number of decays including $B^0 \to K_S^0 K_S^0 K_S^0$. In this algorithm the decay time difference $\Delta t$ is extracted from a single vertex fit to the $\Upsilon(4S) \to B^0\overline{B}^0$ decay tree, using all reconstructed particles associated with $B_{\rm rec}$ and $B_{\rm tag}$ and knowledge of the average interaction point and $\Upsilon(4S)$ momentum. The particles missing from the $B_{\rm tag}$ vertex are parameterized as a single unconstrained four-vector at the $B_{\rm tag}$ vertex. This algorithm maximally exploits all available information from reconstruction and beam parameter calibration. It is interesting that it obtains a competitive resolution only if a constraint on the $B$ decay time sum is applied. The latter is implemented as a $\chi^2$ constraint $t_1 + t_2 = 2\tau_B$ with (RMS) uncertainty $\sqrt{2}\tau_B$. Note that this approach is similar but not identical to the substitution $t_1 + t_2 = \tau_B + |\Delta t|$ applied in the 'momentum corrected' method described above. It has been verified that such a constraint does not bias the $\Delta t$ measurement. However, since this method does not lead to a significant improvement in resolution, it has only been applied to studies of $B^0 \to K_S^0\pi^0$ and alike.

### 6.5.3 $\Delta t$ resolution function

To account for the finite decay time resolution the *p.d.f.* describing the physical time evolution in a time-dependent analysis is convolved with a *r*esolution function which is the response function that describes the distribution of the observed decay time as a function of the true decay time $\Delta t_{\rm true}$. To first order the resolution function is a Gaussian function with zero mean and a width corresponding to the average resolution. In practice, the deviations from a Gaussian are important. The parameterization and calibration of the resolution function is described in detail in Section 10.4. Here, we briefly emphasize features of the vertex resolution that impact the $\Delta t$ resolution in time-dependent analyses.

The estimated uncertainty in the $B_{\rm tag}$ vertex $z$ position is a function of the number of tracks assigned to the vertex and the direction and momentum of those tracks. It differs substantially between events, leading to a large variation in the estimated uncertainty on $\Delta t$, as shown in

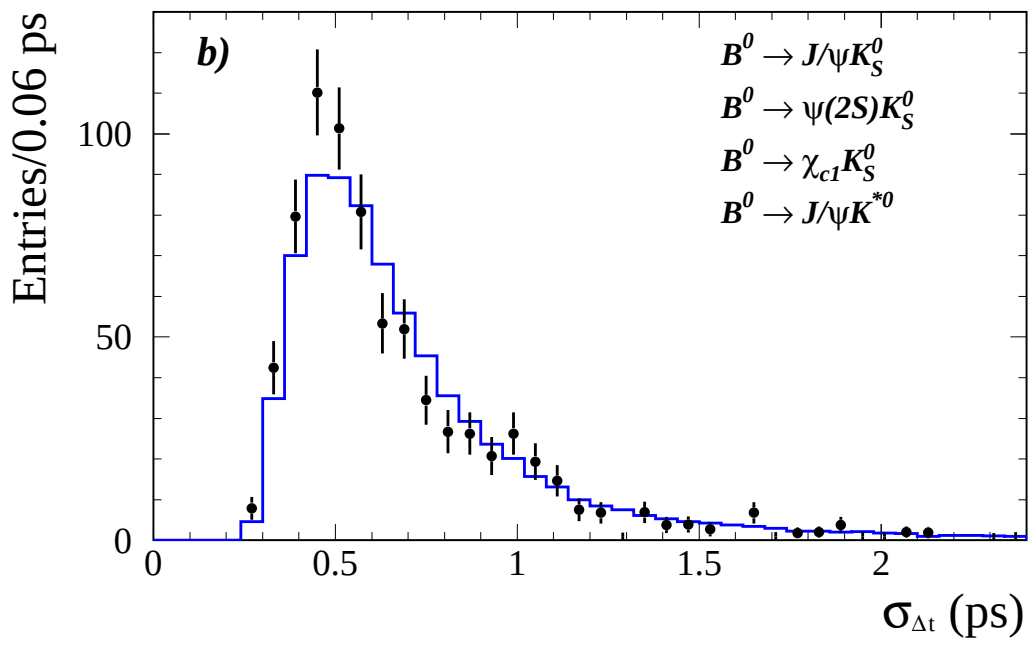

**Figure 6.5.2.** Distribution of event-by-event uncertainty on $\Delta t$ for the $J/\psi K_S^0$, $\psi(2S)K_S^0$, $\chi_{c1}K_S^0$ and $J/\psi K^{*0}$ events. The histogram corresponds to Monte Carlo simulation and the points with error bars to *BABAR* data. From (Aubert, 2002a).

Fig. 6.5.2. The estimated event-by-event uncertainty on $\Delta t$ is denoted by $\sigma_{\Delta t}$.

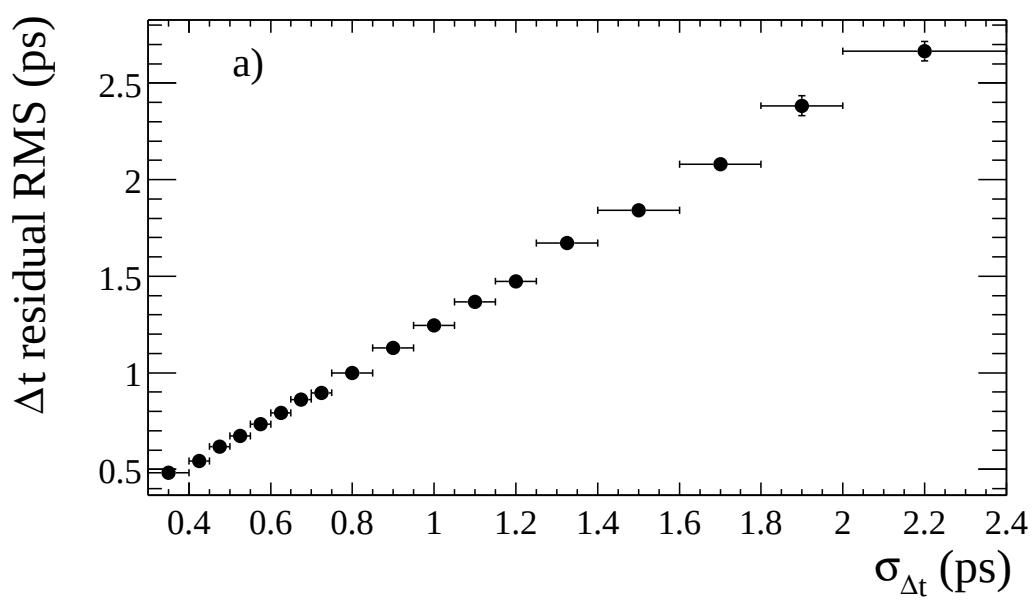

**Figure 6.5.3.** RMS of the $\delta t = \Delta t - \Delta t_{\rm true}$ distribution in *BABAR* simulated events as a function of the estimated event-by-event uncertainty in $\Delta t$. From (Aubert, 2002a).

To benefit statistically from this variation the estimated uncertainty is used in the parameterization of the resolution function. Fig. 6.5.3 shows the actual $\Delta t$ resolution — defined as the RMS of the error distribution — in simulated *BABAR* events as a function of the estimated uncertainty $\sigma_{\Delta t}$. The linear correlation illustrates that $\sigma_{\Delta t}$ is a good measure for the actual resolution, although a scaling factor of approximately 1.1 must be applied to obtain pulls with unit RMS. Therefore, the parameterization of the resolution function typically uses a width that is proportional to $\sigma_{\Delta t}$. The proportionality factor is derived from the data.

The bias due to tracks from $D$ daughters depends on the direction of the $D$ meson in the $B$ rest frame: If the $D$ meson moves approximately perpendicular to the $z$ axis, the $z$ positions of $D$ and $B$ vertices coincide and the bias is small. Due to the boost of the $D$ meson in the $B$ frame, in such events the $D$ daughter trajectories also have a relatively large angle with respect to the beam direction, leading to a small vertex position uncertainty. It is for this reason that both experiments observe that the bias from $D$ daughter tracks is roughly proportional to the per-event estimated uncertainty on the $B_{\rm tag}$ vertex $z$ position, as il-lustrated in Figure 6.5.4. Therefore, the parameterization of the resolution function for $B$ decays often also uses a mean that is proportional to $\sigma_{\Delta t}$.

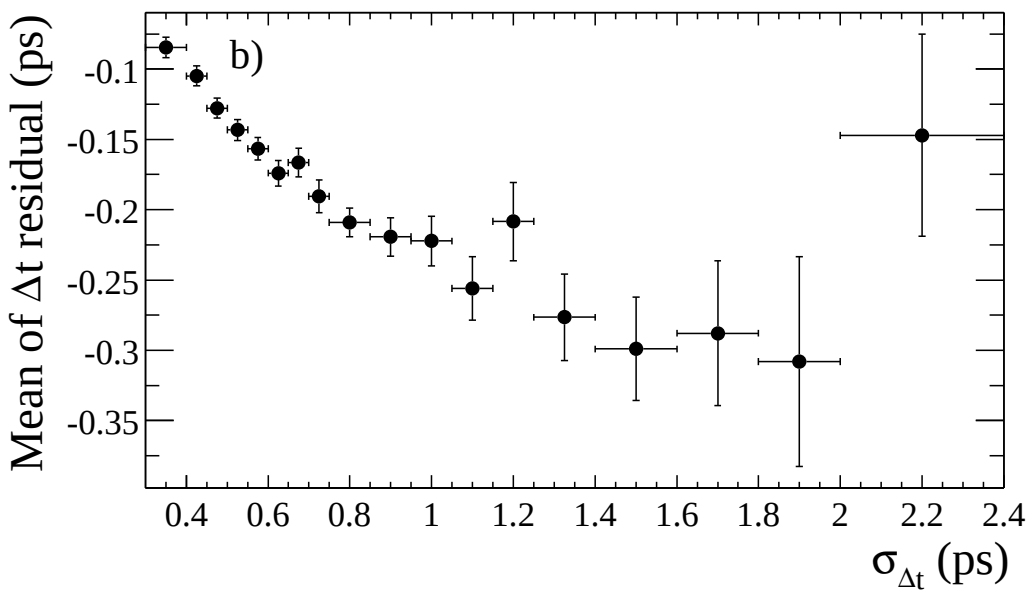

**Figure 6.5.4.** Mean of the $\delta t = \Delta t - \Delta t_{\rm true}$ distribution in *BABAR* simulated events as a function of the estimated event-by-event uncertainty in $\Delta t$. From (Aubert, 2002a).

# Chapter 7
# $B$-meson reconstruction

***Editors:***
*Paul Jackson (BABAR)*
*Anže Zupanc (Belle)*

**Additional section writers:**
*José Ocariz*

The *BABAR* and Belle detectors were designed and built to detect and reconstruct particles produced in $e^+e^-$ collisions and their decay products. Particles with long enough lifetimes or stable particles that deposit signals in subdetectors which in turn allow the measurements of their momenta or energies and consequently their four-momenta (see Chapters 2 and 5 for more details) are: $e^\pm$, $\mu^\pm$, $\pi^\pm$, $K^\pm$, $p$, $\overline{p}$, $\gamma$, and $K^0_L$ and are commonly collectively referred to as final state particles. Particles such as $B$ mesons and charm mesons decay inside the beam pipe close to the interaction point. In order to study the properties of $B$ mesons, or other short-lived particles, they must first be reconstructed from their final state particles.

Reconstruction of $B$ mesons proceeds via summing the momenta of all final state particles to check for consistency with specific exclusive $B$-meson decays. The goal is to measure the four-momentum vector of a reconstructed $B$ meson, or to at least identify particles in an event arising from the same $B$ meson. Candidates are identified by utilizing discriminating variables sensitive to the $B$-meson properties. The building of these candidates from their final state particle momenta is referred to as exclusive $B$-meson reconstruction or also full hadronic reconstruction and is described in detail in Section 7.1. Full reconstruction of (semi-) leptonic $B$-meson decays is not possible because the neutrinos leave the detectors undetected and hence the momentum they carry is not measured directly. However, due to the experimental setup of $B$ Factories additional kinematic constraints can be applied which allow us to infer the neutrino or semi-leptonically decaying $B$-meson momentum indirectly. The constraints and methods are described in more detail in Sections 7.2 and 7.4. As explained in Section 7.3 the unique kinematic properties of $B$-meson decays to $D^{*\pm}$ mesons permit a partial reconstruction approach, without recourse to constraining the entirety of the $B$ decay. As a consequence the partial reconstruction efficiency of $B$ mesons is much higher than that achieved by more exclusive techniques. The choice of the most suitable reconstruction method in any given analysis depends on the studied decay mode and the physics parameters of interest.

The rest of this chapter describes the methods – procedures and main kinematical constraints – used by *BABAR* and Belle to reconstruct and identify decays of $B$ mesons. In each subsection example $B$-meson decay modes are used for illustration of the reconstruction procedures. The techniques relevant to the reconstruction of charm, tau and other events are described in other chapters.

## 7.1 Full hadronic $B$-meson reconstruction

In most of the analyses we wish to extract some physics parameters of interest for a given specific exclusive $B$-meson decay mode, meaning that the entire $B$-meson decay chain from intermediate particles to all final state particles is reconstructed. For example, $B^0 \to D^{*-}\pi^+$ decays can be reconstructed from final state particles produced in the following exclusive decay chain:

$$\begin{aligned} B^0 \to\ & D^{*-}\pi^+ \\ & \hookrightarrow \overline{D}^0\pi^- \\ & \qquad \hookrightarrow K^+\pi^-\pi^0 \\ & \qquad\qquad \hookrightarrow \gamma\gamma. \end{aligned} \tag{7.1.1}$$

In exclusive reconstruction the reconstruction of the decay chain proceeds from bottom up. First the selection of tracks and clusters not associated with any track is performed. The former are used to construct final state charged particle candidates (*i.e.* to determine their four-momentum vector), $K^\pm$ and $\pi^\pm$ in the above example, and the latter to construct photon candidates as described in Chapter 2. In the next stages all decaying particles in the decay chain are reconstructed: two photon candidates are combined to form $\pi^0$ meson candidates; $\overline{D}^0$ candidates are formed by combining $K^+$, $\pi^-$ and $\pi^0$ candidates; $D^{*-}$ by pairing $\overline{D}^0$ candidates from the previous level and a negatively charged pion; and finally the $D^{*-}$ and $\pi^+$ candidates are combined to form the $B^0$ candidates. At each stage the four-momentum of a decaying particle is given by the sum of the four-momenta of its decay products following the momentum conservation rule.

Not all combinations of two or more particles which form the 'mother' particle candidates are correct. Wrong combinations (or background candidates) can be roughly divided into two categories:[23] combinatorial background and physics background. Combinatorial background candidates are random combinations of particles which are not produced in a decay of the same particle. For example, in an event two $\pi^0$ mesons are produced and both decay into two photons. If all four photons are detected then six different $\pi^0$ candidates (two photon combinations) can be reconstructed in total – two of them represent correctly reconstructed $\pi^0$ mesons (signal candidates) while the other four represent combinatorial background candidates. Similarly, the $B^0$ candidate in our example can be a combination of correctly reconstructed $D^{*-}$ and $\pi^+$ candidates, where the former originates from one $B$-meson decay and the latter from the decay of the second $B$ meson produced in the same event. Another large source of combinatorial background are events in which a light quark–anti-quark pair is produced instead of a pair of $B$ mesons – so called continuum events (see Chapter 9). The 'continuum' background is usually the dominant background for rare $B$-meson decay studies (decays of $B$ mesons that do not proceed through the dominant $b \to c$ transition). Much effort has therefore been invested in the development of continuum suppression techniques. They are described in detail in Chapter 9.

[23] Background composition strongly depends on the studied $B$-meson decay mode. Here only a general overview is given.

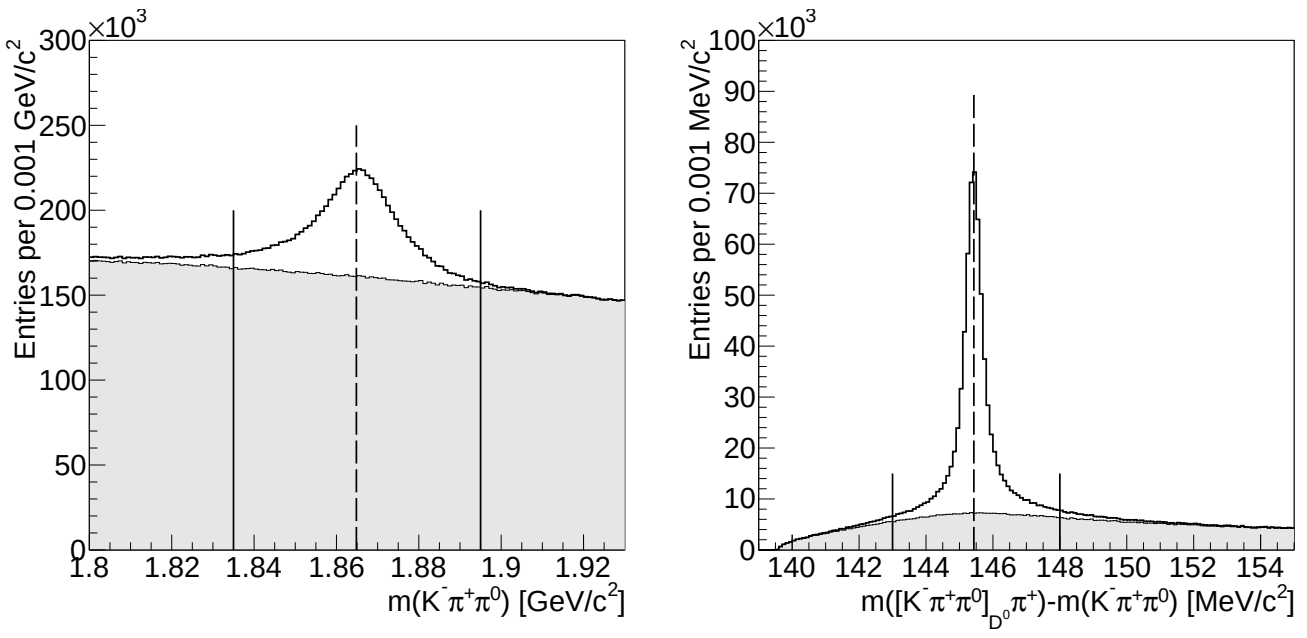

**Figure 7.1.1.** Invariant mass distribution of $D^0$ candidates reconstructed in $K^-\pi^+\pi^0$ (left) and $D^{*+} - D^0$ mass difference for $D^{*+}$ candidates reconstructed in $D^0\pi^+$ and $D^0$ in $K^-\pi^+\pi^0$ decay modes (right) in simulated events. The correctly reconstructed $D^0$ ($D^{*+}$) candidates peak at the nominal $D^0$ mass ($D^{*+} - D^0$ mass difference) indicated by vertical dashed lines. Full histograms show the contribution of background candidates. The signal regions are indicated by the two vertical lines.

The physics background originates from specific $B$-meson decays to final states which can be easily misidentified as the final state under study. For example, consider the charmless $B^+ \to K^+\pi^-\pi^+$ decays. The same or a very similar final state can also be achieved in many other $B$-meson decays, like for example: the $B^+ \to \overline{D}^0\pi^+ \to K^+\pi^-\pi^+$ decay chain leads to the same final state; $B^+ \to \overline{D}^0\pi^+ \to K^+K^-\pi^+$ and $B^+ \to K^+J/\psi \to K^+\mu^-\mu^+$, where in the former case the $K^-$ from the $\overline{D}^0$ decay is mis-identified as $\pi^-$ and in the latter the two muons as pions, respectively; $B^+ \to \overline{D}^0\pi^+ \to K^+\pi^-\pi^+\pi^0$, $B^+ \to \overline{D}^0\rho^+ \to K^+\pi^-\pi^+\pi^0$ and $B^+ \to K^+\eta' \to K^+\pi^-\pi^+\gamma$ decays have four-body final states but can still contaminate signal candidates when the $\pi^0$ or $\gamma$ are not reconstructed. Physics backgrounds are potentially more dangerous than combinatorial background because their distributions often peak around same values as distributions of the signal decay mode.

In the rest of this section most commonly used kinematical constraints which can help to reduce the contribution of combinatorial as well as physics backgrounds are discussed.

### 7.1.1 Kinematical discrimination of $B$ mesons

#### 7.1.1.1 Invariant mass and mass difference

In the case of $B$-meson decays via intermediate resonances, as shown in Equation (7.1.1), the most straightforward way to suppress the contribution of combinatorial background is to select only candidates populating the regions around the nominal masses (signal regions) of the decaying particles in the invariant mass distributions. Figure 7.1.1 shows for example the invariant mass distribution of $D^0$ candidates reconstructed in the $K^-\pi^+\pi^0$ decay mode (charge conjugation is implied). In this example a clear signal peak is visible over the smooth contribution of combinatorial background candidates. By selecting candidates that populate the signal region, indicated by two vertical lines, large amounts of combinatorial background are rejected while retaining almost all signal $D^0$ candidates. The signal region varies for different particles and even for the same particle reconstructed in different decay modes. In general, the invariant mass distribution of signal candidates is given by a convolution of the particle's true lineshape (usually a relativistic Breit-Wigner) and a detector resolution (usually described by the Gaussian function) stemming from the experimental uncertainty in the determination of momenta of the particle's decay products. It therefore depends on the resolution achieved in a given decay mode and the natural width of the reconstructed particle, if it's comparable or larger to the resolution. In case of $D^0$ mesons the natural width is negligible compared to the detector resolution which ranges from around 5-6 MeV/$c^2$ in decay modes to charged final state particles only (*e.g.* $K^-\pi^+$, $K^-\pi^+\pi^+\pi^-$) and up to around 12 MeV/$c^2$ in decay modes with one neutral pion. Composite particles whose natural width is much larger than the invariant mass resolution are for example $K^*(892)$ and $\rho(770)$ with natural widths around 50 and 150 MeV/$c^2$, respectively.

In the example $B$-meson decay the $D^{*+}$ mesons are reconstructed in the $D^0\pi^+$ decay mode. The energy release in the $D^{*+} \to D^0\pi^+$ decay is very small (The $D^{*+}$ mass is only about 6 MeV/$c^2$ above the $D^0\pi^+$ threshold). The $D^{*+}$ momentum measurement is dominated by the $D^0$ momentum. The pion has low momentum, whose magnitude and direction are well measured. Therefore, most of the uncertainty in the $D^*$'s momentum results from the measurement resolution of the $D^0$ momentum. This introduces a correlation between the measured $D^0$ and $D^*$ invariant masses. Due to this correlation, the experimental smearing of the $D^0$ momentum (partly) cancels in the $D^{*+} - D^0$ mass difference, $\Delta m = m(D^{*+}) - m(D^0)$. The mass difference has a much better resolution and discriminates more effectively between signal $D^{*+}$ and background than the $D^{*+}$ invariant mass. Figure 7.1.1 shows the mass difference distribution for $D^{*+} \to D^0\pi^+$ decays, where the $D^0$ is reconstructed in the $K^-\pi^+\pi^0$ mode. As can be seen the mass difference is about an order of magnitude better resolved than the mass of the $D^0$. The mass difference is commonly used to discriminate between the signal and background for particles reconstructed from composite particles with small energy released in the decay; apart from $D^*$ mesons, such cases include also excited charm baryons decaying to $\Lambda_c$, charmonium(-like) states decaying to $J/\psi$, etc.

Kinematic fitting can improve the momentum (invariant mass) resolution of reconstructed particles and therefore also the signal and background discrimination. Details of kinematic fitting and performance improvements that can be achieved are described in Chapter 6.

### 7.1.1.2 Energy difference $\Delta E$ and beam-energy substituted mass $m_{\rm ES}$

In principle, the invariant mass of $B$ mesons could also be used to distinguish between signal and background $B$-meson candidates. However, as it will be explained in what follows, the experimental setup of the $B$ Factories allows one to set additional kinematical constraints which improve the knowledge of the $B$-meson's momentum and hence allow for better signal and background discrimination.

The $\Upsilon(4S)$ decays in two same-mass particles, $B$ and $\overline{B}$, thus imposing two constraints in the CM frame. If the $B$ meson is correctly reconstructed, the energy of its decay products has to be equal to half the CM energy or equal to the beam energy in the $\Upsilon(4S)$ rest frame,[24] and its reconstructed mass has to be equal to that of the $B$ meson:

$$E^{\star}_{\rm rec} = E^{\star}_{\rm beam} = \sqrt{s}/2, \tag{7.1.2}$$

$$m_{\rm rec} = m_B. \tag{7.1.3}$$

In order to exploit the specifics of $B$-meson decay kinematics, two variables are defined, the beam-energy substituted mass, $m_{\rm ES}$, and the energy difference, $\Delta E$. They together exploit in an optimal way the information contained in the equations above.

The energy difference $\Delta E$ can be expressed in a Lorentz-invariant form as

$$\Delta E = \left(2 q_B q_0 - s\right)/2\sqrt{s}, \tag{7.1.4}$$

where $\sqrt{s} = 2E^{\star}_{\rm beam}$ is the total energy of the $e^+e^-$ system in the CM frame, and $q_B$ and $q_0 = (E_0, \boldsymbol{p}_0)$ are the Lorentz four-vectors representing the energy-momentum of the $B$ candidate and of the $e^+e^-$ system, $q_0 = q_{e^+} + q_{e^-}$. In the CM frame, $\Delta E$ takes the more familiar form

$$\Delta E = E^{\star}_B - E^{\star}_{\rm beam}, \tag{7.1.5}$$

where $E^{\star}_B$ is the reconstructed energy of the $B$ meson. The uncertainty of $\Delta E$ originates from the error in the $B$-meson energy measurement, $\sigma^2_{E^{\star}_B}$, and the beam energy spread, $\sigma^2_{E^{\star}_{\rm beam}}$:

$$\sigma^2_{\Delta E} = \sigma^2_{E^{\star}_B} + \sigma^2_{E^{\star}_{\rm beam}}. \tag{7.1.6}$$

The $\Delta E$ resolution receives a sizable contribution from the beam energy spread, but is generally dominated by detector energy resolution (this being the dominant term for modes involving photons). Figure 7.1.2 (a and b) shows the $\Delta E$ distributions for two cases: $B^+ \to K^0_S\pi^+$, $K^0_S \to \pi^+\pi^-$ and $B^+ \to K^+\pi^0$, $\pi^0 \to \gamma\gamma$. A clear difference in the $\Delta E$ resolution is seen between decay modes with and without photons in the final state. The long tail at low $\Delta E$ for the $B^0 \to K^+\pi^0$ signals comes from the photon shower leakage in the calorimeter crystals.

The measurement error $\sigma_{E^{\star}_B}$ receives contributions from the errors in the absolute values of the momenta of the decay products. The momenta of the $B$-meson decay products can be combined in a second variable that is only weakly correlated to $\Delta E$. This is possible if the variable depends on the small three-momentum of the $B$ meson to which the larger momenta of the $B$ decay products contribute with opposing signs in the CM frame. The pioneering experiments invented for this purpose a beam-energy constrained mass. While ARGUS did actually a fit of the $B$-meson four momentum with the $B$-meson energy constrained to the beam energy, CLEO used a simpler approach adopted also at Belle, substituting the $B$ energy with the beam energy, which is what we call the beam-energy substituted mass or beam-energy constrained mass[25]

$$m_{\rm ES}^{\rm CLEO} = m_{\rm bc} = \sqrt{E^{\star 2}_{\rm beam} - \boldsymbol{p}^{\star 2}_B}, \tag{7.1.7}$$

where $p^{\star}_B$ is the CM momentum of the $B$ meson, derived from the momenta of their decay products, and the $B$-meson energy is substituted by $E^{\star}_{\rm beam}$.

The idea behind $\Delta E$ is different and complementary to that of $m_{\rm ES}$. Whereas the latter is by construction independent of the mass hypothesis for each of the particles, $\Delta E$ depends strongly on them. If, for example, a kaon is misidentified as a pion, its energy, and consequently that of the $B$ candidate, will be smaller than its true energy. The event then will be shifted towards negative values of $\Delta E$. In contrast, the distribution for signal events peaks at zero as expected, making $\Delta E$ especially helpful for discriminating from physics background events involving misidentification. On the other hand, $m_{\rm ES}$ will not change if a particle is misidentified, leading to peaking background from true $B$ decays with incorrectly assigned particle identities.

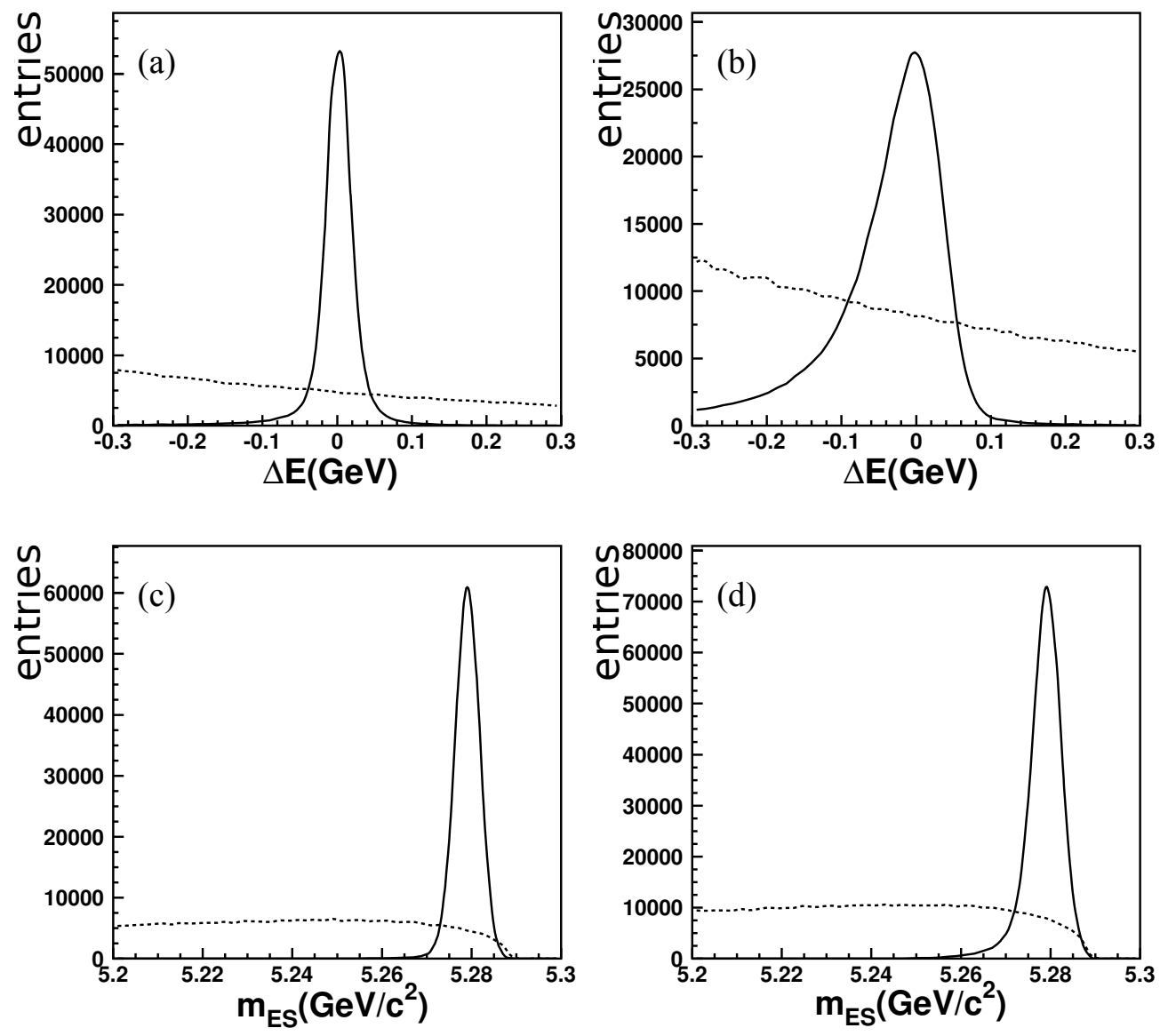

**Figure 7.1.2.** The $\Delta E$ and $m_{\rm ES}$ distributions for (a and c) $B^+ \to K^0_S\pi^+$ and (b and d) $B^+ \to K^+\pi^0$. Solid line histograms are signal events generated using `GEANT` Monte Carlo and dotted histograms are from the continuum MC. The signal resolution in $\Delta E$ is much worse for $B^+ \to K^+\pi^0$, due to the neutral pion present in the final state, but the difference is less pronounced in $m_{\rm ES}$ as explained in the text.

[24] All quantities with a star symbol ($\star$) are estimated in the CM frame unless otherwise stated.

[25] Since only the three-momentum of the $B$-meson candidate is used, this quantity is not Lorentz-invariant.

While this is true for symmetric-energy $e^+e^-$ colliders operating at the Υ(4S) (such as CLEO), where the laboratory system and the CM system are identical, it does not hold for the asymmetric $B$ Factories. The $B$ momentum vector can only be boosted to the CM frame after masses have been assigned, and the result depends on these mass assignments, although much weaker than for $\Delta E$. To strictly keep mass independence, *BABAR* is using a modified variable, which makes use of the three-momenta in the laboratory system and of the beam energy in the CM system:

$$m_{\rm ES} = \sqrt{\left(s/2 + \boldsymbol{p}_B \boldsymbol{p}_0\right)^2 / E_0^2 - \boldsymbol{p}_B^2}. \qquad (7.1.8)$$

where $(E_0, \boldsymbol{p}_0)$ is the four-momentum of the CM system in the laboratory. This definition is identical with Eq. (7.1.7) if the laboratory system is the CM system, *i.e.*, at a symmetric-energy collider. But due to the weak mass dependence, the behavior of $m_{\rm ES}$ and $m_{\rm bc}$ are largely the same even at asymmetric colliders and therefore throughout this book the common notation $m_{\rm ES}$ will be used for both of them. When presenting beam-energy substituted mass or beam-energy constrained mass distributions the reader should keep in mind that Belle uses the definition given in Eq. (7.1.7) while *BABAR* uses the definition given in Eq. (7.1.8).[26]

To appreciate this subtlety, we approximate $m_{\rm ES} \approx m_{\rm bc}$, where the approximation arises from the uncertainty in the $B$ momentum measurement (boosted to the CM frame), $\sigma^2_{p^\star_B}$, and the beam energy spread, $\sigma^2_{E^\star_{\rm beam}}$:

$$\sigma^2_{m_{\rm ES}} \approx \sigma^2_{E^\star_{\rm beam}} + \left(\frac{p^\star_B}{m_B}\right)^2 \sigma^2_{p^\star_B}. \qquad (7.1.9)$$

As the $B$ mesons are almost at rest in the CM frame, $p^\star_B/m_B \approx 0.06$, the second term in the above equation gets small and the resolution in $m_{\rm ES}$ is dominated by the spread in the beam energy. This is illustrated in Figure 7.1.2 (c and d) which shows the $m_{\rm ES}$ distributions for $B^+ \to K^0_S\pi^+$ and $B^+ \to K^+\pi^0$. The signal resolution in $m_{\rm ES}$ is much less affected by the uncertainty in the measured $B$-meson four-momentum compared to $\Delta E$. For signal events, $m_{\rm ES}$ yields the mass of the $B$ meson and shows a clean peak. For continuum events, composed of light quarks, the only way of reaching the $B$ rest mass is by artificially associating random particles. As a consequence, their distribution displays a slowly varying shape, as expected from their combinatorial nature.

The $m_{\rm ES}$ resolution is around 3 MeV/$c^2$ when no neutral particles contribute to the final state. The resolution for $\Delta E$ more strongly depends on the $B$-meson decay mode: it is much larger for low mass final states such as $\pi^+\pi^-$ (Lees (2013b) quotes $\sigma_{\Delta E} \sim 29$ MeV) than for high mass final states such as $\overline{D}^{(*)}D^{(*)}K$ (del Amo Sanchez (2011e) quotes $\sigma_{\Delta E}$ between 6 and 14 MeV for modes with zero or one $D^{*0}$ meson in the final state).

The energy difference and beam substituted mass, defined in Eqs (7.1.5) and (7.1.8), exploit optimally the kinematical constraints from the $\Upsilon(4S)$ decay to two $B$ mesons. A small correlation between the $\Delta E$ and $m_{\rm ES}$ variables follows from their common inputs – the beam energy, measured momentum of charged particles and energy of neutrals. The correlation from the energy measurement becomes severe if the final state particles contain high energy photons, as shown in the top scatter plot in Figure 7.1.3. The correlation coefficient is +18% for $m_{\rm ES}$ and $\Delta E$ in $B^+ \to K^+\pi^0$. The correlation can be reduced by calculating $m_{\rm ES}$ after modifying the magnitude of the $\pi^0$ momentum but retaining its direction to constrain the reconstructed $B$ energy to be the beam energy.[27] The bottom scatter plot in Figure 7.1.3 shows that the correlation between the modified $m_{\rm ES}$ and $\Delta E$ is reduced and the corresponding correlation coefficient is −4% (Duh, 2012). This technique is found useful only for two-body $B$ decays with a hard photon, $\pi^0$ or $\eta \to \gamma\gamma$ meson in the final state. For other $B$ decays with soft photons only, the modified $m_{\rm ES}$ has similar distribution as that of $m_{\rm ES}$ because the $m_{\rm ES}$ resolution is dominated by the beam-energy spread. Furthermore, the modification does not artificially create an enhancement in $m_{\rm ES}$ for the continuum background.

For final states with heavy particles, in particular $B$ decays to baryons, the correlation becomes strong since the beam energy spread $\sigma_{E^\star_{\rm beam}}$ dominates in both variables. The difference between the mean beam energy used in the calculation of $\Delta E$ and $m_{\rm ES}$ and the true beam energy of the event is the same, hence this contribution alone would lead to 100% correlation. Therefore, in these analyses other pairs of variables are preferred. If $\Delta E$ is replaced by the invariant mass

$$m_B = \sqrt{E_B^2 - \boldsymbol{p}_B^2} \qquad (7.1.10)$$

of the reconstructed $B$ candidate, this variable will not depend on the beam energy at all and the correlation with $m_{\rm ES}$ becomes again very small, as shown in Fig. 7.1.4: distributions from simulated events $\overline{B}^0 \to \Lambda_c^+\overline{p}\pi^+\pi^-$ in $\Delta E$ vs $m_{\rm ES}$ with a correlation coefficient of −29% compared to $m_B$ vs $m_{\rm ES}$ with a correlation coefficient of $(-2.3 \pm 0.5)\%$ (Lees, 2013h).

[26] As to any rule there is also an exception to this one: In the measurement reported by Belle in Abe (2001f) the definition Eq. (7.1.8) is used.

[27] In the calculation of the modified $m_{\rm ES}$ (using Eq. 7.1.7) the momentum of the $B$ meson given as $\boldsymbol{p}_B = \boldsymbol{p}_{K^+} + \boldsymbol{p}_{\pi^0}$ is replaced with $\boldsymbol{p}_B = \boldsymbol{p}_{K^+} + \sqrt{(E^2_{\rm beam} - E_{K^+})^2 - M^2_{\pi^0}} \cdot \frac{\boldsymbol{p}_{\pi^0}}{|\boldsymbol{p}_{\pi^0}|}$, where $M_{\pi^0}$ is the nominal mass of $\pi^0$, and $\boldsymbol{p}_{K^+}$ ($\boldsymbol{p}_{\pi^0}$) is the measured $K^+$ ($\pi^0$) momentum.

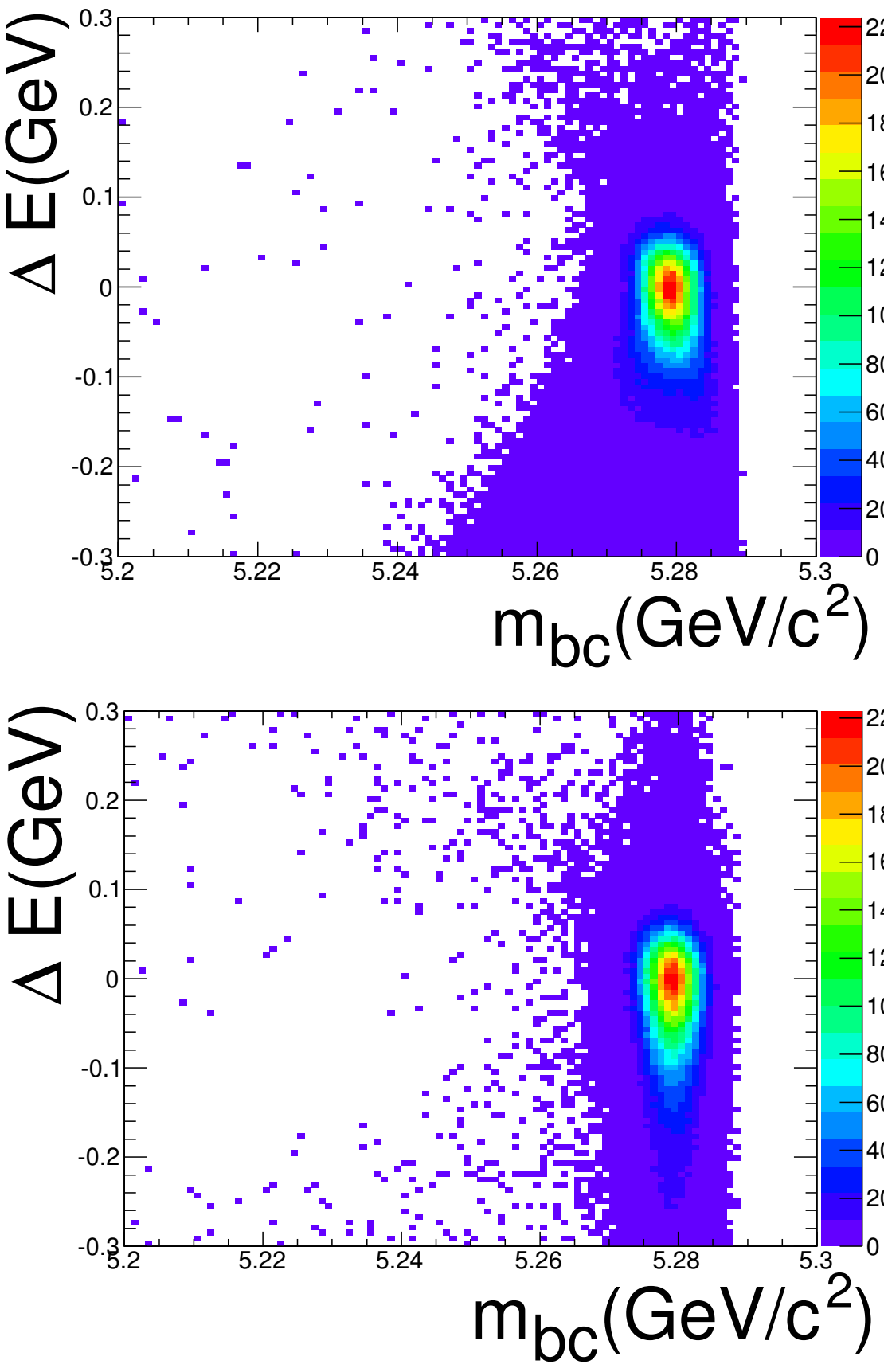

**Figure 7.1.3.** The $\Delta E$ vs $m_{\rm ES}$ $(= m_{\rm bc})$ distributions for the $B^+ \to K^+\pi^0$ signals. The top plot is for the original $m_{\rm ES}$ definition and the bottom is for the modified $m_{\rm ES}$ case. Belle internal, from the Duh (2012) analysis.

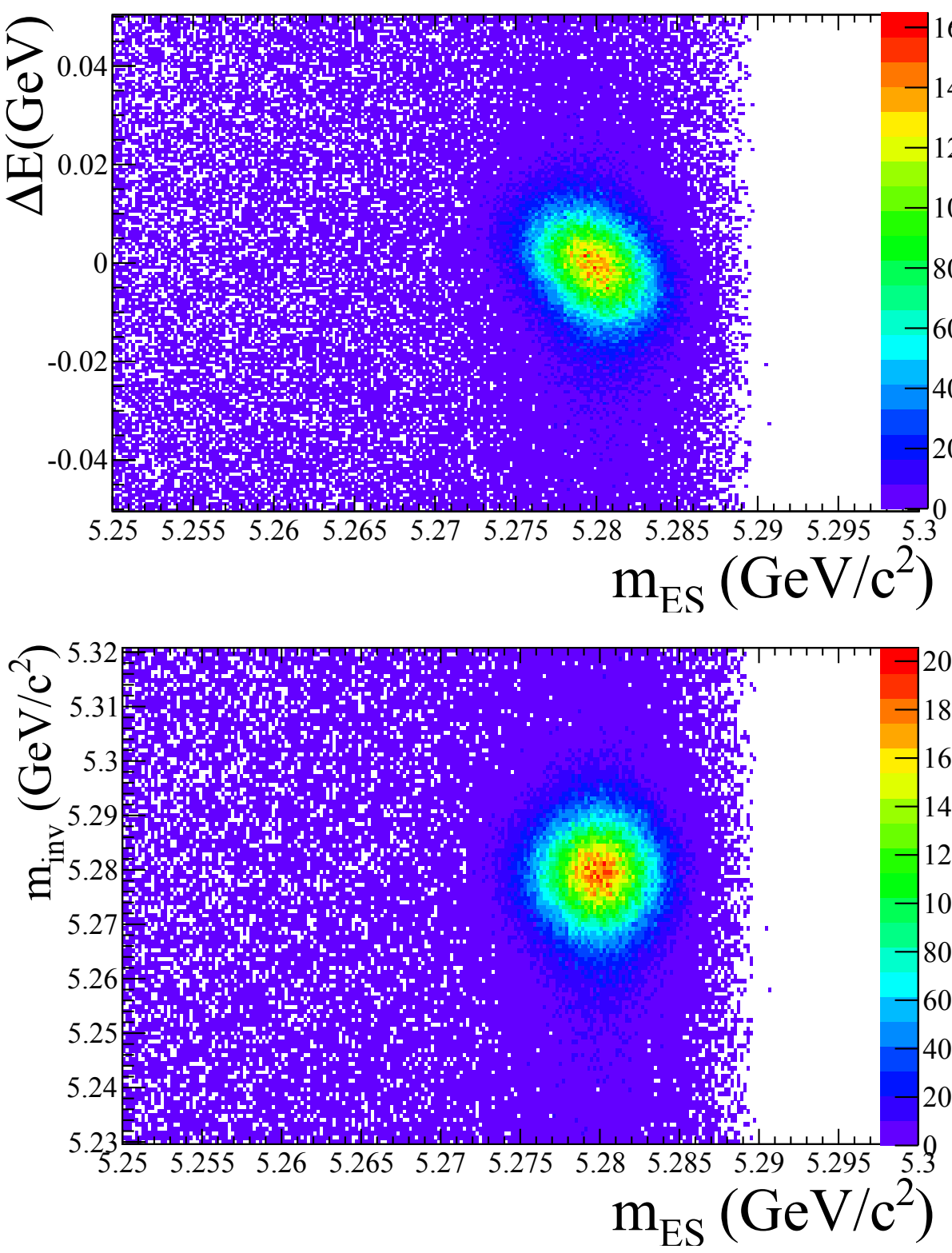

**Figure 7.1.4.** Distributions from $\overline{B}^0 \to \Lambda_c^+\overline{p}\pi^+\pi^-$ events (Monte Carlo). The top plot is for $\Delta E$ vs $m_{\rm ES}$ showing a strong correlation, and the bottom is for the invariant mass $m_B$ vs $m_{\rm ES}$ which is only weakly correlated through measurement errors. *BABAR* internal, from the Lees (2013h) analysis.

#### 7.1.1.3 Signal yield extraction

After the reconstruction and selection of a specific exclusive $B$-meson decay is performed the next step is to determine the number of correctly reconstructed $B$-meson candidates. Most often the signal yield is extracted by performing an extended maximum likelihood fit to the two dimensional $\Delta E$-$m_{\rm ES}$ distribution. In studies in which there is negligible correlation between the two variables the distribution of events can be modeled by a product of two one dimensional probability density functions. The $\Delta E$ and $m_{\rm ES}$ distributions of signal $B$-meson candidates are often modeled with a Gaussian function (or sum of two or more Gaussian functions). The background candidates are modeled in $m_{\rm ES}$ with an empirical function introduced by the ARGUS collaboration (Albrecht et al., 1990a):

$$\begin{aligned}\text{Argus}(m_{\rm ES}|m_{\rm thr}, c) &= m_{\rm ES}\sqrt{1-\left(\frac{m_{\rm ES}}{m_{\rm thr}}\right)^2}\times \\ &\exp\left[-c\left(1-\left(\frac{m_{\rm ES}}{m_{\rm thr}}\right)^2\right)\right], \quad (7.1.11)\end{aligned}$$

where $m_{\rm thr}$ represents the endpoint in $m_{\rm ES}$ distribution and $c$ is a free shape parameter. Background in $\Delta E$ is usually modeled with a polynomial function. The choice of signal and background models given above is very general and depends on the properties of the studied decay mode and background composition. The models used in specific studies are provided in relevant sections and details about maximum likelihood fitting are provided in Chapter 11.

## 7.2 Semileptonic $B$-meson reconstruction

Analyses of $B$-meson decay modes containing leptons present one of the richest means of extracting information about the CKM matrix, along with an understanding of properties of the $b$ quark bound in a meson. These probes are used in a variety of final states where the measurement strategy can be more or less inclusive. Decays of the form: $B \to X\ell\nu$, are used to measure $|V_{cb}|$, $|V_{ub}|$ and to extract branching fractions of $B$ transitions to charm-type and up-type mesons. For semileptonic decays involving charm states (denoted $B \to X_c\ell\nu$), the final state can be reconstructed from the particles produced in a typical exclusive

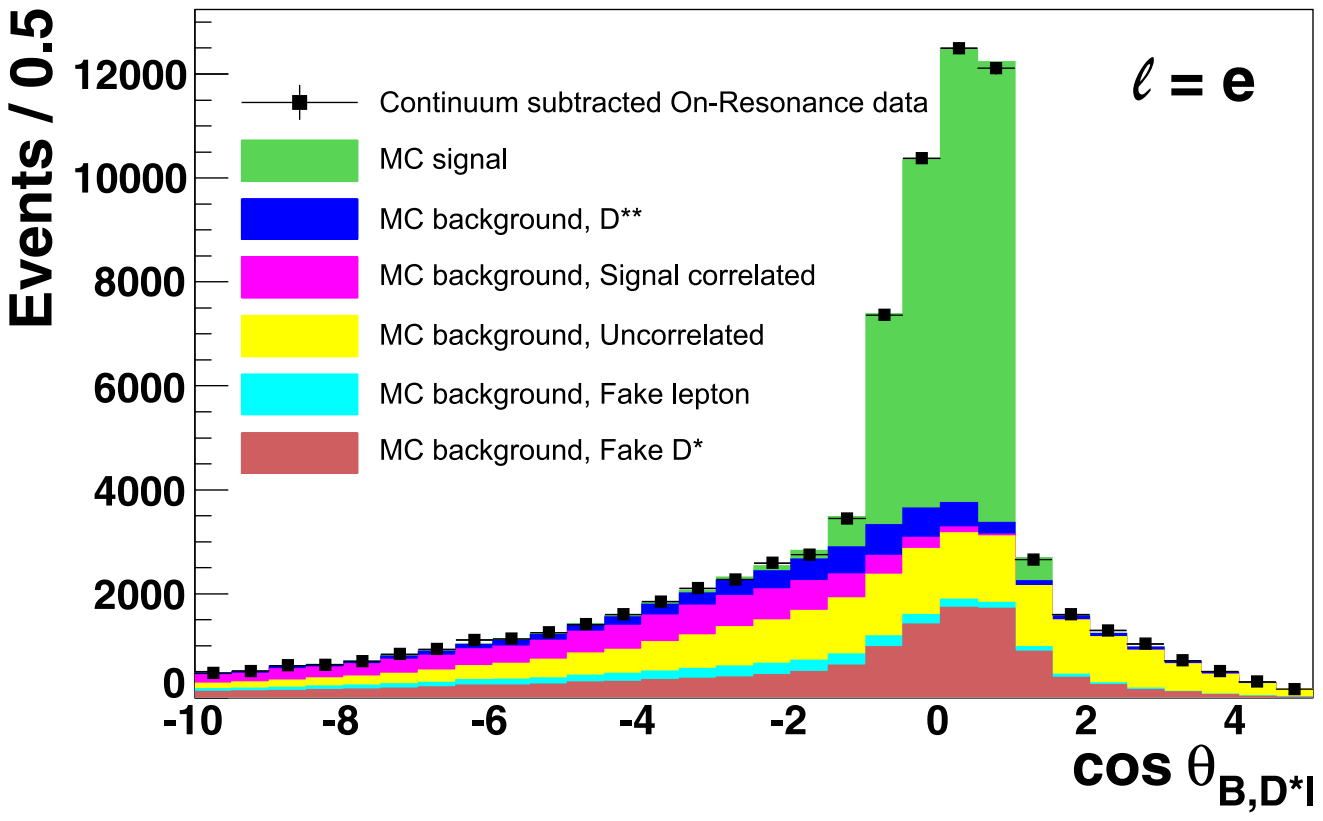

**Figure 7.2.1.** The $\cos\theta_{B,D^*\ell}$ distribution for $B^0 \to D^{*-}e^+\nu_e$ decays (Dungel, 2010). The points with error bars are data and full histograms are, top to bottom, the signal component and different types of background. Signal decays are constrained to lie in the interval $(-1, 1)$, while background decays populate a much wider region.

decay chain:

$$\begin{aligned} B^0 \to\ & D^{*-}\ell^+\nu \\ & \hookrightarrow \overline{D}^0\pi^- \\ & \quad \hookrightarrow K^+\pi^-\pi^0 \\ & \qquad\quad \hookrightarrow \gamma\gamma. \end{aligned} \tag{7.2.1}$$

The reconstruction of the decay chain proceeds from the identification of the charged lepton. In tandem with this, the reconstruction of a $D$ meson occurs, most commonly a suitable ground state neutral or charged meson ($D^0$, $\overline{D}^0$, $D^+$, $D^-$). This ground state $D$ meson may then be combined with soft a $\pi^{\pm}$ or $\pi^0$ in an attempt to form a $D^{*\pm}$ or $D^{*0}$. A tight constraint on $\Delta m$ is applied to evidence such transitions. Higher resonant states of charm mesons (*e.g.* $D^{**}$) are usually examined in a combination of angular and mass distributions.

Under the assumption that the neutrino is the only missing particle, the cosine of the angle between the inferred direction of the reconstructed $B$ and that of the $D^{(*)}\ell$ system is

$$\cos\theta_{B,D^{(*)}\ell} = \frac{2E^*_B E^*_{D^{(*)}\ell} - m^2_B - m^2_{D^{(*)}\ell}}{2|\boldsymbol{p}^*_B||\boldsymbol{p}^*_{D^{(*)}\ell}|}\ , \tag{7.2.2}$$

where $E^*_B$ is half of the CM energy and $|\boldsymbol{p}^*_B|$ is $\sqrt{E^{*2}_B - m^2_B}$. The quantities $E^*_{D^{(*)}\ell}$, $\boldsymbol{p}^*_{D^{(*)}\ell}$ and $m_{D^{(*)}\ell}$ are calculated from the reconstructed $D^{(*)}\ell$ system. This cosine is also a powerful discriminator between signal and background: signal events should strictly lie in the interval $(-1, 1)$, although – due to finite detector resolution – about 5% of the signal is reconstructed outside this interval. The background on the other hand does not have this restriction and populates a much wider region (see Fig. 7.2.1).

The experimental techniques used in reconstruction of semileptonic $B$-meson decays are described in more details in Section 17.1.1.3.

## 7.3 Partial $B$-meson reconstruction

The term partial reconstruction refers to a reconstruction technique in which not all of the final state particles are required to be detected and identified, as is the case in exclusive (full) reconstruction described in Section 7.1. Partial reconstruction of the $B$ meson can therefore result in substantially larger efficiency, albeit with reduced purity resulting from higher backgrounds.

*BABAR* and Belle use the partial reconstruction technique mainly in time-dependent studies of $B^0 \to D^{*-}X^+$ (where $X$ represents some hadronic state like $\pi$, $\rho$ or $D$) and $B^0 \to D^{*-}\ell^+\nu_\ell$ decays. In these measurements the $B$ mesons are reconstructed using only the hadronic state $X$ (or charged lepton) and the soft pion from the $D^{*-} \to \overline{D}^0\pi^-$ decay. The $\overline{D}^0$ decay is not reconstructed which increases the acceptance.

The remainder of this section describes the kinematic constraints and variables used to distinguish between partially reconstructed signal and background $B^0 \to D^{*-}X^+$ and $B^0 \to D^{*-}\ell^+\nu_\ell$ candidates. Physics use cases are described in Sections 17.5 and 17.8.

### 7.3.1 $B \to D^{*\pm}X$ decays

The partial reconstruction technique was originally applied by CLEO (Brandenburg et al., 1998; Giles et al., 1984) to

$$\begin{aligned} B^0 \to\ & D^{*-}\pi^+_f \\ & \hookrightarrow \overline{D}^0\pi^-_s \end{aligned} \tag{7.3.1}$$

decays, where $\pi_f$ and $\pi_s$ are referred to as "fast" and "slow" pions, respectively. *BABAR* and Belle applied this technique to generic $B \to D^{*\pm}X$ decays. In principle, $X$ may be any single-particle state (*e.g.* $\pi$, $\rho$, $D$, $D^{(*)}_s$) as long as it can be exclusively reconstructed. For simplicity the discussion is restricted only to $B \to D^{*\pm}\pi$ decays. In this mode the $D^{*\pm}$ meson is created in a helicity zero state and the characteristic angular distributions of the $D^{*+}$ decay products (see Chapter 12 for more details) can be exploited for background suppression.

#### 7.3.1.1 Kinematic Variables

The decay chain given in Eq. (7.3.1) involves 5 particles ($B^0$, $D^*$, $D^0$, $\pi_s$ and $\pi_f$), each determined by it's four-momentum. There are thus 20 parameters in total which describe the entire decay chain. The experimentally measured inputs to the partial reconstruction are only the three-momenta of the fast and slow pion, $\boldsymbol{p}_{\pi_f}$ and $\boldsymbol{p}_{\pi_s}$, respectively. In principle, it is possible to determine all five four-momenta from the measured $\boldsymbol{p}_{\pi_f}$ and $\boldsymbol{p}_{\pi_s}$ using energy-momentum conservation in the $B^0$ and $D^*$ decays (8 constraints), the known particle masses of $B^0$, $D^{*-}$, $\overline{D}^0$, $\pi_s$ and $\pi_f$ (5 constraints), and the fact that the energy of the $B^0$ in the CM frame is equal to the half of the beam energy (1 constraint). However, since the $B$-meson mass and the beam-energy constraints are imposed to determine the $B$-meson four-momentum the signal and background $B$-meson candidates cannot be separated by kinematic variables used in exclusive studies, like $\Delta E$ and $m_{\rm ES}$ given in Eqs (7.1.5) and (7.1.8), respectively. Instead, variables which can be used to identify signal events from the decay kinematics are utilized. Many different possible kinematic variables have been used in analyses of partially reconstructed $B^0 \to D^*\pi$ decays performed by *BABAR* and Belle.

The measured[28] $\boldsymbol{p}_{\pi_f}$ and $\boldsymbol{p}_{\pi_s}$ represent six independent variables which can be used to distinguish signal events from background. Consider three of these as $\boldsymbol{p}_{\pi_f}$ in spherical polar coordinates: magnitude ($p_{\pi_f}$), polar ($\theta_{\pi_f}$) and azimuthal ($\phi_{\pi_f}$) angle. Since the fast pion has no preferred direction (distribution of signal decays is uniform in $\theta_{\pi_f}$ and $\phi_{\pi_f}$), only the magnitude, $p_{\pi_f}$, is useful. Signal decays are uniformly distributed within a small window in $p_{\pi_f}$, smeared by the $B^0$ momentum in the CM frame, as the fast pion is mono-energetic in the $B$ rest frame. Background events are distributed predominantly outside this window. The three remaining degrees of freedom can be considered as the magnitude of the slow pion momentum, $p_{\pi_s}$, the angle between the slow pion direction and the opposite of the fast pion direction, $\delta_{fs}$, and the azimuthal angle of the slow pion direction around the fast pion direction. The last of these three provides no useful information. The $\cos\delta_{fs}$ peaks sharply at +1 for signal, as the slow pion follows the $D^*$ direction due to the small energy released in the $D^*$ decay, while the background events populate the entire physical region. Instead of the slow pion magnitude the cosine of the angle between the slow pion direction in $D^*$ rest frame and the $D^*$ direction in CM frame, $\cos\theta_{\rm hel}$, is used since the former is correlated with the $p_{\pi_f}$ for signal events, while the latter is not. For partially reconstructed $D^*\pi$ events the $\cos\theta_{\rm hel}$ is given by

$$\cos\theta_{\rm hel} = \frac{1}{p^\star_{\pi_s}}\left(\frac{E_{\pi_s}E_{D^*} - E^\star_{\pi_s}m_{D^*}}{p_{D^*}\gamma_{D^*}} - \beta_{D^*}E^\star_{\pi_s}\right), \tag{7.3.2}$$

where the energy and magnitude of the $D^*$ momentum are given by $E_{D^*} = E_B - \sqrt{|\boldsymbol{p}_{\pi_f}|^2 + m^2_{\pi_f}}$ and $p_{D^*} = \sqrt{E^2_{D^*} - m^2_{D^*}}$, respectively, and $\gamma_{D^*} = E_{D^*}/m_{D^*}$ and $\beta_{D^*} = \sqrt{1 - 1/\gamma^2_{D^*}}$. The $B$-meson energy is taken to be half of the CM energy, $E_B = \sqrt{s}/2$. The quantities denoted with asterisks in the above equation are calculated in the $D^*$ rest frame. The distribution for signal events in $\cos\theta_{\rm hel}$ is proportional to $\cos^2\theta_{\rm hel}$, as the $B^0 \to D^*\pi$ decay is a pseudoscalar to vector pseudoscalar transition. The $\cos\theta_{\rm hel}$ is calculated using kinematic constraints valid only for signal decays so the background events can populate also the unphysical region $|\cos\theta_{\rm hel}| > 1$. Figure 7.3.1 illustrates the discriminating power of the $p_{\pi_f}$, $\cos\delta_{fs}$ and $\cos\theta_{\rm hel}$ kinematic variables for partially reconstructed $B^0 \to D^{*-}\pi^+$ decays (Ronga, 2006).

[28] All momenta in the partial reconstruction section are evaluated in the CM frame unless stated otherwise.

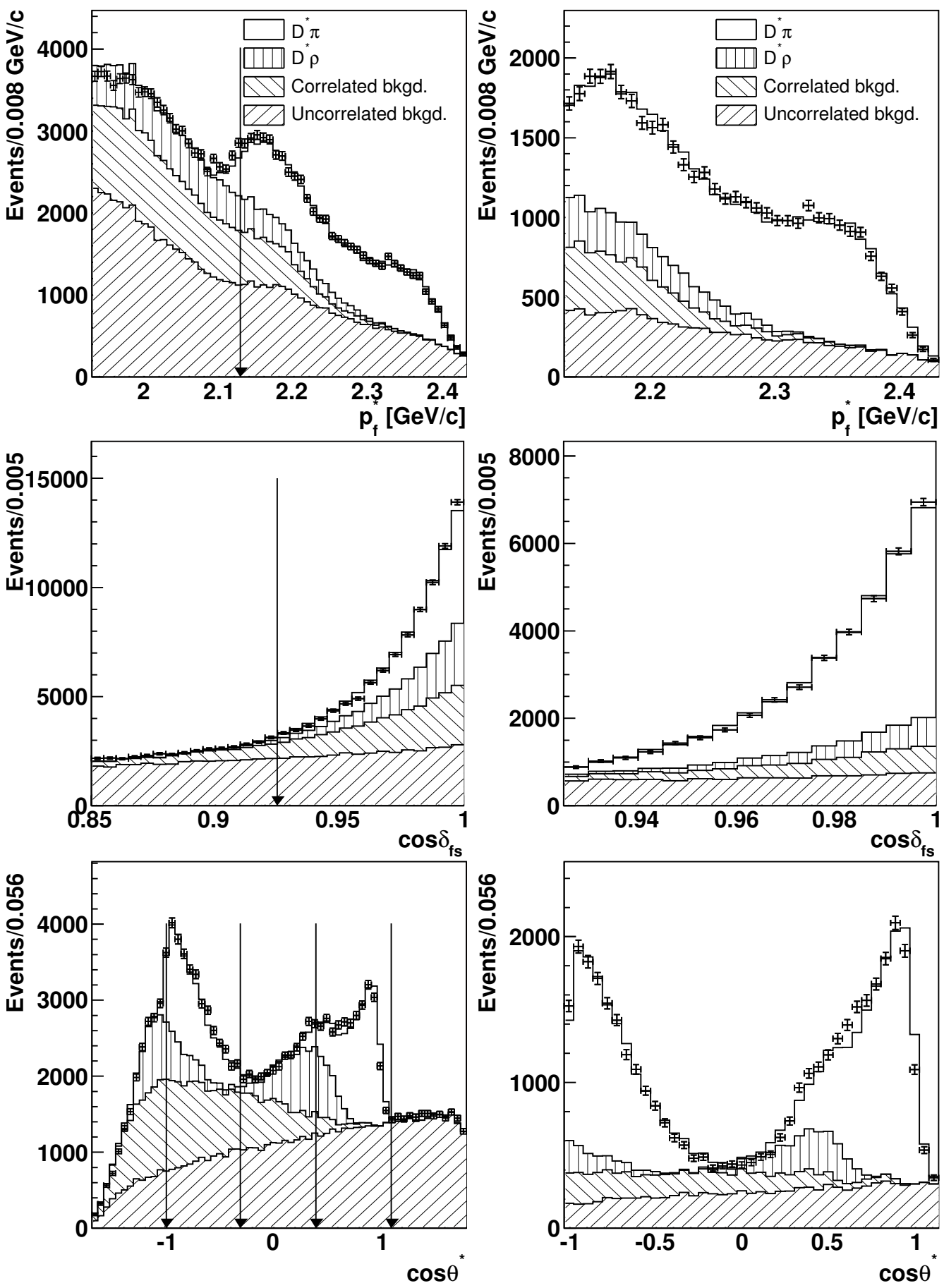

**Figure 7.3.1.** The $p_{\pi_f}$ (top), $\cos\delta_{fs}$ (middle), and $\cos\theta_{\rm hel}$ (bottom) distributions of partially reconstructed $D^*\pi$ candidates showing selection regions (left) and signal region (right). The arrows indicate the borders of the signal region. Points with error bars show the observed data distribution, while the empty histograms show the distribution of signal $D^*\pi$ candidates, and the hatched histograms show the contributions of background candidates originating from different sources (Ronga, 2006).

In quite few measurements, the $\cos\delta_{fs}$ variable is replaced by the 'missing mass',[29] $m_{\rm miss}$, which should be equal to the $D^0$ meson mass for signal $B^0 \to D^*\pi$ decays. The four-momentum of the missing $D^0$, $p_{D^0}$, can be obtained from the four-momentum conservation in the decay of the $B^0$ and $D^*$. The magnitude of the $B$-meson momentum in the CM frame, $p_B$, is given by the known $B$-meson energy, $E_B = \sqrt{s}/2$, and the known $B$-meson mass: $p_B = \sqrt{E^2_B - m^2_B}$. From the angle between the $B$ and $\pi_f$, given by,

$$\cos\theta_{B\pi_f} = \frac{m^2_B + m^2_{\pi^\pm} - m^2_{D^{*\pm}} - 2E_BE_{\pi_f}}{2p_Bp_{\pi_f}}, \tag{7.3.3}$$

and the measured slow and fast pion momenta, the $B$ four-momentum may be calculated up to an unknown azimuthal angle $\phi$ around $\boldsymbol{p}_{\boldsymbol{\pi_f}}$. Depending on the value of

[29] The variables $m_{\rm miss}$ and $\cos\delta_{fs}$ are strongly correlated.

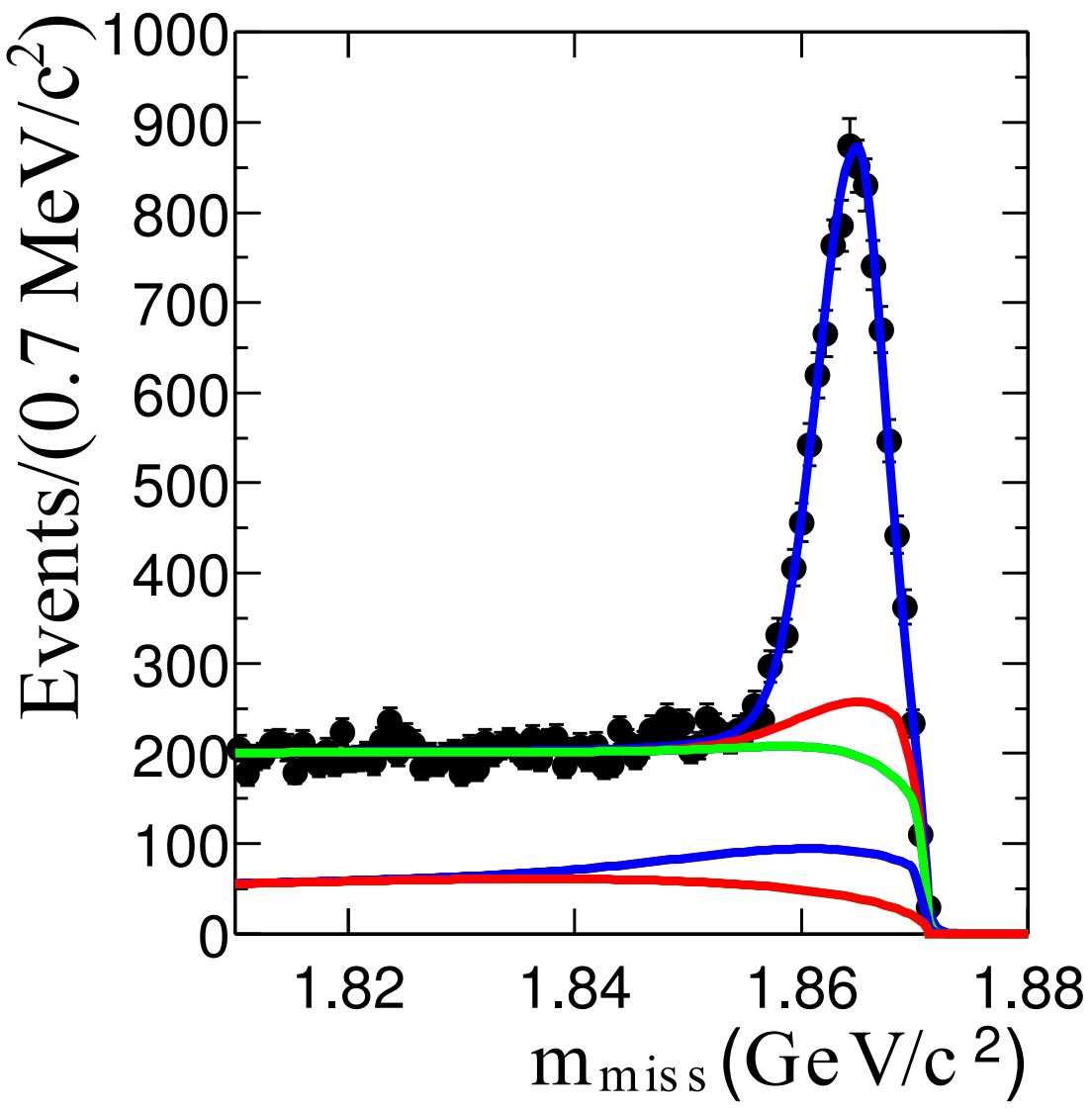

**Figure 7.3.2.** The $m_{\rm miss}$ distributions. The curves show, from bottom to top, the cumulative contributions of continuum, peaking $B\overline{B}$, combinatorial $B\overline{B}$, and $B^0 \to D^{*-}\rho^+$ background, and $B^0 \to D^{*-}\pi^+$ signal events (Aubert, 2004p).

$\phi$, the expected $D^0$ momentum can then be calculated as

$$\begin{aligned} p^2_{D^0}(\phi) = m^2_B + (p_{\pi_f} + p_{\pi_s})^2 - 2E_B(E_{\pi_f} + E_{\pi_s}) \\ +2p_B p_{\pi_f} \cos\theta_{B\pi_f} + 2p_B p_{\pi_s} \cos\theta_{B\pi_f} \cos\theta_{fs} \\ +2p_B p_{\pi_s} \sin\theta_{B\pi_f} \sin\theta_{fs} \cos\phi. \end{aligned} \quad (7.3.4)$$

The $\phi$-dependent missing mass is then calculated as, $m(\phi) = \sqrt{p^2_{D^0}(\phi)}$. The value of $\phi$ is not constrained by kinematics and may be chosen arbitrarily: *BABAR* defines in Aubert (2004p) the missing mass for partially reconstructed $B^0 \to D^{*-}\pi^+$ decays to be $m_{\rm miss} = \frac{1}{2}[m_{\rm max} + m_{\rm min}]$, where $m_{\rm max}$ and $m_{\rm min}$ are the maximum and minimum values of $m(\phi)$, while in analysis of partially reconstructed $B^0 \to D^{*+}D^{*-}$ decays *BABAR* chooses the value for which $\cos\phi = 0.62$, which is the median of the corresponding Monte Carlo distribution for signal events obtained using generated momenta, and defines the missing mass $m_{\rm miss} = m_{\rm miss}(\cos\phi = 0.62)$ (Lees, 2012k). For signal candidates, the $m_{\rm miss}$ variable peaks at the nominal $D^0$ mass $m_{D^0}$, with a spread of about 3 MeV/$c^2$, while the background is smoothly distributed, dropping off just above the $D$ mass due to lack of phase space. The distribution of $m_{\rm miss}$ for partially reconstructed $B^0 \to D^{*-}\pi^+$ decays is shown in Fig. 7.3.2 (Aubert, 2004p).

### 7.3.2 $B \to D^{*\pm}\ell\nu_\ell$ decays

The partial reconstruction technique of semileptonic

$$\begin{aligned} B^0 \to D^{*-}\ell^+\nu_\ell \\ \hookrightarrow \overline{D}^0\pi^-_s \end{aligned} \quad (7.3.5)$$

decays was first applied by ARGUS (Albrecht et al., 1987a, 1994a) and later used by other experiments, including *BABAR* and Belle. The signal events are selected using only the charged lepton from the $B^0$ decay and the slow pion from the $D^*$ decay. Due to the undetected neutrino in the final state, the kinematics of these decays differ from the partial reconstruction of hadronic $B \to D^{*\pm}X$ decays.

As a consequence of the limited phase space available in the $D^*$ decay, the slow pion is emitted within a one-radian wide cone centered about the $D^*$ direction in $\Upsilon(4S)$ rest frame. The $D^*$ four-momentum can therefore be computed by approximating its direction as that of the slow pion, and parameterizing its momentum as a linear[30] function of the slow pion's momentum, $p_{\pi_s}$:

$$p_{D^*} = \alpha + \beta p_{\pi_s}, \quad (7.3.6)$$

$$E_{D^*} = \sqrt{p^2_{D^*} + m^2_{D^*}}, \quad (7.3.7)$$

where the offset and slope parameters $\alpha$ and $\beta$ are taken from the simulation. The approximations used in the determination of the $D^*$ four-momentum result in an uncertainty in the $D^*$ energy of about 400 MeV. The missing momentum carried by the neutrino is then given by energy-momentum conservation in the $B \to D^*\ell\nu_\ell$ decays

$$p_\nu = p_B - p_{D^*} - p_\ell.$$

One requires knowledge of the $B$-meson four-momentum, $p_B$, to solve this equation. The direction of the motion of the $B$ is not known, but it's momentum is sufficiently small (on average 0.34 GeV/$c$) compared to the typical values of the magnitudes of lepton and $D^*$ momenta so that the three-momentum of the $B$ meson can be set to zero. The neutrino invariant mass can then be computed as

$$M^2_\nu = \left(\frac{\sqrt{s}}{2} - E_{D^*} - E_\ell\right)^2 - (\boldsymbol{p}_{D^*} + \boldsymbol{p}_\ell)^2, \quad (7.3.8)$$

where the energy of the $B$ meson is taken to be half of the CM energy. Figure 7.3.3 shows the distribution of partially reconstructed $B^0 \to D^{*-}\ell^+\nu_\ell$ decays ($\pi_s\ell$ combinations) from Aubert (2006s). The signal events produce a prominent peak at $M^2_\nu \approx 0$ with spread around 0.850 GeV$^2/c^4$ while background events are distributed in a wide range, dropping sharply to zero where there is a lack of phase space.

## 7.4 Recoil $B$-meson reconstruction

$B$-meson decays to a final state with one or more neutrinos offer very little or even no kinematic constraints which are usually exploited in $B$ decay searches in order to distinguish these decays from continuum and $B\overline{B}$ backgrounds, as described in Sections 7.1 and 7.3 and Chapter 9. Prominent examples of such decays are:

$$\begin{aligned} B^0 &\to \nu\overline{\nu}, \\ B^+ &\to K^+\nu\overline{\nu}, \\ B^0 &\to D^-\ell^+\nu_\ell. \end{aligned}$$

[30] *BABAR* uses in (Aubert, 2006s) a third order polynomial.

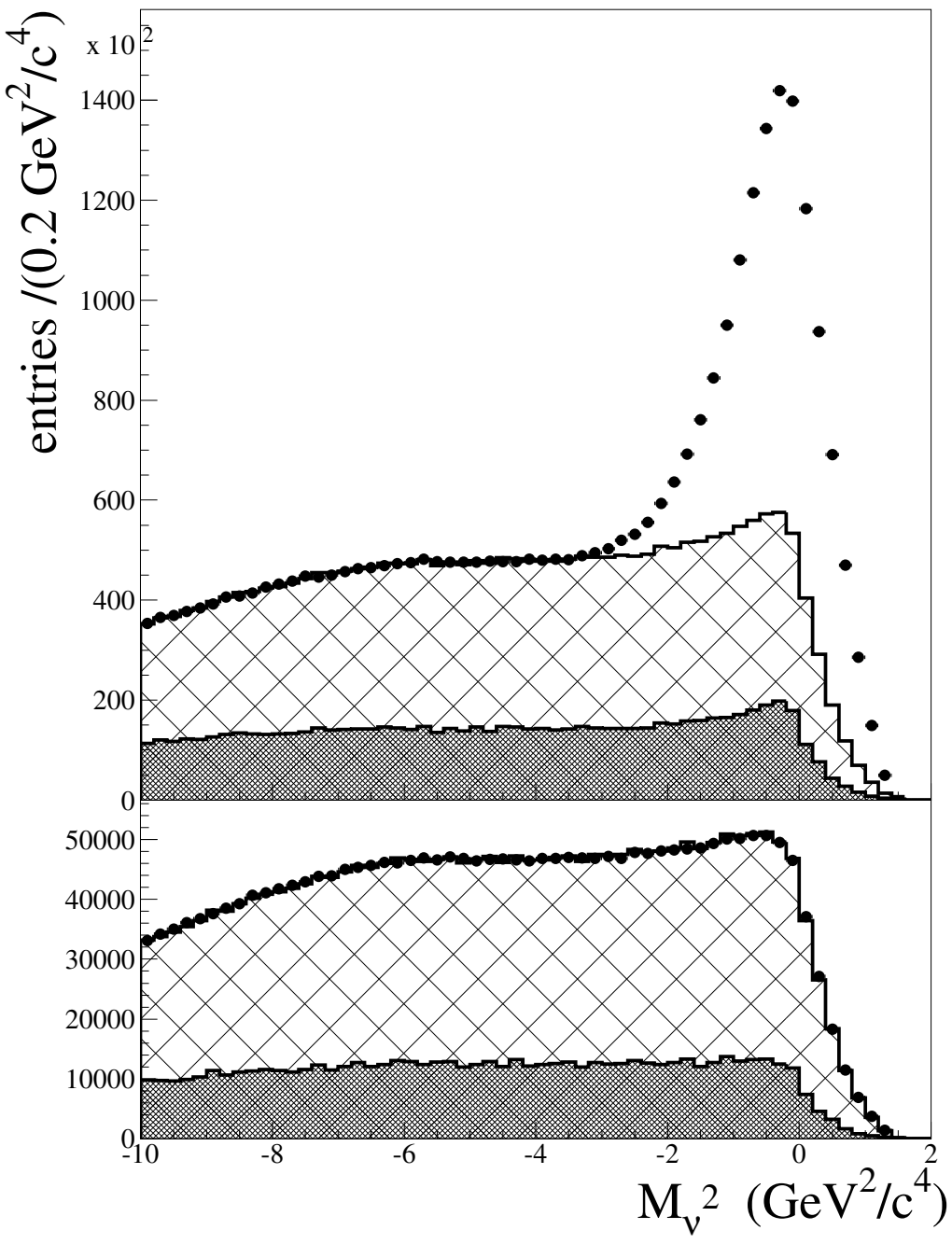

**Figure 7.3.3.** $M_\nu^2$ distribution for right-charge, $\ell^\pm\pi_s^\mp$, (top) and wrong-charge, $\ell^\pm\pi_s^\pm$, (bottom) events. The points correspond to on resonance data. The distributions of continuum events (dark histogram), obtained from luminosity-rescaled off-resonance events, and $B\overline{B}$ combinatorial background events (hatched area), obtained from the simulation, are overlaid. Monte Carlo events are normalized to the difference between on-resonance and rescaled off peak data in the region $M_\nu^2 < -4.5\,\mathrm{GeV}^2/c^4$ (Aubert, 2006s).

The above decays cannot be measured by reconstructing all the decay products since the neutrinos cannot be detected in detectors like *BABAR* and Belle. A different approach is taken instead, which is referred to as recoil $B$-meson reconstruction and is described in detail in the rest of this subsection. Herein, specific reference will be made to the searches for above example decays, to elucidate the necessity of the recoil method, although the techniques of studying the system recoiling against a reconstructed $B$ meson, referred to as the "tag"-$B$ ($B_{\mathrm{tag}}$[31]), can be applied to any analysis. The full list of measurements utilizing the recoil method performed by *BABAR* and Belle is given in Table 7.4.1.

Several different approaches are used in the recoil $B$-meson reconstruction technique. These can be separated according to the method used to reconstruct the decay of the $B$ meson accompanying the signal $B$-meson decay. The accompanying $B$ meson can be reconstructed either inclusively or exclusively. In the exclusive reconstruction the accompanying $B$ meson is reconstructed in several specific decay modes. It is further divided into the hadronic and semileptonic reconstruction, depending whether the decay modes used are hadronic or semileptonic, respectively. In the inclusive reconstruction all detected particles which are not assigned to the signal $B$-meson are used to reconstruct the accompanying $B$-meson, without testing whether the assigned particles are consistent with a specific $B$-meson decay chain. In all cases the recoil $B$-meson reconstruction relies on the following unique properties of experimental setup of the $B$ Factories (see Chapters 1 and 2 for more details):

[31] The same notation, $B_{\mathrm{tag}}$, is also used in Chapter 8 where it represents a flavor tagged $B$-meson.

**Table 7.4.1.** List of measurements performed by *BABAR* and Belle using the $B$ recoil techniques.

| | |
|---|---|
| **Hadronic $B_{\mathrm{tag}}$** | |
| $B\to X_u\ell\nu$ | (Bizjak, 2005; Urquijo, 2010) |
| | (Aubert, 2008ac) |
| $B\to X_c\ell\nu$ | (Schwanda, 2007; Urquijo, 2007) |
| | (Aubert, 2010c,e) |
| $B\to D^{(*)}\pi\ell\nu$ | (Abe, 2005d) |
| $B\to D^{**}\ell\nu$ | (Liventsev, 2008) |
| | (Aubert, 2007z, 2008s) |
| $B\to\pi\ell\nu$ | (Aubert, 2006r) |
| $B\to X_s\gamma$ | (Aubert, 2008q) |
| $B\to\tau\nu$ | (Adachi, 2012b; Ikado, 2006) |
| | (Aubert, 2005ae, 2008c; Lees, 2013a) |
| $B\to h^{(*)}\nu\overline{\nu}$ | (Chen, 2007b) |
| | (Aubert, 2008an) |
| $B\to$ invisible | (Hsu, 2012) |
| $B\to D^{(*)}\ell\nu_\ell$ | (Aubert, 2008h,y, 2010e) |
| $B\to D^{(*)}\tau\nu_\tau$ | (Aubert, 2008al; Lees, 2012e) |
| $B\to K\tau\mu$ | (Aubert, 2007au) |
| $B\to\ell\tau/\ell\nu$ | (Aubert, 2008az) |
| $B\to\tau\tau$ | (Aubert, 2006b) |
| **Semileptonic $B_{\mathrm{tag}}$** | |
| $B\to K\nu\overline{\nu}$ | (Aubert, 2005b, 2008an) |
| | (del Amo Sanchez, 2010p) |
| $B\to$ invisible $(+\gamma)$ | (Aubert, 2004y) |
| $B\to\pi\ell\nu$ | (Hokuue, 2007) |
| $B\to\rho\ell\nu$ | (Hokuue, 2007) |
| $B\to\tau\nu$ | (Hara, 2010) |
| | (Aubert, 2005ae, 2006a, 2007a, 2010a) |
| $B\to\ell\nu$ | (Aubert, 2010a) |
| **Inclusive $B_{\mathrm{tag}}$** | |
| $B\to D^{(*)}\tau\nu_\tau$ | (Bozek, 2010; Matyja, 2007) |
| $B\to D_s^{(*)}K\ell\nu_\ell$ | (Stypula, 2012) |

– the $B\overline{B}$ pairs are produced without any additional particles,
– the detectors enclose the interaction region almost hermetically,
– the collision energy (or initial state energy) is precisely known.

The most commonly used strategy in the recoil $B$-meson reconstruction is to reconstruct exclusively the decay of one of the $B$ mesons ($B_{\mathrm{tag}}$) in the event. The remaining particle(s) in the event (detected as tracks or energy deposits in the calorimeter) must therefore originate from the other $B$-meson decay, referred to as the "recoil"-$B$ ($B_{\text{recoil}}$) or "signal"-$B$ ($B_{\text{sig}}$),[32] and are compared with the signature expected for the signal mode. In studies of the example decay, $B^+ \to K^+\nu\overline{\nu}$ the presence of exactly one charged track (positively identified as a kaon) not used in the reconstruction of the $B_{\text{tag}}$ is required. An additional powerful variable which allows for separation of signal and background is the remaining energy in the calorimeter, denoted as $E_{\text{extra}}$ at *BABAR* or as $E_{\text{ECL}}$ at Belle. It is defined as the sum of the energy deposits in the calorimeter that cannot be directly associated with the reconstructed daughters of the $B_{\text{tag}}$ or the $B_{\text{recoil}}$. Figure 7.4.1 shows a typical distribution of simulated signal and background events. For signal events (*e.g.* example decays given in beginning of this subsection), $E_{\text{extra}}$ must be either zero or a small value arising from beam background hits and detector noise, since neutrinos do not loose any energy in the calorimeter. On the other hand, background events are distributed toward higher $E_{\text{extra}}$ due to the contribution from additional clusters, produced by unassigned tracks and neutrals from the mis-reconstructed tag and recoil $B$ mesons. For signal $B$-meson decays to a final state with only one neutrino (like the example decay $B^0 \to D^-\ell\nu_\ell$) where the $B_{\text{tag}}$ is reconstructed in a hadronic decay mode, the neutrino momentum can be inferred using the momentum conservation relation from the measured momenta of $B_{\text{tag}}$, $D^-$ and $\ell$, and known initial state: $p_{\nu_\ell} = p_{e^-} + p_{e^+} - p_{B_{\text{tag}}} - p_{D^-} - p_\ell$. This allows for the construction of a powerful kinematic constraint – missing mass squared, defined as $MM^2 = |p_\nu|^2$, which peaks at the neutrino mass ($MM^2 = 0$) for correctly reconstructed events.

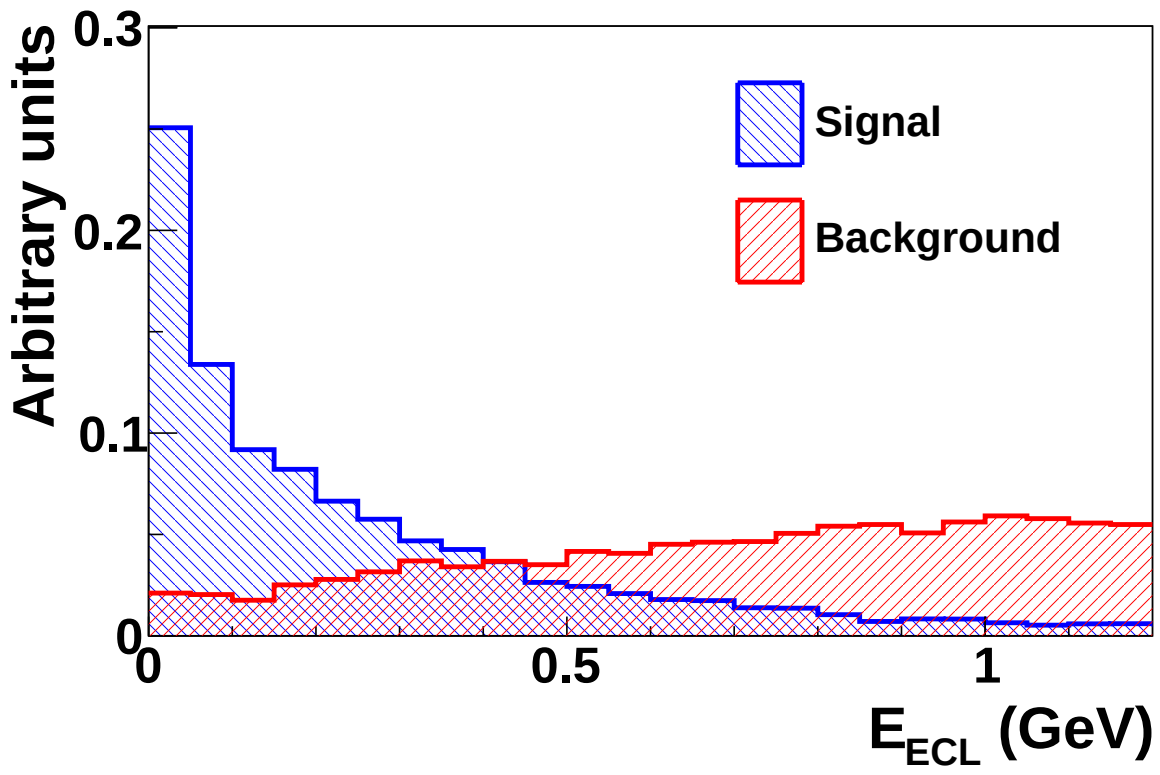

**Figure 7.4.1.** The $E_{\text{extra}}(=E_{\text{ECL}})$ distribution of simulated signal and background events. Belle internal, from the $B^+ \to \tau^+\nu_\tau$ Adachi (2012b) analysis.

In studies of $B$-meson decay modes using the exclusive recoil $B$-meson reconstruction technique the number of reconstructed signal decays is linearly proportional to the efficiency of the $B_{\text{tag}}$ reconstruction, which is given by

$$\varepsilon_{B_{\text{tag}}} = \sum_f \varepsilon_f \mathcal{B}_f, \qquad (7.4.1)$$

and the sum runs over the $B$-meson decays to the exclusively reconstructed final states $f$. The $\varepsilon_f$ are the corresponding reconstruction efficiencies and the $\mathcal{B}_f$ are the branching fractions of the $B \to f$ decays. In order to achieve as high efficiency as possible a large number of $B$-meson decay modes are used for the $B_{\text{tag}}$ reconstruction. On the quark level $B$ mesons decay dominantly via $\overline{b} \to \overline{c}W^+$ transitions, where the virtual $W$ materializes either into a pair of leptons $\ell\nu_\ell$ (semileptonic decay), or into a pair of quarks, $u\overline{d}$ or $c\overline{s}$, which then hadronize. The most common choice for exclusive $B_{\text{tag}}$ reconstruction are therefore semileptonic $B \to \overline{D}^{(*)}\ell^-\nu_\ell$ decays (semileptonic $B_{\text{tag}}$ reconstruction) and hadronic $B \to \overline{D}^{(*)}n\pi$, $\overline{D}^{(*)}D_s^{(*)}$ or $B \to J/\psi K m\pi$ (hadronic $B_{\text{tag}}$ reconstruction), where $n$ and $m$ indicate any number ($n,\ m \leq 10$) of charged or neutral pions and kaons, respectively. The branching fractions of these hadronic decay modes are between $10^{-3}$ and up to $10^{-2}$, and the branching fraction for inclusive semileptonic decays[33] of a $B$ meson to a $D$ meson plus anything else is around 20%. The two analysis techniques are complimentary and non-overlapping and, as such, can be readily combined to improve the sensitivity of any recoil $B$ analysis. This essentially doubles the size of the available $B_{\text{tag}}$ sample.

Many decay modes for which the $B$ meson cannot be exclusively reconstructed rely on these methods to make measurements feasible. For the proposed high luminosity asymmetric $e^+e^-$ super flavor factories, measurements of $B$ decays, not related to *CP* violation or the CKM picture of the Standard Model, will benefit from recoil methods. This corresponds to a wide program of purely leptonic, semileptonic and radiative penguin[34] $B$ decays. Furthermore, with a huge dataset the recoil methods will provide a clean "single $B$ beam" which will permit the extraction of hadronic $B$ decay branching fractions using a missing mass technique.

In this section the general idea behind the recoil $B$-meson reconstruction has been presented. In addition the variables or constraints which can be imposed in studies of $B$-meson decays involving one or more neutrinos with recoil $B$-meson technique have been briefly described. The rest of this section is devoted to the description of different approaches to $B_{\text{tag}}$ reconstruction. More details on analyses of decay modes utilizing the recoil $B$-meson reconstruction (given in Table 7.4.1) can be found in Sections 17.9, 17.10 and 17.11.

### 7.4.1 Hadronic tag $B$ reconstruction

The full reconstruction of one $B$ meson, decaying hadronically, has been utilized in a multitude of analyses by the $B$ Factories (see Table 7.4.1). The approaches of *BABAR* and Belle differ somewhat, providing samples which vary in efficiency and purity. The optimization of these choices depends primarily on the signal mode in the recoil system and the available kinematic constraints which can be imposed.

[32] The terms used for this $B$ meson in the various *BABAR* and Belle papers are not consistent. Elsewhere in this book we use the term $B_{\text{sig}}$.

[33] Semitauonic decays are not included in this case.

[34] A penguin decay is represented by a higher order Feynman diagram including a loop with a $W$ or $Z$ boson; a quark in the loop undergoes a tree process - either a strong interaction one, or electroweak one.

#### 7.4.1.1 *BABAR*

*BABAR* opts for a semi-exclusive approach where hadronic $B$ decays are reconstructed by *seeding* the event with a charm meson, and combining it with a number of pions and kaons. The algorithm underwent a major expansion in 2008 doubling its reconstruction efficiency. The starting point is the creation of a list with all the possible seeds in the event. In the original algorithm, $D^0$, $D^+$, $D^{*0}$ and $D^{*+}$ mesons were used as seeds, reconstructed in the following decay chains: $D^- \to K^+\pi^-\pi^-$, $K^+\pi^-\pi^-\pi^0$, $K^0_S\pi^-$, $K^0_S\pi^-\pi^0$, $K^0_S\pi^-\pi^-\pi^+$; $\overline{D}^0 \to K^+\pi^-$, $K^+\pi^-\pi^0$, $K^+\pi^-\pi^-\pi^+$, $K^0_S\pi^+\pi^-$; $D^{*-} \to \overline{D}^0\pi^-$; and $\overline{D}^{*0} \to \overline{D}^0\pi^0$, $\overline{D}^0\gamma$. The 2008 expansion added the decay chains $D^- \to K^+K^-\pi^-$, $K^+K^-\pi^-\pi^0$; $D^0 \to K^0_S\pi^-\pi^-\pi^0$, $K^+K^-$, $K^0_S\pi^0$, $\pi^+\pi^-\pi^0$, $\pi^+\pi^-$; $D^{*-} \to D^-\pi^0$, and the new seeds $D_s^+ \to \phi\pi^0$, $K^0_S K^+$; $D_s^{*+} \to D_s^+\gamma$ and $J/\psi \to e^+e^-$, $\mu^+\mu^-$.

Subsequently, each one of the reconstructed seeds is combined with up to 5 charmless particles to form a $B_{\rm tag} \to D_{\rm seed}\, Y$ candidate, where $D_{\rm seed}$ refers to the charm meson used to seed events. The $Y$ system represents a collection of hadrons composed of $n_1\pi^\pm + n_2K^\pm + n_3\pi^0 + n_4K^0_S$ ($n_1 = 1, ..., 5$, $n_2 = 0, ..., 2$, $n_3 = 0, ..., 2$ and $n_4 = 0, 1$) and having total charge equal to $\pm1$. In the expansion, four neutral $Y$ systems, $K^+\pi^-$, $\pi^+\pi^-$, $K^+K^-$ and $\pi^0$, were added. Overall, the original algorithm reconstructs $B_{\rm tag}$ candidates in 630 different decay chains, and the expansion in 1768.

The $B_{\rm tag}$ candidates thus formed are accepted if they satisfy some loose requirements that ensure kinematic consistency with a $B$ meson: the beam-energy substituted mass, $m_{\rm ES}$, has to be greater than 5.18 GeV/$c^2$, and $\Delta E$ has to satisfy $-0.12 < \Delta E < 0.12$ GeV. Correctly reconstructed events should have the $m_{\rm ES}$ and $\Delta E$ distributions peak at the $B$-meson mass and at zero, respectively.

These algorithms provide several $B_{\rm tag}$ candidates per event. One of the most extended methods to choose a unique candidate selects the decay chain with the highest *purity*, defined as the fraction of $B$ candidates that are correctly reconstructed for $m_{\rm ES} > 5.27$ GeV/$c^2$ in each particular chain. The purity is determined from a fit to the $m_{\rm ES}$ spectrum of a data sample, where the signal distribution is described by a Crystal Ball function (Skwarnicki, 1986), named after the Crystal Ball collaboration, defined as

$$\mathrm{CB}(m|\alpha, n, m_0, \sigma) = \begin{cases} e^{-(m-m_0)^2/2\sigma^2}, & \text{if } \frac{m-m_0}{\sigma} < -\alpha \\ A\left(B - \frac{m-m_0}{\sigma}\right)^{-n}, & \text{otherwise,} \end{cases} \tag{7.4.2}$$

where $A = (n/|\alpha|)^n e^{-|\alpha|^2/2}$ and $B = n/|\alpha| - |\alpha|$. The background distribution is described by an ARGUS function as defined in Eq. (7.1.11). The purity can also be used to reject combinatorial background by selecting only decay chains with a minimum value of purity, typically between 30% and 55%.

In more recent analyses, the best $B_{\rm tag}$ candidate tends to be selected together with the rest of the event. For instance, in $B \to D^*\ell\nu$, each $B_{\rm tag}$ candidate is combined with $D^*$ and $\ell$ candidates. The best $B_{\rm tag}D^*\ell$ candidate is selected maximizing the energy measured in the calorimeter that is used in the reconstruction.

In the final selection the kinematic requirements on $B_{\rm tag}$ are tightened, candidates are selected with $m_{\rm ES} > 5.27$ GeV/$c^2$ and narrower $\Delta E$ windows ($-90 < \Delta E < 60$ MeV is typically used). Events outside these regions may be used to study the combinatorial background.

When all the $B_{\rm tag}$ decay chains are used in the analysis, the efficiencies of the original algorithm, defined as

$$\varepsilon_{B^0_{\rm tag}} = \frac{N(B^0_{\rm tag})}{N(B\overline{B})}, \tag{7.4.3}$$

$$\varepsilon_{B^+_{\rm tag}} = \frac{N(B^+_{\rm tag})}{N(B\overline{B})}, \tag{7.4.4}$$

reaches typically 0.2% ($B^0\overline{B}^0$) and 0.4% ($B^+B^-$).

#### 7.4.1.2 Belle

Belle developed two versions of hadronic $B_{\rm tag}$ reconstruction algorithms in the course of its history. In both versions the $B_{\rm tag}$ mesons are reconstructed in a set of exclusive final states, although the approach is slightly different from the one used by *BABAR* described above. The difference between the two versions is in the selection of $B_{\rm tag}$ candidates. In the first version a set of rectangular cuts is imposed on $B_{\rm tag}$ candidates (referred to as cut-based selection), while in the second the selection of $B_{\rm tag}$ candidates is made using a NeuroBayes neural network (referred to as NB selection) (Feindt, 2004) (see Section 4.4.4 for more details on neural nets). The latter version is mostly used in the measurements using the full data sample collected by Belle at the $\Upsilon(4S)$. At the end of this section a comparison between the two versions in terms of performance is provided.

In the cut-based approach Belle reconstructs a set of the following exclusive decay modes: $B^+ \to \overline{D}^{(*)0}(\pi, \rho, a_1, D_s^{(*)})^+$ and $B^0 \to D^{(*)-}(\pi, \rho, a_1, D_s^{(*)})^+$. $\overline{D}^0$ mesons are reconstructed in 7 decay modes: $K^+\pi^-$, $K^+\pi^-\pi^0$, $K^+\pi^-\pi^-\pi^+$, $K^0_S\pi^0$, $K^0_S\pi^-\pi^+$, $K^0_S\pi^-\pi^+\pi^0$ and $K^-K^+$. $D^-$ mesons are reconstructed in 6 decay modes: $D^- \to K^+\pi^-\pi^-$, $K^+\pi^-\pi^-\pi^0$, $K^0_S\pi^-$, $K^0_S\pi^-\pi^0$, $K^0_S\pi^-\pi^-\pi^+$ and $K^+K^-\pi^-$, and the $D_s^+$ mesons are reconstructed in two decay modes: $K^0_SK^+$ and $K^+K^-\pi^+$. The $D$ candidates are required to have an invariant mass $m_D$ within $(4-5)\sigma$ of the nominal $D$ mass value depending on the decay mode, where $\sigma$ represents the $D$ mass resolution. The $\overline{D}^{*0}$, $D^{*-}$ and $D_s^{*+}$ mesons are reconstructed in $\overline{D}^{*0} \to \overline{D}^0\pi^0$, $\overline{D}^0\gamma$, $D^{*-} \to \overline{D}^0\pi^-$, $D^-\pi^0$ and $D_s^{*+} \to D_s^+\gamma$ modes, respectively. $D^*_{(s)}$ candidates are required to have a mass difference $\Delta m = m_{D\pi} - m_D$ within $\pm5$ MeV/$c^2$ of its nominal mass or $\Delta m = m_{D_{(s)}\gamma} - m_{D_{(s)}}$ within $\pm20$ MeV/$c^2$.

The $\rho^0$, $\rho^+$ and $a_1^+$ are reconstructed in $\pi^+\pi^-$, $\pi^+\pi^0$ and $\rho^0\pi^+$ modes, respectively. The invariant mass of the $\pi\pi$ pairs is required to be within $\pm225$ MeV/$c^2$ of the nominal $\rho$ mass, and the $\rho\pi$ combinations are required to have invariant mass between 0.7 and 1.6 GeV/$c^2$ ($a_1$ mass region). In order to obtain reasonable purity of the $B_{\rm tag}$ sample (*e.g.* above 20% in the $m_{\rm ES} > 5.27$ GeV/$c^2$ region) the decay chains with a high multiplicity of tracks and neutrals (and hence with a large contribution of combinatorial background) in the final state are excluded. Therefore in the $B \to \overline{D}^{(*)}a_1$ decay modes only the $D^- \to K^+\pi^-\pi^-$, $K_S^0\pi^-$ and $\overline{D}^0 \to K^+\pi^-$ modes are used. The selection of $B_{\rm tag}$ candidates is based on $m_{\rm ES}$ and $\Delta E$. The definition of the signal region in the $\Delta E - m_{\rm ES}$ plane depends on the studied signal decay mode. If an event has multiple $B_{\rm tag}$ candidates the one with the smallest $\chi^2$ is selected based on deviations from the nominal values of $\Delta E$, the $D_{(s)}$ candidate mass and the $D^*_{(s)} - D_{(s)}$ mass difference, if applicable. The efficiencies as defined in Eqs (7.4.4) and (7.4.3) of $B^0_{\rm tag}$ and $B^+_{\rm tag}$ are found to be 0.10% and 0.14%, respectively.

In the second approach Belle increased the number of reconstructed exclusive $B$ decay modes and used a neural network in their selection in order to increase the hadronic $B_{\rm tag}$ reconstruction efficiency (Feindt et al., 2011). In addition to the decay modes used in the cut-based selection the $B_{\rm tag}$ candidates are reconstructed also in the following decay modes: $B^+ \to \overline{D}^{*0}\pi^+\pi^+\pi^-\pi^0$, $D^-\pi^+\pi^+$, $\overline{D}^0K^+$, $J/\psi K^+$, $K^+\pi^0$, $K^0_S\pi^+$, $K^+\pi^+\pi^-$, and for neutral $B$ mesons via $B^0 \to D^{*-}\pi^+\pi^+\pi^-\pi^0$, $\overline{D}^0\pi^0$, $J/\psi K^0_S$, $K^+\pi^-$, and $K^0_S\pi^+\pi^-$.[35] The $D$ meson decay modes used in the reconstruction of $B_{\rm tag}$ are $\overline{D}^0 \to \pi^-\pi^+$, $K^0_SK^-K^+$, $D^- \to K^+K^-\pi^-\pi^0$ and $D_s^+ \to K^+\pi^+\pi^-$, $K^+K^-\pi^+\pi^0$, $K^0_SK^+\pi^+\pi^-$, $K^0_SK^-\pi^+\pi^+$, $K^+K^-\pi^+\pi^+\pi^-$ and $\pi^+\pi^+\pi^-$ in addition to the modes used by Belle in the cut-based reconstruction, given above. The $J/\psi$ is reconstructed in $e^+e^-$ and $\mu^+\mu^-$ modes. The sum of branching ratios of reconstructed decay modes adds up to around 12% for $B^+$, 10% for $B^0$ (not taking into account branching fractions of $D^{(*)}$, $J/\psi$, and other intermediate states), 38% for $D^0$, 29% for $D^+$, 18% for $D_s^+$ and 12% for $J/\psi$. The reconstruction and selection proceeds in four stages. At each stage all available information on a given candidate is used to calculate a single scalar variable (referred to as network output) using the NeuroBayes neural network which can be by construction interpreted as a probability that a given candidate is correctly reconstructed (Feindt, 2004). The network output for each reconstructed particle is used as an input to other neural networks in the later stage(s). In the first stage $\pi^\pm$, $K^\pm$, $K^0_S$, $\gamma$ and $\pi^0$ candidates are reconstructed and classified, in the second $D^0$, $D^\pm_{(s)}$ and $J/\psi$, in the third $D^{*0}$ and $D^{*\pm}_{(s)}$, and finally in the last, fourth stage the $B^\pm$ and $B^0$ candidates are reconstructed and classified. The neural networks of the first stage particles include measurements of time-of-flight, the energy loss in the CDC and Cherenkov light in the ACC for the charged particles, and shower shape variables for photons (see Chapter 2 for subdetectors description). The variables with the largest separation power in the second stage, *e.g.* classification of $D_{(s)}$ mesons, are the network outputs of the daughters (charged or neutral kaons and pions), the invariant masses of daughter pairs (in case of multi-body decay modes), the angle between the momentum of the $D_{(s)}$ meson and the vector joining the $D_{(s)}$ decay vertex and interaction point and the significance of the distance between the decay vertex and the interaction point. In the last stage, the $B$-meson stage, the variables providing good discrimination between correctly reconstructed $B$ mesons and background candidates are again the network outputs of the daughters ($D^{(*)}$, $J/\psi$, pions, kaons), the mass of the $D_{(s)}$ or mass difference between $D^*_{(s)}$ and $D_{(s)}$, $\Delta E$, and the angle between the $B$-meson momentum and the beam. A large fraction of $B_{\rm tag}$ candidates are background candidates from continuum events. As explained in detail in Chapter 9 continuum background can be quite successfully suppressed at $B$ Factories by exploiting event shape variables, such as the reduced second Fox-Wolfram moment, $R_2$, thrust angle and super Fox-Wolfram moments. In the default $B_{\rm tag}$ networks these variables are excluded, but outputs of some additional neural networks, which take also the continuum suppression variables into account (with $R_2$ and thrust angle only, or with $R_2$, thrust angle and super Fox-Wolfram moments), are provided.

In an ideal case, one would reconstruct all possible $B_{\rm tag}$ candidates in the given decay modes without making any selections (cuts) between the stages. No signal candidates would be lost, *i.e.* the efficiency is maximized. Postponing the moment of the selection to the latest possible stage is always the preferred strategy in data analyses, since at the end more information is available which can be used to more successfully separate signal and background candidates. However this procedure is limited by combinatorics and computing resources. Events with many reconstructed particles lead to a large number of possible $B_{\rm tag}$ candidates which of course require more computing time. Loose cuts between the reconstruction stages are therefore required in order to keep computing time at a bearable level. These cuts on the network output for a given candidate are not performed at the end of each stage in which the candidate is reconstructed and classified but it is performed at the next stage and depends on the complexity of the decay mode in the next stage. As an example, the amount of combinations of the decay $B \to \overline{D}\pi\pi\pi$ is much higher then that of the decay $B \to \overline{D}\pi$ given the same number of $\overline{D}$ candidates. Therefore, a tighter cut on the signal probability of $\overline{D}$ candidates is performed only when necessary, *e.g.* when the reconstruction of all candidates would require too many resources, as in the case of $B \to \overline{D}\pi\pi\pi$ decays.

At the end the kinematic consistency of a $B_{\rm tag}$ candidate with a $B$-meson decay is checked using the beam constrained mass, $m_{\rm ES}$, as described previously. Since the network output can be interpreted as signal probability the candidates which are reconstructed in different decay modes can be easily compared to one another. In case multiple $B_{\rm tag}$ candidates are found in an event the one with highest signal probability is taken as the best one. The $m_{\rm ES}$ distributions of $B^+_{\rm tag}$ and $B^0_{\rm tag}$ samples obtained by Belle with cut-based and NB selections are shown in Fig. 7.4.2. In order to compare the performance in terms of $B_{\rm tag}$ efficiencies and purities of the NB and cut-based selections the network output cuts in the NB selection are chosen is such a way that equal purities (in $m_{\rm ES} >$ 5.27 GeV/$c^2$ region) or equal efficiencies are obtained in both selections. As can be seen from Fig. 7.4.2 at the same purity the signal yield (and hence efficiency) is approximately two times larger. The NB selection with efficiency equal to the cut-based selection will result in a much purer sample: nearly 90% versus 25% (reducing the background level by more than a factor of 20). The NB selection used in Fig. 7.4.2 is arbitrary and is chosen only for the purpose of comparing the NB and cut-based $B_{\rm tag}$ selections. The final selection depends on the studied decay mode and can be selected either to give maximal possible $B_{\rm tag}$ efficiency or high purity. Figure 7.4.3 shows purity-efficiency plots for $B^+_{\rm tag}$ and $B^0_{\rm tag}$ for the default NB selection and the one including continuum suppression. The highest possible efficiency that can be achieved with the NB selection at Belle is around 0.18% for $B^0_{\rm tag}$ and 0.28% for $B^+_{\rm tag}$ with around 10% purity. This corresponds to an improvement in efficiency by roughly a factor of two comparing to Belle's cut-based $B_{\rm tag}$ selection.

[35] The $B \to D^{(*)}(\rho, a_1)^+$ modes are reconstructed as $B \to D^{(*)}(\pi^0\pi^+, \pi^+\pi^-\pi^+)$ in the network based $B_{\rm tag}$ selection meaning that there are no explicit restrictions made on the invariant masses of the two or three pion systems.

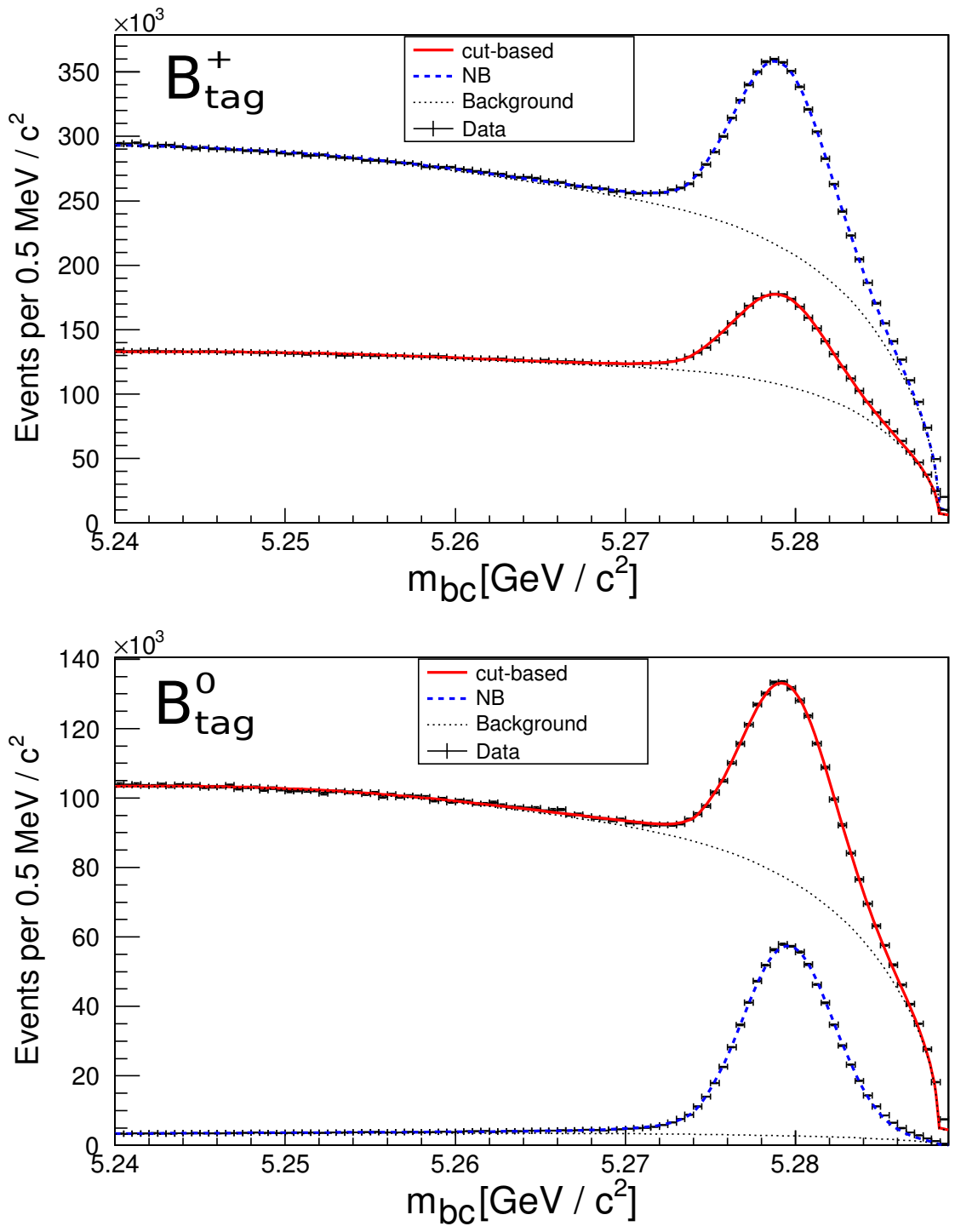

**Figure 7.4.2.** The $m_{\rm ES}$ ($= m_{\rm bc}$) distribution of hadronic $B^+_{\rm tag}$ (top) and $B^0_{\rm tag}$ (bottom) samples obtained by Belle with cut-based (red) and NB selection (blue) (Feindt et al., 2011). In case of the $B^+_{\rm tag}$ sample the cut on the network output in the NB selection is chosen to give equal purity as the cut-based selection in $m_{\rm ES} > 5.27$ GeV/$c^2$. In case of the $B^0_{\rm tag}$ sample the cut on the network output in the NB selection is chosen to give equal $B$-meson signal yield as the cut-based selection. These cuts are arbitrary and are chosen only for the purpose of comparing the NB and cut-based $B_{\rm tag}$ selections.

### 7.4.2 Semileptonic tag $B$ reconstruction

This method of semi–exclusive $B$ reconstruction involves the selection of a $D$ meson and suitable lepton candidate, $\ell$, which are then combined into a $D\ell$ candidate.

The $B_{\rm tag}$ is reconstructed in the set of semileptonic $B$ decay modes $B^- \to D^0\ell^-\overline{\nu}_\ell X$, where $\ell$ denotes an $e$ or $\mu$, and $X$ can be either nothing or a transition particle from a higher mass charm state decay, which one does not necessarily need to reconstruct. This methodology naturally includes the $B^- \to D^0\ell^-\overline{\nu}_\ell$ and $B^- \to D^{*0}\ell^-\overline{\nu}_\ell$ modes and also retains those modes with excited $D$ meson states which decay, via the emission of soft transitions particles, to a $D^0$. The technique can be similarly applied to the tagging of neutral $B$ mesons where one would reconstruct $\overline{B}{}^0 \to D^{(*)+}\ell^-\overline{\nu}_\ell$ for a combination of all possible $\overline{B}{}^0 \to D^+\ell^-\overline{\nu}_\ell$ and $\overline{B}{}^0 \to D^{*+}\ell^-\overline{\nu}_\ell$ states reconstructed exclusively. The main loss in efficiency arises from the $B$ and charm decay branching fractions while further selection criteria must be applied in order to suppress non-$B$ decay backgrounds (continuum) and fakes from hadronic $B$ decays.

The $D^0$ decay is reconstructed by *BABAR* in the four cleanest hadronic modes: $K^-\pi^+$, $K^-\pi^+\pi^-\pi^+$, $K^-\pi^+\pi^0$, and $K^0_S\pi^+\pi^-$. The $K^0_S$ is reconstructed only in the mode $K^0_S \to \pi^+\pi^-$. Belle reconstructs $D^0$ candidates in ten decay modes (Hokuue, 2007): in addition to the four decay modes above, the $K^0_S\pi^0$, $K^0_S\pi^+\pi^-\pi^0$, $K^-\pi^+\pi^+\pi^-\pi^0$, $K^+K^-$, $K^0_SK^+K^-$ and $K^0_SK^-\pi^+$ modes are also included. The added benefit of reconstructing the low momentum transition daughter in $D^{*0}$ decays is to provide a more complete and exclusive tag $B$ selection. Indeed if one neglects to reconstruct these $\pi^0$ or $\gamma$ daughters (from $D^{*0} \to D^0\pi^0/\gamma$) then they will be considered in the reconstruction of the signal $B$ target mode. However, it is observed that the semi-exclusive reconstruction of $B \to D^0\ell\nu X$ provides a higher efficiency with some loss of purity.

For neutral $B$ tags the selection becomes that of either $\overline{B}{}^0 \to D^+\ell^-\overline{\nu}_\ell$ or $\overline{B}{}^0 \to D^{*+}\ell^-\overline{\nu}_\ell$. The $D^+$ decays are reconstructed at Belle in seven decay modes $K^-\pi^+\pi^+$, $K^0_S\pi^+$, $K^-\pi^+\pi^+\pi^0$, $K^0_S\pi^+\pi^0$, $K^0_S\pi^+\pi^+\pi^-$, $K^0_SK^+$ and $K^+K^-\pi^+$ (Hokuue, 2007), while *BABAR* uses only the first two decay modes. The $D^{*+}$ decays can be reconstructed as both $D^0\pi^+$ and $D^+\pi^0$. The mass difference between $D^*$ and $D$ provides a powerful constraint as does the invariant mass of the $D^0$ or $D^+$ candidate.

The center-of-mass lepton momentum ($p^*_\ell$) for both electrons and muons is selected to be greater than 0.8 (1.0) GeV/$c$ at *BABAR* (Belle). This is the lower end of muon identification for the current $B$ Factories and there is commonly non-$B$ background below $p^*_\ell \sim$1 GeV/$c$. The reconstructed $D$ mesons are required to be within $\pm 3\sigma$ ($\pm 2.5\sigma$) at *BABAR* (Belle) of their nominal mass value. As explained in Section 7.2 the cosine of the angle between the $B$ meson and the $D^{(*)}\ell$ candidate momenta, $\cos\theta_{B,D\ell}$ defined in Eq. (7.2.2), is a powerful discriminant. In case the $D\ell$ and the neutrino are the only decay products of the $B$ then $\cos\theta_{B,D\ell}$ must lie in the physical region between $\pm 1$. If additional decay products from the cascade of a higher mass charm state down to the $D^0$ go unreconstructed then this will force the value of $\cos\theta_{B,D\ell}$ to be smaller. In order to keep such candidates events with $\cos\theta_{B,D\ell}$ between $-2.5$ and $+1.1$ are usually accepted. The positive limit is allowed to be slightly outside of the physical region to account for detector and reconstruction effects. Of course, for the reconstruction of exclusive channels ($B^- \to D^0\ell^-\overline{\nu}_\ell$, $B^- \to D^{*0}\ell^-\overline{\nu}_\ell$, $\overline{B}^0 \to D^+\ell^-\overline{\nu}_\ell$ and $\overline{B}^0 \to D^{*+}\ell^-\overline{\nu}_\ell$), the selection is tightened to only consider the physical region.

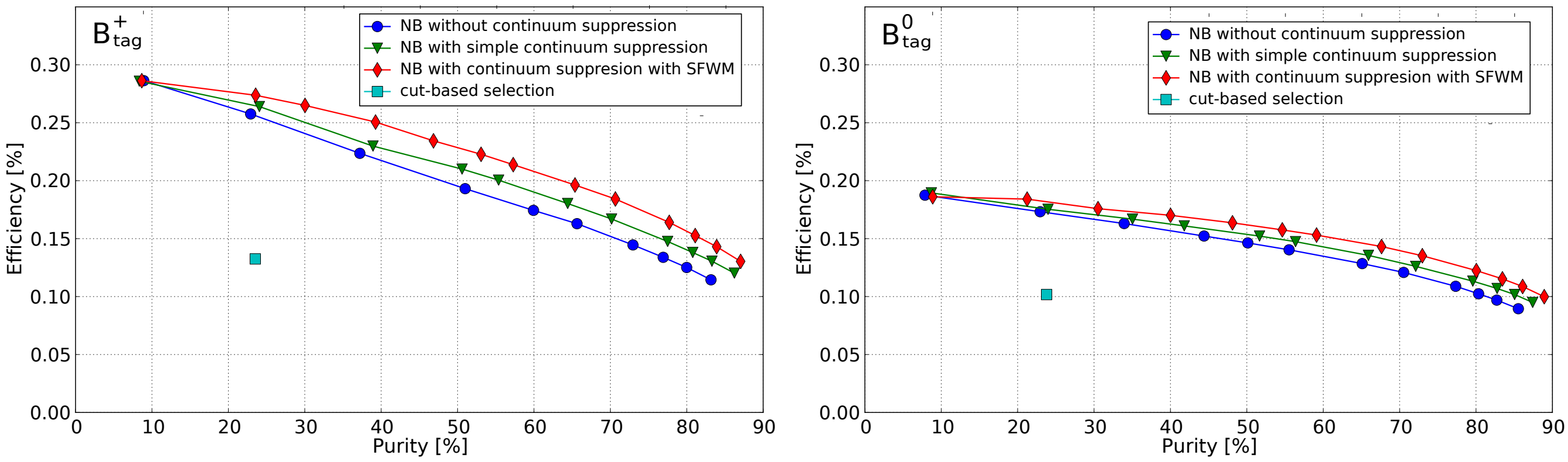

**Figure 7.4.3.** Purity-efficiency plots for hadronic $B^+_{\rm tag}$ (left) and $B^0_{\rm tag}$ (right) as obtained by Belle with neural network based selection (NB) and cut-based selection (Feindt et al., 2011). The network based selection can include no continuum suppression variables (blue), only simple ones (green) or Super-Fox-Wolfram moments (SFWM; red).

A typical $B^- \to D^0\ell^-\overline{\nu}_\ell X$ selection at *BABAR* yields an efficiency of approximately $6\times 10^{-3}$ with a mode dependent purity which averages to $\sim 60\%$. For the neutral $B$ reconstruction the efficiency is typical half that of a similar charged $B$ selection.

The loss of a neutrino in the semileptonic tagging mode limits the constraints that can be imposed compared to the case when all of the $B$ meson decay products are reconstructed. For example the signal $B$ direction cannot be found as is possible for hadronic $B$ reconstruction. However, this constraint is not of paramount importance in the analysis of signal decay modes to final state with more than one neutrino like for example $B^+ \to \tau^+\nu_\tau$ or $B^0 \to \nu\overline{\nu}$). The knowledge of signal $B$ momentum enables calculation of missing mass which is a very powerful variable to separate signal $B$ decays with a single neutrino in the final state from background decays, but becomes weak when multiple neutrinos are present in the signal $B$ decay.

### 7.4.3 Inclusive $B_{\rm tag}$ reconstruction

As discussed in the previous two sections the reconstruction of the recoil $B$ meson using the hadronic and semileptonic $B_{\rm tag}$ samples has many benefits, however suffers from low reconstruction efficiencies. To increase the statistics Belle adopted an inclusive $B_{\rm tag}$ reconstruction (Bozek, 2010; Matyja, 2007) in studies of semitauonic $B \to D^{(*)}\tau^-\nu$ decays (see Section 17.10). In contrast to the measurements utilizing the hadronic or semileptonic recoil $B_{\rm tag}$ reconstruction technique the procedure in this case is first to reconstruct the signal side (pairs of a $D^{(*)}$ and a lepton or pion from tau decay). In the second step the $B_{\rm tag}$ is inclusively reconstructed from all remaining particles passing certain selection criteria however without checking consistency with any specific $B$-meson decays. The number of neutral particles on the tagging side $N_{\pi^0} + N_\gamma < 6$ and $N_\gamma < 3$. The quality of $B_{\rm tag}$ reconstruction and suppression of background is further improved by requiring zero total charge and net proton/antiproton number, no leptons on the tagging side and extra energy to be close to zero (less then 350 MeV). These criteria reject events in which some particles from the signal or tagging side were undetected and suppress events with a large number of spurious showers. The consistency of $B_{\rm tag}$ with a $B$-meson decay is checked using the beam constrained mass, $m_{\rm ES}$, and the energy difference, $\Delta E$. The simulation and reconstruction of the inclusive $B_{\rm tag}$ sample is checked using a control sample of events, where the $B \to D^{*-}\pi^+$ decays (followed by $D^{*-}\pi^-$, $\overline{D}^0 \to K^+\pi^-$) are reconstructed on the signal side. Figure 7.4.4 shows the $m_{\rm ES}$ and $\Delta E$ distributions of the control sample for data and the MC simulation. The good agreement of the shapes and of the absolute normalization demonstrates the validity of the MC-simulations for $B_{\rm tag}$ decays. While the $m_{\rm ES}$ distribution shows a clear peak at the $B$-meson mass, the $\Delta E$ distribution is very broad. On the negative side events with undetected particles contribute and the main source of the events with $\Delta E > 0$ are spurious showers in the electromagnetic calorimeter from secondary interactions of hadrons. These clusters add linearly to $\Delta E$, but tend to average in the vector sum of their momenta that enters the calculation of $m_{\rm ES}$, see Eq. (7.1.8).

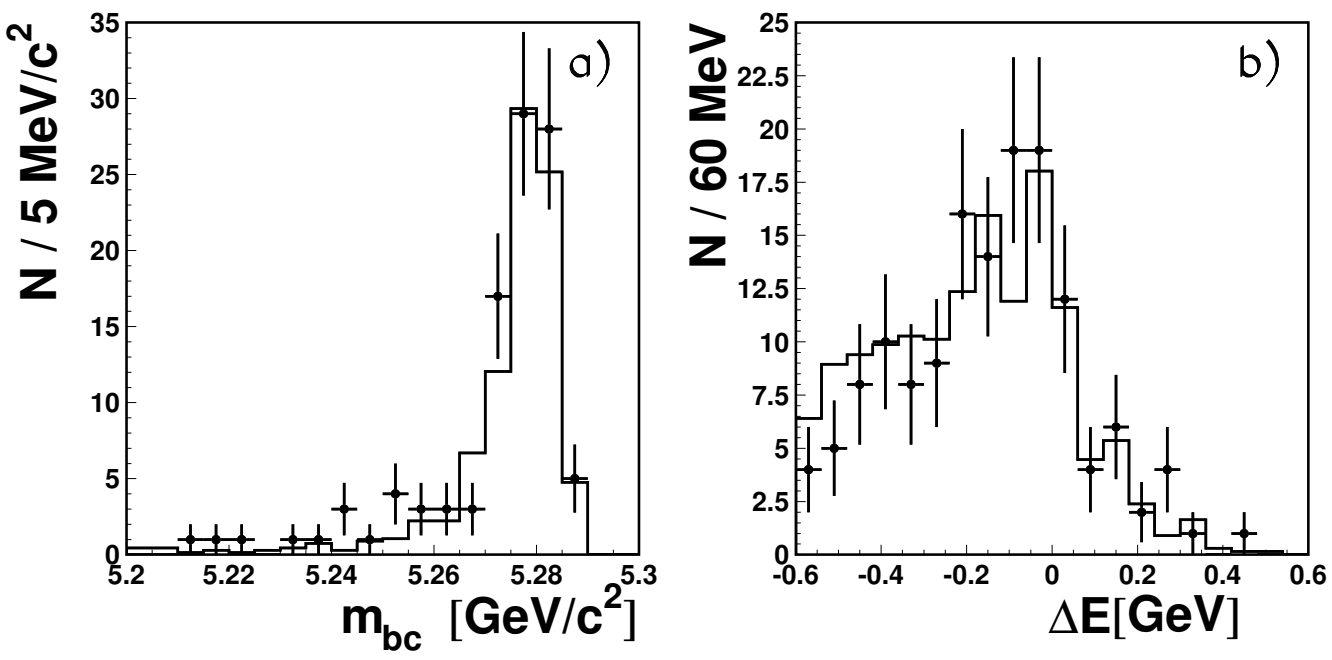

**Figure 7.4.4.** The $m_{\mathrm{ES}}$ $(= m_{\mathrm{bc}})$ (a) and $\Delta E$ (b) distributions for inclusively reconstructed $B_{\mathrm{tag}}$ using a $B^0 \to D^{*-}\pi^+$ recoil control sample from data (points with error bars) and MC (histograms) (Matyja, 2007). The $\Delta E$ ($m_{\mathrm{ES}}$) of $B_{\mathrm{tag}}$ candidates is required to be between $-0.25$ and 0.05 GeV (larger then 5.27 GeV/$c^2$) when plotting $m_{\mathrm{ES}}$ ($\Delta E$).

### 7.4.4 Double tagging

There are two assumptions made when using the recoil method: the first is that the $B$ reconstruction efficiency is well modeled by the Monte Carlo simulations of generic $B$ decays and continuum events. The hadronic $B_{\mathrm{tag}}$ reconstruction efficiencies defined in Eqs (7.4.4) and (7.4.3) depend on the decay rates of $B$-meson decays to final state included in the reconstruction. Some of them are poorly known and hence the $B_{\mathrm{tag}}$ reconstruction efficiencies determined on simulated samples need to be validated or calibrated using the real data sample. The second is that for analyses with few reconstructed particles from the signal $B$, the extra energy used to discriminate signal from background events is also well-modeled. These assumptions can be checked by using control samples which test both the tag $B$ reconstruction efficiency and the description of extra energy in a fully-reconstructed event. Both *BABAR* and Belle use double-tagged samples, in which both $B$ mesons are fully reconstructed either in semileptonic or hadronic final states, as such a control.

The crosscheck using the double-tag approach was first applied by *BABAR* (Aubert, 2004y), using double semileptonic $B$ decays. For the semileptonic $B_{\mathrm{tag}}$ technique described in Section 7.4.2 this means the reconstruction of two oppositely charged and non-overlapping $B \to D^0 \ell \nu_\ell X$ candidates with little other detector activity. Both *BABAR* and Belle have also used "hybrid double-tags", where one $B$ is reconstructed in a hadronic final state while the second $B$ is reconstructed in a semileptonic final state ($B \to D^{(*)} \ell \nu_\ell$). These samples vary in size, depending on the final states used, but given a semileptonic tag reconstruction efficiency (quoted by *BABAR*) of $\sim 0.7\%$ and a hadronic tag efficiency of $\sim 0.2\%$, one expects to find approximately 50 semileptonic double-tagged events per $\mathrm{fb}^{-1}$, 30 hybrid tags per $\mathrm{fb}^{-1}$, and 4 hadronic double-tagged events per $\mathrm{fb}^{-1}$. Given the large datasets of the $B$ Factories, and the expected dataset at future super flavor factories, these are significant samples which can be used as important cross-checks of the assumptions in the recoil method.

The double-tagged events have two important features. The first is that one expects naïvely the yield to be proportional to $\varepsilon^2_{\mathrm{tag}}$, which is the basis of the cross-check of the tag efficiency. The second is that the complete reconstruction of both $B$ mesons creates an environment in which the extra energy in a given event should represent the effect of energy deposits unassociated with the $B$ decays themselves. This latter feature is an important ingredient in the cross-check of the extra energy modeling in signal events, where it is also assumed that all detected particles associated with the $B$ decays have been reconstructed.

The cross-check of the tag efficiency is currently only used in the semileptonic approach, and only by *BABAR*. The early approach to the double-tag sample (Aubert, 2006a) made two assumptions. Given an efficiency, $\varepsilon_{\mathrm{tag}}$, for reconstructing one of the two $B$s in an event in a semileptonic final state, the number of double tags ($N_2$) is given simply by

$$N_2 = \varepsilon^2_{\mathrm{tag}} \times N_{B^+B^-} \tag{7.4.5}$$

where $N_{B^+B^-}$ is the number of charged $B$ pairs originally produced by the $B$ Factory or generated in Monte Carlo simulations. The tag efficiency cross-check was performed by taking the ratio of the above equation in data and in MC simulation and assuming that the double-tag sample is dominated by charged $B$ mesons so that $N_{B^+B^-}$ cancels, yielding the correction factor ($c_{\mathrm{tag}}$) for the tagging efficiency in MC,

$$c_{\mathrm{tag}} = \frac{\varepsilon^{\mathrm{data}}_{\mathrm{tag}}}{\varepsilon^{\mathrm{MC}}_{\mathrm{tag}}} = \sqrt{\frac{N^{\mathrm{data}}_2}{N^{\mathrm{MC}}_2}}. \tag{7.4.6}$$

While MC studies of the double-tags suggest that the contamination from neutral $B$ decays, or other backgrounds, is very small, the second assumption - that the reconstruction of the first $B$ does not bias the reconstruction of the second - is not addressed. The closeness of the correction to 1.0, as cited by *BABAR*, does suggest that also the second assumption is essentially correct.

A second approach to the efficiency correction attempts to address some of the potential deficiencies of the first method outlined above. In the alternative approach (Aubert, 2007a), the data/MC comparison is performed using the ratio of single-tagged to double-tagged events. If the efficiency of reconstructing the first tag is $\varepsilon_{\mathrm{tag},1}$ and the efficiency of reconstructing the second tag is $\varepsilon_{\mathrm{tag},2}$, then the single-tag and double-tag yields, $N_1$ and $N_2$, are given by

$$N_1 = \varepsilon_{\mathrm{tag},1} \times N_{B^+B^-} \tag{7.4.7}$$

$$N_2 = \varepsilon_{\mathrm{tag},1} \times \varepsilon_{\mathrm{tag},2} \times N_{B^+B^-}. \tag{7.4.8}$$

The ratio of the two cancels some of the common factors, yielding the following quantity to be determined in both data and MC simulations,

$$\varepsilon_{\mathrm{tag},2} = \frac{N_2}{N_1} \tag{7.4.9}$$

*BABAR* determines the number of single-tagged events by subtracting the combinatorial component under the $D^0$

mass distribution using an extrapolation of events from the $D^0$ mass sideband. This leaves a sample of events containing correctly reconstructed events, mis-reconstructed events from neutral $B$ semileptonic decay, and events from $e^+e^- \to c\bar{c}$ continuum background events with real $D^0$ mesons paired with a combinatorial lepton. The correction to the tag efficiency is assumed to be equal for either the first or second tag, and is computed from the data and MC as,

$$c_{\rm tag} = \frac{\varepsilon^{\rm data}_{\rm tag,2}}{\varepsilon^{\rm MC}_{\rm tag,2}} = \frac{N_2^{\rm data}/N_1^{\rm data}}{N_2^{\rm MC}/N_1^{\rm MC}} \tag{7.4.10}$$

The correction is computed using only events in which the $D^0$ meson in the first $B_{\rm tag}$ decays into the $K^-\pi^+$ final state. This is cross-checked using a sample in which the $D^0$ meson from the first tag decays into the $K^-\pi^+\pi^-\pi^+$ final state only, yielding complementary results.

In both of the above methods, and across several iterations of semileptonic recoil-based analyses, *BABAR* has found the correction to be very close to 1.0. This suggests both that the assumptions in the above two methods are largely accurate, and also that existing simulations of these and the background decays are adequate for the purposes of modeling the decays. The correction has an associated systematic error, which is typically determined by propagating the statistical uncertainty due to the finite sample sizes of the double-tag and single-tag samples. The uncertainty of the correction is about 4%.

Belle (Sibidanov, 2013) uses fully reconstructed events to calibrate the efficiency of the NB-based $B_{\rm tag}$ reconstruction. One of the produced $B$ mesons is reconstructed as hadronic $B_{\rm tag}$ while the other $B$ meson is reconstructed in the semileptonic decay mode $B_{sl} \to D^{(*)}\ell\nu$. The number of double tagged events is therefore given by:

$$N(B_{\rm tag}B_{sl}) = N_{B\overline{B}} \times \mathcal{B}(B_{\rm tag} \to f)\varepsilon_{B_{\rm tag}\to f} \times \mathcal{B}(B_{sl} \to D^{(*)}\ell\nu)\varepsilon_{B_{sl}}, \tag{7.4.11}$$

where $\mathcal{B}(B_{\rm tag} \to f)\varepsilon_{B_{\rm tag}\to f}$ is the product of branching fraction and reconstruction efficiency of the specific decay $B_{\rm tag} \to f$ and $\mathcal{B}(B_{sl} \to D^{(*)}\ell\nu)\varepsilon_{B_{sl}}$ is the corresponding product for the semileptonically decaying $B$ meson, which is well modeled in the simulation. The correction factor for $B_{\rm tag} \to f$ is then obtained by measuring the ratio of the numbers of reconstructed double tagged events in real data and MC samples

$$c^f_{\rm tag} = \frac{\mathcal{B}^{\rm data}(B_{\rm tag} \to f)\varepsilon^{\rm data}_{B_{\rm tag}\to f}}{\mathcal{B}^{\rm MC}(B_{\rm tag} \to f)\varepsilon^{\rm MC}_{B_{\rm tag}\to f}} = \frac{N^{\rm data}(B_{\rm tag}B_{sl})}{N^{\rm MC}(B_{\rm tag}B_{sl})} \cdot \frac{N^{\rm MC}_{B\overline{B}}\mathcal{B}^{\rm MC}(B_{sl} \to D^{(*)}\ell\nu)}{N^{\rm data}_{B\overline{B}}\mathcal{B}^{\rm data}(B_{sl} \to D^{(*)}\ell\nu)}. \tag{7.4.12}$$

In this method of the $B_{\rm tag}$ efficiency calibration it is assumed that the $B_{sl} \to D^{(*)}\ell\nu$ modes are well modeled in the MC sample and hence the $\varepsilon^{\rm data}_{B_{sl}} = \varepsilon^{\rm MC}_{B_{sl}}$. The overall correction factor (averaged over all $B_{\rm tag}$ modes) is found to be around 0.7 and consistent between different $B_{sl}$ decay modes. The total uncertainty of the calibration is estimated to be 4.2% for $B^+_{\rm tag}$ and 4.5% for $B^0_{\rm tag}$.

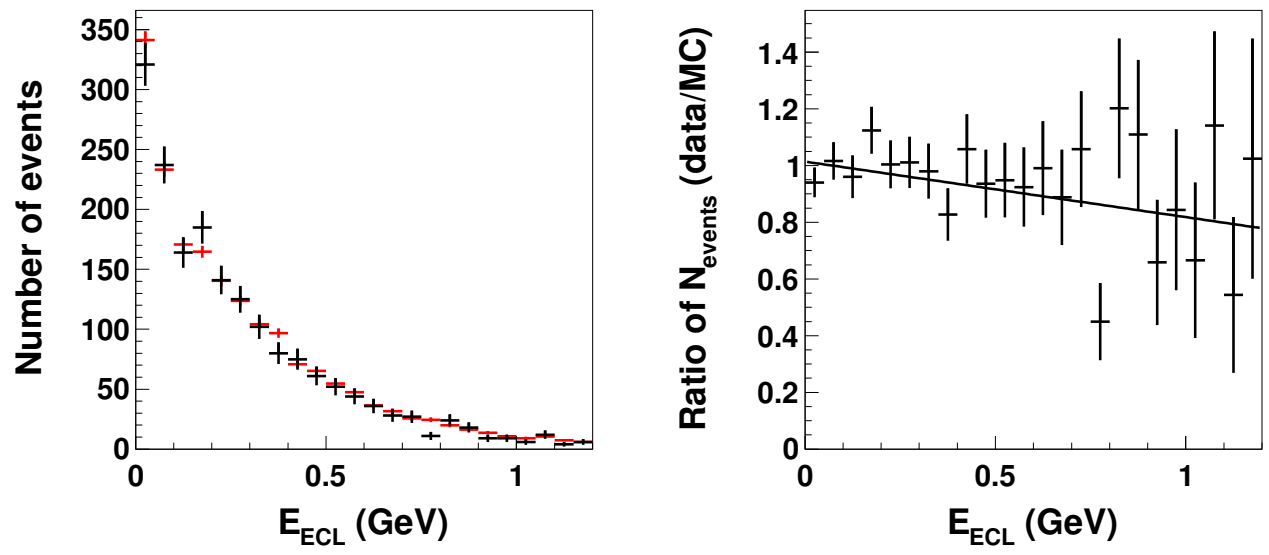

**Figure 7.4.5.** Extra energy distribution for double-tagged $B^+_{\rm tag}B^-_{sl}$ events (left plot), where the semileptonically decaying $B$ meson is reconstructed in the $D^{*0}\ell^-\overline{\nu}_\ell$ decay mode. Black and red data points show the distribution obtained in data and in a sample of simulated events, respectively. The right plot shows the ratio of the two distributions fitted with a linear function. Belle internal, from the Adachi (2012b) analysis.

The second application of the double-tagged sample is to test the modeling of extra particles left in the detector after both $B$ mesons have been reconstructed. In the case of signal events, this typically means that the tag $B$ is reconstructed up to any neutrinos in the final state (as in semileptonic tags), and that the signal $B$ is also reconstructed up to possible neutrinos in its final state. After reconstruction of both $B$ mesons the remaining particles left in the event are assumed to come from several sources: neutrals, such as photons, which arise from the electron-positron beams but not the interaction point; some low momentum charged particles associated with interactions between the beam and the beampipe; neutral clusters from hadronic showering in the calorimeter which fail to associate with a track; and detector noise. These sources would typically lead to a few extra neutral particles left in a signal event in about 20-30% of the reconstructed events.

Double-tagged events are used to test the simulation of these extra neutral particles by fully reconstructing both $B$ mesons either semileptonically, hadronically, or in a hybrid configuration. An example of the use of the double-tags to test the extra energy simulation is the Belle collaboration's hadronic-tagged search for $B^+ \to \tau^+\nu_\tau$. Belle constructs a hybrid double-tag sample (one hadronic $B$ and one semileptonic $B$ per event in the sample), and assumes that the extra neutral clusters remaining in these events comes from the same sources as in signal events. They compare the extra energy in data and MC (Fig. 7.4.5) and use the difference as a variation on their *p.d.f.* model for signal events. Comparisons show that existing detector simulations at the $B$ Factories handle the variety of sources of extra neutral clusters fairly well, even in moderate to high multiplicity final states of $B$ decay.

## 7.5 Summary

$B$-meson reconstruction is crucial for the broad physics program performed at Belle and *BABAR*. All of the techniques presented in this chapter utilize unique constraints provided by the experimental setup of $B$ Factories. They either improve the resolution (*e.g.* $m_{\mathrm{ES}}$ and $\Delta E$ versus $B$-meson invariant mass in full hadronic reconstruction), increase reconstruction efficiency (partial reconstruction) or make possible studies of $B$-meson decays with multiple neutrinos in the final state (recoil reconstruction). Some of the $B$ reconstruction methods presented herein were already used by experiments prior to Belle and *BABAR*. Others, in particular recoil techniques using fully- or semi-exclusive $B$-meson reconstruction, were pioneered in the $B$ Factories era and proved invaluable to access rare processes where the kinematics of the signal $B$ meson could not be fully constrained. Together with background discrimination (see Chapter 9) $B$ reconstruction techniques have been constantly improved over the past ten years which has enabled studies of less clean modes and increased sensitivity to rare decays.

# Chapter 8
# $B$-flavor tagging

***Editors:***
*Juerg Beringer (BABAR)*
*Kazutaka Sumisawa (Belle)*

**Additional section writers:**
*Robert Cahn, Simone Stracka*

## 8.1 Introduction

The goal of $B$-flavor tagging is to determine the flavor of a $B$ meson (*i.e.* whether it contains a $b$ or a $\bar{b}$ quark) at the time of its decay. At the $B$ Factories, flavor tagging is needed for most measurements of time-dependent *CP* asymmetries and $B$ meson mixing. As will be discussed in Chapter 10, these measurements usually require full reconstruction of the decay of one of the $B$ mesons (referred to as $B_{\rm rec}$ or "signal" $B$), measurement of the decay time difference $\Delta t$ between the two $B$ meson decays, and flavor tagging of the other $B$ meson (referred to as $B_{\rm tag}$ in the following).

At the $B$ Factories, in contrast to hadron colliders, $B$ meson pairs are produced in isolation (apart from any initial-state radiation), since there is no "underlying event" and the fraction of events with multiple $e^+e^-$ interactions ("pile-up") is negligible. Therefore, if a $B_{\rm rec}$ decay is fully reconstructed, the remaining tracks in the event can be assumed to come from the $B_{\rm tag}$ decay. In this case flavor tagging is to a good approximation independent of the specific $B_{\rm rec}$ decay mode reconstructed (but of course still depends on whether decays of $B^0/\overline{B}^0$, $B^+/B^-$ or, when running at the $\Upsilon(5S)$, $B^0_s/\overline{B}^0_s$ are tagged), and the flavor tagging performance can be measured using fully reconstructed flavor-specific $B_{\rm rec}$ decays. For inclusive reconstruction of the signal $B$, flavor tagging in general depends on the specific $B_{\rm rec}$ reconstruction since the remaining tracks in the event cannot be unambiguously assigned to either the $B_{\rm rec}$ or $B_{\rm tag}$ meson.

The tagging of neutral $B^0/\overline{B}^0$ mesons from $\Upsilon(4S)$ decays assuming a fully reconstructed $B_{\rm rec}$ decay is the primary use case for flavor tagging at the $B$ Factories. This is the situation considered in the following.

Flavor tagging relies on the fact that a large fraction of $B$ mesons decay to a final state that is flavor specific, *i.e.* to good approximation, can only be reached either through the decay of a $b$ quark, or through the decay of a $\bar{b}$ quark. Because of the large number of decay channels, full reconstruction of a sufficiently large number of flavor-specific $B_{\rm tag}$ decays is not feasible. Instead inclusive techniques are employed that make use of different flavor-specific signatures of $B$ decays. For example, in semileptonic decays $B^0 \to D^{*-}\ell^+\nu_\ell$ the charge of the lepton unambiguously identifies the flavor of the decaying $B$ meson as long as the lepton can be clearly associated with the semileptonic $B$ decay and does not come from a secondary $D$ meson decay.

The flavor tagging algorithms developed by *BABAR* and Belle proceed in two stages. In the first stage, individual flavor-specific signatures are analyzed, each of which provides a signature-specific flavor tag that by itself could be used for flavor tagging. In the second stage, the results from the first stage signatures are combined into a final flavor tag. Both stages rely on multivariate methods in order to optimally combine all available information.

The outline of this chapter is as follows. After defining the relevant quantities characterizing the performance of $B$-flavor tagging and discussing the choice of tagging categories, the different sources of flavor information and the corresponding discriminating variables are reviewed. Section 8.6 describes the specific flavor tagging algorithms used by the *BABAR* and Belle experiments and quotes the performance of these algorithms. The method used to measure the flavor tagging performance is described elsewhere (see Section 10.6).

## 8.2 Definitions

The figure of merit for the performance of a tagging algorithm is the effective tagging efficiency $Q$,

$$Q = \varepsilon_{\rm tag}(1-2w)^2, \quad (8.2.1)$$

where $\varepsilon_{\rm tag}$ denotes the fraction of events to which a flavor tag can be assigned, and the mistag probability $w$ is the fraction of events with an incorrectly assigned tag. The term

$$D = 1 - 2w \quad (8.2.2)$$

is called the dilution and is the factor by which measured *CP* and mixing asymmetries are reduced from their physical values due to incorrectly assigned flavor tags. The definition of $Q$ is motivated by the fact that the statistical uncertainties $\sigma$ on such asymmetry measurements generally scale approximately as (see Section 8.4)

$$\sigma \propto \frac{1}{\sqrt{Q}}. \quad (8.2.3)$$

Tagging efficiencies and mistag fractions are not a priori the same for tagging $B^0$ and $\overline{B}^0$ decays because the detector performance may not be completely charge symmetric. Therefore the averages

$$\varepsilon_{\rm tag} = \frac{\varepsilon_{B^0} + \varepsilon_{\overline{B}^0}}{2} \quad (8.2.4)$$

$$w = \frac{w_{B^0} + w_{\overline{B}^0}}{2} \quad (8.2.5)$$

and differences

$$\Delta\varepsilon_{\rm tag} = \varepsilon_{B^0} - \varepsilon_{\overline{B}^0} \quad (8.2.6)$$

$$\Delta w = w_{B^0} - w_{\overline{B}^0} \quad (8.2.7)$$

are defined where the subscript refers to the true decay. For example, $w_{B^0}$ refers to the fraction of neutral $B_{\rm tag}$ mesons that decay as $B^0$ but are tagged as $\overline{B}^0$.

## 8.3 Tagging categories

The effective tagging efficiency $Q$ can be improved (and hence the statistical uncertainty of a measurement decreased) by grouping events into mutually exclusive tagging categories according to their mistag probabilities $w$ (or dilutions $D$). For tagging categories $c$ with fractions of events $\varepsilon_c$, dilutions $D_c$, total tagging efficiency $\varepsilon = \sum_c \varepsilon_c$ and average dilution $D = \sum_c \varepsilon_c D_c / \varepsilon$ one finds

$$Q = \sum_c \varepsilon_c D_c^2 = \varepsilon D^2 + \sum_c \varepsilon_c (D_c - D)^2. \tag{8.3.1}$$

Thus the resulting $Q$ is always larger or equal to the one obtained when all events are treated as a single category. One gains most from dividing events into categories when the differences in dilution (or mistag fraction) between categories can be made large. However, the characteristics and any systematic effects, such as correlations with the tag vertex resolution, tag-side interference (see Section 15.3.6), or background levels, are expected to be determined by the different flavor-specific signatures. For this reason one would prefer a grouping of events according to different signatures over a category definition based on $w$.

The mistag probability $w$ that can be achieved for a given set of $B_{\rm tag}$ decay modes is determined by the flavor-specific signatures present in these decays. Fortunately, the mistag probabilities of different flavor-specific signatures tend to be different. For example, in semileptonic decays the charge of a reconstructed high-momentum electron or muon gives a much better indication of the correct tag than the charge of a low momentum pion ("slow pion") from a secondary $D^*$ decay.

Therefore a grouping of events into tagging categories according to the mistag probability naturally provides a grouping according to the different signatures of the corresponding $B_{\rm tag}$ decays. Conversely, a grouping according to different signatures leads to an approximate grouping according to mistag probabilities. As a result it is possible to define tagging categories that both optimize the tagging performance and group events according to different signatures.

## 8.4 Dilution factor and effective tagging efficiency

As mentioned above, a *CP* asymmetry $A^{\rm rec}$ measured using flavor tagging is reduced from the physical asymmetry by a factor $D$ due to incorrectly assigned flavor tags. This scaling is easy to see by writing the measured asymmetry $A^{\rm rec}$ as

$$A^{\rm rec} = \frac{N - \overline{N}}{N + \overline{N}}, \tag{8.4.1}$$

where $N$ and $\overline{N}$ denote the number of reconstructed $B$ decays

$$N = \varepsilon_{\rm tag}(1-w)N_0 + \varepsilon_{\rm tag} w \overline{N}_0$$
$$\overline{N} = \varepsilon_{\rm tag}(1-w)\overline{N}_0 + \varepsilon_{\rm tag} w N_0 \tag{8.4.2}$$

tagged as $B^0$ and $\overline{B}^0$, respectively. $N_0$ and $\overline{N}_0$ are the corresponding number of reconstructed $B$ decays of a certain type before tagging is applied. Substituting Eq. (8.4.2) into (8.4.1) one directly obtains

$$A^{\rm rec} = (1-2w)A^0 = DA^0, \tag{8.4.3}$$

where $A^0 = (N_0 - \overline{N}_0)/(N_0 + \overline{N}_0)$ denotes the true physical asymmetry.

The statistical uncertainty in $A^0$ is

$$\sigma_{A^0} = \frac{\sigma_{A^{\rm rec}}}{1-2w} \ . \tag{8.4.4}$$

Using Eq. (8.4.1) and denoting the total number of tagged events by $N_{\rm tag} = N + \overline{N}$, assuming a small asymmetry (*i.e.* $N \approx \overline{N} = N_{\rm tag}/2$), one finds

$$\sigma_{A^{\rm rec}} \propto \frac{1}{\sqrt{N_{\rm tag}}} \ . \tag{8.4.5}$$

Together with Eq. (8.4.4) it follows

$$\sigma_{A^0} \propto \frac{1}{\sqrt{\varepsilon_{\rm tag}}(1-2w)} = \frac{1}{\sqrt{Q}} \ . \tag{8.4.6}$$

In general, this scaling of $\sigma_{A^0}$ with $Q$ is only approximate. For a likelihood-based analysis and assuming a sufficiently large number of events, the expected uncertainty in an estimated *CP* or mixing asymmetry $\widehat{A}$ can be obtained from the maximum-likelihood estimator for the variance on $\widehat{A}$ (see Section 11.1.3),

$$\sigma(\widehat{A})^2 = V(\widehat{A}) = \left(\frac{d^2 \log(L(A))}{d^2 A}\right)^{-1}_{A=\widehat{A}} . \tag{8.4.7}$$

This was calculated (Cahn, 2000; Le Diberder, 1990) for the case of a measurement of a time-dependent *CP* asymmetry with no direct *CP* violation such as *e.g.* the measurement of $A = \sin 2\phi_1$. Using several tagging categories $c$, ignoring effects of resolution and background, and with $x_d = \Delta m / \Gamma$, the approximation

$$\sigma(\widehat{A}) \approx \left[ N \frac{2x_d^2}{1+4x_d^2} \sum_c \epsilon_c D_c^2 \left(1 + \frac{12 x_d^2 D_c^2 A^2}{1+16x_d^2}\right)\right]^{-1/2} , \tag{8.4.8}$$

was derived. This leads to an improved definition $Q'$ of the effective tagging efficiency,

$$Q' = \sum_c \varepsilon_c D_c^2 \left(1 + \frac{12 x_d^2 D_c^2 A^2}{1+16x_d^2}\right) \tag{8.4.9}$$

with $\sigma(\widehat{A}) \propto 1/\sqrt{Q'}$.

$Q'$ depends on the true asymmetry $A$ and reduces to the standard definition of $Q$ for $A = 0$. For large asymmetries ($A \approx 1$) and for the most powerful tagging categories used by the *BABAR* or Belle tagging algorithms with $w_c \approx 2\%$, the factor

$$1 + \frac{12 x_d^2 D_c^2 A^2}{1+16x_d^2} \tag{8.4.10}$$

amounts to a correction of more than 60%. This effect was clearly observed when the scaling of the uncertainties of different *BABAR* $\sin 2\phi_1$ results with effective tagging efficiency was analyzed.

## 8.5 Physics sources of flavor information

In the following the different flavor-specific signatures are discussed in more detail. Since the focus of this chapter is on tagging for fully reconstructed $B_{\rm rec}$ decays, it is assumed that only tracks from the $B_{\rm tag}$ decays are considered in the calculation of any of the discriminating variables described below.

### 8.5.1 Leptons

Electrons and muons produced directly in semileptonic $B$ decays (primary leptons) provide excellent tagging information. The charge of a lepton from a $b \to c\,\ell^-\,\overline{\nu}$ transition is directly associated to the flavor of the $B^0$ meson: a positively charged lepton indicates a $B^0$, a negatively charged lepton indicates a $\overline{B}^0$.

Leptons from cascade decays (secondary leptons) occurring via the transition $b \to W^- c\,(\to s\,\ell^+\,\nu)$ carry tagging information as well: their charge is opposite to that of primary leptons from $B_{\rm tag}$ and they are characterized by a much softer momentum spectrum.

The following kinematical variables are useful to identify primary and secondary leptons:

- $q$, the charge of the track.
- $p^*$, the center-of-mass momentum of the candidate track. Combined with the charge of the track this is the most powerful discriminating variable.
- $\theta_{\rm lab}$, the polar angle in the laboratory frame.
- $E_{90}^W$, the energy in the hemisphere defined by the direction of the virtual $W^\pm$ in the semi-leptonic $B_{\rm tag}$ decay. $E_{90}^W$ is calculated in the center-of-mass frame under the assumption that the $B_{\rm tag}$ is produced at rest. The sum of energies for $E_{90}^W$ extends over all charged and neutral candidates of the recoiling charm system $X$ that are in the same hemisphere (with respect to the direction of the virtual $W^\pm$) as the lepton candidate:

$$\begin{aligned} p_B^\mu &= p_W^\mu + p_X^\mu \approx (m_{B^0}, \mathbf{0}) \\ p_W^\mu &= p_\ell^\mu + p_\nu^\mu \\ p_X^\mu &= \sum_{i \neq \ell} p_i^\mu \\ E_{90}^W &= \sum_{i \in X,\ \boldsymbol{p}_i \cdot \boldsymbol{p}_W > 0} E_i \end{aligned} \tag{8.5.1}$$

- $\boldsymbol{p}_{\rm miss}$, the missing momentum given by:

$$\boldsymbol{p}_{\rm miss} = \boldsymbol{p}_B - \boldsymbol{p}_X - \boldsymbol{p}_\ell \approx -(\boldsymbol{p}_X + \boldsymbol{p}_\ell). \tag{8.5.2}$$

- $\cos\theta_{\rm miss}$, the cosine of the angle between the lepton candidate's momentum $\boldsymbol{p}_\ell$ and the missing momentum $\boldsymbol{p}_{\rm miss}$ is calculated in the $\Upsilon(4S)$ center-of-mass frame (again with the approximation of the $B_{\rm tag}$ being produced at rest).
- $M_{\rm recoil}$, mass recoiling against $\boldsymbol{p}_{\rm miss} + \boldsymbol{p}_\ell$ in the $B_{\rm tag}$ frame. The $M_{\rm recoil}$ distribution for semileptonic $B$ decays peaks around the $D$ mass and has a tail toward the lower side due to missing particles, while that for semileptonic $D$ decays is more broad with a tail up to 5 GeV$/c^2$.

The above kinematical variables can be combined with particle identification (PID) information and applied only to selected electron or muon candidate tracks. Or they can be applied to all tracks in order to recover the tagging information from leptons that fail the PID selection ("kinematically identified leptons").

### 8.5.2 Kaons

The dominant source of charged kaons are $b \to c \to s$ transitions ($B^0 \to \overline{D}(\to K^+X')X$ decays), where the charge of the kaon tags the flavor of $B_{\rm tag}$. Kaons from such decays are referred to as "right sign" kaons (a $K^+$ indicates a $B^0$ decay). The high average multiplicity of charged kaons of $0.78 \pm 0.08$ (Beringer et al., 2012), combined with the higher multiplicity of right sign vs wrong sign kaons of $0.58 \pm 0.01 \pm 0.08$ vs. $0.13 \pm 0.01 \pm 0.05$ (Albrecht et al., 1994b) make kaons overall the most powerful source of tagging information.

The following discriminating variables are useful for flavor tagging with kaons:

- $q$, the charge of the track.
- $\mathcal{L}_K$, the kaon likelihood obtained from PID information.
- If more than one charged kaon is identified, it is useful to combine the information $(q \cdot \mathcal{L}_K)$ from up to three charged kaons.
- $n_{K_S^0}$, the number of $K_S^0$ mesons reconstructed on the tag side. A kaon produced together with one or more $K_S^0$ tends to originate from a strange quark in a $b \to c\overline{c}(d,s)$ decay or from the appearance of $s\overline{s}$ out of the vacuum, while one without an accompanying $K_S^0$ has a higher probability to come from the $b \to c \to s$ cascade decay.
- The sum of the squared transverse momenta of charged tracks on the tag side. A large total transverse momentum squared increases the likelihood that a charged kaon was produced from a $b \to cW^-, c \to s \to K^-$ transition, rather than the transition $b \to XW^-, W^- \to \overline{c}s/d, \overline{c} \to \overline{s} \to K^+$, which would give a "wrong-sign" kaon.
- $p^*$, the center-of-mass momentum of the candidate track.
- $\theta_{\rm lab}$, the polar angle in the laboratory frame.

### 8.5.3 Slow pions

Low momentum $\pi^\pm$ from $D^{*\pm}$ decays (slow pions) provide another source of tagging information. The substan-tial background from low momentum tracks can be reduced by correlating the direction of the slow pion and the remaining tracks from the $B_{\rm tag}$ decay. Since the slow pion and the $D^0$ are emitted nearly at rest in the $D^{*\pm}$ frame, the slow pion direction in the $B_{\rm tag}$ rest frame will be along the direction of the $D^0$ decay products and opposite to the remainder of the $B_{\rm tag}$ decay products. This direction can be approximately determined by calculating the thrust axis of the $B_{\rm tag}$ decay products. The thrust is calculated using both charged tracks and neutral clusters not used in the reconstruction of $B_{\rm rec}$.

The following variables provide useful discriminating power:

- $q$, the charge of the track.
- $p^*$, the momentum of the slow pion candidate in the $\Upsilon(4S)$ center-of-mass frame.
- $p^{\rm lab}$, the momentum of the slow pion candidate in the laboratory frame.
- $\theta_{\rm lab}$, the polar angle in the laboratory frame.
- $\cos\theta_{\pi{\rm T}}$, the cosine of the angle between the slow pion direction and the $B_{\rm tag}$ thrust axis in the $\Upsilon(4S)$ center-of-mass frame.
- $\mathcal{L}_K$, the PID likelihood of the track to be a kaon. PID information helps to reject the contribution from low momentum kaons flying in the thrust direction.
- $\mathcal{L}_e$, the PID likelihood of the track to be a electron. This helps to reject background from electrons produced in photon conversions and $\pi^0$ Dalitz decays.

### 8.5.4 Correlation of kaons and slow pions

In events where both a charged kaon and a slow pion candidate (*e.g.* from a $D^{*+} \to D^0(\to K^-X)\pi^+$ decay) are found, the corresponding flavor tagging information can potentially be improved by using the angular correlation between the kaon and slow pion. A kaon and a slow pion of opposite charge (*i.e.* agreeing flavor tag) that are emitted in approximately the same direction in the $\Upsilon(4S)$ center-of-mass frame can provide a combined tag with a relatively low mistag fraction.

In addition to the information used to identify kaons and slow pions, the following discriminating variable can be used:

- $\cos\theta_{K,\pi}$, the cosine of the angle between the kaon and the slow pion momentum calculated in the $\Upsilon(4S)$ center-of-mass frame.

### 8.5.5 High-momentum particles

A very inclusive tag can be obtained by selecting tracks with the highest momentum in the $\Upsilon(4S)$ center-of-mass frame and using their charge as a tag. Given the other signatures discussed above, the aim of such a tag is to identify fast particles coming from the hadronization of the $W$ boson produced in the decay $b \to c\,W^-$ (for example fast pions from $\overline{B}{}^0 \to D^{*+}\,\pi^-$) as well as high momentum leptons that may have failed the selection for the lepton tag signature. Direct hadrons or leptons with a positive (negative) charge indicate a $B^0$ ($\overline{B}{}^0$) tag. These particles are produced at the $B_{\rm tag}$ decay vertex and, in the $\Upsilon(4S)$ center-of-mass frame, are energetic and fly in a direction opposite to the charm decay products of $B_{\rm tag}$.

Useful discriminating variables are:

- $q$, the charge of the track.
- $p^*$, the momentum of the track in the $\Upsilon(4S)$ center-of-mass frame.
- $d_0$, the impact parameter in the $xy$ plane.
- The angle between the particle and the $B_{\rm tag}$ thrust axis in the $\Upsilon(4S)$ center-of-mass frame.

### 8.5.6 Correlation of fast and slow particles

The angular correlations between slow charged pions from $D^{*\pm}$ decays and fast, oppositely charged particles originating from the $W^\mp$ hadronization in the decay $b \to c\,W$ can be exploited for flavor tagging. Since the $W^\mp$ and the $D^{*\pm}$ are emitted back-to-back in the $B_{\rm tag}$ center-of-mass frame, the slow pion and the fast tracks are expected to be emitted at a large angle.

The following discriminating variables are useful:

- $p^*_{\rm Slow}$, the center-of-mass momentum of the slow track.
- $p^*_{\rm Fast}$, the center-of-mass momentum of the fast track.
- $\cos\theta_{\rm SlowFast}$, the cosine of the angle between the slow and the fast track.
- $\cos\theta_{\rm SlowT}$, the cosine of the angle between the slow track and $B_{\rm tag}$ thrust axis.
- $\cos\theta_{\rm FastT}$, the cosine of the angle between the fast track and $B_{\rm tag}$ thrust axis.
- $\mathcal{L}_{K{\rm Slow}}$, the PID likelihood for the slow track to be a kaon.

### 8.5.7 $\Lambda$ baryons

The flavor of a $\Lambda$ baryon produced in $B_{\rm tag}$ decays carries tagging information because it contains an $s$ quark that was likely produced in the cascade decay $b \to c \to s$. Therefore, the presence of a $\Lambda$ ($\overline{\Lambda}$) will indicate a $\overline{B}{}^0$ ($B^0$).

$\Lambda \to p\pi$ decays on the tag side are reconstructed by combining charged tracks with tracks that are identified as protons (or antiprotons). Although $\Lambda$ candidates are found in a small fraction of events, they provide relatively clean flavor tags that are fully complementary to the other signatures.

Useful discriminating variables include:

- $q$, the flavor of $\Lambda$ ($\Lambda$ or $\overline{\Lambda}$).
- $M_\Lambda$, the reconstructed mass of the $\Lambda$.
- $\chi^2_\Lambda$, the $\chi^2$ probability of the fitted $\Lambda$ decay vertex.
- $\cos\theta_\Lambda$, the cosine of the angle between the $\Lambda$ momentum and the direction from the primary vertex to the $\Lambda$ decay vertex.
- $s_\Lambda$, the flight length of the $\Lambda$ candidate before decay.
- $p_\Lambda$, the momentum of the $\Lambda$ candidate.
- $p_{\rm proton}$, the momentum of the proton candidate used for the $\Lambda$ reconstruction.

– $n_{K_S^0}$, the number of $K_S^0$ mesons reconstructed on the tag side.
– $\Delta z$, difference between the $z$ coordinate of the two tracks at the $\Lambda$ vertex point.

## 8.6 Specific flavor tagging algorithms

In this section the tagging algorithms developed by *BABAR* and Belle are discussed. In both experiments these algorithms have been improved greatly during the lifetime of the experiment, resulting in a substantial performance increase. In the following only the final versions of the tagging algorithms are discussed.

### 8.6.1 Multivariate tagging methods

The *BABAR* and Belle tagging algorithms both use multivariate methods: *BABAR* uses an artificial neural network, while Belle's tagger is based on a multi-dimensional look-up table. Both algorithms provide not only a flavor tag but also an estimated mistag probability for each event.

Both tagging algorithms were trained using large samples of simulated events. Imperfections in the simulation of particle decays (*e.g.* due to incomplete knowledge of branching fractions) or detector response may lead to inaccurate estimates of the per-event mistag probability by the tagging algorithm. Therefore both algorithms use the estimated per-event mistag probabilities only when separating events into tagging categories. For each category, $w$ and $\Delta w$ are measured using a sample of events where the signal $B$ decays into a self-tagging decay mode ($B_{\rm flav}$ control sample, see Section 10.6). As a result, inaccuracies in the simulation of the training sample can only lead to a non-optimal tagging performance but will not introduce any systematic errors. The loss in tagging performance that results from using tagging categories rather than per-event mistag probabilities was found to be small both for the *BABAR* and the Belle tagging algorithms.

### 8.6.2 Systematic effects

Systematic effects associated with tagging are discussed in Chapter 15; only a brief overview is given here. As discussed above, by using tagging categories whose $w$ and $\Delta w$ are measured on data, systematic effects that could arise from imperfections in the tagging algorithm or its training are replaced by the statistical uncertainties of the measurements of $w$ and $\Delta w$. The remaining systematic effects associated with flavor tagging arise from

– potential differences in the tagging performance for signal events and for the $B_{\rm flav}$ control sample used to measure $w$ and $\Delta w$, and
– tag-side interference (see Section 15.3.6).

### 8.6.3 Flavor tagging in BABAR

The *BABAR* tagging algorithm (Aubert, 2005i, 2009z; Lees, 2013c) is a modular, multivariate flavor-tagging algorithm that analyses charged tracks on the tag side in order to provide a flavor tag and a mistag probability $w$. The flavor of $B_{\rm tag}$ is determined from a combination of nine different flavor-specific signatures, which include charged leptons, kaons, pions and $\Lambda$ baryons (see Section 8.5).

For each of these signatures, properties such as charge, momentum, and decay angles are used as input to a specific neural network (NN) or "sub-tagger". Three sub-taggers are dedicated to charged leptons, making use of identified electrons (Electron), muons (Muon) and kinematically identified leptons (Kin. Lepton). The Kaon sub-tagger combines the information from up to three kaons into a single tag. Slow pions are used both by a dedicated slow pion sub-tagger (Slow Pion) and in correlation with kaons (K-Pi). The Max p* sub-tagger analyzes high-momentum particles. The correlation of fast and slow particles is exploited by the FSC sub-tagger. The Lambda sub-tagger looks at $\Lambda$ baryons.

These sub-taggers are combined by a single final neural network (BTagger) that is trained to determine the correct flavor of $B_{\rm tag}$. Based on the output of this NN and the contributing sub-taggers, each event is assigned to one of six mutually exclusive tagging categories. The overall structure of the *BABAR* tagging neural network is shown in Figure 8.6.1.

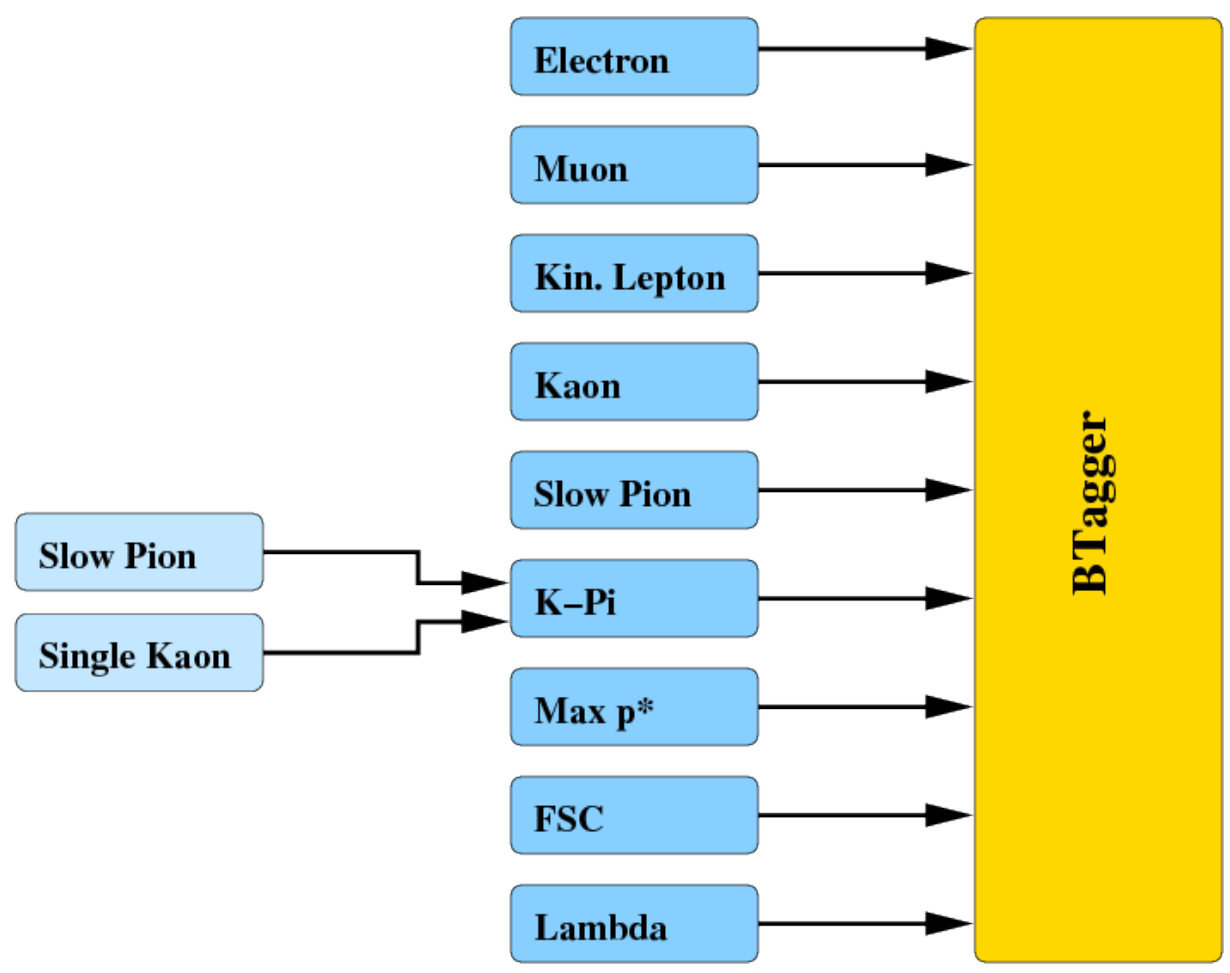

**Figure 8.6.1.** Schematic overview of the *BABAR* tagging algorithm. Each box corresponds to a separate neural network.

The use of sub-taggers dedicated to specific signatures allows one to keep track of the underlying physics of each event and simplifies studies of systematics. For example, events with an identified electron or muon from a semi-leptonic $B_{\rm tag}$ decay can be separated from other decays and assigned to the `Lepton` tagging category. The `Lepton` category does not only have a low $w$ but also more precisely reconstructed $B_{\rm tag}$ vertices, is less sensi-tive to the bias from charm on the tag side, and is immune to the intrinsic mistagging associated with doubly Cabibbo-suppressed decays (see tag-side interference in Section 15.3.6).

The training and validation of each of the sub-tagger NNs is based on the Stuttgart Neural Network Simulator (Zell et al., 1995). Extensive studies have been performed for each sub-tagger, including a wide search for the most discriminating input variables. NN architectures and the number of training cycles are optimized to yield the most efficient flavor assignment. The NNs are feed-forward networks with one hidden layer. The weights and bias values of the logistic activation functions are optimized during training using standard back-propagation.

The NNs are trained using a simulated sample of about 500,000 $B^0\overline{B}^0$ pairs in which one meson ($B_{\rm rec}$ ) decays to a $\pi^+\pi^-$ final state while the other ($B_{\rm tag}$) decays to any possible final state according to known or expected branching fractions. Half of this sample is used for training the NN, while the other half is used as a test sample for an unbiased evaluation of the performance. Each sub-tagger is trained separately before the training of the BTagger network.[36]

Details of the architecture of the different neural networks used by the *BABAR* tagging algorithm are given in Table 8.6.1. For each of the nine sub-taggers and for the final BTagger NN the table lists all input variables and the training target. Some of the sub-taggers are trained to separate $B^0$ from $\overline{B}^0$ decays, while others are trained to discriminate true from fake signatures.

The output $y_{\rm BTagger}$ of the final BTagger NN is mapped to values between $-1$ (for a perfectly tagged $\overline{B}^0$) and $+1$ ($B^0$). The distribution of this output for the $B_{\rm flav}$ control sample is shown in Figure 8.6.2. Excellent agreement is observed between data and simulation.

The estimated probability $p$ of a correct tag assignment is given by the BTagger NN output

$$p = 1 - w = (1 + |y_{\rm BTagger}|)/2, \qquad (8.6.1)$$

and the probability of a given $B_{\rm tag}$ being a $B^0$ is

$$p_{B_{\rm tag}=B^0} = (1 + y_{\rm BTagger})/2. \qquad (8.6.2)$$

The correctness of these probabilities can be checked with the $B_{\rm flav}$ control sample. For example, one can plot the probability of observing a $\overline{B}^0$ on the $B_{\rm flav}$ side as a function of the estimated probability $p_{B_{\rm tag}=B^0}$. Taking into account the time-integrated mixing probability $\chi_d = 0.1862 \pm 0.0023$ (Beringer et al. (2012)), one expects for a perfectly trained tagging algorithm

$$\begin{aligned} p_{B_{\rm flav}=\overline{B}^0} &= (1-2\chi_d)p_{B_{\rm tag}=B^0} + \chi_d \qquad (8.6.3)\\ &= (1-2\chi_d)(1+y_{\rm BTagger})/2 + \chi_d. \qquad (8.6.4)\end{aligned}$$

[36] Simultaneous training of all sub-taggers and the BTagger NN has been shown not to result in a significantly better classification performance.

As can be seen from Figure 8.6.3, the probabilities obtained from the BTagger NN output are in very good agreement with the expectations for both data and simulation. Nevertheless, as discussed in Section 8.6.1, these estimated probabilities are only used to separate events into tagging categories.

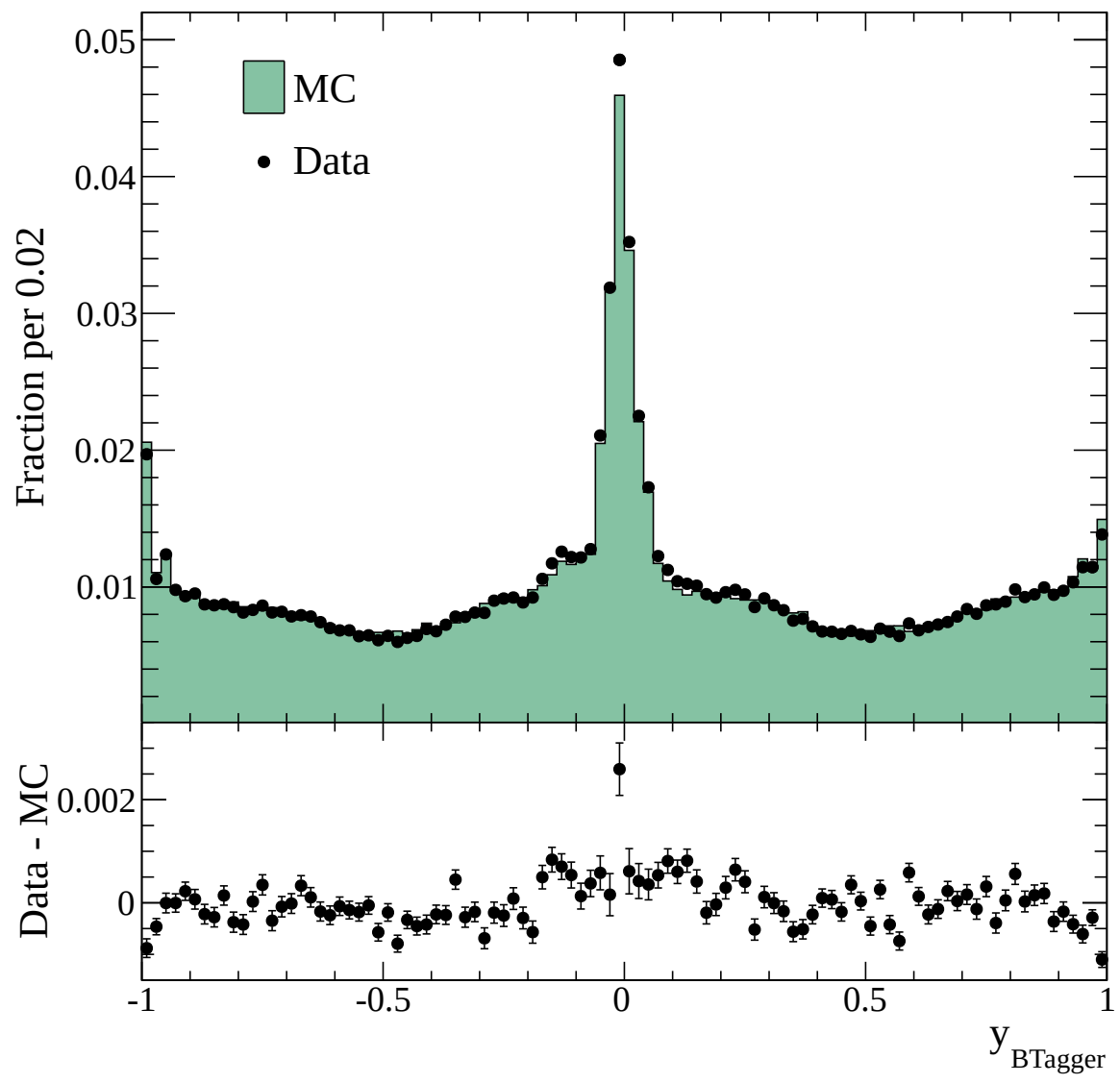

**Figure 8.6.2.** Distribution of the output of the final BTagger NN ($y_{\rm BTagger}$) on the $B_{\rm flav}$ control sample for data and simulation, using the full *BABAR* data sample. A contribution of up to 22% from combinatorial background is subtracted in each bin based on a fit to the $m_{\rm ES}$ distribution. The difference between data and simulation (with statistical uncertainties added in quadrature) is also shown.

The tagging algorithm assigns each event to one of six hierarchical and mutually exclusive tagging categories: `Lepton`, `Kaon I`, `Kaon II`, `Kaon-Pion`, `Pion` or `Other`. The name given to each category indicates the dominant physics processes (or sub-tagger) contributing to the flavor identification. For most categories, this classification is based on $y_{\rm BTagger}$. For the `Lepton` category, which singles out events with a cleanly identified primary lepton, additional cuts are made on the output of the electron or muon sub-taggers. Over 95% of events in the `Lepton` category contain a semileptonic $B_{\rm tag}$ decay. The definition of the tagging categories is summarized in Table 8.6.2.

The final version of the *BABAR* tagging algorithm[37] (Lees, 2013c) achieves an effective tagging efficiency $Q = (33.1 \pm 0.3)\%$ on the full *BABAR* data set. The breakdown of this performance into the different tagging categories is shown in Table 8.6.3.

[37] Improvements in the particle identification algorithms used for the final version of the *BABAR* tagging algorithm (Lees, 2013c) lead to a higher Q value of $(33.1 \pm 0.3)\%$, compared to $Q \approx 31\%$ achieved by the previous version (Aubert, 2005i). The tagging algorithm itself did not change.

**Table 8.6.1.** Overview of the neural networks used by the *BABAR* BTagger and its sub-taggers. For each sub-tagger the network architecture is shown in the second column according to the notation $N_{\text{inputs}} : N_{\text{hidden nodes}} : N_{\text{outputs}}$. The input variables are listed in the third column while the fourth column describes the goal of the NN training.

| (Sub-)Tagger | Network architecture | Discriminating input variables | Training goal |
|---|---|---|---|
| Electron | 4:12:1 | $q$, $p^*$, $E_{90}^W$, $\cos\theta_{\text{miss}}$ | Classify $B^0$ versus $\overline{B}^0$ |
| Muon | 4:12:1 | $q$, $p^*$, $E_{90}^W$, $\cos\theta_{\text{miss}}$ | Classify $B^0$ versus $\overline{B}^0$ |
| Kin. Lepton | 3:3:1 | $p^*$, $E_{90}^W$, $\cos\theta_{\text{miss}}$ | Recognize primary leptons |
| Kaon | 5:10:1 | $(q\mathcal{L}_K)_1$, $(q\mathcal{L}_K)_2$, $(q\mathcal{L}_K)_3$, $n_{K_S^0}$, $\Sigma p_\perp$ | Classify $B^0$ versus $\overline{B}^0$ |
| Slow Pion | 3:10:1 | $p^*$, $\cos\theta_{\pi\text{T}}$, $\mathcal{L}_K$ | Recognize slow pions from $D^{*\pm}$ decays |
| Max $p^*$ | 3:6:1 | $p^*$, $d_0$, $\cos\theta$ | Recognize direct $B$ daughters |
| K–Pi | 3:10:1 | $(q\mathcal{L}_K)$, SlowPion tag, $\cos\theta_{K,\pi}$ | Recognize $K$-$\pi$ pairs from $D^{*\pm}$ decays |
| FSC | 6:12:1 | $\cos\theta_{\text{SlowFast}}$, $p^*_{\text{Slow}}$, $p^*_{\text{Fast}}$, $\cos\theta_{\text{SlowT}}$, $\cos\theta_{\text{FastT}}$, $\mathcal{L}_{K\text{Slow}}$ | Recognize fast-slow correlated tracks |
| Lambda | 6:14:1 | $M_\Lambda$, $\chi^2$, $\cos\theta_\Lambda$, $s_\Lambda$, $p_\Lambda$, $p_{\text{proton}}$ | Recognize $\Lambda$ decays |
| BTagger | 9:20:1 | All of the above tags | Classify $B^0$ versus $\overline{B}^0$ |

**Table 8.6.2.** Definition of tagging categories for the *BABAR* flavor tagging algorithm. Events with $|y_{\text{BTagger}}| < 0.1$ are classified as `Untagged` and are not used to extract time-dependent information from data.

| Category | Definition |
|---|---|
| `Lepton` | $(\lvert y_{\text{Electron}}\rvert > 0.8$ or $\lvert y_{\text{Muon}}\rvert > 0.8)$ and $\lvert y_{\text{BTagger}}\rvert > 0.8$ |
| `Kaon I` | $\lvert y_{\text{BTagger}}\rvert > 0.8$ |
| `Kaon II` | $0.6 < \lvert y_{\text{BTagger}}\rvert < 0.8$ |
| `Kaon-Pion` | $0.4 < \lvert y_{\text{BTagger}}\rvert < 0.6$ |
| `Pion` | $0.2 < \lvert y_{\text{BTagger}}\rvert < 0.4$ |
| `Other` | $0.1 < \lvert y_{\text{BTagger}}\rvert < 0.2$ |

**Table 8.6.3.** Performance of the final *BABAR* tagging algorithm on data.

| Category | $\varepsilon_{\text{tag}}(\%)$ | $\Delta\varepsilon_{\text{tag}}(\%)$ | $w(\%)$ | $\Delta w(\%)$ | $Q(\%)$ | $\Delta Q(\%)$ |
|---|---|---|---|---|---|---|
| `Lepton` | $9.7 \pm 0.1$ | $0.2 \pm 0.2$ | $2.1 \pm 0.2$ | $0.2 \pm 0.5$ | $8.9 \pm 0.1$ | $0.1 \pm 0.4$ |
| `Kaon I` | $11.3 \pm 0.1$ | $-0.1 \pm 0.2$ | $4.1 \pm 0.3$ | $0.2 \pm 0.6$ | $9.6 \pm 0.1$ | $-0.1 \pm 0.4$ |
| `Kaon II` | $15.9 \pm 0.1$ | $-0.1 \pm 0.2$ | $13.0 \pm 0.3$ | $-0.2 \pm 0.6$ | $8.7 \pm 0.2$ | $0.0 \pm 0.5$ |
| `Kaon-Pion` | $13.2 \pm 0.1$ | $0.4 \pm 0.2$ | $23.0 \pm 0.4$ | $-1.3 \pm 0.7$ | $3.9 \pm 0.1$ | $0.5 \pm 0.3$ |
| `Pion` | $16.8 \pm 0.1$ | $-0.3 \pm 0.3$ | $33.3 \pm 0.4$ | $-2.7 \pm 0.6$ | $1.9 \pm 0.1$ | $0.6 \pm 0.2$ |
| `Other` | $10.6 \pm 0.1$ | $-0.5 \pm 0.2$ | $41.8 \pm 0.5$ | $5.9 \pm 0.7$ | $0.28 \pm 0.03$ | $-0.4 \pm 0.1$ |
| Total | $77.5 \pm 0.1$ | $-0.3 \pm 0.5$ | | | $33.1 \pm 0.3$ | $0.7 \pm 0.8$ |

The contribution of each of the nine sub-taggers to the overall tagging performance can be evaluated in two ways:

- the absolute effective tagging efficiency obtained by using only one sub-tagger ($Q_{\text{abs}}$);
- the incremental effective tagging efficiency ($Q_{\text{incr}}$), defined as the improvement in $Q$ associated with adding a single sub-tagger on top of all the others.

Table 8.6.4 shows $Q_{\text{abs}}$ and $Q_{\text{incr}}$ for the nine sub-taggers. In most events multiple flavor tagging signatures are present and contribute to the final tag as can be seen from the fact that $Q_{\text{incr}}$ is small for most sub-taggers. The exception is the Kaon sub-tagger which is the only tagger whose presence is essential to maintain a high tagging performance. The fact that in most cases several sub-taggers contribute to the final tag helps to ensure the robustness of the tagging algorithm.

### 8.6.4 Flavor tagging in Belle

The flavor tagging method used by Belle (Kakuno, 2004) is based on a multi-dimensional look-up table. A schematic diagram of the algorithm is shown in Figure 8.6.4.

The algorithm provides two parameters as the flavor tagging outputs: $q$ denoting the flavor of $B_{\text{tag}}$ ($+1$ for $B^0$, $-1$ for $\overline{B}^0$), and $r$ is an expected flavor dilution factor that ranges from zero for no flavor information ($w \simeq 0.5$) to unity for an unambiguous flavor assignment ($w \simeq 0$). In order to obtain a high overall effective tagging efficiency $Q$, an estimated flavor dilution factor is assigned to each event based on multiple discriminants. Using a multi-dimensional look-up table prepared from a large sample of simulated events and binned by the values of the discriminants, the signed probability, $q \cdot r$, is given by

$$q \cdot r = \frac{N(B^0) - N(\overline{B}^0)}{N(B^0) + N(\overline{B}^0)}, \tag{8.6.5}$$

where $N(B^0)$ and $N(\overline{B}^0)$ are the numbers of $B^0$ and $\overline{B}^0$ in the corresponding bin of the look-up table.

**Table 8.6.4.** Contribution of the nine sub-taggers to the *BABAR* tagging algorithm for the version of the algorithm used in 2004. The final version of the algorithm has the same architecture of the sub-taggers and BTagger but uses an improved kaon identification, leading to a slightly larger tagging performance. The determination of $Q_{\rm abs}$ on data was made using the $B_{\rm flav}$ control sample, assuming a time-integrated mixing probability of $\chi_d = 0.182$ and correcting for background. See text for the definition of $Q_{\rm abs}$ and $Q_{\rm incr}$.

| Sub-tagger | $Q_{\rm abs}$ on MC (%) | $Q_{\rm abs}$ on data (%) | $Q_{\rm incr}$ on MC (%) |
|---|---|---|---|
| Electron | $6.1 \pm 0.1$ | $5.0 \pm 0.2$ | 1.14 |
| Muon | $4.0 \pm 0.1$ | $3.3 \pm 0.2$ | 1.0 |
| Kin. Lepton | $2.9 \pm 0.1$ | $2.6 \pm 0.2$ | 0.36 |
| Kaon | $18.8 \pm 0.1$ | $18.3 \pm 0.4$ | 9.91 |
| Slow Pions | $5.2 \pm 0.1$ | $6.1 \pm 0.4$ | 0.47 |
| K-Pi | $9.3 \pm 0.1$ | $10.0 \pm 0.4$ | 0.25 |
| Max $p^*$ | $11.0 \pm 0.3$ | $9.7 \pm 0.5$ | 0.06 |
| FSC | $6.0 \pm 0.1$ | $6.6 \pm 0.4$ | 0.08 |
| Lambda | $0.3 \pm 0.1$ | $0.2 \pm 0.1$ | 0.38 |

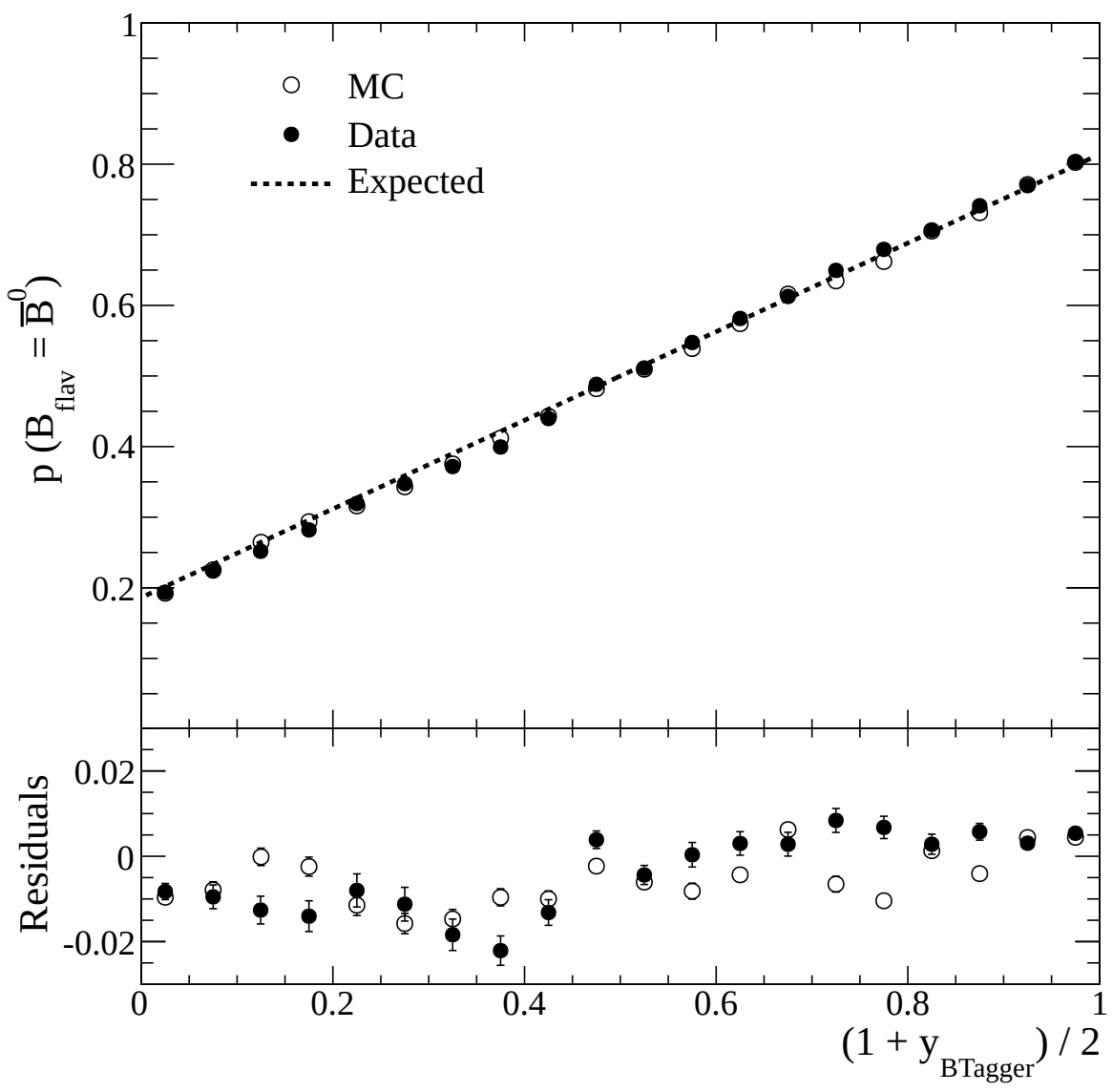

**Figure 8.6.3.** Probability of observing a fully reconstructed $\overline{B}^0$ on the $B_{\rm flav}$ side as a function of the probability $p_{B_{\rm tag}=B^0} = (1 + y_{\rm BTagger})/2$ of having a $B^0$ on the $B_{\rm tag}$ side. The dotted line shows the dependence expected for a perfectly trained tagging algorithm. The solid points are from the full *BABAR* $B_{\rm flav}$ control sample, the open circles are obtained from simulation. A contribution of up to 22% from combinatorial background is subtracted in each bin based on a fit to the $m_{\rm ES}$ distribution. The residuals with respect to the expectation are shown at the bottom.

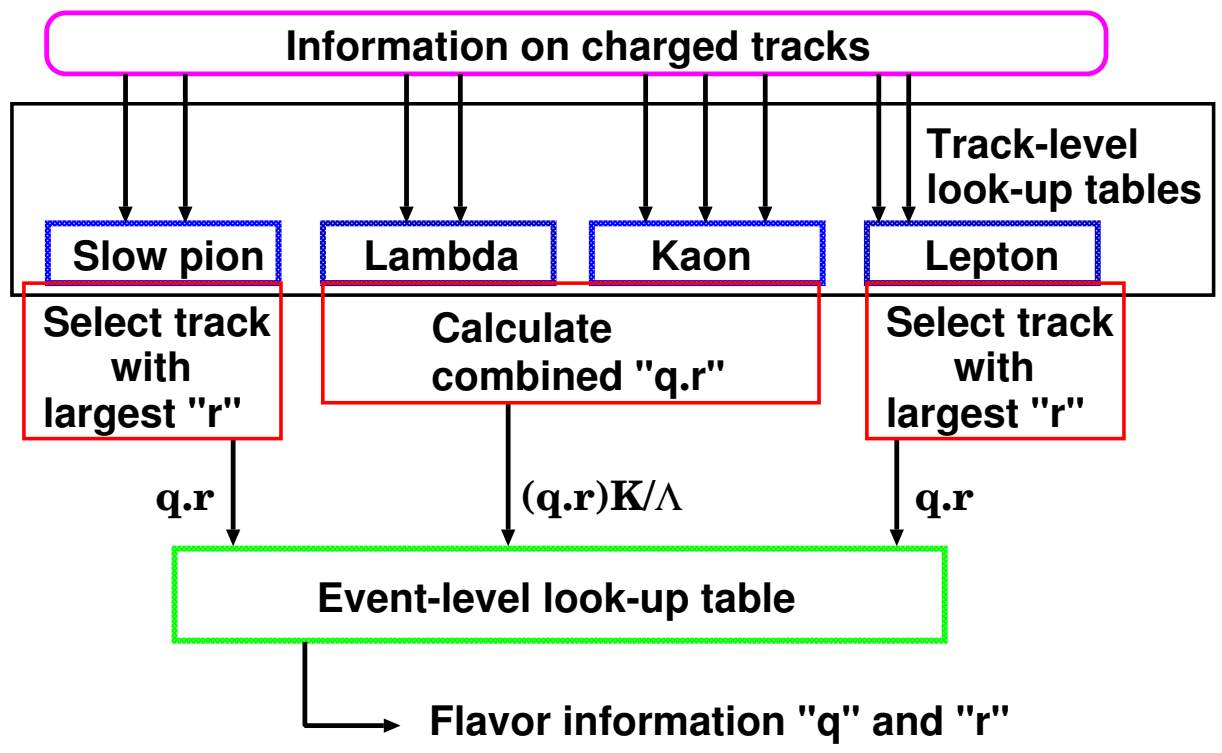

**Figure 8.6.4.** Schematic diagram of Belle's two-stage flavor tagging algorithm. See the text for the definition of the parameters "$q$" and "$r$".

The flavor tagging algorithm proceeds in two stages: the track stage and the event stage. In the track stage, each pair of oppositely charged tracks is examined to satisfy criteria for the $\Lambda$-like particle category. The remaining charged tracks are sorted into slow-pion-like, lepton-like and kaon-like particle categories. The $b$ flavor and its dilution factor of each particle, $q \cdot r$, in the four categories is estimated using the discriminants shown in Table 8.6.5.

In the second stage, the results from the first stage are combined to obtain the event-level value of $q \cdot r$. From the lepton-like and slow-pion-like track categories, the track with the highest $r$ value from each category is chosen as the input to the event level look-up table. The flavor dilution factors of the kaon-like and $\Lambda$-like particle candidates are combined by calculating the product of the flavor dilution factors in order to account for the cases with multiple $s$ quarks in an event. The product of flavor dilution factors gives better effective efficiency than taking the track with the highest $r$. Using the flavor dilution factor $r$ determined from Monte Carlo (MC) simulation as a measure of the tagging quality is a straightforward and powerful way of taking into account correlations among various tagging discriminants.

**Table 8.6.5.** Discriminants used in the Belle tagging algorithm.

| (Sub-)Stage | Variables | Number of bins |
|---|---|---|
| Lepton | $q$, $e$ or $\mu$, $\mathcal{L}_\ell$, $p^*$, $\theta_{\rm lab}$, $M_{\rm recoil}$, $p^*_{\rm miss}$ | 31680 |
| Kaon | $q$, $n_{K^0_S}$, $p^*$, $\theta_{\rm lab}$, $\mathcal{L}_K$ | 19656 |
| Lambda | $q$, $n_{K^0_S}$, $M_\Lambda$, $\theta_\Lambda$, $\Delta z$ | 32 |
| Slow pion | $q$, $p^{\rm lab}$, $\theta_{\rm lab}$, $\cos\theta_{\pi\rm T}$, $\mathcal{L}_e$ | 7000 |
| Event | $(q\cdot r)_\ell$, $(q\cdot r)_{K/\Lambda}$, $(q\cdot r)_{\pi_S}$ | 16625 |

By using two stages, the look-up tables can be kept small enough to provide sufficient statistics for each bin. Four million $B^0\overline{B}^0$ MC events are used to generate the particle-level look-up tables. To reduce statistical fluctuations of the $r$ values in the particle-level look-up tables, the $r$ value in each bin is calculated by including events in nearby bins with small weights. The event-level look-up table is prepared using MC samples that are statistically independent of those used to generate the track-level tables to avoid any bias from a statistical correlation between the two stages. Seven million $B^0\overline{B}^0$ MC events are used to create the event-level look-up table. The performance of individual tagging categories as obtained in MC simulation is shown for illustration in Table 8.6.6.

**Table 8.6.6.** Performance of sub-taggers in the Belle flavor tagging algorithm in terms of effective tagging efficiency $Q_{\rm abs}$ in simulated events.

| Sub-tagger | $Q_{\rm abs}$ on MC |
|---|---|
| Leptons | 12% |
| Kaons and $\Lambda$'s | 18% |
| Slow Pions | 6% |

All tagged events are sorted into seven subsamples according to the value of $r$: $0 \le r \le 0.1$, $0.1 < r \le 0.25$, $0.25 < r \le 0.5$, $0.5 < r \le 0.625$, $0.625 < r \le 0.75$, $0.75 < r \le 0.875$ and $0.875 < r \le 1$. For each subsample $l$, the corresponding average wrong tag fraction $w_l$ is determined. For events with $r \le 0.1$, there is negligible flavor discrimination available and $w_0$ is set to 0.5. For the other six subsamples, the average wrong tag fractions $w_l$ $(l = 1, 6)$ are measured directly from data using samples of semi-leptonic ($B^0 \to D^{*-}\ell^+\nu$) and hadronic ($B^0 \to D^{(*)-}\pi^+$ with $D^{*-}\rho^+$) $B$ meson decays. These decays are fully reconstructed and the flavor of the associated $B$ mesons is tagged. A total of 1461983 events are used to evaluate the performance of the tagging algorithm. An effective tagging efficiency of $Q = (30.1 \pm 0.4)\%$ is obtained. The wrong tag fractions, differences and tagging efficiencies for each subsample are shown in Table 8.6.7.

The average value of $r$ for each region ($r_l$) and the measured wrong tag fraction ($w_l$) should satisfy $r_l \simeq 1 - 2w_l$ if the MC simulation used for constructing the look-up tables simulates generic $B$ decays correctly. The degradation from the subdivision into $r$ bins and use of the corresponding measured wrong tag fractions $w_l$ is estimated to be about $\sim 0.5\%$, according to a Monte Carlo study.

**Table 8.6.7.** Tagging efficiencies ($\varepsilon_{\rm tag}$), wrong tag fractions ($w$) and their differences ($\Delta w$) for each $r$-interval for data taking with the SVD2 by Belle.

| $r$ − interval | $\varepsilon_{\rm tag}$ | $w$ | $\Delta w$ |
|---|---|---|---|
| $0.000 - 0.100$ | $0.222 \pm 0.004$ | 0.5 | 0.0 |
| $0.100 - 0.250$ | $0.145 \pm 0.003$ | $0.419 \pm 0.004$ | $-0.009 \pm 0.004$ |
| $0.250 - 0.500$ | $0.177 \pm 0.004$ | $0.319 \pm 0.003$ | $+0.010 \pm 0.004$ |
| $0.500 - 0.625$ | $0.115 \pm 0.003$ | $0.223 \pm 0.004$ | $-0.011 \pm 0.004$ |
| $0.625 - 0.750$ | $0.102 \pm 0.003$ | $0.163 \pm 0.004$ | $-0.019 \pm 0.005$ |
| $0.750 - 0.875$ | $0.087 \pm 0.003$ | $0.104 \pm 0.004$ | $+0.017 \pm 0.004$ |
| $0.875 - 1.000$ | $0.153 \pm 0.003$ | $0.025 \pm 0.003$ | $-0.004 \pm 0.002$ |

# Chapter 9
# Background suppression for $B$ decays

***Editors:***
*José Ocariz (BABAR)*
*Paoti Chang (Belle)*

**Additional section writers:**
*Jacques Chauveau*

## 9.1 Introduction

While the physics program of the $B$ Factories is not limited to $B$ physics, this chapter focuses on the techniques used to discriminate $B$ decay events from backgrounds: details on specific background-suppression techniques for charm, $\tau$ lepton and other decay modes are described in the relevant chapters of this book. For both *BABAR* and Belle, most analyses of $B$ decays use the kinematical constraints from the $e^+e^-$ collision at the $\Upsilon(4S)$ resonance to identify signal events; additional discrimination can be obtained from information based on the "event shape", that is the phase-space distribution of decay particles detected in the event, and are the main topic of this chapter.

## 9.2 Main backgrounds to $B$ decays

The production cross-section from $e^+e^-$ collisions at the $\Upsilon(4S)$ resonance receives sizable contributions other than $B\overline{B}$, and so the event rate is dominated by non-$B$ events. The identification of specific $B$ decay channels therefore has to deal with a potentially large number of backgrounds from various sources. The dominant source of combinatorial background comes from $e^+e^- \to q\overline{q}$ events, which are usually referred to as "continuum background". To study this background using real data, in addition to using signal sidebands (for example by requiring $m_{\rm ES}$ to lie safely below the $B$ mass peak), the $B$ Factories have also dedicated a significant fraction of off-resonance data-taking, at a center-of-mass energy slightly below the $\Upsilon(4S)$ peak: 40 MeV for *BABAR*, 60 MeV for Belle (see Chapter 3). Also depending on the decay channel under consideration, other backgrounds (either from other $B$ decays or from other processes) may also contribute, and need to be addressed correspondingly. For example, $B$ decay modes with only charged particles suffer backgrounds from QED processes (Bhabha scattering $e^+e^- \to e^+e^-$, $e^+e^- \to \mu^+\mu^-$, and $e^+e^- \to \tau^+\tau^-$) which can usually be suppressed by taking advantage of their clean leptonic signatures.

In the case of charmless $b \to u$ and $b \to s$ decay channels, background rates outnumber the signal by orders of magnitude, so combinatorial background from continuum events is most often the dominant source of background: random combinations of particles in the final state may mimic the kinematical signatures of the signal. Thus background suppression is a crucial issue in the analysis techniques. While the signal-to-background rates are usually more favorable in $b \to c$ decay modes, background suppression can play an important role in controlling potential systematic uncertainties in precision measurements of charmed $B$ decays. Also, rejection of backgrounds from other $B$ decay modes can play a significant role in the analysis results, as decay rates of such backgrounds, or their *CP* nature, can be poorly known.

## 9.3 Topological discrimination

For simplicity, the discussion in this chapter is restricted to fully-reconstructed $B$ decays; while most of the tools and techniques described here can be easily implemented or adapted to partly-reconstructed $B$ modes, for a discussion of specific issues related to such modes, the reader is referred to the relevant chapters.

As discussed in Chapter 7, one fundamental difference between $B$ meson signal and combinatorial background is the kinematics of their underlying production at the $B$ Factories, so essentially all $B$ meson analyses performed by *BABAR* and Belle take advantage of this information to identify the signal decay modes. After kinematic selection, additional background rejection is ensured by exploiting differences in the angular distributions of the particles produced in $e^+e^- \to \Upsilon(4S) \to B\overline{B}$ and background processes. For instance in a $B\overline{B}$ event, both $B$ mesons are produced almost at rest in the $\Upsilon(4S)$ frame, as the $\Upsilon(4S)$ mass is barely above the $B\overline{B}$ production threshold; as a result, the $B$ decay products are distributed isotropically in the $e^+e^- \to \Upsilon(4S) \to B\overline{B}$ rest frame. In contrast for $q\overline{q}$ events, the quarks are produced with a large initial momentum, and yield a back-to-back fragmentation into two jets of light hadrons. For the same reason in $B\overline{B}$ events, the angular distribution of decay products from the two $B$ mesons are uncorrelated, while for continuum a sizeable correlation arises, as the decay particles from each $B$ candidate tend to align with the direction of its jet.

Information based on the phase-space distribution of decay particles can be quantified in many different ways. Early *BABAR* and Belle physics analyses used methods initially developed by the ARGUS and CLEO collaborations; they then moved to develop more refined background-suppression techniques. We recall these methods in this section, and proceed to the description of those developed by *BABAR* and Belle in the next two sections. The *BABAR* Physics Book (Harrison and Quinn, 1998) is a useful reference for background suppression tools and methods available on the eve of $B$ Factories; for consistency, a few definitions and variables inherited prior to the advent of the $B$ Factories are summarized here:

– *Variables related to the $B$ meson direction:* the spin-1 $\Upsilon(4S)$ decaying into two spin-0 $B$ mesons results in a $\sin^2\theta_{\rm B}$ angular distribution with respect to the beam axis; in contrast for $e^+e^- \to f\bar{f}$ events, the spin-1/2 fermions $f$, and its two resulting jets, are distributed following a $1+\cos^2\theta_{\rm B}$ distribution. Using the angle $\theta_{\rm B}$ between the reconstructed momentum of the $B$ candidate (computed in the $\Upsilon(4S)$ reference frame) and the beam axis, the variable $|\cos\theta_{\rm B}|$ allows one to discriminate between signal $B$ decays and the $B$ candidates from continuum background.

- *Thrust and related variables:* for a collection of $N$ momenta $\boldsymbol{p}_i$ $(i = 1, \cdots N)$, the thrust axis $\boldsymbol{T}$ is defined as the unit vector along which their total projection is maximal; the thrust scalar $T$ (or thrust) is a derived quantity defined as

$$T = \frac{\sum_{i=1}^{N} |\boldsymbol{T}\cdot\boldsymbol{p}_i|}{\sum_{i=1}^{N} |\boldsymbol{p}_i|}. \tag{9.3.1}$$

A useful related variable is $|\cos\theta_{\rm T}|$, where $\theta_{\rm T}$ is the angle between the thrust axis of the momenta of the $B$ candidate decay particles (all evaluated in the $\Upsilon(4S)$ rest frame), and the thrust axis of all the other particles in the event (we call the set of those particles not associated with the $B$ candidate, "the rest-of-the-event", or ROE). For a $B\overline{B}$ event, both $B$ mesons are produced almost at rest in the $\Upsilon(4S)$ rest frame, so their decay particles are isotropically distributed, their thrust axes are randomly distributed, and thus $|\cos\theta_{\rm T}|$ follows a uniform distribution in the range $[0, 1]$. In contrast for $q\bar{q}$ events, the momenta of particles follow the direction of the jets in the event, and as a consequence the thrusts of both the $B$ candidate and the ROE are strongly directional and collimated, yielding a $|\cos\theta_{\rm T}|$ distribution strongly peaked at large values. Altogether, these arguments bring a qualitative description of the discriminating power provided by $|\cos\theta_{\rm T}|$.
Another thrust-related variable is $\theta_{T,B}$ the angle between the thrust axis of the $B$ decay particles and the beam axis; for signal, $|\cos\theta_{\rm T,B}|$ is uniformly distributed, while for continuum events, the thrust of particle momenta from the $B$ candidate tends to be aligned with the $1+\cos^2\theta_{\rm T,B}$ distribution followed by the jets.

- *Sphericity and related variables:* sphericity and thrust are strongly correlated concepts, nonetheless both are commonly used. For a collection of momenta $\boldsymbol{p}_i$, the sphericity tensor $S$ is defined as

$$S^{\alpha,\beta} = \frac{\sum_{i=1}^{N} p_i^{\alpha} p_i^{\beta}}{\sum_{i=1}^{N} |\boldsymbol{p}_i|^2}, \tag{9.3.2}$$

(with $\alpha,\beta = x, y, z$) and provides a three-dimensional representation of the spatial distribution of the $\boldsymbol{p}_i$ collection. For an isotropic distribution, its three eigenvalues $\lambda_k$ have similar magnitude; while for a planar distribution, one of the eigenvalues is significantly smaller, with its eigenvector orthogonal to that plane; and finally for a very directional distribution, the eigenvector oriented in that preferred direction has an eigenvalue considerably larger than the two others. Useful quantities derived from sphericity are the sphericity scalar (or sphericity), and the sphericity axis. The sphericity scalar $S$ is defined as

$$S = \frac{3}{2}\left(\lambda_2 + \lambda_3\right), \tag{9.3.3}$$

$\lambda_2$ and $\lambda_3$ being the two lowest eigenvalues; values of $S$ close to 1 correspond to isotropically distributed momentum collections, while very collimated distributions yield sphericity values close to zero. The sphericity axis is collinear with the sphericity eigenvector having the largest eigenvalue. In the same spirit as $|\cos\theta_{\rm T}|$, the variable $|\cos\theta_{\rm S}|$ is often used, where $\theta_{\rm S}$ is the angle between the sphericity axes of the $B$ candidate and the ROE.

- *The Fox-Wolfram moments:* another useful parameterization of phase-space distribution of momentum and energy flow in an event, was introduced in (Fox and Wolfram, 1978): for a collection of $N$ particles with momenta $\boldsymbol{p}_i$, the $k$-th order Fox-Wolfram moment $H_k$ is defined as

$$H_k = \sum_{i,j}^{N} |\boldsymbol{p}_i|\,|\boldsymbol{p}_j|\,P_k\left(\cos\theta_{ij}\right), \tag{9.3.4}$$

where $\theta_{ij}$ is the angle between $\boldsymbol{p_i}$ and $\boldsymbol{p_j}$, and $P_k$ is the $k$-th order Legendre polynomial. Notice that in the limit of vanishing particle masses, $H_0 = 1$; that is why the normalized ratio $R_k = H_k/H_0$ is often used, so that for events with two strongly collimated jets, $R_k$ takes values close to zero (one) for odd (even) values of $k$. These sharp signatures provide a convenient discrimination between events with different topologies.

The variables and tools described in the list above do not necessarily provide the optimal background discriminating power, and for channels suffering from large background rates, additional specific tools are developed. One such example is provided by a multivariate discriminant variable introduced by the CLEO collaboration (Asner et al., 1996) in the context of charmless $B$ decays; it is a Fisher combination (see Chapter 4 for the description of the Fisher discriminant) of nine variables corresponding to the momentum flow around the thrust axis of the $B$ candidate, binned in nine cones of $10^\circ$ around the thrust axis as illustrated in Figure 9.3.1. The linear coefficients assigned to the combination of these nine variables are extracted from MC generated events for the signal, and either $B$ mass sidebands or events collected off-resonance for continuum. The Fisher used by CLEO has often been referred to as "the CLEO Fisher" by the $B$ Factories.

## 9.4 BABAR strategy

For *BABAR*, a typical analysis strategy is based on a two-step approach: first, variables using the complete set of particles in the event are built to reject copious backgrounds while maintaining high efficiency for signal. In the second step, variables are built separately, using information from the decay particles of the signal $B$ candidate and of the ROE, to further reject backgrounds through additional requirements on the selection, and/or as inputs to a maximum-likelihood fit (see Chapter 11 for the description of maximum-likelihood fits) at later stages in the analysis.

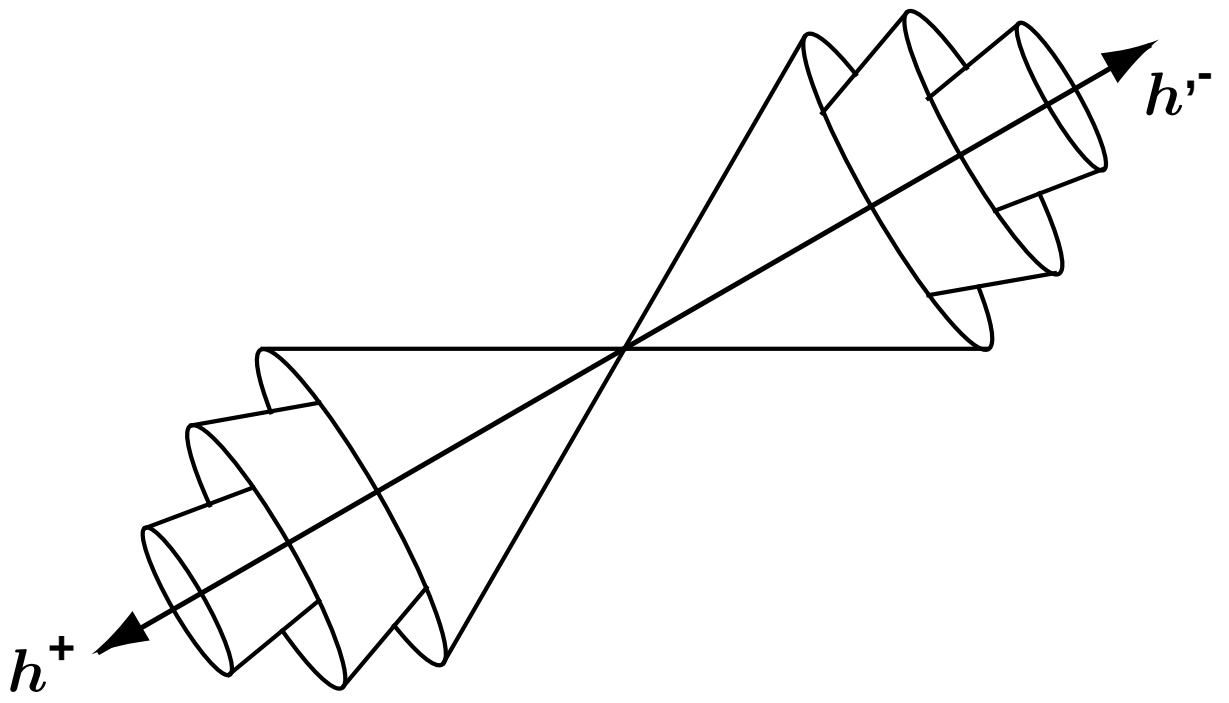

**Figure 9.3.1.** A graphical illustration of the CLEO Fisher discriminant, from (Asner et al., 1996). The $h^+$, $h'^-$ arrows indicate the momenta of the two charged hadronic tracks in a $B^0 \to h^+h'^-$ candidate; the momentum of ROE particles within each cone (the first three cones around its thrust axis being drawn in the figure) are summed and combined to give the Fisher discriminant.

Figure 9.4.1 illustrates two typical variables used in the first step. A simple requirement on the number of charged tracks per event can provide highly efficient background suppression. Also, in this first step, a simple requirement on the normalized second Fox-Wolfram moment ratio $R_2$ is applied; a loose cut on the value of $R_2$ has negligible impact on signal, while efficiently removing a substantial fraction of diphoton or dilepton backgrounds. In this first step, typical *BABAR* analyses also combine information both from the decay particles of the $B$ meson candidate and from the ROE, and use them to achieve additional background rejection. For example, Figure 9.4.2 shows the distributions of $|\cos\theta_S|$, both for signal (from simulated $B$ decays) and for continuum events (from sidebands on data, by requiring $m_{ES}$ to be in the $5.20 - 5.26$ GeV/$c^2$ range). A simple per-event requirement on the value of $|\cos\theta_S|$ is applied to define the final analysis sample.

An important advantage of variables based on the ROE is that for the signal $B$ decays, their correlation is small or negligible with the variables built out of the $B$ candidate observables. Therefore it is appropriate to construct a joint likelihood function from the product of their $p.d.f.$s to use in a fit.

### 9.4.1 Linear discriminants

For typical *BABAR* analyses, several combinations of variables from the ROE are built, and combined in multivariate discriminants. A general description of linear discriminants in the optimization of the analyses is given in

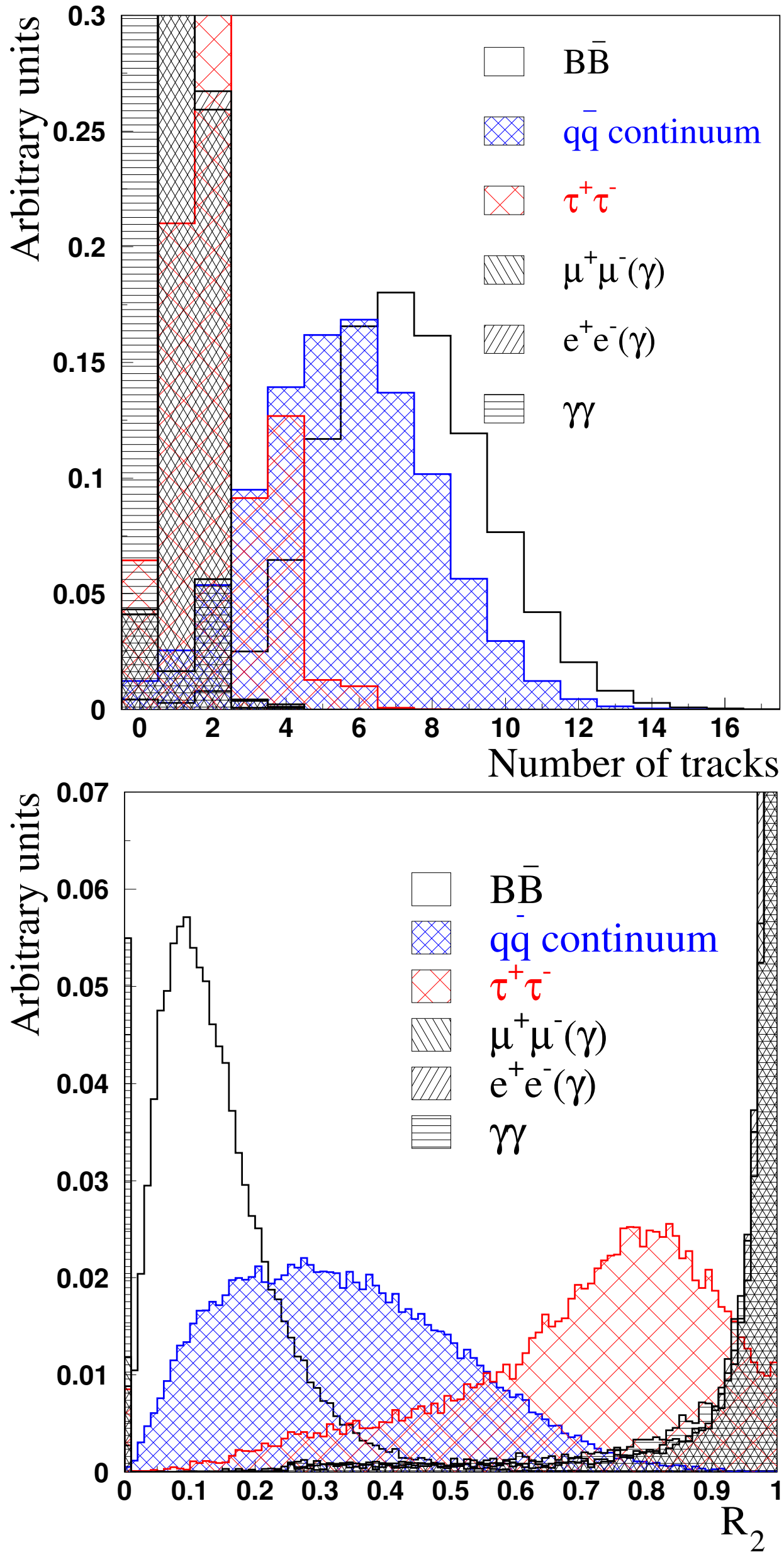

**Figure 9.4.1.** Two examples of global variables, used as a first step in background suppression in most *BABAR* analyses of $B$ decays. The top plot shows the number of charged tracks per event for various processes; the bottom plot is the distribution of the normalized second Fox-Wolfram moment ratio $R_2$, for various processes. The figures are from a *BABAR* Thesis (Rahatlou, 2002).

Chapter 4. Many of these discriminants use the so-called "monomials" $L_n$, defined as

$$L_n = \sum_{i \in ROE} p_i \times |\cos\theta_i|^n \,, \tag{9.4.1}$$

where $p_i$ is the momentum (computed in the $\Upsilon(4S)$ reference frame) of particle $i$ belonging to the ROE, and $\theta_i$ is the angle between its momentum and the thrust axis of the $B$ candidate. Dedicated studies concluded that the $L_0$ and $L_2$ pair provides most of the discriminating power to separate signal from continuum background; for instance, a bi-variate linear (Fisher) combination $\mathcal{F} = c_0 L_0 + c_2 L_2$ (using $L_0$ and $L_2$ only) reaches a signal-to-background separation comparable to a Fisher using the nine variables in the CLEO Fisher. Figure 9.4.3 illustrates the contribution from a single 1 GeV particle to both discriminants, as a function of its angle with respect to the thrust axis. That same figure shows the contribution from a three-variable Fisher discriminant (including also the $L_1$ monomial), that exhibits an almost equivalent angular dependence to the nine-variable CLEO discriminant, thus showing that a comparable discriminating power can be achieved with a smaller number of variables.

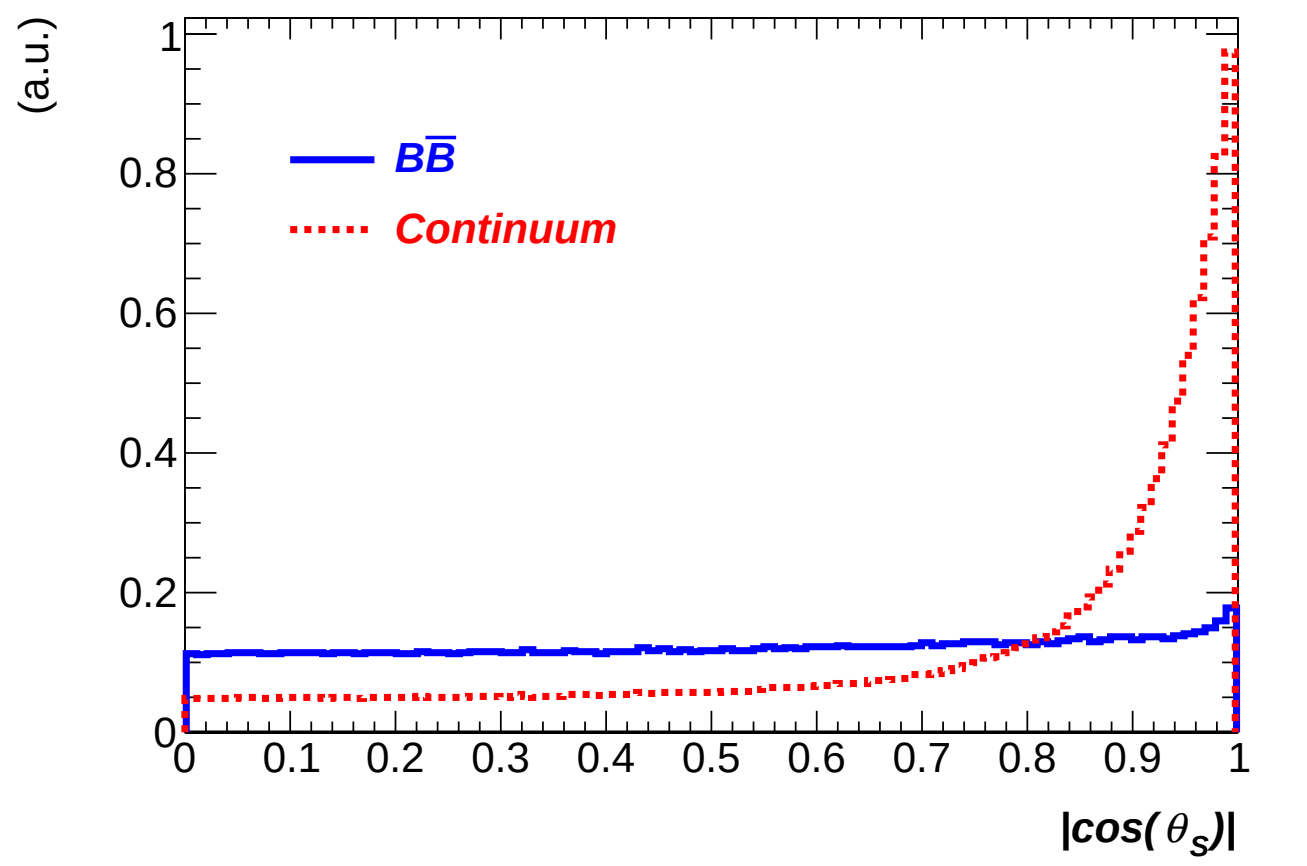

**Figure 9.4.2.** The signal (solid blue line) and continuum background (dashed red line) distributions of $|\cos\theta_{\rm S}|$, a variable often used as a first step in background suppression for charmless two-body $B$ decays. $|\cos\theta_{\rm S}|$ is uniformly distributed for the signal, while for continuum it is sharply peaked at large values. The figure is adapted from a *BABAR* Thesis (Malclès, 2006). The vertical scale is in arbitrary units (a.u.).

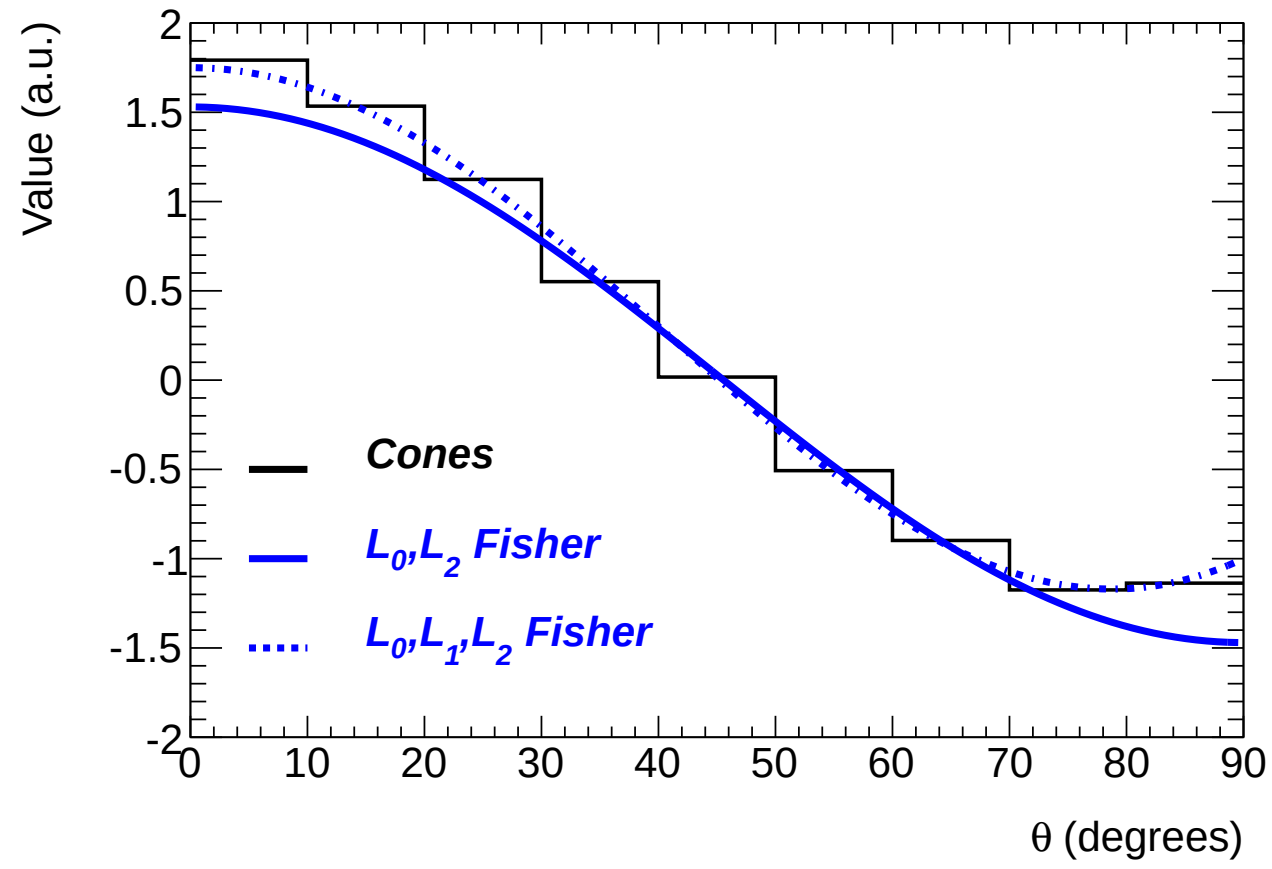

**Figure 9.4.3.** The contribution to the *BABAR* and CLEO Fisher discriminants, for a single 1 GeV particle, as a function of the angle of its momentum and the thrust axis of the $B$ candidate. The nine-step line indicates the values of the nine cone coefficients in $10^\circ$ bins for the CLEO Fisher, while the continuous blue line is the resulting function for the $\mathcal{F}$ used by *BABAR*. The dash-dotted line corresponds to a three-variable Fisher (shown for illustration only, not used in actual *BABAR* analyses). The coefficients for these Fisher discriminants were optimized using samples of charmless two-body $B$ decays for signal, and data events from $m_{\rm ES}$ sidebands for background. The figure is adapted from a *BABAR* Thesis (Pivk, 2003). The vertical scale is in arbitrary units (a.u.).

For most charmless $B$ decay analyses, the optimization algorithm returns values very close to $\mathcal{F} = L_2 - 2 \times L_0$ (*i.e.* $c_0 = -2c_2$, up to arbitrary offset and scale parameters) for the Fisher coefficients. To a certain extent, this two-variable Fisher discriminant can be thought of as a simple, continuous extension of the CLEO discriminant, that can be explained in terms of the relative sign and ratio of the $c_0$ and $c_2$ coefficients described above. For an isotropically distributed collection of particles, the total $\mathcal{F}$ value will be close to zero, as particles with angles collinear/orthogonal to the $B$ candidate thrust axis contribute with opposite signs, and tend to cancel out in the sum. In contrast, contributions from a collection of particles collinear with the thrust axis will mostly sum up to give a positive value.

Figure 9.4.4 shows the distributions of this bi-variate $\mathcal{F}$ discriminant, with coefficients evaluated both before and after a first-step cut on $|\cos\theta_{\rm S}| < 0.8$ (c.f. Figure 9.4.2). Before the first-step selection, the $\mathcal{F}$ discriminant provides a $\sim 1.6\sigma$ separation between signal and background. The first-step cut on $|\cos\theta_{\rm S}|$ rejects $\sim$ 65% of all continuum background, while retaining $\sim$ 80% of signal; for the significantly signal-enriched remaining selected events, $\mathcal{F}$ still provides a $\sim 1.2\sigma$ separation. This remaining discriminating power is efficiently exploited in the maximum-likelihood analysis.

The monomial $L_0$ is the total momentum flow observed in the detector, and $L_2$ is a direction-weighted sum of contributions to the total momentum flow. Hence the ratio $L_2/L_0$ is expected to be rather insensitive to the actual per-event value of the total momentum flow, which largely cancels in the ratio. The relative sign of the $c_0, c_2$ coefficients in $\mathcal{F}$ expresses the same cancellation. As a result, the simulated distributions of both $\mathcal{F}$ and $L_2/L_0$ are found to be in excellent agreement with data. Some *BABAR* analyses have therefore preferred to use the simpler $L_2/L_0$ ratio. Simplicity over complexity (*i.e.* adding $L_1$ or splitting the ROE between charged and neutral particles) has been privileged by most *BABAR* analyses because the discriminating gain was found to be marginal.

### 9.4.2 Nonlinear discriminants

Many *BABAR* analyses combine the information from the monomials with other variables to further enhance their discriminating power and the resulting performance in background suppression. As already mentioned, there are significant correlations among event-shape variables (since they all quantify in different ways the spatial distribution of momentum flow). To better exploit such potentially nonlinear correlations, neural networks (NN, see Chapter 4 for a description of multivariate methods) and other nonlinear discriminant algorithms are used. As an illustration, typical charmless 3-body analyses use, in addition to the $L_0$ and $L_2$ monomials, variables such as $|\cos\theta_{\rm B}|$ and $|\cos\theta_{\rm T}|$ in their final MVA. Figure 9.4.5 illustrates the discriminating power achieved with a NN based on these four variables, used in several Dalitz-plot analyses of charmless 3-body $B$ decays in *BABAR* (see Chapter 13 for a description of Dalitz-plot analyses). In these analyses, the NN output is used both for selection and in the maximum-likelihood fit. At the first stage, this NN provides a $\sim 1.9\sigma$ separation between signal and background. A cut at NN $> -0.4$ is then applied to remove roughly 75% of continuum background while retaining a 90% signal efficiency; on top of enhancing its signal-to-background content, this cut also reduces the sample size to a value that is suitable for the CPU constraints affecting multidimensional amplitude fits in Dalitz-plot analyses. Then, at the amplitude analysis stage, the NN is implemented in the likelihood function, where its remaining $\sim 1.4\sigma$ separation is exploited in the maximum-likelihood fit. Two specific features, relevant to the implementation of a NN in a Dalitz analyses are worth mentioning:

– For continuum background, the NN is correlated with the Dalitz variables. This feature can be qualitatively described as follows: for continuum event candidates passing all selection criteria, and belonging to the center of the Dalitz plot, the angular distribution of particles tends to exhibit a more isotropic distribution, since already the three particles composing the signal candidate have similar momenta and roughly equidistant angular separation. In order to include the NN in the likelihood function, a parameterization of this correlation as a function of Dalitz masses, has to be effectively implemented for its continuum component.
– In light of the aforementioned systematic sensitivity to the simulation of the total momentum flow, some *BABAR* analyses have opted for not allowing the $L_0$ and $L_2$ monomials to be independently optimized in the training stage of the NN, and used instead a linear combination with fixed coefficients or the $L_2/L_0$ ratio in the NN training.

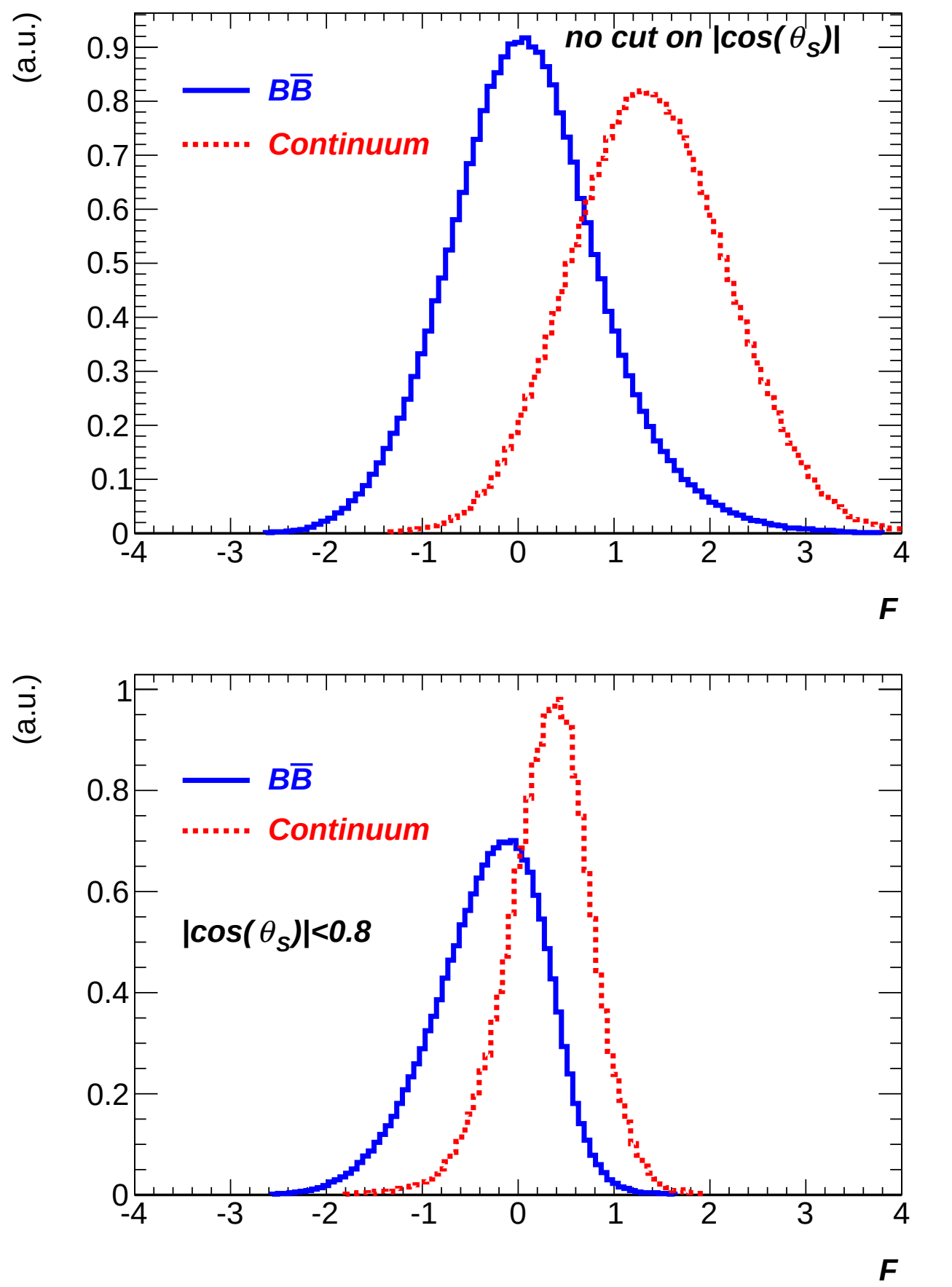

**Figure 9.4.4.** The signal (solid blue line) and background (dashed red line) distributions of the Fisher discriminant $\mathcal{F}$ based on the $L_0$ and $L_2$ monomials, used for continuum background suppression in several *BABAR* charmless $B$ decay analyses. To illustrate the two-step procedure, the distributions are shown both before (top) and after (bottom) a first-step cut of $|\cos\theta_{\rm S}| < 0.8$. The figures are adapted from a *BABAR* Thesis (Malclès, 2006). The vertical scale is in arbitrary units (a.u.).

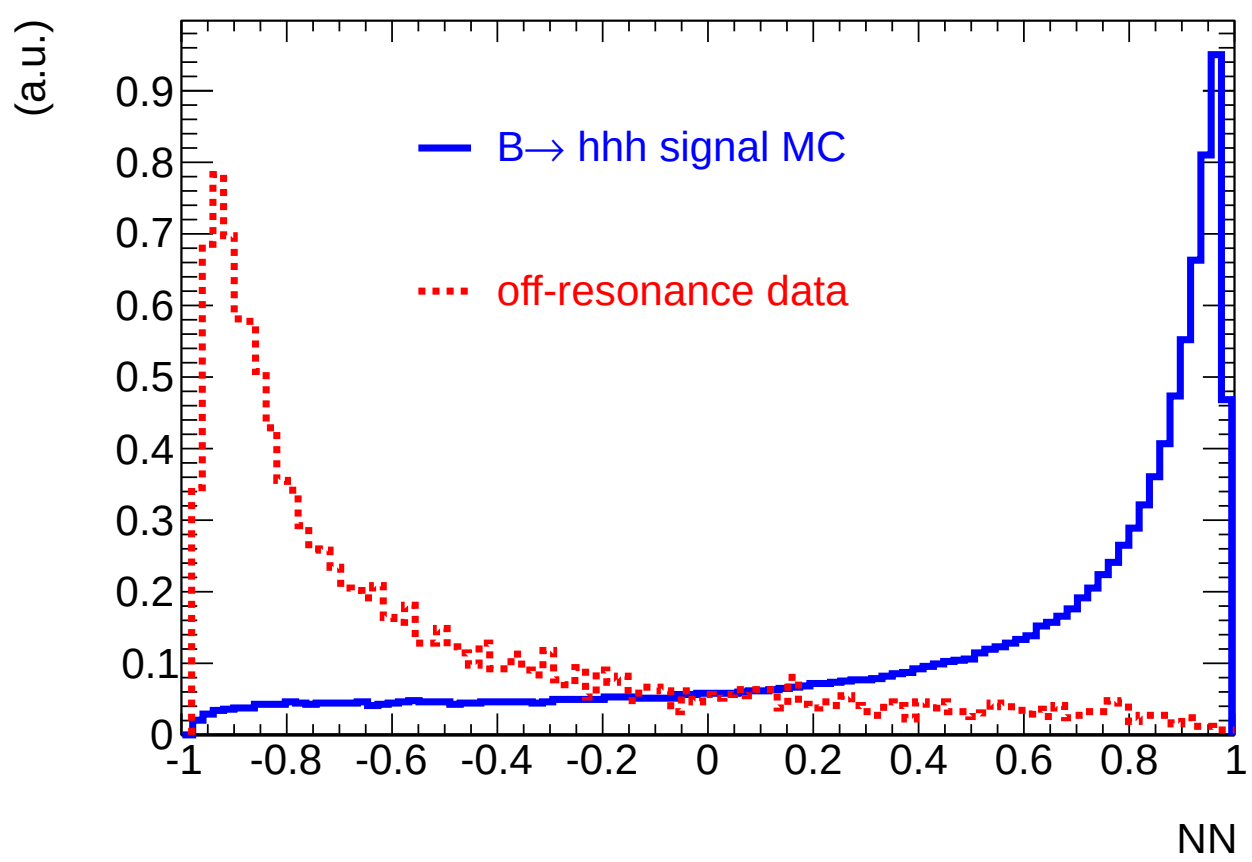

**Figure 9.4.5.** An example of a multilayer perceptron output NN, used to discriminate between the signal $B$ decay and continuum background in the charmless 3-body analysis of $B^0 \to K^0_S\pi^+\pi^-$ decays. The solid blue histogram is the NN output evaluated on signal Monte-Carlo, and the dashed red histogram uses off-resonance data. This neural network uses four variables as inputs : $L_0$, $L_2$, $|\cos\theta_B|$ and $|\cos\theta_T|$. The figure is adapted from a *BABAR* Thesis (Pérez, 2008). The vertical scale is in arbitrary units (a.u.).

### 9.4.3 Including additional sources of background suppression

In addition to the “event-shape” variables discussed in the previous sections, various other sources of discriminating information are also available in $B$ decay analyses: in particular, decay-time information extracted from vertexing (discussed in Chapter 6), kinematical variables extracted from $B$ meson reconstruction (Chapter 7), and the output of $B$-flavor tagging (Chapter 8), can all contribute to background suppression. As described in more detail in Chapter 11, a generic time-dependent analysis combines all this information in a maximum-likelihood analysis.

For specific analyses, only a subsample of this information is effectively used in the likelihood function; for instance, timing information is not necessary to perform a time-independent fit, and analysis of a flavor-specific decay (like charged $B$ modes, or "self-tagging" neutral decay modes), does not require tagging. In such scenarios, some *BABAR* analyses (particularly in searches of rare decay channels) exploit this available background-suppressing power, by combining event-shape variables with the tagging index output and/or the time difference significance $\Delta t/\sigma(\Delta t)$ into a linear Fisher discriminant, which is in turn used in the likelihood function.

## 9.5 Belle strategy

For Belle, the correlated shape variables are first combined to form a Fisher discriminant and then other uncorrelated variables are included with the Fisher variable to form a signal-to-background likelihood ratio $\mathcal{R}$. The numbers of signal and background events can be extracted by either applying a cut on the likelihood ratio and then performing a fit using $m_{\rm ES}$ and $\Delta E$, or by requiring a loose cut on $\mathcal{R}$, and then performing a fit using the variables $m_{\rm ES}$, $\Delta E$ and $\mathcal{R}$. Later in the lifetime of Belle, more analyses employ the neural network technique to combine correlated variables with the Fisher discriminant and other uncorrelated variables. One can make a requirement on the neural network output to suppress the background or include the output after a loose requirement in a multi-dimensional likelihood fit to extract the signal.

### 9.5.1 $SFW$

There are two kinds of Fisher discriminant used to study charmless $B$ decays on Belle. All reconstructed particles in an event are divided into two categories: $B$ candidate daughters (denoted as $s$) and the ROE (denoted as $o$). Two Fisher discriminants are constructed using the energy and momentum of each particle in the $e^+e^-$ center-of-mass frame. The first Fisher discriminant is composed of several Fox-Wolfram moments $h_j^{kl}$ and is defined as

$$SFW = a_2 h_2^{so} + a_4 h_4^{so} + \sum_{j=1}^{4} b_j h_j^{oo}, \qquad (9.5.1)$$

where $a_2, a_4$ and $b_j$ are the Fisher coefficients determined to separate signal and backgrounds in an optimal way using the signal and continuum MC events. The $SFW$ variable is colloquially referred to as the "Super Fox-Wolfram Moment". In order to avoid the data-MC discrepancy in event shapes, data in regions dominated by continuum are often used to determine the coefficients. Variables $h_i^{so}(i = 2, 4)$ and $h_j^{oo}$ are the normalized Fox-Wolfram moments, defined as

$$h_l^k = \frac{\sum_{m,n} |\vec{p_m}||\vec{p_n}| P_l(\cos\theta_{mn})}{\sum_{m,n} |\vec{p_m}||\vec{p_n}|}, \qquad (9.5.2)$$

where $\vec{p_m}$ and $\vec{p_n}$ are the momenta of particles $m$ and $n$; $P_l(\cos\theta_{mn})$ is the $l$-th order Legendre polynomial of cosine of the angle ($\theta_{mn}$) between $\vec{p_m}$ and $\vec{p_n}$; $k$ categorizes the type of Fox-Wolfram moment, $so$ and $oo$, where $m$ is from $B$ signal daughters and $n$ is from the ROE for $so$, and both $m$ and $n$ are from the ROE for $oo$. If $B$ daughter particles themselves decay into several particles, the event shape is more isotropically distributed. However, the $B$ candidates from the continuum are also more isotropically distributed to mimic the $B\overline{B}$ events. For two-body or three-body $B$ decays, the signal-to-background separation is therefore better if the $SFW$ variable is computed using the particles directly from $B$ decays. For instance, in the decay $B \to \omega K$ with $\omega \to \pi^+\pi^-\pi^0$, the Fox-Wolfram moment $h_l^{so}$ in Eq. (9.5.2) is calculated using the $\omega$ momentum instead of the momenta of its daughter pions. The difference of the separation power between the two different treatments is less pronounced for multi-body $B$ decays.

### 9.5.2 $KSFW$

To further improve the continuum suppression, a second Fisher discriminant was developed by Belle:

$$KSFW = \sum_{l=0}^{4} R_l^{so} + \sum_{l=0}^{4} R_l^{oo} + \gamma \sum_{n=1}^{N_t} |(P_t)_n|, \qquad (9.5.3)$$

where $R_l^{so}$ and $R_l^{oo}$ are modified Fox-Wolfram moments similar to $h_l^{so}$ and $h_l^{oo}$ in Eq. (9.5.2), respectively; the third term is the scalar sum of the transverse momentum of each particle multiplied by a free parameter $\gamma$ and $N_t$ is the total number of particles. The expressions of $R_l^{so}$ and $R_l^{oo}$ are described as follows:

– $R_l^{so}$
  In constructing $R_l^{so}$, the missing momentum of an event is treated as an additional particle and the moment is decomposed into three categories: a charged particle part (c), neutral particle part (n), and missing particle part (m). The variable $R_l^{so}$ is expressed as

$$R_l^{so} = \frac{\alpha_{\rm cl} H_{\rm cl}^{so} + \alpha_{\rm nl} H_{\rm nl}^{so} + \alpha_{\rm ml} H_{\rm ml}^{so}}{E^*_{\rm beam} - \Delta E}. \qquad (9.5.4)$$

  For odd $l$, we have

$$H_{\rm nl}^{so} = H_{\rm ml}^{so} = 0 \qquad \text{and} \qquad (9.5.5)$$

$$H_{\rm cl}^{so} = \sum_i \sum_{j\rm x} Q_i Q_{j\rm x} |p_{j\rm x}| P_l(\cos\theta_{i,j\rm x}), \qquad (9.5.6)$$

where $i$ runs over the $B$ daughters; $j$x indexes the ROE in the category x (x = c, n, m); $Q_i$ and $Q_{j\mathrm{x}}$ are the charges of particle $i$ and $j$x, respectively; $p_{j\mathrm{x}}$ is the momentum of particle $jx$; and $P_l(\cos\theta_{i,j\mathrm{x}})$ is the $l$-th order Legendre polynomial of the cosine of the angle between particles $i$ and $j$x.
For even $l$,

$$H^{so}_{\mathrm{x}l} = \sum_i \sum_{j\mathrm{x}} |p_{j\mathrm{x}}| P_l(\cos\theta_{i,j\mathrm{x}}), \qquad (9.5.7)$$

which is similar to Eq. (9.5.6) except for the charge factors. There are two free parameters for $l = 1, 3$ and nine ($3 \times 3$) for $l = 0, 2, 4$.

– $R^{oo}_l$
The definition of the second term of Eq. (9.5.3) is simpler.
For odd $l$, we have

$$R^{oo}_l = \sum_j \sum_k \beta_l Q_j Q_k |p_j||p_k| P_l(\cos\theta_{j,k}), \qquad (9.5.8)$$

where $j$ and $k$ run over the ROE and other variables are the same as used in Eq. (9.5.6).
For even $l$, we have

$$R^{oo}_l = \sum_j \sum_k \beta_l |p_j||p_k| P_l(\cos\theta_{j,k}). \qquad (9.5.9)$$

There are five Fisher coefficients ($\beta_l$) to be determined.

The total number of Fisher coefficients in $KSFW$ is 17, determined using the signal and continuum MC events. To further improve the background suppression, the 17 coefficients are obtained in seven missing mass squared ($M^2_{\mathrm{miss}}$) bins, where $M^2_{\mathrm{miss}}$ is defined as

$$M^2_{\mathrm{miss}} = \left( E_{\Upsilon(4S)} - \sum_{n=1}^{N_t} E_n \right)^2 - \sum_{n=1}^{N_t} |p_n|^2, \qquad (9.5.10)$$

where $E_{\Upsilon(4S)}$ is the energy of $\Upsilon(4S)$ and $E_n$ and $p_n$ are the energy and momentum of particle $n$, respectively. Therefore, there are seven sets of 17 Fisher coefficients in $KSFW$. In general $KSFW$, compared to $SFW$, provides better signal-background separation for charmless two-body and three-body $B$ decays, but the improvement is less pronounced for the $B$ decays into a final state with more than three particles.

Two other variables that can distinguish between signal and continuum are $\cos\theta_{\mathrm{B}}$ (as mentioned in Section 9.3) and $\Delta Z$, where the former is the cosine of the angle between the $B$ momentum and the beam direction in the CM frame and the latter is the distance in the beam direction between the $B$ vertex and the vertex from the ROE. Figure 9.5.1 shows the $\cos\theta_{\mathrm{B}}$ and $\Delta Z$ distributions for the $B^+ \to K^+\pi^0$ signal and the continuum events. Since $\Upsilon(4S)$ produced at $e^+e^-$ resonance is transversely polarized, the $B$ moving distribution behaves as $\sin^2\theta_{\mathrm{B}}$ while it is more or less flat for the continuum background.[38] The $\Delta Z$ distribution is broader for $B\overline{B}$ events due to the relatively longer lifetime of $B$ mesons. Signal $B$ vertices are constructed using the charged tracks of the $B$ daughters. For a decay mode with only one charged track in the final state, for instance $B^+ \to K^+\pi^0$, the $z$ vertex position is obtained by projecting the single track trajectory to the beam axis. Obviously the $\Delta Z$ resolution is better if there is more than one charged particle used to reconstruct the decay vertex. The $\Delta Z$ variable is not applicable for the decay modes with only photons in the final state, for instance $B^0 \to \pi^0\pi^0$. It is possible to use photon conversions to obtain the $B$ vertex in a future super flavor factory. The primary aim for this case is to perform a time-dependent measurement.

Finally all the shape information is combined to form a signal-to-background likelihood ratio ($\mathcal{R}$), defined as

$$\mathcal{R} = \frac{\mathcal{L}_S}{\mathcal{L}_S + \mathcal{L}_B}, \qquad (9.5.11)$$

$$\mathcal{L}_{S/B} = P(KSFW)_{S/B} \times P(\cos\theta_{\mathrm{B}})_{S/B} \times P(\Delta Z)_{S/B}, \qquad (9.5.12)$$

where $P_{S/B}$ is the probability density function for signal ($S$) and background ($B$). Continuum suppression can be achieved by applying a cut selection on $\mathcal{R}$ based on a figure of merit or requiring a loose selection and including $\mathcal{R}$ in a multi-dimensional likelihood fit. To avoid poor modeling of the rising edges as shown in the top plot of Fig. 9.5.2, in some analyses a modified likelihood ratio $\mathcal{R}'$ can be defined as

$$\mathcal{R}' = \log \frac{\mathcal{R} - lb}{ub - \mathcal{R}}, \qquad (9.5.13)$$

where $lb$ is the lower bound of $\mathcal{R}$, which is the loose $\mathcal{R}$ selection value to reduce the background, and $ub$ is the upper bound (usually 1.0). The bottom plot of Fig. 9.5.2 shows the $\mathcal{R}'$ distribution with lower $\mathcal{R}$ bound at 0.2 for $B^+ \to K^+\pi^0$ signal and the continuum background. The signal and background $\mathcal{R}'$ distributions may be described by a single or double Gaussian, which can be used as *p.d.f.* representations of $\mathcal{R}'$ in the multi-dimensional fit.

### 9.5.3 Additional variables and neural network

Additional background discrimination is provided by $B$-flavor tagging. As described in Chapter 8, events with good flavor tags usually contain high momentum leptons and are more likely to be $B\overline{B}$ events. The top plot of Fig. 9.5.3 shows the normalized signed probability ($q \cdot r$) distributions for $B$ signal and the continuum background from MC. Note that the $q \cdot r$ definition for the tag $B$ described in Eq. 8.6.5 is also valid for the charged $B$ meson system by replacing $B^0(\overline{B}^0)$ with $B^+(B^-)$. It is easy to understand that the majority of the continuum events populate the central $q \cdot r$ region, where the flavor information is poorly known, while sizable fractions of $B$ signal events have $q \cdot r \sim \pm 1$. If the signal $B$ decays into a flavor specific state, one can use the product of the signal $B$-flavor ($q_B$) and $q \cdot r$ to distinguish between signal and background. As shown in the bottom plot of Fig. 9.5.3, a large fraction of signal events populate the region around $q_B \cdot q \cdot r = -1$ and the distributions for both signal and the continuum events in the $B^+ \to K^+ l^+ l^-$ study become asymmetric. The asymmetric $q_B \cdot q \cdot r$ distribution for the continuum is due to the correlation of strangeness between the tag and signal sides. To utilize all available information, the quantity $q_B \cdot q \cdot r$ ($q \cdot r$ for the *CP* eigenmodes) can be used in the likelihood for background suppression, or alternatively the original $\mathcal{R}$ selections can be optimized depending on the value of $q_B \cdot q \cdot r$. The latter method has been used in many Belle analyses.

[38] The distribution of the angle between $f$ and the beam axis for $e^+e^- \to f\bar{f}$ (continuum) events has a $1 + \cos^2\theta_{\mathrm{B}}$ shape. However, the reconstructed $\theta_{\mathrm{B}}$ in continuum events is a consequence of random combinations of tracks. The distribution is also affected by acceptance effects. The resulting distribution turns out to be almost uniform.

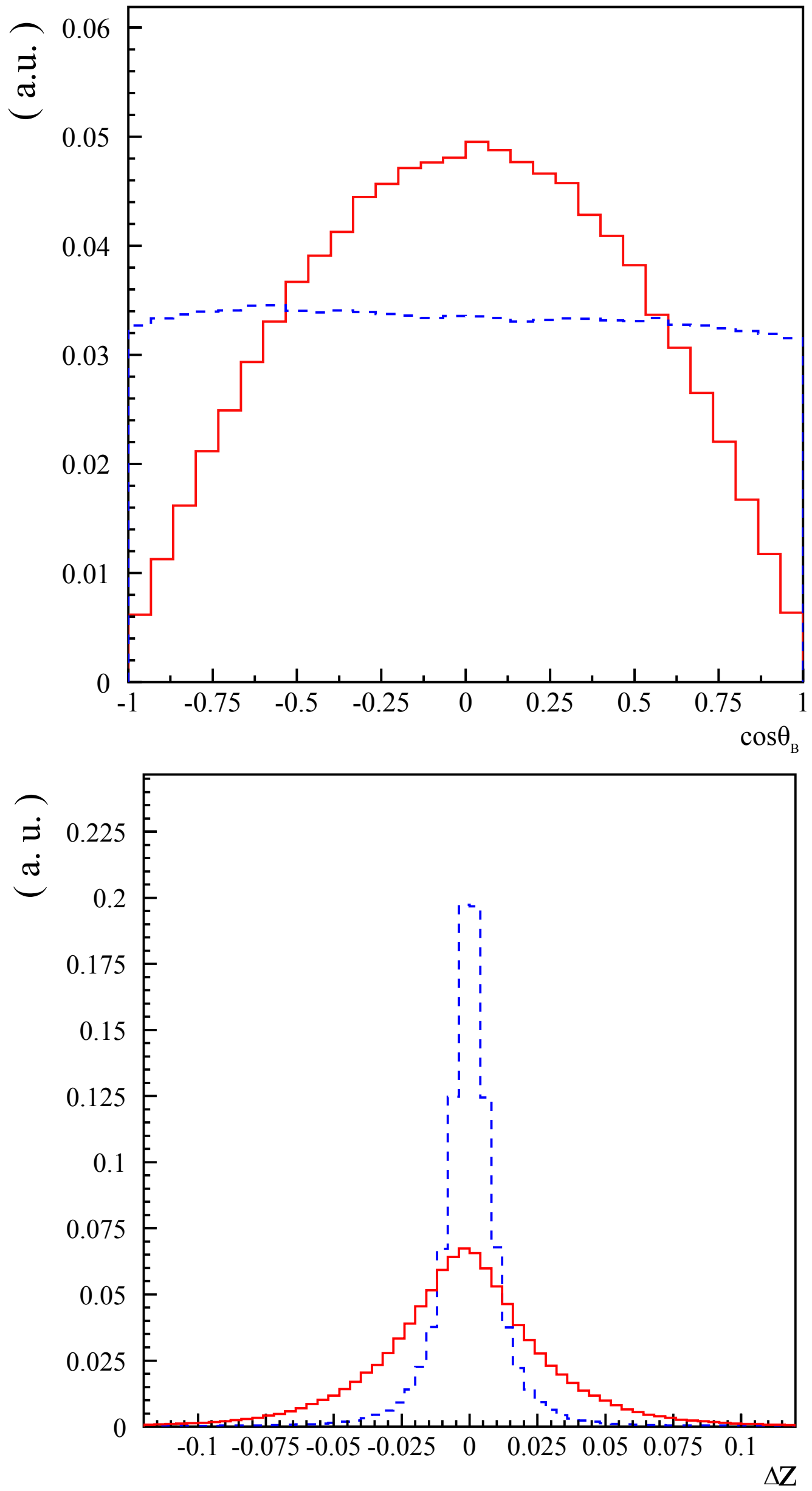

**Figure 9.5.1.** The $\cos\theta_{\rm B}$ (top) and $\Delta Z$ (bottom) distributions for the $B^+ \to K^+\pi^0$ and continuum MC events. Solid red lines are $B$ signal candidates and dashed blue lines are the continuum background. These figures are Belle internal, from the (Duh, 2012) analysis. The vertical scale is in arbitrary units (a.u.).

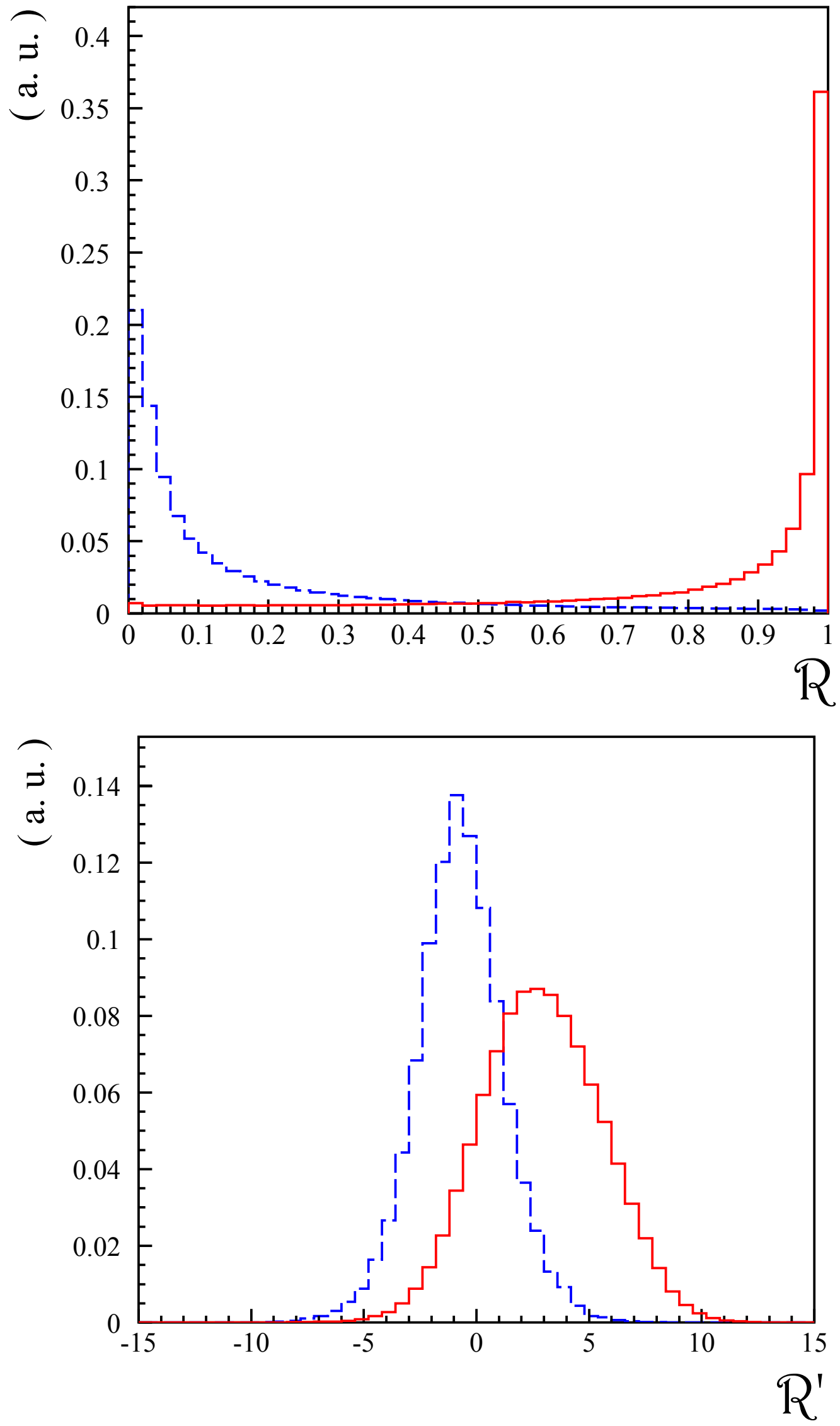

**Figure 9.5.2.** The $\mathcal{R}$ (top) and modified $\mathcal{R}$ (bottom) distributions for the $B^+ \to K^+\pi^0$ and continuum MC events. Solid red lines are $B$ signal candidates and dashed blue lines are the continuum background. The modified $\mathcal{R}$ ($\mathcal{R}'$) is defined after requiring $\mathcal{R} > 0.2$. These figures are Belle internal, from the (Duh, 2012) analysis. The vertical scale is in arbitrary units (a.u.).

To utilize all the available information, in some Belle analyses the variables described above were combined us-ing the neural network technique. One of the popular packages used in Belle is the NeuroBayes package (Feindt and Kerzel, 2006; Phi-T, 2008). For instance, the suppression of the continuum background in the Belle analyses of $B^0 \to D^0 K^{*0}, D^0 \to K^-\pi^+$ (Negishi, 2012) and $B^- \to DK^-, D \to K^+\pi^-$ (Horii, 2011) was achieved using several variables as the NeuroBayes inputs: such as KSFW, $\cos\theta_{\rm T}$, $\cos\theta_{\rm B}$, $\Delta Z$, flavor tagging information $q{\cdot}r$, and the cosine of the angle between the momentum of the kaon candidate from the $D$ decay and the $B$ momentum in the $D$ rest frame. Three more variables are included in the $B^0 \to D^0 K^{*0}$ search: (1) the distance of closest approach between the trajectories of the $K^*$ and $D$ candidates; (2) the difference between the sum of the particle charges in the $D$ hemisphere and the sum in the opposite hemisphere, excluding those used in the reconstruction of the B meson; and (3) the angle between the $D$ and $\Upsilon(4S)$ directions in the rest frame of the $B$ candidate. The advantage of employing the neural network technique is that variables having correlations with each other can be added and their correlations are considered non-linearly. As with the signal-background likelihood ratio, one can make a requirement on the NeuroBayes output to suppress the continuum background or include it in a multi-dimensional likelihood fit to extract the signal yield.

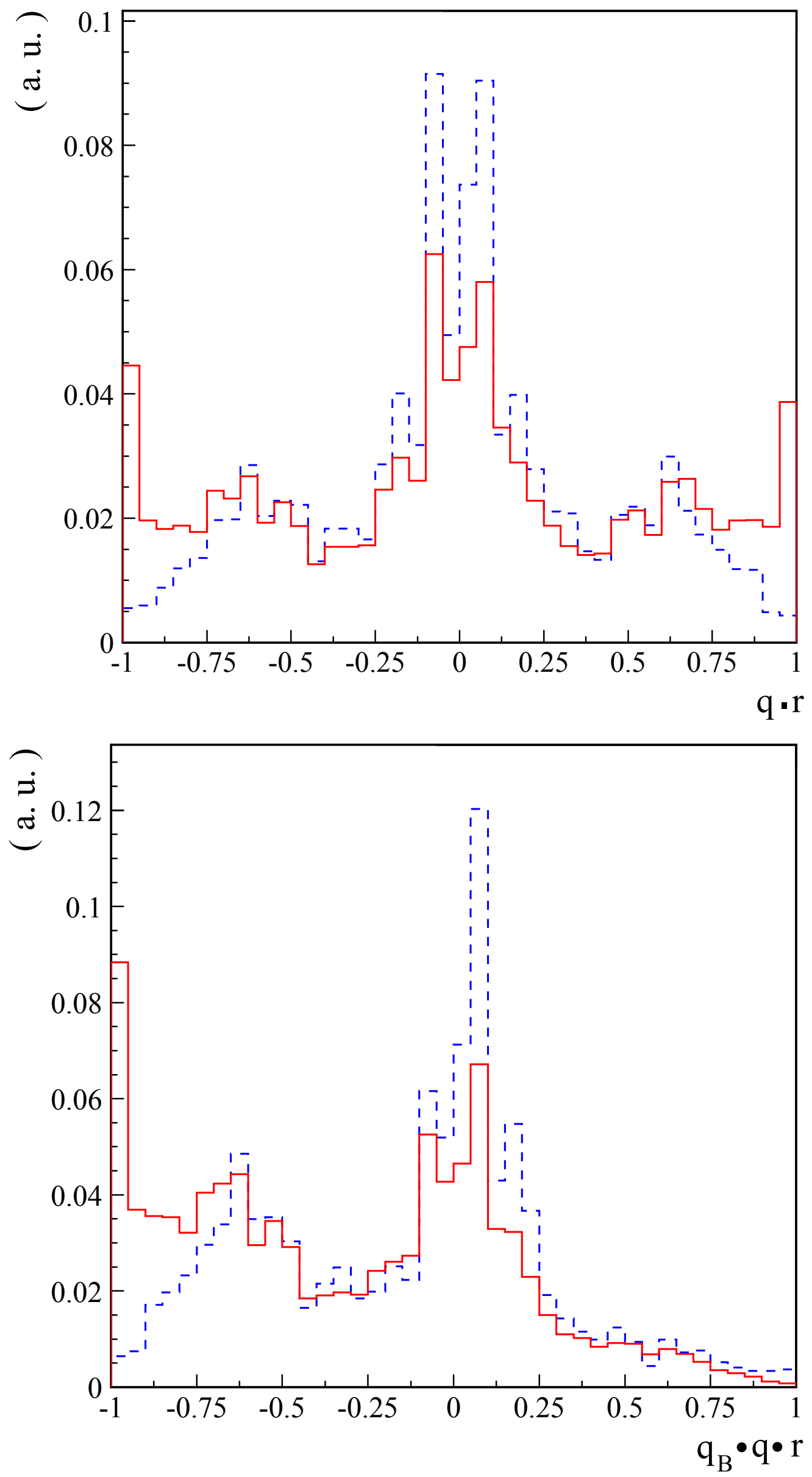

**Figure 9.5.3.** The $q{\cdot}r$ (top) and $q_B{\cdot}q{\cdot}r$ (bottom) distributions for signal (solid red) and continuum MC (dashed blue) events. Signal $B$ events are generated to decay into a flavor specific state. These figures are Belle internal, from the (Wei, 2009) analysis. The vertical scale is in arbitrary units (a.u.).

NeuroBayes is widely used in many high energy experiments. The application, to name a few, ranges from Higgs search (Aaltonen et al., 2009d), studies of single top production (Aaltonen et al., 2010; Chatrchyan et al., 2012a), measuring $B$ and $D$ meson properties (Aaij et al., 2012l; Aaltonen et al., 2011d), and full $B$ meson reconstruction at $B$ factories (Feindt et al., 2011).

## 9.6 Summary

In summary, various techniques of background suppression, mostly inspired by charmless $B$ decay analyses suffering from huge backgrounds, are described in this chapter.

As an illustration, for an analysis of $B \to \eta' h$ ($h = \rho, K^*, \omega, \phi$) (Schumann, 2007) in Belle, the continuum background is suppressed by imposing $q \cdot r$ dependent selections on $\mathcal{R}$. The signal efficiency due to the suppression is (42–88)% and the background is reduced by (98–45)%, depending on the decay mode. The possible improvement by including the variable $\mathcal{R}'$ in the fit for signal extraction is investigated in the $B^+ \to K^+\pi^0$ analysis in Belle. With a lower bound ($lb$) value chosen to be 0.2, the significance (the signal yield from the fit divided by its uncertainty) of the extracted signal is improved by 15%. Note that there may be correlations between $\mathcal{R}'$ and other variables. For instance, the variables $\mathcal{R}'$ and $\Delta E$ for the continuum background is found to be correlated in the $B \to hh'$ analysis (Duh, 2012). Hence, different $\Delta E$ *p.d.f.*s in different $\mathcal{R}'$ regions are implemented in the analysis. Examples of using the NeuroBayes package to include various correlated variables are described in Section 9.5.3. A requirement on the NeuroBayes output in the analysis of $B^- \to DK^-, D \to K^+\pi^-$ (Horii, 2011) retains 96% of the signal and rejects 74% of the background. In the search of $B^0 \to DK^{*0}, D \to K^-\pi^+$, the NeuroBayes output, ranging from $-1$ to 1, is first required to be greater than $-0.6$ to suppress the background, and is then included in the multi-likelihood fit after being transformed using Eq. (9.5.13) with the NeuroBayes output $\mathcal{R}$, $lb = -0.6$ and $ub = 1.0$. The loose cut ($lb = -0.6$) rejects 70.5% of the background, while the signal loss is 3.9%.

For *BABAR*, most analyses of $B \to hh$ channels ($h = \pi, K$) (see Chapter 17.4) followed strategies in line with the generic approach described in Section 9.4.1: a two-step background suppression, starting with simple loose cuts on strongly discriminating variables, then using Fisher discriminants as a discriminating variable in a likelihood fit. At the selection step, signal efficiencies were often adapted to the specific signal-to-background rates for the final state being considered; for example in (Aubert, 2007ay), the cut on the $|\cos\theta_{\rm S}|$ value applied in the $B^+ \to h^+\pi^0$ study was chosen to retain about $\sim 80\%$ of signal while rejecting $\sim 65\%$ of continuum; in contrast, a tighter selection was applied for $B^0 \to \pi^0\pi^0$, as a consequence of its smaller signal-to-background rate. In the same spirit, the final update of the $B^0 \to \pi^+\pi^-$, $K^+\pi^-$ study (Lees, 2013b) applied a looser cut on $|\cos\theta_{\rm S}|$, achieving close to $\sim 90\%$ signal efficiency. Owing to its larger signal purity, in this study both the signal and background parameters of the Fisher *p.d.f.* were extracted from the signal sample itself in the maximum-likelihood fit (instead of being extrapolated from sidebands or simulation control samples), thus minimizing the corresponding systematic uncertainties. The observation of the rare $B^+ \to K^+\overline{K}^0$ and $B^0 \to K^0\overline{K}^0$ decays (Aubert, 2006ai) is another useful illustration of linear discriminants in *BABAR*; the enhancement of signal sensitivity provided by a similar Fisher discriminant was instrumental in establishing the observation of these two rare channels. Concerning nonlinear discriminants, most *BABAR* analyses of charmless $B \to hhh$ decays $(h = \pi, K)$ implemented NN discriminants in line with the generic strategy discussed in Section 9.4.2; at the selection level, typical cuts on the NN value were chosen to retain some $\sim 90\%$ of signal, while rejecting up to $\sim 75\%$ of continuum. For Dalitz-plot analyses such as (Aubert, 2009av), non-negligible correlations between the NN and the Dalitz variables for continuum events were observed, and addressed with a dedicated parameterization; in this way, the $\sim 1.4\sigma$ separation provided by these NN discriminants could be implemented in the likelihood function, and used in the amplitude fits.

# Chapter 10
# Mixing and time-dependent analyses

*Editors:*
*Adrian Bevan (BABAR)*
*Thomas Mannel (theory)*

This Chapter introduces neutral meson mixing, as well as the principles and methods underlying time-dependent analyses in $B$ meson decays. A detailed discussion of experimental concerns for a time-dependent analysis follows on from a theoretical introduction of mixing and time-dependent formalism (Sections 10.1 and 10.2). The experimental aspects discussed here include the use of flavor tagging methods introduced in Chapter 8 and the inevitable dilution of information when the tagging assignment is incorrect (Section 10.3). The impact of the detector resolution on the reconstructed value of the proper time difference between the decays of two neutral mesons and on the measurement of physical observables is raised in Section 10.4. The corresponding time evolution of background events is discussed in Section 10.5. The final part of this chapter discusses how parameters required to describe the mixing and time-evolution of $B$ mesons can be extracted from the data (Section 10.6). Systematic uncertainties common to all time-dependent analyses of $B$ decays are discussed in Section 15.3.

Mixing in the neutral $B$ meson system was discovered by the ARGUS Collaboration (Albrecht et al., 1987b), and Section 17.5 summarizes the measurements of $B$ mixing performed by *BABAR* and Belle. An understanding of mixing in $B$ mesons is one of the ingredients in the study of time-dependent $CP$ asymmetries: in particular, it is crucial for the measurement of the angles of the Unitarity Triangle introduced in Chapter 16, and discussion of measurements of the angles can be found in Sections 17.6 through 17.8. Tests of quantum entanglement, the $CPT$ symmetry, and Lorentz covariance using neutral $B$ mesons, discussed in Sections 17.5.3 through 17.5.5, also rely on a good understanding of mixing. Neutral meson mixing in charm decays was discovered at the $B$ Factories: this is discussed in Section 19.2.

## 10.1 Neutral meson mixing

Meson mixing is a phenomenon that only occurs for the weakly-decaying, open-flavor (*i.e.* not $q\overline{q}$ pairs) neutral $K$, $D$, and $B^0_{d,s}$ mesons. Collectively we can refer to these mesons as $P$ when describing the formalism common to all three systems. The effective Hamiltonian describing neutral meson mixing is given by

$$\mathcal{H}_{\rm eff} = \mathbf{M} - \frac{i\mathbf{\Gamma}}{2} = \left[ \begin{pmatrix} M_{11} & M_{12} \\ M_{21} & M_{22} \end{pmatrix} - \frac{i}{2} \begin{pmatrix} \Gamma_{11} & \Gamma_{12} \\ \Gamma_{21} & \Gamma_{22} \end{pmatrix} \right], \quad (10.1.1)$$

where $\mathbf{M}$ and $\mathbf{\Gamma}$ are two-by-two Hermitian matrices describing the mass and decay rate components of $\mathcal{H}_{\rm eff}$, respectively.

The $CPT$ symmetry imposes that the matrix elements in Eq. (10.1.1) satisfy $M_{11} = M_{22}$ and $\Gamma_{11} = \Gamma_{22}$. In the limit of $CP$ or $T$ invariance in mixing, $\Gamma_{12}/M_{12} = \Gamma_{21}/M_{21}$ is real. Figure 10.1.1 shows the short-distance box diagrams responsible for (top) $D$ and (bottom) $B^0_{d,s}$ mixing transitions in the SM. For the cases of kaons and $D$ mesons these diagrams are dominated by long-distance contributions that are difficult to compute. The long-distance pieces are strongly CKM suppressed only in the case of $B$ mesons for which $M_{12}$ can be computed in perturbation theory. Long-distance contributions are due to real intermediate states whereas the short-distance contributions arise from heavy quark transitions (in particular, the top quark).

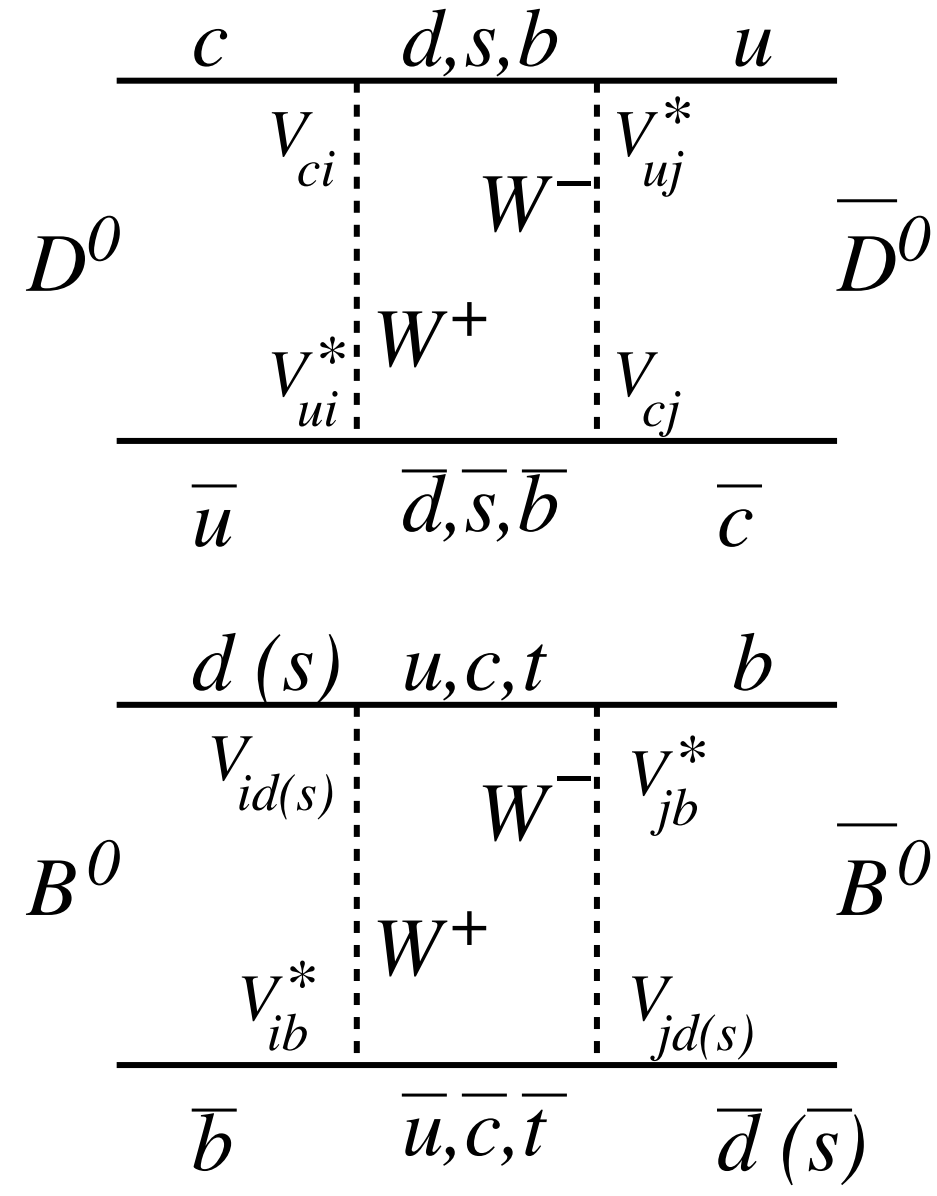

**Figure 10.1.1.** Box diagrams corresponding to the short-distance contributions to neutral meson mixing for (top) $D$ and (bottom) $B^0_{d,s}$ mesons. Each of these contributions is matched by a diagram where the quark triplet, and $W$ bosons are interchanged. The $V_{ij}$ are CKM matrix elements discussed in Chapter 16.

Solving the time evolution represented by the effective Hamiltonian of Eq. (10.1.1) amounts to determining its eigenstates; however, the eigenvalue problem is non-Hermitian, hence the eigenvalues will be complex and the eigenstates will not be orthogonal. This non-Hermiticity and thus the imaginary parts of the eigenvalues lead to a non-unitary time evolution in the two-dimensional subspace spanned by the $B_d$ and the $\overline{B}_d$. As a consequence, probability is not conserved in this subspace, which describes the fact that both mesons will eventually decay and hence disappear from this two-dimensional space.

The eigenstates of the effective Hamiltonian can be represented as an admixture of the flavor eigenstates via

$$|P_{1,2}\rangle = p|P^0\rangle \pm q|\overline{P}^0\rangle, \tag{10.1.2}$$

where $|q|^2 + |p|^2 = 1$ to normalize the wave function, and

$$\frac{q}{p} = \sqrt{\frac{M_{12}^* - \frac{i}{2}\Gamma_{12}^*}{M_{12} - \frac{i}{2}\Gamma_{12}}}, \tag{10.1.3}$$

and the corresponding eigenvalues read

$$m_1 - \frac{i}{2}\Gamma_1 = M_{11} - \frac{i}{2}\Gamma_{11} + \frac{p}{q}\left(M_{12} - \frac{i}{2}\Gamma_{12}\right) \tag{10.1.4}$$

$$m_2 - \frac{i}{2}\Gamma_2 = M_{11} - \frac{i}{2}\Gamma_{11} - \frac{p}{q}\left(M_{12} - \frac{i}{2}\Gamma_{12}\right) \tag{10.1.5}$$

where $m_{1,2}$ are the masses and $\Gamma_{1,2}$ are the widths of the two effective Hamiltonian eigenstates. These states are graphically depicted for various neutral meson systems in Fig. 10.1.2, illustrating their mass and width differences. These two parameters determine the time evolution of a neutral meson that oscillates between the particle and the anti-particle state, as explained in more detail below.

Assuming $m_2 > m_1$ we define $\Delta m = m_2 - m_1 > 0$ and $\Delta\Gamma = \Gamma_2 - \Gamma_1$, and then write the time evolved state that had been a $|P^0\rangle$ at $t = 0$ as

$$|P^0(t)\rangle = g_+(t)|P^0\rangle + \frac{q}{p}g_-(t)|\overline{P}^0\rangle \tag{10.1.6}$$

with

$$g_\pm(t) = e^{-im_1 t}e^{-\frac{1}{2}\Gamma_1 t}\frac{1}{2}\left[1 \pm e^{-i\Delta m\, t}e^{\frac{1}{2}\Delta\Gamma\, t}\right]. \tag{10.1.7}$$

From these relations we can compute the time-dependent decay rates for both $P^0$ and $\overline{P}^0$. If $|f_{CP}\rangle$ is a common final state for both $P^0$ and $\overline{P}^0$, we denote the corresponding decay amplitudes as

$$A_f = \langle f|H_{\Delta F=1}|P^0\rangle \tag{10.1.8}$$

$$\overline{A}_f = \langle f|H_{\Delta F=1}|\overline{P}^0\rangle \tag{10.1.9}$$

where $H_{\Delta F=1}$ is the Hamiltonian for transitions involving a flavor change of one unit. Defining

$$\lambda = \frac{q}{p}\frac{\overline{A}_f}{A_f} \tag{10.1.10}$$

and — following the textbook (Bigi and Sanda, 2000) — the auxiliary variables $K_\pm(t)$ and $L(t)$

$$\begin{aligned} K_\pm(t) &= 4\,e^{\Gamma_1 t}|g_\pm(t)|^2 & (10.1.11)\\ &= 1 + e^{\Delta\Gamma\, t} \pm 2e^{\frac{1}{2}\Delta\Gamma\, t}\cos(\Delta m\, t) \\ L(t) &= 4\,e^{\Gamma_1 t}g_-^*(t)g_+(t) & (10.1.12)\\ &= 1 - e^{\Delta\Gamma\, t} - 2i\,e^{\frac{1}{2}\Delta\Gamma\, t}\sin(\Delta m\, t) \end{aligned}$$

one arrives at

$$\begin{aligned} &\Gamma(P^0(t) \to f) \propto |\langle f|H_{\Delta F=1}|P^0(t)\rangle|^2 & (10.1.13)\\ &= e^{-\Gamma_1 t}|A_f|^2\left[K_+(t) + |\lambda|^2K_-(t) + 2\mathrm{Re}\left\{\lambda L^*(t)\right\}\right] \\ &\Gamma(\overline{P}^0(t) \to f) \propto |\langle f|H_{\Delta F=1}|\overline{P}^0(t)\rangle|^2 & (10.1.14)\\ &= e^{-\Gamma_1 t}|\overline{A}_f|^2\left[K_+(t) + \frac{1}{|\lambda|^2}K_-(t) + 2\mathrm{Re}\left\{\frac{1}{\lambda}L^*(t)\right\}\right]. \end{aligned}$$

These expressions — as well as the resulting *CP* asymmetries — simplify considerably in the cases where some of the parameters are small. For comparison we list the values for the relevant parameters for the various neutral meson systems in Table 10.1.1 The width difference $\Delta\Gamma$ in the kaon system is large compared to the average decay width $\Gamma$ $(= (\Gamma_1 + \Gamma_2)/2 = 1/\tau)$ and the mass difference $\Delta m$; hence, the above expressions are typically expanded in a different way. In the system of neutral $D$ mesons, both the oscillation frequency $\Delta m$ and the width difference $\Delta\Gamma$ are very small compared to the average decay width $\Gamma$. The resulting expressions are given in Section 19.2.

Furthermore, for kaons and $D$ mesons, the expressions for $M_{12}$ and $\Gamma_{12}$ are dominated by long-distance contributions. This makes the theoretical estimates of $\Delta m$ and $\Delta\Gamma$ in these systems difficult to compute.

The situation is simpler for $B$ mesons. The matrix element $\Gamma_{12}$ is strongly CKM suppressed, and thus $\Delta\Gamma$ is small compared to $\Delta m$, and can be set to zero. Furthermore, $\Delta m$ is dominated by the short-distance top quark contribution. We relate $\Delta m$ and $\Delta\Gamma$ to $M_{12}$ and $\Gamma_{12}$ using Eqs (10.1.4) and (10.1.5)

$$\begin{aligned} \Delta m_{d,s}^2 - (\Delta\Gamma_{d,s}/2)^2 &= 4\left[|M_{12}|^2 - |\Gamma_{12}/2|^2\right] \\ \Delta m_{d,s}\Delta\Gamma_{d,s} &= 4\mathrm{Re}(M_{12}\Gamma_{12}^*)\,. & (10.1.15) \end{aligned}$$

Neglecting $|\Gamma_{12}|$ in the above expressions and explicitly calculating the box diagram amplitude for $B_d$ leads to

$$\begin{aligned} \Delta m_d &\simeq 2|M_{12}| & (10.1.16)\\ &= 2\frac{G_F^2M_W^2}{16\pi^2 m_{B_d}}S_0|V_{td}V_{tb}^*|\eta_B\langle B_d|(\bar{b}d)(\bar{b}d)|\overline{B}_d\rangle & (10.1.17) \end{aligned}$$

where $S_0$ is a function of $m_t^2/M_W^2$ whose leading term behaves as $m_t^2/M_W^2$, reflecting the Glashow-Iliopoulos-Maiani (GIM) mechanism (Buras and Fleischer, 1998; Inami and Lim, 1981), $\eta_B$ are the perturbative QCD corrections known to next to leading order (NLO) precision, and $(\bar{b}d)(\bar{b}d)$ is a local $(V - A) \times (V - A)$ operator with $\Delta B = 2$.

For the small width difference $\Delta\Gamma_d$, it follows from Eqs (10.1.15) that

$$\Delta\Gamma_d \simeq 2|M_{12}|\,\mathrm{Re}\left(\frac{\Gamma_{12}}{M_{12}}\right). \tag{10.1.18}$$

Recall that $\Delta m_d$ was defined to be positive; the sign of $\Delta\Gamma_d$ must be determined by experiment.

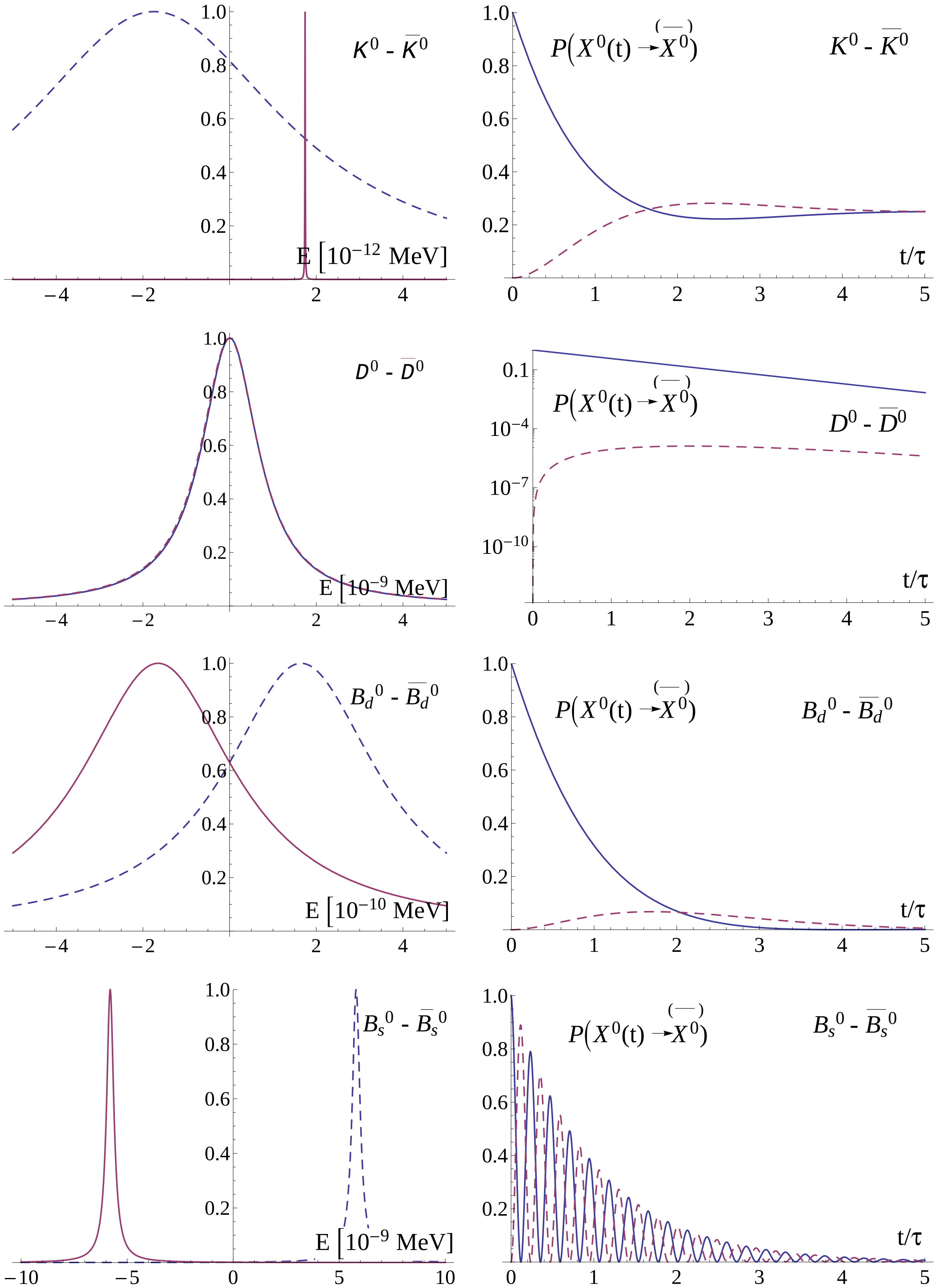

**Figure 10.1.2.** Left: Illustration of mass and width differences of the eigenstates (one denoted by full (red) line and the other by dashed (blue) line) for various neutral meson systems. Right: Probabilities for an initially produced neutral meson to be found after the time $t$ in a particle (full (blue) line) or an anti-particle state (dashed (red) line).

**Table 10.1.1.** Values of the mixing parameters for the different neutral mesons. All numbers are approximate to illustrate the relative sizes.

| Meson | $M$/MeV | $\Delta m$/MeV | $\Gamma$/MeV | $\Delta\Gamma$/MeV |
|---|---|---|---|---|
| $K^0$ | 497.6 | $3.48 \times 10^{-12}$ | $3.68 \times 10^{-12}$ | $7.34 \times 10^{-12}$ |
| $D^0$ | 1864.9 | $9.45 \times 10^{-12}$ | $1.6 \times 10^{-9}$ | $2.57 \times 10^{-11}$ |
| $B_d$ | 5279.6 | $3.34 \times 10^{-10}$ | $4.43 \times 10^{-10}$ | $\sim 0$ |
| $B_s$ | 5366.8 | $1.16 \times 10^{-8}$ | $4.39 \times 10^{-10}$ | $6.58 \times 10^{-11}$ |

With the same assumption $|\Gamma_{12}| \ll |M_{12}|$, it also follows from Eq. (10.1.3) that

$$\left(\frac{q}{p}\right)_d = e^{-i\phi_{M_{12}}}, \tag{10.1.19}$$

where $\phi_{M_{12}}$ is the complex phase of $M_{12}$.

## 10.2 Time-dependent evolution

Neutral $B_d$ mesons (from now on referred to as $B^0$ mesons) are produced via $e^+e^- \to \Upsilon(4S) \to B^0\overline{B}^0$ transitions at *BABAR* and Belle. The wave function for the final state $B$ meson pair is prepared in an anti-symmetric coherent $P$-wave ($L = 1$) state $\Psi$, where

$$\Psi = \frac{1}{\sqrt{2}}\left(|B^0\rangle|\overline{B}^0\rangle - |\overline{B}^0\rangle|B^0\rangle\right). \tag{10.2.1}$$

The $B_d$ mesons remain in this coherent state, where there is always exactly one $B^0$ and one $\overline{B}^0$, until one of them decays. When the first $B$ meson decays, the wave function collapses and the remaining un-decayed $B$ meson will continue to propagate through space-time and oscillate between a $B^0$ and $\overline{B}^0$ state, with a characteristic frequency $\Delta m_d$, until it also decays. This assumes that the $B\overline{B}$ pair is successfully described by quantum mechanics, despite the macroscopic extent of the state; aspects of this assumption can be tested at the $B$ Factories, as discussed in Section 17.5.3.

If one of the $B$ mesons decays into a final state that can be used to unambiguously determine the flavor of the $B$ at the time it decayed, we refer to that as a $B_{\rm tag}$. The set of decay modes of interest as a $B_{\rm tag}$ candidate are referred to as flavor-specific final states. An example of a flavor-specific decay is $B^0 \to D^{(*)-}\ell^+\nu_\ell$, where $\ell = e,\, \mu$. The *CP*-conjugate process has a $\ell^-$ in the final state, so the charge of the final-state lepton is used to identify the flavor of the $B_{\rm tag}$ with a $B^0$ ($\overline{B}^0$) tag originating from a decay with a $\ell^+$ ($\ell^-$). Similarly, if the other $B$ decays into a *CP*-eigenstate or admixture thereof, we refer to that as the $B_{CP}$. Events with one $B_{\rm tag}$ and one $B_{CP}$ are of interest in the study of time-dependent *CP* violation. This sequence is illustrated in Fig. 10.2.1 as seen from the laboratory frame of reference: in this frame, the center-of-mass frame is boosted forward in the direction of the electron (high energy) beam. The $B$ mesons are created almost at rest in the center-of-mass frame.

Having identified the flavor of $B_{\rm tag}$, one can infer the flavor of $B_{CP}$ at the instant the first $B$ meson decays, and the correlated wave function collapses, using the time evolution of the $B^0\overline{B}^0$ system. The detailed study of this system leads to the measurement of so-called time-dependent asymmetries.

The decay times of $B_{CP}$ and $B_{\rm tag}$ in the center-of-mass frame of reference can be labeled as $t_1$ and $t_2$, respectively, and the time evolution of the $B^0\overline{B}^0$ system is a function of $t_1 + t_2$ and $t_1 - t_2$ in general. Assuming a negligible difference between the decay rates of the mass eigenstates (*i.e.* $\Delta\Gamma_d = 0$), the $B_{CP}$ decay rate distribution for $B_{CP}$ decaying into a *CP* eigenstate for a $B^0$ ($\overline{B}^0$) tagged event is given by $f_+$ ($f_-$), following from $g_\pm$ defined in Eq. (10.1.7)

$$f_\pm(\Delta t) = \frac{e^{-|\Delta t|/\tau_{B^0}}}{4\tau_{B^0}}\left[1 \pm \frac{2{\rm Im}\lambda}{1+|\lambda|^2}\sin(\Delta m_d \Delta t) \mp \frac{1-|\lambda|^2}{1+|\lambda|^2}\cos(\Delta m_d \Delta t)\right], \tag{10.2.2}$$

where $\tau_{B^0} \equiv 1/\Gamma_d$ is the $B^0$ meson lifetime and $\lambda$ is given in Eq. (10.1.10). The sign of sine and cosine terms indicated in Eq. (10.2.2) is for a *CP* odd final state such as $J/\psi K^0_S$. *CP* even final states, such as $\pi^+\pi^-$ have the opposite sign conventions for the sinusoidal terms. The proper time difference $t_1 - t_2$ between the decay times of the two $B$ mesons is denoted by $\Delta t$ (see Section 6.5), and terms involving $t_1 + t_2$ drop out. One can compute the time dependence of neutral mesons decaying into flavor-specific final states (so called $B_{\rm flav}$ events), where $\lambda = 0$. These events are used to provide an experimental cross check of the time-dependent measurement and input parameters required to perform time-dependent fits to data (see Section 10.6). Analysis of such decays enables one to measure $\Delta m_d$, where the time dependence becomes

$$h_\pm(\Delta t) = \frac{e^{-|\Delta t|/\tau_{B^0}}}{4\tau_{B^0}}\left[1 \mp \cos(\Delta m_d \Delta t)\right]. \tag{10.2.3}$$

It has been pointed out that, while the assumption $\Delta\Gamma_d = 0$ is valid at the $B$ Factories, improved constraints on this will be required at future experiments in order to verify if one can continue to use this approximation (Bevan, Inguglia, and Meadows, 2011).

The coefficients of the sine and cosine terms in equation (10.2.2) are often referred to in terms of the parameters $S$ and $C$ by the *BABAR* experiment and in terms of

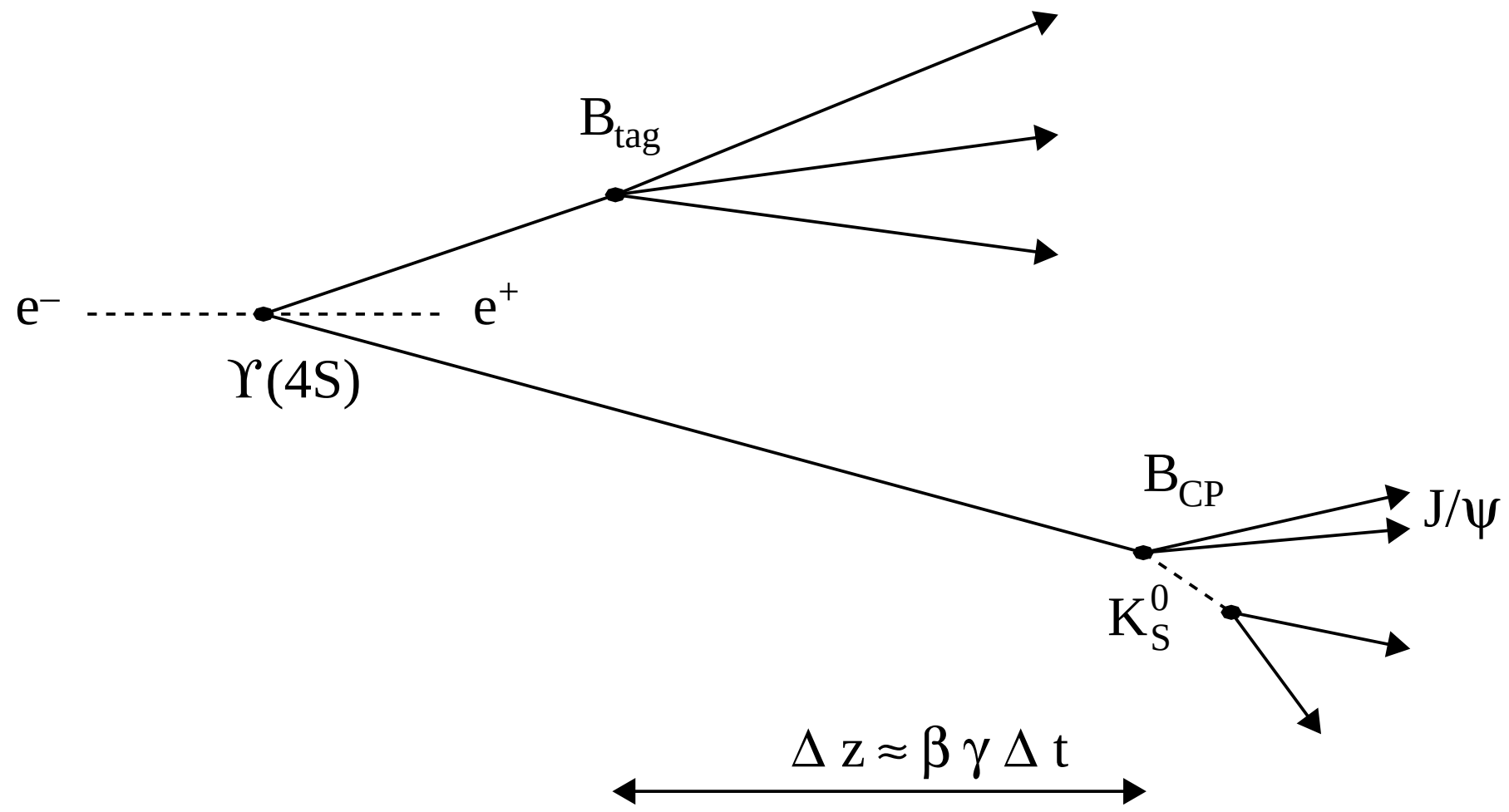

**Figure 10.2.1.** An illustration (not to scale) of a $B$ meson pair decaying in the laboratory frame of reference. On the left hand side of the figure, the initial $e^+e^-$ pair collides producing an $\Upsilon(4S)$. This subsequently decays into two $B$ mesons described by the wave function given in Eq. (10.2.1), one decaying into a $B_{\rm tag}$ final state and the other into a $B_{CP}$ final state. Once the first $B$ meson decays, the remaining one oscillates with the characteristic frequency $\Delta m_d$ before finally decaying. The spatial distance $\Delta z$ between the decay vertices of the $B_{\rm tag}$ and $B_{CP}$ as measured in the laboratory frame of reference is related to the proper time difference $\Delta t$ between the decays of these particles in the center-of-mass frame of reference (see Section 6.5). In this example the $B_{CP}$ final state is $J/\psi K^0_S$.

$S$ and $-A$ by Belle, where

$$S = \frac{2\ {\rm Im}\lambda}{1+|\lambda|^2}, \tag{10.2.4}$$

$$C = -A = \frac{1-|\lambda|^2}{1+|\lambda|^2}. \tag{10.2.5}$$

Note that $S$ and $C$ are related through

$$\Big(\frac{S}{\sin\theta}\Big)^2 + \big(C\big)^2 = 1\ , \tag{10.2.6}$$

where $\theta$ is the phase of $\lambda$.[39] For brevity, we use the notation $S$ and $C$ to refer to these coefficients in the remainder of this book.

An asymmetry between $f_+(\Delta t)$ and $f_-(\Delta t)$ is constructed in order visualize possible *CP* violation. If we neglect experimental effects for the moment, this time-dependent decay-rate asymmetry is given by

$$\mathcal{A}(\Delta t) = \frac{f_+(\Delta t) - f_-(\Delta t)}{f_+(\Delta t) + f_-(\Delta t)}, \tag{10.2.7}$$

which reduces to the form

$$\mathcal{A}(\Delta t) = S\sin(\Delta m_d \Delta t) - C\cos(\Delta m_d \Delta t). \tag{10.2.8}$$

[39] Often the relation between parameters $S$ and $C$ is written in a form of inequality $S^2 + C^2 \leq 1$.

In certain modes, the fitted parameters $S$ and $C$ are related to fundamental parameters of the SM, the angles of the Unitarity Triangle. As discussed in Chapter 16, two notations are used in the literature for these angles. The *BABAR* experiment uses $\beta$, $\alpha$, and $\gamma$ to denote the angles, whereas the Belle experiment reports results in terms of $\phi_1$, $\phi_2$, and $\phi_3$, respectively. In this book, we use the second notation for brevity.

## 10.3 Use of flavor tagging

The purpose of flavor tagging is to classify the $B_{\rm tag}$ either as a $B^0$ or as a $\overline{B}^0$ (see Chapter 8). The performance of the flavor tagging algorithm determines how well the values of $S$ and $C$ can be extracted from the data.

The *BABAR* experiment classifies events according to the information content used in determining the flavor of the $B_{\rm tag}$ meson. These categories of events are ranked in order of decreasing contribution to the total tagging efficiency $Q$ (see Eq. 8.2.1). Thus, the *BABAR* classification is effectively one based on the $B_{\rm tag}$ decay mode. The Belle experiment's algorithm uses the same information but, instead of having distinct categories of events, that algorithm computes a continuous variable that assigns a dilution factor for a given event.

As discussed in Section 8.2, the algorithm for assigning a flavor tag to an event, thus categorizing the tag-side $B$ meson as a $B^0$ or as a $\overline{B}^0$, is not perfect. There is a finite probability to incorrectly tag an event and thus dilute measurements that rely on this information. The mistag probability is denoted by $w_{B^0}$ ($w_{\overline{B}^0}$) for a $B^0$ ($\overline{B}^0$)-tagged event. The value of the mistag probability depends on the $B_{\rm tag}$ final state used, and results in a dilution factor $\langle D\rangle = 1-2\langle w\rangle$ given by Eq. (8.2.2), where $\langle w\rangle$ is the average mistag probability for $B^0$ and $\overline{B}^0$ events (which is often just written as $w$). This dilution factor reduces the amplitude of oscillation from the ideal level (with $D = 1$ when $w_{B^0,\overline{B}^0} = 0$) by some value $D < 1$ for a non-zero mistag probability. The time-dependent formalism developed in Section 10.2 needs to be modified to account for the dilution; indeed one should also account for possible differences in mistag probability between $B^0$- and $\overline{B}^0$-tagged events, denoted by $\Delta w = w_{B^0} - w_{\overline{B}^0}$. Such a difference could be manifest through asymmetries in particle identification, as well as the intrinsic difference in cross section between particles and anti-particles interacting with the matter of the detector. On allowing for dilution effects, the rates of tagged $B^0$ and $\overline{B}^0$ events are given by

$$\begin{aligned} f_+^{\rm Phys} &= (1-w_{B^0})f_+ + w_{\overline{B}^0} f_-, \\ f_-^{\rm Phys} &= (1-w_{\overline{B}^0})f_- + w_{B^0} f_+. \end{aligned} \quad (10.3.1)$$

Taking dilution into account, the time dependence of the physical states given by Eq. (10.3.1) becomes

$$\begin{aligned} f_\pm^{\rm Phys}(\Delta t) = \frac{e^{-|\Delta t|/\tau_{B^0}}}{4\tau_{B^0}} [\, 1 \mp \Delta w & \\ \pm \langle D\rangle S \sin(\Delta m_d \Delta t) & \\ \mp \langle D\rangle C \cos(\Delta m_d \Delta t)]. & \end{aligned} \quad (10.3.2)$$

The observed amplitudes of the sine and cosine terms in the time-dependent asymmetry are suppressed by the average dilution factor $\langle D\rangle$ for $B$ and $\overline{B}$. As $\Delta w$ is small, this factor is sometimes omitted for analyses with a low number of signal events. The analog of the asymmetry given by Eq. (10.2.8) is

$$\mathcal{A}(\Delta t) = \frac{f_+^{\rm Phys}(\Delta t) - f_-^{\rm Phys}(\Delta t)}{f_+^{\rm Phys}(\Delta t) + f_-^{\rm Phys}(\Delta t)} \quad (10.3.3)$$

$$= -\Delta w + \langle D\rangle [S \sin(\Delta m_d \Delta t) - C \cos(\Delta m_d \Delta t)]. \quad (10.3.4)$$

Thus, a non-zero mistag probability $\Delta w$ results in a small offset in $\mathcal{A}(\Delta t)$ at $\Delta t = 0$. Figure 10.3.1 shows the distribution of $\mathcal{A}(\Delta t)$ for $S = 0.7$, $C = 0.0$, and $\Delta w = 0.0$. The amplitude of the sinusoidal oscillation is given by the magnitude of $S$ in the case of a perfectly tagged asymmetry. In reality, dilution effects reduce the measured amplitude relative to the physical one, as illustrated in the figure below with the case of $\langle w\rangle = 0.2$.

The time dependence of events that one typically uses to study mixing ($C = 1$, $S = 0$), allowing for mistagged events, is given by

$$h_\pm^{\rm Phys}(\Delta t) = \frac{e^{-|\Delta t|/\tau_{B^0}}}{4\tau_{B^0}} [1 \mp \Delta w \pm \langle D\rangle \cos(\Delta m_d \Delta t)], \quad (10.3.5)$$

where the $\pm$ index refers to mixed ($-$) and unmixed ($+$) events. Unmixed events have a $B^0\overline{B}^0$ final state whereas mixed events are either $B^0B^0$ or $\overline{B}^0\overline{B}^0$ final states. Given that the distribution is symmetric about $\Delta t = 0$, the modulus of this distribution is shown sometimes when illustrating neutral meson oscillation.

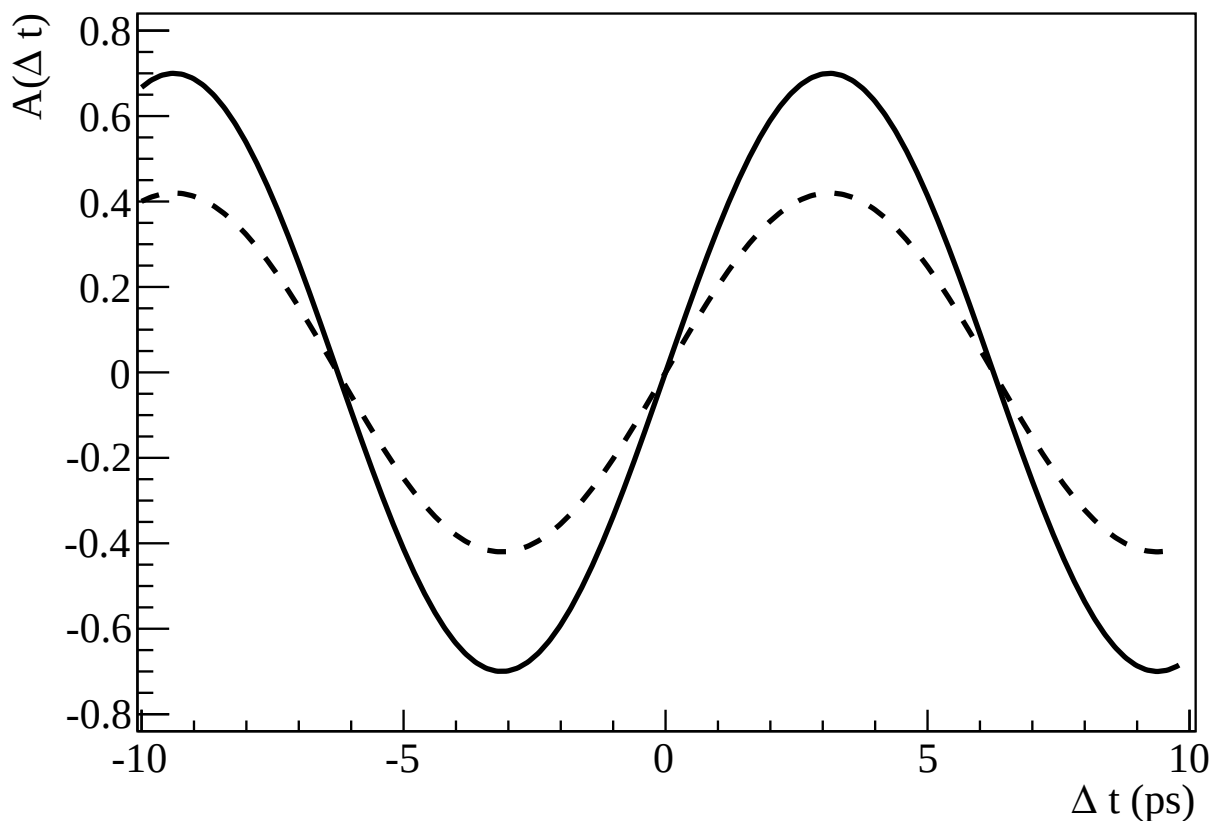

**Figure 10.3.1.** Distributions of the time-dependent *CP* asymmetry with $S = 0.7$, $C = 0$, and $\Delta w = 0$ for (solid) perfect tagging, and (dashed) the corresponding distributions after taking into account dilution with $\langle w\rangle = 0.2$.

## 10.4 Resolution of $\Delta t$

A number of factors contribute to the resolution of the reconstructed value of $\Delta z$, and hence to that of the computed value of $\Delta t \simeq \Delta z/\beta\gamma$. The experimental resolution $R(\delta t, \sigma_{\Delta t})$, as a function of $\delta t = \Delta t - \Delta t_{\rm true}$ and the uncertainty on $\Delta t$, $\sigma_{\Delta t}$, can be accounted for when measuring time-dependent *CP* asymmetry parameters by convoluting $R(\delta t, \sigma_{\Delta t})$ with $f_\pm^{\rm Phys}(\Delta t)$, giving

$$\begin{aligned} F_\pm^{\rm Phys}(\Delta t) &= \int_{-\infty}^{\infty} f_\pm^{\rm Phys}(\Delta t_{\rm true}) R(\delta t, \sigma_{\Delta t}) {\rm d}\Delta t_{\rm true}, \\ &= f_\pm^{\rm Phys}(\Delta t) \otimes R(\delta t, \sigma_{\Delta t}). \end{aligned} \quad (10.4.1)$$

Therefore, one can replace $f_\pm^{\rm Phys}$ with $F_\pm^{\rm Phys}$ in Eqs (10.3.3) and (10.3.4) to obtain the corresponding equations that account for both dilution and resolution effects. Factors contributing to the resolution of $\Delta t$ include:

- $B_{\rm tag}$ vertex resolution, which is a combination of tracking effects and, for a sub-sample of $B_{\rm tag}$ mesons, the finite lifetime of $D$ mesons;
- $B_{CP}$ vertex resolution, which is a superposition of tracking effects; and

– resolution of the measurement of the boost factor $\beta\gamma$ determined from the energy of the $e^+$ and $e^-$ beams.

It is important to understand the $\Delta t$ resolution in detail as this is of a similar magnitude to the average separation between the $B_{CP}$ and $B_{\rm tag}$ proper decay times. Thus, this resolution has a significant effect on the extraction of $S$ and $C$ from a time-dependent analysis.

Different approaches are used to understand resolution effects at the $B$ Factories. *BABAR* adopts a parametric approach to describe the $\Delta t$ resolution, whereas Belle characterizes resolution effects according to their physical source. Both approaches work well and provide a good description of resolution for use in time-dependent analyses.

The nominal *BABAR* $\Delta t$ resolution function has a triple Gaussian form, where the mean $\mu_i$ and width $s_i$ of the two central Gaussian components are scaled by $\sigma_{\Delta t}$ on an event-by-event basis. The three Gaussians are denoted by $G_i$, where $i =$ core, tail, and outlier, in order of increasing width. The resolution function is given by

$$\begin{aligned}\mathcal{R}_{\rm sig}(\delta t, \sigma_{\Delta t}) = {} & f_{\rm core} G_{\rm core}\left(\delta t, \mu_{\rm core}\sigma_{\Delta t}, s_{\rm core}\sigma_{\Delta t}\right) + \\ & f_{\rm tail} G_{\rm tail}\left(\delta t, \mu_{\rm tail}\sigma_{\Delta t}, s_{\rm tail}\sigma_{\Delta t}\right) + \\ & f_{\rm outlier} G_{\rm outlier}\left(\delta t, \mu_{\rm outlier}, s_{\rm outlier}\right) .\end{aligned} \tag{10.4.2}$$

The parameters $s_{\rm tail}$, $s_{\rm outlier}$ and $\mu_{\rm outlier}$ are set to 3.0, 8.0 ps and 0.0 ps, respectively, and the other parameters are determined from reference samples of fully reconstructed $B$ meson decays as described in Section 10.6. The tail width was determined from Monte Carlo simulated data, and the outlier mean was taken as unbiased, with a width varying from $4-12$ ps. The mean of this range was taken as the nominal value for $s_{\rm outlier}$. As the physical tagging categories for *BABAR* have different purities and dilutions, the values of $\mu_i$ and $s_i$ for the core Gaussian contribution to the resolution function depend on the flavor category of an event. This difference is taken into account when analyzing data. For early analyses, each of the *BABAR* flavor tagging categories had a separate value for $\mu_{\rm core}$ and $s_{\rm core}$; in later iterations, the distinction was only made between `Lepton` and non-`Lepton` tagging categories. For *BABAR* data, $s_{\rm core}$ is typically $1.01 \pm 0.04$ $(1.10 \pm 0.02)$ for `Lepton` (non-`Lepton`) events.

The Belle $\Delta t$ resolution function (Tajima, 2004) accounts for four different physical effects

– $B_{\rm tag}$ vertex resolution,
– $B_{CP}$ vertex resolution,
– shift in the $B_{\rm tag}$ vertex position resulting from secondary tracks from charm meson decays, and
– kinematic approximation that the $B$ mesons are at rest in the center-of-mass frame.

The $B_{\rm tag}$ and $B_{CP}$ vertices are described by (i) a Gaussian resolution function in the case of multi-track vertices, and (ii) a sum of two Gaussians in the case of single-track vertices. The widths of these Gaussians are scaled by the uncertainty on the reconstructed vertex being described. The resolution function resulting from non-prompt tracks associated with a decay in flight of charm mesons is described by the sum of a delta function and exponentials. The kinematic approximation is described by a resolution function dependent on the polar angle of $B_{\rm tag}$ as reconstructed in the center-of-mass frame of reference. Given that a $B_{CP}$ or $B_{\rm flav}$ candidate is fully reconstructed, and decays opposite the $B_{\rm tag}$ in the center-of-mass frame of reference, whereas the $B_{\rm tag}$ may not be, the polar angle of the $B_{\rm tag}$ candidate is determined from the fully reconstructed $B_{CP}$ or $B_{\rm flav}$ decay. The physical time dependence $f_\pm^{\rm Phys}$ is convoluted by each of these resolution functions in turn in order to obtain the resultant $F_\pm^{\rm Phys}$.

Figure 10.4.1 shows the $f_\pm^{\rm Phys}$ and $F_\pm^{\rm Phys}$ distributions for $S = 0.7$ and $C = 0.0$, where both dilution and resolution effects are considered. The distribution $f_\pm^{\rm Phys}$ is smeared out considerably as a result of experimental resolution when computing $F_\pm^{\rm Phys}$. The effect of dilution serves to reduce the reconstructed asymmetry between $B^0$- and $\overline{B}^0$-tagged events. This can be seen as a reduction in the asymmetry between $F_+$ and $F_-$ in comparison with the true distributions $f_+$ and $f_-$.

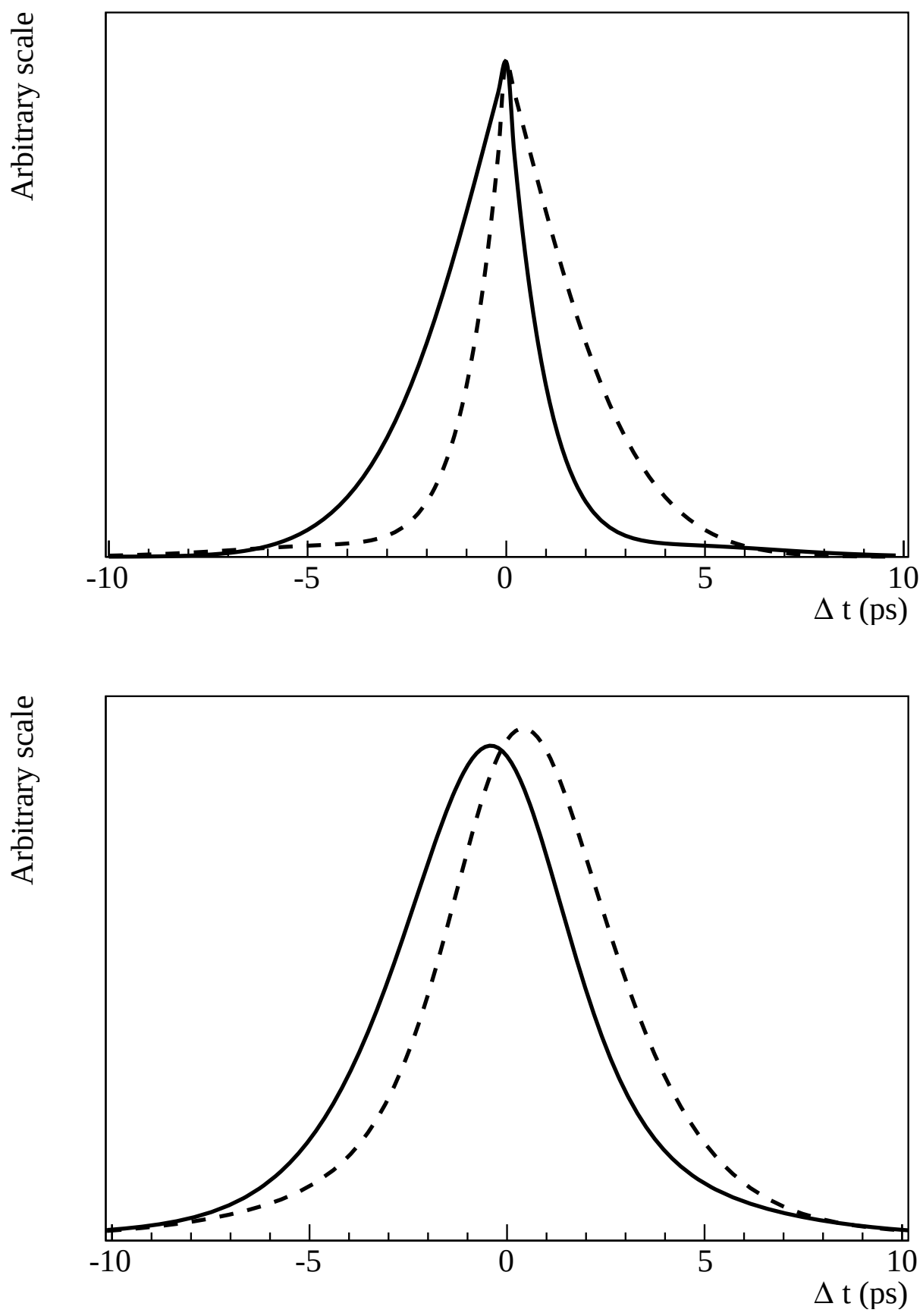

**Figure 10.4.1.** Distributions of (top) $f_\pm^{\rm Phys}(\Delta t)$ with $S = 0.7$, and $C = 0.0$ for (solid) $B^0$- and (dashed) $\overline{B}^0$-tagged events for perfectly reconstructed decays, and (bottom) the corresponding distributions $F_\pm^{\rm Phys}$ after taking into account typical dilution and resolution effects.

## 10.5 Modeling the $\Delta t$ distribution for background events

Generically, one can categorize three types of background that are encountered in time-dependent analyses at the $B$ Factories: (i) continuum events, (ii) $B$ background including charm mesons that decay in flight, and (iii) other $B$ background categories. The effect on the time-evolution of each of these types of events from the resolution of $\Delta t$ needs to be considered. The following describes the general approach adopted for each of these types of background.

- The hadronization processes resulting from continuum $e^+e^- \to q\overline{q}$ background, where $q = u, d, s$, or $c$, occur on a time scale too small to measure. As a result, the time dependence for this type of background is assumed to be a prompt distribution modeled using a $\delta$ function convoluted with the resolution function. The resolution function typically adopted for continuum background is a simplified version of Eq. (10.4.2), where the scale factors $s_{\rm tail}$ and $s_{\rm outlier}$ are set to 2.0 ps and 8.0 ps, respectively, and only the core Gaussian mean and scale factor are weighted by $\sigma_{\Delta t}$. The remaining parameters of the background resolution function are obtained from fits to data.
- The time evolution of $B$ background events that contain charm particles is biased as a result of the assumption that all tracks in the $B_{CP}$ vertex originate from the same point whereas, in reality, the tracks from the charm meson in the event originate from a secondary vertex that is displaced from the $B_{CP}$ vertex. This type of background can occur in the analysis of charmless $B$ decays and, where necessary, the time dependence is assumed to be similar to the signal one, except that the lifetime is taken to be different from $\tau_{B^0}$. An effective lifetime is extracted from samples of Monte Carlo simulated data and used in place of $\tau_{B^0}$ for this type of background. Cross checks using control fits to data validate the approximation of using Monte Carlo simulated data to determine the effective lifetime. A signal resolution function is assumed to be valid for this category of events.
- The time evolution of $B$ background events that do not contain charm particles is assumed to be the same as that for signal. Such backgrounds occur in time-dependent measurements of charmless $B$ decay processes. While these events will be mis-reconstructed as a given hypothesized signal mode, the differences observed between the resolution functions for signal Monte Carlo simulated data and $B$ background Monte Carlo simulated data are small. Some analyses perform systematic cross checks where the time dependence is given by a kernel estimation *p.d.f.* corresponding to the $\Delta t$ distribution observed for Monte Carlo simulated data in order to account for any bias. Such a distribution is formed from the sum of kernels, one for each event in a control sample. In this case Gaussian kernels are used with a mean corresponding to the value of $\Delta t$ of a given event, and a width given by the RMS of the ensemble of data in the control sample. As such a model neglects the per-event uncertainty on $\Delta t$, when this approach is used, a systematic cross check is performed where the kernel estimation *p.d.f.* is replaced with a signal-like time dependence.

Both $B$ Factories categorize continuum background with a prompt distribution as described above. *BABAR* treats background from different types of $B$ decays as indicated above, whereas Belle assigns an exponentially decaying distribution convoluted with the resolution function as the *p.d.f.* for $B$ background events. The lifetime assumed for the Belle $B$ background *p.d.f.* is an effective one determined from Monte Carlo simulated data.

It is possible that background events may themselves be *CP* violating. In such cases, one can account for the level of *CP* violation by ensuring that the time dependence incorporates the asymmetry given in Eq. (10.2.8) for neutral $B$ decays, or the corresponding time-integrated asymmetry for charged $B$ decays. This issue is discussed in Section 15.3.5.

## 10.6 Parameter extraction from data

In order to perform a time-dependent analysis, one needs to determine the values of $w$, $\Delta w$, and the tagging efficiencies, which are collectively referred to as tagging parameters, and the resolution function parameters required to evaluate the convolution of $f_\pm(\Delta t)$ with $R(\delta t, \sigma_{\Delta t})$. A sample of neutral $B$ mesons decaying into flavor-specific final states is used to determine these parameters. Several hundred thousand events were in the control samples used by the $B$ Factories. The set of modes used by *BABAR* for this is $B^0 \to D^{(*)-}(\pi^+, \rho^+, a_1^+)$, whereas Belle uses $B^0 \to D^{(*)-}\pi^+$, $D^{*-}\rho^+$, $D^{*-}\ell^+\nu$ as well as the charmonium decays $J/\psi K^0_S$, and $J/\psi K^*(892)^0$. No flavor tag information is used by Belle when extracting the parameters using the charmonium decays. *BABAR* only uses the $B \to D^*\ell^-\nu$ sample to perform a cross-check as there is a larger background in that mode than the other control sample channels. Collectively, this ensemble of flavor-specific decay modes is referred to as the $B_{\rm flav}$ control sample in the following. In addition to determining tagging and resolution function parameters for use in extracting information on *CP* asymmetries from neutral $B_{\rm flav}$ modes, a set of charged control samples is also used to perform a number of independent validation checks. One of these validations is the determination of $S$ for a sample of charged $B$ decays. As $S$ is physically related to the $B^0-\overline{B}^0$ mixing amplitude, the fitted value for this parameter in a sample of charged $B$ decays should be consistent with zero. The charged $B$ control sample is formed using $B^+ \to J/\psi K^+$, $J/\psi K^*(892)$, $\psi(2S)K^+$, $\chi_{c1}K^+$, and $\eta_c K^+$ in the case of *BABAR*, while $B^+ \to J/\psi K^+$ and $\overline{D}^0\pi^+$ are used by Belle. The corollary of using a set of control modes is that, for each mode used to determine the parameters of interest, one introduces additional parameters relating to the shape of distributions of signal and background events, and the purity of each control channel in the signal region. Having determined the purities for each $B_{\rm flav}$ mode, one can use these events to extract estimates of tagging and resolution parameters. This procedure implicitly assumes that there is no significant interference on the tag side of the event (see Section 15.3.6), so that the mistag probabilities computed from the $B_{\text{flav}}$ sample are the same as those on the $B_{CP}$ side of the event. While this assumption was valid for the $B$ Factories, the precision of measurements at a super flavor factory may require that one formally accounts for tag-side interference in the time dependence of the neutral meson system.

In order to determine tagging efficiencies, one simply needs to determine the fractions of the $B_{\text{flav}}$ sample reconstructed in each of the physical categories; to determine the mistag probabilities and differences, one needs to account for $B^0 - \overline{B}^0$ mixing in the $B_{\text{flav}}$ control sample. The time evolution of these decays, neglecting resolution effects, is given by Eq. (10.3.5). One can account for experimental resolution by convoluting $h_{\pm}$ with a resolution function as described in Section 10.4:

$$\begin{aligned} H_{\pm}^{\text{Phys}}(\Delta t) &= \int_{-\infty}^{\infty} h_{\pm}^{\text{Phys}}(\Delta t_{\text{true}}) R(\delta t, \sigma_{\Delta t}) \mathrm{d}\Delta t_{\text{true}}, \\ &= h_{\pm}^{\text{Phys}}(\Delta t) \otimes R(\delta t, \sigma_{\Delta t}). \end{aligned} \tag{10.6.1}$$

Therefore, it is possible to not only extract the tagging parameters but also the resolution function parameters from the $B_{\text{flav}}$ sample, where one assumes that the $\Delta t$ resolution function is the same for the $B_{\text{flav}}$ and $B_{CP}$ events. There are many more events in the $B_{\text{flav}}$ sample than the $B_{CP}$ sample; hence, a more precise determination of the resolution function parameters can be obtained using the $B_{\text{flav}}$ data. Tagging performance is discussed in Chapter 8, and vertex resolution is discussed in Chapter 6.

Given the complexity of the situation, the extraction of parameters related to the tagging performance and $\Delta t$ resolution is done in a two-step process. The first step involves extracting the purity of each of the $B_{\text{flav}}$ decay modes used. Having done this, one determines the tagging and resolution function parameters from the ensemble of $B_{\text{flav}}$ modes. The result of this process is a set of parameters and the corresponding error matrix that can be subsequently used as input parameters for the time-dependent analyses described in Chapter 17. In a number of cases, the time-dependent asymmetry parameters are extracted from a simultaneous fit to both the $B_{CP}$ and $B_{\text{flav}}$ samples so that tagging and resolution parameters are transparently propagated into the $CP$ analysis.

# Chapter 11
# Maximum likelihood fitting

***Editors:***
*Wouter Verkerke (BABAR)*

## 11.1 Formalism of maximum likelihood fits

The final step in a physics analysis, after appropriate event selection and reconstruction steps have been performed, is extracting a statement on a physics parameter of interest from the observed distribution of events in the data. To make such an estimation, a model must be formulated that describes the expected distribution of the observable quantities $\boldsymbol{x}$ for a given set of physics parameters of interest $\boldsymbol{p}$. Then, given an observed data sample $\boldsymbol{x_0}$ one uses the relation between $\boldsymbol{x}$ and $\boldsymbol{p}$ described by the model to infer a statement on the value $\boldsymbol{p}$ for which the observed data is most likely. A standard technique to make such an inference is a maximum likelihood estimator. In this section the basics of this technique are described, starting with a description of probability density function as a means to model the observed data density, followed by a brief description of the maximum likelihood formalism and a discussion on the structure of typical models used for $B$-physics data modeling.

### 11.1.1 Probability Density Functions

For many analyses, the models of observable distributions are described with a *probability density function* (p.d.f.) for the observable quantities $\boldsymbol{x}$:

$$f(\boldsymbol{x};\boldsymbol{p}). \tag{11.1.1}$$

Such a probability density function is positive definite, and normalized to unity over the allowed range of the observable $\boldsymbol{x}$ for any value of $\boldsymbol{p}$, i. e.

$$\forall \boldsymbol{p} : \int f(\boldsymbol{x};\boldsymbol{p}) d\boldsymbol{x} \equiv 1, \tag{11.1.2}$$

where the integral is over the allowed domain of the observables $\boldsymbol{x}$.

In addition to the parameter(s) of interest $\boldsymbol{p}$, realistic models often incorporate a set of additional 'nuisance parameters' $\boldsymbol{q}$ that represent quantities that affect the relation between $\boldsymbol{p}$ and $\boldsymbol{x}$ that are not a priori known and must be simultaneously inferred from the data. Examples of such nuisance parameters are resolution parameters and flavor tagging efficiencies (see Section 10 for details). The model is thus defined as

$$f(\boldsymbol{x};\boldsymbol{p},\boldsymbol{q}). \tag{11.1.3}$$

### 11.1.2 Maximum Likelihood estimation of model parameters

The basis of parameter inference using a model $F$ and observed data is the *likelihood*, defined as the probability density function evaluated at the measured data point $\boldsymbol{x_0}$:

$$L(\boldsymbol{p},\boldsymbol{q}) = f(\boldsymbol{x_0};\boldsymbol{p},\boldsymbol{q}). \tag{11.1.4}$$

The likelihood is then treated as a function of the parameters $\boldsymbol{p}$ and $\boldsymbol{q}$.

For measurements consisting of an ensemble of data points the likelihood of the ensemble is simply the product of the likelihood of each observation:

$$L(\boldsymbol{p},\boldsymbol{q}) = \prod_{i=0,\ldots,N} f(\boldsymbol{x_i};\boldsymbol{p},\boldsymbol{q}), \tag{11.1.5}$$

where $\boldsymbol{x_i}$ represent independent and identically distributed measurements of the observable $x$. In practice one often uses the negative log-likelihood

$$-\log L(\boldsymbol{p},\boldsymbol{q}) = -\sum_{i=0,\ldots,N} \log f(\boldsymbol{x_i};\boldsymbol{p},\boldsymbol{q}), \tag{11.1.6}$$

instead of the likelihood as this is numerically easier to calculate.

Equation (11.1.5) defines an *unbinned* likelihood - the likelihood is evaluated at each data point and no binning of the data is needed. The (unbinned) maximum likelihood estimator $\widehat{\boldsymbol{p}}$ for a parameter vector $\boldsymbol{p}$ is defined as the value of $\boldsymbol{p}$ for which the likelihood is maximal or, equivalently, the negative log-likelihood is minimal.

For an analysis with a very large number of observed events and a small number of observables, it can be efficient to minimize a binned log-likelihood instead, defined as

$$-\log L(\boldsymbol{p},\boldsymbol{q}) = -\sum_{i=0\ldots N} n_i \cdot \log f(\boldsymbol{x_i};\boldsymbol{p},\boldsymbol{q}), \tag{11.1.7}$$

where $x_i$ and $n_i$ represent the bin center and event count of bin $i$ of a histogram with $N$ bins. The computation time scales with the number of bins $N$ rather than the number of events. A binned likelihood is a priori less precise than an unbinned likelihood as the information of the precise position of the event in each bin is discarded, but at small bin sizes this may be a negligible loss of precision. In practice, the prediction $f(\boldsymbol{x_i};\boldsymbol{p},\boldsymbol{q})$ in each bin is often approximated with the value of the probability density function at the bin center, where the integral of the p.d.f. over the bin volume should be used. This approximation has little impact if the bin size is chosen sufficiently small, but can otherwise result in biases in sharply falling or rising distributions, e. g. in the fitted lifetime of exponential decay distributions.

The traditional $\chi^2$ fit is related to the binned maximum likelihood (ML) fit by inserting the additional assumption that the uncertainty can be interpreted as Gaussian, however, this assumption is a poor approximation of reality for bins with low statistics (roughly $n < 10$).

The properties of likelihood estimators are extensively described in the literature (Edwards, 1992). In the asymptotic limit of infinite statistics maximum likelihood estimators (ML estimators) are so-called ideal estimators: they are *consistent*, meaning that they give the correct answer in the limit of infinite statistics, *unbiased*, meaning that they give the correct answer on average for finite statistics, and *efficient*, meaning that the variance of the estimated parameter values is equal to the bound of the expectation value of the variance predicted by the second derivative of the log-likelihood. On finite samples, ML estimators are not ideal, but nevertheless generally well behaved if samples statistics are sufficiently large. However, some particular care must be exercised when using ML estimators for problems with very small (signal) event counts: in these cases bias terms appear in the likelihood, which are generally proportional to $1/N_{obs}$, where $N_{obs}$ is the number of observed events, and may be non-negligible compared to the statistical uncertainty, which is approximately proportional to $1/\sqrt{N_{obs}}$.

### 11.1.3 Estimating the statistical uncertainty using the likelihood

The simplest way to measure the statistical uncertainty $\sigma(\widehat{p})$ on the estimate of a single parameter $\widehat{p}$ is to estimate the variance $V(\widehat{p})$ of that parameter and calculate the uncertainty as the square-root of the variance. The ML estimator for the variance on $\widehat{p}$ is given by the second derivative of the log-likelihood at $p = \widehat{p}$:

$$\sigma(\widehat{p})^2 = V(\widehat{p}) = \left( \frac{d^2 \log(L(p))}{d^2 p} \right)^{-1}_{p=\widehat{p}}. \qquad (11.1.8)$$

In case there are multiple parameters, the variance of the ensemble of parameters is represented by the covariance matrix defined as

$$V(p,p') = \langle pp' \rangle - \langle p \rangle \langle p' \rangle \,, \qquad (11.1.9)$$

and can be estimated as

$$\widehat{V}(p,p') = \left( \frac{\partial^2 \log(L(p,p'))}{\partial p \partial p'} \right)^{-1}_{p=\widehat{p},p'=\widehat{p}'}, \qquad (11.1.10)$$

A multivariate covariance can also be expressed in terms of scalar variances and a correlation matrix

$$V(p,p') = \sqrt{V(p)V(p')} \cdot \rho(p,p'). \qquad (11.1.11)$$

Here $\rho(p,p')$ expresses the linear correlation between parameters $p$ and $p'$ and has values in the range $[-1,1]$ by construction.

An alternative estimator for the uncertainty on a parameter is based on an interval defined by the log-likelihood ratio

$$\lambda(p) = \log \frac{L(p)}{L(\widehat{p})}, \qquad (11.1.12)$$

where $L(p)$ is the likelihood for a given value $p$, $\widehat{p}$ is the value of $p$ for which the likelihood is maximal and $L(\widehat{p})$ is therefore the maximum value of the likelihood. An interval in $p$ defined by a rise in the log-likelihood-ratio of half a unit from zero corresponds to nominally a 68% confidence interval. Intervals defined this way are related to classic frequentist confidence intervals — under the condition that Wilks' theorem[40] (Wilks, 1938) holds.

When nuisance parameters are present, an interval can be defined for each parameter replacing the likelihood ratio with the profile likelihood ratio

$$\lambda_P(p) = \log \frac{L(p, \widehat{\widehat{\boldsymbol{q}}}(p))}{L(\widehat{p}, \widehat{\boldsymbol{q}})}, \qquad (11.1.13)$$

where $\widehat{p}$ and $\widehat{\boldsymbol{q}}$ represent again the ML estimates of parameters $p$ and $\boldsymbol{q}$ and $\widehat{\widehat{\boldsymbol{q}}}(p)$ represents the *conditional* ML estimate of parameters $\widehat{\boldsymbol{q}}$ for a given value of $p$.

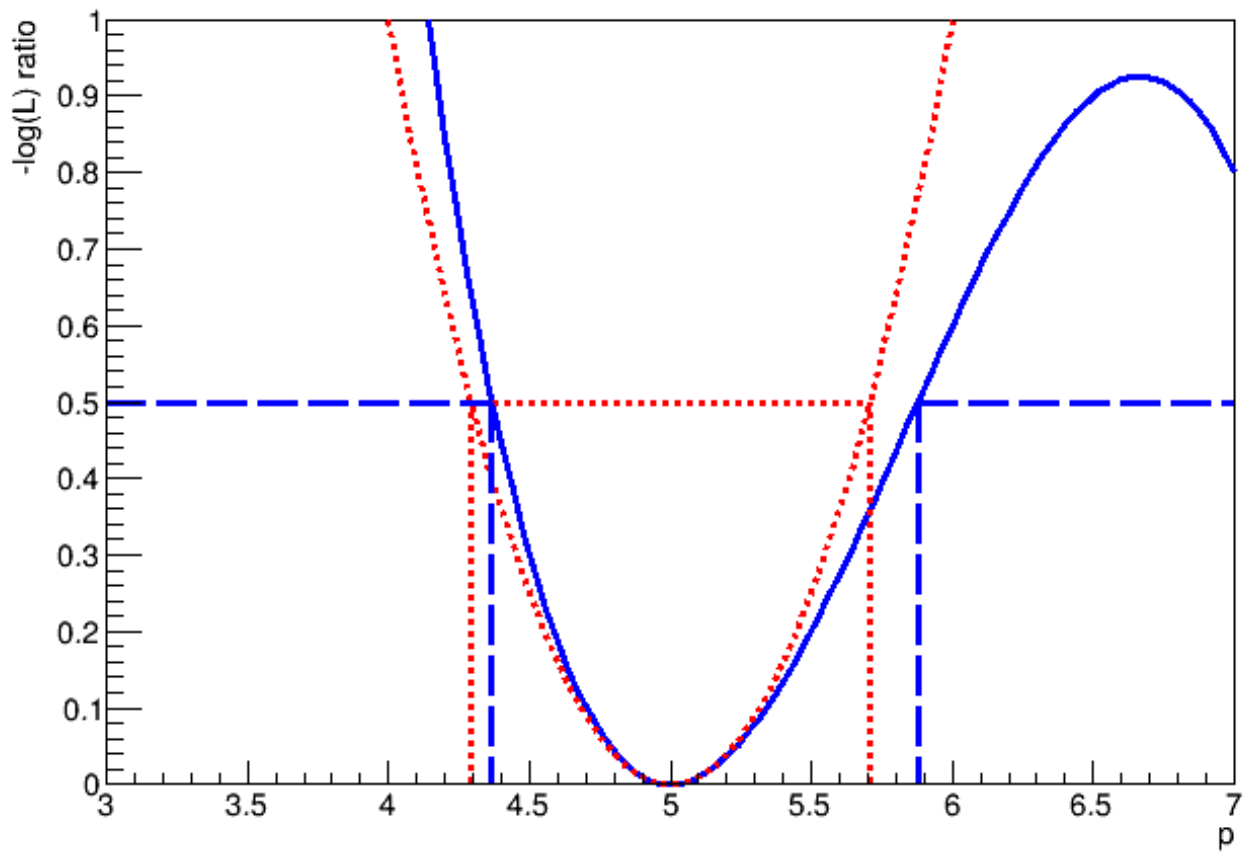

**Figure 11.1.1.** Illustration of the definition of parameter uncertainties in an example log-likelihood ratio (blue solid curve). The variance estimator (`HESSE` , see Section 11.1.5.2) of Eq. (11.1.8) uses the second derivative at $\widehat{p}$ (here $\widehat{p} = 5$) and corresponds to assuming a parabolic log-likelihood ratio shape (red dashed curve) and defining the interval by the intersection points of the parabola with the horizontal line at +0.5 units. The likelihood ratio estimator (`MINOS` , see Section 11.1.5.2) of Eq. (11.1.12) defines the interval using the intersection of the actual log-likelihood ratio curve with a horizontal line at +0.5 units (blue curve, long dashes).

The difference between the variance-based uncertainty and the likelihood-ratio-based uncertainty is visualized in Fig. 11.1.1. If the log-likelihood has a perfectly parabolic shape, as is expected in the limit of infinite statistics (under certain regularity conditions), both uncertainty estimates will give the same interval.[41] At low statistics differences may occur due to the different methods of estimating the uncertainty. In particular, the profile likelihood-based intervals can yield asymmetric intervals around the central values.

---

[40] Wilks' theorem states that the likelihood ratio $\lambda(p)$ will be asymptotically $\chi^2$ distributed under certain regularity conditions as the samples sizes approaches infinity.

[41] The 2nd derivative will perfectly predict the value of the parameter where the log of the likelihood ratio has increased by half a unit from zero in this case.

### 11.1.4 Hypothesis testing and significance

Most measurements of $CP$-violating parameters are expressed as interval estimates. Conversely, the result of a search for a rare signal is usually not expressed as an interval on a signal (strength) parameter, but rather as a test of the background-only hypothesis.

The *significance* of the observation is the probability of the background-only hypothesis to result in the observed signal strength, or larger. This probability is known as the $p$-value. A $p$-value threshold of $1.2{\cdot}10^{-7}$ – corresponding to the probability of a $5\sigma$ Gaussian fluctuation – is conventionally taken to reject the background-only hypothesis, and to declare the discovery of a new signal.

To calculate the $p$-value one must construct a test statistic as function of the data that distinguishes the background-only hypothesis (the 'null hypothesis') from the signal-plus-background hypothesis (the 'alternate hypothesis'). A common choice is $\lambda_P(0)$ of Eq. (11.1.13), where $p$ is the signal strength, so that $\lambda_P(0)$ becomes the ratio of the maximum likelihood of the background-only model and the maximum likelihood of the signal-plus-background model. A dataset that is perfectly consistent with the background-only hypothesis will thus have $\lambda_P(0) = 0$, as the numerator and denominator of Eq. (11.1.13) are equal, whereas datasets with increasing signal strength will result in increasing values of $\lambda_P(0)$. The $p$-value is then calculated as the fraction of experiments sampled from the background-only hypothesis that result in a value $\lambda_P(0)$ that is as large as the observed value or larger:

$$p = \int_{\lambda_P^{obs}(0)}^{\infty} f(\lambda_P(0)|p=0)\,d\lambda_P(0), \qquad (11.1.14)$$

where $\lambda_P^{obs}(0)$ is the value of $\lambda_P(0)$ observed in the data, and $f(\lambda_P(0)|p=0)$ is the expected distribution of $\lambda_P(0)$ values for the background-only hypothesis.

Customarily the significance is re-expressed as a Gaussian fluctuation of $Z\sigma$ that results in the same $p$-value, where $Z$ is defined as

$$p = \int_{-\infty}^{Z\sigma} \frac{1}{\sqrt{2\pi}\sigma} e^{-x^2/(2\sigma^2)}dx, \qquad (11.1.15)$$

and can be calculated from $p$ using the inverse of the error function.[42]

In the asymptotic regime of large statistics, and under certain regularity conditions (Wilks' theorem), $f(\lambda_P(0)|0)$ becomes a $\log(\chi^2)$ distribution with one degree of freedom for each parameter-of-interest. The significance expressed in Gaussian standard deviations can in that case be directly related to the value of $\lambda_P^{obs}(0)$:

$$\lambda_P^{obs}(0) = \frac{1}{2}Z^2. \qquad (11.1.16)$$

Finally, for the specific and simple case of a likelihood describing a counting experiment with an expected signal count $s$ and background count $b$, both with Gaussian uncertainties, the value of $Z$ can be directly expressed as

$$Z_{sb} = \frac{s}{\sqrt{s+b}}, \qquad (11.1.17)$$

but it should be noted that the assumption of Gaussian uncertainties for $s < 10$ or $b < 10$ is poor.

### 11.1.5 Computational aspects of maximum likelihood estimates

For all but a handful of textbook examples, the expression for maximum likelihood estimator for $\widehat{p}$ cannot be expressed analytically, hence the maximum likelihood estimate is computed numerically. The computational problem factorizes into two pieces: definition of the likelihood function for a given problem, and heuristic searches for the maximum of the likelihood function.

#### 11.1.5.1 Likelihood definition

The definition of the likelihood involves coding the definition of the probability density function that is used to model the data, and then evaluating the natural log of this p.d.f. for each observed data point.

The `Root` framework (Brun and Rademakers, 1997) implements definitions of basic functional shapes such as polynomials and Gaussian distributions, but the complexity of models used in typical $B$ Factory analyses is such that they cannot be expressed in terms of this limited set of basic functions. For the first round of $B$ Factory measurements custom software packages were developed that implemented the probability density functions representing the physics models as `Fortran`, `LISP`, or C++ functions.

In the next iteration, the `RooFit` toolkit (Verkerke and Kirkby, 2003) was developed by the *BABAR* collaboration that allowed one to build probability density functions of arbitrary complexity inside the `Root` framework with a minimum amount of custom code. To this end, `RooFit` defines generic software objects that represent observables, probability density functions defining basic shapes as well as $B$-physics specific shapes, and operator objects that allow a user to combine basic shapes through addition, multiplication and convolution. Over time a large number of analyses have migrated to using `RooFit` to encode their likelihood functions. The package has been available in the `Root` framework since 2005. Such models were either coded 'by hand', or for certain complicated models constructed by higher level packages that automate building of `RooFit` p.d.f.s with a certain structure from an configuration file.

[42] In `Root` this calculation is easily accessible as function `RooStats::PValueToSignificance(double pvalue)`

#### 11.1.5.2 Likelihood minimization

The standard tool used by the HEP community for nearly forty years for minimization and uncertainty estimation is the `Minuit` package (James and Roos, 1975), originally written in `Fortran`. A version translated in C++ is now available in the `Root` analysis framework, as well as a new version, `Minuit2`, that was written from scratch in C++ by the original authors. The main components of the `Minuit` package are three algorithms that operate on an user-defined (likelihood) function: `MIGRAD` , `HESSE` , and `MINOS` .

`MIGRAD` is a heuristic algorithm that searches for minima in externally provided multi-variate functions and follows mostly a strategy based on a steepest descent algorithm following a numerically calculated gradient of the input function. Convergence is declared when the input function is within a preset estimated distance from the function value in the nearest minimum assuming a quadratic form. The algorithm has been demonstrated to work well on problems with a very large number of dimensions (> 100), but computational cost increases with the dimensionality.[43] An inherent difficulty with a heuristic search algorithm is distinguishing between local minima and the global minimum. In most cases, the algorithm will settle on the first minimum it finds along its search trajectory, even if this is not the true global minimum. The odds of finding the true global minimum increase if the search is started at a point close to where it is expected to be, putting a premium on an educated guess by the analyzer for the starting values of the algorithm. It is almost impossible to prevent the finding of local minima.

`HESSE` calculates the covariance matrix by sampling the likelihood in small steps around the minimum found by `MIGRAD` and calculating the second derivative from these samples. Its output is the covariance matrix as defined in Eq. (11.1.10). The calculation takes $\frac{1}{2}N^2$ likelihood samplings, where $N$ is the number of parameters allowed to vary, which for large $N$ may exceed the calculation spent in `MIGRAD` minimization.

`MINOS` performs the calculation of the uncertainty interval defined by an increase in the negative log-likelihood of half a unit[44] with respect to the assumed global minimum. When a `MINOS` error calculation is requested for all $N$ parameters of a fit, a $N-1$ dimensional hypersurface is first reconstructed that is defined by $\lambda(\boldsymbol{p}) = 0.5$. The $N$-dimensional hyper-cube that encloses this hypersurface defines the `MINOS` uncertainty on each parameter. Through geometrical arguments it can be shown that the uncertainty defined this way is identical to that of Eq. (11.1.13) for each parameter. `MINOS` calculations can be prohibitively time consuming for a large number of parameters (roughly $N > 30$), but it is also possible to perform a `MINOS` calculation on any subset of the parameters. In such cases `MINOS` uses Eq. (11.1.13) to reduce the parameter space to the desired subset.

[43] The cost of numeric derivative calculations increases linearly with the number of parameters. The number of descent steps required to find the minimum typically increases also with the number of parameters, but is strongly dependent on the shape of the likelihood.

[44] The default `MINOS` value of the increase is 1 unit, as built-in `Root` fitting functions pass two times the value of the negative log-likelihood. Conversely, `Minuit` instances owned by `RooFit` reconfigure the `MINOS` error definition to half a unit.

## 11.2 Structure of models for signal yield measurements and rare decay searches

The probability density functions used as models in $B$ Factory analyses serve two main goals: analysis of the data in terms of a signal and a background component, and if needed inference of the physics parameters of interest. This section covers techniques used to describe the data in terms of signal and background.

The simplest model $M$ to extract a signal yield from the data in the presence of background is a model that describes the data sample as a sum of a signal and background components.

$$m(\boldsymbol{x};\boldsymbol{p},\boldsymbol{q}) = f \cdot s(\boldsymbol{x};\boldsymbol{p}) + (1-f)\cdot b(\boldsymbol{x};\boldsymbol{q}). \qquad (11.2.1)$$

In this equation, $s(\boldsymbol{x};\boldsymbol{p})$ is the model of a signal distribution in the observables $\boldsymbol{x}$, $b(\boldsymbol{x};\boldsymbol{q})$ is a model of the background distribution, and $f$ is the fraction of signal in the data.

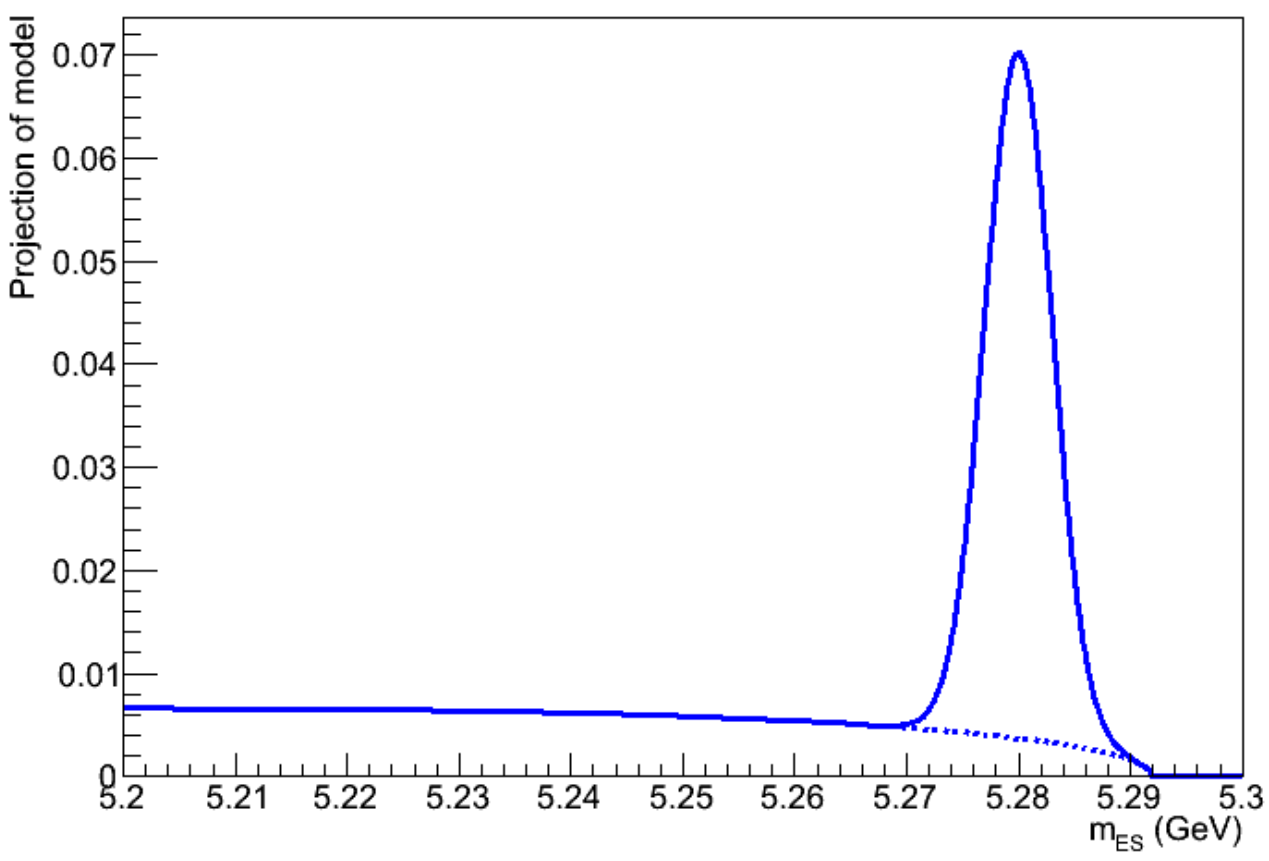

**Figure 11.2.1.** A simple composite probability density model (solid line) consisting of a background component defined by an Argus function (dashed line) plus a signal component defined by a Gaussian function.

Figure 11.2.1 shows an example of a simple version of such a model where the signal is described by a Gaussian distribution of the energy-substituted mass $m_{ES}$ and the background by an Argus function (Albrecht et al., 1990a) that models the kinematics of continuum background events for this observable. See Section 9 for more details on p.d.f. choices to describe signal and background.

With sufficient statistics, the shape parameters $\boldsymbol{p}$ and $\boldsymbol{q}$ of both signal and background can be constrained from the data, in addition to the parameter of interest $f$: the fraction of signal events in the data. The estimate of the number of signal events in data is then $f$ times the total number of observed events.

In the model of Eq. (11.2.1) the p.d.f. only models the shape of the distribution of the observed events and not its count, hence the parameter of interest can only be a fraction, and not a yield. As one is usually interested in the latter in the context of a measurement, the likelihood formalism can be extended to also include the event count of the sample so that a yield can be obtained straight from the fit.

### 11.2.1 Extended ML formalism

In the extended maximum likelihood formalism (EML) (Barlow, 1990) the normalization of the model is not fixed to one, but to a parameter $N_{\rm exp}$, so that the likelihood expression effectively becomes

$$L(\boldsymbol{p},\boldsymbol{q}) = \left(\prod_{i=0\ldots N_{obs}} f(\boldsymbol{x_i};\boldsymbol{p},\boldsymbol{q})\right) \cdot \mathrm{Poisson}(N_{\rm obs}|N_{\rm exp}(\boldsymbol{p},\boldsymbol{q})), \tag{11.2.2}$$

where $N_{\rm obs}$ is the observed event count, modeled by a Poisson distribution with the expected event count $N_{\rm exp}(\boldsymbol{p},\boldsymbol{q})$ as mean. The likelihood of a composite model with a signal and background term can then be rewritten in the EML formalism taking

$$m(\boldsymbol{x};\boldsymbol{p},\boldsymbol{q}) = \frac{N_S}{N_S+N_B}\cdot s(\boldsymbol{x};\boldsymbol{p}) + \frac{N_B}{N_S+N_B}\cdot b(\boldsymbol{x};\boldsymbol{q}), \tag{11.2.3}$$

as the probability density function and

$$N_{\rm exp} = N_S + N_B, \tag{11.2.4}$$

as the expression for the expected event count. A minimization of the extended likelihood will now directly return the estimates for the signal and background event yields $N_S$ and $N_B$.

Often, we may assume that the shapes of the component distributions and the numbers of events are uncorrelated. That is, $N_S$ etc are not dependent on $\boldsymbol{p}$ and $\boldsymbol{q}$. In this case the extended likelihood information does not improve the precision of the measurement of $N_S$ and $N_B$, as the fit can always tune $N_{\rm exp}$ to match $N_{\rm obs}$ exactly for every possible value of $\boldsymbol{p}$, $\boldsymbol{q}$ and $f \equiv N_S/(N_S+N_B)$.

The extended ML formalism in this form is thus mostly used for notational convenience in $B$ Factory analyses, allowing one to directly extract signal event yields from the fits, and to write sums of more than two components in a straightforward form with yield parameters for every component

$$m(\boldsymbol{x};...) = N_S\cdot s(\boldsymbol{x};\boldsymbol{p}) + \sum_i N_B^i b^i(\boldsymbol{x};\boldsymbol{q^i}), \tag{11.2.5}$$

where the index $i$ runs over all background components and $b^i$ denotes the model for background component $i$, with parameters $\boldsymbol{q^i}$.

### 11.2.2 Extending a model to multiple dimensions

In searches for rare decays, a single observable often does not contain sufficient information to distinguish signal from background and the information of multiple observables must be used. Several strategies can be followed to include the information contained in additional observables. One way is to preselect events using cuts in these additional observables in order to obtain a subsample enriched in signal events, and to restrict the signal extraction fit to the original observable. Another strategy – one that is often used for $B$ Factory analyses and which maximizes the statistical precision – is to extend the signal and background models to describe the distributions in these additional observables, effectively constructing a multidimensional probability density function that is fit to the full event sample.

For observables that are uncorrelated, a multidimensional model can be constructed as a simple product of one-dimensional p.d.f.s, *e.g.*

$$f(x,y,z;\boldsymbol{p}) = f_1(x;\boldsymbol{p_1})\cdot f_2(y;\boldsymbol{p_2})\cdot f_3(z;\boldsymbol{p_3}), \tag{11.2.6}$$

where the $f_1, f_2, f_3$ represent normalized one-dimensional probability density functions. In case there are expected correlations between observables, *e.g.* between $x$ and $y$, these must be modeled inside a higher-dimensional p.d.f. $f(x,y)$. This may be accomplished, for example, through the inclusion of conditional probability density functions

$$f(x,y;\boldsymbol{p}) = f_1(x|y;\boldsymbol{p_1})\cdot f_2(y;\boldsymbol{p_2}), \tag{11.2.7}$$

where $f_1(x|y)$ is the conditional probability density in $x$ for a given value of $y$, *i.e.*

$$\forall y,\boldsymbol{p_1} : \int f_1(x|y;\boldsymbol{p_1})dx \equiv 1, \tag{11.2.8}$$

which describes the distribution of $x$ for each given value of $y$, and $f_2(y)$ describes the distribution in $y$. Advantages of the formalism with conditional p.d.f.s are that correlations are often easier to formulate in this way and that all normalization integrals remain one-dimensional. The latter is of particular importance if numeric integration is needed, which is substantially more difficult in two or more dimensions at the level of precision required for `Minuit` minimization. The downside of conditional p.d.f.s is that the normalization integral must be calculated for each value of $y$ separately, which may be computationally expensive, in case the integration needs to be performed numerically.

Apart from their construction, the use of multidimensional probability density functions presents no new technical or conceptual issues in ML estimation, but visualization and validation of multidimensional p.d.f.s introduce some additional issues.

A multi-dimensional model can be most simply visualized by projecting it on one of its observables:

$$P_{yz}(x) = \int f(x,y,z)dydz. \tag{11.2.9}$$

In the case of a factorizing model as defined in Eq. (11.2.6) the projection integral simply reduces to $f_1(x)$ and is trivial to calculate. If correlations are present, the integral must be explicitly calculated.

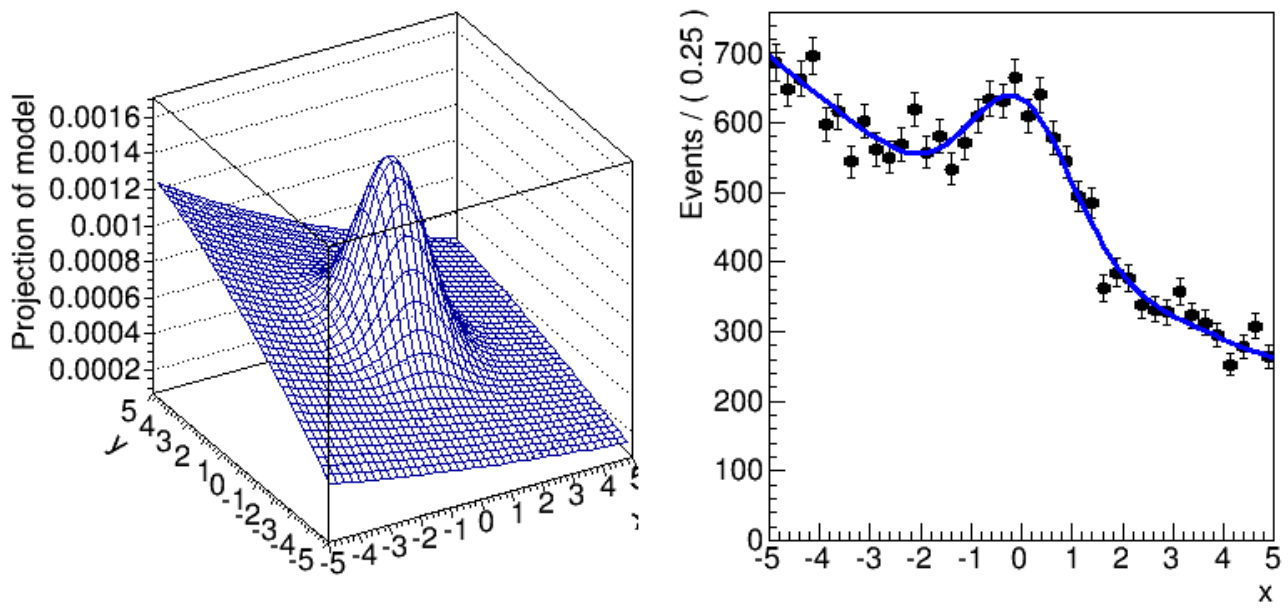

**Figure 11.2.2.** A two-dimensional probability density function consisting of a linear background and a Gaussian signal. On the left the probability density of the model is shown as a function of $x$ and $y$. On the right the projection of the model on the observable $x$ is overlaid on the distribution of a simulated data sample.

A conceptual issue with plain projection plots is that they include the full background and are not suitable to visualize the presence of a small signal in the data that is concentrated in a restricted region of the observable phase space. This is demonstrated in Fig. 11.2.2, which visualizes a two-dimensional model with a linear background and a Gaussian signal concentrated in the central region: while the signal is clearly visible in the central region, it is washed out in the projection plot. This can be mitigated by only projecting a 'signal region' defined in the projected observable $x$

$$P_{y^{SR}}(x) = \int_{y_{SR}^{min}}^{y_{SR}^{max}} F(x, y)dy, \qquad (11.2.10)$$

where the interval $y_{SR}^{min}$ to $y_{SR}^{max}$ represents the region in the observable $y$ that is enhanced in the signal.

**Likelihood ratio plots**. In the search for rare decays many observables are typically used and the signal may not be confined to an easily definable signal region as was possible in the example of Fig. 11.2.2. In these cases, projections of the data and model on a single observable can be defined using a likelihood ratio, rather than a series of cuts on each of the projected observables.

For such a plot, the signal and background models are first integrated over the plotted observable $x$ to obtain the signal and background probabilities according to these models using *only* the information contained in the projected observables $\boldsymbol{y}$ and then combined in a likelihood ratio as follows:

$$LR(\boldsymbol{y}) = \frac{\int S(x, \boldsymbol{y})dx}{\int \left(f \cdot S(x, \boldsymbol{y}) + (1-f)B(x, \boldsymbol{y})\right) dx}. \qquad (11.2.11)$$

A likelihood ratio projection plot is then constructed by taking all parameters ($f$ in the example above) at their estimated values from the data, and by only plotting the data that meet a criterion $LR(\boldsymbol{y}) > \alpha$, where $\alpha$ is a threshold in the predicted signal probability (between 0 and 1), and projecting the model with corresponding selection

$$P_{\boldsymbol{y}}^{LR}(x) = \int_{LR(\boldsymbol{y})>\alpha} F(x, \boldsymbol{y})d\boldsymbol{y}. \qquad (11.2.12)$$

The integral over the region defined by $LR(\boldsymbol{y}) > \alpha$ is clearly not calculable analytically, even if the model itself is, but can be approximated with a Monte Carlo integration technique as follows

$$C(x; \boldsymbol{p}, \boldsymbol{q}) = 1/N_D \sum_{D_{LR}(\boldsymbol{y})} F(x; \boldsymbol{y}, \boldsymbol{p}, \boldsymbol{q}), \qquad (11.2.13)$$

where $D_{LR}(\boldsymbol{y})$ is a pseudo-experiment dataset with $N_D$ events, sampled from the p.d.f. $F(x, \boldsymbol{y})$ from which all events that fail the requirement $LR(\boldsymbol{y}) > \alpha$ have been removed. Figure 11.2.3 shows an example of a likelihood ratio plot defined using a three-dimensional extension of the model shown in Fig. 11.2.2 projecting over the $y$ and $z$ dimensions using a likelihood ratio cut with a value of 0.7.

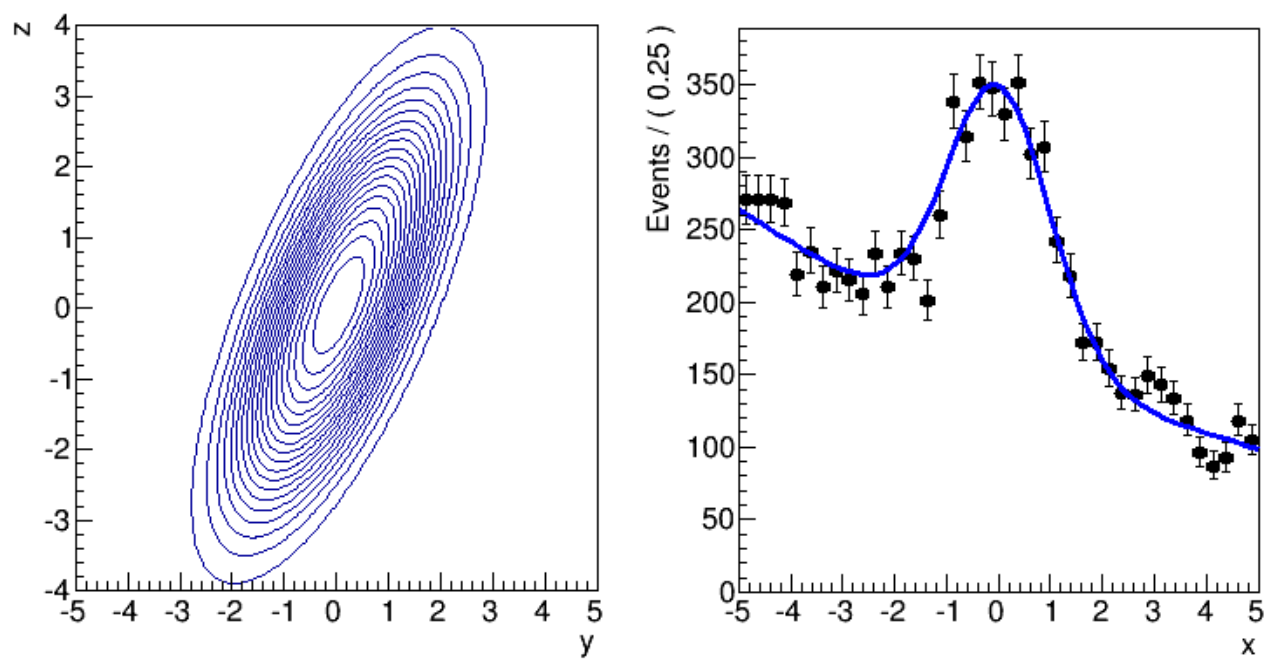

**Figure 11.2.3.** Visualization of a three-dimensional model, similar to that of Fig. 11.2.2. On the left a contour plot with constant values of likelihood ratio defined by Eq. (11.2.11) of a model in the observables $y$ and $z$ is shown. On the right the projection of the model on the observable $x$ is shown, requiring $LR(y, z) > 0.7$ for both data and model to enhance the visibility of the signal.

### 11.2.3 ${}_s\mathcal{P}lots$

A challenge in multi-dimensional models with a large number of observables is to verify that each component describes the data well in all observables. For factorizing p.d.f.s, Eq. (11.2.6), a new technique named ${}_s\mathcal{P}lot$ has been developed at the $B$ Factories (Pivk and Le Diberder, 2005) to facilitate such studies.

In the ${}_s\mathcal{P}lot$ technique the distribution in observable $x$ is predicted using the distribution in all of the other variables, $\boldsymbol{y}$, which must be uncorrelated to $\boldsymbol{y}$, and can be compared to the direct model prediction in $x$. The central concept in ${}_s\mathcal{P}lot$ is the definition of the ${}_sWeight$

$$ {}_s\mathcal{P}_n(\boldsymbol{y}) = \frac{\sum_{j=1}^{n_c} V_{nj}^{-1} \cdot F_j(\boldsymbol{y})}{\sum_{k=1}^{n_c} N_k \cdot F_k(\boldsymbol{y})}, \tag{11.2.14} $$

where $n$ is the selected component of a model consisting of $n_c$ components (*e.g.* signal and one or more backgrounds). In this expression the indices $j, k$ run over the $n_c$ model components, $F_j$ is the p.d.f. for component $j$ in the observables $\boldsymbol{y}$, $N_k$ is the expected number of events for the $k^{\text{th}}$ component, and $V_{nj}^{-1}$ is the inverse of the covariance matrix $V_{nj}$ in these yield parameters. The matrix $V_{nj}$ is obtained from the data, either through a numeric summation over the per-event contributions using Eq. (11.1.10), or from `HESSE` following a maximum likelihood fit to the data. Note that ${}_sWeights$ can be negative, as $V_{nj}$ is not positive definite. The predicted distribution for any component $j$ in observable $x$ is given by the histogram of events in $x$ where each event contributes with a weight ${}_s\mathcal{P}_n(\boldsymbol{y})$.

An example is shown in Figure 11.2.4, where for a 3-dimensional model in observables $m_{ES}, \Delta E, \mathcal{F}$, the p.d.f. in $m_{ES}$ for signal and background are compared with the ${}_s\mathcal{P}lots$ in this observable, calculated using ${}_sWeights$ that use exclusively the data and the model prediction in observables $\Delta E, \mathcal{F}$. In this example the data was simulated and has been sampled from the model itself and perfect agreement is observed between the p.d.f. and the ${}_s\mathcal{P}lot$ prediction. When applied on samples of observed data, discrepancies between the ${}_s\mathcal{P}lot$ and the direct model prediction may occur, which may be indicative of disagreements between data and model.

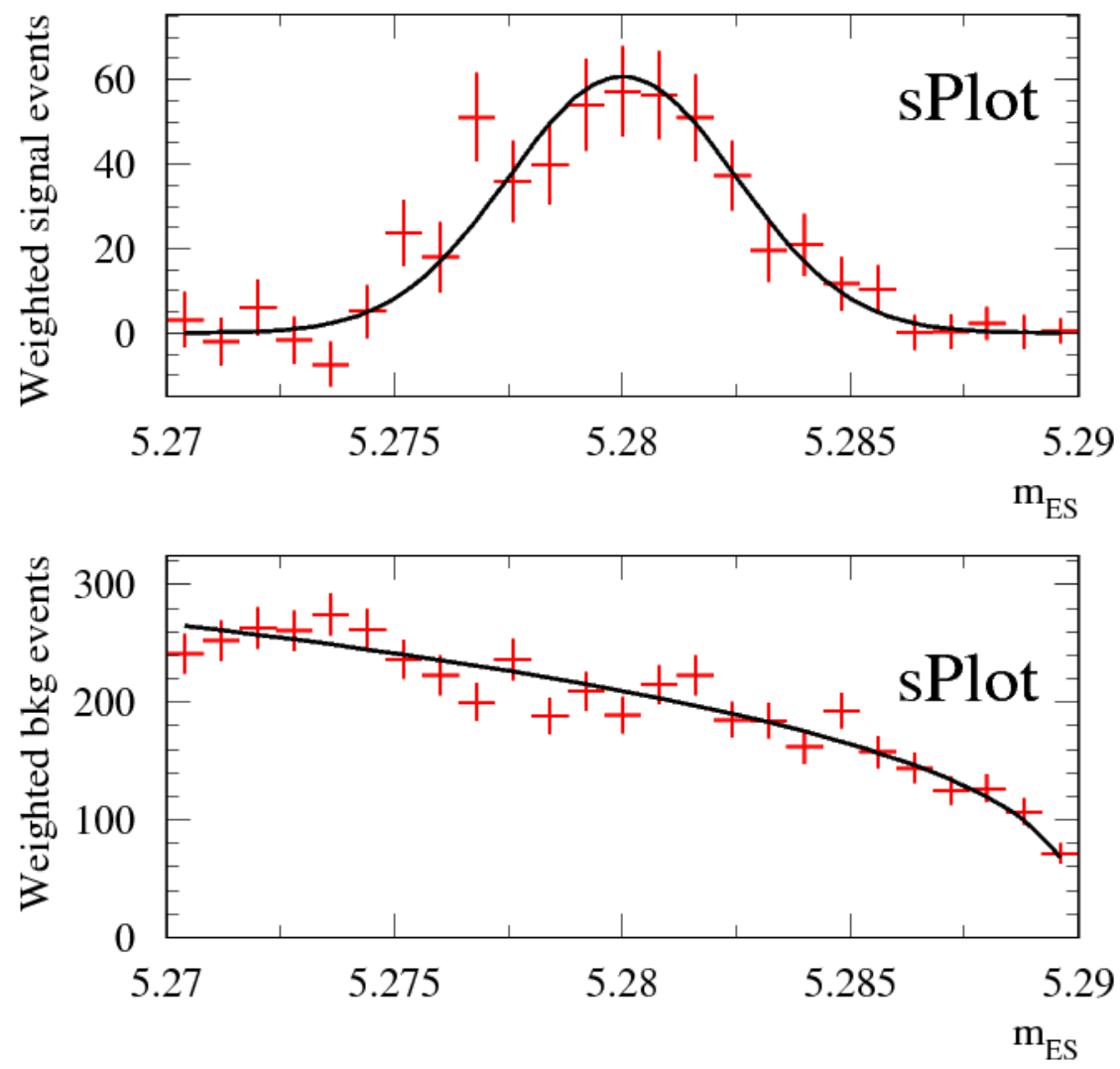

**Figure 11.2.4.** Demonstration of ${}_s\mathcal{P}lot$ concept using a model in three observables $m_{ES}, \Delta E, \mathcal{F}$ with a signal and background component. The top and bottom plot show the estimated signal and background shape in $m_{ES}$, respectively. In either plot the line represents the model prediction in the observable $m_{ES}$, and the histogram is the ${}_s\mathcal{P}lot$ defined as weighted sum over the data using ${}_sWeight$ ${}_s\mathcal{P}_n(\boldsymbol{y})$ calculated from the model prediction using only the observables $\Delta E$ and $\mathcal{F}$.

## 11.3 Structure of models for decay time-dependent measurements

Much of the interesting physics of the $B$-analyses is encoded in the distribution of the decay-time difference $\Delta t$ between $B^0$ and $\overline{B}^0$ mesons, and connected to the phenomena of $B^0 - \overline{B}^0$ flavor oscillations (see also Chapter 10 and Section 6.5). The time scale of flavor oscillations is close to the decay time of $B^0$ mesons and to the experimental resolution of the $B$ Factory detectors. Thus it is important to precisely model both the physics effects encoded in the decay time distribution, as well as the effect of the detector resolution on this distribution, of which the effect may vary on an event-by-event basis.

A priori, the observed inclusive decay-time distribution is expected to be modeled by the convolution of the physics distribution, a pure exponential decay law, and a detector resolution function:

$$ f(\Delta t; \tau, \boldsymbol{q}) = \frac{\exp(-|\Delta t_{\text{tr}}|/\tau) \otimes r(\Delta t - \Delta t_{\text{tr}}; \boldsymbol{q})}{\int \exp(-|\Delta t_{\text{tr}}|/\tau) \otimes r(\Delta t - \Delta t_{\text{tr}}; \boldsymbol{q}) d\Delta t}, \tag{11.3.1} $$

where $\Delta t$ is observed decay time difference, $\Delta t_{\text{tr}}$ is the true decay time difference, which is the integration variable of the convolution integral, and $\tau$ is the lifetime of $B^0$ mesons. Fig. 11.3.1 illustrates the shape of the convoluted p.d.f. of Eq. (11.3.1) and of its components.

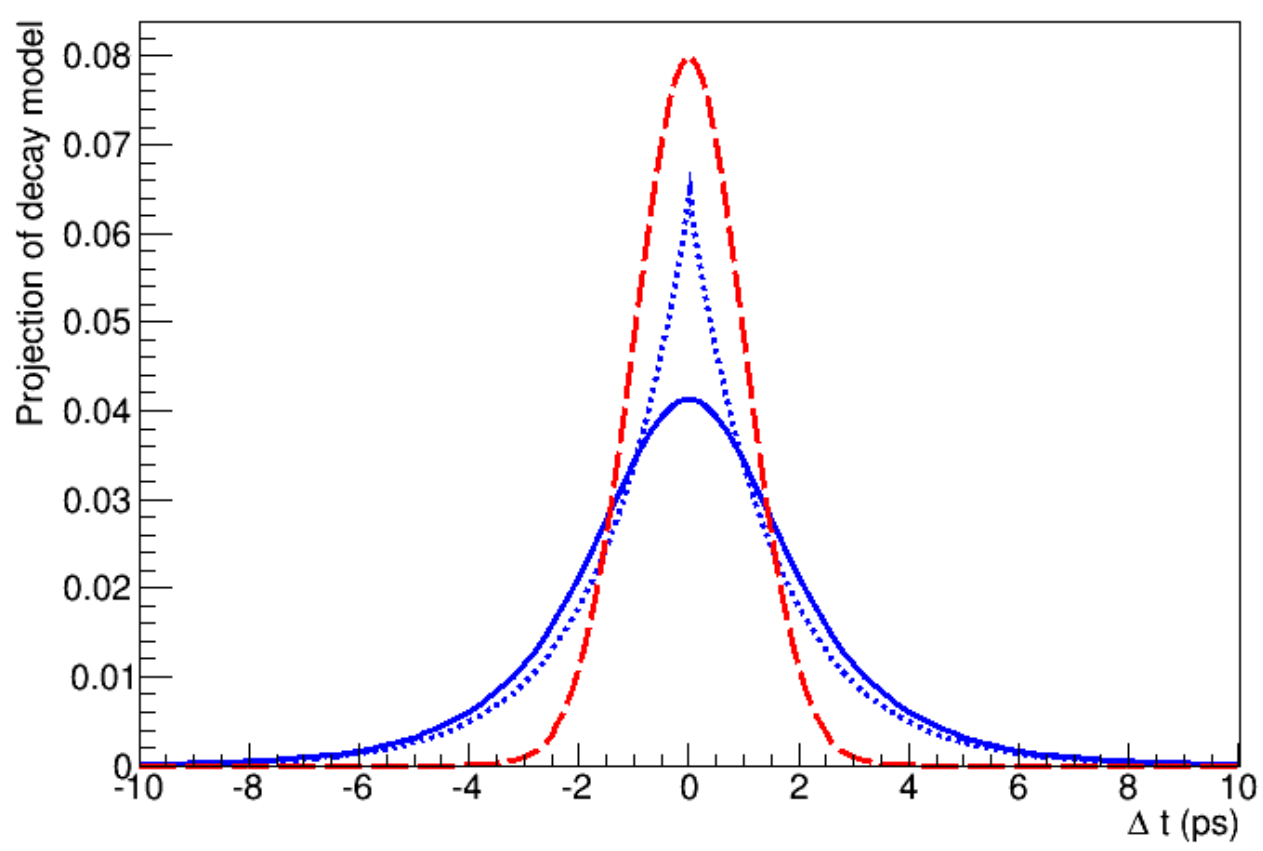

**Figure 11.3.1.** Visualization of exponential decay time difference distribution before (blue dashed) and after (blue solid) convolution with a Gaussian resolution function (red, long dashes).

The resolution model of Eq. (11.3.1) is usually empirically described as a sum of Gaussians, describing a 'core' (C) and a 'tail' (T) resolution, and often includes a very wide 'outlier' (O) term to account for the possibility that outlier events can occur in the data:

$$r(\Delta t; \boldsymbol{\mu}, \boldsymbol{\sigma}) = f_C \cdot \text{Gauss}(\Delta t; \mu_C, \sigma_C) + (1 - f_C - f_O) \cdot \text{Gauss}(\Delta t; \mu_T, \sigma_T) + f_O \cdot \text{Gauss}(\Delta t; 0, \sigma_O), \tag{11.3.2}$$

where $\mu_{C,T}$ and $\sigma_{C,T,O}$ represent the means and widths of the corresponding Gaussian distributions, respectively, and $f_C$ and $f_O$ represent the fraction of events in the core and outlier component, respectively. While very few events are expected that are not described by the convolution of the physics model with a core and tail Gaussian resolution term, it is important to include a wide outlier term in the resolution model, as otherwise a single event that is 'far' from both core and tail models has the potential to contribute disproportionally to the likelihood and can strongly and unduly influence the fit result, even when outliers only contribute at the permille level to the event sample. A common pragmatic choice for the outlier term is a very broad Gaussian distribution, as shown in the example of Eq. (11.3.2), but other shapes have also been used.

The resolution model of the previous example describes the average performance of the decay-time reconstruction. Since the decay-time difference is calculated from the distance between two decay vertices, the resolution in the time difference will depend on the number of tracks used in the vertex fits as well as their configuration, and the vertex fit procedure returns an estimate of the uncertainty on the decay-time difference for each event.

A more precise inference on the physics parameter $\tau$ of the model $f$ can be made by taking into account this per-event uncertainty on the decay time difference – weighting events with a precise measurement of $\sigma_{\Delta t}$ more strongly than those with a poorer measurement by modifying the resolution model as follows

$$r'(\Delta t|\sigma_{\Delta t}; \boldsymbol{\mu}, \boldsymbol{\sigma}) = \text{Gauss}(\Delta t; \mu_C, S \cdot \sigma_{\Delta t}), \tag{11.3.3}$$

where $\sigma_{\Delta t}$ is the estimate of the uncertainty on $\Delta t$ for each event. In this form the mean and width parameters of the resolution model $r'$ describes an a posteriori shift $\mu_C$ and scaling $S$ of the per-event error $\sigma_{\Delta t}$ that is needed to match the model to the data. If the per-event uncertainty estimated by the vertex fit is correct, the mean and width will be 0 and 1, respectively, and $r'$ will be a unit Gaussian. In practice, this is often not the case due to the complexity of the underlying vertex fitting procedure and a more complex p.d.f. is needed to describe the shape of the resolution function. Here one can either take an empirical form for $r'$, *e.g.* a sum of two or three Gaussians, or try to construct a form that parameterizes the effect of the leading underlying causes explicitly. Various choices of resolution models used for time-dependent analyses at the $B$ Factories are described in more detail in Section 6.5. Inserting $r'$ in Eq. (11.3.1) results in a conditional probability density function

$$f(\Delta t|\sigma_{\Delta t}; \tau, \boldsymbol{q}) = \frac{e^{-|\Delta t_{\text{tr}}|/\tau} \otimes r'(\Delta t - \Delta t_{\text{tr}}; \sigma_{\Delta t}, \boldsymbol{q})}{\int e^{-|\Delta t_{\text{tr}}|/\tau} \otimes r'(\Delta t - \Delta t_{\text{tr}}; \sigma_{\Delta t}, \boldsymbol{q}) d\Delta t}, \tag{11.3.4}$$

where $\Delta t_{\text{tr}}$ is again the integration variable of the convolution integral, and which describes the distribution of $\Delta t$ for a *given* value of $\sigma_{\Delta t}$, but not the distribution of $\sigma_{\Delta t}$ itself. Such a conditional p.d.f. can be fit directly to the data, or be multiplied with another (empirical) p.d.f. that describes the distribution of the per-event uncertainty on $\Delta t$:

$$F'(\Delta t, \sigma_{\Delta t}|\tau; \boldsymbol{q}) = F(\Delta t|\sigma_{\Delta t}; \tau, \boldsymbol{q}) \cdot \text{Gauss}(\sigma_{\Delta t}; \boldsymbol{q}). \tag{11.3.5}$$

In realistic models that account for the presence of background in the data, a separate decay-time distribution is defined for signal and background, each multiplied with one or more probability density functions in other observables that primarily serve to distinguish signal from background events. This approach to model building is straightforward except for one aspect related to conditional models: The p.d.f. of Eq. (11.3.4) makes no assumptions on the distribution of $\sigma_{\Delta t}$ in the data, but does assume that signal and background events have the same distribution, whereas the p.d.f. of Eq. (11.3.5) allows for different distributions of $\sigma_{\Delta t}$ for signal and background, but requires an explicit description of both. The most appropriate form depends on the specifics of the analysis. Using Eq. (11.3.4) in cases where it is not appropriate, *e.g.* when distributions of $\sigma_{\Delta t}$ for signal and background are expected to be different, is referred to as the "Punzi problem" (Punzi, 2003a) in HEP statistics literature, and may lead to biased fit results.

Finally, the physics of interest in the decay time distribution is exposed by splitting the event sample in two more categories, *e.g.* same-flavor and opposite-flavor $B^0$ meson pairs to expose flavor oscillations:

$$F(\Delta t, f; \tau, \boldsymbol{q}) = \begin{cases} \frac{\left(e^{-|\Delta t_{\text{tr}}|/\tau} \cos(\Delta m \Delta t_{\text{tr}})\right) \otimes R(\ldots)}{\int \left(e^{-|\Delta t_{\text{tr}}|/\tau} \cos(\Delta m \Delta t_{\text{tr}})\right) \otimes R(\ldots) d\Delta t} & : f \equiv -1 \\ \frac{\left(e^{-|\Delta t_{\text{tr}}|/\tau}(1-\cos(\Delta m \Delta t_{\text{tr}})\right) \otimes R(\ldots)}{\int \left(e^{-|\Delta t_{\text{tr}}|/\tau}(1-\cos(\Delta m \Delta t_{\text{tr}})\right) \otimes R(\ldots) d\Delta t} & : f \equiv +1 \end{cases} \tag{11.3.6}$$

defining a two-dimensional p.d.f. in a continuous observable $\Delta t$ and a discrete observable $f$ that distinguishes same-flavor from opposite-flavor events. The techniques illustrated on inclusive decay time distributions apply transparently to models modified in this way.

### 11.3.1 Visualization of p.d.f.s of decay time distributions

Models describing measurements of time-dependent *CP*-violating decay processes commonly have two or three continuous observables: the decay time and one or two kinematic variables, such as $m_{ES}$ or $\Delta E$, to distinguish $B$ decays from continuum background. These observables are usually uncorrelated and the kinematic variables have a well-defined 'signal' range that allows one to plot the decay time distribution of these events inside this signal range only, using Eq. (11.2.10).

In the case that the p.d.f. contains a conditional observable, such as $\sigma_{\Delta t}$, a different technique is required to project the conditional observable, as the p.d.f. does not contain information on the distribution of that observable. An average curve $C$ is constructed from curves representing the projection over the non-conditional observables taken at the values $\sigma_{\Delta t}$ found in the data:

$$C(\Delta t; \boldsymbol{p}, \boldsymbol{q}) = \frac{1}{n} \sum_{i=0,\ldots,n} \int d\boldsymbol{y} F(\Delta t; \sigma^i_{\Delta t}, \boldsymbol{y}, \boldsymbol{p}, \boldsymbol{q}). \tag{11.3.7}$$

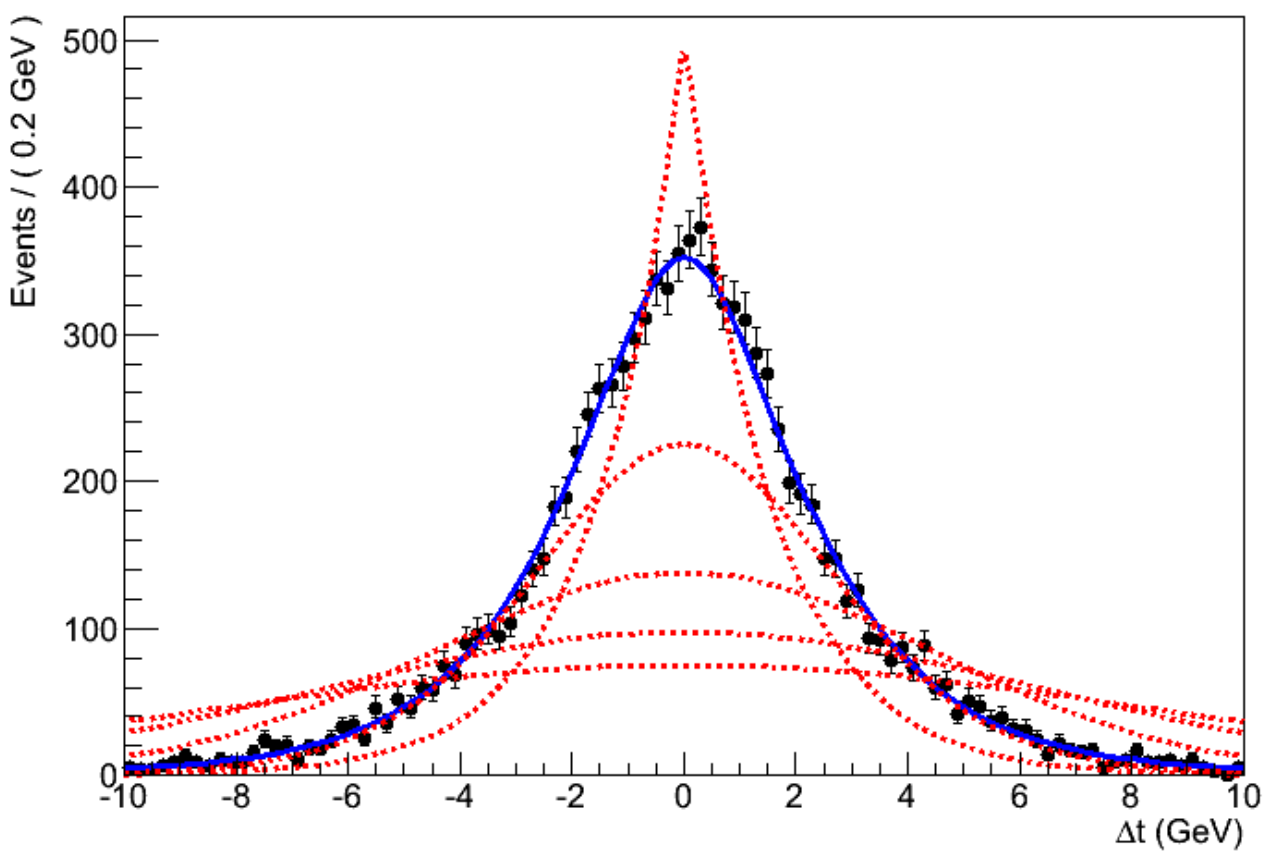

**Figure 11.3.2.** Distribution of the conditional decay time model $F(\Delta t|\sigma_{\Delta t})$ of Eq. (11.3.4) with values of $\sigma_{\Delta t}$ of 0, 2, 4, 6, 8 ps (red dashed, ordered high to low at $\Delta t = 0$) and distribution of the data overlayed with the weighted average of the conditional model using the $\sigma_{\Delta t}$ values of the data sample.

where $n$ is the number of events in the data. Figure 11.3.2 shows an example of a decay-time distribution: the red dashed curves illustrate the shape of the model at various values of $\sigma_{\Delta t}$, and the blue curve represents the weighted average using the $\sigma_{\Delta t}$ values of the dataset. Note that in the limit of the data $\sigma^i_{\Delta t}$ describing the true distribution of $\sigma_{\Delta t}$ Eq. (11.3.7) amounts to the Monte-Carlo integral (given by Eq. 11.2.13) over observable $\sigma_{\Delta t}$. For computational efficiency the summation of the data $\sigma^i_{\Delta t}$ is sometimes approximated by a summation over a histogram of the data.

## 11.4 Techniques used for constraining nuisance parameters from control samples

### 11.4.1 Simultaneous fits to control regions

As a general analysis strategy it is preferable to constrain the nuisance parameters $\boldsymbol{q}$, such as the decay time resolution model parameters, as much as possible from the data itself, instead of inferring them from simulation studies. In many cases this can be accomplished by simply floating the nuisance parameters in the ML fit. This will worsen the estimated uncertainty on the physics parameter of interest, as the values of nuisance parameters are no longer assumed to be known exactly, instead their statistical uncertainties, as inferred by the ML fit from the data, are propagated to the uncertainty on the physics parameter of interest.

In many $B$-physics analyses additional high-statistics control samples exist that can constrain these nuisance parameters with greater precision than the signal sample. For example, for decay-time dependent $CP$ violation measurements high statistics control samples from the $B^0$ flavor tagged samples can be used to measure the nuisance parameters originating from the description of the flavor tagging performance as well as the modeling of the detector decay time resolution.

The most straightforward way to incorporate the knowledge on nuisance parameters – their uncertainties and their correlations – in a measurement is to perform a joint likelihood minimization:

$$-\log L(\boldsymbol{p}, \boldsymbol{q}, \boldsymbol{q'}) = -\log L_{\mathrm{SIG}}(\boldsymbol{p}, \boldsymbol{q}) - \log L_{\mathrm{CTL}}(\boldsymbol{q}, \boldsymbol{q'}), \tag{11.4.1}$$

where $L_{\mathrm{SIG}}(\boldsymbol{p}, \boldsymbol{q})$ is the likelihood for the signal region in terms of parameters of interest $\boldsymbol{p}$ and nuisance parameters $\boldsymbol{q}$ and $L_{\mathrm{CTL}}(\boldsymbol{q}, \boldsymbol{q'})$ is the likelihood for the control region in terms of nuisance parameters $\boldsymbol{q}$ that are shared with the signal region and nuisance parameters $\boldsymbol{q'}$ that are unique to the control region. Equivalently, this construction can be expressed as a joint probability density function

$$F(x|i; \boldsymbol{p}, \boldsymbol{q}, \boldsymbol{q'}) = \begin{cases} F_{\mathrm{SIG}}(x; \boldsymbol{p}, \boldsymbol{q}) & \text{if } (i = \mathrm{SIG}) \\ F_{\mathrm{CTL}}(x; \boldsymbol{q}, \boldsymbol{q'}) & \text{if } (i = \mathrm{CTL}) \end{cases} \tag{11.4.2}$$

that is conditional on a newly introduced discrete observable $i$ that has states SIG and CTL, which label the events in the signal and control samples respectively.

A minimization of a joint likelihood ensures that the full information in both samples is taken into account, and the estimated uncertainty of parameters and their correlations reflect the information from both samples, as is illustrated in Fig. 11.4.1. The p.d.f.s describing the signal and control sample can be very dissimilar in shape and structure, the only requirement is that the common parameters have the same physics interpretation in both models.

### 11.4.2 Simultaneous fits to multiple signal regions

A mathematically similar, but conceptually different application of joint fits is to perform a joint likelihood fit to multiple signal regions, with similar p.d.f.s. If a signal region can be split into regions with different expected signal purities, a split into these regions will exploit this difference in purity without the need to provide an explicit parameterization of the change in purity over the phase space of the original signal sample.

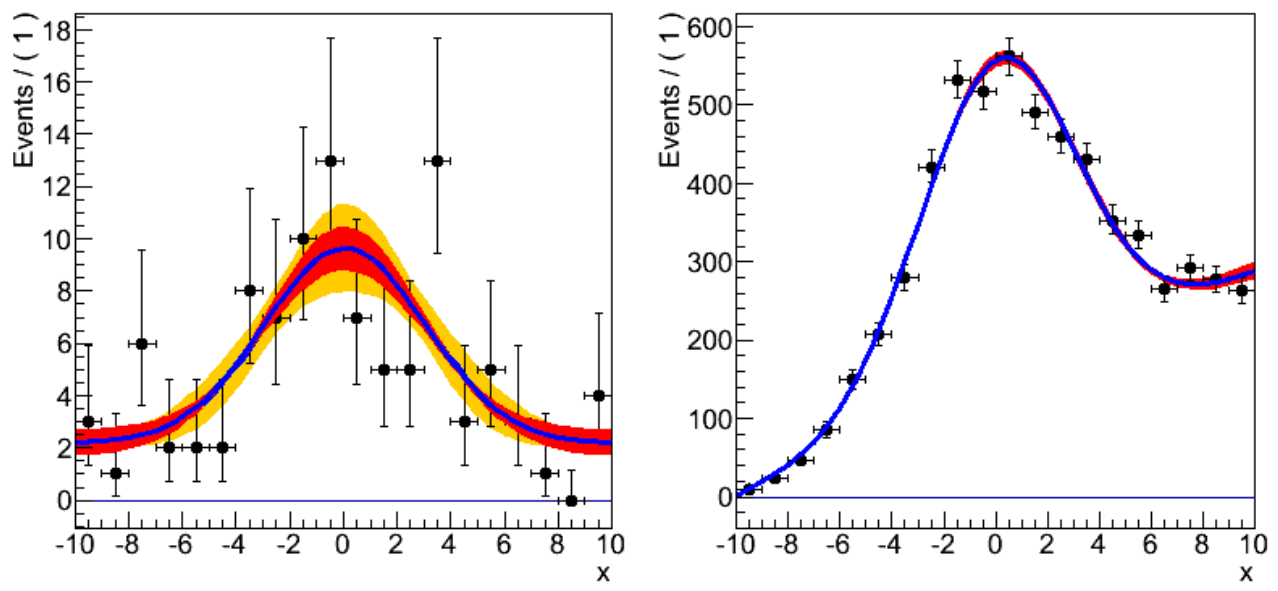

**Figure 11.4.1.** Visualization of the effect of a simultaneous fit. On the left a fictitious low statistics signal sample is shown (modeled by a flat background and a Gaussian signal). The model uncertainty from the fit to the signal sample only is visualized with the light orange band. On the right a fictitious high statistics control sample is shown (modeled by a sloped background and the same Gaussian signal). The uncertainty on the control sample model is visualized with the dark red band. The reduced uncertainty on the signal sample by performing a joint fit with the control sample is shown also in dark red in the left plot.

A prime example of this technique is splitting a signal model according to the flavor tagging technique (see also Chapter 8) that was used to tag a particular event. Different tagging techniques are expected to result in quite different purities. The original model

$$F(x; \boldsymbol{p}, \boldsymbol{q}) = F(x; \boldsymbol{p}, \boldsymbol{q}, w_{tag}) \tag{11.4.3}$$

is substituted with

$$F(x|c; \boldsymbol{p}, \boldsymbol{q}) = \begin{cases} F_{\text{tag1}}(x; \boldsymbol{p}, \boldsymbol{q}, w_{\text{tag1}}) & \text{if } (c = \text{tag1}) \\ F_{\text{tag2}}(x; \boldsymbol{p}, \boldsymbol{q}, w_{\text{tag2}}) & \text{if } (c = \text{tag2}) \\ F_{\text{tag3}}(x; \boldsymbol{p}, \boldsymbol{q}, w_{\text{tag3}}) & \text{if } (c = \text{tag3}) \\ ... & \\ F_{\text{tag}n}(x; \boldsymbol{p}, \boldsymbol{q}, w_{\text{tag}n}) & \text{if } (c = \text{tag}n) \end{cases} , \tag{11.4.4}$$

where $c$ is a discrete observable that labels which flavor tagging technique was used (here these are labeled tag1 through tag$n$, for illustration). The component models $F_{\text{tag}i}(x; \boldsymbol{p}, \boldsymbol{q}, w_{\text{tag}i})$ are structurally identical to the original $F(x; \boldsymbol{p}, \boldsymbol{q}, w_{\text{tag}})$, and expressed in terms of the same observables and parameters, *except* for the parameter $w$ that describes the mistag probability, which is now uniquely defined by $w_{\text{tag}i}$ for each state tag$i$ rather than being a global parameter $w$. At the likelihood level, the original likelihood $L(\boldsymbol{p}, \boldsymbol{q}, w_{\text{tag}})$ is now reparameterized for each region as $L(\boldsymbol{p}, \boldsymbol{q}, w_{\text{tag1}}, w_{\text{tag2}}, w_{\text{tag3}}, ..., w_{\text{tag}n})$.

Since the model of Eq. (11.4.4) is defined conditionally on the discrete split observable $c$, the model makes no assumptions on the distribution of events over the defined subsets, and since each subset is equipped with its own nuisance parameter $w_{\text{tag}i}$, also no assumption is made on the variation of the mistag rates over the subsets. The split likelihood is generally expected to improve the statistical uncertainty on the parameter of interest: in the above example, events with better than average mistag properties will now weigh more strongly in the likelihood than events with less than average mistag properties, when compared to the original likelihood definition.

While the signal splitting technique can quickly increase the number of parameters allowed to vary in the fit, the likelihood tends to be uncorrelated between the 'split' parameters and the minimization stability in `Minuit` is not as strongly impacted as one might a priori expect. The calculation of the covariance matrix by `HESSE` will nevertheless be more time consuming, as `HESSE` is not aware of this block-diagonal form and will simply calculate all covariance matrix elements.

## 11.5 Miscellaneous issues

### 11.5.1 Background subtraction and weighted events

An alternative approach to extracting signal properties from a data sample with a known background contribution is to subtract the background in the data, using an estimate from a sideband region, and then fitting the background subtracted data samples with a signal-only model.

An advantage of the subtraction approach is that no parametric form is needed to describe the distribution of the background. Background subtraction can be applied in both binned and unbinned ML estimates. In the latter case, background events are added to the unbinned dataset with negative weights. Another form of background subtraction is to reweight the data, using the ${}_sWeights$ defined in Eq. (11.2.14), in such a way that the sum of weights reflects only the signal component. ${}_sWeights$ can be either positive or negative.

In all cases of event weighting the distribution of the expected event count in any given region is modified from a Poisson distribution to a distribution that reflects the effect of the subtracted background distribution. In a $\chi^2$ fit, the (squared) uncertainty associated with each bin is the calculated with the sum of the squares of the weights of the events in the bin, using the prescription for the variance with weighted events:

$$\sigma^2 = V = \sum_{i=1}^{N} w_i^2, \tag{11.5.1}$$

where $w_i$ is the weight of the $i$-th of $N$ events contributing to a given bin. For example in a bin containing 20 events with weight $+1$ and 10 events of weight $-1$, the uncertainty on the weighted sum of 10 events is estimated as $\sqrt{30}$, compared to $\sqrt{10}$ for a bin containing 10 events with only positive weights.

In the likelihood formalism, the definition of the likelihood of Eq. (11.1.6) can be modified to include event weights

$$-\log L(\boldsymbol{p}, \boldsymbol{q}) = -\sum_{i=0,...,n} w_i \cdot \log F(\boldsymbol{x_i}; \boldsymbol{p}, \boldsymbol{q}), \tag{11.5.2}$$

so that the ML estimators for the parameters $\boldsymbol{p}, \boldsymbol{q}$ will take the weights into account. The likelihood of Eq. (11.5.2) is however not directly suitable for variance estimators: unlike a $\chi^2$ fit, where the uncertainty associated with each data point can be externally specified according to Eq. (11.5.1), the variance estimator for these parameters will *not* reflect the increased uncertainty and will simply (incorrectly) assume Poisson uncertainties with $\mu = \sum w_i$. and will thus – in case of datasets with events that have weights less than unity – underestimate the uncertainty.

Nevertheless it is possible to extract an approximately correct covariance matrix by combining the ML estimate of variance $V$ given by Eq. (11.5.2) with another ML estimate of the variance ($C$) for which the weight $w_i$ in Eq. (11.5.2) was substituted by $w_i^2$ :

$$V' = V \cdot C^{-1} \cdot V. \qquad (11.5.3)$$

The estimation of errors using Eqs (11.5.2) and (11.5.3) is not restricted to cases where event weights are $\pm 1$, but can also be more generally applied to event samples with arbitrary event weights.

### 11.5.2 Validation of ML fits on complex models

Maximum likelihood fits on complex models can be validated by studying their behavior on simulated data that are sampled from the model itself. A typical study consists of simulating many (of order 1000) data samples according to the model under study, and fitting the model to each of these datasets. For every estimated parameter in the fit, the distribution of its pull, defined as

$$\text{pull}(p) = \frac{\widehat{p} - p_{\text{true}}}{\sigma(\widehat{p})} \qquad (11.5.4)$$

can be examined. If the estimator $\widehat{p}$ is free of bias, *i.e.* it will estimate the true value correctly on average, the mean of the pull distribution will be consistent with zero. If the estimator $\widehat{\sigma}(p)$ represents the uncertainty correctly the variance of the pull distribution will be consistent with one. A too narrow pull distribution indicates that $\widehat{\sigma}(p)$ overestimates the uncertainty. Conversely, a too wide pull distribution indicates that $\widehat{\sigma}(p)$ underestimates the uncertainty. Verification of the absence of bias is of particular importance for estimators of small yields for which ML estimators can be rather imperfect. Studies on simulated data can also be used to determine the expected spread in (statistical) uncertainties on the physics parameter-of-interest, indicating whether the uncertainty obtained on the measured data was 'lucky' or not. For these studies, in particular when studying the aspect of expected statistical uncertainties, it is important to draw event samples from a conditional model evaluated at the observed values of the conditional observables (such as $\sigma_{\Delta t}$ in decay time dependent fits), to get a maximally relevant answer.

Another aspect of validation of ML estimates is to measure the goodness-of-fit of the model $f(x)$ with respect to the data $x$. To do so, a test statistic $T(x)$ must be defined to quantify the agreement between data and model in some way. A common test statistic for this purpose is Pearson's $\chi^2$, which divides the data in bins and measures the distance between the model prediction and the data in each bin. The goodness-of-fit is then expressed by the $p$-value of the hypothesis that model $f(x)$ is true, calculating Eq. (11.1.14) using the chosen test statistic. If the $p$-value is low, one may need to reject the model $f$ as a valid model.[45] The $\chi^2$ test statistic is popular, despite its requirement that the data must be binned, because this test statistic is *distribution-free*: in the limit of sufficient event counts in all bins the distribution of $\chi^2$ values is independent of the distribution of the data predicted by model $f$, simplifying the interpretation of $\chi^2$ values in terms of probabilities.

Estimating the goodness-of-fit for complex likelihood models with multiple observables constitutes a more difficult problem: due to the large number of empty bins that arise in binning multi-dimensional observables distributions the $\chi^2$ test statistic is no longer in the distribution-free regime. Various unbinned multidimensional goodness-of-fit tests have been developed over the years (Aslan and Zech, 2002), but are not as easy to use as the $\chi^2$ test in the high-statistics regime for various reasons, *e.g.* because they are not distribution-free either, and have not been routinely used at the $B$ Factories.

It should be noted that the unbinned maximum likelihood itself is *not* a reliable goodness-of-fit estimator. One reason for this is that this test statistic does not generally provide a sane definition of agreement between data and model. For example, for a likelihood assuming an exponential decay law distribution, the maximum log-likelihood is simply proportional to average lifetime of the events in the data. Thus, all data samples with the same average lifetime will result in the same goodness-of-fit independent of the observed distribution of events. Another problem with the maximum likelihood as goodness-of-fit test statistic is in obtaining the distribution of the test statistic: ML estimates are not distribution free, unlike likelihood ratios, so the expected distribution of ML values under the hypothesis that model $f$ is true, must be obtained from an ensemble of pseudo-experiments. For models $f(\boldsymbol{\theta})$ with parameters $\boldsymbol{\theta}$ this poses a challenge as the true values of the parameters are unknown. Instead, the distribution is usually obtained from pseudo-experiments sampled from the model $f(\widehat{\boldsymbol{\theta}})$ using the ML estimates of the parameters $\widehat{\boldsymbol{\theta}}$. In the example of the decay law distribution, where the maximum likelihood is simply proportional to the estimated lifetime $\widehat{\tau}$, the $p$-value for the hypothesis $f(\widehat{\tau})$ will then be close to 50% by construction and thus not provide meaningful information on the goodness-of-fit.

[45] Note that since a goodness-of-fit test is a hypothesis test in which the alternate hypothesis is the set of all possible alternatives to the hypothesis $f$ being tested, one cannot formulate the alternate hypothesis, and thus not quantify the *power* of the test: the probability that the hypothesis $f$ is false and the alternate hypothesis is true. One should therefore not conclude from a high $p$-value that the hypothesis $f$ is true.

### 11.5.3 Computational optimizations of likelihood calculations

Unbinned maximum likelihood fits are computationally intensive, and the fits underlying many of the $B$ Factory results have taken many hours or even days to complete. Efficient computation of the likelihood is thus important. In this section we discuss a number of the techniques that are applied in many of the $B$ Factory likelihood fits to optimize computational efficiency. The techniques discussed here are applied automatically in all `RooFit`-based likelihood implementations and have often been applied by hand in custom likelihood implementations.

*Constant term pre-calculation.* In many models (partial) expressions occur that do not depend on any floating model parameter. These terms can be identified and precalculated once at the beginning of the fit.

*Caching and lazy evaluation.* Expensive objects such as numeric integrals over functions, may not need to be recalculated every time their value is needed. By explicitly tracking if input variables have changed and caching the value of the previous outcome of the calculation, unnecessary repeated calculations can be prevented. For simultaneous fits, this strategy is also applied to components of the likelihood, so that these are only recalculated if a parameter on which the component actually depends is changed.

*Analytical (partial) integrals.* For many functions, analytical expressions are known for their integrals. By using the analytical forms, expensive numeric integration can be avoided. For multi-dimensional functions, knowledge of partial analytical integrals can be used to reduce the dimensionality of the numeric integration that is needed. This is particularly efficient in cases where the dimension of the numeric integral is reduced to one, as numeric integrals in one dimension can be calculated much more efficiently and with accurate convergence estimates than multi-dimensional integrals.

*Approximation of the complex error function.* Many time-dependent $B$-physics models that involve convolution with Gaussian resolution models are expressed in terms of the complex error function. Standard calculation of the complex error function can take $\mathcal{O}(100)$ complex number multiplications to estimate its value. Instead, inside p.d.f.s interpolation in a 2-dimensional lookup table is used to speed up the calculation.

*Parallelization of the likelihood calculation.* The calculation of the likelihood is by its nature very suitable for parallelized calculation. The wall-time of execution of ML fits can be decreased by roughly a factor $N$ by parallelizing the likelihood calculation over all $N$ available cores on a multi-core host, or alternatively over multiple hosts.

# Chapter 12
# Angular analysis

***Editors:***
*Georges Vasseur (BABAR)*

An angular analysis uses the information coming from the angular distributions of the final state particles. These distributions depend on the spin and polarization of all the particles involved in the decay chain. Consequently an angular analysis may determine the spin of a particle if unknown, and the polarization of the particles in a given decay chain.

Furthermore, the angles of the Unitarity Triangle have been determined, in several $B$-meson decay modes, through the measurement of time-dependent $CP$ asymmetries in vector-vector final states, such as $J/\psi K^*$, $D^* D^*$, and $\rho\rho$, which have both $CP$-even and $CP$-odd components. These components need to be disentangled in order to extract the value of the $CP$ asymmetry. This can be achieved by performing an angular analysis.

In this chapter the angular analysis is described. The formalism is presented in Section 12.1. An overview of the main modes studied at the $B$ Factories that require an angular analysis is given in Section 12.2. Several analysis details are discussed in Section 12.3. Finally angular fits are described in Section 12.4.

## 12.1 Formalism

### 12.1.1 Spin and helicity

The spin is a quantum number characterizing a particle. It is a positive half-integer for particles called fermions (for example, electrons, muons, and protons have a spin of $\frac{1}{2}$) or integer for particles called bosons (for example, mesons). The spin $J$ of a given particle and its parity $P$ are often given using the notation $J^P$. According to the values of $J^P$, particles are referred to as scalars ($0^+$: $f_0$, $a_0$, $K_0^*$, ...), pseudoscalars ($0^-$: $\pi$, $\eta$, $\eta'$, $K$, $D$, $\eta_c$, $B$, ...), vectors ($1^-$: $\rho$, $\omega$, $\phi$, $K^*$, $D^*$, $\psi$, ...), axial vectors ($1^+$: $a_1$, $K_1$, ...), or tensors ($2^+$: $a_2$, $K_2^*$, ...).

The helicity $h$ of a particle of spin $J$ corresponds to the projection of its spin along its momentum. For particles with mass, it can be one of $2J+1$ values: $-J$, $-J+1$, ..., $J-1$, $J$. For massless particles, only two values are allowed: $-J$ and $J$. For example, photons, of spin 1, can have two helicities, $-1$ and $+1$. More information on the helicity formalism can be found in (Jacob and Wick, 1959).

### 12.1.2 Angular bases

Let us consider a spin 0 particle $M_0$ (for example a $B$ meson or a $D$ meson) decaying to two particles $M_1$ and $M_2$. Since the spin of $M_0$ is zero, the spin projection of the final state on the decay axis in the $M_0$ rest frame has to be zero. In other words, $M_1$ and $M_2$ must have the same helicity. For example, if one of the final state particles has spin 0 and hence its helicity is 0, the helicity of the other final state particle must also be equal to 0: it is longitudinally polarized.

Let us now focus on the case where at least one of the direct decay products of $M_0$ has spin 1 (a vector or axial vector particle) and the other a spin greater or equal to 1. If $M_1$ or $M_2$ is of spin 1, $h$ can take three values: $-1$, 0, and $+1$. There is one complex amplitude $A_h$ associated with each case: the longitudinal amplitude $A_0$ and the transverse ones $A_{+1}$ and $A_{-1}$. The three amplitudes ($A_0$, $A_{+1}$, $A_{-1}$) correspond to helicity eigenstates and define the helicity basis.

For a $CP$ eigenstate, the longitudinal amplitude is $CP$-even, while the transverse ones are an admixture of $CP$-even and $CP$-odd components. In the transversity basis ($A_L$, $A_\parallel$, $A_\perp$), the amplitudes correspond to $CP$ eigenstates:

$$\begin{array}{lll} CP\text{-even longitudinal} & : A_L & = A_0 \quad , \\ CP\text{-even transverse} & : A_\parallel & = \frac{A_{+1}+A_{-1}}{\sqrt{2}} \quad , \\ CP\text{-odd transverse} & : A_\perp & = \frac{A_{+1}-A_{-1}}{\sqrt{2}} \quad . \end{array}$$

Both “0” and “$L$” subscripts are commonly used for the longitudinal amplitude. In what follows, the latter notation is used. Additional information on the subject can be found in (Kramer and Palmer, 1992), in (Dunietz, Quinn, Snyder, Toki, and Lipkin, 1991), and in the review of polarization in $B$ decays of (Beringer et al., 2012).

The fractions of each polarization amplitude are defined as $f_{L,\parallel,\perp} = \frac{|A_{L,\parallel,\perp}|^2}{\Sigma |A_h|^2}$, where $h$ runs on the three polarization eigenstates. They satisfy the relation $f_L + f_\parallel + f_\perp = 1$. The phase differences of the two transverse amplitudes with respect to the longitudinal one are defined as $\phi_{\parallel,\perp} = \mathrm{Arg}(A_{\parallel,\perp}/A_L)$. As the decay is described by three independent complex amplitudes, $A_L$, $A_\parallel$, and $A_\perp$, there are six independent real parameters, often chosen as $f_L$, $f_\perp$, $\phi_\parallel$, $\phi_\perp$, the total decay rate $\Gamma$, and an overall phase $\delta_0$.

This overall phase is meaningless in most cases. It is relevant when there exists an external amplitude which can be used as a reference to measure it. This is the case, for example, if one of the $B$-meson daughters is a $K^*$. In addition to the three amplitudes, $A_L, A_\parallel, A_\perp$, describing the decay mode with the $K^*$, there is another amplitude $A_{00}$ associated with the related decay mode where the $K^*$ is replaced by the $J = 0$ $(K\pi)$ wave, $K_0^*$. The overall phase can be defined as $\delta_0 = \mathrm{Arg}(A_{00}/A_L)$. As both the $K^*$ and $K_0^*$ decay to the same final state $K\pi$, the $\delta_0$ phase can be measured through interference between the $B \to M_1 K^*$ and $B \to M_1 K_0^*$ decays (Aubert, 2008bf).

The total amplitude may also be expressed as a function of $S$, $P$, or $D$ partial waves, characterized by the relative orbital angular momentum $L$ between $M_1$ and $M_2$, $L$ being equal to 0, 1, and 2 for $S$, $P$, and $D$ waves respectively. The partial wave basis is used for example in (Chung, 1997).

The expressions for the angular dependence are relatively simple in the helicity and transversity bases. They are given in the next subsections.

### 12.1.3 Angular distributions in the helicity basis

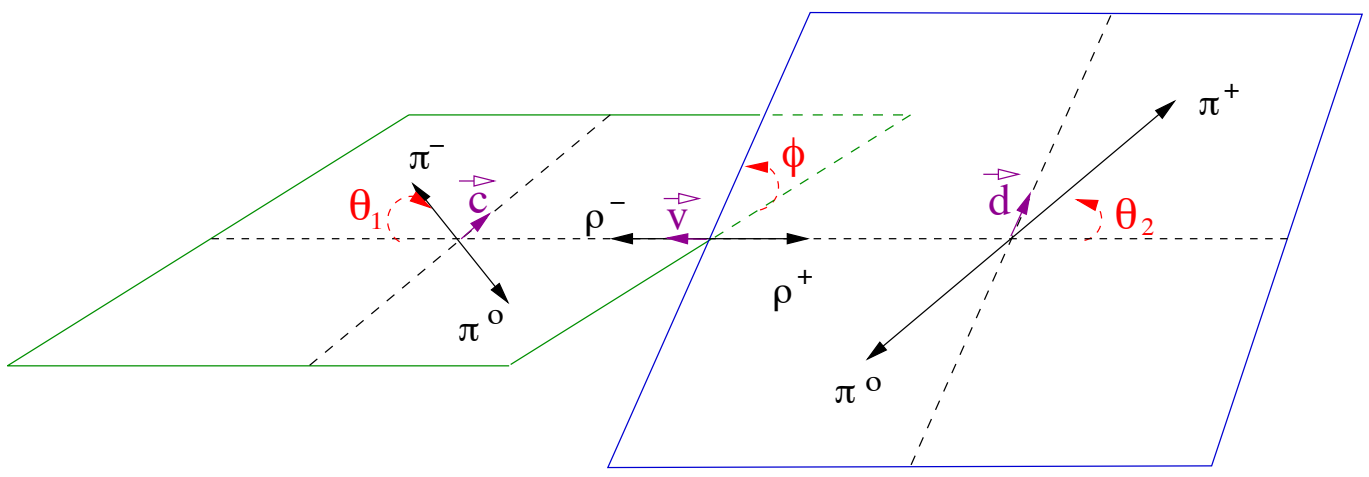

**Figure 12.1.1.** The three angles in the helicity frame: $\theta_1$, $\theta_2$, and $\phi$, shown in the example of $B \to \rho^-\rho^+$ decays. The $B \to \rho^-\rho^+$, $\rho^- \to \pi^-\pi^0$, and $\rho^+ \to \pi^+\pi^0$ decays are represented in the $B$, $\rho^-$, and $\rho^+$ rest frames respectively. The unit vector $\boldsymbol{v}$ defines the direction of the $\rho^-$ in the $B$ rest frame, or equivalently the direction of (opposite to) the line of flight of the $B$ in the $\rho^+$ ($\rho^-$) rest frame. The decay plane of the $\rho^-$ ($\rho^+$) is defined by the $\boldsymbol{c}$ ($\boldsymbol{d}$) and $\boldsymbol{v}$ unit vectors. Here $\phi$ is the angle between the two decay planes and $\theta_1$ ($\theta_2$) is the polar angle of the $\pi^-$ ($\pi^+$) with $\boldsymbol{v}$ ($-\boldsymbol{v}$).

In the helicity frame, in the case of the $M_0 \to M_1 M_2$ decay with $M_1$ and $M_2$ each subsequently undergoing a two-body decay, the relevant angles are the polar angle $\theta_1$ of a decay product of $M_1$ with respect to the direction opposite to the line of flight of $M_0$ in the $M_1$ rest frame, the angle $\theta_2$ for $M_2$ (same as $\theta_1$ for $M_1$), and the angle $\phi$ between the decay planes of $M_1$ and $M_2$ in the $M_0$ rest frame. The choice of the decay product of $M_1$ ($M_2$) used to define $\theta_1$ ($\theta_2$) is arbitrary, but it must be consistent throughout the analysis. Figure 12.1.1 shows the three angles in the case of $B \to \rho^-\rho^+$ decays, with $\rho^- \to \pi^-\pi^0$ and $\rho^+ \to \pi^+\pi^0$. If $M_i$ ($i = 1$ or 2) undergoes a three-body decay, $\theta_i$ is defined as the angle between the normal of the decay plane of $M_i$ with respect to the direction opposite to the line of flight of $M_0$ in the $M_i$ rest frame.

The differential decay rate in the helicity frame can be expressed as:

$$\frac{1}{\Gamma}\frac{d^3\Gamma}{d\cos\theta_1\, d\cos\theta_2\, d\phi} = \frac{9}{8\pi}\,\Sigma\,\alpha_i\, g_i(\cos\theta_1, \cos\theta_2, \phi)\,. \qquad (12.1.1)$$

The $g_i$ functions depend on the quantum numbers of the particles in the decay chain and are given for the most common cases in the next section. The $\alpha_i$ are real parameters, which can be expressed as functions of the fractions $f_L$, $f_\parallel$, $f_\perp$ and of the phase differences $\phi_\parallel$, $\phi_\perp$ Beringer et al. (2012):

$$\begin{aligned}
\alpha_1 &= \frac{|A_L|^2}{\Sigma|A_h|^2} = f_L\,,\\
\alpha_2 &= \frac{|A_\parallel|^2 + |A_\perp|^2}{\Sigma|A_h|^2} = 1 - f_L\,,\\
\alpha_3 &= \frac{|A_\parallel|^2 - |A_\perp|^2}{\Sigma|A_h|^2} = f_\parallel - f_\perp\,,\\
\alpha_4 &= \frac{\mathrm{Im}(A_\perp A_\parallel^*)}{\Sigma|A_h|^2} = \sqrt{f_\perp f_\parallel}\,\sin(\phi_\perp - \phi_\parallel)\,,\\
\alpha_5 &= \frac{\mathrm{Re}(A_\parallel A_L^*)}{\Sigma|A_h|^2} = \sqrt{f_\parallel f_L}\,\cos(\phi_\parallel)\,,\\
\alpha_6 &= \frac{\mathrm{Im}(A_\perp A_L^*)}{\Sigma|A_h|^2} = \sqrt{f_\perp f_L}\,\sin(\phi_\perp)\,.
\end{aligned} \qquad (12.1.2)$$

### 12.1.4 Angular distributions in the transversity basis

The angles used in the transversity frame are illustrated in Figure 12.1.2 for the decay mode $B \to \rho^+\rho^-$. The angle $\theta_1$ has the same definition as in the helicity frame. In the $M_2$ rest frame, the axes ($x$,$y$,$z$) are defined such that the $x$-axis has the direction opposite to the momentum of the $M_1$ particle, the $z$-axis is normal to the decay plane of the $M_1$ particle, and the projection of the momentum along the $y$-axis is positive for the decay product of $M_1$ that is used to define $\theta_1$ (the $\pi^-$ in the example of Figure 12.1.2). Then $\theta_{tr}$ and $\phi_{tr}$ are the polar and azimuthal angles of one decay product of $M_2$. They are called the transversity angles.

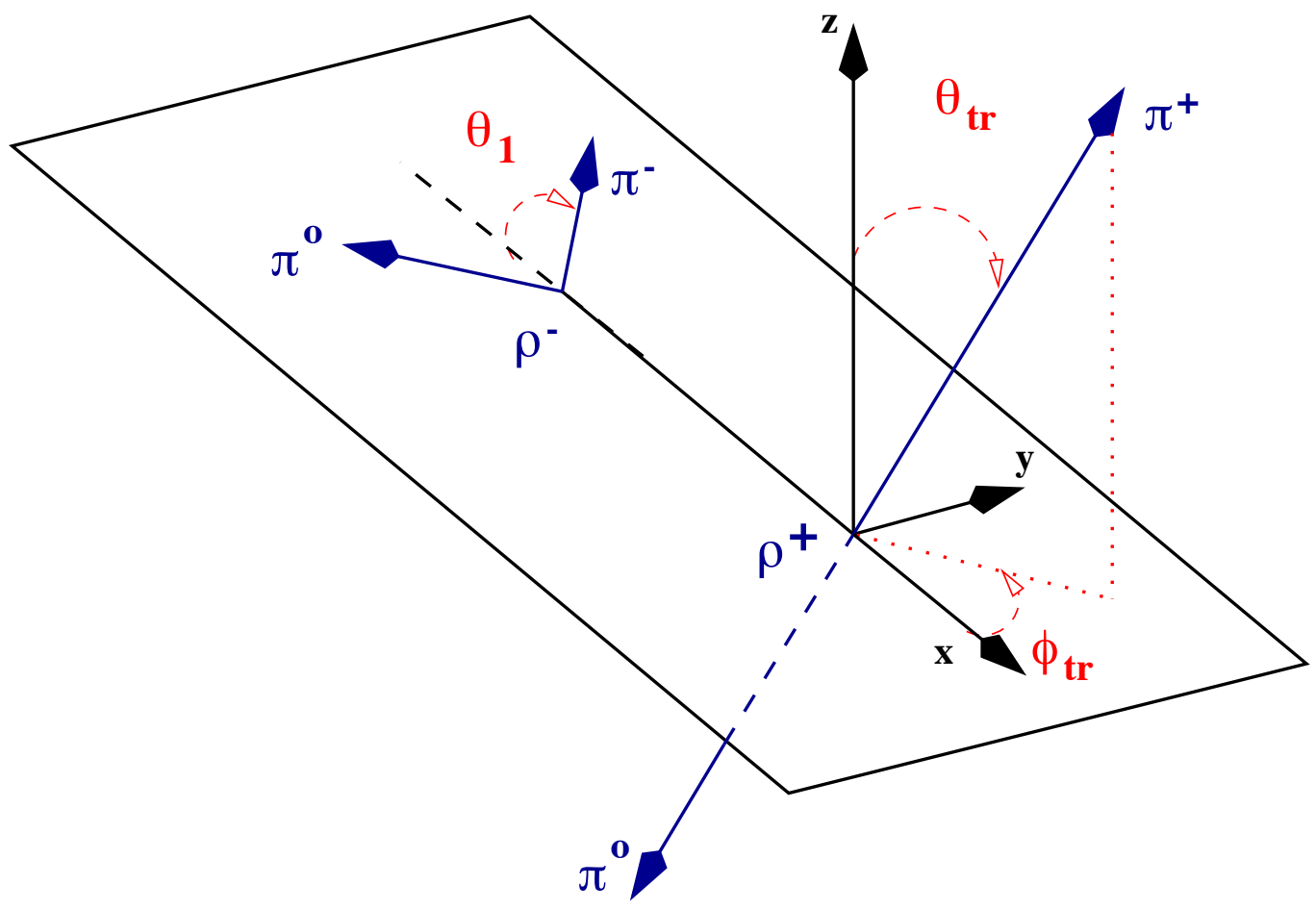

**Figure 12.1.2.** The three angles in the transversity frame: $\theta_1$, $\theta_{tr}$, and $\phi_{tr}$, shown in the example of $B \to \rho^+\rho^-$ decays. Here $\theta_1$ is the angle of the $\pi^-$ with the $\rho^-$ direction. And $\theta_{tr}$ ($\phi_{tr}$) is the polar (azimuthal) angle of the $\pi^+$ in the $\rho^+$ rest frame.

It is convenient to write the differential decay rate, as in the helicity frame, as the sum of six terms:

$$\frac{1}{\Gamma}\,\frac{d^3\Gamma}{d\cos\theta_1\,d\cos\theta_{tr}\,d\phi_{tr}} = \frac{9}{8\pi}\,\Sigma\,\alpha_i^{tr}\,g_i^{tr}(\cos\theta_1,\cos\theta_{tr},\phi_{tr})\,, \qquad (12.1.3)$$

with $\alpha_i^{tr} = \alpha_i$, except for

$$\alpha_2^{tr} = \frac{\alpha_2+\alpha_3}{2} = \frac{|A_\parallel|^2}{\Sigma|A_h|^2} = f_\parallel\,,$$
$$\alpha_3^{tr} = \frac{\alpha_2-\alpha_3}{2} = \frac{|A_\perp|^2}{\Sigma|A_h|^2} = f_\perp\,. \qquad (12.1.4)$$

#### 12.1.5 *CP* violation

If both the $M_0$ decay and its charge conjugate $\overline{M}_0$ decay are considered, there are now six complex amplitudes or twelve real parameters to describe the two decays. They can be chosen as the six parameters, $\Gamma$, $f_L$, $f_\perp$, $\phi_\parallel$, $\phi_\perp$, and $\delta_0$, already given for the $M_0$ decay, and the corresponding ones, $\overline{\Gamma}$, $\overline{f}_L$, $\overline{f}_\perp$, $\overline{\phi}_\parallel$, $\overline{\phi}_\perp$, and $\overline{\delta}_0$, for the $\overline{M}_0$ decay. Alternatively they can be defined as the six averages of the $M_0$ and $\overline{M}_0$ parameters and the six differences between $M_0$ and $\overline{M}_0$ parameters, written below:

$$\begin{aligned}
A_{CP} &= \frac{\overline{\Gamma}-\Gamma}{\overline{\Gamma}+\Gamma}\,,\\
A_{CP}^L &= \frac{\overline{f}_L-f_L}{\overline{f}_L+f_L}\,,\\
A_{CP}^\perp &= \frac{\overline{f}_\perp-f_\perp}{\overline{f}_\perp+f_\perp}\,, \qquad (12.1.5)\\
\Delta\phi_\parallel &= \frac{1}{2}(\overline{\phi}_\parallel-\phi_\parallel)\,,\\
\Delta\phi_\perp &= \frac{1}{2}(\overline{\phi}_\perp-\phi_\perp-\pi)\,,\\
\Delta\delta_0 &= \frac{1}{2}(\overline{\delta}_0-\delta_0)\,.
\end{aligned}$$

The quantity $\pi$, introduced in the definition of $\Delta\phi_\perp$, is the phase difference between $A_\perp$ and $\overline{A}_\perp$ if *CP* were conserved. *CP* violation can be established in an angular analysis if one measures a non-zero value for any of these last six parameters.

#### 12.1.6 Time dependence

The transversity basis is most suited to study *CP* violation in time-dependent asymmetries in neutral $B$ decays. Where $\tau$ is the $B^0$ lifetime, $\Delta m_d$ is the mass difference responsible for the $B^0$-$\overline{B}^0$ oscillations, and $\Delta t$ is the proper time difference between the decay times of the two $B$ mesons (see Section 10), the time-evolution for each amplitude is given by:

$$\begin{aligned}
A_L(\Delta t) &= A_L(0)\,e^{-im\Delta t}\;e^{-|\Delta t|/2\tau}\\
&\quad\times\left(\cos\frac{\Delta m_d\Delta t}{2}+i\eta\lambda_L\sin\frac{\Delta m_d\Delta t}{2}\right),\\
A_\parallel(\Delta t) &= A_\parallel(0)\,e^{-im\Delta t}\;e^{-|\Delta t|/2\tau} \qquad (12.1.6)\\
&\quad\times\left(\cos\frac{\Delta m_d\Delta t}{2}+i\eta\lambda_\parallel\sin\frac{\Delta m_d\Delta t}{2}\right),\\
A_\perp(\Delta t) &= A_\perp(0)\,e^{-im\Delta t}\;e^{-|\Delta t|/2\tau}\\
&\quad\times\left(\cos\frac{\Delta m_d\Delta t}{2}-i\eta\lambda_\perp\sin\frac{\Delta m_d\Delta t}{2}\right).
\end{aligned}$$

The $\eta$ parameter equals 1 for $B$ decays and $-1$ for $\overline{B}$ decays. The parameter $\lambda$, introduced in Section 10, may have three values, $\lambda_L$, $\lambda_\parallel$, and $\lambda_\perp$, which are in general different from each other. The total amplitude is the sum of the three amplitudes, and the time-dependent total neutral $B$-meson decay rate is expressed as:

$$\begin{aligned}
\Gamma(\Delta t) &= |A_L(\Delta t)+A_\parallel(\Delta t)+A_\perp(\Delta t)|^2 \qquad (12.1.7)\\
&= |A_L(\Delta t)|^2+|A_\parallel(\Delta t)|^2+|A_\perp(\Delta t)|^2\\
&\quad+2\mathrm{Re}(A_\parallel(\Delta t)A_L^*(\Delta t)+A_\perp(\Delta t)A_L^*(\Delta t)\\
&\quad+A_\perp(\Delta t)A_\parallel^*(\Delta t))\,.
\end{aligned}$$

Thus the time-dependence of the various terms entering the differential decay rate needs to be obtained. Since equivalent expressions describe the two *CP*-even amplitudes $A_L$ and $A_\parallel$, the "+" subscript is used to denote both "$L$" and "$\parallel$" to minimize the number of relations to be used:

$$\begin{aligned}
|A_+(\Delta t)|^2 = |A_+(0)|^2\ e^{-|\Delta t|/\tau} \Bigg( & \frac{1+|\lambda_+|^2}{2} \\
& + \frac{1-|\lambda_+|^2}{2}\cos(\Delta m_d \Delta t) \\
& - \eta\, \mathrm{Im}\lambda_+ \sin(\Delta m_d \Delta t) \Bigg) , \\
|A_\perp(\Delta t)|^2 = |A_\perp(0)|^2\ e^{-|\Delta t|/\tau} \Bigg( & \frac{1+|\lambda_\perp|^2}{2} \\
& + \frac{1-|\lambda_\perp|^2}{2}\cos(\Delta m_d \Delta t) \\
& + \eta\, \mathrm{Im}\lambda_\perp \sin(\Delta m_d \Delta t) \Bigg) , \\
A_\parallel(\Delta t) A_L^*(\Delta t) = A_\parallel(0) A_L^*(0)\ e^{-|\Delta t|/\tau} \Bigg( & \frac{1+\lambda_\parallel \lambda_L^*}{2} \\
& + \frac{1-\lambda_\parallel \lambda_L^*}{2}\cos(\Delta m_d \Delta t) \\
& + \frac{i\eta}{2}(\lambda_\parallel - \lambda_L^*) \sin(\Delta m_d \Delta t) \Bigg) , \\
A_\perp(\Delta t) A_+^*(\Delta t) = A_\perp(0) A_+^*(0)\ e^{-|\Delta t|/\tau} \Bigg( & \frac{1-\lambda_\perp \lambda_+^*}{2} \\
& + \frac{1+\lambda_\perp \lambda_+^*}{2}\cos(\Delta m_d \Delta t) \\
& - \frac{i\eta}{2}(\lambda_\perp + \lambda_+^*) \sin(\Delta m_d \Delta t) \Bigg) .
\end{aligned} \tag{12.1.8}$$

These general expressions are rather complex. However, they can be simplified under certain assumptions. If the final state interactions can be neglected, the three parameters, $\lambda_L$, $\lambda_\parallel$ and $\lambda_\perp$ are equal to a common value $\lambda$. If direct *CP*-violation effects can also be neglected, $\lambda$ satisfies $|\lambda| = 1$. The expressions for the time-dependent terms then become:

$$\begin{aligned}
& |A_+(\Delta t)|^2 = |A_+(0)|^2\ e^{-|\Delta t|/\tau} \left(1 - \eta\, \mathrm{Im}\lambda \sin(\Delta m_d \Delta t)\right) , \\
& |A_\perp(\Delta t)|^2 = |A_\perp(0)|^2\ e^{-|\Delta t|/\tau} \left(1 + \eta\, \mathrm{Im}\lambda \sin(\Delta m_d \Delta t)\right) , \\
& \mathrm{Re}(A_\parallel(\Delta t) A_L^*(\Delta t)) = \\
& \quad \mathrm{Re}(A_\parallel(0) A_L^*(0))\ e^{-|\Delta t|/\tau}\ \left(1 - \eta\, \mathrm{Im}\lambda \sin(\Delta m_d \Delta t)\right) , \\
& \mathrm{Im}(A_\perp(\Delta t) A_+^*(\Delta t)) = \\
& \quad \mathrm{Im}(A_\perp(0) A_+^*(0))\, e^{-|\Delta t|/\tau} \cos(\Delta m_d \Delta t) \\
& \quad - \mathrm{Re}(A_\perp(0) A_+^*(0))\, e^{-|\Delta t|/\tau} \eta \mathrm{Re}\lambda \sin(\Delta m_d \Delta t) .
\end{aligned} \tag{12.1.9}$$

In the case of the $B \to K^* J/\psi$ decay mode, where $\lambda = e^{2i\beta}$, the first three terms of Eq. (12.1.9) have the usual $\mathrm{Im}\lambda = \sin 2\beta$ coefficient in front of $\sin(\Delta m_d \Delta t)$, while the last term introduces a $\mathrm{Re}\lambda = \cos 2\beta$ coefficient, allowing one to resolve an ambiguity on the measurement of $\beta$ (Aubert, 2005c), as discussed in Section 17.6.

## 12.2 List of modes

The common decay modes are reviewed here according to the type of the particles $M_0$, $M_1$, $M_2$ and, when relevant, the daughters of $M_1$ and $M_2$. In the title of the subsections, $P$, $V$, and $T$ stand for pseudoscalar, vector, and tensor mesons, respectively, while $l(\gamma)$ is for a lepton (photon). When two vector mesons with different decay types are present, they are labeled $V_1$ and $V_2$.

For each mode, the expressions governing the angular distributions are given. The procedure to derive the formulae can be found elsewhere (Chung, 1971; Richman, 1984).

### 12.2.1 $V \to PP$

Let us start with the simple case of a vector meson decaying to two pseudoscalar mesons. The distribution of the helicity angle $\theta_1$ of the vector meson depends upon the polarization of the vector meson.

For longitudinal polarization, the distribution is given by:

$$\frac{1}{\Gamma}\ \frac{d\Gamma}{d\cos\theta_1} = \frac{3}{2}\cos^2\theta_1\ . \tag{12.2.1}$$

For transverse polarization, the expression is the following:

$$\frac{1}{\Gamma}\ \frac{d\Gamma}{d\cos\theta_1} = \frac{3}{4}\sin^2\theta_1\ . \tag{12.2.2}$$

The latter case applies to the decay $\Upsilon(4S) \to B\overline{B}$, as the $\Upsilon(4S)$ vector meson produced in $e^+e^-$ collisions through a virtual photon is transversely polarized. Hence the angle of the $B$-meson direction with respect to the beam axis at the $B$ Factories is governed by Eq. (12.2.2).

### 12.2.2 $P \to VP$ , $V \to PP$

The case of a pseudoscalar meson decaying to a pseudoscalar meson and a vector meson, which then decays to two pseudoscalar mesons, is found for example in the following modes:

- $B \to \rho\pi$, with $\rho \to \pi\pi$,
- $B \to D^*\pi$, with $D^* \to D\pi$,
- $B \to D^*D$, with $D^* \to D\pi$.

This case was briefly mentioned in Section 12.1. Here there is no degree of freedom. As the helicity of the pseudoscalar meson is 0, the helicity of the vector meson must also be 0: it is then known that the vector meson is longitudinally polarized. Hence the distribution of the $\theta_1$ angle is determined: it follows Eq. (12.2.1). In such modes, where the angular distribution is known, the helicity angle can be used in the selection for background rejection.

### 12.2.3 $P \to V\gamma$ , $V \to PP$ and $P \to T\gamma$ , $T \to PP$

Similarly, if a pseudoscalar meson decays to a photon and a vector meson, which then decays to two pseudoscalar mesons, as in the mode:

– $B \to K^*\gamma$, with $K^* \to K\pi$,

the vector meson can only have an helicity which is allowed for the photon, *i.e.* $\pm 1$. So it is transversely polarized. Consequently the distribution of the $\theta_1$ angle is given by Eq. (12.2.2).

When applying the same argument to a pseudoscalar meson decaying to a photon and a tensor meson, which then decays to two pseudoscalar mesons, such as:

– $B \to K_2^*(1430)\gamma$, with $K_2^*(1430) \to K\pi$,

it is found that the tensor meson can only have helicity $\pm 1$. In this case the $\theta_1$ angle is distributed according to:

$$\frac{1}{\Gamma}\,\frac{d\Gamma}{d\cos\theta_1} = \frac{15}{4}\sin^2\theta_1\cos^2\theta_1 \ . \tag{12.2.3}$$

### 12.2.4 $P \to VV$ , $V \to PP$

The case of a pseudoscalar meson decaying to two vector mesons, each of them decaying to two pseudoscalar mesons, can be illustrated by the following decay modes:

– $B \to \rho\rho$, with $\rho \to \pi\pi$,
– $B \to K^*\rho$, with $K^* \to K\pi$ and $\rho \to \pi\pi$,
– $B \to K^*\phi$, with $K^* \to K\pi$ and $\phi \to K^+K^-$,
– $B \to D^*K^*$, with $D^* \to D\pi$ and $K^* \to K\pi$,
– $B \to D^*D^*$, with $D^* \to D\pi$.

Both bases have been used to analyse this type of decay. In the helicity basis, the $g_i$ functions of Eq. (12.1.1) have the following angular dependence:

$$\begin{aligned} g_1 &= \cos^2\theta_1\cos^2\theta_2\,, \\ g_2 &= \frac{1}{4}\sin^2\theta_1\sin^2\theta_2\,, \\ g_3 &= \frac{1}{4}\sin^2\theta_1\sin^2\theta_2\cos 2\phi\,, \\ g_4 &= -\eta\frac{1}{2}\sin^2\theta_1\sin^2\theta_2\sin 2\phi\,, \\ g_5 &= \frac{1}{2\sqrt{2}}\sin 2\theta_1\sin 2\theta_2\cos\phi\,, \\ g_6 &= -\eta\frac{1}{2\sqrt{2}}\sin 2\theta_1\sin 2\theta_2\sin\phi\,. \end{aligned} \tag{12.2.4}$$

When integrating over the $\phi$ angle, the last four terms $g_3 - g_6$ disappear and the differential decay rate reduces to the following expression, with $f_L$ as the single parameter:

$$\frac{1}{\Gamma}\,\frac{d^2\Gamma}{d\cos\theta_1 d\cos\theta_2} = \frac{9}{4}\left(f_L\cos^2\theta_1\cos^2\theta_2 + (1-f_L)\frac{1}{4}\sin^2\theta_1\sin^2\theta_2\right) \ . \tag{12.2.5}$$

In the transversity basis the corresponding $g_i^{tr}$ functions appearing in Eq. (12.1.3) are:

$$\begin{aligned} g_1^{tr} &= \cos^2\theta_1\sin^2\theta_{tr}\cos^2\phi_{tr}\ , \\ g_2^{tr} &= \frac{1}{2}\sin^2\theta_1\sin^2\theta_{tr}\sin^2\phi_{tr}\ , \\ g_3^{tr} &= \frac{1}{2}\sin^2\theta_1\cos^2\theta_{tr}\ , \\ g_4^{tr} &= -\eta\frac{1}{2}\sin^2\theta_1\sin 2\theta_{tr}\sin\phi_{tr}\ , \\ g_5^{tr} &= \frac{1}{2\sqrt{2}}\sin 2\theta_1\sin^2\theta_{tr}\sin 2\phi_{tr}\ , \\ g_6^{tr} &= -\eta\frac{1}{2\sqrt{2}}\sin 2\theta_1\sin 2\theta_{tr}\cos\phi_{tr}\ . \end{aligned} \tag{12.2.6}$$

After integrating over the $\phi_{tr}$ angle, the last three terms disappear and the differential decay rate simplifies to:

$$\frac{1}{\Gamma}\,\frac{d^2\Gamma}{d\cos\theta_1 d\cos\theta_{tr}} = \frac{9}{8}\left(f_L\cos^2\theta_1\sin^2\theta_{tr} + f_\parallel\frac{1}{2}\sin^2\theta_1\sin^2\theta_{tr} + f_\perp\sin^2\theta_1\cos^2\theta_{tr}\right) \ . \tag{12.2.7}$$

This expression allows the extraction of the fraction of the three amplitudes. Figure 12.2.1 illustrates, in the case of the $B^0 \to D^{*+}D^{*-}$ analysis (Miyake, 2005), the sine square (cosine square) dependence on $\theta_{tr}$ of the $A_L$ and $A_\parallel$ amplitudes ($A_\perp$ amplitude) and the cosine square (sine square) dependence on $\theta_1$ of the $A_L$ amplitude ($A_\parallel$ and $A_\perp$ amplitudes).

The following expression, depending only on the fraction $f_\perp$ of the *CP*-odd amplitude, is obtained by integrating also over the $\theta_1$ angle:

$$\frac{1}{\Gamma}\,\frac{d\Gamma}{d\cos\theta_{tr}} = \frac{3}{4}\left((1-f_\perp)\sin^2\theta_{tr} + 2f_\perp\cos^2\theta_{tr}\right) \ . \tag{12.2.8}$$

### 12.2.5 $P \to VV$ , $V_1 \to P\gamma$ , $V_2 \to PP$

Vector-vector final states, where one vector meson decays to a pseudoscalar meson and a photon and the other one to two pseudoscalar mesons, include:

– $B \to D^*K^*$, with $D^* \to D\gamma$ and $K^* \to K\pi$,
– $B \to D_s^*\rho$, with $D_s^* \to D_s\gamma$ and $\rho \to \pi\pi$,
– $B \to D_s^*D^*$, with $D_s^* \to D_s\gamma$ and $D^* \to D\pi$,
– $B_s \to D_s^*\rho$, with $D_s^* \to D_s\gamma$ and $\rho \to \pi\pi$.

Here the helicity basis is used and the differential decay rate, integrated over the $\phi$ angle, is expressed as:

$$\frac{1}{\Gamma}\,\frac{d^2\Gamma}{d\cos\theta_1 d\cos\theta_2} = \frac{9}{4}\left(f_L\sin^2\theta_1\cos^2\theta_2 + (1-f_L)\frac{1}{4}(1+\cos^2\theta_1)\sin^2\theta_2\right) \ . \tag{12.2.9}$$

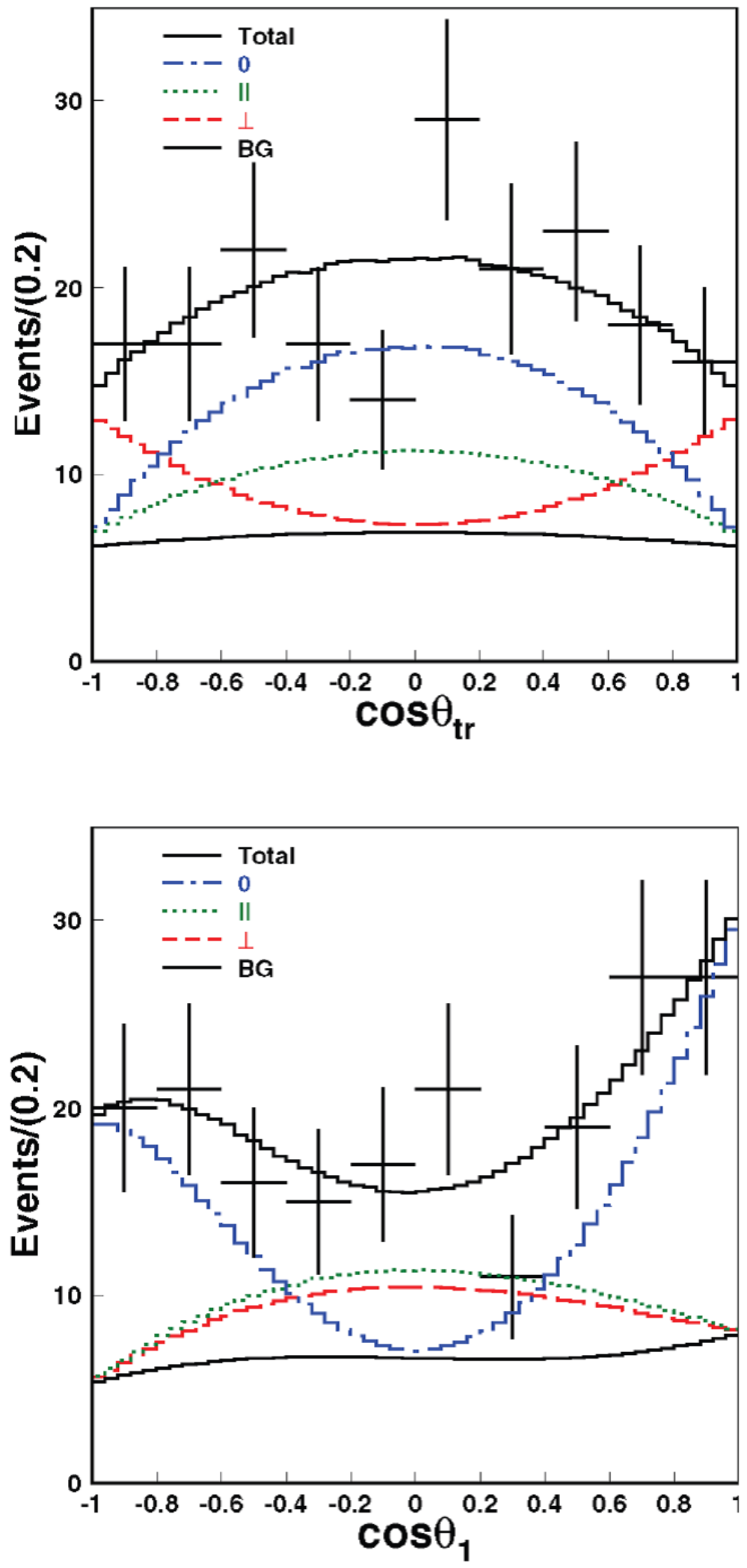

**Figure 12.2.1.** Angular distributions in the transversity frame for (top) $\cos\theta_{tr}$ and (bottom) $\cos\theta_1$, shown in the example of the $B^0 \to D^{*+}D^{*-}$ analysis (Miyake, 2005). The points with error bars represent the data. The dot-dashed, dotted, and dashed lines correspond to the $A_L$, $A_\parallel$, and $A_\perp$ amplitudes respectively. The lower solid line is the background (BG), while the upper solid line shows the sum of all contributions. The asymmetry in the $\cos\theta_1$ distribution is due to an inefficiency for low momentum track reconstruction.

### 12.2.6 $P \to VV\ , V \to P\gamma$

An example of a vector-vector final state, where both vector mesons decay to a pseudoscalar meson and a photon, is:

– $B_s \to D_s^* D_s^*$, with $D_s^* \to D_s\gamma$.

The differential decay rate in the helicity basis, integrated over the $\phi$ angle, is:

$$\frac{1}{\Gamma}\,\frac{d^2\Gamma}{d\cos\theta_1 d\cos\theta_2} = \frac{9}{4}\Big(f_L \sin^2\theta_1\sin^2\theta_2 + (1-f_L)\frac{1}{4}(1+\cos^2\theta_1)(1+\cos^2\theta_2)\Big)\,. \tag{12.2.10}$$

### 12.2.7 $P \to VV\ , V_1 \to PP\ , V_2 \to ll$

In the case of a $B$-meson decay to two vector mesons, where $M_1$ decays to two pseudoscalar mesons and $M_2$ decays to two leptons, the transversity basis is used. An example of this is $B \to K^*\psi$, with $K^* \to K\pi$ and $\psi \to e^+e^-$, where $\psi$ is either $J/\psi$ or $\psi(2S)$. The $g_i^{tr}$ functions have the following angular dependence:

$$\begin{aligned}
g_1^{tr} &= \frac{1}{2}\cos^2\theta_1(1-\sin^2\theta_{tr}\cos^2\phi_{tr})\,,\\
g_2^{tr} &= \frac{1}{4}\sin^2\theta_1(1-\sin^2\theta_{tr}\sin^2\phi_{tr})\,,\\
g_3^{tr} &= \frac{1}{4}\sin^2\theta_1\sin^2\theta_{tr}\,,\\
g_4^{tr} &= \eta\frac{1}{4}\sin^2\theta_1\sin 2\theta_{tr}\sin\phi_{tr}\,,\\
g_5^{tr} &= -\frac{1}{4\sqrt{2}}\sin 2\theta_1\sin^2\theta_{tr}\sin 2\phi_{tr}\,,\\
g_6^{tr} &= \eta\frac{1}{4\sqrt{2}}\sin 2\theta_1\sin 2\theta_{tr}\cos\phi_{tr}\,.
\end{aligned} \tag{12.2.11}$$

### 12.2.8 $P \to VV\ , V_1 \to PP\ , V_2 \to V\gamma$

The case of a $B$-meson decay to two vector mesons, where $M_1$ decays to two pseudoscalar mesons and $M_2$ decays to a vector meson and a photon, is illustrated by the decay $B \to K^*\chi_{c1}$, with $K^* \to K\pi$ and $\chi_{c1} \to J/\psi\gamma$. The transversity basis is used in this case. The $g_i^{tr}$ functions have the following angular dependence:

$$\begin{aligned}
g_1^{tr} &= \frac{1}{2}\cos^2\theta_1(1+\sin^2\theta_{tr}\cos^2\phi_{tr})\,,\\
g_2^{tr} &= \frac{1}{4}\sin^2\theta_1(1+\sin^2\theta_{tr}\sin^2\phi_{tr})\,,\\
g_3^{tr} &= \frac{1}{4}\sin^2\theta_1(2\cos^2\theta_{tr}+\sin^2\theta_{tr})\,,\\
g_4^{tr} &= -\eta\frac{1}{4}\sin^2\theta_1\sin 2\theta_{tr}\sin\phi_{tr}\,,\\
g_5^{tr} &= -\frac{1}{4\sqrt{2}}\sin 2\theta_1\sin^2\theta_{tr}\sin 2\phi_{tr}\,,\\
g_6^{tr} &= -\eta\frac{1}{4\sqrt{2}}\sin 2\theta_1\sin 2\theta_{tr}\cos\phi_{tr}\,.
\end{aligned} \tag{12.2.12}$$

### 12.2.9 $P \to TV$ , $T \to PP$ , $V \to PP$

The mode $B \to K_2^*(1430)\phi$, with $K_2^*(1430) \to K\pi$ and $\phi \to K^+K^-$, is an example of a pseudoscalar meson decaying to a tensor and a vector meson, each of them decaying to two pseudoscalar mesons. In the helicity basis the $g_i$ functions have the following angular dependence (Datta et al., 2008):

$$\begin{aligned} g_1 &= \frac{5}{12}(3\cos^2\theta_1 - 1)^2\cos^2\theta_2\,, \\ g_2 &= \frac{5}{4}\cos^2\theta_1\sin^2\theta_1\sin^2\theta_2\,, \\ g_3 &= \frac{5}{4}\cos^2\theta_1\sin^2\theta_1\sin^2\theta_2\cos 2\phi\,, \\ g_4 &= -\eta\frac{5}{2}\cos^2\theta_1\sin^2\theta_1\sin^2\theta_2\sin 2\phi\,, \\ g_5 &= \frac{5}{8\sqrt{6}}(3\cos^2\theta_1 - 1)\sin 2\theta_1\sin 2\theta_2\cos\phi\,, \\ g_6 &= -\eta\frac{5}{8\sqrt{6}}(3\cos^2\theta_1 - 1)\sin 2\theta_1\sin 2\theta_2\sin\phi\,. \end{aligned} \qquad (12.2.13)$$

After integrating over the $\phi$ angle, the four last terms disappear and the differential decay rate depends simply on the parameter $f_L$:

$$\frac{1}{\Gamma}\,\frac{d^2\Gamma}{d\cos\theta_1 d\cos\theta_2} = \frac{15}{16}\Big(f_L(3\cos^2\theta_1 - 1)^2\cos^2\theta_2 + 3(1-f_L)\cos^2\theta_1\sin^2\theta_1\sin^2\theta_2\Big)\,. \qquad (12.2.14)$$

## 12.3 Analysis details

### 12.3.1 Generators

In order to perform an angular analysis it is important to have simulated data with the correct angular distributions. This allows one to calculate, for example, the correct efficiencies on the signal (see the next subsection) and to study how well (with how much bias) the angular fits described in Section 12.4 can extract the fitted parameters.

Here is a brief explanation of how this is achieved in the `EvtGen` event generator (Lange, 2001) introduced in Chapter 3. The crucial point is that decay amplitudes, and not probabilities, are used for each step in the generation of a decay chain. This allows one to include all angular correlations in the entire decay chain. Each particle is described according to the value of its spin and mass by an object with the corresponding number of degrees of freedom. Each decay in the decay chain is handled by a specific model taking into account the spin of the initial and final state particles. Relevant parameters can be given as arguments to the decay model. For example in the case of the model describing the decay of a scalar to two vector mesons, the six arguments are the magnitude and the phase of the three helicity amplitudes.

### 12.3.2 Experimental effects

A large number of angular analyses require cuts on the helicity angles, $\theta_i$ ($i = 1$ or 2), as the region at high values of $|\cos\theta_i|$, which usually corresponds to soft decay products, has a rapidly changing efficiency and may be dominated by background. The cut may be asymmetric if the decay products are different. For example, if one of the decay products is a $\rho^+$ vector meson, decaying subsequently into $\pi^+\pi^0$, the kinematics of the $\rho^+$-meson decay are strongly correlated with the value of the relevant helicity angle $\theta_i$. Assuming $\theta_i$ was defined with respect to the $\pi^+$, high ($\cos\theta_i \sim 1$), medium ($\cos\theta_i \sim 0$), and low ($\cos\theta_i \sim -1$) values correspond respectively to a decay with a hard $\pi^+$ and a soft $\pi^0$, two pions of similar momentum, and a hard $\pi^0$ and a soft $\pi^+$ in the laboratory frame. Since there is usually a huge background of low momentum $\pi^0$s, an upper cut on $\cos\theta_i$ in this case sould be tighter than a lower cut in order to reduce the soft $\pi^0$ background.

For the same reason, the reconstruction efficiency depends upon the fraction of longitudinal polarization. Defining $\epsilon_L$ ($\epsilon_T$) as the reconstruction efficiency obtained if the signal was completely longitudinally (transversely) polarized, *i.e.* $f_L = 1$ ($f_L = 0$), $\epsilon_L$ and $\epsilon_T$ would be different, usually with $\epsilon_L < \epsilon_T$. This effect has to be taken into account to correct the measured raw value $f_L^{\rm meas}$ of the fraction of longitudinal polarization to obtain the true longitudinal polarization fraction:

$$f_L = \frac{f_L^{\rm meas}}{f_L^{\rm meas} + (1 - f_L^{\rm meas})\frac{\epsilon_L}{\epsilon_T}}\,. \qquad (12.3.1)$$

Similarly, the rate of mis-reconstructed signal events depends on the value of the fraction of longitudinal polarization.

In the various analyses, the efficiency is often modeled as a function of $\cos\theta_i$ ($i = 1$ or 2) with an appropriate function $A(\cos\theta_i)$. Figure 12.3.1 illustrates the efficiency function in the case of the $B^0 \to \phi K^{*0}$ analysis (Aubert, 2008bf). This otherwise smooth function shows some sharp dips due to $D$ meson vetoes (special cuts applied in the analysis in order to reject the background coming from a $D$ meson decay), as seen in Figure 12.3.1(a) near $\cos\theta_1 = 0.8$.

### 12.3.3 Caveats

Here is a discussion of some technical points that should be considered in specific angular analyses.

When studying decays with identical particles in the final state, the formulae need to be symmetrized. For example the $B^0 \to \rho^0\rho^0 \to (\pi^+\pi^-)(\pi^+\pi^-)$ decay has four bosons, identical by pairs, in the final state. In this case the amplitude $A(p_1^+, p_1^-, p_2^+, p_2^-)$, as a function of the four-momenta, $p_1^+$, $p_2^+$, $p_1^-$, and $p_2^-$, of the two $\pi^+$ and the two $\pi^-$, has to be replaced by the symmetrized amplitude under the permutations $p_1^+ \to p_2^+$ and $p_1^- \to p_2^-$.

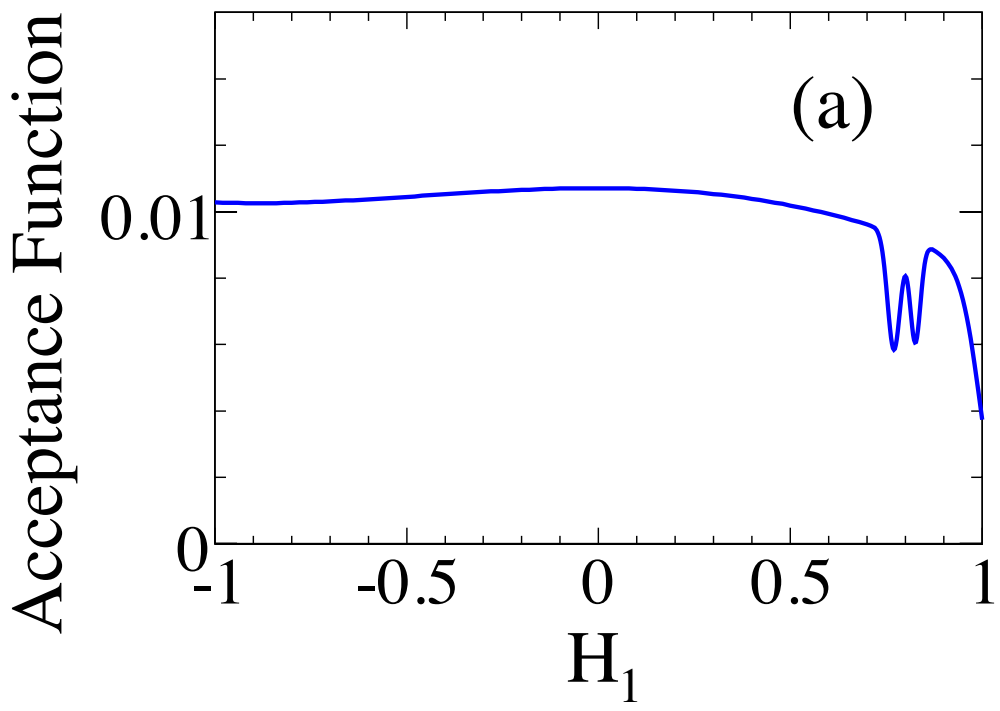

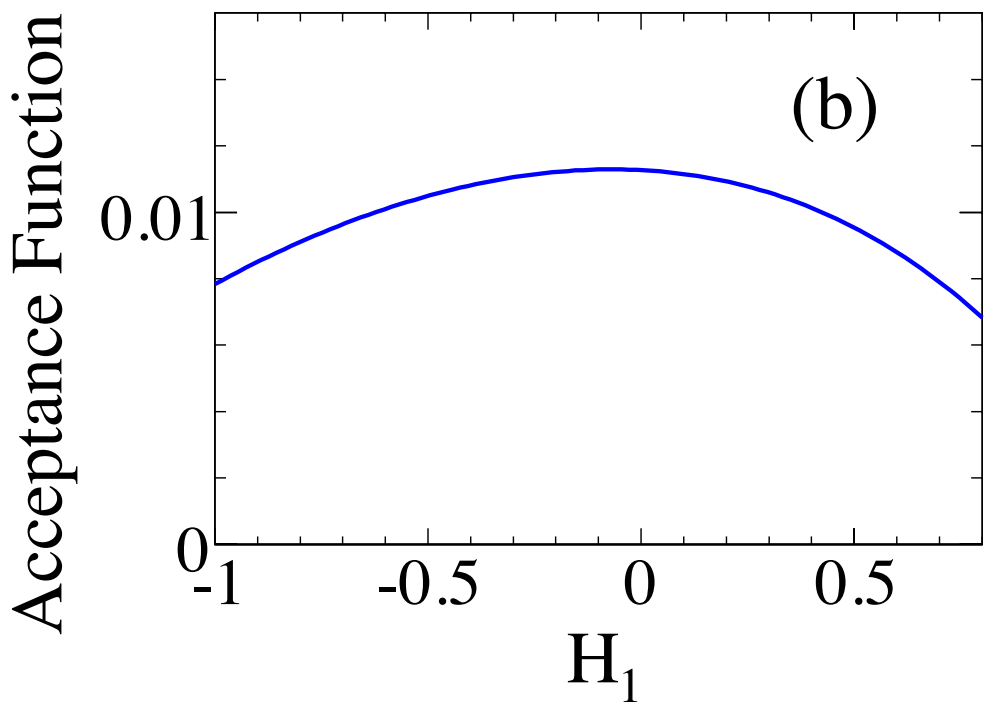

**Figure 12.3.1.** Angular efficiency functions for $H_1 = \cos\theta_1$ (here $\theta_1$ is the angle associated with the $K\pi$ system) in the cases of (a) $B^0 \to \phi K^{\pm}\pi^{\mp}$ and (b) $B^0 \to \phi K^0_s\pi^0$ (Aubert, 2008bf). The wiggles in the upper plot are due to the $D$ meson vetoes.

When performing a multi-variable maximum likelihood fit, care has to be taken for correlations between variables. In particular, continuum events tend to have correlations between the masses of the reconstructed $M_i$ ($i = 1$ or 2) candidates and the cosine of their helicity angles. A solution is to use a two-dimensional probability density function in this case.

## 12.4 Angular fits

In this section, the various types of angular fits which have been performed are quickly reviewed.

### 12.4.1 Dedicated or global fits

Two strategies are possible:

- The signal yield is first extracted using variables such as $m_{\rm ES}$ and $\Delta E$. Second, only the angular variables are fitted to extract the polarization information. Where appropriate, time-dependent information can be obtained in a third step.
- A single maximum likelihood fit is performed using the signal selection variables, as well as the angular variables, and any relevant time-dependence. The polarization parameters are determined in the fit at the same time as other parameters such as signal yields.

The angular parameters can usually be extracted in time-integrated analyses. Numerous results in various decay modes are given in this book, in particular in Section 17.4. The time-dependence, when used, is added essentially to study *CP* violation, as shown in Sections 17.6 and 17.7.

The angular information is also used in Dalitz analyses, through either the helicity formalism or Zemach tensors, as described in Chapter 13.

### 12.4.2 Partial and complete angular analyses

Most angular analyses integrate over the angle $\phi$, for which the acceptance in the $B$ Factory detectors is uniform, to determine the fraction of longitudinally polarized events: $f_L$. The helicity basis is the natural one to use in this case, as the two daughters are treated symmetrically. The formulae to fit have been given in Section 12.2 for different cases:

- Eq. (12.2.5) for $P \to VV\ , V \to PP$,
- Eq. (12.2.9) for $P \to VV\ , V_1 \to P\gamma\ , V_2 \to PP$,
- Eq. (12.2.10) for $P \to VV\ , V \to P\gamma$,
- Eq. (12.2.14) for $P \to TV\ , T \to PP, V \to PP$.

Partial angular analyses have been performed to measure $f_L$ in a large number of decay modes, such as $B \to \rho\rho$, $B \to K^*\rho$, and $B \to D^*K^*$. In some cases, the angular analysis is performed to disentangle the *CP*-even and *CP*-odd components. In that case, $f_\perp$ has to be measured and the transversity basis is more suited for such a partial angular analysis. If the decay is dominated by the *CP*-even longitudinal polarization, however, one can effectively use either basis and deal with the small transverse component when addressing systematic uncertainties. If no attempt is made to disentangle the *CP*-even and *CP*-odd components, the mixture of these two components results in a dilution of the *CP* asymmetry.

Finally, in a limited number of channels, a complete angular fit has been performed to measure not only the fractions of the three amplitudes, but also the relative phases between them. Of course, the complete angular analysis is more difficult than the partial one, as it implies fitting more free parameters. Consequently it requires sufficiently large data samples. Such an analysis has been performed in the $B$-meson decays to $\phi K^*$, both in the vector-vector modes ($B^+ \to \phi K^{*+}$ and $B^0 \to \phi K^{*0}$) using either Eq. (12.2.4) or Eq. (12.2.6), and in the vector-tensor mode ($B^0 \to \phi K_2^{*0}(1430)$) using Eq. (12.2.13). More details can be found in (Chen, 2005a), (Aubert, 2007c), and (Aubert, 2008bf). A complete angular analysis was also performed in the $B$-meson decays to charmonium $K^*$, according to Eq. (12.2.11) when the charmonium decays to two leptons ($B^+ \to J/\psi K^{*+}$, $B^0 \to J/\psi K^{*0}$, and $B^0 \to \psi(2S)K^{*0}$), and to Eq. (12.2.12) when the charmonium decays to a vector meson and a photon ($B^0 \to \chi_{c1}K^{*0}$). They are documented in (Aubert, 2005c), (Itoh, 2005b), and (Aubert, 2007x).

### 12.4.3 Other angular analyses

Not all the types of angular analyses have been covered in this chapter and other kinds of angular analyses have also been performed at $B$ Factories. The goal may be to determine the unknown spin of a particle by studying the angular distribution of its decay products. Examples can be found in charmed meson spectroscopy (Section 19.3) and in charmed baryon spectroscopy (Section 19.4). Angular analyses also allow one to study angular asymmetries or correlations, in particular in the case of baryonic decay modes which are presented in Section 17.12, in order to investigate the underlying dynamics of the decay. Finally, in two-photon physics, described in Chapter 22, the angular dependence of the differential cross section for various processes is studied.

# Chapter 13
# Dalitz-plot analysis

***Editors:***
*Thomas Latham (BABAR)*
*Anton Poluektov (Belle)*

**Additional section writers:**
*Eli Ben-Haim, Mathew Graham, Fernando Martinez-Vidal*

## 13.1 Introduction

Dalitz-plot analysis is a powerful technique that involves studying the amplitude for the decay of a parent particle into a three-body final state. Compared to two-body decays, the three-body decay possesses intrinsic degrees of freedom that permit the determination of the relative magnitudes and phases of interfering amplitudes. The types of measurements that can benefit from using the Dalitz-plot analysis technique include:

– Searches for new states;
– Measurements of properties of resonances — masses, widths, quantum numbers;
– *CP* violation searches and measurements of the associated parameters;
– Studies of flavor mixing.

This chapter starts with a discussion of the kinematics of three-body decays (Sections 13.1.1 and 13.1.2) before describing the formalisms commonly used to model the three-body decay amplitude (Section 13.2). This is followed by an outline of the experimental effects that must also be accounted for in order to successfully describe the distribution of the data over the Dalitz plot (Section 13.3). Technical details of the implementation are presented in Section 13.4 before a discussion of the uncertainties arising from the chosen model (Section 13.5).

### 13.1.1 Three-body decay phase space

In the case of a two-body decay, the energies of the final state particles in the center-of-mass frame are fully determined by the conservation of energy and momentum, up to an overall rotation. In contrast, the kinematics of three-body decays are not similarly constrained: after requiring energy and momentum conservation in the system of three final state particles, there are five remaining degrees of freedom. In the case where the initial and final state particles all have spin zero, after taking into account arbitrary rotations, two degrees of freedom remain. The amplitude of the decay can thus be represented as a function of two parameters; the scatter plot of this pair of parameters is called the Dalitz plot (Dalitz distribution).

There is freedom in the choice of which two parameters one uses to describe the amplitude of a three-body decay. It is often convenient to choose a pair of parameters where the phase-space term is constant within the kinematically allowed region in the two-dimensional space spanned by these variables. In this case the structure of the amplitude becomes apparent. This can be achieved by taking either the kinetic energies of two of the final-state particles, or the squares of the invariant masses of two pairs of final-state particles. The former parameterization is convenient for nonrelativistic decays and was originally proposed by R. H. Dalitz to study the decay of charged kaons to three pions (Dalitz, 1953). The corresponding relativistic formulation was first introduced in Fabri (1954). However, the latter approach is generally more suitable for relativistic decays and has an additional advantage that it allows for easy determination of the masses of intermediate states. For a particle of mass $M$ decaying into three particles denoted as $a$, $b$ and $c$, the differential decay probability is

$$d\Gamma = \frac{1}{(2\pi)^3}\frac{1}{32M^3}|\mathcal{A}|^2 dm^2_{ab}dm^2_{bc}\,, \tag{13.1.1}$$

where $m_{ab}$ and $m_{bc}$ are the invariant masses of the pairs of particles $ab$ and $bc$, respectively. Thus, any nonuniformity observed in the distribution of the variables $m^2_{ab}$ and $m^2_{bc}$ is due to the dynamical structure of the decay amplitude $\mathcal{A}$. Most of the analyses performed at the $B$ Factories deal with the Dalitz plot expressed this way; the exception to this will be considered in Section 13.4.1.

### 13.1.2 Boundaries, kinematic constraints

The invariant masses of pairs of final-state particles are related by the linear dependence:

$$m^2_{ab} + m^2_{bc} + m^2_{ac} = M^2 + m^2_a + m^2_b + m^2_c. \tag{13.1.2}$$

The range of invariant masses $m^2_{bc}$ can be written in terms of either one of the other squared invariant masses (*e.g.*, $m^2_{ab}$):

$$\begin{aligned}(m^2_{bc})_{\max} &= (E^*_b + E^*_c)^2 - (p^*_b - p^*_c)^2\,,\\ (m^2_{bc})_{\min} &= (E^*_b + E^*_c)^2 - (p^*_b + p^*_c)^2\,,\end{aligned} \tag{13.1.3}$$

where

$$E^*_b = \frac{m^2_{ab} - m^2_a + m^2_b}{2m_{ab}}\,,\quad E^*_c = \frac{M^2 - m^2_{ab} - m^2_c}{2m_{ab}}\,, \tag{13.1.4}$$

are the energies of particles $b$ and $c$ in the $ab$ rest frame and

$$p^*_b = \sqrt{E^{*2}_b - m^2_b}\,,\quad p^*_c = \sqrt{E^{*2}_c - m^2_c}\,, \tag{13.1.5}$$

are the corresponding momenta.

The region of kinematically allowed phase space described by these constraints is shown in Fig. 13.1.1. The points on the boundary of the phase space correspond to the configurations where the final state particles are collinear. In particular, three extreme points where $m^2_{ab}$, $m^2_{bc}$, or $m^2_{ac}$ are maximal, correspond to the configurations with one of the particles produced at rest (in the frame of the decaying particle).

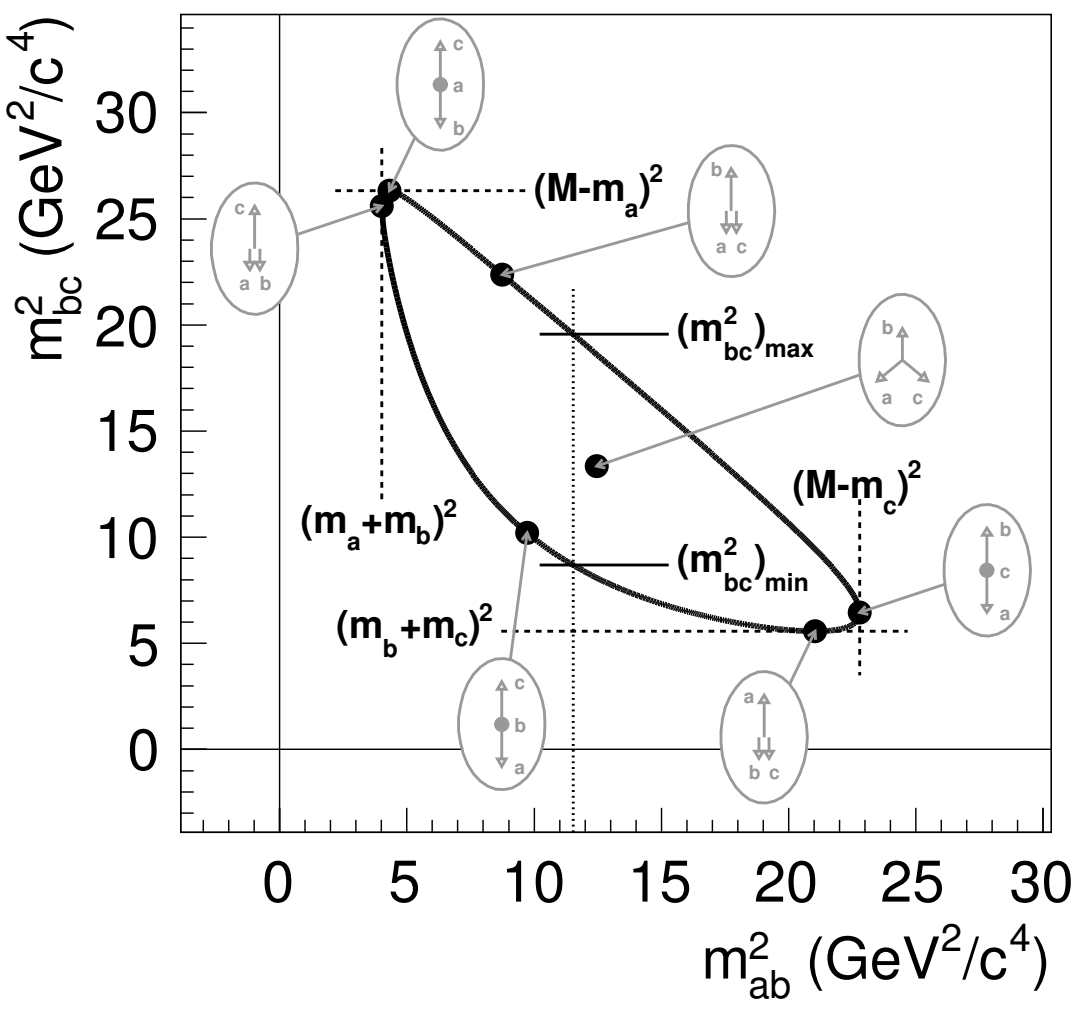

**Figure 13.1.1.** Kinematic boundaries of the three-body decay phase space and illustration of various kinematic configurations of the final-state particles for characteristic Dalitz plot points. In this example, the $B^0 \to \pi^- \overline{D}^0 K^+$ phase space is shown; $a = \pi^-$, $b = \overline{D}^0$, $c = K^+$.

## 13.2 Amplitude description

Experimental data show that nonleptonic three-body $B$ and $D$ decays proceed predominantly through resonant two-body decays. For three-body decays of a spin-zero particle $P$ (*e.g.*, a $D$ or $B$ meson) to pseudoscalar final-state particles $abc$, the baseline model commonly adopted to describe the decay amplitude $\mathcal{A}(m^2_{ab}, m^2_{bc})$ consists of a coherent sum of two-body amplitudes (subscript $r$) and a "nonresonant" (subscript NR) contribution (Beringer et al., 2012),

$$\mathcal{A}(\mathbf{m}) = \sum_r a_r e^{i\phi_r} \mathcal{A}_r(\mathbf{m}) + a_{\rm NR} e^{i\phi_{\rm NR}} \mathcal{A}_{\rm NR}(\mathbf{m}) \,. \quad (13.2.1)$$

The parameters $a_r$ ($a_{\rm NR}$) and $\phi_r$ ($\phi_{\rm NR}$) are the magnitude and phase of the amplitude for component $r$ (NR). The functions $\mathcal{A}_r$ and $\mathcal{A}_{\rm NR}$ are Lorentz-invariant expressions that describe the dynamical properties of the decay into the multi-body final state as a function of position in the Dalitz plot $\mathbf{m} \equiv (m^2_{ab}, m^2_{bc})$. When the final state contains identical particles, *e.g.* $D^+ \to \pi^+\pi^+\pi^-$ or $B^0 \to K^0_S K^0_S K^0_S$, it is important that the total amplitude $\mathcal{A}(\mathbf{m})$ is correctly symmetrized with respect to exchange of those particles.

The most common ways to parameterize the functions $\mathcal{A}_r$ are reviewed in the following Sections 13.2.1 and 13.2.2. The parameterizations of nonresonant amplitude are discussed in Section 13.2.3. Section 13.2.4 discusses a special case of time-dependent amplitude analyses.

### 13.2.1 Isobar formalism

The isobar formalism (or isobar model) is so-called because it was first used to describe pion-nucleon, nucleon-nucleon, and antinucleon-nucleon interactions (Sternheimer and Lindenbaum, 1961). In such reactions the intermediate resonances are isobars of a particular nuclear state. The isobar model was later generalized to any three-body final state (Herndon, Soding, and Cashmore, 1975).

In this formalism, the function $\mathcal{A}_r$ describes the decay through a single intermediate resonance $r$ and takes the form

$$\mathcal{A}_r = F_P \times F_r \times T_r \times W_r \,, \quad (13.2.2)$$

where $T_r \times W_r$ is the resonance propagator ($T_r$ is the dynamical function for the resonance $r$, while $W_r$ describes the angular distribution of the decay), $F_P$ and $F_r$ are the transition form factors of the parent particle and resonance, respectively. In what follows, we assume that the resonance is produced in the $ab$ channel. In that case the particle $c$ will be referred to as the bachelor particle. Naturally, the full amplitude $\mathcal{A}$ may contain contributions of resonances in any of the $ab$, $ac$, and $bc$ channels.

The dynamical function $T_r$ is commonly described using a relativistic Breit-Wigner (BW) parameterization with mass-dependent width (see, *e.g.*, review on Dalitz plot analysis formalism on p. 889 in Beringer et al. (2012))

$$T_r = \frac{1}{m_r^2 - m_{ab}^2 - i m_r \Gamma_{ab}} \,. \quad (13.2.3)$$

Here $m_r$ is the mass of the resonance, and the mass-dependent width $\Gamma_{ab}$ is given by

$$\Gamma_{ab} = \Gamma_r \left(\frac{q_{ab}}{q_r}\right)^{2J+1} \left(\frac{m_r}{m_{ab}}\right) F_r^2 \,, \quad (13.2.4)$$

where $\Gamma_r$ and $J$ are the width and spin of the resonance, $q_{ab}$ is the momentum of the daughter particles in the center-of-mass frame of $a$ and $b$, and $q_r$ is the momentum the decay products would have in the rest frame of a resonance with mass $m_r$.

Strictly speaking, the Breit-Wigner parameterization works well only in the case of narrow states. The use of the mass-dependent width results in the amplitude $T_r$ becoming a non-analytic function. An alternative parametrization proposed by Gounaris and Sakurai (GS) (Gounaris and Sakurai, 1968) recovers the analyticity of the amplitude and provides a better description for broad vector resonances such as $\rho(770)$ and $\rho(1450)$.

For resonances such as the $f_0(980) \to \pi\pi$ that lie close to the threshold of another channel ($f_0(980) \to KK$ in this case), the effect of the opening of the second channel must be taken into account, for example, by employing the Flatté coupled-channel form (Flatte, 1976),

$$T_r = \frac{g_1}{m_r^2 - m_{ab}^2 - i(\rho_1 g_1^2 + \rho_2 g_2^2)} \,, \quad (13.2.5)$$

where $\rho_1$, $\rho_2$ and $g_1$, $g_2$ are the phase-space factors and coupling constants of the $\pi\pi$ and $KK$ channels, respectively.

Values of the mass and width of resonances are in general taken from world averages (Beringer et al., 2012).

Since different parameterizations of the resonance lineshapes, especially for broad resonances, often give different values, one has to make sure that the values used in the fit were extracted using the same parameterization as in the model. If the resonance is apparent and systematic biases (or external errors) of its parameters are expected to be larger than their statistical errors from the fit, the mass and width can be left unconstrained.

The angular dependence $W_r$ is described using either Zemach tensors (Zemach, 1964, 1965), where transversality is enforced, or the helicity formalism (Bonvicini et al., 2008; Jacob and Wick, 1959), which allows for a longitudinal component in the resonance propagator (see Beringer et al. (2012) for a comprehensive summary). The expressions for scalar, vector and tensor states are

$$\begin{aligned}
J=0: \quad & W_r = 1\,, \qquad\qquad (13.2.6)\\
J=1: \quad & W_r = m_{ac}^2 - m_{bc}^2 - \frac{(M^2-m_c^2)(m_a^2-m_b^2)}{m_r^2}\,,\\
J=2: \quad & W_r = \left[m_{bc}^2 - m_{ac}^2 + \frac{(M^2-m_c^2)(m_a^2-m_b^2)}{m_r^2}\right]^2 - \\
& \frac{1}{3}\left[m_{ab}^2 - 2M^2 - 2m_c^2 + \frac{(M^2-m_c^2)^2}{m_r^2}\right]\times \\
& \left[m_{ab}^2 - 2m_a^2 - 2m_b^2 + \frac{(m_a^2-m_b^2)^2}{m_r^2}\right].
\end{aligned}$$

Transversality is enforced by substituting $m_{ab}^2$ for $m_r^2$ in the denominators of the previous expressions. This leads to the alternative expressions

$$\begin{aligned}
J=0: \quad & W_r = 1, \qquad\qquad (13.2.7)\\
J=1: \quad & W_r = -2\,(\boldsymbol{p}\cdot\boldsymbol{q})\,,\\
J=2: \quad & W_r = \frac{4}{3}\left[3\,(\boldsymbol{p}\cdot\boldsymbol{q})^2 - (|\boldsymbol{p}|\,|\boldsymbol{q}|)^2\right]\,,
\end{aligned}$$

where $\boldsymbol{q}$ and $\boldsymbol{p}$ are the momenta of one of the resonance daughters and the bachelor particle, respectively, evaluated in the rest frame of the resonance. The decision as to which daughter to choose is a matter of convention and it is very important that this choice be documented since it affects the interpretation of the relative phases. The angle between $\boldsymbol{q}$ and $\boldsymbol{p}$ is known as the helicity angle (see also Section 12.1) and $\boldsymbol{p}\cdot\boldsymbol{q}$ is proportional to the cosine of the helicity angle $\cos\theta_H$. The Zemach expressions are essentially Legendre polynomials of $\cos\theta_H$ multiplied by coefficients that contain the momenta of the daughter and bachelor particles raised to the power $J$.

The form factors $F_P$ and $F_r$ usually use the Blatt-Weisskopf parameterization for the decay vertex (Blatt and Weisskopf, 1952). The expressions for the Blatt-Weisskopf penetration factors depend on the spin $J$ of the intermediate resonance

$$\begin{aligned}
J=0: \quad & F = 1\\
J=1: \quad & F = \sqrt{\frac{1+R^2q_r^2}{1+R^2q_{ab}^2}} \qquad\qquad (13.2.8)\\
J=2: \quad & F = \sqrt{\frac{9+3R^2q_r^2+R^4q_r^4}{9+3R^2q_{ab}^2+R^4q_{ab}^4}}\,,
\end{aligned}$$

where $R$ is the radial parameter of the decaying meson and typically takes values between 1 and 5 $(\mathrm{GeV})^{-1}$. In this prescription, $F$ is normalized so that $F=1$ for $q_r = q_{ab}$.

While the $P$- and $D$-waves of the decay amplitude are usually well described using a certain number of BW or GS propagators, the actual number depending on the specific decay, the $S$-wave typically contains a number of broad overlapping states, for which the isobar model gives a poor description. In that case, more complex alternatives have been adopted, which are reviewed in Section 13.2.2.

Figure 13.2.1 illustrates how various intermediate two-body states appear in the Dalitz plot. Unlike the uniform distribution of the phase-space decay (Fig. 13.2.1(a)), scalar resonances appear as bands in the Dalitz plot, as shown in Fig. 13.2.1(b-d) for resonances in $bc$, $ac$, and $ab$ channels, respectively. Angular distributions for vector and tensor intermediate states introduce characteristic non-uniformity of the event density along the resonance bands (Fig. 13.2.1(e,f)). Finally, the region where the amplitudes of two resonances overlap is sensitive to the phase difference between the two amplitudes (Fig. 13.2.1(g,h)).

### 13.2.2 K-matrix formalism

The complex $S$-wave dynamics, which can also include the presence of several broad and overlapping scalar resonances, can alternatively be described through the use of a K-matrix formalism (Chung et al., 1995; Wigner, 1946) with the production vector (P-vector) approximation (Aitchison, 1972). Within this formalism, the production process described by the P-vector can be viewed as the initial formation of several states, which are then propagated by the K-matrix term into the final state that is observed. This approach ensures that the two-body scattering matrix respects unitarity, which is not guaranteed in the case of the isobar model. At the $B$ Factories this approach is most commonly used to describe the $\pi^+\pi^-$ $S$-wave contribution to the Dalitz-plot amplitude, *e.g.* in the *BABAR* analyses (Aubert, 2008l, 2009h; del Amo Sanchez, 2010a,b) and Belle analysis (Abe, 2007b). In such cases the amplitude is given by

$$\mathcal{A}_u(s) = \sum_v \left[I - iK(s)\rho(s)\right]^{-1}_{uv} P_v(s)\,, \qquad (13.2.9)$$

where $s \equiv m_{ab}^2$ is the $\pi^+\pi^-$ invariant mass, $I$ is the identity matrix, $K$ is the matrix describing the scattering process, $\rho$ is the diagonal phase-space matrix, and $P$ is the production vector. The indices $u$ and $v$ represent the production and scattering channels, respectively, and take the values 1 to 5, where $1=\pi\pi$, $2=K\overline{K}$, $3=\pi\pi\pi\pi$, $4=\eta\eta$, $5=\eta\eta'$. Hence in the case of describing the $\pi^+\pi^-$ amplitude $u=1$. The propagator can be described using scattering data, provided that the two-body system in the final state is isolated and does not interact with the rest of the final state in the production process.

The parameterizations adopted for the $K$, $\rho$, and $P$ terms in Eq. (13.2.9) by the $B$ Factories are the same as those used by previous analyses (Anisovich and Sarantsev,

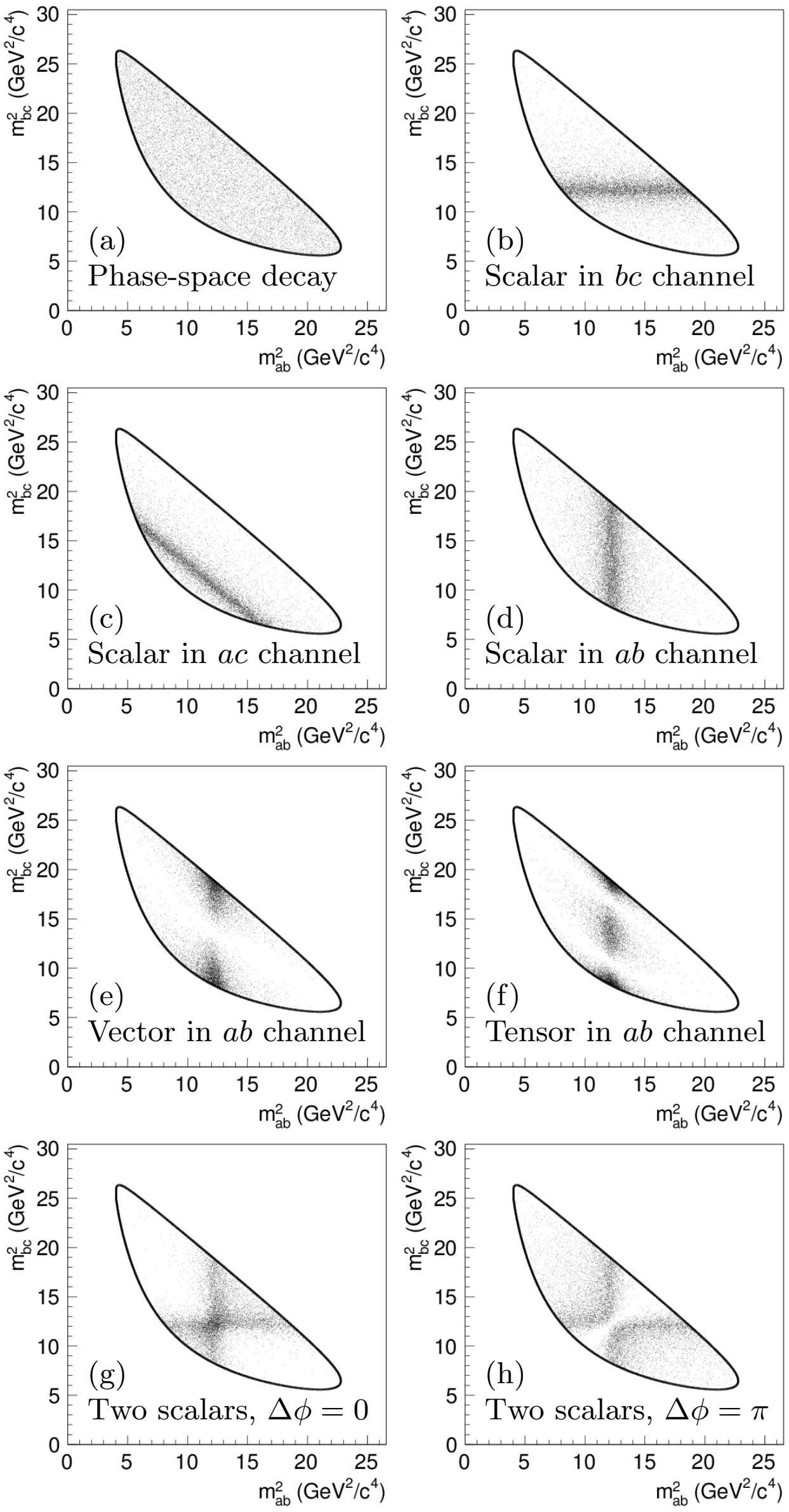

**Figure 13.2.1.** Example Dalitz plots with (a) phase-space decay, (b-d) one scalar resonance appearing in various decay channels, (e, f) vector and tensor resonances, and (g, h) the interference of two scalar resonances with different values of the relative phase $\Delta\phi$.

2003; Link et al., 2004a), up to some sign conventions and constant terms. The K-matrix is formulated as

$$K_{uv}(s) = \left( \sum_{\alpha} \frac{g_u^{\alpha} g_v^{\alpha}}{m_{\alpha}^2 - s} + f_{uv}^{\text{scatt}} \frac{1 - s_0^{\text{scatt}}}{s - s_0^{\text{scatt}}} \right) f_{A0}(s), \tag{13.2.10}$$

where $g_u^{\alpha}$ is the coupling constant of the K-matrix pole at $m_{\alpha}$ to the $u^{\text{th}}$ channel. The parameters $f_{uv}^{\text{scatt}}$ and $s_0^{\text{scatt}}$ describe the slowly varying part of the K-matrix. The factor

$$f_{A0}(s) = \frac{1 - s_{A0}}{s - s_{A0}} \left( s - s_A \frac{m_{\pi}^2}{2} \right) \tag{13.2.11}$$

suppresses the false kinematic singularity at $s = 0$ in the physical region near threshold, the Adler zero (Adler, 1965). For example, the parameter values used in the *BABAR* analysis of $D^0 \to K_S^0 \pi^+ \pi^-$ (Aubert, 2008l) are listed in Table 13.2.1, and are adapted from a global analysis of the available $\pi\pi$ scattering data from threshold up to 1900 MeV/$c^2$ (Anisovich and Sarantsev, 2003). The parameters $f_{uv}^{\text{scatt}}$, for $u \neq 1$, are all set to zero since they are not related to the $\pi\pi$ scattering process. Similarly, the parameterization for the P-vector is

$$P_v(s) = \sum_{\alpha} \frac{\beta_{\alpha} g_v^{\alpha}}{m_{\alpha}^2 - s} + f_{1v}^{\text{prod}} \frac{1 - s_0^{\text{prod}}}{s - s_0^{\text{prod}}} \,. \tag{13.2.12}$$

Note that the P-vector has the same poles as the K-matrix, otherwise the $\mathcal{A}_1$ amplitude would vanish (diverge) at the K-matrix (P-vector) poles. The parameters $\beta_{\alpha}$, $f_{1v}^{\text{prod}}$, and $s_0^{\text{prod}}$ of the initial P-vector depend on the production mechanism and cannot be extrapolated from scattering data. Thus they have to be determined directly from the $D$ or $B$ meson decay data sample. They are complex numbers analogous to the $a_r e^{i\phi_r}$ coefficients in Eq. (13.2.1), hence they can be fitted in the same way.

For the $K\pi$ $S$-wave, the $B$ Factories generally have either used a simple $K_0^*(1430)$ BW that neglects a possible nonresonant contribution or a $K_0^*(1430)$ BW together with an effective-range nonresonant component with a phase shift derived from scattering data (Aston et al., 1988),

$$\mathcal{A}_{K\pi,L=0}(\mathbf{m}) = T_{K\pi,L=0}(s)/\rho(s) \,. \tag{13.2.13}$$

Here $s \equiv m_{K\pi}^2$, $\rho(s) = 2q/\sqrt{s}$ is the phase-space factor, $q$ is the momentum of the kaon and pion in the $K\pi$ rest frame, and

$$\begin{aligned} T_{K\pi,L=0}(s) = B \sin(\delta_B + \phi_B) e^{i(\delta_B + \phi_B)} + \\ R \sin \delta_R e^{i(\delta_R + \phi_R)} e^{i2(\delta_B + \phi_B)} \,, \end{aligned} \tag{13.2.14}$$

where the phases $\delta_B$ and $\delta_R$ have a dependence on $s$ and $q$ given by

$$\begin{aligned} \tan \delta_R &= M\Gamma(s)/(M^2 - s) \,, \\ \cot \delta_B &= 1/(aq) + rq/2 \,. \end{aligned} \tag{13.2.15}$$

The parameters $a$ and $r$ play the role of a scattering length and effective interaction length, respectively, and $B$ ($\phi_B$) and $R$ ($\phi_R$) are the magnitudes (phases) for the nonresonant and resonant terms. $M$ and $\Gamma(s)$ are the mass and mass-dependent width, see Eq. (13.2.4), of the $K_0^*(1430)$ resonance. This parametrization in fact corresponds to a K-matrix approach describing a rapid phase shift coming from the resonant term and a slowly rising phase shift governed by the nonresonant term, with relative strengths $R$ and $B$. The parameters $B$, $\phi_B$, $R$, $\phi_R$, $a$, and $r$ can be determined from the fit to data as with the P-vector parameters and isobar coefficients. Or, in the case of limited data sample, they can be taken from fits to the LASS scattering data (Aston et al., 1988). Other recent experimental efforts to improve the description of the $K\pi$ $S$-wave using K-matrix and model independent parameterizations from large samples of $D^+ \to K^-\pi^+\pi^+$ decays are described in Aitala et al. (2006); Bonvicini et al. (2008); Link et al. (2007).

**Table 13.2.1.** K-matrix parameters used in the *BABAR* analysis of $D^0 \to K^0_S\pi^+\pi^-$ (Aubert, 2008l). They are adapted from the results of a global analysis of the available $\pi\pi$ scattering data from threshold up to 1900 MeV/$c^2$ (Anisovich and Sarantsev, 2003). Masses and coupling constants are given in GeV/$c^2$.

| $m_\alpha$ | $g^\alpha_{\pi^+\pi^-}$ | $g^\alpha_{K\overline{K}}$ | $g^\alpha_{4\pi}$ | $g^\alpha_{\eta\eta}$ | $g^\alpha_{\eta\eta'}$ |
|---|---|---|---|---|---|
| 0.65100 | 0.22889 | −0.55377 | 0.00000 | −0.39899 | −0.34639 |
| 1.20360 | 0.94128 | 0.55095 | 0.00000 | 0.39065 | 0.31503 |
| 1.55817 | 0.36856 | 0.23888 | 0.55639 | 0.18340 | 0.18681 |
| 1.21000 | 0.33650 | 0.40907 | 0.85679 | 0.19906 | −0.00984 |
| 1.82206 | 0.18171 | −0.17558 | −0.79658 | −0.00355 | 0.22358 |
| $s_0^{\rm scatt}$ | $f_{11}^{\rm scatt}$ | $f_{12}^{\rm scatt}$ | $f_{13}^{\rm scatt}$ | $f_{14}^{\rm scatt}$ | $f_{15}^{\rm scatt}$ |
| −3.92637 | 0.23399 | 0.15044 | −0.20545 | 0.32825 | 0.35412 |
| $s_{A0}$ | $s_A$ | | | | |
| −0.15 | 1 | | | | |

### 13.2.3 Nonresonant description

In many analyses the nonresonant amplitude is taken to be a uniform phase-space distribution, *i.e.* a constant magnitude and phase. Indeed, such a constant matrix element is the most strict definition of a nonresonant amplitude. However, final-state interactions and other effects are likely to change this behavior, meaning that a uniform amplitude is not fully motivated. In addition, it is found in many cases not to give a good description of the data. This has been seen both in analyses of charm decays with very large event yields and in analyses of $B$ decays where, although the event yields are generally much smaller, the phase space is considerably larger and so there is greater sensitivity to the nonresonant description. This has led analysts either to adopt various empirical forms or to attempt to use information from scattering data to describe the entire $S$-wave amplitude. The latter approach is described in the previous Section 13.2.2.

An example of one of the empirical forms that has been adopted by Garmash (2005) is

$$\mathcal{A}_{\rm NR} = e^{-\alpha m^2_{ab}}\,, \qquad (13.2.16)$$

where $\alpha$ is a free parameter of the fit. This modification of the uniform amplitude allows for enhancements of the magnitude at lower $m^2_{ab}$ values while the phase remains constant over the Dalitz plot. In most cases more than one such term is employed, often for each neutral or singly charged $m^2_{ij}$ combination. The recent *BABAR* analysis of $B \to KKK$ decays (Lees, 2012y) uses a model that has polynomial dependence on the invariant mass and includes an explicit $P$-wave term.

In recent years there has been an increasing amount of theoretical work towards an understanding of the dynamics of nonresonant three-body amplitudes, see for example Lesniak et al. (2009) and Kamano, Nakamura, Lee, and Sato (2011). In particular, the work focuses on both the effects of final-state interactions and the requirement that two- and three-body prescriptions respect unitarity in the Dalitz-plot model. However, these developments are, in general, yet to be put into practice in the analysis of experimental data.

### 13.2.4 Time-dependent analyses

Performing a time-dependent Dalitz-plot analysis of neutral $B$ decays allows the extraction of the $CP$-violating parameters along with the parameters of the isobar model. A full time- and tag-dependent Dalitz-plot analysis has the following advantages compared to a quasi-two-body analysis:

- determines weak and strong phases simultaneously, alleviating ambiguities from different amplitude contributions;
- provides sensitivity to $\cos 2\phi$ (where $\phi$ is the appropriate weak phase), alleviating the degeneracy of the trigonometric ambiguities;
- correctly accounts for contamination between different resonant contributions.

See Chapters 8 and 10 for details of the techniques of flavor tagging and time-dependent analyses. Here we will give a brief description of how the time dependence and the Dalitz-plot dependence are combined.

With $\Delta t \equiv t_{\rm rec} - t_{\rm tag}$ defined as the proper time interval between the decay of the fully reconstructed $B_{\rm rec}$ and that of the other meson $B_{\rm tag}$ from the $\Upsilon(4S)$ decay, the time-dependent decay rate $|\mathcal{A}^+(\Delta t)|^2$ ($|\mathcal{A}^-(\Delta t)|^2$) when the $B_{\rm tag}$ is a $B^0$ ($\overline{B}^0$) is given by

$$|\mathcal{A}^{\pm}(\Delta t)|^2 = \frac{e^{-|\Delta t|/\tau_{B^0}}}{4\tau_{B^0}}\Bigg[|\mathcal{A}|^2 + |\overline{\mathcal{A}}|^2 \mp \left(|\mathcal{A}|^2 - |\overline{\mathcal{A}}|^2\right)\cos(\Delta m_d \Delta t) \pm 2{\rm Im}\left[\frac{q}{p}\overline{\mathcal{A}}\mathcal{A}^*\right]\sin(\Delta m_d \Delta t)\Bigg], \quad (13.2.17)$$

where $\tau_{B^0}$ is the mean neutral $B$ lifetime and $\Delta m_d$ is the mass difference between $B_H$ and $B_L$. The time distribution is convolved with the $\Delta t$ resolution function in the typical way. Here, we have assumed that *CP* is conserved in $B^0\overline{B}^0$ mixing ($|q/p| = 1$) and that the lifetime difference between $B_H$ and $B_L$ is negligible ($\Delta\Gamma_d = 0$). The decay rate, Eq. (13.2.17), is used as a *p.d.f.* in a maximum-likelihood fit and must therefore be normalized:

$$|\mathcal{A}^{\pm}(\Delta t)|^2 \longrightarrow \frac{1}{\langle|\mathcal{A}|^2 + |\overline{\mathcal{A}}|^2\rangle}|\mathcal{A}^{\pm}(\Delta t)|^2 , \quad (13.2.18)$$

where $\langle ... \rangle$ denotes the value of the integral over the Dalitz plot.

## 13.3 Experimental effects

The amplitude formalisms outlined above provide a model of the underlying physics of the three-body decay. However, these descriptions may have to be modified or augmented to account for the imperfections of experimental measurements. Broadly, these modifications fall into two categories, one accounting for candidates from background processes (see Section 13.3.1) and the other for effects of reconstruction of signal candidates. The latter category incorporates two main effects: efficiency (Section 13.3.2) and misreconstruction (Section 13.3.3).

### 13.3.1 Backgrounds

At the $B$ Factories, the dominant source of background in most three-body analyses is from combinatorics, *i.e.* where three random particles in an event happen to fake the signal decay under consideration. This is largely due to the cross-section for light quark production being two to three times higher than that for charm or bottom. Additionally, in searches for rare decays (such as charmless $B$ decays) the branching fraction of the decay of interest is small, $\mathcal{O}\left(10^{-7} - 10^{-5}\right)$. Therefore the relative rate of particles from other $B$ decays combining to fake the signal is correspondingly greater. While these types of backgrounds can be greatly suppressed using the multivariate techniques described in Chapter 4, the Dalitz-plot distribution of the events that remain must still be modeled. In general, such random combinations of particles tend to populate the edges and corners of the Dalitz plot, since they are most frequently formed from collinear and anti-collinear particles in the predominantly jet-like continuum events.

In addition to the combinatoric backgrounds, there exist fully or partially reconstructed backgrounds that originate from decays of the same class of parent meson to a final state similar to the one under consideration. For example, in an analysis of the decay $B^0 \to K^0_S\pi^+\pi^-$ there are potentially large backgrounds from many other $B$ decays including $B^0 \to \eta'(\to \rho^0\gamma)K^0_S$, $B^0 \to D^-(\to K^0_S\pi^-)K^+$, and $B^+ \to K^0_S\pi^+$. In the first of these examples the decay has been partially reconstructed but the energy of the missing photon is sufficiently small that the reconstructed $B^0$ candidate passes selection criteria. In the second case the decay is fully reconstructed but a kaon/pion misidentification occurs. In the third case the decay is again fully reconstructed and combined with an additional soft pion from the rest of the event to form a signal candidate. Each of these scenarios can lead to very different distributions of events in the Dalitz plot.

In general, the distributions of backgrounds across the Dalitz plot are rather difficult to model with parametric functions. Additionally, the precise nature of the backgrounds can vary dramatically from one analysis to another. Thus, the most common approach for modeling the Dalitz-plot distributions of the backgrounds is to use histograms obtained from either Monte Carlo simulation or sidebands in data. Often some form of smoothing or interpolation is applied to the histograms in order to limit the effect of statistical fluctuations in the input data sample. In $B$ decay analyses, it is found that most backgrounds (particularly the dominant combinatoric backgrounds) preferentially populate the corners and edges of the Dalitz plot. In order to increase the resolution of the histograms in these regions, adaptive binning techniques can be used and/or the histograms can be formed in the so-called "square Dalitz plot", which is discussed in detail in Section 13.4.1.

### 13.3.2 Efficiency

The most obvious effect of detector acceptance is a reduction in the number of events detected. In three-body decays this is complicated by the fact that the kinematic properties of the decay products differ accross the Dalitz plot. Thus, the acceptance as a function of the Dalitz plot variables is, in general, nonuniform.

The typical acceptance function drops at the corners of the phase space, which correspond to the kinematic configuration where one of the final state particles is produced at rest in the frame of the decaying particle. Reconstruction efficiency is typically smaller for such particles, especially if the decaying particle has a small boost in the laboratory frame.

At *BABAR* and Belle, the efficiency profile is usually well modeled by the full detector simulation. The profile is then modeled either by a parameterized form, such as a two-dimensional polynomial, or by a histogram. Either way, this allows the efficiency as a function of the position in phase space, $\varepsilon(\mathbf{m})$, to be included in the signal Dalitz-plot model, where it multiplies the squared absolute value of the amplitude. When histograms are used they often utilize adaptive binning techniques and/or are formed in the "square Dalitz plot" (see Section 13.4.1) to improve the resolution in the areas of most rapidly changing efficiency or of greatest importance for the signal model. Interpolation or smoothing techniques can be employed to reduce the effect of statistical fluctuations.

Another, nonparametric, technique to include the efficiency profile in the Dalitz plot fit was used in some Belle analyses (Abe, 2004f; Kuzmin, 2007). The method uses the fact that in the unbinned maximum likelihood fit the efficiency profile enters only the normalization term. The normalization of the *p.d.f.* over the Dalitz plot is calculated using the Monte-Carlo integration technique, but instead of a uniformly distributed sample in the phase-space variables, a large number of simulated events is used that pass the same selection as applied to data.

### 13.3.3 Misreconstructed signal

Another extremely important effect of reconstruction for a Dalitz-plot analysis is the potential migration of an event from its true coordinate on the Dalitz plot to its reconstructed position. In reality, these effects of reconstruction lie on a continuum, but in order to produce a reasonable model they are most often classified into two types. The first type consists of so-called "correctly reconstructed" events, where the migration is negligible relative to the widths of the resonances under consideration. In this case the amplitude models are used without alteration. In very few cases the migration is not negligible but can be modeled using a simple Gaussian resolution. This class of correctly reconstructed signal events will not be discussed further here. The second type contains events which have more pronounced migration and are sometimes called "self cross feed" in *BABAR* and Belle publications; they form the main topic of this section, and will be referred to as "misreconstructed signal".

For many three-body decay modes, there is a significant fraction of signal events that are incorrectly reconstructed yet still satisfy the selection criteria. Such events typically occur when one low-energy particle from the signal decay is replaced by another in the same event. This behavior is especially prevalent in decays containing neutral pions, where another photon in the event is incorrectly assigned as one of the low-energy photons used to reconstruct the $\pi^0$.

In order to correctly model this behavior, it is necessary to determine both the frequency of the misreconstruction (including the variation of that frequency over the Dalitz plot) and the precise migration effects that occur. This can only be achieved with full detector simulation, where both the generated and reconstructed Dalitz-plot positions are known.

Consider an event that is generated with Dalitz-plot coordinate $\mathbf{m}^t$. The probability that this event passes the selection criteria is given by the efficiency as a function of the true position, $\varepsilon(\mathbf{m}^t)$. If the event is selected then there is a further chance that it is misreconstructed. Since such misreconstructions are dependent on the kinematic configuration, this probability is also a function of the true position, $f_{\rm MR}(\mathbf{m}^t)$. The resulting migration probability from true coordinate $\mathbf{m}^t$ to the reconstructed one, $\mathbf{m}^r$, can be described by the four-dimensional function $R_{\rm MR}(\mathbf{m}^r, \mathbf{m}^t)$, which obeys the unitary condition

$$\iint R_{\rm MR}\left(\mathbf{m}^r, \mathbf{m}^t\right) \, d\mathbf{m}^r = 1 \ \forall\, \mathbf{m}^t \,. \tag{13.3.1}$$

Consequently, for an event reconstructed at $\mathbf{m}^r$ the probability for it to be a well-reconstructed signal event is

$$P_{\rm sig}^{\rm WR} \propto \left[1 - f_{\rm MR}(\mathbf{m}^r)\right] \varepsilon(\mathbf{m}^r) \, \left|\mathcal{A}(\mathbf{m}^r)\right|^2 \,, \tag{13.3.2}$$

while the corresponding probability for a misreconstructed signal event is

$$P_{\rm sig}^{\rm MR} \propto \iint f_{\rm MR}(\mathbf{m}^t)\, \varepsilon(\mathbf{m}^t) \left|\mathcal{A}(\mathbf{m}^t)\right|^2 \times R_{\rm MR}(\mathbf{m}^r, \mathbf{m}^t) \, d\mathbf{m}^t \,. \tag{13.3.3}$$

Typically, this integration is implemented as a summation over binned distributions. Therefore, it is essential to include factors that account for the amount of phase space contained within each bin in both the generated and reconstructed histograms.

## 13.4 Technical details

This part of the chapter describes various technical issues not related to the physics processes involved, but aimed to improve or simplify analyses or presentation of their results. These include the square Dalitz plot transformation (Section 13.4.1), various parameterizations of the complex coefficients for amplitude components (Section 13.4.2), fitting techniques (Section 13.4.3), and the concept of fit fractions, which are used in the presentation of fit results (Section 13.4.4).

### 13.4.1 Square Dalitz plot

A common feature of Dalitz-plot analyses of $B$-meson decays to charmless final states is that both the signal events and the combinatorial $e^+e^- \to q\bar{q}$ ($q = u, d, s, c$) continuum background events populate the kinematic boundaries of the Dalitz plot. This is due to the low masses of the final state particles compared with the $B$ mass. Large variations occurring over small areas of the Dalitz plot are difficult to describe in detail. As a result, the typical representation of the Dalitz plot may be inconvenient when using empirical reference shapes in a maximum-likelihood fit. The boundaries of the Dalitz plot are particularly important since it is here that the interference between light meson resonances occurs. These are the regions with the greatest sensitivity to relative phases. A solution that was adopted by some analyses is to apply a transformation to the kinematic variables that maps the Dalitz plot into a rectangle: the so-called *square Dalitz plot* (SDP). Such a transformation avoids the curved kinematic boundary, which simplifies the use of nonparametric *p.d.f.*s (histograms) to model the distribution of events over the Dalitz plot. Moreover, the transformation is required to expand the regions of interference and simplify parameterization; for instance, the Dalitz plot can be tiled by equally sized bins.

A common definition of the SDP first appeared in the analysis of $B^+ \to \pi^+\pi^+\pi^-$ by *BABAR* (Aubert, 2005d), where the SDP is obtained by the transformation:

$$dm_{ab}^2\, dm_{bc}^2 \;\longrightarrow |\det J|\, dm'\, d\theta'. \tag{13.4.1}$$

The new coordinates are

$$m' \equiv \frac{1}{\pi} \arccos\left(2\frac{m_{ac} - m_{ac}^{\min}}{m_{ac}^{\max} - m_{ac}^{\min}} - 1\right), \tag{13.4.2}$$

$$\theta' \equiv \frac{1}{\pi}\theta_{ac}\,, \tag{13.4.3}$$

where $m_{ac}^{\max} = M - m_b$ and $m_{ac}^{\min} = m_a + m_c$ are the kinematic limits of $m_{ac}$, $\theta_{ac}$ is the helicity angle of the $ac$ combination, and $J$ is the Jacobian of the transformation. Both new variables range between 0 and 1. The determinant of the Jacobian is given by

$$|\det J| \;=\; 4\,|\mathbf{p}_a^*||\mathbf{p}_b^*|\, m_{ac} \cdot \frac{\partial m_{ac}}{\partial m'} \cdot \frac{\partial \cos\theta_{ac}}{\partial \theta'}\,, \tag{13.4.4}$$

where $|\mathbf{p}_a^*| = \sqrt{E_a^* - m_a^2}$, $|\mathbf{p}_b^*| = \sqrt{E_b^* - m_b^2}$, and the energies are defined in the $ac$ rest frame. Figure 13.4.1 shows the determinant of the Jacobian as a function of the SDP parameters $m'$ and $\theta'$. If the events in the nominal Dalitz plot were distributed according to a uniform three-body phase space, their distribution in the SDP would match the plot of $|\det J|$.

The effect of the transformation, Eq. (13.4.1), is illustrated in Fig. 13.4.2, which shows the nominal and square Dalitz plots for Monte Carlo simulated $B^0 \to \pi^+\pi^-\pi^0$ signal events, where the Dalitz-plot model contains only $\rho^+\pi^-$, $\rho^-\pi^+$, and $\rho^0\pi^0$ amplitudes. The benefits of the SDP explained above are clearly visible in this figure. This simulation does not take into account any detector effects and corresponds to a particular choice of the decay amplitudes for which destructive interferences occur in regions where the $\rho$ resonances overlap. To simplify the comparison, hatched areas showing the interference regions between $\rho$ bands and dashed isocontours $m_{ij} = 1.5\,\mathrm{GeV}/c^2$ have been superimposed on both Dalitz plots.

Another transformation of the phase space was used in the recent *BABAR* amplitude analysis of $B^0 \to K_S^0 K_S^0 K_S^0$ decays (Lees, 2012c). In this particular case, due to the presence of identical particles in the final state, symmetrization of the amplitude under exchange of the identical particles is required. The square Dalitz plot transformation described above would result in curved boundaries. On the other hand, mapping the invariant masses to the plane defined by two helicity angles results in a rectangle.

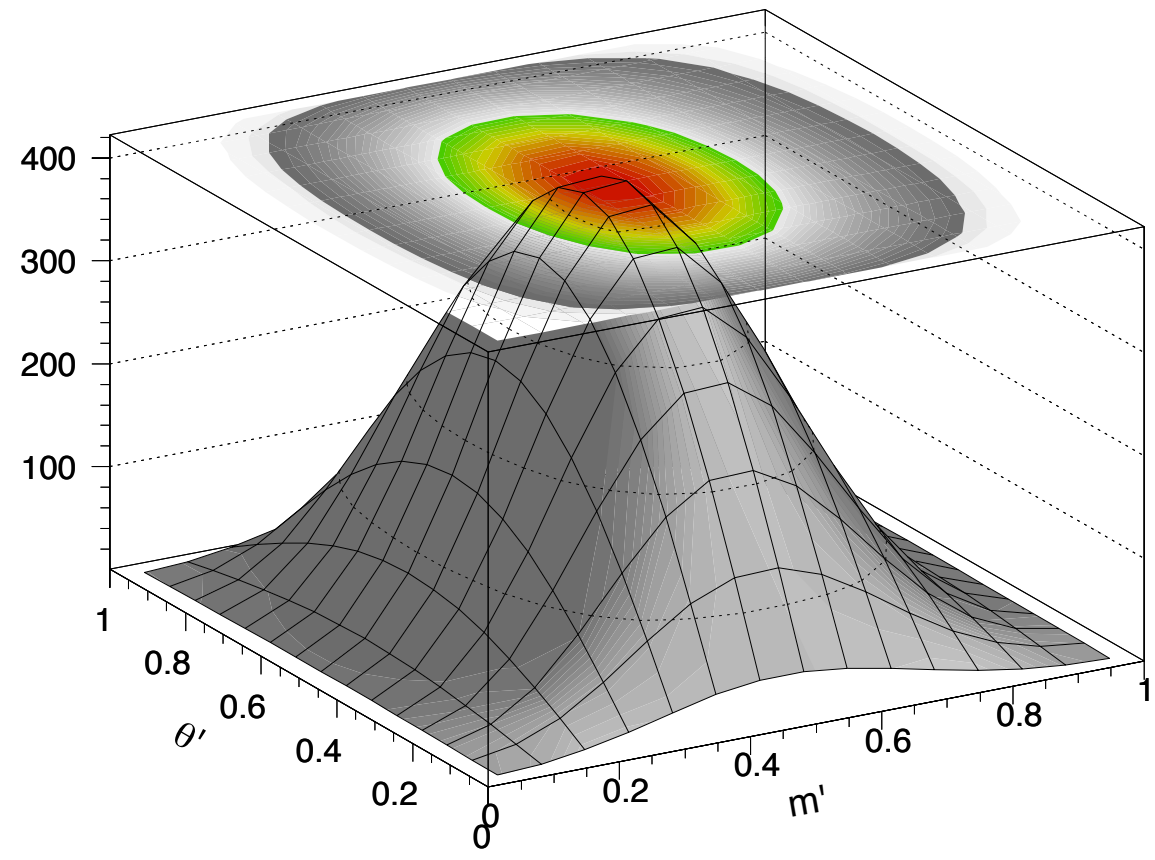

**Figure 13.4.1.** Jacobian determinant, Eq. (13.4.4), of the transformation, Eq. (13.4.1), defining the square Dalitz plot (SDP). Such a distribution would be obtained in the SDP if events were uniformly distributed over the nominal Dalitz plot.

### 13.4.2 Complex coefficients

The complex coefficients of each contribution to the amplitude are expressed in Eq. (13.2.1) in terms of a magnitude and a phase,

$$c_r = a_r e^{i\phi_r}\,, \tag{13.4.5}$$

which is arguably the most intuitive formulation. However, it is also possible to use the real and imaginary parts as the fit parameters

$$c_r = x_r + iy_r\,. \tag{13.4.6}$$

This latter form has the advantage that the parameters are well behaved when the magnitude of the contribution is small, while the former expression can exhibit biases under these circumstances. One caveat is that, conversely, when the magnitude is large the latter form can appear to exhibit bias. Since the magnitude of a contribution is, in general, better constrained than the phase, the fitted values from a group of pseudo experiments tend to lie on an arc in the complex plane. When projecting this arc onto the real and imaginary axes the distributions can appear skewed. This behavior is not generally indicative of a true bias in the fit; indeed the distributions of the magnitudes and phases (calculated from the fitted $x_r$ and $y_r$ parameters) can be perfectly centered on the true values. However, care should be taken when interpreting the errors on the fit parameters due to their large correlation.

The choice of formulations is much broader when parametrizing *CP* violation. Perhaps the simplest approach is to assign the $B$ (or $D$) one set of parameters and the $\overline{B}$ (or $\overline{D}$) another set

$$c_r = a_r e^{i\phi_r} \tag{13.4.7}$$

$$\bar{c}_r = \bar{a}_r e^{i\bar{\phi}_r}\,,$$

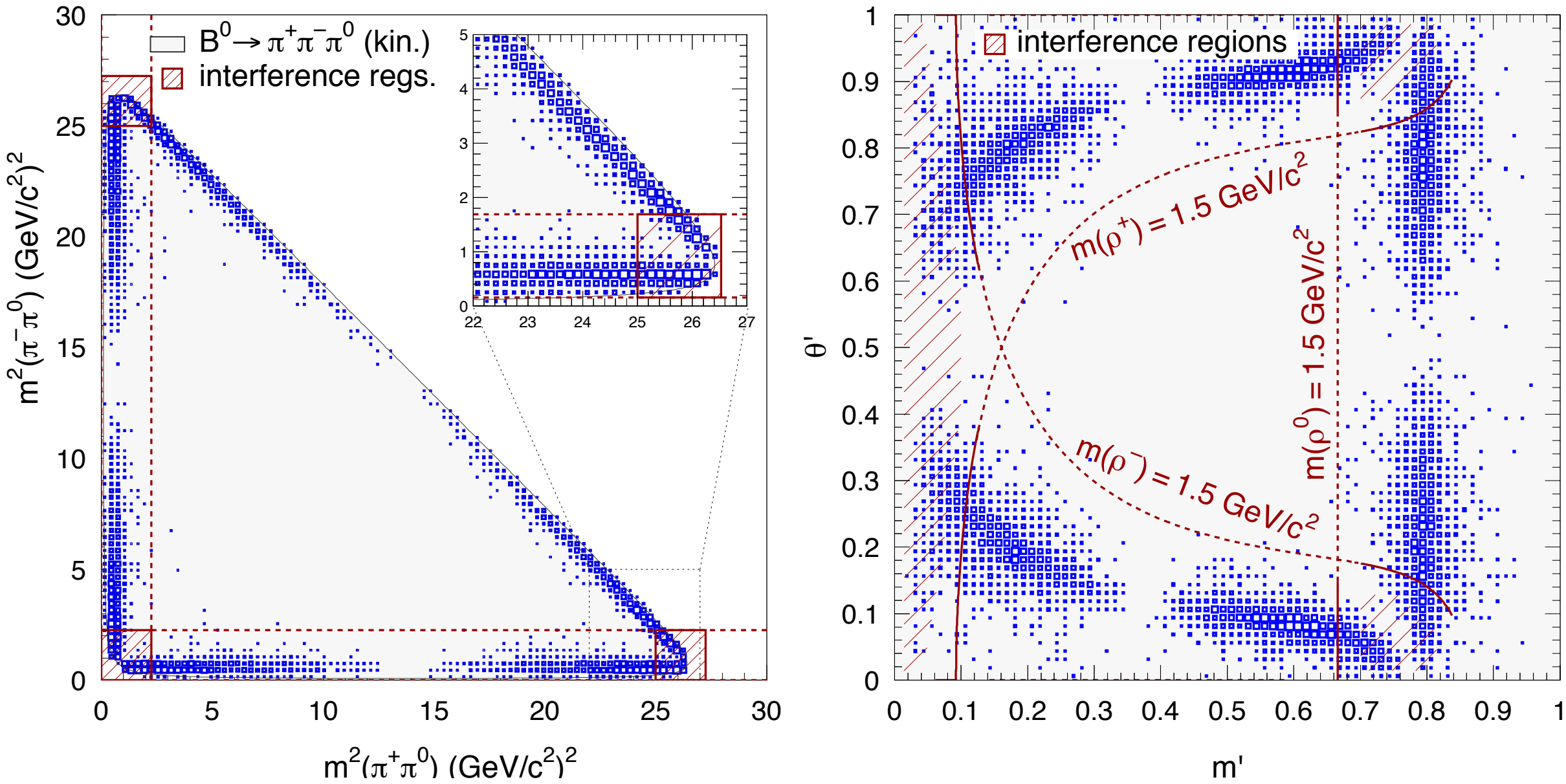

**Figure 13.4.2.** Nominal (left) and square (right) Dalitz plots for Monte Carlo generated $B^0 \to \pi^+\pi^-\pi^0$ decays (Aubert, 2007v). The comparison of the two Dalitz plots shows that the transformation, Eq. (13.4.1), indeed homogenizes the distribution of events, which are no longer near the plot boundaries but rather cover a larger fraction of the physical region. The decays have been simulated without any detector effects and the three $\rho\pi$ amplitudes have been chosen in order to have destructive interference where the $\rho$ bands overlap. The main overlap regions between the $\rho$ bands are indicated by the hatched areas. Dashed lines in both plots correspond to $m_{ij} = 1.5\,\mathrm{GeV}/c^2$.

or

$$\begin{aligned} c_r &= x_r + iy_r \\ \bar{c}_r &= \bar{x}_r + i\bar{y}_r\,. \end{aligned} \tag{13.4.8}$$

Alternatively, one can use sets of *CP*-conserving and *CP*-violating parameters, such as those used in the *BABAR* analysis of $B^+ \to K^+\pi^+\pi^-$ (Aubert, 2008j)

$$\begin{aligned} c_r &= (x_r + \Delta x_r) + i(y_r + \Delta y_r) \\ \bar{c}_r &= (x_r - \Delta x_r) + i(y_r - \Delta y_r)\,, \end{aligned} \tag{13.4.9}$$

or those used in the Belle analysis of the same decay (Garmash, 2006)

$$\begin{aligned} c_r &= a_r e^{i\delta_r}\left(1 + b_j e^{i\phi_j}\right) \\ \bar{c}_r &= a_r e^{i\delta_r}\left(1 - b_j e^{i\phi_j}\right)\,, \end{aligned} \tag{13.4.10}$$

or those used in the CLEO analysis of $D^0 \to K^0_S\pi^+\pi^-$ (Asner et al., 2004b)

$$\begin{aligned} c_r &= a_r e^{i(\delta_r+\phi_r)}\left(1 + \frac{b_j}{a_j}\right) \\ \bar{c}_r &= a_r e^{i(\delta_r-\phi_r)}\left(1 - \frac{b_j}{a_j}\right)\,. \end{aligned} \tag{13.4.11}$$

Each of these formulations has advantages and disadvantages. For example, there can be ambiguities in the phases in the CLEO prescription. While the formulation in terms of real and imaginary parts is generally better behaved when the magnitude of the *CP* violation is small, it is less intuitive in terms of interpretation of the results. Hence it is advisable to try several forms and to choose that which best suits the particular measurement being attempted.

### 13.4.3 Fitting

Once the model of the Dalitz-plot distribution has been formed for all event categories (signal and backgrounds) it is necessary to fit the data to determine the values of the parameters of the model. This is generally achieved using the technique of maximum-likelihood fitting, which is discussed in detail in Chapter 11. As such, only the details specific to Dalitz-plot analyses will be discussed here. Both binned and unbinned fits are used, the former being more common in the analysis of charm decays where the signal yields and purities are greater.

One of the key issues is the normalization of the signal Dalitz-plot *p.d.f.*. There is, in general, no analytic solution to the integral of such a function and so numerical techniques must be employed. The two most commonly used approaches are Monte Carlo and Gauss-Legendre estimation. When a Dalitz-plot model contains narrow resonances such as $\phi(1020)$ or $\chi_{c0}$, it can be useful to perform an integration with higher resolution in the region of those structures. This can involve dividing the Dalitz plot into a number of regions, performing the integration with different resolutions in each region, and finally combining the results.

Since the calculation of the normalization integrals can be computationally expensive, it is desirable to calculate them only once and to cache the values for later use. From Eq. (13.2.1), it is clear that while the complex coefficients factorize from the integral, the parameters of the resonance dynamics, *e.g.*, the mass and width, do not. It is thus possible to cache the integrals only if the parameters of the resonances are fixed in the fit. Under these circumstances, the integrals of each of the $\mathcal{A}_r \mathcal{A}^\star_{r'}$ terms can be calculated prior to the fit. The *p.d.f.* normalization can then be calculated by combining these cached terms and the current values of the complex coefficients at each iteration of the fit. Consequently, it is a common procedure to fix the resonance parameters in the fit. Where necessary, likelihood scans are used to determine the values of any less well-known parameters.

Due to the complexity of the likelihood function and the large numbers of parameters involved in Dalitz-plot fits it is quite common for several local minima to appear in the parameter space. This can cause problems for the minimization routine in finding the global minimum. In addition, these local minima can be almost degenerate with the global minimum, leading to the need to quote multiple solutions. This can occur, for example, when ambiguities arise from broad overlapping states. The data can often be well described by two or more configurations of the magnitudes and phases of these states. The problem of finding the global minimum is usually overcome by performing multiple fits to a given data sample, each with different (often randomized) starting values for the various parameters. One can then choose the case where the best likelihood was obtained as the global solution. This method also permits the exploration of the other local minima, which allows the results from other solutions to be quoted if they are not significantly separated in likelihood from the global minimum.

### 13.4.4 Fit fractions

The choice of normalization, phase convention, and amplitude formalism may not always be the same for different experiments or indeed among the different fitting packages used within a single experiment. Consequently, it is extremely important to provide as much convention-independent information as possible to allow a more meaningful comparison of results. Fit fractions are quite commonly used, both for this purpose and for providing a means to estimate the branching fractions of the various decay modes involved. The fit fraction for a component $j$ is defined as the integral of the square of the decay amplitude for that component divided by the integral of the square of the entire matrix element over the Dalitz plot:

$$FF_j = \frac{\iint_{\rm DP} \left|c_j \mathcal{A}_j(\mathbf{m})\right|^2 d\mathbf{m}}{\iint_{\rm DP} \left|\sum_k c_k \mathcal{A}_k(\mathbf{m})\right|^2 d\mathbf{m}} . \qquad (13.4.12)$$

Similarly, the fit fraction for the conjugate process is defined to be:

$$\overline{FF}_j = \frac{\iint_{\rm DP} \left|\bar{c}_j \overline{\mathcal{A}}_j(\mathbf{m})\right|^2 d\mathbf{m}}{\iint_{\rm DP} \left|\sum_k \bar{c}_k \overline{\mathcal{A}}_k(\mathbf{m})\right|^2 d\mathbf{m}} . \qquad (13.4.13)$$

Furthermore, the fit fraction asymmetry is defined to be

$$A_j^{FF} = \frac{\overline{FF}_j - FF_j}{\overline{FF}_j + FF_j} , \qquad (13.4.14)$$

and the *CP*-conserving (*CP*-violating) fit fraction is given by the sum (difference) of the numerators of Eq. (13.4.12) and Eq. (13.4.13) divided by the sum of the denominators of the same equations. These definitions follow those in Asner et al. (2004b). Note that the sum of the fit fractions is not necessarily unity due to the presence of net constructive or destructive interference.

While the fit fractions can be very useful in comparing results for a given channel, there is additional information in the interference between the contributing decay modes. In order to allow such comparisons one can define interference fit fractions by (del Amo Sanchez, 2010a)

$$FF_{ij} = \frac{\iint_{\rm DP} 2{\rm Re}\left[c_i c_j^* \mathcal{A}_i(\mathbf{m}) \mathcal{A}_j^*(\mathbf{m})\right] d\mathbf{m}}{\iint_{\rm DP} \left|\sum_k c_k \mathcal{A}_k(\mathbf{m})\right|^2 d\mathbf{m}} , \qquad (13.4.15)$$

for $i < j$ only. Note that, with this definition, $FF_{jj} = 2FF_j$.

## 13.5 Model uncertainties

While most of the experimental uncertainties in the measurements involving Dalitz-plot analyses can, in principle, be controlled with Monte Carlo simulation and control samples, there is an essential contribution to the systematic error which is usually hard to quantify. This is the uncertainty on the amplitude arising from model assumptions in its description. This section will describe the possible sources of model uncertainties and outline some methods by which they can be estimated (Section 13.5.1), before discussing the various approaches towards model-independent analysis that have been adopted by the $B$ Factories (Sections 13.5.2 and 13.5.3 ).

### 13.5.1 Estimation of model uncertainties

The sources of model uncertainty and common methods to estimate them are listed below.

- Isobar description:
  The isobar formalism is valid only in the case of narrow and non-overlapping resonances, otherwise the unitarity of the amplitude is violated. In contrast, most of the Dalitz-plot analyses have to deal with wide states that interfere with each other. If the use of the isobar model is not implied by the nature of the measurement, more accurate results can be obtained (or at least, the uncertainty due to the use of the isobar description can be quantified) by using an alternative approach, such as the K-matrix.
- Lineshapes of two-body amplitudes:
  Reasonable theoretical description of broad resonances requires corrections to be applied to the Breit-Wigner lineshape, discussed in Section 13.2. Those corrections (*i.e.* Blatt-Weisskopf form factors and mass-dependent widths) depend on a number of poorly constrained parameters, such as radial parameters of the decaying particle and intermediate resonances. The uncertainty due to these parameters can be estimated by variation within their errors, if known, or otherwise within some reasonable range.

– Identification of intermediate states:
While the presence of narrow states is usually apparent, some broad states can be misinterpreted as reflections of other two-body channels or as nonresonant structures. In addition, a good description of the amplitude requires that broad states beyond the kinematically allowed region of phase space are properly accounted for. Thus, the model uncertainty estimation often involves variation of the list of intermediate resonances.
– Parameters of the intermediate states:
Uncertainty due to the finite precision on, for example, the masses and widths of resonances, can be evaluated in a straightforward way by varying the parameters within their errors.
– Uncertainty of the nonresonant amplitude:
A range of different parameterizations of the nonresonant amplitude is available. Analyses involving $D$ decays, where the phase space of the decay is reasonably small, often parameterize the nonresonant amplitude with the constant complex term, while in $B$ decays more complicated parameterizations, discussed in Section 13.2.3, are necessary. Comparison of the fit results when using alternative parameterizations can give an estimate of the associated uncertainty.

### 13.5.2 Model-independent analysis

Some applications of Dalitz-plot analyses require a model description of the amplitude, such as searches for intermediate states and measurements of their parameters. Other applications need only that the three-body amplitude (or part of it) is described as a certain function of the phase space variables. In the latter case, the model-independent (MI) Dalitz-plot analysis is a possible option. Below we give two examples of MI approaches: binned analysis and MI partial-wave analysis.

One example of the type of analysis that does not require a model description of the amplitude is the search for *CP* violation in the three-body decays of $B$ or $D$ mesons. While the *CP* asymmetry integrated over the phase space can be small, the local asymmetries in some areas of the phase space can be significant. The understanding of these local asymmetries requires a Dalitz-plot analysis. On the other hand, establishing the existence of *CP* violation does not require a full amplitude analysis. One can therefore divide the phase space into a large number of bins and search for asymmetries in the number of events reconstructed in each bin (Bediaga et al., 2009). The drawback of such an approach is that if *CP* violation is observed, its interpretation will require a full amplitude analysis.

There is, however, a quantitative measurement that uses a model-independent binned Dalitz-plot analysis approach — it is the measurement of the angle $\phi_3$ in $B \to DK$, $D \to K^0_S\pi\pi$ decays. In this measurement, the Dalitz-plot analysis is a tool to obtain the parameters of the admixture of $D^0$ and $\overline{D}^0$ states: their relative amplitude and phase difference. This is possible in the binned approach. The average amplitude and $D^0 - \overline{D}^0$ strong phase difference over the bin is described by a few coefficients. The analysis of the binned $D \to K^0_S\pi\pi$ Dalitz plot from $B \to DK$ allows the extraction of $\phi_3$ once the amplitude coefficients are known. These coefficients can be extracted from other measurements: flavor-tagged $D^0 \to K^0_S\pi\pi$ decays, and quantum-correlated decays of pairs of $D$ mesons from $e^+e^- \to \psi(3770) \to D^0\overline{D}^0$ processes. This analysis, performed by the Belle collaboration (Aihara, 2012) using the strong phase parameters measured by CLEO (Libby et al., 2010), is described in detail in Section 17.8.

### 13.5.3 Model independent partial wave analysis

Another kind of model-independent Dalitz-plot analysis is possible in cases when the data sample is large: the (quasi) model-independent partial wave analysis (MI-PWA). The basic idea behind MI-PWA is that most of the model uncertainty in Dalitz-plot analyses usually comes from the scalar component. One can deal with the scalar component in a model-independent way while keeping the model description for the rest of the amplitude. The scalar component can be parameterized as

$$\mathcal{A}_0(s) = f(s)e^{i\phi(s)}\,, \tag{13.5.1}$$

where the functions $f(s)$ and $\phi(s)$ are defined by interpolation of the values $f_j$ and $\phi_j$ in each bin $j$. The values $f_j$ and $\phi_j$ are treated as free parameters in the amplitude fit. The interference with the non-scalar (reference) part of the amplitude allows one to obtain not only the absolute value of the scalar amplitude, but also its phase as a function of $s$. The MI-PWA analysis was proposed in the E791 collaboration (Aitala et al., 2006) and used by *BABAR* for the analysis of the $D_s^+ \to \pi^+\pi^-\pi^+$ Dalitz plot (Aubert, 2009i).

In cases where the size of the data sample is insufficient to use a full MI-PWA, it is still possible to study the contributions of each partial wave using an angular-moments analysis. This can then inform the choice of model to be used. Such an approach can be highly informative when a number of overlapping contributions are present. A recent example of this approach is the *BABAR* analysis of $B \to KKK$ decays (Lees, 2012y).

# Chapter 14
# Blind analysis

***Editors:***
*Aaron Roodman (BABAR)*
*Alan Schwartz (Belle)*

In developing an analysis, it is important not to optimize the analysis procedure on the data that will be used for the measurement (known colloquially as "tuning on the data"). This point is discussed above in Section 4. In this chapter we discuss the method of a blind analysis, which aims to exclude the possibility of even *unintentional* optimization based on the data. Blind analyses have become widespread in particle physics in recent years, and the blind analysis method has been used extensively at the $B$ Factories. Some of the jargon of blind analyses ("opening the box" for a measurement) has also entered into widespread use, even for measurements that are not blind analyses in the strict sense; there has been an increased awareness of the general requirement to avoid tuning on the data.

Here we present the blind analysis method, introducing its definition and history (Section 14.1), and giving pedagogical examples for the cases of upper limits (Section 14.2) and precision measurements (Section 14.3). We then provide some examples of the use of the method at Belle (Section 14.4) and *BABAR* (Section 14.5). For an in-depth discussion of the blind analysis method, see the review article by Klein and Roodman (2005).

## 14.1 Definition and brief history

A *blind analysis* is a measurement such as that of a branching fraction or upper limit that is performed without looking at the data result until most or all analysis criteria are finalized. The purpose is to eliminate the possibility of an experimenter biasing the result in a particular direction. For example, if all previous measurements of a parameter had obtained positive values, then one might be tempted to keep adjusting analysis criteria until a positive value is obtained. This, however, yields a result biased positive. An early example of a blind analysis is the measurement of the $e/m$ ratio of the electron performed by Dunnington (1933). In this measurement, the $e/m$ value was proportional to the angle between the electron source and the detector. Dunnington asked his machinist to arbitrarily label this angle around 340°; only when the analysis was completed did Dunnington accurately measure this angle to obtain the final result.

Within high energy physics, the blind analysis technique was motivated by a number of positive results that were later found to be due to faulty analysis methods (for examples see Harrison (2002)). It was originally championed by rare kaon decay experiments running at Brookhaven National Laboratory (BNL) in the mid-1980s. Probably the first experiment to use this technique was BNL E791 (Arisaka et al., 1993), which searched for the forbidden decay $K_L^0 \to \mu^{\pm}e^{\mp}$. The experiment defined a signal region in two kinematic variables, the $\mu^{\pm}e^{\mp}$ invariant mass ($M_{\mu e}$) and the $K_L^0$ candidate's transverse momentum squared ($P_T^2$). The signal region was subsequently "blinded," *i.e.*, events falling within this region were not selected for viewing, while all selection criteria were finalized. Only after these criteria were finalized was this region unblinded and signal events counted. A similar technique was used by BNL E787 (Adler et al., 1996), which searched for the rare decay $K^+ \to \pi^+\nu\overline{\nu}$, and by BNL E888 (Belz et al., 1996a,b), which searched for a long-lived $H$ dibaryon. The method was subsequently adopted by the Fermilab KTeV experiment (Alavi-Harati et al., 1999), which measured $\epsilon'/\epsilon$ in the $K^0$-$\overline{K}^0$ system; Fermilab E791 (Aitala et al., 1999b, 2001a), which measured rare/forbidden $D$ meson decays; and the CERN NOMAD experiment (Astier et al., 1999), which searched for neutrino oscillations.

As mentioned, the principle of a blind analysis is to *not* look at potential signal events before finalizing analysis criteria in order to avoid biasing the result. There are three main types of measurements this applies to: setting an upper limit, in which one wants to avoid selection criteria that bias one against signal events; measuring a branching fraction, in which one wants to avoid selections that bias one against background events (this can "sculpt" a signal peak); and precision measurements such as that of measuring mixing or *CP*-violation parameters, in which one wants to avoid selections or fitting procedures that bias the result in a preferred direction. Some general examples of these cases are discussed below, followed by specific examples from Belle and *BABAR*. Not every measurement requires a blind analysis: usually when one searches for new particles and does not know *a priori* where to look, one inspects relevant distributions in an unblind manner. However, one still must be careful not to adjust selection criteria to increase or decrease the signal yield while looking at the signal events for feedback. A blind analysis is typically more time-consuming than an unblind one and, in the case of setting an upper limit, can produce a poor result (see below).

## 14.2 Setting upper limits: a quantitative example

An upper limit can become biased when one searches for a decay that is not expected to occur and observes one or more signal candidates; one tends to assume they are background and tighten one or more selection cuts to eliminate them. The problem with this procedure is that one may eliminate a *real* signal event, in which case the upper limit obtained for the rate of the rare process is biased low and has statistical undercoverage.

To illustrate this bias quantitatively, consider the following example. An ensemble of 1000 identical experiments search for the rare decay $D \to X$, which we postulate to have a branching fraction of $2.5 \times 10^{-5}$. If the experiments have a single-event-sensitivity (S.E.S.) of $1.0 \times$ $10^{-5}$, then the expected number of observed events is 2.5 (The S.E.S. of an experiment is the branching fraction that would produce, given the experiment's data set and efficiency, an average over a statistical ensemble of one detected event). From Poisson statistics for $\mu = 2.5$, we calculate that the ensemble obtains the following results:

- about 82 experiments observe no events;
- about 205 experiments observe one event;
- about 257 experiments observe two events;
- the remainder, about 456 experiments, observe $\geq 3$ events.

For simplicity we assume that the experiments observe no background (this is typically the case for rare $K$ and $\tau$ decay searches). This assumption does not change our final conclusions. The experiments that observe no events will set a 90% C.L. upper limit of 2.30 times the S.E.S. [see Section 36.3.2.5 and Table 36.3 of Beringer et al. (2012)] or $2.30 \times 10^{-5}$, which is below the true value. The experiments observing one, two, three, etc., events will set upper limits of 3.89, 5.32, 6.68, etc., times the S.E.S., which are above the true value. In this manner 8.2% of experiments obtain "incorrect" upper limits, which is less than 10% of the ensemble and thus consistent with the definition of a 90% C.L. limit.

Now suppose that each experiment that observed events looks at their candidate(s) and that some find a kinematic or particle identification variable (for at least one of the candidates) that is more than $2\sigma$ away from the value expected for a signal event. These experiments then impose a $2\sigma$ cut on that variable to eliminate the event(s) and adjust the S.E.S. upwards to account for the 4.6% loss in sensitivity. However, if up to 20 variables are potentially considered to be cut on, then the chance of an event surviving this procedure is only $(0.9545)^{20} = 0.394$. Therefore, after experiments observing events adjust a single cut value, approximately $82 + (1 - 0.394)(205) + [1 - (0.394)^2](1 - 0.954)(257) = 216$ experiments observe no events and set an upper limit of either $2.30 \times 10^{-5}$ (no events originally observed) or $2.30 \times (\text{S.E.S.})/0.954 = 2.41 \times 10^{-5}$. Both limits are below the true value. The fraction of experiments is 22%, which is larger than 10% and thus inconsistent with the definition of a 90% C.L. limit. The bias of the procedure has resulted in undercoverage. To avoid such bias, the decision whether to cut on a variable or not must be made *before* looking at signal candidate events.

While a blind analysis does yield unbiased upper limits, it has a serious drawback in that it is possible to miss an obvious background, observe a large number of events in the signal region, and end up setting a poor upper limit. This situation does a disservice to the experiment, as the full "discriminating power" of the detector has not been utilized. Thus in practice, experiments carefully study signal candidates after all cuts have been finalized to check whether there are any due to a trivial background or instrumental problem such as the high voltage having been tripped off. If such events are found, it usually is preferable to eliminate them and set a biased but useful upper limit rather than leave them and set an unbiased but not useful limit.

Here we have discussed only bias introduced in the signal acceptance, not bias potentially introduced when estimating backgrounds. The latter depends upon the background sample used and the method of estimation. For example, if one is estimating background by counting events in a sideband and extrapolating, then to avoid bias one must blind that part of the sideband used to estimate background when finalizing cuts, or at least not "tune" cuts to explicitly remove events from that sideband region.

## 14.3 Precision measurements

For precision measurements of parameters in which one typically performs a fit rather than simply counts events, a different technique for avoiding bias must be used. In this case *hiding the answer* is often the appropriate method. For example, the KTeV experiment used this technique for its measurement of $\epsilon'/\epsilon$. The value of $\epsilon'/\epsilon$ was obtained from a fit to the data, and to avoid bias KTeV inserted an unknown offset into the fitting program such that the fit yielded the "hidden" value

$$\left(\frac{\epsilon'}{\epsilon}\right)_{\text{hidden}} \equiv \left\{\begin{array}{c} +1 \\ -1 \end{array}\right\} \times \left(\frac{\epsilon'}{\epsilon}\right)_{\text{true}} + c\,. \quad (14.3.1)$$

In this expression, $c$ is a hidden random constant, and the choice of the factor $\pm 1$ is also hidden. The values of $c$ and $\pm 1$ were made by a pseudo-random number generator. Thus KTeV could finalize its data samples, analysis cuts, Monte-Carlo corrections, and fitting technique while remaining unaffected by the (hidden) true value of $\epsilon'/\epsilon$. The use of the factor $\pm 1$ prevented KTeV from knowing the direction in which the result moved as changes to the analysis were applied.

When performing a blind analysis using the "hidden answer" technique, one must consider whether there exists figures, tables, or other ancillary results that could inadvertently reveal the blinded result. Only if the measurement result is not readily apparent from such information should the figure, table, etc. be presented.

## 14.4 Examples from Belle

The Belle experiment used blind analysis methods extensively: in measuring branching fractions, *CP* asymmetries, in fitting Dalitz plots, and in searching for rare and forbidden decays. Only after selection criteria and the fitting procedure were finalized, and the background estimated, were the results unblinded. To unblind a result required approval from one's internal review committee. If a committee member felt that more studies were needed before unblinding, then the analyzer could not proceed. If an analysis was an update to a previous Belle result, then before unblinding the analyzer was usually required to run his/her analysis code on the previous data set used and compare the result obtained with that obtained previously. If there were a discrepancy, it had to be understood before continuing. After unblinding, the only steps remaining in the analysis were finalizing the systematic errors and, occasionally, refining the background estimate.

This methodology yielded unbiased results but also occasional surprises such as:

– significant direct *CP* violation in $B^0 \to K^+\pi^-$ decays (Chao, 2004);
– large direct *CP* violation in $B^0 \to \pi^+\pi^-$ decays, and values of *CP* parameters $C_{\pi\pi}$ and $S_{\pi\pi}$ outside the physical region (Abe, 2004b, 2005b);
– the value of $\sin 2\phi_1$ and *CP* asymmetries measured in $b \to sq\overline{q}$ transitions such as $B^0 \to \phi K^0_S$ and $B^0 \to f_0(980)K^0_S$ differed substantially from that expected based on measurements of the $b \to c\overline{c}s$ transition $B^0 \to J/\psi K^0_S$ (Chen, 2005b); and
– the branching fraction for $B^+ \to \tau^+\nu$ measured with 414 fb$^{-1}$ was much larger than that expected based on the value of $|V_{ub}|$ determined from $B$ semileptonic decays (Ikado, 2006).

A typical example of a blind analysis is a search for the lepton-number-violating decays $\tau^- \to \mu^- V^0$ and $\tau^- \to e^- V^0$, where $V^0$ is a neutral vector meson $\rho^0$, $\phi$, $\omega$, or $K^{*0}$ (Miyazaki, 2011). These mesons were reconstructed via $\rho^0 \to \pi^+\pi^-$, $\phi \to K^+K^-$, $\omega \to \pi^+\pi^-\pi^0$, and $K^{*0} \to K^+\pi^-$. The analysis selected candidate events based on the variables $M_{\ell V}$ and $\Delta E$, where $M_{\ell V}$ is the invariant mass of the $\ell^- V^0$ pair ($\ell = e, \mu$), and $\Delta E$ is the difference in energy between the $\ell^- V^0$ system and the beam energy in the $e^+e^-$ center-of-mass frame. Events were first selected by dividing the reconstructed tracks and calorimeter hits for each event into two azimuthal hemispheres and requiring that, in one of the hemispheres, there be only a single track. This topology corresponds to a $\tau^- \to \ell^-\nu\overline{\nu}$, $\tau^- \to \pi^-\nu$, or $\tau^- \to \rho^-(\to \pi^-\pi^0)\nu$ decay; the presence of this "tagging" decay indicates $e^+e^- \to \tau^+\tau^-$ production.

From this tagged sample, events were selected that have three tracks in the "signal hemisphere". Two of the tracks were required to reconstruct to a $\rho^0$, $\phi$, $\omega$, or $K^{*0}$ meson and satisfy particle identification criteria. The third track was required to satisfy muon or electron identification criteria. At this point an elliptical signal region in the $M_{\ell V}$-$\Delta E$ plane was blinded while topological and kinematic selection criteria were optimized using MC-simulated events and applied. The blinded signal ellipse was centered near $M_{\ell V} = m_\tau$ and $\Delta E = 0$ and had semi-major and semi-minor axes equal to $3\sigma$ in resolution.

After the cut optimization procedure, the background in the signal region was estimated by extrapolating from the number of events observed in a larger $M_{\ell V}$-$\Delta E$ region surrounding the blinded ellipse. The backgrounds ranged from 0.06 to 1.5 events. After selection criteria were finalized and the backgrounds estimated, the signal regions were unblinded and the signal yields obtained. The results for four modes are shown in Fig. 14.4.1. From the observed signal yields along with the background estimates, reconstruction efficiencies, and systematic uncertainties, upper limits were calculated using a frequentist approach (Conrad, Botner, Hallgren, and Perez de los Heros, 2003).

It should be noted that in performing a blind analysis one doesn't have to rely solely on the simulated data. Any data sample statistically independent from the data used for the evaluation of the measurement result can be used. This includes (real) data samples with decay modes exhibiting similarities with the studied one, or even samples of the studied decay mode on a distinct (typically smaller) data set. For example, in Belle study of $D^+ \to K^0_S K^+$ and $D^+_s \to K^0_S\pi^+$ decays (Won, 2009) a smaller sample of selected decays obtained in the off-resonance data sample was used to optimize the selection, subsequently applied to the larger on-resonance data sample.

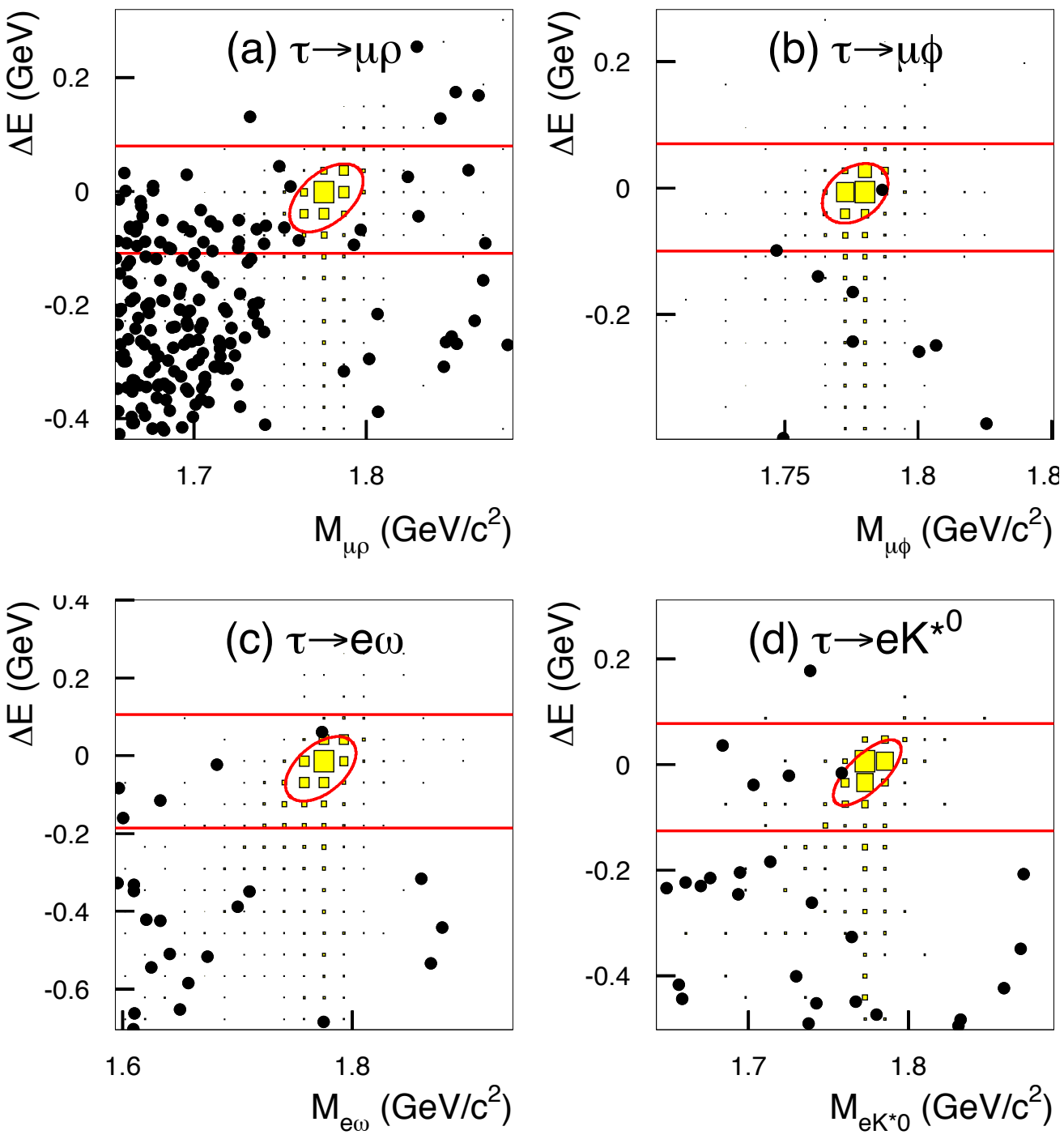

**Figure 14.4.1.** $M_{\ell V}$-$\Delta E$ signal region (see text) for four typical lepton-number-violating decays: *(a)* $\tau^- \to \mu^-\rho^0$ *(b)* $\tau^- \to \mu^-\phi$ *(c)* $\tau^- \to e^-\omega$, and *(d)* $\tau^- \to e^-K^{*0}$ from Miyazaki (2011). Data points are shown as solid circles, and MC signal distributions are shown as yellow boxes (with arbitrary normalization). Red ellipses denote blinded regions, and horizontal lines denote the regions used for estimating background within the blinded ellipses.

## 14.5 Examples from BABAR

The *BABAR* collaboration extensively discussed the use of the blind analysis method prior to data taking, and wrote a document describing possible methods (Ford, 2000) which recommended their use whenever possible. Most *BABAR* results that could make use of a blind analysis technique did in fact do so.

For certain measurements, hiding the answer is not sufficient; it may also be necessary to hide the visual aspect of the measurement. One example is the *CP*-violation measurements performed by *BABAR*. In this case the approximate size and sign of the $CP$ asymmetry can be seen by looking at the $\Delta t$ distributions for $B^0$ and $\overline{B}^0$ decays into $CP$ eigenstates, as shown in Figure 14.5.1a (see also Chapter 10). Before $CP$ violation had been established, and to avoid any chance of bias, a blind analysis was developed to hide both the answer and the visual asymmetry (Roodman, 2000).

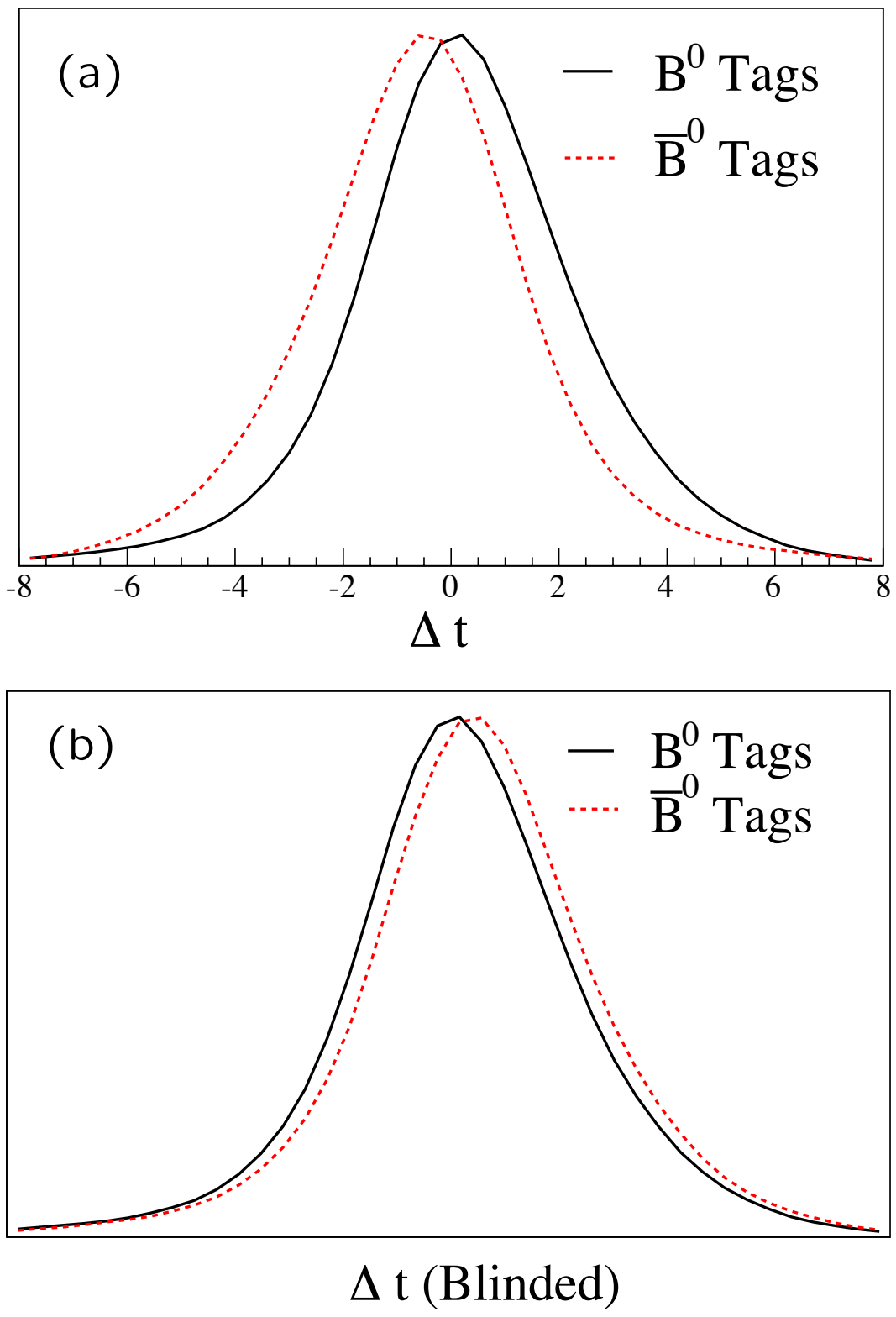

**Figure 14.5.1.** The $\Delta t$ distributions for $B$ decays into $CP$ eigenstates, for $\sin 2\phi_1 = 0.75$ with the $B^0$ flavor tagging and vertex resolution that are typical for *BABAR*. *(a)* The true number of $B^0$-tagged (solid line) and $\overline{B}^0$-tagged (dashed line) decays into $CP$ eigenstates as a function of $\Delta t$. *(b)* The $\Delta t_{\text{Blind}}$ distributions for $B^0$-tagged (solid) and $\overline{B}^0$-tagged (dashed) decays.

In *BABAR*'s initial $CP$-violation measurement (Aubert, 2001a), the result (obtained from fitting the data) was hidden as in Eq. (14.3.1). In addition, the visual asymmetry was hidden by altering the $\Delta t$ distribution used to display the data. This was achieved by using the variable

$$\Delta t_{\text{Blind}} \equiv \begin{Bmatrix} +1 \\ -1 \end{Bmatrix} \times s_{\text{tag}} \times \Delta t \; + \; c\,. \qquad (14.5.1)$$

The parameter $s_{\text{tag}}$ equals $+1$ or $-1$ for $B^0$ or $\overline{B}^0$ flavor tags, respectively. Since the asymmetry is nearly equal and opposite for the two $B$ flavors, *BABAR* hid the asymmetry by flipping one of the distributions. In addition, $CP$-violation can be manifest by the asymmetry about $\Delta t = 0$ of an individual $B^0$ or $\overline{B}^0$ distribution. This feature was hidden by the offset term $c$ in Eq. (14.5.1), which has the affect of hiding the $\Delta t = 0$ point. The result is shown in Fig. 14.5.1b, where the amount of $CP$-violation is no longer visible.

This technique allowed *BABAR* to use the $\Delta t_{\text{Blind}}$ distribution and blinded fit results to validate the analysis and study systematic effects while remaining blind to the presence of any asymmetry. There was one additional restriction: that the fit result could not be superimposed on the data, since the smooth fit curve would show the asymmetry. Instead, to assess the agreement of the fit curve and the data, a distribution of only the residuals was used. In practice, this added only a small complication to the measurement. In fact, after the second iteration of the analysis (Aubert, 2001e), it was realized that the asymmetry would remain blinded if the only $\Delta t$ distribution used was that of the sum of $B^0$ and $\overline{B}^0$ events. Subsequently, no additional checks were done (or needed) using individual $B^0$ and $\overline{B}^0$ $\Delta t$ distributions.

*BABAR* developed other methods for blinding an analysis, depending on its nature (upper limit, branching fraction, or precision measurement). For example, fit results were sometimes blinded directly within the `RooFit` package (Verkerke and Kirkby, 2003), and so `Root`-based fits to data could be subjected to a blind analysis methodology with relative ease. An alternative to a `RooFit`-based blinding method was to set up an analysis chain whereby one performs a fit to data using `Minuit` (James and Roos, 1975) and writes the output to a log file, removing any reference to signal observables while writing the log file. In this manner the output of the fit can be viewed in order to study issues such as the convergence of the fit, the values of ancillary fit parameters, and the covariance matrix.

Lastly, *BABAR* often worked the blind analysis strategy into its internal review process. For many, but not all analyses, the three-person review committee's approval was required before the authors could unblind their analysis (as done in Belle).

# Chapter 15
# Systematic error estimation

***Editors:***
*Wolfgang Gradl (BABAR)*
*Pao-Ti Chang (Belle)*

**Additional section writers:**
*Adrian Bevan, Chih-hsiang Cheng, Andreas Hafner, Kenkichi Miyabayashi*

For most measurements at the $B$ Factories, the estimation of systematic uncertainties is a very important and challenging part of the analysis. There are a number of effects which can systematically influence the result. The ones which are frequently encountered in measurements performed by the $B$ Factories are discussed in the present chapter.

Sources of systematic effects include the difference between data and simulation, the uncertainty on external input needed to convert a directly measured value (*e.g.* the number of signal events) to the desired quantity (*e.g.* a branching fraction), and the analysis procedure chosen to extract the signal (*e.g.* background model, fit bias). In addition, physics processes can introduce discrepancies between the measured value and the parameter of interest. This is often the case because the signal model used is only an approximation of the true, underlying process. An example of this type of systematic uncertainty is the effect of tag-side interference in measurements of time-dependent $CP$ asymmetries.

Where possible, measured values are corrected for such systematic shifts, and there is a systematic uncertainty associated with the correction. Some of the systematic corrections are derived from control sample studies; their associated uncertainty is essentially statistical in nature and scales with the size of the corresponding control sample and therefore with the data sample available for analysis.

Careful design of the analysis strategy can help to minimize the effect of systematic errors on the final result. A particular systematic effect might cancel in the ratio of two observable quantities, such as the total number of produced $B$ mesons in the measurements of rate asymmetries. Similarly, if the branching fraction of a decay is measured relative to a well-known decay mode with similar final state topology, systematic uncertainties due to reconstruction or PID efficiency cancel to a certain extent.

## 15.1 Differences between data and simulation

Most analyses at the $B$ Factories are designed and optimized using simulated data ('Monte Carlo'). Collected data are only looked at after the analysis procedure has been thoroughly tested and validated (see Chapter 14 for a rationale and methods). Quantities such as the event selection efficiency and mis-tag or mis-identification rates are needed for measurements of branching fractions or absolute cross sections, and they are typically obtained from simulated data. If the simulation does not describe the detector perfectly, the efficiency of the selection as applied to real data differs from the efficiency derived from simulated events; this difference needs to be quantified and corrected. The correction factors to be applied to efficiencies obtained from simulation are derived from independent control samples and their simulated counterparts and have their own statistical and systematic uncertainties. The total uncertainty in the correction factor is taken as a systematic uncertainty for the selection efficiency. Correlations between systematic uncertainties need to be taken into account; for example, for a final state with multiple $\pi^0$, the efficiency correction has to be applied for each $\pi^0$ in the final state, and the systematic uncertainties are added linearly.

### 15.1.1 Track reconstruction

Many analyses performed at the $B$ Factories require a precise simulation of the charged track finding and reconstruction efficiency in order to determine absolute rates or cross sections. The way to measure the tracking efficiency is by predicting the presence of a charged particle unambiguously (*e.g.* using kinematic constraints on a series of particle decays) and checking if a reconstructed track matches the prediction. Once the method is validated, one can study the tracking efficiency as a function of the track momentum and polar angle. The same procedure is applied to Monte Carlo events to estimate the tracking efficiency in simulation. From the tracking efficiencies in data and Monte Carlo, one produces a look-up table of correction factors and their uncertainties to correct for the data-MC discrepancy in terms of track momentum and polar angle. This table is used to correct for the signal efficiency estimated from Monte Carlo simulation and to calculate the systematic uncertainty from track reconstruction.

#### 15.1.1.1 Methods at *BABAR*

At *BABAR* several methods are exploited to study possible efficiency differences between the data and simulation over a wide range of particle momenta and production environments relevant to most analyses. They are discussed in detail in (Allmendinger, 2012).

These methods rely on distinct data samples, where additional constraints are applied to select specific event topologies. The primary method to study the charged track reconstruction efficiency in the data and simulation uses $e^+e^- \to \tau^+\tau^-$ events. Events of interest for the efficiency study involve one leptonic $\tau$ decay $\tau^\pm \to \mu^\pm\nu_\mu\nu_\tau$ ('tag side'), $\mathcal{B}(\tau^\pm \to \mu^\pm\nu_\mu\nu_\tau) = (17.36 \pm 0.05)\%$ (Beringer et al., 2012), back-to-back with a semi-leptonic decay $\tau^\mp \to h^\mp h^\mp h^\pm\nu_\tau(\geq 0n)$ ('signal side') with a branching fraction of $\mathcal{B}(\tau^\mp \to h^\mp h^\mp h^\pm\nu_\tau(\geq 0n)) = (14.56 \pm 0.08)\%$, (Beringer et al., 2012). Here, $h$ denotes a charged hadron, and at least two tracks are required to fail a loose electron selection. The presence of one or more neutral particles, denoted by $\geq 1n$, *e.g.* $\pi^0$, but excluding $K^0_S \to \pi^+\pi^-$, is allowed in the final state. This data sample is referred to as 'Tau31' sample. The primary selection for $\tau$ pair candidates requires one isolated muon track in combination with at least two tracks consistent with being hadrons. Through charge conservation, the existence of an additional track is inferred.

Due to the presence of multiple neutrinos in the event, however, the direction of the additional track cannot be determined exactly. Using the measured trajectories of the muon and the two hadrons, kinematic regions can be defined which are correlated with the polar angle $\theta$ and transverse momentum $p_T$ of the missing track. The variation of the agreement between data and simulation as a function of $\theta$ and $p_T$ is conservatively quantified using these estimator quantities.

This variation is the largest uncertainty when applying the results to a physics analysis, where the events typically have distributions in $\theta$ and $p_T$ different from the $\tau$ pair events. The other main uncertainties include both $\tau$ and non-$\tau$ backgrounds: radiative Bhabha events with a converted photon (*i.e.*, $e^+e^- \to e^+e^-\gamma, \gamma \to e^+e^-$), $\tau$ pair events with a converted photon or a $K^0_S \to \pi^-\pi^+$, 2-$\gamma$ events, and continuum events ($q\bar{q}$, with $q = u, d, s, c$). Control samples are used to estimate the levels and/or shapes of the most important backgrounds. This study shows no difference in the track finding efficiency between the data and simulation with an uncertainty of (0.13-0.24)% per track, depending on the exact requirements on the track quality. This method is also used to investigate the stability of track reconstruction over the diverse *BABAR* running periods. No time-dependent effects in the difference between the data and simulation have been observed.

Initial-state radiation (ISR) events in the reaction channel $e^+e^- \to \pi^+\pi^-\pi^+\pi^-\gamma_{\rm ISR}$ are used to cross-check the systematic uncertainties in track reconstruction determined from $\tau^+\tau^-$ events. The absence of neutrinos in this reaction allows to apply a fit with kinematic constraints to events with at least three detected pions. Hereby the kinematic parameters of the possibly missing track are determined using energy and momentum conservation, and the track reconstruction efficiency can be measured as a function of track momentum and angles. In these events, the high-energy ISR photon is emitted back-to-back to the collimated hadronic system in the center-of-mass frame. Because the analysis only selects events with photon energy $E_\gamma > 3\,\rm GeV$, this back-to-back topology is approximately preserved in the laboratory frame. This leads to an environment with a slightly higher track overlap probability. In this environment, the track reconstruction efficiency difference between the data and simulation is found to be $(0.7 \pm 0.4)\%$ per track, compatible with the result of the $\tau$ based study of no significant bias.

Low momentum tracks are studied in $D^{*\pm} \to D^0\pi_s^\pm$ decays, using inclusively selected $D^{*\pm}$. $\pi_s$ denotes the low momentum pion ("slow pion") from the $D^*$ decay. The relative reconstruction efficiency for the slow pions as a function of the pion momentum is measured using their angular distribution, following a method developed by CLEO (Menary, 1992). This method exploits the fact that in the decay of a vector meson to two pseudoscalar mesons the expected distribution of events is an even function of the cosine of the $\pi_s$ helicity angle $\theta^*$. Furthermore, $\cos\theta^*$ is related to the slow pion momentum in the lab frame: $p_{\pi_s} = \gamma(p^*_{\pi_s}\cos\theta^* - \beta E^*_{\pi_s})$. Any observed asymmetry in $dN/d\cos\theta^*$ can be therefore mapped to a relative efficiency difference as a function of $p_{\pi_s}$ (see Allmendinger (2012) for a more complete discussion). Repeating the study on data and simulation, a relative difference between the slow $\pi$ reconstruction efficiencies is extracted, which is then ascribed as a systematic uncertainty. Using the full *BABAR* dataset, this study results in a systematic uncertainty of 1.5% per track with a transverse momentum of $p_T < 180\,{\rm MeV}/c$. This systematic uncertainty includes the effects from both reconstruction efficiency and detector acceptance.

An asymmetry in the track reconstruction efficiency between positively and negatively charged tracks can arise from a charge dependence of the interaction with the detector material; such a detector-induced asymmetry can introduce a bias when measuring small *CP* rate asymmetries. The asymmetry in reconstruction efficiency has to be determined directly from data with a precision of $\mathcal{O}(10^{-3})$. Like the overall tracking efficiency, it can also be measured using the above mentioned Tau31 sample by comparing the number of (2+1)-track events (in which one track was not reconstructed) to the number of (3+1)-track events. The asymmetry in the reconstruction efficiency is found to be $(\varepsilon(\pi^+) - \varepsilon(\pi^-))/(\varepsilon(\pi^+) + \varepsilon(\pi^-)) = (0.10 \pm 0.26)\%$, thus consistent with zero within its uncertainty. This high-statistics measurement is cross-checked and validated with a high purity sample of $D^0 \to \pi^+\pi^-$ events tagged by the decay $D^{*+} \to D^0\pi_s^+$; the charge-dependent reconstruction asymmetry as measured in this decay is also consistent with zero asymmetry, but has a larger uncertainty.

Very sensitive measurements of charge asymmetries, such as $A_{CP}$ in charm meson decays, require a much better control of any detector-induced charge asymmetry. These analyses rely on data-driven methods to determine the charge asymmetry in the track reconstruction with a systematic uncertainty as small as 0.08% (see Section 19.2.6).

Finally, the effect of a vertex of the charged tracks that is displaced from the primary event origin is investigated in $B \to h^+h^-K^0_S$ (with $h = \pi, K$) decays with $K^0_S \to \pi^+\pi^-$. Here the finite lifetime of the $K^0_S$ leads to a displacement of the vertex of the two daughter pions. For these tracks a difference of $(0.5 \pm 0.8)\%$ in the reconstruction efficiency between data and simulation has been observed.

The results of these studies show that at *BABAR* the track finding efficiency in data agrees within uncertainties with the simulated data. Thus, in a *BABAR* analysis, simulated track finding efficiencies can be applied to data. The appropriate systematic errors depending on the number of tracks involved need to be propagated, taking into account that the systematic errors are fully correlated *i.e.*, the systematic uncertainties per track are added linearly.

#### 15.1.1.2 Methods at Belle

The track finding efficiency for charged particles with momenta above 200 MeV/$c$ is studied using the decay chain $D^* \to D^0\pi_s, D^0 \to \pi^+\pi^- K_S^0$ and $K_S^0 \to \pi^+\pi^-$.[46] Partially reconstructed $D^*$ decays provide a clean sample with sufficient statistics to perform the tracking study. The decay chain can be reconstructed without actually detecting one of the pions from the $K_S^0$ decay. The four-momentum of this pion can be inferred from the kinematic constraints of the decay chain.

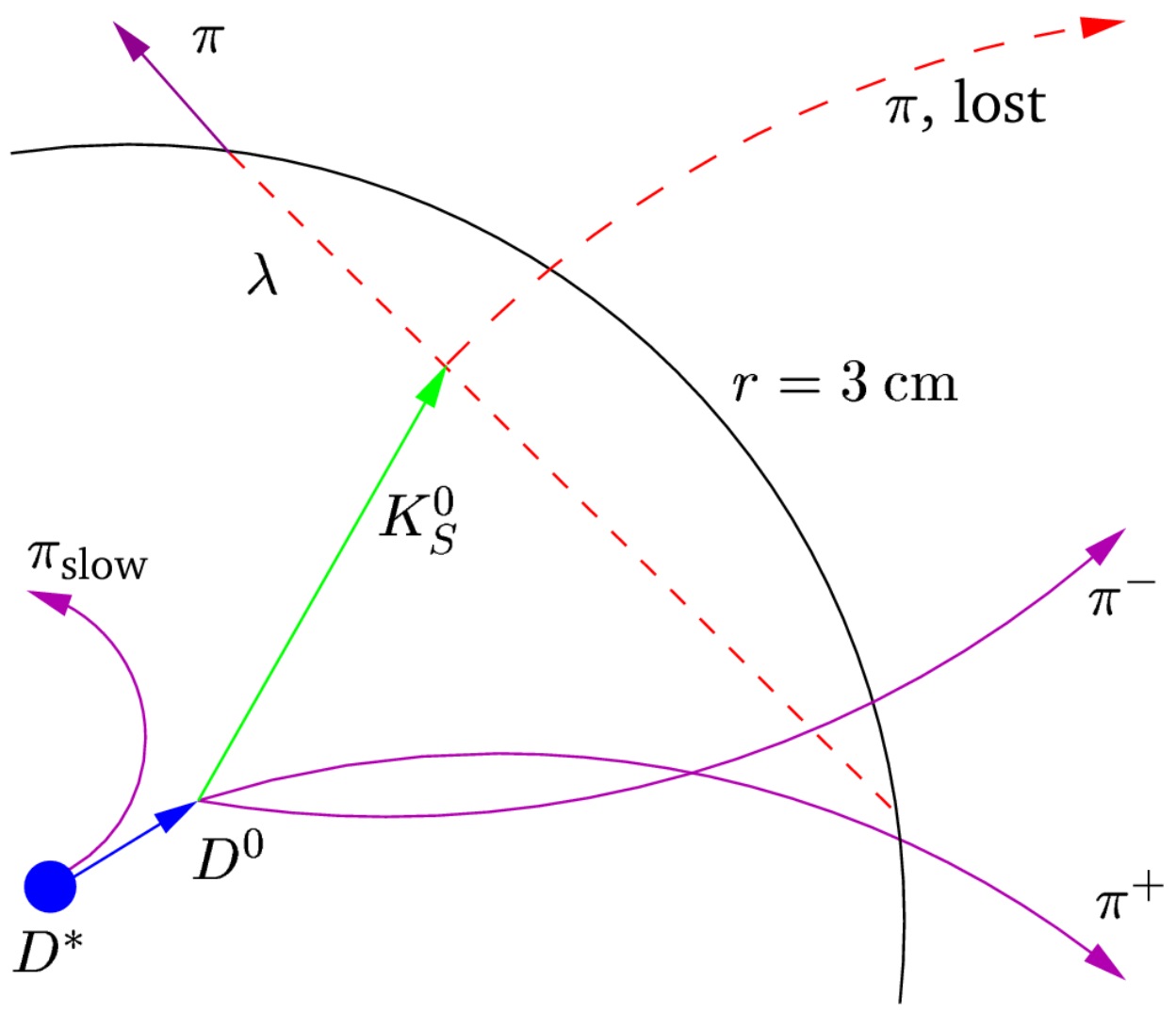

**Figure 15.1.1.** Illustration of the Belle method to determine the efficiency of tracking.

The method is illustrated in Fig. 15.1.1. The $D^*$ meson partial reconstruction starts from the reconstruction of the common vertex of the two charged pions from the $D^0$ decay (the $D^0$ decay vertex; see Chapter 6 about vertexing). Following is the determination of possible $K_S^0$ decay vertex positions, which are constrained to lie on the trajectory of the detected pion and within a certain radius (specifically this is chosen to be 3 cm) from the interaction region to limit the amount of possible points. The segment of the pion track on which the $K_S^0$ vertex is searched for is discretely scanned and for each discrete part of the track the momentum magnitude and direction of the $K_S^0$ is calculated (the latter is determined by the line joining the $D^0$ and $K_S^0$ decay vertices, and the former from the requirement that the $K_S^0$ together with the detected charged pion pair yields the invariant mass of the $D^0$). For each possible value of $K_S^0$ four-momentum (corresponding to each possible $K_S^0$ decay vertex position) a corresponding un-detected pion four-momentum can be calculated by subtracting the momentum of the detected daughter pion. The correct $K_S^0$ momentum (*i.e.* the correct position of the $K_S^0$ decay vertex) is then determined by requiring that the resulting pion four-momentum magnitude corresponds to the pion nominal mass. A slow pion candidate is added to the $D^0$ and the signal of partially reconstructed $D^*$'s is determined from the $D^0\pi_s$ invariant mass distribution (Fig. 15.1.2).

Practically, several selection requirements are implemented to improve the signal-to-background ratio. For instance, the $D^0$ momentum must be larger than 2 GeV/c in the laboratory frame to reduce combinatorial background. The $K_S^0$ vertex should be inside the innermost layer of the silicon vertex detector to ensure silicon hits for the $K_S^0$ daughter tracks, and the missing pions are required to be in the tracking fiducial region. The ratio of the yield of fully reconstructed $D^*$'s to those partially reconstructed with one pion from the $K_S^0$ not required is the track reconstruction efficiency.

Finally the ratio of the tracking efficiencies of data and MC can be obtained as a function of other variables, such as track total momentum and polar angle. The efficiency as a function of particle's transverse momenta for real and simulated data is shown in Fig. 15.1.3. Since the ratio of the data-MC efficiencies is found to be consistent with unity the difference and its uncertainties are assigned as the tracking uncertainty. For Belle, the systematic error for charged-track reconstruction is 0.35% on average for high momentum tracks ($p > 200$ MeV/$c$).

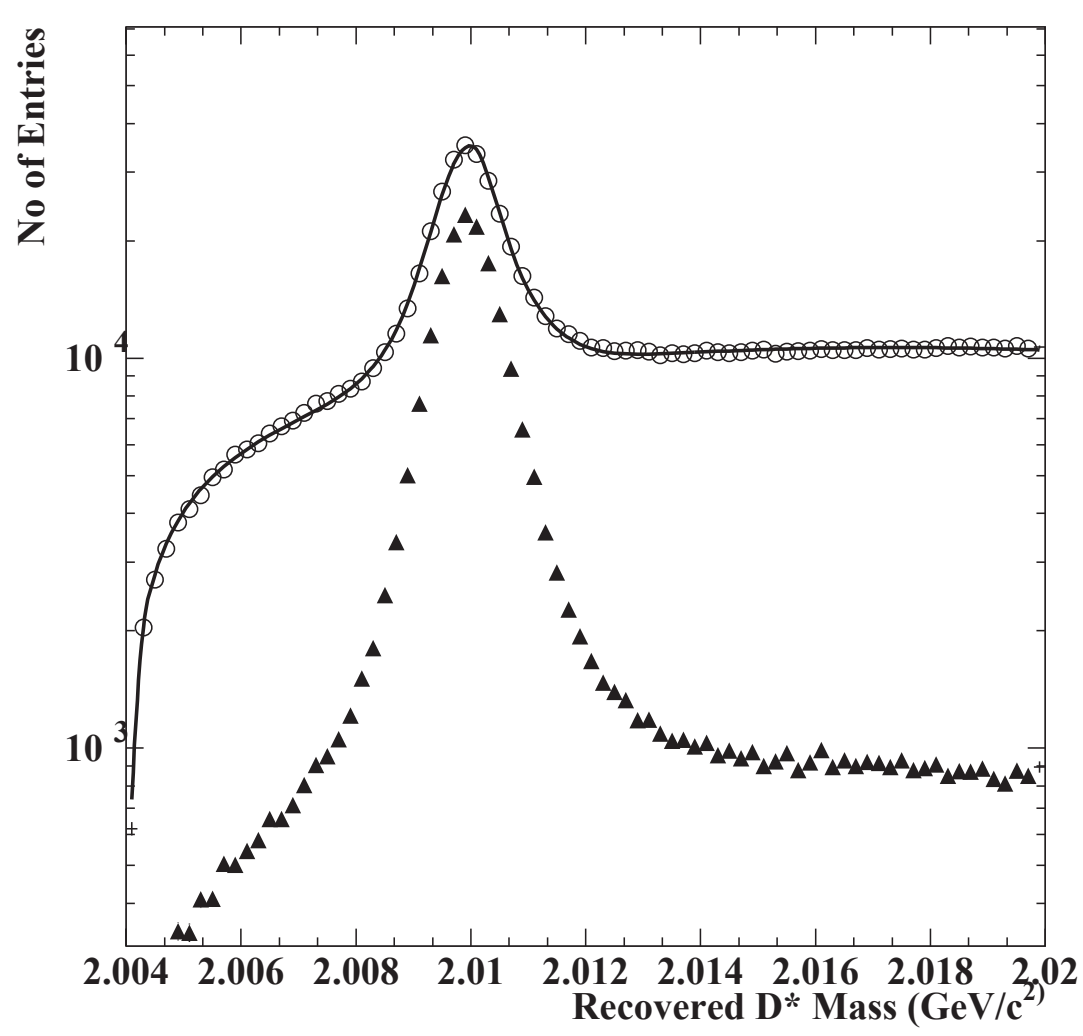

**Figure 15.1.2.** $D^*$ mass distribution for partially (circle) and fully (triangle) reconstructed candidates in Belle data. A similar reconstruction in the simulated data yields the ratio of data-MC simulation tracking efficiencies. The solid line represents a fit to the partially reconstructed candidates.

The efficiency difference of low momentum tracks ($p <$ 200 MeV/c) is studied using the decay chain, $B^0 \to D^{*-}\pi^+$ and $D^{*-} \to \overline{D}^0\pi_s^-$. The large $B^0 \to D^{*-}\pi^+$ branching fraction provides a sample of slow pions large enough to investigate possible track reconstruction discrepancies be-tween data and simulation. Since the tracking difference between data and the MC expectation at higher momenta is known, the data-MC ratios are normalized according to the data-MC ratio obtained using the $D^*$ partial reconstruction method for track momenta above 200 MeV/c. For events with lower $\pi_s$ momentum, the difference between the reconstructed yields in data and MC simulation is ascribed to a difference in the low momentum track reconstruction efficiency. Experimentally $\overline{D}^0$ candidates are reconstructed using several sub-decay modes. A slow pion ($\pi_s$) and a high momentum pion with opposite charge are included to form a $B$ candidate. The sample is divided in terms of the momentum of the slow pion, and the number of $B$ events in each momentum bin can be extracted using a fit to $m_{\rm ES}$ and $\Delta m$ (mass difference between $D^{*-}$ and $\overline{D}^0$). The yield ratio of data and MC is thus obtained in each $\pi_s$ momentum bin. The normalized ratios and their uncertainties at low momentum are used to correct for data-MC differences and to estimate the corresponding uncertainties. In Belle the tracking efficiency in simulation agrees well with that in data for track momenta above 125 MeV/c and the simulation may over-estimate the reconstruction efficiency for tracks with momentum below 100 MeV/c. On average the systematic uncertainty for low momentum tracks is 1.3% per track.

[46] For particles with $p < 200$ MeV/$c$ a different method is used as described below.

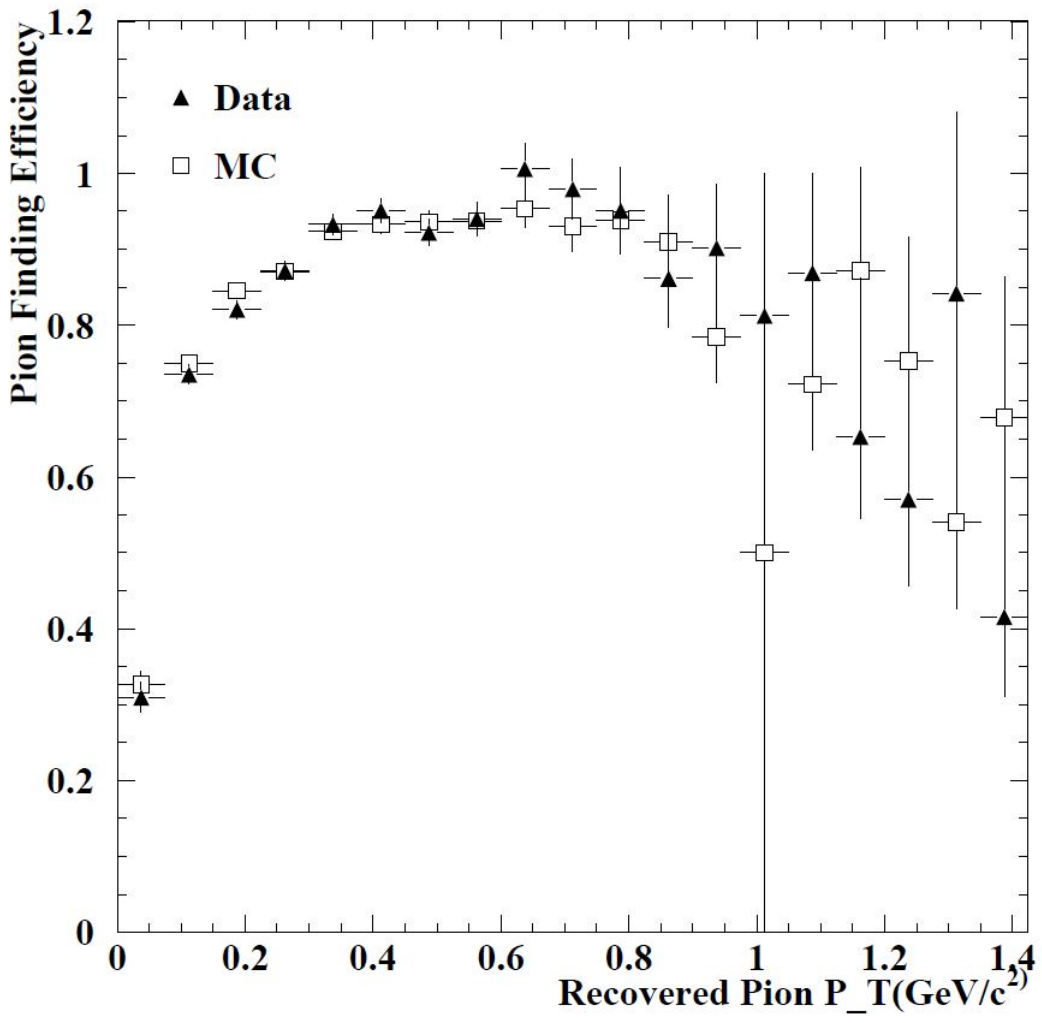

**Figure 15.1.3.** Reconstruction efficiency for charged tracks as a function of the particle's transverse momentum for simulated and real Belle data.

### 15.1.2 $K_S^0$ and $\Lambda$ reconstruction

Experimentally, $K_S^0$ and $\Lambda$ usually are reconstructed through $K_S^0 \to \pi^+\pi^-$ and $\Lambda \to p\pi^-$ decays. Both particles have a long lifetime. They are identified by the requirement that their decay vertex is displaced from the interaction point and that their reconstructed mass is close to their corresponding nominal mass. For long-lived particles systematic uncertainties in addition to the tracking uncertainties of their daughter particles need to be taken into account. The tracks may originate far from the interaction point, and also the reconstruction of the secondary vertex may show differences between simulation and data. Several studies are performed to investigate the $K_S^0/\Lambda$ reconstruction.

#### 15.1.2.1 Exclusive $D^* \to D^0\pi_s, D^0 \to \pi^+\pi^-K_S^0$

Similar to the study of track reconstruction systematics, the $K_S^0$ reconstruction is studied using the exclusive decays of $D^* \to D^0\pi_s, D^0 \to \pi^+\pi^-K_S^0$. One measures the efficiency of the displaced vertex requirement for the $K_S^0$ reconstruction by obtaining the numbers of $K_S^0$ candidates with and without reconstructing a $K_S^0$ vertex. The Belle method is described as follows. Two oppositely charged tracks that are identified as pions are selected and their invariant mass is computed without applying a vertex constraint. A pair with invariant mass close to the nominal $K_S^0$ mass is selected as a $K_S^0$ candidate. Every $K_S^0$ candidate is combined with another $\pi^+\pi^-$ pair to form a $D^0$ candidate, which is required to pair with a slow charged pion to form a $D^*$. To reduce the combinatorial background, a suitable mass range, estimated using simulations, is selected in the $D^0$ mass and $\Delta m' = m_{D^*} - m_D - m_{\pi_s}$. The uncertainty in $\Delta m'$ is significantly reduced with respect to $m_{D^*}$ because the contribution from the $K_S^0$ candidate momentum largely cancels in the subtraction and hence a tighter signal window can be applied due to a better resolution. Finally, the numbers of all $K_S^0$ particles and of those passing a displaced vertex selection are estimated by fitting the candidate $K_S^0$ mass with and without requiring a displaced vertex, respectively.

The control sample has sufficiently high statistics so that the study is extended to measure the efficiency in terms of $K_S^0$ momentum and polar angle similar to the charged track study described in Section 15.1.1. Likewise the efficiency of requiring a displaced vertex for Monte Carlo events can be estimated. Hence, the data-MC efficiency ratio can be obtained. The systematic uncertainty that arises from the reconstruction of the two $K_S^0$ daughter pion tracks has to be added to the efficiency uncertainty of the displaced vertex for the total $K_S^0$ systematic uncertainty. Since $\Lambda$ and $K_S^0$ decays have a similar topology, the $\Lambda$ systematic uncertainty can be estimated using the $K_S^0$ results. For the Belle full data sample, the total systematic uncertainty of the $K_S^0$ reconstruction is on average around 1% including track reconstruction systematic uncertainties.

#### 15.1.2.2 Ratio of two $D$ decays

The performance of the $K_S^0$ reconstruction in data can be checked using the double ratio

$$\eta(K_S^0) = \frac{N(D^+ \to K_S^0\pi^+)^{\rm data}}{N(D^+ \to K^-\pi^+\pi^+)^{\rm data}} \Bigg/ \frac{N(D^+ \to K_S^0\pi^+)^{\rm MC}}{N(D^+ \to K^-\pi^+\pi^+)^{\rm MC}} \,. \qquad (15.1.1)$$

In order to obtain a higher purity sample one can demand a high enough momentum of the $D^+$ candidates.

The disadvantages of this method are: the uncertainty in the $D^+ \to K^+\pi^+\pi^-$ branching fraction is large, the resonant substructure in these $D^+$ decays needs to be properly implemented in the simulation of $K^+\pi^+\pi^-$ decays, and the systematic uncertainty from particle identification needs to be included.

#### 15.1.2.3 $K_S^0$ decay length distribution

Another method to check the data-MC discrepancy in the $K_S^0$ reconstruction is to compare the $K_S^0$ decay length distribution. The $D^*$ decay mode $D^* \to D^0\pi_s, D^0 \to \pi^+\pi^- K_S^0$ used for the $K_S^0$ efficiency study above provides a clean sample to measure the $K_S^0$ decay length. Assuming that decays with short decay length are inside the fiducial region of the silicon vertex detector and are well simulated based on the tracking study, one can compare the fraction of reconstructed $K_S^0$ with longer decay length between data and MC events. The $K_S^0$ data-MC efficiency correction and the corresponding systematic uncertainty are thus obtained.

### 15.1.3 Particle identification

The performance of particle identification (PID) for *BABAR* and Belle is described in Chapter 5, with the related systematic uncertainties briefly discussed in Sections 5.3.2 and 5.4. The PID efficiency and its uncertainty are studied by choosing low-background samples in which the type of charged particles is identified without using the PID information. Then one can examine if the PID gives the correct identification. The PID efficiency and uncertainty can be estimated by counting the number of particles that are correctly identified. For instance, $K_S^0$ and $\Lambda$ are relatively long-lived and can fly a measurable distance before they decay into $\pi^+\pi^-$ or $p\pi^-$; requiring a distinct vertex and the appropriate mass range for the two-track mass provides clean samples of pions and protons. As for kaons, the sample of $D^{*+} \to D^0\pi_s^+$ and $D^0 \to K^-\pi^+$ is used. For electrons and muons, samples of $e^+e^- \to e^+e^-l^+l^-, e^+e^- \to l^+l^-(\gamma)$ and $J/\psi \to l^+l^-$ ($l = e$ or $\mu$) are chosen to study the performance of lepton identification; by positively identifying one of the leptons, the PID efficiency for the other can be studied.

The correction and systematic uncertainty for the signal efficiency due to PID can be estimated using the data-MC ratios of the PID efficiency and their uncertainties in different momentum, polar angle and azimuthal angle bins, similar to what is described for tracking systematics in Section 15.1.1. An alternative way to obtain the PID efficiency and its systematic uncertainty is to use signal MC events without applying any PID selection and weight each event according to the PID efficiency obtained in data. For sufficiently large Monte Carlo samples, the uncertainty due to the size of the sample for understanding the PID performance can be omitted. Typical systematic uncertainties per charged track in *BABAR* and Belle measurements are 0.8%-1.0%. The uncertainty due to the PID efficiency is treated as correlated among several tracks.

### 15.1.4 $\pi^0$ reconstruction

The reconstruction efficiency of $\pi^0$'s in the decay channel $\pi^0 \to \gamma\gamma$ can differ between data and simulation mainly for the following reasons (see Section 2.2.4 for the description of the electromagnetic calorimeters):

- *Imperfect modeling of the material distribution in the detector.* A photon can undergo pair production in the material of the detector before reaching the calorimeter. If the produced tracks are reconstructed in the tracking detectors, the corresponding clusters in the calorimeter, if any, are tagged as being produced by a charged track and the photon candidate is lost. Even if the reconstruction algorithms still find a photon candidate, the energy resolution might be degraded, leading to a $\pi^0$ candidate with an incorrectly reconstructed energy or mass.
- *Imperfect modeling of photon shower shape.* In order to discriminate electromagnetic from hadronic showers, shower shape variables such as the lateral moment,[47] the number of crystals in a shower etc. are used. Showers tend to be somewhat narrower in simulation than in data, creating a small efficiency difference between data and MC.
- *Split-offs.* The particle showers created by hadrons interacting with the material in the calorimeter contain a fraction of neutral hadrons. Such secondary hadrons can travel a sizable distance in the calorimeter before interacting with the material and depositing (a part of) their energy. These so-called *split-offs* leave the signature of a calorimeter cluster without an associated track pointing to it, which is hard to distinguish from a real photon. Cluster split-offs occur close to tracks, and the secondary showers usually have low energies. Detailed modeling of hadronic showers is difficult, thus split-offs present a further potential source of systematic difference between data and simulation.
- *Additional background in data.* Real data events typically contain more (soft) photon candidates, most of which originate from beam-related background. This background consists primarily of electrons and positrons from radiative Bhabha scattering which hit elements of the detector or the beam line, producing neutrons with energies in the MeV range, which then can produce low energy showers in the calorimeter. These additional photon candidates increase the number of $\gamma\gamma$ combinations in data, giving rise to more $\pi^0$ candidates, especially at low $\pi^0$ momentum.

The data-MC efficiency ratio is first measured in very clean events in which the presence of a $\pi^0$ can be predicted with little background. Possible differences between the $\pi^0$ reconstruction efficiency in such events and high-multiplicity events with higher background must then be also estimated. The data-MC efficiency ratio is measured using $\tau$ (Belle, *BABAR*) and $\eta$ (Belle) decays and multi-hadronic events with a photon radiated from the initial state (*BABAR*). An important step is the validation of the efficiency correction which is derived from this class of events and to make sure the correction is applicable to $B$ or charm decays, which tend to produce substantially more activity in the detector. In the following, we present some of the methods used to determine the $\pi^0$ efficiency correction and the associated systematic uncertainty.

[47] The lateral moment of a cluster in the calorimeter is defined as $LAT \equiv \sum_{i=3}^{N} r_{\perp i}^2 E_i/(25(E_1+E_2)+\sum_{i=3}^{N} r_{\perp i}^2 E_i)$, where the $N$ crystals which belong to a cluster are sorted by their energy $E_i$, and $r_{\perp i}$ is the (transverse) distance between the cluster centroid and the $i$th crystal.

### 15.1.4.1 Methods using $\tau$ decays

A clean way to extract the $\pi^0$ reconstruction efficiency is provided by comparing the observed rates of $\tau^- \to \pi^-\pi^0\nu_\tau$ to $\tau^- \to \pi^-\nu_\tau$ with the respective ratio of the branching fractions. The branching fractions of the two decays are known with sub-percent precision, allowing a measurement of the $\pi^0$ reconstruction efficiency with an uncertainty of the order of 1%.

In *BABAR*, $e^+e^- \to \tau^+\tau^-$ events are tagged with one $\tau$ decaying into $e^\pm\nu_e\nu_\tau$ (tag). On the signal side, a charged track incompatible with either the electron or the muon hypothesis is required. $\pi^0$ candidates are reconstructed from two photon candidates; events with more than two photon candidates (*i.e.* those with extra activity in the calorimeter) are removed.

The efficiency correction $\eta \equiv \varepsilon^{\rm data}/\varepsilon^{\rm MC}$ is computed as a function of the $\pi^0$ momentum $p_{\pi^0}$ as the double ratio

$$\eta(p_{\pi^0}) = \frac{N(\tau\to\pi\pi^0\nu)^{\rm data}(p_{\pi^0})}{N(\tau\to\pi\nu)^{\rm data}} \Bigg/ \frac{N(\tau\to\pi\pi^0\nu)^{\rm MC}(p_{\pi^0})}{N(\tau\to\pi\nu)^{\rm MC}} \tag{15.1.2}$$

$$= \frac{N(\tau\to\pi\pi^0\nu)^{\rm data}(p_{\pi^0})}{N(\tau\to\pi\pi^0\nu)^{\rm MC}(p_{\pi^0})} \Bigg/ \frac{N(\tau\to\pi\nu)^{\rm data}}{N(\tau\to\pi\nu)^{\rm MC}} . \tag{15.1.3}$$

In this double ratio, the track reconstruction and PID efficiencies (used on the tag side track) largely cancel provided there are no correlations between the tag and the signal side of the event:

$$\frac{N(\tau\to\pi\pi^0\nu)^{\rm data}}{N(\tau\to\pi\nu)^{\rm data}} = \frac{N_{\tau\tau}\,\mathcal{B}(\tau\to\pi\pi^0\nu)\,\varepsilon^{\rm data}_{\rm tag}\,\varepsilon^{\rm data}_{\rm track}\,\varepsilon^{\rm data}_{\pi^0}}{N_{\tau\tau}\,\mathcal{B}(\tau\to\pi\nu)\,\varepsilon^{\rm data}_{\rm tag}\,\varepsilon^{\rm data}_{\rm track}} \approx \varepsilon^{\rm data}_{\pi^0}\frac{\mathcal{B}(\tau\to\pi\pi^0\nu)}{\mathcal{B}(\tau\to\pi\nu)} \tag{15.1.4}$$

Using the well-measured branching fractions (and the corresponding values for simulated data), the double ratio directly measures the ratio of $\pi^0$ reconstruction efficiencies in data and simulation, modulo a few small corrections for split-offs and the mis-modeling of the high-energy tail of the $\pi^0\pi^-$ mass spectrum. The resulting correction factor depends on the $\pi^0$ momentum in the laboratory frame. For a typical $\pi^0$ momentum spectrum, the correction factor is around 0.97 with a statistical uncertainty well below 1%.

The result of the $\tau$ based study is combined with the results from $\omega$ production in events with hard initial state radiation (see below, Section 15.1.4.2) to obtain an overall momentum dependent $\pi^0$ efficiency correction. A systematic uncertainty of about 1.5% is assigned to cover the systematic differences between the two methods.

Belle also uses $\tau^+\tau^-$ events where one of the $\tau$ leptons decays leptonically and the other into $\pi^\pm\pi^0\nu$ (single $\pi^0$ events), and events where both decay into $\pi^\pm\pi^0\nu$ (double $\pi^0$ events). The ratio of data and MC simulation $\pi^0$ reconstruction efficiencies can be expressed as

$$\frac{\varepsilon^{\rm data}_{\pi^0}}{\varepsilon^{\rm MC}_{\pi^0}} = 2\cdot\frac{N_2^{\rm data}}{N_1^{\rm data}}\cdot\frac{\mathcal{B}(\tau\to\ell\nu\bar\nu)}{\mathcal{B}(\tau\to\pi\pi^0\nu)}\cdot\frac{\varepsilon_1^{\rm MC}}{\varepsilon_2^{\rm MC}}\cdot\frac{(\varepsilon^{\rm data}_{1'}/\varepsilon^{\rm MC}_{1'})}{(\varepsilon^{\rm data}_{2'}/\varepsilon^{\rm MC}_{2'})}, \tag{15.1.5}$$

where $N_{1,2}^{\rm data}$ are the numbers of reconstructed single and double $\pi^0$ events, and $\varepsilon_{1,2}^{\rm MC}$ are the efficiencies to reconstruct these events in the Monte Carlo; writing $\varepsilon_1 = \varepsilon_{\pi^0}\varepsilon_{1'}$ and $\varepsilon_2 = \varepsilon^2_{\pi^0}\varepsilon_{2'}$, Belle separates the efficiency for each class of event into the $\pi^0$ reconstruction efficiency, and a remainder term. The final double-ratio expression in Eq. (15.1.5) is assumed to be unity. Such a study reveals a correction factor of around 0.96 to be applied to the simulated reconstruction efficiency, with an uncertainty of 2.4%.

A comparison of $\eta\to 3\pi^0$ and $\eta\to\pi^+\pi^-\pi^0$ decays also yields the $\pi^0$ reconstruction efficiency directly from the data, and the systematic uncertainty at Belle is found to be 4%.

### 15.1.4.2 Methods using $\omega$-ISR and $\omega\pi^0$-ISR events

Another approach to measure the difference in the $\pi^0$ reconstruction efficiency between data and Monte Carlo is to use the low-background processes $e^+e^-\to\gamma_{\rm ISR}\omega$ and $e^+e^-\to\gamma_{\rm ISR}\omega\pi^0$, where the $\omega$ decays to $\pi^+\pi^-\pi^0$ and the initial state radiation photon is required to have a laboratory energy above 3 GeV. As in the case of the tracking efficiency study, one can exploit the fact that the kinematics of the reaction are fully known: both energy and momentum vector of the $\pi^0$ are predicted by a kinematic fit, using only the information of the initial state particles, the ISR photon and the two pion tracks. In the reaction $e^+e^-\to\gamma_{\rm ISR}\omega\pi^0$ the directly produced $\pi^0$ is required to be reconstructed while the efficiency study is performed with the $\pi^0$ from the $\omega$ decay. This method allows to study the $\pi^0$ efficiency as a function of the $\pi^0$ momentum and flight direction.

This method, in both reaction channels, makes use of the rather narrow width of the $\omega$. Signal events in which the $\pi^0$ momentum and energy were correctly inferred by the kinematic fit peak strongly close to the nominal $\omega$ mass; a fit to this mass spectrum yields the number of produced events which should contain a $\pi^0$. The number of events with a reconstructed $\pi^0$ is extracted from testing all $\pi^0$ candidates in the event with a $5C$ kinematic fit under the hypothesis $e^+e^-\to\omega\gamma\to\pi^+\pi^-\pi^0\gamma$. The classification of events into the categories '$\pi^0$ found' or '$\pi^0$ lost' is quite sensitive to the presence of extra $\pi^0$ candidates due to background photons, which is different in data and simulation.

At *BABAR*, the $\pi^0$ efficiency corrections derived from $\tau$ and $\omega$-ISR events as described above are combined into an overall efficiency correction with an associated systematic uncertainty which also accounts for the remaining differences between the two methods. The $\pi^0$ efficiency correction factors as a function of the $\pi^0$ lab momentum for both analyses as well as the combined correction factor which is recommended for general analyses are shown as a function of the $\pi^0$ lab momentum in Fig. 15.1.4.

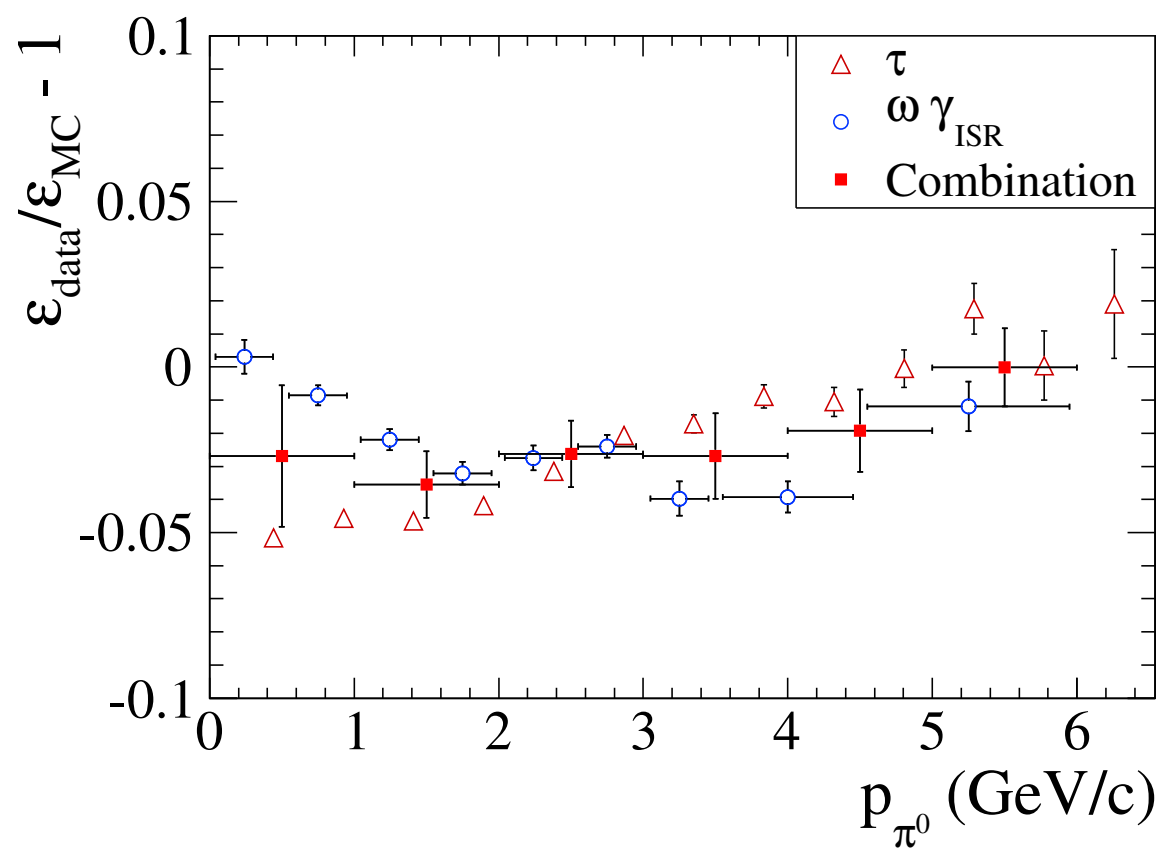

**Figure 15.1.4.** *BABAR* $\pi^0$ efficiency correction factors as a function of the $\pi^0$ lab momentum. The closed squares show the combination of the two analyses described in the text, with the error bars indicating the total systematic uncertainty associated with the efficiency correction (Aubert, 2013).

#### 15.1.4.3 Slow $\pi^0$, and $\pi^0$ efficiency in multi-hadronic decays

As described above, the $\pi^0$ efficiency correction is primarily measured in very clean events with few tracks and few or no additional neutrals. Typical $B$ decays, however, contain more tracks and neutral candidates, which can affect the probability to correctly reconstruct a $\pi^0$. To ensure that the efficiency correction is applicable to this class of multi-hadron events, an inclusive measurement of the ratio of $D^0 \to K\pi\pi^0$ to $D^0 \to K\pi$ decays has been performed at *BABAR*. The $\pi^0$ efficiency correction derived from this analysis suffers from a larger statistical uncertainty due to background subtraction and is less precise than the one derived from $\tau$ or ISR events. Within the given uncertainties, the $\pi^0$ efficiency corrections agree.

A dedicated study of low-momentum $\pi^0$'s has been performed using the decay chain $B^0 \to D^{*-}\pi^+, D^{*-} \to D^-\pi^0$. The method is similar to the one used for low-momentum tracks described in Section 15.1.1.2.

Similar to the study for the tracking efficiency, the data-MC efficiency ratio can also be computed using the double ratio

$$\eta(p_{\pi^0}) = \frac{N(D^0 \to K^+\pi^-\pi^0)^{\rm data}}{N(D^0 \to K^+\pi^-)^{\rm data}} \bigg/ \frac{N(D^0 \to K^+\pi^-\pi^0)^{\rm MC}}{N(D^0 \to K^+\pi^-)^{\rm MC}} \,. \quad (15.1.6)$$

This $\pi^0$ efficiency correction is thus obtained in hadronic events, as opposed to the efficiency in clean $e^+e^- \to \tau^+\tau^-$ events. To reduce the $D^0$ combinatorial background, one can demand a soft $\pi^+$ that combines with a $D^0$ candidate to form a $D^{*+}$ and the reconstruction uncertainty for slow pions cancels in the ratio. The dominant systematic error in the correction is the branching fraction uncertainty of $D^0 \to K^+\pi^-\pi^0$, which results in a common scale factor across the full momentum range. If the data-simulation efficiency ratio for the $\pi^0$ reconstruction is known from other studies in a typical momentum range, one can normalize the double ratio in that momentum range to obtain the correction factors and the corresponding uncertainties in other momentum ranges. For neutral pions with momenta below 200 MeV/$c$ the data-MC simulation correction factor at Belle is found to be $1.023 \pm 0.024$ for the data recorded with the SVD2 vertex detector (see Chapter 2). For *BABAR*, a similar study results in a correction factor for low-momentum $\pi^0$ of $0.98 \pm 0.07$.

### 15.1.5 High-energy photons

The detection efficiency of high energy photons (with typical energies above $E_\gamma \approx 2\,\text{GeV}$) is measured using radiative Bhabha events: $e^+e^- \to e^+e^-\gamma$ (Belle) and $e^+e^- \to \mu^+\mu^-\gamma$ (*BABAR*). After requiring exactly two tracks in an event that are identified as an $e^+e^-$ or $\mu^+\mu^-$ pair, the missing energy direction can be computed. The photon efficiency is estimated from the fraction of events that have a reconstructed photon matching the magnitude and direction of the missing energy, which is required to point to the electromagnetic calorimeter fiducial region. The precise value of the efficiency correction depends on the details of the criteria to select photon candidates and the decision whether a photon candidate matches the prediction from the kinematic fit.

For recent *BABAR* analyses of ISR events (Lees, 2012h), the difference in the reconstruction efficiency of high-energy photons between data and simulation was determined to be $\varepsilon_{\rm data} - \varepsilon_{\rm MC} = (-1.00 \pm 0.02\,(\text{stat}) \pm 0.55\,(\text{syst})) \times 10^{-2}$.

## 15.2 Analysis procedure

A second, important group of systematic uncertainties is related to the analysis procedure. This includes the use of external parameters as well as the use of specific models to separate signal from background and to extract the quantity of interest from the data. In a typical analysis at the $B$ Factories, multi-dimensional maximum likelihood fits are often used to separate signal and background on a statistical basis (see Section 11). This procedure needs to be carefully checked and validated and systematic uncertainties assigned where appropriate. The most important sources of these systematic uncertainties are discussed in this section.

### 15.2.1 External input

In many analyses, the physics observables are extracted from a fit with some of the parameters fixed to values based on external information. Using external information is necessary if for example the statistical power of the selected sample under consideration is not large enough to determine all relevant parameters with sufficient accuracy. For instance, in rare $B$ decay searches the peak positions and resolutions of $m_{\rm ES}$ and $\Delta E$ of signal events are often fixed; in Dalitz plot analyses the masses and natural widths of intermediate resonances are fixed to their PDG values; the mixing parameter $\Delta m_d$ and the $B^0$ meson lifetime are not allowed to vary in fits for time-dependent *CP* asymmetries. The systematic uncertainties that arise from using external input parameters are obtained by checking the deviations in the fitted values after varying the external parameters according to their uncertainties.

Unlike the PDG values used as the external parameters, some of the *p.d.f.* parameters explicitly depend on the detector resolution, and the corresponding uncertainties are determined using data. For instance, the uncertainty of $m_{\rm ES}$ is dominated by the beam energy spread and the $m_{\rm ES}$ peak position and resolution are determined using high-statistics control samples such as $B \to D^0\pi$ and $D^0 \to K^+\pi^-(\pi^0)$ for decay modes without (with) photons in the final state. The corrections between data and simulation and their uncertainties are obtained from these control samples and applied to the decay modes of interest. The same procedure is applied to estimate the correction and uncertainty for the $\Delta E$ *p.d.f.* parameters obtained in simulation. It is preferred to choose a control decay mode with high statistics that has the same numbers of charged and neutral particles in the final state as the mode under study. The same consideration can be applied to estimate systematic uncertainties related to flavor tagging, vertexing, mass resolutions and other external parameters.

Most analyses also rely on external input to derive the quantity of interest from directly measured quantities. Examples of such external parameters are the integrated luminosity (or, alternatively, the number of $B\overline{B}$ pairs produced), branching fractions of daughter decays, particle masses and their lifetimes, etc. These quantities and their uncertainties are typically taken from averages calculated by the Particle Data Group, with the exception of the luminosities, which are measured by the $B$ Factories (see Sections 3.2.1 and 3.6.2). At both experiments, the precision of the luminosity measurement is limited by systematic uncertainties, mainly by uncertainties of the Monte Carlo generator(s) used to calculate the cross-sections of the physics processes used to measure luminosity. At Belle, the luminosity is measured using Bhabha scattering to a precision of about 1.4%. *BABAR* uses both Bhabha scattering and $e^+e^- \to \mu^+\mu^-$ (Lees, 2013i); the systematic uncertainty of the luminosity is about 0.5% for the data collected at the $\Upsilon(4S)$.

The uncertainties from these external parameters are propagated to the final result using either Gaussian error propagation in the simplest cases, or by varying the parameters within their uncertainties and repeating the analysis.

### 15.2.2 Modeling of background

Background distributions are often modeled using events from simulation or sidebands of *e.g.* mass distributions. A typical example is modeling the background distributions in the Dalitz plot for $B$ or $D$ decays. One can assume that the Dalitz plot distributions for the combinatorial background are the same as those obtained using events outside the $m_{\rm ES} - \Delta E$ signal region or in the $D$ mass sideband region. The background model can be cross-checked by comparing the distributions of simulated background events in the signal and sideband regions or by comparing the data distributions in different sideband regions. The systematic uncertainty due to the background modeling is then estimated by using the *p.d.f.*s obtained from different sideband regions and by varying the *p.d.f.* parameters according to the uncertainties.

In many cases the background is sufficiently large so that the background *p.d.f.* parameters can be determined directly from a fit to data. This procedure moves the uncertainty originating from the background *p.d.f.* parameters into the overall statistical uncertainty returned by the fit. However, in many cases the actual shape of the background distribution is not known from first principles, and there may be several different parameterizations which describe, within the given uncertainties of the data, the background shape equally well. The systematic uncertainty related to this is determined by choosing different functions for the background *p.d.f.*s and repeating the fit. For example, $B$ yields in many rare decay searches are extracted with an unbinned maximum likelihood fit to the distributions of $m_{\rm ES}$, $\Delta E$ and other variables (see Chapter 9). The *p.d.f.*s of the $B$ decay background are usually estimated from simulations, while the continuum *p.d.f.*s are modeled as a polynomial function for $\Delta E$ and an ARGUS function (see Eq. 7.1.11) for $m_{\rm ES}$ with their parameters allowed to vary in the fit. Systematic uncertainties of the fit can be evaluated using other function models that provide an acceptable goodness of the fit.

### 15.2.3 Fit bias

The results of multi-dimensional maximum likelihood fits (see Chapter 11) can be systematically biased when the correlations between various discriminating variables are not considered or several components have similar *p.d.f.*s, so that the fit cannot completely distinguish between those components. The fit bias can be examined using large ensembles of simulated experiments ('toy MC', see Section 11.5.2); a bias correction is then derived from these studies. There is no unique method of assigning a systematic uncertainty to this bias correction, and analyst discretion is required. As a conservative approach, the systematic uncertainty associated with the bias correction is often taken to be half or even all of the correction.

## 15.3 Systematic effects for time-dependent analyses

A number of systematic effects need to be understood in order to verify that one is able to correctly extract time-dependent information from fits to data. The general methodology for performing a time-dependent *CP* asymmetry analysis at the $B$ Factories is outlined in Chapter 10. In addition there are special cases that have been considered over the course of these experiments including time-dependent analyses of modes requiring a full angular analysis (Chapter 12), and time-dependent Dalitz plot analyses (Chapter 13).

In the following we discuss systematic uncertainties arising from detector and reconstruction effects (see Sections 15.3.1 through 15.3.3), uncertainties from physics parameters (see Section 15.3.4), and uncertainties arising from approximations made in the analyses (see Sections 15.3.5 through 15.3.6). The systematic uncertainties quoted on $S$ and $C$ (see Chapter 10) in the remainder of this chapter are typical values obtained by the $B$ Factories.

### 15.3.1 Alignment of the vertex detector

In order to precisely reconstruct the decay vertex position of both $B$ mesons in an event and the value of the proper time difference $\Delta t$ between the decays of both mesons (see Chapter 6 for a detailed discussion on these matters), accurate information is required on the position of the reconstructed *hits* that correspond to the signature of charged particles traveling through the tracking volume. The silicon detectors at the $B$ Factories dominate our understanding of the vertex positions by virtue of their proximity to the interaction point, and hence the $B$ decay vertices. The first few measurement points of each track originating from a $B$ decay will be recorded in the silicon detector, and hence one must precisely know the position of the strips embedded in the silicon. This position changes slightly with time, and if not corrected for, will smear out the knowledge of each hit position, and hence fitted track and computed vertex. The purpose of the silicon detector calibration is to correct for variations in the alignment as a function of run period, and in the case of Belle, the differences between the different SVDs installed during operation (see Chapter 2).

While the detector calibration is extremely effective at correcting for variations in detector position as a function of time, there is an uncertainty arising from any residual lack of knowledge in the position and orientation of each double-sided silicon sensor module that provides a measurement of $r$, $\phi$ and $z$ within the detector. The local alignment procedure adopted by *BABAR* is described in detail in (Brown, Gritsan, Guo, and Roberts, 2009). In order to estimate the magnitude of the uncertainty arising from the alignment of the silicon detector, different sets of alignment constants are applied to simulated Monte Carlo data for signal events or equivalently the silicon detector positions are intentionally modified in a plausible range in both global displacement and rotation as well as random misalignment for each silicon sensor, and the change in fitted values of the *CP*-violating parameters $S$ and $C$ (see Section 10.2) from the nominal value is assigned as an uncertainty from this source of systematic. The magnitude of this uncertainty on $S$ and $C$ is at most a few per mille. In extreme cases, for example modes such as $B^0 \to \rho^+\rho^-$ that suffer from a significant contribution from mis-reconstructed signal in the final state, the effect of the silicon detector alignment is somewhat larger: $\sim 0.01$. The reason for this is that some of the mis-reconstructed signal in this final state has a biased reconstructed vertex position, resulting from the inclusion of low-momentum tracks reconstructed at the extremities of the helicity angle distributions (see Chapter 12). Sometimes these low momentum tracks are incorrectly assigned from the rest of the event to a signal $B$ candidate, rather than including the correct tracks from the signal side. Different alignment sets change the reconstruction rate of this component of mis-reconstructed signal, and thus induce a bias on the measured observables $S$ and $C$.

### 15.3.2 Beamspot position, $z$ scale and boost

As discussed in Chapter 6, the beamspot location can be used to improve constraints on vertex reconstruction, and is used when reconstructing the tagging $B$ meson vertex. The dominant contribution to the systematic uncertainty when adding this constraint comes from the limited knowledge of the vertical position of the beamspot. The knowledge of the beamspot is included in the vertex fit via the addition of an extra term in the $\chi^2$ of the track fit. The limitation in the absolute knowledge of the beamspot location therefore translates into a systematic uncertainty on the reconstructed value of $\Delta t$, and hence propagates through onto the measured observables $S$ and $C$ in a time-dependent *CP* asymmetry analysis. Detailed studies of the beam-spot position calibration were performed at the $B$ Factories (see Section 6.4).

Knowledge of the mean vertical position is the dominant systematic uncertainty from the use of the beamspot in *BABAR*, while its spread is found to give a much larger effect in Belle. The corresponding systematic uncertainties in the measured values of $S$ and $C$ are estimated by modifying the position and uncertainty on the vertical beamspot position according to the variations seen in data. For example, *BABAR* varies this position by $\pm 20\mu$m, as well as increase the uncertainty on this quantity by $20\mu$m to evaluate the systematic uncertainty arising from the use of the beamspot in vertex reconstruction. Belle changes the beamspot position uncertainty to a factor of 2 larger or smaller value than the nominal one, $21\mu$m. The relative change in $S$ ($C$) from its nominal value ($S \sim \sin 2\phi_1$, $C \sim 0$) is found to be 0.13% (0.06%) in *BABAR* and 0.3% (0.08%) in Belle for $B$ decays to $c\bar{c}s$ final states.

Other important factors impacting the measurement of $S$ and $C$ are the $z$ scale determined from the vertex detector, and the boost factor. Detailed studies of control samples show that the $z$ scale uncertainty is the dominant of these two effects, and to account for these uncertainties $\Delta t$ and $\sigma_{\Delta t}$ are scaled by 0.6%. This results in negligible systematic shifts, of the order of $4.7\times10^{-4}$ in $S$ and $2.3\times10^{-4}$ in $C$, for $B$ decays to $c\bar{c}s$ final states. These are interpreted as systematic uncertainties from the $z$ scale and boost determination.

### 15.3.3 Resolution function and flavor tagging parameters

Both the $\Delta t$ resolution function parameters and flavor tagging performance parameters are integral inputs to a time-dependent analysis. There are two conceptual ways to incorporate systematic uncertainties from these parameters into the extracted values of $S$ and $C$. Firstly one can perform a simultaneous fit to the so-called $B_{\rm flav}$ sample of events (see Section 10.2) and the selected signal candidates. In this approach the uncertainties on and correlations between resolution function and tagging model parameters are automatically folded into the statistical uncertainty reported for the asymmetry parameters. This approach is adopted by *BABAR*. The second approach is to take the results of a reference fit to the $B_{\rm flav}$ data sample, and incorporate the variations of $S$ and $C$ from the nominal result when varying the resolution and tagging parameters by their uncertainties. This approach results in a number of contributions that are added in quadrature ignoring the correlations that exist between them. Belle uses the second approach as there are only small correlations between the parameters describing the resolution and tagging performance. As a result this second approach provides a conservative and still proper estimation of the systematic uncertainty from the knowledge of these parameters. The typical uncertainty on $S$ and $C$ obtained for the resolution function and tagging parameters using the second approach is $\leq 0.01$.

### 15.3.4 The effect of physics parameters

Time-dependent *CP* asymmetry measurements at the $B$ Factories follow the method described in Chapter 10. In particular, these analyses assume $\Delta\Gamma_d = 0$, unlike the situation for time-dependent measurement for $B_s$ (and eventually $D$) meson decays. No systematic uncertainty is ascribed for the use of this assumption, which is well motivated by theoretical arguments for the statistics available at *BABAR* and Belle. A non-zero value of $\Delta\Gamma_d = 0$ would give rise to hyperbolic sine and cosine terms in the time-dependent asymmetries as discussed in Section 17.5.2.6, and one can estimate the magnitude of any systematic uncertainty from neglecting these hyperbolic terms by comparing results obtained using an ensemble of simulated Monte Carlo experiments with $\Delta\Gamma_d \neq 0$, and observing the bias introduced on the fitted values of $S$ and $C$. If one assumes that $\Delta\Gamma_d \leq 0.01$, the systematic uncertainties in $S$ and $C$ would be negligible, if one were to use the existing experimental limit on the value of $\Delta\Gamma_d$ the bias on $S$ would be 0.005.

The physics parameters $\tau_{B^0}$ and $\Delta m_d$ are required inputs for time-dependent measurements. During the ML fitting procedure used to extract $S$ and $C$ from data, the $B^0$ lifetime and mixing frequency are fixed to their nominal values. The uncertainty on the measured values of $\tau_{B^0}$ and $\Delta m_d$ are propagated through the fitting procedure, assuming that they are uncorrelated, and the resulting variation of $S$ and $C$ from the nominal fitted values is assigned as an uncertainty. This source of uncertainty is found to be at most a few per-mille.

### 15.3.5 *CP* violation in background components

A subtlety raised in Chapter 10 is the issue of correctly accounting for any *CP* asymmetry (time-dependent or time-integrated) in background modes when performing a time-dependent analysis. This issue is not significant for the case of charmonium decays such as $B^0 \to J/\psi K^0_S$, where there is very little background, however it should be considered when analyzing modes with significant levels of background such as $B^0 \to \rho^+\rho^-$.

There are two types of *CP* violating background that may occur (*i.e.* direct and mixing-induced *CP* violation, see Chapter 16) from neutral $B$ mesons, and charged $B$ mesons may only violate *CP* via direct decay. In addition one may need to consider the $B\overline{B}$ background, where the $B$ signal candidates are formed by combining the daughter particles of the true $B_{\rm tag}$ and $B_{\rm rec}$ . In general the reconstructed $|\Delta t|$ values of these background events are smaller than the true ones as the reconstructed $B_{\rm tag}$ and $B_{CP}$ vertices tend to be closer to each other.

Such an effect can be taken into account by replacing the $B$ lifetime in the exponential decay of Eq. (10.2.2) with an effective lifetime. This is particularly relevant for final states with charm mesons in them as discussed in Chapter 10, but is also manifest at a lower level for $B$ backgrounds without charm decays. Generally one assumes that any bias for the latter class of $B$ decays is negligible.

Having corrected for the above reconstruction effects one is faced with having to address the issue of a physical asymmetry in the background decay channel. In the case of a neutral $B$ decay the asymmetry will be of the form of Eq. (10.2.8). One has to account for tagging and resolution effects, and typically it is assumed that it is valid to use the same tagging and resolution parameters for the background channels as for the correctly reconstructed signal. Ideally one should generate samples of Monte Carlo simulated data for each *CP* violating background mode with the values of $S$ and $C$ as measured in data. This way any dilution from mis-reconstructing a given channel is taken into account when setting the values of the effective $S$ and $C$ required to model the *CP* asymmetry of a given background mode. In cases where there is no measurement of the asymmetry parameters, but it is reasonable to expect a non-zero asymmetry, one varies the effective values of $S$ and $C$ between $+1$ and $-1$ to estimate the maximal effect a given background would have on the signal. *CP* violation in charged decay modes can be accounted for in an analogous way, where one uses the time-integrated asymmetry to allow for any possible direct *CP* violation.

Typical systematic uncertainties in the values of the *CP* asymmetry parameters measured for the high-background decay $B \to \rho^+\rho^-$ arising from possible *CP* violation in the background are $\leq 0.2\%$ for $S$ and $1-2\%$ for $C$, see Aubert (2007b). This uncertainty is dominated by contributions from $B \to a_1\pi$ decays, assuming that *CP* violation could be large, as the example discussed predates *CP* asymmetry measurements of $B \to a_1^{\pm}\pi^{\mp}$.

### 15.3.6 Tag-side interference

In order for a decay channel to have non-zero *CP* asymmetry, it must have at least two interfering amplitudes with different weak phases. This is a necessary condition, but it is not sufficient to guarantee that there will be an observable *CP* violation effect in that final state. The discussion so far has focused on interfering amplitudes on the $B_{\rm rec}$ side of the event leading to a measurable *CP* violation effect. However it was pointed out by Long, Baak, Cahn, and Kirkby (2003) that in addition to interference on the $B_{\rm rec}$ side, one has to consider possible effects of interference on the $B_{\rm tag}$ side, where more than one amplitude contributes to the final state. If neglected, interference effects on the $B_{\rm tag}$ side of the event could result in an undesired contribution to the measured *CP* asymmetry for the $B_{\rm rec}$. Many different final states are included in the (inclusive) reconstruction of the $B_{\rm tag}$ with different contributions to the so-called tag-side interference effect.

As discussed in Section 8, the dominant contributions to the tagging efficiency come from semi-leptonic decays with final state leptons, and hadronic decays such as $B \to D^{(*)-}\pi^+$. Since the semi-leptonic decays proceed via a single amplitude in the SM, semi-leptonic tagged decays do not suffer from tag-side interference. However possible interference effects need to be considered when performing a time-dependent analysis, where $B_{\rm tag}$ decays to a hadronic final state as the decay can proceed by more than one amplitude.

If one considers the decay $B \to D^-\pi^+$, with subsequent $D^- \to K^+\pi^-\pi^-$ decay as an example, the final state can be reached via the CKM preferred $b \to c\overline{u}d$ transition of a $\overline{B}^0$. The same final state can also be reached from a $B^0$ through $B^0 - \overline{B}^0$ mixing followed by a doubly-CKM suppressed $\overline{b} \to \overline{u}c\overline{d}$ transition. The ratio of these two amplitudes is given approximately by the ratio of CKM matrix elements $|(V_{ub}^*V_{cd})/(V_{cb}V_{ud}^*)| \simeq 0.02$.

The strength of the amplitude of the doubly-CKM suppressed relative to the allowed decay can be parameterized as

$$\frac{\overline{A}_f}{A_f} = r_f e^{-i\phi_3 + i\delta_f}, \tag{15.3.1}$$

where $r_f$ is the ratio of suppressed to favored decays, and $\delta_f$ is the relative strong phase difference between the $B^0$ and $\overline{B}^0$ decay proceeding via $\overline{b} \to \overline{c}u\overline{d}$ and $b \to u\overline{c}d$ transitions, respectively. In practice a number of modes are summed over on the tag-side of the event, and we replace $r_f$ and $\delta_f$ with primed variants to represent the effective ratio of amplitudes and phase difference of an ensemble of modes.

It is possible to compute a correction on the time-dependent asymmetry parameters $S$ and $C$ resulting from the use of hadronic tag modes, either for a given mode, or an ensemble of modes. These corrections are a function of $\Delta t$ and have the effect of slightly reducing the amplitude and broadening the time distribution, or increasing the amplitude and narrowing the distribution as discussed in Section VI and Fig. 3 of Long, Baak, Cahn, and Kirkby (2003). The effect depends on the values of $r'_f$ and $\delta'_f$. Thus one can expect the measured values of $S$ and $C$ in a time-dependent analysis to differ from the true values for hadronically tagged events.

The semileptonic decay, $B^0 \to D^{*-}\ell^+\nu_\ell$ is a high purity $B_{\rm flav}$ mode and free from doubly-CKM suppressed diagram as already discussed. Thus applying the proper flavor tagging algorithm on the $B_{\rm tag}$ decay products in this sample gives an estimation of the possible range of the effective ratio of the amplitudes and phase difference for an ensemble of the tag-side modes. This estimation is used to see the effects on $S$ and $C$ as described in more detail later.

If a time-dependent analysis were limited by systematic uncertainties arising from tag-side interference, there are two possible approaches that may be considered to mitigate this uncertainty: (i) only use semi-leptonic tagged events, thus removing the affected data from the analysis, and (ii) given sufficient data, to measure the ratio of CKM allowed to suppressed decays, and the corresponding phase difference between the amplitudes using control samples. In the following discussion the true values of these time-dependent asymmetry parameters are represented by $S_0$ and $C_0$, whereas the measured values of these observables are denoted by $S_{\rm fit}$ and $C_{\rm fit}$.

#### 15.3.6.1 The tree dominated $B^0 \to J/\psi K_S^0$ decay

The prime example of a time-dependent measurement made by the $B$ Factories is that of $B^0 \to J/\psi K_S^0$, which is described in Section 17.6. The biases on the true values of measured time-dependent asymmetries in this decay arising from tag-side interference can be treated as a perturbation on the measurement, *i.e.* a systematic shift with an associated uncertainty. It is possible to relate the true values of the *CP* asymmetry parameters $S_0$ and $C_0$ to the fitted values $S_{\rm fit}$ and $C_{\rm fit}$ up to some correction related to the additional amplitudes interfering on the tag side of the event. The correction depends on $\Phi = 2\phi_1 + \phi_3$ resulting from the phase difference between the doubly CKM suppressed and CKM allowed amplitudes on the tag side of the decay and the short distance $B^0 - \overline{B}^0$ mixing box contributions. The corrections to the fitted *CP* asymmetry parameters are related to the magnitude of the effective ratio of CKM suppressed to allowed amplitudes for the tag-side decay given by $r'_f$, as shown in the following

$$
\begin{aligned}
C_{\rm fit} &= C_0 + 2C_0 r'_f \cos\delta'_f \{G\cos(\Phi) - S_0\sin(\Phi)\} \\
&\quad -2r'_f \sin\delta'_f\{S_0\cos(\Phi) + G\sin(\Phi)\}, \qquad (15.3.2)\\
S_{\rm fit} &= S_0 + 2S_0 r'_f \cos\delta'_f G\cos(\Phi) \\
&\quad +2r'_f\sin\delta'_f C_0\cos(\Phi). \qquad (15.3.3)
\end{aligned}
$$

Here the factor $G$ is $2{\rm Re}\lambda_{CP}/(|\lambda_{CP}|^2+1)$, and $\lambda_{CP}$ is the quantity given in Eq. (10.1.10) evaluated for the $B_{\rm rec}$ reconstructed in a *CP* eigenstate.

Using a Monte Carlo simulation based approach, one can estimate the magnitude of the effect on the value of $S_{\rm fit}$ and $C_{\rm fit}$ extracted from data, and hence determine $S_0$ and $C_0$. In order to do this one has to determine $r'_f$ and $\delta'_f$. The value of $r'_f$ is given by $|(V^*_{ub}V_{cd})/(V_{cb}V^*_{ud})|$ and an estimate of the uncertainty on this can be derived from a comparison of rates for allowed to suppressed $D \to K\pi$ transitions. This comparison indicates that the error on $r'_f$ is about 25%. As there is no knowledge of the phase difference, one assumes that this parameter is uniformly distributed in the simulated pseudo-experiments. This approach of evaluating the effect of tag-side interference for $B^0 \to J/\psi K^0_S$ has been broadly applied to $b\to c\overline{c}s$, $c\overline{c}d$, and $q\overline{q}s$ final states. The magnitude of the systematic uncertainty ascribed to the measurement of $S$ ($C$) in this set of channels is typically 0.001 (0.014). The systematic uncertainty is negligible for the extraction of $\sin 2\phi_1$ from the golden $b \to c\overline{c}s$ measurements. However this source of systematic uncertainty is significant for some of the precision measurements of $C$, and in fact dominant for the golden channel $B^0 \to J/\psi K^0_S$ discussed in Section 17.6. For the measurement of $\phi_1$ from an ensemble of *CP*-even and odd states (*i.e.* $J/\psi K^0_L$ and $c\overline{c}K^0_S$) *BABAR* ascribes a systematic uncertainty as described above. However, Belle note that there may be some cancellation between the even and odd states and account for this in their estimation of the systematic uncertainty from this source on the combined measurements of $S$ and $C$ (Adachi, 2012c).

There is no indication of a significant shift in the measured values of $S$ and $C$ found via this Monte Carlo simulation based approach, hence no corrections are applied to the results obtained by the $B$ Factories.

#### 15.3.6.2 The complication of loop amplitudes in $B^0 \to \pi^+\pi^-$

An example of a decay with both tree and loop (penguin) amplitudes used in a time-dependent analysis is $B^0 \to \pi^+\pi^-$ which is discussed further in Section 17.7. The decay amplitude for the reconstructed $B$ meson depends on $\phi_3$, as does the tag-side. Thus the situation encountered with $B^0 \to \pi^+\pi^-$ is therefore much more complicated than the previous case. The uncertainty from tag-side interference can be as large as $2r'_f$. This complication for calculating tag-side interference applies not only to $B^0 \to \pi^+\pi^-$ decays, but more generally to the set of $b \to u\overline{u}d$ transitions related to $\phi_2$ where there are significant penguin contributions. The least problematic of these decays being $B^0 \to \rho^+\rho^-$, which is known to have a small penguin contribution, relative to other $b \to u\overline{u}d$ transitions.

The magnitude of the systematic uncertainty ascribed to the measurement of $S$ ($C$) in this set of channels is typically $0.007 - 0.010$ ($0.016 - 0.04$) depending on the final state. While small, compared to the overall experimental uncertainty, this is the dominant source of systematic uncertainty for the extraction of $C$ from the $B^0 \to \pi^+\pi^-$ and $\rho^+\rho^-$ channels discussed in Section 17.7. The systematic uncertainty is negligible on the extraction of $\phi_2$ for the golden $b \to u\overline{u}d$ measurements given the statistics available at the $B$ Factories.

#### 15.3.6.3 Time-dependent measurement of $\sin(2\phi_1+\phi_3)$

The measurement of $\sin(2\phi_1+\phi_3)$ using $B \to D^{*\pm}\pi^{\mp}$ decays is discussed in Section 17.8. The manifestation of tag-side interference in this time-dependent measurement differs from that discussed for the previous two examples as described below. As with the $b \to u\overline{u}d$ transition case the reconstructed $B$ meson depends on $\phi_3$, so it is not straightforward to extract an estimate of tag-side interference for $B \to D^{*\pm}\pi^{\mp}$ decays. Furthermore, the amplitude of the $\sin(\Delta m_d \Delta t)$ term in the time-evolution of this decay is $2r\sin(2\phi_1+\phi_3)$. Here the parameter $r$ is the ratio of doubly-CKM suppressed to allowed decays for the reconstructed $B$ meson (the $B \to D^{*\pm}\pi^{\mp}$) and has nothing to do with the tag-side of the event.[48] The magnitudes of both $r_f$ and $r'_f$ are expected to be comparable and of the order of 0.02, thus there is the potential for tag-side interference to obscure the signal measurement. It is possible to perform an analysis of the time-dependence of $B \to D^{*\pm}\pi^{\mp}$ explicitly taking into account the effect of tag-side interference while doing so. In contrast to the discussion of $B$ decays to $J/\psi K^0_S$ or $\pi^+\pi^-$ final states where the effect of tag-side interference is treated as a perturbation on a measurement, for $\sin(2\phi_1+\phi_3)$ one attempts to formally incorporate the full time-dependence of both $B$ mesons decaying in an event, allowing for *CP* violation for both the signal and tag sides. A scheme for doing this is outlined by Long, Baak, Cahn, and Kirkby (2003) and this approach has been adopted by the $B$ Factories.

## 15.4 Summary

In order to provide for very precise measurements of various observables the systematic uncertainties of the measurements must be kept under control. In an ideal case the systematic uncertainty should not exceed the statistical one by a large margin. At the $B$ Factories several ingenious methods were developed to estimate the remaining systematic errors as precisely as possible. Whenever possible the uncertainties are obtained using real data control samples, thus avoiding systematic effects due to possible discrepancies between MC simulation and data. For some sources of systematic uncertainties encountered in several measurements performed at the $B$ Factories the estimation methods and representative values are summarized in Table 15.4.1.

[48] The parameter $r$ should not be confused with either the ratio $r_f$ in Eq. (15.3.1), or the effective parameter $r'_f$ for an ensemble of modes on the tag-side of the event.

**Table 15.4.1.** Summary of typical *BABAR* and Belle systematic uncertainties appearing in various measurements. $\Delta\varepsilon$ denotes the difference between the efficiency as estimated in the MC simulation and in the real data, $\sigma_\varepsilon$ denotes the uncertainty on the efficiency. $\sigma_{S,C}$ denotes the uncertainty of *CP* violating parameters $S, C$ (see Chapter 10).

| Measurement | BABAR Method | BABAR Typical value | Belle Method | Belle Typical value | Comment |
|---|---|---|---|---|---|
| Tracking | $e^+e^- \to \tau^\pm\tau^\mp$ , | $\sigma_\varepsilon/\varepsilon = (0.13-0.24)\%$ | $D^{*+} \to D^0(\to \pi^+\pi^- K^0_s)\pi^+$ | $\sigma_\varepsilon/\varepsilon = 0.35\%$ | Uncertainty per charged track |
| | $\tau^\pm \to h^\pm h^\pm h^\mp \nu,\ \tau^\mp \to \mu^\mp\bar\nu$ | $\Delta\varepsilon'/2\varepsilon = (0.10-0.26)\%$ | | | Asymmetry for $h^+/h^-$ |
| | $e^+e^- \to 2\pi^+2\pi^-\gamma_{\rm ISR}$ | $\Delta\varepsilon/\varepsilon = (0.7-0.4)\%$ | | | Uncertainty per charged track |
| | $B \to h^+h^-K^0_s(\to \pi^+\pi^-)$ | $\Delta\varepsilon/\varepsilon = (0.5\pm0.8)\%$ | | | Uncertainty for tracks with displaced vertex |
| | $D^{*+} \to D^0\pi^+_s$ | $\sigma_\varepsilon/\varepsilon = 1.5\%$ | $B^0 \to D^{*-}(\to \overline{D}^0\pi^-_s)\pi^+$ | $\sigma_\varepsilon/\varepsilon = 1.3\%$ | Uncertainty for low momentum tracks ($p \lesssim 200$ MeV/$c$) |
| | | | $D^{*+} \to D^0(\to \pi^+\pi^- K^0_s)\pi^+$ | $\sigma_\varepsilon/\varepsilon = 1\%$ | Uncertainty for $K^0_s$ reconstruction (including tracking uncertainty for $\pi^+\pi^-$) |
| PID | $D^{*+} \to D^0(\to K^-\pi^+)\pi^+$<br>$e^+e^- \to e^+e^-(\gamma)$<br>$e^+e^- \to e^+e^-\ell^+\ell^-$<br>$J/\psi \to \ell^+\ell^-$ | $\sigma_\varepsilon/\varepsilon = (0.8-1.0)\%$ | $D^{*+} \to D^0(\to K^-\pi^+)\pi^+$<br>$e^+e^- \to e^+e^-(\gamma)$<br>$e^+e^- \to e^+e^-\ell^+\ell^-$<br>$J/\psi \to \ell^+\ell^-$ | $\sigma_\varepsilon/\varepsilon = (0.8-1.0)\%$ | Uncertainty per charged track |
| $\pi^0$ reconstruction | $e^+e^- \to \tau^+\tau^-$ ,<br>$\tau^+ \to e^+\nu\bar\nu,\ \tau^- \to \pi^-(\pi^0)\nu$<br>$e^+e^- \to \gamma_{\rm ISR}\omega(\to \pi^+\pi^-\pi^0)$ | $\varepsilon_{\rm data}/\varepsilon_{\rm MC} \sim 0.970 \pm 0.015$ | $e^+e^- \to \tau^+\tau^-$ ,<br>$\tau^+ \to \ell^+\nu\bar\nu$ or $\pi^+\pi^0\bar\nu, \tau^- \to \pi^-\pi^0\nu$<br>$\eta \to 3\pi^0/\eta \to \pi^+\pi^-\pi^0$ | $\varepsilon_{\rm data}/\varepsilon_{\rm MC} = 0.960 \pm 0.024$<br>$\sigma_\varepsilon/\varepsilon = 4\%$ | Efficiency correction per $\pi^0$ ($p \geq 200$ MeV/$c$) |
| | $B^0 \to D^{*-}\pi^+ \to D^0\pi^0\pi^+$ | $\varepsilon_{\rm data}/\varepsilon_{\rm MC} = 0.98 \pm 0.07$ | $B^0 \to D^{*-}\pi^+ \to D^0\pi^0\pi^+$ | $\varepsilon_{\rm data}/\varepsilon_{\rm MC} = 1.024 \pm 0.027$ | Efficiency correction per $\pi^0$ ($p < 200$ MeV/$c$) |
| High energy photons | $e^+e^- \to \mu^+\mu^-\gamma$ | $\Delta\varepsilon = (1.00 \pm 0.55)\cdot 10^{-2}$ | | | |
| $t$-dependent | MC | $\sigma_{S,C} \lesssim 0.01$ | MC | $\sigma_{S,C} \lesssim 0.01$ | vertex detector alignment |
| | changing beam spot position | $\sigma_S/S \sim 0.13\%$<br>$\sigma_C/C \sim 0.06\%$ | changing beam spot position | $\sigma_S/S \sim 0.3\%$<br>$\sigma_C/C \sim 0.08\%$ | beam spot position |
| | varying parameters from $B_{\rm flav}$ sample | $\sigma_{S,C} \lesssim 0.01$ | varying parameters from $B_{\rm flav}$ sample | $\sigma_{S,C} \lesssim 0.01$ | resolution function, flavor tag |
| | varying parameters | $\sigma_{S,C} \sim \mathcal{O}(10^{-3})$ | varying parameters | $\sigma_{S,C} \sim \mathcal{O}(10^{-3})$ | physics parameters ($\Delta\Gamma_d, \Delta m_d, \ldots$) |
| | varying possible $S$, $C$ for background | $\sigma_C/C \lesssim (1-2)\%$ | varying possible $S$, $C$ for background | $\sigma_C/C \lesssim (1-2)\%$ | *CP* violation in background |
| | $B^0 \to D^{*-}\ell^+\nu$ | $\sigma_{S,C} \sim (0.1, 1.4)\cdot 10^{-2}$<br>$\sigma_S \sim 1.0\cdot 10^{-2}$<br>$\sigma_C \sim (1.6-4.0)\cdot 10^{-2}$ | $B^0 \to D^{*-}\ell^+\nu$ | $\sigma_{S,C} \sim (0.1, 0.8)\cdot 10^{-2}$<br>$\sigma_S \sim 1.0\cdot 10^{-2}$<br>$\sigma_C \sim (1.6-4.0)\cdot 10^{-2}$ | tag-side interference in $B \to J/\psi K^0_s$<br>} tag-side interference in $b \to u\bar u d$ |

# Part C
# The results and their interpretation

## Chapter 16
## The CKM matrix and the Kobayashi-Maskawa mechanism

***Editors:***
*Adrian Bevan and Soeren Prell (BABAR)*
*Boštjan Golob and Bruce Yabsley (Belle)*
*Thomas Mannel (theory)*

### 16.1 Historical background

*Fundamentals*

In the early twentieth century the "elementary" particles known were the proton, the electron and the photon. The first extension of this set of particles occurred with the neutrino hypothesis, first formulated by W. Pauli in his famous letter to his "radioactive friends" in 1924. From the theoretical side, the formulation of a theory of weak interactions by Fermi in 1934 marked another milestone in the development of our understanding. This set up for the first time a framework, in which some of the fundamental questions on the role of hadrons versus leptons and on the properties of particles and their interactions could be formulated. This also resulted in a clear formulation of "weak" versus "strong" interactions and the understanding of interactions as an exchange of mediating particles. In particular, Yukawa postulated the existence of such a particle and triggered the search for what we now know as the pion. At about the same time the muon was discovered, and initially called the "$\mu$ meson", however this soon turned out to be distinct from the pion.

Although the the term "flavor" came much later, one may mark the beginning of (quark) flavor physics by the discovery of strange particles (Rochester and Butler, 1947). Their decays into non-strange particles had lifetimes too long to be classified as strong decays: this led to the introduction of the strangeness quantum number (Gell-Mann, 1953), which is conserved in strong decays but may change in a weak decay.

The subsequent proliferation of new particles could nicely be classified and ordered by Gell Mann's "eightfold way" (Gell-Mann, 1962), which was an extension of the isospin symmetry to a symmetry based on the group $SU(3)$. However, none of the particles fitted into the fundamental representation of this group, although there were various attempts such as Sakata's model, in which the proton, the neutron and the $\Lambda$ baryon formed the fundamental representation. Eventually this puzzle was resolved by the postulate of quarks as the fundamental building blocks of matter.

*Strangeness, parity violation, and charm*

The decays of the strange particles, in particular of the kaons, paved the way for the further development of our understanding. Before 1954, the three discrete symmetries C (charge conjugation), P (parity) and T (time reversal) were believed to be conserved individually, a conclusion drawn from the well known electromagnetic interaction. Based on this assumption, the so called $\theta$-$\tau$ puzzle emerged: Two particles (at that time called $\theta$ and $\tau$, where the latter is not to be confused with the third generation lepton) were observed, which had identical masses and lifetimes. However, they obviously had different parities, since the $\theta$ particle decayed into two pions (a state with even parity), and the $\tau$ particle decays into three pions (a state with odd parity).

The resolution was provided by the bold assumption by Lee and Yang (1956) that parity is not conserved in weak interactions, and $\theta$ and $\tau$ are in fact the same particle, which we now call the charged kaon. Subsequently the parity violating $V - A$ structure of the weak interaction was established and, on the experimental side, parity violation was confirmed directly in $\beta$ decays (Garwin, Lederman, and Weinrich, 1957; Wu, Ambler, Hayward, Hoppes, and Hudson, 1957). However, the combination of two discrete transformations, namely $CP$, still seemed to be conserved.

Another puzzle related to kaon decays was the relative coupling strength. It tuned out that the coupling strength of strangeness-changing processes is much smaller than that of strangeness-conserving transitions. This finding eventually led to the parameterization of quark mixing by Cabibbo (1963). In modern language, the up quark $u$ couples to a combination $d \cos\theta_C + s \sin\theta_C$ of the down quark $d$ and the strange quark $s$. The value $\theta_C \sim 13^\circ$ for the Cabibbo angle explained the observed pattern of branching ratios in baryon decays.

Experiments at that time only probed the three lightest quarks, and there was no known reason for the extreme suppression of the flavor changing neutral current (FCNC) decay $K^+ \to \pi^+\ell^+\ell^-$ with respect to the charged current decay $K^+ \to \pi^0\ell^+\overline{\nu}$, $\Gamma(K^+ \to \pi^+\ell^+\ell^-)/\Gamma(K^+ \to \pi^0\ell^+\overline{\nu}) \sim 10^{-6}$. The resolution of this puzzle was found by Glashow, Iliopoulos, and Maiani (1970): one includes the charm quark, with the same quantum numbers as the up quark, and coupling to the orthogonal combination $-d \sin\theta_C + s \cos\theta_C$.

FCNC processes are suppressed by this "GIM mechanism". In fact, FCNC's in the kaon system involve a transition of an $s$ quark into a $d$ quark. This can be achieved by two successive charged current processes involving (in the two-family picture) either the up or the charm quark as an intermediate state. Taking Cabibbo mixing into ac-count, these amplitudes are

$$\mathcal{A}(s \to d) = \mathcal{A}(s \to u \to d) + \mathcal{A}(s \to c \to d) = \sin\theta_C \, \cos\theta_C \, [f(m_u) - f(m_c)], \quad (16.1.1)$$

where $f(m)$ is some smooth function of the mass $m$. Hence, if the up and charm quark masses were degenerate, $K^0 - \overline{K}^0$ mixing and other kaon FCNC processes would not occur.

However, the up and charm masses are not degenerate and thus $K^0 - \overline{K}^0$ mixing can occur. Neglecting the small up-quark mass, the mixing amplitude turns out to be

$$\mathcal{A}(K \to \overline{K}) \propto \sin^2\theta_C \, \cos^2\theta_C \, \frac{m_c^2}{M_W^2}. \quad (16.1.2)$$

This implies that a mass difference $\Delta m_K$ appears in the neutral kaon system. From this mass difference (an expression analogous to Eq. 10.1.17) Gaillard and Lee (1974b) could extract the prediction that the charm-quark mass should be about $m_c \sim 1.5$ GeV, and it was one of the great triumphs of particle physics when narrow resonances with masses of about 3 GeV were discovered a few months later (Aubert et al., 1974; Augustin et al., 1974): these were identified as $c\overline{c}$ bound states. Around this time the term "particle family" was coined, and the discovery of the charm quark completed the second particle family; it also introduced a $2 \times 2$ quark mixing matrix into the phenomenology of weak interactions.

#### *CP violation and the Kobayashi-Maskawa mechanism*

Almost ten years before the discovery of charm, *CP* violation was observed in the study of rare kaon decays by Christenson, Cronin, Fitch, and Turlay (1964). This effect is difficult to accommodate for two families, but an extension to three families allows it to be taken into account naturally. The "six-quark model" was proposed by Kobayashi and Maskawa (1973), extending Cabibbo's $2\times2$ quark mixing matrix into the $3 \times 3$ Cabibbo-Kobayashi-Maskawa (CKM) matrix. The GIM mechanism for the six quark model is implemented by the unitarity of the CKM matrix.

While the observation of decays $K_L^0 \to 2\pi$ meant that *CP* was violated, the data at that time only required *CP* violation in mixing (see Section 16.6 for the classification of *CP*-violating effects). The observed strength of *CP* violation in mixing, $\varepsilon_K \simeq 2.3 \times 10^{-3}$, was consistent with the Kobayashi-Maskawa (KM) mechanism (Ellis, Gaillard, and Nanopoulos, 1976; Pakvasa and Sugawara, 1976). However, this did not constitute a proof that the KM mechanism was really the origin of the observed *CP* violation; the measurement of the single parameter $\varepsilon_K$ could not be used to test the KM mechanism. One alternative explanation was offered by the super-weak model of Wolfenstein (1964), where *CP* violation was due to a new, very weak four-fermion interaction that changed strangeness by 2 units ($\Delta S = 2$). This possibility was ruled out by the observation of direct *CP* violation in $K_L \to \pi\pi$ decays, $\text{Re}(\varepsilon'_K/\varepsilon_K) = (1.65 \pm 0.26) \times 10^{-3}$ (Alavi-Harati et al., 1999; Burkhardt et al., 1988; Fanti et al., 1999). Nonetheless, convincing evidence for the KM mechanism required the measurement of $\sin(2\phi_1)$ at the $B$ Factories.

With the discovery of the $\tau$ lepton in 1975 (Perl et al., 1975) and of the bottom quark in 1977 (Herb et al., 1977) it became clear that there is a third generation of quarks and leptons. Furthermore, the bottom quark turned out to be quite long-lived, indicating a small mixing angle between the first and second generation. This fact is the experimental foundation of using $B$ decays to study *CP* violation, as well as for $b$ tagging in high-$p_t$ physics.

The third generation remained incomplete for many decades, since the top quark turned out to be quite heavy, and a direct discovery had to wait until 1995, when it was discovered at the Tevatron at Fermilab (Abachi et al., 1995a; Abe et al., 1994). However, the first hint of the large top-quark mass was the discovery of $B^0 - \overline{B}^0$ oscillations (also known as mixing) by ARGUS (Albrecht et al., 1987b). The measured $\Delta m_d$ implied a heavy top with a mass $m_t$ above $50 - 70$ GeV, if the standard six quark model was assumed (Bigi and Sanda, 1987; Ellis, Hagelin, and Rudaz, 1987). The phenomenon of neutral meson mixing is discussed in Chapter 10, while Section 17.5 discusses results on $B$ mixing from the $B$ Factories.

In fact, if the top mass had been significantly smaller, ARGUS could not have observed $B^0 - \overline{B}^0$ oscillations. The GIM mechanism for down-type quarks leads generally to suppression factors of the form

$$\text{CKM Factor} \times \frac{1}{16\pi^2} \frac{m_t^2 - m_u^2}{M_W^2} \quad (16.1.3)$$

and hence the GIM suppression for the bottom quark is much weaker than in the up-quark sector, where the corresponding factor is

$$\text{CKM Factor} \times \frac{1}{16\pi^2} \frac{m_b^2 - m_d^2}{M_W^2}. \quad (16.1.4)$$

Hence FCNC decays of $B$-mesons have branching ratios in the measurable region, while FCNC processes for $D$-mesons are heavily suppressed.

The third particle family was completed by the discovery of the $\tau$ neutrino as a particle distinct from the electron and the muon neutrino by the DONUT collaboration (Kodama et al., 2001). Although models with a fourth particle generation are frequently considered as benchmark models for physics beyond the Standard Model, there is no indication of a fourth family. On the contrary, from the width of the $Z$ boson precisely measured at LEP it can be inferred that there is no further family with a neutrino lighter than 40 GeV, and the recent discovery of a Higgs boson in the mass range of 125 GeV (Aad et al., 2012; Chatrchyan et al., 2012b) rules out a large class of fourth-generation models.

## 16.2 $CP$ violation and baryogenesis

Particle physics experiments of the past thirty years have confirmed the Standard Model (SM) even at the quantum level, including quark mixing and $CP$ violation. However, the observed matter-antimatter asymmetry of the universe indicates that there must be additional sources of $CP$ violation, since the amount of $CP$ violation implied by the CKM mechanism is insufficient to create the observed matter-antimatter asymmetry.

In fact, the excess of baryons over antibaryons in the universe

$$\Delta = n_{\mathfrak{B}} - n_{\overline{\mathfrak{B}}} \tag{16.2.1}$$

is small compared to the number of photons: the ratio is measured to be $\Delta/n_\gamma \sim 10^{-10}$. Although it is conceivable that there might be regions in the universe consisting of antimatter, just as our neighborhood consists of matter, no mechanism is known which could, from the Big Bang, produce regions of matter (or antimatter) as large as we observe today. Furthermore, searches have been performed for sources of photons indicative of regions of matter and antimatter colliding. These searches failed to find any large regions of antimatter.

The conditions under which a non-vanishing $\Delta$ can emerge dynamically from the symmetric situation $\Delta = 0$ have been discussed by Sakharov (1967). He identified three ingredients

1. There must be baryon number violating interactions $H_{\text{eff}}(\Delta\mathfrak{B} \neq 0) \neq 0$.
2. There must be $CP$ violating interactions. If $CP$ were unbroken, then we would have for every process $i \to f$ mediated by $H_{\text{eff}}(\Delta\mathfrak{B} \neq 0)$ the $CP$ conjugate one with the same probability
$$\Gamma(i \to f) = \Gamma(\bar{i} \to \overline{f}) \tag{16.2.2}$$
which would erase any matter-antimatter asymmetry.
3. The universe must have been out of thermal equilibrium. Under the assumption of locality, causality, and Lorentz invariance, $CPT$ is conserved. Since in an equilibrium state time becomes irrelevant on the global scale, $CPT$ reduces to $CP$, and the argument of point 2 applies.

In order to illustrate the first two Saharov conditions, we employ a very simplistic example. Assume that in the early universe, there was a particle $X$ that could decay to only two final states $|f_1\rangle$ and $|f_2\rangle$, with baryon numbers $N_{\mathfrak{B}}^{(1)}$ and $N_{\mathfrak{B}}^{(2)}$ respectively, and decay rates

$$\Gamma(X \to f_1) = \Gamma_0 r \quad \text{and} \quad \Gamma(X \to f_2) = \Gamma_0(1-r)\,, \tag{16.2.3}$$

where $\Gamma_0$ is the total width of $X$. Taking the $CP$ conjugate, the particle $\overline{X}$ decays to the state $\overline{f}_1$ with baryon number $-N_{\mathfrak{B}}^{(1)}$ and $\overline{f}_2$ with baryon number $-N_{\mathfrak{B}}^{(2)}$; the rates are

$$\Gamma(\overline{X} \to \overline{f}_1) = \Gamma_0 \overline{r} \quad \text{and} \quad \Gamma(\overline{X} \to \overline{f}_2) = \Gamma_0(1-\overline{r}), \tag{16.2.4}$$

where $\Gamma_0$ is the same as for $X$ due to $CPT$ invariance.

The overall change $\Delta N_{\mathfrak{B}}$ in baryon number induced by the decay of an equal number of $X$ and $\overline{X}$ particles is

$$\begin{aligned}\Delta N_{\mathfrak{B}} &= rN_{\mathfrak{B}}^{(1)} + (1-r)N_{\mathfrak{B}}^{(2)} - \overline{r}N_{\mathfrak{B}}^{(1)} - (1-\overline{r})N_{\mathfrak{B}}^{(2)} \\ &= (r-\overline{r})\left(N_{\mathfrak{B}}^{(1)} - N_{\mathfrak{B}}^{(2)}\right)\end{aligned} \tag{16.2.5}$$

Thus $\Delta N_{\mathfrak{B}} \neq 0$ means that we have to have $CP$ violation ($r \neq \overline{r}$) and a violation of baryon number ($N_{\mathfrak{B}}^{(1)} \neq N_{\mathfrak{B}}^{(2)}$), illustrating the first two conditions.

Sakharov's paper remained mostly unnoticed until the first formulation of Grand Unified Theories (GUTs). In these theories, for the first time, all the necessary ingredients were present. In particular, baryon number violation appears naturally since quarks and leptons appear in the same multiplets of the GUT symmetry group. Furthermore, there are additional sources of $CP$ violation, and a phase transition takes place at the scale $M_{\text{GUT}}$, which has to be quite high to prevent proton decay.

One may also consider electroweak baryogenesis. The electroweak interaction provides $CP$ violation through the CKM mechanism, and the electroweak phase transition has been thoroughly studied. The first ingredient is also present, as the current corresponding to baryon number is conserved only at the classical level: electroweak quantum effects violate baryon number, but still conserve the difference $\mathfrak{B} - L$ of baryon and lepton number. However, although all the ingredients are present, this cannot explain $\Delta$. In particular, the CKM $CP$ violation is too small by several orders of magnitude.

Given the firm evidence for non-vanishing neutrino masses, there could be new sources of $CP$ violation in the lepton sector, and even (although there is no evidence for this as yet) lepton-number violation. This could lead to violation of baryon number via leptogenesis, with the surplus of leptons transferred to the baryonic sector through $(\mathfrak{B} - L)$-conserving interactions.

In any case, an additional source(s) of $CP$ violation is needed, beyond the phase of the CKM matrix (which is explained in the next section), in order to explain the matter-antimatter asymmetry of the universe. The search for this new interaction is one of the main motivations for flavor-physics experiments.

## 16.3 $CP$ violation in a Lagrangian field theory

The SM is formulated as a quantum field theory based on a Lagrangian derived from symmetry principles. To this end, the (hermitian) Lagrangian of the SM is given in terms of scalar operators $\mathcal{O}_i$ with couplings $a_i$

$$\mathcal{L}(x) = \sum_i \left(a_i \mathcal{O}_i(x) + a_i^* \mathcal{O}_i^\dagger(x)\right)\,, \tag{16.3.1}$$

where the $\mathcal{O}_i$ are composed of the SM quark, lepton, and gauge fields. It is straightforward to verify that $CP$ conservation implies that all couplings $a_i$ can be made real by suitable phase redefinitions of the fields composing the $\mathcal{O}_i$. In turn, *CP* is violated in a Lagrangian field theory if there is no choice of phases that renders all $a_i$ real.

In the SM there are in principle two sources of *CP* violation. The so-called "strong *CP* violation" originates from special features of the QCD vacuum, resulting in a contribution of the form

$$\mathcal{L}_{\mathrm{strong}\,CP} = \theta \frac{\alpha_S}{8\pi} G^{\mu\nu,a} \tilde{G}^a_{\mu\nu} \qquad (16.3.2)$$

where $G_{\mu\nu}$ ($\tilde{G}_{\mu\nu}$) is the (dual) strength of the gluon field. This term is *P* and *CP* violating due to its pseudoscalar nature. However, a term such as Eq. (16.3.2) will have a strong impact on the electric dipole moment (EDM) of the neutron, $d_N \sim \theta \times 10^{-15}$ e cm. In combination with the current limit on the neutron EDM of $d_N < 0.29 \times 10^{-25}$ e cm, this yields a stringent limit, $\theta \leq 10^{-10}$. However, the theoretical reason for its smallness has not yet been discovered. This is known as the "strong *CP* problem" (see for example Cheng, 1988; Kim and Carosi, 2010); we shall ignore this in what follows by setting $\theta = 0$.

The second source of *CP* violation is the CKM matrix. It turns out that all terms in the SM Lagrangian are *CP* invariant except for the charged current interaction term

$$H_{\mathrm{cc}} = \frac{g}{\sqrt{2}} \left(\overline{u}_L\, \overline{c}_L\, \overline{t}_L\right) V_{\mathrm{CKM}} \gamma^\mu \begin{pmatrix} d_L \\ s_L \\ b_L \end{pmatrix} W^+_\mu . \qquad (16.3.3)$$

Under a *CP* transformation we have

$$\left(\overline{u}_L\, \overline{c}_L\, \overline{t}_L\right) V_{\mathrm{CKM}} \gamma^\mu \begin{pmatrix} d_L \\ s_L \\ b_L \end{pmatrix} W^+_\mu \qquad (16.3.4)$$

$$\xrightarrow{CP} \left(\overline{d}_L\, \overline{s}_L\, \overline{b}_L\right) V^T_{\mathrm{CKM}} \gamma^\mu \begin{pmatrix} u_L \\ c_L \\ t_L \end{pmatrix} W^-_\mu \qquad (16.3.5)$$

and hence the combination $H_{\mathrm{cc}} + H^\dagger_{\mathrm{cc}}$ appearing in the SM Lagrangian is *CP* invariant, if

$$V^T_{\mathrm{CKM}} = V^\dagger_{\mathrm{CKM}} \quad \text{or} \quad V_{\mathrm{CKM}} = V^*_{\mathrm{CKM}}. \qquad (16.3.6)$$

This statement refers to a specific phase convention for the quark fields; in general terms it implies that in the *CP*-invariant case, the CKM matrix can be made real by an appropriate phase redefinition of the quark fields.

## 16.4 The CKM matrix

The CKM matrix $V_{\mathrm{CKM}}$ appearing in Eq. (16.3.3) is explicitly written as

$$V_{\mathrm{CKM}} = \begin{pmatrix} V_{ud} & V_{us} & V_{ub} \\ V_{cd} & V_{cs} & V_{cb} \\ V_{td} & V_{ts} & V_{tb} \end{pmatrix}. \qquad (16.4.1)$$

Here the $V_{ij}$ are the couplings of quark mixing transitions from an up-type quark $i = u, c, t$ to a down-type quark $j = d, s, b$.

In the SM the CKM matrix is unitary by construction. Using the freedom of phase redefinitions for the quark fields, the CKM matrix has $(n-1)^2$ physical parameters for the case of $n$ families. Out of these, $n(n-1)/2$ are (real) rotation angles, and $((n-3)n+2)/2$ are phases, which induce *CP* violation. For $n = 2$, no *CP* violation is possible, while for $n = 3$ a single phase appears. This is the unique source of *CP* violation in the SM, once the possibility of strong *CP* violation is ignored.

The CKM matrix for 3 families may be represented by three rotations and a matrix generating the phase

$$U_{12} = \begin{bmatrix} c_{12} & s_{12} & 0 \\ -s_{12} & c_{12} & 0 \\ 0 & 0 & 1 \end{bmatrix},$$

$$U_{13} = \begin{bmatrix} c_{13} & 0 & s_{13} \\ 0 & 1 & 0 \\ -s_{13} & 0 & c_{13} \end{bmatrix},$$

$$U_{23} = \begin{bmatrix} 1 & 0 & 0 \\ 0 & c_{23} & s_{23} \\ 0 & -s_{23} & c_{23} \end{bmatrix},$$

$$U_\delta = \begin{bmatrix} 1 & 0 & 0 \\ 0 & 1 & 0 \\ 0 & 0 & e^{-i\delta_{13}} \end{bmatrix}, \qquad (16.4.2)$$

where $c_{ij} = \cos\theta_{ij}$, $s_{ij} = \sin\theta_{ij}$, and $\delta$ is the complex phase responsible for *CP* violation; by convention the mixing angles $\theta_{ij}$ are chosen to lie in the first quadrant so that the $s_{ij}$ and $c_{ij}$ are positive. Then (Chau and Keung, 1984)

$$V_{\mathrm{CKM}} = U_{23} U^\dagger_\delta U_{13} U_\delta U_{12}$$
$$= \begin{pmatrix} c_{12}c_{13} & s_{12}c_{13} & s_{13}e^{-i\delta} \\ -s_{12}c_{23} - c_{12}s_{23}s_{13}e^{i\delta} & c_{12}c_{23} - s_{12}s_{23}s_{13}e^{i\delta} & s_{23}c_{13} \\ s_{12}s_{23} - c_{12}c_{23}s_{13}e^{i\delta} & -c_{12}s_{23} - s_{12}c_{23}s_{13}e^{i\delta} & c_{23}c_{13} \end{pmatrix}. \qquad (16.4.3)$$

This is the representation used by the PDG (Beringer et al., 2012).

The elements of the CKM matrix exhibit a pronounced hierarchy. While the diagonal elements are close to unity, the off-diagonal elements are small, such that *e.g.* $V_{ud} \gg V_{us} \gg V_{ub}$. In terms of the angles $\theta_{ij}$ we have $\theta_{12} \gg \theta_{23} \gg \theta_{13}$. This fact is usually expressed in terms of the Wolfenstein parameterization (Wolfenstein, 1983), which can be understood as an expansion in $\lambda = |V_{us}|$. It reads up to order $\lambda^3$

$$V_{\mathrm{CKM}} = \begin{pmatrix} 1-\lambda^2/2 & \lambda & A\lambda^3(\rho - i\eta) \\ -\lambda & 1-\lambda^2/2 & A\lambda^2 \\ A\lambda^3(1-\rho-i\eta) & -A\lambda^2 & 1 \end{pmatrix} + \mathcal{O}(\lambda^4). \qquad (16.4.4)$$

The parameters $A$, $\rho$ and $\eta$ are assumed to be of order one. When using this parameterization, one has to keep in mind that unitarity is satisfied only up to order $\lambda^4$. As it turns out that both $\rho$ and $\eta$ are also of order $\lambda$, the extension to higher orders becomes non-trivial, and one has to consider redefining the parameters accordingly; this has been studied by Ahn, Cheng, and Oh (2011).

One can obtain an exact parameterization of the CKM matrix in terms of $A$, $\lambda$, $\rho$, and $\eta$, for example, by following the convention of Buras, Lautenbacher, and Ostermaier (1994), where

$$\lambda = s_{12}, \tag{16.4.5}$$
$$A = s_{23}/\lambda^2, \tag{16.4.6}$$
$$A\lambda^3(\rho - i\eta) = s_{13}e^{-i\delta}, \tag{16.4.7}$$

and by substituting Eqs (16.4.5) through (16.4.7) into Eq. (16.4.3), while noting that $\sin^2\theta = 1 - \cos^2\theta$. Such a parameterization is described in Section 19.2.1.3 to illustrate *CP* violation in the charm sector.

Sometimes a slightly different convention for the Wolfenstein parameters is used, with parameters denoted $\overline{\rho}$ and $\overline{\eta}$. These parameters were defined at fixed order by Buras, Lautenbacher, and Ostermaier (1994); the modern definition (Charles et al., 2005),

$$\overline{\rho} + i\overline{\eta} = -\frac{V_{ud}V_{ub}^*}{V_{cd}V_{cb}^*}, \tag{16.4.8}$$

holds to all orders. The difference with the parameterization defined above appears only at higher orders in the Wolfenstein expansion; the relation between this scheme and the one defined in (16.4.5–16.4.7) is given by

$$\rho + i\eta = (\overline{\rho} + i\overline{\eta})\frac{\sqrt{1 - A^2\lambda^4}}{\sqrt{1-\lambda^2}[1 - A^2\lambda^4(\overline{\rho} + i\overline{\eta})]}. \tag{16.4.9}$$

## 16.5 The Unitarity Triangle

The unitarity relations $V_{\rm CKM} \cdot V_{\rm CKM}^\dagger = 1$ and $V_{\rm CKM}^\dagger \cdot V_{\rm CKM} = 1$ yield six independent relations corresponding to the off-diagonal zeros in the unit matrix. They can be represented as triangles in the complex plane; each triangle has the same area, reflecting the fact that (with three families) there is only one irreducible phase. A non-trivial triangle — one with angles other than 0 or $\pi$ — indicates *CP* violation, proportional to the triangles' common area. Bigi and Sanda (2000) provide a detailed discussion of the various triangles, their interpretation, and the possibilities to probe them. Only two triangles have sides of comparable length, which means that they are of the same order in the Wolfenstein parameter $\lambda$. The corresponding relations are

$$V_{ud}V_{ub}^* + V_{cd}V_{cb}^* + V_{td}V_{tb}^* = 0 \tag{16.5.1}$$
$$V_{ud}V_{td}^* + V_{us}V_{ts}^* + V_{ub}V_{tb}^* = 0. \tag{16.5.2}$$

Inserting the Wolfenstein parameterization, both relations turn out to be identical, up to terms of order $\lambda^5$; the apex of the Unitarity Triangle is given by the coordinate $(\overline{\rho}, \overline{\eta})$. The three sides of this triangle (Fig. 16.5.1) — usually referred to as "the" Unitarity Triangle— control semi-leptonic and non-leptonic $B_d$ transitions, including $B_d - \overline{B}_d$ oscillations. In order to obtain the triangle shown in Fig 16.5.1, Eq. (16.5.1) is divided by $V_{cd}V_{cb}^*$ so that the base of the triangle is of unit length. Due to the sizable angles, one expects large *CP* asymmetries in $B$ decays in the SM; this was actually realized *before* the discovery of "long" $B$ lifetimes. Note that in both unitarity-triangle relations CKM matrix elements related to the top quark appear; in particular $V_{td}$ and $V_{ts}$ can be accessed only indirectly via FCNC decays of bottom quarks.

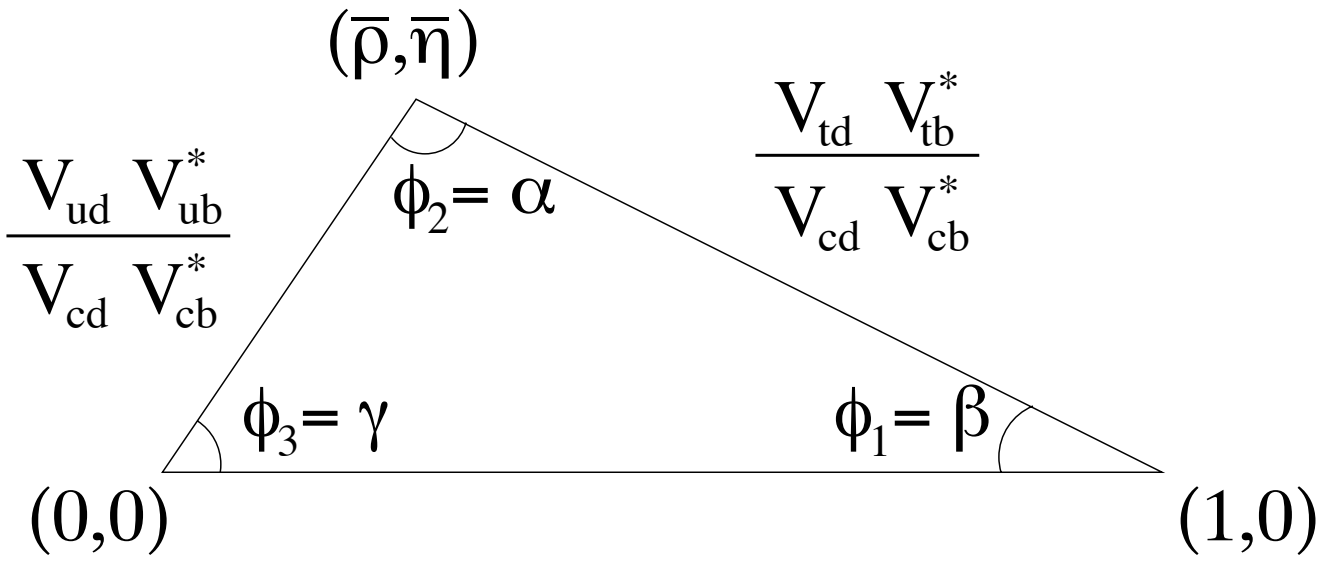

**Figure 16.5.1.** The Unitarity Triangle.

The angles of the Unitarity Triangle are defined as

$$\phi_1 = \beta \equiv \arg\left[-V_{cd}V_{cb}^*/V_{td}V_{tb}^*\right], \tag{16.5.3}$$
$$\phi_2 = \alpha \equiv \arg\left[-V_{td}V_{tb}^*/V_{ud}V_{ub}^*\right], \tag{16.5.4}$$
$$\phi_3 = \gamma \equiv \arg\left[-V_{ud}V_{ub}^*/V_{cd}V_{cb}^*\right], \tag{16.5.5}$$

where this definition is independent of the specific phase choice expressed in Eq. (16.4.3). Different notation conventions have been used in the literature for these angles. In particular the *BABAR* experiment has used $\alpha$, $\beta$, and $\gamma$, whereas the Belle experiment has reported results in terms of $\phi_2$, $\phi_1$, and $\phi_3$, respectively. We use the latter for brevity when discussing results in later sections.

The presence of *CP* violation in the CKM matrix implies non-trivial values for these angles ($\phi_i \neq 0°$, $180°$), corresponding to a non-vanishing area for the Unitarity Triangle. In fact, all the triangles that can be formed from the unitarity relation have the same area, which is proportional to the quantity

$$\Delta = \mathrm{Im}\, V_{cs}^* V_{us} V_{cd} V_{ud}^* \tag{16.5.6}$$

which is independent of the phase convention. Note that all other, rephasing invariant fourth order combinations of CKM matrix elements, which cannot be reduced to products of second order invariants, can be related to $\Delta$, which is thus unique.

Furthermore, the phase in the CKM matrix could also be removed, if the masses of either two up-type quarks or two down-type quarks were degenerate. In summary, the presence of *CP* violation is equivalent to (Jarlskog, 1985)

$$\begin{aligned} J &= \det[M_u\,,\, M_d] \\ &= 2i\Delta \times (m_u - m_c)(m_u - m_t)(m_c - m_t) \\ &\qquad \times (m_d - m_s)(m_d - m_b)(m_s - m_b) \end{aligned} \tag{16.5.7}$$

being non-vanishing.

The SM allows us to construct "the" Unitarity Triangle by measuring its angles or its sides or any combinations of them. Any discrepancy between the observed and predicted values indicates a manifestation of dynamics beyond the SM. Clearly this requires good control of experimental and theoretical uncertainties, both in their *CP* sensitive and insensitive rates.

Measurements of the magnitudes of CKM matrix elements $V_{ub}$ and $V_{cb}$ can be found in Section 17.1, and measurements of $V_{td}$ and $V_{ts}$ in Section 17.2. Measurements of the angles $\phi_1$, $\phi_2$, and $\phi_3$ are discussed in Sections 17.6, 17.7, and 17.8 respectively. It is possible to perform *global fits*, using data from many decay processes to over-constrain our knowledge of the CKM mechanism. Given the lack of knowledge of the determination of the apex of the Unitarity Triangle, these global fits are often expressed in terms of constraints on the $(\overline{\rho}, \overline{\eta})$ plane. Some experimental results require input from Lattice QCD calculations in order to be used in a global fit. These global fits are discussed in Chapter 25, both in the context of the SM (Section 25.1) and allowing for physics beyond the SM (Section 25.2).

It is exactly some of the measurements described in Chapter 17 and further in Section 25.1 which were addressed in (Nobelprize.org, 2010) among experimental verifications of the Kobayashi-Maskawa mechanism in the scientific background to the 2008 Nobel Prize in Physics awarded to M. Kobayashi and T. Maskawa: "The respective collaborations BABAR and BELLE have now measured the *CP* violation in remarkable agreement with the model ... and all experimental data are now in impressive agreement with the model ...".

## 16.6 *CP* violation phenomenology for $B$ mesons

Since *CP* violation is due to irreducible phases of coupling constants, it becomes observable through interference effects. The simplest example is an amplitude consisting of two distinct contributions

$$A_f = \lambda_1 \langle f|O_1|B\rangle + \lambda_2 \langle f|O_2|B\rangle \tag{16.6.1}$$

where $\lambda_{1,2}$ are (complex) coupling constants (in our case combinations of CKM matrix elements) and $\langle f|O_{1,2}|B\rangle$ are matrix elements of interaction operators between the initial and final state.

The *CP* conjugate is the process $\overline{B} \to \overline{f}$, yielding

$$\overline{A}_{\overline{f}} = \lambda_1^* \langle \overline{f}|O_1^\dagger|\overline{B}\rangle + \lambda_2^* \langle \overline{f}|O_2^\dagger|\overline{B}\rangle. \tag{16.6.2}$$

The matrix elements of $\mathcal{O}_{1,2}^{(\dagger)}$ involve only strong interactions, which we assume to be *CP*-invariant. Hence we have

$$\langle \overline{f}|O_1^\dagger|\overline{B}\rangle = \langle f|O_1|B\rangle \quad \text{and} \quad \langle \overline{f}|O_2^\dagger|\overline{B}\rangle = \langle f|O_2|B\rangle. \tag{16.6.3}$$

Thus for the *CP* asymmetry we find

$$\begin{aligned} \mathcal{A}_{CP}(B \to f) &\equiv \frac{\Gamma(B \to f) - \Gamma(\overline{B} \to \overline{f})}{\Gamma(B \to f) + \Gamma(\overline{B} \to \overline{f})} \\ &\propto 2\,\mathrm{Im}[\lambda_1 \lambda_2^*]\,\mathrm{Im}[\langle f|O_1|B\rangle \langle f|O_2|B\rangle^*]. \end{aligned} \tag{16.6.4}$$

Consequently, in order to create *CP* violation, there has to be — aside from the "weak phase" due to the complex phases of the CKM matrix — also a "strong phase", *i.e.* a phase difference between the matrix elements $\langle f|O_1|B\rangle$ and $\langle f|O_2|B\rangle$. In the SM these two contributions correspond to different diagram topologies. In many cases, one can identify tree-level contributions which carry different CKM factors compared to loop (penguin) contributions. *CP* violation then emerges from the interference of "trees" and "penguins".

In the following we are going to consider decays into CP eigenstates $f$ in which case we have $f = \overline{f}$. For a quantum-coherent pair of neutral $B$-mesons (like the color-singlet $B^0\overline{B}^0$ pair from $\Upsilon(4S)$ decay) the time evolution generates a phase difference $\Delta m\, \Delta t$, which acts like the strong phase difference between the amplitudes for $B \to f$ and for $B \to \overline{B} \to f$. Hence we make use of the time-dependent *CP* asymmetry

$$\begin{aligned} \mathcal{A}_{\rm CP}^{B\to f}(\Delta t) &\equiv \frac{\Gamma(B^0(\Delta t) \to f) - \Gamma(\overline{B}^0(\Delta t) \to f)}{\Gamma(B^0(\Delta t) \to f) + \Gamma(\overline{B}^0(\Delta t) \to f)} \\ &= S^{B\to f} \sin(\Delta m_d\, \Delta t) - C^{B\to f} \cos(\Delta m_d\, \Delta t). \end{aligned} \tag{16.6.5}$$

The derivation (see the discussion in Chapter 10 leading to Eq. 10.2.8) neglects the small lifetime difference $\Delta\Gamma$ in the $B_d$ system; the expressions for $S$ and $C$ can be found in Eqs (10.2.4) and (10.2.5).

We may distinguish three different types of *CP* violation according to the various sources from which it emerges. *CP violation in decays*, sometimes referred to as *direct CP violation*, stems from different rates for a process and for its *CP* conjugate: hence we have $|\overline{A}_f/A_f| \neq 1$. This contribution leads to $C^{B\to f} \neq 0$: it is already present at $\Delta t = 0$, and remains in time-integrated measurements. *CP violation in the mixing* emerges in cases where we have $|p/q| \neq 1$.[49] One observable related to this is the semileptonic decay asymmetry $a_{\rm SL}$, which is the asymmetry between the decay rate of $B^0 \to X^-\ell^+\nu_\ell$ and the *CP* conjugate process. Finally, *mixing-induced CP violation*, sometimes also called *CP violation in interference between a decay without mixing and a decay with mixing* occurs for Im$\lambda \neq 0$, in which case interference of the amplitudes $B \to f$ and $B \to \overline{B} \to f$ leads to *CP* violation.[50]

[49] For a definition of the quantities $p$, $q$, and $\lambda$, we refer to Chapter 10, where time evolution is considered.

[50] In kaon physics sometimes the notion *indirect CP violation* is used for saying that the parameter $\epsilon$ is non-vanishing. Comparing this with the definitions given here, non-vanishing $\epsilon$ corresponds to a combination of $|q/p| \neq 1$ and $|\overline{A}_f/A_f| \neq 1$.

In the $B_d$ system we have to a very good approximation[51]

$$\frac{q}{p} = \exp(-2i\phi_1) \ . \tag{16.6.6}$$

This follows from Eq. (10.1.19) and by inspection of the box diagram contributing to the $B_d$ mixing (Fig. 10.1.1), from which it can be seen that the CKM matrix elements appearing in the amplitude yield $\phi_{M_{12}} = 2\phi_1$. Hence in all cases where $A = \overline{A}$, we find $|\lambda| = 1$ and $\text{Im}\lambda = -\sin(2\phi_1)$, leading to

$$C^{B\to f} = 0 \quad \text{and} \quad S^{B\to f} = -\sin(2\phi_1). \tag{16.6.7}$$

This holds for the golden mode $B \to J/\psi K_s$ where there is no relative weak phase between $A$ and $\overline{A}$. However, if there appears a relative weak phase in the decay amplitudes, then we may still have $|A| = |\overline{A}|$ and hence $|\lambda| = 1$, and thus no direct *CP* violation. For example, the tree amplitude in $B \to \pi\pi$ carries a weak phase $e^{-i\phi_3}$ which (neglecting penguin contributions) would lead to

$$\lambda = \exp(-2i(\phi_1 + \phi_3)) = \exp(+2i\phi_2). \tag{16.6.8}$$

However, the penguin contribution in $B \to \pi\pi$ cannot be neglected; in particular it leads to $|\lambda| \neq 1$ and to direct *CP* violation in these decays.

In general we have the "unitarity relation" between the quantities $S^{B\to f}$ and $C^{B\to f}$,

$$\left(C^{B\to f}\right)^2 + \left(S^{B\to f}\right)^2 = 1 - \left(D^{B\to f}\right)^2 \leq 1 \tag{16.6.9}$$

where

$$D^{B\to f} = \frac{2\,\text{Re}\lambda}{1 + |\lambda|^2}. \tag{16.6.10}$$

However, in the limit of vanishing lifetime difference the time-dependent *CP* asymmetry does not depend on $D^{B\to f}$, and hence a direct measurement of this quantity in the $B_d$ system is difficult.

[51] This relation depends on the phase conventions used. It holds in the convention used in (16.4.3).

# Chapter 17
# $B$ physics

The main objective of the $B$ Factories was to perform measurements of the decays and $CP$ asymmetries of $B$ mesons. While the asymmetric set-up and high luminosities of the $B$ Factories allowed us for the first time to perform statistically significant measurements of time-dependent $CP$ asymmetries, the symmetric predecessors of the $B$ Factories, DORIS and CESR, had already produced some data on $B$ decays. Experiments at LEP and the Tevatron had provided a proof of principle of the time-dependent $CP$ asymmetry measurement in the golden mode $B^0 \to J/\psi K^0_s$, and improved our knowledge of $B^0_d$ mixing.

Most of the time, the $B$ Factories took data near the $\Upsilon(4S)$ resonance, which decays almost exclusively into $B^0\overline{B}^0$ and $B^+B^-$ pairs. As a consequence, the overwhelming majority of $B$ Factory measurements relate to these states: these measurements are described in the following sections. However, some data has been taken at the $\Upsilon(5S)$ resonance, which also decays into $B_s^{(*)0}\overline{B}_s^{(*)0}$ pairs. Measurements of $B^0_s$ decays performed with this data are discussed in Chapter 23.

There are many ways to arrange this vast amount of material. The scheme adopted for this book uses the Unitarity Triangle as an organizing principle. We start from a discussion of the ways the sides of the triangle are constrained, including theoretical methods as well as experimental results in the corresponding sections. Hence we start with the measurements determining the magnitude of the CKM matrix elements $V_{cb}$, $V_{ub}$, $V_{ts}$, and $V_{td}$. This is followed by a discussion of the decay rates of charmed and charmless non-leptonic processes, including a comparison with theoretical expectations. The reason for this is that many charmed and charmless non-leptonic decay modes are used in the measurement of $CP$ asymmetries, and therefore should be discussed before moving on to review work related to the angles of the Unitarity Triangle.

Before treating the $CP$ asymmetries related to the angles of the Unitarity Triangle, we discuss measurements of $B$ lifetimes and $B^0-\overline{B}^0$ mixing, which are needed to understand the time-dependent analyses performed for the extraction of the angles. Searches for $CPT$ and other symmetry violations which are based on the lifetime- and mixing-measurement techniques are then presented. Following on from this one will find the description of measurements of $CP$ violation, *i.e.* the extraction of the angles $\phi_1$, $\phi_2$, and $\phi_3$.

The end of this chapter is devoted to special processes. These are either rare decays related to flavor changing neutral current transitions of the $b$ quark, processes involving $\tau$ leptons or baryons in the final state, or decays which are very rare or forbidden in the Standard Model.

## 17.1 $V_{ub}$ and $V_{cb}$

***Editors:***
*Vera Luth (BABAR)*
*Christoph Schwanda (Belle)*
*Paolo Gambino [$V_{cb}$]; Frank Tackmann [$V_{ub}$] (theory)*

**Additional section writers:**
*Christine Davies, Jochen Dingfelder, Alexander Khodjamirian, Andreas Kronfeld, Matthias Steinhauser, and Ruth Van de Water*

### 17.1.1 Overview of semileptonic $B$ decays

#### 17.1.1.1 Motivation

Semileptonic decays of $B^+$ and $B^0$ mesons proceed via leading-order weak interactions. In the following, only decays involving low-mass charged leptons, $\ell = e^\pm$ or $\mu^\pm$, are considered. They are expected to be free of non-Standard Model contributions, and therefore play a critical role in the determination of the magnitudes of the CKM-matrix elements $V_{cb}$ and $V_{ub}$. $|V_{cb}|$ normalizes the Unitarity Triangle, and the ratio $|V_{ub}|/|V_{cb}|$ determines the side opposite to the angle $\phi_1$. Thus, their values impact most studies of flavor physics and *CP*-violation in the quark sector. Leptonic and semileptonic decays involving $\tau^\pm$ leptons are sensitive to couplings to the charged Higgs boson and are discussed in Section 17.10.

There are two methods to determine $|V_{cb}|$ and $|V_{ub}|$, one based on the study of exclusive semileptonic $B$ decays where the hadron in the final state is a $D$, $D^*$, $D^{**}$, $\pi$ or $\rho$ meson, the other based on the study of inclusive decays of the form $B \to X\ell^+\nu_\ell$, where $X$ refers to either $X_c$ or $X_u$, *i.e.*, to any hadronic final state with charm or without charm, respectively.

To extract $|V_{cb}|$ or $|V_{ub}|$ from the measured partial decay rates, both inclusive and exclusive determinations rely on theoretical descriptions of the QCD contributions to the underlying weak decay process. Since both methods rely on different experimental techniques and involve different theoretical approximations, they complement each other and provide largely independent determinations (of comparable accuracy) of $|V_{cb}|$ and $|V_{ub}|$. This in turn provides a crucial cross check of the methods and our understanding of semileptonic $B$ decays in general.

#### 17.1.1.2 Theoretical Overview

Semileptonic decays of $B$ mesons, $B \to X\ell\nu$, proceed through the electroweak transitions $b \to c\ell\nu$ and $b \to u\ell\nu$, as illustrated in Figure 17.1.1. These are governed by the CKM-matrix elements $V_{cb}$ and $V_{ub}$, and since the intermediate $W$ boson decays leptonically, do not involve any other CKM matrix elements. Hence, measurements of the $B \to X\ell\nu$ decay rate can be used to directly measure $|V_{cb}|$ and $|V_{ub}|$.

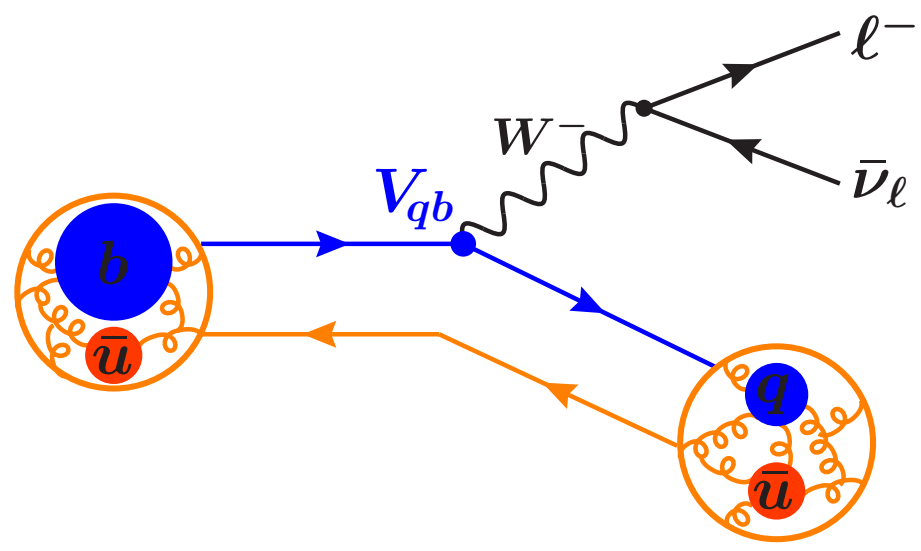

**Figure 17.1.1.** Illustration of semileptonic decay $B^- \to X\ell^-\overline{\nu}_\ell$.

The theoretical description of semileptonic $B$ decays starts from the electroweak effective Hamiltonian,

$$\mathcal{H}_{\rm eff} = \frac{4G_F}{\sqrt{2}} \sum_{q=u,c} V_{qb}\, (\overline{q}\gamma_\mu P_L b)(\ell\gamma^\mu P_L \nu_\ell)\,, \qquad (17.1.1)$$

where $P_L = (1-\gamma_5)/2$, and $G_F$ is the Fermi constant as extracted from muon decay. The $W$ boson has been integrated out at tree level using the hierarchy $m_b \ll m_W$, and higher-order electroweak corrections are suppressed by additional powers of $G_F$ and are thus very small. The differential $B$ decay rates take the form

$$d\Gamma \propto G_F^2 |V_{qb}|^2 \left| L^\mu \langle X|\overline{q}\gamma_\mu P_L b|B\rangle \right|^2 . \qquad (17.1.2)$$

An important feature of semileptonic decays is that the leptonic part in the effective Hamiltonian and the decay matrix element factorizes from the hadronic part, and that QCD corrections can only occur in the $b \to q$ current. The latter do not affect Eq. (17.1.1) and are fully contained in the hadronic matrix element $\langle X|\overline{q}\gamma_\mu P_L b|B\rangle$ in Eq. (17.1.2). This factorization is violated by small electromagnetic corrections, for example by photon exchange between the quarks and leptons, which must be taken into account in situations where high precision is required.

The challenge in the extraction of $|V_{cb}|$ and $|V_{ub}|$ is the determination of the hadronic matrix element of the quark current in Eq. (17.1.2). For this purpose, different theoretical methods have been developed, depending on the specific decay mode under consideration. In almost all cases, the large mass of the $b$-quark, $m_b \sim 5\,\text{GeV}$ is exploited.

In exclusive semileptonic decays, one considers the decay of the $B$ meson into a specific final state $X = D, D^*, \pi$, or $\rho$. In this case, one parameterizes the hadronic matrix element in terms of form factors, which are nonperturbative functions of the momentum transfer $q^2$. This is discussed in Sections 17.1.2 and 17.1.4. The two methods commonly used to determine the form factors are lattice QCD (LQCD) and light-cone sum rules (LCSR). In LQCD, the QCD functional integrals for the matrix elements are computed numerically from first principles. Heavy quark effective theory (HQET), and non-relativistic QCD (NRQCD), were first introduced, at least in part, to enable lattice-QCD calculations with heavy quarks. Even when these formalisms are not explicitly used, heavy-quark dynamics are usually used to control discretization effects. An exception are the most recent determinations of $m_c$ and $m_b$ from lattice QCD, discussed below, which use a fine lattice in combination with a highly improved lattice action such that heavy quarks with masses almost up to $m_b$ can be treated with a light-quark formalism. A complementary method is based on LCSR which use hadronic dispersion relations to approximate the form factor in terms of quark-current correlators and can be calculated in an operator product expansion (OPE).

In inclusive semileptonic decays, one considers the sum over all possible final states $X$ that are kinematically allowed. Employing parton-hadron duality one can replace the sum over hadronic final states with a sum over partonic final states. This eliminates any long-distance sensitivity to the final state, while the short-distance QCD corrections, which appear at the typical scale $\mu \sim m_b$ of the decay, can be computed in perturbation theory in terms of the strong coupling constant $\alpha_s(m_b) \sim 0.2$. The remaining long-distance corrections related to the initial $B$ meson can be expanded in powers of $\Lambda_{\rm QCD}/m_b \sim 0.1$, where $\Lambda_{\rm QCD}$ is the hadronic scale of order $m_B - m_b \sim 0.5\,{\rm GeV}$. This is called *heavy quark expansion* (HQE), and it systematically expresses the decay rate in terms of non-perturbative parameters that describe universal properties of the $B$ meson. This is discussed in Sections 17.1.3 and 17.1.5.

#### 17.1.1.3 Experimental Techniques

As in other analyses of $B\overline{B}$ data recorded at $B$ Factories, the two dominant sources of background for the reconstruction of semileptonic $B$ decays are the combinatorial $B\overline{B}$ and the continuum backgrounds (see Chapter 9).

The suppression of the continuum processes, $e^+e^- \to \ell^+\ell^-(\gamma)$ with $\ell = e, \mu$, or $\tau$, and quark-antiquark pair production, $e^+e^- \to q\overline{q}(\gamma)$ with $q = u, d, s, c$, is achieved by requiring at least four charged particles in the event and by imposing restrictions on several event shape variables, either sequentially on individual variables or by constructing multivariable discriminants. Among these variables are thrust, the maximum sum of the longitudinal momenta of all particles relative to a chosen axis, $\Delta\theta_{\rm thrust}$, the angle between the thrust axis of all particles associated with the signal decay and the thrust axis of the rest of the event, $R_2$, the ratio of the second to the zeroth Fox-Wolfram moments, and $L_0$ and $L_2$, the normalized angular moments.

The separation of semileptonic $B$ decays from $B\overline{B}$ backgrounds is very challenging because they result in one or more undetected neutrinos. The energy and momentum of the missing particles can be inferred from the sum of all other particles in the event,

$$(E_{\rm miss}, \mathbf{p}_{\rm miss}) = (E_0, \mathbf{p}_0) - \left(\sum_i E_i, \sum_i \mathbf{p}_i\right), \quad (17.1.3)$$

where $(E_0, \mathbf{p}_0)$ is the four-vector of the colliding beams. If the only undetected particle in the event is a single neutrino, the missing mass should be close to zero and the missing momentum should be non-zero. Figure 17.1.2 shows examples of missing mass squared distributions, $m^2_{\rm miss} = E^2_{\rm miss} - |\mathbf{p}_{\rm miss}|^2$, for selected $B^- \to X_c\ell^-\overline{\nu}$ candidates. There are narrow peaks at zero for correctly reconstructed decays and in most cases rather small backgrounds from other decays modes. In Figure 17.1.2a there is a broad enhancement above the peak due to $B^- \to D^{*0}\ell^-\overline{\nu}$ decays, in which the low energy pion or photon from the decay $D^{*0} \to D^0\pi^0$ or $D^{*0} \to D^0\gamma$ escaped detection. To reduce the impact of the dependence of the $m^2_{\rm miss}$ resolution on the neutrino energy, the variable $E_{\rm miss} - p_{\rm miss} = m^2_{\rm miss}/(E_{\rm miss} + p_{\rm miss})$ is often preferred.

A variable first introduced by the CLEO Collaboration (Bartelt et al., 1999) to select exclusive semileptonic decays $B \to D\ell\nu$ is

$$\cos\theta_{BY} = (2E_BE_Y - m^2_B - m^2_Y)/2|\mathbf{p}_B||\mathbf{p}_Y|, \quad (17.1.4)$$

where $m_Y$ and $|\mathbf{p}_Y|$ refer to the invariant mass and momentum sum of the hadron $X$ and the charged lepton $\ell$. If the only missing particle is the neutrino, $\theta_{BY}$ corresponds to the angle between the momentum vectors $\mathbf{p}_B$ and $\mathbf{p}_Y = \mathbf{p}_X + \mathbf{p}_\ell$, and the condition $|\cos\theta_{BY}| \leq 1.0$ should be fulfilled, while for background events or incompletely reconstructed semileptonic decays the distribution extends to values well beyond this range, thus enabling a separation from the signal decays.

For the isolation of the exclusive signal decay the kinematic variables

$$\Delta E = E^*_B - E^*_{\rm beam} \ \text{ and } \ m_{\rm ES} = \sqrt{E^{*2}_{\rm beam} - \mathbf{p}^{*2}_B} \quad (17.1.5)$$

are used. A comparison of $\Delta E$ and $m_{\rm ES}$ distributions for selected samples of hadronic and semileptonic $B$ decays is given in Figure 17.1.3. $\Delta E$ is centered on zero and the $m_{\rm ES}$ distribution peaks at the $B$-meson mass. For hadronic decays the $\Delta E$ resolution is dominated by the detector resolution. The resolution in $m_{\rm ES}$ is determined by the spread in the energy of the colliding beams, typically less than 3 MeV. For semileptonic decays both variables are affected by the measurement of the neutrino momentum and energy. The size of the continuum and combinatorial $B\overline{B}$ background depends on the decay mode and the overall event selection. Backgrounds with kinematics very similar to the signal $B$ decays may contribute to the peak region defined as $|\Delta E| < 0.125\,{\rm GeV}, m_{\rm ES} > 5.27\,{\rm GeV}$.

There are several variables that are commonly used to describe the kinematics of semileptonic decays, both for exclusive and inclusive decays: the momentum transfer squared $q^2$, the momentum of the charged lepton $\mathrm{p}_\ell$, and the hadronic mass $m_X$. The last two are of particular importance for analyses of inclusive decays, summing over all possible hadronic states $X$. They are used to separate charmless decays from the dominant decays to charm hadrons.

There are two ways to define and measure $q^2$, either as the invariant mass squared of the four-vector sum of the reconstructed lepton and neutrino, or as the momentum transfer squared from the $B$ meson to the final state hadron $X$, $q^2 = (p_\ell + p_{\rm miss})^2 = (p_B - p_{\rm X})^2$. In the first case, the resolution in $q^2$ is dominated by the measurement of the missing energy which tends to have a poorer resolution than the measured missing momentum, because the missing momentum is a vector sum and contributions from particle losses (or additional tracks and EMC showers) do not add linearly as is the case for $E_{\rm miss}$. Thus it is advantageous to replace $E_{\rm miss}$ by $|\mathbf{p}|_{\rm miss}$, the absolute value of the measured missing momentum, $q^2 = [(E_\ell, \boldsymbol{p}_\ell) + (p_{\rm miss}, \boldsymbol{p}_{\rm miss})]^2$.

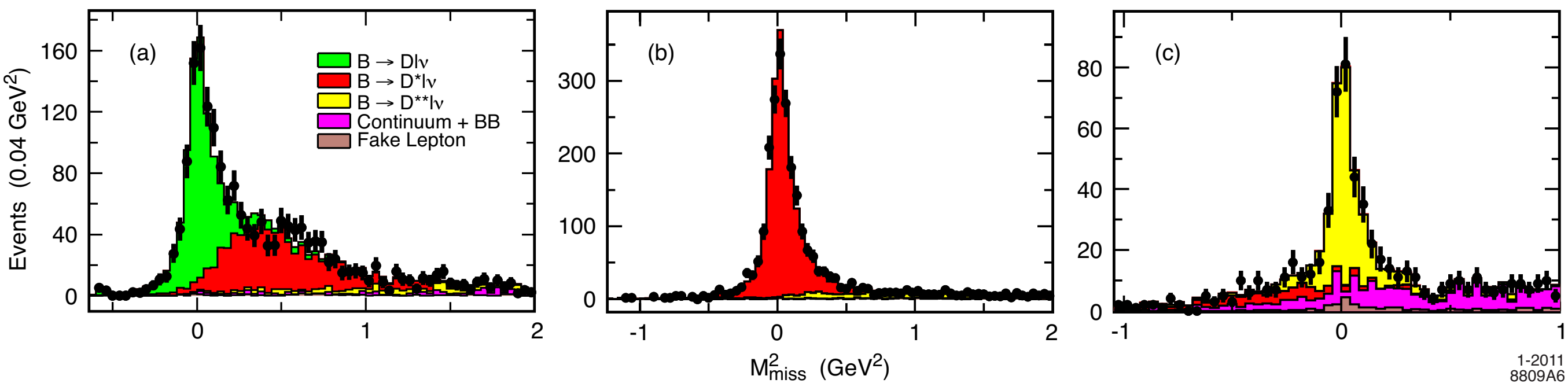

**Figure 17.1.2.** Distributions of the missing mass squared for exclusive $B \to X_c \ell \nu$ candidates in $B\overline{B}$ events tagged by a hadronic decay of the second $B$ meson (Aubert, 2008b), a) $B^- \to D^0 \ell^- \overline{\nu}_\ell$, b) $\overline{B}^0 \to D^{*+} \ell^- \overline{\nu}_\ell$, and c) $B^- \to D^{*+} \pi^- \ell^- \overline{\nu}_\ell$. The contributions from various exclusive decay modes are marked by color shading.

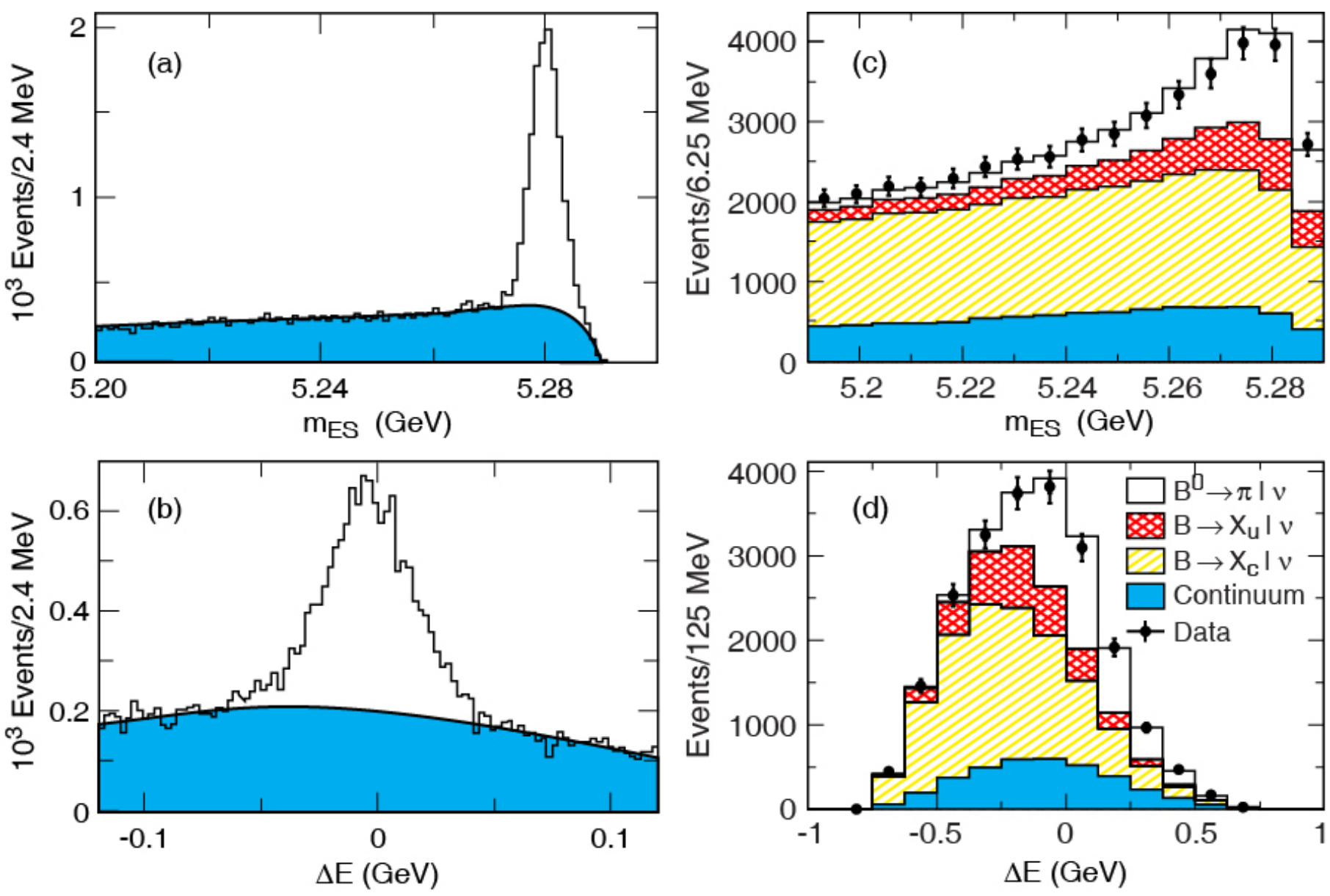

**Figure 17.1.3.** Distributions of $m_{\rm ES}$ and $\Delta E$ for (a, b) hadronic $B$ decays above combinatorial continuum and $B\overline{B}$ background (blue) (Mazur, 2007), and (c, d) selected $B^0 \to \pi^- \ell^+ \nu$ decays (Ha, 2011) in the $q^2$ range of $0 - 16\,{\rm GeV}^2$, above a variety of backgrounds contributions, specifically $B \to X_u \ell \nu$ (red), various $B \to X_c \ell \nu$ decays (yellow), and continuum background (blue). For both samples, the distributions are restricted to events in the signal bands, *i.e.*, $m_{\rm ES}$ is shown for events in the peak region for $\Delta E$, $|\Delta E| < 0.125\,{\rm GeV}$, and $\Delta E$ is restricted to events in the peak region for $m_{\rm ES}$, $m_{\rm ES} > 5.27\,{\rm GeV}$.

In the second case, the $q^2$ measurement is not affected by the measurement of the missing momentum, but the direction of the $B$ meson momentum is not known. Therefore the $B$ momentum vector is estimated as the average over four or more possible directions of the $B$ meson. The two methods have different sensitivity to combinatorial background: the first has the best resolution at high $q^2$, whereas the second method shows the best resolution at low $q^2$ where the hadron background is smaller. The width of the core resolution is in the range $(0.18 - 0.34)\,{\rm GeV}^2$, and the tails can be approximated by a second Gaussian function with widths in the range $(0.6 - 0.8)\,{\rm GeV}^2$. Figure 17.1.4 shows the resolution for selected $B \to \pi \ell \nu$ candidates. The $q^2$ resolution is important for many analyses of semileptonic decays.

With increasing data samples, more recent analyses have employed $B\overline{B}$ tagging techniques to substantially reduce continuum and combinatorial $B\overline{B}$ backgrounds. The detection of the decay of one of the $B$ mesons produced at the $\Upsilon(4S)$ not only identifies the second $B$ decay, but it uniquely determines its momentum, mass, charge and flavor. Furthermore, the kinematics of the final state are constrained such that an undetectable neutrino from the second decay can be identified from the missing momentum and missing energy of the rest of the event.

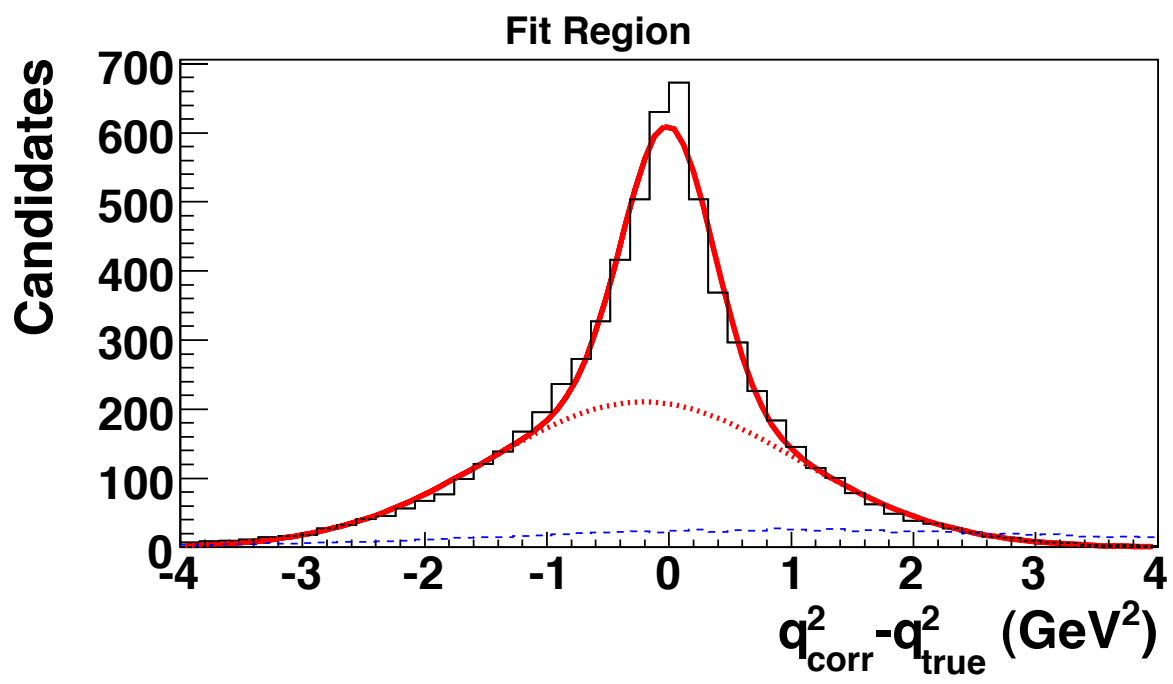

**Figure 17.1.4.** $q^2$ resolution for selected $B^0 \to \pi^- \ell^+ \nu$ decays (del Amo Sanchez, 2011n) for true signal (black, solid histogram) and combinatorial signal (blue, dashed histogram) as obtained from simulation. The result of the fit to the signal with the sum of two Gaussian functions is shown (solid and dotted lines).

The cleanest samples of $B\overline{B}$ events are obtained with hadronic tags. Tag efficiencies and purities vary considerably, depending on the number of charged and neutral particles in the tag decay and the associated signal decay. Given the low branching fractions for individual hadronic decays and their high final-state particle multiplicity, the average achievable tagging efficiency is typically 0.3% for purities of $\simeq 0.5$. Recently, tag efficiencies have been increased as much as a factor of three by the addition of other hadronic decay modes, and by simultaneous constraints on the semileptonic signal decay in a given event, and by effectively selecting the best of several candidates per event (see Chapter 7).

Tag efficiencies in the range of $1-3\%$ can be obtained using semileptonic $B$ decays. As for hadronic tags, the achievable tag efficiencies and purities are strongly dependent on both the tag decay and the decay of the signal $B$ recoiling against the tag. In comparison with fully reconstructed hadronic tags, events tagged by semileptonic decays provide looser kinematic constraints on the recoiling $B$ and result in a less accurate measurement of the missing neutrino and higher combinatorial backgrounds.

### 17.1.2 Exclusive decays $B \to D^{(*)}\ell\nu$

#### 17.1.2.1 Theoretical Overview

In the following, we discuss exclusive $B$ decays to a $D$ or $D^*$ meson. The transition matrix elements of the weak current given in Eq. (17.1.2) are decomposed into Lorentz-covariant forms, built from the independent four-vectors of the decay and form factors. For a pseudoscalar final state, only the vector current contributes,

$$\langle P|\overline{q}\gamma^\mu b|\overline{B}\rangle = f_+(q^2)\left(p_B^\mu + p_P^\mu - \frac{m_B^2 - m_P^2}{q^2}q^\mu\right) + f_0(q^2)\,\frac{m_B^2 - m_P^2}{q^2}q^\mu, \quad (17.1.6)$$

where $p_B$ and $p_P$ denote the four-vector momenta of the mesons, $q = p_B - p_P$ is the momentum transfer, and $f_{+,0}(q^2)$ are two form factors. For a vector final state, both the vector and axial currents contribute:

$$\langle V|\overline{q}\gamma^\mu b|\overline{B}\rangle = V(q^2)\,\varepsilon^{\mu\sigma}{}_{\nu\rho}\epsilon^*_\sigma \frac{2p_B^\nu p_V^\rho}{m_B + m_V}, \quad (17.1.7)$$

$$\langle V|\overline{q}\gamma^\mu\gamma^5 b|\overline{B}\rangle = i\epsilon^*_\nu\left[A_0(q^2)\,\frac{2m_V q^\mu q^\nu}{q^2} + A_1(q^2)\,(m_B + m_V)\eta^{\mu\nu} - A_2(q^2)\,\frac{(p_B + p_V)_\sigma q^\nu}{m_B + m_V}\eta^{\mu\sigma}\right], \quad (17.1.8)$$

where $\epsilon_\nu$ is the polarization vector of the vector meson, $\eta^{\mu\nu} = g^{\mu\nu} - q^\mu q^\nu/q^2$, $\varepsilon^{\alpha\beta\gamma\delta}$ is the Levi-Civita tensor, and $V(q^2)$ and $A_i(q^2)$ are form factors. These form-factor decompositions are general: to determine $|V_{cb}|$, $\overline{q} = \overline{c}$, $P = D$, and $V = D^*$; to determine $|V_{ub}|$, $\overline{q} = \overline{u}$, $P = \pi$, and $V = \rho$.

The key feature of $B \to D^{(*)}$ decays is that the masses of both the charm and bottom quarks are large compared to the energy scale of non-perturbative QCD. Therefore, in both cases the heavy quark is nearly static, surrounded by a cloud of gluons, the light valence quark, and virtual quark-antiquark pairs. In particular, the effects of spin and flavor are suppressed by powers of $\Lambda_{\rm QCD}/m_Q$ ($Q = c, b$). In turn, approximate heavy-quark symmetries impose constraints on the form factors. These constraints become more transparent with a different basis of form factors,

$$\frac{\langle D|\overline{c}\gamma^\mu b|B\rangle}{\sqrt{m_B m_D}} = h_+(w)\,(v_B + v_D)^\mu + h_-(w)\,(v_B - v_D)^\mu, \quad (17.1.9)$$

$$\frac{\langle D^*|\overline{c}\gamma^\mu b|B\rangle}{\sqrt{m_B m_{D^*}}} = h_V(w)\,\varepsilon^{\mu\nu\rho\sigma} v_{B,\nu} v_{D^*,\rho}\epsilon^*_\sigma, \quad (17.1.10)$$

$$\frac{\langle D^*|\overline{c}\gamma^\mu\gamma^5 b|B\rangle}{\sqrt{m_B m_{D^*}}} = ih_{A_1}(w)\,(1+w)\epsilon^{*\mu} - i\left[h_{A_2}(w)v_B^\mu + h_{A_3}(w)v_{D^*}^\mu\right]\epsilon^*\cdot v_B, \quad (17.1.11)$$

where the velocities (for hadrons $H = B, D, D^*$) are $v_H = p_H/m_H$, the velocity transfer is $w = v_B \cdot v_{D^{(*)}} = (m_B^2 + m_{D^{(*)}}^2 - q^2)/2m_B m_{D^{(*)}}$. Again, these decompositions are completely general.

At zero recoil $w = 1$, heavy-quark dynamics requires (Isgur and Wise, 1989, 1990b; Shifman and Voloshin, 1987)

$$h_+(1) = 1 + {\rm O}(\alpha_s) + {\rm O}\left((\Lambda_{\rm QCD}/m_q)^2\right), \quad (17.1.12)$$

$$h_-(1) = 0 + {\rm O}(\alpha_s) + {\rm O}(\Lambda_{\rm QCD}/m_q), \quad (17.1.13)$$

$$h_{A_1}(1) = 1 + {\rm O}(\alpha_s) + {\rm O}\left((\Lambda_{\rm QCD}/m_q)^2\right). \quad (17.1.14)$$

The other zero-recoil form factors are not crucial to the extraction of $|V_{cb}|$. The task is to compute the corrections to heavy-quark symmetry; this is usually done in a way that aims to have the error scale with the deviation from the symmetry limit. The long-distance corrections of order $(\Lambda_{\text{QCD}}/m_q)^n$ must be obtained non-perturbatively; the short-distance corrections of order $\alpha_s^l$ may be obtained perturbatively or non-perturbatively. It is, however, important to ensure that the separation of long- and short-distance effects is done in a consistent way. At nonzero recoil, all form factors receive contributions at first order in $\Lambda_{\text{QCD}}/m_q$. Calculations of the form factors dependence on $w$ require more effort.

The differential decay rates for $B^- \to D^{0(*)}\ell^-\overline{\nu}$ are

$$\begin{aligned}\frac{d\Gamma_{B^-\to D^0\ell^-\overline{\nu}}}{dw} &= \frac{G_F^2 m_D^3}{48\pi^3}(m_B + m_D)^2(w^2-1)^{3/2} \\ &\times |\eta_{\text{EW}}|^2|V_{cb}|^2|\mathcal{G}(w)|^2, \qquad (17.1.15)\\ \frac{d\Gamma_{B^-\to D^{0*}\ell^-\overline{\nu}}}{dw} &= \frac{G_F^2 m_{D^*}^3}{4\pi^3}(m_B - m_{D^*})^2(w^2-1)^{1/2} \\ &\times |\eta_{\text{EW}}|^2|V_{cb}|^2\chi(w)|\mathcal{F}(w)|^2, \qquad (17.1.16)\end{aligned}$$

where $\eta_{\text{EW}} = 1.0066$ is the one-loop electroweak correction (Sirlin, 1982) defined relative to $G_F$ as extracted from muon decay.[52] The form factor $\mathcal{G}(w)$ is a function of $h_+(w)$ and $h_-(w)$ and in $\chi(w)|\mathcal{F}(w)|^2$, $\mathcal{F}(w)$ contains all four $B \to D^*$ form factors that enter Eqs (17.1.10) and (17.1.11). The full expressions can be found in Section 5.2 of Antonelli et al. (2010a). At zero recoil, $\mathcal{G}(1) = 1$ and $\mathcal{F}(1) = h_{A_1}(1)$. For decays of neutral mesons, $\overline{B}^0 \to D^{+(*)}\ell^-\overline{\nu}$, Coulomb attraction in the final state leads to an additional factor $1 + \alpha\pi$ (Atwood and Marciano, 1990; Ginsberg, 1968) on the right-hand sides of Eqs (17.1.15) and (17.1.16).

For the determination of $|V_{cb}|$, the decay $B \to D^*\ell\nu$ is preferred over $B \to D\ell\nu$ for three reasons: First, theoretical predictions are simplest at zero recoil, where the rates are phase-space suppressed, but less so for the $D^*$ final state $[(w^2-1)^{1/2}$ versus $(w^2-1)^{3/2}]$. Second, at zero recoil, the form factor $\mathcal{G}(1)$ receives corrections of order $\Lambda_{\text{QCD}}/m_Q$, instead of $(\Lambda_{\text{QCD}}/m_Q)^2$ for $\mathcal{F}(1)$. On the other hand, for $B \to D\ell\nu$ only the vector current contributes, resulting in a single form factor $\mathcal{G}(1)$, for low-mass leptons. Finally, and less crucially, the three polarization states of $D^*$ increase the rate.

For these reasons let us first consider $\mathcal{F}(1) = h_{A_1}(1)$. One can show that the optical theorem and the OPE imply

$$|h_{A_1}(1)|^2 + \frac{1}{2\pi}\int_0 d\epsilon\, w(\epsilon) = 1 - \Delta_{1/m_Q^2} - \Delta_{1/m_Q^3}, \quad (17.1.17)$$

where $\epsilon = E - m_{D^*}$ is the excess energy of charmed states with $J^{PC} = 1^{-+}$, $w(\epsilon)$ is a structure function, and the upper limit of integration may be considered large for the moment. The contributions $\Delta_{1/m^n}$ describe corrections to the axial vector current for finite-mass quarks. The $\Delta_{1/m_Q^2}$ contributions can be conveniently written as

$$\Delta_{1/m_Q^2} = \frac{\mu_G^2}{3m_c^2} + \frac{\mu_\pi^2(\mu) - \mu_G^2}{4}\left(\frac{1}{m_c^2} + \frac{2/3}{m_c m_b} + \frac{1}{m_b^2}\right), \quad (17.1.18)$$

where $\mu_G^2 \simeq 3(m_{B^*}^2 - m_B^2)/4$ and $\mu_\pi^2(\mu)$ are matrix elements of the chromomagnetic energy and kinetic energy of the $b$ quark in the $B$ meson. The meaning of the scale $\mu$ in $\mu_\pi^2(\mu)$ is explained below. The $1/m_Q^3$ contributions have a similar expression (see, *e.g.*, Gambino, Mannel, and Uraltsev (2010)) with analogs of $\mu_G^2$ and $\mu_\pi^2$ that are related to moments of the inclusive semileptonic distribution.

For $\epsilon \gg \Lambda_{\text{QCD}}$, the hadronic states in the excitation integral are dual to quark-gluon states. Introducing a scale $\mu$ to separate this short-distance part from the long-distance part (which must be treated non-perturbatively), one writes

$$\frac{1}{2\pi}\int_0 d\epsilon\, w(\epsilon) = \frac{1}{2\pi}\int_0^\mu d\epsilon\, w(\epsilon) + [1 - \eta_A(\mu)^2]. \quad (17.1.19)$$

Here the quantity $\eta_A(\mu)$ combines the short-distance ($\epsilon > \mu$) contributions. It has been calculated to two loops in perturbation theory (Czarnecki, Melnikov, and Uraltsev, 1998); its $\mu$ dependence is compensated by $\mu_\pi^2(\mu)$. Rearranging Eq. (17.1.17) results in

$$h_{A_1}(1) \simeq \eta_A(\mu) - \frac{1}{2}\Delta_{1/m_Q^2} - \frac{1}{2}\Delta_{1/m_Q^3} - \frac{1}{4\pi}\int_0^\mu d\epsilon\, w(\epsilon). \quad (17.1.20)$$

The last term from higher hadronic excitations is not directly constrained by data.

Using recent data to compute $\Delta_{1/m_Q^2} + \Delta_{1/m_Q^3}$, Gambino, Mannel, and Uraltsev (2010) find

$$\Delta_{1/m_Q^2} + \Delta_{1/m_Q^3} = 0.11 \pm 0.03 \quad (17.1.21)$$

in the kinetic scheme with $\mu = 0.75\,\text{GeV}$. Combining this with the two-loop result of $\eta_A(0.75\,\text{GeV}) = 0.985 \pm 0.010$, Eq. (17.1.20) implies

$$\mathcal{F}(1) < 0.93, \quad (17.1.22)$$

since the excitation integral is positive. They further estimate the excitation contribution to be (in the notation used here)

$$\frac{1}{4\pi}\int_0^{0.75\ \text{GeV}} d\epsilon\, w(\epsilon) \approx 0.065, \quad (17.1.23)$$

leading to

$$\mathcal{F}(1) = 0.86 \pm 0.02. \quad (17.1.24)$$

One should note, however, that the estimate in Eq. (17.1.23) entails the application of the OPE at scales of 1 GeV or lower, and consequently the error is difficult to assess.

[52] This is just the QED running of the semileptonic form fermion operator from the $W$ mass to the $m_b$ scale. The leading bremsstrahlung part of the QED corrections is subtracted by experiments using approximate methods. Structure dependent corrections are still poorly understood (Becirevic and Kosnik, 2010; Bernlochner and Schonherr, 2010), but are unlikely to give non-negligible corrections.

With lattice QCD, the QCD action is discretized on an Euclidean space-time lattice, and calculations are performed numerically using Monte Carlo methods and importance sampling (see, *e.g.*, Bazavov et al., 2010; DeGrand and Detar, 2006; Hashimoto and Onogi, 2004; Kronfeld, 2002). Physical results are recovered in the limit of zero lattice spacing. Since lattice results are obtained from first principles in QCD, they can be improved to arbitrary precision, given sufficient computing resources. In recent years, LQCD has made substantial progress, particularly in flavor physics. The most computationally demanding part of QCD calculations, namely the treatment of the sea of virtual quark-antiquark pairs, has become feasible.

The Fermilab-MILC calculations (Bernard et al., 2009a) are based on $2+1$ flavors of sea quarks, two corresponding to up and down quarks (Bazavov et al., 2010) and one for the strange sea. The former two have masses larger than in nature, $m_l > 0.1 m_s$, but calculations at a sequence of light-quark masses are guided to the physical limit with chiral perturbation theory (Laiho and Van de Water, 2006).The uncertainties of these calculations can be reliably estimated, because a HQET analysis of the form factor follows through on the lattice (Harada, Hashimoto, Kronfeld, and Onogi, 2002; Kronfeld, 2000), and in this way, the error scales as $1-\mathcal{F}(1)$, rather than as $\mathcal{F}(1)$.

The current value (Bailey et al., 2010),

$$\eta_{\rm EW}\mathcal{F}(1) = 0.9077(51)(88)(84)(90)(30)(33), \quad (17.1.25)$$

includes the electroweak correction $\eta_{\rm EW} = 1.0066$. The stated uncertainties stem, respectively, from Monte-Carlo statistics, the $D^* \to D\pi$ coupling, the chiral extrapolation, discretization errors, perturbative matching, and tuning the bare quark masses. This result is an update of earlier calculations by Bernard et al. (2009a) that were based on a smaller set of LQCD data. Adding the errors in quadrature, we obtain

$$\eta_{\rm EW}\mathcal{F}(1) = 0.908 \pm 0.017, \quad (17.1.26)$$

which agrees well with the bound in Eq. (17.1.22).

The difference between the value in Eq. (17.1.24) and the LQCD result of $\mathcal{F}(1) = 0.902 \pm 0.017$ (without $\eta_{\rm EW}$), though not large, might be due to a breakdown of the OPE in the estimate the low-energy excitation integral, although this appears to be unlikely in view of our present understanding of heavy quark physics. LQCD form-factor calculations have passed several very challenging tests, including predictions of the shapes of $D \to \pi\ell\nu$ and $D \to K\ell\nu$ form factors (Aubin et al., 2005; Bernard et al., 2009b) and agreement to high precision for the normalization of these form factors with experiment (Na, Davies, Follana, Lepage, and Shigemitsu, 2010; Na et al., 2011).

Let us now turn, more briefly, to $B \to D\ell\nu$ and $\mathcal{G}(1)$. Unquenched LQCD calculations (Okamoto et al., 2005) result in

$$\mathcal{G}(1) = 1.074 \pm 0.024 \ , \quad (17.1.27)$$

and are compatible with the HQE calculation (Uraltsev, 2004),

$$\mathcal{G}(1) = 1.04 \pm 0.02 \ , \quad (17.1.28)$$

within the stated uncertainties.

#### 17.1.2.2 Measurements of Branching Fractions and Differential Distributions

The decay $B \to D^*\ell\nu$ was measured at Belle (Dungel, 2010) and *BABAR* (Aubert, 2008h,v, 2009ab) assuming the HQET parameterization of the form factor $\eta_{\rm EW}\mathcal{F}(w)$ given by (Caprini, Lellouch, and Neubert, 1998) in terms of the four quantities: the normalization $\eta_{\rm EW}\mathcal{F}(1)|V_{cb}|$, the slope $\rho^2_{D^*}$, and the form-factor ratios $R_1(1) = R^{*2}V(1)/A_1(1)$ and $R_2(1) = R^{*2}A_2(1)/A_1(1)$, where

$$R^* = (2\sqrt{m_B m_{D^*}})/(m_B + m_{D^*}). \quad (17.1.29)$$

In some analyses (Aubert, 2008v, 2009ab) the partial width $d\Gamma/dw$ was measured as a function of the velocity transfer $w = v_B \cdot v_{D^*}$ to determine the normalization $\eta_{\rm EW}\mathcal{F}(1)|V_{cb}|$ and the slope $\rho^2_{D^*}$, with form-factor ratios $R_1(1)$ and $R_2(1)$ taken as input from other measurements. In the analyses by Dungel (2010) and Aubert (2008h) the differential decay rate of $B \to D^*\ell\nu$ with $D^* \to D\pi$ is measured as a function of four variables, $w$ and the angles $\theta_\ell$, $\theta_V$ and $\chi$ (Figure 17.1.5), where

- $\theta_\ell$ is the angle between the direction of the lepton and the direction opposite the $B$ meson in the rest frame of the virtual $W$,
- $\theta_V$ is the angle between the direction of the $D$ meson and the direction opposite the $B$ meson in the $D^*$ rest frame, and
- $\chi$ is the angle between the decay planes of the $D^*$ and the $W$, defined in the $B$ meson rest frame.

The differential rate in terms of these four kinematic variables gives access to all four HQET parameters of the $B \to D^*\ell\nu$ decay.

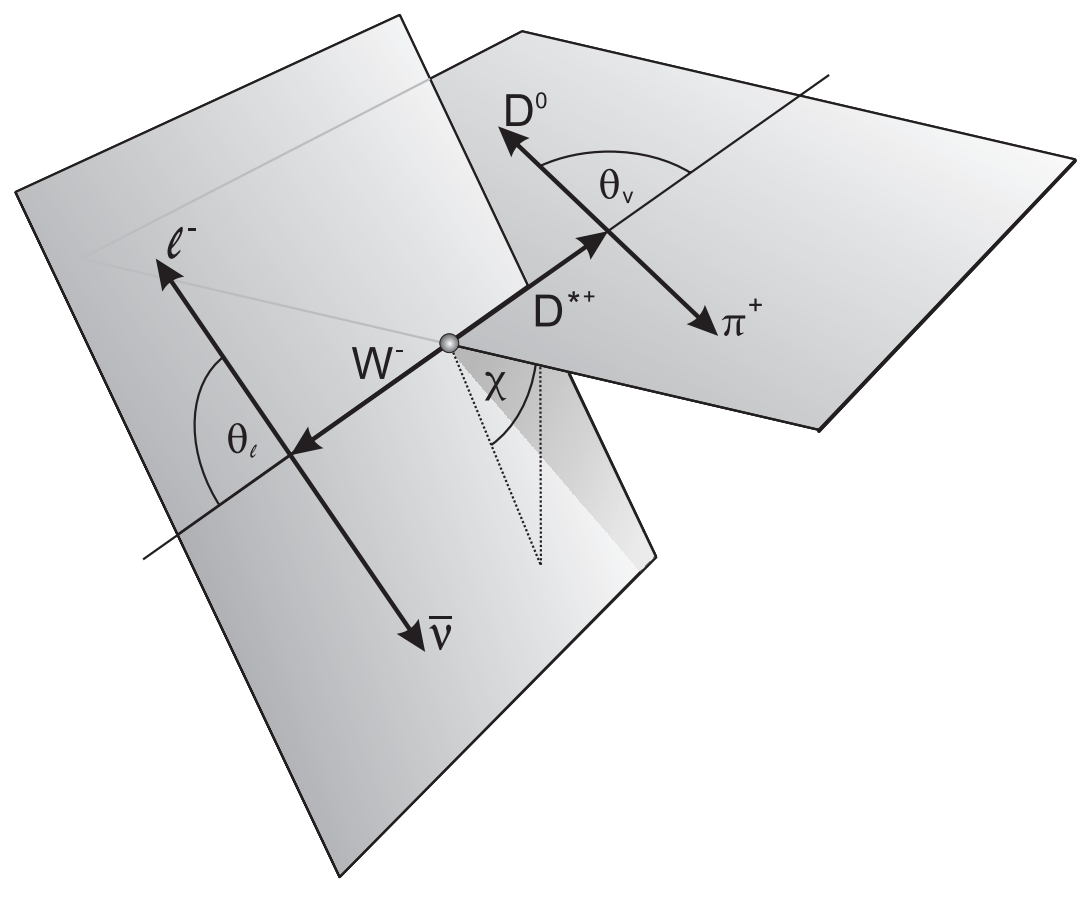

**Figure 17.1.5.** Definition of the angles $\theta_\ell$, $\theta_V$ and $\chi$ for the decay $\overline{B}^0 \to D^{*+}\ell^-\overline{\nu}$ with $D^{*+} \to D^0\pi^+$ (Dungel, 2010).

The Belle measurement (Dungel, 2010) is based on 711 fb$^{-1}$ of $\Upsilon(4S)$ data resulting in about 120,000 reconstructed $B^0 \to D^{*-}\ell^+\nu$ decays. In this analysis the decay chain $D^{*-} \to \overline{D}^0\pi^-$ followed by $\overline{D}^0 \to K^+\pi^-$ is reconstructed and $D^*$ candidates are combined with a charged lepton $\ell$ ($\ell = e, \mu$) with momentum between 0.8 GeV and 2.4 GeV. As the analysis is untagged, the direction of the neutrino is not precisely known. However, using the $\cos\theta_{BY}$ variable with $Y = D^*\ell$ (see Section 17.1.1.3), the $B$ momentum vector is constrained to a cone centered on the $D^*\ell$ direction. By averaging over the possible $B$ directions one can approximate the neutrino momentum and calculate the kinematic variables of the decay, $w$, $\cos\theta_\ell$, $\cos\theta_V$ and $\chi$. The typical $1\sigma$ resolutions for these variables are 0.025, 0.049, 0.050 and 13.5°, respectively. Figure 17.1.6 shows the result of the simultaneous fit to the one-dimensional projections of the four variables for the selected $B^0 \to D^{*-}\ell^+\nu$ sample. A feature of this method is that the same events enter into the four projections and the resulting correlations are accounted for by combining separate covariance matrices for the data and the simulated signal and background distributions.

*BABAR* performed a similar analysis of the decay $B^0 \to D^{*-}\ell^+\nu$ based on a sample of 79 fb$^{-1}$ (Aubert, 2008h). Several $D^0$ decay modes are analyzed and the selected sample contains about 52,800 $B^0 \to D^{*-}\ell^+\nu$ decays. The results extracted from the fit to the four one-dimensional decay distributions were combined with another *BABAR* analysis of $B^0 \to D^{*-}\ell^+\nu$ which performed a fit to the four-dimensional decay rate $\Gamma(w, \theta_\ell, \theta_V, \chi)$ (Aubert, 2006af), thereby enhancing the sensitivity to $R_1(1)$, $R_2(1)$ and $|V_{cb}|$.

*BABAR* also analyzed the isospin conjugated decay mode $B^+ \to \overline{D}^{*0}e^+\nu$ in a sample of 205 fb$^{-1}$, with the neutral $D^*$ meson decaying to $\overline{D}^{*0} \to \overline{D}^0\pi^0$ and $\overline{D}^0 \to K^+\pi^-$ (Aubert, 2008v). The reconstruction of the neutral $D^*$ meson involves a low momentum $\pi^0$ rather than a charged pion, thus it is sensitive to different detection efficiencies and provides an independent check of the $D^*$ reconstruction. In this analysis the HQET form-factor ratios $R_1(1)$ and $R_2(1)$, are taken as external parameters from other measurements.

The results of the $B \to D^*\ell\nu$ form-factor measurements, with common input parameters (mainly $B$ lifetimes and $D$ meson branching ratios) rescaled to the values available by the end of the year 2011 (Beringer et al., 2012), are summarized in Table 17.1.1. The $B^0 \to D^{*-}\ell^+\nu$ branching ratios calculated by using these form-factor parameters are given in Table 17.1.2.

The HQET parameterization of the $B \to D\ell\nu$ form factor $\eta_{\rm EW}\mathcal{G}(w)$ (Caprini, Lellouch, and Neubert, 1998) has only two free parameters: the normalization given by $\eta_{\rm EW}\mathcal{G}(1)|V_{cb}|$ and the slope $\rho_D^2$. These parameters are adopted for the Belle (Abe, 2002c) and *BABAR* (Aubert, 2009ab, 2010e) measurements of this decay.

Untagged analyses of $B \to D\ell\nu$ are limited by large backgrounds and related large irreducible uncertainties. Based on a sample of 417 fb$^{-1}$, *BABAR* performed a study of $B \to D\ell\nu$ decays, in which the second $B$ meson in the event is reconstructed in a hadronic decay mode (Aubert, 2010e). This tagging technique results in a sizable background reduction and a more precise measurement of $w$. With a tagging efficiency of about 0.5%, 16 $D$ meson decay modes and with a lower limit on the lepton momentum at

**Table 17.1.2.** The $B^0 \to D^{*-}\ell^+\nu$ branching ratio, calculated using the HQET parameterization of the form factor $\eta_{\rm EW}\mathcal{F}(w)$ (Caprini, Lellouch, and Neubert, 1998) and the parameter values in Table 17.1.1. For measurements that do not determine $R_1(1)$ and $R_2(1)$, we assume the average values of these parameters (Section 17.1.2.3). The errors quoted correspond to the statistical and systematic uncertainties, respectively.

| Analysis | $\mathcal{B}(B^0 \to D^{*-}\ell^+\nu)$ (%) |
|---|---|
| Belle (Dungel, 2010) | $4.59 \pm 0.03 \pm 0.26$ |
| *BABAR* $D^{*-}\ell^+\nu$ (Aubert, 2008h) | $4.58 \pm 0.04 \pm 0.25$ |
| *BABAR* $\overline{D}^{*0}e^+\nu$ (Aubert, 2008v) | $4.95 \pm 0.07 \pm 0.34$ |
| *BABAR* $DXl\nu$ (Aubert, 2009ab) | $4.96 \pm 0.02 \pm 0.20$ |
| Average | $4.83 \pm 0.01 \pm 0.12$ |

0.6 GeV yields of $2147 \pm 69$ $B^+ \to \overline{D}^0\ell^+\nu$ and $1108 \pm 45$ $B^0 \to D^-\ell^+\nu$ decays are obtained. These signal yields are determined by a fit to the missing-mass-squared distribution, $m^2_{\rm miss} = (p_B - p_D - p_\ell)^2$ (see Figure 17.1.2). The normalization $\eta_{\rm EW}\mathcal{G}(1)|V_{cb}|$ and the slope $\rho_D^2$ are extracted from a fit to the efficiency-corrected signal yields in ten bins of $w$ (see Figure 17.1.7).

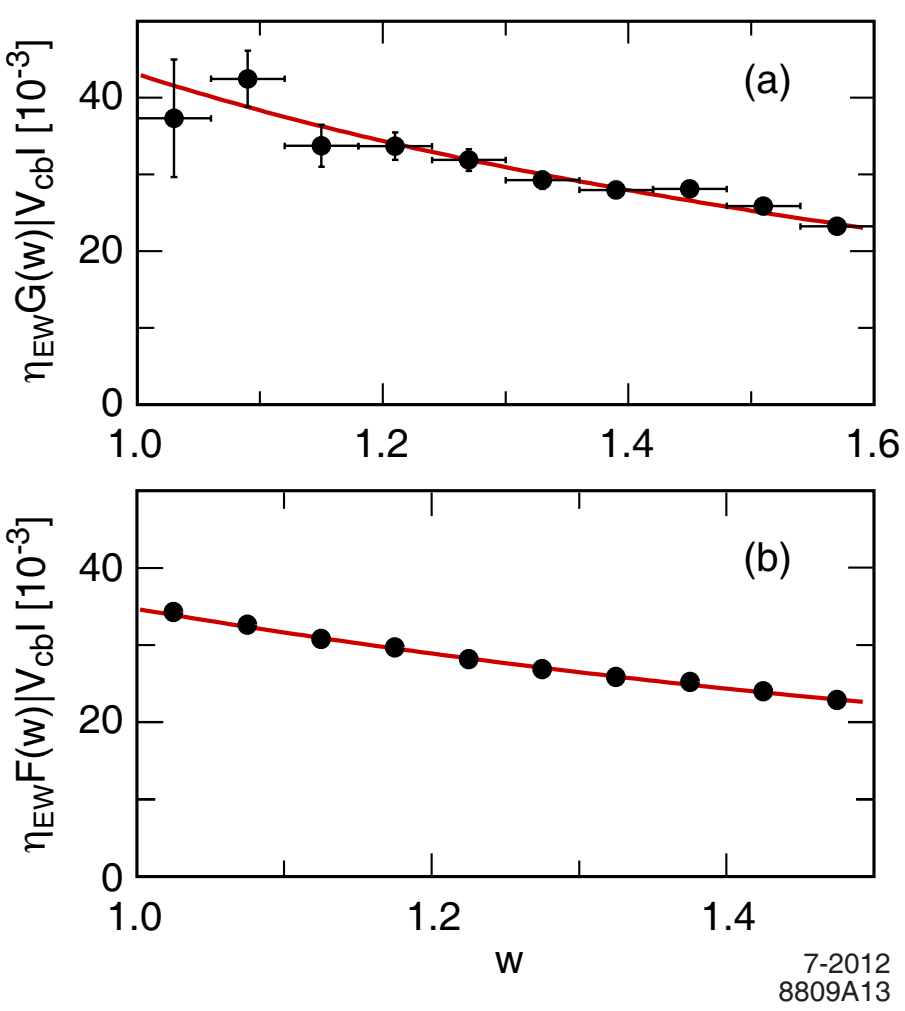

**Figure 17.1.7.** *BABAR* measurements, corrected for the reconstruction efficiency, of the $w$ dependence of the form factors, with fit results superimposed (solid line): (a) $\eta_{EW}\mathcal{G}(w)|V_{cb}|$ for $B \to D\ell\nu$ decays from tagged events (Aubert, 2010e), and for comparison (b) $\eta_{EW}\mathcal{F}(w)|V_{cb}|$ for $B \to D^*\ell\nu$ decays from untagged events (Aubert, 2008h).

The results of the $B \to D\ell\nu$ form-factor measurements at the $B$ Factories, rescaled to common input parameters (Beringer et al., 2012), are summarized in Table 17.1.3. We also calculate the $B^0 \to D^-\ell^+\nu$ branching fraction from these values (Table 17.1.4).

*BABAR* also published a measurement of $B \to D^*\ell\nu$ and $B \to D\ell\nu$ adopting an innovative approach. Using a sample of 207 fb$^{-1}$, this analysis is based on an inclusive selection of $B \to DX\ell\nu$ decays, where only the $D$ meson and the charged lepton are reconstructed (Au-bert, 2009ab). To reduce background from $D^{**}\ell\nu$ decays and other background sources, the lepton momentum is restricted to $p_\ell > 1.2$ GeV, and the $D$ mesons are reconstructed only in the two cleanest decay modes, $D^0 \to K^-\pi^+$ and $D^+ \to K^-\pi^+\pi^+$. The $D^{(*)}\ell\nu$ signal and background yields, the values of $\rho^2_D$, $\rho^2_{D^*}$, $\eta_{\rm EW}\mathcal{G}(1)|V_{cb}|$ and $\eta_{\rm EW}\mathcal{F}(1)|V_{cb}|$ are obtained from a binned $\chi^2$ fit to the three-dimensional distributions of the lepton momentum $p_\ell$, the $D$ momentum $p_D$, and $\cos\theta_{BY}$. The results of this analysis are listed in Tables 17.1.1 and 17.1.3. The statistical errors are less than those of the tagged analysis which was based on a larger overall event sample, but the systematic uncertainty of the $B \to D\ell\nu$ measurement is larger by a factor of two.

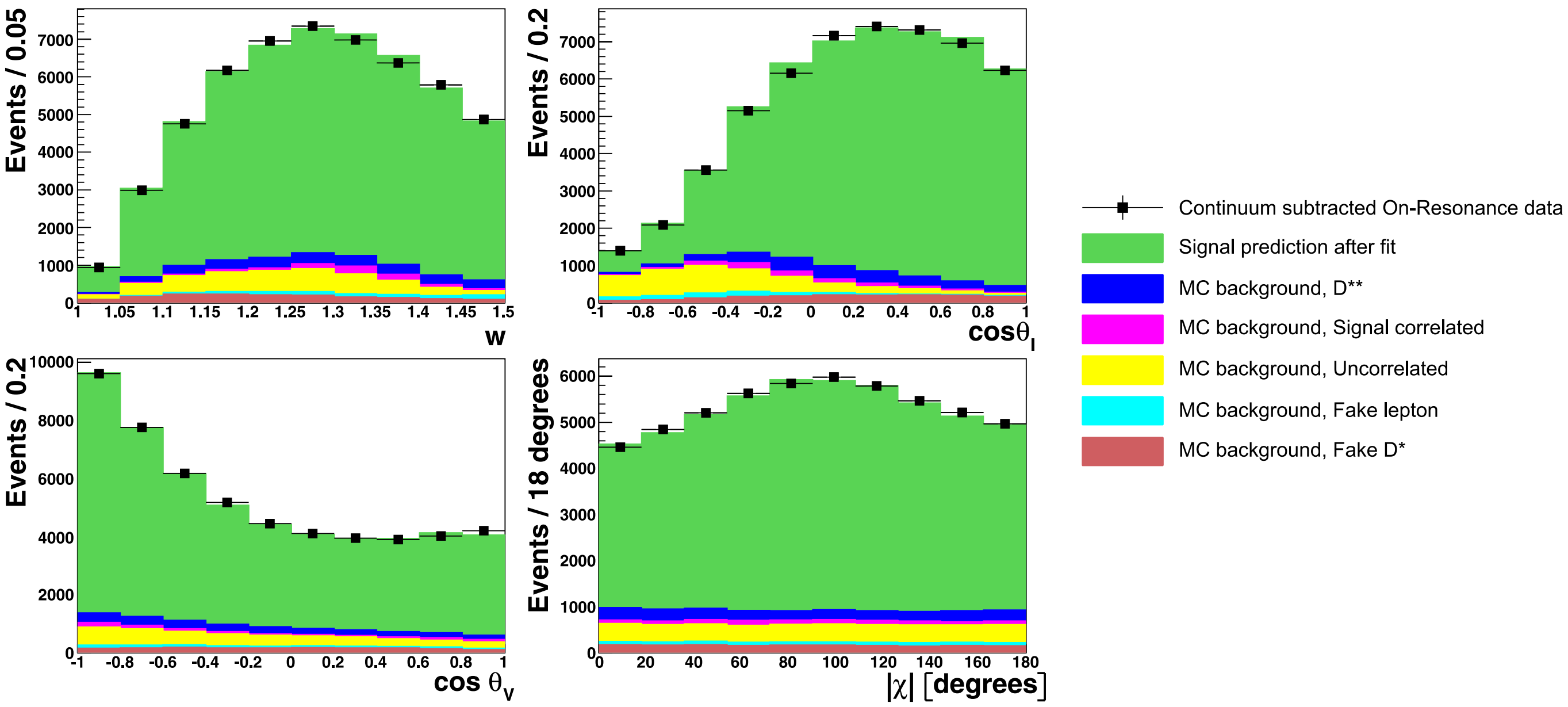

**Figure 17.1.6.** Belle analysis of $B \to D^* \ell \nu$ (Dungel, 2010): Result of the simultaneous fit to four one-dimensional projections of selected $B^0 \to D^{*-} \ell^+ \nu$ events: $w$ (top-left), $\cos\theta_\ell$ (top-right), $\cos\theta_V$ (botton-left) and $\chi$ (bottom-right). The data points represent continuum subtracted event yields. The histograms represent the signal component and different background contributions.

**Table 17.1.1.** Summary of the $B$ Factory results for the $B \to D^* \ell \nu$ form-factor parameters $\eta_{\rm EW}\mathcal{F}(1)|V_{cb}|$, $\rho^2_{D^*}$, $R_1(1)$ and $R_2(1)$. The measurements have been rescaled to the end of year 2011 values of the common input parameters (Beringer et al., 2012). The errors quoted for each parameter correspond to the statistical and systematic uncertainties, respectively. The average is obtained by a four dimensional fit to these values taking into account correlated systematic uncertainties.

| Analysis | $\eta_{\rm EW}\mathcal{F}(1)\|V_{cb}\|$ $(10^{-3})$ | $\rho^2_{D^*}$ | $R_1(1)$ | $R_2(1)$ |
|---|---|---|---|---|
| Belle (Dungel, 2010) | $34.7 \pm 0.2 \pm 1.0$ | $1.21 \pm 0.03 \pm 0.01$ | $1.40 \pm 0.03 \pm 0.02$ | $0.86 \pm 0.02 \pm 0.01$ |
| BABAR $D^{*-}\ell^+\nu$ (Aubert, 2008h) | $34.1 \pm 0.3 \pm 1.0$ | $1.18 \pm 0.05 \pm 0.03$ | $1.43 \pm 0.06 \pm 0.04$ | $0.83 \pm 0.04 \pm 0.02$ |
| BABAR $\overline{D}^{*0}e^+\nu$ (Aubert, 2008v) | $35.1 \pm 0.6 \pm 1.3$ | $1.12 \pm 0.06 \pm 0.06$ | | |
| BABAR $DXl\nu$ (Aubert, 2009ab) | $35.8 \pm 0.2 \pm 1.1$ | $1.19 \pm 0.02 \pm 0.06$ | | |
| Average | $35.5 \pm 0.1 \pm 0.5$ | $1.20 \pm 0.02 \pm 0.02$ | $1.40 \pm 0.03 \pm 0.01$ | $0.86 \pm 0.02 \pm 0.01$ |

**Table 17.1.3.** Summary of the $B$ Factory results for the $B \to D \ell \nu$ form-factor parameters $\eta_{\rm EW}\mathcal{G}(1)|V_{cb}|$ and $\rho^2_D$. The measurements have been rescaled to the end of year 2011 values of the common input parameters (Beringer et al., 2012). The errors quoted for each parameter correspond to the statistical and systematic uncertainties, respectively. The average is obtained by a two dimensional fit to these values taking into account correlated systematic uncertainties.

| Analysis | $\eta_{\rm EW}\mathcal{G}(1)\|V_{cb}\|$ $(10^{-3})$ | $\rho^2_D$ |
|---|---|---|
| Belle (Abe, 2002c) | $40.8 \pm 4.4 \pm 5.0$ | $1.12 \pm 0.22 \pm 0.14$ |
| BABAR $DXl\nu$ (Aubert, 2009ab) | $43.4 \pm 0.8 \pm 2.1$ | $1.20 \pm 0.04 \pm 0.06$ |
| BABAR tagged (Aubert, 2010e) | $42.5 \pm 1.9 \pm 1.1$ | $1.18 \pm 0.09 \pm 0.05$ |
| Average | $42.7 \pm 0.7 \pm 1.5$ | $1.19 \pm 0.04 \pm 0.04$ |

**Table 17.1.4.** The $B^0 \to D^-\ell^+\nu$ branching ratio, calculated using the HQET parameterization of the form factor $\eta_{\rm EW}\mathcal{G}(w)$ (Caprini, Lellouch, and Neubert, 1998) and the parameter values in Table 17.1.3. The errors quoted correspond to the statistical and systematic uncertainties, respectively.

| Analysis | $\mathcal{B}(B^0 \to D^-\ell^+\nu)$ (%) |
|---|---|
| Belle (Abe, 2002c) | $2.07 \pm 0.12 \pm 0.52$ |
| *BABAR* $DXl\nu$ (Aubert, 2009ab) | $2.18 \pm 0.03 \pm 0.13$ |
| *BABAR* tagged (Aubert, 2010e) | $2.12 \pm 0.10 \pm 0.06$ |
| Average | $2.14 \pm 0.03 \pm 0.10$ |

#### 17.1.2.3 Extraction of $|V_{cb}|$ and the Decay Form Factors

We combine the results of four measurements of $B \to D^*\ell\nu$ decays, three obtained by *BABAR* (Aubert, 2008h,v, 2009ab) and one by Belle (Dungel, 2010), by performing a four-dimensional fit to the HQET parameters $\eta_{\rm EW}\mathcal{F}(1)|V_{cb}|$, $\rho^2_{D^*}$, $R_1(1)$ and $R_2(1)$ taking into account systematic error correlations. The results are

$$\begin{aligned} \eta_{\rm EW}\mathcal{F}(1)|V_{cb}| &= (35.45 \pm 0.50) \times 10^{-3} \ , \\ \rho^2_{D^*} &= 1.199 \pm 0.027, \\ R_1(1) &= 1.396 \pm 0.033, \\ R_2(1) &= 0.860 \pm 0.020. \end{aligned} \tag{17.1.30}$$

The correlations between the different fit parameters are

$$\begin{aligned} \rho_{\eta_{\rm EW}\mathcal{F}(1)|V_{cb}|,\rho^2_{D^*}} &= \phantom{-}0.326, \\ \rho_{\eta_{\rm EW}\mathcal{F}(1)|V_{cb}|,R_1(1)} &= -0.084, \\ \rho_{\eta_{\rm EW}\mathcal{F}(1)|V_{cb}|,R_2(1)} &= -0.064, \\ \rho_{\rho^2_{D^*},R_1(1)} &= \phantom{-}0.563, \\ \rho_{\rho^2_{D^*},R_2(1)} &= -0.804, \\ \rho_{R_1(1),R_2(1)} &= -0.761. \end{aligned} \tag{17.1.31}$$

The $\chi^2$ of the combination is 8.0 for 8 degrees of freedom. For $B \to D\ell\nu$, there are three measurements, one by Belle (Abe, 2002c) and two by *BABAR* (Aubert, 2009ab, 2010e). The results of the fit to $\eta_{\rm EW}\mathcal{G}(1)|V_{cb}|$ and $\rho^2_D$ are

$$\begin{aligned} \eta_{\rm EW}\mathcal{G}(1)|V_{cb}| &= (42.68 \pm 1.67) \times 10^{-3}, \\ \rho^2_D &= 1.186 \pm 0.057, \end{aligned} \tag{17.1.32}$$

with a correlation of

$$\rho_{\eta_{\rm EW}\mathcal{G}(1)|V_{cb}|,\rho^2_D} = 0.839. \tag{17.1.33}$$

The $\chi^2$ of the average is 0.3 for 4 degrees of freedom. The measured values and the averages are shown in Figure 17.1.8.

Using the form-factor normalization from the latest LQCD calculation of Eq. (17.1.26), we obtain for $|V_{cb}|$ from $B \to D^*\ell\nu$ decays,

$$|V_{cb}| = (39.04 \pm 0.55_{\rm exp} \pm 0.73_{\rm th}) \times 10^{-3} \ . \tag{17.1.34}$$

Based on an earlier LQCD calculations, Eq. (17.1.27), we derive $|V_{cb}|$ from $B \to D\ell\nu$ decays,

$$|V_{cb}| = (39.46 \pm 1.54_{\rm exp} \pm 0.88_{\rm th}) \times 10^{-3} \ . \tag{17.1.35}$$

On the other hand, we obtain values for $|V_{cb}|$ that are about 5% larger if we rely on heavy flavor sum rule calculations, Eq. (17.1.24), for $B \to D^*\ell\nu$ decays

$$|V_{cb}| = (40.93 \pm 0.58_{\rm exp} \pm 0.95_{\rm th}) \times 10^{-3} \ , \tag{17.1.36}$$

or on HQE calculations, Eq. (17.1.28), for $B \to D\ell\nu$ decays,

$$|V_{cb}| = (40.75 \pm 1.59_{\rm exp} \pm 0.78_{\rm th}) \times 10^{-3} \ . \tag{17.1.37}$$

While the results for the two decay modes agree well, $|V_{cb}|$ measured in $B \to D^*\ell\nu$ decays is more precise and will be considered as the main result.

### 17.1.3 Inclusive Cabibbo-favored $B$ decays

#### 17.1.3.1 Theoretical Overview

Our understanding of inclusive semileptonic $B$ decays rests on a simple idea: since inclusive decays include all possible hadronic final states, the final state quark hadronizes with unit probability and the transition amplitude is sensitive only to the long-distance dynamics of the initial $B$ meson. Thanks to the large hierarchy between the typical energy release, of $O(m_b)$, and the hadronic scale $\Lambda_{\rm QCD}$, and to asymptotic freedom, any residual sensitivity to non-perturbative effects is suppressed by powers of $\Lambda_{\rm QCD}/m_b$.

An OPE allows us to express the non-perturbative physics in terms of $B$ meson matrix elements of local operators of dimension $d \geq 5$, while the Wilson coefficients can be expressed as a perturbative series in $\alpha_s$ (Bigi, Shifman, Uraltsev, and Vainshtein, 1993; Bigi, Uraltsev, and Vainshtein, 1992; Blok, Koyrakh, Shifman, and Vainshtein, 1994; Manohar and Wise, 1994). The OPE disentangles the physics associated with *soft* scales of order $\Lambda_{\rm QCD}$ (parameterized by the matrix elements of the local operators) from that associated with *hard* scales $\sim m_b$ (in the Wilson coefficients). The total semileptonic width and the moments of the kinematic distributions are therefore double expansions in $\alpha_s$ and $\Lambda_{\rm QCD}/m_b$, with a leading term that is given by the free $b$ quark decay. Quite importantly, the power corrections start at $O(\Lambda^2_{\rm QCD}/m_b^2)$ and are comparatively suppressed. At higher orders in the OPE, terms suppressed by powers of $m_c$ also appear, starting with $O(\Lambda^3_{\rm QCD}/m_b^3 \times \Lambda^2_{\rm QCD}/m_c^2)$ (Bigi, Mannel, Turczyk, and Uraltsev, 2010).

The relevant parameters in the double series are the heavy quark masses $m_b$ and $m_c$, the strong coupling $\alpha_s$, and the matrix elements of the local operators. As there are only two dimension five operators, two matrix elements appear at $O(1/m_b^2)$:

$$\mu^2_\pi(\mu) = \frac{1}{2m_B}\langle B|\bar{b}\,\boldsymbol{\pi}^2\, b|B\rangle_\mu, \tag{17.1.38}$$

$$\mu^2_G(\mu) = \frac{1}{2m_B}\langle B|\bar{b}\frac{i}{2}\sigma_{\mu\nu}G^{\mu\nu}b|B\rangle_\mu, \tag{17.1.39}$$

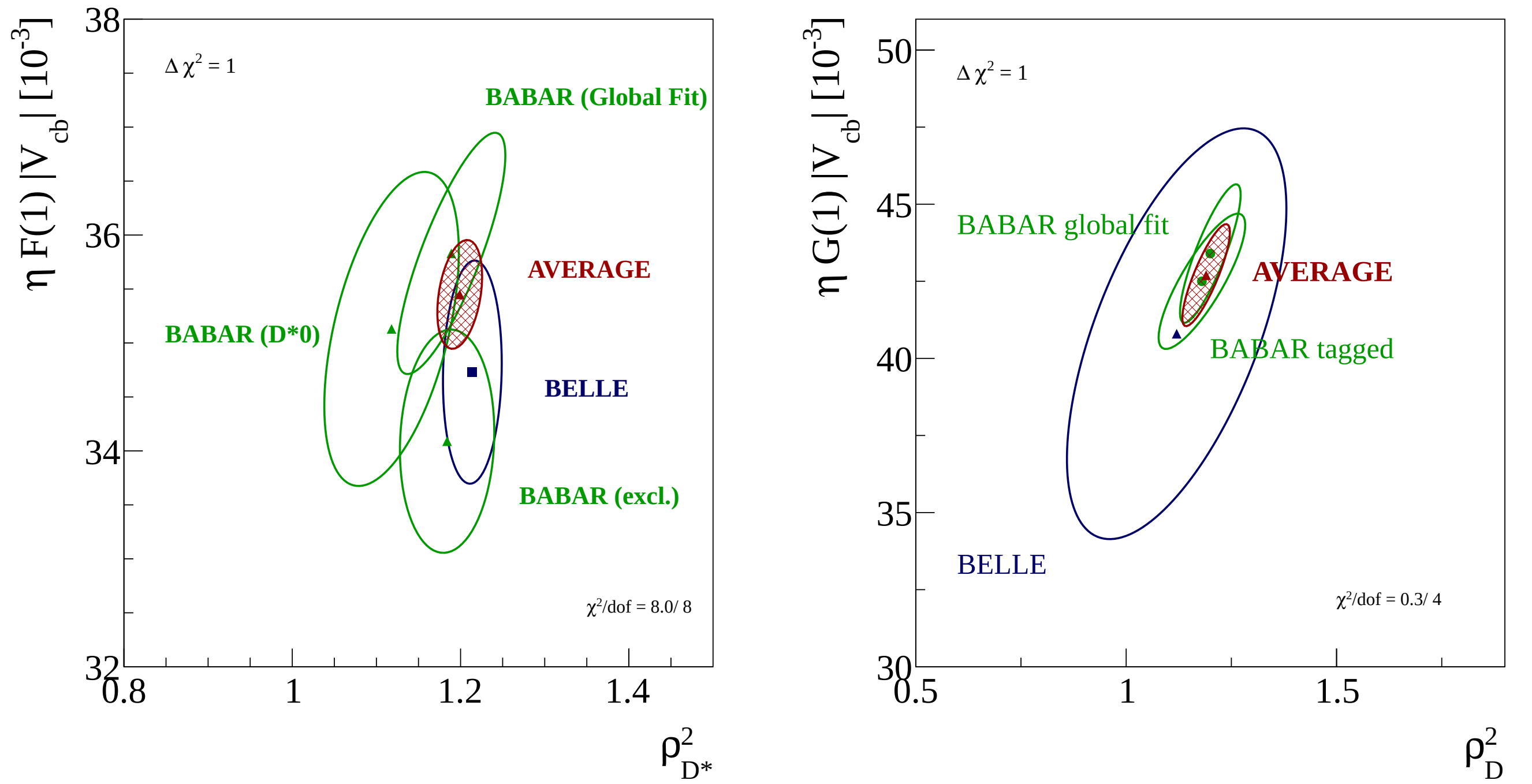

**Figure 17.1.8.** One sigma contour plots of the averages of $\eta_{\rm EW}\mathcal{F}(1)|V_{cb}|$ and $\rho^2_{D^*}$ (left), and of $\eta_{\rm EW}\mathcal{G}(1)|V_{cb}|$ and $\rho^2_D$.

where $\boldsymbol{\pi} = -i\boldsymbol{D}$ with $\boldsymbol{D}$ the space component of the covariant derivative, $\sigma^{\mu\nu} = i/2[\gamma^\mu, \gamma^\nu]$, and $G^{\mu\nu}$ the gluon field tensor. The matrix element of the kinetic operator, $\mu_\pi^2$, is naturally associated with the average kinetic energy of the $b$ quark in the $B$ meson, while that of the chromomagnetic operator, $\mu_G^2$, is related to the $B^*$-$B$ hyperfine mass splitting. They generally depend on a cutoff $\mu = O(1\,{\rm GeV})$ chosen to separate soft and hard physics. The cutoff can be implemented in different ways. In the kinetic scheme (Bigi et al., 1995, 1997), a Wilson cutoff on the gluon momentum is employed in the $b$ quark rest frame: all soft gluon contributions are attributed to the expectation values of the higher dimensional operators, while hard gluons with momentum $|\boldsymbol{k}| > \mu$ contribute to the perturbative corrections to the Wilson coefficients. In the HQET a different notation is usually employed: at leading order in $1/m_Q$ one can identify $\mu_\pi^2$ with $-\lambda_1$ and $\mu_G^2$ with $3\lambda_2$. Most current applications of the OPE involve $O(1/m_b^3)$ effects (Gremm and Kapustin, 1997) as well, parameterized in terms of two additional parameters, generally indicated by $\rho_D^3$ and $\rho_{LS}^3$ or by their HQET counterparts $\rho_{1,2}$. These OPE parameters describe universal properties of the $B$ meson and of the quarks and are useful in several applications.

The interesting quantities to be measured are the total rate and some global shape parameters, such as the first few moments of the lepton energy spectrum or of the hadronic invariant mass distribution. The lepton energy moments are defined as

$$\langle E_\ell^n \rangle = \frac{1}{\Gamma_{E>E_{\rm cut}}} \int_{E>E_{\rm cut}} dE_\ell E_\ell^n \frac{d\Gamma}{dE_\ell}, \quad (17.1.40)$$

where $E_\ell$ is the lepton energy in $B \to X_c \ell\nu$, $\Gamma_{E>E_{\rm cut}}$ is the semileptonic width above the energy threshold $E_{\rm cut}$ and $d\Gamma/dE_\ell$ is the differential semileptonic width as a function of $E_\ell$. The hadronic mass moments are similarly defined as

$$\langle m_X^{2n} \rangle = \frac{1}{\Gamma_{E>E_{\rm cut}}} \int_{E>E_{\rm cut}} dm_X^2 m_X^{2n} \frac{d\Gamma}{dm_X^2}. (17.1.41)$$

Here, $d\Gamma/dm_X^2$ is the differential width as a function of the mass squared of the hadronic system $X$. For both types, $n$ is the order of the moment. For $n > 1$, the moments can also be defined relative to $\langle E_\ell \rangle$ and $\langle m_X^2 \rangle$, respectively, in which case they are called central moments.

The OPE cannot be expected to converge in regions of phase space where the momentum of the final hadronic state is $O(\Lambda_{\rm QCD})$ and where perturbation theory has singularities. This is because what actually controls the expansion is not $m_b$ but the energy release, which is $O(\Lambda_{\rm QCD})$ in those cases. The OPE is therefore valid only for sufficiently inclusive measurements and in general cannot describe differential distributions. The lepton energy moments can be measured very precisely, while the hadronic mass moments are directly sensitive to higher dimensional matrix elements such as $\mu_\pi^2$ and $\rho_D^3$. In most cases, one has to take into account an experimental lower threshold on the lepton momentum. The leptonic and hadronic moments give information on the quark masses and on the non-perturbative OPE matrix elements, while the total rate allows for the extraction of $|V_{cb}|$.

The reliability of the inclusive method rests on our ability to control the higher order contributions in the double series and to constrain quark-hadron duality violation, *i.e.* effects beyond the OPE, which exist but are expected to be rather suppressed in semileptonic decays. The calculation of higher order effects allows us to verify the convergence of the double series and to reduce and properly estimate the residual theoretical uncertainty. Duality violation effects (see Bigi and Uraltsev, 2001a, for a review) can be constrained *a posteriori*, by checking whether the OPE predictions fit the experimental data. This in turn depends on precise measurements and precise OPE predictions. As the experimental accuracy reached at the $B$ Factories is better than the theoretical accuracy for all the measured moments, any effort to improve the latter is strongly motivated.

The main ingredients for an accurate analysis of the experimental data on the moments and the subsequent extraction of $|V_{cb}|$ have been known for some time. Two implementations are currently employed in global analyses; they are based on either the kinetic scheme (Benson, Bigi, Mannel, and Uraltsev, 2003; Gambino and Uraltsev, 2004) or the $1S$ mass scheme for the $b$ quark (Bauer, Ligeti, Luke, Manohar, and Trott, 2004). They both include terms through $O(\alpha_s^2\beta_0)$ and $O(1/m_b^3)$ ($\beta_0 = 11 - 2n_l/3$ is the first coefficient of the QCD beta function) but they use different perturbative schemes, include a somewhat different choice of experimental data under specific assumptions, and estimate the theoretical uncertainties in two distinct ways. Nevertheless, the two methods yield similar results for $|V_{cb}|$.

An important component of the OPE calculation are the purely perturbative contributions. Although the $O(\alpha_s)$ perturbative corrections to various kinematic distributions and to the rate have been computed long ago, the triple differential distribution was first computed at $O(\alpha_s)$ only recently by Aquila, Gambino, Ridolfi, and Uraltsev (2005); Trott (2004). The so-called BLM corrections, *i.e.* those of $O(\alpha_s^2\beta_0)$, are usually the dominant source of two-loop corrections in $B$ decays. They can be found in complete form in Aquila, Gambino, Ridolfi, and Uraltsev (2005). The complete two-loop perturbative corrections to the width and moments of the lepton energy and hadronic mass distributions have been recently computed (Biswas and Melnikov, 2010; Melnikov, 2008; Pak and Czarnecki, 2008) by both numerical and analytic methods. The kinetic scheme implementation for actual observables can be found in Gambino (2011). In general, using $\alpha_s(m_b)$ in the on-shell scheme, the non-BLM corrections amount to about $-20\%$ of the two-loop BLM corrections and give small contributions to normalized moments. In the kinetic scheme with cutoff $\mu = 1\,\text{GeV}$, the perturbative expansion of the total width is

$$\Gamma[\overline{B} \to X_c e\overline{\nu}] \propto 1 - 0.96\,\frac{\alpha_s(m_b)}{\pi} - 0.48\,\beta_0\left(\frac{\alpha_s}{\pi}\right)^2 + 0.82\left(\frac{\alpha_s}{\pi}\right)^2 + O(\alpha_s^3) \approx 0.916. \tag{17.1.42}$$

Higher order BLM corrections of $O(\alpha_s^n\beta_0^{n-1})$ to the width and moments are also known (Aquila, Gambino, Ridolfi, and Uraltsev, 2005; Benson, Bigi, Mannel, and Uraltsev, 2003). The resummed BLM result is numerically very close to that from NNLO calculations (Benson, Bigi, Mannel, and Uraltsev, 2003). The residual perturbative error in the total width is therefore about 1%.

The global fit to moments can be performed to NNLO to extract the OPE parameters and $|V_{cb}|$. In the normalized leptonic moments the perturbative corrections cancel to a large extent, independently of the mass scheme, because hard gluon emission is comparatively suppressed. This pattern of cancellations, crucial for an accurate estimate of the theoretical uncertainties, is confirmed by the complete $O(\alpha_s^2)$ calculation, although the numerical precision of the available results is not sufficient to improve the overall accuracy for the higher central leptonic moments (Gambino, 2011). The non-BLM corrections turn out to be more important for the hadronic moments. Even though it improves the overall theoretical uncertainty only moderately, the complete NNLO calculation leads to the meaningful inclusion of precise mass constraints, such as those discussed in Section 17.1.3.2, in various perturbative schemes (Gambino, 2011).

Sources of significant residual theoretical uncertainty are the perturbative corrections to the Wilson coefficients of the power-suppressed operators. They induce corrections of $O(\alpha_s\Lambda_{\text{QCD}}^2/m_b^2)$ to the width and to the moments. Only the $O(\alpha_s\mu_\pi^2/m_b^2)$ terms are presently known (Becher, Boos, and Lunghi, 2007). A complete calculation of these effects has recently been performed for inclusive radiative decays (Ewerth, Gambino, and Nandi, 2010), where the $O(\alpha_s)$ corrections increase the coefficient of $\mu_G^2$ in the rate by almost 20%. The extension of this calculation to the semileptonic decay rate is in progress. In view of the importance of $O(1/m_b^3)$ corrections, if a theoretical precision of 1% in the decay rate is to be reached, the $O(\alpha_s/m_b^3)$ effects may need to be calculated.

As to the higher order power corrections, a first analysis of $O(1/m_b^4)$ and $O(1/m_Q^5)$ effects is given in Mannel, Turczyk, and Uraltsev (2010). The main problem is the proliferation of non-perturbative parameters: *e.g.* as many as nine new expectation values appear at $O(1/m_b^4)$ and more at the next order. Because they cannot all be extracted from experiment, they are estimated in the ground state saturation approximation, thus reducing them to the known $O(1/m_b^{2,3})$ parameters. In this approximation, the total $O(1/m_Q^{4,5})$ correction to the width is about $+1.3\%$. The $O(1/m_Q^5)$ effects are dominated by $O(1/m_b^3m_c^2)$ intrinsic charm contributions, amounting to $+0.7\%$ (Bigi, Mannel, Turczyk, and Uraltsev, 2010). The net effect on $|V_{cb}|$ also depends on the corrections to the moments. Mannel, Turczyk, and Uraltsev (2010) estimate that the overall effect on $|V_{cb}|$ is a 0.4% increase. While this sets the scale of higher order power corrections, it is as yet unclear how much the result depends on the assumptions made for the expectation values.

It is worth stressing that the semileptonic moments are sensitive to the values of the heavy quark masses and in particular to a specific linear combination of $m_c$ and $m_b$ (Voloshin, 1995), which to a good approximation is the one needed for the extraction of $|V_{cb}|$ (Gambino and Schwanda, 2011). Checking the consistency of the con-straints on $m_c$ and $m_b$ from semileptonic moments with the precise determinations of these quark masses (see Section 17.1.3.2) is an important step in the effort to improve our theoretical description of inclusive semileptonic decays. The inclusion of these constraints in the semileptonic fits will eventually improve the accuracy of the $|V_{ub}|$ and $|V_{cb}|$ determinations. Indeed, the $b$ quark mass and the OPE expectation values obtained from the moments are crucial inputs in the determination of $|V_{ub}|$ from inclusive semileptonic decays (see Section 17.1.5 and Antonelli et al., 2010a). The heavy quark masses and the OPE parameters are also relevant for a precise calculation of other inclusive decay rates such as that of $B \to X_s\gamma$ (Gambino and Giordano, 2008).

The first two moments of the photon energy distribution in $B \to X_s\gamma$ are also often included in the semileptonic fits. They are sensitive to $m_b$ and $\mu_\pi^2$ and play the same role as a loose constraint on $m_b$ ($\delta m_b \sim 90$ MeV). However, as discussed in Section 17.9, experiments place a lower limit on the photon energy, which introduces a sensitivity to the Fermi motion of the $b$-quark inside the $B$ meson and tends to disrupt the OPE. One can still re-sum the higher-order terms into a non-local distribution function and since the lowest integer moments of this function are given in terms of the local OPE parameters, one can parameterize it assuming different functional forms (Benson, Bigi, and Uraltsev, 2005). Another serious problem is that only the leading operator contributing to inclusive radiative decays can be described by an OPE. Therefore, unknown $O(\alpha_s\Lambda_{\rm QCD}/m_b)$ contributions should be expected (Paz, 2010) and radiative moments, though interesting in their own respect, should be considered with care in the context of precision moment analyses.

#### 17.1.3.2 Recent charm and bottom quark mass determinations (other than from semileptonic $B$ decays)

In the following, we discuss recent determinations of $m_c$ and $m_b$, excluding those from semileptonic $B$ decays, and only including results since 2007 (except for the use of non-relativistic sum rules). All quark mass values are presented in the $\overline{\rm MS}$ scheme where the renormalization scale is set to $\mu = m_b$ for the bottom and $\mu = 3$ GeV for the charm quark. For convenience we also provide results for $m_c(m_c)$, even though the scale $\mu = m_c$ is too small considering the current level of precision.

*Low-energy sum rules (LESR)*

The theoretical prediction of moments of the vector current correlator depend on the heavy quark mass and thus the latter can be extracted from the comparison to moments evaluated with the help of experimental data for the total cross section $\sigma(e^+e^- \to \text{hadrons})$. The method is restricted to the first few moments which permits using the fixed-order polarization function. In Kühn, Steinhauser, and Sturm (2007) the charm quark mass has been determined with an uncertainty of 13 MeV. The extraction of the bottom quark mass has been updated (Chetyrkin et al., 2009) using new experimental input. More recently, LESR have also been used to extract the charm quark mass (Dehnadi, Hoang, Mateu, and Zebarjad, 2011).

*Non-relativistic sum rules (NRSR)*

This method requires the evaluation of the polarization function in the non-relativistic limit and is therefore not restricted to lower moments. The most advanced analysis (Pineda and Signer, 2006) uses an almost complete next-to-next-to-leading logarithmic approximation to determine the bottom quark mass.

In Signer (2009) non-relativistic sum rules have been used to extract the charm-quark mass in an approach which combines fixed-order and non-relativistic calculations.

*Finite-energy sum rules (FESR)*

The residue theorem can be used to relate the (appropriately weighted) experimental cross section $\sigma(e^+e^- \to$ hadrons) to a contour integral of the vector current correlation function. The freedom to choose the integration kernel can be used to extract a precise value for the charm quark mass. The most recent analysis was published in Bodenstein, Bordes, Dominguez, Penarrocha, and Schilcher (2011).

*Lattice QCD (LQCD)*

Each quark mass in the lattice QCD Lagrangian must be tuned at each value of the lattice spacing by calibrating to the experimentally-measured value of a 'gold-plated' hadron mass. For $m_c$ and $m_b$ the best choices are ground-state heavy quarkonium or heavy-strange mesons, because they allow very precise tuning. Direct conversion to the $\overline{\rm MS}$ scheme using $m_c(\mu) = Zm_{c,\rm latt}$ is possible using lattice or continuum QCD perturbation theory, but this introduces a significant source of error. The most recent determination of $m_c(m_c)$ using this approach (Blossier et al., 2010) gives a value of 1.28(4) GeV. Since the stated uncertainty includes only the impact of working with only $u$ and $d$ quarks in the sea, it is omitted from Table 17.1.5.

For the $b$ quark, it is also possible to use non-relativistic or even static quark methods to determine the binding energy of a heavy-light meson, and thereby $m_b$. These calculations are currently underway with gluon configurations that include the full effect of sea quarks.

The most precise results from full lattice QCD instead use time-moments of charmonium or bottomonium current-current correlators, extrapolated to the continuum limit and compared to the high-order continuum QCD perturbation theory developed for LESR (Allison et al., 2008; McNeile, Davies, Follana, Hornbostel, and Lepage, 2010). The pseudoscalar current in a highly improved relativistic quark formalism with an exact Partially Conserved

Axial Current (PCAC) relation produces the smallest errors, although reaching bottomonium also requires an extrapolation in the heavy quark mass to the $b$ on current lattices. This method also allows for a completely non-perturbative determination of the ratio of $m_b(\mu)/m_c(\mu) = 4.51(4)$, which can be used to test other determinations.

In Tables 17.1.5 and 17.1.6 the results for $m_c$ and $m_b$ mentioned in the text are listed in chronological order. One observes that the method based on NRSR is not (yet) competitive which is probably due to missing third-order corrections. They are available for the other analyses. For both $m_b$ and $m_c$ the results from LESR and LQCD are very precise and in excellent agreement. They use similar perturbative analyses, but very different input data, with different sources of systematic errors. The recent LESR result (Dehnadi, Hoang, Mateu, and Zebarjad, 2011) gives an error on $m_c$, which is two to four times larger than for other analyses. This has sparked a debate on the theoretical uncertainties of LESR, and in particular on the use of renormalization scales as low as 1 GeV in their estimation, and on the uncertainty of the perturbative QCD prediction for $R(s) = \sigma(e^+e^- \to hadrons, s)/\sigma(e^+e^- \to \mu^+\mu^-, s)$ above 5 GeV. Tables 17.1.5 and 17.1.6 also include the results from Narison (2012), where $m_c$ and $m_b$ have been extracted together with the gluon condensates. In contrast to other determinations based on LESRs and LQCD, a significant influence of the gluon condensate on the quark masses is observed which is quite surprising. Furthermore, the energy region between 3.73 GeV and 4.6 GeV has been parameterized using $\psi$ resonances in the narrow-width approximation, instead of precise experimental data. This might explain the $1.5\sigma$ difference in the central value of $m_c(m_c)$ compared to, for example, the LQCD result. An analogous treatment for bottom quarks seems to have a smaller effect.

We conclude that $m_b$ and $m_c$ can be reliably and precisely extracted using a variety of methods. The results in Tables 17.1.5 and 17.1.6 have correlated errors, so we do not average them. They are well encompassed, however, by $m_c(3\ \text{GeV}) = 0.99(1)$ GeV and $m_b(m_b) = 4.16(2)$ GeV.

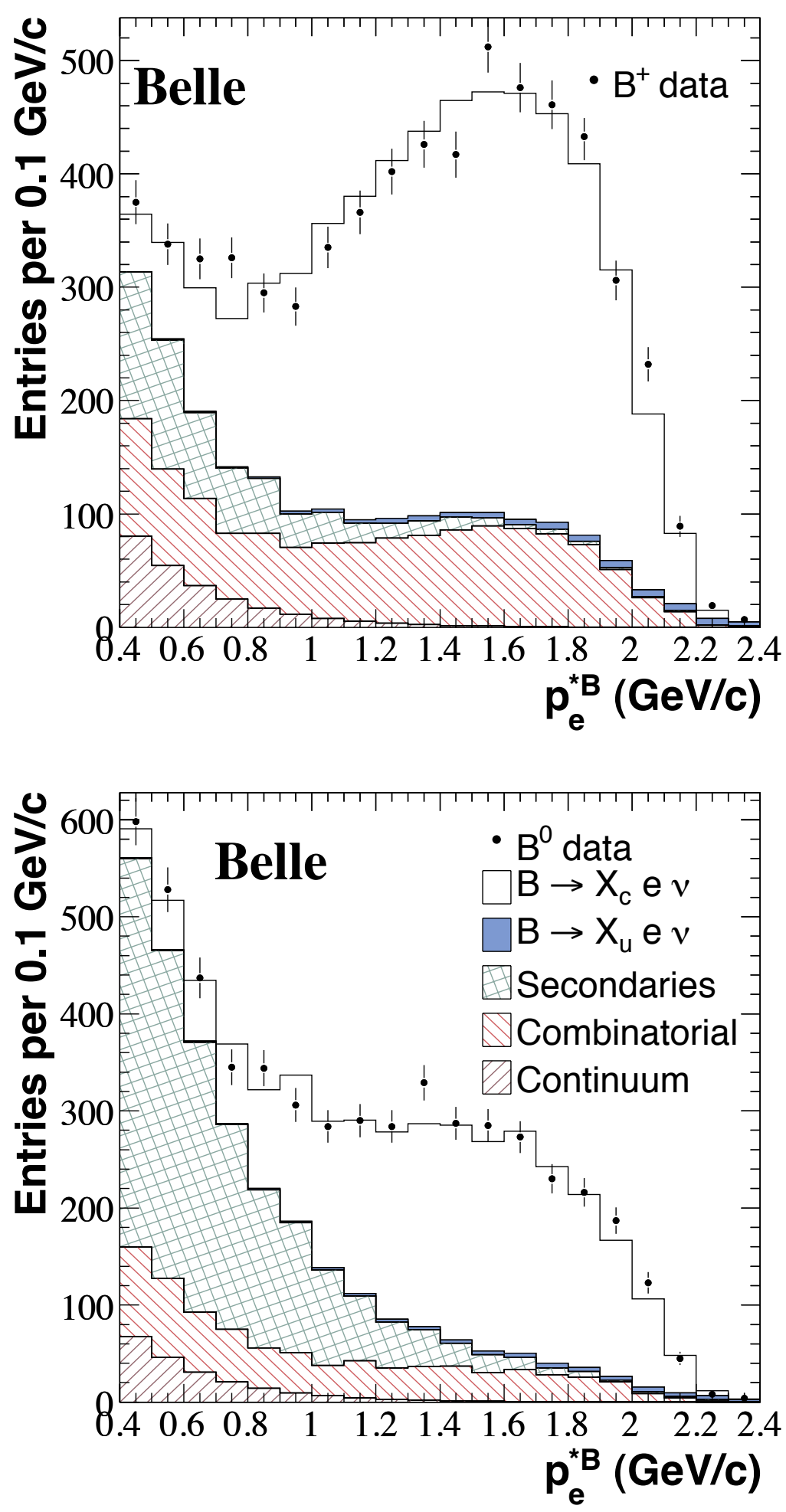

**Figure 17.1.9.** Belle analysis of the electron momentum spectrum for $B^+$ and $B^0$ decays (Urquijo, 2007) before background subtraction, overlaid with the sum of various background contributions and the signal.

### 17.1.3.3 Moment Measurements

Moments of inclusive observables in $B \to X_c \ell \nu$ decays have been measured by the Belle (Schwanda, 2007; Urquijo, 2007) and *BABAR* (Aubert, 2004c,n,r, 2010c) collaborations.

The Belle collaboration has measured spectra of the lepton energy $E_\ell$ and the hadronic mass $m_X$ in $B \to X_c \ell \nu$ using 152 million $\Upsilon(4S) \to B\overline{B}$ events (Schwanda, 2007; Urquijo, 2007). These analyses proceed as follows: first, the decay of one $B$ meson in the event is fully reconstructed in a hadronic mode ($B_{\text{tag}}$). Next, the semileptonic decay of the second $B$ meson in the event ($B_{\text{sig}}$) is identified by searching for a charged lepton among the remaining particles in the event. In Urquijo (2007), the electron momentum spectrum $p_e^*$ in the $B$ meson rest frame is measured down to 0.4 GeV (Figure 17.1.9). In Schwanda (2007), all remaining particles in the event, excluding the charged lepton (electron or muon), are combined to reconstruct the hadronic $X$ system. The $m_X$ spectrum is measured for different lepton energy thresholds in the $B$ meson rest frame (Figure 17.1.10).

The observed spectra are distorted by resolution and acceptance effects and cannot be used directly to obtain the moments. In the Belle analyses, acceptance and finite resolution effects are corrected by unfolding the observed spectra using the Singular Value Decomposition (SVD) algorithm (Höcker and Kartvelishvili, 1996). Belle measures the energy moments $\langle E_\ell^k \rangle$ for $k = 0, 1, 2, 3, 4$ and minimum lepton energies ranging from 0.4 to 2.0 GeV. Moments of the hadronic mass $\langle m_X^k \rangle$ are measured for $k = 2, 4$ and minimum lepton energies from 0.7 to 1.9 GeV.

To determine $|V_{cb}|$, Belle performs fits to 14 lepton energy moments, 7 hadronic mass moments and 4 moments of the photon energy spectrum in $B \to X_s \gamma$ (Schwanda, 2008) based on OPE expressions derived in the kinetic (Benson, Bigi, Mannel, and Uraltsev, 2003; Benson, Bigi, and Uraltsev, 2005; Gambino and Uraltsev, 2004) and 1S schemes (Bauer, Ligeti, Luke, Manohar, and Trott, 2004). Both theoretical frameworks are considered independently and yield very consistent results (see Table 17.1.7).

**Table 17.1.5.** Recent results for the charm-quark mass. An asterisk indicates that we have obtained this number from the value of $m_c$ quoted as the main result of the paper using four-loop accuracy (together with $\alpha_S(m_Z) = 0.1184$ (Nakamura et al., 2010)).

| $m_c$(3 GeV) (GeV) | $m_c(m_c)$ (GeV) | Method | Reference |
|---|---|---|---|
| $0.986 \pm 0.013$ | $1.275 \pm 0.013^*$ | LESR | Kühn, Steinhauser, and Sturm (2007) |
| $0.96 \pm 0.04^*$ | $1.25 \pm 0.04$ | NRSR | Signer (2009) |
| $0.986 \pm 0.006$ | $1.275 \pm 0.006^*$ | LQCD | McNeile, Davies, Follana, Hornbostel, and Lepage (2010) |
| $0.998 \pm 0.029$ | $1.277 \pm 0.026$ | LESR | Dehnadi, Hoang, Mateu, and Zebarjad (2011) |
| $0.987 \pm 0.009$ | $1.278 \pm 0.009$ | FESR | Bodenstein, Bordes, Dominguez, Penarrocha, and Schilcher (2011) |
| $0.972 \pm 0.006^*$ | $1.262 \pm 0.006$ | FESR | Narison (2012) |

**Table 17.1.6.** Recent results for the bottom-quark mass.

| $m_b(m_b)$ (GeV) | Method | Reference |
|---|---|---|
| $4.19 \pm 0.06$ | NRSR | Pineda and Signer (2006) |
| $4.163 \pm 0.016$ | LESR | Chetyrkin et al. (2009) |
| $4.164 \pm 0.023$ | LQCD | McNeile, Davies, Follana, Hornbostel, and Lepage (2010) |
| $4.167 \pm 0.013$ | LESR | Narison (2012) |

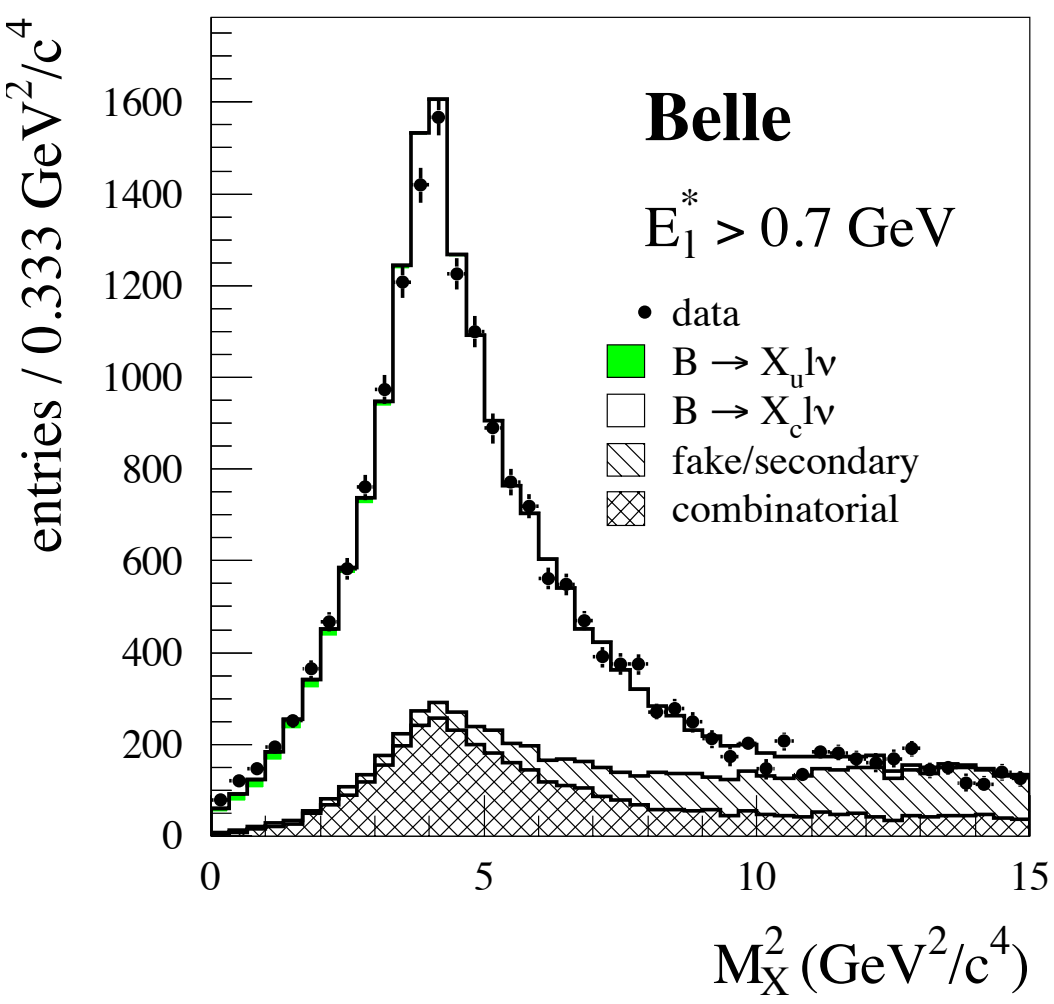

**Figure 17.1.10.** Belle analysis of the hadronic mass distribution for $B \to X_c \ell \nu$ decays (Schwanda, 2007). The data after continuum subtraction are compared with the sum of simulated $X_c \ell \nu$ signal and background contributions.

**Table 17.1.7.** Results of the OPE fits in the kinetic and 1S schemes to moments measured by Belle (Schwanda, 2008): $|V_{cb}|$ and the inclusive branching fractions for $B \to X_c \ell \nu$ decays, plus $\chi^2$ per degree of freedom.

| | Kinetic scheme | 1S scheme |
|---|---|---|
| $\lvert V_{cb}\rvert$ ($10^{-3}$) | $41.58 \pm 0.90$ | $41.56 \pm 0.68$ |
| $\mathcal{B}_{X_c \ell \nu}$ (%) | $10.49 \pm 0.23$ | $10.60 \pm 0.28$ |
| $\chi^2$/ndf. | 4.7/18 | 7.3/18 |

*BABAR* has measured the hadronic mass spectrum $m_X$ in $B \to X_c \ell \nu$ using a data sample of 232 million $\Upsilon(4S) \to B\overline{B}$ events (Aubert, 2010c). The experimental method is similar to the Belle analysis discussed previously, *i.e.*, one $B$ meson is fully reconstructed in a hadronic mode and a charged lepton with momentum above 0.8 GeV in the $B$ meson frame identifies the semileptonic decays of the second $B$. The remaining particles in the event are combined to reconstruct the hadronic system $X$. The resolution in $m_X$ is improved by a kinematic fit to the whole event, taking into account 4-momentum conservation and constraining the missing mass to zero.

To derive the true moments from the reconstructed ones, *BABAR* applies a set of linear corrections. These corrections depend on the charged particle multiplicity of the $X$ system, the normalized missing mass, $E_{\rm miss} - p_{\rm miss}$, and the lepton momentum. In this way, *BABAR* measures the moments of the hadronic mass spectrum up to $\langle m_X^6 \rangle$ for minimum lepton energies ranging from 0.8 to 1.9 GeV.

This study also updates the previous *BABAR* measurement of the lepton energy moments in $B \to X_c \ell \nu$ (Aubert, 2004n) using new branching fraction measurements for background decays and improving the evaluation of systematic uncertainties. Furthermore, first measurements of combined hadronic mass and energy moments of the form $\langle n_X^k \rangle$ with $k = 2, 4, 6$ are presented. They are defined as $n_X^2 = m_X^2 - 2\widetilde{\Lambda} E_X + \widetilde{\Lambda}^2$, where $m_X$ and $E_X$ are the mass and the energy of the $X$ system and the constant $\widetilde{\Lambda}$ is taken to be 0.65 GeV.

*BABAR* performs a simultaneous fit to 12 hadronic mass moments (or 12 combined mass-energy moments), 13 lepton energy moments (including partial branching fractions as zero order moments), and 3 photon energy moments in $B \to X_s \gamma$ (Aubert, 2005x, 2006t), and based on OPE calculations in the kinetic scheme (Benson, Bigi, Mannel, and Uraltsev, 2003; Benson, Bigi, and Uraltsev, 2005; Gambino and Uraltsev, 2004) extracts $|V_{cb}|$, the total branching

**Table 17.1.8.** Results of the OPE fits in the kinetic scheme to moments measured by *BABAR* (Aubert, 2010c). $|V_{cb}|$ and the inclusive branching fractions for $B \to X_c \ell \nu$ decays, plus $\chi^2$ per degree of freedom. The first uncertainty is experimental, the second theoretical.

| | Hadronic moment | Mass-energy moment |
|---|---|---|
| $|V_{cb}|$ $(10^{-3})$ | $42.05 \pm 0.45 \pm 0.70$ | $41.91 \pm 0.48 \pm 0.70$ |
| $m_b$ (GeV) | $4.549 \pm 0.031 \pm 0.038$ | $4.556 \pm 0.034 \pm 0.041$ |
| $\mathcal{B}_{X_c \ell \nu}$ (%) | $10.64 \pm 0.17 \pm 0.06$ | $10.64 \pm 0.17 \pm 0.06$ |
| $\chi^2$/ndf. | 10.9/28 | 8.2/28 |

fractions, $m_b$ and $m_c$, and OPE parameters. The results are given in Table 17.1.8.

#### 17.1.3.4 Global Fit and Determination of $|V_{cb}|$

We perform a global analysis of the $B$ Factory measurements using the full $\mathcal{O}(\alpha_s^2)$ calculations of the moments in the kinetic scheme (Gambino, 2011). This fit combines the 54 moment measurements shown in Table 17.1.9 and determines $|V_{cb}|$, the $b$-quark mass $m_b$ and the higher order parameters in the OPE description of semileptonic decays. The only external input is the average $B^0$ and $B^+$ lifetime, assumed to be $(1.582 \pm 0.007)$ ps (Beringer et al., 2012).

From the fits to moments in $B \to X_c \ell \nu$ we obtain a linear combination of the $b$- and $c$-quark masses. To enhance the precision on $m_b$, we make two choices to gain additional constraints: we either include photon energy moments from $B \to X_s \gamma$ decays in the fit, or use as a precise constraint on the $c$-quark mass $m_c(3\text{ GeV}) = 0.998 \pm 0.029$ GeV, as derived with low-energy sum rules (Dehnadi, Hoang, Mateu, and Zebarjad, 2011). The results for the kinetic scheme based on the Belle and *BABAR* moments are shown in Table 17.1.10 and Figure 17.1.11.

The same moments are fit with expressions derived in the 1S scheme (Bauer, Ligeti, Luke, Manohar, and Trott, 2004). In this framework, we cannot introduce a $c$-quark mass constraint. Results are thus presented for the entire set of 54 moment measurements and for the $X_c \ell \nu$ moments only (Table 17.1.11).

The fit results shown in the first rows of Tables 17.1.10 and 17.1.11 are based on the same set of measurements and can thus be compared directly. For $|V_{cb}|$, the result obtained in the kinetic scheme, $(42.09 \pm 0.75) \times 10^{-3}$, agrees very well with the 1S result, $(42.01 \pm 0.49) \times 10^{-3}$. The uncertainty on $|V_{cb}|$ in the kinetic scheme is 1.8% compared to 1.2% is the 1S scheme. Note however that the assumptions on the dominant theory error are significantly different in the two frameworks. The results for the $b$-quark mass cannot be compared directly due to different mass definitions.

We adopt the results of the fit in the kinetic scheme with the constraint on the $c$-quark mass as currently the most precise result, based on inclusive $B \to X_c \ell \nu$ decays,

$$|V_{cb}|_{\text{incl}} = (42.01 \pm 0.47_{\text{exp}} \pm 0.59_{\text{th}}) \times 10^{-3}. \quad (17.1.43)$$

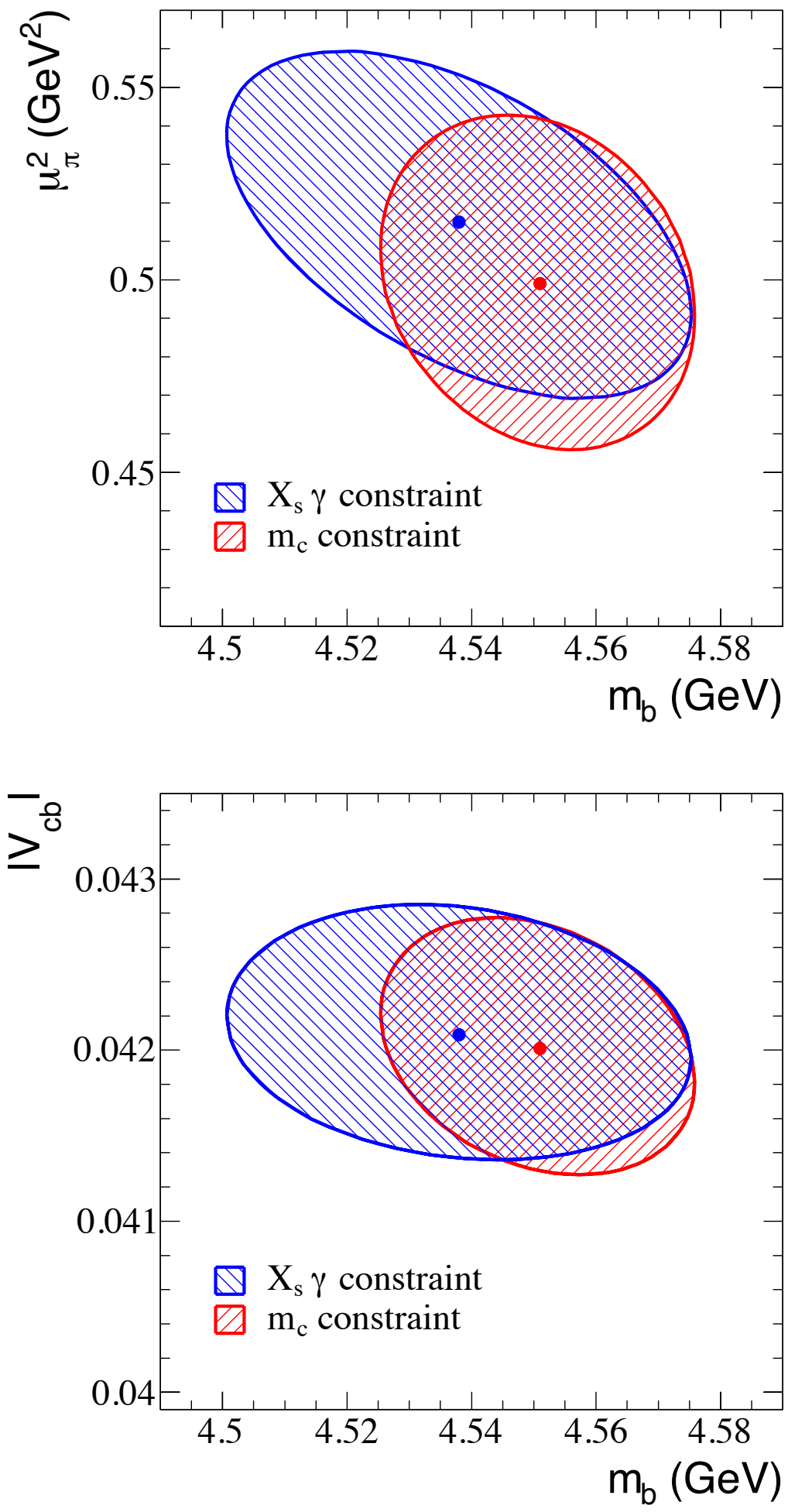

**Figure 17.1.11.** $\Delta\chi^2 = 1$ contours for the global fit to Belle and *BABAR* moments in the kinetic mass scheme, for details see the text.

### 17.1.4 Exclusive decays $B \to \pi \ell \nu$

#### 17.1.4.1 Theoretical Overview

The decay rate for $B \to \pi \ell \nu$ semileptonic decay is given by:

$$\frac{d\Gamma}{dq^2} = \frac{G_F^2 |V_{ub}|^2}{24\pi^3} \frac{(q^2 - m_\ell^2)^2 \, \mathbf{p}_\pi}{q^4 m_B^2} \times \left\{ \left(1 + \frac{m_\ell^2}{2q^2}\right) m_B^2 \, \mathbf{p}_\pi^2 \left[f_+^{B\pi}(q^2)\right]^2 + \frac{3m_\ell^2}{8q^2} (m_B^2 - m_\pi^2)^2 \left[f_0^{B\pi}(q^2)\right]^2 \right\}, \quad (17.1.44)$$

where $q \equiv p_B - p_\pi$ is the 4-momentum transferred to the lepton-neutrino pair and

$$\mathbf{p}_\pi = \left[(m_B^2 + m_\pi^2 - q^2)^2 - 4m_B^2 m_\pi^2\right]^{1/2} / (2m_B)$$

is the pion 3-momentum in the $B$ rest frame. The form factors $f_+^{B\pi}(q^2)$ and $f_0^{B\pi}(q^2)$ are defined in Eq. (17.1.6).

**Table 17.1.9.** Experimental inputs used in the global analysis of $B \to X_c \ell \nu$. $n$ is the order of the moment, $c$ is the threshold value in GeV. In total, there are 29 measurements from BaBar and 25 from Belle.

| Experiment | Hadron moments $\langle m_X^n \rangle$ | Lepton moments $\langle E_\ell^n \rangle$ | Photon moment $\langle E_\gamma^n \rangle$ |
|---|---|---|---|
| BaBar | $n = 2,\ c = 0.9, 1.1, 1.3, 1.5$<br>$n = 4,\ c = 0.8, 1.0, 1.2, 1.4$<br>$n = 6,\ c = 0.9, 1.3$<br>(Aubert, 2010c) | $n = 0,\ c = 0.6, 1.2, 1.5$<br>$n = 1,\ c = 0.6, 0.8, 1.0, 1.2, 1.5$<br>$n = 2,\ c = 0.6, 1.0, 1.5$<br>$n = 3,\ c = 0.8, 1.2$<br>(Aubert, 2004n, 2010c) | $n = 1,\ c = 1.9, 2.0$<br>$n = 2,\ c = 1.9$<br>(Aubert, 2005x, 2006t) |
| Belle | $n = 2,\ c = 0.7, 1.1, 1.3, 1.5$<br>$n = 4,\ c = 0.7, 0.9, 1.3$<br>(Schwanda, 2007) | $n = 0,\ c = 0.6, 1.0, 1.4$<br>$n = 1,\ c = 0.6, 0.8, 1.0, 1.2, 1.4$<br>$n = 2,\ c = 0.6, 1.0, 1.4$<br>$n = 3,\ c = 0.8, 1.0, 1.2$<br>(Urquijo, 2007) | $n = 1,\ c = 1.8, 1.9$<br>$n = 2,\ c = 1.8, 2.0$<br>(Limosani, 2009) |

**Table 17.1.10.** Results of the OPE global fit to $B \to X_c \ell \nu$ moments in the kinetic scheme: the first row refers to the fit including $B \to X_s \gamma$ moments, the second row gives the results obtained with the charm-quark mass constraint. In all cases, the first error is the uncertainty of the global fit. For $|V_{cb}|$ the second error is an additional theoretical uncertainty arising from the calculation of $|V_{cb}|$. The $\chi^2$/ndf. is 17.1/(54 − 7) for the $B \to X_s \gamma$ and 23.3/(44 − 7) for the $m_c$ constrained fit.

| Constraint | $\lvert V_{cb}\rvert$ $(10^{-3})$ | $m_b^{\rm kin}$ (GeV) | $\mu_\pi^2$ (GeV$^2$) | $\rho_D^3$ (GeV$^3$) | $\mu_G^2$ (GeV$^2$) | $\rho_{LS}^3$ (GeV$^3$) |
|---|---|---|---|---|---|---|
| $B \to X_s \gamma$ | $42.09 \pm 0.46 \pm 0.59$ | $4.538 \pm 0.038$ | $0.515 \pm 0.045$ | $0.209 \pm 0.021$ | $0.263 \pm 0.047$ | $-0.121 \pm 0.090$ |
| $m_c$(3 GeV) | $42.01 \pm 0.47 \pm 0.59$ | $4.551 \pm 0.025$ | $0.499 \pm 0.044$ | $0.177 \pm 0.021$ | $0.227 \pm 0.048$ | $-0.081 \pm 0.092$ |

**Table 17.1.11.** Results of the OPE global fit to $B \to X_c \ell \nu$ moments in the 1S scheme: the first row refers to the fit including $B \to X_s \gamma$ moments, the second row gives the results obtained with $B \to X_c \ell \nu$ moments only.

| Input | $\lvert V_{cb}\rvert$ $(10^{-3})$ | $m_b^{1S}$ (GeV) | $\lambda_1$ (GeV$^2$) | $\rho_1$ (GeV$^3$) | $\tau_1$ (GeV$^3$) | $\tau_2$ (GeV$^3$) | $\tau_3$ (GeV$^3$) |
|---|---|---|---|---|---|---|---|
| all moments | $42.01 \pm 0.49$ | $4.696 \pm 0.043$ | $-0.354 \pm 0.072$ | $0.057 \pm 0.060$ | $0.154 \pm 0.122$ | $-0.039 \pm 0.078$ | $0.194 \pm 0.105$ |
| $X_c \ell \nu$ only | $42.58 \pm 0.78$ | $4.595 \pm 0.110$ | $-0.428 \pm 0.099$ | $0.080 \pm 0.062$ | $0.150 \pm 0.124$ | $-0.023 \pm 0.086$ | $0.204 \pm 0.112$ |

In the limit of zero momentum-transfer the form factors must satisfy the kinematic constraint $f_+^{B\pi}(0) = f_0^{B\pi}(0)$. Furthermore, in the limit $m_\ell \to 0$, which is a good approximation for $\ell = e, \mu$, the scalar form factor $f_0^{B\pi}(q^2)$ becomes negligible:

$$\frac{d\Gamma}{dq^2} = \frac{G_F^2 |V_{ub}|^2}{24\pi^3} \mathbf{p}_\pi^3 |f_+^{B\pi}(q^2)|^2. \qquad (17.1.45)$$

Hence precise experimental measurements of the $B \to \pi \ell \nu$ branching fraction along with reliable theoretical calculations of the form factor $f_+^{B\pi}(q^2)$ enable a clean determination of the CKM matrix element $|V_{ub}|$.[53]

The form factors encode the non-perturbative dynamics of binding quarks into hadrons and therefore they cannot be calculated perturbatively. In practice, two methods are available for computing QCD form factors with controlled uncertainties: lattice QCD and light-cone sum rules. As discussed in Section 17.1.2, LQCD is a first-principles approach providing results with steadily improvable errors. LCSR is derived from the correlator of quark currents calculated in terms of the OPE. Matching the result of this calculation to the hadronic dispersion relation yields an analytical expression for the form factor. The precision of LCSR is limited by the accuracy of OPE and by the quark-hadron duality approximation used in the dispersion relation. Lattice QCD and light-cone sum rule form-factor calculations are complementary in that they work in different kinematical regions: LQCD is best at high $q^2$ while LCSR are applicable at low $q^2$-values.

[53] In principle, the exclusive semileptonic decay channel $B \to \rho \ell \nu$ can also be used to determine $|V_{ub}|$ (see, *e.g.*, Flynn, Nakagawa, Nieves, and Toki, 2009). In practice, however, systematic uncertainties are not under control in current lattice QCD calculations of the $\rho$ meson because the $\rho$ is unstable and is not described within the framework of chiral perturbation theory; these concerns will be addressed in future LQCD calculations when more computing resources are available. Light-cone sum rule determinations of the $B \to \rho \ell \nu$ form factor are available, such as in Ball and Zwicky (2005b), but there has not been any recent work on this channel.

#### *Heavy-to-light form-factor parameterizations*

It is useful for comparing different theoretical calculations or theory with experiment to parameterize the form factor $f_+^{B\pi}(q^2)$ as a function of $q^2$. Many parameterizations are available in the literature, but here we focus on the model-independent parameterization of Boyd, Grinstein, and Lebed (1995), hereafter "BGL", and its variants, which is based on the general properties of analyticity, unitarity and crossing-symmetry. All form factors are analytic functions of $q^2$, except at physical poles and threshold branch points. Hence, given an appropriate change of variables, they can be expressed in a particularly useful manner as a convergent power series (see, *e.g.*, Arnesen, Grinstein, Rothstein, and Stewart, 2005; Bourrely, Machet, and de Rafael, 1981; Boyd and Savage, 1997; Lellouch, 1996).

Consider the following change of variables:

$$z(q^2,t_0) = \frac{\sqrt{1-q^2/t_+} - \sqrt{1-t_0/t_+}}{\sqrt{1-q^2/t_+} + \sqrt{1-t_0/t_+}}, \qquad (17.1.46)$$

where $t_+ \equiv (m_B + m_\pi)^2$, and $t_0 < t_+$ is an arbitrary parameter to be discussed later. This transformation maps the semileptonic region of $q^2$ onto a unit circle in the complex $z$ plane. In terms of the new variable $z$, the $B \to \pi$ form factor takes a simple form:

$$P_+(q^2)\phi_+(q^2,t_0)f_+(q^2) = \sum_{k=0}^{\infty} a_k(t_0) z(q^2,t_0)^k. \quad (17.1.47)$$

(A similar function can be derived for the scalar form factor $f_0(q^2)$.) The function $P_+(q^2)$ must be chosen to vanish at the $B^*$ pole in order to preserve the correct analytic structure of $f_+(q^2)$:

$$P_+^{B\pi}(q^2) = z(q^2, m_B^*)\,, \qquad (17.1.48)$$

while the function $\phi_+(q^2,t_0)$ can be any analytic function. It is helpful, however, to choose $\phi_+(q^2,t_0)$ so that the unitarity constraint on the series coefficients ($a_k$'s) obeys a simple form. The choice for $\phi_+(q^2,t_0)$ corresponding to the BGL parameterization is given in Arnesen, Grinstein, Rothstein, and Stewart (2005):

$$\begin{aligned}\phi_+(q^2,t_0) = &\sqrt{\frac{3}{96\pi\chi_J^{(0)}}}\left(\sqrt{t_+-q^2}+\sqrt{t_+-t_0}\right)\\ &\times\left(\sqrt{t_+-q^2}+\sqrt{t_+-t_-}\right)^{3/2}\\ &\times\left(\sqrt{t_+-q^2}+\sqrt{t_+}\right)^{-5}\frac{(t_+-q^2)}{(t_+-t_0)^{1/4}},\end{aligned} \qquad (17.1.49)$$

where the numerical factor $\chi_J^{(0)}$ can be calculated using perturbation theory and the OPE.

Unitarity constrains the size of the BGL series coefficients:

$$\sum_{k=0}^{N} a_k^2 \lesssim 1, \qquad (17.1.50)$$

where this holds for any value of $N$. In the case of the $B \to \pi$ form factor, Becher and Hill (2006) use the heavy-quark power-counting to argue that the sizes of the series coefficients should in fact be much less than one:

$$\sum_{k=0}^{N} a_k^2 \le \left(\frac{\Lambda}{m_Q}\right)^3 \ll 1, \qquad (17.1.51)$$

where $\Lambda$ is a typical hadronic scale; this is consistent with lattice calculations by Bailey et al. (2009) and experimental measurements by *BABAR* in del Amo Sanchez (2011n) and Belle in Ha (2011). The free parameter $t_0$ appearing in Eq. (17.1.46) determines the range of $|z|$ in the semileptonic region, and hence can be chosen to accelerate the series convergence. For example, Arnesen, Grinstein, Rothstein, and Stewart (2005) use the value $t_0 = 0.65t_-$ such that $-0.34 < z < 0.22$ for $B \to \pi l \nu$ decay. The small magnitude of $|z|$, in conjunction with the tight heavy-quark bound on the size of the series coefficients, ensures that only the first few terms in the series are needed to describe the $B \to \pi$ form factor to sub-percent accuracy.

Bourrely, Caprini, and Lellouch (2009) (BCL) use the same series expansion of Eq. (17.1.47), but without an outer function $\phi_+$ and with a different Blashke factor $P_+$:

$$f_+(q^2) = \frac{1}{1-q^2/m_{B^*}^2}\sum_{k=0}^{K} b_k(t_0) z(q^2,t_0)^k. \qquad (17.1.52)$$

Their choice avoids unphysical singularities which are generated at $q^2 = t_+$ by the outer function in a truncated BGL parameterization. Further, Bourrely, Caprini, and Lellouch (2009) optimize the parameter $t_0$ such that the semileptonic domain is mapped onto a symmetric interval in $z$. With the choice $t_0 = (m_B + m_\pi)(\sqrt{m_B} - \sqrt{m_\pi})^2$, the value of $|z| < 0.279$. Although the BCL parameterization has a simpler functional form, the constraint on the series is more complicated than Eq. (17.1.50) in that it is no longer diagonal in the series index $k$. We use the BCL parameterization to obtain $|V_{ub}|$ in Section 17.1.4.3.

A different approach suggested by Flynn and Nieves (2007a,b) uses the Omnès parameterization, allowing one to express the form-factor shape in terms of the elastic $B$-$\pi$ scattering phase shift and the value of $f_+(q^2)$ at a few subtraction points below the $B\pi$ production threshold.

### *Lattice QCD form-factor calculations*

State-of-the-art LQCD computations now regularly include the effects of three light dynamical quarks. Often calculations are done in the isospin limit with two lighter degenerate quarks and one heavier quark with a mass close to the physical strange quark; these are referred to as "2+1" flavor simulations.

In practice, limited computational resources prohibit calculations with simulated values of the $u$- and $d$-quark as light as those in the real world. LQCD calculations must also be done at fixed, nonzero values of the lattice spacing. Hence one generates data with a sequence of light-quark masses (down to $\sim m_{\rm strange}/10$ for current $B \to \pi$ calculations) and a sequence of lattice spacings (down to $a \sim 0.09$ fm for current $B \to \pi$ calculations) and extrapolates the remainder of the way to the physical masses and zero lattice spacing. Because these limits are interrelated, it is now standard to use model-independent functional forms derived in Chiral Perturbation Theory ($\chi$PT) for the specific lattice quark formulation being used (*i.e.* including discretization corrections) to guide the extrapolation (see, *e.g.*, Aubin and Bernard, 2007, for the case of $B \to \pi$). This procedure leaves a remaining systematic uncertainty in the physical matrix element due to truncation of the chiral expansion that is typically included in error budgets as a "chiral extrapolation error." This, in combination with statistical errors, is currently the largest source of uncertainty in lattice calculations of the $B \to \pi$ form factor. Fortunately, increasing computational resources are allowing this error to be reduced in a straightforward manner.

The next-largest uncertainty in current lattice $B \to \pi$ form-factor calculations is due to perturbative operator matching. Numerical lattice simulations evaluate the hadronic matrix element of the vector current $V_\mu = i\overline{u}\gamma_\mu b$ written in terms of the discretized versions of the heavy-quark ($b$) and light anti-quark ($\overline{u}$) fields that appear in the lattice actions. Hence one must compute matching factors to relate the continuum vector current to its lattice counterpart. Current $B \to \pi$ form-factor calculations rely on either a combination of perturbative and non-perturbative methods or on one-loop lattice perturbation theory; the residual uncertainties from neglected 2-loop and higher-order terms in the perturbative series can be approximately as large as the chiral-continuum extrapolation error. Hence new methods are being developed and new actions are being used in order to reduce the renormalization error in the future.

Currently there are two realistic "2+1" flavor LQCD calculations of the $B \to \pi$ form factor – one by the HPQCD Collaboration (Dalgic et al., 2006) and one by the Fermilab Lattice and MILC collaborations (Bailey et al., 2009). These calculations were both performed on gauge configurations made publicly available by the MILC Collaboration (see Aubin et al., 2004) and include the effects of three flavors of dynamical staggered light quarks; hence the statistical errors are somewhat correlated among the two results. The two calculations use different heavy-quark formalisms, however, for the $b$ quark. The Fermilab and MILC collaborations use the Fermilab formalism developed by El-Khadra, Kronfeld, and Mackenzie (1997) in which one uses knowledge of the heavy-quark limit of QCD to systematically remove heavy-quark discretization errors order-by-order in $1/m_b$. The HPQCD Collaboration uses the formulation of the NRQCD action from Lepage, Magnea, Nakhleh, Magnea, and Hornbostel (1992), in which the $b$-quark is a non-relativistic field and the action is expanded in powers of $\boldsymbol{v_b}/c$, where $\boldsymbol{v_b}$ is the spatial velocity of the $b$ quark. Both heavy-quark formulations work well for $b$ quarks at currently available values of the lattice spacing. The Fermilab formalism, however, has two advantages in that it possesses a continuum limit and that it can also be used for $c$ quarks, thereby providing a cross check of the method. Future calculations using other lattice formulations for the light and heavy quarks, such the relativistic heavy-quark action developed by Christ, Li, and Lin (2007) and used by the RBC and UKQCD Collaborations for $B$-meson leptonic decays and mixing (see Van de Water and Witzel, 2010), will provide valuable independent cross checks of the $B \to \pi$ form factor in the next few years.

The Fermilab Lattice and MILC Collaborations present their form-factor results in terms of the BGL series coefficients and the correlation matrix; these are given in Table 17.1.12. The series coefficients can be used to obtain the form factor over the entire $q^2$ range, and are therefore a useful way to present the data, as pointed out by Bernard et al. (2009b). This is particularly helpful for state-of-the art extractions of $|V_{ub}|$ that rely on simultaneous BGL fits of the lattice and experimental data including correlations (see del Amo Sanchez, 2011n; Ha, 2011). Alternatively one can present the integrated decay rate over a $q^2$ range for which the lattice calculation is most reliable, typically from $q^2 = 16\ \text{GeV}^2$ to $q^2_{\text{max}} = (m_B - m_\pi)^2$:

**Table 17.1.12.** Coefficients $a_k$ and correlation matrix $\rho_{kl}$ of a 3-parameter BGL series expansion of $f_+^{B\pi}$ from Bernard et al. (2009b). Statistical and systematic errors are added in quadrature.

| Fit: | 0.0216(27) | −0.0378(191) | −0.113(27) |
|---|---|---|---|
| $\rho$ | $a_0$ | $a_1$ | $a_2$ |
| $a_0$ | 1.000 | 0.640 | 0.475 |
| $a_1$ | 0.640 | 1.000 | 0.964 |
| $a_2$ | 0.474 | 0.964 | 1.000 |

$$\Delta\zeta \equiv \frac{G_F^2}{24\pi^3} \int_{q_i^2}^{q_f^2} dq^2 \mathbf{p}_\pi^3 |f_+^{B\pi}(q^2)|^2 \,. \qquad (17.1.53)$$

The quantity $\Delta\zeta$ is given for both the Fermilab/MILC and HPQCD calculations in Table 17.1.13.

The 2006 HPQCD $B \to \pi$ form-factor calculation relies on the parameterization of Ball and Zwicky (2005a) (BZ) during an intermediate step to interpolate their data to fiducial values of the pion energy before performing the chiral extrapolation. Use of models such as the one in Becirevic and Kaidalov (2000), hereafter "BK"; or the BZ parameterization can lead to an underestimation in the quoted form-factor errors, particularly at low values of $q^2$ where the lattice data are poor or nonexistent and the shape is constrained primarily by the model function. Moreover, any comparisons between different theoretical or experimental determinations of the BK or BZ fit parameters are not necessarily meaningful, since any observed discrepancies could simply be due to limitations of the model. Hence only lattice QCD form factor determinations based on BGL-like series (such as also the BCL parameterization) should be considered model-independent.

### *Light-cone sum rule form-factor calculations*

The method of QCD light-cone sum rules allows one to calculate the $B \to \pi$ form factors at small and intermediate $q^2$ (see, *e.g.*, Bagan, Ball, and Braun, 1998; Ball and Zwicky, 2005c; Belyaev, Khodjamirian, and Ruckl, 1993; Duplancic, Khodjamirian, Mannel, Melic, and Offen, 2008; Khodjamirian, Ruckl, Weinzierl, and Yakovlev, 1997). The key element of the calculational procedure is

**Table 17.1.13.** Results for the integrated decay rate $\Delta\zeta = \Delta\Gamma^{B\to\pi l\nu}/|V_{ub}|^2$ from lattice QCD and light-cone sum rules. Statistical and systematic errors are added in quadrature.

| | $q^2$ (GeV$^2$) | $\Delta\zeta$ (ps$^{-1}$) |
|---|---|---|
| HPQCD (Dalgic et al., 2006) | $> 16$ | $2.02 \pm 0.55$ |
| Fermilab/MILC (Bailey et al., 2009) | $> 16$ | $2.21^{+0.47}_{-0.42}$ |
| LCSR (Khodjamirian, Mannel, Offen, and Wang, 2011) | $< 12$ | $4.59^{+1.00}_{-0.85}$ |

the correlator of the two heavy-light quark currents:

$$i\int d^4x e^{iqx}\langle\pi^+(p)|\,T\{\overline{u}\gamma_\mu b(x), m_b\overline{b}i\gamma_5 d(0)\}\,|0\rangle$$
$$\equiv F((p+q)^2,q^2)p_\mu + \widetilde{F}((p+q)^2,q^2)q_\mu\,, \tag{17.1.54}$$

$$F((p+q)^2,q^2) = \frac{2m_B^2 f_B f_+^{B\pi}(q^2)}{m_B^2-(p+q)^2} + \dots\,, \tag{17.1.55}$$

where Eq. (17.1.55) represents the hadronic dispersion relation for the amplitude $F$, with the ground-state $B$-meson contribution containing the vector $B\to\pi$ form factor multiplied by the $B$ decay constant. The remaining hadronic sum in Eq. (17.1.55) is indicated by ellipses. The amplitude $\widetilde{F}$ is used to calculate the scalar form factor $f_0^{B\pi}(q^2)$. At $(p+q)^2 \ll m_b^2$ and $q^2 \ll m_b^2$, the $T$-product in Eq. (17.1.54) is expanded near the light-cone $x^2\sim 0$, yielding process-independent nonlocal vacuum-pion matrix elements, such as $\langle\pi(p)|\overline{u}_\alpha(x)d_\beta(0)|0\rangle$. The light-cone OPE yields

$$F((p+q)^2,q^2) = \tag{17.1.56}$$
$$\sum_{t=2,3,4,\dots}\int Du_i \sum_{k=0,1,\dots}\left(\frac{\alpha_S}{\pi}\right)^k T_k^{(t)}((p+q)^2,q^2,u_i,m_b,\mu)\varphi_\pi^{(t)}(u_i,\mu)\,,$$

a convolution (at the factorization scale $\mu$) of calculable short-distance coefficient functions $T_k^{(t)}$ and universal pion light-cone distribution amplitudes (DA's) $\varphi_\pi^{(t)}(u_i,\mu)$ of growing twist $t\geq 2$. The integration goes over the momentum fractions $u_i = u_1, u_2, \dots$ of quarks and gluons in the pion. The terms in Eq. (17.1.56) corresponding to higher-twist pion DA's are suppressed by inverse powers of the $b$-quark virtuality $((p+q)^2-m_b^2)\sim\overline{\Lambda}m_b$, where $\overline{\Lambda}\gg\Lambda_{\rm QCD}$ does not scale with $m_b$. Currently Eq. (17.1.54) includes all LO contributions of the twist 2, 3, 4 quark-antiquark and quark-antiquark-gluon DA's of the pion and the NLO, $O(\alpha_S)$ corrections to the twist-2 and twist-3 two-particle coefficient functions.

Furthermore, one uses quark-hadron duality and approximates the sum over excited $B$-states in the hadronic dispersion relation by the quark-gluon spectral density $\mathrm{Im}\,F^{(\rm OPE)}(s,q^2)$ calculated from the OPE, Eq. (17.1.56), introducing the effective threshold parameter $s_0^B$. The final step involves a Borel transformation $(p+q)^2\to m^2\sim\overline{\Lambda}m_b$. The resulting LCSR for the $B\to\pi$ form factor has the following form

$$f_+^{B\pi}(q^2) = \left(\frac{e^{m_B^2/m^2}}{2m_B^2 f_B}\right)\frac{1}{\pi}\int_{m_b^2}^{s_0^B} ds\,\mathrm{Im}\,F^{(\rm OPE)}(s,q^2)\,e^{-s/m^2}. \tag{17.1.57}$$

The uncertainty introduced by the quark-hadron duality approximation is minimized by calculating the $B$ meson mass from the derivative of the same LCSR, thereby fixing $s_0^B$. Details of the method can be found in Duplancic, Khodjamirian, Mannel, Melic, and Offen (2008); an introductory review is in Colangelo and Khodjamirian (2000). The LCSR method and input was also successfully tested for $D\to\pi, K$ form factors by Khodjamirian, Klein, Mannel, and Offen (2009). The input includes $\alpha_S$ and the $b$ quark mass (in the $\overline{\rm MS}$ scheme), as well as the non-perturbative parameters of the pion DA's, *e.g.*, $f_\pi$ and the shape parameters (Gegenbauer moments) for the twist-2 pion DA $\varphi_\pi^{(2)}$. For the decay constant $f_B$ the QCD sum rule for the two-point correlator of $\overline{b}i\gamma_5 q$ currents is employed. More details on the numerical results and their uncertainties can be found in the most recent LCSR analysis by Khodjamirian, Mannel, Offen, and Wang (2011), predicting $f_+^{B\pi}(q^2)$ at $0\leq q^2 < 12\,\mathrm{GeV}^2$, in particular, $f_+^{B\pi}(0) = 0.28\pm 0.03$. Extrapolation to larger $q^2$ reveals a reasonable agreement with the lattice QCD results (see Figure 17.1.12). The most convenient quantity for the $|V_{ub}|$ determination is the integrated decay rate $\Delta\zeta$ defined in Eq. (17.1.53); the most recent LCSR result from Khodjamirian, Mannel, Offen, and Wang (2011) is given in Table 17.1.13. The estimated error corresponds to the quadratic sum of the uncertainties due to variations of the input parameters in LCSR. The largest individual errors originate from the uncertainties of the $\overline{\rm MS}$ quark masses ($m_{u,d}$ and $m_b$) and of the shape parameters in the pion twist-2 DA, as well as from the renormalization scale uncertainty. There is still room for improvement of the OPE in the future, *e.g.*, if one calculates the $O(\alpha_S^2)$ and twist-5, 6 corrections and gains a better control over the pion DA's. On the other hand, the systematic error due to the quark-hadron duality approximation cannot be completely eliminated from the LCSR calculation. Hence, with this method it seems not feasible to reach a precision at a few percent level foreseeable with the future improvements of the lattice QCD calculations.

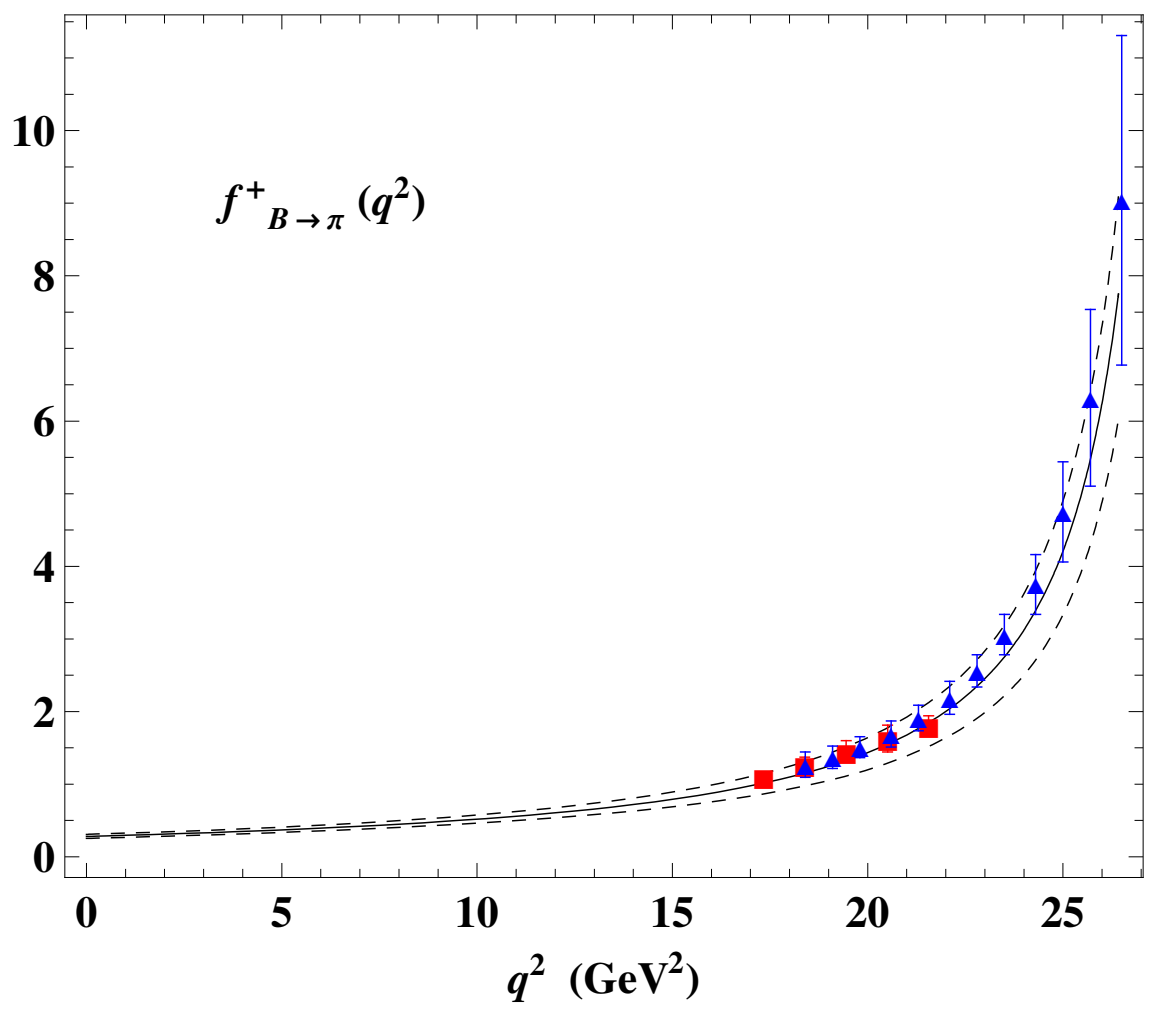

**Figure 17.1.12.** The vector form factor $f_+^{B\pi}(q^2)$ (in arbitrary units) calculated from LCSR (Khodjamirian, Mannel, Offen, and Wang, 2011) and fitted to the BCL parameterization from Bourrely, Caprini, and Lellouch (2009) (solid line) with uncertainties (dashed lines), compared to the LQCD results by HPQCD (Dalgic et al., 2006) (squares) and by FNAL/MILC (Bailey et al., 2009) (triangles with error bars).

#### 17.1.4.2 Measurements of Branching Fractions and $q^2$ Distributions

The semileptonic decay $B \to \pi\ell\nu$ has been studied with different experimental approaches at the $B$ Factories. The goal is a precise measurement of the branching fraction and the spectrum of the squared momentum transfer, $q^2$, to allow for a determination of the $q^2$ dependence of the $B \to \pi$ form factor. The main experimental challenge is the reduction of the much more abundant background from $B \to X_c\ell\nu$ decays, where $X_c$ is any hadronic final state with a charm quark. It is also difficult to separate $B \to \pi\ell\nu$ decays from the other $B \to X_u\ell\nu$ decays, where $X_u$ is a charmless hadronic final state, due to very similar decay kinematics. The $B \to \pi\ell\nu$ analyses are based on event samples with a tagged $B$ meson or on untagged event samples. In the tagged analyses, one of the two $B$ mesons in the $B\overline{B}$ event is either fully reconstructed in a hadronic decay mode or partially reconstructed in a semileptonic decay mode. While the tagged analyses provide a very clean environment, they are statistically limited for the $B$ Factory data samples. At present, untagged analyses, which were first performed by the CLEO collaboration (Athar et al., 2003), still provide the most precise results for $B \to \pi\ell\nu$.

In untagged analyses, the four-momentum of the undetected neutrino is inferred from the missing energy and momentum in the whole event. The reconstructed neutrino is combined with a charged lepton ($\ell = e, \mu$) and a pion to form a $B \to \pi\ell\nu$ candidate. The dominant background at low $q^2$ is due to $e^+e^- \to q\overline{q}$ ($q = u, d, s, c$) continuum events, where the charged lepton originates from a semileptonic decay of a produced hadron (mostly from $e^+e^- \to c\overline{c}$ events) or the misidentification of a charged hadron as a lepton. Continuum events produce jet-like event topologies and can thus be efficiently separated from the more isotropic $B\overline{B}$ events with selection criteria on event shape variables (*e.g.* $R_2$, $L_2$, $\cos\Delta\theta_{\rm thrust}$, see Chapter 9). The overall largest background comes from $B \to X_c\ell\nu$ decays. It is reduced by selection criteria on variables that are related to the neutrino reconstruction, *e.g.* the missing mass squared in the event or the polar angle of the missing momentum vector, or on kinematic variables, *e.g.* the helicity angle of the lepton. These variables also help to partially suppress the $B \to X_u\ell\nu$ background, which has a large uncertainty and limits the measurement at high $q^2$.

Three untagged analyses have been performed by the *BABAR* (del Amo Sanchez, 2011d,n) and Belle collaborations (Ha, 2011). The background suppression based on event shape, neutrino reconstruction and kinematical variables is optimized as a function of $q^2$ to allow for a precise measurement over the full $q^2$ range. While the Belle (Ha, 2011) and one of the *BABAR* (del Amo Sanchez, 2011d) analyses use one-dimensional selection criteria, the other *BABAR* measurement (del Amo Sanchez, 2011n) makes use of neural-network discriminators, which have been trained individually for each background class and $q^2$ interval, yielding an improved background suppression. In contrast to the other two analyses that focus on $B^0 \to \pi^-\ell^+\nu$ decays, this analysis includes a simultaneous measurement of $B^0 \to \pi^-\ell^+\nu$, $B^+ \to \pi^0\ell^+\nu$, $B^0 \to \rho^-\ell^+\nu$ and $B^+ \to \rho^0\ell^+\nu$ decays. By measuring these four decay modes, the uncertainties due to cross feed between these modes and various background contributions are reduced. In all three analyses the signal is extracted from a fit to the two-dimensional $\Delta E$-$m_{\rm ES}$ distribution. The fit is performed in several intervals of $q^2$ to measure the shape of the $q^2$ spectrum. The Belle analysis uses 13 $q^2$ intervals (Ha, 2011), the *BABAR* analyses 6 (del Amo Sanchez, 2011n) or 12 (del Amo Sanchez, 2011d) $q^2$ intervals. The shapes of the signal and background contributions are taken from simulation whereas the yields for the signal and the dominant background contributions are obtained from the fit. Figures 17.1.3 and 17.1.13 show the $m_{\rm ES}$ and $\Delta E$ projections from *BABAR* and Belle for a specific $q^2$ range, indicating the signal above the sum of backgrounds from several sources.

A number of tagged measurements have been performed by *BABAR* (Aubert, 2006r, 2008y) and Belle (Adachi, 2008a; Hokuue, 2007). They have led to a simpler and more precise reconstruction of the neutrino momentum and have low backgrounds and a uniform acceptance in $q^2$. This is achieved at the expense of much smaller signal samples which limit the statistical precision of the form-factor measurement. The semileptonic-tag measurements from *BABAR* and Belle use $B \to D^{(*)}\ell\nu$ decays to partially reconstruct one of the two $B$ mesons. They have a signal-to-background ratio of $\sim 2$ and yield $\sim 0.5$ signal decays per fb$^{-1}$. The signal is extracted from the distribution of the variable $\cos^2\phi_B$, where $\phi_B$ is the angle between the direction of either $B$ meson and the plane containing the momentum vectors of the tag-side $D^*\ell$ system and the signal-side $\pi\ell$ system (Aubert, 2008y; Hokuue, 2007). The hadronic-tag measurements yield fewer signal events, $\sim 0.1$ signal decays per ${\rm fb}^{-1}$, but reach signal-to-background ratios of up to $\sim 10$. The signal is extracted from the missing mass squared distribution, where the signal is expected to be located in a narrow peak near zero, as shown in Figure 17.1.14.

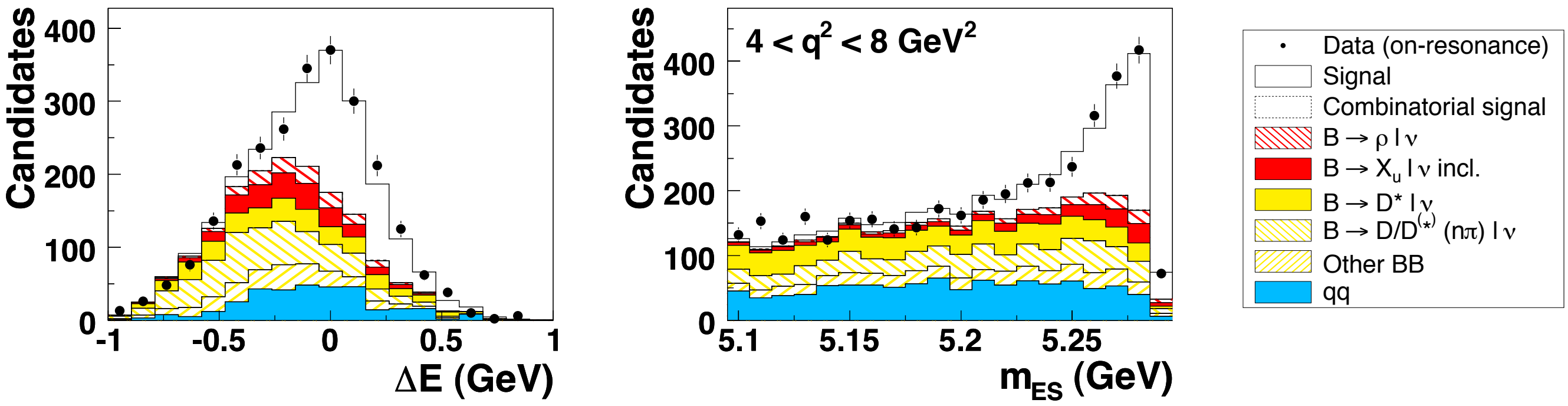

**Figure 17.1.13.** $m_{\rm ES}$ and $\Delta E$ distributions for the $q^2$ interval $4 < q^2 < 8\,{\rm GeV}^2$ from the *BABAR* untagged $B \to \pi\ell\nu$ measurement (del Amo Sanchez, 2011n).

Table 17.1.14 summarizes the signal yields, approximate signal-to-background ratios and integrated luminosities of the various measurements.

The leading experimental systematic uncertainties are associated with the reconstruction of charged and neutral particles, which affect the reconstruction of the missing momentum, with backgrounds from continuum events at low $q^2$ and from $B \to X_u\ell\nu$ decays at high $q^2$. Due to the feed-down from $B \to \rho\ell\nu$ decays, the uncertainties on the branching fraction and form factors for this decay mode also contribute to the systematic uncertainty. For the tagged measurements, the systematic uncertainties are about a factor of two smaller. They contribute to the knowledge of the total branching fraction, but their statistical precision is not yet sufficient to provide significant information on the shape of the $q^2$ spectrum.

Table 17.1.15 summarizes all $B \to \pi\ell\nu$ branching fraction measurements. Shown are the total branching fraction as well as the partial branching fractions for $q^2 < 12$ GeV$^2$ and $q^2 > 16$ GeV$^2$. Overall the individual measurements are in a good agreement, though for the tagged measurements the partial branching fractions at intermediate $q^2$ are somewhat smaller. A combination of all untagged $B \to \pi\ell\nu$ measurements from the $B$ Factories results in an average total branching fraction of $(1.44 \pm 0.03 \pm 0.05) \times 10^{-4}$, with a precision of $3-4\%$ (2% statistical and 3% systematic).

Figure 17.1.15 shows a fit of the $z$-expansion introduced in Section 17.1.4.1 to the measured $q^2$ spectra from all untagged $B \to \pi\ell\nu$ analyses, using the BCL parameterization with three parameters $(b_0, b_1, b_2)$. The results are summarized in Table 17.1.16. The $\chi^2$ probability of this fit is 1.1%. An inclusion of the tagged measurements would decrease the probability to 0.02%. This low probability is mostly due to the lower branching fractions from the tagged measurements. The *BABAR* measurement in 12 $q^2$ bins prefers a larger (negative) quadratic term and a smaller linear term in the $z$ expansion compared to the other two untagged analyses. The fitted function also determines the product $f_+(0)|V_{ub}|$, which for a given value of $|V_{ub}|$ can be compared with LCSR predictions of $f_+(0)$, the $B \to \pi$ form factor at $q^2 = 0$. The largest value of $f_+(0)|V_{ub}|$ from the individual measurements comes from the untagged *BABAR* measurement in 6 $q^2$ bins. For the combination of all untagged measurements, a value of $f_+(0)|V_{ub}| = (0.940 \pm 0.029) \times 10^{-3}$ is obtained. Combining this value with the $|V_{ub}|$ result obtained using the LCSR calculation (see Table 17.1.17) gives $f_+(0) = (0.27 \pm 0.03)$, in good agreement with the LCSR result, $f_+(0) = (0.28 \pm 0.02)$. A comparison of the fitted BCL parameterization with the shapes predicted by form-factor calculations from LQCD, LCSR or quark models like ISGW2. is presented in Figure 17.1.15 (right). It agrees best with the recent LCSR calculation (Khodjamirian, Mannel, Offen, and Wang, 2011) and deviates significantly from the ISGW2 quark model prediction.

### 17.1.4.3 Determination of $|V_{ub}|$

Two different methods have been used to determine $|V_{ub}|$ from the measured $B \to \pi\ell\nu$ differential decay rates. The more traditional approach relates the measured partial branching fractions, $\Delta\mathcal{B}(q^2_{\rm min}, q^2_{\rm max})$, with the normalized partial decay rate, $\Delta\zeta(q^2_{\rm min}, q^2_{\rm max})$, predicted by form-factor calculations integrated over a certain $q^2$ range. For LQCD calculations (Bailey et al., 2009; Dalgic et al., 2006), the range $q^2 > 16\,{\rm GeV}^2$ is used, and for the recent LCSR (Khodjamirian, Mannel, Offen, and Wang, 2011) calculation the range is $q^2 < 12\,{\rm GeV}^2$ . $|V_{ub}|$ is obtained from the relation

$$|V_{ub}| = \sqrt{\frac{\Delta B(q^2_{\rm min}, q^2_{\rm max})}{\tau_0 \Delta\zeta(q^2_{\rm min}, q^2_{\rm max})}}, \tag{17.1.58}$$

where $\tau_0 = (1.519 \pm 0.007)$ ps is the $B^0$ lifetime (Beringer et al., 2012). Table 17.1.17 shows the values of $\Delta B(q^2_{\rm min}, q^2_{\rm max})$, $\Delta\zeta(q^2_{\rm min}, q^2_{\rm max})$ and the $|V_{ub}|$ results for the three untagged $B \to \pi \ell \nu$ measurements and the averages of the untagged and tagged measurements, and for three form-factor calculations. The uncertainty on $|V_{ub}|$ is dominated by the theoretical form-factor uncertainty.

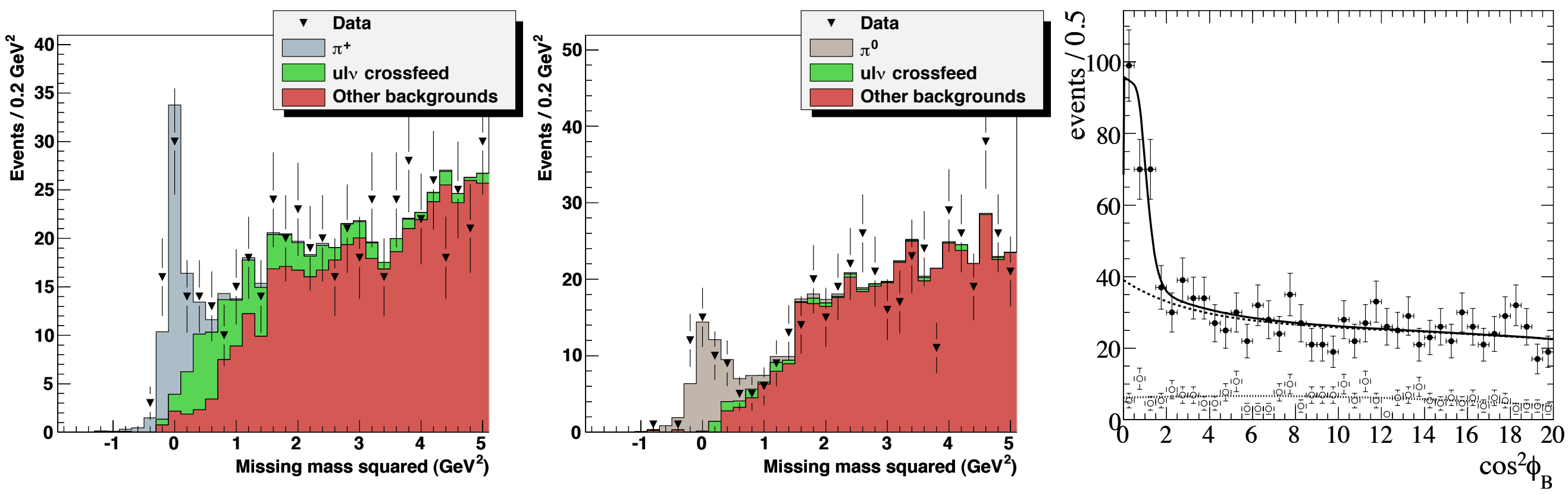

**Figure 17.1.14.** Missing mass squared distributions from the Belle tagged $B^0 \to \pi^-\ell^+\nu$ (left) and $B^+ \to \pi^0\ell^+\nu$ (center) measurements (Adachi, 2008a) and $\cos^2\phi_B$ distribution from the *BaBar* semileptonic-tag $B^0 \to \pi^-\ell^+\nu$ measurement (Aubert, 2008y) (right). In the right figure, the solid line represents the signal and the dotted and dashed lines represent the backgrounds with combinatorial and with correctly reconstructed $D$ mesons in the semileptonic tag, respectively.

**Table 17.1.14.** Integrated luminosity, signal yield and approximate signal-to-background ratio, $S/B$, for the $B \to \pi\ell\nu$ measurements.

| Measurement | Int. lumi. (fb$^{-1}$) | $N_{\rm sig}(B^0 \to \pi^-\ell^+\nu)$ | $N_{\rm sig}(B^+ \to \pi^0\ell^+\nu)$ | $S/B$ |
|---|---|---|---|---|
| *BaBar* untagged (6 bins) (del Amo Sanchez, 2011n) | 349 | 7181 | 3446 | $\sim 0.2$ |
| *BaBar* untagged (12 bins) (del Amo Sanchez, 2011d) | 423 | 11778 | – | $\sim 0.1$ |
| Belle untagged (Ha, 2011) | 605 | 21486 | – | $\sim 0.1$ |
| *BaBar* semileptonic tag (Aubert, 2008y) | 348 | 150 | 134 | $\sim 2$ |
| Belle semileptonic tag (Hokuue, 2007) | 253 | 156 | 69 | $\sim 2$ |
| *BaBar* hadronic tag (Aubert, 2006r) | 211 | 31 | 26 | $\sim 10$ |
| Belle hadronic tag (Adachi, 2008a) | 605 | 59 | 49 | $\sim 10$ |

**Table 17.1.15.** Branching fractions for $B^0 \to \pi^-\ell^+\nu$. The two untagged *BaBar* measurements are assumed to be statistically independent since the selected data samples have less than 1% of the events in common (del Amo Sanchez, 2011d).

| Measurement | $\mathcal{B}_{\rm tot}$ $(10^{-4})$ | $\Delta\mathcal{B}(q^2 < 12\,{\rm GeV}^2)$ $(10^{-4})$ | $\Delta\mathcal{B}(q^2 > 16\,{\rm GeV}^2)$ $(10^{-4})$ |
|---|---|---|---|
| *BaBar* untagged (6 bins) | $1.41 \pm 0.05 \pm 0.07$ | $0.88 \pm 0.03 \pm 0.05$ | $0.32 \pm 0.02 \pm 0.02$ |
| *BaBar* untagged (12 bins) | $1.42 \pm 0.05 \pm 0.07$ | $0.84 \pm 0.03 \pm 0.04$ | $0.33 \pm 0.02 \pm 0.03$ |
| Belle untagged | $1.49 \pm 0.04 \pm 0.07$ | $0.83 \pm 0.02 \pm 0.04$ | $0.40 \pm 0.02 \pm 0.02$ |
| Average untagged | $1.44 \pm 0.03 \pm 0.05$ | $0.84 \pm 0.02 \pm 0.03$ | $0.36 \pm 0.01 \pm 0.02$ |
| Average tagged | $1.31 \pm 0.08 \pm 0.06$ | $0.67 \pm 0.06 \pm 0.03$ | $0.37 \pm 0.04 \pm 0.02$ |
| Average | $1.42 \pm 0.03 \pm 0.05$ | $0.81 \pm 0.02 \pm 0.03$ | $0.36 \pm 0.01 \pm 0.02$ |

**Table 17.1.16.** Results of the fits of the BCL parameterization with 3 parameters to the measured $\Delta B/\Delta q^2$ distribution.

| Measurement | $\chi^2$/ndf | Prob($\chi^2$/ndf) | Fit parameters | $f_+(0)\lvert V_{ub}\rvert$ $(10^{-3})$ |
|---|---|---|---|---|
| *BaBar* (6 bins) | 6.0/3 | 11.2% | $b_1/b_0 = -0.90 \pm 0.45$ | $1.090 \pm 0.055$ |
| | | | $b_2/b_0 = +0.47 \pm 1.49$ | |
| *BaBar* (12 bins) | 4.1/9 | 90.5% | $b_1/b_0 = +0.09 \pm 0.53$ | $0.863 \pm 0.044$ |
| | | | $b_2/b_0 = -4.65 \pm 1.55$ | |
| Belle | 11.9/10 | 29.4% | $b_1/b_0 = -1.31 \pm 0.27$ | $0.914 \pm 0.040$ |
| | | | $b_2/b_0 = -0.79 \pm 0.91$ | |
| *BaBar* +Belle | 48.0/28 | 1.1% | $b_1/b_0 = -0.75 \pm 0.22$ | $0.940 \pm 0.029$ |
| | | | $b_2/b_0 = -1.84 \pm 0.69$ | |

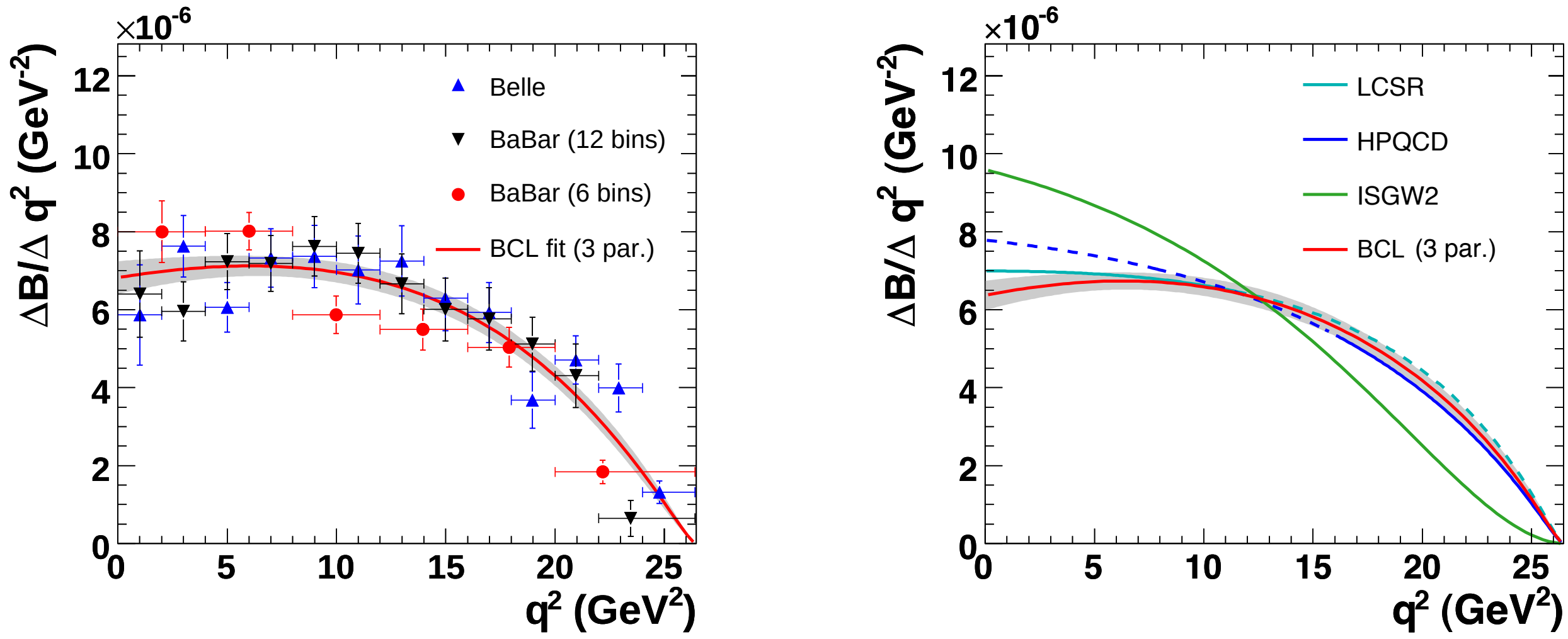

**Figure 17.1.15.** Left: Fit of the BCL parameterization with 3 parameters to the measured $B \to \pi \ell \nu$ $q^2$ distribution. The uncertainty of the fit is shown as shaded error band. Right: Comparison of the fit result with form-factor predictions from HPQCD (Dalgic et al., 2006), LCSR (Khodjamirian, Mannel, Offen, and Wang, 2011) and ISGW2 (Scora and Isgur, 1995). The extrapolations of the predictions to the full $q^2$ range are shown as dashed lines.

The more recent method is based on a simultaneous fit to the measured $q^2$ spectra and the LQCD predictions. The BCL parameterization is used as parameterization for $f_+(q^2)$ over the whole $q^2$ range to minimize the model dependence of the form factor. This method makes use of the full shape information from data and the shape and normalization from theory, which results in a reduced uncertainty on $|V_{ub}|$.

The combined fit to the FNAL/MILC lattice calculations and the data from the three untagged measurements yields $|V_{ub}| = (3.23 \pm 0.30) \times 10^{-3}$. Figure 17.1.16 and Table 17.1.17 show the results of the fit. Only four of the twelve FNAL/MILC points have been included in the fit, avoiding LQCD points with a correlation higher than 80%. This reduction of the theoretical input does not change the $|V_{ub}|$ result but leads to a better agreement of the fitted curve with the lattice points. The fit results for the parameters in the BCL parameterization are $b_1/b_0 = -0.82 \pm 0.20$ and $b_2/b_0 = -1.63 \pm 0.62$, and a value of $f_+(0)|V_{ub}| = 0.945 \pm 0.028$ is obtained. The $\chi^2$ probability of the fit is 2.2% ($\chi^2/ndf = 58.9/31$). The $|V_{ub}|$ values obtained from fits to the individual untagged measurements agree with each other within about one standard deviation. The total uncertainty of $|V_{ub}|$ is about 9%. The contributions to this uncertainty have been estimated to be 3% from the branching fraction measurements, 4% from the shapes of the $q^2$ spectra determined from data, and 8% from the form-factor normalization obtained from theory. Using the HPQCD lattice calculation gives similar fit results. However, at present no information on the correlation of the HPQCD points is available and therefore only one point can be used in the fit to determine the normalization of the decay rate, which results in larger uncertainties.

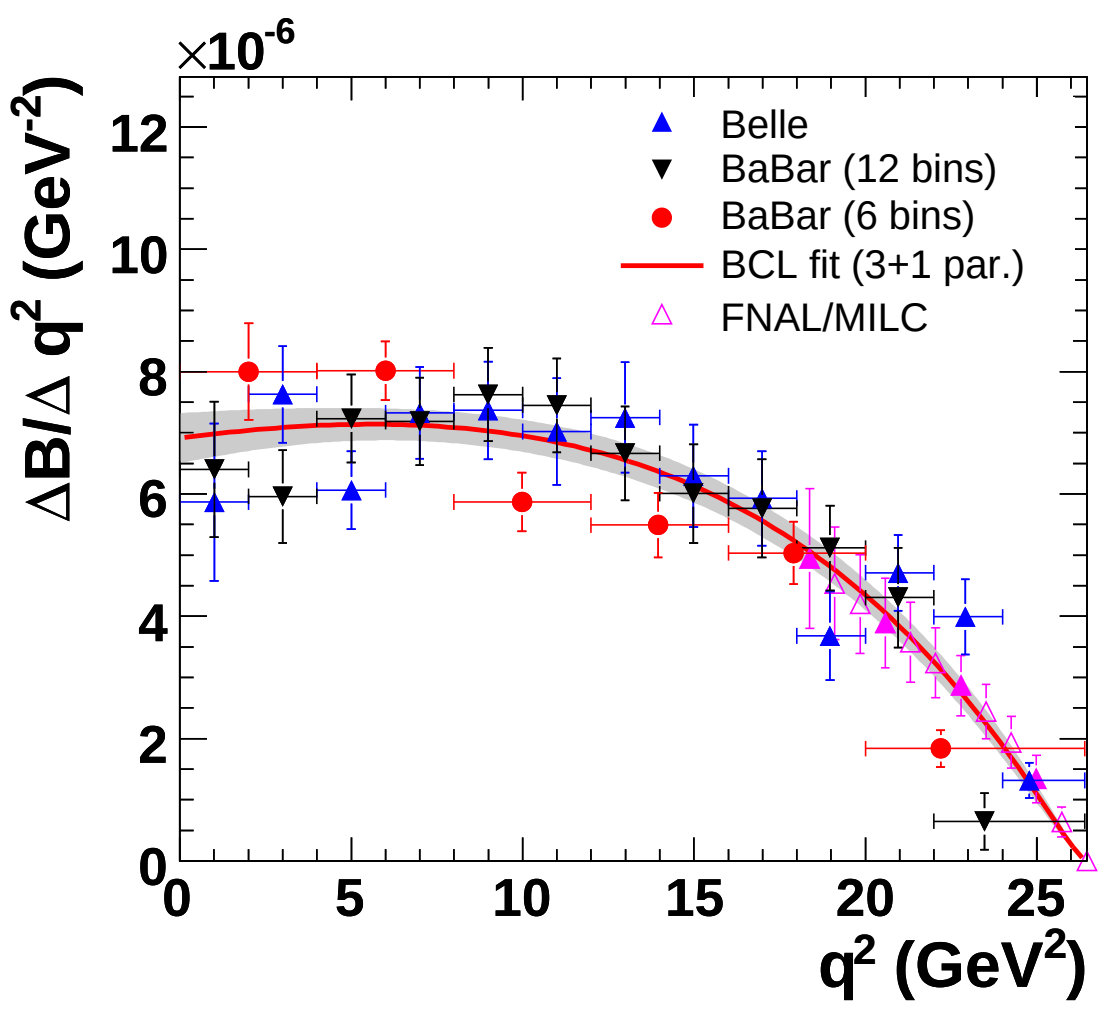

**Figure 17.1.16.** Simultaneous fit of the BCL parameterization to the measured $q^2$ spectra and to four of the twelve points of the FNAL/MILC calculation (magenta, closed triangles). The FNAL/MILC prediction has been rescaled to the data according to the $|V_{ub}|$ value obtained in the fit.

As a final result for $|V_{ub}|$ from $B \to \pi \ell \nu$ decays we quote the value obtained from the simultaneous fit to the three untagged measurements from *BABAR* and Belle, combined with the FNAL/MILC calculation:

$$|V_{ub}|_{\rm excl} = (3.23 \pm 0.30) \times 10^{-3}. \qquad (17.1.59)$$

Future improvements for $|V_{ub}|$ will rely on progress in form-factor calculations based on LQCD or LCSR and on more precise experimental determinations of the $q^2$ spectrum in $B \to \pi\ell\nu$ decays. In particular an improved precision in the high $q^2$ region, where LQCD predictions exist, would be important. This will require a better understanding of the composition and dynamics of the $B \to X_u\ell\nu$ background and significantly larger data samples for tagged event samples expected at the next generation of $B$ Factories.

**Table 17.1.17.** $|V_{ub}|$ derived from $B \to \pi\ell\nu$ decays for various $q^2$ regions and form-factor calculations: LCSR (Khodjamirian, Mannel, Offen, and Wang, 2011), HPQCD (Dalgic et al., 2006), FNAL/MILC (Bailey et al., 2009). The quoted errors on $|V_{ub}|$ are due to experimental uncertainties and theoretical uncertainties on $\Delta\zeta$. The last column shows the $|V_{ub}|$ results of the simultaneous fits to data and the FNAL/MILC prediction. Here the stated error represents the combined experimental and theoretical uncertainty.

| | LCSR | HPQCD | FNAL/MILC | FNAL/MILC fit |
|---|---|---|---|---|
| $\Delta\zeta$ ($\mathrm{ps}^{-1}$) | $4.59^{+1.00}_{-0.85}$ | $2.02\pm0.55$ | $2.21^{+0.47}_{-0.42}$ | $2.21^{+0.47}_{-0.42}$ |
| $q^2$ range ( $\mathrm{GeV}^2$) | $0-12$ | $16-26.4$ | $16-26.4$ | $16-26.4$ |
| Experiment | | $\lvert V_{ub}\rvert$ ($10^{-3}$) | | |
| BABAR (6 bins) | $3.54\pm0.12^{+0.38}_{-0.33}$ | $3.22\pm0.15^{+0.55}_{-0.37}$ | $3.08\pm0.14^{+0.34}_{-0.28}$ | $2.98\pm0.31$ |
| BABAR (12 bins) | $3.46\pm0.10^{+0.37}_{-0.32}$ | $3.26\pm0.19^{+0.56}_{-0.37}$ | $3.12\pm0.18^{+0.35}_{-0.29}$ | $3.22\pm0.31$ |
| Belle | $3.44\pm0.10^{+0.37}_{-0.32}$ | $3.60\pm0.13^{+0.61}_{-0.41}$ | $3.44\pm0.13^{+0.38}_{-0.32}$ | $3.52\pm0.34$ |
| BABAR +Belle | $3.47\pm0.06^{+0.37}_{-0.32}$ | $3.43\pm0.09^{+0.59}_{-0.39}$ | $3.27\pm0.09^{+0.36}_{-0.30}$ | $3.23\pm0.30$ |
| Tagged | $3.10\pm0.16^{+0.33}_{-0.29}$ | $3.47\pm0.23^{+0.60}_{-0.39}$ | $3.32\pm0.22^{+0.37}_{-0.31}$ | $3.33\pm0.39$ |

### 17.1.5 Inclusive Cabibbo-suppressed $B$ decays

#### 17.1.5.1 Theoretical Overview

The theoretical description of inclusive $B \to X_u\ell\nu$ decays rests on the same basic principles as that of inclusive $B \to X_c\ell\nu$ decays described in Section 17.1.3.1. Due to the inclusive nature of the process, the only sensitivity to long-distance dynamics comes from the $B$ meson in the initial state. The total $B \to X_u\ell\nu$ rate is given by an OPE in terms of local operators, which has a similar structure as that for the $B \to X_c\ell\nu$ rate, with non-perturbative corrections first appearing at $O(1/m_b^2)$.

In practice, the experimental sensitivity to $B \to X_u\ell\nu$ and $|V_{ub}|$ is highest in the region of phase space that is less impacted by the dominant $B \to X_c\ell\nu$ background, namely the region where the hadronic $X_u$ system has invariant mass $m_X$ below the mass of the lightest charm meson, $m_X \lesssim m_D$. In this phase-space region non-perturbative corrections are kinematically enhanced, and as a result the non-perturbative dynamics of the decaying $b$ quark inside the $B$ meson becomes an $O(1)$ effect.

In addition to the lepton energy, $E_\ell$, convenient variables to describe the decay kinematics are the hadronic variables

$$p_X^+ = E_X - |\boldsymbol{p}_X|\,, \qquad p_X^- = E_X + |\boldsymbol{p}_X|\,, \qquad (17.1.60)$$

where $E_X$ and $\boldsymbol{p}_X$ are the energy and momentum of the hadronic system in the $B$-meson rest frame. In terms of these variables, the total hadronic and leptonic invariant masses are given by

$$m_X^2 = p_X^+ p_X^-\,, \quad q^2 = (m_B - p_X^+)(m_B - p_X^-)\,. \quad (17.1.61)$$

The fully differential decay rate is given by

$$\frac{d^3\Gamma}{dp_X^+\, dp_X^-\, dE_\ell} = \frac{G_F^2|V_{ub}|^2}{192\pi^3}\int dk\, C(E_\ell, p_X^-, p_X^+, k)\, F(k) + O\Big(\frac{\Lambda_{\mathrm{QCD}}}{m_b}\Big)\,. \qquad (17.1.62)$$

The coefficient $C(E_\ell, p_X^-, p_X^+, k)$ describes the quark decay $b \to u\ell\nu$ and can be computed in QCD perturbation theory. The "shape-function" $F(k)$ is a non-perturbative function. It describes the momentum distribution of the $b$ quark in the $B$ meson (Bigi, Shifman, Uraltsev, and Vainshtein, 1994; Neubert, 1994a). For $p_X^+ \sim k \sim \Lambda_{\mathrm{QCD}}$, which includes a large portion of the small $m_X$ region, the full non-perturbative shape of $F(k)$ is necessary to obtain an accurate description of the differential decay rate. On the other hand, in the limit $p_X^+ \gg k \sim \Lambda_{\mathrm{QCD}}$, only the first few moments of $F(k)$ are needed. Typically, the experimental measurements can lie anywhere between these two kinematic regimes.

There are several sources of uncertainties in the theoretical predictions that must be considered. First, there are perturbative uncertainties in the calculation of $C$ due to unknown higher-order corrections. Second, there are parametric uncertainties due to the imprecise knowledge of inputs, in particular the $b$-quark mass and $F(k)$. The total decay rate scales like $m_b^5$, while partial rates restricted to the small $m_X$ region typically exhibit an even stronger dependence on $m_b$. The first few moments of $F(k)$ are determined by $m_b$ and the expectation values of local operators that are constrained by fits to $B \to X_c\ell\nu$ moments. A substantial part of the $m_b$ dependence enters indirectly via the first moment of $F(k)$. Depending on the kinematic cuts, the shape of $F(k)$ (beyond what is encoded in its first few moments) can also have a significant influence on the predictions. An important consistency check for the overall shape of $F(k)$ is to give a reasonable description of the measured shape of the photon-energy spectrum in inclusive $B \to X_s\gamma$ decays (see Section 17.9), which at leading order in $1/m_b$ is given in terms of the same function $F(k)$ via an expression analogous to Eq. (17.1.62).

In addition to the leading shape function $F(k)$, several additional shape functions appear at $O(\Lambda_{\rm QCD}/m_b)$ (Bauer, Luke, and Mannel, 2003). Apart from their first few moments, very little is known about the form of these subleading shape functions. They thus introduce an uncertainty in the theoretical predictions that is hard to quantify in a systematic fashion. An even larger number of unknown shape functions appears at $O(\alpha_S\Lambda_{\rm QCD}/m_b)$ (Lee and Stewart, 2005).

Weak annihilation contributions could have a large impact at large $q^2$ and might be another source of theoretical uncertainties. However, recent analyses (Bigi, Mannel, Turczyk, and Uraltsev, 2010; Gambino and Kamenik, 2010; Ligeti, Luke, and Manohar, 2010) have used CLEO-c data to constrain contributions from weak annihilation, resulting in a rather small impact. The corresponding uncertainty is below 2% for the total rate, translating into an uncertainty of less than 1% on $|V_{ub}|$ for the most inclusive analyses.

For the determination of $|V_{ub}|$ theoretical predictions by different groups are in use. A more detailed summary and comparison can be found elsewhere (Antonelli et al., 2010a). At their core, the different calculations are all based on Eq. (17.1.62), but they differ in the treatment of the perturbative and non-perturbative contributions.

The BLNP approach (Bosch, Lange, Neubert, and Paz, 2004; Lange, Neubert, and Paz, 2005) preferentially treats the kinematic region $p_X^+ \ll p_X^-$ where the $p_X^+$ and $p_X^-$ dependences of $C$ factorize. This allows for the resummation of Sudakov double logarithms of $p_X^+/p_X^-$ and $p_X^+/m_B$ to NNLL. They also include the full $O(\alpha_S)$ corrections, for which the perturbative expansions are performed using the so-called shape-function scheme for $m_b$, and a subset of the perturbative corrections in $C$ are absorbed into $F(k)$. The subleading shape functions are separately modeled and included in the predictions.

The GGOU approach (Gambino, Giordano, Ossola, and Uraltsev, 2007) treats the $p_X^+ \ll p_X^-$ and $p_X^+ \sim p_X^-$ regions on the same footing. The coefficient $C$ is computed at fixed order to $O(\alpha_S)$ and $O(\alpha_S^2\beta_0)$ (Gambino, Gardi, and Ridolfi, 2006), where the perturbative expansion is performed using the kinetic scheme for $m_b$. In this case no resummation effects at small $p_X^+$ are included. The effect of resummation as well as all contributions from subleading shape functions are absorbed into $F(k)$. This results in three non-universal distribution functions $F_i(k, q^2)$, which have subleading dependence on $q^2$.

In the dressed-gluon exponentiation (DGE) approach (Andersen and Gardi, 2006; Gardi, 2008) the perturbative expansion includes the NNLL resummation in moment space as well as the full $O(\alpha_S)$ and $O(\alpha_S^2\beta_0)$ corrections. It also incorporates an internal resummation of running coupling corrections in the Sudakov exponent. This approach effectively corresponds to a perturbative model for the leading shape function, with non-perturbative corrections only included via its moments. Therefore, it tends to be more predictive than the other approaches, resulting in smaller theoretical uncertainties within the framework. However, the intrinsic uncertainties due the assumptions inherent in the framework are not estimated. Another approach based on Sudakov resummation has been proposed in (Aglietti, Di Lodovico, Ferrera, and Ricciardi, 2009). It employs the so-called analytic coupling in the infrared.

The full $O(\alpha_S^2)$ corrections to the $b \to u\ell\nu$ spectrum are only known in the limit $p_X^+ \ll p_X^-$ (Greub, Neubert, and Pecjak, 2010), and are currently not included in the determination of $|V_{ub}|$. In case of BLNP, their effect turns out to be much larger than expected from the perturbative uncertainties at $O(\alpha_S)$, resulting in an increase of $|V_{ub}|$ by 8%. On the other hand, the $O(\alpha_S^2\beta_0)$ terms often dominate the $O(\alpha_S^2)$ corrections, and their inclusion in the GGOU and DGE approaches does not lead to similarly large corrections. A resolution of this apparent discrepancy will probably have to await a calculation of the complete $O(\alpha_S^2)$ corrections.

All the above approaches choose specific model parameterizations of the shape function(s), and it is unclear to what extent the model variations used to estimate the shape function uncertainties reflect the actual limited knowledge of their form, particularly at subleading order in $1/m_b$. Also, the theoretical uncertainties do not include explicit estimates of the possible size of $O(\alpha_S\Lambda_{\rm QCD}/m_b)$ corrections.

Given all the above, it is possible that the theoretical uncertainties currently quoted for $|V_{ub}|$ might be underestimated. On the other hand, the different theoretical frameworks yield values of $|V_{ub}|$ that are compatible within uncertainties with each other and across a variety of different experimental cuts.

Imposing an additional lower cut on $q^2$ restricts the decay kinematics to the part of the small $m_X$ region where $p_X^+ \sim p_X^-$. Formally, this allows the application of the OPE in terms of local operators (Bauer, Ligeti, and Luke, 2001). In practice, the resulting OPE still has rather large $1/m_b^2$ and higher order corrections, and some residual shape-function effects must be included. Nevertheless, this approach provides an important cross check on the extracted value of $|V_{ub}|$.

In some recent experimental analyses the phase-space restrictions have been relaxed and up to 90% of the total inclusive $B \to X_u\ell\nu$ rate is measured. In principle, this makes it possible to use a simpler theoretical description based on the local OPE only. Consequently, the main theoretical uncertainties are due to $m_b$ and higher-order perturbative corrections. In practice, these analyses still make explicit use of the theoretical description of the signal shape in the shape-function region to determine the experimental reconstruction efficiencies, and the associated theoretical uncertainties contribute via the experimental systematic uncertainties. Nevertheless, the fact that the resulting values of $|V_{ub}|$ are consistent with the other anal-yses enhances the confidence in our current understanding of inclusive $B \to X_u \ell\nu$ decays.

Recently, an improved treatment of the shape function has been developed (Ligeti, Stewart, and Tackmann, 2008), which combines the advantages of the BLNP and GGOU approaches and uses appropriate basis functions to approximate the shape function. It is expected that this procedure will allow for a combined global fit to all available inclusive $B \to X_s\gamma$ and $B \to X_u \ell\nu$ measurements (Bernlochner et al., 2011). As in the determination of $|V_{cb}|$ from inclusive $B \to X_c \ell\nu$ decays, a global fit has the advantage that the input parameters, such as $F(k)$ and $m_b$, are directly constrained by data and are determined together with $|V_{ub}|$.

#### 17.1.5.2 Measurements of Partial Branching Fractions

The observation of charged leptons with momenta exceeding the kinematic limit for $B \to X_c \ell\nu$ decays by the CLEO Collaboration (Bartelt et al., 1993) was the first evidence for charmless semileptonic decays. Since then, a series of measurements near the kinematic limit have been performed (Bornheim et al., 2002; Limosani, 2005; Aubert, 2006x); they differ in the kinematic selection and the size of the data sample. At lower lepton momenta, the background from $B \to X_c \ell\nu$ increases sharply to more than 10 times the signal and the dominant uncertainty arises from the subtraction of the sum of lepton spectra from exclusive $B \to X_c \ell\nu$ decays, for which the branching fractions and form factors are known to different degrees. The signal-to-background ratio can be substantially improved by combining the high energy lepton with a measurement of the missing neutrino in the event, but this can only be achieved with a substantial reduction in the selection efficiency (Aubert, 2005h).

Experimenters simulate the charmless semileptonic $B \to X_u \ell\nu$ decays as a hybrid, *i.e.*, a combination of two components: three-body decays involving a single low-mass charmless meson, $\pi, \rho, \eta, \eta'$, or $\omega$, and decays to non-resonant multi-body hadronic final states. The three-body decays make up about the 20% of the charmless semileptonic decay rate, and their simulation is based on OPE calculations and form-factor measurements and measured branching fractions (Beringer et al., 2012). The generated mass distribution and kinematics of multi-body hadronic states $X_u$ are based on the prescription by De Fazio and Neubert (De Fazio and Neubert, 1999). The fragmentation of $X_u$ into final state hadrons are simulated by using `Jetset` (Sjöstrand, 1994). The two components are combined so that the cumulative distributions of the hadronic mass, the momentum transfer squared, and the lepton momentum reproduce OPE predictions. The generated distributions are often reweighted to accommodate specific choices of the parameters for the inclusive and exclusive decays. The overall normalization is adjusted to reproduce the measured inclusive charmless branching fraction (Beringer et al., 2012).

An example of the extraction of the signal yield is illustrated in Figure 17.1.17 (Aubert, 2006x), showing the observed spectra of the highest momentum electron in events recorded on and below the $\Upsilon(4S)$ resonance. The data collected on the $\Upsilon(4S)$ resonance include contributions from $B\overline{B}$ events and continuum events. The latter is subtracted using off-resonance data, collected below the $B\overline{B}$ production threshold, and on-resonance data with lepton momenta above 2.8 GeV, *i.e.*, well above the endpoint for semileptonic $B$ decays. The principal challenge is the subtraction of the electron spectrum from $B$-meson decays which is dominated by various $B \to X_c \ell\nu$ decays. Hadronic $B$ decays contribute mostly via hadron misidentification and secondary electrons from decays of $D$, $J/\psi$, and $\psi(2S)$ mesons. The signal contribution is determined from a $\chi^2$ fit of the observed inclusive electron spectrum to the sum of Monte Carlo (MC) simulated signal and individual background contributions. The relative normalization factors for signal and background distributions are free parameters of the fit.

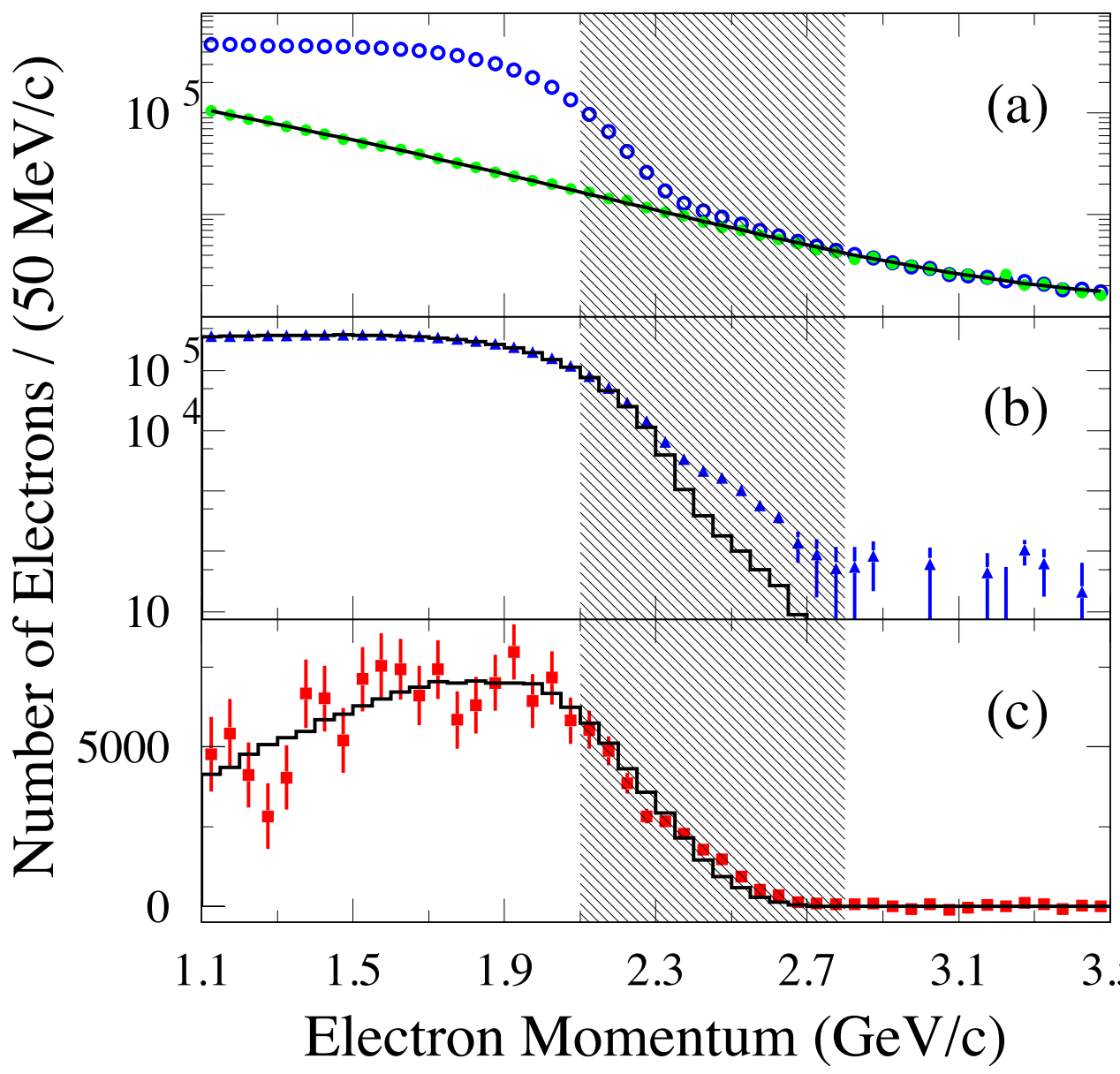

**Figure 17.1.17.** *BABAR* analysis of the electron momentum spectra in the $\Upsilon(4S)$ rest frame (Aubert, 2006x): (a) on-resonance data (open circles - blue), scaled off-resonance data (solid circles - green); the solid line shows the result of the fit to the non-$B\overline{B}$ events using both on- and off-resonance data; (b) on-resonance data after subtraction of the fitted non-$B\overline{B}$ background (triangles - blue) compared to simulated $B\overline{B}$ background (histogram) that is adjusted by a combined fit to the on- and off-resonance data; (c) on-resonance data after subtraction of all backgrounds (data point - red), compared to the simulated $B \to X_u e\nu$ signal spectrum (histogram); the error bars indicate errors from the fit, which include the uncertainties in the fitted yields for continuum and $X_c e\nu$ backgrounds. The shaded area indicates the momentum interval for which the on-resonance data are combined into a single bin for the purpose of reducing the sensitivity of the fit to the simulated shape of the signal spectrum in this region.

In this analysis, a potential bias of the fitted yield from the assumed shape of the signal spectrum is reduced by combining the on-resonance data for the interval from 2.1 to 2.8 GeV in a single bin. The lower limit of this bin is chosen so as to retain the sensitivity to the steeply falling $B\overline{B}$ background distributions, while containing a large fraction of the signal events in a region where the background is low.

In total, the selected sample includes $610 \times 10^3$ electrons, from which roughly 6.5% have been extracted as the signal yield in the momentum interval $2.0 - 2.6$ GeV. This translates to a partial branching fraction of $\Delta\mathcal{B}(B \to X_u e\nu) = (0.572 \pm 0.041 \pm 0.051) \times 10^{-3}$. Here the first error is statistical and the second is the total systematic error. The systematic error includes the uncertainty in the assumed shape of the signal spectrum. The gain in precision compared to earlier analyses of the lepton spectrum near the kinematic endpoint can be attributed to higher statistics, and to improved background estimates. While earlier measurements were restricted to lepton energies close to the kinematic endpoint for $B \to X_c \ell\nu$ decays at 2.3 GeV and covered only 10% of the $B \to X_u \ell\nu$ spectrum, this and other more recent measurements have been extended to lower momenta, thus covering about 25% to 35% of the spectrum (see Table 17.1.18).

More recently, the large data samples accumulated at the $B$ Factories have enabled studies of $B\overline{B}$ event samples tagged by the full reconstruction of the hadronic decays of one of the $B$ mesons. An electron or muon with momentum $p^*_\ell > 1\,\text{GeV}$ in the CM system is taken as a signature for a semileptonic decay of the second $B$ meson. The overall event rate is low due to the low tag efficiency, but the combinatorial backgrounds are substantially reduced allowing the extension of the acceptance for signal events to 90% of the remaining phase space. The tag decay determines the CM momentum and charge of the recoiling signal $B$ decay, and permits the reconstruction of hadronic observables with good resolution. Of particular relevance are $q^2$ and $m_X$, the mass of the hadronic state $X$. The systematic uncertainties related to the tag efficiency largely cancel in the measurement of the ratio of event yields for selected charmless semileptonic decays relative to all $B \to X\ell\nu$ decays. Corrections to the signal yield account for a possible difference in the tagging efficiency in the presence of a signal $B \to X_u\ell\nu$ decay or generic semileptonic decay. The combinatorial background of the tag decay is subtracted by fits to the $m_{\rm ES}$ distributions. Other backgrounds originate from secondary $B \to X \to \ell$ decays and hadron misidentified as leptons, primarily muons. The dominant $B \to X_c\ell\nu$ background is reduced by vetoing kaons from charm particle decays and low-momentum pions from $D^* \to D\pi$ decays. Events with additional missing particles result in large values of the missing mass squared $m^2_{\rm miss}$ and are rejected. This not only reduces the backgrounds, but also improves the resolution of the reconstructed variables describing the signal decays. In particular, the hadronic variable $P_+ = p^+_X$ is sensitive to detector resolution and the background modeling. The normalization of the remaining $B \to X_c\ell\nu$ background is determined from fits to the observed inclusive spectra of different kinematic variables.

Using the hadron-tagged $B\overline{B}$ events, Belle (Bizjak, 2005; Urquijo, 2010) and *BABAR* (Aubert, 2008ac; Lees, 2012x) have measured partial decay rates. The *BABAR* measurements are based on the full dataset of 467 million produced $B\overline{B}$ events, whereas the Belle results are based on 275 million (Bizjak, 2005) and 657 million (Urquijo, 2010) produced $B\overline{B}$ pairs, respectively. Figure 17.1.18 shows *BABAR* data and results of fits (Lees, 2012x) to four different kinematic distributions of $B \to X_u\ell\nu$ decays, performed to extract the partial branching fractions. These branching fractions are listed in Table 17.1.19 for tagged data samples from *BABAR* and Belle. Unless stated otherwise, the minimum lepton momentum is 1 GeV. The listed branching fractions and extraction of $|V_{ub}|$ are based on fits to the distributions of the variables listed in the first column with the specific restrictions imposed. For the *BABAR* and Belle results listed in the last line, no additional restriction is imposed, and the results agree very well within the stated errors.

These most recent analyses by Belle (Urquijo, 2010) and *BABAR* (Lees, 2012x), based on their full data samples, use a two-dimensional fit to $m_X$ versus $q^2$ to extract the branching fraction. Figures 17.1.19 and 17.1.20 show the Belle and *BABAR* data and fit results. The *BABAR* selection of the signal candidates is cut-based, whereas Belle employs a nonlinear multivariate discriminator, a boosted decision tree. For the two analyses, the statistical and systematic errors on the branching fractions are comparable in size ($\simeq 7-9\%$). The systematic uncertainties are dominated by the simulation of the signal decays; in particular, they are sensitive to the shape function and the $b$-quark mass. The average of these two branching fraction measurements, assuming full correlation of the uncertainty in the predicted signal spectrum, is $\Delta\mathcal{B}(p^*_\ell > 1\,\text{GeV}) = (1.87 \pm 0.10 \pm 0.11) \times 10^{-3}$.

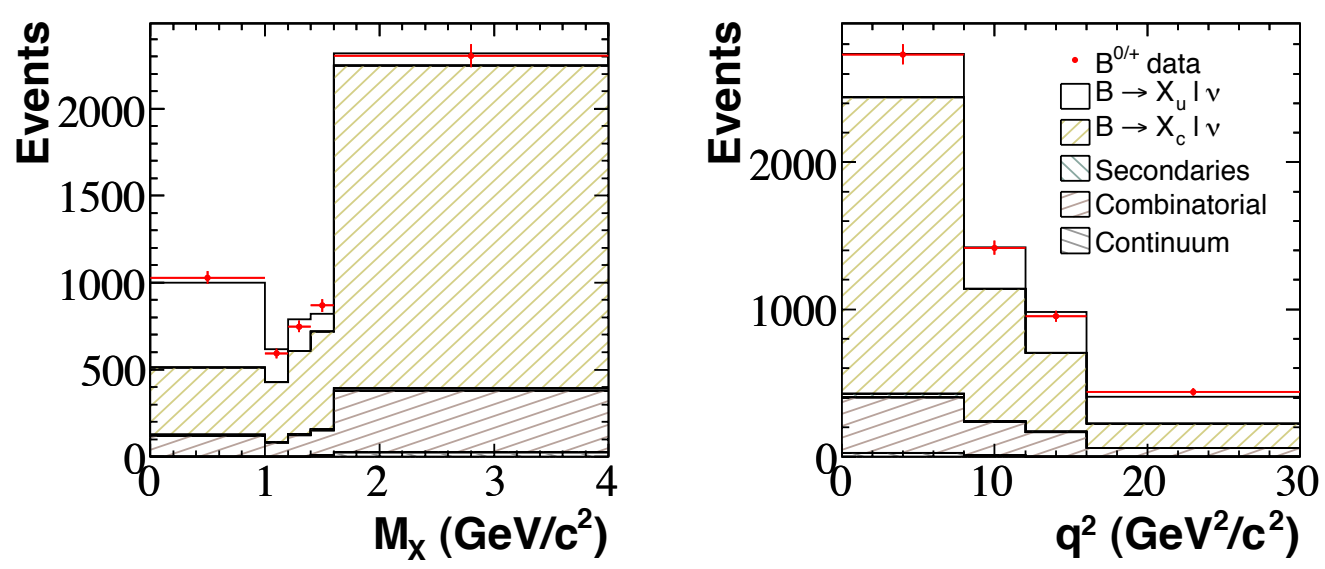

**Figure 17.1.19.** Belle (Urquijo, 2010): Projections of measured distributions (data points) of (a) $m_X$ and (b) $q^2$ with varying bin size, compared to results of a two-dimensional $m_X - q^2$ distribution for the sum of scaled MC contributions. The data are not efficiency corrected.

**Table 17.1.18.** Overview of partial branching fraction measurements with statistical and systematic errors, based on measurements of the inclusive lepton spectrum for $B \to X_u \ell \nu$ decays using untagged data samples. $s_{\rm h}^{\rm max}$ refers to the maximum kinematically allowed hadronic mass squared for a given electron energy and $q^2$.

| Experiment | Selection | $\Delta\mathcal{B}$ $(10^{-3})$ |
|---|---|---|
| CLEO (Bornheim et al., 2002) | $p_\ell^* > 2.1\,\mathrm{GeV}$ | $0.328 \pm 0.023 \pm 0.073$ |
| Belle (Limosani, 2005) | $p_\ell^* > 1.9\,\mathrm{GeV}$ | $0.847 \pm 0.037 \pm 0.153$ |
| BABAR (Aubert, 2006x) | $p_\ell^* > 2.0\,\mathrm{GeV}$ | $0.572 \pm 0.041 \pm 0.051$ |
| BABAR (Aubert, 2005h) | $p_\ell^* > 2.0\,\mathrm{GeV}, s_{\rm h}^{\rm max} > 3.5\,\mathrm{GeV}^2$ | $0.441 \pm 0.042 \pm 0.042$ |

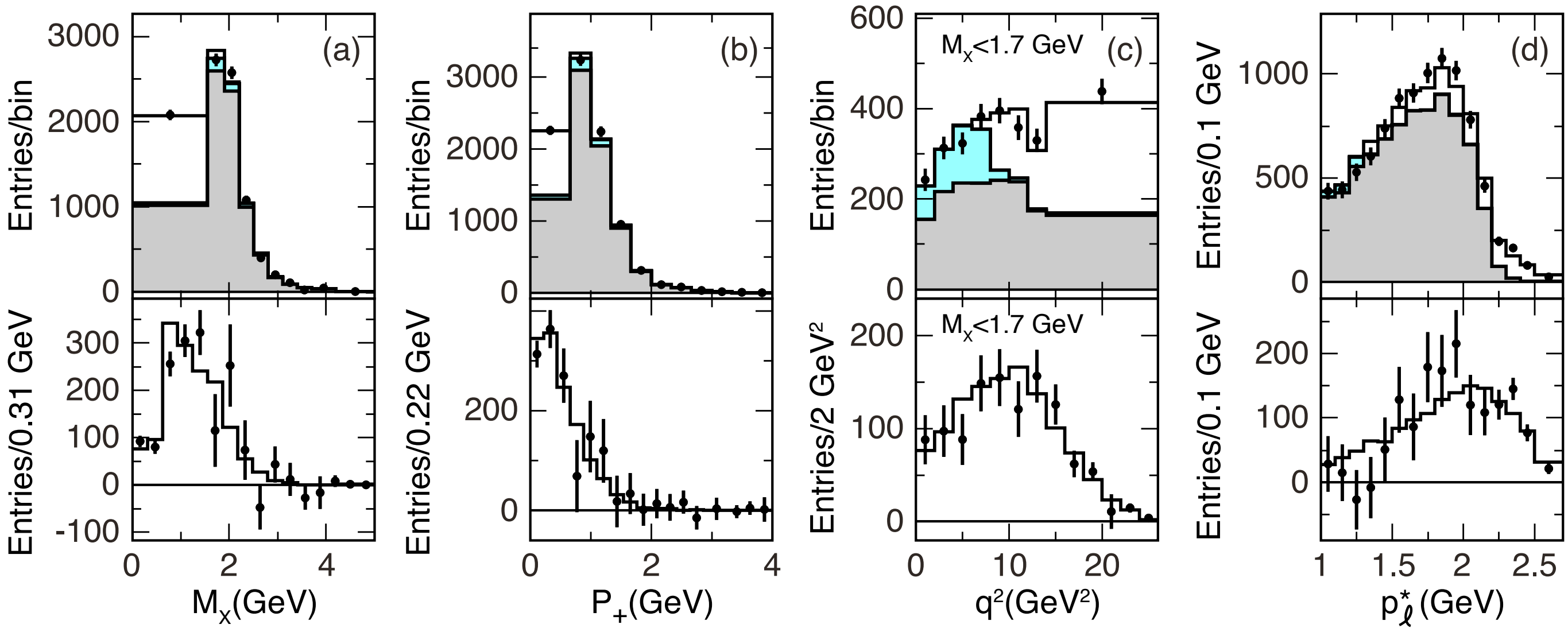

**Figure 17.1.18.** BABAR (Lees, 2012x): Extraction of $|V_{ub}|$ from selected samples of inclusive $B \to X_u \ell \nu$ decays: (a) hadronic mass $m_X$, (b) $P_+$, (c) $q^2$ with restriction $m_X \leq 1.7\,\mathrm{GeV}$, and (d) lepton momentum $p_\ell^*$. upper row: comparison of data (points with statistical errors) with results of $\chi^2$ fit with varying bin size for the sum of scaled MC distributions (histograms) of signal inside (white) and outside (blue) the selected kinematic region and background (gray); lower row: background subtracted distributions, compared to the results of the fit with finer binning. The data are not efficiency corrected.

**Table 17.1.19.** Partial $B \to X_u \ell \nu$ branching fractions (Bizjak, 2005; Urquijo, 2010; Lees, 2012x) and values of $|V_{ub}|$ (Lees, 2012x) based on BLNP calculations for different kinematic regions in tagged $B\overline{B}$ events. The stated errors are statistical and systematic, and for $|V_{ub}|$ the third error refers to the theoretical uncertainty.

| Selection | Belle: $\Delta\mathcal{B}$ $(10^{-3})$ | BABAR: $\Delta\mathcal{B}$ $(10^{-3})$ | BABAR: $\lvert V_{ub}\rvert$ $(10^{-3})$ |
|---|---|---|---|
| $m_X \leq 1.55\,\mathrm{GeV}$ | — | $1.08 \pm 0.08 \pm 0.06$ | $4.17 \pm 0.15 \pm 0.12^{+0.24}_{-0.24}$ |
| $m_X \leq 1.70\,\mathrm{GeV}$ | $1.24 \pm 0.11 \pm 0.12$ | $1.15 \pm 0.06 \pm 0.08$ | $3.97 \pm 0.17 \pm 0.14^{+0.20}_{-0.20}$ |
| $P_+ \leq 0.66\,\mathrm{GeV}$ | $1.11 \pm 0.10 \pm 0.16$ | $0.98 \pm 0.09 \pm 0.08$ | $4.02 \pm 0.18 \pm 0.16^{+0.24}_{-0.23}$ |
| $m_X \leq 1.70\,\mathrm{GeV}, q^2 \geq 8\,\mathrm{GeV}^2$ | $0.84 \pm 0.08 \pm 0.10$ | $0.68 \pm 0.06 \pm 0.04$ | $4.25 \pm 0.19 \pm 0.13^{+0.23}_{-0.25}$ |
| $p_\ell^* > 1.3\,\mathrm{GeV}$ | — | $1.52 \pm 0.16 \pm 0.14$ | $4.29 \pm 0.22 \pm 0.20^{+0.19}_{-0.20}$ |
| $p_\ell^* > 1.0\,\mathrm{GeV}, m_X - q^2$ | $1.96 \pm 0.17 \pm 0.16$ | $1.80 \pm 0.13 \pm 0.15$ | $4.28 \pm 0.15 \pm 0.18^{+0.18}_{-0.20}$ |

#### 17.1.5.3 Determination of $|V_{ub}|$

The measured partial branching fractions $\Delta\mathcal{B}$ can be related to $|V_{ub}|$ in the following way,

$$|V_{ub}| = \sqrt{\Delta\mathcal{B}/(\tau_B\, \Delta\Gamma_{\rm theory})}, \qquad (17.1.63)$$

where $\Delta\Gamma_{\rm theory}$ is the theoretically predicted partial rate (in units of $\mathrm{ps}^{-1}$) for a selected phase space region.

The extracted values of $|V_{ub}|$ are presented in Table 17.1.20 for both untagged and tagged $B\overline{B}$ samples. The $|V_{ub}|$ results have been adjusted by HFAG to include updates of input parameters and reflect the latest understanding of the theoretical uncertainties. The averages of the various available measurements have been obtained by taking correlations into account. In particular, all theoretical uncertainties are considered to be correlated, as are the uncertainties on the modeling of $B \to X_c \ell \nu$ and $B \to X_u \ell \nu$ decays. Experimental uncertainties due to particle identification and reconstruction efficiencies are fully correlated for measurements from the same experi-ment, and uncorrelated for different experiments. Statistical correlations are also taken into account, whenever available. The averaging procedure used is documented by the HFAG Collaboration (Asner et al., 2010). The earlier measurements near the kinematic limit of the lepton spectrum covered limited fractions of the total phase space and had sizable experimental and theoretical uncertainties. The more recent measurements based on the tagged $B\overline{B}$ samples of the full *BABAR* and Belle data sets have reduced backgrounds and cover a much larger fraction of the phase space.

**Table 17.1.20.** Overview of $|V_{ub}|$ measurements based on inclusive $B \to X_u \ell \nu$ decays. The critical input parameters $m_b$ and $\mu_\pi^2$ depend on the different mass schemes and have been obtained from the OPE fits to $B \to X_c \ell \nu$ hadronic mass and lepton energy moments in the kinetic mass scheme. For the BLNP and the DGE calculations, they have been subsequently translated from the kinetic to the shape function and $\overline{\text{MS}}$ schemes, respectively. The additional uncertainties $m_b$ and $\mu_\pi^2$ are due to these scheme translations. The first error is experimental and the second reflects the uncertainties of the QCD calculations and the HQE parameters (Asner et al., 2011).

| | BLNP | GGOU | DGE |
|---|---|---|---|
| $m_b$ scheme | SF scheme | Kinetic scheme | $\overline{\text{MS}}$ scheme |
| $m_b$ (GeV) | $4.588 \pm 0.023 \pm 0.011$ | $4.560 \pm 0.023$ | $4.194 \pm 0.043$ |
| $\mu_\pi^2$ (GeV$^2$) | $0.189^{+0.041}_{-0.040} \pm 0.020$ | $0.453 \pm 0.036$ | — |
| Experiment | | $\lvert V_{ub}\rvert$ $(10^{-3})$ | |
| CLEO (Bornheim et al., 2002) | $4.19 \pm 0.49^{+0.26}_{-0.34}$ | $3.93 \pm 0.46^{+0.22}_{-0.29}$ | $3.82 \pm 0.43^{+0.23}_{-0.26}$ |
| Belle (Limosani, 2005) | $4.88 \pm 0.45^{+0.24}_{-0.27}$ | $4.75 \pm 0.44^{+0.17}_{-0.22}$ | $4.79 \pm 0.44^{+0.21}_{-0.24}$ |
| *BABAR* (Aubert, 2006x) | $4.48 \pm 0.25^{+0.27}_{-0.28}$ | $4.29 \pm 0.24^{+0.18}_{-0.24}$ | $4.28 \pm 0.24^{+0.22}_{-0.24}$ |
| *BABAR* (Aubert, 2005h) | $4.66 \pm 0.31^{+0.31}_{-0.36}$ | — | $4.32 \pm 0.29^{+0.24}_{-0.29}$ |
| Average untagged | $4.65 \pm 0.22^{+0.26}_{-0.29}$ | $4.39 \pm 0.22^{+0.18}_{-0.24}$ | $4.44 \pm 0.21^{+0.21}_{-0.25}$ |
| Belle (Urquijo, 2010) | $4.47 \pm 0.27^{+0.19}_{-0.21}$ | $4.54 \pm 0.27^{+0.10}_{-0.11}$ | $4.60 \pm 0.27^{+0.11}_{-0.13}$ |
| *BABAR* (Lees, 2012x) | $4.28 \pm 0.24^{+0.18}_{-0.20}$ | $4.35 \pm 0.24^{+0.09}_{-0.11}$ | $4.40 \pm 0.24^{+0.12}_{-0.13}$ |
| Average tagged | $4.35 \pm 0.19^{+0.19}_{-0.20}$ | $4.43 \pm 0.21^{+0.09}_{-0.11}$ | $4.49 \pm 0.21^{+0.13}_{-0.13}$ |
| Average all | $4.40 \pm 0.15^{+0.19}_{-0.21}$ | $4.39 \pm 0.15^{+0.12}_{-0.14}$ | $4.45 \pm 0.15^{+0.15}_{-0.16}$ |

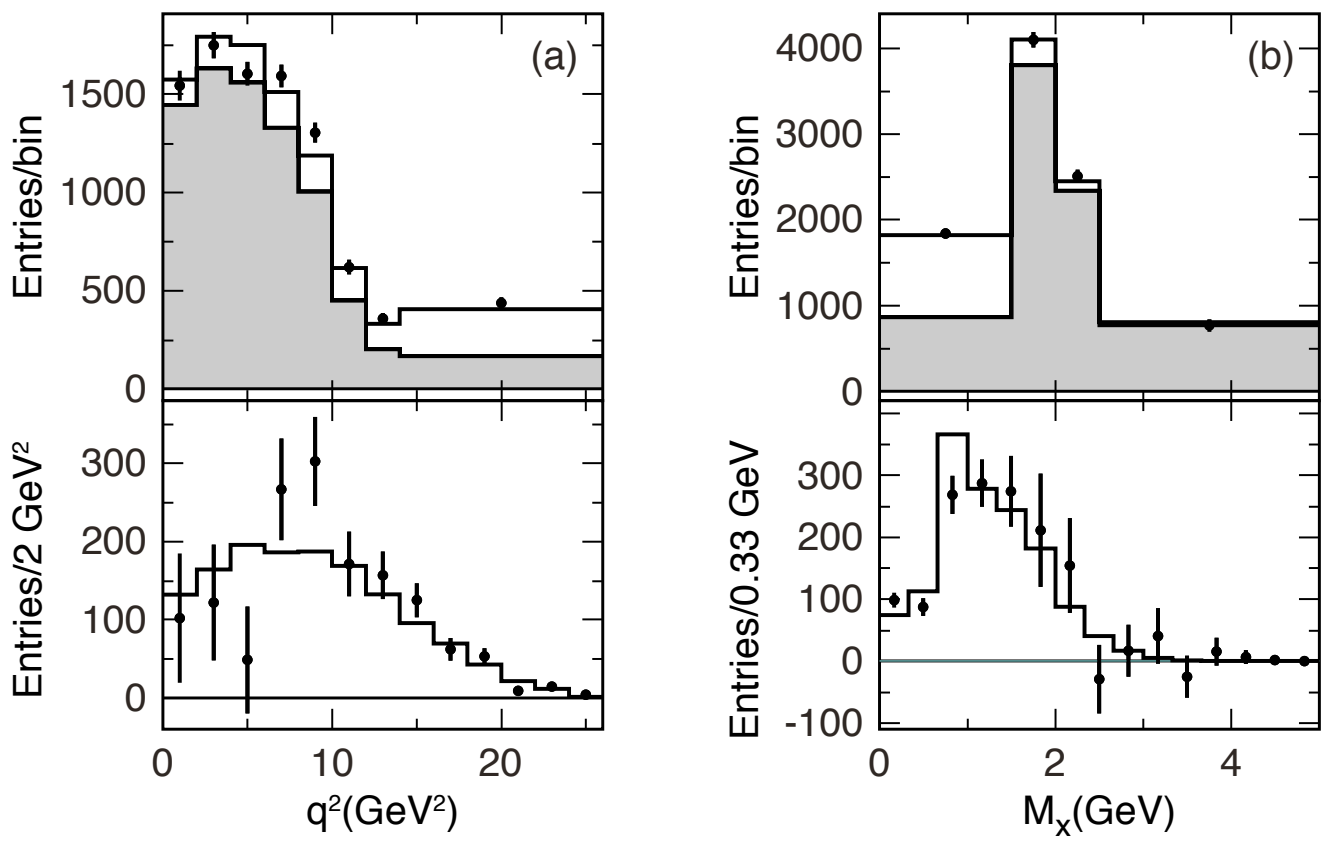

**Figure 17.1.20.** *BABAR* (Lees, 2012x): Projection of measured distributions (data points) of (a) $q^2$ and (b) $m_X$ with varying bin size. Upper row: comparison with the result of the $\chi^2$ fit to the two-dimensional $m_X - q^2$ distribution for the sum of two scaled MC contributions. Lower row: corresponding spectra with equal bin size after background subtraction based on the fit. The data are not efficiency corrected.

The extracted values of $|V_{ub}|$ based on the different QCD calculations agree well. The estimated theoretical errors are dominated by the uncertainty on $m_b$, and by other non-perturbative corrections. For BLNP there are sizable contributions from the leading and subleading shape functions and the matching scales. For GGOU the uncertainties in the parameterization of the different shape functions are important. For the DGE calculation, the main uncertainty comes from $\alpha_S$ and $m_b$ for which the $\overline{\text{MS}}$ renormalization scheme is used. The uncertainty in the weak annihilation process is included. It contributes asymmetrically to the error for the three QCD calculations.

Values of $|V_{ub}|$ based on partial branching fractions (Lees, 2012x) for different regions of phase space are presented in Table 17.1.19 for the BLNP calculation. The resulting uncertainties are highly correlated. For the different kinematic regions, the variations of $|V_{ub}|$ are consistent within the experimental uncertainties. Similar results were also obtained for other QCD calculations. The analysis based on the restricted region $m_X < 1.7\,\text{GeV}$ combined with $q^2 > 8\,\text{GeV}^2$, is expected to be less affected by non-perturbative contributions to the shape functions. Therefore, the use of a more HQE inspired approach (Bauer, Ligeti, and Luke, 2001) is appropriate. It results in a value of $|V_{ub}|$ that is in good agreement with the results based on the three QCD calculations presented here. As discussed in Section 17.1.5, NNLO effects in the BLNP calculation would lead to an increase of about 8% in $|V_{ub}|$ in some of the BLNP values reported above, but not in those related to tagged measurements with looser signal selection criteria. Further investigation is necessary to clarify this unexpected indication.

There is a high degree of consistency among the measurements and results for different QCD calculations show little variation. Based on results in Table 17.1.20, we quote the unweighted arithmetic average of the results and uncertainties from the tagged data analyses as the overall result,

$$|V_{ub}|_{\rm incl} = (4.42 \pm 0.20_{\rm exp} \pm 0.15_{\rm th}) \times 10^{-3}. \quad (17.1.64)$$

### 17.1.6 Evaluation of the results

As a result of joint efforts by theorists and experimentalists our understanding of semileptonic $B$-meson decays has substantially advanced over the last decade. Here we summarize the present situation.

#### 17.1.6.1 Summary on $|V_{cb}|$

Substantial progress has been made in the application of HQE calculations to extract $|V_{cb}|$ and $m_b$ from fits to measured moments from $B \to X_c \ell\nu$ decays. The total error quoted on $|V_{cb}|$ is 1.8% and the introduction of a $c$-quark mass constraint, $m_c(3\,\text{GeV}) = (0.998 \pm 0.029)\,\text{GeV}$, has reduced the overall uncertainty on $m_b$ to only 25 MeV.

The measurement of $|V_{cb}|$ based on the exclusive decay $B \to D^*\ell\nu_\ell$ now has a combined experimental and theoretical uncertainty of 2.3%, still dominated by the form-factor normalization. The measurement based on $B \to D\ell\nu_\ell$ has substantially improved and now provides a very useful cross check on the more precise $B \to D^*\ell\nu_\ell$ determination. However, the values of $|V_{cb}|$ based on the latter differ by about 5%, depending on the choice of the QCD calculation for the normalization of the form factors; lattice calculations lead to lower values of $|V_{cb}|$ than heavy flavor sum rules.

Consequently the comparison of the inclusive and exclusive determinations of $|V_{cb}|$ depends on the choice of the normalization of the form factors. For the LQCD calculations, the values of the inclusive and exclusive determination of $|V_{cb}|$ differ at the level of $2.5\sigma$,

$$\begin{aligned} |V_{cb}|_{\rm excl} &= [39.04\ (1 \pm 0.014_{\rm exp} \pm 0.019_{\rm th})] \times 10^{-3} \\ |V_{cb}|_{\rm incl} &= [42.01\ (1 \pm 0.011_{\rm exp} \pm 0.014_{\rm th})] \times 10^{-3}\ . \end{aligned} \quad (17.1.65)$$

The average has a probability of $P(\chi^2) = 0.015$. We therefore scale the errors by $\sqrt{\chi^2} = 2.51$ and arrive at

$$|V_{cb}| = [40.81\ (1 \pm 0.022_{\rm exp} \pm 0.028_{\rm th})] \times 10^{-3}\ . \quad (17.1.66)$$

For the heavy flavor sum rule calculations, the value is

$$|V_{cb}|_{\rm excl} = [40.93\ (1 \pm 0.014_{\rm exp} \pm 0.023_{\rm th})] \times 10^{-3} \quad (17.1.67)$$

and agrees very well with the inclusive measurement. The average value with unscaled uncertainties is

$$|V_{cb}| = [41.67\ (1 \pm 0.009_{\rm exp} \pm 0.012_{\rm th})] \times 10^{-3}\ . \quad (17.1.68)$$

#### 17.1.6.2 Summary on $|V_{ub}|$

For inclusive measurements of $|V_{ub}|$ experimental and theoretical errors are comparable in size. The dominant experimental uncertainties are related to the limited size of the tagged samples, the signal simulation, and background subtraction. The theoretical uncertainties are dominated by the error on the $b$-quark mass; a 20-30 MeV uncertainty in $m_b$ impacts $|V_{ub}|$ by 2-3%.

Measurements of the differential decay rate as a function of $q^2$ for $B^0 \to \pi^-\ell^+\nu_\ell$ provide valuable information on the shape of the form factor, though with sizable errors due to large backgrounds. Results based on different QCD calculations agree within the stated theoretical uncertainties. While the traditional method of normalizing to QCD calculations in different ranges of $q^2$ results in uncertainties of $^{+17\%}_{-10\%}$, combined fits to LQCD predictions and the measured spectrum using a theoretically motivated ansatz (Becher and Hill, 2006; Bourrely, Caprini, and Lellouch, 2009; Boyd, Grinstein, and Lebed, 1995) have resulted in a reduction of the theoretical uncertainties to about 8%.

The values of the inclusive and exclusive determinations of $|V_{ub}|$ are only marginally consistent, they differ at a level of $3\sigma$,

$$\begin{aligned} |V_{ub}|_{\rm excl} &= [3.23\ (1 \pm 0.05_{\rm exp} \pm 0.08_{\rm th})] \times 10^{-3} \\ |V_{ub}|_{\rm incl} &= [4.42\ (1 \pm 0.045_{\rm exp} \pm 0.034_{\rm th})] \times 10^{-3}. \end{aligned} \quad (17.1.69)$$

This average has a probability of $P(\chi^2) = 0.003$. Thus we scale the error by $\sqrt{\chi^2} = 3.0$ and arrive at

$$|V_{ub}| = [3.95\ (1 \pm 0.096_{\rm exp} \pm 0.099_{\rm th})] \times 10^{-3}. \quad (17.1.70)$$

#### 17.1.6.3 Conclusions and Outlook

While there has been tremendous progress, we have not achieved the precision of 1% for $|V_{cb}|$ or 5% on $|V_{ub}|$, goals many of us had hoped to reach by now, based on the final results of the Belle and *BABAR* experiments. The puzzling differences in the results of exclusive and inclusive measurements of $|V_{ub}|$, and to a lesser extent of $|V_{cb}|$ if we rely on non-lattice calculations, challenge our current understanding of the experimental and theoretical techniques. To resolve this puzzle a major effort will be required. It will take much larger tagged data samples and a more detailed assessment of the detector performance and the background composition to reduce experimental errors. It will also require further progress in QCD calculations, based on lattice or heavy flavor sum rules or other methods, to reduce the uncertainties of form-factor predictions for exclusive decays, to adopt precision determinations of the heavy quark masses, and to improve the detailed predictions of inclusive processes.

## 17.2 $V_{td}$ and $V_{ts}$

***Editors:***
*Kevin Flood (BABAR)*
*Tobias Hurth (theory)*

The CKM matrix elements $|V_{td}|$ and $|V_{ts}|$ are fundamental parameters of the Standard Model that can only be determined experimentally using rare radiative $B$ or $K$ decays (Fig. 17.2.1), or $B^0$ and $\overline{B}^0$ oscillations involving top quarks through a box diagram (Fig. 17.2.2). A discussion of kaon decays is beyond the scope of this article; see, *e.g.*, (Donoghue, Golowich, and Holstein, 1982; Gaillard and Lee, 1974b; Gilman and Wise, 1983). Measurement of the single top quark production cross-section allows for a model-independent direct determination of $|V_{tb}|$, but the magnitudes of $|V_{td}|$ and $|V_{ts}|$ cannot be similarly extracted from tree-level decays. However, a recent paper (Ali, Barreiro, and Lagouri, 2010) speculates that $\sim 10\%$ precision for the signal $t \to Ws$ can be achieved at the LHC with an integrated luminosity of $10\text{fb}^{-1}$, despite the presence of a nearly three orders of magnitude larger background from single top production of $t \to Wb$. Derivation of $|V_{td}|$ and $|V_{ts}|$ from the experimental observables necessarily assumes the SM although the FCNC observables used, *e.g.* from $B_{d,s}$ mixing, $B \to X(s,d)\gamma$, or $\epsilon$ in the kaon sector, may receive new physics contributions from unrelated sources (with the term new physics - NP - one addresses experimentally yet unconfirmed processes and particles beyond those included in the Standard Model). Independent determination of the magnitudes of $|V_{td}|$ and $|V_{ts}|$ from several different sources, along with $V_{tb}$ from single top measurements, can provide a robust model-independent check of the unitarity of the CKM matrix or, conversely, offer a sensitive probe for the possible presence of physics beyond the SM.

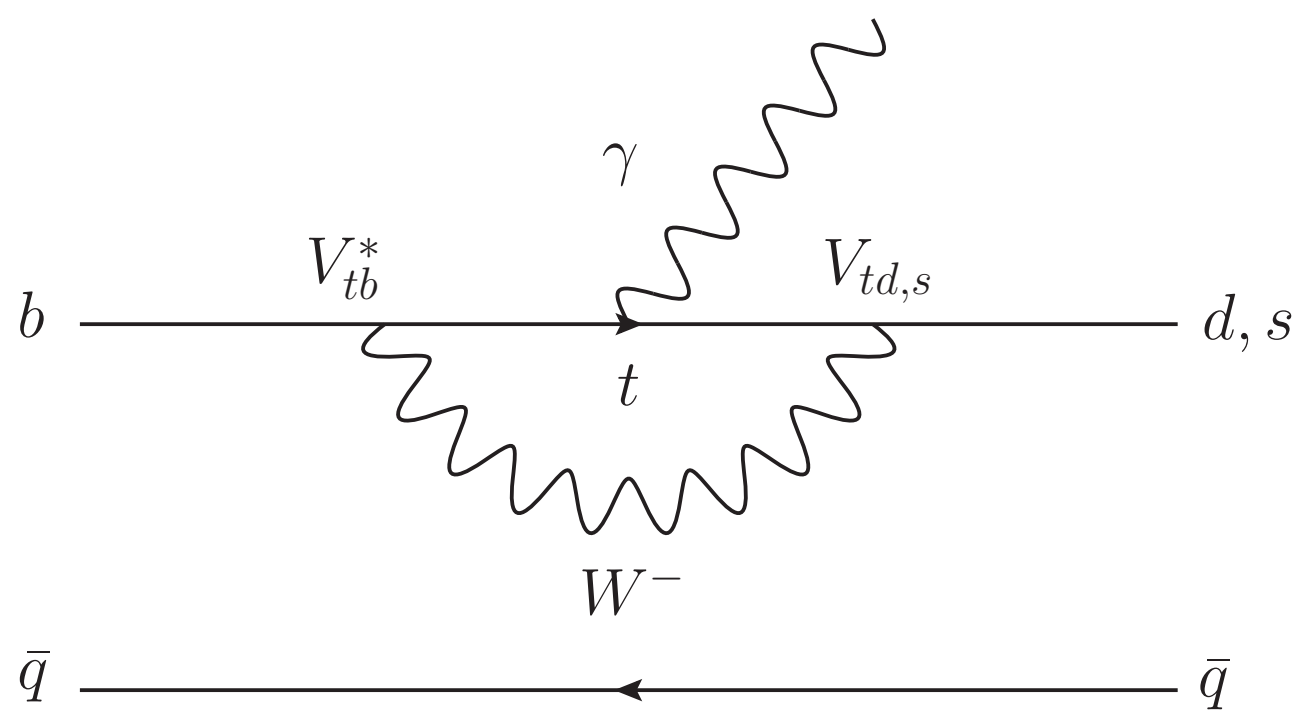

**Figure 17.2.1.** Lowest order SM Feynman diagram for a loop-mediated radiative $B$ decay.

In the past few years, the experimental and lattice QCD inputs necessary to calculate $|V_{td}|$ and $|V_{ts}|$ to good precision have become available. The $B$ Factories have contributed measurements of $\Delta m_d$, the mass difference between the neutral $B_d$ mass eigenstates, and branching fractions from the inclusive and exclusive one-loop radiative penguin processes $B \to X(s,d)\gamma$, while the CDF, DØ and LHCb collaborations have measured $\Delta m_s$, the mass difference between the neutral $B_s$ mass eigenstates, to sub-percent precision. These results have been matched by progress in lattice QCD calculations leading to increased precision in the additional parameters required to extract $|V_{td}|$ and $|V_{ts}|$ from the experimental results.

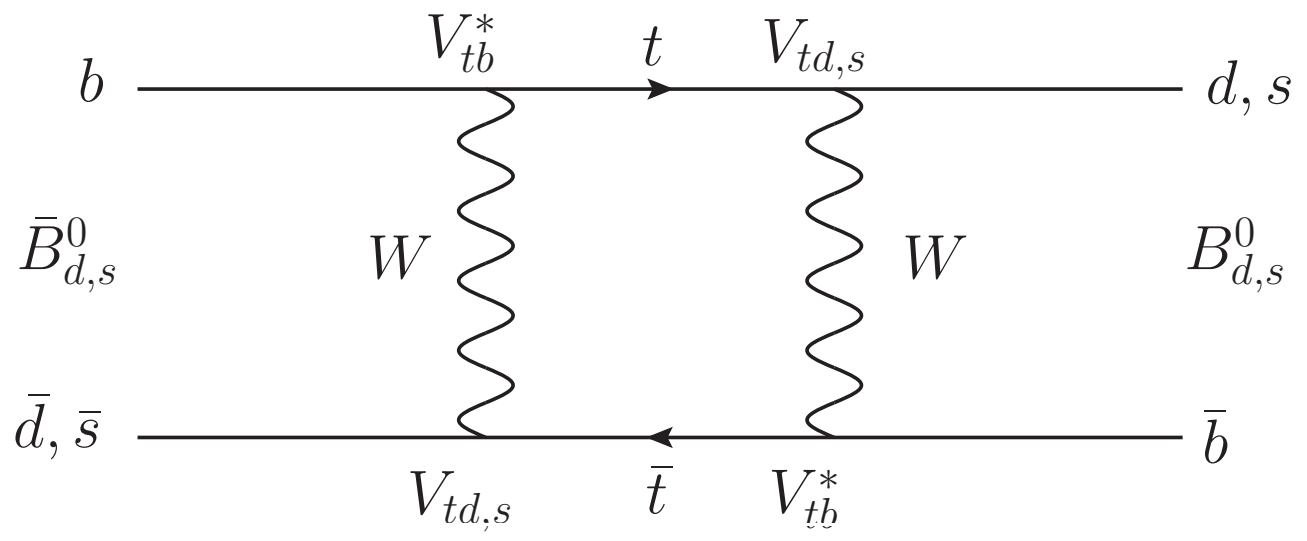

**Figure 17.2.2.** Lowest order SM Feynman diagram describing $B^0$ and $\overline{B}^0$ oscillations.

### 17.2.1 $B_{d,s}$ mixing

Equation (17.2.1) relates $\Delta m_d$ to $|V_{td}|$ (Bigi and Sanda, 2000):

$$\Delta m_d = \frac{G_F^2}{6\pi^2} f_B^2 m_B M_W^2 \eta_B S_0 |V_{tb}^* V_{td}|^2 \widehat{B}_B \ , \qquad (17.2.1)$$

where we have inserted

$$\langle B^0|(\bar{b}d)(\bar{b}d)|\overline{B}^0\rangle = \frac{4}{3} f_B^2 m_{B_d}^2 \widehat{B}_B \qquad (17.2.2)$$

for the hadronic matrix element in Eq. (10.1.17). Here, $m_B$ and $M_W$ are respectively the $B^0$ and $W$ masses; $G_F$ is the Fermi constant; $\eta_B$ is a QCD correction (Buras, Jamin, and Weisz, 1990); $S_0$ is a function of $m_t^2/m_W^2$ (Buras, 1981; Inami and Lim, 1981); $f_B$ is the $B$-meson decay constant; and $\widehat{B}_B$ is the $B$-meson bag parameter (Donoghue, Golowich, and Holstein, 1992). A discussion of the experimental techniques used at the $B$ Factories to measure $\Delta m_d$ is given in Section 17.5.

In order to extract $|V_{td}|$ using Eq. (17.2.1), we adopt the latest combination of lattice QCD results available from “www.latticeaverages.org” (Laiho, Lunghi, and Van de Water, 2010), who report $f_b\sqrt{\widehat{B}_B} = 227 \pm 19\,\text{MeV}$. This result is obtained by combining the average decay constant $f_b$ obtained from the MILC and HPQCD collaborations, along with the HPQCD determination of the bag parameter $\widehat{B}_B$, which reduces the total uncertainty with respect to taking the two parameters separately. Other required inputs are taken from Tables 25.1.2 and 25.1.3, as well as the PDG (Beringer et al., 2012). We additionally assume that $|V_{tb}| = 1$. Using the $B$ Factory results given in Table 17.5.2, which are averaged by the Heavy Flavor Averaging Group (HFAG) to obtain a final value of $\Delta m_d = 0.508 \pm 0.003 \pm 0.003\ \mathrm{ps}^{-1}$, we find $V_{td} = (9.5 \pm 0.7) \times 10^{-3}$.

The uncertainty in $|V_{td}|$ induced by the uncertainty in $f_b\sqrt{\widehat{B}_B}$ can be reduced by rewriting this factor as $f_b\sqrt{\widehat{B}_B} = f_s\sqrt{\widehat{B}_{B_s}}/\xi$, where $\xi = f_s\sqrt{\widehat{B}_{B_s}}/f_b\sqrt{\widehat{B}_B}$. The factor $\xi$ can be more accurately determined in lattice QCD calculations than its individual terms because of the inclusion of $f_s\sqrt{\widehat{B}_{B_s}}$, which is obtained directly at the physical strange quark mass rather than by extrapolation to the down quark mass, and approximate cancellation of some uncertainties in the ratio. Using the values $\xi = 1.237 \pm 0.032$ and $f_s\sqrt{\widehat{B}_{B_s}} = 279 \pm 15\,\mathrm{MeV}$, we find $V_{td} = (9.6 \pm 0.5) \times 10^{-3}$, with a reduction in the uncertainty of $\sim 30\%$ relative to the result based solely on $f_b\sqrt{\widehat{B}_B}$. The lattice parameter uncertainties can be further controlled by taking the ratio $|V_{td}/V_{ts}|$, which directly uses $\xi^{-1}$, and incorporating the PDG combination of the $B^0_s\overline{B}^0_s$ oscillation frequency results from CDF (Abulencia et al., 2006b) and LHCb (Aaij et al., 2012f), $\Delta m_s = 17.69 \pm 0.08\ \mathrm{ps}^{-1}$. Using an expression for $|V_{ts}|$ analogous to Eq. (17.2.1), we obtain $|V_{td}/V_{ts}| = 0.208 \pm 0.005$.

### 17.2.2 $B \to X(s,d)\gamma$

Loop-mediated radiative decays provide a set of processes complementary to $B^0$ and $\overline{B}^0$ oscillations from which the value of $|V_{td}/V_{ts}|$ can be derived using experimental branching fraction results together with inputs from lattice QCD. Since new physics may enter each type of process differently, a comparison of $|V_{td}/V_{ts}|$ extracted from both mixing and radiative decays provides a robust test of the consistency of the SM CKM paradigm or, conversely, offers a powerful probe for the presence of new physics (Descotes-Genon, Ghosh, Matias, and Ramon, 2011; Lenz et al., 2011). Amplitudes for the rare $\Delta F = 1$ decays $b \to d\gamma$ and $b \to s\gamma$, essentially proportional to $V_{td}$ and $V_{ts}$ respectively, have been measured using both inclusive and exclusive final states at the $B$ Factories. These provide the experimental inputs necessary to calculate the ratio of CKM elements $|V_{td}/V_{ts}|$.

The details of the various experimental techniques used to measure the branching fractions for the radiative penguin $B \to X(s,d)\gamma$ processes are addressed in Section 17.9. Here, we discuss the calculation of the ratio $|V_{td}/V_{ts}|$ using a combination of the latest branching fraction results from *BABAR* (Aubert, 2008z, 2009r) and Belle (Nakao, 2004; Taniguchi, 2008) in the exclusive $B \to (\rho, \omega, K^*)\gamma$ modes, followed by calculation of the ratio using *BABAR*'s latest $B \to X_d\gamma$ semi-inclusive results (del Amo Sanchez, 2010q). Belle has no comparable semi-inclusive analysis as of the time of publication of this review. The exclusive and inclusive *BABAR* results use the same *BABAR* dataset as well as a similar event selection, and are thus highly correlated; they cannot be averaged easily. Since there are correlated inputs to both the inclusive and exclusive calculations, as well as non-trivial correlations in the theory assumptions, we forego any attempt here to make any combination of the exclusive and inclusive results.

In their measurements of combinations of the exclusive mode branching fractions, both *BABAR* (Aubert, 2008z) and Belle (Taniguchi, 2008) assume an exact isospin symmetry, *i.e.* $\Gamma(B^\pm \to \rho^\pm\gamma) \equiv 2\overline{\Gamma}(B^0 \to \rho^0\gamma)$, as well as $2\overline{\Gamma}(B^0 \to \rho^0\gamma) \equiv 2\overline{\Gamma}(B^0 \to \omega\gamma)$ However, these relations are not exact and symmetry-breaking corrections have been calculated (Ball, Jones, and Zwicky, 2007; Ball and Zwicky, 2006b). The asymmetry expected between $\rho^0$ and $\omega$ predominantly arises from the different form factors for these decays, while the principal contribution to symmetry breaking between neutral and charged $\rho$ mesons is the presence of a weak annihilation diagram with photon emission from the spectator quark. Both collaborations report *CP*- and isospin-averaged results for $B \to (\rho,\omega)\gamma$ and $B \to \rho\gamma$, as well as branching fractions for contributing individual modes. *BABAR* and Belle have searched for isospin asymmetries in these modes, and no statistically significant asymmetry is observed in either the $\rho\gamma$ or $(\rho,\omega)\gamma$ modes. A discussion of the experimental measurements themselves, as well as related theoretical background, can be found below in Section 17.9.

Belle (Taniguchi, 2008) calculates the ratio of branching fractions from products of likelihoods for each of the individual $B \to (\rho,\omega)\gamma$ and $B \to K^*\gamma$ final states, which are convolved with residual systematics that do not cancel in the ratio of branching fractions, and finds

$$R_{\rho^0} = \frac{\mathcal{B}(B^0 \to \rho^0\gamma)}{\mathcal{B}(B^0 \to K^{*0}\gamma)} = 0.0206^{+0.0045+0.0014}_{-0.0043-0.0016}, \quad (17.2.3)$$

$$R_{\rho} = \frac{\mathcal{B}(B \to \rho\gamma)}{\mathcal{B}(B \to K^{*}\gamma)} = 0.0302^{+0.0060+0.0026}_{-0.0055-0.0028}, \quad (17.2.4)$$

$$R_{\rho/\omega} = \frac{\mathcal{B}(B \to (\rho,\omega)\gamma)}{\mathcal{B}(B \to K^{*}\gamma)} = 0.0284 \pm 0.0050^{+0.0027}_{-0.0029}, \quad (17.2.5)$$

where the first and second errors are statistical and systematic, respectively.

The *BABAR* result for the exclusive modes (Aubert, 2008z) employs a different strategy, first concatenating all $B \to (\rho,\omega)\gamma$ final states into a single dataset which is then simultaneously fit over all modes with an isospin constraint applied in order to extract the isospin-averaged $B \to (\rho,\omega)\gamma$ branching fraction. A similar procedure omitting the $\omega\gamma$ final state is used to produce the $B \to \rho\gamma$ branching fraction. The $B \to K^*\gamma$ branching fraction used in *BABAR*'s calculation of the ratio is taken from HFAG, and thus it is not possible to account for systematic experimental effects which may be common to both numerator and denominator in the ratio of branching fractions, and they quote only a total uncertainty for the branching frac-tion ratio results,

$$R_{\rho^+} = \frac{\mathcal{B}(B^+ \to \rho^+\gamma)}{\mathcal{B}(B^+ \to K^{*+}\gamma)} = 0.030^{+0.012}_{-0.011}, \quad (17.2.6)$$

$$R_{\rho^0} = \frac{\mathcal{B}(B^0 \to \rho^0\gamma)}{\mathcal{B}(B^0 \to K^{*0}\gamma)} = 0.024 \pm 0.006, \quad (17.2.7)$$

$$R_{\omega} = \frac{\mathcal{B}(B^0 \to \omega\gamma)}{\mathcal{B}(B^0 \to K^{*0}\gamma)} = 0.012^{+0.007}_{-0.006}, \quad (17.2.8)$$

$$R_{\rho} = \frac{\mathcal{B}(B \to \rho\gamma)}{\mathcal{B}(B \to K^{*}\gamma)} = 0.042 \pm 0.009, \quad (17.2.9)$$

$$R_{\rho/\omega} = \frac{\mathcal{B}[B \to (\rho/\omega)\gamma]}{\mathcal{B}(B \to K^{*}\gamma)} = 0.039 \pm 0.008. (17.2.10)$$

We use a weighted average of the common central values reported by each collaboration, given the total uncertainty for each measurement and symmetrizing uncertainties where applicable, to arrive at averaged values subsequently used in the calculation of $|V_{td}/V_{ts}|$,

$$R_{\rho^0} = \frac{\mathcal{B}(B^0 \to \rho^0\gamma)}{\mathcal{B}(B^0 \to K^{*0}\gamma)} = 0.0219 \pm 0.0037,$$
$$R_{\rho} = \frac{\mathcal{B}(B \to \rho\gamma)}{\mathcal{B}(B \to K^{*}\gamma)} = 0.0341 \pm 0.0052,$$
$$R_{\rho/\omega} = \frac{\mathcal{B}(B \to (\rho,\omega)\gamma)}{\mathcal{B}(B \to K^{*}\gamma)} = 0.0320 \pm 0.0047. \quad (17.2.11)$$

Both collaborations adopt similar formalisms to derive the ratio of CKM elements from the underlying experimental results, with the ratio $R_{\rm th}(\rho\gamma/K^*\gamma)$ (and similarly $R_{\rm th}(\omega\gamma/K^*\gamma)$) given by (Ali, Lunghi, and Parkhomenko, 2004; Ball, Jones, and Zwicky, 2007; Beneke, Feldmann, and Seidel, 2005; Bosch and Buchalla, 2005):

$$R_{\rm th}(\rho\gamma/K^*\gamma) = \frac{\mathcal{B}_{\rm th}(B \to \rho\gamma)}{\mathcal{B}_{\rm th}(B \to K^*\gamma)} \quad (17.2.12)$$
$$\equiv S_\rho \left|\frac{V_{td}}{V_{ts}}\right|^2 \frac{(M_B^2 - m_\rho^2)^3}{(M_B^2 - m_{K^*}^2)^3}$$
$$\zeta^2 \left[1 + \Delta R(\rho/K^*)\right], \quad (17.2.13)$$

where $m_\rho$ is the mass of the $\rho$ meson, $\zeta$ is the ratio of the transition form factors, $\zeta = \overline{T}_1^{\rho}(0)/\overline{T}_1^{K^*}(0)$ and $S_\rho = 1$ and $1/2$ for the $\rho^\pm$ and $\rho^0$ mesons, respectively. A similar expression applies for $B \to (\rho,\omega)\gamma$ with the substitution $\rho \to (\rho,\omega)$ based on the symmetries defined above. These theoretical relations are based on the method of QCD factorization; the application of this method to radiative decays is discussed in Section 17.9. Within such factorization formulae, process-independent non-perturbative functions like form factors are separated from perturbatively calculable functions. Here, the main sources of theoretical uncertainties are the form factors and the $\Lambda/m_b$ corrections. The former is expected to be reduced by taking ratios of the observables. The $\alpha_S$ corrections to the hard kernels and the power corrections, both included in the ratio in Eq. (17.2.13) via the factor $(1 + \Delta R)$, introduce further dependences on the CKM matrix elements, namely $\phi_2$ as given in Eq. (16.5.4) and $R_{ut} = |V_{ud}V_{ub}^*/V_{td}V_{tb}^*|$, and one finds numerically (Beneke, Feldmann, and Seidel, 2005):

$$\Delta R(\rho^\pm/K^{*\pm}) = \{1 - 2R_{ut}\cos\phi_2\,[0.24^{+0.18}_{-0.18}] + R_{ut}^2\,[0.07^{+0.12}_{-0.07}]\}, \quad (17.2.14)$$
$$\Delta R(\rho^0/K^{*0}) = \{1 - 2R_{ut}\cos\phi_2\,[-0.06^{+0.06}_{-0.06}] + R_{ut}^2\,[0.02^{+0.02}_{-0.01}]\}. \quad (17.2.15)$$

These results are consistent with the predictions given in the literature (Ali, Lunghi, and Parkhomenko, 2004; Ball, Jones, and Zwicky, 2007; Bosch and Buchalla, 2005). The neutral mode is better suited for the determination of $|V_{td}/V_{ts}|$ than the charged mode, in which the function $\Delta R$ is dominated by the weak annihilation contribution, which leads to a larger error. The most recent determination of the ratio $\zeta$ within the light-cone QCD sum rule approach (Ball and Zwicky, 2006b), $1/\zeta = 1.17 \pm 0.09$, leads to the determination of $|V_{td}/V_{ts}|$ via Eq. (17.2.13). However, the experimental data on the branching fractions of $B \to K^*\gamma$ and $B \to \rho\gamma$ calls for a larger error on $\zeta$, if one assumes no large power corrections beyond the known annihilation terms (Beneke, Feldmann, and Seidel, 2005) (see also Section 17.9.4.1).

Using the combined results from both experiments, we obtain

$$\left|\frac{V_{td}}{V_{ts}}\right|_{\rho^0} = 0.26 \pm 0.02 \pm 0.03, \quad (17.2.16)$$
$$\left|\frac{V_{td}}{V_{ts}}\right|_{\rho} = 0.22 \pm 0.02 \pm 0.02, \quad (17.2.17)$$
$$\left|\frac{V_{td}}{V_{ts}}\right|_{\rho,\omega} = 0.21 \pm 0.02 \pm 0.02, \quad (17.2.18)$$

where the first error is the total experimental uncertainty and the second is the theory uncertainty. *BABAR* additionally reports the ratio for the two exclusive modes not measured by Belle:

$$\left|\frac{V_{td}}{V_{ts}}\right|_{\rho^+} = 0.198^{+0.039}_{-0.035} \pm 0.016, \quad (17.2.19)$$
$$\left|\frac{V_{td}}{V_{ts}}\right|_{\omega} = 0.202^{+0.058}_{-0.050} \pm 0.016. \quad (17.2.20)$$

Although experimental uncertainties on the exclusive branching fractions may be substantially reduced in the future, irreducible theory uncertainties can complicate interpretation of any observed discrepancy in $|V_{td}/V_{ts}|$ with values from other processes. Such uncertainties are generally under better control for inclusive radiative penguin decays, where $|V_{td}/V_{ts}|$ has been calculated to next-to-leading-log (NLL) precision (Ali, Asatrian, and Greub, 1998). Following this formalism, the ratio of the inclusive branching fractions can be written as a function of the Wolfenstein parameters $\lambda$, $\overline{\rho}$, $\overline{\eta}$

$$R(d\gamma/s\gamma) = \lambda^2[1+\lambda^2(1-2\overline{\rho})]\,[(1-\overline{\rho})^2+\overline{\eta}^2 + \frac{D_u}{D_t}(\overline{\rho}^2+\overline{\eta}^2)+\frac{D_r}{D_t}(\overline{\rho}(1-\overline{\rho})-\overline{\eta}^2)]\,, \simeq 0.046\ \ [\text{for } (\rho,\eta)=(0.11,0.33), \text{ or } (\overline{\rho},\overline{\eta})=(0.107,0.322)]\,, \quad (17.2.21)$$

where the quantities $D_i$, which depend on several input parameters such as $m_t$, $m_b$, $m_c$, must be calculated numerically. As with the exclusive decays, care must be taken to use a set of input parameters determined independently from $|V_{td}|$ and $|V_{ts}|$. For the *BABAR* result, this was done by re-expressing the Unitarity Triangle apex ($\overline{\rho}$, $\overline{\eta}$) as a function of $\phi_1$ and using the HFAG world-average for $\phi_1$. Given the current HFAG world-average values of the CKM inputs, the theory uncertainty on the ratio $R(d\gamma/s\gamma)$ is expected to be $< 0.2\%$, an order of magnitude smaller than the uncertainty for the exclusive modes prediction.

The *BABAR* analysis (del Amo Sanchez, 2010q) of the $b \to d\gamma$ and $b \to s\gamma$ inclusive rates used in the calculation of $|V_{td}/V_{ts}|$ are extrapolated from measurements of the partial decay rates to seven exclusive hadronic final states, shown in Table 17.9.6, in the mass ranges $0.5 < M(X_d) < 1.0\,\mathrm{GeV}/c^2$ and $1.0 < M(X_d) < 2.0\,\mathrm{GeV}/c^2$. The low-mass region contains contributions that are highly correlated with the dataset used for the *BABAR* exclusive modes analysis and, in the inclusive analysis, it is assumed that there is no non-resonant signal component in this mass range.

To obtain the inclusive rates, the experimentally determined partial rates must be corrected for the fraction of missing final states, as well as for hadronic systems with $M(X) > 2.0\,\mathrm{GeV}/c^2$. Well-characterized corrections for final states with neutral kaons and non-reconstructed $\omega$ final states are made in the low-mass region. In the high-mass region, the missing fractions depend on the details of the fragmentation of the hadronic system, which is modeled using `Jetset` (Sjöstrand, 1995) and expected to be different for $X_d$ and $X_s$. The Kagan-Neubert photon spectrum model (Kagan and Neubert, 1998) is used to correct for the mass region above $2.0\,\mathrm{GeV}/c^2$ that is not measured. The photon spectra for $b \to d\gamma$ and $b \to s\gamma$ are expected to be nearly identical, and the uncertainty in the extrapolation is mainly from lack of knowledge of the details of the underlying fragmentation process. In the high-mass region, this is the largest contribution to the total systematic uncertainty. *BABAR* finds

$$\frac{\mathcal{B}(b \to d\gamma)}{\mathcal{B}(b \to s\gamma)} = 0.040 \pm 0.009 \pm 0.010\,, \qquad (17.2.22)$$

and determines

$$\left|\frac{V_{td}}{V_{ts}}\right| = 0.199 \pm 0.022 \pm 0.024 \pm 0.002\,, \qquad (17.2.23)$$

where the first error is purely statistical, the second accounts for systematic effects including the uncertainty in the extrapolation for the missing mass and final states, and the third uncertainty is purely from theory considerations.

There is good agreement among the values of $|V_{td}/V_{ts}|$ obtained from exclusive and inclusive analyses of radiative penguin processes. The farthest outlier from the central value of $|V_{td}/V_{ts}|$ is obtained from the average of the $\rho^0$ mode. However, all results are in reasonable agreement with each other. While the total uncertainty in the current results for the exclusive and inclusive approaches is comparable, the relatively very small inclusive theory uncertainty will make it a more sensitive observable at future flavor facilities that plan to integrate much larger datasets than available at Belle or *BABAR*. Comparing these results with the $|V_{td}/V_{ts}|$ value from mixing, there is also good agreement, albeit with substantially larger uncertainties for the radiative decays results. For any future Belle inclusive analysis, it seems reasonable to assume that the uncertainty will be similar to that for their exclusive analysis, just as at *BABAR*. This would allow for more precise comparisons between $|V_{td}/V_{ts}|$ from rare radiative decays and from mixing.

### 17.2.3 Summary

A direct determination of $|V_{ts}|$ and $|V_{td}|$ from a measurement of the decays $t \to s$ and $t \to d$ at LHC is difficult, and will likely remain so at least in the near future. Indirect methods involving virtual top quarks are therefore required to measure these CKM matrix elements. At the $B$ Factories, the FCNC transitions $b \to s$ and $b \to d$ in radiative penguin processes have been used to obtain measurements of the ratio $|V_{td}/V_{ts}|$, while the value of $|V_{td}|$ has been obtained from measurements of $B_d$ mixing. Extracting the values of the CKM elements from these processes necessarily assumes there are no contributions from physics beyond the SM and it is difficult to distinguish possible NP contributions, which may enter at the same order as the lowest order SM processes.

The major uncertainties in the existing measurements originate from ignorance of the hadronic matrix elements. The current method for extracting $|V_{td}|$ and $|V_{ts}|$ from $\Delta B = \pm 2$ processes relies heavily on lattice calculations, and any further experimental improvements in $\Delta m_{s/d}$ measurements will need to be matched by corresponding improvements in the lattice calculations. Likewise, for improvement in the precision of $|V_{td}|$ and $|V_{ts}|$ extracted from radiative penguin processes, significant advances in the theoretical methods will be necessary.

Experimentally, it may be possible at future super flavor factories to make a fully inclusive branching fraction measurement of $b \to d\gamma$, as well as $b \to s\ell^+\ell^-$ and $b \to d\ell^+\ell^-$, which will help to reduce theory and model dependences. In $b \to s\ell^+\ell^-$ and $b \to d\ell^+\ell^-$ decays, additional amplitudes arise from diagrams similar to Fig. 17.2.1 but with a $Z$ boson replacing the photon (see Section 17.9 for a discussion of these modes). Because the contribution of these additional electroweak amplitudes becomes greater, and the contribution from the photon pole decreases, with increasing invariant mass of the di-lepton final state, extracting $|V_{td}/V_{ts}|$ as a function of dilepton mass using these decays may allow one to disentangle any underlying new physics contributions from those of the SM CKM matrix elements. Finally, if such future facilities obtain enough data at the $\Upsilon(5S)$, it may also be possible to very cleanly determine $|V_{td}/V_{ts}|$ from the ratio of branching fractions for the annihilation penguin processes $B_d \to \gamma\gamma$ and $B_s \to \gamma\gamma$ (Bosch and Buchalla, 2002a). These di-photon modes are further discussed below in Section 17.11.

## 17.3 Hadronic $B$ to charm decays

***Editors:***
*Richard Kass (BABAR)*
*Martin Beneke (theory)*

**Additional section writers:**
*Justin Albert, Vincent Poireau, Stephen Schrenk*

### 17.3.1 Introduction

$B$ meson decays into all hadronic final states containing open charm or charmonium account for almost three quarters of all $B$ decays. Despite constituting the majority of final states, these decays pose a challenge to both experiment and theory. The large available phase space in a $B$ meson decay means that there are hundreds of possible final states all with rather small branching fractions, typically a few tenths of a percent. Therefore to study in detail any particular final state a very large sample of $B$ mesons is necessary as well as a detector capable of measuring the energy, momentum, and identity of the final state particles to high precision. Since these are all hadronic final states, decay rate calculations must be done using non-perturbative QCD. For the majority of final states, a quantitative prediction with controlled theoretical uncertainties remains out of reach. Only the decay rates of the simplest hadronic decays to charm, such as $\overline{B}^0 \to D^+\pi^-$, can be calculated from first principles using QCD.

In spite of the above drawbacks, hadronic $B$ decays to charm play an important role in the more glamorous aspects of $B$ physics, *i.e.* the determination of the CKM parameters, measurements of *CP* violation, and search for physics beyond the Standard Model. If for no other reason these decay modes must be measured in order to understand the possible backgrounds involved in a measurement of a CKM parameter. Although the branching fractions here are small, it is still possible to collect very clean samples of $B$ events using modes such as $B \to D\pi$, $B \to D^*\pi$, etc. Two-body decays such as $\overline{D}^0\pi^+$ and $D^-\pi^+$ provide important detector calibration tools for determining momentum resolution ($\pi^\pm$, $K^\pm$; see Sections 2.2.2, 6.2), electromagnetic energy resolution ($\pi^0$, $\eta$; see Section 2.2.4), mass resolution ($D$, $B$; see Chapter 7), secondary vertex location ($D$, $K_s^0$; see Chapter 6), and particle identification efficiency and rejection ($\pi/K$; see Chapter 5). Finally, precision measurements of modes such as $B \to D^{(*)}\pi$, $D^{(*)}\pi\pi$ may serve as standard candles for QCD calculations.

In this section we are mainly concerned with decay rates and not the specifics of how the final states are reconstructed and the techniques involved. These techniques are described in detail in Chapters 7 ($B$ reconstruction), 12 (angular analysis), and 13 (Dalitz analysis).

### 17.3.2 Theory overview

A first principles calculation of the decay rate of the full set of $B$ decays to charm, and even the two-body final states only, is still beyond our capabilities. Instead a variety of approaches to these calculations have been tried with various levels of success. An excellent and still relevant discussion of these techniques can be found in Chapters 2 and 10 of (Harrison and Quinn, 1998). In the following overview we cover the generalized factorization approach of Bauer, Stech, and Wirbel (1987) (BSW) and Neubert and Stech (1998) (NS), and the QCD factorization approach (Beneke, Buchalla, Neubert, and Sachrajda, 2000), which provides a first principles calculation for a limited class of final states.

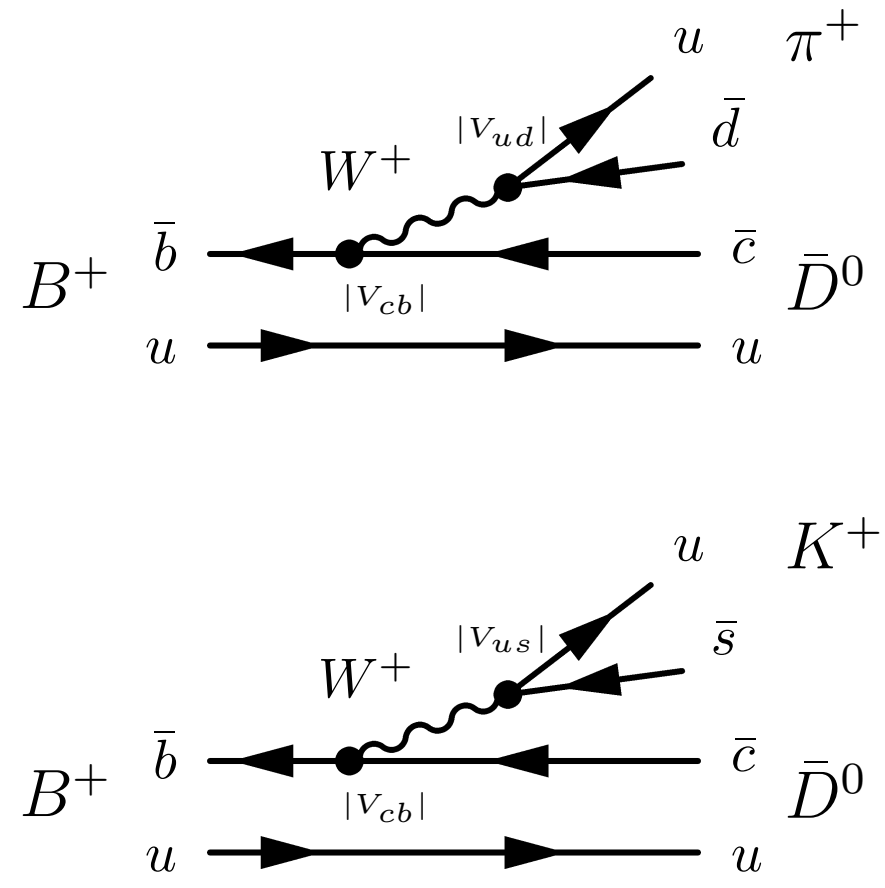

**Figure 17.3.1.** Dominant Feynman diagrams contributing to the decays $B^+ \to \overline{D}^0\pi^+$ (top) and $B^+ \to \overline{D}^0 K^+$ (bottom).

We begin our discussion with $b \to c\bar{u}d$ transitions. The case of $b \to c\bar{u}s$ is completely analogous. Examples for these transitions are shown in Fig. 17.3.1. A popular and useful approach to calculate decay rates (especially for two-body $B$ decays) is the factorization ansatz. To understand this technique, consider the decays that are shown in Fig. 17.3.2. In this figure only the electroweak contributions to the decay amplitudes are shown. A naïve attempt to calculate the decay rate would write the matrix element in terms of the usual currents, *e.g.*, $\bar{c}\gamma_u(1-\gamma_5)b$. However, this is clearly a drastic approximation as it neglects the all important role of gluons in the production of the final state hadrons. Nevertheless, at this early stage of calculation an important distinction becomes apparent. The decay $B^+ \to \overline{D}^0\pi^+$ can proceed through two amplitudes as shown in Figs 17.3.2a) and b). Since all final state particles must be color singlets, diagram b) will be suppressed due to color matching relative to a) by $1/N_c$, with $N_c$ the number of colors. Amplitudes such as Fig. 17.3.2 a) are known as "color-allowed" while an amplitude such as Fig. 17.3.2 b) is often called "color-suppressed". The decay $B^0 \to D^-\pi^+$ shown in Fig. 17.3.2 c) is color-allowed, while $B^0 \to \overline{D}^0\pi^0$, Fig. 17.3.2 d), is color-suppressed. Although not suitable for a quantitative prediction, the notion of color suppression provides a useful guide to the hierarchies in the branching fractions of $B$ to charm decays, in addition to the hierarchies caused by the CKM elements.

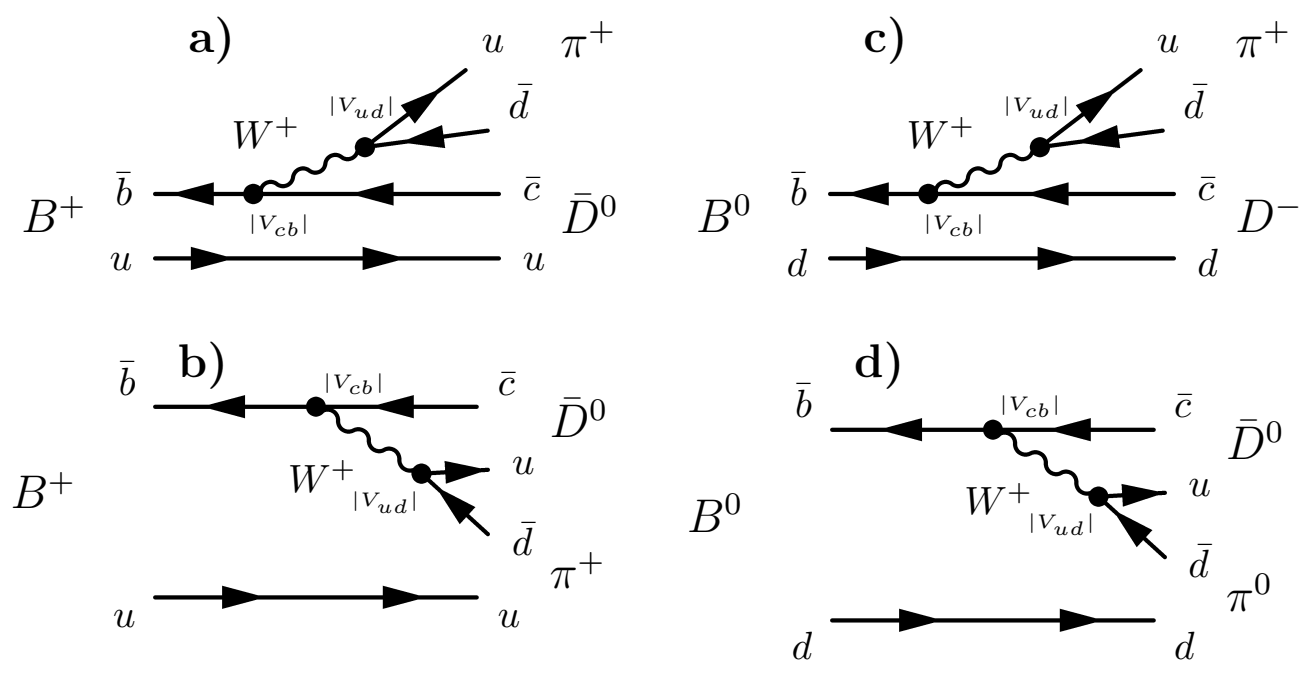

**Figure 17.3.2.** Two-body Feynman diagrams contributing to the $B^+ \to \overline{D}^0\pi^+$ (a, b) and $B^0 \to D\pi$ (c, d) decays.

For a more detailed discussion we recall the effective Hamiltonian

$$H_{\rm eff} = \frac{G_F}{\sqrt{2}} V_{cb} V_{ud}^* \Bigg\{ \left( C_1 + \frac{C_2}{N_c} \right) [\bar{c}_i b_i]_{\rm V-A} [\bar{d}_k u_k]_{\rm V-A} + 2C_2\, [\bar{c}_i T^a_{ij} b_j]_{\rm V-A} [\bar{d}_k T^a_{kl} u_l]_{\rm V-A} \Bigg\} \tag{17.3.1}$$

for the $b \to c\overline{u}d$ transition. Here $C_1$ and $C_2$ are Wilson coefficients that account for short-distance QCD effects and Eq. (17.3.1) includes the color indices $i, j, k, l$. In the naïve factorization approach the $\langle D^+\pi^-|H_{\rm eff}|\overline{B}^0\rangle$ matrix element is separated into currents by inserting the QCD vacuum state, which ignores all long-distance QCD interactions between the currents. Applied to $\overline{B}^0 \to D^+\pi^-$ ($B^0 \to D^-\pi^+$ in Fig. 17.3.2c) the "factorized" matrix element is now:

$$\frac{G_F}{\sqrt{2}} V_{cb} V_{ud}^*\, a_1 \, \langle D^+|\bar{c}\gamma_\mu(1-\gamma_5)b|\overline{B}^0\rangle \langle \pi^-|\bar{d}\gamma^\mu(1-\gamma_5)u|0\rangle \tag{17.3.2}$$

with $a_1 = C_1 + C_2/N_c$. The matrix element of the color-octet operator is set to zero in the factorization approximation. Decays which involve this combination of Wilson coefficients are often called color-allowed or Type I transitions. In addition there are also color-suppressed (or Type II) transitions. As an example $B^0 \to \overline{D}^0\pi^0$ is illustrated in Fig. 17.3.2 d). Here one first uses a so-called Fierz identity $[\overline{\psi}_1\psi_2]_{\rm V-A}[\overline{\psi}_3\psi_4]_{\rm V-A} = [\overline{\psi}_3\psi_2]_{\rm V-A}[\overline{\psi}_1\psi_4]_{\rm V-A}$ to rearrange the four-fermion operators in $H_{\rm eff}$ into the form $[\bar{d}b]_{\rm V-A}[\bar{c}u]_{\rm V-A}$. Then the factorized amplitude similar to Eq. (17.3.2) for this process is

$$\frac{G_F}{\sqrt{2}} V_{cb} V_{ud}^*\, a_2 \, \langle \pi^0|\bar{d}\gamma_\mu(1-\gamma_5)b|\overline{B}^0\rangle \langle D^0|\bar{c}\gamma^\mu(1-\gamma_5)u|0\rangle, \tag{17.3.3}$$

where now $a_2 = C_2 + C_1/N_c$. Finally there are decay modes such as $B^+ \to \overline{D}^0\pi^+$ (Fig. 17.3.2 a) and b)) which are a combination of color-allowed and color-suppressed amplitudes. These decays are called "Type III" processes.

In the absence of any QCD effects, $C_1 = 1$ and $C_2 = 0$, and we recover the estimate based on color-matching. Short-distance QCD effects renormalize the Wilson coefficients, such that at the mass scale $\mu = m_b = 4.8$ GeV we have $a_1 \approx 1$ and $a_2 \approx 0.2$. The value of $a_2$ is strongly scale-dependent. The uncanceled scale-dependence of the physical amplitude is a clear manifestation of the shortcomings of the naïve factorization approach. As we discuss below, factorization is expected to work more reliably for the color-allowed amplitude.

In applying Eqs (17.3.2) and (17.3.3) the matrix elements with the quarks are usually written in the familiar forms:

$$\langle \pi|\bar{d}\gamma_\mu\gamma_5 u|0\rangle = -if_\pi q_\mu \tag{17.3.4}$$

$$\langle D|\bar{c}\gamma_\mu b|B\rangle = f_+(q^2)(p_B + p_D)_\mu + f_-(q^2)q_\mu. \tag{17.3.5}$$

Here $q = p_B - p_D$ where $p_D$ and $p_B$ are the $D$ and $B$ 4-momentum respectively and $q^2 = m_\pi^2$. The parameterization of the matrix elements in terms of the pion decay constant $f_\pi$ and two $B \to D$ transition form factors follows from the spin and parity transformations of the meson states and current operators, and Lorentz invariance. Thus using the factorization approach, the amplitude Eq. (17.3.2) for $\overline{B}^0 \to D^+\pi^-$ can now be written conveniently as:

$$-i\frac{G_F}{\sqrt{2}} V_{cb} V_{ud}^* a_1 f_\pi f_+(m_\pi^2)(m_B^2 - m_D^2). \tag{17.3.6}$$

The pion decay constant and $B \to D$ form factor must be determined by other methods or from data (for the latter see Section 17.1.2).

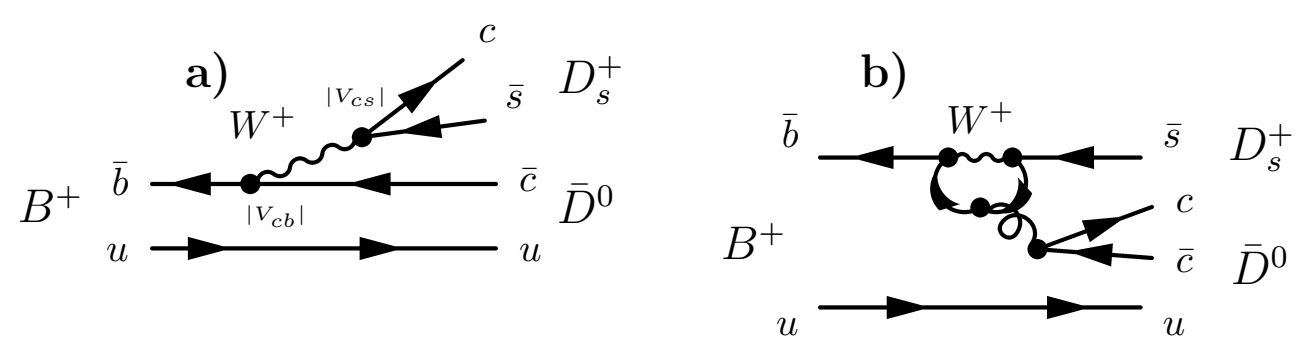

**Figure 17.3.3.** Spectator a) and penguin b) diagrams contributing to $B^+ \to D_s^+\overline{D}$.

This formalism can also be applied to $b \to c\bar{c}s$ (and the Cabibbo-suppressed $b \to c\bar{c}d$) transitions. The color-allowed amplitude leads to final states such as $B \to DD$ and $B \to DD_s$ with two charmed mesons. The color-suppressed amplitude produces a charmonium. The momentum transfer $q^2$ is now large, approximately $m_D^2$, and therefore both form factors ($f_+$, $f_-$) appear in the decay amplitude, which, *e.g.*, for $B^+ \to \overline{D}^0 D_s^+$ is now given by:

$$A_{\rm tree} = -i\frac{G_F}{\sqrt{2}} V_{cb}V_{cs}^* a_1\, f_{D_s} f_+(m_{D_s}^2)(m_B^2 - m_D^2)\,\mathcal{F}$$
$$\text{with} \quad \mathcal{F} = 1 + \frac{f_-(m_{D_s}^2)m_{D_s}^2}{(m_B^2 - m_D^2)f_+(m_{D_s}^2)}\,. \qquad (17.3.7)$$

As shown in Fig. 17.3.3, these decays include contributions from penguin diagrams, since they contain two quarks with identical flavor. However, the penguin operator coefficients ($C_3$, $C_4$, $C_5$, $C_6$) in the effective Hamiltonian are all small, of the order of a few percent. The amplitude can be written as the sum of two pieces, $A_{\rm tree}$ and $A_{\rm peng}$:

$$\mathcal{A}(B \to DD) = A_{\rm tree} + A_{\rm peng}, \qquad (17.3.8)$$

with an estimate $|A_{\rm peng}| < 0.1|A_{\rm tree}|$. It is important to note that while the decay rate is hardly changed by including the penguin contributions they are essential for the observation of direct *CP*-violating asymmetries (see Section 16.6).

A phenomenological approach to predict the branching fractions of hadronic $B$ decays that incorporates factorization is followed in Bauer, Stech, and Wirbel (1987) (BSW) and Neubert and Stech (1998) (NS). In this approach the QCD effects and Wilson coefficients are captured by two phenomenological parameters, $a_1$ and $a_2$. Here $a_1$ represents the factor for decay modes that proceed via Type I (color-favored) amplitudes while $a_2$ is the corresponding factor for Type II (color-suppressed) amplitudes. Decay amplitudes that have contributions from both Type I and II amplitudes (Type III) contain a linear combination of $a_1$ and $a_2$. The values of $a_1$ and $a_2$ are determined from fits to measured $B$ decay rates. For $B$ meson decays the relative phase between $a_1$ and $a_2$ turns out to be positive, which implies constructive interference in the Type III decays. These constants, once determined, are assumed to apply universally to all two-body hadronic $B$ final states.

Table 17.3.1 gives predictions from this model for several Type I, II, and III $B$ decay modes as well as the current PDG (Beringer et al., 2012) values (dominated by *BABAR* and Belle results) for the corresponding branching fractions. The model reproduces well the Type I (color-favored) measurements as well as the Type III where the $a_1$ term dominates the amplitude. Not surprisingly, the Type II predictions differ considerably for some of the decay modes. In particular, the NS model predictions for the $K^{(*)}\psi'$ differ by a factor of two from the experimental measurements. For the $K\psi'$ modes, the prediction is half the measurement while for the $K^*\psi'$ modes the prediction is twice the measurement.

A generalization of factorization can indeed be rigorously derived from the first principles of QCD for the color-allowed amplitude of final states with one charmed meson (Beneke, Buchalla, Neubert, and Sachrajda, 2000). The physical picture is that of color transparency (Bjorken, 1989): in the heavy-quark mass limit, the light meson (*e.g.* the pion) is emitted as a compact color-singlet object with large momentum from the $B \to D$ transition region. In the QCD factorization approach of BBNS (Beneke, Buchalla, Neubert, and Sachrajda, 2000) the coefficient $a_1$ is written as

$$a_1(M) = \sum_{i=1,2} C_i \int_0^1 du\, T_i(u) \varPhi_M(u), \qquad (17.3.9)$$

up to $1/m_b$ corrections, where $\varPhi_M$ denotes the light-cone distribution amplitude of the light meson, which, roughly speaking, describe how the longitudinal momentum of the energetic meson $M$ is shared between the quark and antiquark in the meson, and $T_i(u)$ is a function that can be calculated order by order in the strong coupling $\alpha_s(m_b)$. At tree level, the QCD factorization result reproduces naïve factorization. At the one-loop order, the previously neglected matrix element of the color-octet operator in Eq. (17.3.1) is now non-zero, and leads to a consistent cancellation of the renormalization scale dependence. A consequence of this is that $a_1$ is non-universal, and depends on the light final state meson $M$. However, the non-universality is small, a few percent, as is the correction to naïve factorization. In Table 17.3.1, the decay modes labeled "Type I" receive small corrections to factorization, see (Beneke, Buchalla, Neubert, and Sachrajda, 2000).

Unfortunately, the color-suppressed amplitude $a_2$ in heavy-light final states and the color-allowed amplitude in all final states with two charmed mesons, are not accessible to a rigorous factorization treatment. Counting powers of the small quantity $\Lambda_{\rm QCD}/m_b$ shows that the color-suppressed amplitude in $B \to D\pi$ and related decays is $1/m_b$ suppressed, but the parametric suppression from the form factors and decay constants is not operative in practice. This implies that contrary to the Type I decays, there are no first-principles calculations of Type II and III modes. The same statement applies to the calculation of *CP*-violating charge asymmetries in decays to two charmed mesons.

It is instructive to compare the Type I, II, and III amplitudes for the $B \to D\pi$ final states. In complete generality, we may write

$$\mathcal{A}(\overline{B}^0 \to D^+\pi^-) = T + A, \qquad (17.3.10)$$
$$\sqrt{2}\,\mathcal{A}(\overline{B}^0 \to D^0\pi^0) = C - A, \qquad (17.3.11)$$
$$\mathcal{A}(B^- \to D^0\pi^-) = T + C, \qquad (17.3.12)$$

where $T$ stands for the "color-allowed tree topology", $C$ for "color-suppressed tree topology", and $A$ for "annihilation topology". Since the three final states are related by exchanges of up and down quark, and since the corresponding $SU(2)$ isospin symmetry is a very good approximate symmetry of the QCD Lagrangian, only two of the three amplitudes are independent. The isospin relation $\mathcal{A}(\overline{B}^0 \to D^+\pi^-) + \sqrt{2}\,\mathcal{A}(\overline{B}^0 \to D^0\pi^0) + \mathcal{A}(B^- \to D^0\pi^-) = 0$ allows one to regard $(T + A)$ and $(C - A)$ as the two independent amplitudes. These amplitudes are complex due to strong-interaction phases from final-state interactions. Only the relative phase of the two independent amplitudes is an observable. We define $\delta_{TC}$ to be the relative phase of $(T + A)$ and $(C - A)$. The QCD factorization formula implies that (Beneke, Buchalla, Neubert, and Sachrajda, 2000)

$$\left|\frac{C - A}{T + A}\right| = O(\Lambda_{\text{QCD}}/m_b), \qquad \delta_{TC} = O(1). \quad (17.3.13)$$

**Table 17.3.1.** Predictions of branching fractions of the Neubert & Stech (NS) model (Neubert and Stech, 1998) using $a_1 = 0.98$ and $a_2 = 0.29$ and comparisons with the PDG (Beringer et al., 2012) values.

| Decay mode | NS Model | $\mathcal{B}_{\text{theo}}(\times 10^{-3})$ | PDG $\mathcal{B}(\times 10^{-3})$ |
|---|---|---|---|
| | Type I | | |
| $D^-\pi^+$ | $0.318a_1^2$ | 3.0 | 2.68±0.13 |
| $D^-K^+$ | $0.025a_1^2$ | 0.2 | 0.197±0.021 |
| $D^-\rho^+$ | $0.778a_1^2$ | 7.5 | 7.8±1.3 |
| $D^-K^{*+}$ | $0.041a_1^2$ | 0.4 | 0.45±0.07 |
| $D^-a_1^+$ | $0.844a_1^2$ | 8.1 | 6.0±2.2±2.4 |
| $D^{*-}\pi^+$ | $0.296a_1^2$ | 2.8 | 2.76±0.13 |
| $D^{*-}K^+$ | $0.022a_1^2$ | 0.2 | 0.214±0.016 |
| $D^{*-}\rho^+$ | $0.870a_1^2$ | 8.4 | 6.8±0.9 |
| $D^{*-}K^{*+}$ | $0.049a_1^2$ | 0.5 | 0.33±0.06 |
| $D^{*-}a_1^+$ | $12.17a_1^2$ | 11.6 | 13.0±2.7 |
| | Type II | | |
| $\overline{D}^0\pi^0$ | $0.084a_2^2$ | 0.07 | 0.263±0.014 |
| $K^0J/\psi$ | $0.800a_2^2$ | 0.7 | 0.871±0.032 |
| $K^+J/\psi$ | $0.852a_2^2$ | 0.7 | 1.013±0.034 |
| $K^0\psi'$ | $0.326a_2^2$ | 0.3 | 0.62±0.05 |
| $K^+\psi'$ | $0.347a_2^2$ | 0.3 | 0.639±0.033 |
| $K^{*0}J/\psi$ | $2.518a_2^2$ | 2.1 | 1.33±0.06 |
| $K^{*+}J/\psi$ | $2.680a_2^2$ | 2.3 | 1.43±0.08 |
| $K^{*0}\psi'$ | $1.424a_2^2$ | 1.2 | 0.61±0.05 |
| $K^{*+}\psi'$ | $1.516a_2^2$ | 1.3 | 0.67±0.14 |
| $\pi^0J/\psi$ | $0.018a_2^2$ | 0.02 | 0.0176±0.0016 |
| $\pi^+J/\psi$ | $0.038a_2^2$ | 0.03 | 0.049±0.004 |
| $\rho^0J/\psi$ | $0.050a_2^2$ | 0.04 | 0.027±0.004 |
| $\rho^+J/\psi$ | $0.107a_2^2$ | 0.09 | 0.050±0.008 |
| | Type III | | |
| $\overline{D}^0\pi^+$ | $0.338(a_1 + 0.729a_2(f_D/200\,\text{MeV}))^2$ | 4.8 | 4.84±0.15 |
| $\overline{D}^0\rho^+$ | $0.828(a_1 + 0.450a_2(f_D/200\,\text{MeV}))^2$ | 10.2 | 13.4±1.8 |
| $\overline{D}^{*0}\pi^+$ | $0.315(a_1 + 0.886a_2(f_{D^*}/230\,\text{MeV}))^2$ | 4.8 | 5.19±0.26 |
| $\overline{D}^{*0}\rho^+$ | $0.926(a_1^2 + 0.456a_2^2(f_{D^*}/230\,\text{MeV})^2 + 1.291a_1a_2(f_{D^*}/230\,\text{MeV}))$ | 12.6 | 9.8±1.7 |
| $\overline{D}^{*0}a_1^+$ | $1.296(a_1^2 + 0.128a_2^2(f_{D^*}/230\,\text{MeV})^2 + 0.269a_1a_2(f_{D^*}/230\,\text{MeV}))$ | 13.6 | 19±5 |

Treating the charm meson as a light meson compared to the scale $m_b$, one finds that it is not difficult to accommodate $|C - A|/|T + A| \sim 0.2 - 0.3$ and a large phase $\delta_{TC} \sim 40^\circ$, which is in qualitative agreement with experimental results. The large phase shows that large corrections to naïve factorization must be expected for the color-suppressed amplitude in heavy-light decays.

The situation for $B$ decays to charmonium is ambiguous. QCD factorization formally holds for these decays despite their color suppression, since the "emitted" charmonium is a compact object (Beneke, Buchalla, Neubert, and Sachrajda, 2000). However, various corrections from soft gluon reconnections (Melic, 2004) and color-octet contributions (Beneke and Vernazza, 2009) turn out to be very large relative to the formally dominant color-suppressed amplitude, and prevent a reliable prediction. One should therefore expect large corrections to the naïve factorization and generalized factorization (BSW) estimates of these decay modes, as is indeed observed. Again, these uncertainties prevent a reliable calculation of the (small) $CP$-violating charge asymmetries for final states such as $K\psi$.

While the BSW/NS approach to factorization and the QCD factorization approach (where applicable) provides estimates in agreement with many measured branching fractions, extending this technique to decays with more than two particles in the final state (*e.g.* $B \to D\pi\pi$ or $B^0 \to D^{*-}\pi^+\pi^+\pi^-\pi^0$) is not nearly as successful. A fundamental problem here is that some of the final state particles are the result of gluons and therefore the role of QCD can not be ignored. In (Reader and Isgur, 1993) the problem of multi-body decays with $D$'s and $\pi$'s in the final state is discussed using results from heavy-quark symmetry and factorization. Here the decay process proceeds through intermediate states such as $D\rho$ or $D_2^*\pi$ and the contributions are summed to obtain the total branching fraction. Unfortunately, for many of the modes mentioned in (Reader and Isgur, 1993) and Table 17.3.2, precision measurements are lacking, making a detailed comparison not possible. In Table 17.3.2 the entries with a ">" indicate modes where only an intermediate state and not the explicit final state has been measured. In these cases the measured branching fraction of the intermediate state is taken as the lower limit of the branching fraction of the mode of interest. An example of such a mode is $\overline{D}^0\pi^+\pi^0$ where only the intermediate state $\overline{D}^0\rho^+$ has been measured. For this mode we note that their model's prediction for $B^+ \to \overline{D}^0\pi^+\pi^0$ of 0.59% is significantly lower than the measured $1.34 \pm 0.18\%$ for $\overline{D}^0\rho^+$.

**Table 17.3.2.** Branching fraction predictions of the RI model (Reader and Isgur, 1993) and comparisons with the PDG (Beringer et al., 2012) values. The entries with a ">" indicate modes where only an intermediate state and not the explicit final state has been measured. The measured branching fraction of the intermediate state is taken as the lower limit of the branching fraction of the mode of interest.

| Decay mode | RI Model $\mathcal{B}$ ($\times10^{-3}$) | PDG $\mathcal{B}$ ($\times10^{-3}$) |
|---|---|---|
| $D^-\pi^+\pi^0$ | 5.9 | $> 7.8\pm0.13$ |
| $D^-\pi^+\pi^-$ | 0.7 | $0.84\pm0.09$ |
| $\overline{D}^0\pi^+\pi^0$ | 5.9 | $> 13.4\pm1.8$ |
| $D^-\pi^+\pi^+$ | 0.7 | $1.07\pm0.05$ |
| $D^{*-}\pi^+\pi^0$ | 7.5 | $15\pm5$ |
| $D^{*0}\pi^+\pi^-$ | 1.1 | $0.62\pm0.22$ |
| $D^{*0}\pi^+\pi^0$ | 7.5 | $> 9.8\pm1.7$ |
| $D^{*-}\pi^+\pi^+$ | 1.1 | $1.35\pm0.22$ |
| $D^-\pi^+\pi^+\pi^-$ | 2.1 | $8.0\pm2.5$ |
| $\overline{D}^0\pi^+\pi^+\pi^-$ | 2.1 | $11\pm4$ |
| $D^{*-}\pi^+\pi^+\pi^-$ | 2.9 | $> 13\pm3$ |
| $D^{*-}\pi^+\pi^+\pi^0$ | 2.2 | $15\pm7$ |

### 17.3.3 Decays with a single $D$ decay ($D$, $D^*$, $D_s$)

Due to the experiments at the $B$ Factories there has been an enormous increase in both the number of single charm modes reconstructed and the precision of their branching fractions. As shown in Tables 17.3.3 and 17.3.4 the typical branching fractions for decay modes in this category are in the few tenths of a percent for the modes with a $W \to u\overline{d}$ transition and an order of magnitude smaller for modes with a $W \to u\overline{s}$ transition. In Fig. 17.3.1 the simplest diagrams for $B^+ \to \overline{D}^0\pi^+$ and $B^+ \to \overline{D}^0K^+$ are shown. Including the CKM factors $V_{ud}$ and $V_{us}$ at the relevant vertices explains the dominance of the pion modes over the kaon modes. Other mechanisms such as color-suppression can play an important role in simple two-body final states such as $\overline{D}^0\pi^0$ (Fig. 17.3.2 d). It is important to note that although these diagrams contain only pseudoscalars in the final state it is also likely that the quarks will hadronize into vector particles. Thus the $D$'s can be replaced with $D^*$'s, $\pi$'s with $\rho$'s, $K$'s with $K^*$'s, etc. Finally, the hadronization process also allows for more complicated final states such as $\overline{D}^0K^+K^*$, $D^{*-}3\pi^+\pi^-$, etc.

#### 17.3.3.1 Two body final states

In this section we do not consider the kaon final states (*e.g.* $\overline{D}^0K^+$) as they are discussed in detail in Section 17.8 due to their important role in determining $\phi_3$.

Color-favored two-body decay modes, $D^{(*)-}\pi^+$, $\overline{D}^{(*)0}\pi^+$, were studied in (Aubert, 2007g) using approximately one quarter of the final *BABAR* $\Upsilon(4S)$ data sample. These final states are such that even with relatively simple selection criteria (*e.g.* only using $\overline{D}^0 \to K^+\pi^-$ and $D^- \to K^+\pi^-\pi^-$), high purity samples are obtained. To illustrate the quality (*i.e.* very large signal to background) possible in hadronic $B$ decays into charm we show the beam-energy-substituted mass plots ($m_{\rm ES}$) from (Aubert, 2007g) in Fig. 17.3.4. In all modes the systematic errors are at least a factor of two larger than the statistical errors. In general, there is good agreement between the model predictions in Table 17.3.1 and the branching fraction measurements from this study.

Color-suppressed two-body decay modes have been extensively studied in (Lees, 2011b; Blyth, 2006; Kuzmin, 2007; Schumann, 2005). In the most comprehensive study (Lees, 2011b) eight modes ($\overline{D}^{(*)0}X$, $X = \pi^0$, $\eta$, $\omega$, $\eta'$) are analyzed and their branching fractions measured. The results of this study are in agreement with previous *BABAR* and Belle measurements, although with higher precision. The improved precision in the branching fractions allows for a detailed comparison with predictions from factorization models (Chua, Hou, and Yang, 2002; Deandrea and Polosa, 2002; Eeg, Hiorth, and Polosa, 2002; Neubert and Stech, 1998) and perturbative QCD (pQCD) (Keum, Kurimoto, Li, Lu, and Sanda, 2004; Lu, 2003). There is poor agreement with the factorization predictions; in most cases the measurements are significantly larger than the expectation. In contrast, with the exception of $D^0\omega$ where the measurement is significantly lower than the prediction, experiment and pQCD are close. These differences should not come as a surprise since there is no rigorous QCD approach to the color-suppressed amplitude, as discussed in the theory overview of this chapter. The experimental results along with the model predictions are given in Table VII of Lees (2011b).

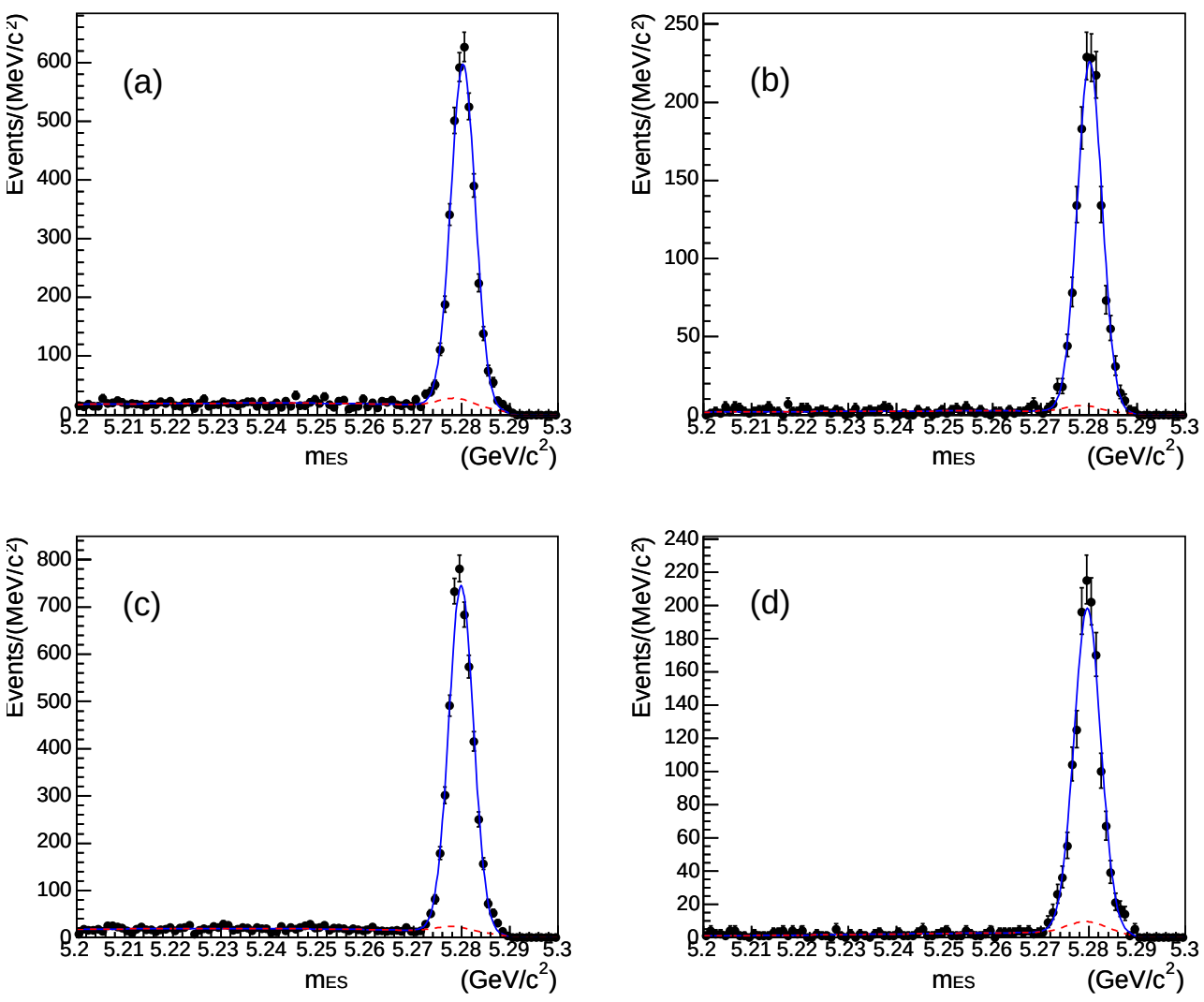

**Figure 17.3.4.** The $m_{\rm ES}$ distributions for (a) $B^0 \to D^-\pi^+$, (b) $B^0 \to D^{*-}\pi^+$, (c) $B^+ \to \overline{D}^0\pi^+$, and (d) $B^+ \to \overline{D}^{*0}\pi^+$ (from Aubert, 2007g). In the figures the solid line is the fit to the data while the background component (including peaking background) is shown as a dashed line.

$B$ meson decay provides a convenient laboratory to study orbitally excited states of the $D$ meson. For the case where a light quark is bound to a $c$ quark, heavy quark effective theory (HQET) suggests that $\boldsymbol{j} = \boldsymbol{L} + \boldsymbol{s}_l$ with $\boldsymbol{s}_l$ the total angular momentum of the light quark and $\boldsymbol{L}$ the orbital angular momentum of the $c\overline{q}$ system will be a good quantum number. As a result, four $L=1$ states are expected with total angular momentum and parity ($J^P$), and $j$ values of $0^+$ ($j=1/2$), $1^+$ ($j=1/2$), $1^+$ ($j=3/2$), and $2^+$ ($j=3/2$). These states are known as the $D_0^*$, $D_1'$, $D_1$, and $D_2^*$ respectively. The states in the mass range of 2.2-2.8 GeV/$c^2$ are often collectively referred to as the $D^{**}$. Both *BABAR* and Belle have studied these states in detail using both specific decay channels (Aubert, 2006p; Abe, 2005i) and Dalitz plot analyses of $B^+ \to D^-\pi^+\pi^+$ and $\overline{B}^0 \to D^0\pi^+\pi^-$ (Abe, 2004f; Kuzmin, 2007; Aubert, 2009g). In addition to branching fraction measurements these studies have also determined the masses and widths of these states. The results are in good agreement with the expectations of HQET. More details on those measurements can be found in Section 19.3.

A variety of final states with a $D_s$ or $D_s^*$ in addition to a scalar or vector meson were the subject of several studies by *BABAR* (Aubert, 2007l, 2008u) and Belle (Das, 2010; Joshi, 2010). These decays are of interest as they can proceed via color-suppressed $W$ exchange (*e.g.* $B^0 \to D_s^{(*)-}K^{(*)+}$), and assuming $SU(3)$ flavor symmetry can be used to calculate the amplitude ratio $r(D^{(*)}\pi) = |A(B^0 \to D^{(*)+}\pi^-)|/|A(B^0 \to D^{(*)-}\pi^+)|$, an important parameter for the determination of $\sin(2\phi_1 + \phi_3)$ using $B^0 \to D^{\mp}\pi^{\pm}$. As shown in Tables 17.3.3 and 17.3.4 the branching fractions into $D_s^{(*)}X$ states are small, a few times $10^{-5}$, as expected from CKM factors and the evident lack of rescattering in the $W$ exchange modes. As predicted in Mantry, Pirjol, and Stewart (2003) the ratios $\mathcal{B}(B^0 \to D_s^-K^+)/\mathcal{B}(B^0 \to D_s^{*-}K^+)$ and $\mathcal{B}(B^0 \to D_s^-K^{*+})/\mathcal{B}(B^0 \to D_s^{*-}K^{*+})$ are consistent with one within the experimental uncertainties.

#### 17.3.3.2 Three or more body final states

Given the large phase space available and mean charged multiplicity of almost six in $B$ meson decay, final states with three or more particles make up a sizable amount of hadronic $B$ decays. The branching fractions for many of the modes of the form $B \to \overline{D}^{(*)}(n\pi)$, $n = 2-5$ charged pions have been measured in (Aubert, 2009g) and (Abe, 2005i; Majumder, 2004). The analysis in (Majumder, 2004) illustrates a difficulty with final states involving a large number of particles, *i.e.* systematic errors from track finding dominate in such high multiplicity decays. It is also interesting to note the absence of branching fraction measurements with multiple $\pi^0$s (*i.e.* not a decay product of a $D^{(*)}$) in the final state.

The three-body decay, $B^0 \to D^{*-}\omega\pi^+$, has been used to study factorization in Aubert (2006ay). As discussed in Reader and Isgur (1993) and Ligeti, Luke, and Wise (2001) the factorization approach allows data from $\tau \to X\nu$ to be used to predict the properties of decays such as $B \to D^*X$, where $X$ is the same hadronic system in both decays. The invariant mass spectrum of the $\omega\pi$ system was found to be in good agreement with the theoretical expectations based on factorization and $\tau$ decay data. In addition, a Dalitz plot analysis shows a non uniform distribution with a preference for $\omega\pi$ at low mass. A broad enhancement in the $D^*\pi$ system at about 2.5 GeV/$c^2$ may indicate the presence of $B^0 \to D_1'\omega$. Finally, the longitudinal polarization of the $D^*$ was found to be in agreement with expectations of HQET.

Three-body decays with charged and neutral kaons as well as $K^*$s in the final state were studied in (Drutskoy, 2002). Even though only 29.4 fb$^{-1}$ of data was used here (a small fraction of Belle's final data sample) five modes of the form $B \to D^{(0)}KK^{(*)0}$ were observed for the first time. An angular analysis of the $KK^*$ system is consistent with the assignment $J^P = 1^+$ and that the decay mainly proceeds through an $a_1(1260)$ intermediate state.

The branching fraction and resonant substructure of the CKM-favored mode $B^0 \to \overline{D}^0K^+\pi^-$ (not including the $D^*$) was determined in (Aubert, 2006n). A motivation for studying this decay was to gain access to $\phi_3$ through the interference of the $\overline{b} \to \overline{c}u\overline{s}$ and $\overline{b} \to \overline{u}c\overline{s}$ amplitudes and use the Dalitz plot to reduce the ambiguity in the strong phase. Unfortunately, the branching fraction turned out to be too small to be of practical use in determining $\phi_3$ with the final *BABAR* and Belle data samples.

Decays of the type $B \to D_s^{(*)} K\pi$ can proceed through the production of an $s\bar{s}$ pair "popping" out of the vacuum. Three such modes ($D_s^- K^+\pi^+$, $D_s^{*-} K^+\pi^+$, and $D_s^- K_s^0\pi^+$) as well as the CKM suppressed $D_s^- K^+K^+$ mode were observed in (Aubert, 2008ai). The first two modes were also studied by Belle (Wiechczynski, 2009). Both groups find that the invariant mass distributions of the $D_s^{(*)} K^+$ subsystem are incompatible with three-body phase space and with enhancements near 2.7 GeV/$c^2$, suggestive of charm resonances below the $D_s^{(*)} K^+$ threshold.

### 17.3.4 Decays with 2 $D$'s

#### 17.3.4.1 $W \to c\bar{d}$

In the neutral $B \to D^{(*)+}D^{(*)-}$ decays, the interference of the dominant tree diagram (see Fig. 17.3.5 a) with the $B^0\overline{B}{}^0$ mixing diagram is sensitive to the CKM phase $\phi_1$ . However, the theoretically uncertain contributions of penguin diagrams (Fig. 17.3.5 b) with different weak phases are potentially significant and may shift both the observed *CP* asymmetries and the branching fractions by amounts that depend on the ratios of the penguin to tree contributions and their relative phases.

The penguin-tree interference in neutral and charged $B \to D^{(*)}\overline{D}{}^{(*)}$ decays can also provide some sensitivity to the angle $\phi_3$, with additional information on the branching fractions of $B \to D_s^{(*)}\overline{D}{}^{(*)}$ decays, assuming $SU(3)$ flavor symmetry between $B \to D^{(*)}\overline{D}{}^{(*)}$ and $B \to D_s^{(*)}\overline{D}{}^{(*)}$.

The color-suppressed decay modes $B^0 \to D^{(*)0}\overline{D}{}^{(*)0}$, if observed, would provide evidence of $W$-exchange or annihilation contributions (see Fig. 17.3.5 c, 17.3.5 d). In principle, these decays could also provide sensitivity to the CKM phase $\phi_1$, if sufficient data were available.

The most precise published results on $D^{(*)}\overline{D}{}^{(*)}$ decays from the $B$ Factories use exclusive reconstruction of these decays: all tracks and neutral energy from each of the decay chain products is reconstructed, and the reconstructed $B$ meson is ultimately composed from these charged tracks and clusters. The $D$ mesons are reconstructed in their decays to some or all of the following: $D^0 \to K^-\pi^+$, $K^-\pi^+\pi^0$, $K^-\pi^+\pi^+\pi^-$, $K^+K^-$, $K_S^0\pi^+\pi^-$, $K_S^0\pi^+\pi^-\pi^0$; and $D^+ \to K_S^0\pi^+$, $K_S^0\pi^+\pi^0$, $K_S^0K^+$, $K^-\pi^+\pi^+$, $K^-K^+\pi^+$. The $D^{*+}$ mesons are then reconstructed in their decays to $D^0\pi^+$ or $D^+\pi^0$. Charge conjugate decays are of course implied throughout. As the product branching fractions of these decays are small ($\mathcal{O}(10^{-7} - 10^{-6})$) particular attention must be paid to particle identification as well as background rejection, details of which can be found in Chapters 5 (charged particle identification) and 9 (background suppression). An example of an $m_{\rm ES}$ distribution with good signal-to-background is shown in Fig. 17.3.6.

Both *BABAR* and Belle have several results for these decays. The branching fraction results, as well as corresponding theoretical predictions, are summarized in Table 17.3.5.

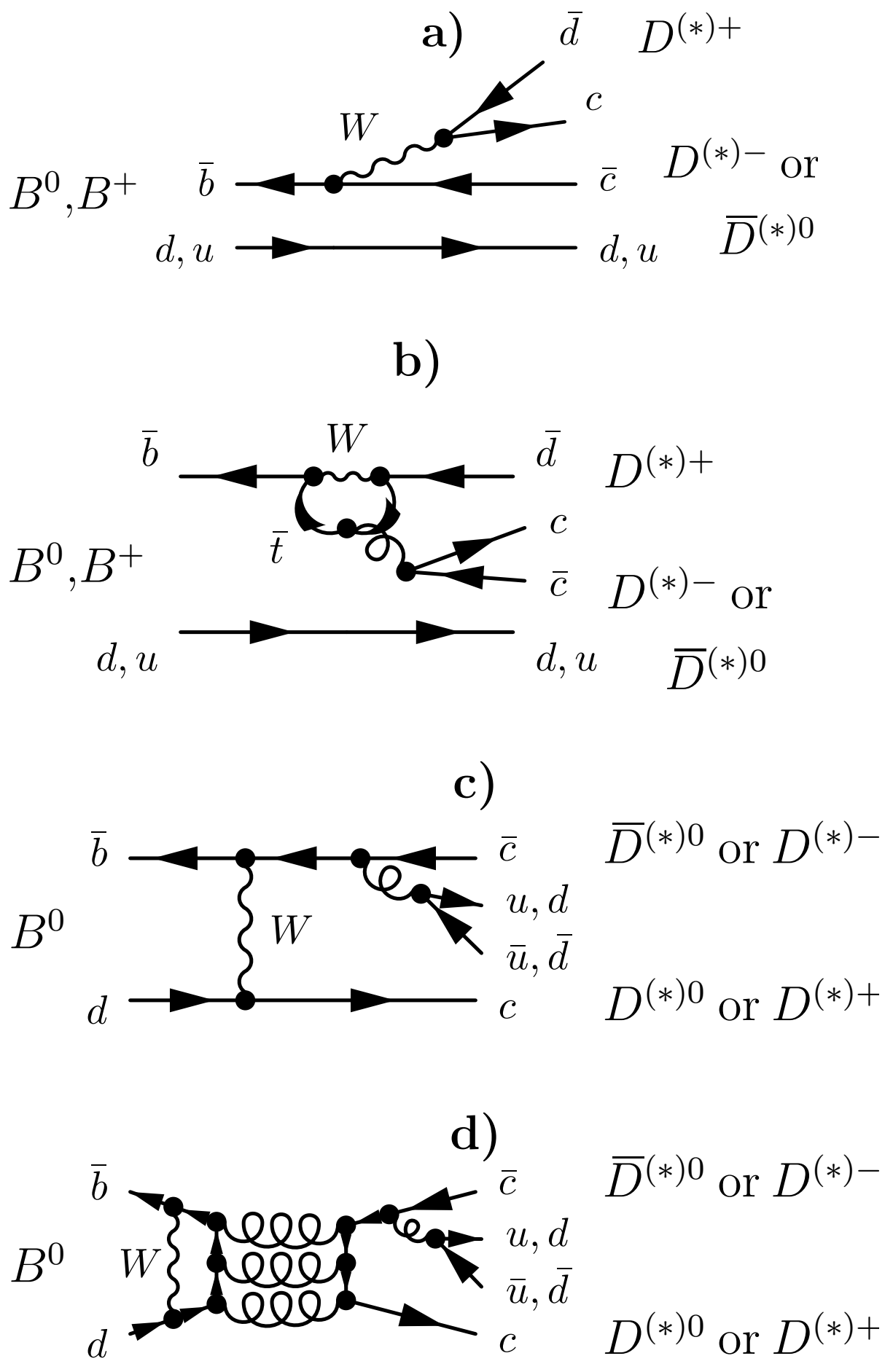

**Figure 17.3.5.** Feynman graphs for $B \to D^{(*)}\overline{D}{}^{(*)}$ decays: the tree (a) and penguin (b) diagrams are the leading terms for both $B^0 \to D^{(*)+}D^{(*)-}$ and $B^+ \to D^{(*)+}\overline{D}{}^{(*)0}$ decays, whereas the exchange (c) and annihilation (d) diagrams (the latter of which is OZI-suppressed) are the lowest-order terms for $B^0 \to D^{(*)0}\overline{D}{}^{(*)0}$ decays.

In addition to the branching fractions (and to the time-dependent *CP* asymmetries, detailed in Section 17.6), *CP*-violating charge asymmetries can be measured in the four charged $B \to D^{(*)}\overline{D}{}^{(*)}$ decays as well as in $B^0 \to D^{*\pm}D^{\mp}$, and also polarization can be measured in the vector-vector decays $B^0 \to D^{*+}D^{*-}$ and $B^+ \to D^{*+}\overline{D}{}^{*0}$. Those results are summarized in Tables 17.3.6 and 17.3.7 respectively.

The Cabibbo-favored $D_s^{(*)}\overline{D}{}^{(*)}$ decays (which can occur via tree and penguin diagrams analogous to those in Fig. 17.3.5a and b, each with the upper $\bar{d}$ replaced with an $\bar{s}$) typically have branching fractions an order of magnitude higher than their $D^{(*)}\overline{D}{}^{(*)}$ analogues, *i.e.* in the $\mathcal{O}(10^{-3} - 10^{-2})$ range rather than $\mathcal{O}(10^{-4} - 10^{-3})$. The measured and predicted branching fractions for these modes can be found in Table 17.3.8, and measured and predicted polarizations can be found in Table 17.3.9.

**Table 17.3.3.** Measured single charm $B^+$ branching fractions ($\mathcal{B}$) from *BABAR*, Belle, and the PDG (Beringer et al., 2012) (average). The PDG value may use measurements from other experiments when calculating the average.

| | *BABAR* results | | Belle results | | PDG Averages |
|---|---|---|---|---|---|
| Final state | $\mathcal{B}$ ($\times 10^{-3}$) | Ref. | $\mathcal{B}$ ($\times 10^{-3}$) | Ref. | $\mathcal{B}$ ($\times 10^{-3}$) |
| $\overline{D}^0\pi^+$ | $4.90 \pm 0.07 \pm 0.22$ | (Aubert, 2007g) | | | $4.84 \pm 0.15$ |
| $\overline{D}^0K^+/\mathcal{B}(\overline{D}^0\pi^+)$ | $83.1 \pm 3.5 \pm 2.0$ | (Aubert, 2004l) | $67.7 \pm 2.3 \pm 3.0$ | (Horii, 2008) | $76 \pm 6$ |
| $\overline{D}^0K^*(892)^+$ | $0.529 \pm 0.030 \pm 0.034$ | (Aubert, 2006q) | | | $0.53 \pm 0.04$ |
| $\overline{D}^0K^+\overline{K}^0$ | | | $0.55 \pm 0.14 \pm 0.08$ | (Drutskoy, 2002) | $0.55 \pm 0.14 \pm 0.08$ |
| $\overline{D}^0K^+\overline{K}^*(892)^0$ | | | $0.75 \pm 0.13 \pm 0.11$ | (Drutskoy, 2002) | $0.75 \pm 0.13 \pm 0.11$ |
| $D^*(2010)^-\pi^+\pi^+$ | | | $1.25 \pm 0.08 \pm 0.22$ | (Abe, 2004f) | $1.35 \pm 0.22$ |
| $D^-\pi^+\pi^+$ | $1.08 \pm 0.03 \pm 0.05$ | (Aubert, 2009g) | $1.02 \pm 0.04 \pm 0.15$ | (Abe, 2004f) | $1.07 \pm 0.05$ |
| $\overline{D}^*(2007)^0\pi^+$ | $5.52 \pm 0.17 \pm 0.42$ | (Aubert, 2007g) | | | $5.18 \pm 0.26$ |
| $\overline{D}^*(2007)^0K^+$ | $0.421^{+0.030}_{-0.026} \pm 0.021$ | (Aubert, 2005t) | $0.40 \pm 0.11 \pm 0.02$ | (Abe, 2001f) | $0.420 \pm 0.034$ |
| $\overline{D}^*(2007)^0K^*(892)^+$ | $0.83 \pm 0.11 \pm 0.10$ | (Aubert, 2004k) | | | $0.81 \pm 0.14$ |
| $\overline{D}^*(2007)^0K^+\overline{K}^*(892)^0$ | | | $1.53 \pm 0.31 \pm 0.29$ | (Drutskoy, 2002) | $1.53 \pm 0.31 \pm 0.29$ |
| $\overline{D}^*(2007)^0\pi^+\pi^+\pi^-$ | | | $10.55 \pm 0.47 \pm 1.29$ | (Majumder, 2004) | $10.3 \pm 1.2$ |
| $\overline{D}^{*0}3\pi^+2\pi^-$ | | | $5.67 \pm 0.91 \pm 0.85$ | (Majumder, 2004) | $5.67 \pm 0.91 \pm 0.85$ |
| $D^*(2010)^-3\pi^+\pi^-$ | | | $2.56 \pm 0.26 \pm 0.33$ | (Majumder, 2004) | $2.56 \pm 0.26 \pm 0.33$ |
| $\overline{D}^{**0}\pi^+$ | $5.9 \pm 1.3 \pm 0.2$ | (Aubert, 2006p) | | | $5.9 \pm 1.3 \pm 0.2$ |
| $\overline{D}_1(2420)^0\pi^+ \times \mathcal{B}(\overline{D}_1^0 \to \overline{D}^0\pi^+\pi^-)$ | | | $0.185 \pm 0.029^{+0.035}_{-0.055}$ | (Abe, 2005i) | $0.185 \pm 0.029^{+0.035}_{-0.055}$ |
| $\overline{D}_1(2421)^0\pi^+ \times \mathcal{B}(\overline{D}_1^0 \to D^{*-}\pi^+)$ | | | $0.68 \pm 0.07 \pm 0.13$ | (Abe, 2004f) | $0.68 \pm 0.07 \pm 0.13$ |
| $\overline{D}_2^*(2462)^0\pi^+ \times \mathcal{B}(\overline{D}_2^{*0} \to D^-\pi^+)$ | $0.35 \pm 0.02 \pm 0.04$ | (Aubert, 2009g) | $0.34 \pm 0.03 \pm 0.072$ | (Abe, 2004f) | $0.35 \pm 0.04$ |
| $\overline{D}_2^*(2462)^0\pi^+ \times \mathcal{B}(\overline{D}_2^{*0} \to D^{*-}\pi^+)$ | | | $0.18 \pm 0.03 \pm 0.04$ | (Abe, 2004f) | $0.18 \pm 0.03 \pm 0.04$ |
| $\overline{D}_0^*(2400)^0\pi^+ \times \mathcal{B}(\overline{D}_0^{*0} \to D^-\pi^+)$ | $0.68 \pm 0.03 \pm 0.2$ | (Aubert, 2009g) | $0.61 \pm 0.06 \pm 0.18$ | (Abe, 2004f) | $0.64 \pm 0.14$ |
| $\overline{D}_1'(2427)^0\pi^+ \times \mathcal{B}(\overline{D}_1'^0 \to D^{*-}\pi^+)$ | | | $0.50 \pm 0.04 \pm 0.11$ | (Abe, 2004f) | $0.50 \pm 0.04 \pm 0.11$ |
| $D_s^+\pi^0$ | $0.016^{+0.006}_{-0.005} \pm 0.001$ | (Aubert, 2007l) | | | $0.016^{+0.006}_{-0.005} \pm 0.001$ |
| $D_s^-\pi^+K^+$ | $0.202 \pm 0.013 \pm 0.038$ | (Aubert, 2008ai) | $0.171^{+0.008}_{-0.007} \pm 0.025$ | (Wiechczynski, 2009) | $0.180 \pm 0.022$ |
| $D_s^{*-}\pi^+K^+$ | $0.167 \pm 0.016 \pm 0.035$ | (Aubert, 2008ai) | $0.131^{+0.013}_{-0.012} \pm 0.028$ | (Wiechczynski, 2009) | $0.145 \pm 0.024$ |
| $D_s^-K^+K^+$ | $0.011 \pm 0.004 \pm 0.002$ | (Aubert, 2008ai) | | | $0.011 \pm 0.004 \pm 0.002$ |

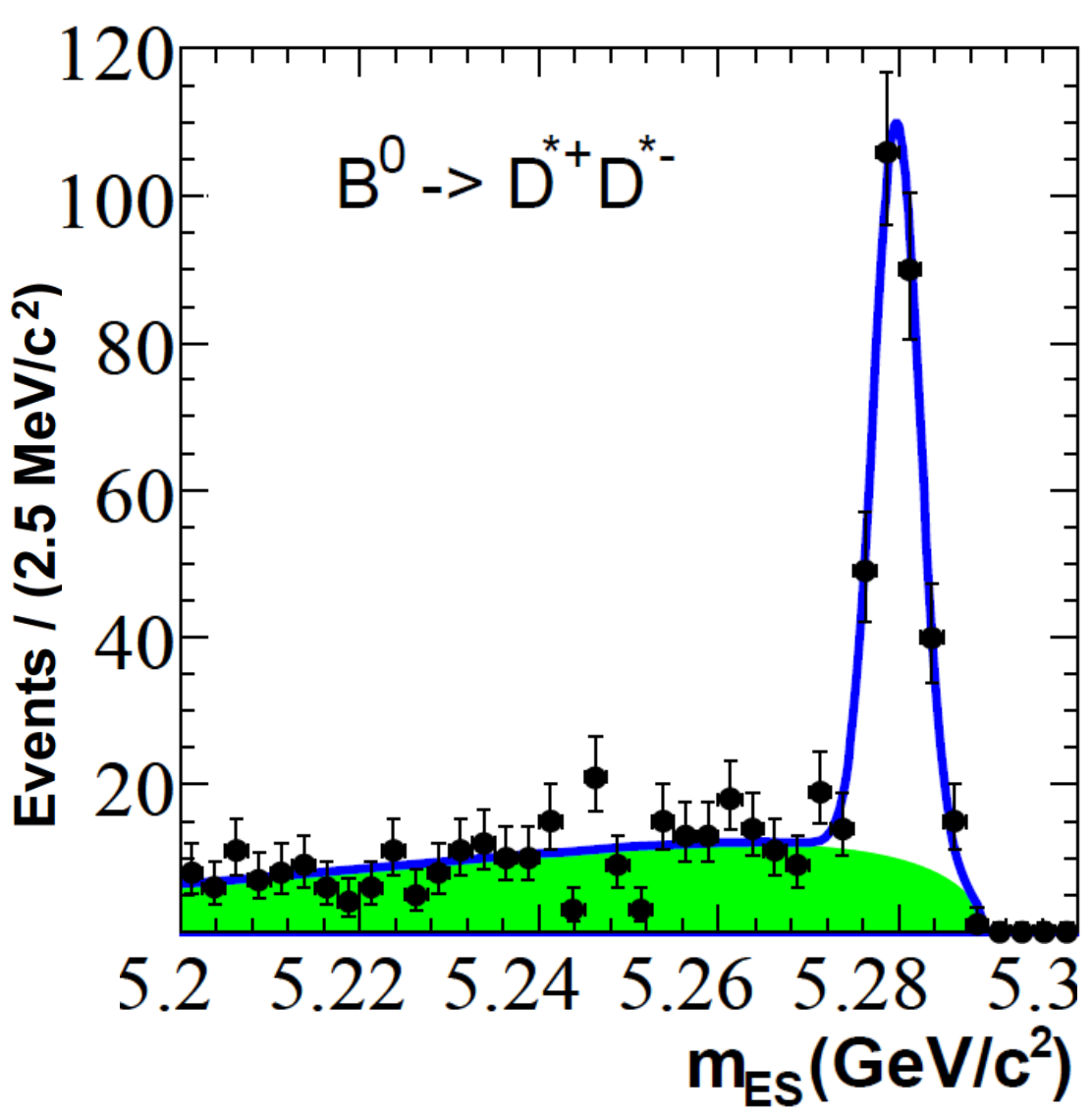

**Figure 17.3.6.** The $m_{\rm ES}$ distribution for $D^{*+}D^{*-}$ candidates from (Aubert, 2006m).

$B$ mesons can also decay to $D_{sJ}^{(*)}\overline{D}^{(*)}$ states, with the multiple $D_{sJ}^{(*)}$ states having been discovered at the $B$ Factories since the original observation of $D_{sJ}^*(2317)^+$ at *BABAR* in 2002. These decays are described in Section 19.3 of this Book, and specifically $B$ decays to $D_{sJ}^{(*)}\overline{D}^{(*)}$ are described in Section 19.3.4.

### 17.3.4.2 $W \to c\bar{s}$

Diagrams similar to Fig. 17.3.2 with $W \to c\bar{s}$ and $u\bar{u}/d\bar{d}$ popping lead to $B \to \overline{D}^{(*)}D^{(*)}K$ final states. These final states play a substantial role in the $B$ decays since they account for about 4% of their total branching fraction. Here, $D^{(*)}$ is either a $D^0$, $D^{*0}$, $D^+$ or $D^{*+}$, $\overline{D}^{(*)}$ is the charge conjugate of $D^{(*)}$ and $K$ is either a $K^+$ or a $K^0$. Twenty-two decay modes are possible with this configuration. The decays of $B$ mesons to $\overline{D}^{(*)}D^{(*)}K$ final states are interesting for many different reasons. For example, in the past (*i.e.* early 1990's), the hadronic decays of the $B$ meson were in theoretical conflict with the $B$ semileptonic branching fraction due to the inconsistency originating from the number of charmed hadrons per $B$ decay (Bigi, Blok, Shifman, and Vainshtein, 1994). At the time, the measured semileptonic branching fraction, $\approx 10\%$, was in

**Table 17.3.4.** Measured single charm $B^0$ branching fractions ($\mathcal{B}$) from *BaBar*, Belle, and the PDG (Beringer et al., 2012) (average). The PDG value may use measurements from other experiments when calculating the average.

| | *BaBar* results | | Belle results | | PDG Averages |
|---|---|---|---|---|---|
| Final state | $\mathcal{B}$ ($\times 10^{-3}$) | Ref. | $\mathcal{B}$ ($\times 10^{-3}$) | Ref. | $\mathcal{B}$ ($\times 10^{-3}$) |
| $D^-\pi^+$ | $2.55 \pm 0.05 \pm 0.16$ | (Aubert, 2007g) | | | $2.68 \pm 0.13$ |
| $D^-K^0\pi^+$ | $0.49 \pm 0.07 \pm 0.05$ | (Aubert, 2005aa) | | | $0.49 \pm 0.07 \pm 0.05$ |
| $D^-K^*(892)^+$ | $0.46 \pm 0.06 \pm 0.05$ | (Aubert, 2005aa) | | | $0.45 \pm 0.07$ |
| $D^-K^+$ | | | $0.18 \pm 0.04 \pm 0.01$ | (Abe, 2001f) | $0.197 \pm 0.021$ |
| $D^-K^+\overline{K}^*(892)^0$ | | | $0.88 \pm 0.11 \pm 0.15$ | (Drutskoy, 2002) | $0.88 \pm 0.11 \pm 0.15$ |
| $\overline{D}^0\pi^+\pi^-$ | | | $0.84 \pm 0.04 \pm 0.08$ | (Kuzmin, 2007) | $0.84 \pm 0.04 \pm 0.08$ |
| $D^*(2010)^-\pi^+$ | $2.79 \pm 0.08 \pm 0.17$ | (Aubert, 2007g) | | | $2.76 \pm 0.13$ |
| $D^*(2010)^-K^+$ | $0.214 \pm 0.012 \pm 0.010$ | (Aubert, 2006n) | $0.20 \pm 0.04 \pm 0.01$ | (Abe, 2001f) | $0.214 \pm 0.016$ |
| $D^*(2010)^-K^0\pi^+$ | $0.30 \pm 0.07 \pm 0.03$ | (Aubert, 2005aa) | | | $0.30 \pm 0.07 \pm 0.03$ |
| $D^*(2010)^-K^*(892)^+$ | $0.32 \pm 0.06 \pm 0.03$ | (Aubert, 2005aa) | | | $0.33 \pm 0.06$ |
| $D^*(2010)^-K^+\overline{K}^*(892)^0$ | | | $1.29 \pm 0.22 \pm 0.25$ | (Drutskoy, 2002) | $1.29 \pm 0.22 \pm 0.25$ |
| $D^*(2010)^-\pi^+\pi^+\pi^-$ | | | $6.81 \pm 0.23 \pm 0.72$ | (Majumder, 2004) | $7.0 \pm 0.8$ |
| $D^{*-}3\pi^+2\pi^-$ | | | $4.72 \pm 0.59 \pm 0.71$ | (Majumder, 2004) | $4.72 \pm 0.59 \pm 0.71$ |
| $\overline{D}^*(2010)^-\omega\pi^+$ | $2.88 \pm 0.21 \pm 0.31$ | (Aubert, 2006k) | | | $2.89 \pm 0.30$ |
| $D_1(2430)^0\omega$ | $0.41 \pm 0.12 \pm 0.11$ | (Aubert, 2006k) | | | $0.41 \pm 0.12 \pm 0.11$ |
| $\times\mathcal{B}(D_1(2430)^0 \to D^{*+}\pi^+)$ | | | | | |
| $\overline{D}^{**-}\pi^+$ | $2.1 \pm 1.0 \pm 0.1$ | (Aubert, 2006p) | | | $2.1 \pm 1.0 \pm 0.1$ |
| $D_1(2420)^-\pi^+$ | | | $0.089 \pm 0.015^{+0.017}_{-0.032}$ | (Abe, 2005i) | $0.100^{+0.021}_{-0.025}$ |
| $\times\mathcal{B}(\overline{D}_1^- \to D^-\pi^+\pi^-)$ | | | | | |
| $\overline{D}_2^*(2460)^-\pi^+$ | | | $0.215 \pm 0.017 \pm 0.031$ | (Kuzmin, 2007) | $0.215 \pm 0.017 \pm 0.031$ |
| $\times\mathcal{B}(D_2^*(2460) \to D^0\pi^-)$ | | | | | |
| $\overline{D}_0^*(2400)^-\pi^+$ | | | $0.060 \pm 0.013 \pm 0.027$ | (Kuzmin, 2007) | $0.060 \pm 0.013 \pm 0.027$ |
| $\times\mathcal{B}(D_0^*(2400) \to D^0\pi^-)$ | | | | | |
| $D_{s0}(2317)^-K^+$ | | | $0.042^{+0.014}_{-0.013} \pm 0.004$ | (Drutskoy, 2005) | $0.042^{+0.014}_{-0.013} \pm 0.004$ |
| $\times\mathcal{B}(D_{s0}(2317) \to D_s\pi^0)$ | | | | | |
| $D^+\pi^-$ | | | $(7.8 \pm 1.3 \pm 0.4) \times 10^{-4}$ | (Das, 2010) | $(7.8 \pm 1.3 \pm 0.4) \times 10^{-4}$ |
| $D_s^+\pi^-$ | $0.025 \pm 0.004 \pm 0.002$ | (Aubert, 2008u) | $0.0199 \pm 0.0026 \pm 0.0018$ | (Das, 2010) | $0.0216 \pm 0.0026$ |
| $D_s^{*+}\pi^-$ | $0.026^{+0.005}_{-0.004} \pm 0.002$ | (Aubert, 2008u) | $0.0175 \pm 0.0034 \pm 0.0020$ | (Joshi, 2010) | $0.021 \pm 0.004$ |
| $D_s^{*+}\rho^-$ | $0.041^{+0.013}_{-0.012} \pm 0.004$ | (Aubert, 2008u) | | | $0.041^{+0.013}_{-0.012} \pm 0.004$ |
| $D_s^-K^+$ | $0.029 \pm 0.004 \pm 0.002$ | (Aubert, 2008u) | $0.0191 \pm 0.0024 \pm 0.0017$ | (Das, 2010) | $0.022 \pm 0.005$ |
| $D_s^{*-}K^+$ | $0.024 \pm 0.004 \pm 0.002$ | (Aubert, 2008u) | $0.0202 \pm 0.0033 \pm 0.0022$ | (Joshi, 2010) | $0.0219 \pm 0.0030$ |
| $D_s^-K^*(892)^+$ | $0.035^{+0.01}_{-0.009} \pm 0.004$ | (Aubert, 2008u) | | | $0.035^{+0.01}_{-0.009} \pm 0.004$ |
| $D_s^{*-}K^*(892)^+$ | $0.032^{+0.014}_{-0.012} \pm 0.004$ | (Aubert, 2008u) | | | $0.032^{+0.014}_{-0.012} \pm 0.004$ |
| $D_s^-\pi^+K^0$ | $0.110 \pm 0.026 \pm 0.020$ | (Aubert, 2008ai) | | | $0.110 \pm 0.026 \pm 0.020$ |
| $\overline{D}^0K^0$ | $0.053 \pm 0.007 \pm 0.003$ | (Aubert, 2006k) | $0.050^{+0.013}_{-0.012} \pm 0.006$ | (Krokovny, 2003a) | $0.052 \pm 0.007$ |
| $\overline{D}^0K^+\pi^-$ | $0.088 \pm 0.015 \pm 0.009$ | (Aubert, 2006n) | | | $0.088 \pm 0.015 \pm 0.009$ |
| $\overline{D}^0K^*(892)^0$ | $0.040 \pm 0.007 \pm 0.003$ | (Aubert, 2006k) | $0.048^{+0.013}_{-0.010} \pm 0.005$ | (Krokovny, 2003a) | $0.042 \pm 0.006$ |
| $D_2^*(2460)^-K^+$ | $0.0183 \pm 0.0040 \pm 0.0031$ | (Aubert, 2006n) | | | $0.0183 \pm 0.0040 \pm 0.0031$ |
| $\times\mathcal{B}(D_2^*(2460)^- \to \overline{D}^0\pi^-)$ | | | | | |
| $\overline{D}^0\pi^0$ | $0.269 \pm 0.009 \pm 0.013$ | (Lees, 2011b) | $0.225 \pm 0.014 \pm 0.035$ | (Blyth, 2006) | $0.263 \pm 0.014$ |
| $\overline{D}^0\rho^0$ | | | $0.319 \pm 0.020 \pm 0.045$ | (Kuzmin, 2007) | $0.319 \pm 0.020 \pm 0.045$ |
| $\overline{D}^0f_2$ | | | $0.120 \pm 0.018 \pm 0.038$ | (Kuzmin, 2007) | $0.120 \pm 0.018 \pm 0.038$ |
| $\overline{D}^0\eta$ | $0.253 \pm 0.009 \pm 0.011$ | (Lees, 2011b) | $0.177 \pm 0.016 \pm 0.021$ | (Blyth, 2006) | $0.236 \pm 0.032$ |
| $\overline{D}^0\eta\prime$ | $0.148 \pm 0.013 \pm 0.007$ | (Lees, 2011b) | $0.114 \pm 0.020^{+0.010}_{-0.013}$ | (Schumann, 2005) | $0.138 \pm 0.016$ |
| $\overline{D}^0\omega$ | $0.257 \pm 0.011 \pm 0.014$ | (Lees, 2011b) | $0.237 \pm 0.023 \pm 0.028$ | (Blyth, 2006) | $0.253 \pm 0.016$ |
| $\overline{D}^*(2007)^0\pi^0$ | $0.305 \pm 0.014 \pm 0.028$ | (Lees, 2011b) | $0.139 \pm 0.018 \pm 0.026$ | (Blyth, 2006) | $0.22 \pm 0.06$ |
| $\overline{D}^*(2007)^0\eta$ | $0.269 \pm 0.014 \pm 0.023$ | (Lees, 2011b) | $0.140 \pm 0.028 \pm 0.026$ | (Blyth, 2006) | $0.23 \pm 0.06$ |
| $\overline{D}^*(2007)^0\eta\prime$ | $0.148 \pm 0.022 \pm 0.013$ | (Lees, 2011b) | $0.121 \pm 0.034 \pm 0.022$ | (Schumann, 2005) | $0.140 \pm 0.022$ |
| $\overline{D}^*(2007)^0\pi^+\pi^-$ | | | $0.62 \pm 0.012 \pm 0.018$ | (Satpathy, 2003) | $0.62 \pm 0.012 \pm 0.018$ |
| $\overline{D}^*(2007)^0K^0$ | $0.036 \pm 0.012 \pm 0.003$ | (Aubert, 2006k) | | | $0.036 \pm 0.012 \pm 0.003$ |
| $\overline{D}^*(2007)^0\pi^+\pi^+\pi^-\pi^-$ | | | $2.60 \pm 0.47 \pm 0.37$ | (Majumder, 2004) | $2.7 \pm 0.5$ |
| $\overline{D}^*(2007)^0\omega$ | $0.455 \pm 0.024 \pm 0.0039$ | (Lees, 2011b) | $0.229 \pm 0.039 \pm 0.040$ | (Blyth, 2006) | $0.36 \pm 0.11$ |

**Table 17.3.5.** Results of the measured branching fractions for the ten $B \to D^{(*)}\overline{D}^{(*)}$ decay modes from *BABAR* and Belle: the number of events for fitted signal $N^{\rm sig}$, the branching fractions $\mathcal{B}$ (and where appropriate 90% C.L. upper limits on branching fractions), as compared with theoretical predictions. All *BABAR* measurements are from (Aubert, 2006m). (Empty entries indicate no measurement from the given experiment, or no prediction.)

| Mode | $N^{\rm sig}_{BABAR}$ | $N^{\rm sig}_{\rm Belle}$ | $\mathcal{B}_{BABAR}$ $(10^{-4})$ | $\mathcal{B}_{\rm Belle}$ $(10^{-4})$ | $\mathcal{B}^{\rm theory}_{\rm predict}$ $(10^{-4})$ |
|---|---|---|---|---|---|
| $B^0 \to D^{*+}D^{*-}$ | 270±19 | 1225±59 | 8.1 ±0.6± 1.0 | 7.82 ±0.38± 0.63 (Kronenbitter, 2012) | 6.0 (Rosner, 1990) |
| $B^0 \to D^{*\pm}D^{\mp}$ | 156±17 | 887±39 | 5.7 ±0.7± 0.7 | 6.14 ±0.29± 0.50 (Rohrken, 2012) | |
| $B^0 \to D^+D^-$ | 63±9 | 221±19 | 2.8 ±0.4± 0.5 | 2.12 ±0.16± 0.18 (Rohrken, 2012) | |
| $B^0 \to D^{*0}\overline{D}^{*0}$ | 0±6 | | −1.3 ±1.1± 0.4 (< 0.9) | | |
| $B^0 \to D^{*0}\overline{D}^0$ | 10±8 | | 1.0 ±1.1± 0.4 (< 2.9) | | |
| $B^0 \to D^0\overline{D}^0$ | −11±12 | 0±25 | −0.1 ±0.5± 0.2 (< 0.6) | < 0.43 (Adachi, 2008b) | |
| $B^+ \to D^{*+}\overline{D}^{*0}$ | 185±20 | | 8.1 ±1.2± 1.2 | | 7.1 (Sanda and Xing, 1997) |
| $B^+ \to D^{*+}\overline{D}^0$ | 115±16 | 74±12 | 3.6 ±0.5± 0.4 | 4.57 ±0.71± 0.56 (Majumder, 2005) | 3.7 (Sanda and Xing, 1997) |
| $B^+ \to D^+\overline{D}^{*0}$ | 63±11 | | 6.3 ±1.4± 1.0 | | 3.1 (Sanda and Xing, 1997) |
| $B^+ \to D^+\overline{D}^0$ | 129±20 | 370±29 | 3.8 ±0.6± 0.5 | 3.85 ±0.31± 0.38 (Adachi, 2008b) | 5.3 (Sanda and Xing, 1997) |

**Table 17.3.6.** Results of measured *CP*-violating charge asymmetries $\mathcal{A}_{CP}$ for $D^{*\pm}D^{\mp}$ and the four charged $B$ modes, as compared with theoretical predictions (where $\mathcal{A}_{CP}$ is defined as $(\Gamma^- - \Gamma^+)/(\Gamma^- + \Gamma^+)$, where the superscript refers to the sign of the $B^{\pm}$ meson in the case of the charged $B$ decays, and for $D^{*\pm}D^{\mp}$, $\Gamma^+$ refers to $D^{*-}D^+$ and $\Gamma^-$ to $D^{*+}D^-$. Empty entries indicate no measurement from the given experiment, or no prediction.)

| Mode | $\mathcal{A}_{CP}^{BABAR}$ | $\mathcal{A}_{CP}^{\rm Belle}$ | Theoretical predictions |
|---|---|---|---|
| $B^0 \to D^{*\pm}D^{\mp}$ | $0.008 \pm 0.048 \pm 0.013$ (Aubert, 2009ad) | $0.06 \pm 0.05 \pm 0.02$ (Rohrken, 2012) | |
| $B^+ \to D^{*+}\overline{D}^{*0}$ | $-0.15 \pm 0.11 \pm 0.02$ (Aubert, 2006m) | | 0.012 (Xing, 2000) |
| $B^+ \to D^{*+}\overline{D}^0$ | $-0.06 \pm 0.13 \pm 0.02$ (Aubert, 2006m) | | 0.012 (Xing, 2000) |
| $B^+ \to D^+\overline{D}^{*0}$ | $0.13 \pm 0.18 \pm 0.04$ (Aubert, 2006m) | | 0.002 (Xing, 2000) |
| $B^+ \to D^+\overline{D}^0$ | $-0.13 \pm 0.14 \pm 0.02$ (Aubert, 2006m) | $0.00 \pm 0.08 \pm 0.02$ (Adachi, 2008b) | 0.030 (Xing, 2000) |

**Table 17.3.7.** Results of measured polarization parameters for the two $D^*\overline{D}^*$ vector-vector decays, as compared with theoretical predictions. Here $R_L$ is the fraction of longitudinal polarization and $R_\perp$ is the *CP*-odd fraction. (Empty entries indicate no measurement from the given experiment, or no prediction. There are presently no published measurements of, or predictions for, polarization in the $D^{*+}\overline{D}^{*0}$ mode.)

| Mode | $\begin{pmatrix} R_\perp \\ R_L \end{pmatrix}^{BABAR}$ | $\begin{pmatrix} R_\perp \\ R_L \end{pmatrix}^{\rm Belle}$ | Theoretical predictions |
|---|---|---|---|
| $B^0 \to D^{*+}D^{*-}$ | $0.158 \pm 0.028 \pm 0.006$ (Aubert, 2009ad) | $0.138 \pm 0.024 \pm 0.006$ (Kronenbitter, 2012) | 0.06 (Rosner, 1990) |
| | | $0.624 \pm 0.029 \pm 0.011$ (Kronenbitter, 2012) | 0.55 (Rosner, 1990) |

conflict with expectations from parton model calculations, 15 − 16%. It was realized (Buchalla, Dunietz, and Yamamoto, 1995) that an enhancement in the $b \to c\bar{c}s$ transition was needed to resolve the theoretical discrepancy with the $B$ semileptonic branching fraction. Buchalla *et al.* predicted sizable branching fractions for decays of the form $B \to \overline{D}^{(*)}D^{(*)}K\,(X)$. Furthermore, the $\overline{D}^{(*)}D^{(*)}K$ events have been used to investigate isospin relations and to extract a measurement of the ratio of $\Upsilon(4S) \to B^+B^-$ and $\Upsilon(4S) \to B^0\overline{B}^0$ decays (Poireau and Zito, 2011). Likewise, the mode $B^0 \to D^{*-}D^{*+}K^0_S$ has been used to perform a time-dependent *CP* asymmetry measurement to determine the sign of $\cos 2\phi_1$ (see Section 17.6). It is also worth recalling that many $D^{(*)}K$ and $\overline{D}^{(*)}D^{(*)}$ resonant processes are at play in the studied decay channels. Using $B \to \overline{D}^{(*)}D^{(*)}K$ final states, *BABAR* and Belle observed and measured properties of the resonances $D^+_{s1}(2536)$ (see Section 19.3), $D_{sJ}(2700)$ (see also Section 19.3), $\psi(3770)$ (see Section 18.2), and $X(3872)$ (see Section 18.3).

*BABAR* reconstructs the $B^0$ and $B^+$ mesons in the 22 $\overline{D}^{(*)}D^{(*)}K$ modes using 429 fb$^{-1}$ (del Amo Sanchez, 2011e), while Belle studies only the modes $B^0 \to D^{*-}D^{*+}K^0$ and $B^+ \to \overline{D}^0D^0K^+$ with 414 fb$^{-1}$ (Brodzicka, 2008; Dalseno, 2007). The collaborations use the decays of particles into $K^0_S \to \pi^+\pi^-$, $D^0 \to K^-\pi^+$, $K^-\pi^+\pi^0$, and $K^-\pi^+\pi^-\pi^+$, $D^+ \to K^-\pi^+\pi^+$, $D^{*+} \to D^0\pi^+$, and $D^+\pi^0$, $D^{*0} \to D^0\pi^0$, and $D^0\gamma$ final states. Additionally, Belle uses the decays $D^0 \to K^0_S\pi^+\pi^-$ and $D^0 \to K^-K^+$. The selection of these particles is based on mass cuts, energies of the decay products, vertexing and particle identification to name a few. The $B$ candidates are reconstructed by combining a $\overline{D}^{(*)}$, a $D^{(*)}$ and a $K$ candidate in a subset of the 22 modes. To suppress the background, topological variables are used which dis-criminate against continuum background (see Chapter 4). Signal events have $m_{\rm ES}$ compatible with the known $B$ meson mass, and a difference between the candidate energy and the beam energy in the center-of-mass, $\Delta E$ (see Chapter 9), compatible with zero.

**Table 17.3.8.** Results of the measured branching fractions for the eight $B \to D_s^{(*)}\overline{D}^{(*)}$ decay modes from *BABAR* and Belle: the number of events for fitted signal $N^{\rm sig}$, and the branching fractions $\mathcal{B}$, as compared with theoretical predictions. (Empty entries indicate no measurement from the given experiment, or no prediction.)

| Mode | Analysis technique | $\mathcal{B}_{BABAR}$ $(10^{-3})$ | $\mathcal{B}_{\rm Belle}$ $(10^{-3})$ | $\mathcal{B}^{\rm theory}_{\rm predict}$ $(10^{-3})$ |
|---|---|---|---|---|
| $B^0 \to D_s^{*+}D^{*-}$ | Semi-exclusive tag | $17.3 \pm 1.8 \pm 1.5$ (Aubert, 2006aw) | | $24.0 \pm 6.7$ (Luo and Rosner, 2001) |
| | $D_s^*$ partial reco. | $18.8 \pm 0.9 \pm 1.7$ (Aubert, 2005q) | | |
| | $D^*$ partial reco. | $15.8 \pm 1.7 \pm 1.4$ (Aubert, 2003d) | | |
| $B^0 \to D_s^{*+}D^{-}$ | Semi-exclusive tag | $7.1 \pm 1.6 \pm 0.6$ (Aubert, 2006aw) | | $10.0 \pm 2.8$ (Luo and Rosner, 2001) |
| $B^0 \to D_s^{+}D^{*-}$ | Semi-exclusive tag | $7.3 \pm 1.3 \pm 0.7$ (Aubert, 2006aw) | | $8.6 \pm 2.4$ (Luo and Rosner, 2001) |
| | $D^*$ partial reco. | $8.3 \pm 1.5 \pm 0.7$ (Aubert, 2003d) | | |
| $B^0 \to D_s^{+}D^{-}$ | Semi-exclusive tag | $6.6 \pm 1.4 \pm 0.6$ (Aubert, 2006aw) | | $14.9 \pm 4.1$ (Luo and Rosner, 2001) |
| | Full reconstruction | | $7.3 \pm 0.4 \pm 0.7$ (Zupanc, 2007) | |
| $B^+ \to D_s^{*+}\overline{D}^{*0}$ | Semi-exclusive tag | $16.7 \pm 1.9 \pm 1.5$ (Aubert, 2006aw) | | |
| $B^+ \to D_s^{*+}\overline{D}^{0}$ | Semi-exclusive tag | $7.9 \pm 1.7 \pm 0.7$ (Aubert, 2006aw) | | |
| $B^+ \to D_s^{+}\overline{D}^{*0}$ | Semi-exclusive tag | $7.8 \pm 1.8 \pm 0.7$ (Aubert, 2006aw) | | |
| $B^+ \to D_s^{+}\overline{D}^{0}$ | Semi-exclusive tag | $9.5 \pm 2.0 \pm 0.8$ (Aubert, 2006aw) | | |

**Table 17.3.9.** Results of measured polarization parameters for the $D_s^*\overline{D}^*$ vector-vector decays, as compared with theoretical predictions. Here $R_L$ is the fraction of longitudinal polarization and $R_\perp$ is the *CP*-odd fraction. (Empty entries indicate no measurement from the given experiment, or no prediction. There are presently no published measurements of, or predictions for, polarization in the $D_s^{*+}\overline{D}^{*0}$ mode.)

| Mode | $\begin{pmatrix} R_\perp \\ R_L \end{pmatrix}^{BABAR}$ | Theoretical Predictions |
|---|---|---|
| $B^0 \to D_s^{*+}D^{*-}$ | | 0.06 (Rosner, 1990) |
| | $0.519 \pm 0.050 \pm 0.028$ (Aubert, 2003d) | 0.55 (Rosner, 1990) |

For each mode, *BABAR* fits the $m_{\rm ES}$ distribution to get the signal yield. According to their physical origin, four categories of events with differently shaped $m_{\rm ES}$ distributions are separately considered: $\overline{D}^{(*)}D^{(*)}K$ signal events, "cross-feed" events, combinatorial background events, and peaking background events. To determine the yields and the branching fractions, the shape of each of these contributions are determined. The cross-feed events are from all the $\overline{D}^{(*)}D^{(*)}K$ modes, except the one we reconstruct, that pass the complete selection, and which are reconstructed in the signal mode; the peaking background is the part of the combinatorial background that is peaking in the signal region. *BABAR* observes from the analysis of simulated samples that most of the cross-feed originates from the combination of an unrelated soft $\pi^0$ or $\gamma$ with the $D^0$ from a $D^{*+}$ decay to form a wrong $D^{*0}$ candidate. A part of the combinatorial $B\overline{B}$ background is peaking in the signal region, and is fitted separately from generic MC samples $e^+e^- \to q\overline{q}$ $(q = u, d, s, c, b)$ satisfying the $\overline{D}^{(*)}D^{(*)}K$ selection. For the modes $B^+ \to \overline{D}^{*0}D^{*0}K^+$ and $B^+ \to \overline{D}^0D^0K^+$, the cross-feed events and the peaking background are negligible, and Belle performs a two dimensional fit on $m_{\rm ES}$ and $\Delta E$ to obtain the signal yield.

Due to the presence of cross-feed events, the fit for the branching fraction for any one channel uses as inputs the branching fractions from the other channels. Since these branching fractions are not a priori known, *BABAR* employs an iterative procedure to obtain the 22 branching fractions. It has been shown that $\overline{D}^{(*)}D^{(*)}K$ events contain resonant contributions (Aubert, 2008bd). In order to measure the branching fractions inclusively without any assumptions on the resonance structure of the signal, *BABAR* estimates the efficiency as a function of location in the Dalitz plane of the data. *BABAR* uses this efficiency at the event position in the Dalitz plane to reweight the signal contribution. To isolate the signal contribution event-per-event, *BABAR* uses the ${}_s\mathcal{P}lots$ technique (Pivk and Le Diberder, 2005) (see Chapter 11). The ${}_s\mathcal{P}lots$ technique exploits the result of the $m_{\rm ES}$ fit (yield and covariance matrix) and the *p.d.f.*s of this fit to compute an event-per-event weight for the signal category and background category.

Both experiments consider several sources of systematic uncertainties on the branching fraction measurements: signal shape, cross-feed determination, peaking background, combinatorial background, fit bias, iterative procedure, limited MC statistics, efficiency mapping, difference between data and MC, number of $B$ mesons in the data sample, and secondary branching fractions.

The combination from the *BABAR* and Belle results can be found in Table 17.3.10. Summing the 10 neutral modes and the 12 charged modes, the $\overline{D}^{(*)}D^{(*)}K$ events represent $(3.65\pm0.10\pm0.24)\%$ of the $B^0$ decays and $(4.06\pm 0.11 \pm 0.28)\%$ of the $B^+$ decays.

### 17.3.5 Decays to charmonium

Decays of $B$ mesons to charmonium modes are color suppressed. In all they consist of a few percent of $B$ decays. Despite their small branching fractions, these decays play a major role in *CP* studies due to the ability to reconstruct many charmonium modes cleanly with little background as well as the simplicity in interpreting the results theoretically. $B^0$ meson decays to charmonium are used to measure the *CP* violation parameter $\sin 2\phi_1$ as well as $\cos 2\phi_1$ (see Sections 17.6.3 and 17.6.8). The relevant decay diagrams for charmonium modes are shown in Fig. 17.3.7.

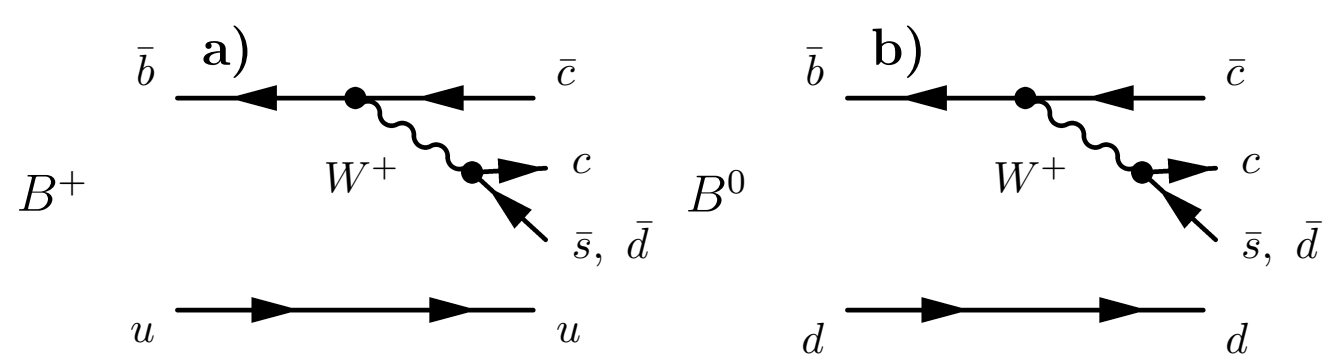

**Figure 17.3.7.** Color-suppressed Feynman diagrams for $B$ meson decays to charmonium.

The easiest way to reconstruct decays to charmonium is via the dileptonic decays of $J/\psi$ or $\psi(2S)$ to electrons or muons. The relatively high energy and topology of the leptons allows a clean sample of charmonium to be reconstructed (which helped in earning the decay to the *CP* state $B^0 \to J/\psi K^0_S$ the title of "Golden Mode"). The $\chi_{c1}$ and $\chi_{c2}$ states can be reconstructed through their radiative decays to $J/\psi\gamma$. $\psi(2S)$ can also be reconstructed through the decay $\psi(2S) \to J/\psi\pi\pi$. The lower mass states $\eta_c$ and $\chi_{c0}$ do not decay to two leptons and the $\chi_{c0}$ branching fraction to $J/\psi\gamma$ is small thus these states must be reconstructed through their decay to hadrons. The higher mass "exotic" charmonium-like $X$ states are reconstructed through decays to $J/\psi\pi\pi$, radiative decays to $J/\psi$ or $\psi(2S)$, or through decays that include two $D$ mesons. They are covered in Section 18.3. Inclusive decays of $B$ mesons to charmonium are covered in Section 18.2.4.1.

#### 17.3.5.1 Reconstruction of charmonium via dileptons

There are several factors that must be taken into account when reconstructing $B$ decays to charmonium where the charmonium are reconstructed via dileptons. The first is that the invariant mass of the two leptons is often significantly below the nominal $J/\psi$ or $\psi(2S)$ mass. This is the result of both final state radiation and energy loss in the detector via bremsstrahlung. This is particularly true for the dielectron mode. Analyses often correct for this energy loss by adding in the energy of photon showers that are within a small angle (typically 50 mrad) of the initial electron direction (*e.g.* Aubert, 2009m; Guler, 2011) to the invariant mass calculation.

The second is that for fully reconstructed $B$ mesons, it is important to perform a mass-constrained fit of the $J/\psi$ or $\psi(2S)$. This improves the energy resolution of the reconstructed $B$ significantly as most of the energy of a charmonium meson coming from a $B$ decay is in its mass.[54] This fit or a global fit for the $B$ meson usually includes the well-measured dilepton vertex.

For charmonium states with radiative decay to $J/\psi$ or $\psi(2S)$, radiative $\gamma$ candidates must pass a minimum energy cut, typically 30 MeV. A common additional requirement is that the $\gamma$ candidate not be a part of a $\pi^0 \to \gamma\gamma$ candidate.

#### 17.3.5.2 Reconstruction of charmonium via hadrons

The $\eta_c$ is reconstructed via $KK\pi$ modes (Fang, 2003; Aubert, 2008ba) in addition to the $p\overline{p}$ mode (Fang, 2003; Aubert, 2007k). The $\eta_c(2S)$ is reconstructed via $KK\pi$ (Vinokurova, 2011; Aubert, 2008ba), as well as via $\eta_c\gamma$.

#### 17.3.5.3 Reconstruction of $B$ candidates

$B$ mesons are reconstructed by combining charmonium candidates with the appropriate other particle candidates. Typically a vertex-constrained fit is done at this point. Both Belle and *BABAR* use kinematic variables to discriminate signal candidates from background. These variables are discussed in Section 7.1.

#### 17.3.5.4 $W \to c\overline{d}$

Decays of $B$ mesons to charmonium $c\overline{d}$ are Cabibbo suppressed and thus are expected to have decay rates of about 5% of the equivalent Cabibbo-allowed $c\overline{s}$ modes. In these modes, the tree and penguin contributions have different phases (unlike the Cabibbo-allowed modes where they are the same) and thus charge asymmetries of a few percent may occur. See Section 17.6.4 for more details.

[54] A key variable for $B$ reconstruction is $\Delta E = E^*_B - E^*_{\rm beam}$ where * refers to the center-of-mass system, $E_B$ is the $B$ candidate's energy, and $E^*_{\rm beam}$ is the beam energy.

**Table 17.3.10.** Branching fractions of $B \to \overline{D}^{(*)} D^{(*)} K$ decays in units of $10^{-4}$. The first uncertainties are statistical and the second are systematic. The results from the modes $B^0 \to D^{*-} D^{*+} K^0$ and $B^+ \to \overline{D}^0 D^0 K^+$ are a combination between the *BABAR* (del Amo Sanchez, 2011e) and Belle (Brodzicka, 2008; Dalseno, 2007) measurements.

| Mode | $\mathcal{B}$ $(10^{-4})$ | Mode | $\mathcal{B}$ $(10^{-4})$ |
|---|---|---|---|
| | *B* decays through external *W*-emission amplitudes | | |
| $B^0 \to D^- D^0 K^+$ | $10.7 \pm 0.7 \pm 0.9$ | $B^+ \to \overline{D}^0 D^+ K^0$ | $15.5 \pm 1.7 \pm 1.3$ |
| $B^0 \to D^- D^{*0} K^+$ | $34.6 \pm 1.8 \pm 3.7$ | $B^+ \to \overline{D}^0 D^{*+} K^0$ | $38.1 \pm 3.1 \pm 2.3$ |
| $B^0 \to D^{*-} D^0 K^+$ | $24.7 \pm 1.0 \pm 1.8$ | $B^+ \to \overline{D}^{*0} D^+ K^0$ | $20.6 \pm 3.8 \pm 3.0$ |
| $B^0 \to D^{*-} D^{*0} K^+$ | $106.0 \pm 3.3 \pm 8.6$ | $B^+ \to \overline{D}^{*0} D^{*+} K^0$ | $91.7 \pm 8.3 \pm 9.0$ |
| | *B* decays through external+internal *W*-emission amplitudes | | |
| $B^0 \to D^- D^+ K^0$ | $7.5 \pm 1.2 \pm 1.2$ | $B^+ \to \overline{D}^0 D^0 K^+$ | $14.0 \pm 0.7 \pm 1.2$ |
| $B^0 \to D^{*-} D^+ K^0$ | $64.1 \pm 3.6 \pm 3.9$ | $B^+ \to \overline{D}^0 D^{*0} K^+$ | $63.2 \pm 1.9 \pm 4.5$ |
| $+D^- D^{*+} K^0$ | | $B^+ \to \overline{D}^{*0} D^0 K^+$ | $22.6 \pm 1.6 \pm 1.7$ |
| $B^0 \to D^{*-} D^{*+} K^0$ | $79.3 \pm 3.8 \pm 6.7$ | $B^+ \to \overline{D}^{*0} D^{*0} K^+$ | $112.3 \pm 3.6 \pm 12.6$ |
| | *B* decays through internal *W*-emission amplitudes | | |
| $B^0 \to \overline{D}^0 D^0 K^0$ | $2.7 \pm 1.0 \pm 0.5$ | $B^+ \to D^- D^+ K^+$ | $2.2 \pm 0.5 \pm 0.5$ |
| $B^0 \to \overline{D}^0 D^{*0} K^0$ | $10.8 \pm 3.2 \pm 3.6$ | $B^+ \to D^- D^{*+} K^+$ | $6.3 \pm 0.9 \pm 0.6$ |
| $+\overline{D}^{*0} D^0 K^0$ | | $B^+ \to D^{*-} D^+ K^+$ | $6.0 \pm 1.0 \pm 0.8$ |
| $B^0 \to \overline{D}^{*0} D^{*0} K^0$ | $24.0 \pm 5.5 \pm 6.7$ | $B^+ \to D^{*-} D^{*+} K^+$ | $13.2 \pm 1.3 \pm 1.2$ |

**Table 17.3.11.** Measured $B^0$ to charmonium $c\overline{d}$ branching fractions ($\mathcal{B}$) from *BABAR*, Belle, and the PDG (Beringer et al., 2012) (average). The PDG value may use measurements from other experiments when calculating the average.

| | *BABAR* results | | Belle results | | PDG Averages |
|---|---|---|---|---|---|
| Final state | $\mathcal{B}$ $(\times 10^{-6})$ | Ref. | $\mathcal{B}$ $(\times 10^{-6})$ | Ref. | $\mathcal{B}$ $(\times 10^{-6})$ |
| $J/\psi\,\pi^0$ | $16.9 \pm 1.4 \pm 0.7$ | (Aubert, 2008i) | $23 \pm 5 \pm 2$ | (Abe, 2003c) | $17.6 \pm 1.6$ |
| $J/\psi\,\eta$ | | | $12.3^{+1.8}_{-1.7} \pm 0.7$ | (Chang, 2012) | $12.3 \pm 1.9$ |
| $J/\psi\,\pi^+\pi^-$ | $46 \pm 7 \pm 6$ | (Aubert, 2003a) | | | $46 \pm 9$ |
| $J/\psi\,\rho^0$ | $27 \pm 3 \pm 2$ | (Aubert, 2007e) | | | $27 \pm 4$ |
| $\chi_{c1}\pi^0$ | | | $11.2 \pm 2.5 \pm 1.2$ | (Kumar, 2008) | $11.2 \pm 2.8$ |

**Table 17.3.12.** Measured $B^+$ to charmonium $c\overline{d}$ branching fractions ($\mathcal{B}$) from *BABAR*, Belle, and the PDG (Beringer et al., 2012) (average). The PDG value may use measurements from other experiments when calculating the average. Note: in (Aubert, 2004ae) *BABAR* measures the ratio $\mathcal{B}(J/\psi\pi^+)/\mathcal{B}(J/\psi K^+)$.

| | *BABAR* results | | Belle results | | PDG Averages |
|---|---|---|---|---|---|
| Final state | $\mathcal{B}$ $(\times 10^{-6})$ | Ref. | $\mathcal{B}$ $(\times 10^{-6})$ | Ref. | $\mathcal{B}$ $(\times 10^{-6})$ |
| $J/\psi\,\pi^+$ | | | $38 \pm 6 \pm 3$ | (Abe, 2003c) | $49 \pm 4$ |
| $J/\psi\,\rho^+$ | $50 \pm 7 \pm 3$ | (Aubert, 2007e) | | | $50 \pm 8$ |
| $\psi(2S)\pi^+$ | | | $24.4 \pm 2.2 \pm 2.0$ | (Bhardwaj, 2008) | $24.4 \pm 3.0$ |
| $\chi_{c1}\pi^+$ | | | $22 \pm 4 \pm 3$ | (Kumar, 2006) | $22 \pm 5$ |

Measured branching fractions for these modes are given in Tables 17.3.11 and 17.3.12. In Table 17.3.1 the measured branching fractions of the $J/\psi\pi$ and $J/\psi\rho$ modes are compared with the predictions from the NS model. Among the four measured modes only the $J/\psi\pi^0$ is consistent with the model's prediction. Both of the $\rho$ modes are overestimated by the model while the $J/\psi\pi^+$ is underestimated.

#### 17.3.5.5 $W \to c\overline{s}$

Measured branching fractions for these modes are given in Tables 17.3.13 and 17.3.14.

The measured branching fractions of the $J/\psi K$ and $J/\psi K^*$ modes are compared in Table 17.3.1 with the predictions from the NS model. All of the $J/\psi K$ measurements are higher than the predictions from the model while for the $K^*$ modes the situation is reversed. The mea-surements are all lower than the predictions. The level of disagreement is typically about a factor of two.

**Table 17.3.13.** Measured $B^0$ to charmonium $c\bar{s}$ branching fractions ($\mathcal{B}$) from *BABAR*, Belle, and the PDG (Beringer et al., 2012) (average). The PDG value may use measurements from other experiments when calculating the average.

| | *BABAR* results | | Belle results | | PDG Averages |
|---|---|---|---|---|---|
| Final state | $\mathcal{B}$ ($\times 10^{-3}$) | Ref. | $\mathcal{B}$ ($\times 10^{-3}$) | Ref. | $\mathcal{B}$ ($\times 10^{-3}$) |
| $\eta_c K^0$ | $0.64^{+0.22}_{-0.20} \pm 0.20$ | (Aubert, 2007k) | $1.23 \pm 0.23^{+0.40}_{-0.41}$ | (Fang, 2003) | $0.83 \pm 0.12$ |
| $\eta_c K^{*0}$ | $0.57 \pm 0.07 \pm 0.8$ | (Aubert, 2007k) | $1.62 \pm 0.32^{+0.55}_{-0.60}$ | (Fang, 2003) | $0.64 \pm 0.09$ |
| $J/\psi K^0$ | $0.869 \pm 0.022 \pm 0.030$ | (Aubert, 2007k) | $0.79 \pm 0.04 \pm 0.09$ | (Abe, 2003c) | $0.874 \pm 0.032$ |
| $J/\psi K^{*0}$ | $1.309 \pm 0.026 \pm 0.077$ | (Aubert, 2005k) | $1.29 \pm 0.05 \pm 0.013$ | (Abe, 2002d) | $1.34 \pm 0.06$ |
| $J/\psi K_1(1270)^0$ | | | $1.30 \pm 0.34 \pm 0.32$ | (Abe, 2001e) | $1.30 \pm 0.5$ |
| $J/\psi \eta K^0_S$ | $0.084 \pm 0.026 \pm 0.027$ | (Aubert, 2004v) | | | $0.08 \pm 0.04$ |
| $J/\psi \phi K^0$ | $0.102 \pm 0.038 \pm 0.010$ | (Aubert, 2003l) | | | $0.094 \pm 0.026$ |
| $J/\psi \omega K^0$ | $0.23 \pm 0.03 \pm 0.03$ | (del Amo Sanchez, 2010c) | | | $0.23 \pm 0.04$ |
| $\psi(2S) K^0$ | $0.646 \pm 0.065 \pm 0.051$ | (Aubert, 2005k) | $0.67 \pm 0.011$ | (Abe, 2003c) | $0.62 \pm 0.05$ |
| $\psi(2S) K^{*0}$ | $0.592 \pm 0.085 \pm 0.089$ | (Aubert, 2005k) | $0.552^{+0.035+0.053}_{-0.032-0.058}$ | (Mizuk, 2009) | $0.61 \pm 0.05$ |
| $\chi_{c0} K^0$ | $0.142^{+0.055}_{-0.044} \pm 0.022$ | (Aubert, 2009av) | | | $0.14^{+0.06}_{-0.04}$ |
| $\chi_{c0} K^{*0}$ | $0.17 \pm 0.03 \pm 0.02$ | (Aubert, 2008ag) | | | $0.17 \pm 0.04$ |
| $\chi_{c1} K^0$ | $0.42 \pm 0.03 \pm 0.03$ | (Aubert, 2009m) | $0.351 \pm 0.033 \pm 0.045$ | (Soni, 2006) | $0.393 \pm 0.027$ |
| $\chi_{c1} K^{*0}$ | $0.25 \pm 0.02 \pm 0.02$ | (Aubert, 2009m) | $0.173^{+0.015+0.034}_{-0.012-0.022}$ | (Mizuk, 2008) | $0.222^{+0.040}_{-0.031}$ |
| $\chi_{c2}\ K^{*0}$ | $0.066 \pm 0.018 \pm 0.005$ | (Aubert, 2009m) | | | $0.066 \pm 0.019$ |

**Table 17.3.14.** Measured $B^+$ to charmonium $c\bar{s}$ branching fractions ($\mathcal{B}$) from *BABAR*, Belle, and the PDG (Beringer et al., 2012) (average). The PDG value may use measurements from other experiments when calculating the average.

| | *BABAR* results | | Belle results | | PDG Averages |
|---|---|---|---|---|---|
| Final state | $\mathcal{B}$ ($\times 10^{-3}$) | Ref. | $\mathcal{B}$ ($\times 10^{-3}$) | Ref. | $\mathcal{B}$ ($\times 10^{-3}$) |
| $\eta_c K^+$ | $0.87 \pm 0.15$ | (Aubert, 2006ae) | $1.25 \pm 0.14^{+0.39}_{-0.40}$ | (Fang, 2003) | $0.96 \pm 0.12$ |
| $\eta_c K^{*+}$ | $1.1^{+0.5}_{-0.4} \pm 0.1$ | (Aubert, 2007k) | | | $1.1^{+0.5}_{-0.4}$ |
| $\eta_c(2S) K^+$ | $0.34 \pm 0.18 \pm 0.03$ | (Aubert, 2007k) | | | $0.34 \pm 0.18$ |
| $J/\psi K^+$ | $1.061 \pm 0.015 \pm 0.048$ | (Aubert, 2007k) | $1.01 \pm 0.02 \pm 0.07$ | (Abe, 2003c) | $1.016 \pm 0.033$ |
| $J/\psi K^+\pi^+\pi^-$ | $1.16 \pm 0.07 \pm 0.09$ | (Aubert, 2008d) | $0.716 \pm 0.010 \pm 0.060$ | (Guler, 2011) | $0.81 \pm 0.013$ |
| $J/\psi K^{*+}$ | $1.454 \pm 0.047 \pm 0.097$ | (Aubert, 2005k) | $1.28 \pm 0.07 \pm 0.014$ | (Abe, 2002d) | $1.43 \pm 0.08$ |
| $J/\psi K_1(1270)^+$ | | | $1.80 \pm 0.34 \pm 0.39$ | (Abe, 2001e) | $1.80 \pm 0.5$ |
| $J/\psi \eta K^+$ | $0.108 \pm 0.023 \pm 0.024$ | (Aubert, 2004v) | | | $0.108 \pm 0.033$ |
| $J/\psi \phi K^+$ | $0.044 \pm 0.014 \pm 0.005$ | (Aubert, 2003l) | | | $0.052 \pm 0.017$ |
| $J/\psi \omega K^+$ | $0.32 \pm 0.01^{+0.06}_{-0.03}$ | (del Amo Sanchez, 2010c) | | | $0.320^{+0.060}_{-0.032}$ |
| $\psi(2S) K^+$ | $0.617 \pm 0.032 \pm 0.044$ | (Aubert, 2005k) | $0.665 \pm 0.017 \pm 0.055$ | (Guler, 2011) | $0.639 \pm 0.033$ |
| $\psi(2S) K^{*+}$ | $0.592 \pm 0.085 \pm 0.089$ | (Aubert, 2005k) | | | $0.67 \pm 0.14$ |
| $\psi(2S) K^+\pi^+\pi^-$ | | | $0.431 \pm 0.020 \pm 0.050$ | (Guler, 2011) | $0.43 \pm 0.05$ |
| $\psi(3370) K^+$ | $3.5 \pm 2.5 \pm 0.3$ | (Aubert, 2006ae) | $0.48 \pm 0.11 \pm 0.07$ | (Chistov, 2004) | $0.49 \pm 0.13$ |
| $\chi_{c0} K^+$ | $0.123^{+0.027}_{-0.025} \pm 0.006$ | (Aubert, 2008l) | $0.112 \pm 0.012^{+0.030}_{-0.020}$ | (Garmash, 2006) | $0.134^{+0.019}_{-0.016}$ |
| $\chi_{c1} K^+$ | $0.45 \pm 0.01 \pm 0.03$ | (Aubert, 2009m) | $0.449 \pm 0.019 \pm 0.053$ | (Garmash, 2006) | $0.479 \pm 0.023$ |
| $\chi_{c1} K^{*+}$ | $0.26 \pm 0.05 \pm 0.04$ | (Aubert, 2009m) | $0.405 \pm 0.059 \pm 0.095$ | (Soni, 2006) | $0.30 \pm 0.06$ |
| $\chi_{c2} K^+$ | | | $0.0111^{+0.0036}_{-0.0034} \pm 0.0009$ | (Bhardwaj, 2011) | $0.011 \pm 0.004$ |

As discussed in Colangelo, De Fazio, and Pham (2002) naïve factorization would predict a zero branching fraction for decays such as $B \to \chi_{c0} K^{(*)}$ and $B \to \chi_{c2} K^{(*)}$ [55]. However, as seen in Tables 17.3.13 and 17.3.14 this is not the case. There are non-zero branching fraction measurements for five of the eight possible final states. In fact, the $B \to \chi_{c0} K^{(*)}$ branching fractions are the same order of magnitude as the factorization allowed $B \to \chi_{c1} K^{(*)}$ decays. In Beneke and Vernazza (2009) it is shown that including color-octet contributions leads to a "correction" to naïve factorization that may even dominate the entire decay amplitude. The calculation is, however, highly uncertain and formally valid only, when the charmonium is a truly non-relativistic bound state. It qualitatively describes correctly the hierarchies of charmonium branching fractions with a sizable $\chi_{c0} K$ one, and a suppression of

[55] Note that in the amplitude for $\overline{B} \to \chi_{c0(2)} \overline{K}$ decays one encounters the $\langle \chi_{c0(2)} | (\bar{c}c)_{\mathrm{V-A}} | 0 \rangle$ matrix element, as can be seen following the examples given in Eqs 17.3.2 and 17.3.3. This matrix element includes the (axial-)vector operator between states with spin 0 and 0 (2) and hence equals to zero.

$\chi_{c2}K$ and $h_cK$, although the suppression of the latter two is not as strong as seen in the data.

### 17.3.6 Summary

Hadronic decays of $B$ mesons into charm make up the largest category of final states. From the point of view of an experimentalist many of these final states are easy to reconstruct as the hardware and software capabilities of both Belle and *BABAR* are well matched to the demands made by final states such as $D\pi$, $D^*\pi$, and $DDK$ to name a few. A glance at the PDG (Beringer et al., 2012) reveals the enormous progress made by *BABAR* and Belle in the number of final states observed and the precision of the measurement of their branching fractions. However there are still some challenges left for experimentalists in this area. To date, no radiative $B$ decays have been observed, there is only an upper limit for $B^0 \to \overline{D}^{*0}\gamma$ (Aubert, 2005ad). There is also much work to be done reconstructing final states with multiple $\pi^0$s.

Since all of these final states rely on QCD to turn quarks into hadrons, gluons play an important role in the dynamics of the decay. At the moment a comprehensive theoretical picture capable of first principle calculations of decay rates is still an elusive goal. The precision data now available on a large number of decay modes will make it easier to achieve this goal. It is also important to keep in mind that the glory in $B$ physics lies not with the QCD component of these decay modes but with the electroweak role in the transition to the final state. Much of what we have learned about *CP* violation in the $b$ sector of the CKM model has come from hadronic final states with charm such as $\psi K^0_S$ ($\phi_1$) and $DK^-$ ($\phi_3$). Looking to the upcoming era of super flavor factories it is clear that this category of final state will continue to play an important role in many aspects of the physics program.

## 17.4 Charmless $B$ decays

***Editors:***
*Fergus Wilson (BABAR)*
*Peter Krizan (Belle)*
*Martin Beneke (theory)*

### 17.4.1 Introduction

In 1964, indirect *CP* violation was discovered in the mixing of the neutral kaon system (Christenson, Cronin, Fitch, and Turlay, 1964) with a value that is currently $|\epsilon| = (2.228 \pm 0.011) \times 10^{-3}$ (Beringer et al., 2012). It took another 30 years before direct *CP* violation was fully established in the kaon system. The absence of direct *CP* violation in the meantime led to the super-weak theory that suggested that *CP* violation would only occur in mixing with a change of two units of flavor ($\Delta S = 2$) and that *CP* violation in the $B$ system would be negligible. At the same time, the discovery of neutral currents in the 1970s and the suggestion that there were six quarks meant that a *CP* violating phase could be introduced into what became the CKM matrix. This would allow flavor to change by one unit ($\Delta S = 1$) and lead to direct *CP* violation in decays. It wasn't until 1999, six years after the start of the construction of the $B$ Factories, that direct *CP* in the kaon system was finally found to be non-zero, $\mathrm{Re}(\epsilon'/\epsilon) = (1.65 \pm 0.26) \times 10^{-3}$ (Fanti et al. (1999)). This result appeared just as the $B$ Factories hoped to establish *CP* violation in the $B$ meson sector. This was achieved through the observation of the angle $\phi_1$ in $B^0 \to J/\psi K^0_S$ in 2001 (see Section 17.6).

Although *CP* violation was initially measured in a $b \to c\bar{c}s$ quark transition, the decays of $B$ mesons to final states without a charm quark are equally as important for the thorough understanding of *CP* violation. The study of the branching fractions and angular distributions (see Chapter 12) probes the dynamics of both weak and strong interactions. In many cases, the measurement of the weak phases can be directly related to the CKM angles ($\phi_1$, $\phi_2$, $\phi_3$). Since the CKM element $|V_{ub}|$ is much smaller than $|V_{cb}|$, the branching fractions for these charmless modes are typically less than $10^{-5}$, and so are only feasible in the era of large integrated luminosities. The accumulated datasets has made it possible to measure branching fractions, direct and indirect *CP* asymmetries, G-parity conservation, longitudinal polarization $f_L$, weak and strong phases. This has enabled comprehensive comparison with theoretical predictions and models.

Figure 17.4.1 shows six of the main amplitudes that contribute to the hadronic $B$ meson decays (there are a number of other less important diagrams that are not shown). The color-allowed tree diagram (T) dominates in $b \to c$ decays but the color-suppressed diagram (C) can also contribute. If the $c$ quark is replaced by a $u$ quark, the tree diagrams are suppressed and the one-loop flavor-changing neutral current (FCNC) penguin diagrams (P) become more or equally important. For example, decays such as $B \to \pi\pi, \pi\rho, \rho\rho$, proceed through $b \to u$ tree diagram but also have a non-negligible $b \to d$ penguin loop contribution. Transitions of $b \to s$ can only occur through the penguin diagrams (P) and CKM suppressed tree decays ($b \to u\bar{u}s$). The former have approximately the same weak phase $\phi_1$ as the $b \to c\bar{c}s$ modes (see Chapter 17.6). Penguin diagrams in $B$ meson decays can be relatively large as they involve the CKM elements $|V_{tb}|$ and $|V_{ts}|$. This is in contrast to $D$ meson decays which require $|V_{cb}|$ and $|V_{ub}|$. As a result $D$ meson decays are a good place to study tree-level, SM-dominated *CP* violation (such as $\phi_3$) while $B$ meson charmless decays have the potential to reveal non-SM physics through heavy virtual particles in the penguin loops.

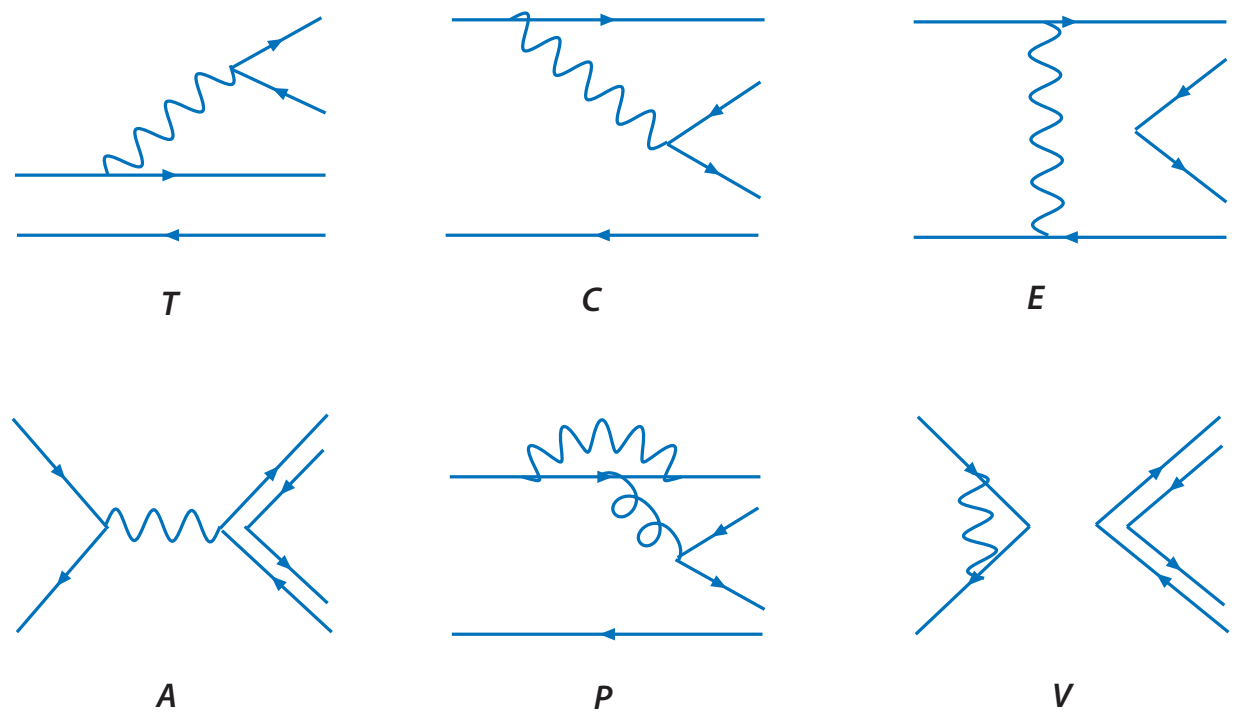

**Figure 17.4.1.** The dominant amplitudes contributing to charmless $B$ meson decays: T) color-allowed external $W$-emission tree diagram; C) color-suppressed internal $W$-emission tree diagram; E) $W$-exchange diagram; A) $W$-annihilation diagram; P) penguin diagram with gluon exchange; and V) $W$-loop diagram.

In $B$ meson decays with an odd number of kaons, the penguin loop (P) will dominate as the $b \to u$ tree diagram is suppressed by the $|V_{ub}|$ coupling. If there are an even number of kaons, the $b \to u$ color-allowed tree diagram (T) again becomes possible and start to contribute a noticeable level.

In the search for indirect *CP* violation, any decay with a $b \to d u \bar{u}$ transition is useful as it provides a possible source of measurement of $\phi_2$, through interference between the decay and the $B$ meson mixing. Examples include $B \to \pi\pi, \pi\rho, \rho\rho$ as discussed in Chapter 17.7. However the presence of the penguin loop as an alternative decay channel complicates the interpretation. Similarly, transitions $b \to q\bar{q}s$ (where $q$ is not a charm quark) provide a precise measurement of $\phi_1$ but in this case there is one dominant penguin decay. Since penguin loops are sensitive to new virtual heavy particles, discrepancies in the value of $\phi_1$ measured in different decay modes could be a sign of new physics. This chapter does not explicitly discuss the CKM angles and more information on the extraction of $\phi_1$ (*e.g.* $B \to \eta' K^0_S$), $\phi_2$ (*e.g.* $B \to \rho\rho$), and $\phi_3$ (*e.g.* $B \to K\pi\pi$) from charmless decays can be found in Chapters 17.6, 17.7, and 17.8, respectively. For information on charmless baryonic decays, please see Chapter 17.12.

Direct $CP$ violation is observed as an asymmetry in the yields between a decay and its $CP$ conjugate when at least two contributing decay amplitudes $A_i$ carry different weak $\phi_i$ and strong phases $\delta_i$ as explained in detail in Chapter 16:

$$A_{CP} = \frac{2\sin(\phi_i - \phi_j)\sin(\delta_i - \delta_j)}{R + R^{-1} + \cos(\phi_i - \phi_j)\cos(\delta_i - \delta_j)}, R \equiv \left|\frac{A_i}{A_j}\right| \tag{17.4.1}$$

Neutral and charged $B$ meson decays involving both tree and penguin amplitudes are a natural place to look for this effect and charmless meson decays have provided evidence for direct $CP$ violation in $B^0 \to K^+\pi^-$, $B^0 \to \pi^+\pi^-$, $B^0 \to \eta K^{*0}$, and $B^+ \to \rho^0 K^+$ (see below).

The diagrams in Fig. 17.4.1 give a simplistic view of the decays. The weak decays of the $B$ meson are subject to both short and long distance QCD effects. The calculation of these properties is challenging as it involves both short-distance perturbative and long-distance non-perturbative QCD. The various models, techniques and successes are the subject of the next section.

### 17.4.2 Theoretical overview

Theoretical calculations of charmless decays of $B$ mesons are based on an effective description of the weak interaction valid at scales below the scale $M_W$. Extracting the CKM elements $\lambda_p^{(D)} \equiv V_{pb}V_{pD}^*$ ($p = u, c$, $D = d, s$), the effective Hamiltonian for $\Delta B = 1$ transitions is

$$H_{\rm eff} = \frac{G_F}{\sqrt{2}} \sum_{p=u,c} \lambda_p^{(D)} \sum_i C_i\, Q_i^p, \tag{17.4.2}$$

where $Q_i^p$ denotes the so-called tree, QCD and electroweak penguin, and dipole operators. The Wilson coefficients $C_i$ include the physics from the highest scales, including $M_W$, down to the scale $m_b$, and their calculation is under complete theoretical control, provided the underlying short-distance physics is known. Eq. 17.4.2 assumes the Standard Model, and the convention that $\lambda_t^{(D)}$ is eliminated by the unitarity relation $\lambda_u^{(D)} + \lambda_c^{(D)} + \lambda_t^{(D)} = 0$. The structure of the operators $Q_i$, the values of their Wilson coefficients, and the flavor structures can be modified in extensions of the SM.

It is sufficient to work to first order in the weak interaction. The decay amplitude $A(\bar{B} \to f) = \langle f|H_{\rm eff}|\bar{B}\rangle$ can be written as

$$A(\bar{B} \to f) = \lambda_u^{(D)} A_f^u + \lambda_c^{(D)} A_f^c\,. \tag{17.4.3}$$

The larger of the two partial amplitudes determines the branching fraction, while the interference with the sub-leading one causes the direct $CP$ asymmetry, provided there is a relative strong phase between the hadronic amplitudes $A_f^u$ and $A_f^c$. For a first estimate, the size of an amplitude is governed by three factors:

- the size of the Wilson coefficients, which divides the amplitudes into tree ($C_i \sim 1$) and penguin ($C_i \sim 0.1$) which are loop-suppressed. Tree amplitudes can be color-allowed or color-suppressed (see the introduction to the section on $B$ decays to charm, 17.3).
- the size of the CKM factors is $\lambda_u^{(d)} \sim \lambda_c^{(d)} \sim \lambda^3$ for $b \to d$ transitions. For these transitions the penguin amplitude $A_f^c$ is typically sub-leading on account of its smaller Wilson coefficient. For $b \to s$ transitions $\lambda_c^{(s)} \sim \lambda^2 \gg \lambda_u^{(s)} \sim \lambda^4$, hence these transitions are dominated by the loop-induced penguin amplitude despite their smaller Wilson coefficient.
- the size of the hadronic matrix elements $\langle f|Q_i^p|\bar{B}\rangle$, which can vary substantially depending on the spin and parity of the final state particles, and whether the final state can only be reached by annihilation of the $B$ meson constituents. The direct $CP$ asymmetry depends crucially on the phases of these matrix elements.

The three factors in combination lead to a fascinating variety of decay patterns, which are summarized in this section.

From the theoretical point of view, the basic problem for the quantitative prediction of charmless $B$ decays is the computation of the hadronic matrix elements $\langle f|Q_i^p|\bar{B}\rangle$. The difficulty resides in the strong interaction, which cannot be treated perturbatively at the hadronic scale $\Lambda \approx 0.5\,\text{GeV}$ relevant to the formation of the hadronic final state $f$, and to the initial bound state. An extreme point of view ("non-perturbative anarchy") would declare the matrix elements to be non-perturbative and unpredictable. In this case, large phases and large direct $CP$ asymmetries in charmless $B$ decays would be expected. The other extreme is the assumption of naïve factorization. The operators $Q_i$ can mostly be written as local products of two bilinear quark currents $J_i^a J_i^b$. In the decay of a $B$ meson to two light mesons M, naïve factorization sets

$$\langle M_1 M_2|Q_i|\bar{B}\rangle \approx \langle M_1|J_i^a|\bar{B}\rangle\langle M_2|J_i^b|0\rangle \tag{17.4.4}$$

(with $M_1 \leftrightarrow M_2$ added where appropriate). With this assumption all direct $CP$ asymmetries vanish.

A direct computation of the matrix elements $\langle f|Q_i^p|\bar{B}\rangle$ with numerical simulations of QCD is neither conceptually nor practically within reach. The available theoretical methods therefore exploit (approximate) flavor symmetries of QCD, or the existence of several scales, which allows for an expansion in $\Lambda/m_b$. The two methods are complementary to a large extent. While the $SU(3)$ approach does not allow the computation of any individual decay from first principles of QCD, its virtue lies in relating groups of decays by expressing them in terms of only a few reduced matrix elements. The second method, the factorization approach, begins with the identification of $m_b$, $\sqrt{m_b\Lambda}$, and $\Lambda$ as relevant scales in $\langle f|Q_i^p|\bar{B}\rangle$. Only the scale $\Lambda$ requires a non-perturbative treatment of the strong interaction. By computing the strong interaction effects at the other two scales perturbatively, a great deal of simplification of the matrix elements can be achieved. Most of the analytical progress in the theory of hadronic $B$ decays achieved over the past few years can be attributed to a systematic implementation of factorization and the heavy-quark expansion. The conclusion is that the truth for $B$ decays lies in between the two above extremes, but closer to naïve factorization than non-perturbative anarchy. In the remainder of this section we provide a brief overview of the different methods and some generic results.

#### 17.4.2.1 $SU(3)$ approach

The $SU(3)$ approach is based on an approximation to QCD, where the up, down, and strange quark masses are equal. In practice, this amounts to an expansion in $m_s/\Lambda$, or, since only the first term is kept, to the approximation $m_s \simeq 0$. In this approximation, QCD acquires an $SU(3)$ flavor symmetry. The quark fields, meson states and the weak interaction Hamiltonian are decomposed into $SU(3)$ representations, and the matrix elements $\langle f|H_{\rm eff}|\bar{B}\rangle$ are expressed in terms of reduced matrix elements and $SU(3)$ Clebsch-Gordan coefficients (Zeppenfeld, 1981). The generic accuracy of this approach is determined by the size of $SU(3)$-breaking corrections, which cannot be calculated. A typical estimate for the ratio of $K$ and $\pi$ decay constants $f_K/f_\pi - 1 \simeq 25\%$ at the amplitude level, though it appears that the non-factorizable $SU(3)$-breaking effects may be smaller than those in decay constants and form factors.

For applications it is more intuitive to work with topological or flavor amplitudes rather than the abstract reduced matrix elements, and hence this notation is widely used. These amplitudes arise naturally in factorization-based calculations of $\langle f|H_{\rm eff}|\bar{B}\rangle$ as well. The "color-allowed tree amplitude" $T$ stands for an amplitude $\langle M_1M_2|\bar{B}\rangle$ with quark flavors $\langle[\bar{q}_s u][\bar{u}D]|[\bar{q}_s b]\rangle$ ($q_s = u, d, s$ the spectator quark, $D = d, s$); the "color-suppressed tree amplitude" $C$ is related to $\langle[\bar{q}_s D][\bar{u}u]|[\bar{q}_s b]\rangle$. The terminology comes from the structure of the effective Hamiltonian $H_{\rm eff}$ and the naïve factorization approximation, where $T$ ($C$) contains a large (small) combination of Wilson coefficients, giving rise to the naïve expectation that $C/T \simeq 0.2$. The "tree" amplitudes are distinguished from the QCD and electroweak "penguin" amplitudes, in which $u\bar{u}$ is replaced by $\sum_q q\bar{q}$ ($q = u, d, s$) and $\sum_q e_q q\bar{q}$, respectively .

The amplitudes for a given set of $B$ decays are written in terms of the independent $SU(3)$ (or topological) amplitudes and CKM parameters, all of which are then fitted to the relevant data. There are often too many amplitude parameters to carry out this program to completion. Possible ways to proceed consist of marginalizing over the phases of amplitudes, resulting in "$SU(3)$ bounds" for the other parameters, or of making further simplifying assumptions beyond the $SU(3)$ limit. The most common additional assumptions are a particular implementation of meson-mixing for $\eta$, $\eta'$ ($\omega$, $\phi$), and neglecting weak annihilation amplitudes.

For instance, with the latter assumption, the $B \to \pi\pi$ and $B \to \pi K$ decay amplitudes are parameterized as follows

$$
\begin{aligned}
\sqrt{2}\,\mathcal{A}_{B^-\to\pi^-\pi^0} &= \lambda_u^{(d)}[T + C + P_u^{EW} + P_u^{C,EW}] \\
&\quad + \lambda_c^{(d)}[P_c^{EW} + P_c^{C,EW}] \\
\mathcal{A}_{\bar{B}^0\to\pi^+\pi^-} &= \lambda_u^{(d)}[T + P_u + \frac{2}{3}P_u^{C,EW}] \\
&\quad + \lambda_c^{(d)}[P_c + \frac{2}{3}P_c^{C,EW}] \\
-\mathcal{A}_{\bar{B}^0\to\pi^0\pi^0} &= \lambda_u^{(d)}[C - P_u + P_u^{EW} + \frac{1}{3}P_c^{C,EW}] \\
&\quad + \lambda_c^{(d)}[-P_c + P_c^{EW} - \frac{1}{3}P_c^{C,EW}], \\
\\
\mathcal{A}_{B^-\to\pi^-\bar{K}^0} &= \lambda_c^{(s)}[P_c - \frac{1}{3}P_c^{C,EW}] \\
&\quad + \lambda_u^{(s)}[P_u - \frac{1}{3}P_u^{C,EW}] \\
\sqrt{2}\,\mathcal{A}_{B^-\to\pi^0K^-} &= \lambda_c^{(s)}[P_c + P_c^{EW} + \frac{2}{3}P_c^{C,EW}] \\
&\quad + \lambda_u^{(s)}[T + C + P_u + P_u^{EW} + \frac{2}{3}P_u^{C,EW}] \\
\mathcal{A}_{\bar{B}^0\to\pi^+K^-} &= \lambda_c^{(s)}[P_c + \frac{2}{3}P_c^{C,EW}] \\
&\quad + \lambda_u^{(s)}[T + P_u + \frac{2}{3}P_u^{C,EW}] \\
\sqrt{2}\,\mathcal{A}_{\bar{B}^0\to\pi^0\bar{K}^0} &= \lambda_c^{(s)}[-P_c + P_c^{EW} + \frac{1}{3}P_c^{C,EW}] \\
&\quad + \lambda_u^{(s)}[C - P_u + P_u^{EW} + \frac{1}{3}P_u^{C,EW}],
\end{aligned}
\tag{17.4.5}
$$

in terms of $T$, $C$, the two penguin amplitudes $P_p$, and four electroweak (EW) penguin amplitudes (the superscript "C" indicates color-suppressed). Since $T$, $C$, $P_u$, $P_u^{EW}$ and $P_u^{C,EW}$ appear only as $T + P_u^{C,EW}$, $C + P_u^{EW}$, and $P_u - P_u^{C,EW}/3$ the parameterization contains six complex strong interaction amplitudes. Assuming only $SU(2)$ isospin symmetry, the most general parameterization of the $\pi\pi$ ($\pi K$) amplitudes requires four (six) complex numbers, so $SU(3)$ symmetry has eliminated four of the 10 independent amplitudes. The full power of $SU(3)$ symmetry becomes apparent, when one adds the analogous decomposition of the $B \to KK$ decays and all the $B_s \to \pi\pi$, $\pi K$, $KK$ decays. The parameterization can be extended to include $\eta$ and $\eta'$ (requiring two singlet penguin amplitudes $S_p$ and an assumption on meson-mixing), and to final states including vector mesons (requiring a larger number of new parameters).

The $SU(3)$ approach is primarily data-driven. No attempt is undertaken to predict the decay amplitudes from QCD dynamics. Where enough experimental information is available, $SU(3)$ relations can give direct access to CKM angles. In particular, if only the more accurate relations of $SU(2)$ isospin are required, this leads to strategies to determine angles almost free of theoretical uncertainties, as discussed elsewhere in this book. $SU(3)$ fits of large sets of final states have been performed (Chiang, Gronau, Luo, Rosner, and Suprun, 2004; Chiang, Gronau, Rosner, and Suprun, 2004; Chiang and Zhou, 2006, 2009; Soni and Suprun, 2007).

#### 17.4.2.2 QCD-based factorization

The factorization approach is more ambitious than the $SU(3)$ approach as it attempts the calculation of individual decays directly from the Lagrangian of the theory in terms of only a few remaining hadronic parameters. In the following, we outline the factorization structure of the matrix elements of hadronic two-body decays, and discuss some general results. The discussion applies to (quasi) two-body final states of mesons. A theoretical description of multi-body final states with similar rigour is not yet available.

The concept of factorization has a long history in $B$ physics as an approximation of $\langle f|H_{\rm eff}|\bar{B}\rangle$ as a product of a decay constant, form factor and a Wilson coefficient (Bauer, Stech, and Wirbel, 1987; Wirbel, Stech, and Bauer, 1985). The term "QCD factorization" refers to a systematic separation of scales in $\langle f|H_{\rm eff}|\bar{B}\rangle$. Contrary to the (useful but ad-hoc) approximation of "naïve" factorization, QCD factorization implies an expansion of the matrix element in the small parameters $\alpha_S(\mu)$ and $\Lambda/m_b$, with $\mu = m_b$ or $\sqrt{m_b\Lambda}$ one of the perturbative scales. Since the $\alpha_S$ series can be calculated (with some effort), but only the leading term in the $1/m_b$ expansion assumes a simple form, the generic accuracy of this approach is limited by power corrections $\Lambda/m_b \simeq 20\%$ at the amplitude level.

The QCD factorization approach developed in (Beneke, Buchalla, Neubert, and Sachrajda, 1999, 2000, 2001) replaces the naïve factorization ansatz by a factorization formula that includes radiative corrections and spectator-scattering effects. Where it can be justified, the naïve factorization ansatz emerges in the simultaneous limit, when $m_b$ becomes large and when radiative corrections are neglected. The basic formula for the hadronic matrix elements is

$$\begin{aligned} \langle M_1 M_2|Q_i|\bar{B}\rangle &= F^{BM_1}(0)\int_0^1 du\, T_i^I(u)\Phi_{M_2}(u) \\ &+ \int_0^1 d\xi du dv\, T_i^{II}(\xi,u,v)\,\Phi_B(\xi)\Phi_{M_1}(v)\Phi_{M_2}(u) \\ &= F^{BM_1}\, T_i^{\rm I}\star\Phi_{M_2} + \Phi_B\star[H_i^{\rm II}\star J^{\rm II}]\star\Phi_{M_1}\star\Phi_{M_2}\,, \end{aligned} \qquad (17.4.6)$$

where $F^{BM_1}(0)$ is a (non-perturbative) $B$ to light-meson transition form factor, $\Phi_{M_i}$ and $\Phi_B$ are light-cone distribution amplitudes, and $T_i^{\rm I,II}$ are perturbatively calculable hard-scattering kernels. $M_1$ is the meson that picks up the spectator quark from the $B$ meson, as illustrated in Fig. 17.4.2. The third line uses a short-hand notation $\star$ for convolutions and indicates that the spectator-scattering effect in the second line is a convolution of physics at the hard scale $m_b$, encoded in $H_i^{\rm II}$, and the hard-collinear scale $\sqrt{m_b\Lambda}$, encoded in the jet function $J^{\rm II}$. Eq. 17.4.6 shows that there is no long-distance interaction between the constituents of the meson $M_2$ and the $(BM_1)$ system at leading order in $1/m_b$. This is the precise meaning of factorization. Strong interaction scattering phases are generated at leading order in the heavy-quark expansion only by perturbative loop diagrams contributing to the kernels $T_i^{\rm I}$ and $H_i^{\rm II}$. Thus the phases are of order $\delta \sim \mathcal{O}(\alpha_S(m_b), \Lambda/m_b)$.

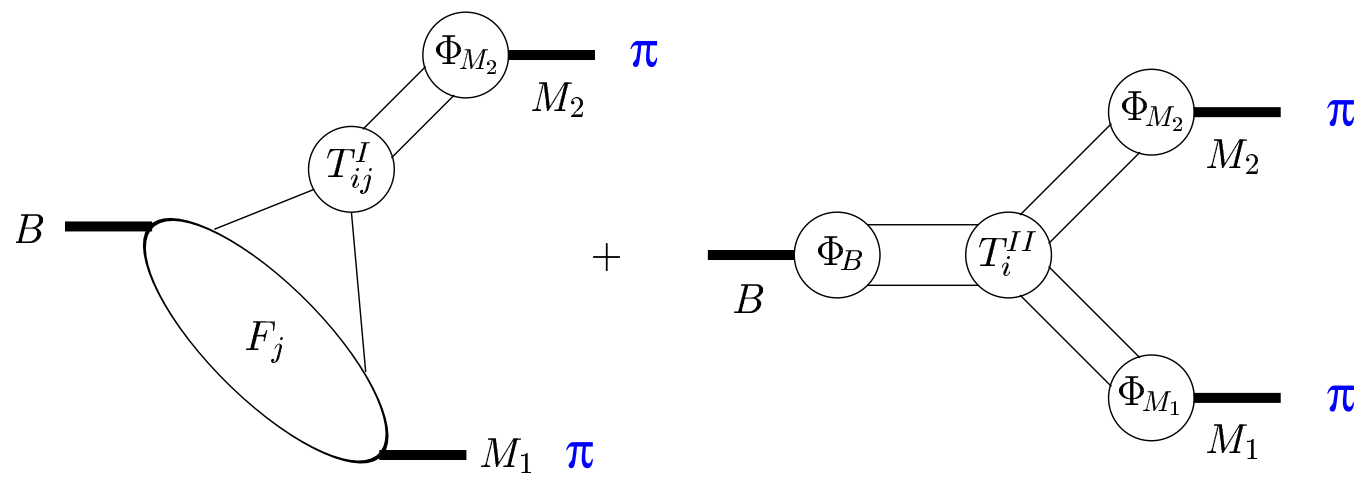

**Figure 17.4.2.** Graphical representation of the factorization formula given Eq. 17.4.6 (Beneke, Buchalla, Neubert, and Sachrajda, 2000).

Factorization as embodied by Eq. 17.4.6 is not expected to hold at sub-leading order in $1/m_b$. Some power corrections related to scalar currents are enhanced by factors such as $m_\pi^2/((m_u+m_d)\Lambda)$. Some corrections of this type, in particular those related to scalar penguin amplitudes, nevertheless appear to be calculable and turn out to be numerically important. On the other hand, attempts to compute sub-leading power corrections to hard spectator-scattering in perturbation theory usually result in infrared divergences, which signal the breakdown of factorization. These effects are usually estimated and included into the error budget. All weak annihilation contributions belong to this class of effects and often constitute the dominant source of theoretical error, in particular for the direct $CP$ asymmetries. Factorization as above applies to pseudoscalar flavor-non-singlet final states and to the longitudinal polarization amplitudes for vector mesons. Final states with $\eta$ and $\eta'$ require additional considerations, but can be included (Beneke and Neubert, 2003a). The transverse helicity amplitudes for vector mesons are formally power-suppressed but can be sizeable, and do not factorize in a simple form (Beneke, Rohrer, and Yang, 2007; Kagan, 2004). The description of polarization is therefore more model-dependent than branching fractions and $CP$ asymmetries. QCD factorization results are available for a variety of complete sets of final states. (Beneke and Neubert, 2003b; Beneke, Rohrer, and Yang, 2007) contain the theoretical predictions for pseudoscalar and vector meson final states (PP, PV, VV). A similar analysis has been performed for final states with a scalar meson (Cheng, Chua, and Yang, 2008), axial-vector mesons (Cheng and Yang, 2007, 2008), and a tensor meson (Cheng and Yang, 2011).

Several variations of factorization have been considered in the literature and applied to the calculation of branching fractions, $CP$ asymmetries and polarization observables. The perturbative QCD (PQCD) framework (Keum, Li, and Sanda, 2001; Lu, Ukai, and Yang, 2001) makes the stronger (and controversial) additional assumption that the $B$ meson transition form factors $F^{B\to M_1}(0)$ are also dominated by short-distance physics and factorize into light-cone distribution amplitudes. Both terms in Eq. 17.4.6 can then be combined to

$$\langle M_1 M_2 | Q_i | \bar{B} \rangle = \phi_B \star [T^{\mathrm{PQCD}} \star J^{\mathrm{PQCD}}] \star \phi_{M_1} \star \phi_{M_2} . \tag{17.4.7}$$

PQCD needs fewer non-perturbative input parameters, but there is a larger dependence on unknown light-cone distribution amplitudes. Since the approach relies on regularizing the infrared sensitivity by intrinsic transverse momentum, there is a larger sensitivity to perturbative corrections at low scales, where the strong coupling is large and perturbation theory is potentially unreliable. From a phenomenological perspective, the principal difference between the PQCD and all other approaches is the relative importance of the weak annihilation mechanism. In QCD factorization the strong interaction phases arise at the scale $m_b$ from loop diagrams, that have yet to be included in the PQCD approach, and from the model for weak annihilation. In the most widely used implementation of PQCD, the strong phases originate only from a weak annihilation tree diagram. As a consequence, the predicted direct *CP* asymmetries can be rather different in the two approaches. There is a large literature covering individual or few decay modes in PQCD. Large sets of final states were analyzed in (Ali et al., 2007; Li and Mishima, 2006). We note that the PQCD factorization formula Eq. 17.4.7 was recently revised due to infrared divergences in loop effects (Li and Mishima, 2011), which weakens its predictive power. Most phenomenological analyses predate this revision.

Alternative to the diagrammatic arguments put forward in the BBNS approach in (Beneke, Buchalla, Neubert, and Sachrajda, 1999, 2000, 2001), factorization of charmless $B$ decays can be elegantly derived in the framework of soft-collinear effective theory (SCET) (Bauer, Pirjol, Rothstein, and Stewart, 2004; Beneke and Feldmann, 2004; Chay and Kim, 2004). It is important to stress that the theoretical basis of QCD factorization and SCET is exactly the same. However, the phenomenological implementation of factorization put forward in (Bauer, Pirjol, Rothstein, and Stewart, 2004) differs in two respects from the BBNS approach. First, perturbation theory at the intermediate scale $\sqrt{m_b \Lambda}$ is avoided by not factorizing the spectator-scattering term into a hard and jet function. Eq. 17.4.6 then takes the form

$$\langle M_1 M_2 | Q_i | \bar{B} \rangle = F^{BM_1} \, T_i^{\mathrm{I}} \star \Phi_{M_2} + \Xi^{BM_1} \star H_i^{\mathrm{II}} \star \Phi_{M_2} \, , \tag{17.4.8}$$

where $\Xi^{BM_1}$ is a generalized, non-local $B$ meson form factor related to the matrix element $\langle M_1 | \bar{q} A_\perp b | \bar{B} \rangle$ (Beneke and Feldmann, 2004), which depends on momentum transfer $q^2$ and an additional convolution variable. Second, penguin diagrams with charm loops (Ciuchini, Franco, Martinelli, Pierini, and Silvestrini, 2001) are supposed to be non-factorizable, hence non-perturbative. From the phenomenological perspective, the principal difference to the BBNS approach concerns again the generation of strong interaction phases. Since the non-local form factor is unknown, Eq. 17.4.8 can be used only at the tree level, hence the amplitudes, including the color-suppressed tree amplitude C, have no phases. The only exception is the charm penguin amplitude $P_c$, which is considered as an unknown complex number and is therefore the only source of direct *CP* violation. The approach proposed in (Bauer, Pirjol, Rothstein, and Stewart, 2004) assumes that scalar penguin and weak annihilation power corrections are zero, but since $P_c$ is a phenomenological parameter, this has no effect on the analysis. Because of the need to fit the dominant penguin amplitudes to data, the "SCET" approach, unlike the QCD factorization or PQCD approach, shares many features of other data-driven approaches such as the $SU(3)$ amplitude approach. It uses the fewest theoretical assumptions of the three factorization-based methods, at the price of having less predictive power. Large sets of final states have been analyzed with this method in (Bauer, Rothstein, and Stewart, 2006; Wang, Wang, Yang, and Lu, 2008; Williamson and Zupan, 2006). We mention that the question whether the penguin loops with charm factorize or not, which for some time has been a point of controversy, has meanwhile been resolved in favor of factorization (Beneke, Buchalla, Neubert, and Sachrajda, 2009).

For a more detailed comparison of the various QCD-based factorization approaches we refer to the short review in (Artuso et al., 2008). This review also provides an overview of the status of the calculation of radiative corrections, which up to now are computed at next-to-leading order (NLO), and partly even at next-to-next-to-leading order (NNLO) (Bell, 2008, 2009; Beneke, Huber, and Li, 2010; Beneke and Jager, 2006, 2007), only in the QCD factorization (BBNS) approach.

#### 17.4.2.3 Generic results

To conclude this overview we summarize a few general results that emerged from comparing theoretical calculations to data. The remainder of this section contains a more specific mode-by-mode analysis. The comparison still suffers from a lack of precise knowledge of quantities such as $|V_{ub}|$, $B$ meson form factors, and light-cone distributions amplitudes, which cause a significant theoretical uncertainty.

1. The *color-allowed tree amplitude* T that governs the branching fractions of decays to final states such as $\pi^+\pi^-$ and its vector-meson relatives is well described by factorization, and even close to its naïve factorization value. The main uncertainty in color-allowed tree-dominated decays comes from $F^{BM_1}(0)$, the $B$ meson form factor.
2. The *color-suppressed tree amplitude* C that governs branching fractions of decays to final states such as $\pi^0\pi^0$ and its vector-meson relatives is often underestimated. Its value depends strongly on the precise magnitude of the spectator-scattering effect. This can be seen from the numerical representation (Beneke, Huber, and Li, 2010) of the NNLO color-suppressed tree amplitude:

$$\alpha_2(\pi\pi) = 0.220 - [0.179 + 0.077\, i]_{\mathrm{NLO}}$$

$$- [0.031 + 0.050\, i]_{\rm NNLO} + \left[\frac{r_{\rm sp}}{0.445}\right] \Big\{ [0.114]_{\rm LOsp} + [0.049 + 0.051 i\,]_{\rm NLOsp} + [0.067]_{\rm tw3} \Big\} = 0.240^{+0.217}_{-0.125} + (-0.077^{+0.115}_{-0.078}) i\,. \quad (17.4.9)$$

Here 0.220 represents the naïve factorization value. Loop corrections to the form-factor-like term in the first line of Eq. 17.4.6 and the first two lines of Eq. 17.4.9 almost cancel this number, but generate a sizable imaginary part, *i.e.* scattering phase. The real part of the amplitude is regenerated by spectator-scattering in the second line of Eq. 17.4.6 and the third and fourth line of Eq. 17.4.9. It is evident that the strong interaction dynamics of the color-suppressed tree amplitude is far from the naïve factorization picture, and is governed by quantum effects. The theoretical uncertainty is correspondingly large.

3. The *QCD penguin amplitude* P that governs branching fractions of decays to final states such as $\pi K$ and its vector-meson relatives is certainly underestimated in leading order in the heavy-quark expansion. The power-suppressed but chirally-enhanced scalar penguin amplitude, and perhaps a (difficult to disentangle) weak annihilation contribution, is required to explain the penguin-dominated PP final states. While the scalar penguin amplitude is calculable, some uncertainty remains. An important observation is the smaller size of the PV, VP and VV penguin amplitudes as compared to PP final states, which can be inferred from the measured branching fractions of hadronic $b \to s$ transitions. This is a clear indication of the relevance of factorization, which predicts this pattern as a consequence of the quantum numbers of the operators $Q_i$. If the penguin amplitude were entirely non-perturbative, no pattern of this form would be expected. A similar statement applies to the $\eta^{(\prime)} K^{(*)}$ final states, where factorization explains naturally the strikingly large differences in branching fractions, including the large $\eta' K$ branching fraction, through the interference of penguin amplitudes, although sizeable theoretical uncertainties remain. A flavor-singlet penguin amplitude seems to play a sub-ordinate role in these decays.
4. The situation is much less clear for the *strong phases and direct CP asymmetries.* A generic qualitative prediction is that the strong phases are small, since they arise through either loop effects ($\alpha_s(m_b)$) or power corrections ($\Lambda/m_b$). Enhancements may arise, when the leading-order term is suppressed, for instance by small Wilson coefficients. This pattern is indeed observed. Quantitative predictions have met only partial success. The observed direct *CP* asymmetry in the decay to $\pi^+\pi^-$, and the asymmetry difference in the decays to $\pi^0 K^+$ and $\pi^- K^+$ are prominently larger than predicted. A comparison of all *CP* asymmetry results shows a pattern of quantitative agreements and disagreements that are not presently understood. Since $\alpha_s(m_b)$ and $\Lambda/m_b$ are roughly of the same order, it is quite possible that power corrections are $\mathcal{O}(1)$ effects relative to the perturbative calculation, preventing a reliable quantitative estimate. However, the direct *CP* asymmetry calculations are still LO calculations, contrary to the branching fractions, so the final verdict must await the completion of the NLO asymmetry calculation. Contrary to direct *CP* asymmetries, the $S$ parameter that appears in time-dependent *CP* asymmetries is predicted more reliably, since it does not require the computation of a strong phase. This is exploited in computations of the difference between $\sin 2\phi_1$ from $b \to s$ penguin dominated and $b \to c\bar{c}s$ tree decays (Beneke, 2005; Cheng, Chua, and Soni, 2005b).
5. *Polarization in $B \to VV$ decays* was expected to be predominantly longitudinal, since the transverse helicity amplitudes are $\Lambda/m_b$ suppressed due to the V-A structure of the weak interaction and helicity conservation in short-distance QCD. While this is parametrically true (with one exception (Beneke, Rohrer, and Yang, 2006)), a closer inspection shows that the parametric suppression is hardly realized in practice for the penguin amplitudes (Beneke, Rohrer, and Yang, 2007; Kagan, 2004). This leads to the qualitative prediction (or rather, in this case, postdiction) that the longitudinal polarization fraction should be close to 1 in tree-dominated decays, but can be much less, even less than 0.5, in penguin-dominated decays, as is indeed observed. However, quantitative predictions of polarization fractions for penguin-dominated decays must be taken with a grain of salt, since they rely on model-dependent or universality-inspired assumptions of the non-factorizing transverse helicity amplitudes.

The remainder of this chapter is devoted to an overview of experimental techniques of importance to charmless $B$ decay measurements and provides a summary of two-body and three-body final state data collected by the *BABAR* and Belle experiments. A detailed comparison and interpretation of the data in the light of theoretical approaches as discussed above is beyond the scope of this review. For this reason we will generally refrain from making reference to specific theoretical papers in the following.

### 17.4.3 Experimental techniques

The decays of $B$ mesons to final states with two or three hadrons without a charm quark are loosely broken down into “two-body”, “quasi-two-body” and “three-body” decays. The “two-body” analyses concentrate on long-lived final states such as $\pi\pi$, $K\pi$, $KK$, etc. As these modes can be used to access the CKM angle $\phi_2$, they are covered in Chapter 17.7; only the observation of direct *CP* violation is discussed here. The “quasi-two-body” category includes decays where one or both of the decay products is a resonance. Final state particles that have been measured include scalar (S) particles ($a_0$ (980), $f_0(980)$, $f_0(1370)$, $f_0(1500)$, $K_0^*(1430)$); pseudoscalar (P) particles ($K^{\pm}$, $K^0$, $\pi^{\pm}$,$\pi^0$, $\eta$, $\eta'$); vector (V) particles ($\rho$, $\phi$, $\omega$, $K^*$); tensor (T) particles ($K_2^*(1430)$, $f_2(1270)$); and axial-vector (A) mesons, which can be classified into two groups as the $^3P_1$ nonet ($a_1(1260)$, $f_1(1285)$, $f_1(1420)$, $\mathrm{K}_{1A}$) and the $^1P_1$ nonet ($b_1$, $b_1(1170)$, $b_1(1380)$, $\mathrm{K}_{1B}$).[56] Three-body charmless decays concentrate on final states with $\pi$ or $K$ but can sometimes branch out to include protons and resonances such as $K^*$ *e.g.* $B^+ \to \overline{p}pK^+$ (see Chapter 17.12) and $B^+ \to K^{*0}K^{*0}K^+$.

The "quasi-two-body" decays are traditionally reconstructed assuming that the resonances decaying to the same final state (such as $\rho$ and $f_0(980)$ decaying to $\pi\pi$) do not interfere. This has the advantage that branching fractions can be compared to measurements from earlier experiments but the effect of interference is then considered as a systematic. The main differences between the ways that decays are analysed are usually dictated by the extent and nature of the background, as the $B$ meson charmless decays have a low signal-to-background ratio. This can be compared to $D$ meson decays which are typically selected with very high purity.

Whatever the final state, the candidate selection process follows a broadly similar path (see Chapter 7 for more details on $B$ meson reconstruction). The $B$ meson candidates are reconstructed through their decays. The intermediate resonance will be formed first and then combined with a third particle to form the $B$ meson. The reconstructed mass will usually be required to be less than $\sim 3$ times the width from the nominal central value. If the natural widths of the resonances are smaller than the detector resolution, the resonance masses (including $\pi^0$'s) are constrained to their nominal PDG values in the fit for the $B$ meson candidate (Beringer et al., 2012); this improves the precision of the parameters obtained in the fit. Quality criteria are applied to the tracks before fitting, such as demanding the tracks are well-measured, have a minimum $p_T$, and originate from close to the beam spot. The momenta of the charged tracks will usually be extracted assuming a particular mass hypothesis determined by the particle type (*e.g.* pion versus kaon, see Chapter 5). However, in some analyses, such as $B^0 \to h^+h^-$ (with $h = K, \pi$), the $B$ meson will be fitted under one mass hypothesis (usually a pion), and any shift in the value of $\Delta E$ is used to differentiate between decays with one or more kaons. The shift is of the order of $\sim 50\,\mathrm{MeV}$ per kaon. The vertexing will apply various constraints to improve the resolution (see Chapter 6) and to take into account the flight distance of long lived particles such as the $K^0_S$ meson. These constraints become more important as the number of neutral particles in the decay increases. A further criterion that is sometimes applied is to require that there is at least one additional charged track from the beam spot region; this is a crude indicator that there has been at least one other decay in the event, which is assumed to be the other $B$ meson.

Two kinematic variables, $m_{\mathrm{ES}}$ and $\Delta E$, are used to select the events (see Eqs 7.1.8 and 7.1.5 for definitions). Any linear correlation between these variables can be removed by rotating them in the ($m_{\mathrm{ES}}$, $\Delta E$) plane or a two dimensional *p.d.f.* can be used in the maximum likelihood (ML) fit. Events with $|\Delta E| < 300\,\mathrm{MeV}$ are typically accepted, although an asymmetric acceptance region is used if there is a chance of energy loss from photon emission or $\pi^0$ reconstruction. The minimum value of $m_{\mathrm{ES}}$ is set to allow a good fit to the $m_{\mathrm{ES}}$ background distribution and is rarely set less than $5.220\,\mathrm{GeV}/c^2$ (below this value, other selection criteria start to distort the selection efficiency).

The ($m_{\mathrm{ES}}$, $\Delta E$) plane is divided into regions to aid analysis. A signal region is defined around the point $m_{\mathrm{ES}} = m_B, \Delta E = 0$ with a width roughly 3 times the resolution on $m_{\mathrm{ES}}$ ($\sim 3\,\mathrm{MeV}/c^2$) and $\Delta E$ ($\sim 20-50\,\mathrm{MeV}$, depending on the number of neutral particles). Although the signal region is usually rectangular in shape, elliptical signal regions have been used *e.g.* Fig. 3 in (Garmash, 2005). Two sidebands are defined above and below the $\Delta E$ signal region, the upper region allowing for the study of two-body decays that have been combined with a random track and the lower region to study four-body decays that have lost a track. A further sideband below the signal region in $m_{\mathrm{ES}}$ can be used to study the continuum background, although care must be taken to account for any decays from $B$ mesons.

In the center-of-mass (CM) frame, the continuum background is characterized by a jet-like, back-to-back structure while the $B\overline{B}$ events have a more spherical distribution since they are produced close to rest (see Chapter 9 for details). Therefore, event shape variables are used to separate signal from this background. Many different criteria have been used over the years including sphericity, spherocity, planarity, acoplanarity and thrust (see the Glossary and Chapter 9). In addition, angles are often measured between the direction of the $B$ meson decay and a reference axis, such as the beam line or the direction of the rest of the event (ROE). An important example is the thrust angle in the CM frame, defined as the angle $\theta_T$ between the thrust axis of the $B$ meson candidate and that of the rest of the particles in an event. Signal events are uniformly distributed in $\cos\theta_T$, while continuum events are peaked near $\cos\theta_T = \pm 1$. A requirement on $\cos\theta_T$ or $|\cos\theta_T|$ of less than $0.7-0.9$ is usually applied.

Any remaining event shape variables are combined into a multivariate discriminant that can either be used as selection criteria or as a *p.d.f.* observable in a ML fit. Fisher discriminants and neural networks are popular but Boosted Decision Trees (or Forests) have also been applied (see Chapter 4 for details). The number of variables is typically about six. Although discriminants with many more variables have been tried, they rarely bring any additional discrimination. The choice of variables depends on the mode under consideration, consistency with previously used discriminants, and ultimately on the prejudice of the analyst. It is important to check for correlations between the input variables and any other variables used in the ML fit. Variables that have been used over the years include (see Chapter 9 for many definitions): CLEO cones (momentum distribution in nine angular cones about the thrust vector); modified Fox-Wolfram moments (Abe, 2001c); the variable $S_T$, the scalar sum of the transverse momenta, calculated with respect to the thrust axis, of particles outside a $45^{\circ}$ cone around the $B$ thrust axis, divided by the scalar sum of their momenta (Jen, 2006); the polar angles of the $B$ meson momentum vector and the $B$ meson thrust axis with respect to the beam axis; the angle between the $B$ meson thrust axis and the thrust axis of the rest of the event; and the ratio of the second- and zeroth-order momentum-weighted polynomial moments of the energy flow around the $B$ meson thrust axis (Aubert, 2004a). Although not strictly event shape variables, some success has been achieved by using two additional inputs to the neural network: the flavor of the other $B$ meson as reported by a multivariate tagging algorithm (Aubert, 2005i); and the boost-corrected proper-time difference between the decay vertices of the two $B$ mesons divided by its error. The multivariate discriminant can be trained with Monte Carlo (MC) simulation for the signal, and $q\bar{q}$ continuum MC, off-resonance data or sideband data for the background. The discriminant can sometimes be used as a selection criterion as well as a *p.d.f.* as a simple cut on the output can eliminate a substantial part of the background (of the order of 20%-40%) with little signal loss. Instead of using the tagging information in the event-shape, Belle have sometimes used the $B$ meson flavor tagging output (Kakuno, 2004) to calculate a figure of merit; signal retention of greater than 60% with background rejection greater than 90% has been achieved (Jen, 2006).

[56] $\mathrm{K}_1(1270)$ and $\mathrm{K}_1(1400)$ are admixtures of $\mathrm{K}_{1A}$ and $\mathrm{K}_{1B}$

$B$ meson decays to charm have large branching fractions and final states that are either the same as the mode under consideration or easily mis-reconstructed. These charm backgrounds can be suppressed by reconstructing the charm candidate from combinations of tracks and applying a veto around the nominal mass (typically $\sim 40\,\mathrm{MeV}/c^2$ for the $D$ meson).

The helicity distribution is an important variable that can be used to identify particles of a particular spin, extract the longitudinal polarization $f_L$, or simply as a selection criterion. The helicity angle $\theta_H$ of the resonance is defined as the angle between the momentum vector of one of the resonance's daughter particles and the direction opposite to the $B$ meson momentum in the resonance rest frame (Kramer and Palmer (1992)). The choice of daughter must be consistent from event to event (either based on charge or flavor) and care must be taken to avoid any unexpected ordering in momentum or azimuthal angle introduced by the track finding algorithms.

It is often necessary to limit the range of the helicity angle. At values of $|\cos\theta_H| > 0.9$ the signal reconstruction efficiency starts to fall off, as one of the daughter tracks of the resonance has a low momentum. At the same time, backgrounds created from combinations of tracks start to increase. If the resonance decays to particles of differing mass, then the momentum selection criteria on the daughter particles will cause the $\cos\theta_H$ distribution to be skewed, requiring careful compensation for the change in efficiency. The allowed range of $\cos\theta_H$ is mode dependent but typically events are rejected if $\cos\theta_H$ is greater than $0.7 - 0.9$, with different ranges for negative and positive $\cos\theta_H$. In Vector-Vector (VV) decays, the longitudinal component is typically dominant and causes the helicity angle distribution to be enhanced near $\pm 1$. For these decays, careful consideration of the $\cos\theta_H$ rejection criterion is required to optimize the signal and background ratio. The value of the longitudinal component is an important measurement (see Section 17.4.5.3) so care must be taken to limit any bias in the acceptance.

There can be multiple $B$ meson candidates in an event. The average number of candidates per accepted event ranges up to $\sim 1.5$, with more mis-reconstructed candidates expected in decays with more neutrals (such as $\pi^0$) due to low-energy photons or noise in the electromagnetic calorimeter. Resonances with large widths (such as $\rho^0$) also have more mis-reconstructed candidates due increased combinatorics. One approach to dealing with multiple candidates is to accept all $N$ candidates in an event with a weight $1/N$ applied to each, but usually a criterion is used to select the best one. This is sometimes a random choice but more common methods rely on a $\chi^2$ based on the pull of the fitted resonance mass from the nominal value or the $B$ meson vertex probability. The accuracy of the selection depends on the mass width of the resonance and the number of neutral particles in the fit. The true candidate is selected with an accuracy that is rarely below 75% and often greater than 95%, based on MC simulation. If the number of mis-reconstructed candidates is large then these "self-crossfeed" candidates are sometimes included as a separate hypothesis category in the ML fit. There is no agreed point at which this happens but it is typically considered as an option when the true candidate selection accuracy falls below $\sim 85\%$.

After the application of all the selection criteria, there will still be a number of backgrounds from $B$ meson decays either from decays via a charm particle that have not been rejected by the $D$ meson mass requirement or $B$ meson decays that have been mis-reconstructed. Unlike the continuum background, these $B\overline{B}$ backgrounds are likely to have a peaking distribution in one or more of the observables used in a ML fit. The contribution to the background from $B\overline{B}$ decays is identified by running the selection on generic $B\overline{B}$ background MC decays, where all the known decay channels have been included. Decays that have not been observed are often included assuming some estimated branching fraction ($10^{-6} - 10^{-5}$) that allows a small number to be selected and characterized. If a decay is observed to pass the selection, the analysis is rerun on the exclusive MC events to extract an estimate of the number of events expected in the final sample. If there are many modes ($\sim 20$ are not uncommon) an attempt is often made to group them into a smaller number based on similarities in the distributions of the observables. The combined sample must be correctly weighted by the expected branching fractions and reconstruction efficiencies for each individual mode. This is a problem for decays that have not been measured yet and in these cases it is typical to assume a branching fraction that is about half the reported branching fraction upper limit. If no upper limit has been reported, a branching fraction is chosen such that only a few events can be expected to appear in the data.

Higher mass resonances that peak outside the invariant mass selection region can still feed-down to the signal region because: they have a large width, such as $f_0(1370)$; through reflection, where a daughter particle is mis-identified, such as in $B \to \omega\pi^+$; or where a resonance has a long range component *e.g.* the S-wave component of the $K_0^*(1430)$. These backgrounds are often treated in a separate analysis that looks in the mass region above the resonance under consideration (since this is still blinded) and performs a ML fit to the higher mass region using $m_{\rm ES}$, $\Delta E$, the multivariate discriminant, and the reconstructed mass. Once the yield is extracted, the number of events in the resonance signal region is estimated by extrapolating the fitted mass *p.d.f.* (or a fit to the extracted ${}_sWeights$, see Chapter 4) down to the low mass region and integrating.

A further category of background occurs when the $B$ meson decays to the same final state without passing through a resonance, such as $B^+ \to \pi^+ \pi^- \pi^+$ when looking for $B^+ \to \rho^0 \pi^+$ or $B^0 \to \pi^+ \pi^- \rho^0$ when looking for $B^0 \to \rho^0 \rho^0$. These backgrounds also become important when $D$ mesons are used as calibration channels as these "non-resonant" decays can be responsible for a significant number of the events underneath the calibration channel of interest. Strictly speaking, "non-resonant" means a decay in a Dalitz Plot that is uniformly distributed in phase space (see later and Chapter 13). However, this distinction is generally ignored and any final state which cannot be represented by a peaking structure is usually categorized as non-resonant. This has practical benefits when performing a fit as it is often difficult to identify the source of smoothly varying distributions. A fit which uses more than one such distribution is likely to find that the background events flow between the different distributions without affecting the significance of the signal. As a result, some papers will report a non-resonant measurement while others will simply consider it as part of the background.

The signal modes, mis-reconstructed signal modes (if used), continuum background and $B\overline{B}$ backgrounds distributions are used in a ML fit to extract the signal yield, branching fraction, $A_{CP}$, and longitudinal polarization $f_L$. The observables used are usually $m_{\rm ES}$, $\Delta E$, the multivariate discriminant and the intermediate resonance masses. If an angular analysis is required, the helicity $\cos\theta_H$ of the resonances is also used. In this later case, the reconstruction efficiency as a function of $\cos\theta_H$ must be taken into account, often by multiplying the expected true distribution by a polynomial of a suitable order. The efficiency for the other variables is usually treated as uniform.

The observables used in the *p.d.f.*s are usually assumed to be uncorrelated and the total *p.d.f.* is taken to be the product of the separate individual *p.d.f.*s. However, in some cases this assumption is invalid and the correlations need to be taken into account explicitly. If the correlation only exists between two observables and is reasonably linear, then the correlation can be reduced by using rotated variables derived from the observables in the *p.d.f.*s. Sometimes, a two-dimensional *p.d.f.* is used. A third option is to create a *p.d.f.* based on one of the observables, where the *p.d.f.* parameters (*e.g.* means and widths) are dependent on the other observable.

A standard set of cross-checks on the fit is performed. The *p.d.f.*s are used to generate a series of simulated data samples that are then fitted with the ML method. This reveals any problems with minimization, pulls and biases. The tests are repeated with data samples generated from the full MC simulated data; this can reveal problems with correlations between observables. The ML fit is sometimes performed on a calibration channel taken from the data, such as a charm decay to the same or similar final state as the $B$ meson decay under consideration. In this case, the ML model is simplified (*e.g.* no angular observables are used), any charm vetoes are removed, and all the model parameters are floated, if possible. This can reveal any differences between the MC simulation and data in the $m_{\rm ES}$ and $\Delta E$ signal distributions, which can then be corrected for in the final fit.

The systematic uncertainties for the result are often separated into two categories and will depend on the measurement under consideration. Additive systematics affect the fit yield and hence the significance of a branching fraction measurement. Multiplicative systematics affect the central value of the result but not the significance. In the additive category, we place uncertainties on the accuracy of the fixed parameters in the *p.d.f.*s, any ML fit biases in extracting the yields, model-dependent parameters (such as the mean and width of poorly known resonances), the presence or absence of uncertain resonances (such as the $\sigma(600)$), interference, $B\overline{B}$ background yields, and uncertainty on the longitudinal polarization $f_L$. In a large number of modes, the uncertainty on the fixed parameters extracted from the MC simulation is the dominant systematic (see Chapter 15 for more details on systematic error estimation). In the multiplicative category falls the reconstruction efficiency uncertainties arising from differences between data and MC simulation from tracking, uncertainties in the branching fractions of any intermediate decays, charged particle identification, neutral particle ($\pi^0$) identification, and long-lived particle ($K_S^0$) identification. Also, the accuracy of the known $B\overline{B}$ cross-section, luminosity and limited MC statistics can contribute. If various sub-decays are combined (*e.g.* $K^{*+} \to K_S^0 \pi^+$ or $K^+ \pi^0$) to form an overall measurement, the multiplicative systematics are correlated and must be added linearly.

Many of the systematic errors associated with $f_L$ and $A_{CP}$ cancel since these two measurements are based on ratios of signal yields. The systematic uncertainty on $A_{CP}$ caused by the detector responding differently to positive and negative tracks or the presence of $s$ and $\overline{s}$ in $K^-$ and $K^+$ respectively is generally considered to be 0.5% at most.

Once calculated, the systematic error is convolved with the likelihood function with a Gaussian distribution with a variance equal to the total systematic error (see Chapter 15). The signal significance is then defined as $\sqrt{2\Delta \ln \mathcal{L}}$, where $\Delta \ln \mathcal{L}$ is the change in log-likelihood $\ln \mathcal{L}$ from the maximum value to the value when the number of signal events is set to zero. If multiple signal resonances are extracted in the same fit it is often helpful to state the linear correlation coefficient between the results.

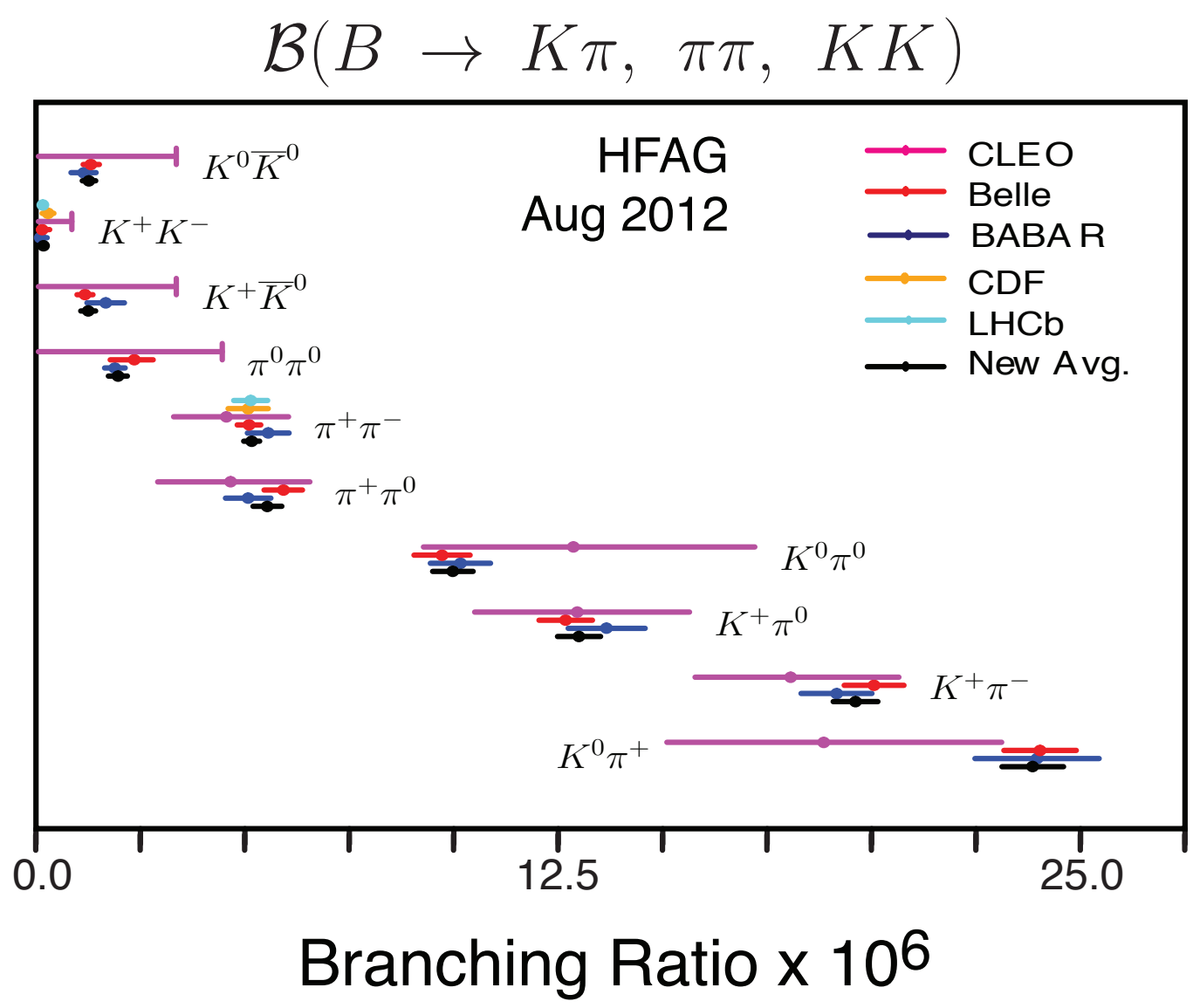

**Figure 17.4.3.** Summary of branching fraction measurements ($\times 10^{-6}$) and HFAG averages for two-body decays to $K\pi$,$\pi\pi$ and $KK$ (Amhis et al. (2012)).

### 17.4.4 Two-body decays

In this section, we just report on branching fractions and $A_{CP}$ measurements; further details are covered in Chapter 17.7. Table 17.4.1 summarises the branching fraction and direct $CP$ measurements made for charmless two-body decays, while Figure 17.4.3 illustrates the branching fraction measurements made so far.

The final state particles in $B$ meson decays to two long-lived particles benefit from having relatively larger momenta than most $B$ decays, leading to a cleaner analysis environment. Decays such as $B \to K\pi$ and $B \to \pi\pi$ are therefore good places to look for new physics and $CP$ violation, both direct and indirect. The first observations of QCD penguin $b \to d$ transitions were made in $B^+ \to \overline{K}^0 K^+$ (Aubert, 2006ai) and $B^0 \to K^0\overline{K}^0$ (Abe, 2005e).

The decay $B^0 \to K^+\pi^-$ proceeds via both $b \to u$ tree and $b \to s$ transitions, which can interfere, leading to a direct $CP$ violating asymmetry (Lin, 2008). The two dominant decay diagrams are shown in Fig. 17.4.4. The world average is now $A_{CP}(K^+\pi^-) = -0.098 \pm 0.013$. The four $K\pi$ asymmetries can be related through sum rules. From Eq. 17.4.5 it follows that

$$\mathcal{A}_{B^- \to \pi^- \bar{K}^0} - \sqrt{2}\,\mathcal{A}_{B^- \to \pi^0 K^-} + \mathcal{A}_{\bar{B}^0 \to \pi^+ K^-} + \sqrt{2}\,\mathcal{A}_{\bar{B}^0 \to \pi^0 \bar{K}^0} = 0 \qquad (17.4.10)$$

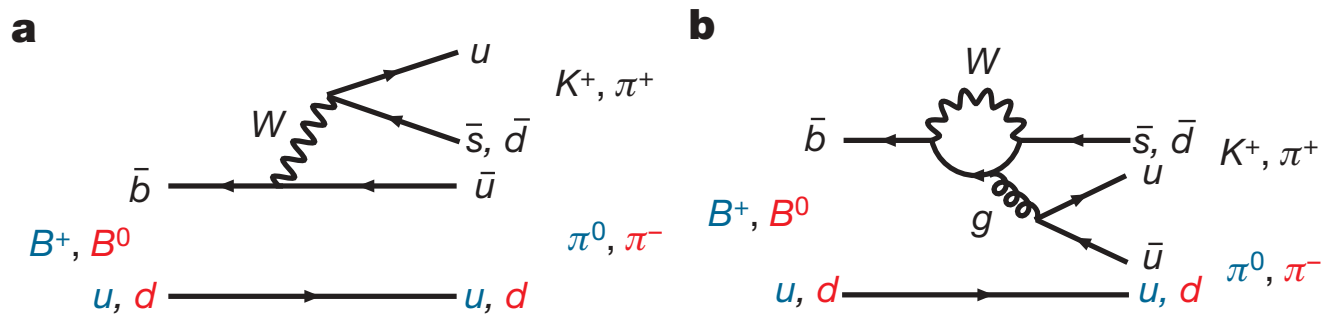

**Figure 17.4.4.** The dominant Tree-level (a) and Penguin-loop (b) Feynman diagrams in the two-body decays $B \to K\pi$ and $B \to \pi\pi$ (Lin, 2008).

This, together with the fact that penguin decays dominate, leads to the prediction (Gronau, 2005)

$$\Delta(K^+\pi^-) + \Delta(K^0\pi^+) = 2\,(\Delta(K^+\pi^0) + \Delta(K^0\pi^0)) \times (1 + \mathcal{O}(5\%)) \qquad (17.4.11)$$

where $\Delta(K\pi) = \Gamma(\overline{B} \to \bar{K}\bar{\pi}) - \Gamma(B \to K\pi)$. Consequently, it is expected that:

$$A_{CP}(K^+\pi^-) + A_{CP}(K^0\pi^+) \approx A_{CP}(K^+\pi^0) + A_{CP}(K^0\pi^0) \qquad (17.4.12)$$

The sum rule prediction for the width agrees quite well with experimental results. If $A_{CP}(K^0\ \pi^+)$ and $A_{CP}(K^0\ \pi^0)$ are small then the predicted values for $A_{CP}(K^+\ \pi^0)$ and $A_{CP}(K^+\ \pi^-)$ are very similar. However the measured values for $A_{CP}(K^+\ \pi^0)$ and $A_{CP}(K^+\ \pi^-)$ differ by about five standard deviations. This is shown in Fig. 17.4.5 where the difference in the number of events is clearly visible and the sign of the difference between the number of events in $B^0 \to K^+\pi^-$ is opposite to that of $B^+ \to K^+\pi^0$. This could be a sign of new physics but other effects, including enhancements in sub-dominant decay diagrams or strong interaction effects, have also been suggested as an explanation.

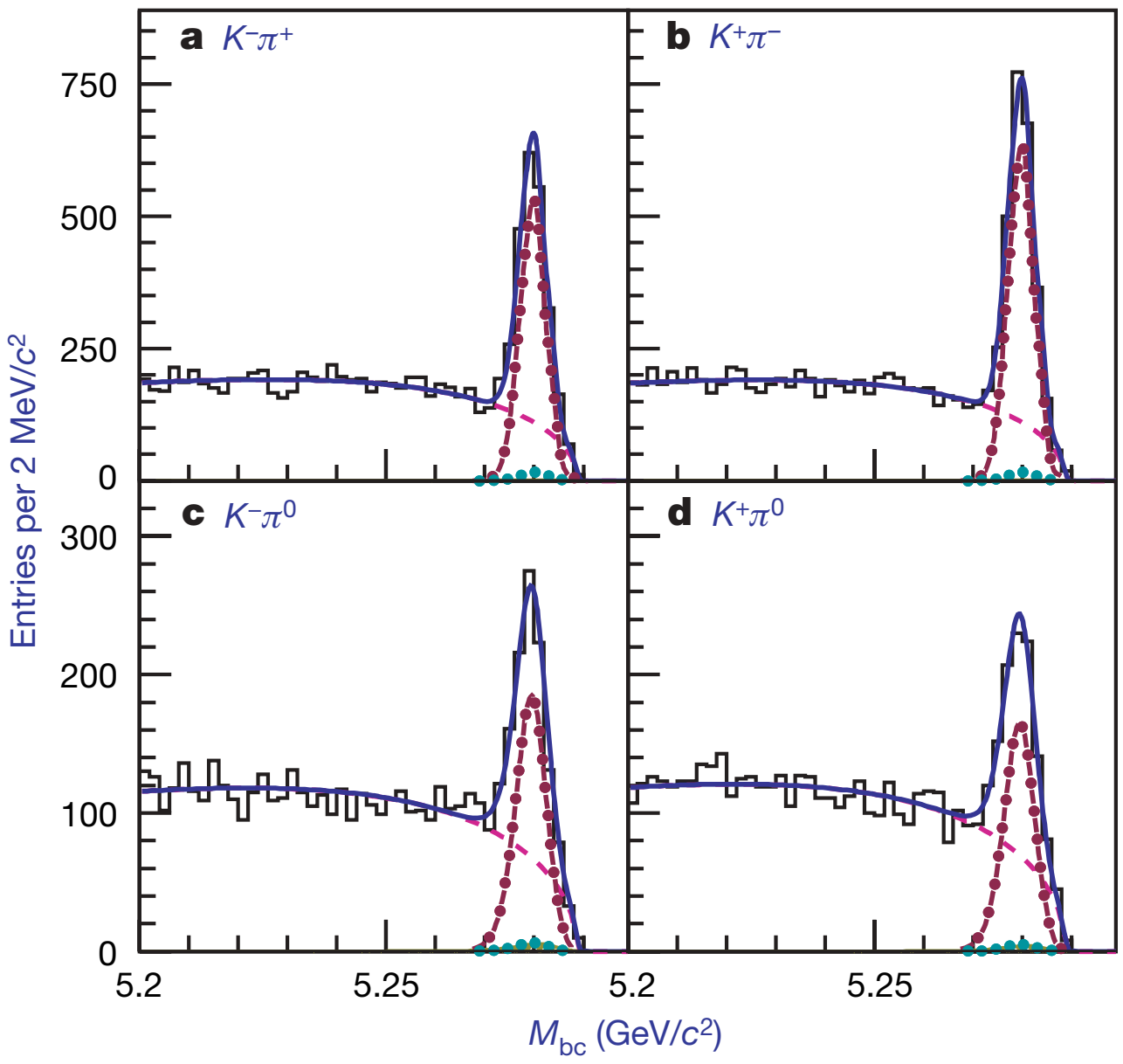

**Figure 17.4.5.** The direct $CP$ violation in $B \to K^{\mp}\pi^{\pm}$ (top) and $B^{\pm} \to K^{\pm}\pi^0$ (bottom) can be seen in the difference between the heights of signal distributions (red/points) in the left and right plots (Lin, 2008).

**Table 17.4.1.** Charmless $B$ decays branching fractions $\mathcal{B}$ and $CP$ asymmetries $A_{CP}$ for *BABAR* and Belle for decays $B \to KK, K\pi, \pi\pi$. The averages come from HFAG and may include measurements from other experiments such as CLEO and CDF (Amhis et al. (2012)).

| | *BABAR* results | | | Belle results | | | Averages | |
|---|---|---|---|---|---|---|---|---|
| Final state | $\mathcal{B}$ ($\times 10^{-6}$) | $A_{CP}$ | Ref. | $\mathcal{B}$ ($\times 10^{-6}$) | $A_{CP}$ | Ref. | $\mathcal{B}$ ($\times 10^{-6}$) | $A_{CP}$ |
| $K^+K^-$ | $0.04 \pm 0.15 \pm 0.08$ | | (Aubert, 2007o) | $0.09^{+0.18}_{-0.13} \pm 0.01$ | | (Abe, 2007g) | $0.15^{+0.11}_{-0.10}$ | |
| $K^+\overline{K}^0$ | $1.61 \pm 0.44 \pm 0.09$ | $0.10 \pm 0.26 \pm 0.03$ | (Aubert, 2006ai) | $1.11^{+0.19}_{-0.18} \pm 0.05$ | $0.017 \pm 0.168 \pm 0.002$ | (Chang, 2011) | $1.19 \pm 0.18$ | $0.041 \pm 0.141$ |
| $K^+\pi^-$ | $19.1 \pm 0.6 \pm 0.6$ | $-0.107 \pm 0.016^{+0.006}_{-0.004}$ | (Aubert, 2007o) | $20.0 \pm 0.34 \pm 0.63$ | $-0.069 \pm 0.014 \pm 0.007$ | (Chang, 2011) | $19.55^{+0.54}_{-0.53}$ | $-0.086 \pm 0.007$ |
| $K^+\pi^0$ | $13.6 \pm 0.6 \pm 0.7$ | $0.030 \pm 0.039 \pm 0.010$ | (Aubert, 2007ay) | $12.62 \pm 0.31 \pm 0.56$ | $0.043 \pm 0.024 \pm 0.002$ | (Chang, 2011) | $12.94^{+0.52}_{-0.51}$ | $0.037 \pm 0.021$ |
| $K^0\overline{K}^0$ | $1.08 \pm 0.28 \pm 0.11$ | | (Aubert, 2006ai) | $1.26^{+0.19}_{-0.18} \pm 0.06$ | | (Chang, 2011) | $1.21 \pm 0.16$ | |
| $K^0\pi^+$ | $23.9 \pm 1.1 \pm 1.0$ | $-0.029 \pm 0.039 \pm 0.010$ | (Aubert, 2006ai) | $23.97^{+0.53}_{-0.52} \pm 0.69$ | $-0.014 \pm 0.012 \pm 0.006$ | (Chang, 2011) | $23.80 \pm 0.74$ | $-0.015 \pm 0.012$ |
| $K^0\pi^0$ | $10.1 \pm 0.6 \pm 0.4$ | | (Aubert, 2008m) | $9.66 \pm 0.46 \pm 0.49$ | | (Chang, 2011) | $9.92^{+0.49}_{-0.48}$ | |
| $\pi^+\pi^-$ | $5.5 \pm 0.4 \pm 0.3$ | | (Aubert, 2007o) | $5.04 \pm 0.21 \pm 0.19$ | | (Chang, 2011) | $5.11 \pm 0.22$ | |
| $\pi^+\pi^0$ | $5.02 \pm 0.46 \pm 0.29$ | $0.03 \pm 0.08 \pm 0.01$ | (Aubert, 2007ay) | $5.86 \pm 0.26 \pm 0.38$ | $0.025 \pm 0.043 \pm 0.007$ | (Chang, 2011) | $5.48^{+0.35}_{-0.34}$ | $0.026 \pm 0.039$ |
| $\pi^0\pi^0$ | $1.83 \pm 0.21 \pm 0.13$ | $0.43 \pm 0.26 \pm 0.05$ | (Aubert, 2008m) | $2.3^{+0.4+0.2}_{-0.5-0.3}$ | $0.44^{+0.53}_{-0.52} \pm 0.17$ | (Abe, 2005h) | $1.91^{+0.22}_{-0.23}$ | $0.43 \pm 0.24$ |

The branching fractions of two-body decays to $\pi\pi$ final states are of interest to understand the so-called penguin pollution in $B^0 \to \pi^+\pi^-$ (discussed in Section 17.7). It was observed that in the $\pi^+\pi^-$ final state there appeared to be evidence for a significant tail in the $\Delta E$ distribution for selected events that was not apparent in the Monte Carlo simulation used at that time. After some investigation it was realised that the tail in $\Delta E$ was the result of final state radiation (FSR) which needed to be accounted for properly in the simulation in order to continue to improve the precision of branching fraction measurements in an un-biased way. The first $B^0 \to \pi^+\pi^-$ branching fraction measurement that attempted to account for this FSR effect appropriately was performed by *BABAR* (Aubert, 2007o). Subsequent results account for this effect.

Initial expectations for the decay $B^0 \to \pi^0\pi^0$ were that the branching fraction would be small, led in part by theoretical calculations indicating that this process would be dominated by a color suppressed tree. In the summer of 2002, preliminary results from the $B$ Factories started to show hints of a relatively large signal with a branching fraction central value of a few $10^{-6}$. Subsequent results published by *BABAR* and Belle led to the observation of this channel. The world average branching fraction is currently $(1.91^{+0.22}_{-0.23}) \times 10^{-6}$.

### 17.4.5 Quasi-two-body decays

In the sections that follow, the quasi-two-body decays have been grouped according to the spins of their final state particles. For each grouping of spin, the results for the branching fractions are itemized in tables and are shown in the plots to enable convenient comparison.

#### 17.4.5.1 $B \to$ two Pseudoscalars, Pseudoscalar Vector, Pseudoscalar Scalar, Pseudoscalar Tensor with $\eta^{(\prime)}$

A number of searches have been performed with a Pseudoscalar $\eta$ or $\eta'$ in the final state together with one other particle. For Pseudoscalar-Pseudoscalar (PP) modes, the other particle is an $\eta^{(\prime)}$, $K$, or $\pi$; for Pseudoscalar-Vector (PV) modes, a $K^*$, $\rho$, $\omega$, or $\phi$; for Pseudoscalar-Scalar (PS) modes, an $f_0(980)$ or $K_0^*(1430)$; for Pseudoscalar-Tensor (PT), $K_2^*(1430)$. The branching fractions and asymmetries reported by Belle and *BABAR*, and their HFAG averages, are given in Table 17.4.2. Figure 17.4.6 shows the branching fractions. The HFAG averages represent a snapshot of the field in late 2012 (Amhis et al. (2012)) but are being annually updated on the website (see Asner et al. (2011)).

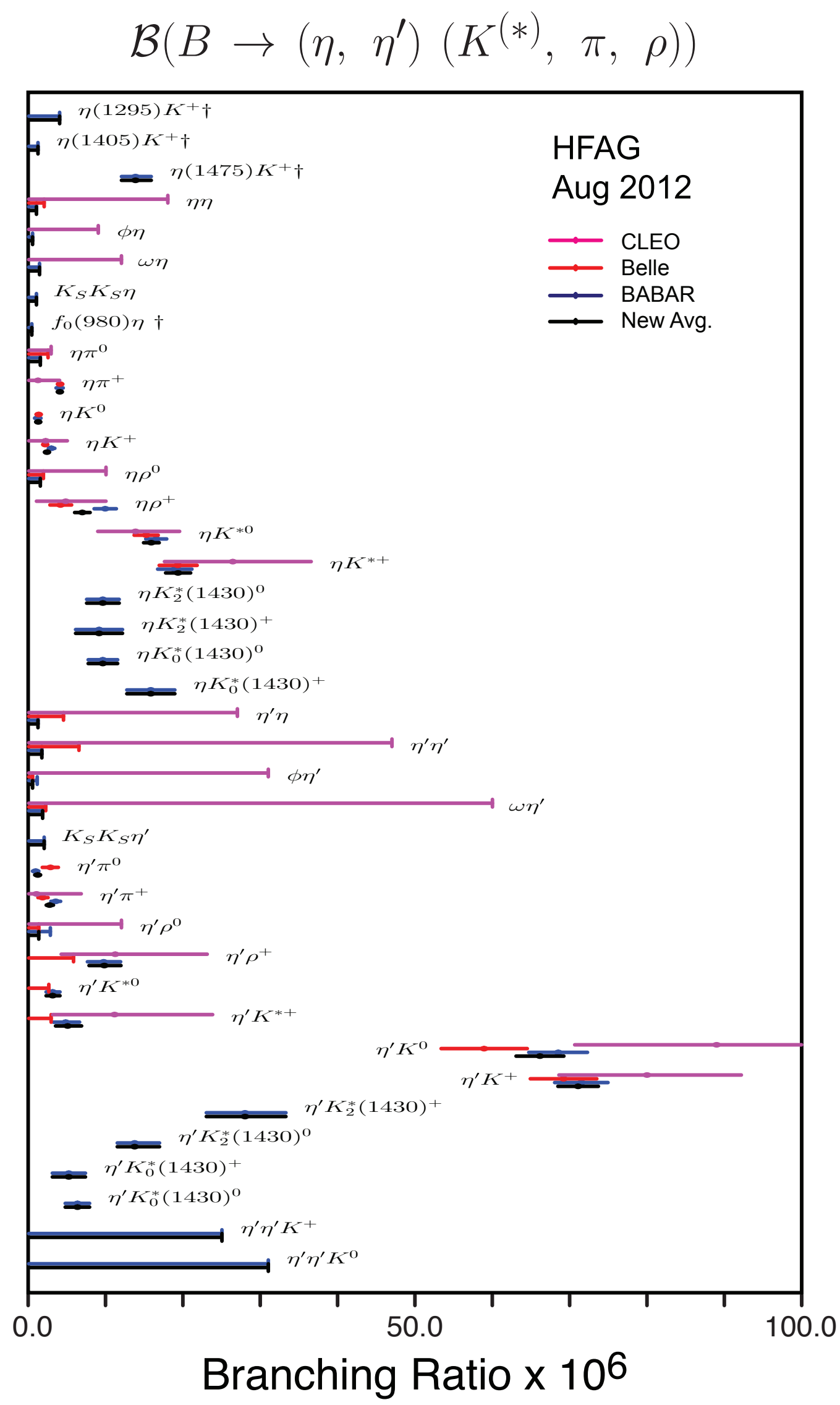

**Figure 17.4.6.** Summary of branching fraction measurements ($\times 10^{-6}$) and HFAG averages for decays with an $\eta$ or $\eta'$ meson combined with a pseudoscalar, vector, scalar or tensor particle (Amhis et al. (2012)).

**Table 17.4.2.** Charmless $B$ decays branching fractions $\mathcal{B}$ and $CP$ asymmetries $A_{CP}$ for BABAR and Belle for decays with an $\eta$ or $\eta'$ in the final state. The averages come from HFAG. The decays are arranged from top to bottom, as Pseudoscalar-Pseudoscalar (PP), Pseudoscalar-Vector (PV), Pseudoscalar-Scalar (PS), Pseudoscalar-Tensor (PT), and three-body decays (Amhis et al. (2012)).

| | BABAR results | | | Belle results | | | Averages | |
|---|---|---|---|---|---|---|---|---|
| Final state | $\mathcal{B}$ ($\times 10^{-6}$) | $A_{CP}$ | Ref. | $\mathcal{B}$ ($\times 10^{-6}$) | $A_{CP}$ | Ref. | $\mathcal{B}$ ($\times 10^{-6}$) | $A_{CP}$ |
| $\eta K^+$ | $2.94^{+0.39}_{-0.34} \pm 0.21$ | $-0.36 \pm 0.11 \pm 0.03$ | (Aubert, 2009d) | $2.12 \pm 0.23 \pm 0.11$ | $-0.38 \pm 0.11 \pm 0.01$ | (Hoi, 2012) | $2.36^{+0.22}_{-0.21}$ | $-0.37 \pm 0.09$ |
| $\eta K^0$ | $1.15^{+0.43}_{-0.38} \pm 0.09$ | | (Aubert, 2009d) | $1.27^{+0.33}_{-0.29} \pm 0.08$ | | (Hoi, 2012) | $1.23^{+0.27}_{-0.24}$ | |
| $\eta\eta$ | $< 1.0$ | | (Aubert, 2009d) | $< 2.0$ | | (Chang, 2005) | $< 1.0$ | |
| $\eta\pi^+$ | $4.00 \pm 0.40 \pm 0.24$ | $-0.03 \pm 0.09 \pm 0.03$ | (Aubert, 2009d) | $4.07 \pm 0.26 \pm 0.21$ | $-0.19 \pm 0.06 \pm 0.01$ | (Hoi, 2012) | $4.02 \pm 0.27$ | $-0.13 \pm 0.10$ |
| $\eta\pi^0$ | $< 1.5$ | | (Aubert, 2008af) | $< 2.5$ | | (Chang, 2005) | $< 1.5$ | |
| $\eta(1295)K^+$ | $< 4.0$ | | (Aubert, 2008bb) | | | | $< 4.0$ | |
| $\eta(1405)K^+$ | $< 1.2$ | | (Aubert, 2008bb) | | | | $< 1.2$ | |
| $\eta(1475)K^+$ | $13.8^{+1.8+1.0}_{-1.7-0.6}$ | | (Aubert, 2008bb) | | | | $13.8^{+2.1}_{-1.8}$ | |
| $\eta' K^+$ | $71.5 \pm 1.3 \pm 3.2$ | $0.008^{+0.017}_{-0.018} \pm 0.009$ | (Aubert, 2009d) | $69.2 \pm 2.2 \pm 3.7$ | $0.028 \pm 0.028 \pm 0.021$ | (Schumann, 2006) | $71.1 \pm 2.6$ | $0.013 \pm 0.017$ |
| $\eta' K^0$ | $68.5 \pm 2.2 \pm 3.1$ | | (Aubert, 2009d) | $58.9^{+3.6}_{-3.5} \pm 4.3$ | | (Schumann, 2006) | $66.1 \pm 3.1$ | |
| $\eta'\eta$ | $< 1.2$ | | (Aubert, 2008af) | $< 4.5$ | | (Schumann, 2007) | $< 1.2$ | |
| $\eta'\eta'$ | $< 1.7$ | | (Aubert, 2009d) | $< 6.5$ | | (Schumann, 2007) | $< 1.7$ | |
| $\eta'\pi^+$ | $3.5 \pm 0.6 \pm 0.2$ | $0.03 \pm 0.17 \pm 0.02$ | (Aubert, 2009d) | $1.8^{+0.7}_{-0.6} \pm 0.1$ | $0.20^{+0.37}_{-0.36} \pm 0.04$ | (Schumann, 2006) | $2.7^{+0.5}_{-0.4}$ | $0.06 \pm 0.16$ |
| $\eta'\pi^0$ | $0.9 \pm 0.4 \pm 0.1$ | | (Aubert, 2008af) | $2.8 \pm 1.0 \pm 0.3$ | | (Schumann, 2006) | $1.2 \pm 0.4$ | |
| $\eta K^{*+}$ | $18.9 \pm 1.8 \pm 1.3$ | $0.01 \pm 0.08 \pm 0.02$ | (Aubert, 2006l) | $19.3^{+2.0}_{-1.9} \pm 1.5$ | $0.03 \pm 0.10 \pm 0.01$ | (Wang, 2007a) | $19.3 \pm 1.6$ | $0.02 \pm 0.06$ |
| $\eta K^{*0}$ | $16.5 \pm 1.1 \pm 0.8$ | $0.21 \pm 0.06 \pm 0.02$ | (Aubert, 2006l) | $15.2 \pm 1.2 \pm 1.0$ | $0.17 \pm 0.08 \pm 0.01$ | (Wang, 2007a) | $15.9 \pm 1.0$ | $0.19 \pm 0.05$ |
| $\eta\rho^+$ | $9.9 \pm 1.2 \pm 0.8$ | $0.13 \pm 0.11 \pm 0.02$ | (Aubert, 2008af) | $4.1^{+1.4}_{-1.3} \pm 0.4$ | $-0.04^{+0.34}_{-0.32} \pm 0.01$ | (Wang, 2007a) | $6.9 \pm 1.0$ | $0.11 \pm 0.11$ |
| $\eta\rho^0$ | $< 1.5$ | | (Aubert, 2007as) | $< 1.9$ | | (Wang, 2007a) | $< 1.5$ | |
| $\eta' K^{*+}$ | $4.8^{+1.6}_{-1.4} \pm 0.8$ | $-0.26 \pm 0.27 \pm 0.02$ | (del Amo Sanchez, 2010h) | $< 2.9$ | | (Schumann, 2007) | $5.0^{+1.8}_{-1.6}$ | $-0.30^{+0.33}_{-0.37} \pm 0.02$ |
| $\eta' K^{*0}$ | $3.1^{+0.9}_{-0.8} \pm 0.3$ | $0.02 \pm 0.23 \pm 0.02$ | (del Amo Sanchez, 2010h) | $< 2.6$ | | (Schumann, 2007) | $3.1 \pm 0.9$ | $0.08 \pm 0.25 \pm 0.02$ |
| $\eta'\rho^+$ | $9.7^{+1.9}_{-1.8} \pm 1.1$ | $0.26 \pm 0.17 \pm 0.02$ | (del Amo Sanchez, 2010h) | $< 5.8$ | | (Schumann, 2007) | $9.8^{+2.1}_{-2.0}$ | $0.04 \pm 0.28 \pm 0.02$ |
| $\eta'\rho^0$ | $< 2.8$ | | (del Amo Sanchez, 2010h) | $< 1.3$ | | (Schumann, 2007) | $< 1.3$ | |
| $\omega\eta$ | $< 1.4$ | | (Aubert, 2009d) | | | | $< 1.4$ | |
| $\omega\eta'$ | $< 1.8$ | | (Aubert, 2009d) | $< 2.2$ | | (Schumann, 2007) | $< 1.8$ | |
| $\phi\eta$ | $< 0.5$ | | (Aubert, 2009d) | | | | $< 0.5$ | |
| $\phi\eta'$ | $< 1.1$ | | (Aubert, 2009d) | $< 0.5$ | | (Schumann, 2007) | $< 0.5$ | |
| $f_0(980)\eta$ | $< 0.4$ | | (Aubert, 2007as) | | | | $< 0.4$ | |
| $f_0(980)\eta'$ | $< 0.9$ | | (del Amo Sanchez, 2010h) | | | | $< 0.9$ | |
| $\eta K_0^*(1430)^+$ | $15.8 \pm 2.2 \pm 2.2$ | $0.05 \pm 0.13 \pm 0.02$ | (Aubert, 2006l) | | | | $15.8 \pm 3.1$ | $0.05 \pm 0.13 \pm 0.02$ |
| $\eta K_0^*(1430)^0$ | $9.6 \pm 1.4 \pm 1.3$ | $0.06 \pm 0.13 \pm 0.02$ | (Aubert, 2006l) | | | | $9.6 \pm 1.9$ | $0.06 \pm 0.13 \pm 0.02$ |
| $\eta' K_0^*(1430)^+$ | $5.2 \pm 1.9 \pm 1.0$ | $0.06 \pm 0.20 \pm 0.02$ | (del Amo Sanchez, 2010h) | | | | $5.2 \pm 2.1$ | |
| $\eta' K_0^*(1430)^0$ | $6.3 \pm 1.3 \pm 0.9$ | $-0.19 \pm 0.17 \pm 0.02$ | (del Amo Sanchez, 2010h) | | | | $6.3 \pm 1.6$ | |
| $\eta K_2^*(1430)^+$ | $9.1 \pm 2.7 \pm 1.4$ | $-0.45 \pm 0.30 \pm 0.02$ | (Aubert, 2006l) | | | | $9.1 \pm 3.0$ | $-0.45 \pm 0.30 \pm 0.02$ |
| $\eta K_2^*(1430)^0$ | $9.6 \pm 1.8 \pm 1.1$ | $-0.07 \pm 0.19 \pm 0.02$ | (Aubert, 2006l) | | | | $9.6 \pm 2.1$ | $-0.07 \pm 0.19 \pm 0.02$ |
| $\eta' K_2^*(1430)^+$ | $28.0^{+4.6}_{-4.3} \pm 2.6$ | $0.15 \pm 0.13 \pm 0.02$ | (del Amo Sanchez, 2010h) | | | | $28.0^{+5.3}_{-5.0}$ | |
| $\eta' K_2^*(1430)^0$ | $13.7^{+3.0}_{-1.9} \pm 1.2$ | $0.14 \pm 0.18 \pm 0.02$ | (del Amo Sanchez, 2010h) | | | | $13.7^{+3.2}_{-2.2}$ | |
| $K_S^0 K_S^0 \eta$ | $< 1.0$ | | (Aubert, 2009am) | | | | $< 1.0$ | |
| $K_S^0 K_S^0 \eta'$ | $< 2.0$ | | (Aubert, 2009am) | | | | $< 2.0$ | |
| $\eta'\eta' K^+$ | $< 25$ | | (Aubert, 2006al) | | | | $< 25$ | |
| $\eta'\eta' K^0$ | $< 31$ | | (Aubert, 2006al) | | | | $< 31$ | |

Theory predictions for PP and PV branching fractions are typically in the low parts per million. In leading order SM calculations, the time dependent *CP* violation asymmetry parameter $S = \sin 2\phi_1$ in decays such as $B^0 \to \eta' K_S^0$, $B^0 \to KKK_S^0$ and $B^0 \to \phi K_S^0$ is expected to be the same as in the golden mode $B^0 \to J/\psi K_S^0$ and provide a useful comparison between decays mediated by $b \to s\bar{s}u$, $b \to su\bar{u}$, $b \to sd\bar{d}$ and $b \to c\bar{c}s$ (see Section 17.6 for details), provided the decays are dominated by a single weak phase. Within the standard model (SM), the decay $B \to \eta' K$ proceeds through $b \to s$ penguin loops with only a small contribution from $b \to u$ tree diagrams (Chen, 2002). Corrections can be estimated in QCD factorization and turn out to be small. Therefore a significant deviation would be a sign of new physics (Abe, 2003e). The decay rates of $\eta\eta$, $\eta'\eta'$, $\eta\phi$ and $\eta'\phi$ can be related to any deviation in $\Delta S$ from the charmonium measured $\phi_1$ via $SU(3)$ flavor symmetry (Aubert, 2006av, 2007am).

In charged decays such as $B^+ \to \eta' K^+$ and $B^+ \to \eta' \pi^+$ (Abe, 2001d; Schumann, 2006), the *CP* charge asymmetry $A_{CP}$ is expected to be small in $\eta' K^+$. A large direct *CP* asymmetry is expected in $B^+ \to \eta K^+$ but not in $B^+ \to \eta' K^+$ because the overall penguin amplitudes in $B^+ \to \eta K^+$ are of the same order as the tree amplitude, while in $B^+ \to \eta' K^+$ the penguin dominates. This is confirmed by the experiments, which measure $A_{CP} = -0.37 \pm 0.09$ for $B^+ \to \eta K^+$ but only $A_{CP} = 0.013 \pm 0.017$ for $B^+ \to \eta' K^+$. In $\eta\rho^+$, $\eta\pi^+$ and $\eta'\pi^+$, the $b \to u$ and $b \to s$ amplitudes are of similar size possibly leading to large direct *CP* violation (Aubert, 2005l).

Any sub-leading terms in $B^0 \to \eta' K_S^0$ can be constrained by measuring the decays $\eta'\eta$, $\eta\pi^0$ and $\eta'\pi^0$. $B^0 \to \eta\pi^0$ and $B^0 \to \eta'\pi^0$ may also constrain isospin breaking effects on the value of $\sin 2\phi_2$ in $B^0 \to \pi^+\pi^-$ decays. The branching fractions are a useful test of predictions from QCD factorization, perturbative QCD (for $\eta'\pi^0$) and flavor-$SU(3)$ symmetry (Aubert, 2006g). These limit the deviation $\Delta S$ of the measured $S$ from the value seen in charmonium decays, with bounds on $|\Delta S| < 0.05$.

Mixing-induced *CP* violation has been observed in $B^0 \to \eta' K^0$ (Aubert, 2007am) and (Chen, 2007a). The $\eta' K^*$ and $\eta K$ branching fractions are suppressed while $\eta' K$ and $\eta K^*$ are enhanced, since the two $b \to s$ penguins that contribute interfere constructively in $\eta' K$ decays and destructively in $\eta K$, while the situation is reversed for $\eta' K^*$ and $\eta K^*$ (Beneke and Neubert, 2003a; Lipkin, 1991) as is observed (Abe, 2007a; Chang, 2005). Searches for $\eta' h$ can improve the understanding of flavor-singlet penguin amplitudes with intermediate up-type ($u$,$c$,$t$) quarks (Schumann, 2007).

Searches for excited $\eta$ and $\eta'$ mesons (*e.g.* the $J^P = 0^-$ states $\eta(1295)$, $\eta(1405)$, $\eta(1475)$) with a kaon have also been performed (Aubert, 2008bb). They decay strongly to at least three pseudoscalar mesons but their exact nature is uncertain and they could be gluonium admixtures (*i.e.* a state with additional $gg$ components). Partial wave analysis suggests that the meson spectrum is a linear combination of the resonant state and a non-resonant phase-space contribution. The $J^P = 1^+$ states $f_1(1285)$ and $f_1(1420)$, and the $J^P = 1^-$ state $\phi(1680)$ have a similar mass and final decay states as the $J^P = 0^-$ states so have to be included in any search for excited $\eta$ and $\eta'$ mesons.

Penguin (tree) diagrams dominate in the $B$ decay to $\eta K^*$ ($\eta\rho$) (Wang, 2007a). The decays $\eta'\rho$ are suppressed due to the small value of the CKM matrix element, even though they proceed via tree diagrams (Aubert, 2007ak). The expected branching fraction for $B^0 \to \eta'\rho^0$ is of the order $10^{-8} - 10^{-7}$ and a few times $10^{-6}$ for $B^+ \to \eta'\rho^+$. The measured values for $\mathcal{B}\,(B^+ \to \eta^{(\prime)}\rho^+)$ are $\sim 10 \times 10^{-6}$ while only upper limits (UL) of $< (1.5 - 2.8) \times 10^{-6}$ have been placed on $B^0 \to \eta^{(\prime)}\rho^0$ decays.

Upper limits of $0.4 \times 10^{-6}$ and $0.9 \times 10^{-6}$ have been found for $B^0 \to f_0(980)\eta$ and $B^0 \to f_0(980)\eta'$, respectively (Aubert, 2007as; del Amo Sanchez, 2010h). Measurements also exist for $B \to \eta^{(\prime)} K_0^*(1430)$ and $B \to \eta^{(\prime)} K_2^*(1430)$; the measured values of $A_{CP}$ are compatible with zero.

Decays involving two identical neutral spin zero particles and another spin zero particle can be used to add important information on time-dependent *CP* violation and hadronic $B$ decays (Aubert, 2006al). Examples of such decays include $B \to \eta'\eta' K$ and $B^0 \to K_S^0 K_S^0 \eta^{(\prime)}$. There are no theoretical predictions for the branching fractions for these SM-suppressed modes.

#### 17.4.5.2 $B \to$ PV excluding $\eta^{(\prime)}$

The branching fractions and asymmetries for the remaining PV modes without an $\eta$ or $\eta'$ are given in Table 17.4.3 and the hierarchy of the branching fraction values are shown in Fig. 17.4.7. The decays $B \to \omega\pi^-$ and $B \to \omega K^-$ are dominated by $b \to u$ tree and $b \to s$ QCD penguin diagrams. They can therefore give an insight into gluonic penguin diagrams (Lu, 2002; Wang, 2004a), (Aubert, 2006ad) and direct *CP* (Lu, 2002).

Charmless $B$ meson decays to final states with an odd number of kaons are usually expected to be dominated by $b \to s$ penguin loops while $b \to u$ tree amplitudes are typically large for final states with $\pi$ and $\rho$ but $\eta K$ decays are suppressed relative to the abundant $\eta' K$.

The $B \to \omega K^0$ decay is a $b \to u\bar{u}s$ process dominated by a single penguin loop amplitude with the same weak phase $\phi_1$ as $\phi K^0$, $K^+K^-K^0$, $\eta' K^0$, $\pi^0 K^0$, and $f_0(980)K^0$, but additional amplitudes and multiple particles in the loop complicate the situation by introducing non-negligible weak phases. $B$ meson decays to *CP* eigenstates $\omega K_S^0$ (together with $\eta' K_S^0$, $\eta' K_L^0$ and $\pi^0 K_S^0$) can be used to extract $S$ and $C$ (Aubert, 2009aa). The maximum deviation $\Delta S$ from the value of $S = \sin 2\phi_1$ measured in charmonium $K_S^0$ decays is $\sim 0.1$. The charged decay modes are expected to have a direct *CP* violation value consistent with zero (Aubert, 2006ad).

In the Standard Model, $B \to K^* K$ decays are dominated by $b \to d s\bar{s}$ gluonic penguin diagrams; for the charged $B^\pm$ decay, the spectator $\bar{d}$ is replaced with $\bar{u}$. Such transitions provide a valuable tool with which to test the quark-flavor sector of the SM. The mode $B^+ \to \overline{K}^{*0} K^+$ is also relevant for the interpretation of the time dependent *CP* asymmetry obtained with the $B^0 \to \phi K^0_S$ mode. To leading order, the *CP* asymmetry equals $\sin 2\phi_1$ for this mode. However, sub-dominant amplitudes proportional to $V^*_{ub}V_{us}$ could produce a deviation $\Delta S_{\phi K^0_S}$ from $\sin 2\phi_1$. Bounds can be placed on $\Delta S_{\phi K^0_S}$ by exploiting $SU(3)$ flavor symmetry and combining measured rates for relevant $b \to s$ and $b \to d$ processes (including $B^+ \to \overline{K}^{*0}K^+$). Measurements yielding a significant deviation in excess of such a bound would be a strong indication of physics beyond the SM. Furthermore, $B^+ \to \overline{K}^{*0}K^+$ is one of several charmless decays that can be used, together with U-spin symmetry, to extract the angle $\phi_3$ (Aubert, 2007av). Only upper limits exist on $B^0 \to \overline{K}^{*0}K^0$ (Aubert, 2006au), which can be used to constrain certain extensions of the Standard Model.

**Table 17.4.3.** Charmless $B$ decays branching fractions $\mathcal{B}$ and $CP$ asymmetries $A_{CP}$ for BABAR and Belle for Pseudoscalar-Vector (PV) final states. The averages come from HFAG and may include measurements from other experiments such as CLEO and CDF (Amhis et al. (2012)).

| | BABAR results | | | Belle results | | | Averages | |
|---|---|---|---|---|---|---|---|---|
| Final state | $\mathcal{B}$ ($\times10^{-6}$) | $A_{CP}$ | Ref. | $\mathcal{B}$ ($\times10^{-6}$) | $A_{CP}$ | Ref. | $\mathcal{B}$ ($\times10^{-6}$) | $A_{CP}$ |
| $K^*(1410)^+\pi^-$ | | | | $< 86$ | | (Garmash, 2007) | $< 86$ | |
| $K^*(1410)^0\pi^+$ | | | | $< 45$ | | (Garmash, 2005) | $< 45$ | |
| $K^*(1680)^+\pi^-$ | $< 25$ | | (Aubert, 2008g) | $< 10.1$ | | (Garmash, 2007) | $< 10.1$ | |
| $K^*(1680)^0\pi^+$ | $< 15$ | | (Aubert, 2005g) | $< 12$ | | (Garmash, 2005) | $< 12$ | |
| $K^*(1680)^0\pi^0$ | $< 7.5$ | | (Aubert, 2008g) | | | | $< 7.5$ | |
| $K^{*+}\pi^-$ | $8.3^{+0.9}_{-0.8} \pm 0.8$ | $-0.24 \pm 0.07 \pm 0.02$ | (Aubert, 2009av) | $8.4 \pm 1.1^{+1.0}_{-0.9}$ | $-0.21 \pm 0.11 \pm 0.07$ | (Garmash, 2007) | $8.6 \pm 0.9$ | $-0.19 \pm 0.07$ |
| $K^{*+}\pi^0$ | $8.2 \pm 1.5 \pm 1.1$ | $-0.06 \pm 0.24 \pm 0.04$ | (Lees, 2011g) | | | | $8.2 \pm 1.8$ | $0.04 \pm 0.29 \pm 0.05$ |
| $K^{*0}\overline{K}^0$ | $< 1.9$ | | (Aubert, 2006au) | | | | $< 1.9$ | |
| $K^{*0}\pi^+$ | $10.8 \pm 0.6^{+1.2}_{-1.4}$ | $0.032 \pm 0.052^{+0.016}_{-0.013}$ | (Aubert, 2008j) | $9.7 \pm 0.6^{+0.8}_{-0.9}$ | $-0.149 \pm 0.064 \pm 0.022$ | (Garmash, 2006) | $9.9^{+0.8}_{-0.9}$ | $-0.04 \pm 0.09$ |
| $K^{*0}\pi^0$ | $3.3 \pm 0.5 \pm 0.4$ | $-0.15 \pm 0.12 \pm 0.04$ | (Lees, 2011a) | $0.4^{+1.9}_{-1.7} \pm 0.1$ | | (Chang, 2004) | $2.5 \pm 0.6$ | $-0.09^{+0.23}_{-0.26}$ |
| $\omega K^+$ | $6.3 \pm 0.5 \pm 0.3$ | $-0.01 \pm 0.07 \pm 0.01$ | (Aubert, 2007f) | $8.1 \pm 0.6 \pm 0.6$ | $0.05^{+0.08}_{-0.07} \pm 0.01$ | (Jen, 2006) | $6.7 \pm 0.5$ | $0.02 \pm 0.05$ |
| $\omega K^0$ | $5.4 \pm 0.8 \pm 0.3$ | | (Aubert, 2007f) | $4.4^{+0.8}_{-0.7} \pm 0.4$ | | (Jen, 2006) | $5.0 \pm 0.6$ | |
| $\omega\pi^+$ | $6.7 \pm 0.5 \pm 0.4$ | $-0.02 \pm 0.08 \pm 0.01$ | (Aubert, 2007f) | $6.9 \pm 0.6 \pm 0.5$ | $-0.02 \pm 0.09 \pm 0.01$ | (Jen, 2006) | $6.9 \pm 0.5$ | $-0.04 \pm 0.06$ |
| $\omega\pi^0$ | $< 0.5$ | | (Aubert, 2008af) | $< 2.0$ | | (Jen, 2006) | $< 0.5$ | |
| $\overline{K}^{*0}K^+$ | $< 1.1$ | | (Aubert, 2007av) | | | | $< 1.1$ | |
| $\phi K^+$ | $9.2 \pm 0.4^{+0.7}_{-0.5}$ | $0.128 \pm 0.044 \pm 0.013$ | (Lees, 2012y) | $9.60 \pm 0.92^{+1.05}_{-0.84}$ | $0.01 \pm 0.12 \pm 0.05$ | (Garmash, 2005) | $8.8 \pm 0.5$ | $-0.01 \pm 0.06$ |
| $\phi K^0$ | $7.1 \pm 0.6^{+0.4}_{-0.3}$ | $-0.05 \pm 0.18 \pm 0.05$ | (Lees, 2012y) | $9.0^{+2.2}_{-1.8} \pm 0.7$ | | (Chen, 2003) | $7.3^{+0.7}_{-0.6}$ | |
| $\phi\pi^+$ | $< 0.24$ | | (Aubert, 2006am) | $< 0.33$ | | (Kim, 2012) | $< 0.24$ | |
| $\phi\pi^0$ | $< 0.28$ | | (Aubert, 2006am) | | | | $< 0.28$ | |
| $\phi(1680)K^+$ | | | | $< 0.8$ | | (Garmash, 2005) | $< 0.8$ | |
| $\rho(1450)^-K^+$ | $2.4 \pm 1.0 \pm 0.6$ | | (Lees, 2011a) | | | | $2.4 \pm 1.2$ | |
| $\rho(1450)^0\pi^+$ | $1.4 \pm 0.4^{+0.5}_{-0.8}$ | $-0.06 \pm 0.28^{+0.23}_{-0.32}$ | (Aubert, 2009h) | | | | $1.4^{+0.6}_{-0.9}$ | $-0.06^{+0.36}_{-0.42}$ |
| $\rho(1700)^-K^+$ | $0.6 \pm 0.6 \pm 0.4$ | | (Lees, 2011a) | | | | $0.6 \pm 0.7$ | |
| $\rho^+K^0$ | $8.0^{+1.4}_{-1.3} \pm 0.6$ | $-0.12 \pm 0.17 \pm 0.02$ | (Aubert, 2007al) | | | | $8.0^{+1.5}_{-1.4}$ | $-0.12 \pm 0.17 \pm 0.02$ |
| $\rho^+\pi^0$ | $10.2 \pm 1.4 \pm 0.9$ | $-0.01 \pm 0.13 \pm 0.02$ | (Aubert, 2007y) | $13.2 \pm 2.3^{+1.4}_{-1.9}$ | $0.06 \pm 0.17^{+0.04}_{-0.05}$ | (Zhang, 2005) | $10.9^{+1.4}_{-1.5}$ | $0.02 \pm 0.11$ |
| $\rho^-K^+$ | $6.6 \pm 0.5 \pm 0.8$ | $0.20 \pm 0.09 \pm 0.08$ | (Lees, 2011a) | $15.1^{+3.4+2.4}_{-3.3-2.6}$ | $0.22^{+0.22+0.06}_{-0.23-0.02}$ | (Chang, 2004) | $7.2 \pm 0.9$ | $0.15 \pm 0.13$ |
| $\rho^0K^+$ | $3.56 \pm 0.45^{+0.57}_{-0.46}$ | $0.44 \pm 0.10^{+0.06}_{-0.14}$ | (Aubert, 2008j) | $3.89 \pm 0.47^{+0.43}_{-0.41}$ | $0.30 \pm 0.11^{+0.11}_{-0.05}$ | (Garmash, 2006) | $3.81^{+0.48}_{-0.46}$ | $0.37 \pm 0.10$ |
| $\rho^0K^0$ | $4.4 \pm 0.7 \pm 0.3$ | | (Aubert, 2009av) | $6.1 \pm 1.0^{+1.1}_{-1.2}$ | | (Garmash, 2007) | $4.7 \pm 0.7$ | |
| $\rho^0\pi^+$ | $8.1 \pm 0.7^{+1.3}_{-1.6}$ | $0.18 \pm 0.07^{+0.05}_{-0.15}$ | (Aubert, 2009h) | $8.0^{+2.3}_{-2.0} \pm 0.7$ | | (Gordon, 2002) | $8.3^{+1.2}_{-1.3}$ | $0.18^{+0.09}_{-0.17}$ |
| $\rho^0\pi^0$ | $1.4 \pm 0.6 \pm 0.3$ | | (Aubert, 2004g) | $3.0 \pm 0.5 \pm 0.7$ | | (Kusaka, 2008) | $2.0 \pm 0.5$ | |
| $\rho^0(1450)K^+$ | $< 11.7$ | | (Aubert, 2005g) | | | | $< 11.7$ | |
| $\rho^{\mp}\pi^{\pm}$ | $22.6 \pm 1.8 \pm 2.2$ | | (Aubert, 2002g) | $22.6 \pm 1.1 \pm 4.4$ | | (Kusaka, 2008) | $23.0 \pm 2.3$ | |

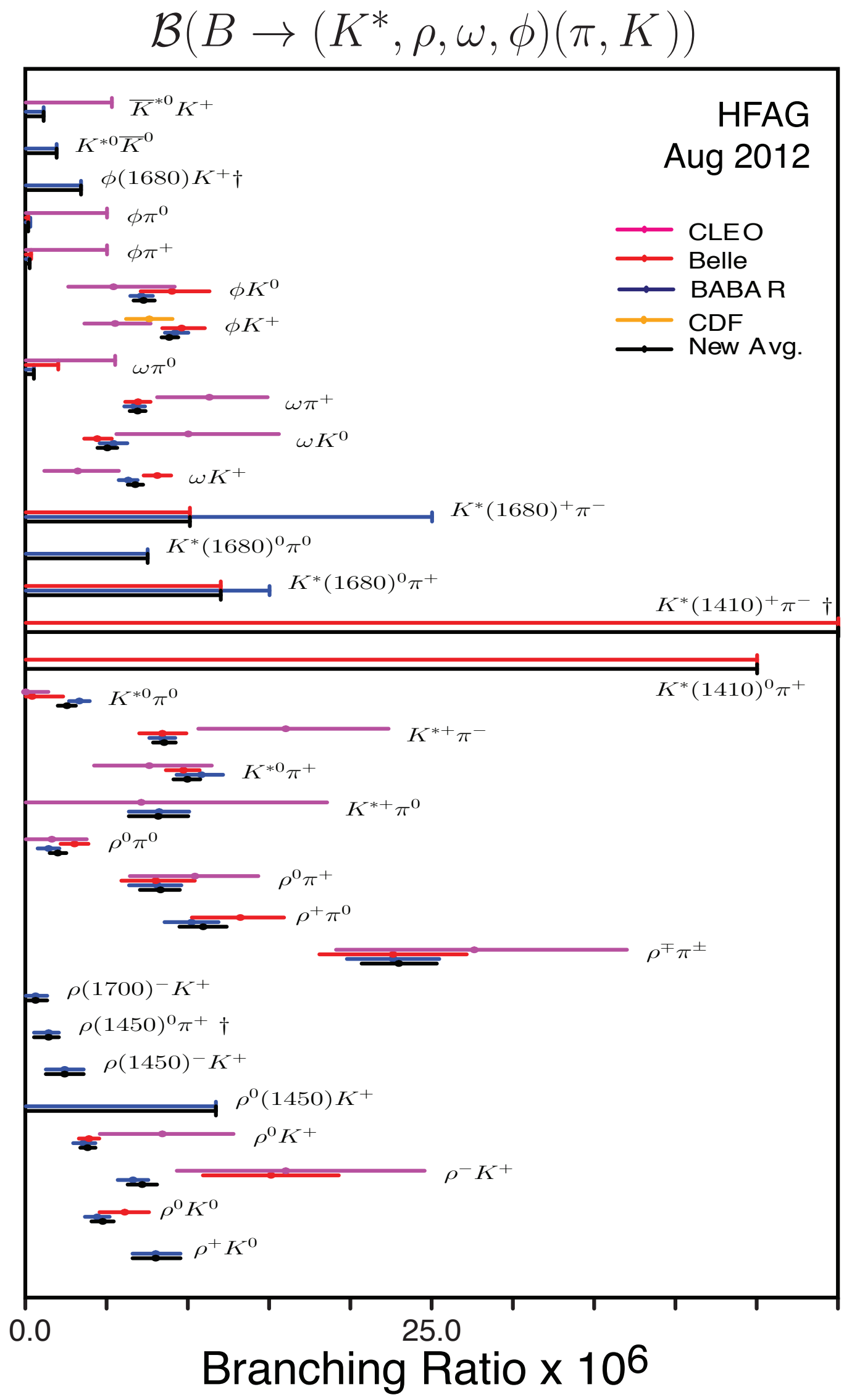

**Figure 17.4.7.** Summary of branching fraction measurements ($\times 10^{-6}$) and HFAG averages for Pseudovector-Vector (PV) decays (Amhis et al. (2012)).

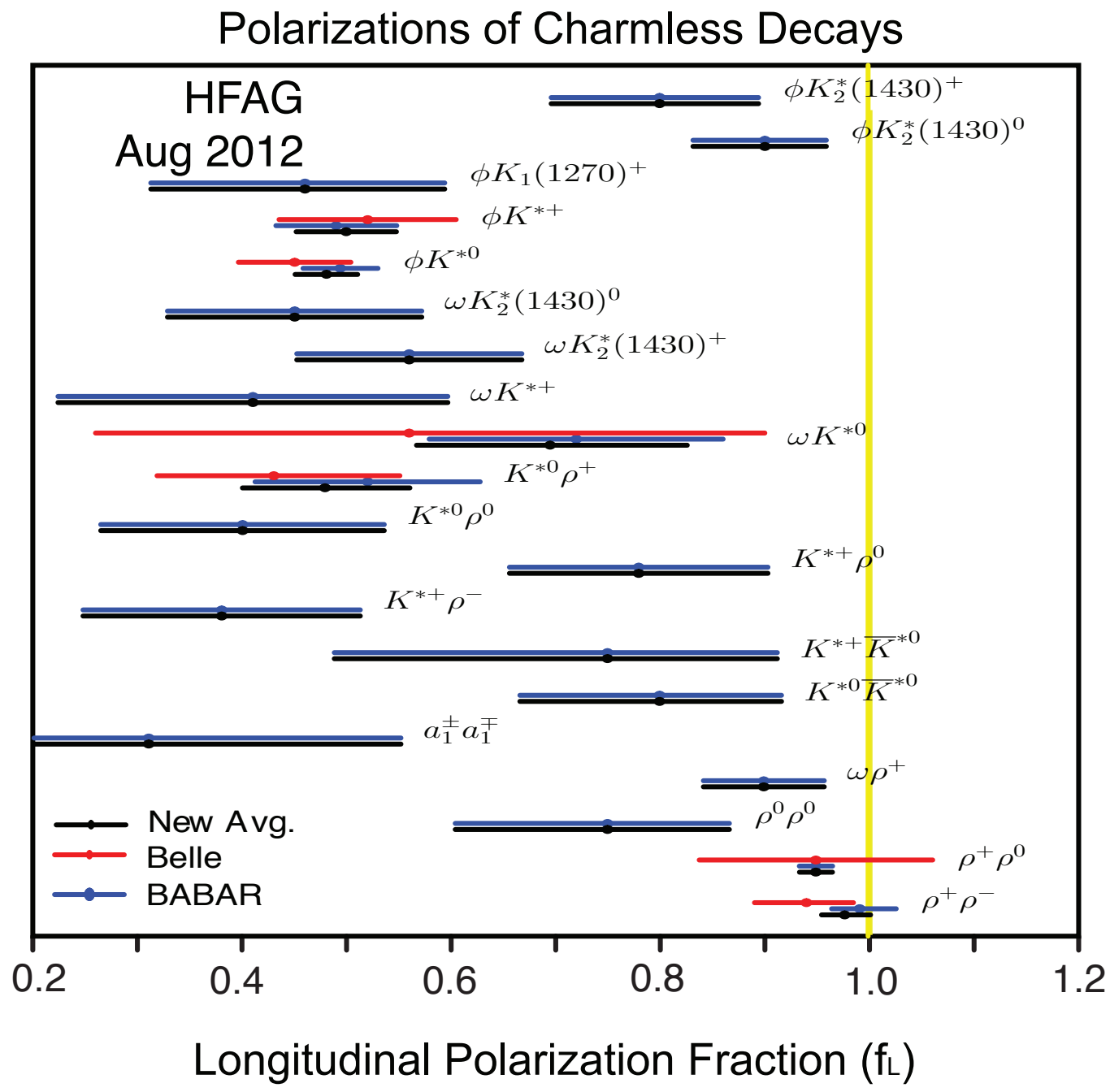

**Figure 17.4.8.** The longitudinal polarization fractions $f_L$ for charmless $B$ decays at *BABAR* and Belle. The average is from the HFAG group (Amhis et al. (2012)).

#### 17.4.5.3 $B \to$ VV

Decays to a Vector-Vector (VV) final state with pairs formed from $\omega$, $K^*$, $\rho$, and $\phi$ can, in principle, be used to determine the helicity amplitudes of the decay. However, this requires a complete angular analysis and in general the number of reconstructed events currently restricts any analysis to integrating over two of the helicity angles and simply reporting the longitudinal polarization $f_L$. A full angular analysis *has* been done for low-background decays such as $B^0 \to \phi K^{*0}$. Further details of the angular analysis process can be found in Chapter 12, where the angular distributions for the VV final states is given in Eq. 12.2.5.

As discussed in Section 17.4.2, the $B \to VV$ decays are naïvely predicted to be dominated by the longitudinal polarization since $f_L \approx 1 - 4m_V/m_B \sim 0.9$, but the naïve factorization expectation is not born out by the QCD factorization analysis for the penguin-dominated decays.

The measured $f_L$ from a number of VV decays are given in Table 17.4.4. Figure 17.4.8 shows the reported results from Belle and *BABAR* and their HFAG averages. There is an apparent hierarchy with $\rho\rho$ modes near $f_L = 1$, $K^*K^*$ and $\phi K_2^*(1430)$ near 0.75, and $\phi K^*$, $\omega K^*$, and $a_1^\pm(1260)a_1^\mp(1260)$ near 0.5. Modes dominated by tree decays have $f_L \sim 1$ while penguin-dominated decays are closer to 0.5. There is also a hierarchy based on the masses of the vector mesons, with larger masses having smaller values of $f_L$. However, this is more evident when comparing decays with a $D^*$ as one or both of the daughter vector mesons.

**Table 17.4.4.** Longitudinal Polarization fractions $f_L$ for *BaBar* and Belle. The average is from the HFAG group (Amhis et al. (2012)).

| | *BaBar* results | | Belle results | | Average |
|---|---|---|---|---|---|
| Final state | $f_L$ | Ref. | $f_L$ | Ref. | $f_L$ |
| $K^{*+}\overline{K}^{*0}$ | $0.75^{+0.16}_{-0.26} \pm 0.03$ | (Aubert, 2009k) | | | $0.75^{+0.16}_{-0.26} \pm 0.03$ |
| $K^{*0}\overline{K}^{*0}$ | $0.80^{+0.10}_{-0.12} \pm 0.06$ | (Aubert, 2008ah) | | | $0.80^{+0.10}_{-0.12} \pm 0.06$ |
| $K^{*+}\rho^-$ | $0.38 \pm 0.13 \pm 0.03$ | (Lees, 2012l) | | | $0.38 \pm 0.13 \pm 0.03$ |
| $K^{*+}\rho^0$ | $0.78 \pm 0.12 \pm 0.03$ | (del Amo Sanchez, 2011g) | | | $0.78 \pm 0.12 \pm 0.03$ |
| $K^{*0}\rho^+$ | $0.52 \pm 0.10 \pm 0.04$ | (Aubert, 2006ab) | $0.43 \pm 0.11^{+0.05}_{-0.02}$ | (Abe, 2005f) | $0.48 \pm 0.08$ |
| $K^{*0}\rho^0$ | $0.40 \pm 0.08 \pm 0.11$ | (Lees, 2012l) | | | $0.40 \pm 0.08 \pm 0.11$ |
| $\omega K^{*+}$ | $0.41 \pm 0.18 \pm 0.05$ | (Aubert, 2009af) | | | $0.41 \pm 0.18 \pm 0.05$ |
| $\omega K^{*0}$ | $0.72 \pm 0.14 \pm 0.02$ | (Aubert, 2009af) | $0.56 \pm 0.29^{+0.18}_{-0.08}$ | (Goldenzweig, 2008) | $0.69 \pm 0.13$ |
| $\omega K_2^*(1430)^+$ | $0.56 \pm 0.10 \pm 0.04$ | (Aubert, 2009af) | | | $0.56 \pm 0.10 \pm 0.04$ |
| $\omega K_2^*(1430)^0$ | $0.45 \pm 0.12 \pm 0.02$ | (Aubert, 2009af) | | | $0.45 \pm 0.12 \pm 0.02$ |
| $\omega\rho^+$ | $0.90 \pm 0.05 \pm 0.03$ | (Aubert, 2009af) | | | $0.90 \pm 0.06$ |
| $\phi K^{*+}$ | $0.49 \pm 0.05 \pm 0.03$ | (Aubert, 2007c) | $0.52 \pm 0.08 \pm 0.03$ | (Chen, 2005a) | $0.50 \pm 0.05$ |
| $\phi K^{*0}$ | $0.494 \pm 0.034 \pm 0.013$ | (Aubert, 2008bf) | $0.45 \pm 0.05 \pm 0.02$ | (Chen, 2005a) | $0.480 \pm 0.030$ |
| $\phi K_1(1270)^+$ | $0.46^{+0.12+0.06}_{-0.13-0.07}$ | (Aubert, 2008ad) | | | $0.46^{+0.12+0.06}_{-0.13-0.07}$ |
| $\phi K_2^*(1430)^+$ | $0.80^{+0.09}_{-0.10} \pm 0.03$ | (Aubert, 2008ad) | | | $0.80^{+0.09}_{-0.10} \pm 0.03$ |
| $\phi K_2^*(1430)^0$ | $0.901^{+0.046}_{-0.058} \pm 0.037$ | (Aubert, 2008bf) | | | $0.901^{+0.046}_{-0.058} \pm 0.037$ |
| $\rho^+\rho^-$ | $0.992 \pm 0.024^{+0.026}_{-0.013}$ | (Aubert, 2007b) | $0.941^{+0.034}_{-0.040} \pm 0.030$ | (Somov, 2006) | $0.977^{+0.028}_{-0.024}$ |
| $\rho^+\rho^0$ | $0.950 \pm 0.015 \pm 0.006$ | (Aubert, 2009p) | $0.95 \pm 0.11 \pm 0.02$ | (Zhang, 2003) | $0.950 \pm 0.016$ |
| $\rho^0\rho^0$ | $0.75^{+0.11}_{-0.14} \pm 0.04$ | (Aubert, 2008r) | | | $0.75^{+0.11}_{-0.14} \pm 0.04$ |
| $a_1^\pm a_1^\mp$ | $0.31 \pm 0.22 \pm 0.10$ | (Aubert, 2009ae) | | | $0.31 \pm 0.22 \pm 0.10$ |

The branching fractions and asymmetries are given in Table 17.4.5 and the hierarchy of measured branching fractions is shown in Fig. 17.4.9.

The decay to $\omega K^*$ is penguin dominated but the tree diagrams are more important for the other decays (Aubert (2006f) and Goldenzweig (2008)). The branching fraction hierarchy of the decays to $\omega K^*$ and $\omega\phi$ is a useful determination of the contribution of electro-weak penguins and so potentially helpful for the understanding of $\phi_2$. The $\omega K^*$ final state can also be used to look at branching fractions and $f_L$ in Vector-Tensor (VT) decays ($B \to \omega K_2^*(1430)$) and Scalar-Vector (SV) decays ($B \to \omega K^{*0}(1430)$) and compared to other VT decays such as $B \to \phi K_2^*(1430)$ (Aubert, 2009af).

Decays proceeding via electro-weak and gluonic $b \to d$ penguin diagrams have been measured in the decays $B \to \rho\gamma$ and $B^0 \to K^0\overline{K}^0$. The charmless decay $B^0 \to K^{*0}\overline{K}^{*0}$ proceeds through both electro-weak and gluonic $b \to d$ penguin loops to two vector particles (VV). The standard model suppressed decay $B^0 \to K^{*0}K^{*0}$ could appear via an intermediate heavy boson (Aubert (2008ah,ao, 2009k) and Chiang (2010)).

### 17.4.5.4 $B \to$ SP, SV, SS

The modes involving a $B$ meson decay to Pseudoscalar-Scalar (PS), Vector-Scalar (VS) and Scalar-Scalar (SS) are summarized in Table 17.4.6 with branching fractions plotted in Fig. 17.4.10.

**Table 17.4.5.** Charmless $B$ decays branching fractions $\mathcal{B}$ and $CP$ Asymmetries $A_{CP}$ for *BABAR* and Belle for mode Vector-Vector (VV) final states. The averages come from HFAG and may include measurements from other experiments such as CLEO (Amhis et al. (2012)).

| | *BABAR* results | | | Belle results | | | Averages | |
|---|---|---|---|---|---|---|---|---|
| Final state | $\mathcal{B}$ ($\times 10^{-6}$) | $A_{CP}$ | Ref. | $\mathcal{B}$ ($\times 10^{-6}$) | $A_{CP}$ | Ref. | $\mathcal{B}$ ($\times 10^{-6}$) | $A_{CP}$ |
| $K^{*+}K^{*-}$ | $< 2.0$ | | (Aubert, 2008ao) | | | | $< 2.0$ | |
| $K^{*+}\overline{K}^{*0}$ | $1.2 \pm 0.5 \pm 0.1$ | | (Aubert, 2009k) | | | | $1.2 \pm 0.5$ | |
| $K^{*0}K^{*0}$ | $< 0.41$ | | (Aubert, 2008y) | $< 0.2$ | | (Chiang, 2010) | $< 0.2$ | |
| $K^{*0}\overline{K}^{*0}$ | $1.28^{+0.35}_{-0.30} \pm 0.11$ | | (Aubert, 2008y) | $0.26^{+0.33+0.10}_{-0.29-0.08}$ | | (Chiang, 2010) | $0.81 \pm 0.23$ | |
| $K^{*+}\rho^-$ | $10.3 \pm 2.3 \pm 1.3$ | $0.21 \pm 0.15 \pm 0.02$ | (Lees, 2012l) | | | | $10.3 \pm 2.6$ | |
| $K^{*0}\rho^+$ | $9.6 \pm 1.7 \pm 1.5$ | $-0.01 \pm 0.16 \pm 0.02$ | (Aubert, 2006ab) | $8.9 \pm 1.7 \pm 1.2$ | | (Abe, 2005f) | $9.2 \pm 1.5$ | $-0.01 \pm 0.16$ |
| $K^{*0}\rho^0$ | $5.1 \pm 0.6^{+0.6}_{-0.8}$ | $-0.06 \pm 0.09 \pm 0.02$ | (Lees, 2012l) | $2.1^{+0.8+0.9}_{-0.7-0.5}$ | | (Kyeong, 2009) | $3.9 \pm 0.8$ | $-0.06 \pm 0.09$ |
| $K^{*+}\rho^0$ | $4.6 \pm 1.1 \pm 0.4$ | $0.31 \pm 0.13 \pm 0.03$ | (del Amo Sanchez, 2011g) | | | | $4.6 \pm 1.1$ | $0.31 \pm 0.13$ |
| $\omega K^{*+}$ | $< 7.4$ | $0.29 \pm 0.35 \pm 0.02$ | (Aubert, 2009af) | | | | $< 7.4$ | $0.29 \pm 0.35$ |
| $\omega K^{*0}$ | $2.2 \pm 0.6 \pm 0.2$ | $0.45 \pm 0.25 \pm 0.02$ | (Aubert, 2009af) | $1.8 \pm 0.7^{+0.3}_{-0.2}$ | | (Goldenzweig, 2008) | $2.0 \pm 0.5$ | $0.45 \pm 0.25$ |
| $\omega\omega$ | $< 4.0$ | | (Aubert, 2006f) | | | | $< 4.0$ | |
| $\omega\phi$ | $< 1.2$ | | (Aubert, 2006f) | | | | $< 1.2$ | |
| $\omega\rho^+$ | $15.9 \pm 1.6 \pm 1.4$ | $-0.20 \pm 0.09 \pm 0.02$ | (Aubert, 2009af) | | | | $15.9 \pm 2.1$ | $-0.20 \pm 0.09$ |
| $\omega\rho^0$ | $< 1.6$ | | (Aubert, 2009af) | | | | $< 1.6$ | |
| $\phi K^*(1410)^+$ | $< 4.3$ | | (Aubert, 2008ad) | | | | $< 4.3$ | |
| $\phi K^*(1680)^0$ | $< 3.5$ | | (Aubert, 2007ap) | | | | $< 3.5$ | |
| $\phi K^{*+}$ | $11.2 \pm 1.0 \pm 0.9$ | $0.00 \pm 0.09 \pm 0.04$ | (Aubert, 2007c) | $6.7^{+2.1+0.7}_{-1.9-1.0}$ | $-0.02 \pm 0.14 \pm 0.03$ | (Chen, 2003) | $10.0 \pm 1.1$ | $-0.01 \pm 0.08$ |
| $\phi K^{*0}$ | $9.7 \pm 0.5 \pm 0.6$ | $0.01 \pm 0.06 \pm 0.03$ | (Aubert, 2008bf) | $10.0^{+1.6+0.7}_{-1.5-0.8}$ | $0.02 \pm 0.09 \pm 0.02$ | (Chen, 2003) | $9.8 \pm 0.7$ | $0.01 \pm 0.05$ |
| $\phi\phi$ | $< 0.2$ | | (Aubert, 2008ay) | | | | $< 0.2$ | |
| $\phi\rho^+$ | $< 3.0$ | | (Aubert, 2008ay) | | | | $< 3.0$ | |
| $\phi\rho^0$ | $< 0.33$ | | (Aubert, 2008ay) | | | | $< 0.33$ | |
| $\rho^+\rho^-$ | $25.5 \pm 2.1^{+3.6}_{-3.9}$ | | (Aubert, 2007b) | $22.8 \pm 3.8^{+2.3}_{-2.6}$ | | (Somov, 2006) | $24.2^{+3.1}_{-3.2}$ | |
| $\rho^+\rho^0$ | $23.7 \pm 1.4 \pm 1.4$ | $-0.054 \pm 0.055 \pm 0.010$ | (Aubert, 2009p) | $31.7 \pm 7.1^{+3.8}_{-6.7}$ | $0.00 \pm 0.22 \pm 0.03$ | (Zhang, 2003) | $24.0^{+1.9}_{-2.0}$ | $-0.05 \pm 0.05$ |
| $\rho^0\rho^0$ | $0.92 \pm 0.32 \pm 0.14$ | | (Aubert, 2008r) | $0.4 \pm 0.4^{+0.2}_{-0.3}$ | | (Chiang, 2008) | $0.73^{+0.27}_{-0.28}$ | |

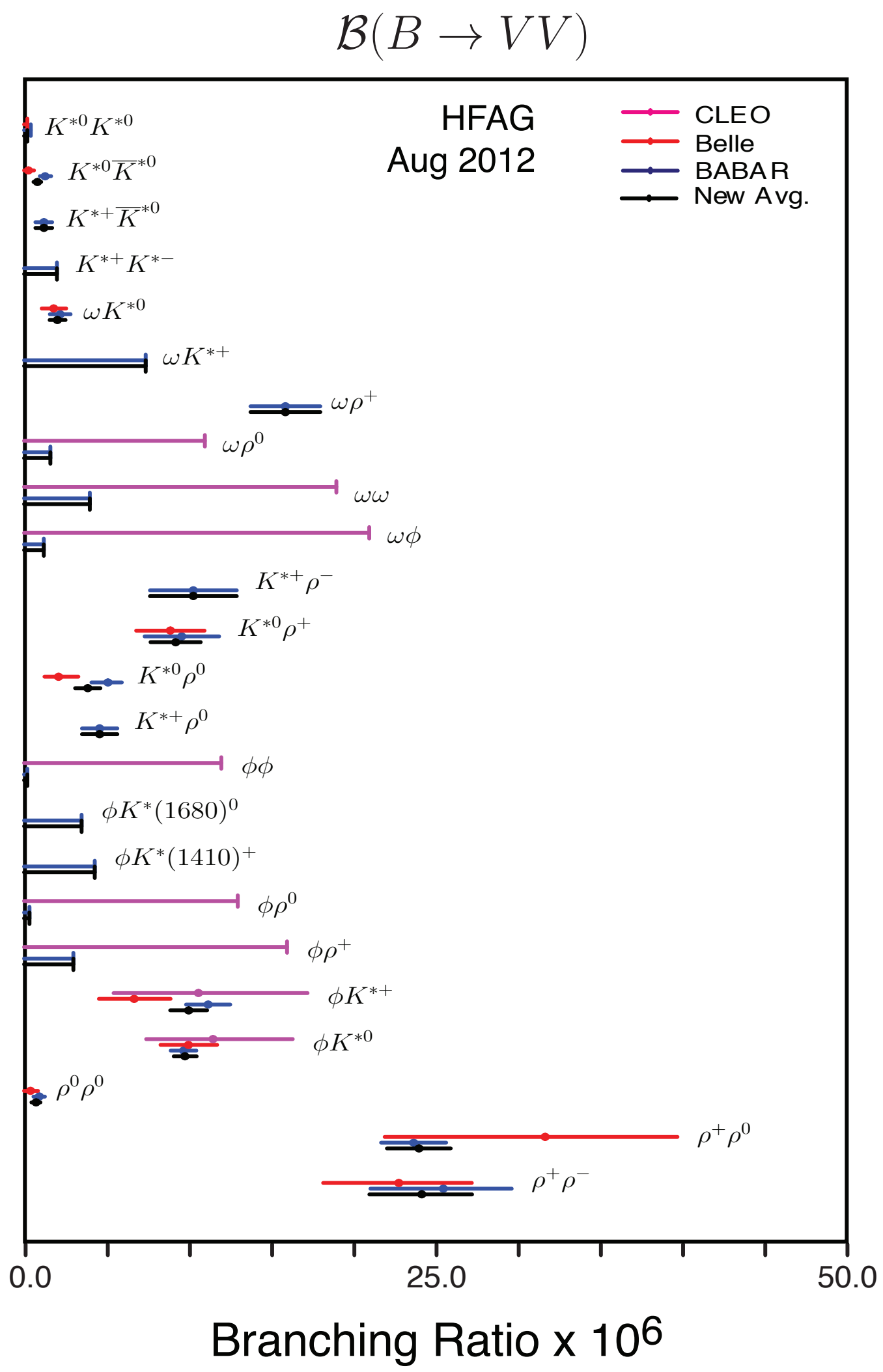

**Figure 17.4.9.** Summary of branching fraction measurements ($\times 10^{-6}$) and HFAG averages for Vector-Vector (VV) decays (Amhis et al. (2012)).

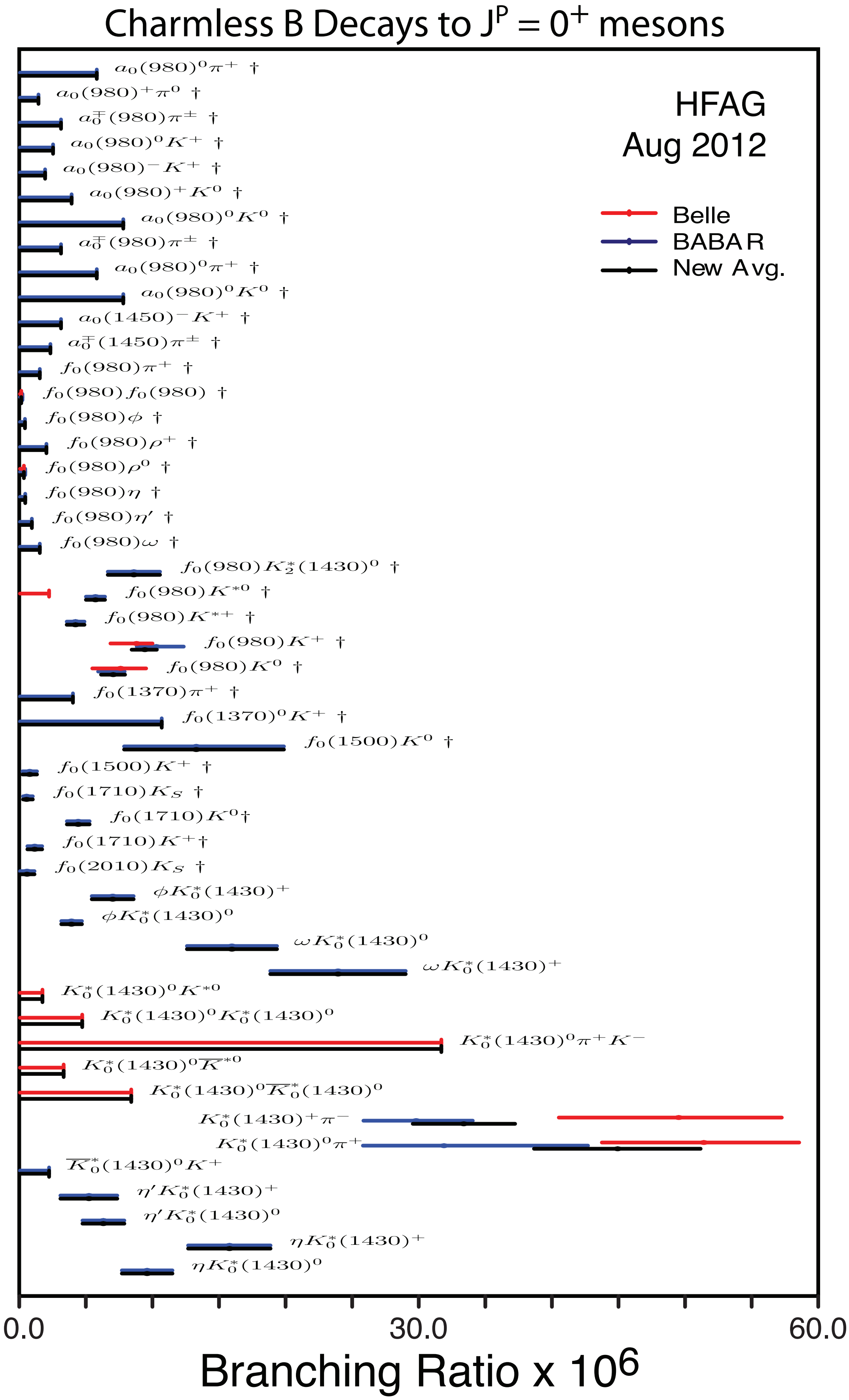

**Figure 17.4.10.** Summary of branching fraction measurements ($\times 10^{-6}$) and HFAG averages for $J^P = 0^+$ final states, including Scalar-Pseudoscalar (SP), Scalar-Vector (SV) and Scalar-Scalar (SS) decays (Amhis et al. (2012)).

**Table 17.4.6.** Charmless $B$ decays branching fractions $\mathcal{B}$ and $CP$ asymmetries $A_{CP}$ for BABAR and Belle for $J^P = 0^+$ final states, including Scalar-Pseudoscalar (SP), Scalar-Vector (SV) and Scalar-Scalar (SS) decays. The averages come from HFAG (Amhis et al. (2012)).

| | BABAR results | | | Belle results | | | Averages | |
|---|---|---|---|---|---|---|---|---|
| Final state | $\mathcal{B}$ ($\times 10^{-6}$) | $A_{CP}$ | Ref. | $\mathcal{B}$ ($\times 10^{-6}$) | $A_{CP}$ | Ref. | $\mathcal{B}$ ($\times 10^{-6}$) | $A_{CP}$ |
| $\overline{K}_0^*(1430)^0 K^+$ | $< 2.2$ | | (Aubert, 2007av) | | | | $< 2.2$ | |
| $K_0^*(1430)^+ \pi^-$ | $29.9^{+2.3}_{-1.7} \pm 3.6$ | $0.07 \pm 0.14 \pm 0.01$ | (Aubert, 2009av) | $49.7 \pm 3.8^{+6.8}_{-8.2}$ | | (Garmash, 2007) | $33.5^{+3.9}_{-3.8}$ | $0.10 \pm 0.07$ |
| $K_0^*(1430)^0 \pi^+ K^-$ | | | | $< 31.8$ | | (Chiang, 2010) | $< 31.8$ | |
| $K_0^*(1430)^0 \pi^+$ | $32.0 \pm 1.2^{+10.8}_{-6.0}$ | $0.032 \pm 0.035^{+0.034}_{-0.028}$ | (Aubert, 2008j) | $51.6 \pm 1.7^{+7.0}_{-7.5}$ | $0.076 \pm 0.038^{+0.028}_{-0.022}$ | (Garmash, 2006) | $45.1 \pm 6.3$ | $0.55 \pm 0.33$ |
| $\eta K_0^*(1430)^+$ | $15.8 \pm 2.2 \pm 2.2$ | $0.05 \pm 0.13 \pm 0.02$ | (Aubert, 2006l) | | | | $15.8 \pm 3.1$ | $0.05 \pm 0.13$ |
| $\eta K_0^*(1430)^0$ | $9.6 \pm 1.4 \pm 1.3$ | $0.06 \pm 0.13 \pm 0.02$ | (Aubert, 2006l) | | | | $9.6 \pm 1.9$ | $0.06 \pm 0.13$ |
| $\eta' K_0^*(1430)^+$ | $5.2 \pm 1.9 \pm 1.0$ | $0.06 \pm 0.20 \pm 0.02$ | (del Amo Sanchez, 2010h) | | | | $5.2 \pm 2.1$ | |
| $\eta' K_0^*(1430)^0$ | $6.3 \pm 1.3 \pm 0.9$ | $-0.19 \pm 0.17 \pm 0.02$ | (del Amo Sanchez, 2010h) | | | | $6.3 \pm 1.6$ | |
| $a_0(1450)^- K^+$ | $< 3.1$ | | (Aubert, 2007as) | | | | $< 3.1$ | |
| $a_0(980)^+ K^0$ | $< 3.9$ | | (Aubert, 2004x) | | | | $< 3.9$ | |
| $a_0(980)^+ \pi^0$ | $< 1.4$ | | (Aubert, 2008ax) | | | | $< 1.4$ | |
| $a_0(980)^- K^+$ | $< 1.9$ | | (Aubert, 2007as) | | | | $< 1.9$ | |
| $a_0(980)^0 K^+$ | $< 2.5$ | | (Aubert, 2004x) | | | | $< 2.5$ | |
| $a_0(980)^0 K^0$ | $< 7.8$ | | (Aubert, 2004x) | | | | $< 7.8$ | |
| $a_0(980)^0 \pi^+$ | $< 5.8$ | | (Aubert, 2004x) | | | | $< 5.8$ | |
| $a_0^{\mp}(1450)\pi^{\pm}$ | $< 2.3$ | | (Aubert, 2007as) | | | | $< 2.3$ | |
| $a_0^{\mp}(980)\pi^{\pm}$ | $< 3.1$ | | (Aubert, 2007as) | | | | $< 3.1$ | |
| $f_0(1370)\pi^+$ | $< 4.0$ | | (Aubert, 2009h) | | | | $< 4.0$ | |
| $f_0(1370)^0 K^+$ | $< 10.7$ | | (Aubert, 2005g) | | | | $< 10.7$ | |
| $f_0(1500) K^+$ | $0.74 \pm 0.18 \pm 0.52$ | | (Lees, 2012y) | | | | $0.74 \pm 0.55$ | |
| $f_0(1500) K^0$ | $13.3^{+5.8}_{-4.4} \pm 3.2$ | | (Lees, 2012y) | | | | $13.3^{+6.6}_{-5.4}$ | |
| $f_0(1710) K_S^0$ | $0.50^{+0.46}_{-0.24} \pm 0.11$ | | (Lees, 2012c) | | | | $0.5^{+0.5}_{-0.3}$ | |
| $f_0(1710) K^+$ | $1.12 \pm 0.25 \pm 0.50$ | | (Lees, 2012y) | | | | $1.12 \pm 0.56$ | |
| $f_0(2010) K_S^0$ | $0.54^{+0.21}_{-0.20} \pm 0.52$ | | (Lees, 2012c) | | | | $0.54 \pm 0.56$ | |
| $f_0(980) K^+$ | $10.3 \pm 0.5^{+2.0}_{-1.4}$ | $-0.106 \pm 0.050^{+0.036}_{-0.015}$ | (Aubert, 2008j) | $8.8 \pm 0.8^{+0.9}_{-1.8}$ | $-0.077 \pm 0.065^{+0.046}_{-0.026}$ | (Garmash, 2006) | $9.4^{+0.9}_{-1.0}$ | $-0.10^{+0.05}_{-0.04}$ |
| $f_0(980) K^0$ | $6.9 \pm 0.8 \pm 0.6$ | $-0.28 \pm 0.24 \pm 0.09$ | (Lees, 2012y) | $7.6 \pm 1.7^{+0.9}_{-1.3}$ | | (Garmash, 2007) | $7.0 \pm 0.9$ | |
| $f_0(980) K^{*+}$ | $4.2 \pm 0.6 \pm 0.3$ | $-0.15 \pm 0.12 \pm 0.03$ | (del Amo Sanchez, 2011g) | | | | $4.2 \pm 0.7$ | $-0.15 \pm 0.12$ |
| $f_0(980) K^{*0}$ | $5.7 \pm 0.6 \pm 0.4$ | $0.07 \pm 0.10 \pm 0.02$ | (Lees, 2012l) | $< 2.2$ | | (Kyeong, 2009) | $5.7 \pm 0.7$ | $0.07 \pm 0.10$ |
| $f_0(980)\eta$ | $< 0.4$ | | (Aubert, 2007as) | | | | $< 0.4$ | |
| $f_0(980)\eta'$ | $< 0.9$ | | (del Amo Sanchez, 2010h) | | | | $< 0.9$ | |
| $f_0(980)\omega$ | $< 1.5$ | | (Aubert, 2009af) | | | | $< 1.5$ | |
| $f_0(980)\phi$ | $< 0.38$ | | (Aubert, 2008ay) | | | | $< 0.38$ | |
| $f_0(980)\pi^+$ | $< 1.5$ | | (Aubert, 2009h) | | | | $< 1.5$ | |
| $K_0^*(1430)^0 K^{*0}$ | | | | $< 1.7$ | | (Chiang, 2010) | $< 1.7$ | |
| $K_0^*(1430)^0 \overline{K}^{*0}$ | | | | $< 3.3$ | | (Chiang, 2010) | $< 3.3$ | |
| $K_0^*(1430)^0 \rho^0$ | $27 \pm 4 \pm 2 \pm 3$ | | (Lees, 2012l) | | | | | |
| $K_0^*(1430)^+ \rho^-$ | $28 \pm 10 \pm 5 \pm 3$ | | (Lees, 2012l) | | | | | |
| $K_0^*(1430)^+ \omega$ | $24.0 \pm 2.6 \pm 4.4$ | $-0.10 \pm 0.09 \pm 0.02$ | (Aubert, 2009af) | | | | $24.0 \pm 5.1$ | $-0.10 \pm 0.09$ |
| $K_0^*(1430)^0 \omega$ | $16.0 \pm 1.6 \pm 3.0$ | $-0.07 \pm 0.09 \pm 0.02$ | (Aubert, 2009af) | | | | $16.0 \pm 3.4$ | $-0.07 \pm 0.09$ |
| $K_0^*(1430)^+ \phi$ | $7.0 \pm 1.3 \pm 0.9$ | $0.04 \pm 0.15 \pm 0.04$ | (Aubert, 2008ad) | | | | $7.0 \pm 1.6$ | $0.04 \pm 0.15$ |
| $K_0^*(1430)^0 \phi$ | $3.9 \pm 0.5 \pm 0.6$ | $0.20 \pm 0.14 \pm 0.06$ | (Aubert, 2008bf) | | | | $3.9 \pm 0.8$ | $0.20 \pm 0.15$ |
| $f_0(980)\rho^+$ | $< 2.0$ | | (Aubert, 2009p) | | | | $< 2.0$ | |
| $f_0(980)\rho^0$ | $< 0.40$ | | (Aubert, 2008r) | $< 0.3$ | | (Chiang, 2008) | $< 0.3$ | |
| $f_0(980) f_0(980)$ | $< 0.19$ | | (Aubert, 2008r) | $< 0.1$ | | (Chiang, 2008) | $< 0.1$ | |
| $f_0(980) K_2^*(1430)^0$ | $8.6 \pm 1.7 \pm 1.0$ | | (Lees, 2012l) | | | (Kyeong, 2009) | $8.6 \pm 2.0$ | |
| $K_0^*(1430)^0 K_0^*(1430)^0$ | | | | $< 4.7$ | | (Chiang, 2010) | $< 4.7$ | |
| $K_0^*(1430)^0 \overline{K}_0^*(1430)^0$ | | | | $< 8.4$ | | (Chiang, 2010) | $< 8.4$ | |

The exact structure of scalar mesons is not clear with various models proposed such as two-quark and four-quark states with potential contributions from glueballs and molecules (compare with the search for exotic states in Chapter 18.3). The experimental measurement of scalars is also complicated as they are often quite broad, decay to pions (and so can be faked by combining the relatively large number of unrelated pions), and have an angular decay structure that is very similar to the non-resonant background. The $a_0(980)$ (along with the $a_1(1260)$ and $b_1$) is an ideal candidate for a four-quark structure as it lies near the $K\bar{K}$ threshold and so could be a $q\bar{q}$ state with a $K\bar{K}$ admixture. As an example, the decay $B^+ \to a_0^+\pi^0$ can differentiate between two- and four-quark models as the two-body branching fraction could be as high as $2 \times 10^{-7}$ while the four-quark model is an order of magnitude lower. The branching fraction however is only measured to a precision of $\mathcal{B}(B^+ \to a_0^+\pi^0) < 1.4 \times 10^{-6}$ (Aubert, 2008ax). The current experimental upper limits on $\mathcal{B}(B^0 \to a_0^\pm\pi^\mp)$ and $\mathcal{B}(B^0 \to a_0^- K^+)$ are $3.1\times10^{-6}$ and $1.9\times10^{-6}$, respectively (Aubert, 2007as). Vector-current considerations and G-parity conservation suppress the color-allowed electroweak tree decay, leading to the small predicted branching fractions. G-parity $G = Ce^{i\pi I_2}$ is a product of charge conjugation $C$ and a rotation about the second Isospin axis $I_2$; it is expected to be conserved by strong interactions (as the strong force conserves both $C$ and Isospin) but not in electro-weak interactions.

The $a_0$, including the $a_0(980)$ and $a_0(1450)$, decays to $\eta\pi$ but the exact branching fraction is not well known (roughly 85%). The decay $B^\pm \to a_0\pi^\pm$ has the benefit of being self-tagging as the pion charge identifies the $B$ meson flavor (Aubert, 2004x).

The averaged decay rates for $B^+ \to f_0(980)K^+$ and $B^0 \to f_0(980)K^0$ have been measured to be $9.4\times10^{-6}$ (Aubert, 2008j), (Garmash, 2006) and $7.0 \times 10^{-6}$ (Aubert, 2009av) (Garmash, 2007), respectively. These are compatible with expectations that the $b \to ss\bar{s}$ penguin dominates over the $b \to su\bar{u}$ penguin.

The SV mode $\phi K_0^*(1430)^0$ has been measured as part of a time-dependent and time-integrated analysis of $B \to \phi K_S^0\pi^0$ and $B \to \phi K^\pm\pi^\mp$ decays (Aubert, 2008bf), which also include VV and VT decays (see Tables 17.4.5 and 17.4.8). The decay $B \to \omega f_0(980)$ naturally forms part of a search for $\omega\rho$.

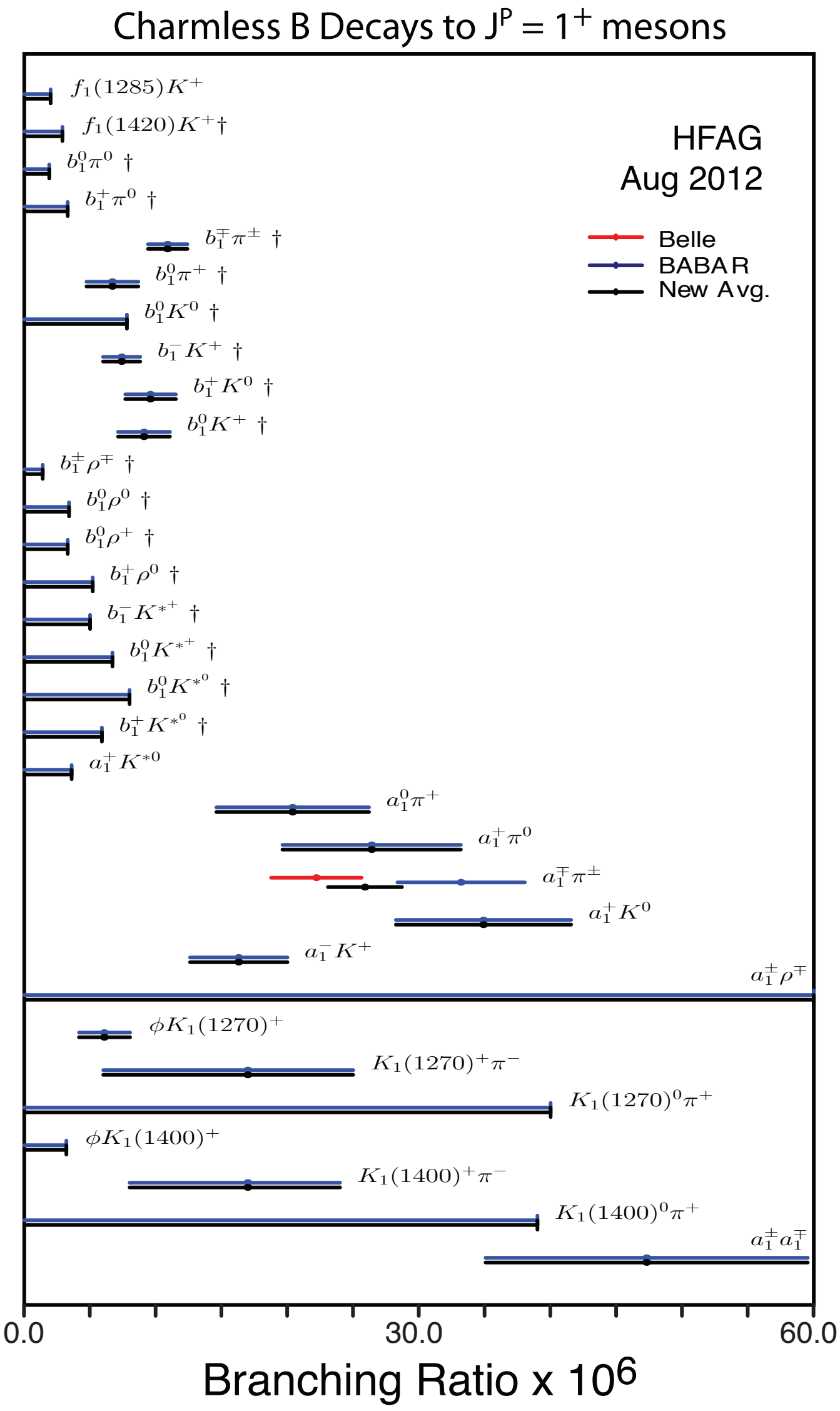

**Figure 17.4.11.** Summary of branching fraction measurements ($\times10^{-6}$) and HFAG averages for $J^P = 1^+$ final states, including Axial-Pseudovector (AP), Axial-Vector (AV) and Axial-Axial (AA) decays (Amhis et al. (2012)).

### 17.4.5.5 $B \to$ AP, AV, AA

Figure 17.4.11 and Table 17.4.7 show the reported results from Belle and *BABAR*, and their HFAG averages, for modes involving Axial-Pseudovector (AP), Axial-Vector (AV) and Axial-Axial (AA) decays.

**Table 17.4.7.** Charmless $B$ decays branching fractions $\mathcal{B}$ and $CP$ asymmetries $A_{CP}$ for BABAR and Belle for $J^P = 1^+$ final states, including Axial-Pseudovector (AP), Axial-Vector (AV) and Axial-Axial (AA) decays. The averages come from HFAG (Amhis et al. (2012)).

| | BABAR results | | | Belle results | | | Averages | |
|---|---|---|---|---|---|---|---|---|
| Final state | $\mathcal{B}$ ($\times 10^{-6}$) | $A_{CP}$ | Ref. | $\mathcal{B}$ ($\times 10^{-6}$) | $A_{CP}$ | Ref. | $\mathcal{B}$ ($\times 10^{-6}$) | $A_{CP}$ |
| $K_1(1270)^+\pi^-$ | $17^{+8}_{-11}$ | | (Aubert, 2010d) | | | | $17^{+8}_{-11}$ | |
| $K_1(1270)^0\pi^+$ | $< 40$ | | (Aubert, 2010d) | | | | $< 40$ | |
| $K_1(1400)^+\pi^-$ | $17^{+7}_{-9}$ | | (Aubert, 2010d) | | | | $17^{+7}_{-9}$ | |
| $K_1(1400)^0\pi^+$ | $< 39$ | | (Aubert, 2010d) | | | | $< 39$ | |
| $a_1^+K^0$ | $34.9 \pm 5.0 \pm 4.4$ | $0.12 \pm 0.11 \pm 0.02$ | (Aubert, 2008ae) | | | | $34.9 \pm 6.7$ | $0.12 \pm 0.11 \pm 0.02$ |
| $a_1^+\pi^0$ | $26.4 \pm 5.4 \pm 4.1$ | | (Aubert, 2007i) | | | | $26.4 \pm 6.8$ | |
| $a_1^-K^+$ | $16.3 \pm 2.9 \pm 2.3$ | $-0.16 \pm 0.12 \pm 0.01$ | (Aubert, 2008ae) | | | | $16.3 \pm 3.7$ | $-0.16 \pm 0.12 \pm 0.01$ |
| $a_1^0\pi^+$ | $20.4 \pm 4.7 \pm 3.4$ | | (Aubert, 2007i) | | | | $20.4 \pm 5.8$ | |
| $a_1^\mp\pi^\pm$ | $33.2 \pm 3.8 \pm 3.0$ | | (Aubert, 2006aj) | | | | $33.2 \pm 4.8$ | |
| $b_1^+K^0$ | $9.6 \pm 1.7 \pm 0.9$ | $-0.03 \pm 0.15 \pm 0.02$ | (Aubert, 2008aj) | | | | $9.6 \pm 1.9$ | $-0.03 \pm 0.15$ |
| $b_1^+\pi^0$ | $< 3.3$ | | (Aubert, 2008aj) | | | | $< 3.3$ | |
| $b_1^-K^+$ | $7.4 \pm 1.0 \pm 1.0$ | $0.07 \pm 0.12 \pm 0.02$ | (Aubert, 2007aj) | | | | $7.4 \pm 1.4$ | $0.07 \pm 0.12 \pm 0.02$ |
| $b_1^0K^+$ | $9.1 \pm 1.7 \pm 1.0$ | $-0.46 \pm 0.20 \pm 0.02$ | (Aubert, 2007aj) | | | | $9.1 \pm 2.0$ | $-0.46 \pm 0.20 \pm 0.02$ |
| $b_1^0K^0$ | $< 7.8$ | | (Aubert, 2008aj) | | | | $< 7.8$ | |
| $b_1^0\pi^+$ | $6.7 \pm 1.7 \pm 1.0$ | $0.05 \pm 0.16 \pm 0.02$ | (Aubert, 2007aj) | | | | $6.7 \pm 2.0$ | $0.05 \pm 0.16 \pm 0.02$ |
| $b_1^0\pi^0$ | $< 1.9$ | | (Aubert, 2008aj) | | | | $< 1.9$ | |
| $b_1^\mp\pi^\pm$ | $10.9 \pm 1.2 \pm 0.9$ | $-0.05 \pm 0.10 \pm 0.02$ | (Aubert, 2007aj) | | | | $10.9 \pm 1.5$ | $-0.05 \pm 0.10 \pm 0.02$ |
| $b_1^\pm\rho^\mp$ | $< 1.4$ | | (Aubert, 2009ak) | | | | $< 1.4$ | |
| $f_1(1285)K^+$ | $< 2.0$ | | (Aubert, 2008bb) | | | | $< 2.0$ | |
| $f_1(1420)K^+$ | $< 2.9$ | | (Aubert, 2008bb) | | | | $< 2.9$ | |
| $\phi K_1(1270)^+$ | $6.1 \pm 1.6 \pm 1.1$ | $0.15 \pm 0.19 \pm 0.05$ | (Aubert, 2008ad) | | | | $6.1 \pm 1.9$ | $0.15 \pm 0.20$ |
| $\phi K_1(1400)^+$ | $< 3.2$ | | (Aubert, 2008ad) | | | | $< 3.2$ | |
| $a_1^+K^{*0}$ | $< 3.6$ | | (del Amo Sanchez, 2010l) | | | | $< 3.6$ | |
| $a_1^\pm\rho^\mp$ | $< 61$ | | (Aubert, 2006as) | | | | $< 61$ | |
| $b_1^+K^{*0}$ | $< 5.9$ | | (Aubert, 2009ak) | | | | $< 5.9$ | |
| $b_1^-K^{*+}$ | $< 5.0$ | | (Aubert, 2009ak) | | | | $< 5.0$ | |
| $b_1^0K^{*+}$ | $< 6.7$ | | (Aubert, 2009ak) | | | | $< 6.7$ | |
| $b_1^0K^{*0}$ | $< 8.0$ | | (Aubert, 2009ak) | | | | $< 8.0$ | |
| $b_1^0\rho^+$ | $< 3.3$ | | (Aubert, 2009ak) | | | | $< 3.3$ | |
| $b_1^0\rho^0$ | $< 3.4$ | | (Aubert, 2009ak) | | | | $< 3.4$ | |
| $b_1^+\rho^0$ | $< 5.2$ | | (Aubert, 2009ak) | | | | $< 5.2$ | |
| $a_1^\pm a_1^\mp$ | $47.3 \pm 10.5 \pm 6.3$ | | (Aubert, 2009ae) | | | | $47.3 \pm 12.2$ | |

The $b_1$ is the $I^G = 1^+$ member of the $J^{PC} = 1^{+-}$, ${}^1P_1$ nonet while the $a_1(1260)$ is the $I^G = I^-$ state in the $J^{PC} = 1^{++}$, ${}^3P_1$ nonet. The decays that happen via a tree diagram favor final states with a pion due to Cabibbo-favored coupling ($B^+ \to b_1^0\pi^+$, $B^0 \to b_1^-\pi^+$) while penguin loop decays favor the kaon final states ($B^+ \to b_1^0K^+$, $B^0 \to b_1^-K^+$). The even G-parity of the $b_1$ means only amplitudes in which the $b_1$ contains the spectator quark from the $B$ meson are allowed (apart from isospin-breaking and radiative correction effects). This is because the weak current has a G-parity even vector part and a G-parity odd axial-vector part. Neither part can produce a G-parity odd scalar meson such as the $a_1^0(1260)$. The $W^+$ is constrained to decay to states of even G-parity. As a result, the decay $B^0 \to b_1^+\pi^-$ is suppressed with respect to $B^0 \to b_1^-\pi^+$. The $B^0 \to b_1^-K^+$ decays can be used to measure $A_{CP}$ while $B^0 \to b_1\pi^\pm\pi^\mp$ can also measure $C$ and *CP*-conserving $\Delta C$ (Aubert, 2007aj, 2008aj). The dominant decay of the $b_1$ is through $\omega\pi$.

$B$ decays involving an $a_1(1260)$ are similarly of interest to the $b_1$ but with the added distinction that decays to $a_1(1260)$ with a $\pi^+$ proceed via a $b \to u\overline{u}d$ transition and the angle $\phi_2$ can be measured through the time-dependent decay rate asymmetry caused by interference between the direct decay and the decay after $B\overline{B}$ mixing. The branching fraction, when combined together with decays of the $a_1(1260)$ and $K_1$ can be used to differentiate between QCD and naïve factorization model predictions for branching fractions and branching fraction ratios, as well as $B \to a_1(1260)$ transition form factors calculations. These decays can also be an important background to other $\phi_2$ measurements, such as $\rho\pi$ and $\rho\rho$. The measurements can be combined with $SU(3)$ symmetry arguments to place bounds on the deviation $\Delta\phi_2$ of the measured $\phi_2$ from the true value. The $a_1(1260)$ decays predominantly to $\pi\pi\pi$ via intermediate states involving a vector P-wave $\rho$ or scalar S-wave $\sigma$ but most analyses assume a pure $\rho\pi$ intermediate decay.

The branching fractions $\mathcal{B}(B^0 \to b_1^-\pi^+)$ are expected to be much greater than $\mathcal{B}(B^0 \to b_1^+\pi^-)$ and that of $\mathcal{B}(B^0 \to a_1^+(1260)\pi^-)$ to be much greater than $\mathcal{B}(B^0 \to a_1^-(1260)\pi^+)$ and this has been confirmed (Aubert, 2006aj, 2007aj). The branching fractions for charged and neutral decays $B \to b_1K$ and $B \to b_1\pi$ are also in line with expectations (Aubert, 2006am, 2007aj, 2008aj). $A_{CP}$ has also been successfully measured in $B^+ \to b_1^+K^0$, $B^0 \to b_1^-K^+$, $B^+ \to b_1^0K^+$, $B^+ \to b_1^0\pi^+$, $B^0 \to b_1^\pm\pi^\mp$ and is compatible with zero (see Table 17.4.11).

For the $B \to a_1(1260)K$ and $B \to a_1(1260)\pi$ decays, both the neutral $B^0$ decays (Aubert, 2006aj, 2008ae) and the charged $B^+$ modes (Aubert, 2007i, 2008ae) have been measured as well as the asymmetries $A_{CP}$ and S in $B^+ \to a_1^+(1260)\ K_S^0$ and $B^0 \to a_1^-(1260)\ K^+$ (Aubert, 2008ae). There is strong evidence for $B^+ \to a_1^+(1260)\pi^0$ and $B^+ \to a_1^0(1260)\pi^\pm$ (Aubert, 2007i). The neutral decay $B^0 \to a_1^\pm(1260)\pi^\mp$ has been observed (Aubert, 2006aj) and a separate paper later measured $A_{CP}$, the mixing induced *CP* asymmetry, and the direct *CP* asymmetry (Aubert, 2007ae); as a result the angle $\phi_2$ was extracted (see Chapter 17.7). Belle have recently published their results and report the first evidence for mixing-induced *CP* violation in $B^0 \to a_1^\pm(1260)\pi^\mp$ (Dalseno, 2012).

The $B$ meson decay $B \to K_1\pi$, which changes the strangeness by one unit $\Delta S = 1$, is sensitive to the presence of penguin amplitudes because its CKM couplings are larger than the corresponding $\Delta S = 0$ penguin amplitudes. Therefore, measurements of the decay rate for $\Delta S = 1$ transitions sharing the same $SU(3)$ flavor multiplet as $a_1(1260)$ can be used to put constraints on $\phi_2$ (Aubert, 2010d). This is similar to the $SU(3)$-based approach to measuring $\phi_2$ in $\pi^+\pi^-$, $\rho^\pm\pi^\mp$ and $\rho^+\rho^-$ channels. The decay rate to $K_{1A}\pi$ (where $K_{1A}$ is the $SU(3)$ partner of the $a_1(1260)$ and a nearly equal admixture of $K_1(1270)$ and $K_1(1400)$ with the quantum numbers $I^{JP} = 1/2^{1+}$) can be derived from the decay rates to $K_1(1270)\pi$ and $K_1(1400)\pi$. The $K_1$ is reconstructed through its predominant decay to $K\pi\pi$ final states.

There are a number of results for the branching fractions of $B$ meson decays to Axial-Vector (AV) and Axial-Axial (AA) final states. Decays to a $b_1$ and a vector meson ($\rho$ or $K^*$) have been searched for as a possible measurement of longitudinal polarization $f_L$, but only upper limits on the branching fractions $\mathcal{B}(B \to b_1K^*) \leq 8 \times 10^{-6}$ and $\mathcal{B}(B \to b_1\rho) \leq\ (3.3 - 5.2) \times 10^{-6}$ have been measured (Aubert, 2009ak). $B^0 \to a_1^\pm(1260)\rho^\mp$ has also been searched for as it is both a background to $\phi_2$ measurements in $B \to \rho\rho$ and a possible place to measure $\phi_2$ itself. An upper limit of $< 61 \times 10^{-6}$ has been obtained (Aubert, 2006as). However this was only performed with $100\,\mathrm{fb}^{-1}$.

The $B^+ \to \phi K_1(1270)^+$, $B^+ \to \phi K_1(1400)^+$, and $B^+ \to a_1^+(1260)K^{*0}$ modes have been searched for (Aubert, 2008ad; del Amo Sanchez, 2010l) and $f_L$ in $B^+ \to \phi K_1(1270)^+$ has been measured.

AA modes such as $a_1^+(1260)\ a_1^-(1260)$, $a_1^+(1260)\ a_1^0(1260)$, $a_1^+(1260)\ b_1^-$, and $a_1^+(1260)\ b_1^0$ should have branching fractions in the range $(20-40)\times10^{-6}$. Although all the branching fractions have been measured, only $\mathcal{B}(B^0 \to a_1^+(1260)a_1^-(1260)) = (47.3 \pm 10.5 \pm 6.3) \times 10^{-6}$ has been observed (Aubert, 2009ae).

#### 17.4.5.6 $B \to$ VT, TP

Table 17.4.8 summarizes the reported branching fractions $\mathcal{B}$ and $A_{CP}$ asymmetries from Belle and *BABAR* and their HFAG averages for Tensor-Pseudoscalar (TP) and Vector-Tensor (VT) states. The hierarchy of branching fractions is shown in Figure 17.4.12. There are as yet very few predictions for these modes.

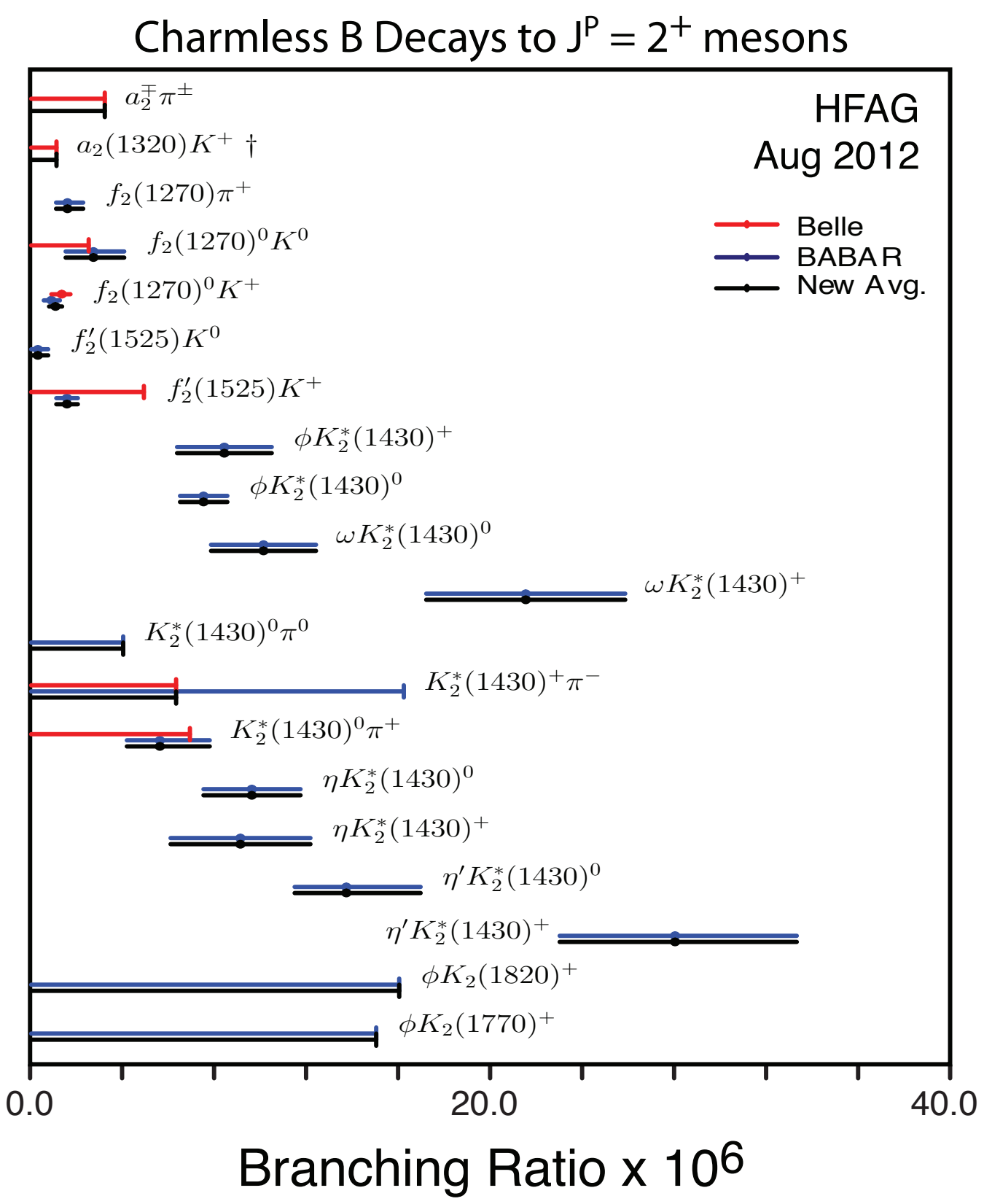

**Figure 17.4.12.** Summary of branching fraction measurements ($\times 10^{-6}$) and HFAG averages for $J^P = 2^+$ final states, including Tensor-Pseudoscalar (TP) and Tensor-Vector (TV) states (Amhis et al. (2012)).

**Table 17.4.8.** Charmless $B$ decays branching fractions $\mathcal{B}$ and $CP$ asymmetries $A_{CP}$ for BABAR and Belle for $J^P = 2^+$ final states, including Tensor-Pseudoscalar (TP) and Tensor-Vector (TV) states. The averages come from HFAG (Amhis et al. (2012)).

| | BABAR results | | | Belle results | | | Averages | |
|---|---|---|---|---|---|---|---|---|
| Final state | $\mathcal{B}$ $(\times 10^{-6})$ | $A_{CP}$ | Ref. | $\mathcal{B}$ $(\times 10^{-6})$ | $A_{CP}$ | Ref. | $\mathcal{B}$ $(\times 10^{-6})$ | $A_{CP}$ |
| $K_2^*(1430)^+\pi^-$ | $< 16.2$ | | (Aubert, 2008g) | $< 6.3$ | | (Garmash, 2007) | $< 6.3$ | |
| $K_2^*(1430)^0\pi^+$ | $5.6 \pm 1.2^{+1.8}_{-0.8}$ | $0.05 \pm 0.23^{+0.18}_{-0.08}$ | (Aubert, 2008j) | $< 6.9$ | | (Garmash, 2005) | $5.6^{+2.2}_{-1.4}$ | $0.05^{+0.29}_{-0.24}$ |
| $K_2^*(1430)^0\pi^0$ | $< 4.0$ | | (Aubert, 2008g) | | | | $< 4.0$ | |
| $\eta K_2^*(1430)^+$ | $9.1 \pm 2.7 \pm 1.4$ | $-0.45 \pm 0.30 \pm 0.02$ | (Aubert, 2006l) | | | | $9.1 \pm 3.0$ | $-0.45 \pm 0.30$ |
| $\eta K_2^*(1430)^0$ | $9.6 \pm 1.8 \pm 1.1$ | $-0.07 \pm 0.19 \pm 0.02$ | (Aubert, 2006l) | | | | $9.6 \pm 2.1$ | $-0.07 \pm 0.19$ |
| $\eta' K_2^*(1430)^+$ | $28.0^{+4.6}_{-4.3} \pm 2.6$ | $0.15 \pm 0.13 \pm 0.02$ | (del Amo Sanchez, 2010h) | | | | $28.0^{+5.3}_{-5.0}$ | |
| $\eta' K_2^*(1430)^0$ | $13.7^{+3.0}_{-1.9} \pm 1.2$ | $0.14 \pm 0.18 \pm 0.02$ | (del Amo Sanchez, 2010h) | | | | $13.7^{+3.2}_{-2.2}$ | |
| $a_2(1320)K^+$ | | | | $< 1.1$ | | (Garmash, 2005) | $< 1.1$ | |
| $f_2(1270)\pi^+$ | $1.57 \pm 0.42^{+0.55}_{-0.25}$ | $0.41 \pm 0.25^{+0.18}_{-0.15}$ | (Aubert, 2009h) | | | | $1.57^{+0.69}_{-0.49}$ | $0.41^{+0.31}_{-0.29}$ |
| $f_2(1270)^0K^+$ | $0.88 \pm 0.26^{+0.26}_{-0.21}$ | | (Aubert, 2008j) | $1.33 \pm 0.30^{+0.23}_{-0.34}$ | | (Garmash, 2006) | $1.06^{+0.28}_{-0.29}$ | |
| $f_2(1270)^0K^0$ | $2.7^{+1.0}_{-0.8} \pm 0.9$ | | (Aubert, 2009av) | $< 2.5$ | | (Garmash, 2007) | $2.7^{+1.3}_{-1.2}$ | |
| $f_2'(1525)K^+$ | $1.56 \pm 0.36 \pm 0.30$ | $0.14 \pm 0.10 \pm 0.04$ | (Lees, 2012y) | $< 4.9$ | | (Garmash, 2005) | $1.56 \pm 0.47$ | $0.14 \pm 0.11$ |
| $f_2'(1525)K^0$ | $0.29^{+0.27}_{-0.18} \pm 0.36$ | | (Lees, 2012y) | | | | $0.29^{+0.45}_{-0.40}$ | |
| $\omega K_2^*(1430)^+$ | $21.5 \pm 3.6 \pm 2.4$ | $0.14 \pm 0.15 \pm 0.02$ | (Aubert, 2009af) | | | | $21.5 \pm 4.3$ | $0.14 \pm 0.15$ |
| $\omega K_2^*(1430)^0$ | $10.1 \pm 2.0 \pm 1.1$ | $0.37 \pm 0.17 \pm 0.02$ | (Aubert, 2009af) | | | | $10.1 \pm 2.3$ | $0.37 \pm 0.17$ |
| $\phi K_2(1770)^+$ | $< 15$ | | (Aubert, 2008ad) | | | | $< 15$ | |
| $\phi K_2(1820)^+$ | $< 16$ | | (Aubert, 2008ad) | | | | $< 16$ | |
| $\phi K_2^*(1430)^+$ | $8.4 \pm 1.8 \pm 1.0$ | $-0.23 \pm 0.19 \pm 0.06$ | (Aubert, 2008ad) | | | | $8.4 \pm 2.1$ | $-0.23 \pm 0.20$ |
| $\phi K_2^*(1430)^0$ | $7.5 \pm 0.9 \pm 0.5$ | $-0.08 \pm 0.12 \pm 0.05$ | (Aubert, 2008bf) | | | | $7.5 \pm 1.0$ | $-0.08 \pm 0.13$ |

The angular distributions for the VT final states is given in Eq. 12.2.14. The longitudinal polarization $f_L$ for the VT mode $\phi K_2^*(1430)$ is close to $0.8-0.9$ (Aubert, 2008ad,bf) but there is a large transverse component in the VA mode $\phi K_1(1270)^+$ with $f_L \sim 0.46$ (Aubert, 2008ad). This lower value of $f_L$ is also seen in $\omega K_2^*(1430)$ (Aubert, 2009af).

Table 17.4.9 itemizes a few measurements that have been a by-product of the analyses described above. In a number of cases, the non-resonant component of $B$ meson decays has been measured, primarily by Belle (Chiang, 2008, 2010; Kyeong, 2009). *BABAR* has extended their analysis of $B \to \phi K^*$ to include the higher mass and higher spin resonances $K^*(1680)^0$, $K_3^*(1780)^0$, and $K_4^*(2045)^0$ (Aubert, 2007ap). Rather than look at individual modes, the partial branching fractions of the inclusive charmless decays $B \to K^+\ X$, $B \to K^0\ X$, and $B \to \pi^+\ X$ have been measured. The inclusive branching fraction of $B$ mesons to charmless final states is about 2%. Here $X$ represents any accessible final state above the endpoint for $B$ meson decays to charmed mesons and the branching fractions and $A_{CP}$ are reported for a restricted range of $K$ and $\pi$ momentum range.

#### 17.4.5.7 $A_{CP}$ summary

A subset of the most precise $A_{CP}$ measurements currently available are shown graphically in Fig. 17.4.13. Figures 17.4.14, 17.4.15, and 17.4.16 show the $A_{CP}$ $CP$ asymmetries for kaonic modes, separated into final states with a kaon or pion (both quasi-two-body and three-body), final states with an $\eta$ or $\phi$, and final states with an $\rho$, $\omega$, $f$, $a_1$, or $b_1$, respectively.

### 17.4.6 Dalitz experimental techniques

A quasi-two-body approach to extracting CKM parameters is not ideal as these modes often interfere with other resonances as well as non-resonant decays to the same final state. As a result, quasi-two-body measurements have an unknown uncertainty in their reported results that requires careful consideration. In principle, these effects can be taken into account by a Dalitz Plot (also known as a Dalitz Plane) analysis. The major advantage to the Dalitz Plot is that it gives access to the phases as well as the magnitudes of the resonances. Since the weak phase changes sign under $CP$ but the strong phase does not, the weak and strong phase components can be extracted by subtracting or adding together the $B$ meson flavor-tagged Dalitz Plots. In some Dalitz Plots, the weak phase can often be directly interpreted as one of the Wolfenstein angles *e.g.* Dalseno (2009). The mathematical formalism for a Dalitz Plot analysis is given in Chapter 13. In this section we consider the experimental problems in its implementation.

The extension of quasi-two-body charmless decays to three-body charmless decays brings with it greater complexity but provides a deeper understanding of the decays and their $CP$ properties. As the integrated luminosity increases, the analyses have started with inclusive measurements of branching fractions and charge asymmetries, integrated over the three-body phase space (*e.g.* $B \to \pi\pi\pi$). This has been followed by exploring intermediate states ignoring interference (*e.g.* $B \to \rho\pi$) before finally performing a full Dalitz Plot analysis taking into account interference between all intermediate resonance states. And finally, time-dependent asymmetries can be extracted from individual resonances. The choice is dictated by the luminosity, expected signal and background, and the understanding of the intermediate resonances (such as the presence or absence of poorly known states such as $\sigma/\kappa$, and higher mass $f_0$ and $K^*$).

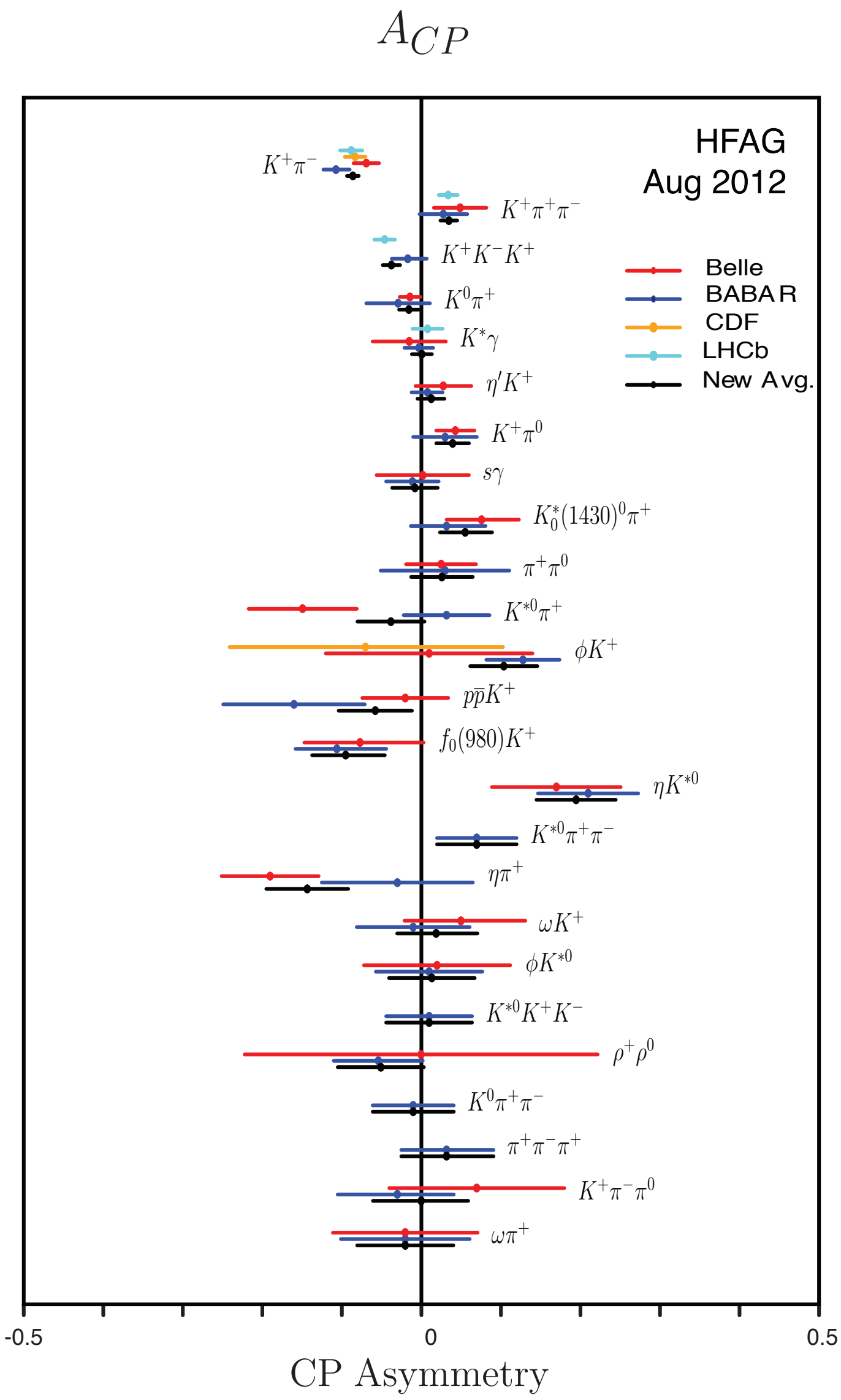

**Figure 17.4.13.** Summary of the most precise $A_{CP}$ measurements (Amhis et al. (2012)).

The Dalitz Plots of $B$ meson decays are usually interpreted in the scattering matrix (S-matrix) or isobar model (see Section 13.2.1). If a more detailed understand-ing of the amplitude properties is required, for instance the spin, the scattering amplitude can be expressed in terms of partial-wave amplitudes. The drawback of the S-matrix formalism is that it is not unitary and as a result the sum of the amplitudes of the resonances in the Dalitz Plot can be greater or less than the inclusive Dalitz Plot amplitude depending on whether the overall interference is constructive or destructive. The individual branching fractions are therefore often reported as fit fractions (FF), defined as the integral of a single amplitude squared divided by the coherent matrix element squared for the whole Dalitz Plot (Section 13.4.1). An alternative parameterization uses the K-matrix formalism which is unitary by construction but has a drawback that the masses and widths can be different to the S-matrix results. The K-matrix formalism is more commonly used in Dalitz Plot analyses of $D$ meson decays (section 13.2.2). This is because many of the resonances in the $D$ meson Dalitz Plot contain a large number of events and the S-matrix approximation of a Breit-Wigner or similar shape for the decay of the resonance is no longer adequate, especially when the resonances overlap in the Dalitz Plot.

**Table 17.4.9.** Charmless $B$ decays branching fractions $\mathcal{B}$ and $CP$ asymmetries $A_{CP}$ for BABAR and Belle for non-resonant decays and other unclassified modes. The averages come from HFAG and may include measurements from other experiments such as CLEO, CDF and DØ (Amhis et al. (2012)).

| | BABAR results | | | Belle results | | | Averages | |
|---|---|---|---|---|---|---|---|---|
| Final state | $\mathcal{B}$ $(\times 10^{-6})$ | $A_{CP}$ | Ref. | $\mathcal{B}$ $(\times 10^{-6})$ | $A_{CP}$ | Ref. | $\mathcal{B}$ $(\times 10^{-6})$ | $A_{CP}$ |
| $K^+X$ | $< 187$ | $0.57 \pm 0.24 \pm 0.05$ | (del Amo Sanchez, 2011c) | | | | | |
| $K^0X$ | $195^{+51}_{-45} \pm 50$ | | (del Amo Sanchez, 2011c) | | | | | |
| $\pi^+X$ | $372^{+50}_{-47} \pm 59$ | $0.10 \pm 0.16 \pm 0.05$ | (del Amo Sanchez, 2011c) | | | | | |
| $K^+X(1812)$ | | | | $< 0.32$ | | (Liu, 2009) | $< 0.32$ | |
| $\phi K_3^*(1780)^0$ | $< 2.7$ | | (Aubert, 2007ap) | | | | $< 2.7$ | |
| $\phi K_4^*(2045)^0$ | $< 15.3$ | | (Aubert, 2007ap) | | | | $< 15.3$ | |
| $K^+\pi^-K^+\pi^-$ | | | | $< 6.0$ | | (Chiang, 2010) | $< 6.0$ | |
| $K^+\pi^-\pi^+K^-$ | | | | $< 72$ | | (Chiang, 2010) | $< 72$ | |
| $K^+\pi^-\pi^+\pi^-$ | | | | $< 2.1$ | | (Kyeong, 2009) | $< 2.1$ | |
| $\pi^+\pi^-\pi^+\pi^-$ | $< 23.1$ | | (Aubert, 2008r) | $< 19.3$ | | (Chiang, 2008) | $< 19.3$ | |

The selection criteria for three-body decays are very similar to that employed for quasi-two-body analyses. An obvious exception is that the $B$ meson decay is treated as a decay to the three final state particles and no intermediate resonance vertex is formed when reconstructing the $B$ meson. As the number of neutral final state particles increases the importance of any constraint from the beam spot on the $B$ meson vertex position also increases.

In quasi-two-body analyses, event shape variables and multivariate discriminants can be used to extract the signal yield because the reconstruction efficiency is flat in the small volume of phase space under consideration. In Dalitz Plot analyses, this is no longer true and variables that depend on momentum vectors are correlated with position in phase space. Even variables like $m_{\rm ES}$ and $\Delta E$ need to be treated carefully. Some analyses deal with the problem using an elliptical selection region in ($m_{\rm ES}$, $\Delta E$). Others rotate ($m_{\rm ES}$, $\Delta E$) about a point to eliminate the linear correlation component. If the event-by-event resolution on $\Delta E$ changes significantly, this can be compensated for by using a derived observable such as $\Delta E/\sigma(\Delta E)$. For similar reasons, multivariate discriminants need to be carefully constructed from variables that are as independent as possible from the position of the event in the Dalitz Plot.

As in quasi-two-body analyses, care must be taken with charm mesons that either decay to the same final state or are mis-reconstructed *e.g.* where a lepton is mistaken for a pion or kaon. This is particularly important in searches for highly suppressed modes such as $B^- \to K^+\pi^-\pi^-$ Aubert (2008aw). The charm background can usually be much reduced by applying mass range criteria about known resonances such as $D$ mesons, $J/\psi$ and $\psi(2S)$. This will result in empty bands in the Dalitz Plot that must be carefully considered when calculating efficiencies and migrations. Alternatively, some charm decays are deliberately kept in the Dalitz Plot. A motivation for this comes from resonances such as the $\chi_{c0}$ that have no weak phase and so can be used in an interference analysis to extract the weak phase from the Dalitz Plot. Unfortunately, the branching fraction for $B \to \chi_{c0}h$ is too small to be useful currently.

When the Dalitz Plot is represented as a Cartesian coordinate system, with the square of the mass of pairs of final state particles as the $x$ and $y$ axes, the phase space is roughly triangular in shape. Figure 17.4.17 illustrates the distribution of events extracted from data in the decay of $B^0 \to K^0_S\pi^+\pi^-$. The distribution of events on the Dalitz Plot is plotted after applying a constraint on the $B$ meson mass ($m_{\rm ES} = m_B$). This improves the resolution and ensures that all events fall within the kinematic boundaries of the Dalitz Plot. An alternative often used is a "square" Dalitz Plot where one of the axes is transformed into a "helicity-like" variable *e.g.* (Aubert, 2007v) or see Chapter 13. Although this transforms the distribution of resonances from simple bands parallel to the axes to more complex hyperboloids, the 'square" Dalitz Plot has a number of benefits. It can expand the region near areas where large variations are occurring such as in narrow resonances like the $\phi$. Bands near the Dalitz Plot edges also get expanded, enabling finer control over regions where the efficiency is changing (such as the $\rho$ meson in $B \to \pi\pi\pi$). However attention must be paid to the Jacobian as equal areas in the "square" Dalitz Plot no longer correspond to equal areas of phase-space.

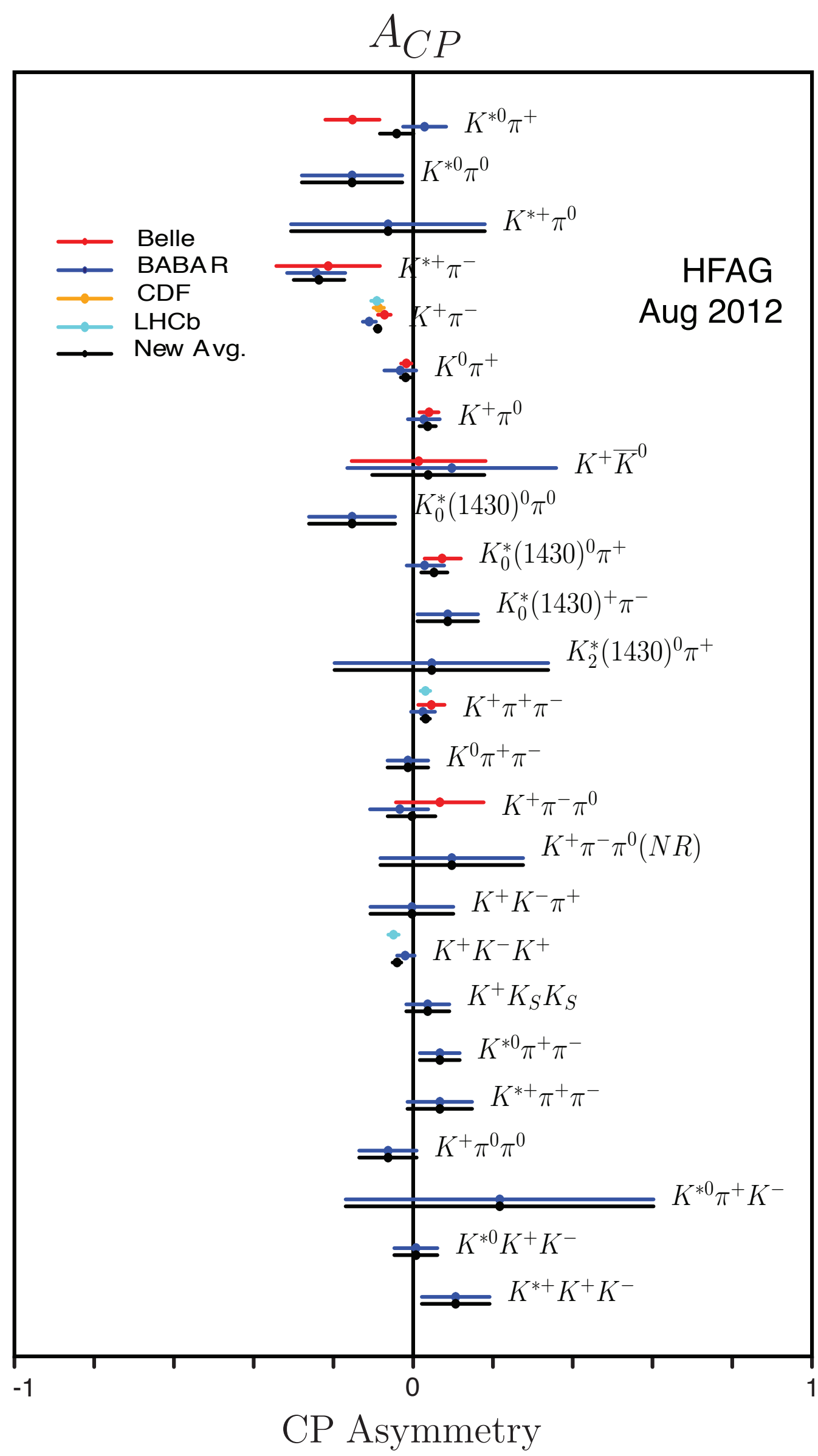

**Figure 17.4.14.** $A_{CP}$ measurements for kaonic modes with kaons or pions (Amhis et al. (2012)).

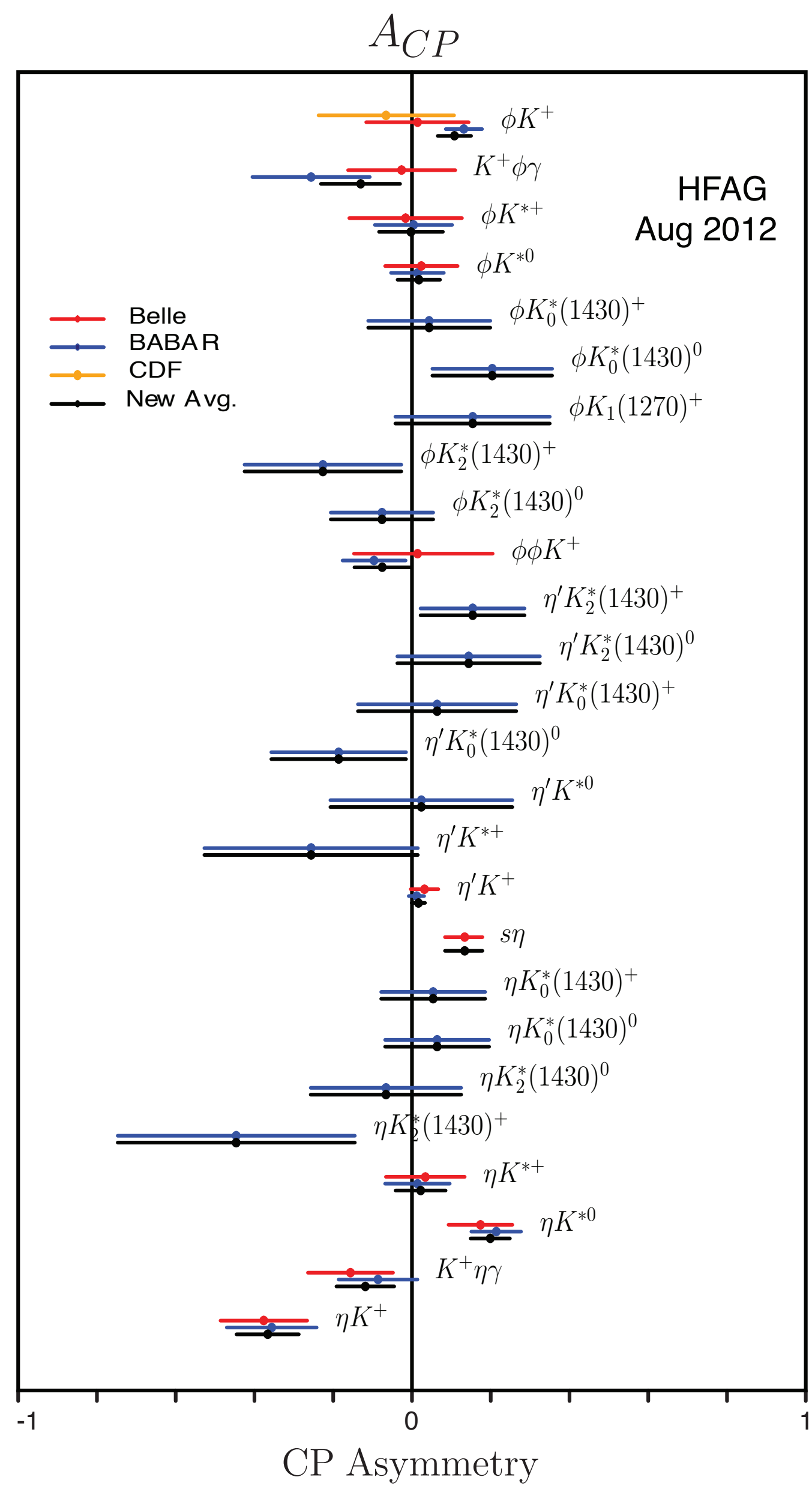

**Figure 17.4.15.** $A_{CP}$ measurements for kaonic modes with $\eta$ or $\phi$ (Amhis et al. (2012)).

Whatever the choice of Dalitz Plot, care must be taken in plotting the candidates, especially in three-body states which have two or more final state identical particles of the same mass and sign (*e.g.* $B^+ \to \pi^+ \pi^+ \pi^-$). Typical choices are to randomly select one of the pair, to fold the Dalitz Plot about the diagonal, or to consistently plot the higher mass pair on one of the axes. Even so, artificial ordering of the candidates must be eliminated or controlled. Such effects can be introduced by, for example, reconstruction tracking software that, through its track-finding algorithm, can result in momentum ordering.

The reconstruction efficiency over the Dalitz Plot can be modeled with a two-dimensional histogram, a technique that benefits from the "square" Dalitz Plot. All selection criteria are applied apart from any mass vetoes. A ratio is taken between the histogram of reconstructed events and a histogram of the true Dalitz Plot distribution of all generated MC simulated events. The reconstructed events are re-weighted to take into account any known differences between MC simulation and data such as particle identification and tracking efficiencies. The ratio can be used to provide event-by-event weighting, with linear interpolation between histogram bins where needed. The efficiency can be calculated from phase-space generated MC simulated events, but this will result in poor accuracy for narrow resonances such as the $\phi$. Better accuracy can be obtained by generating the MC with a model that contains the expected resonances in the Dalitz distribution, perhaps guided by previous quasi-two-body measurements or theory. Interference is a secondary effect but full Dalitz Plot MC simulation models which include interference effects can be used to achieve a more uniform accuracy on the efficiency. Narrow resonances pose an additional problem since their reconstructed width is dominated by the detector resolution.

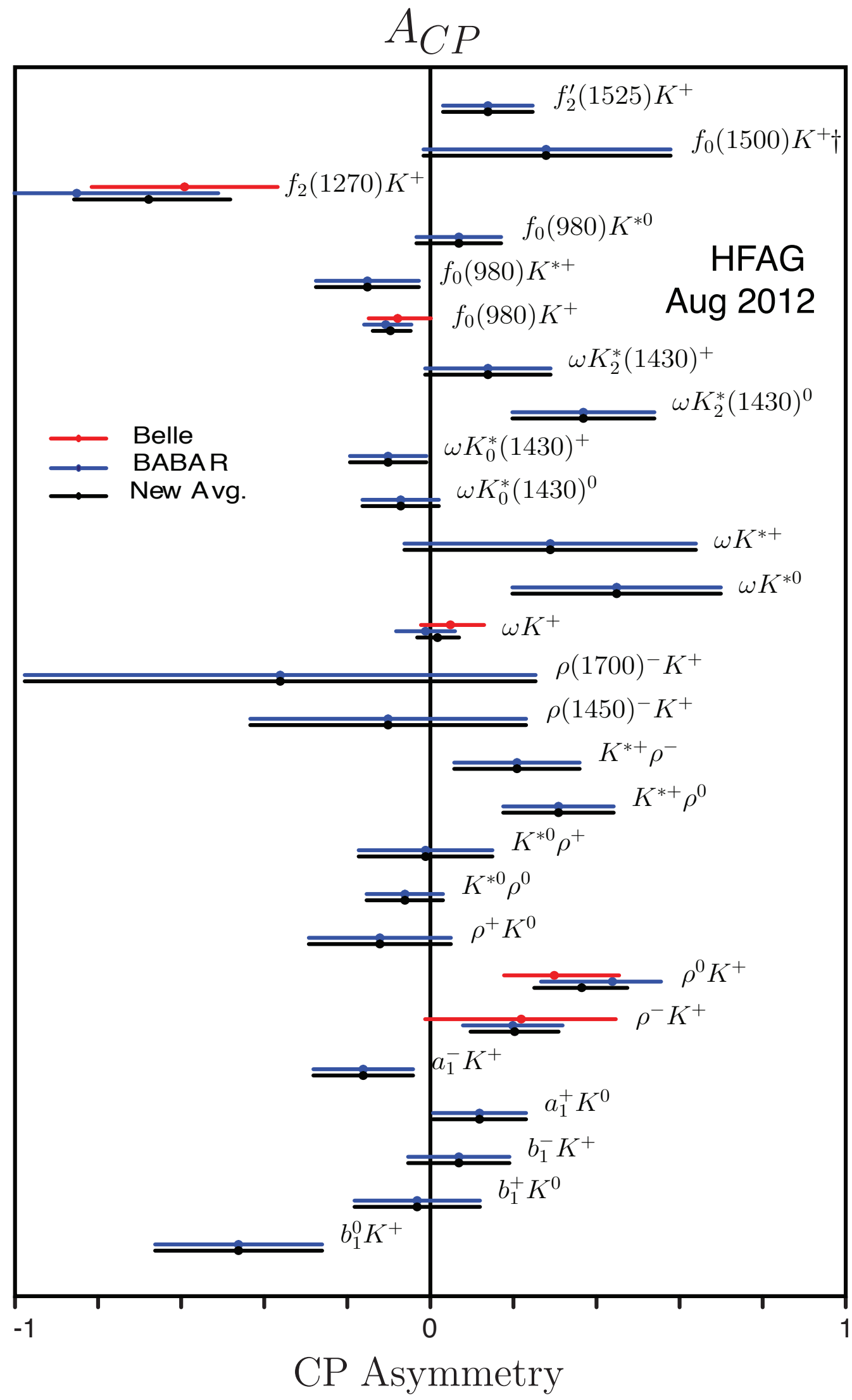

**Figure 17.4.16.** $A_{CP}$ measurements for kaonic modes with $\rho$, $\omega$, $f$, $a_1$, or $b_1$ (Amhis et al. (2012)).

If the reconstruction resolution is poor compared to the size of the histogram bin then it is necessary to take into account migrations from the true Dalitz Plot position to the reconstructed position. This becomes more important as the number of neutral particles in the final state increases. Care needs to be taken near the Dalitz Plot edges where migrations can be systematically in one direction, and also near mass regions that are close to any region that is excluded by the selection criteria *e.g.* $D$ meson mass vetoes.

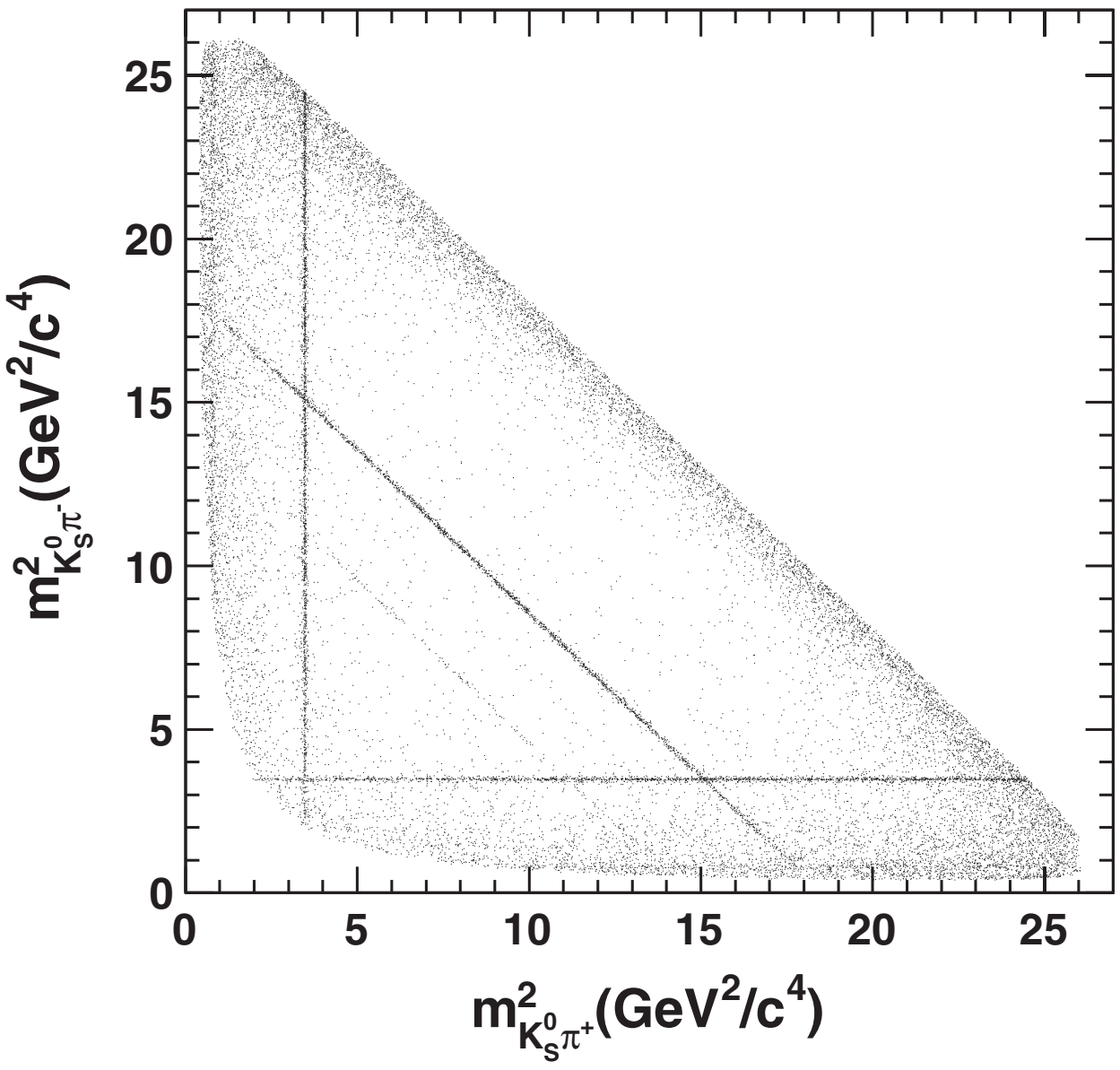

**Figure 17.4.17.** Dalitz Plot of data selected from $B^0 \to K^0_S\pi^+\pi^-$ decays (Aubert, 2009av). The narrow bands correspond to $D^{\pm}\pi^{\mp}$, $J/\psi K^0_S$, and $\psi(2S)K^0_S$ background events. As in many charmless $B$ meson decay Dalitz Plots, the events of interest are often at the edges of the allowed kinematic region.

The identification of the $B\overline{B}$ backgrounds is an intensive task. These backgrounds arise from combinations of unrelated tracks; three- and four-body decays involving intermediate $D$ mesons; charmless two- and four-body decays with an extra or missing track; and three-body decays with one or more particles misidentified. The number of such decays can be large ($\sim 50$). For fitting purposes, modes are combined that have similar behavior in the discriminating variables such as $m_{\rm ES}$ and $\Delta E$. The relative contributions are estimated from the reconstruction efficiency and estimates of the branching fractions either from measurement or theory. In some cases, the $B\overline{B}$ backgrounds are included in the maximum likelihood (ML) fit through the use of two-dimensional histograms rather than *p.d.f.*s.

The term “non-resonant” is used quite loosely by experimentalists and is often used as a short-hand for continuum background. In Dalitz Plot analyses, it should strictly refer to decays that are uniformly distributed in phase-space. In principle, this allows phenomenological predictions of the distribution to be used in the fits. These typically involve decaying exponential distributions as a function of the invariant mass-squared of the pairs of particles (*e.g.* $Ae^{-c_1 m^2}$). These functions attempt to describe the increase in the number of background events near the borders and corners of the Dalitz Plot. This increase originates from the jet-like structure of the continuum background (Garmash, 2007). However, these distri-butions have turned out not to be very satisfactory and other more complex functions are called upon. This can partly be explained as the influence of poorly understood resonances (such as the $\sigma/\kappa$ or the higher mass resonances mentioned above). As a result "non-resonant" has come to mean anything that is not modeled by a resonance. In practical terms, this means the distributions often have to come from MC simulations, off-resonance data or sideband data, or a combination of all three. In the case of sideband data, MC samples must be used to remove events from $B$ meson decays that are also present and to determine possible differences in the background shape between the sideband and signal regions. Linear interpolation between bins can be used where needed.

The backgrounds are constructed separately for both the $B^0$ and $\overline{B}^0$ events and a *p.d.f.* or histogram is formed taking into account any asymmetry that might be present in the background distributions (see, for example Eq. 20 in Aubert (2009h)).

The observables that are used in the ML depend on the analysis under consideration. Typically, a combination of $\Delta E$, $m_{\rm ES}$, multivariate discriminant, position in the Dalitz Plot and charge (flavor) of the $B$ meson candidate is used. Sometimes a cut is applied to the observable first (*e.g.* on the multivariate discriminant) and then this observable is excluded from the fit. This usually happens for observables that are correlated with position in the Dalitz Plot.

As in two-body and quasi-two-body decays, certain $D$ meson decays to the same or similar final state can be used as a calibration channel and allow for correction to fitted parameters derived just from MC simulation.

Although many of the resonances in the Dalitz Plot can be predicted from previous quasi-two-body measurements, there is still a large uncertainty in the number and type of resonances that should be included in any particular model. Examples include the exact parameterization of the non-resonant three-body decay component, the $\sigma/\kappa$ with masses in the region $400-600\,{\rm MeV}/c^2$ and widths that are large and uncertain, the $\omega(782)$, the $\chi_{c0}$ and $\chi_{c2}$, and the higher mass partners of the $\rho$, $f_0(980)$, and $K^*$. The addition of a resonance to the model that is not present in the data can be just as problematic as any exclusion of a resonance that is present. The problem is exacerbated if a blind fit is being performed. One technique is to use the log-likelihood reported by a particular model fitted to the data or to calculate a $\chi^2$ statistic based on the number of events predicted from a fit and the number of real events in a bin in the Dalitz Plot. The statistical significance of the presence of a component can be estimated by evaluating the difference $\Delta \ln \mathcal{L}$ between the negative log-likelihood of the nominal fit and that of a fit where the amplitude and $A_{CP}$ is set to zero. This is then used to evaluate a $p$ value which is the integral from $2\Delta \ln \mathcal{L}$ to infinity of the *p.d.f.* of the $\chi^2$ distribution.

An important goal of the Dalitz Plot analysis is the extraction of *CP* asymmetries either from a time-integrated or time-dependent analysis. Consequently, the resonances are parameterized not just in terms of their widths and masses but as functions of the decay dynamics, angular distributions, and the transition form factors for the $B$ meson and the resonances (see Chapter 13.2.1). As explained in more detail in Chapter 13.4.2, complex coefficients are used to parameterize the $B$ and $\overline{B}$ meson decay. The same parameterization is not consistently used between papers or experiments, although they are all mathematically related. As a specific example from (Dalseno, 2009), the intermediate resonances $i$ in $B$ and $\overline{B}$ meson decay are parameterized respectively as:

$$
\begin{aligned}
a_i' &= a_i(1+c_i)e^{i(b_i+d_i)} \\
\bar{a}_i' &= a_i(1-c_i)e^{i(b_i-d_i)}
\end{aligned}
\tag{17.4.13}
$$

where $b_i$ and $d_i$ represent the strong and weak phase respectively (notice the strong phase does not change sign). Consequently, the *CP* asymmetry for each resonance $i$ can be written as:

$$
A_{CP}(i) = \frac{|\bar{a}_i'|^2 - |a_i'|^2}{|\bar{a}_i'|^2 + |a_i'|^2} = \frac{-2c_i}{1+c_i^2} \tag{17.4.14}
$$

In the case of time-dependent Dalitz plot analyses, the resonance parameterizations above are combined with the equation describing the time-dependent decay properties of the $B$ and $\overline{B}$ meson as given in Equation 13.2.17. In this case, a great deal of attention has to be given to the tagging and resolution functions.

Charmless $B$ decays, especially those without access to tree decay diagrams, may have a large non-resonant contribution. This can be as high as 90% for $B \to KKK$. The contribution is not uniform across the Dalitz diagram and so a parameterization must be adopted that depends on position in the Dalitz Plot. In some analyses, *BABAR* and Belle have adopted the same non-resonant parameterization but in most cases they differ, which can complicate comparisons.

The statistical errors on the measured fit fractions and *CP* parameters are often derived from fits to a large number of MC experiments generated with the fitted parameters obtained from the data. These MC experiments are also vital for understanding the minimization process. With a large number of floating parameters, the fit can sometimes have more than one local minimum. There can be systematic shifts in the fit caused by the starting values of the floating parameters. A number of techniques for investigating this effect have been applied, including using different minimizers, scanning through a set of starting values, randomly initializing the starting values, and the use of genetic algorithms. Each has its benefits and drawbacks but there is no one method that works better than the others in all circumstances.

The systematic uncertainties that affect the final result are very similar to those seen in other charmless $B$ decays. However their effects can be modified since there are more opportunities for correlations between parameters and the fitted results are often reported as ratios rather than absolute numbers. Although the magnitude and phase of the complex coefficients of the amplitude are sometimes transformed to a more orthogonal set of parameters, this does not wholly eliminate the correlations. Systematic uncertainties that are unique to the Dalitz Plot are: the asymmetries in the background; limited statistics from the sidebands used to form the continuum histograms (if histograms are used); the mass rejection regions; differences in the continuum shape between the sideband and the signal region; and charge bias introduced either by the detector response or the selection criteria. A model dependent error derived from performing fits with an alternative set of resonances is sometimes quoted either in quadrature with the systematic error or its own. As with quasi-two-body modes, an important systematic is associated with uncertainty on the parameters that are fixed in the fit. If a resonance is deemed to be significant, the mass and width may still not be well known. Rather than float the mass and width, a series of fits can be performed with the mass and width fixed at different values and the change in the likelihood used as a guide to the best values. Even so, it may be necessary to modify a model after unblinding, particularly to remove resonances that are not significant.

### 17.4.7 Three-body and Dalitz decays

Approximately seven $B^0$ and eleven $B^{\pm}$ Dalitz Plots have been investigated by *BABAR* and Belle. It is impossible to do justice to the wealth of information available. Decays involving three pions, particularly $B \to \rho\pi$, are important for the measurement of $\phi_2$ and are considered in Chapter 17.6. Decays with an $\eta$, $\eta'$, $\omega$, $f_0(980)$, or $K^*$ in the final three-body state are itemized in the tables and figures of this section but are not described in detail. Instead, this section concentrates on modes with one or more kaons in the final state.

$B$ meson decays to three-body final states $B \to Khh$ proceed predominantly via $b \to u$ tree-level diagrams (T and C diagrams in Fig. 17.4.1) and $b \to s(d)$ penguin diagrams (P in Fig. 17.4.1). The other diagrams can contribute but are expected to be much smaller. Final states with an odd number of kaons ($s$-quarks) are expected to proceed dominantly via $b \to s$ penguin transitions as the $b \to u$ transition is color-suppressed. If there are two kaons, the decay proceeds through the color-allowed $b \to u$ tree diagram and the $b \to d$ penguin decay with no $b \to s$ penguin contribution. As a result, these Dalitz decays provide an excellent opportunity to understand the relative contribution of tree and penguin amplitudes in charmless decays. This is shown in Fig. 17.4.18 where the extracted values of $\sin 2\phi_1$ in $b \to s$ penguin transitions are compared to $b \to c\bar{c}s$ decays.

Table 17.4.10 summarizes the reported branching fractions and asymmetries. In many cases, no resonances have been found in a Dalitz Plot and so consequently it has only been possible to give a branching fraction (or upper limit) and a *CP* asymmetry for the whole Dalitz Plot. Figure 17.4.19 shows the relative values of the reported branching fractions so far measured.

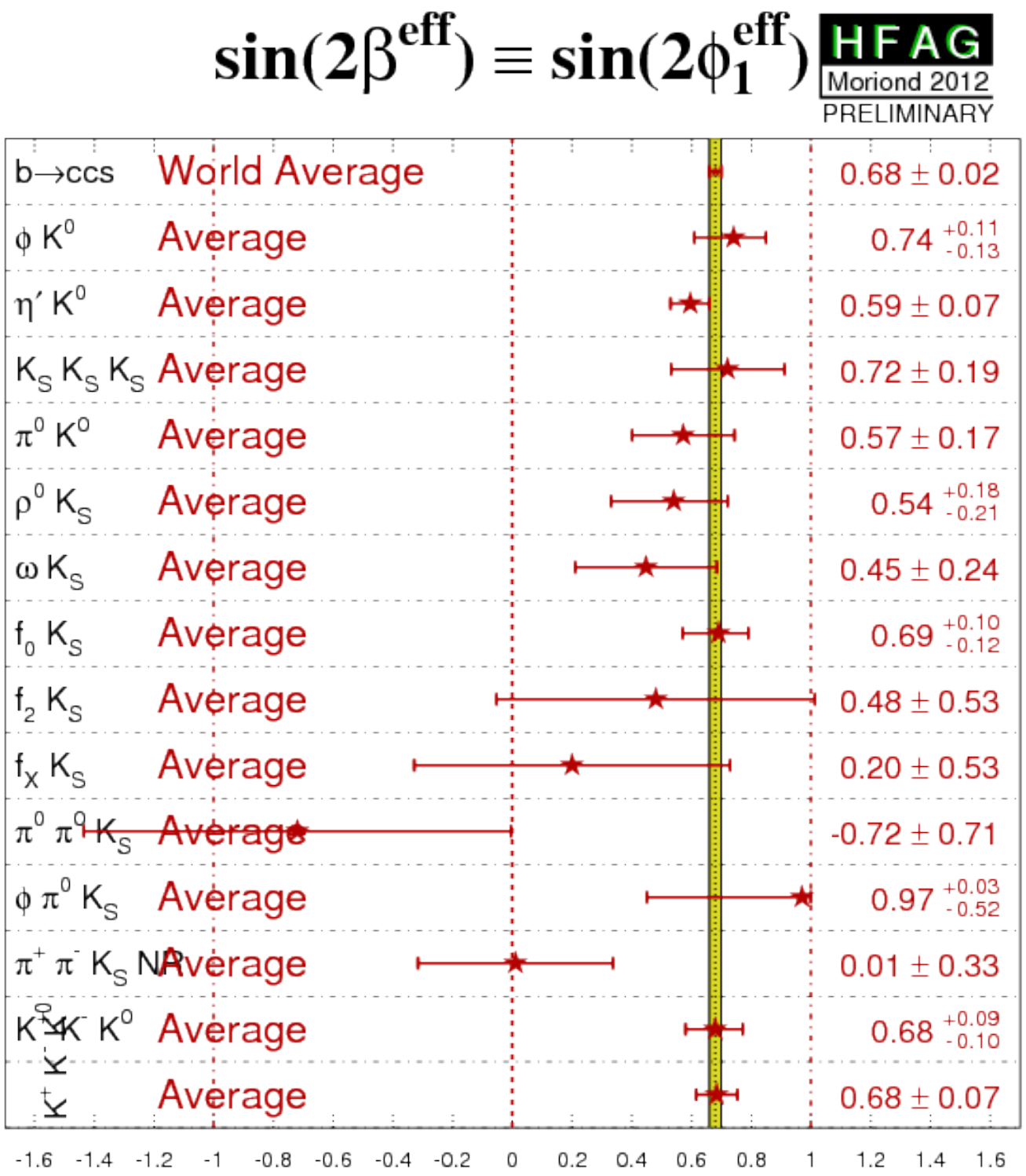

**Figure 17.4.18.** Comparison between the value of $\sin 2\phi_1$ from $b \to c\bar{c}s$ decays such as $B^0 \to J/\psi K^0$ (indicated by "World Average") and strange charmless $b \to u\bar{u}s$ decays (Amhis et al. (2012)).

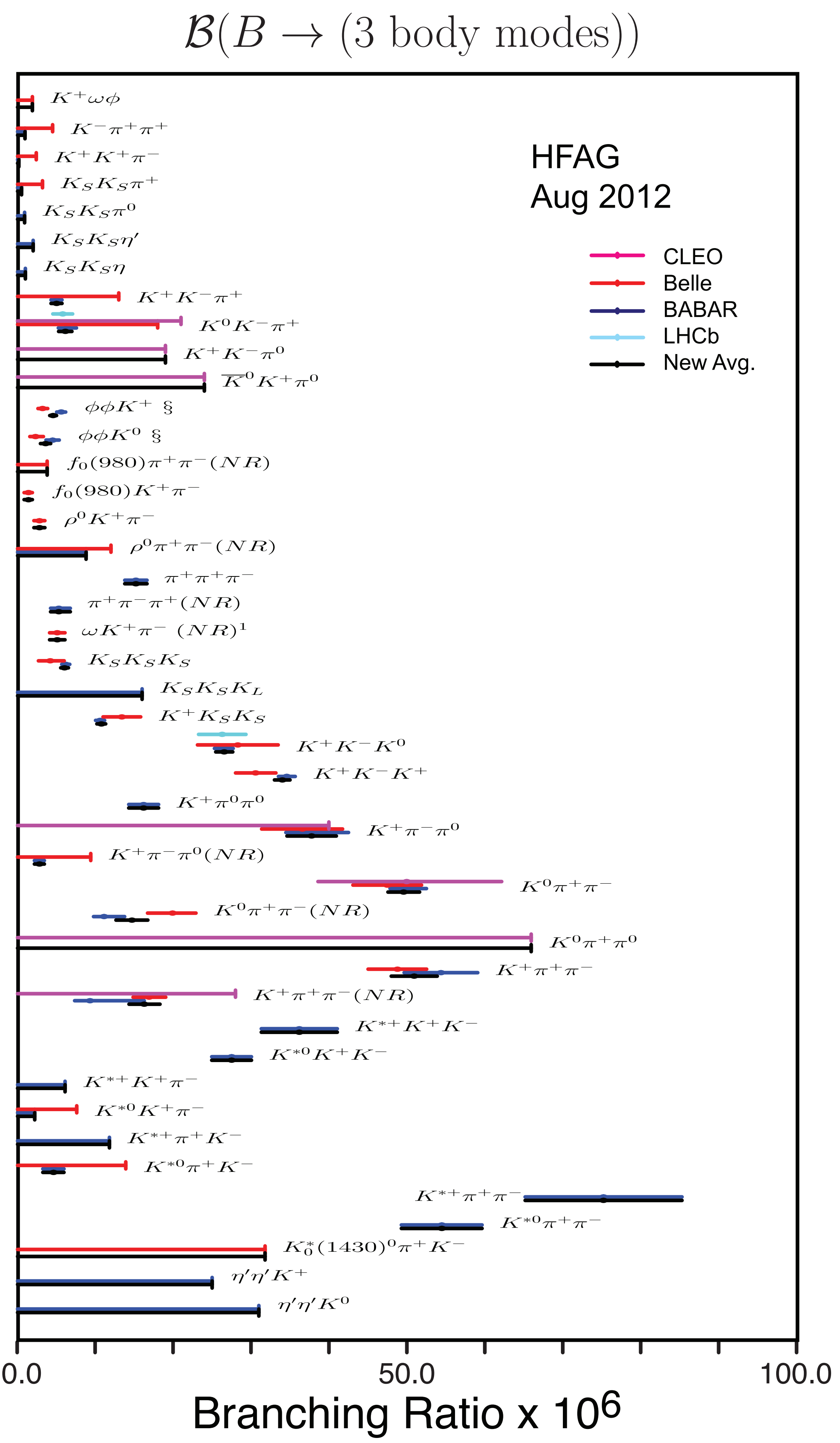

**Figure 17.4.19.** Summary of branching fraction measurements ($\times 10^{-6}$) and HFAG averages for decays with three mesons in the final state (Amhis et al. (2012)).

**Table 17.4.10.** Charmless $B$ decays branching fractions $\mathcal{B}$ and $CP$ asymmetries $A_{CP}$ for BABAR and Belle for decays with three mesons in the final state. The averages come from HFAG and may include measurements from other experiments such as CLEO and LHCb (Amhis et al. (2012)).

| | BABAR results | | | Belle results | | | Averages | |
|---|---|---|---|---|---|---|---|---|
| Final state | $\mathcal{B}$ ($\times 10^{-6}$) | $A_{CP}$ | Ref. | $\mathcal{B}$ ($\times 10^{-6}$) | $A_{CP}$ | Ref. | $\mathcal{B}$ ($\times 10^{-6}$) | $A_{CP}$ |
| $K^+K^+\pi^-$ | $< 0.16$ | | (Aubert, 2008aw) | $< 2.4$ | | (Garmash, 2004) | $< 0.16$ | |
| $K^+K^-K^+$ | $34.6 \pm 0.6 \pm 0.9$ | $-0.017^{+0.019}_{-0.014} \pm 0.014$ | (Lees, 2012y) | $30.6 \pm 1.2 \pm 2.3$ | | (Garmash, 2005) | $34.0 \pm 1.0$ | $-0.017 \pm 0.026$ |
| $K^+K^-K^0$ | $26.5 \pm 0.9 \pm 0.8$ | | (Lees, 2012y) | $28.3 \pm 3.3 \pm 4.0$ | | (Garmash, 2004) | $26.6 \pm 1.1$ | |
| $K^+K^-\pi^+$ | $5.0 \pm 0.5 \pm 0.5$ | $0.00 \pm 0.10 \pm 0.03$ | (Aubert, 2007an) | $< 13$ | | (Garmash, 2004) | $5.0 \pm 0.7$ | $0.00 \pm 0.10$ |
| $K^+K^-\pi^0$ | | | | | | | $< 19$ | |
| $K^+K^0_SK^0_S$ | $10.6 \pm 0.5 \pm 0.3$ | $0.04 \pm 0.05 \pm 0.02$ | (Lees, 2012y) | $13.4 \pm 1.9 \pm 1.5$ | | (Garmash, 2004) | $10.8 \pm 0.6$ | $0.04 \pm 0.05$ |
| $K^+\pi^+\pi^-(NR)$ | $9.3 \pm 1.0^{+6.9}_{-1.7}$ | | (Aubert, 2008j) | $16.9 \pm 1.3^{+1.7}_{-1.6}$ | | (Garmash, 2006) | $16.3 \pm 2.0$ | |
| $K^+\pi^+\pi^-$ | $54.4 \pm 1.1 \pm 4.6$ | $0.028 \pm 0.020 \pm 0.023$ | (Aubert, 2008j) | $48.8 \pm 1.1 \pm 3.6$ | $0.049 \pm 0.026 \pm 0.020$ | (Garmash, 2006) | $51.0 \pm 3.0$ | $0.038 \pm 0.022$ |
| $K^+\pi^-\pi^0(NR)$ | $2.8 \pm 0.5 \pm 0.4$ | $0.10 \pm 0.16 \pm 0.08$ | (Lees, 2011a) | $< 9.4$ | | (Chang, 2004) | $2.8 \pm 0.6$ | $0.23^{+0.22}_{-0.28}$ |
| $K^+\pi^-\pi^0$ | $38.5 \pm 1.0 \pm 3.9$ | $-0.030^{+0.045}_{-0.051} \pm 0.055$ | (Lees, 2011a) | $36.6^{+4.2}_{-4.3} \pm 3.0$ | $0.07 \pm 0.11 \pm 0.01$ | (Chang, 2004) | $37.8 \pm 3.2$ | $0.00 \pm 0.06$ |
| $K^+\pi^0\pi^0$ | $16.2 \pm 1.2 \pm 1.5$ | $-0.006 \pm 0.006 \pm 0.004$ | (Lees, 2011g) | | | | $16.2 \pm 1.9$ | |
| $K^+\omega\phi$ | | | | $< 1.9$ | | (Liu, 2009) | $< 1.9$ | |
| $K^-\pi^+\pi^+$ | $< 0.95$ | | (Aubert, 2008aw) | $< 4.5$ | | (Garmash, 2004) | $< 0.95$ | |
| $K^0K^-\pi^+$ | $6.4 \pm 1.0 \pm 0.6$ | | (del Amo Sanchez, 2010j) | $< 18$ | | (Garmash, 2004) | $6.4 \pm 1.2$ | |
| $K^0\pi^+\pi^-(NR)$ | $11.1^{+2.5}_{-1.0} \pm 0.9$ | | (Aubert, 2009av) | $19.9 \pm 2.5^{+1.7}_{-2.0}$ | | (Garmash, 2007) | $14.7 \pm 2.0$ | |
| $K^0\pi^+\pi^-$ | $50.2 \pm 1.5 \pm 1.8$ | $-0.01 \pm 0.05 \pm 0.01$ | (Aubert, 2009av) | $47.5 \pm 2.4 \pm 3.7$ | | (Garmash, 2007) | $49.6 \pm 2.0$ | $-0.01 \pm 0.05$ |
| $K^0\pi^+\pi^0$ | | | | | | | $< 66$ | |
| $K^{*+}K^+K^-$ | $36.2 \pm 3.3 \pm 3.6$ | $0.11 \pm 0.08 \pm 0.03$ | (Aubert, 2006h) | | | | $36.2 \pm 4.9$ | $0.11 \pm 0.09$ |
| $K^{*+}K^+\pi^-$ | $< 6.1$ | | (Aubert, 2006h) | | | | $< 6.1$ | |
| $K^{*+}\pi^+K^-$ | $< 11.8$ | | (Aubert, 2006h) | | | | $< 11.8$ | |
| $K^{*+}\pi^+\pi^-$ | $75.3 \pm 6.0 \pm 8.1$ | $0.07 \pm 0.07 \pm 0.04$ | (Aubert, 2006h) | | | | $75.3 \pm 10.1$ | $0.07 \pm 0.08$ |
| $K^{*0}K^+K^-$ | $27.5 \pm 1.3 \pm 2.2$ | $0.01 \pm 0.05 \pm 0.02$ | (Aubert, 2007ah) | | | | $27.5 \pm 2.6$ | $0.01 \pm 0.05$ |
| $K^{*0}K^+\pi^-$ | $< 2.2$ | | (Aubert, 2007ah) | $< 7.6$ | | | $< 2.2$ | |
| $K^{*0}\pi^+K^-$ | $4.6 \pm 1.1 \pm 0.8$ | $0.22 \pm 0.33 \pm 0.20$ | (Aubert, 2007ah) | $< 13.9$ | | | $4.6 \pm 1.4$ | $0.22 \pm 0.39$ |
| $K^{*0}\pi^+\pi^-$ | $54.5 \pm 2.9 \pm 4.3$ | $0.07 \pm 0.04 \pm 0.03$ | (Aubert, 2007ah) | $4.5^{+1.1+0.9}_{-1.0-1.6}$ | | | $54.5 \pm 5.2$ | $0.07 \pm 0.05$ |
| $K^*_0(1430)^0\pi^+K^-$ | | | | $< 31.8$ | | (Chiang, 2010) | $< 31.8$ | |
| $K^0_SK^0_SK_L$ | $< 16$ | | (Aubert, 2006at) | | | | $< 16$ | |
| $K^0_SK^0_SK^0_S$ | $6.19 \pm 0.48 \pm 0.19$ | | (Aubert, 2005e) | $4.2^{+1.6}_{-1.3} \pm 0.8$ | | (Garmash, 2004) | $6.2 \pm 0.9$ | |
| $K^0_SK^0_S\eta$ | $< 1.0$ | | (Aubert, 2009am) | | | | $< 1.0$ | |
| $K^0_SK^0_S\eta'$ | $< 2.0$ | | (Aubert, 2009am) | | | | $< 2.0$ | |
| $K^0_SK^0_S\pi^+$ | $< 0.51$ | | (Aubert, 2009ar) | $< 3.2$ | | (Garmash, 2004) | $< 0.51$ | |
| $K^0_SK^0_S\pi^0$ | $< 0.9$ | | (Aubert, 2009am) | | | | $< 0.9$ | |
| $\eta'\eta'K^+$ | $< 25$ | | (Aubert, 2006al) | | | | $< 25$ | |
| $\eta'\eta'K^0$ | $< 31$ | | (Aubert, 2006al) | | | | $< 31$ | |
| $\omega K^+\pi^-$ $(NR)$[1] | | | | $5.1 \pm 0.7 \pm 0.7$ | | (Goldenzweig, 2008) | $5.1 \pm 1.0$ | |
| $\overline{K}^0K^+\pi^0$ | | | | | | | $< 24$ | |
| $\phi\phi K^+$ | $5.6 \pm 0.5 \pm 0.3$ | $-0.10 \pm 0.08 \pm 0.02$ | (Lees, 2011e) | $3.2^{+0.6}_{-0.5} \pm 0.3$ | $0.01^{+0.19}_{-0.16} \pm 0.02$ | (Abe, 2008b) | $4.6 \pm 0.4$ | $-0.08 \pm 0.07$ |
| $\phi\phi K^0$ | $4.5 \pm 0.8 \pm 0.3$ | | (Lees, 2011e) | $2.3^{+1.0}_{-0.7} \pm 0.2$ | | (Abe, 2008b) | $3.6 \pm 0.7$ | |
| $\pi^+\pi^+\pi^-$ | $15.2 \pm 0.6 \pm 1.3$ | | (Aubert, 2009h) | | | | $15.2 \pm 1.4$ | |
| $\pi^+\pi^-\pi^+(NR)$ | $5.3 \pm 0.7^{+1.3}_{-0.8}$ | $-0.14 \pm 0.14^{+0.18}_{-0.08}$ | (Aubert, 2009h) | | | | $5.3^{+1.5}_{-1.1}$ | $-0.14^{+0.23}_{-0.16}$ |
| $\rho^0K^+\pi^-$ | | | | $2.8 \pm 0.5 \pm 0.5$ | | (Kyeong, 2009) | $2.8 \pm 0.7$ | |
| $\rho^0\pi^+\pi^-(NR)$ | $< 8.8$ | | (Aubert, 2008r) | $< 12$ | | (Chiang, 2008) | $< 8.8$ | |
| $f_0(980)K^+\pi^-$ | | | | $1.4 \pm 0.4^{+0.3}_{-0.4}$ | | (Kyeong, 2009) | $1.4^{+0.5}_{-0.6}$ | |
| $f_0(980)\pi^+\pi^-(NR)$ | | | | $< 3.8$ | | (Chiang, 2008) | $< 3.8$ | |

$CP$ asymmetries are expected in $b \to s\bar{s}s$ decays consistent with asymmetries measured in $b \to c\bar{c}s$. The tree contributions are small and the amplitude is dominated by loop contributions, where new virtual particles can contribute. In $B^0 \to K^+K^-K^0$, both the direct $CP$ asymmetry $A_{CP}$ and $\phi_1^{eff}$ ($\phi_1 = arg(-V_{cd}V_{cb}^*/V_{td}V_{tb}^*)$) have been measured for the whole Dalitz Plot and the dominant individual resonances. *BABAR* find two equally likely solutions for $B^0 \to \phi K^0$ and $B^0 \to f_0(980)K^0$, the first consistent with the SM and the second with a significantly different phase $\phi_1^{eff}$ for $B^0 \to f_0(980)K^0$. In the high mass region, the $CP$-conserving case $\phi_1^{eff} = 0$ is excluded at the 5.1 standard deviations level. Across the whole Dalitz Plot the $CP$ asymmetry is $A_{CP} = -0.015 \pm 0.077 \pm 0.053$ and $\phi_1^{eff} = 0.352 \pm 0.076 \pm 0.026$ (Aubert, 2007af). Belle, with approximately twice the data size, find four solutions for $\phi_1^{eff}$ but solution 1 is preferred when external constraints, such as known branching fraction ratios, are included (Nakahama, 2010). Belle see no evidence for $A_{CP}$ in $B^0 \to \phi K_S^0$ nor in $B^0 \to f_0(980)K_S^0$ and measure $\phi_1^{eff}$ to be $(33.2 \pm 9.0 \pm 2.6 \pm 1.4)^\circ$ for $B^0 \to \phi K_S^0$ and $(31.3 \pm 9.0 \pm 3.4 \pm 4.0)^\circ$ for $B^0 \to f_0(980)K_S^0$ (solution 1). These are consistent with $\phi_1^{eff}$ measurements from other $b \to c\bar{c}s$ transitions, such as $B^0 \to J/\psi K^0$.

*BABAR* performed a binned fit to the $B^+ \to K^+K^+K^-$ Dalitz Plot and found no evidence for $CP$ violation, neither for the whole plane ($A_{CP} = -0.017 \pm 0.026 \pm 0.015$) nor for any resonance (Aubert, 2006i). Belle in their analysis (Garmash, 2005) do not report asymmetries but their results for branching and fit fractions do not agree well with *BABAR*. This is primarily due to the fact that *BABAR* report a broad scalar resonance, which they label $X_0(1550)$, while Belle include only the $f_0(980)$ in their model.

The Dalitz Plot structure of $B^0 \to K_S^0K_S^0K_S^0$ has been investigated and the inclusive branching fractions measured. The product branching fractions of $f_0(980)K_S^0$, $f_0(1270)K_S^0$ and $f_2(2010)K_S^0$ have been measured and there are hints of $f_2'(1525)$ and $f_0(1500)$ (Lees, 2012c). The mixing-induced $CP$-violation parameters for $B^0 \to K^0K^0K^0$ are measured to be $S = -0.94^{+0.24}_{-0.21} \pm 0.06$ and $C = -0.17 \pm 0.18 \pm 0.04$. These are compatible within 2 standard deviations with those measured in tree-dominated $B^0 \to J/\psi K_S^0$ decays. As a result $CP$ conservation is excluded at the 3.8 standard deviation level. Belle have looked at $B^0 \to K_S^0K_S^0K_S^0$ and intermediate resonances that decay to the final state $K^+K^-K_S^0$ (Chen, 2007a).

Belle measure $\sin 2\phi_1$ in $B^0 \to \eta' K^0$ to be $0.64 \pm 0.10 \pm 0.04$ with a significance of 5.6 standard deviations and find no evidence for direct $CP$ violation. *BABAR* also measure a significant value of $\sin 2\phi_1 = 0.58 \pm 0.10 \pm 0.03$ (5.5 standard deviations significance) in $B^0 \to \eta' K^0$ (Aubert, 2007am). However, in this case, the direct $CP$ result $A_f = -0.16 \pm 0.07 \pm 0.03$ is 2.1 standard deviation from zero.

For $B^+ \to K^0K^0K^+$, Belle report branching fractions (Garmash, 2004), while *BABAR* has also extracted the $CP$ charge asymmetry $A_{CP} = -0.04 \pm 0.1 \pm 0.02$ (Aubert, 2004b).

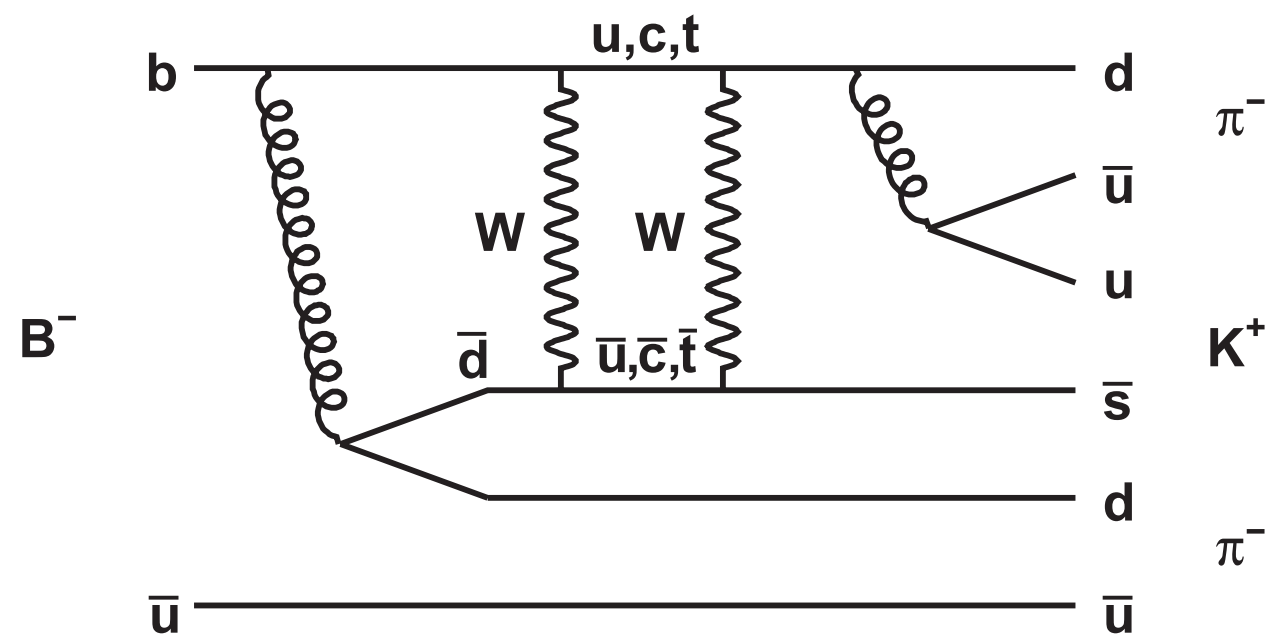

**Figure 17.4.20.** Example of the SM suppressed decay diagram for the decay $B^- \to K^+\pi^-\pi^-$.

Modes with just two kaons in the final state are important as they proceed through $b \to d$ penguin loops and are suppressed. Consequently, small branching fractions are expected and the opportunities for measuring asymmetries are few.

There have been no measurements of the decay $B^0 \to K^+K^-\pi^0$ by *BABAR* or Belle. The decay $B^+ \to K^+K^-\pi^+$ has been observed by *BABAR* (Aubert, 2007an) with $\mathcal{B} = (5.0 \pm 0.5 \pm 0.5) \times 10^{-6}$ and $A_{CP} = 0.00 \pm 0.10 \pm 0.03$; Belle have placed upper limits (UL) on the branching fraction $< 13 \times 10^{-6}$ (Garmash, 2004). The mode $B^+ \to K^+K^+\pi^-$ is additionally suppressed by a factor $|V_{td}V_{ts}^*| \sim 3 \times 10^{-4}$ but could be enhanced in SM extensions with extra $Z'$ bosons. *BABAR* finds for this decay a branching fraction UL of $0.16 \times 10^{-6}$ (Aubert, 2008aw). The decay $B^0 \to K_S^0K^\pm\pi^\mp$ has been observed by *BABAR* with branching fraction $(3.2 \pm 0.5 \pm 0.3) \times 10^{-6}$ at 5.2 standard deviation significance (del Amo Sanchez, 2010j).

Both Belle and *BABAR* have made significant progress in measuring $B \to K_S^0K_S^0h$ where $h$ includes mesons such as $\pi^+$, $\pi^0$, $\eta$, and $\eta'$. The $b \to d$ transition has been measured in $B^0 \to \pi^+\pi^-\pi^0$ where the beauty flavor changes by $\Delta F = 2$ (due to mixing) but in $B^+ \to K_S^0K_S^0\pi^+$ by $\Delta F = 1$ (due to decay). Both *BABAR* and Belle have placed upper limits of $\mathcal{B}(B^+ \to K_S^0K_S^0\pi^+) < 0.51 \times 10^{-6}$ (Aubert, 2009ar) and $< 3.2 \times 10^{-6}$ (Garmash, 2004), respectively. *BABAR* find ULs on $\mathcal{B}(B^0 \to K_S^0K_S^0\pi^0)$, $\mathcal{B}(B^0 \to K_S^0K_S^0\eta)$, and $\mathcal{B}(B^0 \to K_S^0K_S^0\eta')$ of $2 \times 10^{-6}$ and below (Aubert, 2009am).

Large $CP$ asymmetries are expected in $B^+ \to \rho^0K^+$. *BABAR* find evidence of direct $CP$ violation in $B^+ \to \rho^0K^+, \rho^0 \to \pi^+\pi^-$ with $A_{CP} = (0.44 \pm 0.10 \pm 0.04^{+0.06}_{-0.13})$ (Aubert, 2008j) at the $3.7\sigma$ level and Belle report very similar results, with $A_{CP} = (0.30 \pm 0.11 \pm 0.02^{+0.11}_{-0.04})$ with $3.9\sigma$ significance (Garmash, 2006) A Dalitz analysis is essential due to the possibility of interference of the wide $\rho^0$ width with neighboring resonances. $CP$ asymmetries in $B^+ \to K^{*0}\pi^+$, $B^+ \to K_0^{*0}(1430)\pi^+$, and $B^+ \to K_2^{*0}(1430)\pi^+$, on the other hand, are small. The SM-suppressed mode $B^- \to K^+\pi^-\pi^-$ has also been investigated by both experiments and the decay diagram is shown in Fig. 17.4.20. *BABAR* and Belle place UL on the branching fraction of $0.95 \times 10^{-6}$ (Aubert, 2008aw) and $< 4.5 \times 10^{-6}$ (Garmash, 2004), respectively.

The decay $B^+ \to K^+\pi^-\pi^+$ is important for searching for direct *CP* violation in $B \to K^*\pi$ decays. *BABAR* find four compatible solutions of the Dalitz Plot (Lees, 2011a). When combined with the time-dependent analysis of $B^0 \to K^0_S\pi^-\pi^+$ (Aubert, 2009av), *BABAR* report $A_{CP} = -0.24 \pm 0.07 \pm 0.02$ with a significance of $3.1\sigma$ for $B \to K^{*+}\pi^-$ decays. A similar Belle analysis has half the number of events and is restricted to branching fraction measurements and ranges for $A_{CP}$ (Chang, 2004).

In $B^0 \to \pi^+\pi^-K^0_S$, the decay $B^0 \to f_0(980)K^0_S$ is expected to be dominated by $b \to s$ transitions. The $f_0(980)$ can overlap with nearby resonances, requiring a Dalitz analysis to extract a robust estimate of $\sin 2\phi_1$, taking interference into account. Belle find no evidence for direct *CP* violation in $B^0 \to \rho^0 K^0_S$, $B^0 \to f_0(980)K^0_S$, and $B^0 \to K^{*+}\pi^-$ and measure $A_{CP}(K^{*+}\pi^-) = -0.21 \pm 0.11 \pm 0.05 \pm 0.05$ (Dalseno, 2009; Garmash, 2007). The $\sin 2\phi_1$ measurements for $B^0 \to \rho^0 K^0_S$ and $B^0 \to f_0(980)K^0_S$ are consistent with $\sin 2\phi_1$ from $b \to c\bar{c}s$ decays. The phase difference between $B^0 \to K^{*+}\pi^-$ and $\overline{B}^0 \to K^{*-}\pi^+$, which could lead to a measurement of $\phi_3$, is reported as $\Delta\phi(K^{*+}\pi^-) = (-0.7^{+23.5}_{-22.8} \pm 11.0 \pm 17.6)^\circ$. *BABAR* has also looked at this mode but only report ranges for $\phi_1$ in $B^0 \to \rho^0 K^0_S$ and $B^0 \to f_0(980)K^0_S$ but they measure $A_{CP}(K^{*+}\pi^-)$ consistent with Belle (Aubert, 2009av).

The $B$ Factories have started to look at Dalitz Plots involving short-lived particles such as the $K^*$. The branching fractions of the decays $B^0 \to K^{*0}\pi^+K^-$ and $B^+ \to K^{*+}\pi^+K^-$ are sensitive to the CKM matrix elements $V_{td}$ and $V_{ub}$. Additionally, a branching fraction of the Standard Model suppressed decay $B^0 \to K^{*0}K^+\pi^-$ comparable or larger than that of $B^0 \to K^{*0}\pi^+K^-$ would be an indication of new physics (Aubert, 2006h, 2007ah). There is no evidence for this in the current data with branching fraction measurements of $\mathcal{B}(B^0 \to K^{*0}K^+\pi^-) = (4.6 \pm 1.1 \pm 0.8) \times 10^{-6}$ and $\mathcal{B}(B^0 \to K^{*0}\pi^+K^-) < 2.2 \times 10^{-6}$.

As an example of the detail of information that can be extracted from a Dalitz Plot analysis, Table 17.4.11 shows the branching fractions, charged asymmetries, fit fractions, and phases for the decay $B^+ \to K^+K^+K^-$. Similar results exist for a number of the Dalitz Plots listed in Table 17.4.10.

### 17.4.8 Summary

Together *BABAR* and Belle have collected well over $1\,\text{ab}^{-1}$ of $B$ meson decays. Even with low branching fractions, the study of charmless hadronic B decays have enabled the measurement of: the CKM angles $\phi_1$, $\phi_2$, $\phi_3$; the discovery of many new decay modes with a measured branching fraction; new branching fraction upper limits placed on many rare decays; direct and indirect *CP* asymmetries; G-parity conservation tests; longitudinal polarization; interference effects; and weak and strong phases. This has enabled a comprehensive comparison with theoretical predictions and models. These theoretical models continue to progress, with more precise calculations over a wider range of observables. Yet despite this, the study of charmless hadronic decays is still only partially complete. Work is still on-going in understanding the hierarchy of the longitudinal polarization. Some measured branching fractions do not agree with predictions. The prediction, understanding and interpretation of the phases and amplitudes in three-body Dalitz Plots are still in their infancy.

**Table 17.4.11.** An illustration of the results that can be extracted from a full Dalitz Plot analysis of $B^+ \to K^+K^+K^-$ for BABAR (Aubert, 2006i) and Belle (Garmash, 2005). The extracted parameters are: the branching fraction $\mathcal{B}$ or product branching fraction $\mathcal{B}\times\mathcal{B}_f$ ($\times 10^{-6}$); the charged $CP$ asymmetry $A_{CP}$ (%); the fit fraction $FF$ (%); the phase $\delta$ (°) relative to the reference decay; mass $M$ and width $\Gamma$ (GeV/$c^2$); $NR$ is the non-resonant component and some errors have been rounded.

| Decay | Param. | BABAR | Belle |
|---|---|---|---|
| $K^+ K^+ K^-$ | $\mathcal{B}$ | $33.5 \pm 0.9 \pm 1.6$ | $30.6 \pm 1.2 \pm 2.3$ |
| | $A_{CP}$ | $-0.02 \pm 0.03 \pm 0.02$ | |
| $\phi K^+$ | $\mathcal{B}$ | $8.4 \pm 0.7 \pm 0.7$ | $9.60 \pm 0.92 \pm 0.71$ |
| | $A_{CP}$ | $0 \pm 8 \pm 2$ | |
| | $FF$ | $11.8 \pm 0.9 \pm 0.8$ | $14.7 \pm 1.3$ |
| | $\delta$ | $-7 \pm 0.11 \pm 3$ | $-123 \pm 10$ |
| $\phi(1680)K^+$ | $\mathcal{B}\times\mathcal{B}_f$ | | $< 0.8$ |
| $f_0(980)K^+$ | $\mathcal{B}\times\mathcal{B}_f$ | $6.5 \pm 2.5 \pm 1.6$ | $< 2.9$ |
| | $A_{CP}$ | $-31 \pm 25 \pm 8$ | |
| | $FF$ | $19 \pm 7 \pm 4$ | |
| | $\delta$ | $28 \pm 9 \pm 5$ | |
| $f_X(1500)K^+$ | $\mathcal{B}\times\mathcal{B}_f$ | $43 \pm 6 \pm 3$ | |
| | $A_{CP}$ | $-4 \pm 7 \pm 2$ | |
| | $FF$ | $121 \pm 19 \pm 6$ | $63.4 \pm 6.9$ |
| | $\delta$ | $74 \pm 5 \pm 2$ | 0 (fixed) |
| | $M$ | $1.539 \pm 0.020$ | $1.524 \pm 0.014$ |
| | $\Gamma$ | $0.257 \pm 0.033$ | $0.136 \pm 0.023$ |
| $f_0(1710)K^+$ | $\mathcal{B}\times\mathcal{B}_f$ | $1.7 \pm 1.0 \pm 0.3$ | |
| $f'(1525)K^+$ | $\mathcal{B}\times\mathcal{B}_f$ | | $< 4.9$ |
| $a_2(1320)K^+$ | $\mathcal{B}\times\mathcal{B}_f$ | | $< 1.1$ |
| $NR$ | $\mathcal{B}$ | $50 \pm 6 \pm 4$ | $24.0 \pm 1.5 \pm 1.8$ |
| | $FF$ | $141 \pm 16 \pm 9$ | $74.8 \pm 3.6$ |
| | $\delta$ | 0 (fixed) | $-68 \pm 9$ |

## 17.5 $B$-meson lifetimes, $B^0 - \overline{B}^0$ mixing, and symmetry violation searches

***Editors:***
*Soeren Prell (BABAR)*
*Bruce Yabsley (Belle)*

**Additional section writers:**
*Thomas Mannel*

The charged and neutral $B$ meson lifetimes, $\tau_{B^+}$ and $\tau_{B^0}$, and the $B^0 - \overline{B}^0$ oscillation frequency $\Delta m_d$, are fundamental parameters of $B$ meson decays. They provide important input for the determination of the CKM matrix elements $|V_{cb}|$ and $|V_{td}|$ (discussed in Sections 17.1 and 17.2). In addition, precise knowledge of $\tau_{B^0}$ and $\Delta m_d$ is necessary for the extraction of $CP$ asymmetries from the neutral $B$ decay-time distributions. Here we describe precision measurements of $\tau_{B^+}$, $\tau_{B^0}$ (Section 17.5.1), and $\Delta m_d$ (Section 17.5.2); measurements of $\Delta\Gamma_d$ are also discussed (Section 17.5.2.6). By relaxing the assumptions behind standard mixing analyses, it is also possible to test the quantum-mechanical nature of $B^0 - \overline{B}^0$ oscillations (Section 17.5.3), search for violations of $CP$, $T$, or even $CPT$ symmetry in mixing (Section 17.5.4), and search for violations of Lorentz symmetry (Section 17.5.5).

### 17.5.1 $B$-meson lifetimes

In 1983 the MAC and MARK II Collaborations (Fernandez et al., 1983; Lockyer et al., 1983) discovered, in 29 GeV center-of-mass energy $e^+e^-$ collisions recorded at the PEP storage ring at SLAC, that the impact parameters of high-momentum leptons in hadronic final states were largely positive. From the measured impact parameter distributions and assuming these leptons originated mostly from $b$ hadron decays, the collaborations estimated a $b$ hadron lifetime of the order of one picosecond. Such a long lifetime was unexpected. At the time, the phenomenological guidance on the strength of weak $b$ hadron decays was the mixing between the first and second quark generation, characterized by the Cabibbo angle $\theta_C$ (Section 16). If quark mixing between the second and the third generation was similar, the expected $b$ lifetime would be around 0.1 ps (Barger, Long, and Pakvasa, 1979). The long lifetime of $b$ hadrons was the first evidence that the magnitude of the CKM matrix element $V_{cb}$ is much smaller than $\sin\theta_C$. Along with first limits on the branching fractions of semileptonic $b \to u$ transitions, and thus $|V_{ub}/V_{cb}|$, from experiments at Cornell around the same time (Chen et al., 1984; Klopfenstein et al., 1983) and unitarity constraints, the measurement of $|V_{cb}|$ led to the first complete picture of the magnitudes of all the CKM matrix elements (Ginsparg and Wise, 1983). Soon after, it was realized that due to its long lifetime the $B^0$ can oscillate into a $\overline{B}^0$ before it decays, allowing for measurements of $B^0 - \overline{B}^0$ mixing and time-dependent $CP$ asymmetries.

At the time when the $B$ Factories started to record their first data, the Particle Data Group listed in their 2000 Review of Particle Physics (Groom et al., 2000) the averages of the $B^0$ and $B^+$ lifetimes and their ratio as: $\tau_{B^0} = (1.548 \pm 0.032)$ ps, $\tau_{B^+} = (1.653 \pm 0.028)$ ps, and $\tau_{B^+}/\tau_{B^0} = 1.062 \pm 0.029$, with relative uncertainties of 2.1%, 1.7%, and 2.7%, respectively.

While the first measurements of the magnitude of the CKM matrix element $V_{cb}$ were provided by the initial $b$ hadron lifetime measurements, the most precise determination of $|V_{cb}|$, based on advances in the theoretical descriptions of $B$-meson decays, now comes from semileptonic branching ratios (see Section 17.1).

In the following, we briefly discuss the theory of $B$ meson lifetimes (Section 17.5.1.1), and the motivation and principles of lifetime measurements (Section 17.5.1.2), before reviewing lifetime measurements at the $B$ Factories using fully-reconstructed (Section 17.5.1.3) and partially-reconstructed final states (Section 17.5.1.4). Averages of the $B$ lifetimes and their ratio are presented in Section 17.5.1.5.

#### 17.5.1.1 Theory of $B$ meson lifetimes

From the theoretical side the lifetime (or equivalently the total decay rate $\Gamma$) of a heavy quark hadron is a fully inclusive quantity for which a systematic expansion in powers of $\Lambda_{\rm QCD}/m_Q$ can be performed (Bigi, 1996; Neubert and Sachrajda, 1997). Schematically one obtains an expression of the form

$$\Gamma = \Gamma_0 + \Gamma_1 \left(\frac{\Lambda_{\rm QCD}}{m_Q}\right) + \Gamma_2 \left(\frac{\Lambda_{\rm QCD}}{m_Q}\right)^2 + \Gamma_3 \left(\frac{\Lambda_{\rm QCD}}{m_Q}\right)^3 + \cdots \qquad (17.5.1)$$

*The leading term in the decay rate*

It turns out that the leading term of this expansion does not depend on any hadronic matrix element and is simply the decay of a free quark. This is illustrated in Fig. 17.5.1: It depicts the square of the amplitude of a heavy quark decaying via a four quark operator into three final state fermions, *i.e.* the internal lines should not be interpreted as propagators, but rather as the corresponding phase space integration. Since only the heavy quark is involved, to this level of the expansion, the lifetime of all charm and bottom hadrons, respectively, are predicted to be identical. Neglecting CKM suppressed contributions and the masses of the electron, the muon, the up and the down quark, the leading term for charm hadrons (*i.e.* without QCD corrections) can be written as,

$$\Gamma_c = |V_{cs}|^2 \left[ N_c \Gamma(c \to s\overline{u}d) + 2\Gamma(c \to s\ell\overline{\nu}_\ell) \right], \quad (17.5.2)$$

where $N_c$ is the number of colors, $\ell = e, \mu$, and

$$\Gamma(c \to s\overline{f}f') = \frac{G_F^2 m_c^5}{192\pi^3} f_{\mathrm{PS}}, \quad (17.5.3)$$

where $f_{\mathrm{PS}}$ is a phase space factor depending on the mass of the charm and the strange quarks.

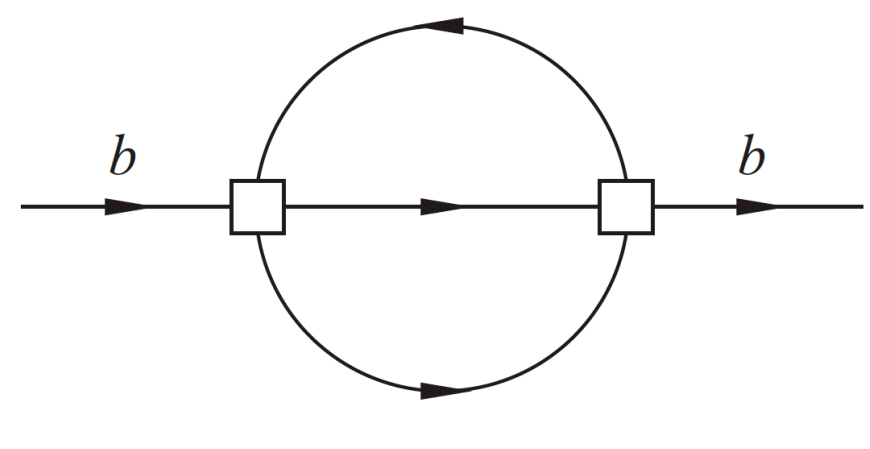

**Figure 17.5.1.** Illustration of the leading term of the heavy quark expansion for the total rate.

For bottom hadrons this expression is slightly more complicated since more final states are involved. For the leading term, neglecting again CKM suppressed contributions, setting $|V_{cs}| = |V_{ud}| \approx 1$, and neglecting the $e$, $\mu$, $u$, and $d$ masses, one obtains

$$\Gamma_b = |V_{cb}|^2 \Bigg[ N_c[\Gamma(b \to c\overline{c}s) + \Gamma(b \to c\overline{u}d)] + 2\Gamma(b \to c\ell\overline{\nu}_\ell) + \Gamma(b \to c\tau\overline{\nu}_\tau) \Bigg], \quad (17.5.4)$$

where now

$$\Gamma(b \to c\overline{f}f') = \frac{G_F^2 m_b^5}{192\pi^3} f_{(\overline{f}f')}, \quad (17.5.5)$$

and $f_{(\overline{f}f')}$ is a phase space function depending on the bottom and the charm mass as well as on the masses of the two additional fermions $f$ and $f'$.

Although the analytic expression for the phase space functions are not complicated, we give here only a simple numerical consideration. Putting in the phase space functions, one obtains

$$\Gamma_c \approx 3.5 \times \frac{G_F^2 m_c^5}{192\pi^3} |V_{cs}|^2 = [1.1 \times 10^{-12}\,\mathrm{s}]^{-1}, \quad (17.5.6)$$

$$\Gamma_b \approx 2.9 \times \frac{G_F^2 m_b^5}{192\pi^3} |V_{cb}|^2 = [1.2 \times 10^{-12}\,\mathrm{s}]^{-1}. \quad (17.5.7)$$

Being the first term of a systematic expansion, it is reassuring that these numbers are in the right ballpark. Note that the rates have to be proportional to $m_Q^5$ to compensate the dimension of the Fermi coupling $G_F$; however the full dependence on the heavy quark mass is not as strong due to the phase space factors. The fact that the bottom and charm lifetimes are still comparable is due to the small magnitude of the CKM element $|V_{cb}|$ relative to $|V_{cs}|$.

The prediction that the heavy-hadron lifetimes are identical was considered a problem in the early days of the heavy quark expansion. In fact, we have for example $\tau(D^+)/\tau(D^0) = 2.52 \pm 0.09$, indicating large corrections from higher-order terms in the expansion. Furthermore, the leading term depends on a high power of $m_Q$, such that any uncertainty in $m_Q$ would be amplified so much that it was originally believed that no precise predictions could be made. However, including QCD corrections in combination with suitable mass definitions, this could be remedied.

We note in passing that the naïve spectator model also predicts the semileptonic branching ratios. Taking into account only the Cabibbo-allowed contributions and neglecting the masses of the final state fermions we obtain

$$\mathcal{B}(D \to X\ell\overline{\nu}) = \frac{\Gamma(c \to s\ell\overline{\nu})}{N_c \Gamma(c \to s\overline{d}u) + N_{\mathrm{lept}} \Gamma(c \to s\ell\overline{\nu})}, \quad (17.5.8)$$

where $\ell = e$ or $\mu$, $N_c = 3$ is the number of colors, and $N_{\mathrm{lept}} = 2$ is for the two leptons that can appear as a final state in a $D$ decay. With the approximation $|V_{cs}| = |V_{ud}| \sim 1$, and final state masses neglected, the partial widths are equal,

$$\Gamma(c \to s\overline{d}u) = \Gamma(c \to s\ell\overline{\nu}) = \frac{G_F^2 m_c^5}{192\pi^3}, \quad (17.5.9)$$

so we find

$$\mathcal{B}(D \to X\ell\overline{\nu}) = \frac{1}{3+2} = 0.2. \quad (17.5.10)$$

For bottom we can perform the same calculation, however here one has to take into account phase space factors, since the phase space for *e.g.* $b \to c\overline{c}s$ is significantly different from that for $b \to c\overline{u}d$. Taking this effect into account one arrives at

$$\mathcal{B}(B \to X\ell\overline{\nu}) = 0.17\,. \quad (17.5.11)$$

Again these predictions are in the right ballpark, but depend strongly on the quark masses and the definitions used for these masses. Including the higher order terms in $\alpha_S$ as well as in the heavy quark expansion improves the precision of the predictions dramatically. In particular, the determination of $|V_{cb}|$ is performed on the basis of the total semileptonic rate, which is computed at the percent level of precision.

### *Higher-order terms*

The higher order terms in the heavy quark expansion have been investigated in detail. The term of order $\Lambda_{\mathrm{QCD}}/m_Q$ vanishes due to heavy quark symmetries, so the first non-perturbative input to the lifetimes appears at the second order of the expansion. To this order, the kinetic energy parameter $\mu_\pi^2$ and the chromo-magnetic moment $\mu_G^2$ appear as non-perturbative input (for the precise definition of these parameters see Section 17.1). However, assuming light-quark flavor symmetry, one obtains $\mu_\pi^2(B^0) = \mu_\pi^2(B^\pm) = \mu_\pi^2(B_s^0)$ and hence the second order in the expansion still does not induce a lifetime difference between $B^0$, $B^\pm$, and $B_s^0$.

A lifetime difference between the bottom mesons needs to involve the spectator quark. Contributions of this kind are illustrated in Fig. 17.5.2. However, such contributions are induced only at order $(\Lambda_{\mathrm{QCD}}/m_Q)^3$, which was taken as an embarrassment at the time this was derived, since the lifetimes in the $D$ meson system differ by a factor as large as 2.5. However, subsequently it has been found that the coefficient $\Gamma_3$ can be enhanced by a loop factor $16\pi^2$, which at least qualitatively explains this large effect. This can actually be seen by comparing Figs 17.5.1 and 17.5.2: the leading term shown in Fig 17.5.1 is a two-loop diagram, leading to a factor $(1/(16\pi^2))^2$, whereas the spectator contributions shown in Fig 17.5.2 are one-loop diagrams with only a single power of $1/(16\pi^2)$ (Neubert and Sachrajda, 1997).

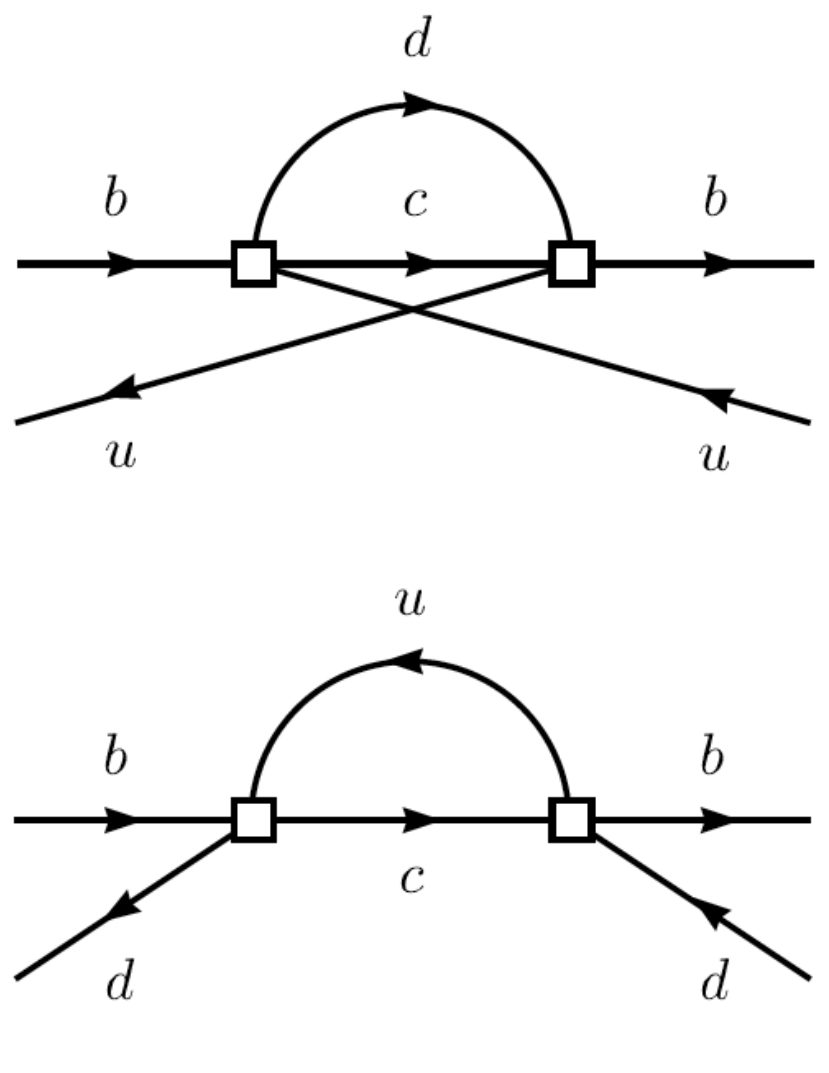

**Figure 17.5.2.** Spectator contributions.

Over the past ten years lifetime calculations have been refined by adding higher order terms in the $1/m_b$ expansion as well as QCD corrections. A recent review can be found in Lenz (2008).

#### 17.5.1.2 Motivation and principles of lifetime measurements

There were both theoretical and experimental reasons for the $B$ Factories to measure the $B$-meson lifetimes more precisely:

- Predictions for lifetime ratios based on a naïve estimate of the hadronic matrix elements yielded $\tau_{B^+}/\tau_{B^0} = 1.067 \pm 0.027$ (Becirevic, 2001). While in agreement with this prediction, the pre-$B$ Factory data were not conclusive on whether the charged or neutral $B$ lifetime was longer, motivating a more precise measurement of $\tau_{B^+}/\tau_{B^0}$ to provide a stronger test of these calculations.
- The $B^0$ meson lifetime provides an essential input to the measurements of the $B^0 - \overline{B}^0$ oscillation frequency (see Section 17.5.2) and time-dependent *CP* asymmetries including the angles $\phi_1$ and $\phi_2$ of the Unitarity Triangle (see Sections 17.6 and 17.7). Accurate values of $\tau_{B^0}$ and $\Delta m_d$ reduce the systematic uncertainties in these analyses of time-dependent *CP* asymmetries.

The most precise measurements of the $B$-meson lifetimes before the first $B$ Factory results became available were from experiments at the $Z^0$ resonance and CDF. These experiments measured the distance $l$ the $B$ meson travels from its production point to its decay vertex. The production point is, respectively, the $e^+e^-$ or $p\overline{p}$ interaction point and the decay vertex is determined from the $B$ decay products. From this decay distance $l$, the measured $B$ momentum $p_B$, and the known $B$ mass $m_B$, they determined the proper time of the $B$-meson decay $t = l/c(\beta\gamma)_B = m_B l/(p_B c)$. The proper-time distribution of the $B$-meson candidates is given by $\Gamma(t) = \frac{1}{\tau_B}\exp(-t/\tau_B)$ before accounting for detector resolution and backgrounds. The experiments extracted the $B$-meson lifetimes from fits to the measured proper-time spectra. While the ARGUS and CLEO experiments had collected large samples of $B$ mesons at the $\Upsilon(4S)$ resonance, their $B$ mesons were essentially produced at rest in the laboratory frame, rendering a proper-time method through decay-length measurements impossible.

These earlier $B$-lifetime measurements are characterized by high-precision measurements of the relative decay length of the $B$ mesons ($\sigma_l/\langle l\rangle \approx 10\%$), but typically suffered from a combination of relatively small signal samples, large backgrounds, and in the case of partially-reconstructed $B$ mesons, a poor measurement of the $B$ momentum. In contrast, the measurements from *BABAR* and Belle have worse $\sigma_l/\langle l\rangle$ resolution, but their high-statistics $B$ samples have little background and excellent knowledge of the $B$ momentum.

A principal difference between the $B$-meson lifetime measurements at previous experiments and at the asymmetric-energy $B$ Factories is the knowledge of the $B$ production point. At all experiments the $B$ mesons are produced in the luminous region of the particle beams (beam spot). The coordinates of the beam spot are well known. The beam spot size is much smaller in the plane transverse to the beam direction than along the beam direction. At the LEP and Tevatron experiments and at SLD most $B$ mesons travel a measurable distance in the transverse plane before they decay, and the $B$ meson proper time is derived from this distance. In fits to the proper-time distributions, events with measured $t < 0$ provide valuable information about the proper-time resolution function. Since there are no true negative proper times, all events with measured $t < 0$ are due to resolution effects. In contrast, at the $B$ Factories the $B$ mesons are barely moving in the center-of-mass frame. Thus their transverse momentum and transverse flight distance are close to zero and cannot be used for a precise proper-time measurement. The length of the beam spot in the $z$ direction is about a centimeter in *BABAR* and Belle and there are no fragmentation tracks coming from the $B$ production point (as only a $B\overline{B}$-pair is produced in the decay of the $\Upsilon(4S)$). Therefore the $z$ coordinate of the $B$ production vertex cannot be reconstructed with good precision. Instead, at the $B$ Factories the distance $\Delta z$ between the decay vertices of the two $B$ mesons is measured. The proper-time difference is then given to good approximation by

$$\Delta t \approx \Delta z/(c(\beta\gamma)_B), \qquad (17.5.12)$$

where $(\beta\gamma)_B$ is the Lorentz boost factor of the $B$ meson in the lab frame (see Section 6.5). The $\Delta t$ distribution is given by

$$\Gamma(\Delta t) = \frac{1}{2\tau_B} \exp\left(-|\Delta t|/\tau_B\right). \qquad (17.5.13)$$

It is symmetric around $\Delta t = 0$. Detector resolution effects will smear this distribution, but there is no region in $\Delta t$ that allows a similarly clean access to the $\Delta t$ resolution function as in the experiments at the $Z^0$ and CDF and DØ (see Fig. 17.5.3). One of the challenges of the $B$-lifetime measurements at the $B$ Factories is to disentangle the underlying true $\Delta t$ distribution from the resolution function.

Both *BABAR* and Belle use multiple samples of $B$ mesons to determine the $B^0$ and $B^+$ lifetimes and their ratio. One of the $B$ mesons, $B_{\rm rec}$, is typically reconstructed in an exclusive final state. The various samples differ in their $B$ meson yield per inverse femtobarn and in their signal purity.[57] More exclusive samples have less background, but also a smaller yield. In the lifetime analyses, the $z$ position of the $B_{\rm rec}$ decay vertex $z_{\rm rec}$ is determined from its decay products. The $z$ position $z_{\rm oth}$ of the decay vertex of the other $B$ meson, $B_{\rm oth}$, is reconstructed from the tracks not belonging to $B_{\rm rec}$. The proper-time difference $\Delta t$ is then calculated from $\Delta z = z_{\rm rec} - z_{\rm oth}$ using Eq. (17.5.12). It turns out that the uncertainty in $\Delta t$ is dominated by the uncertainty in $z_{\rm oth}$ and is almost the same for all lifetime analyses at the $B$ Factories. The $B$ lifetimes are extracted from a fit to the $\Delta t$ distributions of the selected candidates after accounting for detector resolution effects and background. In the following, we will briefly describe the various measurements of the $B$-meson lifetimes by the $B$ Factories. The results of these analyses are summarized in Table 17.5.1; averages are discussed in Section 17.5.1.5.

### 17.5.1.3 Fully-reconstructed final states

$B$ lifetime measurements with samples in which one $B$ decays to an exclusive hadronic final state have the lowest background. *BABAR* measures the $B^0$ and $B^+$ lifetimes with the hadronic decays $B^0 \to D^{(*)-}\pi^+$, $D^{(*)-}\rho^+$, $D^{(*)-}a_1^+$, $J/\psi K^{*0}$ and $B^+ \to \overline{D}^{(*)0}\pi^+$, $J/\psi K^+$, $\psi(2S)K^+$ in a data sample of $20.6\,{\rm fb}^{-1}$ (Aubert, 2001c). Belle performs an analysis combining the exclusive hadronic final states $B^0 \to D^{(*)-}\pi^+$, $D^{*-}\rho^+$, $J/\psi K_S^0$, $J/\psi K^{*0}$ to measure the $B^0$ lifetime and the modes $B^+ \to \overline{D}^0\pi^+$, $J/\psi K^+$ to measure the $B^+$ lifetime in a sample of $29.1\,{\rm fb}^{-1}$ (Abe, 2002m). The decay channels $K^+\pi^-$, $K^+\pi^-\pi^0$, $K^+\pi^-\pi^+\pi^-$, and $K_S^0\pi^+\pi^-$ are used to reconstruct $\overline{D}^0$ candidates, while the modes $K^+\pi^-\pi^+$ and $K_S^0\pi^-$ are used for $D^-$ candidates (Belle does not use the $D$ decay modes involving a $K_S^0$). Charged $D^{*-}$ candidates are formed by combining a $\overline{D}^0$ with a soft $\pi^-$.

[57] The signal purity is the fraction of signal events in the selected candidates (see also Section 4.3). It is often defined for a region of about $\pm 2$ standard deviations around the signal peak (for example, in $m_{\rm ES}$ or $\Delta E$).

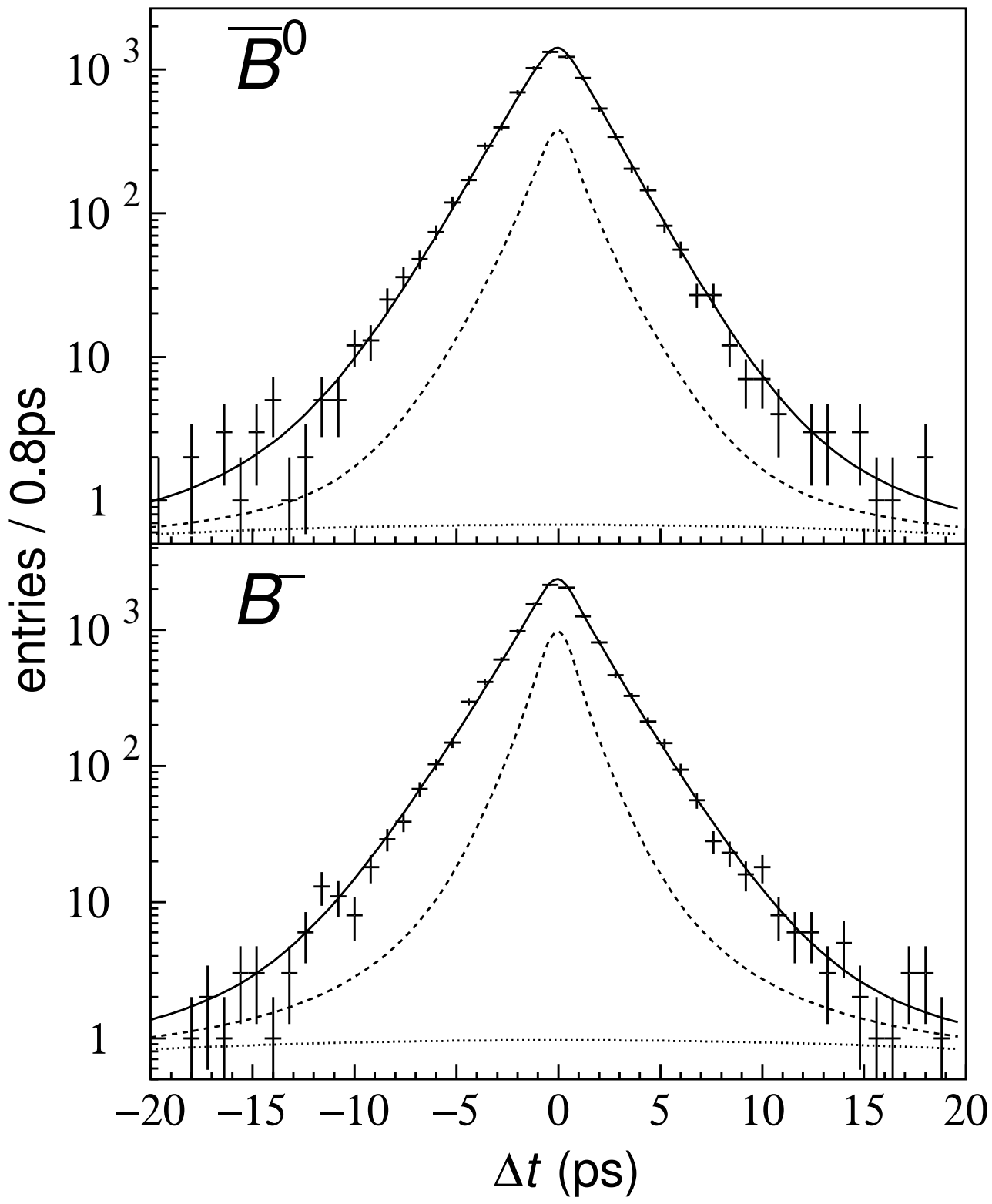

**Figure 17.5.3.** The $\Delta t$ distributions of $\overline{B}^0$ (top) and $B^-$ (bottom) candidates (plus *c.c.*) for fully-reconstructed $B$ decays to hadronic final states. The dashed lines represent the sum of the background and outlier components, and the dotted lines represent the outlier component (Abe, 2002m).

The $B^0$ candidates are formed by combining a $D^{*-}$ or $D^-$ with a $\pi^+$, $\rho^+$ ($\rho^+ \to \pi^+\pi^0$) or $a_1^+$ ($a_1^+ \to \pi^+\pi^-\pi^+$). The $B^0 \to J/\psi K^{*0}$ and $B^0 \to \psi(2S)K^{*0}$ candidates are reconstructed from combinations of a $J/\psi$ or a $\psi(2S)$ candidate, in the decay modes $e^+e^-$ and $\mu^+\mu^-$, with a $K^{*0}$ ($K^{*0} \to K^+\pi^-$). The $\psi(2S)$ candidates are reconstructed in their decays to $J/\psi\pi^+\pi^-$. In these measurements, the collaborations impose constraints on the $B$ candidates requiring them to be compatible with one of the final states mentioned above. The corresponding branching fractions for the $B$ and $D$ decays to these final states are at most a few percent. Therefore, the selected signal samples have relatively small $B^0$ and $B^+$ yields (for example, 291 $B^0$ and 304 $B^+$ per inverse femtobarn of data for the *BABAR* analysis). Due to the tight selection criteria, a main background present in other analyses that arises from incorrect combinations of tracks is highly suppressed, leading to event samples with high signal purities of 80%–90%.

The $z$ position of the decay vertex, $z_{\rm rec}$, of the fully-reconstructed $B$ meson, $B_{\rm rec}$ , is measured with high precision, typically of the order of $\sigma(z_{\rm rec}) \sim 50$ $\mu$m. The decay vertex position of the other $B$, $z_{\rm oth}$, is determined from all tracks not belonging to $B_{\rm rec}$ as described in Chapter 6. For these samples, the $z_{\rm rec}$ resolution is 100–200 $\mu$m with an RMS value of about 170 $\mu$m. Thus, the $\Delta z$ resolution is dominated by the resolution of $z_{\rm oth}$. It is similar for all decay modes ($\sigma(\Delta z) = 180 - 190\ \mu$m). Belle converts the measured $\Delta z$ into a $\Delta t$ value according to Eq. (17.5.12), whereas in fully-reconstructed decays *BABAR* uses a more precise approximation by exploiting the precise knowledge of the $B$ flight direction to correct for the $B$ momentum in the $\Upsilon(4S)$ frame (Eq. 6.5.5). The $\Delta t$ distributions of the selected $B^0$ and $B^+$ candidates are then fit to a likelihood function that describes the true $\Delta t$ distribution of the signal events (Eq. 17.5.13), convoluted with a $\Delta t$ signal resolution function $\mathcal{R}_{\rm sig}$ to account for the uncertainty in the $\Delta t$ measurements; and to an empirical $\Delta t$ distribution describing background events. *BABAR* uses a signal $\Delta t$ resolution function $\mathcal{R}_{\rm sig}$ consisting of the sum of a Gaussian distribution with zero mean and its convolution with an exponential decay that models the bias of $z_{\rm oth}$ due to tracks originating from a displaced decay vertex of a charm meson. Charged and neutral $B$ decays are described with the same $\Delta t$ resolution function. Belle's signal $\Delta t$ resolution function $\mathcal{R}_{\rm sig}$ is formed by the convolution of four components: the detector resolutions for $z_{\rm rec}$ and $z_{\rm oth}$, the bias in $z_{\rm oth}$ due to tracks originating from the decay of a charm meson, and the kinematic approximation that the $B$ mesons are at rest in the center-of-mass frame (Tajima, 2004). Both resolution functions have a term accounting for a small number of poorly reconstructed vertices, so-called $\Delta t$ outliers. Both experiments describe the background $\Delta t$ distribution with a prompt term (*i.e.* zero lifetime) and a term with an effective background lifetime. The background $\Delta t$ resolution functions for the component with effective lifetime is of the same form as the signal resolution functions, but with separate parameters in order to minimize correlations with the signal resolution parameters. The $\Delta t$ resolution function is discussed further in Chapter 10.

**Table 17.5.1.** $B$ Factory measurements of $\tau_{B^0}$, $\tau_{B^+}$, and $\tau_{B^0}/\tau_{B^+}$ along with the journal paper, selected final state, signal purity $f_{\rm signal}$, $B$ meson signal yield, and integrated luminosity for each measurement. The purity and yield values marked with an asterisk * are approximate.

| Experiment | Method | $f_{\rm signal}$ | Yield $[B/{\rm fb}^{-1}]$ | $\int \mathcal{L}\, dt$ $[{\rm fb}^{-1}]$ | |
|---|---|---|---|---|---|
| *Neutral $B$ meson lifetime $\tau_{B^0}$:* | | | | | $\tau_{B^0}$ [ps] |
| BABAR (Aubert, 2001c) | Excl. hadronic modes | 90% | 291 | 21 | $1.546 \pm 0.032 \pm 0.022$ |
| BABAR (Aubert, 2003e) | Incl. $D^*\pi$, $D^*\rho$ | 55% | 603 | 21 | $1.533 \pm 0.034 \pm 0.038$ |
| BABAR (Aubert, 2003m) | Excl. $D^*l\nu$ | 76% | 680 | 21 | $1.523^{+0.024}_{-0.023} \pm 0.022$ |
| BABAR (Aubert, 2002f) | Incl. $D^*l\nu$ | 53% | 4430 | 21 | $1.529 \pm 0.012 \pm 0.029$ |
| BABAR (Aubert, 2006s) | Incl. $D^*l\nu$ | 64% | 605 | 81 | $1.504 \pm 0.013^{+0.018}_{-0.013}$ |
| Belle (Abe, 2002m) | Excl. hadronic modes | 82%* | 220* | 29 | $1.554 \pm 0.030 \pm 0.019$ |
| Belle (Abe, 2005c) | Excl. had. modes + $D^*l\nu$ | 81% | 707 | 140 | $1.534 \pm 0.008 \pm 0.010$ |
| BABAR-Belle average | | | | | $1.530 \pm 0.005 \pm 0.009$ |
| *Charged $B$ meson lifetime $\tau_{B^+}$:* | | | | | $\tau_{B^+}$ [ps] |
| BABAR (Aubert, 2001c) | Excl. hadronic modes | 93% | 304 | 21 | $1.673 \pm 0.032 \pm 0.023$ |
| Belle (Abe, 2002m) | Excl. hadronic modes | 75%* | 310* | 29 | $1.695 \pm 0.026 \pm 0.015$ |
| Belle (Abe, 2005c) | Excl. hadronic modes | 81% | 319 | 140 | $1.635 \pm 0.011 \pm 0.011$ |
| BABAR-Belle average | | | | | $1.640 \pm 0.010 \pm 0.010$ |
| *$\tau_{B^+}/\tau_{B^0}$:* | | | | | $\tau_{B^+}/\tau_{B^0}$ |
| BABAR (Aubert, 2001c) | Excl. hadronic modes | 93%, 90% | 304, 291 | 21 | $1.082 \pm 0.026 \pm 0.012$ |
| Belle (Abe, 2002m) | Excl. hadronic modes | 75%, 82%* | 310, 220* | 29 | $1.091 \pm 0.023 \pm 0.014$ |
| Belle (Abe, 2005c) | Excl. had. modes + $D^*l\nu$ | 81%, 81% | 319, 707 | 140 | $1.066 \pm 0.008 \pm 0.008$ |
| BABAR-Belle average | | | | | $1.068 \pm 0.009 \pm 0.007$ |

*BABAR* and Belle determine the values of $\tau_{B^0}$ and $\tau_{B^+}$ from a simultaneous fit to the samples of $B^0$ and $B^+$ candidates. *BABAR* measures $\tau_{B^0} = (1.546 \pm 0.032 \pm 0.022)$ ps and $\tau_{B^+} = (1.673 \pm 0.032 \pm 0.023)$ ps, while Belle measures $\tau_{B^0} = (1.554 \pm 0.030 \pm 0.019)$ ps and $\tau_{B^+} = (1.695 \pm 0.026 \pm 0.015)$ ps. The measurements of the $B^0$ and the $B^+$ lifetimes share the same sources of systematic uncertainty. Some of these uncertainties cancel in the ratio of the lifetimes $r_\tau \equiv \tau_{B^+}/\tau_{B^0}$. In a separate fit the parameter $\tau_{B^+}$ is replaced with $r_\tau \cdot \tau_{B^0}$ to estimate the statistical error of the lifetime ratio. *BABAR* and Belle measure, respectively, $\tau_{B^+}/\tau_{B^0} = 1.082 \pm 0.026 \pm 0.012$ and $\tau_{B^+}/\tau_{B^0} = 1.091 \pm 0.023 \pm 0.014$. The largest contribu-tions to the systematic uncertainties in the measured lifetimes come from the modeling of the signal $\Delta t$ resolution function (0.009 – 0.014 ps) and the background $\Delta t$ distribution (0.005 – 0.012 ps), the alignment of the vertex detector (0.008 ps), the knowledge of the $z$ scale of the detector (0.008 ps), and limited statistics of the MC simulation (0.007 – 0.009 ps). The dominant contributions to the systematic error in $r_\tau$ come from limited MC statistics (0.005 – 0.006), uncertainties in the background $\Delta t$ distributions (0.005 – 0.011), and the signal $\Delta t$ resolution function (0.006 – 0.008).

In another analysis *BABAR* uses events in which $B_{\rm rec}$ is reconstructed in the semileptonic decay $B^0 \to D^{*-}l^+\nu$ ($l = e, \mu$) to determine the $B^0$ lifetime (Aubert, 2003m). The $B$ yield is larger than for the hadronic final state analysis due to the large $B$ semileptonic branching fraction. They reconstruct 680 $B/\,{\rm fb}^{-1}$. Due to the missing neutrino the background level is higher than in the sample of fully-reconstructed hadronic $B$ decays. The combinatorial $D^{*-}$ background is about 18% and the sum of the backgrounds from events where the $D^{*-}$ and the lepton come from different $B$ decays, events with a fake lepton candidate and events from continuum $c\overline{c} \to D^{*-}X$ processes add up to 5 – 8% depending on the lepton flavor. In this analysis, *BABAR* simultaneously fits for $\tau_{B^0}$ and the $B^0 - \overline{B}^0$ mixing frequency $\Delta m_d$ (see also Section 17.5.2). Because of the different $\Delta t$ distributions for mixed ($B^0B^0$ or $\overline{B}^0\overline{B}^0$) and unmixed ($B^0\overline{B}^0$) events, separately fitting the two $\Delta t$ distributions enhances the sensitivity to the common signal $\Delta t$ resolution function. As a result the uncertainty of $\tau_{B^0}$ is reduced by approximately 15%. *BABAR* measures the $B^0$ lifetime to be $\tau_{B^0} = (1.523^{+0.024}_{-0.023} \pm 0.022)$ ps. The dominant systematic error sources are the same as for the analyses of the hadronic final states and similar in size. A large additional systematic uncertainty in the $\tau_{B^0}$ measurement comes from the limited statistical precision in determining the bias due to the background modeling. By comparing the fitted $\tau_{B^0}$ in simulated events, *BABAR* observes a shift of (0.022 ± 0.009) ps between a signal-only sample and a signal-plus-background sample. The measured $B^0$ lifetime is corrected for the observed bias from the fit to the MC sample with background; the full statistical uncertainty in $\tau_{B^0}$ from this fit (±0.018 ps) is assigned as systematic uncertainty.

Belle also performs a measurement of the $B$ lifetimes and their ratio in a larger sample of 140 fb$^{-1}$ (Abe, 2005c). In this analysis they reconstruct $B^0$ and $B^+$ candidates in the same hadronic decay modes as in their previous analysis. In addition they reconstruct $B^0$ candidates in the semileptonic decay $B^0 \to D^{*-}l^+\nu$. Using a fit to the $\Delta t$ distributions of the signal candidates, they determine the $B^0$ and $B^+$ lifetimes and the $B^0 - \overline{B}^0$ mixing frequency $\Delta m_d$ simultaneously. The analysis of the neutral $B$ decays is described in more detail in Section 17.5.2. Belle measures $\tau_{B^0} = (1.534 \pm 0.008 \pm 0.010)$ ps, $\tau_{B^+} = (1.635 \pm 0.011 \pm 0.011)$ ps and $\tau_{B^+}/\tau_{B^0} = 1.066 \pm 0.008 \pm 0.008$. The largest contributions to the systematic uncertainties in the measured lifetimes come from uncertainties in the vertex reconstruction (0.005 – 0.007 ps) and the modeling of the background (0.007 ps). The dominant contributions to the systematic error in $r_\tau$ come from uncertainties in the background $\Delta t$ distributions (0.005) and the signal $\Delta t$ resolution function (0.004).

#### 17.5.1.4 Partially-reconstructed final states

*BABAR* also measures the $B^0$ meson lifetime in a sample of 21 fb$^{-1}$ using the decay modes $B^0 \to D^{*-}l^+\nu$ (Aubert, 2002f) and $B^0 \to D^{*-}\pi^+$, $B^0 \to D^{*-}\rho^+$ (Aubert, 2003e) with a partially-reconstructed $D^{*-}$ in the final state. These measurements also serve as a proof-of-principle for the analyses of the time-dependent *CP* asymmetries in $B \to D^{(*)\mp}\pi^\pm$ to extract $\sin(2\phi_1+\phi_3)$ (see Section 17.8.5).

In the measurement of $\tau_{B^0}$ with $B^0 \to D^{*-}l^+\nu$ decays, *BABAR* requires a high-momentum lepton ($1.4 < p^*_l < 2.3\,{\rm GeV}/c$) and an opposite-charge soft pion ($\pi_s$) consistent with coming from the decay $D^{*-} \to \overline{D}^0\pi^-_s$ ($p^*_{\pi_s} < 0.19\,{\rm GeV}/c$). The $D^{*-}$ momentum is inferred from the $\pi_s$ momentum without reconstructing the $\overline{D}^0$ (see Eq. 7.3.6). The analysis of this inclusive final state does not suffer from the small $\overline{D}^0$ branching fractions to exclusive final states and consequently has a large $B$ yield (4430 $B/\,{\rm fb}^{-1}$). However, without the additional constraints from the $\overline{D}^0$ reconstruction the signal purity of the selected $B$ candidates is only 53%. The $B_{\rm rec}$ decay vertex is calculated from the lepton and $\pi_s$ tracks, and the beam spot. The decay point of the $B_{\rm oth}$ is determined from the remaining tracks in the event. In events that have another high-momentum lepton ($p^*_l > 1.1$ GeV/$c$), the $B$ vertex is calculated from this lepton track constrained to the beam spot in the transverse plane. Otherwise, all tracks with a center-of-mass angle greater than 90° with respect to the $\pi_s$ direction are considered. This requirement removes most of the tracks from the decay of the $\overline{D}^0$ daughter of the $D^{*-}$, which would otherwise bias the reconstruction of the $B_{\rm oth}$ vertex position. Tracks are also removed if they contribute more than 6 to the vertex $\chi^2$. *BABAR* measures the $B^0$ lifetime with a binned maximum likelihood fit to the $\Delta t$ and $\sigma_{\Delta t}$ distributions of the selected $B$ candidates to be $\tau_{B^0} = (1.529 \pm 0.012 \pm 0.029)$ ps. For this result, the fitted $B^0$ lifetime is multiplied by a correction factor $\mathcal{R}_{\overline{D}^0} = 1.032 \pm 0.007 \pm 0.007$ to account for daughter tracks of the $\overline{D}^0$ included in the calculation of the $B_{\rm oth}$ decay vertex. The largest systematic uncertainties in $\tau_{B^0}$ are due to the knowledge of the fractions and parameterizations of the background types (0.015 ps), the $\Delta t$ resolution model (0.017 ps) and $\mathcal{R}_{\overline{D}^0}$ (0.015 ps).

In a more recent analysis with 81 fb$^{-1}$, *BABAR* uses $B^0 \to D^{*-}l^+\nu$ decays with a partially-reconstructed $D^{*-}$ to measure $\tau_{B^0}$ and the $B^0 - \overline{B}^0$ oscillation frequency $\Delta m_d$ (Aubert, 2006s). They require the other $B^0$ in the event $B_{\rm oth}$ also to decay semileptonically and determine its decay vertex by constraining the high-energy lepton to the beam spot. After correcting for a small bias (−0.006 ps) observed in MC-simulated events they measure $\tau_{B^0} = (1.504 \pm 0.013^{+0.018}_{-0.013})$ ps. The dominant contributions to the systematic error in $\tau_{B^0}$ come from uncertainties in the alignment ($^{+0.013}_{-0.004}$ ps) and $z$ scale (0.007 ps) of the SVT, and from MC statistics (0.007 ps).

*BABAR* also measures the $B^0$ lifetime with a partially-reconstructed $D^{*-}$ in the decays $B^0 \to D^{*-}h^+$, where $h^+$ is either a $\pi^+$ or a $\rho^+$ (Aubert, 2003e). Similarly to the partial reconstruction of the semi-leptonic final state, they reconstruct only the soft pion $\pi_s$ from the decay $D^{*-} \to \overline{D}^0\pi_s^-$ and the $D^{*-}$ momentum is inferred from the $\pi_s$ momentum. The main variable to suppress background in this analysis is the missing $\overline{D}^0$ mass $m_{\rm miss}$, which peaks at the nominal $D^0$ mass with a spread of 3 MeV/$c^2$ for $B^0 \to D^{*-}\pi^+$ and 3.5 MeV/$c^2$ for $B^0 \to D^{*-}\rho^+$. Additional variables to suppress backgrounds include the angle between $h$ and the $B^0$, the $D^{*-}$ and $\rho^+$ helicity angles, and event shape variables. After all selection requirements are applied, the signal purity is approximately 55%. The dominant background comes from continuum events. The remaining background from $B\overline{B}$ events is due to random $h$ and $\pi_s$ combinations and feed-down from $B \to D^{**}\pi$, $B^0 \to D^{*-}\rho^+$ (for $B^0 \to D^{*-}\pi^+$), and $B^0 \to D^{*-}a_1^+$ (for $B^0 \to D^{*-}\rho^+$). The $z$ position of the $B^0$ decay vertex is determined from the $h$ and $\pi_s$ tracks constrained to the nominal beam spot. The decay vertex of $B_{\rm oth}$ is determined in the same way as in *BABAR*'s early analysis of $B^0 \to D^{*-}l^+\nu$ (Aubert, 2002f). For the mode $B^0 \to D^{*-}\pi^+$ they calculate an event-by-event $\Delta z$ correction to account for tracks from the $\overline{D}^0$ included in the vertex of $B_{\rm oth}$. In both modes a small additional correction to the fitted $B^0$ lifetime is applied. *BABAR* uses several data control samples to determine the different background fractions in the signal sample and their *p.d.f.* parameters. These parameters are fixed in the fit to the signal sample. The fitted lifetimes are $\tau_{B^0} = (1.510 \pm 0.040 \pm 0.041)$ ps in $B^0 \to D^{*-}\pi^+$ and $\tau_{B^0} = (1.616 \pm 0.064 \pm 0.075)$ ps in $B^0 \to D^{*-}\rho^+$. The combined result accounting for correlated errors is $\tau_{B^0} = (1.533 \pm 0.034 \pm 0.038)$ ps. The dominant uncertainties in the measurements with the modes $B^0 \to D^{*-}\pi^+$ and $B^0 \to D^{*-}\rho^+$ come from the knowledge of the composition of the background and its *p.d.f.* parameters (0.024 ps, 0.050 ps), limited MC statistics (0.021 ps, 0.042 ps), and the $\overline{D}^0$ track bias (0.017 ps, 0.026 ps).

#### 17.5.1.5 Averages of $\tau_{B^0}$, $\tau_{B^+}$ and $\tau_{B^+}/\tau_{B^0}$

The world averages of the $B^0$ and $B^+$ lifetimes and their ratio are calculated by HFAG from the *BABAR* measurements in Aubert (2001c, 2002f, 2003e,m, 2006s), the Belle measurements in Abe (2005c) and measurements from CDF, DØ, ALEPH, DELPHI, L3, OPAL, SLD and ATLAS (Beringer et al., 2012) to be, respectively, $\tau_{B^0} = (1.519 \pm 0.007)$ ps, $\tau_{B^+} = (1.641 \pm 0.008)$ ps and $\tau_{B^+}/\tau_{B^0} = (1.079 \pm 0.007)$ ps. The most precise measurements contributing to these averages come from the $B$ Factories and a recent set of measurements from CDF using fully-reconstructed $B \to J/\psi K^{(*)}$ events (Aaltonen et al., 2011a). DØ provides a precise measurement of $\tau_{B^+}/\tau_{B^0}$ from samples of $B \to D^{*+}\mu\overline{\nu}X$ and $B \to D^0\mu\overline{\nu}X$ (Abazov et al., 2005). By using only the $B$ Factories measurements, one obtains the averages $\tau_{B^0} = (1.530 \pm 0.010)$ ps, $\tau_{B^+} = (1.640 \pm 0.014)$ ps and $\tau_{B^+}/\tau_{B^0} = (1.068 \pm 0.011)$ ps.

The measurements of the charged and neutral $B$ lifetimes and their ratio now have errors of about half a percent, and the $B^+$ lifetime is now measured to be larger than the $B^0$ lifetime by many standard deviations. The precision in these measurements exceeds that of existing theoretical calculations. Thus, with the original motivations fully addressed by the current set of measurements and the multitude of relevant systematic error sources that come with sub-percent precision measurements, it is unlikely that there will be improved measurements using the full data set of the $B$ Factories or the even larger data sets of future super flavor factories.

### 17.5.2 $B^0 - \overline{B}^0$ mixing

Neutral meson-antimeson oscillations were predicted by Gell-Mann and Pais (1955) and first observed in 1956 in the $K^0 - \overline{K}^0$ system (Lande, Booth, Impeduglia, Lederman, and Chinowsky, 1956). Mixing in the $B^0 - \overline{B}^0$ system was discovered in 1987 by the ARGUS collaboration (Albrecht et al., 1987b). It was clear from the first $B_s^0$ measurements that mixing was an important effect in the $B_s^0 - \overline{B}_s^0$ system (see for example the review of Danilov, 1993), although the mixing frequency was not resolved until much later by the CDF collaboration (Abulencia et al., 2006b) as previous results established only lower limits on $x_s = \Delta m_s/\Gamma_s$. Finally, $D^0 - \overline{D}^0$ mixing was first observed by the $B$ Factories and is described in detail in Section 19.2.

Meson-antimeson oscillations proceed in general through both long distance effects (common decay modes) and second order weak interactions as described by box diagrams containing virtual quarks (Figure 10.1.1). $B^0 - \overline{B}^0$ mixing is predominantly a short-distance phenomenon; among the various box diagrams, those containing the top quark dominate due to the large top mass. The observation of mixing, in fact, provided the first indication that the top quark was very heavy: see the discussion in Section 16.1. The mixing frequency $\Delta m_d$ is sensitive to the CKM matrix element $V_{td}$ (see Section 17.2). In the neutral $K$, $D$ (see Section 19.2), and $B_s^0$ meson systems, mixing also has contributions from real intermediate states accessible to both the meson and the antimeson. Real intermediate states lead to a difference in the decay rate for the two mass eigenstates of the neutral meson system. However, for the $B_d^0$ system, the decay rate difference $\Delta\Gamma$ is expected to be of $O(10^{-2} - 10^{-3})$ times smaller than the average decay rate and the mixing frequency (Lenz and Nierste, 2011), and is typically ignored in the measurements of $\Delta m_d$.

In the following, we briefly review the principles of $\Delta m_d$ measurements (Section 17.5.2.1), and then summarize the techniques and results of the $B$ Factory measurements using dilepton (Section 17.5.2.2), partially-reconstructed (Section 17.5.2.3), and fully-reconstructed final states (Section 17.5.2.4). The average of these results is discussed in Section 17.5.2.5. Throughout, we set $\Delta\Gamma = 0$; the specialized analyses allowing $\Delta\Gamma \neq 0$, and setting constraints on its value, are discussed in Section 17.5.2.6 below.

#### 17.5.2.1 Principles of $\Delta m_d$ measurements

The time-evolution of the $B^0 - \overline{B}^0$ system is given by a phenomenology-based $2 \times 2$ Hamiltonian matrix (for details see Chapter 10). Solving this system of equations gives the time-dependent probabilities for $B^0 - \overline{B}^0$ oscillations. For a $\overline{B}^0$ decay to a flavor eigenstate that is not accessible from a $B^0$ decay (*e.g.* the semileptonic decay $\overline{B}^0 \to D^{*+} l^- \nu_l$), the parameter $\lambda$ in Eqs (10.1.10–10.1.15) and (10.2.2–10.2.5) is zero. Neglecting *CP* violation in mixing, the probability that a $B^0$ produced at time $t = 0$ decays as $\overline{B}^0$ at time $t$ is given by

$$P_{B^0 \to \overline{B}^0}(t) = \frac{e^{-t/\tau_{B^0}}}{2\tau_{B^0}} \times [1 - \cos(\Delta m_d t)], \qquad (17.5.14)$$

where $\Delta m_d$ is the $B^0 - \overline{B}^0$ oscillation frequency and $\tau_{B^0}$ is the neutral $B$ lifetime. Similarly, the probability that a produced $B^0$ decays as $B^0$ (for example through $B^0 \to D^{*-} l^+ \overline{\nu}_l$) is given by

$$P_{B^0 \to B^0}(t) = \frac{e^{-t/\tau_{B^0}}}{2\tau_{B^0}} \times [1 + \cos(\Delta m_d t)]. \qquad (17.5.15)$$

Likewise the probabilities for a produced $\overline{B}^0$ to decay as a $B^0$ or a $\overline{B}^0$ are given, respectively, by

$$P_{\overline{B}^0 \to B^0}(t) = \frac{e^{-t/\tau_{B^0}}}{2\tau_{B^0}} \times [1 - \cos(\Delta m_d t)], \qquad (17.5.16)$$

and

$$P_{\overline{B}^0 \to \overline{B}^0}(t) = \frac{e^{-t/\tau_{B^0}}}{2\tau_{B^0}} \times [1 + \cos(\Delta m_d t)]. \qquad (17.5.17)$$

The first measurements of $B^0 - \overline{B}^0$ oscillations were time-integrated measurements by ARGUS (Albrecht et al., 1987b) and CLEO (Artuso et al., 1989). They measured the time-integrated probability $\chi_d$ that a $B^0$ ($\overline{B}^0$) produced in $\Upsilon(4S) \to B^0\overline{B}^0$ decays as a $\overline{B}^0$ ($B^0$),

$$\chi_d = \frac{x_d^2}{2\,(1 + x_d^2)}, \qquad (17.5.18)$$

where $x_d = \Delta m_d / \Gamma_d = \Delta m_d \tau_{B^0}$. In 1993 the LEP experiments started to provide the first time-dependent measurements of $\Delta m_d$, made possible through their precision vertex detectors and highly-boosted $B^0$ mesons from $Z^0$ decays (Abreu et al., 1994; Acciarri et al., 1996; Akers et al., 1994b; Buskulic et al., 1993b). The CDF collaboration published their first $\Delta m_d$ measurement in 1998 (Abe et al., 1998). In the 2000 Review of Particle Physics (Groom et al., 2000) the PDG calculated an average $B^0 - \overline{B}^0$ oscillation frequency from time-dependent measurements by the LEP experiments and CDF of $\Delta m_d = (0.478 \pm 0.018)$ ps$^{-1}$. Including the measurements of the time-integrated mixing probability $\chi_d = 0.156 \pm 0.024$ by CLEO and ARGUS, they obtained $\Delta m_d = (0.472 \pm 0.017)$ ps$^{-1}$.

The experimental strengths and weaknesses in $\Delta m_d$ measurements, when comparing these older experiments to the $B$ Factories, are the same as for measurements of the $B$ lifetimes. The former benefit from high-precision proper-time measurements, whereas the latter have the advantage of low-background, high-statistics $B$ samples, and excellent $B$-momentum resolution.

The experimental methods of the $\Delta m_d$ analyses are very similar to those used in the measurement of time-dependent *CP* asymmetries in $B$ decays to *CP* eigenstates (see the measurement of $\sin 2\phi_1$ in Section 17.6). In particular, fully-reconstructed $B$ decays to flavor final states $B_{\rm flav}$, such as $\overline{B}^0 \to D^{(*)+}\pi^-$, have the same $B$ vertex resolutions and thus $\Delta t$ resolution function as $B$ decays to $(c\overline{c})s$ *CP* eigenstates, $B_{CP}$ (see Chapter 6 and Section 17.6). The same $B$ flavor-tagging algorithms are used to determine the flavors of $B_{\rm flav}$ and $B_{CP}$ at the time of their production (see Chapter 8). In both cases, maximum likelihood fits are used to extract the parameters of the time-dependent asymmetries from the measured $\Delta t$ distributions. By-products of the $\Delta m_d$ measurement with fully-reconstructed final states are the $B$ flavor-tagging mistag rates, which cannot be determined with *CP* eigenstates. In addition, a confirmation of the $\Delta m_d$ results of previous experiments served as a convincing proof-of-principle of this novel technique for measuring time-dependent *CP* asymmetries at the asymmetric beam-energy $B$ Factories. So, it is no coincidence that one of the first measurements from the $B$ Factories was the precise time-dependent measurement of $\Delta m_d$. On the other hand, improving the knowledge of $\Delta m_d$ has been and still is interesting in its own right. The oscillation frequency $\Delta m_d$ is proportional to $|V_{td}|^2$ (Eq. 17.2.1). Thus, a precise $\Delta m_d$ measurement along with a measurement of the $B_s^0 - \overline{B}_s^0$ oscillation frequency $\Delta m_s$ from hadron colliders, combined with lattice QCD calculations of the decay constants and QCD bag parameters of $B^0$ and $B_s^0$ mesons (for details see Section 17.2) provide strong constraints on the Unitarity Triangle (see Section 25.1).

The time-dependent $\Delta m_d$ measurements by the $B$ Factories all follow the same basic idea. In the $\Upsilon(4S) \to B^0\overline{B}^0$ decay the two neutral $B$ mesons are produced in a coherent $P$-wave state. If one of the $B$ mesons, referred to as $B_{\rm tag}$, can be ascertained to decay to a state of known flavor (*i.e.* $B^0$ or $\overline{B}^0$) at a certain time $t_{\rm tag}$, the other $B$, referred to as $B_{\rm rec}$, at that time must be of the opposite flavor as a consequence of Bose symmetry. Consequently, the probabilities to observe *unmixed* (+) $B^0\overline{B}^0$, or *mixed* (−) $B^0B^0/\overline{B}^0\overline{B}^0$ events, are functions of the proper-time difference $\Delta t = t_{\rm rec} - t_{\rm tag}$ and of $\Delta m_d$:

$$P_{B^0\overline{B}^0 \to B^0\overline{B}^0}(\Delta t) \equiv P_+(\Delta t) = \frac{e^{-|\Delta t|/\tau_{B^0}}}{4\tau_{B^0}} \times [1 + \cos(\Delta m_d \Delta t)]\,,$$

$$P_{B^0\overline{B}^0\to B^0B^0/\overline{B}^0\overline{B}^0}(\Delta t) \equiv P_-(\Delta t) = \frac{e^{-|\Delta t|/\tau_{B^0}}}{4\tau_{B^0}} \times [1-\cos(\Delta m_d \Delta t)]\,. \tag{17.5.19}$$

From these two equations one can define the so-called $B^0\overline{B}^0$ mixing asymmetry as

$$A_{\rm mix}(\Delta t) \equiv \frac{P_+(\Delta t) - P_-(\Delta t)}{P_+(\Delta t) + P_-(\Delta t)} = \cos(\Delta m_d \Delta t). \tag{17.5.20}$$

The functions $P_\pm(\Delta t)$ are illustrated in Fig. 17.5.4. The mixed event function ($P_-$) rises slowly from zero at $\Delta t = 0$ until it reaches a maximum at around $\Delta t = 2.6$ ps.

The $B$ Factories have measured the $B^0-\overline{B}^0$ oscillation frequency with various final states and $B$-reconstruction techniques. In the analyses of dilepton inclusive final states, the flavors of both $B$ mesons are identified only through high-momentum leptons from semileptonic decays. In all other $\Delta m_d$ measurements one $B$ is reconstructed through its decay to an exclusive flavor final state, $B_{\rm rec}$, while the remaining charged particles in the event are used to identify (or "tag") the flavor of the other $B$ (referred to as $B_{\rm tag}$), as a $B^0$ or $\overline{B}^0$. The proper-time difference $\Delta t = \Delta z/\langle\beta\gamma\rangle c$ is determined from the $z$ positions of the $B$ decay vertices $\Delta z = z_{\rm rec} - z_{\rm tag}$ and the average boost of the $\Upsilon(4S)$ frame in the lab frame $\langle\beta\gamma\rangle$. The boost is known to good precision from the $e^+$ and $e^-$ beam energies, so that the $\Delta z$ measurement dominates the $\Delta t$ resolution (see Chapter 6). The value of $\Delta m_d$ is then extracted from a simultaneous fit to the $\Delta t$ distributions of the unmixed and mixed events. There are two principal experimental complications to the probability distributions in Eq. (17.5.19). First, the flavor tagging algorithm sometimes incorrectly identifies the $B_{\rm tag}$ flavor. The probability to incorrectly identify the flavor of $B_{\rm tag}$, $w$, reduces the observed amplitude for the oscillation by a factor $(1-2w)$. Second, the resolution of $\Delta t$ is comparable to the oscillation period and must be accounted for. The *p.d.f.*s for the unmixed and mixed signal events $\mathcal{H}_{\pm,\rm sig}$ can be expressed as the convolution of the underlying $\Delta t$ distribution,

$$h_{\pm,\rm sig}(\Delta t; \Delta m_d, w) = \frac{e^{-|\Delta t|/\tau_{B^0}}}{4\tau_{B^0}}\left[1 \pm (1-2w)\cos(\Delta m_d \Delta t)\right], \tag{17.5.21}$$

with a signal $\Delta t$ resolution function $\mathcal{R}_{\rm sig}$ containing parameters $\hat{a}_j$:

$$\mathcal{H}_{\pm,\rm sig}(\Delta t; \Delta m_d, w, \hat{a}_j) = h_{\pm,\rm sig}(\Delta t; \Delta m_d, w) \otimes \mathcal{R}_{\rm sig}(\Delta t; \hat{a}_j). \tag{17.5.22}$$

The functions $\mathcal{H}_{\pm,\rm sig}$ are shown in Fig. 17.5.4. The impact of typical mistag and $\Delta t$ resolution effects is clearly visible in the comparison with the functions $P_\pm(\Delta t)$ that represent ideal detector performance.

A fit is then performed to simultaneously extract the mistag rates $w$, the resolution function parameters $\hat{a}_j$, and the mixing frequency $\Delta m_d$. In the following sections we give brief descriptions of the various $\Delta m_d$ measurements by the $B$ Factories, in dilepton (Section 17.5.2.2), partially-reconstructed (17.5.2.3), and fully-reconstructed (17.5.2.4) final states. The results are summarized in Table 17.5.2; their average is discussed in Section 17.5.2.5.

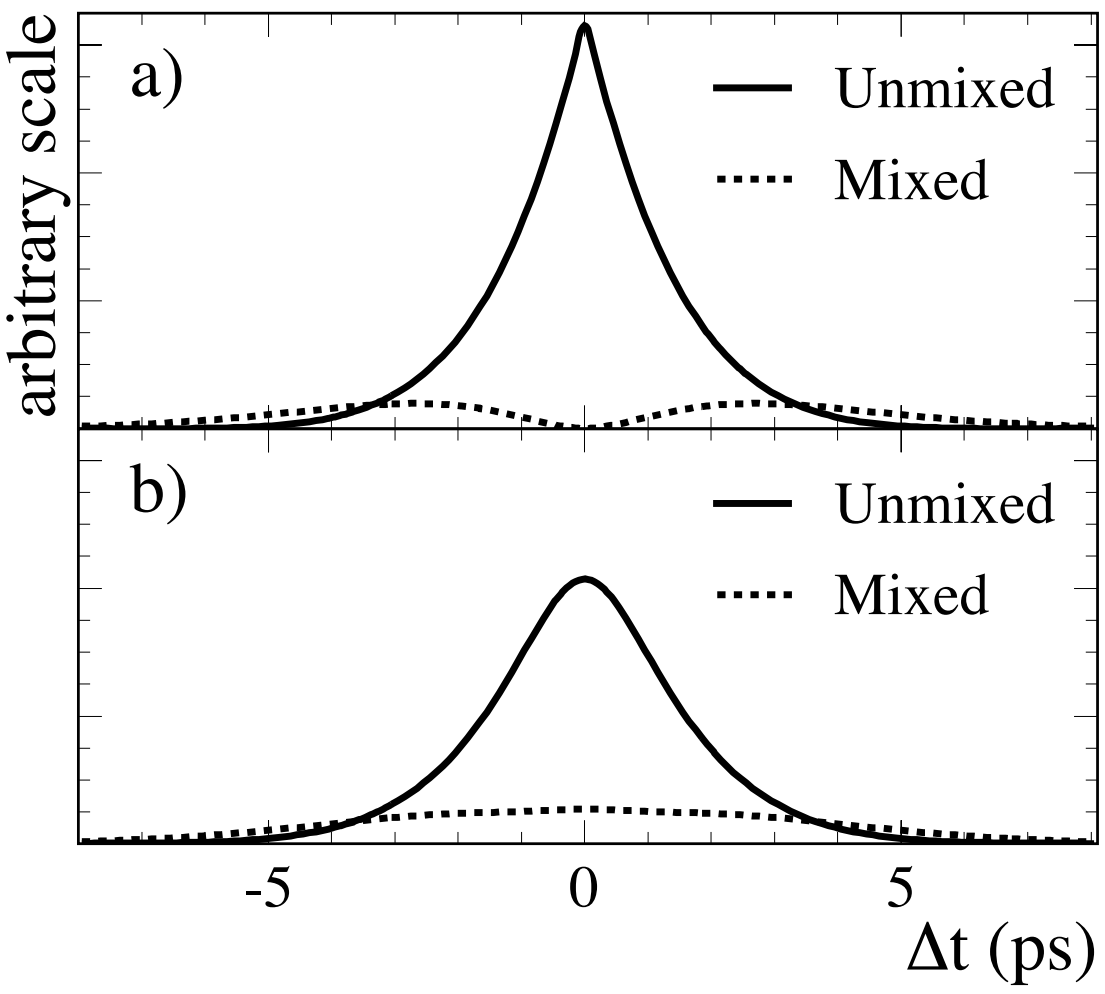

**Figure 17.5.4.** The $\Delta t$ distributions of mixed and unmixed events (a) with perfect tagging and $\Delta t$ resolution ($P_\pm(\Delta t)$) (b) with mistag rates and $\Delta t$ resolution typical at the $B$ Factories ($\mathcal{H}_{\pm,\rm sig}(\Delta t)$). From Aubert (2002a).

### 17.5.2.2 Dilepton final states

Belle published their first measurement of $\Delta m_d$ using dilepton events in a sample of $5.9\,{\rm fb}^{-1}$ (Abe, 2001b). In a later analysis of the same final state, they used a sample of $29\,{\rm fb}^{-1}$ (Hastings, 2003). *BABAR* published one measurement of $\Delta m_d$ with dilepton events using a sample of $21\,{\rm fb}^{-1}$ (Aubert, 2002e).

The inclusive nature of the dilepton final state provides large event samples. The measurements are based on the identification of events containing pairs of high-momentum leptons ($ee$, $\mu\mu$ and $e\mu$) from semileptonic decays of $B$ mesons. The flavors of the $B$ mesons at the time of their decay are determined by the charges of the leptons in the final state. For $\Upsilon(4S)$ resonance decays into $B^0\overline{B}^0$ pairs, opposite-sign charge (OS) and same-sign charge (SS) lepton pairs correspond to unmixed and mixed events, respectively. Both experiments apply selection requirements on the lepton momenta, overall event shape, and track quality to ensure a well-measured $\Delta t$ and to suppress backgrounds from fake leptons, continuum events, $J/\psi$ decays, and so-called $B$ cascade decays. In the latter, one lepton originates from the semileptonic decay of a charm meson, which can come from the same or the opposite $B$ as the other lepton. An irreducible background comes from semileptonic decays of $B^+B^-$ pairs.

Belle determines the $z$ coordinates of the $B$ decay vertices from the intersections of the lepton tracks with the profile of the beam interaction point (IP) convoluted with the average $B^0$ flight length ($\sim 20\,\mu$m in the $\Upsilon(4S)$ rest frame). The mean position and width of the IP are determined on a run-by-run basis using hadronic events (see Chapter 6). The proper-time difference $\Delta t$ is calculated from the $z$ positions of the two lepton vertices using Eq. (17.5.12), where $\Delta z = z_1 - z_2$ is the distance along the beam axis between the two vertices. For OS events, the positively charged lepton is taken as the first lepton ($z_1$). For SS events Belle uses the absolute value of $\Delta z$. *BABAR* applies a beam spot constraint to the two lepton tracks to find the primary vertex of the event in the transverse plane. The positions of closest approach of the two tracks to this vertex in the transverse plane are computed and their $z$ coordinates are denoted $z_1$ and $z_2$, where the subscripts refer to the highest and second highest momentum leptons in the $\Upsilon(4S)$ rest frame. The vertex fit constrains the lepton tracks to originate from the same point in the transverse plane, thereby neglecting the nonzero transverse flight length for $B^0$ mesons. As a consequence, the $\Delta t$ resolution function is $\Delta z$ dependent, becoming worse at higher $|\Delta z|$. Neglecting this dependence introduces a small bias that *BABAR* accounts for in the systematic uncertainty.

**Table 17.5.2.** $B$ Factory measurements of $\Delta m_d$ along with the journal paper, selected final state, signal purity $f_{\rm signal}$, $B$ meson signal yield, and integrated luminosity for each measurement. The $\Delta m_d$ measurements in Hara (2002) and Tomura (2002b) have been superseded by Abe (2005c), and those in Abe (2001b) by Hastings (2003); the superseded measurements are not separately included in the $B$ Factories average (Asner et al., 2010).

| Experiment | Method | $f_{\rm signal}$ | Yield $[B/\,{\rm fb}^{-1}]$ | $\int \mathcal{L}\, dt$ | $\Delta m_d$ [ps$^{-1}$] |
|---|---|---|---|---|---|
| *BABAR* (Aubert, 2002a,b) | Excl. hadronic modes | 86% | 214 | 30 fb$^{-1}$ | $0.516 \pm 0.016 \pm 0.010$ |
| *BABAR* (Aubert, 2002e) | Incl. dilepton | | | 21 fb$^{-1}$ | $0.493 \pm 0.012 \pm 0.009$ |
| *BABAR* (Aubert, 2006s) | $D^* l\nu$ (partial) | 64% | 605 | 81 fb$^{-1}$ | $0.511 \pm 0.007 \pm 0.007$ |
| *BABAR* (Aubert, 2003m) | Excl. $D^* l\nu$ | 76% | 680 | 21 fb$^{-1}$ | $0.492 \pm 0.018 \pm 0.014$ |
| Belle (Abe, 2001b) | Incl. dilepton | | | 6 fb$^{-1}$ | $0.463 \pm 0.008 \pm 0.016$ |
| Belle (Hastings, 2003) | Incl. dilepton | | | 29 fb$^{-1}$ | $0.503 \pm 0.008 \pm 0.010$ |
| Belle (Zheng, 2003) | $D^*\pi$ (partial) | 70% | 118 | 29 fb$^{-1}$ | $0.509 \pm 0.017 \pm 0.020$ |
| Belle (Hara, 2002) | Excl. $D^* l\nu$ | 80% | 453 | 29 fb$^{-1}$ | $0.494 \pm 0.012 \pm 0.015$ |
| Belle (Tomura, 2002b) | Excl. hadronic modes | 80% | 229 | 29 fb$^{-1}$ | $0.528 \pm 0.017 \pm 0.011$ |
| Belle (Abe, 2005c) | Excl. hadronic modes, $D^* l\nu$ | 81% | 707 | 140 fb$^{-1}$ | $0.511 \pm 0.005 \pm 0.006$ |
| *BABAR*-Belle average | | | | | $0.508 \pm 0.003 \pm 0.003$ |

*BABAR* and Belle use binned maximum likelihood fits to the $\Delta t$ and $\Delta z$ distributions, respectively, of the selected dilepton candidates to extract $\Delta m_d$. *BABAR* fits the shapes of the $\Delta t$ distributions with the *p.d.f.*s for OS and SS dilepton events as given in Eq. (17.5.22). Belle fits the $\Delta z$ distributions and constrains the integrated mixing probability to $\chi_d$. Their $\Delta z$ distributions are described by converting the constrained signal $\Delta t$ distributions $\mathcal{P}_\pm(\Delta t)$ to $\Delta z$ distributions using Eq. (17.5.12) and convolving them with the $\Delta z$ resolution function. The constrained signal $\Delta t$ distributions are given by

$$\mathcal{P}_\pm(\Delta t) = N_{\Upsilon(4S)} f_0 b_0^2 \epsilon_{ll}^\pm \frac{e^{-|\Delta t|/\tau_{B^0}}}{4\tau_{B^0}} \left[1 \pm \cos(\Delta m_d \Delta t)\right],$$
$$\mathcal{P}_{\rm ch}(\Delta t) = N_{\Upsilon(4S)} f_{\rm ch} b_{\rm ch}^2 \epsilon_{ll}^{\rm ch} \frac{e^{-|\Delta t|/\tau_{B^+}}}{2\tau_{B^+}}, \qquad (17.5.23)$$

where $N_{\Upsilon(4S)}$ is the total number of $\Upsilon(4S)$ events, $f_0$ and $f_{\rm ch}$ are the branching fractions of the $\Upsilon(4S)$ to neutral and charged $B$ pairs (assuming $f_0 + f_{\rm ch} = 1$), $b_0$ and $b_{\rm ch}$ are the semileptonic branching fractions for neutral and charged $B$ mesons, and $\epsilon_{ll}^\pm$ are the efficiencies for selecting dilepton events of unmixed and mixed origins. Belle determines the ratio $\epsilon_{ll}^+ : \epsilon_{ll}^-$ from MC simulation and fixes it in the fit to the data assuming detector effects that are not simulated correctly equally affect events with these origins. The $\Delta z$ distributions are obtained for these distributions by conversion from $\Delta t$ and convolution with the $\Delta z$ resolution function.

*BABAR* describes the $\Delta t$ resolution function for dilepton events as the sum of three Gaussian distributions. The resolution function parameters are free parameters in the fit. Belle determines the signal $\Delta z$ resolution function from $J/\psi$ decays in data. For these decays the true $\Delta z$ is equal to zero and the measured $\Delta z$ distribution, after the contributions of backgrounds are subtracted, yields the $\Delta z$ resolution function. A comparison between data and MC simulation shows that after convolving the MC $\Delta z$ distribution of $J/\psi$ decays with a Gaussian of width $\sigma = (50 \pm 18)\,\mu$m, the MC distribution agrees with data.

The $\Delta t$ and $\Delta z$ distributions of background events are determined from MC-simulated events and data control samples. The large background from semileptonic $B^+B^-$ events has the same resolution function as the signal events. The numbers of selected OS and SS dilepton pairs along with the corresponding mixing asymmetry as a function of $\Delta t$ from the *BABAR* analysis are shown in Fig. 17.5.5. Due to the small mixing frequency, OS signal events are much more abundant than SS events. Most of the background events are also OS (for example from $B^+B^-$ events). Therefore, even a small mistag probability will blur the characteristic features of the SS $\Delta t$ distribution. This is particularly evident at $\Delta t = 0$ where the OS $\mathcal{P}_+$ distribution has its maximum and the SS $\Delta t$ distribution $\mathcal{P}_-$ is zero: the measured $\Delta t$ distribution of selected SS events does not have a dip at zero. However, the mixing asymmetry still shows the expected cosine shape as mis-tagging and non-oscillating backgrounds, respectively, only reduce the amplitude and shift the baseline of the asymmetry curve.

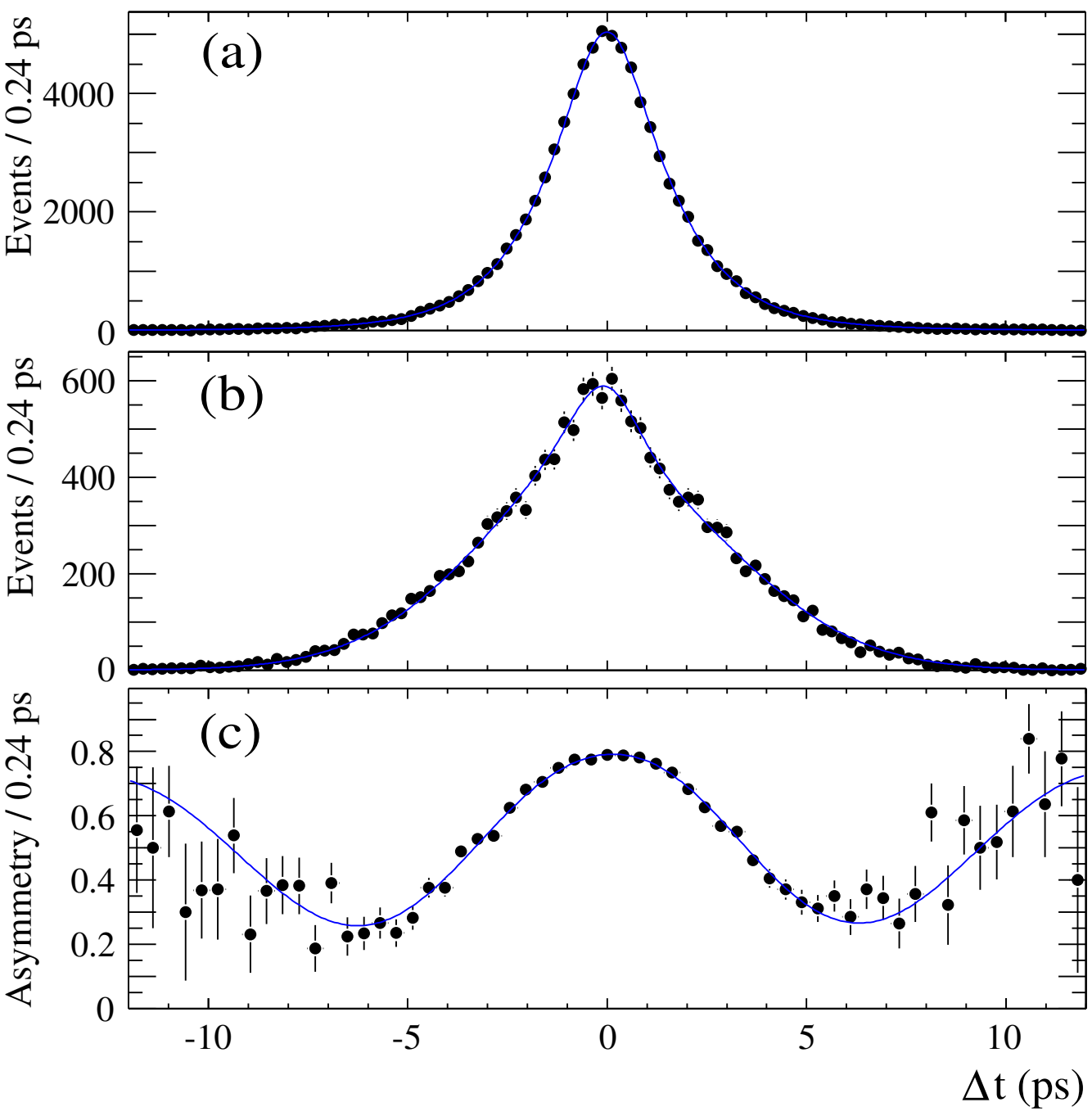

**Figure 17.5.5.** The $\Delta t$ distributions for (a) opposite-sign and (b) same-sign charge dilepton events; (c) mixing asymmetry between opposite-sign and same-sign dilepton events. The points are the data and lines correspond to the projection of the likelihood fit (Aubert, 2002e).

From dilepton events *BABAR* measures $\Delta m_d = (0.493 \pm 0.012 \pm 0.009)\,\text{ps}^{-1}$ in a sample of $21\,\text{fb}^{-1}$ (Aubert, 2002e) while Belle measures $\Delta m_d = (0.503 \pm 0.008 \pm 0.010)\,\text{ps}^{-1}$ in a sample of $29\,\text{fb}^{-1}$ (Hastings, 2003), where the first errors are statistical and the second are systematic. The largest contributions to the systematic errors come from the uncertainties in the $B^0$ and $B^+$ lifetimes ($\sim 0.006\,\text{ps}^{-1}$) and in the $\Delta z$ and $\Delta t$ resolution functions ($\sim 0.006\,\text{ps}^{-1}$).

#### 17.5.2.3 Partially-reconstructed final states

*BABAR* measures $\Delta m_d$ with a sample of partially-reconstructed $\overline{B}^0 \to D^{*+} l^- \overline{\nu}_l$ events in $81\,\text{fb}^{-1}$ (Aubert, 2006s). They select $\overline{B}^0 \to D^{*+} l^- \overline{\nu}_l$ ($l = e$ or $\mu$) events with partial reconstruction of the decay $D^{*+} \to D^0 \pi_s^+$, using only the charged lepton from the neutral $B$ decay ($l_{\text{rec}}$) and the soft pion ($\pi_s^+$) from the $D^{*+}$ decay. This decay mode has a large selection efficiency since the $D^0$ decay is not reconstructed and the branching fraction of $\overline{B}^0 \to D^{*+} l^- \overline{\nu}_l$ is about half of the semileptonic branching ratio of the $\overline{B}^0$. The other $B$ in the event is identified through a second high-momentum lepton ($l_{\text{tag}}$).

Events are required to have at least four charged tracks. The normalized second Fox-Wolfram moment $R_2$ (see Chapter 9) must be less than 0.5 to reduce background from light quark production in continuum events. The lepton from the $B$ decay must have a momentum in the range 1.3–2.4 GeV/$c$, and the soft pion momentum must be between 60 and 200 MeV/$c$. By approximating the $D^{*+}$ momentum from the $\pi_s^+$ momentum, they calculate the square of the missing neutrino mass $(m^{\nu}_{\text{miss}})^2$. The $(m^{\nu}_{\text{miss}})^2$ distribution peaks at zero for signal events, while it is spread over a wide range of mostly negative values for background events.

*BABAR* determines the $B_{\text{rec}}$ decay vertex from a vertex fit of the $l_{\text{rec}}$ and $\pi_s$ tracks, constrained to the beam spot position in the transverse plane, but accounting for the average $\overline{B}^0$ flight distance. The decay point of $B_{\text{tag}}$ is determined from $l_{\text{tag}}$ and the beam spot following a procedure similar to that of the $B_{\text{rec}}$ decay vertex. The flavor of $B_{\text{rec}}$ is determined from the $l_{\text{rec}}$ and soft pion charges. The flavor of the other $B$ in the event is determined from the charge of $l_{\text{tag}}$.

After all selection criteria *BABAR* finds 49,000 signal events over a background of 28,000 events in the region $(m^{\nu}_{\text{miss}})^2 > -2.5\,\text{GeV}^2/c^4$. Background studies are done with events in the region $(m^{\nu}_{\text{miss}})^2 < -2.5\,\text{GeV}^2/c^4$ if no signal candidate is found in the event.

*BABAR* simultaneously fits the distributions of $(m^{\nu}_{\text{miss}})^2$, $\Delta t$, and its uncertainty $\sigma_{\Delta t}$, for mixed and unmixed events, with a binned maximum-likelihood method. Probabilities for a given event to belong to any of the identified background sources ($e^+e^- \to q\overline{q}$ continuum, $B\overline{B}$ combinatorial, and $B^+$ peaking background) are calculated based on the background $(m^{\nu}_{\text{miss}})^2$ distributions. Signal is considered to be any combination of a lepton and a charged $D^{*+}$ produced in the decay of a single $\overline{B}^0$ meson. They further divide their signal events according to the origin of the tag lepton into primary, cascade, and decay-side lepton tags. A primary lepton tag is produced in the direct decay $B^0 \to X l^+ \overline{\nu}_l$, a cascade lepton tag is produced in the process $B^0 \to \overline{D}X$, $\overline{D} \to l^- Y$, and a decay-side tag is produced by the semi-leptonic decay of the unreconstructed $D^0$. The relative normalization between mixed and unmixed signal events is constrained based on the time-integrated mixing rate $\chi_d$. The $\Delta t$ signal *p.d.f.* for both unmixed and mixed events consists of the sum of *p.d.f.*s for primary, cascade, and decay-side tags each convoluted with its own resolution function. They use the standard three Gaussian resolution function with event-by-event $\Delta t$ uncertainties.

From the fit *BABAR* obtains $\tau_{B^0} = (1.504 \pm 0.013^{+0.018}_{-0.013})\,\text{ps}$ and $\Delta m_d = (0.511 \pm 0.007^{+0.007}_{-0.006})\,\text{ps}^{-1}$, where the first errors are statistical and the second are systematic. The statistical correlation between $\tau_{B^0}$ and $\Delta m_d$ is 0.7%. The results include corrections of $-0.006\,\text{ps}$ on $\tau_{B^0}$ and $+0.007\,\text{ps}^{-1}$ on $\Delta m_d$ due to biases from event selection, boost approximation, $B^-$ peaking background, and combinatorial $B\overline{B}$ background based on MC studies. The systematic error in $\Delta m_d$ is dominated by uncertainties in the SVT alignment ($^{+0.0038}_{-0.0033}\,\text{ps}^{-1}$), the selected range of $\Delta t$ and $\sigma_{\Delta t}$ ($0.0033\,\text{ps}^{-1}$), and analysis bias ($0.0035\,\text{ps}^{-1}$), whereas the largest systematic error sources in the $\tau_{B^0}$ measurement are the SVT alignment ($^{+0.0132}_{-0.0038}$ ps), the $z$ scale of the detector (0.0070 ps), and analysis bias (0.0070 ps).

Belle measures $\Delta m_d$ with a sample of partially-reconstructed $\overline{B}^0 \to D^{*+}\pi^-$ events in 29.1 fb$^{-1}$ (Zheng, 2003). They select $\overline{B}^0 \to D^{*+}\pi_h^-$ events with partial reconstruction of the decay $D^{*+} \to D^0\pi_s^+$, using only the hard pion ($\pi_h^-$) from the $\overline{B}^0$ decay and the soft pion ($\pi_s^+$) from the $D^{*+}$ decay. Using this partial reconstruction method, Belle obtains an order of magnitude more events compared to the full reconstruction of the $D^{*+}$. The flavor of the other $B$ in the event is identified through a high-momentum lepton $l_{\rm tag}$ from semileptonic decay.

Hadronic events are selected by applying requirements on track multiplicity and total energy variables. The hard pion from the $B$ decay must have a momentum in the range 2.05–2.45 GeV/$c$ and the soft pion momentum must be below 450 MeV/$c$. Belle applies impact parameter requirements on $\pi_h^-$ and $\pi_s^+$ to suppress backgrounds from interactions of beam particles with residual gas in the beam pipe or the beam pipe wall. They require both tracks to have SVD information and to not be identified as leptons.

The event kinematics are fully constrained by four-momentum conservation in the decays $\overline{B}^0 \to D^{*+}\pi_h^-$ and $D^{*+} \to D^0\pi_s^+$, the masses of all particles in these decays, the $\overline{B}^0$ energy, and the $\pi_h^-$ and $\pi_s^+$ momenta. Belle uses two variables, the missing $D^0$ mass, $M_{D_{\rm miss}}$, and the cosine of the angle between the soft pion in the $D^{*+}$ rest frame and the momentum of the $D^{*+}$ in the center-of-mass frame $\cos\theta^*_{\pi_s}$. The $M_{D_{\rm miss}}$ distribution for signal events peaks sharply at the nominal $D^0$ mass, while background events spread towards smaller values. Signal events are required to have $M_{D_{\rm miss}} > 1.85$ GeV/$c$ and $0.3 < |\cos\theta^*_{\pi_s}| < 1.05$.

The flavor of $B_{\rm rec}$ is determined from the $\pi_h$ charge. The flavor of the other $B$ in the event is determined from the charge of $l_{\rm tag}$. The tag lepton is required to have momentum greater than 1.1 GeV/$c$ and to pass similar requirements on SVD hits and impact parameter as the $B_{\rm rec}$ pions. Tag leptons are rejected if when combined with any other lepton in the event the pair has an invariant mass consistent with a $J/\psi$. Belle determines the $B_{\rm rec}$ ($B_{\rm tag}$) decay vertex from the intersection of the $\pi_h$ ($l_{\rm tag}$) track with the beam spot accounting for the $B$ meson flight distance.

After all selection criteria Belle finds 3433 signal events over a background of 1466 events which are used in the $\Delta m_d$ measurement. Studies of MC-simulated events show that a significant fraction of the selected events come from $\overline{B}^0 \to D^{*+}\rho^-$ decays.

Belle simultaneously fits the $\Delta t$ distributions of the mixed and unmixed events with an unbinned maximum-likelihood method. The $B^0\overline{B}^0$ mixing frequency $\Delta m_d$ is the only free parameter in the fit. The $B^0$ lifetime is fixed to the world average. The signal $\Delta t$ resolution function uses a triple-Gaussian *p.d.f.* (see Eq. 10.4.2) in the $\Delta t$ residuals. The resolution function parameters are determined from decays of $J/\psi$ to $e^+e^-$ and $\mu^+\mu^-$. Backgrounds are divided into peaking and non-peaking categories. Non-peaking background is dominated by random combinations of $\pi_h^-$ and $\pi_s^+$ with primary leptons from $B^0$ and $B^\pm$ decays, and combinatorial background from continuum. Peaking background is dominated by the following sources: $\overline{B}^0 \to D^{*+}\pi^-$ and $\overline{B}^0 \to D^{*+}\rho^-$ with secondary-lepton or fake lepton tags; $B^0 \to D^{**-}\pi^+$, $B^+ \to \overline{D}^{**0}\pi^+$, and $B^0 \to D^{*-}\pi^+\pi^0$ decays with primary-lepton, secondary-lepton, or fake lepton tags. Peaking and non-peaking background *p.d.f.*s are convolved with their own resolution functions.

From the fit Belle obtains $\Delta m_d = (0.509 \pm 0.017 \pm 0.020)\,{\rm ps}^{-1}$, where the first error is statistical and the second is systematic. The systematic error in $\Delta m_d$ is dominated by uncertainties in the background fractions (0.014 ps$^{-1}$) and the signal $\Delta t$ resolution function (0.012 ps$^{-1}$).

#### 17.5.2.4 Fully-reconstructed final states

*Hadronic decay modes*

*BABAR* reconstructs neutral $B$ mesons in the decay modes $B^0 \to D^{(*)-}\pi^+$, $D^{(*)-}\rho^+$, $D^{(*)-}a_1^+$, $J/\psi K^{*0}$ using a data sample of 29.7 fb$^{-1}$ (Aubert, 2002a,b). Belle uses the $B$ decays to the hadronic final states $D^-\pi^+$, $D^{*-}\pi^+$, and $D^{*-}\rho^+$ in a data sample of 29.1 fb$^{-1}$ (Tomura, 2002b). The $B^0\overline{B}^0$ mixing analyses with fully-reconstructed final states reconstruct the same decay modes of the $B^0$ daughters as in the $B^0$ lifetime measurements described in Section 17.5.1.3 (Aubert, 2001c; Abe, 2002m). Both experiments reduce background from continuum events by applying requirements on the normalized second Fox-Wolfram moment $R_2$ and the angle between the thrust axis of the particles that form the reconstructed $B$ candidate and the thrust axis of the remaining tracks and unmatched calorimeter clusters in the event, computed in the $\Upsilon(4S)$ frame. Neutral $B$ candidates are identified by their $\Delta E$ and $m_{\rm ES}$ values. *BABAR* selects events with $m_{\rm ES} > 5.2\,{\rm GeV}/c^2$ and $|\Delta E|$ within $\pm 2.5\sigma$ of zero. They use the events in the background-dominated region $m_{\rm ES} < 5.27\,{\rm GeV}/c^2$ to determine the parameters of the background $\Delta t$ distributions. Belle requires $m_{\rm ES}$ and $\Delta E$ to be within $\pm 3\sigma$ around their expected means. They use candidates from a sideband region in the $m_{\rm ES} - \Delta E$ plane to determine the background parameters.

Events with a reconstructed $B^0$ are then analyzed to determine the flavor of the other $B$ using the $B$ flavor tagging algorithms described in detail in Chapter 8. Belle assigns 99.5% of the events to a flavor tag category, while *BABAR* rejects the 30% of events with marginal flavor discrimination.

The decay time difference $\Delta t$ between $B$ decays is determined from the measured separation $\Delta z = z_{\rm rec} - z_{\rm tag}$ along the $z$ axis between the vertices of the reconstructed $B_{\rm rec}$ and the flavor-tagging $B_{\rm tag}$ according to Eq. (17.5.13). *BABAR* applies an event-by-event correction for the directions of the $B$ meson momenta with respect to the $z$ direction in the $\Upsilon(4S)$ frame. A description of this correction and details of the calculation of $z_{\rm rec}$ and $z_{\rm tag}$ and their respective resolutions for fully-reconstructed $B$ decays are given in Chapter 6. In its paper, *BABAR* notes a correlation between the $\Delta t$ residual $\delta\Delta t = \Delta t - \Delta t_{\text{true}}$ and $\sigma_{\Delta t}$ (see Fig. 6.5.4). It is due to the fact that, in $B$ decays, the vertex error ellipse for the $D$ decay products is oriented with its major axis along the $D$ flight direction, leading to a correlation between the $D$ flight direction and the calculated uncertainty on the vertex position in $z$ of the $B_{\text{tag}}$. In addition, the flight length of the $D$ in the $z$ direction is correlated with its flight direction. Therefore, the bias in the measured $B_{\text{tag}}$ vertex position due to including the $D$ decay products is correlated with the $D$ flight direction. Taking into account these two correlations, *BABAR* concludes that $D$ mesons that have a flight direction perpendicular to the $z$ axis in the laboratory frame will have the best $z$ resolution and will introduce the least bias in a measurement of the $z$ position of the $B_{\text{tag}}$ vertex, while $D$ mesons that travel forward in the laboratory will have poorer $z$ resolution and will introduce a larger bias in the measurement of the $B_{\text{tag}}$ vertex.

After all selection criteria are applied, *BABAR* (Belle) finds 6300 (5300) signal events with an average purity of 86% (80%). Both experiments use an unbinned maximum likelihood fit to extract $\Delta m_d$ from the $\Delta t$ distributions of the selected candidates. The *p.d.f.* describing the data accounts for the presence of backgrounds with terms added to the signal description of Eq. (17.5.22):

$$\mathcal{H}_{\pm,i} = f_{\text{sig},i}\mathcal{H}_{\pm,\text{sig},i} + \sum_{j=\text{bkgd}} f_{i,j}\mathcal{B}_{\pm,i,j}(\Delta t, \hat{b}_{\pm,i,j}). \tag{17.5.24}$$

The background $\Delta t$ *p.d.f.*s $\mathcal{B}_{\pm,i,j}(\Delta t, \hat{b}_{\pm,i,j})$ provide an empirical description for the $\Delta t$ behavior of background events in each tagging category $i$. The background $\Delta t$ types considered are a prompt component and an exponentially decaying component with an effective lifetime. The prompt term is modeled with a delta function $\delta(\Delta t)$. Both experiments describe the background resolution *p.d.f.* with the same function as the signal resolution *p.d.f.*, but with separate parameters to minimize correlations. Both experiments determine the signal probability $f_{\text{sig},i}$ for each $B$ candidate $i$ from its $m_{\text{ES}}$ and $\Delta E$ values (*BABAR* only uses $m_{\text{ES}}$) based on separate fits to the $m_{\text{ES}}$ and $\Delta E$ distributions.

In the likelihood fit *BABAR* approximates the signal $\Delta t$ resolution function by a sum of three Gaussian distributions (core, tail, and outlier) with different means and different widths (see Chapter 10). The resolution is determined separately for each signal candidate depending on the uncertainty of its $\Delta t$ value. *BABAR* uses separate resolution function parameters for each tagging category, while Belle uses a common parameterization.

In the final fit Belle lets only $\Delta m_d$ and the mistag rates $w_i$ ($i$ = 1–6) vary. *BABAR*'s likelihood fit has 44 free parameters: $\Delta m_d$, average mistag rate and difference between $B^0$ and $\overline{B}^0$ for each tagging category (8), signal resolution function parameters (16), and parameters for background time dependence (5), $\Delta t$ resolution (6), and effective mistag rates (8).

In fully-reconstructed $B$ decays to hadronic final states *BABAR* measures in a sample of 29.7 fb$^{-1}$ $\Delta m_d = (0.516 \pm 0.016 \pm 0.010)\,\text{ps}^{-1}$, where the first error is statistical and the second is systematic. The central value has been corrected by $(-0.002\pm0.002)\,\text{ps}^{-1}$ to account for a small variation of the background composition as a function of $m_{\text{ES}}$. An additional correction of $(-0.007\pm0.003)\,\text{ps}^{-1}$ has been applied to account for a bias observed in fully-simulated MC events due to correlations between the mistag rate and the $\Delta t$ resolution that are not explicitly included in the likelihood function. Belle measures $\Delta m_d = (0.528 \pm 0.017 \pm 0.011)\,\text{ps}^{-1}$ in a sample of 29.1 fb$^{-1}$.

The largest contributions to the systematic uncertainty in the Belle measurement come from the uncertainties in the signal $\Delta t$ resolution function parameters ($0.008\,\text{ps}^{-1}$) and limited MC statistics ($0.005\,\text{ps}^{-1}$). In the *BABAR* fit the parameters of the signal and background $\Delta t$ resolutions functions are allowed to vary, and their contribution to the uncertainty on $\Delta m_d$ is included as part of the statistical error. The largest remaining systematic uncertainties come from uncertainties in the $B^0$ lifetime ($0.006\,\text{ps}^{-1}$) and in the alignment of the SVT ($0.005\,\text{ps}^{-1}$).

#### *Semileptonic decays* $B^0 \to D^{*-}l^+\nu_l$

*BABAR* performs a simultaneous measurement of the $B^0$ lifetime and $\Delta m_d$ with a sample of semileptonic $B^0 \to D^{*-}l^+\nu_l$ decays using 21 fb$^{-1}$ of data (Aubert, 2003m). The $D^{*-}$ candidates are selected in the decay mode $D^{*-} \to \overline{D}^0\pi^-$, and the $\overline{D}^0$ candidates are reconstructed in the modes $K^+\pi^-$, $K^+\pi^-\pi^+\pi^-$, $K^+\pi^-\pi^0$, and $K^0_S\pi^+\pi^-$. Candidate $B^0 \to D^{*-}l^+\nu_l$ events are rejected if they fail selection criteria required to suppress backgrounds and ensure a well-measured $\Delta t$. These requirements include lepton and kaon identification, momenta of the lepton and the $D^{*-}$ and $\overline{D}^0$ daughter tracks and $\pi^0$, the $\overline{D}^0$ invariant mass, the $D^{*-} - \overline{D}^0$ mass difference, vertex probabilities, $\cos\theta^*_{\text{thrust}}$, the absolute value of $\Delta z$, and the calculated error on $\Delta t$. Furthermore they use two angular variables. The first angle is $\theta_{D^*,l}$, the angle between the $D^{*-}$ and the lepton candidate in the $\Upsilon(4S)$ frame. The second is $\theta_{B,D^*l}$, the inferred angle between the direction of the $B^0$ and the vector sum of the $D^{*-}$ and the lepton candidate momenta, calculated in the $\Upsilon(4S)$ frame.

The $B$ yield is larger than for the hadronic final state analysis due to the large $B$ semileptonic branching fraction. They reconstruct 680 $B/\text{fb}^{-1}$. Due to the missing neutrino the background level is higher than in the sample of fully-reconstructed hadronic $B$ decays. The combinatorial $D^{*-}$ background is about 18%, and the sum of the backgrounds from events where the $D^{*-}$ and the lepton come from different $B$ decays, events with a fake lepton candidate, and events from continuum $c\overline{c} \to D^{*-}X$ processes add up to 5–8%, depending on the lepton flavor.

The measurements of the decay vertex of the $B^0 \to D^{*-}l^+\nu_l$ candidate and that of the other $B$ in the event in this analysis is similar to *BABAR*'s $\Delta m_d$ analysis of fully-reconstructed hadronic final states. The decay time difference is determined from the $z$ positions of these vertices according to Eq. (17.5.12). The flavor of $B_{\text{tag}}$ is determined from the charged tracks in the event that do not belong to the $B^0 \to D^{*-}l^+\nu_l$ candidate using the algorithms described in Chapter 8. About 30% of the selected signal candidates have a mistag rate close to 50%. These events are not sensitive to $\Delta m_d$, but they increase the sensitivity to the $B^0$ lifetime. In this paper, *BABAR* describes an interesting correlation between the mistag rate and the $\Delta t$ resolution for the tagging category based on identified charged kaons.[58] Both the mistag rate for kaon tags and the calculated $\sigma_{\Delta t}$ depend inversely on $\sqrt{\sum p_t^2}$, where $p_t$ is the transverse momentum with respect to the $z$ axis of tracks from the $B_{\rm tag}$ decay. The mistag rate dependence originates from the kinematics of the physical sources for wrong-charge kaons. The three major sources of mis-tagged events in the kaon tag category are wrong-sign $D^0$ mesons from $B$ decays to double charm ($b \to c\bar{c}s$), wrong-sign kaons from $D^+$ decays, and kaons produced directly in $B$ decays. All these sources produce a spectrum of tracks that have smaller $\sqrt{\sum p_t^2}$ than $B$ decays that produce a correct tag. The $\sigma_{\Delta t}$ dependence originates from the $1/p_t^2$ dependence of $\sigma_z$ for the individual contributing tracks due to multiple scattering in the SVT and the beam pipe.

After all selection requirements are applied, the $B^0 \to D^{*-}l^+\nu_l$ selected event sample contains contributions from the following types of background: events with a misreconstructed $D^{*-}$ candidate, events from continuum $c\bar{c} \to D^{*-}X$ processes, events with a fake lepton candidate, events with a charged $B$, and events in which the lepton does not come from the primary $B$ decay. They model the $\Delta t$ distributions of each background with combinations of prompt, exponential, and oscillatory functions convolved with background resolution functions. The parameters of background *p.d.f.*s are obtained from fits to control samples and simulated events. *BABAR* split their data into two signal samples and ten control samples depending on whether the data was taken on or off the $\Upsilon(4S)$ resonance, whether the lepton candidate was on the same side or opposite side to the $D^{*-}$ candidate, and whether the lepton candidate was an electron, muon, or fake lepton. Furthermore they split each of these samples into subsamples according to the reconstruction of the soft pion, the $\overline{D}^0$ decay mode, and the $B$ flavor-tagging category for a total of 360 subsamples.

They extract the $B^0$ lifetime and $\Delta m_d$ from a simultaneous fit to the $\Delta t$ and $\sigma_{\Delta t}$ values of the events of the 360 event samples. The fit has 70 additional free parameters to describe the signal and background $\Delta t$ resolution functions and mistag rates, and the background $\Delta t$ shapes. From the fit they determine $\tau_{B^0} = (1.523^{+0.024}_{-0.023} \pm 0.022)\,\text{ps}$ and $\Delta m_d = (0.492 \pm 0.018 \pm 0.013)\,\text{ps}^{-1}$. The statistical correlation coefficient between $\tau_{B^0}$ and $\Delta m_d$ is $-0.22$. Dominant systematic error sources in the $\Delta m_d$ measurement are the SVT alignment and the signal and background probabilities. An additional systematic uncertainty in the $\Delta m_d$ measurement comes from the limited statistical precision in determining the bias due to the background modeling. By comparing the fitted $\Delta m_d$ in simulated events, *BABAR* observes a shift of $(0.020 \pm 0.005)\,\text{ps}^{-1}$ between a signal-only sample and a signal-plus-background sample. The measured $\Delta m_d$ is corrected for the observed bias from the fit to the MC sample with background, and the full statistical uncertainty in $\Delta m_d$ of $\pm 0.012\,\text{ps}^{-1}$ is assigned as a systematic uncertainty.

Belle also measures $\Delta m_d$ with a sample of semileptonic $B^0 \to D^{*-}l^+\nu_l$ decays corresponding to $29.1\,\text{fb}^{-1}$ of data (Hara, 2002). They select $D^{*-}$ candidates in the decay mode $D^{*-} \to \overline{D}^0\pi^-$ and the $\overline{D}^0$ candidates are reconstructed in the modes $K^+\pi^-$, $K^+\pi^-\pi^+\pi^-$, and $K^+\pi^-\pi^0$. Candidate events are rejected if they fail selection criteria required to suppress backgrounds and ensure a well-measured $\Delta t$. The applied requirements are similar to those in the *BABAR* $\Delta m_d$ measurement described above.

Belle uses its standard algorithms for the $\Delta t$ measurements and $B$ flavor tagging in this analysis, which are the same as in its measurement of $\sin\phi_1$ (Abe, 2002k). The algorithms are described in more detail in Chapters 6 and 8. After all selection criteria, including flavor tagging and vertex reconstruction, are applied, Belle reconstructs 453 $B^0/\,\text{fb}^{-1}$ with a signal purity of 80.4%. The background consists of misreconstructed $D^*$ mesons (7.8%), $B \to D^{**}l\nu$ events (7.4%), random combinations of $D^*$ mesons with leptons with no angular correlation (2.6%), and continuum events (1.8%).

Belle measures $\Delta m_d$ from a simultaneous fit to the $\Delta t$ and $\sigma_{\Delta t}$ distributions of the mixed and unmixed events. The fit has a total of ten free parameters including $\Delta m_d$, six flavor mistag rates, the fraction of the $D^{**}$ background coming from charged $B$ decays, its effective lifetime, and the fraction of charged $B$ decays. All other parameters are determined from MC simulation and data control samples. The likelihood fit gives $\Delta m_d = (0.494 \pm 0.012 \pm 0.015)\,\text{ps}^{-1}$. Dominant systematic error sources in the $\Delta m_d$ measurement are due to uncertainties in the $D^{**}$ branching fractions ($0.007\,\text{ps}^{-1}$), the selected $|\Delta t|$ range ($0.007\,\text{ps}^{-1}$), the background $\Delta t$ *p.d.f.* parameters ($0.006\,\text{ps}^{-1}$), the signal $\Delta t$ resolution function ($0.006\,\text{ps}^{-1}$), and the $B^0$ lifetime ($0.005\,\text{ps}^{-1}$).

### *Belle hadronic and semileptonic combination*

Belle's most recent measurement of $\Delta m_d$ comes from a simultaneous analysis of $B$ decays to the exclusive hadronic final states $B^0 \to D^{(*)-}\pi^+$, $D^*\rho^+$, $J/\psi K_S^0$, $J/\psi K^{*0}$, and the semileptonic decay $B^0 \to D^{*-}l\nu$ in a sample of $140\,\text{fb}^{-1}$ (Abe, 2005c). In the same analysis, they also determine the $B^0$ lifetime and, using the decays $B^+ \to \overline{D}^0\pi^+$ and $J/\psi K^+$, the $B^+$ lifetime.

The signal modes and selection criteria of the hadronic final states are similar to the ones used in Tomura (2002b), while the $B^0 \to D^{*-}l^+\nu$ selection follows that described in Zheng (2003). The $\Delta t$ reconstruction uses the algorithm described in Tajima (2004). The $B$ flavor tagging algorithm is similar to the one used in Belle's previous analyses of fully-reconstructed final states, but they allow for separate mistag rates for $B^0$ and $\overline{B}^0$ tagged events. The overall $B^0$ signal purity after all selection criteria are applied is 80.9%, and the $B^0$ signal yield is 707 $B/\mathrm{fb}^{-1}$.

[58] This correlation is already observed and accounted for in Aubert (2002b), but is not described in that paper.

Belle performs an unbinned maximum likelihood fit to the $\Delta t$ distributions of the selected $B^0$ and $B^+$ candidates to simultaneously obtain values of the $B^0$ and $B^+$ lifetimes (2), $\Delta m_d$ (1), the mistag fractions (12), the signal $\Delta t$ resolution function parameters (14), and parameters to describe the $B^+$ background in $B^0$ decays (3). The signal resolution function has two parameters added to the ones described in Tajima (2004) to better describe the effect of charmed particle decays on the $B_{\mathrm{tag}}$ vertex. The same $\Delta t$ resolution function is used for $B^0$ and $B^+$ signal candidates. The background for the hadronic $B$ decay modes is described by the convolution of the sum of a prompt term and a term with an effective background lifetime with a background $\Delta t$ resolution function. The background for the $B^0 \to D^{*-} l^+ \nu$ decays is the same as in the earlier study of this mode (Hara, 2002) described above. The $\Delta t$ behavior of the backgrounds is modeled with prompt and lifetime terms. The backgrounds due to $D^{**}$ and misreconstructed $D^*$ candidates also have an oscillatory component.

Belle extracts the $B^0$ lifetime and $\Delta m_d$ to be, respectively, $\tau_{B^0} = (1.534 \pm 0.008 \pm 0.010)\,\mathrm{ps}$ and $\Delta m_d = (0.511 \pm 0.005 \pm 0.006)\,\mathrm{ps}^{-1}$. Dominant systematic error sources in the $\Delta m_d$ measurement are the $B$ vertex reconstruction ($0.004\,\mathrm{ps}^{-1}$) and the $D^{**}$ background parameters ($0.003\,\mathrm{ps}^{-1}$).

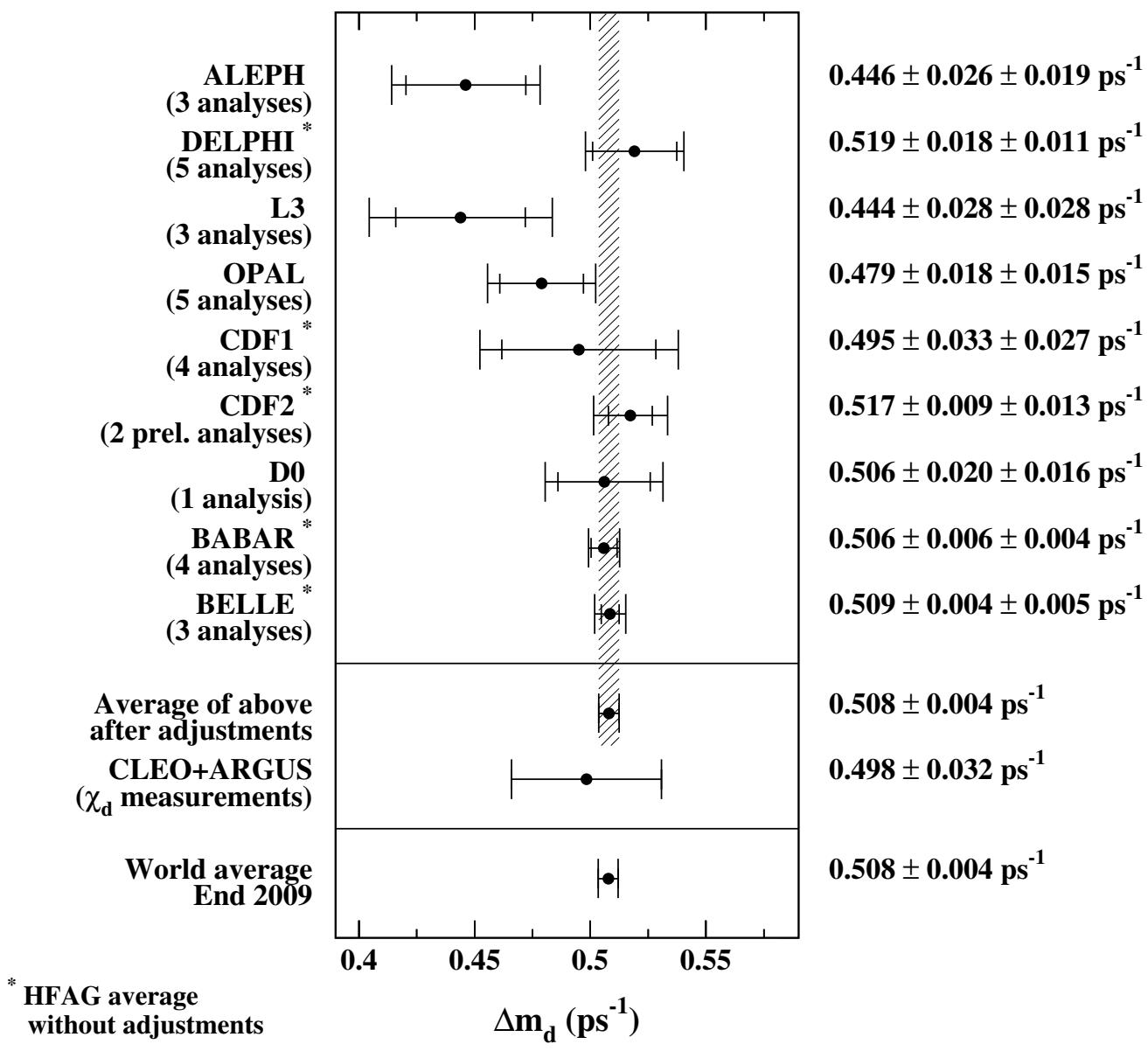

**Figure 17.5.6.** The $B^0\overline{B}^0$ oscillation frequency $\Delta m_d$ as measured by the different experiments along with their average. Averages are also given separately for the time-dependent measurements by the $B$ Factories and the LEP and Tevatron experiments, and the time-integrated measurements by CLEO and ARGUS (Asner et al., 2010).

#### 17.5.2.5 Average of $\Delta m_d$

The various measurements of $\Delta m_d$ by the $B$ Factories listed in Table 17.5.2 have been averaged by the Heavy Flavor Averaging Group (HFAG; Asner et al., 2010), where results superseded by more recent ones have been omitted from the average. Before being combined, the $\Delta m_d$ measurements have been adjusted to a common set of input values, including the $B$ meson lifetimes. The total systematic uncertainty in $\Delta m_d$ is of the same size as the statistical uncertainty, although only a small fraction of the total $B$ Factories' data sets have been used in the measurements. Systematic correlations arise from common physics sources (*e.g.* $B$ lifetimes and branching fractions) and common experimental techniques and algorithms (*e.g.* flavor tagging, $\Delta t$ resolution, and background description). Combining the $B$ Factories $\Delta m_d$ measurements and accounting for all identified correlations, HFAG quotes

$$\Delta m_d = (0.508 \pm 0.003 \pm 0.003)\,\mathrm{ps}^{-1}, \qquad (17.5.25)$$

where the first error is statistical and the second is systematic (Asner et al., 2010). Combining the $B$ Factories $\Delta m_d$ average with time-dependent measurements from the LEP and Tevatron experiments, and time-integrated measurements from CLEO and ARGUS, gives the same value. The values of $\Delta m_d$ as measured by different experiments along with the time-dependent and time-integrated averages are shown in Fig. 17.5.6. Two recent measurements by the LHCb Collaboration, $\Delta m_d = (0.516 \pm 0.005 \pm 0.003)\,\mathrm{ps}^{-1}$ (Aaij et al., 2013b) and $\Delta m_d = (0.499 \pm 0.032 \pm 0.003)\,\mathrm{ps}^{-1}$ (Aaij et al., 2012f), are not included in the HFAG average and the figure. The 2013 PDG world average including these results is $\Delta m_d = (0.510 \pm 0.004)\,\mathrm{ps}^{-1}$ (Beringer et al., 2012). The world average of the $B^0\overline{B}^0$ oscillation frequency $\Delta m_d$ is an input to the calculation of the magnitude of the CKM matrix element $V_{td}$. Along with $\Delta m_s$, the $B_s^0\overline{B}_s^0$ oscillation frequency as measured by CDF and DØ, $\Delta m_d$ is used to calculate the ratio of CKM matrix elements $|V_{td}/V_{ts}|$ (see Section 17.2).

#### 17.5.2.6 Measurements of $\Delta\Gamma_d$

Transitions between a $B^0$ state and a $\overline{B}^0$ state can be mediated by a box diagram involving virtual top quarks (see Fig. 10.1.1) or by real intermediate states accessible to both $B^0$ and $\overline{B}^0$. The former process determines the magnitude of $\Delta m_d$, while the latter gives rise to a difference in decay width of the neutral $B$ mass eigenstates. The decay width difference is defined as $\Delta\Gamma_d \equiv \Gamma_{H,d} - \Gamma_{L,d}$,[59] where $H$ and $L$ refer to the heavy and light $B^0$ states, respectively. In the $B_s^0$ system the corresponding relative decay width difference is large, $\Delta\Gamma_s/\Gamma_s = (15 \pm 2)\%$ (Beringer et al., 2012), due to the significant branching fractions of $B_s^0$ and $\overline{B}_s^0$ to $D_s^{(*)+}D_s^{(*)-}$. Since the decays of $B^0$ and

[59] Note, the Particle Data Group (Beringer et al., 2012) uses the definition $\Delta\Gamma_d = \Gamma_{L,d} - \Gamma_{H,d}$.

$\overline{B}^0$ to common final states are strongly suppressed, $\Delta\Gamma_d$ is expected to be much smaller than $\Delta\Gamma_s$. A recent SM calculation predicts $\Delta\Gamma_d/\Gamma_d = (-4.2 \pm 0.8) \times 10^{-3}$ (Lenz and Nierste, 2011). The best limit prior to the $B$ Factories measured by DELPHI was $\Delta\Gamma_d/\Gamma_d < 0.18$ at 95% C.L. (Abdallah et al., 2003). The small value of $\Delta\Gamma_d$ in the SM makes it a sensitive parameter in the search for new physics (Dighe, Hurth, Kim, and Yoshikawa, 2002).

In the derivation of the time-dependent decay rates in Chapter 10 we have neglected the case of non-zero $\Delta\Gamma_d$. This is justified because of the small values of $\Delta\Gamma_d/\Gamma_d$ and $\Delta\Gamma_d/\Delta m_d$ predicted by the SM. The decay rates given in Eqs (10.2.2) and (10.2.3) are sufficient for all measurements of $\Delta m_d$ (Section 17.5.2) and *mixing-induced CP* asymmetries in $B$ decays (Sections 17.6–17.8). However, physics from processes beyond the SM can lead to a sizable $\Delta\Gamma_d$. *BABAR* (Aubert, 2004e,f) and Belle (Higuchi, 2012) have both measured $\Delta\Gamma_d$ as part of analyses that search for *CP*, *T*, and *CPT* violation in $B^0\overline{B}^0$ mixing. The analyses are described in detail below (see Section 17.5.4). Here we describe the sensitivity to $\Delta\Gamma_d$ and the results from the $B$ Factories.

The time-dependence of the decay rates for $B$ decays to *CP* eigenstates and flavor-specific final states in the absence of *CP* and *CPT* violation in $B^0\overline{B}^0$ mixing, but including additional terms due to $\Delta\Gamma_d$, are given by

$$f_{\pm}^{\Delta\Gamma_d}(\Delta t) \propto e^{-|\Delta t|/\tau_{B^0}} \left[ \cosh\left(\frac{\Delta\Gamma_d \Delta t}{2}\right) \mp C\cos(\Delta m_d \Delta t) + A^{\Delta\Gamma_d}\sinh\left(\frac{\Delta\Gamma_d \Delta t}{2}\right) \pm S\sin(\Delta m_d \Delta t)\right], \tag{17.5.26}$$

with

$$C = \frac{1-|\lambda|^2}{1+|\lambda|^2},\ S = \frac{2\,\mathrm{Im}\lambda}{1+|\lambda|^2},\ A^{\Delta\Gamma_d} = \frac{2\,\mathrm{Re}\lambda}{1+|\lambda|^2}. \tag{17.5.27}$$

The parameter $\lambda = \frac{q}{p}\frac{\overline{A}_f}{A_f}$ has been introduced in Chapter 10, where $A_f$ ($\overline{A}_f$) represents the amplitude for the decay of a $B^0$ ($\overline{B}^0$) to the final state $f$, and $q/p$ is the weak phase in $B^0\overline{B}^0$ mixing. Note that $(S)^2 + (C)^2 + \left(A^{\Delta\Gamma_d}\right)^2 = 1$ by definition. The time-dependence for a $B^0$ ($\overline{B}^0$) tagged event is given by $f_+^{\Delta\Gamma_d}$ ($f_-^{\Delta\Gamma_d}$).

For $B$ decays to *CP* eigenstates that proceed through a single weak amplitude, $|\lambda| = 1$ and thus $C = 0$, $S = \mathrm{Im}\lambda$, and $A^{\Delta\Gamma_d} = \mathrm{Re}\lambda$. For example, the time-dependence for the *golden CP* mode $B \to J/\psi K_S^0$ (see Chapter 10 and Section 17.6) simplifies to

$$f_{\pm}^{J/\psi K_S^0,\Delta\Gamma_d}(\Delta t) \propto e^{-|\Delta t|/\tau_{B^0}} \left[ \cosh\left(\frac{\Delta\Gamma_d \Delta t}{2}\right) + \cos(2\phi_1)\sinh\left(\frac{\Delta\Gamma_d \Delta t}{2}\right) \pm \sin(2\phi_1)\sin(\Delta m_d \Delta t)\right], \tag{17.5.28}$$

where $\phi_1$ is one of the angles of the Unitarity Triangle.

For $B$ decays to flavor-eigenstates which are only accessible from either a $B^0$ or a $\overline{B}^0$, $|\lambda|$ is zero or infinite and thus $C = 1$, $S = A^{\Delta\Gamma_d} = 0$. The corresponding time-dependence is given by

$$h_{\pm}^{\Delta\Gamma_d}(\Delta t) \propto e^{-|\Delta t|/\tau_{B^0}} \left[ \cosh\left(\frac{\Delta\Gamma_d \Delta t}{2}\right) \pm \cos(\Delta m_d \Delta t)\right], \tag{17.5.29}$$

where $h_+^{\Delta\Gamma_d}$ and $h_-^{\Delta\Gamma_d}$ refer to unmixed and mixed events, respectively.

*BABAR* and Belle both use samples of $B$ decays to *CP* eigenstates and flavor-specific final states in their measurement of $\Delta\Gamma_d$. The *BABAR* analysis is performed with $88 \times 10^6$ $B\overline{B}$ pairs and uses the $B_{\text{flav}}$ decays to $D^{(*)-}\pi^+(\rho^+, a_1^+)$, $J/\psi K^{*0}(\to K^+\pi^-)$ and $B_{CP}$ decays to $J/\psi K_S^0$, $\psi(2S)K_S^0$, $\chi_{c1}K_S^0$, and $J/\psi K_L^0$. The Belle analysis is performed with $535 \times 10^6$ $B\overline{B}$ pairs and uses the $B_{\text{flav}}$ decays to $D^{(*)-}\pi^+$, $D^{*-}\rho^+$, and $D^{*-}\ell^+\nu_l$ and $B_{CP}$ decays to $J/\psi K_S^0$ and $J/\psi K_L^0$. The cosh and sinh terms in Eqs (17.5.28) and (17.5.29) do not change sign with the flavor of $B_{\text{tag}}$. This allows the experiments to also use events without a flavor-tagged $B$ in their analyses. The time-dependence of the $B_{CP}$ samples include a $\sinh\left(\frac{\Delta\Gamma_d\Delta t}{2}\right)$ term, which is practically linear in $\Delta\Gamma_d$. The $B_{\text{flav}}$ sample is only sensitive to $\Delta\Gamma_d$ through a $\cosh\left(\frac{\Delta\Gamma_d\Delta t}{2}\right)$ term and thus effectively to $\mathcal{O}(\Delta\Gamma_d^2)$. Therefore, even though the $B_{CP}$ events represent only 8% (4%) of the selected signal events in the *BABAR* (Belle) analysis, they dominate the $\Delta\Gamma_d$ measurement.

The experiments perform unbinned likelihood fits to the $\Delta t$ distributions of the flavor-tagged and untagged $B_{CP}$ and $B_{\text{flav}}$ samples (after accounting for experimental effects such as the $\Delta t$ resolution and $B$ flavor-tagging) to extract parameters that violate *CP*, *T*, or *CPT* symmetries and $\Delta\Gamma_d$. In the fit the sign of $A^{\Delta\Gamma_d}$ is fixed to the value obtained from global CKM fits (see Section 25.1). *BABAR* measures $\Delta\Gamma_d/\Gamma_d = -0.008 \pm 0.037 \pm 0.018$ and Belle measures $\Delta\Gamma_d/\Gamma_d = -0.017 \pm 0.018 \pm 0.011$. The dominant systematic error contributions arise from uncertainties in the reconstruction of the $B$ vertices and the $\Delta t$ resolution function. The results are consistent with each other. The $B$ Factories average value is $\Delta\Gamma_d/\Gamma_d = -0.015 \pm 0.019$ (Beringer et al., 2012), consistent with the small predicted value. Larger $B$ samples at LHCb and future super flavor factories should allow the measurement of $\Delta\Gamma_d$ at the SM value or find discrepancies as evidence of new physics (Gershon, 2011), if the systematic uncertainties can be kept under control.

### 17.5.3 Tests of quantum entanglement

The $B$-lifetime and $B^0 - \overline{B}^0$ mixing results of the previous sections rely on certain assumptions about the physics of $B$-meson production and decay (see the discussion in Chapter 10). Some of these assumptions can be tested by performing an extended analysis including symmetry-breaking parameter(s) in the final fit. This approach is used to test discrete symmetries, including *CPT* (Sections 17.5.4 and 17.5.5 below); if the assumption of Lorentz invariance is also relaxed, qualitatively new phenomena are expected and the $B^0 - \overline{B}^0$ mixing analysis method of Section 17.5.2 must be heavily modified (Section 17.5.5), even when a standard mixing event selection is retained.

Quantum mechanical principles governing the entangled $B^0\overline{B}^0$ state may also be tested. The careful conceptual treatment required in this case is reviewed in Section 17.5.3.1. One such analysis has been performed by Belle (Go, 2007): the event selection and background treatment are both straightforward modifications of those in the $D^*\ell\nu$ mixing analysis of Abe (2005c), as discussed in Section 17.5.3.2. The final analysis is presented in Section 17.5.3.3.

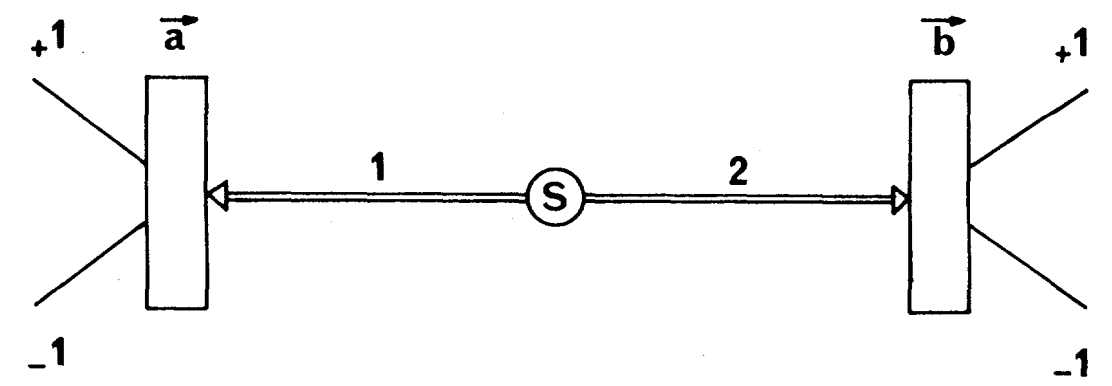

(a) Aspect: freely-chosen analyzer orientations $\boldsymbol{a}$, $\boldsymbol{b}$.

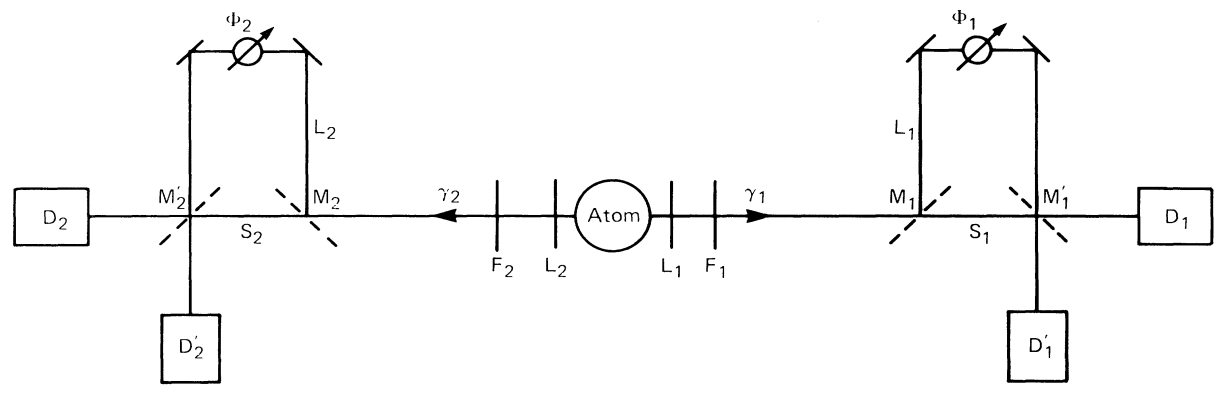

(b) Franson: freely-chosen phase shifts $\phi_1$, $\phi_2$.

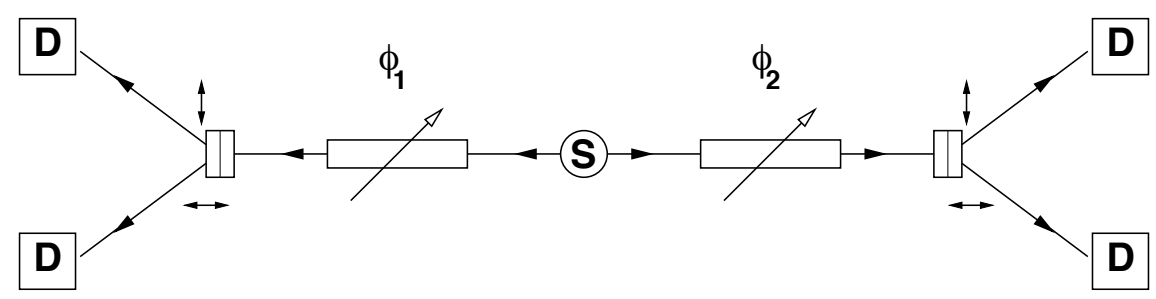

(c) Go: fixed analyzers; variable phase shifts $\phi_1$, $\phi_2$.

**Figure 17.5.7.** Schematics of the Bell inequality tests with photons by (a) Aspect, Grangier, and Roger (1982); (b) the position-time test proposed by Franson (1989); and (c) an optical analog of the Go (2007) analysis of $B^0\overline{B}^0$ pairs (from Yabsley, 2008). To perform a Bell test, projective measurements must be performed onto axes determined outside the system under study. In (a), the analyzer orientations can be freely chosen; in (b) the projections recorded by the detectors $D_i$ are fixed, but phase shifts imposed on the photons can be chosen; in (c), neither the projection axes ($\updownarrow \equiv B^0$ or $\leftrightarrow \equiv B^0$) nor phase shifts ($\phi_i = \Delta m_d t_i$) are subject to experimental control.

### 17.5.3.1 $B^0 - \overline{B}^0$ mixing and entanglement tests

As discussed in Section 10.2, $\Upsilon(4S)$ decay prepares a neutral $B$ meson pair in the coherent state

$$\Psi = \frac{1}{\sqrt{2}} \left[ |B^0(\boldsymbol{p})\rangle|\overline{B}^0(-\boldsymbol{p})\rangle - |\overline{B}^0(\boldsymbol{p})\rangle|B^0(-\boldsymbol{p})\rangle \right] \tag{17.5.30}$$

given there in a more compact notation as Eq. (10.2.1). The formulae for the time-dependent evolution of the $B$ pair in the remainder of that section follow from this expression. Such a state is *entangled*: it cannot be represented as a product of states of the first $B$ (with momentum $\boldsymbol{p}$) and the second $B$ (with momentum $-\boldsymbol{p}$); it is a flavor analog of the spin-singlet state for a photon pair,

$$\Psi = \frac{1}{\sqrt{2}} \left( |\Uparrow\rangle_1 |\Downarrow\rangle_2 - |\Downarrow\rangle_1 |\Uparrow\rangle_2 \right), \tag{17.5.31}$$

familiar from Bohm's version of the thought experiment on "EPR correlations" (Bohm, 1951; Einstein, Podolsky, and Rosen, 1935). Powerful tests of such correlations are possible (Bell, 1964), and have been carried out on photon pairs by Aspect, Grangier, and Roger (1982) and many subsequent investigators. Subject to certain experimental "loopholes", such tests exclude the hypothesis that the individual photons have definite physical states at all times (a feature of so-called "local realistic" models). Quantum entanglement thus appears to be an experimental fact, which would persist even if quantum mechanics (QM) itself were replaced by future developments.

Bell tests using photons rely on experimental choice of the orientation (polarization axis) of analyzers, in experiments of the Aspect type; or on fixed analyzers, and experimental choice of phase shifts imposed on the photons in flight (following Franson, 1989); see Fig. 17.5.7(a) and (b) for schematics of both arrangements. $B^0 - \overline{B}^0$ mixing is analogous to the latter case, as a flavor-tagging decay projects a neutral $B$ meson onto one of two fixed axes: $B^0$, equivalent to spin-up for a fermion or vertical polarization for a photon; or $\overline{B}^0$, equivalent to spin-down or to horizontal polarization. For discussion of this *quasi-spin* analogy, see Lee and Wu (1966), Lipkin (1968), and Bertlmann and Hiesmayr (2001); the assignment of spin and polarization states to flavors is arbitrary. If we ignore $B$-meson decay, the state $|B^0\rangle$ at production evolves to the state

$$\frac{1}{2} \left[ \{1 + \cos(\Delta m_d t)\}|B^0\rangle + \{1 - \cos(\Delta m_d t)\}|\overline{B}^0\rangle \right] \tag{17.5.32}$$

at a later time $t$, from Eqs (10.1.6) and (10.1.7); *cf.* Eqs (17.5.14)–(17.5.17) above. For a $B^0\overline{B}^0$ pair undergoing two flavor-tagging decays, the product $\Delta m_d \Delta t$ therefore corresponds to the difference in phase shifts $\Delta\phi$ imposed in a Franson-type experiment, or the angle between polarization analyzers chosen in an Aspect-type experiment (see Fig. 17.5.7(c)).

An early attempt to re-interpret $B$ Factory mixing results as Bell inequality tests was presented by Go (2004). In fact, no such test is possible using $B$ Factory measurements of the $B^0 - \overline{B}^0$ system (Bertlmann, Bramon, Garbarino, and Hiesmayr, 2004):

1. *Flavor measurements at the B Factories are passive,* relying on spontaneous decay of the $B$ mesons rather

than (say) interactions with converters placed in the path of each $B$. It is therefore not possible to exclude local models where EPR-like decays of the two $B$ mesons have been determined in advance, as $\Delta m_d \Delta t$ is not subject to experimental control. Schematic diagrams comparing this case and entangled-photon experiments are shown in Fig. 17.5.7; note that in photon experiments, control of analyzer orientations has been demonstrated for spacelike-separated measurements (for example Weihs, Jennewein, Simon, Weinfurter, and Zeilinger, 1998).

2. *The rate of $B_d^0$-mixing is too low,* relative to the rate of decay, to construct a Bell test even in the case of active measurements. The crucial value of $x = \Delta m/\Gamma$ is found to be 2.0; *cf.* $x_d = \Delta m_d/\Gamma_d = (0.775 \pm 0.007)$. Note that as $x_s = \Delta m_s/\Gamma_s = (26.82 \pm 0.23) \gg 2.0$, a Bell test using active measurements of the $B_s^0 - \overline{B}_s^0$ system is possible in principle, although not practical with foreseeable technology. Values are taken from the 2013 update of Beringer et al., 2012.

Artificial local models which reproduce QM predictions for $B$ Factory results have been constructed by Bertlmann, Bramon, Garbarino, and Hiesmayr (2004), following Kasday (1971); and by Santos (2007), to further demonstrate that such models cannot be excluded as a class.

It is however possible to compare $B$ Factory results with the predictions of both quantum mechanics and various local models. The Belle analysis (Go, 2007) tested both decoherence models (following Bertlmann, Grimus, and Hiesmayr, 1999), and a broad class of models that reproduce the QM predictions for uncorrelated $B$ decays (Pompili and Selleri, 2000). Predictions for the $B^0\overline{B}^0$ mixing asymmetry $A_{\rm mix}(\Delta t)$ of Eq. (17.5.20) are shown in Fig. 17.5.8 for QM and for spontaneous disentanglement (SD), an extreme form of decoherence corresponding to $\zeta = 1$ in the $\{B^0, \overline{B}^0\}$ basis in Bertlmann et al., or the hypothesis of Furry (1936); asymmetries for models in the Pompili and Selleri class must lie between the two curves $\mathrm{PS_{max}}$ and $\mathrm{PS_{min}}$. With sufficient resolution, $A_{\rm mix}(\Delta t)$ measurements can discriminate between these models; with reconstruction of individual decay times (not just $\Delta t$), much stronger discrimination would be possible at a next-generation flavor factory (see Eqs (2)–(5) of Go, 2007, and Figure 4 of Yabsley, 2008).

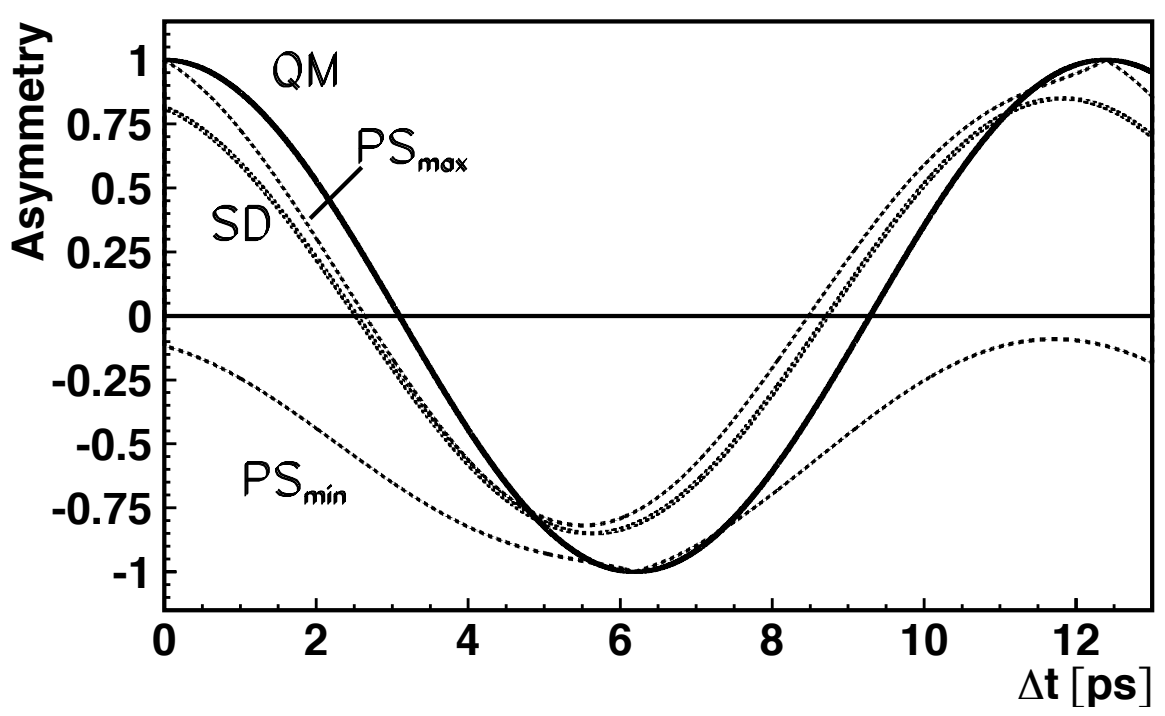

**Figure 17.5.8.** Time dependent asymmetry predictions for (QM) quantum mechanics, (SD) spontaneous disentanglement, and ($\mathrm{PS_{max}}$ to $\mathrm{PS_{min}}$) the allowed range for models in the class described by Pompili and Selleri (2000). See the discussion in the text. From Go (2007).

#### 17.5.3.2 Event selection and background treatment

The Belle analysis (Go, 2007) uses a sample of $B^0\overline{B}^0$ events where one $B$ is reconstructed as $B^0 \to D^{*-}\ell^+\nu$ (or charge conjugate), and the remaining tracks are subjected to the Belle flavor-tagging algorithm (Section 8.6.4). Taken from $140\,\mathrm{fb}^{-1}$ of data, the sample is a subset of the $D^*\ell\nu$ sample of Abe (2005c) discussed in Sections 17.5.1.3 and 17.5.2.4 above. To perform the entanglement analysis, the event selection and background treatment of Abe (2005c) are modified in the following ways:

1. *Only events with the highest-purity flavor tag are used* (*i.e.* $0.875 < r < 1.000$; see Section 8.6.4), with the further restriction that the tag is based on a reconstructed lepton. This reduces the sample from 84823 to 8565 events.
2. *The data are binned*, separately for opposite-flavor (OF) and same-flavor (SF) events, into 11 variable-width bins in $\Delta t$.
3. *Backgrounds are subtracted*, in both OF and SF samples, using the same background categorization as Abe (2005c): $e^+e^- \to q\overline{q}$ continuum (found to be negligible), non-$D^*$ events, wrong $D^*$-lepton combinations, and $B^+ \to \overline{D}^{**0}\ell^+\nu$ events; $B^0 \to \overline{D}^{**-}\ell^+\nu$ events, which undergo mixing, are retained.
4. *Remaining reconstruction effects are unfolded* using deconvolution with single value decomposition (Höcker and Kartvelishvili, 1996) separately on the OF and SF samples, based on $11 \times 11$ response matrices built from MC $D^*\ell\nu$ events; see Go (2007) for the details.

To avoid potential bias due to the MC events underlying the response matrices, the deconvolution procedure is validated on Monte Carlo samples generated according to each of the QM, SD, and PS models. Differences between results and inputs are averaged over the three models, and subtracted from the measured asymmetry; the largest remaining deviation in each $\Delta t$ bin, over all three models, is then assigned as a contribution to the systematic uncertainty.

The resulting asymmetry $A_{\rm mix} = (N_{\rm OF} - N_{\rm SF})/(N_{\rm OF} + N_{\rm SF})$ in bins of the time difference $\Delta t$, with statistical and four categories of systematic uncertainties, is given in Table 1 of Go (2007); systematics become comparable to statistical uncertainties for $\Delta t > 4.0$ ps, with the uncertainties due to background subtraction and deconvolution dominant in the final $[13.0, 20.0]$ ps bin. These results can be directly compared with theoretical models that lie outside the analysis discussed in the following section.

#### 17.5.3.3 Analysis and interpretation

For each model, a weighted least-squares fit is performed, to the asymmetries $A_{\rm mix}$ and their total uncertainties as data, and the function shown in Fig. 17.5.8 as the prediction. The mass difference $\Delta m_d$ appears as a parameter in each model, however the world-average value of $\Delta m_d$ is dominated by $B$ Factory measurements, which assume time evolution according to QM in their analysis (see Section 17.5.2.1). An average of results then available (Barberio et al., 2006), excluding $B$ Factory measurements, was therefore performed, yielding $\langle \Delta m_d \rangle = (0.496 \pm 0.014)\,\text{ps}^{-1}$. The uncertainty was treated by including $\Delta m_d$ as a parameter in the fit, and adding an additional term $[(\Delta m_d - \langle \Delta m_d \rangle)/\sigma_{\Delta m_d}]^2$ to the least-squares statistic; this technique is now in common use for treating systematic uncertainties *e.g.* at LHC experiments).

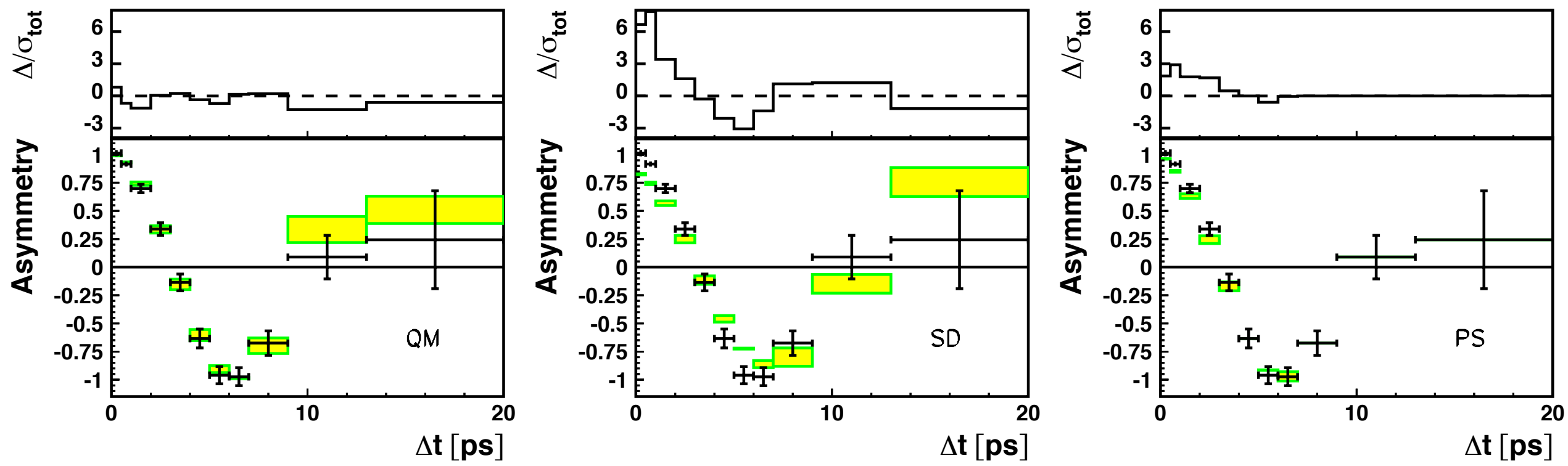

**Figure 17.5.9.** Asymmetry $A_{\rm mix}$ and its total uncertainty (crosses) in bins of $\Delta t$, and the results of fits to predictions from (left: QM) quantum mechanics, (middle: SD) spontaneous decomposition, and (right: PS) the Pompili and Selleri (2000) class of models. The shaded boxes show the variation in the predictions as the fitted value of $\Delta m_d$ is allowed to vary by $\pm 1\sigma$; see the text, in particular for handling of the PS case. The upper panels show normalized residuals in each bin. From Go (2007).

The results of the fits are shown in Fig. 17.5.9. The predictions of quantum mechanics are favored over spontaneous disentanglement at $13\sigma$; more general decoherence models are treated by fitting the data with a function $(1-\zeta)A_{\rm QM} + \zeta A_{\rm SD}$, equivalent to modifying the interference term in the $\{B^0, \overline{B}^0\}$ basis, or assuming disentanglement into $B^0$ and $\overline{B}^0$ of a fraction of neutral $B$ pairs (Bertlmann, Grimus, and Hiesmayr, 1999). The result, $\zeta = 0.029 \pm 0.057$, is consistent with no decoherence.

The analysis of Pompili and Selleri (2000) constrains the relevant models to have an asymmetry within a range ($\text{PS}_{\max}$ to $\text{PS}_{\min}$; see Fig. 17.5.8 and Section 17.5.3.1). If the data fall within this range, a null deviation is assigned; otherwise, the nearest boundary is treated as the PS prediction. Even with this conservative treatment, this class of models is disfavored at $5.1\sigma$. The discrepancy with data is concentrated at $\Delta t < 4.0\,\text{ps}$, where statistical uncertainty dominates. In summary the Belle results are consistent with a QM description of entangled neutral $B$ meson pairs created via $\Upsilon(4S)$ decay.

### 17.5.4 Violation of *CP*, *T*, and *CPT* symmetries in $B^0 - \overline{B}^0$ mixing

The phenomenological description introduced in Section 10.1 of $B^0 - \overline{B}^0$ mixing with a $2 \times 2$ matrix effective Hamiltonian already allows for the possibility of *CP*, *T*, and *CPT* symmetry violations. Section 17.5.4.1 discusses the parameterization of the Hamiltonian including new variables that represent the magnitudes of the symmetry violations. In Section 17.5.4.2 we summarize the $B$ Factory measurements of these variables. In the case of *CPT* violation, one would expect on general grounds that violation of Lorentz invariance would also occur; an extended formalism is required to treat this consistently. Such an approach, and the $B$ Factory analysis taking this into account, are presented in Section 17.5.5.

The *CP* symmetry violations discussed in this section pertain to *CP violation in mixing*.[60] These differ from asymmetries due to *mixing-induced CP violation* that result from non-trivial values of the angles of the Unitarity Triangle, $\phi_1$, $\phi_2$, and $\phi_3$, discussed in Sections 17.6–17.8. The recent observation of $T$ violation by *BABAR* (Lees, 2012m) is discussed in Section 17.6. The large observed $T$ asymmetry is expected in the Standard Model and can be understood as a consequence of the CKM phase. If *CPT* symmetry is conserved, there is a direct correspondence between the $T$ asymmetry resulting from the CKM phase and the magnitude of the corresponding *CP* asymmetry (here $\sin 2\phi_1$). As such one could call this violation of $T$ symmetry *mixing-induced T violation*. In this section we discuss searches for *T violation in mixing*.

#### 17.5.4.1 Parameterization of mixing with *CP*, *T*, *CPT* violation

The effective Hamiltonian of $B^0 - \overline{B}^0$ mixing, $\mathcal{H}_{\rm eff} = \mathbf{M} - i\mathbf{\Gamma}/2$, defined in Eq. (10.1.1), is completely described by only eight independent real quantities. Four of them are the masses and decay rates of the eigenstates. These four quantities are sufficient to describe the $B^0 - \overline{B}^0$ oscillations expected in the Standard Model accurately enough for the sensitivity of the $B$ Factories.

To allow for *CP*, *T* and *CPT*-violating effects in mixing, it is necessary to extend the treatment presented in

[60] For a brief overview of the types of *CP* violation relevant for $B$ mesons see Section 16.6.

Section 10.1; there are many different conventions in the literature. For consistency with $B$ Factory papers we follow the notation of Aubert (2004f).

The quantity $q/p$ (Eq. 10.1.3) is given by

$$\frac{q}{p} \equiv \sqrt{\frac{M_{12}^* - \frac{i}{2}\Gamma_{12}^*}{M_{12} - \frac{i}{2}\Gamma_{12}}}. \tag{17.5.33}$$

Its magnitude is expected to be very close to unity:

$$\left|\frac{q}{p}\right|^2 \approx 1 - \mathrm{Im}\left(\frac{\Gamma_{12}}{M_{12}}\right). \tag{17.5.34}$$

If $|q/p| - 1$ differs from zero, *CP* and *T* symmetries are broken, but *CPT* symmetry can still hold. In the Standard Model, $|q/p| - 1$ is small because $|\Gamma_{12}| \ll |M_{12}|$ and because $\mathrm{Im}(\Gamma_{12}/M_{12})$ is suppressed with respect to $|\Gamma_{12}/M_{12}|$. The size of this suppression is $(m_c^2 - m_u^2)/m_b^2 \approx 0.1$. The suppression reflects the fact that *CP* violation is not possible if two of the quark masses are identical. In that case one could redefine the quark states such that one of them does not mix with the other two, and mixing between two quark generations is insufficient to allow for *CP* violation. The phase of $q/p$ is convention-dependent and unobservable.[61] Therefore, the physics of $\mathcal{H}_{\text{eff}}$ is determined by only seven real parameters.

To allow for *CPT*-violating effects in mixing, we introduce the complex parameter

$$z \equiv \frac{\delta m - \frac{i}{2}\delta\Gamma}{\Delta m - \frac{i}{2}\Delta\Gamma}, \tag{17.5.35}$$

where $\delta m \equiv M_{11} - M_{22}$ and $\delta\Gamma \equiv \Gamma_{11} - \Gamma_{22}$ are the differences of the diagonal terms of $\mathcal{H}_{\text{eff}}$. If $z \neq 0$ *CP* and *CPT* symmetries are broken, but *T* symmetry can be conserved. In the Standard Model, $z$ is zero.[62]

With these definitions, the mass eigenstates of Eq. (10.1.2) are replaced by

$$|B_{1,2}\rangle = p\sqrt{1 \mp z}\,\left|B^0\right\rangle \pm q\sqrt{1 \pm z}\,\left|\overline{B}^0\right\rangle, \tag{17.5.36}$$

and the time-evolved states given in Eq. (10.1.6) are replaced by

$$\begin{aligned}|B^0(t)\rangle &= [f_+(t) + zf_-(t)]\left|B^0\right\rangle + \sqrt{1-z^2}\frac{q}{p}f_-(t)\left|\overline{B}^0\right\rangle, \\ |\overline{B}^0(t)\rangle &= [f_+(t) - zf_-(t)]\left|\overline{B}^0\right\rangle + \sqrt{1-z^2}\frac{p}{q}f_-(t)\left|B^0\right\rangle,\end{aligned} \tag{17.5.37}$$

where the functions $f_\pm(t)$ are defined in Eq. (10.1.7). In Section 17.5.2.6, Eq. (17.5.26), we have already introduced a non-zero $\Delta\Gamma_d$ in the time-dependent decay rate of $B$ meson pairs from $\Upsilon(4S)$ decays. We extend this expression to include the *CP*, *T*, and *CPT*-violating parameters defined above:

$$\begin{aligned}N(\Delta t) \propto\, & e^{-\Gamma|\Delta t|} \times \\ & \left\{\frac{1}{2}c_+ \cosh(\Delta\Gamma_d\Delta t/2) + \frac{1}{2}c_- \cos(\Delta m_d\Delta t)\right. \\ & \left. -\mathrm{Re}(s)\sinh(\Delta\Gamma_d\Delta t/2) + \mathrm{Im}(s)\sin(\Delta m_d\Delta t)\right\},\end{aligned} \tag{17.5.38}$$

where

$$c_\pm = |a_+|^2 \pm |a_-|^2, \quad s = a_+^* a_-. \tag{17.5.39}$$

The complex expressions $a_\pm$ depend on the decay amplitudes for a set of specific final states of $B_{\text{tag}}$ and $B_{\text{rec}}$ and on the symmetry-violating parameters $q/p$ and $z$:

$$\begin{aligned}a_+ &= -A_{\text{tag}}\overline{A}_{\text{rec}} + \overline{A}_{\text{tag}}A_{\text{rec}}, \\ a_- &= \sqrt{1-z^2}\left[\frac{p}{q}A_{\text{tag}}A_{\text{rec}} - \frac{q}{p}\overline{A}_{\text{tag}}\overline{A}_{\text{rec}}\right] \\ &\quad + z\left[A_{\text{tag}}\overline{A}_{\text{rec}} + \overline{A}_{\text{tag}}A_{\text{rec}}\right].\end{aligned} \tag{17.5.40}$$

The amplitudes $A_{\text{tag}}$ ($A_{\text{rec}}$) and $\overline{A}_{\text{tag}}$ ($\overline{A}_{\text{rec}}$) represent the cases where $B_{\text{tag}}$ ($B_{\text{rec}}$) is reconstructed, respectively, as a $B^0$ or a $\overline{B}^0$.

We can write Eq. (17.5.38) explicitly for the cases where the flavors of the two $B$ mesons from a $\Upsilon(4S)$ decay are reconstructed as $B^0B^0$, $\overline{B}^0\overline{B}^0$, or $B^0\overline{B}^0$:[63]

$$N^{BB} \propto \frac{e^{-|\Delta t|/\tau}}{2}\left|\frac{p}{q}\right|^2\left\{\cosh\left(\frac{\Delta\Gamma\Delta t}{2}\right) - \cos(\Delta m_d\Delta t)\right\}$$

$$N^{\overline{B}\overline{B}} \propto \frac{e^{-|\Delta t|/\tau}}{2}\left|\frac{q}{p}\right|^2\left\{\cosh\left(\frac{\Delta\Gamma\Delta t}{2}\right) - \cos(\Delta m_d\Delta t)\right\}$$

$$N^{B\overline{B}} \propto \frac{e^{-|\Delta t|/\tau}}{2}\left\{\cosh\left(\frac{\Delta\Gamma\Delta t}{2}\right) + 2\mathrm{Re}(z)\sinh\left(\frac{\Delta\Gamma\Delta t}{2}\right)\right.$$

---

[61] In addition, different conventions for the sign of the phase of $q/p$ are used in the literature. Here we set the phase of $q/p$ using Eqs (10.1.3) and (17.5.33). The convention in Aubert (2004f) differs by $e^{i\pi} = -1$. As a result their Eqs (9) and (10) have a negative sign in front of the $\sqrt{1-z^2}$ term relative to our Eq. (17.5.37). The same comment applies to Eq. (12.30) of the 2013 PDG review on *CP* violation in meson decays (Beringer et al., 2012), where the phase of $q/p$ is not explicitly stated. Our expressions otherwise agree with those of Aubert (2004f).

[62] In Hastings (2003), Belle uses a different parameterization, following Mohapatra, Satpathy, Abe, and Sakai (1998), based on complex parameters $\theta$ and $\phi$. The relationship between these and several other notations is discussed by Kostelecký (2001): in particular, $\cos\theta = -z = \xi$, and $|\exp(i\phi)| = |q/p| = w$, where $\xi$ (complex) and $w$ (real) are the parameters preferred by Kostelecký. While we rely heavily on this and related references in the Lorentz-violation discussion below (Section 17.5.5), we use the notation of Eq. (17.5.35) throughout. In neutral kaon mixing, *CPT* violation is described by $\delta_K = -z/2$.

[63] In Eqs (17.5.41)–(17.5.43) we assume that the $B$ transition to a flavor eigenstate $f$ has a single weak amplitude $A_f$ and that the $B$ decay does not violate *CPT* symmetry, *i.e.* $A_f = \overline{A}_{\overline{f}}$. We also assume that the amplitude for the $B$ decay to the *CP* conjugate final state is zero, *i.e.* $A_{\overline{f}} = \overline{A}_f = 0$.

$$+\cos(\Delta m_d \Delta t) - 2\mathrm{Im}(z)\sin(\Delta m_d \Delta t)\Big\}; \quad (17.5.41)$$

here we have ignored terms quadratic in $z$. The first (second) $B$ in the superscript denotes the flavor of $B_{\rm tag}$ ($B_{\rm rec}$).

If one of the $B$ mesons decays through a $b \to (c\bar{c})s$ transition to a *CP* eigenstate $B_{CP}$ such as $J/\psi K^0_S$ or $J/\psi K^0_L$, and the other $B$ meson decays to a flavor state ($B^0$ or $\overline{B}^0$), the time-dependent decay rates are given by

$$\begin{aligned} N^{BB_{CP}} &\propto \frac{e^{-|\Delta t|/\tau}}{2} \times \\ &\Big\{ [1 \mp \mathrm{Re}(z)\cos 2\phi_1 \mp \mathrm{Im}(z)\sin 2\phi_1]\cosh\left(\frac{\Delta\Gamma_d \Delta t}{2}\right) \\ &+ [\pm\mathrm{Re}(z)\cos 2\phi_1 \pm \mathrm{Im}(z)\sin 2\phi_1]\cos(\Delta m_d \Delta t) \\ &+ [\mp\cos 2\phi_1 + \mathrm{Re}(z)]\sinh\left(\frac{\Delta\Gamma_d \Delta t}{2}\right) \\ &+ [\pm\sin 2\phi_1 - \mathrm{Im}(z)]\sin(\Delta m_d \Delta t)\Big\}, \end{aligned} \quad (17.5.42)$$

$$\begin{aligned} N^{\overline{B}B_{CP}} &\propto \frac{e^{-|\Delta t|/\tau}}{2} \times \\ &\Big\{ [1 \pm \mathrm{Re}(z)\cos 2\phi_1 \mp \mathrm{Im}(z)\sin 2\phi_1]\cosh\left(\frac{\Delta\Gamma_d \Delta t}{2}\right) \\ &+ [\mp\mathrm{Re}(z)\cos 2\phi_1 \pm \mathrm{Im}(z)\sin 2\phi_1]\cos(\Delta m_d \Delta t) \\ &+ [\mp\cos 2\phi_1 - \mathrm{Re}(z)]\sinh\left(\frac{\Delta\Gamma_d \Delta t}{2}\right) \\ &+ [\mp\sin 2\phi_1 + \mathrm{Im}(z)]\sin(\Delta m_d \Delta t)\Big\}, \end{aligned} \quad (17.5.43)$$

where the upper (lower) sign in front of terms with $\cos 2\phi_1$ or $\sin 2\phi_1$ represents final states with $\eta_{CP} = -1$ (+1). In Eqs (17.5.42) and (17.5.43), we assume $|q/p| = 1$.

The decay rates defined in Eqs (17.5.41)–(17.5.43) can be used to construct asymmetries sensitive to $T$, *CP*, and *CPT* violation. The *same-flavor* asymmetry $A_{T/CP}$ between the two oscillation probabilities $P(\overline{B}^0 \to B^0)$ and $P(B^0 \to \overline{B}^0)$ depends on $|q/p|$ and probes both $T$ and *CP* symmetries:

$$\begin{aligned} A_{T/CP} &= \frac{P(\overline{B}^0 \to B^0) - P(B^0 \to \overline{B}^0)}{P(\overline{B}^0 \to B^0) + P(B^0 \to \overline{B}^0)} \\ &= \frac{N^{BB} - N^{\overline{B}\overline{B}}}{N^{BB} + N^{\overline{B}\overline{B}}} \\ &= \frac{1 - |q/p|^4}{1 + |q/p|^4}. \end{aligned} \quad (17.5.44)$$

The *opposite-flavor* asymmetry, $A_{CPT/CP}$, depends on $z$ and probes both *CP* and *CPT* symmetries. Defining the decay-time difference for such events as $\Delta t = t^+ - t^-$, where $t^+$ ($t^-$) corresponds to $B^0$ ($\overline{B}^0$), the asymmetry

$$\begin{aligned} &A_{CPT/CP}(\Delta t;\, \Delta t > 0) \\ &= \frac{P(B^0 \to B^0) - P(\overline{B}^0 \to \overline{B}^0)}{P(B^0 \to B^0) + P(\overline{B}^0 \to \overline{B}^0)} \\ &= \frac{N^{B\overline{B}}(\Delta t) - N^{B\overline{B}}(-\Delta t)}{N^{B\overline{B}}(\Delta t) + N^{B\overline{B}}(-\Delta t)} \\ &\simeq 2\frac{\mathrm{Im}(z)\sin(\Delta m_d \Delta t) - \mathrm{Re}(z)\sinh(\Delta\Gamma_d \Delta t/2)}{\cos(\Delta m_d \Delta t) + \cosh(\Delta\Gamma_d \Delta t/2)} \\ &\simeq \frac{2\mathrm{Im}(z)\sin(\Delta m_d \Delta t) - \mathrm{Re}(z)\Delta\Gamma_d \Delta t}{\cos(\Delta m_d \Delta t) + \cosh(\Delta\Gamma_d \Delta t/2)}, \end{aligned} \quad (17.5.45)$$

where the approximation in the third line is the neglect of terms of higher order in $z$; in the fourth line, as $|\Delta\Gamma_d/\Gamma| \ll 1$, we take $\sinh(\Delta\Gamma_d \Delta t/2) \simeq \Delta\Gamma_d \Delta t/2$.

By comparing the rates of a $B$ versus a $\overline{B}$ decaying to a *CP* eigenstate, we can define another asymmetry that is sensitive to $z$ using Equations (17.5.42) and (17.5.43):

$$\begin{aligned} A'_{CPT/CP}(\Delta t) &= \frac{P(\overline{B}^0 \to B_{CP}) - P(B^0 \to B_{CP})}{P(\overline{B}^0 \to B_{CP}) + P(B^0 \to B_{CP})} \\ &= \frac{N^{BB_{CP}}(\Delta t) - N^{\overline{B}B_{CP}}(\Delta t)}{N^{BB_{CP}}(\Delta t) + N^{\overline{B}B_{CP}}(\Delta t)} \\ &\simeq \{\pm\mathrm{Re}(z)\cos 2\phi_1[-1 + \cos(\Delta m_d \Delta t)] \\ &\quad + [\pm\sin 2\phi_1 - 2\mathrm{Im}(z)]\sin(\Delta m_d \Delta t)\}/ \\ &\quad \{1 \pm \mathrm{Im}(z)\sin 2\phi_1[-1 + \cos(\Delta m_d \Delta t)] \\ &\quad \mp \cos 2\phi_1 \Delta\Gamma_d \Delta t/2\}. \end{aligned} \quad (17.5.46)$$

In the last step we have again neglected terms of higher than linear order in $z$ and $\Delta\Gamma_d$, and the upper (lower) sign in front of terms with $\cos 2\phi_1$ or $\sin 2\phi_1$ refers to final states with $\eta_{CP} = -1$ (+1).

In the expressions of the decay rates (Eqs 17.5.41–17.5.43) and associated asymmetries (Eqs 17.5.44–17.5.46) we have assumed that a $B$ flavor state can unambiguously be identified by its decay products. In the quark model, this is true for semi-leptonic decays. For hadronic $B$ decays to flavor final states, however, the presence of doubly-CKM-suppressed decays (DCS) makes it impossible to determine the original flavor of the $B$ meson without ambiguity. Accounting for DCS decays leads to more complicated expressions for $c_\pm$, $s$, and the resulting decay rates and asymmetries. A complete list of general expressions of $c_\pm$ and $s$ can be found in Aubert (2004f).

In the SM the asymmetry $A_{T/CP}$ is expected to be very small (Beneke, Buchalla, Lenz, and Nierste, 2003; Ciuchini, Franco, Lubicz, Mescia, and Tarantino, 2003). A recent calculation predicts $A_{T/CP} = (-0.40 \pm 0.06) \times 10^{-3}$ or correspondingly $|q/p| - 1 = (0.20 \pm 0.03) \times 10^{-3}$ (Nierste, 2012). A measurement significantly different from zero with the data samples of the $B$ Factories would be evidence for new physics. In the Standard Model *CPT* symmetry is conserved and $\mathrm{Re}(z)$ and $\mathrm{Im}(z)$ as well as $A_{CPT/CP}$ are expected to be zero. The asymmetry $A'_{CPT/CP}$ reduces to the *CP*-violating, but *CPT*-conserving asymmetry, $-\eta_{CP}\sin 2\phi_1 \sin(\Delta m_d \Delta t)$. It is a measurement of mixing-induced *CP* violation.

### 17.5.4.2 Results on $CP$, $T$ and $CPT$ violation in $B^0 - \overline{B}^0$ mixing

*BABAR* and Belle published several papers on searches for violation of $T$, $CP$, and $CPT$ symmetries in $B^0 - \overline{B}^0$ mixing. The analyses are performed with inclusive dilepton final states or fully-reconstructed hadronic and semileptonic final states. Here we review the measurements of $|q/p| - 1$, $\mathrm{Re}(z)$, and $\mathrm{Im}(z)$.

#### *Measurements of* $|q/p| - 1$

Earlier measurements of the asymmetry $A_{T/CP}$ have been performed by CLEO (Behrens et al., 2000; Jaffe et al., 2001), ALEPH (Barate et al., 2001), and OPAL (Abbiendi et al., 2000b; Ackerstaff et al., 1997a) before the $B$ Factories took their data. In the 2002 Review of Particle Physics (Hagiwara et al., 2002) the PDG calculated an average[64] of $A_{T/CP} = (0 \pm 16) \times 10^{-3}$ corresponding to a value of $|q/p| - 1 = (0 \pm 8) \times 10^{-3}$.

*BABAR* and Belle both measure $|q/p| - 1$ with inclusive samples of semileptonic $B^0$ decays. In these events only the two leptons from semileptonic decays $B \to Xl\nu$ ($l = e, \mu$) are reconstructed. The charge of the lepton $l^+$ ($l^-$) unambiguously identifies the flavor of the parent $B$ meson to be a $B^0$ ($\overline{B}^0$). The asymmetry between the numbers of same-sign dilepton pairs, $N^{BB} = N(l^+l^+)$ and $N^{\overline{B}\overline{B}} = N(l^-l^-)$, is related to the two oscillation probabilities $P(\overline{B}^0 \to B^0)$ and $P(B^0 \to \overline{B}^0)$ as given in Eq. (17.5.44). Although $A_{T/CP}$ is a time-independent asymmetry, both experiments use the decay time difference $\Delta t$ between the two $B$ decays to discriminate signal events from background.

*BABAR* and Belle measure $|q/p| - 1$ in samples of 211 $\mathrm{fb}^{-1}$ (Aubert, 2006aq) and 78 $\mathrm{fb}^{-1}$ (Nakano, 2006), respectively. The *BABAR* result supersedes an earlier measurement with 21 $\mathrm{fb}^{-1}$ (Aubert, 2002i). The event selections in these analyses are similar to the ones for measurements of $\Delta m_d$ with dilepton events described above (see Section 17.5.2.2), but with attention to keeping charge-dependent asymmetries in reconstruction and PID efficiencies small and under control. Events with two identified leptons in a momentum range (0.8 GeV/$c$ (*BABAR*) or 1.2 GeV/$c$ (Belle) $< p^* < 2.3$ GeV/$c$) and with topology consistent with that of semileptonic $B$ decays are selected. Leptons that originate from photon conversions in the detector or from $J/\psi$ and $\psi(2S)$ decays are explicitly vetoed. Both experiments carefully determine detector-induced charge asymmetries in their lepton identification (Chapter 5). *BABAR* uses a control sample of radiative Bhabha events and Belle uses two photon production of $e^+e^-$ to study electron ID asymmetries. The charge asymmetry in muon ID is determined with $e^+e^- \to \mu^+\mu^-\gamma$ events by *BABAR* and with events in which a simulated muon track is embedded in a hadronic event by Belle. Hadron fake rates are determined with pions from $K^0_S \to \pi^+\pi^-$, kaons from $D^{*+} \to D^0\pi^+_s \to (K^-\pi^+)\pi^+_s$ (*BABAR*) and from $\phi \to K^+K^-$ (Belle), and protons from $\Lambda \to p\pi^-$. The distance between the two $B$ decay vertices $\Delta z$ is measured as described in Section 17.5.2.2. *BABAR* fits the $\Delta t$ distributions of the $l^+l^+$ and $l^-l^-$ events and randomly assigns the sign of $\Delta t$ for each event. Belle fits the $\Delta z = z_1 - z_2$ distribution, where $z_1$ ($z_2$) is the $z$ coordinate of the higher- (lower-) momentum lepton. The majority of the events are events where both leptons originate from a $B$ decay. *BABAR* distinguishes background from events in which one lepton is a primary lepton from a $B$ decay and the other lepton comes from a secondary charm decay ($b \to c \to l$), events with one direct lepton and one lepton from a tau ($b \to \tau \to l$) or charmonium ($b \to (c\overline{c}) \to l$) cascade decay, and events from the light quark $e^+e^- \to q\overline{q}$ continuum. Belle subtracts the contribution from light quark continuum candidates to their dilepton sample. The background events from $B\overline{B}$ decays are separated into correctly tagged and wrongly tagged events. The wrongly tagged sample is dominated by events where one lepton is from a primary $B$ decay and the other one from a cascade charm decay. The correctly tagged background sample consists mainly of events in which both leptons come from secondary charm decays. *BABAR* extracts $|q/p| - 1$ from a binned likelihood fit to the $\Delta t$ distribution of the selected dilepton sample.[65] The likelihood function

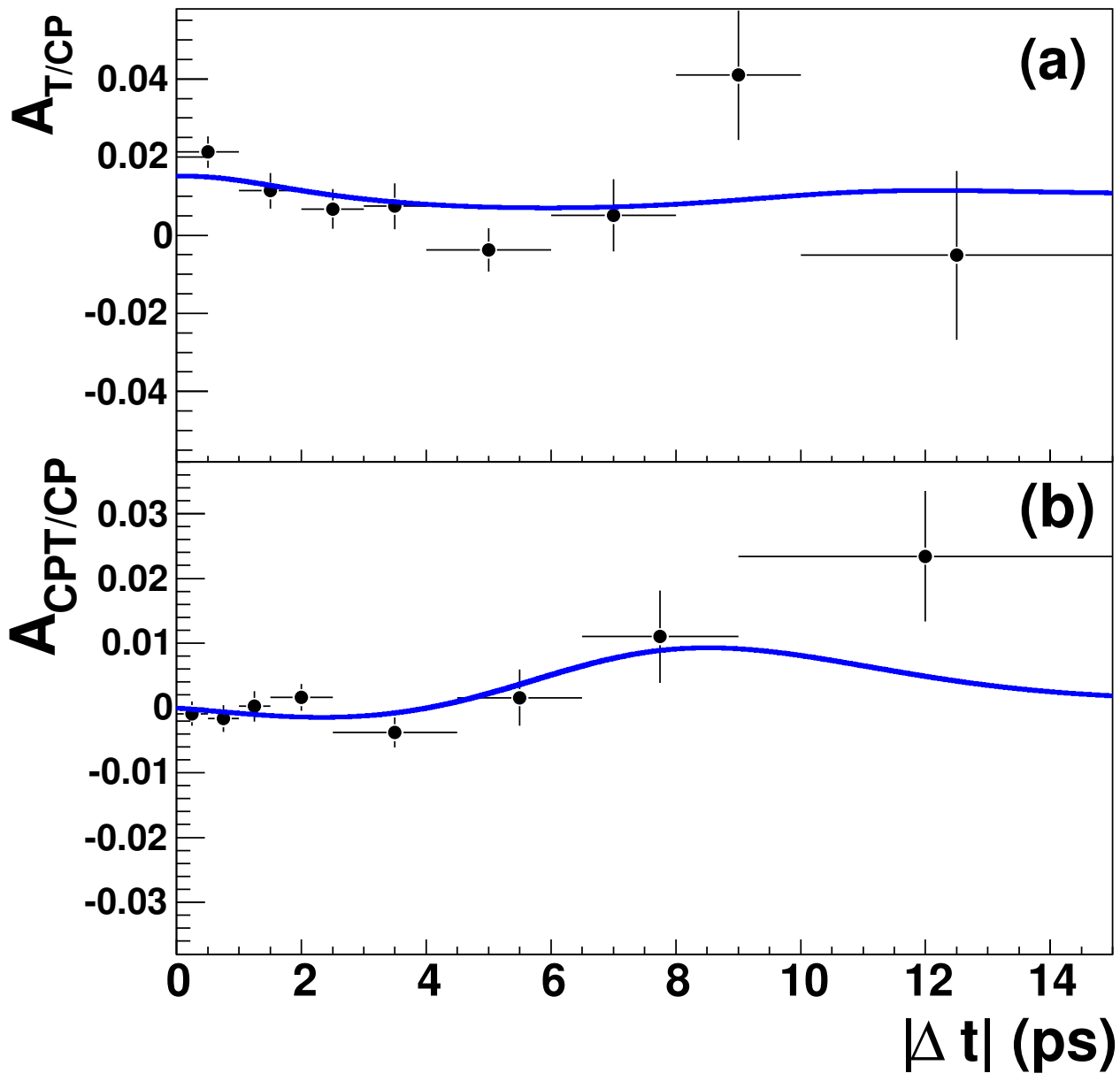

**Figure 17.5.10.** The measured asymmetries (a) $A_{T/CP}$ and (b) $A_{CPT/CP}$ from inclusive dilepton events, as functions of $|\Delta t|$ (Aubert, 2006aq). The deviation from zero in $A_{T/CP}$ is a result of background from cascade muons that is dominant at small $|\Delta t|$.

[64] The PDG quotes $\mathrm{Re}(\epsilon_B)/(1+|\epsilon_B|^2) = (0 \pm 4) \times 10^{-3}$, where $\epsilon_B = (p-q)/(p+q)$ corresponds to the parameter $\epsilon_K$ describing the corresponding asymmetry in the neutral kaon system and $A_{T/CP} \approx 4\mathrm{Re}(\epsilon_B)/(1+|\epsilon_B|^2)$ for small $A_{T/CP}$.

[65] In the same analysis *BABAR* measures the $CP$- and $CPT$-violating parameter $z$ (see below). They employ a simultaneous

combines detector-related charge asymmetries and time-dependent *p.d.f.*s for signal and background events. The measured asymmetry $A_{T/CP}$ as a function of $|\Delta t|$ is shown in Fig. 17.5.10. Belle determines a raw dilepton asymmetry from their selected events as a function of $\Delta z$, applies a bin-wise background correction, and calculates an average $A_{T/CP}$ in a range $0.15\,\mathrm{mm} < |\Delta z| < 2\,\mathrm{mm}$. *BABAR* measures $|q/p| - 1 = (-0.8 \pm 2.7 \pm 1.9) \times 10^{-3}$ and Belle measures $A_{T/CP} = (-1.1 \pm 7.9 \pm 8.5) \times 10^{-3}$, which corresponds to $|q/p| - 1 = (0.5 \pm 4.0 \pm 4.3) \times 10^{-3}$. The largest contributions to the systematic error in the *BABAR* measurement come from potential charge asymmetries in track reconstruction ($1.0 \times 10^{-3}$) and electron identification ($1.0 \times 10^{-3}$). Belle's systematic error in $|q/p| - 1$ is dominated by potential charge asymmetries in the track finding efficiency ($2.6 \times 10^{-3}$) and uncertainties in the continuum background subtraction ($2.4 \times 10^{-3}$).

*BABAR* also measures $|q/p| - 1$ in an analysis of fully-reconstructed $B$ decays to hadronic final states in a sample of $88 \times 10^6$ $B\overline{B}$ pairs (Aubert, 2004e,f). One $B$ is reconstructed either in a flavor state or a *CP* eigenstate. In the same analysis *BABAR* measures $\Delta\Gamma_d$ (Section 17.5.2.6) and the *CP* and *CPT*-violating parameters $\mathrm{Re}(z)$ and $\mathrm{Im}(z)$ (Section 17.5.4.2) from the $\Delta t$ distributions of the selected events. The sensitivity to $|q/p| - 1$ comes from events in which both $B$ mesons have the same flavor. Because the branching fractions to exclusive flavor states are much smaller than the inclusive semileptonic branching fraction, the signal sample is comparatively small compared to event samples in the dilepton analyses. From the analysis of fully-reconstructed hadronic final states *BABAR* quotes $|q/p| - 1 = (29 \pm 13 \pm 11) \times 10^{-3}$.

The average of the measurements of $|q/p| - 1$ published by the $B$ Factories is $(0.3 \pm 2.8) \times 10^{-3}$ (see Table 17.5.3). While we were finishing the writing of this Book, *BABAR* submitted another measurement of $|q/p| - 1$ for publication (Lees, 2013g). In that analysis one $B$ meson is reconstructed as $B^0 \to D^{*-} l^+ \nu_l$ (and the $D^{*-}$ is partially-reconstructed using only the slow pion) and the flavor of the other is tagged with a charged kaon. The value of $|q/p| - 1 = (0.29 \pm 0.84^{+1.88}_{-1.61}) \times 10^{-3}$ measured in this analysis represents the most precise single measurement of $|q/p| - 1$ by the $B$ Factories. However, due to the overlap of events used in this analysis with those used for the $|q/p| - 1$ measurement in Aubert (2006aq), a simple average could not be calculated for this Book. The PDG in their 2013 partial update of the Review of Particle Physics quotes a world average of $\mathrm{Re}(\epsilon^0_B)/(1 + |\epsilon^0_B|^2 = (0.6 \pm 0.7) \times 10^{-3}$ corresponding to $|q/p| - 1 = (-1.2 \pm 1.4) \times 10^{-3}$ (Beringer et al., 2012), which includes two recent measurements from DØ corresponding to $|q/p| - 1 = (0.6 \pm 2.6) \times 10^{-3}$ (Abazov et al., 2011) and $|q/p| - 1 = (-3.4 \pm 2.2 \pm 0.8) \times 10^{-3}$ (Abazov et al., 2012).

There has been significant interest in the measurement of $|q/p|$ or the corresponding asymmetry $A_{T/CP}$ since DØ announced evidence for an anomalous like-sign dimuon charge asymmetry in $p\overline{p}$ collisions (Abazov et al., 2010a,b). The asymmetry $A^b_{\mathrm{sl}}$ is defined similarly to Eq. (17.5.44), but has contributions from the charge asymmetries of $B^0_d$ mesons ($A_{T/CP}$) and $B^0_s$ mesons ($A^s_{T/CP}$):

$$A^b_{\mathrm{sl}} = C_d A_{T/CP} + C_s A^s_{T/CP} \tag{17.5.47}$$

with $C_d = 0.594 \pm 0.022$, $C_s = 0.406 \pm 0.022$ (Abazov et al., 2011). With their latest measurement of $A^b_{\mathrm{sl}} = (-7.87 \pm 1.72 \pm 0.93) \times 10^{-3}$ DØ claims a $3.9\sigma$ discrepancy with the Standard Model prediction of $A^b_{\mathrm{sl}}(\mathrm{SM}) = (-0.28^{+0.05}_{-0.06}) \times 10^{-3}$ (Abazov et al., 2011). However, no significant measurement has yet been observed in either $A_{T/CP}$ or $A^s_{T/CP}$. The world average for $|q/p| - 1$ corresponds to $A_{T/CP} = (2.4 \pm 2.8) \times 10^{-3}$. Based on two DØ measurements HFAG calculates the corresponding average for $B^0_s$ mesons of $A^s_{T/CP} = (-11 \pm 6) \times 10^{-3}$ (Amhis et al., 2012). It is important to improve the measurements of $A_{T/CP}$ or $A^s_{T/CP}$ to understand if there is *CP* and *T* violation in the mixing of $B^0_d$ or $B^0_s$ mesons or both. LHCb is expected to improve the measurement of $A^s_{T/CP}$ in the near future. The $B$ Factories have only analyzed fractions of their full datasets for $|q/p|$. Some improvement in $A_{T/CP}$ will be possible by using more data, but systematic uncertainties are already of comparable size to the statistical errors. LHCb may be able to reduce the error in $A_{T/CP}$ using fully-reconstructed semileptonic $B^0$ decays similar to the DØ analysis described in (Abazov et al., 2012). The large data set of a future super flavor factory paired with analyses of fully-reconstructed semi-leptonic decays could substantially reduce the overall uncertainty in $A_{T/CP}$.

**Table 17.5.3.** $B$ Factory measurements of $|q/p| - 1$ along with the journal paper and selected final state for each measurement. The measurement in Aubert (2002i) has been superseded by Aubert (2006aq) and is not included in the $B$ Factories average.

| Experiment | Method | $\|q/p\| - 1$ $[10^{-3}]$ |
|---|---|---|
| *BABAR* (Aubert, 2006aq) | Incl. dilepton | $-0.8 \pm 2.7 \pm 1.9$ |
| *BABAR* (Aubert, 2002i) | Incl. dilepton | $-2 \pm 6 \pm 7$ |
| *BABAR* (Aubert, 2004e,f) | Hadr. modes | $29 \pm 13 \pm 11$ |
| Belle (Nakano, 2006) | Incl. dilepton | $0.5 \pm 4.0 \pm 4.3$ |
| *BABAR*-Belle average | | $0.3 \pm 2.8$ |

### *Measurements of* $\mathrm{Re}(z)$ *and* $\mathrm{Im}(z)$

Prior to the $B$ Factories a search for *CPT* violation in $B^0 - \overline{B}^0$ mixing was performed by the OPAL collaboration (Ackerstaff et al., 1997a). They quote their result in terms of the *CPT* parameter $\delta_B$, a variable with a definition equivalent to $\delta_K$, which is used to characterize *CPT* violation in kaon mixing (Beringer et al., 2012). OPAL's result $\mathrm{Im}(\delta_B) = -0.020 \pm 0.016 \pm 0.006$ corresponds to fit to both same-sign and opposite-sign dilepton pairs to determine the $\Delta t$ resolution in addition to the symmetry-violating parameters.

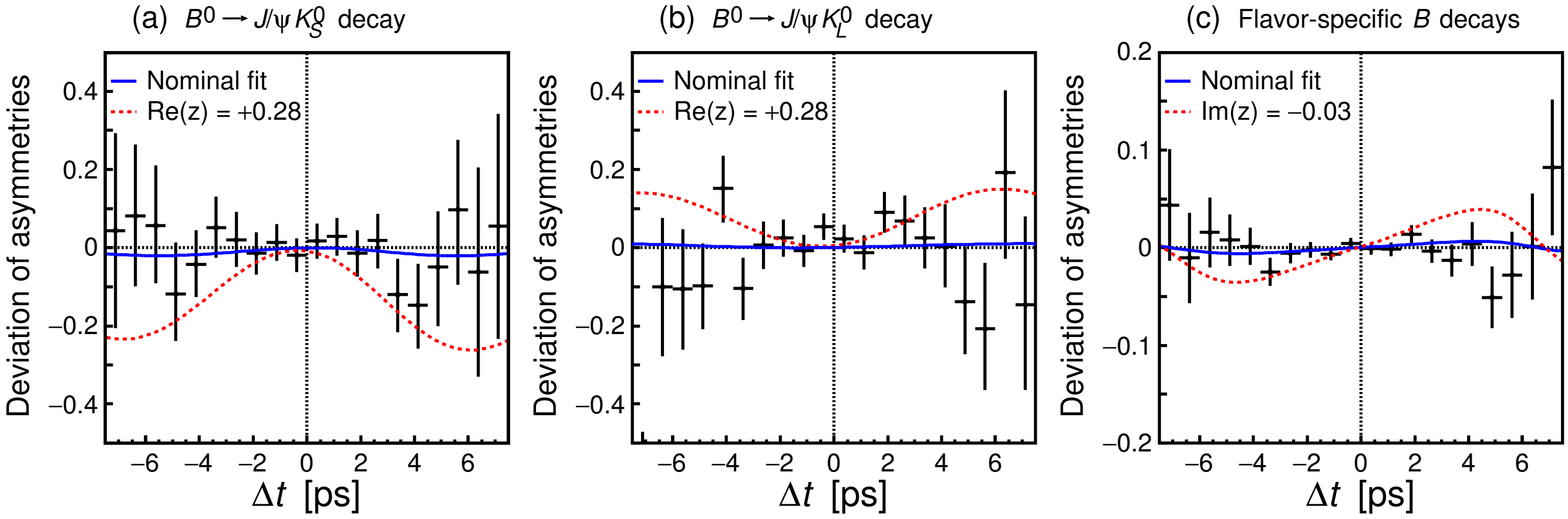

**Figure 17.5.11.** Deviations of the asymmetries from the reference asymmetry [$\mathrm{Re}(z) = \mathrm{Im}(z) = \Delta\Gamma_d/\Gamma_d = 0$] in fully-reconstructed hadronic and semileptonic final states (Higuchi, 2012). Shown are raw asymmetries uncorrected for backgrounds, flavor mis-tagging, and $\Delta t$ resolution. The underlying asymmetries for (a) and (b) corresponding to $\eta_{CP} = -1$ and $+1$, respectively, are given by Eq. (17.5.46). The asymmetry tested in (c) is given in Eq. (17.5.45). The crosses with error bars are data. The solid curves are deviations for the nominal fits. The dashed curves are for illustration only: they represent scenarios where either $\mathrm{Re}(z) = +0.28$ or $\mathrm{Im}(z) = -0.03$. These values are equal to approximately 5× the total uncertainty of the corresponding parameter.

$\mathrm{Im}(z) = 0.040 \pm 0.032 \pm 0.012$. The world averages for the real and imaginary parts of $\delta_K$ determined from kaon experiments are $\mathrm{Re}(\delta_K) = (2.5 \pm 2.3) \times 10^{-4}$ and $\mathrm{Im}(\delta_K) = (-1.5 \pm 1.6) \times 10^{-5}$ (Beringer et al., 2012). The corresponding limit on the mass difference between $K^0$ and $\overline{K}^0$ normalized to the mass average is $|m_{K^0} - m_{\overline{K}^0}|/m_{\mathrm{average}} < 6 \times 10^{-19}$, assuming $\Gamma_{K^0} = \Gamma_{\overline{K}^0}$.

*BABAR* and Belle have studied samples of inclusive dilepton events and fully-reconstructed hadronic and semileptonic events to measure $\mathrm{Re}(z)$ and $\mathrm{Im}(z)$ in $B^0 - \overline{B}^0$ mixing. In dilepton events the $\Delta t$ distribution of opposite-sign dilepton pairs has been used to search for violation of *CP* and *CPT* asymmetry. The asymmetry $A_{CP/CPT}$ between events with positive and negative *true* $\Delta t$ is related to the two oscillation probabilities $P(B^0 \to B^0)$ and $P(\overline{B}^0 \to \overline{B}^0)$ as given in Eq. (17.5.45). Here the decay-time difference is defined as $\Delta t = t^+ - t^-$, where $t^+$ ($t^-$) corresponds to $l^+$ ($l^-$).

*BABAR* measures the *CP*- and *CPT*-violating parameters $\mathrm{Re}(z)$ and $\mathrm{Im}(z)$ with opposite-sign dilepton pairs in a sample of $211\,\mathrm{fb}^{-1}$ (Aubert, 2006aq). The selection criteria are the same as for the measurement of $|q/p| - 1$ with same-sign dilepton pairs in the same paper (see above). The parameter $\mathrm{Im}(z)$ appears as coefficient to the $\sin(\Delta m_d \Delta t)$ term in the $\Delta t$ distribution of opposite-sign dilepton pairs $N^{B\overline{B}}$ given by Eq. (17.5.41) and the corresponding asymmetry $A_{CPT/CP}$ of Eq. (17.5.45). Thus a measurement of the shape of the $\Delta t$ distribution is sensitive to $\mathrm{Im}(z)$. On the other hand, sensitivity to $\mathrm{Re}(z)$ only comes from the $\sinh(\Delta\Gamma_d \Delta t/2)$ term. Since $\Delta\Gamma_d$ is a small quantity and has not been measured, *BABAR* substitutes $\sinh(\Delta\Gamma_d \Delta t/2) \simeq \Delta\Gamma_d \Delta t/2$ and quotes only the product $\Delta\Gamma_d \times \mathrm{Re}(z)$ in their paper. In the $\cosh(\Delta\Gamma_d \Delta t/2)$ term they use $|\Delta\Gamma_d| = (5 \pm 3) \times 10^{-3}\,\mathrm{ps}^{-1}$. Although *BABAR* fits the $\Delta t$ distributions of the same-sign and opposite-sign dilepton pairs in a single fit, they do not constrain the ratio between the numbers of events of the two types. *BABAR* quotes $\mathrm{Im}(z) = (-13.9 \pm 7.3 \pm 3.2) \times 10^{-3}$ and $\Delta\Gamma_d \times \mathrm{Re}(z) = (-7.1 \pm 3.9 \pm 2.0) \times 10^{-3}\,\mathrm{ps}^{-1}$. The statistical correlation between the measurements of $\mathrm{Im}(z)$ and $\Delta\Gamma_d \times \mathrm{Re}(z)$ is 76%. The systematic errors in $\mathrm{Im}(z)$ and $\Delta\Gamma_d \times \mathrm{Re}(z)$ are dominated by uncertainties in the *p.d.f.* modeling ($2.5\times$ and $1.2 \times 10^{-3}$), the external parameters $\tau_{B^0}$, $\tau_{B^-}$, $\Delta m_d$, and $\Delta\Gamma_d$ ($1.9\times$ and $1.1 \times 10^{-3}$) and SVT alignment ($0.6\times$ and $1.2 \times 10^{-3}$). Assuming $\Delta\Gamma_d = 0$, *BABAR* obtains $\mathrm{Im}(z) = (-3.7 \pm 4.6 \pm 2.9) \times 10^{-3}$. The measured asymmetry $A_{T/CP}$ as a function of $|\Delta t|$ is shown in Fig. 17.5.10.

Belle's results on $\mathrm{Re}(z)$ and $\mathrm{Im}(z)$ with $29.4\,\mathrm{fb}^{-1}$ of data are published in Hastings (2003). The analysis uses the same selection criteria as in their measurement of $\Delta m_d$ with dilepton pairs (see Section 17.5.2.2) that is described in the same paper. A major difference to the *BABAR* analysis is that Belle constrains the time-integrated fractions of same-sign and opposite sign events to

$$\chi_d = \frac{|1 - z^2| x_d^2}{|1 - z^2| x_d^2 + 2 + x_d^2 + |z|^2 x_d^2}, \qquad (17.5.48)$$

where $x_d = \tau_{B^0} \Delta m_d$. $N^{B\overline{B}}$ ($N^{BB}$, $N^{\overline{B}\overline{B}}$) is proportional to opposite-sign (same-sign) dilepton efficiency. Belle determines the ratio of the efficiencies from MC simulation. They quote $\mathrm{Re}(z) = (0 \pm 12 \pm 1) \times 10^{-2}$ and $\mathrm{Im}(z) = (-3 \pm 1 \pm 3) \times 10^{-2}$. These measurements supersede the results from an earlier Belle paper (Abe, 2001b). The dominant systematic uncertainties in $\mathrm{Im}(z)$ come from data/MC agreement of the $\Delta t$ *p.d.f.* and the requirement that the polar angle of the lepton tracks be in the fiducial volume. The largest contribution to the systematic error in $\mathrm{Re}(z)$ comes from the MC-modeling of the $\Delta t$ resolution.

**Table 17.5.4.** Measurements of Re($z$) and Im($z$) and, if given in the paper, the corresponding limits (at 90% C.L.) on the mass difference and width difference between $B^0$ and $\overline{B}^0$. In Aubert (2006aq), *BABAR* measures $\Delta\Gamma_d \times \mathrm{Re}(z) = (-0.71 \pm 0.39 \pm 0.20) \times 10^{-2}$ ps$^{-1}$, but does not quote a value for Re($z$).

| Experiment | Method | Re($z$) [$10^{-2}$] | Im($z$) [$10^{-2}$] | $\lvert\delta m/m\rvert$ [$10^{-14}$] | $\delta\Gamma/\Gamma$ |
|---|---|---|---|---|---|
| *BABAR* (Aubert, 2006aq) | Incl. dilepton | — | $-1.39 \pm 0.73 \pm 0.32$ | — | — |
| Belle (Hastings, 2003) | Incl. dilepton | $0.0 \pm 12 \pm 1$ | $-3 \pm 1 \pm 3$ | $< 1.16$ | $\lvert\delta\Gamma/\Gamma\rvert < 0.11$ |
| Belle (Abe, 2001b) | Incl. dilepton | $0 \pm 15 \pm 6$ | $-3.5 \pm 2.9 \pm 5.1$ | $< 1.6$ | $\lvert\delta\Gamma/\Gamma\rvert < 0.161$ |
| *BABAR* (Aubert, 2004e,f) | Hadronic | $2.0 \pm 5.1 \pm 4.9$ | $3.8 \pm 2.9 \pm 2.5$ | $< 1.0$ | $-0.156 < \delta\Gamma/\Gamma < 0.042$ |
| Belle (Higuchi, 2012) | Hadr. + semilep. | $1.9 \pm 3.7 \pm 3.3$ | $-0.57 \pm 0.33 \pm 0.33$ | — | — |

*BABAR* and Belle also measure Im($z$) and Re($z$) in samples of fully-reconstructed hadronic and semi-leptonic final states. *BABAR* reconstructs $B^0$ decays to the flavor final states $D^{(*)-}\pi^+/\rho^+/a_1^+$ and $J/\psi K^{*0}$ and *CP*-eigenstates $J/\psi K_S^0$, $\psi(2S)K_S^0$, $\chi_{c1}K_S^0$, and $J/\psi K_L^0$ in 88 million $B\overline{B}$ events (Aubert, 2004e,f). Belle reconstructs signal events in the decays $B^0 \to D^{(*)-}\pi^+$, $D^{*-}\rho^+$, $D^{*-}l\nu$, $J/\psi K_S^0$, and $J/\psi K_L^0$ in a sample of 535 million $B\overline{B}$ events (Higuchi, 2012). Raw asymmetries as function of $\Delta t$ for $J/\psi K_S^0$, $J/\psi K_L^0$ and flavor-specific final states overlaid with curves representing the nominal fit result and scenarios with significant *CPT* violation are shown in Fig. 17.5.11. In addition to $z$ both analyses also measure $\Delta\Gamma_d$ and $|q/p|-1$. These results are described above. The measurements use the $\Delta t$ reconstruction and flavor tagging methods of the standard time-dependent analyses of the $B$ Factories described in earlier Chapters (6, 8, 10). The time-dependent *p.d.f.*s for events to final states that contain a fully-reconstructed $B_{\mathrm{rec}}$ identified either as $B^0$, $\overline{B}^0$, or $B_{CP}$ , and a $B_{\mathrm{tag}}$ with identified flavor as $B^0$ or $\overline{B}^0$ are given in Eqs (17.5.42) and (17.5.43). However, interference effects between the amplitudes for dominant decays of flavor-eigenstates (*e.g.* $B^0 \to D^-\pi^+$) and for doubly-CKM-suppressed decays (*e.g.* $B^0 \to D^+\pi^-$) lead to more complicated *p.d.f.*s (Aubert, 2004f). These interference effects are present when either $B_{\mathrm{rec}}$ or $B_{\mathrm{tag}}$ is reconstructed in a flavor state. In principle, the ratio of favored and DCS decay amplitudes is different for each mode. *BABAR* shows that an effective ratio can be defined for ensembles of final states as long as terms linear in $|z|$, $|q/p|-1$, and in the amplitude ratios of the contributing modes can be neglected. Belle treats the effects of DCS decays as part of the systematic error. The dominant contribution of Im($z$) to the time-dependence is through the coefficient of $\sin(\Delta m_d \Delta t)$ for flavor final states, while Re($z$) contributes primarily to the coefficients of the $\cosh(\Delta\Gamma_d \Delta t/2) \approx 1$ and $\cos(\Delta m_d \Delta t)$ terms for *CP* eigenstates. The main physics parameters extracted in *BABAR*'s analysis are $\mathrm{sgn}(\mathrm{Re}\lambda_{CP})$, $\Delta\Gamma_d/\Gamma_d$, $|q/p|$, Im($z$), and $(\mathrm{Re}\lambda_{CP}/|\lambda_{CP}|) \times \mathrm{Re}(z)$. The parameters $(\mathrm{Im}\lambda_{CP}/|\lambda_{CP}|)$ and $\Delta m_d$ are determined together with the main parameters as cross checks against earlier measurements. *BABAR* measures $(\mathrm{Re}\lambda_{CP}/|\lambda_{CP}|) \times \mathrm{Re}(z) = 0.014 \pm 0.035 \pm 0.034$ and $\mathrm{Im}(z) = (3.8 \pm 2.9 \pm 2.5) \times 10^{-2}$. Using *BABAR*'s measurement of $\sin 2\phi_1$ ($\mathrm{Im}\lambda_{CP}$) on the same data set (Aubert, 2005i) and assuming $|\lambda_{CP}| = 1$, we calculate a value of $\mathrm{Re}(z) = (+2.0 \pm 5.1 \pm 4.9) \times 10^{-2}$. Belle quotes the physics parameters $\mathrm{Re}(z) = (+1.9 \pm 3.7 \pm 3.3) \times 10^{-2}$ and $\mathrm{Im}(z) = (-5.7 \pm 3.3 \pm 3.3) \times 10^{-3}$. The fit has a twofold ambiguity in the sign of $\mathrm{Re}\lambda_{CP}$. The sign of Re($z$) has been determined assuming $\mathrm{Re}\lambda_{CP} > 0$, which is a result of global fits of the Unitarity Triangle (see Section 25.1). The largest systematic uncertainty in Re($z$) comes from the knowledge of tag-side interference (0.028). The error in Im($z$) is dominated by uncertainties in vertex reconstruction (0.0028).

The results of all Re($z$) and Im($z$) measurements by the $B$ Factories are summarized in Table 17.5.4. The measurements are still mostly statistically limited and many of the systematic uncertainties are statistical in nature. Using the full data sets will allow one to further improve the constraints on these *CP* and *CPT*-violating parameters. Future super flavor factories should be able to improve current limits even further. How much they can improve will depend on how well the systematic uncertainties can be controlled.

### 17.5.5 Lorentz invariance violation in $B^0 - \overline{B}^0$ mixing

If we go beyond a purely phenomenological treatment of *CPT*-violating effects, Lorentz violation should also be considered. *CPT* invariance follows from assumptions that are currently understood to hold in the low-energy (Standard Model-like) domain: point particles, the applicability of quantum field theory, and in particular, Lorentz invariance (Jost, 1957; Luders, 1954; Pauli, 1955; Streater and Wightman, 2000). If *CPT* symmetry is broken, then one or more of these conditions must be violated.

In particle physics, Lorentz violation is usually studied in the framework of the Standard Model Extension. This theory and its application to $B$ mixing is briefly reviewed in Section 17.5.5.1. We then describe the analysis carried out within this framework by *BABAR* (Section 17.5.5.2), and discuss the implications for work at future facilities (Section 17.5.5.3).

### 17.5.5.1 The Standard Model Extension and mixing

Simplifying assumptions are required to make an effective search for Lorentz violation in data. If we posit a fundamental theory whose dynamics are both *CPT*- and Poincaré-invariant,[66] and a low-energy effective theory that exhibits *spontaneous CPT*- and Lorentz-symmetry breaking, we obtain the so-called Standard Model Extension (SME) of Colladay and Kostelecký (1997, 1998); see also Kostelecký (2004). In this theory *nature remains invariant under translations, and is covariant under changes in the inertial frame of the observer*: the usual kinematic expressions may consistently be used to analyse particle motion, reconstruct invariant masses, and so on. However under boosts of individual particles, *CPT* is broken and certain Lorentz-violating terms appear, due to the (constant) expectation values of one or more Lorentz tensors (*cf.* the mass terms due to particles coupling to the [scalar] Higgs field in the Standard Model). The Lorentz-violating coefficients due to these background fields vary from particle to particle in general; the resulting parameters are similar in number to the supersymmetric couplings, and have been exhaustively tabulated, together with current bounds from experimental and observational tests, by Kostelecký and Russell (2011). As neutral-meson oscillation is flavor-changing, *CPT*-violation measurements in mixing provide access to couplings that are not constrained by other experimental tests (Kostelecký, 1998).

For a neutral meson, the Lorentz-violating parameters are given by the four-vector $\Delta a_\mu \equiv r_{q_1} a_\mu^{q_1} - r_{q_2} a_\mu^{q_2}$, the difference in the couplings of the two valence quarks $q_i$ (Kostelecký, 2001; Kostelecký and Potting, 1995).[67] These parameters are constant in any inertial frame. We then find, for the *CPT*-violating parameter defined in Eq. (17.5.35),

$$z \equiv \frac{\delta m - \frac{i}{2}\delta\Gamma}{\Delta m_d - \frac{i}{2}\Delta\Gamma} \simeq \frac{\beta^\mu \Delta a_\mu}{\Delta m_d - \frac{i}{2}\Delta\Gamma}, \qquad (17.5.49)$$

where $\beta^\mu = \gamma(1, \boldsymbol{\beta})$ is the meson four-velocity. The approximation in Eq. (17.5.49) is due to the neglect of higher-order effects in the SME, and does not otherwise rely on the size of $z$ (Kostelecký, 2001; Kostelecký and Potting, 1995).[68] Note that the relative values of the imaginary and real parts of $z$ are fixed by the $B$-mixing parameters (Kostelecký and Potting, 1995),[69]

$$\frac{\mathrm{Im}\, z}{\mathrm{Re}\, z} = \frac{\Delta\Gamma}{2\Delta m_d}, \qquad (17.5.50)$$

[66] That is, invariant under translations, as well as rotations and boosts.

[67] The factors $r_{q_i}$, which represent the effect of binding the quarks $q_i$ within the meson, are not used consistently in the literature, disappearing (for example) in Kostelecký (1998).

[68] *BaBar* (Aubert, 2008ar) cites Kostelecký (1998), where a further approximation exists due to the use of another parameter, $\delta \approx -z/2$ in the case of small $T$- and *CPT*-violating effects.

[69] The *BaBar* analysis (Aubert, 2008ar) derives this condition from Eq. (17.5.49) using $\Delta\Gamma \ll \Delta m_d$, but it is derived from fundamental considerations by Kostelecký and Potting (1995), assuming only that $T$- and *CPT*-violating effects are small.

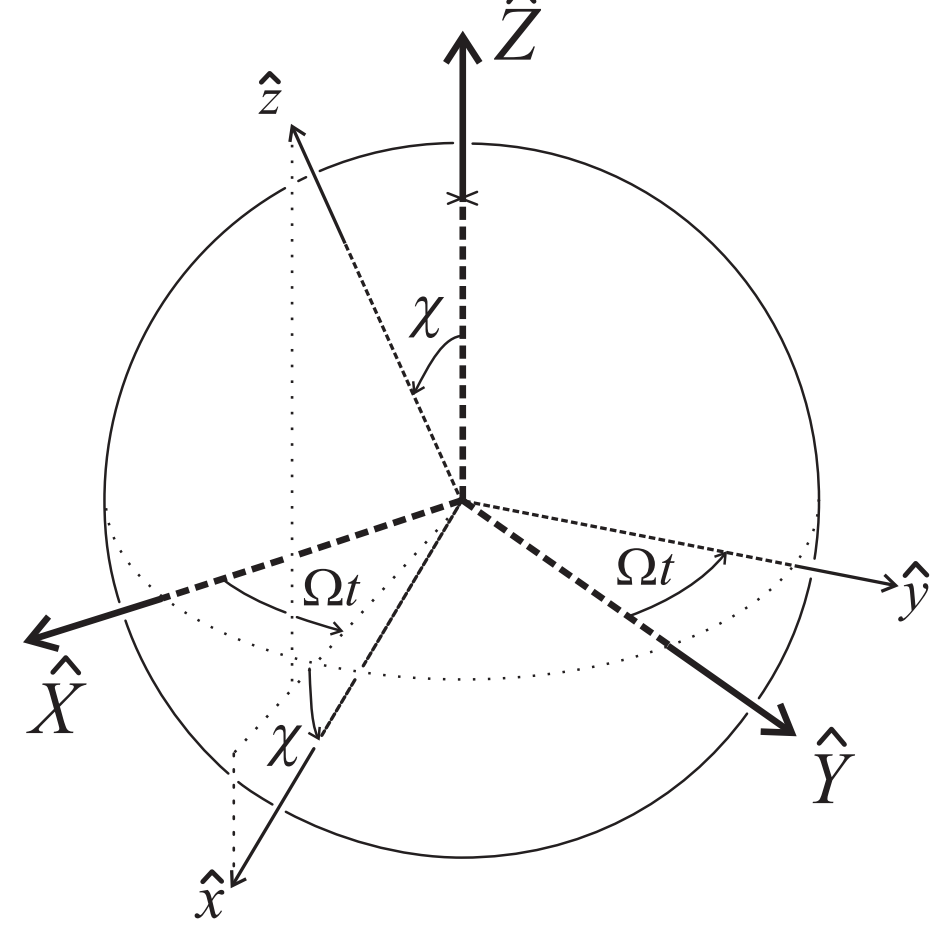

**Figure 17.5.12.** Transformation between non-rotating and laboratory (rotating) reference frames for the Lorentz-violation analysis: from Kostelecký and Lane (1999).

providing a distinctive signature for *CPT*-violating effects within this scheme.

The motion of the laboratory must be taken into account: the (non-relativistic) velocity may be neglected, but the earth's rotation changes the relative orientation of the detector coordinate system and the spatial components $\Delta\boldsymbol{a}$. *BABAR* (Aubert, 2008ar) chooses a non-rotating frame $(\hat{X}, \hat{Y}, \hat{Z})$ following Kostelecký and Lane (1999), with $\hat{Z}$ parallel to the earth's rotation axis, and $\hat{X}$ ($\hat{Y}$) at right ascension 0° (90°). With the further choices that the laboratory coordinate $\hat{z}$ lies along $-\boldsymbol{\beta}$, and $\hat{y}$ lies in the equatorial plane (declination 0°), it follows from Eq. (14) of Kostelecký (2001) that

$$\begin{aligned}\beta^\mu \Delta a_\mu = \gamma\,[\Delta a_0 &- \beta \Delta a_Z \cos\chi \\ &- \beta \sin\chi\,(\Delta a_Y \sin \Omega\, t_{\rm sid} \\ &\qquad + \Delta a_X \cos \Omega\, t_{\rm sid})]\,, \end{aligned} \qquad (17.5.51)$$

where $\cos\chi \equiv \hat{z} \cdot \hat{Z} = 0.628$ for *BABAR*, $t_{\rm sid}$ is the sidereal time, and $\Omega = 2\pi/d_{\rm sid}$ the sidereal frequency; $d_{\rm sid} \simeq 0.99727$ solar days. The transformation between laboratory and non-rotating coordinates is illustrated in Fig. 17.5.12. The sidereal time $t_{\rm sid}$ is given by the right ascension of $\hat{z}$; this will become important when comparing results from different experiments (Section 17.5.5.3 below).

From Eqs (17.5.49) and (17.5.51) it is clear that in the general case, the measured *CPT*-violating parameter $z$ will vary with a period of one sidereal day ($d_{\rm sid}$); a value $z$ obtained from data without time-binning will depend on the latitude of the experiment and the distribution of meson momenta in the laboratory frame.

### 17.5.5.2 The *BABAR* analysis

The Lorentz violation study in Aubert (2008ar) is an extension of the *CPT*-violation search of Aubert (2006aq),

**Table 17.5.5.** Parameters from fits to the asymmetry $A_{CPT}$ as a function of sidereal time, assuming constant ($z_0$) and sinusoidal ($z_1$) contributions according to the Standard Model Extension. Statistical and total systematic uncertainties are shown; a breakdown of systematic contributions is given in Table I of Aubert (2008ar). Results are shown without (center) and with (right) the SME constraint of Eq. (17.5.50) on the real and imaginary parts of $z$.

| $A_{CPT}$ parameter | | unconstrained | SME constraint |
|---|---|---|---|
| $\mathrm{Im}\, z_0$ | $[10^{-3}]$ | $-14.2 \pm 7.3 \pm 2.2$ | $-5.2 \pm 3.6 \pm 1.9$ |
| $\mathrm{Re}\, z_0\, \Delta\Gamma$ | $[10^{-3}/\mathrm{ps}]$ | $-7.3 \pm 4.1 \pm 1.8$ | |
| $\mathrm{Im}\, z_1$ | $[10^{-3}]$ | $-24 \pm 11 \pm 3.3$ | $-17.0 \pm 5.8 \pm 1.9$ |
| $\mathrm{Re}\, z_1\, \Delta\Gamma$ | $[10^{-3}/\mathrm{ps}]$ | $-18.5 \pm 5.6 \pm 1.7$ | |
| $\phi$ | [rad] | $2.63 \pm 0.31 \pm 0.21$ | $2.56 \pm 0.36 \pm 0.15$ |

discussed in Section 17.5.4.2 above, using the same sample of opposite-sign dilepton events to measure the *CP*- and *CPT*-violating asymmetry between $B^0 \to B^0$ and $\overline{B}^0 \to \overline{B}^0$ rates,

$$A_{CPT/CP}(\Delta t) = \frac{2\mathrm{Im}\, z \sin(\Delta m_d \Delta t) - \mathrm{Re}\, z \Delta\Gamma \Delta t}{\cos(\Delta m_d \Delta t) + \cosh(\Delta\Gamma \Delta t/2)}; \tag{17.5.52}$$

see Eq. (17.5.45) for the full expression. As in Aubert (2006aq), same-sign dilepton events are used to provide additional information on the fractions of the various signal and background components.

The analysis is extended to include the sidereal time $t_{\mathrm{sid}}$, allowing for variations in $z$ of the form

$$z = z_0 + z_1 \cos(\Omega t_{\mathrm{sid}} + \phi); \tag{17.5.53}$$

the discrete ambiguity ($z_1 \to -z_1, \phi \to \phi + \pi$) does not affect the physical parameters $\Delta a_\mu$ of Eq. (17.5.51). A two-dimensional maximum likelihood fit is performed, with opposite- and same-sign events separately binned in $(\Delta t, t_{\mathrm{sid}})$; 24 sidereal-time bins are used. The values obtained for the parameters $z_{0,1}$ and $\phi$ are shown in Table 17.5.5. Individual systematic uncertainties are itemized in Table I of Aubert (2008ar): the dominant terms are due to alignment of the *BABAR* SVT and the absolute $z$ scale (especially for $\phi$), and modeling of the resolution.

Deviations from zero are seen for both the constant and sidereal-time-dependent *CPT*-violating terms in Eq. (17.5.53). The constant terms $\mathrm{Re}\, z_0\, \Delta\Gamma$ and $\mathrm{Im}\, z_0$ are almost identical to those in the time-independent analysis (Aubert, 2006aq), where a $\chi^2$ of 3.25 for 2 degrees of freedom is quoted (the results have a correlation coefficient of 0.76 in both analyses): consistent with *CPT* invariance at 19.7% confidence. The sidereal-time dependence of $A_{CPT}$ is shown in Figure 17.5.13; events at small time differences $|\Delta t| < 3$, while included in the fit, are suppressed in the figure as their predicted asymmetry is small.

Results are consistent with the SME condition of Eq. (17.5.50), so a further fit is performed with this expression used as a constraint, to improve the precision of the measurement: these results are also shown in the table. Consistent results are found if second-order terms $|z|^2 = \rho^2 \cos^2(\Omega t_{\mathrm{sid}} + \phi)$ are added to the fit (*cf.* the derivation of Eq. (17.5.41) above).

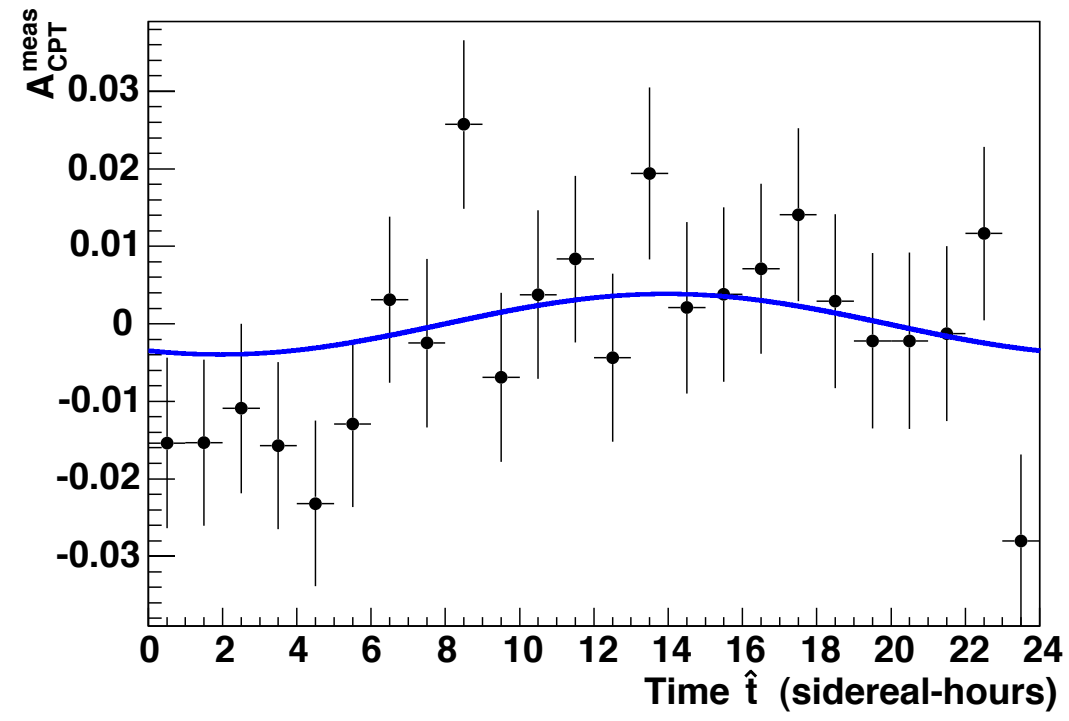

**Figure 17.5.13.** Measured asymmetry $A_{CPT}$ for opposite-sign dilepton events with 3 ps $< |\Delta t| <$ 15 ps, as a function of sidereal time $\hat{t} = t_{\mathrm{sid}}$. The curve shows a projection of the full two-dimensional $|\Delta t| <$ 15 ps fit, also requiring 3 ps $< |\Delta t|$. From Aubert (2008ar).

Two kinds of significance estimates are quoted for the sidereal-time-dependent results. Based on the likelihood fit, ($\mathrm{Re}\, z_1\, \Delta\Gamma$, $\mathrm{Im}\, z_1$) differ from zero at $2.8\sigma$, with or without the SME constraint. Based on the periodogram method (Lomb, 1976; Scargle, 1982) measuring the spectral power $\mathcal{P}(\nu)$ for variations in $z$ at test frequencies $\nu$, a value $\mathcal{P}(1/d_{\mathrm{sid}}) = 5.28$ is found; the probability to exceed this value in the absence of an oscillatory signal is

$$\begin{aligned} P\left[\mathcal{P}(\nu) > S\right] &= e^{-S} \\ &= 5.1 \times 10^{-3} \ \text{ for } S = 5.28, \end{aligned} \tag{17.5.54}$$

also corresponding to $2.8\sigma$. This is significantly stronger than the result at the solar-day frequency, where effects due to diurnal variations in detector response would occur: $\mathcal{P}(1/d_{\mathrm{solar}}) = 1.47$. The largest spectral power among the $M = 9500$ independent frequencies tested is $\mathcal{P}(0.46312/d_{\mathrm{sid}}) = 8.78$; the probability of finding a larger spectral power than this is

$$\begin{aligned} P\left[\mathcal{P}(\nu)|_{\max} > S;\, M\right] &= 1 - \left(1 - e^{-S}\right)^M \\ &= 76\% \ \text{ for } S = 8.78. \end{aligned} \tag{17.5.55}$$

However we note that the frequency expected within the SME is unambiguous: $\nu = 1/d_{\mathrm{sid}}$.

Final results are quoted for the SME quantities:

$$\begin{aligned} \Delta a_0 - 0.30 \Delta a_Z &= (-3.0 \pm 2.4)(\Delta m_d/\Delta\Gamma) \times 10^{-15} \text{ GeV}, \\ \Delta a_X &= (-22 \pm 7)(\Delta m_d/\Delta\Gamma) \times 10^{-15} \text{ GeV}, \\ \Delta a_Y &= (-14^{+10}_{-13})(\Delta m_d/\Delta\Gamma) \times 10^{-15} \text{ GeV}. \end{aligned} \tag{17.5.56}$$

#### 17.5.5.3 Implications for future measurements

A study of Lorentz covariance violation has not been performed by Belle, nor has the full available *BABAR* dataset been used to update the Aubert (2008ar) analysis. The results of that analysis thus remain untested. If confirmed, a non-zero measurement would be a result of the utmost importance; the burden of proof for such a measurement is correspondingly high.

At face value, the Aubert (2008ar) analysis provides weak evidence for *CPT* violation together with departures from Lorentz covariance, consistent with the Standard Model Extension. Combining the confidence levels of the time-independent ($\alpha_1 = 0.197$) and sidereal-time-dependent results ($\alpha_2 = 5.1\times10^{-3}$) discussed above, using Eq. (11.35) of James (2006), we find an overall result compatible with zero at $\alpha = \alpha_1\alpha_2[1-\ln(\alpha_1\alpha_2)] = 7.9\times10^{-3}$: a $2.66\sigma$ effect. While there is greater spectral power at some frequencies $\nu \neq 1/d_{\rm sid}$, and even the largest such signal is within expectations in the absence of oscillation, this adds little new information beyond the relative weakness of the sidereal-time dependence; within the SME, the predicted signal is not at some undetermined frequency but at $\nu = 1/d_{\rm sid}$ — there is no "look-elsewhere effect".

The current results are statistically limited. While much larger datasets are foreseen at super flavor factories, even the final *BABAR* and Belle samples exceed those used by Aubert (2008ar) by factors of 2.0 and 3.3 respectively, allowing for both a repetition of the analysis on independent data, and a test on different equipment with statistical errors reduced by a factor $\sqrt{3.3} = 1.82$. Assuming no change in central value, systematics, or intrinsic power, a hypothetical Belle result with statistical uncertainty of $3.2 \times 10^{-3}$ and systematic uncertainty of $1.9 \times 10^{-3}$ (*cf.* Table 17.5.5) would have $4.6\sigma$ significance for the sidereal-time-dependent measurement alone; the probability to exceed the corresponding spectral power, at *any* frequency, would be 5% (from Eqs (17.5.54) and (17.5.55), assuming the same number of frequencies tested by Aubert, 2008ar).

The latitude of Belle is similar to that of *BABAR*, and by chance the compass orientations of the $\Upsilon(4S)$ boost are also similar for the two experiments. The longitudes are substantially different at the two sites: this leads to a difference in the right ascension of $\hat{z}$, and thus an offset in $t_{\rm sid}$ in Eq. (17.5.51) at a given clock time. The phase $\phi$ of sidereal-time dependence in Eq. (17.5.53) predicted in the SME for Belle therefore differs from that at *BABAR* by a fixed amount, whereas for results due to statistical fluctuation, the phase would be arbitrary.

The dominant systematic uncertainties — alignment, the $z$-scale, and the modeling of resolution — are amenable to improvement at a redesigned experiment, although the underlying time-dependent analysis techniques (Chapter 10) would need to be mature. Even without a significant reduction in systematics, a super flavor factory could perform a measurement of overwhelming statistical power.

## 17.6 $\phi_1$, or $\beta$

***Editors:***
*Chih-hsiang Cheng (BABAR)*
*Yoshihide Sakai (Belle)*
*Ikaros Bigi (theory)*

**Additional section writers:**
*Tagir Aushev, Eli Ben-Haim, Adrian Bevan, Bob Cahn, Chunhui Chen, Ryosuke Itoh, Alfio Lazzaro, Owen Long, Fernando Martinez-Vidal, Vincent Poireau, Klaus Schubert*

Precision measurement of the *CP* asymmetries in $B \to J/\psi K_S^0$ decays was the principal motivation for building the $B$ Factories. With the accumulation of data samples larger than anticipated, the *BABAR* and Belle experiments at the $B$ Factories are able to study *CP* asymmetries in a wide range of related channels. This section describes measurements of the Unitarity Triangle angle $\phi_1$, also known as $\beta$ in the literature. An overview of $\phi_1$ measurements and their motivation is presented in Section 17.6.1, followed by a review of the quark transitions and the formalism of $\phi_1$ measurements in Section 17.6.2. The various channels for $\phi_1$ measurement, and the $B$ Factories results, are then described in Sections 17.6.3–17.6.7. Resolution of discrete ambiguities is discussed in Section 17.6.8, and a summary of $\phi_1$ results is presented in Section 17.6.10.

In the Standard Model, non-zero asymmetries measured in these analyses reflect violation of both the *CP* and $T$ symmetries. Performing the measurement in a way that directly demonstrates $T$ violation, without assuming (for example) *CPT* symmetry, requires special care. Such an analysis has been performed at *BABAR*, and is presented in Section 17.6.9. Tests of *CPT* symmetry are presented in Section 17.5.

### 17.6.1 Overview of $\phi_1$ measurement at the $B$ Factories

Initially, *CP* violation seemed isolated from the mainstream of particle physics. Since it was seen only in the $K_S^0$-$K_L^0$ system, it was possible to imagine that it was due entirely to a $\Delta S = 2$ operator as postulated in the superweak theory (Wolfenstein, 1964). Two developments put *CP* violation at center stage. The first was A. D. Sakharov's demonstration (Sakharov, 1967) that *CP* violation was one of the three requirements for the existence of the baryon anti-baryon asymmetry of the universe (see Section 16.2). The second was Kobayashi's and Maskawa's demonstration that *CP* violation was natural if there were three generations of quarks (see Chapter 16). With the subsequent discovery of the last three quarks, testing the CKM model became urgent.

The $K_S^0$-$K_L^0$ system was not sufficient by itself to test the CKM picture. The measured parameters, $\Delta m_K$, $\epsilon_K$ and $\epsilon'_K$, depended not just on the fundamentals of the weak interactions, but on non-perturbative hadronic matrix elements. Moreover, *CP* violation in the kaon system was feeble. Since $\epsilon_K$ was measured in 1964, it took until 1973 before Kobayashi and Maskawa provided a real theory for *CP* violation. It needed many years to demonstrate that the parameter $\epsilon'$ was non-zero. Even *before* the unexpected 'long' lifetime of $B$ mesons was discovered, the $B$ meson system was recognized as the ideal testing ground for *CP* violation (Bigi and Sanda, 1981) and the decay $B \to J/\psi K_S^0$ as ideal for the purpose. Detection of the final state is especially clean because the $J/\psi$ decays to lepton pairs and the $K_S^0$ is sufficiently long-lived to decay into pairs of oppositely charged pions at a secondary vertex displaced from the interaction region.

Unlike neutral kaons, the neutral $B$ mesons start oscillating just after their production, since their mixing rate $\Delta m_d$ is comparable to their natural widths $\Gamma$ (see Section 17.5). If we begin with a $B^0$, at a later time the state will be a superposition of $B^0$ and $\overline{B}^0$. The decay to $J/\psi K_S^0$ will occur through both components and the interference pattern will depend on the relative phases between the $B^0$ and $\overline{B}^0$ components, which is directly calculable in the CKM model. The interference pattern depends on the two decay amplitudes to the final state. Because the final state is a *CP* eigenstate and because there is only one significant pathway to it from $B^0$ or $\overline{B}^0$, the two decay amplitudes are identical, up to another calculable phase. As a result, the oscillation pattern can be predicted simply in terms of the phases due to the CKM matrix without any dependence on hadronic physics. The time-dependent formalism required for the measurement of $\sin 2\phi_1$ can be found in Chapter 10.

In order to test the CKM paradigm we need to know if we are starting with a $B^0$ or with a $\overline{B}^0$. The $\Upsilon(4S)$ is very near the threshold for $B\overline{B}$ so if one $B$ is observed, the remaining particles must come from another $B$. Moreover, by Bose symmetry, if a $\overline{B}^0$ is observed the other particle must be a $B^0$ at that instant, since the two mesons must be in an antisymmetric state to produce the unit of angular momentum carried by the $\Upsilon(4S)$. Thus "tagging" one $B$ meson tells us both, when to start the clock and the type of $B$ at that time (see Chapter 8).

The decay $B^0 \to J/\psi K_S^0$ is just one of a large family of related decays due to a $b \to c\bar{c}s$ transition. Of particular interest is the decay to $J/\psi K_L^0$ because the final state has the opposite *CP* eigenvalue, and we expect exactly the opposite oscillation. Other charmonia can take the place of $J/\psi$, including $\psi(2S), \eta_c$, and $\chi_{c1}$. The decay $B^0 \to J/\psi K^{*0}$ is more complex because the spins of the final state particles can be combined to produce an overall spin equal to 0, 1, or 2, and correspondingly the orbital angular momentum will be 0, 1, or 2. This complexity has the advantage that it can help resolve the ambiguity inherent in determining the angle $\phi_1$ when only $\sin 2\phi_1$ is known.

At first, the $B$ Factories concentrated on measuring time-dependent asymmetries in the so-called charmonium "golden modes" concentrating on $B^0 \to J/\psi K_S^0$, $\psi(2S)K_S^0$, $\chi_{c1}K_S^0$, $J/\psi K_L^0$, and $J/\psi \pi^0$. However, it was understood that there were other ways to measure $\phi_1$. Once an understanding of how to do these measurements started to develop, the experiments branched out to study similar final states that were more difficult to isolate from the data. These states either had smaller branching fractions, or were experimentally more challenging to isolate. Studies performed in the *BABAR* physics book (Harrison and Quinn, 1998), prior to the commencement of data taking, assumed that a data sample of $30\,\text{fb}^{-1}$ would be available to use for testing the SM. In reality this data sample was quickly attained on both sides of the Pacific Ocean and the $B$ Factories program of measuring $\phi_1$ expanded, both in terms of the number of measurements and in terms of the complexity of analysis used, to accommodate the rich harvest of $B$ meson pairs. The first results on the measurement of $\sin 2\phi_1$ were shown at the International Conference on High Energy Physics in 2000, which became known colloquially within *BABAR* and Belle as 'the Osaka conference'. The $B$ Factories presented values of $\sin 2\phi_1$ of $0.12 \pm 0.37 \pm 0.09$ (Aubert, 2000) and $0.45^{+0.43}_{-0.44}{}^{+0.07}_{-0.09}$ (Aihara, 2000a) at this conference. A year later Belle and *BABAR* established large *CP* asymmetry in this final state. Since then both $B$ Factories have accumulated much larger data samples, and the final results obtained by *BABAR* and Belle are significantly more precise than these first measurements (see Section 17.6.3).

While the charmonium decays were the primary focus of the $\phi_1$ program, final states mediated by other transitions were also studied in subsequent waves of measurements that quickly followed the first results. In particular the modes $B \to \phi K^0_S$, $B \to \eta' K^0_S$, and $D^{(*)+}D^{(*)-}$ were highlighted. The expectation was that these would provide alternative ways of constraining $\phi_1$, and would complement the constraint on the Unitarity Triangle given by the golden mode measurements. Any measurement of $\phi_1$ that differed significantly from expectations, or any two measurements that disagreed with each other, could reveal physics beyond the Standard Model.

The first few measurements of $S \simeq \sin(2\phi_1)$ in a quasi-two-body analysis of $B \to \phi K^0_S$ decays in 2003 were far from the SM expectation. While these were low statistics studies, with only a handful of high purity events (well tagged events with a low mistag probability, see Chapter 8), the community was tantalized by the possibility that this could herald a new age in modern physics. As a result, the interest in alternative measurements of $\phi_1$ blossomed, and this remains a vibrant area a decade later. Alas, the early deviations from the SM turned out to be statistical fluctuations, and the most recent measured values of $\phi_1$ obtained from the $B$ Factories are compatible with SM expectations within experimental and theoretical uncertainties.

The early fluctuation had several consequences. First, a large number of neutral $B$ meson decays to *CP* eigenstate or admixture final states have been studied in the hope that one or more of them might yield a result incompatible with the SM. Second, both the theoretical and experimental communities started to take possible hadronic uncertainties more seriously in both golden and alternative measurements of $\phi_1$. Today the constraints on hadronic uncertainties in these modes are a mixture of theoretical calculations and data-driven constraints obtained via a more phenomenological approach. The golden channels are theoretically clean, up to the extent that analysis at the $B$ Factories would be concerned about. This has been determined via theoretical calculation, and via a data-driven interpretation of results. However other final states, in particular those dominated by penguin loop amplitudes, have non-negligible uncertainties. The cleanest modes are $B \to \eta' K^0_S$, which is the most precisely measured charmless final state, and $B \to \phi K^0_S$. These have hadronic uncertainties of a few percent on the measured value of $S$. In the case of $B \to f_0 K^0_S$ there are only partial calculations where, for example, long distance effects are ignored, and the estimated hadronic uncertainties for this mode provide a lower bound. More details on this part of the $B$ Factory program can be found in Section 17.6.6.

Early time-dependent studies of $B$ decays to charmless final states relied on a simplified analysis paradigm by imposing the quasi-two-body assumption that resonances are particles of definite mass, so that interference between amplitudes could be neglected. As the recorded data samples of the two experiments increased, more sophisticated techniques were incorporated. Just as the measurements of $\phi_2$ ultimately required that the $B$ Factories pioneer the use of time-dependent Dalitz plot techniques, so eventually one had to perform similar analyses in order to constrain $\phi_1$. The ability to study amplitudes in a Dalitz plot leads to the possibility of resolving the four-fold ambiguity in the value of $\phi_1$ obtained from the golden mode measurement, and complements other approaches such as the full angular analysis of the $B^0 \to J/\psi K^*$ final state. Results from three-body charmless decays on $\phi_1$ are discussed in Section 17.6.7, and resolution of discrete ambiguities on the value of this angle using other modes is considered in Section 17.6.8.

The large amounts of data accumulated by the $B$ Factories also required an improvement in understanding the systematic uncertainties involved in the measurements themselves. In particular the concept of flavor tagging as originally conceived, while good enough to describe semileptonic tagged events, turned out to be an approximation for hadronically tagged final states. It is possible to have a small level of *CP* violation manifest on the tag side of the event that would need to be considered as a systematic uncertainty in order to ensure that one reports the correct level of *CP* violation obtained for a given result. In some cases with small expected *CP* violating asymmetries, such as the measurement of $\sin(2\phi_1 + \phi_3)$, this so-called tag-side interference needs to be incorporated into the measurement technique. The main systematic uncertainties for time-dependent measurements at the $B$ Factories, including tag-side interference, are discussed in Chapter 15.

The final measurement of $\sin 2\phi_1 \equiv \sin 2\beta$ obtained by the $B$ Factories has a combined precision of 3%. This can be compared with the estimated relative statistical precision for this measurement estimated in the *BABAR* physics book, 12%, using a foreseen data sample of $30\,\text{fb}^{-1}$ (Harrison and Quinn, 1998). The achieved precision is a nice example of exceeding the initial expectations put forward before the startup of the $B$ Factories. The final result of the $B$ Factories is not systematically limited and may be improved upon by the next generation of experiments.

### 17.6.2 Transitions and formalism

The Unitarity Triangle angle $\phi_1 = \beta$ is defined as

$$\phi_1 \equiv \beta \equiv \arg[-(V_{cd}V_{cb}^*)/(V_{td}V_{tb}^*)]. \quad (17.6.1)$$

It describes $CP$ violation in the interference between decays with and without $B^0$-$\overline{B}^0$ mixing and is best measured in $B^0 \to J/\psi\,(\psi(2S))K_S^0$ transitions, which have $CP$-odd final states (ignoring the small $CP$ violation in $K^0$-$\overline{K}^0$ mixing). As discussed in Section 10.1, $\Delta B = 2$ transitions in the SM are produced by quark box diagrams $O_{\rm box}$ including QCD radiative corrections for $\Delta m_d$.

The most precise technique for measuring $\phi_1$ uses $B^0$ decays to $CP$ eigenstates with quark transitions of the type $b \to c\bar{c}s$ (Fig. 17.6.1). Since the final state $f$ is accessible to both $B^0$ and $\overline{B}^0$, the amplitudes for $B^0 \to f$ (direct decay) and $B^0 \to \overline{B}^0 \to f$ (decay preceded by neutral meson oscillation) will interfere. As described in Section 10.2,[70] the resulting time-dependent $CP$ asymmetry is given as

$$\mathcal{A}(\Delta t) = S\sin(\Delta m_d \Delta t) - C\cos(\Delta m_d \Delta t), \quad (17.6.2)$$

where $S = 2{\rm Im}\lambda/(1+|\lambda|^2)$, $C = (1-|\lambda|^2)/(1+|\lambda|^2)$, and $\lambda = (q/p)(\overline{A}_f/A_f)$. In the SM, $q/p = V_{td}V_{tb}^*/V_{td}^*V_{tb}$ to a good approximation. For the final state $f = J/\psi\,K_S^0$, the $B$ decay is dominated by a tree $b \to c\bar{c}s$ (or its $CP$ conjugate) amplitude[71] followed by $K^0$-$\overline{K}^0$ mixing.[72] The result is $\lambda = \eta_f \frac{V_{td}V_{tb}^*}{V_{tb}V_{td}^*}\frac{V_{cb}V_{cd}^*}{V_{cd}V_{cb}^*}$, which leads to $C = 0$ and $S = -\eta_f \sin 2\phi_1$, where $\eta_f = \eta_{J/\psi\,K_S^0} = -1$ is the $CP$ eigenvalue. $B^0 \to J/\psi\,K_L^0$ has $\eta_f = \eta_{J/\psi\,K_L^0} = +1$ and has the opposite sign for $S$. The same magnitude is expected for the $CP$-even and -odd modes up to a small correction for $CP$ violation in $K^0$-$\overline{K}^0$ oscillations.

To understand the penguin amplitude contributions, one can group tree ($T$) and penguin ($P^q$) amplitudes according to their CKM factors, remove the $V_{tb}V_{ts}^*$ term using the unitarity condition

$$\sum_{q=u,c,t} V_{qb}V_{qs}^* = 0, \quad (17.6.3)$$

and express the $b \to c\bar{c}s$ decay amplitude as

$$A_{c\bar{c}s} = V_{cb}V_{cs}^*(T + P^c - P^t) + V_{ub}V_{us}^*(P^u - P^t), \quad (17.6.4)$$

where the superscripts indicate the quark in the loop. The second term has a different phase but the magnitude is suppressed by $|V_{ub}V_{us}^*/V_{cb}V_{cs}^*| \sim \mathcal{O}(\lambda^2_{\rm Cabibbo})$. Therefore, the effect of the penguin amplitude on $\phi_1$ is expected to be very small.

[70] See in particular Eqs (10.2.2, 10.2.4, 10.2.4, and 10.1.10).

[71] $B$ decay amplitude ratio provides a factor $\eta_f \frac{V_{cb}V_{cs}^*}{V_{cb}^*V_{cs}}$.

[72] $K^0$-$\overline{K}^0$ mixing provides a factor $V_{cd}^*V_{cs}/V_{cd}V_{cs}^*$.

Within the SM the level of $CP$ violation in decay ($|A_f/\bar{A}_{\bar{f}}| \neq 1$) is expected to be inaccessible to existing experiments, and new physics (NP) beyond the SM is unlikely to generate large effects due to the dominance of the tree amplitude in decay. However, NP could modify the time-dependent $CP$ asymmetry across different modes by affecting the phase in $q/p$ and lead to inconsistencies between $\phi_1$ and other observables that determine the Unitarity Triangle.

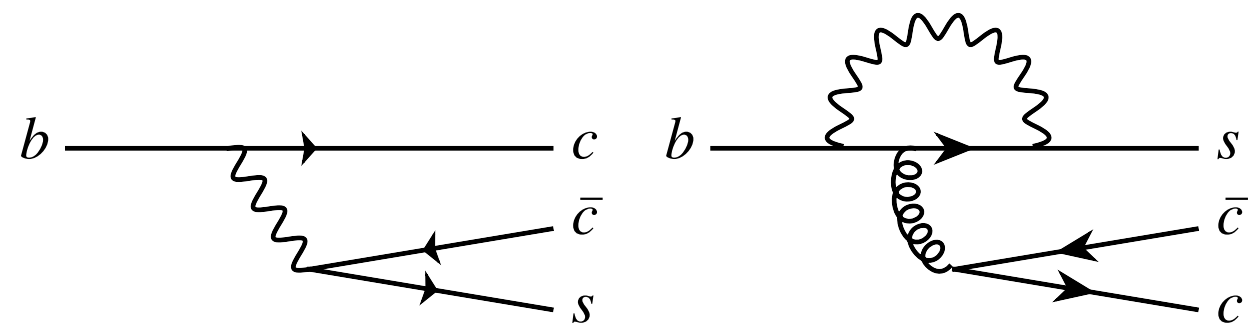

**Figure 17.6.1.** Tree and penguin diagrams of $b \to c\bar{c}s$.

In $b \to c\bar{c}d$ (Fig. 17.6.2) decays, the difference between the CKM phase of the tree diagram and that of $b \to c\bar{c}s$ is negligible. This allows the measurements of $\sin 2\phi_1$ through decays to $CP$ eigenstates of $b \to c\bar{c}d$ (such as $B^0 \to J/\psi\,\pi^0$ and $D^+D^-$) in the same way as $b \to c\bar{c}s$. Unlike $b \to c\bar{c}s$, however, the CKM factors of the penguin diagrams here are of the same order ($\mathcal{O}(\lambda^3_{\rm Cabibbo})$) as the tree diagram. The possible contribution of the $b \to c\bar{c}d$ penguin diagrams, which have a different CKM phase, can alter the measured value of $\sin 2\phi_1$. Any such deviation would be due to the effect of penguin contributions or due to NP.

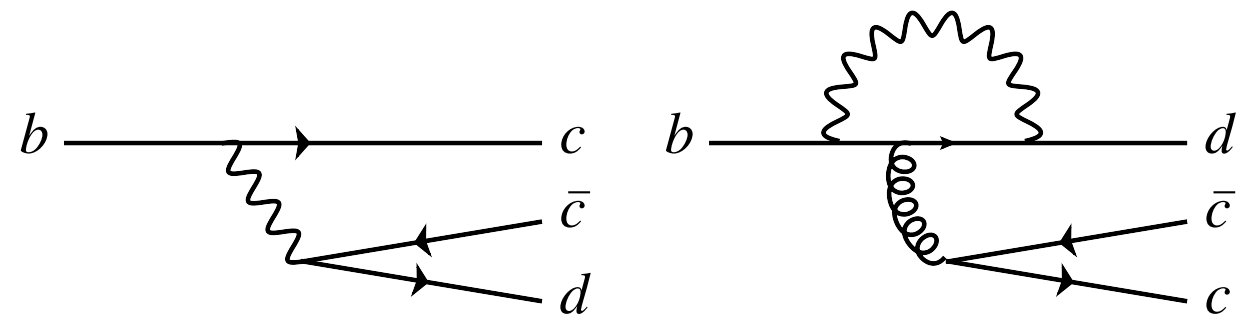

**Figure 17.6.2.** Tree and penguin diagrams of $b \to c\bar{c}d$.

The $b \to c\bar{u}d$ transition (Fig. 17.6.3) proceeds through a tree diagram, and has no penguin contribution. It can again be used to probe $\sin 2\phi_1$ if the final state is accessible to both $B^0$ and $\overline{B}^0$ (*e.g.*, in the case of intermediate $D^0$ and $\overline{D}^0$ decays to the same final state). However, in this case, the process $b \to u\bar{c}d$ also contributes. The relative CKM factor of these two tree diagrams, $V_{ub}V_{cd}^*/V_{cb}V_{ud}^*$, has a large phase and the magnitude is approximately 0.02. Therefore, the deviation from the $b \to c\bar{c}s$ value for $\sin 2\phi_1$ obtained in these decays is expected to be small.

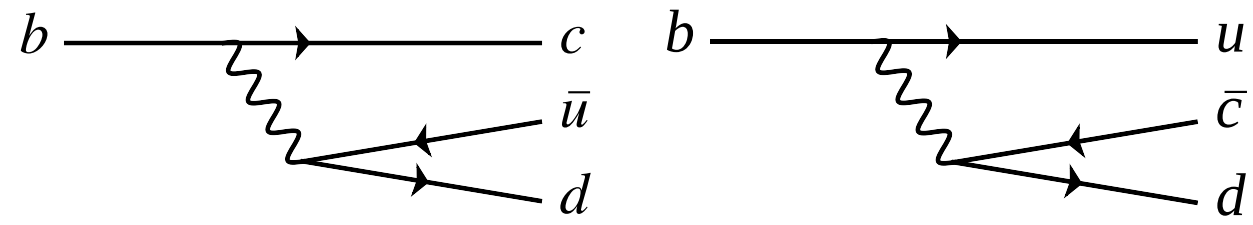

**Figure 17.6.3.** Tree diagrams of $b \to c\bar{u}d$ and $b \to u\bar{c}d$.

The decays to $CP$ eigenstates dominated by $b \to s\bar{q}q$ penguin transitions (Fig. 17.6.4) also can be used for $\sin 2\phi_1$ measurements in the SM. Similar to Eq. (17.6.4), the dominant penguin contribution has the same phase as that in the $b \to c\bar{c}s$ tree diagram, and the sub-dominant term is suppressed. Any deviation of $S$ from the $b \to c\bar{c}s$ decay (beyond theoretical uncertainty) is a clear indication of the effect of NP. The decays proceeding via $b \to s\bar{s}s$ penguin diagrams, such as $B^0 \to \phi K^0$, $K^0_S K^0_S K^0_S$, and $\eta' K^0$, have a small theoretical uncertainty on $S$ due to the lack of a tree amplitude contribution. These decays are particularly promising for future new physics searches.

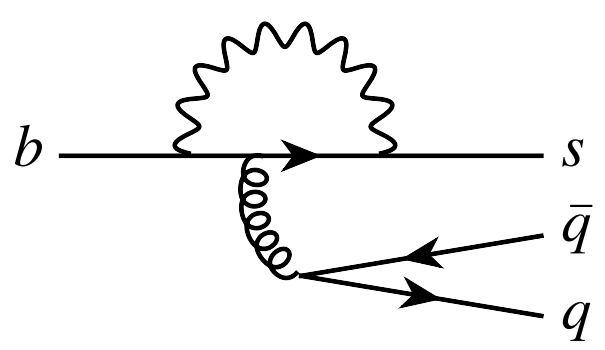

**Figure 17.6.4.** Penguin diagram of $b \to q\bar{q}s$.

Measurements of $\sin 2\phi_1$ have a four-fold ambiguity in $\phi_1$: $\phi_1 \leftrightarrow \pi/2 - \phi_1$, $\phi_1 + \pi$ and $3\pi/2 - \phi_1$ (all these four values result in the same $\sin 2\phi_1$). The $\phi_1 \leftrightarrow \pi/2 - \phi_1$ and $3\pi/2 - \phi_1$ ambiguity can be resolved in one of several ways: the full time-dependent angular analysis of vector-vector final states such as $B^0 \to J/\psi K^{*0}[K^0_S\pi^0]$; time-dependent Dalitz analysis of three-body decays; time-dependent Dalitz analysis of $D^0 \to K^0_S\pi^+\pi^-$ in $B^0 \to D^{(*)0}h^0$; and time-dependent measurements in two separate Dalitz regions in $B^0 \to D^{*+}D^{*-}K^0$. Using these measurements the ambiguity is partially resolved and only the two fold ambiguity $\phi_1 \to \phi_1 + \pi$ remains, which cannot be resolved by a single measurement. When combining with other CKM measurements, one can clearly see which of the two remaining solutions is ruled out. See Chapter 25 for details.

The following sections describe the different measurements of $\phi_1$ made at the $B$ Factories.

### 17.6.3 $\phi_1$ from $b \to c\bar{c}s$ decays

The decays to $CP$ eigenstates via a $b \to c\bar{c}s$ transition include $B^0$ decays to charmonium ($c\bar{c}$) and a $K^0_S$ or $K^0_L$. These modes have experimentally clean signals, and large signal yields are expected due to relatively large branching fractions (they are CKM favored, though color suppressed[73]). These decays are also theoretically very clean for $\phi_1$ determination, *i.e.*, the deviation due to the contribution of penguin diagrams with a different CKM phase is expected to be at the $\leq 1\%$ level (H. Boos and Reuter, 2004, 2007). As a result the $B^0 \to J/\psi K^0_S$ decay is called a "Golden mode".

Since the observation of $CP$ violation in $B$ decays and the precise measurements of $\sin 2\phi_1$ are the primary goals of the asymmetric $B$ Factories, the measurements made using $b \to c\bar{c}s$ modes were performed shortly after data taking commenced, and have been updated several times during the course of data taking. Both $B$ Factories have updated their measurements using the whole data sample collected by each experiment. *BABAR* (Aubert, 2009z) uses $465 \times 10^6$ $B\overline{B}$, while Belle (Adachi, 2012c) uses $772 \times 10^6$ $B\overline{B}$ pairs. For $\phi_1$ measurements with $b \to c\bar{c}s$ decays, the $B^0$ decays to the final states $J/\psi K^0_S$, $J/\psi K^0_L$, $\psi(2S)K^0_S$, $\chi_{c1}K^0_S$, $\eta_c K^0_S$, and $J/\psi K^*(890)^0[K^0_S\pi^0]$ are used. The $J/\psi K^0_L$ state is $CP$-even, and $J/\psi K^*(890)^0$ is an admixture of two $CP$ states. All the others are $CP$-odd states.

The $J/\psi$ and $\psi(2S)$ mesons are reconstructed via their decays to $\ell^+\ell^-$ ($\ell = e, \mu$). For decays to an $e^+e^-$ final state, photons near the direction of the $e^\pm$ are added to recover the energy lost by radiated bremsstrahlung. The $\psi(2S)$ mesons are also reconstructed in the $J/\psi\pi^+\pi^-$ final state. The $\chi_{c1}$ mesons are reconstructed in the $J/\psi\gamma$ final state, and these photons must not be consistent with photons from $\pi^0$ decays. The $\eta_c$ mesons are reconstructed in the $K^0_S K^+\pi^-$ final states, and the regions that contain the dominant intermediate resonant states in $K^+\pi^-$ and $K^0_S K^+$ are selected. Candidate $K^0_S$ mesons are reconstructed via decays to the $\pi^+\pi^-$ final state. For the $B^0 \to J/\psi K^0_S$ decay mode, $K^0_S$ mesons are also reconstructed in the $\pi^0\pi^0$ final state. Inclusion of the $K^0_S \to \pi^0\pi^0$ channel increases a signal yield by about 20% of the $K^0_S \to \pi^+\pi^-$ channel. The masses of $J/\psi$, $\psi(2S)$, $\chi_{c1}$, and $K^0_S$ candidates are constrained to their respective nominal values to improve their momentum resolutions. Candidate $K^0_L$ mesons are identified using information from the electromagnetic calorimeter and IFR/KLM detectors (see Chapter 2), requiring that the signals in these detectors are not associated with any charged tracks. Since the energy of a $K^0_L$ cannot be measured precisely, only the flight direction is used when reconstructing $B^0 \to J/\psi K^0_L$ decay candidates. The $K^{*0}$ candidates are selected by combining $K^0_S$ and $\pi^0$ mesons. *BABAR* uses all of the aforementioned final states for their analysis. While Belle (Abe, 2001g) used the same set of modes for earlier iterations of their analysis, more recent updates do not include the $J/\psi K^0_S(\to \pi^0\pi^0)$, $\eta_c K^0_S$, and $J/\psi K^{*0}$ final states.

Candidate $B^0$ mesons are reconstructed by combining charmonium and $K^0_S$, $K^0_L$, or $K^{*0}$ candidates. Two kinematic variables $\Delta E$ and $m_{\rm ES}$ (see Section 7.1.1) are used to select signal candidates, with the exception of the $B^0 \to J/\psi K^0_L$ channel. For the latter case a kinematic constraint is applied assuming a two-body decay of the $B^0$, and both *BABAR* and Belle use $\Delta E$ and the momentum of the reconstructed $B^0$ in the center-of-mass (CM) system ($p^*_B$) to isolate signal candidates. Figure 17.6.5 shows the $m_{\rm ES}$ and $\Delta E$ distributions for candidates satisfying the flavor tagging and vertex reconstructions in the *BABAR*

[73] Each of the two quarks ($\bar{c}s$) from the virtual $W$ is paired with the quark originating from the initial state ($b\bar{d}$) to form a hadron. Since hadrons have to stay color-neutral, the color of $\bar{c}$ and $s$ must match that of $b$ and $\bar{d}$. Therefore the overall amplitude is 1/number-of-colors smaller than the decays in which $W^* \to \bar{q}q'$ hadronize by themselves.

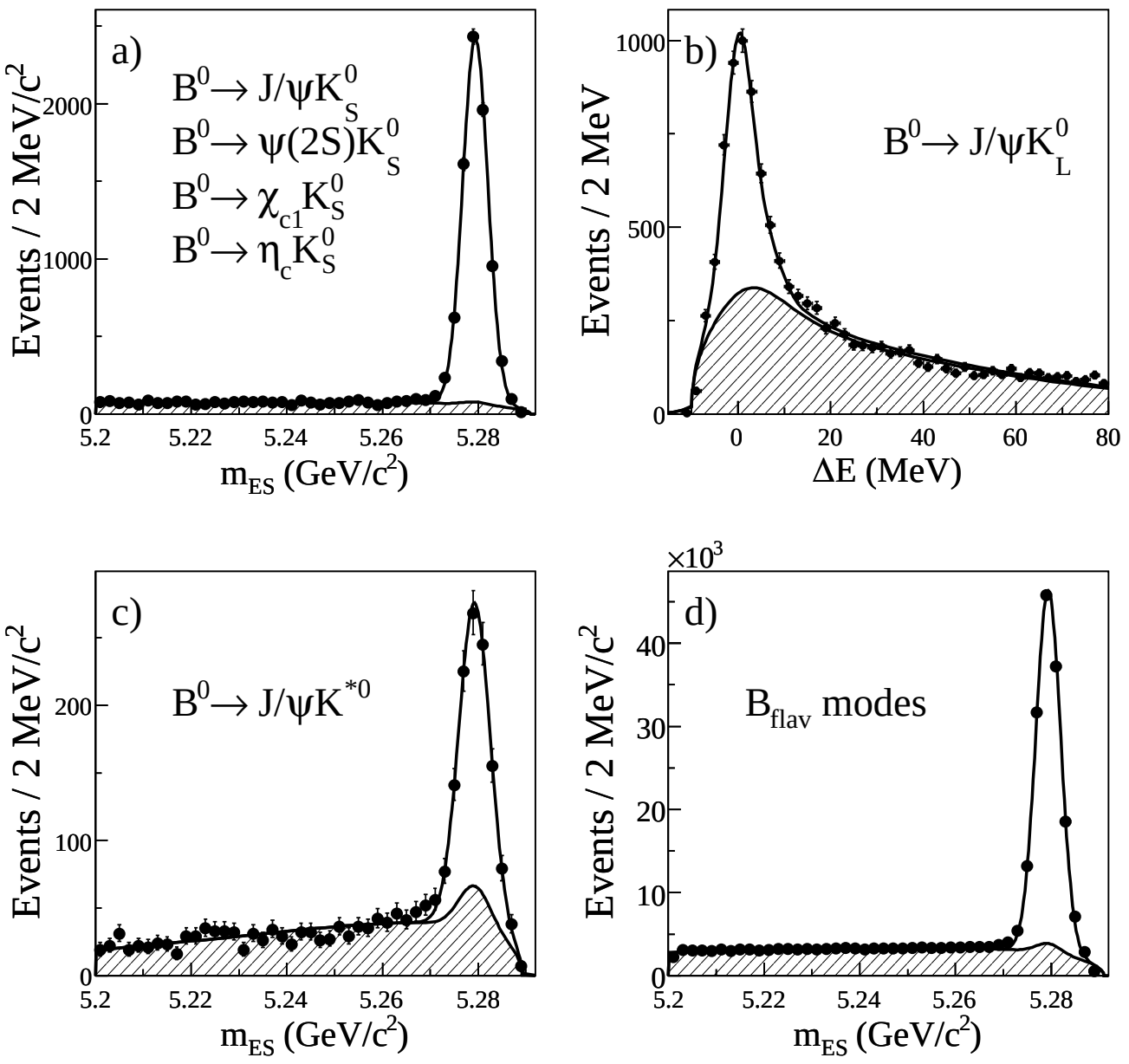

**Figure 17.6.5.** Distributions of $m_{\rm ES}$ or $\Delta E$ for (a) $B^0 \to (c\bar{c})K^0_S$, (b) $B^0 \to J/\psi K^0_L$, (c) $B^0 \to J/\psi K^{*0}$, and (d) $B^0$ decays to flavor-specific final states for the samples used in the *BABAR* measurement (Aubert, 2009z) of $\phi_1$. The shaded regions represent the estimated background, and the solid lines are the projections of the fits to the data.

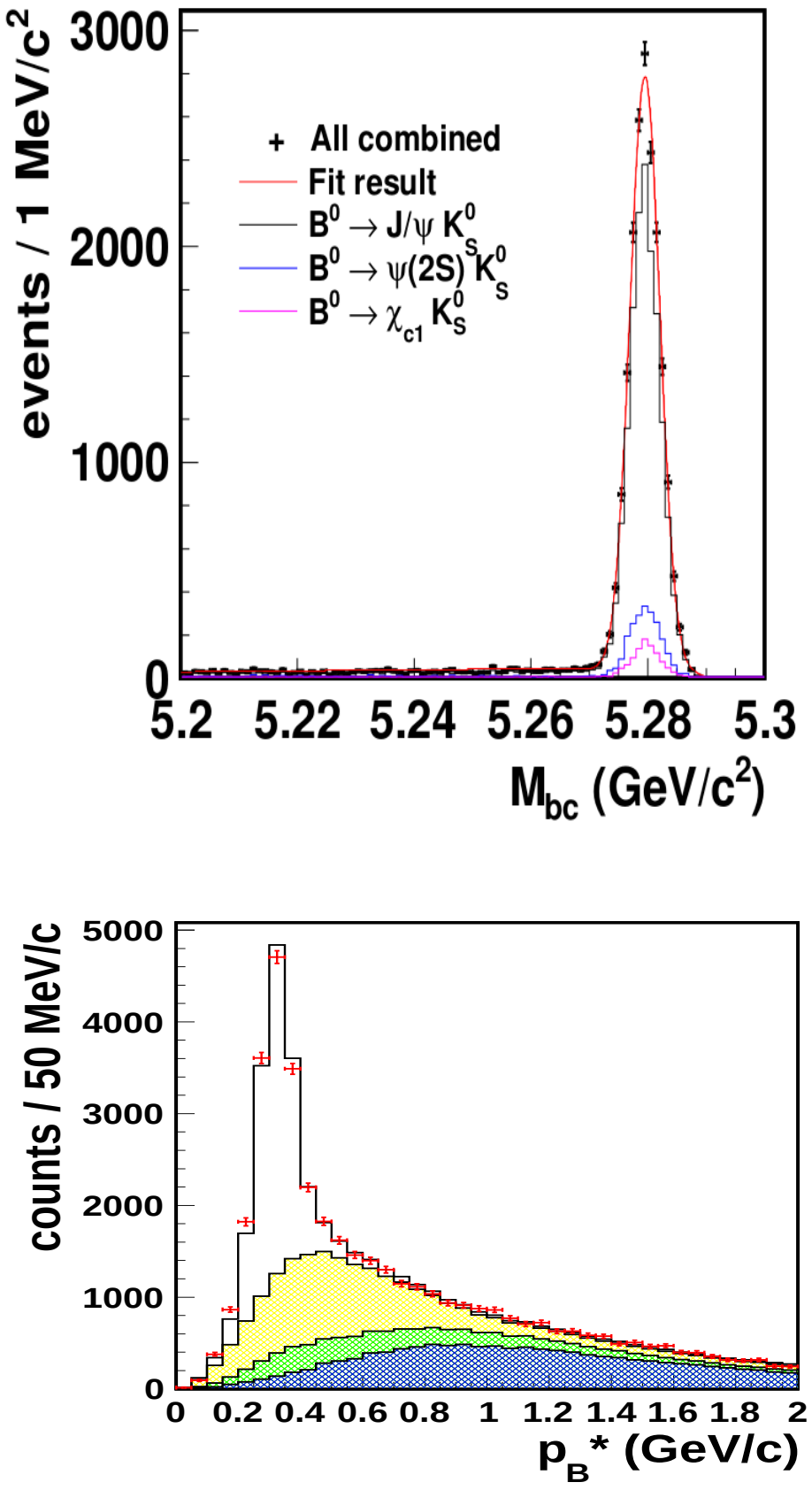

**Figure 17.6.6.** Distributions of $m_{\rm ES}$ ($= M_{\rm bc}$) for $B^0 \to (c\bar{c})K^0_S$ (top) and $p^*_B$ for $B^0 \to J/\psi K^0_L$ (bottom) obtained with the samples used for the Belle measurement (Adachi, 2012c) of $\phi_1$. The shaded regions in the bottom plot represent the estimated background components: (from top to bottom) real $J/\psi$ and real $K^0_L$ (yellow), real $J/\psi$ and fake $K^0_L$ (green), and fake $J/\psi$ (blue).

analysis. Figure 17.6.6 shows the $m_{\rm ES}$ and $p^*_B$ distributions for the Belle analysis.

Vertex reconstruction and $B$ meson flavor tagging algorithms (described in Chapters 6 and 8) are applied to the selected signal candidates. Time-dependent *CP* asymmetry parameters are extracted from fits to the distributions of proper decay time difference between signal and tagged $B$ mesons as described in Chapter 10. *BABAR* extracts the time-dependent asymmetry parameters ($S$ and $C$) from a simultaneous fit to both the $B_{CP}$ and $B_{\rm flav}$ (see Section 10.2) samples with 69 additional free parameters, where tagging and resolution parameters are transparently propagated into the $CP$ analysis as part of the final statistical error. Belle takes a multi-step approach: the final fit includes only $S$ and $C$ as free parameters, and all the fit model parameters, which include signal fractions, flavor tagging performance parameters, and proper time difference resolution function parameters are fixed to the values determined from separate fits to the $B_{\rm flav}$ and $B_{CP}$ samples. Effects arising from the uncertainties of these parameters are included in the final result as systematic errors.

The results of the time-dependent *CP* asymmetry measurements are summarized in Table 17.6.1 for each decay mode, and for the combined set of modes. As described in Section 17.6.8, the time-dependent full angular analysis of the $B^0 \to J/\psi K^{*0}$ decay can provide a value for $\cos 2\phi_1$ in addition to $\sin 2\phi_1$. The angular information presented in the table has been averaged over, resulting in a dilution of the measured *CP* asymmetry by a factor of $1 - 2R_\perp$, where $R_\perp$ is the fraction of the *CP*-odd component. *BABAR* uses the previously measured value $0.233 \pm 0.010 \pm 0.005$ (Aubert, 2007x). Systematic errors on the time-dependent asymmetry parameters are summarized in Table 17.6.2. The dominant sources for $S$ are due to the uncertainties in vertex reconstruction and $\Delta t$ resolutions, flavor tagging, and background in the $J/\psi K^0_L$ mode. The systematic error on $C$ is dominated by tag-side interference. For this source, Belle takes into account a cancellation between *CP*-even and *CP*-odd states, while *BABAR* does not. Chapter 15 discusses the main sources of systematic uncertainty on time-dependent *CP* asymmetry parameter measurements in detail.

The $\Delta t$ distributions and asymmetries obtained from the data for all modes combined are shown in Fig. 17.6.7. The values of $C$ obtained are consistent with zero in accordance with SM expectations, and hence $-\eta_f S$ gives essentially $\sin 2\phi_1$. The average of the two experiments

**Table 17.6.1.** Summary of the time-dependent *CP*-asymmetry measurements using $B^0$ decays to charmonium + $K^0$ final states, for each decay mode and for all modes combined. $N_{\text{tag}}$ and $P$ are the number of candidates and signal purity (in %), respectively, in the signal region after flavor tagging and vertex reconstruction requirements have been applied. $S$ and $C$ are the *CP* asymmetry parameters for the final state with the *CP* eigenvalue $\eta_f$.

| | BABAR (Aubert, 2009z) | | | | Belle (Adachi, 2012c) | | | |
|---|---|---|---|---|---|---|---|---|
| Mode | $N_{\text{tag}}$ | $P$ | $-\eta_f S$ | $C$ | $N_{\text{tag}}$ | $P$ | $-\eta_f S$ | $C$ |
| $J/\psi K_S^0$ | 6750 | 95 | $0.657 \pm 0.036 \pm 0.012$ | $0.026 \pm 0.025 \pm 0.016$ | 13040 | 97 | $0.670 \pm 0.029 \pm 0.013$ | $0.015 \pm 0.021\ ^{+0.023}_{-0.045}$ |
| $J/\psi K_L^0$ | 5813 | 56 | $0.694 \pm 0.061 \pm 0.031$ | $-0.033 \pm 0.050 \pm 0.027$ | 15937 | 63 | $0.642 \pm 0.047 \pm 0.021$ | $-0.019 \pm 0.026\ ^{+0.041}_{-0.017}$ |
| $\psi(2S)K_S^0$ | 861 | 87 | $0.897 \pm 0.100 \pm 0.036$ | $0.089 \pm 0.076 \pm 0.020$ | 2169 | 91 | $0.738 \pm 0.079 \pm 0.036$ | $-0.104 \pm 0.055\ ^{+0.027}_{-0.047}$ |
| $\chi_{c1}K_S^0$ | 385 | 88 | $0.614 \pm 0.160 \pm 0.040$ | $0.129 \pm 0.109 \pm 0.025$ | 1093 | 86 | $0.640 \pm 0.117 \pm 0.040$ | $0.017 \pm 0.083\ ^{+0.026}_{-0.046}$ |
| $\eta_c K_S^0$ | 381 | 79 | $0.925 \pm 0.160 \pm 0.057$ | $0.080 \pm 0.124 \pm 0.029$ | | | | |
| $J/\psi K^{*0}$ | 1291 | 67 | $0.601 \pm 0.239 \pm 0.087$ | $0.025 \pm 0.083 \pm 0.054$ | | | | |
| All | 15481 | 76 | $0.687 \pm 0.028 \pm 0.012$ | $0.024 \pm 0.020 \pm 0.016$ | 32239 | 79 | $0.667 \pm 0.023 \pm 0.012$ | $-0.006 \pm 0.016 \pm 0.012$ |

**Table 17.6.2.** Summary of systematic errors on the time-dependent *CP* asymmetry parameters measured in $B^0$ decays to charmonium + $K^0$ for all modes combined.

| | *BABAR* | | Belle | |
|---|---|---|---|---|
| Source | $S$ | $C$ | $S$ | $C$ |
| Vertex and $\Delta t$ | 0.007 | 0.003 | 0.010 | 0.007 |
| Flavor tagging | 0.006 | 0.002 | 0.004 | 0.003 |
| $J/\psi K_L^0$ background | 0.006 | 0.001 | 0.004 | 0.002 |
| Other signal/background | 0.005 | 0.003 | 0.002 | 0.001 |
| Physics parameters | 0.003 | 0.001 | 0.001 | 0.000 |
| Tag-side interference | 0.001 | 0.014 | 0.001 | 0.008 |
| Possible fit bias | 0.002 | 0.003 | 0.004 | 0.005 |
| Total | 0.012 | 0.016 | 0.012 | 0.012 |

(Amhis et al., 2012) gives

$$\sin 2\phi_1 = 0.677 \pm 0.020 \quad \text{and} \quad C = 0.006 \pm 0.017. \quad (17.6.5)$$

This corresponds to $\phi_1 = (21.30 \pm 0.78)^\circ$ (up to the four-fold ambiguity mentioned above). An accuracy of 3% on $\sin 2\phi_1$ ($0.8^\circ$ on $\phi_1$) is achieved.

The evolution of the measured value of $\sin 2\phi_1$ can be seen in Fig. 17.6.8. Central values for the initial measurements from both experiments were slightly lower than the current world average. A significant milestone in the measurement of $\sin 2\phi_1$ was achieved in the summer of 2001 when both *BABAR* and Belle observed *CP* violation in $B^0$ meson decay.[74] The data samples used for these measurements each consists of about $30 \times 10^6$ $B\overline{B}$ pairs. Since that time, improved measurements have proved to be stable, and the results reported by *BABAR* and Belle have remained consistent with each other.

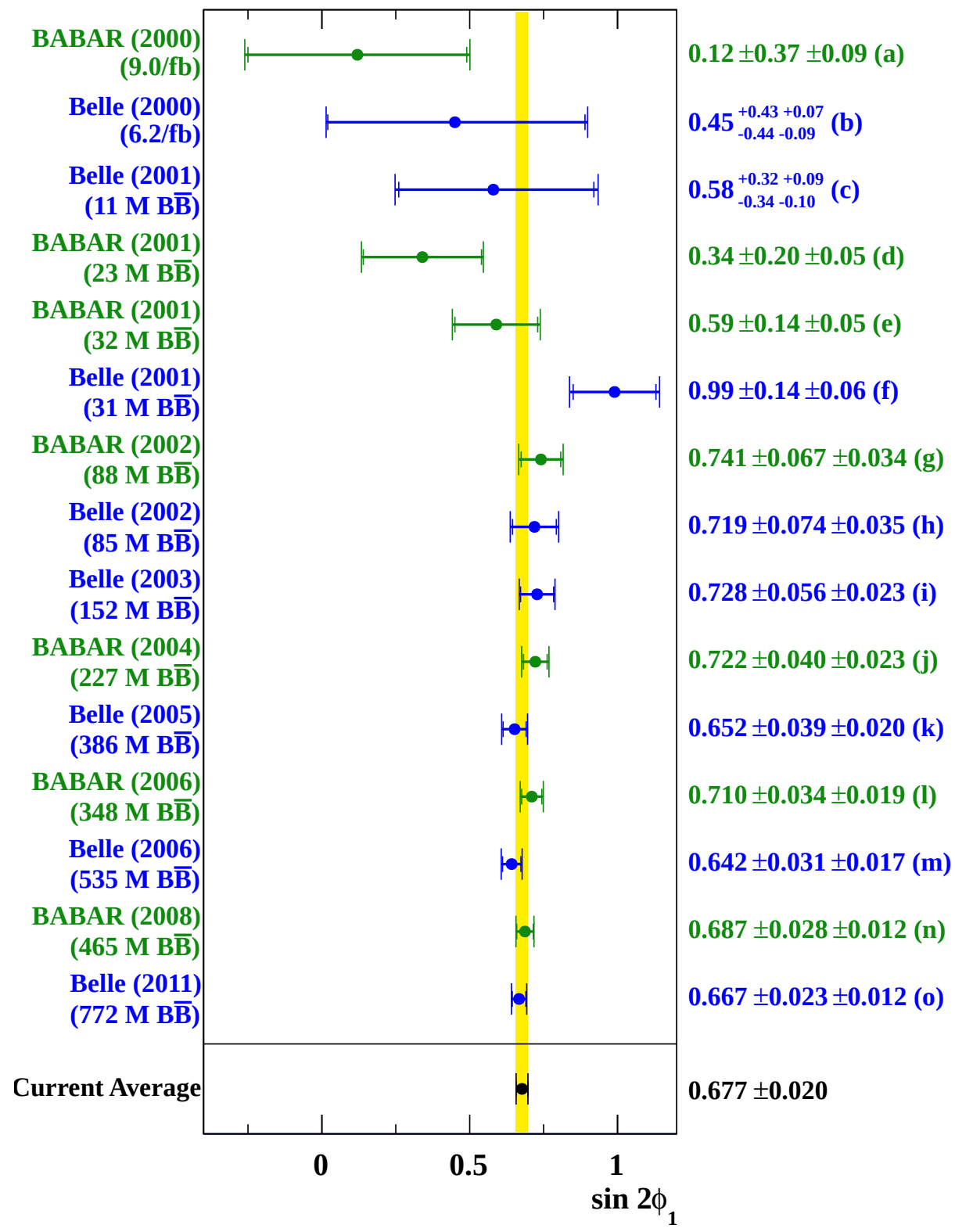

**Figure 17.6.8.** History of the $\sin 2\phi_1$ measurements with $b \to c\bar{c}s$ decays, ordered by the dates they appeared in public. References: (a) (Aubert, 2000), (b) (Aihara, 2000a), (c) (Abashian, 2001), (d) (Aubert, 2001a), (e) (Aubert, 2001e), (f) (Abe, 2001g), (g) (Aubert, 2002g), (h) (Abe, 2002b), (i) (Abe, 2005c), (j) (Aubert, 2005i), (k) (Abe, 2005j), (l) (Aubert, 2006j), (m) (Chen, 2007a), (n) (Aubert, 2009z), (o) (Adachi, 2012c).

### 17.6.4 $\phi_1$ from $b \to c\bar{c}d$ decays

#### 17.6.4.1 $B^0 \to J/\psi \pi^0$

The decay $B^0 \to J/\psi \pi^0$ is a $b \to c\bar{c}d$ transition into a *CP*-even final state. The final state has contributions from both a color- and Cabibbo-suppressed tree amplitude, and penguin amplitudes with different weak phases. In the absence of penguin contributions one can measure the Unitarity Triangle angle $\phi_1$ using this decay. If there are significant penguin contributions, the measured value of $\phi_1$, called the "effective phase" $\phi_1^{\rm eff}$, may differ from that obtained from the tree-dominated $B \to J/\psi K^0$ decays. There are two motivations for such a measurement; firstly it is possible to constrain theoretical uncertainties in $B \to J/\psi K^0$ decays using $B^0 \to J/\psi \pi^0$ (Ciuchini, Pierini, and Silvestrini, 2005), and secondly one may be able to probe, or constrain, possible new physics contributions to $b \to c\bar{c}d$ transitions manifesting via loop diagrams.

[74] A commonly accepted definition of "observation" is a result with a statistical significance of at least five standard deviations if the uncertainties are treated as Gaussian.

Unlike $b \to c\bar{c}s$ decays, which are experimentally clean, one has to consider significant background contributions when trying to extract information from $B^0 \to J/\psi \pi^0$ signal events. These background contributions include events from $B$ decays to $J/\psi \rho^0$, $J/\psi K_S^0$, $J/\psi K^{*0}$, $J/\psi K^{*\pm}$, and $J/\psi \rho^\pm$ final states as well as smaller contributions from other $B$ decays to final states including a $J/\psi$. The aforementioned backgrounds populate the negative $\Delta E$ region (peak $\sim -0.2$ GeV) and have a tail in the signal region around $\Delta E \sim 0$ (see Fig. 17.6.9). Since these modes are well measured, the $B$ Factories have relied on existing branching fraction measurements from the Particle Data Group (Yao et al., 2006) in order to fix the normalization of background contributions while extracting signal yields and *CP* asymmetry parameters. The normalization of the combinatorial background is allowed to vary in the fit.

Both experiments perform an unbinned maximum likelihood fit to data using discriminating variables: $m_{\rm ES}$, $\Delta E$, and $\Delta t$. In order to suppress background from light-quark continuum events, *BABAR* also includes a Fisher discriminant as one of the discriminating variables in their fit to data. This is computed using three variables: $L_0$, $L_2$

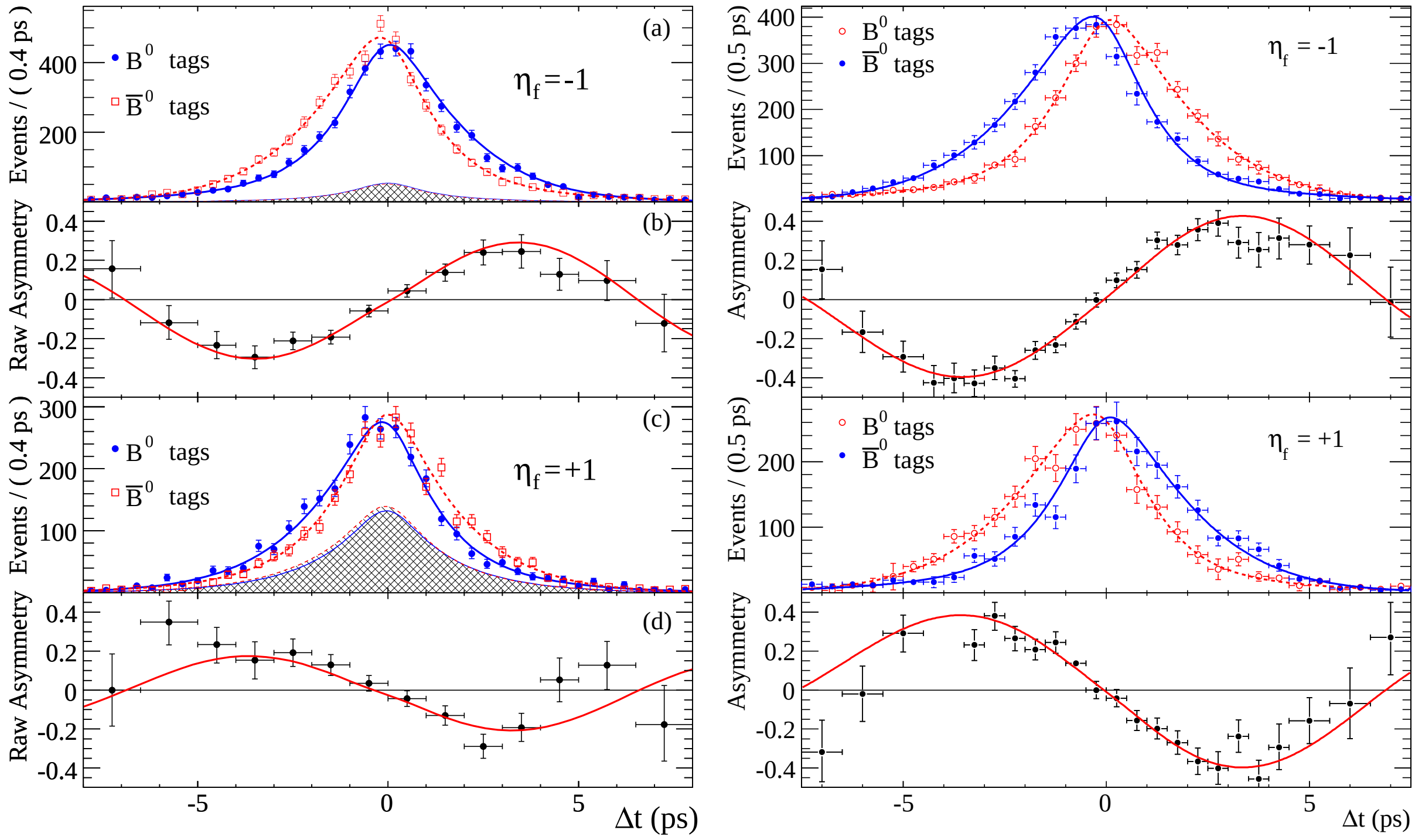

**Figure 17.6.7.** Flavor-tagged $\Delta t$ distributions (a,c) and raw *CP* asymmetries (b,d) for the *BABAR* (left, (Aubert, 2009z)) and Belle (right, (Adachi, 2012c)) measurements of $\sin 2\phi_1$. The top two plots show the $B \to (c\bar{c})K_S^0$ ($\eta_f = -1$) samples, and the bottom two show the $B \to J/\psi K_L^0$ ($\eta_f = +1$) sample. The shaded regions for *BABAR* represent the fitted background, while the Belle distributions are background subtracted. The two experiments adopt the opposite color code in $\Delta t$ distribution plots.

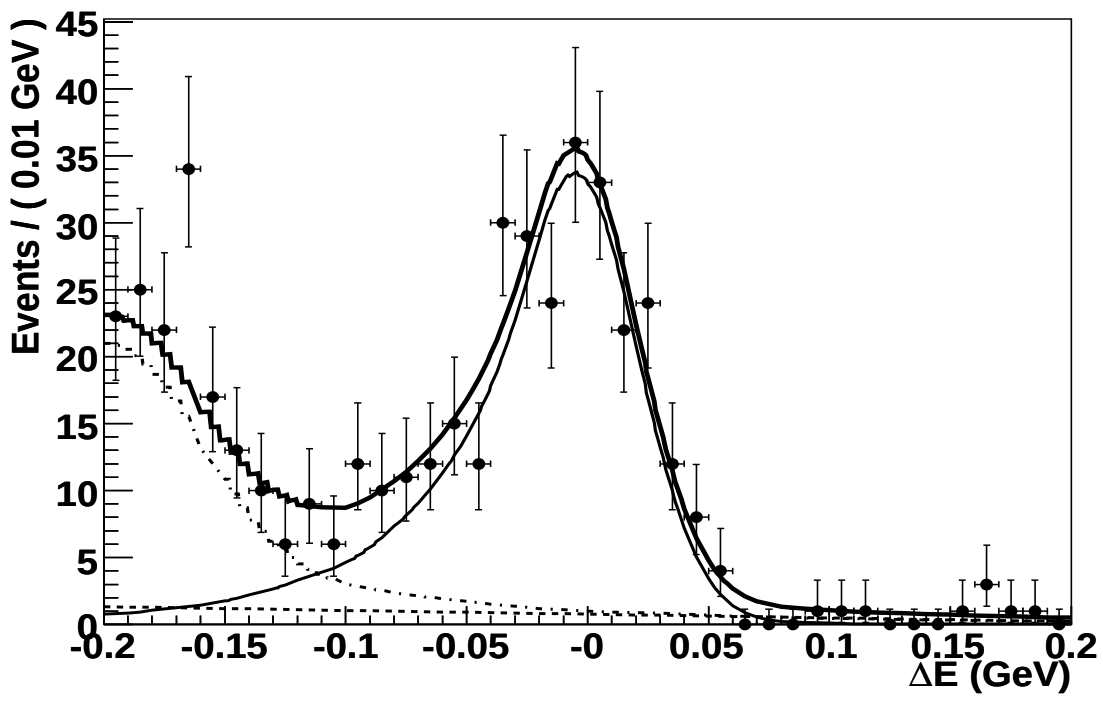

**Figure 17.6.9.** Distributions of $\Delta E$ for $B^0 \to J/\psi \pi^0$ samples used in the Belle measurement (Lee, 2008) of $\phi_1$. The superimposed curves show the signal (solid line), $B \to J/\psi X$ background (dot-dashed line), combinatorial background (dashed line) and the sum of all the contributions (thick solid line).

The most recent results obtained by *BABAR* (Aubert, 2008i) and Belle (Lee, 2008) use 465 $\times 10^6$ and 535 $\times 10^6$ $B\overline{B}$ pairs, respectively, and are summarized in Table 17.6.3. *BABAR* finds *CP* violation with 4.0$\sigma$ significance, and Belle finds 2.4$\sigma$ significance. Both results, and their average, are consistent with the value of $S$ measured in $b \to c\bar{c}s$ decays. The obtained value of $C$ is consistent with zero.

**Table 17.6.3.** The time-dependent *CP* asymmetry parameters $-\eta_f S$ and $C$ for the decay $B^0 \to J/\psi \pi^0$. The first quoted uncertainty is statistical, and the second is systematic. The averages are obtained by HFAG (Amhis et al., 2012).

| Experiment | $-\eta_f S$ | $C$ |
|---|---|---|
| *BABAR* | $1.23 \pm 0.21 \pm 0.04$ | $-0.20 \pm 0.19 \pm 0.03$ |
| Belle | $0.65 \pm 0.21 \pm 0.05$ | $-0.08 \pm 0.16 \pm 0.05$ |
| Average | $0.93 \pm 0.15$ | $-0.10 \pm 0.13$ |

(Eq. (9.4.1)), and $\cos\theta_H$, where $\theta_H$ is the angle between the positively charged lepton and the $B$ candidate momenta in the $J/\psi$ rest frame. In contrast, Belle achieves continuum background rejection by applying a cut on the ratio of zeroth to second Fox-Wolfram moments, $R_2 < 0.4$. Details on these background suppression techniques can be found in Chapter 9.

### 17.6.4.2 $B^0 \to D^{(*)\pm}D^{(*)\mp}$

The decay $B^0 \to D^{(*)\pm}D^{(*)\mp}$ is dominated by a color-favored tree-diagram in the SM. When neglecting the penguin (loop) diagram, the mixing induced *CP* asymmetry of $B^0 \to D^{(*)\pm}D^{(*)\mp}$ is also determined by $\sin 2\phi_1$. The effect of neglecting the penguin amplitude has been estimated in models based on factorization and heavy quark symmetry, and the corrections are expected to be a few percent (Xing, 1998, 2000). Significant deviation of $S$ in $B^0 \to D^{(*)\pm}D^{(*)\mp}$ decays with respect to $\sin 2\phi_1$ determined from $b \to c\bar{c}s$ transitions, or a large non-zero value of $C$, could indicate physics beyond the SM (Grossman and Worah, 1997; M. Gronau and Pirjol, 2008; Zwicky, 2007).

The $B^0 \to D^{(*)\pm}D^{(*)\mp}$ candidates are formed from oppositely charged $D^{(*)}$ mesons reconstructed in the following channels: $D^{*+} \to D^0\pi^+$, $D^{*+} \to D^+\pi^0$, $D^0 \to K^-\pi^+$, $D^0 \to K^-\pi^+\pi^0$, $D^0 \to K^-\pi^+\pi^-\pi^+$, $D^0 \to K^0_S\pi^+\pi^-$, and $D^+ \to K^-\pi^+\pi^+$. Belle also uses the $D^0 \to K^+K^-$, $D^+ \to K^0_S\pi^+$ and $K^0_S\pi^+\pi^0$ channels. In the $B^0 \to D^{*+}D^{*-}$ mode, $B^0$ candidates where both $D^*$ mesons decay to $D\pi^0$ are excluded because of its smaller branching fraction and larger backgrounds. At least one $D$ meson is required to decay via $D^+ \to K^-\pi^+\pi^+$ for the $B^0 \to D^+D^-$ decay.

Both *BABAR* and Belle also analyze these decays using partially reconstructed events. However, while Belle includes these events in their analysis of fully reconstructed events, *BABAR* performs a separate $B^0 \to D^{*+}D^{*-}$ analysis of partially reconstructed events. For the partial reconstruction method one $D^{*-}$ (or a $D^- \to K^+\pi^-\pi^-$) is fully reconstructed as described in the previous paragraph. For the other $D^{*+}$, only a slow pion $\pi^+_{\rm slow}$ from the decay $D^{*+} \to D^0\pi^+_{\rm slow}$, is reconstructed. The details of the partial reconstruction technique are described in Section 7.3. Due to low $B$ meson CM momentum and small energy release in the $D^{*+}$ decay, the momenta of $\pi^+_{\rm slow}$ and $D^{(*)-}$ are almost back-to-back. This signature is used as a discriminator in Belle's analysis. *BABAR* on the other hand exploits the kinematics of the event and calculates the $B$ four-momentum up to an unknown azimuthal angle around the direction of the fully reconstructed $D^*$. *BABAR* uses the median value for this angle based on simulation to calculate the recoil mass of the unreconstructed $D^0$ and uses this recoil mass as a fit variable to separate signal and background. Belle requires the CM momenta of the reconstructed mesons in the $D^{*+}D^-$ mode to satisfy $1.63\ \text{GeV}/c < p^*_{D^{(*)-}} < 1.97\ \text{GeV}/c$ and $p^*_{\pi^+_{\rm slow}} < 0.18\ \text{GeV}/c$. *BABAR* selects events with $1.3\ \text{GeV}/c < p^*_{D^{*-}} < 2.1\ \text{GeV}/c$ and $p^*_{\pi^+_{\rm slow}} < 0.6\ \text{GeV}/c$.

In the partial reconstruction technique used by both experiments, a lepton $\ell_{\rm tag}$ is used to provide flavor tagging, suppress continuum background to a negligible level, and reduce combinatorial $B\overline{B}$ background. In addition to leptons *BABAR* also uses kaons for flavor tagging. The vertex of the reconstructed $B$ ($B_{\rm rec}$) is determined by a fit with the fully reconstructed $\overline{D}^0$ or $D^-$ mesons to the interaction region. On the tagging side, the $\ell_{\rm tag}$ is fitted to the interaction region to provide the $B_{\rm tag}$ vertex information. For the kaon-tagged events (*BABAR*), all tracks that do not belong to $B_{\rm rec}$ and are outside of a cone of $\cos\theta = 0.5$ around the missing $D^0$ direction are used for $B_{\rm tag}$ vertexing. A kinematic cut is applied to remove a large fraction of the background events from $B \to D^{(*)-}\ell^+X$ decays or other sources where the tagging track originates from the same $B$ as the fully reconstructed $D^*$ or $D^-$. In Belle's analysis, the calculated angle between the $B$ and $D^{(*)}\ell_{\rm tag}$ combination is required to be outside the physical region of $B \to D^{(*)-}\ell^+X$, *i.e.*,

$$\cos\theta_{B,D\ell} = \frac{(E_{\rm beam} - E^*_{D\ell})^2 - p^{*2}_B - p^{*2}_{D\ell}}{2p^*_B p^*_{D\ell}} < -1.1. \tag{17.6.6}$$

In *BABAR*'s analysis, the angle between the tagging lepton (kaon) and the missing $D^0$ is required to be larger than arccos 0.75 (arccos 0.5). This kind of background (tagging and reconstructed particles originating from the same $B$) cannot be completely eliminated, and care is taken to evaluate the mistag effects.

For each $B^0 \to D^{(*)\pm}D^{(*)\mp}$ candidate, *BABAR* constructs a likelihood function $\mathcal{L}_{\rm mass}$ from the masses and mass uncertainties of the $D$ and $D^*$ candidates. The values of $\mathcal{L}_{\rm mass}$ and $\Delta E$ are used to reduce the combinatorial background. From the simulated events, the minimum allowed values of $-\ln\mathcal{L}_{\rm mass}$ and $|\Delta E|$ for each individual final state are optimized to obtain the highest expected signal significance.

The technique used to fit the $\Delta t$ distribution is analogous to the one used in $b \to c\bar{c}s$ decays. Since the $B^0 \to D^{*+}D^{*-}$ final state contains two vector mesons, it is an admixture of *CP*-even and *CP*-odd states depending on the orbital angular momentum of the decay products (Chapter 12). In the partial reconstruction approach, the helicity angles are calculated ignoring the $B$ meson momentum. In the fit to data, two scenarios are considered. In the first scenario, the *CP*-even amplitude is allowed to have different *CP*-violating parameters ($C_+$ and $S_+$) from those of the *CP*-odd amplitude ($C_\perp$ and $S_\perp$). In the second scenario, we assume that $C_+ = C_\perp = C_{D^{*+}D^{*-}}$ and $S_+ = -S_\perp = S_{D^{*+}D^{*-}}$.[75] In the absence of penguin contributions, $S_{D^+D^-} = S_+ = -S_\perp = -\sin 2\phi_1$ and $C_{D^+D^-} = C_+ = C_\perp = 0$.

As $B^0 \to D^{*\pm}D^\mp$ is not a *CP* eigenstate, the expressions for the different $S$ and $C$ parameters are related, $S_{D^{*\pm}D^\mp} = -\sqrt{1 - C^2_{D^{*\pm}D^\mp}}\sin(2\phi_1^{\rm eff} \pm \delta)$ (see Eq. 10.2.6) , where $\delta$ is the strong phase difference between the $D^{*+}D^-$ and $D^{*-}D^+$ amplitudes. Neglecting the penguin contributions, $\phi_1^{\rm eff} = \phi_1$ and $C_{D^{*+}D^-} = -C_{D^{*-}D^+}$. It is convenient to express the *CP* asymmetry parameters as

$$\begin{aligned} S_{D^*D} &= \frac{1}{2}(S_{D^{*+}D^-} + S_{D^+D^{*-}}), \\ C_{D^*D} &= \frac{1}{2}(C_{D^{*+}D^-} + C_{D^+D^{*-}}), \\ \Delta S_{D^*D} &= \frac{1}{2}(S_{D^{*+}D^-} - S_{D^+D^{*-}}), \\ \Delta C_{D^*D} &= \frac{1}{2}(C_{D^{*+}D^-} - C_{D^+D^{*-}}). \end{aligned} \tag{17.6.7}$$

[75] In some literature, the opposite sign convention of $S_\perp$ is used, *i.e.*, $S_+ = +S_\perp = S_{D^{*+}D^{*-}}$.

**Table 17.6.4.** Summary of *CP* asymmetry parameter measurements for $B^0 \to D^{(*)\pm}D^{(*)\mp}$ decays. Signal yields quoted here include tagged and untagged events. Reference: (a) (Aubert, 2009ad); (b) Lees (2012k); (c) (Kronenbitter, 2012); (d) (Rohrken, 2012).

| | *BABAR* | Belle |
|---|---|---|
| $B^0 \to D^{*+}D^{*-}$ | $934 \pm 40$ (a) | $1225 \pm 59$ (c) |
| $S_+$ | $-0.76 \pm 0.16 \pm 0.04$ | $-0.81 \pm 0.13 \pm 0.03$ |
| $C_+$ | $+0.00 \pm 0.12 \pm 0.02$ | $-0.18 \pm 0.10 \pm 0.05$ |
| $-S_\perp$ | $-1.80 \pm 0.70 \pm 0.16$ | $-1.52 \pm 0.62 \pm 0.12$ |
| $C_\perp$ | $+0.41 \pm 0.49 \pm 0.08$ | $+0.05 \pm 0.39 \pm 0.08$ |
| $S_{D^{*+}D^{*-}}$ | $-0.70 \pm 0.16 \pm 0.03$ | $-0.79 \pm 0.13 \pm 0.03$ |
| $C_{D^{*+}D^{*-}}$ | $+0.05 \pm 0.09 \pm 0.02$ | $-0.15 \pm 0.08 \pm 0.04$ |
| $B^0 \to D^{*+}D^{*-}$ | $4972 \pm 453$ (b) | - |
| (partial rec.) | | |
| $S_{D^{*+}D^{*-}}$ | $-0.49 \pm 0.18 \pm 0.08$ | - |
| $C_{D^{*+}D^{*-}}$ | $+0.15 \pm 0.09 \pm 0.04$ | - |
| $B^0 \to D^*D$ | $724 \pm 37$ (a) | $887 \pm 39$ (d) |
| $S_{D^*D}$ | $-0.68 \pm 0.15 \pm 0.04$ | $-0.78 \pm 0.15 \pm 0.05$ |
| $C_{D^*D}$ | $+0.04 \pm 0.12 \pm 0.03$ | $-0.01 \pm 0.11 \pm 0.04$ |
| $\Delta S_{D^*D}$ | $+0.05 \pm 0.15 \pm 0.02$ | $-0.13 \pm 0.15 \pm 0.04$ |
| $\Delta C_{D^*D}$ | $+0.04 \pm 0.12 \pm 0.03$ | $-0.12 \pm 0.11 \pm 0.03$ |
| $B^0 \to D^+D^-$ | $152 \pm 17$ (a) | $269 \pm 21$ (d) |
| $S_{D^+D^-}$ | $-0.63 \pm 0.36 \pm 0.05$ | $-1.06\ ^{+0.21}_{-0.14} \pm 0.08$ |
| $C_{D^+D^-}$ | $-0.07 \pm 0.23 \pm 0.03$ | $-0.43 \pm 0.16 \pm 0.05$ |

The parameters $S_{D^*D}$ and $C_{D^*D}$ characterize mixing induced *CP* violation and flavor-dependent direct *CP* violation, respectively. $\Delta S_{D^*D}$ and $\Delta C_{D^*D}$ are insensitive to *CP* violation. In the case of *BABAR*'s $B^0 \to D^{*+}D^{*-}$ partial reconstruction method, the fit parameter $S$ is $(1 - 2R_\perp)S_{D^{*+}D^{*-}}$, where $R_\perp$ is the *CP*-odd fraction measured from fully reconstructed $D^{*+}D^{*-}$ events.

The most recent measurements of the *CP* violation in $B^0 \to D^{(*)\pm}D^{(*)\mp}$ decays by *BABAR* are based on the full data sample, $467 \times 10^6$ $B\overline{B}$ pairs (Aubert, 2009ad), while Belle measurements are based on $772 \times 10^6$ $B\overline{B}$ pairs (Kronenbitter, 2012; Rohrken, 2012). The results are summarized in Table 17.6.4. These supersede the previous *BABAR* (Aubert, 2003g, 2005u, 2007n,u) and Belle (Aushev, 2004; Fratina, 2007; Miyake, 2005; Vervink, 2009) measurements, except for the Belle result based on the $B^0 \to D^*D$ partial reconstruction (Aushev, 2004).

The averages of *BABAR* and Belle results (Amhis et al., 2012) are $S_{D^{*+}D^{*-}} = -0.77 \pm 0.10$, $S_{D^*D} = -0.73 \pm 0.11$, and $S_{D^+D^-} = -0.98 \pm 0.17$, and other parameters are consistent with zero within uncertainties. All three modes have significant *CP* violation asymmetries ($> 5\sigma$), which are consistent with the SM expectation with small penguin amplitude contributions ($S$ parameters are consistent with the $\sin 2\phi_1$ value from $b \to c\bar{c}s$ decays).

### 17.6.5 $\phi_1$ from $b \to c\bar{u}d$ decays

#### 17.6.5.1 $B^0 \to D^{(*)}h^0$

The decay $B^0 \to D^{(*)}h^0$, where $h^0$ is a light, unflavored neutral meson, is dominated by a $b \to c\bar{u}d$ color-suppressed tree diagram in the SM. The final state $D^{(*)}h^0$ is a *CP* eigenstate if the neutral $D$ meson decays to a *CP* eigenstate as well. In this case, the time-dependent asymmetry in $B^0$ decays is similar to that of $b \to c\bar{c}s$ decays but with a small correction from the $b \to u\bar{c}d$ amplitude. This amplitude is suppressed by $V_{ub}V_{cd}^*/V_{cb}V_{ud}^* \simeq 0.02$, and therefore the deviation is expected to be small in the SM (Fleischer, 2003a,b; Grossman and Worah, 1997). $R$-parity violating ($\not{R}_p$) supersymmetric processes (Grossman and Worah, 1997) could enter at the tree level in these decays, leading to a deviation from the SM prediction.

In *BABAR*'s analysis (Aubert, 2007ad) with $383 \times 10^6$ $B\overline{B}$ pairs, the $B^0$ meson is fully reconstructed in the following channels: $D^{(*)}\pi^0$ ($D \to K^+K^-$, $K^0_S\omega$) and $D^{(*)}\eta$ ($D \to K^+K^-$), where $D^{*0} \to D^0\pi^0$, and $D\omega$ ($D \to K^+K^-$, $K^0_S\omega$, $K^0_S\pi^0$). The $\eta$ mesons are reconstructed via $\gamma\gamma$ and $\pi^+\pi^-\pi^0$ final states, and the $\omega$ candidates are reconstructed from the $\pi^+\pi^-\pi^0$ decay mode. The event selection criteria are determined by maximizing the expected signal significance using Monte Carlo simulated signal events and simulated samples of generic $B\overline{B}$ and $e^+e^- \to q\overline{q}$ ($q = u, d, s, c$) continuum events.

Angular distributions of the $D \to K^0_S\omega$ decay mode are exploited to take advantage of the polarization in the decay. The background from continuum $q\overline{q}$ production is suppressed by a Fisher discriminant constructed using several event shape variables and angular distributions (see Chapter 9).

The signal and combinatorial background yields are determined by a fit to the $m_{\rm ES}$ distribution using a Gaussian and a threshold function (ARGUS, see Eq. (7.1.11)) for the signal and combinatorial background components, respectively. The contribution from each mode is shown in Table 17.6.5. Peaking background contributions are studied using both simulation and $D^0$ sideband data. The contributions to *CP*-even and *CP*-odd modes are $(0.8 \pm 2.6)\%$ and $(11 \pm 6)\%$, respectively.

The fit technique adopted to extract the *CP* violating parameters $S$ and $C$ is similar to that used in $b \to c\bar{c}s$ decays. The mistag parameters and the resolution function are determined from a large data control sample of $B^0 \to D^{(*)-}h^+$ decays, where $h^+$ is a $\pi^+$, $\rho^+$, or $a_1^+$ meson. An exponential decay is used to model the $\Delta t$ p.d.f. of the peaking background and accounts for possible *CP* asymmetries in the systematic uncertainty. In addition to the fit to the entire sample, fits to *CP*-even and *CP*-odd subsamples are performed to check consistency. As the SM corrections due to the sub-leading-order $b \to u\bar{c}d$ diagram are different for $D_{CP+}$ and $D_{CP-}$ (Fleischer, 2003a,b), a fit is also performed allowing different *CP* asymmetries for $D_{CP+}$ and $D_{CP-}$. The results are summarized in Table 17.6.5, and the $\Delta t$ distribution projections and the asymmetry of the events in the signal region are shown in Fig. 17.6.10. The result is consistent with the world

**Table 17.6.5.** Summary of the $B^0 \to D^{(*)0}_{CP} h^0$ analysis from *BABAR* (Aubert, 2007ad). The *CP* eigenvalue of the $D^0$ final state is indicated in the column '$D_{CP}$'.

| $\eta_f = +1$ (*CP* even) | | | $\eta_f = -1$ (*CP* odd) | | |
|---|---|---|---|---|---|
| Mode | $D_{CP}$ | $N_{\rm signal}$ | Mode | $D_{CP}$ | $N_{\rm signal}$ |
| $D^0_{K^0_S\omega}\pi^0$ | $-$ | $26.2 \pm 6.3$ | $D^0_{KK}\pi^0$ | $+$ | $104 \pm 17$ |
| $D^0_{K^0_S\pi^0}\omega$ | $-$ | $40.0 \pm 8.0$ | $D^0_{KK}\eta_{\gamma\gamma}$ | $+$ | $28.9 \pm 6.5$ |
| $D^0_{K^0_S\omega}\omega$ | $-$ | $23.2 \pm 6.8$ | $D^0_{KK}\eta_{3\pi}$ | $+$ | $14.2 \pm 4.7$ |
| $D^{*0}_{KK}\pi^0$ | $+$ | $23.2 \pm 6.3$ | $D^0_{KK}\omega$ | $+$ | $51.2 \pm 8.5$ |
| $D^{*0}_{KK}\eta_{\gamma\gamma}$ | $+$ | $9.8 \pm 3.5$ | $D^{*0}_{K^0_S\omega}\pi^0$ | $-$ | $5.5 \pm 3.3$ |
| $D^{*0}_{KK}\eta_{3\pi}$ | $+$ | $6.8 \pm 2.9$ | | | |
| Combined | | $131 \pm 16$ | | | $209 \pm 23$ |
| $\eta_f S$ | | $-0.17 \pm 0.37$ | | | $-0.82 \pm 0.28$ |
| $C$ | | $-0.21 \pm 0.25$ | | | $-0.21 \pm 0.21$ |
| $\eta_f S$ (combined) | | $-0.56 \pm 0.23 \pm 0.05$ | | | |
| $C$ (combined) | | $-0.23 \pm 0.16 \pm 0.04$ | | | |
| | $D_{CP+}$ | | | $D_{CP-}$ | |
| $\eta_f S$ | $-0.65 \pm 0.26 \pm 0.06$ | | | $-0.46 \pm 0.45 \pm 0.13$ | |
| $C$ | $-0.33 \pm 0.19 \pm 0.04$ | | | $-0.03 \pm 0.28 \pm 0.07$ | |

average of $-\sin 2\phi_1$, and is $2.3\sigma$ from the *CP*-conserving hypothesis $S = C = 0$.

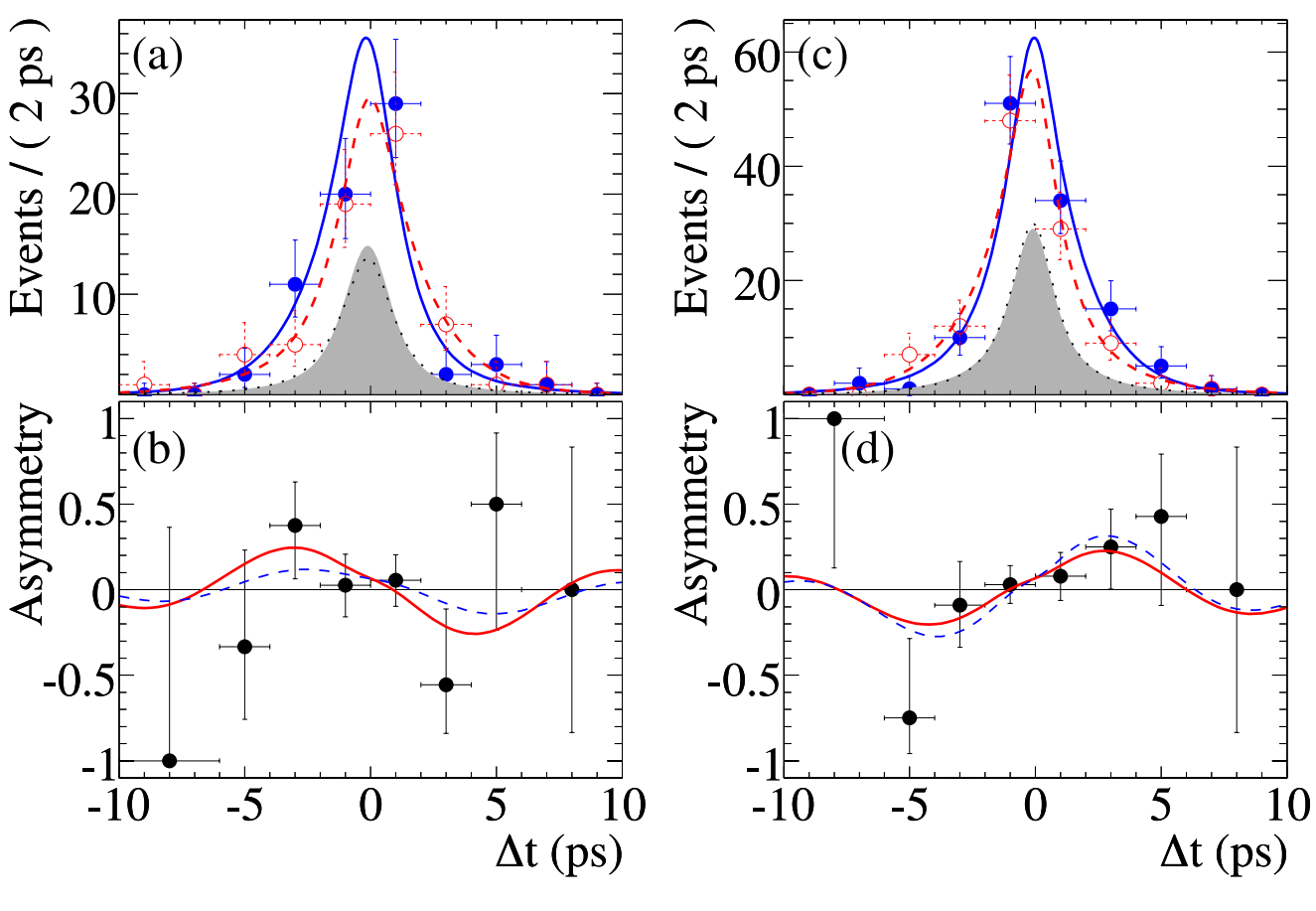

**Figure 17.6.10.** $\Delta t$ distributions and asymmetries of $B^0 \to D^{(*)0}_{CP} h^0$ candidates from *BABAR* (Aubert, 2007ad) for (a, b) *CP*-even and (c, d) *CP*-odd candidates in the signal region ($m_{\rm ES} > 5.27\,{\rm GeV}/c^2$). In (a) and (c), the solid points and curve (open circles and dashed curve) are $B^0$-tagged ($\overline{B}^0$-tagged) candidates and $\Delta t$ projection curves. Shaded areas (dotted lines) are background distributions for the $B^0$-tagged ($\overline{B}^0$-tagged) candidates. In (b) and (d), the solid curve represents the combined fit result, and the dashed curve represents the result of the fits to *CP*-even and *CP*-odd modes separately.

### 17.6.6 $\phi_1$ from charmless quasi-two-body $B$ decays

The time-dependent *CP* asymmetry parameter $S$ measured in charmless decays to *CP* eigenstates via $b \to s\bar{q}q$ penguin transitions is also equal to $S = -\eta_f \sin 2\phi_1$ in the SM. These decays are particularly sensitive to new physics because any unobserved heavy particle could contribute an additional penguin loop and alter the value of the measured weak phase. If the measured $S$ in one or a group of charmless decays deviates significantly from that in tree-dominated processes, it could be a signature of new physics effects. The comparison between loop and tree-dominated decays, however, must to be made with careful estimates of the SM corrections from higher order topologies. The key issue in the theoretical understanding of these *CP* asymmetries is the tree-to-penguin ratio, both in short- and long-distance interactions. The typical deviations in theoretical calculations are below a few percent, and the corresponding uncertainty can be as small as one or two percent. The modes that benefit from the least theoretical uncertainties are $\eta' K^0_S$, $\phi K^0_S$, and $K^0_S K^0_S K^0_S$ (Beneke, 2005; Cheng, Chua, and Soni, 2005a,b).

Of the charmless decays of interest, two-body and quasi-two-body final states are the simplest states to study experimentally. The term "quasi-two-body" refers to a final state that includes a resonance whose interference with any other amplitude is ignored (details in Section 17.4.5). The experiments at the $B$ Factories have studied the *CP*-odd states $B^0 \to \eta' K^0_S$, $\omega K^0_S$, $\pi^0 K^0_S$ and the *CP*-even states $B^0 \to \eta' K^0_L$ and $\pi^0 K^0_L$. Measurements of time-dependent asymmetries in three-body decays are discussed in Section 17.6.7.

Due to the similarity between the experimental techniques used to reconstruct the $B^0 \to \pi^0 K^0_S$ and $B^0 \to K^0_S K^0_S$ decays, the latter measurement is included in this section. The $2K^0_S$ mode is dominated by a $b \to d\bar{s}s$ penguin transition. Assuming top-quark dominance in the virtual loop, the time-dependent *CP* asymmetry parameters in this decay are expected to vanish, *i.e.* $S_{K^0_S K^0_S} = C_{K^0_S K^0_S} = 0$ (Fleischer, 1994). If a significant discrepancy is observed, this would be a clear signature of new physics (Giri and Mohanta, 2004).

Measurements of time-dependent asymmetry parameters of $B$ mesons decaying into $\eta' K^0$, $\omega K^0_S$, $\pi^0 K^0$, and $K^0_S K^0_S$ are described in the following.

#### 17.6.6.1 $B^0 \to \eta' K^0$

The branching fraction of the $B^0 \to \eta' K^0$ decay was first measured by CLEO (Behrens et al., 1998) and was surprisingly large compared to naïve expectations. This result is confirmed by both Belle (Abe, 2001d) and *BABAR* (Aubert, 2001d). Because of the large branching fraction, this mode provides the most precise time-dependent *CP* asymmetry parameter measurement of any $b \to s\bar{q}q$ decay mode. The first measurements were made in 2002 by Belle (Chen, 2002) and in 2003 by *BABAR* (Aubert, 2003i). For these measurements, the $\eta'$ candidates were reconstructed via $\eta' \to \eta\pi^+\pi^-$ and $\eta' \to \rho^0\gamma$ decays, with $\eta \to \gamma\gamma$ and $\rho^0 \to$ $\pi^+\pi^-$. Only the $B^0 \to \eta' K^0_S$ mode was considered, using $K^0_S \to \pi^+\pi^-$. The measured values of $S$ were consistent between the two experiments but the uncertainties were large. Over the years both experiments have improved the measurements method and increased the available data sample. The decays $\eta \to \pi^+\pi^-\pi^0$ and $K^0_S \to \pi^0\pi^0$ are added to the reconstructed sub-decays listed above. All the combinations of the sub-decays are used except for the $\eta' \to \pi^+\pi^-\pi^0$, $K^0_S \to \pi^0\pi^0$ combination. Belle also excludes the $\eta' \to \rho^0\gamma$, $K^0_S \to \pi^0\pi^0$ combination. A tension between these results and the SM expectation at a level of $3\sigma$ was reported by *BABAR* in 2005 (Aubert, 2005w), but was not confirmed by Belle (Chen, 2005b). In the 2007 update of the measurements (Aubert (2007am); Chen (2007a)), the decay $B^0 \to \eta' K^0_L$ with $\eta' \to \eta\pi^+\pi^-$ (both sub-decays of the $\eta$ considered) is also added. With these measurements, both experiments are able to establish the existence of *CP* violation in the $B^0 \to \eta' K^0$ mode, obtained from the combination of the $B^0 \to \eta' K^0_S$ and $B^0 \to \eta' K^0_L$ decays. This is the first observation of *CP* violation (with a significance greater than $5\sigma$) in $b \to s\bar{q}q$ transitions. These measurements are consistent with the SM expectation.

In the most recent measurements, *BABAR* and Belle use data samples of $467 \times 10^6$ and $535 \times 10^6$ $B\overline{B}$ pairs (Aubert (2009aa); Chen (2007a)), respectively. The kinematic variables used to identify $B^0$ candidates are $m_{\rm ES}$ and $\Delta E$ for $\eta' K^0_S$; $\Delta E$ (*BABAR*) or $p^*_B$ (Belle) for $\eta' K^0_L$. As with other charmless $B$ decays, the dominant background comes from $e^+e^- \to q\bar{q}$ $(q = u, d, s, c)$ continuum events. Loose cuts are applied to continuum suppression variables. These variables are also used together with the aforementioned kinematic variables in the fit to extract signals. *BABAR* uses a Fisher discriminant formed from shape variables, while Belle uses a likelihood ratio formed from a Fisher discriminant with modified Fox-Wolfram moments (see Chapter 9). The flavor tagging, vertex reconstruction, and fit procedures used to extract the *CP* asymmetry parameters are essentially the same as for $b \to c\bar{c}s$ decays. The results obtained are shown in Table 17.6.6. The time-dependent event yields and asymmetry from Belle are shown in Fig. 17.6.11. Both experiments measure asymmetry parameters consistent with results from $b \to c\bar{c}s$ decays. These measurements are limited by statistical uncertainties. Most of the systematic uncertainties are in common with the $b \to c\bar{c}s$ modes, and summarized in Section 15.3. The main contributions to the systematic uncertainty arise from the *CP* content of the $B\overline{B}$ background and the likelihood fit model used.

In the course of the book preparation the final Belle result in this mode became available, using the integrated luminosity of 711 fb$^{-1}$ (Santelj, 2013). The measurement mainly profits from the increased statistical power of the sample due to both, the increase in the luminosity as well as the reprocessing of data (see Section 3.3). The result including $K^0_S$ and $K^0_L$ final states is in agreement with the SM prediction,

$$S = 0.68 \pm 0.07 \pm 0.03$$
$$C = 0.03 \pm 0.05 \pm 0.03\ . \qquad (17.6.8)$$

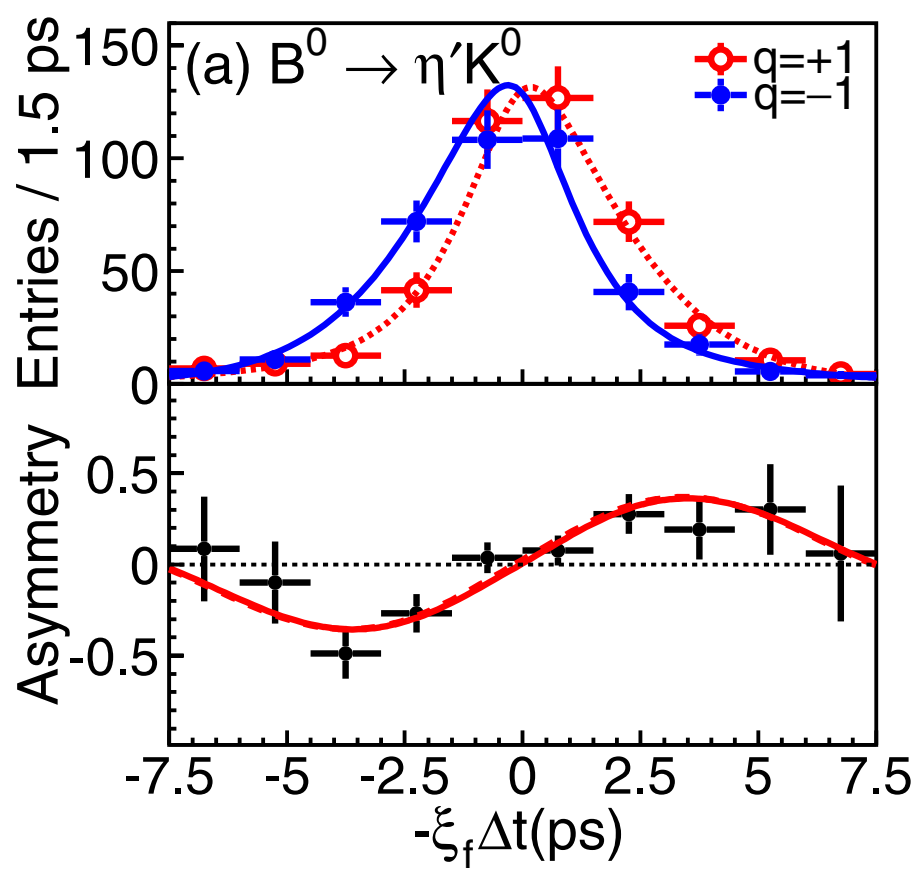

**Figure 17.6.11.** Background subtracted $\Delta t$ distributions and time-dependent asymmetry for $B^0 \to \eta' K^0$ events with a good flavor tag from Belle (Chen, 2007a).

#### 17.6.6.2 $B^0 \to \omega K^0_S$

$B^0 \to \omega K^0_S$ candidates are reconstructed via $\omega \to \pi^+\pi^-\pi^0$ and $K^0_S \to \pi^+\pi^-$ sub-decay channels. The $\omega$ candidates are selected by requiring the $\pi^+\pi^-\pi^0$ invariant mass to be within a window around the nominal mass. As for $B^0 \to \eta' K^0_S$, $m_{\rm ES}$, $\Delta E$, and continuum suppression variables are used to extract signals from background. *BABAR* also includes the invariant mass of $\pi^+\pi^-\pi^0$ and $\mathcal{H}$ in the fit to data to improve signal to background discrimination. The variable $\mathcal{H}$ is the cosine of the angle between the opposite direction of the $B$ meson and the normal to the decay plane in the $\omega$ rest frame. *BABAR* (Aubert, 2009aa) and Belle (Abe, 2007e) analyze data samples of $467 \times 10^6$ and $535 \times 10^6$ $B\overline{B}$ pairs, respectively. Results are shown in Table 17.6.6, where the number of signal events obtained is small ($\sim$ 100) and the uncertainties on $S$ and $C$ are large.

#### 17.6.6.3 $B^0 \to \pi^0 K^0$

Since the $B^0 \to \pi^0 K^0_S$ decay does not produce charged tracks at the $B^0$ decay vertex, it is experimentally challenging to perform a time-dependent analysis. The decay position is determined from the intersection of the $K^0_S$ trajectory, which is determined from the $\pi^+$ and $\pi^-$ tracks and the profile of the interaction point. *BABAR* imposes the constraint that the sum of the two $B$ decay times $(t_{CP} + t_{\rm tag})$ is equal to $2\tau_{B^0}$ with an uncertainty $\sqrt{2}\tau_{B^0}$ in order to further improve the accuracy of the reconstructed value of $\Delta t$. The $\pi^+$ and $\pi^-$ tracks are required to be well measured in the silicon vertex detector. Since $c\tau$ of a $K^0_S$ is 2.84 cm, about 60% and 30% of $K^0_S$ candidates satisfy this condition at *BABAR* and Belle, respectively. Flavor tagged signal events can contribute to the precision obtained on $C$. Events that fail to satisfy the requirement are also used in the fit with a p.d.f which is obtained by integrating the time-dependent p.d.f. with respect to $\Delta t$. *BABAR* (Aubert, 2009aa) and Belle (Fujikawa, 2010) analyze $467 \times 10^6$ and $657 \times 10^6$ $B\overline{B}$ pairs, respectively. The continuum background suppression method adopted by the two experiments is discussed in more detail in Chapter 9.

**Table 17.6.6.** Summary of time-dependent asymmetry parameter measurements for charmless two-body and quasi-two-body decays. Signal yields quoted here are for tagged and untagged events for *BABAR* and only tagged events for Belle. The $B^0 \to K_S^0 K_S^0$ mode is expected to have $S = C = 0$ in the SM, and $S = -\eta_f \sin 2\phi_1$ and $C = 0$ for the other modes.

| | *BABAR* | Belle | Average (Amhis et al., 2012) |
|---|---|---|---|
| | | $\eta' K^0$ | |
| Ref. | Aubert (2009aa) | (Chen, 2007a) | |
| Yield | $2515 \pm 69$ | $1875 \pm 60$ | |
| $-\eta_f S$ | $0.57 \pm 0.08 \pm 0.02$ | $0.64 \pm 0.10 \pm 0.04$ | $0.59 \pm 0.07$ |
| $C$ | $-0.08 \pm 0.06 \pm 0.02$ | $0.01 \pm 0.07 \pm 0.05$ | $-0.05 \pm 0.05$ |
| | | $\omega K_S^0$ | |
| Ref. | (Aubert, 2009aa) | (Abe, 2007e) | |
| Yield | $163 \pm 18$ | $118 \pm 18$ | |
| $-\eta_f S$ | $0.55^{+0.26}_{-0.29} \pm 0.02$ | $0.11 \pm 0.46 \pm 0.07$ | $0.45 \pm 0.24$ |
| $C$ | $-0.52^{+0.22}_{-0.20} \pm 0.03$ | $0.09 \pm 0.29 \pm 0.06$ | $-0.32 \pm 0.17$ |
| | $\pi^0 K_S^0$ | $\pi^0 K^0$ | |
| Ref. | (Aubert, 2009aa) | (Fujikawa, 2010) | |
| Yield | $556 \pm 32$ | $919 \pm 62$ | |
| $-\eta_f S$ | $0.55 \pm 0.20 \pm 0.03$ | $0.67 \pm 0.31 \pm 0.08$ | $0.57 \pm 0.17$ |
| $C$ | $0.13 \pm 0.13 \pm 0.03$ | $-0.14 \pm 0.13 \pm 0.06$ | $0.01 \pm 0.10$ |
| | | $K_S^0 K_S^0$ | |
| Ref. | (Aubert, 2006ai) | (Nakahama, 2008) | |
| Yield | $32 \pm 9$ | $58 \pm 11$ | |
| $S$ | $-1.28^{+0.80+0.11}_{-0.73-0.16}$ | $-0.38^{+0.69}_{-0.77} \pm 0.09$ | $-1.08 \pm 0.49$ |
| $C$ | $-0.40 \pm 0.41 \pm 0.06$ | $0.38 \pm 0.38 \pm 0.05$ | $-0.06 \pm 0.26$ |

Belle also includes $B^0 \to \pi^0 K_L^0$ decays. Here $m_{\rm ES}$ is calculated using the direction of the $K_L^0$ meson assuming that the parent $B^0$ is at rest in the CM system. The signal is extracted using $m_{\rm ES}$ and a likelihood ratio variable for continuum suppression. Since the vertex position cannot be calculated, $B^0 \to \pi^0 K_L^0$ only contributes to the determination of $C$. The signal yield obtained for the $K_L^0$ mode is $285 \pm 52$ events compared to $634 \pm 34$ for the $K_S^0$ mode.

The *CP* asymmetry parameters $S$ and $C$ are obtained by fitting the events with and without the vertex position information. The results are shown in Table 17.6.6. While the $C$ values measured by *BABAR* and Belle have opposite signs, they are consistent at the level of $\sim 1.5\sigma$.

#### 17.6.6.4 $B^0 \to K_S^0 K_S^0$

As with the $B^0 \to \pi^0 K_S^0$ case, prompt charged tracks from the $B$ vertex are absent in $B^0 \to K_S^0 K_S^0$ decays. Therefore, the study of time-dependent *CP* asymmetry parameters uses the same technique developed for $B^0 \to \pi^0 K_S^0$. In this case both charged pions from at least one of the $K_S^0$ mesons are required to have been well reconstructed using hits in the silicon vertex detector. The efficiency is approximately 82% and 61% for *BABAR* and Belle, respectively. Events in which both $K_S^0$ mesons decay outside the silicon vertex detector do not have a well reconstructed $B$ vertex; they are only used to determine $C$.

Data samples of $348 \times 10^6$ and $657 \times 10^6$ $B\overline{B}$ pairs are used for the *BABAR* (Aubert, 2006ai) and Belle (Nakahama, 2008) measurements, respectively. The suppression of the continuum background is achieved in the same way as for the $B^0 \to \pi^0 K_S^0$ measurement. The results obtained for the time-dependent asymmetry parameters are shown in Table 17.6.6. The dominant sources of systematic uncertainty are due to the fit model parameterization. These results are consistent with the SM prediction of no *CP* asymmetry in $b \to d\bar{s}s$ penguin modes.

### 17.6.7 $\phi_1$ from charmless three-body decays

Charmless three-body decays through $b \to s\bar{q}q$ penguin transitions also provide measurements of $\phi_1$. In general, three-body decays are not *CP* eigenstates and also often include intermediate resonances. These resonances complicate the extraction of useful *CP* violation parameters. However, for $B^0 \to P^0 P^0 X^0$ decays, where $P^0$ and $X^0$ are any spin-0 neutral particles, the final state has a definite *CP* eigenvalue, that of the $X^0$ (Gershon and Hazumi, 2004), regardless of intermediate states. The decay $B^0 \to K_S^0 K_S^0 K_S^0$ is of particular interest since it proceeds only through a $b \to s$ penguin transition and is free from any $b \to u$ contribution. The $\pi^0\pi^0K_S^0$ final state also has a definite *CP* (even) eigenvalue but has a $b \to u$ tree contribution, similar to $B^0 \to K_S^0\pi^0$, discussed in Section 17.6.6. As with similar loop-dominated transitions, the deviations of measured *CP* asymmetry parameters from those found in $b \to c\bar{c}s$ decays are expected to be quite small in the SM. If a large deviation were to be measured, then this could indicate the presence of new physics.

In general, analysis of the Dalitz plane for three-body decays can be used to extract the amplitude of each contribution. Time-dependent Dalitz plot analysis, therefore, can be used to extract *CP* asymmetry parameters of each intermediate two-body *CP* eigenstate and also those of any non-resonant *CP* eigenstate components (see Chapter 13). This method is applied to $B^0 \to K^+K^-K^0$ and $B^0 \to \pi^+\pi^-K^0$ decays.

#### 17.6.7.1 $B^0 \to K_S^0K_S^0K_S^0$

The first measurement of *CP* asymmetry parameters for $B^0 \to K_S^0K_S^0K_S^0$ decays is made by Belle (Sumisawa, 2005) with $275 \times 10^6$ $B\overline{B}$ pairs. The latest measurements reported by *BABAR* (Lees, 2012c) and Belle (Chen, 2007a) use $468 \times 10^6$ and $535 \times 10^6$ $B\overline{B}$ pairs, respectively.

The $K_S^0$ candidates are reconstructed in the $K_S^0 \to \pi^+\pi^-$ and $K_S^0 \to \pi^0\pi^0$ modes. $B^0 \to K_S^0K_S^0K_S^0$ decays are reconstructed with all $K_S^0$ mesons decaying into a $\pi^+\pi^-$ final state ($B_{3K_S^0(+-)}$) and also with one of the $K_S^0$ mesons decaying into a $\pi^0\pi^0$ final state ($B_{3K_S^0(00)}$). Signal is extracted by fitting the distributions of kinematic variables ($m_{\rm ES}$ and $\Delta E$) and a continuum suppression variable.

Since $B^0 \to \chi_{c0,2}K_S^0$ ($\chi_{c0,2} \to K_S^0K_S^0$) decays give the same final states but proceed through a $b \to c\bar{c}s$ transition, vetoes are applied for candidates with a $K_S^0K_S^0$ mass combination within a window around the nominal $\chi_{c0}$ mass. The contribution from $\chi_{c2}$ is found to be negligible. Belle also applies a veto based on the measured $D^0$ mass to remove the decays $B^0 \to D^0K_S^0$ ($D^0 \to K_S^0K_S^0$). In case of multiple candidates in an event, a single candidate is selected based on the reconstructed $K_S^0$ mass or the quality of a fit with a constraint on the $D^0$ mass.

The decay vertex position of the reconstructed $B$ is obtained using the trajectories of the $K_S^0$ mesons in the $\pi^+\pi^-$ channels constraining the reconstructed $K_S^0$ mesons to come from the beam spot. As is the case for $\pi^0K_S^0$ and $K_S^0K_S^0$ decays, these measurements use $K_S^0 \to \pi^+\pi^-$ candidates reconstructed from tracks that are well measured in the silicon vertex detectors.

The usual flavor tagging and fitting procedure are applied to extract the *CP* asymmetry parameters. The results obtained are summarized in Table 17.6.7. The $\Delta t$ distribution and the time-dependent asymmetry from *BABAR* is shown in Fig. 17.6.12.

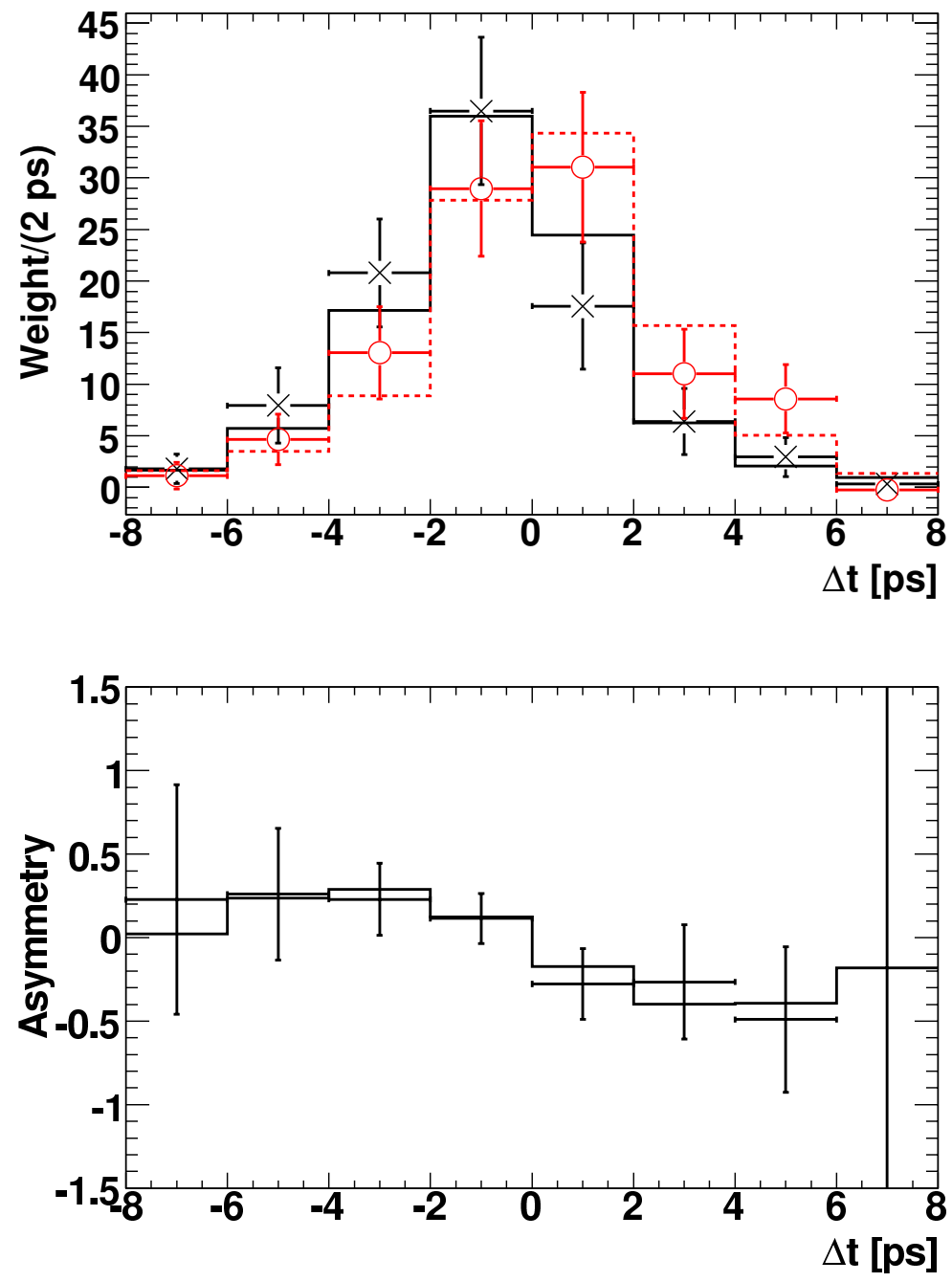

**Figure 17.6.12.** (Top) $\Delta t$ distribution and (bottom) *CP* asymmetry as a function of $\Delta t$, for the $B^0 \to K_S^0K_S^0K_S^0$ signal (points) obtained by *BABAR* (Lees, 2012c) using the $_s\mathcal{P}lot$ technique (see Section 11.2.3), superimposed on the fit results (histograms). The data points marked with crosses (circles) and solid (dashed) histograms correspond to $B^0$ ($\overline{B}^0$) tagged events.

**Table 17.6.7.** Summary of *CP* asymmetry measurements for charmless $B^0 \to K_S^0K_S^0K_S^0$ decays, (Lees, 2012c) and (Chen, 2007a). The signal yield includes both tagged and untagged events.

| | *BABAR* | Belle | Average |
|---|---|---|---|
| Signal | $263\ ^{+21}_{-19}$ | $185 \pm 17$ | |
| $S$ | $0.94\ ^{+0.24}_{-0.21} \pm 0.06$ | $0.30 \pm 0.32 \pm 0.08$ | $0.74 \pm 0.17$ |
| $C$ | $-0.17 \pm 0.18 \pm 0.04$ | $-0.31 \pm 0.20 \pm 0.07$ | $-0.23 \pm 0.13$ |

#### 17.6.7.2 $B^0 \to \pi^0\pi^0K_S^0$

The event reconstruction of $B^0 \to \pi^0\pi^0K_S^0$ is similar to that of $B^0 \to K_S^0\pi^0$ (Section 17.6.6) with an additional $\pi^0$. Even though no charged tracks come directly from the interaction point, the intersection of the $K_S^0$ trajectory and the beamspot provide adequate measurement of the $B^0$ decay vertex. Approximately 70% of the candidates at *BABAR* have well measured $K_S^0 \to \pi^+\pi^-$ tracks in the silicon vertex detector. This is higher than that in $B^0 \to K_S^0\pi^0$ because the $K_S^0$ momentum spectrum is softer in $B^0 \to \pi^0\pi^0K_S^0$ decays. *BABAR* uses a neural network (Section 4.4.4) with event shape variables to discriminate against continuum background. Events consistent with $B^0 \to K_S^0\pi^0$, $D^0\pi^0$, $\eta^{(\prime)}K_S^0$, and $\chi_{c0,2}K_S^0$ decays are vetoed. In case of multiple candidates in an event, the candidate with the smallest value of $\sum_{i=1}^{2}(m_{\gamma\gamma}^{(i)} - m_{\pi^0})^2$ is selected, where $m^{(1)}_{\gamma\gamma}$ and $m^{(2)}_{\gamma\gamma}$ are the invariant masses of the two $\pi^0 \to \gamma\gamma$ candidates.

The fit uses $m_{\rm ES}$, $\Delta E/\sigma(\Delta E)$, the neural-network output, $\Delta t$, $\sigma(\Delta t)$, and flavor tagging as variables. Using a sample of $227 \times 10^6$ $B\overline{B}$ pairs, *BABAR* (Aubert, 2007t) finds $117 \pm 27$ signal events, and $S = 0.72 \pm 0.71 \pm 0.08$ and $C = 0.23 \pm 0.52 \pm 0.13$. Belle has not measured this channel.

#### 17.6.7.3 $B^0 \to K^+K^-K^0$ time-dependent Dalitz plot analysis

$B^0 \to \phi K^0$ decays proceed almost purely through a $b \to s\bar{s}s$ penguin transition. This is one of the most promising modes to search for new physics. In general, the decays $B^0 \to K^+K^-K^0$ have a contribution from the $b \to u\bar{u}s$ tree transition. Therefore, its theoretical uncertainty must be taken into account when comparing asymmetry parameter results with those obtained from charmonium decays.

The measurements were originally made treating this decay in terms of the quasi-two-body process $B^0 \to \phi K^0$, where $\phi \to K^+K^-$ (Abe (2003e,f); Chen (2007a); Aubert (2004o, 2005m)). Other analyses measuring *CP* asymmetry parameters in $B^0 \to K^+K^-K^0_S$ decays excluded the $\phi$ mass region in the $K^+K^-$ invariant mass spectrum (Abe, 2003e,f, 2007e) (Aubert, 2005m), where they found the phase space was dominantly *CP*-even.

There can be nonresonant $B^0 \to K^+K^-K^0$ contributions and also $B^0 \to f_0(980)K^0$ that may interfere with the $B^0 \to \phi K^0$ decay. Therefore, the measurement of $B^0 \to \phi K^0$ as a quasi-two-body decay would ultimately have limited precision. This problem can be resolved via the use of a time-dependent amplitude analysis of the three-body final state. The amplitudes and time-dependent asymmetry parameters can be extracted for each intermediate state (including any nonresonant component) while simultaneously accounting for interference between amplitudes as discussed in Chapter 13. With increasingly large data samples this became feasible, and such a measurement was first made by *BABAR* (Aubert, 2007af) using $383 \times 10^6$ $B\overline{B}$ pairs. The latest measurements are made using $470 \times 10^6$ $B\overline{B}$ pairs by *BABAR* (Lees, 2012y) and $657 \times 10^6$ $B\overline{B}$ pairs by Belle (Nakahama, 2010).

The $B^0 \to K^+K^-K^0$ decays are reconstructed in $K^0_S \to \pi^+\pi^-$ and $K^0_S \to \pi^0\pi^0$ channels (in the first *BABAR* measurement, the $K^+K^-K^0_L$ channel was also used). Belle uses only $K^0_S \to \pi^+\pi^-$ decay. Signal components are extracted using kinematic variables ($m_{\rm ES}$, $\Delta E$) and an $e^+e^- \to q\bar{q}$ continuum suppression variable (a Fisher discriminant, a neural network, or a flavor-tagging quality, see Chapter 9).

In the Dalitz plot analysis, each amplitude of an intermediate resonant or nonresonant state $r$ (called "isobar") is parameterized as

$$a_r = c_r(1+b_r)e^{i(\phi_r+\delta_r)}, \quad \bar{a}_r = c_r(1-b_r)e^{i(\phi_r-\delta_r)}, \tag{17.6.9}$$

for $B^0$ and $\overline{B}^0$ decays respectively, where $c_r$ is the magnitude of the amplitude. Only weak phase of the two amplitudes is written in the above equation, and the *CP* violating weak phase difference is $2\delta_r$. The magnitudes of $B^0$ and $\overline{B}^0$ decay amplitudes are also allowed to be different, parameterized by $b_r$. With this parameterization and following Eqs (10.2.4 and 10.2.5), the direct *CP* asymmetry, effective phase $\phi_1^{\rm eff}$, and time-dependent *CP* coefficient are given, respectively, as

$$C_r \approx -A^r_{CP} = -\frac{|\bar{a}_r|^2 - |a_r|^2}{|\bar{a}_r|^2 + |a_r|^2} = \frac{2b_r}{1+b_r^2}, \tag{17.6.10}$$

$$\phi_1^{{\rm eff},r} = \phi_1 + \delta_r, \tag{17.6.11}$$

$$-\eta_r S_r \approx \frac{1-b_r^2}{1+b_r^2}\sin[2\phi_1^{{\rm eff},r}]. \tag{17.6.12}$$

The measured phase is referred to as "effective" because one measures $\phi_1$ up to theoretical uncertainties related to higher order contributions, which can be significant.

Belle vetoes events consistent with a $B^0$ decaying into the following final states using appropriate mass windows: $\overline{D}^0 K^0_S$, $D^-_{(s)}K^+$, and $J/\psi K^0_S$, where $\overline{D}^0 \to K^+K^-$, $K^+\pi^-$, $D^- \to K^0_S K^-$, $K^0_S\pi^-$, $D^-_s \to K^0_S K^-$, and $J/\psi \to K^+K^-$. The $B^0 \to \chi_{c0}K^0_S$ amplitude is included in the fit. On the other hand, *BABAR* includes $B^0 \to J/\psi K^0_S$, $D^-K^+$, $D^-_s K^+$, and $\overline{D}^0 K^0_S$ as background components in the fit. The latest *BABAR* analysis finds $1419 \pm 43$ $K^+K^-K^0_S[\pi^+\pi^-]$ signal events and $160 \pm 17$ $K^+K^-K^0_S[\pi^0\pi^0]$ signal events. Belle obtains $1176 \pm 51$ signal events.

Both experiments perform a time-dependent fit to the whole Dalitz plane, using three sets of $\phi_1^{\rm eff}$ and $A_{CP}$ parameters; the first two are for $\phi(1020)K^0_S$ and $f_0(980)K^0_S$, and the third is shared by all the other charmless isobars.

Due to the possible presence of multiple solutions, the same fit is performed many times with different starting parameter values to ensure the global minimum of the likelihood is reached. Scans of log-likelihood values are done to study the behavior of the p.d.f. near the minimum and the statistical uncertainties. The latest *BABAR* analysis finds five local minima within 9 units in $-2\ln\mathcal{L}$; the second solution is 3.9 larger than the global minimum. Belle finds four solutions, separated by approximately 10 units in $-2\ln\mathcal{L}$; Solution 1 is taken as the preferred one based on external information though it has the second lowest $-2\ln\mathcal{L}$ value, which is 3.1 units larger than the lowest one (Solution 2). The results are summarized in Table 17.6.8.

It should be noted that the discrete ambiguities on the value of $\phi_1$ can be resolved using the time-dependent Dalitz plot fit method because the log-likelihood values can be compared for multiple solutions. In both $B^0 \to \phi K^0_S$ and $f_0(980)K^0_S$ decays the $\phi_1^{\rm eff} < \pi/2$ solution is clearly preferred. *BABAR* excludes the $\pi/2 - \phi_1^{\rm eff}$ value at 4.8 standard deviations.

#### 17.6.7.4 $B^0 \to \pi^+\pi^-K^0_S$ time-dependent Dalitz plot analysis

The decay $B^0 \to \pi^+\pi^-K^0_S$ includes transitions via $B^0 \to \rho^0 K^0_S$, $B^0 \to f_0(980)K^0_S$, and $B^0 \to K^{*+}\pi^-$. The measurements of time-dependent asymmetry parameters for the first two of these decays were initially made using a quasi-two-body approach (Aubert (2007aa); Abe (2007e)), similar to the $B^0 \to K^+K^-K^0_S$ ($\phi K^0$) case above. Observation of direct *CP* asymmetry in $B^0 \to K^+\pi^-$ and evidence of *CP* asymmetry in resonances in other similar three-body decays such as $B^+ \to K^+\pi^+\pi^-$ (see Chapter 17.4) suggest possible large *CP* asymmetry in resonances in $B^0 \to K^0_S\pi^+\pi^-$ decays. Time-dependent *CP* asymmetry measurements of $B^0 \to \pi^+\pi^-K^0_S$ may shed light on the $A_{CP}(K\pi)$ puzzle together with the *CP* asymmetry of other $B \to K^*\pi$ decays (see for example (Li and Mishima, 2011)). In addition, the phase difference between $B^0 \to K^{*+}\pi^-$ and $\overline{B}^0 \to K^{*-}\pi^+$ decays can be used to determine $\phi_3$ (Ciuchini, Pierini, and Silvestrini, 2006; Deshpande, Sinha, and Sinha, 2003; Gronau, Pirjol, Soni, and Zupan, 2007). Time-dependent Dalitz plot analysis of $B^0 \to \pi^+\pi^-K^0_S$ decays can provide all these measurements simultaneously.

**Table 17.6.8.** Results for time-dependent asymmetry parameters for $B^0 \to K^+K^-K^0$ decays. The three uncertainties are statistical, systematic and Dalitz plot model uncertainty (for BABAR the latter is included in the systematic uncertainty). The solutions with the (three) smallest $-2\ln\mathcal{L}$ value(s) are shown for BABAR (Belle).

| | BABAR (Lees, 2012y) | Belle (Nakahama, 2010) | | |
|---|---|---|---|---|
| | Solution 1 | Solution 1 | Solution 2 | Solution 3 |
| $A_{CP}(\phi K^0_S)$ | $-0.05 \pm 0.18 \pm 0.05$ | $+0.04 \pm 0.20 \pm 0.10 \pm 0.02$ | $+0.08 \pm 0.18 \pm 0.10 \pm 0.03$ | $-0.01 \pm 0.20 \pm 0.11 \pm 0.02$ |
| $\phi_1^{\rm eff}(\phi K^0_S)$ | $(21 \pm 6 \pm 2)^\circ$ | $(32.2 \pm 9.0 \pm 2.6 \pm 1.4)^\circ$ | $(26.2 \pm 8.8 \pm 2.7 \pm 1.2)^\circ$ | $(27.3 \pm 8.6 \pm 2.8 \pm 1.3)^\circ$ |
| $A_{CP}(f_0(980)K^0_S)$ | $-0.28 \pm 0.24 \pm 0.9$ | $-0.30 \pm 0.29 \pm 0.11 \pm 0.09$ | $-0.20 \pm 0.15 \pm 0.08 \pm 0.05$ | $+0.02 \pm 0.21 \pm 0.09 \pm 0.09$ |
| $\phi_1^{\rm eff}(f_0(980)K^0_S)$ | $(18 \pm 6 \pm 4)^\circ$ | $(31.3 \pm 9.0 \pm 3.4 \pm 4.0)^\circ$ | $(26.1 \pm 7.0 \pm 2.4 \pm 2.5)^\circ$ | $(25.6 \pm 7.6 \pm 2.9 \pm 0.8)^\circ$ |
| $A_{CP}$(others) | $-0.02 \pm 0.09 \pm 0.03$ | $-0.14 \pm 0.11 \pm 0.08 \pm 0.03$ | $-0.06 \pm 0.15 \pm 0.08 \pm 0.04$ | $-0.03 \pm 0.09 \pm 0.08 \pm 0.03$ |
| $\phi_1^{\rm eff}$(others) | $(20.3 \pm 4.3 \pm 1.2)^\circ$ | $(24.9 \pm 6.4 \pm 2.1 \pm 2.5)^\circ$ | $(29.8 \pm 6.6 \pm 2.1 \pm 1.1)^\circ$ | $(26.2 \pm 5.9 \pm 2.3 \pm 1.5)^\circ$ |

*BABAR* (Aubert, 2009av) analyzes a sample of $383\times10^6$ $B\overline{B}$ pairs and Belle (Dalseno, 2009) analyzes $657\times10^6$ $B\overline{B}$ pairs. The $B^0 \to \pi^+\pi^-K^0_S$ candidates are identified using the kinematic variables $m_{\rm ES}$ and $\Delta E$. The $e^+e^- \to q\bar{q}$ continuum background is suppressed by a loose requirement on the continuum suppression variable. This requirement retains about 90% of the signal. *BABAR* uses the neural-network output from various shape parameters while Belle uses a likelihood ratio. Belle applies vetoes for $B^0 \to D^-\pi^+$ decays and $B^0 \to (c\bar{c})K^0_S$ decays, while *BABAR* includes them as a background in the fit. Belle finds that 20–30% of events have multiple candidates in quasi-two-body modes. By selecting the $B$ candidate with $m_{\rm ES}$ closest to the nominal $B$ mass, the fraction of misreconstructed events is reduced to the level of a few percent. The signal yield is found to be $1944 \pm 98$ events using the $\Delta E$ distribution. *BABAR* finds that 1–8% of the events have multiple candidates and selects single events randomly The fraction of misreconstructed candidates is 4–8% depending on the intermediate states. The signal yield is extracted using $m_{\rm ES}$, $\Delta E$, and neural-network output information; $2182 \pm 64$ signal events are obtained.

Both groups use square Dalitz plot variables (Section 13.4.1) in the fit. The phase difference for flavor specific decays is given as

$$\Delta\phi_r = 2\delta_r. \tag{17.6.13}$$

As in $B^0 \to K^+K^-K^0_S$ decays, the fits lead to multiple solutions: two for *BABAR* and four for Belle. Table 17.6.9 shows the two most likely solutions in each experiment. *CP* violation parameters for the $B^0 \to f_0(980)K^0_S$ and $\rho^0(770)K^0_S$ decays are similar for two solutions in the Belle result, while they differ in the *BABAR* measurement (note that the statistical uncertainties between different solutions are correlated). In both cases, the $\phi_1^{\rm eff}$ values are consistent with the value of $\phi_1$ measured in $b \to c\bar{c}s$ decays.

#### 17.6.7.5 Summary of $\phi_1$ from charmless decays

Figure 17.6.13 (17.6.14) shows a summary of measurements of $\sin 2\phi_1^{\rm eff}$ (vs. $C$) from charmless decays including both quasi-two-body and three-body decays. The favored solutions are shown for $B^0 \to K^+K^-K^0$ and $B^0 \to \pi^+\pi^-K^0$ decays.

The measured $\sin 2\phi_1^{\rm eff}$ values for all of the individual modes are consistent with the $\sin 2\phi_1$ value measured in $b \to c\bar{c}s$ decays within statistical and theoretical uncertainties. However, the current statistical precision is not enough to draw definite conclusions about the presence of new physics; a much larger data sample is necessary.

### 17.6.8 Resolving discrete ambiguities in $\phi_1$

Since the time-dependent *CP* asymmetry parameter measurements described so far usually provide a value for $\sin 2\phi_1$, there is a four-fold ambiguity on the angle, $\phi_1 \to \pi/2 - \phi_1$, $\phi_1 + \pi$ and $3\pi/2 - \phi_1$. As mentioned in Section 17.6.7, time-dependent Dalitz plot analyses of charmless three-body decays measure (effective) values of $2\phi_1$, rather than $\sin 2\phi_1$, and can resolve the $\phi_1 \to \pi/2 - \phi_1$ ambiguity. However, charmless decays are dominated by penguin transitions, which can be affected by NP entering in loops. Resolving the ambiguity using decays dominated by a $b \to c$ tree transition can avoid such complication Several tree level $b \to c$ measurements are possible, and those performed at the $B$ Factories are described in the following.

**Table 17.6.9.** Results of *CP* asymmetry parameters for $B^0 \to \pi^+\pi^- K^0$ decays. The first uncertainty is statistical, the second is systematic, and the third represents the Dalitz plot signal model dependence.

| | BaBar (Aubert, 2009av) | | Belle (Dalseno, 2009) | |
|---|---|---|---|---|
| | Solution 1 | Solution 2 | Solution 1 | Solution 2 |
| $A_{CP}(f_0(980)K^0_S)$ | $-0.08 \pm 0.19 \pm 0.03 \pm 0.04$ | $-0.23 \pm 0.19 \pm 0.03 \pm 0.04$ | $-0.06 \pm 0.17 \pm 0.07 \pm 0.09$ | $+0.00 \pm 0.17 \pm 0.06 \pm 0.09$ |
| $\phi_1^{\rm eff}(f_0(980)K^0_S)[^\circ]$ | $36.0 \pm 9.8 \pm 2.1 \pm 2.1$ | $56.2 \pm 10.4 \pm 2.1 \pm 2.1$ | $12.7\ ^{+6.9}_{-6.5} \pm 2.8 \pm 3.3$ | $14.8\ ^{+7.3}_{-6.7} \pm 2.7 \pm 3.3$ |
| Fraction [%] | $13.8\ ^{+1.5}_{-1.4} \pm 0.8 \pm 0.6$ | $13.5\ ^{-1.4}_{-1.3} \pm 0.8 \pm 0.6$ | $14.3 \pm 2.7$ | $14.9 \pm 3.3$ |
| $A_{CP}(\rho^0(770)K^0_S)$ | $0.05 \pm 0.26 \pm 0.10 \pm 0.03$ | $0.14 \pm 0.26 \pm 0.10 \pm 0:03$ | $+0.03\ ^{+0.23}_{-0.24} \pm 0.11 \pm 0.10$ | $-0.16 \pm 0.24 \pm 0.12 \pm 0.10$ |
| $\phi_1^{\rm eff}(\rho^0(770)K^0_S)[^\circ]$ | $10.2 \pm 8.9 \pm 3.0 \pm 1.9$ | $33.4 \pm 10.4 \pm 3.0 \pm 1.9$ | $+20.0\ ^{+8.6}_{-8.5} \pm 3.2 \pm 3.5$ | $+22.8 \pm 7.5 \pm 3.3 \pm 3.5$ |
| Fraction [%] | $8.6\ ^{+1.4}_{-1.3} \pm 0.5 \pm 0.2$ | $8.5\ ^{+1.3}_{-1.2} \pm 0.5 \pm 0.2$ | $6.1 \pm 1.5$ | $8.5 \pm 2.6$ |
| $A_{CP}(K^{*-}\pi^+)$ | $-0.21 \pm 0.10 \pm 0.01 \pm 0.02$ | $-0.19\ ^{+0.10}_{-0.11} \pm 0.01 \pm 0.02$ | $-0.21 \pm 0.11 \pm 0.05 \pm 0.05$ | $-0.20 \pm 0.11 \pm 0.05 \pm 0.05$ |
| $\Delta\phi(K^{*-}\pi^+)[^\circ]$ | $72.2 \pm 24.6 \pm 4.1 \pm 4.4$ | $-175.1 \pm 22.6 \pm 4.1 \pm 4.4$ | $-0.7\ ^{+23.5}_{-22.8} \pm 11.0 \pm 17.6$ | $+14.6\ ^{+19.4}_{-20.3} \pm 11.0 \pm 17.6$ |
| Fraction [%] | $45.2 \pm 2.3 \pm 1.9 \pm 0.9$ | $46.1 \pm 2.4 \pm 1.9 \pm 0.9$ | $9.3 \pm 0.8$ | $9.0 \pm 1.3$ |

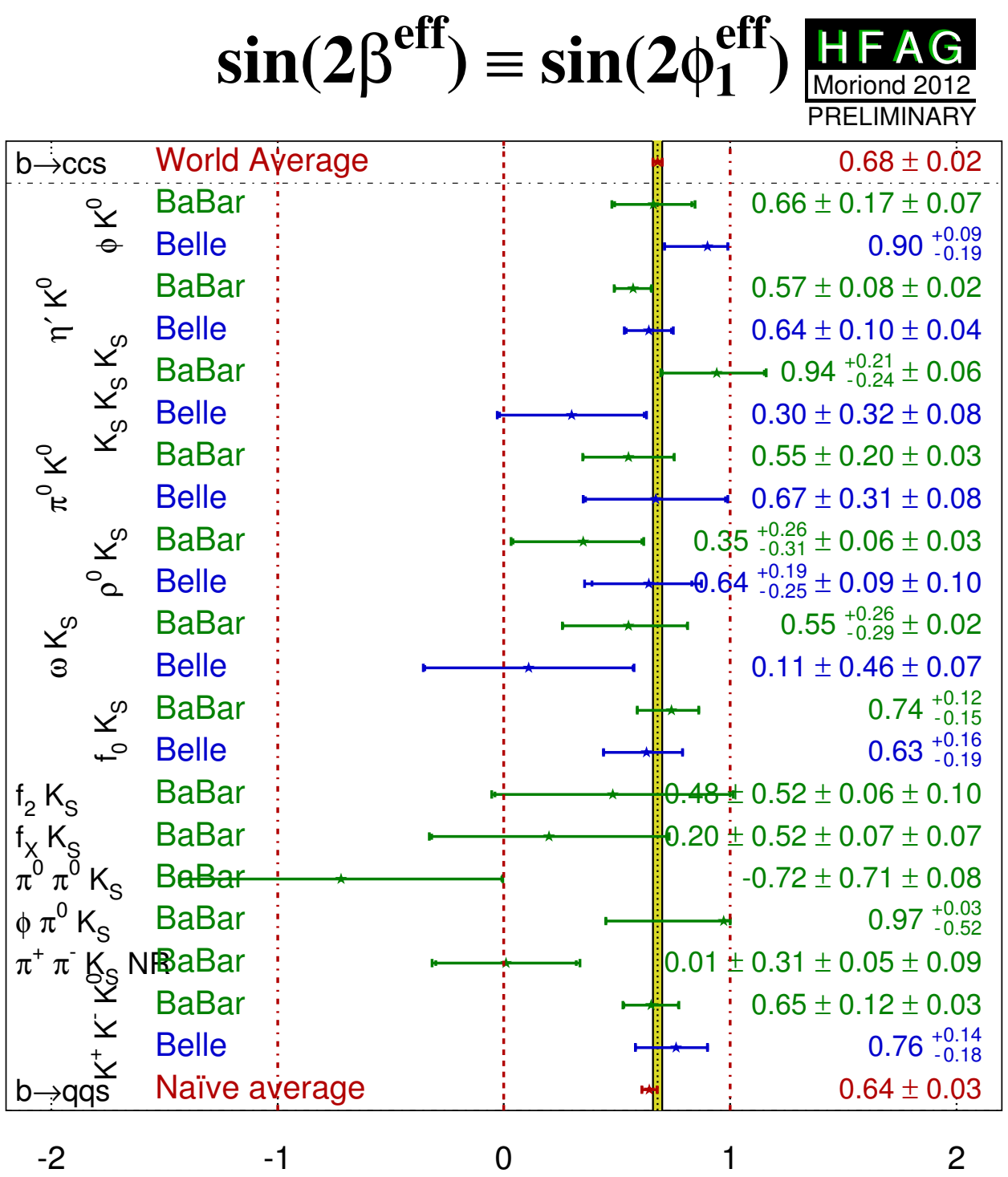

**Figure 17.6.13.** Summary of $\sin 2\phi_1^{\rm eff}$ measurements from charmless $B^0$ decays (Amhis et al., 2012).

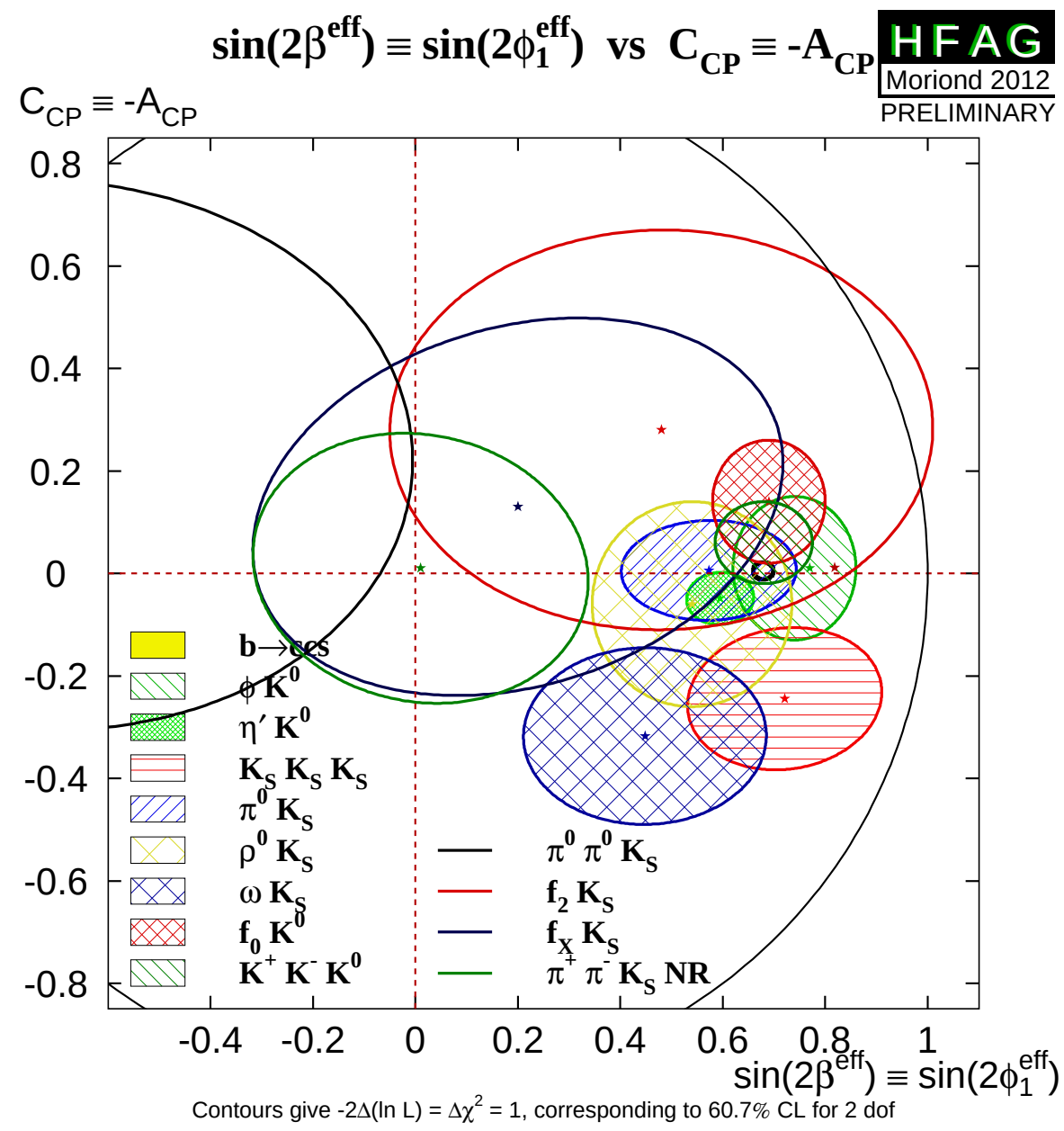

**Figure 17.6.14.** Summary of $C$ vs. $\sin 2\phi_1^{\rm eff}$ measurements from charmless $B^0$ decays (Amhis et al., 2012).

### 17.6.8.1 Time-dependent angular analysis in $B^0 \to J/\psi K^{*0}$

There are two classes of parameters obtained through the angular analysis of the $B$ meson decay to the two vector mesons $J/\psi$ and $K^{*0}$. The first is the measurement of the decay amplitudes of the three angular states. These can be obtained using a time-integrated angular analysis to flavor-specific decays (see Chapter 12). The second class comprises the *CP* parameters ($\sin 2\phi_1$ and $\cos 2\phi_1$) that are measured through a time-dependent angular analysis. In particular, the measurement of $\cos 2\phi_1$, which appears in the time-dependent interference terms (Eq. (12.1.9)), is important both to solve the two-fold ambiguity in $2\phi_1$ and to test the consistency of this determination with the more precise value from other $b \to c\bar{c}s$ decays.

The decay of a pseudo-scalar to vector-vector final state can be described with three angles defined in the transversity basis (Dunietz, Quinn, Snyder, Toki, and Lipkin, 1991), where the three amplitudes, $A_0$, $A_\parallel$, and $A_\perp$ have well-defined *CP* eigenvalues. The amplitudes are determined by a time-integrated angular analysis of $B^0 \to$ $J/\psi K^{*0}[K^+\pi^-]$ and $B^+ \to J/\psi K^{*+}[K^0_S\pi^+, K^+\pi^0]$ decays. Belle (Itoh, 2005b) and *BABAR* (Aubert, 2007x) analyze the data samples of 275 and $232 \times 10^6$ $B\overline{B}$ pairs, respectively.

**Table 17.6.10.** Measured decay amplitudes for $B^0 \to J/\psi K^{*0}$. The first uncertainty is statistical, and the second is systematic.

| | BABAR (Aubert, 2007x) | Belle (Itoh, 2005b) |
|---|---|---|
| $\lvert A_0\rvert^2$ | $0.556 \pm 0.009 \pm 0.010$ | $0.574 \pm 0.012 \pm 0.009$ |
| $\lvert A_\parallel\rvert^2$ | $0.211 \pm 0.010 \pm 0.006$ | $0.231 \pm 0.012 \pm 0.008$ |
| $\lvert A_\perp\rvert^2$ | $0.233 \pm 0.010 \pm 0.005$ | $0.195 \pm 0.012 \pm 0.008$ |
| $\arg(A_\parallel)$ | $-2.93 \pm 0.08 \pm 0.04$ | $-2.89 \pm 0.09 \pm 0.01$ |
| $\arg(A_\perp)$ | $2.91 \pm 0.05 \pm 0.03$ | $2.94 \pm 0.06 \pm 0.01$ |

Figure 17.6.15 shows the projected angular distributions for $B^0 \to J/\psi K^{*0}$ decays, where $K^{*0} \to K^+\pi^-$, from Belle. The decay amplitudes determined from the fit are summarized in Table 17.6.10.

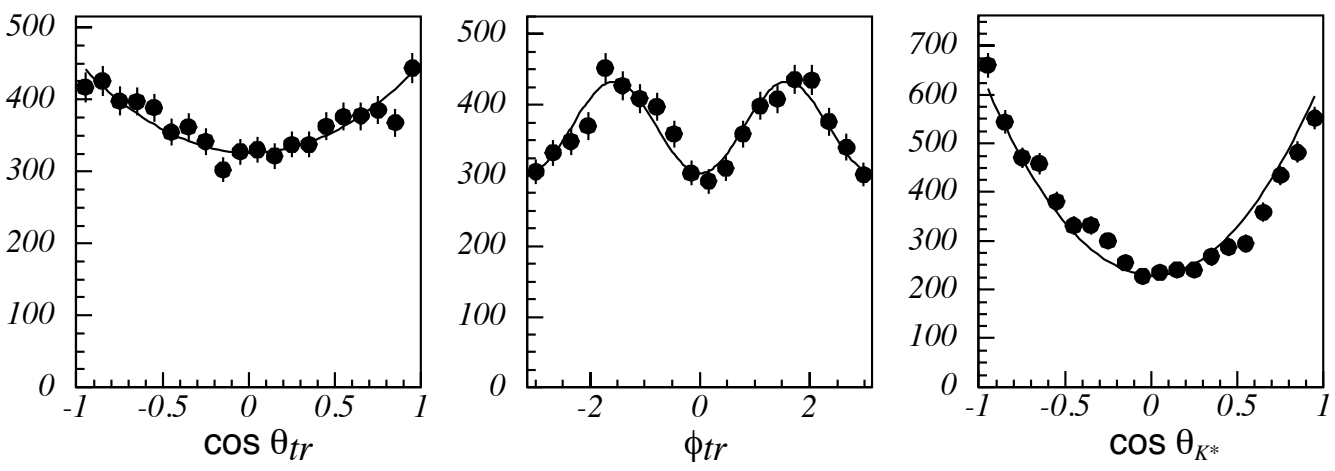

**Figure 17.6.15.** Angular distributions of $B^0 \to J/\psi K^{*0}(K^+\pi^-)$, as obtained by Belle (Itoh, 2005b). The angles are defined in Eq. (12.2.6), where $\theta_1 = \theta_{K^*}$. The curves show the fit results.

There is a two-fold ambiguity in the choice of the phases. *BABAR* resolves this ambiguity by extending the formalism to include a $K\pi$ $S$-wave amplitude and then measuring the $K\pi$ invariant mass dependence of its phase difference with respect to the dominant $K^*(892)$ $P$-wave around its mass peak (Aubert, 2005c). The result agrees with the prediction where the $s$-quark helicity is conserved as predicted by Suzuki (Suzuki, 2001). Belle adopts this choice in their analysis as well. The phases shown in Table 17.6.10 are given for this choice.

The values of $\sin 2\phi_1$ and $\cos 2\phi_1$ are determined by the time-dependent angular analysis of the decays to the $CP$ eigenstate $B^0 \to J/\psi K^{*0}$, where $K^{*0} \to K^0_S\pi^0$, from the same data set of $275 \times 10^6$ $B\overline{B}$ pairs by Belle (Itoh, 2005b), and a sample of $88 \times 10^6$ $B\overline{B}$ pairs by *BABAR* (Aubert, 2005c). The $P$-wave amplitudes are fixed to the results obtained from the time-independent analysis of the flavor-definite final states described above. Figure 17.6.16 shows the $\Delta t$ distributions for $B^0$ and $\overline{B}^0$ tags and the raw asymmetry between them by *BABAR*. Since $\sin 2\phi_1$ and $\cos 2\phi_1$ are independent parameters in the analysis, they can be obtained simultaneously using a fit. However, since the precision of the $\sin 2\phi_1$ measurement using only $B^0 \to J/\psi K^{*0}$ decays is limited by statistics, the value of $\cos 2\phi_1$ is also obtained by fixing $\sin 2\phi_1$ to the world average at that time, 0.726 (Belle) or 0.731 (*BABAR*). The results are summarized in Table 17.6.11.

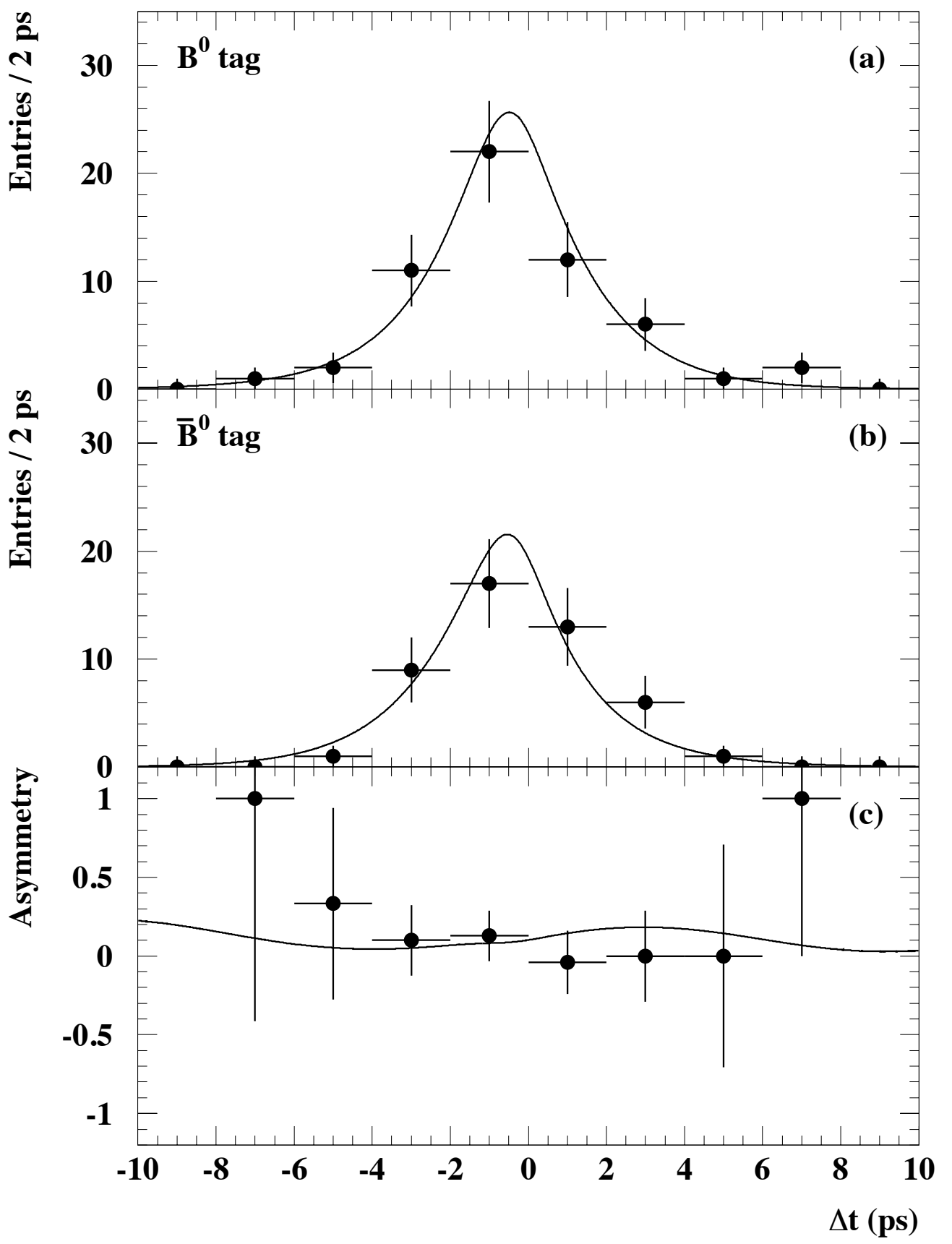

**Figure 17.6.16.** $\Delta t$ distributions for (a) $B^0$ and (b) $\overline{B}^0$ tagged $B^0 \to J/\psi K^{*0}$ events, and (c) raw asymmetry between them by BABAR (Aubert, 2005c).

The sign of $\cos 2\phi_1$ is positive in both measurements, which is consistent with the value of $\phi_1$ predicted by global CKM fits obtained using other measurements (see Section 25.1).

### 17.6.8.2 Time-dependent Dalitz analysis in $B^0 \to D^{(*)0}[K^0_S\pi^+\pi^-]h^0$

Another method to resolve discrete ambiguities uses a time-dependent Dalitz plot analysis with $B^0 \to D^{(*)0}h^0$, $D^0 \to K^0_S\pi^+\pi^-$ decays, where $h^0$ is a light neutral meson, such as $\pi^0$, $\eta$, $\eta'$, and $\omega$ (Bondar, Gershon, and Krokovny, 2005). As described in Section 17.6.5, the $B^0 \to D^{(*)0}h^0$ decay is dominated by a color-suppressed $b \to c\overline{u}d$ tree amplitude. Neglecting a small contribution from $b \to u\overline{c}d$, the decay amplitude for $B^0 \to \overline{D}^0[K^0_S\pi^+\pi^-]h^0$ can be factorized as $A_f = A_B A_{\overline{D}^0}$ and for $\overline{B}^0$ as $\overline{A}_f = A_{\overline{B}} A_{D^0}$, where $A_{D^0} = f(m_+^2, m_-^2)$ and $A_{\overline{D}^0} = f(m_-^2, m_+^2)$ with $m_\pm^2 = M^2_{K_S^0\pi^\pm}$. The $\Delta t$ distribution is given as

$$f_\pm(\Delta t) \propto \frac{e^{-|\Delta t|/\tau_{B^0}}}{2}|A_B|^2[(|A_{\overline{D}^0}|^2 + |\lambda|^2|A_{D^0}|^2) \quad (17.6.14)$$
$$\mp(|A_{\overline{D}^0}|^2 - |\lambda|^2|A_{D^0}|^2)\cos(\Delta m_d \Delta t)$$
$$\pm 2|\lambda|\xi_{h^0}(-1)^L \mathrm{Im}(e^{-2\phi_1}A_{D^0}A^*_{\overline{D}^0})\sin(\Delta m_d \Delta t)],$$

where $\xi_{h^0}$ is the *CP* eigenvalue of $h^0$ and $L$ is the orbital angular momentum of the $Dh^0$ system. One notices that if $A_{D^0} = A_{\overline{D}^0}$ (*e.g.*, if the final state is a *CP* eigenstate), this equation reduces to Eq. (10.2.2). An additional factor of $-1$ is required in the $\sin(\Delta m_d \Delta t)$ term for the $D^{*0}[D^0\pi^0]$ mode to take into account the *CP* eigenvalue of the $\pi^0$. The $\sin(\Delta m_d \Delta t)$ term can be written as

$$\mathrm{Im}(e^{-2\phi_1}A_{D^0}A^*_{\overline{D}^0}) = \mathrm{Im}(A_{D^0}A^*_{\overline{D}^0})\cos 2\phi_1 \quad (17.6.15)$$
$$-\mathrm{Re}(A_{D^0}A^*_{\overline{D}^0})\sin 2\phi_1.$$

**Table 17.6.11.** $\sin 2\phi_1$ and $\cos 2\phi_1$ determined for $B^0 \to J/\psi K^{*0}, K^{*0} \to K_S^0\pi^0$. The first two numbers show the result of the simultaneous fit with both $\sin 2\phi_1$ and $\cos 2\phi_1$ treated as free parameters. The final set of values for $\cos 2\phi_1$ are obtained with $\sin 2\phi_1$ fixed at the world average at the time of the analysis.

| | *BABAR* (Aubert, 2005c) | Belle (Itoh, 2005b) |
|---|---|---|
| $\sin 2\phi_1$ | $-0.10 \pm 0.57 \pm 0.14$ | $+0.24 \pm 0.31 \pm 0.05$ |
| $\cos 2\phi_1$ | $+3.32^{+0.76}_{-0.96} \pm 0.27$ | $+0.56 \pm 0.79 \pm 0.11$ |
| $\cos 2\phi_1$ (fixed value of $\sin 2\phi_1$) | $+2.72^{+0.50}_{-0.79} \pm 0.27$ (0.731) | $+0.87 \pm 0.74 \pm 0.12$ (0.726) |

Therefore, $\cos 2\phi_1$ and $\sin 2\phi_1$ can be independently determined by fitting the time-dependent Dalitz plot distribution.

Belle (Krokovny, 2006) and *BABAR* (Aubert, 2007s) perform the measurements using $386 \times 10^6$ and $383 \times 10^6$ $B\overline{B}$ pairs, respectively. They use $D\pi^0$, $D\eta$, $D\omega$, $D^*\pi^0$, and $D^*\eta$ decay modes. *BABAR* also uses $D\eta'$. The reconstruction includes the decay chains $D^{*0} \to D^0\pi^0$, $D^0 \to K_S^0\pi^+\pi^-$, $K_S^0 \to \pi^+\pi^-$, $\eta \to \gamma\gamma$ and $\pi^+\pi^-\pi^0$, $\eta' \to \eta\pi^+\pi^-$, and $\omega \to \pi^+\pi^-\pi^0$. The $B^0$ signal candidates are identified by $m_{\rm ES}$ and $\Delta E$. The reconstruction of the tag-side $B$ meson and flavor tagging are performed in the same way as other time-dependent *CP* asymmetry measurements.

The parameters $\sin 2\phi_1$ and $\cos 2\phi_1$ are obtained by fitting the Dalitz plot ($m_+^2$, $m_-^2$) and $\Delta t$ distributions for the events in the signal region in $m_{\rm ES}$ and $\Delta E$. The isobar model described in Chapter 13 is used for the $D^0 \to K_S^0\pi^+\pi^-$ decay amplitude. The results are summarized in Table 17.6.12. Belle fixes $|\lambda|$ to unity as expected in the SM, while *BABAR* measures $|\lambda| = 1.01 \pm 0.08(\mathrm{stat.}) \pm 0.02(\mathrm{syst.})$. Belle and *BABAR* determine the sign of $\cos 2\phi_1$ to be positive at 98.3% and 86% C.L., respectively.

#### 17.6.8.3 Time-dependent *CP* asymmetry in $B^0 \to D^{*+}D^{*-}K_S^0$

Another way to resolve the $\phi_1 \to \pi/2 - \phi_1$ ambiguity is to study the decay channel $B^0 \to D^{*+}D^{*-}K_S^0$. No direct *CP* violation is expected in this mode since the penguin contributions are negligible. It is shown (Browder, Datta, O'Donnell, and Pakvasa, 2000) that a time-dependent analysis can be performed in this channel, where in principle the values of $\sin 2\phi_1$ and $\cos 2\phi_1$ can be extracted. The time-dependent $\Delta t$ distribution, considering the mistag probability $w$ and the difference $\Delta w = w(B^0) - w(\overline{B}^0)$, is given by

$$f_\pm(\Delta t) \equiv \frac{\mathrm{e}^{-|\Delta t|/\tau_{B^0}}}{4\tau_{B^0}}\Bigg\{(1 \mp \Delta w) \pm (1-2w).$$
$$\times \left[\eta_y \frac{J_c}{J_0}\cos(\Delta m_d \Delta t) - \left(\frac{2J_{s1}}{J_0}\sin 2\phi_1\right.\right.$$
$$\left.\left. +\eta_y \frac{2J_{s2}}{J_0}\cos 2\phi_1\right)\sin(\Delta m_d \Delta t)\right]\Bigg\}, \quad (17.6.16)$$

where $f_+$ and $f_-$ correspond respectively to a $B^0$ and $\overline{B}^0$ tag. This equation is defined in the half Dalitz plane $s^+ < s^-$ or $s^+ > s^-$, where $s^+ \equiv m^2(D^{*+}K_S^0)$ and $s^- \equiv m^2(D^{*-}K_S^0)$. The parameter $\eta_y$ is equal to $+1$ or $-1$ for $s^- < s^+$ or $s^- > s^+$, respectively. The parameters $J_0$, $J_c$, $J_{s1}$, and $J_{s2}$ are the integrals over the half Dalitz phase space with $s^+ < s^-$ of the functions $|A|^2 + |\overline{A}|^2$, $|A|^2 - |\overline{A}|^2$, $\mathrm{Re}(\overline{A}A^*)$, and $\mathrm{Im}(\overline{A}A^*)$, where $A$ and $\overline{A}$ are the amplitudes of $B^0 \to D^{*+}D^{*-}K_S^0$ and $\overline{B}^0 \to D^{*+}D^{*-}K_S^0$ decays, respectively. The values of these parameters depend strongly on the intermediate resonances present in this final state. The presence of the $D_{s1}(2536)$ resonance is well established (Section 19.3) in this decay mode, but this meson is narrow and does not contribute much to $J_{s2}$. Although it had not been studied specifically in $B^0 \to D^{*+}D^{*-}K_S^0$ decays, the $D^*_{s1}(2700)$ meson (Section 19.3) is expected to have a large contribution due to its large width. $D^*_{s1}(2700)$ decays to $D^*K$ and has a large width, $125 \pm 30$ MeV. This implies that $J_{s2}$ is nonzero and that $J_c$ may be large.

*BABAR* (Aubert, 2006u) and Belle (Dalseno, 2007) study this decay mode using $230 \times 10^6$ and $449 \times 10^6$ $B\overline{B}$ pairs, respectively. The mode $B^0 \to D^{*+}D^{*-}K_S^0$ is reconstructed from $D^{*+} \to D^0\pi^+$ and $D^{*+} \to D^+\pi^0$, requiring at least one $D^0\pi^+$ decay. Candidate $D$ mesons are reconstructed in the modes $D^0 \to K^-\pi^+$, $K^-\pi^+\pi^0$, $K^-\pi^+\pi^-\pi^+$, and $D^+ \to K^-\pi^+\pi^+$. Belle also includes the modes $D^0 \to K_S^0\pi^+\pi^-, K^-K^+$, and $D^+ \to K^-K^+\pi^+$, rejecting cases with two $D^0 \to K_S^0\pi^+\pi^-$ decays. When multiple $B$ mesons are reconstructed in an event, *BABAR* selects the one with the smallest $|\Delta E|$ value; Belle chooses the best candidate by using a $\chi^2$ test based on the mass differences from the world averages of the particles present in the final state.

**Table 17.6.12.** Results of the time-dependent Dalitz plot analysis for $B^0 \to D^{(*)0}[K_S^0\pi^+\pi^-]h^0$ decays. $N_{\rm sig}$ is a signal yield obtained from the fit to data. The uncertainties are statistical, systematic, and those due to the Dalitz model, respectively. The uncertainties in the averages include all sources.

| | *BABAR* (Aubert, 2007s) | Belle (Krokovny, 2006) | Average |
|---|---|---|---|
| $N_{\rm sig}$ | $335 \pm 32$ | $325 \pm 31$ | |
| $\sin 2\phi_1$ | $0.29 \pm 0.34 \pm 0.03 \pm 0.05$ | $0.78 \pm 0.44 \pm 0.20 \pm 0.1$ | $0.45 \pm 0.28$ |
| $\cos 2\phi_1$ | $0.42 \pm 0.49 \pm 0.09 \pm 0.13$ | $1.87\ ^{+0.40}_{-0.53}\ ^{+0.20}_{-0.30} \pm 0.1$ | $1.01 \pm 0.40$ |

In *BABAR*, the signal yield is extracted from a fit to the $m_{\rm ES}$ distribution with an additional peaking component to account for misreconstructed events from $B^+ \to \overline{D}^{*0}D^{*+}K_S^0$ decays ($\sim 1.4\%$ of the signal yield). The unbinned maximum likelihood fit yields $201 \pm 17$ signal events. In Belle, the signal yield is extracted from a simultaneous fit to the $m_{\rm ES}$ and $\Delta E$ distributions. The fit result from Belle, shown in Fig. 17.6.17, has a signal yield of $131 \pm 15$ events.

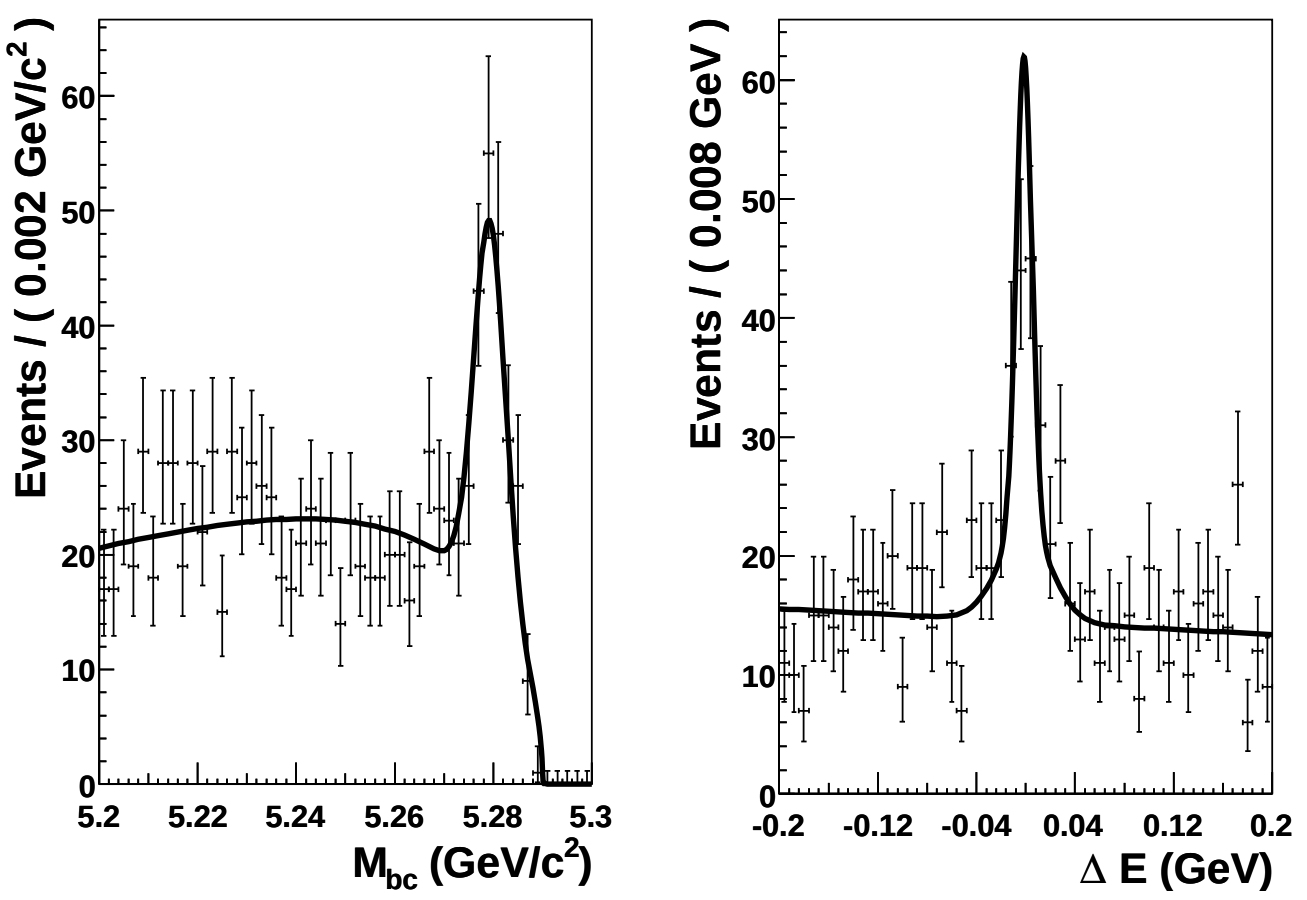

**Figure 17.6.17.** The (left) $m_{\rm ES}$ and (right) $\Delta E$ distributions of $B^0 \to D^{*+}D^{*-}K_S^0$ candidates in Belle (Dalseno, 2007). The curves show the fit projections.

A time-dependent analysis is performed using the event samples described previously. *BABAR* rejects events in which the invariant mass of the $D^{*\pm}K_S^0$ pair is less than $2.55\,{\rm GeV}/c^2$ in order to exclude the $D_{s1}(2536)$ meson, while Belle accounts for this resonance in the systematic uncertainties. Table 17.6.13 shows the results of both experiments and their averages using the half Dalitz plane to fit the coefficients as described in Eq. (17.6.16). Figure 17.6.18 shows the projections in $\Delta t$ of the fits in *BABAR*'s analysis. Belle also uses the whole Dalitz plane to determine the *CP* asymmetry parameters to be:

$$C = +0.01^{+0.28}_{-0.28}\ ({\rm stat}) \pm 0.09\ ({\rm syst}) \qquad (17.6.17)$$

$$D\sin 2\phi_1 = +0.06^{+0.45}_{-0.44}\ ({\rm stat}) \pm 0.06\ ({\rm syst}), \qquad (17.6.18)$$

where $D$ is the dilution factor defined by $D = 1 - 2w$. No evidence for either mixing-induced or direct *CP* violation is found.

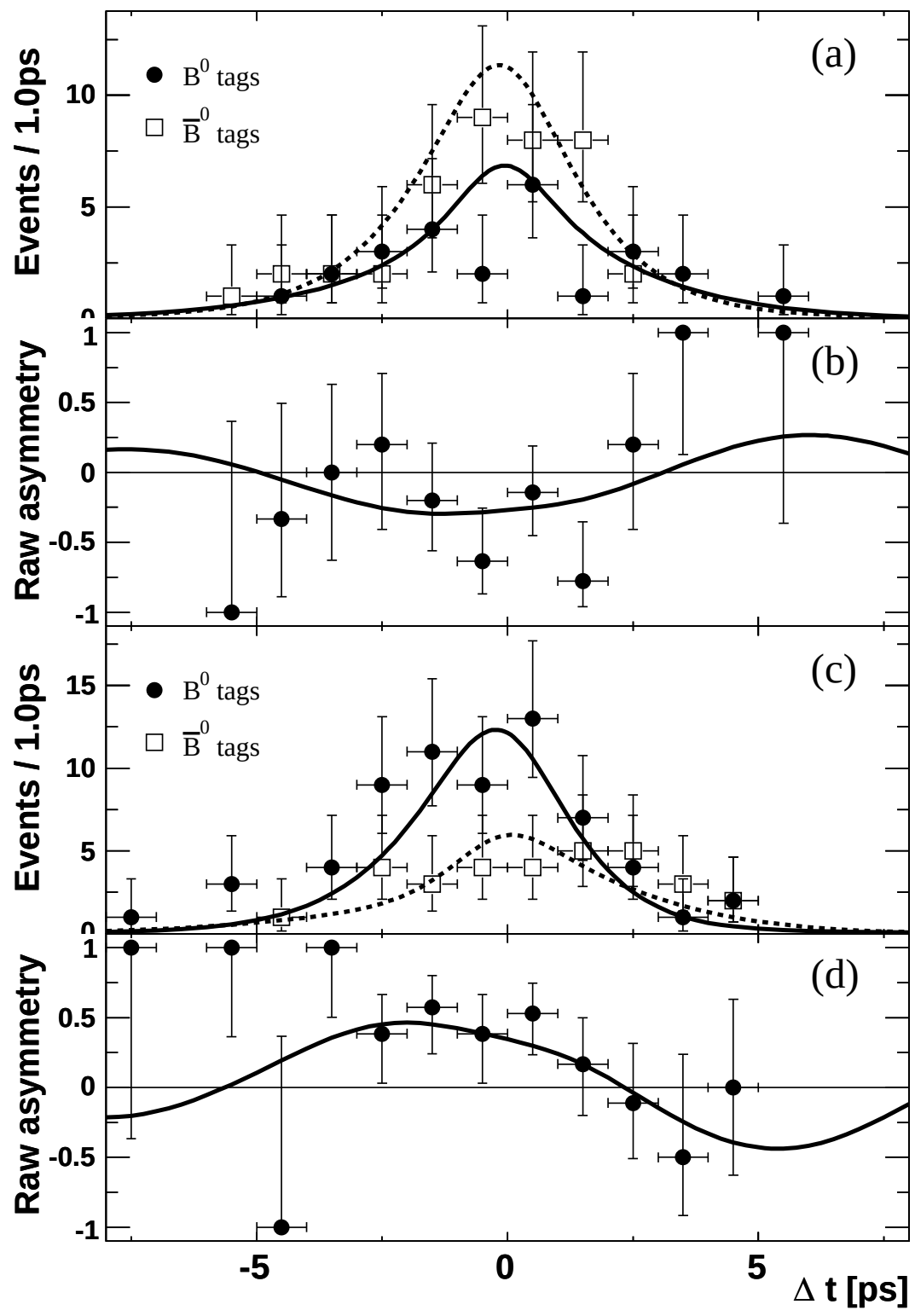

**Figure 17.6.18.** Fit results from *BABAR* for $B^0 \to D^{*+}D^{*-}K_S^0$ (Aubert, 2006u). (a) Distribution of $\Delta t$ in the region $m_{\rm ES} > 5.27$ GeV/$c^2$ for $B^0$ ($\overline{B}^0$) tag candidates in the half Dalitz space $s^+ < s^-$ ($\eta_y = -1$). The solid (dashed) curve represents the fit projections in $\Delta t$ for $B^0$ ($\overline{B}^0$) tags. (b) Raw asymmetry $(N_{B^0} - N_{\overline{B}^0})/(N_{B^0} + N_{\overline{B}^0})$, as a function of $\Delta t$, where $N_{B^0}$ ($N_{\overline{B}^0}$) is the number of candidates with a $B^0$ ($\overline{B}^0$) tag. (c) and (d) contain the corresponding information for the $B^0$ candidates in the other half Dalitz space $s^+ > s^-$ ($\eta_y = +1$).

The main sources of systematic uncertainties, listed here in decreasing order of magnitude, consist of non-uniform acceptance over the Dalitz plane, vertex resolution, mistag fraction, $\Delta t$ resolution function, fit bias, misreconstructed signal events, limited MC statistics, knowledge of the background, and tag-side interference.

The ratio $J_c/J_0$ is found to be significantly different from zero, which confirms a sizable contribution of a broad resonance in the decay $B^0 \to D^{*+}D^{*-}K^0_S$. Since $(2J_{s2})/J_0$ is predicted to be positive when a wide resonance is present (Browder, Datta, O'Donnell, and Pakvasa, 2000), the sign of $\cos 2\phi_1$ can be deduced, in principle, from the measurements presented here. These results are not precise enough to allow one to conclusively determine the sign of $\cos 2\phi_1$. However, the *BABAR* data do prefer a value of $\cos 2\phi_1$ that is positive at the 94% confidence level.

As described above, all of the three independent measurement methods, which use different decay modes and rather different techniques, indicate that confidence levels for $\cos 2\phi_1 > 0$ are around 90% or higher. Therefore, $\cos 2\phi_1$ is experimentally proved to be positive with relatively high confidence. Furthermore, the global fit results discussed in Section 25.1 prove that $\cos 2\phi_1 > 0$ with a large confidence.

### 17.6.9 Time-reversal violation in $b \to c\bar{c}s$ decays

Entangled pairs of neutral $B$ mesons from $\Upsilon(4S)$ decays have been used for establishing *CP* violation in the interference between amplitudes with and without $B^0 - \overline{B}^0$ mixing in decays into $c\bar{c}s$ states (as discussed above, see Sections 17.6.1-17.6.3), and also for demonstrating time-reversal violation in this interference. Just as one $B$ meson in a pair is prepared in the $\overline{B}^0$ state at the time when the other $B$ is identified as a $B^0$ by a decay into a flavor-specific decay such as $e^+\nu_e X$, the decay of one $B$ into $c\bar{c}K^0_S$ prepares the other $B$ in the well defined state $B_+$, which does not decay into $c\bar{c}K^0_S$. Similarly, when the first $B$ decays into $c\bar{c}K^0_L$, the second $B$ is prepared in the state $B_-$, which does not decay into $c\bar{c}K^0_L$.

Violation of *CP* symmetry has been established by observing the difference between the transition rates of $B^0 \to c\bar{c}K^0_S$ and $\overline{B}^0 \to c\bar{c}K^0_S$. In the same way, as proposed by Bañuls and Bernabeu (1999), the difference between the rates of the transitions $B^0 \to B_-$ and $B_- \to B^0$ probes time-reversal symmetry. Such an analysis has been performed by *BABAR* (Lees, 2012m). In the following we define the states $B_+$ and $B_-$ as linear combinations of $B^0$ and $\overline{B}^0$ and show their relevance for time reversal. We then describe the analysis and its results, which are independent of Standard Model or other model assumptions and are only based on quantum mechanics and entanglement.

The time-reversal transformation, usually called $T$, consists of changing the sign of the time coordinate $t$ in the equations of motion. In quantum mechanics, this transformation involves changing the sign of all odd variables under $t \to -t$ in the Hamiltonian $\mathcal{H}$, such as velocities, momenta and spins (called $\widehat{T}$ in the following), and the exchange of final and initial states (Branco, Lavoura, and Silva, 1999; Sachs, 1987). Since it is difficult to prepare the time-reversed process, methods based on $\widehat{T}$-odd observables for non-degenerate stationary states (*e.g.* electric dipole moments for particles), or for final states after weak decay, have been used. The latter, however, require detailed understanding of final-state interactions (FSI), since they may lead to $\widehat{T}$ symmetry violation without the occurrence of $T$ violation (Wolfenstein, 1999).

For $T$-symmetric processes, the probability of an initial state $i$ being transformed into a final state $f$ is the same as the probability that an initial state identical to $f$, but with all momenta and spins reversed, transforms into the state $i$ with all momenta and spins reversed,

$$|\langle f|\mathcal{S}|i\rangle|^2 = |\langle i_T|\mathcal{S}_T|f_T\rangle|^2, \qquad (17.6.19)$$

where $\mathcal{S}$ is the transition matrix given by the Hamiltonian $\mathcal{H}$. This is referred to as detailed balance (Sachs, 1987). In Eq. (17.6.19), $|i\rangle \equiv |\mathbf{p}_i, \mathbf{s}_i\rangle$ and $\langle f| \equiv \langle \mathbf{p}_f, \mathbf{s}_f|$ are the initial and final states, $\langle i_T|$ and $|f_T\rangle$ are the $T$-transformed states of $|i\rangle$ and $\langle f|$, respectively, $\langle i_T| \equiv T|i\rangle = \langle -\mathbf{p}_i, -\mathbf{s}_i|$ and $|f_T\rangle \equiv T\langle f| = |-\mathbf{p}_f, -\mathbf{s}_f\rangle$, and $\mathcal{S}_T = \mathcal{S}^\dagger = T\mathcal{S}T^{-1}$. It should be noted that $T$ invariance is a sufficient, but not necessary, condition for detailed balance. Therefore, detailed-balance breaking is an unambiguous signal for $T$ violation. If $\mathcal{S}$ is Hermitian, $|\langle f|\mathcal{S}|i\rangle| = |\langle i_T|\mathcal{S}|f_T\rangle| = |\langle f_T|\mathcal{S}|i_T\rangle|$; in this case, $T$ invariance implies $\widehat{T}$ invariance, and vice versa. This occurs, for instance, to first order in the weak interactions when FSI may be neglected (Branco, Lavoura, and Silva, 1999; Sachs, 1987).

Within the framework of the Wigner-Weisskopf approximation (Weisskopf and Wigner, 1930a,b), the two contributions to *CP* violation in $K^0 \leftrightarrow \overline{K}^0$ transitions are described by the parameters Re$\epsilon$ (violation of *CP* and $T$ symmetry) and Re$\delta$+$i$Im$\delta$ (violation of *CP* and *CPT* symmetry). Here, *CP* and $T$ symmetry is known to be violated since 1970, when a Bell-Steinberger unitarity analysis determined Re$\epsilon \neq 0$ ($|q_K/p_K| \neq 1$) with a significance of about $5\sigma$ (Schubert et al., 1970). Direct evidence for the violation of *CP* and $T$, however, has been found only 28 years later (Angelopoulos et al., 1998), through the measurement of detailed-balance breaking in $K^0 \leftrightarrow \overline{K}^0$ transitions with a significance of about $4\sigma$, leading to a value of Re$\epsilon$ consistent with that obtained using Bell-Steinberger unitarity.

*CP* violation in $B \to c\bar{c}K^0$ decays is described by the parameter $\lambda = q\overline{A}/pA$, where $A = \langle c\bar{c}K^0|\mathcal{D}|B^0\rangle$, $\overline{A} = \langle c\bar{c}\overline{K}^0|\mathcal{D}|\overline{B}^0\rangle$, and the operator $\mathcal{D}$ is the $B$ decay contribution to $\mathcal{S}$ (Section 10.2). Assuming that the amplitude $A$ can be described by a single weak phase with only one FSI phase shift, the two parts of $\lambda$ (*CP* with $T$ violation, and *CP* with *CPT* violation) are easily identified by separating it into its modulus and phase: $\lambda = |\lambda| \exp(i\phi)$. *CPT* invariance in the decay requires $|\overline{A}/A| = 1$ (Lee, Oehme, and Yang, 1957). With $|q/p| = 1$, which is observed to be well fulfilled (see Section 17.5.4), it follows that $|\lambda| = 1$. $T$ invariance of $\mathcal{S}$ requires $\phi = 0$ or $\pi$, *i.e.* Im$\lambda = 0$ (Enz and Lewis, 1965). Conversely, if $A$ is the sum of two (or more) amplitudes, $|\overline{A}/A| \neq 1$ when both the strong and weak phase differences between the two decay amplitudes do not vanish, even if $\mathcal{D}$ is $CPT$ symmetric (direct $CP$ violation, see Section 16.6). Therefore, if $|\overline{A}/A| = 1$ then we either have both $CPT$ symmetry in decay and a single amplitude, or an unlikely "accidental" cancellation of $T$ and $CPT$ violation in the decay.

**Table 17.6.13.** Time-dependent $CP$ parameters obtained from *BABAR* (Aubert, 2006u) and Belle (Dalseno, 2007) for the decay $B^0 \to D^{*+}D^{*-}K^0_S$. The first uncertainty is statistical and the second is systematic. The averages of the two experiments and their total uncertainty are also shown.

| | *BABAR* | Belle | Average |
|---|---|---|---|
| $\frac{J_c}{J_0}$ | $0.76 \pm 0.18 \pm 0.07$ | $0.60^{+0.25}_{-0.28} \pm 0.08$ | $0.71 \pm 0.16$ |
| $\frac{2J_{s1}}{J_0}\sin 2\phi_1$ | $0.10 \pm 0.24 \pm 0.06$ | $-0.17 \pm 0.42 \pm 0.09$ | $0.03 \pm 0.21$ |
| $\frac{2J_{s2}}{J_0}\cos 2\phi_1$ | $0.38 \pm 0.24 \pm 0.05$ | $-0.23^{+0.43}_{-0.41} \pm 0.13$ | $0.24 \pm 0.22$ |

The first significant observations of large $CP$ violation in $B \to c\bar{c}K^0$ decays (Aubert, 2001e; Abe, 2001g) (see Sections 17.6.2 and 17.6.3) found $C = (1-|\lambda|^2)/(1+|\lambda|^2)$ to be consistent with zero ($|\lambda| = 1$) and $S = 2\mathrm{Im}\lambda/(1+|\lambda|^2) \neq 0$. These results are obtained from the $\Delta t = t_\beta - t_\alpha$ distributions of events $\Upsilon(4S) \to B^0\overline{B}^0 \to (c\bar{c}K^0_S$ or $c\bar{c}K^0_L)$ and $(e^+\nu_e X$ or $e^-\overline{\nu}_e X)$ at times $t_\beta$ and $t_\alpha$, respectively, parameterized according to Eq. (10.2.2) in Section 10.2. This expression assumes a negligible difference between the decay rates of the mass eigenstates (*i.e.* $\Delta\Gamma_d = 0$), $|q/p| = 1$ and $\mathrm{Re}z + i\mathrm{Im}z = 0$ (see Section 17.5.4); *i.e.* $CP$ symmetry in $B^0 - \overline{B}^0$ mixing. However, it is valid for both signs of $\Delta t$ and neither requires $T$ nor $CPT$ symmetry in decay. Within the framework of the Wigner-Weisskopf approximation, the results are compatible with $CP$ and $CPT$ symmetry in decay, and violate $CP$ and $T$ symmetry in the interference between decay and mixing (Fidecaro, Gerber, and Ruf, 2013). 11 years later, time-reversal violation has been directly observed in the measurement of detailed-balance breaking (Lees, 2012m), as described in the following.

Experimentally we know to a sufficiently good approximation that $K^0_S$ and $K^0_L$ are orthogonal states. Adopting an arbitrary sign convention, we have

$$K^0_S = \left(K^0 - \overline{K}^0\right)/\sqrt{2},$$
$$K^0_L = \left(K^0 + \overline{K}^0\right)/\sqrt{2}, \qquad (17.6.20)$$

within $\mathcal{O}(10^{-3})$ due to $CP$ violation in $K^0 - \overline{K}^0$ mixing.[76] Furthermore, assuming the absence of wrong strangeness $B$ decays, *i.e.* the $\overline{B}^0$ does not decay into $c\bar{c}K^0$ and the $B^0$ does not decay into $c\bar{c}\overline{K}^0$, $\langle c\bar{c}K^0|\mathcal{D}|\overline{B}^0\rangle = \langle c\bar{c}\overline{K}^0|\mathcal{D}|B^0\rangle = 0$, we have

$$\lambda_S = q\overline{A}_S/pA_S = -\lambda,$$
$$\lambda_L = q\overline{A}_L/pA_L = \lambda, \qquad (17.6.21)$$

where

$$A_{S,L} = \langle c\bar{c}K^0_S, c\bar{c}K^0_L|\mathcal{D}|B^0\rangle,$$
$$\overline{A}_{S,L} = \langle c\bar{c}K^0_S, c\bar{c}K^0_L|\mathcal{D}|\overline{B}^0\rangle. \qquad (17.6.22)$$

With the aforementioned approximations, the normalized states

$$B_+ = \mathcal{N}\left(B^0 + \frac{A}{\overline{A}}\overline{B}^0\right),$$
$$B_- = \mathcal{N}\left(B^0 - \frac{A}{\overline{A}}\overline{B}^0\right), \qquad (17.6.23)$$

with $\mathcal{N} = |\overline{A}|/\sqrt{|A|^2 + |\overline{A}|^2}$, have the property that the former decays into $c\bar{c}K^0_L$, but not into $c\bar{c}K^0_S$, and the latter into $c\bar{c}K^0_S$, but not into $c\bar{c}K^0_L$ (Alvarez and Szynkman, 2008; Bernabeu, Martinez-Vidal, and Villanueva-Perez, 2012). Like the two mixing eigenstates $B_H$ and $B_L$, the two states $B_+$ and $B_-$ are well defined and phase-convention-free physical states, but all four are not $CP$ eigenstates. In contrast to the $K^0$, $D^0$ and $B^0_s$ systems, where the mass eigenstates are approximate $CP$ eigenstates, none of the linear combinations of $B^0$ and $\overline{B}^0$ has this approximate property because of large $CP$ violation in the system. The states $B_+$ and $B_-$ are orthogonal, *i.e.* $\langle B_+|B_-\rangle = 0$, if $|\overline{A}/A| = 1$. An extended discussion, including wrong strangeness and wrong sign (*i.e.* $\langle e^+\nu_e X\mathcal{D}|\overline{B}^0\rangle \neq 0$, $\langle e^-\overline{\nu}_e X\mathcal{D}|B^0\rangle \neq 0$) $B$ decays has been very recently presented by Applebaum, Efrati, Grossman, Nir, and Soreq (2013).

Preparing the four initial states $B^0$, $\overline{B}^0$, $B_+$ and $B_-$ by entanglement, the *BABAR* analysis (Lees, 2012m) determines the four differences

$$|\langle c\bar{c}K^0_S|\mathcal{S}|B^0\rangle|^2 - |\langle e^+\nu_e X|\mathcal{S}|B_-\rangle|^2,$$
$$|\langle c\bar{c}K^0_L|\mathcal{S}|B^0\rangle|^2 - |\langle e^+\nu_e X|\mathcal{S}|B_+\rangle|^2,$$
$$|\langle c\bar{c}K^0_S|\mathcal{S}|\overline{B}^0\rangle|^2 - |\langle e^-\overline{\nu}_e X\mathcal{S}|B_-\rangle|^2,$$

[76] In general $K^0_{S(L)} \propto K^0(1+\epsilon) - (+)\overline{K}^0(1-\epsilon)$, where $|\epsilon| = (2.228 \pm 0.011) \times 10^{-3}$(Beringer et al., 2012).

$$|\langle c\bar{c}K^0_L|\mathcal{S}|\overline{B}^0\rangle|^2 - |\langle e^-\overline{\nu}_e X|\mathcal{S}|B_+\rangle|^2, \quad (17.6.24)$$

where $\mathcal{S} = \mathcal{D}U(t)$ and $U(t)$ describes the time evolution of $B^0 \leftrightarrow \overline{B}^0$ transitions, given by $\mathbf{M}$ and $\mathbf{\Gamma}$, the two-by-two mass and decay Hermitian matrices of the effective Hamiltonian, as introduced in Section 10.1, and $t > 0$ is the elapsed time between the first and second $B$ decay of the entangled pair. If $|\overline{A}/A| = 1$ (Schubert, Gioi, Bevan, and Di Domenico, 2014), the four differences in Eq. (17.6.24) are equal to the differences

$$\begin{aligned}
&|\langle B_-|U(t)|B^0\rangle|^2 - |\langle B^0|U(t)|B_-\rangle|^2,\\
&|\langle B_+|U(t)|B^0\rangle|^2 - |\langle B^0|U(t)|B_+\rangle|^2,\\
&|\langle B_-|U(t)|\overline{B}^0\rangle|^2 - |\langle \overline{B}^0|U(t)|B_-\rangle|^2,\\
&|\langle B_+|U(t)|\overline{B}^0\rangle|^2 - |\langle \overline{B}^0|U(t)|B_+\rangle|^2, \quad (17.6.25)
\end{aligned}$$

respectively. The observation that these differences are non-zero, with a $\sin \Delta m_d t$ time dependence, is a clear demonstration of detailed-balance breaking.

Within the same approximation, differences like

$$|\langle c\bar{c}K^0_S|\mathcal{S}|B^0\rangle|^2 - |\langle e^-\overline{\nu}_e X|\mathcal{S}|B_-\rangle|^2, \quad (17.6.26)$$

demonstrate *CPT* symmetry.

The experimental analysis (Lees, 2012m) uses the same data sample as the most recent *CP*-violation study in $B \to c\bar{c}K^0$, consisting of 426 $\text{fb}^{-1}$ of integrated luminosity (Aubert, 2009z) (see Section 17.6.3). The analysis relies on identical reconstruction algorithms, selection criteria and calibration techniques. Events are selected in which one $B$ candidate is reconstructed in a $c\bar{c}K^0_S$ or $c\bar{c}K^0_L$ state, and the other $B$ in a flavor eigenstate. We denote generally as $\ell^- X$ ($\ell^+ X$) final states that identify the flavor of the $B$ as $\overline{B}^0$ ($B^0$), which can be either semileptonic decays such as $B^0 \to e^+\nu_e X$ or flavor-specific hadronic decays. The selection leads to event classes $(f_1, f_2)$ where the final state $f_1$ is reconstructed at time $t_1$, and the final state $f_2$ is reconstructed at time $t_2 > t_1$. Thus, only the eight event classes given in Table 17.6.14 are used for further analysis. Within the same approximations and if $|\overline{A}/A| = 1$ can experimentally be proven, these eight classes correspond to the transitions reported on the right column of the table. For example, the event class $(\ell^+ X, c\bar{c}K^0_L)$ involves the decay of one $B$ meson at time $t_1$ into a $\ell^+ X$ final state, thus at this time the $B$ is in a $B^0$ state. It then follows that the still living (second) $B$ meson is, at that time, in a $\overline{B}^0$ state. If this same $B$ meson decays and is reconstructed at time $t_2 > t_1$ as $c\bar{c}K^0_L$, it is a $B_+$ state at $t_2$. Hence, it undergoes a transition $\overline{B}^0 \to B_+$ in the elapsed time $t = t_2 - t_1$. Each of the four time-reversal symmetry differences in Eq. (17.6.24) uses a pair of event classes involving four different final states, $\ell^+ X$ and $\ell^- X$ at times $t_1$ (or $t_2$) and $t_2$ (or $t_1$), and $c\bar{c}K^0_S$ and $c\bar{c}K^0_L$ at times $t_2$ (or $t_1$) and $t_1$ (or $t_2$), respectively.

Assuming $\Delta\Gamma_d = 0$, each of the eight transitions has a time-dependent rate $g^{\pm}_{\alpha,\beta}(t)$ given by

$$e^{-\Gamma t}[1 + S^{\pm}_{\alpha,\beta}\sin(\Delta m_d t) + C^{\pm}_{\alpha,\beta}\cos(\Delta m_d t)], \quad (17.6.27)$$

**Table 17.6.14.** Event classes $(f_1, f_2)$ and their corresponding transitions between $B$ meson states, assuming that $K^0_S$ and $K^0_L$ are orthogonal states, the $\overline{B}^0$ ($B^0$) does not decay into $c\bar{c}K^0$ ($c\bar{c}\overline{K}^0$), and $|\overline{A}/A| = 1$. The effect of the first two assumptions is well below the statistical sensitivity, whereas the third is directly demonstrated in the same analysis (see text).

| Event class | Transition |
|---|---|
| $(\ell^+ X, c\bar{c}K^0_L)$ | $\overline{B}^0 \to B_+$ |
| $(\ell^+ X, c\bar{c}K^0_S)$ | $\overline{B}^0 \to B_-$ |
| $(\ell^- X, c\bar{c}K^0_L)$ | $B^0 \to B_+$ |
| $(\ell^- X, c\bar{c}K^0_S)$ | $B^0 \to B_-$ |
| $(c\bar{c}K^0_L, \ell^+ X)$ | $B_- \to B^0$ |
| $(c\bar{c}K^0_S, \ell^+ X)$ | $B_+ \to B^0$ |
| $(c\bar{c}K^0_L, \ell^- X)$ | $B_- \to \overline{B}^0$ |
| $(c\bar{c}K^0_S, \ell^- X)$ | $B_+ \to \overline{B}^0$ |

where the lower indices $\alpha = \ell^+, \ell^-$ and $\beta = K^0_S, K^0_L$ stand for the final reconstructed states $\ell^+ X$, $\ell^- X$ and $c\bar{c}K^0_S$, $c\bar{c}K^0_L$, respectively, and the upper indices indicate if the flavor eigenstate (+) or the *CP* eigenstate (−) is reconstructed first. The coefficients $S^{\pm}_{\alpha,\beta}$ and $C^{\pm}_{\alpha,\beta}$ are model-independent; the eight pairs of $S$ and $C$ coefficients can be written in terms of eight complex $\lambda$ parameters, as $2\text{Im}\lambda/(1+|\lambda|^2)$ and $(1-|\lambda|^2)/(1+|\lambda|^2)$, respectively. The state $c\bar{c}K^0_S$ is identified by the final states with $c\bar{c} = J/\psi$, $\psi(2S)$ or $\chi_{c1}$, while $c\bar{c}K^0_L$ only by $J/\psi K^0_L$. As in Aubert (2009z), the flavor eigenstates labeled $\ell^+ X$ and $\ell^- X$ are identified by prompt leptons, kaons, pions from $D^*$ mesons, and high-momentum charged particles, combined in a neural network. The final sample contains 7796 $c\bar{c}K^0_S$ events, with purities ranging between 87% and 96%, and 5813 $J/\psi K^0_L$ events with a purity of 56%.

The coefficients $S^{\pm}_{\alpha,\beta}$ and $C^{\pm}_{\alpha,\beta}$ are determined by a simultaneous, unbinned maximum likelihood fit to the four measured $\Delta t = t_\beta - t_\alpha$ distributions. The time difference $\Delta t$ is determined as described in Section 6.5 and used in the *CP*-violation studies based on the same decay modes (see Section 17.6.3). Neglecting time resolution, the elapsed time between the first and second decay is $t = \Delta t$ if the first $B$ decays into a flavor eigenstate, and $t = -\Delta t$ if it decays into a *CP* eigenstate. Time resolution mixes events with positive and negative true $\Delta t$, *i.e.*, a true event class $(\ell^+ X, c\bar{c}K^0_L)$, corresponding to a $\overline{B}^0 \to B_+$ transition, could appear reconstructed as $(c\bar{c}K^0_L, \ell^+ X)$, corresponding to a $B_- \to B^0$ transition, and vice versa. Therefore, the fit cannot be performed with eight event classes but only with four. The separate determination of the coefficients for the event classes with flavor before *CP* eigenstates and those with *CP* before flavor eigenstates, *i.e.*, the unfolding of time ordering and $\Delta t$ resolution, is accomplished by using a signal *p.d.f.* for the four distributions of the form

$$\mathcal{H}_{\alpha,\beta}(\Delta t) = g^+_{\alpha,\beta}(\Delta t_{\text{true}})H(\Delta t_{\text{true}}) \otimes \mathcal{R}(\delta t; \sigma_{\Delta t}) +$$

$$g^{-}_{\alpha,\beta}(-\Delta t_{\rm true})H(-\Delta t_{\rm true}) \otimes \mathcal{R}(\delta t;\sigma_{\Delta t}), \quad (17.6.28)$$

where $\Delta t_{\rm true} \equiv \pm t$ is the signed difference of proper time between the two $B$ decays in the limit of perfect $\Delta t$ resolution, $H$ is the Heaviside step function, $\mathcal{R}(\delta t;\sigma_{\Delta t})$ is the resolution function with $\delta t = \Delta t - \Delta t_{\rm true}$, and $\sigma_{\Delta t}$ is the estimate of the $\Delta t$ uncertainty obtained by the reconstruction algorithms (Bernabeu, Martinez-Vidal, and Villanueva-Perez, 2012). A total of 27 parameters are varied in the likelihood fit: eight pairs $(S^{\pm}_{\alpha,\beta}, C^{\pm}_{\alpha,\beta})$ of signal coefficients and 11 for describing possible $CP$ and $T$ violation in the background. All remaining signal and background parameters are treated in an identical manner as done in the $CP$ violation analysis (see Section 17.6.3). From the 16 signal coefficients, reported in Table 17.6.15, we construct six pairs of independent asymmetry parameters $(\Delta S^{\pm}_T, \Delta C^{\pm}_T)$, $(\Delta S^{\pm}_{CP}, \Delta C^{\pm}_{CP})$, and $(\Delta S^{\pm}_{CPT}, \Delta C^{\pm}_{CPT})$, as shown in Table 17.6.16. The asymmetry parameters have the advantage that the breaking of time-reversal symmetry would directly manifest itself through any nonzero value of $\Delta S^{\pm}_T$ or $\Delta C^{\pm}_T$, or any difference between $\Delta S^{\pm}_{CP}$ and $\Delta S^{\pm}_{CPT}$, or between $\Delta C^{\pm}_{CP}$ and $\Delta C^{\pm}_{CPT}$.

**Table 17.6.15.** Measured values of the $S^{\pm}_{\alpha,\beta}$ and $C^{\pm}_{\alpha,\beta}$ coefficients (Lees, 2012m). The first uncertainty is statistical and the second systematic. The indices $\ell^-$, $\ell^+$, $K^0_S$, and $K^0_L$ stand for reconstructed final states that identify the $B$ meson state as $\overline{B}^0$, $B^0$ and $B_-$, $B_+$, respectively.

| Transition | Parameter | Result |
|---|---|---|
| $\overline{B}^0 \to B_+$ | $S^{+}_{\ell^+,K^0_L}$ | $-0.69 \pm 0.11 \pm 0.04$ |
| | $C^{+}_{\ell^+,K^0_L}$ | $-0.02 \pm 0.11 \pm 0.08$ |
| $\overline{B}^0 \to B_-$ | $S^{+}_{\ell^+,K^0_S}$ | $0.55 \pm 0.09 \pm 0.06$ |
| | $C^{+}_{\ell^+,K^0_S}$ | $0.01 \pm 0.07 \pm 0.05$ |
| $B^0 \to B_+$ | $S^{+}_{\ell^-,K^0_L}$ | $0.51 \pm 0.17 \pm 0.11$ |
| | $C^{+}_{\ell^-,K^0_L}$ | $-0.01 \pm 0.13 \pm 0.08$ |
| $B^0 \to B_-$ | $S^{+}_{\ell^-,K^0_S}$ | $-0.76 \pm 0.06 \pm 0.04$ |
| | $C^{+}_{\ell^-,K^0_S}$ | $0.08 \pm 0.06 \pm 0.06$ |
| $B_- \to B^0$ | $S^{-}_{\ell^+,K^0_L}$ | $0.70 \pm 0.19 \pm 0.12$ |
| | $C^{-}_{\ell^+,K^0_L}$ | $0.16 \pm 0.13 \pm 0.06$ |
| $B_+ \to B^0$ | $S^{-}_{\ell^+,K^0_S}$ | $-0.66 \pm 0.06 \pm 0.04$ |
| | $C^{-}_{\ell^+,K^0_S}$ | $-0.05 \pm 0.06 \pm 0.03$ |
| $B_- \to \overline{B}^0$ | $S^{-}_{\ell^-,K^0_L}$ | $-0.83 \pm 0.11 \pm 0.06$ |
| | $C^{-}_{\ell^-,K^0_L}$ | $0.11 \pm 0.12 \pm 0.08$ |
| $B_+ \to \overline{B}^0$ | $S^{-}_{\ell^-,K^0_S}$ | $0.67 \pm 0.10 \pm 0.08$ |
| | $C^{-}_{\ell^-,K^0_S}$ | $0.03 \pm 0.07 \pm 0.04$ |

**Table 17.6.16.** Measured values of the asymmetry parameters, defined as the differences in $S^{\pm}_{\alpha,\beta}$ and $C^{\pm}_{\alpha,\beta}$ between symmetry-transformed transitions (Lees, 2012m). The parameters $\Delta S^{\pm}_T$, $\Delta C^{\pm}_T$ and the differences $\Delta S^{\pm}_{CP} - \Delta S^{\pm}_{CPT}$, $\Delta C^{\pm}_{CP} - \Delta C^{\pm}_{CPT}$ are all $T$ violating. The first uncertainty is statistical and the second systematic.

| Parameter | Result |
|---|---|
| $\Delta S^{+}_T = S^{-}_{\ell^-,K^0_L} - S^{+}_{\ell^+,K^0_S}$ | $-1.37 \pm 0.14 \pm 0.06$ |
| $\Delta S^{-}_T = S^{+}_{\ell^-,K^0_L} - S^{-}_{\ell^+,K^0_S}$ | $1.17 \pm 0.18 \pm 0.11$ |
| $\Delta C^{+}_T = C^{-}_{\ell^-,K^0_L} - C^{+}_{\ell^+,K^0_S}$ | $0.10 \pm 0.14 \pm 0.08$ |
| $\Delta C^{-}_T = C^{+}_{\ell^-,K^0_L} - C^{-}_{\ell^+,K^0_S}$ | $0.04 \pm 0.14 \pm 0.08$ |
| $\Delta S^{+}_{CP} = S^{+}_{\ell^-,K^0_S} - S^{+}_{\ell^+,K^0_S}$ | $-1.30 \pm 0.11 \pm 0.07$ |
| $\Delta S^{-}_{CP} = S^{-}_{\ell^-,K^0_S} - S^{-}_{\ell^+,K^0_S}$ | $1.33 \pm 0.12 \pm 0.06$ |
| $\Delta C^{+}_{CP} = C^{+}_{\ell^-,K^0_S} - C^{+}_{\ell^+,K^0_S}$ | $0.07 \pm 0.09 \pm 0.03$ |
| $\Delta C^{-}_{CP} = C^{-}_{\ell^-,K^0_S} - C^{-}_{\ell^+,K^0_S}$ | $0.08 \pm 0.10 \pm 0.04$ |
| $\Delta S^{+}_{CPT} = S^{-}_{\ell^+,K^0_L} - S^{+}_{\ell^+,K^0_S}$ | $0.16 \pm 0.21 \pm 0.09$ |
| $\Delta S^{-}_{CPT} = S^{+}_{\ell^+,K^0_L} - S^{-}_{\ell^+,K^0_S}$ | $-0.03 \pm 0.13 \pm 0.06$ |
| $\Delta C^{+}_{CPT} = C^{-}_{\ell^+,K^0_L} - C^{+}_{\ell^+,K^0_S}$ | $0.14 \pm 0.15 \pm 0.07$ |
| $\Delta C^{-}_{CPT} = C^{+}_{\ell^+,K^0_L} - C^{-}_{\ell^+,K^0_S}$ | $0.03 \pm 0.12 \pm 0.08$ |

All eight $C^{\pm}_{\alpha,\beta}$ coefficients are compatible with zero. Since $C = (1-|\lambda|^2)/(1+|\lambda|^2)$, $\lambda = q\overline{A}/pA$ and $|q/p| \approx 1$, $C = 0$ implies $|\overline{A}/A| = 1$. Therefore, the time dependence with only a $\sin \Delta m_d t$ function proves experimentally the approximation $|\overline{A}/A| = 1$ required for the demonstration of time-reversal violation. With this observation (*i.e.* the absence of all eight $\cos \Delta m_d t$ terms) the two states $B_+$ and $B_-$ are orthogonal; we have the association between event classes and $B$ meson transitions given in Table 17.6.14, and the differences in Eqs (17.6.24) and (17.6.25) become identical.

For visualizing the $T$-violating differences of the transition rates, the fit results are shown in Fig. 17.6.19 in the form of asymmetries such as (for the transition $\overline{B}^0 \to B_-$)

$$A_T(\Delta t) = \frac{\mathcal{H}^{-}_{\ell^-,K^0_L}(\Delta t) - \mathcal{H}^{+}_{\ell^+,K^0_S}(\Delta t)}{\mathcal{H}^{-}_{\ell^-,K^0_L}(\Delta t) + \mathcal{H}^{+}_{\ell^+,K^0_S}(\Delta t)}, \quad (17.6.29)$$

where $\mathcal{H}^{\pm}_{\alpha,\beta}(\Delta t) = \mathcal{H}_{\alpha,\beta}(\pm\Delta t)H(\Delta t)$. With this construction, $A_T(\Delta t)$ is defined only for positive $\Delta t$ values. Neglecting reconstruction effects,

$$A_T(t) \approx \frac{\Delta S^{+}_T}{2}\sin(\Delta m_d t) + \frac{\Delta C^{+}_T}{2}\cos(\Delta m_d t). \quad (17.6.30)$$

The three other asymmetries in Fig. 17.6.19 are constructed in an analogous way and have the same time dependence, with $\Delta S^{+}_T$ replaced by $\Delta S^{-}_T$, $\Delta S^{-}_{CP} - \Delta S^{-}_{CPT}$, and $\Delta S^{+}_{CP} - \Delta S^{+}_{CPT}$, respectively, and equally for $\Delta C^{+}_T$.

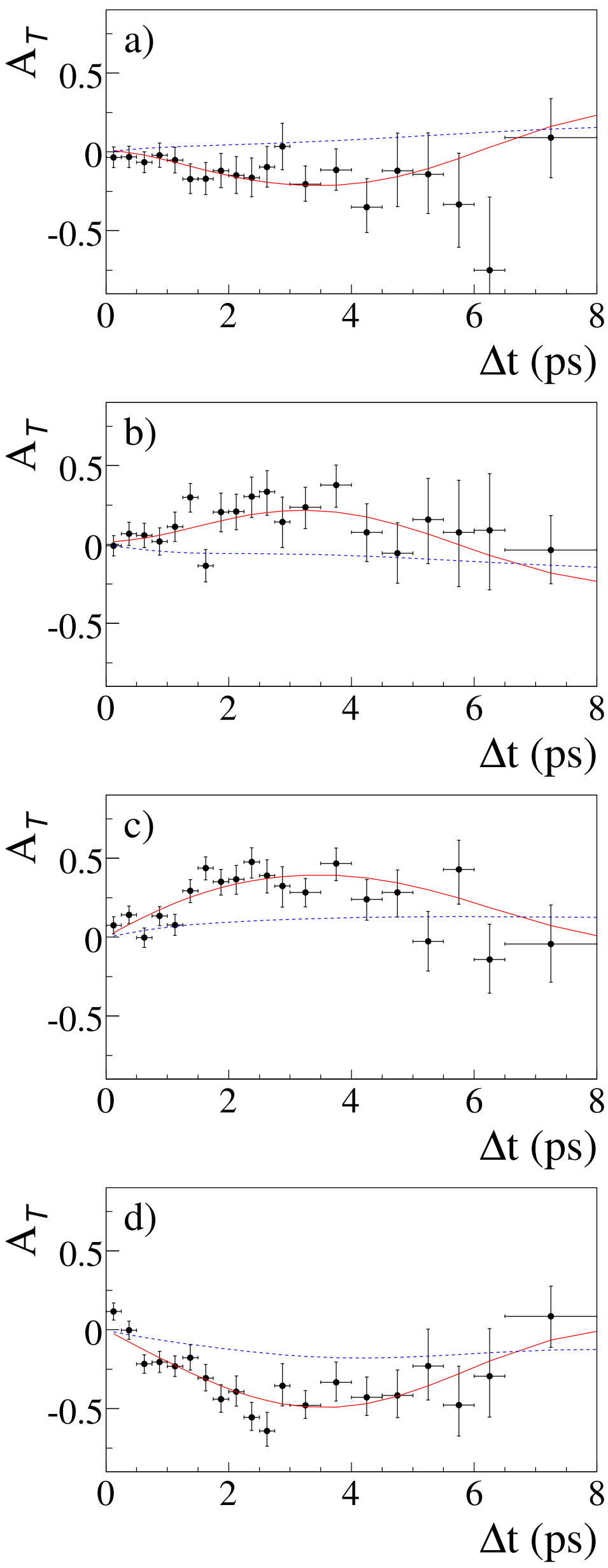

**Figure 17.6.19.** The four independent time-reversal violating asymmetries (Lees, 2012m) for transition a) $\overline{B}^0 \to B_-$ $(\ell^+ X, c\bar{c}K_S^0)$, b) $B_+ \to B^0$ $(c\bar{c}K_S^0, \ell^+ X)$, c) $\overline{B}^0 \to B_+$ $(\ell^+ X, J/\psi K_L^0)$, d) $B_- \to B^0$ $(J/\psi K_L^0, \ell^+ X)$, for combined flavor categories with low misidentification (leptons and kaons), in the signal region ($5.27 < m_{\rm ES} < 5.29$ GeV/$c^2$ for $c\bar{c}K_S^0$ modes and $|\Delta E| < 10$ MeV for $J/\psi K_L^0$). The points with error bars represent the data, the red solid and dashed blue curves represent the projections of the best fit results with and without time-reversal violation, respectively.

The evaluation of systematic uncertainties, reported in Table 17.6.16, follows closely that of the *CP* analysis based on the same final states, discussed in Section 17.6.3. A possible *CP* violation in right- and wrong-sign flavor-specific $B$ decays (denoted $\ell^\pm X$) is found to have an impact on the measurement well below the statistical uncertainty.

As seen in Fig. 17.6.19, time-reversal symmetry is clearly violated in all four transition comparisons. The significance of the observed $T$ violation is obtained from the log-likelihood value $\ln\mathcal{L}$. The difference $2\Delta\ln\mathcal{L}$ between the best fit and the fit without $T$ violation, including systematic errors, is 226 with 8 d.o.f., which corresponds, assuming Gaussian errors, to $14\sigma$. Using the same procedure for the *CPT*-symmetry differences such as in Eq. (17.6.26), no *CPT* violation is observed. The difference $2\Delta\ln\mathcal{L}$ between the values for the best fit and the fit with *CPT* symmetry is 5, equivalent to $0.3\sigma$. The analysis also determines four *CP* asymmetries; the results are compatible with those obtained from the standard *CP* violation analysis based on the same *CP* final states (Aubert, 2009z); the observed significance of *CP* violation is equivalent to $17\sigma$. This is larger than $14\sigma$ for $T$ violation since the comparison between two $(\ell^\pm, c\bar{c}K_S)$ rates has a higher statistical and systematic significance than the comparison of the rates $(\ell^\pm, c\bar{c}K_S)$ and $(\ell^\pm, c\bar{c}K_L)$.

In the Standard Model, the eight coefficients $S^\pm_{\alpha,\beta}$ are measurements of $\sin 2\phi_1$. Hence the four measured $T$-violating asymmetries, $\Delta S_T^\pm$ and $\Delta S_{CP}^\pm - \Delta S_{CPT}^\pm$, can be seen as four measurements of $2\sin 2\phi_1$. The results in Table 17.6.16 lead to a mean value $\phi_1 = (21.8 \pm 2.0)^\circ$, which is of course completely correlated with the $\phi_1$ value obtained from the *CP*-asymmetry measurements discussed in Sections 17.6.3 and 17.6.10.

In conclusion, the *BABAR* experiment (Lees, 2012m) has demonstrated with a large significance of $14\sigma$ that detailed balance and therefore time-reversal symmetry are violated. In $b \to c\bar{c}s$ decays, $T$ and *CP* symmetry breakings are seen in two different observations, are time dependent with only a $\sin\Delta m t$ term, are of order $\mathcal{O}(10^{-1})$, and are induced by the interference between $q\overline{A}$ and $pA$, *i.e.* the interference between decay and mixing. All these properties are different from those of the earlier observed flavor mixing asymmetry in $K^0 - \overline{K}^0$ transitions, where *CP* and $T$ transformations lead to the same observation, the asymmetry is time independent, is of order $\mathcal{O}(10^{-3})$, and is produced by the interference of absorptive $(\Gamma_{12})$ and dispersive $(M_{12})$ contributions to mixing.

### 17.6.10 $\phi_1$ summary

Establishing *CP* violation in $B^0$ meson decays by measuring $\sin 2\phi_1$ was the most important initial goal of the $B$ Factories. Both experiments achieved this goal after two years of operation through time-dependent analyses of $b \to c\bar{c}s$ transitions. This represents the first observation of *CP* violation outside of the neutral kaon system (Christenson, Cronin, Fitch, and Turlay, 1964). With a combined final data set of 1.2 billion $B\overline{B}$ pairs, the achieved precision on $\sin 2\phi_1$ is 0.020. *BABAR* have also demonstrated $T$ violation in $b \to c\bar{c}s$ transitions which provides an additional test of the CKM matrix (this is statistically completely correlated with the *CP* violation result). The ambiguity between $\phi_1$ and $\pi/2 - \phi_1$ is resolved by several measurements. They all use interference with known or measured strong phases (transversity states and $K^* \to K\pi$ phases for $B^0 \to J/\psi K^*$, and Dalitz plot phases for $B^0 \to D^{(*)0}[K_S^0\pi^+\pi^-]h^0$ and $D^*D^*K_S^0$ and other three-body decays). The result in terms of angle is $\phi_1 \equiv \beta = (21.30 \pm 0.78)^\circ$. The direct *CP* asymmetry parameter $C$ is found to be consistent with zero in these channels, as expected in the SM. The consistency between $\phi_1$ and other CKM angles and sides of the Unitarity Triangle demonstrates that the KM mechanism is the dominant source of *CP* violation in the SM. Kobayashi and Maskawa shared the 2008 Nobel Prize in physics for their work on the KM mechanism presented in (Kobayashi and Maskawa, 1973). The test of the CKM matrix by examining the agreement between different measurements is discussed in Section 25.1.

A number of other channels have been studied by the experiments at the $B$ Factories. These are suppressed to various degrees in the SM compared to $b \to c\bar{c}s$ transitions. They are either tree dominated modes with a penguin (or another tree) contribution that has a different weak phase ($J/\psi\pi^0$, $D^{(*)}D^{(*)}$ or $D^{(*)}h^0$), or charmless modes ($b \to sq\bar{q}$). The penguin-dominated modes are particularly sensitive to the presence of any postulated new heavy particles that could contribute to such a loop transition.

The most precisely determined time-dependent asymmetry parameters from a loop dominated $b \to sq\bar{q}$ channel come from $B^0 \to \eta' K^0$ and $K^+K^-K^0$ with a precision of 0.07 on $\sin 2\phi_1$. The uncertainties of other modes range from around 0.2 to 0.7. The $\sin 2\phi_1$ results obtained from these measurements are consistent with the value measured in the $b \to c\bar{c}s$ golden channels. The naïve average of charmless decays is within one sigma of $b \to c\bar{c}s$ results. However, it should be noted that the naïve average is not a good observable to use when searching for NP, as the hadronic uncertainties vary from mode to mode. The most recent measurements of these decays are consistent with the SM.

No significant direct *CP* asymmetry is found in the channels discussed in this section. However some of these channels exhibit central values that are more than $2\sigma$ from $C = 0$ (*e.g.*, $D^+D^-$ for Belle and $\omega K_S^0$ for *BABAR*). The global $\chi^2$ among the different channels studied is consistent with the interpretation that these measurements are the result of a statistical fluctuation.

## 17.7 $\phi_2$, or $\alpha$

***Editors:***
*Yury Kolomensky (BABAR)*
*Tagir Aushev (Belle)*
*Ikaros Bigi (theory)*

**Additional section writers:**
*Adrian Bevan, Cheng-Chin Chiang, Jeremy Dalseno, J. William Gary, Mathew Graham, Akito Kusaka, Fernando Palombo, Kolja Prothmann, Aaron Roodman, Abner Soffer, Alexander Somov, Alexandre Telnov, Karim Trabelsi, Pit Vanhoefer, Georges Vasseur, Fergus Wilson*

In the Standard Model of particle physics the CKM matrix results in a set of nine unitary relationships, six of which are triangles in a complex plane (Chapter 16). The imaginary components of these triangles are manifestations of a single complex phase that dictates the amount of *CP* violation in the theory. The measurement of $\phi_1$ described in Chapter 17.6 establishes one of the angles of the Unitarity Triangle associated with $B_d$ decays, introduced in Section 16.5. To check the self-consistency of the triangle one has to measure its other two angles and the sides. The second angle of the Unitarity Triangle to be measured is $\phi_2$, which is the subject of this chapter. Together the measurements of $\phi_1$ and $\phi_2$ are sufficient to test the predictions of the SM. Constraints on the third angle, $\phi_3$ , are discussed in Chapter 17.8 and how one typically interprets these results in the context of the SM is reviewed in Chapter 25.1.

A probe that can be used to determine $\phi_2$ is the measurement of the time-dependent *CP* asymmetry in $B^0 \to \pi^+\pi^-$ transitions. This *CP* violation is produced by the interference of the dominant box diagram for $B^0-\overline{B}^0$ mixing with the tree diagram $b\overline{d} \to u\overline{u}d\overline{d}$. If these were the only contributing diagrams, the resulting *CP* asymmetry parameters would be $S = \sin 2\phi_2$ and $C = 0$. However the situation is not so simple: this final state is also produced by higher order weak transitions, and of particular relevance is the one-loop diagram usually called the 'penguin' diagram (Shifman, Vainshtein, and Zakharov, 1977). The presence of penguin contributions with weak phases that differ from the leading order tree results in theoretical uncertainties on $\phi_2$ that are sometimes referred to as 'penguin pollution' in the literature. The penguin contribution affects $\mathcal{B}(B^0 \to \pi^+\pi^-)$ and its *CP* asymmetry (Bigi, Khoze, Uraltsev, and Sanda, 1989). Gronau and London (1990) suggested using isospin symmetry to correct for the effect of penguin contributions when extracting $\phi_2$. At first it was thought that penguin contributions are very small in the SM for $B^0 \to \pi\pi$, since the amplitude $b \to dq\overline{q}$ is suppressed by a factor of $|\lambda| = |V_{us}|$ relative to $b \to sq\overline{q}$. However, data showed that the $B^0 \to \pi^0\pi^0$ rate is larger than had been initially expected, and this is explained by the presence of a sizable penguin contribution; therefore penguin amplitudes can significantly affect the extraction of $\phi_2$. Thus the measurement of $\phi_2$ with $B \to \pi\pi$ requires a more complicated approach than the measurement of $\phi_1$ with $B \to J/\psi K^0_S$. The theoretical issues associated with this approach are described in Section 17.7.1.1, and the corresponding experimental treatment is summarized in Section 17.7.3.1. It is worth noting that new physics (NP) could enhance penguin contributions significantly. Experimentally one could identify such contributions by observing a significant difference between values of $\phi_2$ obtained using different decay modes.

The impact of penguin amplitudes in general is different for different final states, and as $b\overline{d} \to u\overline{u}d\overline{d}$ decays can be used to measure $\phi_2$, it became necessary to explore experimentally and theoretically more difficult scenarios in the hope that nature was kind enough to permit measurement of this angle in one or another way. Having determined that the measurement of $\phi_2$ via $B^0 \to \pi\pi$ would be less sensitive than anticipated, the $B$ Factories approached the problem using a rather different technique: a time-dependent analysis of the Dalitz plot of $B^0 \to \pi^+\pi^-\pi^0$. The theoretical issues related to this measurement are introduced in Section 17.7.1.2, while the corresponding experimental discussion can be found in Section 17.7.4. The resulting constraints obtained from *BABAR* and Belle data do not add a significant amount of information to improve the accuracy of the SM solution for $\phi_2$, however they suppress the discrete ambiguities arising from the interpretation of other measurements. For the future higher statistics experiments, the decay $B^0 \to \pi^+\pi^-\pi^0$ is expected to dominate the experimental determination of $\phi_2$ and to provide a sensitive probe for the impact of NP and its features as a non-leading source of *CP* violation.

After several years of data taking it became apparent that extraction of $\phi_2$ from the $B$ Factories is a difficult enterprise. Thus it was realized that one has to think about other final states and *BABAR* started to investigate other related options such as $B \to \rho\rho$ decays. They were previously dismissed by the community as experimentally and theoretically too challenging to be a viable alternative compared with the already ambitious attempts to study $B \to \pi\pi$ and $B \to \rho\pi$. When the experimental work commenced, the outcome of this endeavor was not entirely clear; however, there were hints that indicated these modes could be more promising than originally thought. The presence of two vector particles in the final state meant that one would have to perform a full angular analysis of the final state (see Chapter 12) in addition to constraining penguin contributions. However, it was possible to piece together sufficient information from various sources in order to motivate attempting the measurement of $\phi_2$ with $B \to \rho\rho$ decays. Ultimately a full angular analysis was not required to constrain $\phi_2$ as the fraction of longitudinal polarization in $B \to \rho\rho$ decays was found to almost completely dominate (Section 17.7.3.2). The result of this approach turned out to provide the most stringent constraint on $\phi_2$, where the efforts of *BABAR* and Belle are summarized in Section 17.7.3.2. The time-dependent analysis of $B^0 \to \rho^0\rho^0$ promises to help resolve some of the discrete ambiguities inherent in the isospin analysis and is discussed in Section 17.7.3.3. An additional cross-check using $SU(3)$ for $B \to \rho\rho$ and $K^*\rho$ decays is discussed in Section 17.7.6.

As a further development one constrains $\phi_2$ using final states including vector and axial-vectors particles, in particular using $B \to a_1(1260)\pi$ decays, where one can determine the impact of penguin contributions with the aid of $SU(3)$ flavor symmetry. This theoretical approach is discussed in Section 17.7.1.3. Time-dependent measurements of $B \to a_1(1260)\pi$ and the complementary studies of $B \to K_1\pi$ decays are used to control penguin pollution as discussed in Section 17.7.5.

In contrast to the initial expectations of the $B$ Factories where it was anticipated that $\pi\pi$ final states would provide a measurement of $\phi_2$ and $\rho\pi$ would be used to resolve ambiguities, 'reality' told a different story. Measurements of $B \to \rho\rho$ decays dominate the determination of the angle $\phi_2$ and $B \to a_1(1260)\pi$ decays provide additional precision on the overall measurement of this angle. The study of $\rho\pi$ final states provides additional discrimination: the power to resolve some of the discrete ambiguities, as originally expected. The $B$ Factories have been able to make an accurate measurement of $\phi_2$, using $B$ decays to $\pi\pi$, $\rho\pi$, $\rho\rho$, and $a_1\pi$ final states, as discussed in Section 17.7.7.

### 17.7.1 Introduction

The angle $\phi_2$ can be inferred from time-dependent $CP$ asymmetries in charmless $b \to u$ transitions. Feynman diagrams describing these decays, such as $B^0 \to \pi\pi$ and $B^0 \to \rho\rho$, are shown in Fig. 17.7.1. Interference between the leading tree amplitude and the amplitude of $B^0 - \overline{B}^0$ mixing (Fig. 10.1.1) provides access to the observable $\phi_2$. As explained above, if the tree amplitude was the only decay amplitude (as is the case for $B^0 \to J/\psi K^0_S$), the $S$ parameter in $B^0 \to \pi^+\pi^-$ would be equal to $\sin 2\phi_2$ and $C$ zero (see Eq. 16.6.8). However, the penguin contributions to charmless $B$ decays cannot be ignored. In general $S$ measures $\sin 2\phi_2^{\rm eff}$ instead $\sin 2\phi_2$, where $\phi_2^{\rm eff}$ is related to $\phi_2$ up to a shift $\Delta\phi_2$ resulting from penguin amplitudes with a different weak phase to that of the leading order tree contribution, *i.e.* $\Delta\phi_2 = \phi_2^{\rm eff} - \phi_2$. We have to understand how to control the penguin contributions and determine the difference between $\phi_2^{\rm eff}$ and $\phi_2$.

In the following, we discuss four complementary techniques to extract the angle $\phi_2$ from time-dependent $CP$ asymmetry measurements in $B \to 2\pi$, $3\pi$ and $4\pi$ decays.

- Isospin analysis in $B \to \pi\pi$ and $B \to \rho\rho$;
- Dalitz analysis in $B \to \rho\pi$;
- $SU(3)$ analysis of $B \to a_1(1260)\pi(K)$;
- $SU(3)$ constraints in charmless $B$ decays to two vector meson final states.

The analysis methodology outlined in the remainder of this chapter in terms of the study of four body final states relies on the quasi-two-body approximation (see Section 17.4.3), which is sufficient for work at the $B$ Factories. However, it should be borne in mind that in the future one will want to probe the impact of NP in $4\pi$ and $2\pi K\overline{K}$ final states to search for possible non-leading sources of $CP$ violation. Amplitude analyses, introduced in Section 13, will be required for such searches and future super flavor factory will have to adopt a more general approach for such analyses.

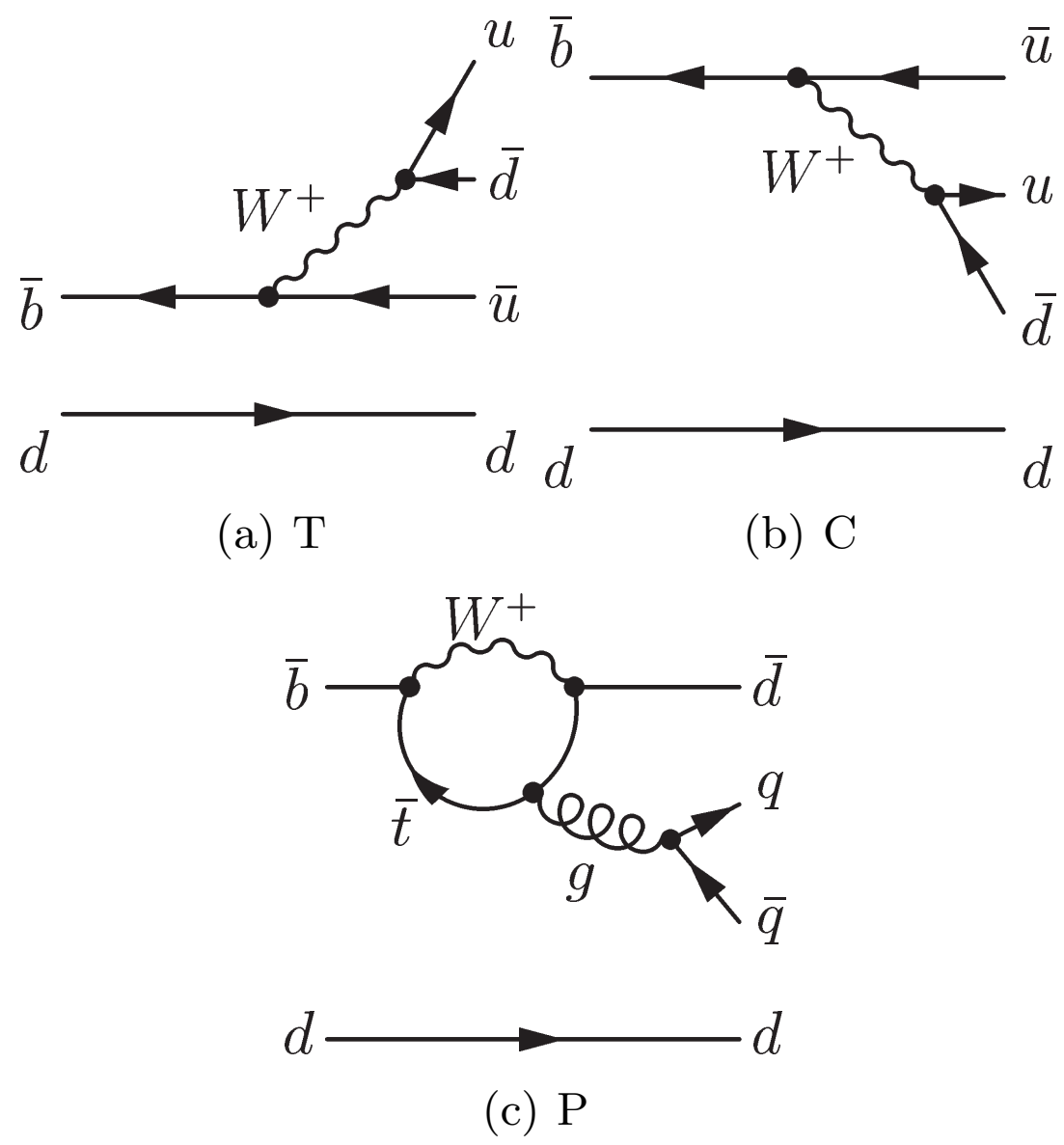

**Figure 17.7.1.** Feynman diagrams contributing to the charmless $B$ decays $B^0 \to \pi\pi$ or $B^0 \to \rho\rho$: (a) external tree ($T$), (b) internal (or color suppressed) tree ($C$), and (c) gluonic penguin ($P$). Nearby quarks are implied to be grouped into mesons.

The combined accuracy on $\phi_2$ obtained by the $B$ Factories is discussed in Section 17.7.7. Some time-integrated measurements are required to constrain penguin contributions in various decays; those are discussed in Section 17.4.

#### 17.7.1.1 Isospin analysis of $B \to \pi\pi$ and $B \to \rho\rho$

The $CP$ asymmetry in $B^0 \to \pi^+\pi^-$ depends on $\phi_2$. However (unlike $B^0 \to J/\psi K^0_S$), because of the two contributing amplitudes (tree and penguin, see Section 16.6) in the SM one expects direct $CP$ violation to be manifest in $B^0 \to \pi^+\pi^-$, hence:

$$\begin{aligned}\frac{\Gamma(\overline{B} \to \pi^+\pi^-) - \Gamma(B \to \pi^+\pi^-)}{\Gamma(\overline{B} \to \pi^+\pi^-) + \Gamma(B \to \pi^+\pi^-)} \\ = C\cos\Delta m_d\Delta t - S\sin\Delta m_d\Delta t\end{aligned} \tag{17.7.1}$$

with

$$\begin{aligned} C &= \frac{1-|\lambda|^2}{1+|\lambda|^2}, \\ S &= \frac{2\,{\rm Im}\lambda}{1+|\lambda|^2}\,, \end{aligned} \tag{17.7.2}$$

where $\lambda = (q/p)\overline{A}/A$ as noted in Chapter 10, and from Chapter 16 we recall that

$$0 \le C^2 + S^2 \le 1. \tag{17.7.3}$$

$CP$ violation is manifest if $0 < C^2 + S^2$, *i.e.* if either of the asymmetry parameters are non-zero. Here $\lambda = (q/p)\overline{R}(\pi^+\pi^-)$ where $\overline{R}(\pi^+\pi^-)$ refers to the amplitude of the $\overline{B}^0$ decay to the final state normalized by the $B^0$ decay to the same one (Eq. 17.7.4):

$$\overline{R}(\pi^+\pi^-) \equiv \frac{A(\overline{B}^0 \to \pi^+\pi^-)}{A(B^0 \to \pi^+\pi^-)} \,. \qquad (17.7.4)$$

Without penguin contributions one predicts in the SM (Bigi, Khoze, Uraltsev, and Sanda, 1989)

$$\left|\frac{q}{p}\overline{R}(\pi^+\pi^-)\right| \simeq 1,$$
$$\mathrm{Im}\left[\frac{q}{p}\overline{R}(\pi^+\pi^-)\right] \simeq \sin 2\phi_2. \qquad (17.7.5)$$

However, penguin amplitudes do contribute. In this case one finds $|\frac{q}{p}\overline{R}(\pi^+\pi^-)| \neq 1$ and therefore $C^2 \neq 0$. Our knowledge of the quantitative impact of penguin amplitudes and in general non-perturbative QCD is rather limited.

One technique for measuring $\phi_2$ is to study time-dependent $CP$ asymmetries in $B^0 \to \pi^+\pi^-$ decays (Aubert (2002h); Abe (2003b)). The data show (see Section 17.7.3.1) that:

$$S = -0.66 \pm 0.07,$$
$$C = -0.30 \pm 0.05, \qquad (17.7.6)$$

which are consistent with expectation from the SM. The non-zero value of $C$, indicating direct $CP$ violation in this mode, arises from the interference of tree and penguin amplitudes with different weak and strong phases. It is not possible to determine if the penguin amplitudes are consistent with the SM expectation, or include contributions from physics beyond the SM. The SM level of contribution to these decays can be determined using the isospin analysis described below, the results of which can be found in Section 17.7.7. Therefore one cannot directly obtain $\phi_2$ from the time-dependent analysis and use this with the value of $\phi_1$ from $B^0 \to J/\psi K^0_S$, discussed in Section 17.6, to construct the SM Unitarity Triangle. A complementary study in $B^0 \to \rho^+\rho^-$ was pioneered by *BABAR* (Aubert, 2004ag) with the hope of being able to contribute to the measurement of $\phi_2$. However once again the extraction of this angle is complicated by the presence of both tree and penguin amplitudes, with different weak phases. An isospin analysis of the $\pi\pi$ or $\rho\rho$ system is necessary (Gronau and London, 1990) to disentangle the tree contribution, and hence determine $\phi_2$ as explained in the following.

The all-charged modes ($B^0 \to \pi^+\pi^-$ and $B^0 \to \rho^+\rho^-$) are dominated by the external tree ($T$) and gluonic penguin ($P$) amplitudes, while the all-neutral modes ($B^0 \to \pi^0\pi^0$ and $B^0 \to \rho^0\rho^0$) are very sensitive to the $P$ contribution, since the internal tree diagram ($C$) is color-suppressed. The amplitudes $A^{00} \equiv A(B^0 \to h^0h^0)$, $A^{+-} \equiv A(B^0 \to h^+h^-)$, and $A^{+0} \equiv A(B^+ \to h^+h^0)$, where $h = \pi, \rho$, and their complex conjugates, obey the Gronau-London isospin relation (Gronau and London, 1990). Here we note that $I = 1/2$ for $u$ and $d$ quarks only and as a result the isospin decomposition of $B \to \pi\pi$ decays follows the corresponding $K \to \pi\pi$ case. Bose statistics forbids $I = 1$ $\pi\pi$ final states, which simplifies the isospin construction of these decays. The tree topologies shown in Fig 17.7.1 come from operators that describe $\Delta I = 1/2$ or $\Delta I = 3/2$ transitions, and so as a $B$ meson has $I = 1/2$, the corresponding final states can either be $I = 0$ or 2. In contrast the gluonic penguin contributions come from a $\Delta I = 1/2$ operator, hence these can only be $I = 0$. Both $h^+h^-$ and $h^0h^0$ final states can be $I = 0$ or 2, and so in general these decays may proceed via both tree and penguin transitions. In contrast, the $h^+h^0$ final state is $I = 2$ and therefore can only have tree contributions. The resulting isospin relations obtained by Gronau and London are:

$$A^{+-}/\sqrt{2} + A^{00} = A^{+0},$$
$$\overline{A}^{+-}/\sqrt{2} + \overline{A}^{00} = \overline{A}^{-0}, \qquad (17.7.7)$$

each of which can be represented by a triangle in a complex plane (Fig. 17.7.2). The $CP$ conjugate relation is usually shown with a tilde replacing the bar to denote that the bases of the two isospin triangles have been aligned such that $A^{+0} = \widetilde{A}^{+0}$, which explicitly neglects any effect coming from electroweak (EW) penguins.[77] The relative sizes and phases of each amplitude can be extracted from the complete isospin analysis of the three decay rates and corresponding $CP$ asymmetries (Gronau and London, 1990). The angle between the sides of lengths $A^{+-}/\sqrt{2}$ and $\widetilde{A}^{+-}/\sqrt{2}$ is $2\Delta\phi_2$.

Experimentally, the complete isospin analysis of the $B \to \pi\pi$ system is complicated by the need to measure time-dependent $CP$ asymmetry of the all-neutral final state decay of $B^0$ mesons to $\pi^0\pi^0$. This is not possible at the present level of statistics, although high luminosity super flavor factory may be able to constrain the decay vertex of the $B^0 \to \pi^0\pi^0$ candidate using Dalitz decays of one or both $\pi^0$ mesons, or events where one or more photons convert in the detector material. The situation is further exacerbated by the relatively large observed branching fraction of $B^0 \to \pi^0\pi^0$ decays (Aubert (2003k); Abe (2003a)); this implies a large penguin contribution, which results in a significant uncertainty in the extraction of $\phi_2$. The branching fraction measurements of the decays $B^0 \to \pi^0\pi^0$ and $B^+ \to \pi^+\pi^0$ are described in Section 17.4.

The isospin analysis of the vector-vector modes $B \to \rho\rho$ is more complicated than that for $B \to \pi\pi$. The $\rho\rho$ final states include three contributions: one longitudinal and two transverse amplitudes following the discussion in Chapter 12. As a result there are three isospin analyses that can be performed, one for each of the transversity amplitudes. Naïve factorization expectations (Suzuki, 2002) indicated that one would expect the longitudinal polarization (*CP*-even) to dominate over the transverse one (a *CP* admixture), which had the implication that analysis of these decays could be simplified from a full angular treatment to a partial angular one where only the fraction of longitudinally polarized events needed to be extracted from data (see Chapter 12). However, the polarization measurements of charmless $B$ decays available at the time were not straightforward and did not all support the expectation of nearly a 100% longitudinal polarization contribution (see Section 17.4). It had also been noted in (Aleksan et al., 1995) that using naïve factorization calculations one obtains a definite hierarchy of penguin contributions in $B \to \pi\pi$, $\rho\pi$ and $\rho\rho$ final states. The results of these calculations implied that the penguin contributions would be largest for $B \to \pi\pi$ decays and smallest for $B \to \rho\rho$. However at the time the $B$ Factories started taking data this message had not been widely appreciated by the community. Given that one expects the ratio of amplitudes of $\rho\rho$ to $\pi\pi$ decays to be $\mathcal{O}(f_\rho^2/f_\pi^2) \sim 2.5$, one could piece together a credible theoretically motivated scenario that indicated the $\rho\rho$ final states might be an attractive alternative way to measure $\phi_2$, if one could overcome the experimental challenges. Fortunately, the longitudinal polarization in $B^0 \to \rho^+\rho^-$ final state has been found to be consistent with unity (Aubert (2004w); Somov (2006)). Moreover, the neutral branching fraction $B^0 \to \rho^0\rho^0$ was found to be relatively small (Aubert (2007h, 2008r); Chiang (2008)), which constrains the penguin uncertainty in the $B \to \rho\rho$ system significantly (see Section 17.4 for the details of the branching fraction measurements of the decays $B^0 \to \rho^+\rho^-$, $B^0 \to \rho^0\rho^0$ and $B^+ \to \rho^+\rho^0$). Shortly after the observation of $B^0 \to \rho^+\rho^-$, *BABAR* performed a time-dependent *CP* asymmetry measurement as a proof of principle that one could indeed constrain $\phi_2$ (Aubert, 2004ag) using larger data samples. It has been noted by Falk et al. (2004) that there could be a small $I = 1$ component to $B \to \rho\rho$, which could be tested by measuring $S$ as a function of the difference between the mass of the two $\rho$'s. Any departure from uniformity would indicate that there is an $I = 1$ component, in which case the isospin construct required to correct for penguins would require some modification.

[77] Electroweak penguins have the same topology as the gluonic penguin shown in Fig 17.7.1, but are mediated by a photon or $Z^0$ boson. It is expected that EW penguins are small and can be neglected. This assumption can be tested by constraining the level of direct $CP$ violation found in $B^+ \to h^+h^0$ decays, which is predicted to be zero in the absence of any EW penguin contribution.

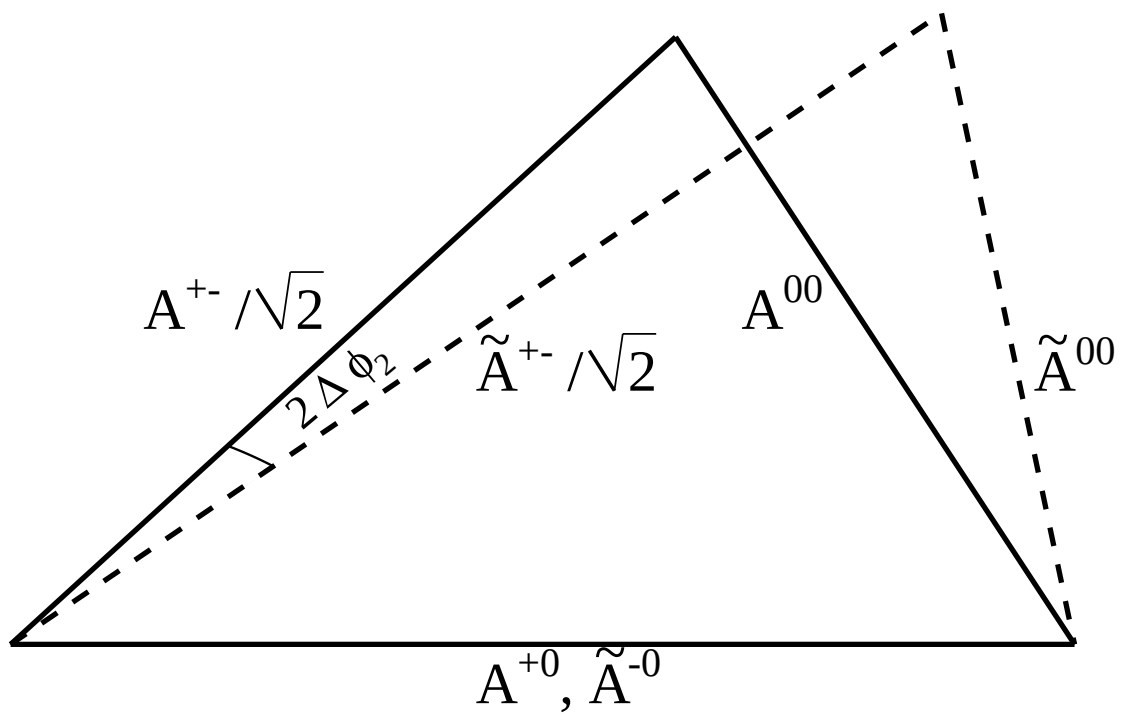

**Figure 17.7.2.** Gronau-London isospin triangles for $B^0 \to hh$ (solid lines) and $\overline{B}^0 \to hh$ (dashed lines), drawn for illustration purposes (not to scale). The $\overline{B}^0$ amplitudes are denoted with a tilde to highlight that the two triangles have been rotated relative to each other so that the $h^+h^0$ and $h^-h^0$ amplitudes are aligned.

#### 17.7.1.2 Dalitz analysis of $B \to \rho\pi$

The $B$ Factories have performed analyses of the quasi-two-body final states $B \to \rho\pi$ to check our theoretical control over extracting $\phi_2$.

A proposed analysis of quasi-two-body final states (Lipkin, Nir, Quinn, and Snyder, 1991; Snyder and Quinn, 1993) relies on the isospin symmetry of the rates of all $B \to \rho\pi$ modes. The decay channels $B^+ \to \rho^0\pi^+$ and $B^0 \to \rho^\pm\pi^\mp$ have been observed first by Belle (Gordon, 2002) and then by *BABAR* (Aubert, 2003h). Evidence for the $B^0 \to \rho^0\pi^0$ mode, which was expected to be small, has been reported by Belle (Dragic, 2004) with a rate higher than an upper bound obtained by *BABAR* (Aubert, 2004g). However, these two results are in agreement at the level of $1.5\sigma$. The remaining mode $B^+ \to \rho^+\pi^0$ has two neutral pions in the final state that makes it a challenging measurement. *BABAR* has reported the observation of this mode (Aubert, 2004g). These analyses are described in detail in Section 17.4.

A better approach uses a Dalitz-plot analysis of $B \to 3\pi$ final states, which relaxes the quasi-two-body approximation and uses information from the interference between resonances in the corners of the Dalitz plot. Snyder and Quinn (1993) pointed out that a time-dependent Dalitz-plot analysis (TDPA) of $B^0 \to \rho\pi \to \pi^+\pi^-\pi^0$ offers a unique way to determine the angle $\phi_2$ without discrete ambiguities. The TDPA uses isospin symmetry and takes into account contamination from $b \to d$ penguin transitions. Additional information to constrain $\phi_2$ can be provided by the measurements of $B^+ \to \rho^+\pi^0$ and $\rho^0\pi^+$ (Gronau, 1991; Lipkin, Nir, Quinn, and Snyder, 1991). Technicalities required to perform a TDPA can be found in Chapter 13.

A preliminary TDPA was reported by *BABAR* at ICHEP in 2004 using a data sample of 213 million $B\overline{B}$ pairs. Subsequent analyses have been published by both Belle (Kusaka, 2007) and *BABAR* (Aubert, 2007v; Lees, 2013c) using larger data samples.

Future Flavor Factories should study the Dalitz plots for $B \to 3\pi$ states beyond intermediate $\rho\pi$ contributions and also explore $B \to K\overline{K}\pi$ to search for possible signs of NP as a non-leading source of *CP* violation.

#### 17.7.1.3 $B \to a_1(1260)\pi$, $B^0 \to a_1(1260)K$ constraints

The last set of decay modes considered at the $B$ Factory experiments for the extraction of $\phi_2$ is $B^0 \to a_1^\pm(1260)\pi^\mp$,

with $a_1^\pm(1260) \to \pi^\mp\pi^\pm\pi^\pm$. As with the previous examples these decays proceed mainly via $b \to u\overline{u}d$ tree amplitudes which can be used to measure time-dependent *CP* asymmetries and allow one to extract the angle $\phi_2$. As with the other modes discussed in this section the existence of non-trivial penguin amplitudes complicates the extraction of $\phi_2$.

Similar to $B \to \rho\pi$ decays, $B$ meson decays to $a_1^\pm(1260)\pi^\mp$ final states are not *CP* eigenstate decays, so to extract $\phi_2$ from these channels one needs to simultaneously consider $B^0(\overline{B}^0) \to a_1^+(1260)\pi^-$ and $B^0(\overline{B}^0) \to a_1^-(1260)\pi^+$ transitions (Aleksan, Dunietz, Kayser, and Le Diberder, 1991). One might cope with the difficulty due to the contribution of penguin amplitudes by using isospin symmetry (Gardner, 1999; Gronau, 1991; Gronau and London, 1990; Gronau and Zupan, 2004; Lipkin, Nir, Quinn, and Snyder, 1991) or a TDPA (Quinn and Silva, 2000; Snyder and Quinn, 1993) or approximate $SU(3)$ flavor symmetry (Charles, 1999; Gronau, London, Sinha, and Sinha, 2001; Grossman and Quinn, 1998).

A full isospin analysis requires the precise measurement of the branching fractions and time-dependent asymmetries in the five modes (and their *CP* conjugates) $B^0 \to a_1^+(1260)\pi^-$, $a_1^-(1260)\pi^+$, $a_1^0(1260)\pi^0$, $B^+ \to a_1^+(1260)\pi^0$, $a_1^0(1260)\pi^+$. Currently the poor precision of most of these measurements (Aubert, 2007i,ae) does not permit the application of this method.

As pointed out in the references (Quinn and Silva, 2000; Snyder and Quinn, 1993) the angle $\phi_2$ may be extracted without ambiguities from a TDPA. This method has been successfully applied to the decay $B^0 \to \pi^+\pi^-\pi^0$ by both experiments. This approach could also be applied to the decay $B^0 \to \pi^+\pi^-\pi^0\pi^0$ with contributions from $a_1^+(1260)\pi^-$, $a_1^-(1260)\pi^-$, $a_1^0(1260)\pi^0$, and $\rho^+\rho^-$ amplitudes or to the decay $B^0 \to \pi^+\pi^-\pi^+\pi^-$ with contributions from $a_1^+(1260)\pi^-$, $a_1^-(1260)\pi^+$, and $\rho^0\rho^0$ amplitudes. Such analyses would be difficult because of the four particles in the final state, the small overlapping region of the phase space of the pions from the $a_1^\pm(1260)$ and $a_1^0(1260)$ mesons, uncertainties in the $a_1(1260)$ meson parameters and line shape, the small number of signal events and the large expected background.

Gronau and Zupan (2006) proposed an $SU(3)$-based procedure for extracting $\phi_2$ in the presence of penguin contributions that the $B$ Factories have followed. This procedure requires measurements of $B$ meson decays into the axial-vector plus pseudoscalar final states $a_1\pi$, $a_1K$, and $K_1\pi$.

*BABAR* (Aubert, 2006aj) and Belle (Dalseno, 2012) measure the branching fraction of the $B^0$ meson decay to $a_1^\pm(1260)\ \pi^\mp$ to be relatively large ($\sim 3 \times 10^{-5}$, see Section 17.4). Following on from the observation of this decay mode *BABAR* performed a set of measurements of $a_1\pi$ and $a_1K$ decays to extract the angle $\phi_2$. This includes the time-dependent asymmetry measurement of $B$ decays to $a_1^\pm(1260)\pi^\mp$ (Aubert, 2007ae), and observation of both $B^+ \to a_1^+(1260)K^0$ and $B^0 \to a_1^-(1260)K^+$ decays (Aubert, 2008ae). The final piece of information required to constrain $\phi_2$ using this approach is the branching fraction of $B$ decays to $K_1\pi$, which was also measured by *BABAR* (Aubert, 2010d), where $K_1$ denotes the axial vector excited $K$ meson states.

The method chosen by *BABAR* for the study of $B^0 \to a_1^\pm(1260)\pi^\mp$ decays follows the quasi-two-body approximation. The decays $B^0(\overline{B}^0) \to a_1^\pm(1260)\pi^\mp$ have been reconstructed with $a_1^\pm(1260) \to \pi^\mp\pi^\pm\pi^\pm$. The other subdecay modes with $a_1^\pm(1260) \to \pi^\pm\pi^0\pi^0$ could be used to enhance statistics, however these are ignored as they have low reconstruction efficiency and large background. From a time-dependent *CP* analysis one extracts an effective angle $\phi_2^{\rm eff}$ which, in analogy with the approaches described above, is an approximate measure of the angle $\phi_2$. Details on this approach for the decays $B^0 \to a_1^\pm(1260)\pi^\mp$ are discussed by Gronau and Zupan (2006). Applying flavor $SU(3)$ symmetry one can determine an upper bound on $\Delta\phi_2 = |\phi_2 - \phi_2^{\rm eff}|$ by relating the $B^0 \to a_1^\pm(1260)\pi^\mp$ decay rates with those of the $\Delta S = 1$ transitions involving the same $SU(3)$ multiplet of $a_1(1260)$, $B \to a_1(1260)K$ and $B \to K_{1A}\pi$. The $K_{1A}$ meson is a nearly equal admixture of the $K_1(1270)$ and $K_1(1400)$ resonances (Amsler et al., 2008). The rates of $B \to K_{1A}\pi$ decays can be derived from the decay rates of $B \to K_1(1270)\pi$ and $B \to K_1(1400)\pi$.

Motivated by the $B \to a_1\pi$ study *BABAR* performed a search for the related decay $B \to a_1^\pm\rho^\mp$ using a data sample of $100\,{\rm fb}^{-1}$, but were unable to establish the presence of a significant signal (Aubert, 2006as). Future experiments may have sufficient data to isolate a clean sample of $a_1^\pm\rho^\mp$ and augment the list of channels used in the determination of $\phi_2$.

#### 17.7.1.4 $SU(3)$ constraints on $\phi_2$ using $B^0 \to \rho^+\rho^-$ and $B^+ \to K^{*0}\rho^+$ decays

A way to constrain penguin contributions to $B^0 \to \rho^+\rho^-$ decays using $SU(3)$ flavor symmetry was proposed by Beneke, Gronau, Rohrer, and Spranger (2006), and is referred to here as the BGRS method. The amplitude of this decay has SM contributions from both tree and penguin topologies, so may be written as

$$A(B^0 \to \rho^+\rho^-) = Te^{i\phi_3} + Pe^{i\delta_{PT}}, \qquad (17.7.8)$$

where $T$ and $P$ are the magnitudes of the tree and penguin contributions to the decay, $\phi_3$ is the Unitarity Triangle angle introduced in Chapter 16, and $\delta_{PT}$ is the strong phase difference between the tree and penguin contributions. Interference between the amplitudes in Eq. (17.7.8) and those responsible for $B^0 - \overline{B}^0$ mixing results in the time-dependent asymmetry of this decay being sensitive to $\phi_2$ as discussed above. The $SU(3)$ related decay $B^+ \to K^{*0}\rho^+$ only proceeds via a penguin transition, so one can use knowledge of the branching fraction and longitudinal polarization fraction of this decay to constrain the corresponding penguin contribution in $\rho^+\rho^-$ up to $SU(3)$ breaking corrections.[78]

[78] As the fraction of longitudinally polarized events is near one it is possible to neglect information contained in the *CP*

In practice in order to constrain $\phi_2$ using this approach one needs to have seven experimental inputs in total: the branching fractions and fraction of longitudinally polarized events for the two decays, as well as $S$ and $C$ measured for $\rho^+\rho^-$, and finally the value of $\phi_1$ obtained from $b \to c\bar{c}s$ transitions (see Section 17.6). These experimental inputs can be used to constrain the three unknowns: $\phi_2$, $\delta_{PT}$, and $r_{PT} = |P/T|$ using

$$C = \frac{2r_{PT}\sin\delta_{PT}\sin(\phi_1+\phi_2)}{1-2r_{PT}\cos\delta_{PT}\cos(\phi_1+\phi_2)+r_{PT}^2}, \tag{17.7.9}$$

$$S = \frac{\sin 2\phi_2 + 2r_{PT}\cos\delta_{PT}\sin(\phi_1-\phi_2) - r_{PT}^2\sin 2\phi_1}{1-2r_{PT}\cos\delta_{PT}\cos(\phi_1+\phi_2)+r_{PT}^2}, \tag{17.7.10}$$

and

$$\left(\frac{|V_{cd}|f_\rho}{|V_{cs}|f_{K^*}}\right)^2 \frac{\Gamma_{\rm L}(B^+\to K^{*0}\rho^+)}{\Gamma_{\rm L}(B^0\to\rho^+\rho^-)} = \frac{Fr_{PT}^2}{1-2r_{PT}\cos\delta_{PT}\cos(\phi_1+\phi_2)+r_{PT}^2}, \tag{17.7.11}$$

where the coefficient $F$ is not equal to one in case of $SU(3)$ breaking, and $f_\rho$ ($f_{K^*}$) is the $\rho$ ($K^*$) decay constant. The factor $F$ is estimated to be $0.9 \pm 0.6$ (Beneke, Gronau, Rohrer, and Spranger, 2006). In fact it turns out that $SU(3)$ breaking has little effect on the overall constraint obtained for $\phi_2$, and one can obtain a precision comparable to the isospin analysis approach even with 100% $SU(3)$ breaking uncertainty. The decay widths $\Gamma_L$ in Eq. (17.7.11) can be replaced by the corresponding branching fractions multiplied by the ratio of $B^0$ to $B^\pm$ lifetimes. This approach provides a stringent constraint on $\phi_2$ that can be used as a cross-check of the traditional $SU(2)$ isospin analysis. Results of using this approach can be found in Section 17.7.6, but given that the same inputs are used for this approach and the isospin analysis (Section 17.7.7), one should take care not to combine the results obtained from the two methods when computing a global average for $\phi_2$.

### 17.7.2 Event reconstruction

The reconstruction of the charmless $B$ decays and event selection follows a similar sequence in both *BABAR* and Belle. First, a sample of charged tracks and photons is selected. Typically, charged tracks are required to originate from the interaction region and to be identified as pions (Chapter 5). In most of the cases an electron veto is applied. After an initial $\pi^0$ selection based on two-photon candidates, the $\pi^0$ candidates are kinematically constrained to the nominal $\pi^0$ mass. Tracks and $\pi^0$ candidates are combined to produce composite candidates (*e.g.* $\rho$, $a_1(1260)$), and finally, signal $B$ candidates are formed. The beam energy substituted mass $m_{\rm ES}$ and the energy difference $\Delta E$ are calculated for these candidates (see Chapter 7).

For time-dependent *CP* analyses, the flavor of the $B$ candidates is determined using a flavor-tagging algorithm (Chapter 8) and the proper time difference $\Delta t$ between the signal $B$ and the accompanying $B$ meson ($B_{\rm tag}$), is measured (Chapter 6). Finally, the signal yields and other decay properties (polarization, *CP* asymmetries) are determined in a multi-variate maximum likelihood fit.

The continuum process $e^+e^- \to q\bar{q}$ ($q = u, d, s, c$) is the main source of background for the charmless $B$ decays. In order to suppress this background, charmless analyses employ multi-variate discriminants based on event topology, which tends to be isotropic for $B\overline{B}$ events and jet-like for $q\bar{q}$ events. A detailed description of these methods is given in Chapter 9.

Additional discrimination against the continuum background is provided by the output of the $B$-flavor tagging algorithms (Chapter 8). In Belle, the tag parameter $r$ ranges from 0 to 1 and is a measure of the likelihood that the $b$ flavor of the accompanying $B$ meson is correctly assigned by the Belle flavor-tagging algorithm. Events with high values of $r$ are well-tagged and are less likely to originate from continuum production. It is found that there is no strong correlation between $r$ and any of the topological variables used above to separate signal from continuum. In *BABAR*, the background discrimination power arises from the difference between the tag efficiencies for signal and continuum background in seven tag categories ($c_{\rm tag} = 1\ldots7$) which is manifest in terms of a different signal purity in each tag category (Section 9.5.3).

After the signal candidates are identified, the proper time difference $\Delta t$ between the signal $B$ and $B_{\rm tag}$ can be determined from the spatial separation between their decay vertices. The $B_{\rm tag}$ vertex is reconstructed from the remaining charged tracks in the event and its uncertainty dominates the $\Delta t$ resolution $\sigma_{\Delta t}$. The typical proper time resolution is $\langle\sigma_{\Delta t}\rangle \approx 0.7$ ps. The distribution of the proper times provides further discrimination against the continuum backgrounds, which are characterized by smaller values of $|\Delta t|$ (Section 9.5.3). The parameters of the proper-time distributions for signal modes, as well as the tagging efficiencies and mistag fractions, are obtained in dedicated fits to events with identified exclusive $B$ decays as discussed in Section 10.6.

### 17.7.3 $B \to \pi\pi$ and $B \to \rho\rho$

An isospin analysis needs as ingredients the measurements of branching fractions, given in Section 17.4, and *CP* violation parameters, described in this section.

#### 17.7.3.1 $B^0 \to \pi^+\pi^-$

The $B$ Factories started performing time-dependent analyses of $B^0 \to \pi^+\pi^-$ early in their lifetime, and these results were updated on a number of occasions. The following describes only the most recent publications by *BABAR* admixture of the transverse polarization without significantly affecting the overall precision on the constraint obtained for $\phi_2$ given the data samples available at the $B$ Factories.

(Lees, 2013b) and Belle (Adachi, 2013), which use data samples of $467 \times 10^6$ and $772 \times 10^6$ $B\overline{B}$ pairs, respectively.

The candidates for the *CP*-eigenstate decay $B^0 \to \pi^+\pi^-$ are constructed from two oppositely charged tracks originating from a common vertex. A good quality of the vertex fit is required (Chapter 6). The $m_{\rm ES}$ and $\Delta E$ requirements are loose relative to the signal resolution ($m_{\rm ES}$ $> 5.24$ GeV/$c^2$ and $-0.2$ GeV $< \Delta E < 0.15$ GeV at Belle, and $m_{\rm ES} > 5.2$ GeV/$c^2$ and $|\Delta E| < 0.15$ GeV at *BABAR*), leaving sidebands for accurate determination of the background level.

The dominant background for this decay mode is continuum. To discriminate between signal and continuum background, multivariate discriminants composed of event-shape variables are used. The definitions of these variables appear in Chapter 9. Belle applies a loose requirement $|\cos\theta_T| < 0.9$, rejecting 50% of the continuum background while keeping 90% of the signal events, and uses a Fisher discriminant $\mathcal{F}_{B\bar{B}}$ composed of $\cos\theta_{\rm T}$, $\cos\theta_{\rm B}$, $\cos\theta_{\rm T,B}$, $\sum p_t^*$ and moments $L_0$ and $L_2$.

*BABAR* requires $|\cos\theta_S| < 0.91$ and $R_2 < 0.7$, rejecting 65% of the continuum background while keeping 90% of the signal events, and then constructs a Fisher discriminant from the $L_0$ and $L_2$ moments.

The selected samples contain not only $B^0 \to \pi^+\pi^-$ signal events but also $B \to K^\pm\pi^\mp$, $q\overline{q}$, and background from higher multiplicity $B$ decays. The signal and background yields and the *CP* parameters are determined from unbinned extended maximum likelihood fits.

The Belle fit is performed using the variables $\Delta E$, $m_{\rm ES}$, $\mathcal{L}(\pi^\pm)$, $\mathcal{F}_{B\bar{B}}$, the $B_{\rm tag}$ flavor $q$ ($q = +1$ for $B_{\rm tag} = B^0$ and $q = -1$ for $B_{\rm tag} = \overline{B}^0$) and $\Delta t$, where $\mathcal{L}(\pi^\pm)$ are the identification likelihoods for each of the pions from the $\pi\pi$ candidate. *BABAR* uses the fit variables $\Delta E$, $m_{\rm ES}$, and Fisher discriminant composed of $L_0$ and $L_2$, the DIRC Cherenkov angles and $dE/dx$ values for each track, $q$ and $\Delta t$. The decay rate as a function of $\Delta t$ of the signal events is described by

$$F_q(\Delta t) = \frac{1}{4\tau_{B^0}} e^{-|\Delta t|/\tau_{B^0}} \cdot (1 \; - qC\cos\Delta m_d \Delta t + qS\sin\Delta m_d\Delta t), \quad (17.7.12)$$

up to vertex position resolution and flavor mistag effects, which are included in the fit *p.d.f.*s.[79] The $\Delta t$ *p.d.f.*s for all other event types in the sample are *CP* conserving.

The two collaborations use different methods for presenting the event distributions and fit functions. Belle applies cuts on discriminating variables and plots the distributions of signal and background of the remaining events, while *BABAR* uses $_s\mathcal{P}lots$ (Section 11.2.3). As an example, Fig. 17.7.3 shows the $\Delta E$ distributions of the *BABAR* data overlaid with components of the fit functions. The $\Delta t$ distributions and time-dependent *CP* asymmetries are shown in Fig. 17.7.4 for the Belle results.

[79] Note the similarity with Eq. (10.2.2), the sign differences resulting from the fact that the *CP* eigenvalue of the decay is opposite that of $B^0 \to J/\psi K^0_S$.

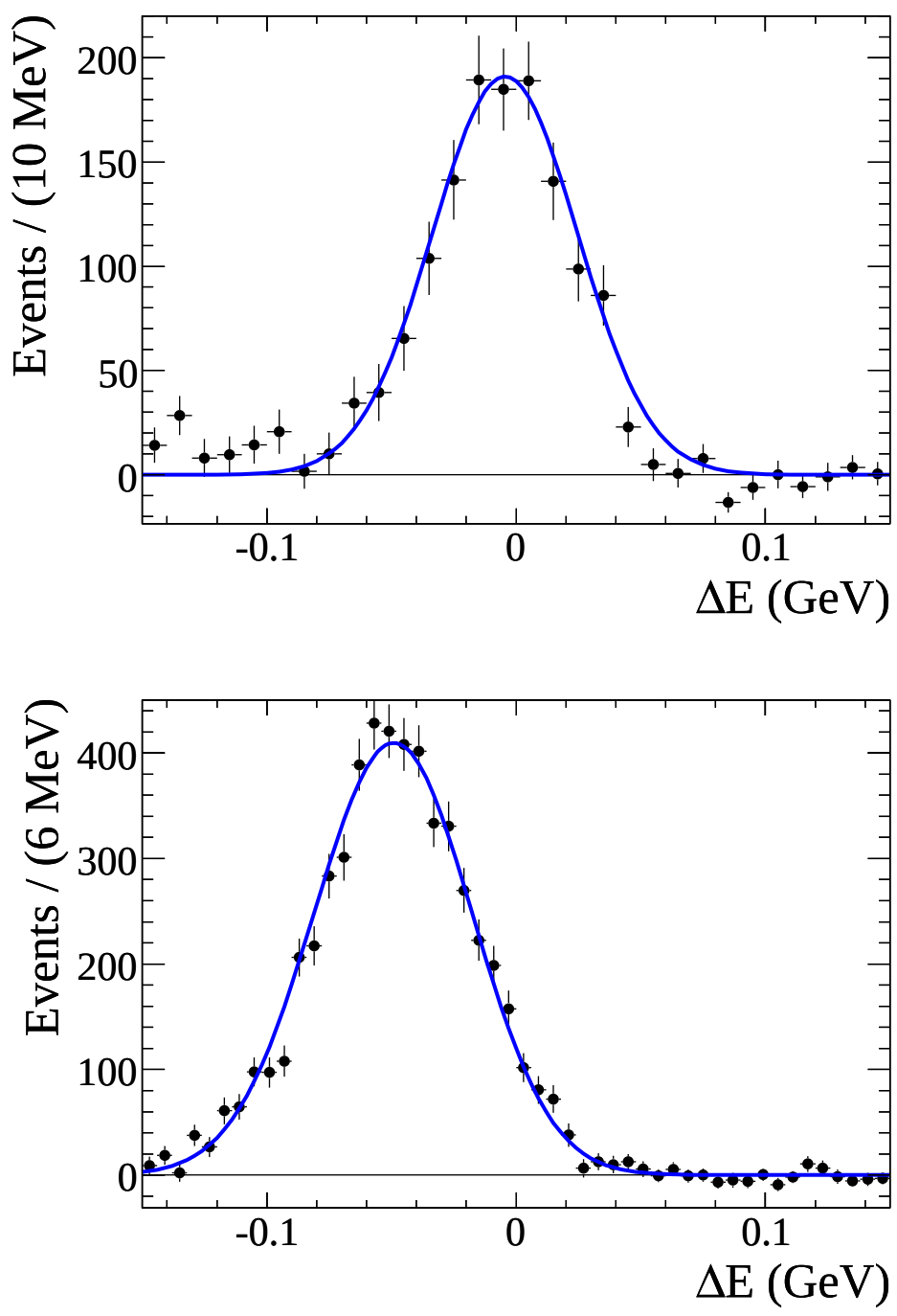

**Figure 17.7.3.** $_s\mathcal{P}lots$ of $\Delta E$ for the *BABAR* $B \to \pi^+\pi^-$ analysis, showing the data distributions (data points with errors) and overlaid fit functions (curves) for the $\pi^+\pi^-$ (top) and background $K^+\pi^-$ (bottom) components, from Lees (2013b). Schematically, the $_s\mathcal{P}lots$ use the information mainly from $m_{\rm ES}$ and the Fisher discriminant to separate the $hh$ (both $\pi^+\pi^-$ and $K\pi$) signal from the continuum background and the information from the Cherenkov angles to separate the $\pi^+\pi^-$ signal from the $K\pi$ signal.

The fit to the Belle data yields $2964 \pm 88$ $B^0 \to \pi^+\pi^-$ events, $9205 \pm 124$ $B^0 \to K^\pm\pi^\mp$ events, and $23 \pm 35$ $B^0 \to K^+K^-$ events. In both analyses, most of the selected candidates (almost 98%) are from continuum background. The *BABAR* fit finds $1394 \pm 54$ $\pi^+\pi^-$ events, $5410 \pm 90$ $K^\pm\pi^\mp$ events, and $7 \pm 17$ $K^+K^-$ events. The *CP* violation parameters obtained by Belle are

$$\begin{aligned} S &= -0.64 \pm 0.08 \pm 0.03, \\ C &= -0.33 \pm 0.06 \pm 0.03, \end{aligned} \quad (17.7.13)$$

and those obtained by *BABAR* are

$$\begin{aligned} S &= -0.68 \pm 0.10 \pm 0.03, \\ C &= -0.25 \pm 0.08 \pm 0.02, \end{aligned} \quad (17.7.14)$$

where the first error is statistical and the second is systematic.

The systematic uncertainties account for a variety of systematic effects. These include biases in $\Delta t$ due to detector misalignment and beam profile; uncertainties on parameters that are fixed in the fit; uncertainties on the parameterization of the detector $\Delta t$ resolution function (main one for $S$), particle-identification performance, and flavor tagging performance and *CP* violation in the $B_{\rm tag}$ decay (main one for $C$). Since $S$ and $C$ are extracted from an asymmetry, the procedure is insensitive to many effects and the systematic uncertainties are quite small.

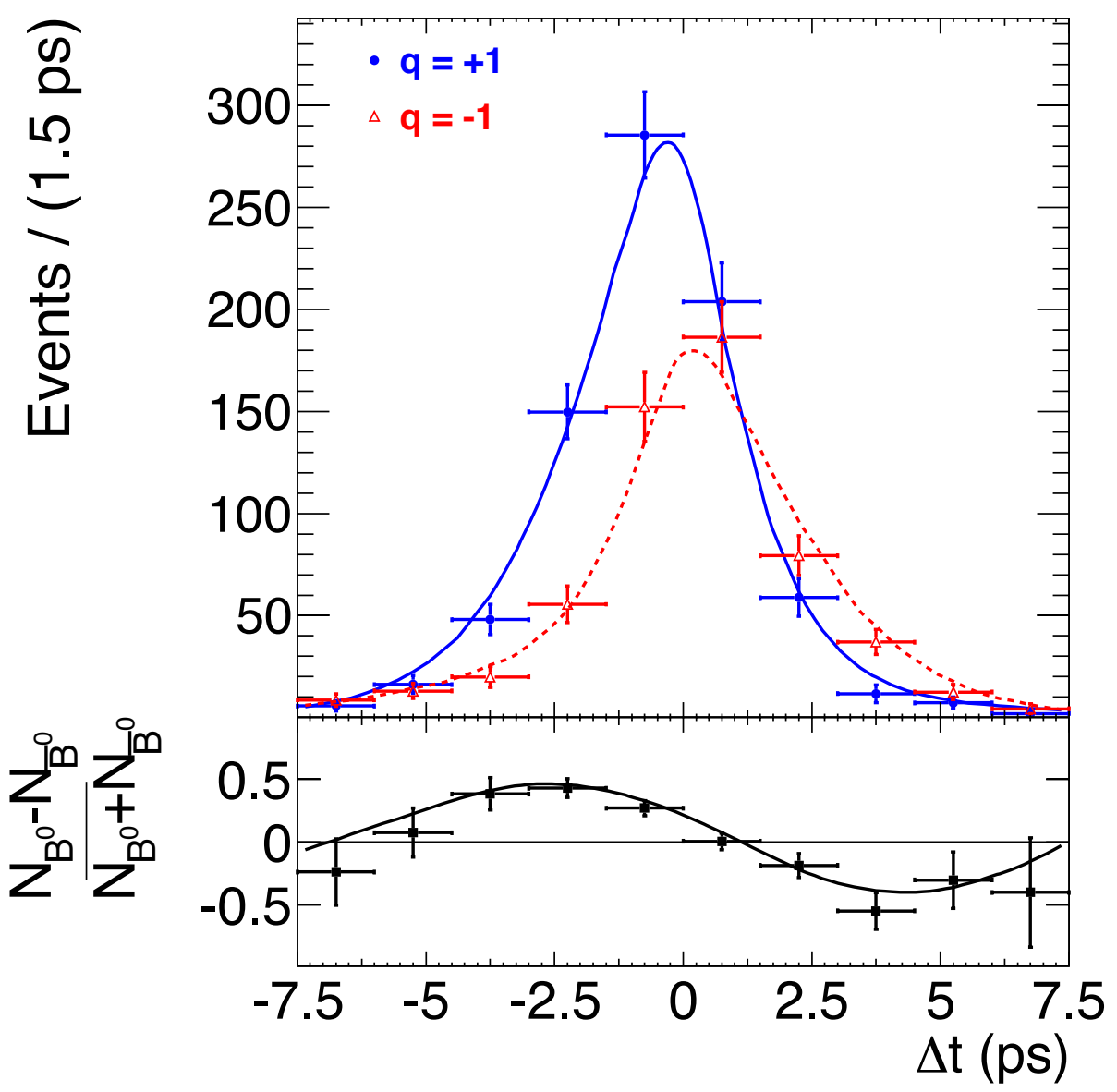

**Figure 17.7.4.** Background subtracted time-dependent fit result for $B^0$ ($\overline{B}^0$)$\rightarrow$ $\pi^+$ $\pi^-$. The top plot shows the time-dependent decay rate for each $B_{\rm tag}$ flavor $q$, where $q = +1(-1)$ refers to a $B^0$ ($\overline{B}^0$) tag. The bottom plot shows the asymmetry between the plots above, $(N_{B^0} - N_{\overline{B}^0})/(N_{B^0} + N_{\overline{B}^0})$. $N_{B^0}(N_{\overline{B}^0})$ is the measured signal yield in each bin of $\Delta t$ for $B^0$ ($\overline{B}^0$) tagged events, from (Adachi, 2013).

Historically there was indication of some level of disagreement between the two experiments on the measurements of $S$ and $C$ in this decay mode. While this difference was never large enough to claim a significant discrepancy between the results, the evolution of the parameters from one conference season to the next remained of wide interest to the community. Over time these measurements slowly regressed toward a common mean, and as one can see from the results presented here – results from the two experiments agree with each other within uncertainties. The $B$ Factory average of these results is

$$\begin{aligned} S &= -0.66 \pm 0.07, \\ C &= -0.30 \pm 0.05, \end{aligned} \qquad (17.7.15)$$

where the resulting correlation between $S$ and $C$ for the average is $-8.1\%$.

#### 17.7.3.2 $B^0 \rightarrow \rho^+\rho^-$

The $B^0 \rightarrow \rho^+\rho^-$ candidates are reconstructed by combining pairs of oppositely charged $\rho$ mesons, which in turn are selected using $\pi^\pm$ candidates and $\pi^0$ candidates. As the $\rho$ meson is a wide resonance, when reconstructing the signal final state some events contain multiple reconstructed $B$ candidates. Most of these candidates arise from combinations of fake $\pi^0$ mesons with tracks from the signal side. In such events the $B$ candidate with the smallest sum $\sum_{\pi^0_{1,2}}(m_{\gamma\gamma} - m^0_\pi)^2$ is selected. Other incorrectly reconstructed $B$ candidates appear from events with mis-reconstructed $\pi^\pm$ tracks. These events contain mis-reconstructed vertices and may bias time-dependent *CP* measurements if not accounted for appropriately. The fraction of signal decays in data samples selected for the time-dependent measurements of *BABAR* and Belle that have at least one $\pi^\pm$ track incorrectly identified but pass all selection criteria is 13.8 % and 6.5 %, respectively. Signal decays that have at least one $\pi$ meson incorrectly identified are referred to as mis-reconstructed signal. The two types of mis-reconstructed signals described above, where the fake pion is either neutral or charged, are dealt with separately.

Similarly to other charmless $B$ decays the dominant background for the $B^0 \rightarrow \rho^+\rho^-$ channel originates from $e^+e^- \rightarrow q\overline{q}$ ($q = u, d, s, c$) continuum events. The procedures adopted for background suppression are described briefly in the following paragraphs.

The Belle analysis uses a Fisher discriminant formed from modified Fox-Wolfram moments and $\theta_B$, the polar angle in the CM frame between the $B$ direction and the beam axis. These two variables are combined into a signal to background likelihood ratio, $\mathcal{R}$. The *p.d.f.*s for signal and $q\overline{q}$ components are obtained from MC simulation and the data $m_{\rm ES}$ sideband, respectively, and used to fit the selected $B \rightarrow \rho^+\rho^-$ candidates.

In the *BABAR* analysis $q\overline{q}$ background is reduced by requiring $|\cos\theta_T| < 0.8$, where $\theta_T$ is the angle between the thrust axis of the candidate and that of the remaining detected particles in the event. Further signal to background separation is performed by using a multi-layer perceptron (neural network, NN, see Chapter 4), which is trained and validated using off-resonance data (background) and MC simulated events (signal). Eight topological variables, see Aubert (2007b), are included into this neural network, and the output is transformed by a 1 : 1 mapping that broadens the peaking contribution for the signal target type (NN $\sim 1$) to facilitate *p.d.f.* parameterization so that the neural network output can be used in a maximum likelihood fit to data.

The following components are distinguished in both Belle and *BABAR* analyses: signal and $\rho\pi\pi$ non-resonant decays, signal events with a mis-reconstructed $\pi^0$, signal events with a mis-reconstructed $\pi^\pm$, continuum background ($q\overline{q}$), charm $B$ background ($b \rightarrow c$), and charmless ($b \rightarrow u$) background. The fitted yield for the non-resonant $4\pi$ component was found to be consistent with zero by both experiments. The ($b \rightarrow u$) background is dominated by $B \rightarrow (\rho\pi, a_1\pi, a_1\rho, \rho^\pm\rho^0)$ decays.

The latest *BABAR* analysis is based on a data sample of 383.6 million $B\overline{B}$ pairs (Aubert, 2007b) and supersedes two previous *BABAR* analyses (Aubert, 2004ag, 2005j). The signal yield, longitudinal polarization fraction $f_L$, and *CP* asymmetry parameters $C$ and $S$ are obtained simultaneously from an unbinned extended ML fit to 37424 events. The background discriminating variables are $m_{\rm ES}$, $\Delta E$, $\Delta t$, $m_{\pi^\pm\pi^0}$, $\cos\theta_\pm$, and NN. Part of the $b \to u$ background has distributions similar to the signal for one or more of the discriminating variables. Therefore, even if it is smaller than the $b \to c$ background, it is important to account for it correctly. A total of 165 different possible background contributions are considered for the $b \to u$ background. However, only components where more than one event is expected to contribute to the selected data sample are modeled individually. Those modes where less than one event is expected are collected together and modeled using an inclusive component (split into neutral and charged contributions). As a result the *BABAR* analysis incorporates 22 components for different background types explicitly, whose yields are fixed to expectations, except for the non-resonant $\rho\pi\pi$ final state yield left free in the fit as a secondary signal. The fit results are $N_{\rho\rho} = 729 \pm 60^{+94}_{-102}$ events, $f_L = 0.992 \pm 0.024^{+0.026}_{-0.013}$, $C = 0.01 \pm 0.15 \pm 0.06$, and $S = -0.17 \pm 0.20^{+0.05}_{-0.06}$. Distributions of $m_{\rm ES}$, $\Delta E$, $\cos\theta_\pm$, and $m_{\pi^\pm\pi^0}$, for the highest purity tagged events are shown in Fig. 17.7.5. The $\Delta t$ distribution for $B^0$ and $\overline{B}^0$ tagged events and the time-dependent decay-rate asymmetry are presented in Fig. 17.7.6.

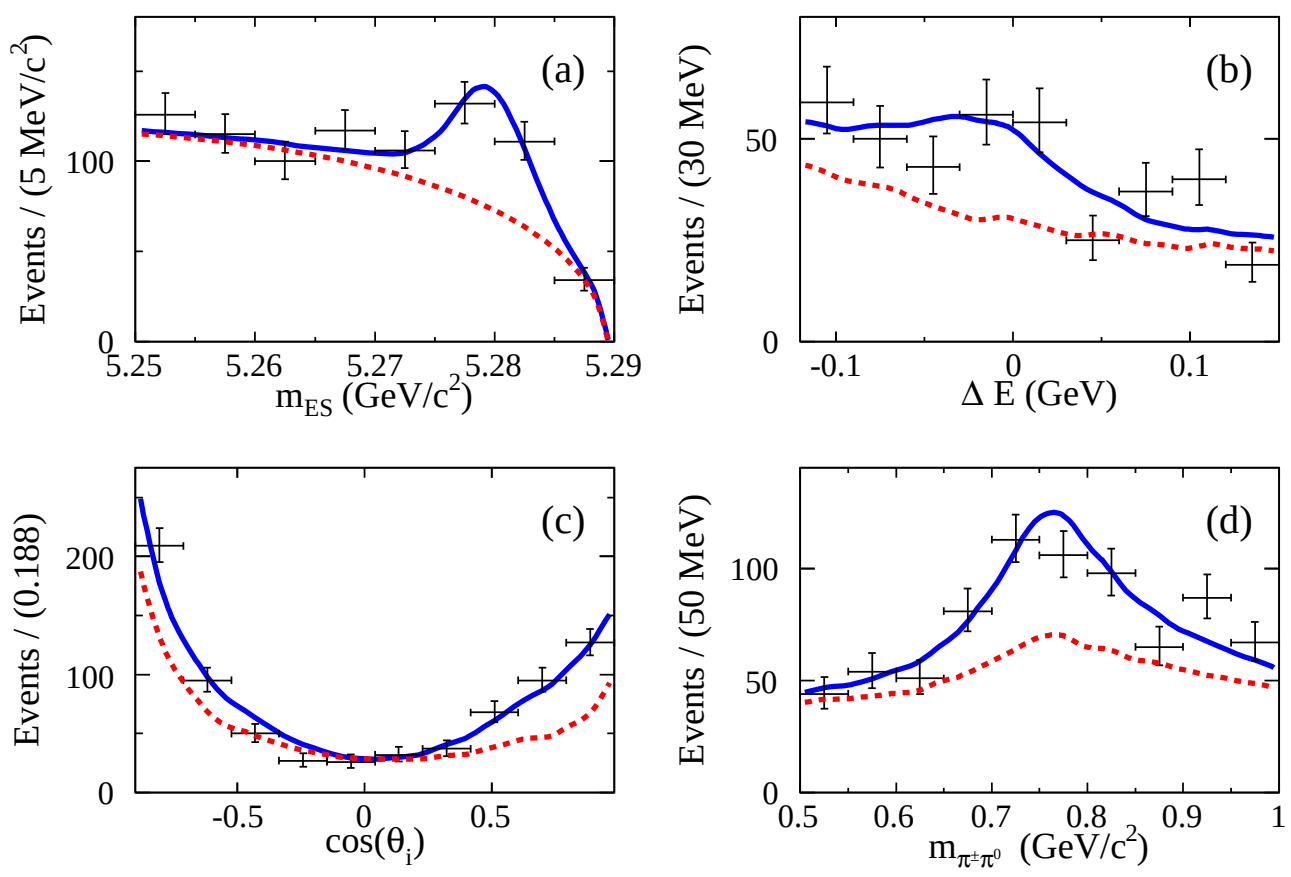

**Figure 17.7.5.** (a) $m_{\rm ES}$, (b) $\Delta E$, (c) cosine of the $\rho$ helicity angle, and (d) $m_{\pi^\pm\pi^0}$ for the highest purity tagged $B \to \rho^+\rho^-$ events. For the plots (b), (c) and (d), $m_{\rm ES}$ is required to be larger than 5.27 GeV/$c^2$. The dashed lines are the sum of backgrounds and the solid lines are the total *p.d.f.* (from Aubert (2007b)).

Measurements of the polarization fraction and the fraction of $\rho\pi\pi$ non-resonant events were performed by Belle in (Somov, 2006) and found to be $0.941^{+0.034}_{-0.040} \pm 0.030$ and $(6.3 \pm 6.7)\%$, respectively. The latest Belle measurements of the *CP* asymmetry parameters are based on a data sample of 535 million $B\overline{B}$ pairs (Somov, 2007). The analysis is organized into two steps. During the first step the yields of signal and background components are obtained using an unbinned extended ML fit to the three-dimensional ($m_{\rm ES}$, $\Delta E$, $\mathcal{R}$) distribution. A total of 176843 events are selected for the analysis. The fit yields $N_{\rho\rho+\rho\pi\pi} = 576 \pm 53$ events. During the second step the *CP* asymmetry parameters $S$ and $C$ are determined from a fit to the $\Delta t$ distributions. The signal region used for the $\Delta t$ fit is 5.27 GeV/$c^2 < m_{\rm ES} <$ 5.29 GeV/$c^2$, $-0.12$ GeV $< \Delta E <$ 0.08 GeV, and $\mathcal{R} > 0.15$, a soft requirement, which eliminates the background dominated events. The fractions of mis-reconstructed signal events are fixed to the expectation from MC simulation, the fraction of non-resonant $B \to \rho\pi\pi$ decays is fixed from the result obtained in Somov (2006), and the fractions of correctly reconstructed signal, $b \to c$ background, $q\overline{q}$ background and $b \to u$ background are normalized to the event fractions obtained from the ($m_{\rm ES}$, $\Delta E$, $\mathcal{R}$) fit and are fixed in the $\Delta t$ fit. The event fractions and $\Delta t$ *p.d.f.*s depend on the flavor tag category assigned to an event. A fit to the 18016 events in the ($m_{\rm ES}$, $\Delta E$, $\mathcal{R}$) signal region gives $C = -0.16 \pm 0.21 \pm 0.08$ and $S = 0.19 \pm 0.30 \pm 0.08$.

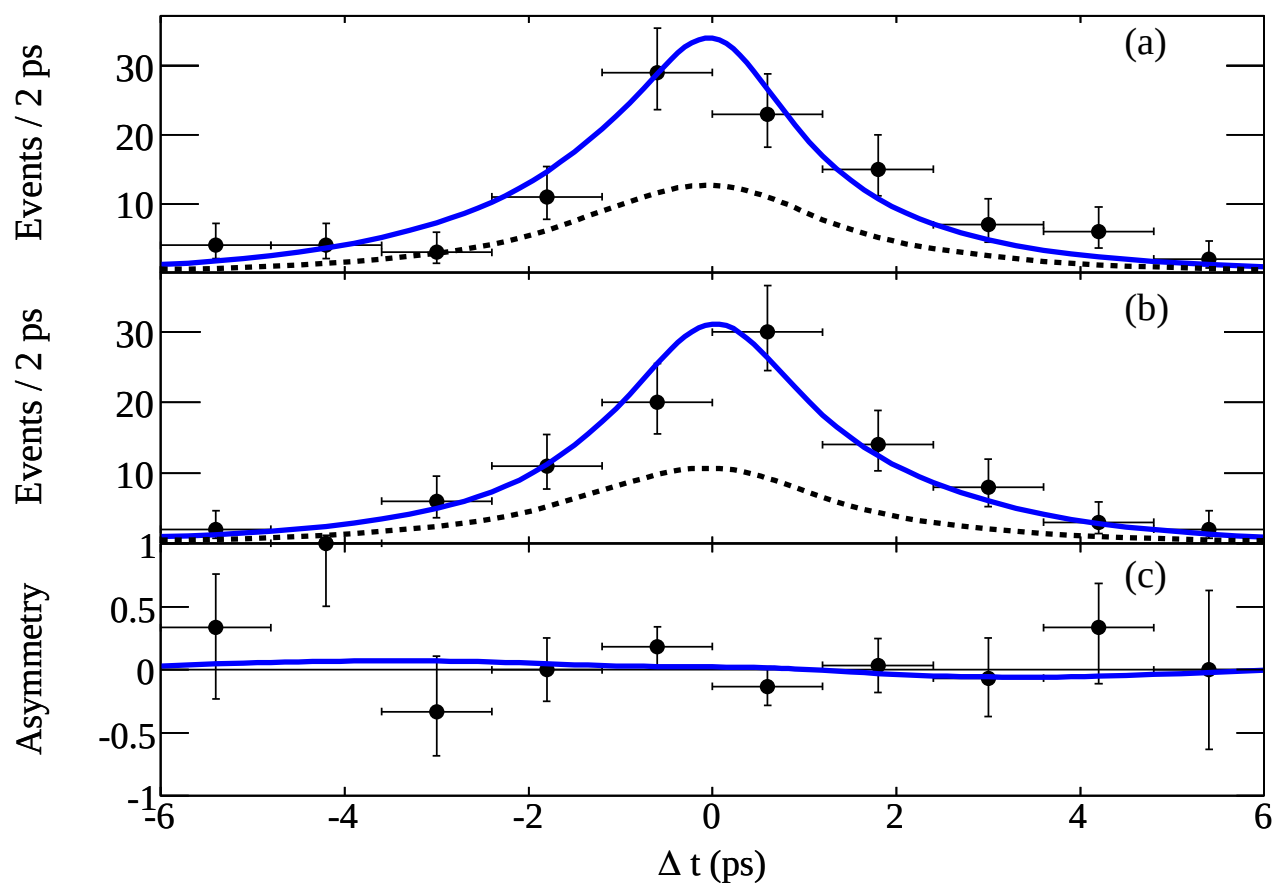

**Figure 17.7.6.** The $\Delta t$ distributions of events enriched in signal for (a) $B^0$ and (b) $\overline{B}^0$ tagged $B \to \rho^+\rho^-$ events. The solid lines are the sum of signal and backgrounds and the dashed lines are the sum of backgrounds. The time-dependent *CP* asymmetry is presented in (c). (As there is no entry in the second bin from the left in (b), the corresponding point is at 1 in (c).) The curve corresponds to the measured asymmetry (from Aubert (2007b)).

The $B$ Factory combined values for $f_L$, $C$ and $S$ are

$$\begin{aligned} S &= -0.05 \pm 0.17, \\ C &= -0.06 \pm 0.13, \\ f_L &= 0.978 \pm 0.023, \end{aligned} \tag{17.7.16}$$

where the correlation between $S$ and $C$ is small. The measured $B^0 \to \rho^+\rho^-$ branching fraction is given in Section 17.4.

#### 17.7.3.3 $B^0 \to \rho^0\rho^0$

The analyses reported here are based on the full data sample of both experiments, containing respectively $465 \times 10^6$ $B\overline{B}$ pairs recorded with the *BABAR* detector (Aubert, 2008r) and $772 \times 10^6$ $B\overline{B}$ pairs collected with the Belle detector (Adachi, 2014). Two other analyses on a partial data sample were also published by *BABAR* (Aubert, 2007h) and Belle (Chiang, 2008).

$B^0$ meson candidates are reconstructed from two $\rho^0$ candidates, each reconstructed from two oppositely charged pions. The analyses use six kinematic variables to reconstruct the signal: $m_{\mathrm{ES}}$, $\Delta E$, the invariant $\pi^+\pi^-$ masses $m(\pi^+\pi^-)_{1,2}$, and the helicity angles $\theta_{1,2}$, defined as the angles between the $\pi^+$ and the $B$ flight direction in each $\rho^0$ rest frame. The *BABAR* analysis applies the following kinematic selection: $5.245 < m_{\mathrm{ES}} < 5.290$ GeV/$c^2$, $|\Delta E| < 85$ MeV, $0.55 < m(\pi^+\pi^-)_{1,2} < 1.05$ GeV/$c^2$, and $|\cos\theta_{1,2}| < 0.98$. The Belle analysis requires that the invariant $\pi^+\pi^-$ masses lie within the signal window $m(\pi^+\pi^-)_{1,2} \in [0.52, 1.15]$ GeV/$c^2$, $|\Delta E| < 0.1$ GeV, and $m_{\mathrm{ES}} > 5.27$ GeV/$c^2$. In both cases, the $\pi^+\pi^-$ mass window is chosen to accept $\rho^0 \to \pi^+\pi^-$, $f_0(980) \to \pi^+\pi^-$, and non-resonant modes, and to exclude $K^0_S \to \pi^+\pi^-$ and charm meson decays such as $D^0 \to \pi^+\pi^-$. Furthermore, to remove the peaking backgrounds from the $D^+$ (especially $D^+ \to K^-\pi^+\pi^+$), $D^+_s$, $D^0$, $J/\psi$, and $K^0_S$ decays, corresponding mass vetoes are applied to any combinations of the two or three final state particles. To remove events from the decay $J/\psi \to \mu^+\mu^-$, the muon mass hypothesis is assigned to the selected pion candidates and the mass veto is applied.

To distinguish $B\overline{B}$ events from the dominant jet-like continuum background *BABAR* uses a neural network-based discriminant, NN, which combines the same eight topological variables used in the $B^0 \to \rho^+\rho^-$ analysis, while Belle uses a Fisher discriminant, $\mathcal{F}$, constructed from seven variables (Chapter 4). These discriminants are used as inputs to the ML fits described below. In addition Belle places a loose requirement on $\mathcal{F}$, which removes about 60% of the continuum background and 10% of the signal.

The $B$ meson can decay to $\rho^0\rho^0$ via two polarizations, longitudinal or transverse. These polarizations have different angular distributions and therefore, as described in Section 12.3, significantly different kinematics and average multiplicities of reconstructed candidates per event. For simulated signal events Belle (*BABAR*) finds 1.17 and 1.03 (1.15 and 1.03) $B$ candidates per event for the longitudinal and transverse polarization, respectively. In case of multiple $B$ candidates, the one whose $m_{\mathrm{ES}}$ is closest to the nominal $B$ mass is chosen for Belle, and the one that has the smallest $\chi^2$ for the four-pion vertex is selected for *BABAR*. The reconstruction efficiency for the signal is calculated from MC to be 21.1% (26.5%) for Belle and 22.3% (26.1%) for *BABAR* for the longitudinal (transverse) polarization.

The branching ratio, $\mathcal{B}(B^0 \to \rho^0\rho^0)$, as well as the fraction of longitudinal polarization, $f_L$, are extracted in Belle from an unbinned extended ML fit, using six discriminating variables ($\Delta E$, $m(\pi^+\pi^-)_1$, $m(\pi^+\pi^-)_2$, $\cos\theta_1$, $\cos\theta_2$, $\mathcal{F}$), where $\theta_{1,2}$ allows one to measure the polarization according to Eq. (12.2.5). In *BABAR* the extended ML fit is used to extract not only $\mathcal{B}(B^0 \to \rho^0\rho^0)$ and $f_L$, but also the coefficients of the time-dependent *CP* asymmetry for the longitudinal signal, $C_L^{\rho^0\rho^0}$ and $S_L^{\rho^0\rho^0}$. Hence it uses ten variables: $m_{\mathrm{ES}}$, $\Delta E$, $m(\pi^+\pi^-)_1$, $m(\pi^+\pi^-)_2$, $\cos\theta_1$, $\cos\theta_2$, NN, $c_{\mathrm{tag}}$, $\Delta t$, and $\sigma_{\Delta t}$, where the tagging category $c_{\mathrm{tag}}$ of the $B$-flavor tagging algorithm, introduced in Chapter 8, provides additional background discrimination power, and $\Delta t$ and its error $\sigma_{\Delta t}$ are added in order to include the time-dependent information.

Four categories of signal are distinguished in the fits: the longitudinally and transversely polarized signals and their respective mis-reconstructed components.[80] The fractions of the mis-reconstructed signal are fixed according to MC expectations. Several kinds of background, including a variety of four pion final states, have to be dealt with. While $\Delta E$ and $m_{\mathrm{ES}}$ are powerful in discriminating $B$ decays into four charged pions from other decays, $m(\pi^+\pi^-)$ helps to distinguish signal from non-resonant final states with four charged pions. Belle considers 17 different event types in its fit. In addition to the four signal types, these are continuum, neutral or charged $B$'s decaying into charm or charmless final states and eight peaking background modes ($a_1^\pm(1260)\pi^\mp$, $a_2^\pm\pi^\mp$, $b_1^\pm\pi^\mp$, $\rho^0\pi^+\pi^-$, non-resonant $4\pi^\pm$, $\rho^0 f_0$, $f_0 f_0$ and $f_0\pi^+\pi^-$). The branching fraction of $B^0 \to a_1^\pm(1260)\pi^\mp$ is fixed to the published value $(33.2 \pm 3.0 \pm 3.8) \times 10^{-6}$ (Aubert, 2006aj) and the ones of $B^0 \to a_2^\pm\pi^\mp$ and $B^0 \to b_1^\pm\pi^\mp$ to values based on measured upper limits and theoretical expectation. All other branching fractions, yields, and the continuum shape are allowed to vary in the fit. In the fit from *BABAR*, the following background categories are considered: continuum, $B$ decays into final states containing at least one charm meson, $B$ decays into charmless final states, $\rho^0 f_0$, $f_0 f_0$, $\rho^0\pi^+\pi^-$, non-resonant $4\pi^\pm$, $a_1^\pm(1260)\pi^\mp$, $\rho^0 K^{*0}$, and $f_0 K^{*0}$. All yields are allowed to vary in the fit, except the last two which are fixed to the expected values. Belle and *BABAR* use slightly different sets of decay modes considered in the fit. The modes which are different have quite small contributions to the final sample.

The *p.d.f.*s are obtained from MC for all $B$ decay components. For the $m_{\mathrm{ES}}$, $\Delta E$, and $\mathcal{F}$ or NN *p.d.f.*s, possible differences between real data and the MC modeling are calibrated using a large control sample of $B^0 \to D^-(K^+\pi^-\pi^-)\pi^+$ decays. To obtain the continuum *p.d.f.*, Belle uses off-resonance data and *BABAR* on-resonance sideband data ($m_{\mathrm{ES}} < 5.27\,\mathrm{GeV}/c^2$), with parameters of most *p.d.f.*s left free in the final fit. When possible, the *p.d.f.* for each component is taken to be the product of analytical one-dimensional functions, but correlations as small as 2% between the fit variables are also accounted for using different techniques such as multidimensional *p.d.f.*s or different *p.d.f.*s for different slices of another discriminating variable. For example in Belle, since the shape of $\mathcal{F}$ depends on the flavor tagging quality $r$, it is described in seven bins of the variable $r$.

Figure 17.7.7 shows the projections of the fit results onto $m_{\mathrm{ES}}$ and $m_{\pi^+\pi^-}$ (where the peaks of the $\rho^0$ and $f_0$ can be seen). Table 17.7.1 summarizes the measurements of the branching fraction, $f_L$, and the time-dependent asymmetries for $B^0 \to \rho^0\rho^0$. The results obtained by the two experiments are in agreement with each other. This mode is seen with a significance of 3.1 $\sigma$ by *BABAR* and 3.4 $\sigma$ by Belle, taking into account systematic uncertainties.

[80] At least one track from the signal decay is replaced by one from the accompanying tag $B$ meson in the event for a signal event to be mis-reconstructed.

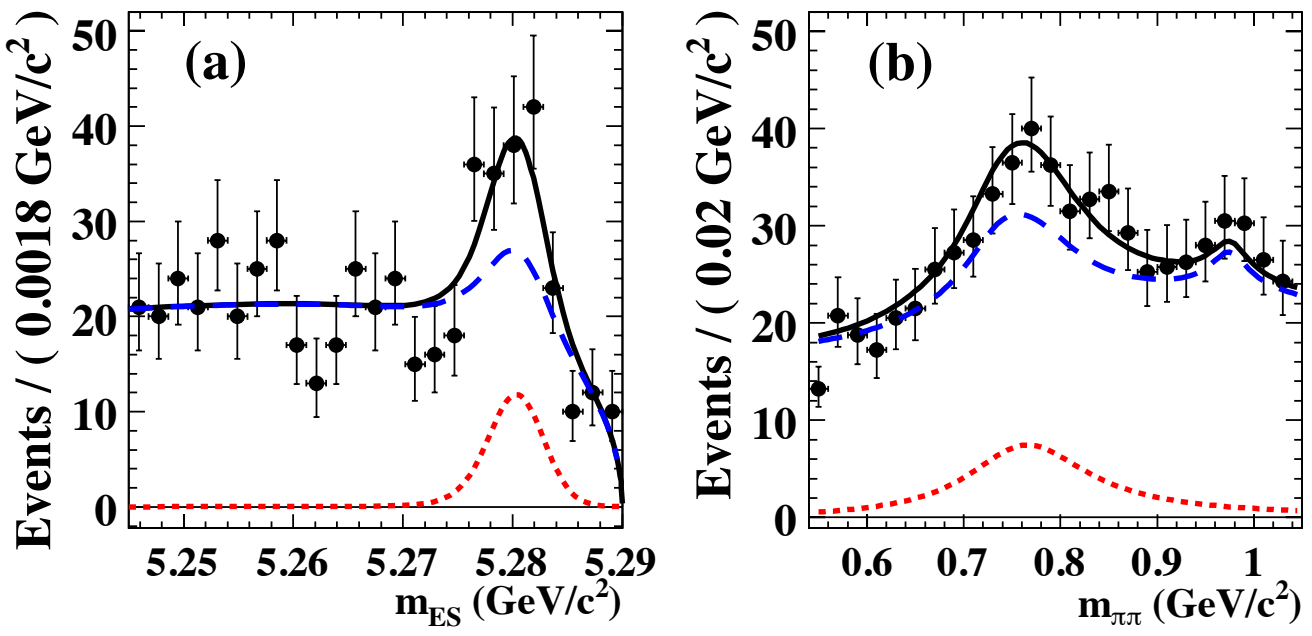

**Figure 17.7.7.** Projections of the multidimensional fit of the *BABAR* $\rho^0\rho^0$ analysis (Aubert, 2008r) onto the (a) $m_{\rm ES}$, and (b) di-pion invariant mass (combining the $m(\pi\pi)_1$ and $m(\pi\pi)_2$ distributions), after a requirement on the signal-to-background probability ratio calculated using all discriminating variables excluding the one plotted, which enhances the fraction of signal events in the sample. This selection has 40% (60%) efficiency for signal for the $m_{\rm ES}$ ($m_{\pi\pi}$) projection. The data points are overlaid by the full *p.d.f.* projection (solid black line). Also shown are the $B^0 \to \rho^0\rho^0$ *p.d.f.* component (dotted line) and the sum of all other *p.d.f.*s (dashed line).

**Table 17.7.1.** Branching fraction, fraction of longitudinal polarization, branching fraction to the longitudinal polarized state, and coefficients of the time-dependent *CP* asymmetry in $B^0 \to \rho^0\rho^0$.

| | *BABAR* (Aubert, 2008r) | Belle (Adachi, 2014) |
|---|---|---|
| $\mathcal{B}_{\rho^0\rho^0}[10^{-6}]$ | $0.92 \pm 0.32 \pm 0.14$ | $1.02 \pm 0.30 \pm 0.15$ |
| $f_L^{\rho^0\rho^0}$ | $0.75\,^{+0.11}_{-0.14} \pm 0.05$ | $0.21\,^{+0.18}_{-0.22} \pm 0.15$ |
| $\mathcal{B}^L_{\rho^0\rho^0}[10^{-6}]$ | $0.69 \pm 0.25 \pm 0.11$ | $0.2 \pm 0.2 \pm 0.1$ |
| $S_L^{\rho^0\rho^0}$ | $0.3 \pm 0.7 \pm 0.2$ | |
| $C_L^{\rho^0\rho^0}$ | $0.2 \pm 0.8 \pm 0.3$ | |

The branching fraction of longitudinally polarized $B^0 \to \rho^0\rho^0$ is an input to the isospin analysis in $B \to \rho\rho$. As pointed out in Section 17.7.1.1, the small branching fraction $\mathcal{B}(B^0 \to \rho^0\rho^0) \approx 10^{-6}$ in comparison with the relatively large branching fraction $\mathcal{B}(B^0 \to \rho^+\rho^-) = (24.2^{+3.1}_{-3.2})$ $\times 10^{-6}$ significantly constrains the penguin uncertainty. The results of the constraint on $\phi_2$ are presented in Section 17.7.7. The time-dependent analysis performed by *BABAR* is a proof of principle of this technique which, with significantly larger data sets, would help determine $\phi_2$ with high precision. The effect of using the measured values of $S$ and $C$ from the *BABAR* analysis in the $\rho\rho$ isospin analysis is evident as the shoulder above $\phi_2 = \alpha \sim 100°$ in Fig. 17.7.12. With larger statistics this input will help to resolve some of the discrete ambiguities inherent in the determination of $\phi_2$ using an isospin analysis.

### 17.7.4 $B^0 \to (\rho\pi)^0$

The decays $B \to \rho\pi$ are similar to $B \to \pi\pi$ and $B \to \rho\rho$ in that they are also dominated by the $b \to u$ tree amplitude and contain $b \to d$ penguin pollution. The *CP*-violation in the $\rho\pi$ final state is related to the Unitarity Triangle angle $\phi_2$. The main complicating factor for measuring $\phi_2$ in $B \to \rho\pi$ is that it is not a *CP*-eigenstate; one can still measure *CP*-violation in $B^0 \to \rho^\pm\pi^\mp$ but it is more challenging than the case for $B^0 \to \pi^+\pi^-$ decays. A measurement of $\phi_2$ is possible in $\rho\pi$ but instead of a triangular isospin relationship between the amplitudes it is pentagonal, requiring measurements of all of the rate and *CP*-violation parameters of the decays $B^0 \to \rho^+\pi^-$, $B^0 \to \rho^-\pi^+$, $B^0 \to \rho^0\pi^0$, $B^+ \to \rho^+\pi^0$, and $B^+ \to \rho^0\pi^+$. Early attempts to constrain $\phi_2$ using these decays in an isospin pentagon analysis were based on a quasi-two-body analysis methodology, measuring decay rates and *CP* asymmetry parameters sensitive to *CP* violating and *CP* conserving observables $S$, $C$, $\Delta S$, $\Delta C$ and $A_{CP}$ (c.f. the $B \to a_1\pi$ time-dependent analysis discussed in Section 17.7.5). Due to the number of parameters involved, extracting $\phi_2$ from $\rho\pi$ via the pentagon relationship yields poor results with many discrete ambiguities.

However, the $B^0 \to \rho^+\pi^-$, $B^0 \to \rho^-\pi^+$, and $B^0 \to \rho^0\pi^0$ all decay to the same final state, $\pi^+\pi^-\pi^0$ and there are regions in the 3-body phase space where there is interference between decay amplitudes. Because of this interference, it is possible to perform an amplitude analysis of the Dalitz plot to extract the complex decay amplitudes. Snyder and Quinn (1993) pointed out that there are enough observables in a full time- and tag-dependent amplitude analysis of the $B^0 \to \pi^+\pi^-\pi^0$ Dalitz plot to simultaneously extract the tree and penguin amplitudes along with the weak phase, $\phi_2$.

The time-dependent rate for $B^0$ ($\overline{B}^0$) decays, $|\mathcal{A}^+_{3\pi}(\Delta t)|^2$ ($|\mathcal{A}^-_{3\pi}(\Delta t)|^2$), is given by:

$$|\mathcal{A}^\pm_{3\pi}(\Delta t)|^2 = \frac{e^{-|\Delta t|/\tau_{B^0}}}{4\tau_{B^0}} \Bigg[ |A_{3\pi}|^2 + |\overline{A}_{3\pi}|^2 \mp \left(|A_{3\pi}|^2 - |\overline{A}_{3\pi}|^2\right)\cos(\Delta m_d \Delta t) \pm 2{\rm Im}\left[\frac{q}{p}\overline{A}_{3\pi}A^*_{3\pi}\right]\sin(\Delta m_d\Delta t)\Bigg] . \quad (17.7.17)$$

The $B^0 \to \pi^+\pi^-\pi^0$ Dalitz plot is dominated by the $\rho$ resonances; it was checked that other contributions ($f_0(980)$, non-resonant, etc.) can safely be neglected at the size of the current data samples. The amplitudes can be written as a sum of terms:

$$A_{3\pi} = f_+A^+ + f_-A^- + f_0A^0 \ , \quad (17.7.18)$$
$$\overline{A}_{3\pi} = f_+\overline{A}^+ + f_-\overline{A}^- + f_0\overline{A}^0 \ , \quad (17.7.19)$$

where the $f_\kappa$ are the Dalitz-plot position-dependent $\rho$ lineshapes and the $A^{+,-,0}$ are Dalitz-plot independent complex amplitudes for the $\rho^+\pi^-$, $\rho^-\pi^+$, and $\rho^0\pi^0$ final states respectively, which contain information on the strong and weak phases. They are indeed related to $\phi_2$ through an isospin relation:

$$e^{2i\phi_2} = \frac{\overline{A}^+ + \overline{A}^- + 2\overline{A}^0}{A^+ + A^- + 2A^0}. \tag{17.7.20}$$

Inserting the amplitudes of Eqs (17.7.18) and (17.7.19) and assuming no *CP* violation in $B^0 - \overline{B}^0$ mixing ($|q/p| = 1$), one obtains the following for the terms appearing in Eq. (17.7.17)

$$\begin{aligned}
|A_{3\pi}|^2 \pm |\overline{A}_{3\pi}|^2 &= \sum_{\kappa\in\{+,-,0\}} |f_\kappa|^2 U_\kappa^\pm + \\
&\sum_{\kappa<\sigma\in\{+,-,0\}} 2\left(\mathrm{Re}\left[f_\kappa f_\sigma^*\right] U_{\kappa\sigma}^{\pm,\mathrm{Re}} - \mathrm{Im}\left[f_\kappa f_\sigma^*\right] U_{\kappa\sigma}^{\pm,\mathrm{Im}}\right) , \\
\mathrm{Im}\left(\frac{q}{p}\overline{A}_{3\pi}A_{3\pi}^*\right) &= \sum_{\kappa\in\{+,-,0\}} |f_\kappa|^2 I_\kappa \; + \\
&\sum_{\kappa<\sigma\in\{+,-,0\}} \left(\mathrm{Re}\left[f_\kappa f_\sigma^*\right] I_{\kappa\sigma}^{\mathrm{Im}} + \mathrm{Im}\left[f_\kappa f_\sigma^*\right] I_{\kappa\sigma}^{\mathrm{Re}}\right),
\end{aligned} \tag{17.7.21}$$

where the coefficients of the bi-linear terms ("the $U$'s and $I$'s") are related to the amplitudes by:

$$\begin{aligned}
U_\kappa^\pm &= |A^\kappa|^2 \pm |\overline{A}^\kappa|^2 , \\
U_{\kappa\sigma}^{\pm,\mathrm{Re}} &= \mathrm{Re}\left[A^\kappa A^{\sigma *} \pm \overline{A}^\kappa \overline{A}^{\sigma *}\right] , \\
U_{\kappa\sigma}^{\pm,\mathrm{Im}} &= \mathrm{Im}\left[A^\kappa A^{\sigma *} \pm \overline{A}^\kappa \overline{A}^{\sigma *}\right] , \\
I_\kappa &= \mathrm{Im}\left[\overline{A}^\kappa A^{\kappa *}\right] , \\
I_{\kappa\sigma}^{\mathrm{Re}} &= \mathrm{Re}\left[\overline{A}^\kappa A^{\sigma *} - \overline{A}^\sigma A^{\kappa *}\right] , \\
I_{\kappa\sigma}^{\mathrm{Im}} &= \mathrm{Im}\left[\overline{A}^\kappa A^{\sigma *} + \overline{A}^\sigma A^{\kappa *}\right] .
\end{aligned} \tag{17.7.22}$$

The above coefficients are used as the fit parameters in the time-dependent maximum likelihood fit to the $B^0 \to \pi^+\pi^-\pi^0$ Dalitz plot. In order to extract the amplitude-level information (*e.g.* the phase $\phi_2$) from the bi-linear coefficients, one can construct a $\chi^2$ quantity using the 27 measured $U$'s and $I$'s and the full correlation matrix and minimize it to find the 12 best fit amplitude parameters.

Both *BABAR* (Lees, 2013c) and Belle (Kusaka, 2008) have performed the time-dependent analysis to the $B^0 \to \pi^+\pi^-\pi^0$ Dalitz plot using the above method. The resulting values, including correlation matrices, for the bi-linear coefficients can be found in the references. The resulting $\phi_2$ contour, with systematic uncertainties taken into account, is shown in Figure 17.7.13 where one can see that the individual measurements are not able to constrain $\phi_2$ significantly. The combined power of the $B$ Factory results is sufficient to start providing important information in terms of our knowledge on $\phi_2$. Solutions appear with values of $\sim 50°$ and $\sim 120°$ that can be used to suppress non-SM solutions when combined with the results from $\pi\pi$ and $\rho\rho$ isospin analyses. A discussion on the extraction of $\phi_2$ using these decays can be found in (Lees, 2013c), where *BABAR* conclude that while the the $U$ and $I$ parameters (and derived quasi-two-body parameters) can be reliably extracted from data, the $1-$C.L. scan for $\phi_2$ is not robust using the full *BABAR* data set.

A significant increase in the available statistics for this mode, which is expected from the next generation of Flavor Factories, will result in the $B^0 \to \pi^+\pi^-\pi^0$ time-dependent Dalitz-plot analysis playing a more prominent role in the determination of $\phi_2$. Given the available level of precision it is not possible to make concrete predictions on the possible sensitivity that one might be able to achieve in the future as there are too many parameters that feed into the determination of $\phi_2$, and many of these are only weakly constrained by the current data.

### 17.7.5 $B^0 \to a_1^\pm(1260)\pi^\mp$

The following describes analyses of $B^0(\overline{B}^0) \to a_1^+(1260)\pi^-$ and $B^0(\overline{B}^0) \to a_1^-(1260)\pi^+$ performed using the quasi-two-body approximation (which is directly analogous to the corresponding early analyses of $B \to \rho\pi$ decays). The $\Delta t$ distributions[81] are given by (Gronau and Zupan, 2006)

$$\begin{aligned}
F_q^{a_1^\pm\pi^\mp}(\Delta t) = (1 \pm A_{CP})\frac{e^{-|\Delta t|/\tau_{B^0}}}{4\tau_{B^0}}\Big\{1 - q\Delta w + \\
q\cdot(1-2w)\Big[(S \pm \Delta S)\sin(\Delta m_d \Delta t) - \\
(C \pm \Delta C)\cos(\Delta m_d \Delta t)\Big]\Big\} ,
\end{aligned} \tag{17.7.23}$$

where $q = +1(-1)$ when the tagging meson $B^0_{\rm tag}$ is a $B^0(\overline{B}^0)$. The time- and flavor-integrated charge asymmetry $A_{CP}$ is the rate asymmetry between $a_1^+(1260)\pi^-$ and $a_1^-(1260)\pi^+$ final states including contributions from both $B^0$ and $\overline{B}^0$ decays. The quantities $S$ and $C$ parameterize mixing-induced *CP* violation related to the angle $\phi_2$, and flavor-dependent direct *CP* violation, respectively. The parameter $\Delta C$ describes the difference in $B^0 - \overline{B}^0$ asymmetries between the $a_1^+(1260)\pi^-$ and $a_1^-(1260)\pi^+$ final states, while $\Delta S$ is related to the strong phase difference between the amplitudes contributing to $B^0 \to a_1^\pm(1260)\pi^\mp$ decays. The parameters $\Delta C$ and $\Delta S$ are insensitive to *CP* violation. Defining the *CP*-averaged rate into a final state $f$ as $\overline{\mathcal{B}}(f) = (\mathcal{B}(B^0 \to f) + \mathcal{B}(\overline{B}^0 \to \overline{f}))/2$, the asymmetry between the *CP*-averaged rates $\overline{\mathcal{B}}(a_1^+(1260)\pi^-)$ and $\overline{\mathcal{B}}(a_1^-(1260)\pi^+)$ is $\Delta C + A_{CP}C$.

The following effective value $\phi_2^{\rm eff}$ of the weak phase $\phi_2$ is obtained (Gronau and Zupan, 2006)

$$\phi_2^{\rm eff} = \frac{1}{4}\Bigg[\arcsin\left(\frac{S+\Delta S}{\sqrt{1-(C+\Delta C)^2}}\right) +$$

[81] Note that the form of the $\Delta t$ distributions involves more parameters than the example described in Eq. 10.2.2. The reason for this is that the $a_1^\pm(1260)\pi^\mp$ final state is not a *CP* eigenstate.

$$\arcsin\left(\frac{S-\Delta S}{\sqrt{1-(C-\Delta C)^2}}\right)\Bigg] . \qquad (17.7.24)$$

A bound on $|\Delta\phi_2|$ is derived (Gronau and Zupan, 2006) from the *CP* asymmetry parameters and from the ratios $R^0_\pm$ and $R^+_\pm$ of *CP*-averaged rates:

$$R^0_+ \equiv \frac{\overline{\lambda}^2 f^2_{a_1}\overline{\mathcal{B}}(K_1^+\pi^-)}{f^2_{K_1}\overline{\mathcal{B}}(a_1^+\pi^-)}, \quad R^0_- \equiv \frac{\overline{\lambda}^2 f^2_{\pi}\overline{\mathcal{B}}(a_1^-K^+)}{f^2_{K}\overline{\mathcal{B}}(a_1^-\pi^+)},$$
$$R^+_+ \equiv \frac{\overline{\lambda}^2 f^2_{a_1}\overline{\mathcal{B}}(K_1^0\pi^+)}{f^2_{K_1}\overline{\mathcal{B}}(a_1^+\pi^-)}, \quad R^+_- \equiv \frac{\overline{\lambda}^2 f^2_{\pi}\overline{\mathcal{B}}(a_1^+K^0)}{f^2_{K}\overline{\mathcal{B}}(a_1^-\pi^+)}. \qquad (17.7.25)$$

The constant $\overline{\lambda}$ is $|V_{us}|/|V_{ud}| = |V_{cd}|/|V_{cs}|$ while $f_K$, $f_\pi$, $f_{a_1(1260)}$, and $f_{K_1}$ are the decay constants of $K$, $\pi$, $a_1(1260)$, and $K_1$ mesons, respectively. The branching fraction measurements are described in detail in Section 17.4.5.5.

The $a_1(1260) \to 3\pi$ decay proceeds mainly through the intermediate states $(\pi\pi)_\rho\pi$ and $(\pi\pi)_\sigma\pi$ (Amsler et al., 2008). Because of the limited number of signal events, *BABAR* and Belle made no attempt to separate the contributions of the dominant $P$-wave $(\pi\pi)_\rho$ and the $S$-wave $(\pi\pi)_\sigma$. The $a_1(1260)$ meson is reconstructed in its $\rho\pi$ decay. A systematic uncertainty is then estimated due to the difference in the selection efficiency. The $B^0 \to a_1^\pm(1260)\pi^\mp$ candidates are reconstructed by combining an $a_1(1260)$ candidate and a charged pion.

Both experiments use the same selection strategy. As for other charmless $B$ decays studied at the $B$ Factories, for this decay mode the dominant background is continuum. To suppress it, the angle $\theta_T$ between the thrust axis of the $B$ candidate and that of the rest of the tracks and neutral clusters in the event, calculated in the CM frame is used. To suppress further combinatorial background in the modes containing an $a_1(1260)$ meson, it is required that the absolute value of the cosine of the angle between the direction of the $\pi$ meson from $a_1(1260) \to \rho\pi$ with respect to the flight direction of the $B$ in the $a_1(1260)$ meson rest frame is less than 0.85. Peaking backgrounds from $B$ decays to final states involving $D^+(\to K^-\pi^+\pi^+, K^0_S\pi^+)$, $D^0(\to K^-\pi^+, K^-\pi^+\pi^0)$, $J/\psi\,(\to \mu^+\mu^-)$ and $K^0_S(\to \pi^+\pi^-)$ mesons are removed by applying the corresponding mass vetoes.

The *CP* asymmetry parameters are measured at *BABAR* using an unbinned extended ML fit using the selected sample of $B^0 \to a_1^\pm(1260)\pi^\mp$ with the input observables $\Delta E$, $m_{\rm ES}$, a Fisher discriminant $\mathcal{F}$ (described in Chapter 9), $m_{a_1(1260)}$, a helicity angle $\mathcal{H}$, and $\Delta t$ (Aubert, 2007ae).

At Belle a two step procedure is used. Firstly the signal yield (thus the branching fraction measurement) in $a_1(1260)\pi$ decay mode is obtained from an extended ML fit to the selected sample $B \to a_1^\pm(1260)\pi^\mp$ with the input observables $\Delta E$, $m_{a_1(1260)}$, $\mathcal{H}$, and $\mathcal{F}$. The $m_{\rm ES}$ distribution is not included in the fit, as it is used to select the best $B$ candidate in cases of multiple candidates per event. After that a one-dimensional unbinned maximum likelihood fit is performed to the $\Delta t$ distribution of events in the signal region $|\Delta E| < 0.04$ GeV, using the branching fraction measurement from the first step to provide the event-dependent signal probability for each component (Dalseno, 2012).

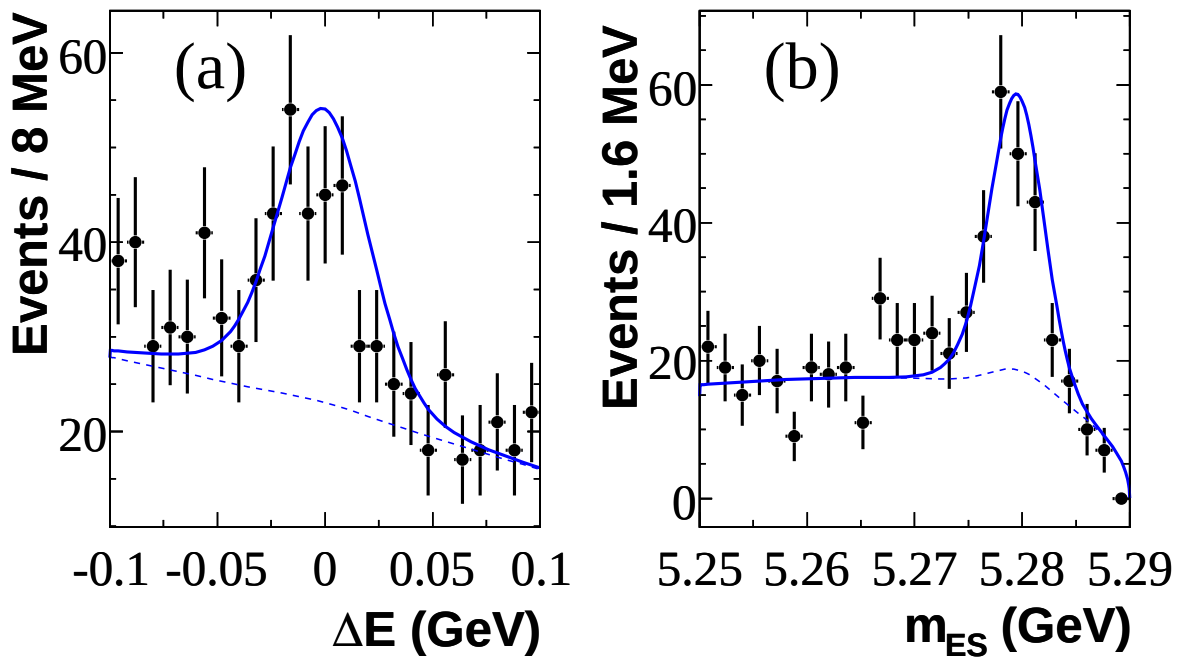

**Figure 17.7.8.** Signal enhanced projections of a) $\Delta E$, b) $m_{\rm ES}$ from the *BABAR* $B \to a_1\pi$ analysis. Points represent on-resonance data, dotted lines the sum of all backgrounds, and solid lines the full fit function. These plots are made with a cut on the signal-to-continuum likelihood ratios excluding the variable being plotted (from Aubert (2007ae)).

In both experiments the *p.d.f.*s for signal and $B\overline{B}$ backgrounds are determined from MC distributions in each observable. For the continuum background the functional forms and initial parameter values of the *p.d.f.*s are established with off-resonance data. The *p.d.f.* of the $a_1(1260)$ meson invariant mass distribution of signal events is parameterized as a relativistic Breit-Wigner line-shape with a mass-dependent width (Armstrong et al., 1990).

At *BABAR*, the maximum likelihood fit to a sample of 29300 events results in a signal yield of $608 \pm 53$, of which $461 \pm 46$ have their flavor identified. Fig. 17.7.8 shows distributions of $m_{\rm ES}$ and $\Delta E$, enhanced in signal content by requirements on the signal-to-continuum likelihood ratios using all discriminating variables other than the one plotted.

With a data sample of $304 \times 10^6$ $B\overline{B}$ pairs, the following results for the *CP* asymmetries are obtained (Aubert, 2007ae)

$$\begin{aligned} S &= \phantom{-}0.37 \pm 0.21 \pm 0.07, \\ C &= -0.10 \pm 0.15 \pm 0.09, \\ A_{CP} &= -0.07 \pm 0.07 \pm 0.02, \end{aligned} \qquad (17.7.26)$$

where the errors are statistical and systematic in nature, respectively. For the *CP* conserving parameters one obtains

$$\begin{aligned} \Delta S &= -0.14 \pm 0.21 \pm 0.06, \\ \Delta C &= \phantom{-}0.26 \pm 0.15 \pm 0.07. \end{aligned} \qquad (17.7.27)$$

At Belle (Dalseno, 2012), the fit to 83799 $a_1^\pm(1260)\pi^\mp$ candidates in the signal region results in

$$\begin{aligned} S &= -0.51 \pm 0.14 \pm 0.08, \\ C &= -0.01 \pm 0.11 \pm 0.09, \\ A_{CP} &= -0.06 \pm 0.05 \pm 0.07, \end{aligned} \qquad (17.7.28)$$

and the *CP* conserving parameters obtained are

$$\Delta S = -0.09 \pm 0.14 \pm 0.06,$$

$$\Delta C = \quad 0.54 \pm 0.11 \pm 0.07. \qquad (17.7.29)$$

Figure 17.7.9 shows the $\Delta t$ and *CP* asymmetry distributions obtained for the Belle analysis. From a likelihood scan, the significance of mixing-induced *CP* violation is found by Belle to be $3.1\sigma$ including systematic uncertainties. There is a discrepancy at the level of $3.2\ \sigma$ between the *BABAR* and Belle values of $S$, which requires more data to resolve.

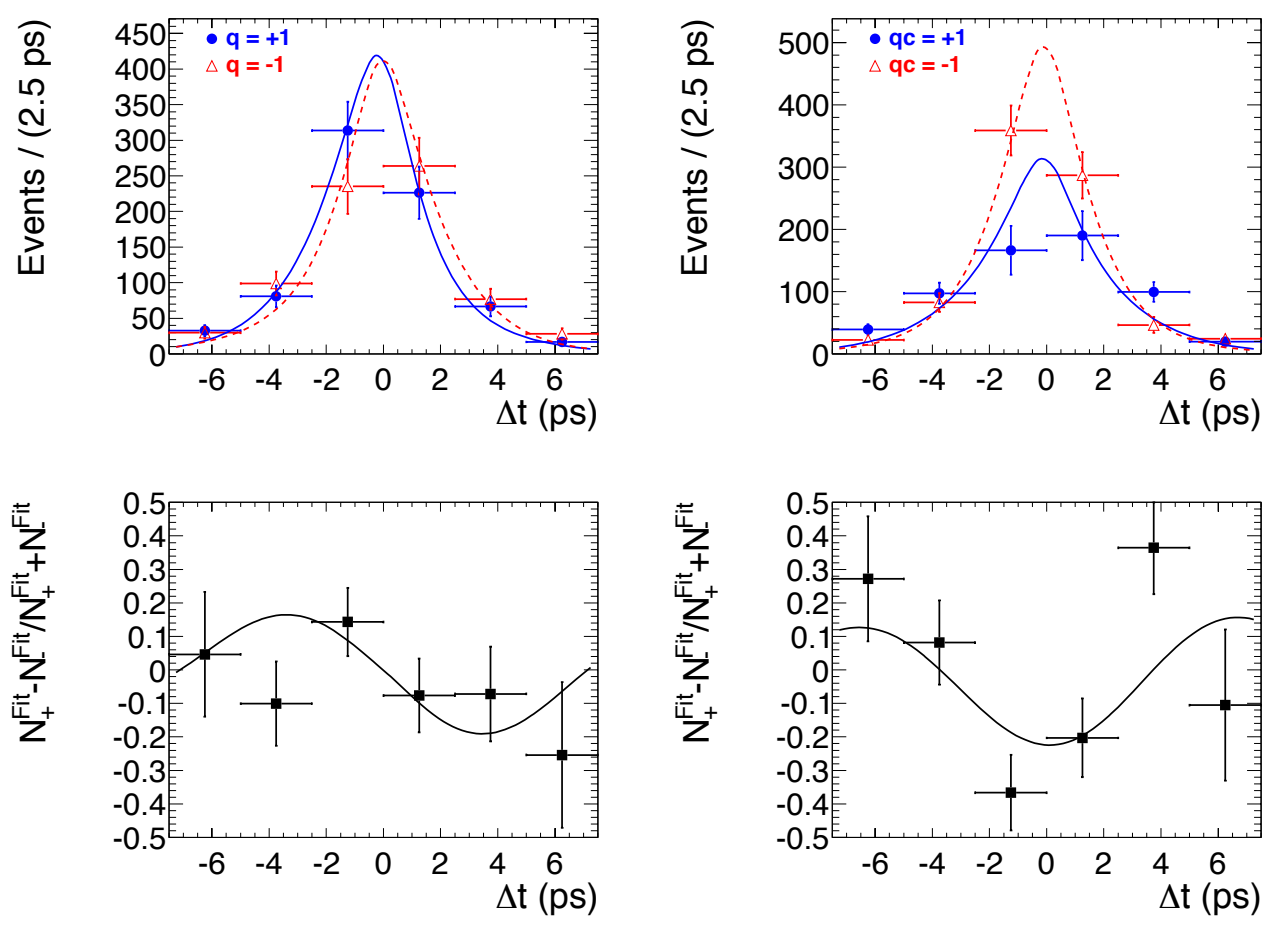

**Figure 17.7.9.** Background subtracted time-dependent fit results for $B^0 \to a_1^{\pm}(1260)\pi^{\mp}$(Dalseno, 2012). (a) and (b) show the $\Delta t$ distributions for the $B^0_{\rm tag}$ flavor $q$ and the product of the $B^0_{\rm tag}$ flavor and $a_1$ charge $qc$, respectively. The left plot shows the effect of flavor-dependent *CP* violation while the plot on the right shows the effects of the *CP* conserving parameters. The solid blue and dashed red curves represent the $\Delta t$ distributions for $q = +1$ and $q = -1$, respectively. (c) and (d) show the asymmetry of the plots immediately above them, $(N_+^{\rm Fit} - N_-^{\rm Fit})/(N_+^{\rm Fit} + N_-^{\rm Fit})$, where $N_+^{\rm Fit}$ ($N_-^{\rm Fit}$) is the measured signal yield of positive (negative) quantities in bins of $\Delta t$. All curves are the expected ones for the values in Eqs (17.7.28) and (17.7.29).

The main contributions to the systematic error on the signal parameters come from the *p.d.f.* parameterization, uncertainty due to *CP* violation present in the $B\overline{B}$ background and uncertainty due to the interference between $B^0 \to a_1^{\pm}(1260)\pi^{\mp}$ and other $4\pi$ final states.

$\phi_2^{\rm eff}$ can be determined from Eq. (17.7.24) up to a four-fold discrete ambiguity in the range $[0°, 180°]$. This can be reduced to a two-fold ambiguity with the assumption that the relative strong phase of the tree amplitudes of the $B^0$ decays to $a_1^-(1260)\pi^+$ and $a_1^+(1260)\pi^-$ is much smaller than 90° (Gronau and Zupan, 2006), as predicted by QCD factorization (Beneke and Neubert, 2003b). This assumption is valid to leading order in $1/m_b$ (Bauer and Pirjol, 2004; Bauer, Pirjol, Rothstein, and Stewart, 2004). Under this assumption, the two solutions at *BABAR* are $(78.6 \pm 7.3)°$ and $(11.4 \pm 7.3)°$. At Belle the four obtained solutions for $\phi_2^{\rm eff}$ are $(-17.3 \pm 6.6 \pm 4.8)°$, $(41.6 \pm 6.2 \pm 3.4)°$, $(48.4 \pm 6.2 \pm 3.4)°$, and $(107.3 \pm 6.6 \pm 4.8)°$. Multiple solutions appear in the fits for reasons analogous to the $\pi\pi$ and $\rho\rho$ constraints on $\phi_2$. The first of the two solutions quoted for *BABAR* is compatible with the results of Standard Model based fits to the Unitarity Triangle to be presented in Section 25.1. In Belle the solution most compatible with the results of Standard Model based fits is the last quoted one.

In *BABAR* a MC technique is used to estimate a probability region for the bound on $|\Delta\phi_2|$. The *CP*-averaged rates and *CP* asymmetry parameters used in estimating the bounds are generated according to the experimental distributions. The input values of branching fractions are those presented in Section 17.4.5.5 while *CP* asymmetry parameters are from this section (Eq. 17.7.26). For the decay constants the following values are used: $f_\pi = (130.4 \pm 0.2)$ MeV (Amsler et al., 2008), $f_K = (155.5 \pm 0.9)$ MeV (Amsler et al., 2008), $f_{a_1} = (203 \pm 18)$ MeV (Cheng and Yang, 2007), and $f_{K_1} = 207$ MeV (Bloch, Kalinovsky, Roberts, and Schmidt, 1999). For $f_{K_1}$ an uncertainty of 20 MeV is assumed. For the constant $\overline{\lambda}$ the value 0.23 (Amsler et al., 2008) is used.

For each set of generated values, the bound on $|\Delta\phi_2|$ is evaluated. The limits on $|\Delta\phi_2|$ are obtained by counting the fraction of bounds within a given value and the results are $|\Delta\phi_2| < 11°(13°)$ at 68% (90%) probability (Aubert, 2010d). Combining the solution near 90°, consistent with the results of global CKM fits, with the bound on $|\phi_2 - \phi_2^{\rm eff}|$ we measure the weak phase $\phi_2 = (79 \pm 7 \pm 11)°$.

This solution is in agreement with the value of $\phi_2$ found in the analyses of $B \to \pi\pi$, $B \to \rho\rho$, and $B \to \rho\pi$ decays. This measurement is currently limited by statistics and a substantial improvement of its precision may come from a future super flavor factory.

### 17.7.6 $SU(3)$ constraint using $B^0 \to \rho^+\rho^-$, and $B^+ \to K^{*0}\rho^+$

Table 17.7.2 summarizes the measurements of the time-dependent asymmetries with their correlation coefficient $\rho_{S,C}$, branching fractions, and $f_L$ for the $B$ Factory results on $B^0 \to \rho^+\rho^-$ and $B^+ \to K^{*0}\rho^+$. The effect of correlated systematic uncertainties on the ratio of branching fractions of these two decay channels is negligible. The constraint obtained from the $B$ Factory data on $\phi_2$ following the BGRS procedure outlined in Section 17.7.1.4 is shown in Figure 17.7.10. There are two overlapping solutions consistent with the SM. These two solutions differ by the magnitude of $\delta_{PT}$, where values of $\delta_{PT} \sim 0$ correspond to the left of the two central peaks in the figure. The determination of $\phi_2$ using this approach only weakly constrains this phase difference given the data from the $B$ Factories, however QCD factorization calculations favor a small phase difference and the solution with $|\delta_{PT}| \sim 0$ is preferred (Beneke, Buchalla, Neubert, and Sachrajda, 1999, 2000, 2001). This favored solution is summarized in Table 17.7.2 for the individual and combined $B$ Factory results where the requirement that $|\delta_{PT}| < 90°$ is imposed. While the *BABAR* data give a more precise value of $\phi_2$ with this method, the Belle data provide a more stringent constraint on $\delta_{PT}$ than *BABAR*. This can be seen in Figure 17.7.10 as the mirror solution at $\phi_2 \sim 5°$ is less probable than the other possible solutions. The ratio of penguin to tree amplitudes (see Section 17.7.1.4) obtained from the data is $r_{PT} = 0.10 \pm 0.04$.

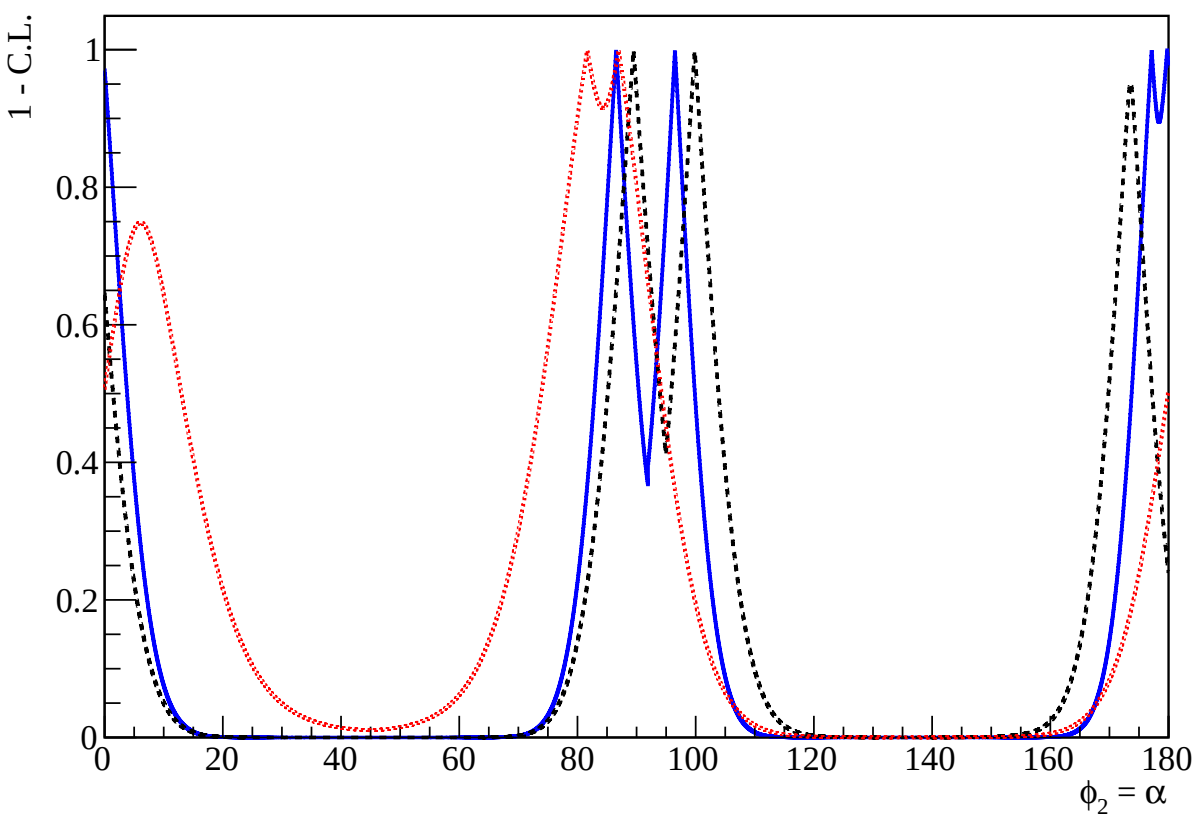

**Figure 17.7.10.** The $1 -$ C.L. function obtained on $\phi_2 = \alpha$ from the $B$ Factories (solid blue) using $B^0 \to \rho^+\rho^-$, and $B^+ \to K^{*0}\rho^+$ decays and the BGRS method discussed in the text. The constraints from *BABAR* (dashed black) and Belle (dotted red) are also shown. The curves shown include solutions for all possible values of $\delta_{PT}$.

### 17.7.7 Summary

First we discuss the constraints on $\phi_2$ obtained by the $B$ Factories mode-by-mode and then on the 'average' of the $B \to \pi\pi/\rho\rho$ and the $B \to \rho\pi \to 3\pi$ Dalitz results. The mode $B \to a_1(1260)\pi$ can play an important role in constraining $\phi_2$; however, at the time of writing this book the $SU(3)$ constraint from $B \to a_1(1260)\pi$ is typically not included in global averages together with the $SU(2)$ constraints, as it is believed that the former measurements have to account for $SU(3)$ breaking effects which may not be straight forward to compute, whereas the latter are agreed to have small theoretical uncertainties. A 'naïve' weighted average of the two sets of measurements is given at the end. The $SU(3)$ constraint from $B^0 \to \rho^+\rho^-$ and $B^+ \to K^{*0}\rho^+$ discussed in Section 17.7.6 is completely correlated with inputs for the isospin analysis and this result is used only as a cross check of the underlying theoretical interpretation of $S$ and $C$. The averaging procedure adopted here is to combine the $\chi^2(\phi_2)$ distributions obtained from each individual constraint ($B \to \pi\pi$, $\pi^+\pi^-\pi^0$ and $\rho\rho$), and from this determine the value of $1 -$ C.L. as a function of $\phi_2$. The value of $\phi_2$ obtained near 90° is reported as the SM value of this angle. The naïve average including $B \to a_1(1260)\pi$ is simply the weighted average obtained from the $\chi^2(\phi_2)$ combination described in this section with the result of the $SU(3)$ constraint from Section 17.7.5.

The constraint on $\phi_2$ obtained from $B \to \pi\pi$ decays from Belle and *BABAR* is shown in Figure 17.7.11. The eight solutions corresponding to the eight-fold ambiguity inherent in the isospin analysis are visible. The solution near the zero value is suppressed by physical constraints on possible magnitudes of penguin contributions within the SM (Bona et al., 2007a). Furthermore one can exclude allowed values around $\phi_2 \sim 40 - 50°$ as can be seen from the figure.

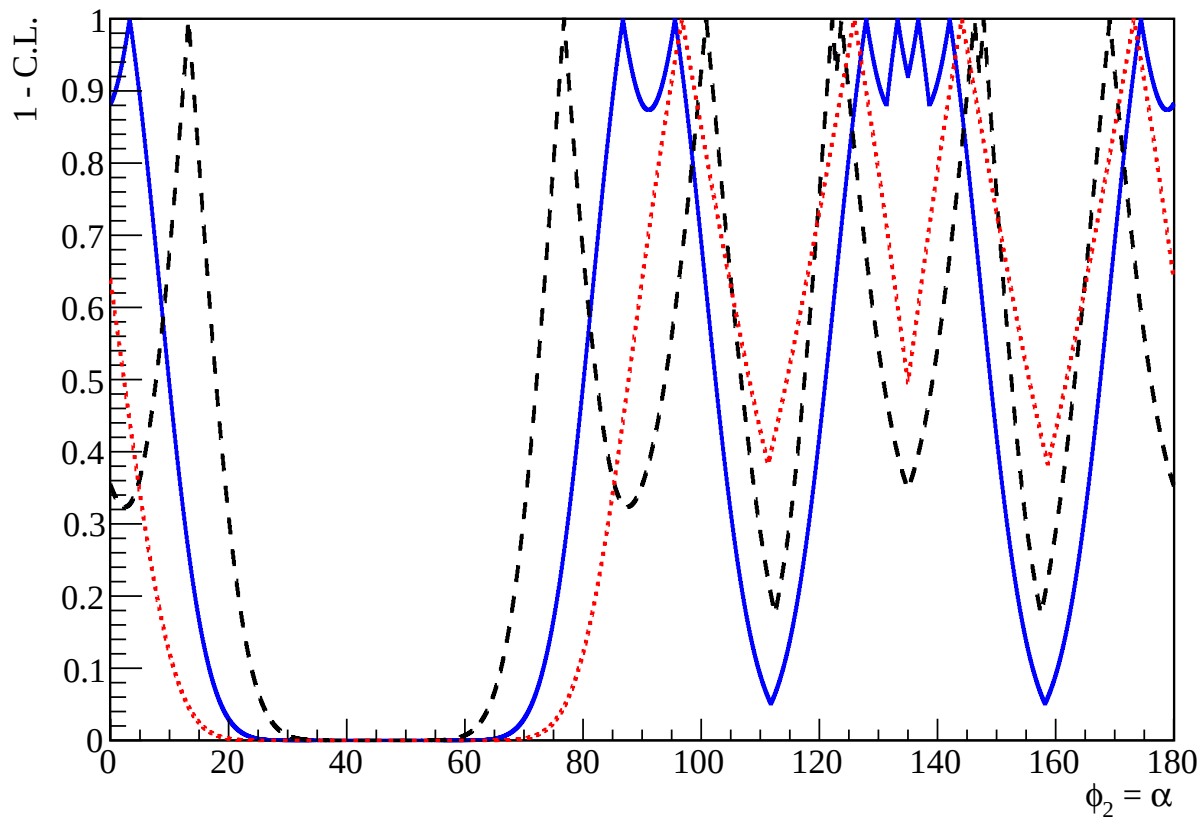

**Figure 17.7.11.** The solid (blue) curve shows the $1 -$ C.L. function obtained on $\phi_2 = \alpha$ from the $B$ Factories using the Gronau-London isospin analysis for $B \to \pi\pi$ decays. The dashed (black) curve shows the *BABAR* only constraint, the dotted (red) curve shows the Belle only constraint. As discussed in the text, the solutions near $\phi_2 = 0$ are unphysical and can be removed by extending the isospin analysis.

The constraint on $\phi_2$ obtained from the *BABAR* and Belle analyses of $B \to \rho\rho$ is shown in Figure 17.7.12. One gets two values, one near zero and the other near 90°. The underlying eight-fold ambiguity expected for the isospin analysis is not visible, since the isospin triangles, illustrated in Figure 17.7.2, are essentially flat as a result of the equally large branching fractions of the $\rho^+\rho^-$ and $\rho^+\rho^0$ modes, and the small value of $\rho^0\rho^0$. The value obtained for the solution consistent with the SM expectation is $\phi_2 = (89.9^{+5.4}_{-5.6})°$. The accurate value obtained using isospin symmetry with $\rho\rho$ decays is similar to that attained using the BGRS $SU(3)$ flavor based approach discussed in Section 17.7.6. The individual *BABAR* and Belle results are $(92.7 \pm 6.3)°$ and $(84.3^{+12.4}_{-12.8})°$, respectively. The better accuracy of *BABAR* with respect to Belle is partly due to the time-dependent *CP* asymmetry measurement of $\rho^0\rho^0$ final state, the more accurate measurement of $\rho^+\rho^-$, and partly due to the accuracy on the $\rho^\pm\rho^0$ branching fraction from *BABAR*. One can see from the figure that the $\rho\rho$ modes exclude values of $\phi_2$ around $\sim 50°$ and $\sim 140°$. These decays help to discard unphysical solutions resulting from the $B \to \pi\pi$ isospin analysis.

**Table 17.7.2.** A summary of experimental inputs used for the BGRS method, along with the constraints on $\phi_2$ derived using this method. ‡The solution given for $\phi_2$ is the one in agreement with the SM with the additional requirement that $|\delta_{PT}| < 90°$ (consistent with theoretical expectations (Beneke, Buchalla, Neubert, and Sachrajda, 1999, 2000, 2001)).

| | | BABAR | Belle | Average |
|---|---|---|---|---|
| $S(\rho^+\rho^-)$ | | $-0.17 \pm 0.20^{+0.05}_{-0.06}$ | $+0.19 \pm 0.30 \pm 0.08$ | $-0.05 \pm 0.17$ |
| $C(\rho^+\rho^-)$ | | $+0.01 \pm 0.15 \pm 0.06$ | $-0.16 \pm 0.21 \pm 0.08$ | $-0.06 \pm 0.13$ |
| $\rho_{S,C}$ | | $-0.035$ | $0.10$ | $0.009$ |
| $\mathcal{B}(\rho^+\rho^-)$ | $[10^{-6}]$ | $25.5 \pm 2.1^{+3.6}_{-3.9}$ | $22.8 \pm 3.8\,^{+2.3}_{-2.6}$ | $24.2 \pm 3.1$ |
| $\mathcal{B}(K^{*0}\rho^+)$ | $[10^{-6}]$ | $9.6 \pm 1.7 \pm 1.5$ | $8.9 \pm 1.7 \pm 1.2$ | $9.2 \pm 1.5$ |
| $f_L\,(\rho^+\rho^-)$ | | $0.992 \pm 0.024^{+0.026}_{-0.013}$ | $0.941\,^{+0.034}_{-0.040} \pm 0.030$ | $0.978 \pm 0.023$ |
| $f_L\,(K^{*0}\rho^+)$ | | $0.52 \pm 0.10 \pm 0.04$ | $0.43 \pm 0.11^{+0.05}_{-0.02}$ | $0.48 \pm 0.08$ |
| $\phi_2 = \alpha$ (SM solution‡) (°) | | $89.5^{+6.7}_{-6.5}$ | $81.7^{+11.4}_{-11.3}$ | $86.5^{+7.3}_{-5.4}$ |

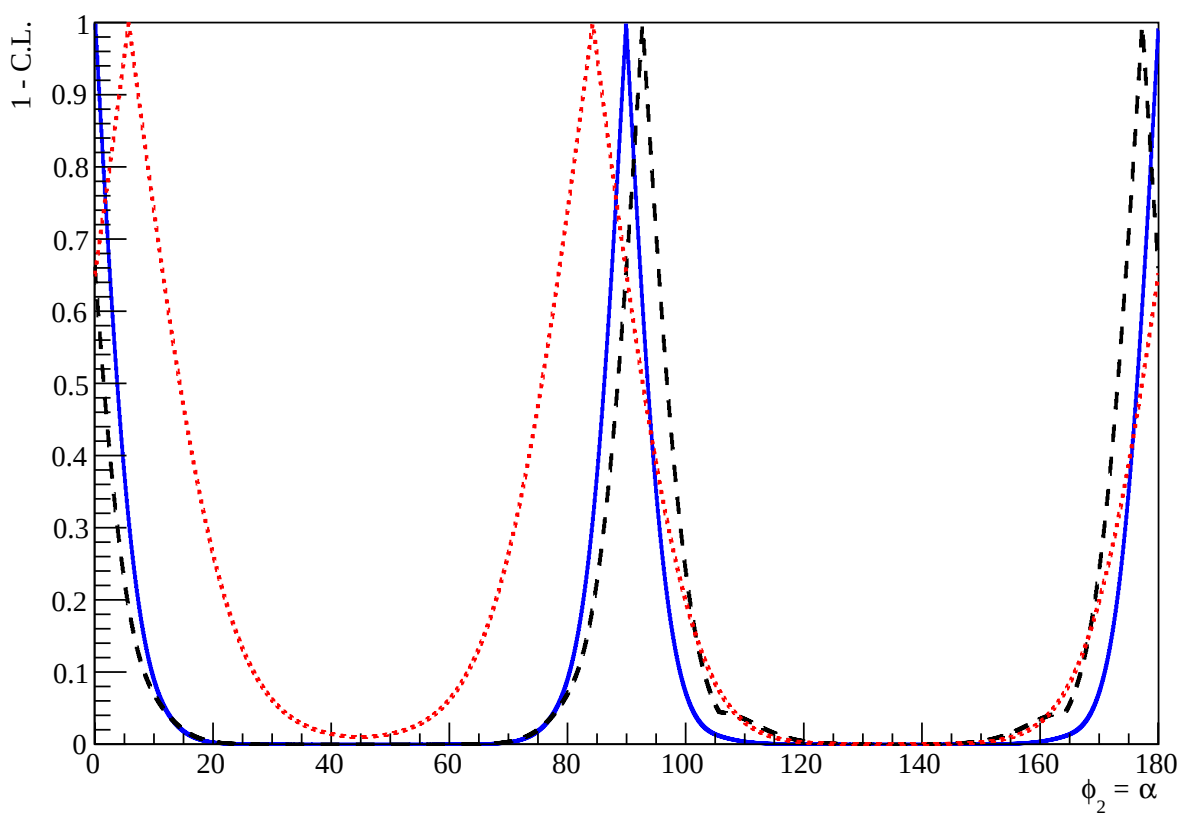

**Figure 17.7.12.** The solid (blue) curve shows the $1 -$ C.L. function obtained on $\phi_2 = \alpha$ from the $B$ Factories using the Gronau-London isospin analysis for $B \to \rho\rho$ decays. The dashed (black) curve shows the BABAR only constraint, the dotted (red) curve shows the Belle only constraint.

The constraint on $\phi_2$ from $B^0 \to \pi^+\pi^-\pi^0$ decays obtained by the $B$ Factories is shown in Figure 17.7.13. While these results do not completely exclude any interval, they do strongly suppress unphysical values. It is the ability of these results to resolve ambiguities that makes the analysis of $B^0 \to \pi^+\pi^-\pi^0$ very important. In the longer term it is expected that the measurement of $\phi_2$ with the time-dependent Dalitz-plot method will provide the most precise value of this angle, however it is not yet possible to make estimates of how much more data will be required for the precision on $\phi_2$ from these decays to surpass the theoretical uncertainty limits of $\sim 1°$ found in the other modes. At the time of compiling this book, it was not straightforward to perform the joint isospin and Dalitz-plot analysis as originally published by Belle. As a result we resorted to using Dalitz-plot results from each experiment to compute an average of $\phi_2$ from $B^0 \to \pi^+\pi^-\pi^0$ decays. Looking to the future, it will be important to probe the Dalitz plot beyond $\rho\pi$ intermediate states in order to maximize our sensitivity to possible second order NP effects that may be manifest in nature. Theoretical work on how one might identify underlying non-SM contributions from three body final states is ongoing.

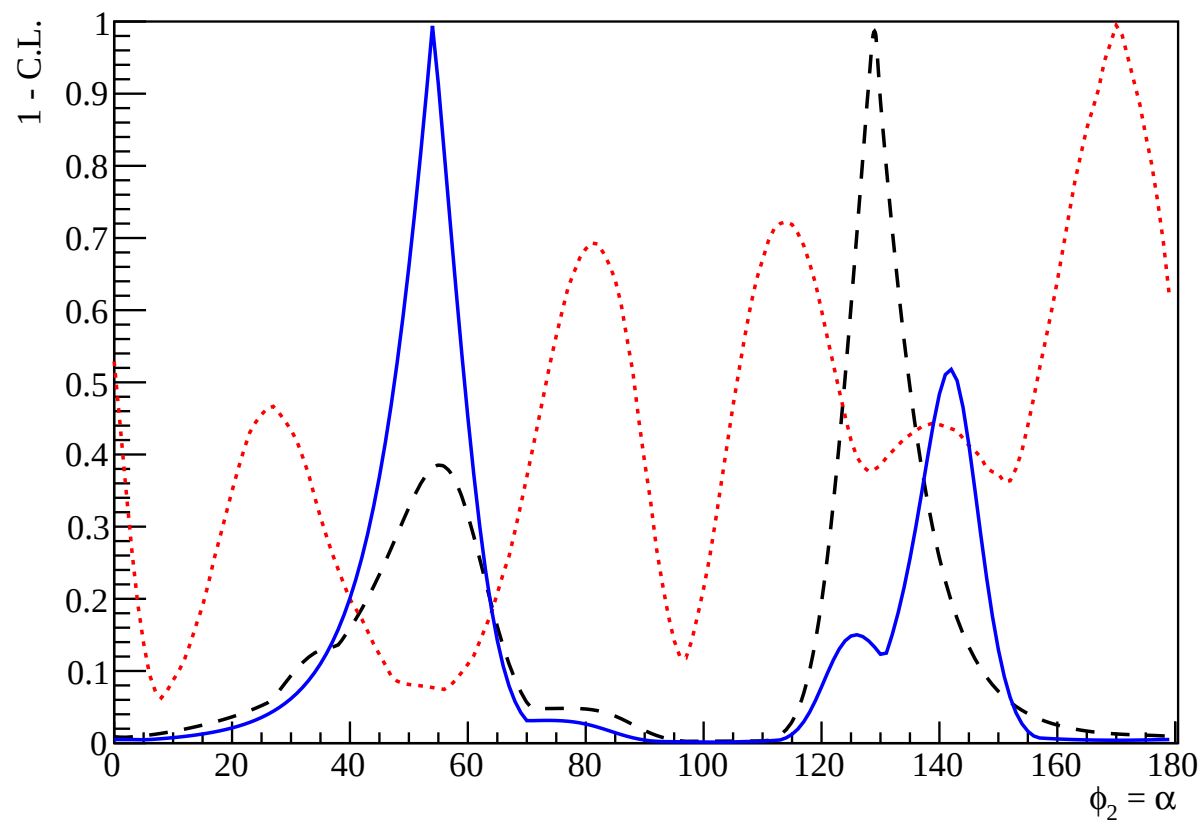

**Figure 17.7.13.** The solid (blue) curve shows $1-$C.L. function obtained on $\phi_2 = \alpha$ from the $B$ Factories using the Dalitz-plot analysis of $B \to \pi^+\pi^-\pi^0$ decays. The dashed (black) curve shows the BABAR only constraint, the dotted curve (red) shows the Belle only constraint.

The $B$ Factories results for $\phi_2$ from $B \to \pi\pi$, $\rho\rho$, and $\pi^+\pi^-\pi^0$ decays are summarized in Figure 17.7.14. Initial measurements from $B \to \pi\pi$ decays excluded the range of $40 - 50°$ while at the same time highlighted the issue of understanding penguin contributions in detail. As can be seen from the combined $B$ Factories result (the dashed line in the figure), it was not possible to measure the angle precisely with this channel alone even with full data sample obtained by both experiments. The measurement of $\phi_2$ using $B \to \rho\rho$ decays (the dashed-dotted line in the figure) provides the most accurate value; it also suppresses the unphysical solution near $\sim 140°$ as well as those in the range $40° - 50°$. Early quasi-two-body analyses of $B \to \rho\pi \to \pi^+\pi^-\pi^0$ provided an interesting third alternative to constrain this angle. However, as the data sets of $B$ Factories increased, it became apparent that a time-dependent Dalitz analysis was required. The value of $\phi_2$ from $\pi^+\pi^-\pi^0$ is also shown (the dotted line). This result plays an important role in suppressing those values inconsistent with the SM solution for the Unitarity Triangle. Noting that physical penguin contributions to $B \to \pi\pi$ suppress $\phi_2 \sim 0$ leaves

$$\phi_2 = \alpha = (88 \pm 5)°, \qquad (17.7.30)$$

which is consistent with the SM expectations as discussed in Section 25.1. While $B \to \rho\rho$ dominates the measured value of the angle, there is a theoretical uncertainty from isospin breaking currently of the order of $\sim 1°$. It is worth noting that $B \to \pi\pi$ decays also have a theoretical limit from isospin breaking at a similar level. A high statistics study of $B \to 4\pi$ will require an amplitude analysis in order to fully utilize interference between the various intermediate states and remove experimental limitations of the quasi-two-body approximation.

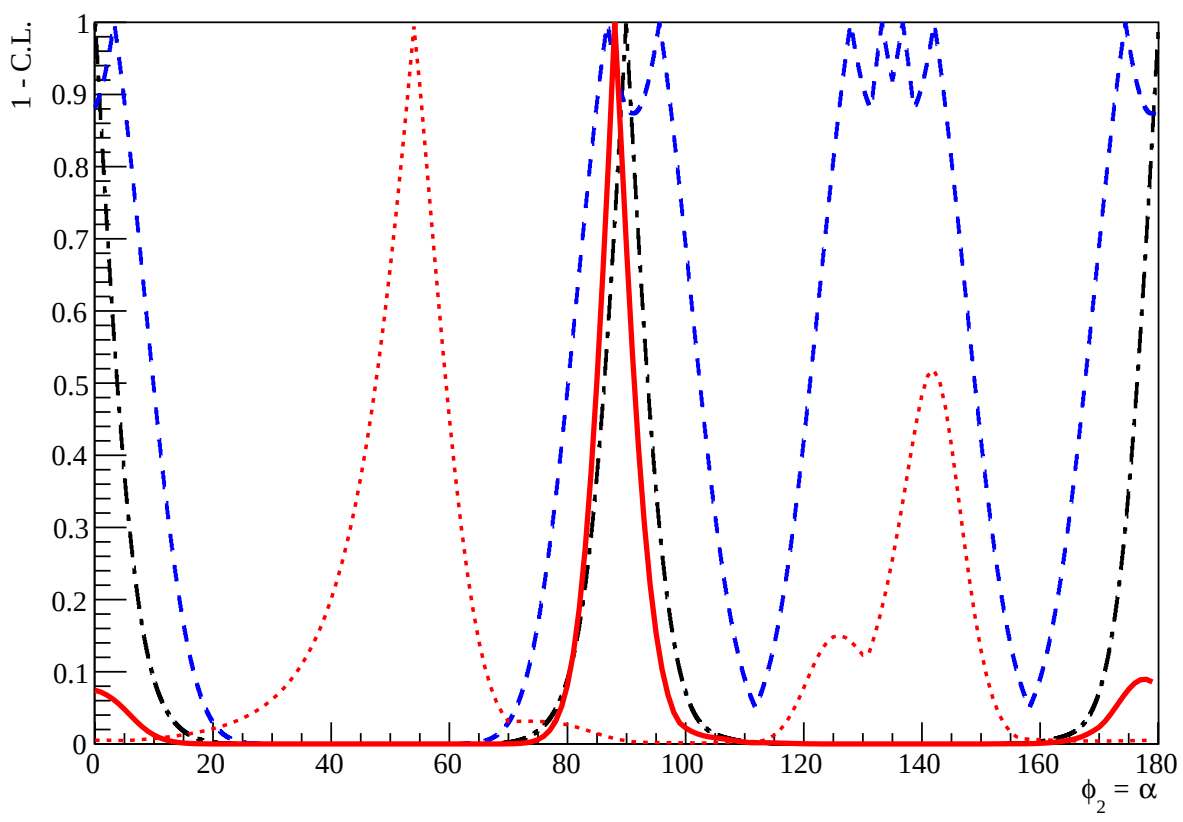

**Figure 17.7.14.** The $1 -$ C.L. function obtained on $\phi_2 = \alpha$ from the $B$ Factories (solid thick red curve). Constraints obtained using the Gronau-London isospin analysis for (dashed blue) $B \to \pi\pi$ and (dash-dotted black) $\rho\rho$ decays as well as the Dalitz analysis for (dotted red) $\pi^+\pi^-\pi^0$ decays are shown.

It is interesting to note that the most probable value from the $B^0 \to \pi^+\pi^-\pi^0$ Dalitz-plot analysis peaks around 55° which is somewhat smaller than the SM expectation. The difference is not significant, however a higher statistics study is called for in order to improve the precision of the $B^0 \to \pi^+\pi^-\pi^0$ Dalitz result.

The last constraint on $\phi_2$ comes from time-dependent measurements of $B \to a_1(1260)\pi$ decays with input from $SU(3)$ related modes in order to constrain penguins. The accuracy obtained using this method on $\phi_2^{\rm eff}$ is $\pm 7°$ for each experiment, with an additional $\pm 11°$ uncertainty from penguin contributions. Usually the community does not include $a_1(1260)\pi$ results in the global average of $\phi_2$ to avoid the complexity of the $SU(3)$ analysis. On the other hand one can compute a 'naïve' weighted average for $\phi_2$ from the combination of $B \to \pi\pi$, $\rho\rho$, $\pi^+\pi^-\pi^0$ decays with the $B \to a_1(1260)\pi$ $SU(3)$ value of $(79 \pm 7 \pm 11)°$ (from *BABAR*). This gives an average of

$$\phi_2 = \alpha = (87 \pm 5)°. \qquad (17.7.31)$$

The measurement of this angle of the Unitarity Triangle is a direct constraint to test the CKM matrix description of quark mixing and thus the KM description of *CP* violation in the SM. The use of $\phi_2$ in this context is discussed in Chapter 25. The presence of penguin contributions, while a nuisance in terms of the determination of $\phi_2$, means that these decays are also sensitive to NP effects manifest in loops. High precision measurements of $\phi_2$ in the different modes can be compared in order to search for NP in analogy with the ongoing searches via penguin measurements of $\phi_1$ discussed in Section 17.6.

One of the great successes of the *BABAR* and Belle $B$ Factories has been the confirmation that the CKM matrix provides the leading order description of *CP* violation in the quark sector. The next generation of experiments will focus on searches for possible second order *CP* violation effects beyond the SM. This will require detailed studies of the interference patterns manifest in three and four body final states.

In the future it may be possible to include constraints on $\phi_2$ obtained using additional modes such as $B^0 \to K\overline{K}\pi\pi$ and $B \to a_1\rho$. The precision of the next generation super flavor factory will approach theoretical limits on isospin breaking for $B \to \pi\pi$ and $\rho\rho$, so it will be important for experimentalists and theorists alike to continue to explore all possible avenues to improve the precision on this weak angle.

## 17.8 $\phi_3$, or $\gamma$

***Editors:***
*Fernando Martinez-Vidal (BABAR)*
*Karim Trabelsi (Belle)*
*Ikaros Bigi (theory)*

**Additional section writers:**
*Giovanni Marchiori, Gagan Mohanty, Anton Poluektov, Matteo Rama, Abner Soffer*

### 17.8.1 Introduction

While $\phi_1$ and $\phi_2$ have been determined to a good level of precision, knowledge of $\phi_3 = \gamma \equiv \arg\left[-V_{ud}V_{ub}^*/V_{cd}V_{cb}^*\right]$ (see Section 16.4) is still limited by the small branching fractions of the processes used in its measurement. The most powerful methods for measuring this angle in a theoretically clean way are based on the interference between $b \to c\bar{u}s$ and $b \to u\bar{c}s$ tree amplitudes in the charged-$B$ meson decays to open-charm final states, $B^- \to D^{(*)}K^{(*)-}$ (charge-conjugate modes are implied here and throughout the text unless otherwise specified). The interference is between $B^- \to DK^-$ followed by a $D \to f$ decay and $B^- \to \overline{D}K^-$ followed by a $\overline{D} \to f$ decay, where $f$ is any common final state of $D$ and $\overline{D}$ mesons (Fig. 17.8.1).

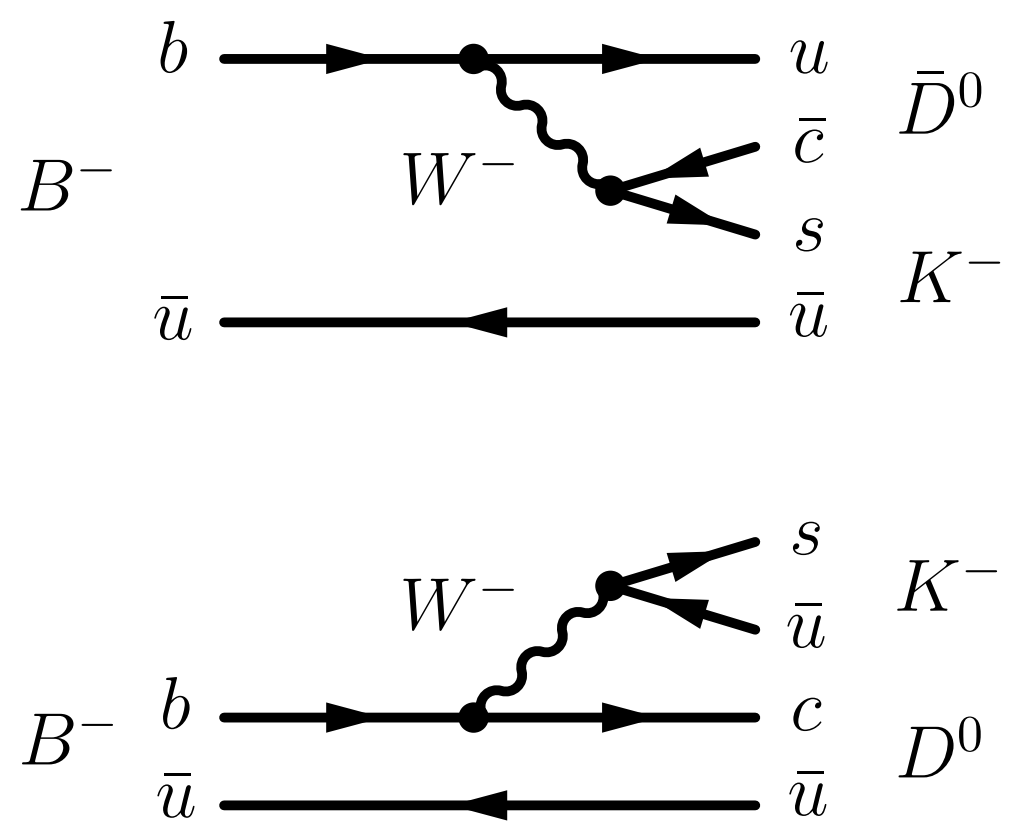

**Figure 17.8.1.** Dominant Feynman diagrams contributing to the $B^- \to DK^-$ decay. The top diagram proceeds via a $b \to u\bar{c}s$ transition, and is suppressed by both the small value of $|V_{ub}|$, and color considerations; the bottom diagram proceeds via a $b \to c\bar{u}s$ transition, and is only singly Cabibbo-suppressed.

Since there is no penguin contribution for these decays and consequently no theoretical uncertainty involved, all the hadronic unknowns are obtainable from experiment. They are $r_B$, the magnitude of the ratio of the amplitudes for the processes $B^- \to \overline{D}{}^0K^-$ and $B^- \to D^0K^-$, and $\delta_B$, the relative strong phase between these two amplitudes. For charged $B$ decays, $r_B \sim c_f|V_{cs}V_{ub}^*/V_{us}V_{cb}^*| \sim 0.1$, where $c_f$ is a color suppression factor ($\sim 0.3$). There is no theoretical guidance for the strong phase difference $\delta_B$. Typically, effects due to neutral $D$ mixing and *CP* violation are neglected, since these are expected (and measured) to be small (see the text on $D$-mixing and *CP* violation in Section 19.2). In general however, such effects can also be taken into account (Grossman, Soffer, and Zupan, 2005). There is also an irreducible error coming from electroweak corrections estimated to be $\delta\phi_3/\phi_3 \sim 10^{-6}$ (Zupan, 2011).

The possibility of observing direct *CP* violation in $B^- \to DK^-$ was first discussed by Bigi, Carter, and Sanda (Bigi and Sanda, 1988; Carter and Sanda, 1980). It was suggested to use charged $B$ decays to final states with $D^0/\overline{D}{}^0 \to K_S^0$ plus pion(s), where the presence of the $K_S^0$ generated by $K^0 - \overline{K}{}^0$ mixing was the essential element for making the interference. Since then, several methods have been proposed which can be grouped according to the choice of the final state: the "GLW" method (Gronau and London, 1991; Gronau and Wyler, 1991), based on Cabibbo-suppressed $D$ decays to *CP* eigenstates, such as $K^+K^-$ or $K_S^0\pi^0$ (Section 17.8.2); the "ADS" method (Atwood, Dunietz, and Soni, 1997, 2001), where the neutral $D$ is reconstructed in Cabibbo-favored (CF) and doubly Cabibbo-suppressed (DCS) final states such as $K^\pm\pi^\mp$ (Section 17.8.3); and the "GGSZ" method (Giri, Grossman, Soffer, and Zupan, 2003b), which uses the Dalitz-plot distribution of the products of $D$ decays to multibody self-conjugate final states, such as $K_S^0\pi^+\pi^-$ (Section 17.8.4). The main issue with these methods is the small overall branching fractions of the decays involved, which range from $5\times10^{-6}$ to $5\times10^{-9}$. Therefore a precise determination of $\phi_3$ requires a very large data sample. The various methods are combined in Section 17.8.6 to provide a determination of $\phi_3$ from $B$ Factory data. The study of the time-dependent decay rates of $B \to D^{(*)\mp}h^\pm$, providing a measure of $\sin(2\phi_1 + \phi_3)$, is discussed separately in Section 17.8.5.

### 17.8.2 GLW method

In the method proposed by Gronau and London (1991) and Gronau and Wyler (1991), the neutral $D$ meson is reconstructed in decays to *CP*-even eigenstates such as $K^+K^-$ (denoted as $D_{CP+}$) or *CP*-odd eigenstates such as $K_S^0\pi^0$ ($D_{CP-}$). Although a $B^0$ may decay weakly to either a $D^0$ or to a $\overline{D}{}^0$, when looking for a *CP*-even decay product one is actually selecting the *CP*-even superposition $(D^0 + \overline{D}{}^0)/\sqrt{2}$. The measurements are of the ratio

$$R_{CP\pm} = 2\frac{\Gamma(B^-\to D_{CP\pm}K^-) + \Gamma(B^+\to D_{CP\pm}K^+)}{\Gamma(B^-\to D_{fav}K^-) + \Gamma(B^+\to D_{fav}K^+)}, \tag{17.8.1}$$

where $D_{fav}$ indicates that the neutral $D$ meson is reconstructed in a favored hadronic decay mode such as

**Table 17.8.1.** Compilation of $R_{CP}$ and $A_{CP}$ results for $CP$-even and $CP$-odd $D$ decay modes. The three horizontal blocks refer to $B^{\pm} \to DK^{\pm}$ (top), $B^{\pm} \to D^{*}K^{\pm}$ (center), and $B^{\pm} \to DK^{*\pm}$ (bottom).

| $B$ decay | | BABAR | Belle | Average |
|---|---|---|---|---|
| $B^{\pm} \to DK^{\pm}$ | | (del Amo Sanchez, 2010e) | (Trabelsi, 2013) | |
| | $R_{CP+}$ | $1.18 \pm 0.09 \pm 0.05$ | $1.03 \pm 0.07 \pm 0.03$ | $1.08 \pm 0.06$ |
| | $R_{CP-}$ | $1.03 \pm 0.09 \pm 0.04$ | $1.13 \pm 0.09 \pm 0.05$ | $1.08 \pm 0.07$ |
| | $A_{CP+}$ | $+0.25 \pm 0.06 \pm 0.02$ | $+0.29 \pm 0.06 \pm 0.02$ | $+0.27 \pm 0.04$ |
| | $A_{CP-}$ | $-0.08 \pm 0.07 \pm 0.02$ | $-0.12 \pm 0.06 \pm 0.01$ | $-0.10 \pm 0.05$ |
| $B^{\pm} \to D^{*}K^{\pm}$ | | (Aubert, 2008o) | (Trabelsi, 2013) | |
| | $R_{CP+}$ | $1.31 \pm 0.13 \pm 0.04$ | $1.19 \pm 0.13 \pm 0.03$ | $1.25 \pm 0.09$ |
| | $R_{CP-}$ | $1.10 \pm 0.12 \pm 0.04$ | $1.03 \pm 0.13 \pm 0.03$ | $1.06 \pm 0.09$ |
| | $A_{CP+}$ | $-0.11 \pm 0.09 \pm 0.01$ | $-0.14 \pm 0.10 \pm 0.01$ | $-0.12 \pm 0.07$ |
| | $A_{CP-}$ | $+0.06 \pm 0.10 \pm 0.02$ | $+0.22 \pm 0.11 \pm 0.01$ | $+0.13 \pm 0.07$ |
| $B^{\pm} \to DK^{*\pm}$ | | (Aubert, 2009t) | | |
| | $R_{CP+}$ | $2.17 \pm 0.35 \pm 0.09$ | – | – |
| | $R_{CP-}$ | $1.03 \pm 0.27 \pm 0.13$ | – | – |
| | $A_{CP+}$ | $+0.09 \pm 0.13 \pm 0.06$ | – | – |
| | $A_{CP-}$ | $-0.23 \pm 0.21 \pm 0.07$ | – | – |

$D^0 \to K^-\pi^+$, and $A_{CP\pm}$, defined as

$$A_{CP\pm} = \frac{\Gamma(B^-\to D_{CP\pm}K^-) - \Gamma(B^+\to D_{CP\pm}K^+)}{\Gamma(B^-\to D_{CP\pm}K^-) + \Gamma(B^+\to D_{CP\pm}K^+)}. \quad (17.8.2)$$

These four observables can be expressed in terms of the physics parameters $\phi_3$, $\delta_B$, and $r_B$,

$$\begin{aligned} R_{CP\pm} &= 1 + r_B^2 \pm 2r_B \cos\delta_B \cos\phi_3, \\ A_{CP\pm} &= \pm 2r_B \sin\delta_B \sin\phi_3 / R_{CP\pm}. \end{aligned} \quad (17.8.3)$$

Effects due to interference would result in $R_{CP\pm} \neq 1$, while $CP$ violation would show up as $A_{CP\pm} \neq 0$.

Both *BABAR* (Aubert, 2008o; del Amo Sanchez, 2010e) and Belle (Trabelsi, 2013) have reconstructed $B^- \to DK^-$ and $B^- \to D^*K^-$ decays with $D^* \to D\pi^0$ and $D^* \to D\gamma$. The data samples used by *BABAR* and Belle consist of 467 and $772 \times 10^6$ $B\overline{B}$ pairs respectively for the $B^- \to DK^-$ decay, whereas 383 and $772 \times 10^6$ $B\overline{B}$ pairs are used respectively for the $B^- \to D^*K^-$ decays. *BABAR* has also included the decay $B^- \to DK^{*-}$ with $K^{*-} \to K_S^0\pi^-$ (Aubert, 2009t). For both experiments, the reconstructed $CP$-even final states are $K^+K^-$ and $\pi^+\pi^-$, whereas the $CP$-odd eigenstates used by *BABAR* are $K_S^0\pi^0$, $K_S^0\omega$ ($\omega \to \pi^+\pi^-\pi^0$) and $K_S^0\phi$ ($\phi \to K^+K^-$); Belle uses $K_S^0\pi^0$ and $K_S^0\eta$. The $K_S^0$ candidates are reconstructed through their decays to charged pion pairs, while $\eta$ and $\pi^0$ mesons are identified through their decays to photon pairs.

The $B$ decay final states are fully reconstructed, with efficiencies between 40% (for low-multiplicity, low-background decay modes, *e.g.* $D \to K^+K^-$) and 5% (for high-multiplicity decays, *e.g.* $D \to K_S^0\omega$). The selection is optimized to maximize the figure of merit $S/\sqrt{S+B}$ (an estimator of the statistical precision; Section 4.3), where the numbers of expected signal ($S$) and background ($B$) events are estimated from simulated and data control samples respectively. Signal $B$ decays are distinguished from $B\overline{B}$ and continuum $q\overline{q}$ background by means of maximum likelihood fits to the energy-substituted invariant mass $m_{\rm ES}$ and/or the energy difference $\Delta E$, as described in Section 7.1.1. Additional continuum background discrimination is in some cases achieved by including in the likelihood a Fisher discriminant $\mathcal{F}$ for *BABAR*, or a non-linear neural network $NB$ for Belle, based on several event-shape quantities that distinguish nearly isotropic $B\overline{B}$ events from more jet-like $q\overline{q}$ events and exploit the different angular correlations in the two event categories (Section 9.3). $B^- \to D^{(*)}\pi^-$ decays, which are 12 times more abundant than $B^- \to D^{(*)}K^-$ and are expected to show negligible $CP$-violating effects ($r_B \approx 0.01$ in such decays), are distinguished using charged hadron identification variables and $\Delta E$, and are used as control samples.

The results are summarized in Table 17.8.1. The $B^{\pm} \to DK^{\pm}$, $D \to K_S^0\phi$ results of *BABAR* are omitted here, as they are included in the Dalitz $B^{\pm} \to DK^{\pm}$, $D \to K_S^0KK$ result (Section 17.8.4). *BABAR* and Belle results are in good agreement. The results are also compatible with expectations from Eqs (17.8.1) and (17.8.2) using the values of $r_B$, $\delta_B$, and $\phi_3$ measured with the Dalitz method (Section 17.8.4). Both experiments find evidence ($> 3\sigma$ and $> 4\sigma$, respectively) of direct $CP$ violation in the $B^- \to D_{CP+}K^-$ decay (Figs 17.8.2 and 17.8.3). The combined result has a significance larger than $6\sigma$.

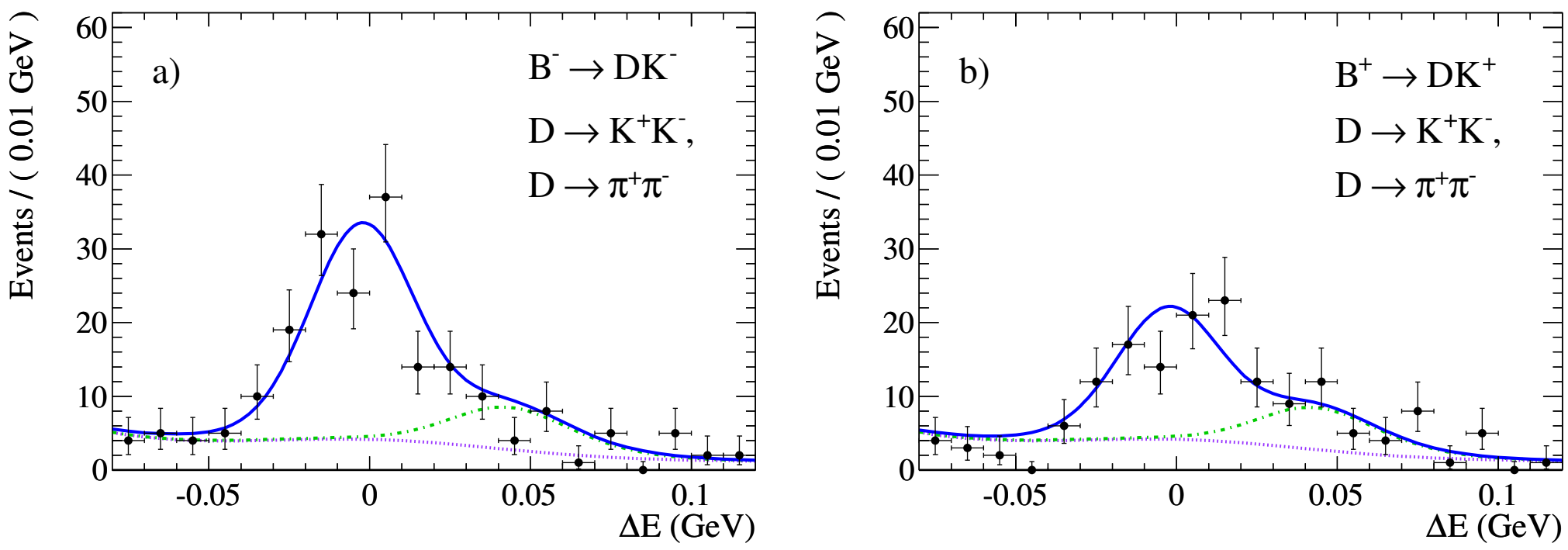

**Figure 17.8.2.** $\Delta E$ projections of the fits to the $B^{\pm} \to D_{CP+}K^{\pm}$ candidates selected in the full *BABAR* data sample (Aubert, 2008o), split into subsets of definite charge of the $B$ candidate: a) $B^- \to D_{CP+}K^-$, and b) $B^+ \to D_{CP+}K^+$. The curves are the full probability density function (solid, blue), and $B^{\pm} \to D\pi^{\pm}$ (dash-dotted, green) stacked on the remaining backgrounds (dotted, purple). The region between the solid and the dash-dotted lines represents the $B^{\pm} \to DK^{\pm}$ contribution. We show the subset of the data sample in which the track from the $B$ decay is identified as a kaon and a signal-enriching selection is applied to the other variables used in the fit ($m_{\rm ES}$ and $\mathcal{F}$).

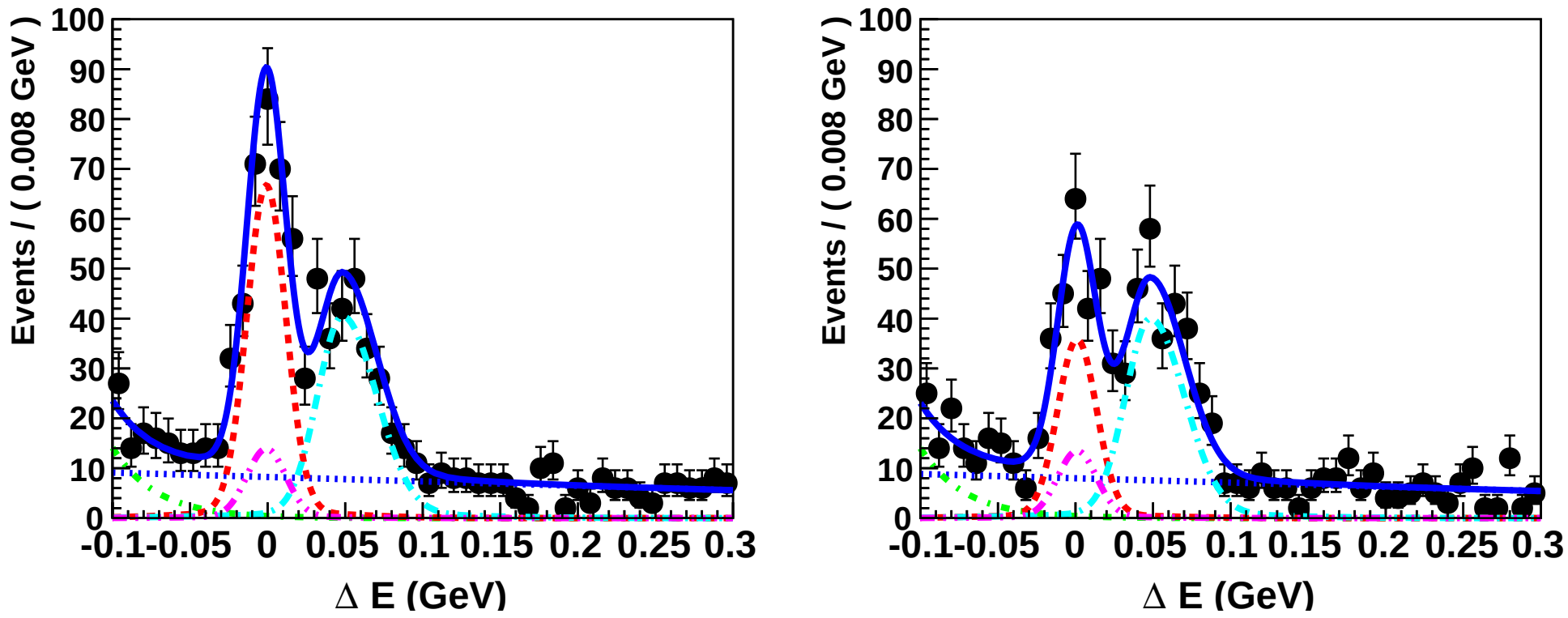

**Figure 17.8.3.** $\Delta E$ projections of the fits to the $B^{\pm} \to D_{CP+}K^{\pm}$ candidates selected in the full Belle data sample (Trabelsi, 2013), split into subsets of definite charge of the $B$ candidate: (left) $B^- \to D_{CP+}K^-$, and (right) $B^+ \to D_{CP+}K^+$. The curves are the full *p.d.f.* (solid, blue), $B^{\pm} \to DK^{\pm}$ (dashed, red), $B^{\pm} \to D\pi^{\pm}$ (dash-dotted, light blue), the charmless peaking background (dash-dot-dotted, magenta), the $B\overline{B}$ background (dash-dotted, green) and the remaining combinatorial background (dotted, blue).

### 17.8.3 ADS method

This idea was extended further by Atwood, Dunietz, and Soni (1997, 2001), who showed that additional neutral $D$ decay modes, in particular DCS $D$ decays, could also be used to measure $\phi_3$. For example, $B^- \to [K^+\pi^-]_D K^-$ can be reached via favored $B^- \to D^0 K^-$ followed by DCS $D^0 \to K^+\pi^-$ decay, or via suppressed $B^- \to \overline{D}^0 K^-$ followed by favored $\overline{D}^0 \to K^+\pi^-$ decay. In the case that the neutral $D$ meson decays to a non-*CP* eigenstate, one also has to consider the ratio of the magnitudes of the suppressed and favored decays to the particular final state ($r_D$) as well as the strong phase difference between them ($\delta_D$). This information on the hadronic parameters of the $D$ meson can be obtained from a charm factory (CLEO-c and BES III). Large *CP* violation effects are possible when the overall ratio of magnitudes of amplitudes is close to unity ($r_D \sim r_B$). There is an effective strong phase shift of 180° between the cases where a $D^*$ is reconstructed in the $D\pi^0$ and $D\gamma$ final states, which in principle allows $\phi_3$ to be measured using the ADS technique with $B^{\pm} \to D^* K^{\pm}$ alone (Bondar and Gershon, 2004).

#### 17.8.3.1 $B^{\pm} \to D^{(*)}K^{(*)\pm}$, $D \to K^+\pi^-$ decays

The observables in this method are the charge-averaged partial decay width ratio

$$R_{ADS} = \frac{\Gamma(B^- \to [K^+\pi^-]K^-) + \Gamma(B^+ \to [K^-\pi^+]K^+)}{\Gamma(B^- \to [K^-\pi^+]K^-) + \Gamma(B^+ \to [K^+\pi^-]K^+)}, \tag{17.8.4}$$

and the *CP* asymmetry

$$A_{ADS} = \frac{\Gamma(B^- \to [K^+\pi^-]K^-) - \Gamma(B^+ \to [K^-\pi^+]K^+)}{\Gamma(B^- \to [K^+\pi^-]K^-) + \Gamma(B^+ \to [K^-\pi^+]K^+)}.$$

**Table 17.8.2.** Compilation of $R_{ADS}$ and $A_{ADS}$ results for (from top to bottom) $B^{\pm} \to DK^{\pm}$, $B^{\pm} \to D^{*}K^{\pm}$ with $D^{*} \to D\pi^{0}$ and $D\gamma$, and $B^{\pm} \to DK^{*\pm}$.

| $B$ decay | | *BABAR* | Belle | Average |
|---|---|---|---|---|
| $B^{\pm} \to DK^{\pm}$ | | (del Amo Sanchez, 2010m) | (Horii, 2011) | |
| | $R_{ADS}$ | $0.011 \pm 0.006 \pm 0.002$ | $0.0163^{+0.0044+0.0007}_{-0.0041-0.0013}$ | $0.015 \pm 0.004$ |
| | $A_{ADS}$ | $-0.86 \pm 0.47^{+0.12}_{-0.16}$ | $-0.39^{+0.26+0.04}_{-0.28-0.03}$ | $-0.51 \pm 0.25$ |
| $B^{\pm} \to D^{*}[D\pi^{0}]K^{\pm}$ | | (del Amo Sanchez, 2010m) | (Trabelsi, 2013) | |
| | $R_{ADS}$ | $0.018 \pm 0.009 \pm 0.004$ | $0.010^{+0.008+0.001}_{-0.007-0.002}$ | $0.013 \pm 0.006$ |
| | $A_{ADS}$ | $+0.77 \pm 0.35 \pm 0.12$ | $+0.4^{+1.1+0.2}_{-0.7-0.1}$ | $+0.72 \pm 0.34$ |
| $B^{\pm} \to D^{*}[D\gamma]K^{\pm}$ | | (del Amo Sanchez, 2010m) | (Trabelsi, 2013) | |
| | $R_{ADS}$ | $0.013 \pm 0.014 \pm 0.008$ | $0.036^{+0.014}_{-0.012} \pm 0.002$ | $0.027 \pm 0.010$ |
| | $A_{ADS}$ | $+0.36 \pm 0.94^{+0.25}_{-0.41}$ | $-0.51^{+0.33}_{-0.29} \pm 0.08$ | $-0.43 \pm 0.31$ |
| $B^{\pm} \to DK^{*\pm}$ | | (Aubert, 2009t) | | |
| | $R_{ADS}$ | $0.066 \pm 0.031 \pm 0.010$ | – | – |
| | $A_{ADS}$ | $-0.34 \pm 0.43 \pm 0.16$ | – | – |

(17.8.5)

In terms of physics parameters, these can be written as

$$R_{ADS} = r_B^2 + r_D^2 + 2r_B r_D \cos(\delta_B + \delta_D)\cos\phi_3,$$
$$A_{ADS} = 2r_B r_D \sin(\delta_B + \delta_D)\sin\phi_3 / R_{ADS}. \quad (17.8.6)$$

The original and first objective for these analyses was to observe $R_{ADS} \neq 0$.

Both *BABAR* (del Amo Sanchez, 2010m) and Belle (Horii, 2011) have reconstructed the decays $B^{-} \to DK^{-}$ and $B^{-} \to D^{*}K^{-}$ with $D^{*} \to D\pi^{0}$ and $D^{*} \to D\gamma$ (del Amo Sanchez, 2010m; Trabelsi, 2013) followed by $D \to K^{+}\pi^{-}$ on datasets of 467 and 772 $\times 10^{6}$ $B\overline{B}$ pairs respectively. *BABAR* (Aubert, 2009t) has also selected $B^{-} \to DK^{*-}$ with $K^{*-} \to K_S^0\pi^{-}$ using $379 \times 10^{6}$ $B\overline{B}$ pairs. As in the GLW analysis, the $B$ decay final states are fully reconstructed. The selection criteria are usually tighter in order to achieve a higher signal purity, given that the signal rate is typically $\mathcal{O}(10^{-2})$ weaker than in the case of $D$ decaying to *CP* eigenstates. Particular care is taken over the suppression of "peaking" backgrounds from misidentified $B^{-} \to D^{(*)}\pi^{-}$ or $B^{-} \to D^{(*)}K^{-}$ decays. After the selection, the main background is due to $q\overline{q}$ events. In order to achieve a better continuum background suppression, a non-linear neural network (Section 4.4.4), $NN$ (*BABAR*) or $NB$ (Belle), of several event-shape quantities (Section 9.3) is used. The final yields are extracted from maximum likelihood fits to $m_{\rm ES}$ and $NN$ (*BABAR*) or $\Delta E$ and $NB$ (Belle) for $B^{-} \to D^{(*)}K^{-}$, and to $m_{\rm ES}$ (after a tight selection criteria on $NN$) for $B^{-} \to DK^{*-}$ (*BABAR*).

These results are summarized in Table 17.8.2. The strongest evidence for suppressed signals have been reported by Belle with a significance (including systematic uncertainties) of 4.1$\sigma$ for the $B^{-} \to DK^{-}$ mode and of 3.5$\sigma$ for the $B^{-} \to D^{*}[D\gamma]K^{-}$ decay (Fig. 17.8.4).

The observables $R^{+}$ and $R^{-}$, defined as

$$R^{+} = \Gamma(B^{+}\to[K^{-}\pi^{+}]K^{+})/\Gamma(B^{+}\to[K^{+}\pi^{-}]K^{+}),$$
$$R^{-} = \Gamma(B^{-}\to[K^{+}\pi^{-}]K^{-})/\Gamma(B^{-}\to[K^{-}\pi^{+}]K^{-}), \quad (17.8.7)$$

have been noted recently as more suitable to use than $R_{ADS}$ and $A_{ADS}$, since the former are better behaved. They are statistically independent observables, while the uncertainty on $A_{ADS}$ depends on the central value of $R_{ADS}$. $R^{+}$ and $R^{-}$ are related to $R_{ADS}$ and $A_{ADS}$ by

$$R_{ADS} = (R^{+} + R^{-})/2,$$
$$A_{ADS} = (R^{-} - R^{+})/(R^{-} + R^{+}). \quad (17.8.8)$$

Recent *BABAR* $B^{-} \to D^{(*)}K^{-}$ measurements have been reported in terms of $R^{+}$ and $R^{-}$ (Table 17.8.3), as well as $R_{ADS}$, $A_{ADS}$ (Table 17.8.2).

**Table 17.8.3.** Compilation of $R^{+}$ and $R^{-}$ results reported by *BABAR* for $B^{\pm} \to DK^{\pm}$ and $B^{\pm} \to D^{*}K^{\pm}$, with $D^{*} \to D\pi^{0}$ and $D\gamma$ (del Amo Sanchez, 2010m).

| | $R^{+}$ $(10^{-2})$ | $R^{-}$ $(10^{-2})$ |
|---|---|---|
| $B^{\pm} \to DK^{\pm}$ | $2.2 \pm 0.9 \pm 0.3$ | $0.2 \pm 0.6 \pm 0.2$ |
| $B^{\pm} \to D^{*}[D\pi^{0}]K^{\pm}$ | $0.5 \pm 0.8 \pm 0.3$ | $3.7 \pm 1.8 \pm 0.9$ |
| $B^{\pm} \to D^{*}[D\gamma]K^{\pm}$ | $0.9 \pm 1.6 \pm 0.7$ | $1.9 \pm 2.3 \pm 1.2$ |

#### 17.8.3.2 $B^{\pm} \to DK^{\pm}$, $D \to K^{+}\pi^{-}\pi^{0}$ decay

*BABAR* has also presented ADS results for the $D \to K^{+}\pi^{-}\pi^{0}$ decay, quoting a measurement of $R^{+}$ and $R^{-}$

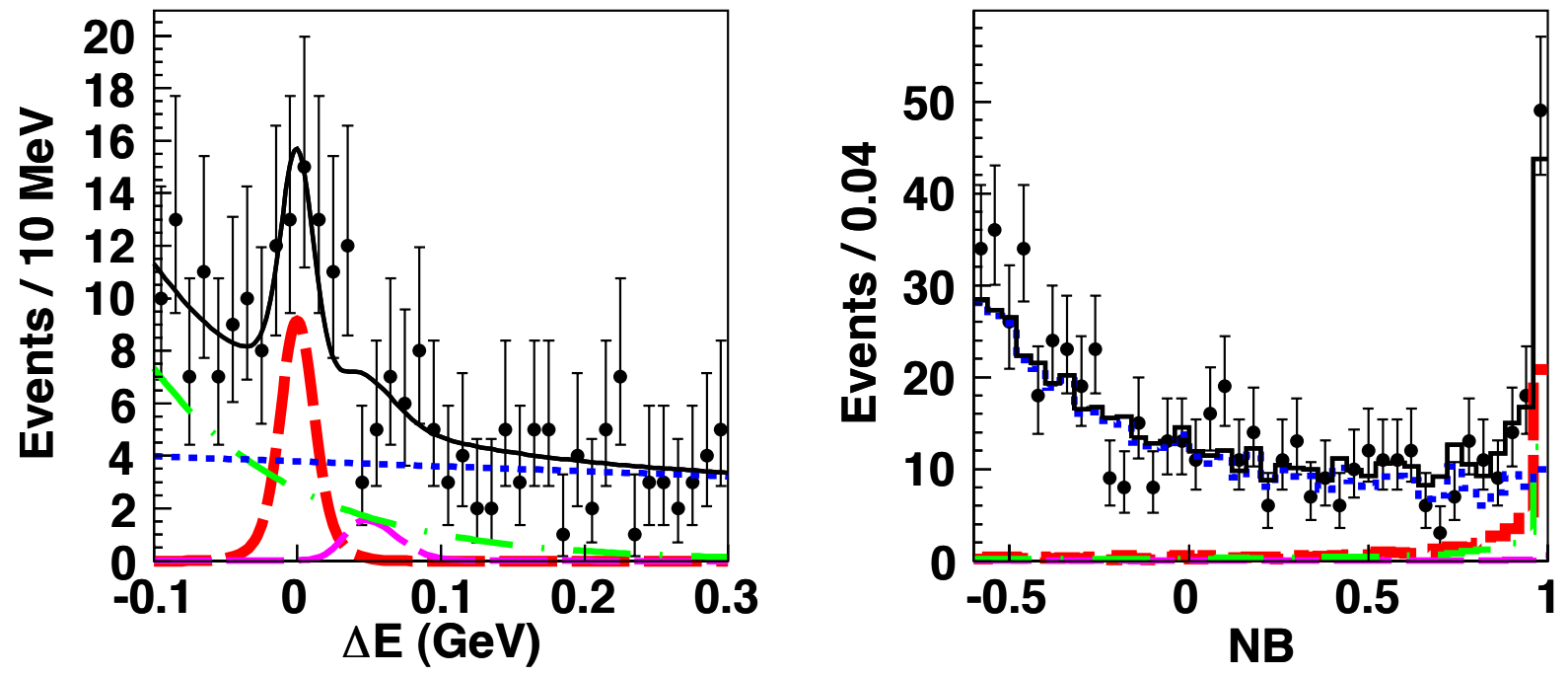

(a) $B^- \to DK^-$, $D \to K^+\pi^-$

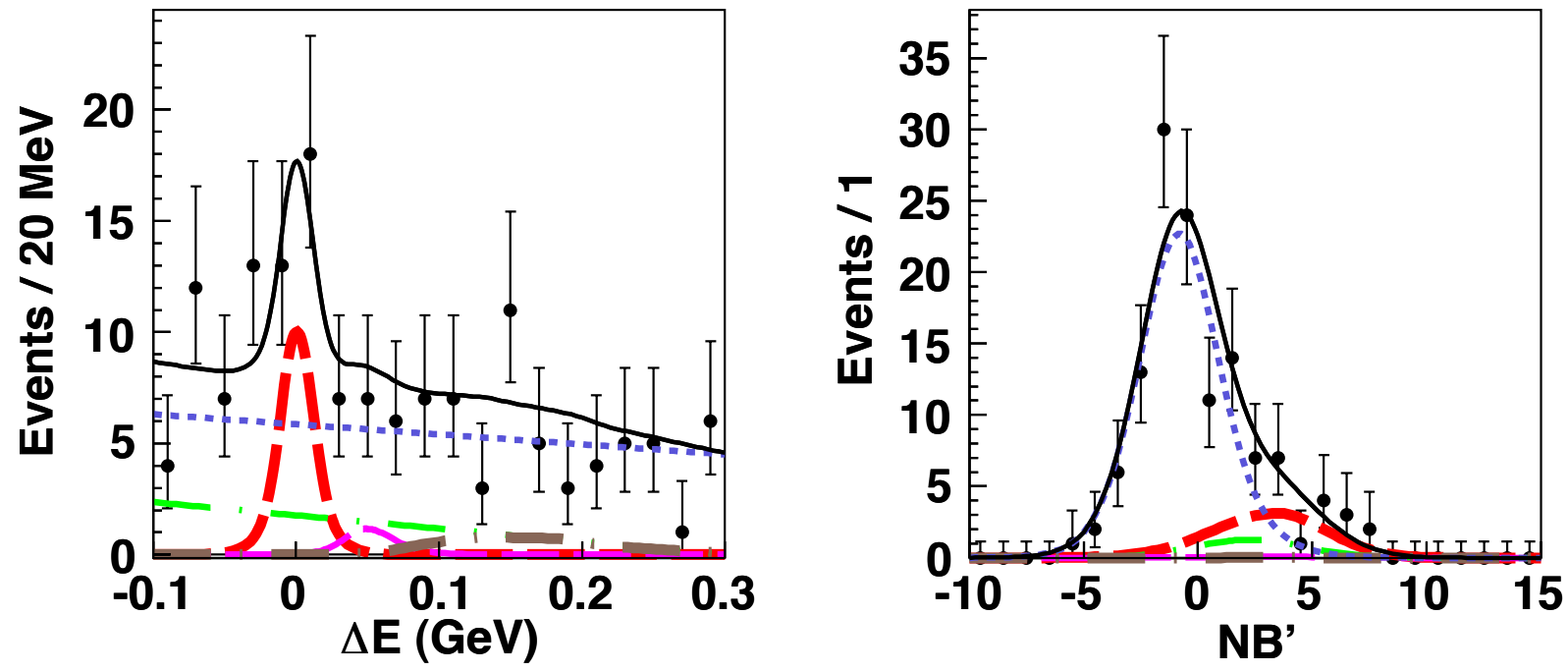

(b) $B^- \to D^*K^-$, $D^* \to D\gamma$, and $D \to K^+\pi^-$

**Figure 17.8.4.** $\Delta E$ and neural-network $NB$ (or $NB'$, a modified $NB$ variable) distributions in a signal-enriched region for $B^- \to DK^-$ where $D \to K^+\pi^-$ (top; Horii, 2011) and for $B^- \to D^*K^-$ where $D^* \to D\gamma$ and $D \to K^+\pi^-$ (bottom; Trabelsi, 2013) from the Belle collaboration. In these plots, $DK^-$ components are shown by thicker dashed curves (red), and $D\pi^-$ components are shown by thinner curves (magenta). $B\overline{B}$ backgrounds are shown by dash-dotted curves (green) while $q\bar{q}$ backgrounds are shown by dotted curves (blue). The sum of all components are shown by solid curves (black).

on a data sample of $474 \times 10^6$ $B\overline{B}$ pairs (Lees, 2011h). The relation between $R^\pm$ and the physics parameters is similar to the two-body case:

$$\begin{aligned} R^+ &= r_B^2 + r_D^2 + 2r_B r_D k_D \cos(\delta_B + \delta_D + \phi_3), \\ R^- &= r_B^2 + r_D^2 + 2r_B r_D k_D \cos(\delta_B + \delta_D - \phi_3), \end{aligned} \tag{17.8.9}$$

with

$$\begin{aligned} r_D^2 &= \frac{\Gamma(D^0 \to f)}{\Gamma\left(D^0 \to \overline{f}\right)} = \frac{\int d\boldsymbol{m} A_{DCS}^2(\boldsymbol{m})}{\int d\boldsymbol{m} A_{CF}^2(\boldsymbol{m})}\,, \\ k_D e^{i\delta_D} &= \frac{\int d\boldsymbol{m} A_{DCS}(\boldsymbol{m}) A_{CF}(\boldsymbol{m}) e^{i\delta(\boldsymbol{m})}}{\sqrt{\int d\boldsymbol{m} A_{DCS}^2(\boldsymbol{m}) A_{CF}^2(\boldsymbol{m})}}\,, \end{aligned} \tag{17.8.10}$$

where $A_{CF}(\boldsymbol{m})$ and $A_{DCS}(\boldsymbol{m})$ are the magnitude of the CF and the DCS $D \to K^\mp\pi^\pm\pi^0$ amplitudes, respectively, $\delta(\boldsymbol{m})$ is the relative strong phase, and $\boldsymbol{m}$ indicates the position in the $D$ decay Dalitz plot of squared invariant masses $(m_{K\pi}^2, m_{K\pi^0}^2)$ (Atwood and A. Soni, 2003). The parameter $k_D$ is called the coherence factor and takes a value in the interval [0,1] depending on the Dalitz structure of the decay. Both $k_D$ and $\delta_D$ have been measured by the CLEO-c collaboration, who find $k_D = 0.84 \pm 0.07$ and $\delta_D = (47^{+14}_{-17})^\circ$ (Lowrey et al., 2009). The ratio $r_D$ has been measured in different experiments with an average value $r_D^2 = (2.2 \pm 0.1) \times 10^{-3}$ (Beringer et al., 2012). *BABAR* found

$$\begin{aligned} R^+ &= \left(5^{+12}_{-10}\,{}^{+1}_{-4}\right) \times 10^{-3}\,, \\ R^- &= \left(12^{+12}_{-10}\,{}^{+2}_{-4}\right) \times 10^{-3}\,. \end{aligned} \tag{17.8.11}$$

From these measurements a limit of $r_B < 0.13$ at the 90% confidence level (C.L.) is obtained.

#### 17.8.3.3 $B^0 \to DK^{*0}$, $D \to K^+\pi^-$ decay

*BABAR* and Belle have also performed an ADS analysis of neutral $B^0 \to DK^{*0}$, with $K^{*0} \to K^+\pi^-$ decays. In these modes, the flavor of the $K^*$ unambiguously determines the flavor of the neutral $B$ so no time-dependent measurement is needed. Here $r_B$ is expected to be approximately 0.3, due to CKM factors only, since both the interfering amplitudes are color-suppressed. The CF charge-conjugate final states are used as normalization and control samples. *BABAR* (Aubert, 2009al) uses a sample of $465 \times 10^6$ $B\overline{B}$ pairs and reconstructs the neutral $D$ mesons in the following DCS $D^0$ final states: $K^+\pi^-$, $K^+\pi^-\pi^0$, and $K^+\pi^-\pi^-\pi^+$. Signal yields are extracted from fits to the $m_{\rm ES}$ distribution of the selected candidates. Due to the small size of the final sample after the selection (24 signal candidates in total for all $D^0$ final states), the *CP* asymmetries $A_{ADS}$ have not been measured. Instead, 95% C.L. limits on $R_{ADS}$ have been set: $R_{ADS}^{K\pi} < 0.244$; $R_{ADS}^{K\pi\pi^0} < 0.181$; $R_{ADS}^{K\pi\pi\pi} < 0.391$. From the combination of these three results the ratio between the $b \to u$ and the $b \to c$ mediated decay amplitudes has been estimated to be $0.07 < r_B < 0.41$ at the 95% confidence level. Belle (Negishi, 2012), with a sample of $772 \times 10^6$ $B\overline{B}$ pairs, reconstructs only the neutral $D$ mesons in the DCS $D^0$ final state $K^+\pi^-$. No signal is found and the most stringent limit to date is set, $R_{ADS}^{K\pi} < 0.16$ at the 95% confidence level.

### 17.8.4 Dalitz plot (GGSZ) method

The measurement of $\phi_3$ using Dalitz plot analysis of three-body decays of the $D$ meson from the $B^\pm \to DK^\pm$ process was proposed by Giri, Grossman, Soffer, and Zupan (2003a) (and is therefore often referred to as the GGSZ method) and independently by Bondar (2002). The basic idea behind this method is to use final states accessible to both $D^0$ and $\overline{D}^0$ and to measure the phase of the interference between them in the decay of $D$ mesons produced in $B^\pm \to DK^\pm$ transitions.

The most convenient decay for this kind of measurement is $D \to K_S^0\pi^+\pi^-$. This mode has a unique combination of three advantages:

1. Large branching fraction.
2. Significant overlap between $D^0 \to K_S^0\pi^+\pi^-$ and $\overline{D}^0 \to K_S^0\pi^+\pi^-$ amplitudes which gives a large interference term sensitive to $\phi_3$.
3. Rich resonant structure which provides large variations of the strong phase in $D$ decay and results in sensitivity to $\phi_3$ that is only weakly dependent on the values of $\phi_3$ and strong phase $\delta_B$.

However, other decay modes can also be used. $D^0 \to K_S^0K^+K^-$ is another convenient mode which has a smaller branching ratio than $D^0 \to K_S^0\pi^+\pi^-$, but is generally cleaner due to the presence of two kaons. The mode $D^0 \to \pi^+\pi^-\pi^0$ has a comparable rate, but is more affected by the background. Modes that are not self-conjugate, such as $K^+\pi^-\pi^0$, can also be used: this requires two amplitudes, $D^0 \to K^+\pi^-\pi^0$ and $\overline{D}^0 \to K^+\pi^-\pi^0$, to be studied separately. However, given the large coherence factor in this mode (Lowrey et al., 2009), there would be little to gain from a GGSZ-like approach. The description of the technique below uses the mode $D \to K_S^0\pi^+\pi^-$ as an example.

The amplitude of the $B^+ \to DK^+$ decay as a function of the two $D$ Dalitz plot variables $m_+^2 \equiv m_{K_S^0\pi^+}^2$ and $m_-^2 \equiv m_{K_S^0\pi^-}^2$ is

$$A_{B^+}(m_+^2, m_-^2) = \overline{A}_D + r_B e^{i(\delta_B+\phi_3)} A_D\,, \qquad (17.8.12)$$

where $A_D = A_D(m_+^2, m_-^2)$ is the complex amplitude of $D^0 \to K_S^0\pi^+\pi^-$ decay, and $\overline{A}_D = \overline{A}_D(m_+^2, m_-^2)$ is the amplitude of $\overline{D}^0 \to K_S^0\pi^+\pi^-$ decay. Similarly, for $B^- \to DK^-$ decay, the amplitude is

$$A_{B^-}(m_+^2, m_-^2) = A_D + r_B e^{i(\delta_B-\phi_3)} \overline{A}_D\,. \qquad (17.8.13)$$

In the case of *CP* conservation in $D^0$ decay and neglecting $D^0 - \overline{D}^0$ mixing (as the mixing parameters $x$, $y$ are 1% or less; see the text on $D$-mixing and *CP* violation in Section 19.2), $\overline{A}_D(m_+^2, m_-^2) = A_D(m_-^2, m_+^2)$. The unknown quantities $\phi_3$, $r_B$, and $\delta_B$ can be obtained from a fit to the $D$-decay Dalitz distributions for $B^\pm \to DK^\pm$ decays once the complex amplitude $A_D$ is known.

Although the original proposal of Giri *et al.* was to use a binned analysis, in the case of limited statistics an unbinned fit is used in order to optimally extract information from the data, at the price of introducing model dependence. Section 17.8.4.1 describes the unbinned analyses using $D^0 \to K_S^0\pi^+\pi^-$ performed by Belle and *BABAR*. Other decay modes studied by *BABAR* are presented in Section 17.8.4.2. A binned approach used by Belle has its own advantages, and is discussed in Section 17.8.4.3.

#### 17.8.4.1 Model-dependent technique

Measurements of $\phi_3$ using the model-dependent unbinned Dalitz plot analysis technique have been performed by Belle (Poluektov, 2004, 2006, 2010) and *BABAR* (Aubert, 2005o, 2008l; del Amo Sanchez, 2010b), their latest analyses using 657 and $468 \times 10^6$ $B\overline{B}$ pairs, respectively. Both experiments use $B^\pm \to DK^\pm$, $B^\pm \to D^*K^\pm$ (with $D^* \to D\pi^0$ and $D^* \to D\gamma$) and $B^\pm \to DK^{*\pm}$ with $K^{*\pm} \to K_S^0\pi^\pm$ decays. While Belle reconstructs the neutral $D$ in the $K_S^0\pi^+\pi^-$ final state, *BABAR* uses both the $K_S^0\pi^+\pi^-$ and $K_S^0K^+K^-$ final states.

This method requires a model to describe the amplitude as a function of the Dalitz plot variables. Both collaborations use $D^{*+} \to D^0\pi^+$ and $D^{*-} \to \overline{D}^0\pi^-$ decays to flavor-tag $D^0$ and $\overline{D}^0$ mesons, and fit the neutral $D$ decay amplitude in the resulting sample. The Dalitz plots obtained by *BABAR* for $D^0 \to K_S^0\pi^+\pi^-$ and $D^0 \to K_S^0K^+K^-$ decays are shown in Fig. 17.8.5 (del Amo Sanchez, 2010b,f).

#### *D decay amplitudes*

The description of amplitudes differs in the Belle and *BABAR* analyses. Belle uses an isobar model for $S$-, $P$-, and $D$-waves in their $D^0 \to K_S^0\pi^+\pi^-$ analysis in each of the $K_S^0\pi^+$, $K_S^0\pi^-$, and $\pi^+\pi^-$ channels, and a flat non-resonant term. The isobar model involves describing amplitudes with relativistic Breit-Wigner (BW) propagators, or Gounaris-Sakurai in the case of $\rho^0 \to \pi^+\pi^-$, with Blatt-Weisskopf centrifugal factors and angular terms (see the Isobar formalism text in Section 13.2.1). The resonance composition measured by Belle is shown in Table 17.8.4. Note that $\sigma_1$ and $\sigma_2$ states, with masses and widths allowed to vary in the fit, are introduced as an effective description of structure in the $\pi\pi$ $S$-wave.

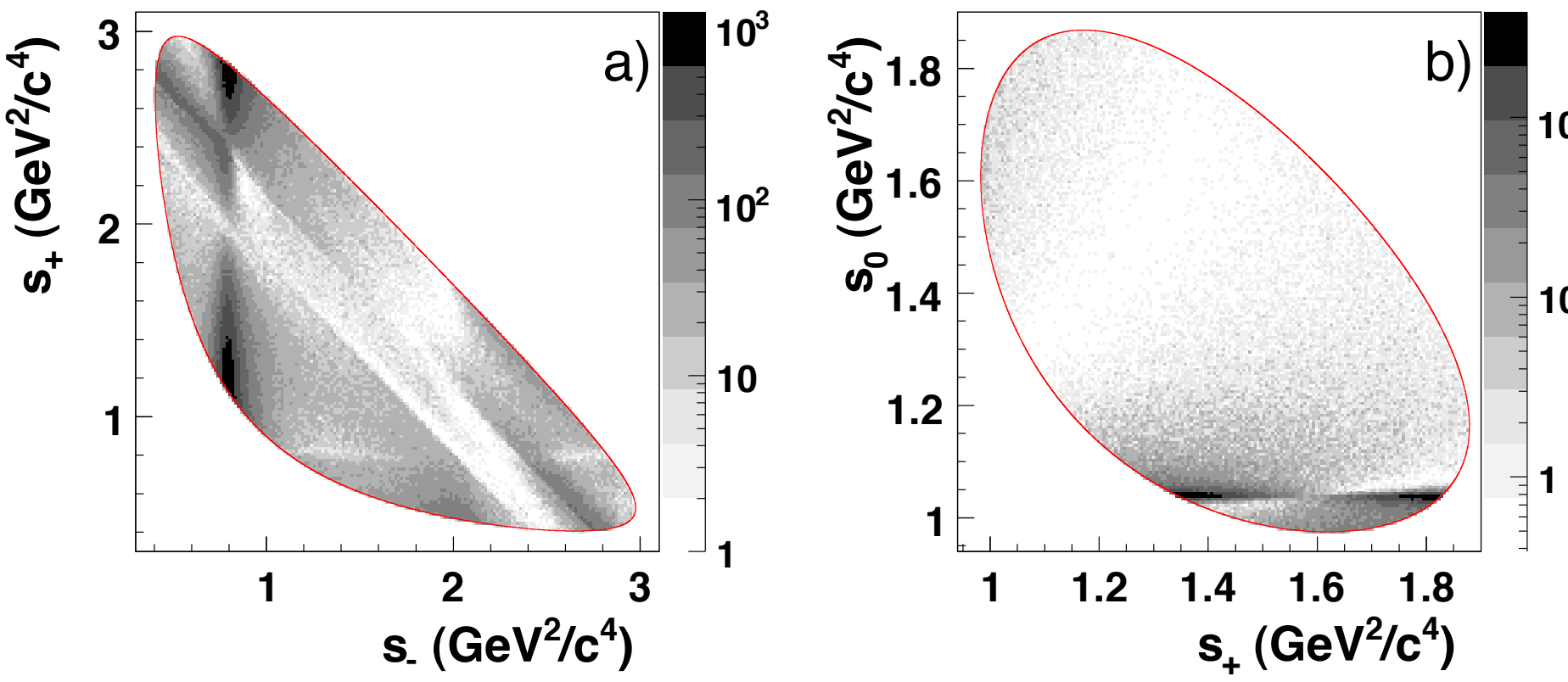

**Figure 17.8.5.** *BABAR* Dalitz plots of (a) $D^0 \to K_S^0\pi^+\pi^-$ and (b) $D^0 \to K_S^0 K^+K^-$ decays (del Amo Sanchez, 2010b,f), where $s_\pm \equiv m_\pm^2 \equiv m_{K_S^0 h^\pm}^2$ and $s_0 \equiv m_{h^+h^-}^2$, with $h = \pi, K$.

**Table 17.8.4.** Belle fit results for the $D^0 \to K_S^0\pi^+\pi^-$ decay (Poluektov, 2010). Errors are statistical only. The phases are given in the interval $[0, 360]^\circ$. The fit fraction for each mode is defined as the ratio of the integrals of the square absolute value of the amplitude for that mode and the squared absolute value of the total amplitude. The fit fractions do not sum to one due to interference effects.

| Intermediate state | Amplitude | Phase (°) | Fit fraction (%) |
|---|---|---|---|
| $K_S^0\sigma_1$ | $1.56 \pm 0.06$ | $214 \pm 3$ | $11.0 \pm 0.7$ |
| $K_S^0 f_0(980)$ | $0.385 \pm 0.006$ | $207.3 \pm 2.3$ | $4.72 \pm 0.05$ |
| $K_S^0\sigma_2$ | $0.20 \pm 0.02$ | $212 \pm 12$ | $0.54 \pm 0.10$ |
| $K_S^0 f_0(1370)$ | $1.56 \pm 0.12$ | $110 \pm 4$ | $1.9 \pm 0.3$ |
| | | | |
| $K_S^0\rho(770)^0$ | 1.0 (fixed) | 0 (fixed) | $21.2 \pm 0.5$ |
| $K_S^0\omega(782)$ | $0.0343 \pm 0.0008$ | $112.0 \pm 1.3$ | $0.526 \pm 0.014$ |
| $K_S^0 f_2(1270)$ | $1.44 \pm 0.04$ | $342.9 \pm 1.7$ | $1.82 \pm 0.05$ |
| $K_S^0\rho^0(1450)$ | $0.49 \pm 0.08$ | $64 \pm 11$ | $0.11 \pm 0.04$ |
| | | | |
| $K_0^*(1430)^-\pi^+$ | $2.21 \pm 0.04$ | $358.9 \pm 1.1$ | $7.93 \pm 0.09$ |
| $K_0^*(1430)^+\pi^-$ | $0.36 \pm 0.03$ | $87 \pm 4$ | $0.22 \pm 0.04$ |
| | | | |
| $K^*(892)^-\pi^+$ | $1.638 \pm 0.010$ | $133.2 \pm 0.4$ | $62.9 \pm 0.8$ |
| $K^*(892)^+\pi^-$ | $0.149 \pm 0.004$ | $325.4 \pm 1.3$ | $0.526 \pm 0.016$ |
| $K^*(1410)^-\pi^+$ | $0.65 \pm 0.05$ | $120 \pm 4$ | $0.49 \pm 0.07$ |
| $K^*(1410)^+\pi^-$ | $0.42 \pm 0.04$ | $253 \pm 5$ | $0.21 \pm 0.03$ |
| $K_2^*(1430)^-\pi^+$ | $0.89 \pm 0.03$ | $314.8 \pm 1.1$ | $1.40 \pm 0.06$ |
| $K_2^*(1430)^+\pi^-$ | $0.23 \pm 0.02$ | $275 \pm 6$ | $0.093 \pm 0.014$ |
| $K^*(1680)^-\pi^+$ | $0.88 \pm 0.27$ | $82 \pm 17$ | $0.06 \pm 0.04$ |
| $K^*(1680)^+\pi^-$ | $2.1 \pm 0.2$ | $130 \pm 6$ | $0.30 \pm 0.07$ |
| | | | |
| non-resonant | $2.7 \pm 0.3$ | $160 \pm 5$ | $5.0 \pm 1.0$ |

*BABAR*, in contrast, uses the K-matrix formalism with the P-vector approximation to describe the $\pi^+\pi^-$ $S$-wave, while the $K\pi$ $S$-wave description uses a BW for the $K_0^*(1430)^\pm$ state and a non-resonant contribution parameterized by a scattering length and effective range, as described in Section 13.2.2. The resonance composition, $P$-vector, and $K\pi$ $S$-wave parameters measured by *BABAR* are shown in Table 17.8.5.

**Table 17.8.5.** *BABAR* fit results for the $D^0 \to K^0_S\pi^+\pi^-$ decay (del Amo Sanchez, 2010b,f). Errors are statistical only. The phases are given in the interval $[-\pi, +\pi]$ rad. The description of the $\pi\pi$ and $K\pi$ $S$-wave parameters can be found in Section 13.2.2; we follow the notation of that section. The $\pi\pi$ $S$-wave parameters $\beta_5$, $f^{\rm prod}_{14}$, and $f^{\rm prod}_{15}$ are fixed to zero due to the lack of sensitivity. We report the mass and the width of the $K^*(892)^\pm$ resonance, which are also determined.

| Intermediate state or component | Parameter value | | Fit fraction (%) |
|---|---|---|---|
| | Amplitude | Phase (rad) | |
| *ππ S-wave* | | | 15.4 |
| $\beta_1$ | $5.54 \pm 0.06$ | $-0.054 \pm 0.007$ | |
| $\beta_2$ | $15.64 \pm 0.06$ | $-3.125 \pm 0.005$ | |
| $\beta_3$ | $44.6 \pm 1.2$ | $+2.731 \pm 0.015$ | |
| $\beta_4$ | $9.3 \pm 0.2$ | $+2.30 \pm 0.02$ | |
| $f^{\rm prod}_{11}$ | $11.43 \pm 0.11$ | $-0.005 \pm 0.009$ | |
| $f^{\rm prod}_{12}$ | $15.5 \pm 0.4$ | $-1.13 \pm 0.02$ | |
| $f^{\rm prod}_{13}$ | $7.0 \pm 0.7$ | $+0.99 \pm 0.11$ | |
| $s^{\rm prod}_0$ | $-3.92637$ | | |
| *$K^0_S$ modes* | | | |
| $K^0_S\rho(770)^0$ | 1 | 0 | 21.1 |
| $K^0_S\omega(782)$ | $0.0420 \pm 0.0006$ | $+2.046 \pm 0.014$ | 0.6 |
| $K^0_S f_2(1270)$ | $0.410 \pm 0.013$ | $+2.88 \pm 0.03$ | 0.3 |
| *Kπ S-wave* | | | |
| $K^*_0(1430)^-\pi^+$ | $2.650 \pm 0.015$ | $+1.497 \pm 0.007$ | 6.1 |
| $K^*_0(1430)^+\pi^-$ | $0.145 \pm 0.014$ | $+1.78 \pm 0.10$ | $< 0.1$ |
| $M_{K^*_0(1430)}$ (MeV/$c^2$) | $1421.5 \pm 1.6$ | | |
| $\Gamma_{K^*_0(1430)}$ (MeV/$c^2$) | $247 \pm 3$ | | |
| $B$ | $0.62 \pm 0.04$ | | |
| $\phi_B$ (rad) | $-\ 0.100 \pm 0.010$ | | |
| $R$ | 1 | | |
| $\phi_R$ (rad) | $+\ 1.10 \pm 0.02$ | | |
| $a$ ([GeV/$c$]$^{-1}$) | $0.224 \pm 0.003$ | | |
| $r$ ([GeV/$c$]$^{-1}$) | $-\ 15.01 \pm 0.13$ | | |
| *$K^*$ modes* | | | |
| $K^*(892)^-\pi^+$ | $1.735 \pm 0.005$ | $+2.331 \pm 0.004$ | 57.0 |
| $K^*(892)^+\pi^-$ | $0.164 \pm 0.003$ | $-0.768 \pm 0.019$ | 0.6 |
| $K^*_2(1430)^-\pi^+$ | $1.303 \pm 0.013$ | $+2.498 \pm 0.012$ | 1.9 |
| $K^*_2(1430)^+\pi^-$ | $0.115 \pm 0.013$ | $+2.69 \pm 0.11$ | $< 0.1$ |
| $K^*(1680)^-\pi^+$ | $0.90 \pm 0.03$ | $-2.97 \pm 0.04$ | 0.3 |
| *$K^*(892)$ parameters* | | | |
| $M_{K^*(892)}$ (MeV/$c^2$) | $893.70 \pm 0.07$ | | |
| $\Gamma_{K^*(892)}$ (MeV/$c^2$) | $46.74 \pm 0.15$ | | |

The description of the $D^0 \to K^0_S K^+K^-$ decay amplitude adopted by *BABAR* is based on an isobar model containing five distinct resonances leading to 8 two-body decays (see Table 17.8.6). The $\phi(1020)$ resonance is described using a relativistic BW, with mass and width allowed to vary in the fit in order to account for mass resolution effects. The use of this approach, rather than the technically challenging convolution of the relativistic BW with a Gaussian-like resolution function, has been studied with simulated data and shown to have a negligible systematic effect. Since the $a_0(980)$ resonance has a mass very close to the $K\overline{K}$ threshold and decays mostly to $\eta\pi$, it is described using a coupled channel BW, as described in Section 13.2.1, where the pole mass and coupling constant to $\eta\pi$ are taken from Abele et al. (1998), and the coupling constant to $K\overline{K}$, $g_{K\overline{K}}$, is directly obtained from the fit.

Both experiments estimate the quality of their amplitude models using $\chi^2$ tests. *BABAR* employs a two-dimensional adaptive binning that requires at least 30 observed events per bin, obtaining $\chi^2/n_{\rm dof} = 1.21$ for 8585 degrees of freedom (dof) for $D^0 \to K^0_S\pi^+\pi^-$, and 1.28 for 1178 dof for $D^0 \to K^0_S K^+K^-$. Belle divides the region bounded by $m^2_\pm = 0.3$ and $3.0\,{\rm GeV}^2/c^4$ into $54 \times 54$ bins; bins with an expected population of less than 50 events are then combined with adjacent ones, finding $\chi^2/n_{\rm dof} = 2.35$ for 1065 dof. The values are large, but both experiments find that the main features of the Dalitz plot are well reproduced, with some significant but numerically small discrepancies at the peaks and dips of the distribution, which are used later to assign systematic uncertainties. *BABAR* has estimated that most of their excess in $\chi^2/n_{\rm dof}$, $\Delta\chi^2/n_{\rm dof} \approx 0.16$, arises from imperfections in modeling experimental effects — mostly efficiency variations at the boundaries of the Dalitz plot, and invariant mass resolution — rather than the amplitude model (Aubert, 2008l).

**Table 17.8.6.** *BABAR* fit results for the $D^0 \to K^0_S K^+ K^-$ decay (del Amo Sanchez, 2010b,f). Errors are statistical only. The phases are given in the interval $[-\pi, +\pi]$ rad. We also report the mass and the width of the $\phi(1020)$ resonance, and the $a_0(980)$ coupling constant to $K\overline{K}$ introduced in Section 13.2.2, as determined from the fit.

| Intermediate state or component | Parameter value | | Fit fraction (%) |
|---|---|---|---|
| | Amplitude | Phase (rad) | |
| $K^0_S a_0(980)^0$ | 1 | 0 | 51.8 |
| $a_0(980)^+ K^-$ | $0.635 \pm 0.006$ | $-2.91 \pm 0.02$ | 19.5 |
| $a_0(980)^- K^+$ | $0.125 \pm 0.008$ | $+2.47 \pm 0.04$ | 0.7 |
| $K^0_S f_0(1370)$ | $0.16 \pm 0.05$ | $+0.2 \pm 0.2$ | 1.7 |
| $K^0_S a_0(1450)^0$ | $0.83 \pm 0.10$ | $-1.93 \pm 0.12$ | 19.3 |
| $a_0(1450)^+ K^-$ | $0.93 \pm 0.03$ | $+1.66 \pm 0.07$ | 25.6 |
| $K^0_S \phi(1020)$ | $0.2313 \pm 0.0011$ | $-0.977 \pm 0.008$ | 44.1 |
| $K^0_S f_2(1270)$ | $0.385 \pm 0.015$ | $+0.06 \pm 0.04$ | 0.7 |
| *$\phi(1020)$ and $a_0(980)$ parameters* | | | |
| $M_{\phi(1020)}$ (MeV/$c^2$) | $1019.55 \pm 0.02$ | | |
| $\Gamma_{\phi(1020)}$ (MeV/$c^2$) | $4.60 \pm 0.04$ | | |
| $g_{K\overline{K}}$ (MeV/$c^2$) | $537 \pm 9$ | | |

### Selection of $B$ decays

Event selection for $B^\pm \to D^{(*)} K^{(*)\pm}$ decays is performed using the $m_{\rm ES}$ and $\Delta E$ variables. Additional suppression of background from $e^+e^- \to q\overline{q}$ ($q = u, d, s, c$) events is provided by using $\cos\theta_{\rm T}$, where $\theta_{\rm T}$ is the angle between the thrust axes of the $B$ signal candidate and the rest of the event, and a Fisher discriminant $\mathcal{F}$ combining 11 parameters that describe the momentum flow in the event relative to the $B$ thrust axis (Belle; see Section 9.3) or combining the monomials $L_0$, $L_2$, and the variables $\cos\theta_{\rm S}$ and $\cos\theta_{\rm B}$ (*BABAR*; see Sections 9.3 and 9.4). All topological variables are optimized to separate continuum events from signal.

The fit of the event distributions differs for the two collaborations. In the Belle approach, the fit is performed in two stages. At the first stage, the distributions of the event selection variables ($m_{\rm ES}$, $\Delta E$, $\cos\theta_{\rm T}$, and $\mathcal{F}$, shown in Fig. 17.8.6, top row, for $B^\pm \to DK^\pm$ decays) are fitted to obtain the relative fractions of signal and backgrounds. In the second stage, the Dalitz plot fit is performed (separately for $B^+$ and $B^-$ data) with the event-by-event background fractions based on the information obtained at the first stage. *BABAR* uses a simultaneous combined fit to $m_{\rm ES}$, $\Delta E$, $\mathcal{F}$, shown in Fig. 17.8.6 (bottom row) for $B^\pm \to DK^\pm$, $D \to K^0_S K^+ K^-$ decays, and to the Dalitz plot variables. The signal event yields and purities obtained by the two experiments, in signal-enriched regions, are given in Table 17.8.7. While Belle has a larger sample of $B\overline{B}$ pairs, the final signal yields from *BABAR* are larger. This difference in $D^0 \to K^0_S \pi^+\pi^-$ reconstruction efficiencies between the two experiments is mostly due to the enhanced tracking performance of *BABAR* for high-multiplicity (low $p_T$) events, reflecting differences in their low momentum pattern recognition and the silicon detector. As discussed in Section 2.2.1, the *BABAR* SVT performs stand-alone efficient low-momentum tracking, while the Belle SVD is employed to extrapolate tracks reconstructed in the CDC to the interaction region.

### Fit results

Instead of directly using the physical observables, both analyses use Cartesian variables

$$z_\pm = x_\pm + i y_\pm, \qquad (17.8.14)$$

first proposed in Aubert (2005o), which are expressed in terms of the physical observables as

$$z_\pm = r_B \exp[i(\delta_B \pm \phi_3)]. \qquad (17.8.15)$$

These observables have better statistical behavior (small correlation, minimal dependence of their uncertainties on the actual values) and allow for easier combination of several measurements into a single result. The obvious disadvantage is the necessary conversion required to obtain the values of $\phi_3$ and other related quantities. The strong phase in the $D^* \to D\pi^0$ and $D^* \to D\gamma$ modes differs by 180°, thus the observables $x^*_\pm$ and $y^*_\pm$ for $B^\pm \to D^* K^\pm$ have opposite sign (Bondar and Gershon, 2004). For $B^\pm \to DK^{*\pm}$ decays, following the suggestion in Gronau (2003),

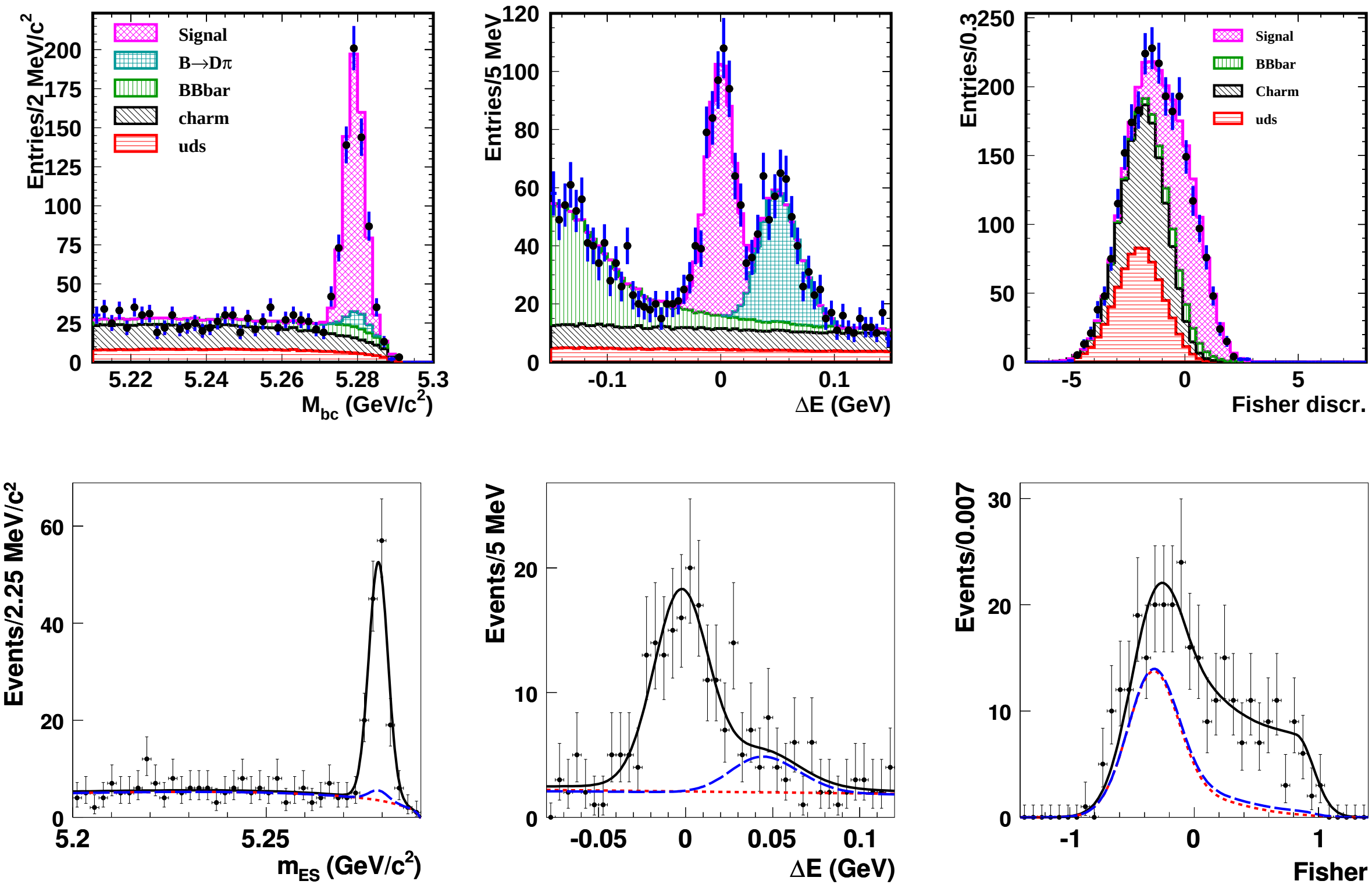

**Figure 17.8.6.** The $m_{\mathrm{ES}}$, or $M_{\mathrm{bc}}$ (left column), $\Delta E$ (middle column), and $\mathcal{F}$ (right column) distributions for $B^{\pm} \to DK^{\pm}$, $D \to K_S^0\pi^+\pi^-$ decays from Belle (Poluektov, 2010; top row) and for $B^{\pm} \to DK^{\pm}$, $D \to K_S^0 K^+K^-$ decays from *BABAR* (del Amo Sanchez, 2010b; bottom row). Points with error bars are the data. In the top row the histograms are fitted contributions due to signal, misidentified $B^{\pm} \to D\pi^{\pm}$ events, and $B\overline{B}$, charm, and continuum background. In the bottom row the curves superimposed represent the projections of the *BABAR* fit: signal plus background (solid black lines), the continuum plus $B\overline{B}$ background contributions (dotted red lines), and the sum of the continuum, $B\overline{B}$, and misidentified $B^{\pm} \to D\pi^{\pm}$ events (dashed blue lines). The distributions are for events in the signal region defined through the requirements $m_{\mathrm{ES}} > 5.272\,\mathrm{GeV}/c^2$, $|\Delta E| < 30\,\mathrm{MeV}$ (common to the two experiments), $|\cos\theta_{\mathrm{T}}| < 0.8$ and $\mathcal{F} > -0.7$ by Belle, and $\mathcal{F} > -0.1$ by *BABAR*, except the one on the plotted variable.

**Table 17.8.7.** Event yields in modes used for Dalitz plot analyses. The numbers in parenthesis indicate the signal purity in the signal region. This region is defined through the requirements $m_{\mathrm{ES}} > 5.272$ GeV/$c^2$, $|\Delta E| < 30$ MeV (common to the two experiments), $|\cos\theta_{\mathrm{T}}| < 0.8$ and $\mathcal{F} > -0.7$ by Belle, and $\mathcal{F} > -0.1$ by *BABAR*.

| Mode | Belle, $D^0 \to K_S^0\pi^+\pi^-$ (Poluektov, 2010) | *BABAR*, $D^0 \to K_S^0\pi^+\pi^-$ (del Amo Sanchez, 2010b) | *BABAR*, $D^0 \to K_S^0K^+K^-$ (del Amo Sanchez, 2010b) |
|---|---|---|---|
| $B^{\pm} \to DK^{\pm}$ | 756 (71%) | $896 \pm 35$ (68%) | $154 \pm 14$ (82%) |
| $B^{\pm} \to D^*K^{\pm}$, $D^* \to D\pi^0$ | 149 (78%) | $255 \pm 21$ (81%) | $56 \pm 11$ (87%) |
| $B^{\pm} \to D^*K^{\pm}$, $D^* \to D\gamma$ | 141 (42%) | $193 \pm 19$ (55%) | $30 \pm 7$ (78%) |
| $B^{\pm} \to DK^{*\pm}$ | $54 \pm 8$ (65%) (Poluektov, 2006) | $163 \pm 18$ (58%) | $28 \pm 6$ (81%) |

*BABAR* measures the effective Cartesian parameters $z_{\pm}^s = x_{\pm}^s + iy_{\pm}^s = \kappa r_B^s \exp[i(\delta_B^s \pm \phi_3)]$, where $0 < \kappa < 1$ is an effective hadronic parameter that accounts for the interference between $B^{\pm} \to DK^{*\pm}$ and other $B^{\pm} \to DK_S^0\pi^{\pm}$ decays, as a consequence of the $K^{*\pm}$ natural width. This effective parameterization also accounts for efficiency variations as a function of the kinematics of the $B$ decay. Belle measures $z_{\pm}^s$ assuming $\kappa = 1$. Both experiments finally assign an additional source of systematic uncertainty due to non-resonant decays.

Results for $z_{\pm}$, $z_{\pm}^*$, and $z_{\pm}^s$ for $B^{\pm} \to DK^{\pm}$, $B^{\pm} \to D^*K^{\pm}$, and $B^{\pm} \to DK^{*\pm}$ decays, respectively, are pre-sented in Table 17.8.8. Belle reports $z^*_{\pm}$ values separately for $D^* \to D\pi^0$ and $D^* \to D\gamma$ modes and combines them in the $\phi_3$ fit (Poluektov, 2010), while *BABAR* reports $D^* \to D\pi^0$ and $D^* \to D\gamma$ combined values inverting the sign for the latter (del Amo Sanchez, 2010b). Belle results for $B^{\pm} \to DK^{*\pm}$ are reported in Poluektov (2006). The model uncertainties in the case of the Belle analysis are reported for the physics parameters $\phi_3$, $r_B$, and $\delta_B$; the uncertainties on $z_{\pm}$, $z^*_{\pm}$, and $z^s_{\pm}$ quoted in Table 17.8.8 are based on information provided by Belle and published in Asner et al. (2011). Figure 17.8.7 shows the corresponding one-, two-, and three-standard deviation contours in two dimensions in the $z_{\pm}$, $z^*_{\pm}$, and $z^s_{\pm}$ planes, together with the HFAG combination (Asner et al., 2011). These averages take into account the effect of correlations within each experiment's set of measurements, both statistical and systematic (excluding effects of the amplitude model), and are performed assuming that both experiments use the same decay amplitude model and that the model uncertainty is fully correlated between the two experiments (thus this source of error is not used in the averaging procedure). It is also assumed that the selection of $B^{\pm} \to DK^{*\pm}$ decays is the same in both experiments, neglecting the Belle model uncertainty due to possible non-resonant decays.

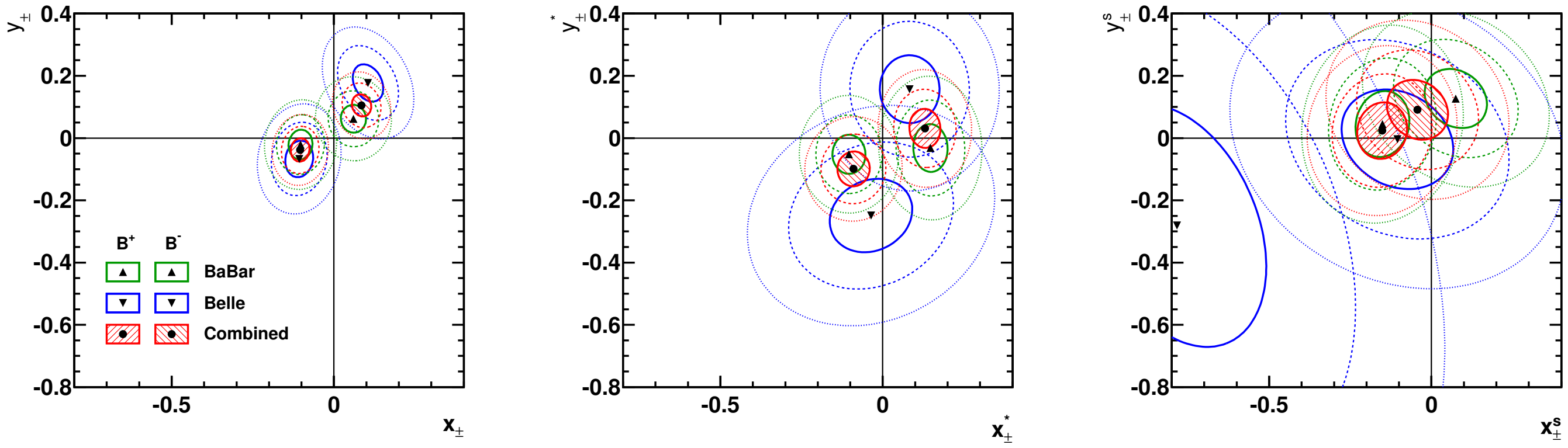

**Figure 17.8.7.** Contours at the 60.7% confidence level for 2 degrees of freedom (corresponding to $-2\Delta \ln L = \Delta\chi^2 = 1$, *i.e.*, one standard deviation in two dimensions assuming Gaussian errors, solid lines), and two- and three-standard deviation contours (dashed lines) in the (left) $z_{\pm}$, (center) $z^*_{\pm}$, and (right) $z^s_{\pm}$ planes, for *BABAR* and Belle separately (including all errors other than model uncertainties), and their HFAG combination (Asner et al., 2011).

**Table 17.8.8.** Fit results for $B^{\pm} \to DK^{\pm}$, $B^{\pm} \to D^*K^{\pm}$, and $B^{\pm} \to DK^{*\pm}$ modes using the Dalitz analysis technique in Cartesian variables $z_{\pm}$, $z^*_{\pm}$, and $z^s_{\pm}$, respectively. The first error is statistical, the second is the experimental systematic uncertainty, and the third reflects the uncertainty in the description of the neutral $D$ decay amplitudes.

| | Real part (%) | Imaginary part (%) | Real part (%) | Imaginary part (%) |
|---|---|---|---|---|
| | *BABAR* (del Amo Sanchez, 2010b) | | Belle (Poluektov, 2010) | |
| $z_-$ | $+6.0 \pm 3.9 \pm 0.7 \pm 0.6$ | $+6.2 \pm 4.5 \pm 0.4 \pm 0.6$ | $+10.5 \pm 4.7 \pm 1.1 \pm 6.4$ | $+17.7 \pm 6.0 \pm 1.8 \pm 5.4$ |
| $z_+$ | $-10.3 \pm 3.7 \pm 0.6 \pm 0.7$ | $-2.1 \pm 4.8 \pm 0.4 \pm 0.9$ | $-10.7 \pm 4.3 \pm 1.1 \pm 5.5$ | $-6.7 \pm 5.9 \pm 1.8 \pm 6.3$ |
| $z^*_-$ $[D\pi^0]$ | $-10.4 \pm 5.1 \pm 1.9 \pm 0.2$ | $-5.2 \pm 6.3 \pm 0.9 \pm 0.7$ | $+2.4 \pm 14.0 \pm 1.8 \pm 9.0$ | $-24.3 \pm 13.7 \pm 2.2 \pm 4.9$ |
| $z^*_+$ $[D\pi^0]$ | $+14.7 \pm 5.3 \pm 1.7 \pm 0.3$ | $-3.2 \pm 7.7 \pm 0.8 \pm 0.6$ | $+13.3 \pm 8.3 \pm 1.8 \pm 8.1$ | $+13.0 \pm 12.0 \pm 2.2 \pm 6.3$ |
| $z^*_-$ $[D\gamma]$ | Included in $z^*_-$ $(D\pi^0)$ | | $+14.4 \pm 20.8 \pm 2.5 \pm 9.0$ | $+19.6 \pm 21.5 \pm 3.7 \pm 4.9$ |
| $z^*_+$ $[D\gamma]$ | Included in $z^*_+$ $(D\pi^0)$ | | $-0.6 \pm 14.7 \pm 2.5 \pm 8.1$ | $-19.0 \pm 17.7 \pm 3.7 \pm 6.3$ |
| | *BABAR* (del Amo Sanchez, 2010b) | | Belle (Poluektov, 2006) | |
| $z^s_-$ | $+7.5 \pm 9.6 \pm 2.9 \pm 0.7$ | $+12.7 \pm 9.5 \pm 2.7 \pm 0.6$ | $-78.4^{+24.9}_{-29.5} \pm 2.9 \pm 9.7$ | $-28.1^{+44.0}_{-33.5} \pm 4.6 \pm 8.6$ |
| $z^s_+$ | $-15.1 \pm 8.3 \pm 2.9 \pm 0.6$ | $+4.5 \pm 10.6 \pm 3.6 \pm 0.8$ | $-10.5^{+17.7}_{-16.7} \pm 0.6 \pm 8.8$ | $-0.4^{+16.4}_{-15.6} \pm 1.3 \pm 9.5$ |

### *Interpretation of fit results*

The fit results expressed in terms of the $z_{\pm}$ variables are then converted into physical parameters $\phi_3$, $r_B$, and $\delta_B$. $\phi_3$ is constrained to be the same in all modes used, while $r_B$ and $\delta_B$ are allowed to be different for different $B$ decay modes. Note that, due to this constraint and the fact that $r_B$ is expected to be the same for $B^+$ and $B^-$, the number of physical parameters is smaller than the number of experimental observables. Thus, the statistical treatment has to take into account the mathematical mismatch between the experimental results $z_{\pm}$ and the set of physical observables $(\phi_3, r_B, \delta_B)$ due to statistical fluctuations (Yabsley, 2006). Both collaborations use a frequentist approach to obtain $\phi_3$, although the details of the treatment differ. In both cases the method requires knowledge of the probability density function (p.d.f.) $p(\boldsymbol{z}|\boldsymbol{\mu})$ of the vector $\boldsymbol{z}$ of measured parameters $z_{\pm}$, $z^{*}_{\pm}$, and $z^{s}_{\pm}$ as a function of the true parameters $\bar{\boldsymbol{z}}$, which can easily be expressed in terms of the vector $\boldsymbol{\mu} = (\phi_3,\, r_B,\, \delta_B,\, r^{*}_B,\, \delta^{*}_B,\, \kappa r^{s}_B,\, \delta^{s}_B)$.

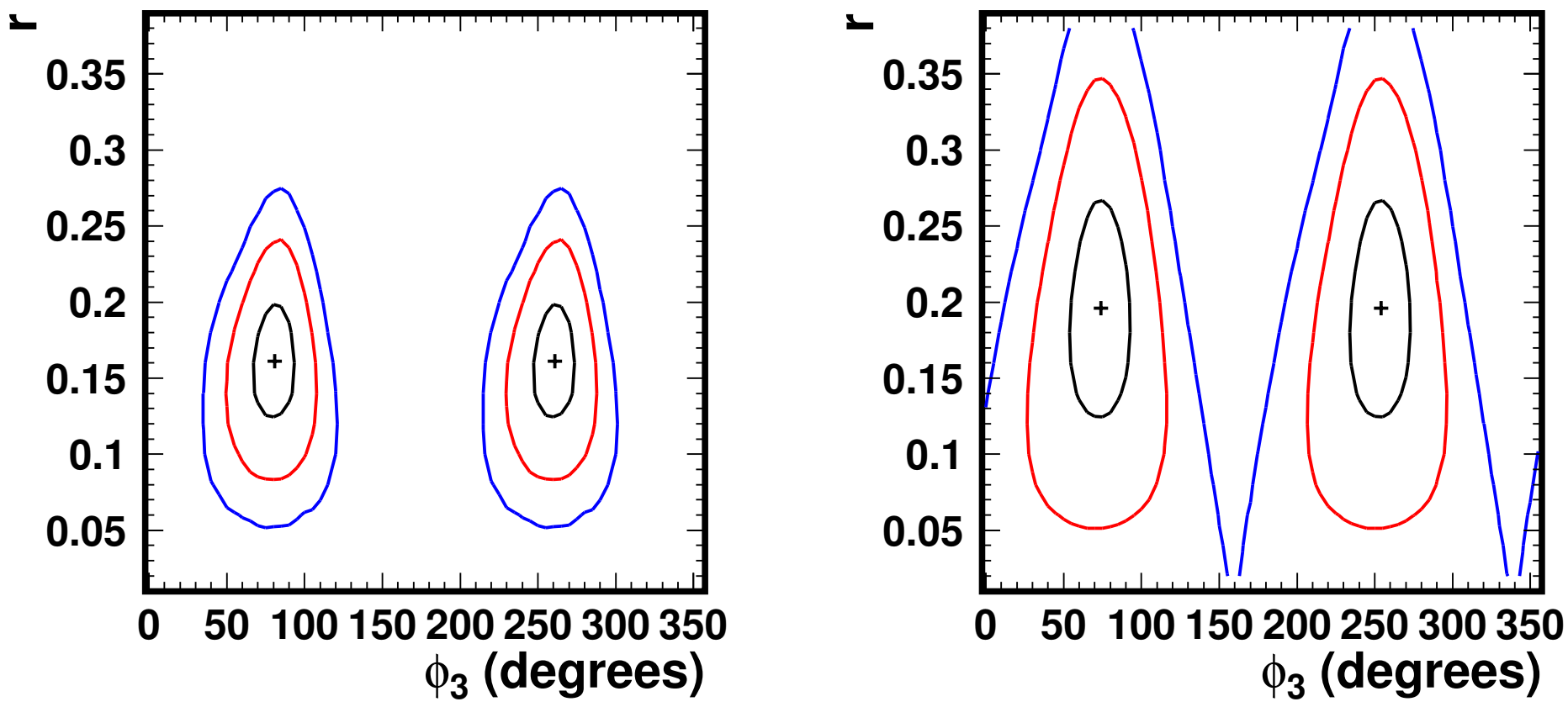

**Figure 17.8.8.** Belle projections of the confidence regions for $B^{\pm} \to DK^{\pm}$ (left) and $B^{\pm} \to D^{*}K^{\pm}$ (right) decays onto the $(\phi_3, r_B)$ and $(\phi_3, r^{*}_B,)$ planes (Poluektov, 2010). Contours indicate projections of one-, two-, and three-standard deviation regions.

To obtain this p.d.f., Belle uses a simplified Monte Carlo (MC) simulation of the experiment which incorporates the same efficiencies, resolution, and backgrounds as used in the fit to the experimental data. Belle constructs three-dimensional regions in the $\boldsymbol{\mu}$ space, using the unified approach of Feldman and Cousins (1998). The confidence level $\alpha$ is calculated as $\alpha(\boldsymbol{\mu}) = \int_{\mathcal{D}(\boldsymbol{\mu})} p(\boldsymbol{z}|\boldsymbol{\mu}) d\boldsymbol{z}$, where the integration domain $\mathcal{D}$ is given by the likelihood ratio ordering

$$\frac{p(\boldsymbol{z}|\boldsymbol{\mu})}{p(\boldsymbol{z}|\boldsymbol{\mu}_{\rm best}(\boldsymbol{z}))} > \frac{p(\boldsymbol{z}_0|\boldsymbol{\mu})}{p(\boldsymbol{z}_0|\boldsymbol{\mu}_{\rm best}(\boldsymbol{z}_0))}, \qquad (17.8.16)$$

where $\boldsymbol{\mu}_{\rm best}(\boldsymbol{z})$ stands for the best parameters $\boldsymbol{\mu}$ such that $p(\boldsymbol{z}|\boldsymbol{\mu})$ is maximized for the given measurement $\boldsymbol{z}$, and $\boldsymbol{z}_0$ is the measurement from the fit to the experimental data.

*BABAR* instead constructs directly one-dimensional intervals calculating the confidence level as a function of the true value of a given parameter $\mu$ from $\boldsymbol{\mu} \equiv \{\mu, \boldsymbol{q}\}$ as $\alpha(\mu) = 1 - F[\Delta\chi^2(\mu)]$, where $F[\Delta\chi^2(\mu)]$ is the cumulative expected distribution of $\Delta\chi^2(\mu)$, with

$$\Delta\chi^2(\mu) = -2\ln\frac{p(\boldsymbol{z}_0|\mu, \boldsymbol{q}(\mu))}{p(\boldsymbol{z}_0|\boldsymbol{\mu}_{\rm best}(\boldsymbol{z}_0))}. \qquad (17.8.17)$$

Here $\boldsymbol{q}(\mu)$ stands for the parameters $\boldsymbol{q}$ that maximize $p(\boldsymbol{z}_0|\boldsymbol{\mu})$ for the given $\mu$, and $\boldsymbol{\mu}_{\rm best}(\boldsymbol{z}_0)$ is as defined above. The $p(\boldsymbol{z}|\boldsymbol{\mu})$ p.d.f. is approximated by a correlated, multidimensional Gaussian in the vector $\boldsymbol{z}$ of measurements, previously validated using a simplified MC simulation, similar to that performed by Belle. The distribution $F[\Delta\chi^2(\mu)]$ is obtained using a large number of MC simulated samples, by counting of the number of experiments generated with true values $\boldsymbol{\mu} = \{\mu, \boldsymbol{q}(\mu)\}$ that have better $\Delta\chi^2(\mu)$ than the actual experiment.

Figure 17.8.8 shows the Belle projection of the three-dimensional region onto the $(r_B, \phi_3)$ plane, for each of the $B^{\pm} \to DK^{\pm}$ and $B^{\pm} \to D^{*}K^{\pm}$ modes, for 20%, 74%, and 97% confidence level regions, which correspond to one-, two-, and three-standard deviations for a three-dimensional Gaussian distribution. Similarly, Fig. 17.8.9 shows the *BABAR* confidence level as a function of $\phi_3$ and $r_B$ for each of the three $B$ decay channels, as well as their combination for the case of the weak phase. Table 17.8.9 reports the corresponding central values with their one- and two-standard deviation intervals. Using these frequentist procedures, each of the experiments obtains a departure from $\phi_3 = 0$ equivalent to 3.5 standard deviations, providing evidence for direct *CP* violation in $B^{\pm} \to D^{(*)}K^{(*)\pm}$ decays.

### *Model uncertainty*

While the experimental systematic uncertainties in the GGSZ measurements with the model-dependent technique can be understood using control samples and Monte Carlo simulation, the uncertainty arising from the description of the amplitude model is more difficult to quantify. Both experiments follow the general guidelines discussed in Section 13.5, although with differences in the details. *BABAR* uses alternative models that give similar $D^0 \to K^0_S\pi^+\pi^-$ fit quality to that of the default model, where BW parameters are varied according to their uncertainties, the reference K-matrix solution is replaced by other solutions (Anisovich and Sarantsev, 2003), and the standard parameterizations are replaced by other related choices, for example, replacing the Gounaris-Sakurai and $K\pi$ $S$-wave parameterizations by BWs, removing the mass dependence in the P vector, changing form factors, and adopting the helicity formalism instead of Zeemach tensors to describe the angular dependence. Other models are built by removing or adding resonances with small or negligible fractions, or accounting explicitly for $D^0 - \overline{D}^0$ mixing effects. Belle performs model variations that employ a reduced number of resonances while keeping the absolute value of the amplitude the same as in the default model. Models excluding the $\sigma_1$ and $\sigma_2$ states; or using only the largest Cabibbo-favored term $D^0 \to K^*(892)^-\pi^+$, the narrow resonances, $D^0 \to K_S^0 f_0(980)$, and $D^0 \to K_0^*(1430)^-\pi^+$, and a large flat non-resonant term; or the model used by CLEO (Muramatsu et al., 2002), are more conservative model variations than those performed by *BABAR*, where these extreme models have been discarded on the basis of their significantly poorer fit quality. Other variations used by Belle include removal of the form factors for the $D$ meson and intermediate resonances, and of the momentum dependence of the resonance width.

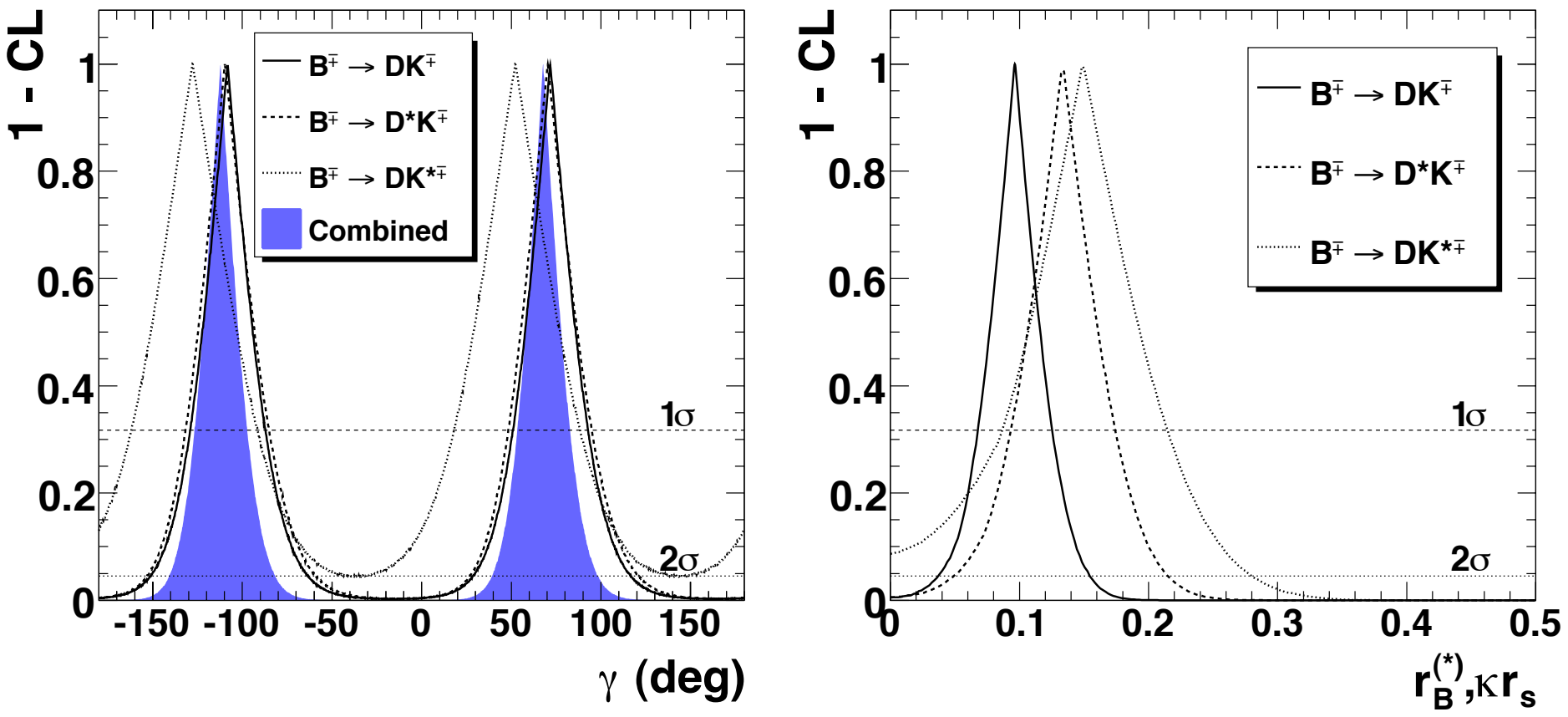

**Figure 17.8.9.** The *BABAR* $1 - \text{C.L.}$ distributions as a function of $\gamma = \phi_3$ (left) and $r_B$ (right) for $B^\pm \to DK^\pm$, $B^\pm \to D^*K^\pm$, and $B^\pm \to DK^{*\pm}$ decays separately, and their combination, including statistical and systematic uncertainties (del Amo Sanchez, 2010b). The dashed (upper) and dotted (lower) horizontal lines correspond to the one- and two-standard deviation intervals, respectively.

**Table 17.8.9.** The 68.3% and 95.4% one-dimensional C.L. regions, equivalent to one- and two-standard deviation intervals, for $\phi_3$, $\delta_B$, $r_B$, $\delta_B^s$, and $\kappa r_B^s$, including all sources of uncertainty. The 68.3% C.L. regions show separate contributions from statistical, experimental systematic, and model uncertainties. The 95.4% regions include statistical and systematic uncertainties for *BABAR*, but only statistical for Belle. $B^\pm \to DK^{*\pm}$ results from Belle (Poluektov, 2006), shown in the bottom panel of the table, are obtained assuming the effective hadronic parameter $\kappa = 1$, and are not included in the combined value for $\phi_3$.

| Parameter | 68.3% C.L. | 95.4% C.L. | 68.3% C.L. | 95.4% C.L. |
|---|---|---|---|---|
| | *BABAR* (del Amo Sanchez, 2010b) | | Belle (Poluektov, 2010) | |
| $\phi_3$ (°) | $68^{+15}_{-14} \pm 4 \pm 3$ | $[39, 98]$ | $78.4^{+10.8}_{-11.6} \pm 3.6 \pm 8.9$ | $[54.2, 100.5]$ |
| $r_B$ (%) | $9.6 \pm 2.9 \pm 0.5 \pm 0.4$ | $[3.7, 15.5]$ | $16.0^{+4.0}_{-3.8} \pm 1.1^{+5.0}_{-1.0}$ | $[8.4, 23.9]$ |
| $r_B^*$ (%) | $13.3^{+4.2}_{-3.9} \pm 1.3 \pm 0.3$ | $[4.9, 21.5]$ | $19.6^{+7.2}_{-6.9} \pm 1.2^{+6.2}_{-1.2}$ | $[6.1, 27.1]$ |
| $\kappa r_B^s$ (%) | $14.9^{+6.6}_{-6.2} \pm 2.6 \pm 0.6$ | $< 28.0$ | – | – |
| $\delta_B$ (°) | $119^{+19}_{-20} \pm 3 \pm 3$ | $[75, 157]$ | $136.7^{+13.0}_{-15.8} \pm 4.0 \pm 22.9$ | $[102.2, 162.3]$ |
| $\delta_B^*$ (°) | $-82 \pm 21 \pm 5 \pm 3$ | $[-124, -38]$ | $341.9^{+18.0}_{-19.6} \pm 3.0 \pm 22.9$ | $[296.5, 382.7]$ |
| $\delta_B^s$ (°) | $111 \pm 32 \pm 11 \pm 3$ | $[42, 178]$ | – | – |
| | | | Belle (Poluektov, 2006) | |
| $\kappa r_B^s$ (%) | | | $56.4^{+21.6}_{-15.5} \pm 4.1 \pm 8.4$ | $[23.1, 1.106]$ |
| $\delta_B^s$ (°) | | | $242.6^{+20.2}_{-23.2} \pm 2.5 \pm 49.3$ | $[186.0, 300.2]$ |

### 17.8.4.2 Model-dependent technique with other final states

*BABAR* has carried out similar analyses using the decay $B^\pm \to DK^\pm$ with the $D \to \pi^+\pi^-\pi^0$ final state (Aubert, 2007w), and the neutral $B$ decay $B^0 \to DK^{*0}$, $K^{*0} \to K^+\pi^-$, with $D \to K_S^0\pi^+\pi^-$ (Aubert, 2009f).

In the study of the $B^{\pm} \to DK^{\pm}$, $D \to \pi^+\pi^-\pi^0$ decay, *BABAR* measures from $324 \times 10^6$ $B\overline{B}$ pairs $\rho_- = 0.815 \pm 0.034$, $\theta_- = (186 \pm 7)^\circ$, $\rho_+ = 0.854 \pm 0.035$, $\theta_+ = (192 \pm 7)^\circ$, where the polar parameterization $\rho_\pm \equiv |z_\pm - x_0|$, $\theta_\pm = \tan^{-1} y_\pm/(x_\pm - x_0)$ (with $x_0 = 0.850$) is chosen to reflect the symmetry properties of the measurement: studies show that this removes nonlinear correlations (and consequent bias) in the fit, and improves the sensitivity of the result. These results are consistent with $\rho_\pm = x_0$, $\theta_\pm = 180^\circ$, which corresponds to $z_\pm = 0$.

For the neutral $B$ decay $B^0 \to DK^{*0}$, $K^{*0} \to K^+\pi^-$, $D \to K^0_S\pi^+\pi^-$, $r_B$ is naïvely expected to be larger, $\sim 0.3$ (Section 17.8.3), although the overall rate of events is significantly smaller than for $B^\pm \to DK^{*\pm}$ decays. The flavor of the neutral $B$ meson is tagged by the charge of the kaon produced in the $K^*(892)^0$ decay ($K^+\pi^-$ or $K^-\pi^+$). The analysis finds $39 \pm 9$ signal events from $371 \times 10^6$ $B\overline{B}$ pairs, and using a Bayesian analysis with external inputs yields $\phi_3 = (162 \pm 56)^\circ$ and $r_B < 0.55$ at the 90% confidence level.

Nevertheless, in both cases the errors on the experimental measurements are too large for a meaningful determination of $\phi_3$, or $\gamma$, and have not been included in the combined determination of $\phi_3$.

#### 17.8.4.3 Binned model-independent technique

In the binned fit approach to $\phi_3$ determination using $B^\pm \to DK^\pm$, $D^0 \to K^0_S\pi^+\pi^-$ decays, it is possible to avoid dependence on a detailed model of the $D^0$ amplitude across the Dalitz plot. Instead, if the plot is divided into bins, the amplitude in each bin can be described by quantities averaged over that bin. These quantities can be extracted from analyses of charm data, thus allowing for a completely model-independent measurement of $\phi_3$. This approach is particularly attractive for precision measurement at a super flavor factory where the model uncertainty would otherwise dominate the precision. The approach was first proposed in Giri, Grossman, Soffer, and Zupan (2003a), and further developed by Bondar and Poluektov (2006, 2008), where the experimental feasibility of the method was shown and an optimization procedure for the analysis was proposed. The analysis has been performed by Belle as a proof of principle using the final data sample of $772 \times 10^6$ $B\overline{B}$ pairs (Aihara, 2012) and based on results of the measurement of strong phase parameters by the CLEO collaboration (Briere et al., 2009; Libby et al., 2010).

##### *Procedure*

In the model-independent approach, the Dalitz plot is divided into $2\mathcal{N}$ bins symmetric under the exchange $m^2_- \leftrightarrow m^2_+$. The bin index "$i$" ranges from $-\mathcal{N}$ to $\mathcal{N}$ (excluding zero); the exchange $m^2_+ \leftrightarrow m^2_-$ corresponds to the exchange $i \leftrightarrow -i$. The expected number of events in the bin "$i$" of the Dalitz plot of the $D$ from a $B^+ \to DK^+$ decay is

$$N^+_i = h_B \left[K_i + r_B^2 K_{-i} + 2\sqrt{K_i K_{-i}}(x_+ c_i + y_+ s_i)\right], \tag{17.8.18}$$

where $h_B$ is a normalization constant and $K_i$ is the number of events in the $i^{\text{th}}$ bin of the Dalitz plot of the $D$ meson decaying into a flavor eigenstate (obtained using a $D^{*\pm} \to D\pi^\pm$ sample). The terms $c_i$ and $s_i$ include information about the cosine and sine of the phase difference $\delta_D(m^2_+, m^-)$ between $D^0$ and $\overline{D}^0$ averaged over the bin region:

$$c_i = \frac{\int\limits_{\mathcal{D}_i} |A_D||\overline{A}_D| \cos\delta_D \, d\mathcal{D}}{\sqrt{\int\limits_{\mathcal{D}_i} |A_D|^2 d\mathcal{D} \int\limits_{\mathcal{D}_i} |\overline{A}_D|^2 d\mathcal{D}}}. \tag{17.8.19}$$

Here $\mathcal{D}$ represents the Dalitz plot phase space and $\mathcal{D}_i$ is the bin region over which the integration is performed. The terms $s_i$ are defined similarly with sine substituted for cosine.

Neglecting effects due to neutral $D$ mixing and *CP* violation (which are measured or constrained at the 1% level or less; see the text on $D$-mixing and *CP* violation in Section 19.2), the strong phase difference $\delta_D$ is antisymmetric ($\delta_D(m^2_+, m^2-) = -\delta_D(m_-, m^+)$) and thus the relations $c_i = c_{-i}$ and $s_i = -s_{-i}$ hold. The values of the $c_i$ and $s_i$ terms can be measured using quantum correlated pairs of $D$ mesons created at charm-factory experiments operated at the threshold of $D\overline{D}$ pair production. The wave function of the two mesons is antisymmetric,

$$A_{\text{corr}} = A^{(1)}_D \overline{A}^{(2)}_D - A^{(2)}_D \overline{A}^{(1)}_D, \tag{17.8.20}$$

where the indices "(1)" and "(2)" correspond to the two decaying $D$ mesons. The four-dimensional probability density for the two correlated $D \to K^0_S\pi^+\pi^-$ Dalitz plots is sensitive to the strong phase difference. In the case of the binned analysis, the number of events where one $D$ meson lies in the $i$-th bin of the Dalitz plot and the other D meson in the $j$-th bin is

$$\begin{aligned} M_{ij} =& K_i K_{-j} + K_{-i} K_j \\ & -2\sqrt{K_i K_{-i} K_j K_{-j}}(c_i c_j + s_i s_j). \end{aligned} \tag{17.8.21}$$

In addition to the process where both $D$ mesons decay into $K^0_S\pi^+\pi^-$, CLEO (Briere et al., 2009; Libby et al., 2010) use decays $(K^0_L\pi^+\pi^-)_D(K^0_S\pi^+\pi^-)_D$ to increase the available data sample, although weak model assumptions are made to constrain the $c_i$ and $s_i$ values in this case, since the amplitudes with $K^0_L$ and $K^0_S$ differ.

Additional information about the values of $c_i$ is obtained from the process where one $D$ meson decays in a *CP* eigenstate, and the second — in the *CP* eigenstate of opposite sign — decays to $K^0_S\pi^+\pi^-$. The amplitude of this decay is

$$A_\pm = A_D \pm \overline{A}_D, \tag{17.8.22}$$

and the number of events in bins of the $D_{CP} \to K^0_S\pi^+\pi^-$ decay is

$$M_i = K_i + K_{-i} \pm 2\sqrt{K_i K_{-i}} c_i. \tag{17.8.23}$$

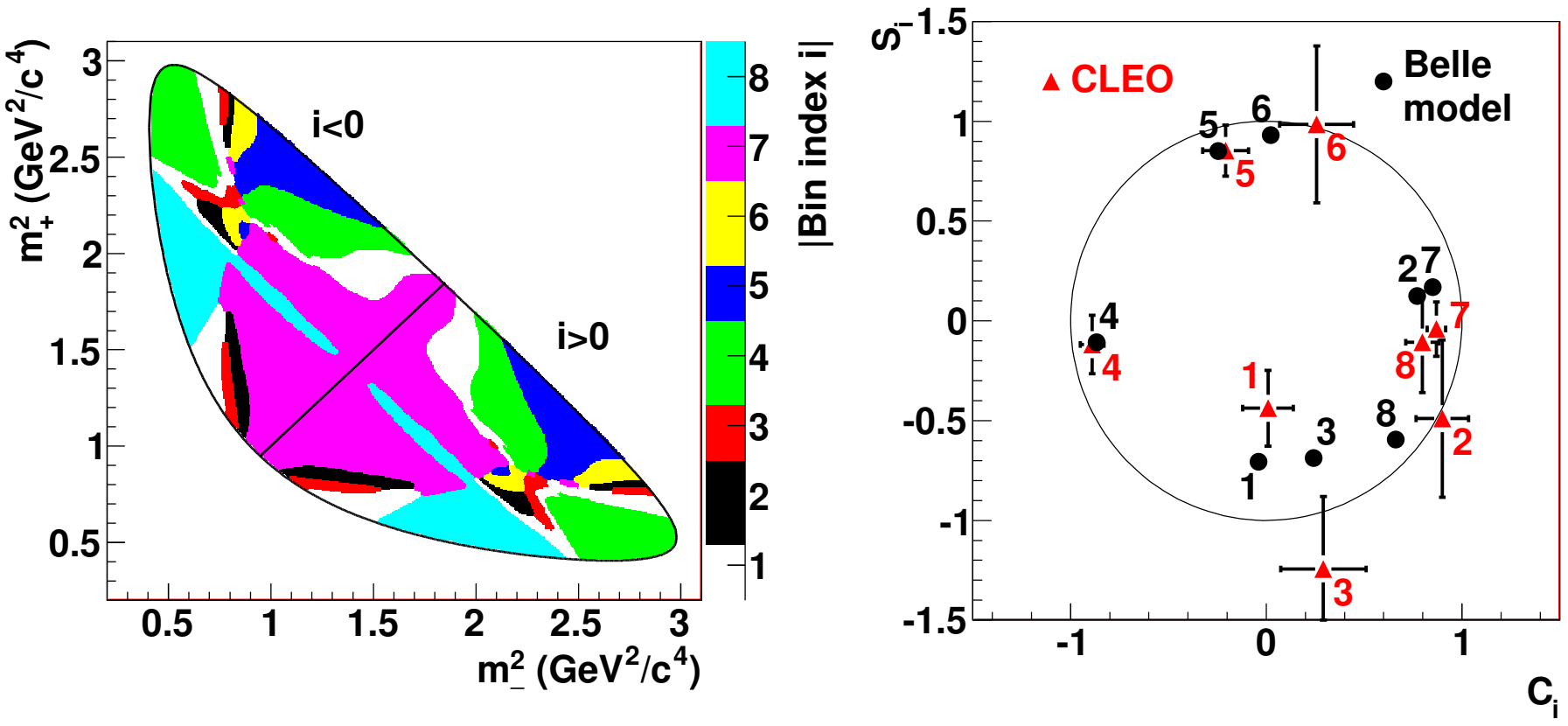

**Figure 17.8.10.** (a) Optimal binning of the $D^0 \to K^0_S\pi^+\pi^-$ Dalitz plot and (b) comparison of phase terms $c_i, s_i$ for the optimal binning measured by CLEO, and calculated from the Belle $D^0 \to K^0_S\pi^+\pi^-$ amplitude model, taken from Aihara (2012).

Note that the use of a *CP* eigenstate allows one to resolve an ambiguity in the measurement of $c_i$ and $s_i$ from correlated $K^0_S\pi^+\pi^-$ decays, as Eq. (17.8.21) is invariant under the simultaneous change of the signs of all $c_i$ or $s_i$, while Eq. (17.8.23) provides the signs of $c_i$. The ambiguity under the simultaneous change of signs of $s_i$ remains (corresponding to complex conjugation of the amplitude $A_D$), however this can be resolved with a weak model assumption. The solution that best fits the isobar model with BW amplitudes is preferred, since the other one, corresponding to the complex-conjugated parameterization, is unphysical: the complex-conjugated BW amplitude corresponds to the converging spherical wave in the quantum-mechanical scattering problem, and violates causality.

### *Optimal binning*

The statistical precision of the binned procedure depends strongly on the chosen binning. If the amplitude varies significantly across the bin area, the integral over the bin averages over the interference term, discarding information and reducing the sensitivity of the analysis. An optimal binning of the Dalitz plot, that takes into account both the variations of the strong phase difference and the absolute value of the $D^0 \to K^0_S\pi^+\pi^-$ amplitude, was proposed by Bondar and Poluektov (2008). The optimization uses the amplitude of $D^0 \to K^0_S\pi^+\pi^-$ decay from the model-dependent analysis. However, although the choice of binning is model-dependent, a bad choice of model results only in poorer statistical precision of the measurement, and not in systematic bias. It has been shown that as few as 16 bins are sufficient to reach statistical precision comparable to that of the unbinned fit.

Measurements of the phase terms $c_i$ and $s_i$ have been performed by CLEO (Briere et al., 2009; Libby et al., 2010), with various binnings of the Dalitz plot. The Belle analysis uses the binning shown in Fig. 17.8.10(a) optimized for the best statistical accuracy under the assumption that the background in $B^\pm \to DK^\pm$ decays is small. This optimization uses the *BABAR* amplitude measurement (Aubert, 2008l). The results of the CLEO measurement of $c_i$ and $s_i$ for this binning are presented in Fig. 17.8.10(b). Comparison with $c_i$ and $s_i$ calculated from the Belle model (Poluektov, 2010) shows reasonable agreement between the model and measurement: $\chi^2/n_{\rm dof} = 18.6/16$.

Once the values of the terms $c_i$ and $s_i$ are known, the system of equations (17.8.18) contains only three free parameters ($x$, $y$, and $h_B$) for each $B$ charge, and can be solved using the maximum likelihood method to extract the values of $r_B$, $\phi_3$, and $\delta_B$. The numbers of events $K_i$ and $N_i$ are extracted from $D^{*\pm} \to D\pi^\pm$ and $B^\pm \to DK^\pm$ samples respectively. To minimize the systematic error coming from the difference in reconstruction efficiency across the phase space for the two samples, the flavor-tagged results $K_i$ are obtained by choosing $D$ mesons in the momentum range $1.8\ {\rm GeV}/c < p_D < 2.8\ {\rm GeV}/c$, *i.e.*, with the same average momentum $p_D$ as for $B^\pm \to DK^\pm$ decays. Momentum resolution is taken into account by using a migration matrix to describe the cross-feed between bins.

### *Fit results and interpretation*

The parameters $x_\pm$ and $y_\pm$ are determined by a simultaneous fit over the 16 bins, using signal selection variables to determine the yield $N_i$ in each bin: $\Delta M$ and $M_D$ for the $D^{*\pm} \to D\pi^\pm$ sample, and $m_{\rm ES}$, $\Delta E$, $\cos\theta_{\rm T}$, and the same Fisher discriminant $\mathcal{F}$ as used in Belle's model-dependent analysis (Section 17.8.4) for the $B^\pm \to DK^\pm$ sample. Figure 17.8.11 shows the binned signal yield separately for $B^+$ and $B^-$ data, its charge asymmetry, and the results of the fit using Eq. (17.8.18) for each bin in the Dalitz plot. Different binned yields for positive and negative $B$ charges in Figs 17.8.11(a,b) suggest significant *CP* asymmetry, while the pattern in Figs 17.8.11(c,d) shows that this *CP* asymmetry is well described by the model involving a nonzero value of $\phi_3$ (Eq. 17.8.18).

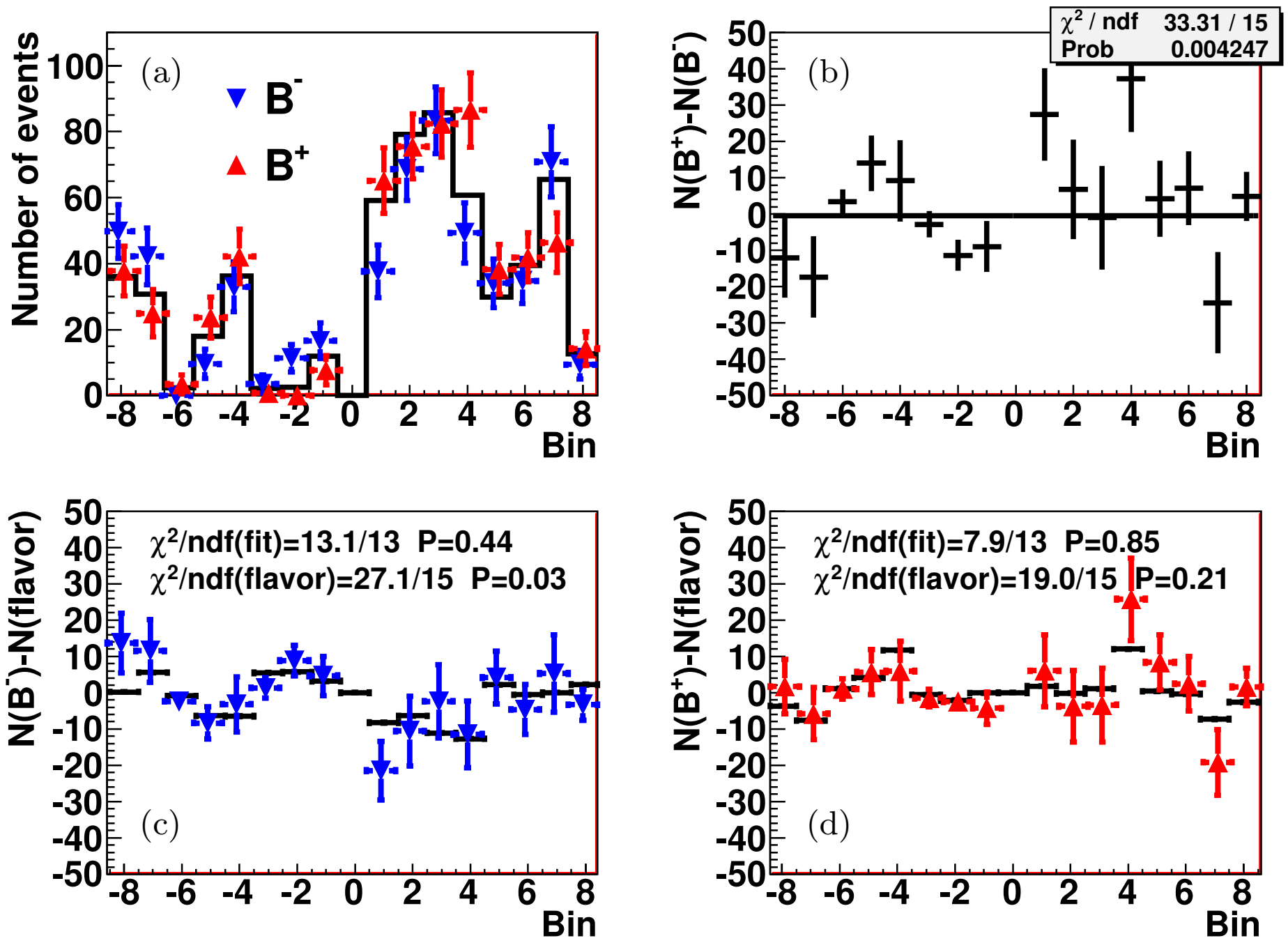

**Figure 17.8.11.** Results of the model-independent binned fit of the $B^{\pm} \to DK^{\pm}$ sample (Aihara, 2012). (a) Numbers of events in bins of the $D^0 \to K_S^0\pi^+\pi^-$ Dalitz plot: from $B^- \to DK^-$ (blue triangle downwards), $B^+ \to DK^+$ (red triangle upwards) and the flavor-tagged sample (histogram). (b) Difference of the number of events from $B^+ \to DK^+$ and $B^- \to DK^-$ decays. (c) Difference of the number of events from $B^- \to DK^-$ and flavor-tagged sample (normalized to the total number of $B^- \to DK^-$ decays): data (points with vertical and horizontal error bars), and as a result of the $(x, y)$ fit (horizontal bars). (d) Same as (c) for $B^+ \to DK^+$ data.

The values of the $z_{\pm}$ parameters obtained from the binned fit to the $B^{\pm} \to DK^{\pm}$ sample are

$$\begin{aligned} x_- &= +0.095 \pm 0.045 \pm 0.014 \pm 0.010, \\ y_- &= +0.137^{+0.053}_{-0.057} \pm 0.015 \pm 0.023, \\ x_+ &= -0.110 \pm 0.043 \pm 0.014 \pm 0.007, \\ y_+ &= -0.050^{+0.052}_{-0.055} \pm 0.011 \pm 0.017. \end{aligned} \quad (17.8.24)$$

Here the first error is statistical, the second error is the systematic uncertainty, and the third error is the uncertainty due to the errors on $c_i$ and $s_i$ terms coming from the CLEO analysis. This translates to

$$\begin{aligned} \phi_3 &= (77.3^{+15.1}_{-14.9} \pm 4.1 \pm 4.3)^{\circ}, \\ r_B &= 0.145 \pm 0.030 \pm 0.010 \pm 0.011, \\ \delta_B &= (129.9 \pm 15.0 \pm 3.8 \pm 4.7)^{\circ}. \end{aligned} \quad (17.8.25)$$

These results are consistent with the *CP*-conservation hypothesis at the 99.35% C.L., which corresponds to a 2.7 standard deviation discrepancy. They are also in good agreement with those obtained with the model-dependent approach, given in Tables 17.8.8 and 17.8.9.

It is important to note that, unlike the model uncertainty of the unbinned analysis, which is difficult to quantify, the error due to $c_i$ and $s_i$ is statistical in nature, since the measurements of these quantities are largely dominated by statistical uncertainties. It is expected that a precision measurement of $\phi_3$ at the 1° level (or better) with the binned model-independent Dalitz plot analysis will be possible at a super flavor factory, using data from the BES III experiment. There are no other critical systematic uncertainties in this analysis that would dominate the measurement at the 1° level — the most significant uncertainties are determined by the finite size of the auxiliary samples (flavor-tagged $D^0 \to K_S^0\pi^+\pi^-$ and $B^{\pm} \to D\pi^{\pm}$), which will also increase in future analyses.

### 17.8.5 $\sin(2\phi_1 + \phi_3)$

#### 17.8.5.1 Method involving $B \to D^{(*)}h$ $(h = \pi, \rho)$ decays

The study of the time-dependent decay rates of $B \to D^{(*)\mp}h^{\pm}$ provides a measure of $\sin(2\phi_1 + \phi_3)$, where $h$ denotes a pion, a $\rho$, or an $a_1$ meson (Dunietz, 1998). As shown in Fig. 17.8.12, these decays proceed through CF and DCS transitions, whose amplitudes are proportional to the CKM matrix element products $V_{cb}^*V_{ud}$ and $V_{ub}^*V_{cd}$, respectively. Thus, the weak phase difference between these amplitudes in the usual Wolfenstein (1983) convention is $\phi_3$ (see Eq. 16.4.4 and Fig. 16.5.1). Interference between the two contributing diagrams also involves $B^0\overline{B}^0$ mixing (see Eq. 16.6.6), hence resulting in a total weak phase difference $2\phi_1 + \phi_3$.

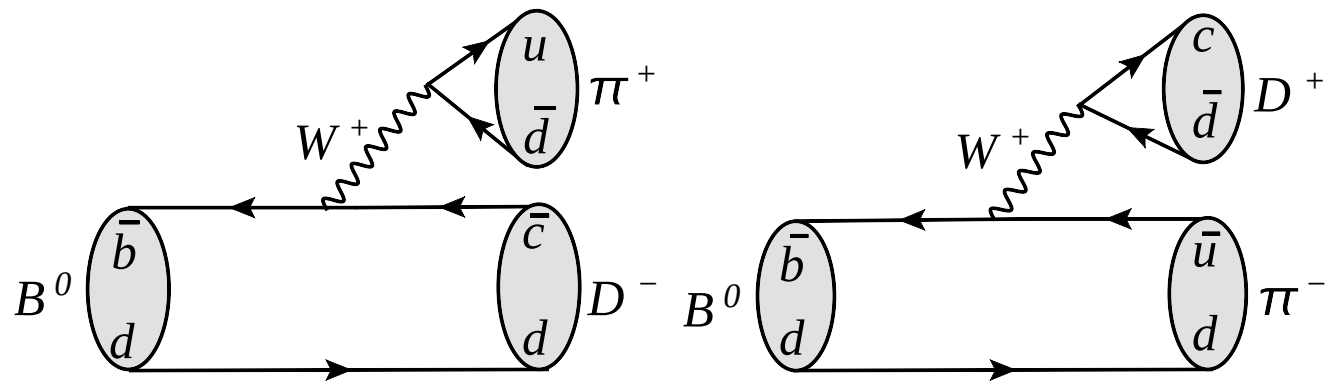

**Figure 17.8.12.** Typical leading order Feynman diagrams for the CF decay $B^0 \to D^- h^+$ (left) and the DCS decay $B^0 \to D^+ h^-$ (right).

In $\Upsilon(4S) \to B\overline{B}$ decays, the observed decay rate distribution of $B \to D^{(*)\mp} h^\pm$ is (see Section 10.2)

$$f_\pm(\Delta t) = \frac{e^{-|\Delta t|/\tau_B}}{4\tau_B} \times [1 \mp S^\xi \sin(\Delta m_d \Delta t) \mp \eta\, C \cos(\Delta m_d \Delta t)], \quad (17.8.26)$$

where $\tau_B$ is the neutral $B$ meson lifetime averaged over the two mass eigenstates, $\Delta m_d$ is the $B^0\overline{B}^0$ mixing frequency, $\Delta t$ is the proper time difference between the $B \to D^{(*)\mp} h^\pm$ decay ($B_{\rm rec}$) and the decay of the other $B$ in the event ($B_{\rm tag}$), the upper (lower) sign on $\pm$ or $\mp$ indicates the flavor of the $B_{\rm tag}$ as a $B^0$ ($\overline{B}^0$), and the parameters $\xi$ and $\eta$ have the values $\xi = +(-)$ and $\eta = +1(-1)$ for the $B_{\rm rec}$ final state $D^{(*)-}h^+$ ($D^{(*)+}h^-$). The coefficients $S^\pm$ and $C$ are

$$S^\pm = \frac{2R}{1+R^2} \sin(2\phi_1 + \phi_3 \pm \delta), \qquad C = \frac{1-R^2}{1+R^2}, \quad (17.8.27)$$

where $R$ is the ratio of the magnitudes of the DCS and CF amplitudes (in the SM, their magnitudes are the same for $B^0$ and $\overline{B}^0$ decays), and $\delta$ is the strong phase difference between the two amplitudes. The values of $R$ and $\delta$ are not necessarily the same for different $D^{(*)}h$ final states, making $S^\pm$ and $C$ mode dependent. For instance,

$$R_{D^{(*)}h} = \frac{|\mathcal{A}(B^0 \to D^{(*)+}h^-)|}{|\mathcal{A}(B^0 \to D^{(*)-}h^+)|} = \left| \frac{V_{ub}^* V_{cd}}{V_{cb}^* V_{ud}} \right| r, \quad (17.8.28)$$

could be different owing to possible distinct values for $r$, where $r$ is the ratio of decay constants and form factors involved with the two diagrams shown in Fig. 17.8.12. Assuming $r \approx 1$ in the above equation, we can estimate $R$ purely in terms of the CKM matrix elements to be 2%.

It follows from Eq. (17.8.27) that the value of $R_{D^{(*)}h}$ dictates the sensitivity of *CP* violation measurement in $B \to D^{(*)\mp} h^\pm$, as the sine term containing weak phases is essentially weighted by the factor $R$. Now because $R$ is predicted to be small, the experimental precision on $2\phi_1 + \phi_3$ expected from these measurements is poor. Furthermore, these measurements are susceptible to potential model uncertainties caused by the assumptions used in the calculation of $R$. However, when the decay proceeds through several interfering amplitudes such as the three helicity amplitudes in $B \to D^{*\mp}\rho^\pm$, it is possible to extract $R$ directly from the data (London, Sinha, and Sinha, 2000; Sinha, Sinha, and Soffer, 2005), eliminating these uncertainties.

### 17.8.5.2 Determination of $R_{D^{(*)}h}$

Unfortunately, we cannot directly measure the $R$ values with the current $B$ Factory dataset as the DCS decay $B^0 \to D^{(*)+}h^-$ is overwhelmed by the copious background from $\overline{B}^0 \to D^{(*)+}h^-$. They can be, however, indirectly obtained from self-tagging neutral $B$ decays involving a charmed-strange meson such as $B^0 \to D_s^+\pi^-$, assuming $SU(3)$ flavor symmetry, or from suppressed charged $B$ decays (*e.g.*, $B^+ \to D^+\pi^0$) with an isospin relation. In the former case, $R$ is extracted using the following relation (Dunietz, 1998; Dunietz and Sachs, 1988; Suprun, Chiang, and Rosner, 2002),

$$R_{D^{(*)}h} = \frac{|V_{cd}|}{|V_{cs}|} \frac{f_{D^{(*)}}}{f_{D_s^{(*)}}} \sqrt{\frac{\mathcal{B}(B^0 \to D_s^{(*)+}h^-)}{\mathcal{B}(B^0 \to D^{(*)-}h^+)}}, \quad (17.8.29)$$

where $f_x$ denotes the decay constant of the meson $x$, and $\mathcal{B}$ denotes the branching fraction of the mode shown. This relation can be inferred from the DCS decay diagram of Fig. 17.8.12, where by replacing the $\overline{d}$ quark with an $\overline{s}$ quark one can get $B^0 \to D_s^{(*)+}h^-$. In the first case the virtual $W^+$ boson hadronizes into a $D^{(*)+}$ meson, with the decay constant $f_{D^{(*)}}$ and CKM matrix element $V_{cd}$, and in the second it forms a $D_s^{(*)+}$ meson, with the decay constant $f_{D_s^{(*)}}$ and CKM matrix element $V_{cs}$. The theory errors on $R$ due to possible $SU(3)$ breaking effects are difficult to quantify, but are estimated to be in the range 10–15% (Baak, 2007). Furthermore, the above relation assumes that internal $W$-exchange amplitudes contribute much less than tree amplitudes to the $B^0 \to D^{(*)\mp}h^\pm$ decays. We can verify this assumption by measuring the branching fraction for $B^0 \to D_s^{(*)-}K^{(*)+}$, because in the absence of re-scattering the exchange diagram is the lone contributor to these decays. Therefore, their branching fractions can provide a measure of the $W$-exchange contribution to $B^0 \to D^{(*)\mp}h^\pm$.

### 17.8.5.3 Results from $B \to D^{(*)}h$ ($h = \pi, \rho$) decays

Both *BABAR* (Aubert, 2005v, 2006aa) and Belle (Bahinipati, 2011; Ronga, 2006) have performed these measurements using both full as well as partial reconstruction of $D^{(*)\mp}\pi^\pm$. In the case of partial reconstruction, signal $D^{*\mp}\pi^\pm$ decay candidates are identified using information solely from the high momentum pion originating from the $B$ decay, and the low momentum pion from the subsequent decay of the $D^*$ meson, without reconstructing the neutral $D$. This results in increased efficiency at the cost of a larger background. *BABAR* (Aubert, 2006aa) has extended the study to include the $B \to D^\mp\rho^\pm$ decays.

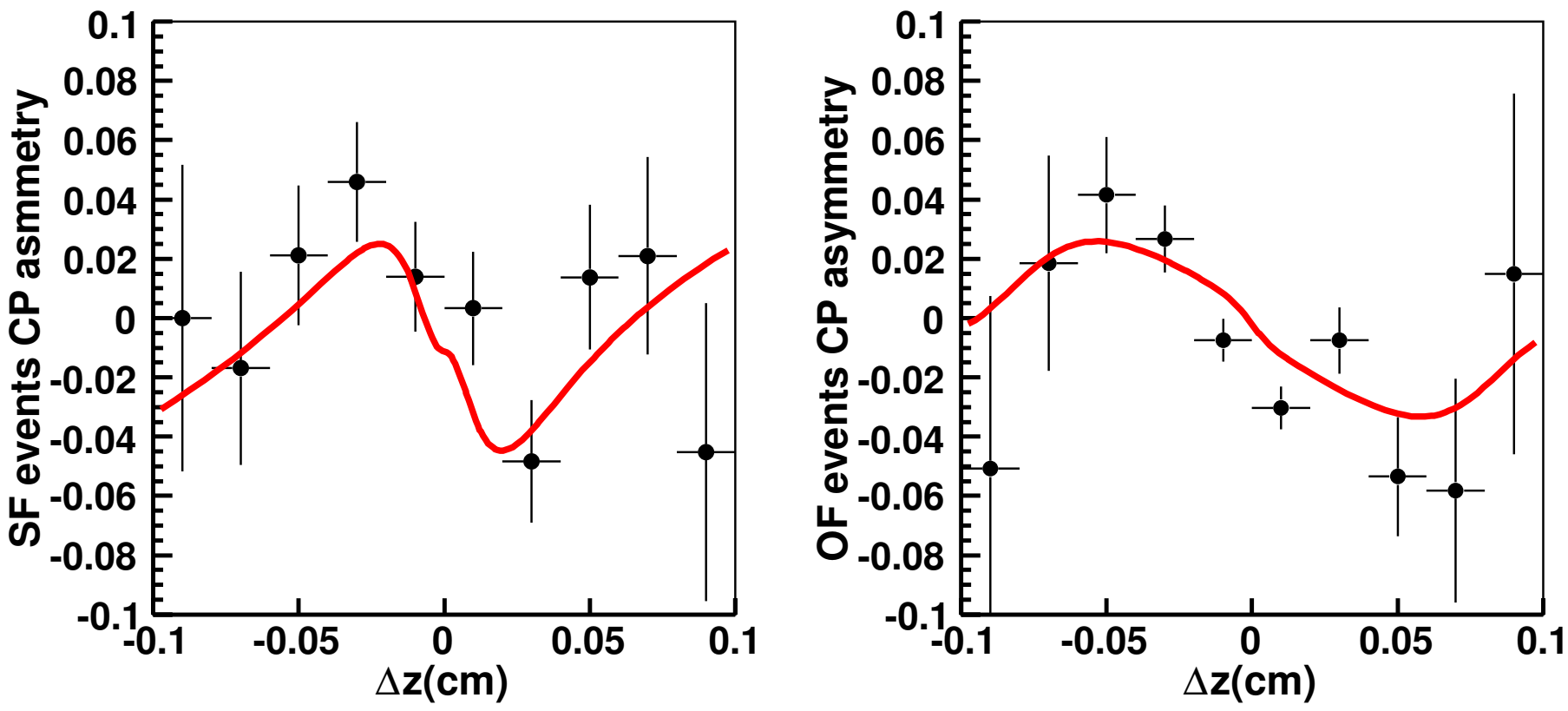

**Figure 17.8.13.** Belle's measurement of the distance between $B_{\rm rec}$ and $B_{\rm tag}$ vertices along the $z$ axis for lepton-tagged events, where the lepton has either the (left) same or (right) opposite charge as the low-momentum pion. The fit results (solid curves) are superimposed on the data (points with error bars). *CP* violation is characterized by a nonzero amplitude of the sinusoidal oscillation. These plots show the central regions of those presented in Bahinipati (2011), with an expanded vertical scale.

**Table 17.8.10.** Time-dependent *CP* violation parameters measured by Belle and *BABAR* in $B \to D^{(*)\mp} h^{\pm}$ decays.

| | *BABAR* | | Belle | |
|---|---|---|---|---|
| | Partial reconstruction (Aubert, 2005v) $N(B\bar{B}) = 232 \times 10^6$ | Full reconstruction (Aubert, 2006aa) $N(B\bar{B}) = 232 \times 10^6$ | Partial reconstruction (Bahinipati, 2011) $N(B\bar{B}) = 657 \times 10^6$ | Full reconstruction (Ronga, 2006) $N(B\bar{B}) = 386 \times 10^6$ |
| $a_{D^*\pi}$ | $-0.034 \pm 0.014 \pm 0.009$ | $-0.040 \pm 0.023 \pm 0.010$ | $-0.046 \pm 0.013 \pm 0.015$ | $-0.039 \pm 0.020 \pm 0.013$ |
| $c_{D^*\pi}$ | $-0.019 \pm 0.022 \pm 0.013$ | $+0.049 \pm 0.042 \pm 0.015$ | $-0.015 \pm 0.013 \pm 0.015$ | $-0.011 \pm 0.020 \pm 0.013$ |
| $a_{D\pi}$ | – | $-0.010 \pm 0.023 \pm 0.007$ | – | $-0.050 \pm 0.021 \pm 0.012$ |
| $c_{D\pi}$ | – | $-0.033 \pm 0.042 \pm 0.012$ | – | $-0.019 \pm 0.021 \pm 0.012$ |
| $a_{D\rho}$ | – | $-0.024 \pm 0.031 \pm 0.009$ | – | – |
| $c_{D\rho}$ | – | $-0.098 \pm 0.055 \pm 0.018$ | – | – |

In Table 17.8.10 we summarize results on *CP* violation parameters sensitive to $2\phi_1 + \phi_3$ obtained with $B \to D^{(*)\pm}\pi^{\mp}$ by the two experiments. Results are given in terms of two parameters $a$ and $c$, defined as

$$a = (S^+ + S^-)/2, \qquad c = (S^+ - S^-)/2. \tag{17.8.30}$$

These parameters were introduced by *BABAR* (Aubert, 2005v, 2006aa) in both partial and full reconstruction analyses in an attempt to disentangle the results from possible *CP* violation effects on the $B_{\rm tag}$ side. The parameter $a$ is always independent of tag-side *CP* violation; the same also holds true for $c$ in the case of semileptonic $B_{\rm tag}$ decays since those decays are dominated by a single amplitude. In the partial reconstruction analysis, Belle (Bahinipati, 2011) uses only lepton tags for $B_{\rm tag}$, while *BABAR* employs kaon- and lepton-tagged events. Both experiments use the $a$ and $c$ notation in the partial reconstruction analyses, whereas full-reconstruction results of Belle (Ronga, 2006) are presented in terms of $S^+$ and $S^-$. To compare results from the two experiments, we convert $S^+$ and $S^-$ into $a$ and $c$ after taking into account the relative factor $(-1)^L$ between Belle and *BABAR* in the definition of $S^{\pm}$, where the orbital angular momentum $L$ equals 0 (1) for the $D\pi$ ($D^*\pi$) final state. The search for *CP* violation in these decays has provided results with significance at the level of 2.5 (2.0) standard deviations from Belle (*BABAR*). Figure 17.8.13, for instance, provides an illustration of *CP* violation results obtained in the partial reconstruction analysis of Belle.

#### 17.8.5.4 Results from $B \to D_s^{(*)} h$ ($h = \pi, K$) decays

Among charmed-strange meson final states, *BABAR* (Aubert, 2008u) and Belle (Das, 2010; Joshi, 2010) have studied $B^0 \to D_s^{(*)+}\pi^-$ and $B^0 \to D_s^{(*)-}K^+$ (see Section 17.3.3). As mentioned earlier, the former decay constitutes an independent measurement of the small parameter $R$ and the latter provides a measure of the $W$-exchange contribution in $B \to D^{(*)\pm}\pi^{\mp}$.

In Table 17.8.11 we present the branching fraction measurement of $B^0 \to D_s^{(*)+} h^-$, where $h = \pi$ or $\rho$, from the two experiments. By substituting these numbers along with world-average values for $|V_{cd}|$, $|V_{cs}|$ and $\mathcal{B}(B^0 \to D^{(*)-} h^+)$ (Beringer et al., 2012) as well as for lattice QCD estimates of the decay constants of the $D^{(*)}$ and

**Table 17.8.11.** Measured branching fractions for $B^0 \to D_s^{(*)+}h^-(h=\pi,\rho)$ and $B^0 \to D_s^{(*)-}K^{(*)+}$ with the corresponding average values. All are in units of $10^{-5}$.

| | *BABAR* (Aubert, 2008u) $N(B\overline{B}) = 381 \times 10^6$ | Belle (Das, 2010; Joshi, 2010) $N(B\overline{B}) = 657 \times 10^6$ | Average |
|---|---|---|---|
| $\mathcal{B}(B^0 \to D_s^+\pi^-)$ | $2.5 \pm 0.4 \pm 0.2$ | $1.99 \pm 0.26 \pm 0.18$ | $2.16 \pm 0.26$ |
| $\mathcal{B}(B^0 \to D_s^{*+}\pi^-)$ | $2.6^{+0.5}_{-0.4} \pm 0.3$ | $1.75 \pm 0.34 \pm 0.20$ | $2.02 \pm 0.33$ |
| $\mathcal{B}(B^0 \to D_s^+\rho^-)$ | $1.1^{+0.9}_{-0.8} \pm 0.3$ | – | $1.10 \pm 0.95$ |
| $\mathcal{B}(B^0 \to D_s^{*+}\rho^-)$ | $4.1^{+1.3}_{-1.2} \pm 0.8$ | – | $4.10 \pm 1.53$ |
| $\mathcal{B}(B^0 \to D_s^-K^+)$ | $2.9 \pm 0.4 \pm 0.2$ | $1.91 \pm 0.24 \pm 0.17$ | $2.21 \pm 0.25$ |
| $\mathcal{B}(B^0 \to D_s^{*-}K^+)$ | $2.4 \pm 0.4 \pm 0.2$ | $2.02 \pm 0.33 \pm 0.22$ | $2.19 \pm 0.30$ |
| $\mathcal{B}(B^0 \to D_s^-K^{*+})$ | $3.5^{+1.0}_{-0.9} \pm 0.4$ | – | $3.50 \pm 1.08$ |
| $\mathcal{B}(B^0 \to D_s^{*-}K^{*+})$ | $3.2^{+1.4}_{-1.2} \pm 0.4$ | – | $3.20 \pm 1.46$ |

$D_s^{(*)}$ mesons (Laiho, Lunghi, and Van de Water, 2010) in Eq. (17.8.29), we determine

$$\begin{aligned} R_{D\pi} &= (1.73 \pm 0.15 \pm 0.04)\%, \\ R_{D^*\pi} &= (1.65 \pm 0.18 \pm 0.04)\%, \\ R_{D\rho} &= (0.74 \pm 0.33 \pm 0.02)\%, \\ R_{D^*\rho} &= (1.50 \pm 0.31 \pm 0.04)\%, \end{aligned} \tag{17.8.31}$$

where the second errors are due to those on $f_{D^{(*)}}/f_{D_s^{(*)}}$. Note that here we have assumed the ratio $f_{D^*}/f_{D_s^*}$ to be the same as $f_D/f_{D_s}$. The $R$ values obtained are somewhat smaller than the naïve expectations of 2%: in particular, $R_{D\rho}$ is significantly below 2%. Table 17.8.11 also summarizes the branching fractions for $B^0 \to D_s^{(*)-}K^{(*)+}$ measured by the two experiments. These branching fractions are two orders of magnitude smaller than those of the CF decays $B^0 \to D^{(*)-}\pi^+$, implying insignificant contributions from the internal $W$ exchange diagram (or a CF $\overline{B}^0 \to \overline{D}^0 d\overline{d}$ diagram followed by $d\overline{d} \to s\overline{s}$ re-scattering). This justifies neglecting contributions from similar diagrams in the determination of $R$ (Eq. 17.8.29).

#### 17.8.5.5 Results from the decay $B^+ \to D^{*+}\pi^0$

Belle (Iwabuchi, 2008) has performed a search for the DCS decay $B^+ \to D^{*+}\pi^0$. No significant signal is found, and a 90% confidence-level upper limit is set on the branching fraction, $\mathcal{B}(B^+ \to D^{*+}\pi^0) < 3.6 \times 10^{-6}$. This limit is used to constrain $R_{D^*\pi}$ to be less than 5.1% at the 90% confidence level. The upper limit on $R$ is consistent with the values obtained from $B \to D_s^{(*)}h$.

#### 17.8.5.6 Constraint on $2\phi_1 + \phi_3$ from $B \to D^{(*)}h$

One can derive a combined constraint on $2\phi_1 + \phi_3$ using relevant observables measured in the $B \to D^{(*)\mp}h^\pm$ decays. There are two measurements, $a$ and $c$ (or $S^+$ and $S^-$), and three unknown quantities, $R$, $\delta$, and $2\phi_1 + \phi_3$, of which the first two are different for each decay channel. To find a solution, we can use the $R_{D^{(*)}h}$ values extracted with the $SU(3)$ relation of Eq. (17.8.29) as an additional input. Combining results on $a$ and $c$ with $R_{D^{(*)}h}$ (see Table 17.8.10 and Eq. 17.8.31) using a frequentist method described in Charles et al. (2005), we obtain a constraint on $2\phi_1 + \phi_3$. The confidence level as a function of $|\sin(2\phi_1 + \phi_3)|$ is shown in Fig. 17.8.14. We set a lower limit $|\sin(2\phi_1 + \phi_3)| > 0.74$ (0.51) at 68% (90%) confidence level.

#### 17.8.5.7 Constraint on $2\phi_1 + \phi_3$ from $B \to DK\pi$

*BABAR* (Aubert, 2008bg) has performed a time-dependent Dalitz plot analysis of $B^0 \to D^\mp K^0\pi^\pm$. Since both $b \to c$ and $b \to u$ diagrams involved in these decays are color suppressed, the $R$ value is expected to be larger than that found in $B \to D^{(*)\mp}h^\pm$. Assuming $R$ is 30% and constant across the Dalitz plot, *BABAR* finds $2\phi_1 + \phi_3 = (83 \pm 53 \pm 20)^\circ$ along with an equivalent solution at a value 180° larger than this, where the first error is statistical and the second is systematic.

### 17.8.6 Determination of $\phi_3$ and discussion

We combine the available *BABAR* ($B^\pm \to DK^\pm$, $B^\pm \to D^*K^\pm$, and $B^\pm \to DK^{*\pm}$) and Belle ($B^\pm \to DK^\pm$, $B^\pm \to D^*K^\pm$) observables obtained for the GLW method (Table 17.8.1), the ADS method (Table 17.8.2), and the GGSZ method (model-dependent results as shown in Table 17.8.8) using the frequentist procedure (plug-in method) exploited in Charles et al. (2005). The $p$-value (1 − C.L.) curves for the angle $\phi_3$ as well as the hadronic parameters ($\delta_B$ and $r_B$) of the $B \to DK$ mode are shown in Fig. 17.8.15 and the 68% C.L. intervals are summarized in Table 17.8.12. The results obtained are in very good agreement with individual constraints available for each experiment (Lees, 2013e and Trabelsi, 2013): the combined $B$ Factory $\phi_3$ average is $(67 \pm 11)^\circ$.

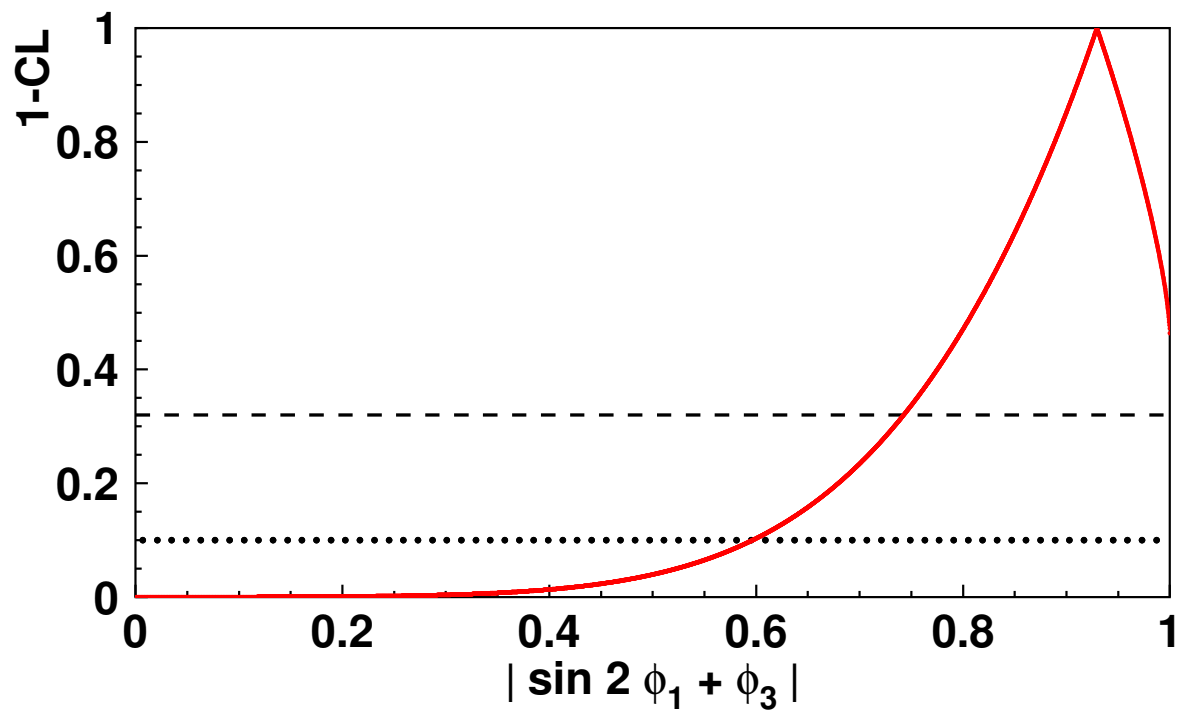

**Figure 17.8.14.** Combined constraint on $2\phi_1 + \phi_3$ using relevant observables measured in the $B \to D^{(*)}h$ decays. The dashed (dotted) line indicates the 68% (90%) confidence-level lower limit.

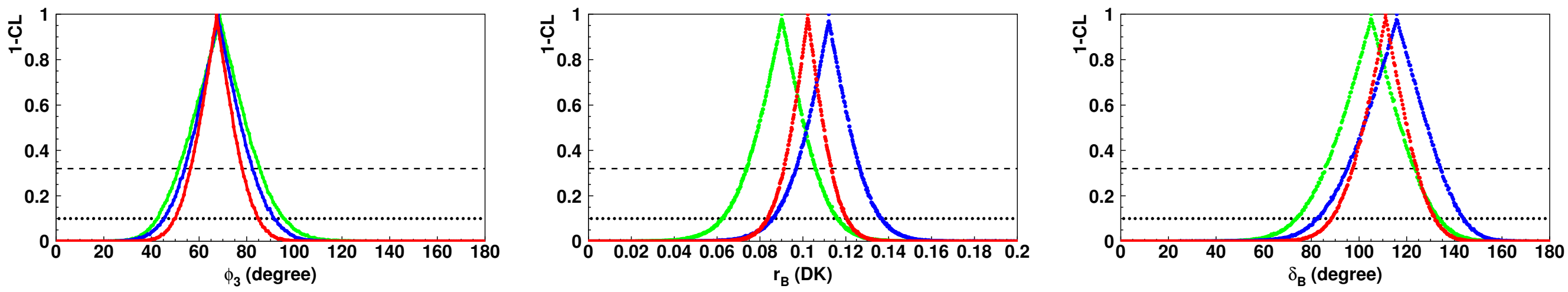

**Figure 17.8.15.** Combined constraint (red curve) on $\phi_3$ (left), $r_B(DK)$ (middle) and $\delta_B(DK)$ (right) using relevant *BaBar* and Belle observables measured in the $B \to D^{(*)}K^{(*)}$ decays. The green (blue) curve represents the results using only the *BaBar* (Belle) observables, the dashed (dotted) line indicates the 68% (90%) confidence-level lower limit.

**Table 17.8.12.** Confidence intervals for $\phi_3$, $r_B(DK)$ and $\delta_B(DK)$ obtained from the combination of the relevant *BaBar* and Belle observables measured in the $B \to D^{(*)}K^{(*)}$ decays.

| | $\phi_3$ (°) | $r_B(DK)$ | $\delta_B(DK)$ (°) |
|---|---|---|---|
| *BaBar* | $69 \pm 17$ | $0.090^{+0.016}_{-0.017}$ | $105 \pm 19$ |
| Belle | $68 \pm 14$ | $0.112 \pm 0.015$ | $116^{+19}_{-21}$ |
| $B$ Factories | $67 \pm 11$ | $0.102 \pm 0.011$ | $111^{+13}_{-14}$ |

## 17.9 Radiative and electroweak penguin decays

***Editors:***
*Al Eisner, Stephen Playfer (BABAR)*
*Mikihiko Nakao (Belle)*
*Tobias Hurth (theory)*

**Additional section writers:**
*John Walsh*

This section discusses radiative penguin $B$ meson decays with $b \to s\gamma$ and $b \to d\gamma$ transitions, and electroweak penguin[82] $B$ meson decays with $b \to s\ell^+\ell^-$, $b \to d\ell^+\ell^-$ and $b \to s\nu\bar{\nu}$ transitions. These $B$ decay modes are considered to be among the most sensitive probes for physics beyond the SM, because they occur at loop level, and their rates can be accurately predicted. In the SM the decays proceed at lowest order through penguin loop and box diagrams involving heavy virtual top quarks and weak $W$ or $Z$ bosons as shown in Figure 17.9.1. Beyond the SM these could also contain hypothetical heavy particles, *e.g.* supersymmetric partners of quarks and bosons, or charged Higgs bosons.

At the $B$ Factories many of these decays have been studied. Inclusive and exclusive branching fractions have been accurately determined for $b \to s\gamma$, and measured for the first time for $b \to d\gamma$ and $b \to s\ell^+\ell^-$. Here, an inclusive decay is denoted for example as $B \to X_s\gamma$, where $X_s$ is the sum of the hadronic final states formed by the recoiling $s$ quark from $b \to s\gamma$ and the spectator $\overline{u}$ or $\overline{d}$ quark, whereas an exclusive decay specifies the final state hadron(s), for example, $B \to K^*(892)\gamma$. Time-integrated and time-dependent $CP$ asymmetries have also been measured. For $b \to s\ell^+\ell^-$, the decay amplitude depends on $q^2$, which is the invariant mass squared of the di-lepton system, or the virtual momentum squared of the electroweak boson in the case of the lowest order penguin diagram. In addition, angular analyses, which are sensitive to the interference between different terms in the decay amplitudes, have been performed as functions of $q^2$.

Theoretically, the SM predictions are at a similar level of accuracy to the experimental precision for the inclusive branching fractions. This is due to the presence of leptons and photons in the final state which reduces the size of non-perturbative QCD corrections.

This section (17.9) is organized into a short review of the theoretical aspects (17.9.1), then discussions of inclusive and exclusive $b \to s\gamma$ (17.9.2 and 17.9.3) and $b \to d\gamma$ (17.9.4) decays, separate subsections on rate asymmetries (17.9.5) and time-dependent $CP$ asymmetry measurements (17.9.6), followed by subsections on $b \to s(d)\ell^+\ell^-$ (17.9.7) and $b \to s\nu\bar{\nu}$ (17.9.8) decays.

[82] In the literature these decays are also called *semileptonic rare decays*. We do not adopt this term here to avoid confusion with semileptonic $B$ meson decays with $b \to c\ell\overline{\nu}$ and $b \to u\ell\overline{\nu}$.

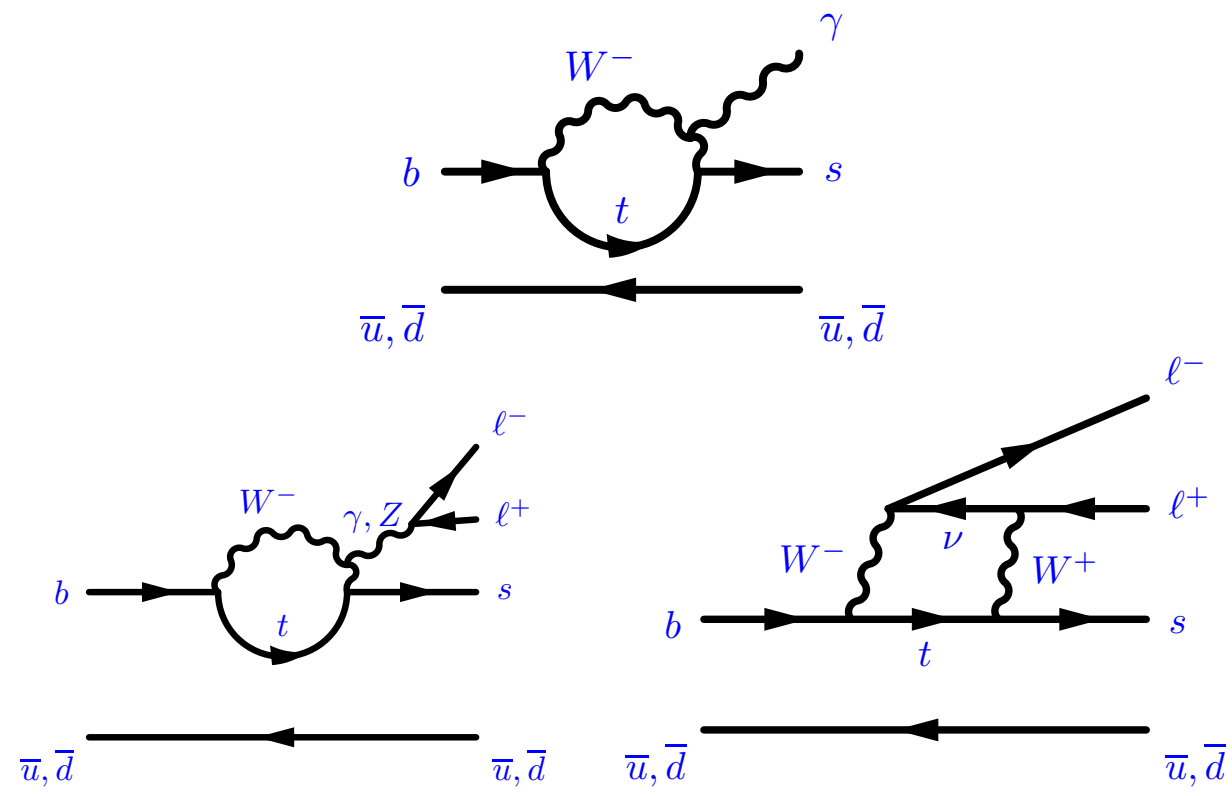

**Figure 17.9.1.** Examples of penguin loop diagram for $b \to s\gamma$ (top), and loop and box diagrams for $b \to s\ell^+\ell^-$ (bottom).

### 17.9.1 Theoretical framework

#### 17.9.1.1 Effective electroweak Hamiltonian

Rare $B$ decays are governed by an interplay between the weak and strong interactions. This is especially the case for inclusive $B$ decay modes, where short-distance QCD effects are very important. In the decay $B \to X_s\gamma$ these effects lead to a rate enhancement by a factor of greater than two. Such effects are induced by hard-gluon exchanges between the quark lines of the one-loop electroweak diagrams.

The perturbative QCD corrections that arise from hard gluon exchange bring in large logarithms of the form

$$\alpha_S^n(m_b)\,\log^m(m_b/M), \tag{17.9.1}$$

where $m \leq n$ (with $n = 0, 1, 2, ...$). $M$ is the top or $W$ mass and $m_b$ the $b$ quark mass. These large logarithms are a natural feature in any process in which two different mass scales are present. To obtain a reasonable result, one must re-sum at least all the leading-log (LL) terms with $m = n$, or $\alpha_S^n(m_b)\log^n(m_b/M)$, with the help of renormalization group techniques (Grinstein, Savage, and Wise, 1989; Grinstein, Springer, and Wise, 1988, 1990). Working to next-to-leading-log (NLL) or next-to-next-to-leading-log (NNLL) precision means that one re-sums all the terms with $m = n - 1$ or $m = n - 2$, too (Buchalla, Buras, and Lautenbacher, 1996; Misiak, 1993).

A suitable framework in which to achieve the necessary re-summations of the large logarithms is an effective low-energy theory with five quarks; this framework is obtained by integrating out the heavy particles *i.e.* by removing them from the theory as dynamical fields (Buchalla, Buras, and Lautenbacher, 1996). These are the electroweak bosons and the top quark in the SM. This effective field theory approach serves as a theoretical framework for both inclusive and exclusive modes. The standard method of the operator product expansion (OPE) (Wilson and Zimmermann, 1972) allows for a separation of the $B$ meson decay amplitude into two distinct parts, the long-distance contributions contained in the operator matrix elements and the short-distance physics described by the Wilson coefficients. The electroweak effective Hamiltonian can schematically be written as (Altarelli and Maiani, 1974; Gaillard and Lee, 1974a; Witten, 1977)

$$\mathcal{H}_{\rm eff} = \frac{4G_F}{\sqrt{2}} \sum_i \lambda_{\rm CKM} C_i(\mu, M)\, \mathcal{O}_i(\mu), \qquad (17.9.2)$$

where $\mathcal{O}_i(\mu)$ are operators of dimension six, $C_i(\mu, M)$ are the corresponding Wilson coefficients, $\lambda_{\rm CKM}$ are products of CKM matrix elements, and $\mu$ denotes the factorization scale. As the heavy fields are integrated out, the complete top and $W$ mass dependence is contained in the Wilson coefficients. Within the observable $\mathcal{H}_{\rm eff}$ the scale dependence ($\mu$) should cancel out.

The effective electroweak Hamiltonian relevant to $b \to s(d)\,\gamma$ and $b \to s(d)\,\ell^+\ell^-$ transitions in the SM reads

$$\mathcal{H}_{\rm eff} = -\frac{4G_F}{\sqrt{2}} \left[ \lambda_q^t \sum_{i=1}^{10} C_i \mathcal{O}_i + \lambda_q^u \sum_{i=1}^{2} C_i (\mathcal{O}_i - \mathcal{O}_i^u) \right], \qquad (17.9.3)$$

where the explicit CKM factors are $\lambda_q^t = V_{tb} V_{tq}^*$ and $\lambda_q^u = V_{ub} V_{uq}^*$ ($q = s, d$). The unitarity relations $\lambda_q^c = -\lambda_q^t - \lambda_q^u$ have already been used. The numerically significant dimension-six operators are:[83]

$$\begin{aligned}
\mathcal{O}_1 &= (\bar{s}_L \gamma_\mu T^a c_L)(\bar{c}_L \gamma^\mu T^a b_L)\,, \qquad (17.9.4)\\
\mathcal{O}_2 &= (\bar{s}_L \gamma_\mu c_L)(\bar{c}_L \gamma^\mu b_L)\,,\\
\mathcal{O}_1^u &= (\bar{s}_L \gamma_\mu T^a u_L)(\bar{u}_L \gamma^\mu T^a b_L)\,,\\
\mathcal{O}_2^u &= (\bar{s}_L \gamma_\mu u_L)(\bar{u}_L \gamma^\mu b_L)\,,\\
\mathcal{O}_7 &= \frac{e}{16\pi^2} m_b (\bar{s}_L \sigma^{\mu\nu} b_R) F_{\mu\nu}\,,\\
\mathcal{O}_8 &= \frac{g_s}{16\pi^2} m_b (\bar{s}_L \sigma^{\mu\nu} T^a b_R) G^a_{\mu\nu}\,,\\
\mathcal{O}_9 &= \frac{e^2}{16\pi^2} (\bar{s}_L \gamma_\mu b_L) \sum_\ell (\bar{\ell} \gamma^\mu \ell)\,,\\
\mathcal{O}_{10} &= \frac{e^2}{16\pi^2} (\bar{s}_L \gamma_\mu b_L) \sum_\ell (\bar{\ell} \gamma^\mu \gamma_5 \ell)\,,
\end{aligned}$$

where $T^a$ are $SU(3)$ color generators, $F_{\mu\nu}$ and $G_{\mu\nu}$ are electromagnetic and chromomagnetic fields, and the subscripts $L$ and $R$ refer to the left- and right-handed components of the fermion fields. In $b \to s$ transitions the contributions proportional to $\lambda_s^u$ are rather small, while in $b \to d$ decays, where $\lambda_d^u$ is of the same order as $\lambda_d^t$, these contributions play an important role in $CP$ and isospin asymmetries. The operators $\mathcal{O}_9$ and $\mathcal{O}_{10}$ only occur in the $b \to s(d)\ell^+\ell^-$ and $b \to s\nu\overline{\nu}$ modes.

It is worth noting that among the four-quark operators, only the effective couplings for $i = 1, 2$ are large at the low scale $\mu = m_b$ where $C_{1,2}(m_b) \approx 1$. The so-called QCD penguin operators

$$\begin{aligned}
\mathcal{O}_3 &= (\bar{s}_L \gamma_\mu b_L) \sum_{q=u,d,c,s,b} (\bar{q}_L \gamma^\mu q_L)\,, \qquad (17.9.5)\\
\mathcal{O}_4 &= (\bar{s}_L \gamma_\mu T^a b_L) \sum_{q=u,d,c,s,b} (\bar{q}_L \gamma^\mu T^a q_L)\,, \qquad (17.9.6)\\
\mathcal{O}_5 &= (\bar{s}_L \gamma_\mu b_L) \sum_{q=u,d,c,s,b} (\bar{q}_R \gamma^\mu q_R)\,, \qquad (17.9.7)\\
\mathcal{O}_6 &= (\bar{s}_L \gamma_\mu T^a b_L) \sum_{q=u,d,c,s,b} (\bar{q}_R \gamma^\mu T^a q_R)\,, \qquad (17.9.8)
\end{aligned}$$

have very small coefficients $C_3, \ldots, C_6$ and hence can safely be neglected. The electromagnetic penguin with $C_7(m_b) \approx -0.3$, and the chromomagnetic penguin with $C_8(m_b) \approx -0.15$, play a significant role in both $b \to s(d)\gamma$ and $b \to s(d)\ell^+\ell^-$. Finally the vector and axial-vector contributions to $b \to s(d)\ell^+\ell^-$ have $C_9(m_b) \approx 4$, $C_{10}(m_b) \approx -4$.

There are three principal calculational steps that lead to the LL (NNLL) result within the effective field theory approach:

1. At the scale $\mu = m_W$ the full SM theory is matched with the effective theory. This means that the calculation of the amplitude in the full SM is expanded in inverse powers of the large masses ($m_W$, $m_Z$, $m_t$) and the result is compared to the corresponding amplitude in the effective theory. In this way the Wilson coefficients $C_i(m_W)$ are extracted by comparison. At the high scale $\mu = m_W$ the $C_i$ pick up only small QCD corrections, which can be calculated within fixed-order perturbation theory. In the LL (NNLL) calculation, the matching has to be worked out at the $\mathcal{O}(\alpha_s^0)$ [$\mathcal{O}(\alpha_s^2)$] level.
2. The evolution of these Wilson coefficients from $\mu = m_W$ down to $\mu \approx m_b$ must then be performed with the help of the renormalization group. In this way the large logarithms (Eq. 17.9.1) are shifted from the matrix elements of the operators into the Wilson coefficients, and the matrix elements of the operators evaluated at the low scale $m_b$ are free of these large logarithms. For the LL (NNLL) calculation, this renormalization step has to be performed up to order $\alpha_s^1$ ($\alpha_s^3$).
3. To LL (NNLL) precision, the corrections to the matrix elements of the operators $\langle s\gamma|\mathcal{O}_i(\mu)|b\rangle$ at the scale $\mu \approx m_b$ must be calculated to order $\alpha_s^0$ ($\alpha_s^2$) precision.

While the Wilson coefficients $C_i$ enter both inclusive and exclusive processes and can be calculated with perturbative methods, the calculational approaches to the matrix elements of the operators differ in the two cases. In inclusive modes, one can use quark-hadron duality in order to derive a well-defined heavy mass expansion of the decay rates in powers of $\Lambda_{\rm QCD}/m_b$ (Heavy Quark Expansion, HQE)[84] (Bigi, Blok, Shifman, Uraltsev, and Vainshtein, 1992; Bigi, Uraltsev, and Vainshtein, 1992; Chay, Georgi, and Grinstein, 1990; Manohar and Wise, 1994). In particular, it turns out that the decay width of $B \to X_s\gamma$ is well approximated by the partonic decay rate, which can be calculated in renormalization group improved perturbation theory (Ali, Hiller, Handoko, and Morozumi, 1997; Falk, Luke, and Savage, 1994):

$$\Gamma(B \to X_s\gamma) = \Gamma(b \to X_s^{\mathrm{parton}}\gamma) + \mathcal{O}(\Lambda/m_b) \quad (17.9.9)$$

[83] There are also operators $\mathcal{O}_7'$ and $\mathcal{O}_8'$ where $m_b$ is replaced by $m_s$ (or $m_d$, respectively), and here these are suppressed by factors $m_{s/d}/m_b$ and are usually omitted.

[84] In the following text the symbol $\Lambda/m_b$ is also used to denote $\Lambda_{\rm QCD}/m_b$.

In exclusive processes one cannot rely on quark-hadron duality, and face the difficult task of estimating matrix elements between meson states. A promising approach is the method of QCD-improved factorization (QCDF) which has been systematically formalized for non-leptonic decays in the heavy quark limit $m_b \to \infty$ (Beneke, Buchalla, Neubert, and Sachrajda, 1999, 2000, 2001). This method allows for a perturbative calculation of QCD corrections to naïve factorization, and is the basis for the up-to-date predictions for exclusive rare $B$ decays. However, within this approach, a general, quantitative method to estimate the important $1/m_b$ corrections to the heavy quark limit is missing.

#### 17.9.1.2 Power corrections to inclusive decays

The inclusive decay rate is defined as (see also Section 17.1)

$$\Gamma = \frac{1}{2m_{H_b}} \sum_X (2\pi)^4 \delta^4(p_i - p_f) \mid \langle X \mid \mathcal{H}_{\mathrm{eff}} \mid H_b \rangle \mid^2 , \quad (17.9.10)$$

where the sum runs over all possible states $X$. In order to set up a systematic approach, we use the optical theorem which relates the inclusive decay rate of a hadron $H_b$ to the imaginary part of the forward scattering amplitude

$$\Gamma(H_b \to X) = \frac{1}{2m_{H_b}} \mathrm{Im} \langle H_b \mid \mathbf{T} \mid H_b \rangle , \quad (17.9.11)$$

where $\mathbf{T}$ is the time-ordered product of two effective Hamiltonians $\mathbf{T} = i \int d^4x \, T[\mathcal{H}_{\mathrm{eff}}(x)\mathcal{H}_{\mathrm{eff}}(0)]$.

From this it is possible to construct an OPE of the operator $\mathbf{T}$, which is expressed as a series of local operators that are suppressed by powers of the $b$ quark mass and written in terms of the $b$ quark field (Bigi, Blok, Shifman, Uraltsev, and Vainshtein, 1992; Bigi, Uraltsev, and Vainshtein, 1992; Chay, Georgi, and Grinstein, 1990; Manohar and Wise, 1994):

$$T[\mathcal{H}_{\mathrm{eff}}\mathcal{H}_{\mathrm{eff}}] \overset{\mathrm{OPE}}{=} \frac{1}{m_b} \Big( \sum_i c_i^{(0)} \mathcal{P}_i^{(0)} + \frac{1}{m_b} \sum_i c_i^{(1)} \mathcal{P}_i^{(1)} + \frac{1}{m_b^2} \sum_i c_i^{(2)} \mathcal{P}_i^{(2)} + ... \Big) , \quad (17.9.12)$$

where $P_i^{(n)}$ are local operators of dimension $n+3$ and $c_i^{(n)}$ are the Wilson coefficient of the OPE.

Taking the forward matrix element (Eq. 17.9.11) generates an expansion in inverse powers of the heavy quark mass. Note that the matrix elements $\langle H_b \mid \mathcal{P}_i^{(n)} \mid H_b \rangle$ are of the order $\Lambda_{\mathrm{QCD}}$ to some appropriate power, and hence this expansion is expected to converge sufficiently well as long as the energy release in the decay is large with respect to the QCD scale, $\Lambda_{\mathrm{QCD}} \ll m_b$. With the help of heavy quark effective theory (HQET), where new heavy quark spin-flavor symmetries arise in the heavy quark limit $m_b \to \infty$ (Isgur and Wise, 1992; Shifman and Voloshin, 1988), the hadronic matrix elements within the OPE, $\langle H_b \mid \mathcal{P}_i^{(n)} \mid H_b \rangle$, can be further simplified. In this well-defined expansion, the free quark model is the first term in the constructed expansion in powers of $1/m_b$, and therefore the dominant contribution. In inclusive rare $B$ decays, one finds no correction of order $\Lambda/m_b$ to the free quark model approximation. The corrections to the partonic decay rate begin with $1/m_b^2$ only, which implies a rather small numerical impact of the non-perturbative corrections on the decay rate of inclusive modes. However, there are more subtleties to consider if other than the leading operators are taken into account (see below).

One can directly apply these methods to the inclusive decay mode $B \to X_s\gamma$. If one neglects perturbative QCD corrections and assumes that the decay $B \to X_s\gamma$ is due to the leading electromagnetic dipole operator $\mathcal{O}_7$ alone, then the photon would always be emitted directly from the hard process of the $b$ quark decay. One has to consider the time-ordered product $T[\mathcal{O}_7^+(x)\,\mathcal{O}_7(0)]$. Using the OPE for $T[\mathcal{O}_7^+(x)\,\mathcal{O}_7(0)]$ and HQET methods, as discussed above, the decay width $\Gamma(B \to X_s\gamma)$ reads (up to and including terms of order $1/m_b^2$):

$$\Gamma_{B\to X_s\gamma}^{(\mathcal{O}_7,\mathcal{O}_7)} = \frac{\alpha_{\mathrm{EM}} G_F^2 m_b^5}{32\pi^4} |V_{tb}V_{ts}|^2 \, C_7^2(m_b) \times \left(1 - \frac{1}{m_b^2}\left[\frac{1}{2}\mu_\pi^2 + \frac{3}{2}\mu_G^2\right]\right) , \quad (17.9.13)$$

where $\mu_\pi^2$ and $\mu_G^2$ are the HQE parameters for the kinetic energy and the chromomagnetic energy, respectively (see Eqs 17.1.38 and 17.1.39 in Section 17.1). If the $B \to X_s\gamma$ decay width is normalized to the charmless semileptonic decays, the non-perturbative corrections of order $1/m_b^2$ cancel out within the ratio $\mathcal{B}(B \to X_s\gamma)/\mathcal{B}(B \to X_u\ell\nu)$. However, in practice the branching fraction of inclusive rare decays are often normalized to the well measured $B \to X_c\ell\nu$ semileptonic branching fraction with which the $m_b^5$ dependence in Eq. (17.9.13) cancels.

The OPE for the inclusive decay $B \to X_s\gamma$ breaks down if one considers operators beyond the leading electromagnetic dipole operator $\mathcal{O}_7$ (Buchalla, Isidori, and Rey, 1998; Ligeti, Randall, and Wise, 1997; Voloshin, 1997). For example, one finds a contribution to the total decay rate due to the interference between the electromagnetic dipole operator $\mathcal{O}_7$ and the charming penguin amplitude due to the current-current operator $\mathcal{O}_2$. This is an example of a so-called *resolved* photon contribution. These contributions contain subprocesses in which the photon couples to light partons instead of connecting directly to the effective weak interaction vertex. A systematic analysis of all resolved photon contributions related to other operators in the weak Hamiltonian establishes this breakdown of the local OPE within the hadronic power corrections as a generic result (Benzke, Lee, Neubert, and Paz, 2010). Estimating such nonlocal matrix elements is very difficult, and leads to an irreducible theoretical uncertainty of $\pm(4-5)\%$ for the total $CP$ averaged decay rate, defined with a photon-energy cutoff $E_\gamma = 1.6\,\text{GeV}$ (Benzke, Lee, Neubert, and Paz, 2011). This result indicates that the theoretical efforts for the $B \to X_s\gamma$ mode have reached the non-perturbative boundaries.

The non-perturbative contributions in the decay $B \to X_d\gamma$ can be treated analogously to those in the decay $B \to X_s\gamma$. The local corrections that scale as $1/m_b^2$ are the same for the two modes (up to CKM factors). Also, the analysis of resolved contributions can be applied to this case. On the other hand, the long-distance contributions from the intermediate $u$ quark in the penguin loops are critical. While they are suppressed in the $B \to X_s\gamma$ mode by the CKM matrix elements, there is no such CKM suppression in $B \to X_d\gamma$, and one must account for the non-perturbative contributions that arise from the operator $\mathcal{O}_1^u$. However, this interference contribution vanishes in the total $CP$-averaged rate of $B \to X_d\gamma$ at order $\Lambda/m_b$. Other interference terms from the double resolved contributions, involving $\mathcal{O}_1^u$ and $\mathcal{O}_8$, or $\mathcal{O}_1^u$ and $\mathcal{O}_1^u$, arise first at order $1/m_b^2$. Thus, there is no power correction due to the operator $\mathcal{O}_1^u$ in the total rate of $B \to X_d\gamma$ at order $\Lambda/m_b$, which implies that the $CP$-averaged decay rate of $B \to X_d\gamma$ is as theoretically clean as the decay rate of $B \to X_s\gamma$ (Benzke, Lee, Neubert, and Paz, 2010).

Local hadronic power corrections due to the leading operator $\mathcal{O}_9$ in the decay $B \to X_s\ell^+\ell^-$ that scale with $1/m_b^2$, $1/m_b^3$, and $1/m_c^2$ have also been considered. They can be calculated analogously to those in the decay $B \to X_s\gamma$. However, a systematic analysis of hadronic power corrections including all relevant operators has yet to be performed. Thus, an additional uncertainty of $\pm 5\%$ should be added to all theoretical predictions for this mode on the basis of a simple dimensional estimate.

In the high-$q^2$ region of the decay $b \to s\ell^+\ell^-$, one encounters a breakdown of the heavy quark expansion at the end point of the di-lepton mass spectrum. Whereas the partonic contribution vanishes, the $1/m_b^2$ and $1/m_b^3$ corrections tend towards non-zero values. In contrast to the end point region of the photon energy spectrum in the $B \to X_s\gamma$ decay (see below), no partial all-order resummation into a shape function is possible. However, for an *integrated* high-$q^2$ spectrum an effective expansion is found in inverse powers of $m_b^{\text{eff}} = m_b \times (1 - \sqrt{s_{\text{min}}})$ rather than $m_b$. The expansion converges less rapidly, depending on the lower dilepton-mass cut $s_{\text{min}} = q^2_{\text{min}}$. The large theoretical uncertainties could be significantly reduced by normalizing the $B \to X_s\ell^+\ell^-$ decay rate to the semileptonic $B \to X_u\ell\overline{\nu}$ decay rate with the same $q^2$ cut:

$$\mathcal{R}(s_0) = \frac{\displaystyle\int_{s_0}^{1} \mathrm{d}s\, \frac{\mathrm{d}\Gamma(B \to X_s\ell^+\ell^-)}{\mathrm{d}s}}{\displaystyle\int_{s_0}^{1} \mathrm{d}s\, \frac{\mathrm{d}\Gamma(B \to X_u\ell\overline{\nu})}{\mathrm{d}s}}. \qquad (17.9.14)$$

In this way, the relative uncertainty in this ratio due to the dominating $1/m_b^3$ term would be reduced to 9%, whereas the relative uncertainty in the numerator alone is about 19%.

### 17.9.1.3 Shape functions and kinematical cuts

In the measurements of the inclusive mode $B \to X_s\gamma$ one needs cuts in the photon energy spectrum to suppress the background from other $B$ decays. A threshold of 1.6 GeV is also required for theoretical predictions to remove $c\overline{c}$ bound states.

In order to deal with these cuts, one needs a theoretical description of the photon energy spectrum. In principle, this can be computed along the same lines as the total rates by using the heavy quark expansion. However, at leading order in $\alpha_S$ and $1/m_b$, the spectrum is simply a $\delta$-function expressing the fact that the photon recoils against a single quark and hence $E_\gamma = m_b/2$ (for a massless $s$ quark). Without $\alpha_S$ corrections, the spectrum remains concentrated at this single energy and the heavy quark expansion takes the form

$$\frac{d\Gamma}{dx} = \frac{G_F^2 \alpha m_b^5}{32\pi^4} |V_{ts}V_{tb}^*|^2 |C_7|^2 \Bigg(\delta(1-x) + \frac{\mu_\pi^2 - \mu_G^2}{2m_b^2}\delta'(1-x) + \frac{\mu_\pi^2}{6m_b^2}\delta''(1-x) + \cdots\Bigg) \qquad (17.9.15)$$

with $x = 2E_\gamma/m_b$.

It has been shown in (Bigi, Shifman, Uraltsev, and Vainshtein, 1994; Mannel and Neubert, 1994; Neubert, 1994a) that the leading terms can be resummed into a *shape function* defined as

$$2M_B f(k_+) = \langle B(v)|\bar{b}_v \delta(k_+ - iD_+) b_v |B(v)\rangle\ , \qquad (17.9.16)$$

which has a moment expansion according to

$$f(\omega) = \delta(\omega) + \frac{\mu_\pi^2}{6}\delta''(\omega) - \frac{\rho_D^3}{18}\delta'''(\omega) + \cdots\ . \qquad (17.9.17)$$

In terms of the shape function, the spectrum takes the form

$$\frac{d\Gamma}{dx} = \frac{G_F^2 \alpha m_b^6}{32\pi^4} |V_{ts}V_{tb}^*|^2 |C_7|^2 f(m_b(1-x))\ . \qquad (17.9.18)$$

The shape function is a non-perturbative quantity, which is universal for all heavy-to-light transitions. It either needs to be modeled or it can be extracted from other heavy-to-light decays such as $b \to u\ell\overline{\nu}$. However, at the sub-leading level, several new shape functions need to be defined, spoiling the simple relation between $B \to X_s\gamma$ and $B \to X_u\ell\overline{\nu}$ (Bauer, Luke, and Mannel, 2002, 2003).

The fact that the shape function is not well known induces uncertainties in experimental branching fraction results in two ways. First, the form (as well as the scheme) chosen for the shape function affects efficiencies, and hence affects the measured integrated branching fractions above $E_\gamma$ thresholds of 1.7 to 2.0 GeV. Second, the need for such thresholds in the measurements leads to further shape-function effects which are taken into account when the branching fractions are extrapolated down to an $E_\gamma$ threshold of 1.6 GeV, in order to compare to theoretical predictions. Both stages result in "model-dependence" uncertainties in the experimental results.

The shape functions have been represented using three different theoretical approaches: the "kinetic" scheme, the "shape function" scheme and "dressed gluon exponentiation" (DGE).

- The kinetic scheme is frequently used in the context of the determination of $V_{cb}$ and $V_{ub}$ from inclusive semileptonic decays and is described in some detail in Section 17.1.3.1.
- In the shape function scheme (Neubert, 2005) a multi-scale OPE with three short-distance scales $m_b$, $\sqrt{m_b \Delta}$, and $\Delta = m_b - 2E_\gamma$ has been proposed to connect the shape function and the local OPE region (Becher and Neubert, 2007). Additional perturbative effects related to the kinematic cutoff have been calculated to NNLL precision by the use of SCET methods. Further work is needed to clarify the applicability of these numerical results (Misiak, 2008).
- An alternative approach to the effects of the cutoff in the photon energy spectrum is based on DGE, which incorporates Sudakov and renormalon re-summations (Andersen and Gardi, 2005). The greater predictive power of this approach is related in part to the assumption that non-perturbative power corrections associated with the shape function follow the pattern of ambiguities present in the perturbative calculation.

In the inclusive decay $B \to X_s \ell^+ \ell^-$, the hadronic and di-lepton invariant masses are independent kinematical quantities. An upper hadronic invariant-mass cut is imposed by the experiments to reduce backgrounds. The high di-lepton mass region is not affected by this cut, since at high di-lepton mass the hadronic invariant mass is constrained to small values due to kinematics. In the low di-lepton mass region the kinematics with a jet-like $X_s$ and $m_X^2 \leq m_b \Lambda$ implies the need to include the effects of a shape function. A recent SCET analysis shows that to leading order, using the universality of the shape function, the form of the di-lepton mass spectrum at small di-lepton masses remains unchanged, but the differential rate becomes smaller by an overall factor of $0.7-0.9$. Nevertheless, the effects of sub-leading shape functions lead to an additional uncertainty of 5% (Lee, Ligeti, Stewart, and Tackmann, 2006; Lee and Stewart, 2006). Another analysis estimates the uncertainties due to sub-leading shape functions more conservatively. By scanning over a range of models of these functions, one finds corrections in the rates relative to the leading-order result to be between $-10\%$ to $+10\%$ with equally large uncertainties (Lee and Tackmann, 2009). In the future it may be possible to decrease such uncertainties significantly by constraining both the leading and sub-leading shape functions using the combined data from $B \to X_s \gamma$, $B \to X_u \ell \overline{\nu}$ and $B \to X_s \ell^+ \ell^-$ (Lee and Tackmann, 2009).

#### 17.9.1.4 Soft Collinear Effective Theory (SCET)

The Wilson coefficients of the weak effective Hamiltonian are process independent and can be used for both inclusive and exclusive modes. However, exclusive final states require the computation of hadronic matrix elements between meson states, which is difficult and limits the theoretical precision. The naïve approach is to write the amplitude $A \simeq C_i(\mu_b) \langle \mathcal{O}_i(\mu_b) \rangle$ and parameterizing $\langle \mathcal{O}_i(\mu_b) \rangle$ in terms of form factors. A substantial improvement can be obtained by using the QCDF method (Beneke, Buchalla, Neubert, and Sachrajda, 1999, 2000, 2001) and its field-theoretical formulation, SCET (Bauer, Fleming, and Luke, 2000; Bauer, Fleming, Pirjol, and Stewart, 2001; Bauer, Pirjol, and Stewart, 2002; Bauer and Stewart, 2001; Beneke, Chapovsky, Diehl, and Feldmann, 2002; Hill and Neubert, 2003). These methods form the basis of the up-to-date predictions of exclusive $B$ decays. Within this framework one can show that, even if the form factors were known with infinite precision, the description of exclusive decays would be incomplete due to the existence of non-factorizable strong interaction effects that cannot be represented by form factors.

The QCDF and SCET methods were first systematized for exclusive non-leptonic decays in the heavy quark limit. In contrast to the HQET, SCET does not correspond to a local operator expansion. Whereas HQET is applicable to $B$ decays if the energy transfer to light hadrons is small, *e.g.* in $B \to D$ transitions at small recoil, HQET is not applicable to rare decays where light particles have momenta of order $m_b$. One faces a multi-scale problem that can be tackled within SCET. There are three relevant scales: (a) $\Lambda = \text{few} \times \Lambda_{\rm QCD}$, the soft scale set by the typical energies and momenta of the light degrees of freedom in the hadronic bound states; (b) $m_b$, the hard scale set by both the heavy $b$ quark mass and the energy of the final state hadrons in the $B$ meson rest frame; and (c) the hard-collinear scale $\mu_{\rm hc} = \sqrt{m_b \Lambda}$, which appears through interactions between the soft and energetic modes in the initial and final states. The dynamics of the hard and hard-collinear parts can be described perturbatively in the heavy quark limit $m_b \to \infty$. In this limit SCET describes $B$ decays to light hadrons with energies much larger than their masses, assuming that their constituents have momenta collinear to the hadron momenta.

#### 17.9.1.5 Application to the modes $B \to K^* \gamma$ and $B \to \rho\gamma$

The QCDF formalism can be applied to exclusive radiative and electroweak penguin decays (Beneke and Feldmann, 2001). For $B \to K^* \gamma$, or more generally for $B \to V\gamma$, where $V$ is a light vector meson, the QCDF formula for the hadronic matrix element of each operator of the effective Hamiltonian in the heavy quark limit and to all orders in $\alpha_S$ reads

$$\langle V\gamma | \, \mathcal{O}_i \, | B \rangle = T_i^I \, F^{B \to V_\perp} + \int_0^\infty \frac{d\omega}{\omega} \, \phi_B(\omega) \int_0^1 du \, \phi_{V_\perp}(u) \, T_i^{II}(\omega, u). \tag{17.9.19}$$

This formula separates out the process independent non-perturbative quantities into $F^{B\to V_\perp}$, a form factor evaluated at maximum recoil ($q^2 = 0$), and the light-cone distribution amplitudes (LCDA), $\phi_B$ and $\phi_{V_\perp}$, for the heavy and light mesons. This leaves the quantities $T^I$ and $T^{II}$, known as hard-scattering kernels, which can be calculated perturbatively. These correspond to vertex and spectator corrections, respectively, and have been calculated to $\mathcal{O}(\alpha_s^1)$ (Ali and Parkhomenko, 2002; Beneke, Feldmann, and Seidel, 2001; Bosch and Buchalla, 2002b; Descotes-Genon and Sachrajda, 2004), and recently in some cases to $\mathcal{O}(\alpha_s^2)$ (Ali, Pecjak, and Greub, 2008).

The LCDA of light pseudoscalar and vector mesons that enter the factorization formula have been studied in detail through the use of light-cone QCD sum rules (Ball and Braun, 1999; Ball, Braun, Koike, and Tanaka, 1998; Braun and Filyanov, 1989, 1990). However, not much is known about the $B$ meson LCDA, whose first moment enters the factorized amplitude at $\mathcal{O}(\alpha_s)$. Because this moment also enters the factorized expression for the $B \to \gamma$ form factor, it might be possible to extract its value from measurements of decays such as $B \to \gamma e\nu$, if the power corrections are under control.

The QCDF formula introduces an important simplification in the form factor description. The $B \to V_\perp$ form factors at large recoil have been analyzed in SCET and are independent of the Dirac structure of the current in the heavy quark limit (Charles, Le Yaouanc, Oliver, Pène, and Raynal, 1999). As a consequence of this, all the form factors reduce to a single form factor up to factorizable corrections in the heavy quark and large energy limits.

Field-theoretical methods such as SCET make it possible to reach a deeper understanding of the QCDF approach. The various momentum regions are represented by different fields, and the hard-scattering kernels $T^I$ and $T^{II}$ can be shown to be Wilson coefficients of effective field operators. Using SCET one can prove the factorization formula to all orders in $\alpha_s$ and to leading order in $\Lambda/m_b$ (Becher, Hill, and Neubert, 2005). QCD is matched on SCET in a two-step procedure that separates the hard scale $\mu \sim m_b$ and then the hard-collinear scale $\mu \sim \sqrt{\Lambda m_b}$ from the hadronic scale $\Lambda$. The vertex correction term $T^I$ involves the hard scales, whereas the spectator scattering term $T^{II}$ involves both the hard and the hard-collinear scales. This is why large logarithms have to be resummed, which can be done most efficiently in SCET.

In principle, the field-theoretical framework of SCET allows one to go beyond the leading-order result in $\Lambda/m_b$. However, a breakdown of factorization is expected at that order. For example, in the analysis of $B \to K^*\gamma$ decays at sub-leading order, an infrared divergence is encountered in the matrix element of $\mathcal{O}_8$ (Kagan and Neubert, 2002). In general, power corrections involve convolutions, which turn out to be divergent. Currently, no solution to this well-analyzed problem of end-point divergences within power corrections is available (Arnesen, Ligeti, Rothstein, and Stewart, 2008; Becher, Hill, and Neubert, 2004; Beneke and Feldmann, 2004). Thus, within the QCDF/SCET approach, a general, quantitative method to estimate the important $\Lambda/m_b$ corrections to the heavy quark limit is missing, which significantly limits the precision in phenomenological applications.

Nevertheless, some very specific power corrections are still computable and are often numerically important. Indeed, this is the case for the annihilation and weak exchange amplitudes in $B \to \rho\gamma$, where the annihilation diagram represents the leading contribution to the isospin asymmetry (Kagan and Neubert, 2002). These corrections are included in recent theoretical predictions of these decays. The method of light-cone QCD sum rules can also help to provide estimates of such unknown sub-leading terms. For example, power corrections to the indirect *CP* asymmetries in $B \to V\gamma$ decays have been analyzed in this manner (Ball, Jones, and Zwicky, 2007).

#### 17.9.1.6 Application to the mode $B \to K^*\ell^+\ell^-$

There is a similar factorization formula for the exclusive electroweak penguin $B$ decays, such as $B \to K^*\ell^+\ell^-$, but the simplification due to form factor relations is even more drastic. The hadronic form factors can be expanded in the small ratios $\Lambda/m_b$ and $\Lambda/E$, where $E$ is the energy of the light meson. If corrections of order $1/m_b$ and $\alpha_s$ are neglected, the *seven* a priori independent $B \to K^*$ form factors reduce to *two* universal form factors $\xi_\perp$ and $\xi_\parallel$ (Charles, Le Yaouanc, Oliver, Pène, and Raynal, 1999). This reduction makes it possible to design interesting ratios of observables in which any soft form factor dependence cancels out for all di-lepton masses $q^2$ at leading order in $\alpha_s$ and $\Lambda/m_b$ (Bobeth, Hiller, van Dyk, and Wacker, 2012; Egede, Hurth, Matias, Ramon, and Reece, 2008, 2010).

The theoretical simplifications of the QCDF/SCET approach are restricted to the kinematic region in which the energy of the $K^*$ is of the order of the heavy quark mass, $q^2 \ll m_B^2$. However, in the limit $q^2 \to 0$ the longitudinal amplitude in the QCDF/SCET approach generates a logarithmic divergence, which indicates problems in the theoretical description. The presence of light resonances below $1\,\mathrm{GeV}/c^2$ may also call into question the QCDF results in this region. Thus, the factorization formula only applies well in the di-lepton mass range $1\,\mathrm{GeV}^2/c^2 < q^2 < 6\,\mathrm{GeV}^2/c^2$.

The QCDF and SCET methods are also applicable to other phenomenologically important electroweak penguin decays such as $B \to K\ell^+\ell^-$ (Bobeth, Hiller, and Piranishvili, 2007), $B \to \rho\ell^+\ell^-$ (Beneke, Feldmann, and Seidel, 2005), and $B_s \to \phi\ell^+\ell^-$. Note that the decay into a pseudoscalar meson is analogous to the decay into a longitudinally polarized vector meson.

### 17.9.2 Inclusive $b \to s\gamma$

The transition $b \to s\gamma$ was first observed by CLEO II through the exclusive decay $B \to K^*\gamma$ (Ammar et al., 1993). This was followed by the first measurement of the inclusive rate for $b \to s\gamma$ using a combination of a fully inclusive photon spectrum, and a "pseudoreconstruction" (see Section 17.9.2.4) of a sum of exclusive final states in $B \to X_s\gamma$ (Alam et al., 1995).

The detailed computation described in Sections 17.9.1.1 and 17.9.1.2 results in the NNLL prediction for a photon-energy ($E_\gamma$) threshold of $E_\gamma > 1.6\,\text{GeV}$ (Misiak et al., 2007)

$$\mathcal{B}(B \to X_s\gamma)_{\text{NNLL}} = (3.15 \pm 0.23) \times 10^{-4}. \quad (17.9.20)$$

The overall uncertainty is the quadratic sum of non-perturbative (5%), parametric (3%), perturbative scale (3%) and $m_c$ interpolation ambiguity (3%) uncertainties. An additional scheme dependence has since been found (Gambino and Giordano, 2008), but it is within the perturbative uncertainty of 3% (Misiak, 2008).

However, in experimental measurements, a large background from non-signal $B\overline{B}$ events at low values of photon energy limits the minimum useful $E_\gamma$. The final $B \to X_s\gamma$ CLEO publication (Chen et al., 2001b) reports results for $E_\gamma$ above 2.0 GeV using $9\,\text{fb}^{-1}$ of $\Upsilon(4S)$ data and $4.4\,\text{fb}^{-1}$ of off-resonance data. They made an extrapolation down to the full energy range, and quoted an inclusive branching fraction of $(3.21 \pm 0.43 \pm 0.27^{+0.18}_{-0.10}) \times 10^{-4}$, where the errors are statistical, systematic and model-dependence, respectively.

The much larger data samples of the $B$ Factories and to some extent their improved detectors have allowed for a number of significant advances in the analysis techniques, leading to large reductions in the systematic uncertainties as well as the statistical uncertainties of measured branching fractions. They have also made it possible to reduce the photon energy threshold, in one case down to 1.7 GeV, and detailed studies of the $E_\gamma$ spectrum are used to help constrain the model-dependent extrapolation. The world-average extrapolated branching fraction now has an uncertainty comparable to that on the theoretical prediction.

In the following, four measurements of the inclusive $B \to X_s\gamma$ branching fraction are described. Three of these are fully inclusive, while the fourth builds up the branching fraction as a sum of exclusive final states. The hallmark of a fully inclusive measurement is that for a signal $B$ it requires only the detection of a high-energy photon with $E_\gamma$ close to half the $b$-quark mass. Because of this the processes $B \to X_s\gamma$ and $B \to X_d\gamma$ are not separated. For branching fractions, the $B \to X_d\gamma$ contribution is easily subtracted using

$$\frac{\mathcal{B}(B \to X_d\gamma)}{\mathcal{B}(B \to X_s\gamma)} = (|V_{td}|/|V_{ts}|)^2 = 0.044 \pm 0.003\ . \quad (17.9.21)$$

As the measurement is based on the photon, the result is not much affected by uncertainties in the hadronization process of the $s$ quark. However, the measured value of $E_\gamma$ is subject to electromagnetic-calorimeter resolution. Also, because (for two of the three measurements) the $B$ rest frame is not known, there is Doppler smearing due to the motion of the $B$ in the $\Upsilon(4S)$ center-of-mass frame. Inclusiveness is not compromised by imposing requirements on the non-signal $B$ ($\overline{B}$) meson in the event. Such requirements can significantly reduce the large background from continuum processes (*i.e.*, $e^+e^- \to q\overline{q}$ or $\tau^+\tau^-$, with $q = u, d, s, c$), which dominates the statistical uncertainty on the extracted signal. One such requirement is lepton tagging: for $B\overline{B}$ events a high-momentum electron or muon can arise from the semileptonic decay of the non-signal $B$. The Belle analysis described in Section 17.9.2.1 combines separate samples of untagged and lepton-tagged events, while the *BABAR* analysis in Section 17.9.2.2 relies on lepton tagging. The *BABAR* analysis in Section 17.9.2.3 fully reconstructs the non-signal $B$ in hadronic decay modes. This has the advantage that the signal-$B$ frame is known, but at a great cost in statistics.

The sum-of-exclusive-modes method in Section 17.9.2.4 specifically reconstructs $X_s$ final states, and determines the photon energy in the $B$ rest frame, using

$$E_\gamma = \frac{m_B^2 - m_{X_s}^2}{2m_B}\ , \quad (17.9.22)$$

with a resolution that is much better than that of the direct photon energy measurement with the calorimeter. On the other hand, there are substantial systematic uncertainties from the hadronization model and unmeasured modes, and signal efficiency decreases significantly with increasing $m_{X_s}$.

The branching-fraction results from all the methods are summarized at the end in Table 17.9.3. The first two fully-inclusive approaches provide by far the best precision on the branching fraction above any given energy threshold. (The sum-of-exclusive-modes approach is systematically limited by uncertainties in the $X_s$ hadronization, which affect both the efficiency for the selected decay modes and the contribution of the unmeasured modes.) Results are also presented for the photon energy spectrum (Section 17.9.2.5) and, later, for direct *CP* asymmetries (Section 17.9.5.3). Extrapolating the branching fraction measurements down to a 1.6 GeV threshold, for which the theoretical SM prediction is made, can provide useful constraints on new physics (Section 17.9.2.7).

#### 17.9.2.1 Belle fully inclusive (untagged and lepton-tagged)

The untagged inclusive method was first applied by Belle with $140\,\text{fb}^{-1}$ of $\Upsilon(4S)$ data (Koppenburg, 2004). High-energy photons leave a clear signal in the CsI (Tl) electromagnetic calorimeter, with good energy resolution. The main challenge for this method is the subtraction of the large background from other sources of photons and photon-like signals.

That initial analysis is superseded by the latest measurement (Limosani, 2009), with $605\,\text{fb}^{-1}$ of $\Upsilon(4S)$ data (657M $B\overline{B}$) and $68\,\text{fb}^{-1}$ of data collected 60 MeV below the $\Upsilon(4S)$. In the updated analysis, a separate measurement with the lepton tag method (Section 17.9.2.2) is also performed on the same sample. The sizes of the statistical errors are comparable between the two methods. Since the events that pass the selection criteria are not fully overlapping between the two methods, the photon spectra are separately measured and then combined (taking correlations into account) to increase the sensitivity.

The photons need to be isolated from other clusters in the calorimeter. They are then matched to other low energy photons to see if they form either a $\pi^0$ or $\eta$ meson. If they do they are rejected from further analysis ("$\pi^0/\eta$ veto"). There is a systematic error of 3% from the photon selection efficiency, which mainly comes from the isolation requirement and the understanding of the $\pi^0$ and $\eta$ veto.

Event shape information is used to suppress a large fraction of the background from continuum events. Then the remainder, which for the untagged sample is still huge, as illustrated in Figure 17.9.2, is subtracted by using a sample of off-resonance data, which is free from $B$-meson decays. The $B$ Factories took off-resonance data at a fraction of ∼10% to 11% of their $\Upsilon(4S)$ data (much lower than CLEO's 50%). The subtraction of the continuum background is the largest statistical error source in this method. There is also a systematic error of up to 7.5% (depending on the photon energy threshold) of the subtracted value,[85] originating from the anti-correlated small uncertainties in the scaling factor for the continuum subtraction (0.3%) and in the number of $B\overline{B}$ for the $B$ background subtraction (1.4%). The continuum scaling factor is the luminosity ratio corrected for the change in cross-section and photon energy spectrum as a function of the center-of-mass energy.

Once the non-$B$ backgrounds are subtracted, the dominant background source are the $B$ decay modes that produce photons through secondary meson decays, with the main contribution coming through $\pi^0 \to \gamma\gamma$ decays, and the next largest through $\eta \to \gamma\gamma$ as shown in Figure 17.9.3. These photons are on average lower in energy, but follow a steeply rising spectrum as the photon energy threshold is reduced. This background is simulated by a generic $B\overline{B}$ Monte Carlo sample, in which the $\pi^0$ and $\eta$ momentum spectra are calibrated using their distributions from $B$ decays measured in the data. Other sources of photons in $B$ decays then dominate the uncertainty on the background, giving a systematic error of 2–7% (depending on the photon energy threshold). These include real photons from $\omega$, $\eta'$ and charmonia decays, and fake photons from electrons, anti-neutrons and $K^0_L$. An artifact due to remnant energy clusters of out-of-time electrons from QED processes, which is not fully subtracted by the off-resonance sample as its rate depends on the instantaneous luminosity, is also subtracted. These contributions are evaluated using data as much as possible.

The combined (untagged and lepton-tagged) photon energy spectrum is shown in Figure 17.9.4, after background subtraction, efficiency correction and unfolding of the calorimeter resolution. The $B \to X_d\gamma$ contribution is then subtracted, using Eq. (17.9.21), to give the $B \to X_s\gamma$ branching fraction. The photon energy spectrum and the reconstruction efficiency are considered to be the same for $B \to X_s\gamma$ and $B \to X_d\gamma$.

[85] The quoted systematic errors in this section are for the combined untagged and lepton-tag results for the integrated branching fraction.

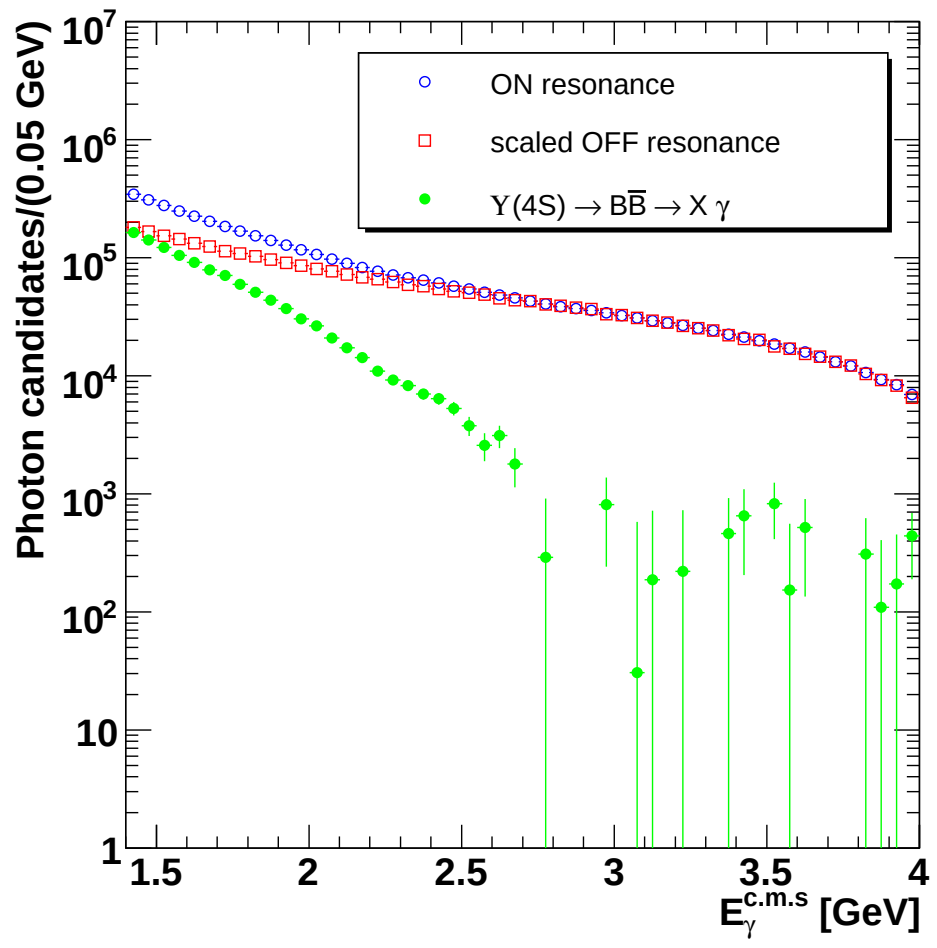

**Figure 17.9.2.** Photon energy spectrum before background subtraction, continuum background estimated from the off-resonance events, and the continuum-subtracted spectrum, all for the untagged selection in Belle's 657M $B\overline{B}$ data.

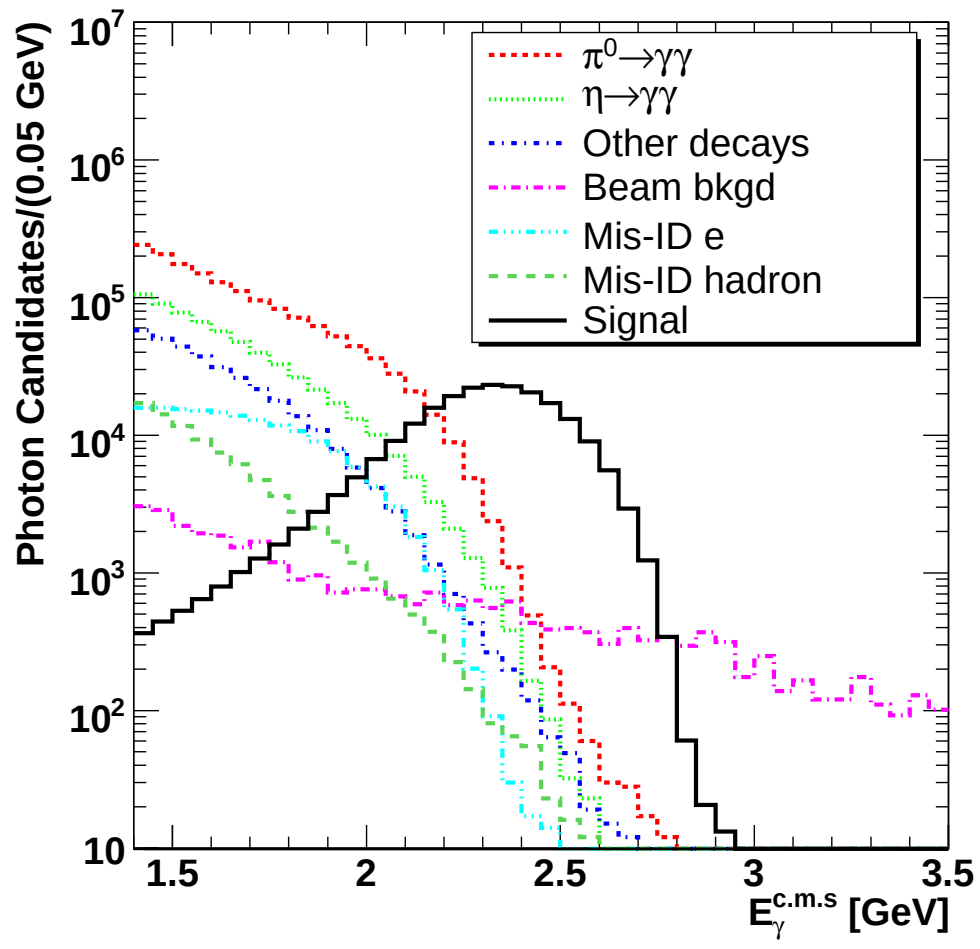

**Figure 17.9.3.** Expected $B \to X_s\gamma$ signal and background contributions as functions of the center-of-mass photon energy from Belle's Monte Carlo simulation for 657M $B\overline{B}$. This illustration is for the untagged selection.

Using a photon energy threshold $E_\gamma > 1.7\,\mathrm{GeV}$, where the photon energy is defined in the $B$-meson rest frame, the $B \to X_s\gamma$ branching fraction is measured to be

$$(3.45 \pm 0.15 \pm 0.40) \times 10^{-4} \quad (\text{Belle}, E_\gamma > 1.7\,\mathrm{GeV}), \tag{17.9.23}$$

where the errors are statistical and systematic. The small correction due to the boost of the $B$ meson in the center-of-mass system is calculated using a Monte Carlo simulation. Results for higher energy thresholds are tabulated in Table 17.9.3 of Section 17.9.2.6.

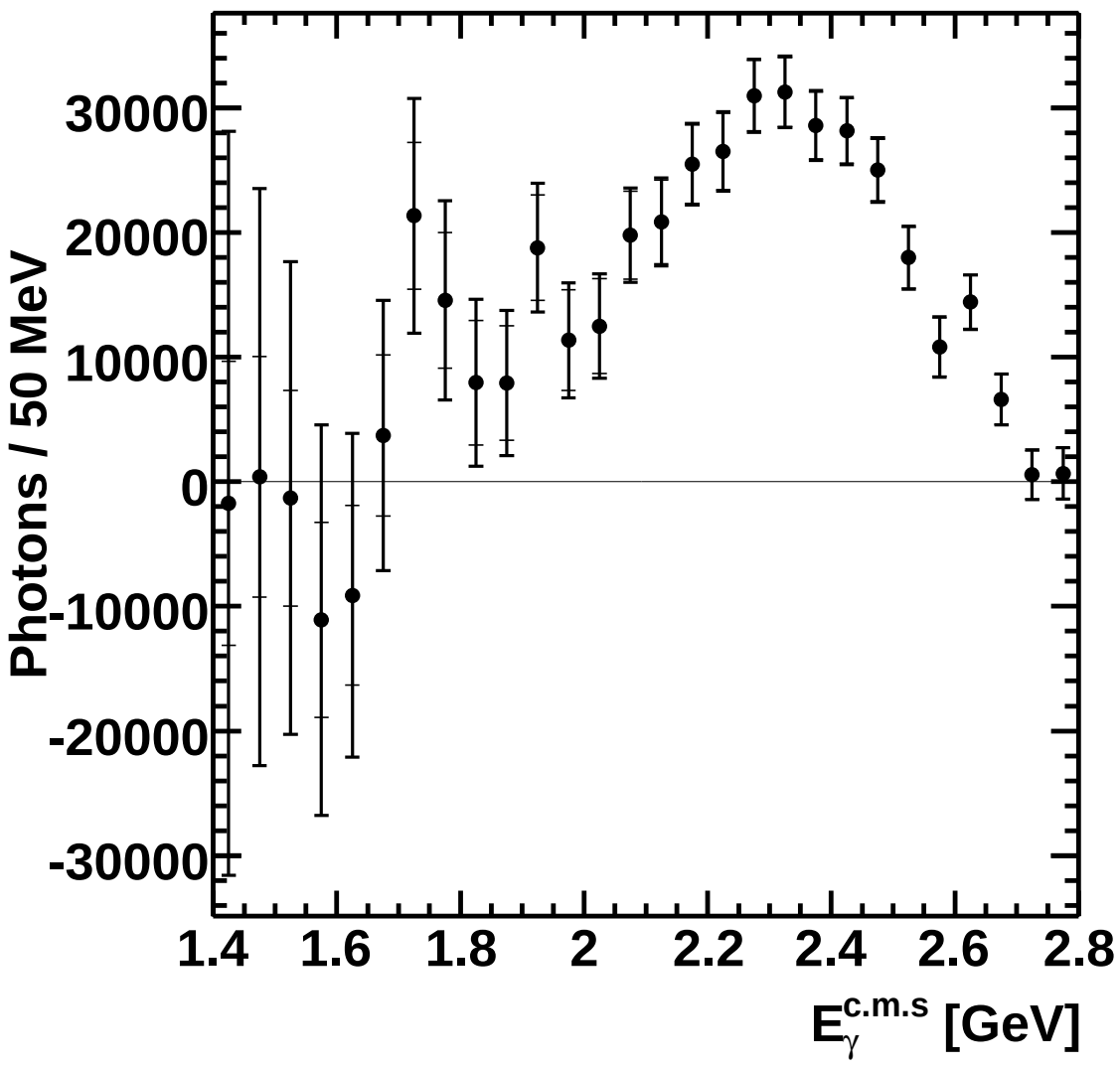

**Figure 17.9.4.** From Limosani (2009). Combined untagged and lepton-tagged inclusive photon spectrum from Belle in the rest frame of the $\Upsilon(4S)$ after background subtraction, efficiency correction and unfolding of calorimeter resolution.

### 17.9.2.2 *BABAR* fully inclusive with lepton tagging

The lepton tag method was first used by the *BABAR* collaboration (Aubert, 2006t). That initial measurement has been superseded by an analysis based on an integrated luminosity of $347\,\text{fb}^{-1}$ (383M $B\overline{B}$ pairs) collected on the $\Upsilon(4S)$ resonance, plus $36\,\text{fb}^{-1}$ collected 40 MeV below the resonance (Lees, 2012j,o). For the signal it relies on the detection of high energy photons — with photon-quality requirements analogous to those used by Belle including isolation, and a veto if the high-energy photon is part of a reconstructed $\pi^0$ or $\eta \to \gamma\gamma$ decay — in association with a lepton tag from a semileptonic decay of the other $B$. The lepton momentum threshold is $p^* > 1.05\,\text{GeV}/c$ for both electrons and muons, where $p^*$ is measured in the center-of-mass frame. There are additional requirements on the angle between the photon and the lepton (near back-to-back configurations are rejected) and on the missing energy in the event (since a semileptonic decay entails a missing neutrino).

These preliminary lepton-tag requirements remove 98% of the continuum events, at a cost of retaining only 12% of signal events. The tag variables are then combined with topological (event-shape) information in a multivariate selector to further suppress continuum background. The subtraction of the remaining continuum background using the off-resonance data still dominates the statistical uncertainty, but at a lower level than for an untagged analysis. Lepton tagging introduces an additional small systematic error of up to 2.4% (decreasing as the photon energy threshold increases) due to lepton identification and uncertainties in $b \to c\ell\nu$ branching fractions.

The largest background is now from other $B$ decays, which have a lepton-tag efficiency slightly below that for signal events. The composition of the $B$ background is similar to that for Belle. It consists of high-energy photons from unvetoed $\gamma\gamma$ decays of $\pi^0$'s (by far the largest component) and $\eta$'s, radiative decays of other mesons, electrons which are misidentified as photons (due to tracking inefficiency or bremsstrahlung), antineutrons which annihilate in the detector, final-state radiation, and other small effects. Each significant component of the MC-predicted $B$ background is corrected by comparisons of data and MC control samples. The uncertainties on these corrections give the main contribution to the systematic error on the branching fraction for lower $E_\gamma$ thresholds (7.8% for 1.8 GeV, decreasing as the threshold rises). The largest uncertainties arise from unvetoed $\pi^0$'s and from electrons without reconstructed tracks. The signal efficiency uncertainty is 3.0%, independent of the threshold; its largest component is from the photon isolation requirement. Correlations between common sources of signal-efficiency and $B\overline{B}$-background uncertainties are additionally taken into account.

The measured photon energy spectrum in the center-of-mass frame, after subtracting both continuum and corrected $B\overline{B}$ backgrounds, is shown in the top plot of Fig. 17.9.5. After correcting for efficiency, adjusting the result to the $B$ rest frame (both of which steps have a small dependence on the spectral shape, *i.e.* model-dependence, as noted in Section 17.9.1.3), and removing the small $B \to X_d\gamma$ contribution (using Eq. 17.9.21), the resulting $\mathcal{B}(B \to X_s\gamma)$ is

$$(3.21 \pm 0.15 \pm 0.29 \pm 0.08) \times 10^{-4} \qquad (\textit{BABAR}, E_\gamma > 1.8\,\text{GeV}), \quad (17.9.24)$$

where the errors are statistical, systematic and model-dependence, respectively. Results for higher energy thresholds are tabulated in Table 17.9.3 of Section 17.9.2.6.

The photon energy spectrum in the top plot of Fig. 17.9.5 is also corrected bin by bin for efficiency, and the effects of calorimeter resolution and Doppler smearing are unfolded using a technique adapted from (Malaescu, 2009). For a binned energy spectrum, the method starts from an assumed model (for the unfolded spectrum) and computes the bin-by-bin difference between the initial (not-yet-unfolded) data and the predictions of this model. A regularization function is then defined to ascribe a fraction of the difference in each bin to fluctuations, with the remainder applied as a correction to the model. The function is optimized using Monte Carlo studies with different starting models. The entire procedure is then iterated, and converges quickly. The resulting spectrum in terms of true photon energy in the $B$ rest frame is shown in the lower plot of Fig. 17.9.5.

An advantage of the lepton tag is that it provides a $CP$-flavor tag for the combined $B \to X_{s+d}\gamma$ decays and can be used to determine the direct $CP$ asymmetry. The measurement is described in Section 17.9.5.3.

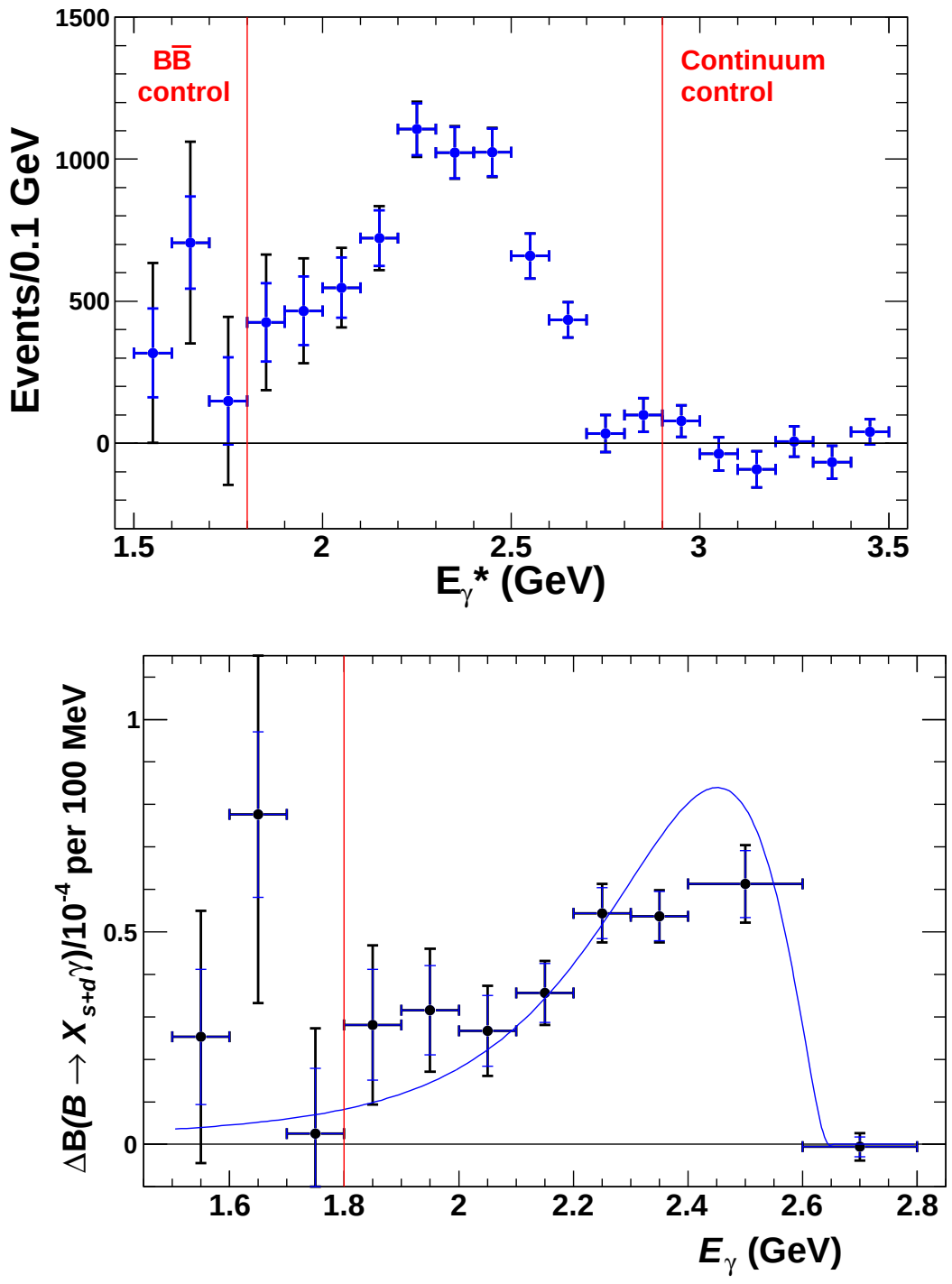

**Figure 17.9.5.** From (Lees, 2012j,o). Photon spectra in the BABAR lepton-tagged fully-inclusive measurement: (top) in the center-of-mass frame after background subtraction, uncorrected for efficiency and resolution smearing, and (bottom) in the $B$ rest frame, after correcting for efficiency and unfolding the effects of both calorimeter resolution and Doppler smearing. The control regions (delineated by vertical red lines) were used to check the backgrounds; all $B\overline{B}$ background corrections were determined and applied before unblinding the signal region (1.8 to 2.9 GeV). The curve in the lower plot is based on the kinetic-scheme computation of (Benson, Bigi, and Uraltsev, 2005), using the HFAG world-average determination of HQE parameters (see Section 17.9.2.5).

#### 17.9.2.3 *BABAR* fully inclusive with reconstructed-$B$ tagging

As an alternative to the lepton-tag approach, *BABAR* (Aubert, 2008q) has also used the recoil-$B$ technique (Section 7.4), in which the signal ("recoil") $B$ meson is tagged by fully reconstructing the non-signal $B$ meson in a hadronic decay mode. This technique (along with reconstruction in semileptonic decay modes) has been widely used at the $B$ Factories to study rare decays with multiple neutrinos, *e.g.*, $B \to \tau\nu$ and $B \to K\nu\overline{\nu}$.

In *BABAR*'s reconstructed-$B$-tag analysis, based on 210 fb$^{-1}$ of $\Upsilon(4S)$ data, more than 1000 different hadronic final states are reconstructed, representing 5% of the decay width of the $B$ meson. This large number of final states is essential in order to reach a maximal signal efficiency for $B \to X_s\gamma$, although it is still only 0.3%. A high-energy photon is required among the remaining particles in the event. After applying a $\pi^0$ veto, suppressing continuum background using event-topology criteria, and selecting events with $\Delta E$ of the hadronic $B$ candidates in a $\pm$60 MeV window, fits are made to the $m_{\rm ES}$ distribution of those candidates in bins of photon energy. These fits remove all the continuum and combinatorial $B$ backgrounds, leaving only signal events and those $B$ decays with a similar topology. These mostly contain photons from $\pi^0$ decays which have survived the $\pi^0$ veto. The remaining $B$ backgrounds are estimated using similar techniques to the untagged and lepton-tagged $B \to X_s\gamma$ analyses.

The hadronic tag has the advantage that it measures the momentum of the tag $B$, which makes it possible to calculate the photon energy in the recoiling signal-$B$ rest frame. It also identifies both the flavor and the charge of the $B$ in the $B \to X_s\gamma$ decay (apart from $B^0-\overline{B}^0$ mixing). Resulting asymmetries are presented in Section 17.9.5. Finally, the rest of the event that has not been used to form the tagging $B$ can be used to study the hadronic $X_s$ system associated with the $b \to s\gamma$ decay. It may eventually be possible to separate out the 4% of $B \to X_d\gamma$ decays using this information.

The *BABAR* analysis (Aubert, 2008q) obtains a $B \to X_s\gamma$ branching fraction

$$(3.66 \pm 0.85 \pm 0.60) \times 10^{-4}\,(\textit{BABAR}, E_\gamma > 1.9\text{ GeV}), \tag{17.9.25}$$

where the errors are statistical and systematic (including some small model-dependence). Although this analysis is currently statistically limited, it is a promising method for the future, *i.e.*, at a high-luminosity $B$ Factory. The dominant systematic uncertainties (*e.g.*, from $B\overline{B}$ backgrounds) may also be reduced significantly with a larger data sample. (This is equally true for the untagged and lepton-tag methods, despite the use in those cases of off-resonance data.) An improvement might also be possible by including semileptonic tags.

#### 17.9.2.4 Sum of exclusive modes

An alternative technique to measure the inclusive branching fraction is to reconstruct the $B \to X_s\gamma$ decay chain with the $X_s$ final state as the sum of as many exclusive modes as possible. This method is a development of the "pseudoreconstruction" method used by CLEO for the continuum background suppression, a $\chi^2$ technique in which an $X_s$ candidate is required to have $m_{\rm ES}$ and $\Delta E$ within broad ranges around expected values.

An early analysis by Belle (Abe, 2001a) has used 6 fb$^{-1}$, and explicitly reconstructed a set of 16 final states with one kaon and 1 to 4 pions of which only one is allowed to be a $\pi^0$. The $X_s$ mass is restricted to be below 2.05 GeV/$c^2$, which corresponds to an $E_\gamma$ threshold of 2.24 GeV. The quoted branching fraction:

$$(3.36 \pm 0.53 \pm 0.42\,^{+0.50}_{-0.54}) \times 10^{-4}\,(\text{Belle},\ E_\gamma > 2.24\,\text{GeV}) \tag{17.9.26}$$

is for the full energy range, where the errors are statistical, systematic and model-dependence, respectively. This method has not been used by Belle with a larger dataset for the branching fraction measurement, but it was used to measure the direct *CP* asymmetry with a 152M $B\overline{B}$ sample as discussed in Section 17.9.5.

In an improved version of this method adopted by *BABAR* (Aubert, 2005x; Lees, 2012f), a set of 38 exclusive final states is explicitly reconstructed. Multiple candidates in an event are resolved using a signal-selecting classifier based on $\Delta E$, and a fit is made to the $m_{\rm ES}$ distributions for the sum of the final states in bins of photon energy. All of the continuum and almost all the $B$ backgrounds are thereby subtracted, apart from a small component which peaks in $m_{\rm ES}$, primarily due to $\pi^0$'s that survive the veto. The photon energy in the $B$ rest frame is precisely deduced from the measured $X_s$ mass (Eq. 17.9.22).

The main limitation of this analysis is the understanding of the hadronization of the $s$ quark into different $X_s$ final states, and the estimation of the fraction of missing final states that have not been included in the analysis. The most prominent exclusive signal is from $B \to K^*\gamma$, but it covers only 12% of the total $B \to X_s\gamma$ branching fraction. The rest of the decay width is covered by modes with higher mass $X_s$ final states. Of the 38 final states considered by *BABAR*, most of them are of the form $B \to Kn(\pi)\gamma$, where $K$ stands for $K^+$ or $K^0_S$, and $n(\pi)$ stands for 1 to 4 pions of which up to two can be a $\pi^0$. In addition they include $B \to Kn(\pi)\eta\gamma$ modes with 0 to 2 pions of which up to one can be a $\pi^0$, and $B \to KK^+K^-(\pi)\gamma$ with 0 or 1 pion. This set of 38 final states accounts for approximately half of the rate for $B \to X_s\gamma$. A further quarter of the rate is due to modes with a $K^0_L$; this contribution can be accurately estimated using the observed $K^0_S$ modes. The remaining 25% of the total rate is mostly in high multiplicity final states, and is associated with lower photon energies and higher $X_s$ mass. The largest $X_s$ mass considered in the analysis is 2.8 GeV, which corresponds to a minimum photon energy of 1.9 GeV. At this mass the missing fraction is 70%, or just over 50% if the $K^0_L$ part is accounted for.

To evaluate the systematic errors associated with the $X_s$ hadronization and the missing fractions, *BABAR* compares the distribution of final states observed in data with a prediction from a MC simulation using `Jetset` (Sjöstrand, 1994). Within the often-large uncertainties, they find agreement except for the low multiplicity $K\pi$ final states where the data are only about 30% of the MC prediction. These modes are dominated by the $K^*(892)$ and $K_2^*(1430)$ resonances. Since `Jetset` performs a non-resonant hadronization, this disagreement is not surprising. For higher multiplicity final states the agreement is better, but the statistical accuracy of the comparisons is limited. With larger data samples it is desirable to add more final states, and make more detailed comparisons with the simulation to understand the hadronization more accurately.

*BABAR*'s first analysis (Aubert, 2005x) used 82 fb$^{-1}$ (89M $B\overline{B}$), and was updated to 429 fb$^{-1}$ (471M $B\overline{B}$) (Lees, 2012f). The latest branching fraction is

$$(3.29 \pm 0.19 \pm 0.48) \times 10^{-4}\,(BABAR,\ E_\gamma > 1.9\,\mathrm{GeV}), \qquad (17.9.27)$$

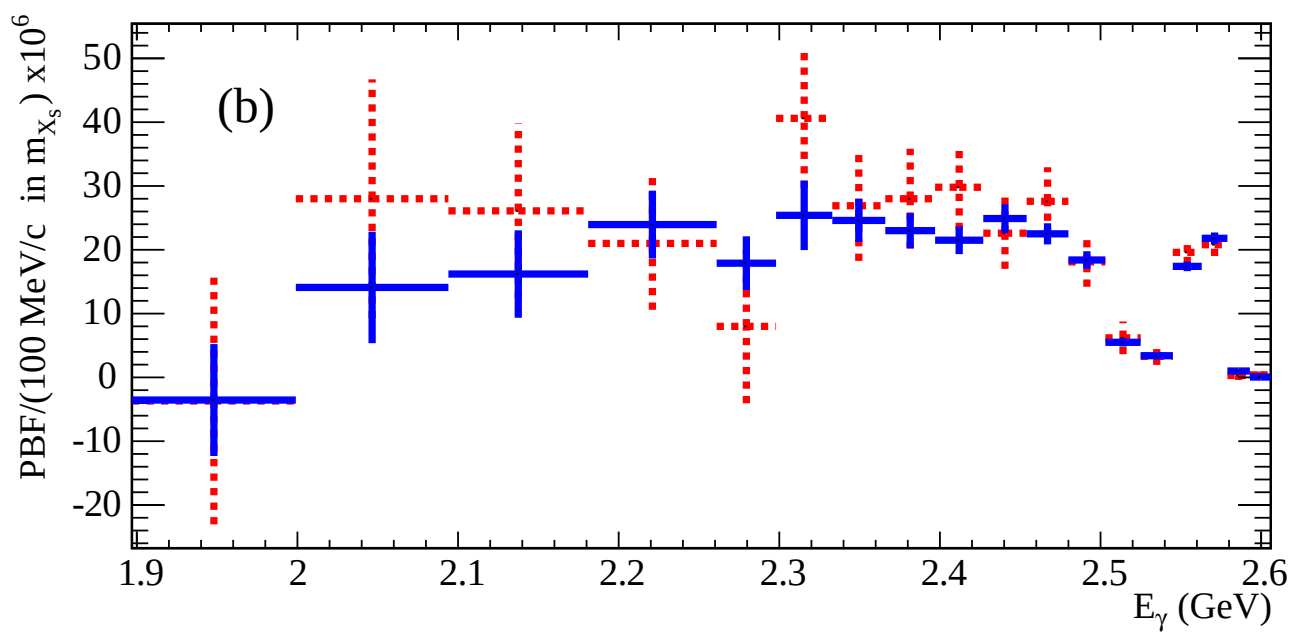

**Figure 17.9.6.** From (Lees, 2012f). Photon spectrum in the $B$ rest frame from *BABAR*'s sum of exclusive final states analysis (blue solid lines), compared to the results from the older (superseded) similar *BABAR* analysis (red dashed lines). The bin widths are defined in terms of the $X_s$ mass, and converted to $E_\gamma$ using Eq. (17.9.22). The peak at 2.56 GeV is from exclusive $B \to K^*\gamma$ decays.

where the errors are statistical and systematic, respectively. The photon spectrum is shown in Figure 17.9.6.

The systematic error from hadronization already dominated with only 82 fb$^{-1}$, and adding statistics did not reduce the overall error in the branching fraction with five times more data. However, this analysis is an important cross-check of the other inclusive analyses, where the systematic error is dominated by a different source, the background from other $B$ decays. And as can be seen in Figure 17.9.6, the spectrum measurement has been significantly improved with more data.

A key point to note about this method is that it is the only one that distinguishes $b \to s\gamma$ from $b \to d\gamma$, and hence is the only method used to measure inclusive $b \to d\gamma$ (see Section 17.9.4). The method also determines the flavor and charge of the $b \to s\gamma$ decay, allowing measurements of direct *CP* and isospin asymmetries in inclusive $b \to s\gamma$ decays. These asymmetry measurements are not expected to be sensitive to the hadronization of the $X_s$.

### 17.9.2.5 Photon energy spectrum and moments

The photon energy spectrum in $B \to X\gamma$ is insensitive to NP (Kagan and Neubert, 1998); rather, it reflects the motion of the $b$ quark inside the $B$ meson, as detailed in Section 17.9.1.3. However, uncertainties in the true spectrum result in a small model-dependence in fully-inclusive measurements of the branching fraction, a quantity which is sensitive to NP. Such uncertainties also enter when extrapolating the measured branching fraction down to an $E_\gamma$ threshold of 1.6 GeV, where the theoretical prediction is made.

A general way to quantify the $E_\gamma$ spectrum is to extract the moments of the spectrum, where the first two moments are equivalent to the mean and width. These quantities depend on the minimum photon energy; or put another way, moments with different minimum photon energies can be treated as additional information about the spectrum, although they are strongly correlated. The results are given in Tables 17.9.1 and 17.9.2. *BABAR* (Lees, 2012j) has also measured third moments, which are several standard deviations away from zero. For all four measurements detailed above, correlation matrices between all measured moments are provided in the published papers or in EPAPS material cited therein.

**Table 17.9.1.** Measured first moments $\langle E_\gamma \rangle$ (in GeV) of the $B \to X_s\gamma$ photon energy spectrum in the $B$ rest frame, for several different photon energy thresholds. The methods are in the order they are described in the text. The errors are statistical and systematic (including model-dependence). SoE refers to sum-of-exclusive modes.

| Measurement | $E_\gamma > 1.7$ GeV | $E_\gamma > 1.8$ GeV | $E_\gamma > 1.9$ GeV | $E_\gamma > 2.0$ GeV |
|---|---|---|---|---|
| Belle | $2.282 \pm 0.015 \pm 0.051$ | $2.294 \pm 0.011 \pm 0.028$ | $2.311 \pm 0.009 \pm 0.015$ | $2.334 \pm 0.007 \pm 0.009$ |
| BABAR lepton tag | | $2.267 \pm 0.019 \pm 0.032$ | $2.304 \pm 0.014 \pm 0.017$ | $2.342 \pm 0.010 \pm 0.009$ |
| BABAR reco.-$B$ tag | | | $2.289 \pm 0.058 \pm 0.027$ | $2.315 \pm 0.036 \pm 0.019$ |
| BABAR SoE | | | $2.346 \pm 0.018\,^{+0.027}_{-0.022}$ | $2.338 \pm 0.010\,^{+0.020}_{-0.017}$ |

**Table 17.9.2.** Measured second moments $\langle (E_\gamma^2 - \langle E_\gamma \rangle^2) \rangle$ (in GeV$^2$) of the $B \to X_s\gamma$ photon energy spectrum in the $B$ rest frame, for several different photon energy thresholds. The methods are in the order they are described in the text. The errors are statistical and systematic (including model-dependence). SoE refers to sum-of-exclusive modes.

| Measurement | $E_\gamma > 1.7$ GeV | $E_\gamma > 1.8$ GeV | $E_\gamma > 1.9$ GeV | $E_\gamma > 2.0$ GeV |
|---|---|---|---|---|
| Belle | $0.043 \pm 0.005 \pm 0.020$ | $0.037 \pm 0.003 \pm 0.008$ | $0.030 \pm 0.002 \pm 0.003$ | $0.023 \pm 0.001 \pm 0.002$ |
| BABAR lep. tag | | $0.0484 \pm 0.0053 \pm 0.0077$ | $0.0362 \pm 0.0033 \pm 0.0033$ | $0.0251 \pm 0.0021 \pm 0.0016$ |
| BABAR $B$ tag | | | $0.033 \pm 0.012 \pm 0.006$ | $0.027 \pm 0.006 \pm 0.002$ |
| BABAR SoE | | | $0.0211 \pm 0.0057\,^{+0.0055}_{-0.0069}$ | $0.0239 \pm 0.0018\,^{+0.0023}_{-0.0030}$ |

**Table 17.9.3.** Measured $B \to X_s\gamma$ inclusive branching fractions (in $10^{-6}$) for several photon energy ($E_\gamma$) thresholds, 1.7 GeV and larger. Errors are statistical, systematic and model-dependence (if applicable); if there is no third error, the model dependence is included in the systematic error. The column with $E_\gamma > 1.6$ GeV contains HFAG's extrapolations (see text) from the lowest measured threshold (HFAG's reciprocal factors are shown in the bottom row.) and their computed world average. The measurements are in the order they are described in the text. The CLEO result is taken from HFAG, who corrected CLEO's published value for the entire spectrum to the value at the listed threshold. The Belle sum-of-exclusive result, which is obtained with $E_\gamma > 2.24$ GeV and is corrected by HFAG, is listed only for the sake of the HFAG extrapolated value and average. All averages above measured threasholds assume errors are uncorrelated. For the HFAG world average of extrapolated values the first error combines statistics and systematics (assumed uncorrelated between experiments), while the second error is from shape function systematics (assumed fully correlated).

| Measurement | ($E_\gamma > 1.6$ GeV) | $E_\gamma > 1.7$ GeV | $E_\gamma > 1.8$ GeV | $E_\gamma > 1.9$ GeV | $E_\gamma > 2.0$ GeV |
|---|---|---|---|---|---|
| CLEO | $328 \pm 44 \pm 28 \pm 6$ | | | | $306 \pm 41 \pm 26$ |
| Belle un- & lepton tag | $350 \pm 15 \pm 41 \pm 1$ | $345 \pm 15 \pm 40$ | $336 \pm 13 \pm 25$ | $321 \pm 11 \pm 16$ | $302 \pm 10 \pm 11$ |
| BABAR lepton tag | $332 \pm 16 \pm 31 \pm 2$ | | $321 \pm 15 \pm 29 \pm 8$ | $300 \pm 14 \pm 19 \pm 6$ | $280 \pm 12 \pm 14 \pm 4$ |
| BABAR reco.-$B$ tag | $390 \pm 91 \pm 64 \pm 4$ | | | $366 \pm 85 \pm 60$ | |
| Belle sum-of-excl. | $369 \pm 58 \pm 46 \pm 60$ | | | | |
| BABAR sum-of-excl. | $352 \pm 20 \pm 51 \pm 4$ | | | $329 \pm 19 \pm 48$ | |
| Average | $343 \pm 21 \pm 7$ | $345 \pm 15 \pm 40$ | $330 \pm 10 \pm 19$ | $315 \pm 8 \pm 12$ | $294 \pm 9 \pm 11$ |
| (Extrapolation factor) | | $(0.985 \pm 0.004)$ | $(0.967 \pm 0.006)$ | $(0.936 \pm 0.010)$ | $(0.894 \pm 0.016)$ |

With sufficient experimental precision, the universal shape function could be extracted from the photon energy spectrum. Fits to the photon spectrum are also relevant to the extraction of $V_{ub}$ from inclusive $B \to X_u \ell \overline{\nu}$ decays (Section 17.1.5). Global fits to spectra from both processes are the goal of the SIMBA collaboration; recent progress on their fits to the $B \to X_s\gamma$ spectrum can be found, for example, in (Bernlochner et al., 2013).

Moments of the photon energy spectrum are related to the parameters of the heavy quark expansion (HQE), such as $m_b$, $\mu_\pi^2$ and $\rho_D$. Thus these parameters could be extracted from the spectrum. Note that the precise meanings and values of such parameters differ somewhat between different schemes. Parameterized computations of the $E_\gamma$ moments (and spectrum) are described for the kinetic scheme in (Benson, Bigi, and Uraltsev, 2005) and for the shape function scheme in (Lange, Neubert, and Paz, 2005). To date, HFAG has found that the HQE parameters are not adequately constrained from fits to $B \to X_s\gamma$ moments alone, but that spectral moments from $B \to X_c \ell \overline{\nu}$ must be included. Such a fit to both sets of moments was first carried out in the kinetic scheme by (Buchmüller and Flächer, 2006). Recent HFAG results can be found in (Amhis et al., 2012). The fit values of $m_b$ and $\mu_\pi^2$ are $(4.57 \pm 0.03)\,\mathrm{GeV}/c^2$ and $(0.46 \pm 0.04)\,\mathrm{GeV}^2/c^2$, respectively. (This fit predates inclusion of the *BABAR* lepton-tagged $B \to X_s\gamma$ results described in Section 17.9.2.2.)

A different approach is DGE (Andersen and Gardi, 2007), in which the moments are predicted as functions of the energy threshold. As pointed out in Section 17.9.1.3, there are additional assumptions in this approach concerning the structure of the non-perturbative contributions. Measurements of moments including energy cuts thus allow for testing these assumptions.

#### 17.9.2.6 Branching fraction summary and extrapolation

All measured results for the inclusive $B \to X_s\gamma$ branching fraction are summarized in Table 17.9.3. For each of the four measurements detailed above, correlations between results at different thresholds are provided in the published papers or in EPAPS material cited therein. The measurements have $E_\gamma$ thresholds ranging from 1.7 to over 2.0 GeV, while theoretical predictions are usually made with a minimum $E_\gamma$ of 1.6 GeV. The approach adopted by HFAG and the PDG to produce a world average has been to extrapolate the experimental results down to the 1.6 GeV threshold from the lowest measured experimental threshold in each case. The extrapolation factors are taken from the initial HQE analysis of $B \to X_c\ell\overline{\nu}$ and $B \to X_s\gamma$ decays by (Buchmüller and Flächer, 2006) (rather than from the latest HFAG HQE fits). A current HFAG summary can be found at (HFAG, 2013). The extrapolation factors and extrapolated branching fractions are included in Table 17.9.3. The quoted world average of extrapolated values is

$$\mathcal{B}(B \to X_s\gamma) = (3.43 \pm 0.21 \pm 0.07) \times 10^{-4} \quad (E_\gamma > 1.6), \tag{17.9.28}$$

where the first error is the combined statistical and systematic error, and the second is from the model dependence of the extrapolations.

We note that because experimental systematic uncertainties decrease strongly with increasing $E_\gamma$ threshold (a consequence of the large $B\overline{B}$ backgrounds at lower $E_\gamma$ values), the uncertainty on an extrapolated branching fraction decreases if one starts with a measurement at a higher threshold. This seems contrary to the theoretical prejudice that the result should be most reliable if one begins at the lowest possible measured threshold, so there may be more unaccounted uncertainties in the extrapolation factors. (For example, the systematic uncertainty of the extrapolated CLEO result seems to be artificially low, a consequence of the high photon energy threshold used.) HFAG's chosen method minimizes the overall dependence on the extrapolation factors when the branching fracion is evaluated at 1.6 GeV.

Eventually, perhaps moments from the different threshold could be combined, taking into account their correlations, to provide a better extrapolation. This is related to spectrum-fitting goals described in Section 17.9.2.5.

#### 17.9.2.7 Constraints on new physics from $\mathcal{B}(B \to X_s\gamma)$

The SM prediction for the extrapolated branching fraction (Eq. 17.9.20) and the latest experimental average now have similar levels of uncertainty, and are consistent at the $1\sigma$ level. This finding implies very stringent constraints on NP models (Section 25.2). As examples we quote

- In the type-II two-Higgs doublet model (THDM) the bound on the charged Higgs mass is $M_{H^+} > 380\,\mathrm{GeV}/c^2$ at 95% C.L. See (Misiak et al., 2007) and the improved computation in (Hermann, Misiak, and Steinhauser, 2012). [Note a slightly older world-average branching fraction was used in the latter work.]
- In the minimal universal extra-dimension model (Haisch and Weiler, 2007), the bound on the inverse compactification radius is $1/R > 600\,\mathrm{GeV}$ at 95% C.L. [This limit should be recomputed to reflect improvements in the world-average branching fraction since 2007.]

In both cases, the bounds are much stronger than those previously derived from other measurements — but see Section 17.10.2.2 regarding an even stronger constraint on the THDM. Constraints on various supersymmetric models have been reviewed in (Altmannshofer, Buras, Gori, Paradisi, and Straub, 2010; Hurth, 2003). Bounds on the little Higgs model with $T$-parity have also been presented (Blanke, Buras, Duling, Recksiegel, and Tarantino, 2010). Finally, model-independent analyses in the effective field theory approach with the assumption of minimal flavor violation (D'Ambrosio, Giudice, Isidori, and Strumia, 2002; Hurth, Isidori, Kamenik, and Mescia, 2009) also show the strong constraining power of the $B \to X_s\gamma$ branching fraction.

### 17.9.3 Exclusive $b \to s\gamma$

As already discussed, $B \to K^*\gamma$ was the first $b \to s\gamma$ decay to be observed, and the $K^*(892)$ resonance is the only one clearly visible in the $X_s$ mass spectrum of $B \to X_s\gamma$ (Fig. 17.9.6). Contributions to $B \to X_s\gamma$ with $X_s$ heavier than $K^*$ have also been studied in detail. Some of the heavier $X_s$ final states are from resonances, but these are harder to disentangle, and non-resonant contributions also seem to be large, as the sum of the measured resonant contributions is far from saturating the total $B \to X_s\gamma$ decay width.

#### 17.9.3.1 $B \to K^*\gamma$

The $B \to K^*\gamma$ signal is reconstructed in the four $K^*$ final states, $K^+\pi^-$, $K_S^0\pi^0$, $K^+\pi^0$, $K_S^0\pi^+$, and their charge conjugate modes; $K_S^0\pi^0$ is a $CP$ eigenstate that can be used to study time-dependent $CP$ violation.

The signal is identified by the kinematic variables $m_{\mathrm{ES}}$ and $\Delta E$. There is a combinatorial background that is suppressed by event shape variables, and then has to be subtracted. There are also small "peaking" backgrounds from $B \to K\pi\pi\gamma$ and $B \to K\pi^0$ which are separated in $\Delta E$ from the signal (see Figure 17.9.7).

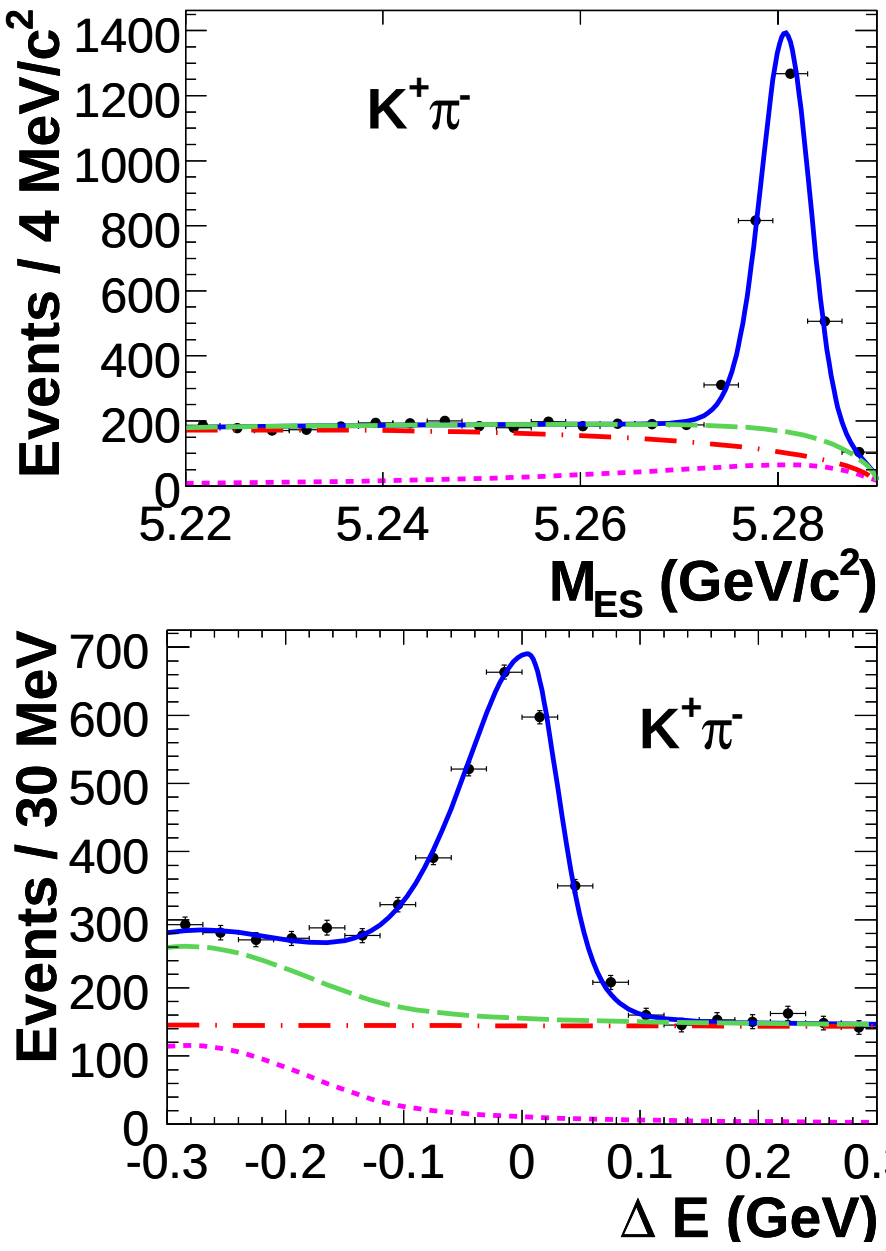

**Figure 17.9.7.** From (Aubert, 2009r). Signals from *BABAR* for $B \to K^*\gamma$ with $K^{*0} \to K^+\pi^-$ in $m_{ES}$ (top) and $\Delta E$ (bottom). Short-dashed (magenta) line is for peaking background, dot-dashed (red) for continuum, long-dashed (green) for the total background, solid (blue) for the total, overlaid on (black) data points.

As a two-body decay of a quite narrow resonance, the photon energy is almost monochromatic in the $B$ rest frame, and ranges between 2.40 and 2.74 GeV in the center-of-mass frame. The energy resolution for this high energy photon as well as the $\Delta E$ resolution is determined by the CsI calorimeter system, and shows a low energy tail due to the energy leakage from the crystals and energy loss in the material of the particle identification device in front of it. The $m_{ES}$ resolution is not much affected because the photon energy is rescaled by using a $\Delta E = 0$ constraint. The final state is a vector-vector state, but as the photon is massless, the $K^*$ is always transversely polarized, and the helicity angle $\theta_K$, the angle of the kaon with respect to the $B$ meson in the $K^*$ rest frame, follows a $1 - \cos^2\theta_K$ distribution.

The branching fraction is now measured to a precision of a few %. This is much more accurate than the theoretical predictions, which are limited by knowledge of the $B \to K^*$ form factor at $q^2 = 0$ (see Section 17.9.3.3). For rate asymmetries, the form factor uncertainties partly cancel, and the theoretical predictions are rather precise. Measurements of *CP* and isospin asymmetries are discussed later in Section 17.9.5.

The results for the branching fractions are summarized in Table 17.9.4. There is also a CLEO measurement (Coan et al., 2000) which is included in the world average. The *BABAR* measurement is based on 347 fb$^{-1}$ of data (Aubert, 2009r), and the Belle measurement on 78 fb$^{-1}$ (Nakao, 2004). Instead of assuming equal $B^+B^-$ and $B^0\overline{B}^0$ production, *BABAR* used the measured production rates $\mathcal{B}(\Upsilon(4S) \to B^0\overline{B}^0) = 0.484 \pm 0.006$ and $\mathcal{B}(\Upsilon(4S) \to B^+B^-) = 0.516 \pm 0.006$ (the 2008 PDG values (Amsler et al., 2008)).[86] With these values the quoted branching fractions are

$$\begin{aligned} \mathcal{B}(B^0 \to K^{*0}\gamma) &= (44.7 \pm 1.0 \pm 1.6) \times 10^{-6}, \\ \mathcal{B}(B^+ \to K^{*+}\gamma) &= (42.2 \pm 1.4 \pm 1.6) \times 10^{-6}. \end{aligned} \quad (17.9.29)$$

**Table 17.9.4.** Summary of measurements of exclusive $b \to s\gamma$ branching fractions (in $10^{-6}$). For the $K\pi\pi\gamma$ final states, Belle (*BABAR*) integrates over masses up to 2.0 (1.8) GeV/$c^2$. Several slightly asymmetric errors are symmetrized. The averages for $K^*(892)\gamma$ are recalculated (see text) also including CLEO results. Otherwise the averages are only from Belle and *BABAR* and are identical to PDG values. Averages assume that systematics are uncorrelated between Belle and *BABAR*.

| Mode | Belle | *BABAR* | Average |
|---|---|---|---|
| $K^*(892)^0\gamma$ | $40.1 \pm 2.1 \pm 1.7$ | $43.3 \pm 1.0 \pm 1.6$ | $42.4 \pm 1.5$ |
| $K^*(892)^+\gamma$ | $42.5 \pm 3.1 \pm 2.4$ | $43.6 \pm 1.4 \pm 1.6$ | $43.1 \pm 1.8$ |
| $K_2^*(1430)^0\gamma$ | $13.0 \pm 5.0 \pm 1.0$ | $12.2 \pm 2.5 \pm 1.0$ | $12.4 \pm 2.4$ |
| $K_2^*(1430)^+\gamma$ | | $14.5 \pm 4.0 \pm 1.5$ | $14.5 \pm 4.3$ |
| $K_1(1270)^+\gamma$ | $43.0 \pm 9.0 \pm 9.0$ | | $43.0 \pm 13.0$ |
| $K^+\pi^-\pi^+\gamma$ | $25.0 \pm 1.8 \pm 2.2$ | $29.5 \pm 1.3 \pm 1.9$ | $27.6 \pm 2.2$ |
| $K^0\pi^+\pi^-\gamma$ | $24.0 \pm 4.0 \pm 3.0$ | $18.5 \pm 2.1 \pm 1.2$ | $19.5 \pm 2.2$ |
| $K^+\pi^-\pi^0\gamma$ | | $40.7 \pm 2.2 \pm 3.1$ | $40.7 \pm 3.9$ |
| $K^0\pi^+\pi^0\gamma$ | | $45.6 \pm 4.2 \pm 3.0$ | $45.6 \pm 5.1$ |
| $K^+\phi\gamma$ | $2.5 \pm 0.3 \pm 0.2$ | $3.5 \pm 0.6 \pm 0.4$ | $2.8 \pm 0.3$ |
| $K^0\phi\gamma$ | $2.7 \pm 0.6 \pm 0.3$ | | $2.7 \pm 0.7$ |
| $K^+\eta\gamma$ | $8.4 \pm 1.5 \pm 1.1$ | $7.7 \pm 1.0 \pm 0.4$ | $7.9 \pm 0.9$ |
| $K^0\eta\gamma$ | $8.7 \pm 2.9 \pm 1.8$ | $7.1 \pm 2.1 \pm 0.4$ | $7.6 \pm 1.8$ |
| $K^+\eta'\gamma$ | $3.6 \pm 1.2 \pm 0.4$ | $1.9 \pm 1.4 \pm 0.1$ | $2.9 \pm 1.0$ |
| $\overline{\Lambda}p\gamma$ | $2.5 \pm 0.4 \pm 0.2$ | | $2.5 \pm 0.5$ |
| $B_s^0 \to \phi\gamma$ | $57 \pm 17 \pm 12$ | | $57 \pm 21$ |

In Table 17.9.4 the published *BABAR* results have been adjusted so that all the results are based on the same assumption of equal $B^+$ and $B^0$ production at the $\Upsilon(4S)$. This also leads to an adjustment of the world averages compared to those given by the Particle Data Group (Beringer et al., 2012).

### 17.9.3.2 Other exclusive $b \to s\gamma$ modes

Many other $b \to s\gamma$ decay modes have been searched for by Belle and *BABAR*. The branching fractions for the observed decays are summarized in Table 17.9.4. The higher $K\pi$ mass region is dominated by the $K_2^*(1430)$ resonance, seen first by CLEO (Coan et al., 2000) and confirmed by both Belle (Nishida, 2002) and *BABAR* (Aubert, 2004i) with relatively small datasets. Non-resonant $B \to K\pi\gamma$ contributions seem to be small and have not been observed so far. Note that an $S$-wave $K\pi$ system is forbidden by angular momentum conservation.

[86] The 2012 PDG value for $B^0\overline{B}^0$ is $0.487 \pm 0.006$. See Section 18.4.6.8 for a detailed discussion.

The decays $B \to K\pi\pi\gamma$ have been suggested as place to measure the photon polarization (Gronau, Grossman, Pirjol, and Ryd, 2002). In the SM the photon in $b \to s\gamma$ decays is left-handed up to small corrections of order $m_s/m_b$. In the $K\pi\pi$ hadronic system Belle (Yang, 2005) reports a signal for $B^+ \to K_1(1270)^+\gamma$ where $K_1(1270)^+ \to K^+\rho^0$. There is no branching fraction measurement for $B^0 \to K_1(1270)^0\gamma$, but this mode is clearly visible and is the dominant contribution in the time-dependent *CP* violation study of $B^0 \to K^0_S\rho^0\gamma$ by Belle (Li, 2008), as illustrated in Fig. 17.9.10 below. Both Belle (Nishida, 2002) and *BABAR* (Aubert, 2007r) measure inclusive rates for the $K\pi\pi$ final states without restricting the final state to a resonance. Belle separates out the $K^*\pi$ and $K\rho$ contributions, but neither experiment has performed a full Dalitz plot analysis of the $K\pi\pi$ system.

In addition, Belle has reported upper limits for radiative branching fractions to other resonant states, such as $K^*(1410)$, $K_1(1400)$, and $K^*_3(1780)$. These were searched for in $K\pi$, $K\pi\pi$ and $K\eta$ final states but no significant signals were found. A summary of these upper limits is given by the Particle Data Group (Beringer et al., 2012).

The $X_s$ final states $K\eta$, $K\eta'$ and $K\phi$ have all been measured for the first time at the $B$ Factories. They are primarily of interest because the neutral decay modes can be used to measure time-dependent *CP* violation (discussed later in Section 17.9.5). The decays $B \to \phi K\gamma$ have clear signals seen by Belle (Drutskoy, 2004; Sahoo, 2011) and *BABAR* (Aubert, 2007q). Observations of the decays $B \to K\eta\gamma$ are also reported by *BABAR* (Aubert, 2009e) and Belle (Nishida, 2005). Only Belle has reported evidence for $B^+ \to K^+\eta'\gamma$ (Wedd, 2010), and neither experiment has seen the neutral counterpart $B^0 \to K^0\eta'\gamma$ with upper limits on its branching fraction of $6.4 \times 10^{-6}$ from Belle (Wedd, 2010) and $6.6\times10^{-6}$ from *BABAR* (Aubert, 2006o).

The baryonic radiative decay $B^+ \to \overline{\Lambda}p\gamma$ has been measured by Belle (Wang, 2007b) with 449M $B\overline{B}$. One would expect similar branching fractions for other baryonic radiative decays. $B^+ \to \overline{\Sigma}^0 p\gamma$ was searched for as a by-product, where the signal would show up at a shifted $\Delta E$ compared to $\overline{\Lambda}p\gamma$. An upper limit of $5\times10^{-6}$ has been obtained by Belle (Wang, 2007b).

Finally, the first radiative $B^0_s$ decay $B^0_s \to \phi\gamma$ was observed by Belle using a data sample of $23\,\text{fb}^{-1}$ taken at the $\Upsilon(5S)$ resonance (Wicht, 2008). The analysis is almost identical to the study of $B \to K^*\gamma$ at the $\Upsilon(4S)$, except that there are three possible $m_{\text{ES}}$-$\Delta E$ peaks because of production through $B^0_s$-$\overline{B}^0_s$, $B^0_s$-$\overline{B}^{*0}_s$ and $B^{*0}_s$-$\overline{B}^{*0}_s$ at the $\Upsilon(5S)$. The measured branching fraction is similar to $B \to K^*\gamma$ as would be expected if exchange diagrams are small. This decay mode has now also been seen by LHCb (Aaij et al., 2012j), and is going to be useful in the search for time-dependent *CP* asymmetry due to new physics in the $B^0_s$ system. Although such a measurement looks similar to those in the $B^0$ system (see Section 17.9.6), it has a different implication due to the different CKM parameters and the non-negligible value of $\Delta\Gamma_s$ (Muheim, Xie, and Zwicky, 2008).

#### 17.9.3.3 Theoretical predictions for exclusive $b \to s\gamma$ modes

Up-to-date theoretical predictions for exclusive radiative decays are based on the method of QCD factorization. Large hadronic uncertainties are due to the non-perturbative input of the QCDF approach, namely light-cone wave functions and form factors, and our limited knowledge of power corrections. These uncertainties do not allow precise predictions of the branching fractions of exclusive modes. For example the branching fraction of $B \to K^*\gamma$ is directly proportional to the soft form factor at $q^2 = 0$, which can only be determined by QCD sum rules with an uncertainty of about 20%. The decay rate is given by (Ali and Parkhomenko, 2002)

$$\Gamma(B \to K^*\gamma) = \frac{G_F^2\alpha|V_{tb}V_{ts}^*|^2}{32\pi^4} m_b^2 M_B^3 |\xi_\perp(0)|^2 \left(1 - \frac{m_{K^*}^2}{M_B^2}\right)\left[C_7^{\text{eff}} + A\right]^2, \quad (17.9.30)$$

where $\xi_\perp(0)$ is the soft form factor at $q^2 = 0$ and $A$ is the contribution of NLO terms such as spectator interactions. Using the value of the form factor from a QCD sum rule calculation, the branching fraction becomes (Ali and Parkhomenko, 2002)

$$\mathcal{B}(B \to K^*\gamma) = (7.3 \pm 2.7) \times 10^{-5}. \quad (17.9.31)$$

This prediction is consistent with the experimental measurements, but, because the form factor input results in by far the largest error, Eq. (17.9.30) is often used to determine the form factor via the experimental data.

However, within ratios of branching fractions of exclusive modes such as *CP* asymmetries, parts of the uncertainties cancel out. This way, exclusive modes also provide valuable constraints, for example on the ratio of CKM elements $|V_{td}/V_{ts}|$ (see Section 17.2).

### 17.9.4 Exclusive and inclusive $b \to d\gamma$

The $b \to d\gamma$ transition from the third generation to the first, proceeds through a penguin loop diagram very similar to $b \to s\gamma$, except that the transition rate is suppressed by the ratio of the CKM matrix elements squared, $|V_{td}/V_{ts}|^2$ (see Eq. 17.9.21). It is thus one of the possible means to extract $|V_{td}/V_{ts}|$ (see Section 17.2). If $|V_{ts}|$ is identical to $|V_{cb}|$ from unitarity to the required precision. this in turn allows a determination of $|V_{td}|$, the length of the least-known side of the Unitarity Triangle.

There are some differences with respect to $b \to s\gamma$, because the suppression of the penguin transition amplitude increases the relative importance of other contributions, as indicated in the detailed theoretical discussion of Sections 17.9.1.1 and 17.9.1.2. The contribution to the penguin diagram from the $u$ quark is no longer small compared to the $t$ quark, since it now involves $V_{ud}$ rather than $V_{us}$. This is also true for the contributions from the four-quark operators containing $u$ quarks. Finally, the annihilation diagram for charged $B$ and the exchange diagram for neutral $B$ become significant, leading to isospin asymmetries and potentially large $CP$ asymmetries. All these corrections have to be taken into account in the extraction of $|V_{td}|$.

Experimentally, $b \to s\gamma$ processes are large backgrounds to similar $b \to d\gamma$ processes, so the former have to be suppressed using the particle identification capabilities of the detectors. The $b \to s\gamma$ decays provide a good control sample, and many systematic uncertainties cancel in the ratios of $b \to d\gamma$ to $b \to s\gamma$.

#### 17.9.4.1 Exclusive modes $B \to \rho\gamma$ and $B \to \omega\gamma$

As was the case in the early days of $b \to s\gamma$, it was the exclusive modes that were used to make the first observation of the $b \to d\gamma$ process at the $B$ Factories. The three decay modes $B^+ \to \rho^+\gamma$, $B^0 \to \rho^0\gamma$, and $B^0 \to \omega\gamma$ have all been seen, although the $B^0 \to \omega\gamma$ signals are only $2\sigma$ significant in each experiment. Based on naïve quark counting, $B^+ \to \rho^+\gamma$ should have twice the decay width (hence approximately twice the branching fraction) of the other two modes, whose branching fractions should be the same:

$$\mathcal{B}(B^+ \to \rho^+\gamma) = 2\frac{\tau_{B^+}}{\tau_{B^0}}\mathcal{B}(B^0 \to \rho^0\gamma) = 2\frac{\tau_{B^+}}{\tau_{B^0}}\mathcal{B}(B^0 \to \omega\gamma). \tag{17.9.32}$$

Due to the wide $\rho$ mass window, the $B \to \rho\gamma$ modes have large backgrounds from similar $B \to K^*\gamma$ modes. These are separated using $K/\pi$ particle identification and $\Delta E$. Both *BABAR* and Belle use a $K/\pi$ likelihood ratio selection, with similar pion efficiencies of 85%, but with significantly different probabilities of misidentifying a kaon as a pion, 1% and 8.5%, respectively (for relevant momenta), due to differences in the particle identification systems. The $K^*\gamma$ background is more pronounced in $B^0 \to \rho^0\gamma$ than in $B^+ \to \rho^+\gamma$ because Eq. (17.9.32) predicts about twice the rate for $\rho^+\gamma$ than $\rho^0\gamma$, while the $(K^+\pi^0)\gamma$ rate is only about 3/8 of that for $(K^+\pi^-)\gamma$. Also, there are two charged pions that could be misidentified kaons in $\rho^0\gamma$ while there is only one in $\rho^+\gamma$. To constrain the background from $K^{*0}\gamma$ Belle additionally uses the $K\pi$ mass in their fit to $\rho^0\gamma$. Both experiments also take into account another peaking background at lower $\Delta E$ from $B \to \rho\pi^0$ events, which survive a $\pi^0$ veto on the high energy photon. The Belle signal for $\rho^0\gamma$ is shown in Figure 17.9.8, in which it is seen that $\Delta E$ and $M(K\pi)$ are used to separate the signal from $K^{*0}\gamma$ and other peaking backgrounds, while $m_{\rm ES}$ is essential to fix the size of the continuum background.

The backgrounds from other $B$ decays are lower in the $B \to \omega\gamma$ mode because the $\omega$ resonance is narrower, $B \to \omega\pi^0$ is color-suppressed and has not been observed, and there is very little $K\pi\pi\gamma$ background at low $X_s$ mass. The branching fractions are summarized in Table 17.9.5, where the *BABAR* measurements (Aubert, 2008z) are from 465M $B\overline{B}$, and the Belle measurements (Taniguchi, 2008) are from 657M $B\overline{B}$.

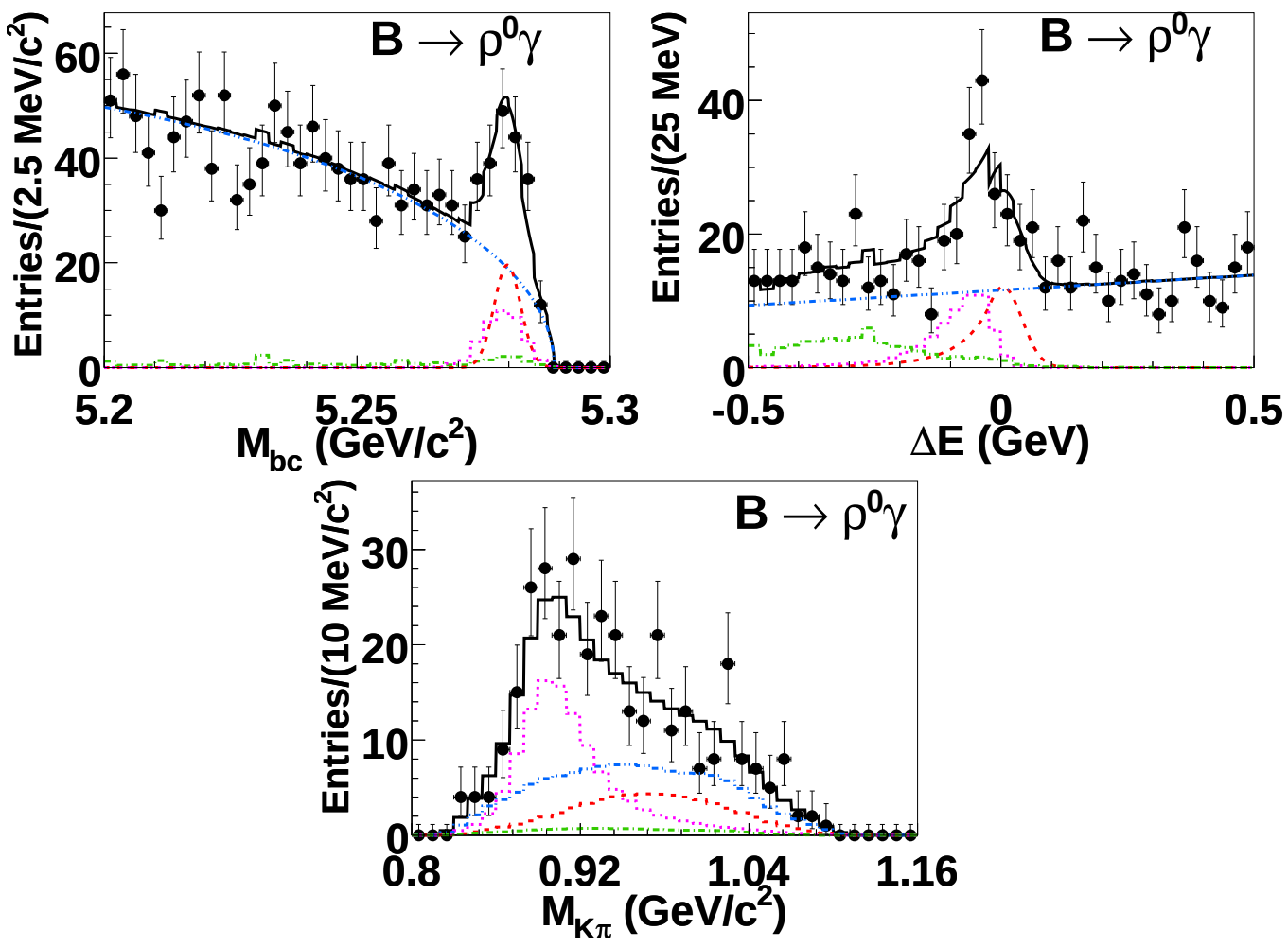

**Figure 17.9.8.** From (Taniguchi, 2008). Signal for $B^0 \to \rho^0\gamma$ from Belle in $m_{\rm ES}$ (=$M_{\rm bc}$), $\Delta E$ and $M(K\pi)$ (dashed, red) and backgrounds from $B^0 \to K^{*0}\gamma$ (dotted, magenta), other $B$ decays (dot-dashed, green) and continuum (dot-dot-dashed, cyan). Each spectrum is made with signal-region cuts on the other plotted quantities.

**Table 17.9.5.** Summary of measurements of exclusive $B \to X_d\gamma$ branching fractions (in $10^{-6}$). The combined results in the last two lines assume naïve quark counting, quoting values of $\mathcal{B}(B^+ \to \rho^+\gamma)$ from constrained fits to the two $\rho\gamma$ modes or all three modes. Errors are statistical and systematic, respectively.

| Mode | Belle | *BABAR* | Average |
|---|---|---|---|
| $\rho^+\gamma$ | $0.87 \pm 0.28 \pm 0.10$ | $1.20 \pm 0.40 \pm 0.20$ | $0.98 \pm 0.25$ |
| $\rho^0\gamma$ | $0.78 \pm 0.17 \pm 0.10$ | $0.97 \pm 0.23 \pm 0.06$ | $0.86 \pm 0.15$ |
| $\omega\gamma$ | $0.40 \pm 0.18 \pm 0.13$ | $0.50 \pm 0.25 \pm 0.09$ | $0.44 \pm 0.17$ |
| $\rho\gamma$ | $1.21 \pm 0.33 \pm 0.17$ | $1.73 \pm 0.33 \pm 0.17$ | $1.39 \pm 0.25$ |
| $\rho/\omega\gamma$ | $1.14 \pm 0.20 \pm 0.11$ | $1.63 \pm 0.29 \pm 0.16$ | $1.30 \pm 0.23$ |

Naïve quark counting breaks down due to the isospin asymmetries from the additional diagrams mentioned above, and due to differences in the hadronic form factors. It can be seen from the branching fractions in Table 17.9.5 that the naïve isospin factor of two between $\rho^+$ and $\rho^0$ is not very consistent with the experimental results (the isospin asymmetry will be discussed in Section 17.9.5 in more detail). Nevertheless each experiment has combined its measurements by using Eq. (17.9.32) to do constrained fits to $\mathcal{B}(B^+ \to \rho^+\gamma)$ using either just the two $\rho$ modes or all three modes. These combined results are quoted in Table 17.9.5 for completeness. The ratios of the measured branching fractions to those for corresponding $K^*\gamma$ final states are consistent with the SM predictions, with the value of $|V_{td}|/|V_{ts}|$ extracted from measurements of $\Delta m_d$ in $B^0 - \overline{B}^0$ mixing using precise lattice calculations. Nu-merical results and detailed discussion can be found in Section 17.2.2.

**Table 17.9.6.** Final states used to measure the ratio of branching fractions of inclusive $B \to X_d\gamma$ to inclusive $B \to X_s\gamma$ decays.

| $B \to X_d\gamma$ | $B \to X_s\gamma$ |
|---|---|
| $B^0 \to \pi^+\pi^-\gamma$ | $B^0 \to K^\pm\pi^\mp\gamma$ |
| $B^\pm \to \pi^\pm\pi^0\gamma$ | $B^\pm \to K^\pm\pi^0\gamma$ |
| $B^\pm \to \pi^\pm\pi^+\pi^-\gamma$ | $B^\pm \to K^\pm\pi^+\pi^-\gamma$ |
| $B^0 \to \pi^+\pi^-\pi^0\gamma$ | $B^0 \to K^\pm\pi^\mp\pi^0\gamma$ |
| $B^0 \to \pi^+\pi^-\pi^+\pi^-\gamma$ | $B^0 \to K^\pm\pi^\mp\pi^+\pi^-\gamma$ |
| $B^\pm \to \pi^\pm\pi^+\pi^-\pi^0\gamma$ | $B^\pm \to K^\pm\pi^+\pi^-\pi^0\gamma$ |
| $B^\pm \to \pi^\pm\eta\gamma$ | $B^\pm \to K^\pm\eta\gamma$ |

#### 17.9.4.2 Inclusive $b \to d\gamma$

An inclusive $B \to X_d\gamma$ measurement could provide a better sensitivity to extract $|V_{td}|/|V_{ts}|$, as it is free from the theoretical uncertainty of the form factors that appear in the exclusive modes $B \to \rho\gamma$ and $B \to K^*\gamma$. As discussed earlier, the fully inclusive approaches cannot separate $X_s$ and $X_d$, and hence cannot be used to measure $B \to X_d\gamma$. With the sum-of-exclusive modes approach it is possible to discriminate between $B \to X_d\gamma$ and $B \to X_s\gamma$ final states using particle identification requirements on the charged pions and kaons. A comparison of the ratio $|V_{td}|/|V_{ts}|$ to that extracted from $B$ mixing studies is a test of the SM, because different diagrams are involved.

The first inclusive $B \to X_d\gamma$ measurement was reported by *BABAR* in (Aubert, 2009q). This analysis has been updated to include the full dataset with 471M $B\overline{B}$ pairs (del Amo Sanchez, 2010q). The $X_d$ and $X_s$ are each reconstructed as the sum of seven final states (Table 17.9.6). The pion identification algorithm in this analysis has an efficiency of 95%, but also accepts kaons with a probability of 4%. Hence, because of the considerably larger $B \to X_s\gamma$ branching fraction, there is significant misidentification of $B \to X_s\gamma$ as $B \to X_d\gamma$. This $B \to X_s\gamma$ background is included as an additional component in the $B \to X_d\gamma$ fits to distributions of the kinematic variables $m_{\rm ES}$ and $\Delta E$; such misidentified events have a displaced and broader $\Delta E$ distribution.

Separate results are quoted for a low mass range $0.5 < m_{X_{d,s}} < 1.0\,{\rm GeV}/c^2$, which is dominated by the $\rho$, $\omega$ and $K^*$ resonances, and for a high mass range $1.0 < m_{X_{d,s}} < 2.0\,{\rm GeV}/c^2$. In the low-mass region, missing final states are readily accounted for by the known resonant decays. However, in the high-mass region there are significant uncertainties due to missing final states. The seven final states given in Table 17.9.6 account for only 43% of $b \to d\gamma$ and 36% of $b \to s\gamma$ in the high mass range. A further 37% of $b \to s\gamma$ is accounted for using isospin to relate neutral and charged kaon modes. The hadronization of a non-resonant $X_{d,s}$ is modeled using `Jetset`, and constrained using the observed distribution among the $X_s$ final states. Two alternative hadronization models are considered: replacing 50% of the inclusive hadronization by known resonances, and setting the $b \to d\gamma$ hadronization fractions to be the same as for the corresponding $b \to s\gamma$ states, instead of allowing for the differences predicted by `Jetset` . The resulting missing fractions in $B \to X_d\gamma$ vary by up to 40% compared to the nominal model. There is a partial cancellation of this uncertainty in the ratio of $B \to X_d\gamma$ to $B \to X_s\gamma$, but it remains the dominant systematic error.

*BABAR* quotes inclusive branching fractions for the low mass range $0.5 < m_{X_{d,s}} < 1.0\,{\rm GeV}/c^2$:

$$\begin{aligned}\mathcal{B}(B \to X_d\gamma) &= (1.3 \pm 0.3 \pm 0.1) \times 10^{-6},\\ \mathcal{B}(B \to X_s\gamma) &= (38 \pm 2 \pm 2) \times 10^{-6},\end{aligned} \qquad (17.9.33)$$

which are consistent with the measurements of exclusive $B \to \rho(\omega)\gamma$ and $B \to K^*\gamma$ decays. In the high mass range $1.0 < m_{X_{d,s}} < 2.0\,{\rm GeV}/c^2$ they measure:

$$\begin{aligned}\mathcal{B}(B \to X_d\gamma) &= (7.9 \pm 2.0 \pm 2.2) \times 10^{-6},\\ \mathcal{B}(B \to X_s\gamma) &= (192 \pm 9 \pm 29) \times 10^{-6}.\end{aligned} \qquad (17.9.34)$$

The results in these two mass ranges do not combine to give fully inclusive branching fractions, because they have not been extrapolated to include higher $X_{d,s}$ masses (lower $E_\gamma$). The ratio of the inclusive decays over the combined mass range $0.5 < m_{X_{d,s}} < 2.0\,{\rm GeV}/c^2$ is

$$\frac{\mathcal{B}(B \to X_d\gamma)}{\mathcal{B}(B \to X_s\gamma)} = 0.040 \pm 0.009 \pm 0.010. \qquad (17.9.35)$$

This ratio can be regarded as fully inclusive, because the photon energy spectra for $b \to d\gamma$ and $b \to s\gamma$ are expected to be almost identical at higher $X_{d,s}$ mass. The measured ratio of 4% is completely consistent with the SM predictions with $|V_{td}|/|V_{ts}|$ derived from $B^0 - \overline{B}^0$ mixing (see Section 17.2).

#### 17.9.4.3 Theoretical predictions for $b \to d\gamma$

The theoretical prediction for the branching fraction $\mathcal{B}(B \to X_d\gamma)$ for photon energies $E_\gamma > 1.6\,{\rm GeV}$ is (Hurth, Lunghi, and Porod, 2005)

$$\begin{aligned}&\mathcal{B}(B \to X_d\gamma) \times 10^5\\ &= \left(1.38\ {}^{+0.14}_{-0.21}\Big|_{\frac{m_c}{m_b}} \pm 0.15_{\rm CKM} \pm 0.09_{\rm param.} \pm 0.05_{\rm scale}\right),\end{aligned} \qquad (17.9.36)$$

and when normalized to $\mathcal{B}(B \to X_s\gamma)$ it is

$$\begin{aligned}&\frac{\mathcal{B}(B \to X_d\gamma)}{\mathcal{B}(B \to X_s\gamma)} \times 10^2\\ &= \left(3.82\ {}^{+0.11}_{-0.18}\Big|_{\frac{m_c}{m_b}} \pm 0.42_{\rm CKM} \pm 0.08_{\rm param.} \pm 0.15_{\rm scale}\right).\end{aligned} \qquad (17.9.37)$$

Scaling by the ratio of $|V_{td}/V_{ts}|^2$ between the current PDG value (Beringer et al., 2012) and that used by (Hurth, Lunghi, and Porod, 2005),[87] the ratio in Eq. (17.9.37) increases to $3.98\times10^{-2}$, and the CKM uncertainty contribution decreases to $0.22\times10^{-2}$. That reduced uncertainty is based in part on lattice QCD calculations of $\Delta m_d/\Delta m_s$.

These predictions are of NLL order. They are fully consistent with previous results presented in (Ali, Asatrian, and Greub, 1998). Due to the fact that a good part of the uncertainties cancel out in the ratio, the CKM uncertainties are an important component. In principle, measurements of $\mathcal{B}(B\to X_d\gamma)$ could be used to constrain the CKM parameters, and thus crosscheck their PDG ratio, but present experimental precision is far from adequate for that task. Such measurements are also of interest with respect to new physics, because the CKM suppression by the factor $|V_{td}/V_{ts}|^2$ in the SM may not hold in extended models. As discussed in the general theory section (Section 17.9.1) the *CP*-averaged decay rate of $B\to X_d\gamma$ is, in principle, as theoretically clean as the decay rate of $B\to X_s\gamma$, but the analogous NNLL QCD calculation is still missing.

### 17.9.5 Rate asymmetries in $b\to s(d)\gamma$

In many cases, signal events can be divided into two halves based on the flavor or charge, and the asymmetry in their decay rate provides information in addition to the branching fractions. In this section, we discuss the isospin asymmetry and direct *CP* asymmetry for $b\to s\gamma$ and $b\to d\gamma$ processes.

The isospin asymmetry in $B$ decays into a final state $X$ (asymmetry between $\overline{B}^0\to\overline{X}^0$ and $B^-\to X^-$) is usually defined as

$$\Delta_{0-}=\frac{\Gamma(\overline{B}^0\to\overline{X}^0)-\Gamma(B^-\to X^-)}{\Gamma(\overline{B}^0\to\overline{X}^0)+\Gamma(B^-\to X^-)}. \tag{17.9.38}$$

A related quantity often used for $B\to\rho\gamma$ is

$$\Delta_\rho=\frac{\Gamma(B^-\to\rho^-\gamma)}{2\Gamma(\overline{B}^0\to\rho^0\gamma)}-1. \tag{17.9.39}$$

The direct *CP* asymmetry in the time-integrated rates is defined as

$$A_{CP}=\frac{\Gamma(\overline{B}\to\overline{X})-\Gamma(B\to X)}{\Gamma(\overline{B}\to\overline{X})+\Gamma(B\to X)}, \tag{17.9.40}$$

We first summarize theoretical predictions in the SM for these asymmetries.

#### 17.9.5.1 Theoretical predictions for rate asymmetries

The theoretical prediction for the isospin breaking ratio $\Delta_{0-}(B\to K^*\gamma)$ based on the QCDF/SCET approach is given by (Beneke, Feldmann, and Seidel, 2005):

$$\Delta_{0-}(B\to K^*\gamma)=(0.28/T_1^{K^*}(0))\,(5.8\,^{+3.3}_{-2.9})\times10^{-2} \tag{17.9.41}$$

where $0.28/T_1^{K^*}(0)$ is a quantity of $\mathcal{O}(1)$, and the partial decay rates are *CP*-averaged. In the SM, spectator-dependent effects enter only at order $\Lambda/m_b$ while isospin-breaking in the form factors is expected to be a negligible effect, and the SM prediction is $\mathcal{O}(5\%)$ (Ali, Lunghi, and Parkhomenko, 2004; Ball, Jones, and Zwicky, 2007; Beneke, Feldmann, and Seidel, 2005; Bosch and Buchalla, 2005; Kagan and Neubert, 2002). The ratio is especially sensitive to new physics effects in the penguin sector, namely to the ratio of the two effective couplings $C_6/C_7$. The isospin ratio in the $\rho\gamma$ decay strongly depends on CKM parameters (again an average over *CP*-conjugate decay modes is made), and predicted for example (Beneke, Feldmann, and Seidel, 2005) to be:

$$\Delta_\rho=(-4.6\,^{+5.4}_{-4.2}|_{\rm CKM}\,^{+5.8}_{-5.6}|_{\rm had})\times10^{-2} \tag{17.9.42}$$

The hadronic error is mainly due the weak annihilation contribution to which a 50% error is assigned. Other predictions (Ali and Lunghi, 2002; Ball, Jones, and Zwicky, 2007; Lu, Matsumori, Sanda, and Yang, 2005) are similarly small.

For direct *CP* asymmetries, in exclusive decays, the uncertainties due to form factors cancel out to a large extent. But both the scale dependence and the dependence on the charm quark mass of the NLO predictions are rather large because the *CP* asymmetries arise at $\mathcal{O}(\alpha_S)$. While the direct *CP* asymmetry in $B\to K^*\gamma$ is doubly Cabibbo suppressed and expected to be very small, with QCDF and SCET one finds $-10\%$ predictions for the direct *CP* asymmetries in $B\to\rho\gamma$ (Ali, Lunghi, and Parkhomenko, 2004; Beneke, Feldmann, and Seidel, 2005; Bosch and Buchalla, 2002b). Since the weak annihilation contribution does not contribute significantly here, the neutral and charged mode are of similar size (Beneke, Feldmann, and Seidel, 2005):

$$A_{CP}(\overline{B}^0\to\rho^0\gamma)=\Big(-10.4\,^{+1.6}_{-2.4}|_{\rm CKM}\,^{+3.0}_{-3.6}|_{\rm had}\Big)\times10^{-2}, \tag{17.9.43}$$

$$A_{CP}(B^-\to\rho^-\gamma)=\Big(-10.7\,^{+1.5}_{-2.0}|_{\rm CKM}\,^{+2.6}_{-3.7}|_{\rm had}\Big)\times10^{-2}. \tag{17.9.44}$$

Finally, one should emphasize again that all predictions of exclusive observables with QCDF and SCET may receive further uncertainties due to the unknown power corrections. This might be specifically important in the case of *CP* asymmetries.

The theoretical situation is significantly better for direct *CP* asymmetries in the inclusive modes, as first noted in (Kagan and Neubert, 1998). (There are apparently no theoretical predictions for isospin asymmetries in the inclusive radiative processes ) The NP sensitivities of direct *CP* asymmetries in these modes have been analyzed in Kagan and Neubert (1998), Hurth, Lunghi, and Porod (2005), and Benzke, Lee, Neubert, and Paz (2011).

[87] The published version of (Hurth, Lunghi, and Porod, 2005) does not present the numerical CKM values used, but they can be found in the arXiv version 2.

A SM computation (Hurth, Lunghi, and Porod, 2005) yielded (for $E_\gamma > 1.6\,\mathrm{GeV}$)

$$A_{CP}(B \to X_s\gamma) = \left(0.44\,{}^{+0.15}_{-0.10}\big|_{\frac{m_c}{m_b}} \pm 0.03_{\mathrm{CKM}}\,{}^{+0.19}_{-0.09}\big|_{\mathrm{scale}}\right) \times 10^{-2}, \quad (17.9.45)$$

$$A_{CP}(B \to X_d\gamma) = \left(-10.2\,{}^{+2.4}_{-3.7}\big|_{\frac{m_c}{m_b}} \pm 1.0_{\mathrm{CKM}}\,{}^{+2.1}_{-4.4}\big|_{\mathrm{scale}}\right) \times 10^{-2}. \quad (17.9.46)$$

Note the very small uncertainty on $A_{CP}(B \to X_s\gamma)$.

However, recent theoretical work (Benzke, Lee, Neubert, and Paz, 2011) has shown that previously unaccounted long-distance (resolved-photon) effects shift the predicted central values of $A_{CP}$in the SM to 0.011 for $B \to X_s\gamma$ and $-0.24$ for $B \to X_d\gamma$. The new contributions greatly increase the uncertainties, and the resulting $A_{CP}$predictions have what the authors term "irreducible" ranges

$$-0.006 < A_{CP}(B \to X_s\gamma) < +0.028\,, \quad (17.9.47)$$

$$-0.62 < A_{CP}(B \to X_d\gamma) < +0.14\,. \quad (17.9.48)$$

These ranges were computed for an $E_\gamma$ threshold of 1.9 GeV, but might be larger for higher thresholds. The implication is that $A_{CP}(B \to X_s\gamma)$ is not as sensitive a probe for new physics as had once been thought. (The authors note that the long-distance effects are essentially isospin-independent, so that the difference between $A_{CP}$for $B^0$ and $B^+$ decays has much better sensitivity.)

The two inclusive *CP* asymmetries are connected by the relative CKM factor $\lambda^2\,((1-\rho)^2+\eta^2)$. The small SM prediction for the *CP* asymmetry in the decay $B \to X_s\gamma$ is a result of three factors: (a) a strong phase can only appear through QCD radiative corrections, so the *CP* asymmetry is $\mathcal{O}(\alpha_s(m_b))$; (b) there is a CKM suppression of order $\lambda^2$; (c) there is a GIM suppression, leading to a factor $(m_c/m_b)^2$, which reflects the fact that in the limit $m_c = m_u$ any *CP* asymmetry in the SM would vanish.

Using CKM unitarity one can derive the following U-spin relation between the un-normalized *CP* asymmetries (Soares, 1991):

$$\Delta\Gamma(B \to X_s\gamma) + \Delta\Gamma(B \to X_d\gamma) = 0, \quad (17.9.49)$$

where $\Delta\Gamma(B \to X_q\gamma) = \Gamma(\overline{B} \to X_q\gamma) - \Gamma(B \to X_{\overline{q}}\gamma)$ and $q = s, d$. U-spin breaking effects can be estimated within the heavy mass expansion (even beyond the partonic level) and one finds that the total $A_{CP}(B \to X_{s+d}\gamma)$ is zero to order $10^{-6}$ (Hurth and Mannel, 2001a,b). This precision is preserved even in the presence of the long-distance effects (Benzke, Lee, Neubert, and Paz, 2011). Since the prediction is based on CKM unitarity, this null test is a clear probe for new *CP* phases beyond the CKM phase.

#### 17.9.5.2 Measurements of Isospin asymmetries

In order to measure isospin asymmetries, the decay widths $\Gamma$ in Eq. (17.9.38) are evaluated as the ratios of branching fractions and $B$ lifetimes, $\mathcal{B}/\tau$, where $\tau_{B^0} = (1.519 \pm 0.007)\,\mathrm{ps}$ and $\tau_{B^-} = (1.641 \pm 0.008)\,\mathrm{ps}$, and the ratio $\tau_{B^-}/\tau_{B^0} = 1.071 \pm 0.009$ from Particle Data Group (Beringer et al., 2012).

There is an inclusive measurement of $\Delta_{0-}(B \to X_s\gamma)$, but only from the sum of exclusive modes (Aubert, 2005x). The recoil-$B$ tag method (Aubert, 2008q) has measured $\Delta_{0-}(B \to X_{s+d}\gamma)$, albeit with limited precision. Other inclusive methods do not distinguish between $\overline{B}^0$ and $B^-$. Both results use a minimum photon $E_\gamma$ of 2.2 GeV, and are given in Table 17.9.7.

The most precise measurement of an isospin asymmetry comes from $B \to K^*\gamma$. Indeed it is so precise that it is important to take into account a possible production asymmetry between $B^+B^-$ and $B^0\overline{B}^0$ at the $\Upsilon(4S)$. *BABAR* chooses to use the measured production fractions $B^+B^- = 0.516 \pm 0.006$ and $B^0\overline{B}^0 = 0.484 \pm 0.006$, and obtains an isospin asymmetry:

$$\Delta_{0-}(B \to K^*\gamma) = 0.066 \pm 0.021 \pm 0.022. \quad (17.9.50)$$

Belle has a measurement based on a smaller data sample, and assumes equal production fractions. For consistency in averaging, and with other decay modes, the *BABAR* result is adjusted to assume equal production fractions in Table 17.9.7. This gives an average isospin asymmetry consistent with zero and with the inclusive isospin asymmetry. If instead the Belle result is changed to use the measured production fractions, the world average becomes

$$\Delta_{0-}(B \to K^*\gamma) = 0.058 \pm 0.025. \quad (17.9.51)$$

The effect of the measured production fractions is a change from no isospin asymmetry to about $2\sigma$ positive asymmetry. The same shift towards a positive asymmetry is expected for the inclusive result, although here it is not significant due to the larger error. A small positive isospin asymmetry is in accordance with SM predictions (Kagan and Neubert, 2002; Keum, Matsumori, and Sanda, 2005) — see Eq. (17.9.41).

The isospin asymmetry for $B \to \rho\gamma$ reported by *BABAR* and Belle is reported in terms of $\Delta_\rho$ (Eq. 17.9.39). To be more consistent with the definition in Eq. (17.9.38), we redefine the isospin asymmetry for $B \to \rho\gamma$ as

$$\Delta_{0-}(B \to \rho\gamma) = \frac{2\Gamma(\overline{B}^0 \to \rho^0\gamma) - \Gamma(B^- \to \rho^-\gamma)}{2\Gamma(\overline{B}^0 \to \rho^0\gamma) + \Gamma(B^- \to \rho^-\gamma)}. \quad (17.9.52)$$

It is straightforward to convert between these different forms. Here in the text the published numbers are given in their original form, since this is what is used by HFAG and many theory papers. In Table 17.9.7 both results are converted to the form of $\Delta_{0-}$.

The published Belle result (Taniguchi, 2008) is $\Delta_\rho = -0.48 \pm 0.20 \pm 0.09$, which converts to $\Delta_{0-}(B \to \rho\gamma) = +0.32 \pm 0.12 \pm 0.05$. The published *BABAR* result (Aubert, 2008z) is $\Delta_\rho = -0.43 \pm 0.24 \pm 0.10$, which converts to $\Delta_{0-}(B \to \rho\gamma) = +0.27 \pm 0.13 \pm 0.05$. The world average provided by HFAG is

$$\Delta_\rho = -0.46 \pm 0.17 \quad (17.9.53)$$

**Table 17.9.7.** Summary of measurements of isospin asymmetries $\Delta_{0-}$ in $b \to s(d)\gamma$ decays (in $10^{-2}$). The *BABAR* $X_s\gamma$ result assumed an older value for the lifetime ratio, $1.086\pm0.017$. The $\rho\gamma$ asymmetry is defined in a consistent fashion with $s\gamma$, which is different from the definition $\Delta_\rho$ found in the literature (see text for discussion). This table assumes equal production of $B^+B^-$ and $B^0\overline{B}^0$ at the $\Upsilon(4S)$ (see Eq. (17.9.50) for the published $\bar{b}$ result for $K^*\gamma$, which uses the measured production fractions). The effect of these on the isospin asymmetry is discussed in the text.

| Mode | *BABAR* | Belle | Average |
|---|---|---|---|
| $X_s\gamma$ | $-0.6 \pm 5.8 \pm 2.6$ | | $-0.6 \pm 6.3$ |
| $X_{s+d}\gamma$ | $-6 \pm 15 \pm 7$ | | $-6 \pm 17$ |
| $K^*\gamma$ | $3.4 \pm 2.1 \pm 2.2$ | $-1.5 \pm 4.4 \pm 1.2$ | $2 \pm 3$ |
| $\rho\gamma$ | $27 \pm 13 \pm 5$ | $32 \pm 12 \pm 5$ | $30 \pm 10$ |

**Table 17.9.8.** Summary of measurements of direct *CP* asymmetries $A_{CP}$in $b \to s(d)\gamma$ decays (in $10^{-2}$). The two values listed for *BABAR* $B \to X_{s+d}\gamma$ are from the lepton-tag and reconstructed-$B$-tag methods, respectively (kept separate because of different $E_\gamma$ thresholds). Uncertainties are statistical and systematic, respectively, combined for the *BABAR* plus Belle average in the last column.

| Mode | *BABAR* | Belle | Average |
|---|---|---|---|
| $X_s\gamma$ | $-1.1 \pm 3.0 \pm 1.4$ | $0.2 \pm 5.0 \pm 3.0$ | $-1 \pm 3$ |
| $X_{s+d}\gamma$ | $+5.7 \pm 6.0 \pm 1.8$ | | $+6 \pm 6$ |
| | $+10 \pm 18 \pm 5$ | | $+10 \pm 19$ |
| $K^*\gamma$ | $-0.3 \pm 1.7 \pm 0.7$ | $1.2 \pm 4.4 \pm 2.6$ | $0 \pm 2$ |
| $K^+\eta\gamma$ | $-9.0\,^{+10.2}_{-9.8} \pm 1.4$ | $-16 \pm 9 \pm 6$ | $-12 \pm 7$ |
| $K^+\phi\gamma$ | $26 \pm 14 \pm 5$ | $-3 \pm 11 \pm 8$ | $-13 \pm 10$ |
| $K_2^*(1430)^0\gamma$ | $-8 \pm 15 \pm 1$ | | $-8 \pm 15$ |
| $\rho^+\gamma$ | | $-11 \pm 32 \pm 9$ | $-11 \pm 33$ |

which converts to:

$$\Delta_{0-}(B \to \rho\gamma) = +0.30 \pm 0.10. \qquad (17.9.54)$$

In either form, there is $3\sigma$ evidence for an isospin asymmetry in $B \to \rho\gamma$, which is much larger than the SM expectation (Eq. 17.9.42); and therefore a larger dataset is needed to clarify the situation.

### 17.9.5.3 Measurements of Direct *CP* asymmetries

The direct *CP* asymmetry of Eq. (17.9.40) has been measured in inclusive $b \to s\gamma$ decays using the sum-of-exclusive-states method. The *BABAR* result (Aubert, 2008a) is based on 383M $B\overline{B}$, and the Belle result (Nishida, 2004) on 152M $B\overline{B}$. The final states are divided into $\bar{b}$ modes ($B^0$ and $B^+$) and $b$ modes ($\overline{B}^0$ and $B^-$) using the kaon charge or total charge. *CP* eigenstates with a $K_S^0$ and an even number of charged pions are excluded from the analysis. Dilution of the asymmetry due to events misreconstructed in the wrong final state is very small and is corrected for. *BABAR* measures the *CP* asymmetry for $X_s$ mass below $2.8\,\mathrm{GeV}/c^2$ (but note that there are relatively few selected events above $\sim 2\,\mathrm{GeV}/c^2$):

$$A_{CP}(B \to X_s\gamma) = -0.011 \pm 0.030 \pm 0.014 \qquad (17.9.55)$$

while Belle measures for $X_s$ mass below $2.1\,\mathrm{GeV}/c^2$:

$$A_{CP}(B \to X_s\gamma) = 0.002 \pm 0.050 \pm 0.030. \qquad (17.9.56)$$

Both measurements are consistent with zero, and are statistically limited. In the SM the direct *CP* asymmetry in $B \to X_s\gamma$ is expected to be small, as explained in Section 17.9.5.1.

In the lepton-tagged fully-inclusive analysis, the lepton charge can be used to flavor-tag the signal-$B$ decay, although there is some dilution due primarily to $B^0 - \overline{B}^0$ mixing. Because this measurement does not separate $B \to X_s\gamma$ from $B \to X_d\gamma$ decays, the resulting *CP* asymmetry is for a sum of $B \to X_{s+d}\gamma$ events. (In the corresponding branching fraction analysis of Section 17.9.2.2, the small $B \to X_d\gamma$ contribution could be removed using a ratio of CKM matrix elements, but this procedure is not applicable to the *CP* asymmetries.) In the SM this asymmetry is predicted to be zero, to a precision of order $10^{-6}$, with the larger $X_d\gamma$ asymmetry (Eq. 17.9.46) exactly compensated by the smaller $X_d\gamma$ branching fraction. *BABAR* has measured this combined asymmetry based on a 383M $B\overline{B}$ sample (Lees, 2012j,o). The photon energy threshold of 2.1 GeV suppresses $B\overline{B}$ background, while being adequately inclusive to preserve the SM prediction. The flavor-mistag fraction (primarily from mixing, with contributions from cascade decays and misidentified leptons) is $0.133\pm0.006$. The measured asymmetry is

$$A_{CP}(B \to X_{s+d}\gamma) = +0.057 \pm 0.060 \pm 0.018\,, \quad (17.9.57)$$

where the errors are statistical and systematic.

Another measurement of the combined $A_{CP}(B \to X_{s+d}\gamma)$ comes from *BABAR*'s use of reconstructed-$B$ tags (Section 17.9.2.3). In this case (Aubert, 2008q) $A_{CP}$is measured by splitting the reconstructed tags into known $B$ and $\overline{B}$ states. The flavor-mistag fraction, due to $B^0 - \overline{B}^0$ mixing, is 0.188 times the fraction of $B^0$ events in the sample, and the measured asymmetry for $E_\gamma > 2.2\,\mathrm{GeV}$ is

$$A_{CP}(B \to X_{s+d}\gamma) = +0.10 \pm 0.18 \pm 0.05\,. \qquad (17.9.58)$$

Using the flavor-specific $B \to K^*\gamma$ final states, precise measurements of direct *CP* violation have been made by both *BABAR* and Belle. The results are given in Table 17.9.8. The world average is consistent with zero with an error of only 2%:

$$A_{CP}(B \to K^*\gamma) = -0.003 \pm 0.017\,. \qquad (17.9.59)$$

There are also measurements of direct *CP* asymmetries in $B^0 \to K_2^*(1430)^0\gamma$ by *BABAR* (Aubert, 2004i), in $B^+ \to \rho^+\gamma$ by Belle (Taniguchi, 2008), and in $B^+ \to K^+\eta\gamma$ (Nishida (2005) and Aubert (2009e)) and $B^+ \to K^+\phi\gamma$ (Sahoo (2011) and Aubert (2007q)) by Belle and *BABAR*, respectively. Direct *CP* asymmetry is also measured for neutral $B$ decay modes with no self-flavor-tagging in time dependent *CP* asymmetry measurements as discussed in Section 17.9.6. All the measurements of direct *CP* asymmetries in $b \to s(d)\gamma$ are consistent with zero.

### 17.9.6 Time-dependent *CP* asymmetries

Exclusive decay modes that have a common final state between $B^0$ and $\overline{B}^0$ are candidates to measure the time-dependent *CP* asymmetry due to interference between mixing and decay amplitudes. The distribution of the proper time difference $\Delta t$ between the decay of a tagging $B^0$ or $\overline{B}^0$ and the decay of the signal $B$ is given by Eq. (10.2.2). The coefficients of the sine and cosine terms are represented as parameters $\mathcal{S}$ and $\mathcal{C}$, respectively. The time-dependent asymmetry is then defined by Eq. (10.2.7), which results in

$$\mathcal{A}_{CP}(B \to f;\ \Delta t) = \mathcal{S}\sin(\Delta m_d \Delta t) - \mathcal{C}\cos(\Delta m_d \Delta t)\,. \tag{17.9.60}$$

For a measurement, $\Delta t$-resolution must be convoluted with the underlying $\Delta t$ distributions, and incorrect $B$-flavor tagging results in a dilution factor multiplying the $\mathcal{S}$ and $\mathcal{C}$ terms. The full effects of this flavor mistagging on the time distributions are given in Eq. (10.3.3). Note that $-\mathcal{C}$ represents the size of the direct *CP* asymmetry discussed above.[88] In contrast to most rare hadronic $B$ decays, the uniquely time-dependent $\mathcal{S}$ term is strongly suppressed in radiative decays because of the left-handedness of the final state photon, as discussed below in Section 17.9.6.1. Studies of these asymmetries are thus considered to be one of the most promising methods to search for non-SM right-handed currents.

#### 17.9.6.1 Theoretical predictions for time-dependent *CP* asymmetries

In the hadronic decay mode $B \to J/\psi K^0_S$, a large value of $\mathcal{S}$ has been measured, its size a consequence of the value of the angle $\phi_1 \equiv \beta = -\arg(V_{td}V^*_{tb}/V_{ud}V^*_{ub})$ of the Unitarity Triangle. Similar large *CP* asymmetries are expected for hadronic penguin decays. However, this asymmetry is suppressed in radiative penguin decays because the photon helicity is opposite between $B^0$ and $\overline{B}^0$ decays as a result of the left-handed current of SM weak decays. In the limit of massless quarks there is no *CP* violation due to interference between mixing and decay amplitudes. This implies a suppression factor of $2m_s/m_b$ in the leading contribution to $\mathcal{S}$ induced by the electromagnetic dipole operator $\mathcal{O}_7$:

$$\mathcal{S}^{\rm SM} = -\sin 2\phi_1 \frac{m_s}{m_b}\left[2 + \mathcal{O}(\alpha_S)\right] + \mathcal{S}^{{\rm SM},s\gamma g} \tag{17.9.61}$$

As noted in Grinstein, Grossman, Ligeti, and Pirjol (2005) and Grinstein and Pirjol (2006), there are additional contributions, $\mathcal{S}^{{\rm SM},s\gamma g}$ induced by the process $b \to s\gamma g$ via other operators than $\mathcal{O}_7$. One example is a contribution of the operator $\mathcal{O}_2 \sim (\bar{b}s)(\bar{c}c)$ where the charm quark forms a loop from which a (right-handed) photon and a gluon is emitted. These corrections are power-suppressed but not helicity-suppressed. A conservative estimate of this contribution in $B \to K^*\gamma$ due to a non-local SCET operator series leads to $|\mathcal{S}^{{\rm SM},s\gamma g}| \approx 0.06$ (Grinstein, Grossman, Ligeti, and Pirjol, 2005; Grinstein and Pirjol, 2006), while within a QCD sum rule calculation the contribution due to soft gluon emission is estimated to be $\mathcal{S}^{{\rm SM},s\gamma g} = -0.005 \pm 0.01$ (Ball, Jones, and Zwicky, 2007; Ball and Zwicky, 2006a) which leads to $\mathcal{S}^{\rm SM} = -0.022 \pm 0.015^{+0}_{-0.01}$ for the process $B \to K^*\gamma$. The QCD sum rule estimates of power corrections, due to long-distance contributions with photon and soft-gluon emission from quark loops (Ball, Jones, and Zwicky, 2007), lead to analogous results for the other radiative decay modes such as $B \to \rho\gamma$ (Ball, Jones, and Zwicky, 2007). If a large value of $\mathcal{S}$ beyond the SM prediction is observed, this will be a clear signal of a new right-handed current beyond the SM.

It was pointed out by (Atwood, Gershon, Hazumi, and Soni, 2007) that, due to the left-handed photon coupling, *CP* asymmetries in the SM are equally small for any decay of the form $B \to P^0Q^0\gamma$, where $P^0$ is a neutral pseudoscalar and $Q^0$ is another neutral pseudoscalar or a neutral vector meson. In the case of the pseudoscalar-vector-photon final states there is also information in the angular distribution of the final state.

**Table 17.9.9.** Summary of measurements of time-dependent *CP* asymmetries in $b \to s(d)\gamma$ decays. $\mathcal{S}$ refers to coefficients of $\sin(\Delta m_d \Delta t)$, and $\mathcal{C}$ to coefficients of $\cos(\Delta m_d \Delta t)$. The entries for $B \to K^0_S\pi^0\gamma$ are for the high $K^0_S\pi^0$ mass region excluding the $K^{*0}$ resonance.

| | BaBar | Belle | Average |
|---|---|---|---|
| $\mathcal{S}(K^{*0}\gamma)$ | $-0.03 \pm 0.29 \pm 0.03$ | $-0.32 \pm 0.35 \pm 0.05$ | $-0.17 \pm 0.20$ |
| $\mathcal{C}(K^{*0}\gamma)$ | $-0.14 \pm 0.16 \pm 0.03$ | $0.20 \pm 0.24 \pm 0.05$ | $0.00 \pm 0.13$ |
| $\mathcal{S}(K^0_S\pi^0\gamma)$ | $-0.78 \pm 0.59 \pm 0.09$ | $-0.10 \pm 0.31 \pm 0.07$ | $-0.40 \pm 0.25$ |
| $\mathcal{C}(K^0_S\pi^0\gamma)$ | $-0.36 \pm 0.33 \pm 0.04$ | $0.20 \pm 0.20 \pm 0.06$ | $0.00 \pm 0.16$ |
| $\mathcal{S}(K^0_S\eta\gamma)$ | $-0.18 \pm 0.48 \pm 0.12$ | | $-0.18 \pm 0.50$ |
| $\mathcal{C}(K^0_S\eta\gamma)$ | $-0.32 \pm 0.40 \pm 0.07$ | | $-0.32 \pm 0.41$ |
| $\mathcal{S}(K^0_S\phi\gamma)$ | | $0.74 \pm 0.90 \pm 0.20$ | $0.74 \pm 0.91$ |
| $\mathcal{C}(K^0_S\phi\gamma)$ | | $-0.35 \pm 0.58 \pm 0.20$ | $-0.35 \pm 0.61$ |
| $\mathcal{S}(K^0_S\rho^0\gamma)$ | | $0.11 \pm 0.33 \pm 0.07$ | $0.11 \pm 0.35$ |
| $\mathcal{C}(K^0_S\pi^+\pi^-\gamma)^\dagger$ | | $0.05 \pm 0.18 \pm 0.06$ | $0.05 \pm 0.20$ |
| $\mathcal{S}(\rho^0\gamma)$ | | $-0.83 \pm 0.65 \pm 0.18$ | $-0.83 \pm 0.68$ |
| $\mathcal{C}(\rho^0\gamma)$ | | $0.44 \pm 0.49 \pm 0.14$ | $0.44 \pm 0.53$ |

$^\dagger$For $m_{K\pi\pi} < 1.8\,{\rm GeV}/c^2$ and $m_{\pi\pi} \in [0.6, 0.9]\,{\rm GeV}/c^2$.

#### 17.9.6.2 Measurements of time-dependent *CP* asymmetries

The decay mode $B^0 \to K^{*0}\gamma$ with $K^{*0} \to K^0_S\pi^0$ has the largest branching fraction and hence has the largest potential for a time-dependent *CP* asymmetry search. However, to measure the time-dependent *CP* asymmetry for $B^0 \to K^0_S\pi^0\gamma$, one has to measure the $B$ meson decay vertex by extrapolating the displaced $K^0_S \to \pi^+\pi^-$ vertex. It is only possible to do this accurately when the $K^0_S$ decays inside the vertex detector volume. This requirement reduces the acceptance by a factor of 0.68 for *BABAR* and 0.55 for Belle. Note that the $K^0_S$ momentum in this decay is lower than in the charmless hadronic decay $B^0 \to K^0_S\pi^0$, so the acceptance is somewhat larger. A control sample of $B \to J/\psi K^0_S$ events is used to demonstrate that it is feasible to make a time-dependent measurement using the vertex reconstruction from the $K^0_S$ extrapolation alone (Ushiroda, 2005).

[88] The symbol $\mathcal{A} = -\mathcal{C}$ is also often used.

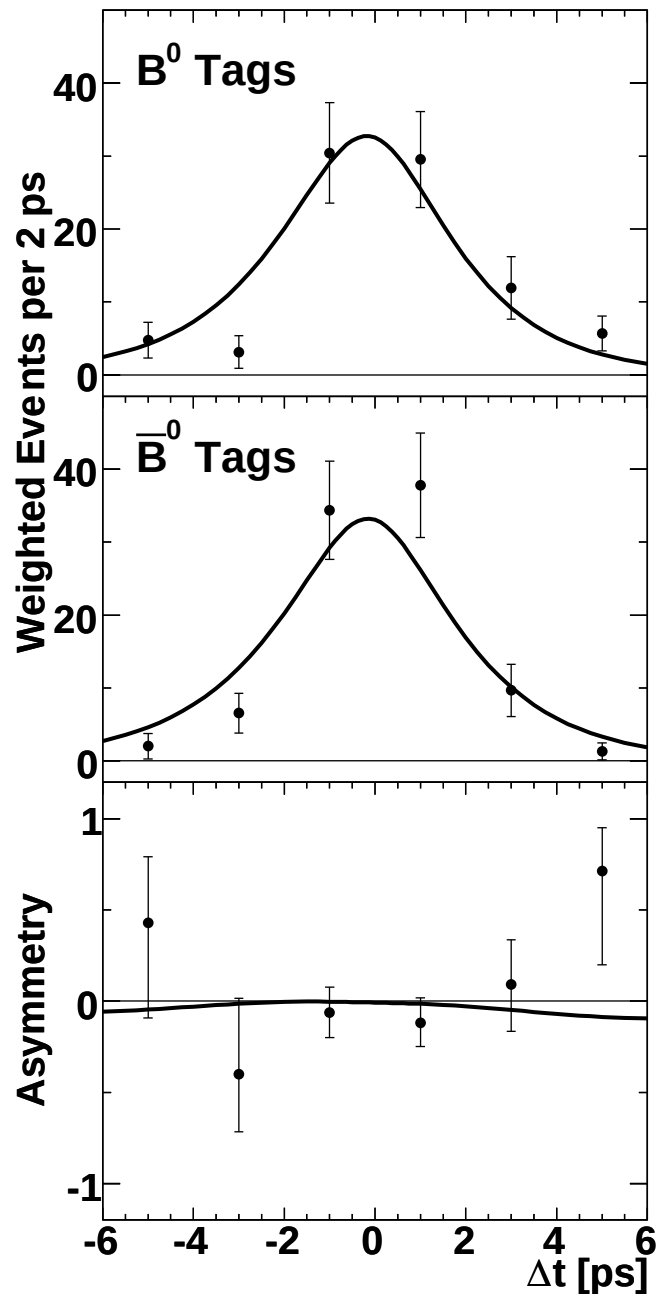

**Figure 17.9.9.** Results of *BABAR* time-dependent *CP* asymmetry fit to $B \to K^*\gamma$ events with $K^* \to K^0_S\pi^0$, from (Aubert, 2008x). The underlying $\Delta t$ distributions (before including resolution effects) are given by Eq. (10.3.3). Shown are background-subtracted distributions for $B^0$ tags (top) $\overline{B}^0$ tags (middle) and their asymmetry (bottom).

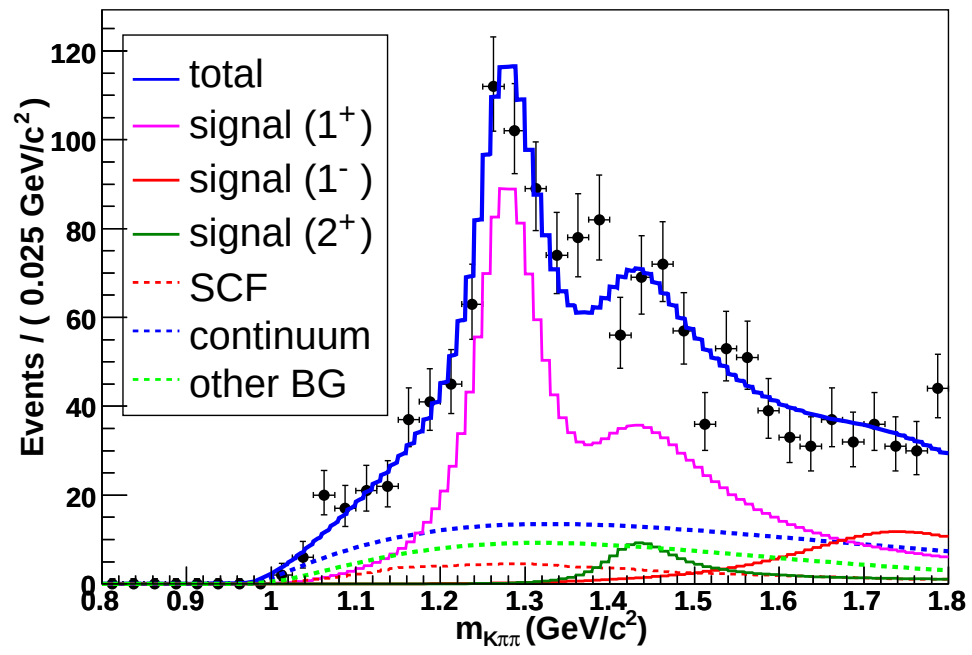

**Figure 17.9.10.** $K^+\pi^+\pi^-$ mass distribution for a $K^*$ enriched $B^+ \to K^+\pi^+\pi^-\gamma$ sample for Belle's $B^0 \to K^0_S\rho^0\gamma$ analysis. Signal components ($1^+$, $1^-$ and $2^+$) are based on known resonances ($K_1(1270)$ and $K_1(1400)$, $K^*(1680)$ and $K^*_2(1430)$), of which the $K^*_2(1430)$ component is fixed. Dashed curves are for self-crossfeed (SCF), continuum and other backgrounds.

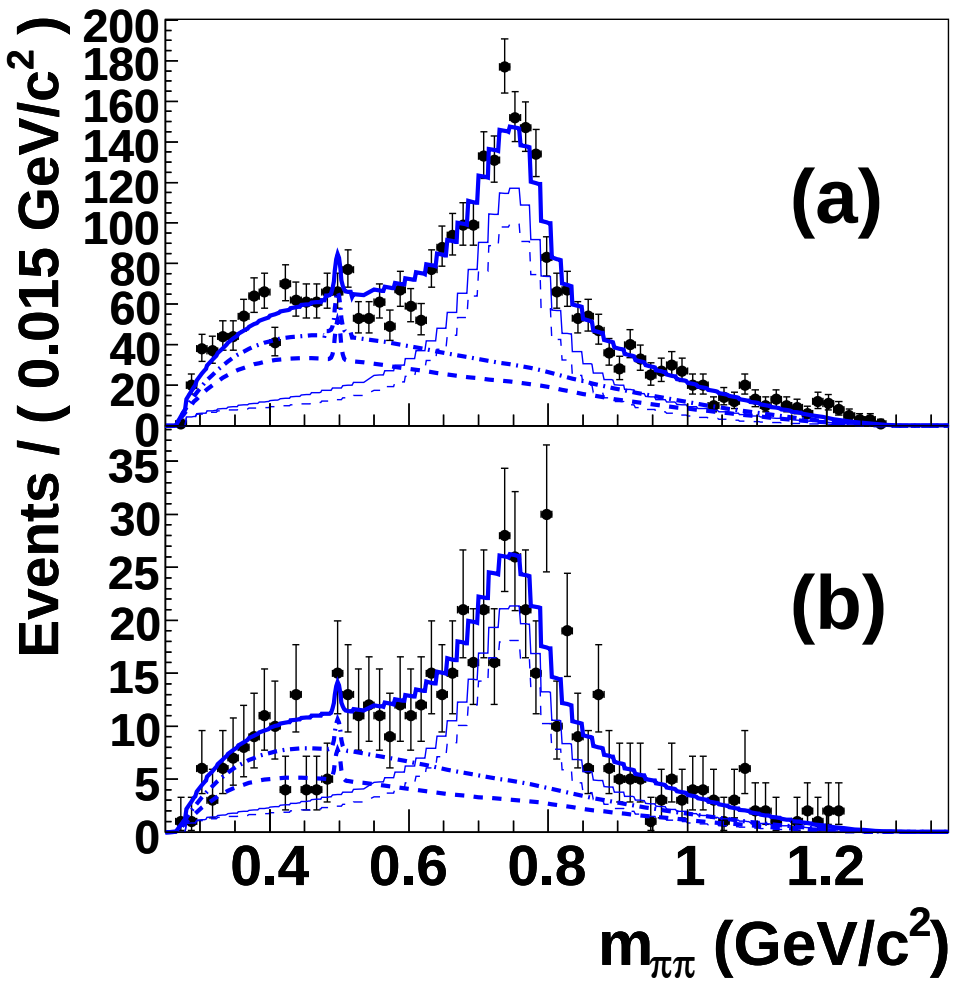

**Figure 17.9.11.** Belle measurements of $\pi^+\pi^-$ mass distributions (with no $K^*$ mass restriction) for (a) $B^+ \to K^+\pi^+\pi^-\gamma$ and (b) $B^0 \to K^0_S\pi^+\pi^-\gamma$, from (Li, 2008). The thin solid (dashed) line corresponds to the $K\rho^0\gamma$ ($K_1(1270)\gamma$ subset) signal, and dot-dashed (dashed) line to the total (continuum) background.

*BABAR* (Aubert, 2008x) and Belle (Ushiroda, 2006) have both made measurements of the amplitudes $\mathcal{S}$ and $\mathcal{C}$ of the $\sin(\Delta m_d\Delta t)$ and $\cos(\Delta m_d\Delta t)$ terms in the time-dependent asymmetry. The results are given in Table 17.9.9. Although they use large data samples of 467M and 535M $B\overline{B}$ respectively, their results are statistically limited, because only 1/9 of $B^0 \to K^{*0}\gamma$ events decay into $K^0_S(\to \pi^+\pi^-)\pi^0\gamma$. The fit results from *BABAR* is shown in Fig. 17.9.9.

Both experiments have also looked at a higher $K^0_S\pi^0$ mass region in $B \to K^0_S\pi^0\gamma$ in a range up to 2 GeV. In this region the final state is no longer dominated by a single resonance, with the largest contribution coming from the $K^*_2(1430)$.

A few other exclusive $b \to s\gamma$ final states have been investigated experimentally (see Table 17.9.9). For $B \to K^0_S\eta\gamma$, the vertex can be reconstructed from charged pions for the final state with $\eta \to \pi^+\pi^-\pi^0$, but for $\eta \to \gamma\gamma$ it is still necessary to reconstruct the vertex from the $K^0_S$. The time-dependent *CP* asymmetry for $B \to K^0_S\eta\gamma$ has been measured by *BABAR* (Aubert, 2009e) with 465M $B\overline{B}$. A similar mode, $B \to K^0\eta'\gamma$, has not been observed yet. The $B \to K^0_S\phi\gamma$ mode has recently been measured by Belle (Sahoo, 2011) with 772M $B\overline{B}$, in which the vertex is determined from $\phi \to K^+K^-$.

The $B \to K^0_S\rho^0\gamma$ mode was measured by Belle with 657M $B\overline{B}$ (Li, 2008). Unlike the other decays discussed here, the *CP* purity of this final state is diluted by the flavor specific $B^0 \to K^{*+}\pi^-\gamma$ mode. (This dilution is in addition to the usual effect of flavor mistags.) In order to fix the contribution of each resonant state, assuming isospin symmetry, the $K^+\pi^+\pi^-$ mass distribution of the $K^*\pi$ enriched sample of more abundant charged mode $B^+ \to K^+\pi^+\pi^-\gamma$ is fitted with known states: those with $1^+$ spin-parity of which $K_1(1270)$ is the dominant resonance with a small $K_1(1400)$ contribution, those with $1^-$ from $K^*(1680)$, and those with $2^+$ from $K_2^*(1430)$ as shown in Figure 17.9.10. As a part of the systematic error study, contributions from other possible modes were tested, and it was found that the inclusion of $B \to K_S^0\sigma\gamma$ causes the largest shift, where $\sigma$ is a controversial scalar state also listed as $f_0(500)$ by the Particle Data Group (Beringer et al., 2012). The absolute values of the amplitudes to the $K^*\pi$ and $K\rho^0$ modes are deduced from the results of the $K^*\pi$ enriched sample and known branching fractions. The relative phase between $K^*\pi$ and $K\rho$ for $K_1(1270)$ is determined from a fit to the $K\pi$ and $\pi\pi$ mass distributions after integrating out the $K\pi\pi$ mass, in which the corresponding phases for $K^*(1680)$ and $K_2^*(1430)$ are fixed to known values and that for $K_1(1400)$ is determined from a scan to give the smallest $\chi^2$.

The dilution factor is calculated by integrating the sum of decay amplitudes for $B^0$ decays, using the parameters obtained for $B^+$ decays. It can be seen in Figure 17.9.11 that the $K\rho^0\gamma$ is the dominant contribution in the $\pi^+\pi^-$ mass range $[0.6, 0.9]\,\mathrm{GeV}/c^2$. The dilution factor in this range is found to be $0.83\,^{+0.19}_{-0.03}$, which is used as a correction factor to the measured *CP* asymmetry in the same mass range. Despite this complication, the $K_S^0\rho^0\gamma$ mode is statistically competitive to the $K^{*0}\gamma$ mode as listed in Table 17.9.9 because the vertex is determined from $\rho^0 \to \pi^+\pi^-$. The dilution effect is corrected for in the $\mathcal{S}$ measurement, but the $\mathcal{C}$ coefficient is not corrected as direct *CP* asymmetry does not necessarily originate only from *CP* eigenstates.

Time dependent *CP* asymmetries can also be measured in $b \to d\gamma$ decay modes. The weak phases due to $V_{td}$ that appear in the $B^0\overline{B}^0$ mixing and $b \to d$ penguin diagrams cancel each other, so a time dependent *CP* asymmetry requires a new contribution to the phase beyond the SM that enters differently in mixing and penguin diagrams. As with the $b \to s\gamma$ decays, this new contribution also has to have a significant right-handed amplitude. Although the rate for $b \to d\gamma$ is only 4% of $b \to s\gamma$, in the case of $B \to \rho^0\gamma$ a large fraction of this suppression is compensated with respect to $B \to K^{*0}\gamma$, because of the 1/9 factor for the $K_S^0\pi^0$ fraction and the $K_S^0$ vertex requirement in the vertex detector volume. Belle has measured time-dependent *CP* asymmetries in $B \to \rho^0\gamma$ with 657M $B\overline{B}$ (Ushiroda, 2008). The results are given in Table 17.9.9.

At present all the measurements of time-dependent *CP* violation are consistent with zero, and dominated by statistical errors that are 0.16 or greater. This is an area where larger samples at a super flavor factory would make a significant improvement.

### 17.9.7 Electroweak penguin decays $b \to s(d)\ell^+\ell^-$

The $b \to s\ell^+\ell^-$ transition, where $\ell^+\ell^-$ is an electron or a muon pair, provides a number of additional probes of the SM and possible new physics contributions. In the language of the effective electroweak Hamiltonian (Section 17.9.1.1), the contributions present for the radiative $b \to s\gamma$ transitions (most importantly the $\mathcal{O}_7$ term) are supplemented by the lepton-current terms, $\mathcal{O}_9$ and $\mathcal{O}_{10}$. The lepton pair can be generated from a virtual photon with the same penguin diagram as $b \to s\gamma$, or the photon can be replaced by a virtual $Z$ boson. There is also a contribution from a box diagram that is formed by virtual $W$ bosons and a neutrino (Figure 17.9.1). The decays $b \to s\ell^+\ell^-$ are suppressed relative to $b \to s\gamma$ by an additional factor of $\alpha$, which results in branching fractions of $\mathcal{O}(10^{-6})$. For this reason they had not been observed at experiments prior to the $B$ Factories. The $b \to d\ell^+\ell^-$ transition has similar properties, but it is further suppressed by $|V_{td}/V_{ts}|^2$ and thus has been beyond the reach of the $B$ Factories.

The theoretical methods to describe the observables in $b \to s(d)\ell^+\ell^-$ are discussed in Section 17.9.1. The exclusive channel $B \to K^*\ell^+\ell^-$ is of particular interest experimentally, but is theoretically more difficult than inclusive channels, since it depends on a set of form factors. A description of the necessary theoretical tools can be found in Section 17.9.1.6.

The $b \to s\ell^+\ell^-$ decays have additional degrees of freedom as compared to $b \to s\gamma$ decays. First, the amplitudes of the different contributions vary as a function of the invariant mass squared $q^2$ of the di-lepton system.[89] The lower end of the $q^2$ distribution has a large contribution from the virtual photon, whereas the higher end is dominated by weak boson transitions. Second, the two final-state leptons provide several additional angular variables (see Section 17.9.7.3 for the specifics in the case of $B \to K^*\ell^+\ell^-$). In particular, the interference between the contributions generates a forward-backward asymmetry in the di-lepton decay angle. The pattern of this asymmetry as a function of $q^2$ is expected to provide a sensitive test of the SM (Ali, Giudice, and Mannel, 1995), especially via the presence in the SM of a zero-crossing point at $q^2 \approx 3$ to $4\,\mathrm{GeV}^2/c^2$ (Ali, Ball, Handoko, and Hiller, 2000).

#### 17.9.7.1 The exclusive modes $B \to K^{(*)}\ell^+\ell^-$

The exclusive modes $B \to K\ell^+\ell^-$ and $B \to K^*\ell^+\ell^-$ have common final states with $B \to \psi K$ and $B \to \psi K^*$. These final states are a source of calibration events for optimizing the search for the $b \to s\ell^+\ell^-$ modes, but they are also a large background in the di-lepton mass ranges around the $J/\psi$ and $\psi'$. There is interference between the electroweak penguin and charmonium amplitudes which is difficult to handle theoretically, although it could eventually provide useful information about relative phases. For both experimental and theoretical reasons the regions around the $J/\psi$ and $\psi'$ are vetoed in the *BABAR* and Belle analyses. For $e^+e^-$ pairs there is a long and large radiative tail on the lower $q^2$ side of the charmonium peaks due to bremsstrahlung from the final state electrons, which also shifts the measured kinematic variables $m_{\rm ES}$ (slightly) and $\Delta E$. These offsets are partially removed by adding photons to the electron momentum if they are found near the electron direction. Details of this bremsstrahlung recovery procedure can be found in (Ishikawa, 2003) and (Aubert, 2006ac). Note that Belle uses wider $J/\psi$ and $\psi'$ veto windows for electron modes than for the muon modes.

[89] Although here we define the $q^2$ in terms of mass, the original meaning is in terms of momentum of the virtual boson and hence we quote the values in the unit of $\mathrm{GeV}^2/c^2$.

Another region that is usually vetoed in $B \to K^{(*)} e^+e^-$ is the very low $q^2$ region where there is a large peak from the virtual photon diagram. Removing the region below the threshold $\mu^+\mu^-$ mass squared from analysis results in similar SM expectations for the inclusive decay rates of $B \to K^* e^+e^-$ and $B \to K^* \mu^+\mu^-$. There is no virtual photon peak in $B \to K e^+e^-$ due to angular momentum suppression of the photons, but there are other background contributions in this region. They arise from charmless hadronic $B$ decays followed by a Dalitz decay of a $\pi^0$, and from pair conversions of photons in the detector, and it is desirable to remove them.

Observation of the decay $B \to K\ell^+\ell^-$ was already reported by Belle with only 31M $B\overline{B}$ data (Abe, 2002l). This was followed by the first observation of $B \to K^*\ell^+\ell^-$ by Belle in 2003 with a data sample of 152M $B\overline{B}$ (Ishikawa, 2003). Both results were supported by *BABAR* with a data sample of 123M $B\overline{B}$ (Aubert, 2003c). There have been frequent updates from *BABAR* using 229M $B\overline{B}$ (Aubert, 2006ac), 384M $B\overline{B}$ (Aubert, 2009c,j), and 471M $B\overline{B}$ (Lees, 2012i), and the latest result from Belle uses 657M $B\overline{B}$ (Wei, 2009).

We also compare in this section the $B$ Factories results on $B \to K^{(*)}\ell^+\ell^-$ modes with recent results from the CDF experiment at the Tevatron (Aaltonen et al., 2011c), and from the LHCb experiment at CERN (Aaij et al., 2012a,b,g,i). CDF has been very competitive with the $B$ Factories in these exclusive modes, and LHCb has recently surpassed the precision of the $B$ Factories in $\mu^+\mu^-$ modes.

### 17.9.7.2 Branching fractions and rate asymmetries in $B \to K^{(*)}\ell^+\ell^-$

Given the limited statistics in these exclusive decay modes, combined branching fractions are determined using all measured final states. In the case of $B \to K\ell^+\ell^-$ there are four final states, with $K^+$ or $K^0_S$ and $\mu^+\mu^-$ or $e^+e^-$. In the case of $B \to K^*\ell^+\ell^-$ there are eight final states with $K^{*+} \to K^0_S\pi^+$ or $K^+\pi^0$, $K^{*0} \to K^+\pi^-$ or $K^0_S\pi^0$ and $\mu^+\mu^-$ or $e^+e^-$. At the hadron colliders the modes with $\pi^0$ or electrons have not been used thus far. The combined branching fractions assume *CP* and isospin symmetries, and lepton universality, all of which are satisfied to good accuracy in the SM, to sum over $K^*$ decay modes and average over lepton flavor and $B$ charge. Regarding lepton flavor, the *BABAR* measurements (Lees, 2012i) are for $q^2 \geq 0.1\,{\rm GeV}^2/c^2$, where the $e^+e^-$ and $\mu^+\mu^-$ branching fractions are expected to be very close. Belle (Wei, 2009), on the other hand, measures in effect to the minimum possible $q^2$ values, hence including the effects of the virtual photon peak in the $B \to K^*e^+e^-$ mode. Then for $B \to K^*\ell^+\ell^-$, Belle averages the lepton flavors using an SM-based constraint of 1.33 on the ratio of $\mathcal{B}(B \to K^*e^+e^-)$ to $\mathcal{B}(B \to K^*\mu^+\mu^-)$, and quotes the latter value for its results.

**Table 17.9.10.** Measurements of exclusive $b \to s\ell^+\ell^-$ branching fractions in $10^{-8}$, integrated over all di-lepton $q^2$ (including the vetoed regions — see text). *BABAR* and Belle results are based on both $e^+e^-$ and $\mu^+\mu^-$ modes, while CDF and LHCb results are based on $\mu^+\mu^-$ modes only. In their papers, *BABAR* does not quote separate results for $B^+$ and $B^0$ decays, while CDF and LHCb do not quote combined results. LHCb does not report the total branching fraction for $B^0 \to K^{*0}\mu^+\mu^-$, although precise $d\mathcal{B}/dq^2$ results are given in (Aaij et al., 2012b). Uncertainties are statistical and systematic, respectively; these errors are combined in some of the LHCb results.

| Model | *BABAR* | Belle | CDF | LHCb |
|---|---|---|---|---|
| $K^+\ell^+\ell^-$ | | $53\,^{+6}_{-5} \pm 3$ | $46 \pm 4 \pm 2$ | $43.6 \pm 1.5 \pm 1.8$ |
| $K^0\ell^+\ell^-$ | | $34\,^{+9}_{-8} \pm 2$ | $32 \pm 10 \pm 2$ | $31\,^{+7}_{-6}$ |
| $K\ell^+\ell^-$ | $47 \pm 6 \pm 2$ | $48\,^{+5}_{-4} \pm 3$ | | |
| $K^{*+}\ell^+\ell^-$ | | $124\,^{+23}_{-21} \pm 13$ | $95 \pm 32 \pm 8$ | $116 \pm 19$ |
| $K^{*0}\ell^+\ell^-$ | | $97\,^{+13}_{-11} \pm 7$ | $102 \pm 10 \pm 6$ | |
| $K^*\ell^+\ell^-$ | $102\,^{+14}_{-13} \pm 5$ | $107\,^{+11}_{-10} \pm 9$ | | |

It has been customary to divide the branching fraction results into a set of six bins in $q^2$, three below the $J/\psi$ mass, two above the $\psi'$ mass, and one in the gap between the two charmonium veto regions. The distributions of $d\mathcal{B}/dq^2$ are in good agreement between experiments and with the SM as shown in Figure 17.9.12 for *BABAR*, Belle and CDF.

The branching fractions integrated over all $q^2$ from *BABAR* (Lees, 2012i), Belle (Wei, 2009), CDF (Aaltonen et al., 2011c) and LHCb (Aaij et al., 2012a,i) are given in Table 17.9.10. For these integrals, the veto regions are filled in by interpolation, using SM-based predictions of the spectral shape *vs.* $q^2$. Figure 17.9.13 compares the isospin-averaged total branching fractions for these three experiments to two SM predictions. The results are consistent with the predicted branching fractions.

Lepton universality is tested by looking at the ratios:

$$R_\ell(K^{(*)}) = \frac{\mathcal{B}(B \to K^{(*)}e^+e^-)}{\mathcal{B}(B \to K^{(*)}\mu^+\mu^-)}, \qquad (17.9.62)$$

which should be equal to one if the region $q^2 < (2m_\mu)^2$ is removed, and about 30% greater than one for $K^*\ell^+\ell^-$ if the low-$q^2$ virtual photon region is included for $e^+e^-$. Deviations from these predictions could be due to enhancements from new physics, *e.g.* a SUSY Higgs. The results from Belle (Wei, 2009) with 657M $B\overline{B}$, and *BABAR* (Aubert, 2009j) with 384M $B\overline{B}$ are given in Table 17.9.11. (The Belle paper defines $R_\ell$ with the $\mu^+\mu^-$ and $e^-e^-$ swapped, as compared to Eq. (17.9.62), so reciprocals have been taken for presentation here.) The measured $R_\ell$ are consistent with SM expectations (1.00, except 1.33 for the

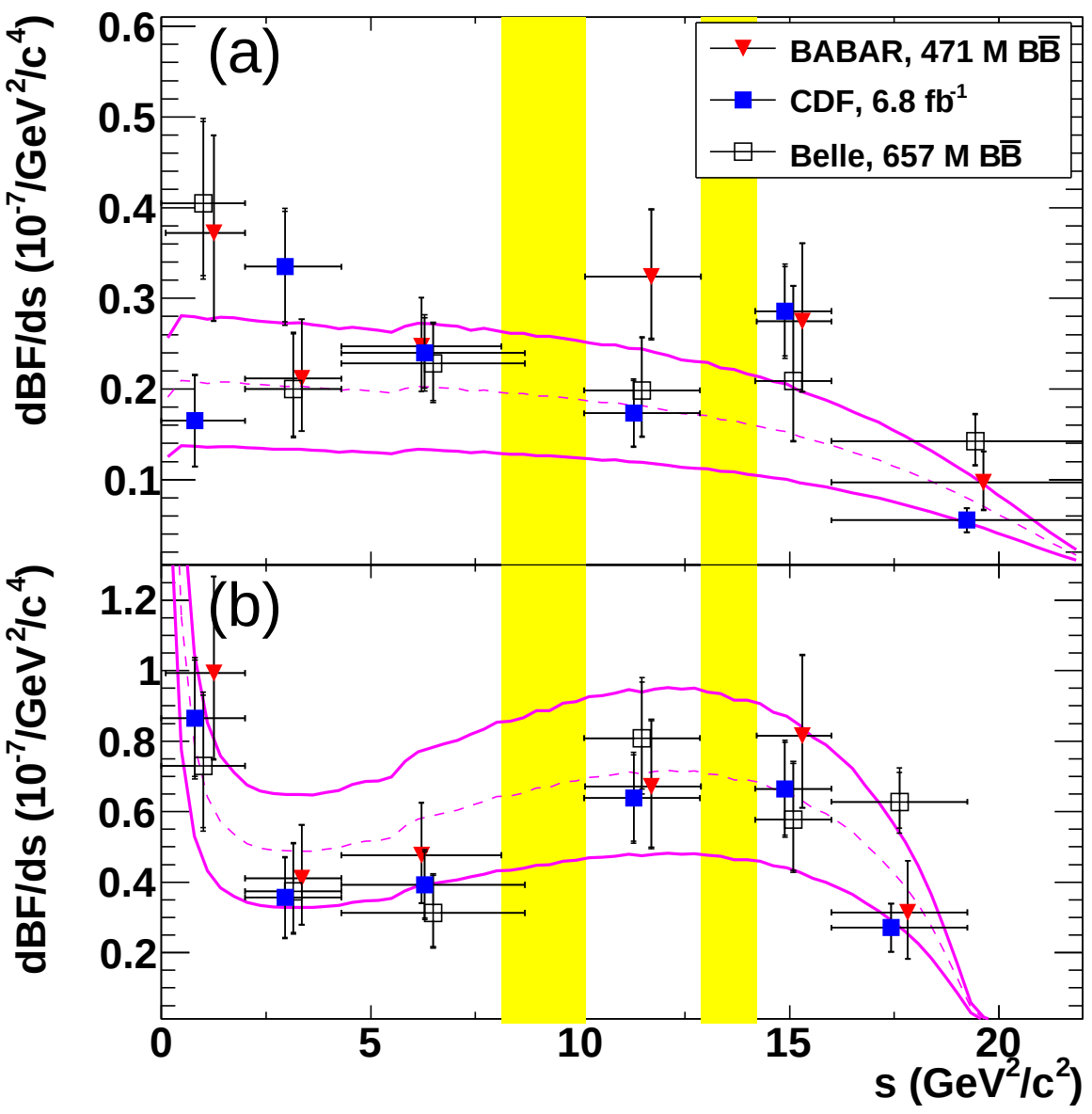

**Figure 17.9.12.** From (Lees, 2012i). Branching fractions $d\mathcal{B}/dq^2$ for (a) $B \to K\ell^+\ell^-$ and (b) $B \to K^*\ell^+\ell^-$ from Belle, CDF and *BABAR* *vs.* $s \equiv q^2$. The yellow bands correspond to the $J/\psi$ and $\psi'$ veto windows used by *BABAR*. The magenta curves show the range of the SM predictions from (Ali, Lunghi, Greub, and Hiller, 2002).

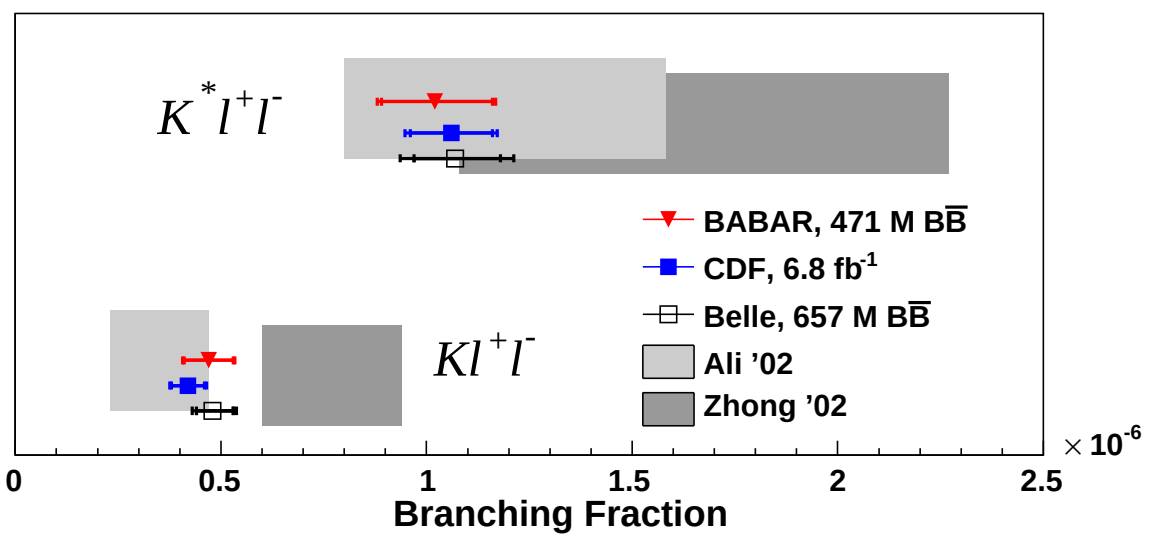

**Figure 17.9.13.** From (Lees, 2012i). Total branching fractions $d\mathcal{B}/dq^2$ for $B \to K\ell^+\ell^-$ and $B \to K^*\ell^+\ell^-$ from *BABAR*, CDF and Belle, compared to SM predictions from the models of (Ali, Lunghi, Greub, and Hiller, 2002) and (Zhong, Wu, and Wang, 2003).

Belle measurement of $R_\ell(K^*)$), but with quite large uncertainties.

Direct *CP* asymmetries, defined as per Eq. (17.9.40), have also been searched for, with results given in Table 17.9.11. They are consistent with zero.

The results for the isospin asymmetries drew more attention. Isospin asymmetry $A_I$ (called $\Delta_{0-}$ in Eq. (17.9.38), which defines its sign) is evaluated as per Section 17.9.5.2. The isospin asymmetry in the mode $B \to K^*\ell^+\ell^-$ is a subleading $\Lambda_{\rm QCD}/m_b$ effect as in the radiative mode, but again the dominant isospin-breaking effects can be calculated perturbatively, while other $\Lambda_{\rm QCD}/m_b$ corrections are just estimated. The exact uncertainty is difficult to estimate due to unknown power corrections, but the observable may still be useful in the NP search because of the high sensitivity to specific Wilson coefficients (Feldmann and Matias, 2003). These authors predict only a rather small isospin asymmetry in the SM, with a positive sign at low $q^2$.

**Table 17.9.11.** Summary of measurements of lepton universality $R_\ell$, direct *CP* asymmetries $A_{CP}$, and isospin asymmetries $A_I$ in $B \to K^{(*)}\ell^+\ell^-$ decays. For $A_I$, "low $q^2$" means below $m^2_{J/\psi}$. Uncertainties are statistical and systematic, respectively, for the separate measurements, combined for the weighted averages. Because the Belle measurements of $R_\ell$ extend to lower $q^2$ than do the *BABAR* measurement, the two $R_\ell(K^*)$ values have different SM expectations, (see text); hence no average is quoted for that quantity.

| | *BABAR* | Belle | Average |
|---|---|---|---|
| $R_\ell(K)$ | $1.00\,^{+0.31}_{-0.25} \pm 0.07$ | $0.97 \pm 0.18 \pm 0.06$ | $0.98 \pm 0.16$ |
| $R_\ell(K^*)$ | $1.13\,^{+0.34}_{-0.26} \pm 0.10$ | $1.20 \pm 0.25 \pm 0.12$ | |
| $A_{CP}(K)$ | $-0.03 \pm 0.14 \pm 0.01$ | $+0.04 \pm 0.10 \pm 0.02$ | $+0.02 \pm 0.08$ |
| $A_{CP}(K^*)$ | $+0.03 \pm 0.13 \pm 0.01$ | $-0.10 \pm 0.10 \pm 0.01$ | $-0.05 \pm 0.08$ |
| $A_I^{\text{low}\,q^2}(K)$ | $-0.58\,^{+0.29}_{-0.37} \pm 0.02$ | $-0.31\,^{+0.17}_{-0.14} \pm 0.08$ | $-0.37 \pm 0.15$ |
| $A_I^{\text{low}\,q^2}(K^*)$ | $-0.25\,^{+0.20}_{-0.17} \pm 0.03$ | $-0.29 \pm 0.16 \pm 0.09$ | $-0.27 \pm 0.13$ |

In (Aubert, 2009j) *BABAR* reported evidence for a large negative isospin asymmetry in the $q^2$ range below the $J/\psi$, with about $3\sigma$ significance in both $B \to K\ell^+\ell^-$ and $B \to K^*\ell^+\ell^-$. However, the latest *BABAR* results (Lees, 2012i) are more consistent with null asymmetry, as listed in Table 17.9.11. Belle's results (Wei, 2009) with higher statistics are compatible with *BABAR* and also with null asymmetry. Note that the average $A_I$ values are still negative, and about $2\sigma$ from zero, so higher-precision measurements are desirable. Neither experiment sees a significant asymmetry in the high-$q^2$ region. See the cited papers for measured isospin asymmetries in all $q^2$ bins. These are illustrated for Belle in Fig. 17.9.14.

### 17.9.7.3 Angular distributions in $B \to K^*\ell^+\ell^-$: formalism and theory

In the $B \to K^*\ell^+\ell^-$ decay mode the angular distribution contains useful information about the different amplitudes. This is in contrast to $B \to K\ell^+\ell^-$ or $B \to K^*\gamma$, where the angular distributions are fully constrained by angular momentum conservation. The decay $B \to K^*\ell^+\ell^-$ ($K^* \to K\pi$) is completely described by four independent kinematic variables:[90] the lepton-pair invariant mass squared, $q^2$, the angle $\theta_K$ of the $K^+$ relative to the $B$ in the $K^*$ rest frame, the angle $\theta_\ell$ of the $\ell^+$ relative to the $B$ in the di-lepton rest frame, and the angle $\phi$ between the $K^*$ decay plane and the di-lepton plane. Summing over the spins of the final particles, the differential decay distribution can be written as (Kruger and Matias, 2005; Kruger, Sehgal, Sinha, and Sinha, 2000)

$$\frac{d^4\Gamma}{dq^2\, d\theta_\ell\, d\theta_K\, d\phi} = \frac{9}{32\pi} I(q^2, \theta_\ell, \theta_K, \phi)\,. \qquad (17.9.63)$$

[90] This discussion assumes a *P*-wave $K^*$ final state with no *S*-wave $K\pi$ background.

A full expression for $I$ contains twelve angular coefficients, $I_{1-9}^{(s,c)}$, which are functions of $q^2$, and may be different for $B$ and $\overline{B}$ decays[91] (Altmannshofer et al., 2009):

$$\begin{aligned} I(q^2,\theta_\ell,\theta_K,\phi) &= I_1^s \sin^2\theta_K + I_1^c \cos^2\theta_K \\ &+ (I_2^s \sin^2\theta_K + I_2^c \cos^2\theta_K)\cos 2\theta_\ell \\ &+ I_3 \sin^2\theta_K \sin^2\theta_\ell \cos 2\phi \\ &+ I_4 \sin 2\theta_K \sin 2\theta_\ell \cos\phi \\ &+ I_5 \sin 2\theta_K \sin\theta_\ell \cos\phi \\ &+ (I_6^s \sin^2\theta_K + I_6^c \cos^2\theta_K)\cos\theta_\ell \\ &+ I_7 \sin 2\theta_K \sin\theta_\ell \sin\phi \\ &+ I_8 \sin 2\theta_K \sin 2\theta_\ell \sin\phi \\ &+ I_9 \sin^2\theta_K \sin^2\theta_\ell \sin 2\phi. \end{aligned} \quad (17.9.64)$$

The coefficients $I_{1-9}^{(s,c)}$ can be expressed in terms of either helicity amplitudes $H_{0,+,-}$, or transversity amplitudes $A_{\perp,||,0}$. These amplitudes contain left and right-handed contributions which can be written in terms of the Wilson coefficients $C_{7,9,10}$ and form-factors. The coefficients $I_{1-3}^{(s,c)}$ are sensitive to amplitudes squared, while $I_{4-9}^{(s,c)}$ are sensitive to interference terms.

In practice a full angular analysis has not yet been done, because it requires a few thousand $B \to K^*\ell^+\ell^-$ signal events. *BABAR* (Aubert, 2009c) and Belle (Wei, 2009) have performed angular fits, in bins of $q^2$, to the $\theta_K$ and $\theta_\ell$ distributions, in each case after integrating over the other two angles. The $\theta_K$ distribution

$$\frac{1}{\Gamma}\frac{\mathrm{d}\Gamma}{\mathrm{d}\cos\theta_K} = \frac{3}{2}F_L\cos^2\theta_K + \frac{3}{4}(1-F_L)(1-\cos^2\theta_K) \quad (17.9.65)$$

is sensitive to the fraction of longitudinal polarization, $F_L = |A_0|^2$. In the SM this varies as a function of $q^2$, going to zero as $q^2 \to 0$, increasing to a maximum of 0.8 at $q^2 \approx 3\,\mathrm{GeV}^2/c^2$, and falling gradually towards higher $q^2$.

The $\theta_\ell$ distribution

$$\begin{aligned} \frac{1}{\Gamma}\frac{\mathrm{d}\Gamma}{\mathrm{d}\cos\theta_\ell} &= \frac{3}{4}F_L(1-\cos^2\theta_\ell) + \frac{3}{8}(1-F_L)(1+\cos^2\theta_\ell) \\ &+ A_{\mathrm{FB}}\cos\theta_\ell \end{aligned} \quad (17.9.66)$$

is sensitive to the forward-backward asymmetry:

$$A_{\mathrm{FB}} = \frac{\int \mathrm{d}\theta_\ell\, \mathrm{sgn}(\theta_\ell)\mathcal{B}(B \to K^*\ell^+\ell^-;\ \theta_\ell)}{\int \mathrm{d}\theta_\ell\, \mathcal{B}(B \to K^*\ell^+\ell^-;\ \theta_\ell)}. \quad (17.9.67)$$

In the SM, $A_{\mathrm{FB}}$ is a strong function of $q^2$. It goes to zero as $q^2 \to 0$, is small and negative at low $q^2$, with a zero-crossing point at $q_0^2 \approx 4\,\mathrm{GeV}^2/c^2$, then it gradually increases to about 0.4 at high $q^2$, where the electroweak $V-A$ contributions dominate; *e.g.*, (Ali, Ball, Handoko, and Hiller, 2000).

Angular fits to the projected distributions of $\theta_K$ and $\theta_\ell$ are used to measure the observables $F_L$ and $A_{\mathrm{FB}}$ in bins of $q^2$. The hadronic uncertainties on these two observables in the SM are large. However, the value of the di-lepton invariant mass $q_0^2$, for which the forward-backward asymmetry vanishes, can be predicted in quite a clean way. In the QCD factorization approach at leading order in $\Lambda_{\mathrm{QCD}}/m_b$, the value of $q_0^2$ is free from hadronic uncertainties at order $\alpha_s^0$. A dependence on the soft form factor and on the light cone wave functions of the $B$ and $K^*$ mesons appears only at order $\alpha_s^1$. At NLO one finds (Beneke, Feldmann, and Seidel, 2005):

$$\begin{aligned} q_0^2[K^{*0}\ell^+\ell^-] &= (4.36\,^{+0.33}_{-0.31})\,\mathrm{GeV}^2/c^2, \\ q_0^2[K^{*+}\ell^+\ell^-] &= (4.15\,^{+0.27}_{-0.27})\,\mathrm{GeV}^2/c^2. \end{aligned} \quad (17.9.68)$$

For all observables from the angular analysis the unknown $\Lambda_{\mathrm{QCD}}/m_b$ power corrections are the source of the largest theoretical uncertainty. The small difference between the two modes is due to isospin-breaking power corrections.

The value of $q_0^2$ is highly sensitive to the ratio of the two Wilson coefficients $C_7$ and $C_9$, and the region near this SM-predicted zero-crossing point is particularly sensitive to the interplay between the terms proportional to $C_7$ and to $C_9$. Using the magnitude of $C_7$ constrained from $B \to X_s\gamma$, this angular distribution can be used to determine the sign of $C_7$, and to constrain the two other Wilson coefficients $C_9$ and $C_{10}$. If the sign of $C_7$ is flipped, there is no zero-crossing point at $q_0^2$ (Ali, Giudice, and Mannel, 1995).

The position of $q_0^2$ can also be moved by new physics contributions (Altmannshofer et al., 2009). Detailed NP analyses of the angular observables have been presented in Bobeth, Hiller, and Piranishvili (2008), Altmannshofer et al. (2009), and Egede, Hurth, Matias, Ramon, and Reece (2008, 2010). They provide sensitivity to various Wilson coefficients, but the sensitivity to new weak phases turns out to be restricted (Egede, Hurth, Matias, Ramon, and Reece, 2010).

#### 17.9.7.4 $B \to K^*\ell^+\ell^-$ angular analysis

*BABAR* (Aubert, 2009c) and Belle (Wei, 2009) have extracted $F_L$ and $A_{\mathrm{FB}}$ values for $B \to K^*\ell^+\ell^-$ by first fitting their measured $\theta_K$ distributions in bins of $q^2$ to Eq. (17.9.65), and then fitting each corresponding $\theta_\ell$ distribution to Eq. (17.9.66) with $F_L$ fixed from the result of the $\theta_K$ fit. Belle measures $A_{\mathrm{FB}}$ in six bins in $q^2$ using a data set of 657M $B\overline{B}$, while *BABAR* measures $A_{\mathrm{FB}}$ in two bins in $q^2$, below and above the $J/\psi$ (with the $\psi(2S)$ window excluded in the higher mass region), using a data set of 384M $B\overline{B}$.

The results for $F_L$ are listed in Table 17.9.12 along with more recent results from CDF (Aaltonen et al., 2011b) and LHCb (Aaij et al., 2012b), with results in good agreement with the SM. Figure 17.9.14 shows the Belle results.

The Belle and *BABAR* results for $A_{\mathrm{FB}}$ are listed in Table 17.9.13, along with the recent results from CDF and LHCb. Figure 17.9.14 illustrates the Belle results. Given the current level of precision, all results are compatible with the SM predictions.

[91] Note that $I_{5,6,8,9}$ change sign between $B$ and $\overline{B}$.

**Table 17.9.12.** Measurements of longitudinal polarization fraction $F_L$ in $B \to K^*\ell^+\ell^-$ as a function of di-lepton $q^2$. (BaBar measures this in only two $q^2$ bins, the other experiments in six bins). Errors are statistical and systematic, respectively.

| $q^2$ (GeV$^2/c^2$) | Belle | BaBar | CDF | LHCb |
|---|---|---|---|---|
| $0.00 - 2.00$ | $0.29 \pm 0.20 \pm 0.02$ | | $0.30 \pm 0.16 \pm 0.02$ | $0.00^{+0.13}_{-0.00} \pm 0.02$ |
| $2.00 - 4.30$ | $0.71 \pm 0.24 \pm 0.05$ | $(0.35 \pm 0.16 \pm 0.04)$ | $0.37^{+0.25}_{-0.24} \pm 0.10$ | $0.77 \pm 0.15 \pm 0.03$ |
| $4.30 - 8.68$ | $0.64 \pm 0.24 \pm 0.07$ | | $0.68^{+0.15}_{-0.17} \pm 0.09$ | $0.60^{+0.06}_{-0.07} \pm 0.01$ |
| $10.09 - 12.86$ | $0.17 \pm 0.16 \pm 0.03$ | | $0.47 \pm 0.14 \pm 0.03$ | $0.41 \pm 0.11 \pm 0.03$ |
| $14.18 - 16.00$ | $-0.15 \pm 0.25 \pm 0.07$ | $(0.71 \pm 0.21 \pm 0.04)$ | $0.29^{+0.14}_{-0.13} \pm 0.05$ | $0.37 \pm 0.09 \pm 0.05$ |
| $16.00 - 19.30$ | $0.12 \pm 0.14 \pm 0.02$ | | $0.20^{+0.19}_{-0.17} \pm 0.05$ | $0.26^{+0.10}_{-0.08} \pm 0.03$ |

**Table 17.9.13.** Measurements of di-lepton forward-backward asymmetry $A_{\rm FB}$ in $B \to K^*\ell^+\ell^-$ as a function of di-lepton $q^2$. (BaBar measures this in only two $q^2$ bins, the other experiments in six bins). Errors are statistical and systematic, respectively.

| $q^2$ (GeV$^2/c^2$) | Belle | BaBar | CDF | LHCb |
|---|---|---|---|---|
| $0.00 - 2.00$ | $+0.47^{+0.26}_{-0.32} \pm 0.03$ | | $-0.35^{+0.26}_{-0.23} \pm 0.10$ | $-0.15 \pm 0.20 \pm 0.06$ |
| $2.00 - 4.30$ | $+0.11^{+0.31}_{-0.36} \pm 0,07$ | $(+0.24^{+0.18}_{-0.23} \pm 0.05)$ | $+0.29^{+0.32}_{-0.35} \pm 0.15$ | $+0.05^{+0.16}_{-0.20} \pm 0.04$ |
| $4.30 - 8.68$ | $+0.45^{+0.15}_{-0.21} \pm 0.15$ | | $+0.01 \pm 0.20 \pm 0.09$ | $+0.27^{+0.06}_{-0.08} \pm 0.02$ |
| $10.09 - 12.86$ | $+0.43 \pm 0.19 \pm 0.03$ | | $+0.38^{+0.16}_{-0.19} \pm 0.09$ | $+0.27^{+0.11}_{-0.13} \pm 0.02$ |
| $14.18 - 16.00$ | $+0.40^{+0.16}_{-0.22} \pm 0.10$ | $(+0.76^{+0.52}_{-0.32} \pm 0.07)$ | $+0.44^{+0.18}_{-0.21} \pm 0.10$ | $+0.47^{+0.06}_{-0.08} \pm 0.03$ |
| $16.00 - 19.30$ | $+0.66^{+0.11}_{-0.16} \pm 0.04$ | | $+0.65^{+0.17}_{-0.18} \pm 0.16$ | $+0.16^{+0.11}_{-0.13} \pm 0.06$ |

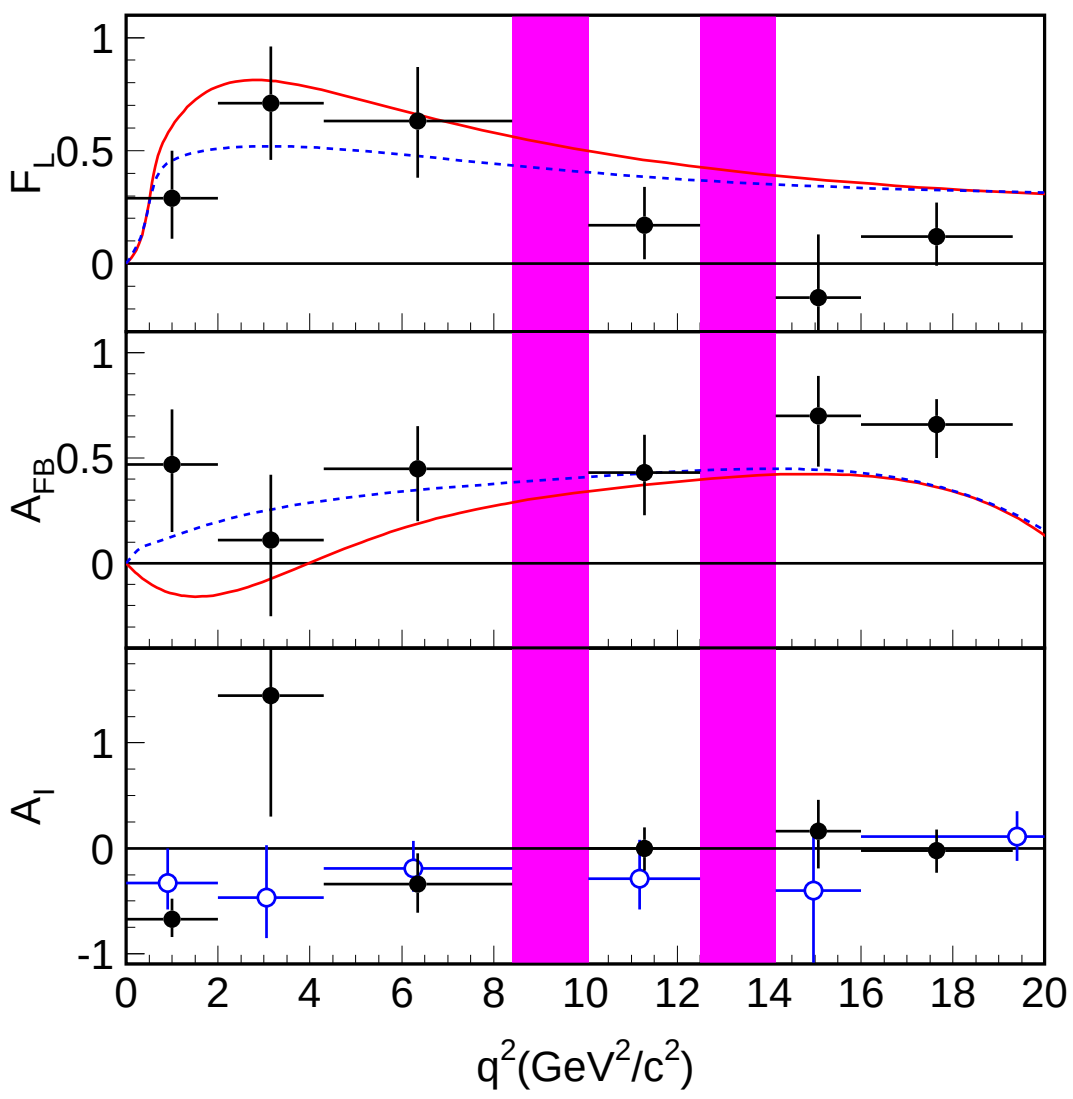

**Figure 17.9.14.** From (Wei, 2009). Measurements of longitudinal polarization fraction $F_L$, and di-lepton forward-backward asymmetry $A_{\rm FB}$ for $B \to K^*\ell^+\ell^-$ from Belle. The solid (red) curves show the SM predictions. The dotted (blue) curves show the effect of reversing the sign of the Wilson coefficient $C_7$. The bottom plot shows the isospin asymmetries for $B \to K^*\ell^+\ell^-$ (filled circles) and $B \to K\ell^+\ell^-$ (open circles).

Belle (Ishikawa, 2006) earlier provided constraints on $C_9$ and $C_{10}$ using 386M $B\overline{B}$ events, from a fit to the $\theta_\ell$ distribution under the assumption that $F_L$ follows the SM distribution.

### 17.9.7.5 Theoretical predictions for inclusive $B \to X_s\ell^+\ell^-$

An inclusive measurement of $b \to s\ell^+\ell^-$ is expected to have reduced theoretical uncertainties in comparison with the exclusive decays $B \to K^{(*)}\ell^+\ell^-$. The situation is similar to $b \to s\gamma$, where understanding of exclusive decays is limited by knowledge of hadronic form factors. In the case of $b \to s\ell^+\ell^-$ these reduced theoretical uncertainties apply not only to the inclusive branching fraction, but also to the angular information, *e.g.* the zero crossing point in the di-lepton forward-backward asymmetry $A_{\rm FB}$.

The angular decomposition of the inclusive decay $B \to X_s\ell^+\ell^-$ provides three independent observables, $H_T$, $H_A$ and $H_L$, from which one can extract the short-distance electroweak Wilson coefficients that test for new physics (Lee, Ligeti, Stewart, and Tackmann, 2007).

$$\frac{d^3\Gamma}{dq^2\,dz} = \frac{3}{8}\left[(1+z^2)H_T(q^2)+2(1-z^2)H_L(q^2)+2zH_A(q^2)\right]. \tag{17.9.69}$$

Here, $z = \cos\theta_\ell$, $H_A$ is equivalent to the forward-backward asymmetry in the exclusive decays, and the di-lepton mass spectrum is given by $H_T + H_L$. The observables depend on the Wilson coefficients $C_7$, $C_9$ and $C_{10}$ in the SM. The present measurements of $B \to X_s\ell^+\ell^-$ already favor the SM sign of the coefficient $C_7$, which is undetermined by the $B \to X_s\gamma$ mode (Gambino, Haisch, and Misiak, 2005).

The observables in inclusive $B \to X_s\ell^+\ell^-$ are dominated by perturbative contributions in a low-$q^2$ region, $1 < q^2 < 6\,{\rm GeV}^2/c^2$, and in the high-$q^2$ region above the $c\bar{c}$ resonances, $q^2 > 14.4\,{\rm GeV}^2/c^2$. The present predictions are based on the perturbative calculations to NNLL precision in QCD and to NLL precision in QED (Section 17.9.1). The branching fraction in the low-$q^2$ region is (Huber, Lunghi, Misiak, and Wyler, 2006):

$$\mathcal{B}(B \to X_s \ell^+\ell^-)_{\rm low} = \begin{cases} (1.59 \pm 0.11) \times 10^{-6} & (\ell = \mu) \\ (1.64 \pm 0.11) \times 10^{-6} & (\ell = e) \end{cases} \tag{17.9.70}$$

and in the high-$q^2$ region (Huber, Hurth, and Lunghi, 2008a):

$$\mathcal{B}(B \to X_s \ell^+\ell^-)_{\rm high} = \begin{cases} 2.40 \times 10^{-7} \times (1^{+0.29}_{-0.26}) & (\ell = \mu) \\ 2.09 \times 10^{-7} \times (1^{+0.32}_{-0.30}) & (\ell = e) \end{cases} \tag{17.9.71}$$

The value of $q_0^2$ for which the inclusive forward-backward asymmetry vanishes,

$$(q_0^2)[X_s \ell^+\ell^-] = \begin{cases} (3.50 \pm 0.12)\,\mathrm{GeV}^2/c^2 & (\ell = \mu) \\ (3.38 \pm 0.11)\,\mathrm{GeV}^2/c^2 & (\ell = e)\,, \end{cases} \tag{17.9.72}$$

is one of the most precise predictions in flavor physics. It determines the relative sign and magnitude of the coefficients $C_7$ and $C_9$ (Huber, Hurth, and Lunghi, 2008a). Unknown subleading non-perturbative corrections of order $\mathcal{O}(\alpha_S \Lambda_{\rm QCD}/m_b)$ are estimated to give an additional uncertainty of order 5%, which has to be added to all $B \to X_s \ell^+\ell^-$ observables. In all predictions it is assumed that there is no cut on the hadronic mass region.

After including the NLL QED matrix elements, the electron and muon channels receive different contributions. This is due to the fact, that the leptons can emit collinear photons, which generate large logarithms of the form $\ln(m_b^2/m_\ell^2)$. This generates a differences between these two channels. We note that in theoretical calculations all collinear photons are assumed to be included in the $X_s$ system. The di-lepton invariant mass does not contain any additional photon, *i.e.* $q^2 = (p_{\ell^+} + p_{\ell^-})^2$. This differs from the experimental analyses which recover bremsstrahlung photons and add them to the di-lepton system, and therefore small modifications to the theoretical predictions are needed – see (Huber, Hurth, and Lunghi, 2008b).

#### 17.9.7.6 Measurements of inclusive $B \to X_s \ell^+\ell^-$

The fully inclusive approach has not been used for $b \to s\ell^+\ell^-$ because the presence of large backgrounds from semileptonic $B$ decays. In these background events the initial $B\overline{B}$ produces two oppositely charged leptons either directly from the two $B$ mesons, or as a cascade from the $b \to c \to s$ decay chain. These backgrounds cannot be removed without further kinematic constraints. The reconstructed-$B$-tag approach (also not yet attempted) should remove the direct two-$B$-decay background component, leaving only the cascade decays, which can then be removed using missing energy variables. The difficulty here is the need for millions of $B$ tags in order to measure an inclusive branching fraction of a few $\times 10^{-6}$. This is a challenging measurement even with the anticipated ultimate dataset of a super flavor factory.

**Table 17.9.14.** Measurements of inclusive $B \to X_s \ell^+\ell^-$ branching fractions (in $10^{-6}$), with $m_{\ell^+\ell^-} > 0.2\,\mathrm{GeV}/c^2$, while the excluded regions around the $J/\psi$ and $\psi'$ are interpolated assuming the SM with no interference. Uncertainties are statistical and systematic, respectively.

| Mode | *BABAR* | Belle | Average |
|---|---|---|---|
| $X_s e^+e^-$ | $6.0 \pm 1.7 \pm 1.3$ | $4.0 \pm 1.3 \pm 0.9$ | $4.7 \pm 1.3$ |
| $X_s \mu^+\mu^-$ | $5.0 \pm 2.8 \pm 1.2$ | $4.1 \pm 1.1 \pm 0.8$ | $4.3 \pm 1.3$ |
| $X_s \ell^+\ell^-$ | $5.6 \pm 1.5 \pm 1.3$ | $4.1 \pm 0.8 \pm 0.8$ | $4.5 \pm 1.0$ |

The sum-of-exclusive method does provide sufficient constraints to discriminate against the semileptonic backgrounds using the $m_{\rm ES}$ and $\Delta E$ kinematic variables. In a similar fashion to the $B \to X_s\gamma$ analysis, the $X_s$ state is reconstructed as one kaon and multiple pions (but including the zero pion case, which corresponds to $B \to K\ell^+\ell^-$). Belle (Iwasaki, 2005) uses up to 4 pions of which one can be a $\pi^0$, and includes the $X_s$ mass range below $2.0\,\mathrm{GeV}/c^2$, while *BABAR* (Aubert, 2004h) uses up to 2 pions of which one can be a $\pi^0$, and includes the $X_s$ mass below $1.8\,\mathrm{GeV}/c^2$. After assuming that modes containing a $K_L^0$ have equal branching fractions to corresponding $K_S^0$ modes, both experiments account for $\sim 70\%$ of $B$ decays in their measured $X_s$ ranges. In order to reduce the semileptonic decay backgrounds, the analyses exploit the fact that energy is carried away by two or more neutrinos. Belle uses the total visible energy and missing mass, while *BABAR* uses the missing energy in the rest of the event (ROE) excluding the $X_s\ell^+\ell^-$ candidate, as well as the $m_{\rm ES}$ of the ROE. If the two leptons originate from semileptonic background, they may have displaced vertices, which is used for further discrimination.

As with the exclusive $B \to K^{(*)}\ell^+\ell^-$ analysis, it is necessary to remove the di-lepton mass ranges around the $J/\psi$ and $\psi'$. In both analyses $e^+e^-$ pairs with masses below $0.2\,\mathrm{GeV}/c^2$ are removed. This makes the di-muon and di-electron samples consistent, and removes the theoretically less interesting region dominated by the virtual photon contribution.

Belle measures $31.8 \pm 10.2$ $X_s e^+e^-$ and $36.3 \pm 9.3$ $X_s\mu^+\mu^-$ signal events with a sample of 152M $B\overline{B}$, while *BABAR* measures $29.2 \pm 8.4$ $X_s e^+e^-$ and $11.2 \pm 6.3$ $X_s\mu^+\mu^-$ events with the smaller sample of 89M $B\overline{B}$. There are experimental systematic uncertainties associated with the background subtraction and the reconstruction efficiency, which total about 10%. However the dominant systematic uncertainties come from the modeling of the $X_s$ system (as they did in $B \to X_s\gamma$). The fractions of exclusive $B \to K\ell^+\ell^-$ and $B \to K^*\ell^+\ell^-$ are varied, as well as the missing fractions of final states in the mass range above $1.1\,\mathrm{GeV}/c^2$. Finally there is an extrapolation to the full $X_s$ mass range, which uses a spectral shape of a Fermi-motion model (Ali, Hiller, Handoko, and Morozumi, 1997) with parameters determined from analyses of inclusive $B \to X_s\gamma$ and $B \to X_c\ell\overline{\nu}$. The $q^2$ distribution is mod-eled based on (Ali, Hiller, Handoko, and Morozumi, 1997; Ali, Lunghi, Greub, and Hiller, 2002; Kruger and Sehgal, 1996) in which the $J/\psi$ and $\psi(2S)$ regions are interpolated as if these contributions do not exist. The model-dependent effects give a total systematic error of 20%, still slightly smaller than the statistical error. The resulting inclusive branching fractions are given in Table 17.9.14. In Figure 17.9.15 Belle's results on the differential branching fractions for $B \to X_s \ell^+ \ell^-$ as functions of $M_{X_s}$ and $q^2$ are shown. With more events the measurements could be broken into bins of di-lepton mass squared, and the forward-backward asymmetry $A_{\rm FB}$ could be studied.

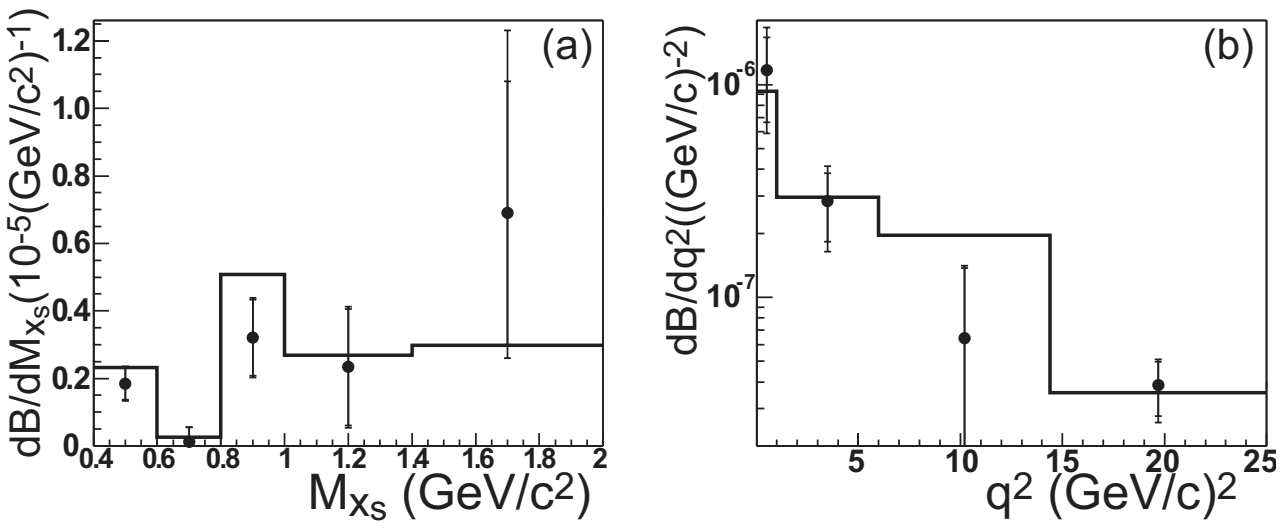

**Figure 17.9.15.** From (Iwasaki, 2005). Belle's results on the differential branching fractions for $B \to X_s \ell^+ \ell^-$ as functions of (a) $M_{X_s}$ and (b) $q^2$. All branching fractions have been interpolated through the vetoed $J/\psi$ and $\psi(2S)$ regions. The $M_{X_s}$ results correspond to all $q^2 > 0.04\,\mathrm{GeV}^2/c^2$, while the $q^2$ results follow an extrapolation to all values of $M_{X_s}$. Inner (outer) error bars are statistical (total) errors. Histograms represent the SM-based predictions described in the text.

The results for the inclusive $b \to s \ell^+ \ell^-$ branching fraction are consistent with the SM prediction (*e.g.*, $\mathcal{B}(B \to X_s \mu^+ \mu^-) = (4.20 \pm 0.70) \times 10^{-6}$ assuming interpolation over the veto region (Ali, Lunghi, Greub, and Hiller, 2002)).[92] They can be interpreted as giving a preference to a negative sign for Wilson coefficient $C_7$, where the inclusive $b \to s\gamma$ branching fraction only determines the magnitude of $C_7$ and not the sign.

Belle has reported an unpublished preliminary result with 657M $B\overline{B}$, with several improvements in the analysis (Iijima, 2010). The largest improvement is that the $X_s$ mass range is divided into the $K$, $K^*$ and high mass range, and the high mass range alone has been measured with $3\sigma$ significance. In addition, several new background sources have been identified. They include a semileptonic $B$ decay background that peaks in the $m_{\rm ES}$ distribution due to one additional misidentified lepton that compensates the missing neutrino, and contributions from higher $c\overline{c}$ resonances that were disregarded in the previous analyses. Although the preliminary results have been used in the HFAG averages and also in some literature, we do not include them in Table 17.9.14.

In addition, *BABAR* has submitted for publication an updated measurement of $B \to X_s \ell^+ \ell^-$ based on 471M $B\overline{B}$ events (Lees, 2014). Along with the results in the usual $q^2$ bins, results are provided for the $1 < q^2 < 6\,\mathrm{GeV}^2/c^2$ range suitable for comparison to the most precise SM prediction.

[92] The more recent theory publications used for Eqs 17.9.70 and 17.9.71 do not quote values for the full $q^2$ range.

#### 17.9.7.7 $B \to \pi \ell^+ \ell^-$

The $B \to \pi \ell^+ \ell^-$ decay mode has been the first exclusive decay mode utilized in the search for the $b \to d \ell^+ \ell^-$ transition. The analysis for $B \to \pi \ell^+ \ell^-$ is almost identical to that for $B \to K \ell^+ \ell^-$ (Section 17.9.7.1). Tight charged-pion identification, similar to that used for the measurement of $B \to \rho\gamma$ (Section 17.9.4.1), is applied to the pion in $B^+ \to \pi^+ \ell^+ \ell^-$. It is necessary to account for misidentified $B^+ \to K^+ \ell^+ \ell^-$ events, which give a peaking background in $m_{\rm ES}$, but are shifted to lower $\Delta E$. Both $B$ Factories also include $B^0 \to \pi^0 \ell^+ \ell^-$ in their searches.

*BABAR* (Aubert, 2007ax) has analyzed 230M $B\overline{B}$ and finds one candidate in the $(m_{\rm ES}, \Delta E)$ signal region in each of $\pi^+ e^+ e^-$, $\pi^+ \mu^+ \mu^-$ and $\pi^0 e^+ e^-$. This is consistent with the expectations from background. Defining the isospin-constrained branching fraction

$$\begin{aligned} \mathcal{B}(B \to \pi \ell^+ \ell^-) &\equiv \mathcal{B}(B^+ \to \pi^+ \ell^+ \ell^-) \qquad (17.9.73) \\ &= 2 \frac{\tau_{B^+}}{\tau_{B^0}} \mathcal{B}(B^0 \to \pi^0 \ell^+ \ell^-)\ , \end{aligned}$$

they set an upper limit of:

$$\mathcal{B}(B \to \pi \ell^+ \ell^-) < 9.1 \times 10^{-8}\,. \qquad (17.9.74)$$

Belle (Wei, 2008a) has analyzed 657M $B\overline{B}$ and also finds a few candidate events. A fit gives a small excess with a significance of $1.2\sigma$. They set an upper limit of:

$$\mathcal{B}(B \to \pi \ell^+ \ell^-) < 6.2 \times 10^{-8}\,. \qquad (17.9.75)$$

The SM prediction of $2 \times 10^{-8}$ is not far below these limits, and the backgrounds are quite manageable, and in fact the first observation for the charged mode $B^+ \to \pi^+ \mu^+ \mu^-$ has been recently reported by LHCb (Aaij et al., 2012e) with a branching fraction consistent with the SM.

### 17.9.8 Electroweak penguin decays $b \to s(d)\nu\overline{\nu}$

The $b \to s\nu\overline{\nu}$ decays are described by an electroweak penguin diagram including a $Z^0$ boson or a $W^+W^-$ box diagram, with Wilson coefficients for the vector and axial-vector parts $C_9$ and $C_{10}$. Unlike $b \to s\ell^+\ell^-$, there is no contribution from a virtual photon penguin diagram ($C_7$). There is also no contribution from $c\overline{c}$ resonances. Measuring the branching fractions of these decays provides a powerful test of new physics complementary to other rare $B$ decays (Altmannshofer, Buras, Straub, and Wick, 2009).

The experimental challenge is to identify a $B$ decay to an $X_s$ system and two missing neutrinos. This is similar to $B^- \to \tau^- \overline{\nu}$ decays, which have been first observed at the $B$ Factories. All that has to be done is to replace the observable $\tau$ decay products ($\pi$, $\rho$, $e$, $\mu$), with a $K$ or $K^*$ meson. However, the SM predictions for $B \to K\nu\overline{\nu}$ or $B \to K^*\nu\overline{\nu}$ branching fractions are rather small, $4 \times 10^{-6}$ and $13 \times 10^{-6}$, respectively. This makes suppression of the background from semileptonic $B$ decays more difficult.

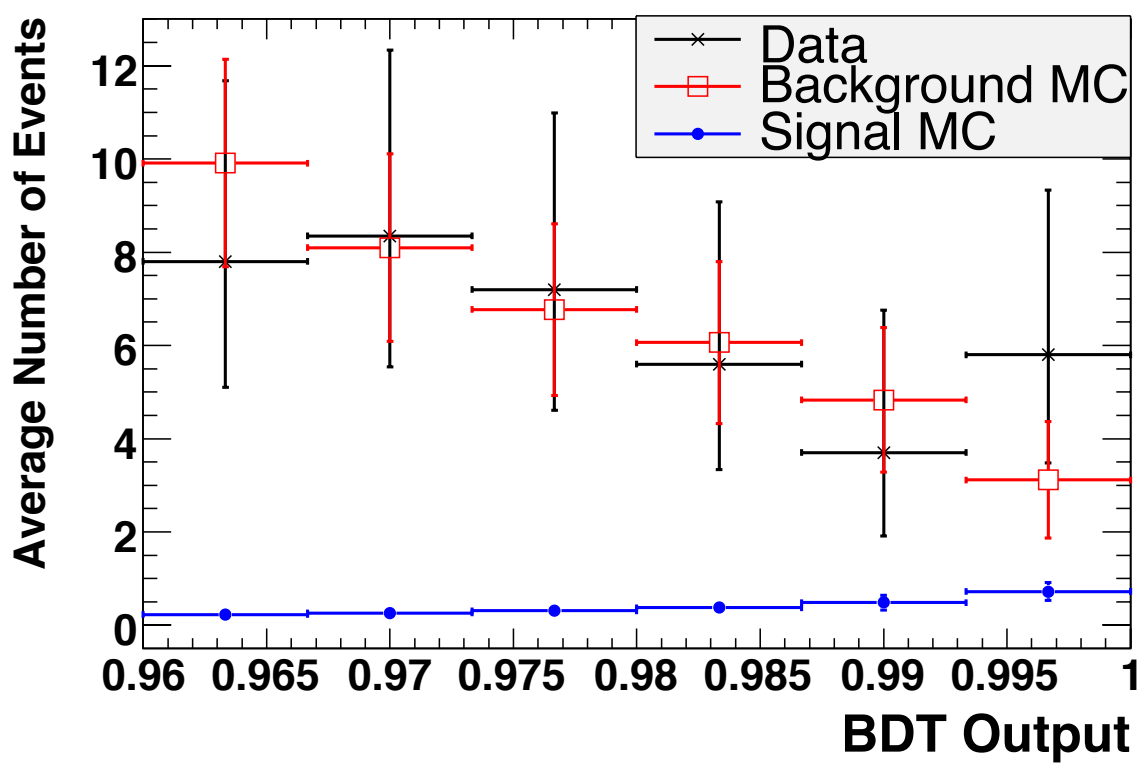

**Figure 17.9.16.** From (del Amo Sanchez, 2010p). Boosted Decision Tree output from the *BABAR* search for $B^+ \to K^+\nu\overline{\nu}$, along with expected SM-signal and background contributions. Errors on the data are statistical, while those on the expected numbers include systematics.

The recoil-$B$ method is used, in which either fully-reconstructed hadronic $B$ decays, or semileptonic $B \to D^{(*)}\ell\nu$ decays, are used as the tag (Section 7.4). The remaining charged tracks and neutral clusters then have to be consistent with the $X_s$ system being searched for. The most powerful variables for suppressing background are associated with the missing energy carried by the neutrinos, and the lack of extra energy in the detector. Information on the momentum, charge and flavor of the recoil tag can be correlated with the $X_s$ system to further reduce the backgrounds.

Belle (Chen, 2007b) uses hadronic $B$ tags to search for $B \to h^{(*)}\nu\overline{\nu}$ where $h^{(*)}$ includes charged and neutral $K$, $K^*$, $\pi$, $\rho$ and $\phi$. They reconstruct 788k charged $B$ and 491k neutral $B$ decays from a sample of 535M $B\overline{B}$ events, with an overall tag efficiency of $2.5 \times 10^{-3}$. They observe between 1 and 30 candidate events in the different $h^{(*)}$ final states. These yields are consistent with the expectations from backgrounds, so they set upper limits at 90% C.L. between $4.4 \times 10^{-4}$ for $\rho^0\nu\overline{\nu}$ and $1.4 \times 10^{-5}$ for $K^+\nu\overline{\nu}$.

The best limit on $K\nu\overline{\nu}$ comes from *BABAR* (del Amo Sanchez, 2010p), also using the hadronic $B$ tag method. This analysis uses bagged decision trees with 26 (38) inputs to separate signal and background in the $K^+$ ($K^0$) modes as illustrated in Figure 17.9.16. The $K^+$ search is separated into high and low-$q^2$ regions, corresponding to low and high kaon momenta. The backgrounds are very large at high-$q^2$, so the sensitivity mainly comes from the low-$q^2$ region. This introduces some model-dependence into the extraction of the upper limit. The quoted upper limits at 90% C.L. are $5.6 \times 10^{-5}$ for $K^0\nu\overline{\nu}$ and $1.3 \times 10^{-5}$ for $K^+\nu\overline{\nu}$. *BABAR* also reports a search using semileptonic $B$ tags (Aubert, 2009aq).

The *BABAR* search for $K^*\nu\overline{\nu}$ uses a combination of hadronic and semileptonic tags (Aubert, 2008an). The efficiency is slightly lower with hadronic tags, but the background with the semileptonic tags is significantly higher. A fit to the distribution of extra energy in the events leads to comparable upper limits from the two types of tags, and yields combined upper limits at 90% C.L. of $12 \times 10^{-5}$ for $K^{*0}\nu\overline{\nu}$ and $8 \times 10^{-5}$ for $K^{*+}\nu\overline{\nu}$. These limits are lower than those reported by Belle (Chen, 2007b).

The experimental limits are already about 3 and 6 times the SM predictions for $K\nu\overline{\nu}$ and $K^*\nu\overline{\nu}$, respectively, but the backgrounds are severe. Initial studies of what could be done at a super flavor factory suggest that a data sample of about $50\,\mathrm{ab}^{-1}$ will be needed to observe either of these decays.

## 17.10 $B^+ \to \ell^+\nu(\gamma)$ and $B \to D^{(*)}\tau\nu$

***Editors:***
*Steven Robertson (BABAR)*
*Toru Iijima (Belle)*

**Additional section writers:**
*Dana Lindemann*

### 17.10.1 Overview

In this section, we review the measurements of purely leptonic decays, $B^+ \to \ell^+\nu$ ($\ell = e, \mu, \tau$), and the semileptonic $B$ decays $B \to D^{(*)}\tau\nu$. As $b \to u$ and $b \to c$ quark transitions, these processes depend on the magnitudes of the CKM matrix elements $V_{ub}$ and $V_{cb}$, respectively, however both have potential sensitivity to physics beyond the SM. In extensions of the SM which include an expanded Higgs sector, in particular the type-II two Higgs doublet model (2HDM) such as in the minimal supersymmetric extension of the Standard Model (MSSM), these processes are potentially sensitive to a charged Higgs boson ($H^\pm$). A number of benchmark new physics models, such as 2HDM and MSSM, are discussed in Section 25.2. The presence of the $H^\pm$ can impact the experimentally observed branching fractions for these decay modes and, in the case of $B \to D^{(*)}\tau\nu$, also the kinematic distributions of final state particles. Figure 17.10.1 shows Feynman diagrams for these tree level processes.

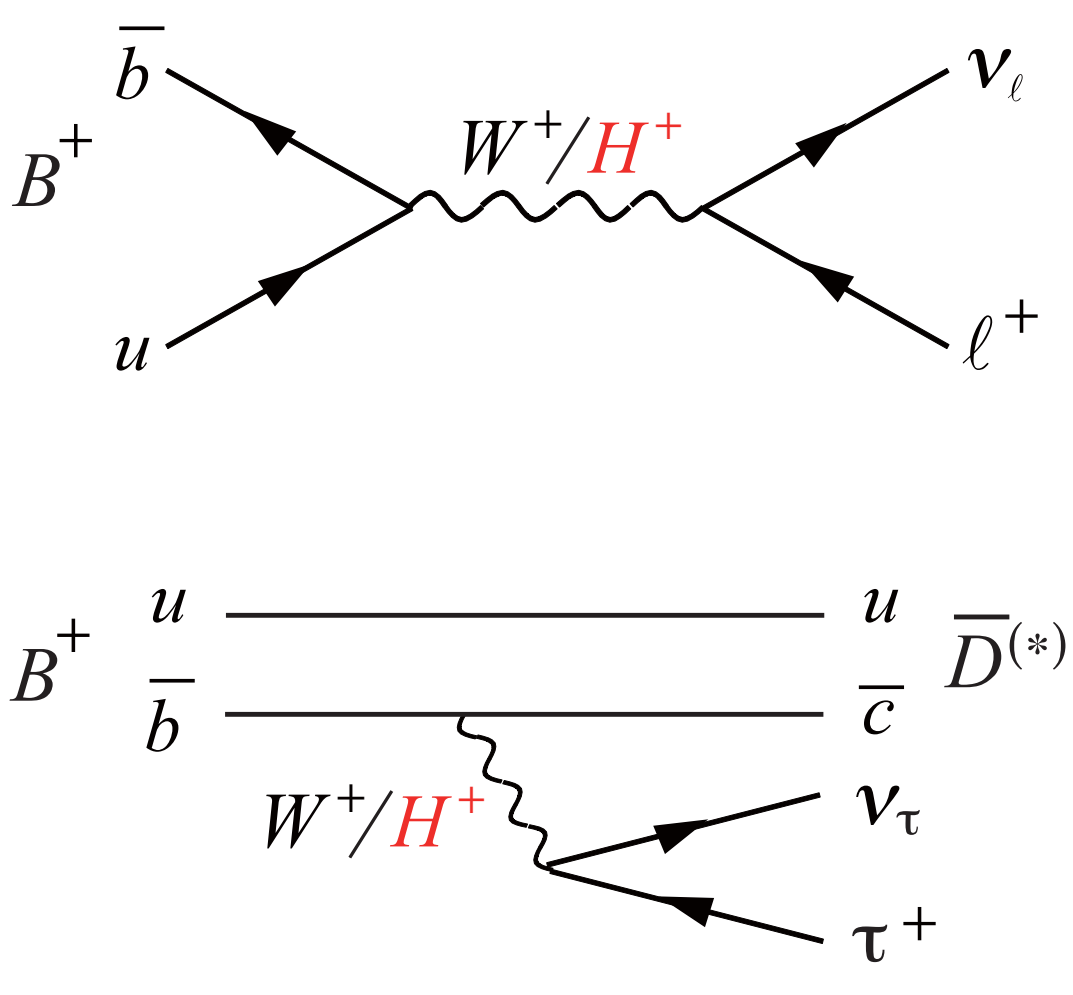

**Figure 17.10.1.** Feynman diagrams of $B^+ \to \ell^+\nu_\ell$ (top) and $B^+ \to \overline{D}^{(*)}\tau^+\nu_\tau$ (bottom).

Since $B^+ \to \ell^+\nu$ proceeds via a quark annihilation process with no hadrons in the final state, all hadronic effects are encapsulated in the $B$ decay constant $f_B$,

$$\langle 0|\overline{b}(0)\gamma_\mu\gamma_5 q(0)|B(p)\rangle = ip_\mu f_B \quad , \qquad (17.10.1)$$

which can be interpreted as the wave function of the light quark at the location of the $b$ quark. In contrast, in semileptonic $B \to D^{(*)}\tau\nu$ decays the hadronic transition is described by form factors, which are functions of the momentum transfer squared, $q^2$, resulting in larger theoretical uncertainties in the decay kinematics and branching fractions.

The effective Hamiltonian describing $B \to D^{(*)}\tau\nu$ and $B \to \tau\nu$ transitions mediated by $W^\pm$ or $H^\pm$ can be written as,

$$\begin{aligned}\mathcal{H}_{eff} = &\frac{G_F}{\sqrt{2}} V_{\rm qb}\{[\overline{q}\gamma^\mu(1-\gamma_5)b][\overline{\tau}\gamma_\mu(1-\gamma_5)\nu_\tau] \\ &- \frac{M_b M_\tau}{M_B^2}\overline{q}[g_S + g_P\gamma_5]b[\overline{\tau}(1-\gamma_5)\nu_\tau]\} \\ &+ \text{h.c.}\,, \end{aligned} \qquad (17.10.2)$$

where $G_F$ is the Fermi coupling constant, $V_{\rm qb}$ is the CKM matrix element and $M_B$ is the $B$ meson mass. The first term corresponds to the SM $W^\pm$ couplings. The second term, which can occur in beyond-SM models, represents scalar couplings to $H^\pm$. Since these couplings are proportional to the fermion masses, which are relatively large for $B$ mesons and tau leptons, it is natural to look for new physics in leptonic or semileptonic $B$ decays involving tau leptons.

In the MSSM, the couplings $g_{S,P}$ in Eq. (17.10.2) are written as

$$g_S = g_P = \frac{M_B^2 \tan^2\beta}{M_H^2}\frac{1}{(1+\epsilon_0\tan\beta)(1-\epsilon_\tau\tan\beta)}\,, \qquad (17.10.3)$$

where $\tan\beta$ is the ratio of the two Higgs vacuum expectation values and $M_H$ is the charged Higgs boson mass. The parameters $\epsilon_{0,\tau}$ arise from sparticle loop contributions and their values depend on other MSSM parameters. These are typically expected to be of $\mathcal{O}(10^{-2})$. Since these contributions to $g_{S,P}$ are relatively small for moderate values of $\tan\beta$, and in the absence of other experimental evidence for SUSY, studies of $B \to \tau\nu$ and $B \to D^{(*)}\tau\nu$ are usually interpreted in the context of the simpler extension of the SM with only the addition of a second Higgs doublet (that is, with only the Type-II 2HDM and no other heavy new physics contributions). In this case, $\epsilon_0 = \epsilon_\tau = 0$, and measurements of $B \to (D^{(*)})\tau\nu$ provide information on $\tan\beta/M_H$.

Experimentally, it is challenging to study these processes due to the presence of neutrinos in the final state. The decays $B^+ \to \tau^+\nu$ and $B \to D^{(*)}\tau\nu$ involve two or more neutrinos and hence cannot be fully constrained kinematically. Although the $B^+ \to e^+\nu$ and $B^+ \to \mu^+\nu$ modes have only a single neutrino in the final state, they have very small branching fractions compared with $B^+ \to \tau^+\nu$, due to helicity suppression. However, the $B^+ \to \tau^+\nu$ branching fraction is roughly two orders of magnitude smaller than the semileptonic $B \to D^{(*)}\tau\nu$ modes due to the relative sizes of $|V_{ub}|$ and $|V_{cb}|$, and the contribution of $f_B$.

At $B$ Factories, one can fully reconstruct one of the $B$ mesons produced in an $\Upsilon(4S) \to B\overline{B}$ event, referred to as the "tag $B$" ($B_{\rm tag}$), and examine the properties of the remaining particles in the event, which are collectively referred to as the "signal $B$" ($B_{\rm sig}$), to look for evidence of a signal decay (see Chapter 7). This method strongly suppresses the combinatorial background and provides a unique identification of the decay daughters of the signal $B$ decay. The disadvantage of this method is the low efficiency of the $B_{\rm tag}$ reconstruction, which is at the level of $\mathcal{O}(0.1)\%$. In spite of this, the high luminosity $B$ Factories have provided a sufficiently large number of events to enable measurements of these decays for the first time.

In this section we describe the theoretical and experimental status of $B^+ \to \ell^+\nu$ and $B \to D^{(*)}\tau\nu$ studies. A theoretical introduction to leptonic decays is presented in Section 17.10.2.1. The experimental status of $B^+ \to \tau^+\nu$ and $B^+ \to \ell^+\nu$ (with $\ell = e, \mu$) is described in Sections 17.10.2.2 and 17.10.2.3, respectively. Searches for radiative leptonic decays are discussed in Section 17.10.2.4. $B \to D^{(*)}\tau\nu$ are likewise presented in the remaining sections, with a brief theory introduction in Section 17.10.3 followed by a description of experimental measurements in Section 17.10.3.1 and interpretation of these results in Section 17.10.3.2. Some comments on the current status and future prospects for studies of these decays conclude this section, while additional interpretation is provided in the context of global fits to the Unitarity Triangle in Chapter 25.

### 17.10.2 $B^+ \to \ell^+\nu(\gamma)$

#### 17.10.2.1 Theory of leptonic decays

In the SM, the purely leptonic decay $B^+ \to \ell^+\nu$ proceeds via the annihilation of $\bar{b}$ and $u$ quarks to a $W^+$ boson (see Figure 17.10.1). The branching fraction is given by

$$\mathcal{B}(B^+ \to \ell^+\nu)_{\rm SM} = \frac{G_F^2 M_B M_\ell^2}{8\pi}\left(1 - \frac{M_\ell^2}{M_B^2}\right)^2 \times f_B^2 |V_{ub}|^2 \tau_B \ , \qquad (17.10.4)$$

where the $M_\ell$ is is the mass of the lepton, and $\tau_B$ is the $B$ meson lifetime. The $B$ meson decay constant $f_B = 0.191 \pm 0.009$ GeV is obtained from the most recent lattice QCD calculations (Na et al., 2012). The helicity suppression of the leptonic decays can be seen in the lepton mass dependence in Eq. (17.10.4). The expected branching fraction for the $\tau$ mode is

$$\mathcal{B}_{\rm SM}(B^+ \to \tau^+\nu) = (1.01 \pm 0.29) \times 10^{-4} \ , \qquad (17.10.5)$$

using $|V_{ub}| = (3.95 \pm 0.38_{\rm exp} \pm 0.39_{\rm th}) \times 10^{-3}$, which is an average of $|V_{ub}|$ values determined using charmless semileptonic $B$ decay data, see Eq. (17.1.70). Due to the relatively small mass of the $e$ and $\mu$ compared with the $\tau$, these modes are suppressed by factors of $1.05 \times 10^{-7}$ and $4.49 \times 10^{-3}$, respectively, relative to the $\tau$ mode.

Within the Type-II 2HDM, the addition of the charged Higgs boson in Eq. (17.10.2) and Eq. (17.10.3) modifies the $B^+ \to \ell^+\nu$ branching fraction (Hou, 1993),

$$\mathcal{B}(B^+ \to \ell^+\nu)_{\rm 2HDM} = \mathcal{B}(B^+ \to \ell^+\nu)_{\rm SM} \times r_H \ , \ (17.10.6)$$

where the ratio $r_H$ is given by

$$r_H = (1 - M_B^2 \tan^2\beta / M_H^2)^2 \ . \qquad (17.10.7)$$

The interference between the SM $W^\pm$ and $H^\pm$ contributions is destructive. Consequently, the charged Higgs contribution suppresses the branching fraction relative to the SM expectation, resulting in $r_H < 1$, unless the $H^\pm$ is sufficiently large that it dominates the SM $W^\pm$ contribution. The case where $M_B^2 \tan^2\beta / M_H^2 = 2$ is indistinguishable from the SM. It is notable that Eq. (17.10.7) applies equally to the other leptonic decay modes, $B^+ \to \mu^+\nu$ and $B^+ \to e^+\nu$. As the $H^\pm$ is expected to decrease the observed branching fraction, much of the present constraint on charged Higgs bosons results from the *lower* bound on the experimental value of the $B^+ \to \tau^+\nu$ branching fractions rather than the *upper* bound on the branching fraction. A much weaker bound is currently obtained from $B^+ \to \mu^+\nu$ decays because only upper limits on its branching fraction have been reported.

The radiative decays $B^+ \to \ell^+\nu_\ell\gamma$ are also of interest since the presence of the radiated photon can remove the helicity suppression of the purely leptonic modes, possibly by coupling the spin-0 $B$ meson to the spin-1 $W^\pm$ boson through an intermediate off-shell state (Burdman, Goldman, and Wyler, 1995). Consequently, the predicted branching fractions of $B^+ \to e^+\nu_\ell\gamma$ and $B^+ \to \mu^+\nu_\ell\gamma$ are considerably larger than the corresponding non-radiative modes, in spite of an additional suppression by the factor $\alpha_{\rm EM}$. The branching fractions for $B^+ \to \ell^+\nu_\ell\gamma$ (with $\ell = e, \mu, \tau$) are predicted to be of order $10^{-6}$ independent of the lepton type, making these modes potentially accessible at the $B$ Factories. They potentially provide an additional method to access $|V_{ub}|$, and they are also a potential background to the non-radiative mode searches.

The decay rate for $B^+ \to \ell^+\nu_\ell\gamma$ is given by

$$\frac{d\mathcal{B}}{dE_\gamma} = \frac{\alpha_{\rm EM} G_F^2 |V_{ub}|^2}{48\pi^2} M_B^5 \tau_B \left[f_A^2(E_\gamma) + f_V^2(E_\gamma)\right](1-y)y^3 \qquad (17.10.8)$$

where $y = 2E_\gamma/M_B$. The axial-vector and vector $B \to \gamma X$ form factors, $f_A$ and $f_V$, respectively, are assumed to be equal in most models. The branching fraction can be approximated as (Korchemsky, Pirjol, and Yan, 2000)

$$\mathcal{B}(B^+ \to \ell^+\nu_\ell\gamma) \approx \frac{\alpha_{\rm EM} G_F^2 |V_{ub}|^2}{288\pi^2} f_B^2 M_B^5 \tau_B \left(\frac{Q_u}{\lambda_B} - \frac{Q_b}{M_b}\right)^2 , \qquad (17.10.9)$$

where $Q_i$ is the quark charge, and $\lambda_B$ is the first inverse moment of the $B$-meson wave function. This last parameter plays an important role in QCD factorization (Descotes-Genon and Sachrajda, 2003; Lunghi, Pirjol, and Wyler, 2003). It also enters into calculations of the $B \to \pi X$ form factor at zero momentum transfer and the branching fractions of two-body hadronic $B$-meson decays such as $B \to \pi\pi$, a benchmark channel for measuring the CKM angle $\phi_2$ (Le Yaouanc, Oliver, and Raynal, 2008). However, $\lambda_B$ has a large theoretical uncertainty, so $B^+ \to \ell^+\nu_\ell\gamma$ is a useful decay for obtaining a clean measurement of $\lambda_B$.

### 17.10.2.2 $B^+ \to \tau^+\nu$ measurements

*Methodology common to Belle and BABAR*

Among the leptonic $B$ decays, $B^+ \to \tau^+\nu$ has the largest branching fraction and, in spite of the difficulties associated with multiple neutrinos in the final state, was the first of these modes to be successfully measured at the $B$ Factories. Both Belle and *BABAR* use a similar analysis method in which they fully reconstruct the accompanying $B$ meson ($B_{\rm tag}$) using either hadronic or semileptonic decays, and examine the rest of the event to search for a $B^+ \to \tau^+\nu$ decay. In both experiments, analyses employing hadronic- and semileptonic-tag methods were performed and published as separate measurements. Since there is essentially no overlap between the tag samples, the analyses are statistically independent $B^+ \to \tau^+\nu$ and the two results from each collaboration can be combined into a single branching fraction measurement.

Details of $B_{\rm tag}$ reconstruction are described in Section 7. Analyses using semileptonic tags have higher efficiency than the hadronic tag searches, but suffer from a lower signal-to-background ratio. This is a consequence of the less stringent kinematic constraints associated with the presence of the undetectable tag-$B$ neutrino. As a result, searches using the hadronic and semileptonic tag methods employ somewhat different optimizations.

Once a $B_{\rm tag}$ has been reconstructed, using either the hadronic or semileptonic tag method, the selection of $B^+ \to \tau^+\nu$ candidates exploits the low multiplicity and missing energy signatures of the signal mode. Since the $\tau^+$ decays into final states in which one or two neutrinos accompany either hadrons or a charged lepton, it is not possible to reconstruct the two-body kinematics of $B^+ \to \tau^+\nu$ from the final state particles. Tau decays to leptons, $\tau^+ \to \ell^+\nu\overline{\nu}$ ($\ell = e, \mu$) comprise approximately 35% of the branching fraction, while decays to $\pi^+\overline{\nu}$, $\pi^+\pi^0\overline{\nu}$, $\pi^+\pi^0\pi^0\overline{\nu}$ and $\pi^+\pi^-\pi^+\overline{\nu}$ contribute approximately 11%, 25%, 9% and 10%, respectively. The $\pi^+\pi^0\overline{\nu}$ and $3\pi\overline{\nu}$ modes proceed primarily through the $\rho(770)$ and $a_1(1260)$ resonances, hence mass constraints can be imposed on the pions to provide background rejection. However, these states are broad and so the suppression is modest. Since modes decaying to kaons (charged or neutral) make up only $\sim 1\%$ of $\tau$ decays, a kaon veto is usually applied to suppress large $B\overline{B}$ backgrounds involving charm mesons. The leptonic modes have a clean signature, but because the lepton has relatively low momentum, efficient and high-purity particle identification is needed. In tau decay modes with charged hadrons the pions are usually efficiently identified but large backgrounds must be overcome. Modes with one or more neutral pions have both low reconstruction efficiency and high backgrounds. The entire $B^+ \to \tau^+\nu$ signal selection is optimized on a mode-by-mode basis so as to maximize the overall sensitivity. Due to differences in detectors and data samples, *BABAR* and Belle do not utilize exactly the same set of tau decay modes in their respective searches.

Charged particles from tau decays are selected as tracks that are not identified as the daughters of the reconstructed $B_{\rm tag}$. It is required that exactly three such tracks are present in the case of the $\pi^+\pi^-\pi^+\overline{\nu}$ final state and exactly one track otherwise. The summed charge of these tracks is required to be consistent with that expected based on the reconstructed $B_{\rm tag}$. Particle identification criteria are applied to the track(s) to distinguish leptons and pions, and to veto kaons. In the case of a single identified charged pion, $\pi^0$ candidates are reconstructed from $\gamma\gamma$ combinations which do not overlap with $B_{\rm tag}$ daughters. These are combined with the $\pi^+$ and constraints are applied to identify $\rho(770)$ or $a_1(1260)$ candidates. Events containing identified leptons are vetoed if a $\pi^0$ is also reconstructed in the event. Once each event has been uniquely classified as one of the candidate tau decay modes and $\pi^0$ candidates have been associated to the tau mode when applicable, there should be no additional energy deposition in the electromagnetic calorimeter in signal events. In practice, a small amount of energy is usually present due to accelerator beam backgrounds and reconstruction effects. In particular, hadronic shower fragments ("split-offs") from pions or kaons interacting in the calorimeter are sometimes reconstructed as separate calorimeter clusters rather than being associated with the originating particle. The most powerful variable for separating signal and background is the sum of the energies of neutral clusters that are not associated with decays of the $B_{\rm tag}$ or the tau. This quantity is denoted as $E_{\rm ECL}$ in Belle and $E_{\rm extra}$ in *BABAR* (and hereafter referred to as $E_{\rm extra}$). The specific definition of $E_{\rm extra}$ depends of the low energy threshold applied to calorimeter clusters and differs between analyses. In *BABAR* analyses, cluster thresholds range from $30-100$ MeV, while in Belle they are chosen to be 50 MeV for the barrel and 100 (150) MeV for the forward (backward) end-cap ECL. Signal events are expected to peak at or near $E_{\rm extra} = 0$. In contrast, many background events contain one or more additional neutral clusters from unreconstructed $\pi^0$ mesons or other particles. Consequently, for backgrounds $E_{\rm extra}$ extends to higher values. The $E_{\rm extra}$ distributions are estimated based on MC simulations. In order to reproduce the effects of beam backgrounds, data recorded with random triggers are overlaid on simulated events in both *BABAR* and Belle analyses. Furthermore, to take into account the possible difference between MC and data description of split-off showers, the signal $E_{\rm extra}$ distribution is calibrated, both for hadronic and semileptonic $B_{\rm tag}$ analyses, using "double tagged" event samples. In these events, a $B_{\rm tag}$ is reconstructed as described above, but a second hadronic or semileptonic $B$ decay is also exclusively reconstructed in the same event using tracks and calorimeter clusters not already assigned to the $B_{\rm tag}$.

Background from $e^+e^- \to \tau^+\tau^-$ and other continuum processes are suppressed using signal-mode-specific criteria relating to event shapes, in particular the ratio of the second and zeroth Fox-Wolfram moments ($R_2$, see Chapter 9), and the angle between the thrust axis computed using the daughters of the $B_{\rm tag}$ and the thrust axis computed using all other track and clusters in the event. $B\overline{B}$ backgrounds in which the $B_{\rm tag}$ has been correctly reconstructed are suppressed by using kinematic variables of the signal track (and $\pi^0$ in the case of $\tau^+ \to \rho^+\overline{\nu}$) and any additional calorimeter clusters. These include the CM frame momentum of the signal track ($p^*_{\rm trk}$) and the angle ($\cos\theta_{miss}$) of the missing momentum vector of the event with respect to the beam axis, computed using the $B_{\rm tag}$ four-vector, the signal track, and any additional calorimeter clusters.

A blind analysis (Chapter 14) procedure was adopted and the signal selection was optimized to obtain the smallest uncertainty on the measured branching fraction. The signal yield is extracted using an extended unbinned maximum likelihood fit to the $E_{\rm extra}$ distribution for the various tau decay signal modes. In the fit, the background yields are allowed to vary independently, while the signal yields in all of the tau modes are constrained to a common $B^+ \to \tau^+\nu$ branching fraction.

### *Belle results*

The Belle collaboration reported the first evidence of the $B^+ \to \tau^+\nu$ decay by applying the hadronic tagging method on a sample of $449 \times 10^6$ $B\overline{B}$ pairs. The extracted signal yield is $N_S = 24.1^{+7.6}_{-6.6}(\rm stat)^{+5.5}_{-6.3}(\rm syst)$ events, corresponding to 3.5 $\sigma$ significance. The branching fraction is measured to be $\mathcal{B}(B^+ \to \tau^+\nu) = (1.79^{+0.56}_{-0.49}(\rm stat)^{+0.46}_{-0.51}(\rm syst))\times 10^{-4}$ (Ikado, 2006).

More recently, Belle has reported an updated result using a similar method on the full $\Upsilon(4S)$ data sample containing $772\times10^6$ $B\overline{B}$ pairs (Adachi, 2012b). This analysis has a number of significant improvements compared to the previous one, improved hadronic tagging efficiency (a factor of 2.2 times larger), and improved signal efficiency due to less restrictive selection criteria (a factor of 1.8 times larger). The $\tau$ lepton is identified in the $\tau^+ \to e^+\nu_e\overline{\nu}_\tau$, $\mu^+\nu_\mu\overline{\nu}_\tau$, $\pi^+\overline{\nu}_\tau$, and $\pi^+\pi^0\overline{\nu}_\tau$ decay channels. Multiple neutrinos in the final state are distinguished using the missing mass squared $M^2_{\rm miss} = (E_{\rm CM} - E_{B_{\rm tag}} - E_{B_{\rm sig}})^2/c^4 - |\boldsymbol{p}_{B_{\rm tag}} + \boldsymbol{p}_{B_{\rm sig}}|^2/c^2$, where $E_{B_{\rm sig}}$ and $\boldsymbol{p}_{B_{\rm sig}}$ are the energy and the momentum, respectively of the $B_{\rm sig}$ candidate in the CM frame. For the $B_{\rm sig}$ selection, the event is required to have no extra $\pi^0$ or $K^0_L$ candidates ("$K^0_L$ veto"). The signal yield is extracted from a two-dimensional extended maximum likelihood fit to $E_{\rm extra}$ and $M^2_{\rm miss}$. By combining the four $\tau$ decay modes, the extracted signal yield is $62^{+23}_{-22}(\rm stat) \pm 6(\rm syst)$ events, corresponding to a significance of 3.0 $\sigma$. The branching fraction is measured to be $\mathcal{B}(B^+ \to \tau^+\nu) = (0.72^{+0.27}_{-0.25}(\rm stat) \pm 0.11(\rm syst)) \times 10^{-4}$. Figure 17.10.2 (a) shows the $E_{\rm extra}$ distribution overlaid with the fit results for the sum of the individual $\tau$ decay modes.

Belle has also reported a result using the semileptonic tagging method, based on a sample of $657\times10^6$ $B\overline{B}$ events (Hara, 2010). In this analysis, $B_{\rm tag}$ candidates were reconstructed via $B^- \to D^{*0}\ell^-\overline{\nu}$ and $B^- \to D^0\ell^-\overline{\nu}$ decays, where $\ell$ is an electron or muon. $D^0$ mesons were reconstructed in the $K^-\pi^+$, $K^-\pi^+\pi^0$, and $K^-\pi^+\pi^-\pi^+$ modes. For the $B_{\rm sig}$, Belle considered $\tau^+$ decays to one charged particle and neutrinos, *i.e.* $\tau^+ \to \ell^+\nu_\ell\overline{\nu}_\tau$ and $\tau^+ \to \pi^+\overline{\nu}_\tau$. Figure 17.10.2 (b) shows the $E_{\rm extra}$ distribution overlaid with the fit results for the sum of the $\tau$ decay modes. Belle reported a clear excess of signal events in the region near zero and obtained a signal yield of $143^{+36}_{-35}$ events, corresponding to a significance of $3.6\sigma$. The branching fraction was determined to be $\mathcal{B}(B^+ \to \tau^+\nu) = (1.54^{+0.38}_{-0.37}(\rm stat)^{+0.29}_{-0.31}(\rm syst)) \times 10^{-4}$.

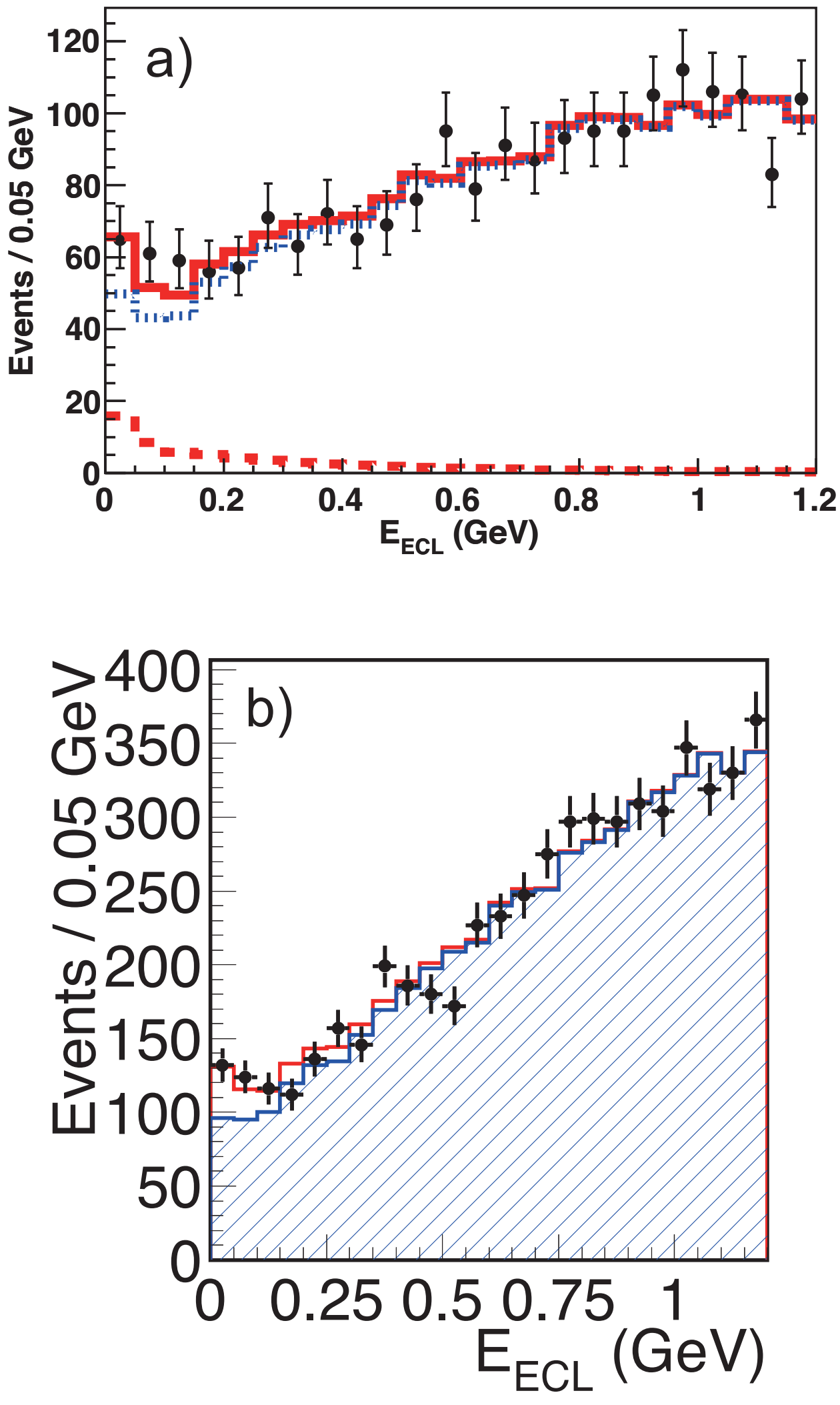

**Figure 17.10.2.** (a) Distribution of residual energy $E_{\rm ECL}$ (referred to as $E_{\rm extra}$ in the text) reported by Belle using hadronic tags (Adachi, 2012b). The solid circles with error bars are data. The solid histograms show the projection of the fit. The dashed and dotted histograms show the signal and background components, respectively. (b) The same distribution obtained using semileptonic tagged events (Hara, 2010). The points with error bars are data. The hatched histogram and solid open histogram are the background and the total signal plus background contributions, respectively.

*BABAR results*

*BABAR* has published searches for $B^+ \to \tau^+\nu$ using both the hadronic and semileptonic tag reconstruction methods. The most recent hadronic tag search is based on the full *BABAR* dataset of $467.8 \times 10^6$ $B\overline{B}$ events (Lees, 2013a). This search utilized the four tau decay channels $\tau^+ \to e^+\nu\overline{\nu}$, $\tau^+ \to \mu^+\nu\overline{\nu}$, $\tau^+ \to \pi^+\overline{\nu}$, and $\tau^+ \to \rho^+(\to \pi^+\pi^0)\overline{\nu}$, totaling approximately 70% of the total branching fraction. This search utilizes an expanded set of hadronic tag reconstruction modes (see Chapter 7) that includes $B^- \to J/\psi X^-$ along with lower purity modes to increase the overall tag reconstruction efficiency by almost a factor of two compared to previous searches by *BABAR*.

Exactly one charged track is required in addition to the reconstructed tag $B$ daughter particles. Events are classified into $\tau$ decay modes according to electron, muon and pion particle identification criteria applied to the track. If the signal track is identified as a $\pi^+$, the event is considered to be a $\tau^+ \to \rho^+\overline{\nu}$ candidate if a $\pi^0$ candidate satisfies the condition $115 < M_{\gamma\gamma} < 155$ MeV/$c^2$. Otherwise it is considered to be a $\tau^+ \to \pi^+\overline{\nu}$ candidate. Multivariate likelihood ratios are constructed, for each signal mode, for signal and background hypotheses. For the $\tau^+ \to e^+\nu\overline{\nu}$, $\tau^+ \to \mu^+\nu\overline{\nu}$ and $\tau^+ \to \pi^+\overline{\nu}$ modes, the likelihoods are constructed from the products of *p.d.f.*s for $p^*_{\rm trk}$ and $\cos\theta_{\rm miss}$. For the $\tau^+ \to \rho^+\overline{\nu}$ mode, the reconstructed invariant mass of the $\gamma\gamma$ combination and of the $\pi^+\pi^0$ combination are also used as inputs. The signal yield is extracted using an unbinned maximum likelihood fit to the $E_{\rm extra}$ distribution for the four signal modes. The combinatorial background *p.d.f.* is obtained from data in the $m_{ES}$ sideband region and combined with a $B^+B^-$ "peaking" component from MC to define an overall background *p.d.f.* to use in the fit. The fit yields a positive signal with a significance of approximately $3.8\sigma$ and a branching fraction of $\mathcal{B}(B^+ \to \tau^+\nu) = (1.83^{+0.53}_{-0.49}(\text{stat}) \pm 0.24(\text{syst})) \times 10^{-4}$. Figure 17.10.3 (a) shows the $E_{\rm extra}$ distribution obtained from this analysis. An earlier version of this search based on $383 \times 10^6$ $B\overline{B}$ events (Aubert, 2008c) and utilizing the same four tau decay modes reported a similar branching fraction central value $\mathcal{B}(B^+ \to \tau^+\nu) = 1.8^{+0.9}_{-0.8}(\text{stat}) \pm 0.4 \pm 0.2(\text{syst})$. However, the statistical significance ($2.2\sigma$) was not sufficient to provide compelling evidence for the signal decay.

*BABAR* has also performed searches for $B^+ \to \tau^+\nu$ using semileptonic tag reconstruction (Aubert, 2006a, 2007a, 2010a). The most recent *BABAR* study (Aubert, 2010a) was based on $458.9 \times 10^6$ $B\overline{B}$ events and analyzed the four tau decay modes $\tau^+ \to e^+\nu\overline{\nu}$, $\tau^+ \to \mu^+\nu\overline{\nu}$, $\tau^+ \to \pi^+\overline{\nu}$, and $\tau^+ \to \rho^+(\to \pi^+\pi^0)\overline{\nu}$. Two likelihood ratios are constructed from kinematic and event shape variables designed, respectively, for suppression of continuum backgrounds and non-signal $B\overline{B}$ backgrounds. Signal and background *p.d.f.*s are obtained from MC for each of the four signal channels and likelihood ratios are constructed from the product of the *p.d.f.*s. The selection is optimized to maximize the expected signal significance for each of the signal modes. Three variables are optimized: the outputs of the continuum and $B\overline{B}$ likelihood ratio selectors and $E_{\rm extra}$. In the optimized selection the cut on $E_{\rm extra}$ ranges from 200 MeV to 350 MeV and predicted background yields range from approximately 60 to 230 events, depending on the signal mode. The overall selection efficiency, including both $B_{\rm tag}$ reconstruction and signal selection, is at the level of $\sim 10^{-3}$. Figure 17.10.3 (b) shows the $E_{\rm extra}$ distribution obtained. A slight excess of events in data compared with the predicted background is found in each of the four signal modes resulting in a combined branching fraction central value of $\mathcal{B}(B^+ \to \tau^+\nu) = (1.7 \pm 0.8(\text{stat}) \pm 0.2(\text{syst})) \times 10^{-4}$ with an overall signal significance of approximately $2.3\sigma$.

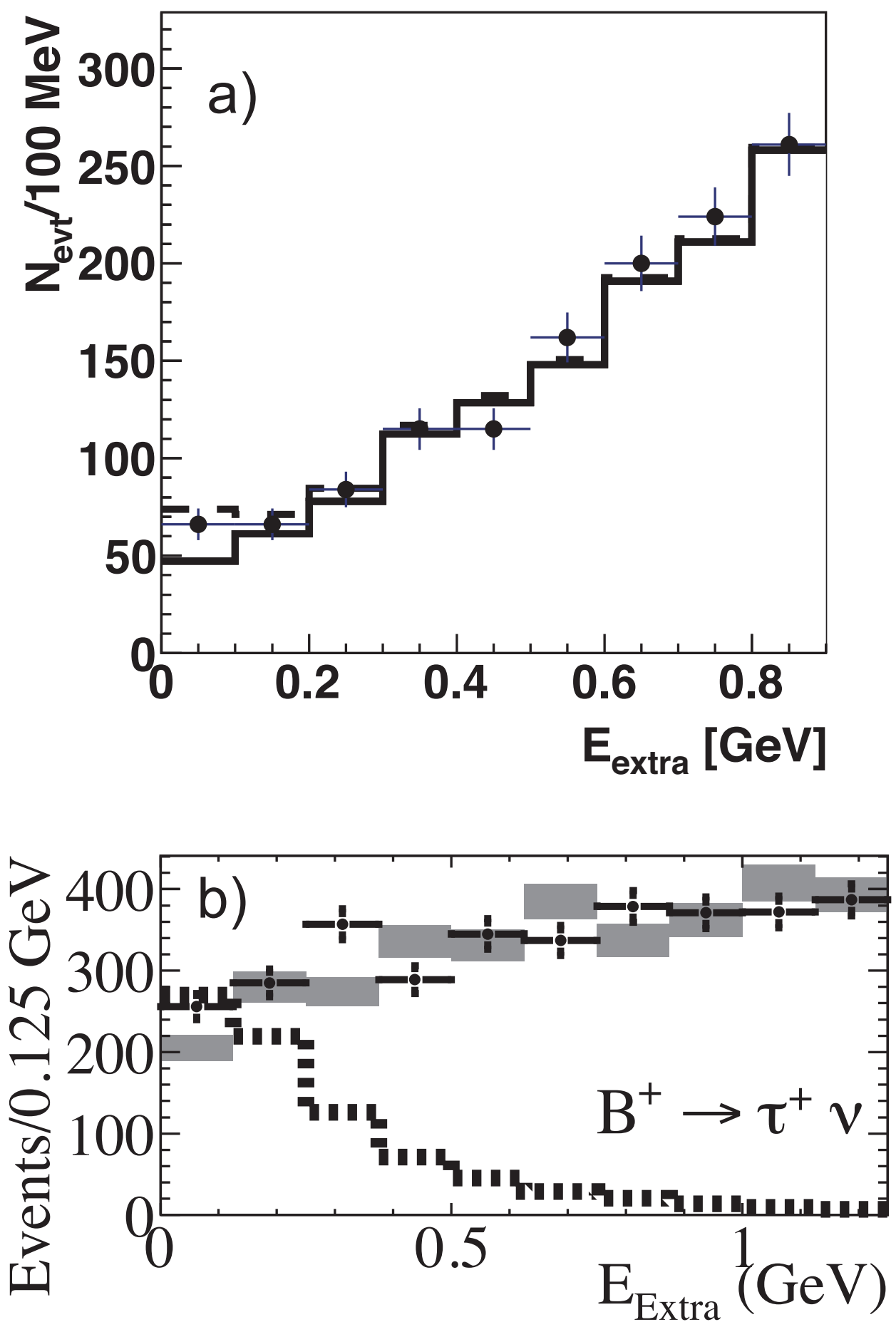

**Figure 17.10.3.** (a) Distribution of $E_{\rm extra}$ reported by *BABAR* using hadronic tags (Lees, 2013a). The points with error bars represent data. The solid histogram shows the background and the dashed component is the best-fit signal excess distribution. (b) The same distribution obtained using semileptonic tagged events (Aubert, 2010a). The points with error bars are data. The gray shaded boxes represent MC simulated backgrounds and the dotted histogram is the signal MC simulation normalized to 10 times the expected branching fraction.

*Summary of $B^+ \to \tau^+\nu$ measurements*

The branching fractions reported by Belle and *BABAR* using the hadronic and semileptonic tagged samples are summarized in Table 17.10.1 and graphically compared in Figure 17.10.4. The errors for $N_{\rm sig}$ are statistical only. For the semileptonic-tag analysis at *BABAR*, we obtain $N_{\rm sig}$ by taking a difference between the total yield of 583 events and the expected background yield of $509 \pm 30$ events, where the error is obtained by taking a quadratic sum of the errors for the above two yields assuming a Poisson error for the total yield. The significance $\Sigma_{\rm sig}$ includes systematic uncertainties. The efficiency $\epsilon_{\rm sig}$ includes the branching ratios of the tau decay modes. The first and second errors for $\mathcal{B}$ are the statistical and systematic uncertainties, respectively.

The four results are consistent within the errors. Both Belle and *BABAR* quote average branching fractions from the combination of their own hadronic and semileptonic tag results. Taking the simple weighted average of these two values one obtains

$$\mathcal{B}(B^+ \to \tau^+\nu)_{\rm AVG} = (1.15 \pm 0.23) \times 10^{-4}. \quad (17.10.10)$$

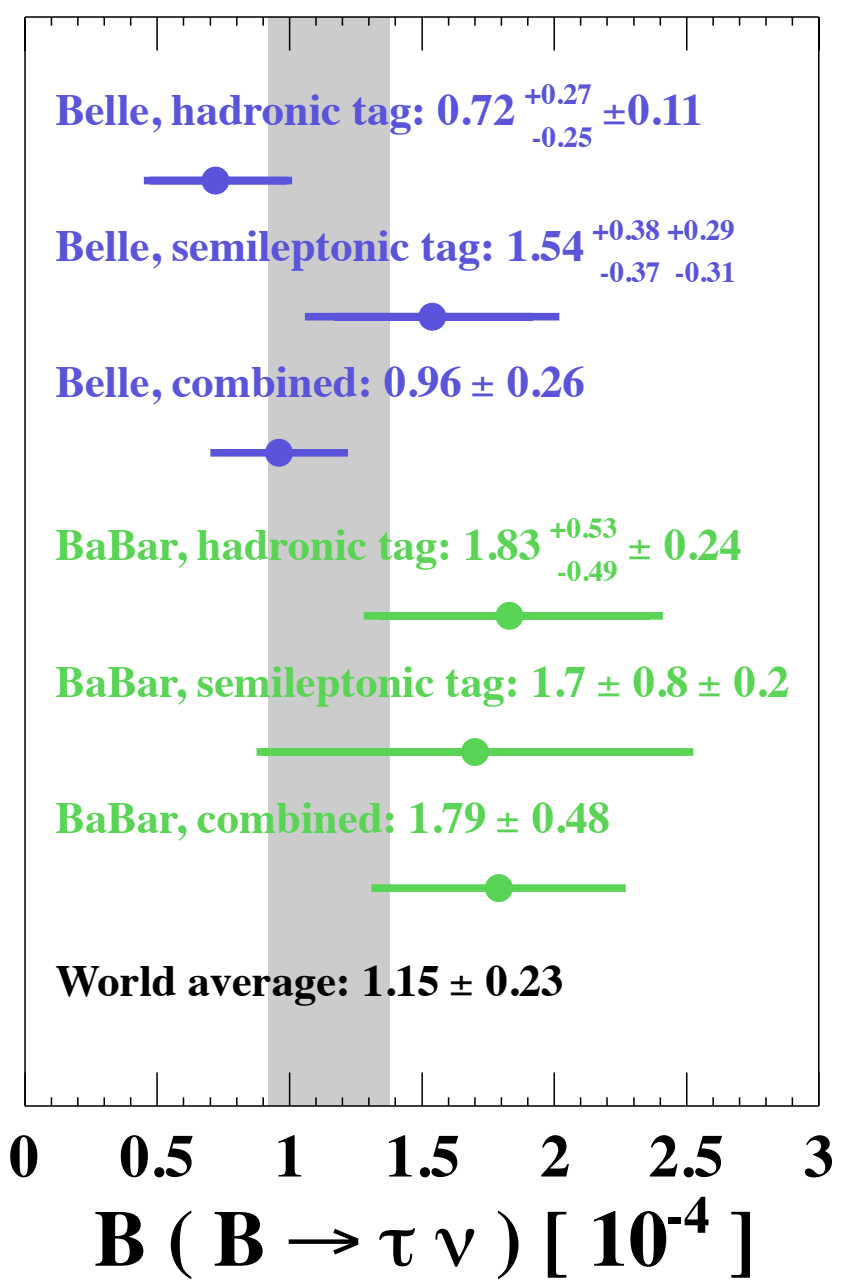

**Figure 17.10.4.** Comparison of the branching fractions reported by Belle and *BABAR* using the hadronic and semileptonic tagged samples of data. The error bars indicate the quadratic sums of the statistical and systematic uncertainties.

*Interpretation of results*

The experimental average for the $B^+ \to \tau^+\nu$ branching fraction is consistent with the SM prediction from Equation (17.10.5). In turn, using the average $B^+ \to \tau^+\nu$ branching fraction, the product of $|V_{ub}|$ and the $B$ meson decay constant $f_B$ is calculated to be,

$$f_B|V_{ub}| = (8.06 \pm 0.81) \times 10^{-4}\,\text{GeV}. \quad (17.10.11)$$

Using the value of $f_B$ above (Section 17.10.2.1), $|V_{ub}|$ is deduced to be,

$$|V_{ub}| = (4.22 \pm 0.47) \times 10^{-3}. \quad (17.10.12)$$

This is consistent, within errors, with the average values of $|V_{ub}|$ obtained from inclusive and exclusive $B$ semileptonic decay data in Eq. (17.1.70). This result is also consistent with the value from inclusive $B$ semileptonic decay data alone, but higher than the value from exclusive $B$ semileptonic decay data by 2.1 $\sigma$.

The obtained branching fraction can also be used to constrain the charged Higgs. The ratio $r_H$, as defined in Eq. (17.10.7), is found to be $r_H = 1.14 \pm 0.40$. Based on this result and Eq. (17.10.6), the charged Higgs can be constrained in the $(\tan\beta, M_H)$ plane, as shown in Figure 17.10.5.

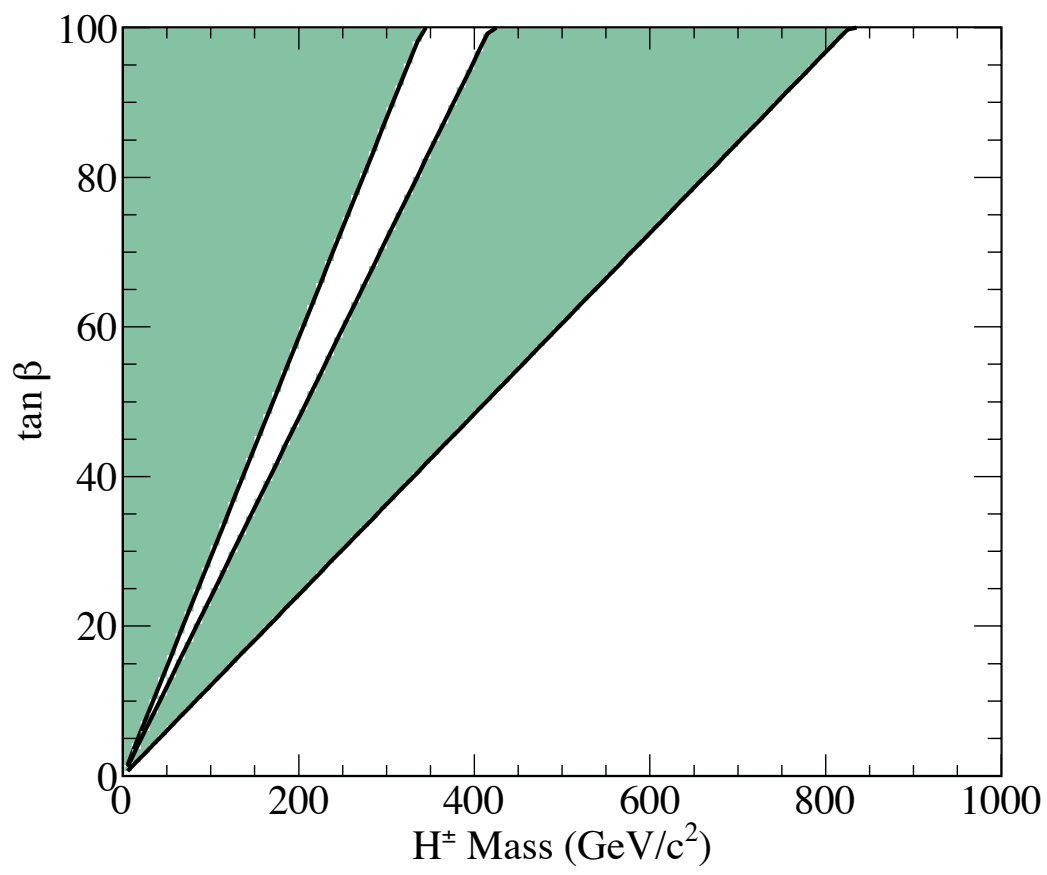

**Figure 17.10.5.** Constraint on the ratio of the two vacuum expectation values $\tan\beta$ and the charged Higgs mass in the type II of two Higgs doublet model. The green regions indicate the excluded regions at a confidence level of 95%.

### 17.10.2.3 $B^+ \to \ell^+\nu$ ($\ell = e, \mu$)

Although the $B^+ \to e^+\nu$ and $B^+ \to \mu^+\nu$ branching fractions are substantially suppressed compared to the $\tau$ mode, these modes are still of considerable interest at the $B$ Factories. While the electron mode, within the SM, is well beyond reach, the $\mu$ mode has a predicted branching fraction of $\sim 5 \times 10^{-7}$, which is potentially detectable by *BABAR* and Belle. It is also notable that the relative enhancement (or suppression) of the leptonic branching fractions due to the existence of a charged Higgs boson is independent of the final state lepton mass as can be seen from Eq. (17.10.7). Consequently, equally precise determinations of experimental branching fractions in any of the three leptonic modes would yield identical constraints on a postulated charged Higgs boson. Additionally, because the $B^+ \to \mu^+\nu$ final state contains only a single neutrino and a high momentum $\mu$, there exist sufficient constraints that the search can be performed without the need for exclusive $B_{\rm tag}$ reconstruction and hence with substantially higher signal efficiency than in the case of $B^+ \to \tau^+\nu$. The higher efficiency and cleaner signature compensates to some degree for the smaller SM branching fraction, however current measurements of $B^+ \to \mu^+\nu$ are not yet sufficiently sensitive to provide evidence of a non-zero signal. Both *BABAR* (Aubert, 2004ac, 2009as) and Belle (Satoyama, 2007) have published the results of searches for $B^+ \to \mu^+\nu$ and $B^+ \to e^+\nu$ using this "inclusive" (*i.e.* un-tagged) approach. These inclusive searches have resulted in branching fraction upper limits that are within about a factor of two of the SM expectation, and that are limited by the finite size of the background event samples. *BABAR* has also performed a search using hadronic $B_{\rm tag}$ reconstruction (Aubert, 2008az) and a search using semileptonic tag reconstruction (Aubert, 2010a).

**Table 17.10.1.** Summary table for the $B^+ \to \tau^+\nu$ analyses. The number of $B\overline{B}$ pairs in the data sample ($N_{B\overline{B}}$), the signal yield ($N_{\rm sig}$), the significance ($\Sigma_{\rm sig}$), the detection efficiency ($\epsilon_{\rm sig}$), and the branching ratio ($\mathcal{B}$) are shown for each of the hadronic-tag and semileptonic-tag analyses. The combined results reported by Belle and *BABAR* for $\mathcal{B}$ are also shown, where the errors are the sum in quadrature of the statistical and systematic uncertainties.

| Experiment | Tagging | $N_{B\overline{B}}$ $(10^6)$ | $N_{\rm sig}$ | $\Sigma_{\rm sig}$ | $\epsilon_{\rm sig}$ $(10^{-4})$ | $\mathcal{B}$ $(10^{-4})$ | Reference |
|---|---|---|---|---|---|---|---|
| | Hadronic | 772 | $62^{+23}_{-22}$ | $3.0\sigma$ | 11.2 | $0.72^{+0.27}_{-0.25} \pm 0.11$ | (Adachi, 2012b) |
| Belle | Semileptonic | 657 | $143^{+36}_{-35}$ | $3.6\sigma$ | 14.3 | $1.54^{+0.38+0.29}_{-0.37-0.31}$ | (Hara, 2010) |
| | Combined | | | | | $0.96 \pm 0.26$ | (Adachi, 2012b) |
| | Hadronic | 468 | $62 \pm 17$ | $3.8\sigma$ | 7.3 | $1.83^{+0.53}_{-0.49} \pm 0.24$ | (Lees, 2013a) |
| *BABAR* | Semileptonic | 459 | $74 \pm 39$ | $2.3\sigma$ | 9.6 | $1.7 \pm 0.8 \pm 0.2$ | (Aubert, 2010a) |
| | Combined | | | | | $1.79 \pm 0.48$ | (Lees, 2013a) |

### *Inclusive searches*

The most stringent limits on $B^+ \to \mu^+\nu$ and $B^+ \to e^+\nu$ are obtained from "inclusive" searches from *BABAR* (Aubert, 2009as) and Belle (Satoyama, 2007). These analyses rely on the distinctive signature of the high-momentum lepton ($e$ or $\mu$) resulting from the two-body $B$ decay. The lepton momentum lies well above the kinematic limit for $b \to c\ell\nu$ and close to the endpoint for $b \to u\ell\nu$. Consequently, backgrounds from $B$ decays with real leptons are relatively limited, but continuum background can be large.

Tight particle identification requirements are imposed in order to cleanly identify the signal candidate electron or muon. Although the lepton is expected to be mono-energetic in the signal $B$ rest frame, the $\Upsilon(4S)$ rest frame (*i.e.* the $e^+e^-$ CM frame) is initially used as an approximation since the rest frame of the parent $B$ is not known. Because the two $B$ mesons have momenta of $\sim 320$ MeV/$c$ in this frame, the signal lepton momentum is smeared out and ranges from 2.4 GeV/$c$ to about 3.2 GeV/$c$.

Since the only other daughter of the signal $B$ is an undetected neutrino, it is expected that all other particles detected in the $\Upsilon(4S) \to B^+B^-$ event originate from the non-signal $B$. Consequently, for signal events, the combination of all particles should yield a four-vector consistent with a $B$ meson, and a total charge that is opposite that of the signal lepton. In order to obtain the best possible resolution for this four-vector, tracks used in this combination are assigned mass hypotheses based on particle identification criteria. Events with any additional identified leptons are vetoed since their presence often implies either additional missing energy from unobserved neutrinos, or that the event is continuum background. Missing energy can also arise due to particles lost outside the detector fiducial acceptance; in particular this can occur for continuum backgrounds, which tend to produce particles in the forward and backward regions of the detector. Belle requires the transverse component of the missing momentum to be greater than 1.75 GeV/$c$ and the cosine of the angle with respect to the beam axis to be less than 0.84 (0.82) for the muon (electron) mode. Continuum background is further suppressed by exploiting the differences in event shapes compared with $B\overline{B}$ events. To this end both *BABAR* and Belle use a Fisher discriminant (see Chapter 4) combining kinematic and angular variables describing the distribution and energies of reconstructed particles in the event.

The four-vector of the non-signal $B$ is obtained by summing the four-vectors of all tracks and clusters in the event other than the signal lepton. The kinematic variables $\Delta E$ and $m_{ES}$ are used to characterize the $B$ candidate. Events in which all non-signal $B$ decay daughters, and no additional particles, are correctly identified and included in the four-vector sum are expected to have $\Delta E \approx 0$ and $m_{ES}$ close to the nominal $B$ mass. Due to the "inclusive" nature of the method, the resolution of both of these quantities is relatively poor compared to what is typically obtained for exclusively reconstructed $B$ decays.

The signal $B$ four-vector can be inferred from the non-signal $B$ four-vector and used to refine the estimate of the $B$ rest frame momentum of the signal lepton. This results in a modest improvement in the resolution of the lepton momentum (see *e.g.* Figure 17.10.6 upper plot), compared with the $e^+e^-$ CM frame.

The signal yield is extracted based on the distributions of the corrected lepton momentum ($p_\ell^B$) and the non-signal $B$ $m_{ES}$ and $\Delta E$ distributions. *BABAR* and Belle use different methods. Belle requires $2.6 < p^B_{\mu(e)} < 2.84\,(2.80)$ GeV/$c$ and $-0.8\,(-1.0) < \Delta E < 0.4$ GeV for the muon (electron) mode (see Figure 17.10.6 upper plot), then defines a fit region, $5.10 < m_{ES} < 5.29$ GeV/$c^2$, and a more constrained signal region, $5.26 < m_{ES} < 5.29$ GeV/$c^2$, which are used for background and signal estimation, respectively. The signal is extracted using an unbinned maximum likelihood fit to $m_{ES}$ in the signal region. The overall signal efficiency is estimated to be $\epsilon_\mu = (2.18 \pm 0.06)\%$ and $\epsilon_e = (2.39 \pm 0.06)\%$ for the muon and electron channels, respectively. Belle observes a total of 12 (15) events with an expected background of $7.4 \pm 1.0$ ($13.4 \pm 1.4$) events in the muon (electron) channel, yielding branching fraction upper limits of $1.7 \times 10^{-6}$ ($0.98 \times 10^{-6}$) at a 90% confidence level (see Table 17.10.2). The SM expectation for the signal yield in the $B^+ \to \mu^+\nu$ channel is 2 – 3 events in this analysis.

In the *BABAR* analysis, $\Delta E$ is required to satisfy $-2.25 < \Delta E < 0$ GeV. A Fisher discriminant is then constructed from the two variables, $p_\ell^{\rm CM}$ and $p_\ell^B$, representing the signal lepton momentum in the CM frame and the inferred signal $B$ rest frame, respectively. The signal yield is then determined using an extended maximum likelihood fit to the Fisher discriminant output and $m_{ES}$. The total signal efficiencies are estimated to be $\epsilon_\mu = (6.1 \pm 0.2)\%$ and $\epsilon_e = (4.7 \pm 0.3)\%$ in the muon and electron channel, respectively. The fit yields $1 \pm 15$ ($18 \pm 14$) events in the muon (electron) channel. In the absence of significant evidence for signal, *BABAR* obtains branching fraction upper limits of $1.0 \times 10^{-6}$ ($1.9 \times 10^{-6}$) at the 90% confidence level (see Table 17.10.2). In a previous inclusive search for $B^+ \to \mu^+\nu$ by *BABAR* (Aubert, 2004ac), a simpler signal extraction method was used in which the signal yield was obtained using a so-called cut-and-count (or rectangular cut) method based on a rectangular signal region defined in the ($m_{ES}$, $\Delta E$) plane. The signal efficiency was estimated to be $\epsilon_\mu = (2.09 \pm 0.06({\rm stat}) \pm 0.13({\rm syst}))\%$ with a total background of $5.2 \pm 0.5$ events in a data sample of $88 \times 10^6$ $B\overline{B}$ pairs.

#### *Searches using tag reconstruction*

*BABAR* has also performed searches for $B^+ \to \mu^+\nu$ and $B^+ \to e^+\nu$ using methods based on hadronic (Aubert, 2008az) and semileptonic (Aubert, 2010a) tag $B$ reconstruction, similar to the methods used for $B^+ \to \tau^+\nu$ and described in Section 17.10.2.2 above. The semileptonic tag search is performed simultaneously with the corresponding $B^+ \to \tau^+\nu$ study, essentially representing a special case of the $\tau^+ \to \ell^+\nu\overline{\nu}$ ($\ell = e, \mu$) signal channels in which the final state lepton has a high momentum.

While the signal efficiency in the "tagged" searches is substantially reduced compared with inclusive searches due to the tag reconstruction procedure, an advantage is gained (particularly in the case of hadronic tags) due to increased continuum background suppression and improved knowledge of the signal event kinematics. In particular, the hadronic $B_{\rm tag}$ four-vector permits the signal $B$ four-vector to be precisely determined, with the consequence that the lepton momentum can be precisely determined in the $B$ rest frame. The improved $p_\mu^B$ resolution is illustrated in the lower plot of Figure 17.10.6, which can be compared with the "inclusive" distribution from Belle shown in the upper plot. The improved resolution allows for a significantly improved separation between signal events and backgrounds from semileptonic $B$ decays, in particular $b \to u\ell\nu$ events. In the *BABAR* study (Aubert, 2008az), the hadronic tag search is essentially background free in the signal region. However, with the number of events available with the *BABAR* and Belle data samples, the tagged approach is statistically limited and the inclusive approach results in a significantly more stringent branching fraction limit than the tagged analyses. However, both the hadronic tag and inclusive methods yield similar sensitivities for a $5\sigma$ signal observation, with the event samples available at the current $B$ Factories, due to the large statistical uncertainty in the background in the inclusive method. It is anticipated that the two methods will provide complementary and precise $B^+ \to \mu^+\nu$ branching fraction measurements with the data expected at future high luminosity $B$ Factories.

#### 17.10.2.4 Radiative decays $B^+ \to \ell^+\nu\gamma$

To date, only a single search for $\mathcal{B}(B^+ \to \ell^+\nu_\ell\gamma)$ has been published (Aubert, 2009a) by the asymmetric $B$ Factories, although CLEO had previously published results using an un-tagged search method (Browder et al., 1997). The *BABAR* result, which is based on a data sample of 465 million $B\overline{B}$ pairs, uses a method based on hadronic-tag reconstruction (see Chapter 7). However, Belle (Abe, 2004d) and *BABAR* (Aubert, 2007aw) have both reported unpublished results using an "inclusive" method similar to that used by CLEO. Although the hadronic-tag technique results in a low signal efficiency (0.3% for signal modes), it compensates by providing a high purity sample of $B$ mesons with comparatively little non-$B\overline{B}$ (continuum) background. This background is very problematic for the inclusive analyses. In addition, by reconstructing the $B_{\rm tag}$ using only detectable hadronic decay modes, the missing four-vector of the signal neutrino is fully determined. Thus the *BABAR* hadronic tag analysis was able to avoid the model-dependent kinematic constraints in the signal selection which had complicated the interpretation of the earlier analyses. However, the inclusive analyses benefit from significantly higher statistical sensitivity. With the available *BABAR* or Belle data it is expected that the inclusive measurements, if they had been published, would have yielded more stringent experimental limits or possibly observation of these decay modes.

**Table 17.10.2.** $B^+ \to \ell^+\nu$ and $B^+ \to \ell^+\nu\gamma$ branching fraction measurements by *BABAR* and Belle.

| Experiment | Decay Mode | Method | $N_{B\overline{B}}$ ($10^6$) | $\mathcal{B}$ upper limit 90% C.L. | Reference |
|---|---|---|---|---|---|
| Belle | $B^+ \to \mu^+\nu$ | inclusive | 253 $\text{fb}^{-1}$ | $1.7 \times 10^{-6}$ | Satoyama (2007) |
| | $B^+ \to e^+\nu$ | | | $0.98 \times 10^{-6}$ | |
| *BABAR* | $B^+ \to \mu^+\nu$ | inclusive | 268 | $1.0 \times 10^{-6}$ | Aubert (2009as) |
| | $B^+ \to e^+\nu$ | | | $1.9 \times 10^{-6}$ | |
| *BABAR* | $B^+ \to \mu^+\nu$ | hadronic | 378 | $5.6 \times 10^{-6}$ | Aubert (2008az) |
| | $B^+ \to e^+\nu$ | | | $5.2 \times 10^{-6}$ | |
| *BABAR* | $B^+ \to \mu^+\nu$ | semileptonic | 459 | $11 \times 10^{-6}$ | Aubert (2010a) |
| | $B^+ \to e^+\nu$ | | | $8 \times 10^{-6}$ | |
| *BABAR* | $B^+ \to \mu^+\nu\gamma$ | hadronic | 465 | $24 \times 10^{-6}$ | Aubert (2009a) |
| | $B^+ \to e^+\nu\gamma$ | | | $16 \times 10^{-6}$ | |
| | $B^+ \to \ell^+\nu\gamma$ | | | $15.6 \times 10^{-6}$ | |

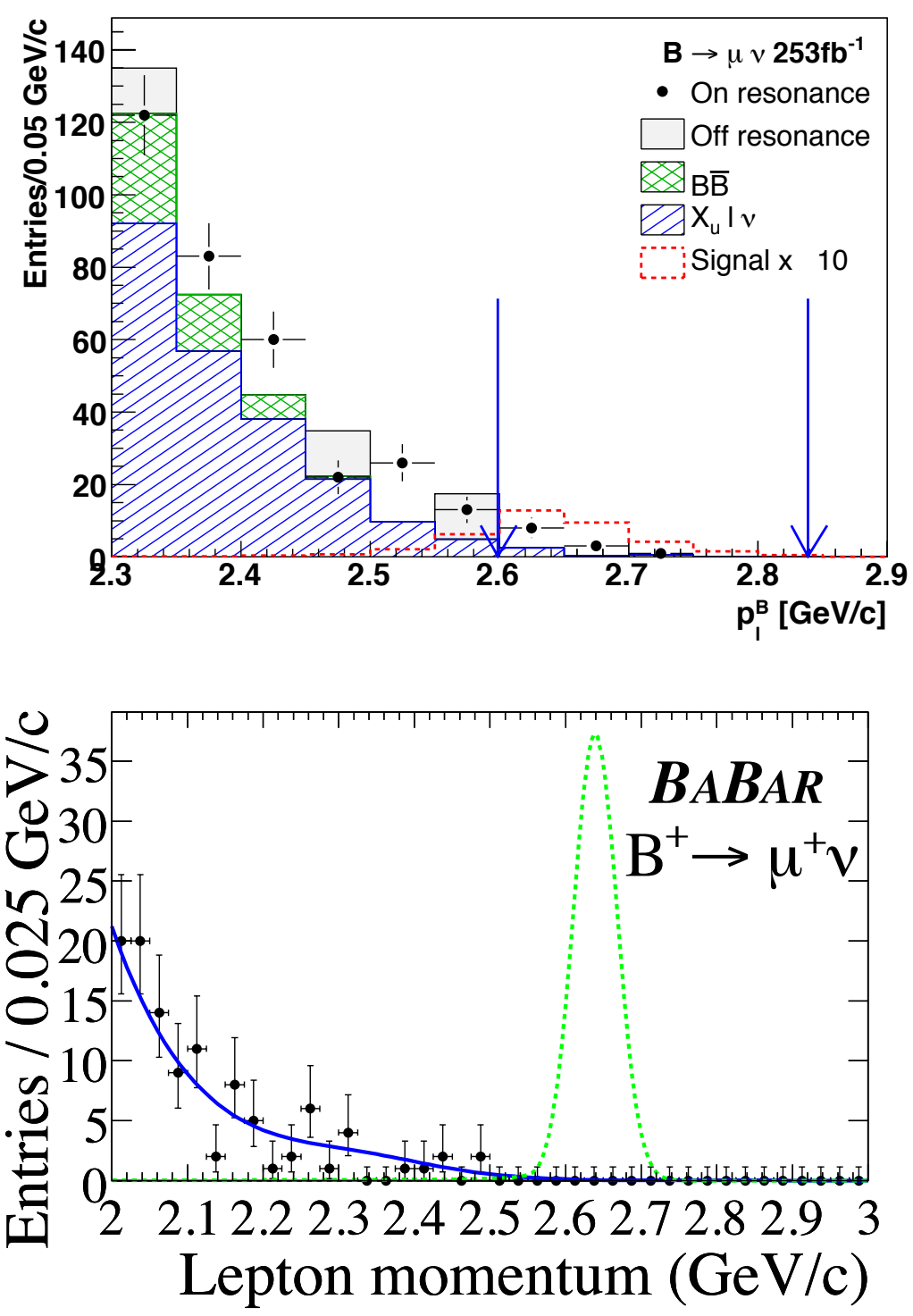

**Figure 17.10.6.** (Top) Momentum of the reconstructed $B^+ \to \mu^+\nu$ signal muon in the Belle "inclusive" search (Satoyama, 2007) and (bottom) in the *BABAR* "hadronic tag" search (Aubert, 2008az).

The hadronic-tag analysis proceeds as follows. After reconstructing a $B_{\text{tag}}$, remaining continuum background is suppressed using a multivariate selector incorporating several event shape variables. From this sample, signal candidate events are required to possess only one "signal-side" track, in addition to those used to reconstruct the $B_{\text{tag}}$. This track is required to satisfy either electron or muon particle identification. In the case of electrons, candidate Bremsstrahlung clusters in the calorimeter are used to correct the momentum of the electron track. The highest energy remaining calorimeter cluster, not associated with the reconstructed $B_{\text{tag}}$, is assumed to be the radiated signal photon. The energy spectrum of radiated photons in signal candidates is expected to peak at $\sim 1\,\text{GeV}$. The energies of any remaining calorimeter clusters are summed to obtain $E_{\text{extra}}$. A loose requirement of $E_{\text{extra}} < 0.8\,\text{GeV}$ is imposed to reject $B$ backgrounds. To ensure that the signal candidates are consistent with a three-body decay, the lepton momentum and the total missing momentum in the event were required to be back-to-back in the frame recoiling against the photon. As the signal $B$ four-vector can be inferred from the $B_{\text{tag}}$ four-vector, the 3-body kinematics can be uniquely determined by combining this information with the signal lepton and photon candidate four-vectors. The most discriminating variable is the reconstructed invariant mass of the neutrino, given as $m_\nu^2 \equiv |p_{\Upsilon(4S)} - p_{B_{\text{tag}}} - p_\ell - p_\gamma|^2$ where $p_i$ is the four-momentum of particle $i$. Figure 17.10.7 shows that the signal peaks at zero, while the background rises with $m_\nu^2$.

The dominant backgrounds arise from $B^+ \to X_u^0\ell^+\nu_\ell$ events, where $X_u$ is a neutral meson containing a $u$-quark. Events in which the signal photon candidate could be combined with another calorimeter cluster to form an invariant mass consistent with the $\pi^0$ or $\eta$ mass, or combined with a $\pi^0$ candidate to form an $\omega$, were rejected. However, $B^+ \to X_u^0\ell^+\nu_\ell$ events can mimic the signal decay kinematics. This can occur especially if only one high-energy photon daughter from the $X_u$ decay is present in the signal-side clusters, or if the two photons from a $B^+ \to \pi^0\ell^+\nu_\ell$ decay are merged into a single calorimeter cluster containing the full energy of the $\pi^0$. Consequently, the combination of this "photon" cluster with the signal lepton results in an $m_\nu^2$ distribution that peaks at zero, mimicking signal. These backgrounds are suppressed by examining the shape of the calorimeter cluster and limiting the lateral moment of the cluster energy deposit.

A cut-and-count method is used to determine the signal yield. The background is divided into two components: events that peak in the $m_{\text{ES}}$ signal region are es-timated from various dedicated $B^+ \to X_u^0\ell^+\nu_\ell$ MC samples, while non-peaking events are extrapolated directly from the data events in the $m_{\rm ES}$ sideband region in order to reduce the dependence on MC simulations. The largest background uncertainties stem, respectively, from the branching fractions and form factors of the various $B^+ \to X_u^0\ell^+\nu_\ell$ decays, and from the limited number of sideband data events.

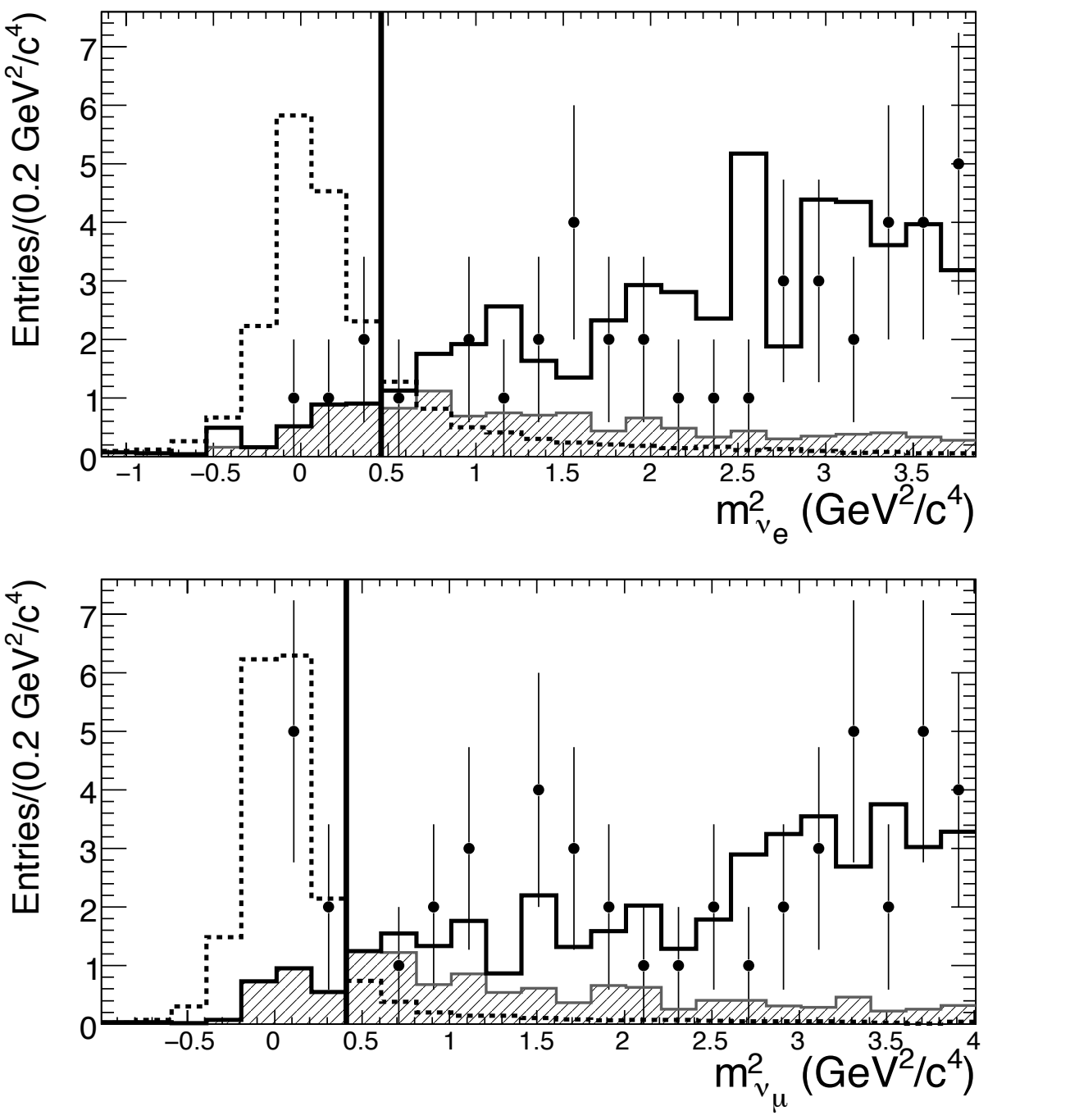

**Figure 17.10.7.** $m_\nu^2$ distribution in $B^+ \to \ell^+\nu\gamma$, from Aubert (2009a), after all selection criteria are applied, in the electron (top) and muon (bottom) modes. The $m_{\rm ES}$-peaking (shaded) and non-peaking (solid) background contributions are shown stacked, along with signal MC (dashed) normalized to $\mathcal{B} = 40 \times 10^{-6}$, and data (points). Events to the left of the vertical lines are selected.

A measurement of $\mathcal{B}(B^+ \to \ell^+\nu_\ell\gamma) = (6.5^{+7.6}_{-4.7}\,{}^{+2.8}_{-0.8}) \times 10^{-6}$ was obtained with a significance of $2.1\sigma$, along with an upper limit of $\mathcal{B}(B^+ \to \ell^+\nu_\ell\gamma) < 15.6 \times 10^{-6}$ at 90% confidence level. These results are the most stringent published limits, and are close to the theoretical predictions for these modes. Effectively no requirements are applied to the lepton or photon kinematics, thus this analysis is essentially independent of the $B \to \gamma$ form factor models and valid over the full kinematic range.

However, the extraction of $\lambda_B$ (see Eq. 17.10.9) can be improved by including a minimum energy requirement on the signal photon (Ball and Kou, 2003; Beneke and Rohrwild, 2011). Therefore, additional branching fraction results were reported in specific kinematic regions. A requirement that the signal photon candidate energy is $>$ 1 GeV results in a partial branching fraction of $\Delta\mathcal{B}(\mathrm{B}^+ \to \ell^+\nu_\ell\gamma) < 14 \times 10^{-6}$ at a C.L. of 90%. More stringent branching fraction limits were determined by introducing a kinematic requirement on the angles between the three daughter particles of the signal decay. In a model in which the two $B \to \gamma$ form-factors, $f_V$ and $f_A$, are equal, the result $\mathcal{B}(B^+ \to \ell^+\nu_\ell\gamma) < 3.0 \times 10^{-6}$ is obtained. In a model with $f_A = 0$, $\mathcal{B}(B^+ \to \ell^+\nu_\ell\gamma) < 18 \times 10^{-6}$ is obtained.

Although a significant $B^+ \to \ell^+\nu_\ell\gamma$ signal has not yet been observed, the sensitivity of the this method is such that it is likely that these decays will be accessible at future high-luminosity $B$ Factories.

### 17.10.3 $B \to D^{(*)}\tau\nu$

In the SM the branching fractions for the semileptonic decays $B \to D^{(*)}\tau\nu_\tau$, which proceed via a tree-level process with an intermediate $W^\pm$, are predicted to be $(0.69 \pm 0.04)\%$ and $(1.41 \pm 0.07)\%$ for $B^0 \to D^-\tau^+\nu_\tau$ and $B^0 \to D^{*-}\tau^+\nu_\tau$, respectively (Chen and Geng, 2006). However, if a charged Higgs boson exists, the branching fraction may differ significantly due to interference from a tree-level $H^\pm$ exchange contribution similar to that for $B^+ \to \tau^+\nu$ (see Figure 17.10.1). Effects of the charged Higgs on $B \to D^{(*)}\tau\nu$ decays are discussed in a number of theoretical papers (Fajfer, Kamenik, and Nisandzic, 2012; Grzadkowski and Hou, 1992; Itoh, Komine, and Okada, 2005; Kiers and Soni, 1997; Nierste, Trine, and Westhoff, 2008; Tanaka, 1995). From a theoretical point of view, the $B \to D^{(*)}\tau\nu_\tau$ decay has a similar sensitivity to $H^\pm$ as the $B^+ \to \tau^+\nu$ decay, but with different theoretical and parametric uncertainties. While the purely leptonic decay depends only on the relatively well known $B$ meson decay constant $f_B$, the semileptonic decays depend on form factors. The formulae for the semileptonic rates can be found in Section 17.1. However, for $\ell = e$, $\mu$ the lepton mass is usually neglected in which case the rates are not sensitive to the "longitudinal" form factors, which are proportional to the momentum transfer $q = (p - p')$

$$\langle D(p')|\bar{c}\gamma_\mu b|B(p)\rangle = (p_\mu - p'_\mu)F_0(q^2) + \cdots \qquad (17.10.13)$$

and similarly for the transition $B \to D^*$. These form factors become relevant for $B \to D^{(*)}\tau\nu$, since their contributions to the differential rates are proportional to $M_\ell^2/M_B^2$ which is sizable for $\ell = \tau$. Detailed results for the differential rates including lepton mass effects can be found, for example, in (Fajfer, Kamenik, and Nisandzic, 2012; Nierste, Trine, and Westhoff, 2008).

The longitudinal form factors for $B \to D^{(*)}\tau\nu$ are not as well known as the ones appearing in the decays with an $e$ or a $\mu$. Nevertheless, one can apply the heavy quark limit for both the bottom and the charm quark in which case all form factors of the $B \to D^{(*)}$ transitions may be related to a single form factor, the Isgur-Wise function (see Section 17.1). This limit is expected to be valid at the level of (20-30)%, however, in combination with the factor $M_\ell^2/M_B^2$ one still arrives at fairly precise predictions for ratios of branching fractions. The ratio

$$\mathcal{R}_{D^{(*)}} = \frac{\mathcal{B}(B \to D^{(*)}\tau\nu)}{\mathcal{B}(B \to D^{(*)}\ell\nu)}\,, \qquad (17.10.14)$$

can provide sensitivity to $H^{\pm}$. Measurement of this quantity has the additional advantage of using two decay modes with very similar experimental signatures, if only the leptonic decay modes of the tau are considered. This method therefore permits the cancellation of form factor uncertainties as well as many experimental systematic uncertainties.

In contrast, sensitivity to $H^{\pm}$ in $B^+ \to \tau^+\nu_\tau$ requires measurement of the absolute branching fraction and comparison with the expected SM prediction, which in turn requires knowledge of the CKM matrix element $V_{ub}$. Given the current discrepancy between inclusive and exclusive $V_{ub}$ measurements (see Section 17.1), the $B \to D^{(*)}\tau\nu_\tau$ modes currently provide a cleaner interpretation of possible $H^{\pm}$ contributions. These modes therefore provide complementary approaches to searching for $H^{\pm}$ signatures in $B$ decays.

The three-body kinematics of the $B \to D^{(*)}\tau\nu_\tau$ decay also potentially permit the study of the $\tau$ polarization via the decay distributions of final state particles. These measurements can in principle discriminate between $H^{\pm}$ and $W^{\pm}$ exchange, however studies performed to date have not been sensitive to these distributions.

### 17.10.3.1 Experimental methodology and results

Like $B^+ \to \tau^+\nu$, the $B \to D^{(*)}\tau\nu$ decay has two or more neutrinos in the final state and so cannot be fully reconstructed using only the observable particles. It therefore relies on exclusive reconstruction of the accompanying $B$ ("$B_{\rm tag}$") to provide the necessary level of background suppression (see Chapter 7). *BABAR* has reported $B \to D^{(*)}\tau\nu$ results using the method of hadronic $B$ tag reconstruction (Aubert, 2008al; Lees, 2012e) while Belle has published results based on another method, referred to as "inclusive tags" (Bozek, 2010; Matyja, 2007), in which $B_{\rm tag}$'s are reconstructed by calculating the four-vector sum of the tracks inclusively without reconstructing the intermediate mesons. This method is similar to the "inclusive" method used in $B^+ \to \ell^+\nu$ ($\ell = e, \mu$) described in Section 17.10.2.3 above. Belle has also produced a preliminary measurement of $B \to D^{(*)}\tau\nu$ based on hadronic tag reconstruction (Adachi, 2009).

The Belle Collaboration reported the first observation ($5.2\sigma$) of a $B^0 \to D^{*-}\tau^+\nu$ using the inclusive tag method with a data sample of 535 M $B\overline{B}$ pairs (Matyja, 2007). The $\tau^+ \to e^+\nu_e\overline{\nu}_\tau$ and $\tau^+ \to \pi^+\overline{\nu}_\tau$ decays were used to reconstruct $\tau$ lepton candidates. A follow-up to this analysis for $B^+ \to \overline{D}^{(*)0}\tau\nu_\tau$ was performed using 657 M $B\overline{B}$ pairs (Bozek, 2010). In this analysis, the signal and combinatorial background yields were extracted using an extended unbinned maximum likelihood fit to the distributions of the $B_{\rm tag}$ $m_{\rm ES}$ (referred to as $M_{\rm tag}$) and the CM frame momentum of the reconstructed $D^0$, $p_{D^0}$. The $\tau^+ \to e^+\nu_e\overline{\nu}_\tau$, $\tau^+ \to \mu^+\nu_\mu\overline{\nu}_\tau$ and $\tau^+ \to \pi^+\overline{\nu}_\tau$ decay modes were used to reconstruct the $\tau^+$ lepton candidates. In total, 13 different decay channels, 8 for $D^{*0}$ and 5 for $D^0$, were considered. The fits were performed simultaneously to all data subsets. In each of the sub-channels, the data were described as the sum of four components; signal, cross-feed between $\overline{D}^{*0}\tau^+\nu_\tau$ and $\overline{D}^0\tau^+\nu_\tau$, combinatorial and peaking backgrounds. Figure 17.10.8 shows the $M_{\rm tag}$ and $p_{D^0}$ distributions and fit results for the two decay modes. The extracted signal yields (significances) are $446^{+58}_{-56}$ ($8.1\ \sigma$) for $B^+ \to \overline{D}^{*0}\tau^+\nu_\tau$ and $146^{+42}_{-41}$ ($3.5\ \sigma$) for $B^+ \to \overline{D}^0\tau^+\nu_\tau$. This was the first evidence for the $B^+ \to \overline{D}^0\tau^+\nu_\tau$ decay. Branching fraction results are given in Table 17.10.3.

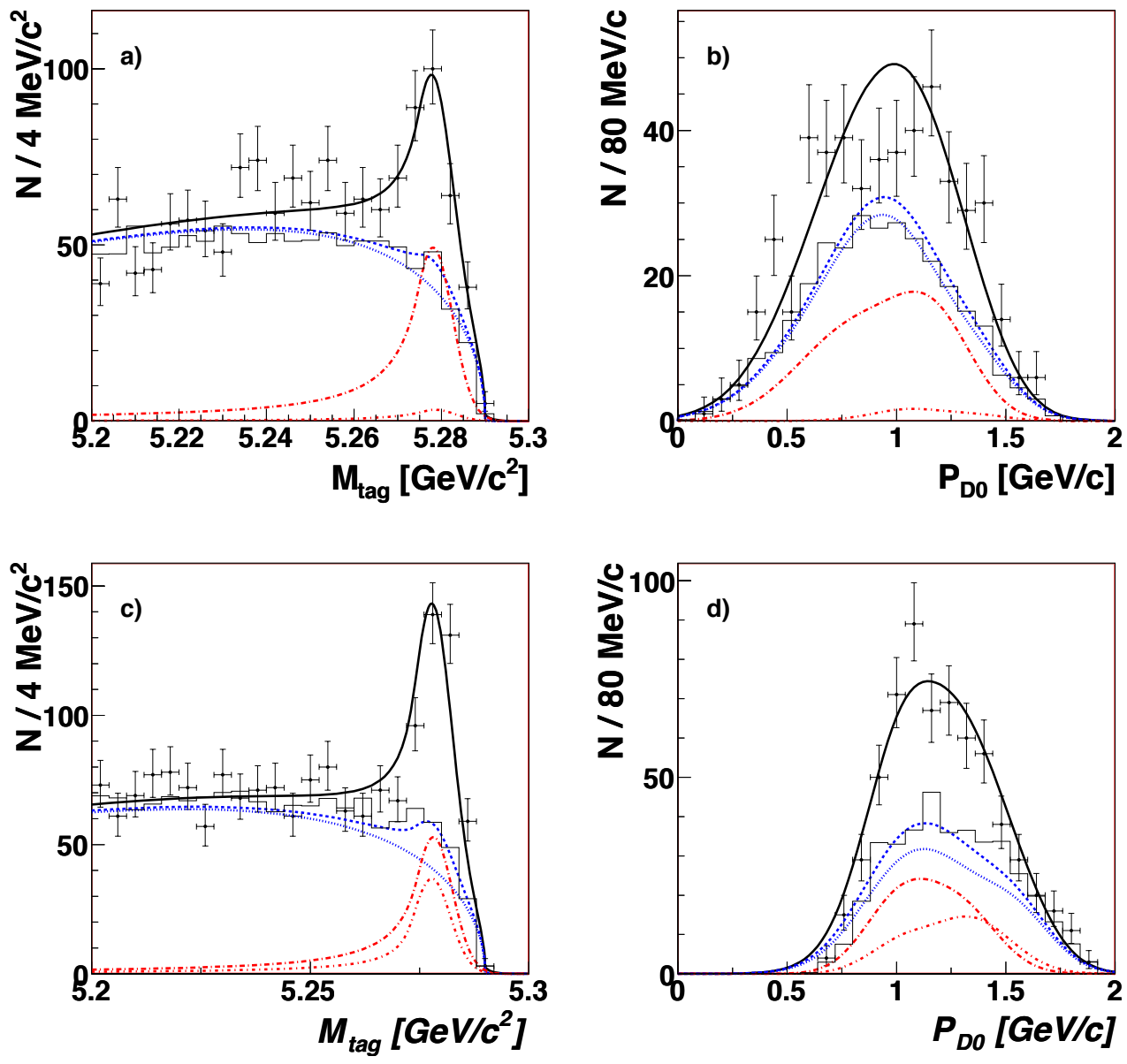

**Figure 17.10.8.** The fit projection to $M_{\rm tag}$ and $p_{D^0}$ for $M_{\rm tag} > 5.26$ GeV/$c^2$ (a, b) for $\overline{D}^{*0}\tau^+\nu_\tau$ and (c, d) for $\overline{D}^0\tau^+\nu_\tau$, from (Bozek, 2010).

*BABAR* reported an observation of $B \to D^*\tau\nu$ as well as first evidence of $B \to D\tau\nu$ in Aubert (2008al) using a method based on exclusive hadronic $B_{\rm tag}$ reconstruction and a data sample of $232M$ $B\overline{B}$ events. This analysis reports measurements of the four modes $B^+ \to \overline{D}^0\tau^+\nu$, $B^+ \to \overline{D}^{*0}\tau^+\nu$, $B^0 \to D^-\tau^+\nu$ and $B^0 \to D^{*-}\tau^+\nu$, as well as the combined modes $B \to D\tau\nu$ and $B \to D^*\tau\nu$. A follow up to this paper, based on the full *BABAR* data sample of $471M$ $B\overline{B}$ events (Lees, 2012e) uses similar methodology. This more recent analysis reports the first observation of $B \to D\tau\nu$ and measures the ratios, $\mathcal{R}(D^{(*)})$ in a simultaneous measurement of $B \to D^{(*)}\tau\nu$ and $B \to D^{(*)}\ell\nu$ ($\ell = e, \mu$), reporting an excess over SM predictions for both $\mathcal{R}(D)$ and $\mathcal{R}(D^*)$. Several improvements are incorporated into the $B_{\rm tag}$ reconstruction and signal selection, which result in a factor of three improvement of the signal efficiency compared to the earlier analysis.

Since only leptonic decay modes of the tau are considered in this analysis, "signal" $B \to D^{(*)}\tau\nu$ decays have identical final states as the "normalization" $B \to D^{(*)}\ell\nu$ modes, differing only in the kinematics of the observed final state. Consequently, many uncertainties associated with the signal reconstruction, including charged particle tracking, particle identification and calorimeter-related reconstruction issues, cancel in the ratio. As $\tau^+ \to \ell^+\nu\overline{\nu}$ results in two additional undetected neutrinos, the missing mass, $M_{\rm miss}$, computed from the signal $B$ (inferred from the $B_{\rm tag}$), the reconstructed $D^{(*)}$ and lepton four-vectors, peaks at zero for the normalization modes but not for signal decays. The final state $e$ or $\mu$ in the $B \to D^{(*)}\tau\nu$ channel also has a softer momentum distribution than the measured primary lepton in the normalization mode. The two quantities $M_{\rm miss}$ and $|p^*_\ell|$, the magnitude of the lepton momentum in signal $B$ rest frame, are used to extract the signal and normalization mode yields, as described below.

Signal events are further distinguished from normalization events and other backgrounds by requiring that the missing momentum magnitude, $|p_{\rm miss}|$ and the square of the magnitude of the exchanged four-momentum in $B \to D^{(*)}\tau\nu$, $q^2$ satisfy $|p_{\rm miss}| > 200$ MeV and $q^2 > 4$ GeV$^2$.

The signal $D^{(*)}$ and lepton candidates are reconstructed from tracks and clusters that are not already associated with the $B_{\rm tag}$. Signal and normalization mode electrons (muons) are required to have laboratory frame momenta greater than 300 MeV (200 MeV) and satisfy particle identification criteria. The $D$ or $D^*$ mesons that are combined with this lepton to form the signal or normalization-mode candidates are reconstructed in the $D^0$ modes $K^-\pi^+$, $K^-K^+$, $K^-\pi^+\pi^0$, $K^-\pi^+\pi^-\pi^+$, $K^0_s\pi^+\pi^-$ and the charged modes $D^+ \to K^-\pi^+\pi^+$, $K^-\pi^+\pi^+\pi^0$, $K^0_s\pi^+$, $K^0_s\pi^+\pi^-\pi^+$, $K^0_s\pi^+\pi^0$, $K^0_sK^+$, where $K^0_s \to \pi^+\pi^-$. $D^*$ mesons are identified by combining reconstructed $D$ candidates with photons or charged or neutral pions to obtain $D^{*+} \to D^0\pi^+$, $D^+\pi^0$ and $D^{*0} \to D^0\pi^0$, $D^0\gamma$ candidates. No additional tracks are permitted in the event after reconstruction of the $D^{(*)}$ and lepton, but additional photons are permitted. If multiple candidates are reconstructed in a single event, the candidate with the lowest $E_{\rm extra}$ is selected, where $E_{\rm extra}$ is the sum of the CM energies of any remaining photons in the event.

Four signal channels are analyzed, corresponding to the final states $D^0\ell$, $D^{*0}\ell$, $D^+\ell$ and $D^{*+}\ell$. Control samples are constructed by requiring the presence of an additional $\pi^0$ in each mode, *i.e.* $D^{(*)}\pi^0\ell$. These control samples are used to estimate background contributions from decays to higher-mass charm states, such as $B \to D^{**}\ell\nu$, which are relatively poorly understood and not reliably modeled in the MC. Additional event shape requirements are imposed on the control samples to suppress large continuum backgrounds in these samples.

Additional background suppression is obtained by using a set of boosted decision tree multivariate selectors (see Chapter 4) trained and optimized for each of the four signal modes to select signal and normalization modes while rejecting backgrounds including $D^{**}$ contributions and cross-feed from other signal and normalization modes. Eight kinematic variables are used as inputs, including $E_{\rm extra}$, the invariant masses of the signal and $B_{\rm tag}$ daughter $D$ mesons, the $D^* - D$ mass difference (when a $D^*$ is present), as well as other quantities related to the quality of the $B_{\rm tag}$ reconstruction and the overall event shape.

The level of agreement between data and MC simulation is verified, and the MC description is improved through use of additional data sideband control samples obtained by requiring $E_{\rm extra} > 0.5$ GeV, $m_{ES} < 5.26$, or $q^2 < 4$ GeV$^2$. Conservative systematic uncertainties on signal efficiencies and background estimates are based on these comparisons.

The signal and normalization mode yields are obtained simultaneously for each of the four signal channels using an unbinned extended maximum likelihood fit to the $M_{\rm miss}$ - $|p^*_\ell|$ distribution for the signal and $D^{(*)}\pi^0\ell$ samples. The fit consists of eight components including signal ($D\tau\nu$, $D^*\tau\nu$), normalization modes ($D\ell\nu$, $D^*\ell\nu$), $D^{**}\ell\nu$, charge cross-feed, other $B\overline{B}$ background and continuum background. The first five of these are allowed to vary in the fit, while the last three are fixed to the values determined from MC and sideband studies. The $D^{**}\ell\nu$ contributions in the signal sample fit are constrained by the fit to the $D^{(*)}\pi^0\ell$ samples. Even with this method to control the $D^{**}\ell\nu$ contributions, this background imposes the dominant systematic uncertainty on the signal yield extraction.

The branching fraction ratios $\mathcal{R}(D^{(*)})$ are determined from the signal and normalization mode yields and selection efficiencies. Fit projections are shown in Fig. 17.10.9 for the four signal modes. Since most of the uncertainties in the signal and normalization mode efficiencies cancel in the ratio of modes, the dominant uncertainty in the efficiency ratio comes from the form factor model uncertainties for $B \to D^{(*)}\tau\nu$ and $B \to D^{(*)}\ell\nu$. The results are presented in Table 17.10.3. In addition to results for the four individual signal channels, two additional results are obtained by combining charged and neutral $B$ results by imposing isospin constraints.

#### 17.10.3.2 Interpretation of results

Table 17.10.3 summarizes the results of the $B \to D^{(*)}\tau\nu$ branching fraction measurements and $\mathcal{R}(D^{(*)})$ results from Belle and *BABAR*. All results are somewhat high compared with the SM expectations, with averages dominated by the 2012 *BABAR* measurements. *BABAR* (Lees, 2012e) has interpreted the results in the context of the SM and the type-II 2HDM, estimating $\mathcal{R}(D)_{\rm SM} = 0.297 \pm 0.017$ and $\mathcal{R}(D^*)_{\rm SM} = 0.252 \pm 0.003$ based on Fajfer, Kamenik, and Nisandzic (2012); Kamenik and Mescia (2008) with updated form factor measurements. The combination of $\mathcal{R}(D)$ and $\mathcal{R}(D^*)$ measurements, including the experimental correlation between the $B \to D\tau\nu$ and $B \to D^*\tau\nu$ measurements, is determined to be inconsistent with the SM at the level of $3.4\sigma$. Interpretation within the context of the 2HDM is shown in Figure 17.10.10. The decay kinematics are sensitive to the presence of a $H^\pm$ contribution, impacting both the momentum spectrum of the final state leptons and the missing mass distribution. This causes the signal efficiency to depend on $\tan\beta/M_H$, with the consequence that the measured value of $\mathcal{R}(D^{(*)})$ also depends on this quantity, as shown in the Figure. In order to correctly estimate the efficiency, *BABAR* re-weights the kinematics of simulated signal events to correspond to representative $\tan\beta/M_H$ values in the range shown in the Figure and repeats the full fit to arrive at measurements of $\mathcal{R}(D^{(*)})$ as a function of $\tan\beta/M_H$. It should be noted that this has implications for averaging the results of the *BABAR* and Belle analyses, since the efficiency dependence of the Belle analysis is not available, a simple average can only be correctly interpreted in the context of the SM. Within the context of the 2HDM, the *BABAR* $\mathcal{R}(D^{(*)})$ measurements imply specific values of $\tan\beta/M_H$ that are incompatible with each other, and hence with the 2HDM type II, at a C.L. of 99.8%.

**Table 17.10.3.** Summary of measurements of $B \to D^{(*)}\tau\nu$. $N_{B\overline{B}}$: number of $B\overline{B}$ pairs in the data sample used for the analysis, $\mathcal{B}$: branching fraction (the first error is statistical, the second systematic, and the third due to the branching fraction uncertainty in the normalization mode), $\Sigma$: significance of the signal including systematic, $\mathcal{R}(D^{(*)})$: the ratio $\mathcal{B}(B \to D^{(*)}\tau\nu)/\mathcal{B}(B \to D^{(*)}\ell\nu)$.

| Experiment | Tag | $N_{B\overline{B}}$ $(10^6)$ | $\mathcal{B}$ $(10^{-4})$ | $\Sigma$ | $\mathcal{R}(D^{(*)})$ | Reference |
|---|---|---|---|---|---|---|
| $B^0 \to D^{*-}\tau^+\nu_\tau$ | | | | | | |
| Belle | inclusive | 535 | $2.02^{+0.40}_{-0.37} \pm 0.37$ | 5.2 | | Matyja (2007) |
| *BABAR* | hadronic | 471 | $1.74 \pm 0.19 \pm 0.12$ | 10.4 | $0.355 \pm 0.039 \pm 0.021$ | Lees (2012e) |
| $B^+ \to \overline{D}^{*0}\tau^+\nu_\tau$ | | | | | | |
| Belle | inclusive | 657 | $2.12^{+0.28}_{-0.27} \pm 0.29$ | 8.1 | | Bozek (2010) |
| *BABAR* | hadronic | 471 | $1.71 \pm 0.17 \pm 0.13$ | 9.4 | $0.322 \pm 0.032 \pm 0.022$ | Lees (2012e) |
| $B^0 \to D^-\tau^+\nu_\tau$ | | | | | | |
| *BABAR* | hadronic | 471 | $1.01 \pm 0.18 \pm 0.12$ | 5.2 | $0.469 \pm 0.084 \pm 0.053$ | Lees (2012e) |
| $B^+ \to \overline{D}^0\tau^+\nu_\tau$ | | | | | | |
| Belle | inclusive | 657 | $0.77 \pm 0.22 \pm 0.12$ | 3.5 | | Bozek (2010) |
| *BABAR* | hadronic | 471 | $0.99 \pm 0.19 \pm 0.13$ | 4.7 | $0.429 \pm 0.082 \pm 0.052$ | Lees (2012e) |
| $B \to \overline{D}\tau^+\nu_\tau$ (isospin constrained) | | | | | | |
| *BABAR* | hadronic | 471 | $1.02 \pm 0.13 \pm 0.11$ | 6.8 | $0.440 \pm 0.058 \pm 0.042$ | Lees (2012e) |
| $B \to \overline{D}^*\tau^+\nu_\tau$ (isospin constrained) | | | | | | |
| *BABAR* | hadronic | 471 | $1.76 \pm 0.13 \pm 0.12$ | 13.2 | $0.332 \pm 0.024 \pm 0.018$ | Lees (2012e) |

### 17.10.4 Discussion and future prospects

Searching for leptonic decays of charged $B$ mesons at the present generation of $B$ Factories has proven to be very challenging, with the light lepton modes $B^+ \to e^+\nu$ and $B^+ \to \mu^+\nu$ remaining beyond the experimental sensitivities and the $B^+ \to \tau^+\nu$ mode observed, but not yet precisely measured. Prior to the most recent Belle hadronic-tag search (Adachi, 2012b), all $B^+ \to \tau^+\nu$ measurements, including previous Belle hadronic-tag studies, had reported branching fractions which were consistently high compared with SM expectations. The most recent *BABAR* hadronic-tag measurement (Lees, 2013a) determines a branching fraction which is approximately a factor of two higher than the corresponding Belle result, although the discrepancy between the two is only about $2\sigma$. Other possible points of concern are the modeling of the experimentally crucial $E_{\rm extra}$ variable, which has shown indications of discrepancies in previous measurements, and the internal consistency of $B^+ \to \tau^+\nu$ branching fraction values obtained with different tau decay modes, see for example Lees (2013a). Leptonic and hadronic tau decay signatures have vastly different background sources and rates, hence inconsistencies in signal yields between tau decay modes is a potential red-flag of experimental problems. As both *BABAR* and Belle have published $B^+ \to \tau^+\nu$ results based on their full data samples, it is unlikely that this situation will be clarified without additional measurements from future $B$ Factory experiments with very large data samples. However, the experimental challenges posed by this decay, in particular low momentum lepton particle identification and modeling of hadronic backgrounds and extra calorimeter energy, will likely be similar (or worse) at high luminosity experiments. Consequently, it is not clear how precisely $B^+ \to \tau^+\nu$ will ultimately be measured. It is notable that precise measurements of $\mathcal{B}(B^+ \to \mu^+\nu)$ at the SM rate will be possible with data samples of $\mathcal{O}(50-100)$ ab$^{-1}$, using both tagged and un-tagged approaches. As the $B^+ \to \mu^+\nu$ searches utilize a cleanly identifiable high momentum muon and do not rely on $E_{\rm extra}$, they will provide an independent test of possible new physics in leptonic $B$ decays. If new physics were present in the form of a Type II 2HDM, then $\mathcal{B}(B^+ \to \mu^+\nu)$ would potentially be sensitive to it. This scenario is, however, disfavored by $B \to D^{(*)}\tau\nu$ measurements discussed above, so it is not clear what the implications might be for the purely leptonic $B$ decay modes. It is, however, likely that these modes will play an important role in elucidating any new physics that might be present in $B \to D^{(*)}\tau\nu$.

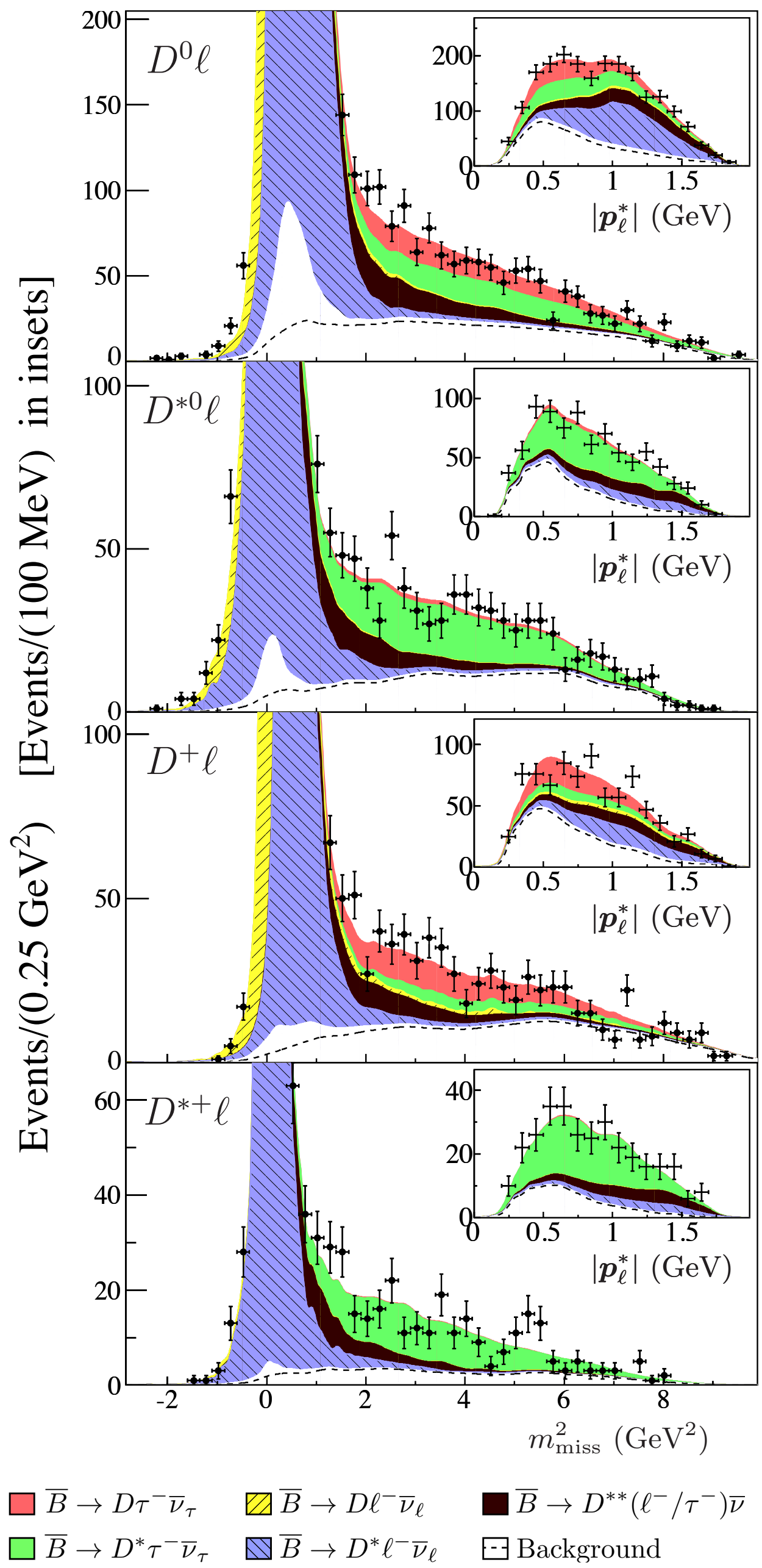

**Figure 17.10.9.** Projections of the *BABAR* (Lees, 2012e) $B \to D^{(*)}\tau\nu$ fit for the four signal modes (a) $D^0\ell$, (b) $D^{*0}\ell$, (c) $D^+\ell$ and (d) $D^{*+}\ell$ in $M^2_{\rm miss}$. Inset figures show the corresponding projection in $|p^*_\ell|$ for $M^2_{\rm miss} > 1$ GeV$^2$ in order to exclude the $B \to D^{(*)}\ell\nu$ normalization component. Stacked in order from the bottom the components are: continuum and $B\overline{B}$ background (below dashed line), charge cross-feed (white above dashed line), $B \to \overline{D}^*\ell^+\nu$ (blue hatched), $B \to \overline{D}\ell^+\nu$ (yellow hatched), $B \to \overline{D}^{**}\ell^+\nu$ (black), $B \to \overline{D}^*\tau^+\nu$ (green), $B \to \overline{D}\tau^+\nu$ (red).

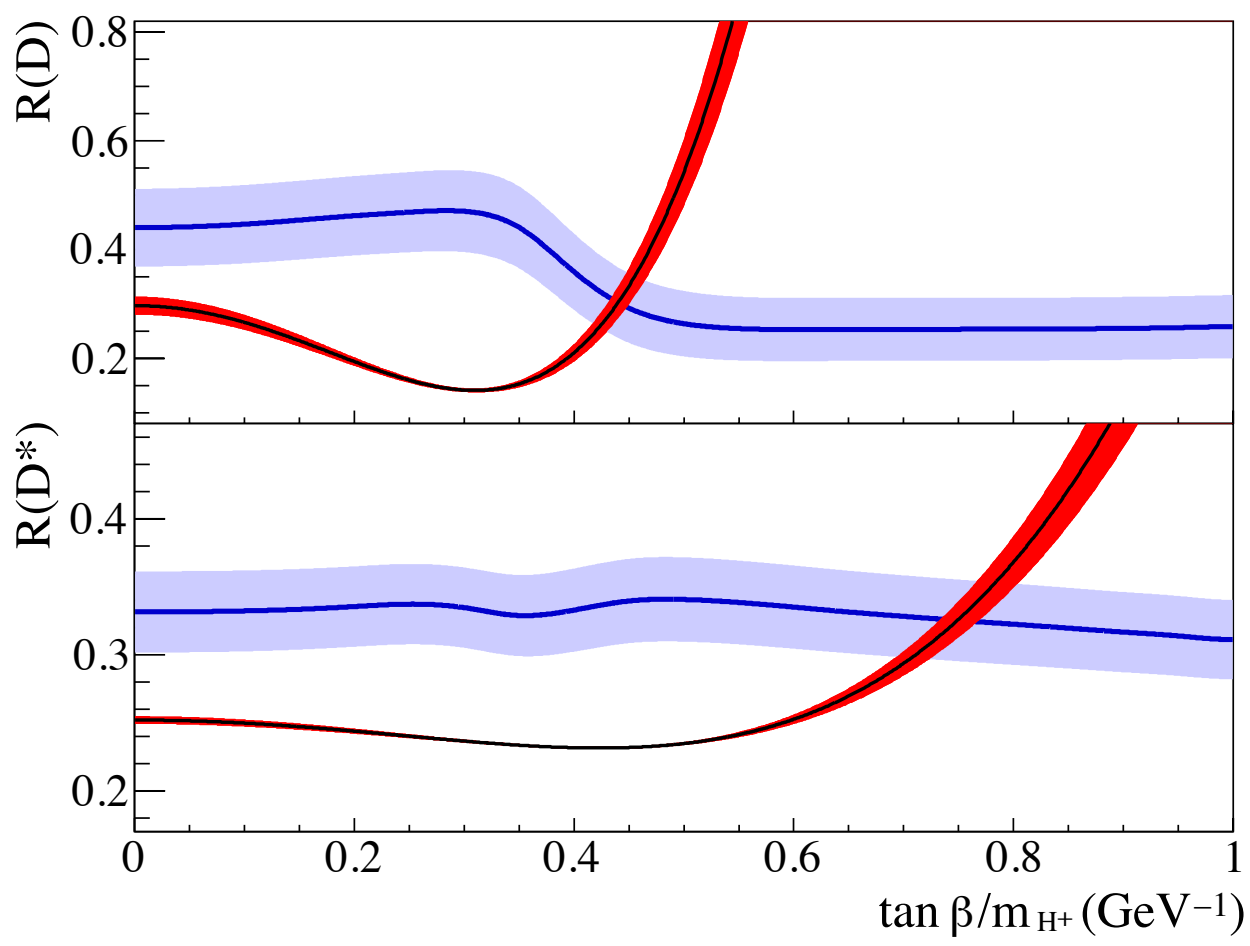

**Figure 17.10.10.** Comparison of the *BABAR* measurements of $\mathcal{R}(D)$ (top) and $\mathcal{R}(D^*)$ (bottom) with the prediction of the Type-II 2HDM (red curves). The $\tan\beta/M_H$ dependence of the *BABAR* $\mathcal{R}(D^{(*)})$ results (blue shaded bands) arises due to changes to the signal kinematics in the case that the process is mediated by a $H^\pm$ rather than a $W^\pm$. The SM case corresponds to the case $\tan\beta/M_H = 0$, and the favored values for $\tan\beta/M_H$ differ for $\mathcal{R}(D)$ and $\mathcal{R}(D^*)$. Plot is taken from Lees (2012e).

Measurement of the radiative decay $B^+ \to \ell^+\nu\gamma$ has proven to be problematic at the current generation of $B$ Factories, but the method using hadronic tag $B$ reconstruction (Aubert, 2009a) appears to offer a solution that can permit observation and measurement of this branching fraction at future high-luminosity $B$ Factories. Although these modes do not provide interesting new physics sensitivity, they may be relevant for the interpretation of precision measurements of $B^+ \to \ell^+\nu$ and of $B \to \pi^+\pi^-$ decays.

Future measurements of $B \to D^{(*)}\tau\nu$ will greatly improve our understanding of this process. The recent *BABAR* study of $\mathcal{R}(D^{(*)})$ (Lees, 2012e) provides a conceptually robust experimental method for measuring these decays using the ratio $\mathcal{R}(D^{(*)})$, but this analysis yields results that are incompatible both with the SM and the "preferred" 2HDM scenario that underlies the MSSM. These results have not as of yet been independently confirmed by Belle, but may provide a tantalizing glimpse of new physics coupling to third-generation leptons. In addition to determining $\mathcal{R}(D^{(*)})$, future measurements of $B \to D^{(*)}\tau\nu$ at high luminosity $B$ Factories also have the potential to precisely study the $q^2$ distributions or angular properties of these decays, providing an additional handle on possible beyond-SM contributions. However, these studies will be confronted with the same experimental challenges as the *BABAR* measurement, in particular the modeling of the missing mass, lepton momentum and $E_{\rm extra}$ distributions, and the understanding of background contributions from semileptonic decays with higher mass open-charm states. All of these issues will need to be addressed with increased precision in order to make substantial improvements to the existing $\mathcal{R}(D^{(*)})$ measurements at future high luminosity experiments.

## 17.11 Rare and forbidden $B$ decays

***Editors:***
*Steve Robertson (BABAR)*
*Youngjoon Kwon (Belle)*

**Additional section writers:**
*Matthew Bellis, Fergus Wilson*

While many $B$ decay modes are "rare" in the sense that they have small branching fractions, the term usually refers specifically to modes which are suppressed within the SM due to some property of the decay, often related to a symmetry or conserved quantum number. The interest in these modes arises from the possibility that contributions from physics beyond the SM may not be similarly suppressed. In this sense, many of the modes discussed in Sections 17.9 and 17.10 would be considered rare decays as well, since for example the tree-level process $B^+ \to \ell^+\nu$ is suppressed by helicity conservation when the decay is mediated by a SM $W^+$ boson. New-physics models permit possible contributions to this process which are mediated by a scalar charged Higgs boson and lead to potentially observable deviations from the SM expectations for the branching fractions. Similarly, the flavor changing neutral current modes in Section 17.9 are suppressed due to the absence of tree-level FCNCs in the SM. Possible new-physics contributions, which also enter at one-loop level, can therefore be comparable in size to the suppressed SM contributions. The decays $B^0 \to \ell^+\ell^-$ and $B^0 \to \nu\bar{\nu}$, discussed below in Sections 17.11.1 and 17.11.2, are in fact exactly such processes: highly suppressed SM electroweak penguin FCNC processes which can be strongly enhanced in beyond-SM models by neutral Higgs boson contributions or other new physics. The distinction as to which modes are considered "rare" is obviously somewhat arbitrary.

In contrast, "forbidden" modes are those which are expected to proceed primarily through processes beyond the SM, which typically violate quantum numbers that are conserved within the SM such as charged lepton flavor and number, and baryon number. These decays are forbidden in the SM in the absence of neutrino masses. However, lepton flavor is not associated with a fundamental conservation law in the SM and in fact the existence of neutrino mixing explicitly requires that lepton flavor is not conserved in the neutrino sector. This in turn implies lepton flavor violation (LFV) in the charged lepton sector as well, via loop processes which contain neutrinos. However, the expected rate for such processes is many orders of magnitude below current or foreseen future experimental sensitivity to these decay modes. Observation of LFV in $B$ decays would therefore be unambiguous evidence for a new source of LFV beyond the SM. Similarly, lepton number is not protected by any fundamental conservation law and in fact is explicitly violated if neutrinos are of Majorana type, *i.e.* if they are their own antiparticles. Consequently, searches for lepton number violation (LNV), discussed in 17.11.5, can provide insight into the nature of neutrinos.

We discuss $B^0 \to \ell^+\ell^-$ along with the radiative mode $B^0 \to \ell^+\ell^-\gamma$ in Section 17.11.1 and the neutrino counterparts to these modes, $B^0 \to \nu\bar{\nu}(\gamma)$, in Section 17.11.2. We describe $B^0_{(s)} \to \gamma\gamma$ in Section 17.11.3. Lepton flavor and lepton number violating modes are presented in Sections 17.11.4 and 17.11.5, respectively, and searches for baryon number violating modes are discussed in Section 17.11.6.

### 17.11.1 $B^0 \to \ell^+\ell^-(\gamma)$

$B^0 \to \ell^+\ell^-$ decays are expected to proceed through the diagrams shown in Figure 17.11.1 within the SM (SM). The branching fraction for the $B^0 \to \ell^+\ell^-$ decays can be written to good accuracy as

$$\mathcal{B}(B^0 \to \ell^+\ell^-) = \frac{G_F^2\alpha^2}{64\pi^3 \sin^4\theta_W} \times \\ |V_{tb}^* V_{td}|^2 \tau_B M_B^3 f_B^2 \sqrt{1 - \frac{4m_\ell^2}{M_B^2}} \cdot \frac{4m_\ell^2}{M_B^2} Y^2(m_t^2/M_W^2) \ . \tag{17.11.1}$$

The equation reveals a high suppression of the decays due to the internal quark annihilation within the $B$ meson involving a $b \to d$ transition (CKM element $|V_{td}|$ and the $B$ meson decay constant $f_B$) and helicity considerations (helicity suppression factor $m_\ell^2/M_B^2$).[93] The SM expected branching fractions are of the order of $10^{-15}$ and $10^{-10}$ for the $e^+e^-$ and $\mu^+\mu^-$ modes, respectively. In some new-physics models, including those with two Higgs doublets and $Z$-mediated FCNC, the branching fractions could be enhanced by two orders of magnitude (Babu and Kolda, 2000; Bobeth, Ewerth, Kruger, and Urban, 2001; Chankowski and Slawianowska, 2001; Choudhury and Gaur, 1999; Hewett, Nandi, and Rizzo, 1989).

$B^0 \to \tau^+\tau^-$ is much less helicity suppressed than $B^0 \to e^+e^-$ and $B^0 \to \mu^+\mu^-$ due to the large tau mass, with a predicted branching fraction of order $10^{-7}$ (Harrison and Quinn, 1998; Grossman, Ligeti, and Nardi, 1997). The large masses of the tau leptons also provide the potential for substantial enhancements due to Higgs couplings in two-Higgs-doublet models (Babu and Kolda, 2000; Logan and Nierste, 2000). However, due to the presence of multiple neutrinos in the experimental final state (between two and four depending on the tau decay modes), the $\tau^+\tau^-$ final state is considerably more difficult to access experimentally than the $e^+e^-$ and $\mu^+\mu^-$ modes. The $e^+e^-$ and $\mu^+\mu^-$ searches are described in 17.11.1.1, and the $\tau^+\tau^-$ decay in 17.11.1.2.

Although $B^0 \to \ell^+\ell^-\gamma$ ($\ell = e$ or $\mu$) decays can occur by emitting a photon from any of the initial or final-state fermions of $B^0 \to \ell^+\ell^-$, the dominant contribution is due to photon emission from one of the initial-state quarks, since this process is free from the helicity suppression associated with $B^0 \to \ell^+\ell^-$. In the SM, the expected $B^0 \to \ell^+\ell^-\gamma$ branching fractions are about $10^{-10}$ (Aliev, Ozpineci, and Savci, 1997; Eilam, Halperin, and Mendel, 1995). Observation of such signals with current sensitivities of *BABAR* and Belle would provide clear evidence for new physics. $B^0 \to \ell^+\ell^-\gamma$ is discussed in Section 17.11.1.3.

[93] The function $Y(m_t^2/M_W^2)$ is known with good accuracy at NLO.

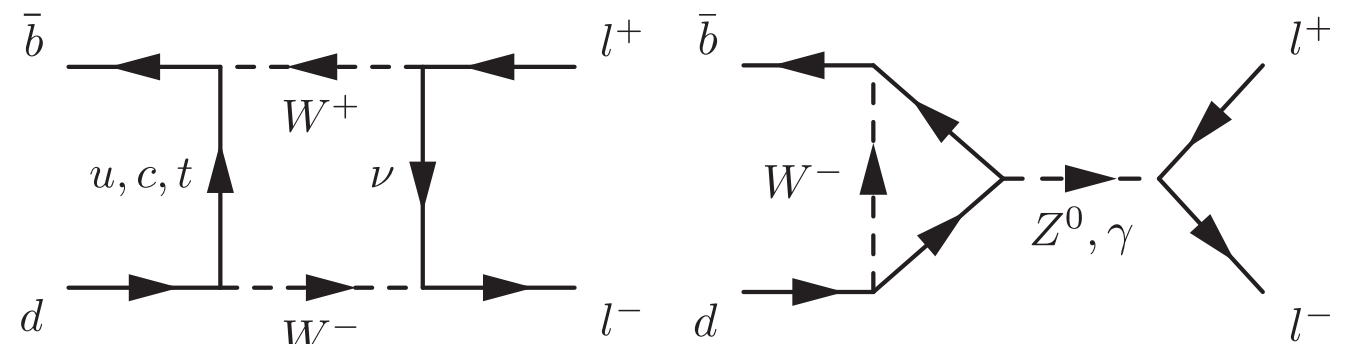

**Figure 17.11.1.** SM diagrams for $B^0 \to \ell^+\ell^-$.

### 17.11.1.1 $B^0 \to \ell^\pm\ell'^\mp$ $(\ell, \ell' = e, \mu)$

Searches for $B^0$ decays to pairs of light leptons, $\ell = e, \mu$, have been performed at both $B$ Factories and at hadron colliders; the latter are able to probe not only $B^0$ but also $B_s^0$ decays. Searches for $B^0 \to e^+e^-$ and $B^0 \to \mu^+\mu^-$ as well as the LFV mode $B^0 \to e^\pm\mu^\mp$ by *BABAR* and Belle provided the most stringent limits on new physics in these modes until around 2008, when they were superseded by results from Run-II at the Tevatron. LHC experiments have now pushed the experimental results considerably beyond the current $B$ Factory sensitivities. The *BABAR* and Belle analyses are described in the following.

*BABAR* has searched for these decays in a data sample of $384 \times 10^6$ $B\overline{B}$ pairs (Aubert, 2008as). The signal candidates are reconstructed by pairing oppositely charged leptons. Leptons are identified with stringent requirements which retain $\sim 93\%$ ($\sim 73\%$) of $e^\pm$ ($\mu^\pm$), while less than $\sim 0.1\%$ ($\sim 3\%$) of pions are misidentified as electrons (muons). The signal candidates are required to satisfy $m_{\rm ES} > 5.2$ GeV/$c^2$ and $|\Delta E| < 0.15$ GeV. To partially recover the energy lost by electrons due to final-state radiation or bremsstrahlung, photons consistent with originating from the $e^+$ or $e^-$ track have their 4-momentum added to the track. Using MC simulations, peaking background contributions from $B^0 \to h^+h'^-$ ($h, h' = \pi$ or $K$) decays are estimated to be of the order of $10^{-4}$ or less. After applying the lepton ID requirements, other $B\overline{B}$ background is found to be negligible. The backgrounds from non-$B\overline{B}$ events, such as $q\bar{q}$ ($q = u, d, s, c$) continuum and $\tau^+\tau^-$ production, are reduced by using event shape variables which are combined into a single Fisher discriminant $\mathcal{F}$. The signal yields for $e^+e^-$, $\mu^+\mu^-$ and $e^\pm\mu^\mp$ (see Section 17.11.4 for further discussion of LFV modes) are independently obtained by maximum likelihood (ML) fits to $m_{\rm ES}$, $\Delta E$ and $\mathcal{F}$, where the *p.d.f.*s used in the likelihood function is composed of an uncorrelated product of the *p.d.f.*s of the individual discriminating variables. As an example the distribution of $\Delta E$ in the search for $B^0 \to \mu^+\mu^-$ is shown in Fig. 17.11.2. No significant excesses of signal were seen in any modes, and the 90% C.L. upper limits on the corresponding branching fractions are calculated utilizing a Bayesian approach assuming a flat positive prior and including systematic uncertainties. The obtained upper limit is, at 90% C.L., 11.3 (5.2) $\times 10^{-8}$ for the $e^+e^-$ ($\mu^+\mu^-$) mode (see Table 17.11.1).

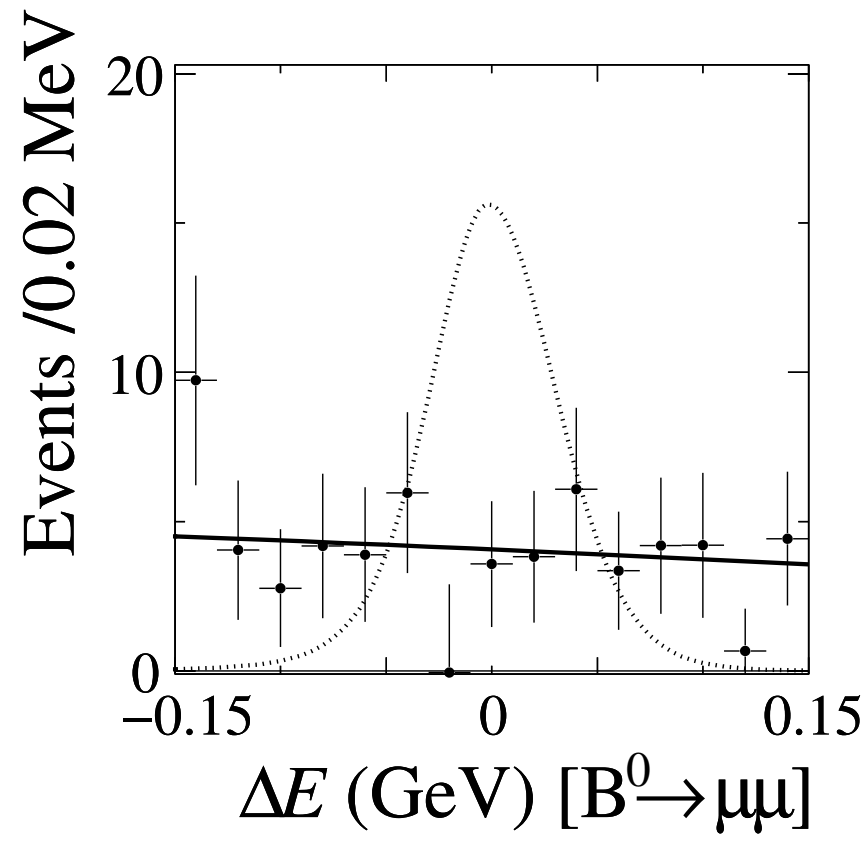

**Figure 17.11.2.** $\Delta E$ distribution of selected $B^0 \to \mu^+\mu^-$ candidates (Aubert, 2008as). The solid curve is the background ${}_s\mathcal{P}lot$ (see Section 11.2.3) and the dashed curve is the expected distribution of signal with an arbitrary normalization.

Using a data sample of 85 million $B\overline{B}$ pairs, Belle has also searched for these modes (Chang, 2003) and obtained 90% C.L. upper limits of 1.9 (1.6) $\times$ $10^{-7}$ for the $e^+e^-$ ($\mu^+\mu^-$) mode. These upper limits are calculated based on the likelihood ratio ordering (Feldman and Cousins, 1998) and including systematic uncertainties using the POLE program (Conrad, Botner, Hallgren, and Perez de los Heros, 2003). [94]

The Belle analysis included missing-momentum-based quantities in its Fisher discriminant, which significantly reduced the important class of background due to double semileptonic charm decays from the continuum. Using the Pati-Salam model (Kuznetsov and Mikheev, 1994), which predicts a vector leptoquark at a mass associated with the scale of the breaking of an $SU(4)$ gauge group to the usual color $SU(3)$ group, along with the assumption that there are no other colored particles between the $t$-quark mass and the mass, $M_{\rm LQ}$, of the Pati-Salam leptoquark, Belle has obtained $M_{\rm LQ} > 46$ TeV/$c^2$ at the 90% C.L..

### 17.11.1.2 $B^0 \to \tau^+\tau^-$

*BABAR* published a search for $B^0 \to \tau^+\tau^-$ (Aubert, 2006b) based on a data sample of $(232 \pm 3) \times 10^6$ $B\overline{B}$ events (210 fb$^{-1}$) and using the method of exclusive hadronic tag reconstruction of the accompanying $B$ meson as described in Section 7.4.1. Evidence for a $B^0 \to \tau^+\tau^-$ decay is sought by considering all charged tracks and clusters which are not associated with a $B$ candidate, "$B_{\rm tag}$", which has been exclusively reconstructed in one of a large number of hadronic decay modes, as illustrated in Figure 17.11.3. Only events with "one-prong" decays of both taus are considered, so signal events are required to contain exactly two charged tracks on the signal side, each of which is identified as an electron, muon or pion. Each of the two tau leptons can potentially decay to $\tau \to e\nu\bar{\nu}$, $\tau \to \mu\nu\bar{\nu}$, $\tau \to \pi\nu$ or $\tau \to \rho(770)\nu$. Signal topologies are therefore defined corresponding to each combination of $\tau^+\tau^-$ decay modes. Charged pion tracks are considered to be $\rho(770)$ candidates if a $\pi^0$ candidate, reconstructed from a pair of photon clusters, can be combined with the $\pi$ track to give $0.6 < m_{\pi\pi^0} < 1.0$ GeV$/c^2$. Events with any additional $\pi^0$ candidates are rejected, and the sum of any remaining calorimeter energy ($E_{\rm extra}$) is required to be less than 110 MeV (summing all clusters with energy exceeding 30 MeV).

[94] The POLE program calculates an upper limit with an extension of the Feldman Cousins method (Feldman and Cousins, 1998) by incorporating systematic uncertainties on the background yields and signal reconstruction efficiency. These uncertainties are incorporated in the calculation by integrating the *p.d.f.*s that parameterize them.

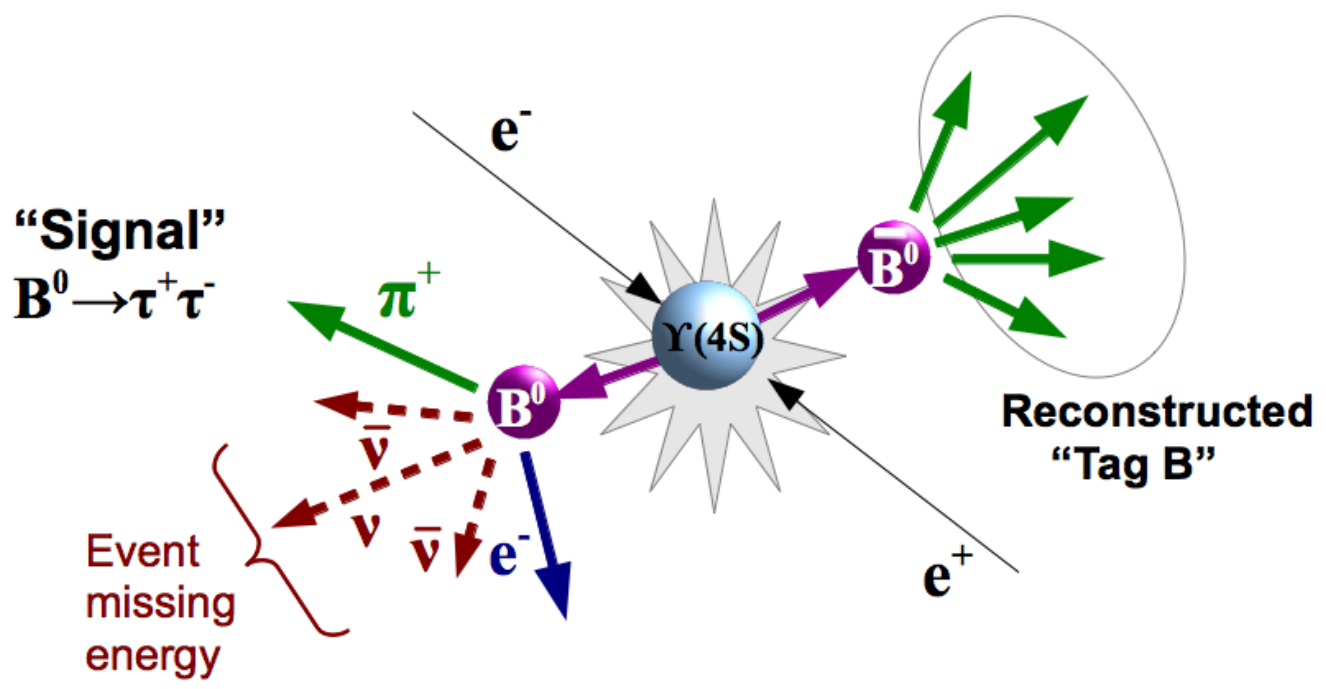

**Figure 17.11.3.** Illustration of a $B^0 \to \tau^+\tau^-$ event in which the associated $\overline{B}^0$ is reconstructed as a hadronic $B_{\rm tag}$. The tau decay modes are depicted as $\tau^- \to e^-\nu\bar{\nu}$ and $\tau^+ \to \pi^+\bar{\nu}$. Since the $\overline{B}^0$ 4-vector is determined by the tag reconstruction, the signal $B^0$ 4-vector can be obtained using the known CM energy, and the event missing energy can be fully attributed to the neutrinos.

Backgrounds from $B$ decays to open charm, which subsequently decay to final states containing a strange quark, are suppressed by vetoing events in which any signal candidate track is identified as a $K^+$, or if the combination of the two tracks is consistent with originating from a $K^0_S \to \pi^+\pi^-$ decay. Note that this background is large due to the Cabibbo favored $b \to c \to s$ transitions. Similarly, events possessing a calorimeter cluster which is identified as a $K^0_L$ candidate, based on cluster energy and event shape information, are rejected.

Additional background suppression is obtained by exploiting correlations between the momenta and angular distributions of the tau decay daughters in the signal $B$ rest frame, which is estimated from the 4-vector of the reconstructed tag $B$. A set of neural networks, one for each of the $\tau^+\tau^-$ decay topologies, are trained to discriminate signal from background based on four inputs: the $B$ rest frame momenta of the positively and negatively charged tau daughters, $p_+$ and $p_-$, respectively, $\cos\theta \equiv \boldsymbol{p}_+ \cdot \boldsymbol{p}_-/p_+p_-$, and $E_{\rm extra}$.

Substantial backgrounds remain following this selection, primarily arising from $b \to c \to s$ processes with significant missing energy and no identified kaon. Typically these are $B$ decays with an undetected $K^0_L$, with one or more particles passing outside of the detector acceptance and/or semileptonic $B$ or charm decays. A total of $281 \pm 48$ background events are expected and $263 \pm 19$ events are observed in data, distributed across all modes. A 90% C.L. branching fraction limit of $\mathcal{B}(B^0 \to \tau^+\tau^-) < 4.1 \times 10^{-3}$ is obtained. Because of the limited sensitivity imposed by the high backgrounds, this analysis has not been repeated, either by *BABAR* with a larger data sample or by Belle.

#### 17.11.1.3 $B^0 \to \ell^+\ell^-\gamma$ ($\ell = e, \mu$)

*BABAR* has searched for the radiative decays $B^0 \to \ell^+\ell^-\gamma$ in the $\ell = e, \mu$ modes in an event sample of $320 \times 10^6$ $B\overline{B}$ pairs (Aubert, 2008av). Signal MC events are simulated using a leading-order calculation of the Wilson coefficients $C_7, C_9$, and $C_{10}$ (Dincer and Sehgal, 2001; see also the discussion in Section 17.9.1 of this book). Events are selected by combining a pair of oppositely-charged leptons and an energetic photon yielding $B$ candidates within the region $|\Delta E| \leq 0.5$ GeV and $5.0 \leq m_{\rm ES} \leq 5.3$ GeV$/c^2$. Signal candidates are required to lie in the smaller signal region defined by $-0.146(-0.112) \leq \Delta E \leq 0.082$ GeV and $5.270 \leq m_{\rm ES} \leq 5.289$ GeV$/c^2$ for the $e^+e^-\gamma$ ($\mu^+\mu^-\gamma$) mode, while the remainder of the larger region is used for background studies. The dominant backgrounds include: (1) un-modelled higher-order QED and two-photon processes for the $e^+e^-\gamma$ mode, (2) $B$ decays where a $\pi^0$ produces the photon or a $J/\psi$ (or $\psi(2S)$) produces one or both of the leptons, and (3) continuum processes. Backgrounds of type (1) are suppressed by imposing fiducial constraints on the electrons, cutting on event shape variables, requiring that the photon energy exceeds 0.3 GeV and requiring that there are at least 5 charged tracks and 10 calorimeter clusters in the event. The $B$ decay backgrounds are suppressed by $\pi^0$ and $J/\psi$ (or $\psi(2S)$) vetoes. Continuum backgrounds are rejected by using a combination of event shape variables (see Chapter 9). A neural network is constructed using event shape, angular and kinematic variables and trained using MC to discriminate between signal events and remaining background events. After applying all selection requirements, the expected number of background events is estimated from data sideband regions to be $1.75 \pm 1.38 \pm 0.36$ and $2.66 \pm 1.40 \pm 1.58$ events for $e^+e^-\gamma$ and $\mu^+\mu^-\gamma$, respectively, where the quoted errors are statistical and systematic. The background is dominated by non-$B$ backgrounds and so is estimated by extrapolation from the signal sideband regions. One event is found in the signal region in data for each mode, consistent with the size of the expected backgrounds. The 90% C.L. upper limits on the corresponding branching fractions are determined using a frequentist method (Barlow, 2002): $\mathcal{B}(B^0 \to e^+e^-\gamma) < 1.2 \times 10^{-7}$ and $\mathcal{B}(B^0 \to$ $\mu^+\mu^-\gamma) < 1.6 \times 10^{-7}$. Belle has not reported any results for these decay modes.

All the results described in this subsection are summarized in Table 17.11.1.

### 17.11.2 $B^0 \to$ invisible

The decay of a $B$ meson into $\nu\bar{\nu}$ pairs is similar from a theoretical point of view to the leptonic decays $B^0 \to \ell^+\ell^-$ described in Section 17.11.1. It is extremely suppressed in the SM due to helicity considerations, and is only permitted, albeit at a rate orders of magnitude below experimental sensitivity, due to the miniscule but non-zero neutrino mass. As is the case with other radiative decays, for example $B^0 \to \ell^+\ell^-\gamma$ (Section 17.11.1) and $B^+ \to \ell^+\nu\gamma$ (Section 17.10.2), the radiation of a photon from an initial-state quark can remove this helicity suppression, resulting in a larger branching fraction than the non-radiative process. SM branching fractions for $B^0 \to \nu\bar{\nu}$ and $B^0 \to \nu\bar{\nu}\gamma$ have been computed to be $\sim 1 \times 10^{-25}$ and $2 \times 10^{-9}$, respectively (Badin and Petrov, 2010). In practice, experimental searches for these modes cannot directly detect the neutrinos, and so are more correctly considered to be searches for $B \to E_{\rm miss}(+\gamma)$, where $E_{\rm miss}$ represents missing energy from all sources including not only neutrinos, but also possible new-physics particles which do not interact in the detector. As such, they are frequently referred to as "$B^0 \to$ invisible$(+\gamma)$".

New stable particles which do not interact in the detector are potential dark matter candidates. Consequently, these decay modes are interesting probes of new physics. Decays to pairs of such particles, $B^0 \to \chi_0\chi_0(\gamma)$ where $\chi_0$ are massive scalars, are not helicity suppressed and hence can occur at rates substantially above the SM rate for $B \to$ invisible decays. A phenomenological model for $B^0 \to \bar{\nu}\chi_1^0$, where $\chi_1^0$ is a neutralino, predicts a branching fraction in the range $10^{-7}$–$10^{-6}$ (Dedes, Dreiner, and Richardson, 2001). Since both decay products would be undetected, the experimental signature would be $B \to$ invisible. Models with large extra dimensions (Agashe, Deshpande, and Wu, 2000; Agashe and Wu, 2001; Davoudiasl, Langacker, and Perelstein, 2002) can also result in significant enhancements to the invisible decay rate.

Because the signature for $B^0 \to$ invisible$(\gamma)$ decays is the absence of detector activity (*i.e.* charged tracks and neutral calorimeter clusters) associated with an identified $B$ meson decay, the analysis strategy relies on exclusive tag-$B$ reconstruction (see Section 7.4). Tag reconstruction serves the dual purpose of identifying the event as an $\Upsilon(4S) \to B^0\overline{B}^0$ transition, and uniquely associating all detector activity with either the tag $B$ or the signal $B$ candidate. Unlike other modes which use this method, for example $B^0 \to \tau^\pm\ell^\mp$ and $B^+ \to h^+\tau^\pm\ell^\mp$ discussed in Section 17.11.4, there is no kinematic advantage to be gained from knowledge of the signal $B$ candidate 4-vector (estimated from the tag $B$ 4-vector). Consequently, tag $B$ reconstruction based on semileptonic $B$ decays, which possess additional missing energy due to the un-reconstructed neutrino, is equally viable to hadronic $B$ tagging for these searches, although signal efficiencies and background rates differ significantly between the two methods.

To date, only Belle has published limits on $B^0 \to$ invisible using hadronic tag reconstruction (Hsu, 2012), while *BABAR* has only published results based on semileptonic tag reconstruction. A recent *BABAR* paper (Lees, 2012g) updated the results of an earlier analysis (Aubert, 2004y) to include the full *BABAR* data sample. *BABAR* also reports limits on the $B^0 \to \nu\bar{\nu}\gamma$ branching fraction for $E_\gamma > 1.2$ GeV and assuming decay kinematics based on a constituent quark model (Lu and Zhang, 1996).

In the *BABAR* analysis, semileptonic $B_{\rm tag}$ candidates are reconstructed in the modes $B^0 \to D^{(*)-}\ell^+\nu$. Details of the *BABAR* semileptonic tag reconstruction procedure can be found in Section 7.4.2. The $D^{(*)-}$ candidates are combined with identified electrons or muons having lab-frame momentum greater than 800 MeV, and the $\ell^+ - D^{(*)-}$ combination is required to be consistent with a common decay vertex. $B^0$ candidates are then selected by requiring the $\ell^+ - D^{(*)-}$ combination to be kinematically consistent with a $B^0 \to D^{(*)-}\ell^+\nu$ event, *i.e.* that the only missing particle is the unobserved neutrino. The quantity $\cos\theta_{B,D^{(*)-}\ell^+}$ is computed as

$$\cos\theta_{B,D^{(*)-}\ell^+} = \frac{2E_B E_{D^{(*)-}\ell^+} - m_B^2 - m_{D^{(*)-}\ell^+}^2}{2|\boldsymbol{p}_B||\boldsymbol{p}_{D^{(*)-}\ell^+}|}\ , \tag{17.11.2}$$

where $E_{D^{(*)-}\ell^+}$, $\boldsymbol{p}_{D^{(*)-}\ell^+}$ and $m_{D^{(*)-}\ell^+}$ are the CM frame energy, momentum 3-vector and the invariant mass of the $\ell^+$-$D^{(*)-}$ combination. The quantity $m_B$ is the nominal $B$ meson mass, while $E_B$ and $|\boldsymbol{p}_B|$ are the expected $B$ energy and momentum magnitude computed from the known CM energy. The quantity $\cos\theta_{B,D^{(*)-}\ell^+}$ represents the cosine of a physical angle only in the case of a correctly reconstructed $B^0 \to D^{(*)-}\ell^+\nu$ decay. For background events, however, it does not relate to a physical angle, and so $\cos\theta_{B,D^{(*)-}\ell^+}$ can assume values outside of the mathematically allowed region $[-1, 1]$. Lees (2012g) accepts events in the region $-5.5 < \cos\theta_{B,D^{(*)-}\ell^+} < 1.5$ in order to retain high signal efficiency while accounting for detector resolution effects which produce values slightly outside of the allowed region. The larger range for negative values was chosen to implicitly include contributions from higher-mass open charm states in which some of the decay products have not been explicitly reconstructed.

After identifying a well-reconstructed $B^0 \to D^{(*)-}\ell^+\nu$ candidate (and an energetic photon in the case of $B^0 \to \nu\bar{\nu}\gamma$), signal events should have little or no additional detector activity. Consequently, the signal selection requires that no additional tracks are present in the event, there is only limited activity in the calorimeter, and the missing momentum vector of the event is required to point within the detector fiducial acceptance.

Additional background suppression is obtained by using a neural network which includes as inputs the CM-frame lepton momentum, $\cos\theta_{B,D^{(*)-}\ell^+}$, and the angle between the event thrust axis and the $D^{(*)-}\ell^+$ momentum direction. A number of additional inputs are included which are specific to the $B^0 \to \nu\bar{\nu}$ and $B^0 \to \nu\bar{\nu}\gamma$ searches.

**Table 17.11.1.** Summary of the results for $B^0 \to \ell^+\ell^-(\gamma)$, $B^0 \to \nu\bar{\nu}(\gamma)$ and $B^0_{(s,d)} \to \gamma\gamma$ modes.

| Experiment | Decay Mode | Method | $N_{B\bar{B}}$ ($10^6$) | $\mathcal{B}$ upper limit (90% C.L.) | Reference |
|---|---|---|---|---|---|
| Belle | $B^0 \to e^+e^-$ | signal recon. only | 85 | $1.9 \times 10^{-7}$ | Chang (2003) |
| BABAR | $B^0 \to e^+e^-$ | signal recon. only | 384 | $1.1 \times 10^{-7}$ | Aubert (2008as) |
| BABAR | $B^0 \to e^+e^-\gamma$ | signal recon. only | 320 | $1.2 \times 10^{-7}$ | Aubert (2008av) |
| Belle | $B^0 \to \mu^+\mu^-$ | signal recon. only | 85 | $1.6 \times 10^{-7}$ | Chang (2003) |
| BABAR | $B^0 \to \mu^+\mu^-$ | signal recon. only | 384 | $0.52 \times 10^{-7}$ | Aubert (2008as) |
| BABAR | $B^0 \to \mu^+\mu^-\gamma$ | signal recon. only | 320 | $1.6 \times 10^{-7}$ | Aubert (2008av) |
| BABAR | $B^0 \to \tau^+\tau^-$ | hadronic tag | 232 | $4.1 \times 10^{-3}$ | Aubert (2006b) |
| Belle | $B^0 \to \nu\bar{\nu}$ | hadronic tag | 657 | $13 \times 10^{-5}$ | Hsu (2012) |
| BABAR | $B^0 \to \nu\bar{\nu}$ | semileptonic tag | 471 | $2.4 \times 10^{-5}$ | Lees (2012g) |
| BABAR | $B^0 \to \nu\bar{\nu}\gamma$ | semileptonic tag | 471 | $1.7 \times 10^{-5}$ | Lees (2012g) |
| Belle | $B^0 \to \gamma\gamma$ | signal recon. only | 111 | $6.2 \times 10^{-7}$ | Villa (2006) |
| BABAR | $B^0 \to \gamma\gamma$ | signal recon. only | 226 | $3.2 \times 10^{-7}$ | del Amo Sanchez (2011k) |
| Belle | $B^0_s \to \gamma\gamma$ | signal recon. at $\Upsilon(5S)$ | 23 fb$^{-1}$ | $8.7 \times 10^{-6}$ | Wicht (2008) |

The quantity $E_{\text{extra}}$ is constructed by summing the CM-frame energies of any remaining calorimeter clusters with a laboratory-frame energy greater than 30MeV. Signal and background *p.d.f.*s are constructed from MC simulation for $E_{\text{extra}}$, and an extended ML fit is performed on data to extract the signal and background yields. No evidence of signal is seen in either the $B^0 \to \nu\bar{\nu}$ or $B^0 \to \nu\bar{\nu}\gamma$ search. Branching fraction upper limits at are obtained using a Bayesian method which assumes a positive prior distribution (*i.e.* the observed negative signal yield does not result in a more stringent branching fraction limit than if zero signal yield had been obtained). Upper limits of $\mathcal{B}(B^0 \to \nu\bar{\nu}) < 2.4 \times 10^{-5}$ and $\mathcal{B}(B^0 \to \nu\bar{\nu}\gamma) < 1.7 \times 10^{-5}$ are obtained at 90% C.L..

The Belle search (Hsu, 2012) utilizes hadronic $B$ tag reconstruction based on $B^0 \to D^{(*)-}h^+$ decays, where $h^+$ can be $\pi^+$, $\rho(770)^+$, $a_1(1260)^+$, or $D_s^{(*)+}$. Details of the reconstruction procedure can be found in Section 7.4.1. Compared with the semileptonic tag reconstruction method used in the BABAR analysis, the hadronic tag method yields a somewhat lower reconstruction efficiency, but, since it does not have to deal with an unobserved neutrino, it also provides more stringent kinematic constraints on the reconstructed $B_{\text{tag}}$. As a consequence, backgrounds arising from $B_{\text{tag}}$ misreconstruction are inherently lower and a simpler signal selection procedure can be used. In the Belle analysis, after hadronic $B_{\text{tag}}$ events are selected, $B^0 \to$ invisible candidate events are required to have no additional tracks and no $\pi^0$ or $K^0_L$ candidates in the rest of the event. Continuum backgrounds are suppressed by considering two quantities: the cosine of the angle of the $B_{\text{tag}}$ flight direction (in the CM frame) relative to the beam axis, $\cos\theta_B$, and the cosine of the angle of the $B_{\text{tag}}$ thrust axis relative to the beam axis, $\cos\theta_T$. Signal decays peak at zero in both of these variables. Events are retained in the region $-0.9 < \cos\theta_B < 0.9$ and $-0.6 < \cos\theta_T < 0.6$. Belle defines the variable $E_{\text{ECL}}$ analogously to $E_{\text{extra}}$ for BABAR by summing the energies of remaining calorimeter clusters. However, different cluster energy thresholds are applied in different regions of the calorimeter: 50 MeV in the barrel region, 100 MeV in the forward endcap and 150MeV in the backward endcap. MC modeling of efficiencies and kinematic distributions is verified by studying $B^0 \to D^{(*)-}\ell^+\nu$ decays in events with a hadronic $B_{\text{tag}}$. The signal yield is extracted using a two-dimensional, unbinned ML fit to the $E_{\text{ECL}}$ and $\cos\theta_B$ distributions, where the *p.d.f.*s for the two distributions are treated as uncorrelated. A slight excess of signal events is obtained in the fit, with a significance of approximately $1.5\sigma$. A branching fraction upper limit of $\mathcal{B}(B^0 \to \text{invisible}) < 1.3 \times 10^{-4}$ at 90% C.L. is obtained, which is slightly worse than the expected sensitivity of $1.1 \times 10^{-4}$. The difference in sensitivity between the BABAR and Belle analyses ($2.4 \times 10^{-5}$ and $13 \times 10^{-5}$, respectively), roughly a factor of five, is thought to be primarily due to the difference in tag efficiencies between the hadronic and semileptonic methods, but also reflects different optimizations of the level of background between the two experiments.[95] The observed distributions of $E_{\text{extra}}$ ($E_{\text{ECL}}$) are shown in Fig. 17.11.4 for both measurements. It is not clear at this point how the sensitivities of the two tag methods compare for $B^0 \to$ invisible, since neither experiment has performed these searches using both methods. The details of the signal selection, in particular the detector acceptance and extra energy environment, differ sufficiently between BABAR and Belle that it is difficult to draw firm conclusions regarding future $B$ Factory sensitivities.

### 17.11.3 $B^0 \to \gamma\gamma$ and $B^0_s \to \gamma\gamma$

The $B^0 \to \gamma\gamma$ mode is related to the $b \to d\gamma$ process, as the $\bar{b}$ and $d$ quarks in the initial state $B^0$ annihilate through a penguin loop. Moreover, it involves radiation of one additional photon. Therefore this mode is highly suppressed in the SM, with a branching fraction which is estimated to be $(3.1^{+6.4}_{-1.6}) \times 10^{-8}$(Bosch and Buchalla, 2002a).

[95] The BABAR measurement optimizes the selection in order to achieve the most stringent upper limit, assuming no signal events to be found.

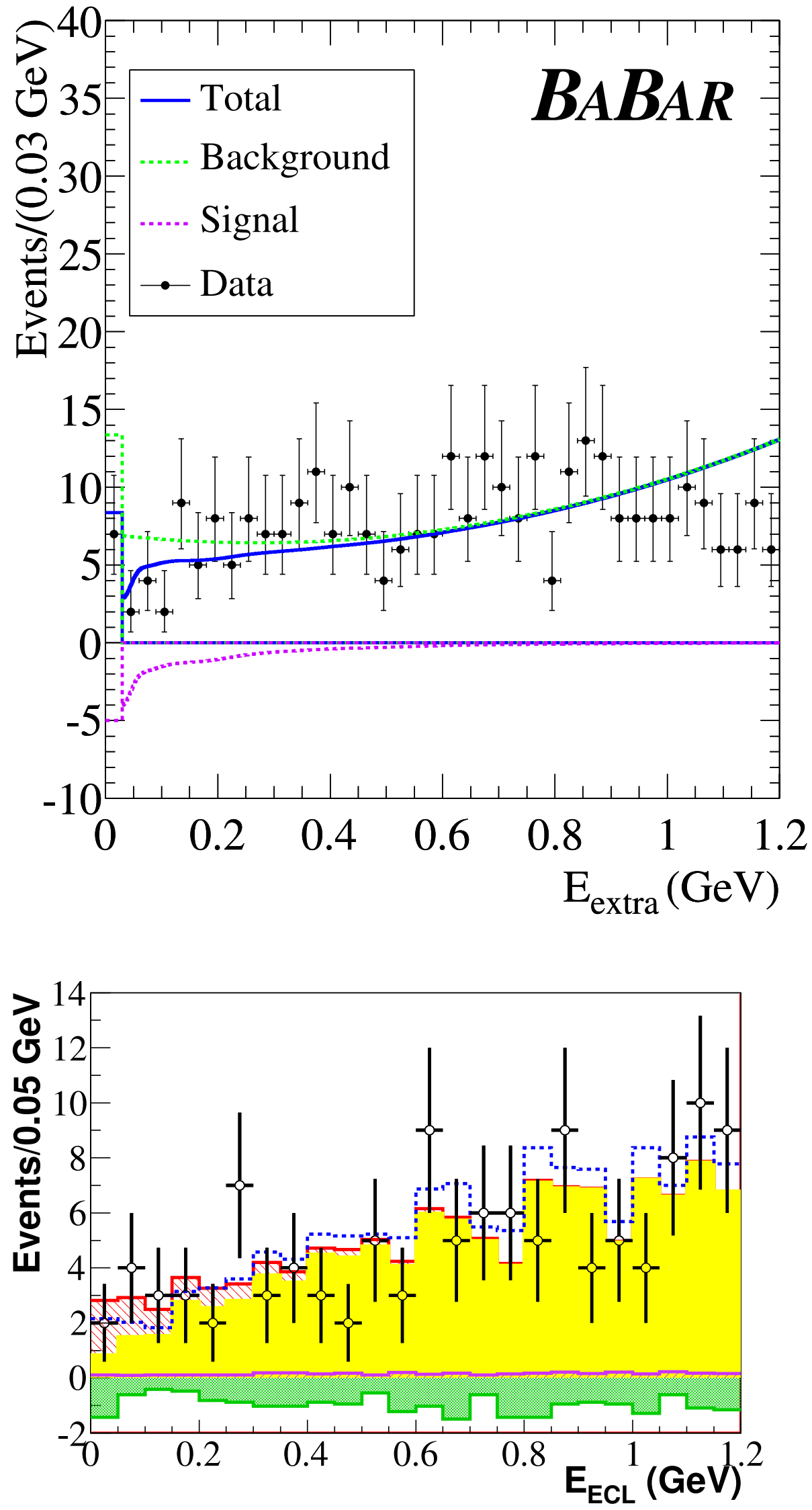

**Figure 17.11.4.** Distribution of $E_{\rm extra}$ in the *BABAR* search for $B^0 \to$ invisible using semileptonic tagged events (Lees, 2012g). Lines represent the result of the fit. Distribution of a similar quantity (bottom, called $E_{\rm ECL}$ at Belle) for the search in the same decay mode using hadronic tagged events (Hsu, 2012). Lines are the result of the fit (lightly hatched yellow region is the background contribution and the solid red line the total).

*BABAR* (del Amo Sanchez, 2011k) used 226M $B^0\overline{B}^0$ (426 fb$^{-1}$) to search for $B^0 \to \gamma\gamma$. Two photons with CM energies within $1.15 \le E^*_\gamma \le 3.50$ GeV are selected and required to satisfy $m_{\rm ES} > 5.1$ GeV$/c^2$ and $|\Delta E| \le 0.50$ GeV. Background from $e^+e^- \to q\bar{q}$ continuum and $e^+e^- \to \tau^+\tau^-$ events, respectively, are suppressed by using the Fox-Wolfram moment ratio, $R_2$ (see Section 9.3), and by requiring the number of reconstructed charged tracks to be larger than two. The dominant sources of background, at this stage, are $\pi^0$ and $\eta$ decays to $\gamma\gamma$. Discrimination against these backgrounds is obtained by combining each of the candidate photons ($\gamma$) with other photons in the event ($\gamma'$), and using the $\gamma\gamma'$ invariant mass and the energy $E_{\gamma'}$ of the other photon as input variables to a likelihood ratio.

Backgrounds due to merged photons from $\pi^0$ decays are suppressed by the energy distribution shape of the photon candidate in the calorimeter. Further suppression of the remaining continuum events is performed with a neural network (see Chapter 4). After all selections are applied, the peaking background contribution from rare $B$ decays is estimated to be $1.18 \pm 0.22$ events. Using an unbinned extended ML fit to $m_{\rm ES}$ and $\Delta E$, the signal yield $N_{\rm sig}$ is determined to be $N_{\rm sig} = 21.3^{+12.8}_{-11.8}$ events, with statistical significance of $1.8\sigma$. The systematic error on the branching fraction, 12.1%, is dominated by the fitting uncertainty (9.9%), and is included by convolution with the likelihood function. A branching fraction upper limit $\mathcal{B}(B^0 \to \gamma\gamma) < 3.2 \times 10^{-7}$ (at the 90% C.L.) is obtained.

In an earlier analysis (Villa, 2006) based on data collected prior to 2004, Belle used a limited data sample of 111M $B\overline{B}$ events to search for $B \to \gamma\gamma$. This search set a branching fraction upper limit $\mathcal{B}(B \to \gamma\gamma) < 6.2 \times 10^{-7}$ at the 90% C.L., but the analysis was notable for the fact that it suffered due to calorimeter backgrounds arising from out-of-time signals from previous bunch crossings. As a result of the relatively long decay time of the scintillation light from the CsI(Tl) crystals in the Belle calorimeter, there is a non-negligible probability that a residual calorimeter signal from a previous QED event, typically $e^+e^- \to e^+e^-$, can persist long enough to produce a "fake" photon cluster in a later $\Upsilon(4S) \to B\overline{B}$ event. If two back-to-back clusters from such a Bhabha event are present, and the reduced energy of the pair happens to match the $B$ mass, then it resembles the $B$ signal in the $m_{\rm ES}$ distribution. This background can be mostly removed using the timing information of the calorimeter signals, however, this information was not available in the reduced data format for data processed before summer 2004, and in particular for the sample used for the Villa (2006) analysis. In subsequent reprocessings of this data sample (see Section 3.3), this information was made available however, at the time of writing, the analysis has not been updated. In the existing measurement the background composed of photons from the continuum events (mainly decays of $\pi^0$ and $\eta$ mesons) is suppressed by the selection based on the polar angle of the more energetic photon. The issue of out-of-time calorimeter clusters has also been studied by *BABAR*, as a potential background for other $B$ decay modes which rely on neutral clusters, in particular $b \to s\gamma$ (Section 17.9) and $B^0 \to \pi^0\pi^0$. While this background was not an issue for the *BABAR* $B^0 \to \gamma\gamma$ study, it is potentially a concern for future high-luminosity experiments in which the Bhabha rate is much higher than the present generation of experiments.

In comparison to $B^0 \to \gamma\gamma$, the $B_s^0 \to \gamma\gamma$ decay is favored by $\sim |V_{ts}/V_{td}|^2$, with the SM prediction $\mathcal{B}(B_s^0 \to \gamma\gamma) \sim (0.5-1.0)\times 10^{-6}$. Production of $B_s$ mesons does not occur at the $\Upsilon(4S)$, hence a large sample of $\Upsilon(5S)$ events is required. *BABAR* did not collect significant data at this energy. However, Belle obtained a substantial sample (see Table 3.2.1). Using 23.6 $\text{fb}^{-1}$ of such data, Belle (Wicht, 2008) performed a search for $B_s^0 \to \gamma\gamma$ decays. Details of the measurement can be found in Section 23.3.6, only the main results are presented at this place. The signal yield is determined by an unbinned extended ML fit to $m_{\rm ES}$ and $\Delta E$. Figure 17.11.5 shows the projections to the fit variables, $m_{\rm ES}$ and $\Delta E$. No signal is observed. Including the systematic error of $^{+21}_{-19}\%$ which is dominated by the uncertainties in the number of $B_s^0$ events in the $\Upsilon(5S) \to b\bar{b}$ process ($^{+16}_{-13}\%$), the 90% C.L. branching fraction upper limit is determined to be $\mathcal{B}(B_s^0 \to \gamma\gamma) < 8.7 \times 10^{-6}$. This limit is about an order of magnitude larger than the SM prediction.

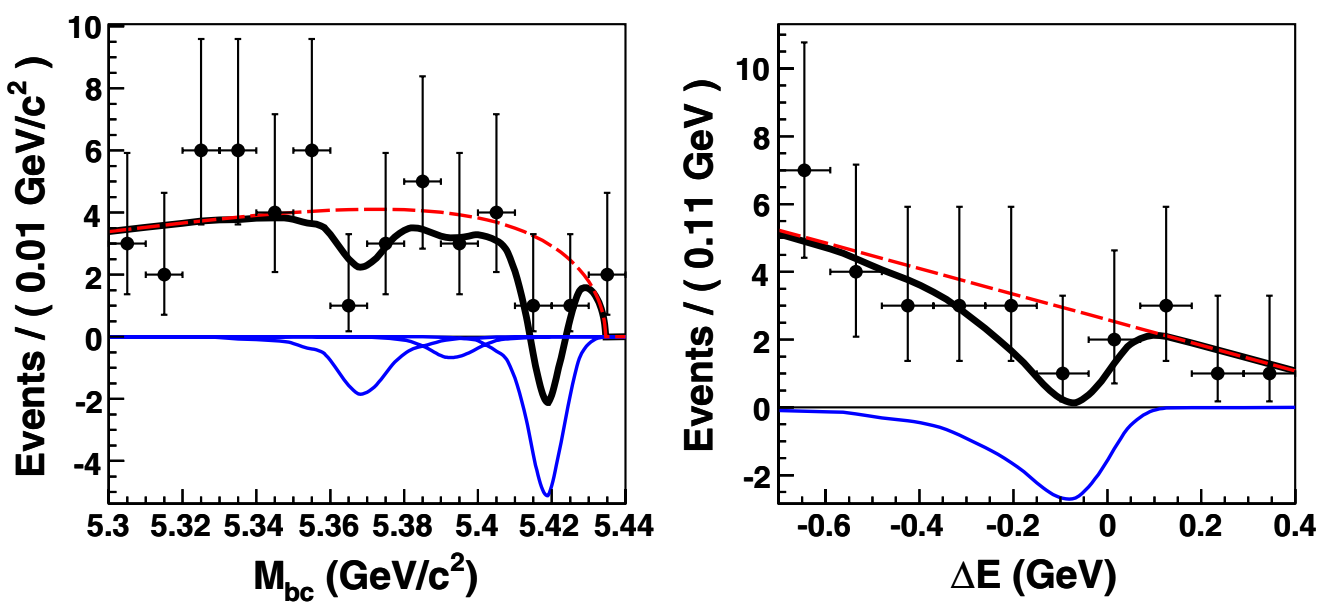

**Figure 17.11.5.** $m_{\rm ES}$ (left; called $M_{\rm bc}$ on the plot) and $\Delta E$ (right) data yield (points) and fit results for $B_s^0 \to \gamma\gamma$ in the Belle analysis (Wicht, 2008). The solid black curve represents the overall fit, while the dashed red and solid blue curves represent the continuum/combinatorial background and signal components of the fit, respectively. See Chapter 23 for an explanation of the structure of the $m_{\rm ES}$ distribution. On the $\Delta E$ projection, $m_{\rm ES} > 5.4$ GeV/$c^2$ is required to select only the $B_s^* \overline{B}{}_s^*$ contributions from $\Upsilon(5S)$. The fit returns a slight negative yield, consistent with zero within the data statistical uncertainties.

### 17.11.4 Lepton flavor violating modes

Lepton flavor violation is permitted within the SM if nonzero neutrino masses are included, since mixing of neutrino generations can then occur. LFV in the charged lepton sector can then occur in processes which contain one or more neutrinos as internal lines in any contributing Feynman diagram. However, the expected rates for LFV $B$ decays via this mechanism are far beyond current or expected future experimental sensitivity. Potentially measurable rates can however result from non-SM contributions, since most models do not explicitly conserve lepton flavor. In models with Higgs-mediated LFV, modes with heavier leptons generally are expected to exhibit larger LFV than modes with lighter leptons. However, experimental searches for modes containing tau leptons in the final state tend to be more difficult due to the multiple decay modes of the tau and missing energy resulting from the presence of one or more neutrinos. Consequently, experimental limits on $\mu$ - $e$ LFV modes tend to be more stringent than $\tau$ - $e$ or $\tau$ - $\mu$. As many of the proposed mechanisms for LFV tend to have couplings which favor heavier-generation leptons, the experimental limits from LFV modes with tau leptons can still provide interesting constraints on the parameters of these models.

Many lepton flavor violation searches, particularly those containing only first and second generation leptons, are performed as “incidental” studies along with related non-LFV modes. This is the case, for example, for $B^0 \to \mu^\pm e^\mp$ and $B \to K^{(*)}\mu^\pm e^\mp$, which have been published by both *BABAR* and Belle along with the corresponding $B^0 \to \ell^+\ell^-$ modes (discussed in Section 17.11.1) and $B \to K^{(*)}\ell^+\ell^-$ modes (see Section 17.9), respectively, as well as for $D^0 \to e^\pm\mu^\mp$ searched for together with $D^0 \to \ell^\pm\ell^\mp$ decays (Section 19.1.8). In general, these analyses have few unique features which distinguish them from the related non-LFV modes, hence we do not discuss them further in this section. For completeness, we tabulate the results in Table 17.11.2. In several instances however, *BABAR* and Belle have published dedicated searches for specific lepton flavor or lepton number violating decays.

Searches involving final state tau leptons generally require special techniques to overcome the challenges presented by the missing neutrinos and lack of a distinctive tau signature. In particular, tau decays to leptonic final states, $\tau \to \ell\nu\bar{\nu}$, are three-body final states with two unobserved neutrinos, providing essentially no kinematic constraints that can be exploited experimentally. Tau decays to hadronic final states are largely indistinguishable from $B$ and continuum backgrounds containing charged and neutral pions. To overcome these limitations, hadronic tag reconstruction (see Section 7.4.1) has been used in *BABAR* searches for $B^+ \to h^+\tau^\pm\ell^\mp$ (17.11.4.1) and $B^0 \to \tau^\pm\ell^\mp$ (17.11.4.2). As is the case with other studies which use this method, the resulting sensitivity is limited primarily by the very low signal efficiency. Hadronic tag reconstruction provides a number of kinematic advantages for these particular searches, since knowledge of the signal $B$ 4-vector (inferred from the $B_{\rm tag}$ 4-vector) allows the 2-body kinematics of $B^0 \to \tau^\pm\ell^\mp$ to be exploited and, in both $B^0 \to \tau^\pm\ell^\mp$ and $B^+ \to h^+\tau^\pm\ell^\mp$, permits the 4-vector of the daughter tau to be uniquely determined from the observed non-tau decay daughters.

Because they can potentially proceed via a “CKM-favored” $b$ - $s$ FCNC process (see Section 17.9) rather than a $b$ - $d$ FCNC process in which the quarks annihilate, $B^+ \to K^+\tau^\pm\ell^\mp$ are generally predicted to have larger branching fractions in potential new-physics models than $B^0 \to \tau^\pm\ell^\mp$. The primary difference from the experimental point of view is the presence of the additional charged kaon in $B^+ \to K^+\tau^\pm\ell^\mp$. This has two consequences: that the signal has 3-body (rather than 2-body) dynamics, and that the final states all topologically resem-ble various $b \to c\ell\nu$ modes, resulting in potentially very large backgrounds from these high-branching-fraction processes.

#### 17.11.4.1 $B^+ \to h^+\tau^\pm\ell^\mp$ ($h = K, \pi$, $\ell = e, \mu$)

Searches for $B^+ \to h^+\tau^\pm\ell^\mp$ (Aubert, 2007au; Lees, 2012b) use a methodology which exploits hadronic tag reconstruction (Section 7.4.1) to enhance the available kinematic constraints and to suppress continuum and combinatorial $B\overline{B}$ backgrounds. Searches for the corresponding neutral modes $B^0 \to h^0\tau^\pm\ell^\mp$ were not performed due to the lower efficiency for $K^0$ reconstruction via $K_s^0 \to \pi^+\pi^-$ and the fact that the tag reconstruction yield is somewhat higher for charged $B$ mesons than for neutral $B^0$ mesons due to the branching fractions of the available tag modes. In the more recent study, $h = K, \pi$, $\ell = e, \mu$ and the decays $B^+ \to h^+\tau^+\ell^-$ and $B^+ \to h^+\tau^-\ell^+$ are considered separately, for a total of eight distinct decay modes (charge conjugate modes are implied, but are not treated as distinct decay modes).

Since details of the specific new physics which could result in a signal for $B^+ \to h^+\tau^\pm\ell^\mp$ are not known *a priori*, a 3-body phase space model is assumed for the signal modes. This is in contrast to studies of the SM $B \to K^+\ell^+\ell^-$ channels described in Section 17.9.

Signal events are required to contain exactly three tracks, with total charge opposite that of the tag $B$. The primary hadron, $h$, is required to be one of the two tracks having charge opposite the tag $B$ and can be identified either as a kaon or pion. The two remaining tracks are then inferred to be the primary lepton $\ell$ and a charged tau decay daughter, which is identified as $e$, $\mu$ or $\pi$.

The tau decay 4-vector is uniquely specified using the tag $B$, primary lepton and primary hadron 4-vectors, independent of the tau decay daughters. The tau invariant mass, $m_\tau$,

$$m_\tau^2 = \left[(E_{CMS}, 0) - (E^*_{B_{tag}}, \mathbf{p}^*_{B_{tag}}) - (E^*_\ell, \mathbf{p}^*_\ell) - (E^*_h, \mathbf{p}^*_h)\right]^2 , \quad (17.11.3)$$

is obtained from this 4-vector and is used to extract the final signal yield as it peaks strongly for signal and is non-peaking for background. In the cases where the primary lepton and the tau daughter are identified as leptons of the same type, vetoes are imposed on the di-lepton invariant mass to reject $J/\psi$ and $\psi(2S)$ decays to $\ell^+\ell^-$: $3.03 < m_{\ell^+\ell^-} < 3.14$ GeV/$c^2$ and $3.60 < m_{\ell^+\ell^-} < 3.75$ GeV/$c^2$. In the di-electron case, a photon conversion veto of $m_{e^+e^-} > 0.1$ GeV/$c^2$ is also applied.

The dominant background sources depend on the relative charge of the primary lepton and the primary hadron in the signal mode. In the case of $B^+ \to h^+\tau^-\ell^+$, the dominant background is from semileptonic $B$ decays, $B^+ \to \overline{D}^{(*)0}\ell^+\nu$ with $\overline{D}^0 \to K^+X^-$(where $X$ is a hadronic system), in which some or all of the $X^-$ system is incorrectly reconstructed as the signal tau decay daughters.

In the case of $B^+ \to h^+\tau^+\ell^-$ however, the dominant background is from $B^+ \to \overline{D}^{(*)0}X^+$ with the charm system decaying semileptonically. In both cases, the primary hadron and the track of opposite charge originate from the $\overline{D}^{(*)0}$. The quantity $m(K\pi)$, the invariant mass of the combination of these two tracks computed assuming appropriate kaon and pion mass hypotheses, is required to be greater than 1.95 GeV/$c^2$, *i.e.* to exceed the $D^0$ mass, effectively suppressing these decays although with a significant loss of signal efficiency (see Figure 17.11.6). The remaining background is mainly from continuum $q\bar{q}$ production and is further suppressed by using a multivariate likelihood selector based on event shape and PID quality criteria, and the scalar sum of any remaining energy in the calorimeter.

Signal branching fractions are determined relative to high-branching fraction decays with similar topologies, specifically $B^+ \to \overline{D}^{(*)0}\ell^+\nu$ with $\overline{D}^0 \to K^+\pi^-$. Signal yields in each of eight signal modes are determined within a mass window of $\pm 60$ MeV/$c^2$ in $m_\tau$, centered around the nominal tau mass. For each signal mode, a likelihood function is obtained from the products of the Poisson *p.d.f.*s representing the yields in the $e$, $\mu$ and $\pi$ decay channels. Since models of LFV can produce signatures in which the charge of the tau is either correlated or uncorrelated with that of the charged hadron, *BABAR* reports results on both the signed (*e.g.* $B^+ \to h^+\tau^-\mu^+$) and unsigned (*e.g.* $B^+ \to h^+\tau^\mp\mu^\pm$) branching fractions. No significant signals were found in any modes and upper limits (see Table 17.11.2) were obtained at the level of a few times $10^{-5}$.

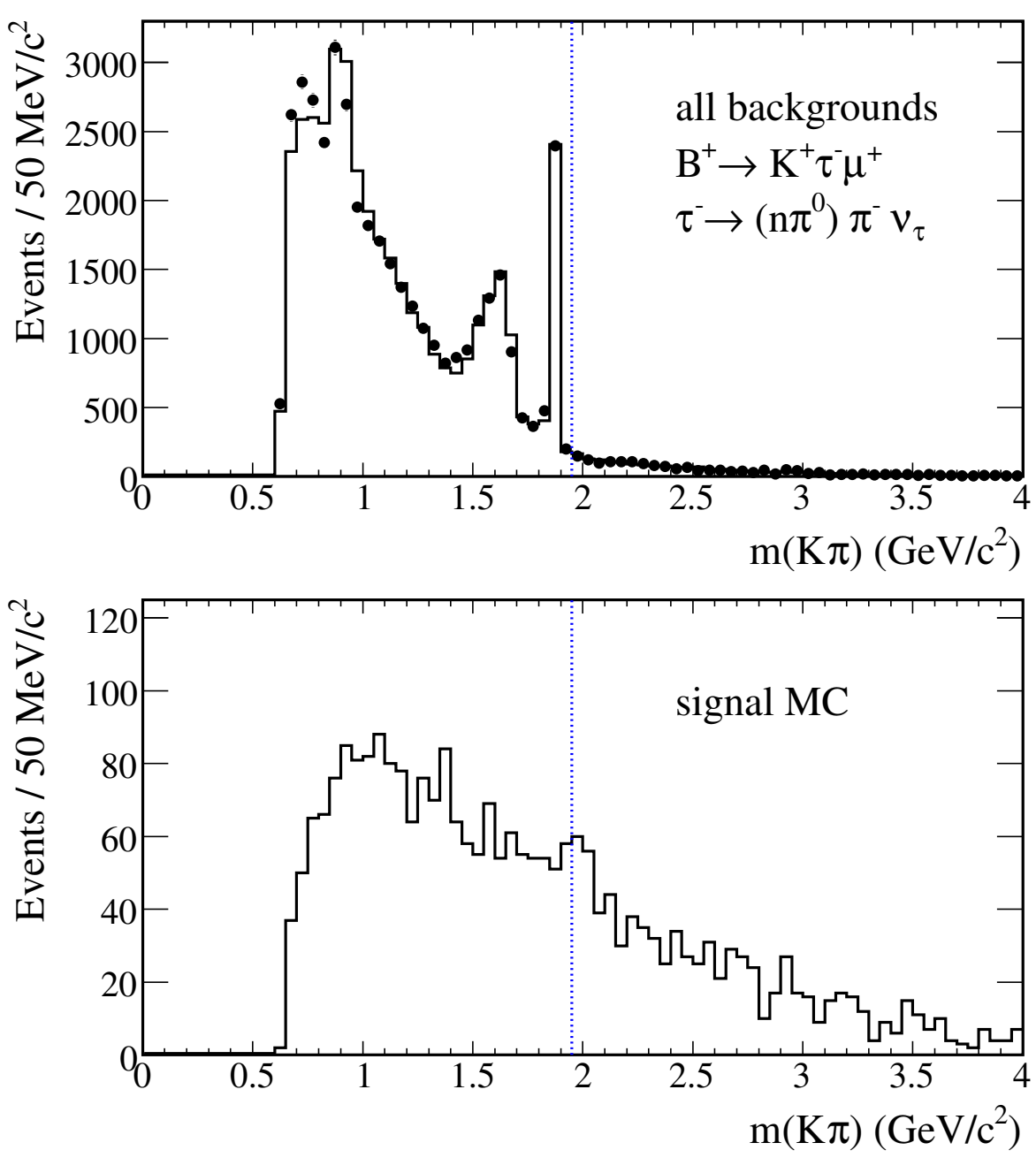

**Figure 17.11.6.** The reconstructed invariant mass, $m(K\pi)$, for the $B^+ \to K^+\tau^-\mu^+$ and $\tau^- \to (n\pi^0)\pi^-\nu_\tau$ (Lees, 2012b). The upper plot shows the full data distribution and the bottom the simulated expectation for the signal. The vertical line denotes the selection requirement just above the $D^0$ peak at approximately 1.86 GeV/$c^2$.

### 17.11.4.2 $B^0 \to \tau^\pm \ell^\mp$ ($\ell = e, \mu$)

*BABAR* performed a search (Aubert, 2008az) for $B^0 \to \tau^\pm \ell^\mp$ (with $\ell = e, \mu$) based on a data sample of $378 \times 10^6$ $B\overline{B}$ pairs using methodology similar to $B^+ \to K^+ \tau^\pm \ell^\mp$ (Section 17.11.4.1). Due to the 2-body kinematics and the large tau mass, the light lepton is expected to possess a momentum (in the signal $B$ rest frame) of $\sim 2.34$ GeV/$c$, near the kinematic endpoint for leptons from $B$ decays. Since the $B$ rest frame is inferred from the $B_{\rm tag}$, the resolution of the signal peak is dictated primarily by the resolution of the $B_{\rm tag}$ 4-vector. This 2-body decay is somewhat similar kinematically to the charged $B$ decay $B^+ \to \ell^+ \nu$ ($\ell = e, \mu$) and hence a search is performed simultaneously (see Section 17.10.2).

The signal and background distribution of the electron momentum in $B^0 \to \tau^\pm e^\mp$, from Aubert (2008az), is shown in Figure 17.11.7. After reconstructing the $B_{\rm tag}$ and high-momentum lepton, all remaining particles in the event are then assumed to be the decay daughters of the tau lepton. Hence there should be either one or three additional tracks with total charge opposite that of the high-$p$ lepton. Six tau decay modes are considered: $e^- \nu \bar{\nu}$, $\mu^- \nu \bar{\nu}$, $\pi^- \nu$, $\rho(770)^- (\to \pi^- \pi^0)\nu$, $a_1(1260)^- (\to \pi^- \pi^0 \pi^0)\nu$ and $a_1(1260)^- (\to \pi^- \pi^+ \pi^-)\nu$, where $\rho(770)$ or $a_1(1260)$ mass constraints are imposed in the latter three cases. In the hadronic tau decay modes only a single neutrino is present. Consequently, the neutrino 4-vector can be uniquely determined from the combination of the reconstructed tag $B$, the high-$p$ lepton and the hadronic tau daughter 4-vectors. The neutrino mass therefore provides an additional kinematic constraint on these modes. Aubert (2008az) exploits this by defining the quantity $\Delta E_\tau$, representing the difference between the expected tau energy and the total energy of the hadronic tau daughters combined with the neutrino (assuming zero mass). Computed in the tau rest frame, this quantity should peak at zero if the missing energy vector is consistent with a single massless neutrino. $\Delta E_\tau$ is used to select a "best" tau candidate from possible $\pi^\pm$, $\rho(770)^\pm$ and $a_1(1260)^\pm$ candidates in the case that one or more $\pi^0 \to \gamma\gamma$ candidates have been reconstructed.

In signal events, the combination of the high-$p$ lepton with the tau decays daughter(s) should account for all particles in the event which are not associated with the reconstructed tag $B$, while in background events other particles may be present. A loose $E_{\rm extra}$ (defined as the scalar sum of energies of any remaining tracks or clusters with energy $> 50$ MeV) requirement is imposed of $E_{\rm extra} < 1.0$ GeV. Signal yields are extracted from unbinned ML fits to the the high-$p$ lepton momentum spectrum in the signal $B$ rest frame, as shown for $B^0 \to \tau^\pm e^\mp$ in Figure 17.11.7. No significant signal is seen in either mode and limits of $\mathcal{B}(B^0 \to \tau^\pm e^\mp) < 2.8 \times 10^{-5}$ and $\mathcal{B}(B^0 \to \tau^\pm \mu^\mp) < 2.2 \times 10^{-5}$ are obtained. Due to the very low backgrounds, this search is statistically limited.

Interpretation of these results in a new-physics context is model-dependent. However as an example we can consider a SUSY seesaw model with degenerate right-handed neutrino masses $M_N = 10^{14}$ GeV (Babu and Kolda, 2002; Dedes, Ellis, and Raidal, 2002). In such a model $B^0 \to \tau^\pm \mu^\mp$ is mediated by a SUSY neutral Higgs with effective LFV couplings, which would also lead to a potentially observable signal in $\tau^- \to \mu^- \mu^+ \mu^-$ (see Section 20.4). In this model $\mathcal{B}(B^0 \to \tau^\pm \mu^\mp) \propto (\tan^2 \beta / M_A)^4$, leading to a lower bound on the $A^0$ mass ($M_A$) of $\sim 30$ GeV for $\tan\beta = 100$. Although not currently very stringent, improved experimental limits on these modes from future experiments could place significant constraints on new-physics models.

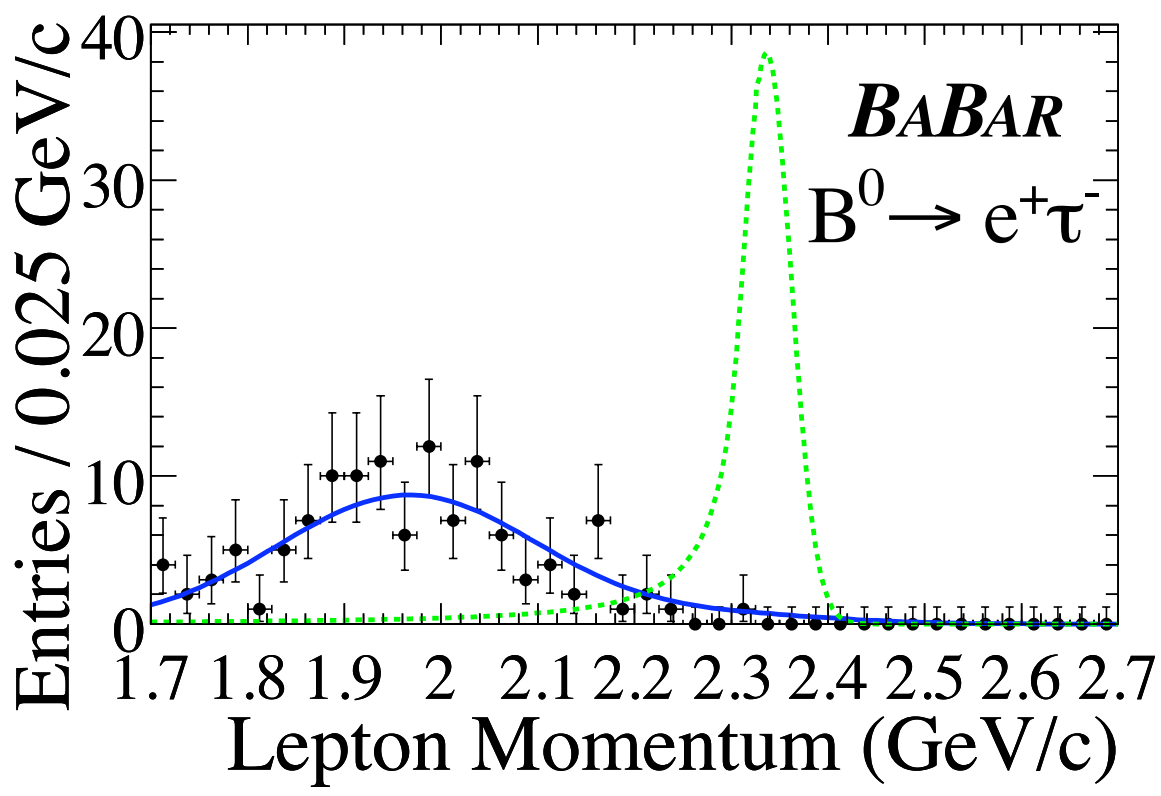

**Figure 17.11.7.** Signal electron candidate momentum in the signal $B$ rest frame for $B^0 \to \tau^\pm e^\mp$ (Aubert, 2008az). The points with errors are *BABAR* data, the solid blue curve is the fitted background *p.d.f.* and the dashed green curve shows the expected signal shape.

### 17.11.5 Lepton number violating modes

LNV processes are possible if neutrinos are of the Majorana type. As for the case of neutrino-less double beta decay, lepton number must change by $\Delta L = 2$. Two possible diagrams for such decays are shown in Figure 17.11.8. For a heavy sterile Majorana neutrino with a mass of a few GeV/$c^2$, the $s$-channel process (Figure 17.11.8(b)) is expected to give the dominant contribution.

*BABAR* reports a measurement of the LNV decays $B^+ \to h^- \ell^+ \ell^+$ (where $h = K, \pi$ based on $471 \times 10^6$ $B\overline{B}$ decays (Lees, 2012r). The experimental technique is very similar to that used for studies of $B^+ \to h^+ \ell^+ \ell^-$ described in Section 17.9, and the analysis sensitivity is similar. A three-body phase space model is assumed for the signal simulation. Events are required to possess at least four charged tracks, including two same-sign charged leptons each with momentum greater than 0.3 GeV/$c$. The leptons are required to originate from a common vertex and to satisfy $m_{\ell^+\ell^+} < 5.0$ GeV/$c^2$. Leptons from identified photon conversions are not permitted, and a bremsstrahlung recovery procedure is applied to electron and positron tracks to provide the best possible 4-vector for these particles.

For consistency with the $B^+ \to h^+ \ell^+ \ell^-$ studies, mass vetoes are imposed on the $J/\psi$ and $\psi(2S)$ mass regions, rejecting events with $2.85 < m_{\ell^+\ell^-} < 3.15$ GeV/$c^2$ and $3.59 < m_{\ell^+\ell^-} < 3.77$ GeV$/c^2$, respectively, although no actual peaking contribution is expected in this case. In the $B^+ \to \pi^-\mu^+\mu^+$ mode, an additional veto is imposed on the combination of the $\pi^-$ with each of the two oppositely charged muons in order to reject $J/\psi$ events in which a muon has been misidentified as a pion. Events are rejected if the $\pi^-\mu^+$ combination is within the range $3.05 < m_{\pi^-\mu^+} < 3.13$ GeV$/c^2$.

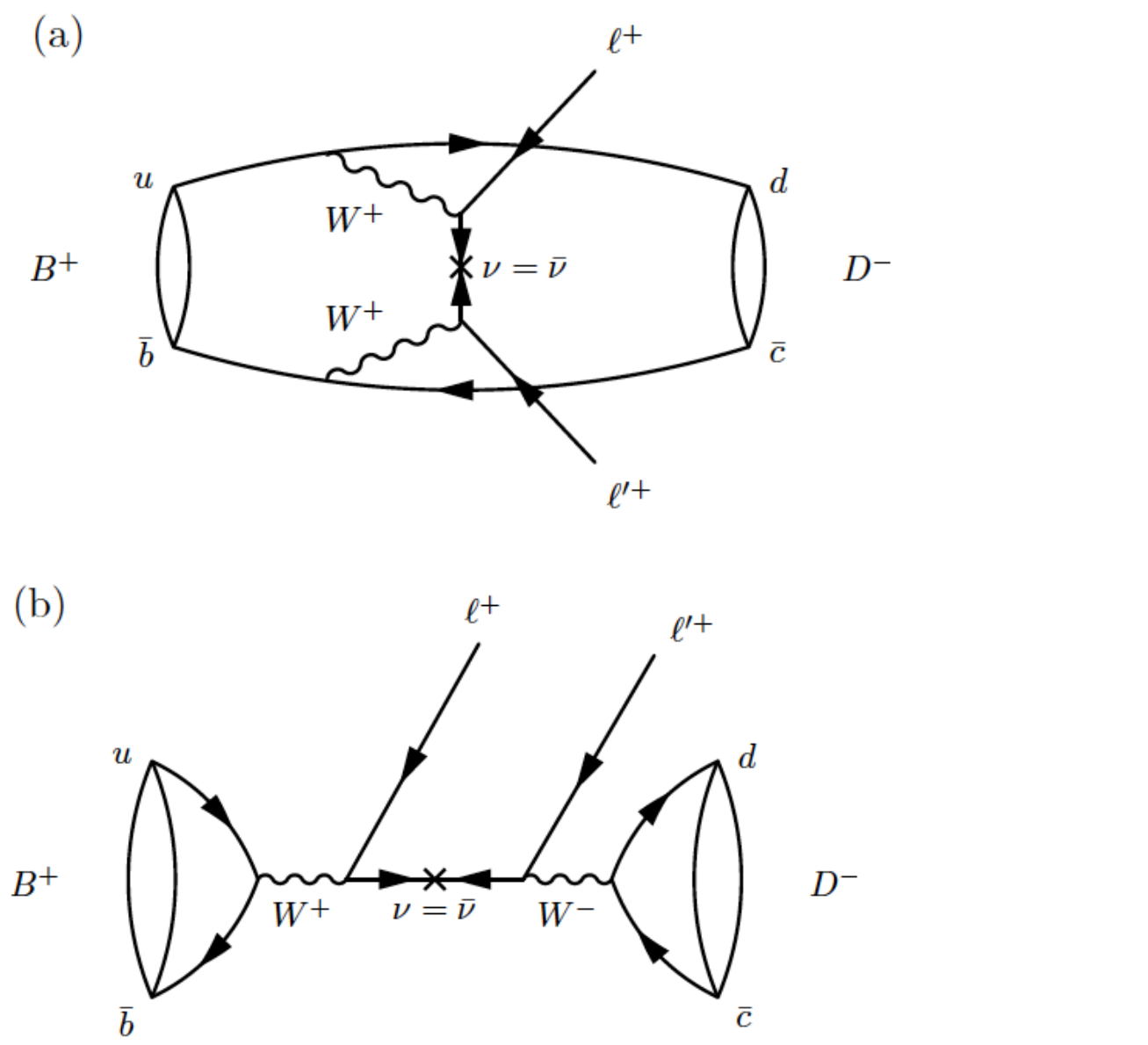

**Figure 17.11.8.** Diagrams for a LNV $B$ decay $B^+ \to D^- \ell^+ \ell^+$ in (a) $t$-channel and (b) $s$-channel processes.

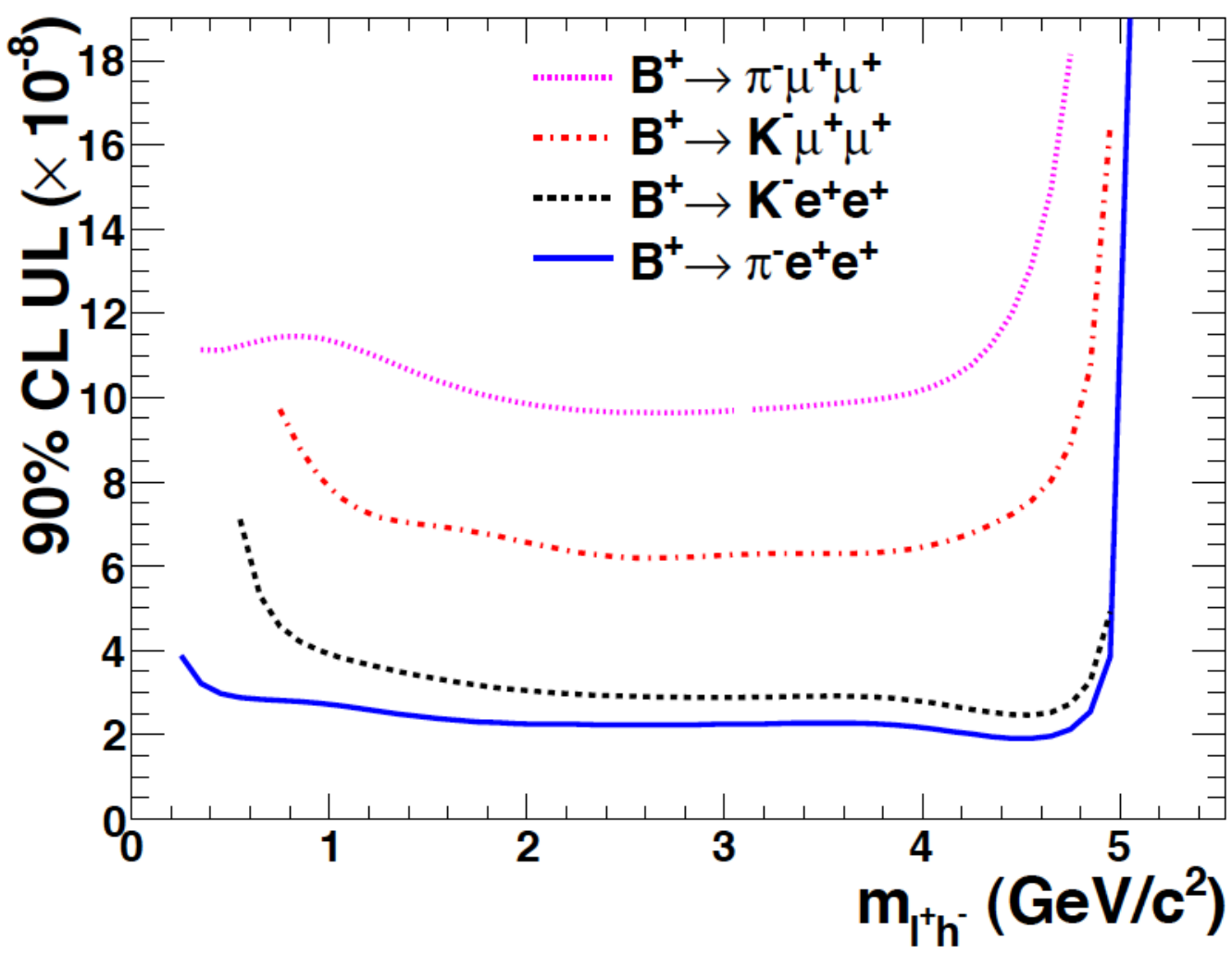

**Figure 17.11.9.** Branching fraction upper limits (UL) as a function of the mass $m_{\ell^+h^-}$ for the *BABAR* (Lees, 2012r) search modes $B^+ \to \pi^-\mu^+\mu^+$ (dotted magenta line), $B^+ \to K^-\mu^+\mu^+$ (dash-dotted red line), $B^+ \to K^-e^+e^+$ (dashed black line) and $B^+ \to \pi^-e^+e^+$ (solid blue line).

The di-lepton pair is then combined with an identified charged kaon or pion track of sign opposite that of the leptons, requiring the combined $B$ candidate to lie within $5.200 < m_{ES} < 5.289$ GeV$/c^2$ and $-0.10 < \Delta E < 0.05$ GeV. Backgrounds from $q\bar{q}$ and $B\overline{B}$ are suppressed using a set of Boosted Decision Trees based on 18 inputs representing event shape and kinematic variables and trained on MC signal and background samples (see Chapter 4 for a description of Boosted Decision Tree classifiers). A likelihood ratio, $\mathcal{L}_R$, is defined using the Boosted Decision Tree outputs as input *p.d.f.*s. The signal yield in each mode is extracted from an unbinned ML fit to $m_{ES}$ and $\mathcal{L}_R$. No significant signals are observed, and branching fraction upper limits are determined in the range $[2, 11] \times 10^{-8}$ at the 90% C.L. as shown in Table 17.11.2.

If $B^+ \to h^-\ell^+\ell^+$ is the result of the exchange of a Majorana neutrino, then the reconstructed invariant mass of the hadron $h$ with the opposite-sign lepton, $m_{\ell^+h^-}$, can be related to the Majorana neutrino mass $m_\nu$ (Atre, Han, Pascoli, and Zhang, 2009; Han and Zhang, 2006; Zhang and Wang, 2011). The *BABAR* results are presented as a function of $m_{\ell^+h^-}$ in Figure 17.11.9.

Since $b \to c$ decays are in general favored over charmless $B$ decays, it is interesting to extend the search for LNV processes to $B^+ \to X_c^-\ell^+\ell^+$ decays, where $X_c^-$ is any charmed hadron that has the opposite charge to the leptons. Using a sample of $772 \times 10^6$ $B\overline{B}$ pairs, Belle reports a measurement of the $B^+ \to D^-\ell^+\ell'^+$ decays (Seon, 2011), where $\ell, \ell' = e$ or $\mu$ in any combination. Since we have no prior knowledge nor widely accepted model for these decays, a 3-body phase-space model is assumed for the signal simulation. To find signal candidates, first an energetic same-sign lepton pair is chosen. The lepton momentum in the lab frame is required to be greater than 0.5(0.8) GeV$/c$ for electrons (muons). Particle identification requirements select electrons (muons) with an efficiency of approximately 90% and a misidentification rate of 0.1% (1%) for pions in the kinematic region of interest. The energy sum of the dilepton system in the CM frame is required to exceed 1.3 GeV: this has minimal effect on the signal efficiency in the phase-space model. The lepton pair is then combined with a $D^- \to K^+\pi^-\pi^-$ decay candidate. Kaons (pions) are discriminated from pions (kaons) with an efficiency of approximately 91% (95%) and a misidentification rate below 4% (6%) in the kinematic region of interest. The $K^+\pi^-\pi^-$ invariant mass ($M_{K\pi\pi}$) is required to be within $\pm 10$ MeV$/c^2$ from the nominal $D^-$ mass. The $B$ candidates are further required to lie within $m_{\rm ES} > 5.2$ GeV$/c^2$ and $|\Delta E| < 0.3$ GeV (the 'analysis region'). The major background sources are from continuum processes and to a lesser degree from semileptonic $B$ decays such as $B \to D^-\ell^+\nu_\ell X$, in which a same-sign lepton from the decay products of the other $B$ is combined with the signal $B$. These backgrounds are suppressed by a single likelihood ratio $\mathcal{R}$ using the four variables: the Fisher discriminant $\mathcal{F}$ of the modified Fox-Wolfram moments (see Section 9.3), the cosine of the polar angle of the $B$ candidate flight direction in the CM frame, $\cos\theta_B$, the missing energy $E_{\rm miss}$ of the event, and the difference $\delta z$ between the impact parameters of the two leptons in the beam direction. The requirement on $\mathcal{R}$, determined mode-by-mode by a MC study, eliminates more than 99% of the background while retaining 11–26% of the signal, depending on the mode. Signal yield is estimated in the 'signal region', $5.27 < m_{\rm ES} < 5.29$ GeV/$c^2$ and $-0.055(-0.035) < \Delta E < 0.035$ GeV for the $e^+e^+$ and $e^+\mu^+$ modes ($\mu^+\mu^+$ mode). The background region is defined as the complement of the analysis region excluding the signal region. The amount of background is determined by fitting the 2-dimensional ($\Delta E$, $m_{ES}$) p.d.f. to the data in the background region and then integrating the fitted p.d.f. over the signal region. There was no event observed in the signal region of any mode. Figure 17.11.10 shows the ($\Delta E$, $m_{ES}$) distribution of selected $B^+ \to D^-\mu^+\mu^+$ candidates. Upper limits (at the 90% C.L.) are calculated based on a frequentist approach (Feldman and Cousins, 1998) including systematic uncertainties using the POLE program, (Conrad, Botner, Hallgren, and Perez de los Heros, 2003). The results are summarized in Table 17.11.2, and are in the range $[1.1, 2.6] \times 10^{-6}$.

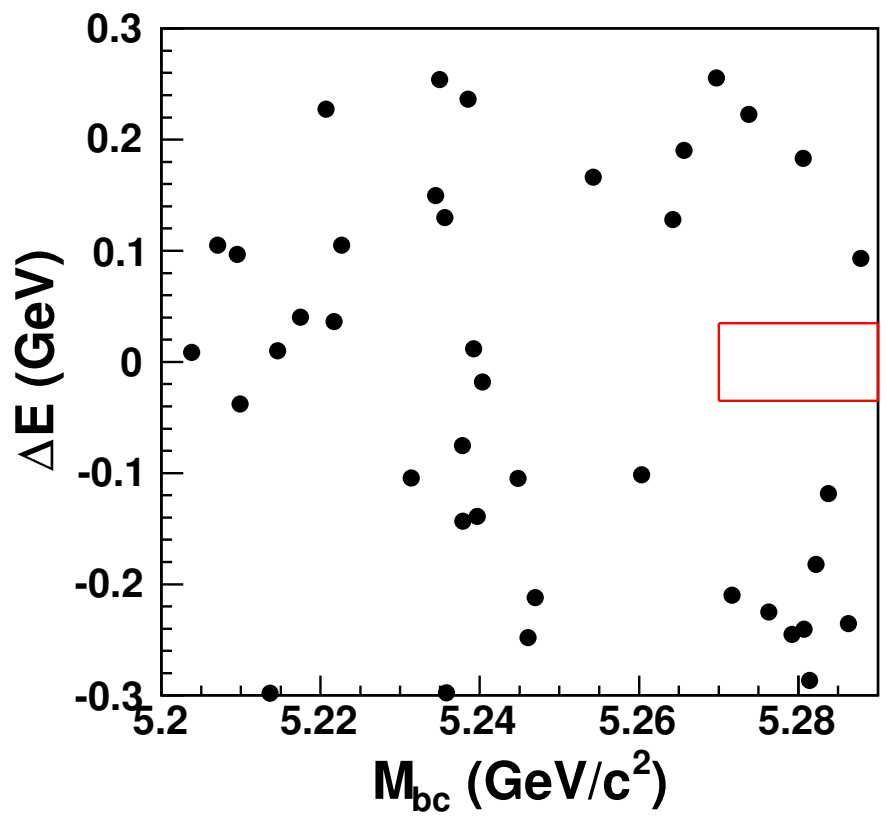

**Figure 17.11.10.** ($\Delta E$, $m_{ES}$) distribution of $B^+ \to D^-\mu^+\mu^+$ candidates from (Seon, 2011). The solid red line denotes the signal region.

## 17.11.6 Lepton/baryon number violating modes

Although baryon number is approximately conserved within the SM, it is predicted to be violated in many models of grand unification (Fritzsch and Minkowski, 1975; Georgi and Glashow, 1974), and indeed baryon number violation is one of Sakharov's conditions for cosmological baryogenesis (see Section 16.2). Most GUT models however conserve $\mathfrak{B} - L$, the difference of baryon and lepton number, implying that lepton number is also violated. In these models the proton is predicted to be unstable, albeit with a very long lifetime, decaying for example into a positron and $\pi^0$. However, the proton decay rates predicted by many of these models have not been observed and very stringent experimental limits have been placed on the proton lifetime (Nakamura et al., 2010). These limits, based on measurements of first-generation quarks, have been used to estimate the potential for baryon number violation in decays involving second- and

**Table 17.11.2.** Summary of the results for lepton flavor, lepton number and baryon number violating modes.

| Decay Mode | $N_{B\overline{B}}$ ($10^6$) | $\mathcal{B}$ upper limit (90% C.L.) | Reference |
|---|---|---|---|
| *Lepton flavor violating modes (light flavors):* | | | |
| $B^0 \to \mu^\pm e^\mp$ | 85 | $17 \times 10^{-8}$ | Chang (2003) |
| $B^0 \to \mu^\pm e^\mp$ | 384 | $9.2 \times 10^{-8}$ | Aubert (2008as) |
| $B^+ \to \pi^+\mu^\pm e^\mp$ | 230 | $17 \times 10^{-8}$ | Aubert (2007ax) |
| $B^0 \to \pi^0\mu^\pm e^\mp$ | | $14 \times 10^{-8}$ | |
| $B \to \pi\mu^\pm e^\mp$ | | $9.2 \times 10^{-8}$ | |
| $B^+ \to K^+\mu^- e^+$ | 229 | $9.1 \times 10^{-8}$ | Aubert (2006ac) |
| $B^+ \to K^+\mu^+ e^-$ | | $13 \times 10^{-8}$ | |
| $B^+ \to K^+\mu^\mp e^\pm$ | | $9.1 \times 10^{-8}$ | |
| $B^0 \to K^0\mu^\mp e^\pm$ | | $27 \times 10^{-8}$ | |
| $B \to K\mu^\mp e^\pm$ | | $3.8 \times 10^{-8}$ | |
| $B^+ \to K^{*0}\mu^- e^+$ | | $53 \times 10^{-8}$ | |
| $B^+ \to K^{*0}\mu^+ e^-$ | | $34 \times 10^{-8}$ | |
| $B^+ \to K^{*0}\mu^\mp e^\pm$ | | $58 \times 10^{-8}$ | |
| $B^+ \to K^{*+}\mu^- e^+$ | | $130 \times 10^{-8}$ | |
| $B^+ \to K^{*+}\mu^+ e^-$ | | $99 \times 10^{-8}$ | |
| $B^+ \to K^{*+}\mu^\mp e^\pm$ | | $140 \times 10^{-8}$ | |
| $B \to K^*\mu^\mp e^\pm$ | | $51 \times 10^{-8}$ | |
| *Lepton flavor violating modes (including $\tau$):* | | | |
| $B^0 \to \tau^\pm e^\mp$ | 378 | $2.8 \times 10^{-5}$ | Aubert (2008az) |
| $B^0 \to \tau^\pm \mu^\mp$ | | $2.2 \times 10^{-5}$ | |
| $B^+ \to K^+\tau^-\mu^+$ | 472 | $4.5 \times 10^{-5}$ | Lees (2012b) |
| $B^+ \to K^+\tau^+\mu^-$ | | $2.8 \times 10^{-5}$ | † |
| $B^+ \to K^+\tau^\mp\mu^\pm$ | | $4.8 \times 10^{-5}$ | |
| $B^+ \to K^+\tau^- e^+$ | | $4.3 \times 10^{-5}$ | |
| $B^+ \to K^+\tau^+ e^-$ | | $1.5 \times 10^{-5}$ | |
| $B^+ \to K^+\tau^\mp e^\pm$ | | $3.0 \times 10^{-5}$ | |
| $B^+ \to \pi^+\tau^-\mu^+$ | | $6.2 \times 10^{-5}$ | |
| $B^+ \to \pi^+\tau^+\mu^-$ | | $4.5 \times 10^{-5}$ | |
| $B^+ \to \pi^+\tau^\mp\mu^\pm$ | | $7.2 \times 10^{-5}$ | |
| $B^+ \to \pi^+\tau^- e^+$ | | $7.4 \times 10^{-5}$ | |
| $B^+ \to \pi^+\tau^+ e^-$ | | $2.0 \times 10^{-5}$ | |
| $B^+ \to \pi^+\tau^\mp e^\pm$ | | $7.5 \times 10^{-5}$ | |
| *Lepton number violating modes:* | | | |
| $B^+ \to \pi^- e^+ e^+$ | 471 | $2.3 \times 10^{-8}$ | Lees (2012r) |
| $B^+ \to K^- e^+ e^+$ | | $3.0 \times 10^{-8}$ | |
| $B^+ \to \pi^-\mu^+\mu^+$ | | $10.7 \times 10^{-8}$ | |
| $B^+ \to K^-\mu^+\mu^+$ | | $6.7 \times 10^{-8}$ | |
| $B^+ \to D^- e^+ e^+$ | 772 | $2.6 \times 10^{-6}$ | Seon (2011) |
| $B^+ \to D^-\mu^+ e^+$ | | $1.8 \times 10^{-6}$ | |
| $B^+ \to D^-\mu^+\mu^+$ | | $1.1 \times 10^{-6}$ | |
| *Baryon and lepton number violating modes:* | | | |
| $B^0 \to \Lambda_c^+\mu^-$ | 471 | $1.8 \times 10^{-6}$ | del Amo Sanchez† |
| $B^0 \to \Lambda_c^+ e^-$ | | $5.2 \times 10^{-6}$ | |
| $B^- \to \Lambda\mu^-$ | | $6.2 \times 10^{-8}$ | |
| $B^- \to \Lambda e^-$ | | $8.1 \times 10^{-8}$ | |
| $B^- \to \overline{\Lambda}\mu^-$ | | $6.1 \times 10^{-8}$ | |
| $B^- \to \overline{\Lambda} e^-$ | | $3.2 \times 10^{-8}$ | |

del Amo Sanchez (2011l)

third-generation quarks (Hou, Nagashima, and Soddu, 2005). In particular for $B^0 \to \Lambda_c^+ \ell^-$, which violates both lepton number and baryon number, the branching fraction is estimated to be less than $4 \times 10^{-29}$. Although this is far beyond any expected experimental sensitivity, searches have still been performed to the precision permitted by current data samples.

*BABAR* has performed a search for the decays $B^0 \to \Lambda_c^+ \ell^-$, $B^- \to \Lambda \ell^-$, and the $\mathfrak{B} - L$ violating mode $B^- \to \overline{\Lambda} \ell^-$, where the lepton is a muon or an electron (del Amo Sanchez, 2011l). This is the first experimental search for these decays, and any positive signal would be evidence of new physics.

$B$-meson candidates are formed by combining a $\Lambda_c^+$, $\Lambda$ or $\overline{\Lambda}$ candidate with an identified muon or electron. The $\Lambda_c^+$ candidates are reconstructed in the decay mode $\Lambda_c^+ \to pK^-\pi^+$, which has a branching fraction of about 5%. The $\Lambda$ candidates are reconstructed in the decay $\Lambda \to p\pi^-$, which has a branching fraction of about 64%.

The final state hadron $(p, K, \pi)$ and lepton $(\mu, e)$ candidates are all required to be consistent with the candidate particle hypothesis according to PID criteria based on $\mathrm{d}E/\mathrm{d}x$ , DIRC, EMC and IFR information. The 4-momenta of photons that are consistent with bremsstrahlung radiation from the electron candidate are added to that of the electron.

$\Lambda_c^+$ candidates are required to have $pK^-\pi^+$ invariant mass within $\pm 15$ MeV/$c^2$ of the nominal $\Lambda_c^+$ mass. Similarly, $\Lambda$ candidates must have $p\pi^-$ mass within $\pm 4$ MeV/$c^2$ of the nominal $\Lambda$ mass. The final state tracks which form the decay daughters of the the $\Lambda_c^+$ ($\Lambda$) are constrained to a common spatial vertex, and their invariant mass is constrained to the $\Lambda_c^+$ ($\Lambda$) mass. This has the effect of improving the 4-momentum resolution for true $B \to \Lambda_{(c)} \ell$ candidates. The baryon and lepton candidates are also constrained to originate from a common vertex.

As the $\Lambda$ has $c\tau = 7.89$ cm, the purity of the $\Lambda$-candidate sample is further improved by selecting candidates for which the reconstructed decay point of the $\Lambda$ candidate is at least 0.2 cm from the reconstructed decay point of the $B$ candidate in the plane perpendicular to the $e^+e^-$ beams. Particle mis-ID backgrounds from $e^+e^- \to e^+e^-\gamma$ events in which the photon converts to an $e^+e^-$ pair are eliminated by requiring that there are more than four tracks in the events.

$B$-meson candidates are selected within the kinematic region $|\Delta E| < 0.2$ GeV and $5.2 < m_{\mathrm{ES}} < 5.3$ GeV/$c^2$ are fitted to extract the signal yield. The signal yield is extracted using an unbinned extended ML fit in which the total *p.d.f.* is a sum of *p.d.f.*s for signal and background. The signal and background *p.d.f.*s are each a product of *p.d.f.*s describing the dependence on $m_{\mathrm{ES}}$ and $\Delta E$. For the $\Lambda_c^+ \ell^-$ modes, additional discriminating power is gained from a three dimensional *p.d.f.*, where the output from a neural network discriminator is used as the third variable.

No significant signal is observed for any of the decay modes, and branching fraction upper limits are determined, ranging from $5.2 \times 10^{-6}$ to $3.2 \times 10^{-8}$ at the 90% C.L. (see Table 17.11.2). Less stringent limits are obtained for the $B^0 \to \Lambda_c^+ \ell^-$ modes than the $\Lambda$ modes due to the relatively low branching fraction for the studied $\Lambda_c^+$ decay and a higher level of background compared with the $\Lambda$ modes.

### 17.11.7 Summary

Although there is currently no evidence for any of the rare or forbidden decay modes described in this section, they remain useful as probes for physics beyond the SM. In the case of $B_{(s)} \to \ell^+\ell^-$ and $B \to h\ell^+\ell^+$ (with $\ell = e, \mu$), experimental results from hadron colliders have already exceeded the current sensitivity from $B$ Factories and it is unlikely that future $e^+e^-$ facilities will change this situation. In other modes, particularly those with tau leptons or neutrinos, hadron colliders are at a significant disadvantage. It is notable that in some cases, either due to experimental challenges or due to the small size of the expected new-physics effects, current experimental limits on these modes do not yet reach the ranges predicted by most reasonable new-physics models. Consequently, these searches are essentially of the “shot-in-the-dark” variety: essentially probing dark corners of the SM to verify that we see nothing in places where we expect to see nothing. As a rule of thumb one can summarize the achieved sensitivity of $B$ Factories to $\mathcal{O}(10^{-5})$ for $B \to$ invisible and LFV decays with $\tau$'s, $\mathcal{O}(10^{-6})$ for $B_{(s)} \to \gamma\gamma$, and $\mathcal{O}(10^{-7})$ for $B \to \ell^+\ell^-$, LFV decays with light leptons and baryon and/or lepton number violating modes. Improvements in experimental sensitivity with large datasets at the future generation of $B$ Factories could change this picture, with experimental results in some cases directly confronting realistic new-physics models.

## 17.12 $B$ decays to baryons

***Editors:***
*Roland Waldi (BABAR)*
*Min-Zu Wang (Belle)*
*Hai-Yang Cheng (theory)*

**Additional section writers:**
*Thomas Hartmann*

Baryons and antibaryons have to be produced in pairs in the Standard Model, therefore most mesons cannot decay to baryons for lack of energy. The only baryonic $D$ decay is $D_s^+ \to p\overline{n}$, which has only just enough energy to proceed. But the phase space of the two or four quarks in a purely hadronic weak decay of a $B$ meson is much larger than that of charmed or light mesons, and leaves ample freedom for high multiplicities and for the production of baryon-antibaryon pairs.

That $B$ decays to baryons play an important role became evident by the large proton multiplicities found by ARGUS and CLEO at the $\Upsilon(4S)$ (Albrecht et al., 1989b; Crawford et al., 1992).

The interest in decays to baryons was increased when ARGUS claimed the observation of $B$ decays to $p\overline{p}\pi^\pm$ and $p\overline{p}\pi^+\pi^-$ (Albrecht et al., 1988b). Subsequently, baryonic $B$ decays were studied extensively by theorists around the early 1990s with the focus on the tree-dominated two-body decay modes. Experimental studies were first led by CLEO, but with the accumulating data at the $B$ Factories, *BABAR* and Belle came to dominate the field.

The features of $B$ decays to baryons reflect the properties of both the weak interaction, and the hadronization of quarks. One or two $q\overline{q}$ pairs have to be produced out of the vacuum to produce a baryon-antibaryon pair, similar to jet fragmentation.

This section presents the inclusive production of baryons in $B$ meson decays (Section 17.12.1), then exclusive two-body decays (Section 17.12.2), followed by the more frequent multibody final states with a baryon-antibaryon pair plus one or more mesons (Section 17.12.3). Complex phenomena are seen in multibody decays, and our treatment includes dedicated discussions of threshold enhancement (Section 17.12.3.3), multiplicity effects (Section 17.12.3.4), and angular correlations (Section 17.12.3.5). Finally, radiative and semileptonic decays with baryons in the final state are discussed (Sections 17.12.4 and 17.12.5 respectively). Theoretical interpretations and model predictions for each of these topics are discussed within the the corresponding section. Baryon number violating decays have been presented in the previous section, 17.11.6.

### 17.12.1 Inclusive decays into baryons

The inclusive production of protons and antiprotons from $\Upsilon(4S)$ decays, *i.e.*, an admixture of $B^+$, $B^-$, $B^0$, and $\overline{B}^0$, has been measured by ARGUS and CLEO (Albrecht et al., 1993a; Crawford et al., 1992). The combined multiplicity of protons and antiprotons in an average $B$ decay is

$$\langle n_p + n_{\overline{p}} \rangle = 0.080 \pm 0.004. \qquad (17.12.1)$$

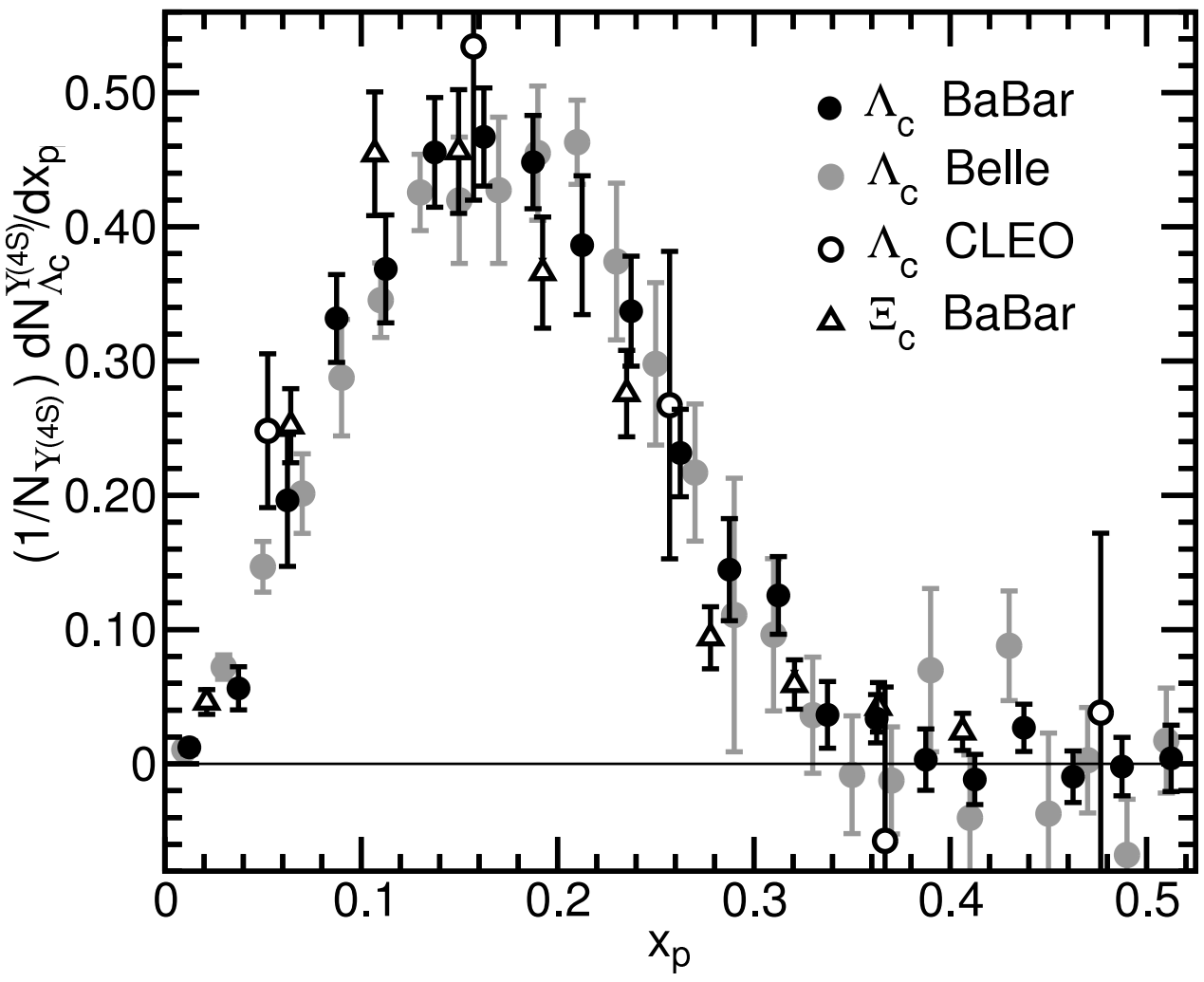

**Figure 17.12.1.** Differential $B \to \Lambda_c^+ X$ production rate per $\Upsilon(4S)$ from *BABAR* (Aubert, 2007p), Belle (Seuster, 2006), and CLEO (Crawford et al., 1992) versus the momentum fraction $x_p = p/p_{\rm max}$ in the $\Upsilon(4S)$ rest frame. Also shown is the differential $\Xi_c^0$ production rate normalized to match the peak of the $\Lambda_c^+$ rate.

Some of these protons come from $\Lambda$ decays; the multiplicity of $\Lambda$ baryon production has been determined to be $\langle n_\Lambda + n_{\overline{\Lambda}} \rangle = 0.040 \pm 0.005$ (Albrecht et al., 1989b; Crawford et al., 1992).

Inclusive particle spectra in (scaled) momentum are obtained from data at the $\Upsilon(4S)$ energy by subtracting the spectra obtained off resonance, scaled to the on resonance luminosity and cross section. The scaled momentum of a particle of mass $m$ is given by $x_p = p/p_{\rm max}$ with the maximum center-of-mass momentum $p_{\rm max} = \sqrt{s/4 - m^2}$. Integration over the extrapolated spectra yields the particle multiplicity.

If protons were the only stable baryons, any baryonic event would have one proton and one antiproton. This would imply a 4% branching fraction into baryon antibaryon + $X$ (ignoring decays with two pairs). But since there are also neutrons, a more sophisticated analysis is required to obtain the total baryonic branching fraction. Such an analysis has been performed by the ARGUS collaboration (Albrecht et al., 1992c) using in addition to the proton and $\Lambda$ multiplicities the fractions of events at the $\Upsilon(4S)$ with baryon-antibaryon pairs, baryon $\ell^+$ pairs, and baryon $\ell^-$ pairs (where $\ell$ is an electron or muon). The baryon-antibaryon fraction allows for the elimination of the unknown contribution of neutrons to the final state, while the remaining fractions help to establish baryon-flavor correlations. The result is

$$\mathcal{B}(B \to \mathfrak{B}_1 \overline{\mathfrak{B}}_2 X) = (6.8 \pm 0.5 \pm 0.3)\% \qquad (17.12.2)$$

where $\mathfrak{B}$ represents a generic baryon.

Since the dominant weak process in $B$ decays is $b \to cX$, most final states with baryons contain either a meson with charm quarks (like $B^+ \to J/\psi p \overline{\Lambda}$ or $\overline{B}^0 \to D^0 p\overline{p}$) or a charmed baryon. While there are a few charmed baryon decays to charmed mesons and non-charmed baryons (such as $\Lambda_c(2880)^+ \to D^0 p$), most decays are to $\Lambda_c^+$, $\Xi_c^0$, $\Xi_c^+$, or $\Omega_c^0$. Inclusive production rates for all of these states, except $\Omega_c^0$, have been determined by ARGUS (Albrecht et al., 1988a) and CLEO (Crawford et al., 1992). Inclusive $\Lambda_c$ production has also been measured at the $B$ Factories (Aubert, 2004m, 2007p,ba; Seuster, 2006): the most precise average multiplicity per $B$ meson has been determined by *BABAR* (Aubert, 2007p),

$$\langle n_{\Lambda_c^+} + n_{\overline{\Lambda}_c^-} \rangle = 0.0456 \pm 0.0009 \pm 0.0031 \pm 0.0118_{\Lambda_c}. \qquad (17.12.3)$$

The scaled momentum spectrum in the $\Upsilon(4S)$ rest frame is shown in Fig. 17.12.1. As in most measurements where a $\Lambda_c^+$ baryon is reconstructed, the decay $\Lambda_c^+ \to pK^-\pi^+$ is used in this analysis. All channels with the $\Lambda_c$ baryon in the final state suffer from a large systematic uncertainty of 26% since the branching fractions are only poorly known, with $\mathcal{B}(\Lambda_c^+ \to pK^-\pi^+) = 0.050 \pm 0.013$ (Beringer et al., 2012). This value is also used to normalize other $\Lambda_c$ branching fractions: it is the source of the dominant third uncertainty on the multiplicity in Eq. (17.12.3). A further discussion of this problem is found in Section 19.4.2.3.[96] The situation is even worse for $\Xi_c$ baryons, where no absolute branching fraction measurement is available. For our multiplicity estimate and also for exclusive branching fractions (*e.g.*, Table 17.12.1) we use $\mathcal{B}(\Xi_c^0 \to \Xi^-\pi^+) \approx 1.2\%$, assuming a 50% error. This value is based on the range of theoretical predictions for the partial width (Cheng and Tseng, 1993) and the lifetime of the $\Xi_c^0$ baryon (Beringer et al., 2012). In the same spirit, we use $\mathcal{B}(\overline{\Xi}_c^- \to \overline{\Xi}^+\pi^-\pi^-) \approx 6.4\%$, assuming a 50% error. This value is obtained from the theoretical prediction $\Gamma(\overline{\Xi}_c^- \to \overline{\Xi}^0\pi^-) \approx 0.8\times10^{11}\,\mathrm{s}^{-1}$ derived from Cheng and Tseng (1993), the experimental ratio $\Gamma(\overline{\Xi}_c^- \to \overline{\Xi}^0\pi^-)/\Gamma(\overline{\Xi}_c^- \to \overline{\Xi}^+\pi^-\pi^-) = 0.55\pm0.16$, and the lifetime $\tau(\overline{\Xi}_c^-) = (4.42\pm0.26)\times10^{-13}\,\mathrm{s}$ (Beringer et al., 2012).

The multiplicity of charged $\Xi_c$ baryons has only been measured at CLEO (Barish et al., 1997), corresponding to $\langle n_{\Xi_c^+} + n_{\overline{\Xi}_c^-} \rangle \sim 0.007$. For neutral $\Xi_c$ baryons the average from the CLEO (Crawford et al., 1992) and *BABAR* (Aubert, 2005z) measurements is $\langle n_{\Xi_c^0} + n_{\overline{\Xi}_c^0} \rangle \sim 0.016$.

Production of $\Omega_c^0$ baryons in $B$ decays is even more rare, and has been observed by *BABAR* (Aubert, 2007ao), with an average multiplicity of the order 0.0005 assuming $\mathcal{B}(\Omega_c \to \Omega^-\pi^+) \sim 1\%$.

[96] While we were finalising this book Belle submitted an absolute branching fraction measurement for publication, with significantly improved precision: $\mathcal{B}(\Lambda_c^+ \to pK^-\pi^+) = (6.84 \pm 0.24^{+0.21}_{-0.27})\%$ (Zupanc, 2013a). See the discussion in Section 19.4.2.3.

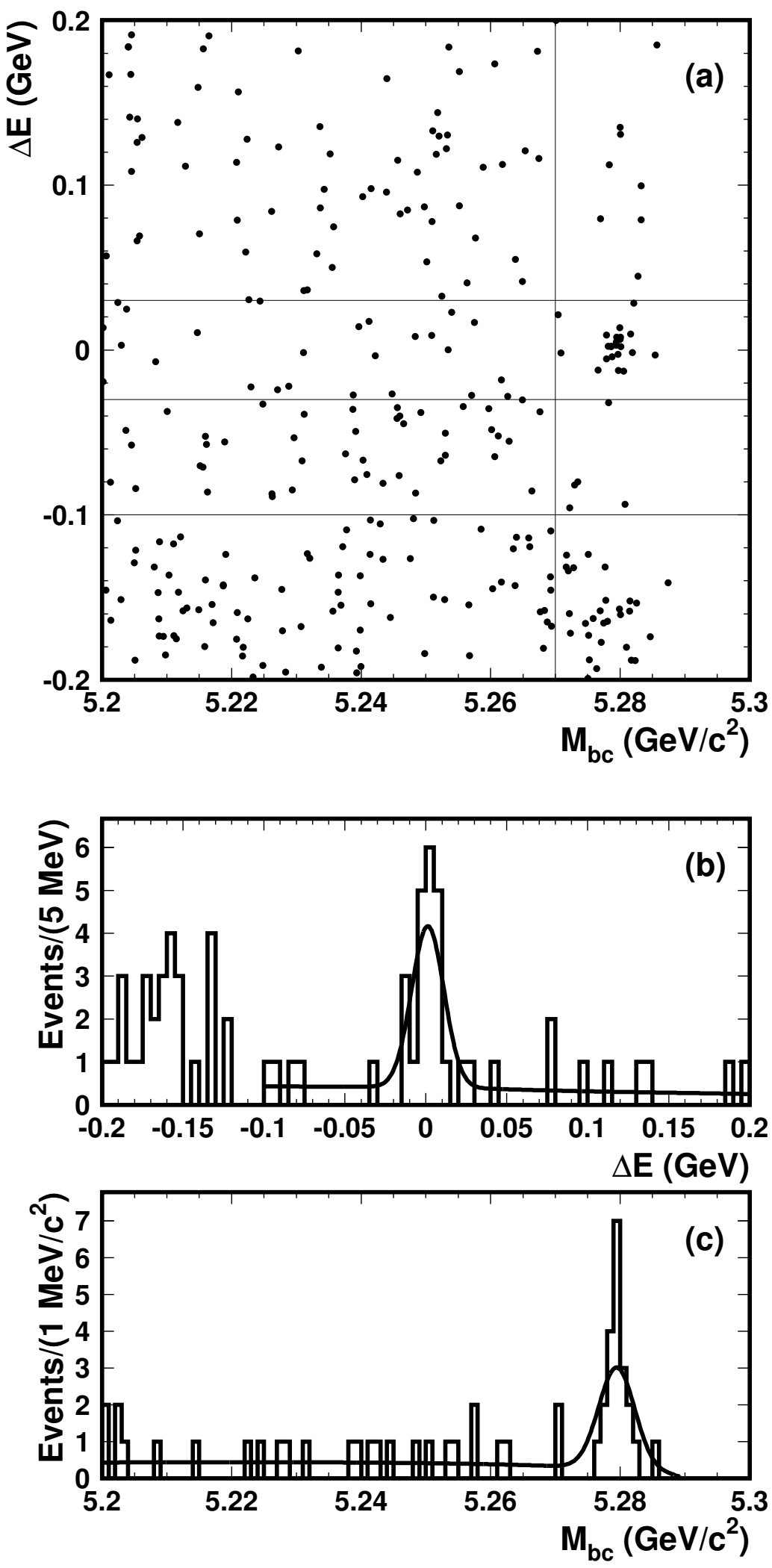

**Figure 17.12.2.** Candidate events for the decay $\overline{B}^0 \to \Lambda_c^+\overline{p}$ from Belle (Gabyshev, 2003): (a) scatter plot of $\Delta E$ versus $m_{\mathrm{ES}} = M_{bc}$, (b) $\Delta E$ distribution for $m_{\mathrm{ES}} > 5.270\,\mathrm{GeV}/c^2$, and (c) $m_{\mathrm{ES}}$ distribution for $|\Delta E| < 0.030\,\mathrm{GeV}$. The curves indicate the result of a two-dimensional fit.

### 17.12.2 Two-body decays

One might expect a large fraction of baryonic decays to proceed via two-body decay channels, since the phase space for heavy particles is rather small. However, from the spectrum in Fig. 17.12.1 it is evident that two-body decays are rare, since they would show up as a peak around $x_p = 0.4$. Indeed, it has been found experimentally that decays of $B$ mesons to just a baryon and an antibaryon have very small branching fractions. First measurements of $B$ decays to baryons were made by CLEO, but no two-body decay could be established, and upper limits of those decays were reported (Bornheim et al., 2003; Dytman et al., 2002; Procario et al., 1994).

**Table 17.12.1.** Branching fractions of observed two-body decays of $B$ mesons to baryons. Upper limits are at the 90% C.L. Channels with the $\Lambda_c^+$ baryon in the final state have an additional $\pm 26\%$ relative error (not included) from the assumption $\mathcal{B}(\Lambda_c^+ \to pK^-\pi^+) = 0.050 \pm 0.013$ and are marked with †. The same holds for $\mathcal{B}(\Xi_c^0 \to \Xi^-\pi^+) = 0.012$ and $\mathcal{B}(\Xi_c^+ \to \Xi^-\pi^+\pi^+) = 0.064$, both with an additional $\pm 50\%$ taken from the range of theoretical predictions (Cheng and Tseng, 1993). Two daggers indicate that two such errors have to be added (linearly due to correlation).

| Decay | BaBar | | Belle | | Average |
|---|---|---|---|---|---|
| *$\mathfrak{B}_c\overline{\mathfrak{B}}_c$ final states* ($\mathcal{B}$: $10^{-3}$) | | | | | |
| $B^+ \to \overline{\Xi}_c^0 \Lambda_c^+$ | $1.73 \pm 0.54 \pm 0.24^{\dagger\dagger}$ | (Aubert, 2008e) | $4.00 \pm 0.83 \pm 0.92^{\dagger\dagger}$ | (Chistov, 2006a) | $2.16 \pm 0.54^{\dagger\dagger}$ |
| $B^0 \to \overline{\Xi}_c^- \Lambda_c^+$ | $0.23 \pm 0.17 \pm 0.03^{\dagger\dagger}$ | (Aubert, 2008e) | $1.5 \pm 0.5 \pm 0.3^{\dagger\dagger}$ | (Chistov, 2006a) | $\sim 0.3$ |
| $B^0 \to \overline{\Lambda}_c^- \Lambda_c^+$ | | | $< 0.062$ | (Uchida, 2008) | $< 0.062$ |
| *singly-charmed final states* ($\mathcal{B}$: $10^{-6}$) | | | | | |
| $B^0 \to \overline{\Lambda}_c^- p$ | $18.9 \pm 2.1 \pm 0.6^{\dagger}$ | (Aubert, 2008aa) | $21.9^{+5.6}_{-4.9} \pm 3.2^{\dagger}$ | (Gabyshev, 2003) | $19 \pm 2^{\dagger}$ |
| $B^+ \to \overline{\Lambda}_c^- \Delta^{++}$ | | | $< 19$ | (Gabyshev, 2006) | $< 19$ |
| $B^+ \to \overline{\Lambda}_c^- \Delta^{++}(1600)$ | | | $59 \pm 10 \pm 6^{\dagger}$ | (Gabyshev, 2006) | $59 \pm 12^{\dagger}$ |
| $B^+ \to \overline{\Lambda}_c^- \Delta^{++}(2420)$ | | | $47 \pm 10 \pm 4^{\dagger}$ | (Gabyshev, 2006) | $47 \pm 11^{\dagger}$ |
| $B^0 \to \overline{\Sigma}_c(2455)^- p$ | $< 30$ | (Aubert, 2010h) | | | $< 30$ |
| $B^+ \to \overline{\Sigma}_c(2455)^0 p$ | $42 \pm 4 \pm 3^{\dagger}$ | (Aubert, 2008aa) | $37 \pm 7 \pm 4^{\dagger}$ | (Gabyshev, 2006) | $40 \pm 4^{\dagger}$ |
| $B^+ \to \overline{\Sigma}_c(2520)^0 p$ | $< 3$ | (Aubert, 2008aa) | $< 27$ | (Gabyshev, 2006) | $< 3$ |
| $B^+ \to \overline{\Sigma}_c(2800)^0 p$ | $40 \pm 8 \pm 8^{\dagger}$ | (Aubert, 2008aa) | | | $40 \pm 11^{\dagger}$ |
| *unflavored final states* ($\mathcal{B}$: $10^{-6}$) | | | | | |
| $B^0 \to p\overline{p}$ | $< 0.27$ | (Aubert, 2004ab) | $< 0.11$ | (Tsai, 2007) | $< 0.11$ |
| $B^+ \to p\overline{\Delta}^0$ | | | $< 1.4$ | (Wei, 2008b) | $< 1.4$ |
| $B^+ \to \Delta^{++}\overline{p}$ | | | $< 0.14$ | (Wei, 2008b) | $< 0.14$ |
| $B^0 \to \Lambda\overline{\Lambda}$ | | | $< 0.32$ | (Tsai, 2007) | $< 0.32$ |
| *strange final states* ($\mathcal{B}$: $10^{-6}$) | | | | | |
| $B^+ \to p\overline{\Lambda}$ | | | $< 0.32$ | (Tsai, 2007) | $< 0.32$ |
| $B^+ \to p\overline{\Lambda}(1520)$ | $< 1.5$ | (Aubert, 2005p) | | | $< 1.5$ |
| $B^+ \to p\overline{\Sigma}(1385)^0$ | | | $< 0.47$ | (Wang, 2007b) | $< 0.47$ |
| $B^0 \to p\overline{\Sigma}(1385)^-$ | | | $< 0.26$ | (Wang, 2007b) | $< 0.26$ |
| $B^+ \to \Delta^+\overline{\Lambda}$ | | | $< 0.82$ | (Wang, 2007b) | $< 0.82$ |
| $B^0 \to \Delta^0\overline{\Lambda}$ | | | $< 0.93$ | (Wang, 2007b) | $< 0.93$ |

#### 17.12.2.1 Results from $B$ Factories

An overview of the results from the $B$ Factories for two-body decays is given in Table 17.12.1.

##### *First observation*

The first two-body baryonic $B$ decay observed was $\overline{B}^0 \to \Lambda_c^+\overline{p}$ with $\Lambda_c^+ \to pK^-\pi^+$ (Gabyshev, 2003) using a $78.2\,\mathrm{fb}^{-1}$ data sample at Belle.

The mass resolution of reconstructed $\Lambda_c$ is very good. One can just select $\Lambda_c$ and apply simple continuum suppression (see Chapter 9) to reject most of the background events. Exclusive reconstruction is described in Section 7.1. Fig. 17.12.2 shows the scatter plot of $\Delta E$ versus $m_{\mathrm{ES}}$ and their projections for selected events. The $\Delta E$ projection is shown for $m_{\mathrm{ES}} > 5.270\,\mathrm{GeV}/c^2$ and the $m_{\mathrm{ES}}$ projection for $|\Delta E| < 0.030\,\mathrm{GeV}$. A two-dimensional binned maximum likelihood fit is performed to determine the signal yield. For this fit, the $\Delta E$ distribution is represented by a double Gaussian for the signal plus a first order polynomial for the background. The $m_{\mathrm{ES}}$ distribution is represented by a single Gaussian for the signal plus the ARGUS function for the background. The signal shapes determined from MC simulation are fixed in the fit. The region $\Delta E < -0.1\,\mathrm{GeV}$ is excluded from the fit to avoid feed-down from modes including extra pions. The measured branching fraction is $(2.19^{+0.56}_{-0.49} \pm 0.32 \pm 0.57) \times 10^{-5}$ where the last error comes from the uncertainty on the sub-decay branching fraction of $\Lambda_c^+ \to pK^-\pi^+$. The branching fraction is thusan order-of-magnitude smaller than that of the three-body decay $B^- \to \Lambda_c^+\overline{p}\pi^-$ (see Table 17.12.6). This suppression is a unique feature of two-body baryonic decays and will be addressed further in Sections 17.12.2.2 and 17.12.3.4. In contrast, the two- and three-body mesonic $B$ decays are comparable.

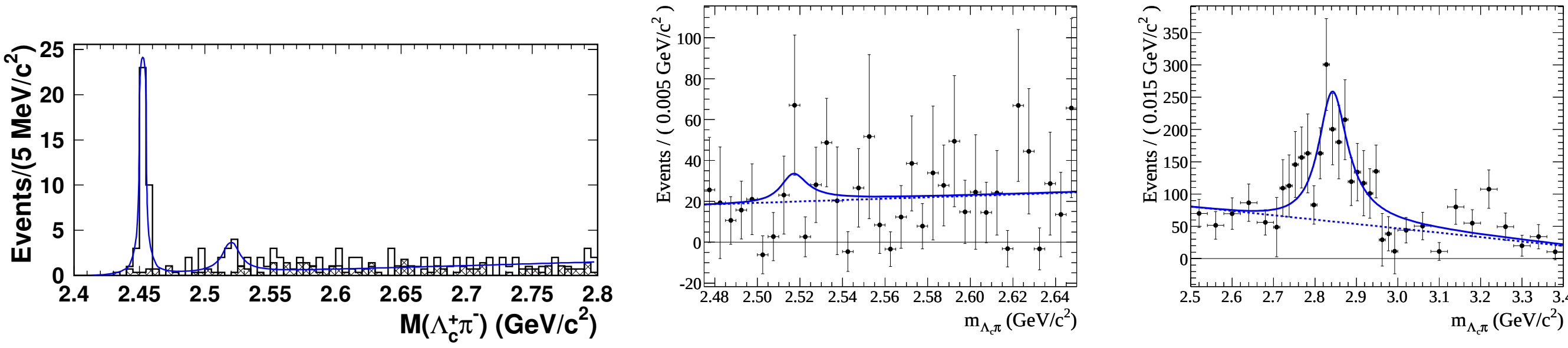

(a) Belle: $B$ signal region (open histogram), fit results (curve) and background from sidebands (hatched).

(b) *BABAR*: ${}_s\mathcal{P}lot$ projections of the $\overline{\Sigma}_c(2520)^0 p$ signal (data points) and fits (curves).

(c) *BABAR*: ${}_s\mathcal{P}lot$ projections of the $\overline{\Sigma}_c(2800)^0 p$ signal (data points) and fits (curves).

**Figure 17.12.3.** $m(\Lambda_c^+\pi^-)$ distributions from (a) Belle (Gabyshev, 2006), and (b,c) *BABAR* (Aubert, 2008aa); for the ${}_s\mathcal{P}lot$ technique, see Section 11.2.3. Three $\Sigma_c^0$ resonances are visible at $2.455\,\mathrm{GeV}/c^2$ (a), $2.52\,\mathrm{GeV}/c^2$ (a and b, not significant) and $2.80\,\mathrm{GeV}/c^2$ (c).

*Quasi-two-body decays*

Following the study of $B^- \to \Lambda_c^+\overline{p}\pi^-$ at CLEO (Dytman et al., 2002) and at Belle (Gabyshev, 2002), an analysis of this final state has been performed using a much larger data sample (containing $\sim 152\times10^6$ $B\overline{B}$ pairs) and extending the $\Lambda_c^+$ reconstruction to the following five decay modes: $\Lambda_c^+ \to pK^-\pi^+$, $p\overline{K}^0$, $\Lambda\pi^+$, $p\overline{K}^0\pi^+\pi^-$, and $\Lambda\pi^+\pi^+\pi^-$ (Gabyshev, 2006). A clear signal peak is seen from the intermediate two-body $B^- \to \Sigma_c(2455)^0\overline{p}$ decay, together with a hint of $B^- \to \Sigma_c(2520)^0\overline{p}$, shown in Fig. 17.12.3a. The open histogram is the distribution from the $B$ signal region ($|\Delta E| < 0.03\,\mathrm{GeV}$ and $m_{\mathrm{ES}} > 5.27\,\mathrm{GeV}/c^2$). The hatched histogram is the distribution from sideband regions ($-0.10\,\mathrm{GeV} < \Delta E < -0.04\,\mathrm{GeV}$ or $0.04\,\mathrm{GeV} < \Delta E < 0.20\,\mathrm{GeV}$) normalized to the $B$ signal region. The curve shows the result of the fit which includes the contributions from $\Sigma_c(2455)^0$ and $\Sigma_c(2520)^0 \to \Lambda_c^+\pi^-$ decays and the background parameterized with a linear function. The $\Sigma_c(2455/2520)^0$ signal shapes are fixed from MC assuming a Breit-Wigner function convolved with the resolution function.

A subsequent *BABAR* analysis (Aubert, 2008aa) using $\sim 383\times10^6$ $B\overline{B}$ pairs confirms the $\Sigma_c(2455)^0$, but with an even weaker signal from $\Sigma_c(2520)^0$ shown in Fig. 17.12.3b. A broad structure $\Sigma_c(2800)^0$, however, is clearly seen (Fig. 17.12.3c). The $\Sigma_c(2455)^0$ is a spin-$\frac{1}{2}$ baryon, while the $\Sigma_c(2520)^0$ has spin $\frac{3}{2}$. A decay of the spin-0 $B$ meson to a spin-$\frac{1}{2}$ antiproton and a spin-$\frac{3}{2}$ $\Sigma_c$ requires one or two units of orbital angular momentum, and hence its suppression is reasonable.

Final states from weak $B$ meson decays $b \to c\overline{u}d$ have three light quarks, therefore the isospin can be $I = \frac{1}{2}$ or $I = \frac{3}{2}$, while the $W$ exchange $b\overline{d} \to c\overline{u}$ has only one and therefore $I = \frac{1}{2}$ (see Fig. 17.12.5a and c below). Since hadronization is a strong interaction process, isospin is conserved and we can classify final states according to their isospin. The average branching fraction $\mathcal{B}(B^- \to \Sigma_c(2455)^0\overline{p}) = (4.0 \pm 0.4 \pm 1.0_{\Lambda_c}) \times 10^{-5}$ (pure isospin $I = \frac{3}{2}$) is about twice that of $\overline{B}^0 \to \Lambda_c^+\overline{p}$ (pure isospin $I = \frac{1}{2}$). Using Clebsch-Gordan coefficients, one would expect $\mathcal{B}(\overline{B}^0 \to \Sigma_c(2455)^+\overline{p}) < 2.5\times10^{-5}$, compatible with the present limit (Table 17.12.1).

*Decays to two charmed baryons*

An unexpectedly large rate is found for $B$ mesons decaying to two charmed baryons, specifically $B^+ \to \overline{\Xi}_c^0\Lambda_c^+$ (Chistov, 2006a). In this analysis, the following sub-decays are reconstructed: $\Xi_c^0 \to \Xi^-\pi^+$ and $\Lambda K^-\pi^+$, $\Lambda_c^+ \to pK^-\pi^+$, $\Xi^- \to \Lambda\pi^-$, and $\Lambda \to p\pi^-$. For $\Xi^- \to \Lambda\pi^-$, one can fit the $p$ and $\pi^-$ tracks to a common vertex in order to get the $\Lambda$ 4-momenta. Then one can fit the $\Lambda$ trajectory and the $\pi^-$ track to a common vertex to reconstruct the long lived $\Xi^-$. Fig. 17.12.4 shows the projection plots of selected candidate events; the maximum likelihood fit results are overlaid. The product of branching fractions $\mathcal{B}(B^+ \to \overline{\Xi}_c^0\Lambda_c^+) \times \mathcal{B}(\overline{\Xi}_c^0 \to \overline{\Xi}^+\pi^-)$ is measured to be $(4.8^{+1.0}_{-0.9} \pm 1.1 \pm 1.2) \times 10^{-5}$. *BABAR* found a somewhat smaller branching fraction (Aubert, 2008e), but still about two orders of magnitude larger than that of $\overline{B}^0 \to \Lambda_c^+\overline{p}$. Using our estimate $\mathcal{B}(\Xi_c^0 \to \Xi^-\pi^+) \approx 1.2\%$ described in Section 17.12.1, the average translates to $\mathcal{B}(B^+ \to \overline{\Xi}_c^0\Lambda_c^+) \approx 0.22\%$. This is quite intriguing and offers an important clue to understand the underlying dynamics for baryonic $B$ decays.

The decay of the neutral $B$ meson $B^0 \to \overline{\Xi}_c^-\Lambda_c^+$ (with $\overline{\Xi}_c^- \to \overline{\Xi}^+\pi^-\pi^-$) has not yet been observed with high significance, however we have averaged the results of Belle and *BABAR*, using the theoretical estimate (described in Section 17.12.1) $\mathcal{B}(\overline{\Xi}_c^- \to \overline{\Xi}^+\pi^-\pi^-) \approx 6.4\%$, assuming a 50% uncertainty.

*Charmless decays*

So far, no charmless two-body baryonic $B$ decays have been observed. The 90% confidence level upper limits have been pushed below the $10^{-6}$ level and are listed in Table 17.12.1. The method to determine upper limit yields is either based on the Feldman-Cousins approach (Conrad, Botner, Hallgren, and Perez de los Heros, 2003; Feldman and Cousins, 1998) or by integration of the likelihood fit function convolved with a Gaussian error function. Since there is no sign of observation, it will be interesting to know the order of magnitude of the branching fractions of these rare decays, and hopefully they can be determined by the LHCb experiment or future super flavor factories.

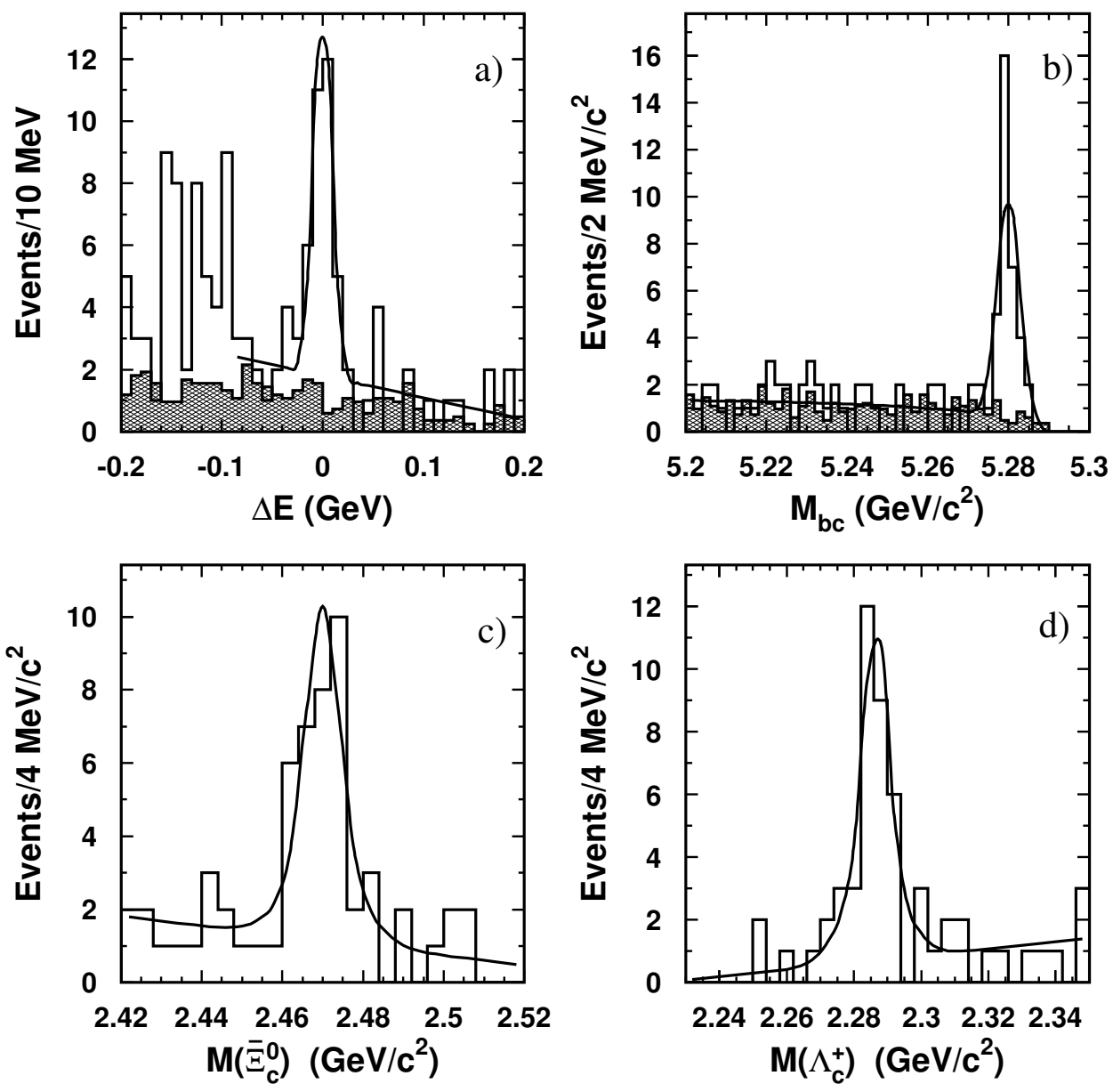

**Figure 17.12.4.** The $\Delta E$ (a) and $m_{\mathrm{ES}} = M_{\mathrm{bc}}$ (b) distributions for the $B^+ \to \overline{\Xi}{}_c^0 \Lambda_c^+$ candidates from Belle (Chistov, 2006a). The hatched histograms show the combined $\overline{\Xi}{}_c^0$ and $\Lambda_c^+$ mass sidebands normalized to the signal region. Also shown are the $\overline{\Xi}{}_c^0$ (c) and $\Lambda_c^+$ (d) mass distributions for the $B^+ \to \overline{\Xi}{}_c^0 \Lambda_c^+$ candidates taken from the $B$ signal region of $|\Delta E| < 0.025\,\mathrm{GeV}$ and $m_{\mathrm{ES}} > 5.272\,\mathrm{GeV}/c^2$. For the $\overline{\Xi}{}_c^0$ ($\Lambda_c^+$) distribution $\Lambda_c^+$ ($\overline{\Xi}{}_c^0$) is required to be within $\pm 15\,\mathrm{MeV}/c^2$ of the nominal mass. The overlaid curves are the fit results.

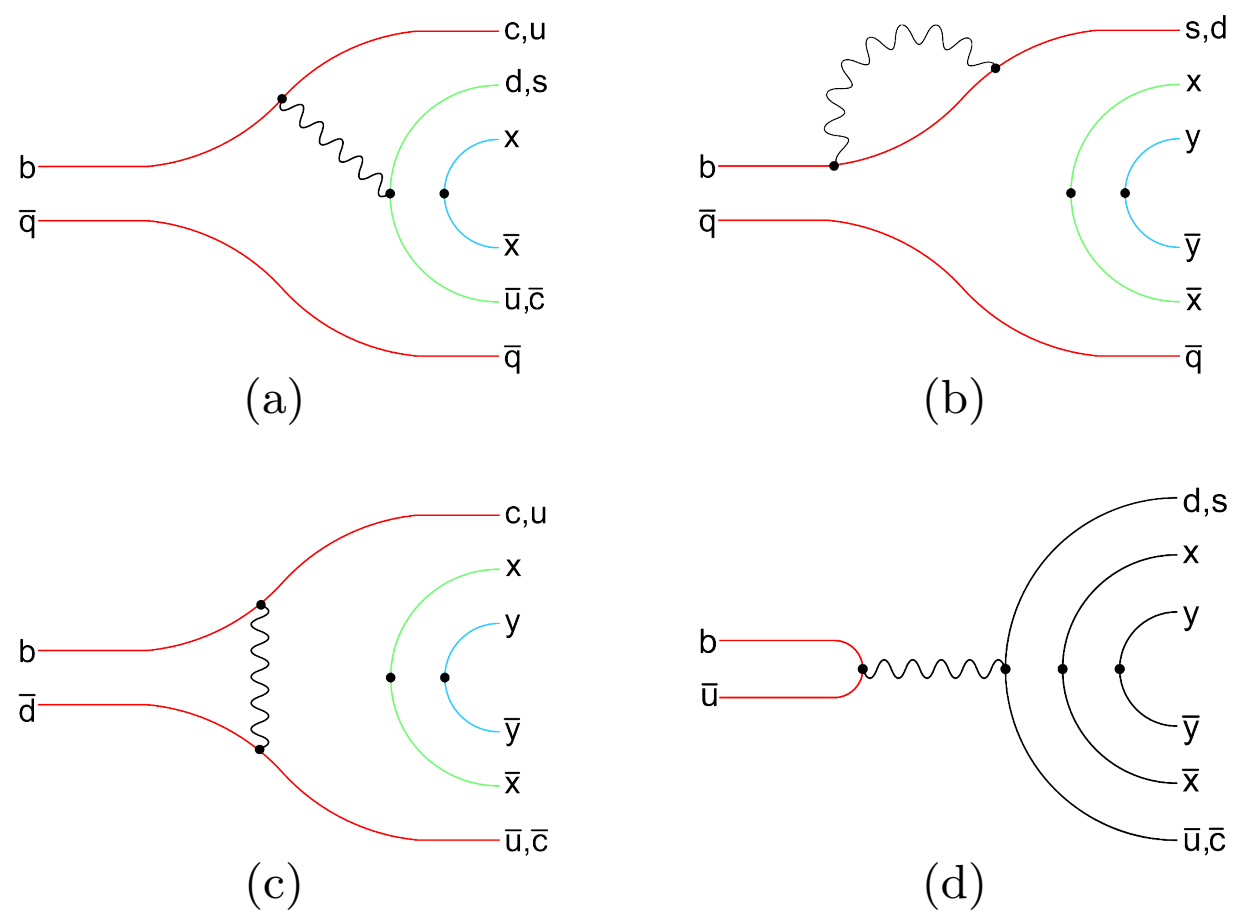

**Figure 17.12.5.** Quark diagrams for two-body baryonic $B$ decay $\overline{B} \to \mathfrak{B}_1 \overline{\mathfrak{B}}_2$. The waveline represents a $W$, while gluon lines are omitted. Quark antiquark pairs from the vacuum are denoted as $x, y$ and determine the flavor of the two final baryons. In this and subsequent figures, the colors of the quark lines are significant: see the discussion in Section 17.12.2.2.

### 17.12.2.2 Theory and interpretation

Since baryonic $B$ decays involve two baryons in the final state, the underlying mechanism is complicated. The quark diagrams for two-body baryonic $B$ decays are shown in Fig. 17.12.5: internal $W$-emission for $b \to c(u)$ (a), the $b \to s(d)$ penguin transition (b), $W$-exchange for the neutral $B$ meson (c), and $W$-annihilation for the charged $B$ (d). As for mesonic $B$ decays, $W$-exchange and $W$-annihilation are expected to be helicity suppressed (Chau, 1983) which can be understood in the same way as for leptonic decays (see Eq. 17.10.4 and the accompanying discussion). Therefore, the main contributions to two-body baryonic $B$ decay $B \to \mathfrak{B}_1 \overline{\mathfrak{B}}_2$ are due to either the internal $W$-emission diagram or the penguin diagram. It should be stressed that, unlike the case of mesonic $B$ decays, internal $W$ emission in baryonic $B$ decays is not necessarily color suppressed. This is because the baryon wave function is totally antisymmetric in the color indices. One component of this wave function is illustrated in Fig. 17.12.5, where the internal $W$-emission (a) needs either of two matching colors (blue or green) to produce a baryon, while exactly one matching color (red) would be required to produce a meson. The black quark lines in (d) indicate that any color will do.

In short, the two-body decay proceeds mainly through the nonfactorizable internal $W$-emission or the $b \to s(d)$ penguin transition. This is why it is difficult to determine theoretical estimates for the rates of two-body decays.

There exist several theoretical models for describing $B$ decays into two baryons: the pole model of Jarfi et al. (1990) and Cheng and Yang (2002a), the diquark model of Ball and Dosch (1991) and the QCD sum rule analysis by Chernyak and Zhitnitsky (1990). The predictions of these models for some selected charmless, singly-charmed and doubly-charmed baryonic $B$ decays are listed in Tables 17.12.2–17.12.4. Evidently, many of the earlier model predictions are either too large compared to or marginally comparable to experiment.

Experimentally, two-body baryonic $B$ decays follow the pattern

$$\begin{aligned} &\mathcal{B}(\overline{B} \to \mathfrak{B}_{1c} \overline{\mathfrak{B}}_{2c}) \sim 10^{-3} \\ \gg\; &\mathcal{B}(\overline{B} \to \mathfrak{B}_c \overline{\mathfrak{B}}) \quad\;\; \sim 10^{-5} \\ \gg\; &\mathcal{B}(\overline{B} \to \mathfrak{B}_1 \overline{\mathfrak{B}}_2) \quad \lesssim 10^{-6}\,. \end{aligned} \tag{17.12.4}$$

where no $c$ subscript indicates a non-charmed baryon.

Since the doubly-charmed baryonic decay $\overline{B} \to \Xi_c \overline{\Lambda}_c$ proceeds via $b \to cs\bar{c}$, while $\overline{B} \to \Lambda_c \bar{p}$ proceeds via a $b \to cd\bar{u}$ quark transition, the CKM matrix elements for the two decays are the same in magnitude but opposite in sign. One may therefore wonder why the $\Xi_c \overline{\Lambda}_c$ mode has a rate two orders of magnitude larger than $\Lambda_c\overline{p}$. Indeed, earlier calculations based on QCD sum rules (Chernyak and Zhitnitsky, 1990) or the diquark model (Ball and Dosch, 1991) all predict that $\mathcal{B}(B \to \Xi_c\overline{\Lambda}_c) \approx \mathcal{B}(\overline{B} \to \mathfrak{B}_c\overline{N})$ (see Table 17.12.3), which is in violent disagreement with experiment. The decay pattern (17.12.4) can be understood as follows. In an energetic heavy baryon, the momentum is mostly carried by the constituent heavy quark. Therefore, in $b \to c\overline{c}s$ decays, the energetic $c$ quark will fragment into $\mathfrak{B}_{1c}$, and $\overline{c}$ into $\overline{\mathfrak{B}}_{2c}$. Consequently, no hard gluon is needed to produce the energetic $\Xi_c\overline{\Lambda}_c$ pair in $B$ decays (see Fig. 17.12.6a). In $\overline{B} \to \Lambda_c\overline{p}$ decay, the three quarks of the energetic proton share the same momentum fraction $\sim 1/3$. Hence, two hard gluons are needed to produce an energetic $\overline{p}$: one hard gluon to kick the spectator quark of the $B$ meson to make it energetic and the other to produce the hard $q\overline{q}$ pair (see Fig. 17.12.6b). Therefore, the decay rate of $\overline{B} \to \Lambda_c\overline{p}$ is suppressed with respect to $\overline{B} \to \Xi_c\overline{\Lambda}_c$ due to a dynamical factor $\mathcal{O}(\alpha_s^4) \sim 10^{-2}$. These qualitative statements have been supported by realistic calculations of the decay rates for $\overline{B} \to \Xi_c\overline{\Lambda}_c$ (Cheng, Chua, and Hsiao, 2009) and $\overline{B}^0 \to \Lambda_c^+\overline{p}$ (He, Li, Li, and Wang, 2007).

**Table 17.12.2.** Branching fractions (in units of $10^{-7}$) for some charmless two-body baryonic $B$ decays classified into two categories: tree-dominated (upper) and penguin-dominated (lower). Branching fractions denoted by "†" are calculated only for the parity-conserving part. Experimental limits are taken from Table 17.12.1. Theoretical predictions are taken from the following references — "CZ": Chernyak and Zhitnitsky (1990); "Jarfi": Jarfi et al. (1990); and "CY": Cheng and Yang (2002a).

| Decay | CZ | Jarfi | CY | Experiment |
|---|---|---|---|---|
| $\overline{B}^0 \to p\overline{p}$ | 12 | 70 | $1.1^\dagger$ | $< 1.1$ |
| $\overline{B}^0 \to n\overline{n}$ | 3.5 | 70 | $1.2^\dagger$ | |
| $B^0 \to n\overline{p}$ | 6.9 | 170 | 5.0 | |
| $\overline{B}^0 \to \Lambda\overline{\Lambda}$ | | 2 | $0^\dagger$ | $< 3.2$ |
| $B^- \to p\overline{\Delta}^{--}$ | 2.9 | 3200 | 14 | $< 1.4$ |
| $\overline{B}^0 \to p\overline{\Delta}^-$ | 0.7 | 1000 | 1.4 | |
| $B^- \to n\overline{\Delta}^-$ | | 1 | 4.6 | |
| $\overline{B}^0 \to n\overline{\Delta}^0$ | | 1000 | 4.3 | |
| $B^- \to \Lambda\overline{p}$ | $\lesssim 30$ | | $2.2^\dagger$ | $< 3.2$ |
| $\overline{B}^0 \to \Lambda\overline{n}$ | | | $2.1^\dagger$ | |
| $\overline{B}^0 \to \Sigma^+\overline{p}$ | 60 | | $0.18^\dagger$ | $< 2.6$ |
| $B^- \to \Sigma^0\overline{p}$ | 30 | | 0.58 | $< 4.7$ |
| $B^- \to \Sigma^+\overline{\Delta}^{--}$ | 60 | | 2.0 | |
| $\overline{B}^0 \to \Sigma^+\overline{\Delta}^-$ | 60 | | 0.63 | |
| $B^- \to \Sigma^-\overline{\Delta}^0$ | 20 | | 0.87 | |

**Table 17.12.3.** Predictions (in units of $10^{-5}$) of singly charmed two-body baryonic $B$ decays in various models. Theoretical references are as in Table 17.12.2. Experimental results are taken from Table 17.12.1.

| Decay | CZ | Jarfi | CY | Experiment |
|---|---|---|---|---|
| $\overline{B}^0 \to \Lambda_c^+\overline{p}$ | 190 | 110 | 1.1 | $1.9 \pm 0.2$ |
| $B^- \to \Sigma_c^0\overline{p}$ | 300 | 1500 | 6.0 | $4.0 \pm 0.4$ |
| $\overline{B}^0 \to \Sigma_c^0\overline{n}$ | 580 | 0.06 | | |
| $B^- \to \Lambda_c^+\overline{\Delta}^{--}$ | 20 | 3600 | 1.9 | $5.9 \pm 1.2$ |

**Table 17.12.4.** Predicted branching fractions (in units of $10^{-4}$) of doubly-charmed two-body baryonic $B$ decays (Cheng, Chua, and Hsiao, 2009). Experimental results are taken from Table 17.12.1.

| Decay | Theory | Experiment |
|---|---|---|
| $B^- \to \Xi_c^0\overline{\Lambda}_c^-$ | $10.4^{+5.7}_{-5.5}$ | $\sim 21.6 \pm 5.4$ |
| $\overline{B}^0 \to \Xi_c^+\overline{\Lambda}_c^-$ | $9.4^{+6.3}_{-4.1}$ | $\sim 3$ |
| $\overline{B}^0 \to \Lambda_c^+\overline{\Lambda}_c^-$ | $0.52^{+0.35}_{-0.19}$ | $< 0.62$ |

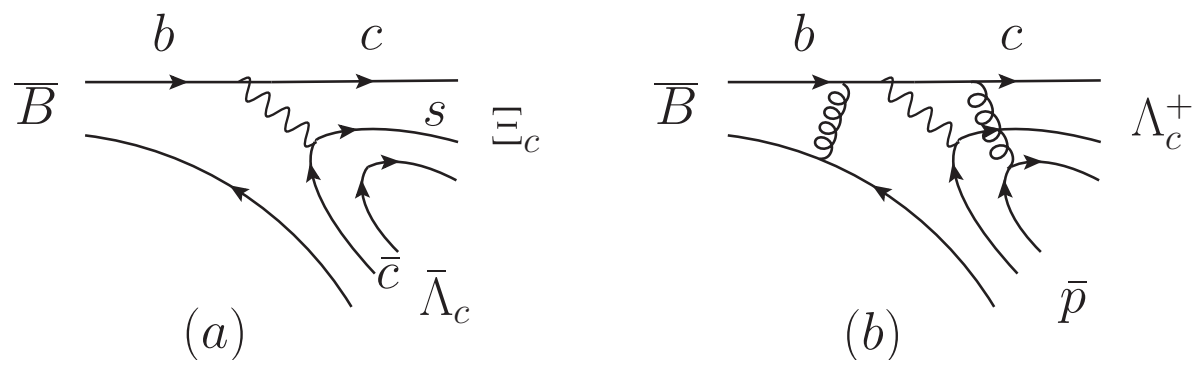

**Figure 17.12.6.** Quark diagrams for two-body baryonic $B$ decays $\overline{B} \to \Xi_c\overline{\Lambda}_c$ and $\overline{B} \to \Lambda_c^+\overline{p}$. At least two hard gluons are needed for $\Lambda_c^+\overline{p}$ production.

The charmless decay $\overline{B}^0 \to p\overline{p}$ is suppressed relative to $\overline{B} \to \Lambda_c\overline{p}$ by the CKM matrix elements $|V_{ub}/V_{cb}|^2$ and is also subject to a possible dynamical suppression:

$$\mathcal{B}(\overline{B} \to p\overline{p}) = \mathcal{B}(\overline{B} \to \Lambda_c^+\overline{p}) \left|\frac{V_{ub}}{V_{cb}}\right|^2 \times f_{\rm dyn} \approx 2{\times}10^{-7} \times f_{\rm dyn}\ . \qquad (17.12.5)$$

In the absence of dynamical suppression $f_{\rm dyn}$, the predicted rate for two-body charmless decays is on the verge of the experimental upper limit.

If the dynamical suppression is of order $10^{-2}$ as in the case of $\overline{B} \to \Lambda_c\overline{p}$, then the branching fraction for charmless two-body decays will be of order $10^{-9}$ and thus beyond the reach even of super flavor factories. In reality, the branching fraction is most likely of order $10^{-8}$, between the extreme cases of $10^{-7}$ and $10^{-9}$. Thus far, there is no clear theoretical prediction for charmless two-body decays. Presumably a reliable prediction based on pQCD can be made as the energy release in charmless two-body decay is very large, justifying the use of pQCD (Cheng and Yang, 2002a).

Most of the previous theoretical predictions are not trustworthy: for example, predictions based on the QCD sum rule, the pole model and the diquark model are too large compared to experiment. The most reliable predictions are based on pQCD, which has been successfully applied to $B \to \Lambda_c\overline{p}$ (He, Li, Li, and Wang, 2007). The pQCD calculation for charmless modes such as $\Lambda\overline{p}$ and $p\overline{p}$ is much more involved and has not yet been carried out.

**Table 17.12.5.** Branching fractions of observed decays of $B$ mesons to charmless baryons plus charmed mesons. Contributing quark diagrams are given in square brackets following each mode: annihilation type A (Fig. 17.12.5c with an extra $q\overline{q}$ pair), external $W$-emission type 1 (Fig. 17.12.7), and internal $W$-emission type 2 (Fig. 17.12.8).

| Decay | | BABAR | Belle | other | Average |
|---|---|---|---|---|---|
| ($\mathcal{B}$ : $10^{-4}$) | | (del Amo Sanchez, 2012) | (Abe, 2002g) | (Anderson et al., 2001) | |
| $\overline{B}^0 \to D^0 p\overline{p}$ | [2c, 2e, A] | $1.02 \pm 0.04 \pm 0.06$ | $1.18 \pm 0.15 \pm 0.16$ | | $1.04 \pm 0.07$ |
| $\overline{B}^0 \to D^{*0} p\overline{p}$ | [2c, 2e, A] | $0.97 \pm 0.07 \pm 0.09$ | $1.20^{+0.33}_{-0.29} \pm 0.21$ | | $1.00 \pm 0.11$ |
| $\overline{B}^0 \to D^{*+} n\overline{p}$ | [2c, 2e, A] | | | $14.5^{+3.4}_{-3.0} \pm 2.7$ | |
| $\overline{B}^0 \to D^+ p\overline{p}\pi^-$ | | $3.32 \pm 0.10 \pm 0.29$ | | | |
| $\overline{B}^0 \to D^{*+} p\overline{p}\pi^-$ | | $4.55 \pm 0.16 \pm 0.39$ | | $6.5^{+1.3}_{-1.2} \pm 1.0$ | $4.67 \pm 0.40$ |
| $B^- \to D^0 p\overline{p}\pi^-$ | | $3.72 \pm 0.11 \pm 0.25$ | | | |
| $B^- \to D^{*0} p\overline{p}\pi^-$ | | $3.73 \pm 0.17 \pm 0.27$ | | | |
| $\overline{B}^0 \to D^0 p\overline{p}\pi^-\pi^+$ | | $2.99 \pm 0.21 \pm 0.45$ | | | |
| $\overline{B}^0 \to D^{*0} p\overline{p}\pi^-\pi^+$ | | $1.91 \pm 0.36 \pm 0.29$ | | | |
| $B^- \to D^+ p\overline{p}\pi^-\pi^-$ | | $1.66 \pm 0.13 \pm 0.27$ | | | |
| $B^- \to D^{*+} p\overline{p}\pi^-\pi^-$ | | $1.86 \pm 0.16 \pm 0.19$ | | | |
| ($\mathcal{B}$ : $10^{-5}$) | | | (Chang, 2009) | | |
| $B^0 \to D^0 \Lambda\overline{\Lambda}$ | [2c, 2e, A] | | $1.05^{+0.57}_{-0.44} \pm 0.14$ | | |
| ($\mathcal{B}$ : $10^{-5}$) | | | (Medvedeva, 2007) | | |
| $\overline{B}^0 \to D_s^+ \Lambda\overline{p}$ | [2c] | | $2.9 \pm 0.7 \pm 0.5 \pm 0.4_{D_s}$ | | |
| ($\mathcal{B}$ : $10^{-5}$) | | | (Chen, 2011) | | |
| $B^- \to D^0 \Lambda\overline{p}$ | [1b, 2e] | | $1.43^{+0.28}_{-0.25} \pm 0.18$ | | |
| $B^- \to D^{*0} \Lambda\overline{p}$ | [1b, 2e] | | $< 4.8$ | | |
| ($\mathcal{B}$ : $10^{-6}$) | | (Aubert, 2003b) | (Xie, 2005) | | |
| $B^- \to J/\psi\, \Lambda\overline{p}$ | [2e] | $12^{+9}_{-6}$ | $11.6 \pm 2.8^{+1.8}_{-2.3}$ | | $11.7 \pm 3.1$ |
| $B^- \to J/\psi\, \Sigma^0\overline{p}$ | [2e] | | $< 11$ | | |
| $B^0 \to J/\psi\, p\overline{p}$ | [2e] | $< 1.9$ | $< 0.83$ | | $< 0.83$ |

## 17.12.3 Decays to baryon antibaryon plus mesons

Before Belle and BABAR investigated baryonic $B$ decays, there were already observations by CLEO of $B^+ \to \overline{\Lambda}_c p\pi^+$ (Fu et al., 1997) and $B^0 \to D^{*-}p\overline{p}\pi^+, D^{*-}p\overline{n}$ (Anderson et al., 2001). These are all generic $b \to c$ transitions, and many more have since been investigated by the $B$ Factories. Their branching fractions are shown in Table 17.12.5 for cases where a $c$ quark hadronizes into a charmed meson, and in Table 17.12.6 where a $c$ quark hadronizes into a charmed baryon. In the following subsections we consider in turn theoretical issues (Section 17.12.3.1), rare decays (Section 17.12.3.2), the threshold enhancement seen in many multibody baryonic decays (Section 17.12.3.3), the role of final state multiplicity (Section 17.12.3.4), and angular correlations of the final state particles (Section 17.12.3.5); we conclude with brief discussions of the role of Cabibbo suppression (Section 17.12.3.6), isospin relations (Section 17.12.3.7), and the suppression of $s\overline{s}$ pairs (Section 17.12.3.8).

### 17.12.3.1 Theoretical models and Feynman diagrams

The complexity of quark diagrams increases with the final state multiplicity. For three-body decays of a $B$ meson to the baryonic final state $\mathfrak{B}_1\overline{\mathfrak{B}}_2M$ there are many distinct quark diagrams: two type-1 or external $W$-diagrams (Figs 17.12.7a and b), and eight type-2 or internal $W$-emission diagrams (Figs 17.12.8a–h); $W$-exchange diagrams (basically Fig. 17.12.5c, inserting another $q\overline{q}$ pair) for the neutral $B$ meson; and $W$-annihilation (Fig. 17.12.5d with another $q\overline{q}$ pair) for the charged $B$. The various possibilities for inserting the extra $q\overline{q}$ pair are only illustrated for the type-2 decay in Fig. 17.12.8, as the same modification to the annihilation diagrams is straightforward. Penguin diagrams can also contribute: these can be obtained from Fig. 17.12.5b by adding the extra $q\overline{q}$ pair in the same way as for $W$ exchange.

**Table 17.12.6.** Branching fractions of observed decays of $B$ mesons to final states with a charmed baryon including three or more final state particles. Channels with a $\Lambda_c^+$ baryon in the final state have an additional ±26% relative uncertainty (not included) from the assumption $\mathcal{B}(\Lambda_c^+ \to pK^-\pi^+) = 0.050 \pm 0.013$ and are marked with †. Two daggers indicate that two such terms have to be added (linearly due to correlation). Decays via intermediate resonances are marked with •. Contributing quark diagrams are given in square brackets following each three-body mode: annihilation type A (Fig. 17.12.5c with an extra $q\overline{q}$ pair), external $W$-emission type 1 (Fig. 17.12.7), and internal $W$-emission type 2 (Fig. 17.12.8).

| Decay | | BaBar | Belle | other | Average |
|---|---|---|---|---|---|
| $(\mathcal{B}: 10^{-4})$ | | (Aubert, 2010h) | | | |
| $\overline{B}^0 \to \Lambda_c^+ \overline{p}\pi^0$ | [2abdfg, A] | $1.94 \pm 0.17 \pm 0.14^\dagger$ | | | |
| • see Table 17.12.1 | | | | | |
| $(\mathcal{B}: 10^{-4})$ | | (Aubert, 2008aa) | (Gabyshev, 2006) | (Dytman et al., 2002) | |
| $B^- \to \Lambda_c^+ \overline{p}\pi^-$ | [1a, 2abdfg] | $3.38 \pm 0.12 \pm 0.12^\dagger$ | $2.01 \pm 0.15 \pm 0.20^\dagger$ | $2.4 \pm 0.6^{+0.19\dagger}_{-0.17}$ | $2.92 \pm 0.14$ |
| • see Table 17.12.1 | | | | | |
| $(\mathcal{B}: 10^{-4})$ | | (Aubert, 2008e) | (Abe, 2006b) | | |
| $B^- \to \Lambda_c^+ \Lambda_c^- K^-$ | [2d] | $11.4 \pm 1.5 \pm 1.7^{\dagger\dagger}$ | $6.5^{+1.0}_{-0.9} \pm 1.1^{\dagger\dagger}$ | | $8.0 \pm 1.3$ |
| $B^0 \to \Lambda_c^+ \Lambda_c^- K^0$ | [2dg] | $3.8 \pm 3.1 \pm 0.5^{\dagger\dagger}$ | $7.9^{+2.9}_{-2.3} \pm 1.2^{\dagger\dagger}$ | | $6.2 \pm 2.0$ |
| $(\mathcal{B}: 10^{-5})$ | | (Lees, 2011f) | | | |
| $\overline{B}^0 \to \Lambda_c^+ \overline{\Lambda} K^-$ | [2abf, A] | $3.8 \pm 0.8 \pm 0.2^\dagger$ | | | |
| $(\mathcal{B}: 10^{-4})$ | | (Lees, 2013h) | (Park, 2007) | (Dytman et al., 2002) | |
| $\overline{B}^0 \to \Lambda_c^+ \overline{p}\pi^+\pi^-$ total | | $12.3 \pm 0.5 \pm 0.7^\dagger$ | $11.2 \pm 0.5 \pm 1.4^\dagger$ | $16.7 \pm 1.9^{+1.9\dagger}_{-1.6}$ | $12.35 \pm 0.72$ |
| • $\overline{B}^0 \to \Lambda_c^+ \overline{p}\pi^+\pi^-$ nonresonant | | $7.9 \pm 0.4 \pm 0.4^\dagger$ | $6.4 \pm 0.4 \pm 0.9^\dagger$ | | $6.66 \pm 0.89$ |
| • $\overline{B}^0 \to \Sigma_c^{++}(2455)\overline{p}\pi^-$ | [1a, 2g, A] | $2.13 \pm 0.10 \pm 0.10^\dagger$ | $2.1 \pm 0.2 \pm 0.3^\dagger$ | $3.7 \pm 0.8 \pm 0.7^\dagger$ | $2.15 \pm 0.13$ |
| • $\overline{B}^0 \to \Sigma_c^{++}(2520)\overline{p}\pi^-$ | [1a, 2g, A] | $1.15 \pm 0.10 \pm 0.05^\dagger$ | $1.2 \pm 0.1 \pm 0.2^\dagger$ | | $1.20 \pm 0.10$ |
| • $\overline{B}^0 \to \Sigma_c^0(2455)\overline{p}\pi^+$ | [2ab, A] | $0.91 \pm 0.07 \pm 0.04^\dagger$ | $1.4 \pm 0.2 \pm 0.2^\dagger$ | $2.2 \pm 0.6 \pm 0.4^\dagger$ | $0.94 \pm 0.08$ |
| • $\overline{B}^0 \to \Sigma_c^0(2520)\overline{p}\pi^+$ | [2ab, A] | $0.22 \pm 0.07 \pm 0.01^\dagger$ | $< 0.38$ | | |
| $B^- \to \Lambda_c^+ \overline{p}\pi^-\pi^0$ | | | | $18.1 \pm 2.9^{+2.2\dagger}_{-1.6}$ | |
| • $B^- \to \Sigma_c^0 \overline{p}\pi^0$ | [2abfg] | | | $4.2 \pm 1.3 \pm 0.4^\dagger$ | |
| $(\mathcal{B}: 10^{-4})$ | | (Aubert, 2009ag) | | | |
| $\overline{B}^0 \to \Lambda_c^+ \overline{p}\pi^+ K^-$ total | | $0.433 \pm 0.082 \pm 0.033^\dagger$ | | | |
| • $\overline{B}^0 \to \Sigma_c^{++}(2455)\overline{p}K^-$ | [1a, 2g] | $0.111 \pm 0.030 \pm 0.009^\dagger$ | | | |
| • $\overline{B}^0 \to \Lambda_c^+ \overline{p}\overline{K}^{*0}$ | [2dg] | $0.160 \pm 0.061 \pm 0.012^\dagger$ | | | |
| $(\mathcal{B}: 10^{-6})$ | | (Grünberg, 2012) | | | |
| $B^- \to \Lambda_c^+ \overline{p}p\overline{p}$ | | $< 6.2$ | | | |
| $(\mathcal{B}: 10^{-4})$ | | (Lees, 2012aa) | | (Dytman et al., 2002) | |
| $B^- \to \Lambda_c^+ \overline{p}\pi^-\pi^+\pi^-$ | | | | $22.5 \pm 2.5^{+2.4\dagger}_{-1.9}$ | |
| • $B^- \to \Sigma_c^0 \overline{p}\pi^-\pi$ | | | | $4.4 \pm 1.2 \pm 0.5^\dagger$ | |
| • $B^- \to \Sigma_c^{++} \overline{p}\pi^-\pi^-$ | | $2.98 \pm 0.16 \pm 0.15^\dagger$ | | $2.8 \pm 0.9 \pm 0.5^\dagger$ | $2.97 \pm 0.21$ |

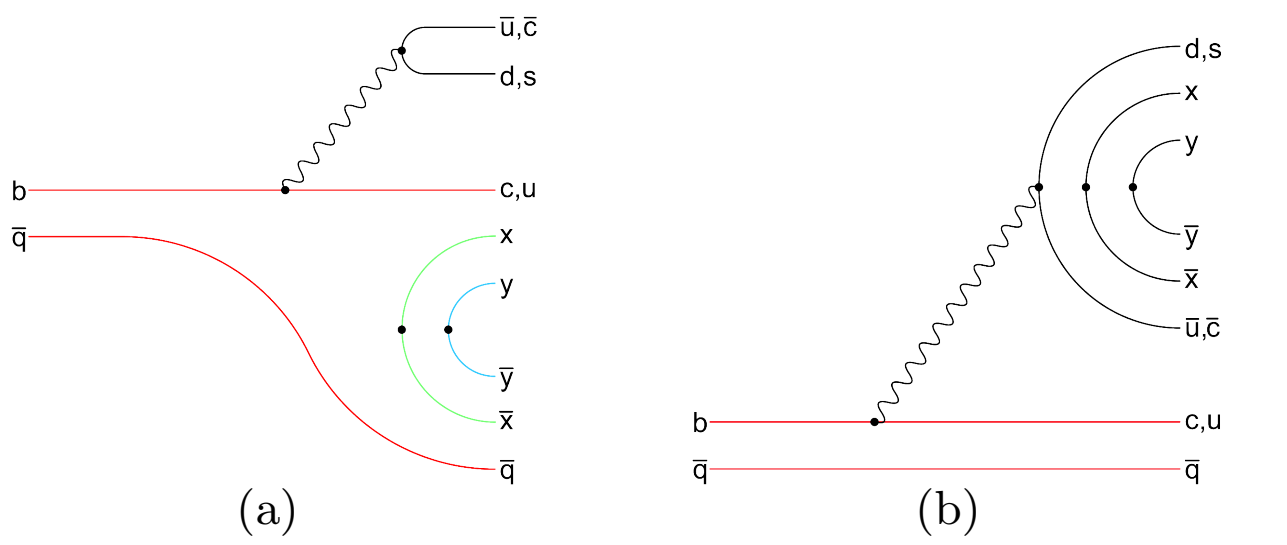

**Figure 17.12.7.** Quark diagrams for three-body baryonic $B$ decay $\overline{B} \to \mathfrak{B}_1\overline{\mathfrak{B}}_2 M$ corresponding to type 1, factorizable external $W$-emission contributions.

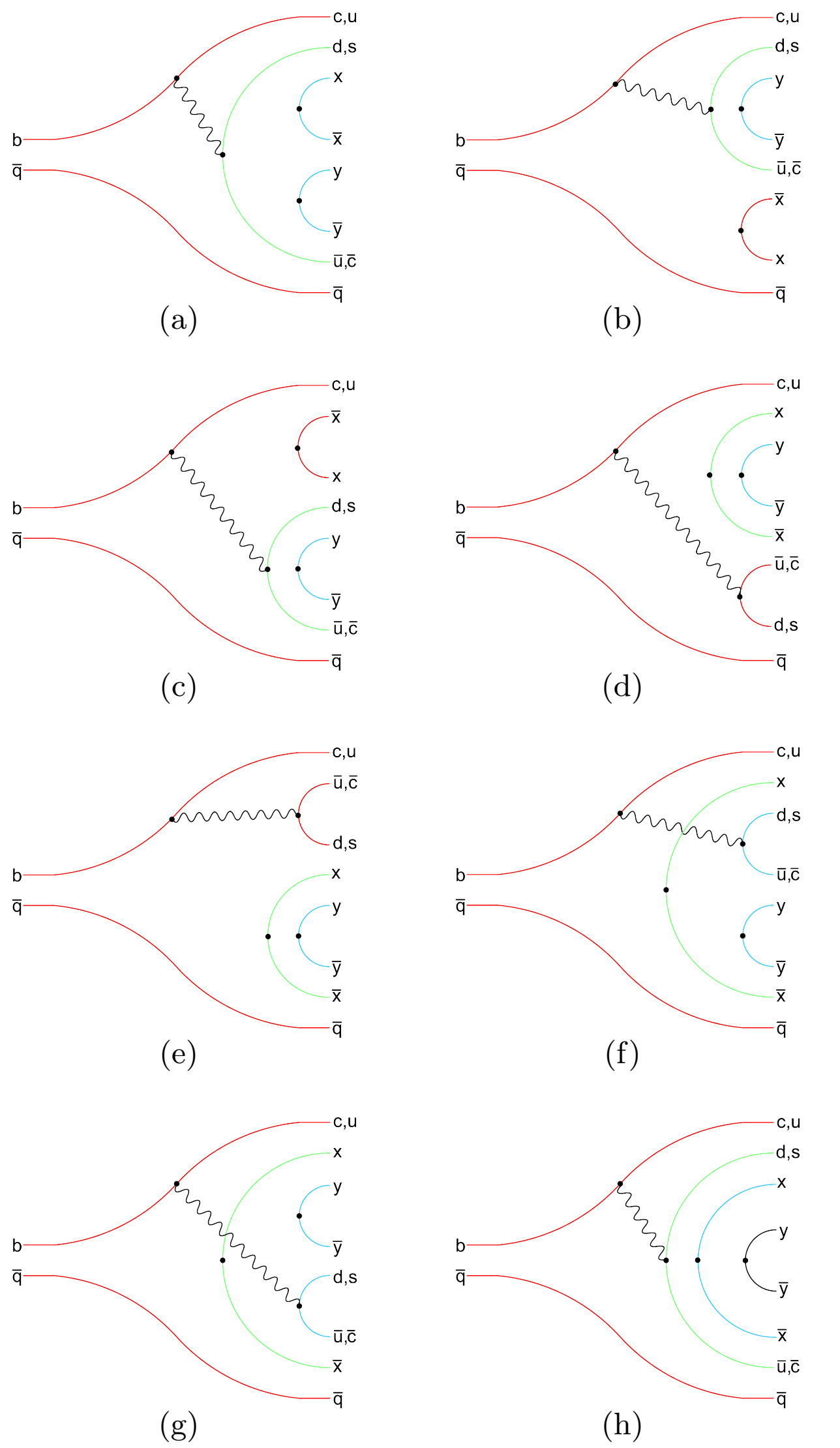

**Figure 17.12.8.** Spectator diagrams of type 2, factorizable (d,e) and non-factorizable (a–c,f–h) internal $W$-emission, for three-body baryonic $B$ decays $\overline{B} \to \mathfrak{B}_1\overline{\mathfrak{B}}_2 M$.

In the Feynman diagrams presented in the figures, the $q\overline{q}$ pairs are shown detached; they can be produced by soft or hard gluons attached to any other quark line in the diagram, and usually involve more than one gluon to accomplish color matching for color neutral mesons and baryons.

It should be stressed that among the internal $W$-emission diagrams, Figs 17.12.8d and e (where a red $q\overline{q}$ pair is created by the $W$) are color suppressed while the remaining diagrams in Fig. 17.12.8 are not, since the baryon wave function is antisymmetric in color indices (Cheng and Yang, 2002a). For example, $B^- \to J/\psi \Lambda\overline{p}$ proceeds via Fig. 17.12.8e, while $\overline{B}^0 \to \Sigma_c^0\overline{p}\pi^+$ receives contributions predominantly from Figs 17.12.8a and b. The experimental observation that $J/\psi \Lambda\overline{p}$ is suppressed by one order of magnitude is due to the color suppression for Fig. 17.12.8e and non-suppression for Figs 17.12.8a and b. The decay to $\Sigma_c^{++}\overline{p}\pi^-$ can also proceed through the (non-suppressed) external $W$-emission Fig. 17.12.7a, and has a higher branching fraction than the internal $W$-emission processes for $\Sigma_c^0\overline{p}\pi^+$. This may be explained by simple color counting: all three colors are possible in color-allowed processes (type 1), only one color in fully color-suppressed processes (type 2d,e) and two colors in the unsuppressed processes (type 2a–c,f–h).

Neglecting the factorizable annihilation contributions, which are helicity suppressed, the factorizable contributions to three-body decays consist of two parts: (i) the transition process with meson emission, $\langle M|(\overline{q}_3 q_2)|0\rangle \times \langle \mathfrak{B}_1\overline{\mathfrak{B}}_2|(\overline{q}_1 b)|\overline{B}\rangle$ where $(\bar{q}_i q_j) \equiv \overline{q}_i\gamma_\mu(1-\gamma_5)q_j$ and $M$ denotes a meson, and (ii) the current-induced process in association with a $B$ to meson transition, $\langle \mathfrak{B}_1\overline{\mathfrak{B}}_2|(\overline{q}_1 q_2)|0\rangle \times \langle M|(\overline{q}_3 b)|\overline{B}\rangle$. The two-body matrix element $\langle \mathfrak{B}_1\overline{\mathfrak{B}}_2|(\overline{q}_1 q_2)|0\rangle$ in the latter process can be either related to some measurable quantities, or calculated using the quark model. Note that while the form factors in the matrix elements $\langle \mathfrak{B}_1\overline{\mathfrak{B}}_2|(\overline{q}_1 q_2)|0\rangle$ and $\langle M|(\overline{q}_3 b)|\overline{B}\rangle$ depend on the dibaryon invariant mass squared $t = (p_1 + p_2)^2$, form factors in $\langle \mathfrak{B}_1(p_1)\overline{\mathfrak{B}}_2(p_2)|(\overline{q}_1 b)|\overline{B}(p_B)\rangle$ are functions not only of $t$ but also of one of the other Mandelstam variables $s = (p_B - p_1)^2$ or $u = (p_B - p_2)^2$. The current-induced contribution to three-body baryonic $B$ decays has been discussed in various publications, *e.g.* Chua, Hou, and Tsai (2002a). By contrast, it is difficult to evaluate the three-body matrix element in the transition process and in this case one can appeal to the pole model (Cheng and Yang, 2002b). Instead of showing various model predictions for three-body decays, we summarize in Table 17.12.7 the references related to the study of doubly-charmed, singly-charmed, and charmless three-body baryonic $B$ decays. The interested reader is referred to the original work for more details.

**Table 17.12.7.** References for the theoretical studies of doubly-charmed, singly-charmed, and charmless three-body baryonic $B$ decays.

| Modes | Reference |
|---|---|
| *Doubly charmed:* | |
| $\Lambda_c\bar{\Lambda}_c\bar{K}^0$, $\Lambda_c\bar{\Lambda}_c K^-$ | Cheng, Chua, and Hsiao (2009) |
| *Singly charmed:* | |
| $n\bar{p}D^{(*)+,0}$, $\Lambda\bar{p}D^{(*)+,0}$, $\Sigma^0\bar{p}D^{(*)+,0}$ | Chua, Hou, and Tsai (2002b) |
| $\Sigma^-\bar{n}D^{(*)+,0}$, $n\bar{n}D^{(*)0}$, $\Lambda\bar{n}D^{(*)0}$<br>$\Sigma^+\bar{p}D^{(*)0}$, $\Xi^-\bar{\Sigma}^-D^{(*)0}$, $\Xi^-\bar{\Sigma}^0D^{(*)0}$<br>$\Xi^-\bar{\Lambda}D^{(*)0}$, $\Xi^0\bar{\Sigma}^+D^{(*)0}$<br>$n\bar{p}J/\psi$, $\Lambda\bar{p}J/\psi$, $\Xi^-\bar{\Sigma}^0J/\psi$, $\Xi^0\bar{\Sigma}^+J/\psi$<br>$n\bar{n}J/\psi$, $\Lambda\bar{n}J/\psi$, $\Xi^0\bar{\Sigma}^0J/\psi$, $\Xi^-\bar{\Sigma}^+J/\psi$ | Chen, Cheng, Geng, and Hsiao (2008) |
| $\Lambda_c^+\bar{p}\pi^-$, $\Sigma_c^{++}\bar{p}\pi^-$, $\Sigma_c^0\bar{p}\pi^+(\pi^0)$ | Cheng and Yang (2003) |
| *Charmless (tree):* | |
| $n\bar{p}\pi^+$, $n\bar{p}\rho^+$, $p\bar{n}\pi^-$, $p\bar{n}\rho^-$ | Cheng and Yang (2002a) |
| $\Sigma^-\bar{\Lambda}\pi^+$, $\Xi^-\bar{\Xi}^0\pi^+$, $\Sigma^0\bar{\Sigma}^-\pi^+$, $p\bar{p}\pi^-$ | Chua and Hou (2003) |
| *Charmless (penguin):* | |
| $p\bar{p}K^-$, $p\bar{p}K^{*-}$, $p\bar{n}K^-$, $p\bar{n}K^{*-}$<br>$p\bar{p}\bar{K}^0$, $p\bar{p}\bar{K}^{*0}$, $n\bar{n}K^-$, $n\bar{n}K^{*-}$<br>$\Lambda\bar{p}\pi^+$, $\Lambda\bar{p}\rho^+$, $\Sigma^0\bar{p}\pi^+$, $\Sigma^0\bar{p}\rho^+$<br>$\Sigma^-\bar{n}\pi^+$, $\Sigma^-\bar{n}\rho^+$, $\Lambda\bar{p}\eta'$ | Cheng and Yang (2002a) |

### 17.12.3.2 Rare decays

The first charmless baryonic $B$ decay observed was $B^+ \to p\bar{p}K^+$, in an analysis of a 29.4 fb$^{-1}$ data sample (Abe, 2002f). One unexpected feature of this rare decay process is that the observed mass distribution of the baryon-antibaryon pair is peaked near threshold as shown in Fig. 17.12.9. To ensure that the measured events are genuine non-$b \to c$ signals, the regions $2.850 < M(p\bar{p}) < 3.128\,\mathrm{GeV}/c^2$ and $3.315 < M(p\bar{p}) < 3.735\,\mathrm{GeV}/c^2$ are excluded to remove background from modes with $\eta_c$ and $J/\psi$ mesons, and $\psi'$, $\chi_{c0}$, and $\chi_{c1}$ mesons, respectively. The mass distribution of vetoed events can be found in the inset plot of Fig. 17.12.9, where a $J/\psi$ peak can be clearly identified. Since the efficiency of particle identification varies with respect to the particle's momentum, the overall reconstruction efficiency is dependent on the mass of the baryon-antibaryon system. The partial branching fractions in bins of baryon-antibaryon mass are then summed to obtain the total branching fraction.

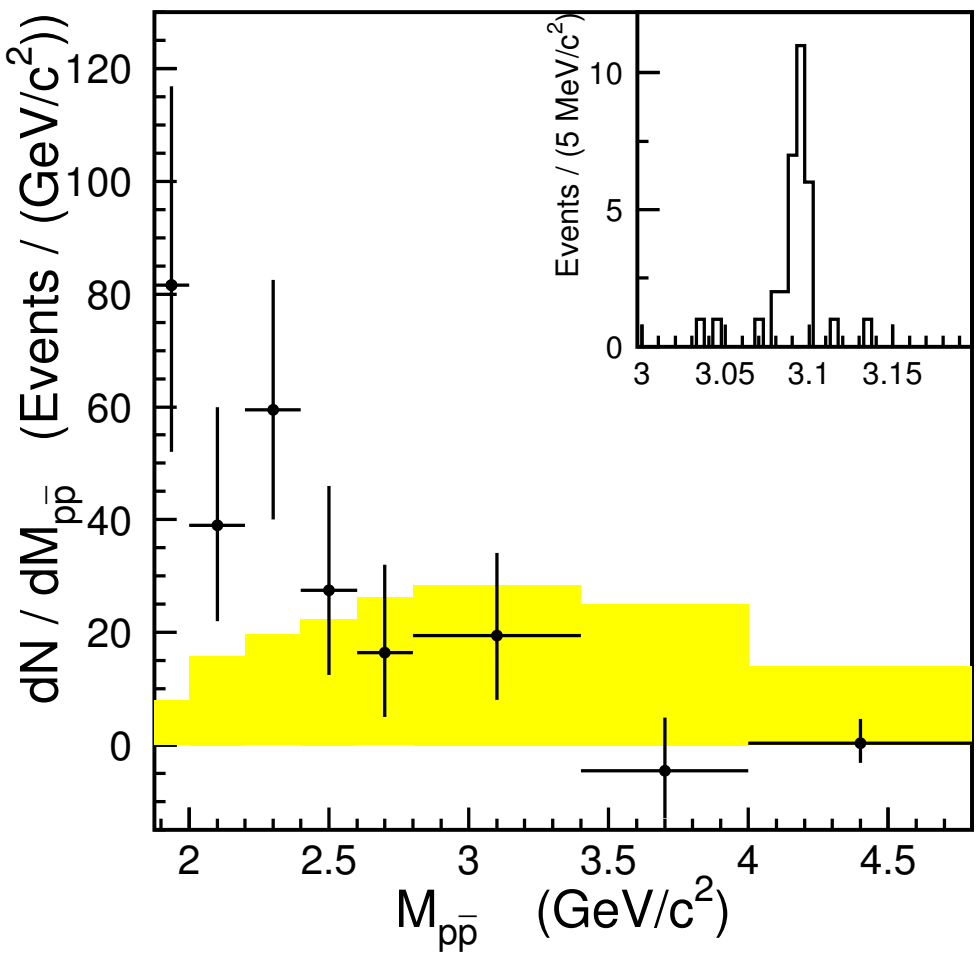

**Figure 17.12.9.** The fitted yield from Belle (Abe, 2002f) divided by the bin size for $B^+ \to p\bar{p}K^+$ as a function of $p\bar{p}$ mass. A charmonium veto is applied. The distribution from non-resonant $B^+ \to p\bar{p}K^+$ MC simulation is superimposed (shaded). The inset shows the $p\bar{p}$ mass distribution for the $J/\psi K^+$ signal region.

The measured $B^+ \to p\bar{p}K^+$ branching fraction is $\sim 4\times10^{-6}$. The decay is difficult to observe due to the large background from continuum events. A good particle identification system is a key for this analysis since both $p$ and $K^+$ should be positively identified in order to reject the background. Another important point is that a more sophisticated pattern recognition method based on event shape information was adopted to discriminate the more isotropic $B$ events from the jet-like continuum events.

Following this first observation, many other three-body charmless baryonic $B$ decays have been found: $p\bar{\Lambda}\pi^-$, $p\bar{\Lambda}\pi^0$, $p\bar{p}\pi^+$, $p\bar{p}K^0$, $p\bar{p}K^{*0}$, $p\bar{p}K^{*+}$, $\Lambda\bar{\Lambda}K^+$, $\Lambda\bar{\Lambda}K^0$, and $\Lambda\bar{\Lambda}K^{*0}$. Except for $B^+ \to p\bar{\Lambda}\pi^0$, all these modes are reconstructed entirely from charged particles in the final state, *i.e.* $\Lambda \to p\pi^-$, $K_S^0 \to \pi^+\pi^-$, $K^{*0} \to K^+\pi^-$, and $K^{*+} \to K_S^0\pi^+$. The signal shape can be well described by a single Gaussian in $m_{\mathrm{ES}}$ and a sum of two Gaussians in $\Delta E$. In the fit to determine signal yield with the $\Delta E$ spectrum, there are feed-across events between similar $B$ decays. For example, $B^+ \to p\bar{p}K^+$ events can form a bump at $-0.05\,\mathrm{GeV}$ in $\Delta E$ in the study of $B^+ \to p\bar{p}\pi^+$ when the $K^+$ is misidentified as a $\pi^+$. There are also feed-down events from similar decays with higher multiplicity, *e.g.* $B^0 \to p\bar{p}K^{*0}$ can form a bump below $-0.01\,\mathrm{GeV}$ in $\Delta E$ in the study of $B^+ \to p\bar{p}K^+$. These structures are useful as a sanity check for the measured branching fractions of related modes. The measured branching fractions of the above modes are all $\sim 10^{-6}$, and are summarized in Table 17.12.8.

### 17.12.3.3 Threshold enhancement

Many of the abovementioned channels also have the special feature that the measured mass spectrum of the baryon-antibaryon pair peaks near threshold. Figure 17.12.10 shows the differential branching fractions in bins of the baryon-antibaryon invariant mass for some representative decays: $B^+ \to p\overline{p}\pi^+$, presumed to proceed via the $b \to u$ tree process, and $B^+ \to p\overline{p}K^+$ and $B^0 \to p\overline{\Lambda}\pi^-$, presumably $b \to s$ strong penguin modes. They will be further discussed below in Section 17.12.3.5 on angular correlations.

**Table 17.12.8.** Branching fractions of observed decays of $B$ mesons to charmless baryons plus charmless mesons. Decays via intermediate resonances are marked with •.

| Decay | BABAR | Belle | Average |
|---|---|---|---|
| $(\mathcal{B}:\ 10^{-6})$ | (Aubert, 2009w) | (Wang, 2003) | |
| $B^0 \to \overline{\Lambda} p \pi^-$ | $3.07 \pm 0.31 \pm 0.23$ | $3.97^{+1.00}_{-0.80} \pm 0.56$ | $3.18 \pm 0.36$ |
| $B^0 \to \overline{\Lambda} p K^-$ | | $< 0.82$ | |
| $B^0 \to \overline{\Sigma}^0 p \pi^-$ | | $< 0.38$ | |
| $(\mathcal{B}:\ 10^{-6})$ | (Aubert, 2007k) | (Chen, 2008a) | |
| $B^0 \to p\overline{p} K^0$ | $3.0 \pm 0.5 \pm 0.3$ | $2.51^{+0.35}_{-0.29} \pm 0.21$ | $2.66 \pm 0.32$ |
| $B^0 \to p\overline{p} K^{*0}$ | $1.5 \pm 0.5 \pm 0.4$ | $1.18^{+0.29}_{-0.25} \pm 0.11$ | $1.23 \pm 0.27$ |
| $B^+ \to p\overline{p} K^{*+}$ | $5.3 \pm 1.5 \pm 1.3$ | $3.38^{+0.73}_{-0.60}$ | $3.57 \pm 0.63$ |
| $(\mathcal{B}:\ 10^{-6})$ | (Aubert, 2005p) | (Wei, 2008b) | |
| $B^+ \to p\overline{p} K^+$ | $6.7 \pm 0.5 \pm 0.4$ | $5.00^{+0.24}_{-0.22} \pm 0.32$ | $5.47 \pm 0.34$ |
| $(\mathcal{B}:\ 10^{-6})$ | (Aubert, 2007k) | (Wei, 2008b) | |
| $B^+ \to p\overline{p} \pi^+$ | $1.7 \pm 0.3 \pm 0.3$ | $1.57^{+0.17}_{-0.15} \pm 0.12$ | $1.59 \pm 0.15$ |
| $(\mathcal{B}:\ 10^{-6})$ | | (Chang, 2009) | |
| $B^+ \to \Lambda\overline{\Lambda}\pi^+$ | | $< 0.94$ | |
| $B^0 \to \Lambda\overline{\Lambda} K^0$ | | $4.76^{+0.84}_{-0.65} \pm 0.61$ | |
| $B^0 \to \Lambda\overline{\Lambda} K^{*0}$ | | $2.46^{+0.87}_{-0.72} \pm 0.34$ | |
| $B^+ \to \Lambda\overline{\Lambda} K^+$ | | $3.38^{+0.36}_{-0.41} \pm 0.41$ | |
| $B^+ \to \Lambda\overline{\Lambda} K^{*+}$ | | $2.19^{+1.13}_{-0.88} \pm 0.33$ | |
| $(\mathcal{B}:\ 10^{-6})$ | | (Chen, 2009) | |
| $B^+ \to \overline{\Lambda} p \pi^- \pi^+$ | | $5.92^{+0.88}_{-0.84} \pm 0.69$ | |
| • $B^+ \to \overline{\Lambda} p \rho^0$ | | $4.78^{+0.67}_{-0.64} \pm 0.60$ | |
| • $B^+ \to \overline{\Lambda} p f_2(1270)$ | | $2.03^{+0.77}_{-0.72} \pm 0.27$ | |

Threshold enhancement has also been found in the $b \to c$ process, although the effect is not as pronounced or dominant as in the charmless case. When the available energy is limited to a small amount, say $\sim 200\,\text{MeV}$, there is no visible peaking feature. Figure 17.12.11 shows the baryon-antibaryon mass for $\overline{B}^0 \to p\overline{p}D^0$, $B^+ \to J/\psi\, p\overline{\Lambda}$, $B^+ \to p\overline{\Lambda}_c\pi^+$, $B^+ \to \Lambda_c\overline{\Lambda}_c K^+$, $\overline{B}^0 \to \Lambda_c\overline{\Lambda}K^-$, and $\overline{B}^0 \to \Sigma_c^0\overline{p}\pi^+$. There are clear threshold enhancement peaks in $\overline{B}^0 \to p\overline{p}D^0$, $B^+ \to p\overline{\Lambda}_c\pi^+$, and $\overline{B}^0 \to \Lambda_c\overline{\Lambda}K^-$ along with non-negligible phase space components. But not all three-body decays show the threshold behavior. For $B^+ \to J/\psi\, p\overline{\Lambda}$ and $B^+ \to \Lambda_c\overline{\Lambda}_c K^+$ (Fig. 17.12.11b and d), where the available phase space is small, there is no clear threshold peak visible, but still a slight enhancement can be detected. The threshold peaking effect is totally absent in $\overline{B}^0 \to \Sigma_c^0\overline{p}\pi^+$ (Fig. 17.12.11f).

The same threshold behavior has also been observed in the baryonic $J/\psi$ decays $J/\psi \to \gamma p\overline{p}$ (Bai et al., 2003) and $J/\psi \to K^- p\overline{\Lambda}$ (Ablikim et al., 2004b). However, it is often argued in the literature (see references in Table 17.12.9 under the item “Final-state interactions”) that threshold enhancement in $J/\psi \to \gamma p\overline{p}$ can be explained in terms

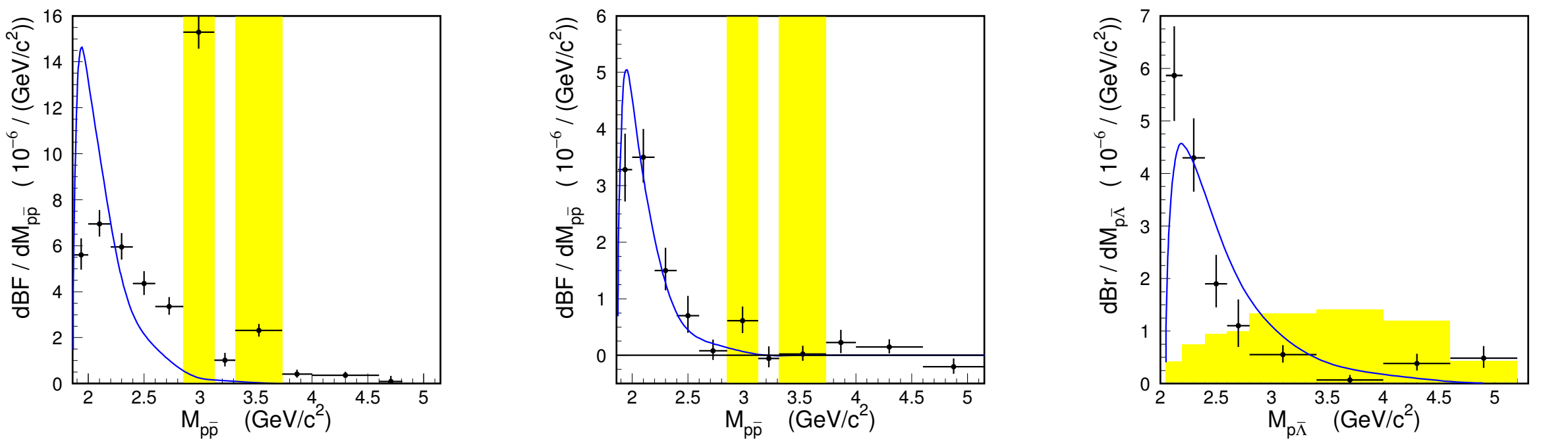

(a) $B^+ \to p\bar{p}K^+$ (Wei, 2008b) (b) $B^+ \to p\bar{p}\pi^+$ (Wei, 2008b) (c) $B^0 \to p\bar{\Lambda}\pi^-$ (Wang, 2007b)

**Figure 17.12.10.** Differential branching fractions in bins of the baryon-antibaryon mass for three representative modes. The two charm veto regions are shown shaded in (a) and (b); in (c), the shaded histogram shows the phase space distribution, which is distinctly different from the measured distribution. The curves indicate the theoretical predictions (Geng and Hsiao, 2006) normalized to the measured charmless branching fractions.

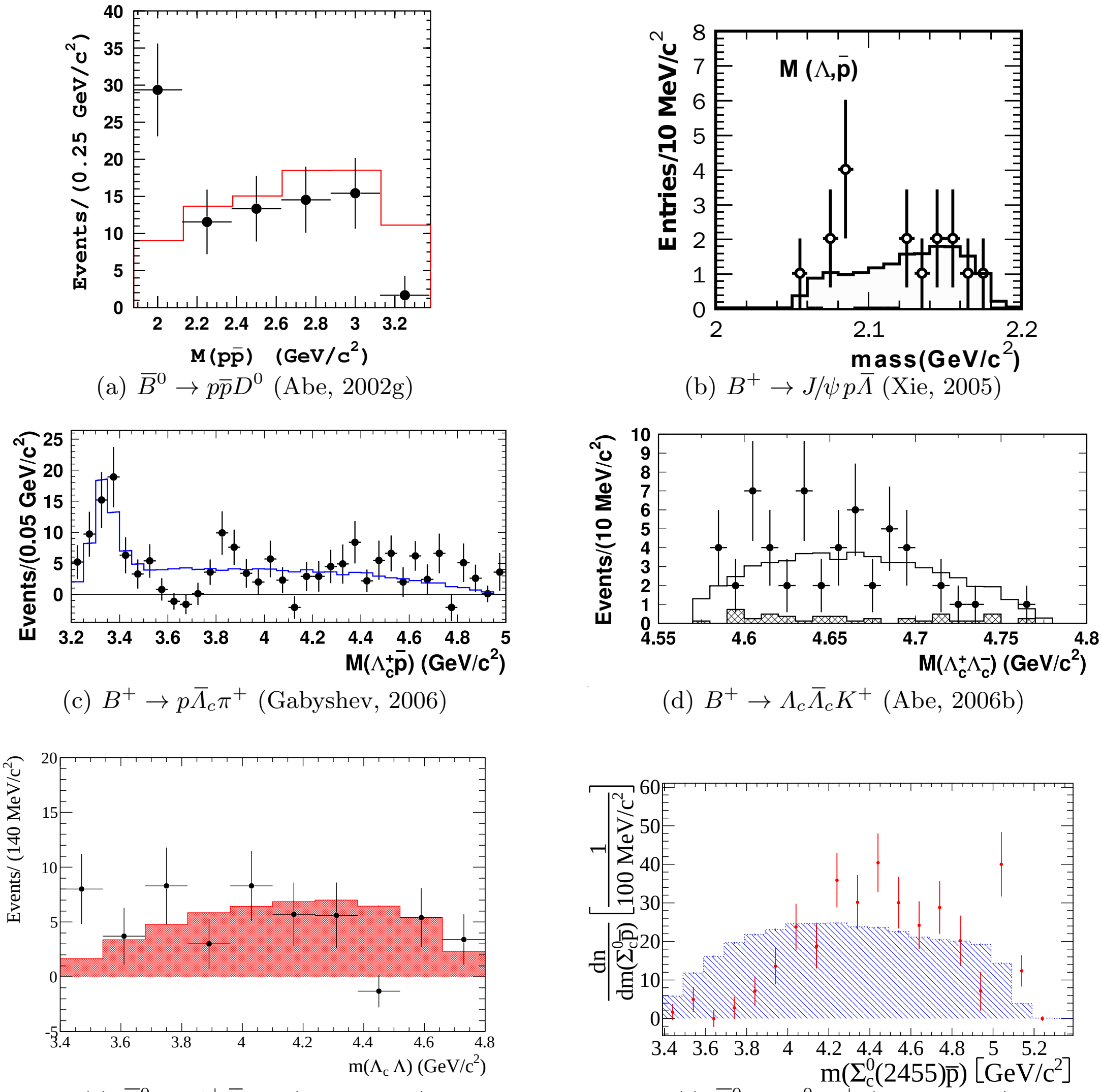

(a) $\overline{B}^0 \to p\bar{p}D^0$ (Abe, 2002g) (b) $B^+ \to J/\psi p\bar{\Lambda}$ (Xie, 2005)

(c) $B^+ \to p\bar{\Lambda}_c\pi^+$ (Gabyshev, 2006) (d) $B^+ \to \Lambda_c\bar{\Lambda}_cK^+$ (Abe, 2006b)

(e) $\overline{B}^0 \to \Lambda_c^+\bar{\Lambda}K^-$ (Lees, 2011f) (f) $\overline{B}^0 \to \Sigma_c^0\bar{p}\pi^+$ (Lees, 2013h)

**Figure 17.12.11.** Signal yields (points with error bars) as a function of the baryon antibaryon mass for various modes.

**Table 17.12.9.** Various interpretations of threshold effects in baryonic $B$ decays.

| Model | Description |
|---|---|
| 1. Pole model | The absence of $1/m_b^2$ suppression of the propagator in pole diagrams and the presence of a $\Lambda_b$ or $\Xi_b$ intermediate state lead to threshold enhancement (Cheng and Yang, 2002a; Hou and Soni, 2001). |
| 2. $\mathfrak{B}_1\overline{\mathfrak{B}}_2$ bound state | The $p\overline{p}$ pair forms a bound state such as baryonium or $X(1835)$ (Datta and O'Donnell, 2003a; Rosner, 2003). |
| 3. Glueball | An isoscalar $p\overline{p}$ pair forms a gluonic state (Chua, Hou, and Tsai, 2002a; Rosner, 2003). |
| 4. Final-state interactions | Enhancement due to final-state interactions between the baryon pair (Haidenbauer, Meissner, and Sibirtsev, 2006; Kerbikov, Stavinsky, and Fedotov, 2004; Laporta, 2007; Sibirtsev, Haidenbauer, Krewald, Meissner, and Thomas, 2005). |
| 5. Baryon form factors | In some approaches such as factorization, fragmentation, etc., the amplitude is governed by dibaryon form factors which fall off rapidly with dibaryon invariant mass as suggested by QCD counting rules (Chua and Hou, 2003; Chua, Hou, and Tsai, 2002a). |

of final-state interactions between the baryon pair, while the same threshold effect in baryonic $B$ decays can be understood in terms of the simple short-distance picture depicted in Fig. 17.12.12.

The so-called "threshold effect" indicates that the $B$ meson prefers to decay into a baryon-antibaryon pair with low invariant mass accompanied by a fast recoil meson. This peaking behavior was quite unexpected, and has lead to various speculations about possible mechanisms, such as a glueball bound state formed by gluons, a baryonium bound state of the baryon-antibaryon pair, etc. Threshold enhancement was first proposed by Hou and Soni (2001), motivated by the CLEO measurement of $B \to D^* p\overline{n}$ and $D^* p\overline{p}\pi$ (Anderson et al., 2001). They argued that in order to enhance baryonic $B$ decay, one has to reduce the energy release and at the same time allow for baryonic ingredients to be present in the final state. In other words, they conjectured that enhanced baryon production is favored by reduced energy release on the baryon side. This is indeed the near threshold effect mentioned above. Hence, the smallness of the two-body baryonic decay $B \to \mathfrak{B}_1\overline{\mathfrak{B}}_2$ has to do with its large energy release.

A heuristic approach to understanding the threshold enhancement in three-body decays can be obtained by looking at the Feynman diagrams in Fig. 17.12.8. In all these diagrams, the weak decay of a $B$ meson produces two quarks and two antiquarks including the spectator. Baryons are formed with additional $q\overline{q}$ pairs produced by strong interaction from the vacuum. The initial arrangement of the primary four quarks determines whether the baryon-antibaryon pair is close in phase space (*i.e.*, at mass threshold) or distant. If the diagram can be converted into a $B \to MM$ diagram by omitting the extra $q\overline{q}$ pairs from the vacuum (as in Figs 17.12.8d and e, the factorizable color-suppressed diagrams) we observe enhancement, while diagrams where the same process leaves a diquark-antidiquark pair would produce no enhancement at threshold. The latter class includes $\overline{B}^0 \to \Sigma_c^0\overline{p}\pi^+$ (see Fig. 17.12.11f) proceeding through diagrams Fig. 17.12.8a and b.

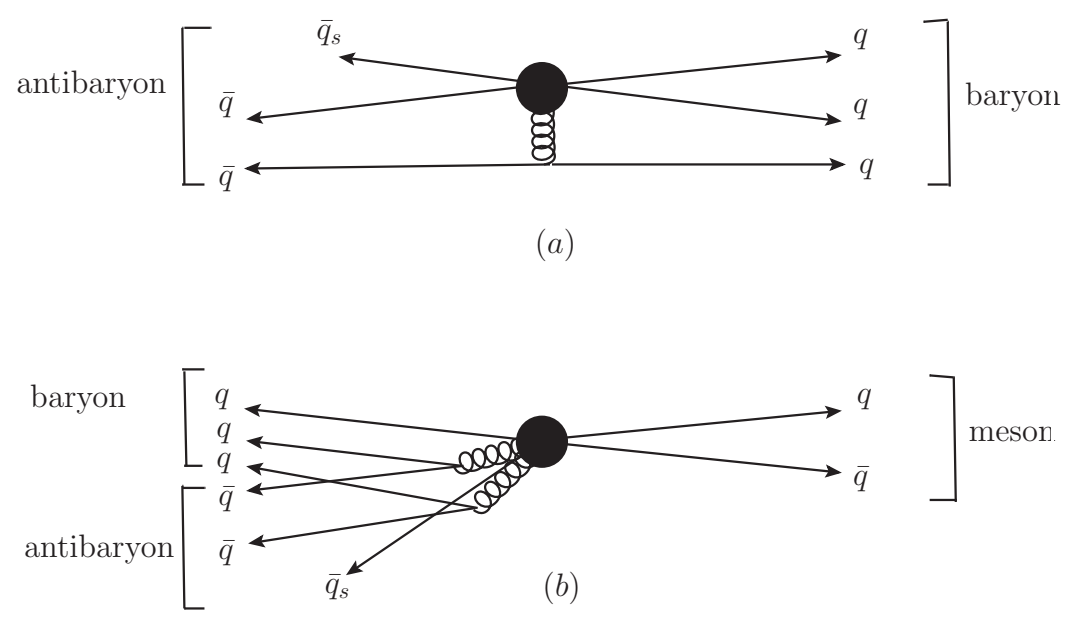

**Figure 17.12.12.** Short-distance picture in terms of quarks and antiquarks for (a) two-body baryonic decay and (b) three-body baryonic decay. The slow spectator antiquark is denoted by the short line $\overline{q}_s$.

This idea is illustrated in Fig. 17.12.13. The diagram (a) would produce two mesons; with the extra $q\overline{q}$ pairs in (b) one meson transforms into a $\mathfrak{B}_1\overline{\mathfrak{B}}_2$ pair ($\Lambda_c^+\overline{p}$) with preferentially low invariant mass. This may be related to a meson pole, as described below. The diagram (c), however, produces a diquark-antidiquark pair, which is transformed by the extra $q\overline{q}$ pairs into a $\mathfrak{B}_1 M \overline{\mathfrak{B}}_2$ state $\Sigma_c^0\overline{p}\pi^+$. Here, no meson pole is possible, and no threshold enhancement is observed.

Of course, one has to understand the underlying origin of the threshold peaking effect. Threshold enhancement is closely linked to the behavior of baryon form factors which fall off sharply with $t$, the invariant mass squared of the dibaryon. While various theoretical ideas, summarized in Table 17.12.9, have been put forward to explain the low mass threshold enhancement, this effect can be understood in terms of a simple short-distance picture illustrated in Fig. 17.12.12 (Suzuki, 2007). To produce a baryon and an antibaryon in the two-body decay, one energetic $q\overline{q}$ pair must be emitted at high invariant mass, *i.e.*, by a hard gluon (high $q^2$). This hard gluon is far off mass shell and hence the two-body decay amplitude is suppressed by a factor of order $\alpha_S/q^2$. In three-body baryonic $B$ decays, a possible configuration is that the $\mathfrak{B}_1\overline{\mathfrak{B}}_2$ pair is emitted collinearly against the meson. The quark-antiquark pair emitted from a gluon is moving in nearly the same direction. Since this gluon is close to mass shell, the corresponding configuration is not subject to the short-distance suppression. This implies that the dibaryon pair tends to have a small invariant mass.

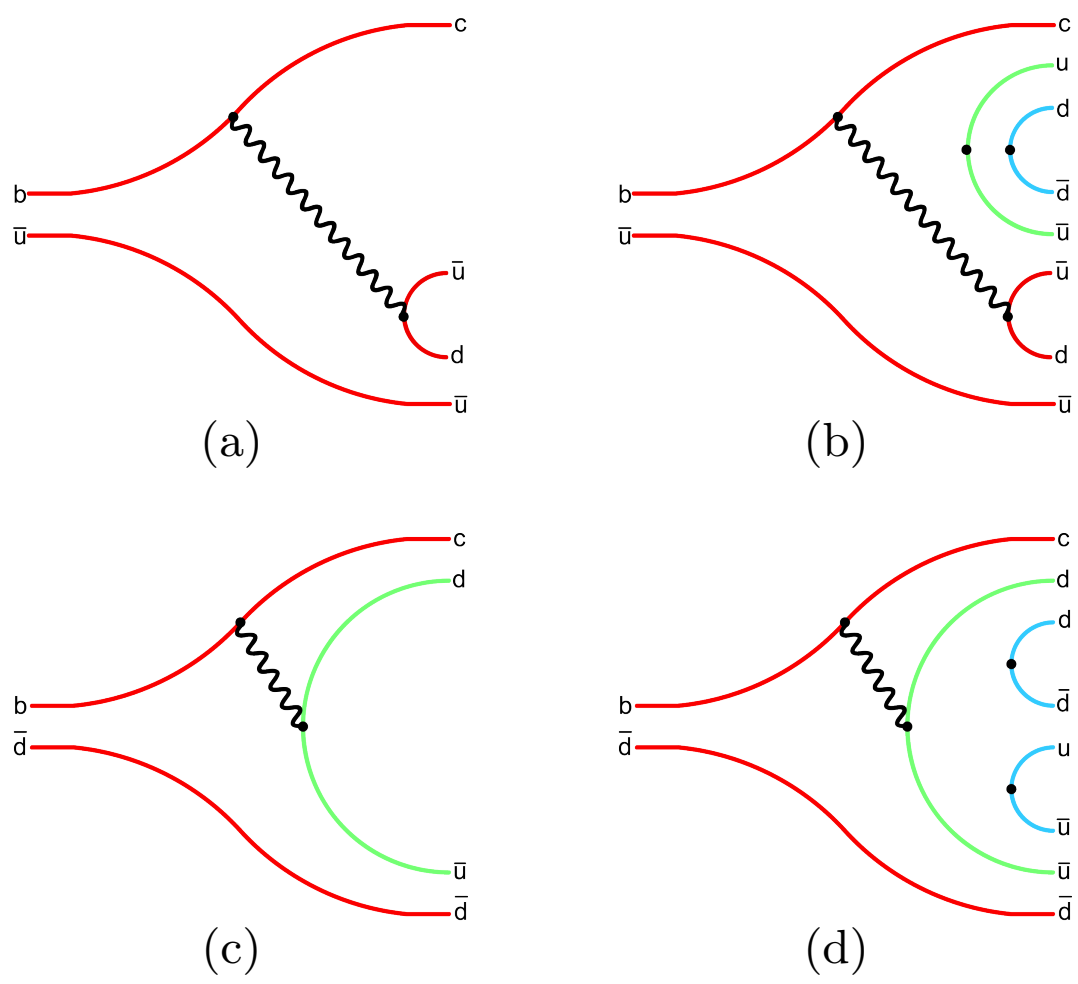

**Figure 17.12.13.** Spectator diagrams of type 2, illustrating the basic picture for a $\mathfrak{B}_1\overline{\mathfrak{B}}_2$ threshold enhancement: (b) shows one diagram for $B^- \to \Lambda_c^+ \overline{p}\pi^-$ with preferentially low $\mathfrak{B}_1\overline{\mathfrak{B}}_2$ mass related to the meson pair diagram (a), and (d) shows $\overline{B}^0 \to \Sigma_c^0 \overline{p}\pi^+$ with preferentially high $\mathfrak{B}_1\overline{\mathfrak{B}}_2$ mass related to the diquark pair diagram (c).

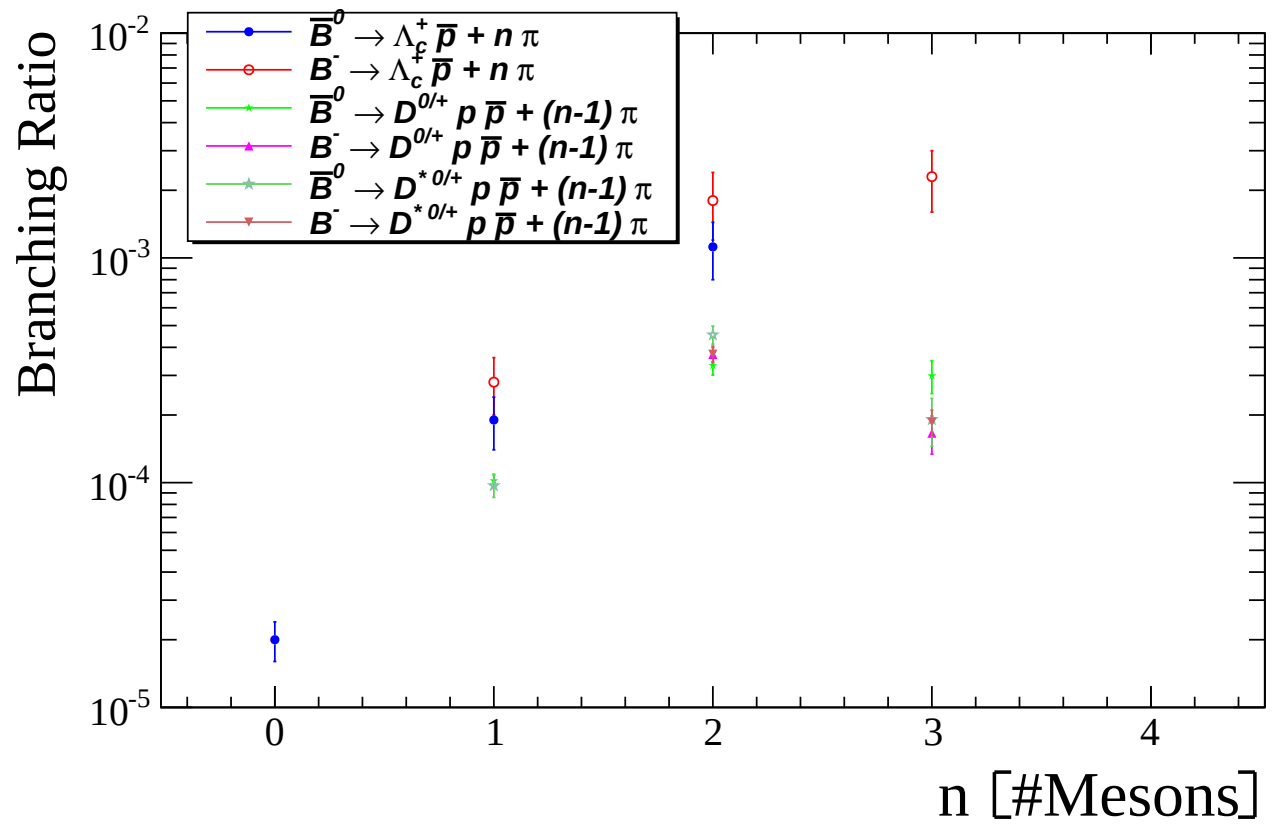

**Figure 17.12.14.** Multiplicities for decays $B \to \mathfrak{B}_1\overline{\mathfrak{B}}_2 + \mathrm{n}M$.

All present explanations for the baryon antibaryon threshold enhancement have in common that the partial rate increases at low values of the baryon-antibaryon invariant mass, while other regions of phase space are poorly populated. Decay channels which have a small phase space would be naturally suppressed through the phase space factor, but this is counteracted by the property of the matrix element to cluster in a small phase space volume anyway, resulting in no or a much smaller suppression of those channels. A first test of this idea has been performed by *BABAR* (Grünberg, 2012), looking for the decay $\overline{B}^0 \to \Lambda_c^+ \overline{p} p \overline{p}$, where two baryon-antibaryon pairs are produced within a small overall phase space region. However, no event of this type has been found, yielding the upper limit shown in Table 17.12.6.

### 17.12.3.4 Multiplicity

There is a noticeable hierarchy in decay rates: three-body decays have substantially larger rates than their two-body counterparts. Likewise, four-body decays are usually more frequent than the three-body ones. For example,

$$\begin{aligned}
\mathcal{B}(B^- \to p\overline{p}\pi^-) &\gg \mathcal{B}(\overline{B}^0 \to p\overline{p}),\\
\mathcal{B}(\overline{B}^0 \to \Lambda\overline{p}\pi^-) &\gg \mathcal{B}(B^- \to \Lambda\overline{p}),\\
\mathcal{B}(\overline{B}^0 \to \Lambda_c^+\overline{p}\pi^+\pi^-) &\gg \mathcal{B}(\overline{B}^0 \to \Lambda_c^+\overline{p}\pi^0)\\
&\gg \mathcal{B}(\overline{B}^0 \to \Lambda_c^+\overline{p}),\\
\mathcal{B}(\overline{B}^0 \to D^{*+}p\overline{p}\pi^-) &\gg \mathcal{B}(\overline{B}^0 \to D^{*0}p\overline{p}), \qquad (17.12.6)
\end{aligned}$$

as shown in Fig. 17.12.14. This phenomenon can be understood in terms of the aforementioned threshold effect, that is, the preference for the invariant mass of the dibaryon to be close to threshold. The configuration of the two-body decay $B \to \mathfrak{B}_1\overline{\mathfrak{B}}_2$ is not favorable since its invariant mass is $m_B$. In $B \to \mathfrak{B}_1\overline{\mathfrak{B}}_2 M$ decays, the effective mass of the baryon pair is reduced as the emitted meson can carry away energy. This explains why $\mathcal{B}(B \to \mathfrak{B}_1\overline{\mathfrak{B}}_2 M) \gg \mathcal{B}(B \to \mathfrak{B}_1\overline{\mathfrak{B}}_2)$. The same reasoning applies to decays with two or more mesons.

However, it is not always true that a larger rate for three-body decays can be ascribed to threshold enhancement. As an example, consider the three-body doubly-charmed baryonic decay $\overline{B} \to \Lambda_c\overline{\Lambda}_c\overline{K}$ which has been observed at the $B$ Factories with a branching fraction of order $10^{-3}$ (see Table 17.12.6). Since this mode is color-suppressed and has a very small phase space, the estimate is $\mathcal{B}(\overline{B} \to \Lambda_c\overline{\Lambda}_c\overline{K}) \sim 10^{-6}$ in naïve factorization. This is too small by two to three orders of magnitude compared to experiment. Possibilities for the enhancement of $\Lambda_c\overline{\Lambda}_c\overline{K}$ rates include final-state interactions and some resonances. There are two possible resonant states: a hidden-charm bound state $X_{c\bar{c}}$ with a mass near the $\Lambda_c\overline{\Lambda}_c$ threshold, $4.6 \sim 4.7$ GeV, and a $\Lambda_c\overline{K}$ resonance. Indeed, Belle has reported a peak, called the $X(4630)$, in the $e^+e^- \to \Lambda_c^+\overline{\Lambda}_c^-$ exclusive cross section (Pakhlova, 2008b; see also the discussion in Sections 21.4.6 and 18.3.5), while *BABAR* has found a resonance in the $\Lambda_c\overline{K}$ invariant mass distribution with mass $\sim 2930$ MeV (Aubert, 2008e; see also Section 19.4.1.3). It is therefore plausible that it is the resonant contribution rather than the threshold effect that renders $\mathcal{B}(\overline{B} \to \Lambda_c^+\overline{\Lambda}_c^-\overline{K}) > \mathcal{B}(\overline{B} \to \Lambda_c^+\overline{\Lambda}_c^-)$.

Also, as has already been pointed out, double charmed two-body decays are enhanced over single charm or charmless decays by the same mechanism: the baryon-antibaryon pair is closer to threshold in the former case. Much softer gluons are employed in double charmed decays. At least two hard gluons are needed for single charm or charmless two-body decays, and they are suppressed by factors of $\alpha_s^4$ relative to double-charm two-body decays as explained above in Section 17.12.2.2.

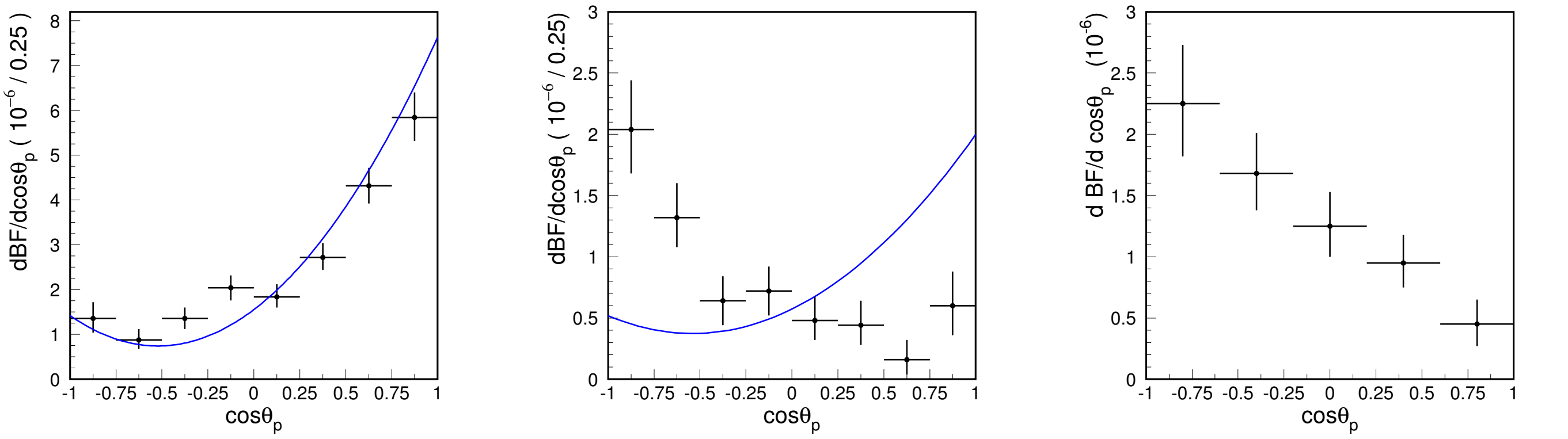

(a) $B^+ \to p\overline{p}K^+$ (Wei, 2008b) (b) $B^+ \to p\overline{p}\pi^+$ (Wei, 2008b) (c) $B^0 \to p\overline{\Lambda}\pi^-$ (Wang, 2007b)

**Figure 17.12.15.** Differential branching fractions as a function of $\cos\theta_p$. The curves show the model of Geng and Hsiao (2006), based on a preliminary version of the $B^+ \to p\overline{p}K^+$ results.

### 17.12.3.5 Baryon-meson angular correlations in three-body decays

The measurement of angular correlations of the outgoing meson in the dibaryon rest frame can provide further insight into the underlying mechanism for three-body decays. From the theoretical point of view, within pQCD it is expected that the meson in $\overline{B} \to \mathfrak{B}_1\overline{\mathfrak{B}}_2M$ decays has a stronger correlation with the antibaryon than with the baryon in the dibaryon rest frame. Hence, the opposite correlation effect seen in $B^- \to p\overline{p}K^-$ and $\Lambda\overline{p}\pi^-$ is astonishing and entirely unexpected.

*Experimental results*

After sufficient data was accumulated at the $B$ Factories, there was an effort to study the threshold region by investigating the angular distribution in the baryon-antibaryon rest frame (Chang, 2009; Wang, 2005, 2007b; Wei, 2008b). In these analyses, $\theta_p$ is defined as the angle between the (anti)proton direction and the oppositely charged meson direction in the baryon antibaryon rest frame for $B^+ \to p\overline{p}K^+$, $B^+ \to p\overline{p}\pi^+$, and $B^0 \to p\overline{\Lambda}\pi^-$. Figure 17.12.15 shows the differential branching fractions as a function of $\cos\theta_p$ for these representative modes with baryon-antibaryon mass $< 2.85\,\mathrm{GeV}/c^2$. These distributions are not symmetric and have some puzzling features. Since the proton and the antiproton move almost collinearly in the $B$ rest frame due to the threshold constraint, the peaking toward $\cos\theta_p = 1$ for $B^+ \to p\overline{p}K^+$ indicates that the baryon containing the spectator quark of the $B$ meson moves faster in the $B$ rest frame. This is opposite to the pQCD expectation for $b \to sg^*$ decays (see below). Similarly, most of the time the protons in $B^0 \to p\overline{\Lambda}\pi^-$ decays move faster in the $p\overline{\Lambda}$ system. An early attempt to account for the $B^+ \to p\overline{p}K^+$ data within pQCD, based on the preliminary result shown at the International Conference on High Energy Physics held in Beijing in 2004, predicted a similar correlation for $B^+ \to p\overline{p}\pi^+$ (Geng and Hsiao, 2006; see the curve in Fig. 17.12.15b). Instead, the opposite effect is seen. The difference between these modes may indicate a crucial difference between the $b \to s$ strong penguin and the $b \to u$ tree processes.

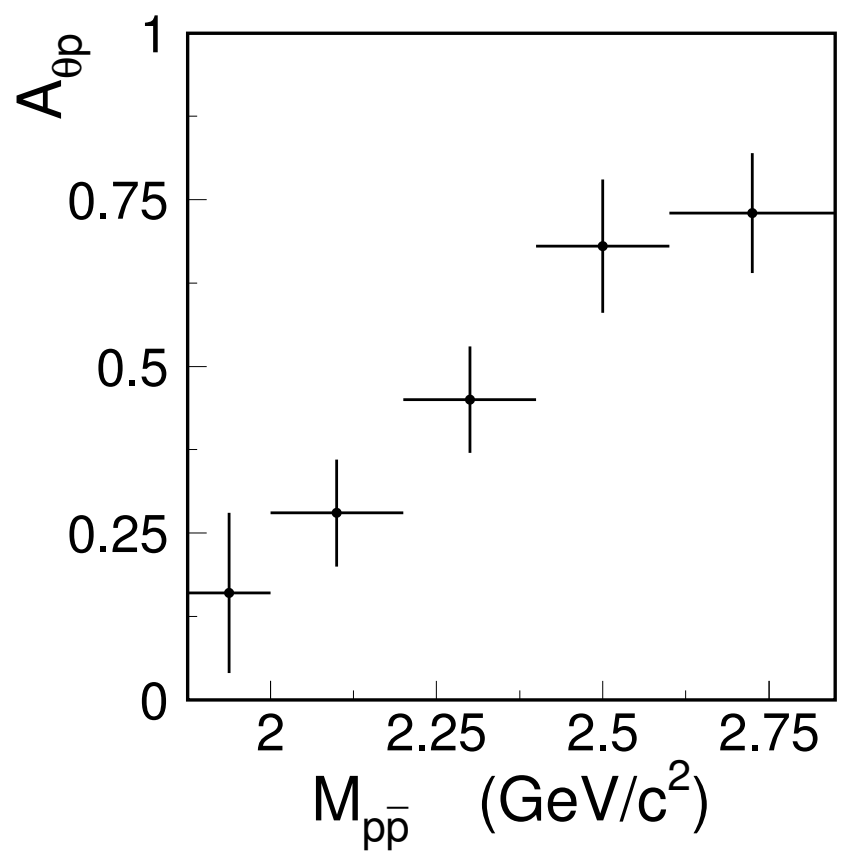

**Figure 17.12.16.** Measured angular asymmetries ($A_{\theta_p}$) as a function of $p\overline{p}$ mass near threshold for $B^+ \to p\overline{p}K^+$ from Belle (Wei, 2008b).

Using $449 \times 10^6$ $B\overline{B}$ pairs (Wei, 2008b), enough data is available for a detailed study of $B^+ \to p\overline{p}K^+$ signal events near threshold. The angular asymmetry

$$A_{\theta_p} = \frac{N_+ - N_-}{N_+ + N_-}, \qquad (17.12.7)$$

where $N_+$ and $N_-$ are the efficiency-corrected $B$ yields with $\cos\theta_p > 0$ and $\cos\theta_p < 0$ respectively, is shown as a function of $m_{p\overline{p}}$ in Fig. 17.12.16. The distribution is not flat, indicating that the relative contributions from differ-ent decay amplitudes are changing in this near-threshold mass range.

*$b \to sg^*$ and other models*

In the short-distance $b \to sg^*$ picture, it is expected that the antibaryon produced in penguin-dominated $\overline{B} \to \mathfrak{B}_1\overline{\mathfrak{B}}_2 M$ decays tends to emerge parallel to the outgoing meson, while the baryon moves antiparallel to the meson in the $\mathfrak{B}_1\overline{\mathfrak{B}}_2$ rest frame. This is also true for tree-dominated three-body decays. Intuitively, this can be understood in the following manner. Since in the $B$ rest frame

$$m_{12}^2 = m_1^2 + m_2^2 + 2(E_1E_2 - |\boldsymbol{p}_1||\boldsymbol{p}_2|\cos\theta_{12}), \quad (17.12.8)$$

threshold enhancement implies that the baryon pair $\mathfrak{B}_1$ and $\overline{\mathfrak{B}}_2$ tends to move collinearly in this frame, *i.e.* $\theta_{12} \to 0$. See Fig. 17.12.5(b) for a comparable penguin diagram, and Fig. 17.12.7a for a comparable three-body decay. From Fig. 17.12.7a we see that the $\mathfrak{B}_1$ is moving faster than $\overline{\mathfrak{B}}_2$ as the former picks up an energetic quark from the $b$ decay. When the system is boosted to the $\mathfrak{B}_1\overline{\mathfrak{B}}_2$ rest frame, $\overline{\mathfrak{B}}_2$ and $M$ are moving collinearly away from the $\mathfrak{B}_1$.

This picture has been tested and confirmed by the measurements of angular correlations in $B^- \to p\overline{p}\pi^-$ (see Fig. 17.12.15b) and $\Lambda_c^+\overline{p}\pi^-$ decays. (For the related $B^- \to \Lambda\overline{p}\gamma$ decay, see Section 17.12.4 and Fig. 17.12.21.) However, from the study of the polar angle distribution of the proton in the $p\overline{p}$ system of $B^- \to p\overline{p}K^-$, it was found by both *BABAR* (Aubert, 2005p) and Belle (Wang, 2004b) that there is a preference for $K^-$ to be collinear with the proton in the $p\overline{p}$ rest frame (see Fig. 17.12.15a, recalling that $\theta_p$ is the angle between $K^+$ and $\overline{p}$ or $K^-$ and $p$). This is against the theoretical prediction based on a short-distance $b \to sg^*$ picture. The fragmentation model by Rosner (2003) implies a large correlation between $K^-$ and $\overline{p}$ from the penguin annihilation diagram. However, this diagram is suppressed by a factor $1/m_b$ relative to the dominant diagram that leads to the opposite correlation. For a detailed discussion of this model on $B^- \to p\overline{p}K^-$, see Cheng (2006). This puzzle may indicate that (i) some long-distance effect enters and reverses the angular dependence, or (ii) the $p\overline{p}$ pair is produced from some intermediate states such as a baryonium, a $p\overline{p}$ bound state, or a glueball. For example, the angular puzzle for $p\overline{p}K^-$ can be resolved if $p\overline{p}$ is produced via a $^1S_0$ or $^3S_1$ baryonium state. However, the same flip mechanism will modify the correct $(1-\cos\theta_p)^2$ distribution for $p\overline{p}\pi^-$ to a wrong one, unless one assumes $D$- and $P$-waves for $p\overline{p}\pi^-$ (Suzuki, 2007). This is the first big surprise.

The second big surprise arises from the experimental findings by Belle (Wang, 2007b) that the $\Lambda$ particle is moving slower than the $\overline{p}$ in the decay of $\overline{B}^0 \to \Lambda\overline{p}\pi^+$ (see Fig. 17.12.15c which shows that $\Lambda$ moves collinearly with $\pi^+$ in the $\Lambda\overline{p}$ rest frame, which in turn implies that $\Lambda$ is moving slower than $\overline{p}$ in the $B$ rest frame). This violates the common idea for $b \to sg^*$ decay since the $\Lambda$ particle inherits the energetic $s$ quark from $b$ decay directly. It is naïvely expected that the pion has no preference for its correlation with $\Lambda$ or $\overline{p}$. The aforementioned baryonium mechanism does not work for this case.

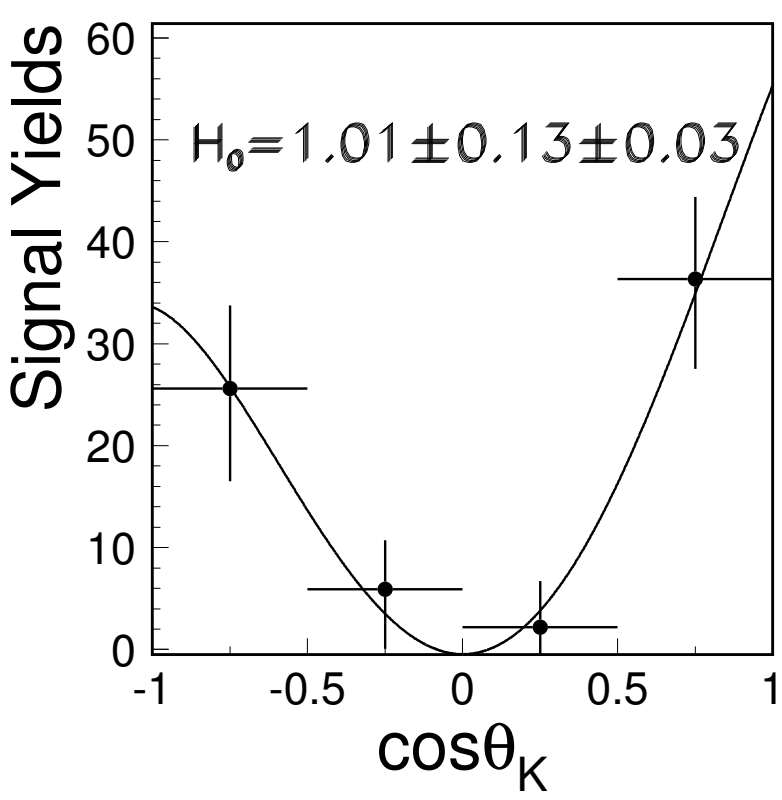

**Figure 17.12.17.** $B$ yield distributions from Belle (Chen, 2008a) as a function of $\cos\theta_K$ with a fit curve overlaid for $B^0 \to p\overline{p}K^{*0}$. The fraction of the signal in the helicity zero state is the fit parameter and is denoted by $H_0$. The asymmetry in the fit curve is due to detection efficiency: the underlying theoretical distribution is symmetric.

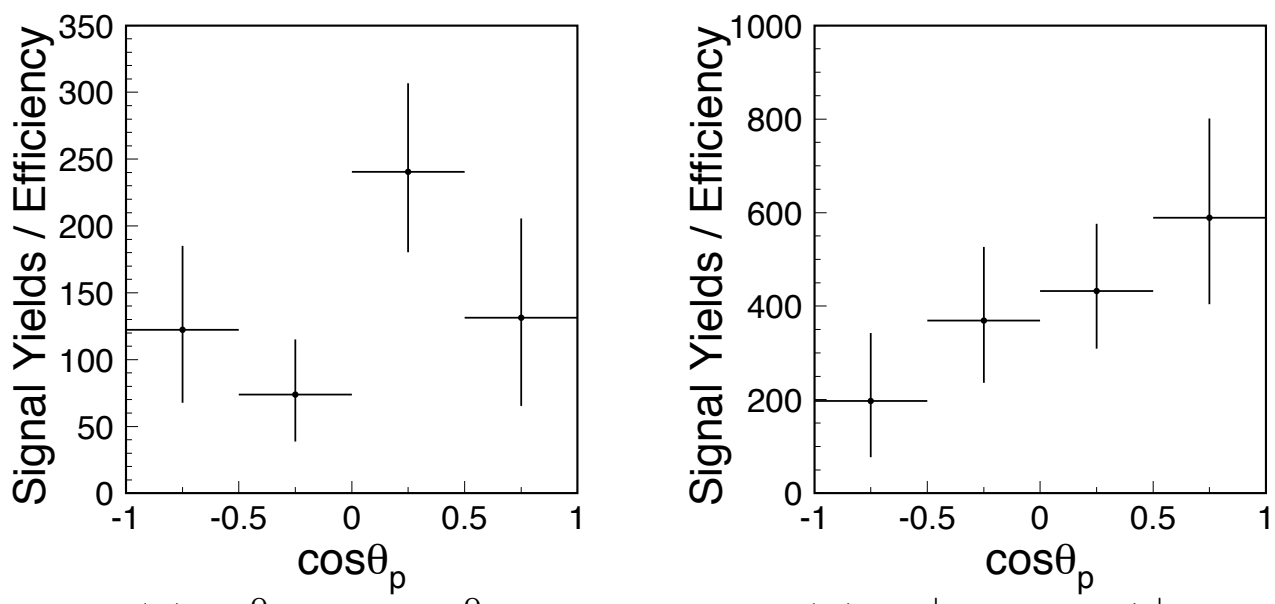

(a) $B^0 \to p\overline{p}K^{*0}$ (b) $B^+ \to p\overline{p}K^{*+}$

**Figure 17.12.18.** Distributions of efficiency corrected signal yields *vs* $\cos\theta_p$ in the proton-antiproton system with $M(p\overline{p}) < 2.85\,\mathrm{GeV}/c^2$ from Belle (Chen, 2008a).

*Further measurements, including $B \to p\overline{p}K^*$*

The abovementioned correlation enigmas are great challenges to theorists. It appears that these puzzles occur only in the penguin-dominated decays $B^- \to p\overline{p}K^-$ and $\overline{B}^0 \to \Lambda\overline{p}\pi^-$. Experimental studies of the angular distributions in charmed decays such as $\overline{B}^0 \to \Lambda\overline{p}D^{*+}$ may help solve the angular correlation puzzle in $\overline{B}^0 \to \Lambda\overline{p}\pi^+$, as the same vacuum to $\Lambda\overline{p}$ transition form factors appear in both cases (Chen, Cheng, Geng, and Hsiao, 2008). It is also very important to study the angular distributions of the baryon for $p\overline{p}X$ with $X = K^{*+}, K^0, K^{*0}, \pi^-$, and the $\Lambda\overline{\Lambda}K^-$ modes.

Both the $p\overline{p}$ threshold enhancement and baryon-meson correlations have been studied for the the isospin-related decays $B^+ \to p\overline{p}K^{*+}$ and $B^0 \to p\overline{p}K^{*0}$ (Chen, 2008a); the analysis relies on a measurement of the helicity of the $K^*$ in the decays. Large MC samples with different helicity states, 0 or $\pm 1$, of $K^*$ mesons are generated in order to obtain the corresponding angular *p.d.f.*s in $\cos\theta_K$, where $\theta_K$ is the polar angle of the $K$ meson in the $K^*$ helicity frame. For events with $m_{p\overline{p}} < 2.85\,\mathrm{GeV}/c^2$, the $B$ yield distribution in bins of $\cos\theta_K$ is used to determine the helicity zero fraction: the result is shown in Fig. 17.12.17 for $B^0 \to p\overline{p}K^{*0}$. It is interesting to note that the $K^{*0}$ meson is likely to be fully polarized in the helicity zero state, whereas the $K^{*+}$ produced in $B^+ \to p\overline{p}K^{*+}$ has only a $(32\pm17\pm9)\%$ fraction in this state (Chen, 2008a). If more than one decay amplitude is important, interference between them could lead to sizeable direct *CP* violation. A theoretical conjecture based on the factorization approach predicts that the *CP* violation in $B^\pm \to p\overline{p}K^{*\pm}$ could be as large as $\sim 22\%$ (Geng, Hsiao, and Ng, 2007). This should be checked experimentally in the future.

With fixed $K^*$ polarization, the $p\overline{p}K^*$ detection efficiency is determined as a function of $M(p\overline{p})$, and the differential cross section is then measured in $M(p\overline{p})$ bins. Threshold enhancements similar to those of $B^+ \to p\overline{p}K^+$ or $p\overline{p}\pi^+$ (Fig. 17.12.10) are seen. Distributions of the efficiency-corrected signal yield as a function of $\cos\theta_p$ are shown in Fig. 17.12.18. For $B^0 \to p\overline{p}K^{*0}$, consistent with a pure helicity state (and presumably dominated by the $b \to s$ penguin transition), the distribution is featureless; for $B^+ \to p\overline{p}K^{*+}$, where the polarization is lower (and both $b \to s$ penguin and external $W$-emission transitions can contribute) an angular correlation comparable to that in $B^- \to p\overline{p}K^-$ or $\overline{B}^0 \to \Lambda\overline{p}\pi^-$ is seen. The statistical power is limited in both cases.

#### 17.12.3.6 Cabibbo suppression

*BABAR* measured the Cabibbo suppressed decay $\overline{B}^0 \to \Lambda_c^+\overline{p}\pi^+K^-$ (Aubert, 2009ag) which can be compared to the Cabibbo favored decay $\overline{B}^0 \to \Lambda_c^+\overline{p}\pi^+\pi^-$ (Park, 2007). From a naïve comparison of the matrix elements one would expect a ratio of $|V_{us}/V_{ud}|^2 = 0.054 \pm 0.002$ if the decay mechanisms would be dominated solely by the CKM matrix elements. Comparing the branching ratios, as given in Table 17.12.6, the ratio is

$$\frac{\mathcal{B}\left(\overline{B}^0 \to \Lambda_c^+\overline{p}K^-\pi^+\right)}{\mathcal{B}\left(\overline{B}^0 \to \Lambda_c^+\overline{p}\pi^-\pi^+\right)} = 0.038 \pm 0.009, \qquad (17.12.9)$$

which implies that additional decay amplitudes (similar to 2b and 2h in Fig. 17.12.8) are only present in $\overline{B}^0 \to \Lambda_c^+\overline{p}\pi^+\pi^-$ and their contribution cannot be neglected.

The resonant subchannels have a ratio

$$\frac{\mathcal{B}\left(\overline{B}^0 \to \Sigma_c^{++}(2455)\overline{p}K^-\right)}{\mathcal{B}\left(\overline{B}^0 \to \Sigma_c^{++}(2455)\overline{p}\pi^-\right)} = 0.048 \pm 0.016, \quad (17.12.10)$$

which is in better agreement with the expectation from Cabibbo suppression. This may be understood by the fact that the spectator amplitudes are the same (1a, 2g, see Table 17.12.6) for these decays.

#### 17.12.3.7 Isospin relations

Isospin relations between two-body decays have already been used in Section 17.12.2 for the ratios of $\Sigma_c^0\overline{p}$, $\Lambda_c^+\overline{p}$, and $\Sigma_c^+\overline{p}$. Similar considerations can be applied to three-body states. The isospin restrictions on the final states in $\overline{B}^0 \to \Lambda_c^+\overline{p}\pi^0$ (Aubert, 2010h) and $B^- \to \Lambda_c^+\overline{p}\pi^-$ (Aubert, 2008aa; Gabyshev, 2006) are different: while $B^- \to \Lambda_c^+\overline{p}\pi^-$ can have only a final isospin of $I_{X\pi^-} = 3/2$ (where $X = \Lambda_c^+\overline{p}$) the neutral decay $\overline{B}^0 \to \Lambda_c^+\overline{p}\pi^0$ can have $I_{X\pi^0} = 1/2,\ 3/2$. If the decay mechanisms were equivalent in both decays one would expect a ratio of the decay rates $\overline{B}^0 \to \Lambda_c^+\overline{p}\pi^0 : B^- \to \Lambda_c^+\overline{p}\pi^-$ of $2:3$ for $I_{X\pi^-} = I_{X\pi^0} = 3/2$. A significant difference would suggest contribution from amplitudes specific to one of the decays, *e.g.* amplitudes where the $\pi^-$ originates from the $W$ in $B^- \to \Lambda_c^+\overline{p}\pi^-$ or $\overline{B}^0 \to \Lambda_c^+\overline{p}\pi^0$ contributions with $I_{X\pi^0} = \frac{1}{2}$. *BABAR* finds the ratio of partial decay widths for all decays to the final state particles

$$\frac{\Gamma\left(\overline{B}^0 \to \Lambda_c^+\overline{p}\pi^0\right)}{\Gamma\left(B^- \to \Lambda_c^+\overline{p}\pi^-\right)} = 0.61 \pm 0.09 \qquad (17.12.11)$$

to be consistent with the expectation of 2/3. When we remove the $\Sigma_c$ resonant states that are only visible in the charged $B$ decay, the ratio

$$\frac{\Gamma\left(\overline{B}^0 \to \Lambda_c^+\overline{p}\pi^0\right)}{\Gamma\left(B^- \to \Lambda_c^+\overline{p}\pi^-\right)_{\text{nonresonant}}} = 0.80 \pm 0.11 \quad (17.12.12)$$

is found to be in agreement with the assumption of similar processes in both decays.

There are, however, penguin decays for which the isospin relations are violated. Using the average of the results reported in Table 17.12.8, and correcting for the different lifetimes $\tau_+/\tau_0 = 1.071 \pm 0.009$ and the different $\Upsilon(4S)$ branching fraction $\mathcal{B}_{+-}/\mathcal{B}_{00} = 1.066 \pm 0.024$ (Beringer et al., 2012) we obtain

$$\frac{\Gamma(B^+ \to p\overline{p}K^+)}{\Gamma(B^0 \to p\overline{p}K^0)} = 1.91 \pm 0.27$$

and

$$\frac{\Gamma(B^+ \to p\overline{p}K^{*+})}{\Gamma(B^0 \to p\overline{p}K^{*0})} = 2.7 \pm 0.3 \qquad (17.12.13)$$

while a ratio of 1 is expected. Note that the helicities of the $K^{*+}$ and $K^{*0}$ mesons in the latter pair differ, as discussed at the end of Section 17.12.3.5 above.

#### 17.12.3.8 $s\overline{s}$ suppression

In fragmentation, $s\overline{s}$-production is suppressed by a factor of three compared to $u\overline{u}$ or $d\overline{d}$. This is attributed to the tunnelling process leading to additional $q\overline{q}$-pairs from the vacuum. In $B$ meson decays, a similar process occurs in hadronization. This can be investigated using pairs of decays such as $\overline{B}^0 \to D^0\Lambda\overline{\Lambda}$ (Chang, 2009) and $\overline{B}^0 \to D^0p\overline{p}$ (Abe, 2002g), where both decays can have contributions from the same diagram types (see Table 17.12.5). The ratio

$$\frac{\mathcal{B}\left(\overline{B}^0 \to D^0 \Lambda\overline{\Lambda}\right)}{\mathcal{B}\left(\overline{B}^0 \to D^0 p\overline{p}\right)} = 0.103^{+0.056}_{-0.043} \pm 0.014 \qquad (17.12.14)$$

does not include the additional possible final states $D^0\Sigma^0\overline{\Lambda}$, $D^0\Lambda\overline{\Sigma}^0$, and $D^0\Sigma^0\overline{\Sigma}^0$ for the $s\overline{s}$ diagrams. Assuming branching fractions of the same order for each of those, the resulting ratio $\sim 0.4$ is compatible with the fragmentation picture.

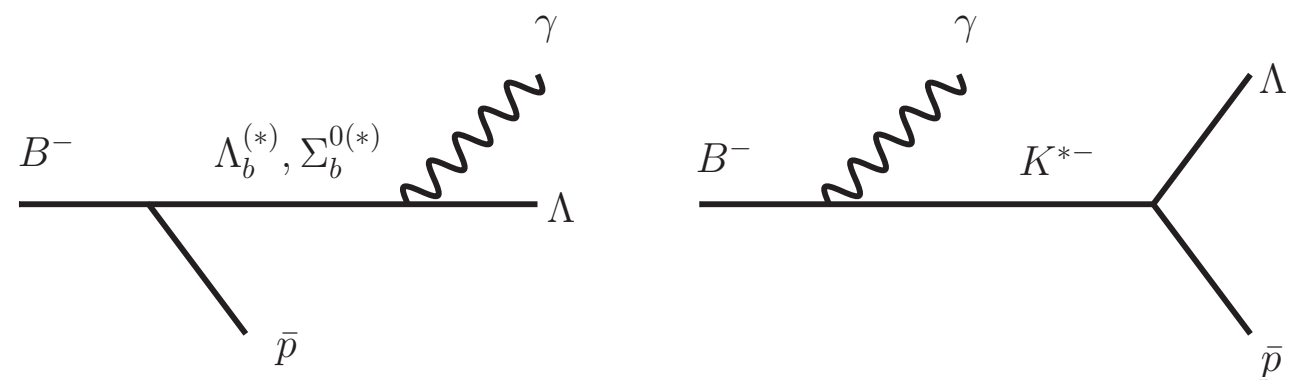

**Figure 17.12.19.** Pole diagrams for $B^- \to \Lambda\overline{p}\gamma$.

One would expect a similar ratio between the penguin decays $B^0 \to K^0\Lambda\overline{\Lambda}$ (Chang, 2009) and $B^0 \to K^0 p\overline{p}$ (Wei, 2008b), however more diagrams can contribute to $B^0 \to K^0\Lambda\overline{\Lambda}$ due to the combinatoric rearrangements of $b \to s + s\overline{s}$ compared to $b \to s + u\overline{u}$ in $B^0 \to K^0 p\overline{p}$. The experimental ratio of

$$\frac{\mathcal{B}\left(\overline{B}^0 \to K^0 \Lambda\overline{\Lambda}\right)}{\mathcal{B}\left(\overline{B}^0 \to K^0 p\overline{p}\right)} = 1.89^{+0.37}_{-0.40} \pm 0.29 \qquad (17.12.15)$$

is larger by a factor of $\sim 10$ than that for the $D^0\mathfrak{B}_1\overline{\mathfrak{B}}_2$ channels.

An $s\overline{s}$ suppression may also be expected between $\overline{B}^0 \to \Lambda_c^+\overline{p}\pi^+\pi^-$ and $\overline{B}^0 \to \Lambda_c^+\overline{p}K^+K^-$. However, differences are not attributable to $s\overline{s}$ suppression alone since possible contributing diagrams differ: Only $\overline{B}^0 \to \Lambda_c^+\overline{p}\pi^+\pi^-$ can have contributions from diagrams of type 1 (Fig. 17.12.7). In decays of type 2 (Fig. 17.12.8) different intermediate resonant states are contributing, and the number of diagrams for $\overline{B}^0 \to \Lambda_c^+\overline{p}K^+K^-$ is smaller.

### 17.12.4 Radiative decays into baryons

It would be very difficult to detect the radiative baryonic $B$ decay $B \to \mathfrak{B}_1\overline{\mathfrak{B}}_2\gamma$ if it proceeded only via bremsstrahlung. Fortunately, there is an important short-distance electromagnetic penguin transition $b \to s\gamma$ which is neither Cabibbo suppressed nor (due to the large top quark mass) loop suppressed. Moreover, it is considerably enhanced by QCD corrections. At the mesonic level, it is well known that the electromagnetic penguin transition $b \to s\gamma$ is represented by the radiative decays $B \to K^*\gamma$. The measurement of $B^- \to \Lambda\overline{p}\gamma$ using a $449 \times 10^6$ $B\overline{B}$ sample in 2005 by Belle (Lee, 2005) provided the first observation of $b \to s\gamma$ in baryonic $B$ decays.

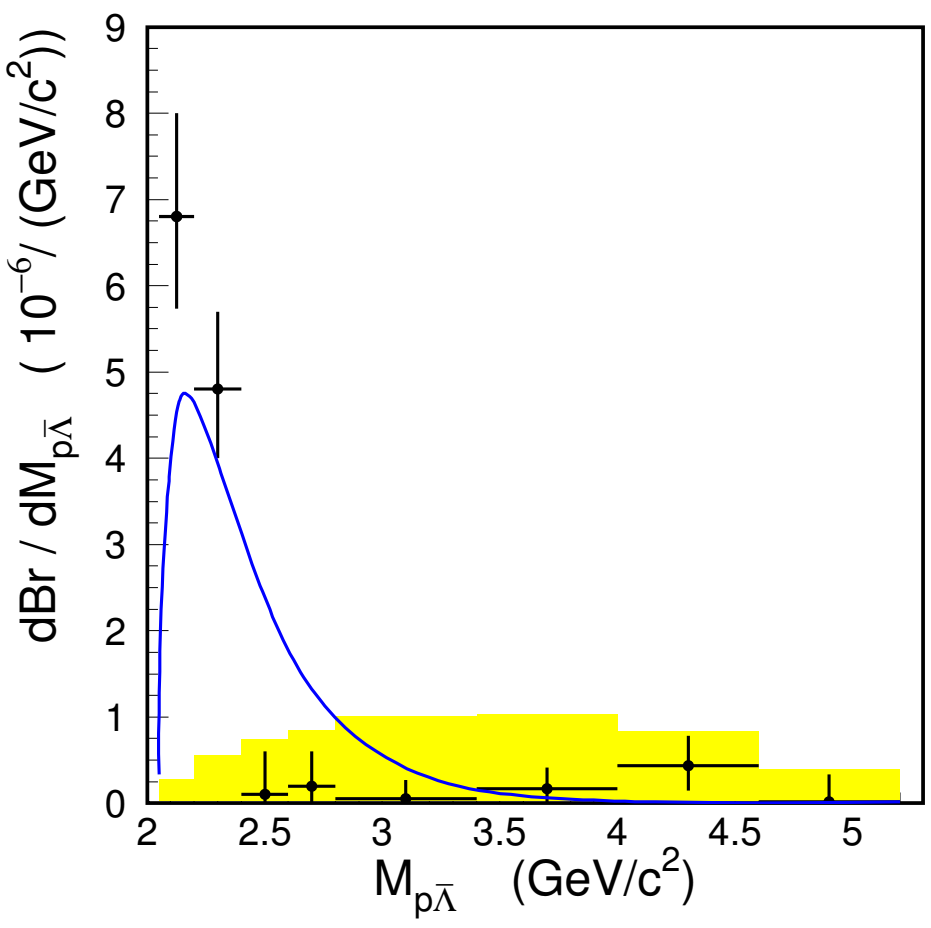

**Figure 17.12.20.** Differential branching fractions for $B^- \to \Lambda\overline{p}\gamma$ as a function of baryon-antibaryon pair mass from Belle (Wang, 2007b). The shaded distribution shows the expectation from a phase-space MC simulation. The theoretical prediction from Geng and Hsiao (2005) is overlaid as a solid line for comparison. The area of the shaded distributions and areas under the theoretical curves are scaled to match the measured branching fractions from data. The uncertainties are statistical only.

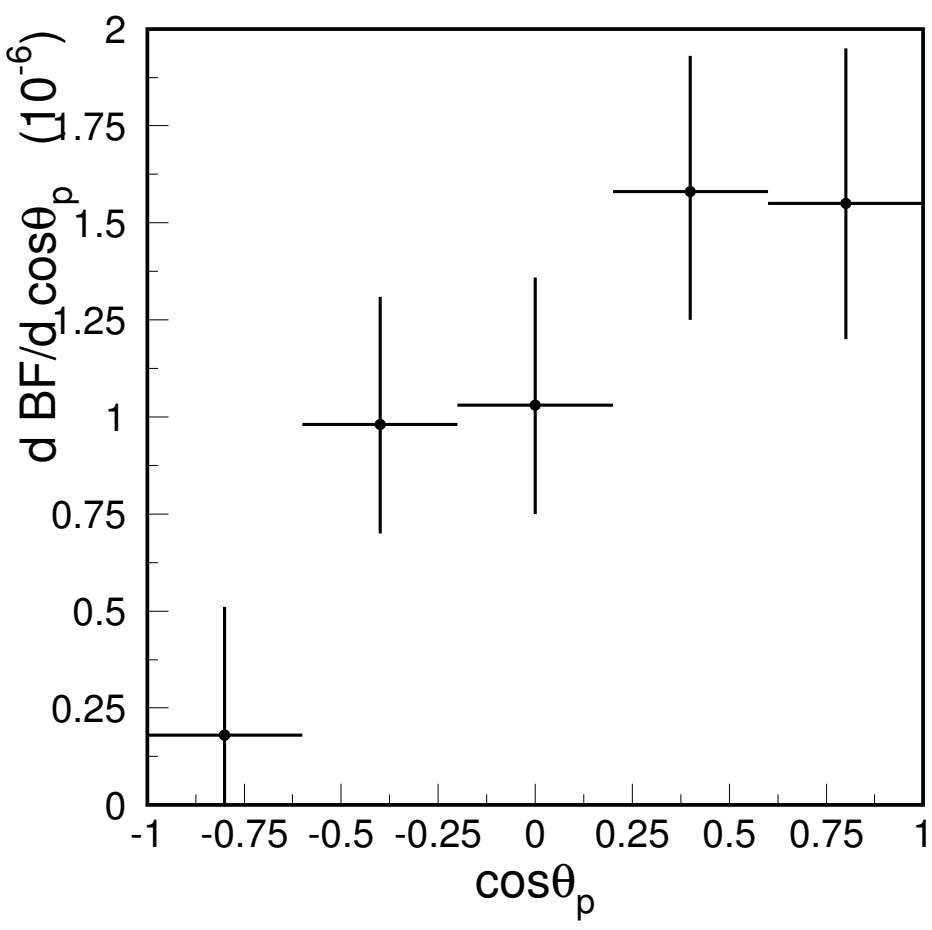

**Figure 17.12.21.** Differential branching fractions *vs* $\cos\theta_p$ for $B^- \to \Lambda\overline{p}\gamma$ in the region near threshold (baryon-antibaryon mass $< 2.8\,\mathrm{GeV}/c^2$) from Belle (Wang, 2007b). The uncertainties are statistical only.

Theoretically, radiative baryonic $B$ decays have been studied in the pole model, in which the dominant contributions are assumed to arise from low-lying baryon and meson intermediate states (Cheng and Yang, 2002b). For example, the relevant intermediate states for $B^- \to \Lambda\overline{p}\gamma$ are $K^*$, $\Lambda_b^{(*)}$ and $\Sigma_b^{(*)0}$ (see Fig. 17.12.19). Predictions for some radiative decay modes are summarized in Table 17.12.10. The model prediction for $\mathcal{B}(B^- \to \Lambda\overline{p}\gamma) =$

**Table 17.12.10.** Predicted branching fractions for radiative baryonic $B$ decays where $A_\theta$ is the angular asymmetry defined in Eq. (17.12.7) and "$\mathcal{B}$(tot)" denotes the sum of the baryon and meson pole contributions (Cheng and Yang, 2006).

| Mode | Baryon | Meson | $\mathcal{B}$(tot) | $A_\theta$ |
|---|---|---|---|---|
| $B^- \to \Lambda\overline{p}\gamma$ | $7.9\times10^{-7}$ | $9.5\times10^{-7}$ | $2.6\times10^{-6}$ | 0.25 |
| $B^- \to \Sigma^0\overline{p}\gamma$ | $4.6\times10^{-9}$ | $2.5\times10^{-7}$ | $2.9\times10^{-7}$ | 0.07 |
| $B^- \to \Xi^0\overline{\Sigma}^-\gamma$ | $7.5\times10^{-7}$ | $1.6\times10^{-7}$ | $5.6\times10^{-7}$ | 0.43 |
| $B^- \to \Xi^-\overline{\Lambda}\gamma$ | $1.6\times10^{-7}$ | $2.4\times10^{-7}$ | $2.2\times10^{-7}$ | 0.13 |

$2.6\times10^{-6}$ is in good agreement with the latest measurement (Wang, 2007b)

$$\mathcal{B}(B^- \to \Lambda\overline{p}\gamma) = (2.45^{+0.44}_{-0.38} \pm 0.22)\times10^{-6} \ . \quad (17.12.16)$$

The measured differential branching fraction as a function of baryon-antibaryon mass for $B^- \to \Lambda\overline{p}\gamma$ is shown in Fig. 17.12.20. This distribution is sharply peaked near threshold and is quite similar to those observed in $B^- \to \Lambda\overline{p}\pi^0$ and $\overline{B}^0 \to \Lambda\overline{p}\pi^+$ (Wang, 2007b; see the discussion in Section 17.12.3.3 above).

Experimentally, the angular correlation in $B^- \to \Lambda\overline{p}\gamma$ is measured by considering the angular asymmetry $A_\theta$ defined in Eq. (17.12.7). Figure 17.12.21 shows the differential branching fraction as a function of $\cos\theta_p$ near the $\Lambda\overline{p}$ mass threshold, where $\theta_p$ is the angle between the photon and the antiproton in the $\Lambda\overline{p}$ rest frame. From Fig. 17.12.20, we know that $\Lambda$ and $\overline{p}$ tend to move in parallel. Since the energetic $s$ quark from the $b \to s\gamma$ process will hadronize into $\Lambda$, it is expected that $\Lambda$ will move faster than $\overline{p}$ in the $B$ rest frame. Indeed, after boosting to the baryon-antibaryon rest frame, the antiproton prefers leaning to the photon direction as shown in Fig. 17.12.21. Again, the predicted angular asymmetry $A_\theta = 0.25$ agrees with the measured value of $0.29 \pm 0.14 \pm 0.03$ (Wang, 2007b). We see from Table 17.12.10 that the decay $B^- \to \Xi^0\overline{\Sigma}^-\gamma$, with an estimated branching fraction of order $6\times10^{-7}$, should be accessible at the $B$ Factories. Penguin-induced radiative baryonic $B$ decays should be further explored both experimentally and theoretically.

### 17.12.5 Semileptonic decays with a baryon-antibaryon pair

Semileptonic decays would proceed only through the type 1 (external) diagram shown in Fig. 17.12.7a, when the $W$ decays into a $\ell^-\overline{\nu}_\ell$ pair instead of a quark antiquark pair. The observation of such decays could therefore establish the relevance of this diagram compared to the type 2 (internal) ones. Unfortunately, no semileptonic decay to baryons has been observed so far. The upper limit from *BABAR* (Lees, 2012p) is

$$\frac{\mathcal{B}(\overline{B} \to \Lambda_c^+ X \ell^- \overline{\nu}_\ell)}{\mathcal{B}(B/\overline{B} \to \Lambda_c^+ X)} < 0.025 \text{ at 90\% CL} \qquad (17.12.17)$$

which, using the inclusive $\Lambda_c^\pm$ multiplicity in $B$ decays from Section 17.12.1, translates into an approximate upper limit $\mathcal{B}(\overline{B} \to \Lambda_c^+ X \ell^- \overline{\nu}_\ell) < 1.2\times10^{-3}$.

### 17.12.6 Summary

The observed pattern in two-body baryonic $B$ decays, namely, $\mathcal{B}(B \to \mathfrak{B}_{1c}\overline{\mathfrak{B}}_{2c}) \gg \mathcal{B}(B \to \mathfrak{B}_c\overline{\mathfrak{B}}) \gg \mathcal{B}(B \to \mathfrak{B}_1\overline{\mathfrak{B}}_2)$, can be understood in pQCD, though there is still no clear theoretical prediction for charmless two-body decays. Given the large energy release in such decays, an estimate based on the pQCD approach should be reliable.

The enhancement of the baryon-antibaryon invariant mass near threshold observed in multi-body baryonic $B$ decays indicates that the $B$ meson preferentially decays into a baryon-antibaryon pair with low invariant mass accompanied by a fast recoil meson. Theoretically, the threshold peaking effect is closely linked to the behavior of baryon form factors which fall off sharply with $t$, the invariant mass squared of the dibaryon. There are two unsolved puzzles in the study of baryon-antibaryon angular correlations. First, the anomalous correlation effect measured in $B^- \to p\overline{p}K^-$ decay is against the theoretical prediction based on the short-distance $b \to sg^*$ picture. Second, the $\Lambda$ particle in the decay of $\overline{B}^0 \to \Lambda\overline{p}\pi^-$ moves collinearly with $\pi^-$ in the $\Lambda\overline{p}$ rest frame, whereas it is naïvely expected that the pion has no preference for its correlation with $\Lambda$ or the antiproton. These correlation enigmas are great challenges to theorists.

The measured branching fraction and the angular correlation in $B^- \to \Lambda\overline{p}\gamma$ are consistent with the theoretical model based on the weak penguin process $b \to s\gamma$. Hence, this radiative baryonic $B$ decay is induced by the electroweak penguin transition. It is important to further explore penguin-mediated radiative baryonic $B$ decays both experimentally and theoretically.

# Chapter 18
# Quarkonium physics

## 18.1 Introduction to quarkonium

***Editors:***
*Nora Brambilla, Thomas Mannel (theory)*

Heavy quarkonia are systems composed of a heavy quark and antiquark of the same flavor (charm, bottom, or top[97]), with mass $m$ much larger than the "QCD confinement scale" $\Lambda_{\rm QCD}$, so that $\alpha_s(m) \ll 1$ holds. Within both the $c\bar{c}$ and $b\bar{b}$ quarkonium spectra, it is evident that the difference in energy levels is much smaller than the quark mass: quarkonia are non-relativistic systems.

Due to the large quark mass and small (relative) quark velocity $|\boldsymbol{v}| = v$ in quarkonia, these states are affected by physical processes at a range of energy scales. They probe all the regimes of QCD, from high energies where an expansion in the coupling constant is possible, to low energies where non-perturbative effects dominate; they also probe intermediate scales. Quarkonium is thus a laboratory where our understanding of non-perturbative QCD and its interplay with perturbative QCD may be tested in a controlled framework. The large mass and the clean and known decay modes also make quarkonia an ideal probes of new physics in some well defined window of beyond Standard Model (BSM) parameters, in particular for some searches for dark matter candidates (Brambilla et al., 2004, 2011; Dermisek, Gunion, and McElrath, 2007; McElrath, 2005; Sanchis-Lozano, 2010).

Belle and *BABAR* have collected a wide range of quarkonium data, including clean samples of charmonia produced in $B$ decays, two-photon fusion, initial state radiation, and $e^+e^-$ annihilation, including the unexpected observation of large associated $(c\bar{c})(c\bar{c})$ production. The final years of datataking have also seen extensive studies of $b\bar{b}$ states. Even if quarkonium studies were not a priority at the start of the running of the $B$ Factories, these facilities have come to function as heavy meson factories, producing many new states and new data on quarkonia, and accumulating large data samples on spectra and decays. In the same period, quarkonia have been studied at BES and BESIII at BEPC and BEPC2, KEDR at VEPP-4M, CLEO-III and CLEO-c at CESR, CDF and DØ at Fermilab, and the PHENIX and STAR experiments at RHIC. New states and exotics, new production mechanisms, new transitions and unexpected states of an exotic nature have been observed. Large new data samples are now being collected at the LHC experiments and new facilities will become operational (PANDA at GSI, a much higher luminosity $B$ Factory at KEK) adding challenges and opportunities to this research field.

In the following, we describe the possible quantum numbers and some features of the spectrum of quarkonium states (Section 18.1.1), and briefly review the potential model approach (Section 18.1.2). The hierarchy of scales required to describe quarkonia, and its implications, are then discussed (Section 18.1.3), followed by an extended review of effective field theory (EFT) methods (Section 18.1.4) and a brief discussion of lattice calculations (Section 18.1.5). We conclude this introduction with examples of results obtained from theoretical studies of quarkonia. In the remainder of this chapter, we describe in turn the $B$ Factory results on conventional charmonium states (Section 18.2), the exotic charmonium-like or "XYZ" states (Section 18.3), and bottomonium states (Section 18.4).

[97] The top quark does not form a proper bound state since it decays weakly on a time scale shorter than that typical of the would-be bound state. However, to calculate the $t\bar{t}$ production cross section, bound state effects have to be taken into account.

### 18.1.1 Quantum numbers and spectroscopy

The term "quarkonium" was coined because of the similarity of heavy quark-antiquark bound states to those of positronium: the bound states of an $e^+$ and an $e^-$. These systems share similar spectroscopy and decays. For example, the parapositronium decays into two photons, while its charmonium analogue $\eta_c$ decays into two gluons; the orthopositronium decays into three photons, while the "orthocharmonium" $(J/\psi)$ decays into and gluons, as these decays are fixed by quantum numbers and related parity and charge conjugation conservation.

In Fig. 18.1.1 our present knowledge of the bottomonium and charmonium energy levels is illustrated: our current understanding has changed dramatically after the $B$ Factory era and we have acquired information about many new energy levels, both below and above threshold, and among several new decays and transitions.

The spectroscopic notation $n\,^{2s+1}\ell_J$ (with the $J^{PC}$ number often given in parenthesis) is conventionally used for quarkonium levels, where $n$ is the radial quantum number (equal to the number of nodes in the wavefunction) plus 1, $\ell$ is the orbital angular momentum between quarks (designated by letters as $S$, $P$, $D$, etc.), $s = 0, 1$ is the total spin of the quarks, and $J$ is the quarkonium spin $(|\ell - s| \le J \le \ell + s)$. We note, that among the above four quantum numbers, only the spin of a state can be measured; the others are merely assigned based on the measured parity $P$, and charge-conjugation $C$. The behavior of a state under parity is dictated by the symmetry of the angular momentum eigenfunctions, the spherical harmonics $Y_l^m$, for which $P = (-1)^\ell$,, and by the opposite parity of the antifermion with respect to the fermion, which finally yields for the quarkonium parity

$$P = (-1)^{\ell+1}. \qquad (18.1.1)$$

Charge conjugation exchanges the two constituents. Because of Fermi-Dirac statistics, the exchange of two identical fermions gives a minus sign. On the other hand, this exchange is performed applying the charge conjugation operator (which gives a factor $C$), exchanging the coordinates (which gives $(-1)^\ell$) and exchanging the spin (which gives a factor $(-1)^{s+1}$). Therefore $C(-1)^\ell(-1)^{s+1} = -1$ and

$$C = (-1)^{\ell+s}. \qquad (18.1.2)$$

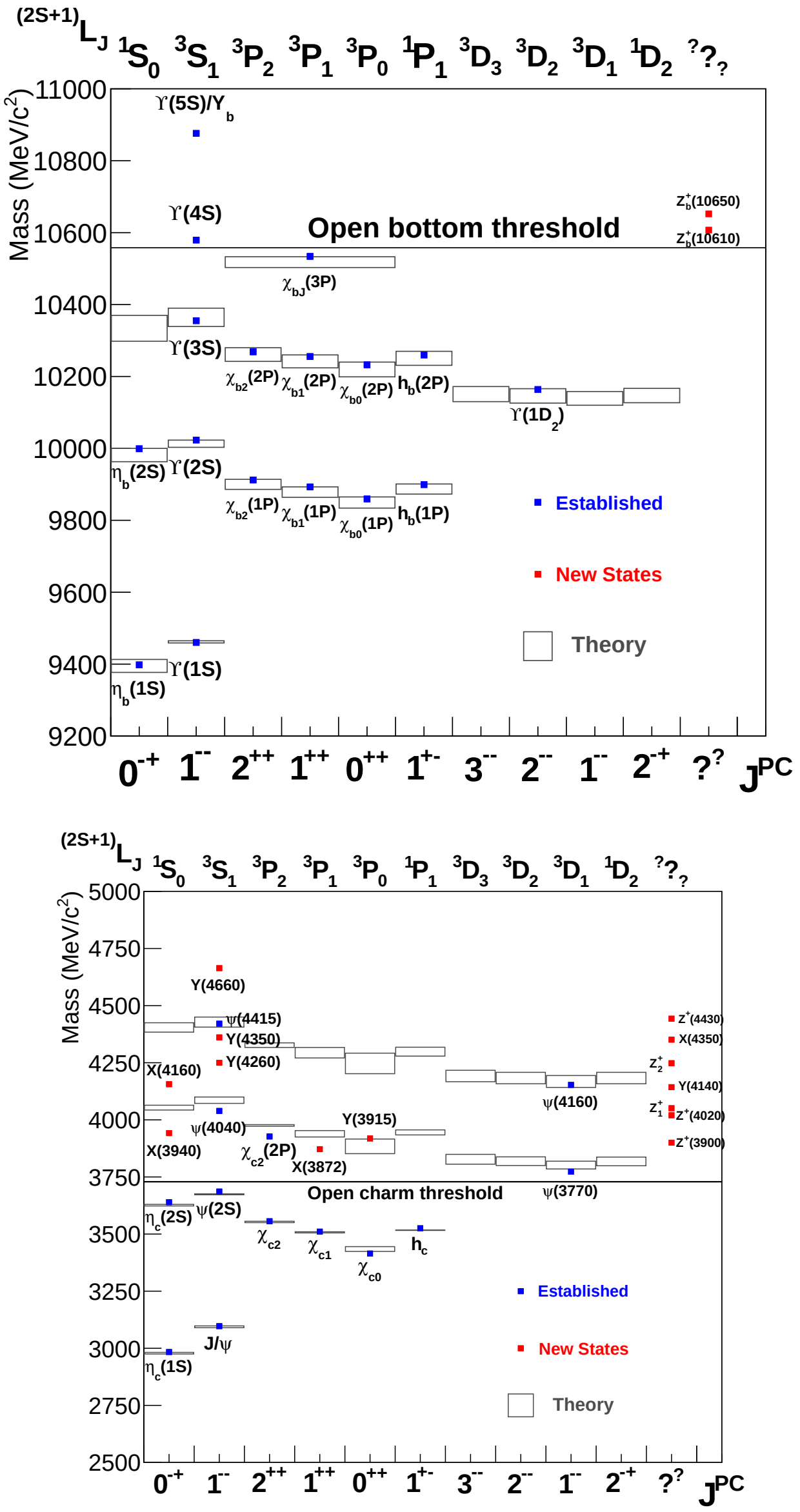

**Figure 18.1.1.** Energy levels of bottomonium (upper plot) and charmonium (lower plot) as known at the end of the $B$ Factory era. "Established" states are those predicted in the theory and whose measured properties are in agreement with predictions. "New states" are unpredicted and/or their measured properties are difficult to accommodate in the theory. In the last column we list states with unknown quantum numbers, and the charged quarkonium-like resonances.

Spin, $P$ or $C$ are often determined from the selection rules both of the production and the decay mechanism. When this is not the case, or if they cannot unambiguously fix $\ell$ and $s$, a quarkonium state assignment can be tried relying on theoretical predictions for the mass, width, decay channels, or production mechanisms.

From its non-relativistic nature some specific features of the quarkonium spectrum can be derived. The separation between levels of different $n$ and same $l$ typically scales like $mv^2$; the spin separation between pseudoscalar mesons $n\,^1S_0(0^{-+})$ and vector mesons $n\,^3S_1(1^{--})$, called hyperfine splitting, scales like $mv^4$; the spin separation between states within the same $\ell \neq 0$ and $S$ multiplets (*e.g.* the splittings in the $1\,^3P_j$ multiplet $\chi_c(1P)$ in charmonium), called fine splitting, scales like $mv^4$; and the hyperfine separation between the spin-singlet state $^1P_1$ and the spin-averaged triplet state $\langle ^3P_j \rangle$, which again scales like $mv^4$.

The fact that all splittings are much smaller than the masses implies that all the dynamical scales of the bound state, such as the kinetic energy or the momentum of the heavy quarks, are small compared to the quark mass. Therefore, the heavy quarkonia are to a good approximation non-relativistic systems. For further discussion of the various energy scales relevant for quarkonium system, see Section 18.1.3.

Another important feature of the spectrum is the presence of an "open flavor threshold" (open charm, or open bottom), where a quarkonium state can undergo strong decay to a pair of mesons carrying the corresponding quark flavor. States above threshold are considerably wider than states below. Excited states below threshold decay either by strong interactions or electromagnetically into lower-lying states; the ground states finally decay by an annihilation process of the heavy quark-antiquark pair. This annihilation is controlled by powers of the strong coupling constant evaluated at the quark mass, which gives a large suppression factor, resulting in a small width.

### 18.1.2 Potential models

To make quantitative predictions of masses and for the the full and partial widths of charmonium states, one has to resort to theory. For many years a phenomenological approach, based on both non-relativistic and relativistic potential model, has been used. Non-relativistic potential models are justified by the fact that the bottom and, to a lesser extent, the charm masses are large in comparison to $\Lambda_{\rm QCD}$, the typical hadronic scale. Hence a quantum mechanical description of the system based on two heavy quarks interacting through a suitable potential appears reasonable. In this approach, the quarks are located in a potential $V(r)$ and the charmonium wave function can be found as a solution of the stationary non-relativistic Schrödinger equation. The potential is usually chosen such that at short distances it coincides with the QCD one-gluon exchange Coulomb potential $-\frac{4}{3}\alpha_S/r$, and at long distances it incorporates confinement by for example including a linearly rising term. Since relativistic effects appear to be sizable for some states, different models incorporate relativistic kinematics appropriately matched to their confinement features. Different models of quark confinement may result in different classes of relativistic corrections. For states close to and beyond the two heavy-light meson threshold, potential models have to be complemented with extra degrees of freedom in order to account for possible mixing effects. Hybrid states which are expected from QCD are also incorporated by hand.

Examples of results obtained for the charmonia in such phenomenological approaches are listed in Table 18.2.1: the states below open flavor threshold are well described.

The problem with this approach is that it is purely phenomenological, with no way to link the model parameters to QCD. In this approach on can neither use the quarkonia to improve our understanding of strong interactions, nor as a tool to extract precise information on the Standard Model and beyond. The modern approach to charmonium physics relies on non-relativistic effective field theories (NRQCD, pNRQCD), discussed in Section 18.1.4, and lattice calculations, discussed in Section 18.1.5 below. These methods are conditioned by the importance of a range of different energy scales in quarkonium physics, to which we now turn.

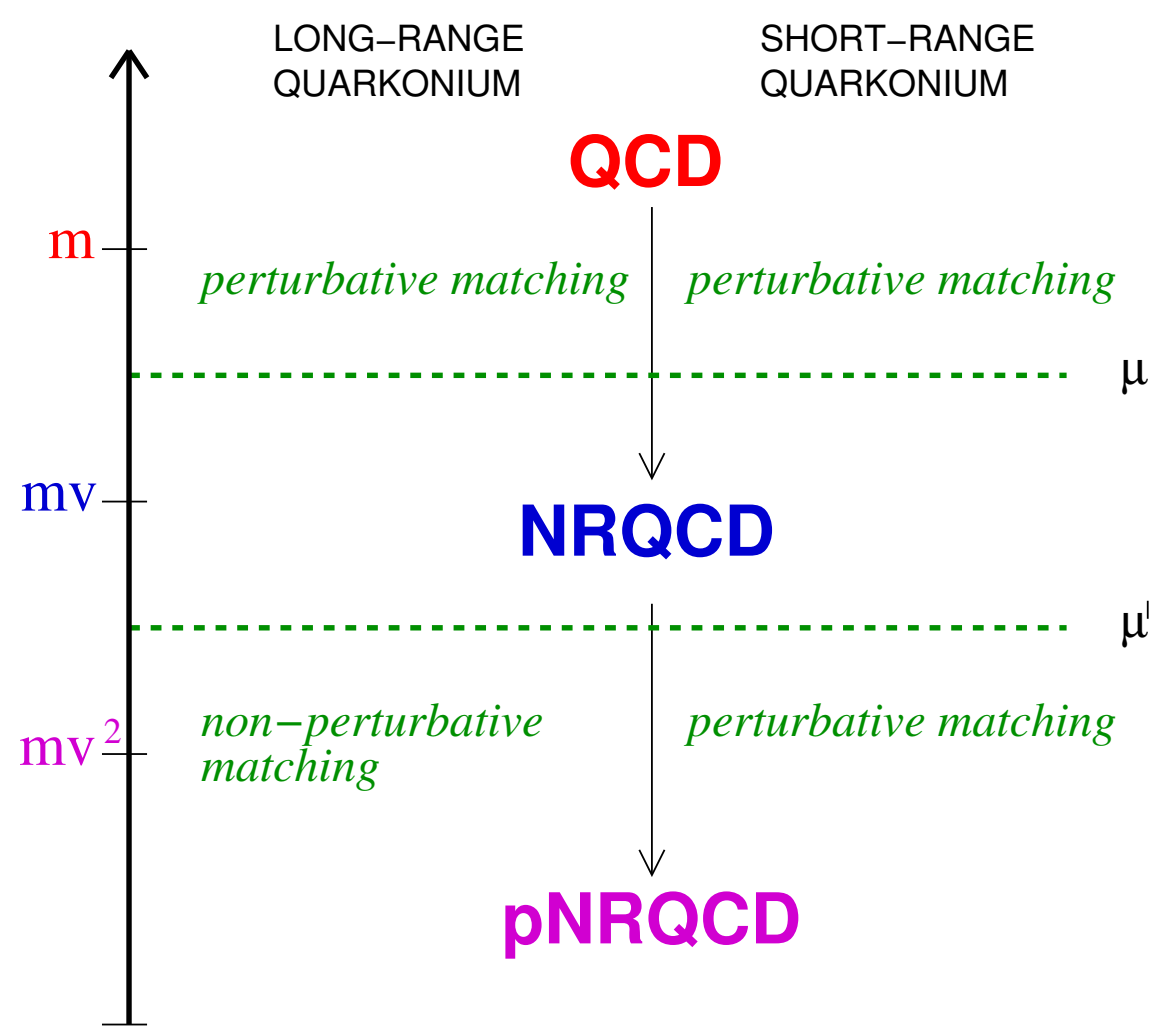

**Figure 18.1.2.** Energy scales and corresponding effective field theories for quarkonium. The scale $\mu$ separates QCD from NRQCD, while $\mu'$ separates NRQCD from pNRQCD.

### 18.1.3 Quarkonium as a multiscale system

As non-relativistic systems, quarkonia are characterized by the heavy-quark velocity $v$ ($v^2 \sim 0.1$ for the $b\bar{b}$, and $\sim 0.3$ for the $c\bar{c}$ systems) and by a hierarchy of energy scales: the mass $m$ of the heavy quark (hard scale), the typical relative momentum $p \sim mv$ (in the meson rest frame) corresponding to the inverse Bohr radius $r \sim 1/(mv)$ (soft scale), and the typical binding (or kinetic) energy $E \sim mv^2$ (ultrasoft scale). This is similar to the energy scales for the hydrogen atom, in which case $v \sim \alpha_{\rm EM}$.

The hierarchy of non-relativistic scales makes the heavy quarkonia qualitatively different from the heavy-light mesons, which are characterized by just two scales: $m$ and $\Lambda_{\rm QCD}$. This makes the theoretical description of quarkonium physics more complicated. There are effects at each of these scales in a typical amplitude involving a quarkonium observable. In particular, quarkonium annihilation and production take place at the scale $m$, quarkonium binding takes place at the scale $mv$, which is the typical momentum exchanged inside the bound state, while very low-energy gluons and light quarks (also called ultrasoft degrees of freedom) live long enough that a bound state has time to form and, therefore, are sensitive to the scale $mv^2$. Ultrasoft gluons are responsible for phenomena similar to the Lamb shift in QCD.

The appearance of a hierarchy of scales calls for the application of effective field theory (EFT) methods. However, "heavy quark effective theory" (HQET), where only an ultraviolet mass scale $m$ and an infrared mass scale $\Lambda_{\rm QCD}$ appear, is not suitable for the description of heavy quarkonia, since HQET is unable to describe the dynamics of binding. The appropriate effective field theories are "non-relativistic QCD" (NRQCD; Section 18.1.4.1) and "potential non-relativistic QCD" (pNRQCD; Section 18.1.4.2) which are far more complicated, since the scales $mv$ and $mv^2$ are generated by the dynamics of the system which determines the velocity of the quarks in the bound state.

The description of the heavy quark-antiquark systems depends on the relation of $\Lambda_{\rm QCD}$ to the above-mentioned scales. Clearly for energy scales close to $\Lambda_{\rm QCD}$ there is no perturbative description and one has to rely on non-perturbative methods. Regardless of this, the non-relativistic hierarchy $m \gg mv \gg mv^2$ persists below the $\Lambda_{\rm QCD}$ threshold, as long as $v$ is small. While the hard scale $m$ is always larger than $\Lambda_{\rm QCD}$, different situations may arise for the other two scales.

In a case with $\Lambda_{\rm QCD} \ll mv^2$ both scales are still perturbative and the system is similar to a Coulombic system: for such a quarkonium we would have — as for the hydrogen atom — $v \sim \alpha_s(mv)$. However, none of the $b\bar{b}$ or $c\bar{c}$ states satisfy this condition. In all realistic quarkonia ($b\bar{b}$ and $c\bar{c}$) the ultrasoft scale is non-perturbative. Only for $t\bar{t}$ threshold states may the ultrasoft scale be considered to lie in the perturbative regime.

The soft scale, proportional to the inverse quarkonium radius $r$, may be either perturbative ($mv \gg \Lambda_{\rm QCD}$) or non-perturbative ($mv \sim \Lambda_{\rm QCD}$) depending on the physical system under consideration. Unfortunately, we do not have any direct information on the radius of the quarkonia, and thus the assignment of some of the lowest bottomonium and charmonium states to the perturbative or the non-perturbative soft regime is at the moment still ambiguous, but it is likely that the lowest bottomonium and possibly also the lowest charmonium states have small enough radii that the scale $mv$ is in fact still perturbative.

In Fig. 18.1.2 we schematically show the various scales. The short-range quarkonia are small enough to allow for a perturbative treatment of the scale $mv$, while for the long range quarkonia already this scale requires a non-perturbative treatment.

The implications of the hierarchy of scales for lattice calculations of quarkonium are outlined in Section 18.1.5 below; effective field theory methods are described in the following section.

### 18.1.4 Effective Field Theories

Effective field theories for the description of quarkonium processes have recently been developed, providing a unifying description as well as a solid and versatile tool giving well-defined, model-independent and precise predictions. They rely on the one hand on high-order perturbative calculations and on the other hand on lattice simulations, the recent progress in both fields having added significantly to the reach of theory. The progress in our understanding of EFTs has made it possible to move beyond phenomenological models (at least for states below open-flavor threshold) and to provide a systematic description, inside QCD, of heavy-quarkonium physics. On the other hand, the recent progress in the measurement of several heavy-quarkonium observables makes it meaningful to address the problem of their precise theoretical determination. Here we will give a brief introduction to EFTs for heavy quarkonium. For a general introduction to the field of quarkonium, a detailed review of quarkonium theory and experiments and a comparison of theory predictions to experiments with a discussion of the most important open problems, see Brambilla et al. (2004, 2011).

The idea of non-relativistic Effective Field Theories (NR EFTs) was pioneered by Caswell and Lepage (1986) and was later refined and cast into the NRQCD effective theory language by Bodwin, Braaten, and Lepage (1995); subsequently, the EFT at the lowest possible energy scale (the ultrasoft scale), potential NRQCD (pNRQCD), was obtained (Brambilla, Pineda, Soto, and Vairo, 2000; Pineda and Soto, 1998). A recent review can be found in Brambilla, Pineda, Soto, and Vairo (2005). The point is to take advantage of the existence of the different energy scales to substitute QCD with simpler but equivalent NR EFTs. A hierarchy of NR EFTs may be constructed by systematically integrating out modes associated with high-energy scales not relevant for the quarkonium system. Such integration is performed in a matching procedure that enforces the equivalence between QCD and the EFT at a given order of the expansion in $v$. The EFT Lagrangian is factorized in matching coefficients, encoding the high-energy degrees of freedom and low-energy operators; relativistic invariance is realized via exact relations among these coefficients (Brambilla, Gromes, and Vairo, 2001, 2003; Manohar, 1997). The EFT displays power counting in the small parameter $v$, *i.e.* we are able to attach a definite power of $v$ to the contribution of each of the EFT operators to the physical observables.

#### 18.1.4.1 Physics at the scale $m$: NRQCD

Quarkonium annihilation and production take place at the scale $m$. The suitable EFT is non-relativistic QCD (Bodwin, Braaten, and Lepage, 1995; Caswell and Lepage, 1986), which follows from QCD after integrating out the scale $m$. As a consequence, the effective Lagrangian is organized as an expansion in $1/m$ and $\alpha_s(m)$:

$$\mathcal{L}_{\text{NRQCD}} = \sum_n \frac{c_n(\alpha_s(m), \mu)}{m^n} \times O_n(\mu, mv, mv^2, ...), \tag{18.1.3}$$

where $c_n$ are Wilson coefficients that contain the contributions from the scale $m$; they can be perturbatively calculated by matching the full QCD result to the effective theory. The $O_n$ are the local operators of NRQCD; the matrix elements of these operators contain the physics of scales below $m$, in particular of the scales $mv$ and $mv^2$ and also of the non-perturbative scale $\Lambda_{\text{QCD}}$. Finally, the parameter $\mu$ is the NRQCD factorization scale, which separates the contributions to be described in QCD from the ones to be described in NRQCD. Matrix elements of $O_n$ depend on the scales $\mu$, $mv$, $mv^2$ and $\Lambda_{\text{QCD}}$ and the power counting is performed in powers of $v$. The low-energy operators $O_n$ are constructed out of two or four heavy quark/antiquark fields plus gluons. The operators with a fermion and an antifermion field are the same ones obtained from the non-relativistic reduction of the QCD Lagrangian. This part of the Lagrangian is equal, in the meson rest frame, to the Lagrangian of Heavy Quark Effective Theory (HQET) which is used to treat heavy-light mesons. The power counting is, however, different: while in HQET there is a strict counting in inverse powers of $m$, in NRQCD the power counting is in powers of $v$, because the energy scales of heavy-light mesons and quarkonia are different. In particular, in a heavy-light meson, the three-momentum and the energy of the heavy quark are both of order $\Lambda_{\text{QCD}}$, in contrast with the situation in a heavy quarkonium, in which the three-momentum is of order $mv$ and the energy is of order $mv^2$.

*Annihilation decays*

To describe the annihilation decays of heavy quarkonia into light hadrons, we have to consider four-fermion operators with four heavy quark/antiquark fields in the effective interaction. Considering operators up to dimension six we have the following contributions

$$\begin{aligned} &\frac{c_1(^1S_0)}{m^2}O_1(^1S_0) + \frac{c_1(^3S_1)}{m^2}O_1(^3S_1) \\ &+\frac{c_8(^1S_0)}{m^2}O_8(^1S_0) + \frac{c_8(^3S_1)}{m^2}O_8(^3S_1), \end{aligned} \tag{18.1.4}$$

where $c_j$ are the matching coefficients: they are calculated as a series in $\alpha_s$ and in this case they acquire also imaginary parts. The dimension-6 operators (*i.e.* the operators containing two quark and two antiquark operators) are

$$\begin{aligned} O_1(^1S_0) &= \psi^\dagger \chi \chi^\dagger \psi, & O_1(^3S_1) &= \psi^\dagger \boldsymbol{\sigma} \chi \chi^\dagger \boldsymbol{\sigma} \psi, \\ O_8(^1S_0) &= \psi^\dagger \mathrm{T}^a \chi \chi^\dagger \mathrm{T}^a \psi, & O_8(^3S_1) &= \psi^\dagger \mathrm{T}^a \boldsymbol{\sigma} \chi \chi^\dagger \mathrm{T}^a \boldsymbol{\sigma} \psi, \end{aligned} \tag{18.1.5}$$

where $\psi$ and $\chi$ are the Pauli fields for the quark and the antiquark, $\boldsymbol{\sigma}$ are the three spin Pauli matrices and $T^a$ are the $SU(3)$ color matrices. The subscript 1 or 8 indicates the color structure: since we consider states made by a quark and an antiquark, from the point of view of color $SU(3)$ these are $3 \times \bar{3}$ states, 3 being the fundamental color representation of $SU(3)$ and $\bar{3}$ the antifundamental. Therefore, since in $SU(3)$ $3 \times \bar{3} = 1 \oplus 8$, $q\bar{q}$ states behave under color transformation as color singlets or color octets. The arguments ${}^{2S+1}L_J$ indicate the angular-momentum state of the $Q\overline{Q}$ pair which is annihilated or created by the operator. Then, the annihilation rate can be calculated by recalling that the decay rate is minus two times the imaginary part of the energy of the state, so that we obtain that the inclusive decays of a heavy quarkonium state $|H\rangle$ induced through annihilation of the heavy quarks can be calculated in NRQCD as (Bodwin, Braaten, and Lepage, 1995)

$$\Gamma(H \to \text{light hadrons}) = \sum_n \frac{2\text{Im}\, c_n}{m^{d_n-4}} \langle H|O_n^{\text{4fermions}}|H\rangle, \tag{18.1.6}$$

$d_n$ being the dimension of the operator $O_n$. This formula realizes a factorization between the physics at the hard scale contained in the imaginary parts of the matching coefficients and the low-energy physics contained in the non-perturbative matrix elements of four-fermion operators. The sum, over operators of increasing dimension, has to be truncated at the desired accuracy counting the contribution of each matrix element in powers of $v$ and the suppression factor in $\alpha_s$ coming from the matching coefficient. Electromagnetic annihilation is treated in a similar way.

### *Color octet contributions*

A quarkonium state $|H\rangle$ in NRQCD is expanded in the number of partons

$$|H\rangle = |\overline{Q}Q\rangle + |\overline{Q}Qg\rangle + |\overline{Q}Q\bar{q}q\rangle + \cdots \tag{18.1.7}$$

where the states including one or more light parton are shown to be suppressed by powers of $v$. In $|\overline{Q}Qg\rangle$ for example the quark-antiquark are in a color octet state, since the in total the state may not carry any color, and thus the two quarks have to compensate the color of the gluon. Then both color singlet and color octet operators contribute in Eq. (18.1.6).

If one instead assumes that only heavy-quarkonium states with quark-antiquark in a color-singlet configuration can exist, then only color-singlet four-fermion operators can contribute and the matrix elements reduce to heavy-quarkonium wave functions (or derivatives of them) calculated at the origin. This assumption is known as the "color-singlet model".

Explicit calculations show that at higher order the color-singlet matching coefficients $c_n$ develop infrared divergences, *e.g.* for $P$-waves this takes place at order $\alpha_s$. In the color-singlet model, these singularities do not cancel in the expression of the decay widths. The first success of NRQCD (Bodwin, Braaten, and Lepage, 1995) was to show that the Fock space of a heavy-quarkonium state may contain a small component of quark-antiquark in a color-octet configuration, bound with some gluonic degrees of freedom. Due to this component, matrix elements of color-octet four-fermion operators contribute, and it is exactly these contributions that absorb the infrared divergences of the color-singlet matching coefficients in the decay widths, giving rise to finite results.

### *Quarkonium production*

The relevant scale for direct quarkonium production is also the hard scale $m$, so this process can be described by a local interaction in NRQCD, as we have done for inclusive decays. As a result, the inclusive cross section for the direct production of a quarkonium state $H$ at large momentum in the center-of-mass frame can be written as a sum of products of NRQCD matrix elements and short-distance coefficients:

$$\sigma[H] = \sum_n \sigma_n \langle \mathcal{K}_n^{\text{4fermions}} \rangle \tag{18.1.8}$$

where the $\sigma_n$ are short-distance coefficients, and the matrix elements $\langle \mathcal{K}_n^{\text{4fermions}} \rangle$ are vacuum-expectation values of objects similar to the four-fermion operators in decays, containing both color singlet and color octet contributions. This factorization formula for production however has been proven only at next-to-next-to-leading order in $\alpha_S$ (Nayak, Qiu, and Sterman, 2005). Interesting new developments are coming from defining fragmentation functions in the NRQCD formalism (Kang, Qiu, and Sterman, 2012) and using soft collinear theory (Fleming, Leibovich, Mehen, and Rothstein, 2012). The short-distance coefficients $\sigma_n(\Lambda)$ are essentially the process-dependent partonic cross sections to make a $Q\overline{Q}$ pair (convoluted with parton distributions if there are hadrons in the initial state). The $Q\overline{Q}$ pair can be produced in a color-singlet state or in a color-octet state. Its spin state can be singlet or triplet, and it can also have orbital angular momentum. The matrix elements $\langle \mathcal{K}_n^{\text{4fermions}} \rangle$ contain all of the non-perturbative physics associated with the evolution of the $Q\overline{Q}$ pair into a quarkonium state. An important property of these matrix elements, which greatly increases the predictive power of NRQCD, is the fact that they should be universal, *i.e.*, process independent. However, this is still an object of study. Again, NRQCD power-counting rules allow one to organize the sum over operators as an expansion in powers of $v$ so that through a given order in $v$, only a finite set of matrix elements contributes.

Aside from NRQCD, which is a QCD-based approach, models have also been used to study quarkonium production. One is the color singlet model, which can be related to NRQCD by retaining in Eq. (18.1.8) only the color singlet contributions at leading order in $v$. Another is the color evaporation model, where the $Q\overline{Q}$ pair only has to have a certain invariant mass close to the quarkonium mass, but the color of the $Q\overline{Q}$ state is assumed to "evaporate". However, both models lead to inconsistencies: see Brambilla et al. (2004, 2011) for a detailed review.

For a review of applications of NRQCD to quarkonium production at the $B$ Factories see Bodwin (2010, 2012) and Brambilla et al. (2004, 2011); for original calculations see Bodwin, Braaten, Lee, and Yu (2006); Bodwin, Kang, and Lee (2006); Bodwin, Lee, and Yu (2008); He, Fan, and Chao (2010); Li, He, and Chao (2009); Wang, Ma, and Chao (2011).

It is important to relate quarkonium production at $B$ Factories and at hadron colliders. For example, recently NLO order NRQCD calculations for the process $e^+e^- \to J/\psi + X(\text{non}-c\bar{c})$ have been carried out by Zhang, Ma, Wang, and Chao (2010) and by Butenschoen and Kniehl (2011); the latter authors rely on the extraction of the NRQCD production matrix elements obtained from a global fit to all production data. For a note on intepretation of $B$ Factory measurements of such cross sections, see Section 18.2.4.3.

Recently, factorization theorems have been obtained in two exclusive heavy-quarkonium production processes: production of two quarkonia in $e^+e^-$ annihilation and production of a quarkonium and a light meson in $B$-meson decays (Bodwin, Garcia i Tormo, and Lee, 2010).

#### 18.1.4.2 Physics at the scales $mv$, $mv^2$: pNRQCD

Quarkonium formation takes place at the scale $mv$. The suitable EFT is potential non-relativistic QCD, pNRQCD (Brambilla, Pineda, Soto, and Vairo, 2000, 2005; Pineda and Soto, 1998), which follows from NRQCD by integrating out the scale $mv \sim r^{-1}$. The soft scale $mv$ may be either larger or smaller than the confinement scale $\Lambda_{\rm QCD}$ depending on the radius of the quarkonium system. When $mv \gg \Lambda_{\rm QCD}$, we speak about weakly-coupled pNRQCD because the soft scale is perturbative and the matching from NRQCD to pNRQCD may be performed in perturbation theory. When $mv \sim \Lambda_{\rm QCD}$, we speak about strongly-coupled pNRQCD because the soft scale is non-perturbative and the matching from NRQCD to pNRQCD is non-perturbative and cannot be calculated with an expansion in $\alpha_s$.

It is generally assumed that the lowest levels of quarkonium, like $J/\psi$ and $\Upsilon(1S)$, may be described by weakly coupled pNRQCD, while the radii of the excited states are larger and presumably need to be described by strongly coupled pNRQCD. All this is valid for states away from open charm (bottom) threshold.

Close to threshold, many additional degrees of freedom become relevant and many more scales, which do not have a clear hierarchy, appear. Hence it will be difficult to devise an effective theory for this situation and thus one has presently to refer to models.

From pNRQCD one can also derive the $Q\overline{Q}$ QCD interaction potentials which may be used as an input for calculations of spectra on the basis of the Schrödinger Equation. In this way one can obtain QCD-based information on the spectra of heavy quarkonium systems.

*The case $mv \gg \Lambda_{\rm QCD}$: weakly-coupled pNRQCD*

The effective Lagrangian is organized as an expansion in $1/m$ and $\alpha_s(m)$, inherited from NRQCD, and an expansion in $r$ (Brambilla, Pineda, Soto, and Vairo, 2000):

$$\mathcal{L}_{\rm pNRQCD} = \int d^3r \sum_n \sum_k \frac{c_n(\alpha_s(m),\mu)}{m^n} \times V_{n,k}(r,\mu',\mu)\, r^k \times O_k(\mu', mv^2, ...), \quad (18.1.9)$$

where $O_k$ are the operators of pNRQCD. The matrix elements of these operators depend on the low-energy scale $mv^2$ and $\mu'$, where $\mu'$ is the pNRQCD factorization scale. The $V_{n,k}$ are the Wilson coefficients of pNRQCD that encode the contributions from the scale $r$ and are non-analytic in $r$. The $c_n$ are the NRQCD matching coefficients as given in Eq. (18.1.3).

The degrees of freedom, which are relevant below the soft scale, and which appear in the operators $O_k$, are $Q\overline{Q}$ states (a color-singlet $S$ and a color-octet $O = O_a T^a$ state) and (ultrasoft) gluon fields, which are expanded in $r$ as well (multipole expanded). Looking at the equations of motion of pNRQCD, we may identify $V_{n,0} = V_n$ with the $1/m^n$ potentials that enter the Schrödinger equation and $V_{n,k\neq 0}$ with the couplings of the ultrasoft degrees of freedom, which provide corrections to the Schrödinger equation. Since the degrees of freedom that enter the Schrödinger description are in this case both $Q\overline{Q}$ color singlet and $Q\overline{Q}$ color octets, both singlet and octet potentials exist. Nonpotential interactions, associated with the propagation of low-energy degrees of freedom are, in general, present as well, and start to contribute at NLO in the multipole expansion. They are typically related to non-perturbative effects.

If the quarkonium system is small ($r \ll \Lambda_{\rm QCD}$), the soft scale is perturbative and the potentials can be calculated in perturbation theory, *i.e.* no non-perturbative quantities enter the potential (Brambilla, Pineda, Soto, and Vairo, 2005). Being matching coefficients of the effective field theory, the potentials undergo renormalization, develop a scale dependence and satisfy renormalization group equations, which allow the resummation of large logarithms having as arguments ratios of physical scales, such as $\log(\frac{mv^2}{mv}) = \log(v)$.

*The case $mv \sim \Lambda_{\rm QCD}$: strongly-coupled pNRQCD*

When $mv \sim \Lambda_{\rm QCD}$ the soft scale is non-perturbative, and matching cannot be performed in perturbation theory any more. Rather the potential matching coefficients $V_{n,k}$ are obtained as non-perturbative quantities in the form of expectation values of gauge invariant Wilson-loop operators. In this case, under certain assumptions, the quarkonium singlet field $S = \overline{Q}Q$ is the only low-energy dynamical degree of freedom in the pNRQCD Lagrangian which reads (Brambilla, Pineda, Soto, and Vairo, 2001, 2005; Pineda

and Vairo, 2001):

$$\mathcal{L}_{\mathrm{pNRQCD}} = \int d^3r\, S^\dagger \left( i\partial_0 - \frac{\mathbf{p}^2}{2m} - V_S(r) \right) S\,. \tag{18.1.10}$$

The singlet potential $V_S(r)$ is a series in the expansion in the inverse of the quark masses; the terms up to $1/m^2$ have been calculated long ago (Brambilla, Pineda, Soto, and Vairo, 2001; Pineda and Vairo, 2001). They involve NRQCD matching coefficients (containing the contribution from the hard scale) and low-energy non-perturbative parts given in terms of static Wilson loops and field strength insertions in the static Wilson loop (containing the contribution from the soft scale).

In this regime, from pNRQCD we recover the quark potential singlet model. However, here the potentials are obtained from QCD by non-perturbative matching and they often appear to have a different form with respect to phenomenological potential models. Their evaluation requires calculations on the lattice or in QCD vacuum models. Recent progress includes new precise lattice calculations of these potentials (Koma, Koma, and Wittig, 2008; Koma and Koma, 2010). Using these potentials, all the masses for heavy quarkonia away from threshold can be obtained by the solution of the Schrödinger equation.

A trivial example of application of this method is the mass of the $h_c$. The lattice data show a vanishing long-range component of the spin-spin potential so that the potential appears to be entirely dominated by its short-range, delta-like part. This suggests that the $^1P_1$ state should be close to the center-of-gravity of the $^3P_J$ system. Indeed, the measurements show consistency between data and this expected value (see experimental results in Table 18.2.1).

### 18.1.5 Lattice calculations

Lattice calculations play a key role for quarkonium physics. For an introduction to the lattice treatment of quarkonia see Brambilla et al. (2004) and for recent results see Brambilla et al. (2011). We already mentioned that the recent progress in this field relies both on high order perturbative calculations and on lattice simulations, the results of the two being often combined inside the EFT framework.

In fact it is difficult to put a multiscale system on the lattice as the lattice step should be smaller than the smallest scale ($m^{-1}$) and the lattice size should be bigger than the biggest scale of the system $\Lambda_{\mathrm{QCD}}^{-1}$, putting prohibitive requirements on the lattice dimensions. This is true in particular for bottomonium, due to its larger mass. In this case one could use direct anisotropic lattice simulations or EFTs. The Lagrangian of NRQCD can be put on the lattice and used to obtain quarkonium energy levels. Recent results can be found in Daldrop, Davies, and Dowdall (2012), Dowdall et al. (2012), Gregory et al. (2011), and Donald et al. (2012). Charmonium spectra may also be calculated on the lattice with relativistic actions. Very recently new lattice techiques have been introduced that will eventually allow the excited charmonium spectroscopy to be obtained from the lattice (Bali, Collins, and Ehmann, 2011; Liu et al., 2012). Another possibility is to evaluate on the lattice the potentials of strongly coupled pNRQCD and use them inside a Schrödinger equation to obtain all the quarkonium energy levels. New precise quenched lattice calculations of these potentials obtained using the Lüscher multilevel algorithm have recently become available (Koma, Koma, and Wittig, 2008; Koma and Koma, 2010).

### 18.1.6 Applications

A large set of phenomenological applications of the EFT framework outlined above to quarkonium spectra, decays, and production has been presented elsewhere (Brambilla, Pineda, Soto, and Vairo, 2005; Brambilla et al., 2004, 2011), and discussed in relation to experimental data. Here we briefly recall some selected results.

In the regime in which the soft scale $mv$ is perturbative the energy levels of quarkonium have been calculated at order $m\alpha_S^5$ (Brambilla, Pineda, Soto, and Vairo, 1999; Kniehl, Penin, Smirnov, and Steinhauser, 2002).

Decay amplitudes (Brambilla, Pineda, Soto, and Vairo, 2005; Brambilla et al., 2011; Kiyo, Pineda, and Signer, 2010) and production and annihilation (Beneke, Kiyo, and Penin, 2007) have been calculated in perturbation theory at high order. Since for systems with a small radius the non-perturbative contributions are power suppressed, it is possible to obtain a good determinations of the masses of the lowest quarkonium resonances with purely perturbative calculations in the cases in which the perturbative series is convergent (after the appropriate subtractions of renormalons have been performed) and large logarithms in the scale ratios are resummed. For example, in Brambilla and Vairo (2000) a prediction of the $B_c$ mass[98] has been obtained: $(6326\,^{+29}_{-9})$ MeV/c$^2$, to be compared to the experimental value of $(6277 \pm 6)$ MeV/c$^2$ (Beringer et al., 2012). An NNLO calculation with finite charm mass effects (Brambilla, Sumino, and Vairo, 2002) predicts a mass that well matches the Fermilab measurement (Brambilla et al., 2011) and the lattice determination (Allison et al., 2005). The same procedure has been applied at NNLO even for higher states (Brambilla, Sumino, and Vairo, 2002). An NLO calculation reproduces in part the $1P$ fine splitting (Brambilla and Vairo, 2005). Including log resummation at NLL, it is possible to obtain a prediction for the $B_c$ hyperfine separation $\Delta = 50 \pm 17(\mathrm{th})^{+15}_{-12}(\delta\alpha_S)$ MeV/c$^2$ (Penin, Pineda, Smirnov, and Steinhauser, 2004) and for the hyperfine separation between the $\Upsilon(1S)$ and the $\eta_b$ the value of $41 \pm 11(\mathrm{th})^{+9}_{-8}(\delta\alpha_S)$ MeV/c$^2$ (where the second error comes from the uncertainty in $\alpha_S$; Kniehl, Penin, Pineda, Smirnov, and Steinhauser, 2004). This last value turned out to considerably undershoot the measurements of *BABAR* (Aubert, 2008ak, 2009l) and Belle (Mizuk, 2012).

[98] The $B_c$ states constitute a separate class of heavy mesons, distinct both from the quarkonia (*e.g.* in lacking electromagnetic and strong decays) and from the simpler heavy mesons (in lacking light valence quarks). Because of their masses they lie outside the scope of research at the $B$ Factories.

NRQCD lattice calculations (Gray et al., 2005) obtained a value close to the experimental one but did not include the calculation of the matching coefficient at one loop. Recent lattice calculations (Hammant, Hart, von Hippel, Horgan, and Monahan, 2011) aim at including the NRQCD matching coefficients in the NRQCD lattice calculation and will help to settle this issue, see for example the result contained on the $\eta_b$ mass in Dowdall et al. (2012). See also the result contained in Meinel (2010). The hyperfine separation of $B_c$ has been calculated on the lattice to be $M_{B_c^*} - M_{B_c} = 54(3)$ MeV/c$^2$ in Dowdall, Davies, Hammant, and Horgan (2012). In the same paper values for the excited energy levels of $B_c$ have been presented.

An EFT of the magnetic dipole transition has been given in Brambilla, Jia, and Vairo (2006), allowing magnetic dipole transitions between $c\bar{c}$ and $b\bar{b}$ ground states to been considered in pNRQCD at NNLO. The results are: $\Gamma(J/\psi \to \gamma\,\eta_c) = (1.5 \pm 1.0)$ keV and $\Gamma(\Upsilon(1S) \to \gamma\,\eta_b) = (k_\gamma/71\ \text{MeV})^3\ (15.1 \pm 1.5)$ eV, where the errors account for uncertainties coming from higher-order corrections. The width $\Gamma(J/\psi \to \gamma\,\eta_c)$ is consistent with the world-average value (Beringer et al., 2012) but bears a large error. Working in the same formalism but exactly incorporating the perturbative static potential in the leading order Hamiltonian and resumming large logarithms in the mass scale in Pineda and Segovia (2013) a number of M1 transitions have been calculated. In particular, the values $\Gamma(J/\psi \to \gamma\eta_c) = 2.12(40)$ keV has been obtained, which is in agreement with the experimental determination with a smaller error, and $\Gamma(\Upsilon(1S \to \gamma\eta_b) = 15.18(51)$ eV and $\Gamma(\Upsilon(2S \to \gamma\eta_b) = 0.668(60)$ eV has been obtained. The transition $\Gamma(J/\psi \to \gamma\eta_c)$ and the $J/\psi$ annihilation constant have been evaluated on the lattice in Becirevic and Sanfilippo (2013) and in Davies et al. (2012). The quarkonium magnetic moment is explicitly calculated in Brambilla, Jia, and Vairo (2006) and turns out to be very small in agreement with a recent lattice calculation (Dudek, Edwards, and Richards, 2006); the M1 transition of the lowest quarkonium states at relative order $v^2$ turn out to be completely accessible in perturbation theory (Brambilla, Jia, and Vairo, 2006). A theory of electric dipole transitions has been given in Brambilla, Pietrulewicz, and Vairo (2012) and Pietrulewicz (2012), reproducing some of the results of the phenomenological potential models with some important differences.

A description of the $\eta_c$ line shape has been given in Brambilla, Roig, and Vairo (2011). Using pNRCD and Soft Collinear EFT (SCET) a good description of the $\Upsilon(1S)$ radiative decay has been obtained (Garcia i Tormo and Soto, 2007).

Concerning decays, substantial progress has recently been made in the evaluation of the NRQCD factorization formula for inclusive decays at order $v^7$ (Brambilla, Mereghetti, and Vairo, 2006, 2009), in the lattice evaluation of the NRQCD matrix elements (Bodwin, Lee, and Sinclair, 2005), and in the higher order perturbative calculation of some NRQCD matching coefficients (Guo, Ma, and Chao, 2011; Jia, Yang, Sang, and Xu, 2011; Li, Ma, and Chao, 2013). The data are clearly sensitive to NLO corrections in the Wilson coefficients and presumably also to relativistic corrections. Improved theory predictability would entail the lattice calculation or data extraction of the NRQCD matrix elements and perturbative resummation of large contribution in the NRQCD matching coefficients. The $J/\psi \to 3\gamma$ decay has been studied in NRQCD in Feng, Jia, and Sang (2012). Inclusive decay amplitudes have been calculated in pNRQCD (Brambilla, Eiras, Pineda, Soto, and Vairo, 2003) and the number of non-perturbative correlators appears to be sizeably reduced with respect to NRQCD so that new model-independent predictions have been made possible (Brambilla, Eiras, Pineda, Soto, and Vairo, 2002). Still, the new data on hadronic transitions and hadronic decays pose interesting challenges to the theory. Exclusive decay modes are more difficult to address in theory (He, Lu, Soto, and Zheng, 2011; Soto, 2011; Vairo, 2004).

For excited states with masses away from threshold, phenomenological applications of the QCD potentials obtained in Brambilla, Pineda, Soto, and Vairo (2001) and Pineda and Vairo (2001) are ongoing (Laschka, Kaiser, and Weise, 2011). For a full phenomenological description of the spectra and decays it would be helpful to have updated, more precise and unquenched lattice calculation of the Wilson loop field strength insertion expectation values and of the local and nonlocal gluon correlators (Brambilla, Pineda, Soto, and Vairo, 2005). For recent lattice results on the spectroscopy see Gregory et al. (2011).

In the most interesting region, the region close to threshold where many new (possibly exotic) states have recently been discovered, a full EFT description has not yet been constructed nor the appropriate degrees of freedom clearly identified (Brambilla, Vairo, Polosa, and Soto, 2008; Brambilla et al., 2011). An exception is the $X(3872)$, which displays universal characteristics related to its being so close to threshold, allowing a beautiful EFT description to be obtained (Braaten, 2009; Braaten and Kusunoki, 2004). The light quark mass dependence in quarkonium has been studied in Guo and Meissner (2012).

The threshold region remains troublesome also for the lattice, although several excited state calculations have recently been pionereed.

## 18.2 Conventional charmonium

***Editors:***
*Riccardo Faccini (BABAR)*
*Pasha Pakhlov (Belle)*
*Nora Brambilla (theory)*

**Additional section writers:**
*Pietro Biassoni, Galina Pakhlova, Antimo Palano, Torsten Schroeder, Korneliy Todyshev, Timofey Uglov, Anna Vinokurova, Bruce Yabsley*

Since its discovery in 1974 the charmonium family has served as a laboratory to test strong interactions.

After the discovery of the first charmonium state, the $J/\psi$, its radial excitation, the $\psi(2S)$, was found just two weeks later, and another eight states were discovered within the subsequent five years. More than half of the charmonium states known by 1980 were observed in $e^+e^-$ annihilation, while the others were found in the decays of $J/\psi$ or $\psi(2S)$. Next, the decays of the ten known charmonia were studied in detail; theoretical frameworks used to compute the masses and widths for charmonium states evolved over the years from purely phenomenological approaches to the effective field and lattice gauge theories that are the current state of the art.

The known conventional charmonium states are listed in Table 18.2.1, together with predictions for their masses from the potential models described in Section 18.1.2 above. For accounts of effective field theory (NRQCD) and lattice approaches to charmonium, see Sections 18.1.4 and 18.1.5 respectively. In the following material, we present the experimental results on charmonia from the $B$ Factories: the observation of four new states (Section 18.2.1), and new decay modes of well known states (Section 18.2.2); more precise determinations of the parameters of some charmonium states (Section 18.2.3); and the various mechanisms of charmonium production at the $B$ Factories (Section 18.2.4). Some concluding remarks are provided in Section 18.2.5.

### 18.2.1 New conventional charmonium states

18.2.1.1 $\eta_c(2S)$

In the heavy quark potential model the $\eta_c(2S)$, the first radial excitation of the charmonium ground state $\eta_c$, is predicted to lie below the $D\overline{D}$ threshold (Buchmüller and Tye, 1981; Ebert, Faustov, and Galkin, 2000; Eichten and Feinberg, 1981; Eichten and Quigg, 1994; Godfrey and Isgur, 1985). Calculations within this model predict a mass splitting $m_{\psi(2S)} - m_{\eta_c(2S)}$ in the range $(42-103)\,\text{MeV}/c^2$.

In 1982 the Crystal Ball Collaboration reported evidence of a signal in $\psi(2S)$ radiative decay attributed to the $\eta_c(2S)$ with a mass of $(3594 \pm 5)\,\text{MeV}/c^2$ (Edwards et al., 1982). This claim remained unconfirmed and unrefuted for about 20 years until the observation of the $\eta_c(2S)$ at the $B$ Factories.

*Exclusive observation in B decays*

The first modern evidence for the $\eta_c(2S)$ was the observation of a significant peak in the $K^0_S K^\pm \pi^\mp$ mass spectrum, near the mass $3.65\,\text{GeV}/c^2$, in $B \to K^0_S K^\pm \pi^\mp K$ decays at Belle (Choi, 2002). In this analysis the $B$ candidates are exclusively reconstructed, and $B$ meson signal events are distinguished from continuum $q\overline{q}$ background by using a likelihood ratio combining event-shape variables (see Chapter 9). To suppress potential backgrounds from $B \to D_{(s)}X$ decays, combinations with any $K\pi$ ($K^0_S K$) pairs lying near the $D$ ($D^+_s$) nominal masses are vetoed. To extract the number of signal $B \to K^0_S K^\pm \pi^\mp K$ decays as a function of $M_{K^0_S K^\pm\pi^\mp}$, Belle performs fits to the $m_{\rm ES}$ and $\Delta E$ distributions in bins of $M_{K^0_S K^\pm\pi^\mp}$ with $40\,\text{MeV}/c^2$ width. The signal yields obtained from these fits are plotted in Fig. 18.2.1, where in addition to a prominent $\eta_c$ signal, another significant peak is evident at higher mass. This spectrum is fitted to a sum of $\eta_c$ and $\eta_c(2S)$ Breit-Wigner signal components and a polynomial background. The signal functions are convolved with a Gaussian representing the detector resolution function. The fitted $\eta_c(2S)$ parameters are reported in Table 18.2.2.

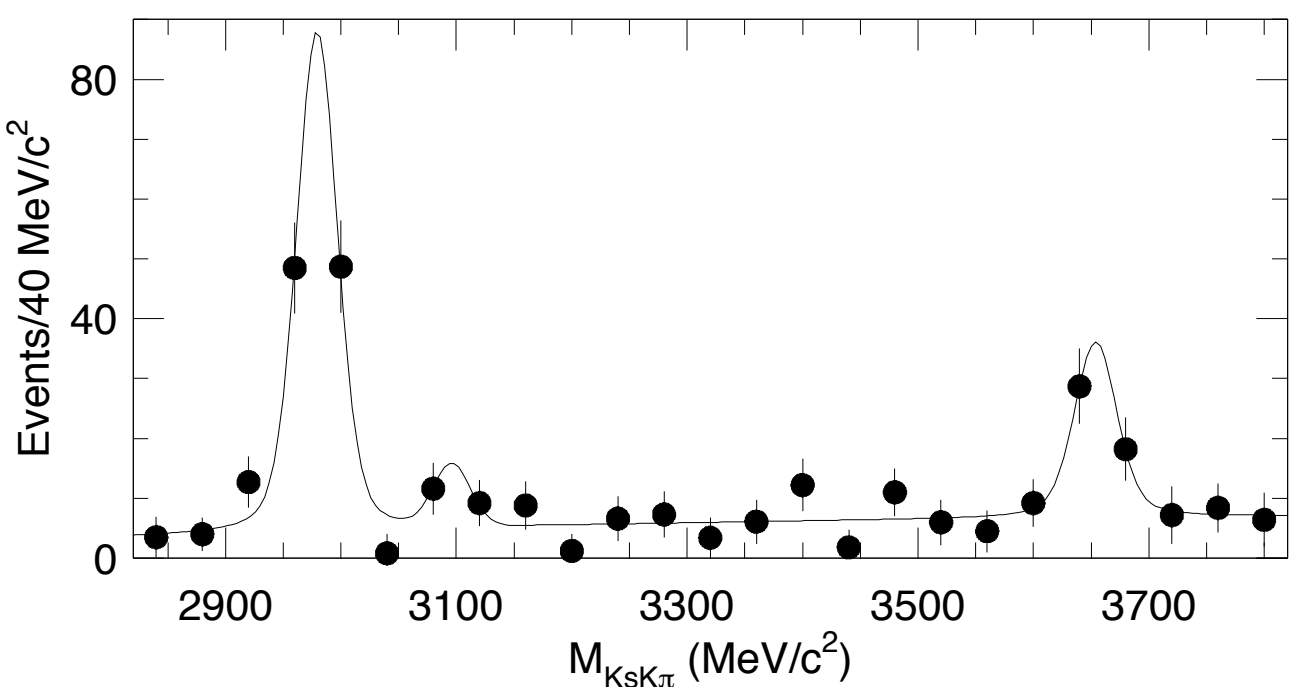

**Figure 18.2.1.** $K^0_S K^\pm \pi^\mp$ invariant mass distribution for $B \to K^0_S K^\pm \pi^\mp K$ signal events with the $\eta_c$ and $\eta_c(2S)$ mass peaks visible (Choi, 2002). The solid line represents the fit function.

*BABAR* follows a similar approach in the analysis of $B$ decays to $K\overline{K}\pi K^{(0)}$ final states (Aubert, 2008ba). The $K\overline{K}\pi$ system is reconstructed in $K^0_S K^\pm \pi^\mp$ and $K^+K^-\pi^0$ final states, and the sum over these two modes is reported in the results. Signal $B$ mesons are selected by applying a tight $\Delta E$ requirement. The $K\overline{K}\pi$ invariant mass distribution from continuum-background events is extrapolated from the $m_{\rm ES}$ sidebands. In the fit to the $K\overline{K}\pi$ background-subtracted distribution the $\eta_c(2S)$ mass and width are fixed to world-average values, 3637 MeV and 14 MeV respectively (Yao et al., 2006). The measured $B^+ \to \eta_c(2S)K^+$ yield is 59±12. By using $\mathcal{B}(B^+ \to \eta_c(2S)K^+) = (3.4 \pm 1.8) \times 10^{-4}$ (Aubert, 2006ae; see the details of this analysis under "Inclusive $B$ decays at *BABAR*" below), *BABAR* measures the absolute branching ratio for $\eta_c(2S) \to K\overline{K}\pi$ to be $\mathcal{B}(\eta_c(2S) \to K\overline{K}\pi) = (1.9 \pm 0.4 \pm 0.5 \pm 1.0)\%$, where the first error is statistical, the second systematic

**Table 18.2.1.** Charmonium masses (MeV/$c^2$) according to potential models, compared with the observed values (Beringer et al., 2012). The states observed by the $B$ Factories and the CLEO collaboration after 2002 are marked with $^*$: see Sections 18.2.1.1 ($\eta_c(2S)$), 18.2.1.2 ($\chi_{c2}(2P)$), and 18.2.1.3 ($\eta_c(3S)$ and $\eta_c(4S)$). The model names are built from the first letters of the authors and the year: GI85 (Godfrey and Isgur, 1985); EG94 (Eichten and Quigg, 1994); F91 (Fulcher, 1991); GJ95 (Gupta and Johnson, 1996); EFG02 (Ebert, Faustov, and Galkin, 2003); ZVR94 (Zeng, Van Orden, and Roberts, 1995); BGS05 (Barnes et al., 2005).

| State | | $J^{PC}$ | Experiment | GI85 | EG94 | F91 | GJ95 | EFG02 | ZVR94 | BGS05 |
|---|---|---|---|---|---|---|---|---|---|---|
| $1\,^1S_0$ | $\eta_c$ | $0^{-+}$ | $2981.0 \pm 1.1$ | 2975 | 2980 | 2987 | 2979 | 2979 | 3000 | 2982 |
| $1\,^3S_1$ | $J/\psi$ | $1^{--}$ | 3096.9 | 3098 | 3097 | 3104 | 3097 | 3096 | 3100 | 3090 |
| $1\,^1P_1$ | $h_c$ | $1^{+-}$ | $3525.41 \pm 0.16$ | 3517 | 3493 | 3529 | 3526 | 3526 | 3510 | 3516 |
| $1\,^3P_0$ | $\chi_{c0}$ | $0^{++}$ | $3414.75 \pm 0.31$ | 3445 | 3436 | 3404 | 3415 | 3424 | 3440 | 3424 |
| $1\,^3P_1$ | $\chi_{c1}$ | $1^{++}$ | $3510.66 \pm 0.07$ | 3510 | 3486 | 3513 | 3511 | 3510 | 3500 | 3505 |
| $1\,^3P_2$ | $\chi_{c2}$ | $2^{++}$ | $3556.20 \pm 0.09$ | 3550 | 3507 | 3557 | 3557 | 3556 | 3540 | 3556 |
| $2\,^1S_0$ | $\eta_c(2S)$ | $0^{-+}$ | $3637 \pm 4$ $^*$ | 3623 | 3608 | 3584 | 3618 | 3588 | 3670 | 3630 |
| $2\,^3S_1$ | $\psi(2S)$ | $1^{--}$ | $3686.09 \pm 0.04$ | 3676 | 3686 | 3670 | 3686 | 3686 | 3730 | 3672 |
| $1\,^1D_2$ | $\eta_{c2}$ | $2^{-+}$ | | 3837 | | 3872 | | 3811 | 3820 | 3799 |
| $1\,^3D_1$ | $\psi(3770)$ | $1^{--}$ | $3772.92 \pm 0.35$ | 3819 | | 3840 | | 3798 | 3800 | 3785 |
| $1\,^3D_2$ | $\psi_2$ | $2^{--}$ | | 3838 | | 3871 | | 3813 | 3820 | 3800 |
| $1\,^3D_3$ | $\psi_3$ | $3^{--}$ | | 3849 | | 3884 | | 3815 | 3830 | 3806 |
| $2\,^1P_1$ | $h_c(2P)$ | $1^{+-}$ | | 3956 | | | | 3945 | 3990 | 3934 |
| $2\,^3P_0$ | $\chi_{c0}(2P)$ | $0^{++}$ | | 3916 | | | | 3854 | 3940 | 3852 |
| $2\,^3P_1$ | $\chi_{c1}(2P)$ | $1^{++}$ | | 3953 | | | | 3929 | 3990 | 3925 |
| $2\,^3P_2$ | $\chi_{c2}(2P)$ | $2^{++}$ | $3927.2 \pm 2.6$ $^*$ | 3979 | | | | 3972 | 4020 | 3972 |
| $3\,^1S_0$ | $\eta_c(3S)$ | $0^{-+}$ | $3942 \pm 9$ $^*$ | 4064 | | | | 4130 | 3991 | 4043 |
| $3\,^3S_1$ | $\psi(3S)$ | $1^{--}$ | $4039 \pm 1$ | 4100 | | | | 4180 | 4088 | 4072 |
| $2\,^3D_1$ | $\psi(2D)$ | $1^{--}$ | $4153 \pm 3$ | 4194 | | | | | | 4142 |
| $4\,^1S_0$ | $\eta_c(4S)$ | $0^{-+}$ | $4156\,^{+29}_{-25}$ $^*$ | 4425 | | | | | | 4384 |
| $4\,^3S_1$ | $\psi(3S)$ | $1^{--}$ | $4421 \pm 4$ | 4450 | | | | | | 4406 |

and the third is due to the uncertainty of the branching fractions used in the calculation.

Recently, Belle has updated the $\eta_c(2S)$ measurement in the decays $B^+ \to (K^0_S K^\pm \pi^\mp)K^+$ by using a much larger data sample (Vinokurova, 2011). Besides improving the statistical accuracy, this analysis accounts for $\eta_c(2S)$ interference with the non-resonant continuum for the first time in a model-independent way, thus providing more reliable measurements of the $\eta_c(2S)$ mass and width, and the branching ratio for the $B^+ \to \eta_c(2S)K^+$ decay. Indeed the decays $B^+ \to K^0_S K^\pm \pi^\mp K^+$ can occur without proceeding via a charmonium state; the amplitude for such decays can interfere with the $\eta_c(2S)$ signal, which has a non-vanishing width. Different values of the interference phase can result in different $\eta_c(2S)$ resonance line shapes, and can lead to significant variations in the number of $\eta_c(2S)$ events while the total number of observed events in the $\eta_c(2S)$ peak remains the same. In this study Belle jointly analyzes the $M_{K^0_S K^\pm \pi^\mp}$ spectrum and the distribution of the angle $\theta$ between the $K^0_S$ and $K^+$ in the rest frame of the $K^0_S K^\pm \pi^\mp$ system. The angular analysis provides discrimination between the component of the non-resonant amplitudes that interfere with the signal and the one that does not. As can be seen from Fig. 18.2.2, the interference deforms the Breit-Wigner, making it asymmetric and lengthening its tail; the angular distribution in the $\eta_c(2S)$ signal region is dominated by $S$-wave,[99] while in the $\eta_c(2S)$ sidebands a sum of $S$-, $P$-, and $D$-waves is visible. The fitted $\eta_c(2S)$ mass and width are listed in Table 18.2.2. Taking interference into account has a dramatic effect on the measured $\eta_c(2S)$ parameters: if interference is ignored, Belle finds a $\sim 10\,\mathrm{MeV}/c^2$ upward mass shift, while the $\eta_c(2S)$ width increases by more than a factor 6. The measured product of branching fractions $\mathcal{B}(B^\pm \to K^\pm \eta_c(2S)) \times \mathcal{B}(\eta_c(2S) \to K^0_S K^\pm \pi^\mp)$ is equal to $(3.4\,^{+2.2}_{-1.5}\,^{+0.5}_{-0.4}) \times 10^{-6}$.

[99] Since the $\eta_c(2S)$ is a pseudoscalar, one expects a uniform distribution in $\cos\theta$ for pure $\eta_c(2S)$ decay (pure $S$-wave). The signal region also contains non-resonant background, but the $\eta_c(2S)$ component is much larger, so the $S$-wave contribution here is dominant.

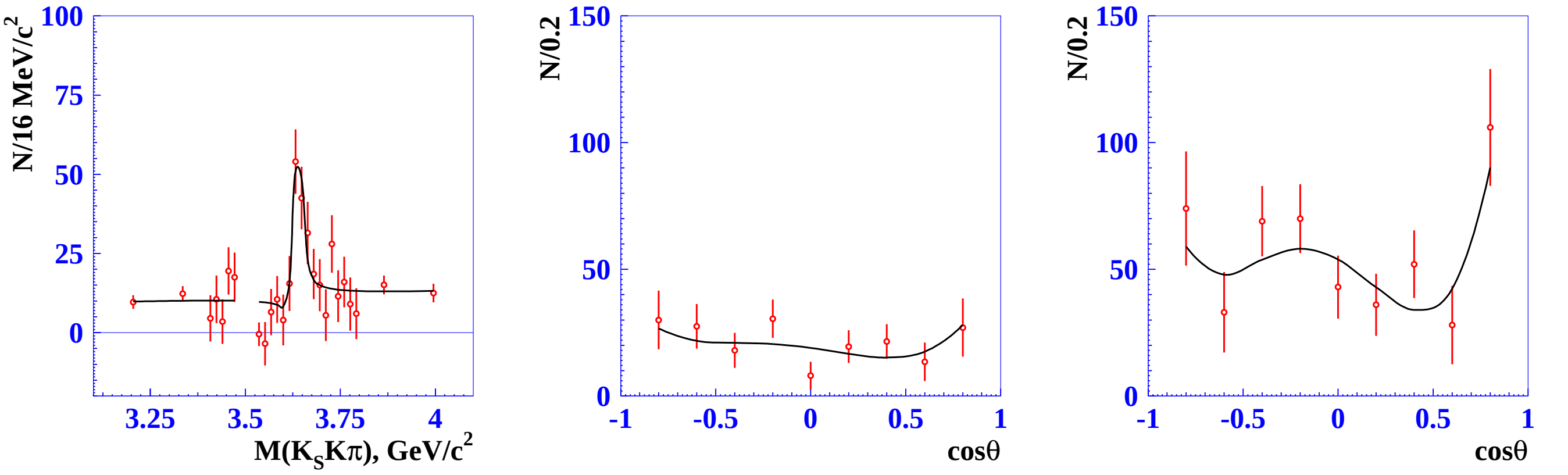

**Figure 18.2.2.** From the Belle $B \to K^0_S K^\pm \pi^\mp K$ analysis Vinokurova (2011): Projections of a 2D-fit onto the $M_{K^0_S K^\pm \pi^\mp}$ axis (left) and onto the $\cos\theta$ axis in the $\eta_c(2S)$ signal (center) and sideband (right) regions. The combinatorial background is subtracted. In the mass plot, the change in binning between the signal ($16\,\mathrm{MeV}/c^2$) and sideband regions ($130\,\mathrm{MeV}/c^2$) is evident; the gap near $3.5\,\mathrm{GeV}/c^2$ is due to a veto of the $\chi_{c1}$ region, $M_{K^0_S K^\pm \pi^\mp} \in [3.48, 3.54]\,\mathrm{GeV}/c^2$.

*Exclusive observation in two-photon fusion*

The $\eta_c(2S)$ decay into the $K^0_S K^\pm \pi^\mp$ final state has also been studied by *BABAR* in the two-photon production process (Aubert, 2004s; del Amo Sanchez, 2011h). In the latter analysis the $\eta_c(2S)$ signal has also been observed in the $K^+K^-\pi^+\pi^-\pi^0$ final state. Events produced via two-photon fusion (see Chapter 22) are selected with requirements on the total number of charged and neutral particles in the event, and transverse momentum $p_T$ of the reconstructed hadronic final state. In addition, to suppress ISR background the missing mass squared is required to be greater than $2\,\mathrm{GeV}^2/c^4$ ISR events are expected to show a peak at $m^2_{\rm miss} \sim 0\,\mathrm{GeV}^2/c^4$, while in two-photon events the missing mass should be large due to the large momentum taken away by the outgoing $e^+e^-$. The invariant mass distributions are fitted to a sum of non-relativistic Breit-Wigner functions to model $\eta_c$, $\chi_{c0}$, $\chi_{c2}$, and $\eta_c(2S)$ signals plus a polynomial shape to describe combinatorial background. In the $K^0_S K^\pm \pi^\mp$ fit a $\chi_{c0}$ component is not included, since a $J^P = 0^+$ resonance cannot decay to this final state, due to angular momentum and parity conservation.[100] The signal shapes are convolved with the detector resolution function obtained from MC simulation. The width of the $\eta_c(2S)$ in the fit to the $K^+K^-\pi^+\pi^-\pi^0$ invariant mass distribution is fixed to the value found in the $K^0_S K^\pm \pi^\mp$ decay mode. Results of the fit are reported in Table 18.2.2 and shown in Fig. 18.2.3. Although the interference of the $\eta_c(2S)$ state with the non-resonant $\gamma\gamma \to K^0_S K^\pm \pi^\mp (K^+K^-\pi^+\pi^-\pi^0)$ continuum may shift the measured parameters, this effect can not be determined in this analysis due to the small signal to background ratio. The $p_T$ distribution is found to be consistent with that expected for two-photon production.

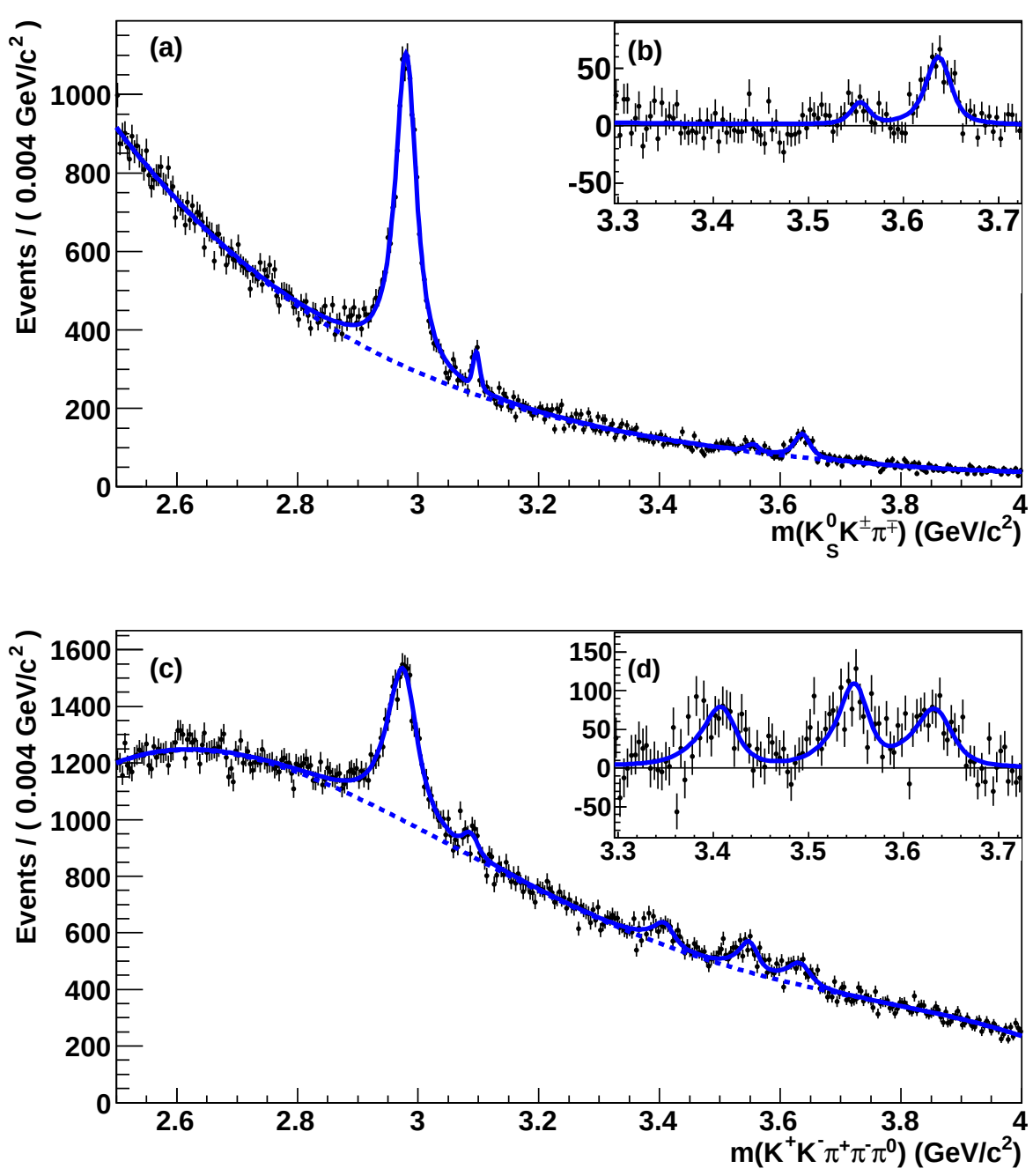

**Figure 18.2.3.** Fit to (a) the $K^0_S K^\pm \pi^\mp$ and (c) the $K^+K^-\pi^+\pi^-\pi^0$ mass spectra in two-photon fusion events (del Amo Sanchez, 2011h). The solid curves represent the total fit functions and the dashed curves show the combinatorial background contributions. The background-subtracted distributions are shown in (b) and (d), where the solid curves indicate the signal components. The prominent peak is the $\eta_c$, while the small peak above $3\,\mathrm{GeV}/c^2$ is due to residual ISR $J/\psi$ production. The peaks in the insets are (left to right) $\chi_{c0}$, $\chi_{c2}$, and $\eta_c(2S)$. The $\chi_{c0}$ peak is not present in (a) and (b), since the decay to $K^0_S K^\pm \pi^\mp$ is not allowed for this state.

[100] The decay proceeds via the strong interaction, so $P$ is conserved. In the $K^0_S K^\pm \pi^\mp$ system, let $l_1$ be the angular momentum between $K^0_S$ and $K^\pm$, and $l_2$ the angular momentum between $\pi^\mp$ and the $K^0_S K^\pm$ system. The final state has $P = (-1)^{1+l_1+l_2}$. Since the final state spin is equal to 0, $\boldsymbol{J} = \boldsymbol{l}_1 + \boldsymbol{l}_2$. Thus $J = 0$ implies $l_1 = l_2$ and $P = -1$.

Angular analysis of the $\eta_c(2S) \to K^0_S K^\pm\pi^\mp$ decay (Aubert, 2004s) confirms that the observed events are consistent with the two-photon production mechanism and inconsistent with production via ISR. This analysis also restricts the allowed $J^{PC}$ values for the final state (Yang, 1950) to be $0^{-+}$ or $J \geq 2$. The measured mass and the probable $J^{PC} = 0^{-+}$ assignment support the interpretation of the observed signal as the $\eta_c(2S)$ resonance. These results are consistent with those previously obtained by CLEO with a similar analysis of the $\gamma\gamma \to K^0_S K^\pm\pi^\mp$ process (Asner et al., 2004a). *BABAR* has also measured the product of the two-photon coupling, $\Gamma_{\gamma\gamma}$, and the final state branching fractions, $\mathcal{B}$. This quantity is related to the ratio of the resonance signal yield and the detection efficiency. In order to reduce the systematic uncertainty due to the unknown resonant substructure of the decays, a weighted fit to the efficiency-corrected mass distribution is performed, taking into account the dependence of the efficiency on the $K^0_S K^\pm\pi^\mp (K^+K^-\pi^+\pi^-\pi^0)$ decay kinematics. The values $\Gamma_{\gamma\gamma}(\eta_c(2S)) \times \mathcal{B}(\eta_c(2S) \to K\overline{K}\pi) = (41 \pm 4 \pm 6)\,\text{eV}$ and $\Gamma_{\gamma\gamma}(\eta_c(2S)) \times \mathcal{B}(\eta_c(2S) \to K^+K^-\pi^+\pi^-\pi^0) = (30 \pm 6 \pm 5)\,\text{eV}$, and the ratio between the $\eta_c(2S)$ branching fractions to the final states $\frac{\mathcal{B}(\eta_c(2S)\to K^+K^-\pi^+\pi^-\pi^0)}{\mathcal{B}(\eta_c(2S)\to K^0_S K^\pm\pi^\mp)} = 2.2 \pm 0.5 \pm 0.5$ are found (del Amo Sanchez, 2011h).

*BABAR* has also searched for the $\eta_c(2S)$ in the process $\gamma\gamma \to \eta_c\pi^+\pi^-$, with the $\eta_c$ decaying to $K^0_S K^\pm\pi^\mp$ (Lees, 2012t). The analysis uses a two-dimensional fit in the variables $m(K^0_S K^\pm\pi^\mp)$ and $m(K^0_S K^\pm\pi^\mp\pi^+\pi^-)$. Signal events peak in both of these quantities. The combinatorial background is expected to be distributed smoothly, with significant correlation between the distributions over these two variables. The correlation, studied with $\eta_c$ sidebands, is found to be consistent with being due to phase space. Utilizing this enables precise determination of the combinatorial background shape from the data. The analysis also accounts for backgrounds that peak in either $m(K^0_S K^\pm\pi^\mp)$ or $m(K^0_S K^\pm\pi^\mp\pi^+\pi^-)$. The fit yields the ratio $\frac{\mathcal{B}(\eta_c(2S)\to\eta_c\pi^+\pi^-)}{\mathcal{B}(\eta_c(2S)\to K^0_S K^\pm\pi^\mp)} = 4.9 \pm 3.5(\text{stat}) \pm 1.3(\text{syst}) \pm 0.8(\mathcal{B})$, where the third error is due to the uncertainty on $\mathcal{B}(\eta_c \to K^0_S K^\pm\pi^\mp)$ (Beringer et al., 2012). No significant signal is found, and an upper limit of 10.0 is set on this ratio of branching fractions at the 90% confidence level.

### *Inclusive observation in double charmonium production*

The $\eta_c(2S)$ has also been observed in the process $e^+e^- \to J/\psi\,\eta_c(2S)$ by both Belle (Abe, 2002j, 2004g, 2007f) and *BABAR* (Aubert, 2005n): see Section 18.2.4.2. The $\eta_c(2S)$ is not reconstructed in these analyses, but inferred from the reconstructed $J/\psi$ by using energy-momentum conservation. The $\eta_c(2S)$ signal is identified as a peak in the mass spectrum of the system recoiling against the reconstructed $J/\psi$; $M_{\text{recoil}}(J/\psi)$ is defined in Eq. (18.2.3), and the technique is discussed in the surrounding text. We illustrate these measurements using the *BABAR* analysis as a representative. The recoil mass spectrum is fitted with signal line shapes determined from MC simulation, taking into account phase space suppression due to the variation of the virtual photon energy in ISR (Fig. 18.2.4). The $J/\psi\,\eta_c(2S)$ system is assumed to be produced in $P$-wave as required by parity conservation. The combinatorial background contribution is estimated from $J/\psi$ sidebands. The $\psi(2S)$ ISR background is estimated by using MC simulations; the other feed-down from $\psi(2S) \to J/\psi\,X$ is estimated by using $\psi(2S)$ events reconstructed in the data. The $M_{\text{recoil}}$ distributions for such backgrounds are structureless and are described by a smooth function in the fit. The main sources of systematic uncertainty in the mass measurement are the uncertainty on the signal lineshape, selection procedure, and mass scale calibration. *BABAR* and Belle results are summarized in Table 18.2.2.

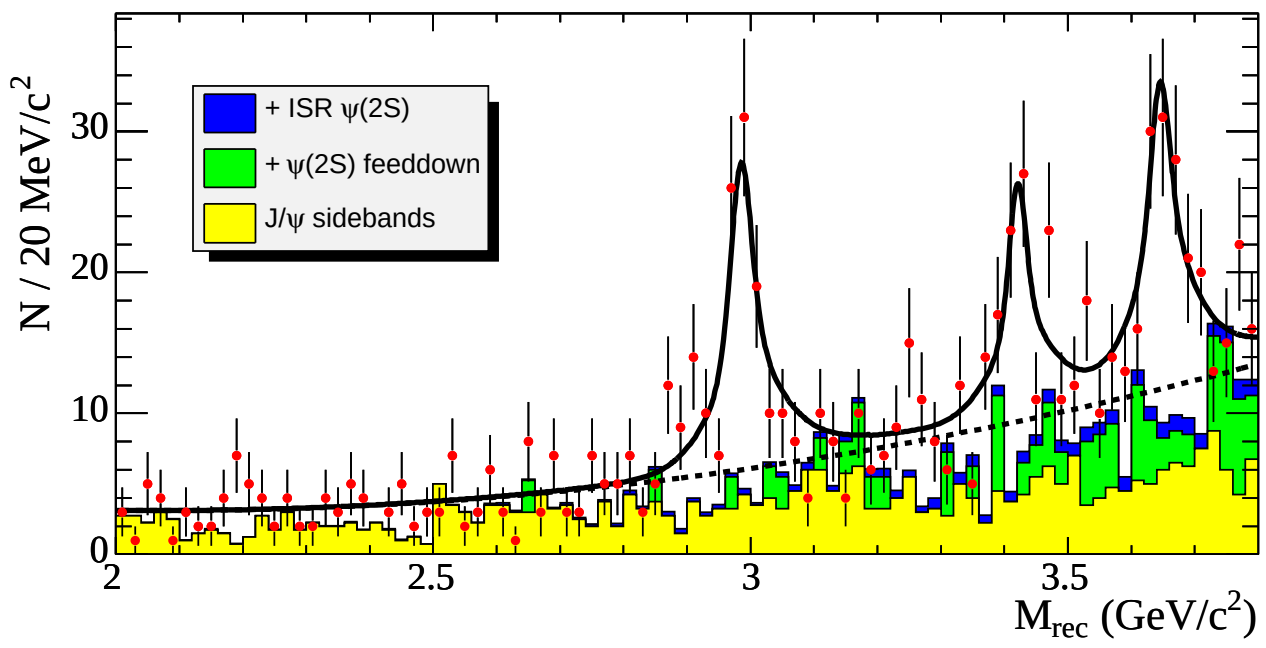

**Figure 18.2.4.** The $M_{\text{recoil}}(J/\psi)$ distribution for the $e^+e^- \to J/\psi\,X$ process (Aubert, 2005n). The solid line represents the total fit function and the dashed line is the background contribution. The histograms represent different sources of backgrounds. Signals from $\eta_c$, $\chi_{c0}$, and $\eta_c(2S)$ are visible.

### *Inclusive B decays at BABAR*

A search for $\eta_c(2S)$ has also been performed by *BABAR* in inclusive $B$-meson decays to $X_{c\overline{c}}K^\pm$ (Aubert, 2006ae). The analysis is carried out by fully reconstructing one $B$ meson ($B_{\text{tag}}$), so the signal $B$-meson ($B_{\text{sig}}$) momentum is known from the $B_{\text{tag}}$ and beam momenta. In events with one charged kaon not associated with $B_{\text{tag}}$, its momentum is calculated in the $B_{\text{sig}}$ rest frame. The mass of $X_{c\overline{c}}$ is $m_X = \sqrt{m_B^2 + m_K^2 - 2E_K m_B}$, where $m_B$ and $m_K$ are the $B^\pm$ and $K^\pm$ masses and $E_K$ is the $K^\pm$ energy in the $B$ rest frame. The resulting $m_X$ spectrum is fitted with a sum of signal and combinatorial background components, to obtain the $\eta_c(2S)$ signal yield. An excess of events with a statistical significance of $1.8\sigma$ is observed at the expected position for the $\eta_c(2S)$ peak. Results of the fit are reported in Table 18.2.2.

### *Summary*

In conclusion, the $\eta_c(2S)$ has been measured in several production processes and final states at the $B$ Factories. The mass values measured in different processes show quite a large spread ranging from 3626 to 3654 $\text{MeV}/c^2$. The large spread among the mass values measured in various processes does not indicate an actual discrepancy: its size is marginally consistent with the experimental uncertainties, and the interference of the $\eta_c(2S)$ resonance with the underlying background is neglected in all but one measurement (Vinokurova, 2011). In that analysis a 10 MeV mass shift due to this effect is estimated. Similar shifts with different values or signs are expected for the various $\eta_c(2S)$ production processes ($B$ decays, two-photon fusion, and double charmonium production). Nonetheless, all the $B$ Factory measurements are in contradiction with the previous result reported by the Crystal Ball Collaboration (Edwards et al., 1982).

**Table 18.2.2.** $\eta_c(2S)$ mass and width as measured by BABAR and Belle. The first three rows refer to measurements performed by using $B$ meson decays, the fourth and fifth rows to two-photon collisions, and the last rows to double charmonium production. These results are discussed in the summary at the end of Section 18.2.1.1. Limits are at 90% C.L.

| Experiment | Process | Luminosity ($\text{fb}^{-1}$) | Mass ($\text{MeV}/c^2$) | Width (MeV) | Reference |
|---|---|---|---|---|---|
| Belle | $B \to (K_S^0 K^\pm \pi^\mp)K$ | 42 | $3654 \pm 6 \pm 8$ | $< 55$ | Choi (2002) |
| BABAR | $B^\pm \to X_{c\bar{c}} K^\pm$ | 211 | $3639 \pm 7$ | $< 23$ | Aubert (2006ae) |
| Belle | $B^\pm \to (K_S^0 K^\pm \pi^\mp)K^\pm$ | 492 | $3636.1\,^{+3.9}_{-4.2}\,^{+0.7}_{-2.0}$ | $6.6\,^{+8.4}_{-5.1}\,^{+2.6}_{-0.9}$ | Vinokurova (2011) |
| BABAR | $\gamma\gamma \to K_S^0 K^\pm \pi^\mp$ | 520 | $3638.5 \pm 1.5 \pm 0.8$ | $13.4 \pm 4.6 \pm 3.2$ | del Amo Sanchez (2011h) |
| BABAR | $\gamma\gamma \to K^+K^-\pi^+\pi^-\pi^0$ | 520 | $3640.5 \pm 3.2 \pm 2.5$ | – | del Amo Sanchez (2011h) |
| BABAR | $e^+e^- \to J/\psi\,\eta_c(2S)$ | 112 | $3645.0 \pm 5.5\,^{+4.9}_{-7.8}$ | $22 \pm 14$ | Aubert (2005n) |
| Belle | $e^+e^- \to J/\psi\,\eta_c(2S)$ | 357 | $3626 \pm 5 \pm 6$ | – | Abe (2007f) |

Hadronic branching fractions of the $\eta_c(2S)$ are expected to be similar to those of $\eta_c$ (Chao, Gu, and Tuan, 1996). However, the measured branching fraction $\mathcal{B}(\eta_c(2S) \to K\overline{K}\pi) = (1.9 \pm 1.2)\%$ (Aubert, 2008ba) is significantly smaller than the corresponding $\mathcal{B}(\eta_c \to K\overline{K}\pi) = (7.0 \pm 1.2)\%$ (Beringer et al., 2012). Furthermore, the $\eta_c$ is observed to decay into $h^+h^-h'^+h'^-$ (with $h^{(\prime)} = K, \pi$) with a branching fraction $\sim 1.6\%$ (Beringer et al., 2012), while the corresponding decays for $\eta_c(2S)$ were searched for, but not observed (*e.g.* $\mathcal{B}(\eta_c(2S) \to 4\pi) < 4.5 \times 10^{-3}$ at 90% C.L.; Uehara, 2008b). The only exclusive $\eta_c(2S)$ decays observed to date are to $K\overline{K}\pi$ and $K^+K^-\pi^+\pi^-\pi^0$.

#### 18.2.1.2 $\chi_{c2}(2P)$

Although the lowest $^3P_J$ charmonium states (the $\chi_{cJ}$) are well established, no experimental information existed about their radial excitations $\chi_{cJ}(2P)$ before the $B$ Factory era. Theory predicts that the masses of these states lie in the region $3.9 - 4.0\,\text{GeV}/c^2$ (Godfrey and Isgur, 1985), which places them well above the $D\overline{D}$ threshold. The $\chi_{c0}(2P)$ and $\chi_{c2}(2P)$ mesons would then decay primarily into $D\overline{D}$; the decay $\chi_{c1}(2P) \to D\overline{D}$ is forbidden by parity conservation, but the $\chi_{c1}(2P)$ could decay into $D\overline{D}^*$, if energetically allowed.

In 2006, the Belle Collaboration reported the observation of a new resonance, provisionally called $Z(3930)$, based on analysis of a data sample of $395\,\text{fb}^{-1}$ (Uehara, 2006). The resonance is observed in two-photon production, a mechanism providing a clean environment for studying resonances in direct formation (see Chapter 22 for the details), both in $\gamma\gamma \to D^0\overline{D}^0$ and $\gamma\gamma \to D^+D^-$ (see Fig. 18.2.5 (a) and (b), respectively). The final state charmed mesons are fully reconstructed. Two-photon events are separated from $e^+e^-$ annihilation and ISR events by requiring that the transverse momentum of the $D\overline{D}$ system be small, as expected for two-photon events in the no-tag mode (*i.e.*, where neither the outgoing electron nor the positron are detected). The resulting combined invariant mass distribution is fitted with a relativistic Breit-Wigner signal function (taking the mass resolution and reconstruction efficiency into account) and a background component (Fig. 18.2.5 (c)). The statistical significance of the $Z(3930)$ peak is $5.3\,\sigma$. The measured mass and total width of the resonance are listed in Table 18.2.3. The systematic uncertainties are dominated by uncertainties in the $D$ mass and the choice of the signal function lineshape.

Belle performs an angular analysis to identify the spin of the observed resonance. If one defines $\theta$ as the angle of a $D$ meson relative to the beam axis in the $\gamma\gamma$ frame (equivalent to the $D\overline{D}$ frame), the $\cos\theta$ distribution for a scalar particle will be flat, while for a spin-2 resonance produced with helicity 2 along the incident axis, a distribution proportional to $\sin^4\theta$ is expected. Spin-1 is largely suppressed in two-photon events with quasi-real photons (Yang, 1950), thus this assignment is not considered. The Belle data significantly favor spin-2 over spin-0 assignment, while the production and decay mechanisms require positive parity and $C$-parity. The resulting quantum numbers, $J^{PC} = 2^{++}$, suggest identifying this particle with the previously unobserved $\chi_{c2}(2P)$ charmonium state. Assuming production of a spin-2 state, Belle calculated the product of its two-photon width and the branching fraction into $D\overline{D}$ (Table 18.2.3). The systematic errors are primarily due to uncertainties in tracking and particle identification efficiencies, the choice of fit lineshapes and the errors of $D$ branching fractions.

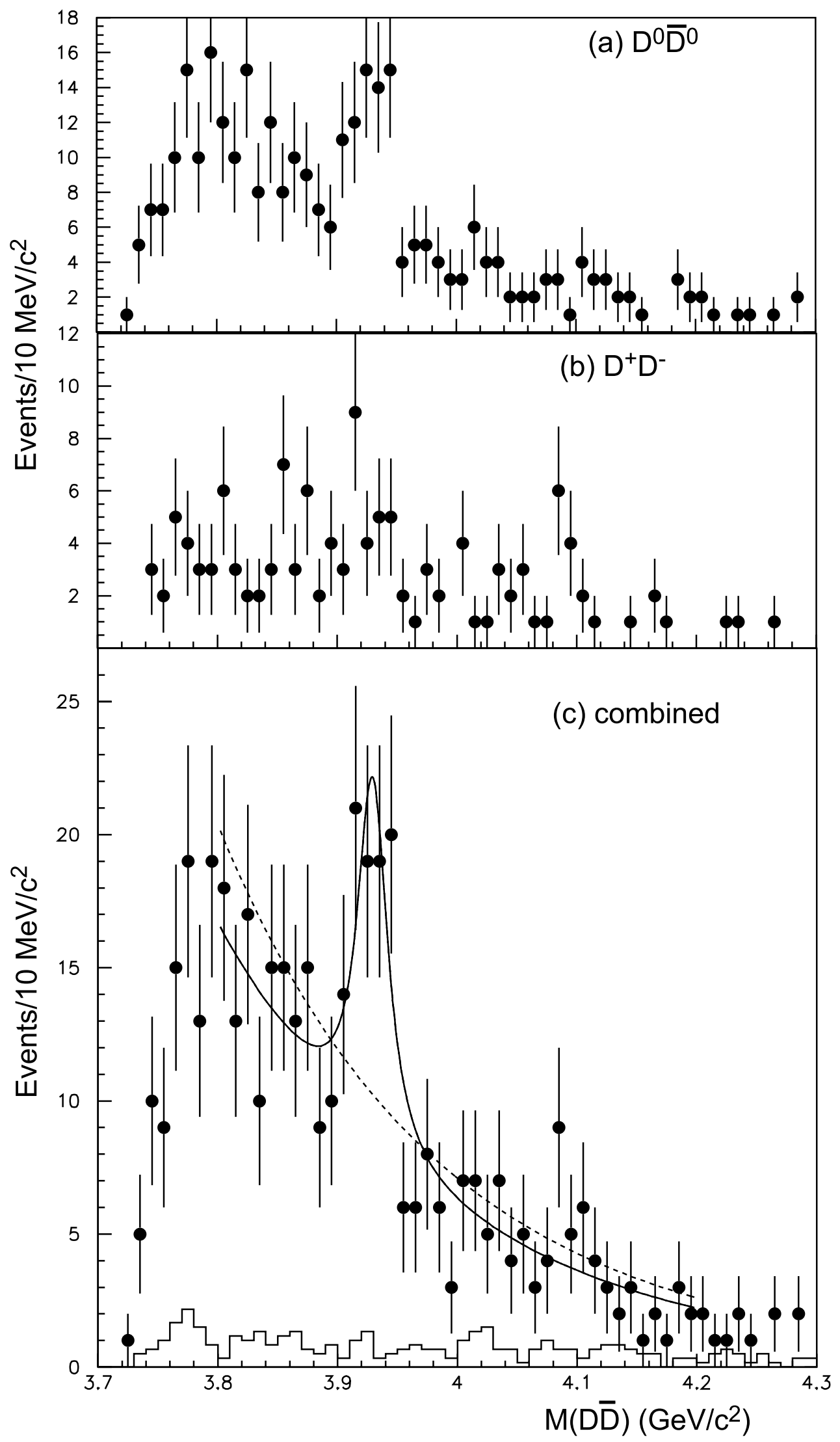

**Figure 18.2.5.** (a) $D^0\overline{D}^0$, (b) $D^+D^-$, and (c) combined $D\overline{D}$ invariant mass distributions in two-photon fusion events (Uehara, 2006). The open histogram shows the combinatorial background distribution estimated from $D$ sidebands. The solid line represents the total fit function, the dashed line the fit without any resonant structure.

Similar results have been obtained by *BABAR*, using a $384\,\text{fb}^{-1}$ data sample (Aubert, 2010g). The $D^0\overline{D}^0$ and $D^+D^-$ final states are fully reconstructed, selecting two-photon events (in the no-tag mode) by requiring a large missing mass ($\sqrt{(p_{e^+e^-} - p_{D\overline{D}})^2}$) and a small transverse momentum of the $D\overline{D}$ system. Additionally, the energy deposited in the calorimeter unmatched to any charged-particle track should not exceed 400 MeV. A peak in the $D\overline{D}$ invariant mass distribution near $3.93\,\text{GeV}/c^2$ is also clearly seen in *BABAR* data. The combined efficiency-corrected $D\overline{D}$ invariant mass spectrum is fitted with a relativistic Breit-Wigner signal function convolved with a mass-dependent Gaussian resolution function and a background lineshape taking the $D\overline{D}$ threshold into account.

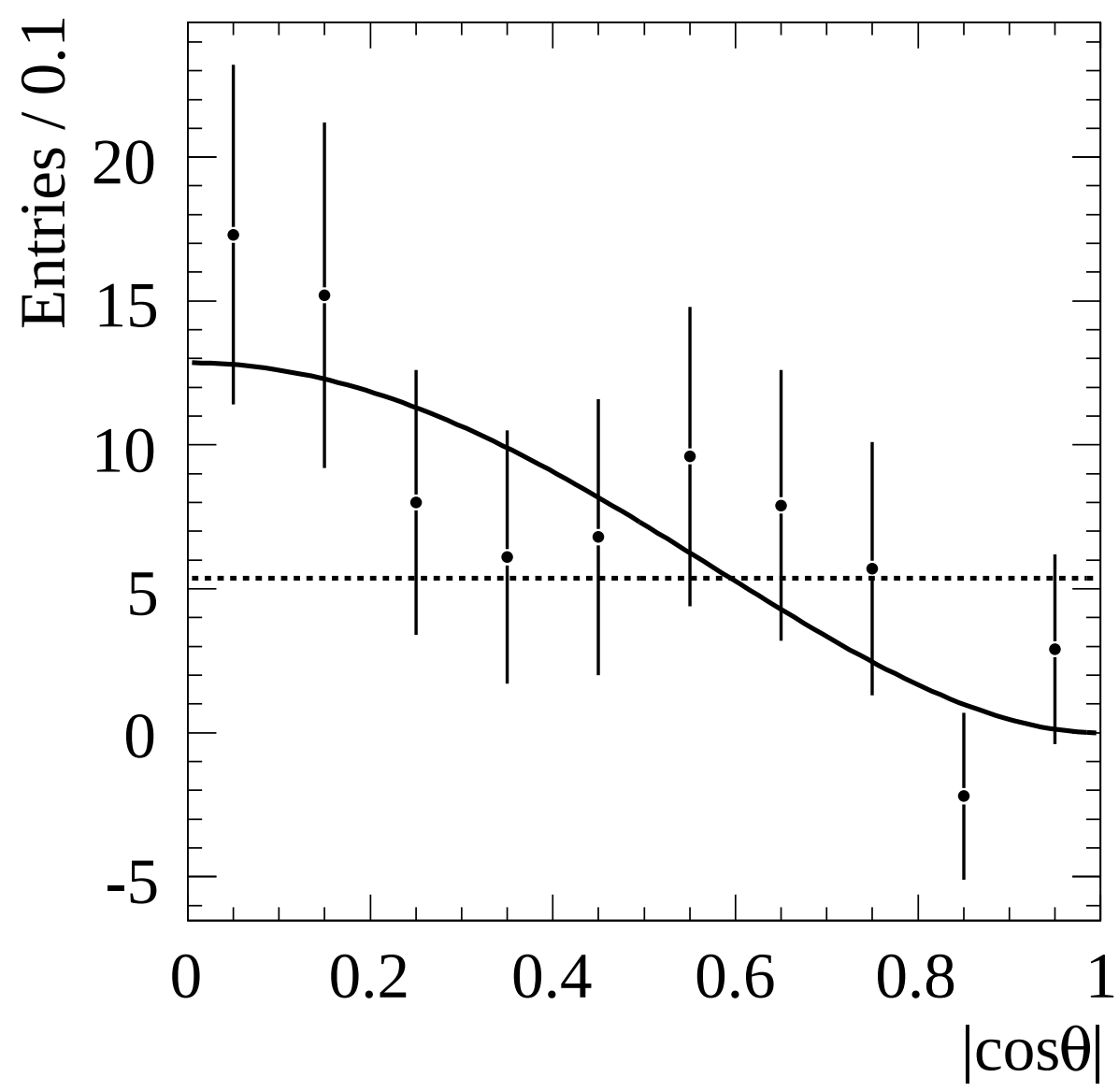

**Figure 18.2.6.** The $\cos\theta$ distribution (see the text) for $\chi_{c2}(2P)$ signal candidates in $D\overline{D}$ events produced in two-photon fusion at *BABAR* (Aubert, 2010g). The results of a fit to the $J = 2$ hypothesis are shown with the solid curve; the dashed curve corresponds to the $J = 0$ hypothesis.

The significance of the $\chi_{c2}(2P)$ observation is $5.8\,\sigma$; the fitted mass and width are listed in Table 18.2.3. The angular distribution for signal entries is obtained from fits to data in 10 bins of $\cos\theta$, with $\theta$ being the angle of a $D$ in the $D\overline{D}$ system relative to the $D\overline{D}$ lab momentum (Fig. 18.2.6). As in the Belle study, the expected distribution for spin-2 is significantly favored over spin-0; taking the production and decay processes into account, $J^{PC} = 2^{++}$ is therefore preferred. The calculated product of the two-photon width and the branching fraction into the $D\overline{D}$ final state is in good agreement with the Belle value (Table 18.2.3). Systematic errors in this analysis address the choice of signal and background lineshapes, tracking and particle identification issues, and uncertainties in $D$ mass and branching fractions.

In summary, the $\chi_{c2}(2P)$ has been observed in two-photon production at both $B$ Factory experiments, decaying into $D^0\overline{D}^0$ and $D^+D^-$. The parameters reported by *BABAR* and Belle are in good agreement. The measured $\chi_{c2}(2P)$ mass is $50\,\text{MeV}/c^2$ lower than potential model predictions (Table 18.2.1); other parameters including the two-photon width are consistent with the model expectations for the $\chi_{c2}(2P)$ state. This state has so far not been seen in any other production mechanism or decay mode. For example, *BABAR* has obtained a 90% C.L. upper limit $\Gamma_{\gamma\gamma}(\chi_{c2}(2P)) \times \mathcal{B}(\chi_{c2}(2P) \to \eta_c\pi^+\pi^-) < 18\,\text{eV}$ (del Amo Sanchez, 2011h). Only two other reported charmonium-like states in the predicted mass region are observed to decay into $D\overline{D}$ or $D\overline{D}^*$: the $X(3872)$, the structure of which is controversial (see Section 18.3.2), and the $X(3940)$, which is likely an excitation of the $\eta_c$ (see

**Table 18.2.3.** Summary of $\chi_{c2}(2P)$ mass and width measurements obtained by Belle and *BaBar* in $\gamma\gamma \to D\overline{D}$.

| Experiment | Luminosity ($\text{fb}^{-1}$) | Mass ($\text{MeV}/c^2$) | Width (MeV) | Spin $J^{PC}$ | $\Gamma_{\gamma\gamma}(\chi_{c2}(2P))\times$ $\mathcal{B}(\chi_{c2}(2P) \to D\overline{D})$ (keV) | Reference |
|---|---|---|---|---|---|---|
| Belle | 395 | $3929 \pm 5 \pm 2$ | $29 \pm 10 \pm 2$ | $2^{++}$ | $0.18 \pm 0.05 \pm 0.03$ | Uehara (2006) |
| *BaBar* | 384 | $3926.7 \pm 2.7 \pm 1.1$ | $21.3 \pm 6.8 \pm 3.6$ | $2^{++}$ | $0.24 \pm 0.05 \pm 0.04$ | Aubert (2010g) |

Sections 18.2.1.3 and 18.3.3). The $\chi_{c2}(2P)$ remains the only confirmed radial excitation of the $^3P_J$ charmonium states.

### 18.2.1.3 $X(3940)$ and $X(4160)$ as candidates for higher radial excitations of $\eta_c$

Double charmonium production in $e^+e^-$ annihilation, first observed in 2002 by Belle (Abe, 2002j) and confirmed by *BABAR* (Aubert, 2005n), can be regarded as a mini-factory of charmonium production. This process, described in detail in Section 18.2.4.2, provides opportunities both to search for new charmonia, and to study the decays of known states. Using two-body kinematics, exclusive final states can be identified by reconstructing a state such as the $J/\psi$, and then studying the spectrum of the recoil mass ($M_{\text{recoil}}(J/\psi)$, as defined in Eq. 18.2.3, Section 18.2.4.2). Both known and new states produced in association with $J/\psi$ appear as peaks in this spectrum.[101] Studies of various double charmonium final states have demonstrated that scalar and pseudoscalar charmonia are copiously produced in recoil against $J/\psi$ or $\psi(2S)$, and there is no significant suppression of the production of radially excited states (Abe, 2004g; Aubert, 2005n).

In the study by Abe (2007f) of the $M_{\text{recoil}}(J/\psi)$ distribution, in addition to previously reported peaks at the $\eta_c$, $\chi_{c0}$, and $\eta_c(2S)$ masses, a fourth enhancement around $3940\,\text{MeV}/c^2$ was found (Fig. 18.2.7). The new state was called $X(3940)$. A fit to this spectrum that includes the three previously seen charmonium states plus a fourth state finds the significance of the new state to be $5.0\sigma$ including systematics. However, in this study it is not possible to prove that the observed peak is due to a single resonance.

The $X(3940)$ mass is above both the $D\overline{D}$ and $D\overline{D}^*$ thresholds, so it is natural to search for $X(3940)$ decays into these final states. Because of the small product of $D^{(*)}$ reconstruction efficiencies and branching fractions, it is not feasible to reconstruct fully the chain $e^+e^- \to J/\psi\, X(3940)$, $X(3940) \to D\overline{D}^{(*)}$. To increase the efficiency, only the $J/\psi$ and one $D$ meson are reconstructed, detecting the other $\overline{D}^{(*)}$ as a peak in the $M_{\text{recoil}}(J/\psi\, D)$ spectrum. The instrumental resolution allows the $\overline{D}$ and $\overline{D}^*$ peaks to be clearly resolved, thus effectively tagging the processes $e^+e^- \to J/\psi\; D\overline{D}$ and $e^+e^- \to J/\psi\; D\overline{D}^*$. A clear $X(3940)$ signal is seen only in the latter process, with $5.0\sigma$ significance. We illustrate the method, and present the measured parameters of both the $X(3940)$ and another new state, using the results of the latest Belle study.

[101] Only states with charge conjugation $C = +1$ can be produced in association with the $J/\psi$, due to the conservation of this quantum number in $e^+e^- \to \gamma^* \to J/\psi\; X$ and to the fact that both the $\gamma^*$ and the $J/\psi$ have $C = -1$.

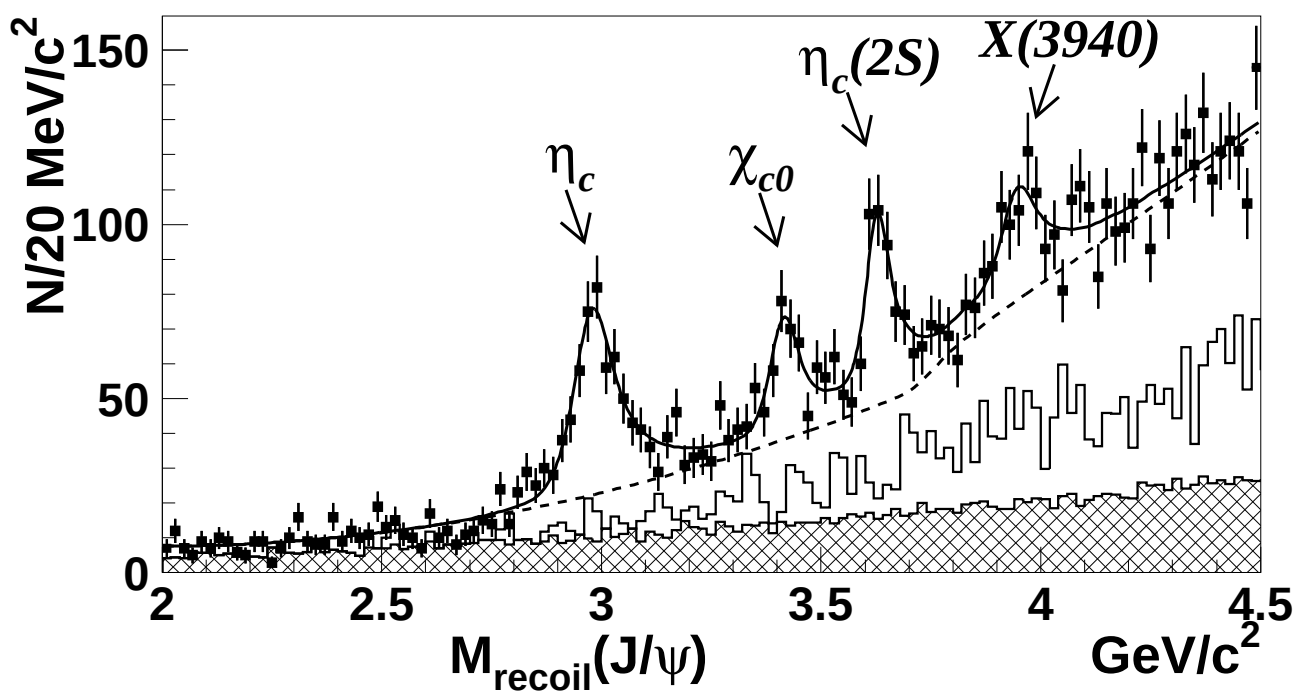

**Figure 18.2.7.** From the Abe (2007f) analysis: The distribution of $M_{\text{recoil}}(J/\psi)$ in inclusive $e^+e^- \to J/\psi\; X$ events (points with error bars). The cross-hatched histogram shows the scaled $J/\psi$ sideband distribution; the open histogram corresponds to the feed-down from $\psi(2S)$ decay. The solid curve is the fit result; the dashed curve shows the background and non-resonant contribution.

Using a dataset twice as large, Pakhlov (2008) performed a detailed study of the processes $e^+e^- \to J/\psi\; D\overline{D}$, $J/\psi\; D\overline{D}^*$, and $J/\psi\; D^*\overline{D}^*$. After reconstruction of a $J/\psi\, D$ combination, signals for all three processes are evident in the spectrum of recoil mass $M_{\text{recoil}}(J/\psi\, D)$ (Fig. 18.2.8 (a)), at the $\overline{D}$ mass, the $\overline{D}^*$ mass, and at $\sim 2.2\,\text{GeV}/c^2$ respectively. The latter peak is shifted and widened due to the missing pion or photon from $\overline{D}^*$ decay. The processes $e^+e^- \to J/\psi\; D\overline{D}^*$ and $J/\psi\; D^*\overline{D}^*$ are also clearly seen following $J/\psi\, D^*$ reconstruction, in the spectrum of recoil mass $M_{\text{recoil}}(J/\psi\, D^*)$ (Fig. 18.2.8 (b)), as distinct peaks around the $\overline{D}$ and $\overline{D}^*$ masses. Selecting $J/\psi\, D$ or $J/\psi\, D^*$ combinations from the proper interval of $M_{\text{recoil}}(J/\psi\, D^{(*)})$, events can be effectively divided into non-overlapping samples corresponding to each of the studied processes. In particular, only the process $e^+e^- \to D\overline{D}$ (and combinatorial background) contributes in the interval $|M_{\text{recoil}}(J/\psi\, D) - M_{\overline{D}}| < 70\,\text{MeV}/c^2$. Events from the adjacent interval $|M_{\text{recoil}}(J/\psi\, D) - M_{\overline{D}^*}| < 70\,\text{MeV}/c^2$ are dominated by the process $e^+e^- \to J/\psi\, D\overline{D}^*$; a small feed-down from the process $e^+e^- \to J/\psi\, D\overline{D}$ appears in this interval due to initial state radiation. $J/\psi\, D^*$ combinations from the interval $|M_{\text{recoil}}(J/\psi\, D^*) - M_{\overline{D}}| < 70\,\text{MeV}/c^2$ provide a very clean sample of the same pro-cess $e^+e^- \to J/\psi\, D\overline{D}^*$, with very small background and free of feed-down. However, as this sample is a small subsample of the previous case, it is used only as a cross check. Finally, $J/\psi\, D^*$ combinations from the interval $|M_{\rm recoil}(J/\psi\, D^*) - M_{\overline{D}^*}| < 70\,{\rm MeV}/c^2$ tag the process $e^+e^- \to J/\psi\, D^*\overline{D}^*$.

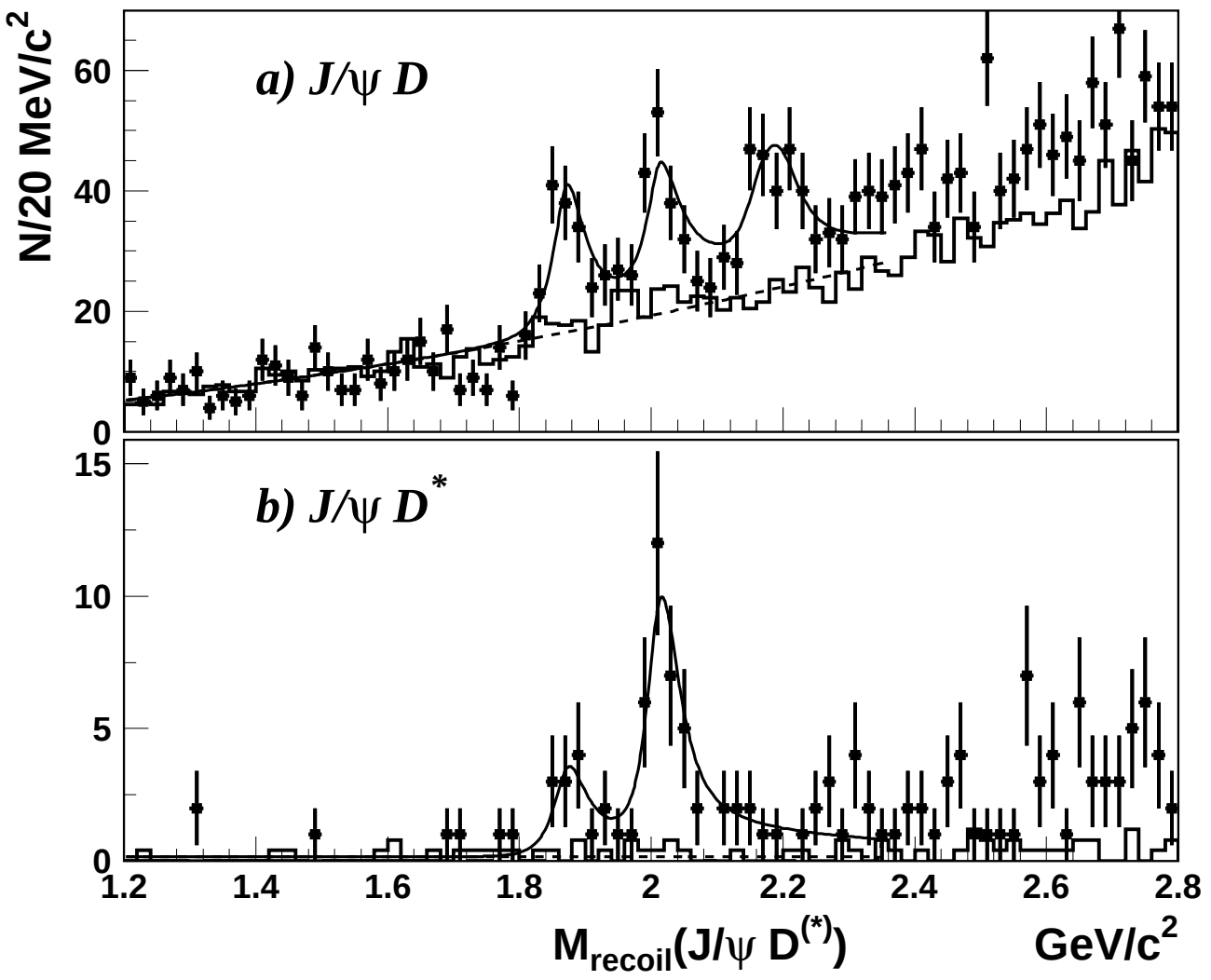

**Figure 18.2.8.** From Pakhlov (2008): The distribution of (a) $M_{\rm recoil}(J/\psi\, D)$ and (b) $M_{\rm recoil}(J/\psi\, D^*)$ in $e^+e^- \to J/\psi\, D^{(*)}X$ events (points with error bars). The histograms show the scaled $D^{(*)}$ sideband distribution. The solid curve is the fit result; the dashed curve shows the background contribution.

The spectra of $M(D^{(*)}\overline{D}^{(*)}) \equiv M_{\rm recoil}(J/\psi)$ are shown in Figs 18.2.9 (a), (b), (c), and (d) for the four selected cases in turn. Enhancements near threshold are evident in each distribution. A fit to the $M(D\overline{D})$ distribution finds a broad resonance-like structure near the threshold, tentatively denoted $X(3880)$. However the significance of the broad peak is low ($3.8\,\sigma$), and the fit is not stable under variation of the background parameterization. Therefore, with the existing sample the resonant structure in this process cannot be reliably determined. The significance of the $X(3940)$ signal found by the fit to the $M(D\overline{D}^*)$ spectrum is $5.7\,\sigma$ (including systematic uncertainties). The $X(3940)$ mass and width are $M = (3942\,^{+\,7}_{-\,6} \pm 6)\,{\rm MeV}/c^2$ and $\Gamma = (37\,^{+\,26}_{-\,15} \pm 8)\,{\rm MeV}$. The insets in Figs 18.2.9 (a) and (b) show the background subtracted spectra with the signal functions superimposed.

The $M(D^*\overline{D}^*)$ spectrum has a clear broad enhancement near threshold, which is seen above the small combinatorial background and the $X(3940)$ reflection. The observed enhancement, which has a significance of $5.1\,\sigma$ (including systematics), was interpreted as a new resonance and denoted $X(4160)$. The $X(4160)$ parameters are $M = (4156\,^{+\,25}_{-\,20} \pm 15)\,{\rm MeV}/c^2$ and $\Gamma = (139\,^{+\,111}_{-\,\ \,61} \pm 21)\,{\rm MeV}$. Although the masses and widths of the $X(4160)$ and $\psi(4160)$ are not inconsistent, the latter cannot be produced in $e^+e^-$ annihilation via a single virtual photon due to $C$-parity conservation, as explained above; annihilation via two virtual photons is strongly suppressed, as demonstrated by the non-observation of $e^+e^- \to J/\psi\, J/\psi$ (Abe, 2004g).

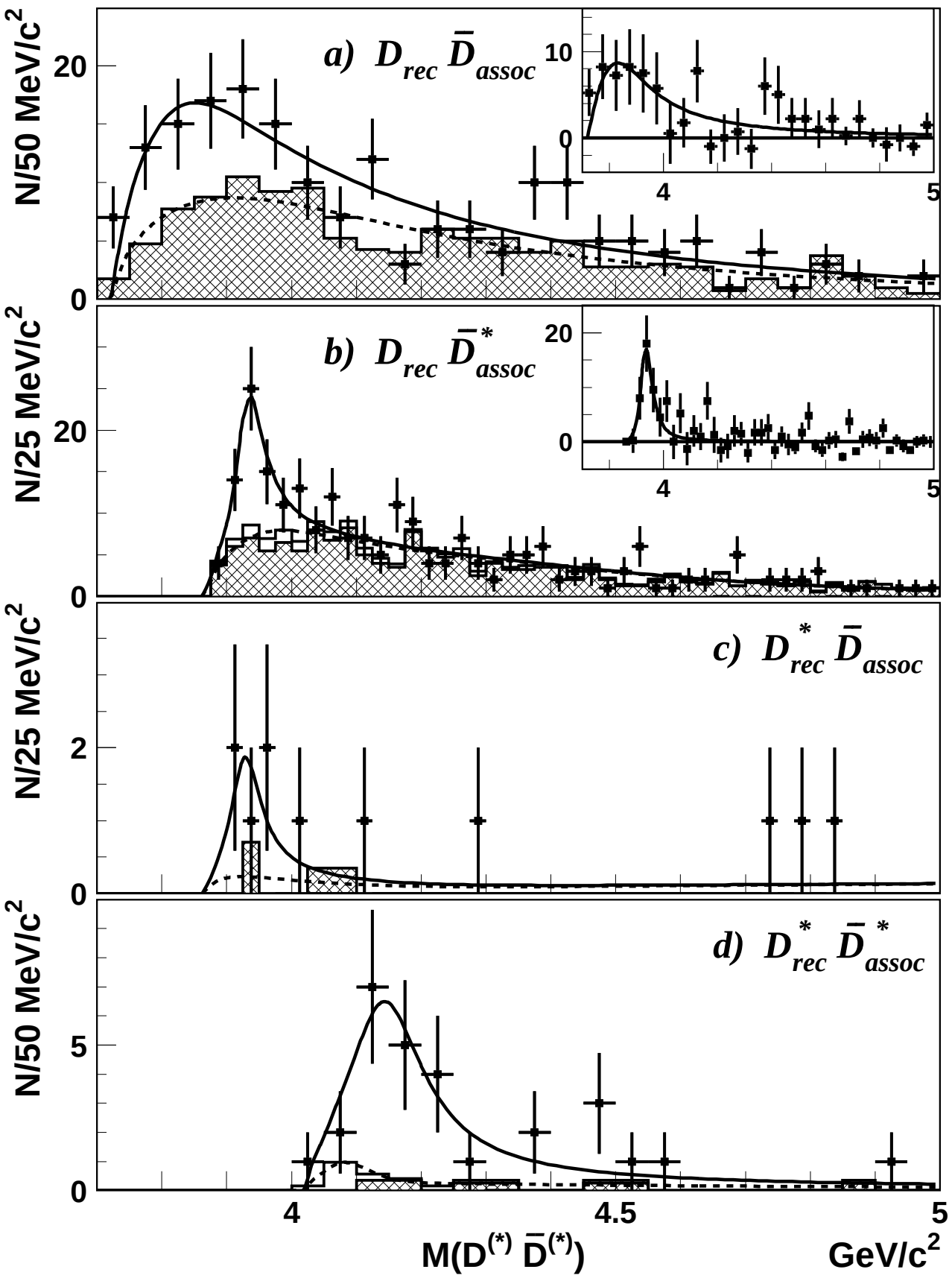

**Figure 18.2.9.** From Pakhlov (2008): The spectra of $M(D^{(*)}\overline{D}^{(*)}) \equiv M_{\rm recoil}(J/\psi)$ for events tagged and constrained as (a) $e^+e^- \to J/\psi\, D\overline{D}$, (b,c) $e^+e^- \to J/\psi\, D\overline{D}^*$, and (d) $e^+e^- \to J/\psi\, D^*\overline{D}^*$ in the data (points with error bars). Hatched histograms show the combinatorial background distributions and open histograms show the feed-down contribution (see text). The solid lines represent the fit results; the dashed lines are background functions.

If the $X(3940)$ has spin equal to 0, like other states produced together with the $J/\psi$, the absence of a $D\overline{D}$ decay mode strongly favors $J^P = 0^-$, for which the most likely charmonium assignment is the $\eta_c(3S)$ (see Table 18.2.1).[102] The fact that the lower-mass $\eta_c$ and $\eta_c(2S)$ are also produced in double charmonium production supports this assignment. However, there is the problem that the measured $X(3940)$ mass is below potential model estimates for the $\eta_c(3S)$ mass of $\sim 4050\,{\rm MeV}/c^2$ or higher (Barnes et al., 2005). A further complication is the observation of the $X(4160)$, which could also be attributed to the $^1S_0$ state, using similar arguments. But the $X(4160)$ mass is well above expectations for the $\eta_c(3S)$ and well be-low those for the $\eta_c(4S)$, which is predicted to be near 4400 MeV/$c^2$ (Barnes et al., 2005). Although either the $X(3940)$ or the $X(4160)$ or both might conceivably fit a charmonium assignment, the final identification of these states will be possible only after angular analysis of the $e^+e^- \to J/\psi\, D^* \overline{D}^{(*)}$ processes has been performed, allowing quantum numbers to be fixed. Such a study requires much larger samples than those collected by the $B$ Factories.

[102] The decay of the pseudoscalar state, $0^-$, into two pseudoscalar mesons is forbidden by parity conservation, as only $S$-wave is allowed. On the contrary, for the scalar state, $0^+$, this decay is allowed and should be dominant.

### 18.2.2 New decay modes of known charmonia

It is difficult for the $B$ Factories to compete with the charm factories (BES and CLEO-c) in searching for new decay modes of the charmonium states below $D\overline{D}$ threshold, as the charm factories have collected large datasets at the $J/\psi$ and $\psi(2S)$ peaks. Automatically, large samples of tagged $\eta_c$, $\chi_{cJ}$, and $h_c$ events are also collected through radiative or hadronic transitions. Nonetheless, one new $\eta_c$ decay mode was observed by the $B$ Factories.

For the states above $D\overline{D}$ threshold the $B$ Factories are competitive: for a study of wide $\psi$ resonances charm factories need to perform an energy scan, with relatively low luminosity at each point, whereas the $B$ Factories can see the whole energy region in many open charm exclusive final states. Here we present the observed new decay mode of the $\eta_c$ (Section 18.2.2.1) and first measurements of the exclusive decays of $\psi$ states above open charm threshold (Section 18.2.2.2).

#### 18.2.2.1 $\eta_c \to \Lambda\overline{\Lambda}$

Belle has studied decays of $\eta_c$ and $J/\psi$ to both $p\overline{p}$ and $\Lambda\overline{\Lambda}$, using two-body $B$ decays $B \to \eta_c K$ and $J/\psi\, K$ (Wu, 2006). The primary goal was to study anisotropy parameters in the decays of $J/\psi$ to baryon-antibaryon pairs (Murgia and Melis, 1995). In addition to a clear $\eta_c \to p\overline{p}$ signal, a significant excess has also been observed in $\eta_c \to \Lambda\overline{\Lambda}$ for the first time.

In this analysis $B$ mesons are reconstructed using the standard procedure, with $\Delta E$ and $m_{\rm ES}$ as discriminating variables (see Section 7.1). The dominant background, continuum events, is suppressed using a Fisher discriminant that combines seven event shape variables (Section 9.3). The $\Lambda\overline{\Lambda}$ mass spectrum from the $B$ meson signal window is presented in Fig. 18.2.10. An unbinned maximum likelihood fit to this spectrum is performed using a relativistic Breit-Wigner function for the $\eta_c$ peak, a Gaussian for the $J/\psi$ peak, and a linear function for the non-resonant background. The Breit-Wigner function is convolved with the detector resolution function, which is taken from the Gaussian width of the $J/\psi$ peak. The fit result is shown in the Fig. 18.2.10 inset. The measured $\eta_c$ mass and width are $(2974 \pm 7\,^{+2}_{-1})$ MeV/$c^2$ and $(40 \pm 19 \pm 5)$ MeV, respectively. The signal yield is $(18.2 \pm 4.8)$ $\eta_c$ events.

A fit to the $m_{\rm ES}$ spectrum from the $\eta_c$ signal window yields $(19.5\,^{+5.1}_{-4.4})$ events, consistent with the result of the former fit. The statistical significance of the observation of the new $\eta_c$ decay mode is estimated to be 7.9 standard deviations. Taking into account reconstruction efficiency, the branching fraction of the $\eta_c \to \Lambda\overline{\Lambda}$ decay is calculated to be $\mathcal{B} = (0.87^{+0.24}_{-0.21}({\rm stat})\,^{+0.09}_{-0.14}({\rm syst}) \pm 0.27\,(\mathcal{B})) \times 10^{-3}$, where the third uncertainty term is due to the poorly known absolute branching fractions of $\eta_c$. This term cancels in the ratio $\mathcal{B}(\eta_c \to \Lambda\overline{\Lambda})/\mathcal{B}(\eta_c \to p\overline{p})$, measured to be $0.67\,^{+0.19}_{-0.16} \pm 0.12$, consistent with the theoretical expectations.

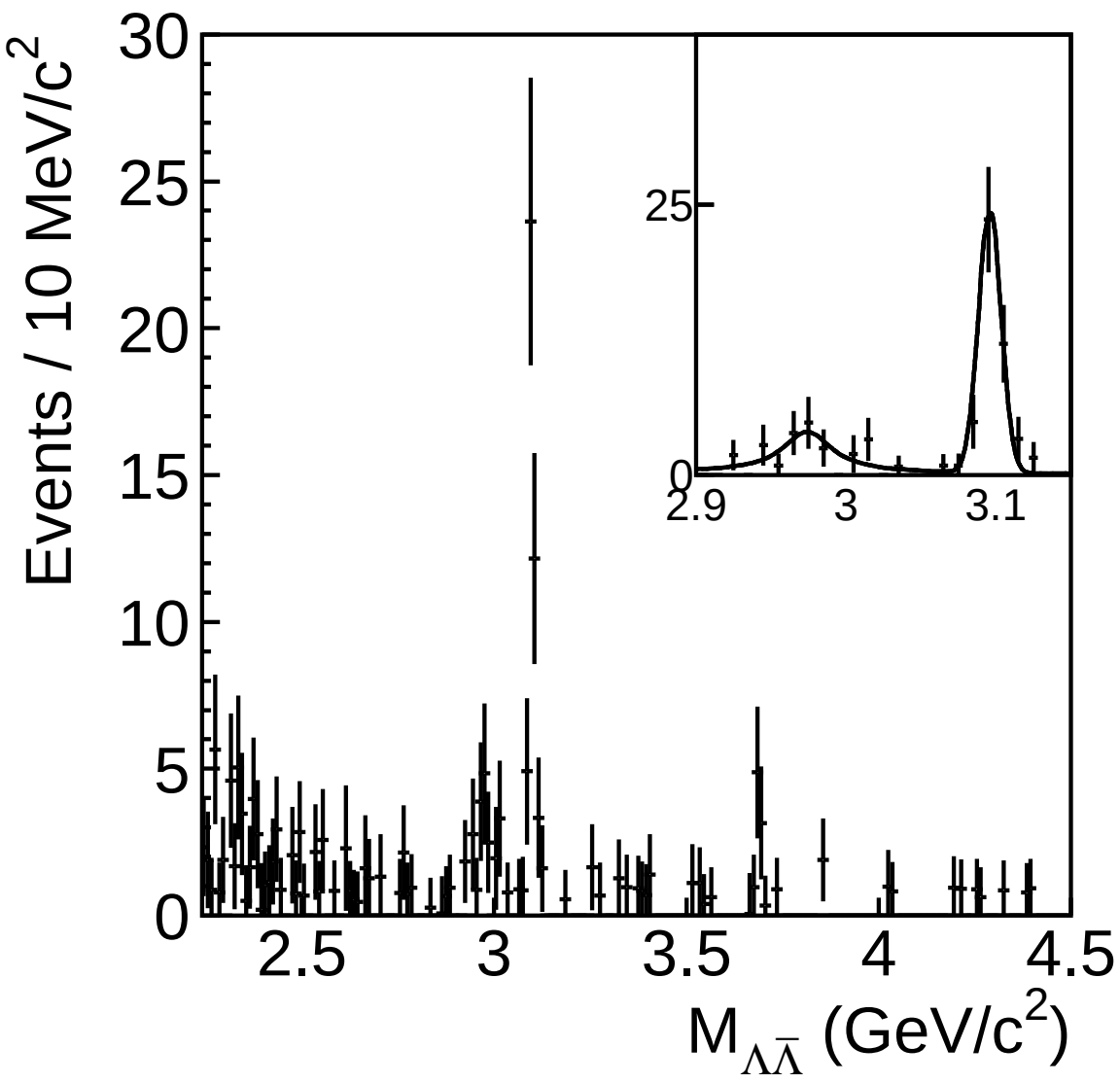

**Figure 18.2.10.** From Wu (2006): The $\Lambda\overline{\Lambda}$ mass spectrum in $B \to \Lambda\overline{\Lambda}K$ events from the $B$ meson signal window. The $\eta_c$ and $J/\psi$ region is shown inset, with the fit results as a solid line. No significant signal is visible in the $\eta_c(2S)$ region.

#### 18.2.2.2 Open charm decays of $J^{CP} = 1^{--}$ charmonium states

The process with a photon radiated from the initial state (ISR), $e^+e^- \to \gamma_{\rm ISR}\, V$ (see Fig. 21.2.2 for the Feynman diagram), generates a state $V$ coupled to the virtual photon, and therefore with the same quantum numbers, $J^{PC} = 1^{--}$. Such events represent an excellent laboratory to study exclusive decays of the vector $V$, with very clean signals observed in most studied final states (see Chapter 21 for a detailed description of the process). The ISR method has been successfully used to measure charmonium decays into open charm final states: their high multiplicity allows for efficient reconstruction with the ISR method, while the small branching fractions of charmed mesons to modes convenient for reconstruction make them difficult to detect in exclusive $B$ decays. The $B$ Factories have provided measurements of the branching fractions of various vector charmonium states for the first time. Here we describe only the procedure that was used to extract the branching fractions, and summarize the results.

Both *BABAR* (Aubert, 2009n) and Belle (Abe, 2007d; Pakhlova, 2008a) have studied the processes $e^+e^- \to$

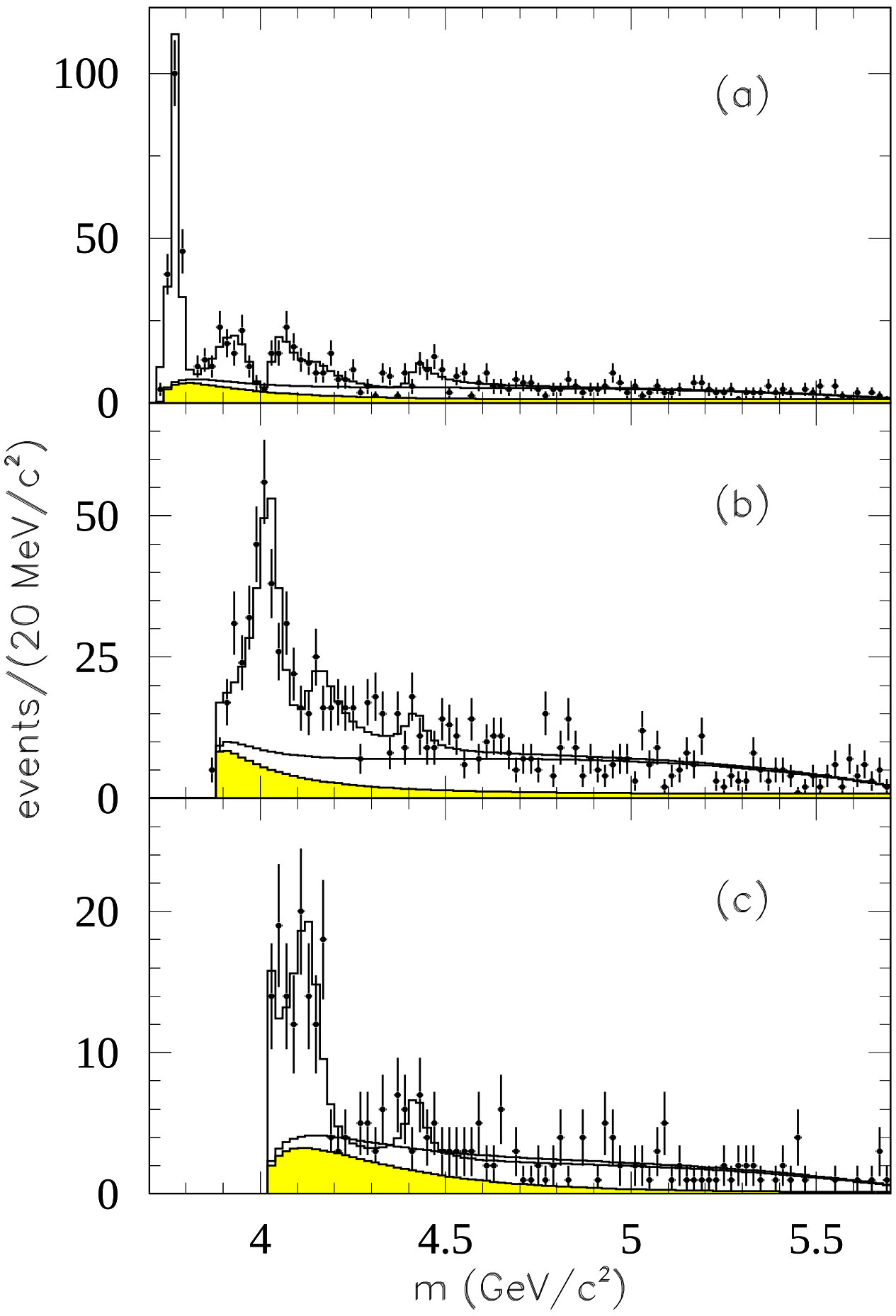

**Figure 18.2.11.** From Aubert (2009n): The (a) $D\bar{D}$, (b) $D\bar{D}^*$, and (c) $D^*\bar{D}^*$ mass spectra in the $e^+e^- \to \gamma_{\rm ISR}\, D^{(*)}\bar{D}^{(*)}$ process. The curves represent the fitted functions as described in the text. The shaded histogram corresponds to the smoothed incoherent background. The second smooth solid line represents the non-resonant contribution.

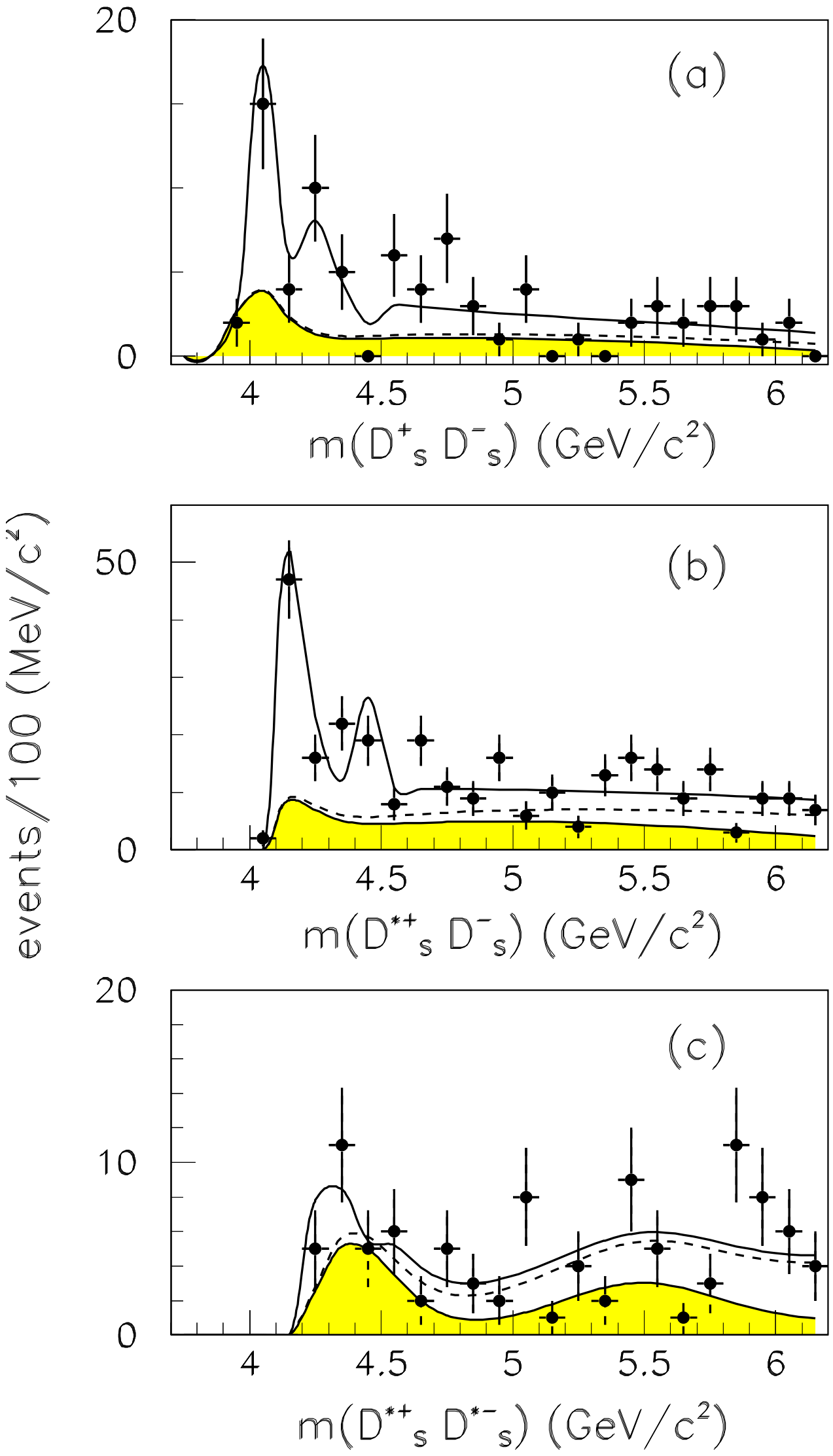

**Figure 18.2.12.** From del Amo Sanchez (2010d): The observed (a) $D_s^+D_s^-$, (b) $D_s^{*+}D_s^-$, and (c) $D_s^{*+}D_s^{*-}$ mass spectra in the processes $e^+e^- \to \gamma_{\rm ISR} D_s^{(*)}\bar{D}_s^{(*)}$ (del Amo Sanchez, 2010d). The shaded areas show the background contribution. The dashed lines indicate the sum of this background and the coherent background. The solid lines are the results from the fit as described in the text.

$\gamma_{\rm ISR}\, D^{(*)}\bar{D}^{(*)}$; their results are in good agreement. The measured cross-sections for these processes around threshold exhibit many structures, which could be attributed to the various $\psi$ states. *BABAR* has performed fits to the measured mass spectra including interference between the resonant terms $(c_i W_i(m) e^{i\phi_i}$, where $W_i(m)$ is a $P$-wave relativistic Breit-Wigner) and the non-resonant contribution. The fitting functions for each channel are computed with their own thresholds, efficiencies, purities, and backgrounds. The fits, summed over the charged and neutral final states, provide a good description of all the data (Fig. 18.2.11). The fraction for each resonant contribution $i$ is defined by

$$f_i = \frac{|c_i|^2 \int |W_i(m)|^2 dm}{\sum_{j,k} c_j c_k^* \int W_j(m) W_k^*(m) dm}; \qquad (18.2.1)$$

the fractions $f_i$ do not necessarily add up to 1 because of interference between amplitudes. The error for each fraction has been evaluated by propagating the full covariance matrix obtained by the fit. The resulting relative branching fractions are listed in Table 18.2.4, and compared with the predictions of theoretical models.

A similar fit was performed by *BABAR* to the $D_s^+D_s^-$, $D_s^{*+}D_s^-$, and $D_s^{*+}D_s^{*-}$ mass spectra (del Amo Sanchez, 2010d). In the fit the mass and width of the $\psi(4040)$, $\psi(4160)$, $\psi(4415)$, and the exotic state $Y(4260)$ (see Section 18.3) are fixed to the PDG values (Amsler et al., 2008); interference with the coherent non-resonant contribution is taken into account. The fit results are shown in Fig. 18.2.12 and measured fit fractions are given in Table 18.2.5.

**Table 18.2.4.** From Aubert (2009n): Ratios of branching fractions for the three $\psi$ resonances. The first error is statistical, the second systematic. Theoretical expectations are from the $^3P_0$ model (Barnes et al., 2005), $C^3$ model (Eichten, Lane, and Quigg, 2006), and $\rho K\rho$ model (Swanson, 2006).

| Ratio | Measurement | $^3P_0$ | $C^3$ | $\rho K\rho$ |
|---|---|---|---|---|
| $\mathcal{B}(\psi(4040) \to D\overline{D})/\mathcal{B}(\psi(4040) \to D^*\overline{D})$ | 0.24 ± 0.05 ± 0.12 | 0.003 | | 0.14 |
| $\mathcal{B}(\psi(4040) \to D^*\overline{D}^*)/\mathcal{B}(\psi(4040) \to D^*\overline{D})$ | 0.18 ± 0.14 ± 0.03 | 1.0 | | 0.29 |
| $\mathcal{B}(\psi(4160) \to D\overline{D})/\mathcal{B}(\psi(4160) \to D^*\overline{D}^*)$ | 0.02 ± 0.03 ± 0.02 | 0.46 | 0.08 | |
| $\mathcal{B}(\psi(4160) \to D^*\overline{D})/\mathcal{B}(\psi(4160) \to D^*\overline{D}^*)$ | 0.34 ± 0.14 ± 0.05 | 0.011 | 0.16 | |
| $\mathcal{B}(\psi(4415) \to D\overline{D})/\mathcal{B}(\psi(4415) \to D^*\overline{D}^*)$ | 0.14 ± 0.12 ± 0.03 | 0.025 | | |
| $\mathcal{B}(\psi(4415) \to D^*\overline{D})/\mathcal{B}(\psi(4415) \to D^*\overline{D}^*)$ | 0.17 ± 0.25 ± 0.03 | 0.14 | | |

**Table 18.2.5.** From del Amo Sanchez (2010d): $D_s^+D_s^-$, $D_s^{*+}D_s^-$, and $D_s^{*+}D_s^{*-}$ fit fractions (in %). Errors are statistical only.

| Resonance | Fraction | | |
|---|---|---|---|
| | $D_s^+D_s^-$ | $D_s^{*+}D_s^-$ | $D_s^{*+}D_s^{*-}$ |
| $\psi(4040)$ | 62 ± 21 | | |
| $\psi(4160)$ | 23 ± 26 | 53 ± 8 | |
| $\psi(4415)$ | 6 ± 11 | 4 ± 2 | 5 ± 12 |
| $Y(4260)$ | 0.5 ± 3.0 | 18 ± 24 | 11 ± 16 |
| non-resonant | 11 ± 5 | 27 ± 5 | 71 ± 20 |
| Sum | 103 ± 36 | 102 ± 26 | 87 ± 28 |

In similar studies of two-body charmed mesons states produced via ISR, Belle does not perform fits to the obtained cross-sections, motivating their choice by the difficulty of taking coupled channel effects into account. However, a fit is performed to the prominent $\psi(4415)$ peak found in the process $e^+e^- \to \gamma_{\rm ISR}\, D^0D^-\pi^+$ (Pakhlova, 2008c). As this peak is observed far from other $\psi$ states, and $\psi(4415)$ decay to this final state turns out to be large, a naïve one-resonance fit is justified in this case. A study of invariant masses of the $D^-\pi^+$ and $D^0\pi^+$ combinations demonstrates that the decay $\psi(4415) \to D^0D^-\pi^+$ is dominated by the $D^0\overline{D}_2^*(2460)^0$ and $D^-\overline{D}_2^*(2460)^+$ intermediate states. Because of their positive interference (due to $C = -1$ of the $\psi(4415)$) Belle does not study them separately, but divides the selected sample into $D\overline{D}_2^*(2460)$ + *c.c.* and non-resonant $D^0D^-\pi^+$ regions. A $\psi(4415)$ peak is seen only in the former region; no sign of $\psi(4415)$ is seen in the second case (Fig. 18.2.13). A fit to the $M_{D^0D^-\pi^+}$ spectrum in the $D\overline{D}_2^*(2460)$ + *c.c.* regions yields 109 ± 25(stat) signal events, and the significance for the $\psi(4415)$ signal is $\sim 10\sigma$. The measured peak mass $M_{\psi(4415)} = (4.411 \pm 0.007(\text{stat}))\,\text{GeV}/c^2$ and total width $\Gamma_{\rm tot} = (77 \pm 20(\text{stat}))\,\text{MeV}$ are in good agreement with the BES results (Ablikim et al., 2007). Belle measures $\mathcal{B}(\psi(4415) \to D\overline{D}_2^*(2460)) \times \mathcal{B}(D_2^*(2460) \to D\pi^+) = (10.5 \pm 2.4 \pm 3.8)\%$ and sets an upper limit on the ratio of the branching fractions of $\psi(4415)$ decays to non-resonant $D^0D^-\pi^+$ and $D\overline{D}_2^*(2460)$ + *c.c.* to be 0.22 at the 90% C.L.

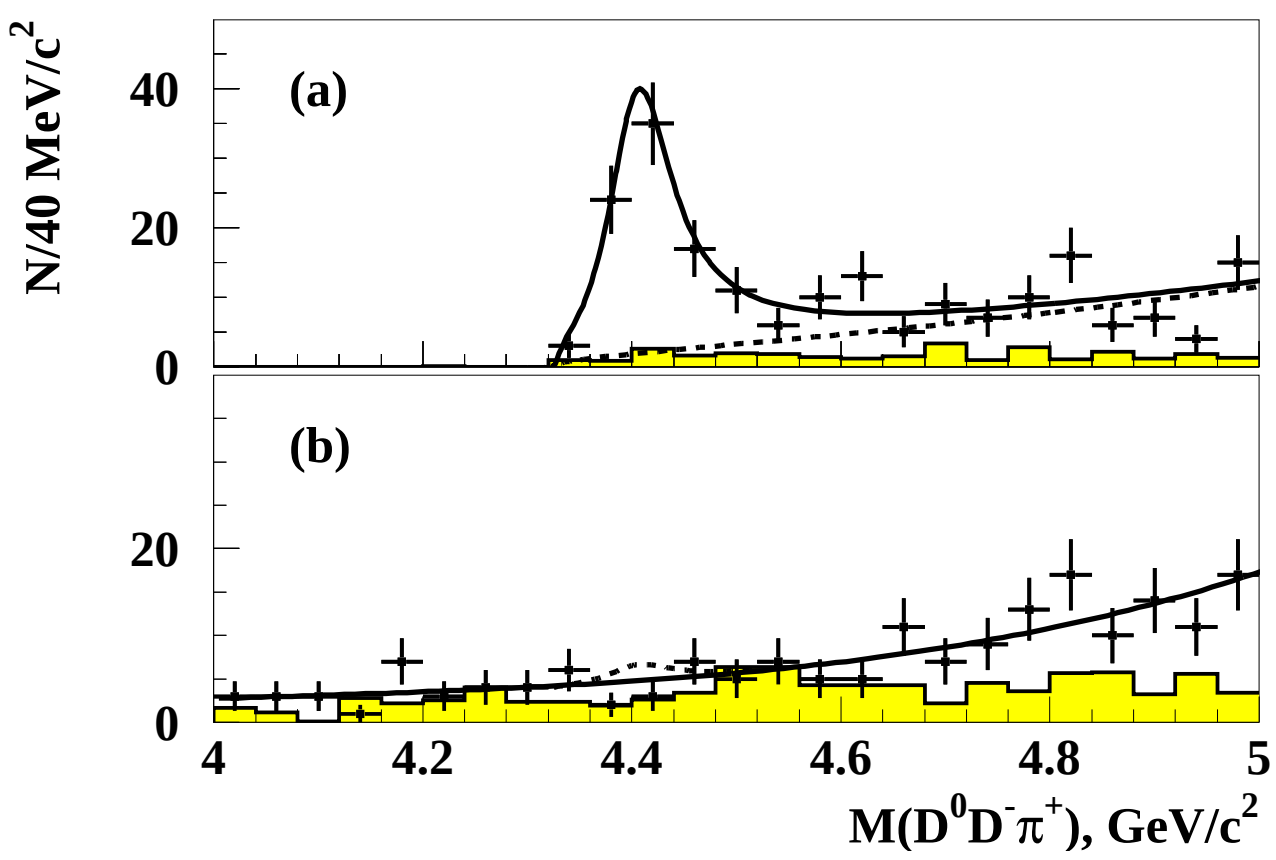

**Figure 18.2.13.** From Pakhlova (2008c): Mass distributions for $D^0D^-\pi^+$ in $e^+e^- \to \gamma_{\rm ISR}D^0D^-\pi^+$ events. Fit results are shown by the solid curve. (a) The $M_{D^0D^-\pi^+}$ spectrum for the $D\overline{D}_2^*(2460)$ signal region; the dashed curve corresponds to the non-$\psi(4415)$ contribution. (b) The $M_{D^0D^-\pi^+}$ spectrum outside the $D\overline{D}_2^*(2460)$ signal region; the dashed curve shows the upper limit on the $\psi(4415)$ yield at the 90% C.L. In both plots, shaded histograms show the normalized contributions from $M_{D^0}$ and $M_{D^-}$ sidebands.

In the study of the process $e^+e^- \to D^0D^{*-}\pi^+$ Belle found only a hint for the $\psi(4415)$ signal with statistical significance $3.1\sigma$ (Pakhlova, 2009) and set an upper limit on the branching fraction $\mathcal{B}(\psi(4415) \to D^0D^{*-}\pi^+) < 10.6\%$ at the 90% C.L.

### 18.2.3 Measurements of parameters

$B$ Factory analyses have contributed to the precision with which various charmonium parameters are known: studies have been performed for the $\eta_c$ (Section 18.2.3.1), $J/\psi$ (Section 18.2.3.2), $\chi_{c0}$ and $\chi_{c2}$ (Section 18.2.3.3), and $\psi(3770)$ (Section 18.2.3.4). The treatment of interference effects is important in a number of these analyses, as discussed below.

**Table 18.2.6.** Summary of $\eta_c$ mass and width measurements obtained by *BaBar* and Belle in different production processes as marked in the second column.

| Experiment | Process | Decay | Mass (MeV/$c^2$) | Width (MeV) | Reference |
|---|---|---|---|---|---|
| Belle | $B \to K\eta_c$ | hadrons | $2979.6 \pm 2.3 \pm 1.6$ | $29 \pm 8 \pm 6$ | Fang (2003) |
| *BaBar* | $B \to K\eta_c$ | inclusive | $2982 \pm 5$ | — | Aubert (2006ae) |
| Belle | $B \to K\eta_c$ | $p\bar{p}$ | $2971 \pm 3^{+2}_{-1}$ | $48^{+8}_{-7} \pm 5$ | Wu (2006) |
| Belle | $B \to K\eta_c$ | $\Lambda\bar{\Lambda}$ | $2974 \pm 7^{+2}_{-1}$ | $40 \pm 19 \pm 5$ | Wu (2006) |
| *BaBar* | $B \to K^{(*)}\eta_c$ | $K\bar{K}\pi$ | $2985.8 \pm 1.5 \pm 3.1$ | $36.3^{+3.7}_{-3.6} \pm 4.4$ | Aubert (2008ba) |
| Belle | $B \to K\eta_c$ | $K^0_S K\pi$ | $2985.4 \pm 1.5^{+0.5}_{-2.0}$ | $35.1 \pm 3.1^{+1.0}_{-1.6}$ | Vinokurova (2011) |
| *BaBar* | $\gamma\gamma \to \eta_c$ | $K\bar{K}\pi$ | $2982.5 \pm 1.1 \pm 0.9$ | $34.3 \pm 2.3 \pm 0.9$ | Aubert (2004s) |
| Belle | $\gamma\gamma \to \eta_c$ | hadrons | $2986.1 \pm 1.0 \pm 2.5$ | $28.1 \pm 3.2 \pm 2.2$ | Uehara (2008b) |
| *BaBar* | $\gamma\gamma \to \eta_c$ | $K^0_S K\pi$ | $2982.2 \pm 0.4 \pm 1.6$ | $31.7 \pm 1.2 \pm 0.8$ | Lees (2010b) |
| Belle | $\gamma\gamma \to \eta_c$ | $\eta'\pi^+\pi^-$ | $2982.7 \pm 1.8 \pm 2.2 \pm 0.3$ | $37.8^{+5.8}_{-5.3} \pm 2.8 \pm 1.4$ | Zhang (2012) |
| Belle | $e^+e^- \to J/\psi\,\eta_c$ | inclusive | $2970 \pm 5 \pm 6$ | — | Abe (2007f) |

### 18.2.3.1 $\eta_c$ mass, width, and transition form factor

The $\eta_c$ is the lightest $S$-wave spin-singlet charmonium state. In spite of a long history of studies, the $\eta_c$ parameters are still not well defined. As detailed in Section 18.1, the $\eta_c$ mass and total width are of particular importance for QCD tests in the charmonium sector, where a set of QCD complications are partially removed due to the large quark mass. While the spin-independent part of the $c\bar{c}$ potential is well-fixed by experimental data, for the study of the spin-dependent part exact knowledge of the $\eta_c$ mass plays an important role. There is a relatively large spread in the $\eta_c$ mass and total width values obtained in different experiments (in Beringer et al., 2012, the fourteen measurements have $\chi^2 = 36$), and no definitive explanation for the discrepancy between the $\eta_c$ parameters measured in $J/\psi$ and $\psi(2S)$ radiative decays, in $\gamma\gamma$ and $p\bar{p}$ production, and in $B$ decays has been suggested to date.

One of the critical issues for the correct measurement of $\eta_c$ parameters in the decays $J/\psi\,(\psi(2S)) \to \eta_c\gamma$ may be the theoretical understanding of the $\eta_c$ line shape in $M1$ radiative transitions. For example, the effect of a distorted $\eta_c$ line shape in these decays was discussed by CLEO (Mitchell et al., 2009b). In other processes where $\eta_c$ parameters are measured, there is another source of systematic uncertainty which was only recently recognized as of possible significance: interference of the $\eta_c$, which is commonly exclusively reconstructed in a multihadron final state, with a non-resonant (continuum) substrate. It is not an easy task to take the effect of interference into account. If the final state contains more than two particles, the resonant structure and the orbital angular momentum between final state hadrons may differ for $\eta_c$ decay and the continuum. Hence the interference may be partial or even absent, and simply adding a coherent continuum amplitude to the $\eta_c$ Breit-Wigner amplitude in the fit does not guarantee results more correct than those obtained when interference is ignored. Over the past decade, both *BABAR* and Belle have carried out many measurements of $\eta_c$ parameters, as summarized in Table 18.2.6. Below we briefly review some recent analyses where the effect of interference has been considered.

In a study of $\eta_c$ production in $\gamma\gamma$ fusion *BABAR* estimates the uncertainty on the $\eta_c$ mass and width due to interference effects (Lees, 2010b). In the baseline fit to the measured $K^0_S K\pi$ mass spectra for the selected $\gamma\gamma$ events the interference term is ignored. To estimate the possible mass shift a fit assuming the maximum (full) interference with the continuum $\gamma\gamma \to K^0_S K\pi$ background is performed. The $\eta_c$ mass value changes by $1.5\,\mathrm{MeV}/c^2$, which is the dominant contribution to the systematic uncertainty. Belle observes a clear $\eta_c$ signal in the mass spectrum of $\eta'\pi^+\pi^-$ combinations produced in two-photon collisions (Zhang, 2012), and measures its mass and width. As in the *BABAR* analysis above, the effect of interference can only be estimated: the differences in the $\eta_c$ parameters with and without interference, $\Delta M = 0.3\,\mathrm{MeV}/c^2$ and $\Delta\Gamma = 1.4\,\mathrm{MeV}$, are taken as model-dependent uncertainties of the measurement. In the Belle study of $B \to K\eta_c$ followed by $\eta_c \to K^0_S K\pi$ (Vinokurova, 2011) an angular analysis is used to distinguish the contributions from the coherent and noncoherent $K^0_S K\pi$ continuum amplitudes from $B \to K(K^0_S K\pi)$ decays, mediated by the penguin diagram. This analysis takes interference into account with no assumptions on its phase or absolute value. If interference is turned off, the fitted mass and width do not vary significantly. Finally, a recent BES paper (Ablikim et al., 2012a) has presented a high statistics measurement of the interference: considering such results in future measurements would help to reduce systematic uncertainties.

Another important $\eta_c$ property, the transition form factor, has been measured by *BABAR* (Lees, 2010b). Such a measurement allows the shape of the charmonium wave function to be probed, and provides a test of the predictions of pQCD as well as calculations that use the lattice QCD approach. *BABAR* studies the process $e^+e^- \to e^+e^-\gamma\gamma^* \to e^+e^-\eta_c$ for the momentum transfer range from 2 to $50\,\mathrm{GeV}^2$. To ensure high virtuality of one of the photons, either the electron or positron is required to be detected, while the other is scattered at a small an-gle and hence escapes detection. The transition form factor $F(Q^2)$ (Fig. 18.2.14) is extracted from the measured differential cross section $\mathrm{d}\sigma/\mathrm{d}Q^2$, where the squared momentum transfer is calculated from the measured ($p'$) and known ($p$) four-momenta of the final and initial state electrons respectively: $Q^2 = -(p' - p)^2$. The obtained distribution is well described by the simple monopole form

$$\left|\frac{F(Q^2)}{F(0)}\right| = \frac{1}{1+Q^2/\Lambda}, \quad \Lambda = (8.5 \pm 0.6 \pm 0.7)\,\mathrm{GeV}^2, \quad (18.2.2)$$

and in fair agreement with the QCD prediction (Feldmann and Kroll, 1997).

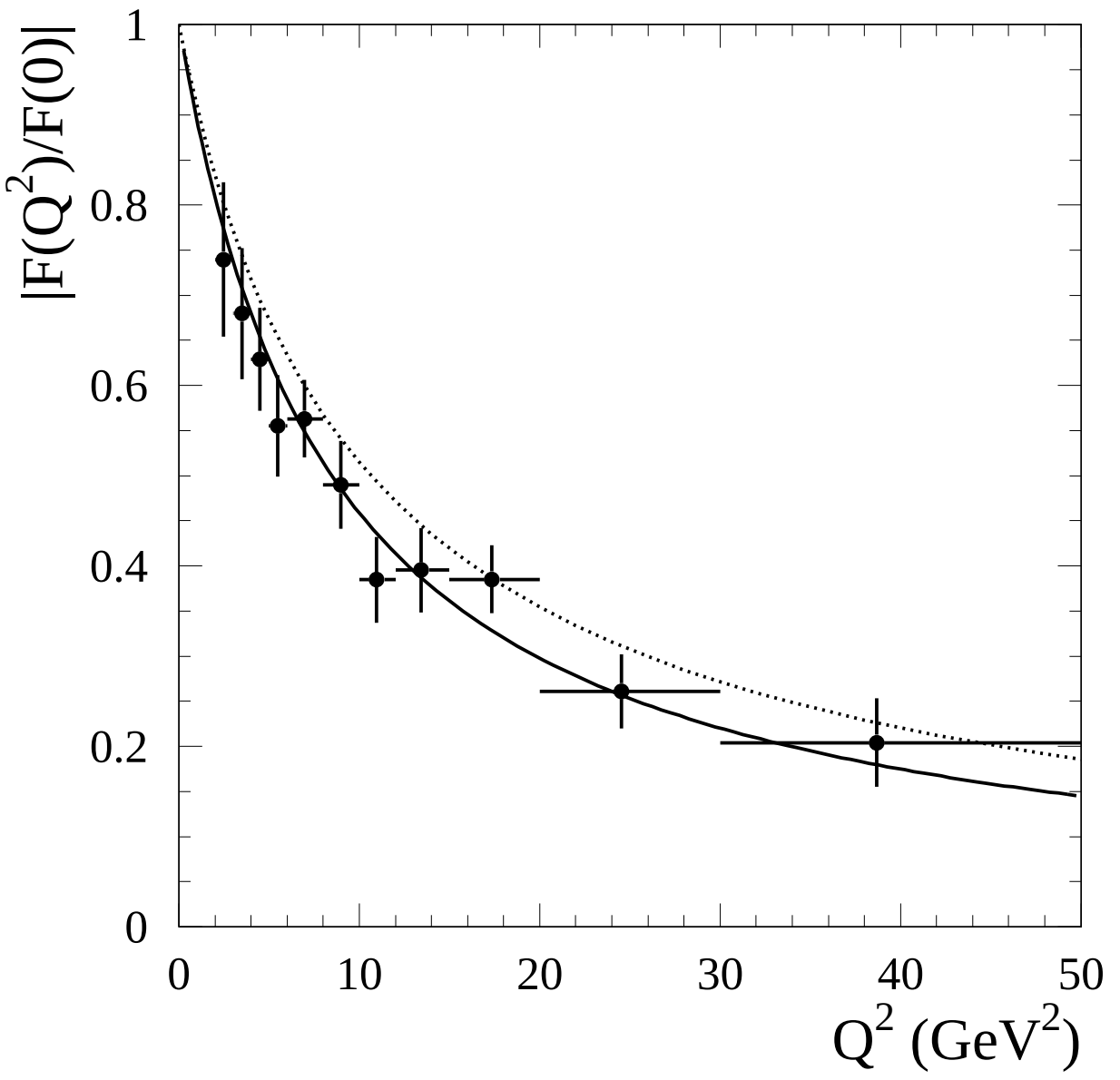

**Figure 18.2.14.** From Lees (2010b): The $F(Q^2)/F(0)$ distribution for the $\eta_c$ (points with error bars). The solid curve shows the fit to the simple monopole shape. The dotted curve shows the leading order pQCD prediction from Feldmann and Kroll (1997).

#### 18.2.3.2 Electronic and total width of the $J/\psi$

The electronic width of $1^{--}$ charmonium resonances is an important characteristic that measures the charmonium wavefunction $\psi(0)$ and helps to fix potential model parameters. Lattice QCD calculations of $\Gamma_{ee}$, which are gradually approaching experimental results in precision, will also soon be put to the test.

Experimentally $\Gamma_{ee}$ can be derived from the resonance peak cross section. This either requires a dedicated energy scan of the resonance at a charm factory, or can be achieved at $B$ Factories via ISR. The latter process provides significant cancellation of systematic uncertainties, as an energy range including both the resonance and nearby regions is simultaneously available. *BABAR* pioneered this method for determination of the $J/\psi$ electronic and total widths (Aubert, 2004d). In this analysis the $J/\psi$ is reconstructed in the dimuon channel only, because the $e^+e^-$ final state has much larger backgrounds from radiative Bhabha events. The directly measured quantity is $\Gamma_{ee} \times \mathcal{B}(J/\psi \to \mu^+\mu^-)$,[103] which is found to be equal to $(0.3301 \pm 0.0077 \pm 0.0073)\,\mathrm{keV}$. Then using the known leptonic branching fractions it is possible to derive both the electronic width $\Gamma_{ee} = (5.61 \pm 0.20)\,\mathrm{keV}$, and the total width $\Gamma_{\mathrm{tot}} = \Gamma_{ee}/\mathcal{B}(J/\psi \to e^+e^-) = (94.7 \pm 4.4)\,\mathrm{keV}$. The statistical and systematic uncertainties are combined in quadrature. The *BABAR* $\Gamma_{ee} \times \mathcal{B}(J/\psi \to \mu^+\mu^-)$ result is one of three measurements contributing to the current world average, which is $(0.334\pm0.005)\,\mathrm{keV}$ (Beringer et al., 2012).

#### 18.2.3.3 $\chi_{c0}$ and $\chi_{c2}$

The $B$ Factories have also made a moderate contribution to the precision measurement of the $P$-wave charmonium masses and widths. Belle has measured charmonium production in two-photon collisions (Uehara, 2008b), observing signals for the three $C$-even charmonia $\eta_c$, $\chi_{c0}$, and $\chi_{c2}$ in the $\pi^+\pi^-\pi^+\pi^-$, $K^+K^-\pi^+\pi^-$, and $K^+K^-K^+K^-$ decay modes. The invariant mass distributions in the vicinity of each charmonium peak are fitted to the sum of charmonium and background components. The combined results for the three decay modes yield a $\chi_{c0}$ mass of $(3414.2 \pm 0.5 \pm 2.3)\,\mathrm{MeV}/c^2$ and a width of $(10.6 \pm 1.9 \pm 2.6)\,\mathrm{MeV}$. The measured $\chi_{c2}$ mass is $(3555.3 \pm 0.6 \pm 2.2)\,\mathrm{MeV}/c^2$.

*BABAR* has also searched for resonances decaying to $\eta_c\pi^+\pi^-$ in two-photon collisions (Lees, 2012t). In this analysis $\eta_c$ is reconstructed in the $K^0_S K\pi$ decay mode, and searches for several known charmonium states, including the $\chi_{c2}$, are performed in the reconstructed $\eta_c\pi^+\pi^-$ mass spectrum. The fit in the $\chi_{c2}$ region yields a central value for the ratio of branching fractions $\frac{\mathcal{B}(\chi_{c2}\to\eta_c\pi^+\pi^-)}{\mathcal{B}(\chi_{c2}\to K^0_S K\pi)} = 14.5 \pm 9.8(\mathrm{stat}) \pm 7.3(\mathrm{syst}) \pm 2.5\,(\mathcal{B})$, where the last uncertainty is due to the uncertainty on $\mathcal{B}(\eta_c \to K^0_S K\pi)$. No significant signal is found, and an upper limit of 31.4 is set on this ratio of branching fractions at the 90% C.L.

#### 18.2.3.4 $\psi(3770)$ mass and width

The $\psi(3770)$ is thought to be a $D$-wave state with a small admixture of $S$-wave, and as it is above $D\overline{D}$ threshold, it is expected to decay mostly to $D\overline{D}$. The $\psi(3770)$ has been investigated in direct formation in $e^+e^-$ annihilation by numerous experiments since its observation by MARK I (Rapidis et al., 1977). Precise measurements of its mass and width have been obtained from energy scans near the resonance. However, as the accuracy of measurements has increased, an anomalous deviation of the $\psi(3770)$ peak lineshape from the Breit-Wigner function has appeared (Ablikim et al., 2008a), and remained puzzling until recently.

[103] The peak Born cross section, which can be extracted from the fit to the data spectrum, is equal to $\sigma^{\mathrm{peak}}_{\mathrm{Born}} = \frac{12\pi^2}{ms}\Gamma_{ee}\mathcal{B}(J/\psi \to \mu^+\mu^-)$

At the $B$ Factories one can use the process with initial state radiation to study $\psi(3770)$ formation, and both *BABAR* and Belle have observed the $\psi(3770)$ in their analysis of $e^+e^- \to \gamma_{\rm ISR} D\overline{D}$ (Pakhlova, 2008a and Aubert, 2009n; see Section 21.4.2). Only *BABAR* has fitted its $D\overline{D}$ mass spectrum to measure $M = (3778.8 \pm 1.9 \pm 0.9)\,{\rm MeV}/c^2$ and $\Gamma = (23.5 \pm 3.7 \pm 0.9)\,{\rm MeV}$. Although the electromagnetic suppression of ISR processes results in small data samples[104] and does not allow the study of the $\psi(3770)$ peak in detail, the access to the large energy range provided by ISR turns out to be extremely important for understanding the $\psi(3770)$ lineshape. Both *BABAR* and Belle observe a structure in the ISR cross section at $\sim 3.9\,{\rm GeV}/c^2$ (Fig. 21.4.3), known as $G(3900)$,[105] which must be taken into account to describe the cross section in the region below 4 GeV. This observation suggests that resonance-continuum interference is essential for determination of the $\psi(3770)$ parameters. A recent KEDR analysis of $e^+e^-$ scan data (Anashin et al., 2012), which includes interference with the tail of the $\psi(2S)$ resonance, concludes that the interference causes a significant shift in the fitted $\psi(3770)$ peak and can explain the nontrivial $\psi(3770)$ lineshape. The Particle Data Group (Beringer et al., 2012), when determining the $\psi(3770)$ mass, now uses only those analyses which take interference into account.

The $B$ Factories have also observed the $\psi(3770)$ in $B$ decays (Brodzicka, 2008; Chistov, 2004; Aubert, 2008bd). The measured mass and width are in good agreement with the parameters obtained from the direct formation analysis that accounts for interference.

### 18.2.4 Production

The $B$ Factories also provide useful information on charmonium production mechanisms. Measurement of the charmonium production rates in different processes, as well as kinematic characteristics of produced charmonia, help to test models numerically and to determine charmonium properties. At the $B$ Factories charmonia are produced in $\gamma\gamma$ fusion, via resonant direct production in $e^+e^-$ annihilation with initial state radiation, in the decays of $B$ mesons, and in the fragmentation of $c\overline{c}$ pairs produced in $e^+e^-$ annihilation.

The former two processes provide a direct measurement of important charmonium parameters, namely the two-photon and dielectron widths. Both are related to the charmonium wave function, and are used to fix the parameters of the potential models that describe charmonium spectroscopy. Chapters 21 and 22 describe in detail the numerous experimental results obtained at the $B$ Factories, in ISR and two-photon physics respectively.

When describing charmonium formation from $c\overline{c}$ pairs produced either in $B$ decays or in $e^+e^-$ annihilation, effective field theories are used. The EFT most often exploited is non-relativistic QCD (NRQCD), which assumes factorization of the production of charmonium partons (*e.g.* a $c\overline{c}$ pair) in the given process, and the formation of charmonium from those partons (Bodwin, Braaten, and Lepage, 1995; Caswell and Lepage, 1986; Thacker and Lepage, 1991). The former part contains a partonic level cross section generally calculated in perturbative QCD, in which the $c\overline{c}$ pair may be produced in a color singlet or color octet state (Braaten and Fleming, 1995; Cho and Leibovich, 1996a). The latter part, which describes the evolution of the $c\overline{c}$ pair with the quantum numbers of the final charmonium state, cannot be calculated in perturbation theory, and the relevant parameters are usually extracted from the data. A signature of the NRQCD approach is the universality of the long distance production matrix elements, which are assumed to be independent of the hard process of parton production. For a more extensive account of EFTs and quarkonia, see Section 18.1.4.

Before presenting the experimental results it is worth emphasizing that the $B$ Factories provide the cleanest processes for calculation of charmonium production, as $c\overline{c}$-pairs in $B$ decays (Section 18.2.4.1) and $e^+e^-$ annihilation (Section 18.2.4.2) are produced via weak and electromagnetic processes, which can be calculated exactly. However in some cases, the test of such predictions requires a careful treatment of the details of the experimental measurements (Section 18.2.4.3).

#### 18.2.4.1 $B$ decays

$B$ mesons can decay into almost all possible charmonium states, with typical inclusive branching fractions $\sim 1\%$, although some states are dynamically suppressed. The first example of charmonium production in $B$ decays, $B \to J/\psi X$, was discovered in 1985 by the ARGUS and CLEO collaborations (Albrecht et al., 1985b; Haas et al., 1985). At present, the Particle Data Group lists branching ratios for 31 charmonium modes, while upper limits are set for a further 26 modes. Although the $B$ Factories have made a formidable contribution to the majority of these measurements, in this section we limit ourselves to the first observations of inclusive and exclusive $B$ to charmonium decays that are interesting for the theory of charmonium production.

*Inclusive decays*

*Two-body $B$-decays*

Inclusive $B$ decays to charmonia provide a very good opportunity to test charmonium production models. The well measured inclusive $J/\psi$ production rate (after subtraction of the contribution from cascade decays $\psi(2S)$ and $\chi_c \to J/\psi\, X$) is a factor $5-10$ larger than the predicted color singlet contribution: inclusion of the color octet mechanism to resolve this discrepancy therefore makes the octet the dominant contribution. One of the cleanest ways to check whether this conclusion is correct is to measure the $\chi_{c2}$-to-$\chi_{c1}$ production ratio in $B$ decays: the only contribution to $\chi_{c2}$ production comes from the color octet model, which favors $\chi_{c2}$ over $\chi_{c1}$ production, the rate being proportional to $2J+1$, the number of spin states.[106] Experimentally, $B \to \chi_{c1}\, X$ was measured many years ago by ARGUS (Albrecht et al., 1992b), while only an upper limit had been set on $\chi_{c2}$ production before the $B$ Factories began operation (Chen et al., 2001c).

[104] Note that the emission of the ISR photon is suppressed by the electromagnetic coupling constant $\alpha_{\rm EM}$.

[105] The $G(3900)$ is not considered to be a real resonance, as the appearance of a bump in this region is qualitatively consistent with predictions of the coupled-channel model of Eichten, Gottfried, Kinoshita, Lane, and Yan (1980).

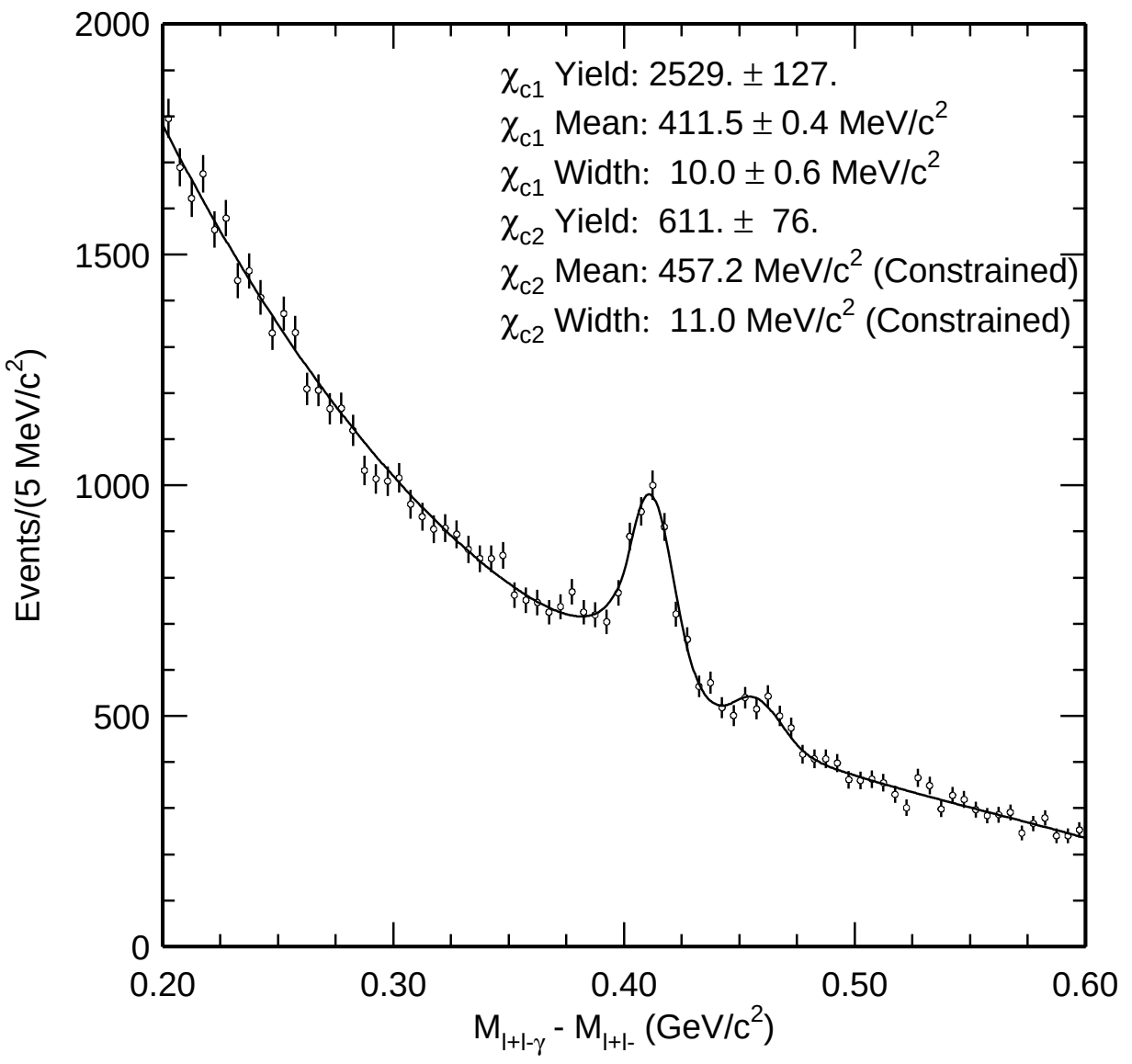

**Figure 18.2.15.** Mass difference between $J/\psi\gamma$ and $J/\psi$ candidates in $B$ decays (Abe, 2002i).

Inclusive $B \to \chi_{c2}X$ decays were first observed by Belle in 2002 using $29.4\,\mathrm{fb}^{-1}$ of data (Abe, 2002i). $\chi_c$ candidates are reconstructed in the $J/\psi\gamma$ mode. In addition to the prominent $\chi_{c1}$ peak, a $\chi_{c2}$ signal is clearly seen in the $J/\psi\gamma - J/\psi$ mass difference spectrum (Fig. 18.2.15). Signal yields are extracted by fitting the distribution with the sum of two Crystal Ball functions representing the $\chi_{c1}$ and $\chi_{c2}$ contributions, and a third-order Chebyshev polynomial parameterizing the background. After subtraction of the $\psi(2S) \to \chi_{cJ}\gamma$ feed-down the direct branching fractions are found to be $\mathcal{B}(B \to \chi_{c1}X) = (3.32 \pm 0.22 \pm 0.34) \times 10^{-3}$ and $\mathcal{B}(B \to \chi_{c2}X) = (1.80^{+0.23}_{-0.28} \pm 0.26) \times 10^{-3}$. A similar analysis was performed by *BABAR* (Aubert, 2003n), with results in good agreement with those of Belle: $\mathcal{B}(B \to \chi_{c1}X) = (3.41 \pm 0.35 \pm 0.42) \times 10^{-3}$ and $\mathcal{B}(B \to \chi_{c2}X) = (1.90 \pm 0.45 \pm 0.29) \times 10^{-3}$. As can be seen, the ratio of production rates of $\chi_{c2}$ and $\chi_{c1}$ is roughly $1:2$, between the pure color singlet and pure color octet predictions ($0:1$ and $5:3$ respectively).

Two-body decays of the type $B \to (c\bar{c})_{\mathrm{res}}K^{(*)}$ have been extensively studied, because of their extremely clean experimental environment and their importance for *CP*-violation measurements. Theoretical calculations for these decays are more difficult than those for inclusive charmonium production, as they have to include the fragmentation of light quarks into $K^{(*)}$ mesons, introducing an additional uncertainty. However, for such decays it is justified to use the factorization hypothesis, since a charmonium state (which does not pick up the spectator quark from the $B$ meson) is an object of small size and escapes the decay region; only the kaon partner is affected by soft-gluon exchange. The factorization approach predicts large suppression in the production of $\chi_{c0}$, $h_c$, and $\chi_{c2}$ in comparison with $\chi_{c1}$ in $B \to (c\bar{c})_{\mathrm{res}}K$ decays (Beneke and Vernazza, 2009). By the start of $B$ Factory data taking, only the $B \to \chi_{c1}K$ decay had been observed.

The decay $B \to \chi_{c0}K$ was seen for the first time at the $B$ Factories; this process was difficult to observe, due to the small $\chi_{c0}$ branching fractions to modes suitable for reconstruction. In 2001 Belle (Abe, 2002e) observed a $B^+ \to \chi_{c0}K^+$ signal in two $\chi_{c0}$ decay modes: $\pi^+\pi^-$ and $K^+K^-$. A more substantial study of this decay, that takes into account interference of the $\chi_{c0}$ resonance with a large variety of possible intermediate hadron resonances in the $K^+\pi^-$, $\pi^+\pi^-$, and $K^+K^-$ systems, was performed by Belle (Garmash, 2005) with a larger data set using the Dalitz analysis technique (see Chapter 13). In this analysis, signal events are selected from an ellipse around the nominal $\Delta E$ and $m_{\mathrm{ES}}$ values in the $\Delta E - m_{\mathrm{ES}}$ plane. The regions with dipion mass around the $J/\psi$ or $\psi(2S)$ nominal masses contain a large background from $B^+ \to J/\psi\,(\psi(2S))K^+$ decays followed by $J/\psi\,(\psi(2S)) \to \mu^+\mu^-$, where both muons are misidentified as pions. Similarly, the region in the $K^+\pi^-$ mass corresponding to $\overline{D}^0 \to K^+\pi^-$ decay is contaminated by the $B^+ \to \overline{D}^0K^+$ process. These three regions are excluded from further analysis. The Dalitz plot for the signal region is shown in Fig. 18.2.16 (a) and (b) for $B^+ \to \pi^+\pi^-K^+$ and $K^+K^-K^+$ decays respectively. The $B^+ \to \chi_{c0}K^+$ signal can be seen as a horizontal band at $M^2(\pi^+\pi^-)$ and $M^2(K^+K^+) \sim 11.6\,\mathrm{GeV}^2/c^4$. The $\chi_{c0}$ signal yield in $B^+ \to \pi^+\pi^-K^+$ is extracted by an unbinned maximum-likelihood fit to the Dalitz distribution with a coherent sum of all known intermediate quasi-two-body processes ($\chi_{c0}K^+$, $K^*(892)^0\pi^+$, $K_0^*(1430)^0\pi^+$, $\rho(770)^0K^+$, $f_0(980)K^+$, $f(1300)K^+$, $\kappa\pi^+$), a non-resonant three-body $K^+\pi^+\pi^-$ contribution, and a background shape fixed from sideband studies. A similar procedure is used for the $K^+K^-K^+$ final state. Significant signals for $\chi_{c0}$ are observed in both modes, and the combined branching fraction is found to be $\mathcal{B}(B^+ \to \chi_{c0}K^+) = (1.96 \pm 0.35 \pm 0.33^{+1.97}_{-0.26}) \times 10^{-4}$, where the first error is statistical, the second is systematic, and the third is the model error due to Dalitz plot parameterization. Subsequent, similar measurements based on larger samples by Belle ($\mathcal{B}(B^+ \to \chi_{c0}K^+) = (1.12 \pm 0.12\,^{+0.30}_{-0.20}) \times 10^{-4}$; Garmash, 2006) and *BABAR* ($\mathcal{B}(B^+ \to \chi_{c0}K^+) = (1.23\,^{+0.27}_{-0.25} \pm 0.06) \times 10^{-4}$; Aubert, 2008j) are in good agreement.

[106] If the $c\bar{c}$ pair is produced in the singlet state, it can directly form a meson, but this one can only have $J = 0, 1$. In order to be able to produce the $\chi_{c2}$ state, which has $J = 2$, a gluon needs to be emitted; gluon emission is also necessary when the $c\bar{c}$ is produced in a color octet state. Note also that in the naïve factorization approach $\Gamma(B \to \chi_{c0(2)}P)$ is expected to vanish as explained in Section 17.3.5.4.

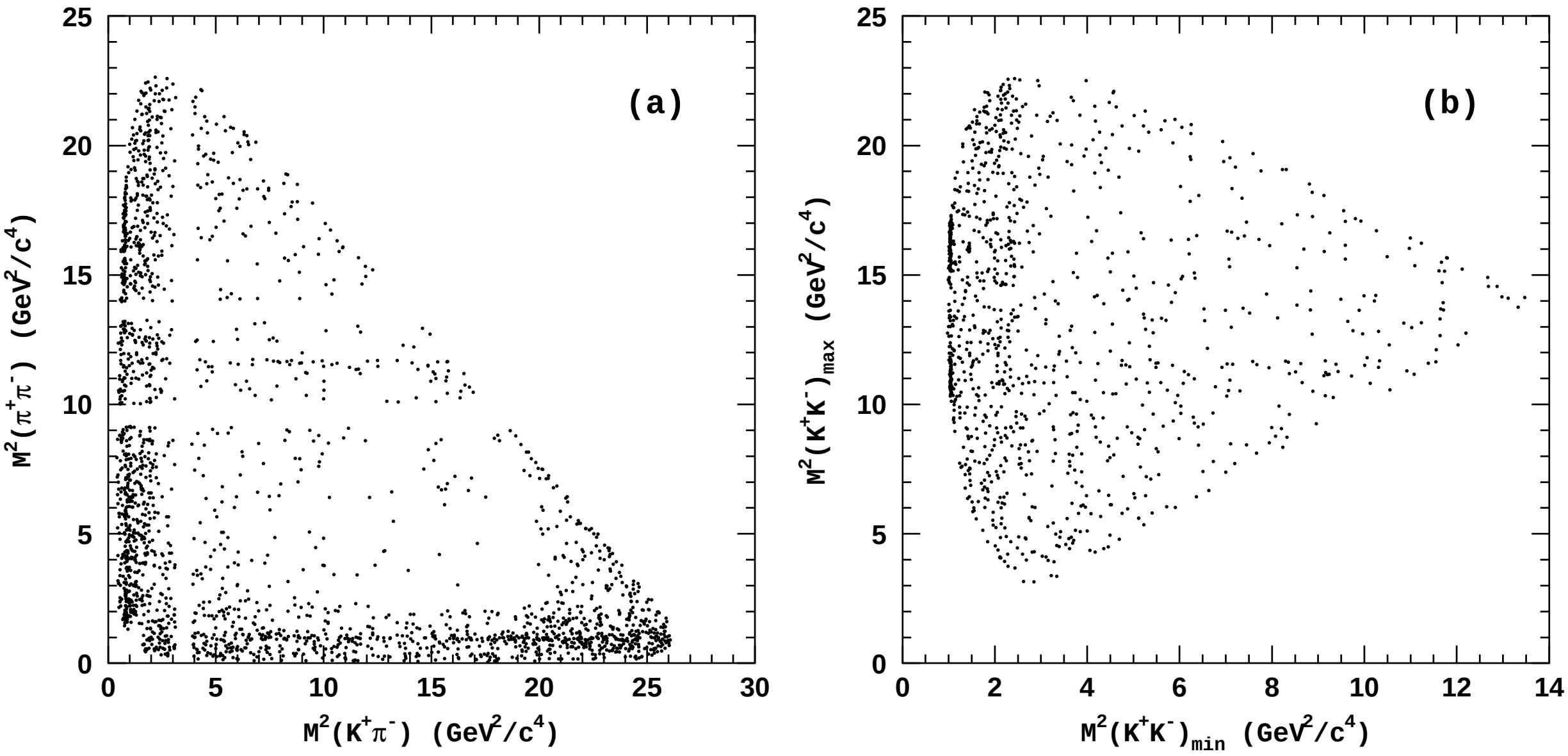

**Figure 18.2.16.** From Garmash (2005): Dalitz plot for events in the signal region for the (a) $B^+ \to \pi^+\pi^-K^+$ and (b) $B^+ \to K^+K^-K^+$ processes.

Two-body $B$ decay into $\chi_{c2}$ (such as $B \to \chi_{c2}K^{(*)}$) has not yet been observed with high statistical significance. The upper limit obtained by *BABAR* is $\mathcal{B}(B^+ \to \chi_{c2}K^+) < 1.8 \times 10^{-5}$ at 90% C.L. (Aubert, 2009m). Belle has found $3.6\sigma$ evidence for the $B^+ \to \chi_{c2}K^+$ decay, with $\mathcal{B}(B^+ \to \chi_{c2}K^+) = (1.11\,^{+0.36}_{-0.34} \pm 0.09) \times 10^{-5}$ (Bhardwaj, 2011), *i.e.* almost 40 times smaller than the branching fraction for the $B^+ \to \chi_{c1}K^+$ decay. There is also an upper limit from Belle $\mathcal{B}(B^+ \to h_cK^+) < 3.8 \times 10^{-5}$ (Fang, 2006). Such a large suppression of production of $\chi_{c2}$ and $h_c$ with respect to $\chi_{c1}$ in two-body $B$ decays is anticipated by theory, as discussed above.

#### 18.2.4.2 $e^+e^-$ annihilation

Prompt charmonium production in $e^+e^-$ annihilation was first observed in 1990 by the CLEO collaboration (Alexander et al., 1990), which found $15.2 \pm 4.9$ events with reconstructed $J/\psi$ above the kinematical limit for $B$-decays ($p_{J/\psi} > 2\,\mathrm{GeV}/c$) in the $\Upsilon(4S)$ data. In the first analysis this observation was misinterpreted as non-$B\overline{B}$ decays of $\Upsilon(4S)$, but later a $J/\psi$ signal was also seen in the CLEO continuum data.

For more than ten years after this observation, there were attempts by theoreticians to explain the estimated cross section ($\sigma \sim 2\,\mathrm{pb}$) without new experimental inputs. Due to the lack of experimental information, all possible production mechanisms had to be considered. The dominant contribution to prompt $J/\psi$ production was expected to be due to color singlet and color octet diagrams $e^+e^- \to c\overline{c}\,g(g)$. In the color singlet $e^+e^- \to c\overline{c}\,gg$ process, two hard gluons are emitted, pushing the mass of the $c\overline{c}$ pair into the charmonium region. Although the radiation of two gluons is suppressed by $\alpha_S^2$, the contribution of this diagram is comparable with single gluon production because it provides a colorless $c\overline{c}$ pair, which can be directly projected into a physical charmonium state, *e.g.* $J/\psi$ (Fig. 18.2.17 (a)). NRQCD, based on leading-order perturbative QCD calculations, predicted that the cross section of the color singlet process $e^+e^- \to c\overline{c}gg \to J/\psi\,X$ might be as high as 0.8 pb (Cho and Leibovich, 1996b; Yuan, Qiao, and Chao, 1997b). The color octet $e^+e^- \to c\overline{c}g$ diagram leads to the formation of a color $(c\overline{c})^8$ state, which is required to be "decolorized" by emission of another soft gluon before it can be transformed into a physical charmonium state (Fig. 18.2.17 (b)). Due to the large value of $\alpha_S$ at low energy, such emission is both large and impossible to compute perturbatively. According to theoretical estimates, the color singlet and color octet contributions may be of the same order (Schuler, 1999; Yuan, Qiao, and Chao, 1997a,b), although the uncertainty of this estimate is large due to poorly-constrained color octet matrix elements. Another color singlet diagram $e^+e^- \to c\overline{c}\,c\overline{c}$ (Fig. 18.2.17 (c)) that can contribute to prompt charmonium production was estimated to be so small ($\sim 0.05\,\mathrm{pb}$; Kiselev, Likhoded, and Shevlyagin, 1994), that detection of this process was considered hardly possible.

##### *Initial cross section measurements*

In 2001 both *BABAR* and Belle performed much more precise measurements of the $e^+e^- \to J/\psi\,X$ cross section using the data sets obtained in the first year of their operation ($\mathcal{L} \sim 20\,\mathrm{fb}^{-1}$). In both collaborations the $J/\psi$ production was studied in the full momentum interval: the region below $2\,\mathrm{GeV}/c$ was studied using continuum data. *BABAR* obtained $(2.52 \pm 0.21 \pm 0.21)\,\mathrm{pb}$ (Aubert, 2001b), while Belle obtained $(1.47 \pm 0.10 \pm 0.13)\,\mathrm{pb}$ (Abe, 2002n). The discrepancies between the two measurements are likely due to differences in the selection criteria for $J/\psi$ events that were used to suppress contributions from the huge QED background. Corrections for the selection efficiency are model dependent and may result in poorly controlled systematic uncertainty. (See also the discussion in Section 18.2.4.3 below.) While the measured cross section is not in contradiction with the NRQCD predictions (color singlet + color octet) of 1.1–1.6 pb (Yuan, Qiao, and Chao, 1997a,b), the expected sole color singlet contribution is too small to describe the data. On the other hand, the $J/\psi$ momentum spectrum measured by Belle and *BABAR* does not show any indication of the sizable color octet contribution, that was expected to result in an enhancement at the maximum momentum value. *BABAR* and Belle also measured the $J/\psi$ production and helicity angle distributions, which roughly agree with NRQCD expectations.

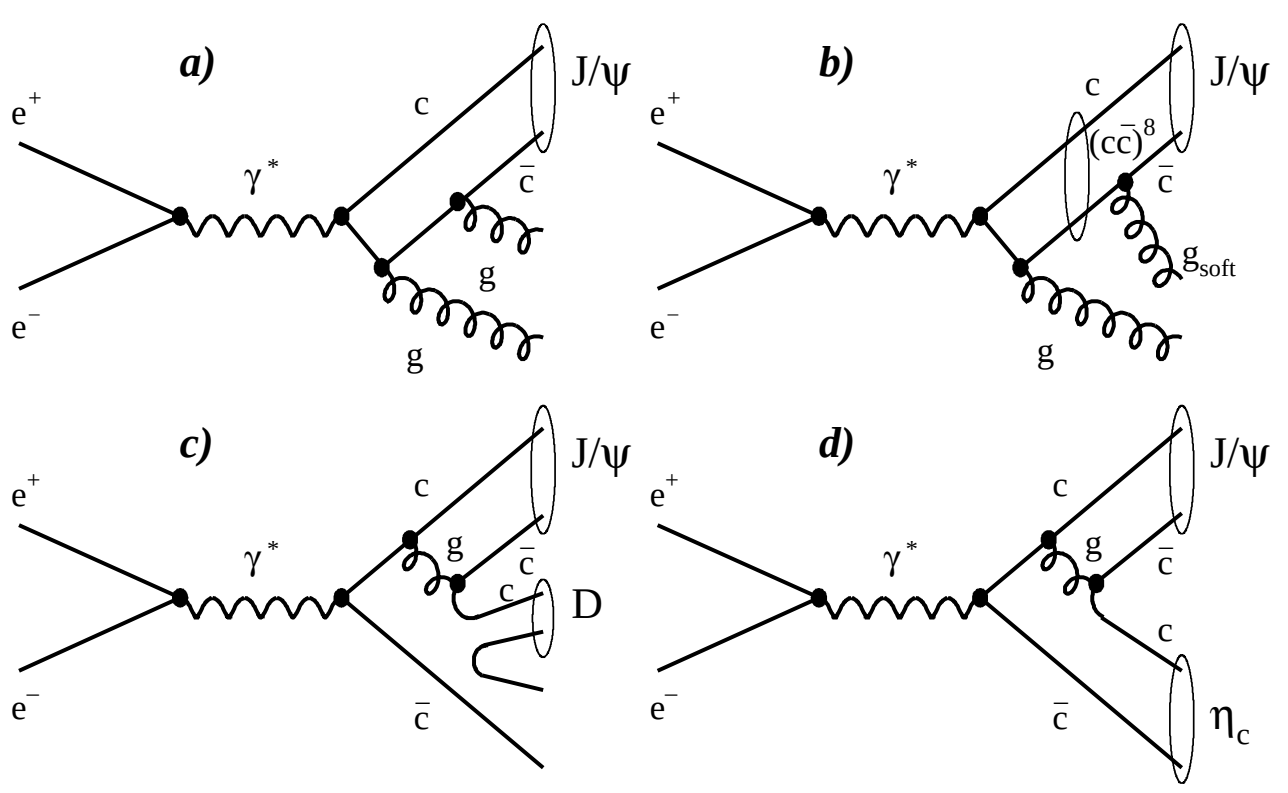

**Figure 18.2.17.** Feynman diagrams describing $J/\psi$ production in $e^+e^-$ annihilation: see the text for details.

Belle and *BABAR* performed searches for other charmonium states produced in $e^+e^-$ annihilation. In addition to the $J/\psi$ production study, Belle also measured $\sigma(e^+e^- \to \psi(2S)\,X) = (0.67 \pm 0.09^{+0.09}_{-0.11})\,\mathrm{pb}$ (Abe, 2002n) and set upper limits on the production of $\chi_{c1}$ and $\chi_{c2}$. Later *BABAR*, using a much larger data sample, improved these limits: $\sigma^{\rm prompt}_{N_{ch}\geq 3}\,(e^+e^- \to \chi_{c1(2)}\,X) < 77(79)\,\mathrm{fb}$ at the 90% confidence level (Aubert, 2007at). Upper limits were set for events where the charmonium momentum exceeds $2.0\,\mathrm{GeV}/c$ and there are at least three additional charged tracks. These limits are consistent with NRQCD predictions.

### *The recoil mass analyses*

In 2002, contrary to NRQCD expectations, Belle observed that most of the prompt $J/\psi$'s are accompanied by charmed hadrons (Abe, 2002j). The $J/\psi$ is reconstructed in its $\ell^+\ell^-$ ($\ell = e, \mu$) decays, with its mass constrained to the nominal value to improve the momentum resolution. $B\overline{B}$ background is suppressed by requiring $p^*_{J/\psi} > 2\,\mathrm{GeV}/c$, where $p^*_{J/\psi}$ is the $J/\psi$ momentum in the CM system. Backgrounds from QED $e^+e^- \to \ell^+\ell^-(\gamma)$ processes are suppressed by requiring the number of tracks in each event to be larger than 4. Unexpectedly, Belle found that a significant fraction of $J/\psi$'s are produced together with another charmonium state in two-body reactions of the type $e^+e^- \to J/\psi\,\eta_c$ (Fig. 18.2.17 (d)). More often, charmed mesons are found in the events with reconstructed $J/\psi$ (Fig. 18.2.17 (c)). To identify the first type of process (with the double charmonium final state) it is not necessary to reconstruct both mesons, which inevitably leads to substantial efficiency loss. Relying on four-momentum conservation, the mass of the system produced together with $J/\psi$ can be calculated using only the measured $J/\psi$ momentum. The mass of the system recoiling against the $J/\psi$ candidate—the "recoil mass"— is defined as

$$M_{\rm recoil}(J/\psi) = \left[(\sqrt{s} - E^*_{J/\psi})^2 - p^{*\,2}_{J/\psi}\right]^{1/2}, \qquad (18.2.3)$$

where $E^*_{J/\psi}$ is the $J/\psi$ energy in the CM system. A clear peak was observed around the $\eta_c$ mass; two more peaks were seen around the masses of $\chi_{c0}$ and $\eta_c(2S)$.

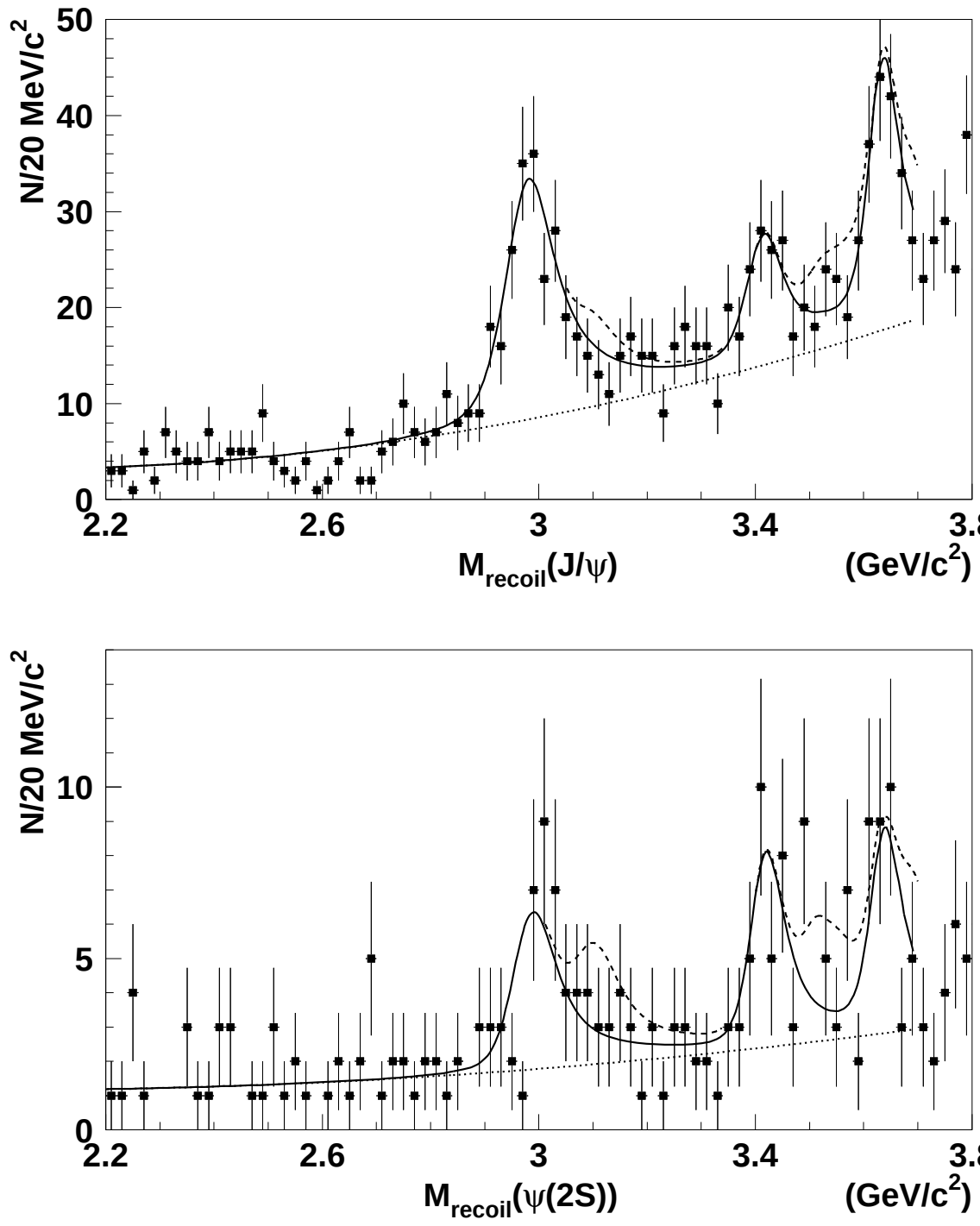

**Figure 18.2.18.** From Abe (2004g): the mass of the system recoiling against the reconstructed (top) $J/\psi$ and (bottom) $\psi(2S)$ in inclusive $e^+e^- \to J/\psi\,(\psi(2S))\,X$ events. The solid curve is the result of a fit that includes $\eta_c$, $\chi_{c0}$, and $\eta_c(2S)$; the dashed curve is the background contribution.

**Table 18.2.7.** Comparison of experimental cross sections ($\sigma \times \mathcal{B}_{>2}$ in fb, see text for symbol definition) with theoretical expectations that do not include the $\mathcal{B}_{>2}$ factor.

| $J/\psi\ c\bar{c}$ | $\eta_c$ | $\chi_{c0}$ | $\eta_c(2S)$ | Reference |
|---|---|---|---|---|
| Belle | $25.6\ \pm 2.8 \pm 3.4$ | $6.4\ \pm 1.7 \pm 1.0$ | $16.5\ \pm 3.0 \pm 2.4$ | Abe (2004g) |
| *BABAR* | $17.6\ \pm 2.8^{+1.5}_{-2.1}$ | $10.3\ \pm 2.5^{+1.4}_{-1.8}$ | $16.4\ \pm 3.7^{+2.4}_{-3.0}$ | Aubert (2005n) |
| NRQCD | $3.78 \pm 1.26$ | $2.40 \pm 1.02$ | $1.57 \pm 0.52$ | Braaten and Lee (2003) |
| NRQCD | 5.5 | 6.9 | 3.7 | Liu, He, and Chao (2003) |
| Light cone | $14.4\ ^{+11.2}_{-9.8}$ | – | $13.0\ ^{+12.2}_{-11.0}$ | Braguta (2009) |
| NLO NRQCD | $17.6\ ^{+10.7}_{-8.3}$ | – | – | Bodwin, Lee, and Yu (2008) |

This observation was later confirmed by subsequent Belle (Fig. 18.2.18; Abe, 2004g) and *BABAR* analyses (Fig. 18.2.4; Aubert, 2005n) using larger samples. The most recent results are summarized in Table 18.2.7. Because of the selection criteria applied by both collaborations the results are given in terms of the product of the cross section and the branching fraction of the recoil charmonium state into more than 2 charged tracks, $\sigma \times \mathcal{B}_{>2}$. The similar process involving $\psi(2S)$ was also observed by Belle (Fig. 18.2.18; Abe, 2004g). Surprisingly, the cross sections of double charmonium production with $\psi(2S)$ are close to those with $J/\psi$.

Following the observation of double charmonium production, the corresponding cross sections were calculated using NRQCD to be an order of magnitude smaller than experimental values (Braaten and Lee, 2003; Liu, He, and Chao, 2003). Later the importance of relativistic corrections was recognized by Ma and Si (2004) and Bondar and Chernyak (2005); the relative momentum of the heavy quarks in the charmonium was taken into account using the light cone approximation. As a result, the calculated cross sections are now close to the experimental values though within a large uncertainty (Braguta, 2009). Alternatively, other authors (Bodwin, Lee, and Yu, 2008; He, Fan, and Chao, 2007) suggested to resolve the discrepancy within the NRQCD approach by the resummation of the corrections of next-to-leading order (NLO) in $\alpha_S$, relativistic corrections, and contributions from pure QED diagrams. The theoretical expectations (not including the $\mathcal{B}_{>2}$ factor) are summarized in comparison with the Belle and *BABAR* measurements in Table 18.2.7.

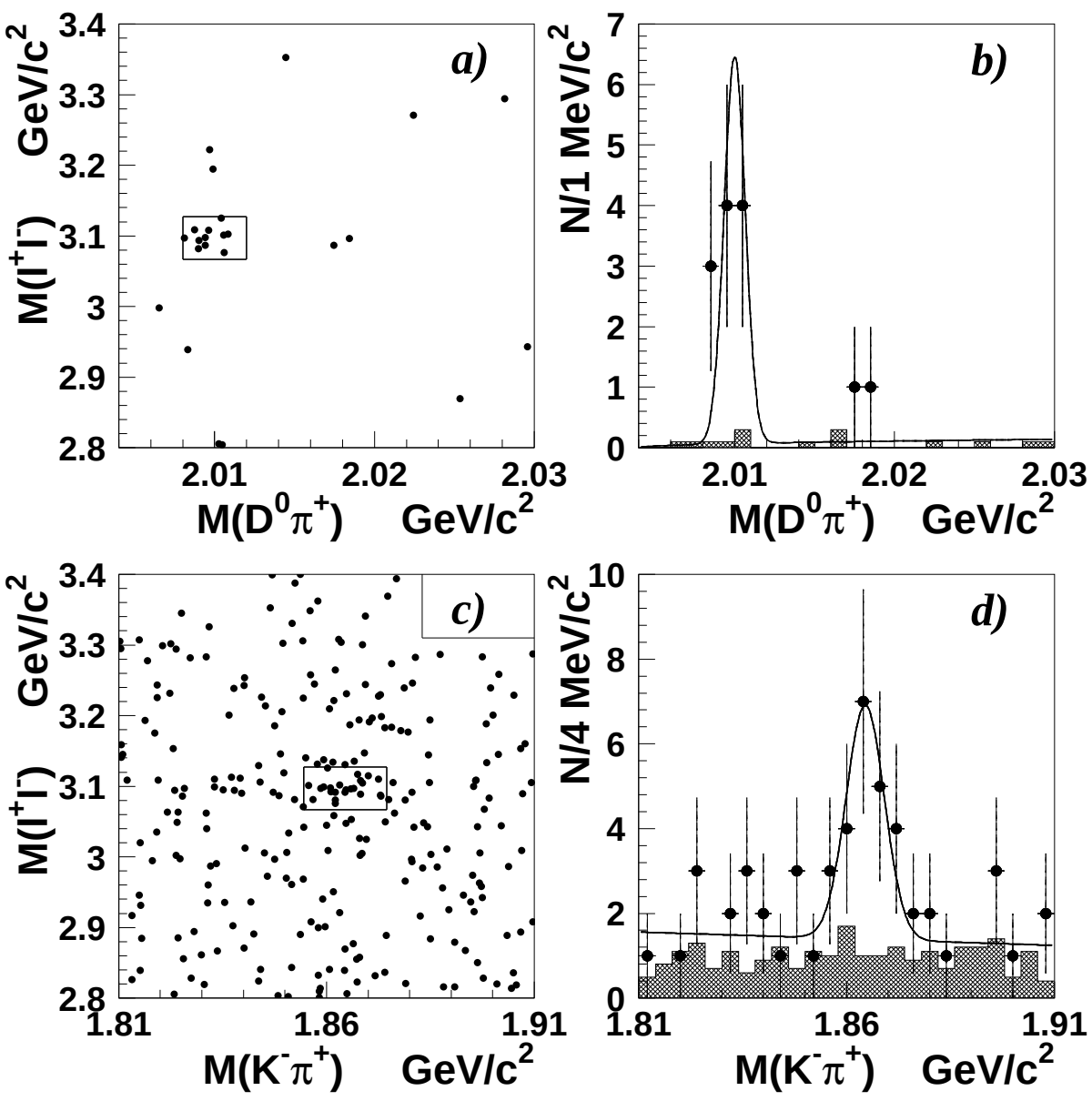

**Figure 18.2.19.** Results of a search for associated production of $J/\psi$ and charm mesons (Abe, 2002j): (a) the scatter plot $M(\ell^+\ell^-)$ vs $M(D^0\pi^+)$; (b) projection onto the $M(D^0\pi^+)$ axis; (c) the scatter plot $M(\ell^+\ell^-)$ vs $M(K^-\pi^+(K^+K^-))$; (d) projection onto the $M(K^-\pi^+(K^+K^-))$ axis. Points with error bars show the $J/\psi$ signal region and the hatched histograms show the scaled sidebands.

### *Measuring the large double-$c\bar{c}$ fraction*

In 2002 Belle also observed that $J/\psi$'s are often accompanied by $D^{*+}$ and $D^0$-mesons (Abe, 2002j). Scatter plots of the invariant mass of $J/\psi$ candidates versus masses of $D^{*+}$ and $D^0$ candidates are shown in Fig. 18.2.19 (a), (c). The charmed meson mass projections for the $J/\psi$ signal and sideband regions are shown in Fig. 18.2.19 (b), (d). A significant excess of $D^{*+}$ ($N_{D^{*+}J/\psi} = 10.1\,^{+3.6}_{-3.0}$, with significance $5.3\,\sigma$), and $D^0$ mesons ($N_{D^0 J/\psi} = 14.9\,^{+5.4}_{-4.8}$, with significance $3.7\,\sigma$) in the events with $J/\psi$'s above the kinematical limit for $\Upsilon(4S)$ decays demonstrates that another $c\bar{c}$ pair is present. To calculate the $e^+e^- \to J/\psi\, c\bar{c}$ cross section, one needs to know how often the second $c\bar{c}$-pair fragments into $D^{*+}$ or $D^0$-mesons. Using the Lund fragmentation model (Sjöstrand, 1994) Belle calculated the ratio of the $J/\psi\ c\bar{c}$ and inclusive $J/\psi\ X$ production cross sections to be equal to $0.59\,^{+0.15}_{-0.13} \pm 0.12$. This result clearly demonstrates that, contrary to NRQCD predictions, the dominant diagram for $J/\psi$ production is $e^+e^- \to J/\psi\ c\bar{c}$.

In 2009, using an order of magnitude larger data sample ($673\,\mathrm{fb}^{-1}$) Belle measured the cross sections for the processes $e^+e^- \to J/\psi\ c\bar{c}$ in a model-independent way (Pakhlov, 2009). In the study of associated production of a $J/\psi$ with charmed hadrons, all the ground state charmed mesons ($D^0$, $D^+$, $D_s^+$) and the $\Lambda_c$-baryon were used. As two charmed hadrons are produced in $c\bar{c}$ fragmentation, the $e^+e^- \to J/\psi\ c\bar{c}$ cross section is given by the sum of double-charmonium $e^+e^- \to J/\psi\ (c\bar{c})_{\rm res}$ cross sections (for $(c\bar{c})_{\rm res}$ states below open-charm threshold) plus half the sum of the cross sections for production of $J/\psi$ with any of ground state charmed hadrons. Production of the $J/\psi$ via mechanisms other than $e^+e^- \to J/\psi\, c\bar{c}$ was also studied: the $e^+e^- \to J/\psi$ non-$c\bar{c}$ cross section was calculated as the difference between inclusive $e^+e^- \to J/\psi\, X$ and $e^+e^- \to J/\psi\, c\bar{c}$ cross sections. Belle found $\sigma(e^+e^- \to J/\psi\, c\bar{c}) = (0.74 \pm 0.08\,^{+0.09}_{-0.08})$ pb and $\sigma(e^+e^- \to J/\psi\, \text{non-}c\bar{c}) = (0.43 \pm 0.09 \pm 0.09)$ pb, respectively, thus confirming the dominance of the $e^+e^- \to J/\psi\, c\bar{c}$ production mechanism. It should be noted that in this analysis (unlike that of Abe, 2002n) no correction for the charged track multiplicity ($N_{\rm ch} > 4$) requirement was applied for any of the processes. For $e^+e^- \to J/\psi$ non-$c\bar{c}$, such corrections are only possible by relying on a model, while for $e^+e^- \to J/\psi\, c\bar{c}$ they are close to unity. A note on the interpretation of these results follows in Section 18.2.4.3.

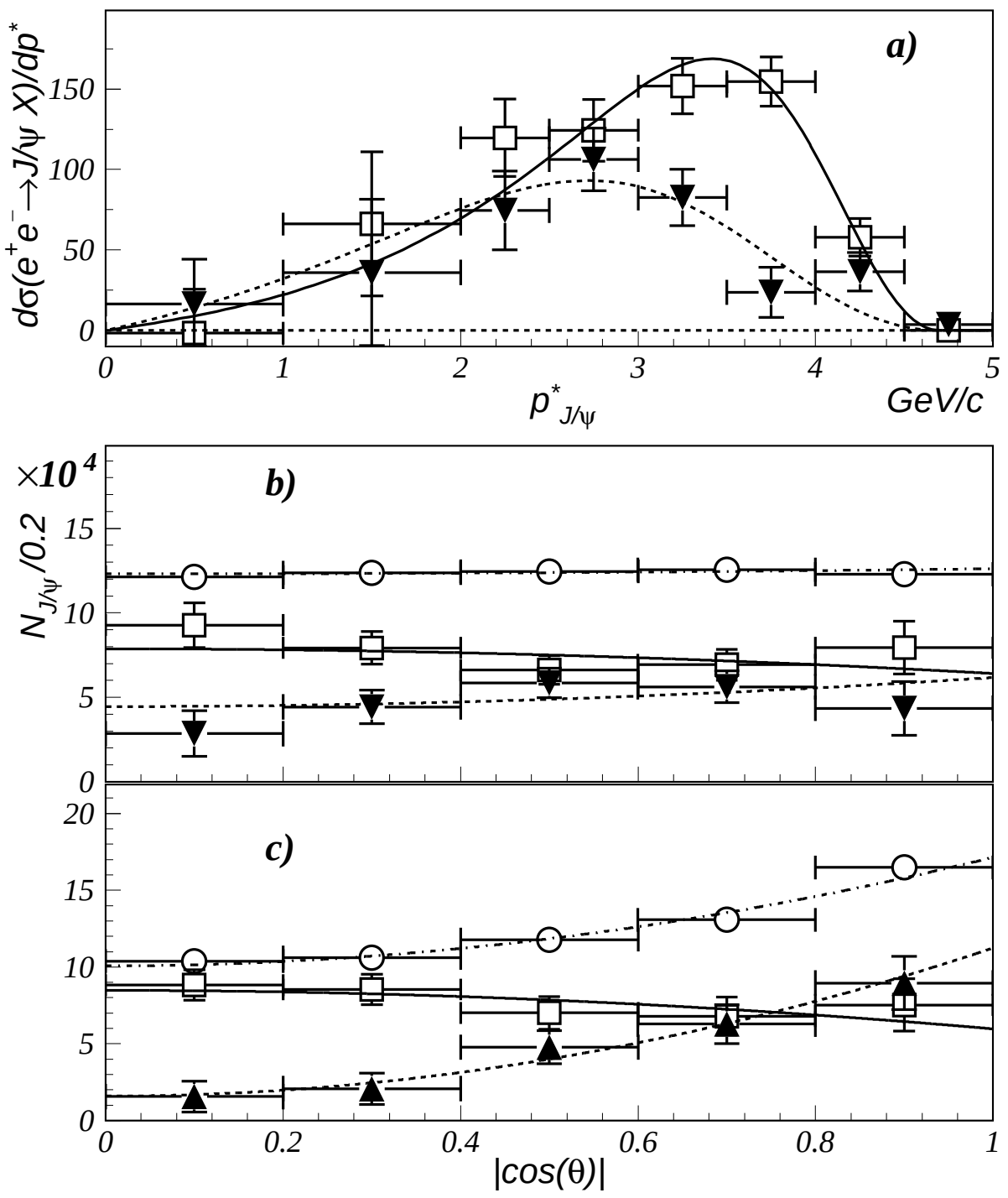

**Figure 18.2.20.** From Pakhlov (2009): (a) Differential cross section for the $e^+e^- \to J/\psi\, c\bar{c}$ (open squares) and $e^+e^- \to J/\psi$ non-$c\bar{c}$ processes (filled triangles). The curves represent a fit to the Peterson fragmentation function (Peterson, Schlatter, Schmitt, and Zerwas, 1983). Angular distributions (b) $|\cos\theta_{\rm helicity}|$ and (c) $|\cos\theta_{\rm production}|$ for inclusive (open circles), $e^+e^- \to J/\psi\, c\bar{c}$ (open squares), and $e^+e^- \to J/\psi$ non-$c\bar{c}$ processes (filled triangles).

With the same technique Belle measured the $J/\psi$ momentum (Fig. 18.2.20 (a)) and $J/\psi$ helicity and production angle distributions (Fig. 18.2.20 (b) and (c), respectively) for both $e^+e^- \to J/\psi\, c\bar{c}$ (open squares) and $e^+e^- \to J/\psi$ non-$c\bar{c}$ (filled triangle) processes. For the $e^+e^- \to J/\psi$ non-$c\bar{c}$ process, the $J/\psi$ momentum spectrum is significantly softer than that for $e^+e^- \to J/\psi\, c\bar{c}$, and the production angle distribution peaks along the beam axis.

Recently, both $e^+e^- \to J/\psi\, gg$ and $J/\psi\, c\bar{c}$ cross sections have been recalculated including NLO corrections (Gong, Wang, and Zhang, 2011; He, Fan, and Chao, 2010; Li, Song, Zhang, and Ma, 2011) and are in better agreement with the experimental data than the first leading-order calculations. A complete discussion can be found in Brambilla et al. (2011).

### 18.2.4.3 Special note: the $e^+e^- \to J/\psi\, X$ cross section

It is both important and difficult to compare measurements of the $e^+e^- \to J/\psi\, X$ cross section with theory. This is especially true in the case where the recoil system $X$ does not include open or hidden charm ("$e^+e^- \to J/\psi$ non-$c\bar{c}$"), as this allows the NRQCD framework—with universal matrix elements describing production in $e^+e^-$, $p\bar{p}$, and other environments—to be tested (Section 18.1.4.1). The measurements are described in detail in Section 18.2.4.2; here we treat problems of interpretation.

The main pitfall is the selection requiring more than four reconstructed tracks. As described above, $B$ Factory $e^+e^- \to J/\psi\, X$ analyses impose such a requirement to suppress low-multiplicity events of QED origin, which are numerous and poorly understood. While the physics of QED events is straightforward, practical measurement requires control of cases where tracks are missed or misreconstructed, and where beam-background tracks are added to the event, together with photon conversions and bremsstrahlung. The lack of coverage close to the beamlines, and the trigger conditions, are key limitations: see Chapter 2 for the design of the experiments; for the forward-peaked cross-section of QED processes, in a simple case (initial state radiation), see the discussion in Section 21.2.1.

The requirement of more than four reconstructed tracks must be taken into account when comparing measurements with theoretical predictions:

1. For double charmonium production, $e^+e^- \to J/\psi\, c\bar{c}$, this is straightforward: both collaborations quote results for $\sigma \times \mathcal{B}_{>2}$ (Table 18.2.7), where the factor $\mathcal{B}_{>2}$ describes the fraction of $c\bar{c}$ decays to final states with more than two charged particles. ($J/\psi$ is reconstructed only in the decay to a lepton pair $\ell^+\ell^-$.)
2. The first $B$ Factory measurements of $e^+e^- \to J/\psi\, X$ quoted the cross section directly, attempting to correct for the effect of track requirements. The initial *BABAR* analysis (Aubert, 2001b) required more than four tracks in the $J/\psi \to e^+e^-$ case (*i.e.* not for $J/\psi \to \mu^+\mu^-$), with additional selections to suppress $e^+e^- \to \gamma_{\rm ISR} J/\psi$ and $\gamma_{\rm ISR}\psi(2S)$. The initial Belle analysis (Abe, 2002n) required more than four tracks in all events (the same condition used by more recent analyses), with additional selections to suppress $e^+e^- \to \gamma\psi(2S)(\to \pi^+\pi^- J/\psi)$. Both experiments incorporated these requirements into their efficiency calculations, making assumptions about the angular distribution and polarization of these events, the fraction of recoil systems $X$ containing charm, and the mix of hadronic final states within the system $X$ in the light-quark case. All of these were poorly known at the time;

in the case of the fraction containing charm, prevailing assumptions were incorrect. These early measurements are thus subject to a significant and poorly-controlled model dependence.

3. The latest Belle measurement (Pakhlov, 2009) seeks to minimize such problems, reconstructing a list of states that exhausts most of the possibilities for $X$ systems containing charm: the omissions are systems including the $\Xi_c$, $\Omega_c$, and their excitations. There is thus only weak dependence on modelling of the system $X$. Because the intrinsic reliance on models is so low, Belle quotes cross-sections *without* correcting for the number-of-tracks cut. As noted above, this has little impact on the $e^+e^- \to J/\psi\, c\bar{c}$ measurement: the correction approaches unity. For the important $e^+e^- \to J/\psi$ non-$c\bar{c}$ cross-section, however, the Belle result is an *underestimate* of the true value.

The remaining issue is the comparison of Belle and *BABAR* results. The disagreement between the initial measurements was much larger than their reported uncertainties (Aubert, 2001b; Abe, 2002n); while the Belle result has been superseded by Pakhlov (2009), there has been no update of the *BABAR* cross section. (The successor analysis Aubert, 2005n concentrated on the then-controversial production of double $c\bar{c}$ final states.) The systematic limitations of the early measurements have been listed at point 2 above: while the two collaborations' results formally disagree, experimentalists do not interpret Aubert (2001b) as casting doubt on the Pakhlov (2009) cross sections.

### 18.2.5 Concluding remarks

The last decade saw both an experimental and a theoretical revival in charmonium physics due to the $B$ Factories, with their large enriched charm sample, playing a leading role with the observation and study of dozens new charmonium-like states. For most of them a charmonium assignment has not been found so far: these states are reviewed in Section 18.3. Only a few of the new states match the conventional charmonium level scheme and have been discussed in this section. However even this selection of new states reveals problems in the quantitative description of the charmonium spectrum, since potential models can not accurately predict masses above the $D\overline{D}$ threshold. This suggests that the coupling between the charmonium and two-charmed-meson sectors is not well described, and $B$ Factory measurements provide a stimulating input for the development of theoretical models. In a complementary development, rigorous work at the $B$ Factories on accurate description of broad charmonium states in their interference with non-resonant background has helped to measure properly their masses and widths, which are also of great importance for theory.

The majority of the results presented in this section are illustrated by Fig. 18.2.21, which shows with colors the significant contribution of the $B$ Factories to the study of the charmonium spectrum.

Charmonium production is another case where the $B$ Factories managed to obtain surprising results. Observation of unexpectedly large double $c\bar{c}$ continuum production stimulated new methods to calculate charmonium production. The importance of relativistic corrections and large NLO contributions were recognized in attempts to resolve this puzzling discrepancy.

In conclusion, the numerous results obtained by the $B$ Factories in the charmonium sector have triggered theoretical developments for better descriptions of the spectroscopy, decay and production of charmonium states.

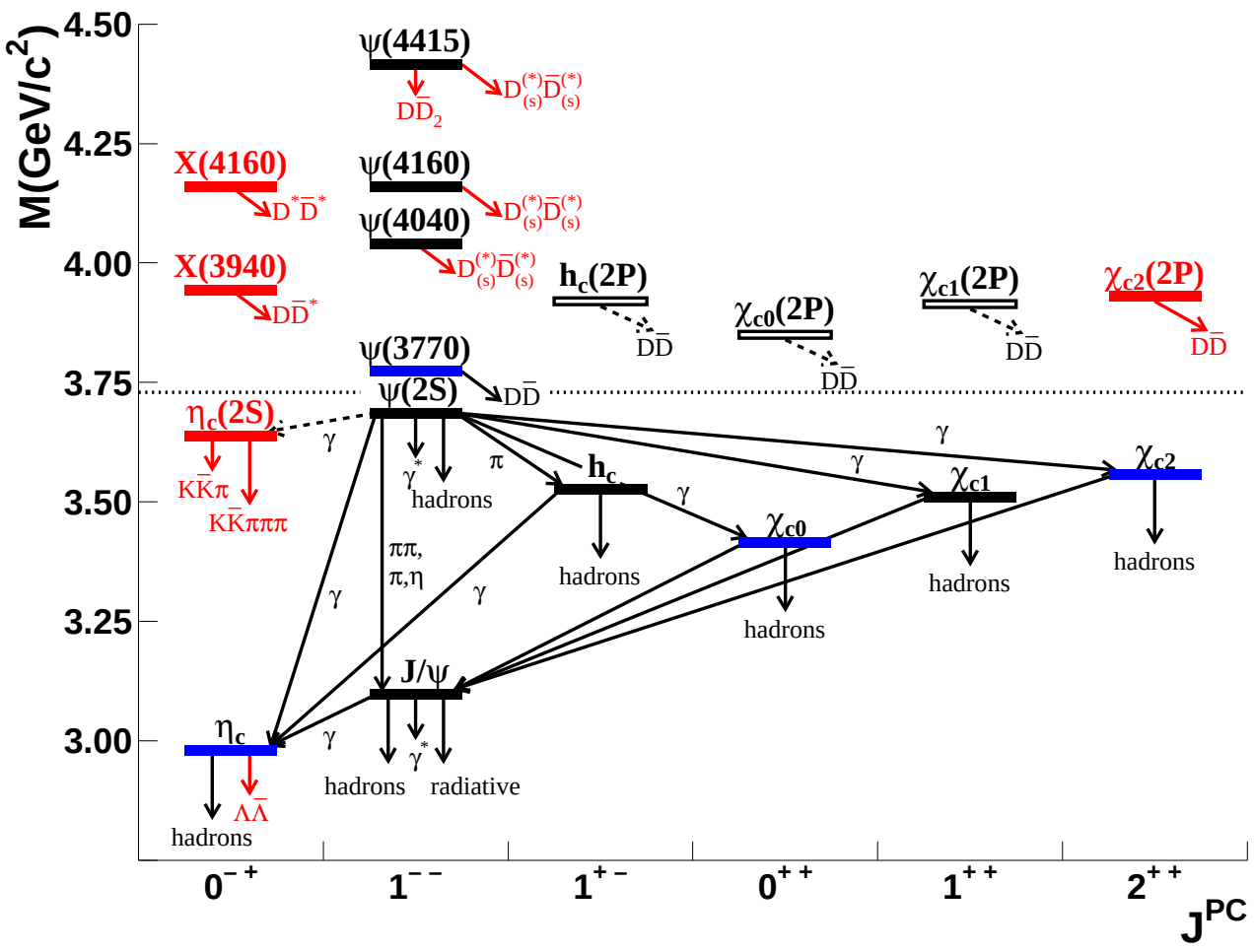

**Figure 18.2.21.** The charmonium spectrum and scheme of charmonium transitions and decays. The red bands correspond to states newly observed at the $B$ Factories, blue bands show states where the $B$ Factories have made a substantial contribution to the accurate measurement of parameters, while the white bands represent yet unobserved states. The arrows show charmonium transitions and decays: decay modes newly observed at the $B$ Factories are shown in red.

## 18.3 Exotic charmonium-like states

***Editors:***
*Riccardo Faccini (BABAR)*
*Stephen Lars Olsen (Belle)*
*Eric Swanson (theory)*

**Additional section writers:**
*Bryan Fulsom, Arafat Gabareen Mokhtar, Alessandro Pilloni, Bruce Yabsley, Shuwei Ye*

As discussed in Section 18.1, the theory of bound states of heavy quarks, such as charmonium, provides quantitative predictions for masses and other properties of the physically observable states with minimal ambiguity, primarily because the velocities of the heavy quarks in these bound states are low enough for relativistic effects to be treated as small perturbations to non-relativistic calculations. The quantum numbers that are most appropriate to characterize a realizable state are, in decreasing order of the energy-splitting among eigenstates: the radial excitation $n$, the orbital angular momentum $\ell$, the spin $s$, and the total angular momentum $J$. Given this set of quantum numbers, the parity and charge conjugation of $c\bar{c}$ states[107] are given by $P = (-1)^{\ell+1}$ and $C = (-1)^{\ell+s}$. States are designated by the usual spectroscopic notation: $n\,^{2s+1}\ell_J$. Figure 18.1.1 shows the mass and quantum number assignments of the experimentally well established charmonium states (see also Table 18.2.1).

All of the predicted $c\bar{c}$ states with mass below the open-charm threshold (*i.e.*, $M < 2m_D$) have been observed with measured masses and other properties that are in good agreement with theoretical predictions. This suggests that the charmonium system is a good environment to search for "exotic" states, *i.e.* states containing a $c\bar{c}$ quark pair, as evidenced from its decay products, but with properties that deviate from theoretical expectations for $c\bar{c}$ spectroscopy. Before the advent of the $B$ Factories no evidence for deviations from standard charmonium expectations was found.

In this section, we first summarize the existing models that describe possible exotic states, then review the experimental observations, reporting both the final states where the states have been observed and those where they have not. Since the easiest quantum number to assign is the charge-conjugation parity $C$, which is uniquely determined either by the production method or decay final state (see Section 18.2), we first examine the $C = +1$ states (Sections 18.3.2, 18.3.3, and 18.3.4) and then discuss the $J^{PC} = 1^{--}$ states (Section 18.3.5). In addition, we discuss the evidence for candidates for states with non-zero electric charge that contain a $c\bar{c}$ pair among their constituents. These play a crucial role since they can by no means be regular charmonium states, which, by definition, contain only a $c\bar{c}$ pair and are, therefore, electrically neutral. We conclude by summarizing the observations and the remaining open issues (Section 18.3.7).

[107] A complete discussion of quantum numbers can be found in Section 18.1.1.

### 18.3.1 Theoretical models

Although the Standard Model is well established, QCD, the fundamental theory of strong interactions, is only amenable to analytic computation at very high energy scales, where perturbation theory is effective due to asymptotic freedom. Lattice gauge theory has recently reached the level where it is able to provide precision predictions of simple hadronic properties (Durr et al., 2008; see also the discussion in Section 18.1.5). Nevertheless, a comprehensive understanding of low energy phenomena remains elusive.

Systems that include heavy quark-antiquark pairs (quarkonia) are a unique and, in fact, ideal laboratory for probing both the high energy regime of QCD and the low energy regime, where non-perturbative effects dominate. For this reason, quarkonia have been the subject of detailed experimental study for several decades. The accuracy of current models of quarkonia is such that a particle which mimics quarkonia but does not fit in the model spectrum is a likely candidate for a nonconventional, "exotic" state.

Indeed, in the past years the $B$ Factories and the Tevatron have provided evidence for states that do not admit a conventional mesonic interpretation and that instead could be made of a larger number of constituents. While this possibility has been considered since the beginning of the quark model (Gell-Mann, 1964), the actual identification of such states would represent a major revolution in our understanding of elementary particles. It would also imply the existence of a possibly large number of additional states that have not yet been observed.

Finally, the study of strong bound states could be of relevance to understanding the Higgs boson. It could transpire, for example, that the Higgs is a bound state, as predicted by several technicolor models, with or without extra dimensions (Contino, Kramer, Son, and Sundrum, 2007; Dietrich, Sannino, and Tuominen, 2005).

A short list of possible "exotic" bound states is:

**hybrids:** bound states of a quark-antiquark pair and a number of constituent gluons. A signature of such states is that they can have quantum numbers that cannot be assumed by quarkonium states (*e.g.* $J^{PC} = 0^{+-}$ or $1^{-+}$). Model and lattice computations indicate that the $1^{-+}$ states are the lightest hybrid states and thus should be easy to distinguish from conventional quarkonia. Additional signatures are the predicted preference for decays to either a pair of open-charm mesons, one $S$- and one $P$-wave (Kokoski and Isgur, 1987) or to quarkonium plus pions; see *e.g.* Kou and Pene (2005) and Close and Page (2005).

**molecules:** bound states of two mesons, usually represented as $[Q\bar{q}][q'\overline{Q}]$, where $Q$ is the heavy quark. The system would be stable if the binding energy were sufficient to place the mass below all meson-meson continua that couple to the molecule. It is expected that this can happen readily when $Q = b$. For $Q = c$ model computations indicate that resonant states are possible in certain channels. These states can decay strongly via configuration mixing (Braaten and Kusunoki, 2004; Braaten and Lu, 2009; Close and Page, 2004; Fleming, Kusunoki, Mehen, and van Kolck, 2007; Swanson, 2006; Tornqvist, 2004; Voloshin, 2006).

**tetraquarks:** a bound quark pair, neutralizing its color with a bound antiquark pair, usually represented as $[Qq][\overline{q}'\overline{Q}]$. A full nonet of states is predicted for each spin-parity, *i.e.* a large number of states are expected. There is no need for these states to be close to any threshold (Maiani, Piccinini, Polosa, and Riquer, 2006).

In addition, before the panorama of states is fully clarified, there is always the lurking possibility that some of the observed states are misinterpretations of threshold effects: a given amplitude might be enhanced when new hadronic final states become energetically possible, even in the absence of resonances.

### 18.3.2 The X(3872)

The $X(3872)$ was the first exotic charmonium-like state discovered. As shown in Fig. 18.3.1, it was initially observed decaying into $J/\psi\pi^+\pi^-$ by the Belle experiment in $B \to XK$ decays (Choi, 2003), and subsequently confirmed both in $B$ decays (Aubert, 2005af) and in inclusive $p\bar{p}$ production (Abazov et al., 2004; Acosta et al., 2004). Far more information is available on the $X(3872)$ than on any other state; this information will be reviewed here by topic: quantum numbers, mass and width, and production and decay.

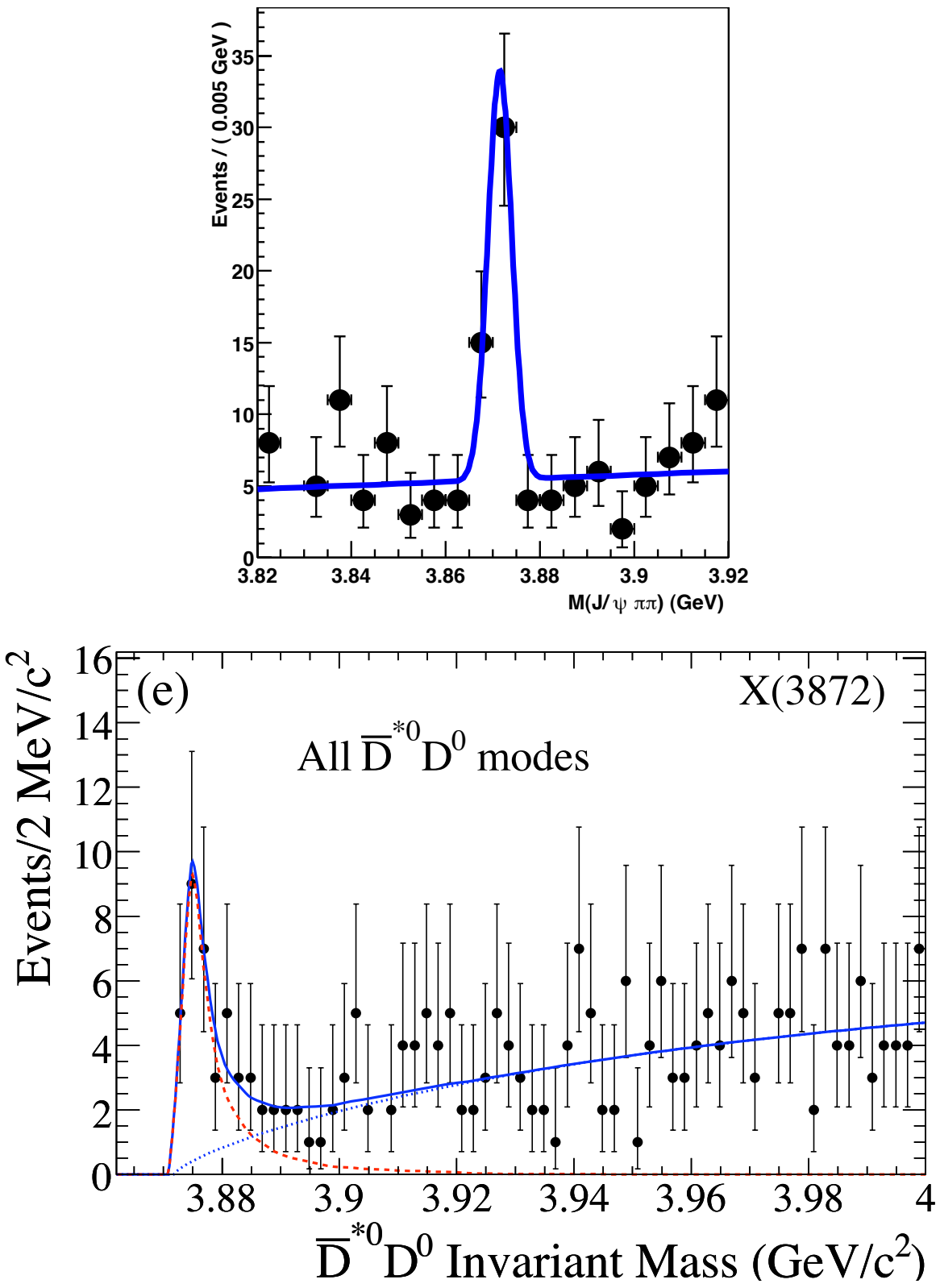

**Figure 18.3.1.** Invariant mass spectrum of the $J/\psi\pi^+\pi^-$ system in $B \to J/\psi\pi\pi K$ decays as observed by Belle (Choi, 2003; upper plot) and of the $\overline{D}^{*0}D^0$ system in $B \to \overline{D}^*DK$ decays as published by *BABAR* (Aubert, 2008bd; lower plot).

#### 18.3.2.1 Quantum numbers

The exotic nature of this state was initially signalled by the narrowness of its width, $\Gamma_{X(3872)} < 2.3\,\mathrm{MeV}/c^2$ at 90% confidence (Choi, 2003), despite being above threshold for decay to a pair of charmed mesons. Furthermore, the $\pi^+\pi^-$ invariant mass distribution first (Choi, 2003; Abulencia et al., 2006a), and a detailed angular analysis next (Abulencia et al., 2007), showed that the dominant decay is $X(3872) \to J/\psi\rho$, which would be isospin violating if the $X(3872)$ were a conventional charmonium state.[108]

The above-mentioned angular analysis from the CDF experiment (Abulencia et al., 2007) was able to discriminate among the possible $J^{PC}$ assignments, excluding all except $J^{PC} = 1^{++}$ and $2^{-+}$. Positive intrinsic charge conjugation had already been established, with the evidence for the decay $X \to J/\psi\gamma$ (Abe, 2005a; Aubert, 2006an) and an upper limit on the branching fraction of the decay $X \to \chi_{c1}\gamma$ (Choi, 2003), thus confirming positive intrinsic charge conjugation.[109]

For years the most favored option has been to assume that the $X$ has $J^{PC} = 1^{++}$; a $D^{*0}\overline{D}$ molecule with $L = 0$ would have these quantum numbers. Such a deuteron-like state, bound by pion exchange, was discussed by Tornqvist (1994), and proposed as a model of the $X(3872)$ structure by Swanson (2004b), and many subsequent investigators. However, definitive arguments against the $2^{-+}$ assignment have been lacking.

The spin-2 hypothesis has been considered implausible from early studies onwards: the $2^{-+}$ decay to $\gamma J/\psi$ is not an electric dipole transition, and so should be suppressed;[110] for the charmonium state with these quantum numbers, the $1\,^1D_2$ or $\eta_{c2}$, the isospin-violating transition $\eta_{c2} \to \pi^+\pi^- J/\psi$ would be expected to have a small rate, relative to isospin-conserving $\eta_{c2} \to \pi^+\pi^-\eta_c$ (Olsen, 2005). However the *BABAR* study of $X(3872) \to J/\psi\omega$ (del Amo Sanchez, 2010c) reported a (relatively weak) preference for $J^{PC} = 2^{-+}$, and there has since been a renewed discussion of this possibility (*e.g.* Burns, Piccinini, Polosa, and Sabelli, 2010; Faccini, Pilloni, and Polosa, 2012; Hanhart, Kalashnikova, Kudryavtsev, and Nefediev, 2012). If the $X(3872)$ has quantum numbers $J^{PC} = 2^{-+}$, it should be produced by two-photon fusion (see Chapter 22).

[108] In this case, an isosinglet particle (the $X(3872)$) would be decaying into an isovector state: the combination of an isosinglet ($J/\psi$) and an isovector ($\rho$) particle.

[109] The $\chi_{c1}$ has $C = +1$, while the $\gamma$ has $C = -1$: the final state therefore has $C = -1$. If the $X$ has positive charge conjugation it cannot decay via electromagnetic interactions into a $C = -1$ final state.

[110] This straightforward point is part of the commonly-accepted wisdom about the $X(3872)$. We are not aware of who first brought it to general attention.

*BABAR* has searched for $\gamma\gamma \to X(3872) \to \eta_c(1S)\pi^+\pi^-$ with the $\eta_c(1S)$ decaying to $K_S^0 K^\pm \pi^\mp$ (Lees, 2012t). No signal events were found, and a 90% confidence-level upper limit $\sigma(\gamma\gamma \to X(3872)) \times \mathcal{B}(X(3872) \to \eta_c(1S)\pi^+\pi^-) <$ 48 fb was set on the product of the $\gamma\gamma \to X(3872)$ cross section and $X(3872) \to \eta_c(1S)\pi^+\pi^-$ branching fraction. CLEO has searched for $\gamma\gamma \to X(3872) \to J/\psi\pi^+\pi^-$ and has found no significant signal (Dobbs et al., 2005).

The updated Belle analysis of $X(3872) \to \pi^+\pi^- J/\psi$, using the full 711 fb$^{-1}$ $\Upsilon(4S) \to B\overline{B}$ data sample (Choi, 2011), confirmed two key CDF results: the $M(\pi^+\pi^-)$ spectrum is consistent with $X(3872) \to \rho^0 J/\psi$ with either $L = 0$ or $L = 1$, consistent with $J^{PC} = 1^{++}$ and $2^{-+}$ respectively (*cf.* Abulencia et al., 2006a); and the angular distribution of the decay allows both $1^{++}$ and $2^{-+}$ interpretations (*cf.* Abulencia et al., 2007). Decays $B \to KX(3872)[\to \pi^+\pi^- J/\psi\{\to \ell^+\ell^-\}]$ are described in general by a five-dimensional angular distribution; under certain assumptions, the $1^{++}$ distribution is fixed, while there are two free parameters for $2^{-+}$: the relative magnitude and phase of two complex amplitudes.[111] Due to this extra freedom for $2^{-+}$, and complementary limitations of the Belle and CDF analyses, it is not possible to exclude $2^{-+}$. The Belle analysis, due to the limited size of the sample, considers three different one-dimensional projections of the full distribution; CDF works with a binned three-dimensional distribution, as the other two quantities are unmeasurable on the sample used (inclusive $X(3872)$ production from $p\overline{p}$, without requiring $B \to KX$).

While we were preparing this book, LHCb published an analysis of a large and clean $B^+ \to K^+X(3872)$ sample ($313 \pm 26$ events), based on an event-by-event likelihood ratio test of $1^{++}$ and $2^{-+}$ hypotheses on the full five-dimensional angular distribution (Aaij et al., 2013a). This study favored $1^{++}$ over $2^{-+}$ by more than eight standard deviations; the complex ratio of amplitudes for the $2^{-+}$ hypothesis, which is treated as a nuisance parameter, is found to be consistent with both the Belle result (Choi, 2011) and with the expectation for decays of a $1^{++}$ state. It therefore appears that the $J^{PC} = 1^{++}$ assignment has finally been established.

#### 18.3.2.2 Mass, width, and hypothetical partner states

Measurements of the mass and width of the $X(3872)$ have been complicated by discussion of two further questions: is the $X$ a single particle, or a pair of neutral states; and, even in the case of a single state, what lineshape do we expect to observe in any given decay mode? These questions have been particularly important for the analysis and interpretation of the decays $X(3872) \to \pi^0 D^0\overline{D}^0$ and $\gamma D^0\overline{D}^0$; at the time of writing, the default approach is to exclude results of these decays from averages of mass and width measurements (see for example Beringer et al., 2012). In the following, we will briefly sketch the history, before presenting a summary of the results from the $B$ Factories and other experiments.

*Partner states, and nontrivial lineshape*

Models in which the $X(3872)$ is a compact four-quark state (a "tetraquark"; see Section 18.3.1) predict partner states, in particular an additional neutral state. The discovery process $B^+ \to K^+X(3872)$ and the isospin-related decay $B^0 \to K_S^0 X(3872)$ in general produce both states with different branching ratios, see for example Maiani, Piccinini, Polosa, and Riquer (2005). However, it could happen that each $B$ decays into a different $X$ mass eigenstate. Both *BABAR* (Aubert, 2006ax, 2008d) and Belle (Adachi, 2008c; Choi, 2011) have performed analyses that distinguish the two samples, in order to test this idea. The most recent results (Aubert, 2008d and Choi, 2011; see Fig. 18.3.2) set the mass difference of the states produced in $B^+$ and $B^0$ decay at

$$\begin{aligned}\delta M &\equiv M(X \mid B^+ \to K^+X) - M(X \mid B^0 \to K^0X)\\ &= (+2.7 \pm 1.6 \pm 0.4)\,\mathrm{MeV}/c^2\ (BABAR),\\ &= (-0.7 \pm 1.0 \pm 0.2)\,\mathrm{MeV}/c^2\ (\mathrm{Belle}),\\ &= (+0.2 \pm 0.8)\,\mathrm{MeV}/c^2 \qquad (\mathrm{mean}). \qquad (18.3.1)\end{aligned}$$

A complementary analysis by CDF (Aaltonen et al., 2009c), fitting the inclusive $\pi^+\pi^- J/\psi$ spectrum, yields no evidence for any other neutral state and sets a limit on the mass difference of 3.6 MeV/$c^2$ at the 95% C.L., to be compared to the expectation

$$\delta M = (7 \pm 2)/\cos(2\theta)\,\mathrm{MeV}/c^2 \qquad (18.3.2)$$

from Maiani, Piccinini, Polosa, and Riquer (2005), where $\theta$ is a (small) angle describing mixing between flavor eigenstates.

The same analyses provide measurements of the ratio of product branching fractions,

$$R \equiv \frac{\mathcal{B}(B^0 \to K^0X) \times \mathcal{B}(X \to \pi^+\pi^- J/\psi)}{\mathcal{B}(B^+ \to K^+X) \times \mathcal{B}(X \to \pi^+\pi^- J/\psi)},$$

finding

$$\begin{aligned}R &= 0.41 \pm 0.24 \pm 0.05\ (BABAR),\\ &= 0.50 \pm 0.14 \pm 0.04\ (\mathrm{Belle}). \qquad (18.3.3)\end{aligned}$$

The expectation in the case of molecular models of the $X$ has been a source of confusion: an extensively-cited study by Braaten and Kusunoki (2005) predicted $R < 0.1$, and together with $\delta M$ measurements prior to Choi (2011), this led to a widespread interpretation that $B$ Factory $B^+$-versus-$B^0$ production results favored the tetraquark picture. Other estimates of $R$ for $D^{*0}\overline{D}^0$ molecules exist: for example, Swanson (2006) found 0.06–0.29. Braaten and Lu (2008) subsequently described the $R < 0.1$ prediction as the result of a "conceptual error" and found that in the molecular model $R$ could be studied only together with the lineshapes of $X(3872)$ decays.

[111] A publicly available LHC note (Mangiafave, Dickens, and Gibson, 2010), cited by Belle (Choi, 2011), lists the various angular distributions but incorrectly represents the normalized ratio of amplitudes for the $2^{-+}$ hypothesis, $\alpha = B_{11}/(B_{11} + B_{12})$, as a real number with $0 \le \alpha \le 1$. The corresponding LHCb thesis (Mangiafave, 2011) notes that $\alpha$ is complex in general.

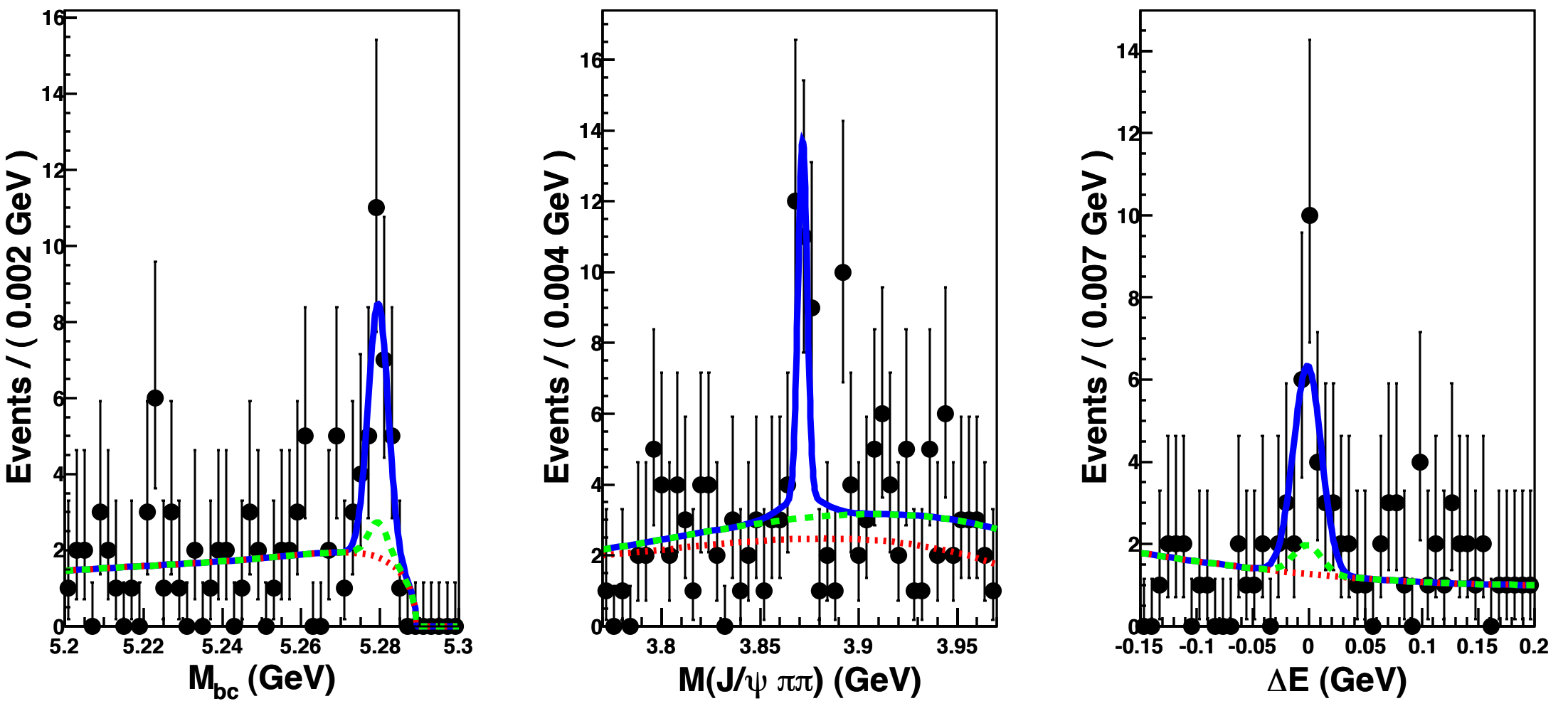

**Figure 18.3.2.** Events from the $B^0 \to K^0_S\pi^+\pi^- J/\psi$ analysis of Belle (Choi, 2011), with the results of a three-dimensional fit to $M_{\rm bc} \equiv m_{\rm ES}$, $M(J/\psi\pi^+\pi^-)$, and $\Delta E$ shown (blue solid curve); the fitted contributions of combinatorial background (red dotted), and combinatorial plus peaking background (green dashed) are also shown.

As no significant separation is seen between the masses of the peaks in the $K^+\pi^+\pi^- J/\psi$ and $K^0_S\pi^+\pi^- J/\psi$ final states, we assume in what follows that both are due to the decay of a single $X(3872)$ state. In the mass and width measurements presented below, the results from $K^+\pi^+\pi^- J/\psi$ dominate.

Belle observed the decay $X(3872) \to D^{*0}\overline{D}^0$ in the $\pi^0 D^0\overline{D}^0$ final state at the higher mass $M = (3875.2 \pm 0.7^{+0.3}_{-1.6} \pm 0.8)\,{\rm MeV}/c^2$ (Gokhroo, 2006; the final error reflects the then-current uncertainty in the $D^0$ mass). Subsequent analyses by both *BABAR* (Aubert, 2008bd) and Belle (Aushev, 2010) confirmed the observation, also adding the $\gamma D^0\overline{D}^0$ final state, and finding a mass of $M = (3873.8 \pm 0.5)\,{\rm MeV}/c^2$ if the two results are averaged. As this is significantly larger than the value observed in the discovery mode $\pi^+\pi^- J/\psi$ (see below) there has been some speculation that $D^{*0}\overline{D}^0$ and $\pi^+\pi^- J/\psi$ are produced by the decay of two distinct parent particles (see for example the discussion in Aubert, 2008bd). While this is possible *a priori*, there are two related problems with using this model to interpret the data:

- *Expected lineshape:* In a decay $X(3872) \to D^{*0}\overline{D}^0$ the $D^{*0}$ will in general be off-shell, because of the proximity of the $D^{*0}\overline{D}^0$ threshold. The effect on the decays is pronounced if the $X$ is below threshold, and study of the $\pi^0 D^0\overline{D}^0$ and $\gamma D^0\overline{D}^0$ lineshapes (which can have a complicated structure in general) is required to distinguish between an $X$ state which is below threshold and an above-threshold "virtual state" (see for example the discussions by Artoisenet, Braaten, and Kang, 2010, and Hanhart, Kalashnikova, and Nefediev, 2010, 2011).
- *Analysis technique:* The mature analyses of both collaborations impose a $D^*$ mass constraint on one of the $\pi^0 D^0$ (or $\gamma D^0$) combinations, to improve the resolution of the resulting $B \to KX$ candidates, and hence the suppression of the background. This yields a reconstructed $X(3872)$ mass that is above threshold by construction, and complicates the task of extracting the $\pi^0 D^0\overline{D}^0$ (or $\gamma D^0\overline{D}^0$) lineshape.

Within a model where a state above threshold is decaying to on-shell $D^{*0}$ and $\overline{D}^0$, both collaborations resolve a nonzero width for that state, with an average of $\Gamma = (3.4\pm1.5)\,{\rm MeV}/c^2$. Some care is taken with the simulation and fitting in both measurements, including (in the Aubert, 2008bd analysis for example) simulations of $X(3872)$ with a range of masses and widths, rather than relying on parameterization of the reconstructed mass. Aubert (2008bd) also finds $\delta M = (0.7 \pm 1.9 \pm 0.3)\,{\rm MeV}/c^2$ for the difference between masses observed in $B^+$ and $B^0$ decays to $D^{*0}\overline{D}^0$.

For the reasons quoted above, and in common with other recent reviews (*e.g.* Beringer et al., 2012), we exclude these $D^{*0}\overline{D}^0$ results from averages of the $X(3872)$ mass and width below.

Searches for charged partner states have also been conducted by both *BABAR* (Aubert, 2005aa) and Belle (Choi, 2011). No evidence for such a state is seen, with limits from Belle (*BABAR*) on the product branching fractions of

$$\mathcal{B}(\overline{B}^0 \to K^- X^+) \times \mathcal{B}(X^+ \to \rho^+ J/\psi) < 4.2(5.4) \times 10^{-6},$$
$$\mathcal{B}(B^+ \to K^0 X^+) \times \mathcal{B}(X^+ \to \rho^+ J/\psi) < 6.1(22) \times 10^{-6}, \quad (18.3.4)$$

to be compared with

$$\mathcal{B}(B^+ \to K^+X) \times \mathcal{B}(X \to \rho^0 J/\psi)$$
$$= (8.4 \pm 1.5 \pm 0.7) \times 10^{-6} \text{ (BABAR)},$$
$$= (8.6 \pm 0.8 \pm 0.5) \times 10^{-6} \text{ (Belle)} \quad (18.3.5)$$

for the discovery mode, from Aubert (2008d) and Choi (2011) respectively. This excludes models in which the $X(3872)$ is the neutral member of an isospin triplet, where decays to the charged states would be favored by a factor of two. However, the tetraquark model of Maiani, Piccinini, Polosa, and Riquer (2005) provides lower limits for the rates in Eq. (18.3.4) which are still allowed by the $X(3872)$ rate in Eq. (18.3.5).

### *Mass measurements in $J/\psi$ final states*

A summary of all available mass measurements is shown in Fig. 18.3.3. The current world average, considering only $X(3872)$ decays to final states including the $J/\psi$, is $M = (3871.68 \pm 0.17)\,\text{MeV}/c^2$ (Beringer et al., 2012). The most precise measurements are those of CDF (Aaltonen et al., 2009c), Belle (Choi, 2011), the new measurement from LHCb (Aaij et al., 2012k), and *BABAR* (Aubert, 2008d), all effectively $\pi^+\pi^- J/\psi$ measurements; the hadron machines measure inclusive production in $p\overline{p}$ and $pp$ respectively, while the $B$ Factory measurements are dominated by $B^+ \to K^+\pi^+\pi^- J/\psi$.

The $D^{*0}\overline{D}^0$ threshold is at $(3871.84 \pm 0.27)\,\text{MeV}/c^2$ (using the $D^0$ mass and $m_{D^{*0}} - m_{D^0}$ difference values from Beringer et al., 2012). If the $X(3872)$ is interpreted as a $D^{*0}\overline{D}^0$ "molecule" (see Section 18.3.1), bound by pion exchange, then the binding is exceptionally weak, $m_X - m_{D^{*0}} - m_{D^0} = (0.16 \pm 0.31)\,\text{MeV}$, to be compared to 2.2 MeV for the deuteron. Although the $B$ Factory (and LHCb) mass measurements for the $X(3872)$ are statistically limited, the precision of the comparison with $D^{*0}\overline{D}^0$ threshold will only significantly improve with better measurements of the $D^0$ mass, the $D^{*0}$ mass (or $m_{D^{*0}} - m_{D^0}$ difference), or the use of some new techniques.

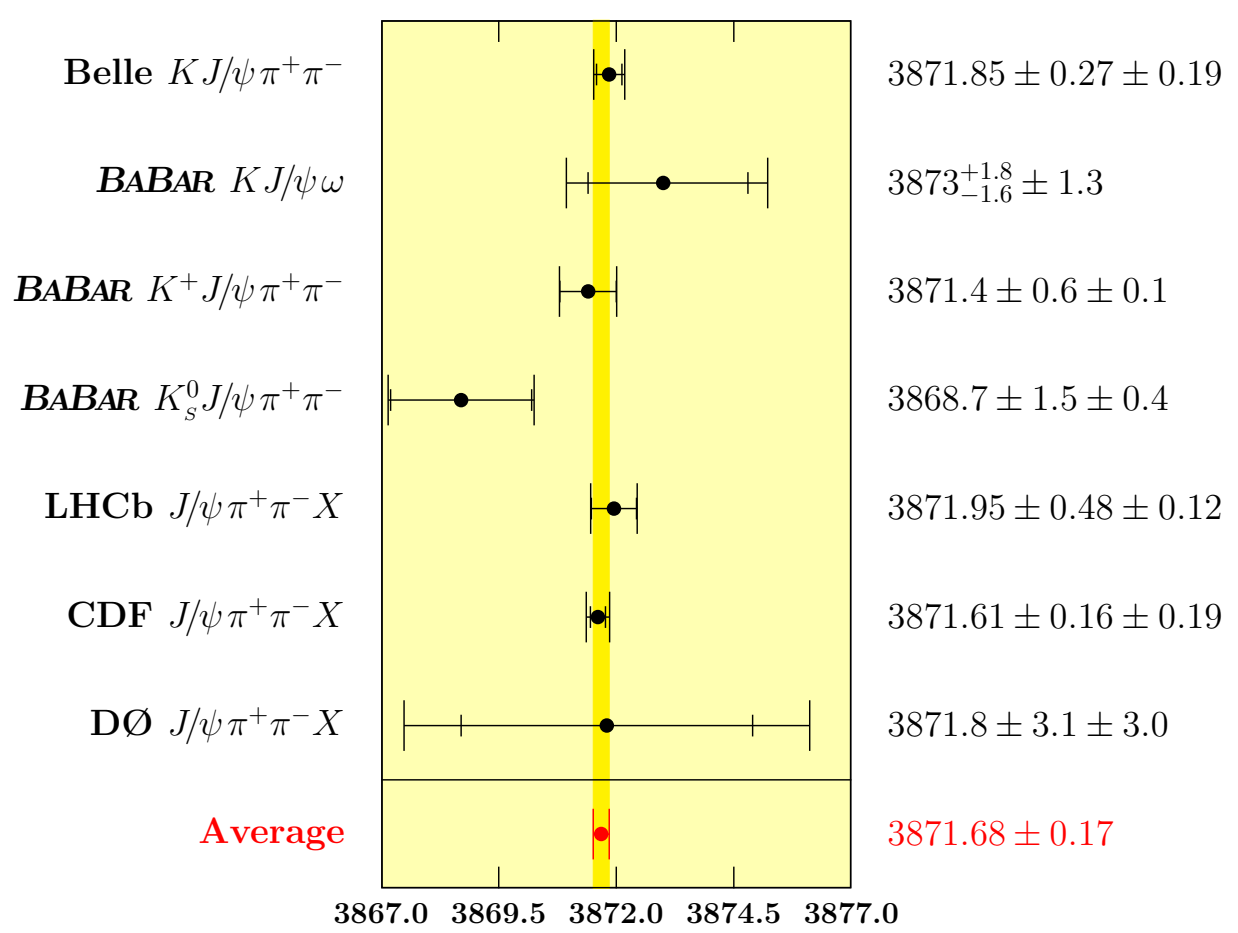

**Figure 18.3.3.** Measured mass of the $X(3872)$. We show the measurements which contribute to the average in Beringer et al. (2012).

### *Width measurements in $J/\psi$ final states*

The $X(3872)$ was known to be relatively narrow from the discovery analysis, with a limit $\Gamma < 2.3\,\text{MeV}/c^2$ at 90% C.L. (Choi, 2003). The confirmations by *BABAR* (Aubert, 2005af), CDF (Acosta et al., 2004), and DØ (Abazov et al., 2004) each found a peak width consistent with the measurement resolution, but did not present explicit width measurements; subsequent analyses by *BABAR* (Aubert, 2006ax, 2008d) set upper limits on the width ($4.1\,\text{MeV}/c^2$ and $3.3\,\text{MeV}/c^2$ respectively). The CDF analysis that set limits on the two-neutral-state hypothesis, and provides the best single measurement of the mass, used an $X(3872)$ intrinsic width of $\Gamma = 1.34\,\text{MeV}/c^2$ (Aaltonen et al., 2009c), based on an average of the central values of the (not statistically significant) width measurements from Belle (Choi, 2003) and *BABAR* (Aubert, 2008d); no independent determination of the width was performed.

The best current estimate of the width comes from the recent Belle analysis (Choi, 2011), which finds $\Gamma < 1.2\,\text{MeV}/c^2$ at 90% C.L. based on a three-dimensional fit to $m_{\text{ES}}$, $\Delta E$, and $M(\pi^+\pi^- J/\psi)$. This is below the experimental resolution. Simulation studies show that natural widths in this range can be recovered; Belle attributes this to constraints on the area of the peak in $M(\pi^+\pi^- J/\psi)$ provided by the distributions in $m_{\text{ES}}$ and $\Delta E$, which make the peak height in $M(\pi^+\pi^- J/\psi)$ sensitive to the natural width. Improved precision will presumably be possible if this technique is applied in the future.

### 18.3.2.3 Production and decay

The $X(3872)$ has been sought in a range of possible decays $B \to KX$, $X \to f$: these are listed, together with the measured product branching fractions (or upper limits) in Table 18.3.1. Plots from a number of important decay modes have been shown above in Figs 18.3.1 and 18.3.2; results in two modes where no signal is seen are shown in Fig. 18.3.4

The most important unresolved case is $X(3872) \to \gamma\psi(2S)$, shown in Fig. 18.3.5, where *BABAR* (Aubert, 2009m) finds a signal with

$$\mathcal{B}(B^+ \to K^+X) \times \mathcal{B}(X \to \gamma\psi(2S))$$
$$= (9.5 \pm 2.7 \pm 0.6) \times 10^{-6}, \quad (18.3.6)$$

while Belle (Bhardwaj, 2011) sees no significant signal and finds

$$= (0.8^{+2.0}_{-1.8} \pm 0.4) \times 10^{-6}. \quad (18.3.7)$$

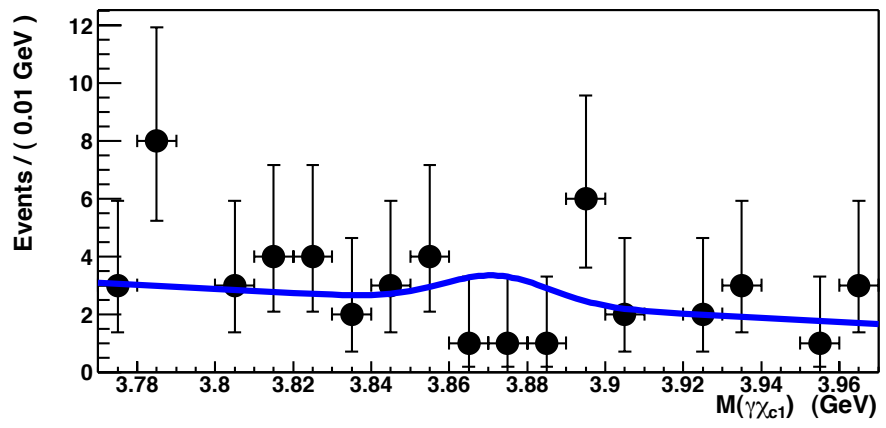

(a) $\gamma\chi_{c1}(1P)$ at Belle (Choi, 2003).

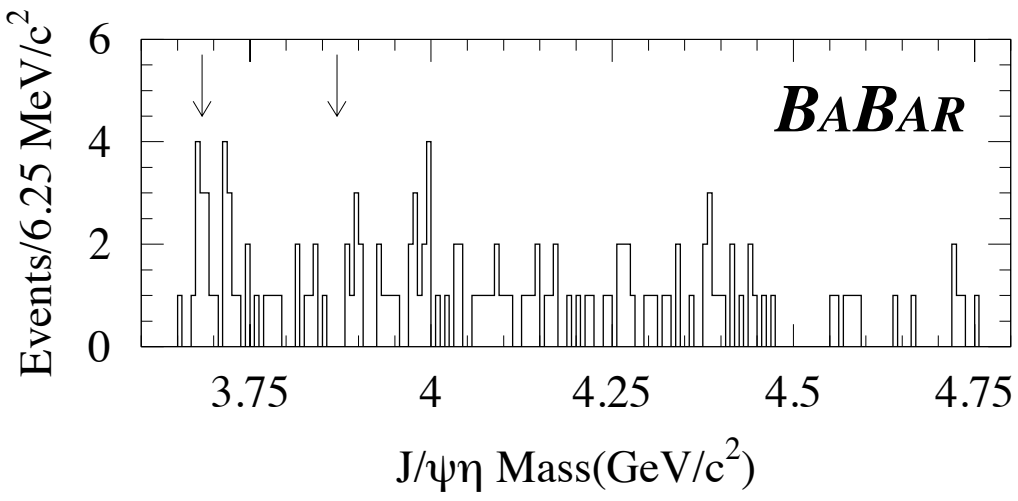

(b) $\eta J/\psi$ at *BABAR* (Aubert, 2004v).

**Figure 18.3.4.** Invariant mass plots for representative $X(3872)$ searches where no signal is seen.

This is in contrast to the $X(3872) \to \gamma J/\psi$ decay, where the same two analyses both find a signal, with consistent product branching fractions

$$\begin{aligned}\mathcal{B}(B^+ \to K^+X) \times \mathcal{B}(X \to \gamma J/\psi) \\ &= (2.8 \pm 0.8 \pm 0.1) \times 10^{-6}\ (\textit{BABAR}), \\ &= (1.8^{+0.5}_{-0.4} \pm 0.1) \times 10^{-6}\ (\text{Belle}).\end{aligned} \tag{18.3.8}$$

The radiative decays are crucial for understanding the structure of the $X(3872)$: while the $\gamma J/\psi$ decay is expected, the $\gamma\psi(2S)$ decay should be heavily suppressed for a molecular state; by contrast, decays of the $2\,^3P_1$ charmonium state ($\chi'_{c1}$) to $\gamma\psi(2S)$ should be enhanced over those to $\gamma J/\psi$ (Barnes and Godfrey, 2004; Suzuki, 2005; Swanson, 2004a, 2006). And unlike other disputed exotic charmonium measurements, where the two experiments disagree on the significance of a signal but make statistically compatible measurements of the rate, the measurements in Eqs (18.3.6) and (18.3.7) are in apparent contradiction. During the final editing of this book, LHCb found evidence for this decay (Aaij et al., 2014a), with $\mathcal{B}(X \to \gamma\psi(2S))/\mathcal{B}(X \to \gamma J/\psi) = 2.46 \pm 0.64 \pm 0.29$. Critical study of the radiative decay modes will therefore be an urgent priority for a super flavor factory.

Measured product branching fractions can be translated into absolute branching fractions of the $X(3872)$ by exploiting the upper limit on $B \to KX(3872)$ measured by *BABAR* from the spectrum of the kaons recoiling against fully reconstructed $B$ mesons (Aubert, 2006ae), $\mathcal{B}(B^\pm \to K^\pm X(3872)) < 3.2 \times 10^{-4}$ at 90% C.L.. Such an analysis has been performed by Drenska et al. (2010), who combine likelihoods for the various product branching fractions (using results available up to mid-2010), the $B \to KX(3872)$ upper limit, and the $X$ width measurement in the $D^{*0}\overline{D}^0$ channel of Aubert (2008bd) using a Bayesian procedure. The resulting 68% confidence intervals are summarized in Table 18.3.1 for each of the decay modes. The same analysis finds a $B \to KX(3872)$ branching fraction in the range $(0.1\text{–}0.2) \times 10^{-3}$, to be compared to the corresponding branching fractions for conventional charmonium states, which are at least $5 \times 10^{-3}$ .

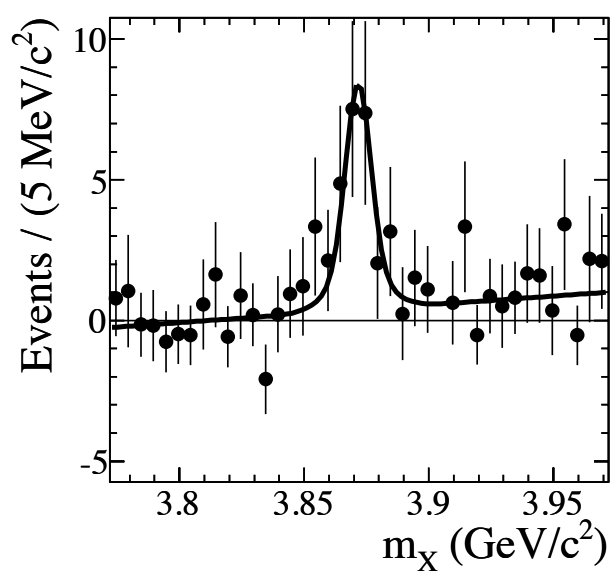

(a) $\gamma J/\psi$ at *BABAR*

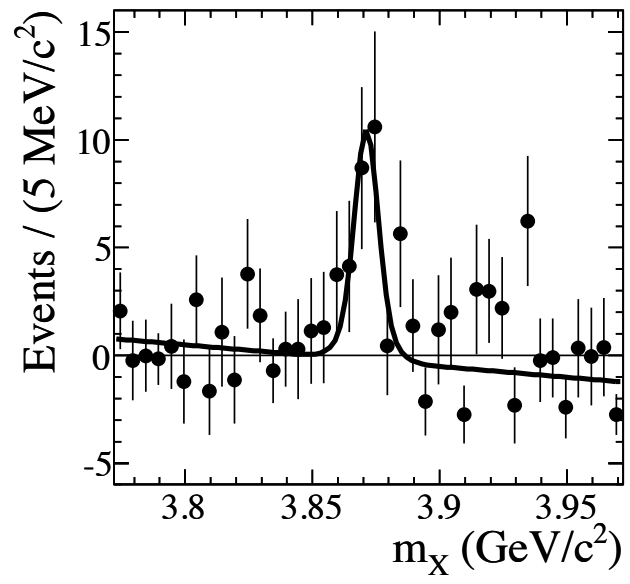

(b) $\gamma\psi(2S)$ at *BABAR*

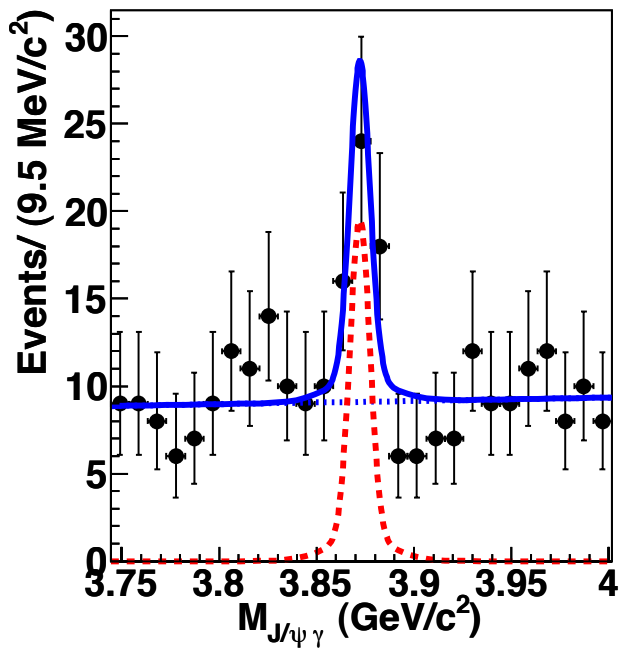

(c) $\gamma J/\psi$ at Belle

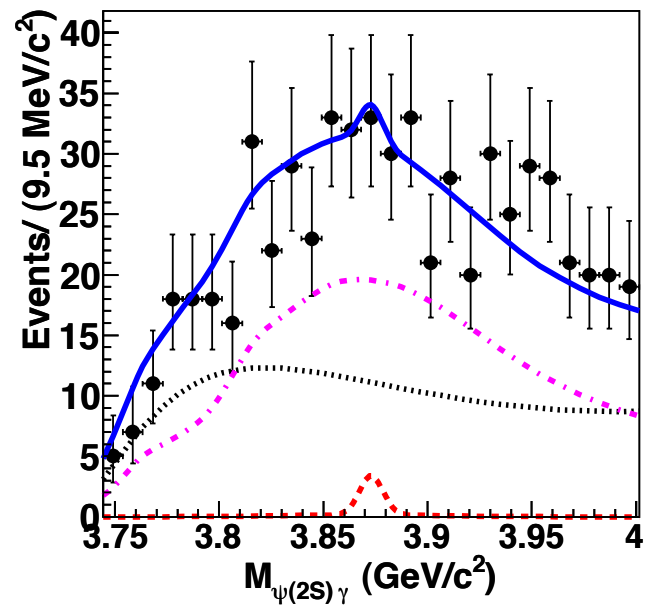

(d) $\gamma\psi(2S)$ at Belle

**Figure 18.3.5.** Invariant mass plots for radiative decays of the $X(3872)$: (a,c) $B^+ \to K^+X[\to \gamma J/\psi]$ and (b,d) $B^+ \to K^+X[\to \gamma\psi(2S)]$, at (a,b) *BABAR* (Aubert, 2009m) and (c,d) Belle (Bhardwaj, 2011). See the discussion in the text. The *BABAR* analyses in (a,b) use the ${}_s\mathcal{P}lot$ technique (Pivk and Le Diberder, 2005) to extract the number of signal events. The curves in the Belle plots show: in (c,d) the fit to data (blue solid) and fitted yields for the signal (red dashed); in (c) the background component (blue dotted); and in (d) the combinatorial background (black dotted), and background from $B \to K^*\psi(2S)$ and $B \to K\psi(2S)$ (pink dot-dashed).

As far as other production mechanisms are concerned, $B^0 \to K^+\pi^-X(3872)$ decays have also been studied. Such decays are seen, but with a smooth distribution in $K^+\pi^-$ invariant mass; an upper limit is set on $\mathcal{B}(B^0 \to K^*(892)^0X(3872))$ (Adachi, 2008c; see the results in Table 18.3.1). This is in contrast to other charmonium states, where $B \to K^*c\overline{c}$ and $Kc\overline{c}$ branching fractions are comparable, and $K^*$ dominates over nonresonant $K\pi$.

### 18.3.2.4 Summary

In summary, the $X(3872)$ is the most studied of the exotic hidden-charm states, and the only one observed in several decay modes; estimates of its width and absolute branching fractions are also available. Some of the early questions about the state — such as its quantum numbers, and whether charged or neutral partners exist — seem to have been resolved. However the structure of the $X(3872)$ is still not fully understood. Outstanding experimental questions are whether the disputed decay $X(3872) \to \gamma\psi(2S)$ takes place (and if so at what rate), and whether the $X(3872)$ lies above or below the $D^{*0}\overline{D}^0$ threshold. For the latter, large and clean $B^+ \to K^+\pi^0 D^0\overline{D}^0$ and $K^+\gamma D^0\overline{D}^0$ samples will be required; a super flavor factory provides some hope of performing these measurements.

**Table 18.3.1.** Measured $X(3872)$ product branching fractions, separated by production and decay mechanism. The combined results and fitted values are taken from Drenska et al. (2010); see the text. When more than one publication is present, the combination is performed assuming Gaussian uncorrelated errors. The last two columns report the results in terms of absolute $X(3872)$ branching fraction ($B_{fit}$) and in terms of the branching fraction normalized to $J/\psi\pi\pi$ ($R_{fit}$) as obtained from the global likelihood fit described in the text. Ranges and limits are provided at 68% and 90% C.L., respectively. Averages marked with a dagger$^\dagger$ include Belle results that have been superseded by subsequent publications: those from Adachi (2008c) by Choi (2011); those from Gokhroo (2006) by Aushev (2010); and those from Abe (2005a) by Bhardwaj (2011). The $\gamma\psi(2S)$ results in particular are controversial: see the text. Concerning $\pi\pi\pi^0$ results (marked with a double dagger$^\ddagger$): the $B$ Factories find that the $X(3872) \to \pi\pi\pi^0 J/\psi$ process is dominated by $\omega J/\psi$, but set no limits on the nonresonant $\pi\pi\pi^0 J/\psi$ rate; the unpublished Belle result Abe (2005a) quotes only the ratio of $\pi\pi\pi^0 J/\psi$ and $\pi\pi J/\psi$ branching fractions.

| $B$ Decay mode | $X$ decay mode | product branching fraction ($\times 10^5$) | | $B_{fit}$ | $R_{fit}$ |
|---|---|---|---|---|---|
| $K^\pm X$ | $X \to \pi\pi J/\psi$ | $0.82 \pm 0.09^\dagger$ | (Aubert, 2008d; Adachi, 2008c) | $[0.035, 0.075]$ | N/A |
| | | $0.84 \pm 0.15 \pm 0.07$ | (Aubert, 2008d) | | |
| | | $0.86 \pm 0.08 \pm 0.05$ | (Choi, 2011) | | |
| $K^0 X$ | $X \to \pi\pi J/\psi$ | $0.53 \pm 0.13^\dagger$ | (Aubert, 2008d; Adachi, 2008c) | | |
| | | $0.35 \pm 0.19 \pm 0.04$ | (Aubert, 2008d) | | |
| | | $0.43 \pm 0.12 \pm 0.04$ | (Choi, 2011) | | |
| $(K^+\pi^+)_{NR} X$ | $X \to \pi\pi J/\psi$ | $0.81 \pm 0.20^{+0.11}_{-0.14}$ | (Adachi, 2008c) | | |
| $K^{*0} X$ | $X \to \pi\pi J/\psi$ | $< 0.34$, 90% C.L. | (Adachi, 2008c) | | |
| $KX$ | $X \to \pi\pi\pi^0 J/\psi$ | $\{R = 1.0 \pm 0.4 \pm 0.3\}^\ddagger$ | (Abe, 2005a) | $[0.015, 0.075]$ | $[0.42, 1.38]$ |
| $K^+ X$ | $X \to \omega J/\psi$ | $0.6 \pm 0.2 \pm 0.1^\ddagger$ | (del Amo Sanchez, 2010c) | | |
| $K^0 X$ | | $0.6 \pm 0.3 \pm 0.1^\ddagger$ | (del Amo Sanchez, 2010c) | | |
| $K^\pm X$ | $X \to D^{*0}\overline{D}^0$ | $13 \pm 3^\dagger$ | (Aubert, 2008bd; Gokhroo, 2006) | $[0.54, 0.8]$ | $[7.2, 16.2]$ |
| | | $16.7 \pm 3.6 \pm 4.7$ | (Aubert, 2008bd) | | |
| | | $7.7 \pm 1.6 \pm 1.0$ | (Aushev, 2010) | | |
| $K^0 X$ | $X \to D^{*0}\overline{D}^0$ | $19 \pm 6^\dagger$ | (Aubert, 2008bd; Gokhroo, 2006) | | |
| | | $22 \pm 10 \pm 4$ | (Aubert, 2008bd) | | |
| | | $9.7 \pm 4.6 \pm 1.3$ | (Aushev, 2010) | | |
| $KX$ | $X \to \gamma J/\psi$ | $0.22 \pm 0.05^\dagger$ | (Aubert, 2009m; Abe, 2005a) | $[0.0075, 0.0195]$ | $[0.19, 0.33]$ |
| $K^+ X$ | | $0.28 \pm 0.08 \pm 0.01$ | (Aubert, 2009m) | | |
| | | $0.18^{+0.05}_{-0.04} \pm 0.01$ | (Bhardwaj, 2011) | | |
| $K^0 X$ | | $0.26 \pm 0.18 \pm 0.02$ | (Aubert, 2009m) | | |
| | | $0.12^{+0.08}_{-0.06} \pm 0.01$ | (Bhardwaj, 2011) | | |
| $KX$ | $X \to \gamma\psi(2S)$ | $1.0 \pm 0.3^\dagger$ | (Aubert, 2009m) | $[0.03, 0.09]$ | $[0.75, 1.55]$ |
| $K^+ X$ | | $0.95 \pm 0.27 \pm 0.06$ | (Aubert, 2009m) | | |
| | | $0.08^{+0.20}_{-0.18} \pm 0.04$ | (Bhardwaj, 2011) | | |
| $K^0 X$ | | $1.14 \pm 0.55 \pm 0.10$ | (Aubert, 2009m) | | |
| | | $0.11^{+0.36}_{-0.29} \pm 0.06$ | (Bhardwaj, 2011) | | |
| $K^+ X$ | $X \to \gamma\chi_{c1}$ | $< 0.19$ | (Bhardwaj, 2013) | | |
| $K^+ X$ | $X \to \gamma\chi_{c2}$ | $< 0.67$ | (Bhardwaj, 2013) | | |
| $KX$ | $X \to \gamma\gamma$ | $< 0.024$ | (Abe, 2008a) | $< 0.0004$ | $< 0.0078$ |
| $KX$ | $X \to \eta J/\psi$ | $< 0.77$ | (Aubert, 2004v) | $< 0.098$ | $< 1.9$ |

**Table 18.3.2.** Measured $J^{PC}$, masses, and widths of the "3940 family" of states. The first error is statistical, the second systematic.

| State | Reference | $J^{PC}$ | Mass (MeV) | Width (MeV) | conventional assignment |
|---|---|---|---|---|---|
| $X(3940)$ | (Pakhlov, 2008) | $0^{\pm+}$ | $3942^{+7}_{-6} \pm 6$ | $37^{+26}_{-15} \pm 8$ | $\eta_c(3S)$ |
| $Y(3940)$ | (Abe, 2005g) | $[0,1,2]^{\pm+}$ | $3943 \pm 11 \pm 13$ | $87 \pm 22 \pm 26$ | $\chi_{c0}(2P)$? |
| $Y(3940)$ | (Aubert, 2008am) | $[0,1,2]^{\pm+}$ | $3914.6^{+3.8}_{-3.4} \pm 1.9$ | $33^{+12}_{-8} \pm 5$ | $\chi_{c0}(2P)$? |
| $Y(3915)$ | (Uehara, 2010b) | $[0,1,2]^{\pm+}$ | $3915 \pm 3 \pm 2$ | $17 \pm 10 \pm 3$ | $\chi_{c0}(2P)$? |
| $Y(3915)$ | (Lees, 2012ad) | $0^{++}$ | $3919 \pm 2 \pm 2$ | $13 \pm 6 \pm 3$ | $\chi_{c0}(2P)$? |
| $Z(3930)$ | (Uehara, 2006) | $2^{++}$ | $3929 \pm 5 \pm 2$ | $29 \pm 10 \pm 2$ | $\chi_{c2}(2P)$ |
| $Z(3930)$ | (Aubert, 2010g) | $2^{++}$ | $3926 \pm 2.7 \pm 1.1$ | $21.3 \pm 6.8 \pm 3.6$ | $\chi_{c2}(2P)$ |

### 18.3.3 The 3940 family

A number of resonances have been reported by the Belle Collaboration with masses near 3940 MeV/$c^2$: the $X(3940)$, $Y(3940)$, $Z(3930)$, and $Y(3915)$. These states have some possible, albeit not certain, interpretations as regular charmonium states and are discussed in Section 18.2. We concentrate here on aspects related to possible exotic assignments. The measured masses and widths of these states are summarized in Table 18.3.2 and Fig. 18.3.6.

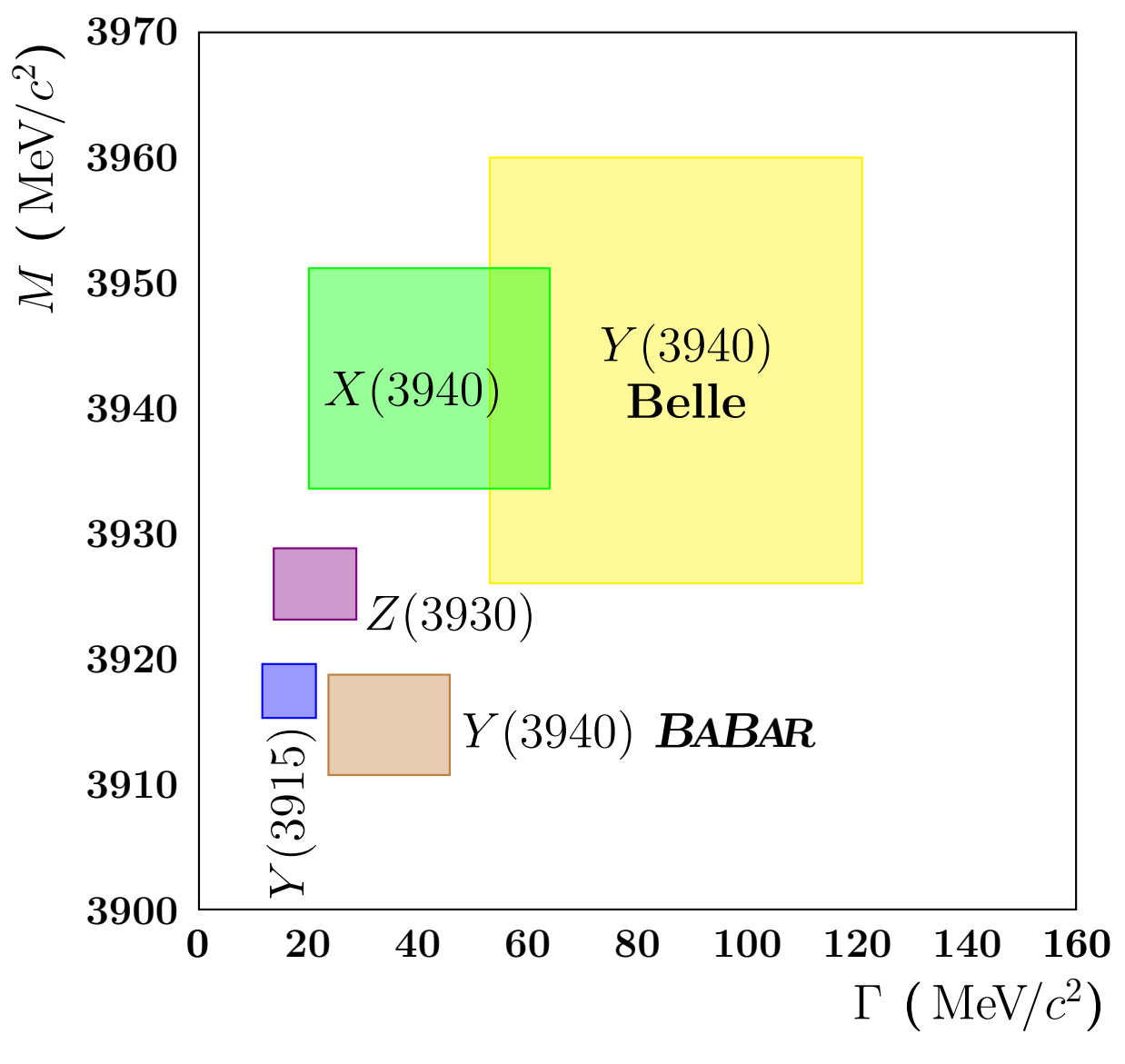

**Figure 18.3.6.** Measured masses and widths of the "3940 family" of states; the boxes represent $\pm 1\sigma$ ranges of the measurements.

#### 18.3.3.1 The $X(3940)$

The $X(3940)$ state was observed in association with a $J/\psi$ meson in double-charmonium production events (*i.e.* not in $\Upsilon(4S)$ decays) (Abe, 2007f; Pakhlov, 2008). Subsequently, by applying a partial reconstruction technique to the same production channel, Belle measured the absolute production rate and established that $X(3940) \to D^*\overline{D}$ is a prominent decay mode; searches were made for $X(3940) \to D\overline{D}$ and $J/\psi\omega$ without evidence for any signals. The lower and upper limits on the branching fractions for these modes set in Abe (2007f) were withdrawn by Belle in Pakhlov (2008), as the inclusive peak in the earlier analysis, used to provide the denominator of the fraction, may have contributions from more than one state. Since the $X(3940)$ is a candidate for a conventional charmonium state, it is also discussed in Section 18.2.1.3. The production mechanism constrains it to have positive charge conjugation since this is an electromagnetic process and the initial state virtual photon and the accompanying $J/\psi$ have the same $C$. Furthermore, all of the known states that are observed via this production mechanism have $J = 0$. Although the reason for this is not understood, it is plausible that this state also has $J = 0$. Thus, the most likely quantum numbers of the $X(3940)$ are $J^{PC} = 0^{++}$ or $0^{-+}$; the apparent absence of the $D\overline{D}$ decay channel favors $0^{-+}$.

#### 18.3.3.2 The $Y(3940)$ and $Y(3915)$

A second state, named $Y(3940)$, was observed as a near-threshold peak in the $J/\psi\omega$ invariant mass spectrum in $B \to J/\psi\omega K$ decays (Abe, 2005g). The $Y(3940)$ state is not seen in $B \to D^*\overline{D}K$ decays and a lower limit has been set on the ratio $\mathcal{B}(Y(3940) \to J/\psi\omega)/\mathcal{B}(Y(3940) \to D^*\overline{D}) > 0.71$ at 90% C.L. (Aushev, 2010); as the $X(3940)$ (discussed above) is seen in $D^*\overline{D}$ but not in $J/\psi\omega$, this strongly suggests that these two states are not the same. The $Y(3940)$ must have positive charge conjugation, while $J = 0, 1, 2$ with either sign parity are possible. The *BABAR* Collaboration confirmed the $Y(3940) \to J/\psi\omega$ observation in $B \to J/\psi\omega K$ decays (Aubert, 2008am), but measure a lower mass and narrower width, which are only marginally consistent with the Belle results (see Table 18.3.2); there are various differences between the two analyses, *e.g.* in the assumptions made about the shape of the background. A Belle study of the $\gamma\gamma \to J/\psi\omega$ reaction (Uehara, 2010b) observed a state, named the $Y(3915)$, with mass and width values consistent with those for the $Y(3940)$ measured by *BABAR* (see Fig. 18.3.7). This observation was confirmed by a *BABAR* analysis of $\gamma\gamma \to J/\psi\omega$ events in which a peak with similar mass and width is seen and an angular correlation study favors a $J^{PC} = 0^{++}$ assignment (Lees, 2012ad).

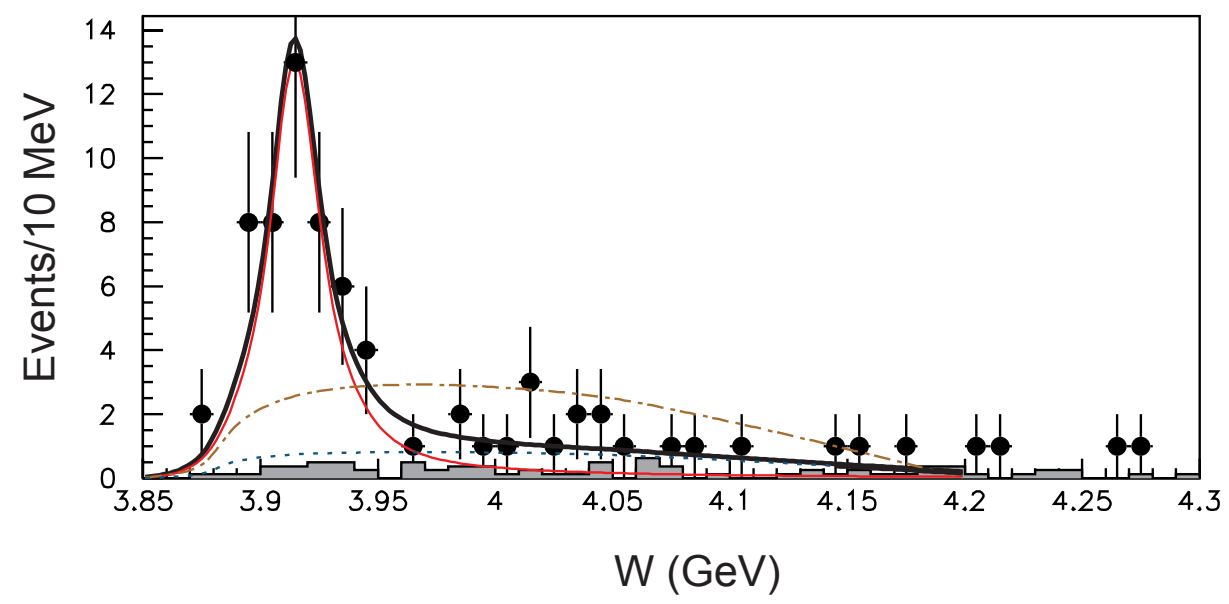

**Figure 18.3.7.** The $J/\psi\omega$ distribution in $\gamma\gamma$ events as measured by Belle (Uehara, 2010b). The black curve is the result of a fit to the data to determine the $Y(3915)$ mass and width. The dot-dashed curve (brown) represents the result of the fit without the resonant contribution.

A likely hypothesis is that the $Y(3940)$ and $Y(3915)$ states coincide, in which case this is the only candidate for an exotic state other than the $X(3872)$ to be seen in two different production mechanisms. The weighted averages of the mass and width measurements listed for $Y(3940)$ and $Y(3915)$ in Table 18.3.2 are $M_{Y(3915)} = (3918.4 \pm 1.9)\,\mathrm{MeV}/c^2$ and $\Gamma_{Y(3915)} = (20 \pm 5)\,\mathrm{MeV}/c^2$. A *BABAR* search for $\gamma\gamma \to Y(3915) \to \eta_c\pi^+\pi^-$ with $\eta_c \to K^0_S K^\pm \pi^\mp$ found no signal events and established an upper limit of $\Gamma_{\gamma\gamma}(Y(3915)) \times \mathcal{B}(Y(3915) \to \eta_c\pi^+\pi^-) < 91$ eV (Lees, 2012t).

#### 18.3.3.3 The $Z(3930)$

Another state, named $Z(3930)$, was seen by Belle in $\gamma\gamma$ fusion into $D\overline{D}$ (Uehara, 2006). This state has been confirmed by the *BABAR* Collaboration (Aubert, 2010g), which has also performed an angular analysis of the decay products that favors a $J^{PC} = 2^{++}$ assignment. This state is generally accepted as the $\chi'_{c2}$, the $2\,^3P_2$ charmonium state. Details of the relevant analyses can be found in Section 18.2.1.2.

#### 18.3.3.4 Charmonium assignments for the $X(3940)$, $Y(3915)$ and $Z(3930)$?

The $X(3940)$'s prominent decay to $D^*\overline{D}$ taken together with the lack of any evidence for $D\overline{D}$ is strongly suggestive of $J^{PC} = 0^{-+}$ quantum numbers, which implies that the most likely charmonium assignment is the $3\,^1S_0$, commonly known as the $\eta_c(3S)$. This assignment is somewhat problematic because the hyperfine partner state, the $3\,^3S_1$, or $\psi(4040)$, has already been established with a measured mass of $(4040 \pm 4)\,\mathrm{MeV}/c^2$ (Ablikim et al., 2007); this would imply an $n = 3$ hyperfine splitting of $(98 \pm 8)\,\mathrm{MeV}/c^2$, almost twice as large as the $n = 2$ splitting of $(47 \pm 1)\,\mathrm{MeV}/c^2$.

The *BABAR* study of the $Y(3915)$ state favors a $0^{++}$ quantum number assignment, for which the closest charmonium level is the $2\,^3P_0$ state, the so-called $\chi'_{c0}$. In this case, its $2\,^3P_2$ multiplet partner is likely the $Z(3930)$ with a measured mass of $(3927{\pm}3)\,\mathrm{MeV}/c^2$, implying an anomalously small $^3P_2$-$^3P_0$ fine splitting of only $\simeq 10\,\mathrm{MeV}/c^2$ for the $n = 2$ triplet $P$-wave multiplet (an order of magnitude smaller than the corresponding $n = 1$ splitting). Moreover, the $\chi'_{c0}$ is expected to have a partial decay width to $D\overline{D}$ of order $30\,\mathrm{MeV}/c^2$ (Barnes et al., 2005), which, by itself, is substantially wider than the measured total width of the $Y(3915)$. Even though no experimental limits on $\mathcal{B}(Y(3915) \to D\overline{D})$ have been reported to date, no signs of a signal for $Y(3915) \to D\overline{D}$ are evident in the measured $D\overline{D}$ invariant mass distributions for $B \to D\overline{D}K$ decays published by *BABAR* (Aubert, 2008bd) or Belle (Brodzicka, 2008), even though both studies see prominent signals for $B \to \psi(3770)K$, $\psi(3770) \to D\overline{D}$.

The $Z(3930)$ has measured properties that match well to the expectations for the $2\,^3P_2$ charmonium state and has no need for an exotic interpretation.

### 18.3.4 Other $C = +1$ states

We review now the remaining $C = +1$ resonances. The first is called $X(4160)$, and was discovered by Belle in double charmonium events. It is produced in association with a $J/\psi$ meson and decays into $D^{*+}D^{*-}$ (Pakhlov, 2008; see top panel of Fig. 18.3.8). The fitted mass and width are $M = (4156^{+25}_{-20}{\pm}15)\,\mathrm{MeV}/c^2$ and $\Gamma = (139^{+111}_{-61} \pm 21)\,\mathrm{MeV}/c^2$. The charge conjugation $C = +1$ is constrained by the production mechanism, which favors also $J = 0$. Hence, this state is a good candidate for a radial excitation of the pseudoscalar charmonium, a $\eta_c(nS)$ state. The identification is discussed in Section 18.2.1.3.

The CDF experiment announced a resonance close to threshold in $J/\psi\,\phi$ invariant mass, in the channel $B \to J/\psi\,\phi K$ (Aaltonen et al., 2009a). This state is called $Y(4140)$, and has mass and width $M = (4143.0 \pm 2.9 \pm 1.2)\,\mathrm{MeV}/c^2$ and $\Gamma = (11.7^{+8.3}_{-5.0} \pm 3.7)\,\mathrm{MeV}/c^2$. The natural quantum number would be $J^{PC} = 0^{++}$, but the exotic assignment $J^{PC} = 1^{-+}$ is not excluded. If the latter hypothesis were confirmed, this could be the hybrid ground state. The measured mass is indeed close to lattice calculations for the lightest hybrid meson (for instance, see Bernard et al., 1997). The search for states in this production mechanism at the $B$ Factories suffers from poor acceptance, and thus does not have sufficient statistical power to be conclusive. Some models (Branz, Gutsche, and Lyubovitskij, 2009) predict a copious production of such state in $\gamma\gamma$ fusion. Belle searched in this channel, but found no evidence for a $Y(4140)$. A limit $\Gamma_{\gamma\gamma} \times \mathcal{B}(\phi J/\psi) < 41\ (6)$ eV for $J^P = 0^+\ (2^+)$ was set at 90% C.L. for the $Y(4140)$ (Shen, 2010a).

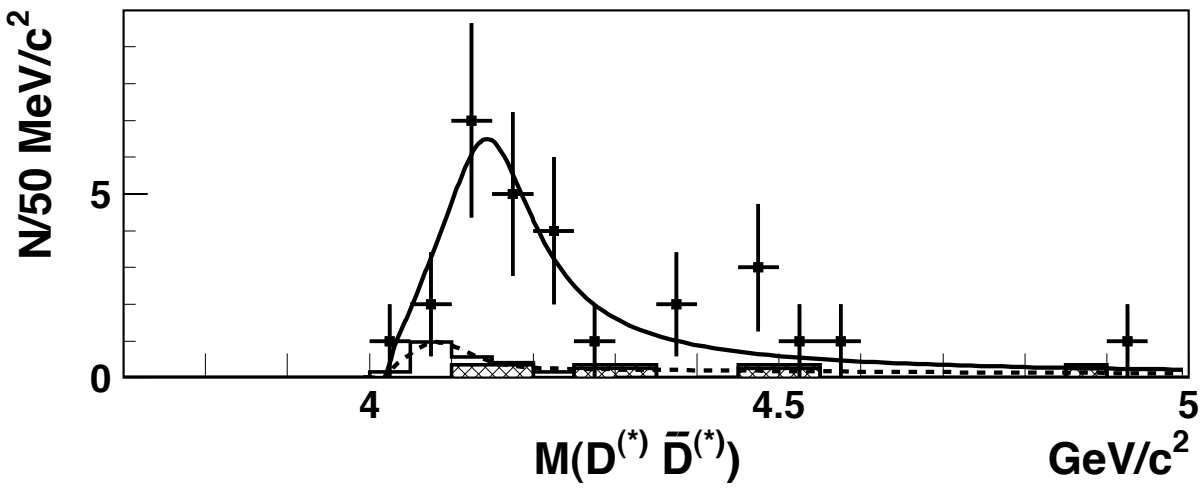

(a) $X(4160) \to D^{*+}D^{*-}$ from Belle (Pakhlov, 2008)

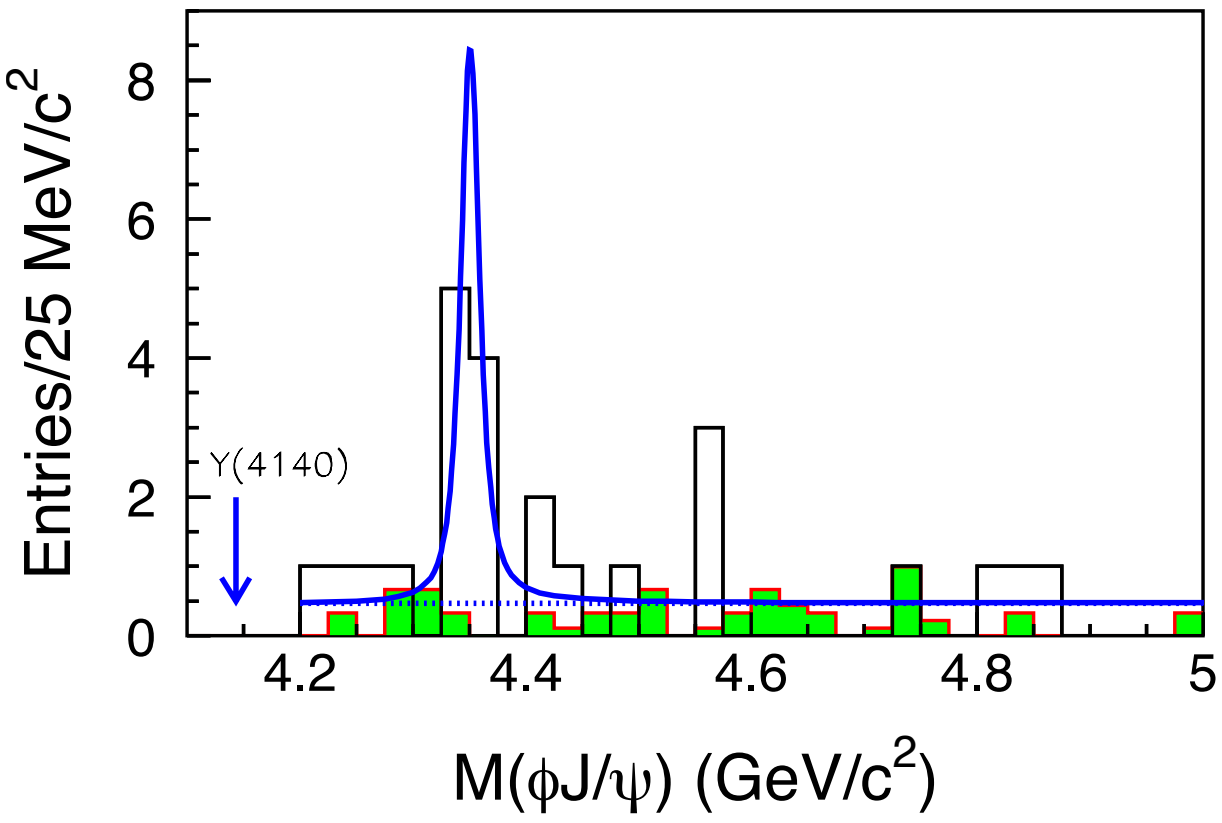

(b) $X(4350) \to J/\psi\phi$ from Belle (Shen, 2010a).

**Figure 18.3.8.** Invariant mass distributions of the most significant observations of states with $C = +1$ and mass above 4 GeV.

During the search for $\gamma\gamma \to X \to J/\psi\,\phi$, Belle found instead a $3.2\sigma$ peak with $M = (4350.6^{+4.6}_{-5.1} \pm 0.7)\,\mathrm{MeV}/c^2$ and $\Gamma = (13^{+18}_{-9} \pm 4)\,\mathrm{MeV}/c^2$ (see Fig. 18.3.8, bottom plot). This is possibly another state called $X(4350)$, with $C = +1$ and close in mass to one of the vector states we will discuss in Section 18.3.5.

### 18.3.5 The $1^{--}$ family

The most unambiguous way to assign $J^{PC}$ quantum numbers a particle is when it is produced in $e^+e^-$ annihilation, so that its quantum numbers must be the same as the photon ones: $J^{PC} = 1^{--}$. The $B$ Factories can investigate a large range of masses for such particles by looking for events where the emission of an energetic photon by the initial state reduces the $e^+e^-$ center-of-mass energy down to the particle's mass (so-called "ISR" events). Such analyses are discussed in more detail in Chapter 21. Alternatively, dedicated $e^+e^-$ machines, like CESR and BEPC scan directly the center-of-mass energies of interest.

The first new state to be observed via these processes was the $Y(4260) \to J/\psi\pi^+\pi^-$ resonance seen by *BABAR* (Aubert, 2005y) and promptly confirmed by CLEO both in ISR events (He et al., 2006) and in direct production (Coan et al., 2006). The latter paper also reported evidence for $Y(4260) \to J/\psi\pi^0\pi^0$ and some events of $Y(4260) \to J/\psi K^+K^-$.

A *BABAR* search for the $Y(4260)$ in the $\psi(2S)\pi^+\pi^-$ decay channel found no evidence of a signal (Aubert, 2007m), but, instead, saw a peak at a different mass, the $Y(4350)$. While the absence of $Y(4260) \to \psi(2S)\pi^+\pi^-$ decays might be understood if the pion pair in the $J/\psi\pi^+\pi^-$ decay were concentrated in an intermediate state (such as $f_0(980) \to \pi\pi$) that is too massive to be produced with a $\psi(2S)$, the absence of any sign of the $Y(4350) \to J/\psi\pi^+\pi^-$ is not so easily understood. Cotugno, Faccini, Polosa, and Sabelli (2010) have shown this absence to be significant: $\mathcal{B}(Y(4350) \to J/\psi\pi^+\pi^-)/\mathcal{B}(Y(4350) \to \psi(2\mathrm{S})\pi^+\pi^-) < 3.4 \times 10^{-3}$ at the 90% C.L..

Belle subsequently confirmed both of these $1^{--}$ states (Wang, 2007c; Yuan, 2007), and observed another state in the $\psi(2S)\pi^+\pi^-$ channel that was not visible in *BABAR* data due to the size of the data sample: the $Y(4660)$. Figure 18.3.9 shows Belle's published invariant mass spectra for both the $J/\psi\pi^+\pi^-$ and the $\psi(2S)\pi^+\pi^-$ channels.

An important question is whether or not the pion pair comes from one or more resonant states. Figure 18.3.10 shows the di-pion invariant mass spectra from Belle for events in the $J/\psi\pi^+\pi^-$ and $\psi(2S)\pi^+\pi^-$ invariant mass peaks that correspond to each of the three resonances. Although a subtraction of the continuum background has not been performed, there is some indication that only the $Y(4660)$ has a well defined intermediate state (most likely the $f_0(980)$); the other two peaks have a more complex structure. In addition, the *BABAR* analysis of the $J/\psi\pi^+\pi^-$ channel (Lees, 2012ab) finds some evidence of a $J/\psi f_0(980)$ component. The observation of decays involving an $f_0$ is particularly interesting because the scalar mesons have long been considered tetraquark candidates (Jaffe, 1977a).

The relative decay rate of these new states into lower-mass charmonium states and into two charm mesons can be used to distinguish between identifications as regular charmonium states and other possibilities. Searches for $Y \to D^{(*)}\overline{D}^{(*)}$ decay channels carried out by Belle (Abe, 2007d; Pakhlova, 2008a) and *BABAR* (Aubert, 2007u) found no evidence for a signal; 90% C.L. limits from *BABAR* are:

$$\begin{aligned}\mathcal{B}(Y(4260) \to D\overline{D})/\mathcal{B}(Y(4260) \to J/\psi\pi^+\pi^-) < 1.0,\\ \mathcal{B}(Y(4260) \to D^*\overline{D})/\mathcal{B}(Y(4260) \to J/\psi\pi^+\pi^-) < 34,\\ \mathcal{B}(Y(4260) \to D^*\overline{D}^*)/\mathcal{B}(Y(4260) \to J/\psi\pi^+\pi^-) < 40.\end{aligned} \tag{18.3.9}$$

*BABAR* also set 90% C.L. limits for the $Y \to D_s^{(*)+}D_s^{(*)-}$ decay channels (del Amo Sanchez, 2010d):

$$\begin{aligned}\mathcal{B}(Y(4260) \to D_s^+D_s^-)/\mathcal{B}(Y(4260) \to J/\psi\pi^+\pi^-) < 0.7,\\ \mathcal{B}(Y(4260) \to D_s^-D_s^{*-})/\mathcal{B}(Y(4260) \to J/\psi\pi^+\pi^-) < 44,\\ \mathcal{B}(Y(4260) \to D_s^{*+}D_s^{*-})/\mathcal{B}(Y(4260) \to J/\psi\pi^+\pi^-) < 30.\end{aligned} \tag{18.3.10}$$

Analogously, Belle studied $Y \to D^0D^{*-}\pi^+$ decays in ISR events as described in Section 21.4.5. The significances for the $Y(4260)$, $Y(4350)$, $Y(4660)$, and $X(4630)$ signals are $0.9\sigma$, $1.4\sigma$, $0.1\sigma$, and $1.8\sigma$, respectively, and the corresponding upper limits on the peak cross sections for $e^+e^- \to X(Y) \to DD^{*-}\pi^+$ processes are presented in Table 18.3.3. The upper limits presented in the Table are at the 90% C.L. and include systematic uncertainties.

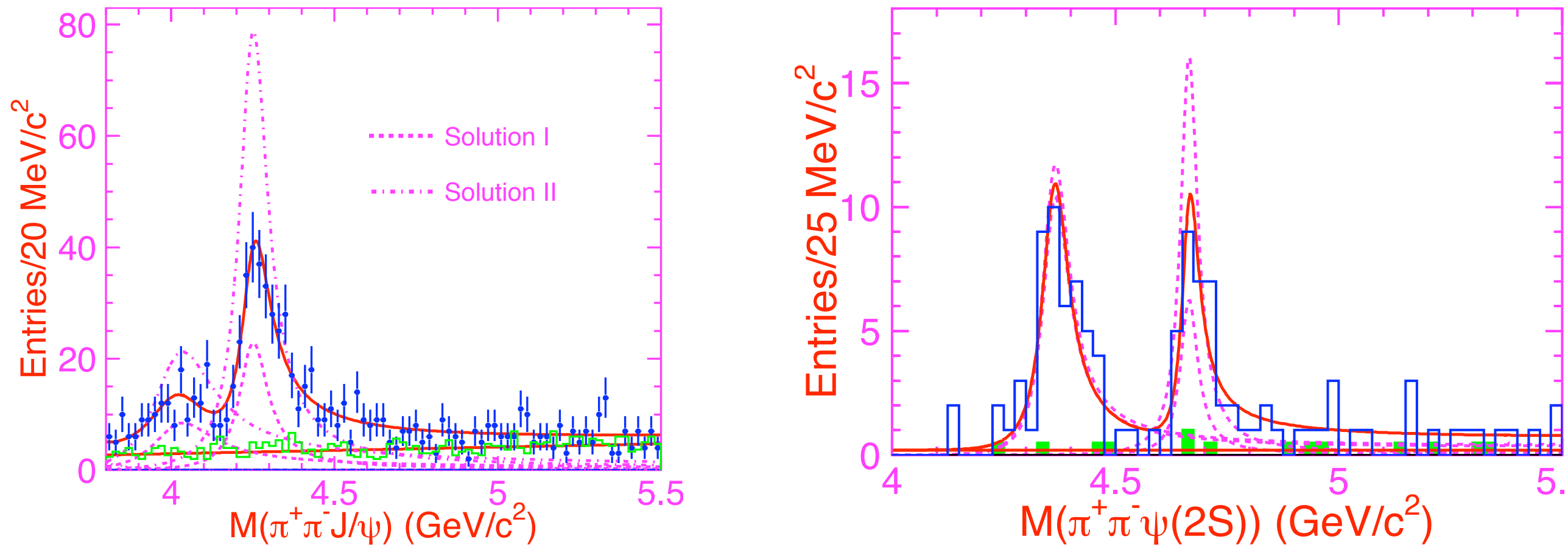

**Figure 18.3.9.** Distributions of $J/\psi\pi^+\pi^-$ (left; Yuan, 2007) and $\psi(2S)\pi^+\pi^-$ (right; Wang, 2007c) invariant masses in ISR production. The data points (left) and open histogram (right) show the data while the green histograms show the normalized sidebands of the charmonium candidates. The curves show the best fit with two coherent resonances together with a background term; the dashed curve shows the contribution from each component. The interference between the two resonances is not shown. In both cases the likelihood fits to the spectra return two solutions of equally good quality as indicated by the two dashed curves.

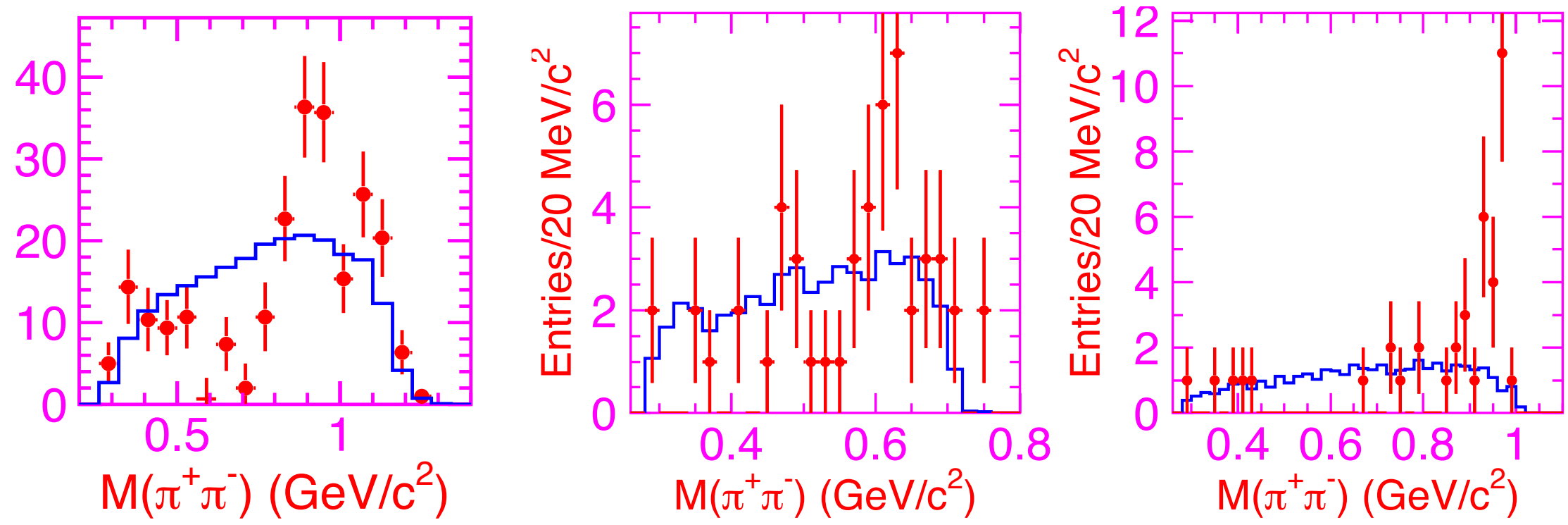

**Figure 18.3.10.** The di-pion invariant mass distribution in $Y(4260) \to J/\psi\pi^+\pi^-$ (left; Yuan, 2007), $Y(4350) \to \psi(2S)\pi^+\pi^-$ (center; Wang, 2007c), and $Y(4660) \to \psi(2S)\pi^+\pi^-$ decays (right; Wang, 2007c). Points are pure signal events, histograms are MC simulations of phase space distributions.

A distinctive signature of tetraquarks would be the observation of $1^{--}$ states decaying into two baryons since it is easier to form two baryons starting from four constituent quarks than from two (see Cotugno, Faccini, Polosa, and Sabelli, 2010 for details). This motivated a Belle search for the ISR production of resonant structures decaying into $\Lambda_c\bar{\Lambda}_c$ (Pakhlova, 2008b). A structure is seen near threshold (see Fig. 18.3.11) that, when fitted with a Breit-Wigner line shape, has $M = (4634^{+8+5}_{-7-8})\,\mathrm{MeV}/c^2$ and $\Gamma_{tot} = (92^{+40+10}_{-24-12})\,\mathrm{MeV}/c^2$, values that are close to those of the $Y(4660)$. An analysis performed by using the same model for the line-shape of the two measured spectra concluded that the two structures are consistent with the hypothesis that they are the same state with a strong preference for the baryonic decay mode: $\mathcal{B}(Y(4660) \to \Lambda_c\bar{\Lambda}_c)/\mathcal{B}(Y(4660) \to \psi(2S)\pi\pi) = 25 \pm 7$ (Cotugno, Faccini, Polosa, and Sabelli, 2010).

#### 18.3.5.1 Charmonium assignments for the $Y(4260)$, $Y(4350)$ and $Y(4660)$?

The reasons that at least some of the $Y(4260)$, $Y(4350)$, and $Y(4660)$ are considered as exotic candidates are the lack of unassigned $1^{--}$ charmonium levels below $4500\,\mathrm{MeV}/c^2$, and the large apparent partial widths of these states in $\pi^+\pi^-$ transitions to the $J/\psi$ or $\psi(2S)$. A comparison of the measured cross section for $e^+e^- \to Y(4260) \to J/\psi\pi^+\pi^-$ with limits on resonance production in the hadron cross section at the same mass leads to the 90% C.L. lower limit $\Gamma(Y(4260) \to J/\psi\pi^+\pi^-) > 1$ $\mathrm{MeV}/c^2$ (Mo et al., 2006). This is much larger than the typical for $1^{--}$ charmonium states: for instance, the corresponding partial width for the $\psi(3770)$ is $52 \pm 8\,\mathrm{keV}/c^2$.

**Table 18.3.3.** Upper limits on the peak cross section for the processes $e^+e^- \to Y \to D^0 D^{*-}\pi^+$, $\mathcal{B}_{ee} \times \mathcal{B}(Y \to D^0 D^{*-}\pi^+)$ and $\mathcal{B}(Y \to D^0 D^{*-}\pi^+)/\mathcal{B}(Y \to \pi^+\pi^- J/\psi\,(\psi(2S)))$ at the 90% C.L., where $Y = Y(4260)$, $Y(4350)$, $Y(4660)$, $X(4630)$. From Pakhlova (2009); this analysis is also briefly discussed in Section 21.4.5 (especially Fig. 21.4.9) and Section 18.2.2.2.

| | | $Y(4260)$ | $Y(4350)$ | $Y(4660)$ | $X(4630)$ |
|---|---|---|---|---|---|
| $\sigma(e^+e^- \to Y) \times \mathcal{B}(Y \to D^0 D^{*-}\pi^+))$ | [nb] | 0.36 | 0.55 | 0.25 | 0.45 |
| $\mathcal{B}_{ee} \times \mathcal{B}(Y \to D^0 D^{*-}\pi^+))$ | $[\times 10^{-6}]$ | 0.42 | 0.72 | 0.37 | 0.66 |
| $\mathcal{B}(Y \to D^0 D^{*-}\pi^+)/\mathcal{B}(Y \to \pi^+\pi^- J/\psi)$ | | 9 | | | |
| $\mathcal{B}(Y \to D^0 D^{*-}\pi^+)/\mathcal{B}(Y \to \pi^+\pi^- \psi(2S))$ | | | 8 | 10 | |

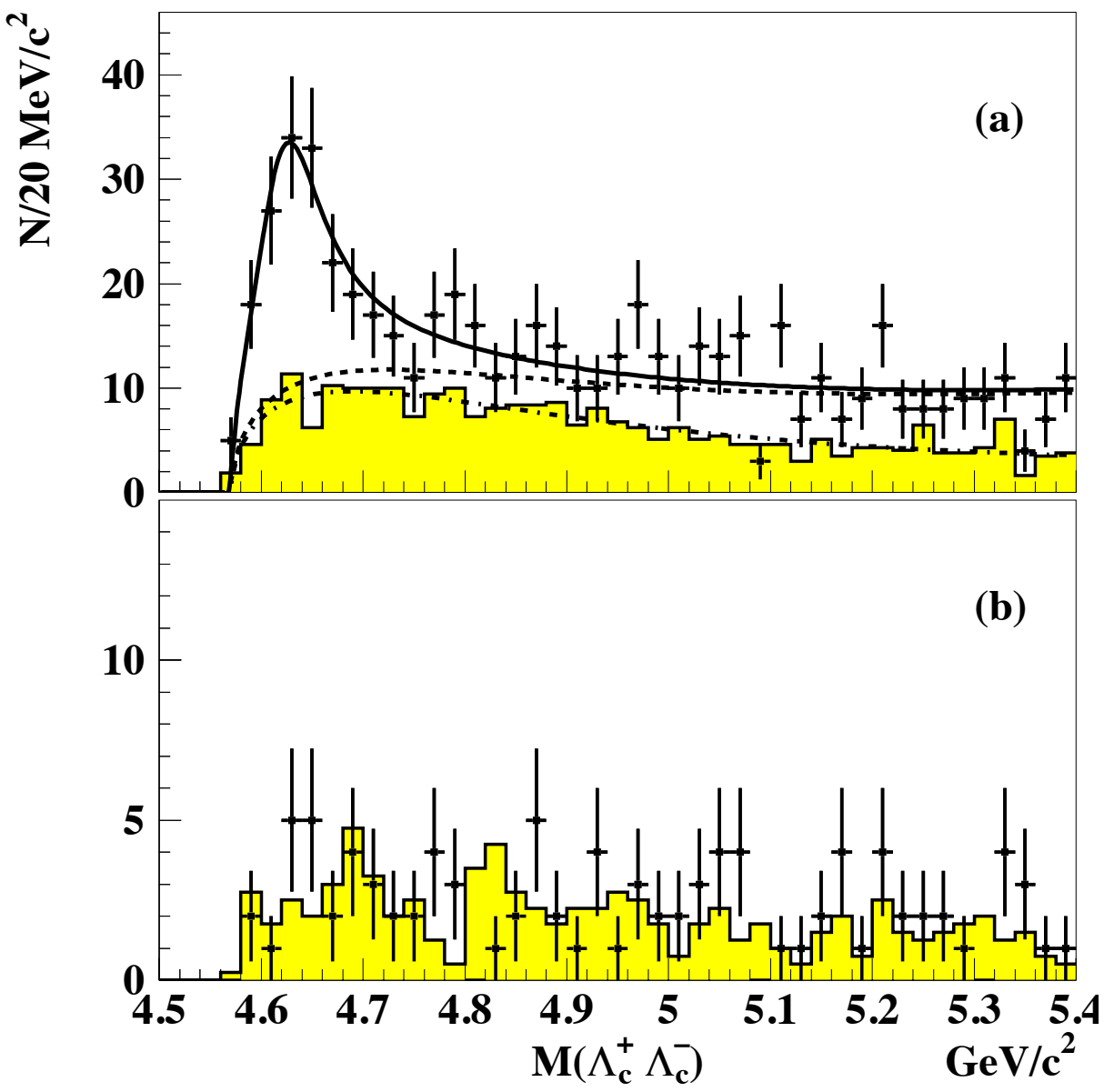

**Figure 18.3.11.** $\Lambda_c\overline{\Lambda}_c$ invariant mass distribution in ISR events from Belle (Pakhlova, 2008b): (a) using partial reconstruction plus proton tags, and (b) background events with wrong-sign proton tags. The histograms in each case show normalized contributions from $\Lambda_c$ sidebands. The superimposed curve is the result of the fit reported in the text.

## 18.3.6 Charged charmonium-like States

A significant turning point in the quest for states beyond the standard charmonium model would be the observation of charged states decaying into charmonium plus accompanying charged hadrons. There is no way to explain such an observation without at least four bound quarks (*e.g.* $c\bar{c}d\bar{u}$). There is evidence for three such charged states, seen by Belle but not by *BABAR*: the $Z(4430)^+$ state decaying into $\psi(2S)\pi^+$ (Choi, 2008; see Section 18.3.6.1), and the $Z_1(4050)^+$ and $Z_2(4250)^+$ states decaying into $\chi_{c1}\pi^+$ (Mizuk, 2008; see Section 18.3.6.2).

These states have been observed in $B$ decays in association with a charged kaon, *i.e.* in three-body $B \to X_{c\bar{c}}\pi K$ decays, where $X_{c\bar{c}} = \psi(2S)$ or $\chi_{c1}$. Three-body decays suffer from interference terms between strong amplitudes mediated by different resonances. In these particular cases the $K\pi$ system has several known resonances that could cause significant reflection effects. Namely, the decays $B \to X_{c\bar{c}}K^*(892)$, $B \to X_{c\bar{c}}K^*(1410)$, and, in particular, their mutual interference constitute irreducible sources of background which are difficult to estimate.

Further developments on charged states (as this book was being finalised), and some possible future studies, are briefly discussed in Section 18.3.6.3

### 18.3.6.1 $Z(4430^+) \to \psi(2S)\pi^+$

In the original Belle paper on the observation of the $Z(4430)^+ \to \psi(2S)\pi^+$ resonance in $B \to \psi(2S)\pi^+K$ decays (Choi, 2008), they report the distinct peak in the $M(\psi(2S)\pi^+)$ invariant mass distribution near $4430\,\mathrm{MeV}/c^2$ that is shown in the upper panel of Fig. 18.3.12. Belle argued that this peak could not be due to interference effects in the $K\pi$ channel because in $B \to \psi(2S)\pi^+K$ decays, events with $M(\psi(2S)\pi^+)$ near $4430\,\mathrm{MeV}/c^2$ correspond to $K\pi$ systems with a decay angle $\theta_{K\pi}$ in the region $\cos\theta_{K\pi} \simeq 0.25$, an angular region where interfering $S$-, $P$- and $D$-waves cannot create a peak without other, much larger structures elsewhere.

The Belle analysis was the subject of scrutiny by the *BABAR* Collaboration, which investigated the same final state by studying in detail the efficiency corrections and the shape of the background, relying for the latter on the data as much as possible (Aubert, 2009at). The search resulted in hints of a structure close to Belle's reported peak, but after estimates of the background they reported an 95% C.L. upper limit on the product branching fraction:

$$\mathcal{B}(B \to Z^+K^-) \times \mathcal{B}(Z^+ \to \psi(2S)\pi^+) < 3.1 \times 10^{-5}, \tag{18.3.11}$$

to be compared with Belle's non-zero value (Choi, 2008)

$$\mathcal{B}(B \to Z^+K^-) \times \mathcal{B}(Z^+ \to \psi(2S)\pi^+) = (4.1^{+1.0}_{-1.4}) \times 10^{-5}. \tag{18.3.12}$$

Subsequent to the *BABAR* analysis, Belle made a detailed Dalitz-plot analyses of $B \to \psi(2S)\pi^+K$ events that included interfering amplitudes for all known resonances in the $K\pi$ channel, both with and without a coherent amplitude for a resonance in the $\psi(2S)\pi^+$ channel (Mizuk, 2009). The results of this Belle analysis confirm those from the original report of a significant resonant structure in the $\psi(2S)\pi^+$ channel near $4430\,\mathrm{MeV}/c^2$. Although both statistical and systematic uncertainties on the $Z(4430)^+$ resonance parameters increased, the significance of the observed resonance signal remained the same; for the default Dalitz distribution model it was found to be $6.4\sigma$; the fitted mass and width of the $Z(4430)^+ \to \psi(2S)\pi^+$ from the Dalitz analysis are $M_{Z(4430)^+} = (4443^{+15+19}_{-12-13})\,\mathrm{MeV}/c^2$ and $\Gamma_{Z(4430)^+} = (109^{+86+74}_{-43-56})\,\mathrm{MeV}/c^2$.

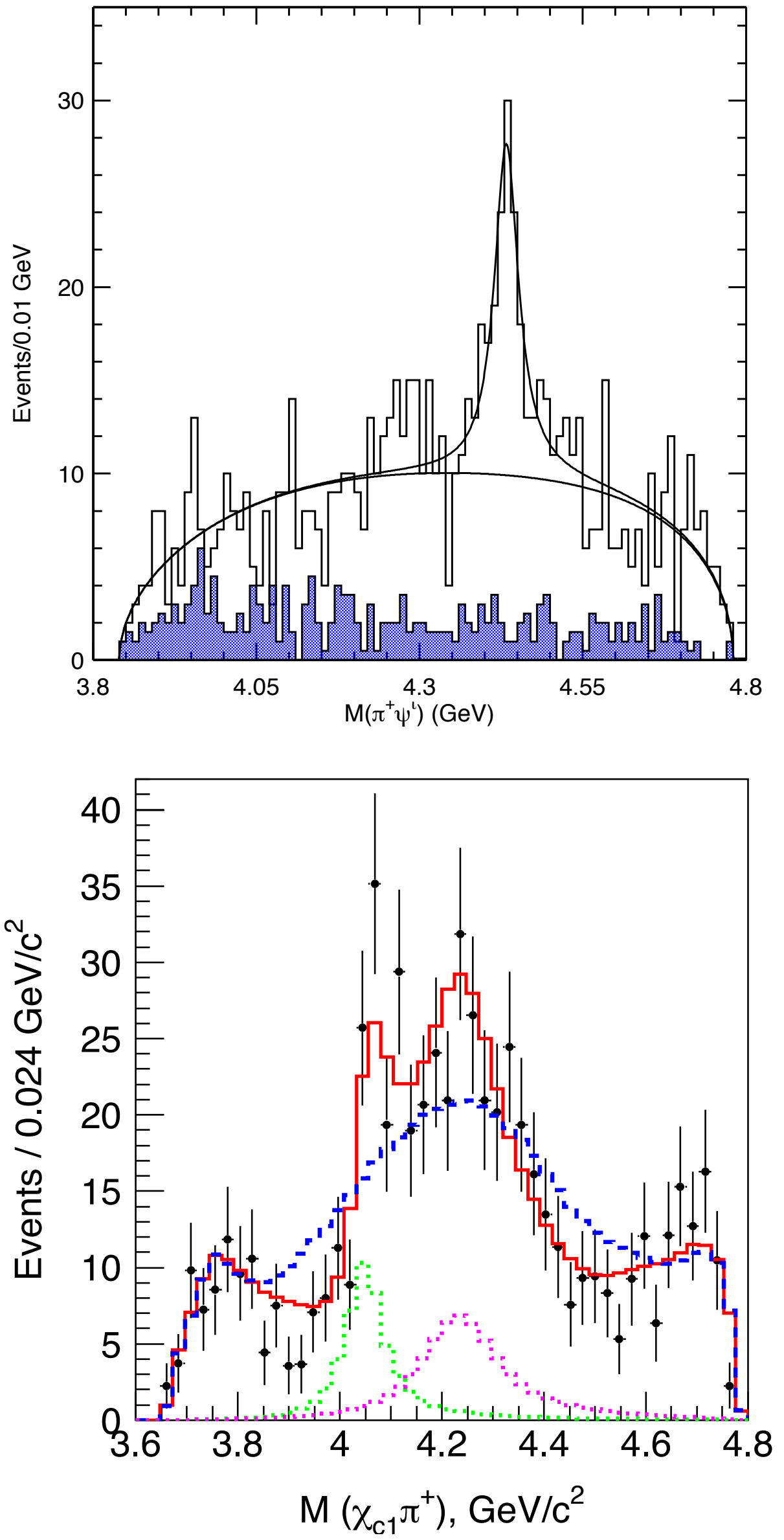

**Figure 18.3.12.** Invariant mass distributions from (top) $\psi(2S)\pi^\pm$ (Choi, 2008), and (bottom) $\chi_{c1}\pi^\pm$ (Mizuk, 2008), superimposed with fit result showing the charged resonances. In both figures, events with $M(K\pi)$ in the region of the $K^*(890)$ and $K^*(1410)$ peaks are removed. The solid red histogram in the lower figure shows the results of the fit that includes coherent $Z_1$ and $Z_2$ amplitudes; the dashed blue curve is the result of the fit using $K\pi$ amplitudes only.

While we were preparing this book, LHCb performed a four-dimensional fit of the decay amplitude (Aaij et al., 2014b). The $Z(4430)^+$ is confirmed with a significance of $13.9\,\sigma$ at least; the fitted mass and width are $M_{Z(4430)^+} = (4475\pm7^{+15}_{-25})\,\mathrm{MeV}/c^2$ and $\Gamma_{Z(4430)^+} = (172\pm13^{+37}_{-34})\,\mathrm{MeV}/c^2$, consistent with Belle measurements. Moreover, an analysis of the Argand diagram confirms the resonant character of the $Z(4430)^+$.

#### 18.3.6.2 States decaying to $\chi_{c1}\pi^+$

In a Dalitz-plot analysis of three-body $B \to \chi_{c1}\pi^+ K$ decays, Belle was unable to get an acceptable fit using only resonances in the $K\pi$ channel (Mizuk, 2008). The inclusion of a single $\chi_{c1}\pi^+$ resonance improved the fit substantially, but still did not reproduce the observed features very accurately. Belle finally settled on a fit that included two resonances in the $\chi_{c1}\pi$ channel: the $Z_1(4050)^+$ and $Z_2(4250)^+$. The $\chi_{c1}\pi^+$ invariant-mass distribution for events in the Dalitz-plot region between the $K^*(980)$ and $K^*(1410)$ bands is shown as data points with the projected final fit shown as a red histogram in the lower panel of Fig. 18.3.12. The fitted masses and widths of the two $\chi_{c1}\pi^+$ resonances are $M_{Z_1^+} = (4051 \pm 14^{+20}_{-41})\,\mathrm{MeV}/c^2$, $M_{Z_2^+} = (4248^{+44+180}_{-29-35})\,\mathrm{MeV}/c^2$, $\Gamma_{Z_1^+} = (82^{+21+47}_{-17-22})\,\mathrm{MeV}/c^2$ and $\Gamma_{Z_2^+} = (177^{+54+316}_{-39-61})\,\mathrm{MeV}/c^2$, respectively.

*BABAR* investigated $B \to \chi_{c1}\pi^+ K$ decays using an analysis that carefully studied the effects of interference between resonances in the $K\pi$ system (Lees, 2012w). They report adequate fits to the data using interfering resonances only in the $K\pi$ channel. They set 95% C.L. upper limits on the product branching fractions to the $Z_1^+$ and $Z_2^+$ states by studying the effects of adding incoherent resonant amplitudes for these two states to their fitting model:

$$\begin{aligned}
&\mathcal{B}(B \to Z_1^+ K^-) \times \mathcal{B}(Z_1^+ \to \chi_{c1}\pi^+) < 1.8 \times 10^{-5},\\
&\mathcal{B}(B \to Z_2^+ K^-) \times \mathcal{B}(Z_2^+ \to \chi_{c1}\pi^+) < 4.0 \times 10^{-5}.
\end{aligned} \tag{18.3.13}$$

For comparison, the non-zero values from Belle for the same quantities are

$$\begin{aligned}
&\mathcal{B}(B \to Z_1^+ K^-) \times \mathcal{B}(Z_1^+ \to \chi_{c1}\pi^+) = (3.0^{+1.2+3.7}_{-0.8-1.6}) \times 10^{-5},\\
&\mathcal{B}(B \to Z_2^+ K^-) \times \mathcal{B}(Z_2^+ \to \chi_{c1}\pi^+) = (4.0^{+2.3+19.7}_{-0.9-0.5}) \times 10^{-5}.
\end{aligned} \tag{18.3.14}$$

Part of the discrepancy between the two experiments may be due to the fact that in the Belle analysis, the $Z_1^+$, $Z_2^+$ and $K\pi$ amplitudes are all coherent and mutually interfere, while in the *BABAR* analysis the $Z_1^+$ and $Z_2^+$ terms are added incoherently and do not interfere with the $K\pi$ amplitudes. In the Belle results shown in Fig. 18.3.12 (lower), significant constructive and destructive interference between the $Z_1^+$ and $Z_2^+$ amplitudes with the $K\pi$ terms is evident (see the dips and peaks of the solid red curve, relative to the dashed blue curve showing the $K\pi$ amplitude fit result).

#### 18.3.6.3 Other candidates for charged charmonium-like states

Both Belle (Liu, 2013) and BESIII (Ablikim et al., 2013a) have claimed the observation of another charged resonance, $Z(3900)^+$, as a peak in the $J/\psi\pi^+$ invariant mass distribution in $Y(4260) \to \pi^+\pi^- J/\psi$ decay; the Belle peak has Breit-Wigner parameters $M = (3894.5 \pm 6.6 \pm 4.5)\,\mathrm{MeV}/c^2$ and $\Gamma = (63 \pm 24 \pm 26)\,\mathrm{MeV}/c^2$. A structure with similar parameters has also been seen by BESIII in $e^+e^- \to (D\overline{D}^*)^+\pi^-$ (Ablikim et al., 2014b). As this book was being finalised, BESIII also presented evidence for a structure $Z(4020)^+$ in the $h_c\pi^+$ invariant mass distribution in $e^+e^- \to \pi^+\pi^- h_c$ (Ablikim et al., 2013b), with parameters similar to those of a structure $Z(4025)^+$ seen in $e^+e^- \to (D^*\overline{D}^*)^\pm\pi^\mp$ (Ablikim et al., 2014a). Note that charged states $Z_b(10610)^+$ and $Z_b(10650)^+ \to h_b\pi^+$ have been seen by Belle in $\Upsilon(5S) \to h_b\pi^+\pi^-$ (see Section 18.4.5).

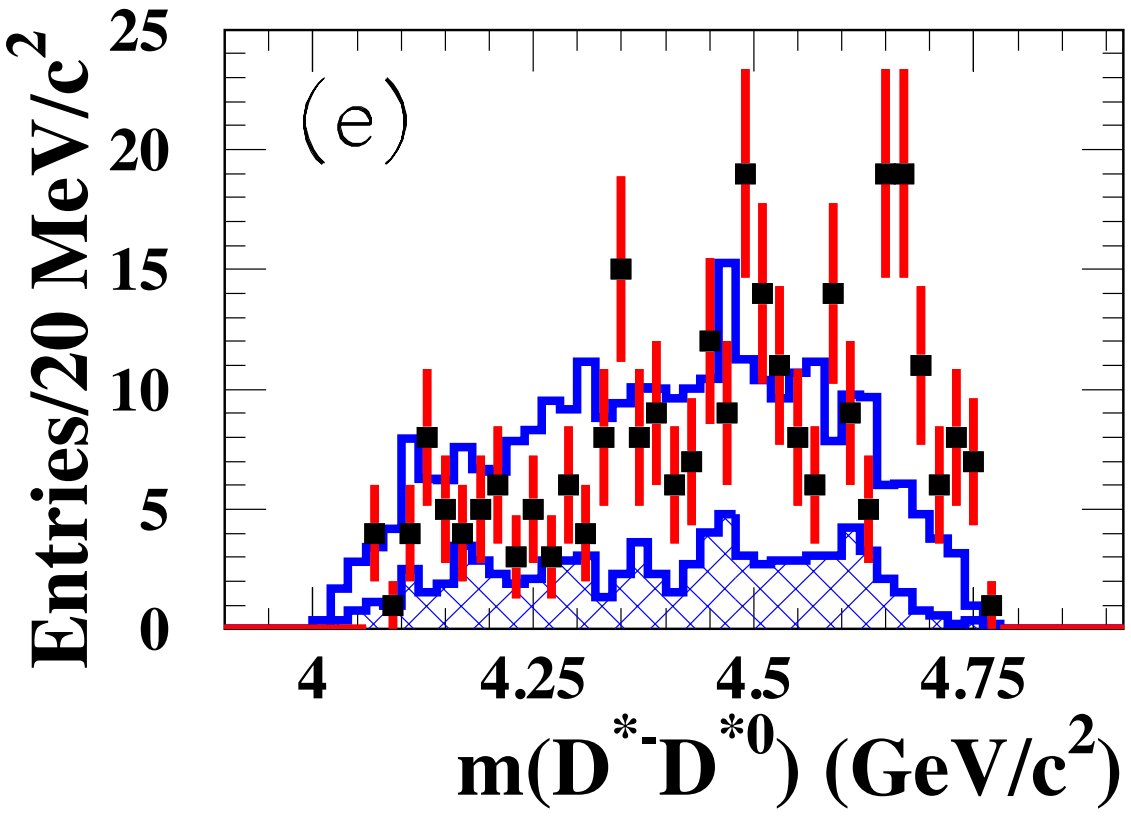

**Figure 18.3.13.** The $D^{*0}D^{*-}$ invariant mass spectrum from $B \to D^{*0}D^{*-}K$ decays from BABAR (Aubert, 2003f). The hatched histograms show the contribution expected from the combinatorial background. The open histograms show the contribution expected for the three-body decay generated with a phase-space model.

A further step in the study of the currently-claimed charged states would be to search for these, or other similar states, in $D^{(*)}\overline{D}^{(*)}$ systems produced in three-body $B \to D^{(*)}\overline{D}^{(*)}K$ decays. For these the experimental efficiencies are lower and detailed three-body amplitude analyses are of limited use with the currently available *BABAR* and Belle data samples. Nevertheless, results based on existing data samples are intriguing, as can be seen, for example, in the $M(D^{*0}D^{*-})$ invariant-mass distribution for $B \to D^{*0}D^{*-}K$ decays in *BABAR* (shown in Fig. 18.3.13). Here signs of structure can be seen, but with limited statistical significance. This will be a promising area of research at future flavor factories.

### 18.3.7 Summary and outlook

The status of experimental studies of exotic charmonium-like states is shown below in tabular form, for the various production mechanisms: $B$ decay (Table 18.3.4), initial state radiation in $e^+e^-$ annihilation (so-called radiative return events; Table 18.3.5), double charmonium production (Table 18.3.6), and two-photon production (Table 18.3.7); studies of charged states are summarized in Table 18.3.8.

As can be seen from the tables, our knowledge is quite fragmentary, despite the large $B$ Factory datasets. Apart from the $X(3872)$ (by far the best-studied state) and the $Y(3915)$, all of the candidate exotic mesons have been observed in only one production mechanism. Most have been observed in only one final state, and none has been the object of a systematic study in a range of final states. Analyses of particular combinations of production mechanism and final state are variously missing, or not performed in the relevant range of invariant mass ("**N**" in the tables), or lacking a fit to the data to test for the presence of exotic states ("**MF**"). To give some examples: the $M(J/\psi\,\gamma)$ spectra in Fig. 18.3.5 are focussed on the $X(3872)$ region and no information is provided for other masses; the $J/\psi\eta$ invariant mass spectrum in Fig. 18.3.4(b) is published, but not fitted for all the candidate new states. To obtain a more complete picture for any candidate would require either an observation or a limit in many final states, allowing quantitative tests of different models to be performed. For a number of particles where the current data is statistically limited, new decay modes would also add to the evidence for the existence of a new state, as opposed to (say) a final state interaction effect.

States produced in $B$ decays (Table 18.3.4) have been the most studied, although many fits are missing, especially of the baryonic final states $p\overline{p}$ and $\Lambda\overline{\Lambda}$ which are predicted to be important for tetraquark states. Among the decays that have never been studied, $B \to \psi(2S)\pi\pi K$ should be relatively clean, whereas $D_s^{(*)+}D_s^{(*)-}$ modes suffer from the low branching fractions of the usable decays of the final-state $D_s$ mesons.

Analysis of states produced in conjunction with an initial state radiation (ISR) photon (Table 18.3.5) is relatively straightforward, due to the unambigous $J^{PC} = 1^{--}$ assignment, although the original exotic candidate of this kind, the $Y(4260)$, is still by far the best studied. A complementary set of states, exclusively $C = +1$, is accessible in recoil from a $J/\psi$ in $e^+e^-$ annihilation (Table 18.3.6) an in production via $\gamma\gamma$ fusion (Table 18.3.7). Many of these analyses are challenging, however, due to large backgrounds and (in the case of $\gamma\gamma$ decays) large missing momentum; as a result, the $C = +1$ states seen only through these production mechanisms are relatively poorly studied. Concerning the recoil analyses, production in recoil against particles other than the $J/\psi$ has not been systematically investigated: the available samples in $B$ Factory data are small. Studies of the system recoiling against the $\chi_{c0}$ and/or the $\chi_{c2}$ would be particularly interesting, given the selection rules.

**Table 18.3.4.** Status of searches for the new states in the process $B \to XK$, $X \to f$, for several final states $f$, adapted from Drenska et al. (2010). Following the discussion in Section 18.3.3.2 we treat the state $Y(3940)$ seen in $B$ decay as a different state to the $X(3940)$ seen in $e^+e^- \to X\, J/\psi$ (see Table 18.3.6 below). Final states where each exotic states were observed (**S**: "seen") or excluded (**NS**: "not seen") are indicated. A final state is marked as **N** ("not performed") if the analysis has not been performed in a given mass range and with **MF** ("missing fit") if the spectra are published but a fit to a given state has not been performed. Finally "—" indicates that, although no search has been performed in this mode, the known quantum numbers or available energy forbid the decay; and "hard" that an analysis is experimentally too challenging.The same labels are used in subsequent tables (18.3.5–18.3.8). In the headings, $\psi$ denotes the $J/\psi$; $\psi' = \psi(2S)$, $2D^* = D^*\overline{D}^*$, and $2D_s^{(*)} = D_s^{(*)+}D_s^{(*)-}$.

| State | $J^{PC}$ | $\psi\pi\pi$ | $\psi\omega$ | $\psi\gamma$ | $\psi\phi$ | $\psi\eta$ | $\psi'\pi\pi$ | $\psi'\omega$ | $\psi'\gamma$ | $\chi_c\gamma$ | $p\overline{p}$ | $\Lambda\overline{\Lambda}$ | $\Lambda_c\overline{\Lambda}_c$ | $D\overline{D}$ | $D\overline{D}^*$ | $2D^*$ | $2D_s^{(*)}$ | $\gamma\gamma$ |
|---|---|---|---|---|---|---|---|---|---|---|---|---|---|---|---|---|---|---|
| X(3872) | $1^{++}$ | **S** | **S** | **S** | — | **NS** | — | — | **S** | **NS** | **MF** | **MF** | — | — | **S** | — | — | **NS** |
| Y(3940) | $J^{P+}$ | **MF** | **S** | **NS** | — | — | — | — | **MF** | — | **MF** | **MF** | — | **MF** | **NS** | — | **N** | **N** |
| Z(3930) | $2^{++}$ | **MF** | **MF** | **NS** | — | — | — | — | **MF** | — | **MF** | **MF** | — | **MF** | **MF** | — | **N** | **N** |
| Y(4140) | $J^{P+}$ | **MF** | **MF** | **N** | **S** | — | **N** | — | **N** | — | **MF** | **MF** | — | **MF** | **N** | **N** | **N** | **N** |
| X(4160) | $0^{P+}$ | **MF** | **MF** | **N** | **MF** | — | **N** | — | **N** | — | **MF** | **MF** | — | **MF** | **N** | **N** | **N** | **N** |
| Y(4260) | $1^{--}$ | **NS** | — | — | — | **MF** | **N** | — | — | **N** | **MF** | **MF** | — | **N** | **N** | **N** | **N** | — |
| X(4350) | $J^{P+}$ | **MF** | **MF** | **N** | **MF** | — | **N** | **N** | **N** | — | **MF** | **MF** | — | **N** | **N** | **N** | **N** | **N** |
| Y(4350) | $1^{--}$ | **MF** | — | — | — | **MF** | **N** | — | — | **N** | **MF** | **MF** | — | **N** | **N** | **N** | **N** | — |
| Y(4660) | $1^{--}$ | **N** | — | — | — | **MF** | **N** | — | — | **N** | **MF** | **MF** | **MF** | **N** | **N** | **N** | **N** | — |

**Table 18.3.5.** Status of searches for the new states in the process $e^+e^- \to \gamma_{ISR}X$, $X \to f$, for several final states $f$, adapted from Drenska et al. (2010). The meaning of the symbols is explained in the caption of Table 18.3.4.

| State | $J^{PC}$ | $\psi\pi\pi$ | $\psi'\pi\pi$ | $\psi\eta$ | $\chi_c\gamma$ | $p\overline{p}$ | $\Lambda\overline{\Lambda}$ | $\Lambda_c\overline{\Lambda}_c$ | $D\overline{D}$ | $D\overline{D}^*$ | $2D^*$ | $2D_s^{(*)}$ |
|---|---|---|---|---|---|---|---|---|---|---|---|---|
| Y(4260) | $1^{--}$ | **S** | **NS** | **NS** | **NS** | **NS** | **MF** | — | **NS** | **NS** | **NS** | **NS** |
| Y(4350) | $1^{--}$ | **NS** | **S** | **MF** | **MF** | **MF** | **MF** | — | **MF** | **MF** | **MF** | **MF** |
| Y(4660) | $1^{--}$ | **NS** | **S** | **MF** | **MF** | **MF** | **MF** | **S** | **MF** | **MF** | **MF** | **MF** |

**Table 18.3.6.** Status of searches for the new states in the process $e^+e^- \to XJ/\psi$, $X \to f$, for several final states $f$, adapted from Drenska et al. (2010). The meaning of the symbols is explained in the caption of Table 18.3.4; as stated there, we treat the $X(3940)$ and $Y(3940)$ as different states.

| State | $J^{PC}$ | $\psi\pi\pi$ | $\psi\omega$ | $\psi\gamma$ | $\psi\phi$ | $\psi'\pi\pi$ | $\psi'\omega$ | $\psi'\gamma$ | $\chi_c\gamma$ | $p\overline{p}$ | $\Lambda\overline{\Lambda}$ | $\Lambda_c\overline{\Lambda}_c$ | $D\overline{D}$ | $D\overline{D}^*$ | $2D^*$ |
|---|---|---|---|---|---|---|---|---|---|---|---|---|---|---|---|
| X(3872) | $1^{++}$ | hard | **N** | hard | — | hard | — | hard | hard | hard | hard | — | **MF** | **MF** | — |
| X(3940) | $0^{-+}$ | hard | **N** | hard | — | hard | — | hard | hard | hard | hard | — | **NS** | **S** | — |
| Z(3930) | $2^{++}$ | hard | **N** | hard | — | hard | — | hard | hard | hard | hard | — | **MF** | **MF** | — |
| Y(4140) | $J^{P+}$ | hard | **N** | hard | **N** | hard | — | hard | hard | hard | hard | — | **MF** | **MF** | **MF** |
| X(4160) | $0^{P+}$ | hard | **N** | hard | **N** | hard | — | hard | hard | hard | hard | — | **MF** | **S** | **MF** |
| X(4350) | $J^{P+}$ | hard | **N** | hard | **N** | hard | **N** | hard | hard | hard | hard | hard | **MF** | **MF** | **MF** |

**Table 18.3.7.** Status of searches for the new states in the process $\gamma\gamma \to X$, $X \to f$, for several final states $f$, adapted from Drenska et al. (2010). The meaning of the symbols is explained in the caption of Table 18.3.4. The identification of the $Y(3915)$ with the $\chi_{c0}(2P)$ is problematic, for the reasons discussed in Section 18.3.3.4, so we retain the former notation. As discussed in Section 18.3.3.2, the $Y(3915)$ and the $Y(3940)$ may be the same state.

| State | $J^{PC}$ | $\psi\pi\pi$ | $\psi\omega$ | $\psi\gamma$ | $\psi\phi$ | $\psi'\pi\pi$ | $\psi'\omega$ | $\psi'\gamma$ | $p\overline{p}$ | $\Lambda\overline{\Lambda}$ | $\Lambda_c\overline{\Lambda}_c$ | $D\overline{D}$ | $D\overline{D}^*$ | $2D^*$ | $2D_s^{(*)}$ |
|---|---|---|---|---|---|---|---|---|---|---|---|---|---|---|---|
| X(3872) | $1^{++}$ | **N** | hard | hard | — | — | — | hard | **MF** | **MF** | — | **MF** | **N** | — | — |
| Y(3915) | $0^{++}$ | **N** | **S** | hard | — | — | — | hard | **MF** | **MF** | — | **MF** | **N** | — | **N** |
| Z(3930) | $2^{++}$ | **N** | **MF** | hard | — | — | — | hard | **MF** | **MF** | — | **S** | **N** | — | **N** |
| Y(4140) | $J^{P+}$ | **N** | **MF** | hard | **NS** | **N** | — | hard | **N** | **N** | — | **MF** | **N** | **N** | **N** |
| X(4160) | $0^{P+}$ | **N** | **MF** | hard | **NS** | **N** | — | hard | **N** | **N** | — | **MF** | **N** | **N** | **N** |
| X(4350) | $J^{P+}$ | **N** | **N** | hard | **S** | **N** | **N** | hard | **N** | **N** | **N** | **N** | **N** | **N** | **N** |

**Table 18.3.8.** Status of searches for the new charged states in several final states, adapted from Drenska et al. (2010). The meaning of the symbols is explained in the caption of Table 18.3.4.

| State | $\psi\pi$ | $\psi\pi\pi^0$ | $\psi'\pi$ | $\psi'\pi\pi^0$ | $\chi_{c1}\pi$ | $h_c\pi$ | $D\overline{D}$ | $D\overline{D}^*$ | $2D^*$ |
|---|---|---|---|---|---|---|---|---|---|
| $X(3872)^+$ | **MF** | **NS** | **MF** | **N** | **MF** | **MF** | **N** | **MF** | — |
| $Z(3900)^+$ | **S** | **MF** | **MF** | **N** | **MF** | **NS** | **N** | **S** | — |
| $Z(3930)^+$ | **MF** | **N** | **MF** | **N** | **MF** | **MF** | **N** | **N** | — |
| $Z(4020)^+$ | **NS** | **N** | **MF** | **N** | **MF** | **S** | **N** | **N** | **S** |
| $Z(4050)^+$ | **MF** | **N** | **MF** | **N** | **S** | **MF** | **N** | **N** | **MF** |
| $Y(4140)^+$ | **MF** | **N** | **MF** | **N** | **MF** | **N** | **N** | **N** | **MF** |
| $Z(4250)^+$ | **MF** | **N** | **MF** | **N** | **S** | **N** | **N** | **N** | **MF** |
| $X(4350)^+$ | **MF** | **N** | **MF** | **N** | **MF** | **N** | **N** | **N** | **MF** |
| $Z(4430)^+$ | **NS** | **N** | **S** | **N** | **MF** | **N** | **N** | **N** | **MF** |
| $Z(4660)^+$ | **MF** | **N** | **MF** | **N** | **MF** | **N** | **N** | **N** | **MF** |

Relatively few searches for charged exotic states — the most striking signature of states made of more than two quarks — have been conducted in $B$ decays; as shown in Table 18.3.8, searches have been accomplished for only five combinations of final states and exotic candidates. Ideally, a search for charged partner states should be performed for each neutral exotic meson. A general spectrum of four-quark bound states would also include mesons containing a single $s$ quark, with distinctive strong decays to charmonium plus a charged kaon. Searches for such states could be conducted in $B$ decays in association with an $s\bar{s}$ state, or inclusively at a hadron collider.

In conclusion, a systematic study of the exotic spectrum is required to form a global and definite picture of these states and their structure. As well as finalizing studies with existing $B$ Factory and Tevatron data, results from newer, even higher luminosity machines such as the LHC and the super flavor factories are needed. The apparent confirmation of the $Z(4430)^+$ by LHCb, just as this book was being completed (Aaij et al., 2014b), is both a welcome clarification of the experimental picture, and a reminder that the exotic charmonium-like states — an unexpected product of the bounty of data from the $B$ Factories— must be studied in larger data samples if they are to be more fully understood.

## 18.4 Bottomonium

***Editors:***
*Stephen Sekula (BABAR)*
*Roberto Mussa (Belle)*
*Nora Brambilla (theory)*

**Additional section writers:**
*Bryan Fulsom, Romulus Godang, Christopher Hearty, Todd Pedlar, Cheng Ping Shen*

### 18.4.1 Introduction

At the advent of the B-factory experiments, measurements of the spectrum of the bottomonium system were limited to only a few states — the $\Upsilon(nS)$ and $\chi_b(nP)$ resonances. However, the theoretical predictions for this spectrum were abundant and the bottomonium system offered open territory for scientific exploration. The spectrum of the bottomonium system is illustrated in Fig. 18.4.1. Of particular interest were the bottomonium ground state, the $\eta_b(1S)$, and its excitations (the $S$-wave singlet states, *e.g.* $\eta_b(2S)$), the discovery of the $h_b(nP)$ $P$-wave singlet states, measurements of their properties and of transitions between bottomonium states as tests of various theoretical frameworks (Lattice QCD, NRQCD, pNRQCD, QCD potential models, etc.). In addition, these measurements allowed for searches for physics beyond the Standard Model (such as violation of universal couplings to leptons, dark matter, and low-mass Higgs bosons).
The bottomonium programs at *BABAR* and Belle yielded a rich assortment of both discoveries and measurements. The ground state of the bottomonium system was first discovered by *BABAR* and later confirmed by Belle, yielding multiple independent mass and branching fraction measurements (Section 18.4.4.2). In addition, Belle discovered the $\eta_b(2S)$ and measured its mass. The *BABAR* Collaboration was the first to show evidence of the existence of the $P$-wave singlet state, $h_b(1P)$, and the Belle Collaboration demonstrated clear discovery of this state and also of its partner, the $h_b(2P)$ (Section 18.4.4.3). Discovery of these states required in parallel the measurement of transitions between states, and these are detailed in the aforementioned sections. Independent measurements (unconnected with searches for new resonances) of transitions between bottomonium states, as well as decays to hadronic final states, are detailed in Section 18.4.6.

The bottomonium system yielded many surprises, however, when compared to the expectations from theoretical predictions. The Belle confirmation of the $\eta_b(1S)$ and discoveries of the $\eta_b(2S)$ and $h_b(1P, 2P)$ were all made possible in part by apparently large and anomalous $\pi^+\pi^-$ transition rates from the $\Upsilon(5S)$ resonance. This is discussed in Section 18.4.4.3. In addition, as a result of their investigations of these anomalous transition rates, two new states were discovered by the Belle Collaboration just above the open-bottom threshold (Fig. 18.4.1) - the charged $Z_b$ states (Section 18.4.5).

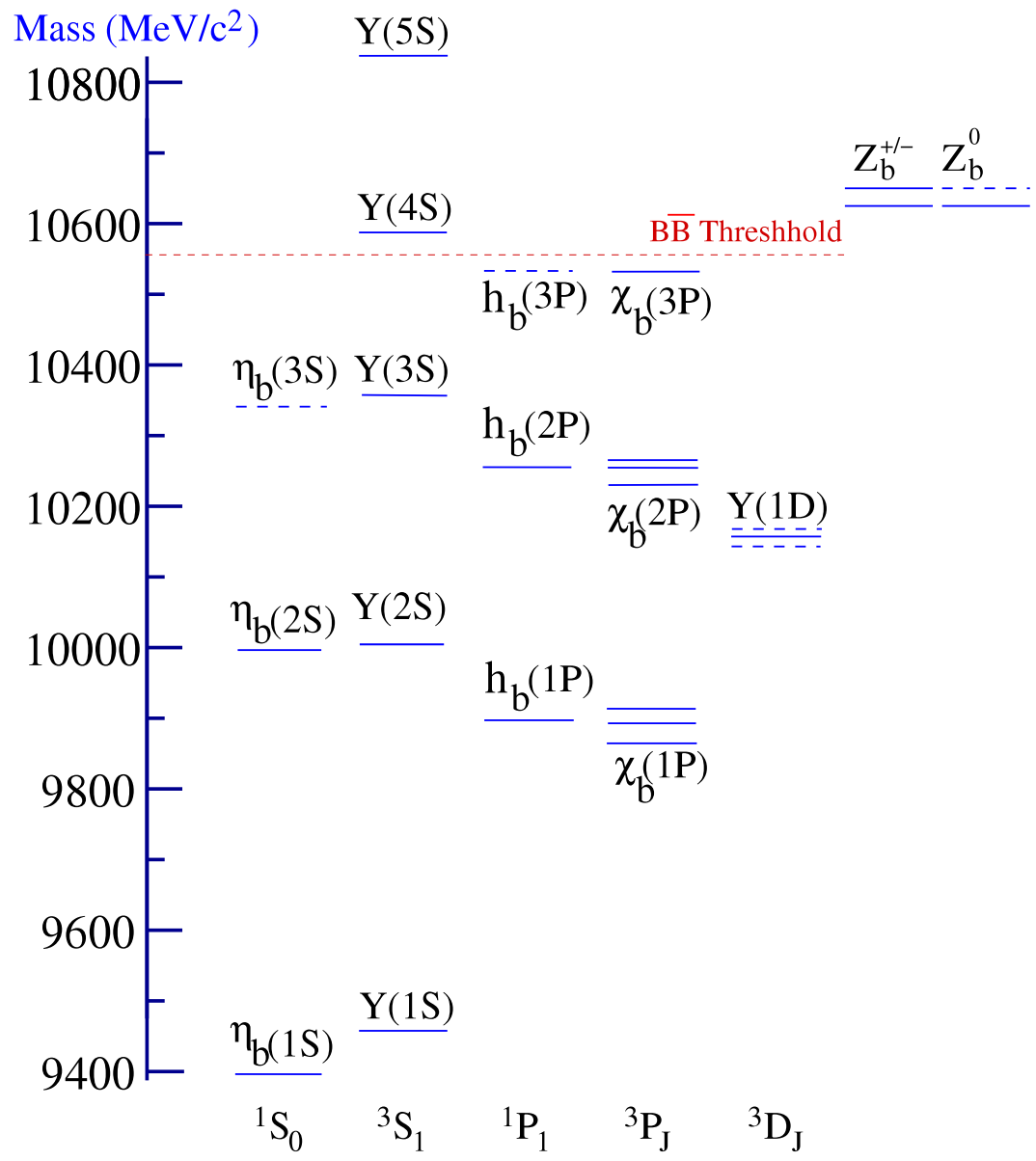

**Figure 18.4.1.** The bottomonium spectrum. Solid lines correspond to observed states while dashed lines indicate the location of predicted ones. See also the upper plot of Fig. 18.1.1, and the accompanying discussion.

Both collaborations pursued physics beyond the Standard Model using their bottomonium samples. The Belle Collaboration, followed by the *BABAR* Collaboration, both searched for invisible decays of the $\Upsilon(1S)$ meson; these searches, and the theoretical work that motivated them, are discussed in Section 18.4.7.2. The *BABAR* Collaboration also searched for evidence of a low-mass Higgs boson (Section 18.4.7.1), as well as lepton-flavor-violating decays (Section 18.4.7.3) and violation of the universality of couplings to leptons (Section 18.4.7.4).

### 18.4.2 Common techniques

#### 18.4.2.1 Transition Recoil Method

The use of transition particles between two resonances to "tag" the presence of one of the resonances is a common technique in the bottomonium analyses. Here, we describe the method and its application.

Consider two resonances with masses $M_1$ and $M_2$, and a transition between them involving the emission of one or more transition particles, $M_1 \to (trans) + M_2$. If the transition particles can be reconstructed, and the mass of either the parent or daughter resonance is known, then the remaining mass, which may be said to "recoil" against the transition particles, can be determined from four-momentum conservation:

$$P(M_1) = P(\mathrm{tr}) + P(M_2) \qquad (18.4.1)$$

where $P(\text{tr})$ is the four-vector describing the entire transition-particle energy and momentum. The mass recoiling against the transition particles, defined here as "recoil mass" (though sometimes the term "missing mass" is also used for the same quantity) is then determined by

$$M^2_{\text{recoil}} \equiv (P(M_1) - P(\text{tr}))^2 = P(M_2)^2 = M_2^2 \quad (18.4.2)$$

where $M_{\text{recoil}}$ is defined as the "recoil mass." This expression is valid in any reference frame. It is usually convenient to then choose a reference frame that simplifies the computation while still giving access to the invariant quantity in question.

The most common use of this technique is to determine the mass of a daughter resonance using the four-momentum of the reconstructed transition particles and that of the parent resonance, in the rest frame of the parent resonance. In this case, the equation simplifies to:

$$M^2_{\text{recoil}} = M_1^2 + m^2_{\text{tr}} - 2M_1E_{\text{tr}} \quad (18.4.3)$$

where $E_{\text{tr}}$ is the total energy of the transition particles in the parent resonance rest frame and $m^2_{\text{tr}}$ is the total invariant mass-squared of the transition particles.

In many instances, the parent resonance is produced at rest in the center-of-mass (CM) frame of the collider, and when this is true, the above calculation holds exactly. When this is not true, the calculated recoil mass using the transition particles' four-momentum in the CM frame will be shifted due to the presence, for instance, of another intermediate resonance before the transition particles are produced. For instance, if the transition particles are produced in the second of two transitions, *e.g.* $M_1 \to (undetected) + M_2$ followed by $M_2 \to (trans) + M_3$, then Lorentz boosting into the CM frame - the frame of the original parent resonance - will yield a shifted recoil mass.

Most often, sequential transitions in which a shifted recoil mass is observed are transitions in which each of states 1, 2 and 3 is a bottomonium state, and as such the emitted transition particles have small masses and momenta. In such cases transition particle energies and momenta are small compared to the masses $M_1, M_2$, and $M_3$. In these cases the amount of the shift is nearly equal to $M_1 - M_2$. When one calculates the recoil mass assuming the transition particles, which were emitted in the second transition, were emitted from the initial state, then the shifted recoil mass obtained is given by

$$M^2_{\text{shifted}} = (P(M_1) - P(\text{tr}))^2. \quad (18.4.4)$$

This is not equal to $M_3^2$, which is properly calculated as $(P(M_2) - P(\text{tr}))^2$, assuming both four-momenta are boosted into the CM frame. In our approximation, then, we find ($E_i$ and $p_i$ are the energies and the spatial components of the four momenta $P(M_i)$ and $P(\text{tr})$)

$$\begin{aligned} M_{\text{shifted}} &= \sqrt{(E_1 - E_{tr})^2 - (p_1 - p_{tr})^2)} \\ &= \sqrt{(M_1 - E_{tr})^2 - (p_{tr})^2)} \\ &= (M_1 - E_{tr})\sqrt{1 - (p_{tr})^2/(M_1 - E_{tr})^2} \\ &\approx (M_1 - E_{tr}) - p^2_{tr}/2(M_1 - E_{tr}) \quad (18.4.5) \end{aligned}$$

and

$$\begin{aligned} M_3 &= \sqrt{(E_2 - E_{tr})^2 - (p_2 - p_{tr})^2)} \\ &= (E_2 - E_{tr})\sqrt{1 - (p_2^2 - p_{tr})^2/(E_2 - E_{tr})^2} \\ &\approx (E_2 - E_{tr}) - (p_2^2 - p_{tr})^2/2(E_2 - E_{tr}). \quad (18.4.6) \end{aligned}$$

We then calculate the shift, $\Delta M = M_{\text{shifted}} - M_3$,

$$\begin{aligned} \Delta M = (M_1 - E_2) &+ (p_2 - p^2_{tr})/2(M_1 - E_{tr}) \\ &- p^2_{tr}/2(E_2 - E_{tr}) \\ &\approx (M_1 - M_2), \quad (18.4.7) \end{aligned}$$

since in the cases we are discussing, $E_2 \approx M_2$ and $p_2$, $E_2$, and $p_{tr}$ are all much smaller than either $M_1$ or $M_2$. Shifted recoil masses of this kind will be observed in the plot of the $M_{\text{recoil}}(\pi^+\pi^-)$ distribution in Figure 18.4.9.

### 18.4.3 $e^+e^-$ energy scans

As is discussed in Section 18.3, the discovery of non-baryonic charmonium states that behave in ways not predicted by two-quark-system models has yielded a renaissance of experimental and theoretical interplay in quarkonium. The observation of such exotic charmonium states suggests that a similar search for exotic bottomonium states is experimentally warranted. Such searches can be conducted in at least a couple of ways: energy scans of the accelerator across a range of center-of-mass energies, or searches at specific center-of-mass energies. The former will be discussed here, while a discussion of the latter approach can be found in Section 18.4.5.

An energy scan can be used to look for anomalous features in the ratio of the $b\bar{b}(\gamma)$ and $\mu^+\mu^-$ production cross-sections. Exotic charmonium states with quantum numbers $J^{PC} = 1^{--}$ have been observed and named the $Y(4260)$, $Y(4350)$, and $Y(4660)$. One can make naïve predictions for the bottomonium system by taking their masses and scaling them up by the mass difference between the $J/\psi$ and the $\Upsilon(1S)$, yielding predicted masses above the $\Upsilon(4S)$ mass and below 11.2 GeV/$c^2$.

During the final two weeks of data taking by the *BABAR* experiment, March 28 – April 7, 2008, the experiment collected 3.9 fb$^{-1}$ of data at more than 300 different center-of-mass energies above the $\Upsilon(4S)$ resonance, with typical spacing of 5 MeV. This is a factor of 30 more data than earlier scans (Besson et al., 1985; Lovelock et al., 1985; Aubert, 2009x). An additional 7.8 fb$^{-1}$ sample recorded at 10.54 GeV was used to study continuum background.

The quantity of interest is $R_b(s) \equiv \sigma_b(s)/\sigma^0_{\mu\mu}(s)$, the ratio of the total cross section for $e^+e^- \to b\bar{b}(\gamma)$ divided by the lowest order cross section for $e^+e^- \to \mu^+\mu^-$. Note that $\sigma_b$ includes Initial State Radiation (ISR) production of $\Upsilon$ states.

The experimental quantities used to calculate $R_b(s)$ are the number of hadronic events and muon pairs at center-of-mass energy $\sqrt{s}$ and in the continuum sample. Hadronic events are selected using criteria that preferentially select events with open bottom ($B$) mesons. Muon pairs are cleanly and efficiently selected using the tracking system and calorimeter only; the muon identification system is not required.

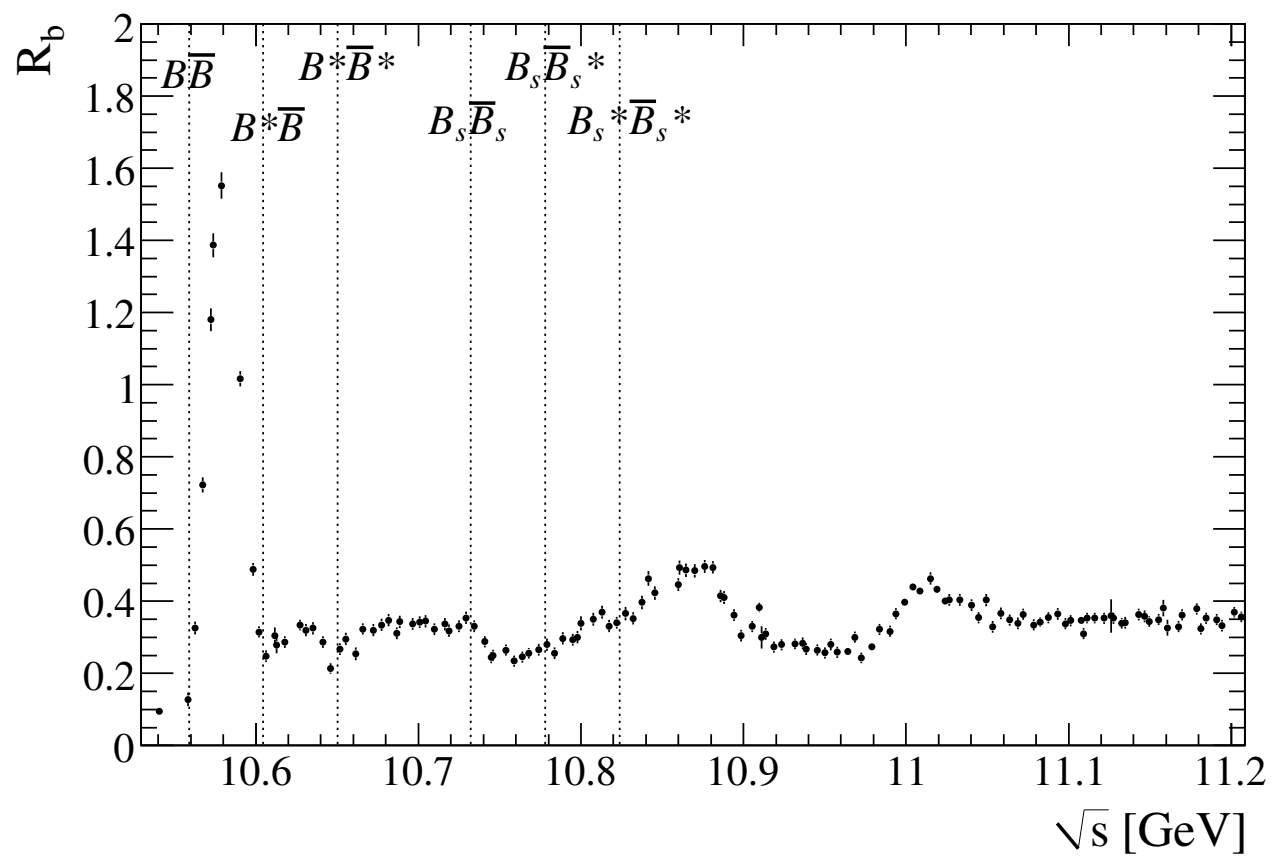

**Figure 18.4.2.** Measured $R_b$ as a function of $\sqrt{s}$ with the position of the opening thresholds of the $e^+e^- \to B^{(*)}_{(s)}\bar{B}^{(*)}_{(s)}$ processes indicated by dotted lines (Aubert, 2009x).

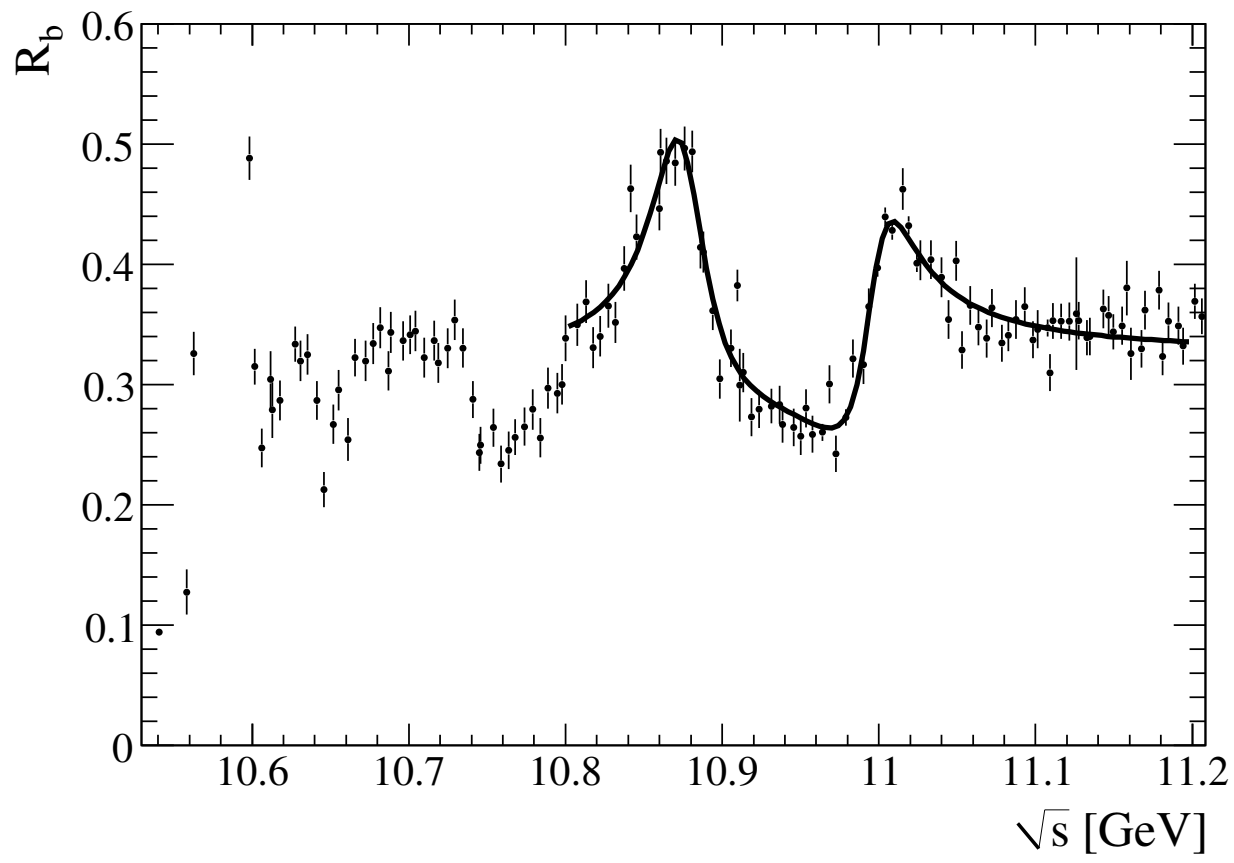

**Figure 18.4.3.** A zoom of Fig. 18.4.2 with the result of the fit superimposed (Aubert, 2009x). The errors on the data represent the statistical and uncorrelated systematic errors added in quadrature.

The 10.54 GeV data includes all event types that satisfy the hadronic selection, other than open bottom: continuum $e^+e^- \to q\bar{q}$ (the dominant background), ISR production of $\Upsilon$, and two photon events. The ISR contribution is calculated using simulated events. The two-photon component, 2% of the continuum sample, is estimated from the direction of the missing-momentum vector. The continuum component is obtained from the 10.54 GeV data by subtracting the other two components.

The efficiency for open bottom events to satisfy the criteria is obtained from simulation. It is taken to be the average of all possible two-body final states. Half the spread is taken as a systematic error.

The resulting values of $R_b(s)$ are shown in Fig. 18.4.2. Note that radiative corrections have not been applied. Not shown in the figure are correlated systematic errors totaling 2.6%, with equal contributions from hadronic event and muon pair efficiencies and $\mu^+\mu^-$ radiative corrections.

The region 10.80–11.20 GeV is fit with a model containing two interfering relativistic Breit Wigner resonances representing the $\Upsilon(10860)$ and the $\Upsilon(11020)$, a flat interfering component, plus an addition flat component representing $b\bar{b}$ continuum not interfering with the resonances (Fig. 18.4.3). The resulting mass and widths are $10.876 \pm 0.002$ GeV/$c^2$ and $43\pm4$ MeV for the $\Upsilon(10860)$ and $10.996\pm 0.002$ GeV/$c^2$ and $37 \pm 3$ MeV for the $\Upsilon(11020)$ (Aubert, 2009x). These widths are considerably narrower than the previous PDG values.

The results are sensitive to the details of the fit model. For example, using a threshold function instead of the flat non-resonant component gives a slightly different mass and a significantly larger width ($74 \pm 4$ MeV) for the $\Upsilon(10860)$. A proper coupled channel approach including the effects of the various thresholds would undoubtedly modify the results.

Finally it is noted that no evidence for exotic states has been found in the scans.

The Belle Collaboration also conducted a scan above the $\Upsilon(4S)$ resonance, in part in order to investigate possible causes for the anomalously large rates for the processes $\Upsilon(10860) \to \Upsilon(nS)\pi^+\pi^-$) with ($n = 1, 2, 3$) that they had observed (Chen, 2008b). These rates, if the $\Upsilon(10860)$ is interpreted as the fourth radial excitation of the $1^{--}$ $\Upsilon(1S)$ state, were up to two orders of magnitude larger than expectations.

Belle undertook a scan similar to that conducted by the *BABAR* Collaboration of the cross section for $e^+e^- \to \Upsilon(nS)\pi^+\pi^-$ ($n = 1, 2, 3$) in the vicinity of the known mass of the $\Upsilon(10860)$, taking data at center-of-mass energies between 10.83 and 11.02 GeV (Chen, 2010). In this study, they observed a peak in $\sigma(e^+e^- \to \Upsilon(nS)\pi^+\pi^-)$ ($n = 1, 2, 3$) at an energy of $(10888^{+2.7}_{-2.6} \pm 1.2)$ MeV with a width of $(30.7^{+8.3}_{-7.0} \pm 3.1)$ MeV. The measured cross section with the result of the fit using the Breit Wigner function for the signal is shown in Figure 18.4.4 (top).

The important thing to note is that this peak differs substantially in mass from the observed maximum in the overall hadronic cross section (shown in the lower pannel of Fig. 18.4.4) , and led to the suggestion that the peak in the $\Upsilon(nS)\pi^+\pi^-$) ($n = 1, 2, 3$) cross section may not, in fact, be due to the $\Upsilon(10860)$ but rather some exotic state (Liu and Ding, 2012).

## 18.4.4 Spectroscopy

### 18.4.4.1 Introduction to Bottomonium Spectroscopy

As discussed in Section 18.4.1, the *BABAR* and Belle Collaborations advanced significantly the experimental measurements of states within the predicted bottomonium

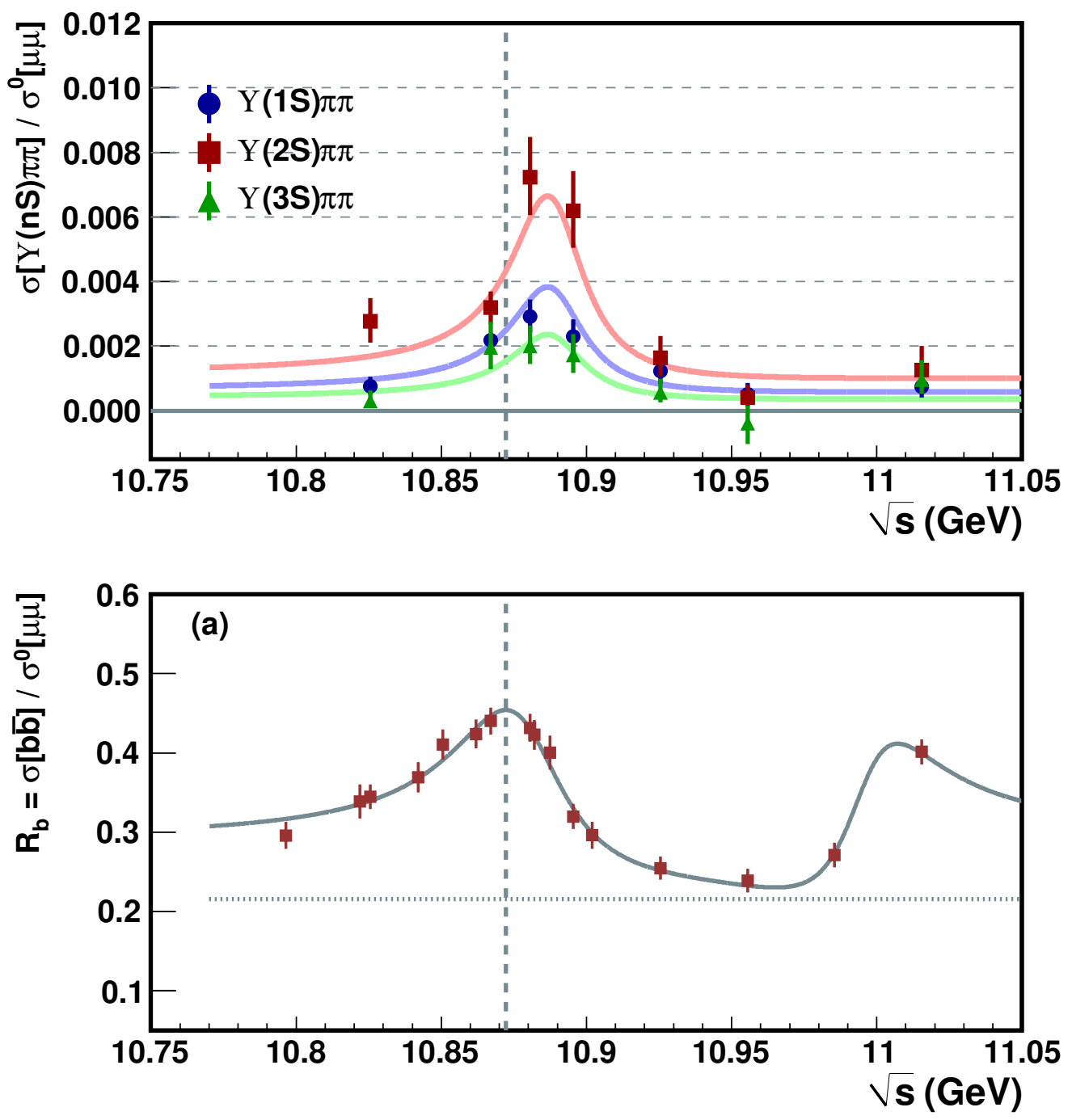

**Figure 18.4.4.** Top: The energy dependence of the cross section for $e^+e^- \to \Upsilon(nS)\pi^+\pi^-$ $(n = 1, 2, 3)$ normalized to the leading order $e^+e^- \to \mu^+\mu^-$ cross section (Chen, 2010). The dashed line shows the energy at which the hadronic cross section is maximal. Bottom: $R_b$ as a function of the energy.

spectrum. At the time that the first $\Upsilon(3S)$ data samples were collected by the Belle and then the *BABAR* Collaborations, predictions of the mass splitting between the lowest-mass $S$-wave pseudoscalar spin-singlet, $\eta_b(1S)$, and the lowest-mass $S$-wave vector spin-triplet, $\Upsilon(1S)$, ranged from 36 MeV/$c^2$ to 100 MeV/$c^2$ (Godfrey and Rosner, 2001). A measurement of this splitting was clearly critical in establishing which theoretical frameworks more accurately predicted the properties of the bottomonium spectrum and to help inform further developments in those frameworks; a discussion on the theoretical framework of quarkonium physics can be found in Section 18.1. For instance, a measurement of the hyperfine mass splitting would further our understanding of non-relativistic QCD bound states and shed light on the contribution of spin-spin interactions in models of the bottomonium system (Burch and Ehmann, 2007; Gray et al., 2005)

This line of argument also extended to the $P$-wave spin-singlet states, the $h_b(nP)$, and their properties as predicted in various theoretical frameworks. Prior to their discovery, it was recognized that measuring the hyperfine splitting for the $P$-wave states was similarly important for assessing the role of spin-spin interactions in potential models for heavy quarks. Treating the bottomonium system non-relativistically, the splitting can be determined from the square of the wave function for the system at the origin. This is expected to be a non-zero quantity for states with $L = 0$; for the $P$-wave states $(L = 1)$, the splitting between the $h_b(1P)$ and the spin-averaged triplet state $\langle\chi_{bJ}(1P)\rangle$ is expected to be approximately zero; therefore, one can approximate the mass of the $h_b(1P)$ as the spin-weighted center-of-gravity of the $\chi_{bJ}(1P)$ system: $M(h_b(1P)) = (9899.87 \pm 0.27)$ MeV/$c^2$ (Beringer et al., 2012). If one takes into account higher-order corrections to this approximation, one would expect to observe deviations from this prediction; however, measured deviations from the prediction that correspond to a hyperfine splitting larger than a few MeV/$c^2$ would indicate a vector component to the confinement potential (Rosner et al., 2005).

In addition to measuring the masses of these states, measuring the branching fractions for the corresponding transitions to and from these states is important for testing theoretical frameworks of heavy quarkonium. The branching fraction for the isospin-violating transition $\Upsilon(3S) \to \pi^0 h_b(1P)$ was expected to be about 0.1% (Godfrey, 2005a; Voloshin, 1986). The branching fraction for the transition $\Upsilon(3S) \to \pi^+\pi^- h_b(1P)$ was expected to range between $\sim 10^{-4}$ (Godfrey, 2005a) and $\sim 10^{-3}$ (Kuang, Tuan, and Yan, 1988; Kuang and Yan, 1981, 1990; Tuan, 1992). The branching fraction for the favored E1 transition $h_b(1P) \to \gamma\eta_b(1S)$ was expected to be large, 41.4% (Godfrey and Rosner, 2002).

#### 18.4.4.2 Observation of the $\eta_b(1S)$ and $\eta_b(2S)$

In the thirty years following the first discovery of the $\Upsilon(nS)$ bottomonium states (Herb et al., 1977), no evidence for the spin-singlet $\eta_b(nS)$ states had been found. The previous best limits for the decays $\Upsilon(3,2S) \to \gamma\eta_b$ were set by the CLEO experiment (Artuso et al., 2005b). *BABAR* first observed the $\eta_b$ in 2008 via the $\Upsilon(3S) \to \gamma\eta_b$ decay channel (Aubert, 2008ak). The discovery was confirmed in 2009 in the *BABAR* $\Upsilon(2S)$ data in decays of $\Upsilon(2S) \to \gamma\eta_b$ (Aubert, 2009l). The CLEO experiment subsequently verified the discovery in a re-analysis of its own $\Upsilon(3S)$ data sample (Bonvicini et al., 2010).

These analyses performed fits to the inclusive photon CM energy $(E^*_\gamma)$ spectrum, searching above the smooth, non-peaking background for evidence of a monochromatic photon associated with a radiative transition to the $\eta_b$. Two other peaking components were expected in the energy region close to this signal: one from photons from $\Upsilon(1S)$ production in ISR $(e^+e^- \to \gamma_{ISR}\Upsilon(1S))$, and a merged triplet from the decays $\Upsilon(nS) \to \gamma\chi_{bJ}(mP)$, $\chi_{bJ}(mP) \to \gamma\Upsilon(1S)$, where $m = n - 1$. The photons that result from the decay $\chi_{bJ}(mP) \to \gamma\Upsilon(1S)$ have energies in the range of the searches and thus serve to contribute background photon energy peaks.

The data samples used in the *BABAR* analyses included 28 (14) fb$^{-1}$ collected the $\Upsilon(3S)$ $(\Upsilon(2S))$ resonance, with approximately 9 (7)% of this data (refered to here as the "test sample") used for preliminary studies and later discarded. A total number of $(109 \pm 1) \times 10^6$ $\Upsilon(3S)$ events and $(91.6 \pm 0.9) \times 10^6$ $\Upsilon(2S)$ events were used in the final analysis. Additionally, "off-resonance" samples of 43.9 (2.4) fb$^{-1}$ were taken approximately 40 (30) MeV below

**Table 18.4.1.** Summary of the optimized variables used in the $\eta_b$ analyses (Aubert, 2008ak, 2009l).

| Variable | $\Upsilon(3S)$ | $\Upsilon(2S)$ |
|---|---|---|
| $\gamma_{LAT}$ | $< 0.55$ | |
| $\cos(\theta_{\gamma,LAB})$ | $-0.762 < \cos(\theta_{\gamma,LAB}) < 0.890$ | |
| $N_{TRK}$ | $> 3$ | |
| $\|\cos\theta_T\|$ | $< 0.7$ | $< 0.8$ |
| $\|m_{\gamma\gamma 2} - m_{\pi^0}\|$ | $> 15\,\mathrm{MeV}$ | |
| $E_{\gamma 2}$ | $> 50\,\mathrm{MeV}$ | $> 40\,\mathrm{MeV}$ |
| Efficiency | 37% | 35.8% |

the $\Upsilon(4S)$ ($\Upsilon(3S)$) resonance energies for studies of ISR production.

Candidate photons were single electromagnetic calorimeter (EMC) bumps not matched to any track, with a minimum lab energy of 30 MeV and a lateral moment (Section 15.1.4) less than 0.8. The selection criteria for this analysis were optimized by maximizing the figure of merit $S/\sqrt{B}$, where $S$ represents the number of signal events from $\Upsilon(nS) \to \gamma\eta_b$ Monte Carlo (MC), and $B$ represents the number of background events taken from the test sample. For optimization purposes, the signal region was restricted to $850 < E^*_\gamma < 950$ ($500 < E^*_\gamma < 700$) MeV for $\Upsilon(3S)$ ($\Upsilon(2S)$). A summary of the optimized selection criteria is given in Table 18.4.1, with the individual variables described below.

To improve photon candidate quality, $\gamma_{LAT} < 0.55$ was required, where "LAT" refers to the lateral moment of the electromagnetic energy deposit. Furthermore, by requiring the photon angle in the lab frame to satisfy $-0.762 < \cos(\theta_{\gamma,LAB}) < 0.890$, only photon candidates with fully-contained electromagnetic showers detected within the barrel of the EMC were used in this analysis. To select hadronic decays of the $\eta_b$, the number of charged tracks ($N_{TRK}$) in the event was required to be greater than or equal to 4, and the ratio of the second to zeroth Fox-Wolfram moments to be less than 0.98. This is typical of the way in which high-multiplicity hadronic final states of $B$ meson decay are selected in the $B$ factories, c.f. Section 9. Background from continuum events was rejected by imposing requirements on $|\cos\theta_T|$, the cosine of the angle in the CM frame between the photon momentum and the thrust axis of the rest of the event. The optimal requirement was found to be $|\cos\theta_T| < 0.7$ (0.8). The dominant background to this analysis, $\pi^0 \to \gamma\gamma$ decays, was reduced by vetoing photon candidates that, when paired with another photon in the event with a lab energy ($E_{\gamma 2}$) greater than 50 (40) MeV, formed an invariant mass ($m_{\gamma\gamma 2}$) within 15 MeV/$c^2$ of the nominal $\pi^0$ mass (Beringer et al., 2012). The values for the thrust angle and $\pi^0$ veto were optimized simultaneously. The signal efficiency resulting from these selection criteria was 37% (35.8%). The efficiency and selection criteria values were independently verified using the signal yield of the nearby $\chi_{bJ} \to \gamma\Upsilon(1S)$ signal peaks as a cross-check.

To extract the $\eta_b$ signal, a binned maximum likelihood fit of the $E^*_\gamma$ spectrum was performed over the range $0.5 < E^*_\gamma < 1.1$ ($0.27 < E^*_\gamma < 0.80$) GeV for the $\Upsilon(3S)$ ($\Upsilon(2S)$) dataset. The fit contained four components: non-peaking background, $\chi_{bJ} \to \gamma\Upsilon(1S)$, $\gamma_{ISR}\Upsilon(1S)$, and the $\eta_b$ signal.

For the $\Upsilon(3S)$ analysis, the non-peaking background was parameterized with a smooth lineshape defined as $f(E^*_\gamma) = A(C + \exp[-\alpha E^*_\gamma - \beta E^{*2}_\gamma])$, where $A$, $C$, $\alpha$, and $\beta$ were empirically-determined variables. The probability density functions (*p.d.f.*s) for the $\chi_{bJ}(2P) \to \gamma\Upsilon(1S)$ transitions were parameterized using the Crystal Ball (CB) function (Gaiser, 1982), a Gaussian distribution with an extended, power-law tail on the low side. The relative rates and peak positions for these three decays were fixed to their PDG values (Beringer et al., 2012). The values of the CB parameters were determined from a fit to the background-subtracted data in $840 < E^*_\gamma < 960$ MeV, and are common to all three peaks. Based on MC-simulated events, the ISR *p.d.f.* was parameterized by a CB function. The $\Upsilon(1S)$ production yield from ISR was measured in the off-resonance $\Upsilon(4S)$ sample, confirmed in $\Upsilon(3S)$ off-resonance data, and extrapolated to fix the size of the contribution in the $\Upsilon(3S)$ on-resonance sample. The $\eta_b$ signal *p.d.f.* was a non-relativistic Breit-Wigner function convolved with a CB function to account for the experimental $E^*_\gamma$ resolution. The CB parameters were fixed from MC events generated with a width of zero, but a natural width of 10 MeV (within the range of theoretical predictions based on expectations for two-photon widths was assumed for the final fit to the data. In the fit, the free parameters were the $\eta_b$ peak position and yield, the total $\chi_{bJ}(2P)$ yield and peak position, and values of the non-peaking background *p.d.f.* parameters.

A similar approach was taken for the analysis of the $\Upsilon(2S)$ decay modes. In this case, the non-peaking background was parameterized using the function $g(E^*_\gamma) = D\exp\left(\Sigma^4_{i=1} c_i E^{*i}_\gamma\right)$, where $D$ and $c_i$ were determined in the fit. A more sophisticated parameterization was used for the $\chi_{bJ}(1P)$ transition peaks. A CB function was analytically convolved with a rectangular function with a width accounting for the Doppler broadening due to the motion of the $\chi_{bJ}(1P)$ relative to the CM frame. The values of the half-width of the rectangular functions were 6.6, 5.5, and 4.9 MeV for the $J = 0, 1, 2$ states, respectively. The CB tail parameters, taken as a free parameter in the final fit, were common to all three peaks, and relative peak positions were fixed to the nominal values (Beringer et al., 2012). The relative yields were fixed to values determined from a control sample of exclusive $\chi_{bJ}(1P) \to \gamma\Upsilon(1S)$, $\Upsilon(1S) \to \mu^+\mu^-$ decays. The ISR *p.d.f.* was determined from MC-simulated events, with the extrapolated yield from the $\Upsilon(4S)$ off-resonance data used only as a cross-check. The $\eta_b$ signal *p.d.f.* was parameterized in a fashion identical to $\Upsilon(3S)$ analysis. For the final fit, the free parameters were the $\eta_b$ peak position and yield, the ISR yield, the total $\chi_{bJ}(1P)$ yield, the $\chi_{b1,2}(1P)$ CB resolutions, the $\chi_{bJ}(1P)$ CB transition point value, an overall energy scale offset based on the $\chi_{bJ}(1P)$ and ISR peak positions, and the power law components for the non-peaking background. The fits to the $E^*_\gamma$ spectrum for both datasets are shown in Fig. 18.4.5 and 18.4.6.

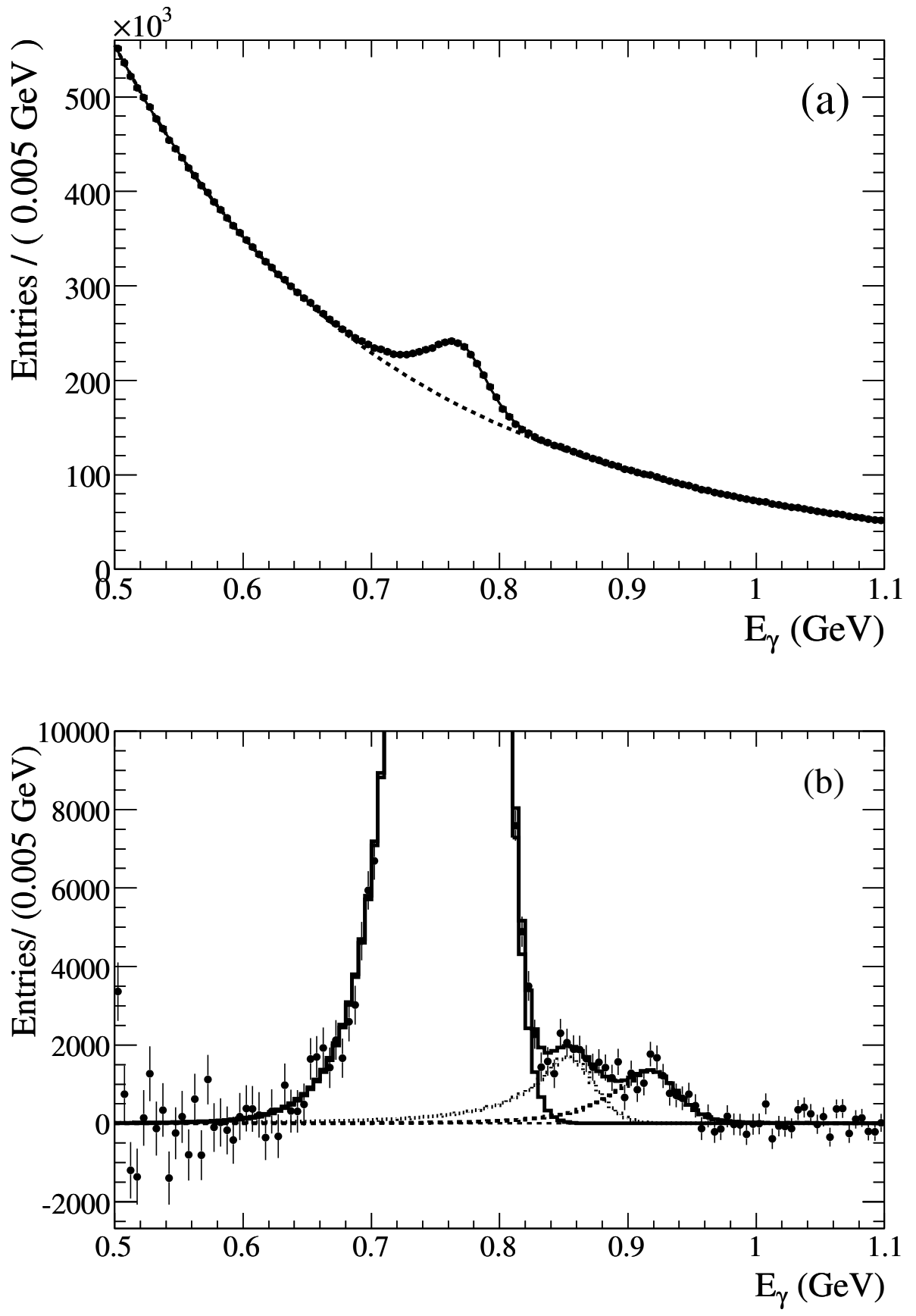

**Figure 18.4.5.** (a) Inclusive photon spectrum from $\Upsilon(3S)$ decays in the region $0.50 < E^*_\gamma < 1.1\,\text{GeV}$ (Aubert, 2008ak). The solid line indicates the total fit to the data; the dotted line indicates the non-peaking background component. (b) The same spectrum after the non-peaking background component has been subtracted, with the $\chi_{bJ}(2P)$, ISR, and $\eta_b$ signal components of the fit indicated from left to right on the plot.

In the $\Upsilon(3S)$ dataset, systematic uncertainties on the yield due to varying the assumed $\eta_b$ Breit-Wigner width, the extrapolated ISR yield, and varying the *p.d.f.* parameters were estimated to produce an 11% effect. By far the largest uncertainty (10%) was due to the $\eta_b$ width assumption. In the $\Upsilon(2S)$ dataset, the largest systematic uncertainties on the yield arise from varying the assumed $\eta_b$ width and the background shape ($\approx 17\%$ total).

The total systematic uncertainty due to assumptions made on the efficiency calculation was estimated to be 5.5%(6.7%) at the $\Upsilon(3S)$ ($\Upsilon(2S)$).

Combining the two *BABAR* results gave a ratio of branching fractions

$$\frac{\mathcal{B}(\Upsilon(2S) \to \gamma\eta_b)}{\mathcal{B}(\Upsilon(3S) \to \gamma\eta_b)} = 0.89^{+0.25+0.12}_{-0.23-0.16}. \qquad (18.4.8)$$

A new, unpredicted pathway to access the $\eta_b(1S)$ and $\eta_b(2S)$ states from $\Upsilon(5S)$ energies allowed Belle to improve substantially the $\eta_b(1S)$ mass measurement, perform the first measurement of its width, and discover the $\eta_b(2S)$.

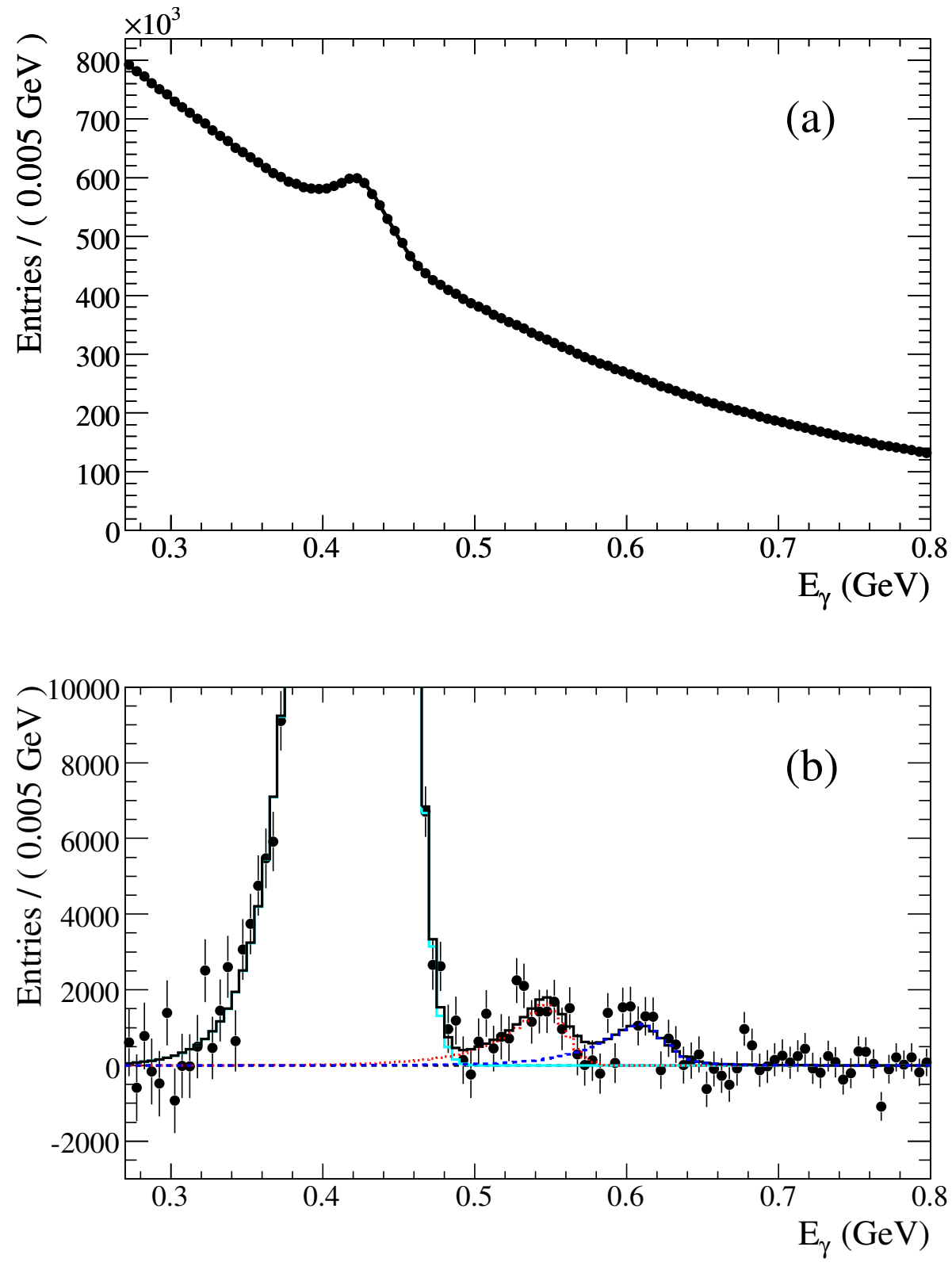

**Figure 18.4.6.** (a) Inclusive photon spectrum from $\Upsilon(2S)$ decays in the region $0.27 < E^*_\gamma < 0.80\,\text{GeV}$ (Aubert, 2009l). The solid line indicates the total fit to the data. (b) The same spectrum after the non-peaking background component has been subtracted, with the $\chi_{bJ}(1P)$ (cyan), ISR (red), and $\eta_b$ (blue) signal components from the fit indicated from left to right on the plot.

This progress followed the discovery of the $h_b(nP)$ and $Z_b$ states, which will be described in the following sections, and allowed Belle to access the $\eta_b(1S)$ and $\eta_b(2S)$ states via E1 transitions from the $h_b(nP)$ states. One key development in Belle's ability to study these transitions was the discovery that the $h_b(nP)$ production in $e^-e^-$ collisions at the $\Upsilon(5S)$ resonance is essentially saturated by charged pion cascades through the $Z_b$ states, By requiring that the single charged pion missing mass be consistent with the mass of the $Z_b$ states, significant improvement in the signal to background ratio for the inclusive $M_{\text{miss}}(\pi^+\pi^-) \equiv M_{recoil}(\pi^+\pi^-)$ spectrum was realizable. Using the resulting cleaner $h_b(nP)$ signals and adding the observation of a photon, Belle was able to report the first evidence for the $\eta_b(2S)$ produced in the $h_b(2P) \to \eta_b(2S)\gamma$ transition and the first observation of the $h_b(1P) \to \eta_b(1S)\gamma$ and $h_b(2P) \to \eta_b(1S)\gamma$ transitions (Mizuk, 2012). The $\eta_b(1S)$ samples obtained via this transition chain enabled to improve the mass measurement with respect to the one obtained via M1 transitions, and enabled the $\eta_b(1S)$ width to be measured for the first time.

In this analysis, Belle used a slightly larger data sample than was used in the two analyses which led to the discovery of the $h_b(nP)$ and of the $Z_b$ states, described in the next sections: 121.4 fb$^{-1}$ at the $\Upsilon(5S)$ resonance and 12.0 fb$^{-1}$ of energy-scan data collected nearby. A pair of charged pions, selected according to the same criteria as described by Adachi (2012a), and then subject to the requirement that the single charged pion missing mass satisfies the relation:

$$10.59\ \mathrm{GeV}/c^2 < M_{recoil}(\pi^\pm) < 10.67\ \mathrm{GeV}/c^2. \tag{18.4.9}$$

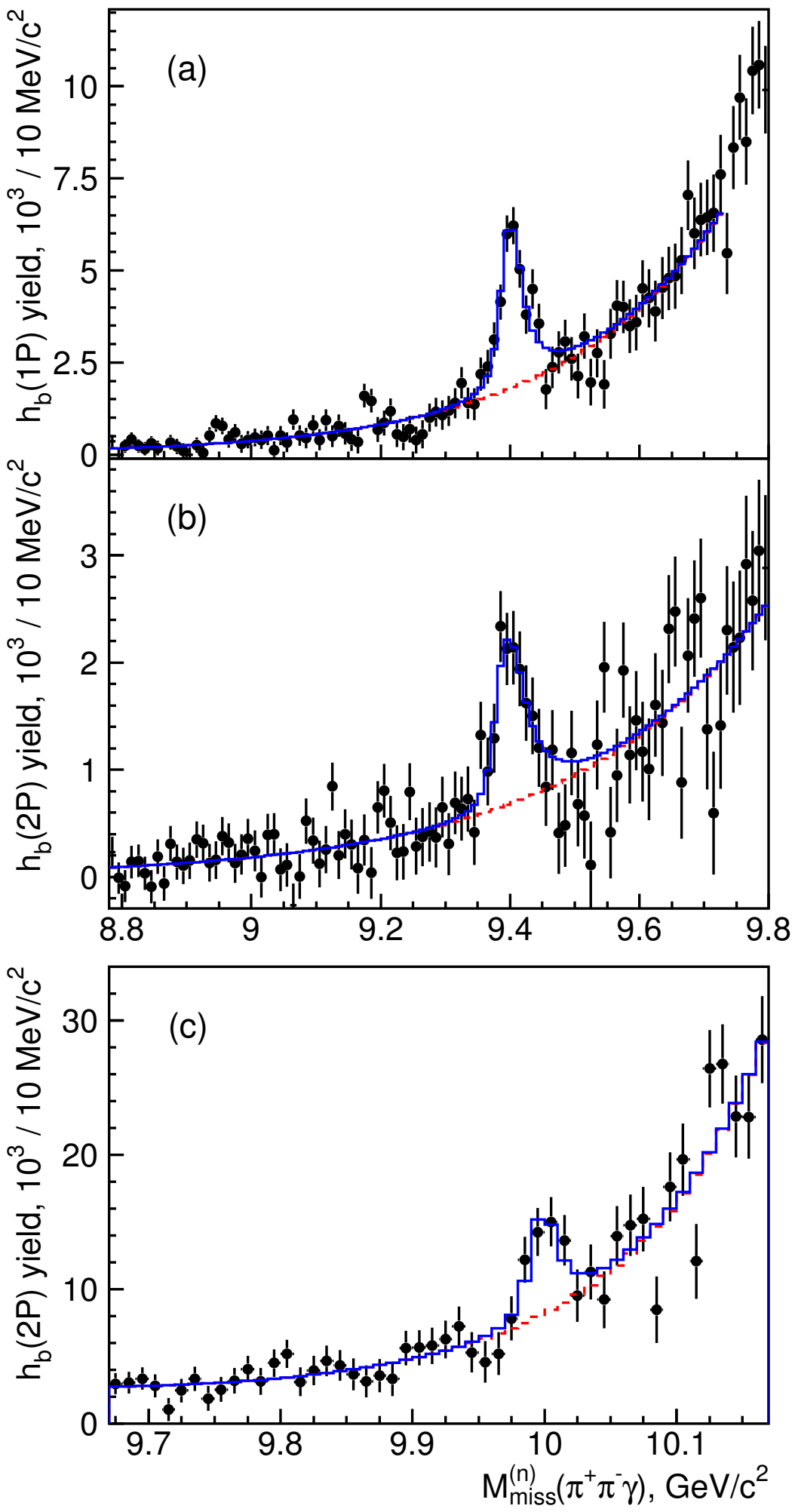

**Figure 18.4.7.** The $h_b(1P)$ yield *vs.* $M^{(1)}_{\rm miss}(\pi^+\pi^-\gamma)$ (a), and $h_b(2P)$ yield *vs.* $M^{(2)}_{\rm miss}(\pi^+\pi^-\gamma)$ in the $\eta_b(1S)$ region (b) and in the $\eta_b(2S)$ region (c). The solid (dashed) histogram presents the fit result (background component of the fit function). From (Mizuk, 2012).

This cut resulted in a reduction of the combinatorial background by a factor of 5 [1.6] for the $h_b(1P)$ [$h_b(2P)$] without any significant loss of the signal.

Clusters in the electromagnetic calorimeter unassociated with any charged track and which could not be paired with another photon in the event to form a $\pi^0$ were identified as photon candidates. The missing mass against $\pi^+\pi^-\gamma$ was then used to form the variable $M^{(n)}_{\rm miss}(\pi^+\pi^-\gamma) \equiv M_{recoil}(\pi^+\pi^-\gamma) - M_{recoil}(\pi^+\pi^-) + m_{h_b(nP)}$, and the yield of $h_b(nP)$ radiative decays to $\eta_b(mS)$ was obtained by determining the yield of $h_b(nP)$ as a function of $M^{(n)}_{\rm miss}(\pi^+\pi^-\gamma)$.

**Table 18.4.2.** Summary of the results on $h_b(1,2P) \to \eta_b$ (Mizuk, 2012).

| Transition | $h_b(1P) \to \eta_b(1S)$ | $h_b(2P) \to \eta_b(1S)$ |
|---|---|---|
| Yield$\times 10^{-3}$ | $23.5 \pm 2.0$ | $10.3 \pm 1.3$ |
| $\mathcal{B} \times 10^2$ | $49.2 \pm 5.7\,^{+5.6}_{-3.3}$ | $22.3 \pm 3.8\,^{+3.1}_{-3.3}$ |
| Significance | $15\sigma$ | $9\sigma$ |
| $m_{\eta_b(1S)}$(MeV/$c^2$) | $9402.4 \pm 1.5 \pm 1.8$ | (joint fit) |
| $\Delta m_{hf}$ (MeV/$c^2$) | $57.9 \pm 2.3$ | (joint fit) |
| $\Gamma(\eta_b(1S))$ (MeV) | $11^{+6}_{-4}$ | (joint fit) |

**Table 18.4.3.** Summary of the results on $h_b(2P) \to \eta_b(2S)$ (Mizuk, 2012).

| Transition | $h_b(2P) \to \eta_b(2S)$ |
|---|---|
| Yield$\times 10^{-3}$ | $25.8 \pm 4.9$ |
| $\mathcal{B} \times 10^2$ | $49.2 \pm 5.7\,^{+5.6}_{-3.3}$ |
| Significance | $4.2\sigma$ |
| $m_{\eta_b(2S)}$(MeV/$c^2$) | $9999.0 \pm 3.5\,^{+2.8}_{-1.9}$ |
| $\Delta m_{hf}$ (MeV/$c^2$) | $24.3^{+4.0}_{-4.5}$ |

Fits to the $M_{\rm miss}(\pi^+\pi^-)$ spectra for each $M^{(n)}_{\rm miss}(\pi^+\pi^-\gamma)$ bin were done using peak shapes for the transitions observed in the inclusive study, keeping the masses of the peaking components fixed at the values given in Table 18.4.5. The combinatorial background was fitted using a polynomial with parameters fixed to the values found in the overall fit, multiplied by a lower-order polynomial with floating coefficients. The resulting $h_b(1P)$ and $h_b(2P)$ yields as a function of $M^{(n)}_{\rm miss}(\pi^+\pi^-\gamma)$ are presented in Fig. 18.4.7. Clear peaks in $M^{(n)}_{\rm miss}(\pi^+\pi^-\gamma)$ at 9.4 GeV/$c^2$ and 10.0 GeV/$c^2$ were identified as signals for the $\eta_b(1S)$ and $\eta_b(2S)$, respectively.

The branching fraction for these radiative transitions was obtained by fitting the $h_b(nP)$ yield as a function of $M^{(n)}_{\rm miss}(\pi^+\pi^-\gamma)$ to the sum of the $\eta_b(nS)$ signal components described by the convolution of a non-relativistic Breit-Wigner function with the resolution function and a background parameterized as $e^{f(x)}$, where $f(x)$ is a first- [second-] order polynomial, in the $\eta_b(1S)$ [$\eta_b(2S)$] region. The two $M^{(n)}_{\rm miss}(\pi^+\pi^-\gamma)$ spectra [from the $h_b(1P)$ and $h_b(2P)$] with $\eta_b(1S)$ signals were fitted simultaneously. In this fit, the width of the $\eta_b(1S)$ Breit-Wigner function was a variable parameter; the width of the $\eta_b(2S)$ was fixed to a value obtained in perturbative calculations (Kwong, Mackenzie, Rosenfeld, and Rosner, 1988):

$$\Gamma_{\eta_b(2S)} = \Gamma_{\eta_b(1S)} \frac{\Gamma_{ee}^{\Upsilon(2S)}}{\Gamma_{ee}^{\Upsilon(1S)}} = (4.9^{+2.7}_{-1.9})\ \mathrm{MeV}, \tag{18.4.10}$$

where the uncertainty is due to the experimental uncertainty in $\Gamma_{\eta_b(1S)}$. If the $\eta_b(2S)$ width was allowed to float in the fit, a value of $\Gamma_{\eta_b(2S)} = (4^{+12}_{-20})$ MeV or $\Gamma_{\eta_b(2S)} < 24$ MeV at 90% C.L. using the Feldman-Cousins approach (Feldman and Cousins, 1998) was obtained. Results are given in Tables 18.4.2 and 18.4.3.

Systematic uncertainties in the $\eta_b(nS)$ parameters were evaluated due to the background fit function choice, fit range and binning choice, as well as signal shape and contributions from the experimental $h_b(nP)$ mass uncertainties and photon energy resolution. The various contributions in quadrature to estimate the total systematic uncertainty.

The efficiencies used to normalize the above radiative transition yields were determined using a combination of Monte Carlo and data-driven studies (Mizuk, 2012)

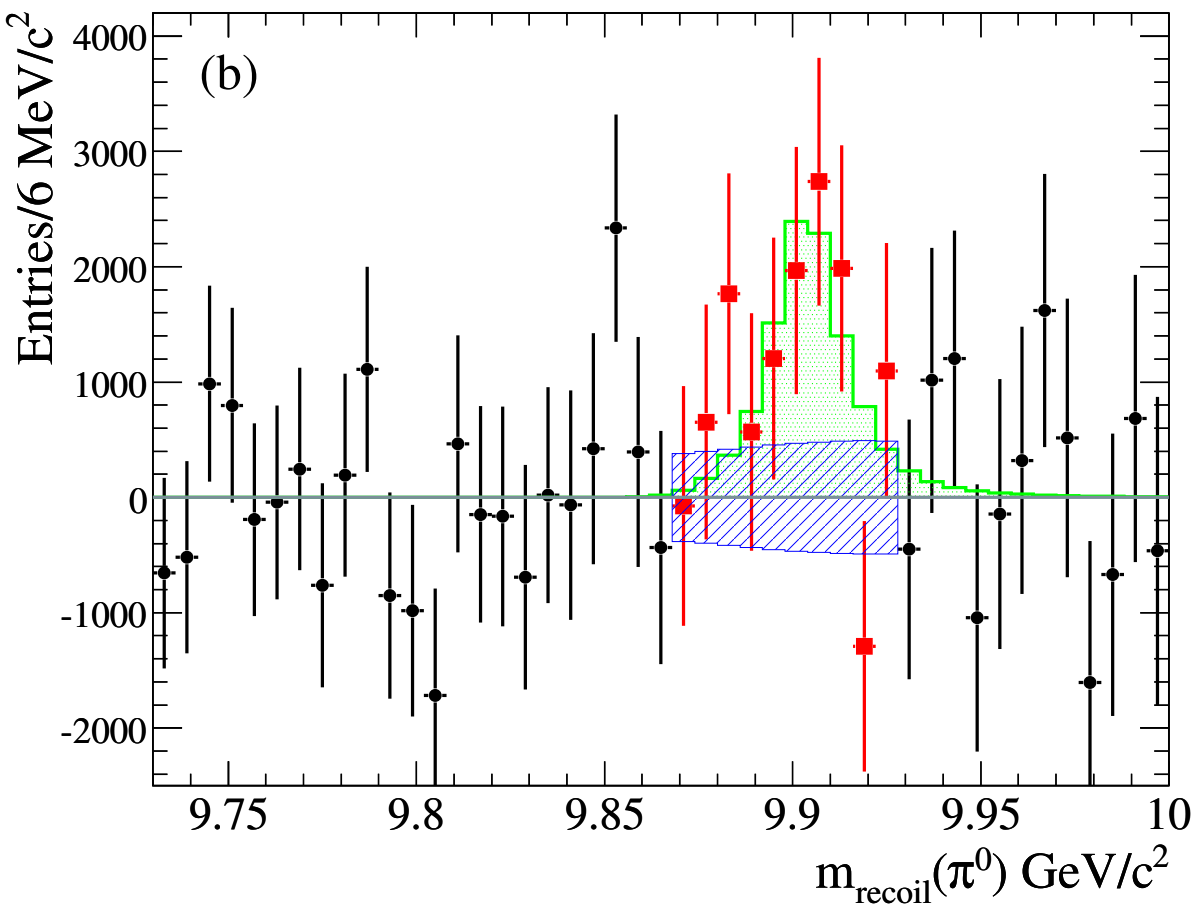

**Figure 18.4.8.** The $\pi^0$ recoil mass spectrum used by the *BABAR* Collaboration (Lees, 2011c) to search for $\Upsilon(3S) \to \pi^0 h_b(1P)$, shown after subtracting the smooth combinatoric background (black points). The green histogram represents the best-fit value from modeling the data with a signal component. The red-colored square points represent the $h_b$ signal region, and the blue-shaded area indicates the uncertainty in that region due to the background.

### 18.4.4.3 Observation of the $h_b(1P, 2P)$

Both the *BABAR* and Belle Collaborations pursued searches for the lowest $P$-wave spin-singlet state, the $h_b(1P)$ (Lees, 2011c) (Adachi, 2012a). These searches and their results are described below.

The *BABAR* Collaboration searched for this state using the experimentally favored transition, $\Upsilon(3S) \to \pi^0 h_b(1P)$, with subsequent decay $h_b(1P) \to \gamma\eta_b(1S)$. This search leveraged the measurement of the $\eta_b$ mass to constrain the expected energy of the photon. The invariant mass of the system recoiling against the $\pi^0$, $M_{\rm recoil}(\pi^0)$ was then used to search for evidence of a resonance consistent with the $h_b$. The distribution of $M_{\rm recoil}(\pi^0)$ was binned, and in each bin a fit was performed to the $\pi^0$ mass spectrum to determine the yield.This resulted in a $M_{\rm recoil}(\pi^0)$ spectrum due only to the recoil against real $\pi^0$ mesons. This distribution was then modeled using a combination of a smooth combinatoric background and a peaking distribution resulting from resonance like the $h_b$ (Fig. 18.4.8). The fit to the data determined that there was 3.3$\sigma$ evidence for a resonance recoiling against the $\pi^0$. The mass of this resonance was determined to be $(9902 \pm 4 \pm 2)\,{\rm MeV}/c^2$, which is consistent with the prediction of the $h_b$ mass from the spin-weighted average of the $\chi_{bJ}(1P)$ states. The product of the branching fractions $\mathcal{B}(\Upsilon(3S) \to \pi^0 h_b(1P)) \times \mathcal{B}(h_b(1P) \to \gamma\eta_b(1S))$ was determined to be $(4.3{\pm}1.1({\rm stat}){\pm}0.9({\rm syst}))\times 10^{-4}$. This measurement established the first evidence for the existence of the $h_b(1P)$ (Lees, 2011c).

Soon thereafter, the Belle Collaboration announced first observations of both $h_b(1P)$ and its radial excitation $h_b(2P)$ in the reaction $e^+e^- \to h_b(nP)\pi^+\pi^-$ using their $121.4\,{\rm fb}^{-1}$ data sample collected at energies near the $\Upsilon(5S)$ resonance (Adachi, 2012a). Among the observations that prompted this search in $\Upsilon(5S)$ data were two anomalous results in data taken above open flavor threshold in both charmonium and bottomonium. First was the observation by CLEO of the process $e^+e^- \to h_c\pi^+\pi^-$ at a rate comparable to that for $e^+e^- \to J/\psi\pi^+\pi^-$ in data taken above open charm threshold (Pedlar et al. (2011)). Such a large rate was unexpected because the production of $h_c$ requires a $c$-quark spin-flip, while production of $J/\psi$ does not. Secondly, Belle had previously observed anomalously high rates for $e^+e^- \to \Upsilon(nS)\pi^+\pi^-$ $(n = 1, 2, 3)$ at energies near the $\Upsilon(5S)$ mass (Chen, 2008b). These observations motivated Belle to undertake a search for the $h_b(nP)$ states in data taken above open-bottom threshold at and near the $\Upsilon(5S)$ resonance.

Belle undertook an inclusive search for the $h_b(nP)$ states using the distribution of the mass recoiling against $\pi^+\pi^-$, denoted $M_{\rm miss}(\pi^+\pi^-)$ in what follows. Rather than relying upon Monte Carlo simulations to determine the shape of the $M_{\rm miss}(\pi^+\pi^-)$ spectrum for signal events, Belle used the $\pi^+\pi^-$ transitions between $\Upsilon(nS)$ states, reconstructed using $\mu^+\mu^-\pi^+\pi^-$ combinations from well-reconstructed four-track events having positively identified $\mu^+\mu^-$ and $\pi^+\pi^-$ pairs. These $\mu^+\mu^-\pi^+\pi^-$ events revealed peaks corresponding to transitions to (and among) the three $\Upsilon(nS)$ states below open-flavor threshold, and masses obtained for each of the $\Upsilon(nS)$ states were consistent within $\pm 1\,{\rm MeV}/c^2$ with the world averages for those states.

The search for the $h_b(nP)$ states was performed inclusively on hadronic events, wherein only $\pi^+\pi^-$ candidate pairs were considered. The inclusive $M_{\rm miss}(\pi^+\pi^-)$ spectrum is dominated by combinatoric $\pi^+\pi^-$ pairs and also, in the region near $M_{\rm miss}(\pi^+\pi^-) = M(\Upsilon(3S))$ a step increase in the $\pi^+\pi^-$ spectrum which occurs because of the opening up of the threshold for $K^0_S$ production. This second background shape was obtained by fitting the $\pi^+\pi^-$ invariant mass corresponding to bins of $M_{\rm miss}(\pi^+\pi^-)$. The fit to the inclusive $M_{\rm miss}(\pi^+\pi^-)$ spectrum included a polynomial term for the combinatoric background, the $K^0_S$ shape as just described, and signal shapes for each of the peaks seen in the $\mu^+\mu^-\pi^+\pi^-$ data as well as those arising from $\pi^+\pi^-$ transitions to $h_b(nP)$ and $\Upsilon(1D)$. The $M_{\rm miss}(\pi^+\pi^-)$ spectrum, after subtraction of both the combinatoric and $K_S^0 \to \pi^+\pi^-$ contributions is shown with the fitted signal functions overlaid in Fig. 18.4.9. The yields and masses obtained in the fits are listed in Table 18.4.4.

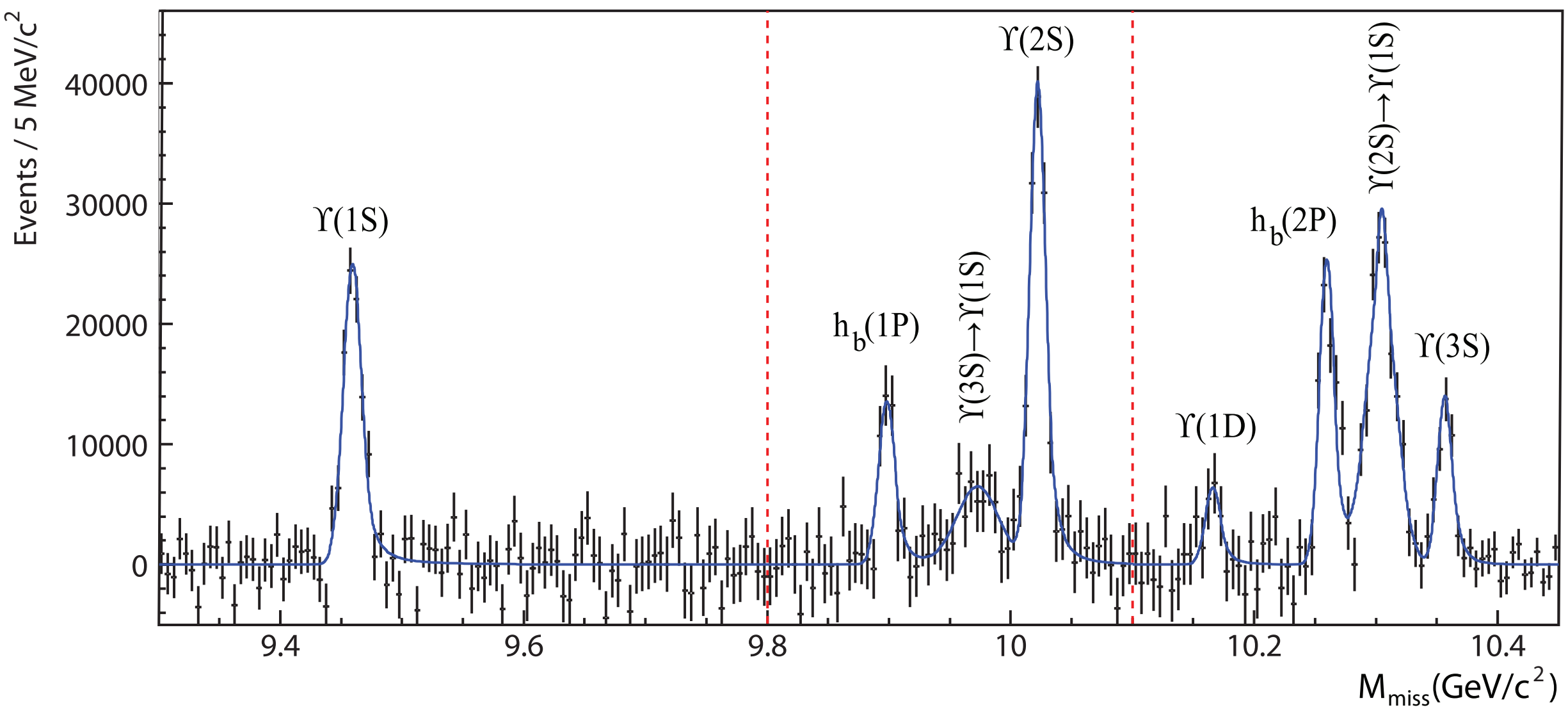

**Figure 18.4.9.** The spectrum of recoil mass $M_{\rm miss} \equiv M_{\rm recoil}(\pi^+\pi^-)$, used by the Belle Collaboration (Adachi, 2012a) to search for $\Upsilon(5S) \to \pi^+\pi^- h_b(nP)$, shown after subtracting the smooth combinatoric background (black points).

**Table 18.4.4.** Yield and mass obtained in the fit (Adachi, 2012a) to the inclusive $M_{\rm miss}(\pi^+\pi^-)$ distribution displayed in Fig. 18.4.9. The first uncertainty is statistical, while the second, if present, is the sum of all systematic uncertainties.

| | Yield, $10^3$ | Mass, MeV/$c^2$ |
|---|---|---|
| $\Upsilon(1S)$ | $105.0 \pm 5.8 \pm 3.0$ | $9459.4 \pm 0.5 \pm 1.0$ |
| $h_b(1P)$ | $50.0 \pm 7.8^{+4.5}_{-9.1}$ | $9898.2^{+1.1+1.0}_{-1.0-1.1}$ |
| $3S \to 1S$ | $55 \pm 19$ | 9973.01 |
| $\Upsilon(2S)$ | $143.8 \pm 8.7 \pm 6.8$ | $10022.2 \pm 0.4 \pm 1.0$ |
| $\Upsilon(1D)$ | $22.4 \pm 7.8$ | $10166.1 \pm 2.6$ |
| $h_b(2P)$ | $84.0 \pm 6.8^{+23.}_{-10.}$ | $10259.8 \pm 0.6^{+1.4}_{-1.0}$ |
| $2S \to 1S$ | $151.3 \pm 9.7^{+9.0}_{-20.}$ | $10304.6 \pm 0.6 \pm 1.0$ |
| $\Upsilon(3S)$ | $45.5 \pm 5.2 \pm 5.1$ | $10356.7 \pm 0.9 \pm 1.1$ |

Systematic uncertainties on the mass and yield of the $h_b(nP)$ states included contributions from the background fit polynomial order, range and bin size used in the fit, variation of selection criteria, and the signal shapes used. By far the most significant source of uncertainty on the yield arose from the choice of signal shape - a relative uncertainty of $^{+9.0\%}_{-18.2\%}$ and $^{+27\%}_{-12\%}$ on the $h_b(1P)$ and $h_b(2P)$ yields, respectively. The most significant source of systematic uncertainty on the $h_b(nP)$ masses ($\pm 1.0$ MeV) is estimated from the differences between the fitted masses of the known $\Upsilon(nS)$ states and the world average values. The signal for the $\Upsilon(1D)$ is marginal ($\sim 2.4\sigma$ statistical significance) and therefore systematic uncertainties on its yield and mass were not evaluated.

The identity of the observed peaks as the $h_b(nP)$ states is established as follows. The observed masses for the $h_b(nP)$ are more than $3\sigma$ from the $\chi_{b1}(nP)$ states, and the $J^{PC}$ for the $h_b(nP)$ candidates can be inferred from two observations. The observation by Belle of the $h_b(nP) \to \eta_b(1S)\gamma$ decays (Section 18.4.4.2) establishes the $C-$parity of the states as odd, while the $\chi_{b1}(nP)$ states have even $C-$parity. Similarly, angular analysis of the $\Upsilon(5S) \to h_b(1P)\pi^+\pi^-$ transition (Adachi, 2011) is consistent with the $h_b(1P)$ candidate having $J^P = 1^+$, as required for $h_b(1P)$ states.

The significances of the $h_b(1P)$ and $h_b(2P)$ signals, with systematic uncertainties accounted for, were, in this measurement, $5.5\sigma$ and $11.2\sigma$, respectively. The measured masses of $h_b(1P)$ and $h_b(2P)$, $M = (9898.2^{+1.1+1.0}_{-1.0-1.1})$ MeV/$c^2$ and $M = (10259.8 \pm 0.6^{+1.4}_{-1.0})$ MeV/$c^2$, respectively, correspond to hyperfine splittings of $\Delta M_{HF} = (+1.7 \pm 1.5)$ MeV/$c^2$ and $(+0.5^{+1.6}_{-1.2})$ MeV/$c^2$, respectively, where statistical and systematic uncertainties are combined in quadrature. As expected, then, the hyperfine splittings for both the 1P and 2P levels are consistent with zero.

The ratios $R \equiv \frac{\sigma(h_b(nP)\pi^+\pi^-)}{\sigma(\Upsilon(2S)\pi^+\pi^-)}$ were determined to be $R = 0.45 \pm 0.08^{+0.07}_{-0.12}$ for the $h_b(1P)$ and $R = 0.77 \pm 0.08^{+0.22}_{-0.17}$ for the $h_b(2P)$. Thus $\Upsilon(5S) \to h_b(nP)\pi^+\pi^-$ and $\Upsilon(5S) \to \Upsilon(2S)\pi^+\pi^-$ proceed at similar rates, despite the fact that the production of $h_b(nP)$ requires a spin-flip of a $b$ quark. The measured rates for $\Upsilon(5S) \to h_b(nP)\pi^+\pi^-$ are much larger than the upper limit for that of $\Upsilon(3S) \to h_b(nP)\pi^+\pi^-$ obtained by the *BABAR* Collaboration (Lees, 2011l). This is consistent with the similarly anomalously high rates for $\Upsilon(5S) \to \Upsilon(mS)\pi^+\pi^-$ with $m = 1, 2, 3$.

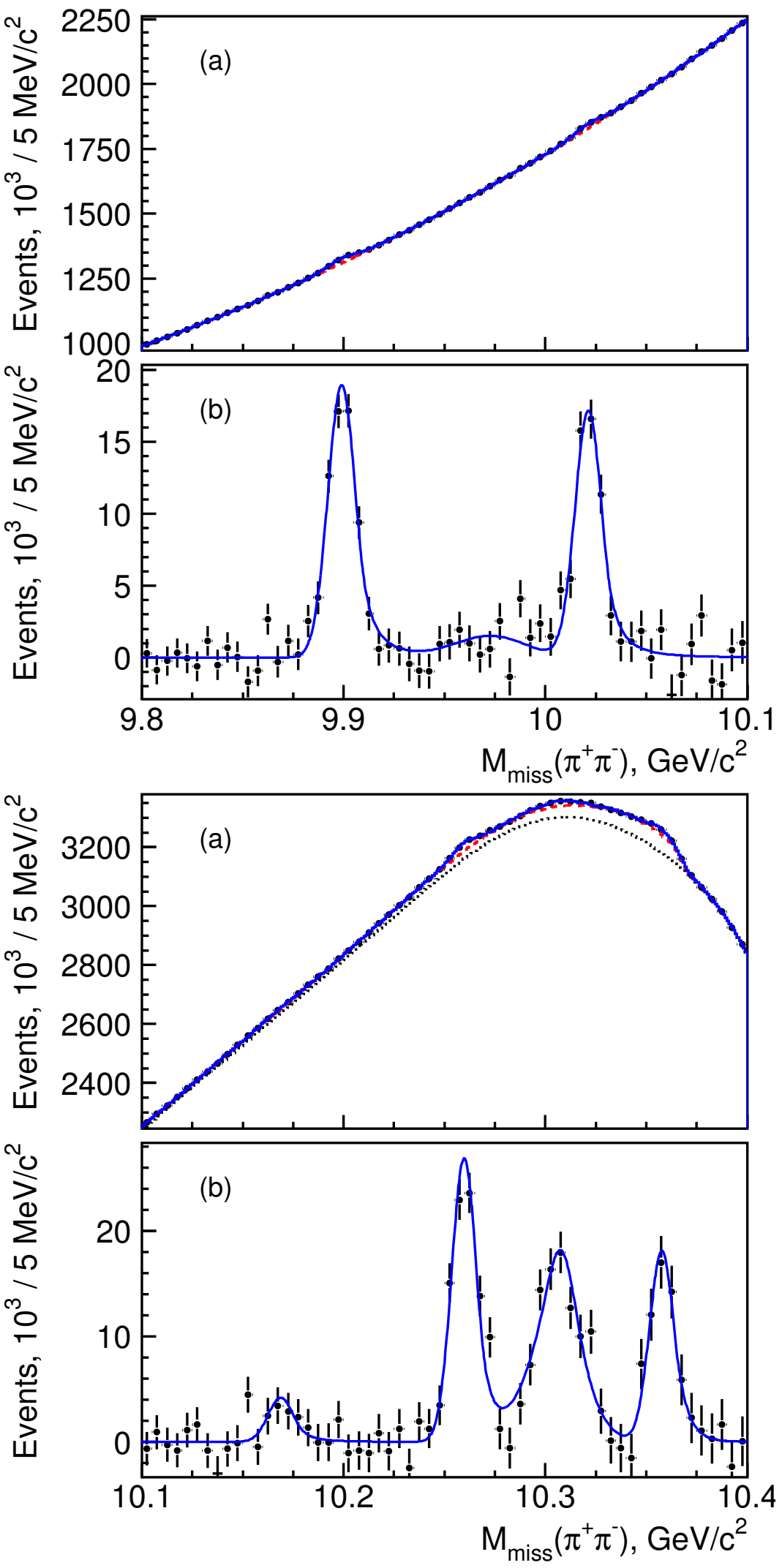

**Figure 18.4.10.** The $M_{\rm miss}(\pi^+\pi^-)$ spectrum with the combinatorial background and $K_S^0$ contribution subtracted (points with errors) and signal component of the fit function overlaid (smooth curve) in the $h_b(1P)$ (a) and $h_b(2P)$ (b) regions (Mizuk, 2012).

Subsequent studies on this anomaly lead to the discovery of the charged $Z_b$ states, which will be described later in this chapter, and which mediate 100% of the $\pi^+\pi^-$ transitions to the $h_b(nP)$ states. This observation allowed a further reduction of the combinatorial background by a factor of 5 [1.6] for the $h_b(1P)$ [$h_b(2P)$], by imposing the requirement:

$$10.59\,{\rm GeV}/c^2 < M_{recoil}(\pi^+\pi^-) < 10.67\,{\rm GeV}/c^2 \tag{18.4.11}$$

on the mass recoiling against the single pion.

The $M_{\rm miss}(\pi^+\pi^-)$ spectra in the $h_b(1P)$ and $h_b(2P)$ regions, defined as $9.8\,{\rm GeV}/c^2 - 10.1\,{\rm GeV}/c^2$ and $10.1\,{\rm GeV}/c^2 - 10.4\,{\rm GeV}/c^2$, are shown in Fig. 18.4.10. The fit procedure was essentially identical to that described before, using a fit function that is the sum of peaking components, a background shape due to the threshold for $K_S^0$ production, and a combinatorial background. The resulting masses and yields are listed in Table 18.4.5.

**Table 18.4.5.** The yield and mass of peaking components from the fits to the $M_{\rm miss}(\pi^+\pi^-)$ (Mizuk, 2012). The first quoted uncertainty is statistical (unless stated otherwise) and the second (if present) is systematic. Parameters without uncertainties were fixed in the fit.

| | $N,\ 10^3$ | Mass, MeV/$c^2$ |
|---|---|---|
| $\Upsilon(5S) \to h_b(1P)$ | $70.3 \pm 3.3^{+1.9}_{-0.7}$ | $9899.1 \pm 0.4 \pm 1.0$ |
| $\Upsilon(3S) \to \Upsilon(1S)$ | $13 \pm 7$ | $9973.0$ |
| $\Upsilon(5S) \to \Upsilon(2S)$ | $61.3 \pm 4.1$ | $10021.3 \pm 0.5$ |
| $\Upsilon(5S) \to \Upsilon(1D)$ | $14 \pm 7$ | $10169 \pm 3$ |
| $\Upsilon(5S) \to h_b(2P)$ | $89.5 \pm 6.1^{+0.0}_{-5.8}$ | $10259.8 \pm 0.5 \pm 1.1$ |
| $\Upsilon(2S) \to \Upsilon(1S)$ | $97 \pm 12$ | $10305.6 \pm 1.2$ |
| $\Upsilon(5S) \to \Upsilon(3S)$ | $58 \pm 8$ | $10357.7 \pm 1.0$ |

Systematic uncertainties in the $h_b(nP)$ parameters arise from the fitting procedure, including polynomial order, fit interval and signal shape. An additional $\pm 1$ MeV/$c^2$ uncertainty in the mass measurements is added, based on the observed deviations of the masses obtained for previously known vector bottomonium states, as in Adachi (2012a).

These updated mass measurements correspond to hyperfine splittings of $\Delta M_{\rm HF}(1P) = (+0.8 \pm 1.1)$ MeV/$c^2$ and $\Delta M_{\rm HF}(2P) = (+0.5 \pm 1.2)$ MeV/$c^2$, where statistical and systematic uncertainties in mass are added in quadrature.

### 18.4.4.4 $\Upsilon(1D)$

The existence of the bottomonium $D$-wave states has been established. The CLEO Collaboration reported observation of the $D$-wave triplet bottomonium state, $\Upsilon(1\,^3D_J)$, where $J = 1, 2, 3$ (Bonvicini et al., 2004). They report observation of a single member of the triplet, $\Upsilon(1\,^3D_2)$, using the decay $\Upsilon(1\,^3D_2) \to \gamma\gamma\Upsilon(1S)$. They identify the state in their analysis as corresponding to the $\Upsilon(1\,^3D_2)$ based on the fact that the mass and branching fractions in question correspond well to the theoretical preductions; however, they were not able to experimentally verify the assignment of quantum numbers $L$ and $J$.

The *BABAR* Collaboration reported in 2010 the observation of the $J = 2$ state of the $\Upsilon(1\,^3D_J)$ triplet using instead the hadronic decay transition $\Upsilon(1\,^3D_2) \to \pi^+\pi^-\Upsilon(1S)$, with subsequent leptonic decay of the $\Upsilon(1S)$ state, $\Upsilon(1S) \to \ell^+\ell^-$ (where $\ell = e, \mu$) (del Amo Sanchez, 2010k). The analysis was performed using a sample of $(121.8 \pm 1.2) \times 10^6$ $\Upsilon(3S)$ mesons. The parent $\Upsilon(3S)$ was then subsequently observed to decay to the $D$-wave state via a two-photon radiative transition, $\Upsilon(3S) \to \gamma\gamma\Upsilon(1\,^3D_J)$. An intermediate $\chi_{bJ'}(2P)$ state is produced between the radiation of the first and second photon, where $J' = 0, 1, 2$. The presence of these intermediate resonances implies a pattern of energies that one can use in the search to reject background and identify candidates for the signal processes in question.

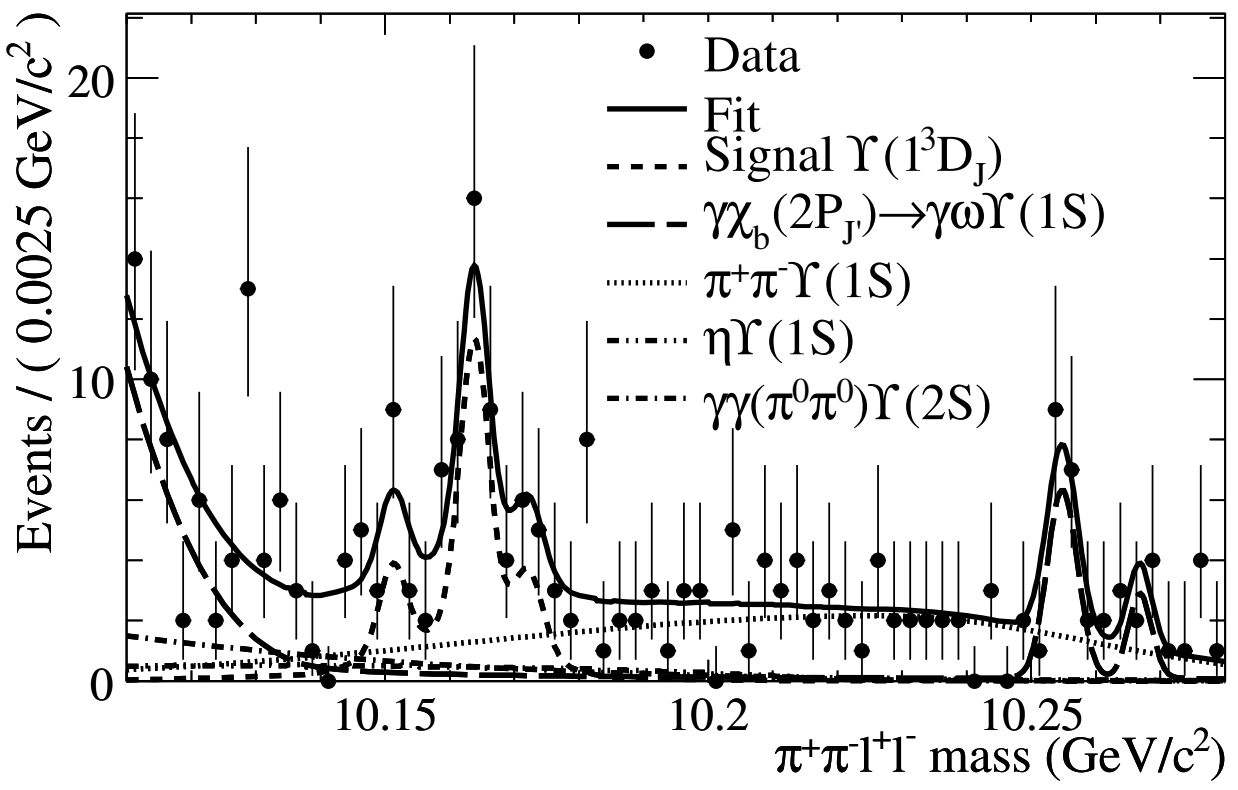

**Figure 18.4.11.** The mass spectrum of the $\Upsilon(1\,^3D_J)$ candidates, and the unbinned maximum likelihood fit to the spectrum. Background peaks from several sources are evident in the spectrum and modeled in the fit. A clear signal from the $D$-wave triplet is also evident (del Amo Sanchez, 2010k).

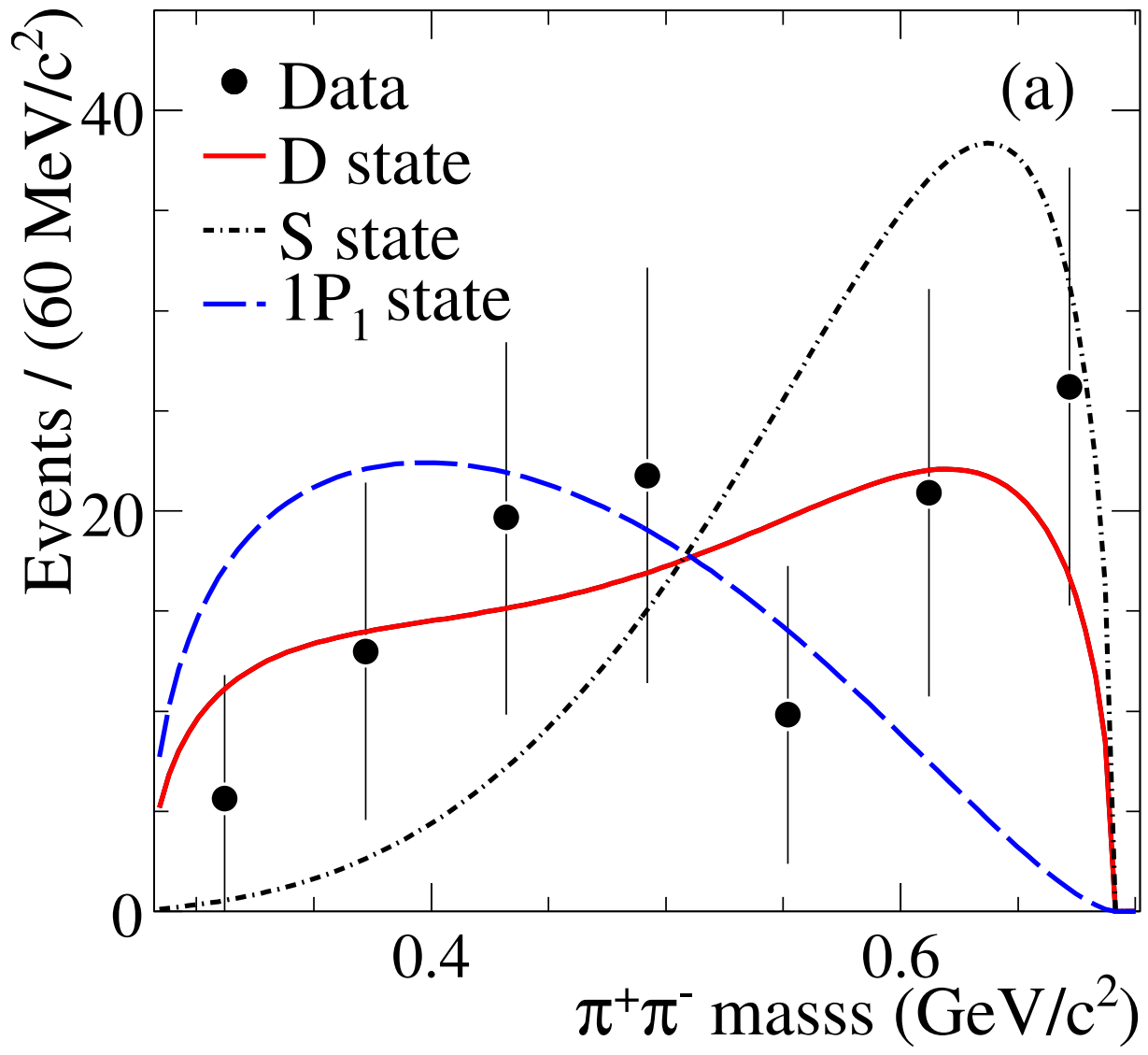

**Figure 18.4.12.** The dipion mass spectrum for the background-subtracted data in the region $10.155 < m_{\pi^+\pi^-\ell^+\ell^-} < 10.68\,\text{GeV}/c^2$. The shapes expected from $S$-wave, $D$-wave, and $1P_1$ states are shown (del Amo Sanchez, 2010k).

Events are required to contain exactly four good charged tracks. Two of the tracks must be identified as same-flavor, opposite-charge leptons. The pion candidates are taken to be the remaining two tracks and must fail an electron requirement. Radiative Bhabha events, a background to this event topology, are rejected by requiring that the electron satisfy a laboratory polar angle requirement, $\cos(\theta) < 0.8$.

The $\Upsilon(1S)$ candidate is selected by making flavor-dependent mass requirements on the lepton pairs: $-0.35 < m_{e^+e^-} - m_{\Upsilon(1S))} < 0.2$ GeV/$c^2$ or $|m_{\mu^+\mu^-} - m_{\Upsilon(1S))}| < 0.2\,\text{GeV}/c^2$. The mass of the dilepton pair is then constrained to the nominal $\Upsilon(1S)$ mass. The pions can be faked by a photon conversion in material. To reject this background, the opening angle between the pions must satisfy $\cos\theta_{\pi^+\pi^-} < 0.95$ if $m_{\pi^+\pi^-} < 0.050\,\text{GeV}/c^2$; for any dipion mass, the angle between the dipion system and either of the leptons must satisfy $\cos\theta_{\pi^+\pi^-,\ell^\pm} < 0.98$.

The events are also required to contain at least two photons, with minimum energy requirements (one with CM energy $> 0.070$ GeV and the other with CM energy $> 0.060$ GeV) consistent with the typical energies expected from the transition photons. Final-state radiation photons are rejected by requiring that $\cos\theta_{\gamma,\ell} < 0.98$. If there is more than one photon pair combination that satisfies these requirements, the pair whose energies minimize a $\chi^2$ constructed from the measured and expected photon energies is chosen as the best pair.

The $\Upsilon(1\,^3D_J)$ candidate is combined with the photon pair to form the $\Upsilon(3S)$ candidate, whose momentum must be $< 0.3\,\text{GeV}/c$. The $\Upsilon(3S)$ candidate mass is then constrained to the nominal mass value.

An extended unbinned maximum likelihood fit is then performed on the mass of the $\pi^+\pi^-\ell^+\ell^-$ system (Fig. 18.4.11). The fit includes components for several expected backgrounds, which were studied using MC simulation: $\Upsilon(3S)$ decays to $\gamma\chi_b(2P_{J'}) \to \gamma\omega\Upsilon(1S)$, $\pi^+\pi^-\Upsilon(1S)$, $\eta\Upsilon(1S)$, and $\gamma\gamma(\pi^0\pi^0)\Upsilon(2S)$. The models for these and signal are obtained from MC simulation. A clear excess exists in the region where $\Upsilon(1\,^3D_J)$, and is fitted with the signal model.

Large data control samples of dipion transitions to $\Upsilon(1S)$ and $\Upsilon(2S)$ final states, directly from the parent $\Upsilon(3S)$, are used to validate the *p.d.f.*s used in the fit to the spectrum. Where shifts are present between the *p.d.f.* parameters determined from MC or data, the shifts are applied as corrections. Only a small shift in the reconstructed mass of the $\Upsilon(2S)$ is observed.

The yield of $D$-wave triplet states is as follows (determined from the fit to the data): $10.6^{+5.7}_{-4.9}\,\Upsilon(1\,^3D_1)$, $33.9^{+8.2}_{-7.5}\,\Upsilon(1\,^3D_2)$, and $9.4^{+6.2}_{-5.2}\,\Upsilon(1\,^3D_1)$. Fit biases for the yield of signal events are determined by applying the data model to 2000 data-sized MC samples with events randomly drawn from the simulation subsamples. The biases are found to typically be at the level of 1-2 events in the signal region, and these biases are subtracted from the signal yields.

Multiplicative systematic uncertainties arise from various sources, with the largest of them being the photon reconstruction efficiency (3.0%) and particle identification (2.0%). Additive systematic uncertainties arise from the *p.d.f.* shapes, and total 1.5-2.0 events in the signal yields.

The statistical significance of the signal yield for the $J = 2$ $D$-wave triplet state is $6.5\sigma$ ($5.8\sigma$) including statistical (statistical and systematic) uncertainties.

The quantum numbers of the state are determined by studying the $\pi^+\pi^-$ mass distribution after subtracting the backgrounds in the region $10.155 < m_{\pi^+\pi^-\ell^+\ell^-} < 10.68\,\text{GeV}/c^2$. The dipion mass distribution is shown in Fig. 18.4.12, compared to the shapes expected from an $S$- wave, $D$-wave, and $1P_1$ state. The data are observed to be most consistent with the $D$-wave hypothesis.

### 18.4.5 Discovery of charged $Z_b$ states

In an effort to explain the large rate of dipion transitions to $\Upsilon(nS)$ and $h_b(nP)$ states in $e^+e^-$ annihilation at energies near $\Upsilon(5S)$, which suggest that exotic mechanisms contribute to $\Upsilon(5S)$ decays, Belle searched for evidence of resonant substructures in these decays (Bondar, 2012). For the analysis of $\pi^+\pi^-$ transitions to $\Upsilon(nS)$ states, the $\Upsilon(nS)$ states were observed in their $\mu^+\mu^-$ decays, which led to a relatively background-free sample for investigation. Transitions to $h_b(nP)$ states were investigated inclusively by examining only the $\pi^+\pi^-$ transition pairs.

$\Upsilon(nS)$ samples were obtained using four-track events, positively identified as a $\pi^+\pi^-$ and $\mu^+\mu^-$ pair, subject to the requirement that $|M_{\rm miss}(\pi^+\pi^-) - M(\mu^+\mu^-)| < 0.2\,{\rm GeV}/c^2$, where $M_{\rm miss}(\pi^+\pi^-)$ is the missing mass recoiling against the $\pi^+\pi^-$ system, and that $|M_{\rm miss}(\pi^+\pi^-) - m_{\Upsilon(nS)}| < 0.05\,{\rm GeV}/c^2$. Sideband regions for the study of background were defined as $0.05\,{\rm GeV}/c^2 < |M_{\rm miss}(\pi^+\pi^-) - m_{\Upsilon(nS)}| < 0.10\,{\rm GeV}/c^2$. The $h_b(nP)$ samples utilized events in which only the $\pi^+\pi^-$ system was selected.

Amplitude analysis of the three-body $\Upsilon(5S) \to \Upsilon(nS)\pi^+\pi^-$ employed unbinned maximum likelihood fits to the two-dimensional $M^2[\Upsilon(nS)\pi^+]$ vs. $M^2[\Upsilon(nS)\pi^-]$ Dalitz distributions. Signal events were found to make up more than 90% of the events in the signal region, and the efficiency-corrected distribution of background events (from $\Upsilon(nS)$ sidebands) was found to be featureless across the Dalitz plot. As an example, the Dalitz distributions of events in the $\Upsilon(2S)$ sidebands and signal regions are shown in Fig. 18.4.13 where, for ease of visualization, the square of the larger of the two $\Upsilon(nS)\pi$ masses is plotted vs. the square of the dipion invariant mass. One-dimensional invariant mass projections for events in each $\Upsilon(nS)$ signal region are shown in Fig. 18.4.14, where two peaks are observed in the $\Upsilon(nS)\pi$ system near 10.61 GeV/$c^2$ and 10.65 GeV/$c^2$, and are subsequently referred to as $Z_b(10610)$ and $Z_b(10650)$, respectively.

The parameterization of the $\Upsilon(5S) \to \Upsilon(nS)\pi^+\pi^-$ three-body decay amplitude includes terms corresponding to the $Z_b$ states as well as $f_0(980)$, $f_2(1270)$ and a non-resonant contribution:

$$M = A_{Z_1} + A_{Z_2} + A_{f_0} + A_{f_2} + A_{\rm nr}. \qquad (18.4.12)$$

In performing the fit, it was assumed that the dominant contributions come from amplitudes that preserve the orientation of the spin of the heavy quarkonium state and, thus, both pions in the cascade decay $\Upsilon(5S) \to Z_b\pi \to \Upsilon(nS)\pi^+\pi^-$ are emitted in an $S$-wave with respect to the heavy quarkonium system. Subsequent angular studies, as outlined in Bondar (2012), support this assumption.

The $Z_b(10610)$ and $Z_b(10650)$ peaks were parameterized with an $S$-wave Breit-Wigner function, and, to allow for the $\Upsilon(5S)$ decay to both $Z_b^+\pi^-$ and $Z_b^-\pi^+$, the amplitudes $A_{Z_1}$ and $A_{Z_2}$ were symmetrized with respect to $\pi^+$ and $\pi^-$ transposition:

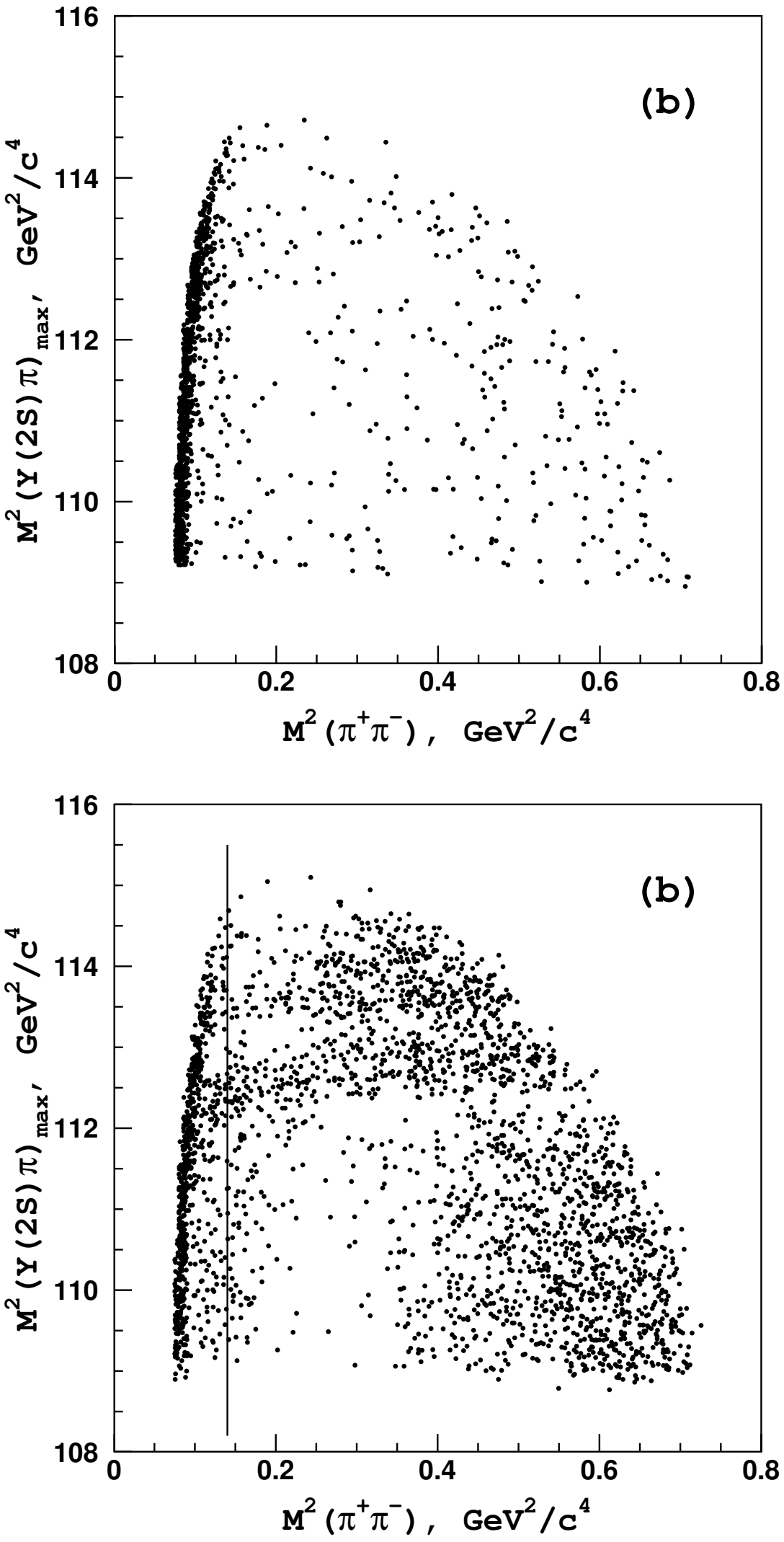

**Figure 18.4.13.** Dalitz plots for $\Upsilon(2S)\pi^+\pi^-$ events in the $\Upsilon(2S)$ sidebands (upper); $\Upsilon(2S)$ signal region (lower). Events to the left of the vertical line are excluded. From (Bondar, 2012).

$$A_{Z_k} = a_{Z_k} e^{i\delta_{Z_k}} (BW(s_1, M_k, \Gamma_k) + BW(s_2, M_k, \Gamma_k)), \qquad (18.4.13)$$

where $s_1 = M^2[\Upsilon(nS)\pi^+]$, $s_2 = M^2[\Upsilon(nS)\pi^-]$. Results of the fits to $\Upsilon(5S) \to \Upsilon(nS)\pi^+\pi^-$ signal events are shown in Fig. 18.4.14, and numerical results are summarized in Table 18.4.6, where the relative normalization is defined as the ratio of amplitudes $a_{Z_2}/a_{Z_1}$ and the relative phase as $\delta_{Z_2} - \delta_{Z_1}$. The systematic uncertainties on the parameters in Table 18.4.6 includes all evaluated sources - the greatest of which is related to the parameterization of the decay amplitude, and was studied by fitting the data with several modifications of the nominal model (Eq. 18.4.12).

For the study of $\Upsilon(5S) \to h_b(nP)\pi^+\pi^-$ resonant substructure (which is naturally a much more background-dominated study given the inclusive nature of the $\pi^+\pi^-$ detection) the yield of $Z_b$ states contributing to $h_b(nP)$ production is measured as a function of the $h_b(1P)\pi^\pm$ invariant mass by fitting the $M_{\rm miss}(\pi^+\pi^-)$ spectra in bins of $M_{\rm miss}(\pi^\mp)$, the mass recoiling against $\pi^\mp$ (which is equivalent to the $h_b(1P)\pi^\pm$ invariant mass). The yields of $\Upsilon(5S) \to h_b(nP)\pi^+\pi^-$ $(n = 1, 2)$ decays as a function of the $M_{\rm recoil}(\pi)$ (both signs of $\pi$ are included) are shown in Fig. 18.4.15. The distributions for the $h_b(nP)$ exhibit a clear two-peak structure without a significant non-resonant contribution. To fit the $M_{\rm recoil}(\pi)$ distributions for $Z_b$ yields, a combination of $P$-wave Breit-Wigner amplitudes is used:

$$|BW_1(s, M_1, \Gamma_1) + ae^{i\phi}BW_1(s, M_2, \Gamma_2) + be^{i\psi}|^2\,\frac{qp}{\sqrt{s}}. \tag{18.4.14}$$

where $\sqrt{s} \equiv M_{\rm recoil}(\pi)$; the variables $M_k$, $\Gamma_k$ $(k = 1, 2)$, $a$, $\phi$, $b$ and $\psi$ are free parameters; $\frac{qp}{\sqrt{s}}$ is a phase-space factor, where $p$ $(q)$ is the momentum of the pion originating from the $\Upsilon(5S)$ $(Z_b)$ decay measured in the rest frame of the corresponding mother particle. The $P$-wave Breit-Wigner amplitude is expressed as $BW_1(s, M, \Gamma) = \frac{\sqrt{M\,\Gamma}\,F\,(q/q_0)}{M^2 - s - iM\,\Gamma}$. Here $F$- is the $P$-wave Blatt-Weisskopf form factor $F = \sqrt{\frac{1+(q_0 R)^2}{1+(qR)^2}}$, $q_0$ is a daughter momentum calculated with pole mass of its mother, $R = 1.6$ GeV$^{-1}$. The function (Eq. 18.4.14) is convolved with the detector resolution function ($\sigma = 5.2$ MeV/$c^2$), integrated over the 10 MeV/$c^2$ histogram bin and corrected for the reconstruction efficiency. The fit results are shown as solid histograms in Fig. 18.4.15 and the numerical results are summarized in Table 18.4.6. The non-resonant contribution to $h_b(nP)$ production is consistent with zero [significance is $0.3\,\sigma$ both for the $h_b(1P)$ and $h_b(2P)$], while the default fit hypothesis is favored over the phase-space fit hypothesis at the $18\,\sigma$ [$6.7\,\sigma$] level for the $h_b(1P)$ [$h_b(2P)$].

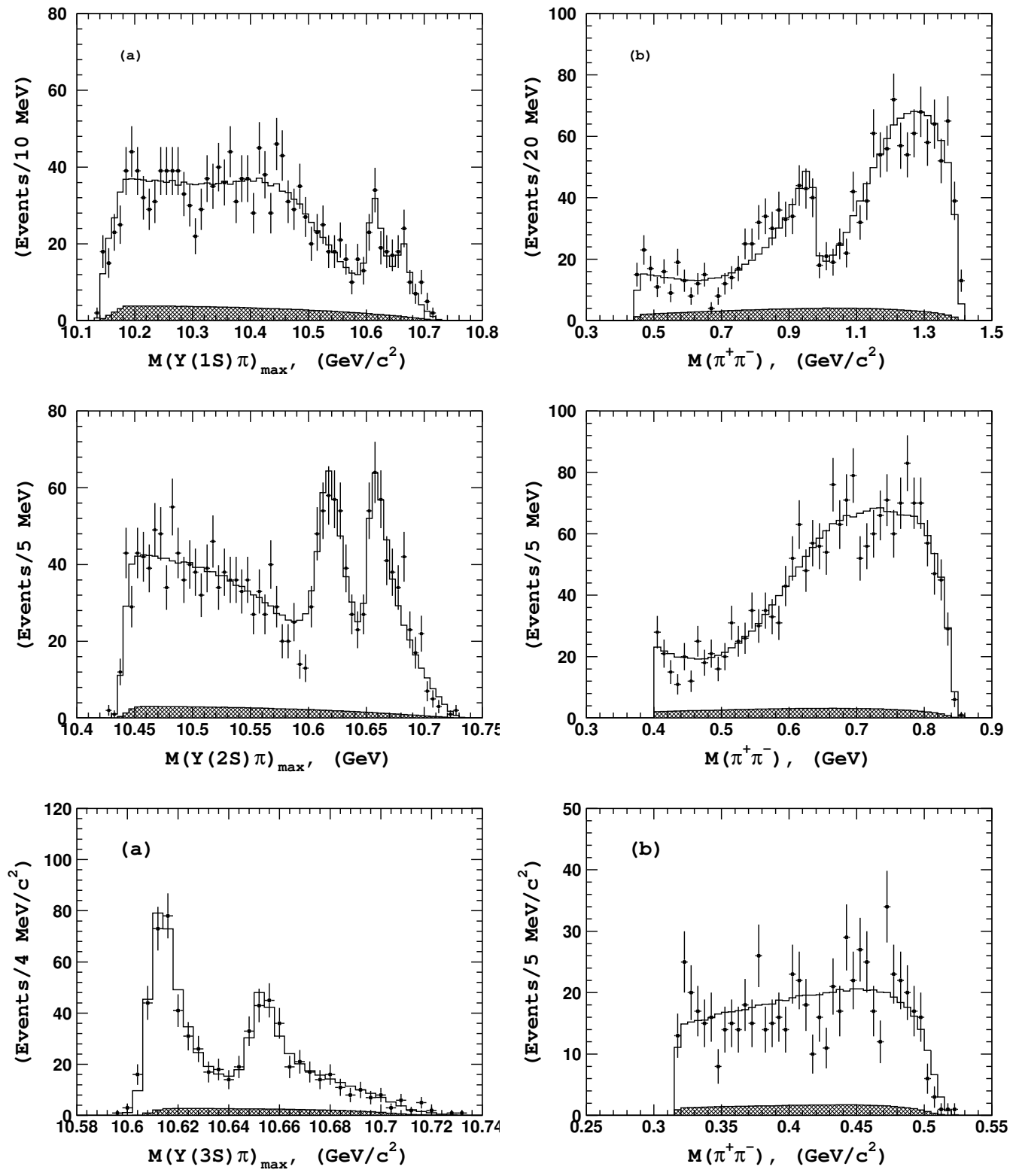

**Figure 18.4.14.** Comparison of fit results (open histogram) with experimental data (points with error bars) for events in the $\Upsilon(1S)$ (upper), $\Upsilon(2S)$ (middle), and $\Upsilon(3S)$ (lower) signal regions. The hatched histogram shows the background component. From (Bondar, 2012).

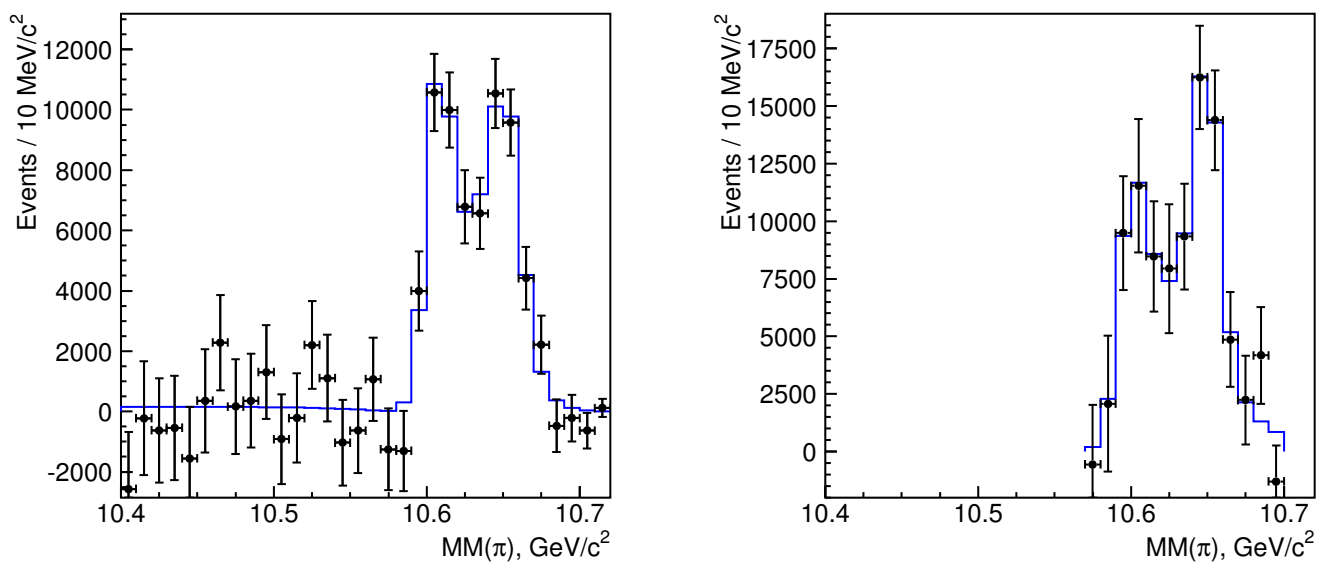

**Figure 18.4.15.** The (a) $h_b(1P)$ and (b) $h_b(2P)$ yields as a function of $M_{\rm recoil}(\pi)$ (points with error bars) and results of the fit (histogram). From (Bondar, 2012).

Systematic uncertainies were studied by evaluating contributions from the background function used in fits to the $M_{\rm miss}(\pi^+\pi^-)$ spectra, effects of finite bin size in the fits, model uncertainties, and data-MC comparisons. An additional 1 MeV/$c^2$ uncertainty in mass measurements was applied, as in the previous analysis, based on the difference between the observed $\Upsilon(nS)$ peak positions and their world averages (Adachi, 2012a). The total systematic uncertainty presented in Table 18.4.6 is the sum in quadrature of contributions from all sources. After inclusion of systematic uncertainties, the significance of the $Z_b(10610)$ and $Z_b(10650)$ including systematic uncertainties was $16.0\,\sigma$ [$5.6\,\sigma$] for the $h_b(1P)$ [$h_b(2P)$].

The two charged bottomonium-like resonances $Z_b(10610)$ and $Z_b(10650)$ are hence firmly established with signals in five different decay channels, $\Upsilon(nS)\pi^\pm$ $(n = 1, 2, 3)$ and $h_b(nP)\pi^\pm$ $(m = 1, 2)$. The weighted averages over all five channels give $M = 10607.2 \pm 2.0$ MeV/$c^2$, $\Gamma = 18.4 \pm 2.4$ MeV for the $Z_b(10610)$ and $M = 10652.2 \pm 1.5$ MeV/$c^2$, $\Gamma = 11.5 \pm 2.2$ MeV for the $Z_b(10650)$, where statistical and systematic errors are added in quadrature. The $Z_b(10610)$ production rate is similar to that of the $Z_b(10650)$ for each of the five decay channels. Their relative phase is consistent with zero for the final states with the $\Upsilon(nS)$ and consistent with 180 degrees for the final states with $h_b(nP)$. Production of the $Z_b$'s saturates the $\Upsilon(5S) \to h_b(nP)\pi^+\pi^-$ transitions and accounts for the high inclusive $h_b(mS)$ production rate reported in Adachi (2012a). Analyses of charged pion angular distributions (Bondar, 2012) favor the $J^P = 1^+$ spin-parity assignment for both the $Z_b(10610)$ and $Z_b(10650)$. Since the $\Upsilon(5S)$ has negative $G$-parity, the $Z_b$ states have positive $G$-parity due to the emission of the pion.

**Table 18.4.6.** Comparison of results on $Z_b(10610)$ and $Z_b(10650)$ parameters obtained from $\Upsilon(5S) \to \Upsilon(nS)\pi^+\pi^-$ $(n = 1, 2, 3)$ and $\Upsilon(5S) \to h_b(nP)\pi^+\pi^-$ $(m = 1, 2)$ analyses (Bondar, 2012).

| Final state | $\Upsilon(1S)\pi^+\pi^-$ | $\Upsilon(2S)\pi^+\pi^-$ | $\Upsilon(3S)\pi^+\pi^-$ | $h_b(1P)\pi^+\pi^-$ | $h_b(2P)\pi^+\pi^-$ |
|---|---|---|---|---|---|
| $M[Z_b(10610)]$, MeV/$c^2$ | $10611 \pm 4 \pm 3$ | $10609 \pm 2 \pm 3$ | $10608 \pm 2 \pm 3$ | $10605 \pm 2^{+3}_{-1}$ | $10599^{+6+5}_{-3-4}$ |
| $\Gamma[Z_b(10610)]$, MeV | $22.3 \pm 7.7^{+3.0}_{-4.0}$ | $24.2 \pm 3.1^{+2.0}_{-3.0}$ | $17.6 \pm 3.0 \pm 3.0$ | $11.4\,^{+4.5+2.1}_{-3.9-1.2}$ | $13\,^{+10+9}_{-8-7}$ |
| $M[Z_b(10650)]$, MeV/$c^2$ | $10657 \pm 6 \pm 3$ | $10651 \pm 2 \pm 3$ | $10652 \pm 1 \pm 2$ | $10654 \pm 3\,^{+1}_{-2}$ | $10651^{+2+3}_{-3-2}$ |
| $\Gamma[Z_b(10650)]$, MeV | $16.3 \pm 9.8^{+6.0}_{-2.0}$ | $13.3 \pm 3.3^{+4.0}_{-3.0}$ | $8.4 \pm 2.0 \pm 2.0$ | $20.9\,^{+5.4+2.1}_{-4.7-5.7}$ | $19 \pm 7\,^{+11}_{-7}$ |
| Rel. normalization | $0.57 \pm 0.21^{+0.19}_{-0.04}$ | $0.86 \pm 0.11^{+0.04}_{-0.10}$ | $0.96 \pm 0.14^{+0.08}_{-0.05}$ | $1.39 \pm 0.37^{+0.05}_{-0.15}$ | $1.6^{+0.6+0.4}_{-0.4-0.6}$ |
| Rel. phase, degrees | $58 \pm 43^{+4}_{-9}$ | $-13 \pm 13^{+17}_{-8}$ | $-9 \pm 19^{+11}_{-26}$ | $187^{+44+3}_{-57-12}$ | $181^{+65+74}_{-105-109}$ |

The minimal quark content of the $Z_b(10610)$ and $Z_b(10650)$ is a four-quark combination. The masses of these new states are a few MeV/$c^2$ above the thresholds for the open beauty channels $B^*\overline{B}$ (10604.6 MeV/$c^2$) and $B^*\overline{B}^*$ (10650.2 MeV/$c^2$), which suggests a "molecular" nature of these new states, which might explain most of their observed properties (Bondar, Garmash, Milstein, Mizuk, and Voloshin, 2011), although other possible interpretations have also been offered (Bugg, 2011; Cui, Liu, and Huang, 2012; Danilkin, Orlovsky, and Simonov, 2012; Guo, Cao, Zhou, and Chen, 2011).

## 18.4.6 Transitions and decays

### 18.4.6.1 Introduction to transitions and decays

Measuring the transitions between bottomonium states, independent of trying to discover new states, also provides important information for theoretical predictions of heavy quarkonium systems. These transitions can be predicted by effective potential models, and for existing measurements of transitions in the bottomonium system the data appeared well described (cf. Brambilla et al., 2004; Eichten, Godfrey, Mahlke, and Rosner, 2008). At leading order, the dominant radiative decays (those involving emission of a photon) are expected to be electric (E1) or magnetic (M1) transitions. If the bottomonium system is treated as a non-relativistic bound state, the predictions are relatively straight-forward and well-characterized. The picture is complicated, however, in transitions such as $\Upsilon(nS) \to \gamma\eta_b(mS)$, where $n > m$, which are referred to as "hindered" M1 transitions between the $S$-wave bottomonium states. In the case of $\Upsilon(3S) \to \gamma\chi_{bJ}(1P)$, there is an overlap between the wave functions of the initial state and the final state; this makes the calculation of such transitions more complex. One experimental goal in measuring such transitions is to improve our understanding of the non-relativistic effects in heavy quarkonium systems, which should in turn inform and improve the theoretical calculations.

Charmonium spectroscopy is a field revived after the operation of the two $B$ Factories. States with $J^{PC} = 1^{--}$ may be studied using ISR in the large $\Upsilon(4S)$ data samples. For a study of charge-parity-even charmonium states, radiative decays of the $\Upsilon$ states below open-bottom threshold may be used.

The production rates of the lowest-lying $P$-wave spin-triplet ($\chi_{cJ}$, $J$=0, 1, or 2) and $S$-wave spin-singlet ($\eta_c$) states in $\Upsilon(1S)$ radiative decays are calculated (Gao, Zhang, and Chao, 2007), where the former is at the part per million level, and the latter is about $5 \times 10^{-5}$. The rates in $\Upsilon(2S)$ decays are estimated to be at the same level.

We know that the OZI-suppressed decays of $J/\psi$ and $\psi(2S)$ to hadrons occur by annihilation of the charm quarks into three gluons or a photon. In either case, pQCD predicts (Appelquist and Politzer, 1975; De Rujula and Glashow, 1975)

$$Q_\psi = \frac{\mathcal{B}_{\psi(2S)\to h}}{\mathcal{B}_{J/\psi\to h}} = \frac{\mathcal{B}_{\psi(2S)\to e^+e^-}}{\mathcal{B}_{J/\psi\to e^+e^-}} \approx 12\% \quad (18.4.15)$$

This relation is referred to as the "12% rule" which is expected to hold to a reasonably good degree for both inclusive and exclusive decays. But the measured experimental data do not follow this rule. The prediction by Eq. 18.4.15 is severely violated in the $\rho\pi$ and several other decay channels. This is the so-called "$\rho\pi$ puzzle". It was first observed by Mark-II Collaboration in 1983 (Franklin et al., 1983). From then on many experimental studies and theoretical explanations have been put forth to decipher this puzzle (Mo, Yuan, and Wang, 2006).

As this so-called "12% rule" in $\psi$ decays is derived from the pQCD and potential models, it is expected to be valid for the bottomonium family, namely, the $\Upsilon$s. Since there are three narrow $\Upsilon$ states below the bottom meson threshold, we expect, using PDG average values of the branching fractions:

$$\begin{aligned} Q_{21} &= \frac{\mathcal{B}_{\Upsilon(2S)\to h}}{\mathcal{B}_{\Upsilon(1S)\to h}} = \frac{\mathcal{B}_{\Upsilon(2S)\to e^+e^-}}{\mathcal{B}_{\Upsilon(1S)\to e^+e^-}} = 0.77 \pm 0.07, \\ Q_{31} &= \frac{\mathcal{B}_{\Upsilon(3S)\to h}}{\mathcal{B}_{\Upsilon(1S)\to h}} = \frac{\mathcal{B}_{\Upsilon(3S)\to e^+e^-}}{\mathcal{B}_{\Upsilon(1S)\to e^+e^-}} = 0.88 \pm 0.09, \\ Q_{32} &= \frac{\mathcal{B}_{\Upsilon(3S)\to h}}{\mathcal{B}_{\Upsilon(2S)\to h}} = \frac{\mathcal{B}_{\Upsilon(3S)\to e^+e^-}}{\mathcal{B}_{\Upsilon(2S)\to e^+e^-}} = 1.14 \pm 0.15. \end{aligned} \quad (18.4.16)$$

These "pQCD rules" should hold better than the 12% rule in $\psi$ decays, since the bottomonium states have higher masses, and pQCD and the potential models should work better, as has been the case for calculations of the bottomonium spectrum.

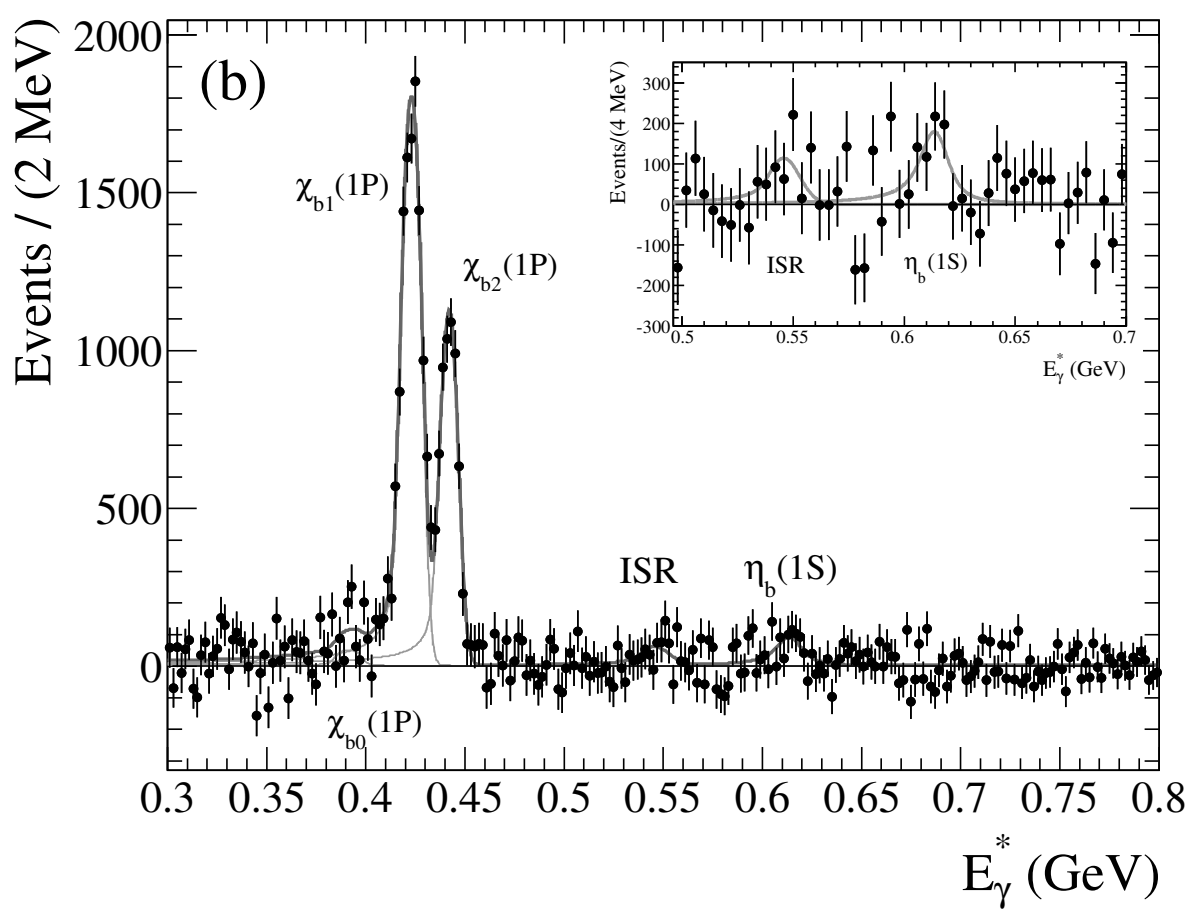

**Figure 18.4.16.** The converted-photon energy spectrum, after subtracting the smooth background (Lees, 2011m). These data were taken at the $\Upsilon(2S)$ resonance.

### 18.4.6.2 Radiative transitions between bottomonium states

The radiative transitions between $\Upsilon$ and $\chi_{bJ}$ states were a background to the discovery of the bottomonium ground state (c.f. Section 18.4.4.2). These transition rates are generally precisely predicted (Section 18.1); however, more precise experimental measurements were needed to determine the accuracy of the methods used to make those predictions. The method used to discover the $\eta_b$ used photons reconstructed using only the *BABAR* electromagnetic calorimeter. The resolution of the $\chi_{bJ}$ transitions is limited by the energy resolution of the calorimeter, which was insufficient to convincingly separate the transitions to and from the three $\chi_{bJ}$ states.

For the *BABAR* Collaboration measurement discussed in this section, photon transitions were reconstructed using photons that had converted in material. This results in a much-improved photon energy resolution, reducing it from 25 MeV using calorimeter-only photons to 5 MeV using converted photons. The improved resolution allows for the separation of many radiative transitions. The photon energy spectrum was analyzed (Lees, 2011m) using data taken at the $\Upsilon(3S)$ in three different energy regions: $E^* = [180, 300]$ MeV, $[300, 600]$ MeV, and $[600, 1100]$ MeV. This was done so that regions expected to contain different kinds of transitions could be separately studied. The data taken at the $\Upsilon(2S)$ resonance was analyzed in a single bin, $E^* = [300, 800]$ MeV. Besides studying prominent transitions, a goal of this approach was to "re-discover" the $\eta_b$ and make an additional measurement of its using an independent technique.

The primary background in these measurements arises due to using a randomly chosen converted photon as the candidate photon from a bottomonium transition. This background is again modeled using a smooth function, and is subtracted (as in Fig. 18.4.16 and 18.4.17). The contributions from monochromatic photons are modeled using functions that describe both their peak location and detector resolution effects.

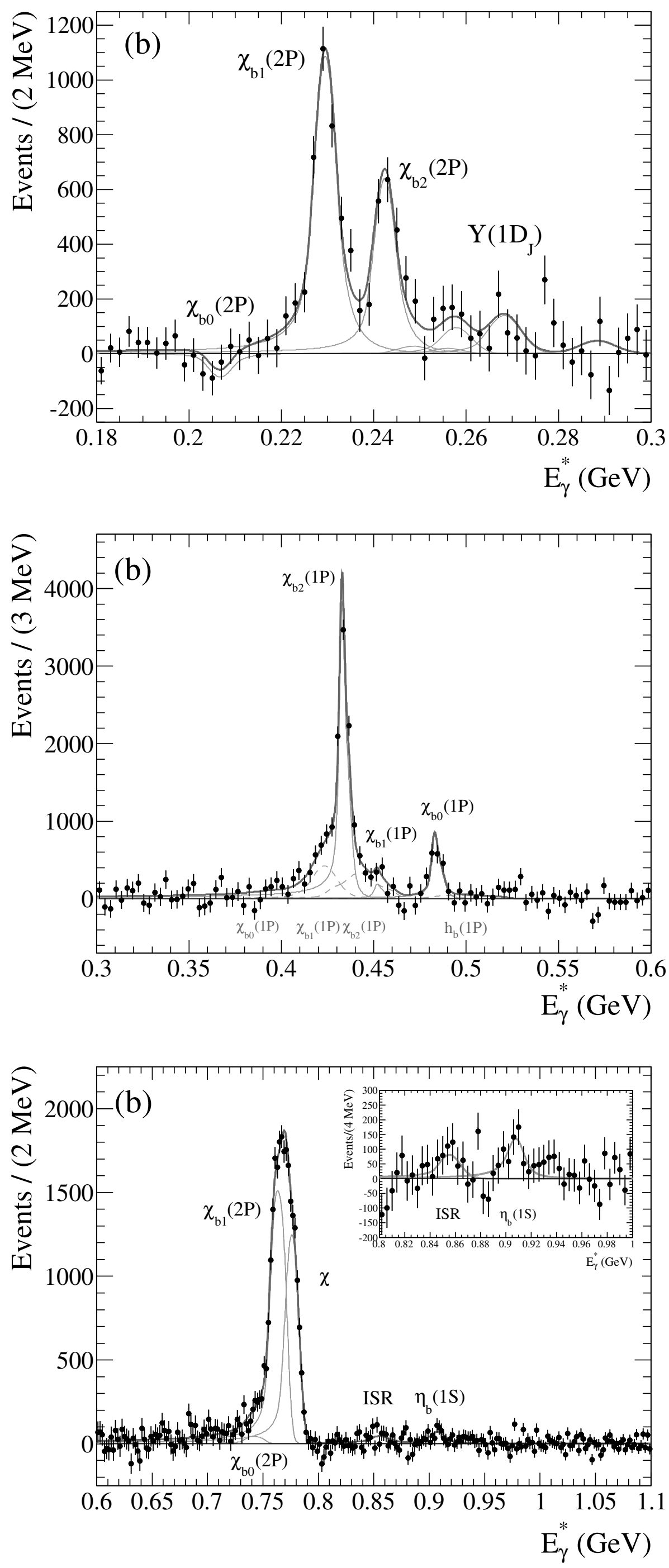

**Figure 18.4.17.** The converted-photon energy spectrum, after subtracting the smooth background (Lees, 2011m). These data were taken at the $\Upsilon(3S)$ resonance. In the lower plot, $\chi$ denotes $\chi_{b2}(2P)$.

The $\Upsilon(3S)$ data in the range $[180, 300]$ MeV were expected to contain three monochromatic peaks due to the transitions $\chi_{bJ}(2P) \rightarrow \gamma\Upsilon(2S)$ and six monochromatic peaks due to the transitions $\Upsilon(1D_J) \rightarrow \gamma\chi_{bJ}(1P)$. Due to the very low rates for these latter transitions, the properties of the models describing these transitions are fixed from existing measurements of the masses of the involved states. The branching fractions for the $\chi_{bJ}(2P) \rightarrow \gamma\Upsilon(2S)$ transitions are determined from the data and shown in Table 18.4.7.

The $\Upsilon(3S)$ data in the range $[300, 600]$ MeV were expected to contain photons from the six transitions $\Upsilon(3S) \rightarrow \gamma\chi_{bJ}(1P)$ and $\chi_{bJ}(1P) \rightarrow \gamma\Upsilon(1S)$. The photons from these transitions are all comparable in energy and were expected to overlap. In addition, this region could also contain photons from the $h_b(1P) \rightarrow \gamma\eta_b(1S)$ transition. No evidence was seen for this latter transition, and the measured rates for the transition $\Upsilon(3S) \rightarrow \gamma\chi_{bJ}(1P)$ are given in Table 18.4.7. The pattern of these transitions is unusual for quarkonium, with the pattern for the relative rates to the different $J = 0, 1, 2$ $\chi_{bJ}$ states being $J = 2 > 0 > 1$.

A comment on the non-observation of $h_b(1P) \rightarrow \gamma\eta_b(1S)$ in this search may be helpful, since in Section 18.4.4.3 a discussion was made of the discovery by the Belle Collaboration of the transition $h_b(1P) \rightarrow \gamma\eta_b(1S)$. From Table 18.4.7, we can see that the smallest branching fraction to which this search method had sensivity at the $\Upsilon(3S)$ resonance was at the level of $0.03 - 0.04\%$ (limited by the statistical uncertainty of the sample). The favored transition for $\Upsilon(3S) \rightarrow h_b(1P)$ is via radiation of a $\pi^0$, and the branching fraction for that process is expected to be at the level of 0.1%, while the branching fraction for $h_b(1P) \rightarrow \gamma\eta_b(1S)$ was measured by the Belle Collaboration to be about 49%, (Mizuk, 2012) as mentioned in Section 18.4.4.1.

The product of these two branching fractions, marking the rate at which $h_b(1P)$ is expected to be produced from the $\Upsilon(3S)$ in the *BABAR* data sample, is therefore about 0.05%, comparable to the statistical uncertainty of this technique. This product of branching fractions is confirmed by the evidence from the *BABAR* Collaboration for $\Upsilon(3S) \rightarrow \pi^0 h_b(1P)$, $h_b(1P) \rightarrow \gamma\eta_b(1S)$, discussed in Section 18.4.4.3.

Another predicted leading process from the $\Upsilon(3S)$ that can produce the $h_b(1P)$ is $\Upsilon(3S) \rightarrow \pi^+\pi^- h_b(1P)$, but this has been found by the *BABAR* Collaboration to occur with a branching fractiom $< 2.5 \times 10^{-4}$ at 90% confidence level (Lees, 2011l). In addition, the contribution of these processes to the photon spectrum overlaps with the much more significant $\chi_{b0}(1P)$ spectral line. Therefore, the non-observation of this photon spectral line in the data from this photon-conversion technique is consistent with independent evidence that the rate is below or comparable to the statistical sensitivity of this technique. In fact, in this study the contribution from these two sources was a fixed component of the fit, and found to be comparable to statistical uncertainties in the region where such a signal would be expected.

**Table 18.4.7.** Measured radiative transition rates, where the first uncertainty is statistical, the second is due to systematic effects, and the third (where present) is due to uncertainties on secondary branching fractions (Lees, 2011m). Numbers in parentheses are the 90% confidence level upper limits.

| Transition | Branching Fraction in % |
|---|---|
| $\Upsilon(3S) \rightarrow \gamma\chi_{b0}(1P)$ | $0.27 \pm 0.04 \pm 0.02$ |
| $\Upsilon(3S) \rightarrow \gamma\chi_{b1}(1P)$ | $0.05 \pm 0.03^{+0.02}_{-0.01}$ $(< 0.10)$ |
| $\Upsilon(3S) \rightarrow \gamma\chi_{b2}(1P)$ | $1.06 \pm 0.03^{+0.07}_{-0.06}$ |
| $\chi_{b0}(1P) \rightarrow \gamma\Upsilon(1S)$ | $2.2 \pm 1.5^{+1.0}_{-0.7} \pm 0.2$ $(< 4.6)$ |
| $\chi_{b1}(1P) \rightarrow \gamma\Upsilon(1S)$ | $34.9 \pm 0.8 \pm 2.2 \pm 2.0$ |
| $\chi_{b2}(1P) \rightarrow \gamma\Upsilon(1S)$ | $19.5 \pm 0.7^{+1.3}_{1.5} \pm 1.0$ |
| $\chi_{b0}(2P) \rightarrow \gamma\Upsilon(2S)$ | $-4.7 \pm 2.8^{+0.7}_{-0.8} \pm 0.5(< 2.8)$ |
| $\chi_{b1}(2P) \rightarrow \gamma\Upsilon(2S)$ | $18.9 \pm 1.1 \pm 1.2 \pm 1.8$ |
| $\chi_{b2}(2P) \rightarrow \gamma\Upsilon(2S)$ | $8.3 \pm 0.8 \pm 0.6 \pm 1.0$ |
| $\chi_{b0}(2P) \rightarrow \gamma\Upsilon(1S)$ | $0.7 \pm 0.4^{+0.2}_{-0.1} \pm 0.1$ $(< 1.2)$ |
| $\chi_{b1}(2P) \rightarrow \gamma\Upsilon(1S)$ | $9.9 \pm 0.3^{+0.5}_{0.4} \pm 0.9$ |
| $\chi_{b2}(2P) \rightarrow \gamma\Upsilon(1S)$ | $7.0 \pm 0.2 \pm 0.3 \pm 0.9$ |

Finally, the $\Upsilon(3S)$ data in the range $[600, 1100]$ MeV were expected to contain photons due to the transitions $\chi_{bJ}(2P) \rightarrow \gamma\Upsilon(1S)$ and $\Upsilon(3S) \rightarrow \gamma\eta_b(1S)$. The latter transition was seen, but only with a significance of $2.7\sigma$ and so no independent measurement of the $\eta_b(1S)$ mass was possible from this sample; more data is needed to fully utilize this technique for measuring that transition. However, the rates for the transitions $\chi_{bJ}(2P) \rightarrow \gamma\Upsilon(1S)$ were measured and are reported in Table 18.4.7.

The $\Upsilon(2S)$ data in the range $[300, 800]$ MeV were expected to contain photons due to the transitions $\chi_{bJ}(1P) \rightarrow \gamma\Upsilon(1S)$ and $\Upsilon(2S) \rightarrow \gamma\eta_b(1S)$. No evidence was seen for the latter transitions, and the rates of the transitions $\chi_{bJ}(1P) \rightarrow \gamma\Upsilon(1S)$ are reported in Table 18.4.7.

### 18.4.6.3 Searches for $\Upsilon(1S)$ and $\Upsilon(2S)$ radiative transitions to charmonium states

The Belle Collaboration searched for radiative transitions from $\Upsilon(2S)$ (Shen, 2010b) and $\Upsilon(1S)$ (Wang, 2011) to charmonium states using the following data samples : on-resonance samples of 5.7 fb$^{-1}$ at the $\Upsilon(1S)$ (102 million $\Upsilon(1S)$ events) and a 24.7 fb$^{-1}$ at the $\Upsilon(2S)$ (158 million $\Upsilon(2S)$ events), and continuum samples of 1.8 fb$^{-1}$ at $\sqrt{s} = 9.43$ GeV and 1.7 fb$^{-1}$ collected at $\sqrt{s} = 9.993$ GeV.

The search for the $\chi_{cJ}$ was conducted using the $\gamma J/\psi$ mode, where $J/\psi$ was observed in both $\mu^+\mu^-$ and $e^+e^-$ decay modes. The $\mu^+\mu^-$ mode shows a clear $J/\psi$ signal, while the $e^+e^-$ mode has some residual radiative Bhabha background. No clear $\chi_{cJ}$ signal is observed in both $\Upsilon(1S)$ and $\Upsilon(2S)$ data samples, as shown in Fig. 18.4.18.

The search for $\eta_c$ was done using full hadronic reconstruction of the $\eta_c$ in the modes: $K_S K^{\pm}\pi^{\mp}$, $\pi^+\pi^- K^+K^-$, $2(K^+K^-)$, $2(\pi^+\pi^-)$, and $3(\pi^+\pi^-)$. The combined mass distributions of the hadronic final states are shown in Fig. 18.4.19 for the five $\eta_c$ decay modes from $\Upsilon(1S)$ and $\Upsilon(2S)$ data, respectively. The large $J/\psi$ signal is due to the ISR process $e^+e^- \to \gamma_{ISR} J/\psi$, while the accumulation of events within the $\eta_c$ mass region is small.

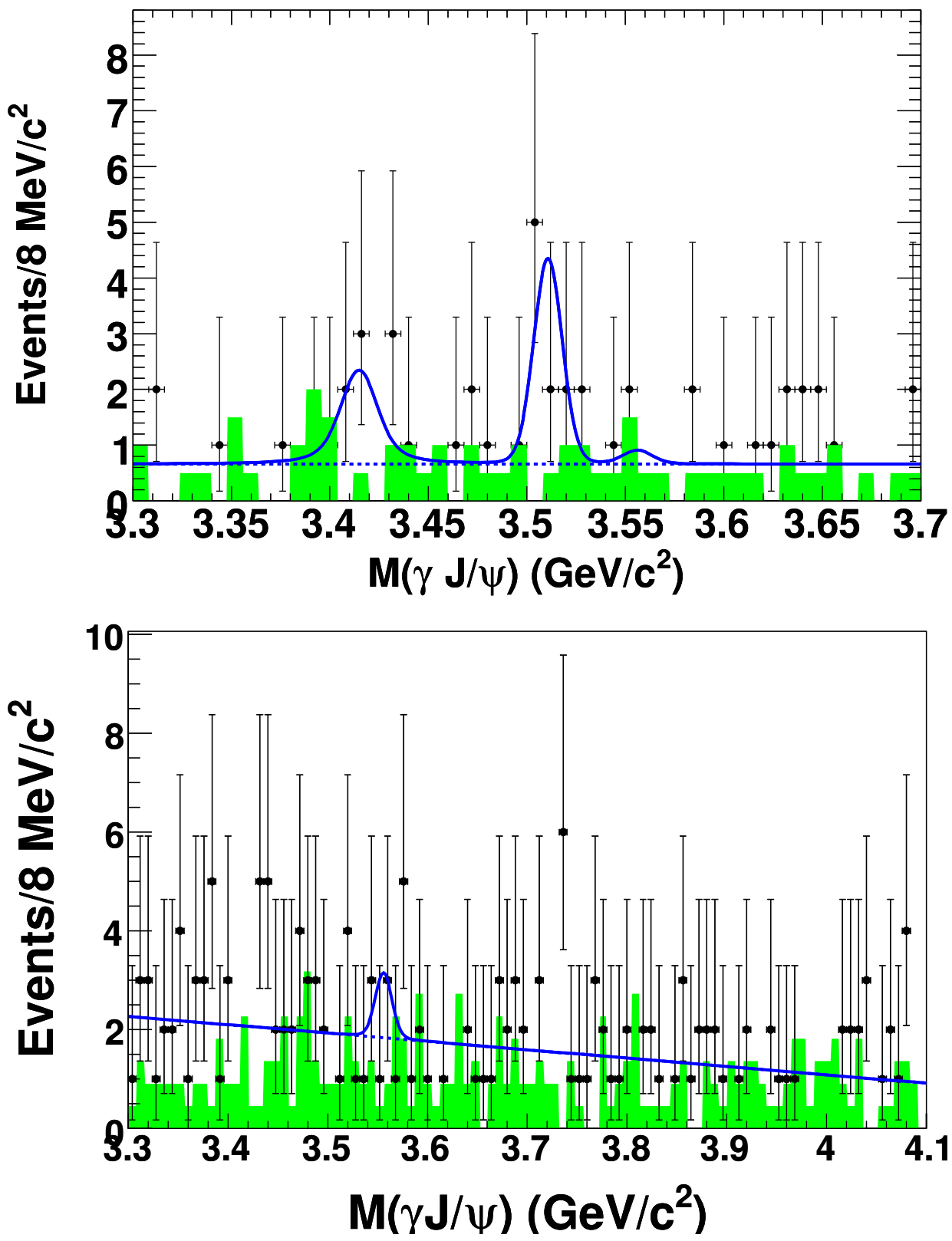

**Figure 18.4.18.** The $\gamma J/\psi$ invariant mass distributions in (top) the $\Upsilon(1S)$ (Shen, 2010b) and (bottom) $\Upsilon(2S)$ (Wang, 2011) data samples. No clear $\chi_{cJ}$ signal is observed. The solid curve is the best fit, the dashed curve is the background, and the shaded histogram is from the normalized $J/\psi$ mass sidebands.

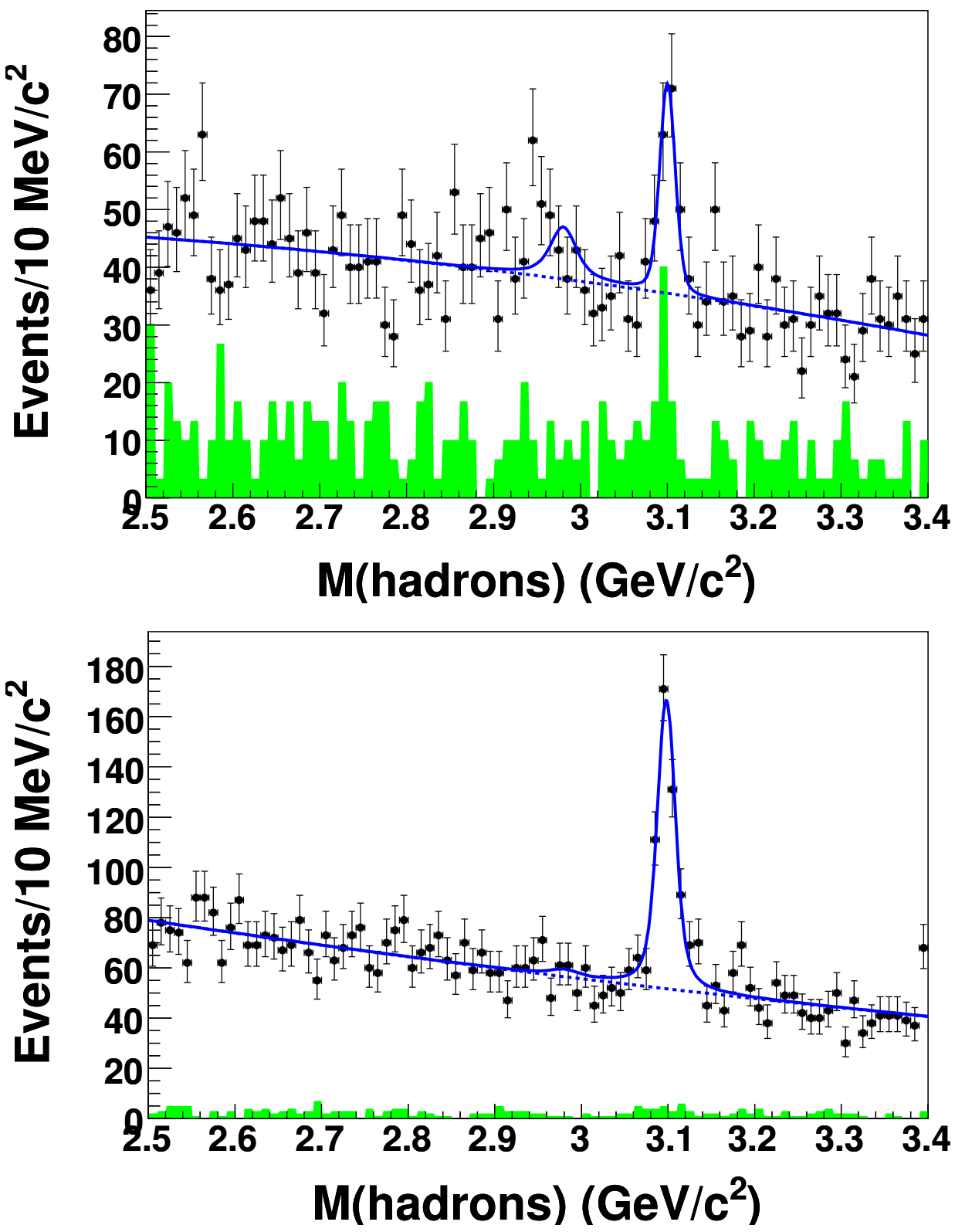

**Figure 18.4.19.** The mass distributions for a sum of the five $\eta_c$ decay modes (top) from $\Upsilon(1S)$ (Shen, 2010b) and (bottom) $\Upsilon(2S)$ (Wang, 2011) data, respectively. The solid curve is a sum of the corresponding functions obtained from a simultaneous fit to all the $\eta_c$ decay modes, and the dashed curve is a sum of the background functions from the fit. The shaded histogram is a sum of the continuum events (not normalized in $\Upsilon(2S)$ data). The $J/\psi$ signal is produced via ISR rather than from a radiative decay of an $\Upsilon(nS)$ resonance.

#### 18.4.6.4 Searches for $\Upsilon(1S)$ and $\Upsilon(2S)$ radiative transitions to charmonium-like states

In addition to many conventional charmonium states, a number of charmonium-like states (the so-called "$XYZ$ particles") have been discovered with unusual properties. These may include exotic states, such as quark-gluon hybrids, meson molecules, and multi-quark states (Brambilla et al., 2011). Many of these new states are established in a single production mechanism or in a single decay mode only. To better understand them, it is necessary to search for such states in more production processes and/or decay modes. For charge-parity-even charmonium-like states, radiative decays of the narrow $\Upsilon$ states below the open bottom threshold can be examined. There are no calculations for radiative decays for "$XYZ$ particles" due to the limited knowledge of their nature.

The Belle Collaboration searched for four of these $XYZ$ states, the $X(3872)$, $X(3915)$, $Y(4140)$ and $X(4350)$, using the $\Upsilon(1S)$(Shen, 2010b) and $\Upsilon(2S)$ (Wang, 2011) data samples described in the previous section.

The $X(3872)$ signal was searched for via $X(3872) \to \pi^+\pi^- J/\psi$ and $\pi^+\pi^-\pi^0 J/\psi$. Except for a few residual ISR produced $\psi(2S)$ signal events, only a small number of events appear in the $\pi^+\pi^- J/\psi$ invariant mass distributions for both $\Upsilon(1S)$ and $\Upsilon(2S)$ decays. Belle observed two events in the $\Upsilon(1S)$ data with masses of 3.67 GeV/$c^2$ and 4.23 GeV/$c^2$; only a few events were observed in the $\Upsilon(2S)$ data.

The search for $X(3915)$ was undertaken in the $\omega J/\psi$ mode. No events were observed within the $X(3915)$ mass region in $\Upsilon(1S)$ data. One event was observed with $m(\pi^+\pi^-\pi^0 J/\psi)$ at 3.923 GeV/$c^2$ and $m(\pi^+\pi^-\pi^0)$ at 0.790 GeV/$c^2$ from $\Upsilon(2S)$ data.

Finally, the $Y(4140)$ in both $\Upsilon(1S)$ and $\Upsilon(2S)$ data, and $X(4350)$ in $\Upsilon(2S)$ data only were searched for using the $\phi J/\psi$ mode. No candidates were observed in either in the $Y(4140)$ or $X(4350)$ mass regions.

Since there is no evidence for charmonium or charmonium-like states signals in the modes studied, Belle placed upper limits on the branching fractions of $\Upsilon(1S)$ and $\Upsilon(2S)$ radiative decays. Table 18.4.8 lists final results for the upper limits on the branching fractions. The results obtained on the $\chi_{cJ}$ and $\eta_c$ production rates are consistent with the theoretical predictions of Gao, Zhang, and Chao (2007). With much larger $\Upsilon(1S)$ and $\Upsilon(2S)$ data samples in the future at super flavor factories, we can obtain better results for charmonium final states which can tell us if experimental results support or disfavor theoretical predictions. If any one of charmonium-like states can be observed, it will do much help to understand its nature.

**Table 18.4.8.** Summary of the limits on $\Upsilon(1S)$ and $\Upsilon(2S)$ radiative decays to charmonium and charmonium-like states $R$. Here $\mathcal{B}_R$ is the upper limit at the 90% C.L. on the decay branching fraction in the charmonium state case, and on the product branching fraction in the case of a charmonium-like state.

| State ($R$) | $\mathcal{B}_R$ ($\Upsilon(1S)$) | $\mathcal{B}_R$ ($\Upsilon(2S)$) |
|---|---|---|
| $\chi_{c0}$ | $6.5 \times 10^{-4}$ | $1.0 \times 10^{-4}$ |
| $\chi_{c1}$ | $2.3 \times 10^{-5}$ | $3.6 \times 10^{-6}$ |
| $\chi_{c2}$ | $7.6 \times 10^{-6}$ | $1.5 \times 10^{-5}$ |
| $\eta_c$ | $5.7 \times 10^{-5}$ | $2.7 \times 10^{-5}$ |
| $X(3872) \to \pi^+\pi^- J/\psi$ | $1.6 \times 10^{-6}$ | $0.8 \times 10^{-6}$ |
| $X(3872) \to \pi^+\pi^-\pi^0 J/\psi$ | $2.8 \times 10^{-6}$ | $2.4 \times 10^{-6}$ |
| $X(3915) \to \omega J/\psi$ | $3.0 \times 10^{-6}$ | $2.8 \times 10^{-6}$ |
| $Y(4140) \to \phi J/\psi$ | $2.2 \times 10^{-6}$ | $1.2 \times 10^{-6}$ |
| $X(4350) \to \phi J/\psi$ | $\cdots$ | $1.3 \times 10^{-6}$ |

#### 18.4.6.5 Search for $\chi_b(1P)$ exclusive decays to double charmonium

The cross sections of the double-charmonium production processes $e^+e^- \to J/\psi\,\eta_c$, $J/\psi\,\eta_c'$, $\psi(2S)\eta_c$, $\psi(2S)\eta_c'$, $J/\psi\,\chi_{c0}$, and $\psi(2S)\chi_{c0}$ measured at the Belle (Abe, 2002j, 2004g) and *BABAR* (Aubert, 2005n) experiments were approximately an order of magnitude larger than the leading order non-relativistic QCD (NRQCD) predictions (Braaten and Lee, 2003; Liu, He, and Chao, 2003, 2008). It was shown that the calculations are very sensitive to the choices of the values of some parameters (Bodwin, Lee, and Braaten, 2003; Bodwin, Lee, and Yu, 2008; Braaten and Lee, 2003; He, Fan, and Chao, 2007; Zhang, Gao, and Chao, 2006) and the agreement between theory and experiment can be achieved if one takes into account radiative and relativistic corrections.

Similar to the production in $e^+e^-$ annihilation, double charmonium final states can also be produced in bottomonium decays, which supplied a new test of the dynamics of hard exclusive processes and the structure of the charmonia. While $\eta_b \to J/\psi\,J/\psi$ has been calculated (Sun, Hao, and Qiao, 2011), the $P$-wave spin-triplet bottomonium states $\chi_{bJ}$ ($J$=0, 1, 2) decays into double charmonium states were calculated using different theoretical models.

Under the NRQCD factorization approach, Zhang, Dong, and Feng (2011) calculated $\chi_{bJ} \to J/\psi\,J/\psi$ to a relativistic correction of the order $v_c^2$ and considered a small pure QED contribution. The branching fraction is predicted to be of order $10^{-5}$ for $\chi_{b0}$ or $\chi_{b2} \to J/\psi J/\psi$, and $10^{-11}$ for $\chi_{b1} \to J/\psi J/\psi$; Sang, Rashidin, Kim, and Lee (2011) considered the corrections to all orders in the charm-quark velocity $v_c$ in the charmonium rest frame, and found decay partial widths that are about a factor of three larger than those determined by Zhang, Dong, and Feng (2011). In the light cone (LC) formalism, however, much larger production rates (with also large uncertainties) are obtained by Braguta, Likhoded, and Luchinsky (2009): $\mathcal{B}(\chi_{bJ} \to J/\psi\,J/\psi) = 9.6 \times 10^{-5}$ or $1.1 \times 10^{-3}$, $\mathcal{B}(\chi_{bJ} \to J/\psi\,\psi(2S)) = 1.6 \times 10^{-4}$ or $1.6 \times 10^{-3}$, and $\mathcal{B}(\chi_{bJ} \to \psi(2S)\psi(2S)) = 6.6 \times 10^{-5}$ or $5.9 \times 10^{-4}$ for $J$=0 or 2, respectively. It is therefore very important to pin down the source of such significant discrepancies between the NRQCD factorization and LC formalisms.

In perturbative QCD theory, the branching fraction of $\mathcal{B}(\chi_{b0} \to J/\psi\,J/\psi) \approx 3 \times 10^{-5}$ and for $\chi_{b1}$ decays, it is even larger (Kartvelishvili and Likhoded, 1984). It has been argued (Braguta, Likhoded, and Luchinsky, 2005) that taking into account the relative motion of quarks in the amplitude of the decay of $\chi_b$ meson to c-quarks increases the branching fractions of the decays of $\chi_{b0}$ ($0^{++}$) and $\chi_{b2}$ ($2^{++}$) into a pair of $J/\psi$ mesons by an order of magnitude.

Based on a 24.7 fb$^{-1}$ $\Upsilon(2S)$ data sample collected by the Belle Collaboration, no significant signals are found for the $\chi_{bJ} \to J/\psi\,J/\psi$, $J/\psi\,\psi(2S)$, or $\psi(2S)\psi(2S)$ final states (Shen, Yuan, Iijima, 2012). The upper limits on the $\chi_{bJ}$ decay branching fractions are lower than the theoretical predictions using LC formalism, while are not in contradiction with other calculations(Zhang, Dong, and Feng, 2011) using NRQCD factorization approach.

#### 18.4.6.6 Two-body Hadronic Transitions: $\pi^0, \eta$

In the charmonium system, single-meson transitions between states have been well-established; the transition $\psi' \to \eta J/\psi$ has a branching ratio of 3.3%, one tenth of the dominant $\pi^+\pi^-$ transition (Beringer et al., 2012). In contrast, the first such transition in the bottomonium system, $\chi_b(2P) \to \omega\Upsilon(1S)$, was observed by the CLEO Collaboration in 2002 (Severini et al., 2004).

Generally, the hadronic transitions between heavy quarkonia are described by the QCD multipole expansion model (QCDME) (Kuang, 2006). In this framework the $\pi\pi$ transitions are mediated by the emission of two gluons in the E1 state, while the $\eta$ transitions proceed either via E1M2 or M1M1 terms. Such terms should be suppressed by a factor which is inversely proportional to the mass of the heavy quark. The suppression should be even larger when the $\eta$ is replaced by $\pi^0$, and the transition violates isospin. In the case of charmonium, the suppression factor is about 1/25. In the proximity of thresholds, coupled channel effects can provide different mechanisms to evade the spin flip suppression.

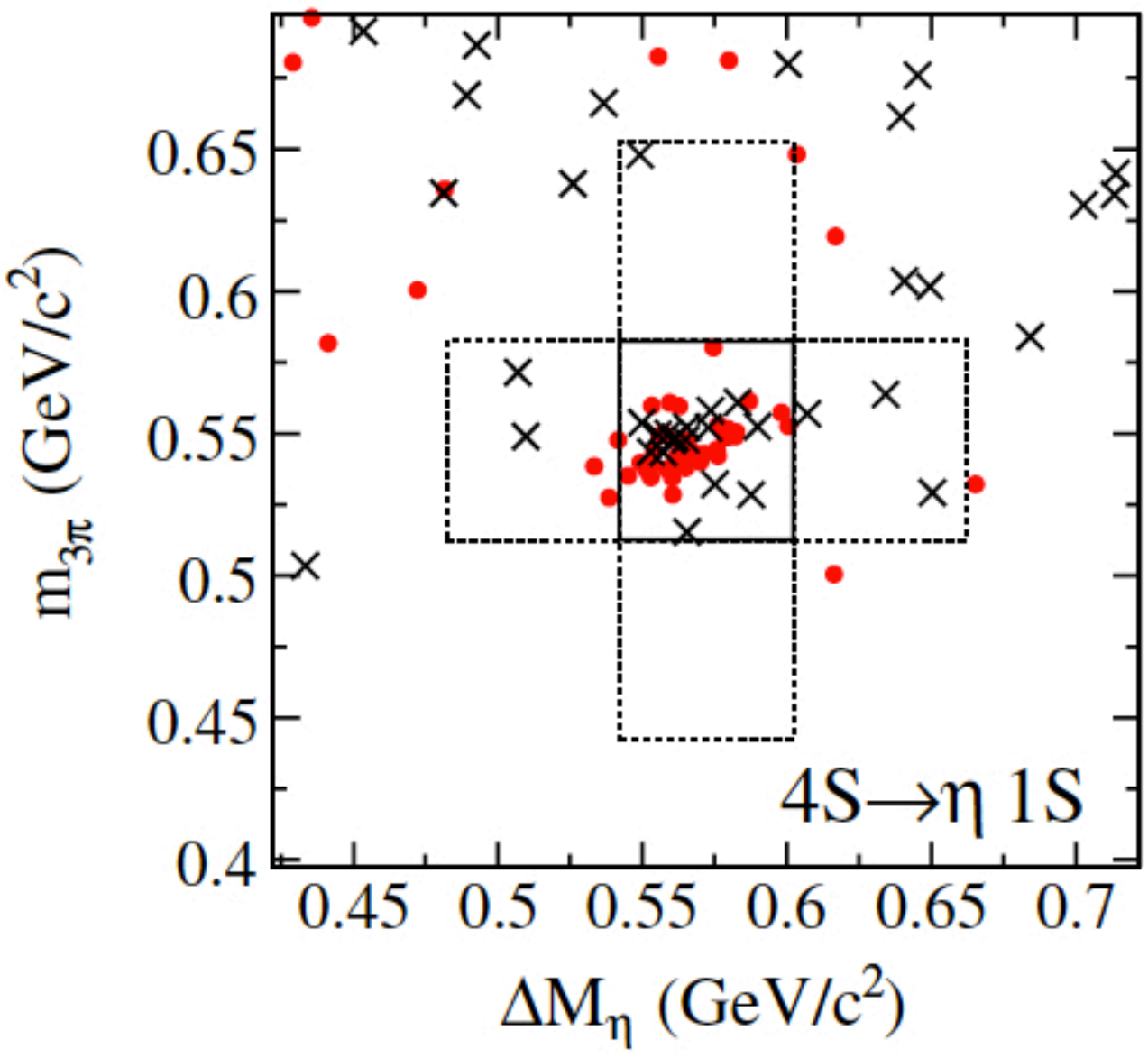

**Figure 18.4.20.** The distribution of $M_{3\pi}$ vs $\Delta M_\eta = M_{3\pi ll} - M_{ll}$ for $\eta\Upsilon(1S)$ candidate events, demonstrating the observed signal (Aubert, 2008bc). Solid lines delimit the signal region, while dashed lines delimit the sideband regions used for the background estimate. Crosses represent the data from the $\Upsilon(1S) \to e^+e^-$ sample and dots the data from the $\Upsilon(1S) \to \mu^+\mu^-$ sample.

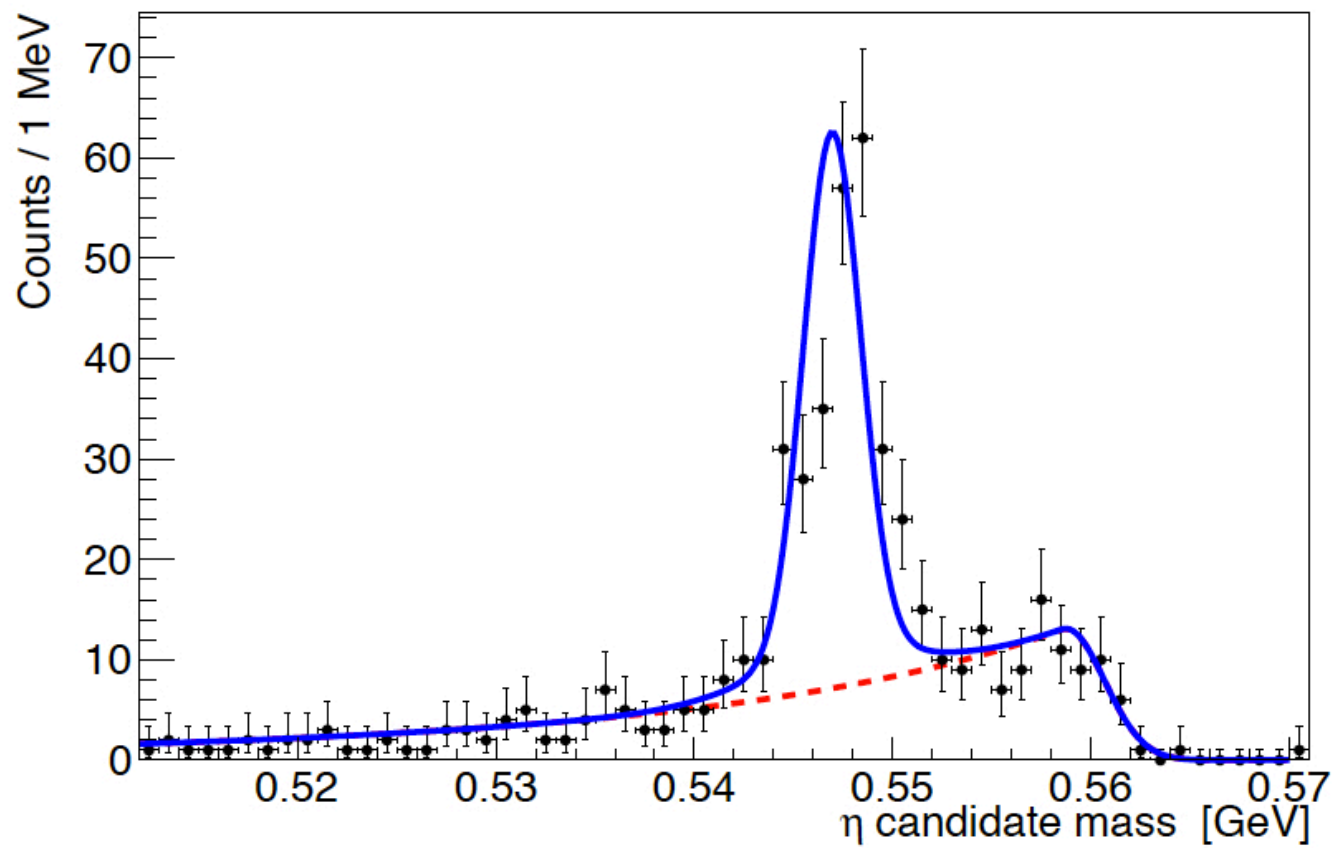

**Figure 18.4.21.** Signal of $\Upsilon(2S) \to \eta\Upsilon(1S)$ from the Belle Collaboration result (Tamponi, 2013).

The *BABAR* Collaboration observed the transition $\Upsilon(4S) \to \eta\Upsilon(1S)$ (Aubert, 2008bc) using $347.5\,\text{fb}^{-1}$ of data taken at the $\Upsilon(4S)$ resonance. The observed rate was 2.5 times larger than that of the $\Upsilon(4S) \to \pi^+\pi^-\Upsilon(1S)$ transition, which is the dominant hadronic transition in both charmonium and bottomonium. The *BABAR* Collaboration performed the search for the hadronic transitions $\pi^+\pi^-\Upsilon(1,2S)$ and $\eta\Upsilon(1S)$ using the $\eta \to \pi^+\pi^-\pi^0$ mode, where the $\Upsilon$ is decaying either to $e^+e^-$ or to $\mu^+\mu^-$. Candidate events were selected requiring at least four charged tracks in acceptance; the lepton candidates are required to have center-of-mass momentum between 4.20 and 5.25 GeV/$c$, and the dilepton pair is required to have invariant mass within $\pm 200$ MeV ($^{+200}_{-350}$ MeV) of the nominal $\Upsilon(1S)$ ($\Upsilon(2S)$) mass. The dipion and dilepton candidates are constrained to have a common vertex.

The dominant background is due to $e^+e^-\gamma$ and $\mu^+\mu^-\gamma$, where a photon converts in material and creates an $e^+e^-$ pair that is misidentified as a pion pair. This background is rejected by requiring the pion pair to have an opening angle above 18° in the laboratory frame; in addition, the invariant mass of the low momentum pair, calculated assuming the $e^\pm$ mass hypothesis, is required to satisfy $m_{e^+e^-} > 100\,\text{MeV}/c^2$.

The two photons used in $\pi^0$ reconstruction are required to have $E_\gamma > 50$ MeV and an invariant mass, $m_{\gamma\gamma}$, in the range [110, 150] MeV/$c^2$. The residual background, from $\Upsilon(mS) \to \pi\pi\Upsilon(nS)$ transitions, is reduced by requiring $\Delta M = M_{\pi\pi ll} - M_{ll}$ to be at least 20 MeV/$c^2$ away from known transitions. The distribution of $M_{3\pi}$ vs $\Delta M_\eta = M_{3\pi ll} - M_{ll}$ for $\eta\Upsilon(1S)$ candidate events is shown in Fig. 18.4.20. Solid lines delimit the signal region, while dashed lines delimit the sideband regions used for the background estimate. The signal observed by *BABAR* has a significance of $11(6.2)\sigma$ in the $\mu\mu(ee)$ channel.

Shortly after the observation of the $\Upsilon(4S) \to \eta\Upsilon(1S)$ transition, the CLEO Collaboration reported the first observation of $\Upsilon(2S) \to \eta\Upsilon(1S)$, with a significance of 5.3 $\sigma$ (He et al., 2008). The observed branching ratio was about two times smaller than the theoretical prediction. Using their samples of $\Upsilon(2,3S)$ decays, the *BABAR* (Lees, 2011n) and Belle Collaborations (Tamponi, 2013) have studied the $\eta$ and $\pi^0$ transitions between narrow bottomonium states.

The dominant peaking backgrounds are due to the favored neutral and charged dipion and diphoton (via $\chi_b$) transitions, and the continuum $e^+e^- \to e^+e^-(n\gamma)$, $\mu^+\mu^-(n\gamma)$ processes. The dominant transition $\Upsilon(2S) \to \pi^-\pi^+\Upsilon(1S)$, which is expected to yield about $\mathbf{O}(10^3)$ more events, can be used as normalization sample. By normalizing the rate of $\eta$ and $\pi^0$ transitions to the rate of the $\pi^+\pi^-$ transition, the systematic error of the measurement is reduced by cancellation of common uncertainties.

Both experiments detect the $\eta$ meson in the $\gamma\gamma$ and the $\pi^+\pi^-\pi^0$ final states. Due to the tighter requirements imposed by the Bhabha veto at trigger level, the *BABAR* Collaboration was only able to use the $\Upsilon \to \mu^+\mu^-$ final state, while the Belle Collaboration uses both leptonic final states.

Charged tracks with momenta in the collider center-of-mass frame are required to have be greater than 4 GeV/$c$; such tracks are selected as candidate leptons from $\Upsilon(1S)$ decay. Particle identification is applied to categorize events as having electrons or muons in the final state.

The momentum of all photons detected in the ECL in the proximity of each leptonic track is added to its momentum, to reduce the effect of final state radiation (FSR) and bremsstrahlung. In order to suppress the contribution from continuum QED processes, a single-constraint kine-matic fit with a constraint on the $\Upsilon(1S)$ mass is then applied on the two lepton candidate momenta, corrected for nearby photons, as described above. The threshold on the confidence level of this fit has been optimized using the Monte Carlo simulation for each specific channel.

**Table 18.4.9.** Branching ratios (in units of $10^{-4}$ ) and upper limits (at 90% C.L.) for $\eta$ and $\pi^0$ transitions from *BABAR* (Aubert, 2008bc; Lees, 2011n) and Belle (Tamponi, 2013), compared to previous results from CLEO (He et al., 2008).

| *transition* | *BABAR* | Belle | *CLEO* |
|---|---|---|---|
| $\Upsilon(2S) \to \eta\Upsilon(1S)$ | $2.39 \pm 0.31 \pm 0.14$ | $3.57 \pm 0.25 \pm 0.21$ | $2.1^{0.7}_{0.6} \pm 0.3$ |
| $\Upsilon(2S) \to \pi^0\Upsilon(1S)$ | – | $< 0.41$ | $< 1.8$ |
| $\Upsilon(3S) \to \eta\Upsilon(1S)$ | $< 1.0$ | – | $< 1.8$ |
| $\Upsilon(3S) \to \pi^0 h_b(1P)$ | $7.4 \pm 2.2 \pm 1.4$ | – | – |
| $\Upsilon(4S) \to \eta\Upsilon(1S)$ | $1.96 \pm 0.06 \pm 0.09$ | – | – |

A requirement on the polar angle of the $e^-$ track with respect to the beam direction, $\cos(\theta^*_{e^-}) < 0.5$, is imposed to further suppress singly or doubly radiative Bhabha events, which represent the dominant QED background. The Bhabha requirement is not included in the $\Upsilon(2S) \to \pi^0\Upsilon(1S)$ analysis, since the $\Upsilon(1S)$ mass constraint provides already a good suppression of the QED processes.

The $\pi^+\pi^-$ candidate for the $\eta \to \pi^+\pi^-\pi^0$ decay (and also for the $\Upsilon(2S) \to \pi^+\pi^-\Upsilon(1S)$ transition used by the Belle Collaboration) is selected requiring the two tracks to be oppositely charged, to originate from the primary interaction point, and to have a large opening angle in the $e^+e^-$ CM frame (*e.g.* $\cos(\theta^*_{ch^+,ch^-}) < 0.6$ in the Belle Collaboraton result), in order to reject the $e^+e^-$ pairs coming from photon conversions in the inner detector. In the search for the $\Upsilon(3S) \to \eta\Upsilon(1S)$ transition, an additional requirement on $\Delta M_{\pi\pi} = M_{\pi\pi ll} - M_{ll}$ is imposed, to avoid contamination from the $\Upsilon(2,3S) \to \pi^+\pi^-\Upsilon(1,2S)$ transitions.

A second kinematic fit is then performed after the $\eta$ or $\pi^0$ selection, constraining all final state particles (*i.e.* the dilepton, the dipion and/or the best photon pair) to have an invariant mass equal to the $\Upsilon(2,3S)$ mass, to generate from the same vertex, and to have a total energy equal to the sum of the beam energies.

An unbinned maximum-likelihood fit to the measured distribution of one (Belle, Fig. 18.4.21) or two (*BABAR*) observables is then performed to extract the signal yields. Each observed distribution is fit to a sum of signal and background components, with functional forms determined from the simulations. The value of the branching fraction for each mode is then extracted. The results of all analyses are summarized in Table 18.4.9.

The branching ratio for the $\Upsilon(3S) \to \pi^0 h_b(1P)$ transition is obtained by combining the product of branching ratio measured by *BABAR* in the study of the reaction $\Upsilon(3S) \to \pi^0 h_b(1P); h_b \to \gamma\eta_b(1S)$, with $\mathcal{B}(h_b(1P) \to \eta_b(1S))$, measured by Belle. The evidence of an isospin violating transition at least one order of magnitude larger than yet unobserved $\eta$ transition from $\Upsilon(3S)$ is even more surprising than the enhanced $\eta\Upsilon(1S)$ rate from $\Upsilon(4S)$.

#### 18.4.6.7 Exclusive $\Upsilon(1S)$ and $\Upsilon(2S)$ decays into light hadrons

A number of channels in $\psi$ decays have been studied, most of which satisfy predictions about their properties to within experimental errors. One example of a property which does not conform to expectation arises from the comparison of $\psi$ decays into vector-pseudoscalar (VP) and vector-tensor (VT) final states: $\rho\pi$, $K^*\bar{K}$, $\rho a_2(1320)$ and $\omega f_2(1270)$. The rates of decay to these final states deviate from expectations, such as those implies by the "12% rule" (Section 18.4.6.1). It is interesting, therefore, to see if similar patterns of deviation occur in the bottomonium system by studying similar final states of $\Upsilon$ decay.

Although 82% of the $\Upsilon(1S)$ and 59% of the $\Upsilon(2S)$ decays are expected to be light-hadron final states, little experimental information exists on exclusive decays of the $\Upsilon$ resonances below the $B\overline{B}$ threshold. This situation is very different in charmonium sector, where numerous channels have been measured and used to perform model tests.

The Belle Collaboration published first observations of exclusive, light-hadron final states of the $\Upsilon(1S)$ and $\Upsilon(2S)$ (Shen, 2012). A large number of final states were studied and the key results are summarized in Table 18.4.10.

The measurements are mostly consistent with the prediction from pQCD (Section 18.4.6.1), $Q_{21} = 0.77 \pm 0.07$. The one measured mode that demonstrates a deviation from the prediction is $\omega\pi^+\pi^-$, which is consistent with the prediction at the level of $2.6\sigma$. For the final states measured so far, the predictions from pQCD appear to be reliable within the experimental uncertainties.

#### 18.4.6.8 $\Upsilon(4S)$ decays to $B\bar{B}$

The $\Upsilon(4S)(10580)$ is a resonance state that has a mass slightly above the $B\overline{B}$ threshold. It decays mostly into $B^0\overline{B}^0$ and $B^+B^-$ pairs which are not available to the lighter resonances due to the phase space.

Given the similar masses of $B^+$ and $B^0$, it is expected that the branching fractions $f_{00} \equiv \mathcal{B}(\Upsilon(4S) \to B^0\overline{B}^0)$ and $f_{+-} \equiv \mathcal{B}(\Upsilon(4S) \to B^+B^-)$ are around 0.5. However, predictions for the ratio $R^{+/0} \equiv f_{+-}/f_{00}$ range from 1.03 to 1.25 (Aubert, 2005r). This is due to the effect of the Coulomb force in the decays of $\Upsilon(4S)$ into $B^0\bar{B}^0$ and $B^+B^-$ pairs. The kinematic aspects of the $\Upsilon(4S)$ decays are treated as non-relativistic. The $B$ meson velocity in the $\Upsilon(4S)$ rest frame is relatively small,

$$\beta = v/c = \sqrt{1 - \frac{4m_B^2}{m_{\Upsilon(4S)}^2}} \approx 0.065, \qquad (18.4.17)$$

where $m_B$ and $m_{\Upsilon(4S)}$ are the masses of the $B$ meson and the $\Upsilon(4S)$ resonance, respectively.

**Table 18.4.10.** Results from the Belle Collaboration (Shen, 2012) in the measurement of exclusive hadronic decays of the $\Upsilon(1S)$ and $\Upsilon(2S)$ mesons. Here, $\mathcal{B}$ is the measured branching fraction (in units of $10^{-6}$), and where the signficance of the result is low $\mathcal{B}^{\rm UP}$ (the 90% confidence level upper limit on the branching fraction) is also reported. $Q_{21}$ is the computed ratio of the $\Upsilon(2S)$ and $\Upsilon(1S)$ branching fractions. Where the significance is small, $Q_{21}^{\rm UP}$ (the upper limit on the value of $Q_{21}$) is also reported. The first error in $\mathcal{B}$ and $Q_{21}$ is statistical, and the second systematic.

| Channel | $\Upsilon$(1S) | | $\Upsilon$(2S) | | | |
|---|---|---|---|---|---|---|
| | $\mathcal{B}$ | $\mathcal{B}^{\rm UP}$ | $\mathcal{B}$ | $\mathcal{B}^{\rm UP}$ | $Q_{21}$ | $Q_{21}^{\rm UP}$ |
| $\phi K^+K^-$ | $2.36 \pm 0.37 \pm 0.29$ | | $1.58 \pm 0.33 \pm 0.18$ | | $0.67 \pm 0.18 \pm 0.11$ | |
| $\omega\pi^+\pi^-$ | $4.46 \pm 0.67 \pm 0.72$ | | $1.32 \pm 0.54 \pm 0.45$ | 2.58 | $0.30 \pm 0.13 \pm 0.11$ | 0.55 |
| $K^{*0}K^-\pi^+$ | $4.42 \pm 0.50 \pm 0.58$ | | $2.32 \pm 0.40 \pm 0.54$ | | $0.52 \pm 0.11 \pm 0.14$ | |
| $\phi f_2'$ | $0.64 \pm 0.37 \pm 0.14$ | 1.63 | $0.50 \pm 0.36 \pm 0.19$ | 1.33 | $0.77 \pm 0.70 \pm 0.33$ | 2.54 |
| $\omega f_2$ | $0.57 \pm 0.44 \pm 0.13$ | 1.79 | $-0.03 \pm 0.24 \pm 0.01$ | 0.57 | $-0.06 \pm 0.42 \pm 0.02$ | 1.22 |
| $\rho a_2$ | $1.15 \pm 0.47 \pm 0.18$ | 2.24 | $0.27 \pm 0.28 \pm 0.14$ | 0.88 | $0.23 \pm 0.26 \pm 0.12$ | 0.82 |
| $K^{*0}\bar{K}_2^{*0}$ | $3.02 \pm 0.68 \pm 0.34$ | | $1.53 \pm 0.52 \pm 0.19$ | | $0.50 \pm 0.21 \pm 0.07$ | |
| $K_1(1270)^+K^-$ | $0.54 \pm 0.72 \pm 0.21$ | 2.41 | $1.06 \pm 0.42 \pm 0.32$ | 3.22 | $1.96 \pm 2.71 \pm 0.84$ | 4.73 |
| $K_1(1400)^+K^-$ | $1.02 \pm 0.35 \pm 0.22$ | | $0.26 \pm 0.23 \pm 0.09$ | 0.83 | $0.26 \pm 0.25 \pm 0.10$ | 0.77 |
| $b_1(1235)^+\pi^-$ | $0.47 \pm 0.22 \pm 0.13$ | 1.25 | $0.02 \pm 0.07 \pm 0.01$ | 0.40 | $0.05 \pm 0.16 \pm 0.03$ | 0.35 |

Table 18.4.11 shows the experimental results on $R^{+/0}$. All measurements assumed the isospin invariance in $\Gamma(B^+ \to x^+) = \Gamma(B^0 \to x^0)$, where $x^+$ and $x^0$ are the charged and neutral final particles.

The only measurement that did not follow the above assumption is the measurement from the Belle experiment, Hastings, 2003. This measurement used dilepton events, but assumed that there is isospin invariance, $\Gamma(B^+ \to \ell^+X) = \Gamma(B^0 \to \ell^+X)$. Therefore, this result is treated slightly differently, described as follows:

- Using the corresponding lifetime ratio (Table 18.4.11), each measurement from CLEO and *BABAR* is converted into its original measurement of $R^{+/0} \times \tau(B^+)/\tau(B^0)$
- No statistical and systematic correlation between the measurements from CLEO and *BABAR* is assumed, and a simple weighted average of $R^{+/0} \times \tau(B^+)/\tau(B^0)$ is computed.
- This weighted average is converted into an average value of $R^{+/0}$ by dividing it by the latest average of the lifetime ratio, $\tau(B^+)/\tau(B^0) = 1.079 \pm 0.007$.
- The measurement of $R^{+/0}$ from the Belle experiment is adjusted using the current values of $\tau(B^+)/\tau(B^0) = 1.079 \pm 0.007$ and $\tau(B^0) = 1.519 \pm 0.007$ ps.
- The weighted-average value of $R^{+/0}$ from CLEO and *BABAR* is then averaged with the adjusted value of the $R^{+/0}$ from Belle, assuming there is 100% correlation of the systematic uncertainty due to the limited knowledge of the lifetime ratio of $\tau(B^+)/\tau(B^0)$.

Most measurements of the $R^{+/0}$ have been made assuming isospin symmetry in specific decay rates and resulting in an average value of $R^{+/0}$,

$$R^{+/0} = 1.056 \pm 0.028 \text{ (total)}. \qquad (18.4.18)$$

This global average of $R^{+/0}$ is in good agreement with isospin invariance in the decay of $\Upsilon(4S) \to B\bar{B}$ pairs at the level of $2\sigma$.

In 2005, the *BABAR* collaboration reported (Aubert, 2005r) the first measurement of the branching fraction of $\mathcal{B}(\Upsilon(4S) \to B^0\overline{B}^0)$, $f_{00}$, using a novel technique: the partial reconstruction method (Section 17.5.1.4). This is a direct measurement of the $f_{00}$ that does not depend on the isospin invariance nor requires the knowledge of the B lifetime ratio, $\tau(B^+)/\tau(B^0)$. The measurement is based on the comparison between the number of events of a single- and double-tag sample using the decay of $\bar{B}^0 \to D^{*+}\ell^-\bar{\nu}_\ell$, and yields

$$f_{00} = 0.487 \pm 0.010(\text{stat}) \pm 0.008(\text{syst}) \qquad (18.4.19)$$

The two results in Equations 18.4.18 and 18.4.19 result from very different approaches and are completely independent. Combining the two results leads to $f_{+-} = 0.514 \pm 0.019$ and the sum of $f_{00}$ and $f_{+-}$ is equal to $1.001 \pm 0.030$ which is consistent with unity.

**Table 18.4.11.** Published measurements of $R^{+/0} = f_{+-}/f_{00}$ values in the decay of $\Upsilon(4S)$ resonance to $B\bar{B}$ pairs. The assumed lifetime ratio for each measurement is included.

| Experiment | Mode B $\rightarrow$ | $R^{+/0}$ Result | $\tau(B^+)/\tau(B^0)$ |
|---|---|---|---|
| CLEO (Alexander et al. (2001)) | $J/\psi K^*$ | $1.04 \pm 0.07 \pm 0.04$ | $1.066 \pm 0.024$ |
| *BaBar* (Aubert, 2002c) | $(c\bar{c})K^*$ | $1.10 \pm 0.06 \pm 0.05$ | $1.062 \pm 0.029$ |
| CLEO (Athar et al. (2002)) | $D^*\ell\nu$ | $1.058 \pm 0.084 \pm 0.136$ | $1.074 \pm 0.028$ |
| Belle (Hastings, 2003) | dilepton events | $1.01 \pm 0.03 \pm 0.09$ | $1.083 \pm 0.017$ |
| *BaBar* (Aubert, 2004j) | $J/\psi K$ | $1.006 \pm 0.036 \pm 0.031$ | $1.083 \pm 0.017$ |
| *BaBar* (Aubert, 2005k) | $(c\bar{c})K^*$ | $1.06 \pm 0.02 \pm 0.03$ | $1.086 \pm 0.017$ |
| Average | | $1.056 \pm 0.028$ (total) | $1.079 \pm 0.007$ |

Assuming $f_{00} + f_{+-} = 1$, the two results in Equations 18.4.18 and 18.4.19 lead to the most precise average values of $f_{00}$ and $f_{+-}$,

$$f_{00} = 0.487 \pm 0.006, \quad (18.4.20)$$
$$f_{+-} = 0.513 \pm 0.006 \quad (18.4.21)$$

and $R^{+/0}$,

$$R^{+/0} = 1.055 \pm 0.025 \quad (18.4.22)$$

where the $R^{+/0}$ ratio differs from unity by $2.2\sigma$.

### 18.4.7 Physics beyond the Standard Model

#### 18.4.7.1 Light Higgs Searches

*Motivation - low-mass, CP-odd Higgs boson*

The existence of a low-mass Higgs boson ($m_{A^0} < m_{b\bar{b}}$) became one of the beyond-the-Standard Model search topics for the bottomonium data sample of the *BABAR* Collaboration. This was motivated by extensions of the Standard Model, such as the next-to-minimal supersymmetric Standard Model, or NMSSM, that was developed to solve problems in the MSSM (cf. Dermisek and Gunion, 2005). While the MSSM requires two Higgs field doublets in order to provide mass to all particles in the theory, the NMSSM adds one more Higgs singlet field for a total of seven physical Higgs bosons. One of these is a *CP*-odd state, and is the lightest of the seven Higgs bosons (henceforth denoted as $A^0$). The purpose of this additional Higgs field singlet is to solve the "naturalness" problem in the MSSM (*e.g.* the apparently fine-tuned value of the $\mu$ parameter). Depending on the couplings and mass of this $A^0$, it was possible that the branching fraction for $\Upsilon \rightarrow \gamma A^0$ could have been as high as $10^{-4}$ and thus easily accessible to the $B$ Factory experiments (Dermisek, Gunion, and McElrath, 2007).

*Experimental searches*

The *BABAR* collaboration has searched for low-mass Higgs bosons produced in bottomonium decay using two decay modes: $A^0 \rightarrow \mu^+\mu^-$, $\tau^+\tau^-$ (Aubert, 2009ai,an). All of these searches assume that the parent Upsilon meson decays to the low-mass Higgs boson by radiating a photon,

$$\Upsilon \rightarrow \gamma A^0. \quad (18.4.23)$$

Prior to the publication of the *BABAR* searches described in this section, the CLEO Collaboration published searches for the same final states (Love et al., 2008) using $21.5 \times 10^6$ $\Upsilon(1S)$ decays. They obtained 90% confidence level limits on the branching fraction for $A^0 \rightarrow \tau^+\tau^-$ decay covering the range $2m_\tau < m_{A^0} < 9.5$ GeV/$c^2$ (where $m_\tau$ is the tau lepton mass) that ranged between and $(1 - 48) \times 10^{-5}$. They obtained 90% confidence level limits on the branching fraction for $A^0 \rightarrow \mu^+\mu^-$ decay covering the range $m_{A^0} < 3.6\,\text{GeV}/c^2$ that ranged between $(1 - 20) \times 10^{-6}$.

The *BABAR* searches use a data sample taken at a collider energy corresponding in the CM frame to the $\Upsilon(3S)$ mass. The sample contains $(121.8 \pm 1.2) \times 10^6$ $\Upsilon(3S)$ mesons.

The *BABAR* searches proceed by selecting events containing a good photon candidate and two opposite electric charge tracks. The selection of these final states diverge after the defintion of the topology due to the differing kinematics of the two di-lepton final states.

*$A^0 \rightarrow \mu^+\mu^-$*

The $A^0 \rightarrow \mu^+\mu^-$ final state is selected by requiring that the photon have a CM energy $E_\gamma \geq 0.5\,\text{GeV}$. Other photons can be present in an event only if their individual energies are below this threshold. The charged tracks are assigned a muon mass hypothesis and are henceforth referred to as "muon candidates" independent of whether additional particle identification is required. The muon candidates must original from a common point in space, and the vertex of the two charged tracks must have a $\chi^2 < 20$ (for 1 degree of freedom) and be displaced by no more than 2 cm from the nominal $e^+e^-$ interaction region in a plane transverse to the beams.

The dimuon system is combined with the highest-energy photon candidate to build an $\Upsilon(3S)$ candidate. A kinematic fit is performed to the three particles, constraining the total energy of the three particles to be within the beam-energy spread of the $e^+e^-$ collision that should have produced the $\Upsilon(3S)$, and the constraint that the particles originate from the primary interaction region. The fit is required to satisfy $\chi^2 < 36$ (for 6 degrees of freedom), which corresponds to a probability of rejecting good kinematic fits that is less than $10^{-6}$.

Particle identification is used in certain regions of photon energy in order to reject specific backgrounds. After the above kinematic selection, the primary background is determined from MC simulation to be $e^+e^- \to \mu^+\mu^-\gamma$. In a photon-energy region corresponding to $m_{A^0} < 1.05\,\text{GeV}/c^2$, contributions from $\phi \to K^+K^-$ (where $K^+ \to \mu^+\nu$) and $\rho \to \pi^+\pi^-$ are suppressed by requiring that both tracks be positively identified as muons using particle identification. In an $m_{A^0}$ region corresponding to the location of the $\eta_b$ mass, events are required to have no additional photons with $E_\gamma > 0.08\,\text{GeV}$; this suppresses radiative transitions of the $\Upsilon(3S)$ to the $\Upsilon(2S)$ through a $\chi_b$ state.

The efficiency of the above selection of $A^0 \to \mu^+\mu^-$ is studied using a signal MC simulation and varies between 24-44%, depending on $m_{A^0}$. The signal yield is extracted from the data in the range $0.212 \le m_{A^0} \le 9.3\,\text{GeV}/c^2$ using a maximum-likelihood fit to the variable,

$$m_R = \sqrt{m_{\mu\mu}^2 - 4m_\mu^2}. \qquad (18.4.24)$$

This equation represents twice the momentum of the muons in the rest frame of the parent particle. This is used instead of just $m_{\mu\mu}$ because it is a smooth function of $m_{\mu\mu}$ across the entire dimuon mass range, including close to the threshold for dimuon production ($m_{\mu\mu} \approx 2m_\mu$ corresponds to $m_R \approx 0$). The background distribution of $m_{\mu\mu}$ turns on sharply near threshold, whereas $m_R$ has a more gradual rise and can be more easily empirically modeled with a simple analytic function. Two functions are developed for use in the maximum likelihood fit: a signal function (determined from signal MC simulation) and a background function (determined from data taken at $\sqrt{s} = M_{\Upsilon(4S)}$). The signal model is constructed from a sum of two Crystal Ball functions; the parameters of this model are determined from many independent simulations of an $A^0$ whose mass varies across the range of interest, and these parameters are cross-checked using a sample of $J/\psi$ mesons obtained from initial-state radiation, $e^+e^- \to \gamma_{ISR}J/\psi$. The background model has alternative parameterizations in different regions of $m_R$; for $m_R < 0.23\,\text{GeV}/c^2$, a threshold (hyperbolic) function is used, while elsewhere the background is described by a first-order ($m_R < 9.3\,\text{GeV}/c^2$) or second-order ($m_R > 9.3\,\text{GeV}/c^2$) polynomial.

The fit for signal and background is performed in steps whose size varies by region. In addition to continuum background, specific backgrounds that contribute peaks to the mass spectrum are accounted using the same model as for signal. The $J/\psi$ and $\psi(2S)$ regions are excluded from the fit since those contributions overwhelm any possible signals in those regions.

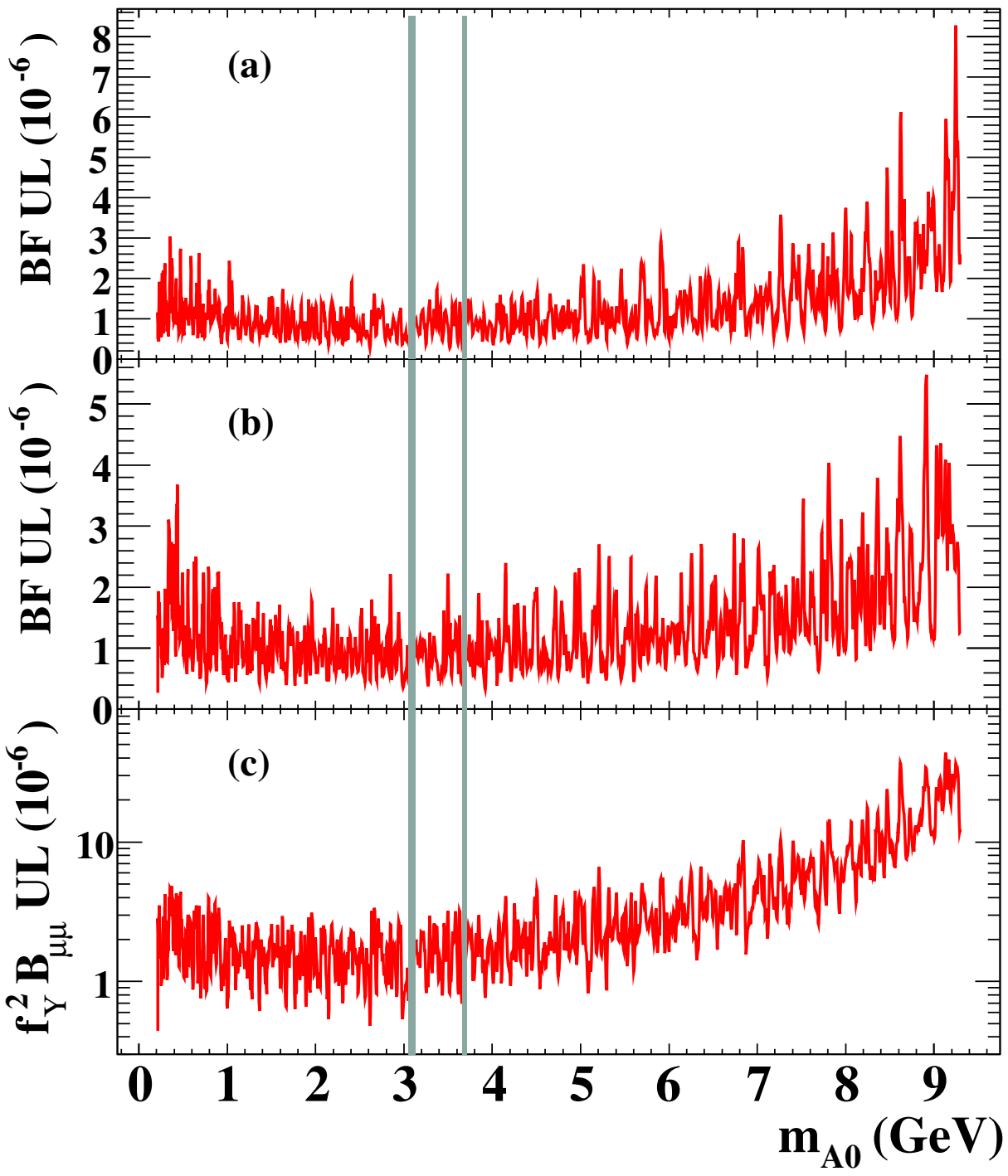

**Figure 18.4.22.** Upper limits on the product of branching fractions (at 90% C.L.) for $\Upsilon(nS) \to \gamma A^0$ and $A^0 \to \mu^+\mu^-$ for (a) $\Upsilon(3S)$ and (b) $\Upsilon(2S)$. The upper limit on the effective coupling, $f_\Upsilon^2 \mathcal{B}_{\mu\mu}$, is shown in (c). From (Aubert, 2009an).

The *BABAR* Collaboration reports no evidence for decay of a low-mass Higgs boson using these data and computes upper limits on the branching fraction for $\Upsilon(3S) \to \gamma A^0 \to \gamma(\mu^+\mu^-)$, shown in Fig. 18.4.22. These upper limits, including systematic uncertainties, range between $(0.25 - 5.2) \times 10^{-6}$ depending on the value of $m_R$ (at the 90% confidence level). These results constrain the branching fraction for $\eta_b \to \mu^+\mu^-$ to be $< 0.8\%$ at the 90% confidence level.

*$A^0 \to \tau^+\tau^-$*

The search for $A^0 \to \tau^+\tau^-$ proceeds similarly to the dimuon search. Events must contain a photon satisfying $E_\gamma > 0.1\,\text{GeV}$; any additional photons (up to nine are allowed) must each satisfy $E < 0.1\,\text{GeV}$. The dominant background processes in this search are due to $e^+e^- \to \gamma\tau^+\tau^-$, $e^+e^- \to e^+e^-e^+e^-$, $e^+e^- \to e^+e^-\mu^+\mu^-$, and $e^+e^- \to q\bar{q}$. These are rejected by using eight discriminating variables: the total CM energy calculated from the two leptons and the most energetic photon; the squared missing mass obtained from the missing four-momentum; the aplanarity of the photon and $A^0$ candidate, which is the cosine of the angle between the photon and the plane of the leptons; the largest cosine between the photon and one of the tracks; the cosine of the polar angle of the highest-momentum track; the transverse momentum of the event calculated in the CM frame; the cosine of the polar angle of the missing momentum vector; and the cosine of the opening angle between the tracks in the photon recoil frame. The selection on these variables is optimized simultaneously in order to achieve the best value of $S/\sqrt{B}$, where $S$ is the number of expected signal and $B$ is the number of expected background. The backgrounds vary depending on the photon energy, so this optimization is performed as a function of photon energy in five regions which slightly overlap in order to reduce discontinuities in performance between the regions.

The energy of the photon in the $\Upsilon(3S)$ rest frame is used to define a range of $A^0$ masses studied in this search, using the relationship

$$m^2_{A^0} = m^2_{\Upsilon(3S)} - 2m_{\Upsilon(3S)}E_\gamma \tag{18.4.25}$$

where $m_{\Upsilon(3S)}$ is the nominal $\Upsilon(3S)$ mass. The range of photon energies studied corresponds to $m_{A^0} = [4.03, 10.10]\,\mathrm{GeV}/c^2$, excluding the region $m_{A^0} = [9.52, 9.61]\,\mathrm{GeV}/c^2$ due to an irreducible background from photons produced in the process $\Upsilon(3S) \to \gamma\chi_{bJ}(2P)$, $\chi_{bJ}(2P) \to \gamma\Upsilon(1S)$, where $J = 0$, 1, 2. The photon energy resolution varies over the range in this search, increasing from 8 MeV at $E_\gamma \approx 0.2$ GeV to 55 MeV at $E_\gamma \approx 4.5$ GeV. The efficiency of event selection also varies with photon energy, ranging as follows: 10-14% for the $\tau\tau \to ee$ final state; 12-20% for $\tau\tau \to e\mu$; and 22-26% for $\tau\tau \to \mu\mu$ (neutrinos are not explicitly written in the final states).

The photon energy spectrum is modeled using the combination of a peaking function for signal and a predominantly smooth function for background. The data are first treated as purely background and fit with only the latter function. This allows the parameters of the background to be determined as initial values for the next stage of the fit to the data.

Backgrounds causing real peaks in the photon spectrum are expected from radiative decays from the $\Upsilon(3S)$ resonance to lower-mass bottomonium states, specifically $\Upsilon(3S) \to \gamma\chi_{bJ}(2P)$, $\chi_{bJ}(2P) \to \gamma\Upsilon(nS)$, and $\Upsilon(nS) \to \tau^+\tau^-$ (J = 0, 1, 2; n = 1, 2). Peaks arise in the photon spectrum when the photon from the $\chi_{bJ}(2P)$ decay is used as the photon radiated by the $\Upsilon(3S)$ when it decays to an $A^0$. Each of the peaks is described using a Crystal Ball function whose means are fixed by the photon energies expected from the PDG values of the bottomonia masses (Beringer et al., 2012) and whose widths are fixed from the MC predictions of the reconstructed widths of the photon peaks. The other parameters of the Crystal Ball functions are also fixed from the MC simulation of these decays.

The results of a fit of the background model to the photon energy spectrum in each final state are shown in Fig. 18.4.23. A complete fit of the data including the signal model is performed by scanning the photon energy in 307 steps and fitting for signal and background yields at each step. This procedure finds no significant yield of signal events anywhere in the spectrum. The branching fraction product $\mathcal{B}(\Upsilon(3S) \to \gamma A^0)\mathcal{B}(A^0 \to \tau^+\tau^-)$ is calculated, along with the upper limit at the 90% C.L., both as a function of Higgs mass, as shown in Fig. 18.4.24. The upper limits on the product branching fraction range between $(1.5 - 16) \times 10^{-5}$ at 90% C.L. for a mass range of $4.03 < m_{A^0} < 10.10\,\mathrm{GeV}/c^2$, excluding $9.52 < m_{A^0} < 9.61$ $\mathrm{GeV}/c^2$ to veto the $\chi_{bJ}(2P)$ with $\chi_{bJ}(2P) \to \gamma\Upsilon(1S)$.

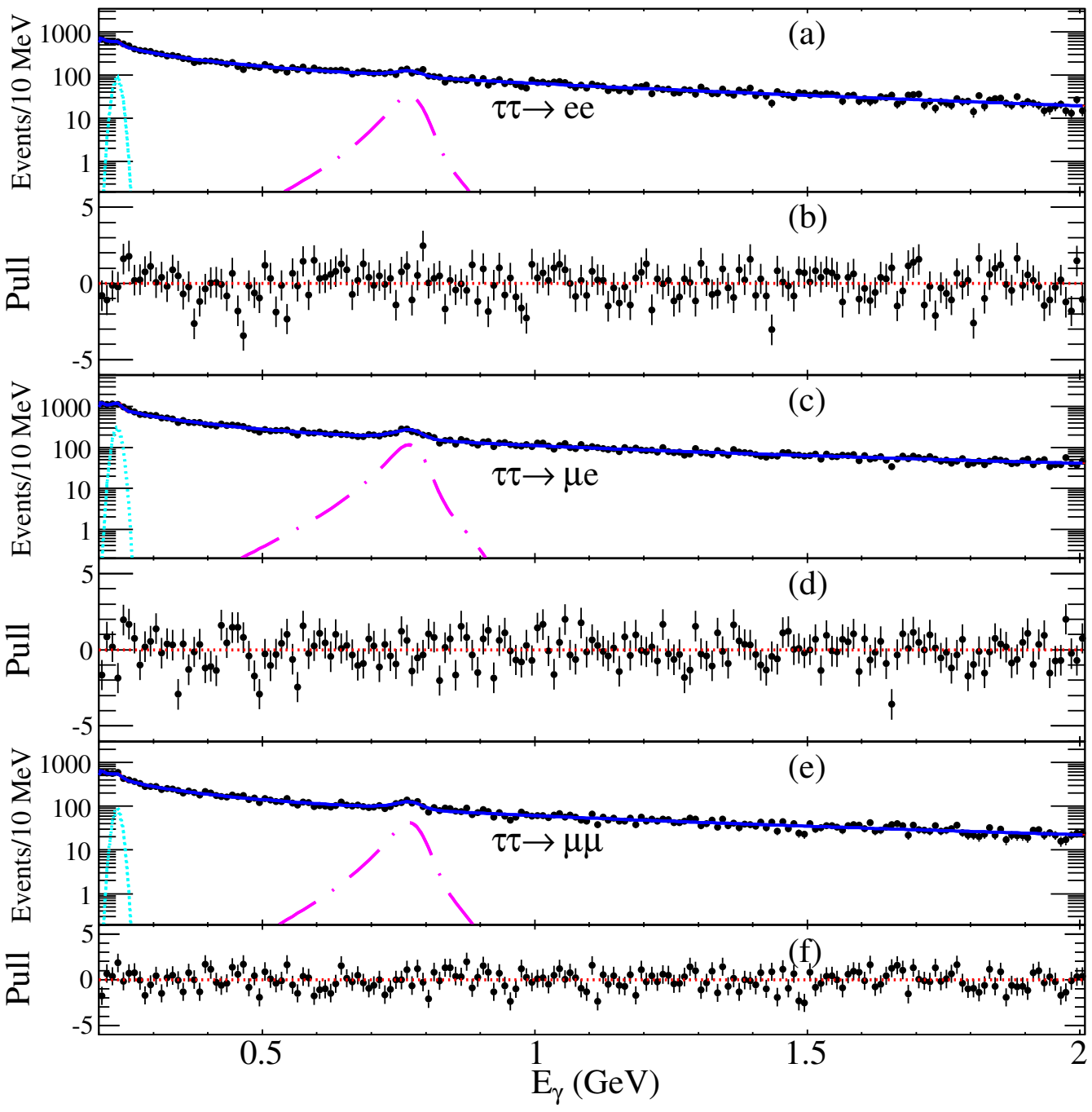

**Figure 18.4.23.** (a), (c), (e): Photon energy distributions for the different $\tau\tau$-decay modes (Aubert, 2009ai). Filled circles show the data; dotted lines represent contributions from $\Upsilon(3S) \to \gamma\chi_{bJ}(2P)$, $\chi_{bJ}(2P) \to \gamma\Upsilon(2S)$; dotted-dashed lines show contributions from $\Upsilon(3S) \to \gamma\chi_{bJ}(2P)$, $\chi_{bJ}(2P) \to \gamma\Upsilon(1S)$; and solid lines show the total background function. For each $\tau\tau$-decay mode, the difference between the background function and the data divided by the uncertainty in the data is shown in (b), (d) and (f).

*Impact of the results*

The low-mass Higgs boson models discussed earlier in this section predicted that the branching fraction for $\Upsilon(1S) \to \gamma A^0$ could range as high as $\sim 10^{-3}$ (the range of possible branching fractions is dependent on the specific NMSSM model used). The measurements from the *BABAR* collaboration put strong constraints on the upper range of this kind of decay down to the level of $10^{-6}$, removing a few orders of magnitude of possible range from the top level of predicted branching fractions.

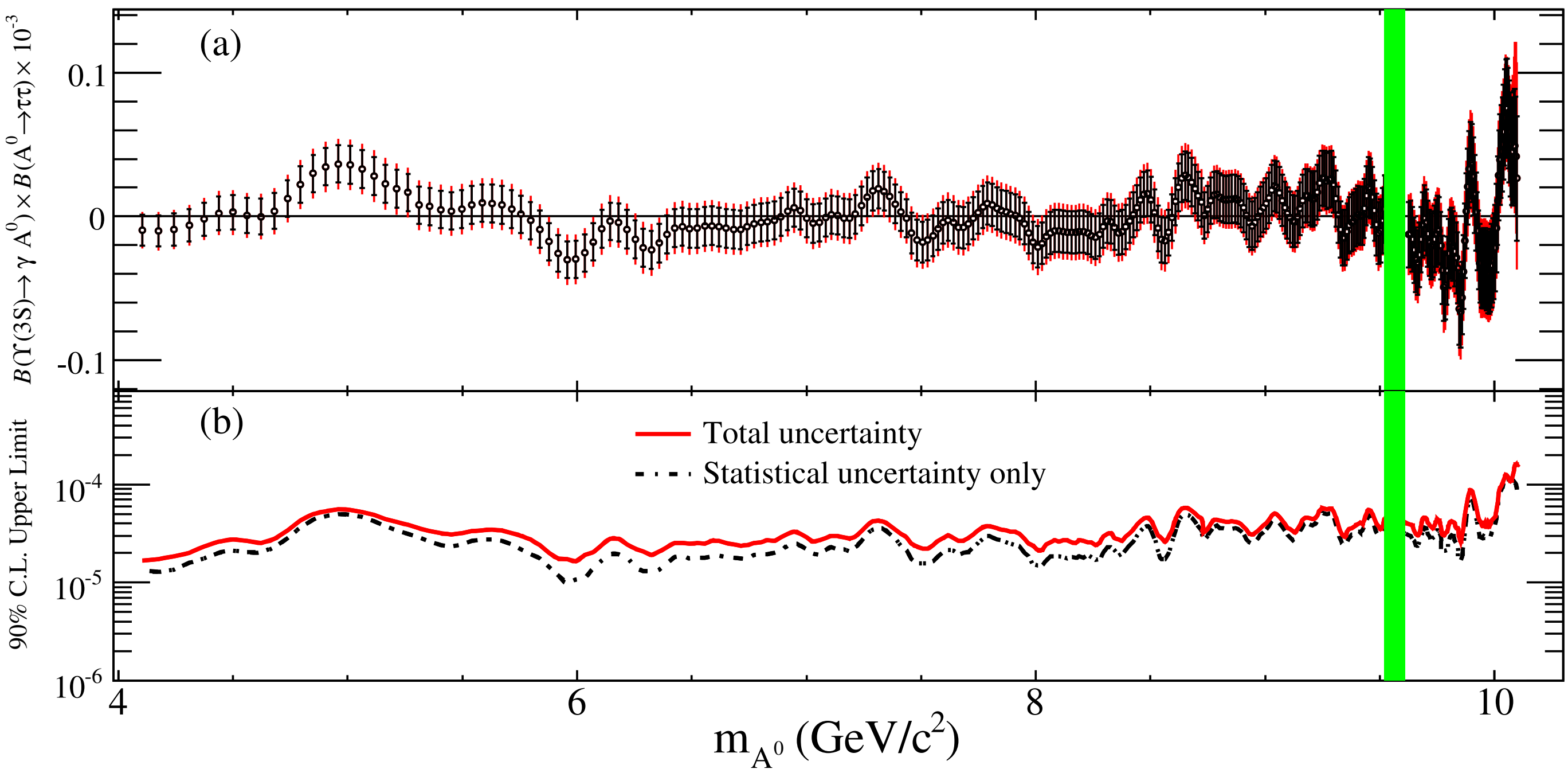

**Figure 18.4.24.** (a) Product branching fractions as a function of the Higgs mass (Aubert, 2009ai). For each point, both the statistical uncertainty (from the central value to the horizontal bar) and the total uncertainty (statistical and systematic added in quadrature) are shown (from the central value to the end of the error bar). In (b), the corresponding 90% C.L. upper limits on the product of the branching fractions versus the Higgs mass values are shown, with total uncertainty (solid line) and statistical uncertainty only (dashed line). The shaded vertical region represents the excluded mass range corresponding to the $\chi_{bJ}(2P) \to \gamma\Upsilon(1S)$ states.

### 18.4.7.2 Invisible Final States of the $\Upsilon(1S)$

*Motivation - low-mass dark matter*

The nature of dark matter is one of the great modern physics puzzles. Assuming dark matter is composed of at least one species of particle, the properties of this particle have not been measured (*e.g.* mass). If the mass of dark matter is small ($< m_{b\bar{b}}$), then there is the possibility of detecting it using rare processes involving undetectable (invisible) final states. One of these, $\Upsilon \to$ invisible, was motived by work by McElrath (2005), where it was suggested that a new interaction that couples Standard Model particles to dark matter particles could mediate the decay of the $\Upsilon$. Based on the interaction cross-section required to achieve the "freeze-out" of dark matter annihilations in the early universe (a process that is required to explain the significant remnant of dark matter in today's universe), it was estimated that the branching fractions for $\Upsilon \to (\gamma+)$ invisible (the dominant decay mechanism depended on the spin of the dark matter constituent) could be as high as 0.41% — easily measured at the B-factories with even a modest sample of $\Upsilon$ mesons.

The *BABAR* and Belle collaborations have both searched for invisible final states of $\Upsilon(1S)$ decay (Tajima, 2007; Aubert, 2009b; del Amo Sanchez, 2011j). Both collaborations produced results in the search for purely invisible final states, $\Upsilon \to$ invisible while the *BABAR* collaboration also produced results for radiative invisible final states, $\Upsilon \to \gamma +$invisible. These searches are sensitive to different possible angular momentum configurations of unknown invisible final states.

*Searches for $\Upsilon \to$ invisible*

The *BABAR* and Belle searches for purely invisible final states proceed similarly. The Belle search appeared first and used a sample of $11 \times 10^6$ $\Upsilon(3S)$ mesons (Tajima, 2007), while the *BABAR* search appeared later and used a sample of $91.4 \times 10^6$ $\Upsilon(3S)$ mesons (Aubert, 2009b). Both searches used the transition

$$\Upsilon(3S) \to \pi^+\pi^-\Upsilon(1S) \tag{18.4.26}$$

to "tag" the presence of the $\Upsilon(1S)$ meson without reconstructing it by using the kinematics of the dipion system. Specifically, if the pions are both produced recoiling against the $\Upsilon(1S)$ state then from four-momentum conservation the mass of the system recoiling against the dipion is given by

$$M^2_{\rm recoil}(\pi^+\pi^-) = s + M^2_{\pi^+\pi^-} - 2\sqrt{s}E^*_{\pi^+\pi^-} \tag{18.4.27}$$

where $s$ is the square of the collider CM energy and $M_{\pi^+\pi^-}$ ($E_{\pi^+\pi^-}$) is the mass (energy) of the dipion system. For a real dipion transition $\Upsilon(3S) \to \pi^+\pi^-\Upsilon(1S)$, $M^2_{\rm recoil}(\pi^+\pi^-) = M^2_{\Upsilon(1S)}$.

The major challenges in this search are the trigger efficiency for signal events and the large background from pions that come from non-transition decays and from real transition decays where the final-state products of the $\Upsilon(1S)$ are simply undetected due to detector effects. The trigger efficiency is a challenge due to the low transverse momentum possessed by the pions; the energy from the transition is shared between the two pions, typically leading to one low-momentum pion and one higher-momentum pion.

Both the Belle and *BABAR* trigger systems require that low-multiplicity events be triggered only when at least one of the tracks has a sufficient $p_T$ to distinguish it from background, and that the opening-angle between the tracks in the plane transverse to the beams satisfy a minimum requirement. The Belle collaboration evaluated their trigger efficiency by studying the efficiency with which events are selected by a single-track trigger and then subsequently by different requirements on a second track in those events. The *BABAR* collaboration evaluated their trigger efficiency by explicitly reconstructing a control sample of events where $\Upsilon(3S) \to \pi^+\pi^-\Upsilon(1S)$ and the $\Upsilon(1S)$ then decays to a pair of leptons (either electrons or muons). The pions in this sample have identical kinematics to those in an equivalent $\Upsilon(1S) \to$ invisible decay, except that these events are triggered by the high-momentum leptons and not the lower-momentum pions. A selection similar to the one applied by the *BABAR* trigger was then used on the pions to evaluate the efficiency with which the dipion system is selected.

The backgrounds to this search come from the two sources mentioned above: events with pions that come from sources other than the $\Upsilon(3S) \to \Upsilon(1S)$ dipion transition (combinatorial) and events with pions that come from a real transition but where the $\Upsilon(1S)$ decay products are simply unreconstructed due to detector effects (peaking). The pions from combinatorial sources have no peak at $M_{\rm recoil}(\pi^+\pi^-) = M_{\Upsilon(1S)}$ but dominate the data samples prior to any rejection after selection the pions. Both collaboration use combinations of kinematic information (Belle uses a Fisher discriminant while *BABAR* uses a Random Forest of Decision Trees - see Chapter 4 for a description of these tools) to reject this source of background. Any remaining background has a smooth distribution through the signal region around the $\Upsilon(1S)$ mass and is easily modeled using a polynomial function whose parameters are determined by fitting the data directly.

The peaking background from real dipion transitions is studied using Monte Carlo simulations and the control sample described above, where $\Upsilon(1S) \to \ell^+\ell^-$ is explicitly reconstructed using electron and muon final states. Both collaborations compare the rate at which both final-state leptons, or just one final-state lepton, are reconstructed as a function of polar angle in the detector. This allows them to correct the MC simulation of these backgrounds using data measurements of the detector acceptance for the $\Upsilon(1S)$ final-state products. In addition to this technique, the *BABAR* collaboration studied the non-leptonic $\Upsilon(1S)$ decay backgrounds by using a control sample where a photon of energy $E_\gamma > 0.250$ GeV is present in addition to the pions; this selection is orthogonal to the nominal signal selection. This sample is enriched in hadronic $\Upsilon(1S)$ decays and was used to correct the modeling of acceptance for such decays in the nominal selection.

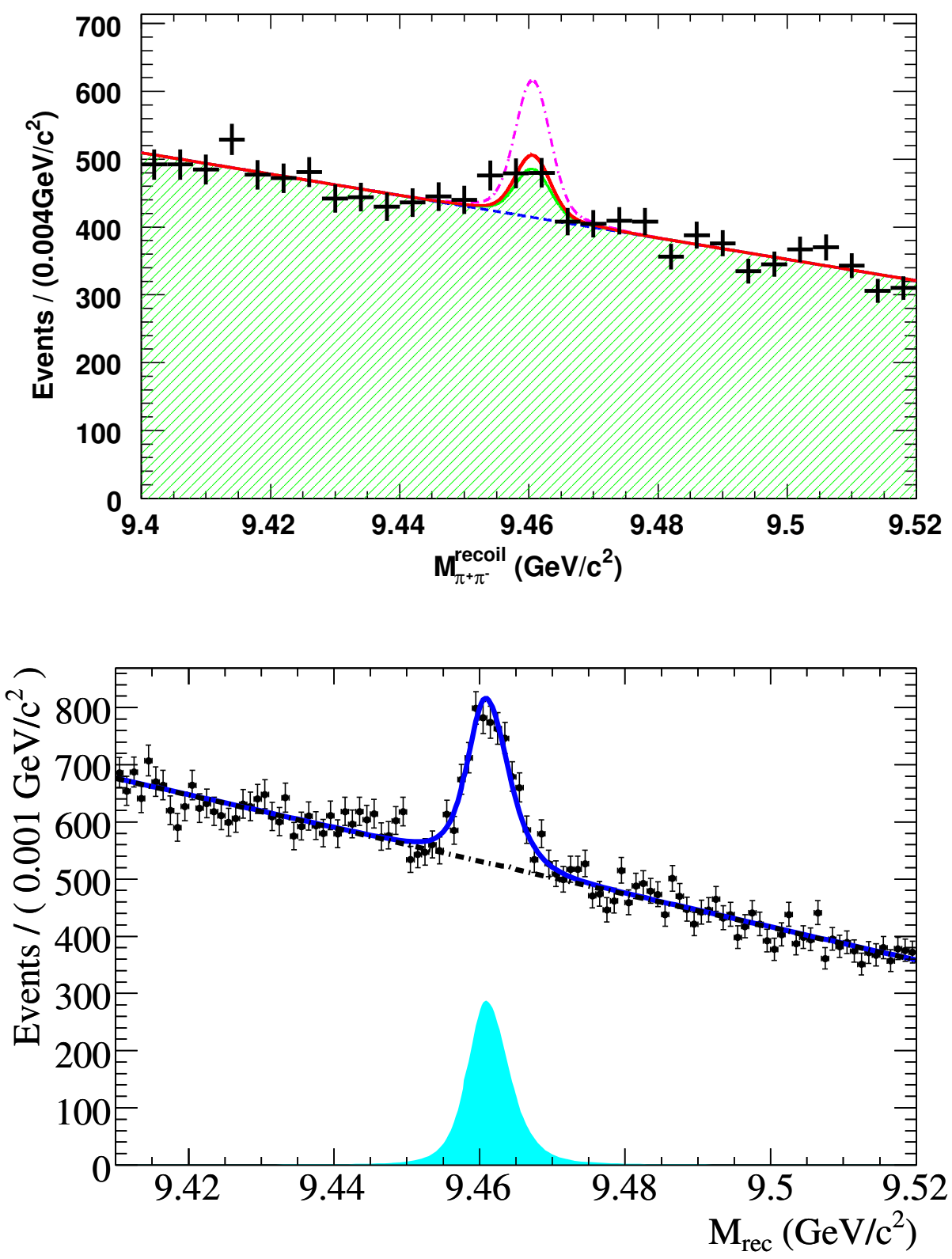

**Figure 18.4.25.** The fits to the recoil mass spectra. (top) The Belle fit, where the solid curve shows the result of the fit to signal plus background distributions, the shaded area shows the total background contribution, the dashed line shows the combinatorial background contribution, and the dot-dashed line shows the expected signal for $\mathcal{B}(\Upsilon(1S) \to \text{invisible}) = 6 \times 10^{-3}$ (Tajima, 2007). (bottom) The *BABAR* fit to $M_{rec} \equiv M_{\rm recoil}(\pi^+\pi^-)$, where the solid line shows the fit including both combinatorial and peaking contributions and the dash-dotted line indicates the contribution only from combinatorial background (Aubert, 2009b).

The shape of the real transition background will be identical to that of the signal because in both cases a real $\Upsilon(1S)$ is recoiling against the dipion system. Thus the recoil mass shape of this background (and that of the signal) can be determined from the $\Upsilon(1S) \to \ell^+\ell^-$ control sample but the yield of the real transition backgrounds must be determined from the aforementioned studies and fixed in the final fit to the data. The Belle collaboration estimated their real transition background to be $133.2^{+19.7}_{-14.69}$ events, while the *BABAR* collaboration measured their peaking background to be $2444 \pm 123$ events, with the dominant contribution coming from $\Upsilon(1S) \to \ell^+\ell^-$ decays where the final-state leptons are undetected.

The fits to the recoil mass spectra from both experiments are shown in Fig. 18.4.25. Neither experiment observed a significant deviation from their expected yield of peaking events due to background sources. Both experiments set upper limits on the branching fraction for $\Upsilon(1S) \to$ invisible. Folding in the measured value for $\Upsilon(3S) \to \pi^+\pi^-\Upsilon(1S)$, the Belle collaboration determined that

$$\mathcal{B}(\Upsilon(1S) \to \text{invisible} < 2.5 \times 10^{-3} \qquad (18.4.28)$$

at 90% C.L. and the *BABAR* collaboration determined that

$$\mathcal{B}(\Upsilon(1S) \to \text{invisible} < 3.0 \times 10^{-4} \qquad (18.4.29)$$

at 90% C.L.

### *Impact of the results*

The measurements described here for $\Upsilon(1S) \to$ invisible leave only about an order-of-magnitude of branching fraction space left before encountering the Standard Model predicted rate for $\Upsilon(1S) \to \nu\overline{\nu}$. This has closed much of the space for low-mass dark matter, with $m_\chi < m_{\Upsilon(1S)}/2$, to be produced in this way.

### *Search for* $\Upsilon \to \gamma +$ invisible

The search for radiative invisible final states, $\Upsilon(1S) \to \gamma +$ invisible also proceeds from a dipion transition sample but uses a sample of $98.3 \times 10^6$ $\Upsilon(2S)$ mesons and the transition $\Upsilon(2S) \to \pi^+\pi^-\Upsilon(1S)$ to tag the presence of the $\Upsilon(1S)$ meson. Such a search has been performed by *BABAR* (del Amo Sanchez, 2011j). The presence of a photon in addition to the dipion system is then used the tag the decay of the $\Upsilon(1S)$ via $\Upsilon(1S) \to \gamma +$ invisible. The analysis assumes that the invisible system recoiling against the photon is a resonance (*e.g.* a low-mass Higgs), denoted $A^0$, that subsequently decays into a two-body invisible final state, $A^0 \to \chi\bar{\chi}$, where $\chi$ denotes an undetectable long-lived particle.

The signal events have a low multiplicity and are triggered in two ways: either by the presence of a pair of tracks each with $p_T > 0.25\,\text{GeV}/c$ or by the presence of a single photon with energy in the CM frame $E^* > 0.8\,\text{GeV}$. Because of the trigger selection depends on the energy of the photon, the analysis is performed in regions corresponding to the mass of the recoiling resonance. The two regions are a high-mass region, $7.5 \leq m_{A^0} \leq 9.2\,\text{GeV}/c^2$ (corresponding to $3.5 \leq m_\chi \leq 4.5\,\text{GeV}/c^2$) and a low-mass region $m_{A^0} \leq 8.0\,\text{GeV}/c^2$ ($m_\chi \leq 4\,\text{GeV}/c^2$). The low-mass region relies entirely on the single-photon trigger, while the high-mass region relies entirely on the track trigger.

The reconstructed dipion system is required to contain two positively identified pion candidates (to reject electron and muon contamination) and have $p_T < 0.5\,\text{GeV}/c$. Neither pion can have momentum $p > 1.0\,\text{GeV}/c$. The photon must have a CM energy satisfying $E^* > 0.15\,\text{GeV}$ and lie well within the central part of the electromagnetic calorimeter. Additional photons can be present as long as their individual energies are less than that of the signal photon and their total energy in the laboratory frame does not exceed 0.14 GeV. A multilayer perceptron neural network is then used to combine kinematic variables from the dipion system into a single discriminant that can reject background. The neural network is trained on data taken at a collider CM energy below the $\Upsilon(2S)$ resonance, and on signal MC simulation. In the low-mass region, this approach retains 87% of simulated signal events while rejecting 96% of events from non-$\Upsilon(2S)$ (continuum) events. In the high-mass region, this approach retains 73% of signal while rejecting 98% of continuum background.

In addition to backgrounds from sources other than the $\Upsilon(2S)$, there could be backgrounds from real $\Upsilon(1S)$ radiative decays where the final-state products are difficult to detect reliably. For instance, $\Upsilon(2S) \to \pi^+\pi^-\Upsilon(1S)$, where the $\Upsilon(1S)$ then decays to either $\Upsilon(1S) \to \gamma n\bar{n}$ or $\Upsilon(1S) \to \gamma K^0_L K^0_L$, are allowed decays where the final-state hadrons are not efficiently reconstructed in the *BABAR* detector. To reject these backgrounds, events are rejected where there is activity in the *BABAR* instrumented flux return within a 20° window opposite the reconstructed signal photon. This requirement is only applied in the low-mass region for $m_{A^0} < 4\,\text{GeV}/c^2$. In the high-mass region there is a potential contamination from the process $e^+e^- \to e^+e^-\gamma^*\gamma^*$ where $\gamma^*\gamma^* \to \eta'$ and $\eta' \to \gamma\pi^+\pi^-$ while the electron and positron escape detection at low-angles to the beams. This is largely rejected by requiring the opening angle between the photon and the dipion system be no more than 160°.

The signal is extracted from a maximum likelihood fit to two variables: the dipion recoil mass $M_{\text{recoil}}(\pi^+\pi^-)$ and the "missing mass", *i.e.* the mass of the system recoiling against the reconstructed dipion and photon,

$$M^2_{\text{recoil}}(\pi^+\pi^-\gamma) = (P_{e^+e^-} - P_{\pi^+\pi^-} - P_\gamma)^2. \qquad (18.4.30)$$

The fit is performed in steps of $M^2_{\text{recoil}}(\pi^+\pi^-\gamma)$. The models contain contributions from multiple sources, including signal (whose shape is determined from MC simulation), continuum background, radiative $\Upsilon(1S)$ decays, and background from real $\Upsilon(3S) \to \Upsilon(1S)$ transitions. Projections of the fits in each of the two dimensions are shown in Fig. 18.4.26. No significant yield of signal events is obtained from any of the scan points in the fits. The 90% C.L. upper limits on the product of branching fraction $\mathcal{B}(\Upsilon(1S) \to \gamma A^0) \times \mathcal{B}(A^0 \to \text{invisible})$ are shown in Fig. 18.4.27 and ranges $(1.9 - 4.5) \times 10^{-6}$ for the low-mass region and $(2.7 - 37) \times 10^{-6}$ in the high-mass region, assuming a scalar $A^0$.

The low-mass dark matter models cited earlier in this section predicted that the branching fraction for the radiative final state could be as high as $10^{-5} - 10^{-4}$. The experimental results discussed here cover well that upper range of the predicted branching fractions and exclude the largest possible rates predicted by these models.

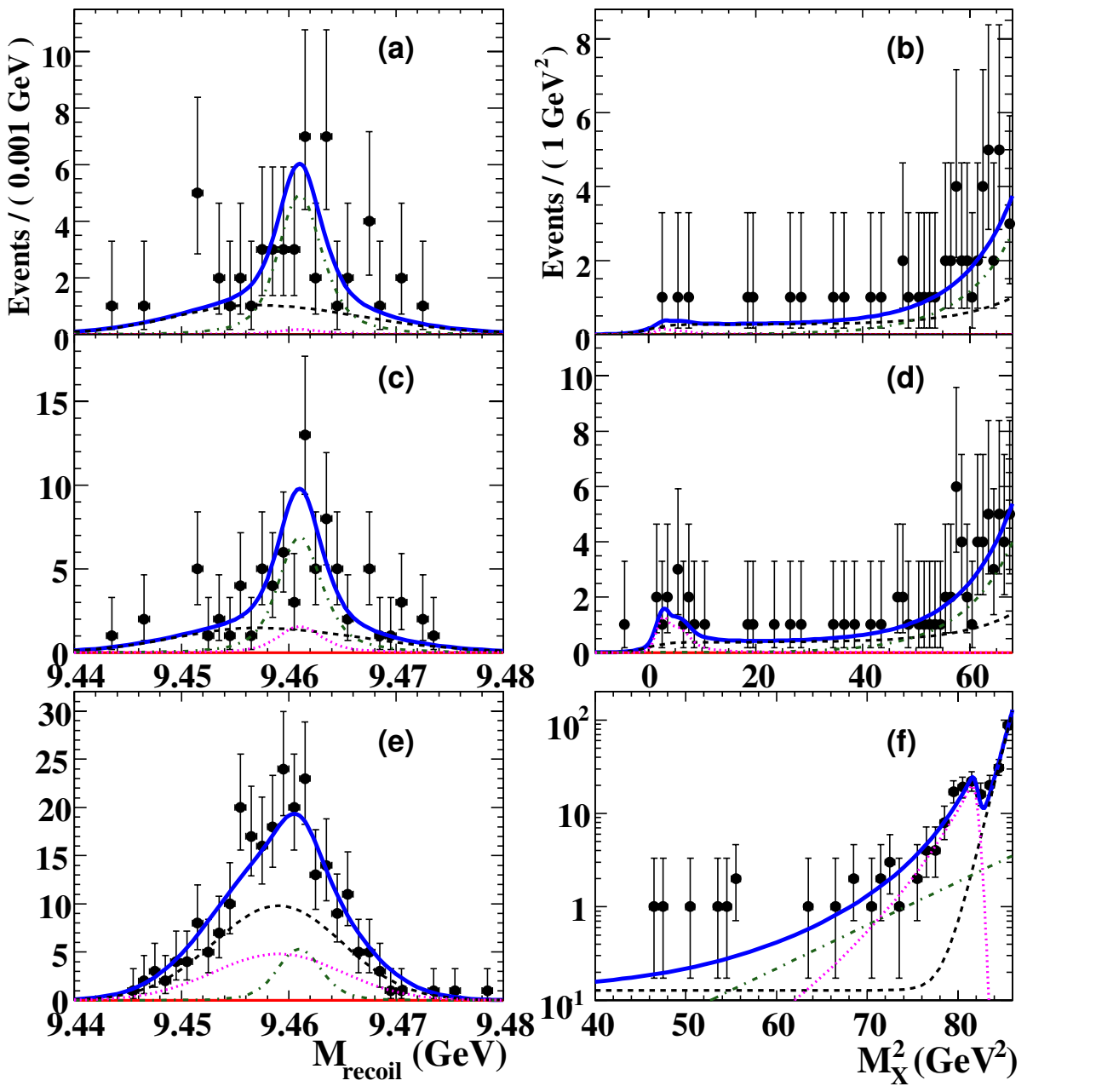

**Figure 18.4.26.** Projection plots from the fit with $N_{\rm sig} = 0$ onto (a,c,e) recoil mass $M_{\rm recoil} \equiv M_{\rm recoil}(\pi^+\pi^-)$ and (b,d,f) missing mass-squared $M_X^2 \equiv M_{\rm recoil}^2(\pi^+\pi^-\gamma)$ (del Amo Sanchez, 2011j). (a,b): low-mass region with a veto on activity in the *BABAR* instrumented flux return (IFR); (c,d): low-mass region without IFR veto; (e,f): high-mass region. Overlaid is the fit with $N_{\rm sig} = 0$ (solid blue line), continuum background (black dashed line), radiative leptonic $\Upsilon(1S)$ decays (green dash-dotted line), and (c,d) radiative hadronic $\Upsilon(1S)$ decays or (e,f) $\eta'$ background (magenta dotted line).

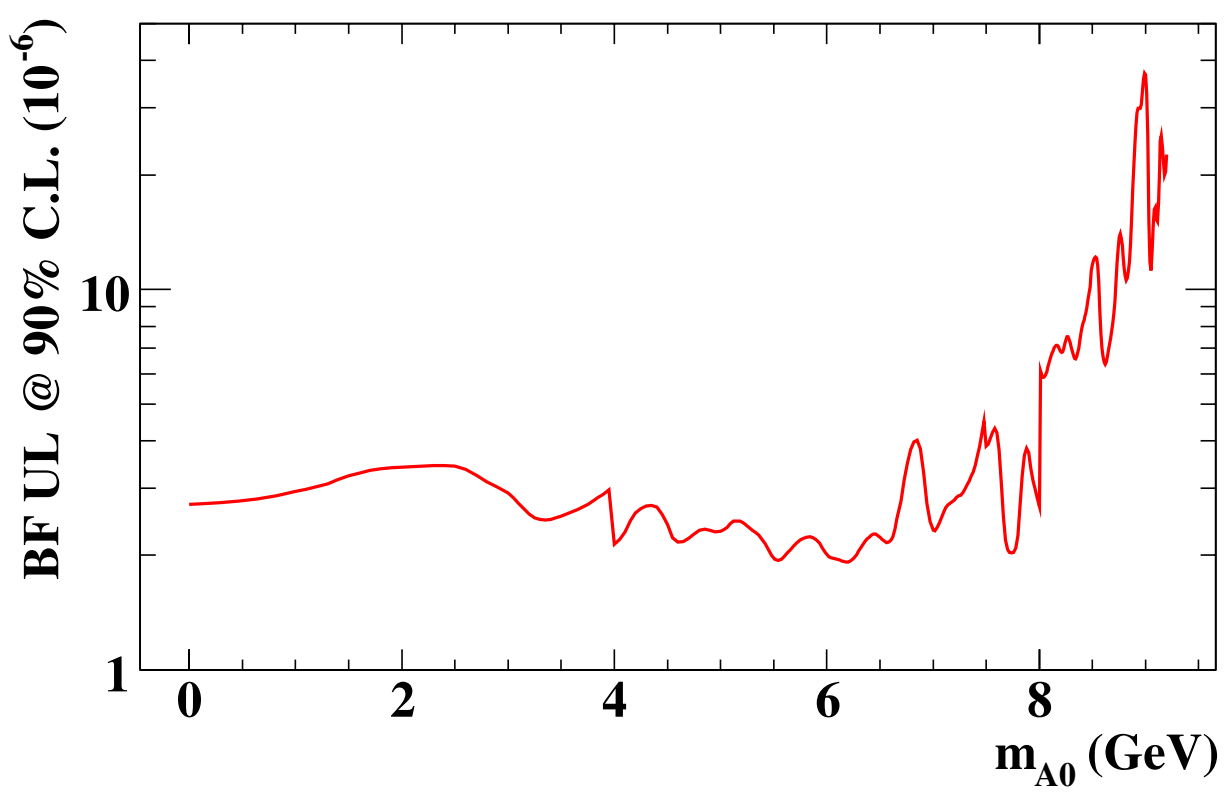

**Figure 18.4.27.** 90% C.L. upper limits for $\mathcal{B}(\Upsilon(1S) \to \gamma A^0) \times \mathcal{B}(A^0 \to \text{invisible})$ (del Amo Sanchez, 2011j).

### 18.4.7.3 Search for Lepton Flavor Violation

*Motivation: new interactions*

Lepton flavor is an accidentally conserved quantum number in the original formulation of the Standard Model; however, the observation of neutrino mixing implies that charged lepton flavor violation should occur, albeit at a scale that is suppressed by a factor of $(\Delta m_\nu^2/M_W^2)^2 \leq 10^{-48}$, as discussed by Feinberg (1958), Bilenky and Pontecorvo (1976), Strumia and Vissani (2006). An experimental observation of such violation at present experimental capabilities would be unambiguous evidence of physics beyond the Standard Model, as discussed by Ellis, Gomez, Leontaris, Lola, and Nanopoulos (2000), Ellis, Raidal, and Yanagida (2004), Pati and Salam (1974), Georgi and Glashow (1974). For instance, supersymmetry includes mechanisms by which such violation naturally occurs.

*Experimental measurement*

The *BABAR* Collaboration performed a search for lepton flavor violation using the decays $\Upsilon(nS) \to (e^\pm/\mu^\pm)\tau^\mp$, where $n = 2, 3$. The searches were performed using $(98.6 \pm 0.9) \times 10^6$ $\Upsilon(2S)$ decays and $(116.7 \pm 1.2) \times 10^6$ $\Upsilon(3S)$ decays (Lees, 2010c). In addition, data from other resonances and from runs taken away from Upsilon resonances were used to characterize and study the backgrounds.

The search uses the fact that the collider is producing the parent Upsilon resonance at rest in the CM frame; the electron or muon that results directly from the Upsilon decay (the "primary lepton") will have energy very close to the single-beam energy in the Upsilon rest frame, $E_B = \sqrt{s}/2$. The primary tau lepton will decay, and a second lepton, or a charged pion, consistent with tau decay is then searched for. If a second lepton is found, it is required to have a different flavor from the primary lepton. If a pion is found, one or two additional neutral pions must also be reconstructed in the same event. These measures are required in order to suppress Bhabha events or $\mu$-pair backgrounds.

The main source of background in this search is from $e^+e^- \to \tau^+\tau^-$ events. The four individual search channels each have other sources of background, arising primarily from lepton and hadron misidentification. Bhabha events are further suppressed by requiring that the visible mass in each candidate event has less than 95% of the collider CM energy (indicating the presence of neutrinos, which are not present in Bhabha events). In addition, the missing momentum in each event must not point close to the beamline. Higher-order QED backgrounds, as from two-photon fusion, are suppressed by requiring that the transverse momentum component of the charge particles' vector sum is more than 20% of the quantity $\sqrt{s} - |\boldsymbol{p}_1| - |\boldsymbol{p}_2|$, where $\boldsymbol{p}_i$ is the three-momentum of charged particle $i$.

The primary lepton momentum is defined by the requirement that $x \equiv |\boldsymbol{p}_1|/E_B > 0.75$. For the hadronic tau decays, the momentum of the tau daughter charged particle is required to satisfy $|\boldsymbol{p}_2|/E_B < 0.8$, and the invariant masses of the track and neutral pion(s) system must be consistent with the mass of either the $\rho^\pm$ or the $a_1^\pm$. The $\mu$-pair background in the $\mu\tau$ channel is suppressed by requiring that the opening angle between the charged tracks in the plane transverse to the beams is less than 172°.

After all selection criteria are applied, the selection efficiency determined from a signal MC simulation range

**Table 18.4.12.** Branching fractions and 90% CL ULs for lepton-flavor violating $\Upsilon(nS) \to \ell^{\pm}\tau^{\mp}$ decays (Lees, 2010c). The first error is statistical and the second is systematic.

| | $\mathcal{B}$ $(10^{-6})$ | UL $(10^{-6})$ |
|---|---|---|
| $\mathcal{B}(\Upsilon(2S) \to e^+\tau^-)$ | $0.6^{+1.5+0.5}_{-1.4-0.6}$ | $< 3.2$ |
| $\mathcal{B}(\Upsilon(2S) \to \mu^+\tau^-)$ | $0.2^{+1.5+1.0}_{-1.3-1.2}$ | $< 3.3$ |
| $\mathcal{B}(\Upsilon(3S) \to e^+\tau^-)$ | $1.8^{+1.7+0.8}_{-1.4-0.7}$ | $< 4.2$ |
| $\mathcal{B}(\Upsilon(3S) \to \mu^+\tau^-)$ | $-0.8^{+1.5+1.4}_{-1.5-1.3}$ | $< 3.1$ |

between $(4-6)\%$, depending on the signal mode, including the tau decay branching fractions.

The signal yield in the data is determined by an unbinned extended maximum likelihood fit to the distribution of $x$, defined earlier. Signal events are expected to peak at $x \approx 0.97$, while $\tau$-pair background exhibits a smoothly falling shape that cuts off at the kinematic endpoint of $x = 0.97$ (the lepton kinematic endpoint for charged leptons produced in tau decay). Bhabha and $\mu$-pair background exhibit a peak at $x = 1$, which is about $(2.5-3)\sigma_x$ above the signal, where $\sigma_x \approx 0.01$ is the detector resolution on $x$.

Probability density functions are obtained for each of these components. The signal and Bhabha/$\mu$-pair *p.d.f.*s are obtained from MC simulation. Signal events are described using a Crystal Ball function, while the Bhabha/$\mu$-pair backgrounds have a smooth component modeled using and ARGUS function and a peaking component modeled using a Gaussian. The $\tau$-pair background is modeled using a polynomial convoluted with a detector resolution function. The fit procedure is validated by using $\Upsilon(4S)$ and off-resonance data that are separated into samples with comparable numbers of events to those expected at the $\Upsilon(3S)$ and $\Upsilon(2S)$. No significant signal yield is obtained in these control tests.

The results of the fits to the $\Upsilon(3S)$ and $\Upsilon(2S)$ data are given in Table 18.4.12. No significant signal yields are obtained. These results are then used to place constraints on physics beyond the Standard Model. For example, Domingo and Ellwanger (2011) use this measurement to put limits on the decay of a Higgs boson to a pair of low-mass *CP*-odd Higgs bosons, interpreting this result as a limit on the mixing of the $\eta_b$ meson with such a *CP*-odd Higgs boson.

### 18.4.7.4 Test of Lepton Universality

*Motivation - mass-dependent couplings*

The Standard Model expresses no preference for the partial width $\Gamma_{\Upsilon(1S)\to\ell\ell}$ that depends on lepton flavor, up to corrections due to phase space (where $\ell = e, \mu, \tau$). This is referred to as "lepton universality" in the decay of the $\Upsilon(1S)$ meson. One can compute the ratio of partial widths to different final states and measure those ratios in data to see whether the Standard Model prediction is correct. For instance, the ratio of the $\tau$-to-$\mu$ partial widths is predicted to be $R_{\tau\mu}(\Upsilon(1S)) \approx 0.992$ in the Standard Model.

*Experimental measurement*

The *BABAR* Collaboration has searched for violation of lepton universality in $\Upsilon$ decay using the $\Upsilon(3S)$ and $\Upsilon(2S)$ data samples (del Amo Sanchez, 2010r). The analysis determines the ratio of branching fractions,

$$\frac{\mathcal{B}(\Upsilon(1S) \to \mu^+\mu^-)}{\mathcal{B}(\Upsilon(1S) \to \tau^+\tau^-)} \tag{18.4.31}$$

whose value is definitively predicted in the Standard Model. For instance, work by Sanchis-Lozano (2004), Fullana and Sanchis-Lozano (2007), and Domingo, Ellwanger, Fullana, Hugonie, and Sanchis-Lozano (2009) calculate the Standard Model rate and then discuss the implications of beyond-the-Standard Model physics on altering this value. Deviation from the SM prediction would indicate the presence of an interaction that couples differently to the two lepton flavors, such as the presence of a low-mass *CP*-odd Higgs Boson like the one searched for in 18.4.7.1. The analysis uses the recoil method to tag the presence of the $\Upsilon(1S)$ in the final state and the presence of leptons to indicate the final-state decay of the $\Upsilon(1S)$.

The *BABAR* Collaboration has measured the above ratio as a test of lepton universality (del Amo Sanchez, 2010r). The measurement uses a sample of $(121.8 \pm 1.2) \times 10^6$ $\Upsilon(3S)$ mesons; 10% of the sample is used to tune the analysis and the remaining 90% is used to obtain the final results. A previous measurement of $R_{\tau\mu}(\Upsilon(1S))$ was performed by the CLEO Collaboration (Besson et al., 2007) and found $R_{\tau\mu}(\Upsilon(1S)) = 1.02 \pm 0.02\,(stat.) \pm 0.05\,(syst.)$.

The *BABAR* measurement uses the transition $\Upsilon(3S) \to \pi^+\pi^-\Upsilon(1S)$ to tag the existence of the $\Upsilon(1S)$ meson, followed by $\Upsilon(1S) \to \ell^+\ell^-$. Only $\tau$ decays to a single charged particle and neutrinos are considered. Event selection is optimized using MC simulations.

Events are required to contain exactly four charged tracks, each with transverse momentum satisfying $0.1 < p_T < 10.0\,\mathrm{GeV}/c$. The four tracks are geometrically constrained to come from the same spatial location, and the distance of closest approach of each track must lie within 10 cm of the interaction region along the beam axis and within 1.5 cm in the plane transverse to the beam axis. The ratio of the second-to-zeroth Fox-Wolfram moments for the events must be $< 0.97$ in order to reject events like radiative Bhabha scatters and $\mu^+\mu^-\gamma$, where the photon converts to a pair of tracks. In addition, the absolute value of the cosine of the polar angle of the event thrust axis must be less than 0.96.

As in previous measurements using this topology, the $\Upsilon(1S)$ candidate is formed from an opposite-charge lepton pair which are constrained to arise from a common spatial point. The $\Upsilon(1S) \to \mu^+\mu^-$ and $\Upsilon(1S) \to \tau^+\tau^-$ final states have different background contributions. Due to the presence of neutrinos in the tau final states, backgrounds from $e^+e^- \to \tau^+\tau^-$ and non-leptonic $\Upsilon(1S)$ decays are possible.

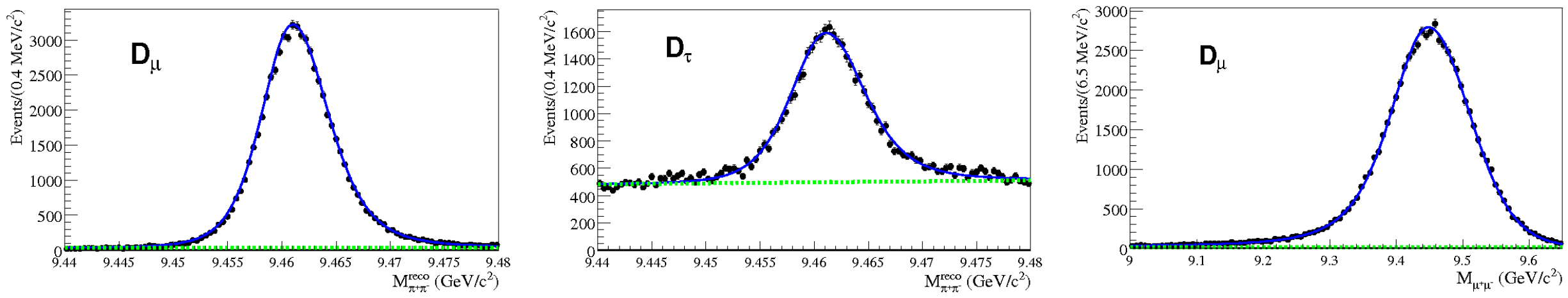

**Figure 18.4.28.** The recoil mass distribution for the dimuon (left) and di-tau (center) final states, and the dilepton mass for the dimuon final state (right) (del Amo Sanchez, 2010r).

The $\Upsilon(1S) \to \mu^+\mu^-$ final state (denoted $D_\mu$) is required to contain same-flavor identified muons. The difference between the intial-state and final-state energies is required to be less than 0.5 GeV, to select events where the four tracks represent all of the final-state particles (*e.g.* no significant missing energy from neutrinos). The magnitude of the dipion momentum in the CM frame is required to satisfy $< 0.875\,\mathrm{GeV}/c$, and the cosine of the angle between the two lepton candidates is required to satisfy $< -0.96$.

For the $\Upsilon(1S) \to \tau^+\tau^-$ candidates (denoted $D_\tau$), a tighter set of restrictions are required to reject backgrounds mentioned above. The difference between the intial-state and final-state energies is required to be greater than 5.0 GeV, to select events with significant missing energy due to neutrinos. The magnitude of the dipion momentum in the CM frame is required to satisfy $< 0.825\,\mathrm{GeV}/c$ and each pion must have a CM momentum satisfying $< 0.725\,\mathrm{GeV}/c$. The measured difference in the energy of the $\Upsilon(3S)$ and $\Upsilon(1S)$ must satisfy $0.835 < \Delta E^* < 0.925\,\mathrm{GeV}$. A boosted decision tree (see Chapter 4 for a description of this tool) is used to further reject background, employing event-shape and kinematic variables. The performance of the classifier is assessed using data taken at a CM energy below the $\Upsilon(3S)$ mass.

In order to select events resulting from the dipion transition, a requirement is placed on the difference in the invariant mass of the $\Upsilon(3S)$ and $\Upsilon(1S)$ states; as determined from the final-state tracks, $\Delta M < 2.5\,\mathrm{GeV}/c^2$. In addition, the dipion mass is required to satisfy $0.28 < M_{\pi^+\pi^-} < 0.90\,\mathrm{GeV}/c^2$. For events with multiple track combinations satisfying these cuts, the combination with the $\Delta M$ closest to the nominal value is chosen.

An unbinned extended maximum likelihood fit is used to extract the signal yields. The fit employs two variables: for the $\Upsilon(1S) \to \mu^+\mu^-$ final state, the fit used the mass recoiling against the dipion system (Equation 18.4.3) and the invariant mass of the lepton pair. Monte Carlo simulations are used to verify these variables are uncorrelated, and the total likelihood is the product of the likelihoods in each dimension. For the $\Upsilon(1S) \to \tau^+\tau^-$ final state, only the dipion recoil mass is used.

The use of a ratio of branching fraction to the two possible final states allows for the cancellation of common systematic uncertainties. The two final-state samples are simultaneously fitted using the likelihood functions, and the result of the fit is $R_{\tau\mu} = 1.006 \pm 0.013$ (statistical uncertainty only). The projections of the fits to the data in the different variables are shown in Fig. 18.4.28.

Residual systematic uncertainties that do not cancel in the ratio are due to the trigger selection efficiency, event selection, and muon selection. The event selection efficiency systematic uncertainty is determined by comparing the shape of each variable in data and MC; the difference in efficiency due to shape differences is determined to be 1.2%. The muon identification systematic uncertainty is determined in situ by comparison the data and MC rates at which only one, or both, tracks are identified as muons. This uncertainty is determined to be 1.2%, and a correction factor of 1.023 on the efficiency is also determined. A correction to the di-tau final-state trigger efficiency is determined to be 1.020; the systematic uncertainty on the efficiency is determined to be 0.10%. For the di-muon mode, the systematic uncertainty on the trigger efficiency is determined to be 0.18%.

The *p.d.f.* shape uncertainty for the signal components is determined by varying the parameters; the model parameters were determined from the 10% of data used for developing the analysis. This uncertainty on the *p.d.f.* shapes is determined to be 0.22%. The recoil mass shape is assumed to be the same for the two final states; shape effects due to trigger efficiency are assumed to be neglectable, and this assumption incurs a systematic uncertainty of 0.6%.

Taking into account correction factors and systematic uncertainties, the final measured ratio of branching fractions is determined to be:

$$R_{\tau\mu} = 1.005 \pm 0.013\,(stat.) \pm 0.022\,(syst.). \qquad (18.4.32)$$

No significant deviation from the Standard Model expectation of one is observed.

# Chapter 19
# Charm physics

In this chapter we proceed to the studies of open charm. It is fair to say that charm physics — more precisely the studies of hadrons containing a charm quark — underwent a revival of interest, both experimentally and theoretically, during the time of the $B$ Factories. The main reasons for this are threefold; the first reason may be called *experimental*, the second *electroweak*, and the third *strong*. The experimental reason is the awareness of the community that $B$ Factories are an abundant source of charm hadrons as well as $B$ mesons. The cross-section for $e^+e^- \to \Upsilon(4S)$ production at $\sqrt{s} \approx 10.58$ GeV is around 1.1 nb, while that for the so-called continuum production of charm quark pairs, $e^+e^- \to \gamma^* \to c\bar{c}$, is around 1.3 nb. For an integrated luminosity of $1\,\mathrm{ab}^{-1}$ this corresponds to about $600 \times 10^6$ $D^{*+}$ mesons produced together with another charmed hadron, available for study. The electroweak reason is due to the first experimental evidence for mixing phenomena in the system of neutral $D$ mesons, which became available in 2007. At about that time it became obvious that — on using the world averages of the measured quantities including results from hadron colliders — the mixing parameters can and will be measured to sub-percent accuracy. Such measurements *per se* represent a possible way to search for processes beyond the SM. When it became clear that with the experimentally determined values of parameters it would be difficult to make specific statements about the presence of New Physics phenomena, experimental as well as theoretical efforts turned to studies of $CP$ violation in the charm sector. This remains the focus of charm physics measurements today. Last but certainly not least there has been a third, strong-sector reason for the increased interest in charm physics. The discovery of the $X(3872)$ particle in 2003 (the observation of this particle has been confirmed and its properties studied by many experiments) with some of its properties similar to the conventional charmonia, but some in obvious disagreement with those, provides strong evidence that QCD has a rich spectrum of states beyond conventional mesons and baryons. Most of these so-called exotic states (although not all of them) bear a resemblance to hadrons composed of charm quarks. Surprises in the spectroscopy of charm hadrons continued also in the open charm sector with the discovery of $D_{s0}^*(2317)^+$ and $D_{s1}(2460)^+$ in 2003 and 2004, with their properties significantly different than expected from the naïve quark model.

The topic of conventional and exotic charmonium-like states is addressed in Chapter 18. This chapter focuses on the studies of open charm: mesons and baryons with a single charm quark. These states yield a number of interesting results which are, in many ways, complementary to results from $B$ mesons. For any study of charm mesons their decay modes must be known. The latter are the subject of the first section in the chapter. The following section discusses the electroweak aspect of the charm physics with the results on mixing and $CP$ violation parameters. The strong-sector is in more details illuminated in the last two sections on the charm meson and charm baryon spectroscopy.

## 19.1 Charmed meson decays

***Editors:***
*Antimo Palano (BABAR)*
*Jolanta Brodzicka (Belle)*

**Additional section writers:**
*Chunhui Chen, Patrick Roudeau, Anže Zupanc*

### 19.1.1 Introduction

The first discovered weak decays of the charm hadrons were $D^0 \to K^-\pi^+$, $K^-\pi^+\pi^+\pi^-$ (Goldhaber et al., 1976) and $D^+ \to K^-\pi^+\pi^+$ (Peruzzi et al., 1976) observed by MARK III at the SPEAR $e^+e^-$ collider operating at charm threshold. They allowed $D$ mass measurements and established the Glashow-Iliopoulos-Maiani (GIM) mechanism (Glashow, Iliopoulos, and Maiani, 1970), which required existence of a charm quark to explain absence of flavor-changing neutral currents (FCNC) at tree level, resulting in large suppression of the strangeness-changing kaon decays like $K^0 \to \mu^+\mu^-$. Later on, the $D_s^+ \to \phi\pi^+$ decay was discovered by CLEO at CESR (Chen et al., 1983), and followed by the color-suppressed $D_s^+ \to \bar{K}^{*0}K^+$ observed by ARGUS at DORIS-II (Albrecht et al., 1986).

Since then, charm decay measurements have been thoroughly performed, and triggered off by searches for non-standard weak interactions, as well as need for understanding of non-perturbative features in strong interactions. Both aspects are significantly different from the $b$-quark sector, making the measurements in general more difficult. Charm sector offers however an unique way to test the flavor physics of up-type quarks,[112] complementary to down-type quarks being investigated through measurements of strange and bottom decays.

Short distance contributions to flavor changing neutral current processes of the charm quark are highly GIM suppressed in the Standard Model, since the mass differences of the down-type quarks are small compared to the weak boson mass. A perturbative calculation of $c \to u$ FCNC processes yield a suppression factor $(m_b^2 - m_d^2)/m_W^2$, whereas FCNC in $B$ and $K$ decays are relatively strong due to the factor $(m_t^2 - m_u^2)/m_W^2$; thus the heavy top quark weakened the GIM suppression mechanism. However, due to the fact that in $D$ decays no particular suppression happens due to CKM factors, there are in general large long distance contributions, making an analysis of the short-distance dynamics difficult. As an example, the short distance contribution to $D^0 - \bar{D}^0$ oscillation is very small, by far exceeded by long distance contributions that are hard to compute.

Consequently, long-distance dynamics, like final state interactions (FSI), play an important role, since they are in general much larger in charm meson decays than for $B_{(s)}$ decays. In many cases they exceed the short-distance contributions even by a few order of magnitudes. Compared to $B$ decays there is a smaller energy release in $D_{(s)}$ decays, resulting in production of slower daughter particles, which thus are more likely to influence each other before they leave interaction region. Any precision electroweak predictions require then theoretical improvement in calculating long-distance QCD effects to remove substantial hadronic uncertainties. Strategies for NP searches and interpretation of measurements highly depend on quantitative information on hadronic effects. Such effects are non-perturbative and their theoretical calculations are still challenging for any approach/method. As charm quark lies inbetween the light flavours ($m_{u,d,s} \leq \Lambda_{QCD}$) described by chiral perturbation theory (ChPT) and heavy quarks ($m_b \gg \Lambda_{QCD}$) treated by heavy-quark effective theory (HQET), charm decays can bring new insight into non-perturbative QCD. Since heavy-quark mass expansion does not work as well for charm decays, thus computation of hadronic effects is more difficult than for corresponding $B$ decays. Despite of this, charm decays can still help to establish theoretical tools and allow their callibration for calculations inevitable for $B_{(s)}$ decays.

[112] Top quarks due to their short lifetime do not hadronize, thus many phenomena cannot be studied.

#### 19.1.1.1 Quark diagrams for weak decays of charm mesons

Quark diagrams underlying hadronic, semileptonic and leptonic decays of charm mesons are shown in Fig. 19.1.1. Taking into account topology of these quark graphs, they are either simple tree-level diagrams (Fig. 19.1.1(a-d,g,h)) or pengiun diagrams (Fig. 19.1.1(e,f)) representing higher-order, loop-level processes.

Tree-level hadronic decays (Fig. 19.1.1(a-c)) and semileptonic ones (Fig. 19.1.1(f)) proceed through $c \to W^+s$ current and thus have amplitudes governed by the CKM matrix element $|V_{cs}| \simeq 0.97$. These decays are Cabibbo-favored ($CF$) processes, while the corresponding Cabibbo-suppressed ($CS$) decays proceed via $c \to W^+d$ and involve $|V_{cd}| \simeq 0.22$. Unless $W^+$ materializes into either $l^+\nu_l$ leptons (Fig. 19.1.1(g,h)) or $u\bar{d}$ pair (Fig. 19.1.1(a,b,d)) inducing the CKM factor of $|V_{ud}|$, Cabibbo suppression may arise from the light-quark $u\bar{s}$ vertex involving $|V_{us}|$. Thus the $CF$ modes at the tree level proceed through $c \to s\bar{d}u$, singly Cabibbo-suppressed ($SCS$) ones through either $c \to d\bar{d}u$ or $c \to s\bar{s}u$, while doubly Cabibbo-suppresed ($DCS$) modes via $c \to d\bar{s}u$.

Figures 19.1.1(a,b,e,g) represent spectator decays, in which a light constituent antiquark does not participate in the weak interaction, contrary to non-spectator decays shown in Fig. 19.1.1(c,d,f,h).

Decays of ground charmed mesons to final states involving leptons (Fig. 19.1.1(g,h)) are the simplest and the cleanest channels and, as such, enable tests of the SM predictions or the Lattice-QCD calculations in the charm sector.

**Semileptonic** $D_{(s)} \to Xl^+\nu_l$ decays (where $l = e$ or $\mu$), comprise significant fractions of $D_{(s)}$ total widths (see Section 19.1.5); up to about for 6% $D^0$, 16% for $D^+$ and 6% for $D_s^+$ mesons. In the underlying $CF$ decay diagram (Fig. 19.1.1(g)), a virtual $W^+$ boson decays to the $l^+\nu_l$ system, while the $s\bar{q}$ pair hadronizes into a strange meson independently of the leptonic current. This hadronization is described by related form factors; single one for spin-0 hadron and three form factors for vector meson. The form factors increase with momentum transfer $q^2 \equiv M^2(l^+\nu_l)$, as for maximal $q^2$ value the final state hadron is at rest in the initial $D$ rest frame and an overlap of wave functions of initial and final states is largest. Measurements of shape of the form factor and its value at maximal $q^2$ allow one to test theoretical calculations performed with either HQET based models or with LQCD. Since semileptonic decays often serve as a reference for hadronic decays, it is important to check whether the $D_s^+$ semileptonic decays mirror the $D$ ones. Precision of $D_s^+$ decays like $D_s^+ \to \eta^{(\prime)} l^+\nu_l$ and $D_s^+ \to \phi l^+\nu_l$ (with uncertainties of $\pm 6\%$-$\pm 20\%$), is still lower than for $CF$ $D \to \bar{K}^{(*)} l^+\nu_l$ and $CS$ $D \to \pi l^+\nu_l$ or $D \to \rho l^+\nu_l$ decays (with uncertainties of $\pm 2\%$-$\pm 3\%$).

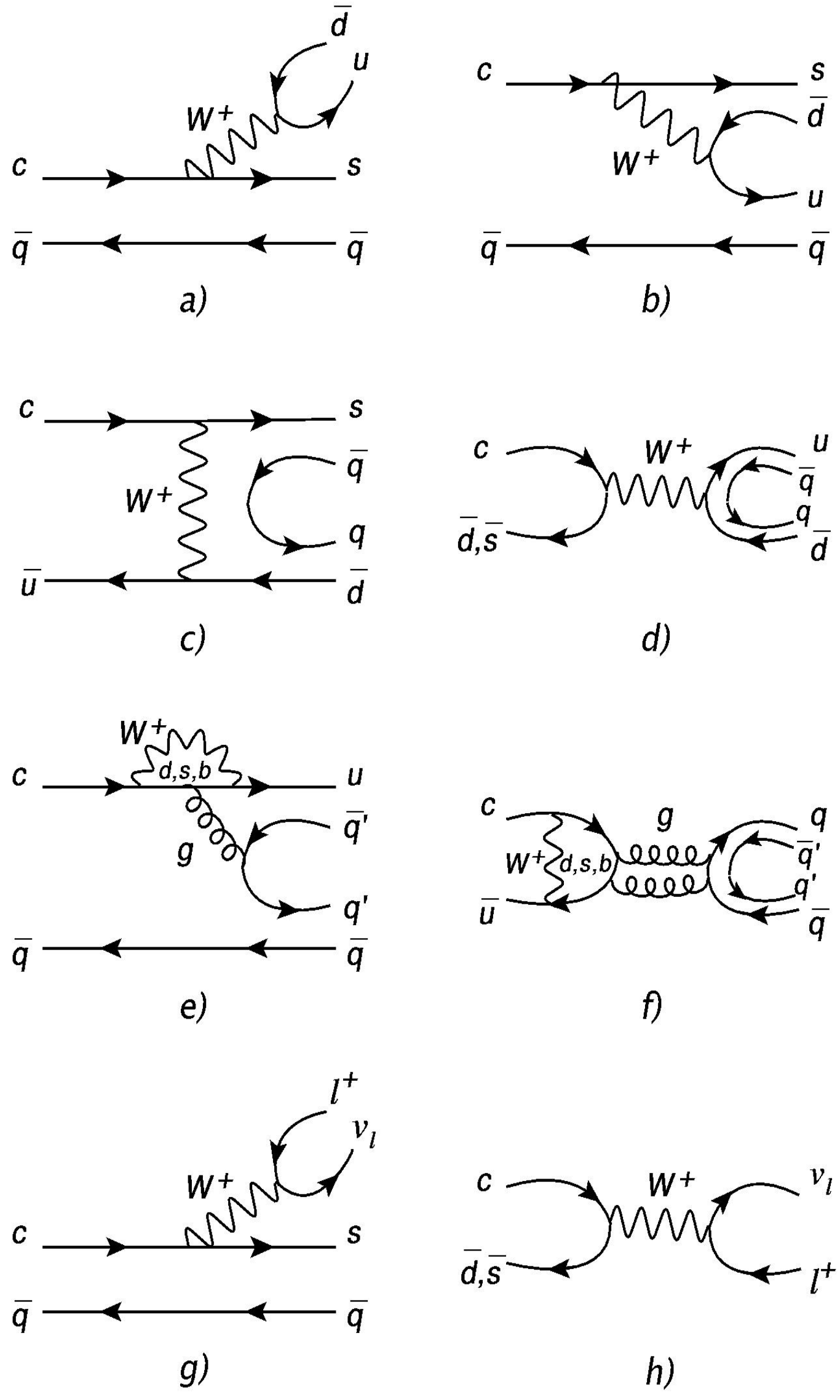

**Figure 19.1.1.** Quark diagrams for charm meson decays: hadronic (a-f), semileptonic (g), leptonic (h). Diagrams underlying hadronic decays: external $W$ emission (a), internal $W$ emission (b), $W$ exchange (c), $W$ annihilation (d), $W$-loop penguin (e) and $W$-loop penguin annihilation (f).

**Hadronic** analogue of the semileptonic process is shown in Fig. 19.1.1(a) where an externally-emitted virtual $W^+$ materializes into quark pair forming the final-state hadron. The corresponding process with an internal $W$ emission (Fig. 19.1.1(b)) has different quark pairings. Such a diagram represents a color-suppressed decay, as it requires a color matching between the quarks originating from different decay vertices. In the simplest approach based on analogy to semileptonic decays, presented hadronic decays are expressed as a product of two hadronic currents describing formation of mesons out of quark-antiquark pairs. In this approach, called naïve factorization, the hadronic currents for a two body decay are expressed by the decay constants of the respective meson and the form factors for the $D \to$ meson transition (see Section 19.1.1.3). Non-perturbative strong interactions, like soft-gluon effects and rescattering of the final-state particles, complicates the simple picture represented by the quark diagrams, and thus renders factorization invalid. Measurements performed so far suggest that naïve factorization for $D$ decays does not work as well as for $B$ decays, where a QCD based factorization has been formulated using a $1/m_b$ expansion. Overall it is fair to say that there is no satisfactory theoretical framework for the description of exclusive non-leptonic charm decays.

**Leptonic** $D_{(s)}^+ \to l^+\nu_l$ decays (see Section 19.1.6), proceed through $W$-annihilation diagram (Fig. 19.1.1(h)). Due to helicity suppression which results in the $m_l^2$ lepton-mass dependence of the decay width, rate for light charged leptons is small. Since $D_s^+ \to l^+\nu_l$ decays involve $V_{cs}$, with respect to $V_{cd}$ underlying $D^+ \to l^+\nu_l$, leptonic $D_s^+$ decay rates are significantly larger than those for $D^+$. All the strong interaction effects, namely hadronic dynamics of the initial meson, are factorized into its decay constant $f_{D_{(s)}^+}$, which is related to an overlap of wave functions of the constituent quark and antiquark. Measurement of $D_{(s)}^+ \to l^+\nu_l$ allows ones to determine the product $f_{D_{(s)}^+}|V_{cd(cs)}|$, thus to extract $f_{D_{(s)}^+}$ one needs the $|V_{cd(cs)}|$ value from other than leptonic decays.

Decay constants are fundamental parameters, and can serve as a test how well we are able to model dynamics of, in general, non-perturbative effects of hadronic dynamics. By measuring $f_{D_{(s)}^+}$ one can verify the Lattice-QCD calculations used for calculation of $B$ decay constants, which are difficult to measure as their leptonic decays are additionally suppressed by the tiny value of $V_{ub}$. Also, having both $f_{D^+}$ and $f_{D_s^+}$ measured one can directly estimate $SU(3)$-flavor symmetry breaking and compare it with theory prediction. Such a test would help to estimate reliably effects of violation of the $SU(3)$-symmetry based $f_{B_s} = f_B$ relation, as $f_{B_s}$ cannot be directly measured and must rely on $f_B$ measurement.

The **W-annihilation** hadronic decays (Fig. 19.1.1(d)), although the helicity suppression is here mitigated by strong interactions, are still strongly suppressed with respect to the tree-level spectator processes; similar applies to **W-exchange** processes (Fig. 19.1.1(c)). First signal for $W$-exchange decay was $D^0 \to \bar{K}^0\phi$ (Albrecht et al., 1985a) owing to the fact that helicity suppression does not apply to spin-0 meson decays with vector particle in final state. Though its large branching ratio of $10^{-2}$ suggests that some QCD effects, like rescattering, are involved in the decay dynamics.

The CKM elements parameterizing the mixing of the first two families into the third are all at least of order $\lambda^2$ in the Wolfenstein parameterization, and hence charm physics can be described to a good approximation by taking into account only the first two families. To this end, we have approximatively $V_{us}V_{cs}^* \approx -V_{ud}V_{cd}^*$ which has the interesting implication that the effective interaction for SCS $c \to u$ transitions takes the form

$$H_{\rm eff} = \frac{G_F}{\sqrt{2}} V_{us}V_{cs}^* (\bar{c}u)_{\rm V-A}[(\bar{s}s)_{\rm V-A} - (\bar{d}d)_{\rm V-A}] \quad (19.1.1)$$

which vanishes in the $SU(3)$ limit, *i.e.* once the strange and the down quark have the same mass. However, $SU(3)$ is severely broken and hence this suppression due to the $SU(3)$ flavor symmetry is not very effective.

#### 19.1.1.2 Hadronic decays; application of symmetries

$D$ mesons decay dominantly (84%) into hadronic final states and, as a charm-quark mass is quite sizable, number of hadronic $D_{(s)}$ decays is quite large. About 63% of the total width are two-body modes, as multibody processes are in fact quasi-two-body ones if intermediate resonances are considered as a single particle. Study of dynamics of multibody $D_{(s)}$ decays via either Dalitz plot or partial-wave analysis (PWA), can bring important information on light-flavor hadron spectroscopy (see Section 19.1.4).

All the two-body hadronic decays of charm mesons can be classified according to the six diagrams shown in Fig. 19.1.1(a-f). Measurements of two-body exclusive $D_{(s)}$ decays allow a quark-diagram analysis in which one determines magnitudes and signs of the amplitudes corresponding to the individual diagrams (Chau and Cheng, 1986) (Cheng and Chiang, 2010) (Bhattacharya and Rosner, 2010) (Bhattacharya and Rosner, 2009) (Bhattacharya, Gronau, and Rosner, 2012). Such decomposition allowed to understand important properties of charm mesons. As an example, external and internal diagrams in Fig. 19.1.1 (a,b) give rise to respectively the $D^0 \to K^-\pi^+$ and $D^0 \to \bar{K}^0\pi^0$ decays, while they both can lead to the $D^+ \to \bar{K}^0\pi^+$ final state. Destructive interference between $CF$ external and internal amplitudes in $D^+$ decays, along with fewer $CF$ $D^+$ channels, increases significantly the $D^+$ lifetime. Somewhat enhanced contribution from the exchange diagram (Fig. 19.1.1(c)) to the $D^0$ width could also reduce the $D^0$ lifetime. A pattern similar to the one for the $D \to \bar{K}\pi$ decays is also preferred by the data for the decays with vector meson in the final state, $D^+ \to \bar{K}^{*0}\pi^+$ and $D^+ \to \bar{K}^0\rho^+$. On the other hand, $\tau(D^0) < \tau(D_s^+)$ lifetime difference can be explained with $W$-annihilation contribution to the $D_s$ decay width, via for example $D_s^+ \to \rho\pi$, as all the spectator diagrams contribute similarly into $D_s^+$ and $D^0$ decays.

These analyses also show that the measured rates of the $D \to \bar{K}\pi$ decays, and many other channels with either two pseudoscalars or pseudoscalar and vector in final state, hardly can be fitted if the contributing quark amplitudes are real. This suggests that strong interactions modify the weak decay amplitudes, so that they carry phases induced by, for example, rescattering effects.

Using the fitted amplitudes one can make predictions for not yet measured decays based on their quark-diagram structure. Precision measurements of certain $D_{(s)}$ channels, especially those involving vector mesons, can help to determine better or/and ambiguously the suppressed amplitudes: exchange amplitudes ($D^0 \to \bar{K}^{*0}K^0$, $\bar{K}^0K^{*0}$, $D^0 \to \bar{K}^{*0}\eta^{(\prime)}$), annihilation amplitudes ($D_s^+ \to \eta^{(\prime)}\pi^+$, $D_s^+ \to \rho^+\pi^0$, $D_s^+ \to \omega\pi^+$) and penguin amplitudes ($D^0 \to \pi^0\pi^0$, $D^0 \to \bar{K}^0K^0$).

A powerful method of studying rescattering is an isospin analysis as isospin invariance holds to a very high accuracy. The isospin analysis requires an isospin decomposition of decay amplitudes for all possible charge states in isospin-related final states. For $D \to \overline{K}\pi$ decays, the decay amplitudes ($\mathcal{A}$) for $D^0 \to K^-\pi^+$, $D^0 \to K^0\pi^0$ and $D^+ \to \bar{K}^0\pi^+$ are linear combinations of the partial-wave isospin amplitudes ($A_I$) with $I = 1/2$ and $I = 3/2$:

$$\begin{aligned}
\mathcal{A}(D^0 \to K^-\pi^+) &= \sqrt{1/3}A_{3/2} + \sqrt{2/3}A_{1/2} \\
\mathcal{A}(D^0 \to \overline{K}^0\pi^0) &= \sqrt{2/3}A_{3/2} - \sqrt{1/3}A_{1/2} \\
\mathcal{A}(D^+ \to \overline{K}^0\pi^+) &= \sqrt{3}A_{3/2}, \qquad (19.1.2)
\end{aligned}$$

decomposition of the $D^0 \to \pi^+\pi^-$, $D^0 \to \pi^0\pi^0$ and $D^+ \to \pi^+\pi^0$ amplitudes into the $I = 0$ and $I = 2$ isospin amplitudes gives:

$$\begin{aligned}
\mathcal{A}(D^0 \to \pi^+\pi^-) &= \sqrt{1/3}A_2 + \sqrt{2/3}A_0 \\
\mathcal{A}(D^0 \to \pi^0\pi^0) &= \sqrt{2/3}A_2 - \sqrt{1/3}A_0 \\
\mathcal{A}(D^+ \to \pi^+\pi^0) &= \sqrt{3/2}A_2, \qquad (19.1.3)
\end{aligned}$$

while the $D^0 \to \overline{K}K$ final states form the isospin amplitudes of $I = 0$ and $I = 1$ as:

$$\begin{aligned}
\mathcal{A}(D^0 \to K^-K^+) &= \sqrt{1/2}A_1 + \sqrt{1/2}A_0' \\
\mathcal{A}(D^0 \to \overline{K}^0K^0) &= \sqrt{1/2}A_1 - \sqrt{1/2}A_0' \\
\mathcal{A}(D^+ \to \overline{K}^0K^+) &= \sqrt{2}A_1. \qquad (19.1.4)
\end{aligned}$$

The amplitudes $A_i^{(\prime)}$ are in general complex numbers; however, only their relative phase is observable. This allows us to introduces relative final-state phase between the isospin amplitudes in Eqs (19.1.2)–(19.1.4), defined as $\delta_{K\pi} = \arg(A_{3/2}/A_{1/2})$, $\delta_{\pi\pi} = \arg(A_2/A_0)$ and $\delta_{KK} = \arg(A_1/A_0)$.

Having measured partial widths for all three final states in each of Eqs (19.1.2)–(19.1.4), one can obtain magnitudes of isospin amplitudes and their relative phase. Old

CLEO analyses (Bishai et al., 1997) (Selen et al., 1993) showed that the isospin-amplitude relative phases for both, $D \to \overline{K}\pi$ and $D \to \pi\pi$ decays were almost 90°, implying large FSI. Adding the isospin amplitudes with no phase allowed estimation of the decay rates without rescattering: $\mathcal{B}(D^0 \to K^-\pi^+)_{\text{no FSI}} \approx 1.3\ \mathcal{B}(D^0 \to K^-\pi^+)$, $\mathcal{B}(D^0 \to \pi^+\pi^-)_{\text{no FSI}} \approx 1.6\ \mathcal{B}(D^0 \to \pi^+\pi^-)$. An isospin phase shift for $D \to \overline{K}K$ decays was however found to be consistent with zero. Thus elastic FSI explains neither enhanced $\mathcal{B}(D^0 \to K^-K^+)$, nor the decay rate of about $10^{-4}$ for $D^0 \to \overline{K}^0K^0$ which, if one neglects $W$-exchange amplitude, should be in the SM forbidden. Therefore either inelastic FSI or large $W$-exchange contribution could explain the latter. Another issue emerging from the isospin analyses is related to a large $I = 2$ amplitude for $D \to \pi\pi$ decays, $|A_2|/|A_0| = 0.72 \pm 0.17$. This indicates that there is no $\Delta I = 1/2$ for the $D$ decays into two pions, while it is well known that for $K \to \pi\pi$ decays $A_2$ is very small.

Each of the Equations 19.1.2-19.1.4 leads to a triangle relation among the decay amplitudes. Nonzero area for the formed triangle would be an evidence for either a difference in phases between the isospin amplitudes or contributions from quark-diagram amplitudes with different weak (CKM) phases. The latter would be a sign of isospin violation, possibly originating form a NP contribution.

Unlike the isospin symmetry, the $SU(3)$-flavor symmetry is heavily broken, and its long-standing indication in charm decays comes from the $D^0 \to K^-K^+$ and $D^0 \to \pi^+\pi^-$ widths, which, without $SU(3)$ violation and with phase-space related factor removed, are expected to be equal. The measured ratio is $\Gamma(D^0 \to K^-K^+)/\Gamma(D^0 \to \pi^+\pi^-) \simeq 3$, although a phase-space allowed in a numerator is smaller. An $SU(3)$-breaking effect in a dominant, external $W$-emission amplitude itself may arise from a difference in decay constants of pion and kaon, $f_K > f_\pi$. It implies a larger external amplitude for $D^0 \to K^-K^+$, but is insufficient to explain the measured ratio. Due to inelastic FSI the $\pi\pi$ mode can be converted into $\bar{K}K$ via for example scalar resonances coupling to both these final states. To confirm this scenario one needs quantitative estimation of inelastic rescattering.

Penguin diagram contributes to $D^0 \to K^-K^+$ and $D^0 \to \pi^+\pi^-$ with opposite relative signs, and for the $KK$ mode is destructive with respect to the tree amplitude, reducing the expected width ratio and thus making the situation even worse. The penguin contribution is expected to be small, however, similarly to the exchange amplitude, the long-distance QCD effects can enhance it significantly and make the theoretical calculations very difficult. On the other hand, knowledge of a size of the penguin amplitude is of great importance for estimation of the $CP$ violation expected in the charm sector within the SM (see Section 19.2). Measurement of a width for $D^0 \to \pi^0\pi^0$, containing the same penguin pollution as $D^0 \to K^-K^+$ and $\pi^+\pi^-$, will allow an estimation of the penguin contribution.

#### 19.1.1.3 Methods for estimating matrix elements

Theoretical description of charm-changing decay, $D \to f$, exploits an (low-energy) effective Hamiltonian constructed with the help of an Operator Product Expansion (OPE) (framework) in terms of local operators $O_i$ and the (couplings) Wilson coefficients $c_i$. Following the same line of argument as for $B$ decays, one obtains an effective Hamiltonian of the form:

$$\langle f|\mathcal{H}_{eff}|D\rangle = \frac{G_F}{\sqrt{2}} V_{CKM} \langle f| \sum_i c_i(\mu) O_i(\mu)|D\rangle, \quad (19.1.5)$$

where $G_F$ is Fermi constant and $V_{CKM}$ is a factor related to the CKM matrix elements involved in the decay. The renormalization scale $\mu$ separates contributions from long-distance dynamics (with length scales above $1/\mu$) and short-distance interactions (with length scales below $1/\mu$). All degrees of freedom with masses above $\mu$ give rise to effectively point-like interactions and are integrated out into the coefficents $c_i$ using perturbation theory. Degrees of freedom having mass scales below $\mu$ remain dynamical and are included in the operators $O_i$. Their hadronic matrix elements (hadronic expectation values), $\langle f|\sum_i c_i(\mu)O_i(\mu)|D\rangle$, involve non-perturbative dynamics. Typically $\mu \approx m_c$ is used as it assures that $\mu \gg \Lambda_{QCD}$ and thus $c_i$, which depends on the strong coupling constant, can be treated perturbatively. Also such a choice provides a resonable momentum cut-off for hadron wave functions used to calculate the hadronic matrix elements. These calculations, especially for nonleptonic transitions, are still challenging.

The effective Hamiltonian for the $\Delta C = 1$ weak decay is (Buchalla, Buras, and Lautenbacher, 1996):

$$\begin{aligned}\mathcal{H}_{eff}^{\Delta C=1} = \frac{G_F}{\sqrt{2}}[&\sum_{q=d,s} V_{uq}V_{cq}^*(c_1O_1 + c_2O_2) \\ &- V_{ub}V_{cb}^* \sum_{i=3}^{6} c_iO_i + c_{8g}O_{8g}] + h.c.,\end{aligned} \quad (19.1.6)$$

where $O_1$ and $O_2$ are the tree-level operators expressed as the current products:

$$O_1 = (\bar{q}c)_{V-A}(\bar{u}q)_{V-A}, O_2 = (\bar{u}c)_{V-A}(\bar{q}q)_{V-A}. \quad (19.1.7)$$

Although the weak decays are mainly driven by the operator $O_1$, QCD effects also induce other operators such as $O_2$ and, in general, both terms contribute to the amplitudes in Fig 19.1.1(a-d). The remaining operators correspond to the penguin contributions in Fig 19.1.1(e-f):

$$\begin{aligned}O_{3,5} &= (\bar{u}c)_{V-A} \sum_{q'=u,d,s} (\bar{q'}q')_{V\mp A}, \\ O_{4,6} &= \sum_{q'=u,d,s} (\bar{u}q')_{V-A} \sum_{q'=u,d,s} (\bar{q'}c)_{V\mp A}, \\ O_{8g} &= -\frac{g_s}{8\pi^2} m_c \bar{u}\sigma_{\mu\nu}(1+\gamma_5)G^{\mu\nu}c,\end{aligned} \quad (19.1.8)$$

where $g_s$ is the strong decay constant, $G^{\mu\nu}$ denotes the QCD field strength tensor, while the $(\bar{q}q)_{V\mp A}$ current structure in Eq. (19.1.7)-(19.1.8) corresponds to $(\bar{q}\gamma(1\mp\gamma_5)q)$. The Wilson coefficients evaluated at $\mu \approx m_c$ are $c_1 \simeq 1.21$, $c_2 = -0.41$, $c_3 = 0.02$, $c_4 = -0.04$, $c_5 = 0.01$, $c_6 = -0.05$ and $c_{8g} = -0.06$ (Buchalla, Buras, and Lautenbacher, 1996).

The remaining task, namely the calculation if the matrix elements of the effective operators, is difficult for charm decays, since charm is neither heavy enough for a reliable $1/m_c$ expansion, nor light enough to be described in terms of chiral perturbation theory. Motivated by the form of the effective Hamiltonian, a pragmatic approach has been suggested some decades ago (Bauer, Stech, and Wirbel, 1987) by introducing two parameters $a_1$ and $a_2$ according to

$$A(D \to f) \sim a_1 \langle f|(\bar{q}c)_H(\bar{u}q)_H|D\rangle + a_2\langle f|(\bar{u}c)_H(\bar{q}q)_H|D\rangle, \tag{19.1.9}$$

where the subscripts $H$ mean that the matrix element is calculated in naïve factorization:

$$\begin{aligned}\langle f_1 f_2|(\bar{q}c)_H(\bar{u}q)_H|D\rangle &\simeq \langle f_1|(\bar{q}c)|0\rangle\langle f_2|(\bar{u}q)|D\rangle,\\ \langle f_1 f_2|(\bar{u}c)_H(\bar{q}q)_H|D\rangle &\simeq \langle f_1|(\bar{u}c)|0\rangle\langle f_2|(\bar{q}q)|D\rangle.\end{aligned} \tag{19.1.10}$$

The quantities $a_1$ and $a_2$ are scale-invariant, phenomenological parameters, which are assumed to be universal for all the decays. Comparing with the QCD calculation of the effective Hamiltonian, we write

$$a_1 = c_1 + \xi c_2,\ a_2 = c_2 + \xi c_1 \tag{19.1.11}$$

where $\xi$ parameterizes the long distance effects that are not correctly treated by naïve factorization.

The matrix elements appearing in the naïve factorization (19.1.10) correspond to the meson decay constants and form factors and can be taken from measurements of the (semi)leptonic $D_{(s)}$ decays. The $a_{1,2}$ can then be fitted from the experimental data and first such a analysis, the Bauer-Stech-Wirbel (BSW) analysis, was performed for the $D \to K\pi$ decays (Bauer, Stech, and Wirbel, 1987) and yielded $a_1 = 1.3 \pm 0.1$ and $a_2 = -0.5 \pm 0.1$. Once compared with the theoretical expectations, $a_1 = 1.25 - 0.48\xi$ and $a_2 = -0.48 + 1.25\xi$, suggested $\xi \simeq 0$.

The non-spectator, exchange and annihilation diagrams (Fig. 19.1.1(c-d)) are within the factorization described using a vacuum insertion:

$$\langle f_1 f_2|(\bar{q}c)_H(\bar{u}q)_H|D\rangle \simeq \langle f_1 f_2|(\bar{q}c)_H|0\rangle\langle 0|(\bar{u}q)_H|D\rangle, \tag{19.1.12}$$

and are in general a small correction with respect to the decays determined by the $a_1$ or/and $a_2$. An impact of the non-spectator diagrams on the two-body charm decays was also studied in the BSW analysis (Bauer, Stech, and Wirbel, 1987).

Naïve factorization as defined by BSW approach is very simply and allows us to parameterize most of the hadronic decays with just two parameters. However, this simple ansatz is not satisfactory for the improved measurements, and thus more sophisticated theoretical approaches need to be developed.

## 19.1.2 Branching ratio measurements

The large statistics accumulated at $B$ Factories and clean environment of the $e^+e^-$ experiment allowed precision measurements of absolute branching fraction of favored weak $D_{(s)}$ decays, previously obtained using sometimes only a few hundred signal events. Such measurements require a knowledge of a total number of produced $D_{(s)}$ mesons to normalize the $D_{(s)}$ signal reconstructed in the decay mode of interest. That can be achieved either via an exclusive reconstruction of $e^+e^- \to c\bar{c}$ events for measurement of directly produced $D_{(s)}$ (Belle technique), or through tagging selected $B$ decays as a source of charmed mesons (*BABAR* technique). The measured decays serve as a reference for measurements of the branching fractions of the $D_{(s)}$ to any other final state, improve our knowledge of most of the decays of the $B_{(s)}$ mesons, and of fundamental parameters of the Standard Model.

The suppressed $D_{(s)}$ decays are measured with respect to the favored processes having a similar topology. Relative branching ratio measurements allow cancelation of numerous systematic uncertainties. Precision measurements of the suppressed charm decays give insight into decay dynamics and allow tests of the various symmetries assumed in the theoretical calculations.

### 19.1.2.1 Absolute Branching Fraction of $D^0 \to K^-\pi^+$

The $D^0 \to K^-\pi^+$ is a reference mode for many measurements of the $D$ decays. CLEO-c (He et al., 2005) published its result on this branching fraction, the most precise at the time, which was widely used (Yao et al., 2006). *BABAR* using $210\,\text{fb}^{-1}$ data, has produced even more precise measurement based on a different technique (Aubert, 2008p). The $D^0 \to K^-\pi^+$ decays are identified in a sample of $D^0$ mesons produced in $D^{*+} \to D^0\pi^+$ decays and obtained with partial reconstruction of $\overline{B}^0 \to D^{*+}(X)\ell^-\bar{\nu}_\ell$ (see also Section 7.3.2). Such $B$ candidates are selected by retaining events containing a charged lepton ($\ell = e,\ \mu$) and a low momentum (soft) pion ($\pi_s^+$) which may arise from the decay $D^{*+} \to D^0\pi_s^+$. Momenta of the lepton and the soft pion must respectively satisfy the $1.4 < p_{\ell^-} < 2.3\,\text{GeV}/c$ and $60 < p_{\pi_s^+} < 190\,\text{MeV}/c$. The minimum $p_{\ell^-}$ and $p_{\pi_s^+}$ are optimized to minimize uncertainties due to charm production in $B$ decays and tracking errors, respectively, while the maximum momenta are determined by the available phase space. The two tracks must be consistent with originating from a common vertex, constrained to the beam-spot in the plane transverse to the beam axis. Then they combine $p_{\ell^-}$, $p_{\pi_s^+}$ and the probability from the vertex fit into a likelihood ratio variable, optimized to reject $B\overline{B}$ background. Using conservation of momentum and energy, the invariant mass squared of the undetected neutrino is calculated as:

$$\mathcal{M}_\nu^2 \equiv (E_{\text{beam}} - E_{D^*} - E_\ell)^2 - (\boldsymbol{p}_{D^*} + \boldsymbol{p}_\ell)^2, \tag{19.1.13}$$

where $E_{\text{beam}}$ is half the total center-of-mass (c.m.) energy and $E_\ell$ ($E_{D^*}$) and $\boldsymbol{p}_\ell$ ($\boldsymbol{p}_{D^*}$) are the energy and momentum of the lepton (the $D^*$ meson) in the c.m. system. Since the magnitude of the $B$ meson momentum in the c.m. system $|\boldsymbol{p}_B| \ll |\boldsymbol{p}_\ell|$, $|\boldsymbol{p}_{D^*}|$, they set $\boldsymbol{p_B} = 0$ in the above equation. As a consequence of the limited phase space available in the $D^{*+}$ decay, the soft pion is emitted nearly at rest in the $D^{*+}$ rest frame. The $D^{*+}$ four-momentum can therefore be computed by approximating its direction as that of the soft pion, and parameterizing its momentum as a linear function of the soft-pion momentum.

All events where $D^{*+}$ and $\ell^-$ originate from the same $B$ meson, producing a peak near zero in the $\mathcal{M}_\nu^2$ distribution, are considered as signal candidates. This sample of events is referred to as the *inclusive* sample. Sample of events with the same charge of $\pi_s$ and lepton is also selected for background studies and is referred to as wrong-charge sample. Number of the inclusive signal events is obtained from a minimum $\chi^2$ fit to the $\mathcal{M}_\nu^2$ distribution in the interval $-10 < \mathcal{M}_\nu^2 < 2.5\,\text{GeV}^2/c^4$. Figure 19.1.2(a) shows the fit result in the $\mathcal{M}_\nu^2$ projection, with signal and background shapes obtained with the MC simulations, while the $\mathcal{M}_\nu^2$ distribution of the wrong-charge sample that contains background candidates only is shown in Fig. 19.1.2(b). The inclusive yield for $\mathcal{M}_\nu^2 > -2\,\text{GeV}^2/c^4$ is $N^{\text{incl}} = (2171 \pm 3 \pm 18) \times 10^3$.

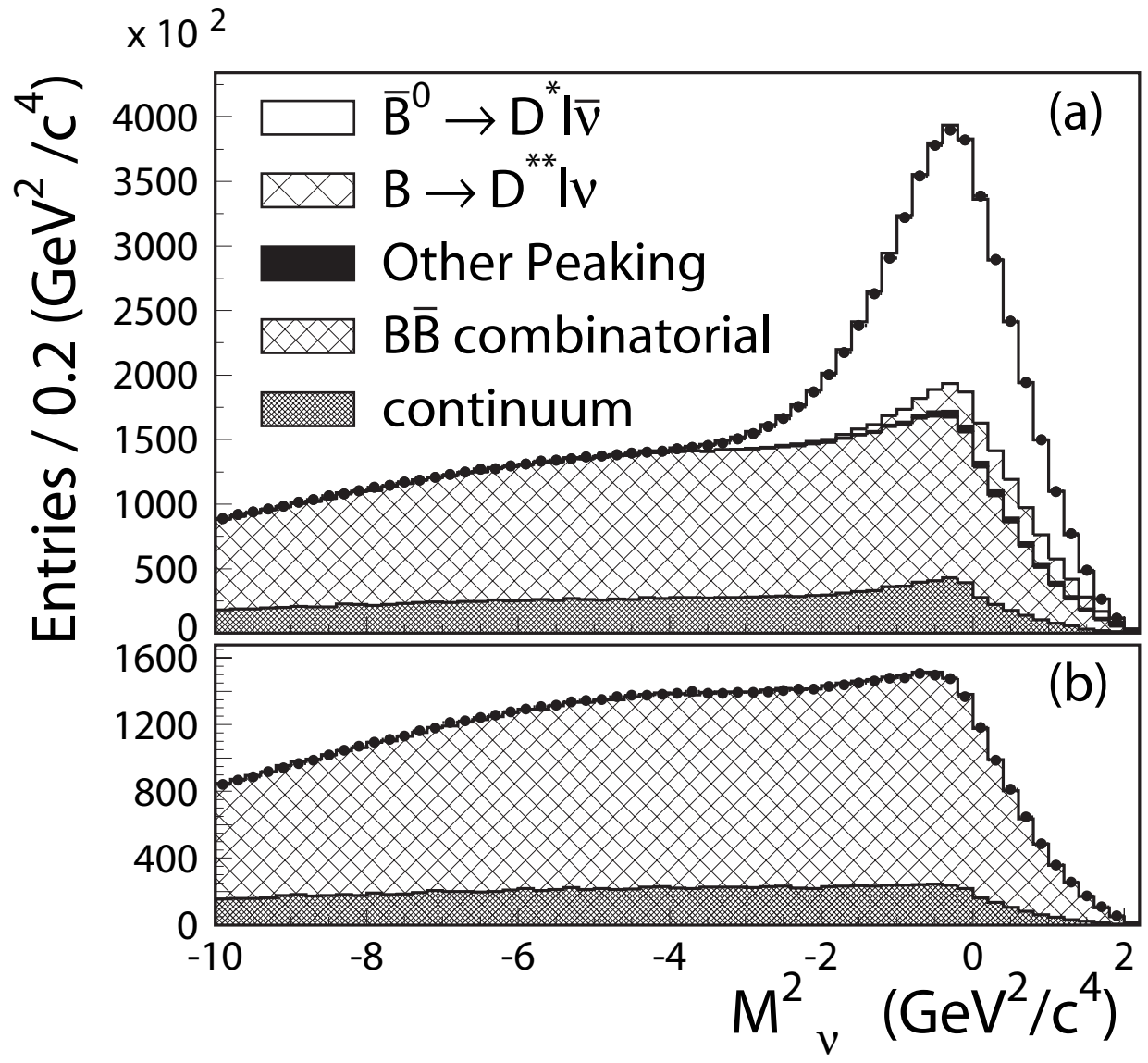

**Figure 19.1.2.** From (Aubert, 2008p). The $\mathcal{M}_\nu^2$ distribution of the inclusive sample, for right-charge (a) and wrong-charge (b) samples. The data are represented by solid points with uncertainty. The MC fit results are overlaid on the data, as explained in the figure.

To obtain the *exclusive* signal sample, $D^0 \to K^-\pi^+$ decays are reconstructed using all tracks in the event, aside from the $\ell^-$ and $\pi_s^+$, with momenta in the direction transverse to the beam axis exceeding $0.2\,\text{GeV}/c$. They combine pairs of tracks with opposite charge and compute the invariant mass $m(K\pi)$ assigning the kaon mass to the track with charge opposite the $\pi_s$ charge. $D^0$ candidates from the mass range $1.82 < m(K\pi) < 1.91\,\text{GeV}/c^2$ are combined with the $\pi_s^+$. Events having the mass difference, $\Delta M = m(K^-\pi^+\pi_s^+) - m(K^-\pi^+)$, in the range of $142.4 < \Delta M < 149.9\,\text{MeV}/c^2$ are selected as the signal candidates (see Fig. 19.1.3). The exclusive selection yields $N^{\text{excl}} = (33.8 \pm 0.3) \times 10^3$ signal events. The branching

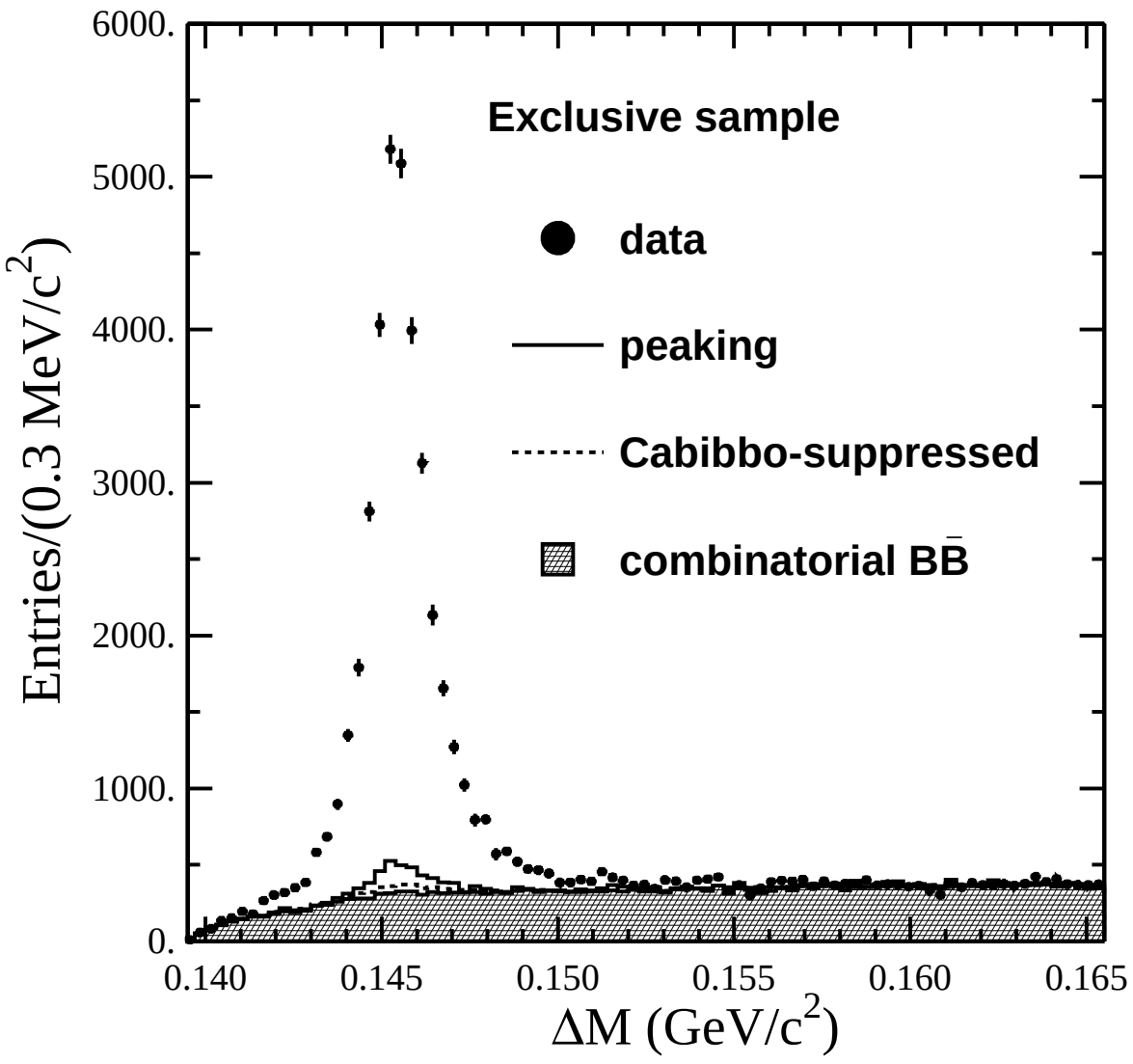

**Figure 19.1.3.** From (Aubert, 2008p). Continuum subtracted $\Delta M$ distribution for data (points with error bars) and backgrounds overlaid as explained in the figure.

fraction is computed as:

$$\mathcal{B}(D^0 \to K^-\pi^+) = N^{\text{excl}}/(N^{\text{incl}}\zeta\varepsilon_{(K^-\pi^+)}), \quad (19.1.14)$$

where $\varepsilon_{(K^-\pi^+)} = (36.96 \pm 0.09)\%$ is the $D^0$ reconstruction efficiency from MC simulation, and $\zeta = 1.033 \pm 0.002$ is the selection bias introduced by the partial reconstruction. The measured result is:

$$\mathcal{B}(D^0 \to K^-\pi^+) = (4.007 \pm 0.037 \pm 0.072)\%, \quad (19.1.15)$$

where the first uncertainty is statistical and the second uncertainty is systematic. This result is comparable in precision with the so far the most precise measurement by Cleo-c (Dobbs et al., 2007b) and is consistent with it within one standard deviation.

#### 19.1.2.2 Absolute Branching Fractions of $D_s^+$ decays

Belle has measured absolute branching fractions of $D_s^+ \to K^+K^-\pi^+$, $D_s^+ \to \bar{K}^0K^+$ and $D_s^+ \to \eta\pi^+$ (Zupanc, 2013b) being reference modes for the leptonic $D_s^+$ decays (see Section 19.1.6). The analysis is based on the $913\,\text{fb}^{-1}$ and the studied $D_s$ mesons are produced in the following reaction:

$$e^+e^- \to c\bar{c} \to D_{tag}KX_{frag}D_s^{*-},\ D_s^{*-} \to D_s^-\gamma, \quad (19.1.16)$$

where one of the charm quarks hadronizes into the $D_s$, while the other into *tagging* charm hadron $D_{tag}$, which is reconstructed as $D^{(*)+}$, $D^{(*)0}$ or $\Lambda_c^+$. The ground charmed hadrons, $D^+$, $D^0$ and $\Lambda_c^+$, are reconstructed in 18 hadronic decay modes in total, with up to one $\pi^0$ in the final state to keep low background level. In order to reject background from $e^+e^- \to B\overline{B}$ events and combinatorial background the $e^+e^-$ center-of-mass $D_{tag}$ momentum is required to be greater than $2.3\,\mathrm{GeV}/c^2$ (or $2.5\,\mathrm{GeV}/c^2$ for less clean $D_{tag}$ modes). To further clean up the reconstructed sample of the ground $D_{tag}$ hadrons several variables being either topological (the $D_{tag}$ decay vertex quality, distance between production and decay vertices, angle between $D_{tag}$ momentum vector and production-to-decay vector), related to the dynamics ($D_{tag}$ decay angle) or quality of the decay products (charged hadron identification, $\pi^0$ quality) are combined into a single neural network output variable (`NBout`), being a probability that a given $D_{tag}$ candidate is a correctly reconstructed signal. The `NBout` selection is optimized using NeuroBayes neural network (see Section B) trained on a small data sample; the optimization procedure maximizes the signal significance measured from the $D_{tag}$ invariant-mass distribution. Within the selected $D_{tag}$ sample, $D^+$ and $D^0$ mesons originating from $D^{*+} \to D^0\pi^+$, $D^+\pi^0$ and $D^{*0} \to D^0\pi^0$, $D^0\gamma$ decays are identified using the invariant-mass difference $\Delta M(D^*) \equiv M(D\pi/\gamma) - M(D)$.

An additional $K$ in Eq. (19.1.16), detected as either $K^+$ or $K^0_S$, assures strangeness conservation in the event,[113] while $X_{frag}$ is a fragmentation system denoting additional particles that can be created in the hadronization. The $X_{frag}$ is formed from the remaining charged pions (up to three) and up to one $\pi^0$ candidate. The $D_{tag}KX_{frag}$ combinations are required to have a common vertex, total electric charge of $\pm1$ (giving inclusively reconstructed $D_s^{*\mp}$) and *right sign* of their charm and strangeness quantum numbers relative to their total charge (charm in the $D_{tag}$ and strangeness of the primary $K$, if specified, are required to be opposite to the charge of the $D_s^*$).

The method uses only $D_s^+$ mesons produced through parent $D_s^{*+}$, so one requires a photon consistent with $D_s^{*+} \to D_s^+\gamma$. This provides a powerful constraint on the $D_s^+$ signal and improves a resolution of the missing mass used to identify $D_s^+$ signal and defined as:

$$M_{miss}(D_{tag}KX_{frag}\gamma) \equiv \sqrt{p^2_{miss}(D_{tag}KX_{frag}\gamma)}, \tag{19.1.17}$$

where $p_{miss}$ is the missing four-momentum in the event

$$p_{miss}(D_{tag}KX_{frag}\gamma){=}p_{e^+}{+}p_{e^-}{-}p_{D_{tag}}{-}p_K - p_{X_{frag}}{-}p_\gamma \tag{19.1.18}$$

For correctly reconstructed events described by Eq. (19.1.16), the $M_{miss}(D_{tag}KX_{frag}\gamma)$ peaks at the nominal $D_s$ mass, while the corresponding $M_{miss}(D_{tag}KX_{frag})$ at the nominal $D_s^*$ mass. To improve the $D_s$ signal resolution, the $D_s^*$ candidates within the $2.0 < M_{miss}(D_{tag}KX_{frag}) < 2.25\,\mathrm{GeV}/c^2$ region are selected and refitted with a $D_s^*$ mass constraint.

To obtain a fully *inclusive* $D_s^+$ sample used for normalization in the branching fraction calculation, there are no requirements on the $D_s^+$ decay products. The inclusive $D_s^+$ signal yield is obtained from the fit to the $M_{miss}(D_{tag}KX_{frag}\gamma)$ spectrum for each $X_{frag}$ mode separately. Figs 19.1.4 show the fitted spectra for $X_{frag}$ modes giving the largest yields, $X_{frag}$ = nothing, $\pi^\pm$, and $\pi^+\pi^-$. Signal component is modeled with histogram from the MC simulations and is convolved with a Gaussian resolution (of about $2\,\mathrm{MeV}/c^2$) measured from real data using fully reconstructed $D_s \to \phi\pi$ decays. Background contributions are: mis-reconstructed signal ($K$ or $X_{frag}$ pion originating from $D_s$ decays), reflections from $D^{*0} \to D^0\gamma$ or $D^*_{(s)} \to D_{(s)} \to D^0\pi^0$ (being sources of the signal photon), wrong $\gamma$ (wrongly reconstructed in the calorimeter) and $\gamma$ coming from $\pi^0$ decays not originating from $D^*_{(s)}$. Their shapes are histograms obtained from the generic MC, while normalizations are mostly kept free in the fit. Total inclusive $D_s^+$ yield is measured to be $(94.4 \pm 1.3 \pm 1.4) \times 10^3$.

To obtain *exclusive* $D_s^+$ sample within the inclusive sample, all the tracks of $D_s^+ \to K^+K^-\pi^+$ decays are fully reconstructed using the events with exactly three such charged tracks remaining. The exclusively reconstructed $D_s^+ \to K^+K^-\pi^+$ events are identified as a peak at $D_s^{*+}$ nominal mass in the $M(K^+K^-\pi^+\gamma)$ mass distribution (Fig. 19.1.5). To increase reconstruction efficiency for $D_s^+ \to \bar{K}^0K^+$ and $D_s^+ \to \eta\pi^+$, only the charged kaon and pion, respectively, are explicitely reconstructed. The $D_s^+ \to \bar{K}^0K^+$ signal is identified in the missing mass squared distribution $M^2_{miss}(D_{tag}KX_{frag}\gamma K)$ (Fig. 19.1.6) and expected at mass squared of $K^0$ nominal mass, while the $D_s^+ \to \eta\pi^+$ in the $M^2_{miss}(D_{tag}KX_{frag}\gamma\pi)$ (Fig. 19.1.7) at $\eta$ nominal mass squared. All these spectra are fitted with signals parameterized using MC simulations with data-based resolution taken into account. In the background part, in addition to a smooth combinatorial background, there are also peaking reflections, modeled in the fits with MC simulations and included in the fits (see Figs 19.1.5-19.1.7 for details). In the $D_s^+ \to \eta\pi^+$ case contribution from $D_s^+ \to \tau^+\nu_\tau \to \pi^+\bar{\nu}_\tau\nu_\tau$ is suppressed by requiring that extra neutral energy in the electromagnetic calorimeter (not associated to the particles used in the inclusive or exclusive $D_s^+$ reconstruction) to be larger than 1 GeV, as significant energy deposit is expected from the $\eta$ decay products while none from the tauonic decays.

The absolute $\mathcal{B}$ of $D_s^+$ to given $f$ final state reconstructed in the exclusive sample is:

$$\mathcal{B}(D_s \to f) = \frac{N^{excl}(D_s^+ \to f)}{N^{incl}(D_s^+) \cdot f_{bias} \cdot \varepsilon(D_s^+ \to f|incl\ D_s^+)}, \tag{19.1.19}$$

where $N^{incl}(D_s)$ is a number of inclusively reconstructed $D_s$ mesons, $N^{excl}(D_s \to f)$ is the number of exclusively reconstructed $D_s \to f$ decays, while $\varepsilon(D_s \to f|incl\ D_s)$ is the efficiency of exclusive $D_s \to f$ reconstruction given inclusively reconstructed $D_s$ and is determined from MC simulations of the events containing the signal decays. The efficiency of inclusive $D_s$ reconstruction depends on the $D_s$-decay mode and drops with increasing multiplicity of the $f$ final state. It is accounted for by introducing a factor $f_{bias}$ which is a ratio of efficiency of inclusive $D_s$ reconstruction for $D_s \to f$ and for $D_s$ decaying generically *i.e.* to all known decay modes. This number is further corrected to account for a different multiplicities of final state particles in $D_s$ decays in MC and real data. Fitted exclusive yields of the hadronic $D_s$ decays and measured absolute branching fractions are summarized in Table 19.1.1. Precision of these measurements is approximately equal to the precission of the curent world average values.

[113] In the case of *tagging* $\Lambda_c^+$ an antiproton is also required to balance the baryon number in the event.

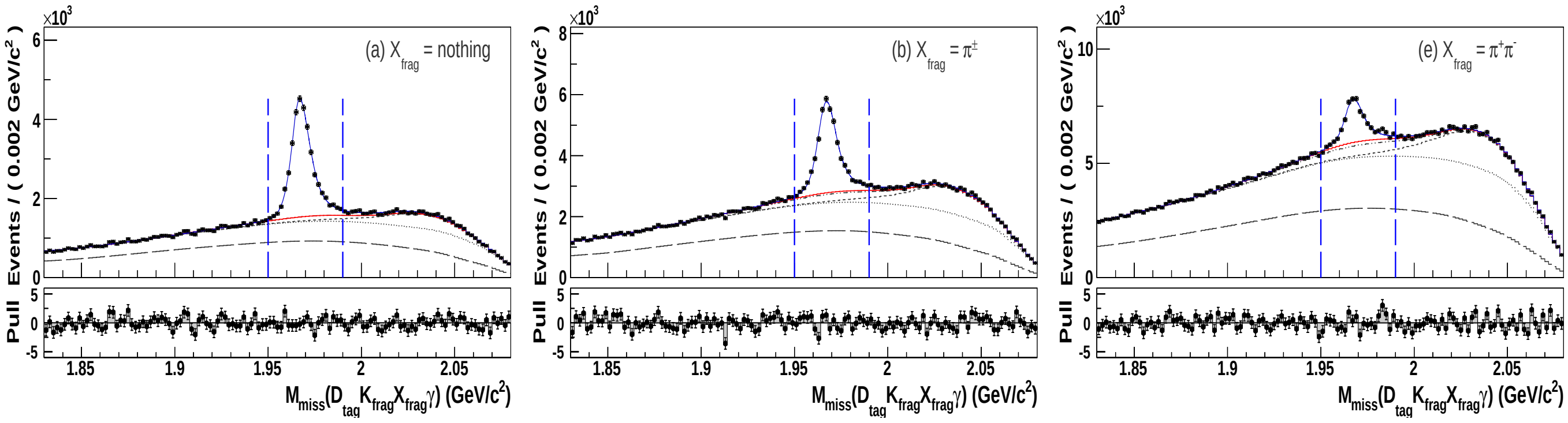

**Figure 19.1.4.** From (Zupanc, 2013b). Inclusive $D_s$ in $M_{miss}(D_{tag}KX_{frag}\gamma)$ for three out of seven possible $X_{frag}$ modes. The solid blue (red) line shows the contribution of signal and background (background only) candidiates. The cumulative contributions of candidates originating from different background sources as described in the text are shown with different gray dashed lines. Dashed vertical lines indicate the signal region considered in further analysis.

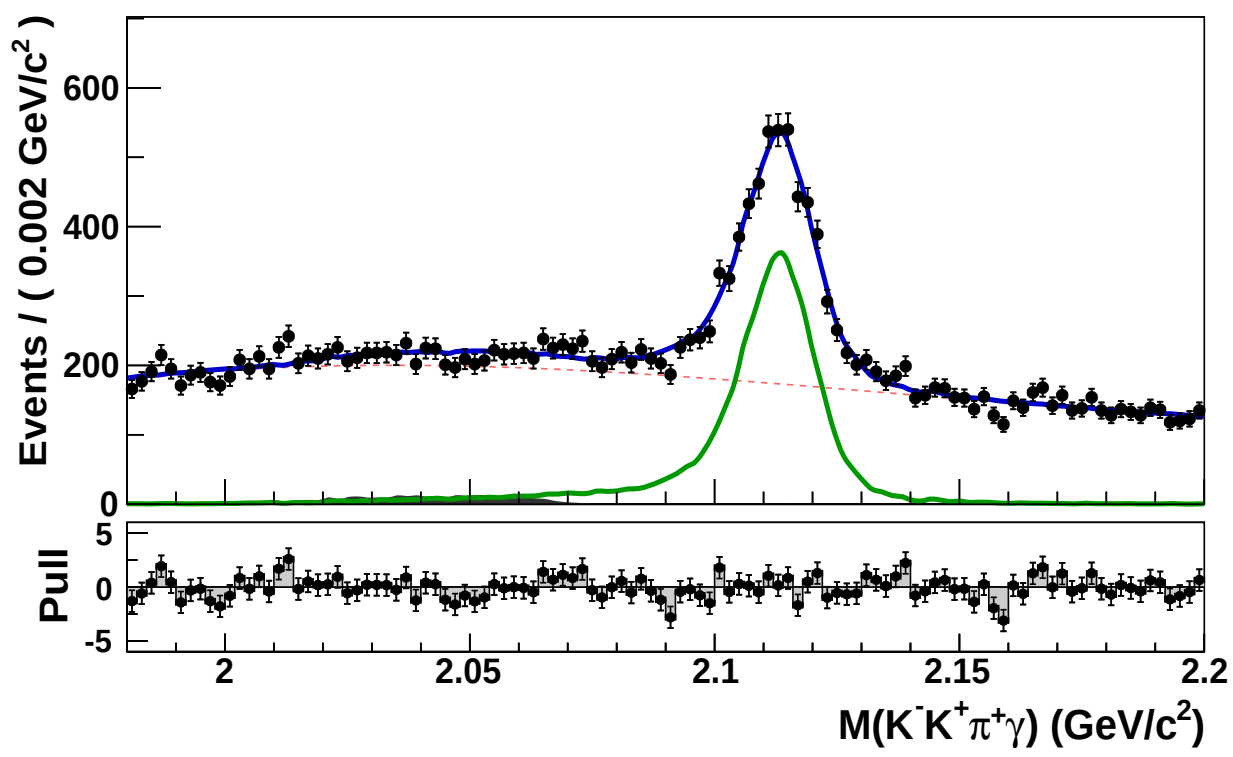

**Figure 19.1.5.** From (Zupanc, 2013b). $M(K^+K^-\pi^+\gamma)$ mass distribution of exclusively reconstructed $D_s^+ \to K^+K^-\pi^+$ decays within the inclusive $D_s$ sample, with the fit result superimposed and including fitted signal contribution (solid green line), reflection from $D_s^* \to D_s\pi^0 \to K^+K^-\pi^+\gamma\gamma$ with one of the photons missing (full dark gray histogram) and combinatorial background (dashed red line).

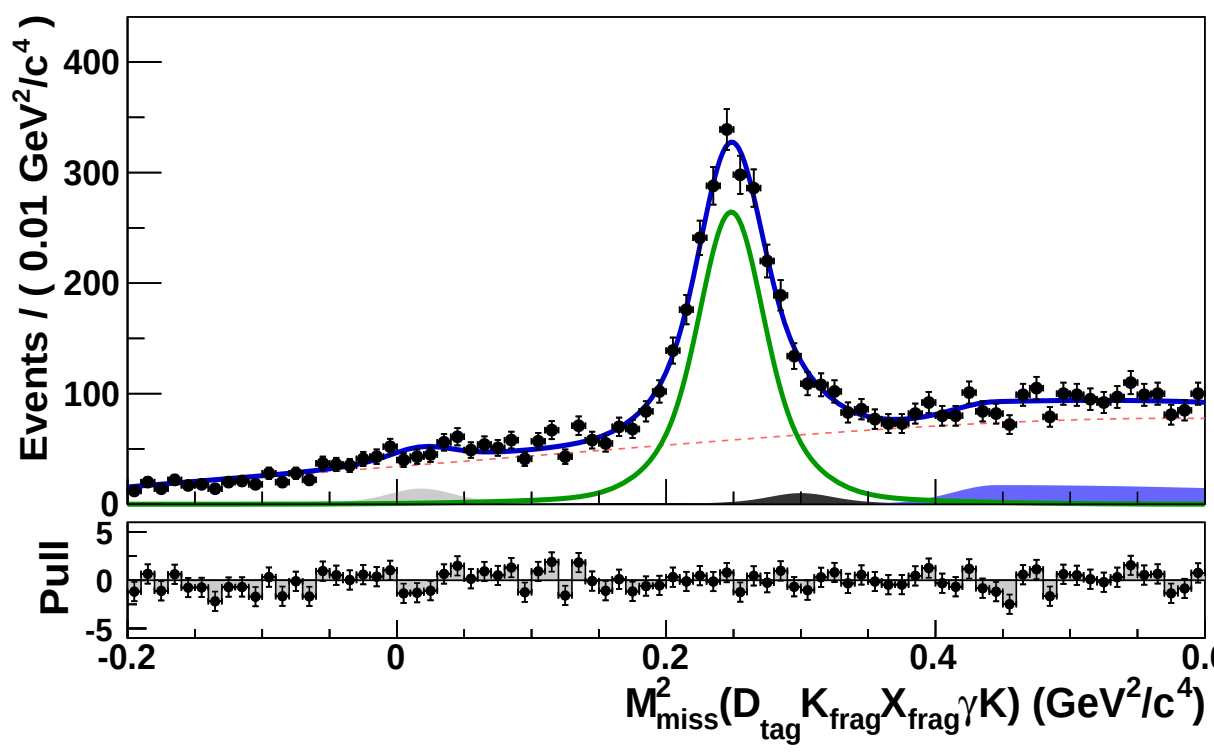

**Figure 19.1.6.** From (Zupanc, 2013b). $M^2_{miss}(D_{tag}KX_{frag}\gamma K)$ distribution of partially reconstructed $D_s^+ \to \bar{K}^0K^+$ decays within the inclusive $D_s$ sample. The fit result is superimposed and includes fitted signal contribution (solid green line), reflections from charged kaon originating from $D_s^+ \to \eta K^+$ and $D_s^+ \to \pi^0K^+$ (full grey histograms) and other true $D_s$ decays (for example $D_s^+ \to \eta\pi^+$ with pion being misidentified as kaon) (full blue histogram), and combinatorial background (dashed red line).

**Table 19.1.1.** Absolute $\mathcal{B}$ of $D_s^+$ from (Zupanc, 2013b)

| Decay mode | *Exclusive* yield | $\mathcal{B}$ [%] |
|---|---|---|
| $D_s^+ \to K^+K^-\pi^+$ | $4094 \pm 123$ | $5.06 \pm 0.15 \pm 0.21$ |
| $D_s^+ \to \bar{K}^0K^+$ | $2018 \pm 75$ | $2.95 \pm 0.11 \pm 0.09$ |
| $D_s^+ \to \eta\pi^+$ | $788 \pm 59$ | $1.82 \pm 0.14 \pm 0.07$ |

All the measured decays are Cabibbo-favored processes. However, since the flavor of neutral kaon in $D_s^+ \to \bar{K}^0K^+$ is not determined, the doubly Cabibbo suppressed decays, $D_s^+ \to K^0K^+$, also contribute to the signal in Fig. 19.1.6. Its expected contribution is at the level of $10^{-4}$, thus, much below the statistical uncertainty of the measured $\mathcal{B}$. The $D_s^+ \to K^+K^-\pi^+$ decays entirely proceed through resonances contributing to either the $KK$ or $K\pi$ systems (see Section 19.1.4.9), which correspond respectively to color-allowed and color-suppressed contributions. A corresponding decay model is assumed in the $D_s^+ \to K^+K^-\pi^+$ MC simulations used for estimation the efficiency for Eq. (19.1.19).

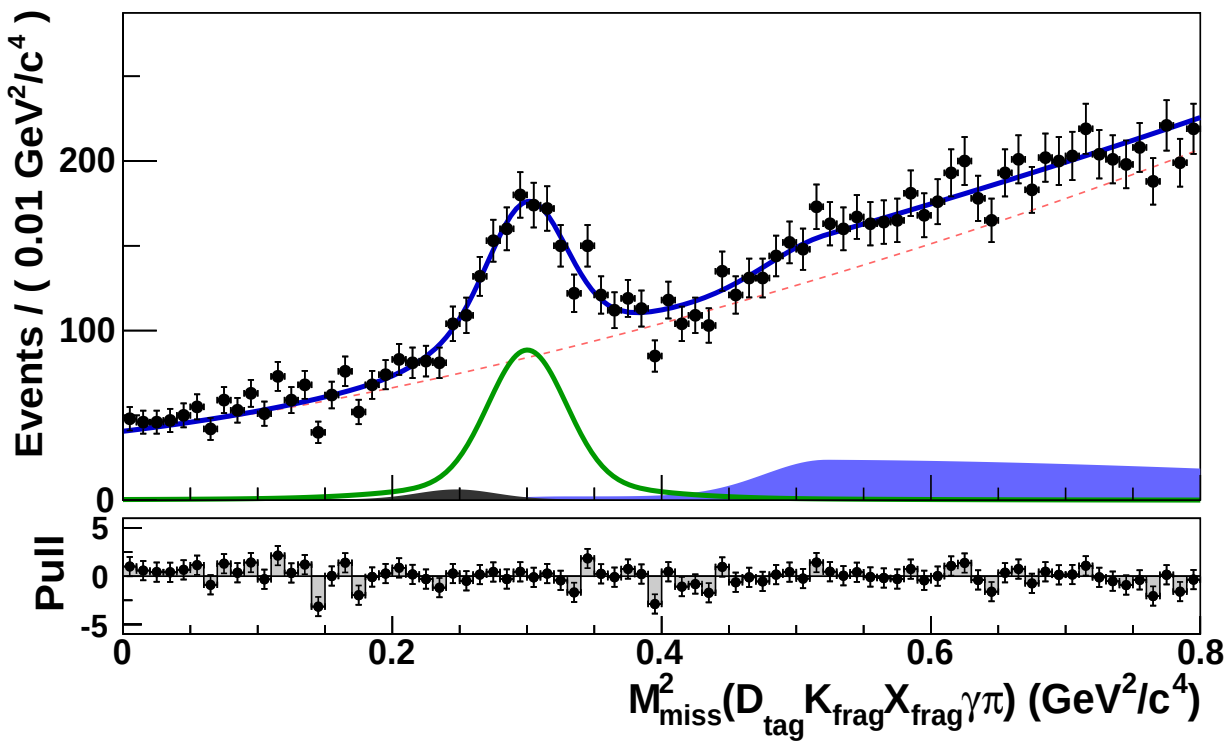

**Figure 19.1.7.** From (Zupanc, 2013b). $M^2_{miss}(D_{tag}KX_{frag}\gamma\pi)$ distribution of partially reconstructed $D_s^+ \to \eta\pi^+$ decays within the inclusive $D_s$ sample. The fit result is superimposed and includes fitted signal contribution (solid green line), reflections from charged pion originating from $D_s^+ \to K^0\pi^+$ (full grey histograms) and other true $D_s$ decays ($D_s^+ \to \rho K^+ \to \pi^+\pi^-K^+$ and $D_s^+ \to \bar{K}^0K^+$ due to $K$-$\pi$ mis-reconstruction) (full blue histogram), and combinatorial background (dashed red line).

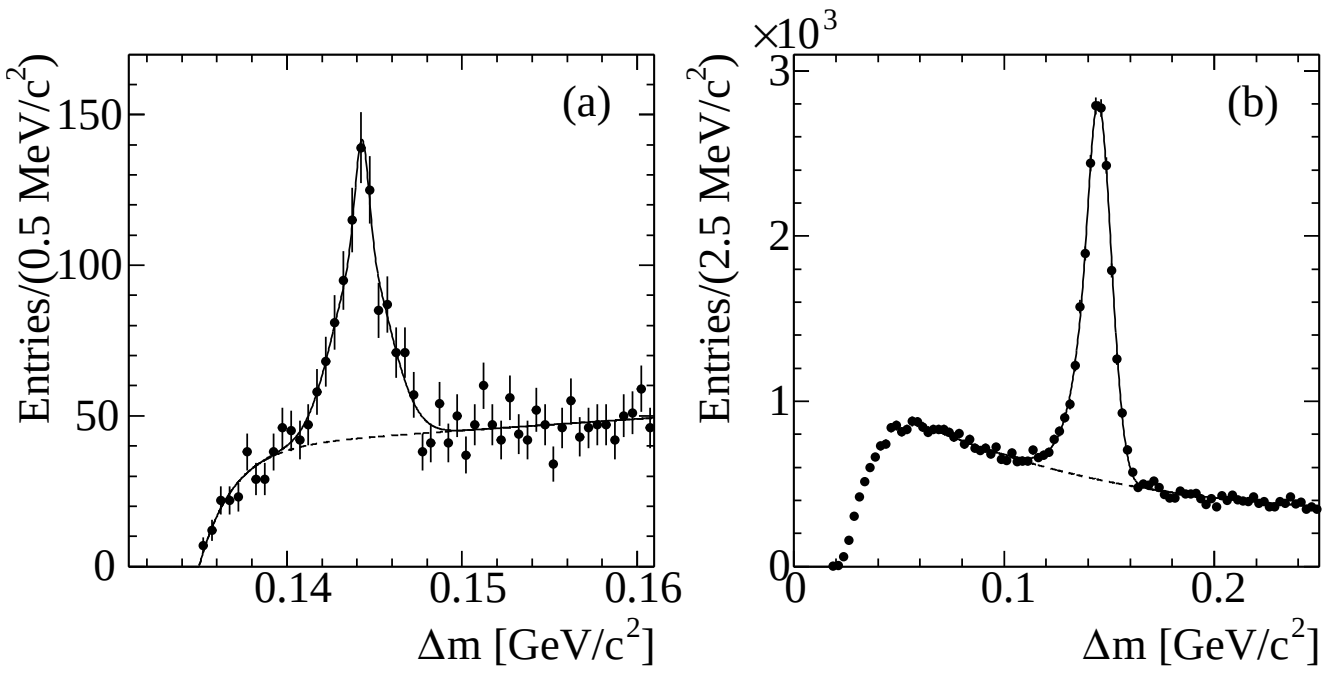

**Figure 19.1.8.** From (Aubert, 2005s). The $D_s^{*+}$ signals in: (a) $\Delta m(D_s^+\pi^0)$ and (b) $\Delta m(D_s^+\gamma)$. The dots represent data points, the solid curve shows the fitted function, the dashed curve indicates the fitted background.

### 19.1.2.3 The $D_s^{*+}$ and $D^{*0}$ Branching Ratios

The decay of any excited $c\bar{s}$ meson into $D_s^+\pi^0$ violates isospin conservation,[114] thus guaranteeing a small partial width. The amount of suppression is a matter of large theoretical uncertainty according to most models of charm-meson radiative decay (Goity and Roberts, 2001). One such a model (Cho and Wise, 1994) suggests that the decay $D_s^{*+} \to D_s^+\pi^0$ may proceed via $\pi^0$-$\eta$ mixing. Even including such considerations, the radiative decay $D_s^{*+} \to D_s^+\gamma$ is still expected to dominate. An existence of isospin-violating decay modes, such as $D_s^{*+} \to D_s^+\pi^0$, is particularly relevant given the observations of the new narrow $D_{sJ}^+$ states decaying dominantly into $D_s^{(*)+}\pi^0$ (see Section 19.3). In particular, in contrast to the $D_s^{*+}$ meson, there is no experimental evidence for the electromagnetic decay of the $D_{sJ}(2317)^+$. Besides the $D_s^+\pi^0$ and $D_s^+\gamma$ final states, no other decay modes of the $D_s^{*+}$ have been observed and none are expected to occur at a significant level.

The decay $D^{*0} \to D^0\pi^0$, in contrast to $D_s^{*+} \to D_s^+\pi^0$, does not violate isospin conservation.[115] As for the $D_s^{*+}$, the $D^0\pi^0$ and $D^0\gamma$ decay modes are expected to saturate the $D^{*0}$ decay width.

The *BABAR* analysis of the $D^*_{(s)}$ decays is based on $90.4\,\text{fb}^{-1}$ (Aubert, 2005s). $D_s^+$ mesons are reconstructed via the decay sequence $D_s^+ \to \phi\pi^+$, $\phi \to K^+K^-$, and the scaled momentum must satisfy $x_p(D_s^+) > 0.6$. The $D_s^+$ signal sample is of $(73.5 \pm 0.3) \times 10^3$ events. In a search for the $D_s^{*+} \to D_s^+\pi^0$, the $D_s^+$ and $\pi^0$ are combined and a fit is applied to the distribution of the mass difference $\Delta m(D_s^+\pi^0) = m(K^+K^-\pi^+\pi^0) - m(K^+K^-\pi^+)$. The result of this fit is shown in Fig. 19.1.8(a), and the obtained signal yield is $560 \pm 40$. To obtain the $D_s^{*+} \to D_s^+\gamma$ signal event yield, a fit is applied to the distribution of the mass difference $\Delta m(D_s^+\gamma) = m(K^+K^-\pi^+\gamma) - m(K^+K^-\pi^+)$, as shown in Fig. 19.1.8(b). The fit function is a sum of a third-order polynomial to model the background plus a Crystal Ball function for the signal.

After correcting for efficiency, the measured branching ratio is:

$$\frac{\Gamma(D_s^{*+} \to D_s^+\pi^0)}{\Gamma(D_s^{*+} \to D_s^+\gamma)} = 0.062 \pm 0.005 \pm 0.006\,, \quad (19.1.20)$$

and is consistent with the previous measurement (Gronberg et al., 1995), but has higher precision.

The ratio $\Gamma(D^{*0} \to D^0\pi^0)/\Gamma(D^{*0} \to D^0\gamma)$, where $D^0 \to K^-\pi^+$, is measured using the same selection criteria for the $\pi^0$ and photon candidates as in the $D_s^{*+}$ reconstruction. The $D^0 \to K^-\pi^+$ signal sample consists of $(996.0 \pm 1.5) \times 10^3$ events. These $D^0$ candidates combined with the $\pi^0$ candidates result in the mass difference $\Delta m(D^0\pi^0) = m(K^-\pi^+\pi^0) - m(K^-\pi^+)$ shown in Fig. 19.1.9(a). A fit, using a double Gaussian for the signal, yields $(69.0 \pm 0.5) \times 10^3$ signal events. The $D^0$ candidates combined with photons produce the distribution of the mass difference $\Delta m(D^0\gamma) = m(K^-\pi^+\gamma) - m(K^-\pi^+)$ shown in Fig. 19.1.9(b). In this case, a peak corresponding to the $D^{*0} \to D^0\gamma$ signal is close to a large reflection from $D^{*0} \to D^0\pi^0$ with one photon from $\pi^0$ decay missing (such a reflection appears also in $D_s^{*+}$ decay, but with a lower rate and less distinctive shape). The $D^{*0} \to D^0\gamma$ signal is modeled by the Crystal Ball function and the fitted signal yield is $(67.9 \pm 0.7) \times 10^3$ events. The resulting branching ratio, corrected for efficiency, is:

$$\frac{\Gamma(D^{*0} \to D^0\pi^0)}{\Gamma(D^{*0} \to D^0\gamma)} = 1.74 \pm 0.02 \pm 0.13. \quad (19.1.21)$$

## 19.1.3 Cabibbo-suppressed decays

Cabibbo-suppressed ($CS$) charm decays offer a good laboratory for studying weak interactions. Branching ratio measurements provide insight into charm decay dynamics and sources of $SU(3)$ flavor symmetry breaking, as well as allow to investigate beyond SM effects affecting decay rates. The $CS$ charm decays are very sensitive probes of the $CP$ violation and $D^0-\bar{D}^0$ mixing, as described in Section 19.2. Understanding the size of the $SU(3)$-violating effects in $D_{(s)}$ decays can help to disantangle New Physics effects from ones having origins in long-distance QCD effects.

[114] The $D_s^{(*)}$ has $I = 0$, while in the final state there is $I = 1$ due to the pion.

[115] The $D^{(*)0}$ has $I = 1/2$, so the final state pion ($I = 1$) can be combined with the $D^0$ to a total of $I = 1/2$.

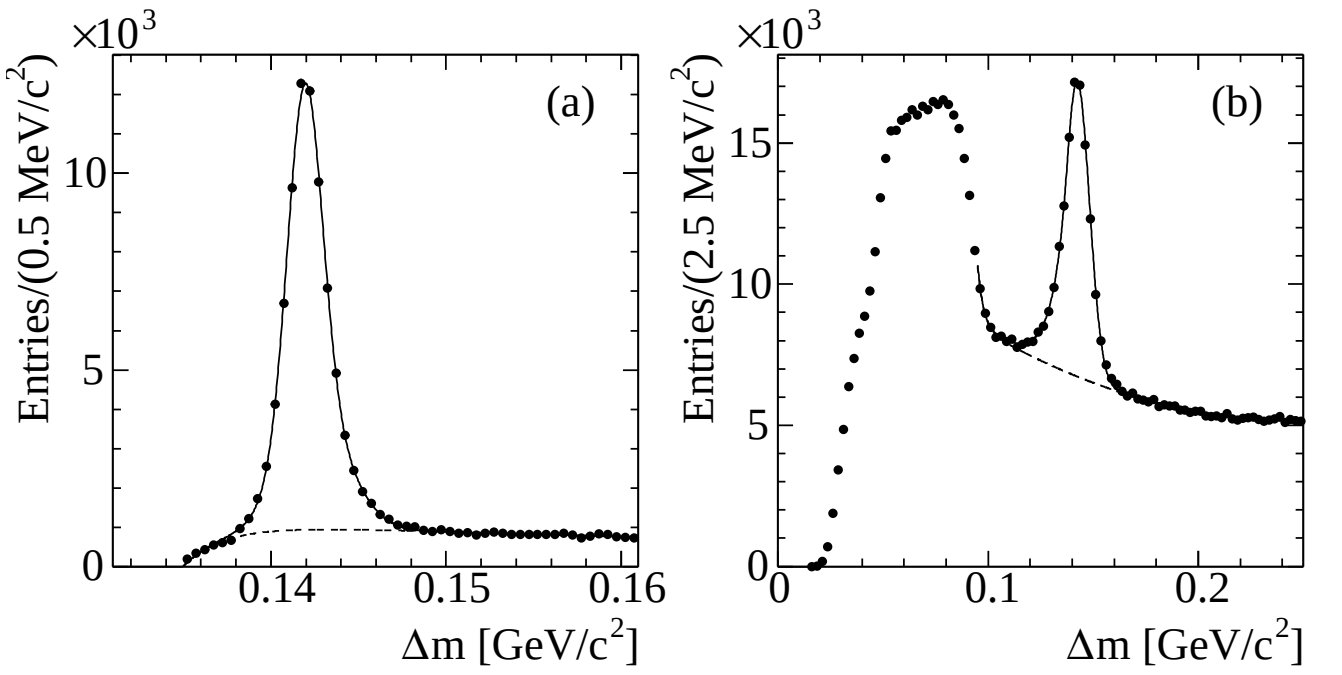

**Figure 19.1.9.** From (Aubert, 2005s). The $D^{*0}$ signals in: (a) $\Delta m(D^0\pi^0)$ and (b) $\Delta m(D^0\gamma)$. The dots represent data points, the solid curve shows the fitted function, the dashed curve indicates the fitted background.

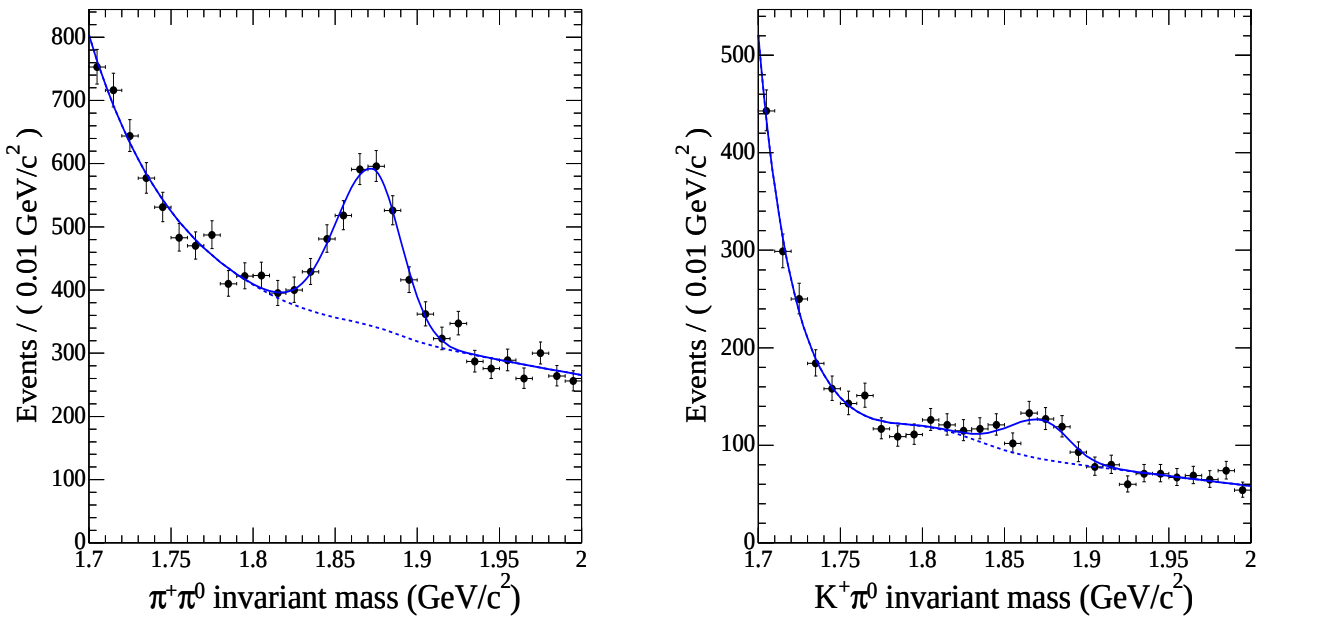

**Figure 19.1.10.** From (Aubert, 2006v). $M(\pi^+\pi^0)$ and $M(K^+\pi^0)$ with the likelihood fit results. The dashed lines show the projected backgrounds in the $D^+$ signal region.

While Cabibbo-favored ($CF$) modes at the tree level proceed through $c \to s\bar{d}u$, the singly Cabibbo-suppressed ($SCS$) decays are mediated by $c \to d\bar{d}u$ or $c \to s\bar{s}u$, and the underlying process for doubly Cabibbo-suppresed ($DCS$) modes is $c \to d\bar{s}u$. The $SCS$ decay rates are naïvely expected to be suppressed relative to $CF$ decay rate by $\tan^2\theta_C$, where $\tan\theta_C = \frac{|V_{cd}|}{|V_{cs}|} \approx 0.23$ and $\theta_C$ is the Cabibbo angle. Correspondingly, the ratio of $DCS$ and $CF$ decay rates is expected to be of $\tan^4\theta_C$. The $SU(3)$ symmetry can be however broken by strong final-state interactions, interference between different contributing amplitudes and leading to the same final states. In particular, the two-body $SCS$ decays of $D^0$ meson have anomalous rates. The $D^0 \to \pi^-\pi^+$ branching fraction is observed to be suppressed relative to the $D^0 \to K^-K^+$ by a factor of almost three (Yao et al., 2006), even though the phase space for the former is larger. On the other hand, three-body decay rates have larger uncertainties but do not exhibit such a suppression. Number of analyses of the $CS$ charm decays were previously performed mainly by MARK III, Argus, CLEO and FOCUS.

#### 19.1.3.1 The $D^+ \to \pi^+\pi^0$ and $D^+ \to K^+\pi^0$ Branching Fractions

*BABAR* has measured branching fractions of the $SCS$ $D^+ \to \pi^+\pi^0$ and the $DCS$ $D^+ \to K^+\pi^0$ decays with respect to the well-measured $D^+ \to K^-\pi^+\pi^+$ decay mode, with a data sample of $124.3\,\text{fb}^{-1}$ (Aubert, 2006v). In order to reduce a large combinatorial background in the $D^+$ signal modes, only $D^+$ mesons that originate from $D^{*+} \to D^+\pi^0$ decays are considered. $D^+$ candidates for the signal modes are obtained by combining a charged track, identified either as a pion or kaon, with a reconstructed $\pi^0$ candidate and requiring $x_p(D^*) > 0.6$. The scaled momentum of the $D^{*+}$, $x_p(D^*)$, is defined as $x_p(D^*) \equiv p^*(D^{*+})/p^*_{max}(D^{*+})$, where $p^*(D^{*+})$ is the momentum of the $D^*$ in the $e^+e^-$ c.m. frame and $p^*_{max}(D^{*+}) \equiv \sqrt{s/4 - m^2_{D^*}}$ is the maximal $D^*$ c.m. momentum allowed, with $s$ being the square of the energy of the initial $e^+e^-$ system. The energy of the $\pi^0$ in the laboratory frame (*lab*) is required to be greater than 0.2 GeV. Requirements on the $D^+$ helicity angle $\theta_h$, $-0.9 < \cos\theta_h < 0.8$ for the $D^+ \to \pi^+\pi^0$ and $-0.9 < \cos\theta_h < 0.7$ for the $D^+ \to K^+\pi^0$, are motivated by uniformly distributed $\cos\theta_C$ expected for signal events and peaking at $\pm 1$ for background. The $\theta_h$ is defined as the angle between the direction of the $D^+$ charged daughter and the direction of the $D^{*+}$ meson evaluated in the $D^+$ rest frame. Figure 19.1.10 shows the measured invariant-mass spectra, $M(\pi^+\pi^0)$ and $M(K^+\pi^0)$, with the fit results superimposed. The signal yields for $D^+ \to \pi^+\pi^0$ and $D^+ \to K^+\pi^0$ are respectively $1229\pm 98$ and $189\pm 35$, while for the $D^+ \to K^-\pi^+\pi^+$ reference mode is $101380 \pm 415$. The branching ratio of the signal and reference modes is obtained as a ratio of the measured yields ($N$) corrected for the reconstruction efficiencies ($\varepsilon$):

$$\frac{\mathcal{B}(D \to \text{signal})}{\mathcal{B}(D \to \text{reference})} = \frac{N(\text{signal})}{N(\text{reference})} \times \frac{\varepsilon(\text{reference})}{\varepsilon(\text{signal})}. \tag{19.1.22}$$

The measured branching ratios:

$$\frac{\mathcal{B}(D^+ \to \pi^+\pi^0)}{\mathcal{B}(D^+ \to K^-\pi^+\pi^+)} = (1.33 \pm 0.11 \pm 0.09) \times 10^{-2},$$
$$\frac{\mathcal{B}(D^+ \to K^+\pi^0)}{\mathcal{B}(D^+ \to K^-\pi^+\pi^+)} = (2.68 \pm 0.50 \pm 0.26) \times 10^{-3}, \tag{19.1.23}$$

combined with the world-average value of $\mathcal{B}(D^+ \to K^-\pi^+\pi^+) = (9.4 \pm 0.3)\%$ yield in the following branching fractions for the $CS$ decays:

$$\mathcal{B}(D^+ \to \pi^+\pi^0) = (1.25 \pm 0.10 \pm 0.09 \pm 0.04) \times 10^{-3},$$
$$\mathcal{B}(D^+ \to K^+\pi^0) = (2.52 \pm 0.47 \pm 0.25 \pm 0.08) \times 10^{-4}, \tag{19.1.24}$$

where the last error is due to the uncertainty in the reference mode branching fraction. This represents the first observation of the $DCS$ $D^+ \to K^+\pi^0$ decays and an improved measurement of the $SCS$ $D^+ \to \pi^+\pi^0$ branching fraction.

#### 19.1.3.2 The $D^+ \to K^0_S K^+$ and $D_s^+ \to K^0_S\pi^+$ Branching Fractions

Both $D^+ \to \bar{K}^0 K^+$ and $D_s^+ \to \bar{K}^0\pi^+$ are $DCS$ decays involving the color-favored tree, penguin and annihilation diagrams, while the related $CF$ modes are $D^+ \to \bar{K}^0\pi^+$ and $D_s^+ \to \bar{K}^0 K^+$.

Belle has measured branching fractions of the corresponding $D^+_{(s)}$ decays including $K^0_S$ in the final states, namely $D^+ \to K^0_S K^+$ and $D_s^+ \to K^0_S\pi^+$ with respect to the $D^+ \to K^0_S\pi^+$ and $D_s^+ \to K^0_S K^+$ decays, using $605\,\mathrm{fb}^{-1}$ data collected at the $\Upsilon(4S)$ resonance (Won, 2009). An additional $60\,\mathrm{fb}^{-1}$ of the off-resonance data collected below the $\Upsilon(4S)$ have been used for the optimization procedures. They select the $K^0_S \to \pi^+\pi^-$ candidates having daughter-pion tracks separated from the interaction point (IP) in the plane perpendicular to the beam axis and $\pi^+\pi^-$ vertex displaced from the IP, whereas the direction of the $K^0_S$ momentum must agree with the direction of the decay vertex point from the IP. Selection criteria on the related distances and angle are optimized, separately for high-momentum (above 1.5 GeV/$c$) and the remaining $K^0_S$ candidates, to maximize a significance of the $K^0_S$ signal identified in the $M(\pi^+\pi^-)$ invariant-mass spectrum. The reconstructed $D^+_{(s)} \to K^0_S h^+$ candidates, where $h^+ = K^+,\ \pi^+$, are required to have a good quality decay vertex and the c.m. momentum greater than 2.6 GeV/$c$ to remove the $D^+_{(s)}$ produced in $B$ meson decays. Removing the $K^0_S h^+$ pairs with an invariant mass close to the nominal $D^+_{(s)}$ mass but having in the laboratory frame (*lab*) highly asymmetrical momenta, has significantly improved significance of the $CS$ $D^+_{(s)}$ signals.

The $M(K^0_S K^+)$ and $M(K^0_S\pi^+)$ invariant mass distributions after the final selections, shown in Fig 19.1.11, exhibit clear signals for both $CF$ and $DCS$ decays in the both decay channels. All the signals are parameterized using double Gaussians with a common mean value. In the maximum-likelihood fits performed to the mass spectra, all the signal parameters are kept free, except for the broad Gaussian fraction and width for the $D_s^+ \to K^0_S\pi^+$; those are fixed to the values obtained for the $D^+ \to K^0_S\pi^+$ mode. In addition to the smooth combinatorial background, there is a peaking background due to particle misidentification. It appears in the $D_s^+ \to K^0_S K^+$ mass region, when $\pi^+$ from $D^+ \to K^0_S\pi^+$ decays is misidentified as $K^+$. Similarly, when $K^+$ is misidentified as $\pi^+$ in $D_s^+ \to K^0_S K^+$ decays, a peaking structure appears under the $D^+ \to K^0_S\pi^+$. The shapes and yields of these peaking backgrounds are obtained from the MC simulations in which hadron momentum scale and resolution are tuned with the data.

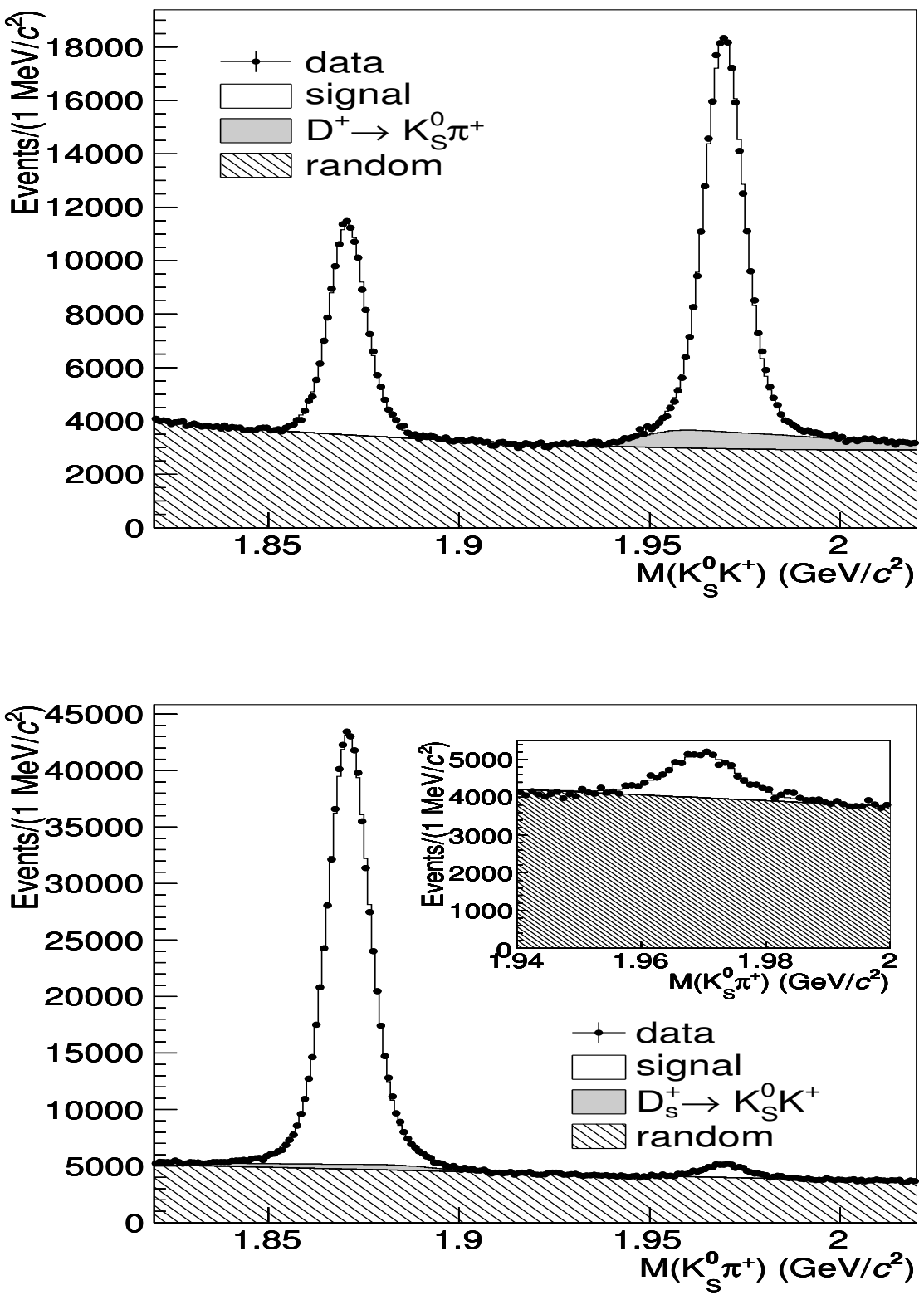

**Figure 19.1.11.** From (Won, 2009). $M(K^0_S K^+)$ (top) and $M(K^0_S\pi^+)$ (bottom) for the selected candidates. Points with error bars show the data, the histograms represent the fit results. Signals for $D^+_{(s)} \to K^0_S K^+$ (top) and $D^+_{(s)} \to K^0_S\pi^+$ (bottom), peaking backgrounds originating from misidentified $D^+ \to K^0_S\pi^+$ (top) and $D_s^+ \to K^0_S K^+$ (bottom) decays, and random combinatorial backgrounds are also shown. The inset in the bottom plot is enlarged view of the $D_s^+$ region.

Based on the fitted signal yields and reconstruction efficiencies (of about $12-15\%$) obtained with the tuned MC, the measured branching ratios are:

$$\frac{\mathcal{B}(D^+ \to K^0_S K^+)}{\mathcal{B}(D^+ \to K^0_S\pi^+)} = (18.99 \pm 0.11 \pm 0.22)\%$$
$$\frac{\mathcal{B}(D_s^+ \to K^0_S\pi^+)}{\mathcal{B}(D_s^+ \to K^0_S K^+)} = (8.03 \pm 0.24 \pm 0.19)\%. \tag{19.1.25}$$

These are the most precise measurements to date and agree with the present WA values of respectively $(20.6 \pm 1.4)\%$ and $(8.4 \pm 0.9)\%$.

The ratio for $D^+$ is larger than naïve expectation of $\tan^2\theta_C$ due to destructive interference between color-favored and color-suppressed tree diagrams contributing to the $CF$ $D^+ \to \bar{K}^0\pi^+$ decay. Increase of the ratio mesured for $D_s^+$ over the $\tan^2\theta_C$ can be due to the color suppression in the tree amplitude in the $D_s^+ \to \bar{K}^0 K^+$. Using the WA values for $CF$ modes, the branching fractions for $CS$ decays are:

$$\mathcal{B}(D^+ \to K^0_S K^+) = (2.75 \pm 0.08) \times 10^{-3}$$
$$\mathcal{B}(D^+_s \to K^0_S \pi^+) = (1.20 \pm 0.09) \times 10^{-3}, \quad (19.1.26)$$

with statistical and systematic uncertainties summed in quadrature. However, experimentaly measured are $D_{(s)}$ decays including $K^0_S$ and converting them to the $\mathcal{B}$ involving $K^0$ or $\bar{K}^0$ is not straightforward, as the corresponding $DCS$ and $CF$ modes can interfere with the unknown interference phase.

#### 19.1.3.3 $D^+ \to K^+\eta^{(')}$ and $D^+ \to \pi^+\eta^{(')}$

$D^+$ decays into two-body final states with $\eta^{(')}$ are all the $CS$ decays with poorly studied $SU(3)$ flavor symmetry structure. Such $DCS$ decays, $D^+ \to K^+\eta^{(')}$, have not been observed before, while their $SCS$ counterparts, $D^+ \to \pi^+\eta^{(')}$, are quite well measured by CLEO.

Belle analysis of the $D^+ \to h^+\eta^{(')}$ decays (Won, 2011), with $h^+ = K^+,\ \pi^+$, is based on the data sample of 791 fb$^{-1}$ and exploits the decays $\eta \to \pi^+\pi^-\pi^0$ and $\eta^{'} \to \eta\pi^+\pi^-$ with $\eta \to \gamma\gamma$. This allows reconstruction of the $D^+$ decay vertex formed using only charged tracks. The $\pi^0$ and $\eta^{(')}$ candidates are selected based on the invariant masses of their decay products. For the final selection the authors optimize criteria to maximize significance of the $D^+ \to K^+\eta^{(')}$ signal studied with the MC simulations. The following variables are considered in the optimization: the $D^+$ c.m. momentum, the $\eta^{(')}$ *lab* momentum, as well as variables related to the decay topology: an angle between momentum vector of the reconstructed $D^+$ and the vector joining its production and decay vertices, and $\chi^2$ of the hypothesis that the candidate tracks forming the $D^+$ are isolated from the primary vertex, defined as a point of intersection of the $D^+$ momentum vector with the IP. Such an isolation is expected due to the finite $D^+$ lifetime. The optimized selection cuts are also applied to the normalization modes, $D^+ \to \pi^+\eta^{(')}$.

The measured $M(\pi^+\eta^{(')})$ and $M(K^+\eta^{(')})$ distributions are shown in Fig. 19.1.12. The signal function is modelled with a sum of Gaussian and bifurcated Gaussian. In the fits for the $DCS$ decays, widths of both Gaussians and bifurcated Gaussian fraction are fixed to the values obtained for the $SCS$ decays and scaled according to the difference obtained with the MC simulations. Observed background is purely combinatorial and smooth, no peaking background has been detected. The signal yields of the $DCS$ modes amount to $166 \pm 23$ for the $D^+ \to K^+\eta$ and $188 \pm 19$ for the $D^+ \to K^+\eta^{'}$, while reconstruction efficiencies are at the level of 1.5%. The measured branching ratios are:

$$\frac{\mathcal{B}(D^+ \to K^+\eta)}{\mathcal{B}(D^+ \to \pi^+\eta)} = (3.06 \pm 0.43 \pm 0.14) \times 10^{-2}$$
$$\frac{\mathcal{B}(D^+ \to K^+\eta^{'})}{\mathcal{B}(D^+ \to \pi^+\eta^{'})} = (3.77 \pm 0.39 \pm 0.10) \times 10^{-2}. \quad (19.1.27)$$

The $D^+ \to K^+\eta^{(')}$ decays are observed for the first time and have completed a class of the $DCS$ $D^+$ decays to pairs of light pseudoscalar mesons. The measured ratios are within errors in agreement with $SU(3)$ based expectations. Using the measurements of the $\mathcal{B}(D^+ \to \pi^+\eta^{(')})$ from (Mendez et al., 2010), the absolute branching fractions for the $DCS$ modes are $\mathcal{B}(D^+ \to K^+\eta) = (1.08 \pm 0.17 \pm 0.08) \times 10^{-4}$ and $\mathcal{B}(D^+ \to K^+\eta^{'}) = (1.76 \pm 0.22 \pm 0.12) \times 10^{-4}$.

Using relations from (Chiang and Rosner, 2002), the measured $\mathcal{B}(D^+ \to K^+\eta^{(')})$ toghether with the WA value $\mathcal{B}(D^+ \to K^+\pi^0) = (1.72 \pm 0.20) \times 10^{-4}$ (Chiang and Rosner, 2002) are used to calculate a relative phase difference between the contributing tree and annihilation amplitudes, $\delta_{TA}$, to be $(72 \pm 9)^\circ$ or $(288 \pm 9)^\circ$. It is an important information for final-state interactions in $D$ decays. Similar measurement for penguin amplitudes is impossible, as they contribute to $DCS$ decays involving $K^0$, while these are overwhelmed by $CF$ decays involving $\bar{K}^0$ in the detected $K^0_S$.

#### 19.1.3.4 Branching Ratios of the Decays $D^0 \to \pi^-\pi^+\pi^0$ and $D^0 \to K^-K^+\pi^0$

The *BABAR* experiment has measuread the rates of three-body $CS$ decays $D^0 \to \pi^-\pi^+\pi^0$ and $D^0 \to K^-K^+\pi^0$ relative to the $CF$ decay $D^0 \to K^-\pi^+\pi^0$, using 232 fb$^{-1}$ (Aubert, 2006ak). To reduce combinatorial backgrounds, the $D^0$ candidates are reconstructed in decays $D^{*+} \to D^0\pi^+_s$. The number of $D^0$ signal events in each decay mode is obtained by fitting the observed $D^0$ candidate mass distributions (Fig. 19.1.13) to the sum of signal and background components, where the latter has combinatorial contributions and reflection contributions from real three-body $D^0$ decays where a kaon (pion) is misidentifed as a pion (kaon). The reconstruction efficiency for each event is calculated as a function of its position in the $D^0$ Dalitz plot. The resulting branching ratios,

$$\frac{\mathcal{B}(D^0 \to \pi^-\pi^+\pi^0)}{\mathcal{B}(D^0 \to K^-\pi^+\pi^0)} = (10.59 \pm 0.06 \pm 0.13) \times 10^{-2}$$
$$\frac{\mathcal{B}(D^0 \to K^-K^+\pi^0)}{\mathcal{B}(D^0 \to K^-\pi^+\pi^0)} = (2.37 \pm 0.03 \pm 0.04) \times 10^{-2}, \quad (19.1.28)$$

combined with $\mathcal{B}(D^0 \to K^-\pi^+\pi^0) = (14.1 \pm 0.5) \times 10^{-2}$ (Yao et al., 2006), give:

$$\mathcal{B}(D^0 \to \pi^-\pi^+\pi^0) = (1.493 \pm 0.008 \pm 0.018 \pm 0.053) \times 10^{-2},$$
$$\mathcal{B}(D^0 \to K^-K^+\pi^0) = (0.334 \pm 0.004 \pm 0.006 \pm 0.012) \times 10^{-2}. \quad (19.1.29)$$

The measured branching ratios contain the phase space factor which enters decay rate as[116] $\Gamma = \int d\Phi|\mathcal{M}|^2 =$

[116] Branching fraction and decay width for given process $D \to f$ are related through $\mathcal{B}(D \to f) = \Gamma(D \to f) \times \tau_D$, where $\tau_D \equiv 1/\Gamma_{total}$ is $D$ lifetime

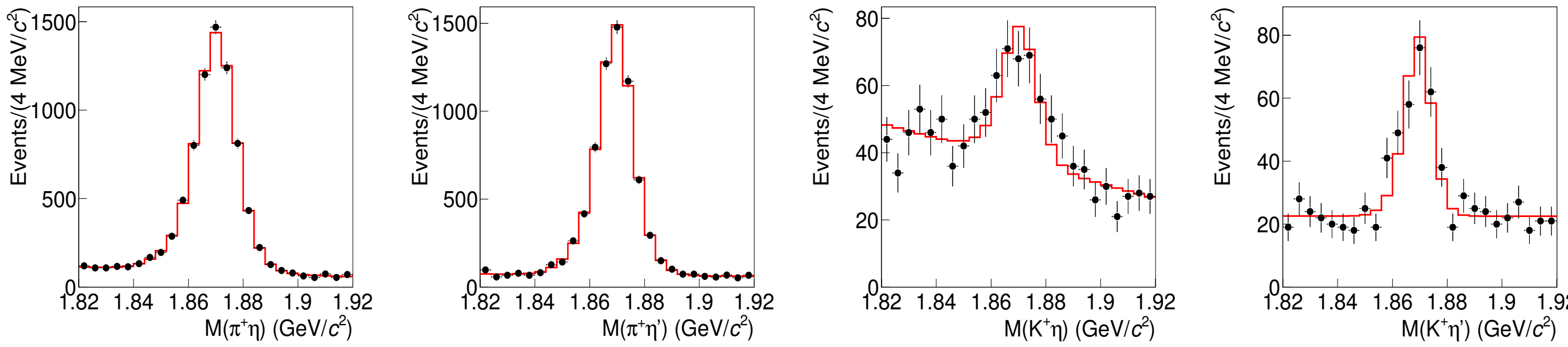

**Figure 19.1.12.** From (Won, 2011). Invariant mass ditributions for the $\pi^+\eta$, $\pi^+\eta^{'}$, $K^+\eta$ and $K^+\eta^{'}$ final states. Points with error bars and histograms correspond respectively to the data and fit result.

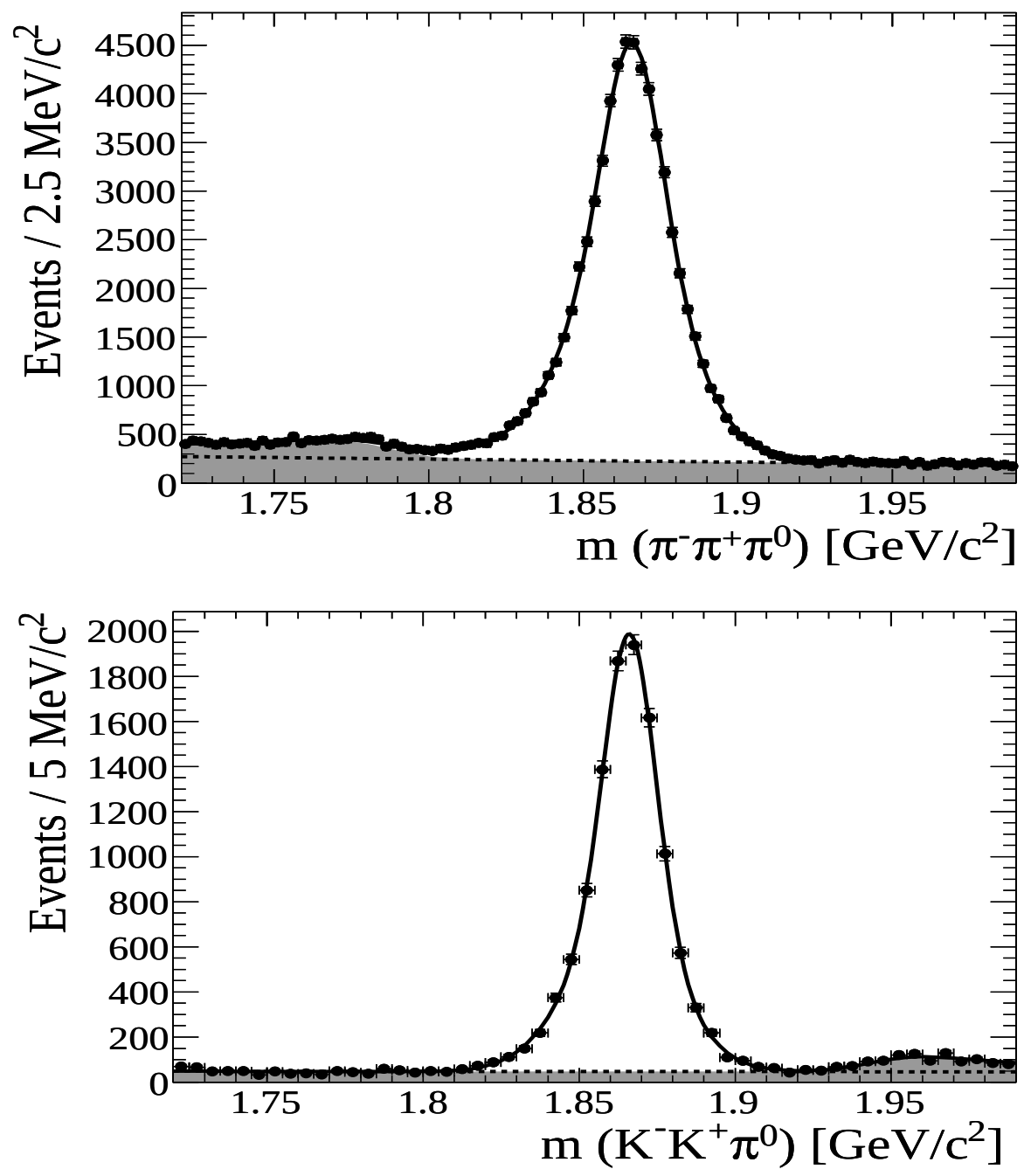

**Figure 19.1.13.** From (Aubert, 2006ak). Fitted mass for the $\pi^-\pi^+\pi^0$, and $K^-K^+\pi^0$ data samples. Dots are data points, the solid curves are the fit. The dot-dashed lines show the level of combinatorial background, the shaded region represents the total background.

$\Phi \times \langle|\mathcal{M}|^2\rangle$, where $\Phi$ is the phase space of a particular final state, $\mathcal{M}$ is the decay matrix element, and $\langle|\mathcal{M}|^2\rangle$ is average $|\mathcal{M}|^2$ value over the Dalitz plot and the three-body phase space. The relative $\Phi$ for the studied decays is $\pi^-\pi^+\pi^0 \ : \ K^-\pi^+\pi^0 \ : \ K^+K^+\pi^- = 5.0 : 3.2 : 1.7$, which gives:

$$\frac{\langle|\mathcal{M}|^2\rangle(D^0 \to \pi^-\pi^+\pi^0)}{\langle|\mathcal{M}|^2\rangle(D^0 \to K^-\pi^+\pi^0)} = (6.68 \pm 0.04 \pm 0.08) \times 10^{-2} \tag{19.1.30}$$

$$\frac{\langle|\mathcal{M}|^2\rangle(D^0 \to K^-K^+\pi^0)}{\langle|\mathcal{M}|^2\rangle(D^0 \to K^-\pi^+\pi^0)} = (4.53 \pm 0.06 \pm 0.08) \times 10^{-2} \tag{19.1.31}$$

$$\frac{\langle|\mathcal{M}|^2\rangle(D^0 \to K^-K^+\pi^0)}{\langle|\mathcal{M}|^2\rangle(D^0 \to \pi^-\pi^+\pi^0)} = (6.78 \pm 0.14 \pm 0.21) \times 10^{-1}. \tag{19.1.32}$$

The deviations from the naïve picture, in which the ratios 19.1.30 and 19.1.31 are of the order $\tan^2\theta_C$, while 19.1.32 is of order unity, are less than 35% for these three-body final states. In contrast, the corresponding ratios may be calculated for the two-body decays $D^0 \to \pi^-\pi^+$, $D^0 \to K^-\pi^+$, and $D^0 \to K^-K^+$. Using the WA values for these two-body branching ratios (Yao et al., 2006), the ratios corresponding to Eqs (19.1.30)–(19.1.32), are, respectively, 0.034±0.001, 0.111±0.002, and $3.53 \pm 0.12$. Thus the naïve Cabibbo-suppression model works well for three-body final states, but not so good for two-body decays.

#### 19.1.3.5 Doubly-Cabibbo Suppressed $D_s^+ \to K^+K^+\pi^-$ and $D^+ \to K^+\pi^+\pi^-$ Decays

The expected branching ratio for the $DCS$ $D_s^+ \to K^+K^+\pi^-$ with respect to its $CF$ counterpart $D_s^+ \to K^+K^-\pi^+$ is about $\frac{1}{2}\tan^4\theta_C$. A factor modifing the $SU(3)$ based expectation arises from the phase space suppresion for the $D_s^+ \to K^+K^+\pi^-$ due to the two identical kaons in the final state. For a similar reason, the ratio of decays rates for corresponding $DCS$ and $CF$ $D^+$ decays, $D^+ \to K^+\pi^+\pi^-$ and $D^+ \to K^-\pi^+\pi^+$, should be about $2\tan^4\theta_C$. Thus within $SU(3)$ symmetry one expects

$$\frac{\mathcal{B}(D_s^+ \to K^+K^+\pi^-)}{\mathcal{B}(D_s^+ \to K^+K^-\pi^+)}\frac{\mathcal{B}(D^+ \to K^+\pi^+\pi^-)}{\mathcal{B}(D^+ \to K^-\pi^+\pi^+)} = \tan^8\theta_C, \tag{19.1.33}$$

as the phase-space related factors cancel out (Lipkin, 2003). In order to test this prediction, Belle has searched for the $DCS$ $D_s^+ \to K^+K^+\pi^-$ decays and studied the other decays entering Eq. (19.1.33) with 605 fb$^{-1}$ data (Ko, 2009). All the $D_{(s)}^+$ candidates are required to have the scaled momentum $x_p$ greater than 0.5 and a good quality decay vertex. In addition, track isolation and consistency of the $D_{(s)}^+$ momentum vector with its production-to-decay vector (see Section 19.1.3.3) are required, with related variables optimized for the $CF$ decays using a small part of the data further discarded from the measurement.

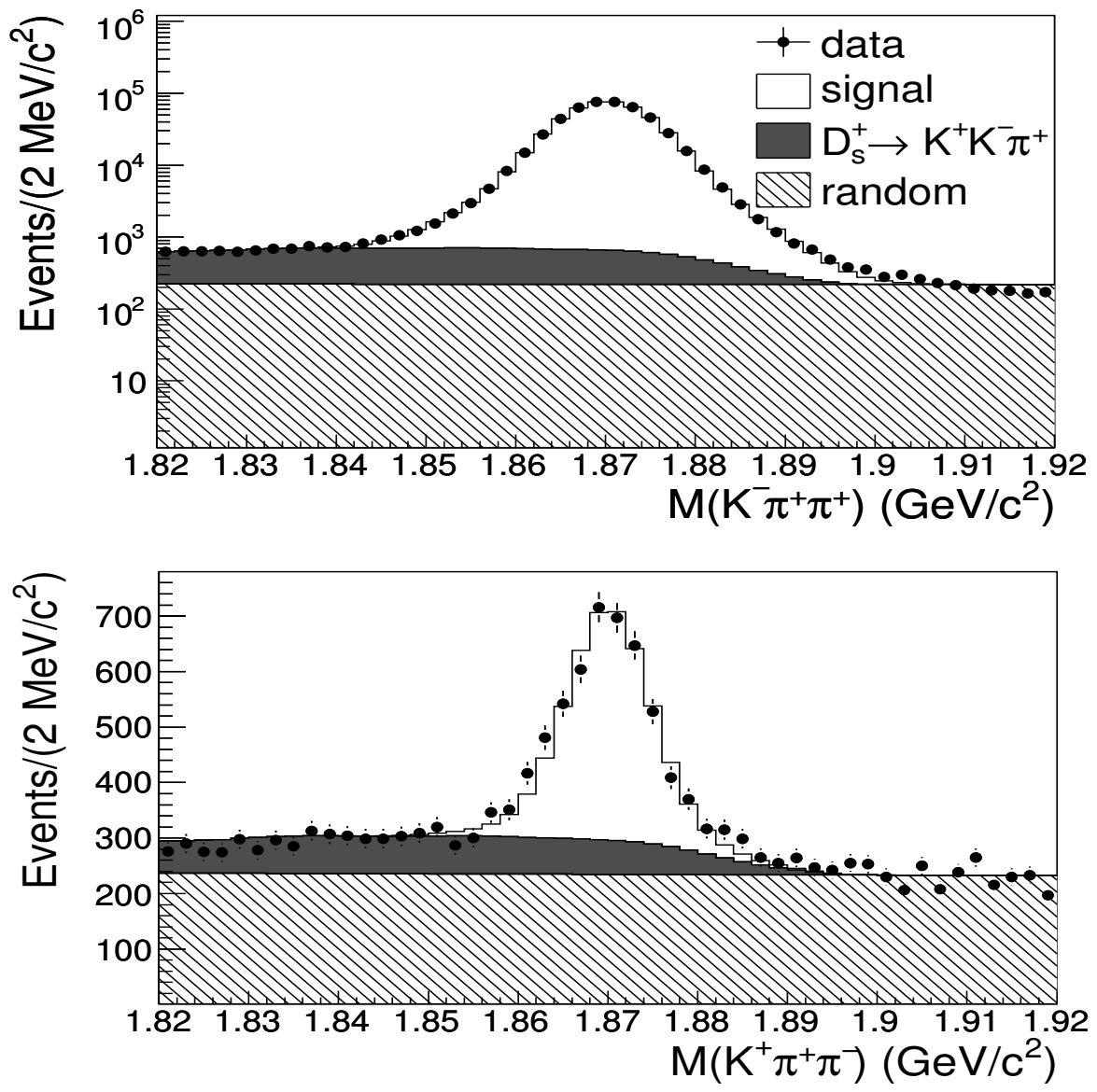

**Figure 19.1.14.** From (Ko, 2009). Invariant mass ditributions for the $K^+\pi^+\pi^-$ and $K^-\pi^+\pi^+$ final states. Points with error bars correspond to the data, histogram is fit result and include signal, random (combinatorial) background and peaking background from $D_s^+ \to K^+K^-\pi^+$.

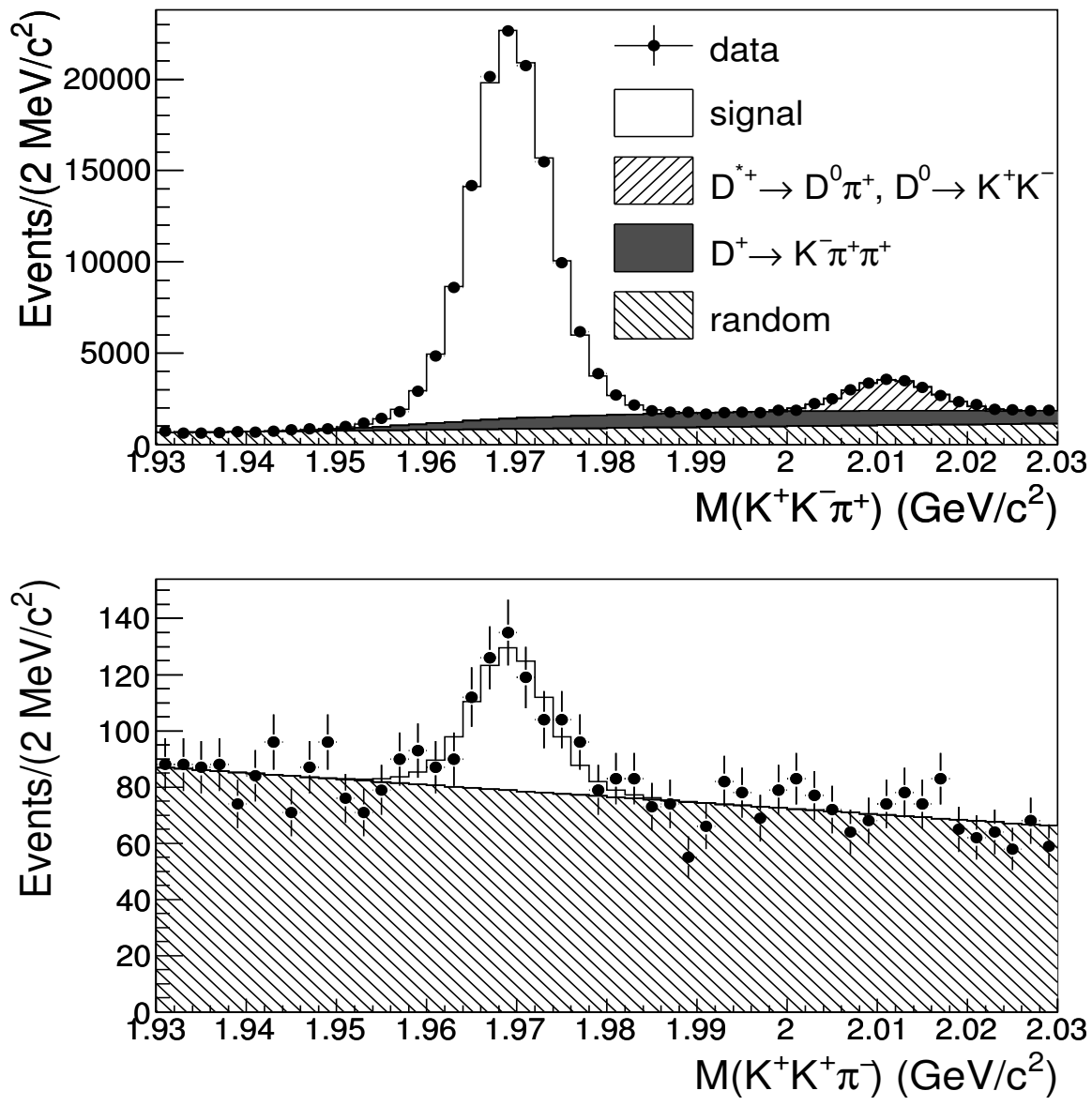

**Figure 19.1.15.** From (Ko, 2009). Invariant mass ditributions for the $K^+K^+\pi^-$ and $K^+K^-\pi^+$ final states. Points with error bars correspond to the data, histogram is fit result and includes signal, random (combinatorial) background and peaking background from $D^+ \to K^-\pi^+\pi^+$ and $D^{*+} \to D^0\pi^+$ with $D^0 \to K^+K^-$.

The measured $M(K^+\pi^+\pi^-)$ and $M(K^-\pi^+\pi^+)$, shown in Fig. 19.1.14, exibit the $D^+$ signals, while $M(K^+K^+\pi^-)$ and $M(K^+K^-\pi^+)$ in Fig. 19.1.15 the $D_s^+$ signals. The signals are parameterized with double Gaussians, with the parameters for the $DCS$ modes fixed to the values fitted for the $CF$ decays. There are peaking backgrounds detected coming from the $K$-$\pi$ misidentification: $D_s^+ \to K^+K^-\pi^+$ reflecting in $D^+ \to K^+\pi^+\pi^-$ and $D^+ \to K^-\pi^+\pi^+$, and $D^+ \to K^-\pi^+\pi^+$ contributing to $D_s^+ \to K^+K^-\pi^+$. Shapes of the reflections are determined from the real data by assigning a nominal kaon (pion) mass to pion (kaon) track, and their yields are kept free in the fits. The measured branching ratios are:

$$\frac{\mathcal{B}(D^+ \to K^+\pi^+\pi^-)}{\mathcal{B}(D^+ \to K^-\pi^+\pi^+)} = (0.569 \pm 0.018 \pm 0.014) \times 10^{-2}$$
$$\frac{\mathcal{B}(D_s^+ \to K^+K^+\pi^-)}{\mathcal{B}(D_s^+ \to K^+K^-\pi^+)} = (0.229 \pm 0.028 \pm 0.012) \times 10^{-2}. \tag{19.1.34}$$

Reconstruction efficiencies used for these $\mathcal{B}$ calculations are based on the MC simulations which include intermediate resonances contributing to the $D^+ \to K^-\pi^+\pi^+$, $D^+ \to K^+\pi^+\pi^-$ and $D_s^+ \to K^+K^-\pi^+$ decays (Amsler et al., 2008; Anjos et al., 1993). As the Dalitz-plot model for the $D_s^+ \to K^+K^+\pi^-$ is not known, it is generated according to the phase space model, while the largest relative differences obtained with various decay models assumed are included in the sytematic uncertainty.

The double branching ratio in Eq. (19.1.33) is measured to be $(1.57 \pm 0.21) \tan^8\theta_C$, with the error being the total uncertainty. Its slight deviation from the expected value can be the effect of different resonant intermediate state, which are not taken into account in the prediction (Lipkin, 2003). Using the world average values for the $CF$ branching fractions (Amsler et al., 2008), the absolute $\mathcal{B}$ for the $DCS$ decays are:

$$\mathcal{B}(D^+ \to K^+\pi^+\pi^-) = (5.2 \pm 0.2 \pm 0.1) \times 10^{-4},$$
$$\mathcal{B}(D_s^+ \to K^+K^+\pi^-) = (1.3 \pm 0.2 \pm 0.1) \times 10^{-4}. \tag{19.1.35}$$

The former is an improvement of the existing measurement, the latter comprises the first significant measurement.

#### 19.1.3.6 Wrong-Sign Decays $D^0 \to K\pi$, $D^0 \to K\pi\pi^0$, $D^0 \to K\pi\pi\pi$

Some $DCS$ $D^0$ decays are measured for the sake of the $D^0-\bar{D}^0$ mixing phenomenon appearing in so called *wrong-sign* ($WS$) decays (see Section 19.2), like $D^0 \to K^+\pi^-$, $D^0 \to K^+\pi^-\pi^0$ or $D^0 \to K^+\pi^-\pi^+\pi^-$, named after a charge of the final state kaon being opposite to the one produced in counterpart $CF$ processes: $D^0 \to K^-\pi^+$, $D^0 \to K^-\pi^+\pi^0$ or $D^0 \to K^-\pi^+\pi^-\pi^-$, and being called *right-sign* ($RS$) decays. The $WS$ process proceeds either through direct $DCS$ decay (for example $D^0 \to K^+\pi^-$) or through $D^0 - \bar{D}^0$ mixing followed by $RS$ $CF$ decays (*i.e.* $D^0 - \bar{D}^0 \to K^+\pi^-$). These two decays can be distinguished by the $D^0$ decay-time distribution (Eq. 19.2.22),

**Table 19.1.2.** $R_{WS}$ measurements from (Tian, 2005; Zhang, 2006) [$10^{-3}$].

| Decay mode | Branching Ratio |
|---|---|
| $\frac{D^0\to K^+\pi^-}{D^0\to K^-\pi^+}$ | $3.77 \pm 0.08 \pm 0.05$ |
| $\frac{D^0\to K^+\pi^-\pi^0}{D^0\to K^-\pi^+\pi^0}$ | $2.29 \pm 0.15^{+0.13}_{-0.09}$ |
| $\frac{D^0\to K^+\pi^-\pi^+\pi^-}{D^0\to K^-\pi^+\pi^+\pi^-}$ | $3.20 \pm 0.18^{+0.18}_{-0.13}$ |

and the $R_D$ entering this formula is the ratio of the $DCS$ and $CF$ decays.

In the Belle studies of the $D^0 \to K^+\pi^-$ (Zhang, 2006) and of the $D^0 \to K^+\pi^-\pi^0$ and $D^0 \to K^+\pi^-\pi^+\pi^-$ (Tian, 2005) the $D^0$ mesons originating from $D^{*+} \to D^0\pi^+$ decays are studied, and both $RS$ and $WS$ signals are obtained from the two-dimensional fit to invariant mass of the $D^0$ decay products, $M(D^0)$, and an energy release in $D^{*+}$ decays, $Q \equiv M(D^{*+}) - M(D^0) - m_\pi$. Measured ratios of the $WS$ to the $RS$ signals are summarized in Table 19.1.2. Efficiencies for both multi-body $D^0$ decays take into account that they are dominated by various intermediate resonances, that can be different for the $RS$ and $WS$ decays. The event yields are corrected for acceptance in multi-dimensional space comprised of the invariant mass squared for various $K\pi$ and $\pi\pi$ subsystems. Given that the mixing is small, $R_{WS} \approx R_D$. Fit to distribution of the $WS$ $D^0 \to K^+\pi^-$ proper decay time yields $R_D = (3.64 \pm 0.17) \times 10^{-3}$, in agreement with the $R_{WS}$. Thus the $R_{WS}$ gives a good estimate of the branching ratio of the $DCS$ to $CF$ rates. In terms of consistency with the $SU(3)$ symmetery based prediction, all the results in Table 19.1.2 are consistent with the expected $\tan^4\theta_C$.

#### 19.1.3.7 Summary of the $CS$ decays

The measured ratios of $CS$ to $CF$ branching ratios are summarized in Table 19.1.3, while branching fractions of $CS$ decays, extracted from measurements relative to reference modes are listed in Table 19.1.4.

**Table 19.1.3.** Branching ratios measured by Belle and BABAR.

| Ratio | type | $\mathcal{B}$ [$10^{-2}$] |
|---|---|---|
| $\frac{D^+\to\pi^+\pi^0}{D^+\to K^-\pi^+\pi^+}$ | $\frac{SCS}{CF}$ | $1.33 \pm 0.11 \pm 0.09$ |
| $\frac{D^+\to K^+\pi^0}{D^+\to K^-\pi^+\pi^+}$ | $\frac{DCS}{CF}$ | $0.268 \pm 0.05 \pm 0.026$ |
| $\frac{D^+\to K^0_S K^+}{D^+\to K^0_S\pi^+}$ | $\frac{DCS}{CF}$ | $18.99 \pm 0.11 \pm 0.22$ |
| $\frac{D^+_s\to K^0_S\pi^+}{D^+_s\to K^0_S K^+}$ | $\frac{DCS}{CF}$ | $8.03 \pm 0.24 \pm 0.19$ |
| $\frac{D^+\to K^+\eta}{D^+\to\pi^+\eta}$ | $\frac{DCS}{SCS}$ | $3.06 \pm 0.43 \pm 0.14$ |
| $\frac{D^+\to K^+\eta'}{D^+\to\pi^+\eta'}$ | $\frac{DCS}{SCS}$ | $3.77 \pm 0.39 \pm 0.10$ |
| $\frac{D^0\to\pi^-\pi^+\pi^0}{D^0\to K^-\pi^+\pi^0}$ | $\frac{SCS}{CF}$ | $6.68 \pm 0.04 \pm 0.08$ |
| $\frac{D^0\to K^-K^+\pi^0}{D^0\to K^-\pi^+\pi^0}$ | $\frac{SCS}{CF}$ | $4.53 \pm 0.06 \pm 0.08$ |
| $\frac{D^+\to K^+\pi^+\pi^-}{D^+\to K^-\pi^+\pi^+}$ | $\frac{DCS}{CF}$ | $0.569 \pm 0.018 \pm 0.014$ |
| $\frac{D^+_s\to K^+K^+\pi^-}{D^+_s\to K^+K^-\pi^+}$ | $\frac{DCS}{CF}$ | $0.229 \pm 0.028 \pm 0.012$ |

**Table 19.1.4.** Branching fractions of $CS$ decays, extracted from measurements relative to reference modes. Errors are respectively statistical, systematic and due to the uncertainty in the reference mode branching fraction. If single error is quoted it includes all these contributions.

| Decay mode | Branching Fraction |
|---|---|
| $D^+ \to \pi^+\pi^0$ | $(1.25 \pm 0.10 \pm 0.09 \pm 0.04) \times 10^{-3}$ |
| $D^+ \to K^+\pi^0$ | $(2.52 \pm 0.47 \pm 0.25 \pm 0.08) \times 10^{-4}$ |
| $D^+ \to K^0_S K^+$ | $(2.75 \pm 0.08) \times 10^{-3}$ |
| $D^+_s \to K^0_S\pi^+$ | $(1.20 \pm 0.09) \times 10^{-3}$ |
| $D^+ \to K^+\eta$ | $(1.08 \pm 0.17 \pm 0.08) \times 10^{-4}$ |
| $D^+ \to K^+\eta'$ | $(1.76 \pm 0.22 \pm 0.12) \times 10^{-4}$ |
| $D^+ \to K^+\pi^+\pi^-$ | $(5.2 \pm 0.2 \pm 0.1) \times 10^{-4}$ |
| $D^+_s \to K^+K^+\pi^-$ | $(1.3 \pm 0.2 \pm 0.1) \times 10^{-4}$ |
| $D^0 \to \pi^-\pi^+\pi^0$ | $(1.493 \pm 0.008 \pm 0.018 \pm 0.053) \times 10^{-2}$ |
| $D^0 \to K^-K^+\pi^0$ | $(0.334 \pm 0.004 \pm 0.006 \pm 0.012) \times 10^{-2}$ |

### 19.1.4 Dalitz analysis of three-body charmed meson decays

#### 19.1.4.1 Introduction

Dalitz plot analyses of three-body charm decays can provide new information on the resonances that contribute to observed three-body final states. In addition, since the intermediate quasi-two-body modes are dominated by light quark meson resonances, new information on light meson spectroscopy can be obtained. Comparison between the production of resonances in decays of differently flavored charmed mesons $D^0(c\bar{u})$, $D^+(c\bar{d})$ and $D^+_s(c\bar{s})$ can yield new information on their possible quark composition. Another benefit of studying charm decays is that, in some cases, partial wave analyses are able to isolate the scalar contribution almost background free. A dedicated description of the Dalitz analysis methods can be found in Chapter 13. Table 19.1.5 gives a list of various three-body charm Dalitz analyses performed by the $B$ Factories together with references and corresponding sections with a detailed description.

#### 19.1.4.2 Dalitz Plot Analysis of $D^0 \to \overline{K}^0 K^+K^-$

The paper from BABAR (Aubert, 2005f) focuses on the study of the three-body $D^0$ meson decay

$$D^0 \to \overline{K}^0 K^+K^-,$$

where the $\overline{K}^0$ is detected via the decay $K^0_S \to \pi^+\pi^-$. The $D^0$ is tagged with a $D^*$. In a first analysis BABAR made use of 91.5 $\text{fb}^{-1}$ collecting N=13536 $\pm$ 116 events with a 97.3% purity. An additional analysis, with increased statistics ($\approx$ 69000 candidates) has been performed in with the aim of measuring $\phi_3$. We first describe the lower statistics analysis and in particular the partial wave analysis of the $K^+K^-$ threshold region.

**Table 19.1.5.** Dalitz analyses of three-body charm decays performed by the $B$ Factories.

| Decay | Reference | Section with description |
|---|---|---|
| $D^0 \to \overline{K}^0 K^+ K^-$ | (Aubert, 2005f, 2008l) | 19.1.4.2 |
| $D^0 \to K^0_S \pi^+ \pi^-$ | (Aubert, 2008l; Zhang, 2006) | 19.1.4.4 |
| $D^0 \to K^+ \pi^- \pi^0$ | (Aubert, 2009u) | 19.1.4.5 |
| $D^0 \to \pi^- \pi^- \pi^0$ | (Aubert, 2007w) | 19.1.4.7 |
| $D^+_s \to \pi^- \pi^+ \pi^+$ | (Aubert, 2009i) | 19.1.4.8 |
| $D^+_s \to K^- K^+ \pi^+$ | (del Amo Sanchez, 2011b) | 19.1.4.9 |

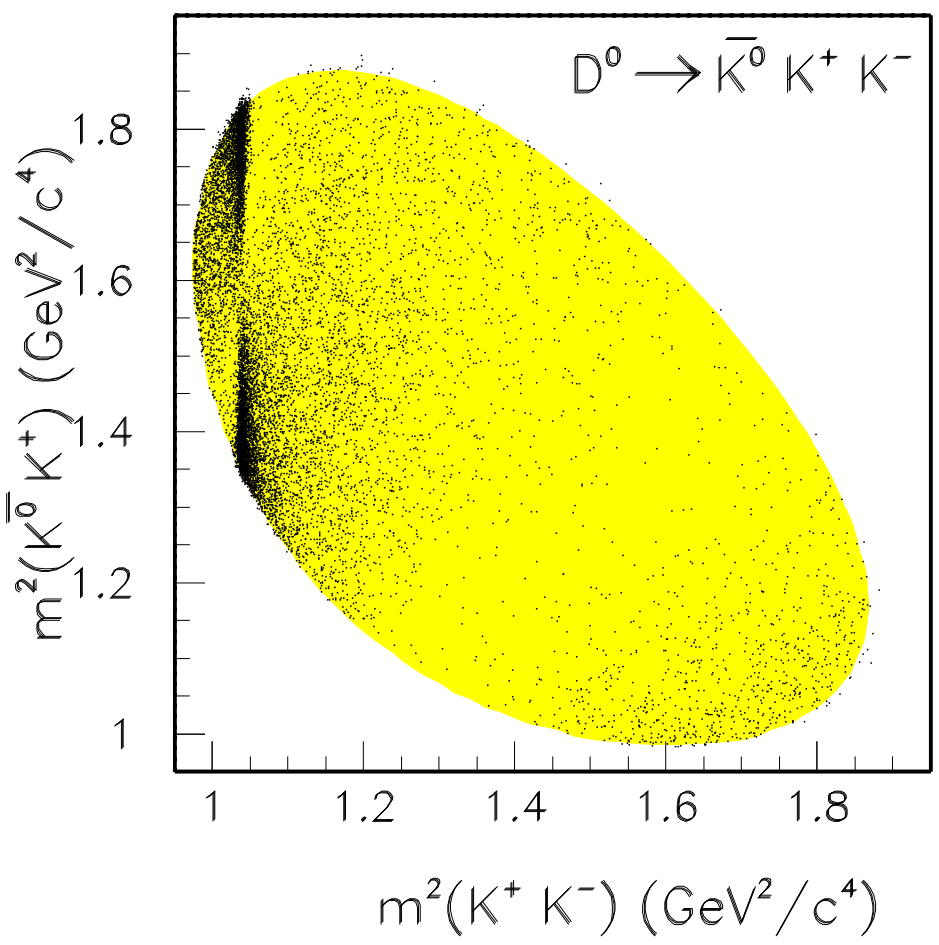

**Figure 19.1.16.** From (Aubert, 2005f). Dalitz plot of $D^0 \to \overline{K}^0 K^+ K^-$.

The Dalitz plot for these $D^0 \to \overline{K}^0 K^+ K^-$ candidates is shown in Fig. 19.1.16.

In the $K^+K^-$ threshold region, a strong $\phi(1020)$ signal is observed, together with a rather broad structure. A large asymmetry with respect to the $\overline{K}^0 K^+$ axis can also be seen in the vicinity of the $\phi(1020)$ signal, which is the result of interference between $S$- and $P$-wave amplitude contributions to the $K^+K^-$ system. The $f_0(980)$ and $a_0(980)$ $S$-wave resonances are, in fact, just below the $K^+K^-$ threshold, and might be expected to contribute in the vicinity of $\phi(1020)$. An accumulation of events due to a charged $a_0(980)^+$ can be observed on the lower right edge of the Dalitz plot. This contribution, however, does not overlap with the $\phi(1020)$ region and this allows the $K^+K^-$ scalar and vector components to be separated using a partial wave analysis in the low mass $K^+K^-$ region.

### 19.1.4.3 Partial Wave Analysis of $D^0 \to \overline{K}^0 K^+ K^-$.

It is assumed that near threshold the production of the $K^+K^-$ system can be described in terms of the diagram shown in Fig. 19.1.17. The helicity angle, $\theta_K$, is then defined as the angle between the $K^+$ for $D^0$ (or $K^-$ for $\overline{D}^0$) in the $K^+K^-$ rest frame and the $K^+K^-$ direction in the $D^0$ (or $\overline{K}^0$) rest frame. The $K^+K^-$ mass distribution has been modified by weighting each $D^0$ candidate by the spherical harmonic $Y^0_L(\cos\theta_K)$, $L = 0-4$, divided by its (Dalitz-plot-dependent) fitted efficiency. The resulting distributions $\langle Y^0_L \rangle$ are shown in Fig. 19.1.18 and are proportional to the $K^+K^-$ mass-dependent harmonic moments. It is found that all the $\langle Y^0_L \rangle$ moments are small or consistent with zero, except for $\langle Y^0_0 \rangle$, $\langle Y^0_1 \rangle$ and $\langle Y^0_2 \rangle$.

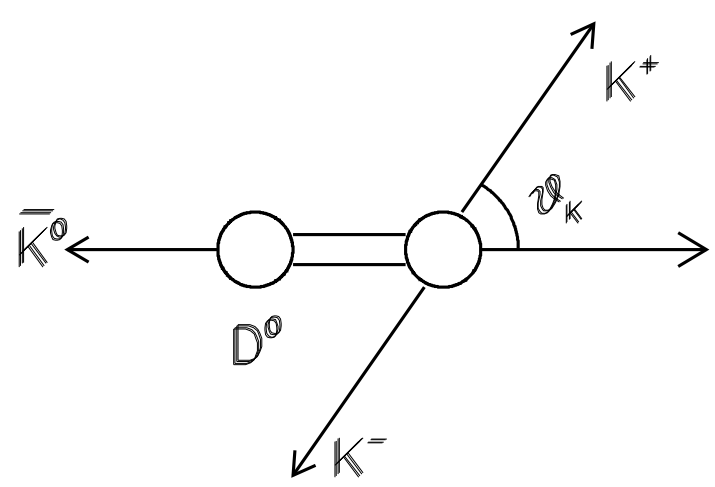

**Figure 19.1.17.** The kinematics describing the production of the $K^+K^-$ system in the threshold region of the decay $D^0 \to \overline{K}^0 K^+ K^-$.

In order to interpret these distributions a simple partial wave analysis has been performed, involving only $S$- and $P$-wave amplitudes. This results in the following set of equations (Chung, 1997):

$$\sqrt{4\pi}\,\langle Y^0_0 \rangle = S^2 + P^2$$
$$\sqrt{4\pi}\,\langle Y^0_1 \rangle = 2\mid S \mid\mid P \mid \cos\phi_{SP}$$
$$\sqrt{4\pi}\,\langle Y^0_2 \rangle = \frac{2}{\sqrt{5}} P^2, \qquad (19.1.36)$$

where $S$ and $P$ are proportional to the size of the $S$- and $P$-wave contributions and $\phi_{SP}$ is their relative phase. Under these assumptions, the $\langle Y^0_2 \rangle$ moment is proportional to $P^2$ so that it is natural that the $\phi(1020)$ appears free of background, as is observed.

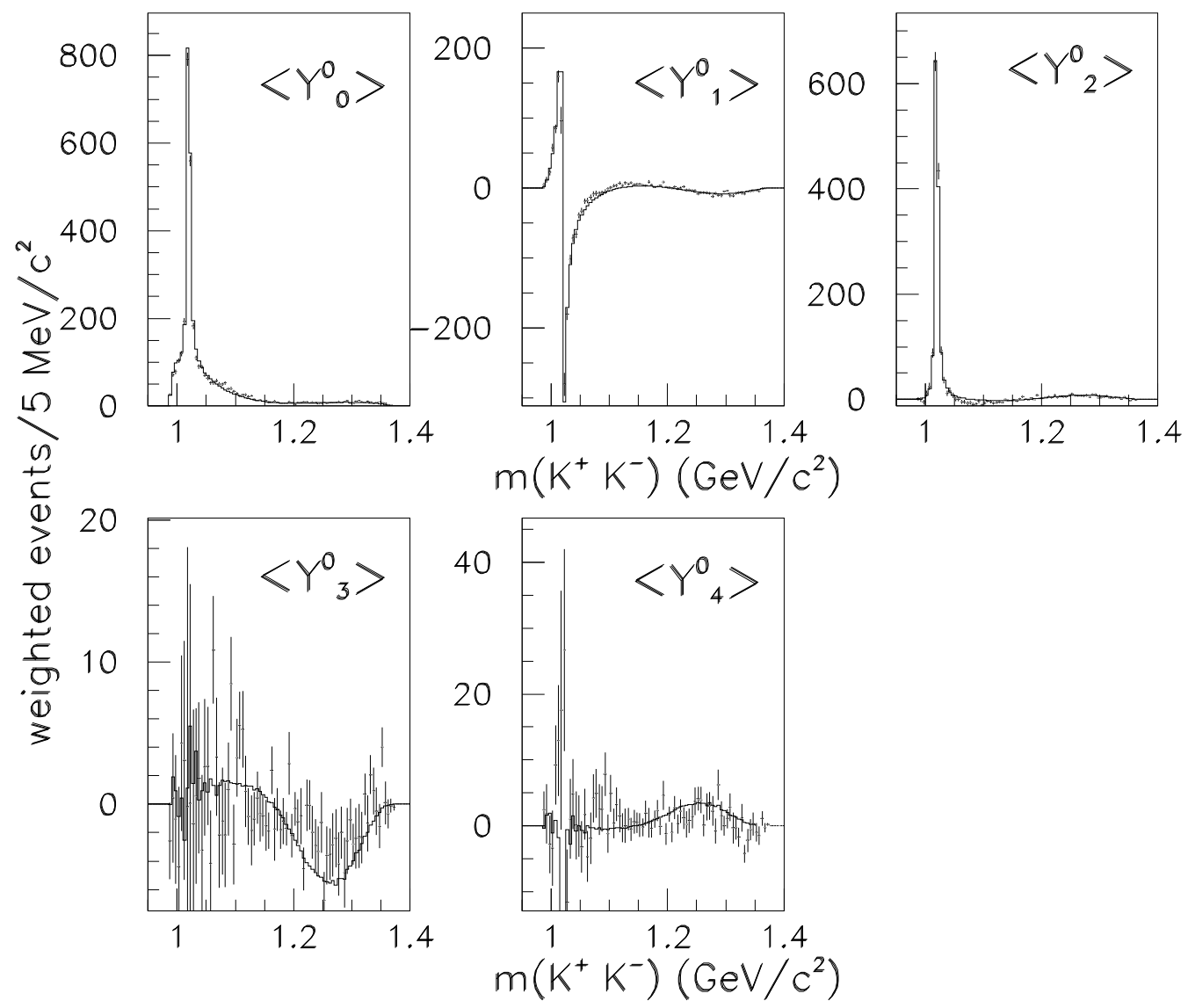

**Figure 19.1.18.** From (Aubert, 2005f). The unnormalized spherical harmonic moments $\langle Y_L^0 \rangle$ as functions of $K^+K^-$ invariant mass. The histograms represent the result of the full Dalitz plot analysis.

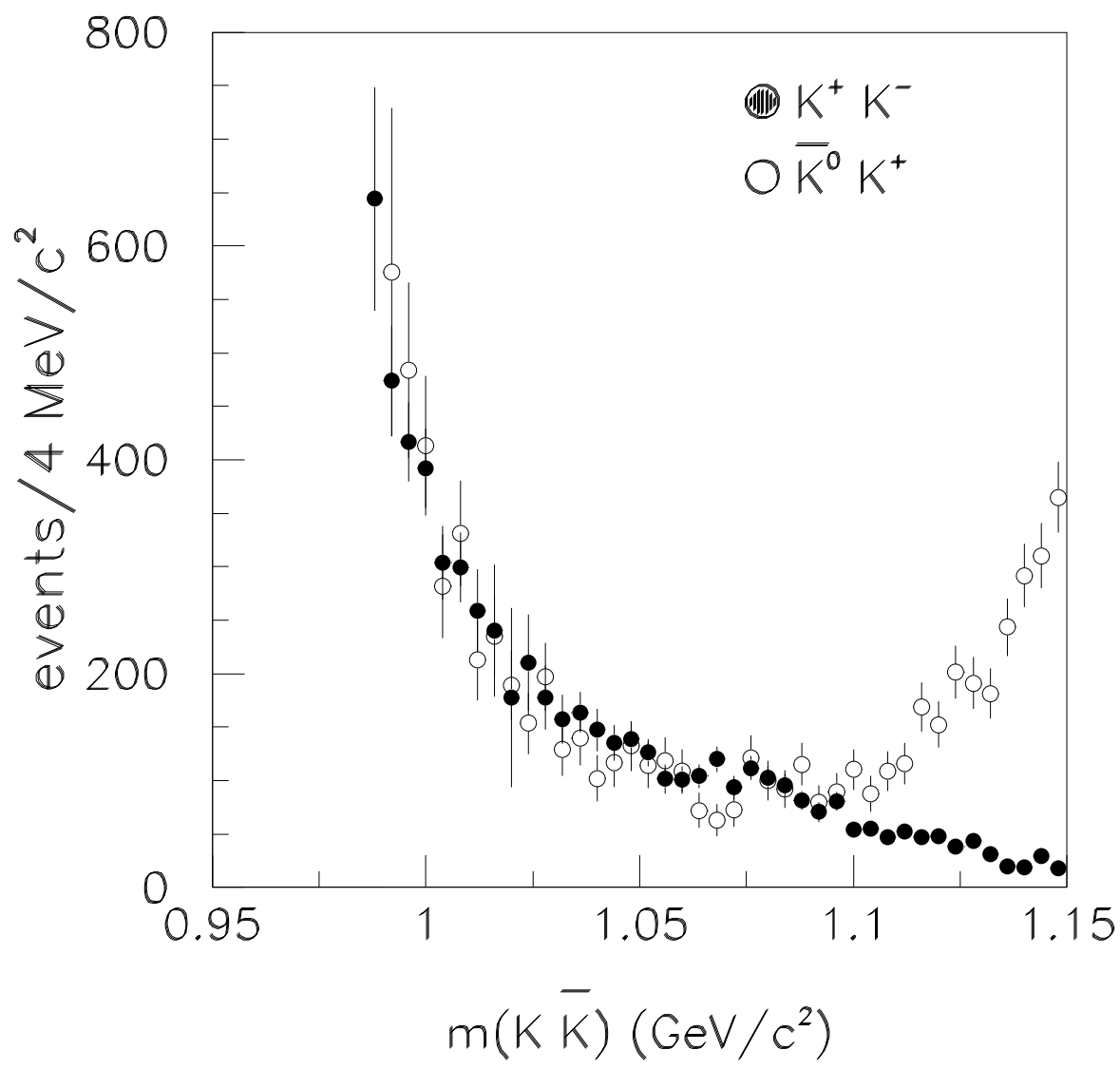

**Figure 19.1.20.** (Aubert, 2005f). Comparison between the phase-space-corrected $K^+K^-$ and $\overline{K}^0K^+$ normalised to the same area in the mass region between 0.992 and 1.05 GeV/$c^2$.

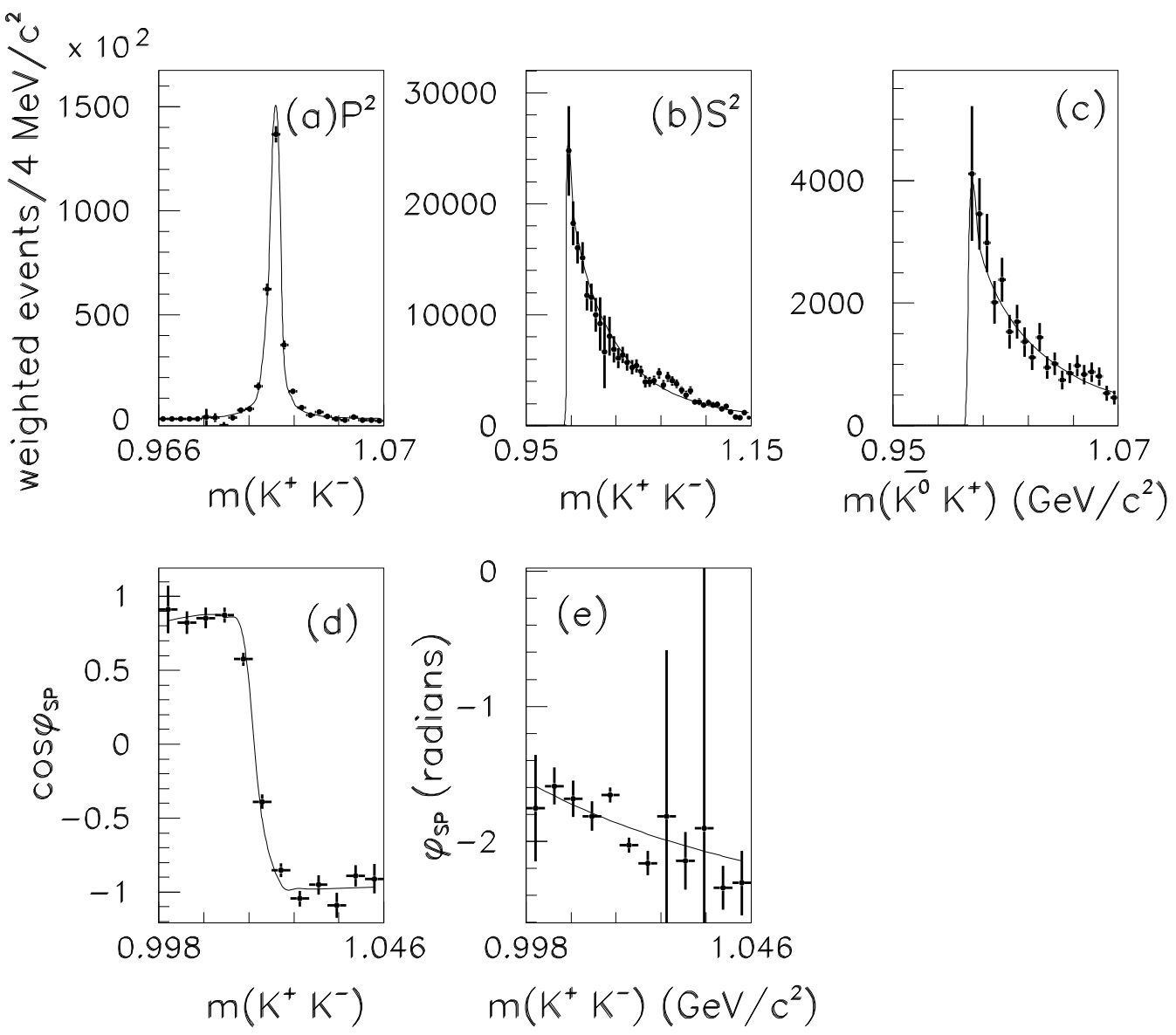

**Figure 19.1.19.** From (Aubert, 2005f). Results from the $K^+K^-$ Partial Wave Analysis corrected for phase space. (a) $P$-wave strength, (b) $S$-wave strength. (c) $m(\overline{K}^0K^+)$ distribution, (d) $\cos\phi_{SP}$ in the $\phi(1020)$ region. (e) $\phi_{SP}$ in the threshold region after having subtracted the fitted $\phi(1020)$ phase motion shown in (d).

A strong $S-P$ interference is evidenced by the rapid motion of the $\langle Y_1^0 \rangle$ moment in Fig. 19.1.18 in the $\phi(1020)$ mass region.

The above system of equations 19.1.36 can be solved directly for $S^2$, $P^2$ and $\cos\phi_{SP}$. However, since these amplitudes are defined in a $D^0$ decay, it is necessary to correct for phase space. The corrected spectra are shown in Fig. 19.1.19.

The distributions have been fitted using the following model:

- The $P$-wave is entirely due to the $\phi(1020)$ meson (Fig.19.1.19(a)).
- The scalar contribution in the $K^+K^-$ mass projection is entirely due to the $a_0(980)^0$ (Fig. 19.1.19(b)).
- The $\overline{K}^0K^+$ mass distribution is entirely due to $a_0(980)^+$ (Fig. 19.1.19(c)).
- The angle $\phi_{SP}$ (Fig. 19.1.19(d)) is obtained fitting the $S$, $P$ waves and $\cos\phi_{SP}$ with $c_{a_0}BW_{a_0} + c_\phi BW_\phi e^{i\alpha}$. Here $BW_{a_0}$ and $BW_\phi$ are the Breit-Wigner functions (BW) describing the $a_0(980)$ and $\phi(1020)$ resonances.

The $a_0(980)$ scalar resonance has a mass very close to the $K\overline{K}$ threshold and decays mostly to $\eta\pi$. It has been described by a coupled channel Breit-Wigner shape of the form:

$$BW_{ch}(a_0)(m) = \frac{g_{K\overline{K}}}{m_0^2 - m^2 - i(\rho_{\eta\pi}g_{\eta\pi}^2 + \rho_{K\overline{K}}g_{K\overline{K}}^2)} \tag{19.1.37}$$

where $\rho(m) = 2q/m$ while $g_{\eta\pi}$ and $g_{K\overline{K}}$ describe the $a_0(980)$ couplings to the $\eta\pi$ and $K\overline{K}$ systems respectively. Fixing $m_0$ and $g_{\eta\pi}$ to the Crystal Barrel measurements (Abele et al., 1998) it is possible to measure:

$$g_{K\overline{K}} = (464 \pm 29)\,\mathrm{MeV})^{1/2}. \tag{19.1.38}$$

Figure 19.1.19(e) shows the residual $a_0(980)$ phase, obtained by first computing $\phi_{SP}$ in the range $(0,\pi)$ and then

**Table 19.1.6.** From (Aubert, 2008l). Complex amplitudes $a_r e^{i\phi_r}$ and fit fractions, obtained from the fit of the $D^0 \to K^0_S K^+ K^-$ Dalitz plot distribution. The mass and width of the $\phi(1020)$, and the $g_{K\overline{K}}$ coupling constant are simultaneously determined in the fit, yielding $M_{\phi(1020)} = 1.01943 \pm 0.00002$ GeV/$c^2$, $\Gamma_{\phi(1020)} = 4.59319 \pm 0.00004$ MeV/$c^2$, and $g_{K\overline{K}} = 0.550 \pm 0.010$ GeV/$c^2$. Errors for amplitudes are statistical only. Uncertainties (largely dominated by systematic contributions) are not estimated for the fit fractions.

| Component | $a_r$ | $\phi_r$ (deg) | Fraction (%) |
|---|---|---|---|
| $K^0_S a_0(980)^0$ | 1 | 0 | 55.8 |
| $K^0_S \phi(1020)$ | $0.227 \pm 0.005$ | $-56.2 \pm 1.0$ | 44.9 |
| $K^0_S f_0(1370)$ | $0.04 \pm 0.06$ | $-2 \pm 80$ | 0.1 |
| $K^0_S f_2(1270)$ | $0.261 \pm 0.020$ | $-9 \pm 6$ | 0.3 |
| $K^0_S a_0(1450)^0$ | $0.65 \pm 0.09$ | $-95 \pm 10$ | 12.6 |
| $K^- a_0(980)^+$ | $0.562 \pm 0.015$ | $179 \pm 3$ | 16.0 |
| $K^- a_0(1450)^+$ | $0.84 \pm 0.04$ | $97 \pm 4$ | 21.8 |
| $K^+ a_0(980)^-$ | $0.118 \pm 0.015$ | $138 \pm 7$ | 0.7 |

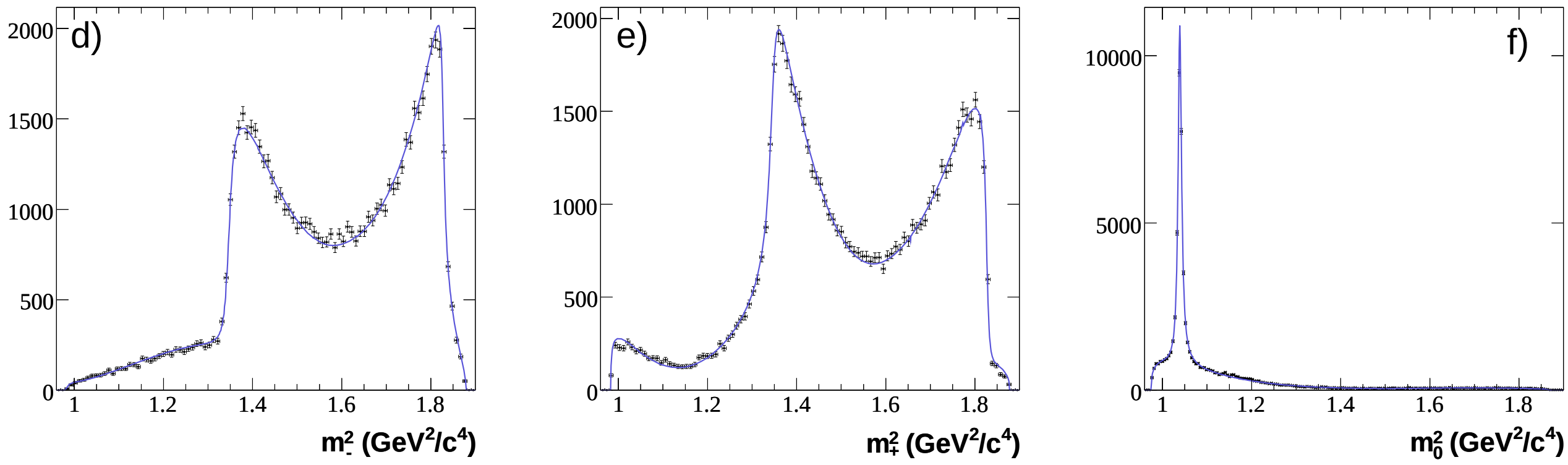

**Figure 19.1.21.** From (Aubert, 2008l). $D^0 \to K^0_S K^+ K^-$ Dalitz plot projections from $D^{*+} \to D^0\pi^+$ events on (d) $m^2_-$, (e) $m^2_+$, and (f) $m^2_0$. The curves are the reference model fit projections.

subtracting the known phase motion due to the $\phi(1020)$ resonance.

In this fit the possible presence of an $f_0(980)$ contribution has not been considered. This assumption can be tested by comparing the $K^+K^-$ and $\overline{K}^0K^+$ phase space corrected mass distributions. Since the $f_0(980)$ has isospin 0, it cannot decay to $\overline{K}^0K^+$. Therefore an excess in the $K^+K^-$ mass spectrum with respect to $\overline{K}^0K^+$ would indicate the presence of an $f_0(980)$ contribution.

Figure 19.1.20 compares the $K^+K^-$ and $\overline{K}^0K^+$ mass distributions, normalised to the same area between 0.992 and 1.05 GeV/$c^2$ and corrected for phase space. It is possible to observe that the two distributions show a good agreement, supporting the argument that the $f_0(980)$ contribution is small.

We now refer to the higher statistics Dalitz plot analysis from *BABAR* (Aubert, 2008l). The description of the $D^0 \to K^0_S K^+ K^-$ decay amplitude consists of five distinct resonances leading to 8 two-body decays: $K^0_S a_0(980)^0$, $K^0_S \phi(1020)$, $K^- a_0(980)^+$, $K^0_S f_0(1370)$, $K^+ a_0(980)^-$, $K^0_S f_2(1270)^0$, $K^0_S a_0(1450)^0$, and $K^- a_0(1450)^+$. This isobar model is essentially identical to that used in the previous analysis, but for the addition of the $a_0(1450)$ scalar, whose contribution is strongly supported by the much larger data sample, as well as of a $D$-wave contribution parameterized with the $f_2(1270)$ tensor. Attempts to improve the model quality by adding other contributions (including the non-resonant term) did not give better results.

Table 19.1.6 summarizes the values obtained for all free parameters of the $D^0 \to K^0_S K^+ K^-$ Dalitz model, the complex amplitudes $a_r e^{i\phi_r}$, the mass and width of the $\phi(1020)$ and the coupling constant $g_{K\overline{K}}$, together with the fit fractions. The value of $g_{K\overline{K}}$ is consistent with the previous result, and differs significantly from the measurement reported in (Abele et al., 1998). All amplitudes are measured with respect to $D^0 \to K^0_S a_0(980)^0$, which gives the largest contribution. The sum of fit fractions is 152.3%, and the reduced $\chi^2$ is 1.09 (with statistical errors only) for 6856 degrees of freedom, estimated from a binning of the Dalitz plot into square regions of size 0.045 GeV$^2$/$c^4$. Figure 19.1.21(d,e,f) shows the fit projections overlaid with the data distributions. The Dalitz plot distributions are well reproduced, with some small discrepancies at the peaks of the $m^2_-$ and $m^2_+$ projections.

### 19.1.4.4 $D^0 \to K^0_S\pi^+\pi^-$ Dalitz model

This analysis from *BABAR* is related to the measurement of $\phi_3$ (Section 17.8) and makes use of 487 000 $D^0 \to K^0_S\pi^+\pi^-$ events tagged with a $D^{*+}$ with 97.7% purity (Aubert, 2008l). The $P$- and $D$-waves of the $D^0 \to K^0_S\pi^+\pi^-$ decay amplitude are described using a total of 6 resonances leading to 8 two-body decay amplitudes: the Cabibbo allowed (CA) $K^*(892)^-$, $K^*(1680)^-$, $K_2^*(1430)^-$, the doubly-Cabibbo suppressed (DCS) $K^*(892)^+$, $K_2^*(1430)^+$, and the *CP* eigenstates $\rho(770)^0$, $\omega(782)$, and $f_2(1270)$. Since the $K\pi$ $P$-wave is largely dominated by the $K^*(892)^\mp$, the mass and width of this resonance are simultaneously determined from the fit to the tagged $D^0$ sample, $M_{K^*(892)^\mp} = 893.61\pm0.08$ MeV/$c^2$ and $\Gamma_{K^*(892)^\mp} = 46.34\pm0.16$ MeV/$c^2$ (errors are statistical only). The mass and width values of the $K^*(1680)^-$ are taken from (Aston et al., 1988), where the interference between the $K\pi$ $S$- and $P$-waves is properly accounted for.

They adopt the same parameterizations for $K$, $\rho$, and $P$ as in Yao et al. (2006), Anisovich and Sarantsev (2003), and Link et al. (2004a). The K matrix is written as

$$K_{uv}(s) = \left( \sum_\alpha \frac{g_u^\alpha g_v^\alpha}{m_\alpha^2 - s} + f_{uv}^{\rm scatt} \frac{1 - s_0^{\rm scatt}}{s - s_0^{\rm scatt}} \right) f_{A0}(s), \tag{19.1.39}$$

where $g_u^\alpha$ is the coupling constant of the K-matrix pole $m_\alpha$ to the $u^{\rm th}$ channel. The parameters $f_{uv}^{\rm scatt}$ and $s_0^{\rm scatt}$ describe the slowly-varying part of the K-matrix. The factor

$$f_{A0}(s) = \frac{1 - s_{A0}}{s - s_{A0}} \left( s - s_A \frac{m_\pi^2}{2} \right), \tag{19.1.40}$$

suppresses the false kinematical singularity at $s = 0$ in the physical region near the $\pi\pi$ threshold (the Adler zero (Adler, 1965)). The parameter values used in this analysis are listed in Section 13, and are obtained from a global analysis of the available $\pi\pi$ scattering data from threshold up to 1900 MeV/$c^2$ (Anisovich and Sarantsev, 2003). The parameters $f_{uv}^{\rm scatt}$, for $u \neq 1$, are all set to zero since they are not related to the $\pi\pi$ scattering process. Similarly, for the P vector we have

$$P_v(s) = \sum_\alpha \frac{\beta_\alpha g_v^\alpha}{m_\alpha^2 - s} + f_{1v}^{\rm prod} \frac{1 - s_0^{\rm prod}}{s - s_0^{\rm prod}}. \tag{19.1.41}$$

Note that the P-vector has the same poles as the K-matrix, otherwise the $F_1$ vector would vanish (diverge) at the K-matrix (P-vector) poles. The parameters $\beta_\alpha$, $f_{1v}^{\rm prod}$ and $s_0^{\rm prod}$ of the initial P-vector are obtained from the fit to the tagged $D^0 \to K^0_S\pi^+\pi^-$ data sample.

For the $K\pi$ $S$-wave contribution to Eq. (19.1.39) they use a parameterization extracted from scattering data (Aston et al., 1988) which consists of a $K_0^*(1430)^-$ or $K_0^*(1430)^+$ BW (for CA or DCS contribution, respectively) together with an effective range non-resonant component with a phase shift,

$$\mathcal{A}_{K\pi\ L=0}(\mathbf{m}) = F \sin\delta_F e^{i\delta_F} + R \sin\delta_R e^{i\delta_R} e^{i2\delta_F}\ , \tag{19.1.42}$$

with

$$\delta_R = \phi_R + \tan^{-1}\left[\frac{M\Gamma(m_{K\pi}^2)}{M^2 - m_{K\pi}^2}\right]\ ,$$
$$\delta_F = \phi_F + \cot^{-1}\left[\frac{1}{aq} + \frac{rq}{2}\right]. \tag{19.1.43}$$

The parameters $a$ and $r$ play the role of a scattering length and effective interaction length, respectively, $F$ ($\phi_F$) and $R$ ($\phi_R$) are the amplitudes (phases) for the non-resonant and resonant terms, and $q$ is the momentum of the spectator particle in the $K\pi$ system rest frame. Note that the phases $\delta_F$ and $\delta_R$ depend on $m_{K\pi}^2$. $M$ and $\Gamma(m_{K\pi}^2)$ are the mass and running width of the resonant term. This parameterization corresponds to a K-matrix approach describing a rapid phase shift coming from the resonant term and a slow rising phase shift governed by the non-resonant term, with relative strengths $R$ and $F$. The parameters $M$, $\Gamma$, $F$, $\phi_F$, $R$, $\phi_R$, $a$ and $r$ are determined from the fit to the tagged $D^0$ sample, along with the other parameters of the model. Other recent experimental efforts to improve the description of the $K\pi$ $S$-wave using K-matrix and model independent parameterizations from high-statistics samples of $D^+ \to K^-\pi^+\pi^+$ decays are described in Aitala et al. (2006), Link et al. (2007), and Bonvicini et al. (2008).

Table 19.1.7 summarizes the values obtained for all free parameters of the $D^0 \to K^0_S\pi^+\pi^-$ Dalitz model: CA, DCS, and *CP* eigenstates complex amplitudes $a_r e^{i\phi_r}$, $\pi^+\pi^-$ $S$-wave P-vector parameters, and $K\pi$ $S$-wave parameters, along with the fit fractions. The non-resonant term of Eq. (19.1.42) has not been included since the $\pi\pi$ and $K\pi$ $S$-wave parameterizations naturally account for their respective non-resonant contributions. The fifth P-vector channel and pole have also been excluded since the $\eta\eta'$ threshold and the pole mass $m_5$ are both far beyond our $\pi\pi$ kinematic range, and thus there is little sensitivity to the associated parameters, $f_{15}^{\rm prod}$ and $\beta_5$, respectively. The amplitudes are measured with respect to $D^0 \to K^0_S\rho(770)^0$ which gives the second largest contribution.

The $K\pi$ and $\pi\pi$ $P$-waves dominate the decay, but significant contributions from the corresponding $S$-waves are also observed (above 6 and 4 standard deviations, respectively). They obtain a sum of fit fractions of $(103.6 \pm 5.2)$%, and the goodness-of-fit is estimated through a two-dimensional $\chi^2$ test performed binning the Dalitz plot into square regions of size 0.015 GeV$^2/c^4$, yielding a reduced $\chi^2$ of 1.11 (including statistical errors only) for 19274 degrees of freedom. The variation of the contribution to the $\chi^2$ as a function of the Dalitz plot position is approximately uniform. Figure 19.1.22(a,b,c) shows the Dalitz fit projections overlaid with the data distributions. The Dalitz plot distributions are well reproduced, with some small discrepancies in low and high mass regions of the $m_0^2$ projection, and in the $\rho(770)^0 - \omega(782)$ interference region.

**Table 19.1.7.** From (Aubert, 2008l). CA, DCS, and $CP$ eigenstates complex amplitudes $a_r e^{i\phi_r}$, $\pi\pi$ $S$-wave $P$-vector parameters, $K\pi$ $S$-wave parameters, and fit fractions, as obtained from the fit of the $D^0 \to K^0_S\pi^+\pi^-$ Dalitz plot distribution from $D^{*+} \to D^0\pi^+$. P-vector parameters $f^{'\rm prod}_{1v}$, for $v \neq 1$, are defined as $f^{\rm prod}_{1v}/f^{\rm prod}_{11}$. Errors for amplitudes are statistical only, while for fit fractions include statistical and systematic uncertainties, largely dominated by the latter. Upper limits on fit fractions are quoted at 95% confidence level.

| Component | $a_r$ | $\phi_r$ (deg) | Fraction (%) |
|---|---|---|---|
| $K^*(892)^-$ | $1.740 \pm 0.010$ | $139.0 \pm 0.3$ | $55.7 \pm 2.8$ |
| $K^*_0(1430)^-$ | $8.2 \pm 0.7$ | $153 \pm 8$ | $10.2 \pm 1.5$ |
| $K^*_2(1430)^-$ | $1.410 \pm 0.022$ | $138.4 \pm 1.0$ | $2.2 \pm 1.6$ |
| $K^*(1680)^-$ | $1.46 \pm 0.10$ | $-174 \pm 4$ | $0.7 \pm 1.9$ |
| $K^*(892)^+$ | $0.158 \pm 0.003$ | $-42.7 \pm 1.2$ | $0.46 \pm 0.23$ |
| $K^*_0(1430)^+$ | $0.32 \pm 0.06$ | $143 \pm 11$ | $< 0.05$ |
| $K^*_2(1430)^+$ | $0.091 \pm 0.016$ | $85 \pm 11$ | $< 0.12$ |
| $\rho(770)^0$ | 1 | 0 | $21.0 \pm 1.6$ |
| $\omega(782)$ | $0.0527 \pm 0.0007$ | $126.5 \pm 0.9$ | $0.9 \pm 1.0$ |
| $f_2(1270)$ | $0.606 \pm 0.026$ | $157.4 \pm 2.2$ | $0.6 \pm 0.7$ |
| $\beta_1$ | $9.3 \pm 0.4$ | $-78.7 \pm 1.6$ | |
| $\beta_2$ | $10.89 \pm 0.26$ | $-159.1 \pm 2.6$ | |
| $\beta_3$ | $24.2 \pm 2.0$ | $168 \pm 4$ | |
| $\beta_4$ | $9.16 \pm 0.24$ | $90.5 \pm 2.6$ | |
| $f^{\rm prod}_{11}$ | $7.94 \pm 0.26$ | $73.9 \pm 1.1$ | |
| $f^{'\rm prod}_{12}$ | $2.0 \pm 0.3$ | $-18 \pm 9$ | |
| $f^{'\rm prod}_{13}$ | $5.1 \pm 0.3$ | $33 \pm 3$ | |
| $f^{'\rm prod}_{14}$ | $3.23 \pm 0.18$ | $4.8 \pm 2.5$ | |
| $s^{\rm prod}_0$ | $-0.07 \pm 0.03$ | | |
| $\pi\pi$ $S$-wave | | | $11.9 \pm 2.6$ |
| $M$ (GeV/$c^2$) | $1.463 \pm 0.002$ | | |
| $\Gamma$ (GeV/$c^2$) | $0.233 \pm 0.005$ | | |
| $F$ | $0.80 \pm 0.09$ | | |
| $\phi_F$ | $2.33 \pm 0.13$ | | |
| $R$ | 1 | | |
| $\phi_R$ | $-5.31 \pm 0.04$ | | |
| $a$ | $1.07 \pm 0.11$ | | |
| $r$ | $-1.8 \pm 0.3$ | | |

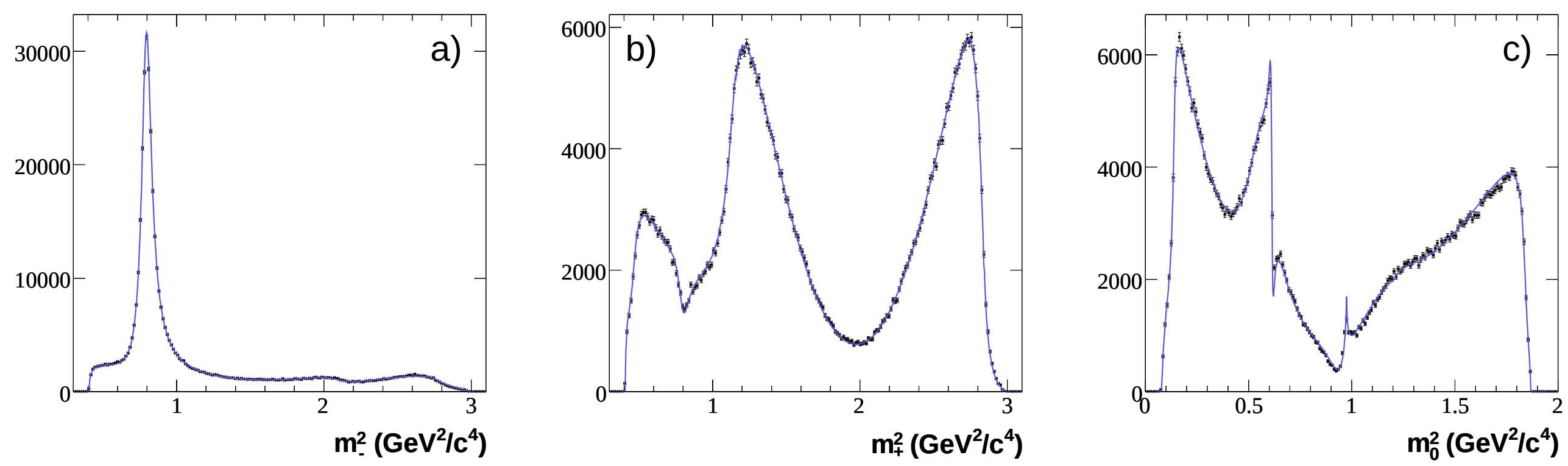

**Figure 19.1.22.** From (Aubert, 2008l). $D^0 \to K^0_S\pi^+\pi^-$ Dalitz plot projections from $D^{*+} \to D^0\pi^+$ events on (a) $m^2_-$, (b) $m^2_+$, and (c) $m^2_0$. The curves are the reference model fit projections.

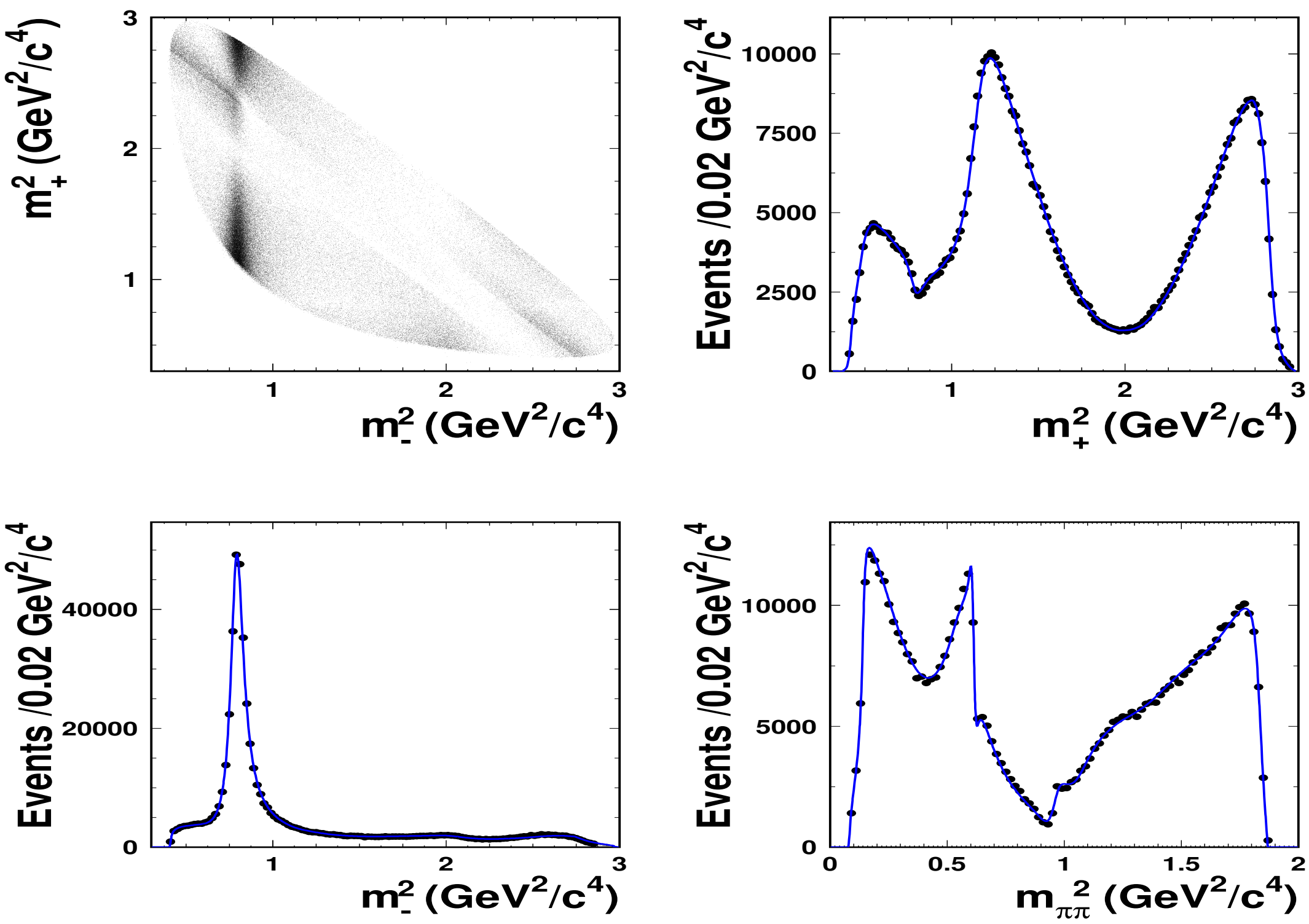

**Figure 19.1.23.** From (Zhang, 2006). Dalitz plot distribution of $D^0 \to K_S^0\pi^+\pi^-$) and the projections for data (points with error bars) and the fit result (curve). Here, $m_\pm^2$ corresponds to $m^2(K_S^0\pi^\pm)$ for $D^0$ decays and to $m^2(K_S^0\pi^\mp)$ for $\bar{D}^0$ decays.

As a cross-check, they alternatively parameterize the $\pi\pi$ and $K\pi$ $S$-waves using the isobar approximation with the following BW amplitudes (plus the non-resonant contribution): the CA $K_0^*(1430)^-$, the DCS $K_0^*(1430)^+$, and the *CP* eigenstates $f_0(980)$, $f_0(1370)$, $\sigma$ and an ad hoc $\sigma'$. Masses and widths of the $\sigma$ and $\sigma'$ scalars are obtained from the fit, $M_\sigma = 528 \pm 5$, $\Gamma_\sigma = 512 \pm 9$, $M_{\sigma'} = 1033 \pm 4$, and $\Gamma_{\sigma'} = 99 \pm 6$, given in MeV/$c^2$. Mass and width values for the $K_0^*(1430)^\mp$, $f_0(980)$, and $f_0(1370)$ are taken from (Aitala et al., 2001b, 2002). They obtain a sum of fit fractions of 122.5%, and a reduced $\chi^2$ of 1.20 (with statistical errors only) for 19274 degrees of freedom, which strongly disfavors the isobar approach in comparison to the K-matrix formalism.

Belle study of the $D^0 \to K_S^0\pi^+\pi^-$ decays, based on 540 fb$^{-1}$ of the data, has been performed for the $D^0-\overline{D}^0$ mixing measurement (Zhang, 2006). The reconstructed signal yield of $D^0$ mesons, tagged with $D^{*+} \to D^0\pi^+$ decays, is of $(534.4 \pm 0.8) \times 10^3$ events and the signal purity amounts to about 95%. The Dalitz distribution for the $D^0 \to K_S^0\pi^+\pi^-$ candidiates is modeled assuming an isobar model in which the total $D^0 \to K_S^0\pi^+\pi^-$ amplitude is a sum of 18 quasi-two-body amplitudes, described with relativistic Breit-Wigner functions, and a constant non-resonant term. The amplitudes and their relative phases, obtained from an unbinned maximum likelihood fit performed to the Dalitz distribution, are summarized in Table 19.1.8. The main features of the Dalitz plot are well reproduced as can be seen from the Dalitz plot and its projections shown, along with projections of the fit result, in Fig. 19.1.23. The goodness-of-fit of the Dalitz plot is estimated to be $\chi^2/ndf = 2.1$ for $3653 - 40$ degrees of freedom ($ndf$). The K-matrix formalism was used in addition to the above mentioned isobar model in estimation of systematic uncertainties for the $D^0 - \overline{D}^0$ mixing measurement.

**Table 19.1.8.** From (Zhang, 2006). Fit results for Dalitz plot parameters for $D^0 \to K_S^0\pi^+\pi^-$. The errors are statistical only.

| Resonance | Amplitude | Phase (°) | Fraction |
|---|---|---|---|
| $K^*(892)^-$ | $1.629 \pm 0.006$ | $134.3 \pm 0.3$ | 0.6227 |
| $K_0^*(1430)^-$ | $2.12 \pm 0.02$ | $-0.9 \pm 0.8$ | 0.0724 |
| $K_2^*(1430)^-$ | $0.87 \pm 0.02$ | $-47.3 \pm 1.2$ | 0.0133 |
| $K^*(1410)^-$ | $0.65 \pm 0.03$ | $111 \pm 4$ | 0.0048 |
| $K^*(1680)^-$ | $0.60 \pm 0.25$ | $147 \pm 29$ | 0.0002 |
| $K^*(892)^+$ | $0.152 \pm 0.003$ | $-37.5 \pm 1.3$ | 0.0054 |
| $K_0^*(1430)^+$ | $0.541 \pm 0.019$ | $91.8 \pm 2.1$ | 0.0047 |
| $K_2^*(1430)^+$ | $0.276 \pm 0.013$ | $-106 \pm 3$ | 0.0013 |
| $K^*(1410)^+$ | $0.33 \pm 0.02$ | $-102 \pm 4$ | 0.0013 |
| $K^*(1680)^+$ | $0.73 \pm 0.16$ | $103 \pm 11$ | 0.0004 |
| $\rho(770)$ | 1 (fixed) | 0 (fixed) | 0.2111 |
| $\omega(782)$ | $0.0380 \pm 0.0007$ | $115.1 \pm 1.1$ | 0.0063 |
| $f_0(980)$ | $0.380 \pm 0.004$ | $-147.1 \pm 1.1$ | 0.0452 |
| $f_0(1370)$ | $1.46 \pm 0.05$ | $98.6 \pm 1.8$ | 0.0162 |
| $f_2(1270)$ | $1.43 \pm 0.02$ | $-13.6 \pm 1.2$ | 0.0180 |
| $\rho(1450)$ | $0.72 \pm 0.04$ | $41 \pm 7$ | 0.0024 |
| $\sigma_1$ | $1.39 \pm 0.02$ | $-146.6 \pm 0.9$ | 0.0914 |
| $\sigma_2$ | $0.267 \pm 0.013$ | $-157 \pm 3$ | 0.0088 |
| NR | $2.36 \pm 0.07$ | $155 \pm 2$ | 0.0615 |

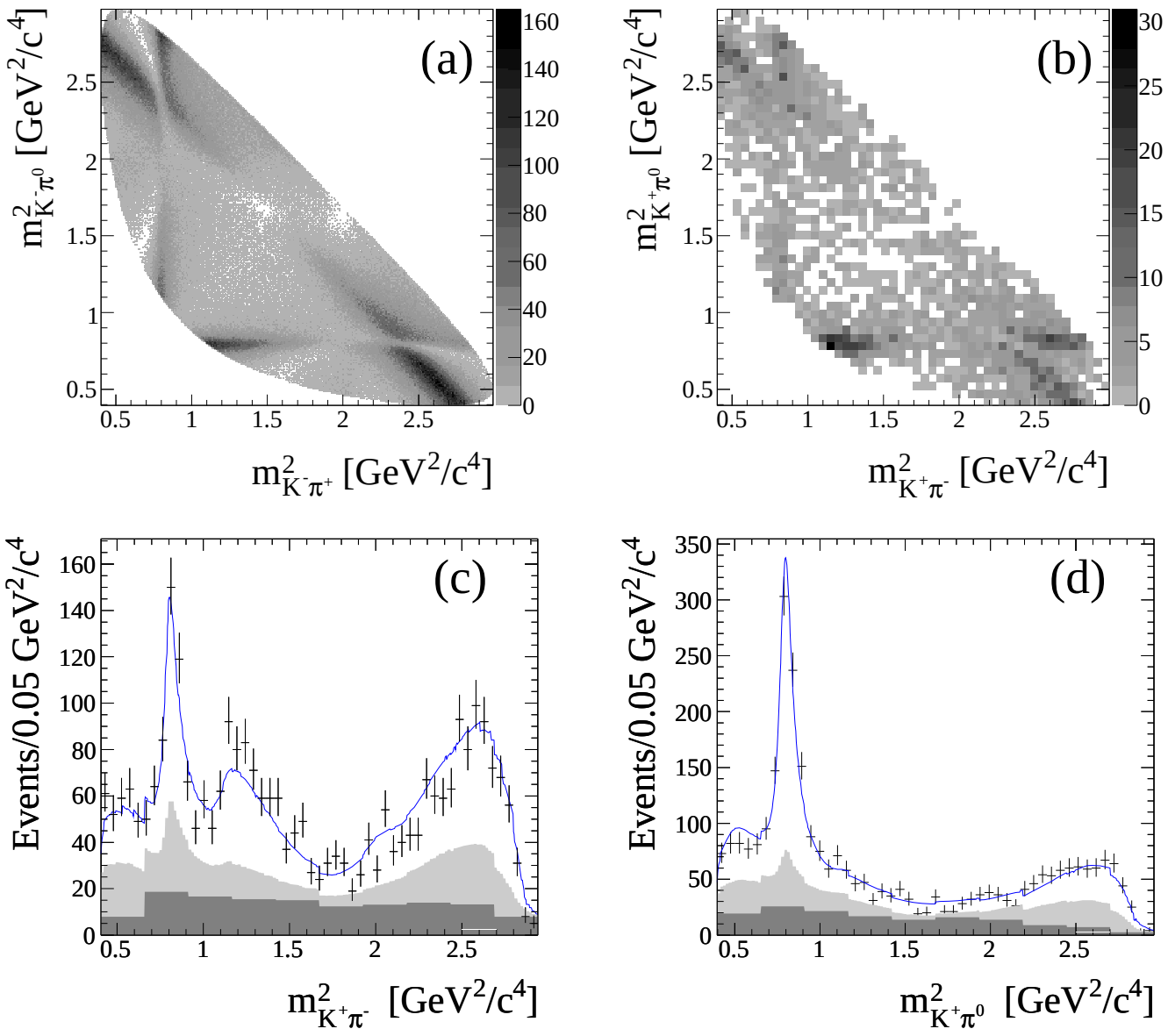

**Figure 19.1.24.** From (Aubert, 2009u). Dalitz plots for the (a) RS $D^0 \to K^-\pi^+\pi^0$ and (b) WS $D^0 \to K^+\pi^-\pi^0$ samples. (c, d) $m^2_{K^+\pi^-}$ and $m^2_{K^+\pi^0}$ projections with superimposed fit results (line). The light histogram represents the mistag background, while the dark histogram shows the combinatoric background;

### 19.1.4.5 Dalitz Plot Analysis of three-body (wrong-sign) $D^0 \to K^+\pi^-\pi^0$ decays.

This analysis from *BABAR* (Aubert, 2009u) has been performed on a data sample of 384 $\text{fb}^{-1}$. The Dalitz analysis is performed on the wrong-sign (WS) dataset only consisting of $D^{*+} \to D^0(\to K^+\pi^-\pi^0)\pi^+$ events. The right sign (RS) sample ($D^{*+} \to D^0(\to K^-\pi^+\pi^0)\pi^+$) is composed of 658,986 events with a purity of 99%, the WS by 3009 events with a purity of 50%. The efficiency of the signal region selection is 54.6%. In the Dalitz plot analysis, the fit fraction of the non-resonant contribution to the $K$-$\pi$ $S$-wave is absorbed into the $K_0^{*+}(1430)$ and $K_0^{*0}(1430)$ fit fractions. Projections of the fit results are shown in Fig. 19.1.24(b-d), amplitudes, phases, and fractions are given in Table 19.1.9.

**Table 19.1.9.** From (Aubert, 2009u). Fit results for the WS $D^0 \to K^+\pi^-\pi^0$ data sample. The total fit fraction is 102% and the $\chi^2/ndof$ is 188/215.

| Resonance | $a_j^{DCS}$ | $\delta_j^{DCS}$ (degrees) | $f_j$ (%) |
|---|---|---|---|
| $\rho(770)$ | 1 (fixed) | 0 (fixed) | $39.8 \pm 6.5$ |
| $K_2^{*0}(1430)$ | $0.088 \pm 0.017$ | $-17.2 \pm 12.9$ | $2.0 \pm 0.7$ |
| $K_0^{*+}(1430)$ | $6.78 \pm 1.00$ | $69.1 \pm 10.9$ | $13.1 \pm 3.3$ |
| $K^{*+}(892)$ | $0.899 \pm 0.005$ | $-171.0 \pm 5.9$ | $35.6 \pm 5.5$ |
| $K_0^{*0}(1430)$ | $1.65 \pm 0.59$ | $-44.4 \pm 18.5$ | $2.8 \pm 1.5$ |
| $K^{*0}(892)$ | $0.398 \pm 0.038$ | $24.1 \pm 9.8$ | $6.5 \pm 1.4$ |
| $\rho(1700)$ | $5.4 \pm 1.6$ | $157.4 \pm 20.3$ | $2.0 \pm 1.1$ |

### 19.1.4.6 Dalitz Plot Analysis of $D^0 \to K^-K^+\pi^0$ decay.

Using 385 $\text{fb}^{-1}$ of $e^+e^-$ collisions, *BABAR* has performed a Dalitz analysis of the singly Cabibbo-suppressed decay $D^0 \to K^-K^+\pi^0$ (Aubert, 2007d). Fig. 19.1.25 shows the $D^0 \to K^-K^+\pi^0$ Dalitz plot and mass projections, together with results from the Dalitz plot analysis. The LASS $K\pi$ $S$-wave amplitude gives the best agreement with data and they use it in the nominal fits. The $K\pi$ $S$-wave modeled by the combination of $\kappa(800)$ (with parameters taken from Aitala et al., 2002), a nonresonant term and $K_0^*(1430)$ has a smaller fit probability ($\chi^2$ probability $< 5\%$). The best fit with this model ($\chi^2$ probability 13%) yields a charged $\kappa$ of mass $(870 \pm 30)$ MeV/$c^2$, and width $(150 \pm 20)$ MeV/$c^2$, significantly different from those reported in Aitala et al. (2002) for the neutral state. This does not support the hypothesis that production of a charged, scalar $\kappa$ is being observed. The E-791 amplitude Aitala et al. (2006) describes the data well, except near threshold ($\chi^2$ probability 23%).

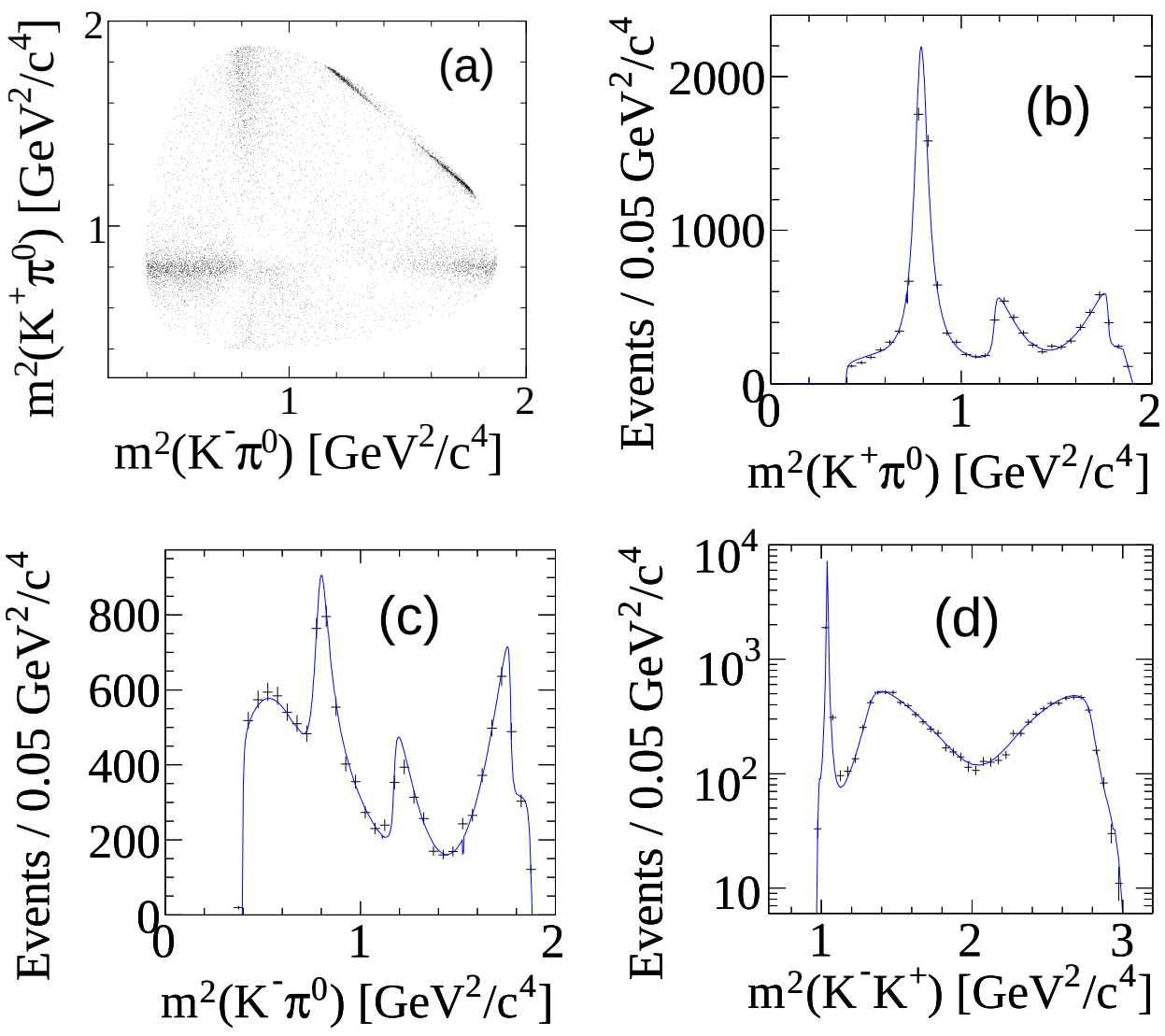

**Figure 19.1.25.** From (Aubert, 2007d). Dalitz plot for $D^0 \to K^-K^+\pi^0$ data (a), and the corresponding squared invariant mass projections (b–d). The three-body invariant mass of the $D^0$ candidate is constrained to the nominal value. In plots (b–d), the dots (with error bars, black) are data points and the solid lines (blue) correspond to the best isobar fit models.

**Table 19.1.10.** From (Aubert, 2007d). The results obtained from the $D^0 \to K^-K^+\pi^0$ Dalitz plot fit. The amplitude coefficients, $a_r$ and $\phi_r$, are computed relatively to those of the $K^*(892)^+$. The $a_0(980)$ contribution, when it is included in place of the $f_0(980)$, is given in square brackets. The $K\pi$ $S$-wave states are incicated as $K^\pm\pi^0(S)$. The LASS amplitude is used to describe the $K\pi$ $S$-wave states in both the isobar models (I and II).

| | Model I | | | Model II | | |
|---|---|---|---|---|---|---|
| State | Amplitude, $a_r$ | Phase, $\phi_r$ (°) | Fraction, $f_r$ (%) | Amplitude, $a_r$ | Phase, $\phi_r$ (°) | Fraction, $f_r$ (%) |
| $K^*(892)^+$ | 1.0 (fixed) | 0.0 (fixed) | 45.2±0.8±0.6 | 1.0 (fixed) | 0.0 (fixed) | 44.4±0.8±0.6 |
| $K^*(1410)^+$ | 2.29±0.37±0.20 | 86.7±12.0±9.6 | 3.7±1.1±1.1 | | | |
| $K^+\pi^0(S)$ | 1.76±0.36±0.18 | -179.8±21.3±12.3 | 16.3±3.4±2.1 | 3.66±0.11±0.09 | -148.0±2.0±2.8 | 71.1±3.7±1.9 |
| $\phi(1020)$ | 0.69±0.01±0.02 | -20.7±13.6±9.3 | 19.3±0.6±0.4 | 0.70±0.01±0.02 | 18.0±3.7±3.6 | 19.4±0.6±0.5 |
| $f_0(980)$ | 0.51±0.07±0.04 | -177.5±13.7±8.6 | 6.7±1.4±1.2 | 0.64±0.04±0.03 | -60.8±2.5±3.0 | 10.5±1.1±1.2 |
| $[a_0(980)^0]$ | [0.48±0.08±0.04] | [-154.0±14.1±8.6] | [6.0±1.8±1.2] | [0.68±0.06±0.03] | [-38.5±4.3±3.0] | [11.0±1.5±1.2] |
| $f_2'(1525)$ | 1.11±0.38±0.28 | -18.7±19.3±13.6 | 0.08±0.04±0.05 | | | |
| $K^*(892)^-$ | 0.601±0.011±0.011 | -37.0±1.9±2.2 | 16.0±0.8±0.6 | 0.597±0.013±0.009 | -34.1±1.9±2.2 | 15.9±0.7±0.6 |
| $K^*(1410)^-$ | 2.63±0.51±0.47 | -172.0±6.6±6.2 | 4.8±1.8±1.2 | | | |
| $K^-\pi^0(S)$ | 0.70±0.27±0.24 | 133.2±22.5±25.2 | 2.7±1.4±0.8 | 0.85±0.09±0.11 | 108.4±7.8±8.9 | 3.9±0.9±1.0 |

Two different isobar models describe the data well. Both yield almost identical behavior in invariant mass (Fig. 19.1.25b–19.1.25d). The results of the best fits (Model I: $\chi^2/\nu = 702.08/714$, probability 61.9%; Model II: $\chi^2/\nu = 718.89/717$, probability 47.3%) are summarized in Table 19.1.10. They find that the $K\pi$ $S$-wave is not in phase with the $P$-wave at threshold as it was in the LASS scattering data. Both fitting models include significant contributions from $K^*(892)$, and each indicates that $D^0 \to K^{*+}K^-$ dominates over $D^0 \to K^{*-}K^+$. This suggests that, in tree-level diagrams, the form factor for $D^0$ coupling to $K^{*-}$ is suppressed compared to the corresponding $K^-$ coupling. While the measured fit fraction for $D^0 \to K^{*+}K^-$ agrees well with a phenomenological prediction (Buccella, Lusignoli, Miele, Pugliese, and Santorelli, 1995) based on a large $SU(3)$ symmetry breaking, the corresponding results for $D^0 \to K^{*-}K^+$ and the color-suppressed $D^0 \to \phi\pi^0$ decays differ significantly from the predicted values.

In a limited mass range, from threshold up to 1.02 GeV/$c^2$, they also measure the scalar amplitude using a model-independent partial-wave analysis. Agreement with similar measurements from $D^0 \to K^-K^+\bar{K}^0$ decay (Aubert, 2005f), and with the isobar models considered here, is excellent.

#### 19.1.4.7 $D^0 \to \pi^+\pi^-\pi^0$

This analysis from *BABAR* makes use of $N_S = 44780 \pm 250$ signal and $N_B = 830 \pm 70$ background events (Aubert, 2007w). Table 19.1.11 summarizes the results of the Dalitz plot analysis. The Dalitz plot distribution of the data is shown in Fig. 19.1.26(a-c). The distribution is marked by three destructively interfering $\rho\pi$ amplitudes, suggesting an $I = 0$-dominated final state (Zemach, 1964).

#### 19.1.4.8 $D_s^+ \to \pi^+\pi^-\pi^+$

*BABAR* has performed a Dalitz plot analysis of $D_s^+ \to \pi^+\pi^-\pi^+$ (Aubert, 2009i) and $D_s^+ \to K^+K^-\pi^+$ (del Amo Sanchez, 2011b) using 380 fb$^{-1}$. The selection of the two channels is similar and it will be described only once.

The three tracks are fitted to a common vertex, and the $\chi^2$ fit probability (labeled $P_1$) must be greater than 0.1 %. A separate kinematic fit which makes use of the $D_s^+$ mass constraint, to be used in the Dalitz plot analysis, is also performed. To help discriminate signal from background, an additional fit which uses the constraint that the three tracks originate from the $e^+e^-$ luminous region (beam spot) is performed. The $\chi^2$ probability of this fit is labeled as $P_2$, and it is expected to be large for background and small for $D_s^+$ signal events, since in general the latter will have a measurable flight distance.

The combinatorial background is reduced by requiring the $D_s^+$ to originate from the decay

$$D_s^{*+} \to D_s^+\gamma \tag{19.1.44}$$

using the mass difference $\Delta m = m(\pi^+\pi^-\pi^+\gamma) - m(\pi^+\pi^-\pi^+)$. Each $D_s^+$ candidate is characterized by three variables: the center-of-mass momentum $p^*$, the difference in probability $P_1 - P_2$, and the signed decay distance $d_{xy}$ between the $D_s^+$ decay vertex and the beam spot projected in the plane normal to the beam collision axis. The distributions for these variables for background are inferred from the $D_s^+ \to \pi^+\pi^-\pi^+$ invariant mass sidebands. Since these variables are (to a good approximation) independent of the decay mode, the distributions for the three-pion invariant mass signal, are inferred from the $D_s^+ \to K^+K^-\pi^+$ decay. These normalized distributions are then combined in a likelihood ratio test. The cut on the likelihood ratio has been chosen in order to obtain the largest statistics with background small enough to perform a Dalitz plot analysis.

The distributions of these variables for the $D_s^+ \to K^+K^-\pi^+$ decay for signal and background are shown in Fig. 19.1.27.

The resulting $D_s^+$ signal region contains 13179 events with a purity of 80%. The resulting Dalitz plot, symmetrized along the two axes, is shown in Fig. 19.1.28. We observe a clear $f_0(980)$ signal, evidenced by the two narrow crossing bands. We also observe a broad accumulation of events in the 1.9 GeV$^2$/$c^4$ region. The efficiency is found to be almost uniform as a function of the $\pi^+\pi^-$ invariant mass with an average value of $\approx 1.6$ %.

In the Dalitz plot analysis spin-1 and spin-2 resonances are described by relativistic Breit-Wigner function. For

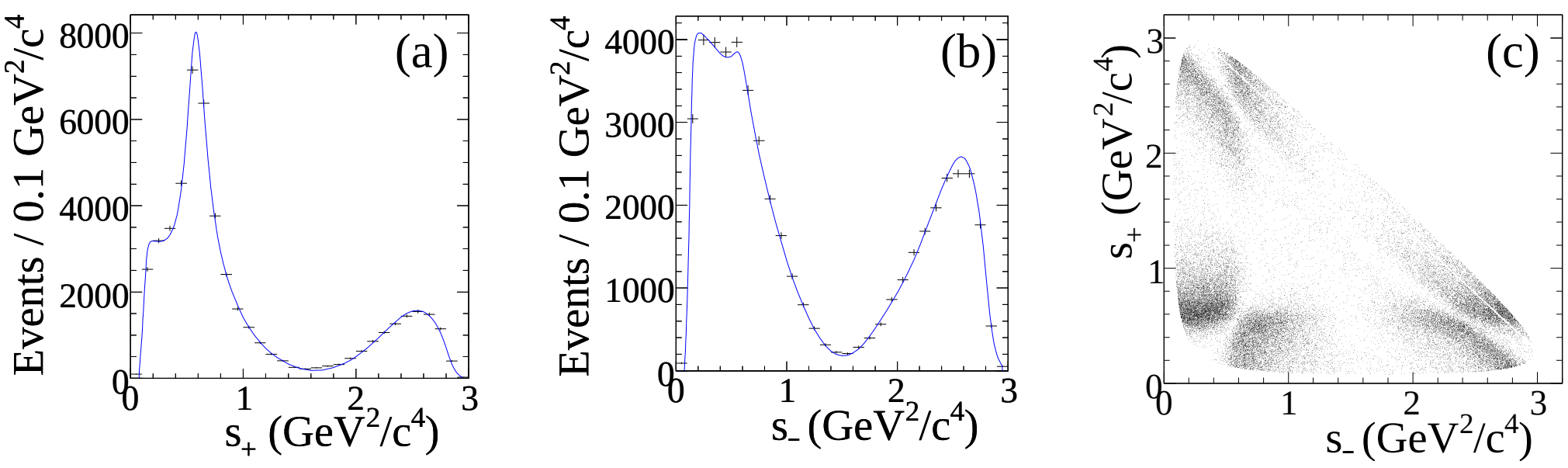

**Figure 19.1.26.** From (Aubert, 2007w). (a,b) Projections of the $D^{*+} \to D^0(\to \pi^+\pi^-\pi^0)\pi^+$ data events and *p.d.f.* onto the Dalitz plot variables $s_+ = m^2(\pi^+\pi^0)$ and $s_- = m^2(\pi^-\pi^0)$. (c) The 2-dimensional $(s_+, s_-)$ distribution of the $D^{*+} \to D^0\pi^+$ data.

**Table 19.1.11.** From (Aubert, 2007w). Result of the fit to the $D^0 \to \pi^+\pi^-\pi^0$ sample, showing the amplitudes ratios $R_r \equiv a_r/a_{\rho^+(770)}$, phase differences $\Delta\phi_r \equiv \phi_r - \phi_{\rho^+(770)}$, and fit fractions $f_r \equiv \int |a_r A_r(s_+, s_-)|^2 ds_- ds_+$ (see 13.4.4). The first (second) errors are statistical (systematic). The mass (width) of the $\sigma$ meson is taken as 400 (600) MeV/$c^2$.

| State | $R_r$ (%) | $\Delta\phi_r$ (°) | $f_r$(%) |
|---|---|---|---|
| $\rho^+(770)$ | 100 | 0 | 67.8±0.0±0.6 |
| $\rho^0(770)$ | 58.8±0.6±0.2 | 16.2±0.6±0.4 | 26.2±0.5±1.1 |
| $\rho^-(770)$ | 71.4±0.8±0.3 | −2.0±0.6±0.6 | 34.6±0.8±0.3 |
| $\rho^+(1450)$ | 21±6±13 | −146±18±24 | 0.11±0.07±0.12 |
| $\rho^0(1450)$ | 33±6±4 | 10±8±13 | 0.30±0.11±0.07 |
| $\rho^-(1450)$ | 82±5±4 | 16±3±3 | 1.79±0.22±0.12 |
| $\rho^+(1700)$ | 225±18±14 | −17±2±3 | 4.1±0.7±0.7 |
| $\rho^0(1700)$ | 251±15±13 | −17±2±2 | 5.0±0.6±1.0 |
| $\rho^-(1700)$ | 200±11±7 | −50±3±3 | 3.2±0.4±0.6 |
| $f_0(980)$ | 1.50±0.12±0.17 | −59±5±4 | 0.25±0.04±0.04 |
| $f_0(1370)$ | 6.3±0.9±0.9 | 156±9±6 | 0.37±0.11±0.09 |
| $f_0(1500)$ | 5.8±0.6±0.6 | 12±9±4 | 0.39±0.08±0.07 |
| $f_0(1710)$ | 11.2±1.4±1.7 | 51±8±7 | 0.31±0.07±0.08 |
| $f_2(1270)$ | 104±3±21 | −171±3±4 | 1.32±0.08±0.10 |
| $\sigma(400)$ | 6.9±0.6±1.2 | 8±4±8 | 0.82±0.10±0.10 |
| Non-Res | 57±7±8 | −11±4±2 | 0.84±0.21±0.12 |

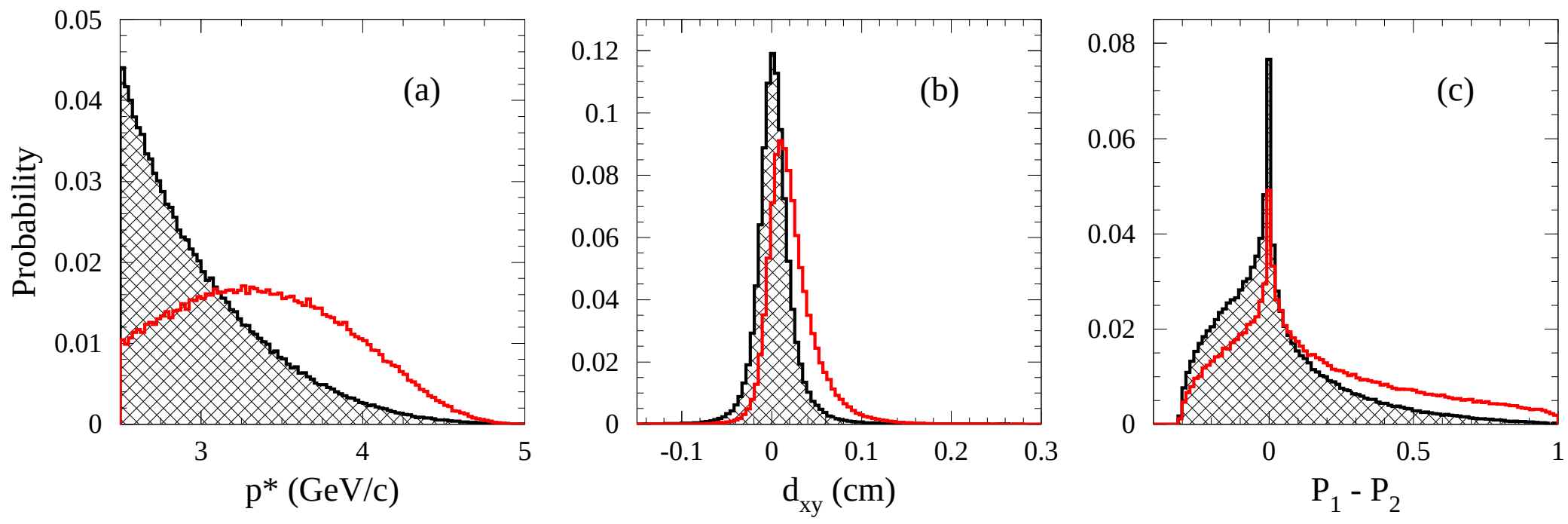

**Figure 19.1.27.** From (del Amo Sanchez, 2011b). Normalized probability distribution functions for signal (solid) and background events (hatched) used in a likelihood-ratio test for the event selection of $D_s^+ \to K^+K^-\pi^+$: (a) the center-of-mass momentum $p^*$, (b) the signed decay distance $d_{xy}$ and (c) the difference in probability $P_1 - P_2$.

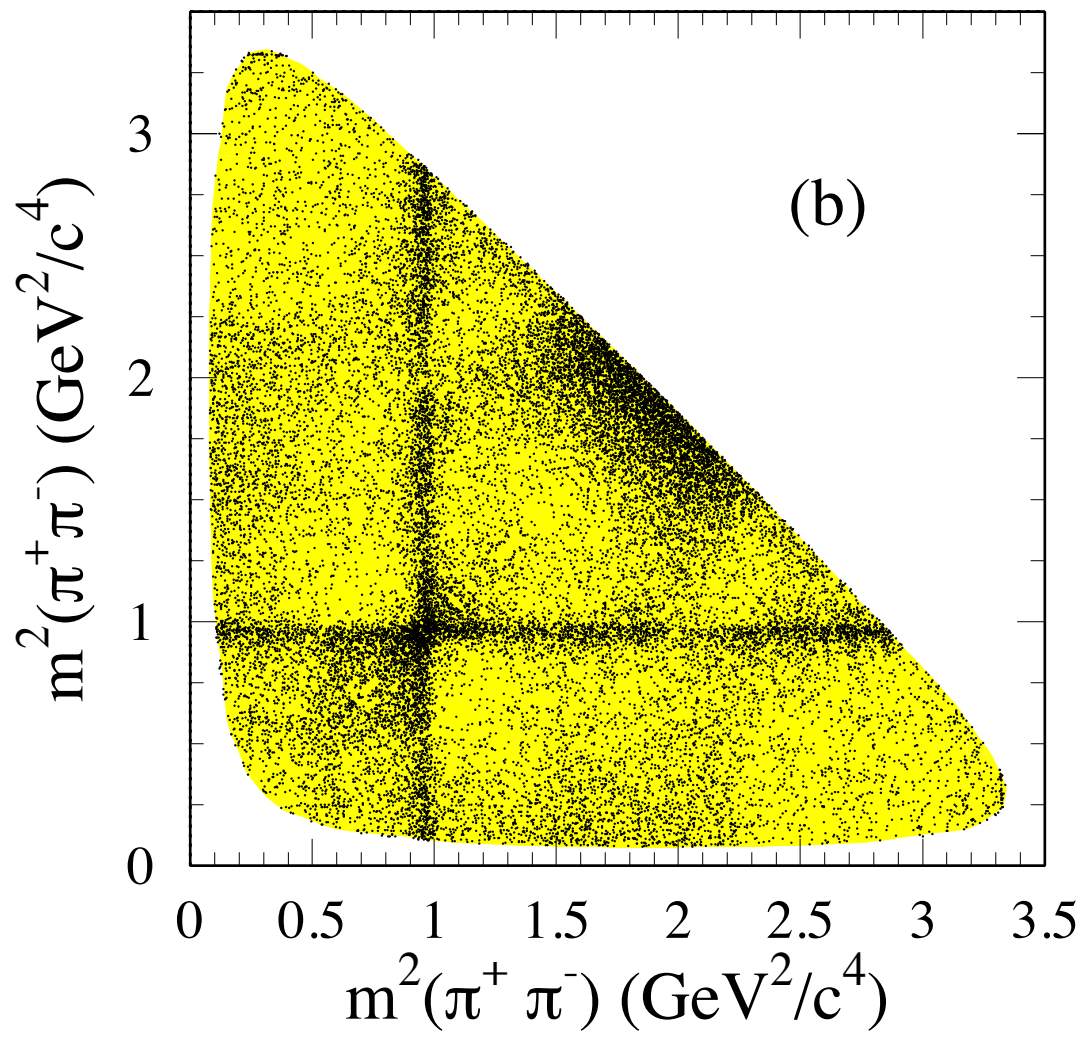

**Figure 19.1.28.** From (Aubert, 2009i). Symmetrized $D_s^+ \to \pi^+\pi^-\pi^+$ Dalitz plot (two entries per event).

the $\pi^+\pi^-$ $S$-wave amplitude, a different approach is used because:

- Scalar resonances have large uncertainties. In addition, the existence of some states needs confirmation.
- Modelling the $S$-wave as a superposition of Breit-Wigner functions is unphysical since it leads to a violation of unitarity when broad resonances overlap.

To overcome these problems, the Model-Independent Partial Wave Analysis introduced by the Fermilab E791 Collaboration (Aitala et al., 2006) has been used. Instead of including the $S$-wave amplitude as a superposition of relativistic Breit-Wigner functions, the $\pi^+\pi^-$ mass spectrum is divided into 29 slices and the $S$-wave is parameterized by an interpolation between the 30 endpoints in the complex plane:

$$A_{S-\text{wave}}(m_{\pi\pi}) = \text{Interp}(c_k(m_{\pi\pi})e^{i\phi_k(m_{\pi\pi})})_{k=1,...,30} \quad (19.1.45)$$

The amplitude and phase of each endpoint are free parameters. The width of each slice is tuned to get approximately the same number of $\pi^+\pi^-$ combinations ($\simeq 13179\times 2/29$). Interpolation is implemented by a Relaxed Cubic Spline. The phase is not constrained in a specific range in order to allow the spline to be a continuous function.

The background shape is obtained by fitting the $D_s^+$ sidebands. In this fit, resonances are assumed to be incoherent, *i.e.* are represented by Breit-Wigner intensity terms only. A good representation of the background includes contributions from $K_S^0$, $\rho^0(770)$ and three ad-hoc scalar resonances with free parameters.

The resulting $S$-wave $\pi^+\pi^-$ amplitude and phase is shown in Fig. 19.1.29(a),(b). The results from the Dalitz analysis are summarized in Table 19.1.12.

The Dalitz-plot projections together with the fit results are shown in Fig. 19.1.30. The labels $m^2(\pi^+\pi^-)_{\text{low}}$ and $m^2(\pi^+\pi^-)_{\text{high}}$ refer to the lower and higher values of the two $\pi^+\pi^-$ mass combinations.

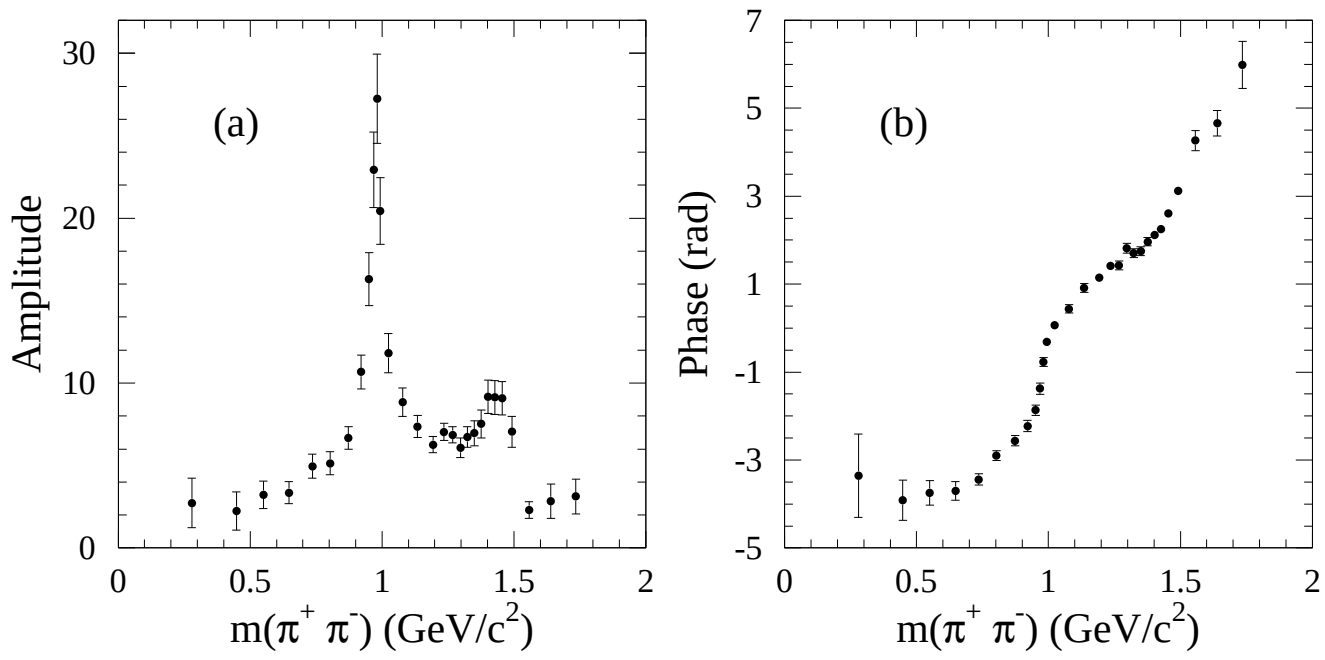

**Figure 19.1.29.** (a) $S$-wave amplitude extracted from the $D_s^+ \to \pi^+\pi^-\pi^+$ Dalitz plot analysis. (b) corresponding $S$-wave phase.

**Table 19.1.12.** From (Aubert, 2009i). Results from the $D_s^+ \to \pi^+\pi^-\pi^+$ Dalitz plot analysis. The table reports the fit fractions, amplitudes and phases. Errors are statistical and systematic respectively.

| Decay mode | Fraction(%) | Amplitude | Phase(rad) |
|---|---|---|---|
| $f_2(1270)\pi^+$ | 10.1±1.5±1.1 | 1.(Fixed) | 0.(Fixed) |
| $\rho(770)\pi^+$ | 1.8±0.5±1.0 | 0.19±0.02±0.12 | 1.1±0.1±0.2 |
| $\rho(1450)\pi^+$ | 2.3±0.8±1.7 | 1.2±0.3±1.0 | 4.1±0.2±0.5 |
| $S$-wave | 83.0±0.9±1.9 | | |
| Total | 97.2±3.7±3.8 | | |
| $\chi^2/NDF$ | $\frac{437}{422-64} = 1.2$ | | |

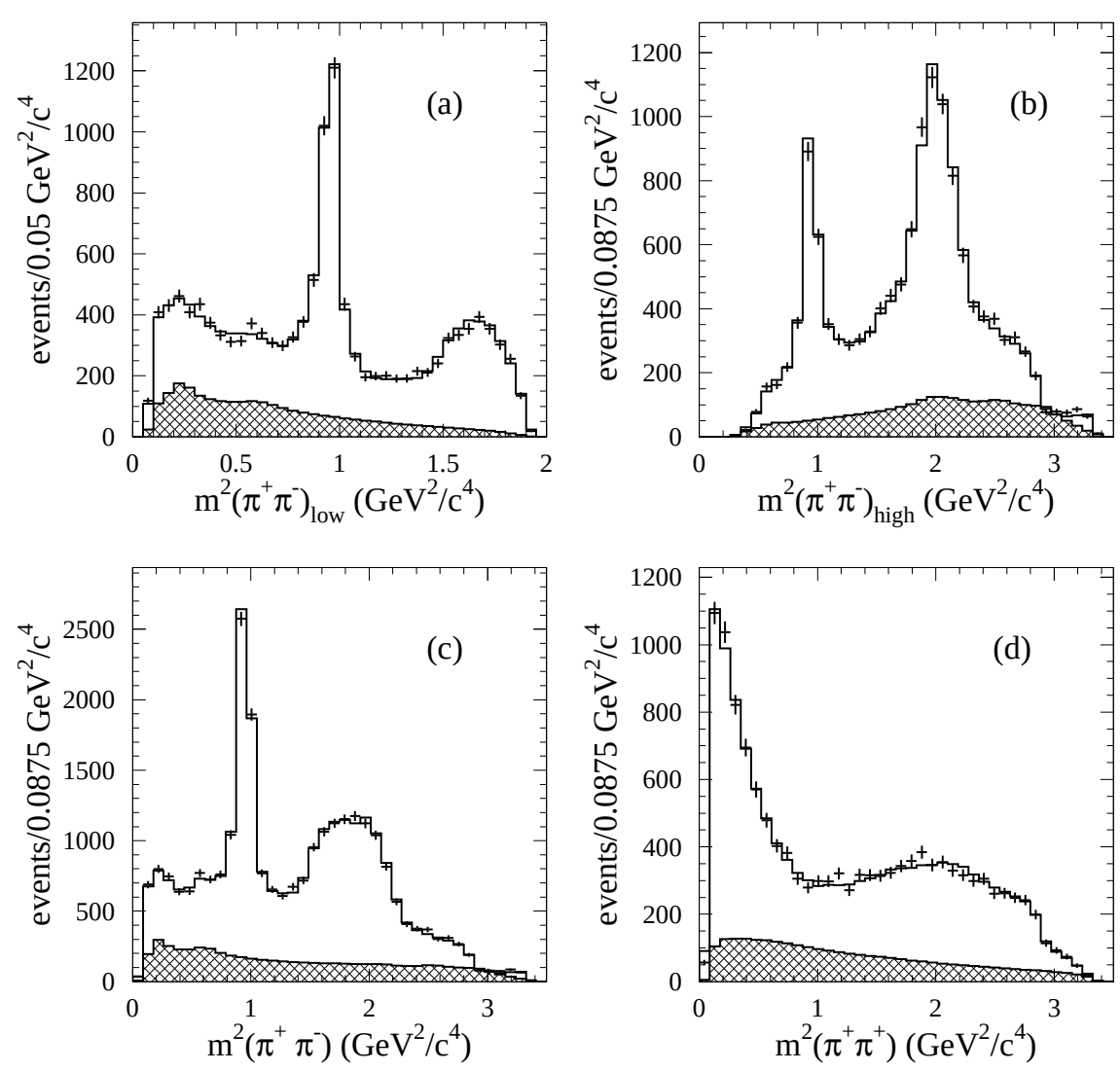

**Figure 19.1.30.** From (Aubert, 2009i). Dalitz plot projections (points with error bars) and fit results (solid histogram) from $D_s^+ \to \pi^+\pi^-\pi^+$ Dalitz plot analysis. (a) $m^2(\pi^+\pi^-)_{\text{low}}$, (b) $m^2(\pi^+\pi^-)_{\text{high}}$, (c) total $m^2(\pi^+\pi^-)$, (d) $m^2(\pi^+\pi^+)$. The hatched histograms show the background distribution.

The fit $\chi^2$ is computed by dividing the Dalitz plot into 30×30 cells with 422 cells having entries.

### 19.1.4.9 Dalitz plot analysis of $D_s^+ \to K^+K^-\pi^+$

The Dalitz analysis of $D_s^+ \to K^+K^-\pi^+$ is described in (del Amo Sanchez, 2011b). The selection of the channel is similar to that of the $D_s^+ \to \pi^+\pi^-\pi^+$ channel (see Section 19.1.4.8). The resulting $K^+K^-\pi^+$ mass distribution contains 96307 ± 369 events with 95% purity. The $D_s^+ \to K^+K^-\pi^+$ Dalitz plot is shown in Fig. 19.1.31.

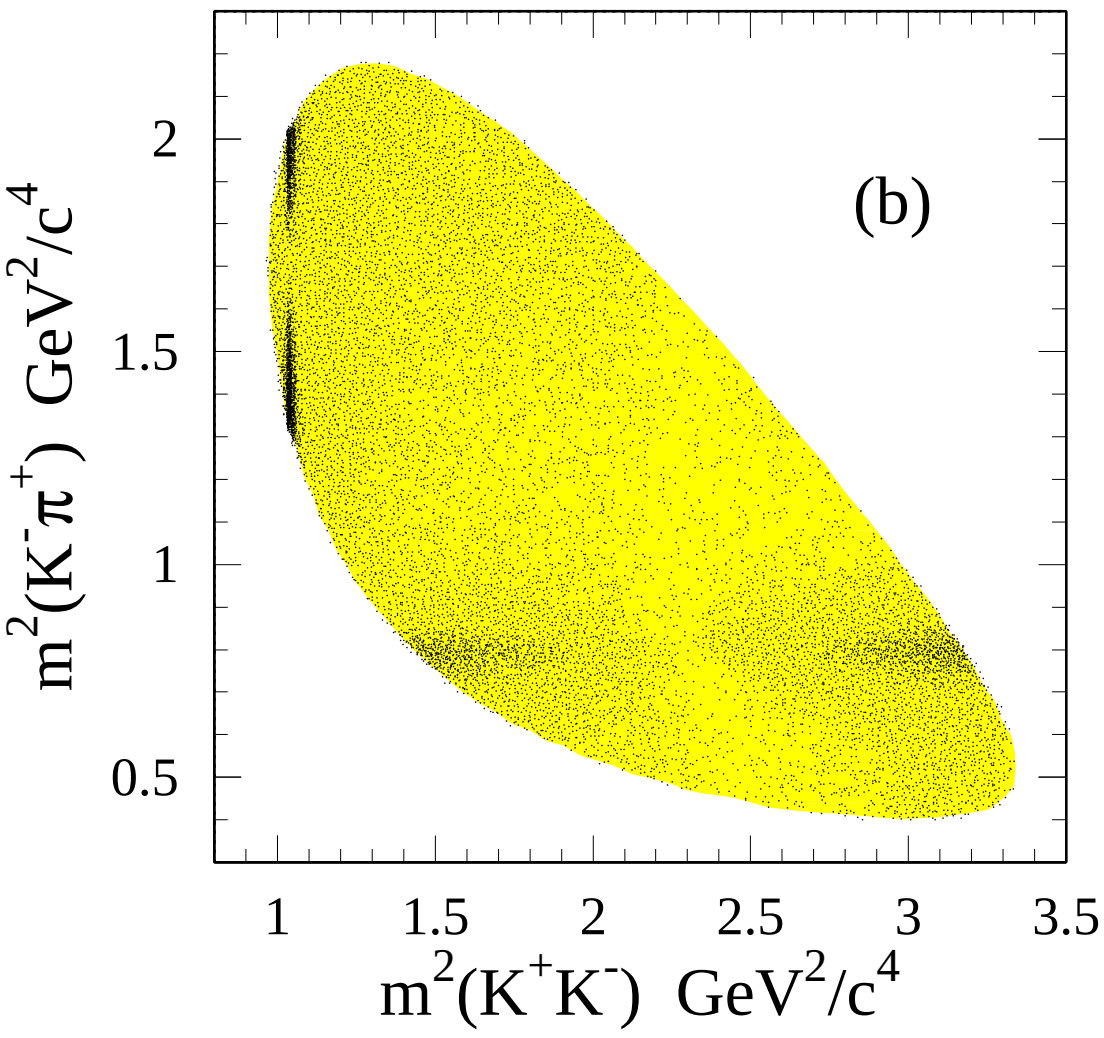

**Figure 19.1.31.** From (del Amo Sanchez, 2011b). $D_s^+ \to K^+K^-\pi^+$ Dalitz plot.

*Partial Wave Analysis of the $K^+K^-$ threshold region for $D_s^+ \to K^+K^-\pi^+$.*

In the $K^+K^-$ threshold region both $a_0(980)$ and $f_0(980)$ can be present, and both resonances have very similar parameters which suffer from large uncertainties. In this section a model-independent information on the $K^+K^-$ $S$-wave is obtained by a performing a partial wave analysis in the $K^+K^-$ threshold region. The procedure is similar to that reported in the analysis of the $D^0 \to \overline{K}^0 K^+K^-$ decay (Section 19.1.4.2).

Figure 19.1.32 shows the $K^+K^-$ mass spectrum up to $1.5\,\mathrm{GeV}/c^2$ weighted by the spherical harmonics moments. These distributions are corrected for efficiency and phase space, and background is subtracted using the $D_s^+$ sidebands.

The results from the Partial Wave Analysis are shown in Fig. 19.1.33. We observe a threshold enhancement in the $S$-wave (Fig. 19.1.33(a)), and the expected $\phi(1020)$ Breit-Wigner (BW) in the $P$-wave (Fig. 19.1.33(b)). We also observe the expected $\mathcal{S}$-$\mathcal{P}$ relative phase motion in the $\phi(1020)$ region (Fig. 19.1.33(c)). In Fig. 19.1.33(c), the $\mathcal{S}$-$\mathcal{P}$ phase difference is plotted twice because of the sign ambiguity associated with the value of $\phi_{\mathcal{SP}}$ extracted from $\cos(\phi_{\mathcal{SP}})$. The lines represent the result from the fit performed using S-P interfering amplitudes.

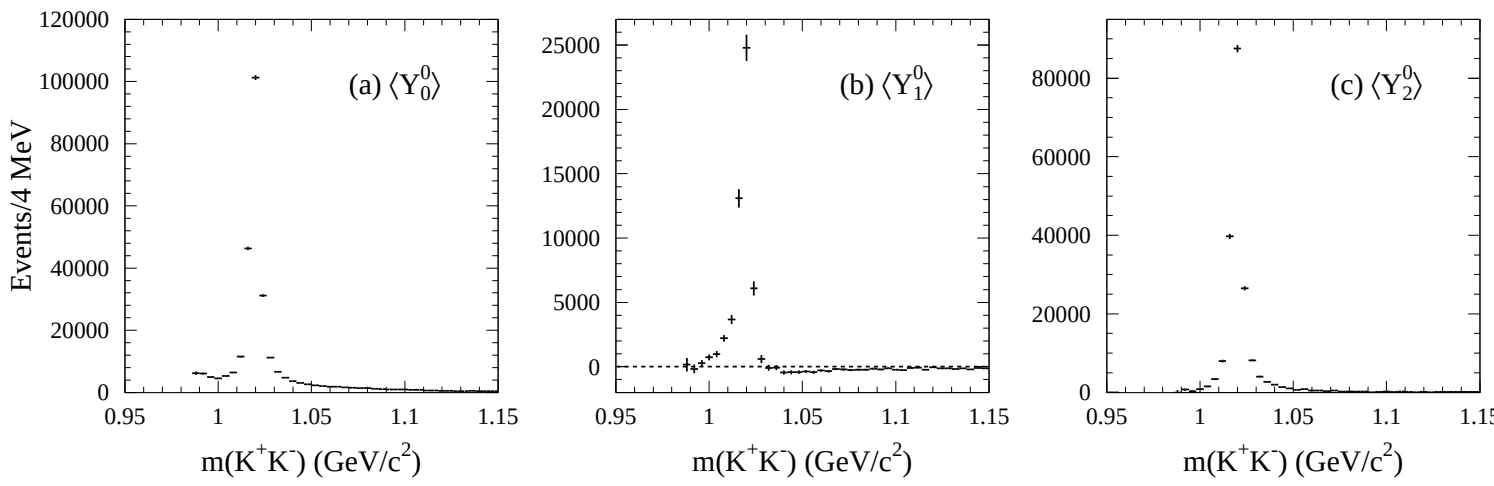

**Figure 19.1.32.** From (del Amo Sanchez, 2011b). $K^+K^-$ mass spectrum from $D_s^+ \to K^+K^-\pi^+$ in the threshold region weighted by (a) $Y_0^0$, (b) $Y_1^0$, and (c) $Y_2^0$, corrected for efficiency and phase space, and background-subtracted.

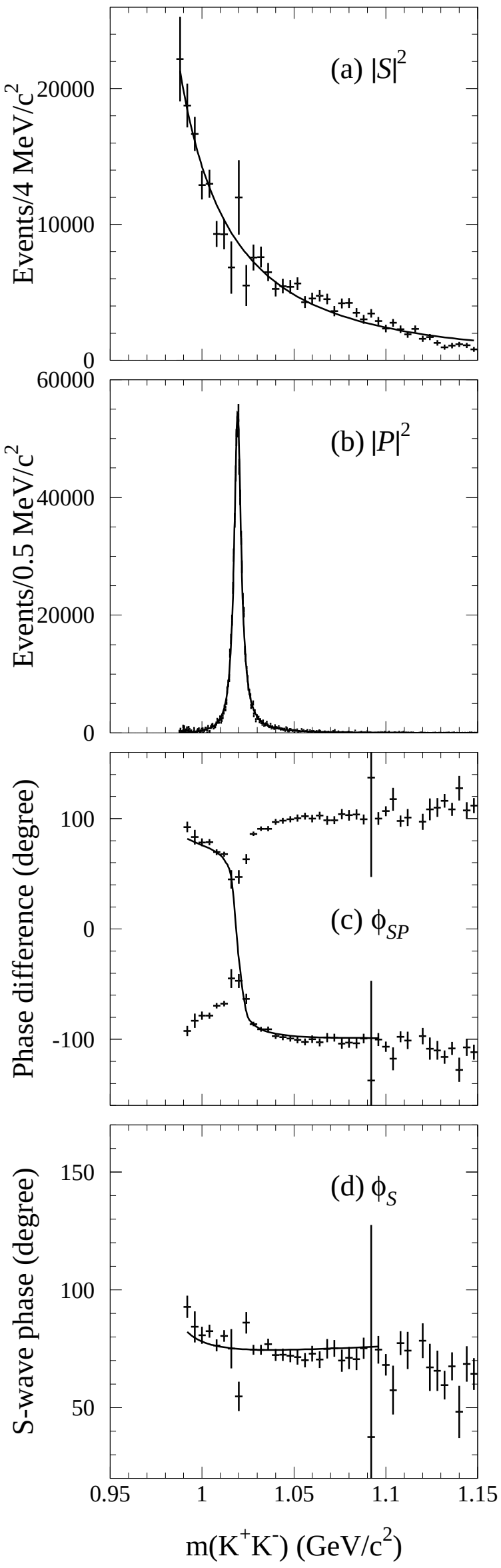

**Figure 19.1.33.** From (del Amo Sanchez, 2011b). Squared (a) $S$- and (b) $P$-wave amplitudes; (c) the phase difference $\phi_{SP}$; (d) $\phi_S$ obtained as explained in the text. The curves result from the fit described in the text.

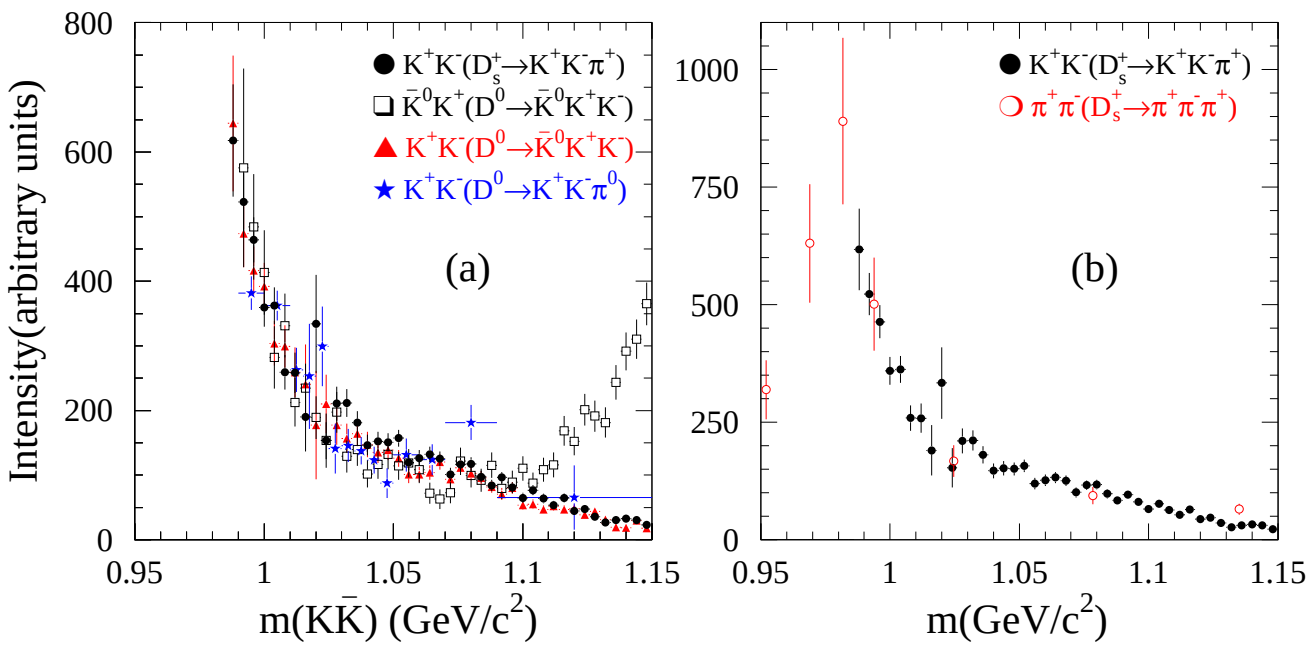

**Figure 19.1.34.** From (del Amo Sanchez, 2011b). (a) Comparison between $K\overline{K}$ $S$-wave intensities from different charmed meson Dalitz plot analyses. (b) Comparison of the $K\overline{K}$ $S$-wave intensity from $D_s^+ \to K^+K^-\pi^+$ with the $\pi^+\pi^-$ $S$-wave intensity from $D_s^+ \to \pi^+\pi^-\pi^+$.

The mass-dependent $f_0(980)$ phase is extracted by adding the mass-dependent $\phi(1020)$ Breit-Wigner phase to the $\phi_{\mathcal{SP}}$ distributions of Fig. 19.1.33(c). The phase ambiguity of Fig. 19.1.33(c) is resolved by choosing as the physical solution the one which decreases rapidly in the $\phi(1020)$ peak region, since this reflects the rapid forward Breit-Wigner-phase motion associated with a narrow resonance. The result is shown in Fig. 19.1.33(d), where we see that the $S$-wave phase is roughly constant, as would be expected for the tail of a resonance.

In Fig. 19.1.34(a) the $S$-wave profile from this analysis is compared with the $S$-wave intensity values extracted for Dalitz plot analyses of $D^0 \to \overline{K}^0K^+K^-$ (Aubert, 2005f) and $D^0 \to K^+K^-\pi^0$ (Aubert, 2007d). The four distributions are normalized in the region from threshold up to 1.05 GeV/$c^2$ and show a substantial agreement. As the $a_0(980)$ and $f_0(980)$ mesons couple mainly to the $u\bar{u}/d\bar{d}$ and $s\bar{s}$ systems respectively, the former is favored in $D^0 \to \overline{K}^0K^+K^-$ and the latter in $D_s^+ \to K^+K^-\pi^+$. Both resonances can contribute in $D^0 \to K^+K^-\pi^0$. We conclude that the $S$-wave projections in the $K\overline{K}$ system for both resonances are consistent in shape. It has been suggested that this feature supports the hypothesis that the $a_0(980)$ and $f_0(980)$ are 4-quark states (Maiani, Polosa, and Riquer, 2007). Figure 19.1.34(b)) also compares the $S$-wave profile from this analysis with the $\pi^+\pi^-$ $S$-wave profile extracted from *BABAR* data in a Dalitz plot analysis of $D_s^+ \to \pi^+\pi^-\pi^+$ Aubert (2009i). The observed agreement supports the argument that only the $f_0(980)$ is present in this limited mass region.

### *Dalitz plot analysis of $D_s^+ \to K^+K^-\pi^+$*

In the full Dalitz plot analysis, the $\overline{K}^*(892)^0$ amplitude is chosen as reference. The decay fractions, amplitudes, and relative phase values are summarized in Table 19.1.13 where the first error is statistical, and the second is systematic. We observe that the decay is dominated by the $\overline{K}^*(892)^0K^+$ and $\phi(1020)\pi^+$ amplitudes and that the fit quality is substantially improved by leaving the $\overline{K}^*(892)^0$ parameters free in the fit. The fitted parameters are:

$$\begin{aligned} m_{\overline{K}^*(892)^0} &= (895.6 \pm 0.2_{\rm stat} \pm 0.3_{\rm sys})\ \mathrm{MeV}/c^2 \\ \Gamma_{\overline{K}^*(892)^0} &= (45.1 \pm 0.4_{\rm stat} \pm 0.4_{\rm sys})\ \mathrm{MeV} \end{aligned} \tag{19.1.46}$$

We notice that the width is about 5 MeV lower than that in Amsler et al. (2008). However this measurement is consistent with results from other Dalitz plot analyses (Mitchell et al., 2009a) and measurements with semileptonic $D^+ \to \overline{K}^*(892)^0e^+\nu_e$ decays (see Section 19.1.5.6).

The $f_0(1370)$ contribution is also left free in the fit, and we obtain the following parameter values:

$$\begin{aligned} m_{f_0(1370)} &= (1.22 \pm 0.01_{\rm stat} \pm 0.04_{\rm sys})\ \mathrm{GeV}/c^2 \\ \Gamma_{f_0(1370)} &= (0.21 \pm 0.01_{\rm stat} \pm 0.03_{\rm sys})\ \mathrm{GeV} \end{aligned} \tag{19.1.47}$$

These values are within the broad range of values measured by other experiments (Amsler et al., 2008).

The nonresonant contribution is consistent with zero.

The results of the best fit are superimposed on the Dalitz plot projections in Fig. 19.1.35. The normalized fit residuals shown under each distribution (Fig. 19.1.35) are defined as $(N_{\rm data} - N_{\rm fit})/\sqrt{N_{\rm data}}$.

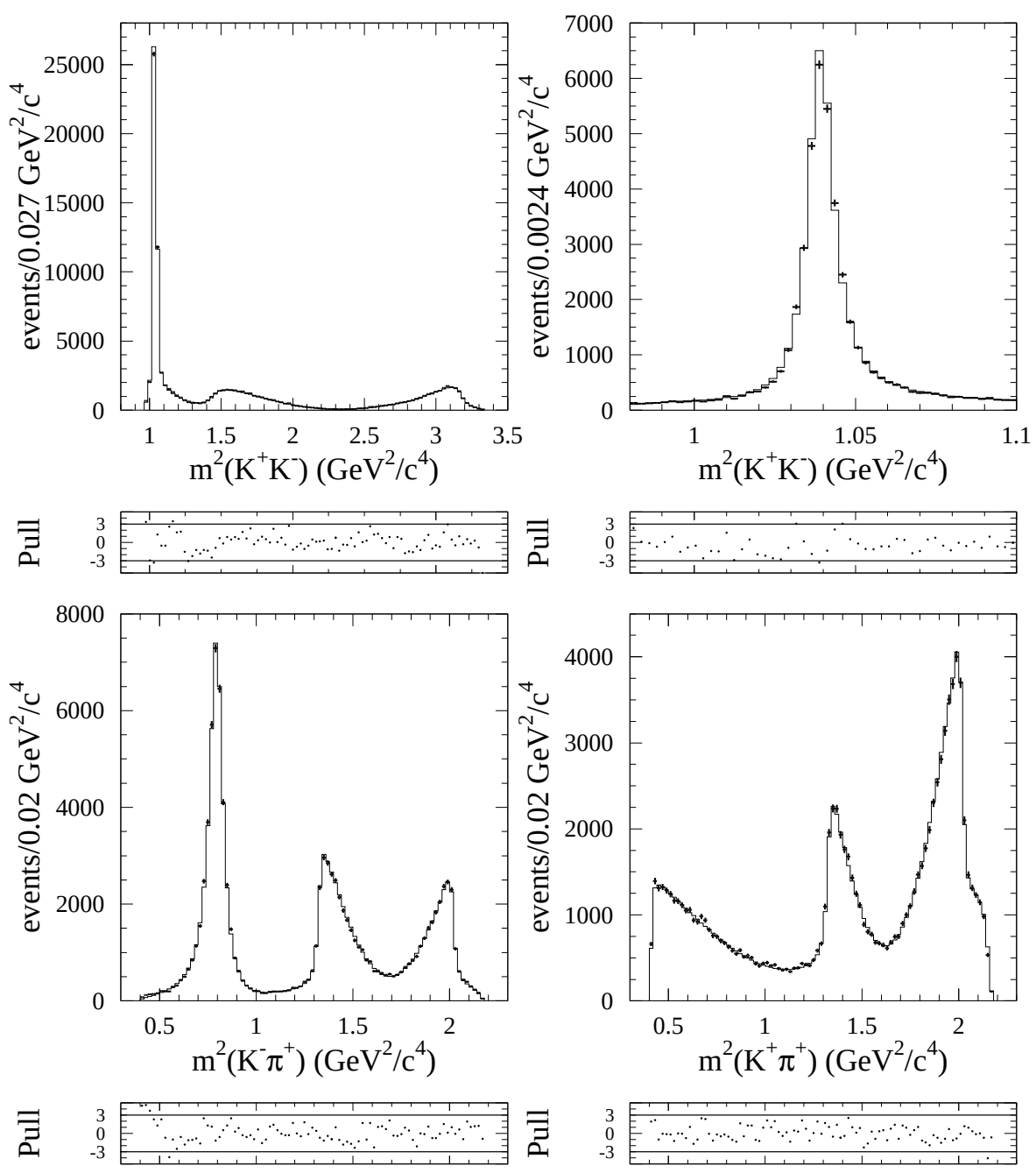

**Figure 19.1.35.** From (del Amo Sanchez, 2011b). $D_s^+ \to K^+K^-\pi^+$: Dalitz plot projections. The data are represented by points with error bars, the fit results by the histograms. The upper right plot shows zoom of the $m^2(K^+K^-)$ in the $\phi$ mass region.

**Table 19.1.13.** From (del Amo Sanchez, 2011b). Results from the $D_s^+ \to K^+K^-\pi^+$ Dalitz plot analysis. The table gives fit fractions, amplitudes and phases from the best fit. Quoted uncertainties are statistical and systematic, respectively.

| Decay Mode | Decay fraction (%) | Amplitude | Phase (radians) |
|---|---|---|---|
| $\overline{K}^*(892)^0 K^+$ | $47.9 \pm 0.5 \pm 0.5$ | 1.(Fixed) | 0.(Fixed) |
| $\phi(1020)\,\pi^+$ | $41.4 \pm 0.8 \pm 0.5$ | $1.15 \pm 0.01 \pm 0.26$ | $2.89 \pm 0.02 \pm 0.04$ |
| $f_0(980)\,\pi^+$ | $16.4 \pm 0.7 \pm 2.0$ | $2.67 \pm 0.05 \pm 0.20$ | $1.56 \pm 0.02 \pm 0.09$ |
| $\overline{K}_0^*(1430)^0 K^+$ | $2.4 \pm 0.3 \pm 1.0$ | $1.14 \pm 0.06 \pm 0.36$ | $2.55 \pm 0.05 \pm 0.22$ |
| $f_0(1710)\,\pi^+$ | $1.1 \pm 0.1 \pm 0.1$ | $0.65 \pm 0.02 \pm 0.06$ | $1.36 \pm 0.05 \pm 0.20$ |
| $f_0(1370)\,\pi^+$ | $1.1 \pm 0.1 \pm 0.2$ | $0.46 \pm 0.03 \pm 0.09$ | $-0.45 \pm 0.11 \pm 0.52$ |
| Sum | $110.2 \pm 0.6 \pm 2.0$ | | |
| $\chi^2/NDF$ | $2843/(2305-14) = 1.24$ | | |

*$P$-wave/$S$-wave ratio in the $\phi(1020)$ region*

The decay mode $D_s^+ \to \phi(1020)\pi^+$ is used often as the normalizing mode for $D_s^+$ decay branching fractions, typically by selecting a $K^+K^-$ invariant mass region around the $\phi(1020)$ peak. The observation of a significant $S$-wave contribution in the threshold region means that this contribution must be taken into account in such a procedure.

*BABAR* has estimated the $P$-wave/$S$-wave ratio in an almost model-independent way. Integrating the distributions of $\sqrt{4\pi}pq' \langle Y_0^0 \rangle$ and $\sqrt{5\pi}pq' \langle Y_2^0 \rangle$ (Fig. 19.1.32), where $p$ is the $K^+$ momentum in the $K^+K^-$ rest frame, and $q'$ is the momentum of the bachelor $\pi^+$ in the $D_s^+$ rest frame, in a region around the $\phi(1020)$ peak yields $\int(|\mathcal{S}|^2 + |\mathcal{P}|^2)pq'\mathrm{d}m_{K^+K^-}$ and $\int|\mathcal{P}|^2pq'\mathrm{d}m_{K^+K^-}$ respectively.

The $S$-$P$ interference contribution integrates to zero, and the $P$-wave and $S$-wave fractions are defined as

$$f_{P-\mathrm{wave}} = \frac{\int |P|^2 pq' \mathrm{d}m_{K^+K^-}}{\int(|S|^2 + |P|^2)pq'\mathrm{d}m_{K^+K^-}} \quad (19.1.48)$$

$$f_{S-\mathrm{wave}} = \frac{\int |S|^2 pq' \mathrm{d}m_{K^+K^-}}{\int(|S|^2 + |P|^2)pq'\mathrm{d}m_{K^+K^-}} = 1 - f_{P-\mathrm{wave}}\,. \quad (19.1.49)$$

The experimental mass resolution is estimated to be $\simeq 0.5$ MeV/$c^2$ at the $\phi$ mass peak. Table 19.1.14 gives the resulting $S$-wave and $P$-wave fractions computed for three $K^+K^-$ mass regions. The last column of Table 19.1.14 shows the measurements of the relative overall rate ($N/N_{\mathrm{tot}}$) defined as the number of events in the $K^+K^-$ mass interval over the number of events in the entire Dalitz plot after efficiency correction and background subtraction.

**Table 19.1.14.** From (del Amo Sanchez, 2011b). $S$-wave and $P$-wave fractions computed in three $K^+K^-$ mass ranges around the $\phi(1020)$ peak. Errors are statistical only.

| $m_{K^+K^-}$ (MeV/$c^2$) | $f_{S-\mathrm{wave}}$ (%) | $f_{P-\mathrm{wave}}$ (%) | $\frac{N}{N_{\mathrm{tot}}}$ (%) |
|---|---|---|---|
| $1019.456 \pm 5$ | $3.5 \pm 1.0$ | $96.5 \pm 1.0$ | $29.4 \pm 0.2$ |
| $1019.456 \pm 10$ | $5.6 \pm 0.9$ | $94.4 \pm 0.9$ | $35.1 \pm 0.2$ |
| $1019.456 \pm 15$ | $7.9 \pm 0.9$ | $92.1 \pm 0.9$ | $37.8 \pm 0.2$ |

## 19.1.5 Semileptonic charm decays

### 19.1.5.1 Introduction

Exclusive semileptonic decays of $B$ and $D$ mesons are a favored means of determining the weak interaction couplings of quarks within the Standard Model because of their relative abundance; in addition, the hadronic uncertainties in their theoretical description are by far better under control than in hadronic decays. Our knowledge of the form factors parameterizing the hadronic current is limiting the precision on extractions of the couplings $|V_{cb}|$ and $|V_{ub}|$. Form factors from $B$ and $D$ meson semileptonic decays have been calculated using lattice QCD techniques whilst heavy quark symmetry relates the two form factors. Measurements of $D \to K/\pi\ell^+\nu_\ell$ are required to confront the theoretical predictions.[117]

Charm meson semileptonic decays with two pseudo scalar mesons ($P_1, P_2$) emitted in the final state allow also to study the strong interaction of the two pseudoscalar mesons in systems with well defined values of isospin and angular momentum, without parasitic effects from the presence of a third hadron as in Dalitz plot analyses. In particular $S$-wave systems can be isolated and properties of $P$-wave resonances can be accurately measured. Large statistics are analyzed, exceeding previous analyses by two orders of magnitude, in particular for the $D_s^+$ meson.

### 19.1.5.2 $D \to K/\pi\ell^+\nu_\ell$ decays

To measure $D$ meson semileptonic decays with a pseudo scalar particle emitted in the final state ($D_{\ell 3}$ decays), Belle and *BABAR* have used completely different techniques. In Belle, events with all particles reconstructed are selected. This allows to isolate signal events over a low background level and a high resolution in $q^2 = (p_\ell + p_{\overline{\nu}_\ell})^2$, at the price of a low efficiency. In *BABAR*, a more inclusive approach allows to have larger statistics at the price of a higher background level and poorer $q^2$ resolution. In this case, additional measurements allow to control distributions from background events.

[117] Inclusion of charge-conjugate states is implied throughout this section.

#### 19.1.5.3 Belle measurement

To achieve good resolution in the neutrino momentum and $q^2$, the $\overline{D}^0$ is tagged by fully reconstructing the remainder of the event (Widhalm, 2006). Events of the type $e^+e^- \to D^{(*)}_{\rm tag} D^{*-}_{\rm sig} X$ with $D^{*-}_{\rm sig} \to \overline{D}^0_{\rm sig}\pi^-_s$ are seeked, where $X$ may include additional $\pi^\pm$, $\pi^0$, or $K^\pm$ mesons. Each candidate is assembled from a fully reconstructed "tag-side" charm meson ($D^{(*)}_{\rm tag}$) which can be $D^{*+} \to D^0\pi^+$, $D^+\pi^0$ or $D^{*0} \to D^0\pi^0$, $D^0\gamma$ with $D^{+/0} \to K^-(\text{n}\pi)^{++/+}$, n = 1, 2, 3. To the ($D^{(*)}_{\rm tag}$) is added a charged pion, that is kinematically consistent with the $\pi^-_s$ from $D^{*-}_{\rm sig}$ decay, and the candidate $X$ is formed from combinations of unassigned $\pi$ and $K^+K^-$ pairs, conserving total event electric charge. The 4-momentum of $D^{*-}_{\rm sig}$ is found by energy-momentum conservation, assuming a $D^{(*)}_{\rm tag} D^{*-}_{\rm sig} X$ event. The candidate $\overline{D}^0_{\rm sig}$ 4-momentum is calculated from that of the $D^{*-}_{\rm sig}$ and $\pi^-_s$. The corresponding $\overline{D}^0_{\rm sig}$ invariant mass distribution, obtained after analyzing an integrated luminosity of 282 fb$^{-1}$, contains $56461 \pm 309 \pm 830$ signal over $39789 \pm 830$ background events, the latter beeing estimated using wrong sign combinations in data and few corrections from the simulation. Within this sample of $\overline{D}^0_{\rm sig}$ tags, the semileptonic decay $\overline{D}^0_{\rm sig} \to K^+/\pi^+\ell^-\overline{\nu}_\ell$ is reconstructed with $K^+/\pi^+$ and $\ell^-$ candidates from among the remaining tracks. The neutrino 4-momentum is reconstructed by energy-momentum conservation, its invariant mass squared, $m^2_\nu$, is required to satisfy $|m^2_\nu| < 0.05\,\text{GeV}^2/c^4$. About 1300 and 150 semileptonic decays are isolated for each lepton flavor ($e$ and $\mu$) in Cabibbo-allowed and Cabibbo-suppressed decays respectively. These numbers have to be corrected for remaining background contributions from other semileptonic and hadronic decays where a hadron is mis-identified as a lepton. Corrections amount typically to 2% and 20% respectively in Cabibbo-allowed and Cabibbo-suppressed decays. The accuracy on branching fraction measurements is limited by systematic uncertainties for Cabibbo-allowed decays whereas systematic and statistical uncertainties are similar in Cabibbo-suppressed events. The main contribution to systematic uncertainties comes from the evaluation of fake $\overline{D}^0_{\rm sig}$ tags. The resolution in $q^2 = (p_\ell + p_{\overline{\nu}_\ell})^2$ is found to be $0.0145 \pm 0.0007_{\rm stat}\,\text{GeV}^2/c^2$ in MC signal events. This is much smaller than statistically reasonable bin widths, which have been chosen as 0.067 (0.3) $\text{GeV}^2/c^2$ for kaon (pion) modes, and hence no unfolding is necessary.

Measured hadronic form factors obtained by Belle are given in Figure 19.1.36 where they are compared with several theoretical expectations.

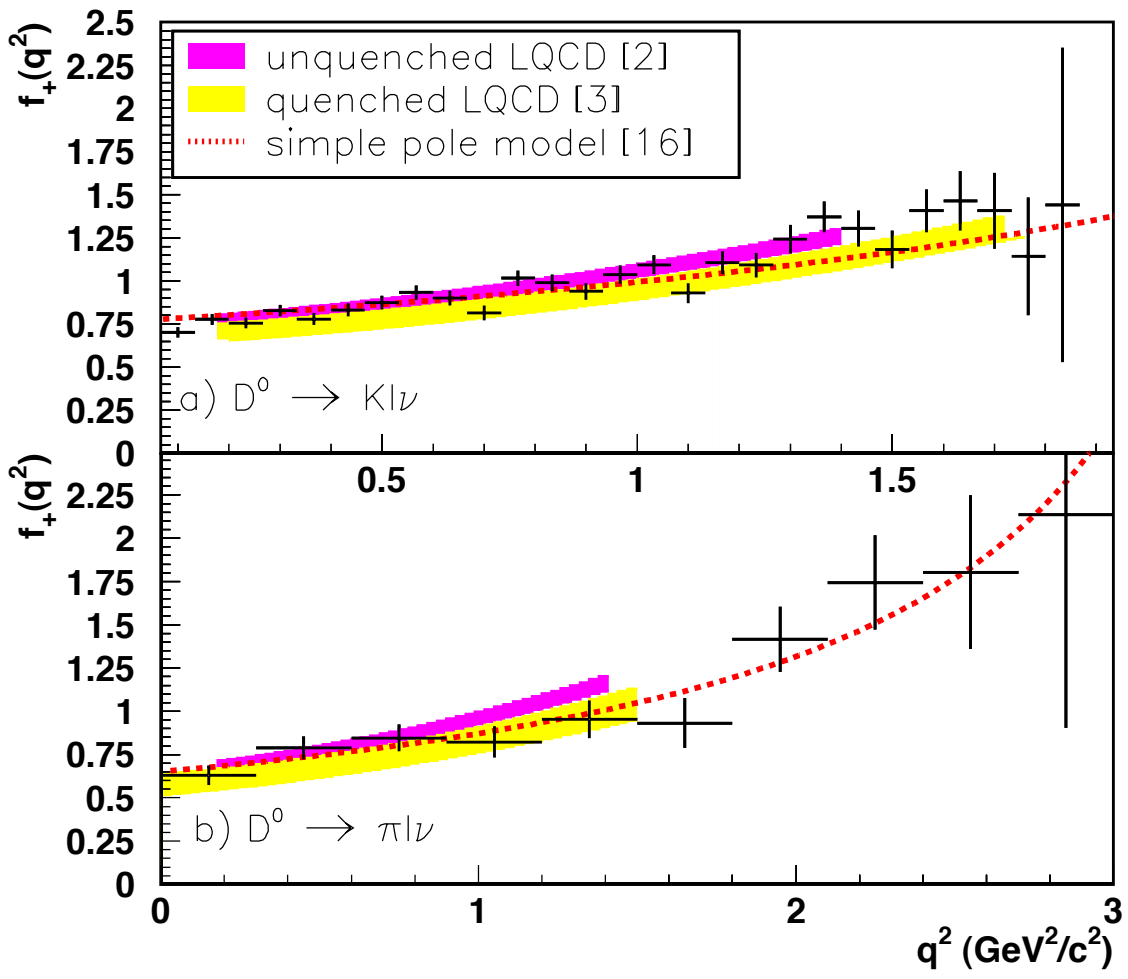

**Figure 19.1.36.** From (Widhalm, 2006). Form factors for (a) $D^0 \to K^-\ell^+\nu_\ell$, in $q^2$ bins of $0.067\,\text{GeV}^2/c^2$ and (b) $D^0 \to \pi^-\ell^+\nu_\ell$, in $q^2$ bins of $0.3\,\text{GeV}^2/c^2$. Overlaid are the predictions of the simple pole model using the physical pole mass (dashed) (Amoros, Noguera, and Portoles, 2003) and a quenched (light grey) (Abada et al., 2003) and unquenched (dark grey) LQCD calculation (Aubin et al., 2005). The shaded band reflects the theoretical uncertainty.

#### 19.1.5.4 *BABAR* measurement

$D^0 \to K^-e^+\nu_e(\gamma)$ decays are reconstructed in $e^+e^- \to c\overline{c}$ events from the continuum where the $D^0$ originates from $D^{*+} \to D^0\pi^+$ (Aubert, 2007ab). The analyzed integrated luminosity is 75 fb$^{-1}$ and semileptonic decays with a muon are not selected. In each event, the direction of the thrust axis is used to define two hemispheres. In each hemisphere, pairs of oppositely charged leptons and kaons are searched. Since the $\nu_e$ momentum is unmeasured, a kinematic fit is performed, constraining the invariant mass of the candidate $K^-e^+\nu_e$ system to the $D^0$ mass. In this fit, the $D^0$ momentum and the neutrino energy are estimated from the other particles measured in the event. Each $D^0$ candidate is retained if the $\chi^2$ probability of the kinematic fit exceeds $10^{-3}$ and it is combined with a charged pion, with the same charge as the lepton, and situated in the same hemisphere. The mass difference $\delta(m) = m(D^0\pi^+) - m(D^0)$ is evaluated and signal events accumulate at low values of this variable. Only events with $\delta(m) < 0.16\,\text{GeV}/c^2$ are used in the analysis. Background events are rejected by applying cuts on Fisher discriminant variables against $B\overline{B}$, other $c\overline{c}$, and light quark events.

To improve the accuracy of the reconstructed $D^0$ momentum, the nominal $D^{*+}$ mass is added as a constraint in the previous fit and only events with a $\chi^2$ probability higher than 1% are kept. There are 85260 selected $D^0$ candidates containing an estimated number of 11280 background events. The 4-momentum squared of the leptonic system is obtained using the fitted $D^0$ and kaon 4-momenta: $q_r^2 = (p_{D^0} - p_K)^2$. The resolution on the reconstructed $q^2$ obtained from the simulation is found to be around $0.16\,\mathrm{GeV}/c^2$. To obtain the true $q^2$ distribution, the measured one is corrected for selection efficiency, detector resolution and radiative effects. This is done using an unfolding algorithm based on MC simulation of these effects. Other data samples are used to validate the MC simulation. Events with $D^{*+} \to D^0\pi^+$, $D^0 \to K^-\pi^+$ allows to control the charm quark hadronization mechanism and the reconstruction accuracy of the $D^0$ 4-momentum from other particles in the event. A sample of $D^{*+} \to D^0\pi^+$, $D^0 \to K^-\pi^+\pi^0$ decays is used also to verify the reconstruction accuracy on $q_r^2$ and to define corrections. To reduce systematic uncertainties on the semileptonic decay branching fraction determination, this quantity is measured relative to the $D^{*+} \to D^0\pi^+$, $D^0 \to K^-\pi^+$ decay channel which is isolated in data and in the simulation using, as much as possible, similar selection criteria. This has also the advantage that future improvements in the $\mathcal{B}(D^0 \to K^-\pi^+)$ determination can be incorporated (this was indeed the case since results have been published). The uncertainty on the branching fraction measurement is dominated by systematic uncertainties which have different origins (reconstruction algorithm, electron identification, background subtraction, control of the discriminant Fisher variable distribution and, counting of $D^{*+}$ events in the normalization channel) of similar importance.

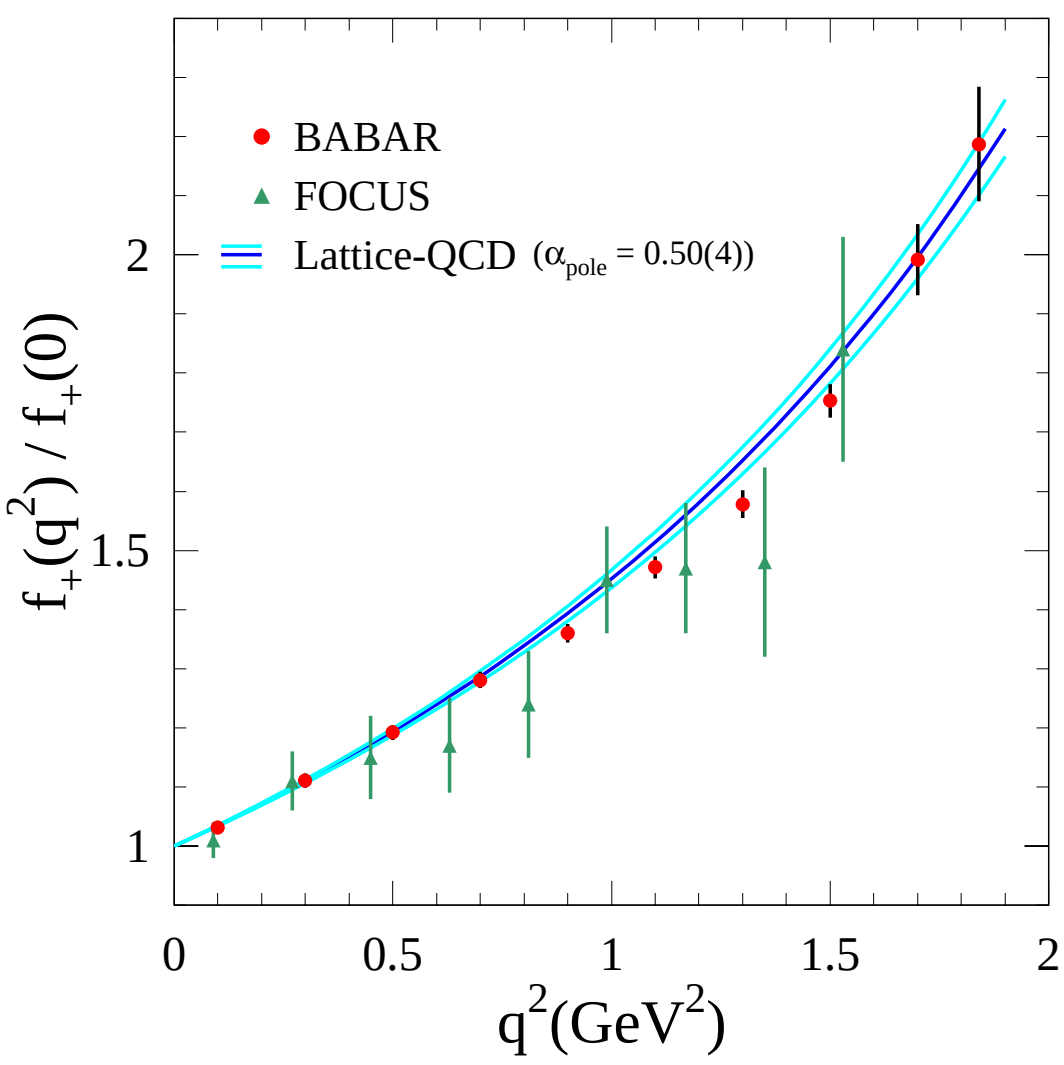

**Figure 19.1.37.** From (Aubert, 2007ab). Comparison of the measured variation of $f_+^K(q^2)/f_+^K(0)$ obtained in *BABAR* and in the FOCUS experiment (Link et al., 2005). The band corresponds to an estimate from LQCD (Aubin et al., 2005).

Figure 19.1.37 gives the measured variation versus $q^2$ of the ratio $f_+^K(q^2)/f_+^K(0)$ obtained in *BABAR*.

#### 19.1.5.5 Comparison with theory and with other experiments

In $D^0 \to K^-/\pi^- e^+\nu_e$ decays, neglecting the electron mass, the differential decay rate depends on only one form factor, $f_+^{K/\pi}(q^2)$:

$$\frac{d\Gamma}{dq^2} = \frac{G_F^2}{24\pi^3} |V_{cq}|^2 \left|\boldsymbol{p}_P(q^2)\right|^3 \left|f_+^P(q^2)\right|^2, \qquad (19.1.50)$$

where $G_F$ is the Fermi constant, $|V_{cq}|$ with $q = s$ or $d$, respectively for $P = K$ or $\pi$, is the absolute value of the corresponding CKM element, and $\boldsymbol{p}_P(q^2)$ is the pseudoscalar ($P$) three-momentum in the $D^0$ rest frame.

The unitarity of the first line of the CKM matrix is verified with high accuracy using beta decays of nuclei, $K_{\ell 3}$, and $K_{\ell 2}$ decays (Antonelli et al., 2010b):

$$|V_{ud}| = 0.97425(22),\ \ |V_{us}| = 0.2253(9), \qquad (19.1.51)$$

giving:

$$|V_{ud}|^2 + |V_{us}|^2 - 1 = -0.0001(8). \qquad (19.1.52)$$

Note the the contribution from $|V_{ub}|^2$ is completely negligible. The unitarity condition for the first column reads:

$$|V_{ud}|^2 + |V_{cd}|^2 + |V_{td}|^2 = 1. \qquad (19.1.53)$$

There could be large effects from new physics in the present value of $|V_{td}|$ but they will have essentially no contribution in the unitarity constraint as $|V_{td}| \sim |V_{ub}|$. So, independently of effects from new physics in $B^0\overline{B}^0$ oscillations, one can use $|V_{cd}| = |V_{us}| = 0.2253(9)$. Using this, together with the unitarity condition of the second line of the CKM matrix, one gets:

$$|V_{cs}| = |V_{ud}| - \frac{|V_{cb}|^2}{2} = 0.97343 \pm 0.00023, \qquad (19.1.54)$$

using the value $|V_{cb}| = (40.6 \pm 1.3) \times 10^{-3}$ (Nakamura et al., 2010). $D_{\ell 3}$ decays depend on the product $|V_{cq}|\left|f_+^P(q^2)\right|$ and, as the CKM elements $|V_{cq}|$ are precisely determined, charm semileptonic decays allow to measure the absolute values of the corresponding hadronic form factors and their variation versus $q^2$.

The most general expressions of the form factor $f_+^P(q^2)$ are analytic functions satisfying the dispersion relation:

$$f_+^P(q^2) = \frac{Res(f_+^P)_{q^2=m^2_{D_q^*}}}{m^2_{D_q^*} - q^2} + \frac{1}{\pi}\int_{t_+}^{\infty} dt \frac{\mathrm{Im} f_+^P(t)}{t - q^2 - i\epsilon}. \qquad (19.1.55)$$

The singularities in the complex $t \equiv q^2$ plane originate from the interaction of the charm and light $s$ or $d$ quarks (for $P = K$ and $\pi$, respectively) forming charmed hadron vector states. They represent a pole, situated at the $D_q^* = D_s^{*+}$ or $D^{*+}$ mass squared and a cut, along the positive real axis, starting at threshold ($t_+ = (m_D + m_P)^2$) for $D^0P$ production.

This cut $t$-plane can be mapped onto the open unit disk with center at $t = t_0$ using the variable:

$$z(t, t_0) = \frac{\sqrt{t_+ - t} - \sqrt{t_+ - t_0}}{\sqrt{t_+ - t} + \sqrt{t_+ - t_0}}. \tag{19.1.56}$$

In this variable, the physical region for the semileptonic decay ($0 < t < t_- = q^2_{\rm max} = (m_D - m_P)^2$) corresponds to a real segment extending between $\pm z_{\rm max} = \pm 0.051$ for $D \to K$ and $\pm 0.17$ for $D \to \pi$. This value of $z_{\rm max}$ is obtained for $t_0 = t_+ \left(1 - \sqrt{1 - t_-/t_+}\right)$, where $t_+ = (m_D + m_P)^2$. The $z$ expansion of $f_+^P$ is thus expected to converge quickly. The most general parameterization (Hill, 2006), consistent with constraints from QCD,

$$f_+(t) = \frac{1}{P(t)\Phi(t, t_0)} \sum_{k=0}^{\infty} a_k(t_0)\ z^k(t, t_0), \tag{19.1.57}$$

is based on earlier considerations by (Boyd and Savage, 1997) and other references quoted therein. For $D^0 \to K^- e^+ \nu_e$, the function $P(t) = z(t, m^2_{D_s^*})$ has a zero at the $D_s^*$ pole mass and $|P| = 1$ along the unit circle. For $D^0 \to \pi^- e^+ \nu_e$, as the $D^{*+}$ mass is higher than $(m_{D^0} + m_{\pi^+})$, $P(t) = 1$. The expression for $\Phi$ can be found in (Boyd and Savage, 1997).

The choice of $P$ and $\Phi$ is such that:

$$\sum_{k=0}^{\infty} a_k^2(t_0) \leq 1. \tag{19.1.58}$$

Having measured the first coefficients of this expansion, Eq. (19.1.58) can constrain the others. This constraint is used in $B$ meson semileptonic decays and found to be quite effective. For charm, this approach is not really justified because the charm quark mass is rather light rendering the perturbative QCD determination of the function $\Phi$ questionable and because the $z$ physical range is quite limited. Numerically it appears that the first measured coefficients are quite small and no useful constraint can be placed on higher order coefficients ($a_k$ with $k \geq 3$).

It seems preferable to consider phenomenological models to describe the $q^2$ variation of $f_+^P(q^2)$ because they have a simple physical interpretation. In the simple pole model,

$$f_+(q^2)_{\rm simple\ pole} = \frac{f_+(0)}{1 - \frac{q^2}{m^2_{\rm pole}}}. \tag{19.1.59}$$

The pole mass value $m_{\rm pole} = m_{D_s^*}$ ($m_{D^{*+}}$) respectively for $D^0 \to K^-(\pi^-)\ell^+\nu_\ell$. When fitting data, $m_{\rm pole}$ is taken as a free parameter and its value can be compared with these expectations.

The $q^2$ dependence of the form factor is nonperturbative and the only first-principles approaches are lattice QCD calculations. Nevertheless, a vector-dominance assumption would lead to a single pole form with the pole located at $q^2 = m^2_{D_s^*}$ for the $D \to K$ vector form factor. However, this does not take into account contributions from other states, and so also double-pole structures have been suggested. The Isgur Scora Grinstein Wise (ISGW2) quark model (Isgur, Scora, Grinstein, and Wise, 1989; Scora and Isgur, 1995) belongs to this category:

$$f_+(q^2)_{\rm ISGW2} = \frac{f_+(q^2_{\rm max})}{\left(1 + \alpha_I (q^2_{\rm max} - q^2)\right)^2},\ \alpha_I = \frac{1}{12} r^2. \tag{19.1.60}$$

This expression is normalized at $q^2 = q^2_{\rm max} = (m_{D^0} - m_P)^2$. For $D \to K$, the predicted values of the parameters are $f_+(q^2_{\rm max}) = 1.23$ and $r = 1.12\ \ {\rm GeV}^{-1}$.

The modified pole model of (Becirevic and Kaidalov, 2000) (BK model) has also two poles. The first pole is at the vector meson mass and the second accounts for higher mass vector states:

$$f_+(q^2)_{\rm BK} = \frac{f_+(0)}{\left(1 - \frac{q^2}{m^2_{D_s^*}}\right)\left(1 - \alpha_{\rm BK}\frac{q^2}{m^2_{D_s^*}}\right)}. \tag{19.1.61}$$

The authors predict $\alpha_{\rm BK} \sim 0.5$.

All proposed ansatze fit well the measured distributions. However, apart for the BK model (which doesn't have precisely determined expectations), predicted values of the parameters differ markedly from the measurements as indicated in Table 19.1.15. Effects from hadronic singularities, in addition to the pole at the $D^*_{s(d)}$ mass, are thus measurable. The simple pole and ISGW2 models, with nominal values of the parameters, are excluded.

QCD based form-factor calculations can also be performed in the framework of QCD sum rules. In particular, light-cone QCD sum rules are well suited for the calculation of form factors especially for heavy-to-light transitions. The input into these sum rules is the light-cone distribution of the quarks in the pion or kaon; a more detailed description of this can be found in Section 17.1 where this is applied to the $B \to \pi$ form factor. A recent discussion of the form factors for semileptonic charm decays can be found in (Khodjamirian, Klein, Mannel, and Offen, 2009). It turns out that the QCD sum rule calculation yields a form factor which is compatible with lattice determinations as well as with the BK parameterization; it can be used to perform an independent and competitive extraction of $V_{cs}$ and $V_{cd}$ from semileptonic charm decays.

**Table 19.1.15.** Values of fitted parameters for different models of the hadronic form factor $q^2$ dependence in $D \to K\ell^+\nu_\ell$ decays. The different labels given in the first column correspond to the following references: Belle (Widhalm, 2006), *BaBar* (Aubert, 2007ab), and CLEO-c (Besson et al., 2009). The simple pole and ISGW2 models, with nominal values of the parameters, are excluded.

| Source | Simple pole | BK | ISGW2 |
|---|---|---|---|
| | $m_{\rm pole}$ (GeV/$c^2$) | $\alpha_{\rm BK}$ | $\alpha_I$ (GeV$^{-2}$) |
| Belle(2006) | 1.82(4)(3) | 0.52(8)(6) | 0.51(3)(3) |
| *BaBar*(2007) | 1.884(12)(15) | 0.38(2)(3) | 0.226(5)(6) |
| CLEOc(2009) | 1.93(2)(1) | 0.30(3)(1) | 0.211(5)(3) |
| Expectation | 2.112 | $\sim 0.5$ | 0.104 |

To compare absolute determinations of the form factors, obtained by several experiments and from LQCD, results are evaluated at $q^2 = 0$ in Table 19.1.16. Measurements at $B$ factories are five times more accurate than previous determinations. Corresponding results from CLEO-c were published later with similar accuracy and central values. The measurement accuracy on $f_+^{K/\pi}(0)$ combined results reaches 1 and 3% respectively. They are in agreement with LQCD recent expectations (Na, Davies, Follana, Lepage, and Shigemitsu, 2010; Na et al., 2011). In spite of spectacular progress from lattice QCD, present computations are still a factor three (two) less accurate than actual measurements for $D \to K$ ($D \to \pi$). In future, detailed comparisons of the measured and expected variations of the hadronic form factors versus $q^2$ are expected. Combining experimental and theoretical uncertainties, one can say also that, at present, the values of $|V_{cs}|$ and $|V_{cd}|$ obtained from the measurements of $D_{l3}$ decays and using the values of the corresponding hadronic form factors from lattice QCD, agree with expectations from unitarity with a relative uncertainty of 2.7% and 5% respectively.

#### 19.1.5.6 The $D^+ \to K^-\pi^+e^+\nu_e$ decay

Detailed study of the $D^+ \to K^-\pi^+e^+\nu_e$ decay channel (del Amo Sanchez, 2011a) is of interest for three main reasons:

- it allows measurements of the different $K\pi$ resonant and non-resonant amplitudes that contribute to this decay. In this respect, *BABAR* has measured the $S$-wave contribution and searched for radially excited $P$-wave and for $D$-wave components.
- high statistics allows accurate measurements of the properties of the $\overline{K}^*(892)^0$ meson, the main contribution to the decay. Both resonance parameters and hadronic transition form factors are precisely measured. The latter can be compared with hadronic model expectations and lattice QCD computations.
- variation of the $K\pi$ $S$-wave phase versus the $K\pi$ mass can be determined, and compared with other experimental determinations.

The approach used to reconstruct $D^+$ mesons decaying into $K^-\pi^+e^+\nu_e$ is similar to that already explained in Section 19.1.5.2. Charged and neutral particles are boosted to the center-of-mass system and the event thrust axis is determined. A plane perpendicular to this axis is used to define two hemispheres.

A candidate $D^+$ is represented by a positron, a charged kaon, and a charged pion present in the same hemisphere. A vertex is formed using these three tracks, and events with the corresponding $\chi^2$ probability larger than $10^{-7}$ are kept. The value of this probability is used with other informations combined in two Fisher discriminant variables to reject, respectively, $\Upsilon(4S)$ decays and continuum background events.

To estimate the neutrino momentum, the $(K^-\pi^+e^+\nu_e)$ system is constrained to the $D^+$ mass. In this fit, estimates of the $D^+$ direction and of the neutrino energy are included from measurements obtained from all tracks registered in the event. The $D^+$ direction estimate is taken as the direction of the vector opposite to the momentum sum of all reconstructed particles but the kaon, the pion, and the positron. The neutrino energy is evaluated by subtracting from the hemisphere energy the energy of reconstructed particles contained in that hemisphere.

Analyzing an integrated luminosity of 347 $\text{fb}^{-1}$, about $244 \times 10^3$ signal events are selected with a ratio Signal / Background= 2.3.

As there are four particles in the final state, the differential decay rate has five degrees of freedom that can be

**Table 19.1.16.** Summary of hadronic form factor measurements at $q^2 = 0$. Measurements from CLEOc(2009) supercede those from CLEOc(2008). Results at $B$ factories have an accuracy similar to CLEOc and results are compatible. All published results have been corrected for normalization branching fractions, lifetimes, and assumed values for CKM matrix elements if needed, using values quoted in (Nakamura et al., 2010). The different labels quoted in the first column correspond to the following references : E691 (Anjos et al., 1989), CLEO (Crawford et al., 1991), CLEOII (Bean et al., 1993), E687(1995) (Frabetti et al., 1995), E687(1996) (Frabetti et al., 1996a), BES II (Ablikim et al., 2004a), CLEO III (Huang et al., 2005), Belle (Widhalm, 2006), *BABAR* (Aubert, 2007ab), CLEO-c (2008) (Dobbs et al., 2008), CLEO-c (2009) (Besson et al., 2009), and HPQCD (Na, Davies, Follana, Lepage, and Shigemitsu, 2010; Na et al., 2011). Combining measurements from Belle , *BABAR* and CLEO-c, and assuming that uncertainties are uncorrelated, the corresponding averaged values are obtained. LQCD results obtained by the HPQCD collaboration are given in the last line. They agree with the measurements.

| Experiment (date) | $f_+^K(0)$ | $f_+^\pi(0)$ |
|---|---|---|
| E691(1989) | 0.70(5)(5) | |
| CLEO(1991) | 0.78(3)(3) | |
| CLEOII(1993) | 0.77(1)(4) | 0.72(13)(5) |
| E687(1995) | 0.70(3)(3) | |
| E687(1996) | | 0.71(7)(2) |
| BESII(2004) | 0.80(4)(3) | |
| CLEOIII(2005) | | 0.62(6)(4) |
| Belle(2006) | 0.695(7)(22) | 0.624(20)(30) |
| *BABAR*(2007) | 0.734(7)(7) | |
| CLEOc(2008) | 0.763(7)(6) | 0.629(22)(7)(3) |
| CLEOc(2009) | 0.739(7)(5) | 0.666(19)(4)(3) |
| Our avg.(2012) | 0.734(6) | 0.657(17)(3) |
| HPQCD(2010 − 11) | 0.747(19) | 0.666(29) |

expressed in the following variables (Cabibbo and Maksymowicz, 1965; Pais and Treiman, 1968):

- $m^2$, the mass squared of the $K\pi$ system;
- $q^2$, the mass squared of the $e^+\nu_e$ system;
- $\cos(\theta_K)$, where $\theta_K$ is the angle between the $K$ three-momentum in the $K\pi$ rest frame and the line of flight of the $K\pi$ in the $D$ rest frame;
- $\cos(\theta_e)$, where $\theta_e$ is the angle between the charged lepton three-momentum in the $e\nu_e$ rest frame and the line of flight of the $e\nu_e$ in the $D$ rest frame;
- $\chi$, the angle between the normals to the planes defined in the $D$ rest frame by the $K\pi$ pair and the $e\nu_e$ pair. $\chi$ is defined between $-\pi$ and $+\pi$.

For the differential decay partial width, we use the formalism given in (Lee, Lu, and Wise, 1992). Apart for the $S$-wave for which there is one hadronic form factor, each higher spin component has three form factors associated. Using the conclusions of the $D^0 \to K^-e^+\nu_e$ analysis, where we find that all usual ansatze give a good parameterization of data, and noting that the $q^2$ range is even more limited, we use the simple pole model to describe the $q^2$ dependence of these form factors. As an example, for the $\overline{K}^*(892)^0$ meson, we use:

$$\begin{aligned} V(q^2) &= \frac{V(0)}{1 - \frac{q^2}{m_V^2}} , \\ A_1(q^2) &= \frac{A_1(0)}{1 - \frac{q^2}{m_A^2}} , \\ A_2(q^2) &= \frac{A_2(0)}{1 - \frac{q^2}{m_A^2}} . \end{aligned} \tag{19.1.62}$$

Hadronic resonances are parameterized using relativistic Breit-Wigner distributions with mass dependent widths and a Blatt-Weisskopf damping factor. For the $S$-wave component we fit the amplitude and measure the phase in several mass intervals.

A binned distribution of data events is analyzed. The expected number of events in each bin depends on signal and background estimates and the former is a function of the values of the fitted parameters. We perform a minimization of a negative log-likelihood distribution. This distribution has two parts. One corresponds to the comparison between measured and expected number of events in bins which span the five dimensional space of the differential decay rate. The other part is used to measure the fraction of background events and corresponds to the distribution of the values of one of the Fisher discriminant variables. There are 2800 bins in total.

#### 19.1.5.7 Measured components

The $K^-\pi^+$ final state is dominated by the $\overline{K}^*(892)^0$ meson (94.1%) and the $S$-wave component (5.8%). There is marginal evidence for the first radial excitation of the $P$-wave (0.3%) and a stringent limit is placed on a $D$-wave contribution. The small $\overline{K}^*(1410)^0$ contribution agrees with the naïve expectation based on corresponding measurements in $\tau$ decays and its phase relative to the $\overline{K}^*(892)^0$ is compatible with zero. Branching fractions for these components are obtained by reference to the $D^+ \to K^-\pi^+\pi^+$ channel which is measured using a similar analysis, and are reported in Table 19.1.17.

**Table 19.1.17.** Comparison between *BaBar* (del Amo Sanchez, 2011a) measurements and present world averages (Nakamura et al., 2010). Values for $\mathcal{B}(D^+ \to \overline{K}^*(1410)^0/\overline{K}_2^*(1430)^0e^+\nu_e)$ are corrected for their respective branching fractions into $K^-\pi^+$.

| Branching fraction | *BaBar* | PDG 2010 |
|---|---|---|
| $\mathcal{B}(D^+ \to K^-\pi^+e^+\nu_e)(\%)$ | 4.00(3)(4)(9) | $4.1 \pm 0.6$ |
| $\mathcal{B}(D^+ \to K^-\pi^+e^+\nu_e)_{\overline{K}^*(892)^0}(\%)$ | 3.77(4)(5)(9) | $3.68 \pm 0.21$ |
| $\mathcal{B}(D^+ \to K^-\pi^+e^+\nu_e)_{S-wave}(\%)$ | 0.232(7)(7)(5) | $0.21 \pm 0.06$ |
| $\mathcal{B}(D^+ \to \overline{K}^*(1410)^0e^+\nu_e)(\%)$ | < 0.6 at 90% C.L. | |
| $\mathcal{B}(D^+ \to \overline{K}_2^*(1430)^0e^+\nu_e)(\%)$ | < 0.05 at 90% C.L. | |

The $S$-wave contribution is dominated by events with a mass below the $\overline{K}_0^*(1430)$ resonance pole and, for the first time, the $S$-wave phase is measured at several values of the $K^-\pi^+$ mass. The $K\pi$ hadronic system can have two isospin components ($I = 1/2,\ 3/2$) however, in charm semileptonic decays, the $c \to s$ transition corresponds to $\Delta I = 0$ and only the $I = 1/2$ component is produced. The $K\pi$ scattering $S$-wave, with isospin $I = 1/2$, remains elastic up to the $K\eta$ threshold, but since the coupling to this channel is weak, it is considered in practice to be elastic up to the $K\eta'$ threshold. In this elastic regime, the Watson theorem (Watson, 1954) implies that, phases measured in $K\pi$ elastic scattering and in a decay channel in which the $K\pi$ system has no strong interaction with other hadrons are equal modulo $\pi$ radians for the same values of isospin and angular momentum. The ambiguity is solved by determining the sign of the $S$-wave amplitude from data. This theorem does not provide any constraint on the corresponding amplitude moduli. In particular, it is not legitimate (though nonetheless frequently done) to assume that the $S$-wave amplitude in a decay is proportional to the elastic amplitude. In Figure 19.1.38 the $K^-\pi^+$ $S$-wave phase with $I = 1/2$ measured in the elastic channel (Aston et al., 1988; Estabrooks et al., 1978) and in $D^+$ semileptonic decays, are compared. We measure that the two amplitudes differ by a negative sign and that phases are compatible within uncertainties, in agreement with the Watson theorem. Similar analyses of the $D^+ \to K^-\pi^+\pi^+$ Dalitz plot (Aitala et al., 2006; Bonvicini et al., 2008; Link et al., 2007, 2009) measure a significant difference between the variation of the $K\pi$ $S$-wave and of the elastic phases versus the $K\pi$ mass. This difference has thus to be attributed to final state interactions with the third hadron.

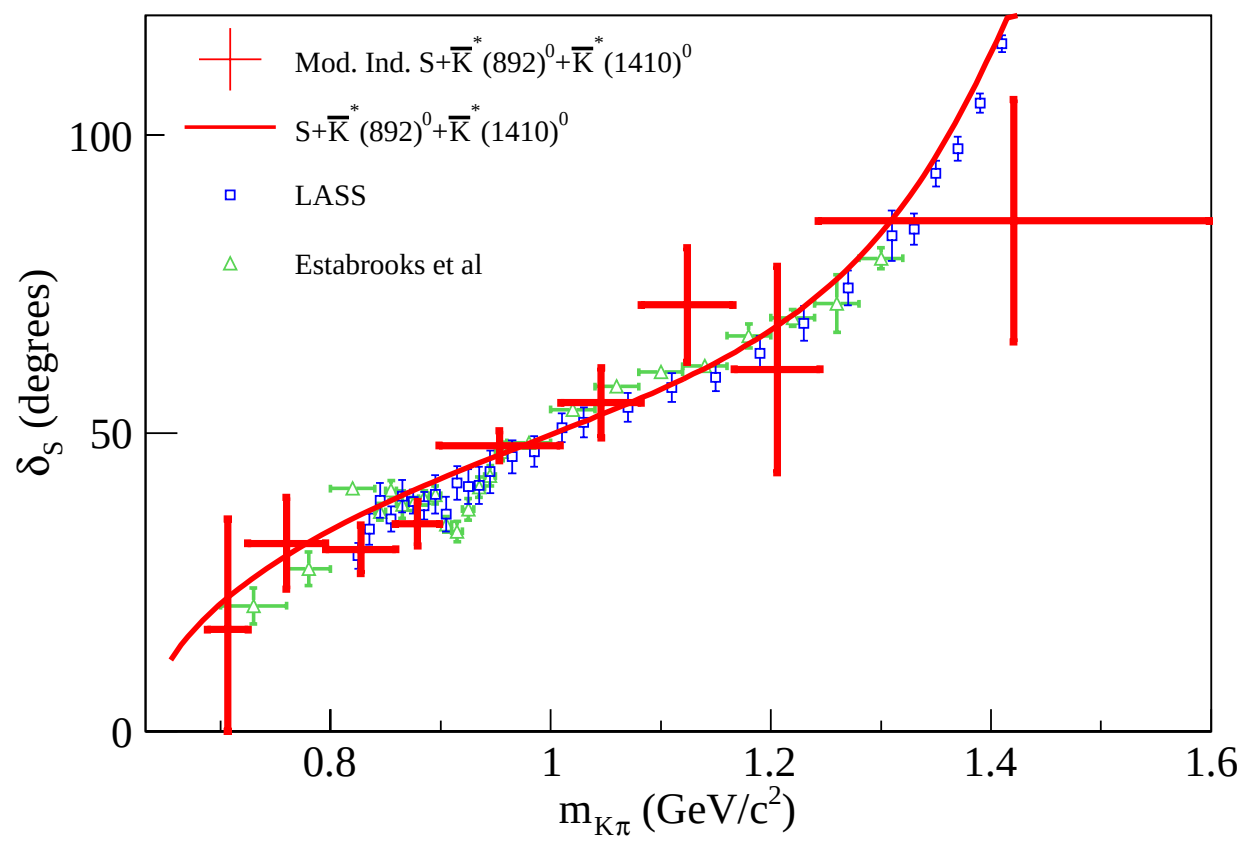

**Figure 19.1.38.** From (del Amo Sanchez, 2011a). Points (red crosses) give the $S$-wave phase variation assuming a signal containing $S$-wave, $\overline{K}^*(892)^0$ and $\overline{K}^*(1410)^0$ components. The $S$-wave phase is assumed to be constant within each considered mass interval. Error bars include systematic uncertainties. The full line corresponds to the fitted parameterized $S$-wave phase variation expected from elastic scattering experiments. The phase variation measured in $K\pi$ scattering by (Estabrooks et al., 1978) (triangles) and LASS (Aston et al., 1988) (squares), after correcting for the $I = 3/2$ isospin component, are given.

#### 19.1.5.8 Detailed measurements of the $\overline{K}^*(892)^0$

Using a model for signal which includes $S$-wave, $\overline{K}^*(892)^0$ and $\overline{K}^*(1410)^0$ contributions, and the simple pole ansatz for the $q^2$ variation of hadronic form factors, parameters of the $\overline{K}^*(892)^0$ component are obtained from a fit to the five-dimensional decay distribution. These parameters, listed in Table 19.1.18, define the $\overline{K}^*(892)^0$ lineshape and the relative contributions of the different form factors: $r_2 = A_2(0)/A_1(0)$ and $r_V = V(0)/A_1(0)$. The analysis is not sensitive to the $q^2$ dependence of the vector form factor $(V(q^2))$ but we measure, for the first time the effective pole mass of the axial vector form factors $(A_{1,2}(q^2))$ and find a value compatible with expectations $m_{\rm pole} \sim m_{D_{s1}}$.

The branching fraction of $D^+ \to \overline{K}^*(892)^0 e^+ \nu_e$ is determined by normalizing the signal yield of $D^+ \to K^-\pi^+ e^+ \nu_e$ to the reconstructed yield of $D^+ \to K^-\pi^+\pi^+$ (with the branching fraction as measured in (Dobbs et al., 2007a)), after subtracting the $S$-wave and $\overline{K}^*(1410)^0$ contributions, and after correcting for the efficiency difference. The decay rate depends only on the value of $A_1(0)$. In the zero-width approximation for the $\overline{K}^*(892)^0$ resonance the measured decay rate corresponds to:

$$A_1(0) = 0.6200 \pm 0.0056 \pm 0.0065 \pm 0.0071. \quad (19.1.63)$$

The last uncertainty includes the uncertainty of $Br(D^+ \to K^-\pi^+\pi^+)$ as well as the systematic uncertainty due to external parameters used in the extraction of $A_1(0)$ (values of $|V_{cs}|$ and $\tau_D^+$).

**Table 19.1.18.** Measured properties (del Amo Sanchez, 2011a) of the $D^+ \to \overline{K}^*(892)^0 e^+ \nu_e$ decay channel and of the $\overline{K}^*(892)^0$ resonance are compared with corresponding world averages (Nakamura et al., 2010). $r_{BW}$ is the Blatt-Weisskopf damping parameter. $m_A$ is the pole mass of the axial vector form factors.

| Measured quantity | BABAR | PDG 2010 |
|---|---|---|
| $m_{K^*(892)^0}$ (MeV/$c^2$) | $895.4 \pm 0.2 \pm 0.2$ | $895.94 \pm 0.22$ |
| $\Gamma^0_{K^*(892)^0}$ (MeV/$c^2$) | $46.5 \pm 0.3 \pm 0.2$ | $48.7 \pm 0.8$ |
| $r_{BW}$ (GeV/$c$)$^{-1}$ | $2.1 \pm 0.5 \pm 0.5$ | $2.72 \pm 0.55$ |
| $r_V$ | $1.463 \pm 0.017 \pm 0.031$ | $1.62 \pm 0.08$ |
| $r_2$ | $0.801 \pm 0.020 \pm 0.020$ | $0.83 \pm 0.05$ |
| $m_A$(GeV/$c^2$) | $2.63 \pm 0.10 \pm 0.13$ | no result |

In the present analysis, the measured distribution in five dimensions is not unfolded and the experimental resolution expected from the simulation is controlled using data. Because angular distributions are completely determined by the kinematics for each helicity component, measurements are giving the decay rate variation versus the remaining two variables, $q^2$ and $m^2$. The $q^2$ dependence of hadronic form factors is smooth and can be parameterized in a simple way as explained in Section 19.1.5.2. For the mass measurement, the experimental resolution is high. It results that the distribution, for which statistical and systematic error matrices on the fitted parameters are provided, can be compared with different theoretical expressions when they will be available.

#### 19.1.5.9 Measured and expected values of the hadronic form factors

The normalization and specific $q^2$ behavior of the hadronic form factors are obtained in some specific limits for which intermediate scales are identified. When a heavy hadron decays semileptonically into another heavy hadron a new symmetry emerges which gives constraints on the behaviour of the form factors. This symmetry is exact in the limit of infinite quark mass values, which can be formulated as an effective field theory, the Heavy Quark Effective Theory (HQET) (Isgur and Wise, 1989; Shifman and Voloshin, 1988). For finite values of quark masses, and depending on the form factors, corrections start at order $1/m_Q$ or $1/m_Q^2$. These properties are used to determine the value of $V_{cb}$ from measurements of $\overline{B} \to D^* \ell \overline{\nu}_\ell$ as corrections are expected to be at order $1/m_Q^2$ for $q^2 \sim q^2_{max}$. For charm, the value of the charm-quark mass is of the order of $2-3$ times $\Lambda_{\rm QCD}$ and the strange quark cannot be described as a heavy quark. Considerations based on HQET are thus expected not to hold for charm semileptonic decays; still they may serve as a starting point, and hence we consider it of interest to review the constraints on form factors, implied by HQET, and to indicate how they are violated in charm to eventually get insights for $B$ decays. For infinite

**Table 19.1.19.** Comparison between measured and expected values of the form factors in $D^+ \to \overline{K}^*(892)^0 e^+\nu_e$ decays, evaluated at $q^2 = 0$. Only measurements from *BABAR* (del Amo Sanchez, 2011a) and FOCUS (Link et al., 2002) which are corrected for the $S$-wave contribution are listed. Results from PDG 2010 (Nakamura et al., 2010) include those from FOCUS and all previous measurements. Expected values are coming from the HM$\chi$T model (Fajfer and Kamenik, 2005) and LQCD (Gill, 2002) computations.

| | FOCUS 2002 | BABAR | PDG 2010 | HM$\chi$T | LQCD |
|---|---|---|---|---|---|
| $r_V$ | 1.504(57)(39) | 1.493(14)(21) | 1.62(8) | 1.6 | 1.23(10) |
| $r_2$ | 0.875(49)(64) | 0.775(11)(11) | 0.83(5) | 0.50 | 0.94(12) |
| $A_1(0)$ | | 0.6200(56)(65)(71) | | 0.62 | 0.65(3) |
| $A_2(0)$ | | 0.480(8)(10) | | 0.31 | 0.61(7) |
| $V(0)$ | | 0.926(12)(19) | | 0.99 | 0.80(5) |

quark masses one has the following equalities (Neubert, 1994b):

$$f_+(q^2) = V(q^2) = A_2(q^2) = \frac{A_1(q^2)}{1 - \frac{q^2}{(m_H+m_V)^2}} = R^{-1}\xi(q^2) \tag{19.1.64}$$

The Isgur-Wise function $\xi(q^2)$ satisfies $\xi(q^2_{max}) = 1$. The parameter $R = 2\sqrt{m_H\, m_{P(V)}}/(m_H+m_{P(V)})$ is equal to $\sim 0.9(0.8)$ for $B(D)$ semileptonic favored decays. $M_H$ is the heavy hadron mass whereas $m_{P(V)}$ are the pseudo scalar and vector meson masses. In this limit, the ratio of form factors are equal for $q^2 = q^2_{max}$: $r_V(q^2_{max}) = r_2(q^2_{max}) = R^{-2}$. For other values of $q^2$ they read:

$$r_{V(2)}(q^2) = \frac{1}{1 - \frac{q^2}{(m_H+m_V)^2}} R_{1(2)}(q^2) \tag{19.1.65}$$

with $R_{1(2)}(q^2_{max}) = 1$. Corrections to these expressions correspond to expansions in $1/m$, where $m$ can take the values of the masses of the two quarks involved in the weak transition, and in the strong coupling constant, from perturbative QCD. It is an unambiguous prediction of HQET that $R_1(q^2) > 1$ as both the QCD and $1/m$ corrections are positive. For $R_2(q^2)$, QCD corrections are small and $1/m$ corrections seem to decrease the value of the ratio. HQET relates the form factors in semileptonic decays to pseudoscalar and vector particles as expressed in Equation (19.1.64). According to (Amundson and Rosner, 1993), QCD corrections alone cannot explain the ratio $\mathcal{B}(D \to \overline{K}^* e^+\nu_e)/\mathcal{B}(D \to K^- e^+\nu_e) = 0.62 \pm 0.02$ as they have rather similar effects on all form factors.

In previous considerations it is assumed that the charm and also the strange quark behave as a heavy quark in $c \to s\ell^+\nu_\ell$ decays. It is also possible to relate form factors in $B$ and $D$ semileptonic decays to light hadrons (Isgur and Wise, 1990a).

Results on absolute values and on ratios of hadronic form factors evaluated at $q^2 = 0$ are compared in Table 19.1.19. In this table, *BABAR* measurements are quoted for fixed values of the pole masses $m_V = 2.1$ GeV/$c^2$ and $m_V = 2.5$ GeV/$c^2$, and only experimental results corrected for the $S$-wave contribution are kept.

The HM$\chi$T model proposed by (Fajfer and Kamenik, 2005, 2006) generalizes the approach of Becirevic and Kaidalov and satisfies the scaling laws of obtained in the infinite mass limit as well as the known scaling at large energy of the outgoing kaon. Values from lattice QCD (Abada et al., 2003; Gill, 2002) are obtained using the quenched approximation. The relative accuracy of experimental measurements is typically 2% whereas it is around 10% for lattice QCD. Theoretical expectations agree with the general picture ($A_2(0) \le A_1(0) < V(0)$) but significant differences are observed.

#### 19.1.5.10 The $D_s^+ \to K^+K^-e^+\nu_e$ decay

Using 214 fb$^{-1}$ of data collected at the $\Upsilon(4S)$ resonance, *BABAR* measure the $D_s^+ \to K^+K^-e^+\nu_e$ channel decay characteristics, for events produced in the continuum (Aubert, 2008be), in a similar manner as described above for $D^+ \to K^-\pi^+e^+\nu_e$ (Section 19.1.5.6).

The $K^+K^-$ mass distribution is displayed in Figure 19.1.39(a). It contains about $25\times10^3$ signal events whereas previous experiments and CLEO-c have collected few hundred events only. The analysis focuses on the $\phi e^+ \nu_e$ final state in the $K^+K^-$ mass range between 1.01 and 1.03 GeV/$c^2$ for which accurate hadronic form factors normalization and $q^2$ dependence are measured. These results are given in Table 19.1.20 where they are compared with previous measurements and those obtained for $D^+ \to \overline{K}^*(892)^0 e^+ \nu_e$.

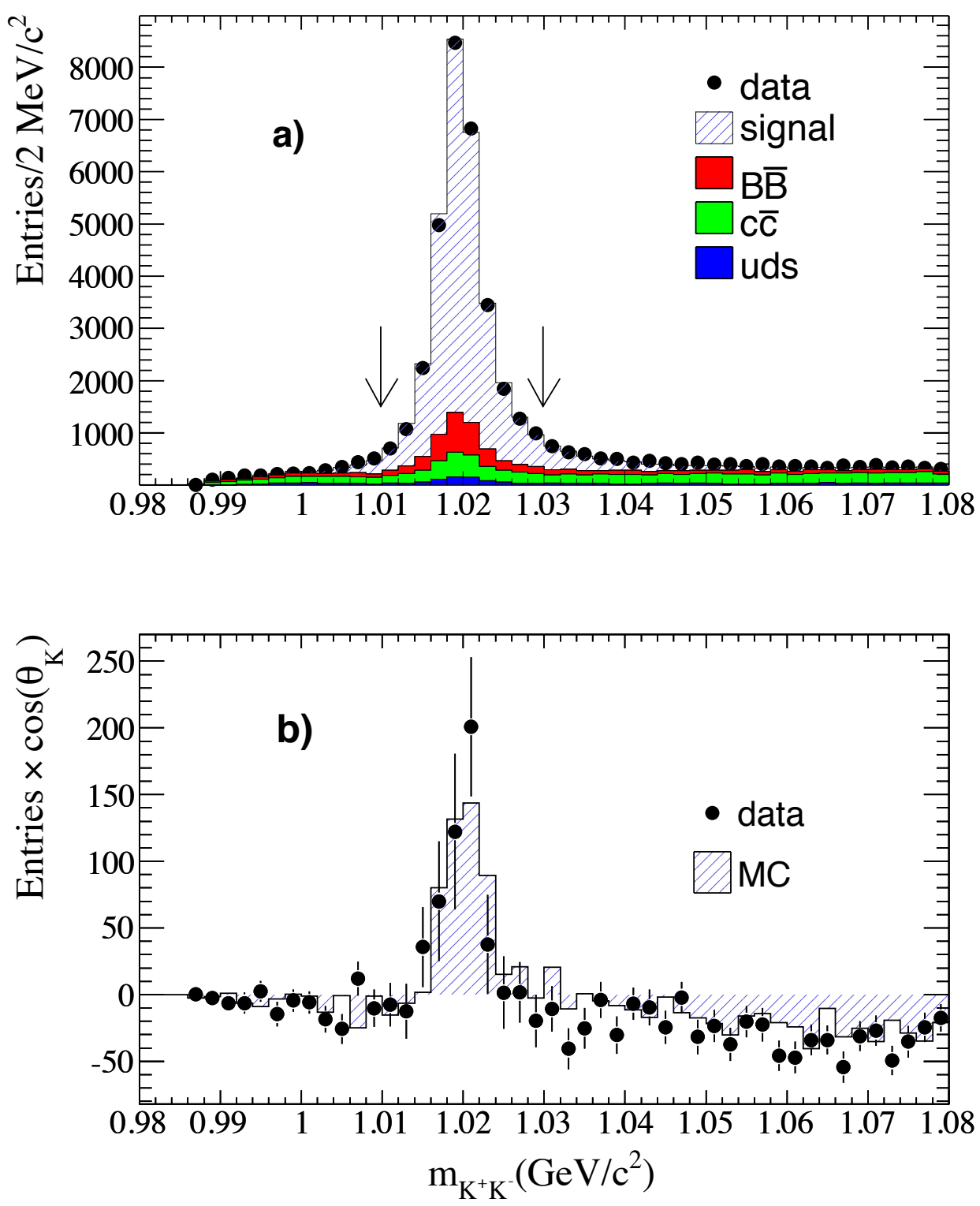

**Figure 19.1.39.** From (Aubert, 2008be). a) $K^+K^-$ invariant mass distribution from $D_s^+ \to K^+K^-e^+\nu_e$ data and simulated events. MC events have been normalized to the data luminosity according to the different cross sections. The arrows indicate the selected $K^+K^-$ mass interval. In b), each event is weighted by the measured value of $\cos\theta_K$. Negative entries are produced by the $c\bar{c}$ background asymmetry in $\cos\theta_K$.

**Table 19.1.20.** Comparison between measured and expected values of the form factors in $D_s^+ \to \phi e^+ \nu_e$ decays, evaluated at $q^2 = 0$. The different labels listed in the first column correspond to the following references: E653 (Kodama et al., 1993), E687 (Frabetti et al., 1994b), CLEO II (Avery et al., 1994a), E791 (Aitala et al., 1999a), FOCUS (Link et al., 2004b), BaBar (Aubert, 2008be), HM$\chi$T (Fajfer and Kamenik, 2005), UKQCD (Gill, 2002), and HPQCD (Donald, Davies, and Koponen, 2011). Results for $K^* e^+ \nu_e$ correspond to the BaBar measurement (del Amo Sanchez, 2011a).

| | $A_1(0)$ | $r_V$ | $r_2$ |
|---|---|---|---|
| E653(1993) | | $2.3^{+1.1}_{-0.9} \pm 0.4$ | $2.1^{+0.6}_{-0.5} \pm 0.2$ |
| E687(1994) | | $1.8 \pm 0.9 \pm 0.2$ | $1.1 \pm 0.8 \pm 0.1$ |
| CLEOII(1994) | | $0.9 \pm 0.6 \pm 0.3$ | $1.4 \pm 0.5 \pm 0.3$ |
| E791(1999) | | 2.27(35)(22) | 1.57(25)(19) |
| FOCUS(2004) | | 1.55(25)(15) | 0.71(20)(28) |
| BaBar (2008) | 0.607(11)(19)(18) | 1.807(46)(65) | 0.816(36)(30) |
| $K^* e^+ \nu_e$ | 0.6200(56)(65)(71) | 1.493(14)(21) | 0.775(11)(11) |
| HM$\chi$T | 0.61 | 1.80 | 0.52 |
| UKQCD(2001) | 0.63(2) | 1.35(7) | 0.98(8) |
| HPQCD(2011) | 0.603(20) | 1.52(12) | 0.62(12) |

The accuracy of experimental measurements compared to previous ones is improved by a factor five. The form factors are defined in an analogous way as for the $D^+ \to \overline{K}^*(892)^0 e^+ \nu_e$ decays in Section 19.1.5.6. Also the angular variables are analogous, with a difference that $\theta_K$ is the angle between the $K^+$ meson direction in the $K^+K^-$ rest frame and the $K^+K^-$ direction in the $D_s^+$ rest frame. The comparison between the values of the form factors measured in the two channels $D^+ \to \overline{K}^*(892)^0 e^+ \nu_e$ and $D_s^+ \to \phi e^+ \nu_e$ shows that they are compatible within uncertainties, apart for the parameter $V(0)$. This indicates that $SU(3)$ violations are small. Recent unquenched LQCD computations (Donald, Davies, and Koponen, 2011) are in better agreement with data than previous unquenched results (Gill, 2002).

The $\phi$ resonance is dominant in this $K^+K^-$ mass region although a small $S$-wave component is measured, for the first time, through its interference with the $\phi$ (see Figure 19.1.39(b). After integration over the $\theta_e$ and $\chi$ angular variables, the differential decay rate is proportional to:

$$|F_1|^2 + \sin^2\theta_K \left(|F_2|^2 + |F_3|^2\right). \qquad (19.1.66)$$

The form factors $F_i$ are function of $q^2$, $m^2$ and, $\theta_K$. Considering contributions from $S-$ and $P-$wave of the $K^+K^-$ system, only the form factor $F_1$ depends on $\theta_K$:

$$F_1 = F_{10} + \cos\theta_K F_{11}. \qquad (19.1.67)$$

Form factors $F_{10}$ and $F_{11}$ correspond respectively to $S-$ and $P$-wave contributions. Using this last expression, the first term in Eq. (19.1.66) generates an interference between the two wave components. The value of the parameter which quantifies this interference, $r_0 = (15.1 \pm 2.6 \pm 1.0)$ GeV$^{-1}$, is obtained with more than $5\sigma$ significance.[118] This is the only experimental result on the contribution from the $S$-wave component, precisely in the $K^+K^-$ mass region of the $\phi$ meson. $D_s^+$ decays can be used in $B_s^0 \to J/\psi\,\phi$ analyses determining the $CP$ violating phase $\beta_s$. They can provide for an evaluation of the $S$-wave contribution in the $\phi$ resonance region, as proposed in (Stone and Zhang, 2009). Using measurements from CLEO-c of $D_s^+ \to f_0 e^+ \nu_e,\ f_0 \to \pi^+\pi^-$, they conclude that there could be, in the $\phi$ mass region, about 10% contribution from the $f_0$ in the $K^+K^-$ final state. Such a large $S$-wave component can add uncertainties in the measurement of $\beta_s$. The direct measurement done in BABAR contradicts these expectations because they obtain, within a range of $\pm 10$ MeV/$c^2$ centered on the nominal $\phi$ meson mass, a relative contribution of the $S$-wave equal to $(0.22^{+0.12}_{-0.08} \pm 0.03)$%. The $S$-wave decay rate in $B_s^0 \to J/\psi\,K^+K^-$ channel, within the same $K^+K^-$ mass interval is thus expected to be below 1%, in agreement with the limit of 6% at 95%C.L. obtained by CDF (Aaltonen et al., 2012c). The recent measurement by LHCb ((4.2 ± 1.5 ± 1.8)% (Aaij et al., 2012h)) is less than $2\sigma$ away form this limit.

### 19.1.6 $D_s^+$ leptonic decays

#### 19.1.6.1 Introduction

The cleanest transitions where a partial decay width can show the manifestation of NP are $D_{(s)}^+ \to \ell^+ \nu_\ell$ ($\ell = \mu,\ \tau$). The decay partial widths depend on a single hadronic parameter, namely the decay constants $f_{D_{(s)}}$:

$$\Gamma(D_{(s)}^+ \to \ell^+ \nu_\ell) \qquad (19.1.68)$$
$$= \frac{G_F^2}{8\pi} f_{D_{(s)}}^2 m_\ell^2 m_{D_{(s)}} \left(1 - \frac{m_\ell^2}{m_{D_{(s)}}^2}\right)^2 \left|V_{cd(s)}\right|^2 ,$$

where $G_F$ is the Fermi coupling constant, $m_\ell$ and $m_{D_{(s)}}$ are the masses of the charged lepton and of the $D$ meson, respectively. $V_{cd(s)}$ is the corresponding CKM matrix element. As these decay channels are suppressed by helicity conservation, corresponding decay rates are proportional to the square of the lepton mass. Decays into electrons are not observable whereas decays into $\tau$ leptons are favored in spite of the reduced phase-space.

[118] The precise definition of the $r_0$ parameter arises from the parameterization of $F_{10}$ assuming $f_0$ production: $F_{10} = r_0[p_{KK} m_{D_s}/(1 - \frac{q^2}{m_A^2})][m_{f_0} g_\pi/(m_{f_0}^2 - m^2 - i m_{f_0}\Gamma_{f_0})]$, where $p_{KK}$ is the momentum of the $K^+K^-$ system in the $D_s^+$ rest frame.

$D^+$ leptonic decays are Cabibbo suppressed and difficult to measure at $B$ factories meanwhile $D_s^+$ leptonic decays are measured in the muon and tau channels.

#### 19.1.6.2 *BABAR* and Belle measurements

The approach pioneered by Belle (Widhalm, 2006) for semileptonic decays of $D^0$ mesons is used by the two $B$-factory experiments to measure absolute leptonic decay branching fractions of $D_s$ mesons. As an example, in the *BABAR* analysis, the $D_s$ meson production is tagged by considering events from the reaction $e^+e^- \to c\bar{c} \to H_c K X D_s^- \gamma$; $H_c$ is $D^0$, $D^+, D^*$ or $\Lambda_c$ exclusively reconstructed, $K$ a $K^+$ or $K_S^0$ and $X$ a system of at most three pions, including at most one $\pi^0$ with a total electric charge appropriate to ensure the neutrality of the overall final state. The number of produced $D_s$ mesons is obtained by considering the distribution in the recoil mass $M_{\rm recoil}(H_c K X \gamma)$ which has a peak at the $D_s$ mass for signal events. The hadron $H_c$ is reconstructed using 15 modes. In addition to the size of the mass window, several other properties of the $H_c$ candidate are used. The center-of-mass (CM) momentum of the $H_c$ must be at least 2.35 GeV/$c$ in order to remove $B$ meson backgrounds. Particle identification requirements are used on the tracks, a cut is applied on the probability of the $H_c$ vertex fit, and a minimum lab energy of $\pi^0$ photons is requested. Only $H_c K X \gamma$ candidates with a total charge, a charm, and a strange quark content consistent with recoiling from a $D_s^-$, are selected from which the signal yield is extracted. A kinematic fit to each $H_c K X$ candidate is performed and the $H_c$ mass is constrained to its nominal value. The 4-momentum of the signal $D_s^{*-}$ is extracted as the missing 4-momentum in the event. It is required that the $D_s^{*-}$ candidate mass be within 2.5$\sigma$ of the signal peak. A similar kinematic fit is performed with the signal $\gamma$ included and with the mass recoiling against the $H_c K X$ constrained to the nominal $D_s^{*-}$ mass in order to determine the $D_s^-$ 4-momentum. It is required that the $D_s^-$ momentum exceeds 3 GeV/$c$ and that its mass be greater than 1.82 GeV/$c^2$. Having reconstructed the inclusive $D_s^-$ sample one proceeds to the selection of $D_s^- \to \mu^- \overline{\nu}_\mu$ events within that sample. The $H_c K X \gamma$ mass range between 1.934 and 2.012 GeV/$c^2$ is used and it is required that there be exactly one more charged particle in the remainder of the event, and that it be identified as a $\mu^-$. In addition it is required that the extra neutral energy in the event, $E_{extra}$, be less than 1 GeV. $E_{extra}$ is defined as the total energy of clusters in the electromagnetic calorimeter with individual energy greater than 30 MeV and not overlapping with the $H_c K X \gamma$ candidate. Since the only missing particle in the event should be the neutrino, the distribution of $E_{extra}$ is expected to peak at zero for signal events. To extract the signal yield, the distribution of the mass squared of the system recoiling against the $H_c K X \gamma \mu^-$ combination, $M^2_{\rm recoil}(H_c K X \gamma \mu^-)$ is used. The method is slightly modified in the updated measurement by Belle (Zupanc, 2013b), as described in Section 19.1.2.2. The result is shown in Fig. 19.1.40. To find $D_s^- \to \tau^- \overline{\nu}_\tau$, $\tau^- \to \mu^-(e^-) \nu_\tau \overline{\nu}_{\mu(e)}$ decays, events associated with $D_s^- \to \mu^- \overline{\nu}_\mu$ decays are removed by requiring $m_r^2 > 0.5\,\mathrm{GeV}^2/c^4$. The $E_{extra}$ distribution is used to extract the yield of signal events as illustrated in Figure 19.1.41 for the *BABAR* (del Amo Sanchez, 2010g) and Belle (Zupanc, 2013b) analysis.

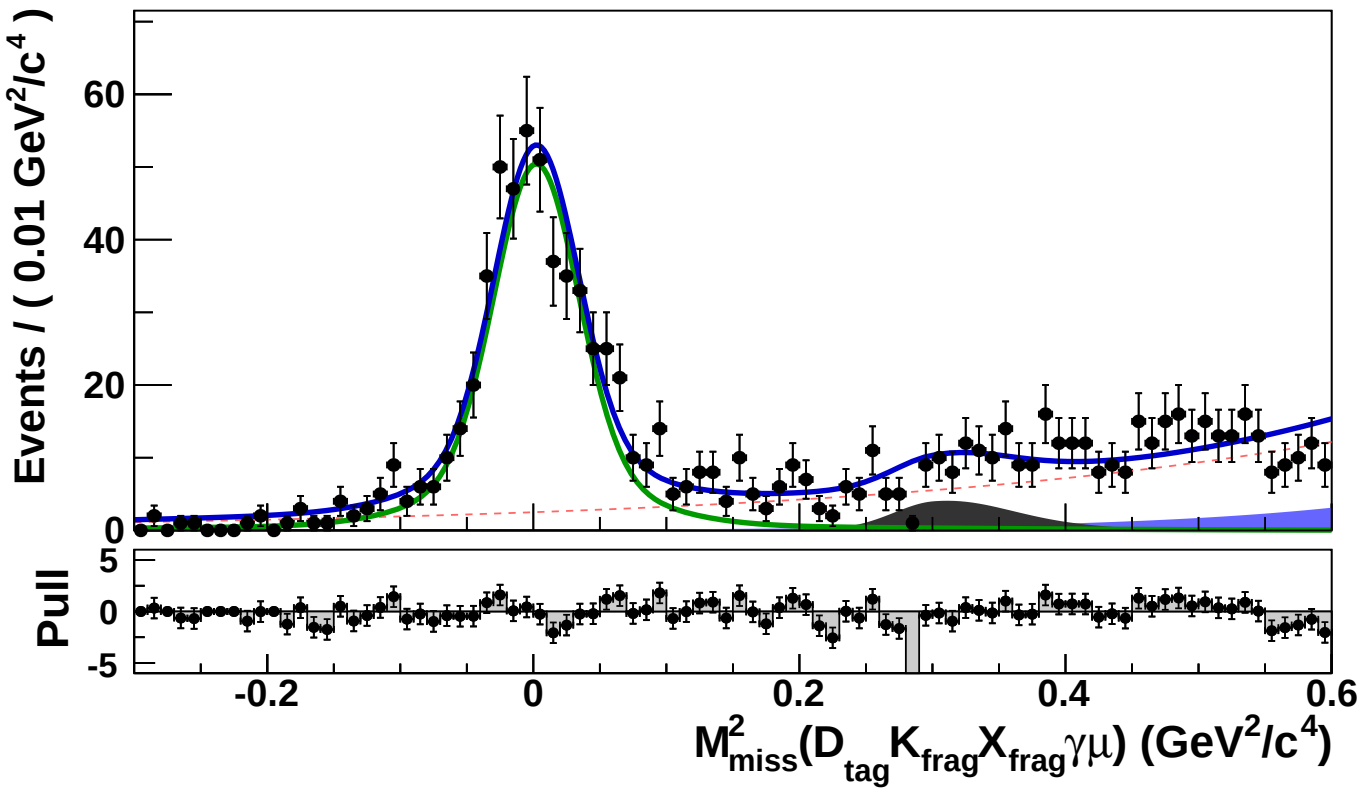

**Figure 19.1.40.** From (Zupanc, 2013b). $M^2_{\rm recoil}(H_c K X \gamma \mu^-)$ spectrum for $D_s^+ \to \mu^+ \nu_\mu$ candidates for the selected data (points with error bars). The solid green line shows the contribution of signal, the red dashed line the contribution of combinatorial background, while the contributions of $D_s^+ \to \tau^+ \nu_\tau$ and $D_s^+ \to \overline{K}^0 K^+$ or $\eta\pi^+$ are indicated by the full blue and dark gray histograms, respectively.

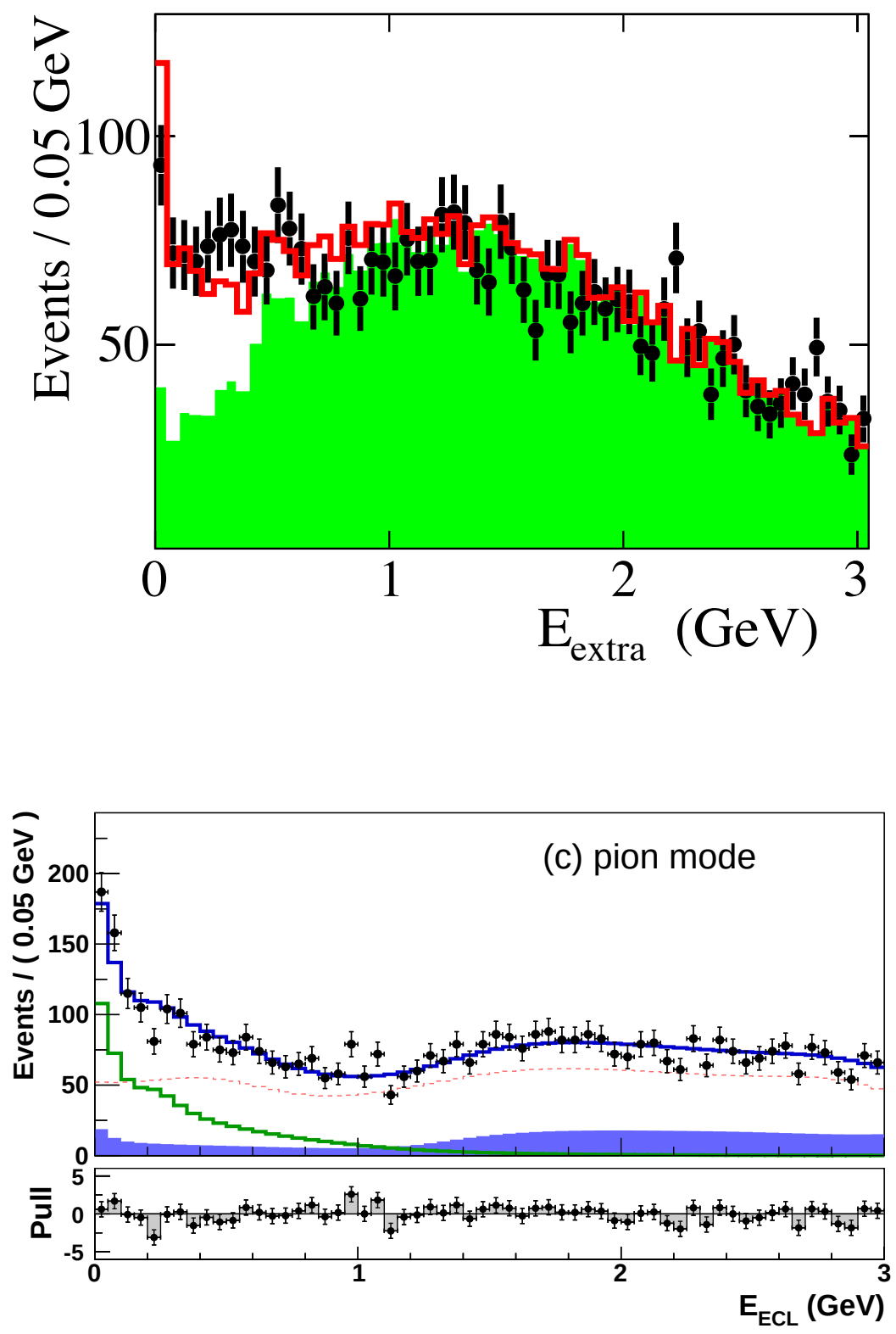

**Figure 19.1.41.** $E_{extra}$ ($E_{ECL}$ in case of Belle) distribution for $D_s^- \to \tau^- \overline{\nu}_\tau$, $\tau^- \to e^- \nu_\tau \overline{\nu}_e$ (del Amo Sanchez, 2010g) (top) and for $D_s^- \to \tau^- \overline{\nu}_\tau$, $\tau^- \to \pi^- \nu_\tau$ (Zupanc, 2013b). The points represent the data with statistical error bars. In the top plot the open histogram is from the fit, and the solid histogram is the background component from the fit. For the bottom plot the lines represent different components of the fit.

**Table 19.1.21.** Statistics of $D_s^{*-}$ tag and signal events measured by Belle (Zupanc, 2013b) and *BaBar* (del Amo Sanchez, 2010g).

| | Belle | *BaBar* |
|---|---|---|
| integrated lumi. | $913\,\mathrm{fb}^{-1}$ | $521\,\mathrm{fb}^{-1}$ |
| $D_s^{*-}$ tag | $(94.4 \pm 1.3 \pm 1.4) \times 10^3$ | $(67.2 \pm 1.5) \times 10^3$ |
| $D_s^- \to \mu^- \overline{\nu}_\mu$ | $492 \pm 26$ | $275 \pm 17$ |
| $D_s^- \to \tau^-_{e\nu\overline{\nu}} \overline{\nu}_\tau$ | $952 \pm 59$ | $408 \pm 42$ |
| $D_s^- \to \tau^-_{\mu\nu\overline{\nu}} \overline{\nu}_\tau$ | $758 \pm 48$ | $340 \pm 32$ |
| $D_s^- \to \tau^-_{\pi\nu} \overline{\nu}_\tau$ | $496 \pm 35$ | |

**Table 19.1.22.** Measured branching fractions of $D_s^+ \to \ell^+ \nu_\ell$ decays by Belle (Zupanc, 2013b) and *BaBar* (del Amo Sanchez, 2010g).

| | Belle | *BaBar* |
|---|---|---|
| $\mathcal{B}(D_s^+ \to \mu^+ \nu_\mu)[10^{-3}]$ | $5.31 \pm 0.28 \pm 0.20$ | $6.02 \pm 0.38 \pm 0.34$ |
| $\mathcal{B}(D_s^+ \to \tau^+ \nu_\tau)[10^{-2}]$ | $5.70 \pm 0.21^{+0.31}_{-0.30}$ | $5.00 \pm 0.35 \pm 0.49$ |

These analyses have a 100 times lower tagging efficiency than CLEO-c data collected at threshold. Yet this small efficiency is compensated by the much higher registered integrated luminosity. Measured statistics for the $D_s^{*-}$ tag and the signals are given in Table 19.1.21. From the obtained $D_s \to \ell \nu_\ell$ signal yields the branching fractions for individual modes are calculated and are given in Table 19.1.22.

### 19.1.6.3 Measured and Expected $f_{D_s}$ Values

Using Eq. (19.1.69), measured branching fractions of $D_s$ leptonic decays are used to extract the value of the decay constant $f_{D_s}^{expt.,SM}$. Results obtained by the different experiments are compared in Table 19.1.23. While measurements at CLEO-c are statistically limited, systematic uncertainties related to the background control dominate the methods developed at $B$ factories which have a total combined accuracy similar to CLEO-c. Having two results with different systematics and similar uncertainty is important to have confidence in the final result as it was already illustrated for the measurement of $f_K^+(0)$ in charm semileptonic decays. The averaged value of all measurements is $f_{D_s} = (257.5 \pm 4.6)\,\mathrm{MeV}$.

New physics can change Eq. (19.1.69) and the value of $f_{D_s}^{expt.,SM}$ extracted previously from data, in the Standard Model framework, may differ from QCD expectations. Several LQCD collaborations using an unquenched formulation of QCD have computed the value of $f_{D_s}^{QCD}$ (see Table 19.1.24). These are in agreement with the measured values within the uncertainties.

Among new physics models which can change the leptonic charm meson decay rate, there are the Two Higgs Doublet Model (2HDM) and the Minimal Supersymmetric Model. (MSSM) In these models it is expected that the leptonic decay partial width, given in Equation (19.1.69) is modified according to expressions given in (Akeroyd and Mahmoudi, 2009) where references to previous studies can be found. For the $D^+$, the correction is negligible, whereas for the $D_s^+$ the relative variation on $f_{D_s}$ is expected to be equal to:

$$k_s = \frac{\delta(f_{D_s})}{f_{D_s}} = -\frac{m_s}{m_c}\left(\frac{m_{D_s}\tan\beta}{m_H}\right)^2 \qquad (19.1.69)$$

for $\tan\beta > \sqrt{m_c/m_s} \simeq 3$. In this expression $m_s$ and $m_c$ are respectively the strange and the charm quark mass, $m_H$ is the charged Higgs boson mass and $\tan\beta$ is the ratio between the vacuum expectations of the two Higgs doublets.

Because $k_s$ is negative, it is expected that $f_{D_s}^{expt.,SM} < f_{D_s}^{QCD}$ in these models. First measurements from CLEO-c ($f_{D_s}^{expt.,SM} = 274(11)\,\mathrm{MeV}$) (Ecklund et al., 2008) and evaluations from the HPQCD collaboration ($f_{D_s}^{QCD} = 241(3)\,\mathrm{MeV}$) (Follana, Davies, Lepage, and Shigemitsu, 2008) were in the opposite direction. These circumstances provide rather stringent limits on the parameters ($m_H$ and $\tan\beta$) of the previous models. At present, measured and expected values of $f_{D_s}$ are more accurate and agree within 1.5 standard deviation. Derived limits on model parameters are thus less impressive. To have evidence ($3\sigma$) for

**Table 19.1.23.** Measured averaged values of the $D_s$ decay constant (averaging $\mu\overline{\nu}$ and $\tau\overline{\nu}$ final states) assuming that corresponding decay rates are given by the Standard Model ($f_{D_s}^{expt.,SM}$). References for the measurements are the following: Belle (Zupanc, 2013b), *BaBar* (del Amo Sanchez, 2010g), and CLEO-c (Naik et al., 2009).

| Belle | *BaBar* | CLEOc |
|---|---|---|
| 255.5(4.2)(5.1) | 258.6(6.4)(7.5) | 259.0(6.2)(3.0) |

**Table 19.1.24.** Expected values of the $D_s$ decay constant from unquenched LQCD ($f_{D_s}^{QCD}$). The different labels correspond to the following references: HPQCD (Davies et al., 2010), PACS-CS (Namekawa et al., 2011), ETMC (Dimopoulos et al., 2012), and Fermilab MILC (Bazavov et al., 2011). Results obtained using QCD sum rules correspond to SR1 (Bordes, Penarrocha, and Schilcher, 2005) and SR2 (Lucha, Melikhov, and Simula, 2011).

| HPQCD | PACS − CS | ETMC | Fermilab MILC | SR1 | SR2 |
|---|---|---|---|---|---|
| 248.0(2.5) | 257(5) | 248(6) | 260.1(10.8) | 205(22) | 245.3(16.3) |

new physics from these models, the following condition must be satisfied:

$$\frac{m_H}{\tan\beta} < m_{D_s}\sqrt{\frac{m_s}{m_c}\frac{1}{3\sigma}}. \quad (19.1.70)$$

The parameter $\sigma$ is the total relative uncertainty on $f_{D_s}$ coming from theory and measurements. At present $\sigma \sim 3\%$ and evidence for new physics can be obtained if $\frac{m_H}{\tan\beta} < 2.1\,\mathrm{GeV}/c^2$. In future, when precision of $\sim 1\%$ can be reached, this condition becomes: $\frac{m_H}{\tan\beta} < 3.6\,\mathrm{GeV}/c^2$. Unless the charged Higgs boson mass is rather low or the value of $\tan\beta$ quite large, no new physics contributions are expected and measurements of leptonic charm decays therefore provide stringent tests of lattice QCD calculations.

It should be noted also that relative uncertainties coming from external parameters as the $\tau$ mass, the $D_s^+$ mass, the value of $|V_{cs}|$, and the $D_s^+$ lifetime have a total contribution of 0.7%. The largest contribution is from the $D_s^+$ lifetime which needs therefore to be more accurately measured.

### 19.1.7 Rare or forbidden charmed meson decays

#### 19.1.7.1 Measurement of the Branching Fractions of the Radiative Charm Decays $D^0 \to \bar{K}^{*0}\gamma$ and $D^0 \to \phi\gamma$

In the $b$-quark sector, radiative decay processes have provided a rich field to study the Standard Model of particle physics. These decays are dominated by short-range electroweak processes, whereas long-range contributions are suppressed. The situation is reversed in the charm sector, where radiative decays are expected to be dominated largely by non-perturbative processes, examples of which are shown schematically in Fig. 19.1.42. Long-range contributions to radiative charm decays are expected to increase the branching fractions for these modes to values of the order of $10^{-5}$, whereas short-range interactions are predicted to yield rates at the $10^{-8}$ level. Given the expected dominance of long-range processes, radiative charm decays provide a laboratory in which these QCD-based calculations can be tested.

Numerous theoretical models have been developed to describe these radiative charm decays. The two most comprehensive studies (Burdman, Golowich, Hewett, and Pakvasa, 1995; Fajfer, Prelovsek, and Singer, 1999) predict very similar amplitudes for the dominant diagrams shown in Fig. 19.1.42.

The first observation of flavor changing radiative decay of charm mesons, $D^0 \to \phi\gamma$, was accomplished by Belle using $78\,\mathrm{fb}^{-1}$ (Abe, 2004c). To reduce the combinatorial background the measurement is performed using $D^{*+} \to D^0\pi^+$ decays, and the $\phi$ meson is reconstructed in decays to $K^+K^-$. The photons are required to have an energy in excess of $450\,\mathrm{MeV}/c^2$ (and not yielding a $\pi^0$ mass with any additional photon in the event - the $\pi^0$ veto). Furthermore, the $|\cos\theta_{hel}| < 0.4$ requirement, where $\theta_{hel}$ is the angle between the $D^0$ and the $K^+$ mesons in the $\phi$ meson rest frame, strongly suppresses contribution of $D^0 \to \pi^0\gamma/\eta\gamma$ decays to the $\phi\gamma$ final state (for the former, due to angular momentum conservation, the distribution in $\cos\theta_{hel}$ is proportional to $\cos^2\theta_{hel}$ while in the latter it is proportional to $1-\cos^2\theta_{hel}$). The yield of $D^0 \to \phi\gamma$ decays, as extracted from the $\phi\gamma$ invariant mass distribution, is $27.6 \pm ^{7.4}_{6.5} \pm ^{0.5}_{1.0}$ with a significance of 5.4 standard deviations. The branching fraction is determined using $D^0 \to K^+K^-$ for normalization and is found to be $\mathcal{B}(D^0 \to \phi\gamma) = (2.60 \pm ^{0.70}_{0.61} \pm ^{0.15}_{0.17}) \times 10^{-5}$. The largest contribution to the systematic uncertainty is due to the uncertainty of the $\mathcal{B}(D^0 \to K^+K^-)$ ($\pm 3.40\%$) and the choice of the fitting model and background estimation (+2.46% -3.99%).

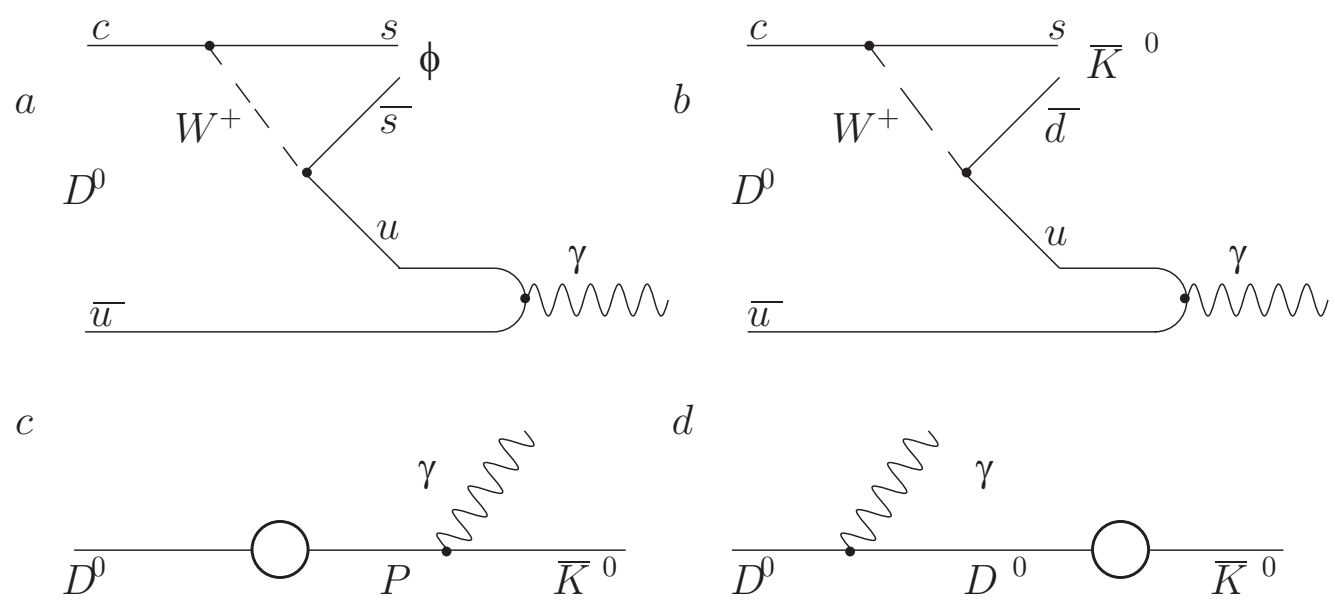

**Figure 19.1.42.** Feynman diagrams for the long-range electromagnetic contributions to $D^0 \to V\gamma$, $V = \bar{K}^{*0}, \phi$. Figures (a) and (b) show sample vector dominance processes, while (c) and (d) are examples of pole diagrams, where the circles signify the weak transition and $P$ represents a pseudoscalar meson.

*BABAR* has performed a measurement of the branching fractions for the Cabibbo-favored radiative decay, $D^0 \to \bar{K}^{*0}\gamma$, and the Cabibbo-suppressed radiative decay, $D^0 \to \phi\gamma$ (Aubert, 2008t).

The analysis is based on $387.1\,\mathrm{fb}^{-1}$ of data. They reconstruct radiative $D^0 \to V\gamma$, $V = \bar{K}^{*0}, \phi$ decays using the charged decay modes of the vector meson, $\bar{K}^{*0} \to K^-\pi^+$ ($\phi \to K^-K^+$). They form $\bar{K}^{*0}$ ($\phi$) candidates from pairs of oppositely charged tracks identified as $K^-\pi^+$ ($K^-K^+$) and accept any $K^-\pi^+$ ($K^-K^+$) candidates with invariant mass in the range 0.848 to $0.951\,\mathrm{GeV}/c^2$ (1.01 to $1.03\,\mathrm{GeV}/c^2$). The significant background from $\pi^0 \to \gamma\gamma$ decays is suppressed by rejecting a photon candidate which, when paired with another photon in the event, results in an invariant mass consistent with the $\pi^0$ mass, $(0.115 < M(\gamma\gamma) < 0.150)\,\mathrm{GeV}/c^2$. Background from random $D^0 \to V\gamma$ candidates is reduced by requiring that the $D^0$ candidate be a product of the decay $D^{*+} \to D^0\pi_s^+$. The mass difference, $\Delta M = M(V\gamma\pi_s^+) - M(V\gamma)$ is required to be in the range $(0.1435 < \Delta M < 0.1475)\,\mathrm{GeV}/c^2$. Combinatoric background from $B\bar{B}$ events is reduced to a negligible level by requiring that the CM momentum of the $D^{*+}$ candidate be greater than $2.62\,\mathrm{GeV}/c$.
The dominant background in the sample of $D^0 \to \bar{K}^{*0}\gamma$ candidates results from $D^0 \to K^-\pi^+\pi^0$ decays, where one of the photons from the $\pi^0$ decay is paired with the kaon and pion from the $D^0$ decay to closely mimic the signal mode. As described above, the $\pi^0$ veto suppresses such

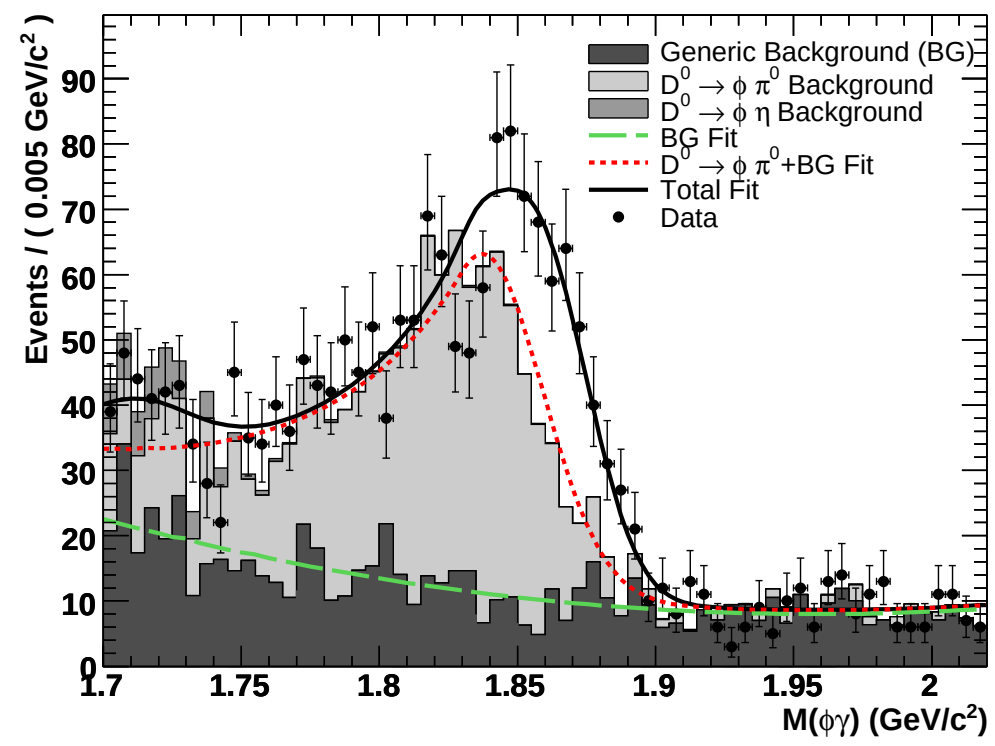

(a) The $\phi\gamma$ invariant mass distribution.

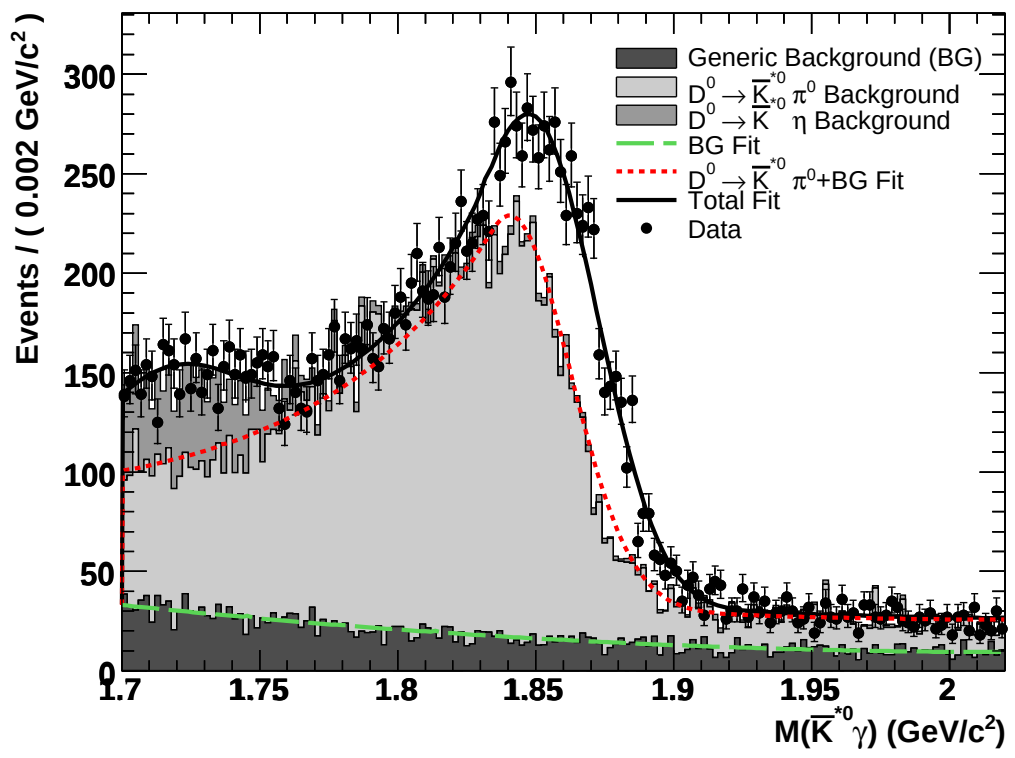

(b) The $\bar{K}^{*0}\gamma$ invariant mass distribution.

**Figure 19.1.43.** From (Aubert, 2008t). Invariant mass distributions for data (points) and simulated events (histograms). The curves show the fit results and the individual signal and background contributions. BG refers to the combinatoric background.

events but, given the large branching fraction of this mode, $\mathcal{B}(D^0 \to K^-\pi^+\pi^0) = (13.5\pm0.6)\%$ (Beringer et al., 2012), a significant number of such candidates survives. It is possible to separate this background from signal on a statistical basis because of differences in the $K^-\pi^+\gamma$ invariant mass distribution. An additional background arises from $D^0 \to \bar{K}^{*0}\eta$ events where the $\eta$ decays to two photons. This contribution peaks well below the nominal $D^0$ mass. The impact of both $D^0 \to \bar{K}^{*0}\pi^0$ and $D^0 \to \bar{K}^{*0}\eta$ is further reduced by using the $\bar{K}^{*0}$ helicity angle $\theta_H$. The helicity angle is defined as the angle between the momentum of the $\bar{K}^{*0}$ meson parent particle ($D^0$) and the momentum of the $\bar{K}^{*0}$ daughter kaon as measured in the $\bar{K}^{*0}$ rest frame. Based on a Monte Carlo study an asymmetric selection of $-0.30 < \cos\theta_H < 0.65$ is chosen to maximize the signal significance. Similarly, but to a lesser extent, the signal of the Cabibbo-suppressed radiative decay $D^0 \to \phi\gamma$ is obscured by backgrounds from $D^0 \to \phi\pi^0$ and $D^0 \to \phi\eta$ decays. The $D^0 \to \bar{K}^{*0}\gamma$ yield is extracted using an unbinned extended maximum likelihood method (E-MLM) 11 to fit the $M(\bar{K}^{*0}\gamma)$ invariant mass spectrum. The yield of $D^0 \to \phi\gamma$ events is extracted using an E-MLM to fit the two dimensional distribution of invariant mass, $M(\phi\gamma)$, and helicity, $\cos\theta_H$.

A Crystal Ball line shape (see Chapter 7) is used to model the invariant mass distributions for $D^0 \to \bar{K}^{*0}\gamma$ ($D^0 \to \phi\gamma$) signal events, and background reflections from $D^0 \to K^-\pi^+\pi^0$ ($D^0 \to \phi\pi^0$) decays. The fit results from data and expected signal and background contributions from MC are shown in Fig. 19.1.43. The resulting branching fractions relative to the well-studied decay $D^0 \to K^-\pi^+$ are $\mathcal{B}(D^0 \to \bar{K}^{*0}\gamma)/\mathcal{B}(D^0 \to K^-\pi^+) = (8.43 \pm 0.51 \pm 0.70) \times 10^{-3}$ and $\mathcal{B}(D^0 \to \phi\gamma)/\mathcal{B}(D^0 \to K^-\pi^+) = (7.15 \pm 0.78 \pm 0.69) \times 10^{-4}$.
This is the first measurement of $\mathcal{B}(D^0 \to \bar{K}^{*0}\gamma)$. In the context of the vector meson dominance (VMD) model the largest contribution to radiative $D^0$ decays is expected to come from a virtual $\rho^0$ coupling directly to a single photon, leading to the prediction that the branching ratios $\mathcal{B}(D^0 \to \phi\gamma)/\mathcal{B}(D^0 \to \bar{K}^{*0}\gamma)$ and $\mathcal{B}(D^0 \to \phi\rho^0)/\mathcal{B}(D^0 \to \bar{K}^{*0}\rho^0)$ should be equal (Burdman, Golowich, Hewett, and Pakvasa, 1995). Comparing these measurements of the radiative $D^0$ decays with the current world averages they find a good agreement with the prediction. Assuming all contributions are from VMD type processes and under the assumption that the $\rho^0$ meson is transversely polarized, as has been confirmed experimentally for $D^0 \to \bar{K}^{*0}\rho^0$, one expects $\mathcal{B}(D^0 \to V\gamma) \approx \alpha_{\rm EM}\mathcal{B}(D^0 \to V\rho^0)$ where $\alpha_{EM} = 1/137$ is the fine structure constant (Burdman, Golowich, Hewett, and Pakvasa, 1995). However they find, for this branching fraction, about a factor of three larger than the VMD prediction. This indicates that enhancements from processes other than VMD are observed, which might be explained by incomplete cancellation between pole diagrams.

### 19.1.8 $D^0 \to \ell^+\ell^-$

In the Standard Model, the flavor-changing neutral current (FCNC) decays $D^0 \to e^+e^-$ and $D^0 \to \mu^+\mu^-$ are highly suppressed by the Glashow-Iliopoulos-Maiani (GIM) mechanism (Glashow, Iliopoulos, and Maiani, 1970) (Chapter 16). Their decay branching fractions have been estimated to be less than $10^{-13}$ even with long-distance processes included. This prediction is orders of magnitude beyond the reach of current experiments. Furthermore, the lepton-flavor-violating (LFV) decay $D^0 \to e^\pm\mu^\mp$ is forbidden in the SM in the limit of vanishing neutrino masses. These decays are in principle allowed due to a non-zero neutrino mass, but branching fractions are expected to be even much smaller than those of $D^0 \to \ell^+\ell^-$.

Some extensions to the Standard Model can enhance the FCNC processes by many orders of magnitude. For example, $R$-parity violating supersymmetry can increase the branching fractions of $D^0 \to e^+e^-$ and $D^0 \to \mu^+\mu^-$ to as high as $10^{-10}$ and $10^{-6}$, respectively (Burdman, Golowich, Hewett, and Pakvasa, 2002). The same model also predicts the $D^0 \to e^\pm\mu^\mp$ branching fraction to be of the order of $10^{-6}$. The upper bounds on the predicted branching fractions of $D^0 \to \mu^+\mu^-$ and $D^0 \to e^\pm\mu^\mp$ are close to the current experimental sensitivities. As a result, searching for the FCNC and LFV decays in the charm sector is a potential way to test the SM and explore new physics. Similar arguments hold for rare $K$ and $B$ decays, but the charm decay is unique since it is sensitive to new physics coupling to the up-quark sector (similar as the $D^0$ mixing, see Section 19.2).

Both *BABAR* and Belle performed a search for the decay of $D^0 \to e^+e^-$, $D^0 \to \mu^+\mu^-$, and $D^0 \to e^\pm\mu^\mp$. A first analysis by *BABAR* was based on $122\,\mathrm{fb}^{-1}$ of data (Aubert, 2004z). Recently, a new anaysis has been performed using $468\,\mathrm{fb}^{-1}$ of data (Lees, 2012v). The measurement by Belle uses $660\,\mathrm{fb}^{-1}$ of data (Petric, 2010).

The $D^0 \to \ell^+\ell^-$ ($\ell = e, \mu$) branching ratio is determined by

$$\mathcal{B}(D^0 \to \ell^+\ell^-) = S(N_{\mathrm{obs}} - N_{\mathrm{bg}}), \qquad (19.1.71)$$

where $N_{\mathrm{obs}}$ is the number of $D^0 \to \ell^+\ell^-$ candidates observed, $N_{\mathrm{bg}}$ is the expected background and $S$ is the sensitivity factor, defined as:

$$S \equiv \mathcal{B}(D^0 \to \pi^+\pi^-)\frac{1}{N_{\pi\pi}}\frac{\epsilon_{\pi\pi}}{\epsilon_{\ell\ell}}. \qquad (19.1.72)$$

Here $\mathcal{B}(D^0 \to \pi^+\pi^-)$ is the $D^0 \to \pi^+\pi^-$ branching fraction, $N_{\pi\pi}$ is the number of reconstructed $D^0 \to \pi^+\pi^-$ decays, $\epsilon_{\ell\ell}$ and $\epsilon_{\pi\pi}$ are the efficiencies for the corresponding decay mode. They choose $D^0 \to \pi^+\pi^-$ as the normalization mode because it is kinematically similar to $D^0 \to \ell^+\ell^-$ and therefore many common systematic uncertainties cancel in the calculation of the efficiency ratio $\epsilon_{\pi\pi}/\epsilon_{\ell\ell}$.

A pair of oppositely charged tracks is selected to form a $D^0 \to \ell^+\ell^-$ or $D^0 \to \pi^+\pi^-$ candidate with particle identification applied 5. In *BABAR*, the average electron and muon efficiencies are about 95 % and 60 %, and their hadron misidentification probabilities are measured from $\tau$ decay control samples to be around 0.2 % and 2.0 %. The corresponding single pion identification efficiency is around 90 %. At Belle, the average muon and electron identification efficiencies are around 90 % with less than 1.5 % and 0.3 % pion misidentification, respectively, whereas the pion identification efficiency is around 83 %. In the analysis, only $D^0$ mesons originating from the fragmentation of charm quark in the continuum $e^+e^- \to c\bar{c}$ are considered. The inclusion of $D^0$ mesons from $B$ meson decays offers no advantage because of their higher combinatorial background. As a result, Belle (*BABAR*) requires the momentum of each $D^0$ ($D^{*+}$) candidate in the center-of-mass frame of the collision to be larger than $2.5\,\mathrm{GeV}/c$ ($2.4\,\mathrm{GeV}/c$). In order to further reduce the background, the $D^0$ candidate is required to originate from a $D^{*+} \to D^0\pi^+$ decay.

Candidate $D^0$ mesons are selected using two kinematic observables: the invariant mass of the $D^0$ decay product, $m_{\ell\ell}$, and the energy released in the $D^{*+}$ decay, $\delta m = m_{D^{*+}} - m_{\ell\ell} - m_\pi$, where $m_{D^{*+}}$ is the invariant mass of the $D^0\pi^+$ combination and $m_\pi$ is the $\pi^+$ mass. Additional experimental observables are exploited in order to increase the search sensitivities. The *BABAR* analysis makes use a linear combination (Fisher discriminant 9) of the following five variables to reduce the combinatorial $B\overline{B}$ background:

- The measured $D^0$ flight length divided by its uncertainty.
- The value of $\cos\theta_{\mathrm{hel}}$, where $\theta_{\mathrm{hel}}$ is defined as the angle between the momentum of the positively-charged $D^0$ daughter and the boost direction from the lab frame to the $D^0$ rest frame, all in the $D^0$ rest frame.
- The missing transverse momentum with respect to the beam axis.
- The ratio of the 2$^{\mathrm{nd}}$ and 0$^{\mathrm{th}}$ Fox-Wolfram moments.
- The $D^0$ momentum in the CM frame.

The Belle analysis uses the maximum allowed missing energy $E_{\mathrm{miss}}$ in the event. MC study shows that the semileptonic $B$ decay, which represent one of the dominant backgrounds, typically have large $E_{\mathrm{miss}}$ due to undetected neutrino.

In order to avoid biases, a blind analysis techniques (see Section 14) has been adopted. All events inside the $D^0$ signal region are blinded until the final event selection criteria are established. The estimate of the number of combinatorial background in the signal window is done using the observed event distribution in the control region. In the *BABAR* analysis they use a sideband region above the signal region in the $D^0$ mass ([1.90, 2.05] GeV) in a wide $\Delta m$ window ([0.141, 0.149] GeV) while the Belle measurement makes use of the region defined by $\Delta m > 1\ \mathrm{MeV}/c^2$. The peaking background in the signal region due to misidentification of $D^0 \to \pi^+\pi^-$ decay is calculated by using the lepton misidentification rates measured from a control data sample. Finally, they determine the optimal selection criteria by maximizing the value $\epsilon_{\ell\ell}/N_{\mathrm{sens}}$, where $N_{\mathrm{sens}}$ is the averaged 90 % confidence level upper limit on the number of observed signal events that would be obtained by an ensemble of experiments with the expected background and no real signal (Feldman and Cousins, 1998).

The number of $D^0 \to \pi^+\pi^-$ candidates in the data, $N_{\pi\pi}$, is extracted by fitting their invariant mass distribution with a binned maximum likelihood fit. The signal efficiencies of $D^0 \to \ell^+\ell^-$ and $D^0 \to \pi^+\pi^-$ are evaluated using a Monte Carlo simulation.

**Table 19.1.25.** From Lees (2012v) and Petric (2010). Summary of the number of expected background events ($N_{\mathrm{bg}}$), the sensitivity factor ($S$), number of observed events ($N_{\mathrm{obs}}$), and the branching fraction upper limits at the 90 % C.L. for each decay mode. The errors quoted include statistical and systematic uncertainties.

| | $D^0 \to e^+e^-$ | $D^0 \to \mu^+\mu^-$ | $D^0 \to e^\pm\mu^\mp$ |
|---|---|---|---|
| | | BABAR | |
| $N_{\mathrm{bg}}$ | $1.01 \pm 0.39$ | $3.88 \pm 0.35$ | $1.42 \pm 0.26$ |
| $N_{\mathrm{obs}}$ | 1 | 8 | 2 |
| $S$ $[10^{-9}]$ | $53.4 \pm 0.22$ | $80.6 \pm 0.44$ | $73.9 \pm 0.4$ |
| UL | $1.7 \times 10^{-7}$ | $[0.6, 8.1] \times 10^{-7}$ | $3.3 \times 10^{-7}$ |
| | | Belle | |
| $N_{\mathrm{bg}}$ | $1.7 \pm 0.2$ | $3.1 \pm 0.1$ | $2.6 \pm 0.2$ |
| $N_{\mathrm{obs}}$ | 0 | 2 | 3 |
| $S$ $[10^{-9}]$ | $64.7(1 \pm 6.4\%)$ | $48.4(1 \pm 5.3\%)$ | $54.8(1 \pm 4.8\%)$ |
| UL | $7.9 \times 10^{-8}$ | $1.4 \times 10^{-7}$ | $2.6 \times 10^{-7}$ |

The branching fraction upper limits (UL) have been calculated including all uncertainties using an extended version (Conrad, Botner, Hallgren, and Perez de los Heros, 2003) of the Feldman-Cousins method (Feldman and Cousins, 1998). Systematic uncertainties are found to have a negligible effect on the limits. The results are listed in Table 19.1.25.

#### 19.1.8.1 Search for the decay $D^0 \to \gamma\gamma$ and Measurement of the branching fraction for $D^0 \to \pi^0\pi^0$

In the Standard Model flavor-changing neutral currents (FCNC) are forbidden at tree level (Glashow, Iliopoulos, and Maiani, 1970). These decays are allowed at higher order (Hurth, 2003) and have been measured in the $K$ and $B$ meson systems. In the charm sector, however, the small mass difference between down-type quarks of the first two families translates to a large suppression at the loop level from the GIM mechanism. To date, measurements of radiative decays of charm mesons are consistent with results of theoretical calculations that include both short-distance and long-distance contributions and predict decay rates several orders of magnitude below the sensitivity of current experiments. While these rates are small, it has been predicted that new physics (NP) processes can lead to significant enhancements (Prelovsek and Wyler, 2001).

*BABAR* has searched for the rare decay of the $D^0$ meson to two photons, $D^0 \to \gamma\gamma$, and measured the branching fraction for a $D^0$ meson decaying to two neutral pions, $\mathcal{B}(D^0 \to \pi^0\pi^0)$ (Lees, 2012u). The data sample analyzed corresponds to an integrated luminosity of 470.5 $\text{fb}^{-1}$. The $D^0 \to K^0_S\pi^0$ decay is chosen for this purpose due to its large branching fraction of $(1.22 \pm 0.05)\%$ (Beringer et al., 2012) and partial cancellation of systematic uncertainties.

The invariant $\gamma\gamma$ mass distribution obtained from the $D^0 \to \gamma\gamma$ analysis is shown in Fig. 19.1.44 together with projections of the likelihood fit and the individual signal and background contributions. The signal yield is $-6 \pm 15$, consistent with no $D^0 \to \gamma\gamma$ events. This result is converted to a branching fraction for $D^0 \to \gamma\gamma$ relative to the $D^0 \to K^0_S\pi^0$ reference mode using

$$\mathcal{B}(D^0 \to \gamma\gamma) = \frac{\frac{1}{\varepsilon_{\gamma\gamma}} N(D^0 \to \gamma\gamma)}{\frac{1}{\varepsilon_{D^0 \to K^0_S\pi^0}} N(D^0 \to K^0_S\pi^0)} \times \mathcal{B}(D^0 \to K^0_S\pi^0), \quad (19.1.73)$$

where $N$ and $\varepsilon$ are the yield and efficiency of the respective modes and $\mathcal{B}(D^0 \to K^0_S\pi^0)$ is the known $D^0 \to K^0_S\pi^0$ branching fraction. In this analysis the $D^0 \to K^0_S\pi^0$ signal yield is $126599 \pm 568$ events. They find $\mathcal{B}(D^0 \to \gamma\gamma) = (-0.49 \pm 1.23 \pm 0.02) \times 10^{-6}$ where the errors are the statistical uncertainty and the uncertainty in the reference mode branching fraction, respectively. Therefore they place an upper limit on the branching fraction for the decay of a $D^0$ meson to two photons, $\mathcal{B}(D^0 \to \gamma\gamma) < 2.2 \times 10^{-6}$, at 90% confidence level.

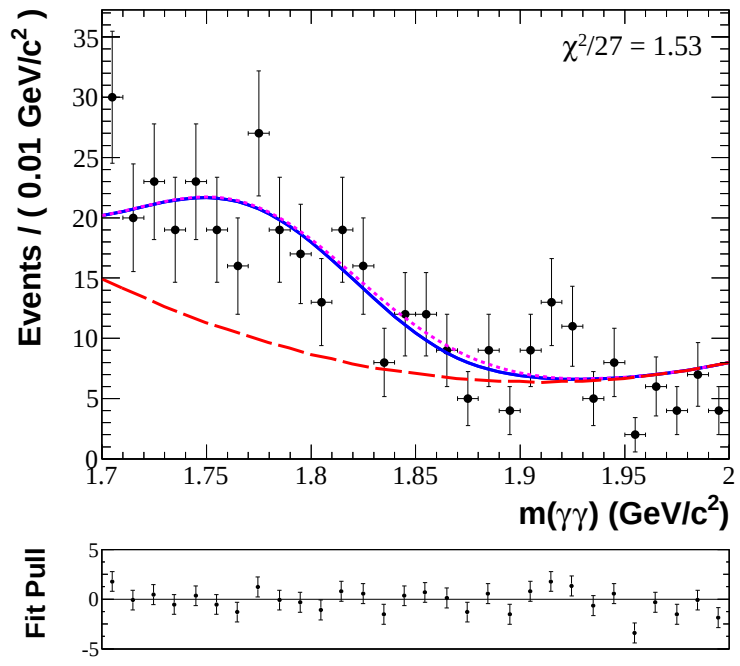

**Figure 19.1.44.** From (Lees, 2012u). The $\gamma\gamma$ mass distribution for $D^0 \to \gamma\gamma$ candidates in data (data points). The curves show the result of an unbinned maximum likelihood fit to the measured mass distribution. The solid (blue) line shows the result of the fit, indicating a slightly negative signal yield (consistent with no signal). The long-dashed (red) curve corresponds to combinatoric background component, and the small-dash pink curve corresponds to the combinatoric background plus $D^0 \to \pi^0\pi^0$ background shape. The $\chi^2$ value is determined from binned data and is provided as a goodness-of-fit measure. The pull distribution shows differences between the data and the solid blue curve with values and errors normalized to the Poisson error.

The invariant mass distribution for events in the $D^0 \to \pi^0\pi^0$ analysis is shown in Fig. 19.1.45. The signal yield for $D^0 \to \pi^0\pi^0$ is $26010 \pm 304$ events. For $D^0 \to K^0_S\pi^0$ the signal yield is $103859 \pm 392$ events. Adjusting Eq. (19.1.73) for the $D^0 \to \pi^0\pi^0$ case this yield can be converted to a branching fraction and obtain $B(D^0 \to \pi^0\pi^0) = (8.4 \pm 0.1 \pm 0.3) \times 10^{-4}$. The first error denotes the statistical uncertainty and the second error reflects the uncertainties in the reference mode branching fraction.

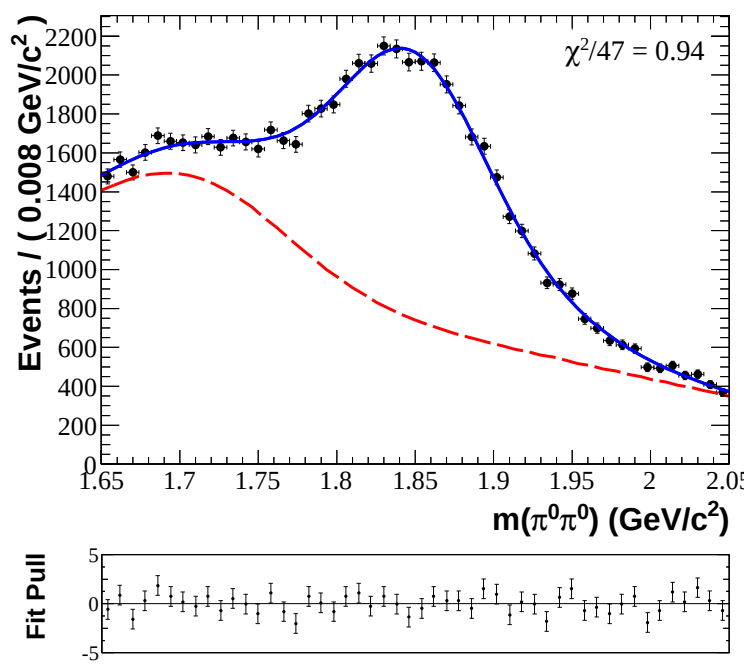

**Figure 19.1.45.** From (Lees, 2012u). The $\pi^0\pi^0$ mass distribution for $D^0 \to \pi^0\pi^0$ candidates in data (data points). The curves show the result of the unbinned maximum likelihood fit to the measured mass distribution. The solid (blue) line shows the result of the fit. The long-dashed (red) curve corresponds to the combinatoric background component. The $\chi^2$ value is determined from binned data and is provided as a goodness-of-fit measure. The pull distribution shows differences between the data and the solid blue curve with values and errors normalized to the Poisson error.

**Table 19.1.26.** From (Lees, 2012u). Summary of predictions and measured values or limits for branching fractions for $D^0 \to \gamma\gamma$ and $D^0 \to \pi^0\pi^0$.

| Theoretical predictions | | |
|---|---|---|
| Mode | Value | Reference |
| $D^0 \to \gamma\gamma$ (SM,VMD) | $\approx (3.5\ ^{+4.0}_{-2.6}) \times 10^{-8}$ | (Burdman, Golowich, Hewett, and Pakvasa, 2002) |
| $D^0 \to \gamma\gamma$ (SM,HQ$\chi$PT) | $(1.0 \pm 0.5) \times 10^{-8}$ | (Fajfer, Singer, and Zupan, 2001) |
| $D^0 \to \gamma\gamma$ (MSSM) | $6 \times 10^{-6}$ | (Prelovsek and Wyler, 2001) |
| **Experimental results** | | |
| Mode | Value | Reference |
| $D^0 \to \gamma\gamma$ (2002) | $< 2.9 \times 10^{-5}$ | (Coan et al., 2003) |
| $D^0 \to \gamma\gamma$ (2012) | $< 2.2 \times 10^{-6}$ | (Lees, 2012u) |
| $D^0 \to \pi^0\pi^0$ (2006) | $(7.9 \pm 0.8) \times 10^{-4}$ | (Rubin et al., 2006) |
| $D^0 \to \pi^0\pi^0$ (2010) | $(8.1 \pm 0.5) \times 10^{-4}$ | (Mendez et al., 2010) |
| $D^0 \to \pi^0\pi^0$ (2012) | $(8.4 \pm 0.1) \times 10^{-4}$ | (Lees, 2012u) |

A summary of the relevant branching fractions is shown in Table 19.1.26.

### 19.1.9 Search for rare or forbidden semileptonic charm decays

*BABAR* has performed a search for charm semileptonic decays that are either forbidden or heavily suppressed in the Standard Model (Lees, 2011k). The decays are of the form $X_c^+ \to h^\pm \ell^\mp \ell^{(\prime)+}$, where $X_c^+$ is a charm hadron ($D^+$, $D_s^+$, or $\Lambda_c^+$), and $\ell^{(\prime)\pm}$ is an electron or muon. For $D^+$ and $D_s^+$ modes, $h^\pm$ can be a pion or kaon, while for $\Lambda_c^+$ modes it is a proton. Decay modes with oppositely charged leptons of the same lepton flavor are examples of flavor-changing neutral current (FCNC) processes, which are expected to be very rare because they cannot occur at tree level in the SM. Decay modes with two oppositely charged leptons of different flavor correspond to lepton-flavor violating (LFV) decays and are essentially forbidden in the SM because they can occur only through lepton mixing. Decay modes with two leptons of the same charge are lepton-number violating (LNV) decays and are forbidden in the SM. Hence, decays of the form $X_c^+ \to h^\pm \ell^\mp \ell^{(\prime)+}$ provide sensitive tools to investigate physics beyond the SM. The most stringent existing upper limits on the branching fractions for $X_c^+ \to h^\pm \ell^\mp \ell^{(\prime)+}$ decays range from 1 to $700 \times 10^{-6}$ and do not exist for most of the $\Lambda_c^+$ decays.

Charm hadron candidates are formed from one track identified as either a pion, kaon, or proton ($h$) and two tracks, each of which is identified as an electron or a muon ($\ell\ell^{(\prime)}$). The total charge of the three tracks is required to be $\pm 1$. For three-track combinations with a pion or kaon track, the $h\ell\ell^{(\prime)}$ invariant mass is required to lie between 1.7 and $2.1\,\mathrm{GeV}/c^2$; for combinations with a proton, the invariant mass is required to lie between 2.2 and $2.4\,\mathrm{GeV}/c^2$.

The combinatorial background at low $p^*$ is very large and they therefore select charm hadron candidates with $p^*$ greater than $2.5\,\mathrm{GeV}/c$. The main backgrounds remaining after this selection are QED events and semileptonic $B$ and charm decays, particularly events with two semileptonic decays.

The QED events are mainly radiative Bhabha, initial-state radiation, and two-photon events, which are all rich in electrons. These events are easily identified by their low multiplicity and/or highly jet-like topology. They strongly suppress this background by requiring at least five tracks in the event and that the hadron candidate be inconsistent with the electron hypothesis. The background from semileptonic $B$ and charm decays is also suppressed by requiring the two leptons to be consistent with a common origin.

For low $e^+e^-$ invariant mass there is a significant background contribution from photon conversions and $\pi^0$ decays to $e^+e^-\gamma$. These are both removed by requiring $m(e^+e^-) > 200\,\mathrm{MeV}/c^2$.

For the $D_{(s)}^+ \to \pi^+\ell^+\ell^-$ decay modes, they exclude events with $0.95 < m(e^+e^-) < 1.05\,\mathrm{GeV}/c^2$ and $0.99 < m(\mu^+\mu^-) < 1.05\,\mathrm{GeV}/c^2$ to reject decays through the $\phi$ resonance.

After the initial event selection, significant combinatorial background contributions remain from semileptonic $B$ decays and other sources. The final candidate selection is performed by forming a likelihood ratio $R_{\mathcal{L}}$ and requiring the ratio to be greater than a minimum value $R_{\mathcal{L}}^{\min}$ 4.

The following three discriminating variables are used in the likelihood ratio: charm hadron candidate $p^*$, total reconstructed energy in the event and flight length significance.

Extended, unbinned, maximum-likelihood fits are applied to the invariant-mass distributions for the $h^\pm \ell^\mp \ell^{(\prime)+}$ candidates. The measured signal yields are converted into branching ratios by normalizing them to the yields of known charm decays. For the $D^+$ and $D_s^+$ mesons, they use decays to $\pi^+\phi$ as normalization modes. For the $\Lambda_c^+$, they use $\Lambda_c^+ \to pK^-\pi^+$ as the normalization mode. The upper limits are set using a Bayesian approach with a flat prior for the event yield in the physical region. The upper limit on the signal yield is defined as the number of signal events for which the integral of the likelihood from zero events to that number of events is 90% of the integral from zero to infinity. The systematic uncertainties are included in the likelihood as additional nuisance parameters. Examples of $h\ell\ell^{(\prime)}$ invariant-mass distributions for signal candidates are in Figs 19.1.46 and 19.1.47. The signal yields obtained from the unbinned likelihood fits are listed in Table 19.1.27 with statistical and systematic uncertainties. Only systematic uncertainties associated with the signal and background *p.d.f.*s are included in the systematic uncertainty for the yields. The curves representing the fits are overlaid in the figures. The most significant signal is seen in the distribution for $\Lambda_c^+ \to p\mu^+\mu^-$; the signal yield has a statistical-only significance of $2.6\sigma$ as determined from the change in log-likelihood with respect to zero assumed signal events. With 35 different measurements, a $2.6\sigma$ deviation is expected with about 25% probability.

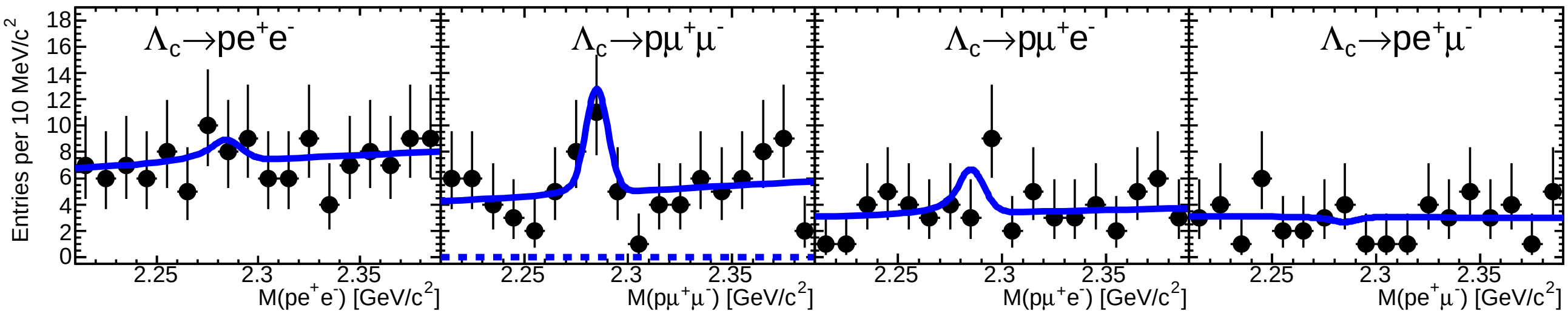

**Figure 19.1.46.** From (Lees, 2011k). Invariant-mass distributions for $\Lambda_c^+ \to p\ell^+\ell^{(\prime)-}$ candidates. The solid lines are the results of the fits. The background component for the dimuon mode in which muon candidates arise from hadrons misidentified is shown as a dashed curve.

**Table 19.1.27.** From (Lees, 2011k). Signal yields for the fits to the 35 $X_c^+ \to h^\pm\ell^\mp\ell^{(\prime)+}$ event samples. The first error is the statistical uncertainty and the second is the systematic uncertainty. The third column lists the estimated signal efficiency. The fourth column gives for each signal mode the 90% C.L. upper limit (UL) on the ratio of the branching fraction of the signal mode to that of the normalization mode (BR). The last column shows the 90% C.L. upper limit on the branching fraction for each signal mode (BF). The upper limits include all systematic uncertainties.

| Decay mode | Yield (events) | Eff. (%) | BR UL 90% C.L. ($10^{-4}$) | BF UL 90% C.L. ($10^{-6}$) |
|---|---|---|---|---|
| $D^+ \to \pi^+e^+e^-$ | $-3.9 \pm 1.6 \pm 1.7$ | 1.56 | 3.9 | 1.1 |
| $D^+ \to \pi^+\mu^+\mu^-$ | $-0.2 \pm 2.8 \pm 0.9$ | 0.46 | 24 | 6.5 |
| $D^+ \to \pi^+e^+\mu^-$ | $-2.9 \pm 3.4 \pm 2.4$ | 1.21 | 11 | 2.9 |
| $D^+ \to \pi^+\mu^+e^-$ | $3.6 \pm 4.3 \pm 1.3$ | 1.54 | 13 | 3.6 |
| $D_s^+ \to \pi^+e^+e^-$ | $8 \pm 34 \pm 8$ | 6.36 | 5.4 | 13 |
| $D_s^+ \to \pi^+\mu^+\mu^-$ | $20 \pm 15 \pm 4$ | 1.21 | 18 | 43 |
| $D_s^+ \to \pi^+e^+\mu^-$ | $-3 \pm 11 \pm 3$ | 2.16 | 4.9 | 12 |
| $D_s^+ \to \pi^+\mu^+e^-$ | $9.3 \pm 7.3 \pm 2.8$ | 1.50 | 8.4 | 20 |
| $D^+ \to K^+e^+e^-$ | $-3.7 \pm 2.9 \pm 3.3$ | 2.88 | 3.7 | 1.0 |
| $D^+ \to K^+\mu^+\mu^-$ | $-1.3 \pm 2.8 \pm 1.1$ | 0.65 | 16 | 4.3 |
| $D^+ \to K^+e^+\mu^-$ | $-4.3 \pm 1.8 \pm 0.6$ | 1.44 | 4.3 | 1.2 |
| $D^+ \to K^+\mu^+e^-$ | $3.2 \pm 3.8 \pm 1.2$ | 1.74 | 9.9 | 2.8 |
| $D_s^+ \to K^+e^+e^-$ | $-5.7 \pm 5.8 \pm 2.0$ | 3.20 | 1.6 | 3.7 |
| $D_s^+ \to K^+\mu^+\mu^-$ | $4.8 \pm 5.9 \pm 1.2$ | 0.85 | 9.1 | 21 |
| $D_s^+ \to K^+e^+\mu^-$ | $9.1 \pm 6.0 \pm 2.8$ | 1.74 | 5.7 | 14 |
| $D_s^+ \to K^+\mu^+e^-$ | $3.4 \pm 6.4 \pm 3.5$ | 2.08 | 4.2 | 9.7 |
| $\Lambda_c^+ \to pe^+e^-$ | $4.0 \pm 6.5 \pm 2.8$ | 5.52 | 0.8 | 5.5 |
| $\Lambda_c^+ \to p\mu^+\mu^-$ | $11.1 \pm 5.0 \pm 2.5$ | 0.86 | 6.4 | 44 |
| $\Lambda_c^+ \to pe^+\mu^-$ | $-0.7 \pm 2.9 \pm 0.9$ | 1.10 | 1.6 | 9.9 |
| $\Lambda_c^+ \to p\mu^+e^-$ | $6.2 \pm 4.6 \pm 1.8$ | 1.37 | 2.9 | 19 |
| $D^+ \to \pi^-e^+e^+$ | $4.7 \pm 4.7 \pm 0.5$ | 3.16 | 6.8 | 1.9 |
| $D^+ \to \pi^-\mu^+\mu^+$ | $-3.1 \pm 1.2 \pm 0.5$ | 0.70 | 7.5 | 2.0 |
| $D^+ \to \pi^-\mu^+e^+$ | $-5.1 \pm 4.2 \pm 2.0$ | 1.72 | 7.4 | 2.0 |
| $D_s^+ \to \pi^-e^+e^+$ | $-5.7 \pm 14. \pm 3.4$ | 6.84 | 1.8 | 4.1 |
| $D_s^+ \to \pi^-\mu^+\mu^+$ | $0.6 \pm 5.1 \pm 2.7$ | 1.05 | 6.2 | 14 |
| $D_s^+ \to \pi^-\mu^+e^+$ | $-0.2 \pm 7.9 \pm 0.6$ | 2.23 | 3.6 | 8.4 |
| $D^+ \to K^-e^+e^+$ | $-2.8 \pm 2.4 \pm 0.2$ | 2.67 | 3.1 | 0.9 |
| $D^+ \to K^-\mu^+\mu^+$ | $7.2 \pm 5.4 \pm 1.6$ | 0.80 | 37 | 10 |
| $D^+ \to K^-\mu^+e^+$ | $-11.6 \pm 4.0 \pm 3.1$ | 1.52 | 6.8 | 1.9 |
| $D_s^+ \to K^-e^+e^+$ | $2.3 \pm 7.9 \pm 3.3$ | 4.10 | 2.1 | 5.2 |
| $D_s^+ \to K^-\mu^+\mu^+$ | $-2.3 \pm 5.0 \pm 2.8$ | 0.98 | 5.3 | 13 |
| $D_s^+ \to K^-\mu^+e^+$ | $-14.0 \pm 8.4 \pm 2.0$ | 2.26 | 2.4 | 6.1 |
| $\Lambda_c^+ \to \overline{p}e^+e^+$ | $-1.5 \pm 4.2 \pm 1.5$ | 5.14 | 0.4 | 2.7 |
| $\Lambda_c^+ \to \overline{p}\mu^+\mu^+$ | $-0.0 \pm 2.1 \pm 0.6$ | 0.94 | 1.4 | 9.4 |
| $\Lambda_c^+ \to \overline{p}\mu^+e^+$ | $10.1 \pm 5.8 \pm 3.5$ | 2.50 | 2.3 | 16 |

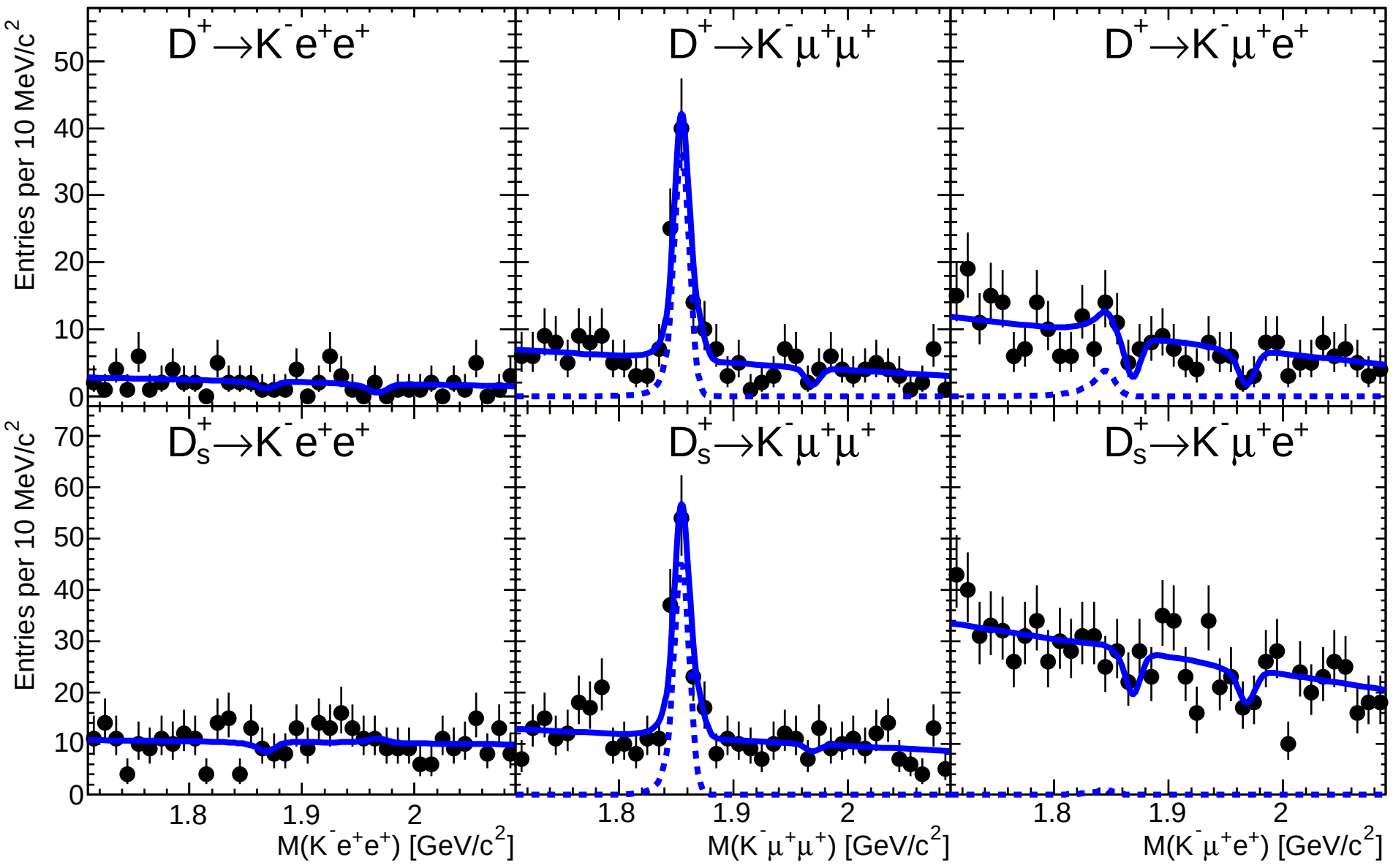

**Figure 19.1.47.** From (Lees, 2011k). Invariant-mass distributions for $D^+ \to K^-\ell^+\ell^{(\prime)+}$ (top) and $D_s^+ \to K^-\ell^+\ell^{(\prime)+}$ (bottom) candidates. The solid lines are the results of the fits. The background components for the dimuon modes and $D_{(s)}^+ \to K^-\mu^+e^+$ in which candidates arise from misidentified hadrons are shown as dashed curves.

### 19.1.10 Summary of charmed meson decays

Charm decays open the road to investigate the flavor physics of up-type quarks, which is complementary to the weak interactions of the (bottom and strange) down type quarks. Since the $B$ decays are dominated by the $b \to c$ transitions, and the $e^+e^- \to c\bar{c}$ cross section is comparatively high, the $B$ factories also generated plenty of charm which allowed detailed measurements with highly competitive precision.

FCNC processes of up-type quarks are predicted to be heavily GIM suppressed, which motivated measurements and searches of rare and even forbidden decays of charm at the $B$ factories. Although for many FCNC processes it is difficult to make a precise theoretical prediction, the $B$ factories added a lot of new information on these decays, constraining significantly the limits on physics beyond the SM.

Aside from the weak interactions also many studies of QCD related issues have been performed. The charm quark is neither heavy enough to be cleanly treated within the heavy quark expansion, nor is it light enough to be sensibly included into chiral Lagrangians, and thus it nicely covers the intermediate region between the two limits. For this reason there has been a lot of effort to measure multibody decays, where a lot of information on QCD has been extracted from *e.g.* Dalitz analyses.

## 19.2 $D$-mixing and $CP$ violation

**Editors:**
*Brian Meadows (BABAR)*
*Boštjan Golob (Belle)*
*Ikaros Bigi (theory)*

**Additional section writers:**
*Ray Cowan, Kevin Flood, Maurizio Martinelli, Alan Schwartz, Marko Starič, Eunil Won, Anže Zupanc*

### 19.2.1 Introduction

#### 19.2.1.1 Brief overview

The mixing phenomenon in $B$, $D$ and $K$ neutral meson system is an example of the flavor changing neutral current (FCNC) process. Within the SM, FCNC's are absent at the tree level (first order). However, mixing can occur through box diagrams (second order), as shown in Fig. 19.2.1.

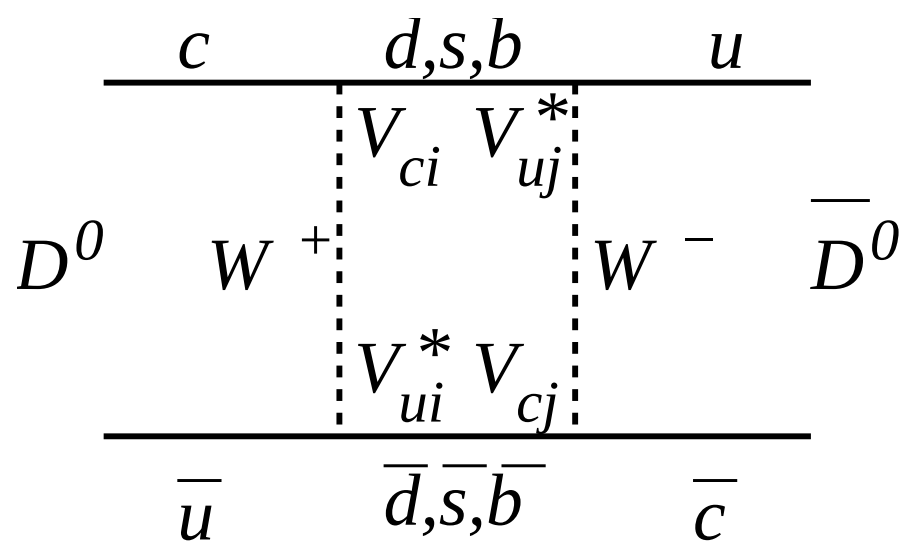

**Figure 19.2.1.** Box diagram leading to $D^0 - \overline{D}^0$ mixing.

The strong suppression of FCNC's is a consequence of the GIM mechanism (Glashow, Iliopoulos, and Maiani, 1970) (see Chapter 16). In the past, measurements of mixing provided a basis for important discoveries. The discoveries of $K^0 - \overline{K}^0$ and $B_d^0 - \overline{B}_d^0$ mixing, for example, enabled predictions of the masses of the charm and top quarks, respectively, before the quarks were first observed at Brookhaven and SLAC (Aubert et al., 1974; Augustin et al., 1974), and at Fermilab (Abachi et al., 1995b; Abe et al., 1995). The probability for any of the above mentioned neutral mesons to transform into its anti-particle in the course of its lifetime is described by the mixing parameters $x$ and $y$. The mixing parameters are defined as

$$\Gamma = \frac{\Gamma_1 + \Gamma_2}{2}$$
$$x = \frac{m_1 - m_2}{\Gamma}$$
$$y = \frac{\Gamma_1 - \Gamma_2}{2\Gamma} \ , \qquad (19.2.1)$$

where $\Gamma_{1,2}$ are the widths of the two mass eigenstates. The time integrated probability for a neutral meson initially produced as $P^0$ to decay at a later time as $\overline{P}^0$ is given by $(x^2+y^2)/2(x^2+1)$. By inspection of approximate values for $x$ and $y$ in Table 19.2.1 it is clear that this probability is by far the smallest for the system of neutral $D$ mesons.

**Table 19.2.1.** Discoveries of neutral mesons and their mixing Approximate values of the mixing parameters are listed as well.

| Meson | Discovery year and place | Mixing parameter |
|---|---|---|
| $K^0$ | 1950 Caltech | |
| Mixing | 1956 Columbia | $x \approx 1, \quad y \approx 1$ |
| $B_d^0$ | 1983 CESR | |
| Mixing | 1987 DESY | $x \approx 0.8, \quad y \sim 0$ |
| $B_s^0$ | 1992 LEP | |
| Mixing | 2006 Fermilab | $x \approx 26, \quad y \sim 0.05$ |
| $D^0$ | 1976 SLAC | |
| Mixing | 2007 KEK, SLAC | $x \sim 0.01, \quad y \sim 0.01$ |

The reason for the small rate of mixing of $D^0$ mesons lies in the fact that they are the only flavored neutral mesons composed of up-type quarks. The GIM mechanism, as explained below, is even more efficient for the case of up-type quark FCNC's. For the same reason measurements of mixing in the $D^0$ system yield complementary constraints on possible contributions from new physics (NP) processes beyond the SM to those arising from the measurements of FCNC's of down-type quarks ($B$ or $K$ mesons). In 2007 the $B$ Factories established evidence for mixing in the neutral charm mesons system, and those results were published back-to-back in Phys. Rev. Lett. as (Aubert, 2007j) and (Staric, 2007). These results are discussed in Sections 19.2.2 and 19.2.3, respectively.

#### 19.2.1.2 Mixing

A general description of oscillations of pseudoscalar neutral mesons is given in Section 10.1. In the following we emphasize some of the specifics of the $D^0$ system. The mixing parameters are defined in Eq. (19.2.1).

In the absence of $CP$ violation ($q = p = 1/\sqrt{2}$ in Eq. 10.1.2), $D_{1(2)}$ is the $CP$-even (odd) state if one adopts the phase convention $CP|D^0\rangle = |\overline{D}^0\rangle$ and $CP|\overline{D}^0\rangle = |D^0\rangle$.[119]

The amplitude for the process of Fig. 19.2.1, $\langle\overline{D}^0|H^{\Delta C=2}|D^0\rangle$, can be schematically written as

$$\sum_{i,j=d,s,b} V_{ui}^* V_{ci} V_{cj} V_{uj}^* \mathcal{F}(m_W^2, m_i^2, m_j^2), \qquad (19.2.2)$$

[119] For the mixing parameter $x$ ($y$) one subtracts the mass (width) of the $CP$-odd state (or in case of $CP$ violation of the state which has a larger $CP$-odd component) from that of the $CP$-even state (or in case of $CP$ violation of the state which has a larger $CP$-even component).

where the kinematic function $\mathcal{F}$ arises from the integration over the momenta of particles exchanged in the loop. The formulation above nicely reflects the GIM mechanism: if masses of all (down-type) quarks exchanged in the loop were equal, *i.e.* $m_i = m_j$, the function $\mathcal{F}$ would be a factor common to all terms in the sum which would make the amplitude zero due to the unitarity of the CKM matrix.

Based on dimensional arguments (Nachtmann, 1990) the form of $\mathcal{F}$ is

$$\mathcal{F}(m_W^2, m_i^2, m_j^2) \propto f_0 m_W^2 + f_1 m_i^2 + f_2 m_j^2 + f_3 m_i m_j + \mathcal{O}(m_W^{-2}) \ . \quad (19.2.3)$$

Due to the unitarity of the CKM matrix, only the $m_i m_j$ terms survives in the sum (19.2.2), and hence

$$\langle \overline{D}^0 | H^{\Delta C=2} | D^0 \rangle \propto \sum_{i,j=d,s,b} V_{ui}^* V_{ci} V_{cj} V_{uj}^* m_i m_j \ . \quad (19.2.4)$$

Furthermore, due to the small magnitude of the $V_{ub}$ coupling, the $b$ quark contribution is negligible. The amplitude is proportional to $V_{us}^* V_{cs} V_{cd} V_{ud}^*$, and vanishes in the limit $m_d \sim m_s$. The $D^0$ mixing thus arises only as a consequence of the $SU(3)_{\rm flavor}$ symmetry breaking resulting from the small differences between the masses of down-type quarks.

The short distance contribution calculated from the box diagrams (like the one in Fig. 19.2.1) is proportional to the local $\Delta C = 2$ operator $\overline{u}\gamma_\mu(1-\gamma^5)c\overline{u}\gamma^\mu(1-\gamma^5)c$ and reads (Bigi and Uraltsev, 2001b; Burdman and Shipsey, 2003; Georgi, 1992)

$$\langle \overline{D}^0 | H^{\Delta C=2} | D^0 \rangle = \frac{G_F^2 m_c^2}{4\pi^2} V_{cs}^* V_{cd}^* V_{ud} V_{us} \frac{(m_s^2 - m_d^2)^2}{m_c^4} \times \langle \overline{D}^0 | \overline{u}\gamma_\mu(1-\gamma^5)c\overline{u}\gamma^\mu(1-\gamma^5)c | D^0 \rangle \ . \quad (19.2.5)$$

Equation (19.2.5) shows that this amplitude is doubly-Cabibbo suppressed. Furthermore, the factor $(m_s^2 - m_d^2)^2/m_c^4$ shows explicitly that $D^0$ mixing vanishes in the limit of exact $SU(3)_{\rm flavor}$ symmetry limit, when $m_s = m_d$.

It has been discussed (Bigi and Uraltsev, 2001b; Georgi, 1992) that the contribution shown in Eq. (19.2.5) is the first term of systematic expansion in $1/m_c$. The peculiar feature of this leading term is the strong suppression by the factor $(m_s^2 - m_d^2)^2/m_c^4$, which is not present any more in the subleading terms of the expansion. In fact, already the first subleading terms exhibits only a factor $m_s^2/m_c^2$. This indicates that the leading term cannot properly describe $D^0$ mixing, and that large long-distance contributions are present.

Taking into account the long-distance contributions by adding the second order terms from $H^{\Delta C=1}$ involving intermediate hadronic states $|n\rangle$ one obtains

$$(M - \frac{i}{2}\Gamma)_{ij} = m_D \delta_{ij} + \frac{1}{2m_D} \langle \overline{D}^0 | H^{\Delta C=2} | D^0 \rangle + \frac{1}{2m_D} \sum_n \frac{\langle \overline{D}^0 | H^{\Delta C=1} | n \rangle \langle n | H^{\Delta C=1} | D^0 \rangle}{m_D - E_n + i\epsilon} \ . \quad (19.2.6)$$

Neglecting, for the moment, the last term in the equation above and evaluating $M_{ij}, \Gamma_{ij}$ from the box diagram ($\langle \overline{D}^0 | H^{\Delta C=2} | D^0 \rangle$), and taking into account the relations for the eigenvalues of the effective Hamiltonian Eq. (10.1.15), it is possible to estimate the expected magnitude of the mixing parameter, $|x| \sim \mathcal{O}(10^{-5})$. Mixing with this rate would be unobservable with the present experimental facilities.

However, the last term in Eq. (19.2.6) also contributes, and represents the long distance contribution to the effective Hamiltonian. The contribution arises from on- and off-shell intermediate states $|n\rangle$ accessible to both $D^0$ and $\overline{D}^0$ (see for example Fig. 19.2.2). Due to the non-perturbative quantum chromodynamic nature of these effects, their contribution is much more difficult to evaluate than the short distance contribution illustrated in Fig. 19.2.1. In general, two methods have been exploited to estimate the magnitude of the mixing parameters that consider these long distance contributions: the exclusive approach (Donoghue, Golowich, Holstein, and Trampetic, 1986; Falk, Grossman, Ligeti, Nir, and Petrov, 2004), considering various possible exclusive intermediate states, and the Operator Product Expansion (OPE) method (Bigi and Uraltsev, 2001b; Georgi, 1992).

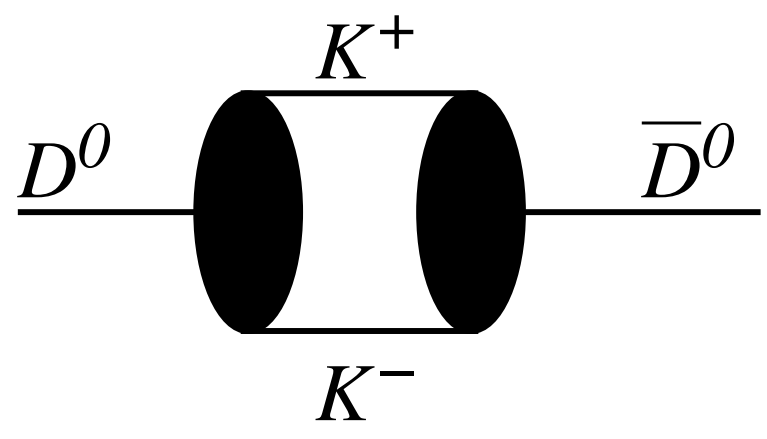

**Figure 19.2.2.** $K^+K^-$ as an example of an intermediate state accessible to both $D^0$ and $\overline{D}^0$, contributing to $D^0-\overline{D}^0$ mixing. Black circles represent quark processes for $D^0 \to K^+K^-$ and its charge conjugate, similar to the process shown in Fig. 19.2.4.

The former method is conceptually straightforward. If we consider only the possible two-body pseudoscalar intermediate states, $K^-\pi^+$, $K^-K^+$, $\pi^-\pi^+$ and $K^+\pi^-$, their contributions are summarized in Table 19.2.2.

The second and third columns of the table show the CKM elements for various states entering the expressions $|\langle \overline{D}^0 | H^{\Delta C=1} | n \rangle|^2$ and $\langle \overline{D}^0 | H^{\Delta C=1} | n \rangle \langle n | H^{\Delta C=1} | D^0 \rangle$, according to the Wolfenstein parameterization (see Chapter 16). The minus signs in the third column are a consequence of a relative sign between the $V_{us}$ and $V_{cd}$ elements of the CKM matrix. Naïvely one expects the contribution to the mixing, when summed over the considered states, to vanish, *i.e.*

$$\sum_n \langle \overline{D}^0 | H^{\Delta C=1} | n \rangle \langle n | H^{\Delta C=1} | D^0 \rangle = 0. \quad (19.2.7)$$

**Table 19.2.2.** Pairs of pseudoscalar mesons accessible to $D^0$ and $\overline{D}^0$. The CKM suppression factors entering two expressions in Eq. (19.2.6) are listed in the second and the third column. Due to the $SU(3)_{\rm flavor}$ symmetry breaking the ratios of branching fractions ($\mathcal{B}$) differ from the naïve CKM expectation (fourth column). From these one can estimate the contributions important for the mixing amplitude, listed in the fifth column.

| State | $\lvert\langle\overline{D}^0\lvert H^{\Delta C=1}\rvert n\rangle\rvert^2 \propto$ | $\langle\overline{D}^0\lvert H^{\Delta C=1}\rvert n\rangle\langle n\lvert H^{\Delta C=1}\rvert D^0\rangle \propto$ | Measured $\mathcal{B}/\mathcal{B}_0$ | Contribution to mixing |
|---|---|---|---|---|
| $K^+\pi^-$ | 1 | $-\lambda^2$ | $r_1$ | $-\sqrt{r_1 r_4}\lambda^2$ |
| $K^-K^+$ | $\lambda^2$ | $\lambda^2$ | $r_2\lambda^2$ | $r_2\lambda^2$ |
| $\pi^-\pi^+$ | $\lambda^2$ | $\lambda^2$ | $r_3\lambda^2$ | $r_3\lambda^2$ |
| $K^-\pi^+$ | $\lambda^4$ | $-\lambda^2$ | $r_4\lambda^4$ | $-\sqrt{r_1 r_4}\lambda^2$ |
| | | $\Sigma = 0$ | | $\Sigma = \lambda^2\big(r_2 + r_3 - 2\sqrt{r_1 r_4}\big)$ |

However, $SU(3)_{\rm flavor}$ symmetry breaking causes differences beyond those in the CKM factors contributing to the various states. These can be estimated from the measured branching fractions as given schematically in the fourth column of the table. If one denotes the expected $\mathcal{B}$ for a final state $i$ by $\lambda^n\mathcal{B}_0^i$, the measured $\mathcal{B}$ is $r_i\lambda^n\mathcal{B}_0^i$. The factors $r_i \neq 1$ point to $SU(3)$ symmetry breaking. The contributions to the mixing amplitude as evaluated using the measured branching fractions is given in the last column. The sum over all the states yields $\lambda^2(r_2 + r_3 - 2\sqrt{r_1 r_4})$ which is in general different from zero. In order to precisely calculate the mixing parameters, one would of course need to take into account other possible intermediate states for some of which the branching fractions are not well known. Nevertheless it is possible to estimate the magnitude of the mixing parameters to be $|x| \lesssim \mathcal{O}(10^{-3})$ and $|y| \lesssim \mathcal{O}(10^{-2})$ (Falk, Grossman, Ligeti, Nir, and Petrov, 2004). These are in rough agreement with the expectation of the OPE method, $|x|, |y| \lesssim \mathcal{O}(10^{-3})$ (Bigi and Uraltsev, 2001b).

In summary, theoretical expectations based on the SM are that the mixing rate in the $D^0$ system is small, arising mainly from long distance contributions that are difficult to estimate. Despite this difficulty, the measurement of mixing in this system over a wide range of decay modes and with sufficient precision can provide important constraints on possible NP parameters. Importantly and uniquely, such constraints will be complementary to those from down-type FCNC processes.

On the experimental side, observations of $D^0$ mixing come predominantly from measurements of the time evolution of neutral $D$ meson decays to final states $f$ that are accessible to both $D^0$ and $\overline{D}^0$. In such cases, as illustrated in Fig. 19.2.3, direct decay and decay preceded by mixing interfere. Mixing can, in principle, also be observed in semi-leptonic decays of $D^0$ mesons where the leptons have the wrong sign. The only way, in the SM, for such decays to occur is through $D^0$-$\overline{D}^0$ mixing, with a tiny rate $\propto (x^2 + y^2)/2 \sim 5 \times 10^{-5}$.

At the $B$ Factories, $c\bar{c}$ pairs are produced in the electroweak annihilation of electrons and positrons. In the fragmentation of primary quarks various species of charmed hadrons together with lighter hadrons are produced. In general, pairs of $D^0$ and $\overline{D}^0$ mesons are not in a quantum correlated state.[120] The time evolution of the mass eigenstates $D_{1,2}$ (*i.e.* of the eigenstates of the effective Hamiltonian, see Section 10.1) follows a simple exponential form

$$|D_{1,2}(t)\rangle = e^{-i(m_{1,2} - i\Gamma_{1,2}/2)t}|D_{1,2}(t=0)\rangle \quad , \qquad (19.2.8)$$

and the form of the experimentally accessible flavor states is

$$\begin{aligned} |D^0(t)\rangle &= \frac{1}{2p}\left[|D_1(t)\rangle + |D_2(t)\rangle\right] \\ |\overline{D}^0(t)\rangle &= \frac{1}{2q}\left[|D_1(t)\rangle - |D_2(t)\rangle\right] \quad . \end{aligned} \qquad (19.2.9)$$

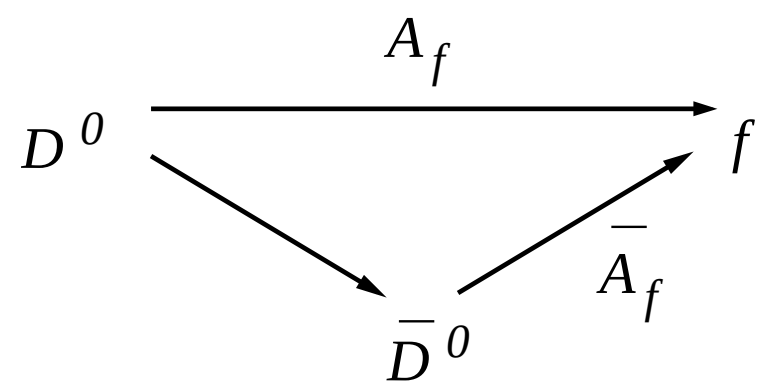

**Figure 19.2.3.** Illustration of interference between direct $D$ meson decays and decays through mixing into the final state $f$, accessible to either $D^0$ or $\overline{D}^0$.

Writing out the time evolution of the mass eigenstates we arrive at

$$\begin{aligned} |D^0(t)\rangle = &\Bigg[|D^0\rangle\cosh\left(\frac{ix+y}{2}\Gamma t\right) - \qquad (19.2.10) \\ &-\frac{q}{p}|\overline{D}^0\rangle\sinh\left(\frac{ix+y}{2}\Gamma t\right)\Bigg]e^{(-im-\Gamma/2)t} \\ |\overline{D}^0(t)\rangle = &\Bigg[|\overline{D}^0\rangle\cosh\left(\frac{ix+y}{2}\Gamma t\right) - \\ &-\frac{p}{q}|D^0\rangle\sinh\left(\frac{ix+y}{2}\Gamma t\right)\Bigg]e^{(-im-\Gamma/2)t} \quad . \end{aligned}$$

[120] This is different from $D^0\overline{D}^0$ production at the charm threshold. There, the pair of $D$ mesons from the decay of a $\psi(3770)$ is in a quantum correlated state similar to a pair of $B$ mesons produced in the decay of an $\Upsilon(4S)$.

In the above equations we use the notation $|D^0\rangle$ and $|\overline{D}^0\rangle$ for the two flavor eigenstates.[121] The time evolutions can be simplified. Since $|x|,\ |y| \ll 1$, the time dependent decay rate of an initially produced $D^0$ meson to a final state $f$ can be written as

$$\frac{d\Gamma(D^0 \to f)}{dt} \propto \left| A_f - \frac{q}{p}\frac{ix+y}{2}\overline{A}_f \Gamma t\right|^2 e^{-\Gamma t} \quad . \quad (19.2.11)$$

$A_f$ and $\overline{A}_f$ denote the instantaneous decay amplitudes $\langle f|D^0\rangle$ and $\langle f|\overline{D}^0\rangle$, respectively. Analogously, for an initially produced $\overline{D}^0$ one finds

$$\frac{d\Gamma(\overline{D}^0 \to f)}{dt} \propto \left| \overline{A}_f - \frac{p}{q}\frac{ix+y}{2}A_f \Gamma t\right|^2 e^{-\Gamma t} \quad . \quad (19.2.12)$$

Further simplification arises if we ignore *CP* violation in mixing and mixing-induced *CP* violation (see further discussion in Section 19.2.1.3), *i.e.* assume that $p = q = 1/\sqrt{2}$:

$$\begin{aligned}\frac{d\Gamma(D^0 \to f)}{dt} &\propto \big[|A_f|^2 - |\overline{A}_f||A_f|(x\sin\delta_f + y\cos\delta_f)(\Gamma t) \\ &+|\overline{A}_f|^2\frac{x^2+y^2}{4}(\Gamma t)^2\big]e^{-\Gamma t} \ , \\ \frac{d\Gamma(\overline{D}^0 \to f)}{dt} &\propto \big[|\overline{A}_f|^2 + |A_f||\overline{A}_f|(x\sin\delta_f - y\cos\delta_f)(\Gamma t) \\ &+|A_f|^2\frac{x^2+y^2}{4}(\Gamma t)^2\big]e^{-\Gamma t},\end{aligned} \quad (19.2.13)$$

where $\delta_f$ is $\arg(A_f/\overline{A}_f)$. In Eqs (19.2.13), the first terms represent direct decay and the third terms describe decay preceded by mixing. The middle terms, linear in $t$, encode the interference between the two processes. They are also linear in the small parameters $x$ and $y$, and it is those terms that make the time dependent decay rates sensitive to the values of the mixing parameters.

The specific dependence on these quantities differs for various final states $f$. When the amplitude for direct decay is Cabibbo-favored (CF), the first term dominates and the decays are, effectively, exponential and mixing can be neglected. When, however, the direct decay amplitude is doubly Cabibbo-suppressed (DCS), all three terms are of the same order of magnitude and such final states offer the maximum sensitivity to the mixing parameters.

#### 19.2.1.3 *CP* violation

For most *BABAR* and Belle measurements of $D^0-\overline{D}^0$ meson mixing, results on the mixing parameters are extracted first assuming that *CP* violation can be neglected, using Eq. (19.2.13). In some measurements a second fit to the data distributions is performed examining the possibility for differences between $D^0$ and $\overline{D}^0$, and *CP* asymmetries are measured.

[121] The experimental identification of these states is possible through decays into flavor specific final states, for example $\langle D^0|K^-\ell^+\nu_\ell\rangle \neq 0$ while $\langle \overline{D}^0|K^-\ell^+\nu_\ell\rangle = 0$.

The decay rates in Eqs (19.2.11) and (19.2.12) depend also on the *CP* violating parameter $q/p$ (cf. Chapter 10). In general, *CP* violation in the SM for processes involving charmed hadrons is expected to be tiny. This is conveniently seen in the parameterization of the CKM matrix given in Eq. (16.4.4). *CP* violation arises from the phase in the CKM matrix, and the elements of the matrix related to the first two generations of quarks, which appear in the charmed hadron processes, are almost real. Hence the magnitude of the *CP* violating effect is expected to be small. As an example, consider the Cabibbo suppressed decay $D^0 \to \pi^+\pi^-$, shown in Fig. 19.2.4. The relevant CKM phase entering the ratio of amplitudes for this decay and its charge conjugate is

$$\arg\frac{\langle\pi^+\pi^-|D^0\rangle}{\langle\pi^+\pi^-|\overline{D}^0\rangle} = 2\ \arg(V_{cd}^*V_{ud}) \quad . \quad (19.2.14)$$

To see the expected magnitude of the *CP* violation due to this weak phase one needs to consider the parameterization of the CKM matrix at least up to the order $\lambda^5$. The usually adopted Wolfenstein parameterization to order $\lambda^3$ is given in Eq. (16.4.4) in Section 16. The parameterization including the order of $\lambda^5$ reads (see for example review "*CP* violation in meson decays" in Beringer et al. (2012))

$$\begin{pmatrix} 1-\lambda^2/2-\lambda^4/8 & \lambda & A\lambda^3(\rho-i\eta) \\ -\lambda+A^2\lambda^5[1-2(\rho+i\eta)]/2 & 1-\lambda^2/2-\lambda^4(1+4A^2)/8 & A\lambda^2 \\ A\lambda^3[1-(1-\lambda^2/2)(\rho+i\eta)] & -A\lambda^2+A\lambda^4[1-2(\rho+i\eta)]/2 & 1-A^2\lambda^4/2 \end{pmatrix} +\mathcal{O}(\lambda^6). \quad (19.2.15)$$

Evaluating Equation (19.2.14) yields

$$2\ \arg(V_{cd}^*V_{ud}) \approx 2A^2\lambda^4\eta = 1.2\times10^{-3} \quad , \quad (19.2.16)$$

where in the last line we use values of parameters from Charles et al. (2005). *CP* violation effects in the charm sector are thus of the order of $10^{-3}$. Recently some authors have argued that the *CP* asymmetries in Cabibbo-suppressed decays can be larger by some factor (Brod, Kagan, and Zupan, 2011).

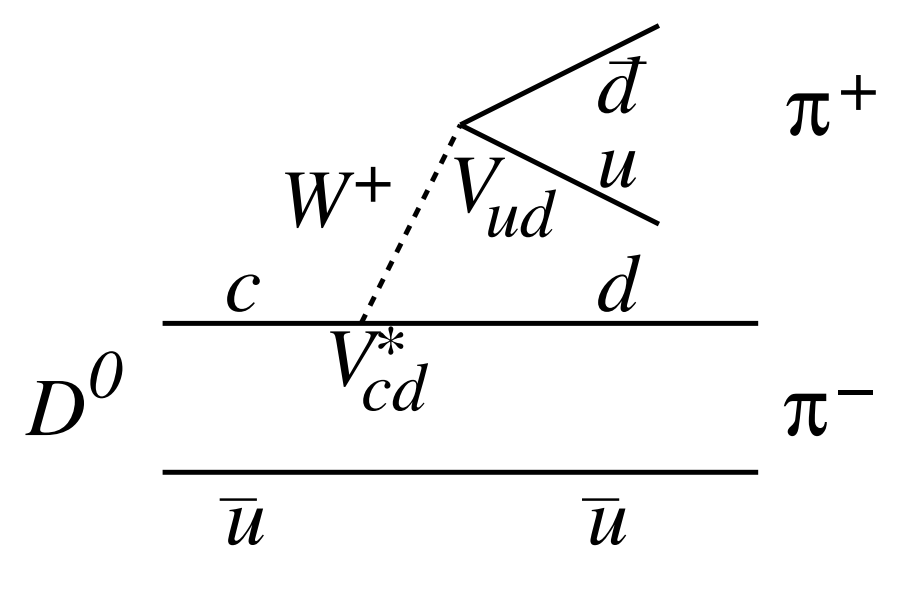

**Figure 19.2.4.** $D^0 \to \pi^+\pi^-$ decay with the corresponding CKM elements.

In measurements of various observables sensitive to *CP* violation, the parameterization

$$\left|\frac{q}{p}\right|^2 \equiv 1 + A_M \qquad (19.2.17)$$

is often used. Three types of *CP* violating effects can be distinguished, as in any other neutral mesons system (see Section 16.6). First, *CP* violation in mixing occurs if $A_M \neq 0$ (alternatively, $|q/p| \neq 1$). *CP* violation in decay is present if $|A_f/\overline{A}_{\overline{f}}| \neq 1$. This effect is sometimes parameterized in terms of

$$\left|\frac{A_f}{\overline{A}_{\overline{f}}}\right|^2 \equiv 1 + A_D^f \quad . \qquad (19.2.18)$$

It should be noted that while the parameter $A_M$ is universal for all $D^0$ decays, the parameter $A_D^f$ depends on the final state $f$. This type of *CP* violation can only occur in decays to which at least two processes with different weak and strong phases contribute (see Eq. (16.6.5) in Chapter 16). For $D$ mesons this is only possible in Cabibbo suppressed decays, where both tree (*e.g.* see Fig. 19.2.4) and penguin (Fig. 19.2.5) diagrams are possible. Finally,

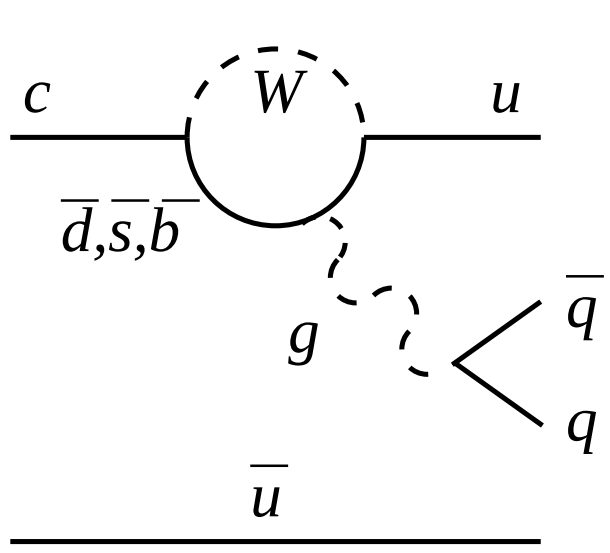

**Figure 19.2.5.** Penguin diagram contributing to Cabibbo suppressed decays of $D$ mesons.

there is a possibility of mixing-induced *CP* violation. This is characterized by

$$\mathrm{Im}\left[\frac{q}{p}\frac{\overline{A}_f}{A_f}\right] \equiv \mathrm{Im}[\lambda_f] \neq 0 \quad . \qquad (19.2.19)$$

Using previous parameterizations (and keeping only linear terms in small quantities $A_M$ and $A_D^f$), $\lambda_f$ is sometimes expressed as

$$\lambda_f = \sqrt{R_D^f}(1 + A_M/2)(1 - A_D^f/2)e^{-i(\delta_f - \phi)} \quad . \quad (19.2.20)$$

In the above expression, the parameter $R_D^f$ is not related to *CP* violation, but to the possible Cabibbo suppression, $R_D^f = |A_{\overline{f}}/A_f|^2 = |\overline{A}_f/\overline{A}_{\overline{f}}|^2$. The phase $\delta_f$ includes a possible strong phase as well as the weak phase difference between the two amplitudes. For decays to *CP* eigenstates (such as $K^+K^-$) $R_D^f = 1$ and the strong phase difference is zero. If we neglect the weak phase differences of the order of $10^{-3}$ then the only source of this type of *CP* violation in decays to *CP* eigenstates is $\phi \equiv \arg(q/p) \neq 0$ which can arise due to some unknown NP processes.

In the following sections we neglect *CP* violation in the neutral $D$ meson system and address this important phenomenon separately in Sections 19.2.6 and 19.2.7.

#### 19.2.1.4 $D^0$ Mixing in New Physics Models

Values of the mixing parameters for the $D^0$ meson system can differ significantly from SM estimates in several NP models. In Golowich, Hewett, Pakvasa, and Petrov (2007) the authors examined a large number of such models and calculated the contributions of new particles and processes to the mixing parameters $x$ and $y$. Due to the large uncertainties in SM calculations, the values are obtained for specific NP contributions alone.[122] In this approach the parameters of a large majority of the models considered are additionally constrained by the measured values of the $D^0$ mixing parameters. An example of the sensitivity of the value of $x$ to the mass and CKM elements of a possible fourth generation $b'$ quark is shown in Fig. 19.2.6.[123]

Not only $x$ but also $y$ can be sensitive to some of the NP models considered. As pointed out in Eq. (19.2.5), the mixing parameters in the SM vanish in the exact $SU(3)_{\text{flavor}}$ limit. Moreover, the contribution to the mixing parameters enters only as a second order effect in the $SU(3)_{\text{flavor}}$ breaking. Hence the NP contributions to $y$ could be significant for the models in which the contributions do not vanish in the $SU(3)_{\text{flavor}}$ symmetry limit (Golowich, Pakvasa, and Petrov, 2007). An example is the $R$-parity violating SUSY model, where the slepton mediated interaction is not suppressed in the $SU(3)_{\text{flavor}}$ symmetry limit and could lead to values as high as $|y| \approx 3.7\%$ for $M_{\tilde{\ell}} = 100\ \mathrm{GeV}/c^2$. Section 25.2 is a more general discussion on how one can constrain benchmark NP models using constraints from the $B$ Factories.

#### 19.2.1.5 General experimental remarks

Two experimental ingredients are necessary to exploit the decay time distributions of Eqs (19.2.11) and (19.2.12) for measuring the mixing parameters. The first one is the determination of the initial neutral $D$ meson flavor ("flavor tagging"), that is used to determine whether a $D^0$ or a $\overline{D}^0$ was produced at $t = 0$. The second ingredient is the determination of the decay time of a the neutral $D$ meson, based on the measurement of its decay length between the production and decay point.

[122] The neglection of the SM contribution leads to less restrictive limits in most cases. If there is also the SM contribution to the magnitudes of the mixing parameters, then the contribution from NP is smaller and hence the constraint gets more severe.

[123] Note that severe lower limits on the $b'$ mass arise also from direct searches at the LHC. For example, the CMS collaboration finds $m(b') > 611$ GeV at 95% C.L., assuming $\mathcal{B}(b' \to Wt) = 100\%$ and hence $|V_{ub'}V_{cb'}| = 0$ (Chatrchyan et al., 2012c). The ATLAS collaboration provides lower mass limits as a function of $\mathcal{B}(b' \to Wt)$ (Gauthier, 2013); for values of the latter between 0.8 and 1.0 the limit is $m(b') > 700$ GeV at 95% C.L. This is complementary to the limit arising from $D^0 - \overline{D}^0$ mixing, which provides an upper limit on the $b'$ mass for a given value of $|V_{ub'}V_{cb'}|$.

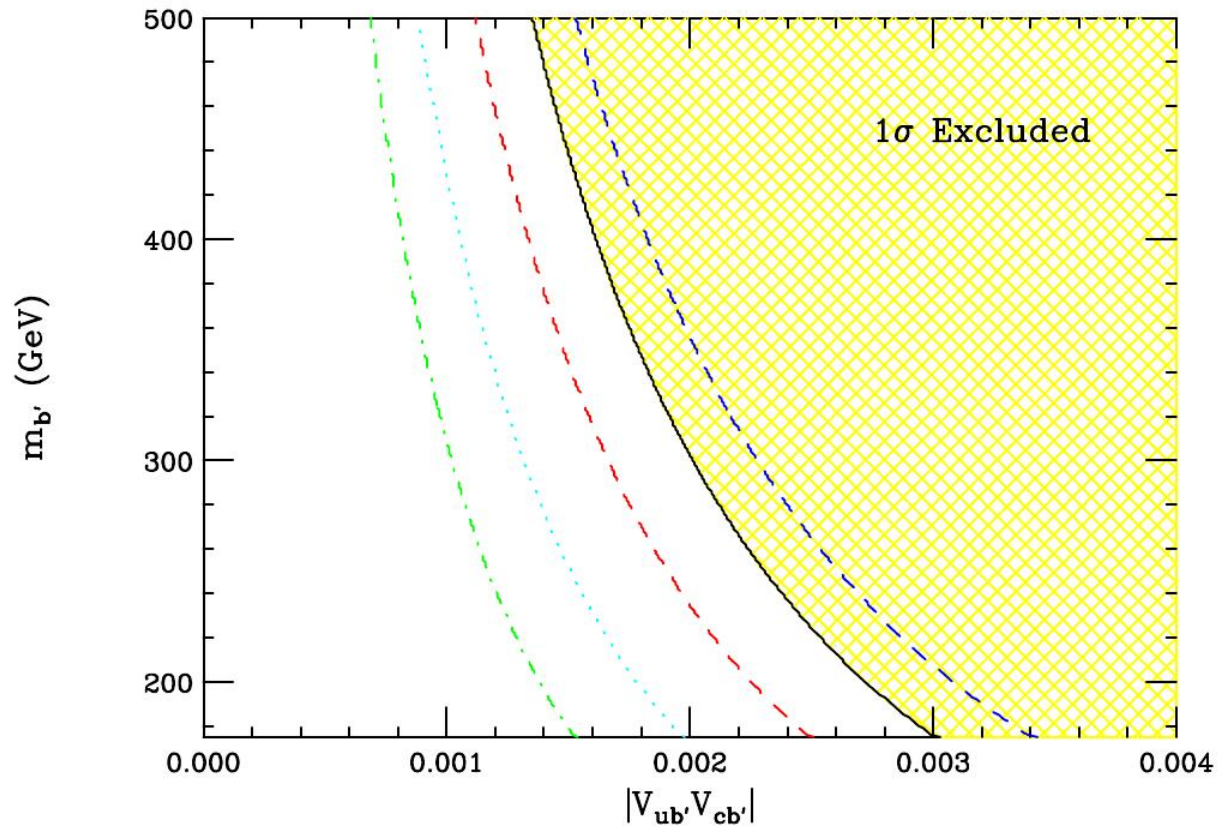

**Figure 19.2.6.** Contours of $x$ in the fourth generation $b'$ quark mass and CKM elements ($|V_{ub'}V_{cb'}|$) plane (see text for explanation). Contours are shown for $x = [15.0, 11.7, 8.0, 5.0, 3.0] \times 10^{-3}$ (from right to left). From (Golowich, Hewett, Pakvasa, and Petrov, 2007).

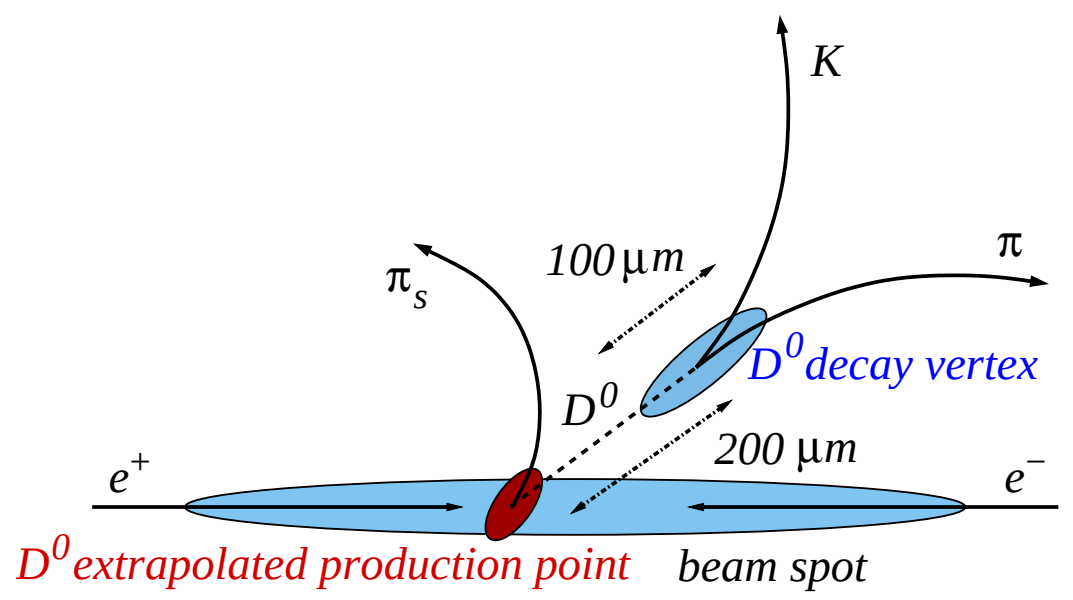

**Figure 19.2.7.** Illustration of $D$ meson decay length determination with some typical dimensions.

The initial flavor tagging is based on the decay chain $D^{*+} \to D^0\pi^+ \to f\pi^+$, and its charge conjugate $D^{*-} \to \overline{D}^0\pi^- \to \overline{f}\pi^-$. The charge of the pion produced in the decay of a $D^*$ determines the flavor of the neutral daughter $D$ meson. It should be noted that the average laboratory momentum of such pions is low, around 400 MeV/$c$, and hence they are usually denoted as "slow pions", $\pi_s$. The use of $D$'s from $D^*$'s also reduces the amount of background in the event samples selected. The difference between the invariant masses of the $D^{*+}$ and the $D^0$ meson, $\Delta m = m(f\pi^+) - m(f)$, is a powerful selection variable that restricts the amount of combinatorial background considerably. The effectiveness is further enhanced by the excellent experimental resolution in $\Delta m$ arising from the cancellation of experimental uncertainty in the determination of the momenta of particles comprising the $D^0$. In some measurements an equivalent variable $Q = \Delta m - m_\pi$ is used instead of $\Delta m$.[124]

The determination of the $D$ meson decay length is illustrated in Fig. 19.2.7 with typical dimensions indicated. The accuracy depends critically on the silicon detectors of the experiments, described in Chapter 2. The decay point is obtained from fitting the $D$ meson decay products to a single vertex. The $D^0$ production point is obtained by intersecting the $D^0$ flight direction, determined by its momentum vector and decay vertex, with the $e^+e^-$ interaction region. The charged ("slow") pion from the $D^*$ decay can be fitted to a common space point together with the $D$ momentum and spatial constraints from the interaction region. This improves the momentum accuracy of the slow pion and, thereby, the experimental resolution in $\Delta m$. The average $D^0$ decay length at the $B$ Factories is about 200 $\mu$m. The precision by which the interaction region is known is given in Chapter 6. The resulting resolution on the decay length is approximately 100 $\mu$m, depending on the final state considered. The proper decay time is obtained from the decay length $l$, momentum $p$ and nominal mass of the $D$ mesons $m_D$ as $m_D \boldsymbol{l} \cdot \boldsymbol{p}/p^2$.

In the measurement of the $D^0$ decay times, it is assumed that the $D^0$'s are produced at the primary $e^+e^-$ production vertex. For $D^0$'s from $B$ meson decays, this assumptions results in a biased measurement due to the finite decay time of the $B$ mesons and the corresponding $D^0$ decay time distribution cannot be described by Eq. (19.2.13). This complication (and the related systematic uncertainties that would result from it) is avoided by requiring the CM momentum of the tagging $D^*$ mesons to exceed the kinematic limit for $B \to D^*X$ decays.[125] The typical minimal value required for the $D^*$ momentum is 2.5 GeV/$c$.

To measure the mixing parameters from the decay time distributions one needs an accurate description of the resolution, to be convolved with the expected distributions of Eqs (19.2.11) and (19.2.12). The simplest way of describing the resolution is to use a Gaussian function with an appropriate width, but this rarely turns out to be accurate enough. In general, the resolution function is more complicated since the resolution in an individual $D$ meson decay time measurement depends on the specific kinematics of the decay. In most cases the resolution function can be successfully parameterized with two or three Gaussian functions of different widths. In an unbinned fit to the decay time distribution the width of the resolution function can be implemented on an event-by-event basis by multiplying the estimated accuracy of the decay time as obtained from the vertex fits by a factor which is either a free parameter in the fit or determined from some control sample of decays, if one exists. For a two Gaussian reso-

[124] Sometimes in the literature the same variable is denoted as $q$, not to be confused with $q$ from Eq. (19.2.9), for example.

[125] The largest momentum of a $D^*$, in a two body $B \to D^*\pi$ decay in the CM system, where the $B$ meson is approximately at rest, is about 2.3 GeV/$c$.

lution parameterization, the likelihood value for the $i$-th signal event is written as

$$\mathcal{L}_i(x,y) = \int_0^\infty dt' \frac{d\Gamma}{dt'}(t';x,y)\Bigg[ f e^{(t_i-t')^2/2S_1\sigma_i^2} + (1-f) e^{(t_i-t')^2/2S_2\sigma_i^2} \Bigg]. \quad (19.2.21)$$

in which $\sigma_i$ is the uncertainty reported by the vertex reconstruction code for this event, and where $S_{1,2}$ are the scale factors mentioned above.

In constructing the resolution function special care should be devoted to possible biases in decay length measurements. These can occur due to a combination of the kinematic properties of decays (mainly arising from dependence on the opening angle of the final state tracks in the laboratory frame) and small residual misalignments of individual detector modules.[126] Such effects could result in a biased measurement of the mixing parameters. These effects can be included in the resolution function through various more sophisticated parameterizations, for example by allowing the mean value of the Gaussian function to deviate from zero. In some cases a further dependence of the mean value on the kinematic properties of the decay is required: for example, the bias can depend on the opening angle of the two tracks in the case of $D^0 \to K^\pm\pi^\mp$ decays; the opening angle is in turn strongly correlated with the invariant mass and hence the bias can vary as the invariant mass changes across the width of the signal peak, as shown in the example below in Fig. 19.2.18. In multi-body decays the bias and hence the mean value of the resolution function can also vary depending on the position of the decay in the Dalitz plane.

The parameterization of the resolution function must be studied to understand what effect such biases have on the determination of the mixing parameters. For this, a number of special simulated samples of events are produced with non-zero values for $x$ and $y$ and then subjected to the same selection, reconstruction and fitting procedures as the data to check for possible bias in these mixing parameters. Another test is to fit the Cabibbo-favored decays (*e.g.* $D^0 \to K^-\pi^+$) where the effects of mixing are negligible to obtain the average decay lifetime. The resulting values are compared to the current world average of $(410.1 \pm 1.5)$ fs (Beringer et al., 2012) to check for possible biases arising from the parameterization of the resolution function.

## 19.2.2 Hadronic wrong-sign decays

The earliest attempts to find evidence for $D^0 - \overline{D}^0$ mixing at the $B$ Factories have used limited samples of "wrong sign" (WS) hadronic decays. The term WS decays is used for decays to flavor specific final states which are either DCS or can proceed through the mixing process. Both *BABAR* and Belle have carefully studied the time dependence of the $D^0 \to K^+\pi^-$ + c.c. decays. These studies initially set limits on the mixing parameters and ultimately provided evidence for charm mixing at the $3.9\sigma$ level, as well as setting limits on $CP$ violation (Zhang, 2006; Aubert, 2007j). These decays, expected to be particularly sensitive to mixing (Bigi and Uraltsev, 2001b; Blaylock, Seiden, and Nir, 1995), were first used to search for this phenomenon by the E791 collaboration (Aitala et al., 1998). This search resulted in the upper limit for the ratio of decays with and without mixing of $r_{\rm mix} < 0.85\%$ at 90% C.L. The CLEO experiment provided limits on the mixing parameters using the same decay mode (see below for the definitions of $x'^2$ and $y'$), $x'^2 < 0.082\%$ and $-5.5\% < y' < 1.0\%$ at 95% C.L. (Godang et al., 2000).

Another type of WS decays that played important role in $D^0 - \overline{D}^0$ mixing measurements is the $D^0 \to K^+\pi^-\pi^+\pi^-$ ($K3\pi$) decay mode. It has been used to search for charm mixing since the earliest days of charm physics. In 1977, the SLAC/LBL magnetic detector at SPEAR was used to search for "same-sign" $K3\pi$ events where the kaon in the $D^0$ decay had the same charge as the kaon in the recoil products opposite the $D^0$ (Goldhaber et al., 1977). The result was that less than 18% (at 90% C.L.) of the observed $D^0$ decays exhibited the same-sign signature, which was consistent with the amount of charged particle mis-identification expected from their time-of-flight system. Other searches for wrong-sign decays also saw no signal but did set limits on the wrong-sign rate. In 1995, E791 (Aitala et al., 1998) reported a wrong-sign measurement of $D^0 \to K3\pi$ showing that the mixing rate is less than 0.85% at 90% C.L. This result was soon followed by evidence for the wrong-sign $K3\pi$ decay from CLEO with a relative rate with respect to the RS decays of $R_{\rm WS}^{K3\pi} = [0.41^{+0.12}_{-0.11} \pm 0.04 \pm 0.10]\%$, where the first error is statistical, the second is systematic, and the third is due to phase space (Dytman et al., 2001). The result is consistent with the WS decays arising solely from DCS decays.

### 19.2.2.1 Method

The time-dependence of mixing in the decays of neutral mesons has been discussed in detail in Section 10.1 and, for charm mesons, in Section 19.2.1. The term WS decays is related to Eq. (19.2.13). If we select $f = K^-\pi^+$ the $D^0$ decays are CF and the second and third term in the expression for $d\Gamma(D^0 \to K^-\pi^+)/dt$ are negligible. On the other hand, the $\overline{D}^0$ decays are DCS. Hence the contribution of direct decay (first term of $d\Gamma(\overline{D}^0 \to K^-\pi^+)/dt$) is not much larger than the decay preceded by mixing ($\overline{D}^0 \to D^0 \to K^-\pi^+$, last term) and the interference between the two (second term). Hence the decays of an initially produced $\overline{D}^0$ to $K^-\pi^+$ (and analogously of an initially produced $D^0$ to $K^+\pi^-$), which are either DCS or can proceed through the mixing process, are called WS decays (as opposed to right sign (RS) decays, $\overline{D}^0 \to K^+\pi^-$ and $D^0 \to K^-\pi^+$). Experimentally, the WS and RS decays are selected based on the correlation between the charge of the slow pion from the tagging $D^{*\pm}$ decay and the charge of the kaon from the $D$ meson decay.

[126] For example, misalignment between the silicon vertex device and the central tracking chamber.

Evaluating the expressions in Eqs (19.2.13), we get the time dependence of the $D^0 \to K\pi$ WS decays for small values of the mixing parameters $x$ and $y$, as defined in Eq. (19.2.1), and assuming that *CP* is conserved:

$$\frac{\Gamma_{\rm ws}}{e^{-\Gamma t}} \propto R_{\rm D} + \sqrt{R_{\rm D}} y' \, \Gamma t + \frac{x'^2 + y'^2}{4} (\Gamma t)^2 , \quad (19.2.22)$$

where we used a short hand notation $R_{\rm D} = R_{\rm D}^{K\pi}$. The parameters $x'$ and $y'$ are the parameters $x$ and $y$ rotated by a strong phase difference $\delta_{K\pi}$ between the CF and DCS decays:

$$\begin{aligned} x' &= x \cos\delta_{K\pi} + y \sin\delta_{K\pi} \\ y' &= y \cos\delta_{K\pi} - x \sin\delta_{K\pi} \; ; \end{aligned} \quad (19.2.23)$$

$\delta_{K\pi}$ is defined through $\overline{A}_{K^-\pi^+}/A_{K^-\pi^+} = -\sqrt{R_{\rm D}^{K\pi}} e^{-i\delta_{K\pi}}$, c.f. Eq. (19.2.20). In the following, often the shorter notation $\delta$ for $\delta_{K\pi}$ is also used. Equation (19.2.22) reveals an exponential decay modulated by terms linear and quadratic in $t$. The differing time development of the three terms — constant, linear, and quadratic — can be used to separate the contributions from DCS decays and decays with mixing. In this approximation, the time-integrated rate, $R_{\rm WS}$, is then

$$R_{\rm WS} = R_{\rm D} + y' \sqrt{R_{\rm D}} + \frac{x^2 + y^2}{2}. \quad (19.2.24)$$

The mixing rate $R_M$ is defined as $R_M = (x^2 + y^2)/2 = (x'^2 + y'^2)/2$. A mixing-only search (not allowing for *CP* violation) combines $D^0$ and $\overline{D}^0$ decays together. To include possible effects from *CP* violation, both *BABAR* and Belle apply Eq. (19.2.22) to $D^0$ and $\overline{D}^0$ decays separately, as discussed further in Section 19.2.7.

#### 19.2.2.2 Measurements of $D^0 \to K^+\pi^-$ decays

Belle and *BABAR* used 400 fb$^{-1}$ and 384 fb$^{-1}$, respectively, in their 2006 and 2007 studies of $K\pi$ WS mixing (Zhang, 2006; Aubert, 2007j). Both experiments used the decay chain $D^{*+} \to \pi_s^+ D^0$, $D^0 \to K^\pm\pi^\mp$, using the large statistics, RS decay $D^0 \to K^-\pi^+$ + c.c. to determine most of the parameters in the *p.d.f.*s used to describe the decay structure in four independent variables: the $D^0$ candidate mass $m_{K\pi}$, the mass difference $\Delta m$ (or $q$), the reconstructed decay time $t$, and the event-by-event decay time error, $\sigma_t$. Both, *BABAR* and Belle perform a "blind" analysis (see Chapter 14), where the analysis procedure is finalized before examining the mixing results.

In both measurements particle identification criteria on charged kaon and pion candidates are imposed, as well as requirements on the quality of selected tracks and/or momentum of slow pions. Belle requires the momentum of the $D^0$ candidate in the center-of-mass frame to be $>$ 2.7 GeV/$c$, which reduces the number of candidates originating in combinatoric background and $B\overline{B}$ events. *BABAR* uses a requirement of 2.5 GeV/$c$ for the same purpose.

Both experiments reconstruct the $D^*$ decay chain by performing a vertex fit of the $K$ and $\pi$ candidate tracks, and extrapolating the flight direction of the resulting $D^0$ candidate back to the interaction region; the resulting intersection is taken as the $D^*$ decay vertex (see Fig. 19.2.3). The $\pi_s$ candidate is constrained to originate from the $D^*$ vertex. Requirements on the $\chi^2$ of all vertex fits are imposed. The resulting decay time $t$ along with its associated uncertainty $\sigma_t$ is calculated from the fitted vertex positions and their uncertainties, and from the reconstructed $D^0$ momentum. A typical value of $\sigma_t$ is (130-160) fs.

Some events have multiple $D^*$ candidates (about 5%).[127] At Belle, the events in which at least two $D^*$ candidates with the opposite charge are found are rejected, which reduces the random $\pi_s$ background (see below) by about 30% and the signal by 1%. In the case of same-sign $D^*$ candidates, the one with the best vertex fit $\chi^2$ was retained. At *BABAR*, if the $D^*$ candidate shared daughter tracks with other $D^*$ candidates, only the one with the largest $P(\chi^2)$ was used.

There are several background components required in order to correctly describe the two-dimensional RS and WS $(m_{K\pi}, \Delta m)$ or $(m_{K\pi}, q)$ distributions in addition to the signal components. *BABAR* defines three background types: "random $\pi_s$" background, where an unassociated $\pi_s$ candidate is paired with a good $D^0$ candidate; "misreconstructed $D^0$" background, where a $\pi_s$ candidate is paired with a $D^0$ that was reconstructed incorrectly, either with an incorrect particle hypothesis for one of the daughter tracks, or a multi-body decay reconstructed as a two-body decay; and combinatoric background. In RS events, misreconstructed $D^0$ candidates are primarily from semileptonic decays; in WS events, from "swapped $D^0$" candidates which are RS $D^0 \to K\pi$ decays with the $K$ and $\pi$ particle identifications interchanged.

Belle defines four background types: random $\pi_s$ (rnd) as above; those with a correct $\pi_s$ but with a misreconstructed $D^0$ decaying to ($\geq$ 3)-body final states (d3b); charged $D^+$ and $D_s^+$ decays (ds3); and combinatoric background (cmb).

Both Belle and *BABAR* determine the shape of the background *p.d.f.*s from MC simulation and only their amplitudes are allowed to vary in the fits. Signal events peak in both $m_{K\pi}$ and $\Delta m$ (or $q$). Random $\pi_s$ events peak in $m_{K\pi}$ but not $\Delta m$ (or $q$). Misreconstructed $D^0$ decays peak in $\Delta m$ (or $q$) but not $m_{K\pi}$. Combinatoric events do not peak in either $m_{K\pi}$ or $\Delta m$ (or $q$).

From the fits to the $(m_{K\pi}, \Delta m)$ or $(m_{K\pi}, q)$ distributions Belle finds $1,073,993 \pm 1108$ RS signal events and $4024 \pm 88$ WS signal events. Fig. 19.2.8 shows $m_{K\pi}$ and $q$ distributions from the Belle analysis for RS and WS data. *BABAR* fits the RS and WS ($m_{K\pi}$,$\Delta m$) plane simultaneously using shared parameters that describe the signal and random $\pi_s$ background. They find $1,141,500 \pm 1200$ RS signal events and $4030 \pm 90$ WS signal from the fits. Projections of the WS fit to data are shown in Fig. 19.2.9.

[127] The largest fraction of multiple $D^*$ candidates results from combinations of a single $D^0$ candidate paired with multiple $\pi_s$ candidates.

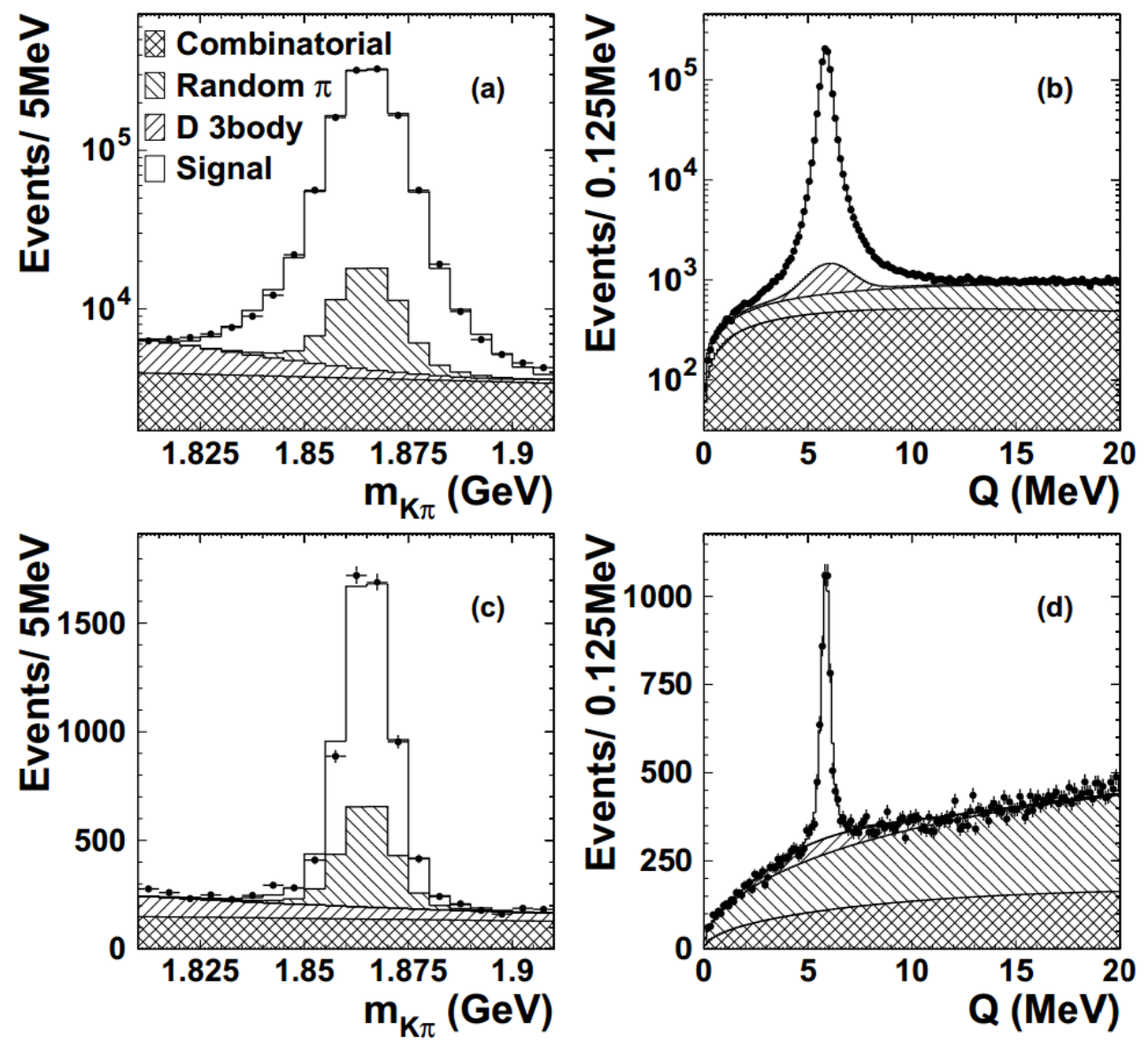

**Figure 19.2.8.** $m_{K\pi}$ and $q$ distributions (Zhang, 2006): (a) RS $m_{K\pi}$ for 0 MeV $< q <$ 20 MeV; (b) RS $q$ for 1.81 GeV/$c^2$ $< m_{K\pi} <$ 1.91 GeV/$c^2$; (c) WS $m_{K\pi}$ for 5.3 MeV $< q <$ 6.5 MeV; and (d) WS $q$ for 1.845 MeV/$c^2$ $< m_{K\pi} <$ 1.885 GeV/$c^2$. Points with error bars represent the data and the histograms different components of the fit.

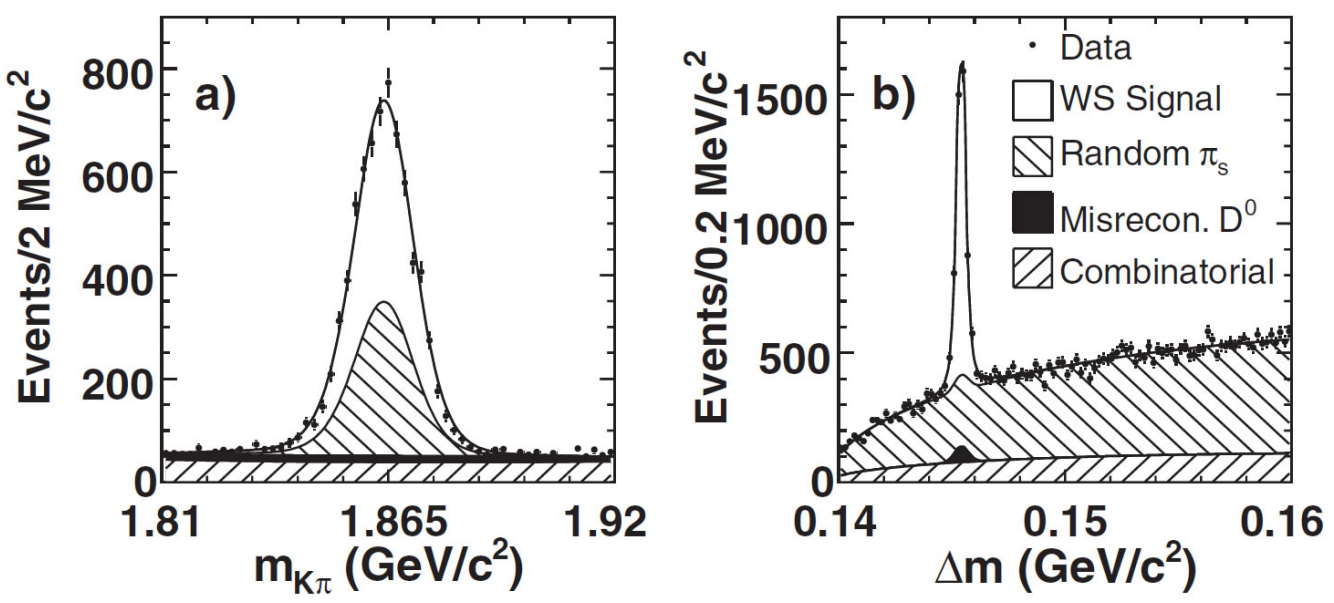

**Figure 19.2.9.** $m_{K\pi}$ and $\Delta m$ distributions (Aubert, 2007j): (a) $m_{K\pi}$ for WS candidates in a signal-enhanced region (0.1445 GeV/$c^2$ $< \Delta m <$ 0.1465 GeV/$c^2$); (b) $\Delta m$ for WS candidates in a signal-enhanced region (0.1843 GeV/$c^2$ $< m_{K\pi} <$ 1.883 GeV/$c^2$). Projections of the fitted signal and background *p.d.f.*s are shown.

The DCS decay parameter $R_{\rm D}$ and the mixing parameters $x'^2$ and $y'$ are determined using unbinned, maximum-likelihood fits to the WS proper decay-time distribution. The fit is done in several stages in order to fix some of the parameters entering the final fit. The RS distribution is fitted first providing the parameters of the resolution functions to be used in the fit to the WS decay-time distribution.

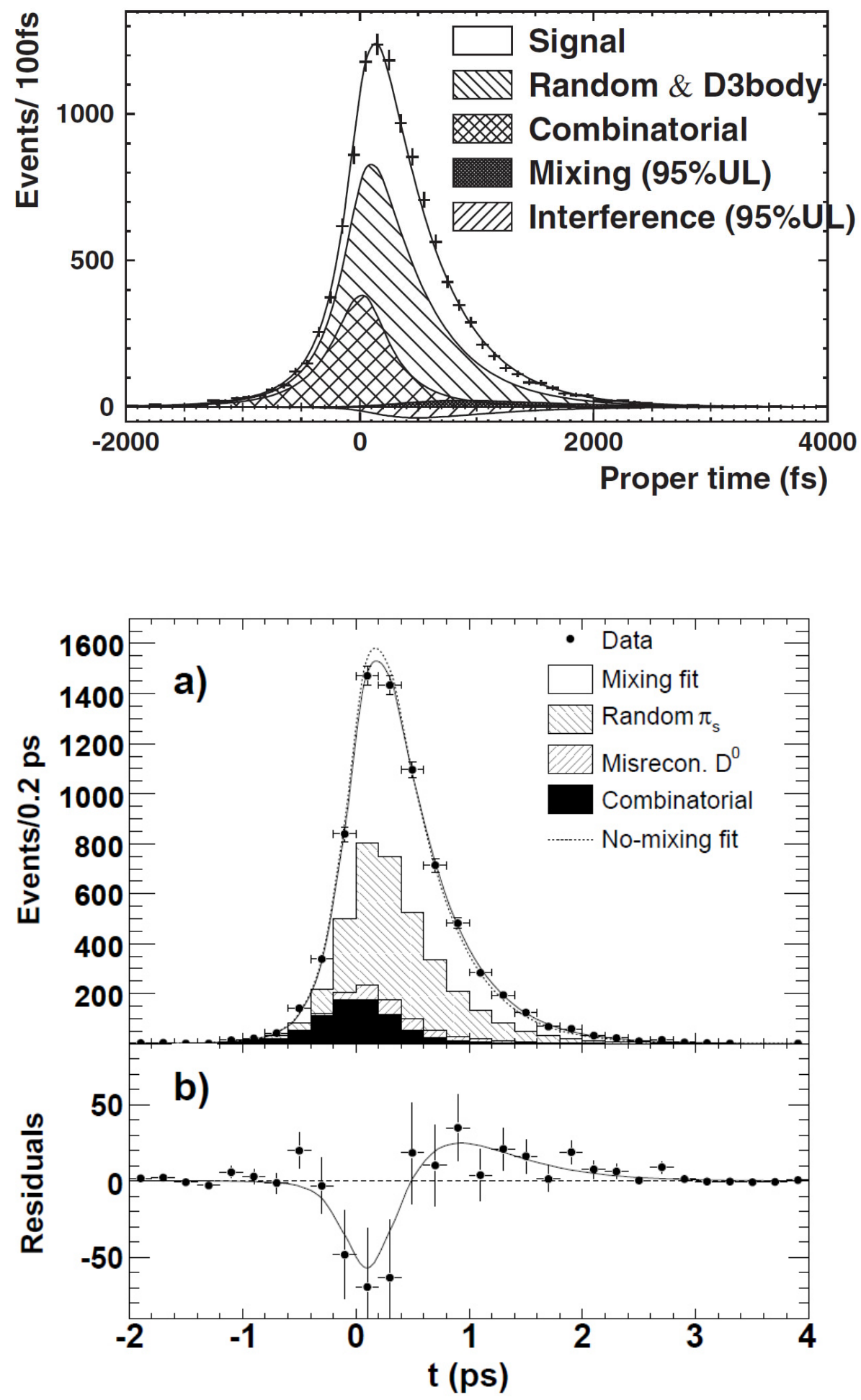

**Figure 19.2.10.** Belle (top, from Zhang, 2006) and *BABAR* (bottom, from Aubert, 2007j) WS decay-time distributions overlaid with projections of fits assuming no *CP* violation. Belle: Data distribution (points with error bars) for WS events in the signal enhanced region $|m_{K\pi} - m_{D^0}| < 22$ MeV/$c^2$ and $|q - 5.9$ MeV$| < 1.5$ MeV. *BABAR*: (a) Data distribution and fit projections for combined $D^0$ and $\overline{D}^0$ candidates in the signal enhanced region 1.843 GeV/$c^2$ $< m_{K\pi} <$ 1.883 GeV/$c^2$ and 0.1445 GeV/$c^2$ $< \Delta m <$ 0.1465 GeV/$c^2$. The fit result allowing (not allowing) mixing is shown as a solid (dashed) line. (b) Points indicate the difference between the data and the no-mixing fit. The solid curve shows the difference between fits with and without mixing.

The Belle decay-time fit uses a likelihood which is a function of the DCS and mixing parameters $R_{\rm D}$, $x'^2$, and $y'$. For event $i$, it is given by

$$\frac{dP_i}{dt'} = \left[f^i_{\rm sig} P_{\rm sig}(t'; R_{\rm D}, x'^2, y') + f^i_{\rm rnd} P_{\rm rnd}(t')\right] \otimes R_{\rm sig}(t_i - t')$$

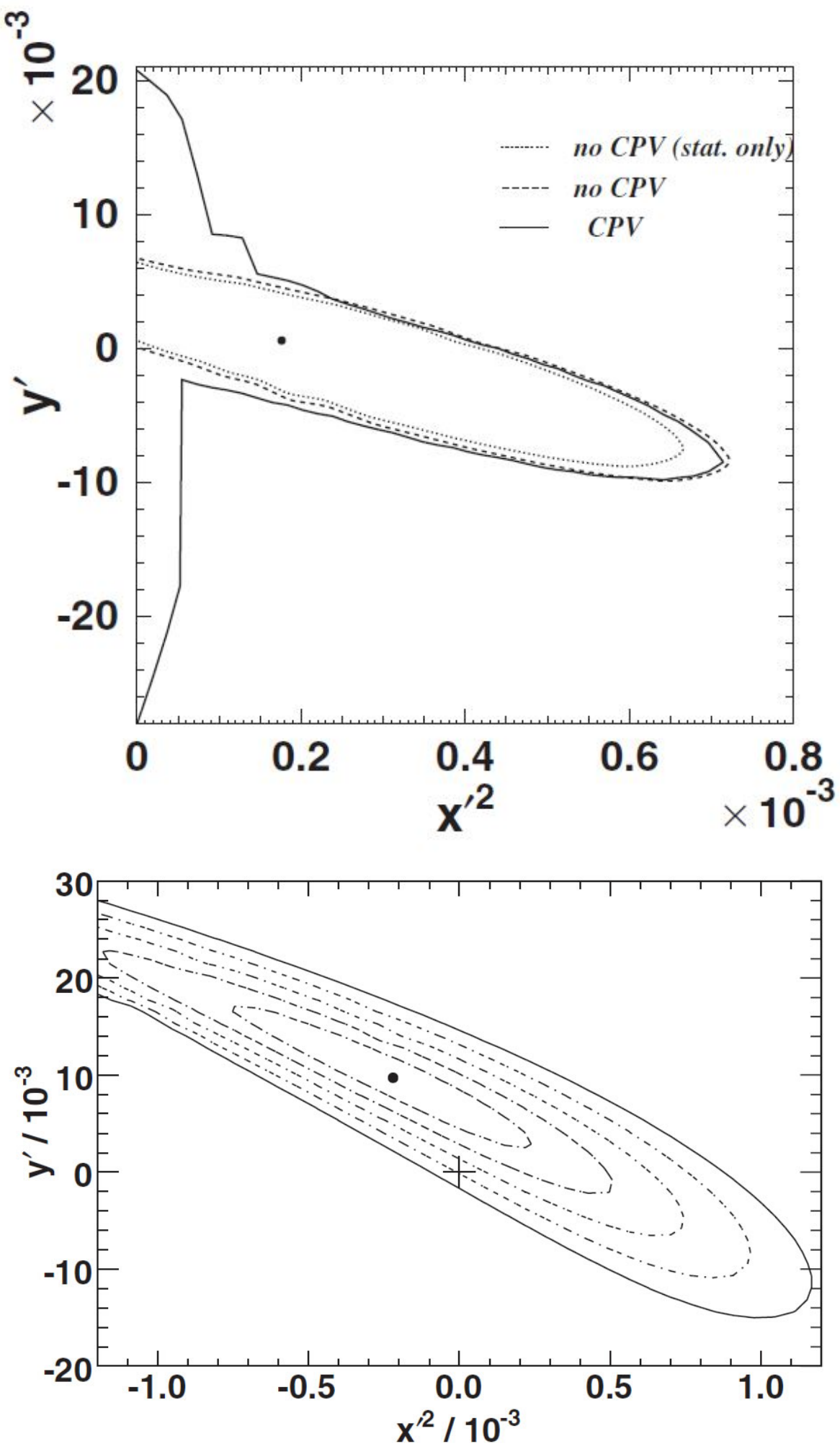

**Figure 19.2.11.** Belle (Zhang, 2006) and *BABAR* (Aubert, 2007j) confidence-level contours from the mixing fits. Belle (top): ($x'^2$, $y'$) 95% confidence-level regions showing the best fit result (point) assuming *CP* conservation. The statistical-only (statistical plus systematic) contour for no *CP* violation is shown as a dotted (dashed) line. The solid line is the statistical plus systematic contour for the *CP*-allowed case. *BABAR* (bottom): Confidence-level contours and fit result (point) for $1 - \text{C.L.} = 0.317(1\sigma)$, $4.55 \times 10^{-2}(2\sigma)$, $2.70 \times 10^{-3}(3\sigma)$, $6.33 \times 10^{-5}(4\sigma)$, and $5.73 \times 10^{-7}(5\sigma)$. The no-mixing point is shown as a "+" sign.

$$
\begin{aligned}
&+f^i_{\rm d3b} P_{\rm d3b}(t') \otimes R_{\rm d3b}(t_i - t') \\
&+f^i_{\rm ds3} P_{\rm ds3}(t') \otimes R_{\rm ds3}(t_i - t') \\
&+f^i_{\rm cmb} \delta(t') \otimes R_{\rm cmb}(t_i - t'). \qquad (19.2.25)
\end{aligned}
$$

The $f^i$ fractions are functions of $m_{K\pi}$, $q$, and $\sigma_t$, and are determined on an event-by-event basis. $P_j$ is the expected decay-time distribution for event category $j$; it is given, for example, by Eq. (19.2.22) for $j = \text{sig}$. $R_j$ is the resolution function for the corresponding events and $\otimes$ denotes the convolution. Signal and random $\pi_s$ background events have the same resolution function since the slow pion is not used in the vertex fit. *BABAR* models the WS decay-time behavior as given by Eq. (19.2.22) convolved with the signal resolution function as determined in the RS decay-time fit. The background distribution is modeled in a similar way as in the Belle case. The WS decay time distributions overlaid with the fit projections for both Belle and *BABAR* are shown in Fig. 19.2.10.

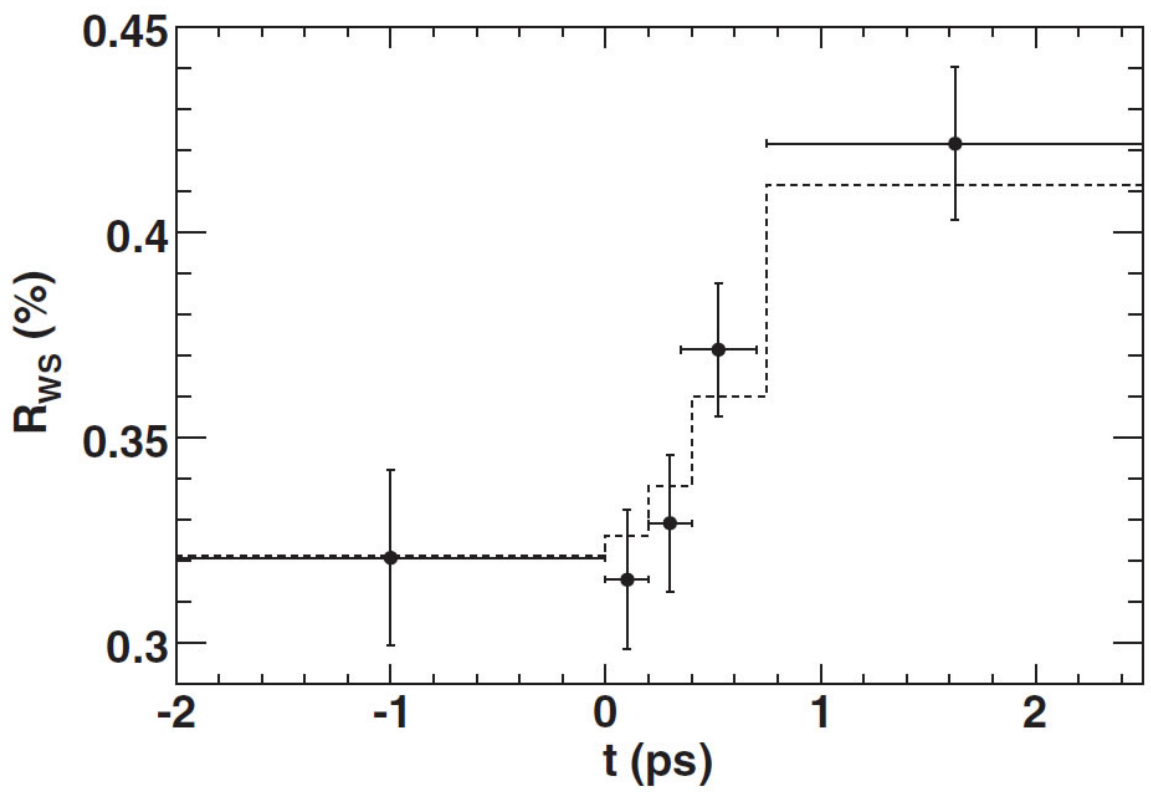

**Figure 19.2.12.** *BABAR* (Aubert, 2007j) WS branching fractions $R_{\rm WS}$ for disjoint regions of measured proper time from fits to the $(m_{K\pi}, \Delta m)$ plane (points with error bars). The dashed line shows the expected values of $R_{\rm WS}$ for each slice in proper time assuming the nominal mixing fit results. The $\chi^2$ with respect to the mixing fit expectation is 1.5; assuming the no-mixing hypotheses (a constant $R_{\rm WS}$ for all time slices), the $\chi^2$ is 24.

Both Belle and *BABAR* present results from fits under three different assumptions: (1) that no mixing or *CP* violation is present; (2) that mixing may be present, but that there is no *CP* violation; and (3) that mixing and *CP* violation may be present. The first two sets of results are shown in Table 19.2.3, while the last is discussed in Section 19.2.7. A large correlation between $x'^2$ and $y'$ is seen in both experiments: $-0.909$ (Belle), $-0.95$ (*BABAR*).

*BABAR* finds the maximum-likelihood point to be in the non-physical region $x'^2 < 0$. Two-dimensional confidence-level contours (see Fig. 19.2.11) in $x'^2$ and $y'$ are calculated based on the change in negative log-likelihood values with respect to the no-mixing point for *BABAR* and using the Feldman-Cousins likelihood ratio ordering (Feldman and Cousins, 1998) for Belle.

Systematic uncertainties are included in the contour evaluation as discussed below.

*BABAR* finds that the fit allowing mixing provides a substantially better result than the one without mixing, as seen in Fig. 19.2.11. The likelihood maximum is at an unphysical value ($x'^2 = -2.2 \times 10^{-4}$, $y' = 9.7 \times 10^{-3}$). The difference in the $\chi^2$ between the most likely point in the physically allowed region and the no-mixing point ($x'^2 = 0$, $y' = 0$) corresponds to a significance for mixing of $3.9\sigma$ ($1 - \text{C.L.} = 10^{-4}$) and provides evidence for mixing.[128]

[128] Here evidence is regarded as a result with a significance of more than three standard deviations. For the explanation about the significance see Chapter 11.

As a cross-check of the results, *BABAR* performs the no-mixing fit in slices in reconstructed decay time with approximately equal numbers of events. The resulting values of $R_D$ increase with increasing decay time as seen in Fig. 19.2.12. The rate of increase is consistent with the measured mixing parameters and inconsistent with the no-mixing hypothesis.

**Table 19.2.3.** $D^0 - \overline{D}^0$ mixing results using $D^0 \to K\pi$ WS decays. When two uncertainties are given, the first is statistical and the second systematic. Results with a single uncertainty have both statistical and systematic components combined.

| Parameter | Fit Results ($\times 10^{-3}$) | |
|---|---|---|
| | *BABAR* (Aubert, 2007j) | Belle (Zhang, 2006) |
| | Assuming no mixing or *CP* violation | |
| $R_D$ | $3.53 \pm 0.08 \pm 0.04$ | $3.77 \pm 0.08 \pm 0.05$ |
| | Assuming mixing but no *CP* violation | |
| $R_D$ | $3.03 \pm 0.16 \pm 0.10$ | $3.64 \pm 0.17$ |
| $x'^2$ | $-0.22 \pm 0.30 \pm 0.21$ | $0.18^{+0.21}_{-0.23}$ |
| $y'$ | $9.7 \pm 4.4 \pm 3.1$ | $0.6^{+4.0}_{-3.9}$ |
| Significance | 3.9 | 2.0 |

Some of the sources of systematic uncertainties investigated are event yields, background modeling, and decay-time *p.d.f.* models. Belle also investigated the effects of changing selection criteria such as particle-identification criteria, vertex fit $\chi^2$ requirement, and $D^*$ momentum selection, which cause the signal-to-background rates to change.

The effect of each individual systematic variation was calculated from the change in $-2\Delta \ln \mathcal{L}$ evaluated in the $(x'^2, y')$ plane between the nominal fit point and the new fit point of the variation under test. This value was scaled by a factor of 2.3, $\chi^{2\prime} \equiv -2\Delta \ln \mathcal{L}/2.3$, to yield the 68% confidence level for a single variable. The largest Belle systematic is from the $D^*$ momentum requirement with $\chi^{2\prime} = 0.083$. When shifts from all systematic checks are added in quadrature, the overall scale factor is $\sqrt{1 + \sum \chi_i^{2\prime}} = 1.12$ which is used to scale the 95% C.L. contours as shown in Fig. 19.2.11. As a cross-check, Belle finds the results for the two SVD subsamples to be within $0.6\sigma$ of each other.

*BABAR* estimated systematic uncertainties and included them in the evaluation of contours in a manner similar to that of Belle, finding the largest contribution to $\sqrt{1 + \sum \chi_i^{2\prime}} = 1.3$ to be 0.06 from the modeling of the long decay-time component of background $D$ decays in the signal region. A non-zero mean value of 3.6 fs for the decay-time was found and is attributed to detector misalignments; this contributes 0.05 to the systematic uncertainty.

*BABAR* validated the fitting procedure on MC data using both the full detector simulation and on ensembles of large parameterized (simulated MC) samples. The fit was found to be unbiased in all cases. A fit allowing for mixing in the RS sample was performed and no significant mixing was observed, as expected. The staged fitting procedure was cross-checked by performing a single simultaneous fit in which all parameters were allowed to vary; results were consistent with the nominal fitting procedure.

#### 19.2.2.3 Measurements of the $D^0 \to K^+\pi^-\pi^+\pi^-$

The advantages of the $K3\pi$ mode relative to $K\pi$ are its larger branching fraction (about 8.1% compared to 3.9% for $K\pi$) (Nakamura et al., 2010) and the improvement obtained in decay vertex resolution with four charged tracks instead of two (0.15–0.16 ps compared to 0.17–0.18 ps for $K\pi$ and $KK$).[129] On the other hand, the strong phase difference of the decay varies over its four-body phase space, making a simple interpretation of an average mixing rate measured over all of the phase space problematic. Possible approaches to this problem include limiting the measurement to specific regions of interest in phase space or performing a four-body, time-dependent fit to the decay using amplitude models. As with mixing studies using the $K\pi$ decay, *CP* violation can also be searched for by analyzing the difference between $D^0$ and $\overline{D}^0$ decays.

*BABAR* has reported preliminary results from a time-dependent analysis of the four-body decay $D^0 \to K^+\pi^-\pi^+\pi^-$ using a 230 fb$^{-1}$ data sample (Aubert, 2006ap). The analysis is very similar to time-dependent analyses of $D^0 \to K\pi$. "Tagged" events (where the decay chain $D^{*+} \to \pi_s D^0$ is reconstructed) are used to determine the production flavor of the $D^0$ candidate via the charge of the slow pion $\pi_s$. The variables of interest are the reconstructed $D^0$ candidate mass $m_{K\pi\pi\pi}$, the mass difference $\Delta m$ between the reconstructed $D^{*+}$ and the $D^0$ candidate, and the decay time $t$ and its uncertainty $\sigma_t$.

The $D^{*+}$ and $D^0$ candidate masses and vertices are obtained from a vertex fit to the decay chain using a beam spot constraint. The fit to the entire chain is required to have large enough probability calculated from the fit $\chi^2$, $P(\chi^2) > 0.01$. The $D^0$ proper decay time $t$ is obtained from the vertex fit along with its uncertainty $\sigma_t$. A requirement is imposed on the decay time uncertainty of $\sigma_t < 0.5$ ps. Unbinned extended maximum-likelihood fits are performed to both right-sign and wrong-sign candidates. Two-dimensional distributions in $(m_{K\pi\pi\pi}, \Delta m)$ are fitted to a combination of *p.d.f.*s that describe the signal shapes and background contributions. The large amount of right-sign signal is used to determine the signal shape parameters for the wrong-sign signal. Approximately $3.5 \times 10^5$ right-sign signal candidates are found are found in each, $D^0$ and $\overline{D}^0$ sample, and about 1100 wrong-sign signal candidates for each. Backgrounds include "bad slow pion" events, where a properly-reconstructed $D^0$ has been paired with an unassociated pion; this background source peaks in $m_{K\pi\pi\pi}$ but not in $\Delta m$. Another background comes from $D^0$'s where the kaon and one of the pions are interchanged, but the $D^{*+}$ is otherwise correctly reconstructed. This source peaks in $\Delta m$ only. The remaining background component is combinatoric, which has no peaking behavior.

[129] The improvement w.r.t. two body decays is less than a factor of $1/\sqrt{2}$ because of the lower momentum of the 4 tracks and thus increased multiple scattering.

Fits to the decay time distributions are performed. Analogously to Eq. (19.2.22) the wrong-sign time-dependence is fitted to

$$\frac{\Gamma_{\rm WS}(t)}{\Gamma_{\rm RS}(t)} = \widetilde{R}_D + \alpha\widetilde{y}'\sqrt{\widetilde{R}_D}(\Gamma t) + \frac{(x^2+y^2)}{4}(\Gamma t^2). \quad (19.2.26)$$

Quantities that are integrated over all or part of phase space are indicated by a tilde. $\alpha$ is a factor describing the suppression due to the strong phase variation over the phase space. Both a *CP*-conserving fit which considers $D^0$ and $\overline{D}^0$ candidates together and a fit that is potentially sensitive to *CP* violation which treats them separately are performed. A description of the latter is given in Section 19.2.7.

Systematics are evaluated by changing various parts of the analysis, including the $\sigma_t$ selection, the *p.d.f.* parameterization of the decay time resolution function, background *p.d.f.* shapes, and the measured $D^0$ lifetime value. In the latter, the fitted lifetime value is fixed in the fit to the PDG value. Combined systematics are smaller than the statistical errors on the measured quantities by about a factor of five.

Assuming *CP* conservation, the *BABAR* preliminary analysis (Aubert, 2006ap) yields a measurement of $R_M$ and the interference term $\alpha\widetilde{y}'$ of

$$\begin{aligned} R_M &= [0.019^{+0.016}_{-0.015}(\text{stat}) \pm 0.002(\text{syst})]\% \\ \alpha\widetilde{y}' &= -0.006 \pm 0.005(\text{stat}) \pm 0.001(\text{syst}) \quad (19.2.27) \end{aligned}$$

which are consistent with the no-mixing hypothesis at the 4.3% confidence level. Results of the fit allowing for *CP* violation are given in Section 19.2.7.

Two-dimensional coverage probabilities of 68.3% and 95.0% ($\Delta\log\mathcal{L} = 1.15$, 3.0, respectively) are shown in Fig. 19.2.13 for the doubly Cabibbo-suppressed rate $\widetilde{R}_D$ vs. the mixing rate $R_M$, and for the interference term $\alpha\widetilde{y}'/\sqrt{x^2+y^2}$ vs. $R_M$.

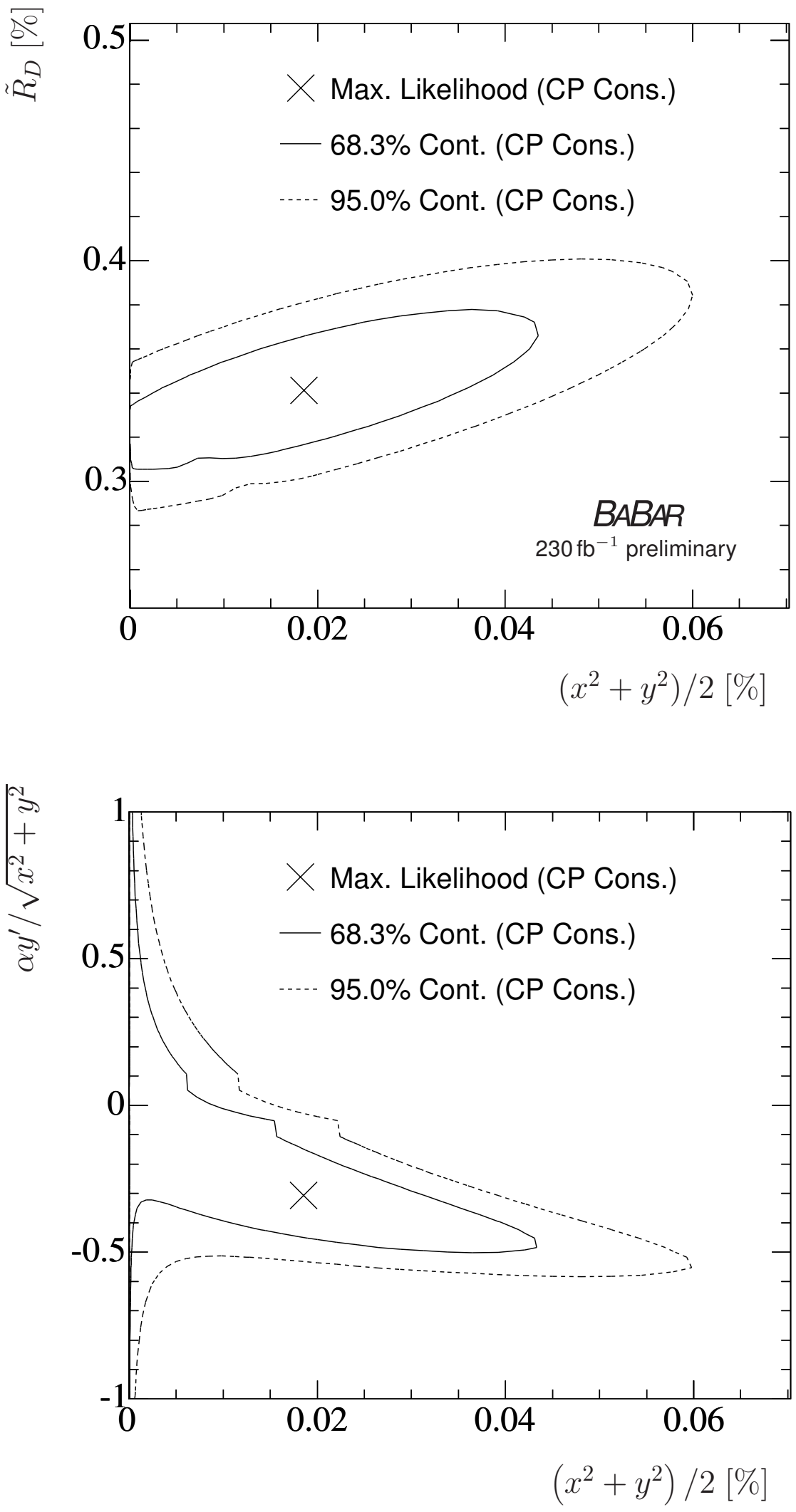

**Figure 19.2.13.** Likelihood contours for the *CP* conserving fit for $\widetilde{R}_D$ (top, from (Aubert, 2006ap)) and for the interference term (bottom, *BABAR* internal, from the (Aubert, 2006ap) analysis) vs. the mixing rate $R_M$. Solid line: $\Delta\ln\mathcal{L} = 1.15$; dotted line: $\Delta\ln\mathcal{L} = 3.0$.

### 19.2.2.4 Summary on hadronic wrong-sign decays

The measurements of decay time distributions in $D^0 \to K^+\pi^-$ decays played an important role in the initial searches and finally in the experimental discovery of the mixing phenomena in the neutral $D$ meson system. It should be noted that neither of the $B$ Factory experiments have performed measurements using their full data set, this remains a task for the future. The uncertainties of determinations of $x'^2$ and $y'$ on measurements made with larger data sets will be, most probably, dominated by experimental systematic uncertainties. The same measurements provide important insights into a possible *CP* violation in the $D^0 - \overline{D}^0$ system as discussed further in Section 19.2.7.

It should be noted that an analysis of wrong-sign decays $D^0 \to K^+\pi^-\pi^0$ has also been made by the *BABAR* collaboration and is described in the section on decay time dependent Dalitz analyses, Section 19.2.4.1.

### 19.2.3 Decays to $CP$ eigenstates

#### 19.2.3.1 Method

If $CP$ is conserved, $q = p \equiv \frac{1}{\sqrt{2}}$, and the mass eigenstates $|D_{1,2}\rangle = \frac{1}{\sqrt{2}}(|D^0\rangle \pm |\overline{D}^0\rangle)$ are $CP$-even and $CP$-odd; they decay with the lifetimes $\tau_1 = 1/\Gamma_1$ and $\tau_2 = 1/\Gamma_2$, respectively, into $CP$-even and $CP$-odd final states, obeying a simple exponential law, see Eq. (19.2.8). On the other hand the time evolution of decays to flavor specific final states is approximately exponential only for the Cabibbo-favored decays, like $D^0 \to K^-\pi^+$. In this case Eqs (19.2.13) simplify due to the fact that $|\overline{A}_f/A_f| \ll 1$ (and $|A_{\overline{f}}/\overline{A}_{\overline{f}}| \ll 1$). Because of typically strong Cabibbo suppression one can neglect the $\Gamma t$ and $(\Gamma t)^2$ terms in Eqs (19.2.13).[130] Hence the decay time distribution is approximately exponential with a lifetime of $\tau_{\rm FS} = 1/\Gamma$, where $\tau_{\rm FS}$ denotes the lifetime for decays into flavor specific final state.

The quantity which represents the relative lifetime difference between decays to $CP$ and flavor specific final states is obtained experimentally as:

$$y_{CP} = \eta_{CP}\left(\frac{\tau_{\rm FS}}{\tau_{CP}} - 1\right), \qquad (19.2.28)$$

where $\eta_{CP} = +1\,(-1)$ for $CP$-even ($CP$-odd) final state, and $\tau_{CP}$ is the lifetime of decays to a $CP$ final state. In the limit of $CP$ conservation $\tau_{CP}$ equals the lifetime of the corresponding mass eigenstate, $\tau_1$ if $\eta_{CP} = +1$ or $\tau_2$ if $\eta_{CP} = -1$, and hence $y_{CP}$ equals the mixing parameter $y$. If $CP$ is violated $y_{CP}$ obtains a contribution from $x$:

$$y_{CP} = y\cos\phi - \frac{1}{2}\left(A_M - A_D^f\right) x\sin\phi, \qquad (19.2.29)$$

with $A_M$ and $A_D^f$ defined in Eqs (19.2.17) and (19.2.18) and $\phi = \arg(q/p)$. We assume here that $CP$ violation is small, *i.e.* $A_M, A_D^f \ll 1$, so that the time evolution is still well described by an exponential law.

To derive equation (19.2.29) one starts with Eqs (19.2.11) and (19.2.12); after squaring the modulus we obtain:

$$\frac{d\Gamma_{D^0\to f}}{dt} \propto (1 - {\rm Re}\left[\lambda_f(ix+y)\right]\Gamma t)\, e^{-\Gamma t} \qquad (19.2.30)$$

and

$$\frac{d\Gamma_{\overline{D}^0\to f}}{dt} \propto \left(1 - {\rm Re}\left[\lambda_f^{-1}(ix+y)\right]\Gamma t\right) e^{-\Gamma t}, \qquad (19.2.31)$$

where

$$\lambda_f = \frac{q}{p}\frac{\overline{A}_f}{A_f} \approx \eta_{CP}\left(1 + \frac{1}{2}\left[A_M - A_D^f\right]\right) e^{i\phi}. \qquad (19.2.32)$$

[130] For example, in $D^0 \to K^-\pi^+$ decays $|\overline{A}_{K^-\pi^+}/A_{K^-\pi^+}| = (5.75 \pm 0.07) \times 10^{-2}$ (Amhis et al., 2012). The decay time distribution $[d\Gamma(D^0 \to K^-\pi^+) + d\Gamma(\overline{D}^0 \to K^+\pi^-)]/dt \propto [1 - |\overline{A}_{K^-\pi^+}/A_{K^-\pi^+}|\; y\cos\delta_{K\pi}\;\Gamma t]e^{-\Gamma t}$, where $|\overline{A}_{K^-\pi^+}/A_{K^-\pi^+}|\; y\cos\delta_{K\pi^+} = 2.3\times10^{-5}$ (Amhis et al., 2012).

Then, Eqs (19.2.30) and (19.2.31) are added together, since in this particular measurement no distinction is made between two possible initial $D^0$ flavors (for measurements of $CP$ violation in such decays, where one tags the flavor of the initial $D^0$, see Section 19.2.7). We obtain:

$$\frac{d\Gamma}{dt} \propto \left(1 - \eta_{CP}\left[y\cos\phi - \frac{1}{2}\left(A_M - A_D^f\right)x\sin\phi\right]\Gamma t\right)e^{-\Gamma t}. \qquad (19.2.33)$$

The expression in front of $e^{-\Gamma t}$ can be regarded as a linear expansion of another exponential function, since $x, y \ll 1$. Denoting the expression in the square brackets by $y_{CP}$ one obtains

$$\frac{d\Gamma}{dt} \propto e^{-\eta_{CP}y_{CP}\Gamma t}e^{-\Gamma t} = e^{-(1+\eta_{CP}y_{CP})\Gamma t}\,. \qquad (19.2.34)$$

Comparison of the decay time for decays into $CP$ eigenstates from the above equation, $\tau_{CP} = 1/[\Gamma(1+\eta_{CP}y_{CP})]$, to the average decay time for decays to flavor specific final states $\tau_{\rm FS} = 1/\Gamma$ yields Eq. (19.2.28).

The measured proper decay time distribution can be written as:

$$\frac{dN}{dt} = \frac{N}{\tau}\int_0^\infty R(t-t')e^{-t'/\tau}dt' + B(t), \qquad (19.2.35)$$

where $R$ is a resolution function and $B(t)$ is a background distribution.

The most suitable decays to measure $y_{CP}$ are the $CP$-even decays $D^0 \to K^+K^-$ and $D^0 \to \pi^+\pi^-$, because of their relatively large branching fractions and since the flavor specific decay $D^0 \to K^-\pi^+$ is kinematically similar. The latter is important in reducing the systematic uncertainty due to resolution function parameterization. Both *BABAR* and Belle have found that up to an overall scale factor in the width, the resolution function has the same shape for all three modes, including its offset $t_0$.

Among the $CP$-odd decays, $D^0 \to K_s^0\omega$ with $\omega \to \pi^+\pi^-\pi^0$ and $D^0 \to K_s^0\phi$ with $\phi \to K^+K^-$ have the largest branching fractions. Both resonances are also narrow. The drawbacks of these decays are smaller reconstruction efficiency due to $K_s^0$ and $\pi^0$ reconstruction, a contribution of other resonances which interfere with the $\omega$ or $\phi$, and large differences in the kinematics compared to $D^0 \to K^-\pi^+$. Up to now only the measurement of $D^0 \to K_s^0\phi$ has been reported (see Section 19.2.3.3).

#### 19.2.3.2 Results for the $D^0 \to KK/\pi\pi$

During the search for $CP$ violation in charm decays, the observable $y_{CP}$ has been measured by a number of experiments. In 2000 the interest of the scientific community was triggered by the result of the FOCUS collaboration, which observed a high value of the parameter, albeit with a rather large statistical uncertainty (Link et al., 2000). The excitement subsided in 2002 following the measurements from CLEO (Csorna et al., 2002) and Belle (Abe, 2002a) providing more precise values consistent with zero. The latter measurement was performed on an untagged sample of $D$ meson decays, the method is explained in more detail below. The year 2007 marks the start of the era of sub-percent accuracy measurements of $y_{CP}$ using the large $B$ Factories data samples resulting in a number of statistically significant results.

The Belle collaboration measured $y_{CP}$ (Staric, 2007) using tagged samples. The $D^{*+} \to D^0\pi_s^+$ decays are reconstructed with a slow pion $\pi_s$, and $D^0 \to K^+K^-$, $K^-\pi^+$, and $\pi^+\pi^-$.

The proper decay time of the $D^0$ candidate is calculated as described in Section 19.2.1.5. Selection based on the the $D^{(*)}$ momentum in the CM system as described there is also applied. The decay time uncertainty $\sigma_t$ is evaluated event-by-event from the covariance matrices of the production and decay vertices, as explained in Section 6. Candidate $D^0$ mesons are selected using the invariant mass of the $D^0$ decay products, $M$, and the energy released in the $D^{*+}$ decay, $q$.

According to Monte Carlo simulated distributions of $t$, $M$, and $q$, background events fall into four categories: (1) combinatorial, with zero apparent lifetime; (2) true $D^0$ mesons combined with random slow pions (this has the same apparent lifetime as the signal) (3) $D^0$ decays to three or more particles, and (4) other charm hadron decays. The apparent lifetime of the latter two categories is 10–30% larger than $\tau_{D^0}$.

The sample of events for the lifetime measurements is selected using $|\Delta M|/\sigma_M$, where $\Delta M \equiv M - m_{D^0}$; $|\Delta q| \equiv q - (m_{D^{*+}} - m_{D^0} - m_\pi)c^2$; and $\sigma_t$. The invariant mass resolution $\sigma_M$ varies from 5.5–6.8 MeV/$c^2$, depending on the decay channel. Selection criteria are chosen to minimize the expected statistical error on $y_{CP}$, using the MC: Belle requires $|\Delta M|/\sigma_M < 2.3$, $|\Delta q| < 0.80$ MeV, and $\sigma_t < 370$ fs. Using 540 fb$^{-1}$ of data, they find $111 \times 10^3$ $K^+K^-$, $1.22 \times 10^6$ $K^-\pi^+$, and $49 \times 10^3$ $\pi^+\pi^-$ signal events, with purities of 98%, 99%, and 92% respectively.

The mixing parameter $y_{CP}$ is determined from the binned maximum likelihood fit performed simultaneously to decay time distributions of all three decay modes. The resolution function of a single event is determined from the estimated accuracy of proper decay time $\sigma_t$ as obtained from the covariance matrices of the vertex fits. Ideally, it is described by a normalized Gaussian distribution with a zero mean and with width equal to $\sigma_t$. However, such a description is not sufficient because of multiple scattering of final state particles in the detector material, which causes the tails of the distribution to increase. To parameterize the tails, one or two additional Gaussian terms are needed, which share a common mean $t_0$ and have widths proportional to $\sigma_t$. The common mean can be offset from zero due to detector misalignment. The parameterization for a single event thus reads:

$$R(t) = \sum_{k=1}^{n_g} w_k G(t; t_0, \sigma_k), \tag{19.2.36}$$

where $G(t; t_0, \sigma_k) = \frac{1}{\sqrt{2\pi}\sigma_k} e^{-(t-t_0)^2/2\sigma_k^2}$ are the normalized Gaussian distributions, $\sigma_k = s_k\sigma_t$ are the widths, $w_k$ are their fractions and $n_g$ is the number of Gaussian terms (usually $n_g = 3$). This parameterization has the following free parameters: $t_0$, $s_k$, $k = 1, ..., n_g$ and $w_k$, $k = 1, ..., n_g - 1$. Parameters $s_k$ and $w_k$ are highly correlated, causing the fit to sometimes have problems converging. In order to ensure stable fitting the fractions $w_k$ can be fixed, using MC simulation, from a fit to the distribution of pulls, *i.e.* the normalized residuals $(t - t_{\text{gen}})/\sigma_t$, where $t_{\text{gen}}$ is the generated proper decay time of an event.

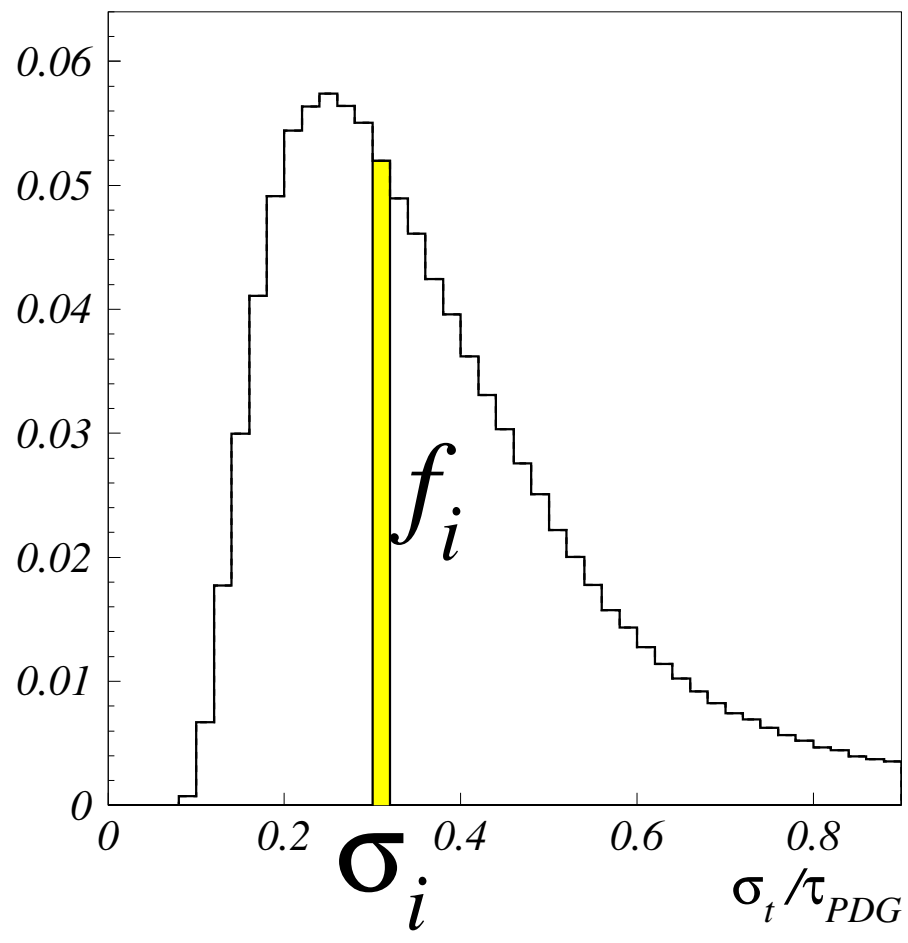

**Figure 19.2.14.** Normalized distribution of errors $\sigma_t$ on the decay time $t$ showing the construction of the resolution function using the fraction $f_i$ in the bin with $\sigma_t = \sigma_i$. Belle internal, from the Staric (2012a) analysis.

The form Eq. (19.2.36) is suitable for use in an unbinned maximum likelihood fit (see Chapter 11). The equivalent parameterization can be derived for a binned maximum likelihood fit. In this case we first construct the normalized distribution of $\sigma_t$ by binning the events in a histogram. Such a distribution is shown in Fig. 19.2.14: a bin $i$ corresponds to a fraction $f_i$ of events with a time resolution $\sigma_t = \sigma_i$. The resolution function for the binned fit is thus:

$$R(t) = \sum_{i=1}^{n} f_i \sum_{k=1}^{n_g} w_k G(t; t_0, \sigma_{ki}), \tag{19.2.37}$$

where $\sigma_{ki} = s_k\sigma_i$ and the first sum runs over bins $i$ of the $\sigma_t$ distribution. Note, that Eq. (19.2.37) has the same free parameters as Eq. (19.2.36).

The resolution function shape including the offset $t_0$ is found to be the same for all three considered decay modes. To account for small differences in the widths of resolution functions among various decay modes, two parameters $S_{KK}$ and $S_{\pi\pi}$ are introduced to scale the overall width of the $KK$ and $\pi\pi$ resolution functions relative to the width of the $K\pi$ resolution function. All other parameters can be shared among the different modes and can be determined by a simultaneous fit to all modes together.

The background term in Eq. (19.2.35) is parameterized assuming two lifetime components: an exponential and a $\delta$ function, each convolved with corresponding resolution functions as parameterized by Eq. (19.2.37). Separate $B(t)$ parameters for each final state are determined by fits to the $t$ distributions of events in $M$ sidebands. The MC is used to select the sideband region that best reproduces the timing distribution of background events in the signal region.

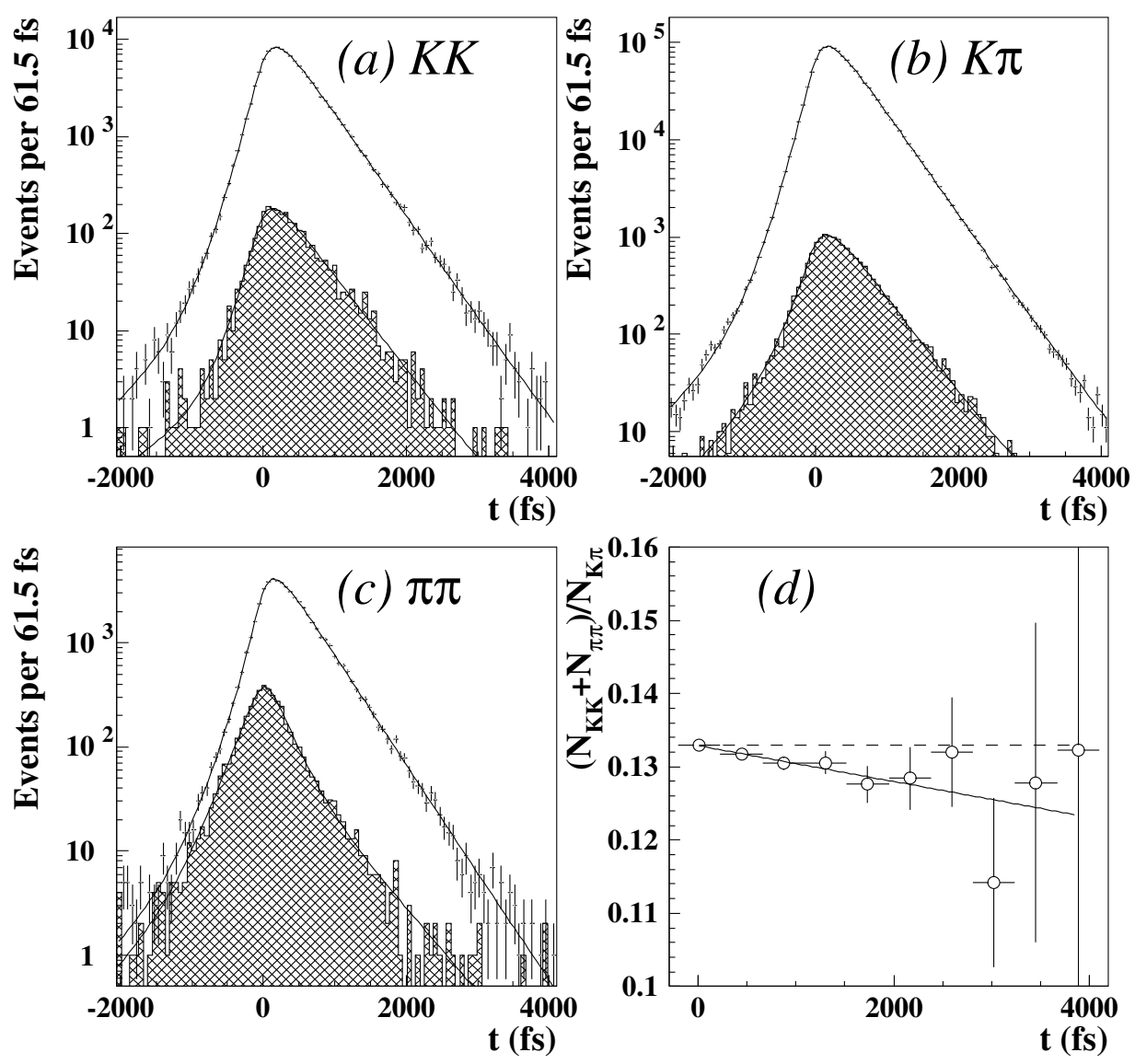

**Figure 19.2.15.** From (Staric, 2007). Belle results of the simultaneous fit to decay time distributions of (a) $D^0 \to K^+K^-$, (b) $D^0 \to K^-\pi^+$ and (c) $D^0 \to \pi^+\pi^-$ decays. The cross-hatched area represents background contributions, the shape of which was fitted using $M$ sideband events. (d) Ratio of decay time distributions between $D^0 \to K^+K^-, \pi^+\pi^-$ and $D^0 \to K^-\pi^+$ decays. The solid line is a fit to the data points and the dashed line represents the no-mixing hypothesis.

The results of a simultaneous fit are shown in Fig. 19.2.15. The fitted lifetime of $D^0 \to K^-\pi^+$, $\tau = (408.7 \pm 0.6(\text{stat}))$ fs, is consistent with the world average of $(410.1 \pm 1.5)$ fs. The value of $y_{CP}$ is determined to be $y_{CP} = (1.31 \pm 0.32(\text{stat}) \pm 0.25(\text{syst}))\%$.

This result and the *BABAR* result in the $D^0 \to K^+\pi^-$ decays, described in Section 19.2.2 represent the first experimental evidence for $D^0 - \overline{D}^0$ mixing.

Belle performed an updated measurement of $D^0 \to KK/\pi\pi$ decay modes using the full available data sample (Staric, 2012a). Using a larger data sample a small bias on the measured lifetime depending on the $D$ meson polar angle in the CM system, $\theta^*$, is observed. It is a consequence of small residual misalignments between the Belle tracking detectors. To reduce the systematic uncertainty due to such effects the measurement is performed in bins of $\cos\theta^*$ and the final value of $y_{CP}$ is obtained as a weighted average of the values in individual bins. The final result is

$$y_{CP} = (1.11 \pm 0.22(\text{stat}) \pm 0.11(\text{syst}))\%, \qquad (19.2.38)$$

the significance of which is above five standard deviations considering the statistical uncertainty alone, and 4.5 $\sigma$ including systematic uncertainties.

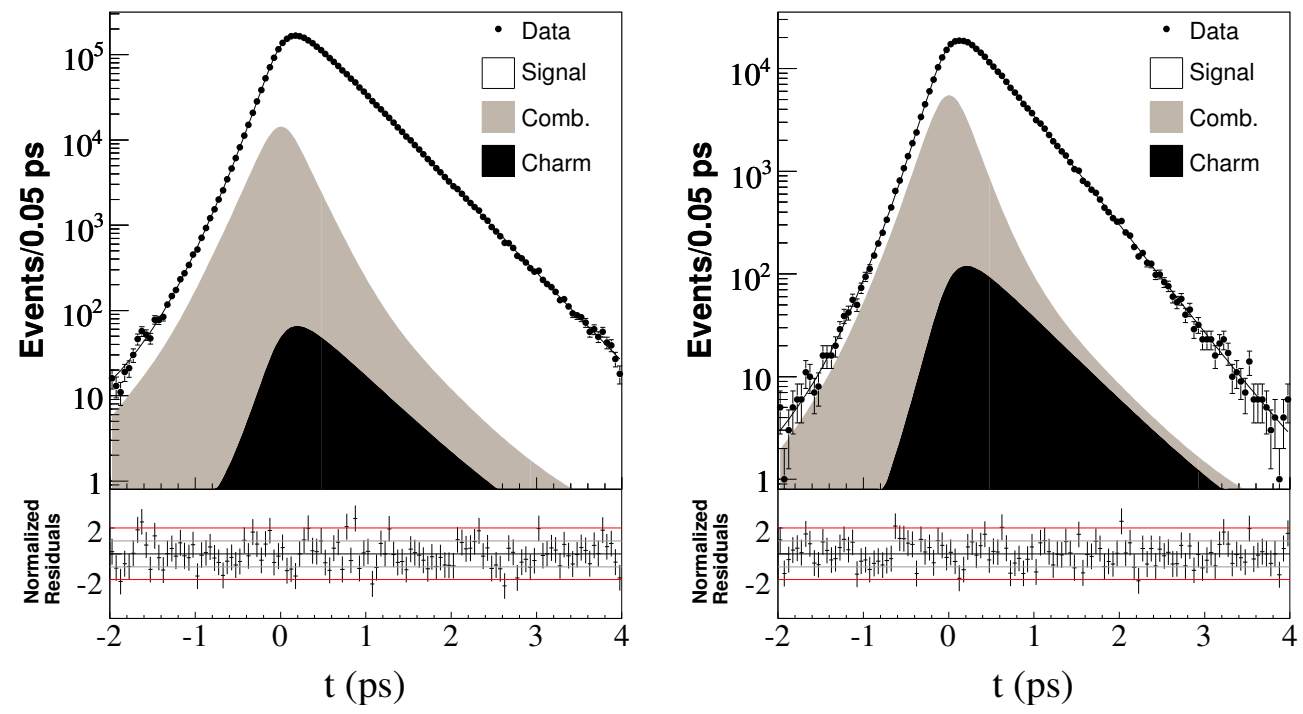

**Figure 19.2.16.** From (Aubert, 2009v). *BABAR* results of the simultaneous fit to decay time distributions of untagged samples: $D^0 \to K^-\pi^+$ (left) and $D^0 \to K^+K^-$ (right).

The *BABAR* analysis of the tagged samples (Aubert, 2008n) is similar to the Belle analysis. *BABAR* has used 384 fb$^{-1}$ of data and an unbinned maximum likelihood fit performed simultaneously to the $K^+K^-$, $K^-\pi^+$ and $\pi^+\pi^-$ decay modes. The main difference with respect to the Belle analysis is the form of the resolution function used, Eq. (19.2.36). The result obtained is

$$y_{CP} = (1.24 \pm 0.39(\text{stat}) \pm 0.13(\text{syst}))\%, \qquad (19.2.39)$$

which is also evidence for $D^0 - \overline{D}^0$ mixing.

*BABAR* has also performed an additional method: here $y_{CP}$ is measured using the untagged samples of $D^0 \to K^+K^-$ and $D^0 \to K^-\pi^+$ (Aubert, 2009v). The event selection is similar to the tagged analysis except that the $D^{*+}$ is not reconstructed. The background is much higher compared to the tagged analysis, as can be seen by comparing Fig. 19.2.15 and Fig. 19.2.16 for the corresponding decay modes. However, the signal yields compared to the tagged analysis are about five times higher. *BABAR* uses the same fitting procedure as in the tagged analysis; the fit is shown in Fig. 19.2.16. The result on 384 fb$^{-1}$ of data is consistent with previous measurements, but with smaller statistical and higher systematic errors: $y_{CP} = (1.12 \pm 0.26(\text{stat}) \pm 0.22(\text{syst}))\%$. The significance of the result is 3.3 $\sigma$.

The above result is superseded by a similar analysis using 468 fb$^{-1}$ (Lees, 2013d) which uses both untagged (for the $y_{CP}$ measurement) and tagged (for $A_\Gamma$; see Section 19.2.7) $D^0$ decays:

$$y_{CP} = (0.72 \pm 0.18(\text{stat}) \pm 0.12(\text{syst}))\%. \qquad (19.2.40)$$

### 19.2.3.3 Results for the $D^0 \to K^0_S\phi$

A large fraction of $D^0 \to K^0_S K^+K^-$ decays proceed via intermediate *CP*-odd $K^0_S\phi$ and *CP*-even $K^0_S a_0$ resonant states. Measurement of the apparent lifetimes $\tau_{K^0_S\phi}$ and $\tau_{K^0_S a_0}$ of candidates populating the $\phi$ and $a_0$ regions in the Dalitz plot, respectively, allows for the extraction of

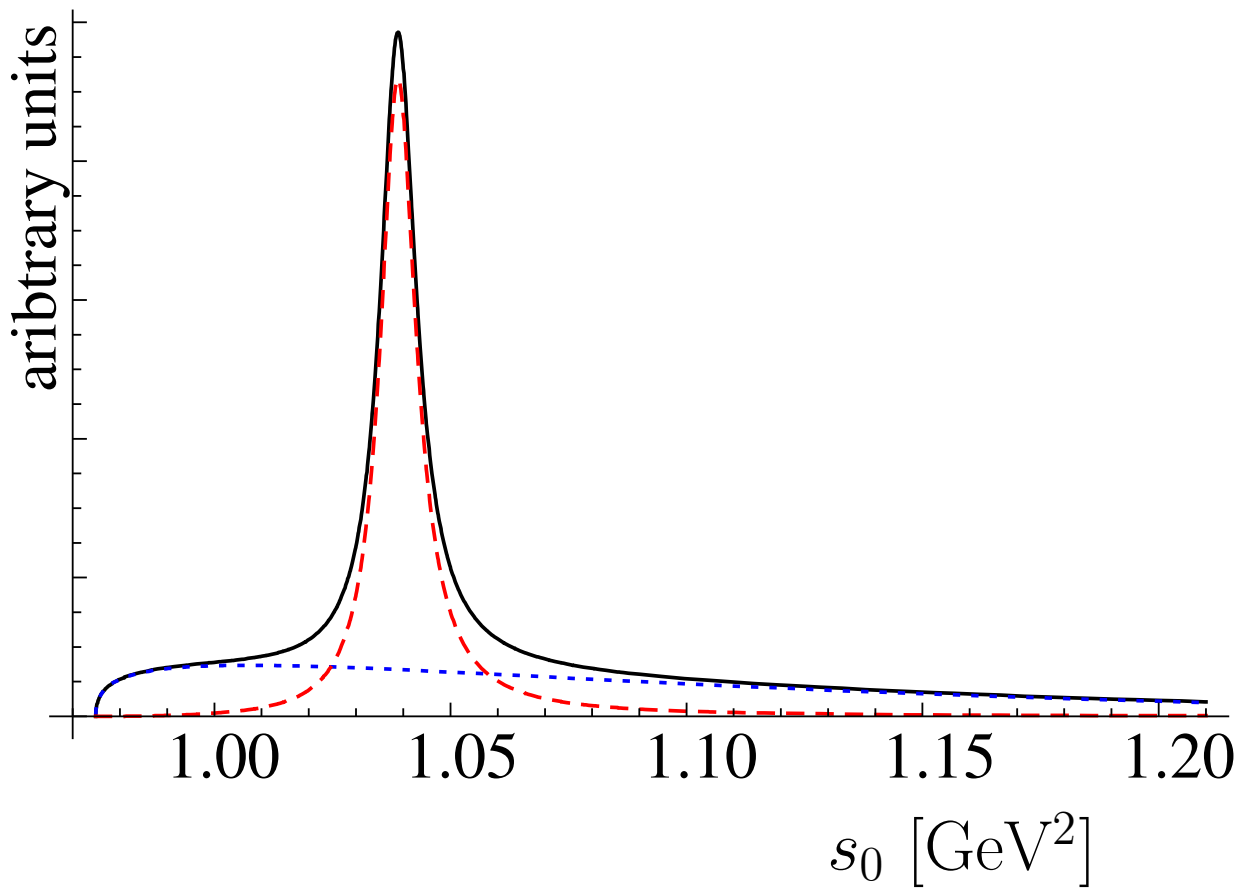

**Figure 19.2.17.** From (Zupanc, 2009). Projections of time integrated Dalitz distribution (solid black line) to $s_0$ and contributions of $CP$-even and $CP$-odd amplitudes, $|\mathcal{A}_1|^2$ (dotted blue line) and $|\mathcal{A}_2|^2$ (dashed red line), respectively.

the mixing parameter $y_{CP}$ as

$$y_{CP} \equiv \frac{\Gamma_{CP-\text{even}} - \Gamma_{CP-\text{odd}}}{\Gamma_{CP-\text{even}} + \Gamma_{CP-\text{odd}}} = \frac{\tau_{K_S^0\phi} - \tau_{K_S^0 a_0}}{\tau_{K_S^0\phi} + \tau_{K_S^0 a_0}}. \quad (19.2.41)$$

The above equation follows directly from the definition of Eq. (19.2.28), noting that $\tau_{CP=\pm 1} = \tau/(1 \pm y_{CP})$.

However, as shown in Fig. 19.2.17, it is impossible to identify the $CP$ value of the final state of an individual $D^0 \to K_S^0 K^+K^-$ decay, since the $a_0$ contribution also populates the region $s_0 \equiv M^2_{K^+K^-}$ below the $\phi$ peak and vice versa. In order to extract the $y_{CP}$ parameter correctly the $CP$ content of each region needs to be estimated.

The time-dependent decay amplitudes of three-body decays of $D^0$ and $\overline{D}^0$ mesons to self-conjugated final states are described in detail in Section 19.2.4 and are given by

$$\begin{aligned}\frac{d\Gamma(D^0)}{ds_0 ds_+ dt} = e^{-t/\tau}\Bigg[ &|\mathcal{A}_1(s_0,s_+)|^2 e^{(1+y)} \\ &+ |\mathcal{A}_2(s_0,s_+)|^2 e^{(1-y)} \\ &+ 2\mathrm{Re}\left[\mathcal{A}_1(s_0,s_+)\mathcal{A}_2^*(s_0,s_+)\right]\cos\left(\frac{xt}{\tau}\right) \\ &+ 2\mathrm{Im}\left[\mathcal{A}_1(s_0,s_+)\mathcal{A}_2^*(s_0,s_+)\right]\sin\left(\frac{xt}{\tau}\right)\Bigg]\end{aligned} \quad (19.2.42)$$

$$\begin{aligned}\frac{d\Gamma(\overline{D}^0)}{ds_0 ds_+ dt} = e^{-t/\tau}\Bigg[ &\left|\overline{\mathcal{A}}_1(s_0,s_+)\right|^2 e^{(1+y)} \\ &+ \left|\overline{\mathcal{A}}_2(s_0,s_+)\right|^2 e^{(1-y)} \\ &+ 2\mathrm{Re}\left[\overline{\mathcal{A}}_1(s_0,s_+)\overline{\mathcal{A}}_2^*(s_0,s_+)\right]\cos\left(\frac{xt}{\tau}\right) \\ &+ 2\mathrm{Im}\left[\overline{\mathcal{A}}_1(s_0,s_+)\overline{\mathcal{A}}_2^*(s_0,s_+)\right]\sin\left(\frac{xt}{\tau}\right)\Bigg],\end{aligned} \quad (19.2.43)$$

where $\tau = 1/\Gamma$ is the $D^0$ lifetime, $s_0$ and $s_+$ are the invariant masses squared of $K^+K^-$ and $K_SK^+$ pairs, respectively. The decay amplitudes $\mathcal{A}_1$ and $\mathcal{A}_2$ can be expressed with $D^0$ and $\overline{D}^0$ decay amplitudes $\mathcal{A}$ and $\overline{\mathcal{A}}$ as $\mathcal{A}_1(s_0,s_+) = [\mathcal{A}(s_0,s_+) + \overline{\mathcal{A}}(s_0,s_+)]/2$ and $\mathcal{A}_2(s_0,s_+) = [\mathcal{A}(s_0,s_+) - \overline{\mathcal{A}}(s_0,s_+)]/2$. In the isobar model (see Chapter 13) the amplitudes $\mathcal{A}$ and $\overline{\mathcal{A}}$ are written as the sum of intermediate decay channel amplitudes (denoted by the subscript $r$) with the same final state, $\mathcal{A}(s_0,s_+) = \sum_r a_r e^{i\phi_r}\mathcal{A}_r(s_0,s_+)$ and $\overline{\mathcal{A}}(s_0,s_+) = \sum_r \overline{a}_r e^{i\overline{\phi}_r}\overline{\mathcal{A}}_r(s_0,s_+) = \sum_r a_r e^{i\phi_r}\mathcal{A}_r(s_0,s_-)$, where $CP$ conservation in decay has been assumed in the final step. If $r$ is a $CP$ eigenstate, then $\mathcal{A}_r(s_0,s_-) = \pm\mathcal{A}_r(s_0,s_+)$, where the sign $+(-)$ holds for a $CP$-even(-odd) eigenstate. Hence the amplitude $\mathcal{A}_1$ is $CP$-even, and the amplitude $\mathcal{A}_2$ is $CP$-odd. According to our current knowledge of the decay dynamics of $D^0 \to K_S^0 K^+K^-$ decays, their Dalitz model includes five $CP$-even intermediate states: $K_S^0 a_0^0(980)$, $K_S^0 f_0(1370)$, $K_S^0 f_2(1270)$, $K_S^0 a_0^0(1450)$, $K_S^0 f_0(980)$); one $CP$-odd intermediate state ($K_S^0\phi(1020)$); and three flavor-specific intermediate states ($K^- a_0^+(980)$, $K^- a_0^+(1450)$, $K^+ a_0^-(980)$ (Aubert, 2008l).

Upon squaring Eqs (19.2.42) and (19.2.43) and integrating over $s_+$, we obtain for the time-dependent decay rates of initially produced $D^0$ and $\overline{D}^0$ (*e.g.* untagged sample):

$$\frac{d\Gamma}{dt ds_0} \propto a_1(s_0)e^{-\frac{t}{\tau}(1+y)} + a_2(s_0)e^{-\frac{t}{\tau}(1-y)}, \quad (19.2.44)$$

where $a_{1,2}(s_0) = \int |\mathcal{A}_{1,2}(s_0,s_+)|^2 ds_+$. When integrating the time-dependent decay rate over $s_+$, all terms depending on the mixing parameter $x$ (*e.g.* Re $[A_1A_2^*]\cos(xt/\tau)$) drop out (see the Appendix in Zupanc, 2009). The two terms in Eq. (19.2.44) have a different time dependence as well as a different $s_0$ dependence (see Fig. 19.2.17). In any given $s_0$ interval, $\mathcal{R}$, and assuming $y \ll 1$, the effective $D^0$ lifetime is

$$\tau_{\mathcal{R}} = \tau\left[1 + (1 - 2f_{\mathcal{R}})y_{CP}\right], \quad (19.2.45)$$

where $f_{\mathcal{R}} = \int_{\mathcal{R}} a_1(s_0)ds_0 / \int_{\mathcal{R}}(a_1(s_0) + a_2(s_0))ds_0$, which represents the effective fraction of the events in the interval $\mathcal{R}$ due to the $\mathcal{A}_1$ amplitude. In Eq. (19.2.45) we introduced the usual notation $y_{CP}$ for the mixing parameter $y$ to indicate that we assumed $CP$ conservation in deriving Eq. (19.2.44).

The mixing parameter $y_{CP}$ can thus be determined from the relative difference in the effective lifetimes of the two $s_0$ intervals, one around the $\phi(1020)$ peak (interval ON) and the other in the sideband (interval OFF). Using Eq. (19.2.45) and taking into account the fact that $[1 - (f_{\rm ON} + f_{\rm OFF})]y_{CP} \ll 1$, we obtain

$$y_{CP} = \frac{1}{f_{\rm ON} - f_{\rm OFF}}\left(\frac{\tau_{\rm OFF} - \tau_{\rm ON}}{\tau_{\rm OFF} + \tau_{\rm ON}}\right). \quad (19.2.46)$$

According to the Dalitz model of $D^0 \to K_S^0 K^+K^-$ decays given in (Aubert, 2008l), the difference in $f_{\rm ON} - f_{\rm OFF}$ is $-0.753 \pm 0.004$ for the ON region given by $M_{K^+K^-} \in$ $[1.015, 1.025]\,\mathrm{GeV}/c^2$ and the OFF region given by the union of intervals $M_{K^+K^-} \in [2m_{K^\pm}, 1.010]\,\mathrm{GeV}/c^2$ and $M_{K^+K^-} \in [1.033, 1.100]\,\mathrm{GeV}/c^2$.

The advantage of this method over the time-dependent Dalitz analysis (described in Section 19.2.4) is that it can be performed on the much larger untagged sample of $D^0$ decays providing better sensitivity on the mixing-related parameter $y_{CP}$. The disadvantage is that the sensitivity to the mixing parameter $x$ is lost.

Belle performed a measurement of $y_{CP}$ using this method on a data sample corresponding to 673 fb$^{-1}$ (Zupanc, 2009). They find $(72.3 \pm 0.4) \times 10^3$ untagged signal $D^0$ candidates in the ON region and $(62.3 \pm 0.7) \times 10^3$ events in the OFF region. The proper decay time of the $D^0$ candidate is reconstructed as described in Section 19.2.1.5. However it is worth noting, that Belle determines the $D^0$ decay position by fitting only one of the charged kaons with the neutral kaon to a common vertex. The reason for using only $K^0_S K^\pm$ pairs for the $D^0$ decay vertex reconstruction, despite the worse resolution, is the strong correlation between the $K^+K^-$ invariant mass $M_{K^+K^-}$ and the mean proper decay time $\bar{t}$ of $D^0$ mesons that has been observed around $m_\phi$ in simulations if the vertex was reconstructed from $K^0_S K^+K^-$ or from $K^+K^-$ pairs (see Fig. 19.2.18). If not accounted for, this correlation could have biased the measurement of $y_{CP}$, and it may be explained as follows. Consider a $K^+K^-$ pair from the decay of a $\phi$ resonance. The reconstructed invariant mass $M_{K^+K^-}$ of the pair is determined from the relation $M^2_{K^+K^-} = 2m^2_{K^\pm} + 2E_{K^+}E_{K^-} - 2p_{K^-}p_{K^+}\cos\alpha_{\rm rec}$, where $p_{K^\pm}$ and $E_{K^\pm}$ are the momenta and energies of $K^\pm$, and $\alpha_{\rm rec}$ is the reconstructed opening angle between $K^+$ and $K^-$. If $\alpha_{\rm rec}$ is bigger (smaller) than the true opening angle $\alpha$ because of, for example, the Coulomb multiple scattering of $K^\pm$ in the detector material, $M_{K^+K^-}$ is shifted to higher (lower) values. Conversely, because of the narrow $\phi$ resonance, $m(K^+K^-) \lesssim m_\phi$ implies $\alpha_{\rm rec} \lesssim \alpha$ for the majority of $K^+K^-$ pairs from $\phi(1020)$. In addition, $\alpha_{\rm rec} \lesssim \alpha$ also implies $L_{\rm rec} \lesssim L$ and thus $t_{\rm rec} \lesssim t$, which then explains the correlation between $M_{K^+K^-}$ and $\bar{t}$.

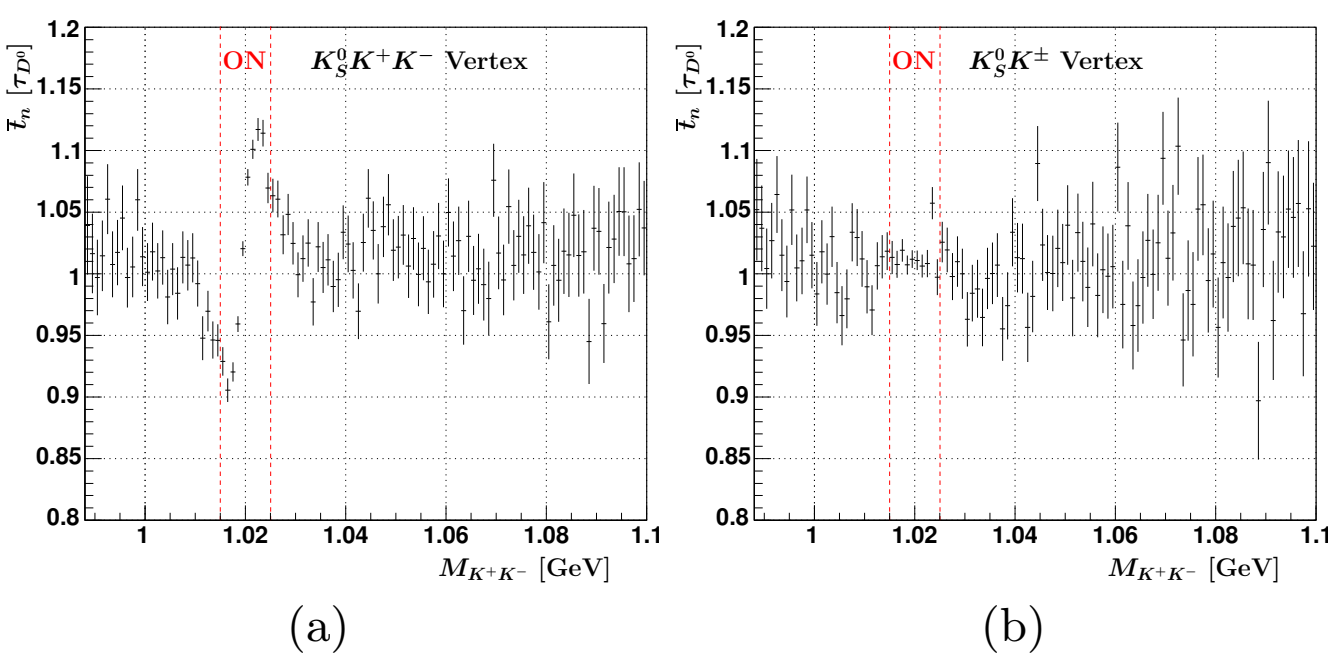

(a) (b)

**Figure 19.2.18.** The mean proper decay time dependence on $M_{K^+K^-}$, where the $D^0$ decay point is determined with a (a) $K^0_S K^+K^-$ and (b) $K^0_S K^\pm$ vertex constrained fit for signal $D^0 \to K^0_S K^+K^-$ decays. The two vertical red dashed lines indicate the borders of the ON $m(K^+K^-)$ interval. Belle internal, from the (Zupanc, 2009) analysis.

The uncertainties on the vertices of the production and decay of $D^0$ mesons are reflected in the uncertainties on the reconstructed decay time $t_{\rm rec}$. The widths (RMS) of the resolution function are $1.35\tau_{D^0}$ and $1.65\tau_{D^0}$ when using the $K^0_S K^+K^-$ vertex and the $K^0_S K^\pm$ vertex, respectively.

In Section 19.2.1.5 a typical p.d.f. used in a likelihood fit to the decay time distribution in charm mixing measurements is given (see Eq. 19.2.21). There exists a simpler and more robust method adopted in the analysis by Belle which has similar statistical sensitivity but does not require detailed knowledge of the resolution function or the time distribution of backgrounds. The average of the convolution is the sum of the averages of the convolved functions. The mean of the proper decay time distribution of a sample consisting of signal decays with lifetime $\tau_{\rm s}$ and signal fraction $p$ and of background is thus given by

$$\langle t \rangle = p(\tau_s + t_0) + (1-p)\langle t \rangle_b, \qquad (19.2.47)$$

where $\langle t \rangle$ and $\langle t \rangle_b$ are the mean decay times of all and of background events, respectively. The latter was estimated using the events in the sideband regions in the plane of $D^0$ and $K^0_S$ invariant masses. The parameter $t_0$ represents a possible non-zero mean of the signal resolution function. Any effect that may cause bias in the lifetime extraction ($t_0 \neq 0$) such as misalignment of the vertex detector is canceled by the use of kinematically equal decays – those in the ON and OFF region.

Belle measured $\tau_{\rm ON} + t_0^{\rm ON} = (413.4 \pm 2.5)$ fs and $\tau_{\rm OFF} + t_0^{\rm OFF} = (412.7 \pm 3.0)$ fs (Zupanc, 2009). The measured values for $\tau + t_0$ are close to the world average for $\tau_{D^0}$, and, since $y_{CP} \ll 1$, this implies $t_0/\tau$ is $\sim 1\%$ or less. Since the topology of events in the ON and OFF intervals is almost identical, Belle assumes $t_0^{\rm ON} = t_0^{\rm OFF}$ and includes a systematic error to account for this assumption. Using Eq. (19.2.46) Belle finds $y_{CP} = (+0.11 \pm 0.61 \pm 0.52)\%$, where the first uncertainty is statistical and the second systematic. This is so far the only measurement of the mixing parameter $y_{CP}$ using a $CP$-odd final state in $D^0$ decays. The value agrees with the measurements using $CP$-even final states described in Section 19.2.3.2.

#### 19.2.3.4 Summary on $y_{CP}$

Measurements of $y_{CP}$ have been at the forefront of the searches and subsequently precise measurements of $D^0 - \overline{D}{}^0$ mixing. Recent individual measurements exhibit significances between three and five standard deviations. Furthermore, they still exhibit larger statistical than systematic uncertainties. However, a good control of the systematic effects will be needed in future measurements at super flavor factories to significantly improve the accuracy of results. Main sources of systematic errors have been identified in the measurements performed at the $B$ Factories and several methods used to reduce the errors were successfully exploited.

Since the main systematic uncertainties are experimental, and method dependent, one can calculate the average value of the parameter $y_{CP}$ performed by the $B$ Factories assuming uncorrelated errors. The result is

$$y_{CP} = (0.86 \pm 0.16)\% \ . \qquad (19.2.48)$$

In the limit of $CP$ conservation $y_{CP} = y$, and hence the result points to a significant decay width difference for the two $D$ meson mass eigenstates.

### 19.2.4 $t$-dependent Dalitz analyses

#### 19.2.4.1 $K^+\pi^-\pi^0$ final state

For WS decays $D^0 \to K^+\pi^-\pi^0$, with an additional $\pi^0$, the phase space of possible final states is greatly increased. Each state can be represented as a point in a Dalitz plot with coordinates $(s_+, s_0)$, where $s_{+,0}$ are the squared invariant masses for the $K\pi^{+,0}$ systems. Ignoring $CP$ violation, each point is populated by decays with a time evolution described by Eq. (19.2.13) with values for $\delta_f$ and the ratio of $|\overline{A}_f/A_f|$ that are unique to that point (*i.e.* they depend on $s_{+,0}$). The interference term in Eq. (19.2.13), linear in $xt$ and $yt$, provides the greatest sensitivity to $x$ and $y$. As for all WS decay channels, the measurement profits from the interference of the doubly-Cabibbo-suppressed (DCS) decay amplitude and the Cabibbo-favored (CF) one preceded by mixing. The interference term is, therefore, comparable in magnitude to each of the other two terms. Furthermore, a model for the Dalitz plane point-to-point variations in $\delta$ can be used that allows, in principle, both $x$ and $y$ to be determined. For the two-body WS decays discussed in Section 19.2.2.1, there is just a single value for $\delta$ that allows a determination only of the combination $y' = y\cos\delta - x\sin\delta$ (and of $x'^2$).

Unfortunately, while a model can be found for the variations in $\delta$ across the separate Dalitz planes of $D^0$ and $\overline{D}^0$ decays, no model exists for the unknown relative strong phase ($\delta_{K\pi\pi}$) between one point in the $D^0$ and another in the $\overline{D}^0$ Dalitz plane. Therefore, only values for $x' = x\cos\delta_{K\pi\pi} + y\sin\delta_{K\pi\pi}$ and $y' = y\cos\delta_{K\pi\pi} - x\sin\delta_{K\pi\pi}$, rotated by this unknown phase, can be measured.

To date only *BABAR* has carried out a mixing analysis of this channel (Aubert, 2009u). The models for the complex decay amplitudes $\mathcal{A}_{DCS}$ and $\overline{\mathcal{A}}_{CF}$, respectively for DCS and CF decays, are based on the isobar model constructed from relativistic Breit-Wigner functions. The $K\pi$ $S$-wave components are described in a way suggested by $K\pi$ scattering in (Aston et al., 1988). This features a Breit-Wigner phase variation for the scalar $K_0^*(1430)$ added to a slowly varying background phase. Parameters (for the $K\pi$ $S$-wave and the complex coefficients for the isobars $K^*\pi$, $K\rho$, etc.) are determined from fits to the RS and WS samples. For $\overline{\mathcal{A}}_{CF}$ a fit to the time-integrated Dalitz plot for RS decays $\overline{D}^0 \to K^+\pi^-\pi^0$ (dominated by this amplitude) is used.

For $\mathcal{A}_{DCS}$, a full time-dependent fit to the WS sample is made. The *p.d.f.* for this has the form given in Eq. (19.2.13) with $\lambda_f$ taken as

$$\lambda(s_+, s_0) = r_0 e^{i\delta_{K\pi\pi}} \frac{\overline{\mathcal{A}}_{CF}(s_+, s_0)}{\mathcal{A}_{DCS}(s_+, s_0)}. \qquad (19.2.49)$$

In this fit, the mixing parameters are defined in the form ($x'/r_0$ and $y'/r_0$), where $r_0$ is the ratio between the CF and DCS amplitudes defined above. These are allowed to vary in the fit.

The time-dependent *p.d.f.* for this fit is convolved with a decay time resolution function derived from a fit to the RS events. The $D^0$ lifetime is also determined from RS events, and is found to agree with the world average (Beringer et al., 2012).

Signal samples consisting of 658,986 RS (purity 99%) and 3,009 WS (purity 50%) candidates are selected. Major sources of background come from a variety of wrongly reconstructed $D^0$ decays, wrongly associated slow pions or from a combination of both. In the WS sample, a small background also comes from events in which both $K^+$ and $\pi^-$ are mis-identified in the PID detectors. Simulated samples of these categories are used to determine the contributions of each in the data. The shape of background events in the RS and WS Dalitz plots are determined from $M$ and $\Delta m$ sideband regions in the data.

For both the RS and WS fits, efficiency variations over the Dalitz plot are estimated from MC samples generated uniformly over the phase space.

The Dalitz plots for the RS and WS samples, together with the distributions of $M$ and $\Delta m$ for the WS sample are shown in Fig. 19.2.19. In each of the Dalitz plots, bands due to charged and neutral states for $K^*(890)$ and for charged $\rho$ are easily seen. CF modes preferentially decay via $K^*\pi$ while DCS modes preferentially decay via $K\rho$ amplitudes.

Values for $x'$ and $y'$ are obtained for the combined $D^0$ and $\overline{D}^0$ samples. Separate values are also obtained from fits to each of the two subsamples (for the latter see more details in Section 19.2.7). In all cases, a value for $r_0$ is required. This is derived from the ratio $N_{\rm WS}/N_{\rm RS}$, where $N_{\rm WS}$ ($N_{\rm RS}$) is the number of wrong-sign (right-sign) signal events observed and their respective time-integrated *p.d.f.*s.

This procedure introduces a correlation between the $x'$ and $y'$ values obtained from the fit. Uncertainties in these mixing parameters are, therefore derived from the values obtained in a similar way for $10^6$ pairs of values for ($x'/r_0$, $y'/r_0$) randomly generated in accordance with the fit covariance matrix (assuming Gaussian errors and including systematic uncertainties).

The major systematic uncertainties arise from uncertainties in resonance masses and widths, and $K\pi$ $S$-wave parameters in the decay amplitude models, variations in the estimates for the numbers of WS and RS signal events and in parameters describing the time resolution.

The mixing parameter results obtained are

$$x' = (+2.61^{+0.57}_{-0.68} \pm 0.39)\% \qquad (19.2.50)$$

$$y' = (-0.06^{+0.55}_{-0.64} \pm 0.34)\% \qquad (19.2.51)$$

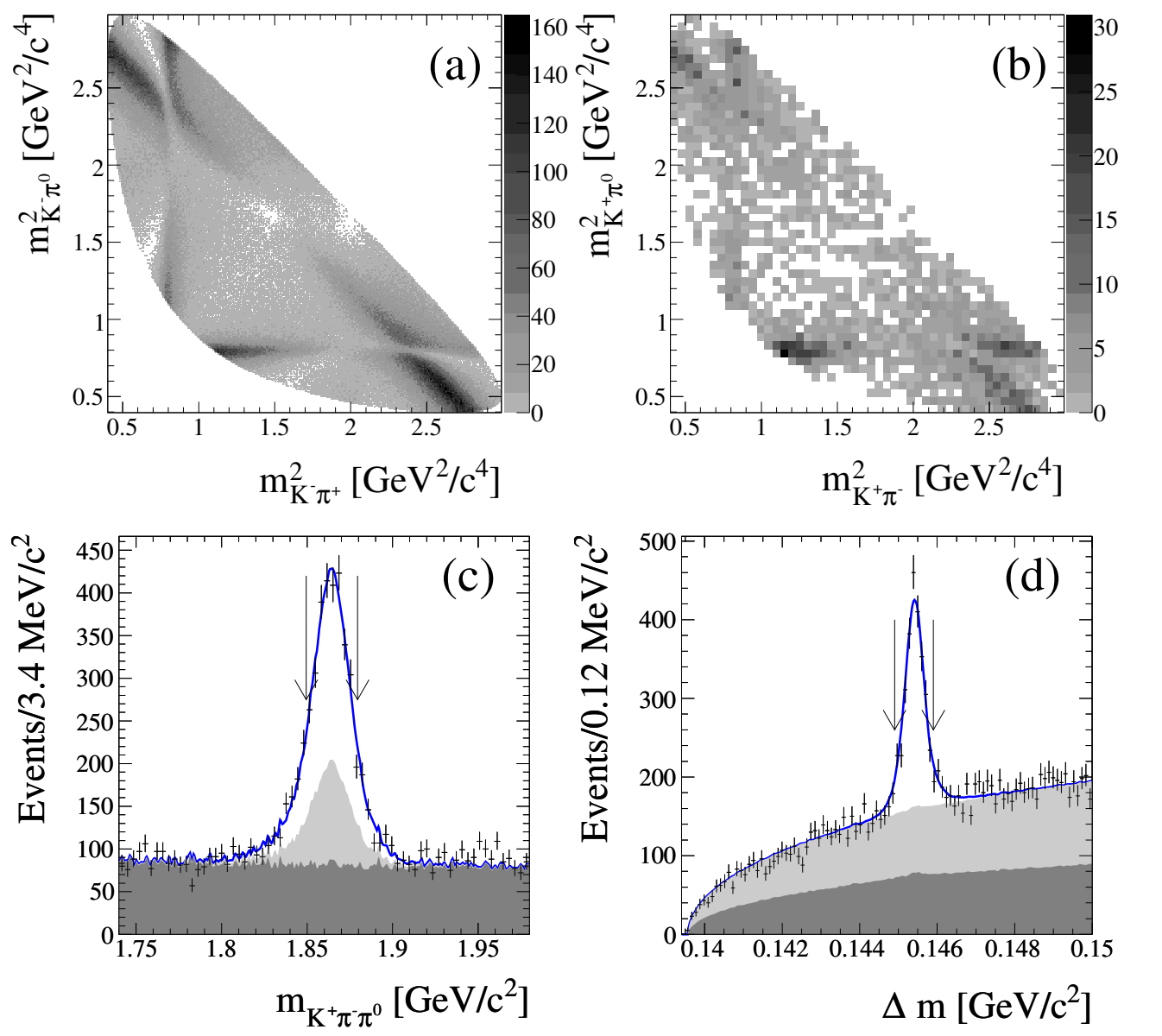

**Figure 19.2.19.** From (Aubert, 2009u). Dalitz plots for (a) RS decays $\overline{D}^0 \to K^+\pi^-\pi^0$ and (b) WS decays $D^0 \to K^+\pi^-\pi^0$. Invariant mass distributions for (c) $M$ ($D^0$ decay products) and (d) $\Delta m = (D^* - D^0)$ mass difference for selected WS event candidates are also shown. Arrows indicate the $M - \Delta m$ region selected for making the Dalitz plot fits described in the text.

where the first uncertainties are statistical and the second are systematic. A Bayesian approach is used to estimate the significance of these values. The data from the $D^0$ and $\overline{D}^0$ samples are consistent with a hypothesis of no mixing ($x' = y' = 0$) at the 3.2$\sigma$ level. This constitutes evidence for mixing.

### 19.2.4.2 $K^0_S h^+ h^-$ final state

Analyses of these channels are important because the final states are *CP* self-conjugate. This implies a vanishing phase between two equal points in the Dalitz planes of $D^0$ and $\overline{D}^0$ decays (analogous to $\delta_{K\pi\pi}$ for $D^0 \to K^+\pi^-\pi^0$ decays, $\delta_{K^0_S h^+h^-} = 0$). This allows a determination (modulo sign) of both $x$ and $y$, and of their relative sign, with no unknown strong phase.

As for WS decays to $D^0 \to K^+\pi^-\pi^0$, the phase space of possible final states is represented by points in a Dalitz plot, with coordinates chosen as $(s_+, s_-)$, where $s_{+,-}$ are the squared invariant masses for the $K^0_S h^{+,-}$ systems, respectively. Ignoring *CP* violation, each point is populated by decays with a time evolution described by Eq. (19.2.13) with values for $\delta_{K^0_S h^+h^-}$ and $|\lambda| = |\overline{\mathcal{A}}_{K^0_S h^+h^-}/\mathcal{A}_{K^0_S h^+h^-}|$ that are unique to that point. The largest sensitivity to $x$ and $y$ arises from the interference term in Eq. (19.2.13) which is linear in the two parameters. Unlike the WS decay channel, however, these decays proceed either by a DCS amplitude after mixing or by a CF one with no mixing at all. They are therefore dominated by the latter process and the interference term is very much smaller. The decay amplitudes for a decay of $D^0$ and $\overline{D}^0$ tagged at $t = 0$ are written as

$$\begin{aligned}\mathcal{M}(s_+, s_-, t) &= \frac{1}{2p}\big[p\big(e_1(t) + e_2(t)\big)\mathcal{A}(s_+, s_-)\\ &\quad + q\big(e_1(t) - e_2(t)\big)\mathcal{A}(s_-, s_+)\big]\\ \overline{\mathcal{M}}(s_+, s_-, t) &= \frac{1}{2q}\big[p\big(e_1(t) - e_2(t)\big)\overline{\mathcal{A}}(s_+, s_-)\\ &\quad + q\big(e_1(t) + e_2(t)\big)\overline{\mathcal{A}}(s_-, s_+)\big], \end{aligned} \tag{19.2.52}$$

where $e_{1,2}(t) = e^{-i(m_{1,2} - (i\Gamma_{1,2}/2))t}$. Both *BABAR* (using 486.5 fb$^{-1}$, (del Amo Sanchez, 2010f)) and Belle (using 540 fb$^{-1}$, (Abe, 2007b)) have analyzed these channels. The model for the complex decay amplitude $\mathcal{A}(s_+, s_-)$, contains CF and DCS terms. As before, these are based on the isobar model constructed from relativistic Breit-Wigner functions. The *BABAR* collaboration also used a $K\pi$ $S$-wave prescription similar to that outlined for the WS channel, and also used a K-matrix parameterization for the $\pi\pi$ $S$-wave. In the Belle analysis the latter was used to estimate systematic uncertainties due to the assumed Dalitz model. Details of these amplitudes are described in Chapter 13. In calculating $\lambda(s_+, s_-)$ in the standard fit, it is assumed that direct *CP* violation can be ignored so that the amplitude $\overline{\mathcal{A}}(s_+, s_-)$ can be taken as $\mathcal{A}(s_-, s_+)$. Belle also performs a fit, in which direct *CP* violation is permitted, where separate isobar coefficients are allowed for the definition of $\overline{\mathcal{A}}(s_+, s_-)$. They find, however, no significant differences in these coefficients.

As in the WS $K^+\pi^-\pi^0$ mode, the time-dependent *p.d.f.* is convolved with a decay time resolution function derived from the combined $D^0$ and $\overline{D}^0$ event samples. This fit is performed for both of these samples, each tagged by the charge of the slow pion from $D^*$ decays.

Belle and *BABAR* each use approximately $540k$ $K^0_S\pi^+\pi^-$ candidates, and *BABAR* also uses $80k$ $K^0_S K^+K^-$ events. Event purities for these samples range from 95-99%. Major sources of the small backgrounds come from a variety of wrongly reconstructed $D^0$ decays, wrongly associated slow pions or from a combination of both. In the $K^0_S\pi\pi$ mode, there is also a small, but significant background from $D^0 \to K^0_S K^0_S$ decays and another from $D^0 \to 4\pi$. Simulated samples of these backgrounds are used to determine the contributions of each in the data. Efficiency variations over the Dalitz plot are estimated from MC samples generated uniformly in phase space.

The decay time distributions are shown in Fig. 19.2.20. Results for the mixing parameters obtained are summarized in Table 19.2.4. They are obtained from fits neglecting *CP* violation, *i.e.* setting $q = p = 1/\sqrt{2}$ in Eq. (19.2.52). Separate fits are performed taking into account the possibility of *CP* violation, as discussed further in Section 19.2.7. The *BABAR* values are the most precise at present. Neither Belle nor *BABAR* show results more than $3\sigma$ from the "no mixing" point $x = y = 0$. The 95% C.L. contour for $(x, y)$ parameters obtained by Belle is shown in Fig. 19.2.21.

Systematic uncertainties in the measurement can be divided into two groups: uncertainties related to experi-mental effects and those related to modeling the Dalitz distribution. The contributions to the systematic error related to the former group are related to the efficiency variations across the DP, modeling of the DP and proper-time distributions for background events, and the selection criteria. The misidentification of the $D^0$ flavor from incorrectly assigned slow pions and variations of the time resolution function, including alternative ways to describe the correlation between time and DP position also contribute to this group of uncertainties. An important source of the experimental systematic uncertainty is also the limited statistics of full detector simulations required to study biases from event selection and instrumental effects arising from the small misalignment of the detector.

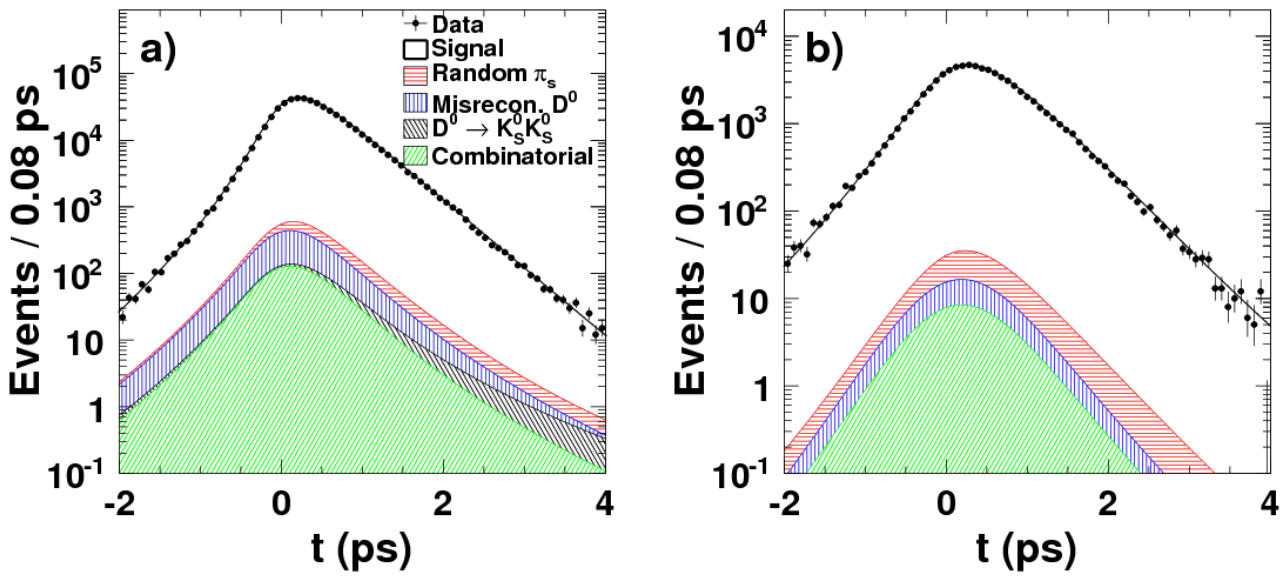

**Figure 19.2.20.** From (del Amo Sanchez, 2010f). Decay time distributions for (a) $D^0 \to K_S^0\pi^-\pi^+$, (b) $D^0 \to K_S^0K^+K^-$ decays with the time-dependent Dalitz plot fits described in the text.

**Table 19.2.4.** Results of fits to $K_S^0h^+h^-$ $CP$ self-conjugate states (del Amo Sanchez, 2010f; Abe, 2007b). The first uncertainties are statistical and the second are systematic. The third uncertainties arise from uncertainties in the model for $\mathcal{A}(s_+, s_-)$.

| Experiment | Sample | Results [$\times 10^3$] |
|---|---|---|
| BaBar | 486.5 fb$^{-1}$ | $x = 1.6 \pm 2.3 \pm 1.2 \pm 0.8$ |
| No $CP$ | Signal: $540 \times 10^3$ | $y = 5.7 \pm 2.0 \pm 1.3 \pm 0.7$ |
| violation | Purity: 98.5% | |
| Belle | 540 fb$^{-1}$ | $x = 8.0 \pm 2.9^{+0.0+1.0}_{-0.7-1.4}$ |
| No $CP$ | Signal: $534 \times 10^3$ | $y = 3.3 \pm 2.4^{+0.8+0.6}_{-1.2-0.8}$ |
| violation | Purity: 98.5% | |

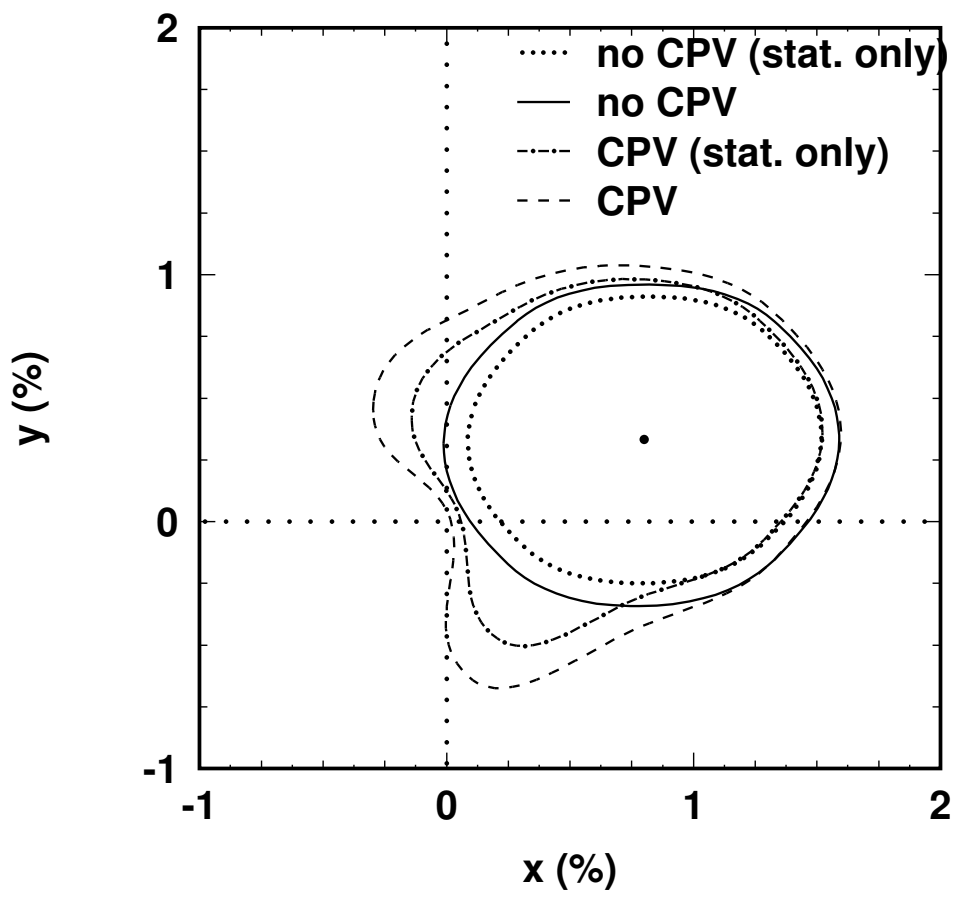

**Figure 19.2.21.** 95% C.L. contour for the mixing parameters $(x, y)$ as obtained from decay-time dependent analysis of Dalitz distribution of $D \to K_S^0\pi^+\pi^-$. From (Abe, 2007b).

The second group of systematic uncertainties is estimated using alternative models for the decay amplitudes $\mathcal{A}(s_+, s_-)$. This introduces one of the largest systematic uncertainties that would ultimately limit the precision of these results. Further improvements should be possible, however, using a model-independent method to analyze the Dalitz distribution. The method is based on the measurements of the strong phase variation over the Dalitz plane as measured from the charm threshold data (Libby et al., 2010) and does not require modelling of the Dalitz distribution.

#### 19.2.4.3 Summary

Study of the decay-time dependence of Dalitz distributions in multibody $D^0$ decays provides an essential tool in studying the charm mesons mixing properties. It is the only method which is sensitive to linear order in both mixing parameters, $x$ and $y$. Especially the time-dependent Dalitz analysis of self-conjugated states ($K_S^0h^+h^-$) enables the determination of the parameters not rotated by an unknown phase. On the other hand such measurements carry a systematic uncertainty arising from modeling the Dalitz distribution. With the increasing statistical power of the data samples the models have been gaining in sophistication. Nevertheless in the future the most accurate measurements of the mixing parameters can be expected using a model independent approach (similar to the measurement of $\phi_3$, see Section 17.8.4.3), which has not yet been used by the $B$ Factories.

### 19.2.5 Semileptonic decays

In the case of semileptonic decays, there are no Doubly Cabibbo Suppressed (DCS) modes as with wrong-sign hadronic decays, and only a pure mixing term modulates the exponential lifetime. In the absence of $CP$ violation, from Eqs (19.2.13) one obtains [131]

$$\frac{d\Gamma(D^0 \to X\ell^+\nu_\ell)}{dt} \sim |A_{X\ell^+\nu_\ell}|^2 e^{-\Gamma t} \; ,$$
$$\frac{d\Gamma(\overline{D}^0 \to X\ell^+\nu_\ell)}{dt} \sim |A_{X\ell^+\nu_\ell}|^2 \left[\frac{x^2+y^2}{4}(\Gamma t)^2\right] e^{-\Gamma t} \; . \quad (19.2.53)$$

[131] By taking $f = X\ell^+\nu_\ell$, $A_{\overline{f}} = \overline{A}_f = 0$, $A_f = \overline{A}_{\overline{f}}$.

Integrating $d\Gamma(\overline{D}^0 \to X\ell^+\nu_\ell)/dt$ over all times $t > 0$ and normalizing to the integrated value of $d\Gamma(D^0 \to X\ell^+\nu_\ell)/dt$, one finds that the relative time-integrated mixing rate is

$$R_M = \frac{\Gamma(\overline{D}^0 \to X\ell^+\nu_\ell)}{\Gamma(D^0 \to X\ell^+\nu_\ell)} = \frac{x^2+y^2}{2}. \qquad (19.2.54)$$

Both *BABAR* and Belle have searched for neutral $D$ mixing in semileptonic $K^{(*)}\ell\nu_\ell$ final states, setting upper limits with two conceptually different analyses using an integrated luminosity of 344 fb$^{-1}$ in the 2007 *BABAR* analysis (Aubert, 2007aq) and 492 fb$^{-1}$ for the 2008 Belle analysis (Bitenc, 2008). The *BABAR* analysis uses the flavor of fully reconstructed hadronic charm decays in the hemisphere opposite the semileptonic signal to provide an additional tag of the production flavor of signal decays (supplemental to the information from the slow pion in the $D^*$ signal decay). This double flavor tag very substantially reduces the rate of incorrectly tagged signal candidates, but it also greatly reduces the overall signal efficiency. Belle's analysis does not use any additional flavor tagging information from the opposite hemisphere, relying instead on a maximum likelihood fit to search for mixed signal events. In both analyses the efficiency corrected ratio of mixed $D^0 \to X\ell^-\overline{\nu}_\ell + \overline{D}^0 \to X\ell^+\nu_\ell$ to un-mixed $D^0 \to \overline{X}\ell^+\nu_\ell + \overline{D}^0 \to \overline{X}\ell^-\overline{\nu}_\ell$ events is used to determine the constraints on $x^2+y^2$.

### 19.2.5.1 Belle

Belle reconstructs the decay chain $D^{*+} \to D^0\pi_s^+$, $D^0 \to K^-l^+\nu_l$, where $l^+$ can be either an electron or a muon. The charge of the slow pion $\pi_s^+$ tags the production flavor of the neutral $D$, with unmixed decays having a lepton and soft pion with identical charge (RS) and mixed decays having a lepton and soft pion with opposite-sign charge (WS). Although the neutrino is not directly detected, the uncertainty due to the missing neutrino four-momentum can be minimized by calculating the mass difference between the $D^0$ and its $D^{*+}$ parent

$$\Delta M \equiv M(K\ell\nu\pi_s) - M(K\ell\nu). \qquad (19.2.55)$$

Since the Belle detector covers nearly the entire solid angle surrounding the interaction point, the signal neutrino four-momentum can be estimated as

$$P_\nu = P_{\rm CM} - P_{K\ell} - P_{\rm ROE}, \qquad (19.2.56)$$

where $P_{\rm CM}$ denotes the CM four-momentum of the initial $e^+e^-$ frame of reference and ROE stands for the "rest of the event", *i.e.* the sum of the CM system four-momenta of all detected neutral and charged candidates in an event except for the signal kaon and lepton. Neutrals with energy less than 70 MeV/$c^2$ and charged tracks with an impact parameter larger than 5 cm (2 cm) in $z$ ($xy$) are not used in this context, although they are used in the computation of event-shape variables. Two kinematic constraints are used to improve the resolution of the neutrino momentum. The invariant mass squared $M^2(K\ell\nu) \equiv (P_\nu + P_{K\ell})^2/c^2$ is computed, and only candidates with $-25 < M^2(K\ell\nu) < 64\,{\rm GeV}^2/c^4$ are kept. For these events, $P_{\rm ROE}$ is rescaled by a correction factor $\xi$ requiring that

$$M^2(K\ell\nu) = (P_{\rm CM} - \xi P_{\rm ROE})^2/c^2 \equiv M^2_{D^0}. \qquad (19.2.57)$$

Having thus determined $\xi$, the neutrino four-momentum is then recalculated as $P_\nu = P_{\rm CM} - P_{K\ell} - \xi P_{\rm ROE}$, and this new neutrino momentum is used in the calculation of $\Delta M$. The most probable value for $\xi$ is close to 1.0, although there is a long tail in the distribution, which leads to an average value of $\sim 1.3$.

The square of the missing mass $M_\nu^2 \equiv P_\nu^2 = 0$ is used as an additional kinematic constraint, where $P_\nu^2$ now includes the $\xi$ correction factor. The squared missing mass can be expressed in terms of the energies and magnitudes of the three-momenta of the final state particles, along with the cosine of the angle between the vector momenta of the $K\ell$ system and $\xi P_{\rm ROE}$. This angle is corrected by rotating the vector $P_{\rm ROE}$ in the plane defined by the vectors $P_{\rm ROE}$ and $p_{K\ell}$ so that the null mass condition is enforced. The final neutrino four-momentum is then calculated using the original expression in Eq. (19.2.56) with a corrected $P_{\rm ROE}$ term on the right-hand side, and this neutrino momentum is used to compute the value of $\Delta M$ which is subsequently used in the fit to the data. The effect of the kinematic constraints on the $\Delta M$ distribution of the simulated WS signal events is shown in Fig. 19.2.22.

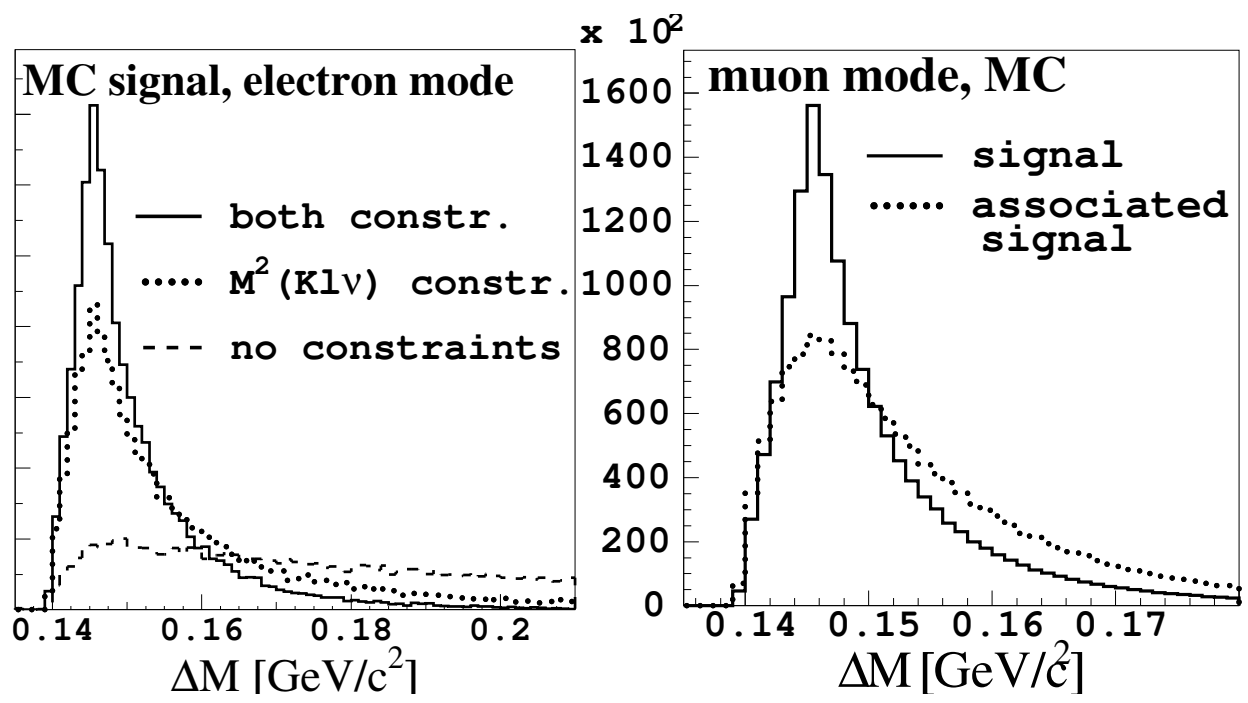

**Figure 19.2.22.** Left: The effect of the kinematic constraints on the $\Delta M$ distribution of the simulated $D^0 \to K^-e^+\nu_e$ decays. Right: $\Delta M$ distribution for simulated associated signal events (see text for the explanation). From (Bitenc, 2008).

The ratio of WS to RS events $N_{\rm WS}/N_{\rm RS} = (x^2+y^2)/2$ is small (n.b. $x^2+y^2 \sim \mathcal{O}(10^{-4})$, see Section 19.2.1), or equivalently $N_{\rm WS} \ll N_{\rm RS}$. Hence the central value and the statistical uncertainty of the ratio is mainly determined by the number of WS events.[132] The selection criteria for the

[132] Note that in the ratio $R_M = N_{\rm WS}/N_{\rm RS}$ the statistical error is given by $\sigma_R = R_M\sqrt{(\sigma_{\rm WS}/N_{\rm WS})^2 + (\sigma_{\rm RS}/N_{\rm WS})^2}$; in case $N_{\rm RS}$ and $N_{\rm WS}$ are simply Poissonian distributed numbers of events $\sigma_R = R_M\sqrt{(1/N_{\rm WS}) + (1/N_{\rm RS})}$ and the smaller of the two numbers determines the uncertainty.

WS events can thus be determined using a much larger sample of kinematically equal RS decays in accordance with blind analysis principles described in Chapter 14.

The selection criteria include the requirements on minimum invariant mass and momentum of the kaon and lepton system, which suppress the backgrounds from various $B$ meson decays. Two important sources of background are $D^0 \to K^+K^-$ and $\pi^+\pi^-$ decays, which in case of misidentification of one or both final state mesons produce a peak in the $\Delta M$ distribution. These are effectively suppressed by calculating the invariant mass of the $K\ell$ system, assigning an appropriate mass to the kaon and lepton candidate,[133] and requiring that the resulting value is not consistent with the nominal $D^0$ mass. The requirement on the magnitude of the CM momentum of the $K\ell$ system, $p^*(K\ell) > 2.0\,\mathrm{GeV}/c^2$, improves the resolution on $\Delta M$ and rejects a majority of $D^0$ mesons produced in $B$ meson decays. Events with $\gamma \to e^+e^-$ conversions are also a source of background for electron and slow pion candidates. These are suppressed by calculating the invariant mass of the electron and the pion candidate, assigning the electron mass to both tracks, and requiring the result to be larger than $140\,\mathrm{MeV}/c^2$.

Apart from the aforementioned backgrounds and genuine signal there are other $D$ meson decays with $\Delta M$ in the signal region. Despite the fact that the measurement method is not specifically aiming to reconstruct those decays the charge correlation between the lepton and the $\pi_s$ is the same as in the signal decays and hence they carry similar information on the possible mixing parameters. For this reason they are referred to as the associated signal:

- $D^0 \to K^-\pi^0\ell^+\nu$,
- $D^0 \to K^{*-}\ell^+\nu_\ell$, followed by $K^{*-} \to K^-\pi^0$,
- $D^0 \to \pi^-\ell^+\nu_\ell$,
- $D^0 \to \rho^-\ell^+\nu_\ell$, followed by $\rho^- \to \pi^-\pi^0$,
- $D^0 \to K^{*-}\ell^+\nu_\ell$, followed by $K^{*-} \to \overline{K}^0\pi^-$.

The mass difference distribution of simulated associated signal events is shown in Fig. 19.2.22.

The mixed events have on average a larger decay time than the background WS events. Hence requiring a larger value of the decay time for selected events can improve the sensitivity. This is illustrated in Fig. 19.2.23 showing the decay time distribution of WS (signal) events.

The proper decay time scaled in units of the nominal PDG $D^0$ lifetime (Yao et al., 2006) is calculated from the $D^0$ momentum $\boldsymbol{p}_{D^0}$ and flight distance $l$:

$$t_{D^0} = \frac{m_{D^0} l}{\tau_{D^0} p_{D^0}}. \quad (19.2.58)$$

The $D^0$ flight distance is the distance between the $D^0$ production and decay vertices, respectively $\boldsymbol{r}_{\rm prod}$ and $\boldsymbol{r}_{\rm dec}$. The $D^0$ momentum is calculated by summing the momenta of the daughter particles. The decay vertex is obtained by fitting the kaon and lepton tracks to a common vertex. The production vertex is obtained by extrapolating the $D^0$ momentum vector to the $e^+e^-$ interaction region. Given the relatively large longitudinal extent of the interaction region as explained in Chapter 6, only the transverse components ($x$, $y$) are used. The radial flight distance $l_{xy}$ is calculated as

$$l_{xy} = \frac{(r^x_{\rm dec} - r^x_{\rm prod}, r^y_{\rm dec} - r^y_{\rm prod}) \cdot (p^x_{D^0}, p^y_{D^0})}{\sqrt{(p^x_{D^0})^2 + (p^y_{D^0})^2}}. \quad (19.2.59)$$

The dimensionless proper decay time is then calculated as

$$t_{xy} = \frac{m_{D^0} l_{xy}}{\tau_{D^0}\sqrt{(p^x_{D^0})^2 + (p^y_{D^0})^2}}. \quad (19.2.60)$$

Because of data recorded by two different Belle SVD detector configurations (Chapter 2) the selected sample is divided into two subsamples. The subsamples are denoted as $\ell - i$, where $\ell = e, \mu$ determines the type of the lepton and $i = 1, 2$ the SVD configuration.

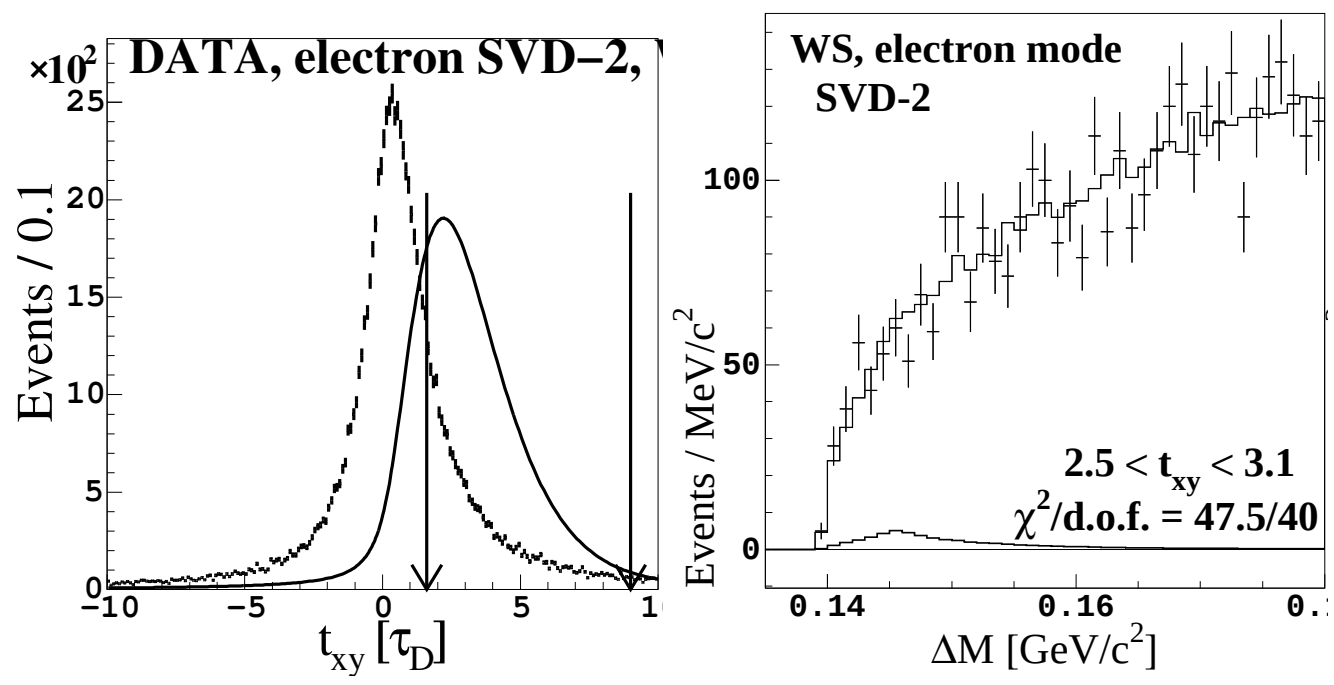

**Figure 19.2.23.** Left: The distribution of the decay time in the transverse plane for selected WS events (dots). The solid line shows the simulated distribution of signal events and the vertical arrows the interval selected for the signal. Right: An example of the fit to the $\Delta M$ data distribution (points) in a single $t_{xy}$ bin for the WS $e-2$ data subsample. The histogram is the fit to the distribution, with the contribution of signal shown along the horizontal axis. From (Bitenc, 2008).

The relative rate of mixed events $N^i_{\rm WS}/N^i_{\rm RS}$ is determined separately for $i = 1, 6$ in six intervals of $t_{xy}$ for decay times $1.6 < t_{xy} < 9.0$. An example of the fit to the $\Delta M$ distribution of WS events in a single $t_{xy}$ bin is shown in Fig. 19.2.23. The shape of the signal is obtained from the simulation. The background is divided into two categories: the correlated and the un-correlated background. The former is defined as a combination in which either the lepton or the kaon candidate or both originate from the same decay chain as the slow pion. The shape of this background is obtained from MC simulation as well. The majority of events in the WS sample belong to the uncorrelated background. The shape of the $\Delta M$ distribution for these events is obtained from real data by embedding a slow pion candidate track into another event and following the same analysis procedure as described above.

[133] Specifically, the invariant mass is calculated assuming the pion mass for the kaon candidate and the pion mass for the lepton candidate, as well as with the kaon mass for the lepton candidate.

Averaging the $(N^i_{\rm WS}/N^i_{\rm RS})(\epsilon^i_{\rm RS}/\epsilon^i_{\rm WS})$ values over the $t_{xy}$ bins, where $\epsilon^i_{\rm RS,WS}$ are the relative acceptances for the RS and WS events in the corresponding bins, one obtains $R_M \equiv (x^2+y^2)/2$ values for $\ell-i$ subsamples. Those values are shown in Fig. 19.2.24.

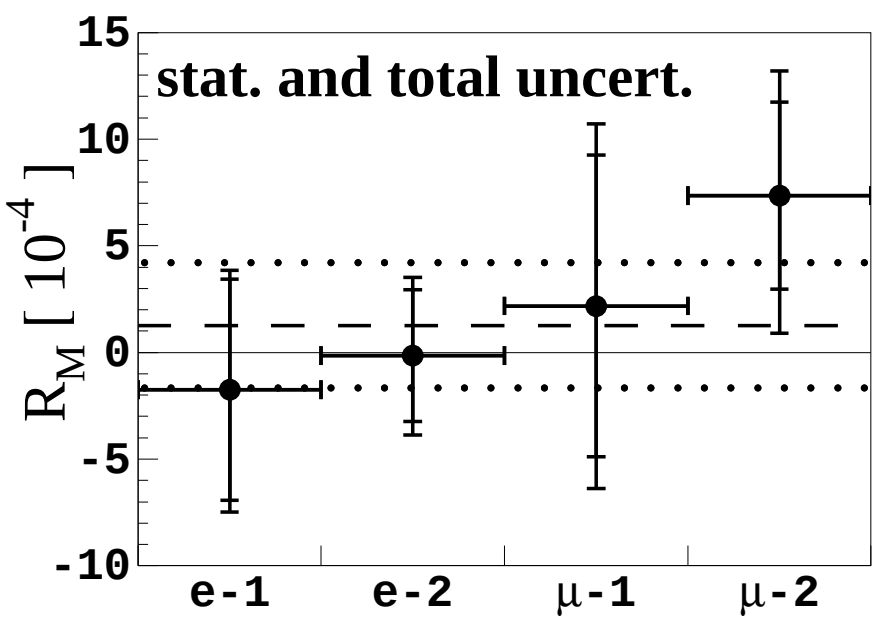

**Figure 19.2.24.** Values of $R_M \equiv (x^2+y^2)/2$ in subsamples of selected $D^0 \to K^-\ell^+\nu_\ell$ events (dots with error bars representing the statistical and systematic uncertainty). The dashed and dotted lines represent the average value and its $\pm 1\sigma$ interval. The solid line corresponds to no mixing. From (Bitenc, 2008).

The average value of the measurement is

$$R_M = (1.3 \pm 2.2 \pm 2.0) \times 10^{-4} , \qquad (19.2.61)$$

including the statistical and systematic uncertainty. The value is consistent with no-mixing and is close to the boundary of the physical region ($R_M > 0$). In order to obtain the upper limit in the vicinity of the physical boundary a Feldman-Cousins (Feldman and Cousins, 1998) method is used. The 90% C.L. upper limit is found to be $R_M < 6.1 \times 10^{-4}$.

One of the largest sources of systematic uncertainty is the finite statistical significance of the samples used to obtain the shapes of the signal and background $\Delta M$ distribution. The uncertainty is estimated by variation of those shapes within their statistical uncertainties. Another important source of systematic error is the amount of the correlated background in the WS sample. It is estimated by conservatively varying the branching fractions of the main decay mode contributions to this type of the background.

### 19.2.5.2 *BABAR*

In the *BABAR* analysis (Aubert, 2007aq), the initial flavor of the neutral $D$ meson is tagged twice, once using the slow pion from a charged $D^*$ decay whose neutral $D$ daughter decays semileptonically, and once using the flavor of a high-momentum $D$ fully reconstructed in the CM hemisphere opposite the semileptonic candidate. Tagging the flavor at production twice, rather than once, highly suppresses the background from false WS slow pions, but it also reduces the signal by an order of magnitude. Additional signal candidate selection criteria similar to that employed above by Belle are used to minimize the remaining sources of background.

Five hadronic tagging samples are used, where three samples explicitly require a reconstructed $D^{*+}$: $D^{*+} \to D^0\pi^+$ with $D^0 \to K^-\pi^+$, $D^0 \to K^-\pi^+\pi^0$, and $D^0 \to K^-\pi^+\pi^+\pi^-$, and the other two samples are CF decays with no $D^{*+}$ requirement: $D^0 \to K^-\pi^+$ and $D^+ \to K^-\pi^+\pi^+$. Candidates from the $D^{*+}$ sample are explicitly excluded from the inclusive $D^0 \to K^-\pi^+$ sample to ensure that the tagging samples are disjoint.

The selection criteria for the tagging samples, such as the $\Delta M$ ranges for the $D^*$ modes or the use of production and decay vertex separation for the $D^+$ mode, vary from channel to channel to balance high purity against high statistical significance. To eliminate candidates from $B\overline{B}$ events, the CM momentum of the tag-side $D$ must be at least 2.5 GeV/$c$.

Using a method similar to that employed above by Belle, the optimal proper decay time range in which to search for mixed decays was similarly found to be $\sim 1.5-9.5$ nominal $D^0$ lifetimes. The *BABAR* fit to the $\Delta M$ distribution for double-flavor-tagged RS data events uses a fit model similar to Belle's, which is shown in Figure 19.2.25 before and after additional kinematic selection exploiting correlations between tag and signal hemispheres are applied. The final RS signal yield is $4780 \pm 94$ events after all event selections are imposed.

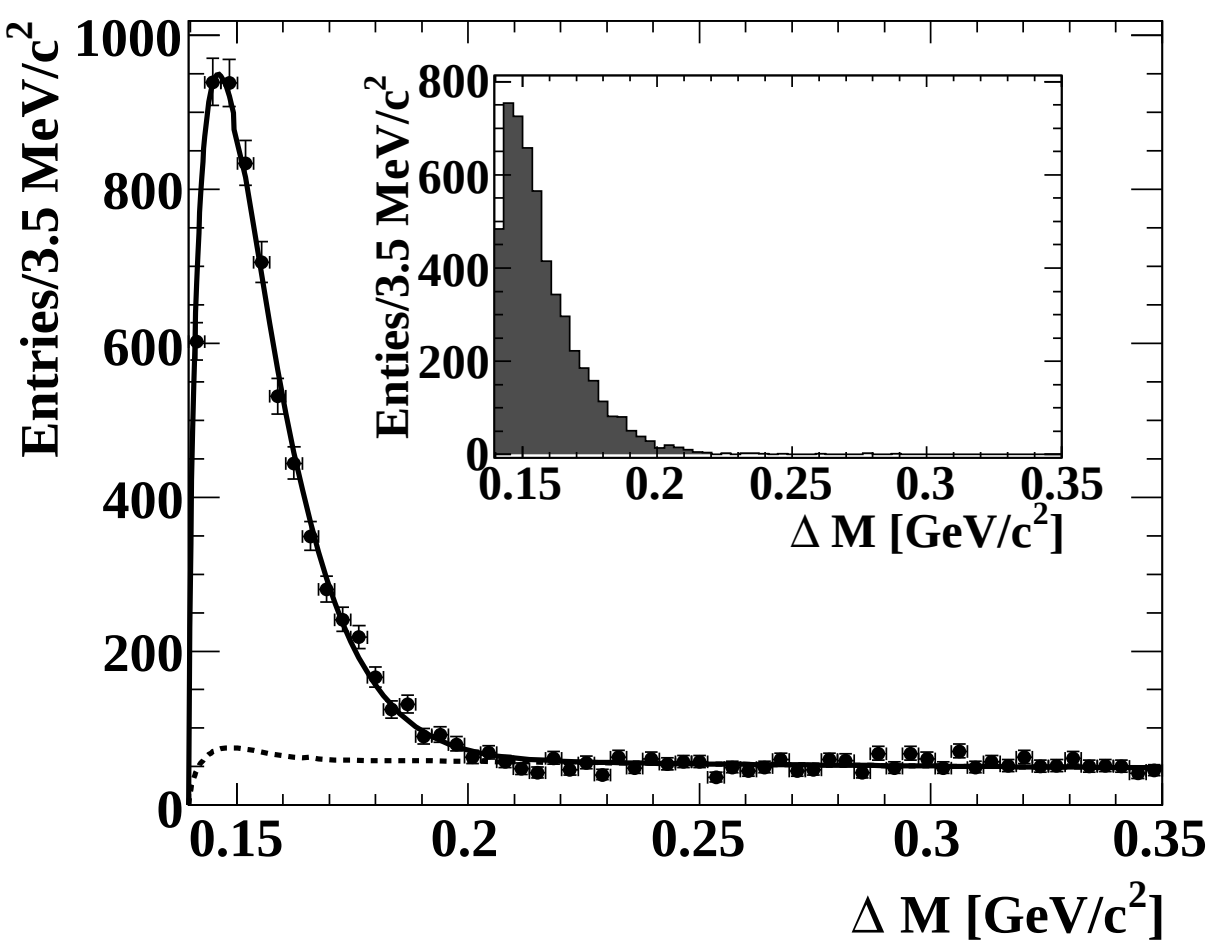

**Figure 19.2.25.** From (Aubert, 2007aq). RS data $\Delta M$ distribution. The main plot shows the RS data (points) before imposing the double-tag kinematic selection, and the projections of the total fit *p.d.f.* (solid line) and the background *p.d.f.* (dashed line). The inset plot shows the RS $\Delta M$ distribution after the double-tag kinematic selection criteria are applied.

Three regions of $\Delta M$ are considered to determine the number of WS mixed events: the signal region, $\Delta M \leq 0.20\,\text{GeV}/c^2$; the near background region, $0.20 < \Delta M \leq 0.25\,\text{GeV}/c^2$; and the far background region, $0.25 < \Delta M \leq 0.35\,\text{GeV}/c^2$. These $\Delta M$ ranges are shown in Fig. 19.2.26, and are respectively labeled "1", "2" and "3" in the plot. A blind analysis was performed where the signal region was not examined until after all details of event selection, fit methodology and statistical procedures for setting upper limits were finalized. An estimated 2.85 background events was expected in the signal region and, as shown in Fig. 19.2.26, three events were found there, yielding a net WS signal of 0.15 events. Note the difference in the yields and purities of the selected samples using the Belle single tag method (Fig. 19.2.23) and the *BABAR* double tag method (Fig. 19.2.26).

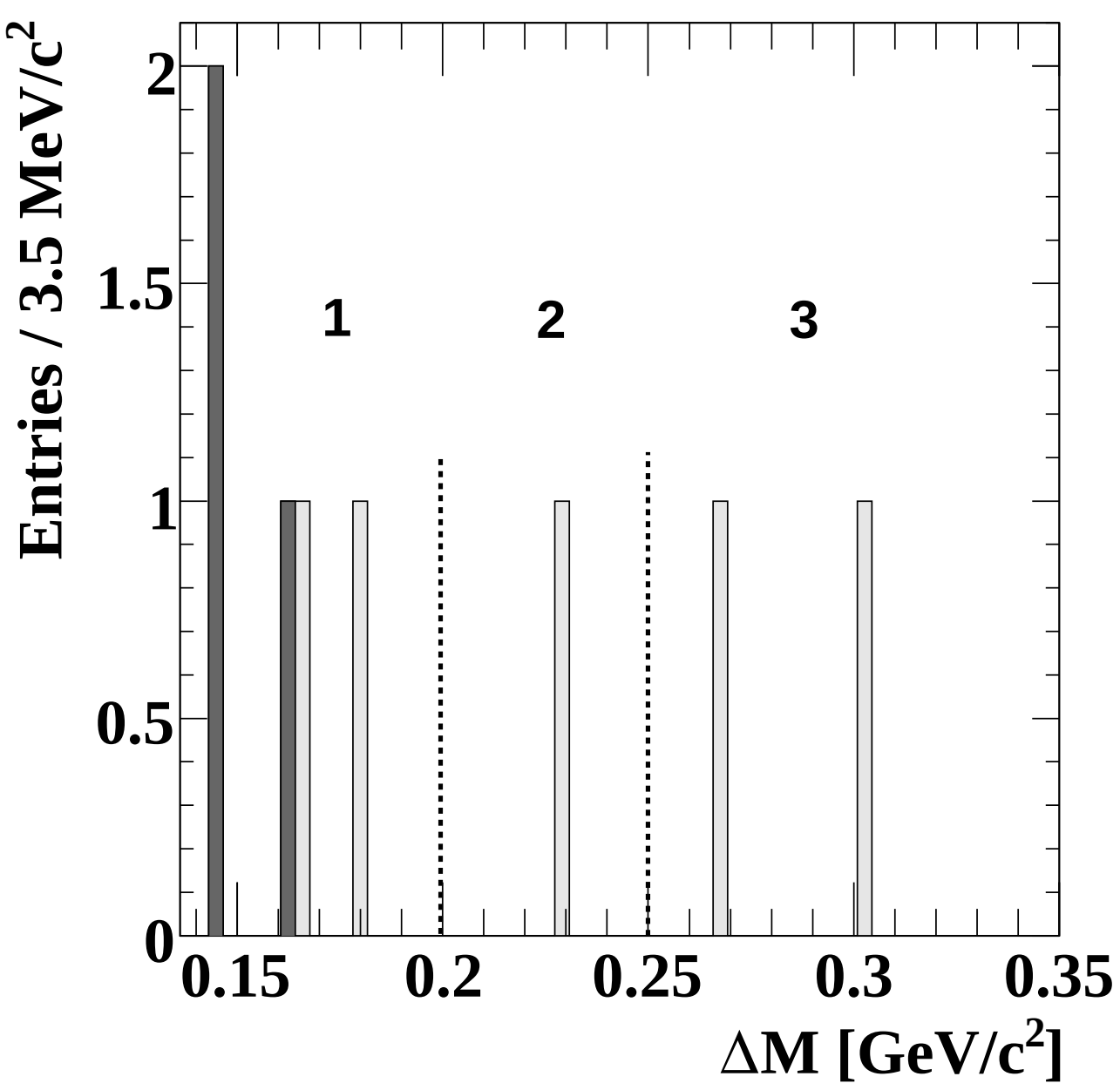

**Figure 19.2.26.** From (Aubert, 2007aq). WS data $\Delta M$ distribution. The dark histogram shows WS events in the data passing all event selections. The light histogram shows WS events passing all selections except the double-tag kinematic selection. Region "1" is the signal region, "2" is the near sideband, and "3" is the far sideband.

To calculate confidence intervals for the number of mixed events observed, a systematic uncertainty associated with the WS background estimate is determined using ten data/MC background control samples. The largest discrepancy between the data and MC rates, 50%, is assigned as the systematic uncertainty associated with the ratio between the MC estimate of the background rate and its true value. To quantify confidence intervals for the number of WS mixed events, a likelihood function is used, $\mathcal{L}(n, n_b; s, b)$, for the number of events observed in the signal region of the WS data sample, $n$, and the corresponding number observed in the MC sample, $n_b$. The likelihood $\mathcal{L}(n, n_b; s, b)$ depends upon the true signal rate $s$ and the true background rate $b$ in the signal region, and also accounts for the systematic uncertainty in the ratio of the true background rate in data to that estimated from MC. The value of $(s, b)$ which maximizes the likelihood function, $\mathcal{L}_{\max}$, is denoted by $(\widehat{s}, \widehat{b})$. As one might naïvely expect, $\widehat{b}$ is equal to $n_b$ times the ratio of data and MC luminosities, while $\widehat{s} = n - \widehat{b}$. A scan is made for the values of $s$ where $-\ln\mathcal{L}(s)$ changes by 0.50 [1.35]; here $\mathcal{L}(s)$ denotes the likelihood at $s$ maximized with respect to $b$. The lower and upper values of $s$ which satisfy this condition define the nominal 68% [90%] confidence interval for $s$, assuming Gaussian uncertainties. The confidence intervals produced using this procedure provide frequentist coverage accurate to within a few percent. A central value of $R_M = 0.4 \times 10^{-4}$ is found, with 68% and 90% confidence intervals $[-5.6, 7.4] \times 10^{-4}$ and $[-13, 12] \times 10^{-4}$, respectively.

#### 19.2.5.3 Summary

In summary, semileptonic decays of $D^0$ mesons can be exploited to determine the time integrated mixing rate. By determining the yield of WS decays $D^0 \to X\ell^-\overline{\nu}_\ell$ relative to the RS decays $D^0 \to X\ell^+\nu_\ell$ one determines $R_M = N_{\text{WS}}/N_{\text{RS}} = (x^2 + y^2)/2$. Both single and double tagged event selections, thereby isolating samples differing in their statistical power and purity, have been used by Belle and *BABAR*. Averaging the two measurements described results in

$$R_M = (0.011 \pm 0.027)\% \qquad (19.2.62)$$

where the error contains statistical and systematic uncertainties (all individual errors are assumed to be uncorrelated).

Though wrong-sign semileptonic decays are a sure sign of mixing, the low mixing rate ($\simeq 10^{-5}$) and the need to apply strong background suppression requirements will continue to limit the use of these decays in the mixing measurements for some time to come.

### 19.2.6 $t$-integrated $CP$ violation measurements

In studies of $CP$ violation in the decays of $D$ mesons one usually measures time-integrated asymmetries of partial decay rates, defined as

$$A^f_{CP} \equiv \frac{\Gamma(D \to f) - \Gamma(\overline{D} \to \overline{f})}{\Gamma(D \to f) + \Gamma(\overline{D} \to \overline{f})} \ . \qquad (19.2.63)$$

The underlying $CP$ violating parameters, see Eqs (19.2.17, 19.2.18, and 19.2.19) upon which such asymmetries depend are determined by the specific final state $f$ and by the type of $D$ meson. For charged $D$ mesons, for example, only mode specific $CP$ violation in the decays is possible.

On the other hand, for neutral $D$ mesons, these asymmetries can include direct ($a^f_{\rm dir}$) and indirect ($a_{\rm ind}$) asymmetry contributions (see Section 19.2.1.3):

$$A^f_{CP} = a^f_{\rm dir} + a_{\rm ind}. \tag{19.2.64}$$

The direct asymmetry term corresponds to the $CP$ violation in decays. The indirect asymmetry in decays to $CP$ eigenstates (like $K^+K^-$ or $\pi^+\pi^-$) consists of the term due to the $CP$ violation in mixing, $-\eta_f(y/2)A_M\cos\phi$, and of the term due to the mixing induced $CP$ violation, $\eta_f x\sin\phi$. $\eta_f$ denotes the $CP$ eigenvalue of the final state ($\eta_f = +1$ for $CP$-even and $\eta_f = -1$ for $CP$ odd states). Hence

$$A^f_{CP} = a^f_{\rm dir} - \eta_f(y/2)A_M\cos\phi + \eta_f x\sin\phi. \tag{19.2.65}$$

As a result of $D^0 - \overline{D}^0$ mixing, $CP$ asymmetries for $D^0$ mesons depend upon the interval of decay time, $t$, over which the asymmetry is integrated. At the $B$ Factories, time resolution is comparable with the $D^0$ lifetime, $\tau_{D^0}$. Therefore, no $D^0$ decay vertex separation requirement is imposed on the samples used in any of the analyses and integration times include $t = 0$. In contrast, hadron collider experiments (CDF, LHCb, etc.) impose decay length based selections to reduce large combinatorial backgrounds. There, the integration times begin at $t = t_{\rm min} > 0$. In all cases, the upper end of the range, $t_{\rm max}$ is large and can be taken as infinite. To first order in the small parameter $y$, the values for $A^f_{CP}$ measured can be approximated (see, for example Gersabeck, Alexander, Borghi, Gligorov, and Parkes, 2012) as combinations of $a^f_{\rm dir}$ and $a_{\rm ind}$ that are linear in $t_{\rm min}$, thereby allowing separate values for $a^f_{\rm dir}$ and $a_{\rm ind}$ to be estimated.[134]

The feasibility of using measurements of $CP$ asymmetries as a function of time (as distinct from values $A_{CP}$ integrated over finite time periods) has also been studied (Bevan, Inguglia, and Meadows, 2011). Many decay modes can be used to study weak phases in $D^0$ meson decays in much the same way that such measurements were used in the $B$ Factory measurements of the CKM phases $\phi_{1-3}$. Such measurements will be feasible using samples about 100 times larger than those of the $B$ Factory's.

### 19.2.6.1 Using data to measure detector induced asymmetries

In the experimental determination of the physics parameter $A^f_{CP}$ other asymmetries not originating from the $CP$ violation may enter and have to be corrected for. These include detector induced asymmetries, for example an asymmetry in the reconstruction efficiencies of positively and negatively charged tracks, as well as the forward-backward ($FB$) asymmetry due to the $\gamma^*-Z^0$ interference in $e^+e^- \to c\bar{c}$. The former may be induced by different cross-sections for the interaction of particles and anti-particles in the material of the detector.[135] Such subtle effects are not described to a sufficient accuracy in the simulated data samples — note that as explained in Section 19.2.1.3 in the search for the $CP$ violation in the charm sector one is interested in effects of the order of $10^{-3}$ — and hence they must be estimated using data control samples.

To explain ideas used in the corrections for the non-$CP$ violating asymmetries mentioned above let us first consider an example of a charged $D$ meson decay, $D^\pm \to Xh^\pm$, where $X$ denotes a neutral hadronic system which is self-conjugated (and hence the same for the $D^+$ and $D^-$) and $h^\pm$ represents a charged hadron. The experimentally determined asymmetry is

$$A_{\rm rec} = \frac{N(D^+ \to Xh^+) - N(D^- \to Xh^-)}{N(D^+ \to Xh^+) + N(D^- \to Xh^-)} \ , \tag{19.2.66}$$

where $N$ denotes the number of observed decays. Taking into account small magnitudes of all $CP$ and non-$CP$ violating asymmetries (*i.e.* neglecting quadratic and higher terms) the measured asymmetry can be expressed as the sum of various contributions

$$A_{\rm rec} = A_{CP} + A_{FB} + A^{h^+}_\epsilon \ , \tag{19.2.67}$$

where $A_{FB}$ and $A^{h^+}_\epsilon$ are the forward-backward asymmetry and the detection efficiency asymmetry between positively and negatively charged tracks.

In order to correct for these one has to use real data as much as possible to minimize systematic uncertainty associated with the correction. The control samples and the technique used vary from mode to mode. For example, if two appropriate control samples can be found, for which $CP$ violation is negligible and the asymmetry due to $A_{FB}$ is equal between the two (both assumptions must be satisfied at least to the level below the $A_{CP}$ measurement sensitivity), then

$$\begin{aligned} A^{\rm cont1}_{\rm rec} &= A^{\rm cont1}_{FB} + A^{h^+}_\epsilon \ , \\ A^{\rm cont2}_{\rm rec} &= A^{\rm cont2}_{FB} \end{aligned} \tag{19.2.68}$$

The above equation assumes that the second control sample does not receive a contribution from the detection efficiency asymmetry (for example the control sample consists of neutral $D$ meson decays). Hence from the difference $A^{\rm cont1}_{\rm rec} - A^{\rm cont2}_{\rm rec}$ one can determine the detection efficiency asymmetry $A^{h^+}_\epsilon$ (with some additional complications as explained below).

After correcting for $A^{h^+}_\epsilon$ using an appropriate control data samples, one arrives at

$$A^{\rm corr}_{\rm rec} = A_{CP} + A_{FB} \ , \tag{19.2.69}$$

where the superscript corr denotes that the measured asymmetry has already been corrected for $A^{h^+}_\epsilon$. The remaining two contributions can be separated using the fact that $A_{FB}$ is antisymmetric with respect to the cosine of the $D$ meson production polar angle in CM system ($\cos\theta^*$), while the intrinsic $CP$ asymmetry $A_{CP}$ is independent of this angle. At tree level the asymmetry in the number of produced fermions ($c$ quarks) and anti-fermions ($\overline{c}$ quarks) as a function of the angle between the fermion and the incoming electron in the CM system, $\theta_c$, is [136]

$$\frac{N_c(\cos\theta_c) - N_{\overline{c}}(\cos\theta_c)}{N_c(\cos\theta_c) + N_{\overline{c}}(\cos\theta_c)} = \frac{8A^0_{FB}\cos\theta_c}{3(1+\cos^2\theta_c)} \ . \qquad (19.2.70)$$

[134] For the final states that are $CP$ eigenstates $a_{\rm ind}$ does not depend on decay mode $f$ while $a^f_{\rm dir}$ does.

[135] Note, for example, that the cross-section for $\pi^-p$ interaction through the $\Delta^0$ resonance at $p_\pi \approx 1\,{\rm GeV}/c$ is three times smaller than the cross-section for the $\pi^+p$ interaction through the $\Delta^{++}$, as can be easily seen using the isospin decomposition.

$A^0_{FB}$ is the forward-backward asymmetry parameter, depending on the axial and vector weak couplings of the electrons and charm quarks (Eidelman et al., 2004). Assuming the fragmentation of the primary quark into a charmed meson does not significantly affect the angular distribution (*i.e.* that $\theta^* \approx \theta_c$) the quark asymmetry above is just the asymmetry $A_{FB}$ of Eq. (19.2.69) induced by the forward backward asymmetry:

$$A^{\rm corr}_{\rm rec}(\cos\theta^*) = A_{CP} + \frac{8A^0_{FB}\cos\theta^*}{3(1+\cos^2\theta^*)} \ . \qquad (19.2.71)$$

Hence one can determine $A_{FB}$ and $A_{CP}$ separately from

$$\begin{aligned} A_{CP} &= [A^{\rm corr}_{\rm rec}(\cos\theta^*) + A^{\rm corr}_{\rm rec}(-\cos\theta^*)]/2 \\ A_{FB} &= [A^{\rm corr}_{\rm rec}(\cos\theta^*) - A^{\rm corr}_{\rm rec}(-\cos\theta^*)]/2. \end{aligned} \qquad (19.2.72)$$

The method of correction for the $A_{FB}$ assumes $\theta^* = \theta_c$ and $A^{\rm cont}_{FB} = A_{FB}$. The expressions can, however, be modified when instead of the quark direction one measures the experimentally accessible polar angle of a $D$ meson (in other words, the assumption $\theta^* = \theta_c$ may not be completely justified). Also, the control data samples often involve decays of different charmed mesons from the ones for which $A_{CP}$ is being measured. Hence another assumption is that $A_{FB}$ is the same for all charmed mesons. Due to slightly different fragmentation when an initial $c$ quark hadronizes into various types of $D$ mesons, the asymmetry can also differ slightly for different $D$ mesons. These assumptions have been tested. The differences between $A_{FB}$'s for different $D$ mesons are small (Staric, 2012b) justifying the method used to correct for the residual asymmetry arising from $A_{FB}$.

For measurements of $A_{CP}$ in $D^0 \to K^+K^-, \pi^+\pi^-$ decays, the flavor of neutral $D$ mesons at production is tagged by reconstructing $D^{*+} \to D^0\pi^+_s$ decays. The measured asymmetry can be written as

$$A_{\rm rec} = A_{CP} + A_{FB} + A^{\pi_s}_{\epsilon}. \qquad (19.2.73)$$

To determine $A^{\pi_s}_{\epsilon}$ (Staric, 2008; Aubert, 2008aq), one reconstructs two $D^0 \to K^-\pi^+$ samples: one consisting of $D$ mesons with tagged initial flavor, and one consisting of untagged candidates. The measured asymmetries for these modes can be written as

$$\begin{aligned} A^{\rm tag}_{\rm rec} &= A^{K\pi}_{CP} + A_{FB} + A^{K\pi}_{\epsilon} + A^{\pi_s}_{\epsilon}, \\ A^{\rm untag}_{\rm rec} &= A^{K\pi}_{CP} + A_{FB} + A^{K\pi}_{\epsilon}. \end{aligned} \qquad (19.2.74)$$

One first uses the difference of the two measurements in Eq. (19.2.74) to determine $A^{\pi_s}_{\epsilon}$.

After correcting for $A^{\pi_s}_{\epsilon}$, one corrects for $A_{FB}$ from Eq. (19.2.73) as explained above. An additional complication arises due to the fact that $A^{\pi_s}_{\epsilon}$, which at least partially arises from the difference of charged particle interactions in the detector material, depends on the momentum and the laboratory polar angle of the pion. Hence $A^{\pi_s}_{\epsilon}$ is examined as a function of $(p_{\pi_s}, \cos\theta_\pi)$ which denote the magnitude of the momentum and the cosine of the polar angle of the slow pion, respectively. The asymmetry values for various decays are thus calculated in bins of $(p_{\pi_s}, \cos\theta_\pi)$, as well as in bins of the $D$ meson polar angle, as explained below.

Graphically the method of correction is illustrated in Fig. 19.2.27. A similar method has been used in the analysis of $D^0 \to K^0_S P^0$ ($P^0 = \pi^0$, $\eta$, or $\eta'$) decays (Ko, 2011). An attentive reader may wonder why the treatment of the asymmetries as a function of the $D$ meson polar angle in the CM system, $\cos\theta^*$, is needed. The forward-backward asymmetry would vanish when integrated over this variable. A subtle reason lies in the fact that the pion polar angle in the laboratory frame, $\theta_\pi$, on which the detector induced asymmetries depend upon [137] is correlated with $\theta^*$. Hence in a given bin of $\cos\theta_\pi$ the integration over $\cos\theta^*$ does not assure a vanishing $A_{FB}$ contribution.

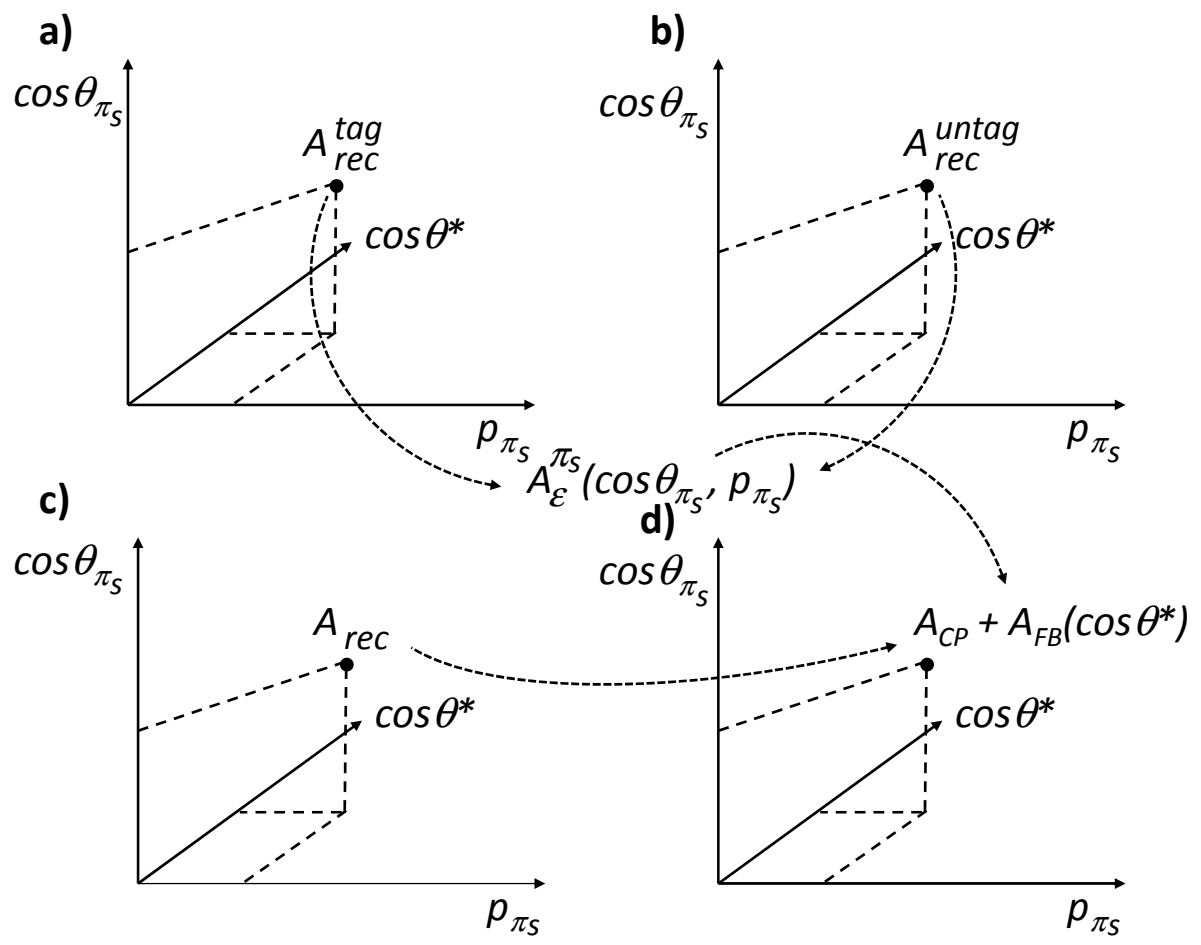

**Figure 19.2.27.** Graphical presentation of the method to correct the measurement of the $CP$ asymmetry in $D$ meson decays. In the case of $D^0 \to h^+h^-$ the difference between the measured asymmetries for $D^{*+} \to D^0(\to K^-\pi^+)\pi^+_s$ (a) and $D^0 \to K^-\pi^+$ (b) decays in a given bin of $\pi_s$ momentum, its polar angle $\theta_{\pi_s}$ and the $D$ meson polar angle in the CM system $\theta^*$, yields the detector induced asymmetry $A^{\pi_s}_{\epsilon}$. This can be used to correct the measured asymmetry for the decays $D^{*+} \to D^0(\to h^-h^+)\pi^+_s$ (c) resulting in the sum $A_{CP} + A_{FB}$ (d). The latter can be distinguished according to their $\cos\theta^*$ dependence.

[136] The asymmetry follows from the angular distribution of produced fermions in the $e^+e^-$ collisions, $d\sigma/d\cos\theta_f \propto 1 + \cos^2\theta_f + (8/3)A^0_{FB}\cos\theta_f$ (Eidelman et al., 2004).

In $D^+_{(s)} \to K^0_S h^+$ decays ($h^+{=}\pi^+$ or $K^+$), the reconstructed asymmetries can be written as

$$A_{\rm rec}^{D^+_{(s)}\to K^0_S h^+} = A_{CP}^{D^+_{(s)}\to K^0_S h^+} + A_{FB}^{D^+_{(s)}} + A_\epsilon^{h^+}. \quad (19.2.75)$$

To correct for $A_{FB}^{D^+_{(s)}}$ and $A_\epsilon^{h^+}$, Belle (Ko, 2010) uses reconstructed samples of $D^+_s \to \phi\pi^+$ and $D^0 \to K^-\pi^+$ decays, assuming that $A_{CP}$ in Cabibbo favored decays is negligibly small at the current experimental sensitivity (note that within the SM *CP* violation in charm decays is expected only for Cabibbo suppressed decays, see Section 19.2.1.3). Another assumption is that $A_{FB}$ is the same for all charmed mesons.

The measured asymmetry for $D^+_s \to \phi\pi^+$ is the sum of $A_{FB}^{D^+_s}$ and $A_\epsilon^{\pi^+}$. Hence one can extract the $A_{CP}$ value for the $K^0_S\pi^+$ final states by subtracting the measured asymmetry for $D^+_s \to \phi\pi^+$ from that for $D^+_{(s)} \to K^0_S\pi^+$. The subtraction is performed in bins of $\pi^+$ momentum, $p_\pi$, and polar angle in the laboratory system, $\cos\theta_\pi$ and the charmed meson's polar angle in the center-of-mass system, $\cos\theta^*_{D^+_{(s)}}$. The three-dimensional (3D) binning is determined in such a way to avoid large statistical fluctuations in each bin. The statistical precision of the $D^+_s \to K^0_S\pi^+$ sample is too low to allow for a 3D correction to $A_{\rm rec}^{D^+_s\to K^0_S\pi^+}$ at present. For this mode one corrects for the forward backward and detection efficiency asymmetries with an inclusive correction obtained by subtracting $A_{\rm rec}^{D^+\to K^0_S\pi^+}$ from $A_{CP}^{D^+\to K^0_S\pi^+}$ after integrating over the entire ($p_\pi$, $\cos\theta_\pi$, $\cos\theta^*_{D^+}$) space. This technique yields a systematic uncertainty of 0.18%, which originates from the statistical uncertainty of the selected $D^+_s \to \phi\pi^+$ sample. Similar methods have been used in the analysis of $D^+ \to \pi^+\eta^{(\prime)}$ decays (Won, 2011).

Recently, the *BABAR* collaboration has developed another data-driven method (del Amo Sanchez, 2011i) to determine the charge asymmetry in the track reconstruction as a function of the magnitude of the track momentum and its polar angle, in the analysis of $D^+ \to K^0_S\pi^+$ decays. $B$ mesons are produced in the process $e^+e^- \to \Upsilon(4S) \to B\overline{B}$. This production mechanism is free of any physics-induced charge or flavor asymmetry. The *CP* violation in the later decays of $B$ mesons must vanish if one takes a completely inclusive sample of $B$ meson decay products. Hence the inclusive $\Upsilon(4S) \to B\overline{B}$ events provide a very large control sample in which any asymmetry is detector induced. However, data recorded at the $\Upsilon(4S)$ resonance also include continuum production $e^+e^- \to q\overline{q}$ ($q = u, d, s, c$), where there is a non-negligible *FB* asymmetry due to the interference between the single virtual photon process and other production processes, as described above. The continuum contribution is estimated using the off-resonance data rescaled to the same luminosity as the on-resonance data sample. Subtracting the number of reconstructed tracks in the rescaled off-resonance sample from the number of tracks in the on-resonance one, *BABAR* obtains the number of tracks corresponding to the $B$ meson decays only. Therefore, the relative detection and identification efficiencies of the positively and negatively charged particles for given selection criteria can be determined using the numbers of positively and negatively reconstructed tracks directly from data. This technique yields a smaller systematic uncertainty of 0.08%. The obtained $\pi^+/\pi^-$ asymmetry map as a function of the pion momentum and polar angle is shown in Fig. 19.2.28. Deviations from unity are largest at low polar angles (due to the amount of the material traversed by the pions) and at relatively low momenta (where the difference between the $\pi^+$ and $\pi^-$ cross-sections for interactions with nucleons is the largest).

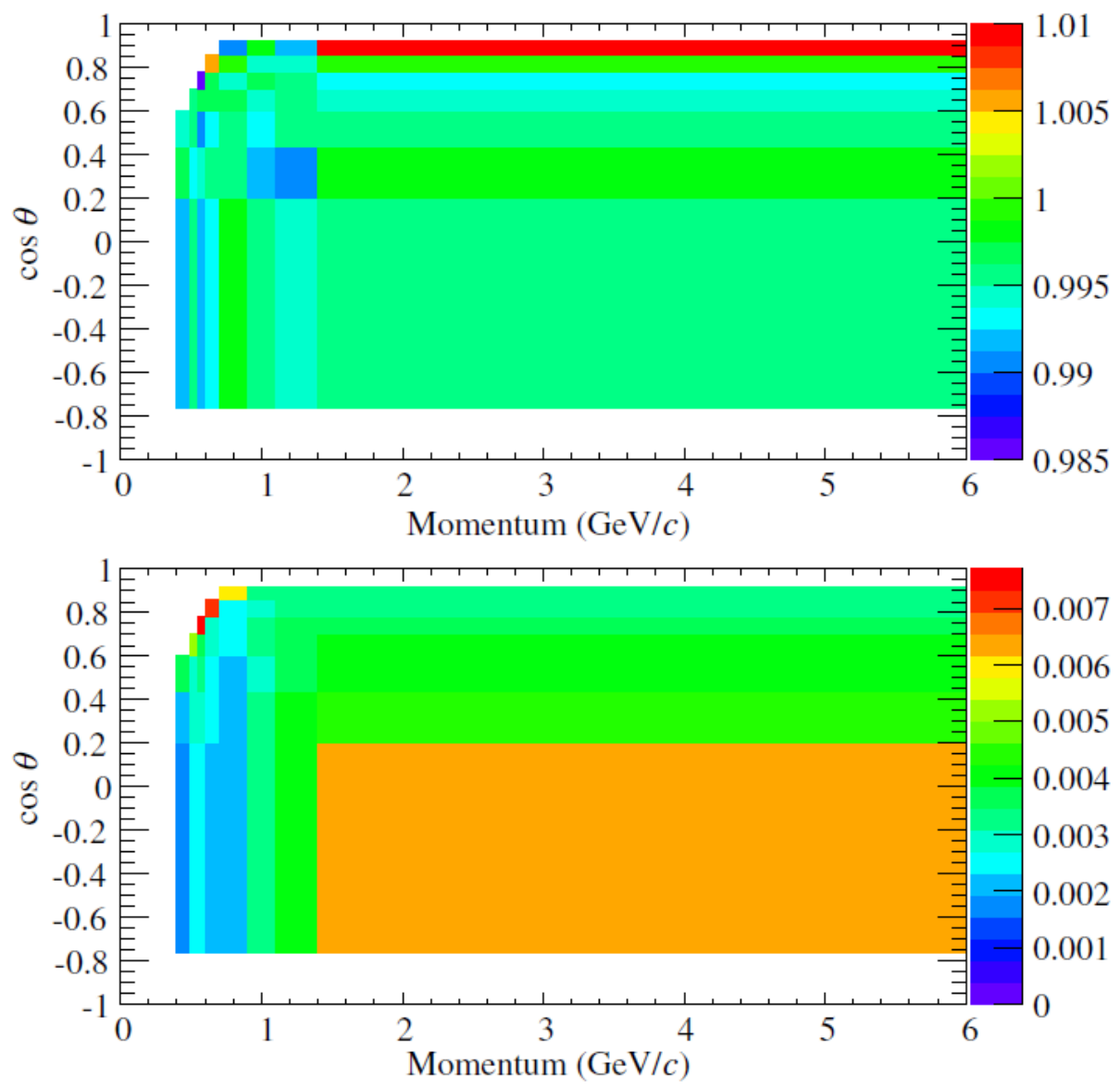

**Figure 19.2.28.** Charged pion reconstruction efficiency ratio $\epsilon(\pi^+)/\epsilon(\pi^-)$ (top) obtained using the method of (del Amo Sanchez, 2011i) as described in the text, and the corresponding statistical errors (bottom).

The method for the measurement of $A_{CP}$ in the $K^0_S K^+$ final state is different from that for the $K^0_S\pi^+$ final state. The $A_{FB}^{D^+_{(s)}}$ and $A_\epsilon^{\pi^+}$ components in $A_{\rm rec}^{D^+_{(s)}\to K^0_S\pi^+}$ are obtained directly from the $D^+_s \to \phi\pi^+$ sample, but there is no corresponding large statistics decay mode that can be used to directly measure the $A_{FB}^{D^+_{(s)}}$ and $A_\epsilon^{K^+}$ components in $A_{\rm rec}^{D^+_{(s)}\to K^0_S K^+}$. Thus, to correct the reconstructed asymmetry in the $K^0_S K^+$ final states, one uses samples of $D^0 \to K^-\pi^+$ and $D^+_s \to \phi\pi^+$ decays.

[137] More precisely, the amount of the material traversed by the pion depends on this angle.

The measured asymmetry for $D^0 \to K^-\pi^+$ is a sum of $A_{FB}^{D^0}$, $A_\epsilon^{K^-}$, and $A_\epsilon^{\pi^+}$. Thus, one can extract $A_\epsilon^{K^-}$ by subtracting the measured asymmetry for $D_s^+ \to \phi\pi^+$ from that for $D^0 \to K^-\pi^+$. An $A_\epsilon^{K^-}$ correction map is obtained as follows; $N_{\rm rec}^{D^0\to K^-\pi^+}$ and $N_{\rm rec}^{\bar{D}^0\to K^+\pi^-}$ are corrected according to the reconstructed asymmetry for $D_s^+ \to \phi\pi^+$ in bins of $(p_\pi, \cos\theta_\pi, \cos\theta^*_{D_{(s)}})$. Subsequently, corrected $N_{\rm rec}^{D^0\to K^-\pi^+}$ and $N_{\rm rec}^{\bar{D}^0\to K^+\pi^-}$ values are determined in bins of $K^\mp$ momentum and polar angle in the laboratory frame, $(p_{K^\mp}, \cos\theta_{K^\mp})$. From the corrected values of $N_{\rm rec}^{D^0\to K^-\pi^+}$ and $N_{\rm rec}^{\bar{D}^0\to K^+\pi^-}$ one obtains an $A_\epsilon^{K^-}$ map that is used to correct for $A_\epsilon^{K^+}$ in the $K_S^0K^+$ final state. By subtracting $A_\epsilon^{K^+}$ from the reconstructed asymmetry of $D_{(s)}^+ \to K_S^0K^+$, one obtains the corrected reconstruction asymmetry $A_{\rm rec,corr}^{D_{(s)}^+\to K_S^0K^+}$ for $D_{(s)}^+ \to K_S^0K^+$:

$$\begin{aligned} A_{\rm rec,corr}^{D_{(s)}^+\to K_S^0K^+} &= A_{\rm rec}^{D_{(s)}^+\to K_S^0K^+} - A_\epsilon^{K^+} \\ &= A_{FB}^{D_{(s)}^+} + A_{CP}^{D_{(s)}^+\to K_S^0K^+} . \end{aligned} \quad (19.2.76)$$

As shown in Eq. (19.2.76), $A_{\rm rec}^{D_{(s)}^+\to K_S^0K^+_{\rm corr}}$ includes not only an $A_{CP}$ component but also the $A_{FB}$ component. Since $A_{CP}$ is independent of all kinematic variables, while $A_{FB}$ is an odd function of $\cos\theta^*_{D_{(s)}^+}$, as explained above, one extracts the two components from $A_{\rm rec}^{D_{(s)}^+\to K_S^0K^+_{\rm corr}}$ as a function of $\cos\theta^*_{D_{(s)}^+}$ through Eq. (19.2.72).

In the most recent measurement of $A_{CP}(D^+ \to K_S^0\pi^+)$ Belle (Ko, 2012) uses two different control samples instead of $D_s^+ \to \phi\pi^+$, composed of selected $D^+ \to K^-\pi^+\pi^+$ and $D^0 \to K^-\pi^+\pi^0$ decays. These decays have larger branching fractions and hence the systematic uncertainty arising from the correction due to the charged pion efficiency asymmetry is reduced. The two aforementioned decay modes are Cabibbo favored and hence one does not expect any observable *CP* violation (see Section 19.2.1.3).[138] Writing out the detector and forward-backward induced contributions to the measured asymmetries in $D^+ \to K^-\pi^+\pi^+$ and $D^0 \to K^-\pi^+\pi^0$ (and assuming the latter is the same for the two modes) it's easy to see that the comparison of the two yields the $A_\epsilon^{\pi^+}$ necessary to correct the measured asymmetry in $D^+ \to K_S^0\pi^+$:

$$\begin{aligned} A_{\rm rec}^{D^+\to K^-\pi^+\pi^+} &= A_{FB}^{D^+} + A_{\epsilon_1}^{\pi^+} + A_{\epsilon_2}^{\pi^+} + A_\epsilon^{K^-} \\ A_{\rm rec}^{D^0\to K^-\pi^+\pi^0} &= A_{FB}^{D^0} + A_{\epsilon_1}^{\pi^+} + A_\epsilon^{K^-} \\ A_{\epsilon_2}^{\pi^+} &= A_{\rm rec}^{D^+\to K^-\pi^+\pi^+} - A_{\rm rec}^{D^0\to K^-\pi^+\pi^0} , \end{aligned} \quad (19.2.77)$$

assuming $A_{FB}^{D^+} = A_{FB}^{D^0}$.

Another correction must be applied to the measured asymmetries with a $K_S^0$ in the final state. In specific decays of charm hadrons either $K^0$ or $\overline{K}^0$ mesons are produced which propagate in time and are at some later time reconstructed as a $K_S^0$ decaying to $\pi^+\pi^-$. The $K^0$ and $\overline{K}^0$ have, however, very different cross-sections for interactions with nucleons in the material of the detector that they traverse (mainly the beam pipe and the material of the silicon vertex detector, see Chapter 2 for the description of the detectors and (Beringer et al., 2012) for the differences in the cross-sections). If a neutral kaon interacts in the material, it normally cannot be reconstructed. Hence the efficiency for the reconstruction of a $K_S^0$ arising in, for example, a $D \to \overline{K}^0X$ decay may differ from the efficiency of the $K_S^0$ reconstruction arising in a $\overline{D} \to K^0\overline{X}$ decay. This asymmetry is not related to the *CP* violation in charm meson decays and must be corrected for. The asymmetry depends on the amount of material traversed by the neutral kaon as well as (due to the energy dependence of the cross-section) on its momentum. The effect is not included in the simulation packages used by Belle and *BABAR* (Agostinelli et al., 2003) and hence a dedicated study has been performed (Ko, Won, Golob, and Pakhlov, 2011) for various detector geometries. The additional contribution to the asymmetry due to this source is found to be of the order of 0.1% and is included in the measurements with $K_S^0$ in the final state either as an additional correction or a systematic uncertainty.

Furthermore, Grossman and Nir (2012) have pointed out a non negligible effect of the $K_S^0 - K_L^0$ interference in decays with neutral kaons in the final state. It depends on the acceptance dependence on the neutral kaon decay time. For the analysis in (Ko, 2012) this effect results in a correction factor of $1.022 \pm 0.007$ for $A_{CP}(D^+ \to K_S^0\pi^+)$.

### 19.2.6.2 Results

All results of $t$-integrated *CP* violation measurements for charm mesons together with the control data samples used for various corrections are listed in Table 19.2.5. In the following some specifics of selected measurements are described.

#### $D^+ \to K_S^0\pi^+,\ D^0 \to K_S^0P^0$

A time-integrated *CP* violation search in $D^+ \to K_S^0\pi^+$ is carried out by both the Belle and *BABAR* collaborations (Ko, 2012; del Amo Sanchez, 2011i). In both experiments corrections for non-*CP* violating asymmetries are performed as explained in the previous section. Specifically, one expects a non-vanishing asymmetry due to the presence of neutral kaons in the final state. The asymmetry measured by Belle is $A_{CP}(D^+ \to K_S^0\pi^+) =(-0.363 \pm$

[138] The *CP* violation in decay is expected only in singly Cabibbo suppressed decays, and this is the only possible type of violation appearing in $D^+$ decays. On the other hand for $D^0$ decays there's also a possibility of *CP* violation in mixing and mixing induced *CP* violation, as explained in Section 16.6. Comparison of the decay-time integrated rates of $D^0 \to K^-\pi^+\pi^0$ and $\overline{D}^0 \to K^+\pi^-\pi^0$ shows that the contribution of this type of *CP* violation to the $A_{CP}$ measurement is $-y\sqrt{R_D^{K\pi\pi^0}}\sin\delta_{K\pi\pi^0}\sin\phi$. This small quantity is included as one of the systematic uncertainties for this result.

0.094 ± 0.067)% where the main systematic uncertainty (0.062%) is due to the statistical uncertainty of the control samples used for the $A_\epsilon^{\pi^+}$ correction. In contrast, *BABAR* uses inclusive on- and off-resonance data for the correction and this enables them to achieve a systematic uncertainty of 0.08% despite the significantly lower total integrated luminosity of the sample used for the measurement. The corrected asymmetry value from *BABAR* is $A_{CP}(D^+ \to K_S^0\pi^+) = (-0.44 \pm 0.13 \pm 0.10)\%$. The *CP* asymmetry from *BABAR* is shown in Fig. 19.2.29 as a function of $|\cos\theta_D^*|$.

If results from the two experiments are combined assuming uncorrelated systematic uncertainties, one obtains $A_{CP}(D^+ \to K_S^0\pi^+) = (-0.389 \pm 0.094)\%$ (where the last error is combined statistical and systematic uncertainty) and this is one of the first hints of *CP* violation in charm decays. It should be noted, however, that this is consistent with *CP* violation due to neutral kaon mixing in the final state. Namely $D^+ \to K_S^0\pi^+$ decays receive a contribution from the CF process $D^+ \to \overline{K}^0\pi^+$ as well as from the DCS process $D^+ \to K^0\pi^+$. Either $\overline{K}^0$ or $K^0$ at a later time decays as a $K_S^0$ reconstructed through the $K_S^0 \to \pi^+\pi^-$ decay. Hence, even in the absence of any *CP* violation in $D$ meson decays, there is an additional contribution to the measured asymmetry due to the *CP* violation in the neutral kaon system. One expects

$$A_{\overline{K}^0} = \left|\frac{\langle\pi^+\pi^-|\overline{K}^0\rangle|^2 - |\langle\pi^+\pi^-|K^0\rangle|^2}{\langle\pi^+\pi^-|\overline{K}^0\rangle|^2 + |\langle\pi^+\pi^-|K^0\rangle|^2}\right|^2 \simeq -\frac{1-|(p/q)_{K^0}|^2}{1+|(p/q)_{K^0}|^2} = -\frac{2\mathrm{Re}(\epsilon)}{1+|\epsilon|^2}. \quad (19.2.78)$$

The expected asymmetry in this decay mode due to the *CP* violation in the neutral kaon system is $(-0.332 \pm 0.006)\%$ (Beringer et al., 2012). The average result for $A_{CP}(D^+ \to K_S^0\pi^+)$ is in good agreement with this expectation, implying no significant *CP* violation in the charged $D$ meson sector. The results from both collaborations also show a remarkable precision achieved despite the need for significant corrections (several corrections of the order of 0.1% to measure the final central value of the same order) in evaluating the intrinsic *CP* asymmetry.

$A_{CP}(D^0 \to K_S^0 P^0)$, where $P^0$ is $\pi^0$ or $\eta^{(\prime)}$ is measured by the Belle experiment (Ko, 2011). Assuming no *CP* violation in decay for the $K_S^0\pi^0$ mode (a mixture of CF and DCS decays) it can be used to test the universality of indirect *CP* violation, as explained further in Section 19.2.7.

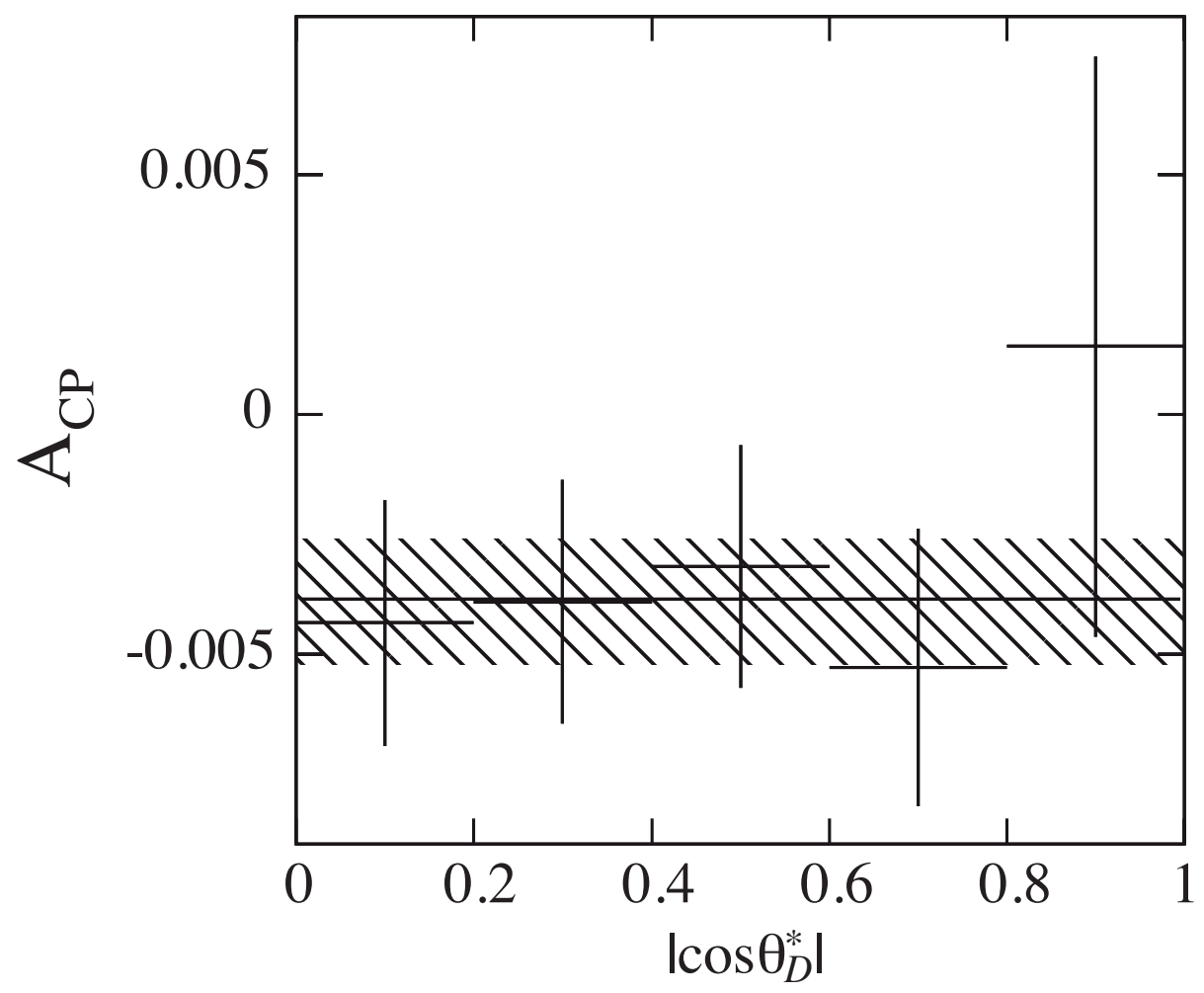

**Figure 19.2.29.** *CP* asymmetries for $D^+ \to K_S^0\pi^+$ candidates as a function of $|\cos\theta_D^*|$. The solid line represents the central value of $A_{CP}$ and the hatched region is the $\pm1\sigma$ interval, obtained from a minimization assuming no dependence on $|\cos\theta_D^*|$ (del Amo Sanchez, 2011i).

As already discussed in Section 19.2.6.1 the measured (or raw) asymmetry can be expressed with a sum of three contributions $A_{\rm rec} = A_{CP} + A_{FB} + A_\epsilon^{\pi_s}$, where the first one is constant, the second one is an odd function of $\cos\theta^*$ and the last one depends on the slow pion phase space $(p_{\pi_s}, \cos\theta_{\pi_s})$, since this asymmetry is due to detector effects. In Section 19.2.6.1 we also discussed that the term $A_\epsilon^{\pi_s}$ can be obtained from data by using tagged and untagged $D^0 \to K^-\pi^+$ decays. Both experiments use this method; *BABAR* has determined the slow pion detection asymmetry $A_\epsilon^{\pi_s}$ in 3 × 3 bins, while Belle uses a 5 × 5 binning in the same momentum range (0.1–0.6 GeV/$c$).

The slow pion asymmetry map is used to correct the raw asymmetry. This is done by weighting the $D^0/\overline{D}^0$ candidates with the following weights:

$$w_{D^0} = 1 - A_\epsilon^{\pi_s}(p_{\pi_s}, \cos\theta_{\pi_s}) \ , \quad w_{\overline{D}^0} = 1 + A_\epsilon^{\pi_s}(p_{\pi_s}, \cos\theta_{\pi_s}) \ . \quad (19.2.79)$$

Note that only candidates in bins with valid $A_\epsilon^{\pi_s}$ measurements are taken into account. This procedure results in a corrected asymmetry $A_{\rm rec}^{\rm cor}$, which is free of the contribution due to the slow pion detection asymmetry. It is calculated as,

$$A_{\rm rec}^{\rm cor}(\cos\theta^*) = \frac{m(\cos\theta^*) - \overline{m}(\cos\theta^*)}{m(\cos\theta^*) + \overline{m}(\cos\theta^*)} \ , \quad (19.2.80)$$

where $m(\overline{m})$ represent the sum of weights of the $D^0(\overline{D}^0)$ candidates in each bin of $\cos\theta^*$.

Finally, taking into account their specific dependence on $\cos\theta^*$, the asymmetries $A_{CP}$ and $A_{FB}$ are extracted by adding or subtracting bins at $\pm\cos\theta^*$, see Eq. (19.2.72).

The systematic uncertainties arise from the signal counting method (*BABAR*: choice of *p.d.f.*, Belle: nonlinear background shape), from slow pion corrections

### $D^0 \to KK/\pi\pi$

Searches for time-integrated *CP*-violating asymmetries in decays $D^0 \to K^+K^-$ and $D^0 \to \pi^+\pi^-$ are carried out by the *BABAR* and Belle collaborations (Aubert, 2008aq; (Staric, 2008)) using 386 fb$^{-1}$ and 540 fb$^{-1}$ of data, respectively. The analysis is similar in both cases, except that the *BABAR* collaboration extracts the signal yields by fitting the mass distributions, while the Belle collaboration uses the method of sideband-subtraction.

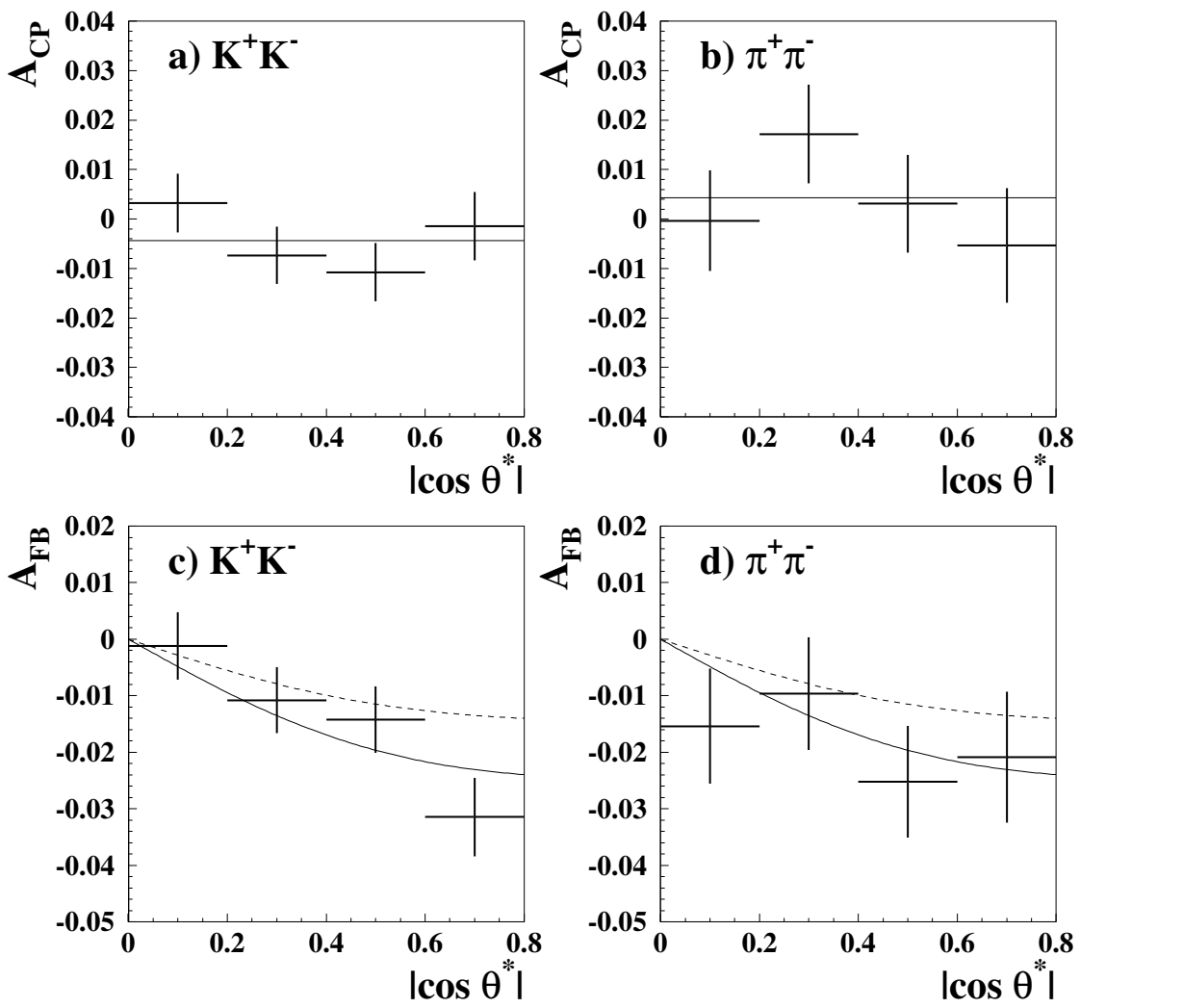

**Figure 19.2.30.** From (Staric, 2008). Measurement of *CP*-violating asymmetries in (a) $KK$ and (b) $\pi\pi$ final states, and forward-backward asymmetries in (c) $KK$ and (d) $\pi\pi$ final states. The solid curves represent the central values obtained from the least square minimizations; the dashed curves in (c) and (d) show the leading order expectation.

(statistics of $K^-\pi^+$ samples, binning) and from the $A_{CP}$ extraction procedure (binning, $\cos\theta^*$ range).

The results of both measurements are presented in Table 19.2.5. Figure 19.2.30 is a graphical representation of the Belle results. For individual decay modes the results agree within the uncertainties. The *BABAR*-Belle averages (assuming uncorrelated systematic uncertainties) are

$$A_{CP}(\pi\pi) = (+0.11 \pm 0.39)\%$$
$$A_{CP}(KK) = (-0.24 \pm 0.24)\% \; . \qquad (19.2.81)$$

The measured values are consistent with no *CP* violation at the level of $\sim 0.3\%$.

Both experiments obtain forward-backward asymmetries that are consistent with each other (see Fig. 19.2.30), but do not agree well with the leading order calculations. Similar is true for measurements of other decay modes. Unfortunately due to lack of reliable predictions at $E_{\rm CM} \approx 10$ GeV it is difficult to quantify the level of disagreement in the $A_{FB}$ measurements.

### *Multibody decays*

Direct *CP* violation is expected to depend on $f$, the final state for each decay. For multibody decays, therefore, the phenomenon will depend on sub modes. *BABAR*, in performing $t$-integrated measurements of $D^0 \to K^+K^-\pi^0$ and $D^0 \to \pi^+\pi^-\pi^0$ decays (Aubert, 2008ap) also exploited the possibility of normalizing $D^0$ and $\overline{D}^0$ rates separately to their total 3-body systems, thereby eliminating the most significant experimental uncertainty (the charge asymmetry in efficiency).

They adopt four different methods, three of which are independent of any decay model. The model-dependent method, introduced by the CLEO collaboration in a study of $D^+ \to K^+K^-\pi^+$ decays (Rubin et al., 2008) compares results of fits to the Dalitz plot distributions of $D^0$ and $\overline{D}^0$ decays to a model based upon the isobar (quasi two-body) approximation with Breit-Wigner parameterizations for the various intermediate resonances (see Chapter 13). Neither magnitude nor phase of any intermediate 2-body mode shows a significant difference between the charge conjugated modes. As an illustration, the intermediate state with the largest contribution to the $\pi^+\pi^-\pi^0$ final state, $\rho^\pm(770)\pi^\mp$, exhibits a relative difference of amplitude magnitudes of $(-3.2 \pm 1.7 \pm 0.8)\%$ and of phases $(-0.8 \pm 1.0 \pm 1.0)^\circ$.

The first model-independent method is, perhaps, the most obvious. Bin by bin differences between efficiency corrected, background subtracted yields of $D^0$ and $\overline{D}^0$ Dalitz plot distributions (see Chapter 13) are examined. The yields are normalized so that there are equal numbers of $D^0$ and $\overline{D}^0$ events. A $\chi^2$ is defined as

$$\chi^2 = \sum_i \chi_i^2 = \sum_i \frac{(n_i^{\overline{D}^0} - Rn_i^{D^0})^2}{(\sigma_i^{\overline{D}^0})^2 + R^2(\sigma_i^{D^0})^2} \; , \qquad (19.2.82)$$

where $\sigma_i^{D^0}$ and $\sigma_i^{\overline{D}^0}$ are the uncertainties of the corresponding yields $n_i^{D^0}$ and $n_i^{\overline{D}^0}$ in the $i$-th bin of the Dalitz plot distribution. The renormalization factor $R$ is introduced to remove the possible overall asymmetry in the total number of reconstructed $D^0$ and $\overline{D}^0$ events. The method is sensitive to possible differences in the shapes of Dalitz distributions but not in the total populations. Using an ensemble of simulated experiments it is found that the value of $\chi^2/\nu$, where $\nu$ is the number of the Dalitz plot bins, is consistent with no *CP* violation at the confidence levels of 33% and 17% for the $D^0 \to \pi^+\pi^-\pi^0$ and $D^0 \to K^+K^-\pi^0$ decays, respectively. The normalized differences in yields for the $D^0 \to \pi^+\pi^-\pi^0$ decays are shown in Fig. 19.2.31.[139]

In the second model independent method the angular distributions for the same, normalized, efficiency corrected, background subtracted yields are studied to extract information on the partial wave content of any differences between $D^0$ and $\overline{D}^0$ decays. The differences $(n_i^{\overline{D}^0} - Rn_i^{D^0})$ are weighted by Legendre polynomial functions $P_\ell(\cos\theta)$ normalized over the range $-1 < \cos\theta \le 1$. For the $AB$ "channel" (quasi two-body mode $D^0 \to r+C, r \to A+B$) $\theta$ is the angle between the momenta of $B$ and $C$ in the $r$ rest frame. The resulting "Legendre difference moments",

$$X_\ell = \frac{(\overline{P}_\ell - R \cdot P_\ell)}{\sqrt{\sigma^2_{\overline{P}_\ell} + R^2 \cdot \sigma^2_{P_\ell}}}, \qquad (19.2.83)$$

(where the $\sigma^2$'s are variances for the indicated quantities) are highly correlated. This is because any partial wave arising from *CP* violation in $D$ decay to any of the resonances in any of the three channels would stimulate non-zero moments in several related values of $\ell$.

[139] An almost identical method is discussed in a later paper (Bediaga et al., 2009) and is commonly referred to as the "Miranda method".

**Table 19.2.5.** Summary of results of time-integrated *CP* violation measurements for charm mesons. The decay modes are grouped to singly Cabibbo suppressed (SCS), Cabibbo favored (CF) and doubly Cabibbo suppressed (DCS) modes. The third column lists the control data samples used to correct for non-*CP* asymmetries appearing in the measurements.

| Decay mode | Reference | Control | $A_{CP}$ [%] | comment |
|---|---|---|---|---|
| | | SCS decays | | |
| $D^0 \to K^+K^-$ | (Staric, 2008) | $D^{*+} \to D^0(\to K^-\pi^+)\pi^+$, $D^0 \to K^-\pi^+$ | $-0.43 \pm 0.30 \pm 0.11$ | |
| | (Aubert, 2008aq) | same as above | $0.00 \pm 0.34 \pm 0.13$ | |
| $D^0 \to \pi^+\pi^-$ | (Staric, 2008) | same as above | $0.43 \pm 0.52 \pm 0.12$ | |
| | (Aubert, 2008aq) | same as above | $-0.24 \pm 0.52 \pm 0.22$ | |
| $D^0 \to K^+K^-\pi^0$ | (Aubert, 2008ap) | same as above | $1.00 \pm 1.67 \pm 0.25$ | |
| $D^0 \to \pi^+\pi^-\pi^0$ | (Aubert, 2008ap) | same as above | $-0.31 \pm 0.41 \pm 0.17$ | |
| | (Arinstein, 2008) | partially reconstructed $D^{*+} \to D^0(\to K_s^0\pi^+\pi^-)\pi^+$ | $0.43 \pm 0.41 \pm 1.23$ | |
| $D^+ \to K_S^0K^+$ | (Ko, 2013) | $D_s^+ \to \phi\pi^+$ , $D^0 \to K^-\pi^+$ | $0.25 \pm 0.28 \pm 0.14$ | |
| $D_s^+ \to K_S^0\pi^+$ | (Ko, 2010) | $D_s^+ \to \phi\pi^+$ | $5.45 \pm 2.50 \pm 0.33$ | |
| | | CF decays | | |
| $D^+ \to K_S^0\pi^+$ | (del Amo Sanchez, 2011i) | inclusive on- and off-resonance data | $-0.44 \pm 0.13 \pm 0.10$ | significant asymmetry due to *CP* violation in the $K^0$ system |
| | (Ko, 2012) | $D^+ \to K^-\pi^+\pi^+$, $D^0 \to K^-\pi^+\pi^0$ | $-0.363 \pm 0.094 \pm 0.067$ | same as above |
| $D_s^+ \to K_S^0K^+$ | (Ko, 2010) | $D_s^+ \to \phi\pi^+$ , $D^0 \to K^-\pi^+$ | $0.12 \pm 0.36 \pm 0.22$ | |
| $D^0 \to K_S^0\pi^0$ | (Ko, 2011) | $D^{*+} \to D^0(\to K^-\pi^+)\pi^+$, $D^0 \to K^-\pi^+$ | $-0.28 \pm 0.19 \pm 0.10$ | |
| $D^0 \to K_S^0\eta$ | (Ko, 2011) | same as above | $0.54 \pm 0.51 \pm 0.16$ | |
| $D^0 \to K_S^0\eta'$ | (Ko, 2011) | same as above | $0.98 \pm 0.67 \pm 0.14$ | |
| | | DCS decays | | |
| $D^0 \to K^+\pi^-\pi^0$ | (Tian, 2005) | | $-0.6 \pm 5.3$ | syst. uncertainty negligible |
| $D^0 \to K^+\pi^-\pi^-\pi^+$ | (Tian, 2005) | | $-1.8 \pm 4.4$ | same as above |

A statistical test for *CP* violation estimates the probability for any moment (which is chosen to be within the range $0-7$) to be inconsistent with no *CP* violation. For this, the quantity

$$\chi^2/\nu = \sum_0^k \sum_{\ell=0}^{7} \sum_{m=0}^{7} (X_\ell \rho_{\ell m} X_m)/\nu \qquad (19.2.84)$$

summed over each of the $k$ invariant mass ranges in various channels is defined. The number of degrees of freedom $\nu = 8k$.

Five hundred Dalitz plots consistent with no *CP* violation are simulated from actual *BABAR* data in which each event is taken randomly as either $D^0$ or $\overline{D}^0$. The resulting moments $X_\ell$ and their correlations $\rho_{\ell m}$ are computed for each sample. The $\chi^2/\nu$ for the actual *BABAR* sample, with $D^*$-tagged assignments as $D^0$ or $\overline{D}^0$ is then compared with the distribution of 500 simulated samples to obtain a one-sided Gaussian C.L. for no *CP* violation. The C.L.'s obtained are 28.2% for the $\pi^+\pi^-$ channel, 28.4% for $\pi^+\pi^0$, 63.1% for $K^+K^-$, and 23.8% for the $K^+\pi^0$ channels, each consistent with no *CP* violation.

The final method, also model independent, consists of comparison of the total number of $D^0$ and $\overline{D}^0$ decays in the quoted modes and was also used by Belle for $D^0 \to \pi^+\pi^-\pi^0$ (Arinstein, 2008). The correction for $A_\epsilon^{\pi_s}$ is, in the *BABAR* case, made as described in Section 19.2.6.1 while Belle uses partially reconstructed $D^{*+} \to D^0(\to K_s^0\pi^+\pi^-)\pi^+$ decays to estimate the tracking efficiency systematics (see Chapter 15), but separately for the negative and positive pions. The values of $A_{CP}$ obtained by this method are quoted in Table 19.2.5.

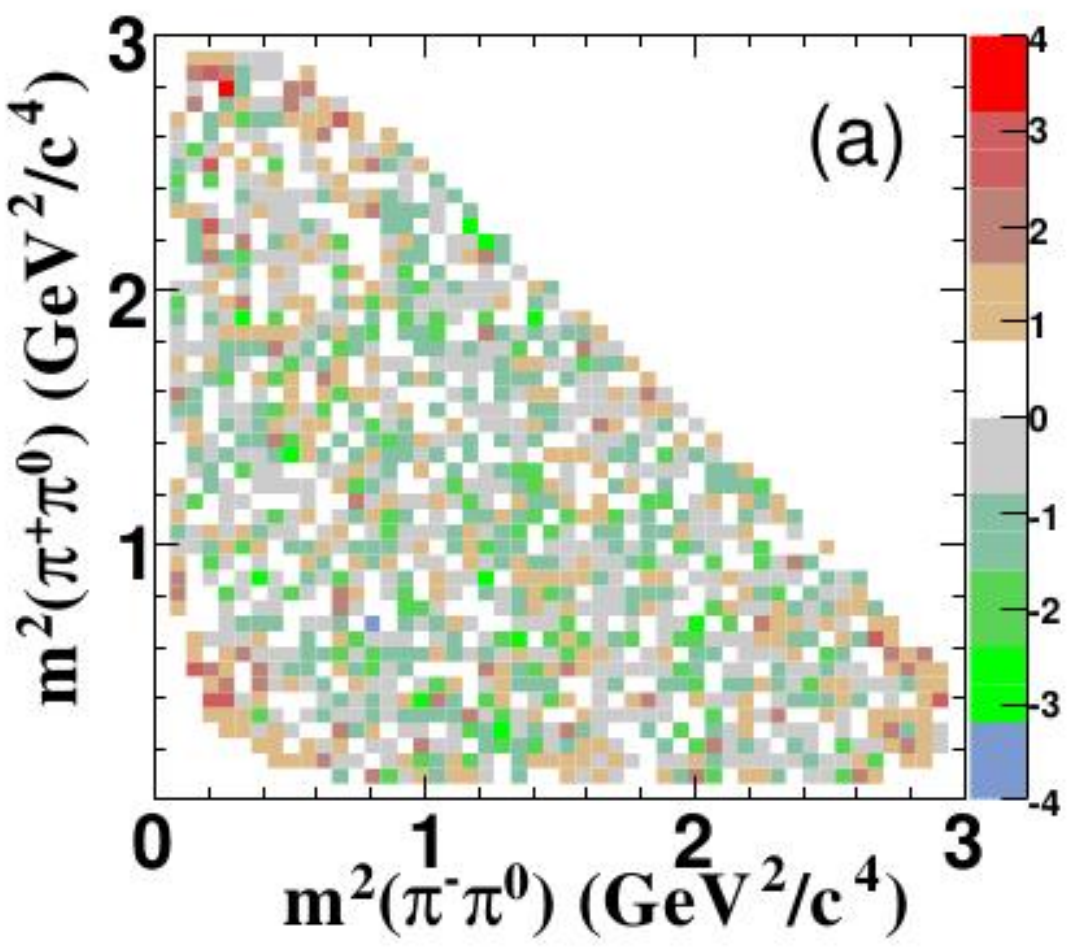

**Figure 19.2.31.** From (Aubert, 2008ap). $\chi_i^2$ defined in Eq. (19.2.82) calculated from efficiency corrected and background subtracted $D^0 \to \pi^+\pi^-\pi^0$ and $\overline{D}^0 \to \pi^+\pi^-\pi^0$ yields across the Dalitz plane.

### 19.2.6.3 $T$-odd correlations

The study of $T$-odd correlations provides a powerful tool to indirectly search for $CP$ violation. It is straight forward to show that any triple product of momenta ($TP$), given by $\boldsymbol{v}_1 \cdot (\boldsymbol{v}_2 \times \boldsymbol{v}_3)$, is odd under the time-reversal symmetry operator $T$. It is clearly also odd under the spatial inversion (parity operator $P$). The kinematics of a four-body decay can be described by a $TP$, and so in the usual way one can construct an asymmetry from $T$ conjugate pairs of triple products, namely $\boldsymbol{v}_1 \cdot (\boldsymbol{v}_2 \times \boldsymbol{v}_3) > 0$ and $\boldsymbol{v}_1 \cdot (\boldsymbol{v}_2 \times \boldsymbol{v}_3) < 0$. The choice of $T$-odd correlations to search for $CP$ violation is proposed by many authors (Bensalem, Datta, and London, 2002a,b; Bensalem and London, 2001; Bigi, 2001; Kayser, 1990; Valencia, 1989) and studies in $D$ decays follow on from similar measurements made for neutral kaon decays.[140] One can construct a $T$-odd observable using the spin or the momentum ($\boldsymbol{v}_i$) of the final state particles in the $D$ CM frame. The $TP$ asymmetry observable of interest is

$$A_{TP} = \frac{\Gamma(\boldsymbol{v}_1 \cdot (\boldsymbol{v}_2 \times \boldsymbol{v}_3) > 0) - \Gamma(\boldsymbol{v}_1 \cdot (\boldsymbol{v}_2 \times \boldsymbol{v}_3) < 0)}{\Gamma(\boldsymbol{v}_1 \cdot (\boldsymbol{v}_2 \times \boldsymbol{v}_3) > 0) + \Gamma(\boldsymbol{v}_1 \cdot (\boldsymbol{v}_2 \times \boldsymbol{v}_3) < 0)} \tag{19.2.85}$$

where $\Gamma$ represents the number of signal events and is measured for $D$ decays only.

However, this is not a true $P$ violating observable, due to final state interaction (FSI) effects that can introduce asymmetries (Bigi and Li, 2009). In order to remove these effects one needs to measure the $CP$ conjugate of this observable ($\overline{A}_{TP}$) using the $\overline{D}$ decays and evaluate the $CP$ violating observable:

$$a_{TP} \equiv \frac{1}{2}(A_{TP} - \overline{A}_{TP}). \tag{19.2.86}$$

This can be explained considering that the asymmetry $A_{TP}$ has a phase that is the sum of a complex $CP$ violating weak phase and a real strong phase (introduced by FSI). Under the operation of $CP$ conjugation the weak phase changes its sign, while the strong phase does not. Thus, the difference between $A_{TP}$ and $\overline{A}_{TP}$ removes the strong phase and the factor 1/2 is required for normalization. The exact definition of $\overline{A}_{TP}$ is given in Eq. (19.2.89).

This method requires three independent momentum vectors, thus at least four different particles are reconstructed in the final state unless the spin of the decaying particle is known. In this last case, the spin of the mother can be used in Eq. (19.2.85).

[140] Some of the literature refers to $CP$ violating $T$-odd observables as $T$-violating observables; however, this is not a correct nomenclature. $T$-violation in kaon decays is discussed in the PDG (Beringer et al., 2012). Section 17.6 discusses $T$-violation measurements by *BABAR* for $B$ decays, and it has been pointed out by Bevan, Inguglia, and Zoccali (2013) that similar measurements are possible using pairs of entangled $D$ mesons produced at the $\psi(3770)$. A recent article discussing $TP$ asymmetries in $K$, $B$ and $D$ decays has been written by Gronau and Rosner (2011).

The search for $CP$ violation by means of the $T$-odd correlations at the $B$ Factories has been performed by *BABAR* in $D^0 \to K^+K^-\pi^+\pi^-$, $D^+ \to K^+K^0_S\pi^+\pi^-$ and $D_s^+ \to K^+K^0_S\pi^+\pi^-$ decays. The latter is a Cabibbo favored decay and no effect is expected, while the others are Cabibbo suppressed decays and the effect could be as large as 0.1%, considering Standard Model processes only (Buccella, Lusignoli, Miele, Pugliese, and Santorelli, 1995). The sensitivity reached at the $B$ Factories for these observables is comparable to the higher SM expectations. Observing an asymmetry would be a signal for processes beyond the SM.

The variable used to build the $T$-odd correlation observable is defined using the momenta of the final state particles in the $D$ CM frame:

$$C_{TP} = \boldsymbol{p}_{K^+} \cdot (\boldsymbol{p}_{\pi^+} \times \boldsymbol{p}_{\pi^-}). \tag{19.2.87}$$

The asymmetry parameters to be measured are then:

$$A_{TP} = \frac{\Gamma(D, C_{TP} > 0) - \Gamma(D, C_{TP} < 0)}{\Gamma(D, C_{TP} > 0) + \Gamma(D, C_{TP} < 0)}, \tag{19.2.88}$$

$$\overline{A}_{TP} = \frac{\Gamma(\overline{D}, -\overline{C}_{TP} > 0) - \Gamma(\overline{D}, -\overline{C}_{TP} < 0)}{\Gamma(\overline{D}, -\overline{C}_{TP} > 0) + \Gamma(\overline{D}, -\overline{C}_{TP} < 0)}. \tag{19.2.89}$$

All three analyses have measured the asymmetry parameters through a simultaneous maximum likelihood fit to the four samples obtained by splitting the data set using the $D$ flavor and the $C_{TP}$ ($\overline{C}_{TP}$) value. Furthermore, a blind analysis has been performed: the asymmetry parameters $A_{TP}$ and $\overline{A}_{TP}$ have been masked adding unknown random offsets, and all the selection criteria and systematic effects have been evaluated before unveiling the final result.

The $D^0 \to K^+K^-\pi^+\pi^-$ analysis (del Amo Sanchez, 2010n) makes use of 471 fb$^{-1}$ data recorded by the *BABAR* detector at the $\Upsilon(4S)$ energy and 40 MeV below. The decay chain

$$e^+e^- \to XD^{*+}; D^{*+} \to \pi_s^+ D^0; D^0 \to K^+K^-\pi^+\pi^-,$$

where $X$ indicates any system composed of charged and neutral particles, has been reconstructed from the sample of events having at least five charged tracks. At first, the $D^0$ has been reconstructed, requiring the momentum in the CM frame $p^*(D^0) > 2.5$ GeV/$c$ to suppress the background. Then, the successful $D^0$ candidate has been combined with any charged track having momentum less than 0.65 GeV/$c$ ($\pi_s^+$) to form the $D^{*+}$ candidate. A contamination from $D^0 \to K^0_S K^+K^-$ has been removed applying a mass veto to the $K^0_S \to \pi^+\pi^-$ candidates. This procedure selects about 50,000 signal events.

The two-dimensional distribution of $m(K^+K^-\pi^+\pi^-)$ vs. $\Delta m = m(K^+K^-\pi^+\pi^-\pi_s^+) - m(K^+K^-\pi^+\pi^-)$ has been described by five components: (i) true $D^0$ signal originating from a $D^{*+}$ decay (signal); and backgrounds comprised of (ii) random $\pi_s^+$ events where a true $D^0$ is combined to an incorrect $\pi_s^+$ ($D^0$-peaking); (iii) misreconstructed $D^0$ decays, where one or more of the $D^0$ decay products are either not reconstructed or reconstructed with the wrong particle hypothesis ($\Delta m$-peaking); (iv) combinatorial; (v) $D_s^+ \to K^+K^-\pi^+\pi^-\pi^+$ contamination. The functional forms of the probability density functions for the signal and background components are based on studies of the generic $e^+e^- \to c\bar{c}$ Monte Carlo (MC) sample. However, all parameters related to these functions are determined from a two-dimensional likelihood fit to data over the full $m(K^+K^-\pi^+\pi^-)$ vs. $\Delta m$ region, shown in Fig. 19.2.32. Combinations of Gaussian and Johnson SU (Johnson, 1949) lineshapes are used for peaking distributions, and polynomials and threshold functions for the non-peaking backgrounds.

Many possible sources of systematic effect have been considered in this analysis, for each of them the related selection criteria have been varied by a small amount to evaluate the deviations with respect to the asymmetries obtained applying the nominal criteria. Among them, the largest contributions to systematic error are due to the particle identification, the selection on $p^*(D^0)$ and the fit bias.

The results are

$$\begin{aligned} A_{TP}(D^0) &= (-68.5 \pm 7.3(\text{stat}) \pm 5.8(\text{syst})) \times 10^{-3}, \\ \overline{A}_{TP}(\overline{D}^0) &= (-70.5 \pm 7.3(\text{stat}) \pm 3.9(\text{syst})) \times 10^{-3}, \\ a_{TP}(D^0) &= (1.0 \pm 5.1(\text{stat}) \pm 4.4(\text{syst})) \times 10^{-3}. \end{aligned} \quad (19.2.90)$$

No $CP$ violation is found, even though $A_{TP}$ and $\overline{A}_{TP}$ are significantly different from zero, indicating the effect of FSIs.

The reconstruction of the $D^+_{(s)} \to K^+K^0_S\pi^+\pi^-$ decays in the other *BABAR* analysis (Lees, 2011i), that used $520\,\text{fb}^{-1}$ recorded around a CM energy of 10.6 GeV, is similar. After the reconstruction of the $K^0_S \to \pi^+\pi^-$ decay, the $K^0_S$ candidates are combined into a vertex with three other charged tracks in the event to reconstruct inclusive $D^+_{(s)}$ decays. A kinematic selection on $p^*(D^+_{(s)}) > 2.5\,\text{GeV}/c$ is required to suppress the background. Possible background contamination from $D^+_{(s)} \to K^+K^0_SK^0_S$ decays have been removed by applying a $K^0_S$ veto.

The signal to background ratio has been then optimized using a likelihood ratio. The probability density functions used to build the likelihood ratio are taken from the distributions of three kinematic variables: (i) $p^*(D^+_{(s)})$; (ii) $\Delta p = P_1 - P_2$, the difference between the probability of the nominal fit ($P_1$) and the probability of the fit obtained constraining the $D^+_{(s)}$ vertex into the interaction region ($P_2$); and (iii) $L_T(D^+_{(s)})$, the flight distance of the $D^+_{(s)}$ meson in the transverse plane. The signal distributions are obtained from the signal regions of two control samples, the Cabibbo favored $D^+ \to K^0_S\pi^+\pi^+\pi^-$ and $D_s^+ \to K^-K^0_S\pi^+\pi^+$ decays, for $D^+$ and $D_s^+$, respectively. The background distributions are taken from data sidebands.

A likelihood ratio, optimized on $S/\sqrt{S+B}$ (see Chapter 4) for the signal region, has been applied to obtain the best significance of the mass peak. About 20,000 (30,000) $D^+$ ($D_s^+$) signal events have been reconstructed.

The model to fit the $K^+K^0_S\pi^+\pi^-$ mass spectrum has been developed and validated on large samples of inclusive MC processed using the same reconstruction and analysis chain as that used for real events. The mass spectrum has been separated in two regions centered on $D^+$ and $D_s^+$ peak, respectively, and the specific requirement on likelihood ratio has been applied.

The model used to simultaneously fit the four samples obtained separating the events by charge and $C_{TP}$ ($\overline{C}_{TP}$) value is composed of two Gaussians for the peak and a second order polynomial for $D^+$ ($D_s^+$) background.

The simultaneous fit on the four data subsamples allows one to measure directly the asymmetries:

$$\begin{aligned} A_{TP}(D^+) &= (+11.2 \pm 14.1(\text{stat}) \pm 5.7(\text{syst})) \times 10^{-3}, \\ \overline{A}_{TP}(D^-) &= (+35.1 \pm 14.3(\text{stat}) \pm 7.2(\text{syst})) \times 10^{-3}, \\ A_{TP}(D_s^+) &= (-99.2 \pm 10.7(\text{stat}) \pm 8.3(\text{syst})) \times 10^{-3}, \\ \overline{A}_{TP}(D_s^-) &= (-72.1 \pm 10.9(\text{stat}) \pm 10.7(\text{syst})) \times 10^{-3}, \end{aligned} \quad (19.2.91)$$

from which the $CP$ violation parameters can be obtained

$$\begin{aligned} a_{TP}(D^+) &= (-12.0 \pm 10.0(\text{stat}) \pm 4.6(\text{syst})) \times 10^{-3}, \\ a_{TP}(D_s^+) &= (-13.6 \pm 7.7(\text{stat}) \pm 3.4(\text{syst})) \times 10^{-3}. \end{aligned} \quad (19.2.92)$$

The sources of systematic error that have been considered are the fit model, the particle identification, the fit bias measured on MC and the selection based on the likelihood ratio. The procedure to evaluate the systematic error follows the same strategy of the $D^0$ analysis. A slightly different behavior is observed for the $A_{TP}$ and $\overline{A}_{TP}$ asymmetries in $D^+$ and $D_s^+$. Even if the final state is the same, a different resonant sub-structure may be responsible for this difference (Gronau and Rosner, 2011).

The $CP$ violation results obtained at the $B$ Factories in four-body $D$ decays are consistent with zero to a precision of 0.5%. These results are in agreement with expectations from SM predictions. Nevertheless, the relative simplicity of the analyses based on $T$-odd correlations and the results that can be obtained allow one to consider this tool as fundamental to search for $CP$ violation in four-body decays. Furthermore, the study of $T$-odd correlations allows one to probe FSI in four-body $D$ decays.

#### 19.2.6.4 Summary

Numerous $t$-integrated searches for $CP$ violation in charm meson decays have been performed by the $B$ Factories and these show no significant intrinsic effect at the levels of sensitivity achieved. These differ from a few percent for the DCS decays to over a few tenths of a percent for SCS decays to 0.1% for CF decays (see Table 19.2.5). Because of the various detector induced and physics (forward-backward) asymmetries the measurements require careful calibration of the data using control samples and hence represent one of the most demanding measurements performed at the $B$ Factories. *BABAR* and Belle developed several methods for such calibrations which assure a satisfactory control of measurement uncertainties that are applicable for the next generation of flavor factories (note that the main systematic uncertainties, arising from limited statistics of control data samples, will be reduced with increased luminosity). It is a remarkable demonstration that both experiments can measure subtle effects such as the asymmetry arising from the *CP* violation in the neutral kaon system in charm meson decays involving $K^0_S$'s. In combination with $t$-dependent measurements, *CP* violating asymmetries represent interesting tests of the SM, as described further in Section 19.2.7.

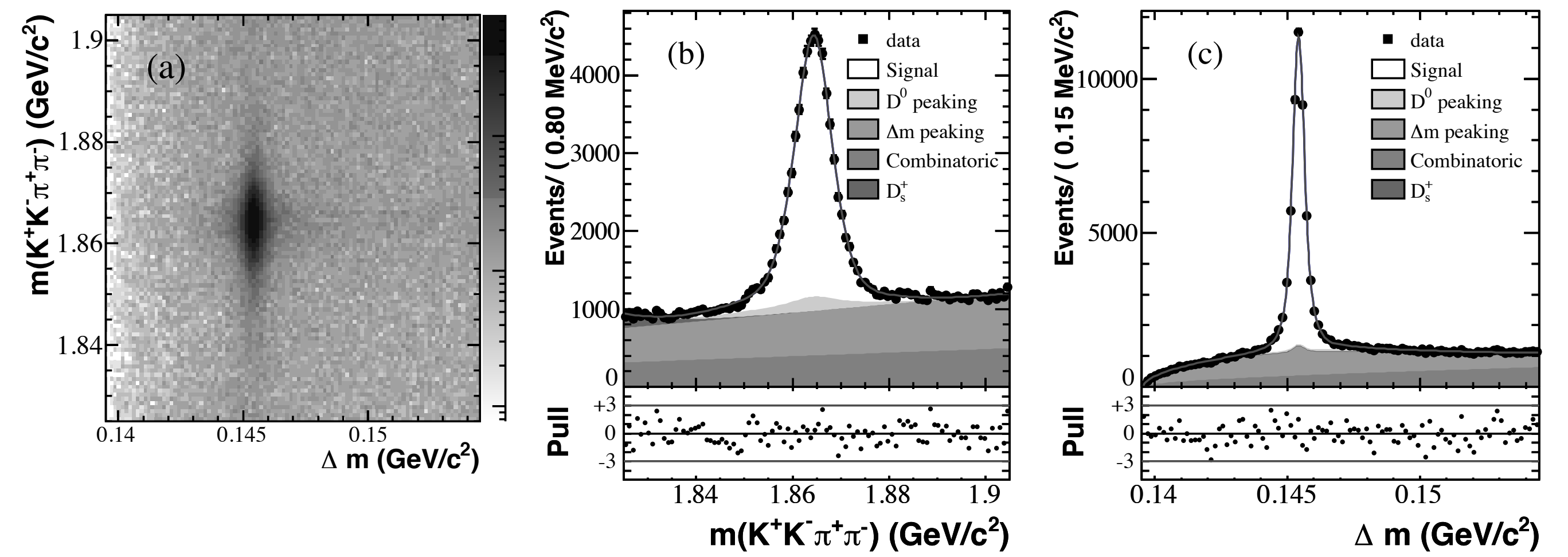

**Figure 19.2.32.** From (del Amo Sanchez, 2010n). The distribution of selected events in the $m(K^+K^-\pi^+\pi^-)$ vs. $\Delta m$ plane (a) is shown together with the projections of the events with overlaid the fit results for $m(K^+K^-\pi^+\pi^-)$ (b) and $\Delta m$ (c), with the shaded areas indicating the different contributions. In the bottom, the distribution of the normalized residuals (Pull), is shown for each fit projection.

### 19.2.7 $t$-dependent *CP* violating asymmetries

A general aspect of the possible *CP* violation effects in the charm sector as well as various parameterizations were discussed in Section 19.2.1.3. In the time dependence of $D^0$ decays one can search for the effects of all three types of *CP* violation.

#### 19.2.7.1 Decays to *CP* eigenstates

Let us start by examining the decays into *CP* eigenstates, like $K^+K^-$. The measurements of the mixing-related parameter $y_{CP}$ in this decay mode are described in Sections 19.2.3.1 and 19.2.3.2. In order to search for *CP* violating effects one has to distinguish between decays of particles and anti-particles. Squaring the modulus in Eqs (19.2.11) and (19.2.12) (keeping the parameter $q/p$, *i.e.* not setting $q = p = 1/\sqrt{2}$), and using the parameterization as defined in Eqs (19.2.17), (19.2.18) and (19.2.20), we arrive at the time evolution of $D^0 \to K^+K^-$ decays

$$|\langle K^+K^-|D^0(\Gamma t)\rangle|^2 = e^{-\Gamma t}|A_{KK}|^2$$

$$[1 - (1 + \frac{A_M - A_D^{KK}}{2})(x\sin\phi - y\cos\phi)\ \Gamma t], \tag{19.2.93}$$

and an equivalent expression for $|\langle K^+K^-|\overline{D}^0(t)\rangle|^2$. It is valid to linear order in the dimensionless decay time $\Gamma t$. One can see that in the above dependence parameters describing all three types of *CP* violation are present: $A_M$ (*CP* violation in mixing), $A_D^{KK}$ (*CP* violation in decay) and $\phi$ (*CP* violation in the interference between decays with and without mixing). If Eq. (19.2.93) is regarded as the first term in the expansion of an exponential function,[141] we arrive at separate expressions for the inverse of the effective lifetime of neutral meson decays in this mode:

$$\frac{1}{\tau_{KK}} = \frac{1}{\tau}\left[1 \mp \left(1 \pm (A_M - A_D^{KK})/2\right)(x\sin\phi \mp y\cos\phi)\right], \tag{19.2.94}$$

where the upper sign corresponds to $D^0$ and the lower one to $\overline{D}^0$ decays. Hence by measuring separately the effective lifetimes of neutral $D$ meson decays (using tagging of the initial $D$ meson flavor with $D^*$ mesons, as described in Section 19.2.1.5) one can determine the lifetime asymmetry:

$$A_\Gamma \equiv \frac{\tau(\overline{D}^0 \to KK) - \tau(D^0 \to KK)}{\tau(\overline{D}^0 \to KK) + \tau(D^0 \to KK)} = \\ = \frac{A_M - A_D^{KK}}{2} y\cos\phi - x\sin\phi \quad . \tag{19.2.95}$$

A non-zero value of the asymmetry $A_\Gamma$ would be a sign of *CP* violation in the $D^0$ system - at least one parameter,

[141] This is equivalent to the derivation of the $y_{CP}$ parameter, see Eq. (19.2.30) – (19.2.34), but now separately for $D^0$ and $\overline{D}^0$ decays.

$A_D^{KK}$, $A_M$ or $\phi$, must be different from zero. The sensitivity of $A_\Gamma$ to the *CP* violating parameters is limited by the small magnitude of the mixing parameters $x$ and $y$.

Measurements of this asymmetry were typically performed together with the measurements of $y_{CP}$. The average of measurements (Staric, 2012a; Lees, 2013d) using the $K^+K^-$ and $\pi^+\pi^-$ final states is found to be

$$A_\Gamma = (0.02 \pm 0.17)\% \quad , \tag{19.2.96}$$

where the error includes statistical and systematic uncertainties.[142]

The main contributions to the systematic uncertainty on $A_\Gamma$ arise from similar sources as in the $y_{CP}$ measurements, from possible biases in the acceptance dependence on the decay time and the assumption of an equal mean of the resolution function in both decay modes used (see Section 19.2.3.1). From the $A_\Gamma$ value one can conclude there is currently no significant sign of *CP* violation at the sensitivity level of 0.25%.

It should be noted that the decay time integrated *CP* violating asymmetry, as defined in Eq. (19.2.63) of Section 19.2.6, also receives contributions from all the types of *CP* violation, and the *CP* violation in decay ($A_D^f$) appears in the linear order. Upon the summation of Eq. (19.2.65) and Eq. (19.2.95) we get

$$A_{CP}^{KK} + A_\Gamma = A_D^{KK} \tag{19.2.97}$$

(in the above equation the *CP* violation in decays to $K^+K^-$ is denoted by $A_D^{KK}$). Hence one can estimate the amount of *CP* violation in decay by summing the two corresponding results. Using the averages of the $A_{CP}^{KK(\pi\pi)}$ measurements (see Section 19.2.6.2), and assuming the systematic uncertainties due to completely different methods of measurements are uncorrelated, one obtains results for the $K^+K^-$ and $\pi^+\pi^-$ final states:

$$A_D^{KK} = (-0.22 \pm 0.29)\% \quad , \\ A_D^{\pi\pi} = (+0.13 \pm 0.43)\% \quad , \tag{19.2.98}$$

including both types of uncertainties.

### 19.2.7.2 Hadronic wrong-sign decays

The measurements of the mixing parameters in this type of decay are described in Section 19.2.2. Since the fit to the decay time dependence of the wrong-sign decays involves parameters $x'^2$ and $y'$, a similar fit is repeated by applying Eq. (19.2.22) separately for $D^0$ and $\overline{D}^0$ decays. One introduces separate parameters $x'^{2\pm}$ and $y'^{\pm}$, where + and − denote parameters for $D^0$ and $\overline{D}^0$ decays, respectively. Any difference between corresponding parameters denoted by + or − represents a sign of *CP* violation. Parameters $x'^{2\pm}$ and $y'^{\pm}$ can of course be related to the *CP* violating parameters in mixing and in the interference introduced in Section 19.2.1.3:

$$x'^{\pm} = \left[\frac{1 \pm A_M}{1 \mp A_M}\right]^{1/4} (x' \cos\phi \pm y' \sin\phi), \\ y'^{\pm} = \left[\frac{1 \pm A_M}{1 \mp A_M}\right]^{1/4} (y' \cos\phi \mp x' \sin\phi). \tag{19.2.99}$$

Apart from the above parameters, $R_D$ is also duplicated to $R_D^\pm$ using an analogous notation. Any difference between $R_D^+$ and $R_D^-$ is related to *CP* violation in decay:

$$A_D = \frac{R_D^+ - R_D^-}{R_D^+ + R_D^-} \quad . \tag{19.2.100}$$

Inclusion of more free parameters to describe the *CP* violation leads to a reduced statistical accuracy of the obtained results. This can be seen in Fig. 19.2.11, comparing the 95% C.L. regions of $x'^2$ and $y'$ as obtained from the fits allowing or neglecting the *CP* violation.

Fits allowing for *CP* violation were performed by Zhang (2006) and Aubert (2007j). They are given in Table 19.2.6.

**Table 19.2.6.** *CP* violation results using $D^0 \to K\pi$ WS decays from *BABAR* (Aubert, 2007j) and Belle (Zhang, 2006). When two uncertainties are given, the first is statistical and the second systematic. Results with a single uncertainty have both statistical and systematic components combined. Limits correspond to 95% C.L.

| Parameter | Fit Results ($\times 10^{-3}$) | |
|---|---|---|
| | *BABAR* | Belle |
| | Assuming both mixing and *CP* violation | |
| $R_D$ | $3.03 \pm 0.16 \pm 0.10$ | — |
| $A_D$ | $-21 \pm 52 \pm 15$ | $23 \pm 47$ |
| $A_M$ | – | $670 \pm 1200$ |
| $x'^{2+}$ | $-0.24 \pm 0.43 \pm 0.30$ | — |
| $y'^{+}$ | $9.8 \pm 6.4 \pm 4.5$ | — |
| $x'^{2-}$ | $-0.20 \pm 0.41 \pm 0.29$ | — |
| $y'^{-}$ | $9.6 \pm 6.1 \pm 4.3$ | — |
| $x'^2$ | – | $< 0.72$ |
| $y'$ | – | $-28 < y' < 21$ |

Taking the averages of the measurements [143] and assuming uncorrelated systematic errors one obtains

$$x'^{2+} - x'^{2-} = (0.011 \pm 0.041)\%$$

[142] Note that $A_\Gamma$ can in principle differ for various different final states due to the $A_D^f$ term in Eq. (19.2.95). Hence the averaging of the $A_\Gamma$ values obtained in $D$ meson decays to $K^+K^-$ and $\pi^+\pi^-$ final states is not justified. However, the difference between $(1/2)(A_M - A_D^{KK})y\cos\phi$ and $(1/2)(A_M - A_D^{\pi\pi})y\cos\phi$, *i.e.* $(1/2)(A_D^{\pi\pi} - A_D^{KK})y\cos\phi$, is only $(1\pm1)\times10^{-5}$ using the average values (Amhis et al., 2012) of parameters involved. As long as the accuracy of $A_\Gamma$ measurement doesn't reach that level the approximation done in the averaging is justified.

[143] Also converting the results from Belle to $x'^{2\pm}$ and $y'^{\pm}$, according to (Amhis et al., 2012).

$$y'^+ - y'^- = (-0.19 \pm 0.64)\% \quad , \qquad (19.2.101)$$

consistent with no *CP* violation.

In the study of the $D^0 \to K^+\pi^-\pi^+\pi^-$ decays (see Section 19.2.2.3) *BABAR* (Aubert, 2006ap) also performs a fit allowing for *CP* violation. The $D^0$ and $\overline{D}^0$ decay-time distributions are treated separately, by making a substitution in Eq. (19.2.26):

$$\begin{aligned} \alpha\widetilde{y}' &\to |p/q|^{\pm 1}(\alpha\widetilde{y}' \cos\widetilde{\phi} \pm \beta\widetilde{x}' \sin\widetilde{\phi}) \\ x^2 + y^2 &\to |p/q|^{\pm 2}(x^2 + y^2), \end{aligned} \qquad (19.2.102)$$

choosing the "+" ("−") sign for $D^0$ ($\overline{D}^0$) candidate decays, respectively. *CP* violation in mixing is parameterized by $|p/q|$ and a *CP*-violating phase $\widetilde{\phi}$ is introduced to account for *CP* violation in the interference between mixing and DCS decay. The resulting values of the *CP* violating parameters are consistent with no *CP* violation:

$$\begin{aligned} R_M &= [0.017^{+0.017}_{-0.016}(\text{stat}) \pm 0.003(\text{syst})]\%, \\ |p/q| &= 1.1^{+4.0}_{-0.6}(\text{stat}) \pm 0.1(\text{syst}), \\ \alpha\widetilde{y}' \cos\widetilde{\phi} &= -0.006^{+0.008}_{-0.006}(\text{stat}) \pm 0.006(\text{syst}), \\ \beta\widetilde{x}' \cos\widetilde{\phi} &= 0.002^{+0.005}_{-0.003}(\text{stat}) \pm 0.006(\text{syst}). \end{aligned} \qquad (19.2.103)$$

### 19.2.7.3 $t$-dependent Dalitz Analyses

Violation of the *CP* symmetry can also be searched for in the analyses of the decay time dependence of the Dalitz distributions in multi-body final states. As for the hadronic wrong-sign decays one can perform a fit to separate $D^0$ and $\overline{D}^0$ samples (see the previous section). Any difference in the resulting mixing parameters can be ascribed to *CP* violation. This is done in the measurement of the $K^+\pi^-\pi^0$ final state (see Section 19.2.4.1) by *BABAR* (Aubert, 2009u). The resulting $x'$, $y'$ parameters are shown in Table 19.2.7. Within the uncertainties they are consistent between the two samples indicating no significant *CP* violation.

**Table 19.2.7.** *CP* violation results using separate $D^0$ and $\overline{D}^0$ samples in the analysis of $K^+\pi^-\pi^0$ final state (Aubert, 2009u).

| | |
|---|---|
| $D^0$ only | $x' = (+2.53^{+0.54}_{-0.63} \pm 0.39)\%$ |
| | $y' = (-0.05^{+0.63}_{-0.67} \pm 0.50)\%$ |
| $\overline{D}^0$ only | $x' = (+3.55^{+0.73}_{-0.83} \pm 0.65)\%$ |
| | $y' = (-0.54^{+0.40}_{-1.16} \pm 0.41)\%$ |

In the Belle search for *CP* violation in the $K^0_S\pi^+\pi^-$ final state (Abe, 2007b) the measurement consists of an extension of the fit procedure described in Section 19.2.4.2. The time evolution of the Dalitz distribution as given in Eq. (19.2.52) is used without setting $q = p = 1/\sqrt{2}$. The possibility of *CP* violation in mixing and mixing induced *CP* violation is parameterized by $q/p = |q/p|e^{i\phi}$ with $|q/p|$ and $\phi$ as free parameters of the fit. The *CP* violation in decay would be manifest as a difference between the magnitudes or phases of the individual intermediate states contributing. This type of the *CP* violation is searched for first by allowing these amplitudes to differ for $D^0$ and $\overline{D}^0$ decays, while not (yet) introducing additional parameters describing the other two types of the *CP* violation. Such an approach is used due to possible complications in the fitting procedure.[144] In this measurement no significant deviations between the amplitudes of intermediate-state contributions to $D^0$ and $\overline{D}^0$ decays were found, and hence one can conclude that no sign of *CP* violation in decay was observed. Following this, the individual amplitudes are fixed to be the same for $D^0$ and $\overline{D}^0$ and the parameters $|q/p|$ and $\phi$ are introduced. The results are listed in Table 19.2.8.

*BABAR* (del Amo Sanchez, 2010f) uses the same approach for the $K^+\pi^-\pi^0$ final state, fitting separately $D^0$ and $\overline{D}^0$ tagged samples with duplicated mixing parameters denoted as $x^+, y^+$ and $x^-, y^-$, respectively. The results are given in Table 19.2.8.

**Table 19.2.8.** Results of search for *CP* violation in time dependent Dalitz analysis of $K^0_S h^+h^-$ final state (del Amo Sanchez, 2010f; Abe, 2007b).

| Experiment | Sample | Results [$\times 10^3$] |
|---|---|---|
| *BABAR* | 486.5 $\text{fb}^{-1}$ | |
| *CP* violation | Purity: 98.5% | $x^+ = 0.0 \pm 3.3$ |
| | | $y^+ = 5.5 \pm 2.7$ |
| | | $x^- = 3.3 \pm 3.3$ |
| | | $y^- = 5.9 \pm 2.8$ |
| Belle | 540 $\text{fb}^{-1}$ | |
| *CP* violation | Purity: 95.0% | $x = 8.1 \pm 3.0^{+1.0+0.9}_{-0.7-1.6}$ |
| | | $y = 3.7 \pm 2.5^{+0.7+0.7}_{-1.3-0.8}$ |
| | | $\lvert q/p\rvert = 0.86^{+0.30+0.06}_{-0.29-0.03} \pm 0.08$ |
| | | $\phi = (-14^{+16+5+2}_{-18-3-4})^\circ$ |

The results show that parameters $x^+, y^+$ are consistent with $x^-, y^-$, $|q/p|$ is consistent with unity and $\phi$ is consistent with 0. Hence the measurements show no significant sign of *CP* violation.

## 19.2.8 Summary

Mixing of neutral charm mesons has been established by the $B$ Factories. The first results were reported in 2007 and published as back-to-back articles. Mixing was established through the study of wrong sign decays by *BABAR* (Aubert, 2007j) (see Section 19.2.2), and via the measurement of $y_{CP}$ by Belle (Staric, 2007) (see Section 19.2.3). The accuracy achieved by the $B$ Factories is well beyond the expectations before the start of data taking in 1999. By the time data taking was well under way, the full potential of $B$ Factories as charm factories was realized and interest in charm physics, especially in the FCNC's of $D$ mesons, was greatly increased. The reason for this was the physics interest in constraints on NP arising from an up-type quark (charm) FCNC's as well as the availability of large samples of reconstructed charm hadrons. Mixing parameters in the $D^0 - \overline{D}^0$ system are now known to an accuracy of $\mathcal{O}(10^{-3})$ from measurements done at the $B$ Factories. The average of results from $B$ Factories on the mixing parameters $x$ and $y$, shown in the upper part of Table 19.2.9, are

$$\begin{aligned} x &= (0.59^{+0.21}_{-0.22})\% \\ y &= (0.78 \pm 0.12)\% \quad . \end{aligned} \tag{19.2.104}$$

[144] Note that by allowing magnitudes and phases of the $D^0$ and $\overline{D}^0$ decays to differ essentially doubles the number of free parameters in the fit.

Graphically they are depicted in Fig. 19.2.33. The no-mixing hypothesis is rejected with a significance of $\mathcal{O}(10)\sigma$.

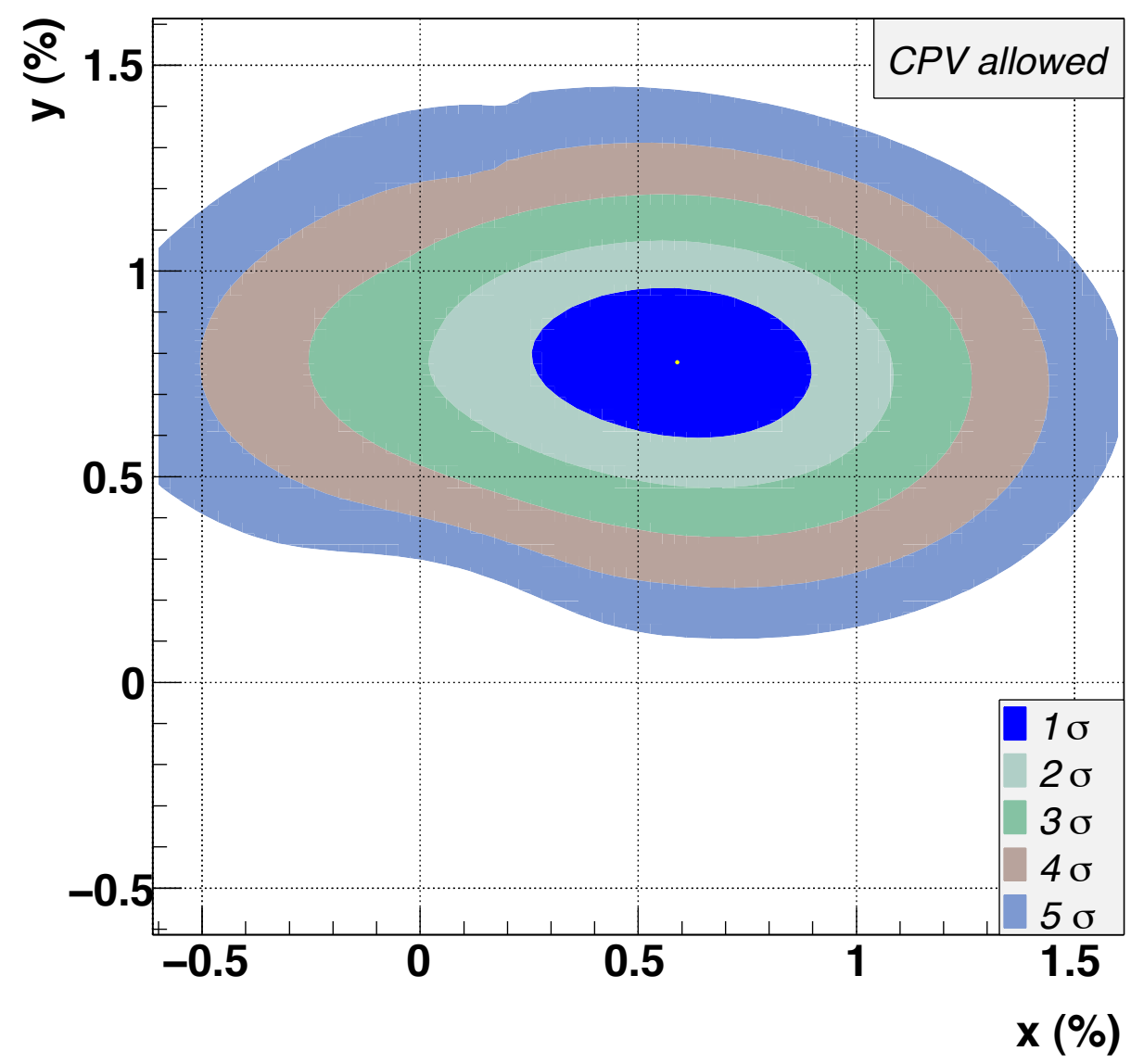

**Figure 19.2.33.** Likelihood contours of the HFAG-like fit to various measurements of mixing and $CP$ violation parameters in the $D^0$ system, in the $(x, y)$ plane.

The average is calculated according to the Heavy Flavor Averaging Group (Amhis et al., 2012) method, using a $\chi^2$ fit assuming uncorrelated systematic errors, but taking into account statistical correlations among the observables as provided by the experiments. Free parameters of the fit are listed in the lower part of Table 19.2.9.

Because the long-distance contributions to the $D^0 - \overline{D}^0$ mixing amplitude are difficult to calculate, the measured values are difficult to interpret. They may receive some contribution from unknown NP processes, although the measured values can be accommodated within the SM.

For several years observation of $CP$ violating asymmetries in the charm sector at the level of $\mathcal{O}(10^{-2})$ was considered as rather clear signature for NP. Measurements performed at the $B$ Factories achieved the sensitivity level of $\mathcal{O}(10^{-3})$. They are probably some of the most demanding measurements as far as the systematic uncertainties are concerned because one needs to control detector induced asymmetries using various control data samples. Several $CP$ violating asymmetries are measured to be at the few per mille level but consistent with no $CP$ violation. Averages of the most important $CP$ violation parameters are given in Table 19.2.9. In Fig. 19.2.34 the likelihood contours of the average are shown in the $(|q/p|, \phi)$ plane. No deviations from $CP$ conservation in the charm sector have been observed. Measurements at the $B$ Factories have triggered an increased activity also on the theoretical side providing more accurate predictions.

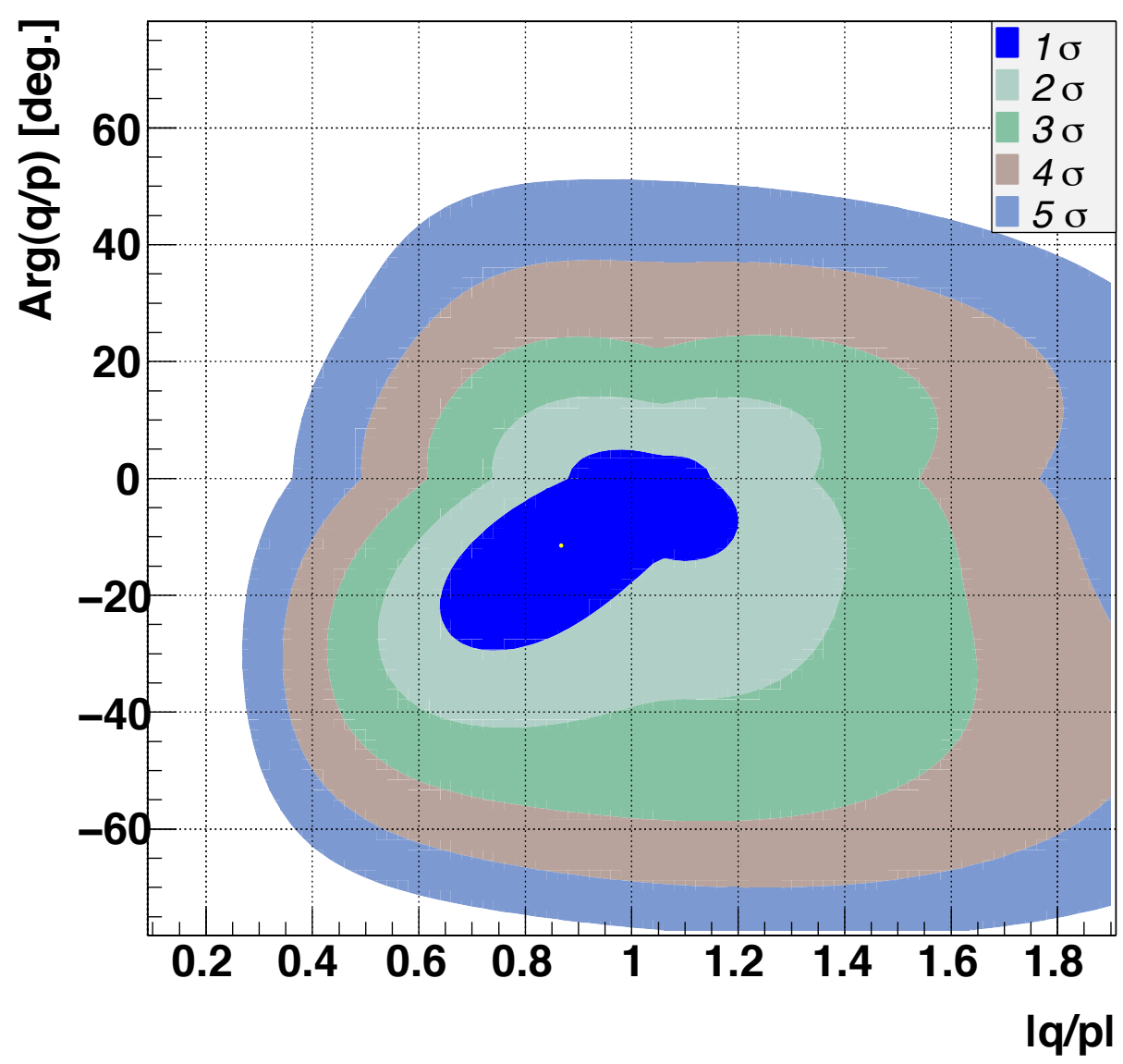

**Figure 19.2.34.** Likelihood contours of the HFAG-like fit to various measurements of mixing and $CP$ violation parameters in the $D^0$ system, in the $(|q/p|, \phi)$ plane.

Results on mixing and especially $CP$ violation in the charm sector have not dried up with the end of the data taking at $B$ Factories. The LHCb collaboration recently presented a very precise measurement of the $CP$ violation in $D^0 \to h^+h^-$ decays (Aaij et al., 2013c). One can hope for more precise (statistically significant) measurements of $CP$ violation in the charm sector from this and future flavor physics experiments.

**Table 19.2.9.** Results on mixing and *CP* violation parameters for neutral charm mesons from *B* Factories. Averages are calculated using the HFAG method assuming uncorrelated systematic errors.

| Results | | |
|---|---|---|
| Decay mode | Parameter | Reference |
| $K^+K^-, \pi^+\pi^-$ | $y_{CP} = (0.72 \pm 0.18 \pm 0.12)\%$ | (Lees, 2013d) |
| | $A_\Gamma = (0.09 \pm 0.26 \pm 0.06)\%$ | |
| | $y_{CP} = (1.11 \pm 0.22 \pm 0.11)\%$ | (Staric, 2012a) |
| | $A_\Gamma = (-0.03 \pm 0.20 \pm 0.08)\%$ | |
| | $A_{CP}^{KK} = (0.00 \pm 0.34 \pm 0.13)\%$ | (Aubert, 2008aq) |
| | $A_{CP}^{\pi\pi} = (-0.24 \pm 0.52 \pm 0.22)\%$ | |
| | $A_{CP}^{KK} = (-0.43 \pm 0.30 \pm 0.11)\%$ | (Staric, 2008) |
| | $A_{CP}^{\pi\pi} = (0.43 \pm 0.52 \pm 0.12)\%$ | |
| $K_S^0\phi$ | $y_{CP} = (0.11 \pm 0.61 \pm 0.52)\%$ | (Zupanc, 2009) |
| $K^\pm\pi^\mp$ | $R_D^{K\pi} = (0.303 \pm 0.0189)\%$ | (Aubert, 2007j) |
| | $A_D^{K\pi} = (-2.1 \pm 5.4)\%$ | |
| | $x'^{2+} = (-0.024 \pm 0.052)\%$ | |
| | $y'^{+} = (0.98 \pm 0.78)\%$ | |
| | $x'^{2-} = (-0.020 \pm 0.050)\%$ | |
| | $y'^{-} = (0.96 \pm 0.75)\%$ | |
| | $R_D^{K\pi} = (0.364 \pm 0.018)\%$ | (Zhang, 2006) |
| | $A_D^{K\pi} = (2.3 \pm 4.7)\%$ | |
| | $x'^{2+} = (0.032 \pm 0.037)\%$ | |
| | $y'^{+} = (-0.12 \pm 0.58)\%$ | |
| | $x'^{2-} = (0.006 \pm 0.034)\%$ | |
| | $y'^{-} = (0.20 \pm 0.54)\%$ | |
| $K^\pm\pi^\mp\pi^0$ † | $x'' = (2.61^{+0.57}_{-0.68} \pm 0.39)\%$ | (Aubert, 2006ap) |
| | $y'' = (-0.06^{+0.55}_{-0.64} \pm 0.34)\%$ | |
| $K_S^0K^+K^-, K_S^0\pi^+\pi^-$ | $x = (0.16 \pm 0.23 \pm 0.12 \pm 0.08)\%$ | (del Amo Sanchez, 2010f) |
| | $y = (0.57 \pm 0.20 \pm 0.13 \pm 0.07)\%$ | |
| | $x = (0.81 \pm 0.30^{+0.13}_{-0.17})\%$ | (Abe, 2007b) |
| | $y = (0.37 \pm 0.25^{+0.10}_{-0.15})\%$ | |
| | $\lvert q/p\rvert = 0.86 \pm 0.30^{+0.10}_{-0.09}$ | |
| | $\phi = (-0.244 \pm 0.31 \pm 0.09)$ rad | |
| $D^0 \to K^+\pi^-\pi^+\pi^-$ | $R_M = (0.017^{+0.017}_{-0.016} \pm 0.003)\%$ | (Aubert, 2006ap) |
| | $\lvert p/q\rvert = 1.1^{+4.0}_{-0.6} \pm 0.1$ | |
| Averages | | |
| | $x = (0.59^{+0.21}_{-0.22})\%$ | |
| | $y = (0.78 \pm 0.12)\%$ | |
| | $\delta_{K\pi} = (26^{+13}_{-14})^\circ$ | |
| | $\delta_{K\pi\pi^0} = (22 \pm 23)^\circ$ | |
| | $R_D^{K\pi} = (0.332 \pm 0.009)\%$ | |
| | $A_D^{K\pi} = (-1.9 \pm 2.4)\%$ | |
| | $\lvert q/p\rvert = 0.87^{+0.18}_{-0.16}$ | |
| | $\phi = (-12^{+10}_{-12})^\circ$ | |
| | $A_D^{KK} = (-0.23 \pm 0.26)\%$ | |
| | $A_D^{\pi\pi} = (0.12 \pm 0.40)\%$ | |

†: $x'' = x\cos\delta_{K\pi\pi^0} + y\sin\delta_{K\pi\pi^0},\ y'' = -x\sin\delta_{K\pi\pi^0} + y\cos\delta_{K\pi\pi^0}$

## 19.3 Charmed meson spectroscopy

***Editors:***
*Antimo Palano (BABAR)*
*Jolanta Brodzicka (Belle)*
*Pietro Colangelo (theory)*

**Additional section writers:**
*Torsten Schroeder*

### 19.3.1 Introduction

In ordinary conditions of temperature and baryon density, quantum chromodynamics (QCD) describes colored quarks permanently bound in colorless hadrons. Accounting for the non-perturbative strong dynamics producing the outstanding phenomenon of confinement represents a difficulty which has been faced in several ways, using methods with various degrees of theoretical soundness, reliability and a variety of results. The main approaches to describe bound states of quarks are the constituent quark models, QCD sum rules and lattice QCD. Moreover, it is possible to formulate effective theories of QCD in limits in which particular symmetries, not present in the full QCD Lagrangian, become apparent and can be exploited. Comparing the outcome of the various calculations (mass spectra, decay rates, etc.) to the measurements has allowed the interpretation of many experimental results. On the other hand, it has been possible to test the accuracy of the theoretical methods and of the procedures adopted to obtain quantitative predictions. The wealth of new information in charm spectroscopy collected at the $B$ Factories has allowed a remarkable progress in the description of the strong dynamics of quarks, as it is briefly described below. A few puzzling features of the observed states deserve further investigations.

#### 19.3.1.1 Constituent quark models

Quark models are traditionally a method to compute properties like hadron masses and couplings. In such approaches the hadrons are approximately described in terms of rest-frame valence quark configurations, the dynamics of which is governed by a Hamiltonian derived from (or inspired by) QCD. In particular, quark confinement is implemented by a flavor-independent, linearly increasing Lorentz-scalar interquark interaction at large distances, while the short-distance quark dynamics is described by a one-gluon exchange interaction. For a system comprising a heavy quark $Q = c$ and a light antiquark $\bar{q} = \bar{u}, \bar{d}, \bar{s}$, the Hamiltonian is written as (Godfrey and Isgur, 1985)

$$H = H_0 + V, \qquad (19.3.1)$$

where

$$H_0 = (p^2 + m_Q^2)^{1/2} + (p^2 + m_{\bar{q}}^2)^{1/2} \qquad (19.3.2)$$

is the kinetic term, with $p$ the modulus of the quark three-momentum in the meson rest frame. The potential $V$ includes spin-independent and spin-dependent terms:

$$V = V_0 + V^{\rm hyp} + V^{\rm so}. \qquad (19.3.3)$$

$V_0$ is the sum of the confining and Coulomb potentials

$$V_0 = -\frac{4}{3}\frac{\alpha_s(r)}{r} + c + \sigma^2 r \qquad (19.3.4)$$

with $c$ and $\sigma^2$ as free parameters. $V^{\rm hyp}$ describes the spin-spin interaction

$$V^{\rm hyp} = \frac{4}{3}\frac{\alpha_S(r)}{m_Q m_{\bar{q}}}\Big[\frac{8\pi}{3}\boldsymbol{s}_Q\cdot\boldsymbol{s}_{\bar{q}}\,\delta^3(\boldsymbol{r}) + \frac{1}{r^3}\left(3\frac{(\boldsymbol{s}_Q\cdot\boldsymbol{r})(\boldsymbol{s}_{\bar{q}}\cdot\boldsymbol{r})}{r^2} - \boldsymbol{s}_Q\cdot\boldsymbol{s}_{\bar{q}}\right)\Big] \qquad (19.3.5)$$

with $\boldsymbol{s}_Q$ and $\boldsymbol{s}_{\bar{q}}$ the heavy quark and light antiquark spin, respectively. $V^{\rm so}$ describes the spin-orbit interaction, expressed as a sum of the chromomagnetic and Thomas-precession contributions:

$$V^{\rm so(cm)} = \frac{4}{3}\frac{\alpha_S(r)}{r^3}\left(\frac{1}{m_Q}+\frac{1}{m_{\bar{q}}}\right)\left(\frac{\boldsymbol{s}_Q}{m_Q}+\frac{\boldsymbol{s}_{\bar{q}}}{m_{\bar{q}}}\right)\cdot\boldsymbol{L}, \qquad (19.3.6)$$

$$V^{\rm so(Tp)} = -\frac{1}{2r}\left(\frac{\partial}{\partial r}V_0\right)\left(\frac{\boldsymbol{s}_Q}{m_Q^2}+\frac{\boldsymbol{s}_{\bar{q}}}{m_{\bar{q}}^2}\right)\cdot\boldsymbol{L}. \qquad (19.3.7)$$

$\boldsymbol{L}$ is the orbital angular momentum. The solution of a Schrödinger-like equation with the Hamiltonian (19.3.1) allows to obtain the mass spectrum and the wave functions for the states $n^{2J+1}L_{2S+1}$ classified according to the orbital angular momentum $\boldsymbol{L}$, the total spin of the quarks $\boldsymbol{S} = \boldsymbol{s}_Q + \boldsymbol{s}_{\bar{q}}$, the total angular momentum $\boldsymbol{J} = \boldsymbol{L} + \boldsymbol{S}$, and the radial quantum number $n$. From the wave functions, other quantities can be computed, *e.g.* the meson decay constants, form factors, the hadron strong couplings.

A few remarks are in order.

- The ("constituent") quark masses in the wave equation are input parameters, and do not coincide with the ("current") masses appearing in the QCD Lagrangian. For the light $u$ and $d$ quarks the constituent masses are fixed to values of $\mathcal{O}$(100 MeV), and of $\mathcal{O}$(300 MeV) for the strange quark, well above the values of the current masses in the QCD Lagrangian.
- The running of the strong coupling $\alpha_S$ can be implemented as a dependence on the interquark distance, $\alpha_S = \alpha_S(r)$; each model is characterized by its implementation of $\alpha_S(r)$.
- Spin dependent terms in the potential present singularities of the type $1/r^n$ with $n > 1$, corresponding to "illegal" operators in the wave equation. This is a consequence of reducing the relativistic quark-antiquark interaction to an instantaneous potential. The treatment of such singularities introduces a model dependence in the calculation of the meson properties.

– The effect of nearby multi-hadron thresholds (or quark unquenching) is not taken into account in the calculation of, *e.g.*, the mass spectrum. Although such an approximation is legitimate in the limit of large number of colors, in real QCD it represents a systematic uncertainty affecting, in particular, the determination of the masses of the orbital and radial excitations.

In a quark model approach the $c\bar{q}$ spectrum ($q = u, d, s$) was already computed long ago with results shown in Table 19.3.1, namely in (Godfrey and Isgur, 1985). These are in rather close agreement (within 20–30 MeV) with the data in the case of the lightest $S$-wave ($L = 0$) states and of two $J^P = 2^+$ and $J^P = 1^+$ $P$-wave ($L = 1$) states, as one can argue considering the experimental measurements reported in Tables 19.3.2 and 19.3.3. As for the state with $J^P = 0^+$ and the second state with $J^P = 1^+$ (both with $L = 1$), in the $c\bar{s}$ case the predicted masses are larger (by about 100 MeV) than the masses of the scalar and axial vector $D_{sJ}$ mesons discussed below in this chapter. Several modifications and improvements have been implemented, namely in models based on the expansion of the Hamiltonian (19.3.1) in the inverse mass of the charm quark, in the spirit of the heavy quark limit described in Section 19.3.1.2 (Di Pierro and Eichten, 2001), and in determinations of the $c\bar{q}$ Regge trajectories (Ebert, Faustov, and Galkin, 2010), but the resulting masses of the scalar and of one of the axial vector $D_{sJ}$ mesons remain larger than in the experiment. A few quark models also predict spin-orbit inversion for the excited states (Godfrey and Kokoski, 1991; Isgur, 1998), which is not observed in data.

To quantitatively assess the accuracy of quark model predictions is not an easy task, due to the assumptions needed to formulate a wave equation for quark-antiquark bound states starting from the QCD Lagrangian; in particular, the effect of quark unquenching is poorly known. Nevertheless, the discrepancy between the predictions of various models and the mass measurements has prompted the idea that some observed states could not be simple quark-antiquark configurations, but more complex structures, like bound state ("molecules") of other mesons (Barnes, Close, and Lipkin, 2003) or mixtures of conventional quark-antiquark with four-quark components (Vijande, Fernandez, and Valcarce, 2006). A discrimination between the different possibilities is feasible considering the results not only for the masses, but also for the widths of the various decay modes. Within the quark models the calculation of the latter quantities presents further uncertainties: in the infinite heavy quark mass limit a different formalism can be developed to study the classification, the spectrum and some decay processes of heavy-light hadrons, as discussed below.

**Table 19.3.1.** Masses (in GeV) of charmed mesons computed in (Godfrey and Isgur, 1985). The corresponding $L = 1$ experimental findings for $D_J$ and $D_{sJ}$ states are reported in Table 19.3.2 and Table 19.3.3 respectively.

| $c\bar{q}$ ($L=0$) | Mass | $c\bar{q}$ ($L=1$) | Mass | $c\bar{q}$ ($L=2$) | Mass |
|---|---|---|---|---|---|
| $D(^1S_0)$ | 1.88 | $D(^3P_0)$ | 2.40 | $D(^3D_1)$ | 2.82 |
| $D(^3S_1)$ | 2.04 | $D(^3P_1)$ | 2.49 | $D(^3D_3)$ | 2.83 |
| | | $D(^3P_2)$ | 2.50 | | |
| | | $D(^1P_1)$ | 2.44 | | |
| $c\bar{s}$ ($L=0$) | Mass | $c\bar{s}$ ($L=1$) | Mass | $c\bar{s}$ ($L=2$) | Mass |
| $D_s(^1S_0)$ | 1.98 | $D_s(^3P_0)$ | 2.48 | $D_s(^3D_1)$ | 2.90 |
| $D_s(^3S_1)$ | 2.13 | $D_s(^3P_1)$ | 2.57 | $D_s(^3D_3)$ | 2.92 |
| | | $D_s(^3P_2)$ | 2.59 | | |
| | | $D_s(^1P_1)$ | 2.53 | | |

#### 19.3.1.2 Exploiting symmetries of QCD in particular limits: the Heavy Quark Chiral Effective Theory

In the limit in which the masses of the heavy quarks (*i.e.* quarks with $m_Q \gg \Lambda_{\rm QCD}$) is sent to infinity, two symmetries emerge in the QCD Lagrangian. The first one is a flavor symmetry, since the dependence on the flavor in QCD is only encoded in the quark mass, and for $m_Q \to \infty$ the heavy flavors are identically described. The second one is a spin symmetry, arising from the decoupling of the spin of a heavy quark from the spin of the light quarks and of the gluons (usually denoted as light degrees of freedom) (Neubert, 1994b). The two symmetries can be recognized at a simple inspection of the Hamiltonian (19.3.1)-(19.3.7); both spin and flavor symmetries are commonly denoted as the "heavy quark symmetry".

A consequence of the heavy quark symmetry is that, in the infinite heavy quark mass limit, heavy-light $Q\bar{q}$ mesons can be classified in doublets labeled by the value of the total angular momentum $\boldsymbol{j}_q$ of the light degrees of freedom with respect to the heavy quark $Q$ (Isgur and Wise, 1991). The spin of each member of the doublet is obtained combining the spin of the heavy quark with the $\boldsymbol{j}_q$: $\boldsymbol{J} = \boldsymbol{s}_Q + \boldsymbol{j}_q$; in the quark model $\boldsymbol{j}_q$ would be given by $\boldsymbol{j}_q = \boldsymbol{s}_{\bar{q}} + \boldsymbol{L}$. Spin symmetry implies that in each doublet the two states are degenerate in mass.

For $L = 0$ the doublet has $j_q = \dfrac{1}{2}$ and consists of two states $(P, P^*)$ ($P$ refers to a generic heavy meson) with spin-parity $J^P_{j_q} = (0^-, 1^-)_{1/2}$. $P$-wave states, with $L = 1$, form two doublets: $(P_0^*, P_1')$ with $J^P_{j_q} = (0^+, 1^+)_{1/2}$, and $(P_1, P_2^*)$ with $J^P_{j_q} = (1^+, 2^+)_{3/2}$. $D$-wave states give rise to two other doublets: $(P_1^*, P_2)$ with $J^P_{j_q} = (1^-, 2^-)_{3/2}$, and $(P_2'^*, P_3)$ with $J^P_{j_q} = (2^-, 3^-)_{5/2}$. Notice that the parity of the doublets has been identified with the parity of the corresponding mesons, $P = (-1)^{L+1}$.

This construction is applied to both open beauty and open charm mesons. In the case of charm, the $(P, P^*)$ doublet is filled by the $D_q$ and $D_q^*$ states, with $q = u, d$ and $s$. The finite heavy quark mass corrections are responsible for removing the mass degeneracy in each doublet (which holds in the light $SU(3)_F$ limit), and are larger in the case of charm than in the case of beauty mesons.

The conservation of angular momentum and parity in strong interactions, together with the heavy quark symmetry, imposes constraints on the transitions between the members of the various doublets with the emission of a light pseudoscalar meson (Isgur and Wise, 1991). In particular, the transitions of the excited states with $j_q^P = \frac{1}{2}^+$ into states with $j_q^P = \frac{1}{2}^-$ and a pion or kaon occur in $S$-wave, while the transitions of the states with $j_q^P = \frac{3}{2}^+$ into $j_q^P = \frac{1}{2}^-$ ones and a pion or kaon are in $D$-wave. The consequence is that, if such transitions are kinematically allowed, the $j_q^P = \frac{1}{2}^+$ resonances are expected to be broader than the $j_q^P = \frac{3}{2}^+$ ones. At the next-to-leading order in the $1/m_Q$ expansion, the axial vector states in the $j_q^P = \frac{3}{2}^+$ doublet can also decay in $S$-wave. An example is provided by the $D_2$ meson, which decays to $D\pi$ in $D$-wave; at the leading order in the heavy quark expansion, its spin partner $D_1$ decays to $D^*\pi$ also in $D$-wave, and their widths, which depend on the three momentum of the emitted pion as $|\boldsymbol{p}|_\pi^5$, are quite narrow.[145] On the other hand, the scalar $D_0^*$ meson decays to $D\pi$ in $S$-wave, which explains its broad width.

The strong transitions between states belonging to the various doublets or within the same doublet, can be studied in an effective field theory formalism. An effective QCD Lagrangian is constructed in the infinite heavy quark mass limit, hence exploiting the heavy quark symmetry, and in the limit in which the light quark masses ($u, d$ and $s$) vanish and another symmetry holds for QCD, the chiral $SU(3)_L \times SU(3)_R$ symmetry. The various heavy meson doublets are represented by fields of $4\times 4$ matrices. The doublets with $j_q^P = \frac{1}{2}^-$ and $j_q^P = \frac{1}{2}^+$ are described by the fields $H_a$ and $S_a$, respectively, while the doublets with $j_q^P = \frac{3}{2}^+$, $j_q^P = \frac{3}{2}^-$ and $j_q^P = \frac{5}{2}^-$ by the fields $T_a^\mu$, $X_a^\mu$ and $X_a'^{\mu\nu}$ ($a$ is a light flavor index):

$$\begin{aligned} H_a &= \frac{1+\not v}{2}[P^*_{a\mu}\gamma^\mu - P_a\gamma_5], \\ S_a &= \frac{1+\not v}{2}[P'^{\mu}_{1a}\gamma_\mu\gamma_5 - P^*_{0a}], \\ T_a^\mu &= \frac{1+\not v}{2} \\ &\times \left\{ P^{*\mu\nu}_{2a}\gamma_\nu - P_{1a\nu}\sqrt{\frac{3}{2}}\gamma_5 \left[ g^{\mu\nu} - \frac{\gamma^\nu}{3}(\gamma^\mu - v^\mu)\right]\right\}, \end{aligned} \tag{19.3.8}$$

and analogous expressions for $X_a^\mu$ and $X_a'^{\mu\nu}$, with $v$ the meson four-velocity. The doublet with $j_q^P = \frac{1}{2}^-$ corresponding to the first radial excitations is described by $H_a'$ with structure identical to $H_a$. The various operators $P_i$ in Eq. (19.3.8) annihilate mesons of four-velocity $v$ which is conserved in strong interaction processes.

The octet of light pseudoscalar mesons is introduced through the fields $\xi = e^{\frac{i\mathcal{M}}{f_\pi}}$, with $\mathcal{M}$ containing $\pi, K$ and $\eta$ fields:

$$\mathcal{M} = \begin{pmatrix} \sqrt{\frac{1}{2}}\pi^0 + \sqrt{\frac{1}{6}}\eta & \pi^+ & K^+ \\ \pi^- & -\sqrt{\frac{1}{2}}\pi^0 + \sqrt{\frac{1}{6}}\eta & K^0 \\ K^- & \bar{K}^0 & -\sqrt{\frac{2}{3}}\eta \end{pmatrix} \tag{19.3.9}$$

and $f_\pi = 132\,\text{MeV}$ the pion decay constant. The strong interaction of the heavy mesons with the octet of light pseudoscalar mesons is described by an effective Lagrangian invariant under chiral transformations of the light fields, and under heavy-quark spin-flavor transformations of the heavy fields. At the leading order in the heavy quark mass and light meson momentum expansion, the transition $F \to HM$ ($F = H, S$ and $T$, and $M$ a light pseudoscalar meson) can be described by the Lagrangian terms (Burdman and Donoghue, 1992; Wise, 1992; Yan et al., 1992)

$$\begin{aligned} \mathcal{L}_H &= g\,\mathrm{Tr}[\bar{H}_a H_b \gamma_\mu\gamma_5 \mathcal{A}^\mu_{ba}], \\ \mathcal{L}_S &= h\,\mathrm{Tr}[\bar{H}_a S_b \gamma_\mu\gamma_5 \mathcal{A}^\mu_{ba}] \; + h.c., \\ \mathcal{L}_T &= \frac{h'}{\Lambda_\chi}\,\mathrm{Tr}[\bar{H}_a T_b^\mu (iD_\mu \mathcal{A} + i\not D \mathcal{A}_\mu)_{ba}\gamma_5] + h.c., \end{aligned} \tag{19.3.10}$$

with $\mathcal{A}_{\mu ba} = \frac{i}{2}\left(\xi^\dagger\partial_\mu\xi - \xi\partial_\mu\xi^\dagger\right)_{ba}$, $D$ the covariant derivative $D_{\mu ba} = -\delta_{ba}\partial_\mu + \mathcal{V}_{\mu ba}$ and $\mathcal{V}_{\mu ba} = \frac{1}{2}\left(\xi^\dagger\partial_\mu\xi + \xi\partial_\mu\xi^\dagger\right)_{ba}$. $\Lambda_\chi$ is a chiral symmetry-breaking scale (which can be set to $\Lambda_\chi = 1\,\text{GeV}$), and $g$ , $h$ and $h'$ are effective couplings, which can be determined from experiment or from theoretical calculations (see Section 19.3.1.4).

A set of other Lagrangian terms for the strong transitions among the various heavy quark doublets can be constructed analogously, including a few $\mathcal{O}(m_Q^{-1})$ corrections (Colangelo, De Fazio, and Ferrandes, 2006), from which the decay widths and ratios of decay branching fractions can be computed and compared to experiment. The results are useful to cast light on the $Q\bar{q}$ spectrum, providing support to the classification of the observed resonances.

In the $c\bar{q}$ ($q = u, d$) system, the mesons $D^{+,0}$ and $D^{*+,0}$ fill the $J^P_{j_q} = (0^-, 1^-)_{1/2}$ doublets. The properties of the positive parity states, collected in Table 19.3.5, follow the expectations based on the classification scheme outlined above, with a mixing between the two $J^P = 1^+$ states. A set of heavier states has been observed in the $D\pi$ and $D^*\pi$ distributions: $D(2550)^0$, $D^*(2600)^0$, $D^*(2600)^{0,+}$, $D(2750)^0$ and $D^*(2760)^{0,+}$ (del Amo Sanchez, 2010i); for them, a tentative assignment is proposed in the following.

Also the observed $c\bar{s}$ mesons fit in the classification scheme based on the heavy quark expansion. The two lightest mesons $D_s(1969)$ and $D_s^*(2112)$ fill the $J^P_{j_q} = (0^-, 1^-)_{1/2}$ doublet. There are four positive parity states:

[145] On very general grounds, the matrix element for the transition involving the orbital momentum $L$ depends on the spatial integration of a $(kr)^L$ term, where $k = p/\hbar$, $p$ is the momentum of the final state particle in the rest frame of the decaying particle, and $L$ is the orbital momentum quantum number. Hence the matrix element is proportional to $p^L$. Furthermore the phase space of a two-body decay is proportional to $p$. Hence the decay width is proportional to $|\mathcal{M}|^2 \propto p^{2L+1}$.

**Table 19.3.2.** Properties of neutral $L = 1$ $D_J$ mesons.

| | $J^P$ | Mass (MeV) | Width (MeV) | Observed decays |
|---|---|---|---|---|
| $D_0^*$ | $0^+$ | $2352 \pm 50$ | $261 \pm 50$ | $D\pi$ |
| $D_1'$ | $1^+$ | $2427 \pm 36$ | $384^{+130}_{-105}$ | $D^*\pi$ |
| $D_1$ | $1^+$ | $2421.3 \pm 0.6$ | $27.1 \pm 2.7$ | $D^*\pi, D^0\pi^+\pi^-$ |
| $D_2^*$ | $2^+$ | $2462.6 \pm 0.7$ | $49.0 \pm 1.4$ | $D^*\pi, D\pi$ |

$D_{s0}^*(2317)$ and $D_{s1}'(2460)$ which can be identified with the members of the doublet $J_{j_q}^P = (0^+, 1^+)_{1/2}$, and $D_{s1}(2536)$ and $D_{s2}^*(2573)$ filling the $J_{j_q}^P = (1^+, 2^+)_{3/2}$ doublet (Becirevic, Fajfer, and Prelovsek, 2004; Colangelo and De Fazio, 2003; Colangelo, De Fazio, and Ozpineci, 2005), with a $\mathcal{O}(\frac{1}{m_Q})$ mixing between the two $1^+$ states. Both the $(0^+, 1^+)_{1/2}$ states have masses below the $DK$ and the $D^*K$ thresholds, respectively, and this explains their very narrow width (Swanson, 2006).[146] Their properties are collected in Tables 19.3.7 and 19.3.8.

The meson $D_{sJ}(2710)$ has been observed in the $DK$ final state, and its spin-parity $J^P = 1^-$ has been determined (Brodzicka, 2008). A resonance $D_{sJ}(2860)$ has also been found in the $DK$ spectrum (Aubert, 2006ag). Since both resonances also appear in the $D^*K$ spectrum (Aubert, 2009au), they have natural parity, $J^P = 1^-, 2^+, 3^-, \cdots$. The decay mode into $D^*K$ excludes the assignment $J^P = 0^+$ for $D_{sJ}(2860)$, and is compatible with the assignment $J^P = 3^-$ with radial quantum number $n = 1$, so that $D_{sJ}(2860)$ could be a member of the doublet $J_{j_q}^P = (2^-, 3^-)_{5/2}$ (Colangelo, De Fazio, and Nicotri, 2006). The rather narrow width of this resonance would be justified by this assignment, since the two-body decay to $DK$ would occur in $F$-wave. The classification of the broad structure $D_{sJ}(3040)$ observed in the $D^*K$ mass spectrum (Aubert, 2009au), has to be done on the basis of the available information on the mass, the width and the decay modes, together with the full set of information about the other doublets (Colangelo and De Fazio, 2010). Important observables are ratios of branching fractions, namely $\mathcal{B}(D_{sJ}(2710) \to D^*K)/\mathcal{B}(D_{sJ}(2710) \to DK)$ and $\mathcal{B}(D_{sJ}(2860) \to D^*K)/\mathcal{B}(D_{sJ}(2860) \to DK)$ (with $D^{(*)}K$ the sum over $D^{(*)0}K^+$ and $D^{(*)^+}K_S^0$): the comparison of the measurement (Aubert, 2009au) with the theoretical results favors the interpretation of $D_{sJ}(2710)$ as the first radial excitation of $D_s^*(2112)$ and a member of the excited doublet $J_{j_q}^P = (0^-, 1^-)_{1/2}$ with radial quantum number $n = 2$ (Colangelo, De Fazio, Nicotri, and Rizzi, 2008; Close, Thomas, Lakhina, and Swanson, 2007). In the case of $D_{sJ}(2860)$ the measured ratio of branching fractions is larger than the theoretical prediction, leaving the classification still an open issue.

[146] Other interpretations of $D_{s0}^*(2317)$ and $D_{s1}'(2460)$ (molecules, multiquarks) are reviewed in (Colangelo, De Fazio, and Ferrandes, 2004) and (Swanson, 2006); the dynamical generation of such states has been proposed in (Guo, Shen, Chiang, Ping, and Zou, 2006; Guo, Shen, and Chiang, 2007).

**Table 19.3.3.** Properties of $L = 1$ $D_{sJ}$ mesons.

| | $J^P$ | Mass (MeV) | Width (MeV) | Observed decays |
|---|---|---|---|---|
| $D_{s0}^*$ | $0^+$ | $2317.8 \pm 0.6$ | $< 3.8$ | $D_s^+\pi^0$ |
| $D_{s1}'$ | $1^+$ | $2459.5 \pm 0.6$ | $< 3.5$ | $D_s^{*+}\pi^0, D_s^+\gamma, D_s^+\pi^+\pi^-$ |
| $D_{s1}$ | $1^+$ | $2535.28 \pm 0.20$ | $< 2.5$ | $D^{*+}K^0, D^{*0}K^+$ |
| $D_{s2}^*$ | $2^+$ | $2572.6 \pm 0.9$ | $20 \pm 5$ | $D^0K^+$ |

The classification of $D_{s0}^*(2317)$ and $D_{s1}'(2460)$ as members of the spin doublet $j_q^P = \frac{1}{2}^+$, together with the observation that the mass splitting $M(D_{s1}') - M(D_{s0}^*)$ coincides with the mass splitting between $D_s(1969)$ and $D_s^*(2112)$ which belong to the negative parity $j_q^P = \frac{1}{2}^-$ doublet, has inspired the notion of chiral heavy-light meson doublets (Bardeen, Eichten, and Hill, 2003; Nowak, Rho, and Zahed, 2004). The idea is that the hadrons comprising a single heavy quark can be considered as "tethered" systems. In a scenario in which (explicitly and spontaneously broken) chiral symmetry is restored in QCD maintaining confinement, the heavy-light hadrons might appear in parity-doubled bound states which transform as linear representations of the chiral symmetry. An effective field theory, constrained by the heavy quark symmetry, can be formulated for such parity-doubled states, and as a consequence the mass difference $\Delta M$ between the $j_q^P = \frac{1}{2}^\pm$ parity doublets can be related to $g_\pi$, the $0^+ \to 0^-\pi$ coupling constant, and to the pion decay constant $f_\pi$ by a relation similar to the Goldberger-Treiman formula: $\Delta M = g_\pi f_\pi$ (Bardeen, Eichten, and Hill, 2003).

The notion of heavy parity doublets needs to be further explored and confirmed, both in the case of the lightest doublets and for the other excited states. It has various consequences for the strong and radiative decay modes. The radiative E1 transitions $(1^+, 0^+) \to (1^-, 0^-)\gamma$, for example, as well as the $(1^+, 1^-) \to (0^+, 0^-)\gamma$ M1 transitions, are governed by similar combinations of the quark masses and electric charges, so that predictions for the various modes can be elaborated and compared to experiment. In Table 19.3.4 a tentative classification of all the observed mesons with open charm in HQ doublets is shown (Colangelo, De Fazio, Giannuzzi, and Nicotri, 2012).

As a last remark, the heavy quark symmetry allows to use the information available in the charm sector to predict properties in the beauty sector. For example, a doublet $J_{j_q}^P = (0^+, 1^+)_{1/2}$ of narrow positive parity $b\bar{s}$ mesons is expected with masses $M(B_{s0}^*) = (5.71 \pm 0.03)$ GeV and $M(B_{s1}') = (5.77 \pm 0.03)$ GeV below the $BK$ and $B^*K$ thresholds and possible decays into $B_s\pi^0$ and $B_s^*\pi^0$, respectively (Colangelo, De Fazio, and Ferrandes, 2006).

**Table 19.3.4.** Tentative classification in HQ doublets of the observed mesons with open charm. States with uncertain assignment are indicated with (∗).

| Doublet | $j_q^P$ | $J^P$ | $c\bar{q}$ $(n=1)$ | $c\bar{q}$ $(n=2)$ | $c\bar{s}$ $(n=1)$ | $c\bar{s}$ $(n=2)$ |
|---|---|---|---|---|---|---|
| H | $\frac{1}{2}^-$ | $0^-$ | $D(1869)$ | $D(2550)$ (∗) | $D_s(1968)$ | |
| | | $1^-$ | $D^*(2010)$ | $D^*(2600)$ (∗) | $D_s^*(2112)$ | $D_{s1}^*(2700)$ |
| S | $\frac{1}{2}^+$ | $0^+$ | $D_0^*(2400)$ | | $D_{s0}^*(2317)$ | |
| | | $1^+$ | $D_1'(2430)$ | | $D_{s1}'(2460)$ | $D_{sJ}(3040)$ (∗) |
| T | $\frac{3}{2}^+$ | $1^+$ | $D_1(2420)$ | | $D_{s1}(2536)$ | $D_{sJ}(3040)$ (∗) |
| | | $2^+$ | $D_2^*(2460)$ | | $D_{s2}^*(2573)$ | |
| X′ | $\frac{5}{2}^-$ | $2^-$ | $D(2750)$ (∗) | | | |
| | | $3^-$ | $D(2760)$ (∗) | | $D_{sJ}(2860)$ (∗) | |

#### 19.3.1.3 Results from QCD sum rules

The masses of the open charm mesons, as well as other hadronic quantities like the decay constants, can be computed in QCD by QCD Sum Rules. Two-point correlation functions of quark currents with the quantum numbers of the mesons of interest (Shifman, Vainshtein, and Zakharov, 1979),

$$\Pi(Q^2) = i \int d^4x \, e^{iq\cdot x} \, \langle T[J(x)J^\dagger(0)]\rangle \qquad (19.3.11)$$

with $J = J_{c\bar{q}}$, are expressed in QCD at short distances ($Q^2 \to \infty$) in terms of the quark masses, the strong coupling constant $\alpha_S$ and of the vacuum matrix elements $\langle O_n\rangle$ of gauge-invariant quark and gluon operators (vacuum condensates). The latter parameters appear in the $1/(Q^2)^n$ corrections to the perturbative expression of the correlation functions $\Pi^{\rm pert}(Q^2)$,

$$\Pi^{\rm QCD}(Q^2) = \Pi^{\rm pert}(Q^2) + \sum_n c_n \frac{\langle O_n\rangle}{(Q^2)^n}. \qquad (19.3.12)$$

The same correlation functions are represented in terms of hadronic states,

$$\Pi(Q^2) = \Pi^{\rm had}(Q^2), \qquad (19.3.13)$$

and the contributions of the lowest lying states are isolated from the excited states and the hadronic continuum. Matching the two (QCD and hadronic) representations on the basis of analyticity and of (global) quark-hadron duality, expressions can be worked out for, *e.g.*, the meson masses in terms of QCD parameters (Colangelo and Khodjamirian, 2000). The method can be formulated also for the effective theories of QCD, namely the Heavy Quark Effective Theory (Neubert, 1994b). The same approach can be applied to investigate multiquark configurations, using correlation functions of currents composed by several quark fields.

The accuracy of the QCD Sum Rule predictions is related to the separation of the various particle contributions in the hadronic representation of the two-point correlation functions, to the procedure of exploiting quark-hadron duality and to the errors in the parameters, in particular the vacuum condensates. Within the uncertainties, the masses of the lightest $c\bar{q}$ and $c\bar{s}$ mesons have been predicted in agreement with experiment (Reinders, Rubinstein, and Yazaki, 1985). Also the masses of the positive parity open charm mesons turn out to be compatible with measurement, and the main features of the radiative and strong decays are reproduced if they are described as ordinary quark-antiquark configurations (Colangelo, De Fazio, and Ozpineci (2005); Dai, Li, Zhu, and Zuo (2008)).

#### 19.3.1.4 Lattice QCD results

The hadron properties can be computed *ab initio* from the QCD Lagrangian by lattice calculations, analyzing correlation functions of quark currents of suitably chosen quantum numbers. The resulting $c\bar{q}$ and $c\bar{s}$ mass spectrum can be compared to the measurement, addressing the issue of the classification of the observed resonances. However, the calculation of hadronic quantities for dynamical light quarks with masses close to the QCD values is still a challenging task. Different results for the meson masses have been found by different groups, leading to a different classification of, *e.g.*, $D_{s0}^*(2317)$ and $D_{s1}'(2460)$. In (Bali, 2003) the mass $M(D_{s0}^*) = 2.57(11)\,\text{GeV}$ is obtained for the scalar $c\bar{s}$ state, hence a value larger than the measured mass of $D_{s0}^*(2317)$, suggesting a non quark-antiquark interpretation for this state. On the other hand, in the calculations in (Dougall, Kenway, Maynard, and McNeile, 2003) and (Lin, Ohta, Soni, and Yamada, 2006) the mass splitting $M(D_s(0^+) - M(D_s(0^-)$ turns out to be compatible with experiment. The results of more recent analyses, with overlap fermions for both light and heavy quarks and only one lattice spacing, are consistent with the experimental masses, in particular for $D_{s0}^*(2317)$ described as a $c\bar{s}$ state (Dong et al., 2009). The same conclusion is drawn in (Gong et al., 2011), where the tetraquark interpretation of $D_{s0}^{*}(2317)$ is tested and found to be inconsistent with data.

## 19.3.2 Production of charmed mesons at $B$ Factories

Charmed mesons are copiously produced at $B$ Factories either directly in $e^{+}e^{-}$ collisions or as products of $B$ meson decays. Both mechanisms allow for complementary measurements of the charmed multiplets.

Charm production in $B$ decays is governed by the CKM-favored $b \to c$ transition, thus $B$ mesons are an abundant source of charmed mesons. The restricted kinematics of $B\overline{B}$ production at the $\Upsilon(4S)$ enable a selection of clean $B$ meson samples, while the zero spin of the parent $B$ constrains possible quantum numbers of the daughter particles and makes their spin-parity measurements easier. Depending on the quantum numbers of charmed mesons, theory predicts certain patterns in their production rates in $B$ meson decays (Section 17.3), which is helpful in clarifying the nature of the produced particles. Charmed mesons bearing high spin or being highly excited are suppressed in $B$ meson decays, thus their studies with $e^{+}e^{-} \to c\bar{c}$ continuum data are more feasible.

At the center-of-mass (CM) energy of the $B$ Factories, the cross-section of prompt $c\bar{c}$ pair production provides a large fraction of the total hadronic cross-section, resulting in large samples of ground and excited charmed mesons from the hadronization of the produced $c$ quarks (see Section 24.1). The produced hadrons are usually studied inclusively *i.e.* without reconstruction of the other particles in the event. Such an approach allows high efficiency but often suffers from large background, while charmed hadrons coming from strongly decaying excited states cannot be distinguished from those originating directly from the $e^{+}e^{-}$ annihilation.

## 19.3.3 Non-strange charm spectroscopy

### 19.3.3.1 Introduction

Charm spectroscopy to this date has still not been fully explored as there are many $D$ meson states predicted in the 1980s, which have not been observed experimentally. Figure 19.3.1 shows the predicted spectrum for a $c\bar{u}$ system (The spectrum of the $c\bar{d}$ system is almost identical.). The ground states, $D^{0,+}$ (Goldhaber et al., 1976; Peruzzi et al., 1976), and spin excitations, $D^{*(0,+)}$ (Feldman et al., 1977), were first observed respectively in 1976 and 1977 by the Mark-I experiment at SLAC. Their properties are quite well known, though they are still being studied with increasing precision. Recent measurements of $D^0$ and $D^+$ masses come from CLEO (Cawlfield et al., 2007) and KEDR (Anashin et al., 2010). As for the total widths, only the $D^{*+}$ width has been measured ($96 \pm 22$ keV), as it is enlarged by strong $D^{*+}$ decays. The widths of the other $D$ and $D^*$ mesons are consistent with zero, because they decay either mainly weakly, or in the case of the $D^{*0}$ mainly radiatively. Therefore, experiments set only upper limits for them. Decays of $D^{(*)}$ mesons are described in detail in Section 19.1. Spin-parities of $D^{(*)}$ quoted in the PDG are assigned based on the quark-model predictions, though many studies performed, especially of $D^{(*)}$ produced in $B$ decays, confirm these assignments (Sections 17.3 and 17.6).

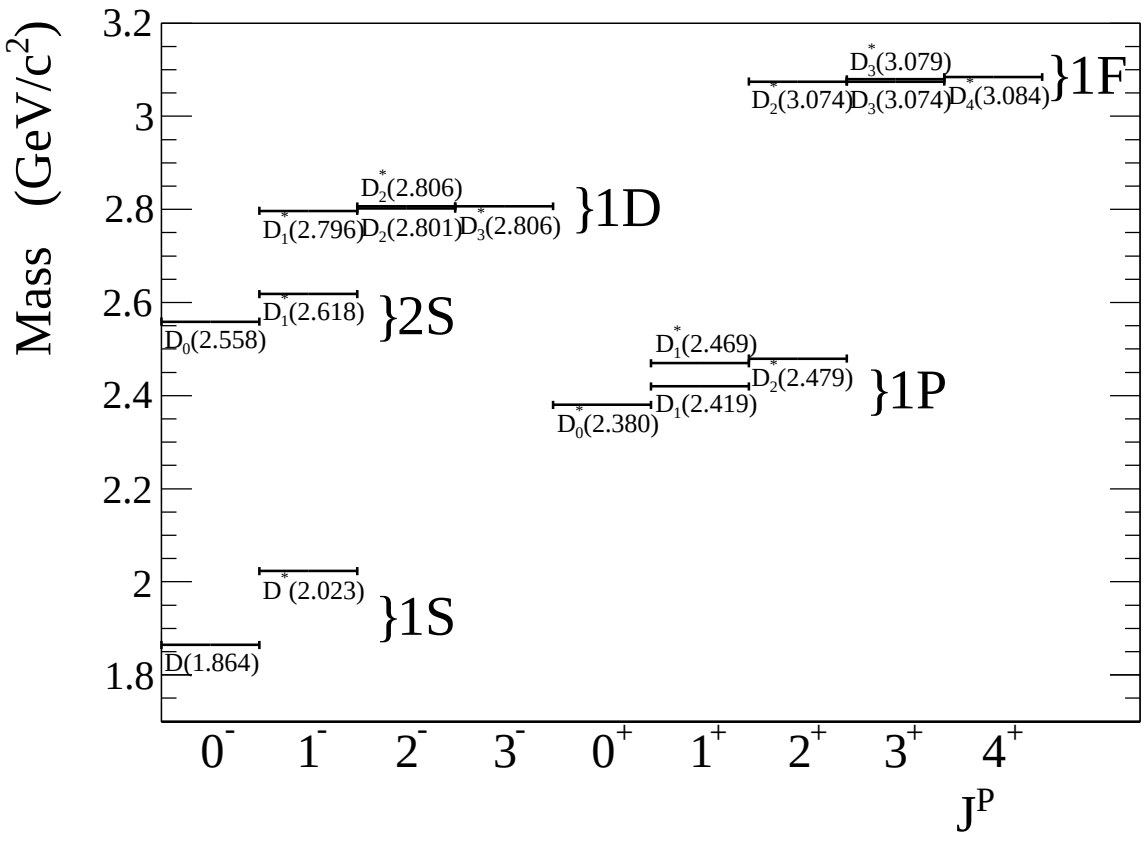

**Figure 19.3.1.** Modified Godfrey-Isgur predictions (Godfrey and Isgur, 1985). The plot shows the $c\bar{u}$ spectrum where the masses have been scaled down so that the ground state matches the $D^0$ mass. Also, the $2^-$ states, not shown in the original paper, have been inserted following the splitting structure of the $1P$ states.

Within the $L = 1$ orbital excitations labeled as $D^{**}$, the narrow doublet was not observed until 1989 owing to its lower production rates and larger widths. Eventually neutral $D_1^0$ and $D_2^{*0}$ mesons were observed in $D^{*+}\pi^-$ final states by ARGUS (Albrecht et al., 1989d) and CLEO (Avery et al., 1990). Their widths of about $20-30\,\mathrm{MeV}/c^2$ and masses around $2.4\,\mathrm{GeV}/c^2$, were in agreement with the model predictions (Rosner (1986); Godfrey and Kokoski (1991); Falk and Peskin (1994)). The broad $L = 1$ states have only recently been observed by Belle and *BABAR* in $B$ decays where they could be separated from the background, as it will be described in the following.

### 19.3.3.2 $D^{**}$ in $B$ decays

$\overline{B}$ decays to the $D\pi$ and $D^*\pi$ final states are two of the dominant hadronic decays and have been measured quite well (Section 17.3). Similar processes, $\overline{B} \to D^{**}\pi$, are expected to be a dominant source of $D^{**}$ mesons. At the quark level such decays proceed through a $b \to cW^- \to c\bar{u}d$ transition for which the underlying diagrams are shown in Fig. 19.3.2. As $D^{**}$ mesons are expected to decay dominantly in the $D^{(*)}\pi$ modes, processes like $\overline{B} \to D^{(*)}\pi\pi$ provide the best way to study them.

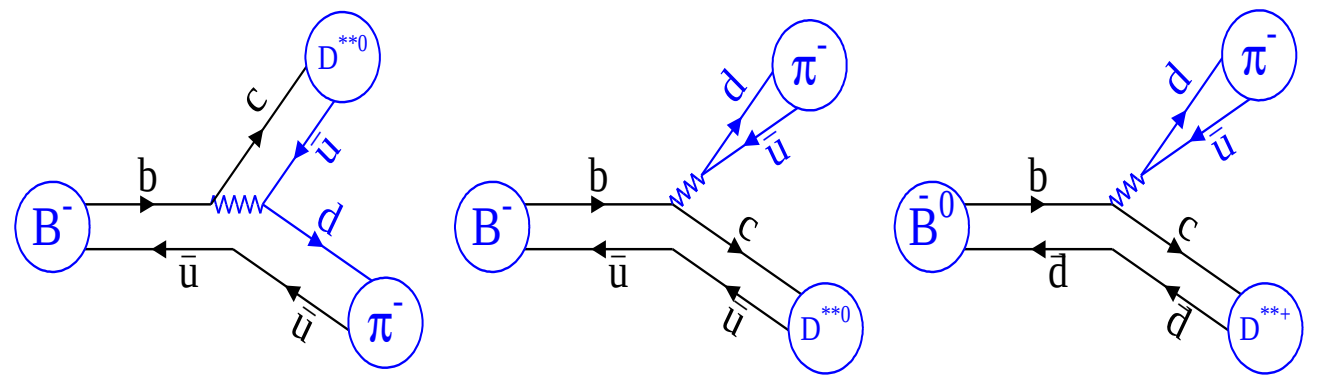

**Figure 19.3.2.** From (Kuzmin, 2007). Quark-line diagrams for charged (left and middle) and neutral (right) $\overline{B} \to D^{**}\pi$ decays.

### 19.3.3.3 Neutral $D^{**}$ via $B^- \to D^{(*)+}\pi^-\pi^-$ Dalitz analysis

The production of neutral $D^{**}$ resonances was studied through a full Dalitz-plot analysis of the three-body decay $B^- \to D^+\pi^-\pi^-$ (Abe, 2004f; Aubert, 2009g) and $B^- \to D^{*+}\pi^-\pi^-$ (Abe, 2004f). Thus, for the first time, interference between intermediate states is taken into account in measuring properties of the $D^{**}$ mesons. First Belle performed such a Dalitz analysis using a data sample of about $65\times10^6$ of $B\overline{B}$ pairs; a *BABAR* analysis, based on $383\times10^6$ $B\overline{B}$ pairs, followed. The $D^{(*)+}$ are reconstructed in the clean decay modes: $D^+ \to K^-\pi^+\pi^+$, $D^{*+} \to D^0\pi^+$ with $D^0 \to K^-\pi^+$ and $K^-\pi^+\pi^+\pi^-$, ensuring good purity over the Dalitz diagram. Reconstructed $B$ candidates are identified by their $\Delta E$ and $m_{\rm ES}$. In addition a thrust angle requirement, $\cos\theta_{\rm T} < 0.8$, is applied to suppress continuum background (see Section 9). Signal yields, obtained from fits to the $\Delta E$ distributions, of about 1100 events (Belle) and 3500 events (*BABAR*) for $B^- \to D^+\pi^-\pi^-$ and 560 events for $B^- \to D^{*+}\pi^-\pi^-$ (Belle) have been obtained.

The Dalitz plots for the $B^- \to D^+\pi^-\pi^-$ candidates within the $\Delta E$-$m_{\rm ES}$ signal region are shown in Fig. 19.3.3. The plot from *BABAR* is symmetric in the $D\pi$ masses because of the two identical pions. Belle used as the Dalitz variables the lower and higher values of the two $D\pi$ mass combinations, denoted as $m^2_{\rm min}(D\pi)$ and $m^2_{\rm max}(D\pi)$; intermediate resonances emerge in $m^2_{\rm min}(D\pi)$. The distributions clearly display the structure of nodes characteristic of the spin-2 resonance $D^*(2460)^0$, while an accumulation of events in the threshold mass region are attributed to the scalar $D_0^{*0}$.

Principles of Dalitz-plot analysis are described in detail in Section 13. The signal density of the decay $B^- \to D^+\pi^-\pi^-$ are parameterized as a coherent sum of amplitudes corresponding to the following intermediate states forming the $D^+\pi^-$ system: $D_2^{*0}$, $D_0^{*0}$ and off-shell vector $D^{*0}$ (labeled as $D_v^{*0}$). Also a virtual $B^{*0}$, contributing as $\overline{B} \to \overline{B}{}_v^{*0}\pi$ with $\overline{B}{}_v^{*0} \to D\pi$, and a constant amplitude for a non-resonant component are included. Such virtual contributions are of phenomenological origin and are introduced to obtain a better description of the Dalitz distributions. Both Belle and *BABAR* analyses employed an isobar model in which resonance amplitudes are parameterized with Breit-Wigner (BW) functions with a mass dependent width and angular dependence related to resonance decay with given orbital momentum ($L = 0,\ 1,\ 2$ respectively for the $D_0^{*0}$, $D_v^{*0}$, $D_2^{*0}$). The latter introduces an amplitude dependence on the helicity angle ($\Theta_h$) defined as the angle between the momentum vectors of the bachelor pion from the $B$ decay and the pion of the $D\pi$ system in the $D\pi$ rest frame. The signal parameterization is convoluted with the experimental mass resolution, typically of order of a few MeV$/c^2$. The background shape is obtained from a fit to the Dalitz distribution for the $\Delta E$ sideband region.

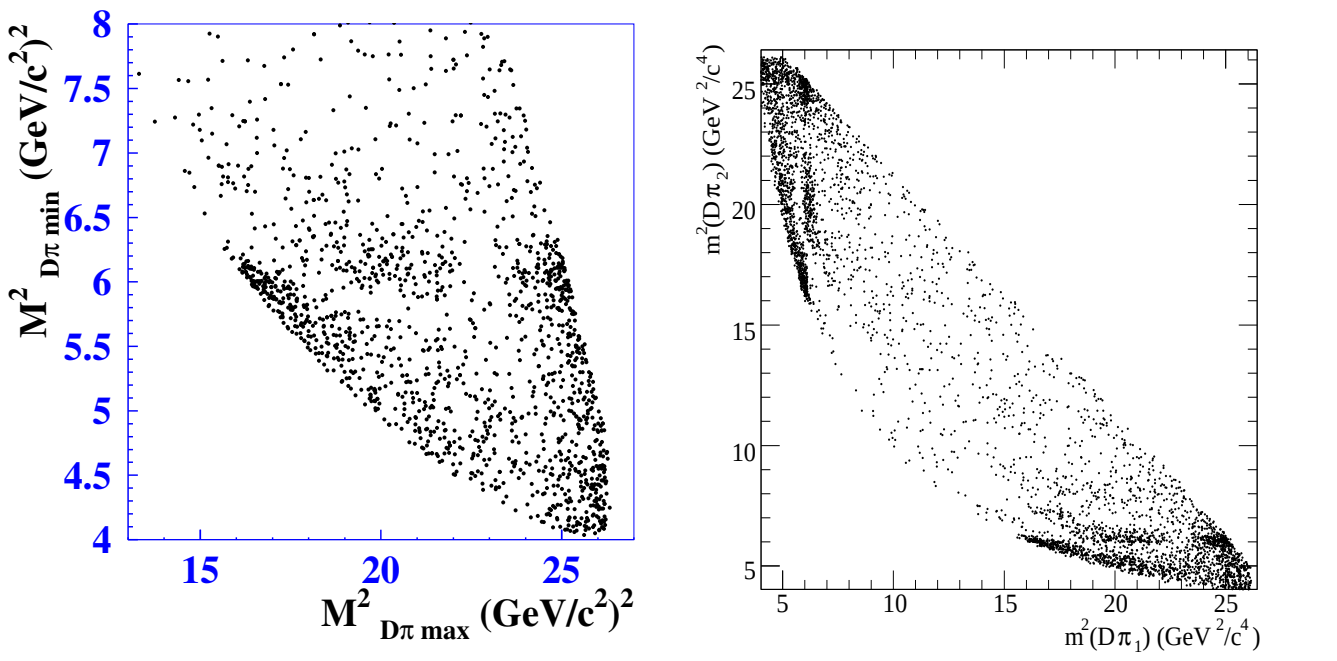

**Figure 19.3.3.** Dalitz plot for $B^- \to D^+\pi^-\pi^-$ from Abe (2004f) (left) and (Aubert, 2009g) (right).

With such models for signal and background densities, an unbinned maximum-likelihood fit to the Dalitz plot is performed. The $D^{**}$ parameters, all the amplitudes and relative phases are free parameters in the fit. The fit likelihood value is significantly improved by the inclusion of the broad scalar resonance, thus Belle claimed the first observation of the $D_0^{*0}$ meson (Abe, 2004f).

Figure 19.3.4 shows the $m^2_{\rm min}(D\pi)$, $m^2_{\rm max}(D\pi)$ and $m^2(\pi\pi)$ projections with the fit result and contributions from the intermediate resonances superimposed, as obtained by *BABAR*. The $D_0^{*0}$ signal and the reflection of $D_2^{*0}$ can easily be distinguished in the $m^2_{\rm min}(D\pi)$ and $m^2_{\rm max}(D\pi)$ projections, respectively. The resonance masses and widths as well as branching ratio products measured in both analyses are very consistent, and are summarized in Table 19.3.5.

Figure 19.3.5 shows $M(D^+\pi^-)_{\rm min}$ (previously labeled also as $m_{\rm min}(D\pi)$) for different helicity angle regions, as measured by Belle. The $D_2^{*0}$ is clearly seen for $|\cos\Theta_h| > 0.67$ where the $D$-wave component peaks, the $D_0^{*0}$ is visible for $0.33 < |\cos\Theta_h| < 0.67$ where the $D$-wave is suppressed with respect to the $S$-wave, the range $|\cos\Theta_h| < 0.33$ demonstrates an interference pattern.

The $B^- \to D^{*+}\pi^-\pi^-$ decay contains a vector particle in the final state, therefore, assuming a negligible $D^*$ width, there are two more variables needed to specify the final state in addition to $M^2(D^*\pi)_{\rm min}$ and $M^2(D^*\pi)_{\rm max}$. The following ones are chosen in the analysis performed by Belle (Abe, 2004f): the $D^*$ helicity angle ($\alpha$) between the momenta of pions from the $D^*$ and $D^{**}$ decays in the $D^*$ rest frame and the azimuthal angle ($\gamma$) of the pion from the $D^*$ relative to the $\overline{B} \to D^*\pi\pi$ decay plane.

**Table 19.3.5.** From Abe (2004f) (A4), Kuzmin (2007) (K), (Abe, 2005i) (A5) and Aubert (2009g) (A9). The fitted parameters of the $D^{**}$ mesons and products of branching ratios $\mathcal{B}(B \to D^{**}\pi) \times \mathcal{B}(D^{**} \to f)$. The first error is statistical, the second one is systematic and the third one is model related. World average (WA) values are taken from (Eidelman et al., 2004). "fixed" indicates parameters which were fixed to values obtained from other fits.

| Ref. | $D^{**}$ | $f$ | Mass [MeV/$c^2$] | Width [MeV] | $\mathcal{B}(B) \times \mathcal{B}(D^{**})$ [$10^{-4}$] |
|---|---|---|---|---|---|
| A4 | $D_0^{*0}$ | $D^+\pi^-$ | $2308 \pm 17 \pm 15 \pm 28$ | $276 \pm 21 \pm 18 \pm 60$ | $6.1 \pm 0.6 \pm 0.9 \pm 1.6$ |
| A4 | $D_1^0$ | $D^{*+}\pi^-$ | $2421.4 \pm 1.5 \pm 0.4 \pm 0.8$ | $23.7 \pm 2.7 \pm 0.2 \pm 4.0$ | $6.8 \pm 0.7 \pm 1.3 \pm 0.3$ |
| A4 | $D_1^{'0}$ | $D^{*+}\pi^-$ | $2427 \pm 26 \pm 20 \pm 15$ | $384^{+107}_{-75} \pm 24 \pm 70$ | $5.0 \pm 0.4 \pm 1.0 \pm 0.4$ |
| A4 | $D_2^{*0}$ | $D^+\pi^-$ | $2461.6 \pm 2.1 \pm 0.5 \pm 3.3$ | $45.6 \pm 4.4 \pm 6.5 \pm 1.6$ | $3.4 \pm 0.3 \pm 0.6 \pm 0.4$ |
| | | $D^{*+}\pi^-$ | 2461.6(fixed) | 45.6(fixed) | $1.8 \pm 0.3 \pm 0.3 \pm 0.2$ |
| A9 | $D_2^{*0}$ | $D^+\pi^-$ | $2460.4 \pm 1.2 \pm 1.2 \pm 1.9$ | $41.8 \pm 2.5 \pm 2.1 \pm 2.0$ | $3.5 \pm 0.2 \pm 0.2 \pm 0.4$ |
| A9 | $D_0^{*0}$ | $D^+\pi^-$ | $2297 \pm 8 \pm 5 \pm 19$ | $273 \pm 12 \pm 17 \pm 45$ | $6.8 \pm 0.3 \pm 0.4 \pm 2.0$ |
| K | $D_0^{*+}$ | $D^0\pi^+$ | 2308(fixed) | 276(fixed) | $0.6 \pm 0.1 \pm 0.1 \pm 0.2$ |
| K | $D_1^+$ | $D^{*0}\pi^+$ | $2428.2 \pm 2.9 \pm 1.6 \pm 0.6$ | $34.9 \pm 6.6^{+4.1}_{-0.9} \pm 4.1$ | $3.7 \pm 0.6^{+0.7}_{-0.4}{}^{+0.6}_{-0.3}$ |
| K | $D_1^{'+}$ | $D^{*0}\pi^+$ | 2427(fixed) | 384(fixed) | $< 0.7$ @ 90% C.L. |
| K | $D_2^{*+}$ | $D^0\pi^+$ | $2465.7 \pm 1.8 \pm 0.8^{+1.2}_{-4.7}$ | $49.7 \pm 3.8 \pm 4.1 \pm 4.9$ | $2.1 \pm 0.2 \pm 0.3 \pm 0.1$ |
| | | $D^{*0}\pi^+$ | 2465.7(fixed) | 49.7(fixed) | $2.4 \pm 0.4^{+0.3}_{-0.4}{}^{+0.4}_{-0.2}$ |
| A5 | $D_1^0$ | $D^0\pi^+\pi^-$ | $2426 \pm 3 \pm 1$ | $24 \pm 7 \pm 8$ | $1.85 \pm 0.29 \pm 0.35^{+0.0}_{-0.43}$ |
| A5 | | $D^{*0}\pi^+\pi^-$ | 2422.2(fixed to WA) | 18.9(fixed to WA) | $< 0.06$ @ 90% C.L. |
| A5 | $D_2^{*0}$ | $D^{*0}\pi^+\pi^-$ | 2458.9(fixed to WA) | 23(fixed to WA) | $< 0.22$ @ 90% C.L. |
| A5 | $D_1^+$ | $D^+\pi^+\pi^-$ | $2421 \pm 2 \pm 1$ | $21 \pm 5 \pm 8$ | $0.89 \pm 0.15 \pm 0.17^{+0.0}_{-0.27}$ |
| A5 | | $D^{*+}\pi^+\pi^-$ | 2422.2(fixed to WA) | 18.9(fixed to WA) | $< 0.33$ @ 90% C.L. |
| A5 | $D_2^{*+}$ | $D^{*+}\pi^+\pi^-$ | 2459(fixed to WA) | 25(fixed to WA) | $< 0.24$ @ 90% C.L. |

Figure 19.3.6 shows the Dalitz distribution, $M^2(D^*\pi)_{\min}$ vs. $M^2(D^*\pi)_{\max}$, for the $\Delta E$-$m_{\rm ES}$ signal region $B$ candidates. The significant increase of the event density in $M^2(D^*\pi)_{\min}$ at about $5.8\,{\rm GeV}^2/c^4$ corresponds to the narrow $D_1$ and $D_2^*$ states. The broad $D_1'$ meson, unobserved at the time the analysis was performed, can also contribute to the $D^*\pi$. The $B^- \to D^{*+}\pi^-\pi^-$ signal is thus parameterized as a coherent sum of the relativistic Breit-Wigner amplitudes of these three intermediate states.

The HQET predicts that the two $1^+$ mesons, with $j_q = \frac{1}{2}$ and $j_q = \frac{3}{2}$, decay into the $D^*\pi$ final state via $S$- and $D$-wave, respectively. Due to the finite $c$-quark mass, the observed (physical) states can be a mixture of such pure states. The mixing can occur for instance via the common $D^*\pi$ decay channel and the resulting $D'_1$ and $D_1$ amplitudes are superpositions of the $S$- and $D$-wave amplitudes:

$$|D'_1\rangle = |1S\rangle \cos\omega - e^{+i\psi}|1D\rangle \sin\omega$$
$$|D_1\rangle = |1S\rangle \sin\omega + e^{-i\psi}|1D\rangle \cos\omega, \quad (19.3.14)$$

where $\omega$ is a mixing angle and $\psi$ is a complex phase. Such an amplitude representation is used in the signal model. Like in the $B^- \to D^+\pi^-\pi^-$ analysis, virtual $D_v^{*0}$ and $B_v^{*0}$ components, as well as the constant term are also included in the signal function, while the background is estimated with events from the $\Delta E$ sidebands.

Amplitudes and phases of the intermediate states are extracted through an unbinned maximum-likelihood fit in the four-dimensional ($M^2(D^*\pi)_{\min}$, $M^2(D^*\pi)_{\max}$, $\alpha$, $\gamma$) phase space. The broad $1^+$ meson significantly improves the fit likelihood and, thus, Belle claimed its discovery. Figure 19.3.6 shows the background-subtracted $M(D^{*+}\pi^-)_{\min}$ distribution with the resonance contributions obtained from the fit. The fitted parameters of the axial mesons are summarized in Table 19.3.5, along with the branching ratio products. The mixing angle between the two $1^+$ mesons and their relative phase were measured as $\omega = (-0.10 \pm 0.03 \pm 0.02 \pm 0.02)$ rad and $\psi = (0.05 \pm 0.20 \pm 0.04 \pm 0.06)$ rad. Such a measurement is performed for the first time for the charmed mesons. Decomposition of the $S$- and $D$-waves was before attempted for the $D_1(2420) \to D^*\pi$ by CLEO. They set a limit on the $S$-wave contribution to the total width as a function of the relative phase (Avery et al., 1994b; Bergfeld et al., 1994).

For better illustration of the fit results, the measured angular distributions along with MC simulations performed with the fitted parameters, are shown in Fig. 19.3.7 for the $M^2(D^*\pi)_{\rm min}$ regions populated by the $D_1^{\prime 0}$ or $D_1^0$.

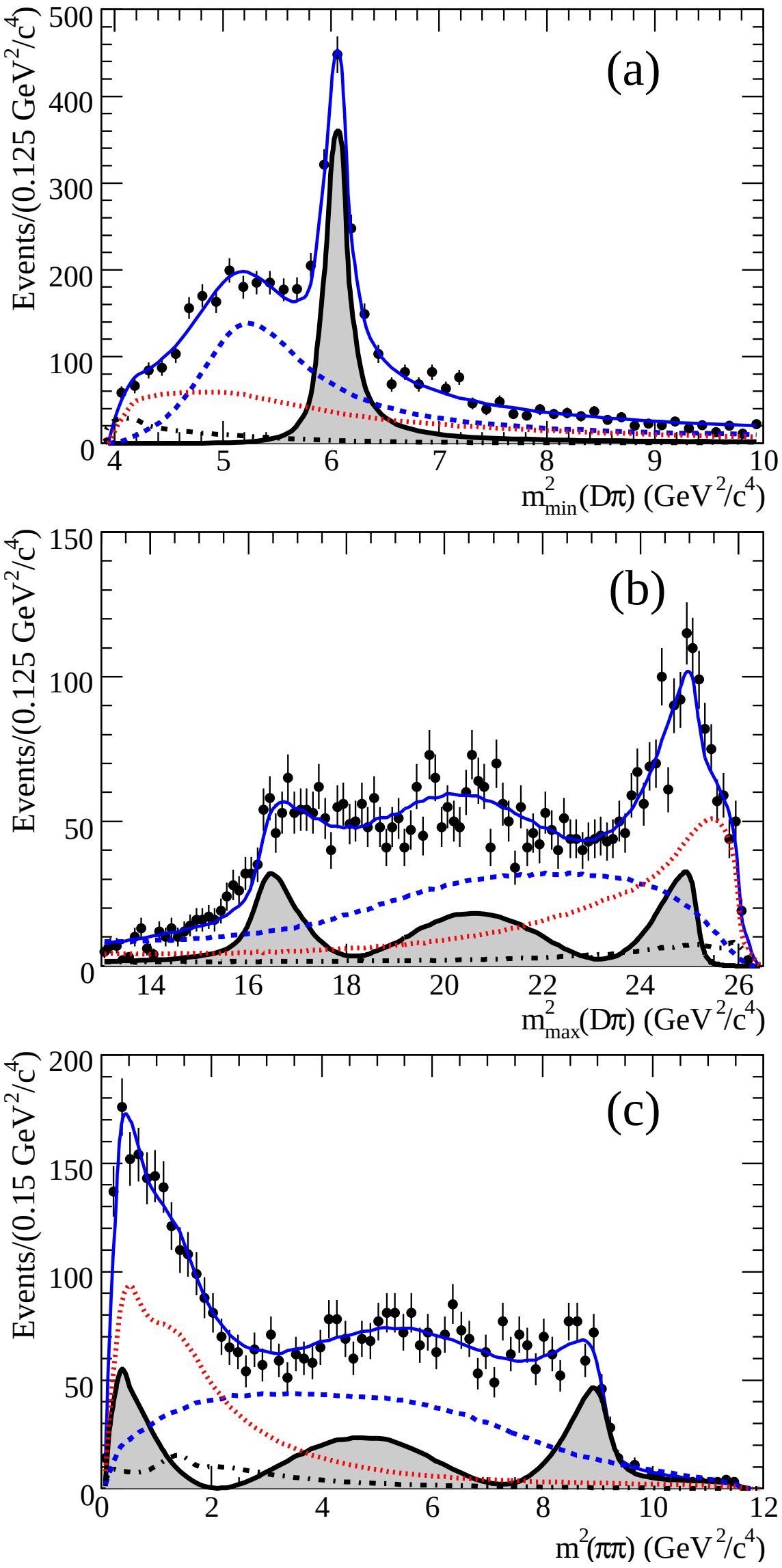

**Figure 19.3.4.** From (Aubert, 2009g). Result of the Dalitz plot fit to the $B^- \to D^+\pi^-\pi^-$ signal candidates: projections on (a) $m^2_{\rm min}(D\pi)$, (b) $m^2_{\rm max}(D\pi)$ and (c) $m^2(\pi\pi)$. The points with error bars are data, the solid curves represent the nominal fit. The shaded areas show the $D_2^{*0}$ contribution, the dashed curves show the $D_0^{*0}$ signal, the dash-dotted curves show the $D_v^*$ and $B_v^*$ signals, and the dotted curves show the background.

### 19.3.3.4 Charged $D^{**}$ via $\overline{B}^0 \to D^{(*)0}\pi^+\pi^-$ Dalitz analysis

The Dalitz analysis of the $\overline{B}^0 \to D^0\pi^+\pi^-$ decay has been performed by Belle (Kuzmin, 2007) using $388 \times 10^6$ $B\overline{B}$ pairs. After excluding a subsample of the $\overline{B}^0 \to D^{*+}\pi^-$ with $D^{*+} \to D^0\pi^+$, Belle obtains a signal yield of about

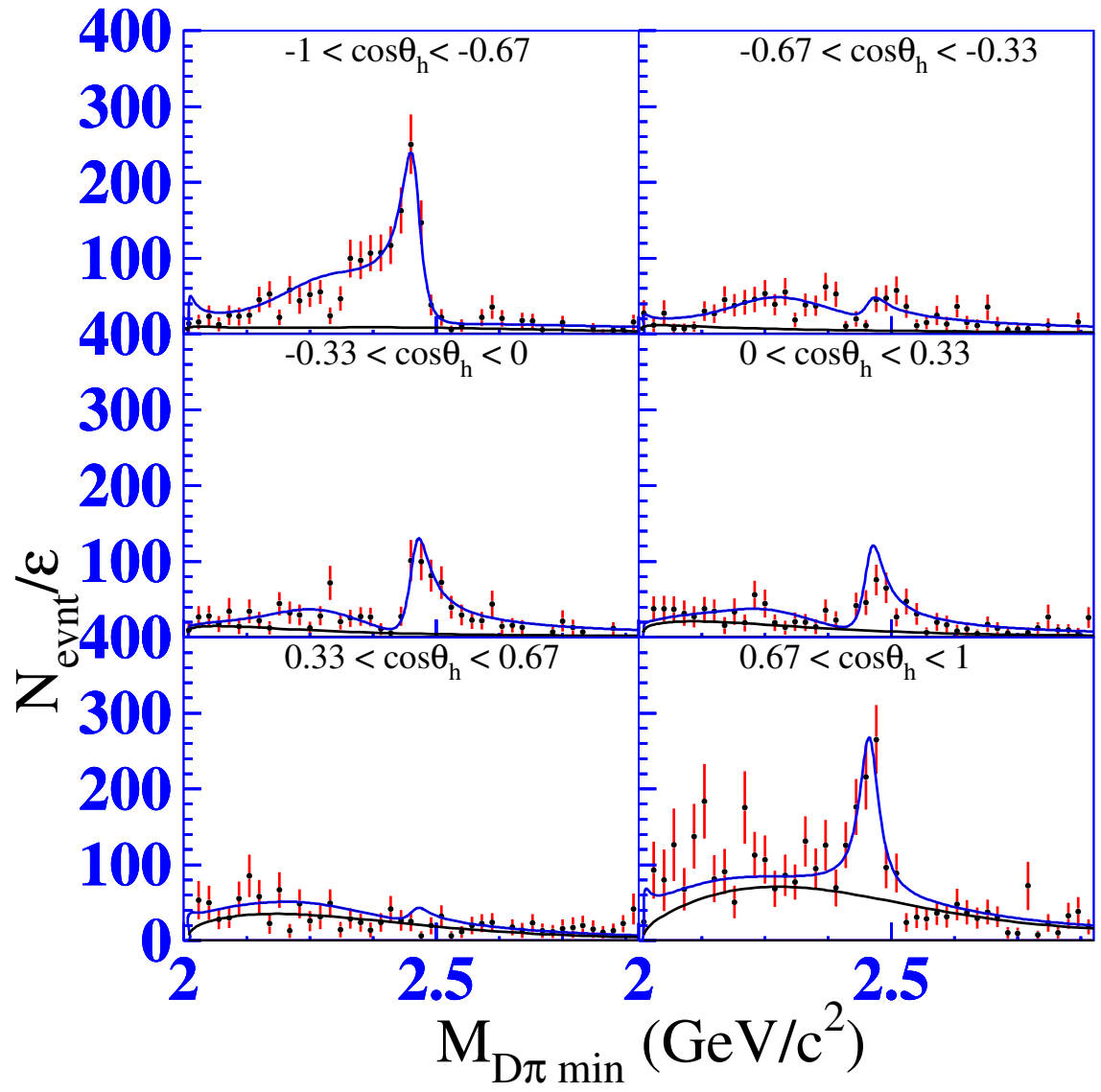

**Figure 19.3.5.** From (Abe, 2004f). Efficiency-corrected $M(D^+\pi^-)_{\rm min}$ distribution in $B^- \to D^+\pi^-\pi^-$ for different helicity angle ranges. Curves correspond to the fit to the total distribution (upper blue) and background (lower black) component.

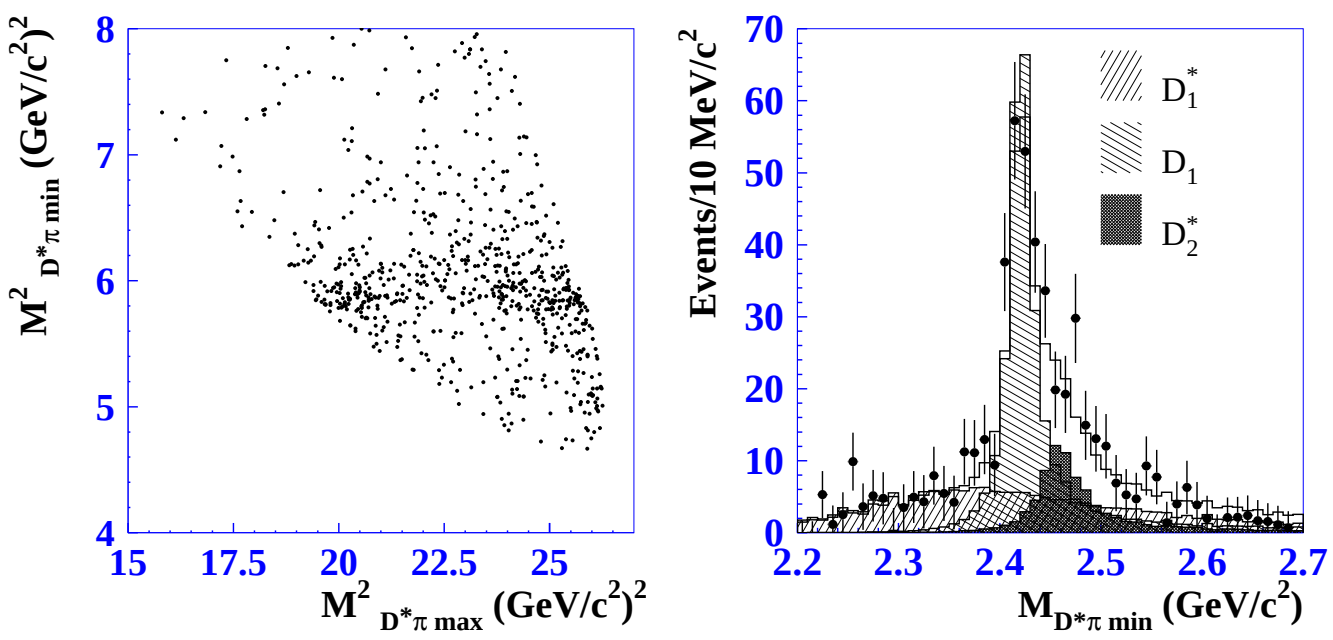

**Figure 19.3.6.** From (Abe, 2004f). Left: Dalitz plot for the $B^- \to D^{*+}\pi^-\pi^-$ signal candidates. Right: Background-subtracted $M(D^*\pi)_{\rm min}$ spectrum. Points with error bars correspond to data, hatched histograms show fitted resonance contributions and the open histogram is a coherent sum of all the contributions. In the figure the $D_1'$ is indicated as $D_1^*$.

2900 events. As intermediate resonances decaying to $D^0\pi^+$ as well as resonances decaying to $\pi^+\pi^-$ can contribute to the $\overline{B}^0 \to D^0\pi^+\pi^-$ reaction, its kinematics is described with the $M^2(D^0\pi^+)$ and $M^2(\pi^+\pi^-)$ invariant masses. The distribution for the reconstructed $\Delta E$-$m_{\rm ES}$ signal region events is shown in Fig. 19.3.8. The *p.d.f.* is comprised of $D_0^{*+}$, $D_2^{*+}$, $D_v^*$ and $B_v^{*+}$ components contributing to the $D^0\pi^+$ system, whereas $\rho(770)$, $\omega$, $f_2(1270)$, $f_0(600)$, $f_0(980)$ and $f_0(1370)$ states can be present in the $\pi^+\pi^-$ projection. Parameters of the $D_0^{*+}$ are fixed to the one from the neutral $D_0^{*0} \to D^-\pi^+$ measurement, while light scalar mesons are included in the fit with their parameters fixed to the PDG values (Beringer et al., 2012). An unbinned fit to the Dalitz distribution gives signifi-cant contributions from both the charged $D_0^{*+}$, observed for the first time, as well as the $D_2^{*+}$. Fig. 19.3.8 shows the $M(D^0\pi^+)$ spectrum with the fitted function superimposed, for the $\cos\Theta_h > 0$ helicity angle region, where the $\pi\pi$ resonance contributions and background are low. Fitted $D^{**+}$ parameters and measured branching ratio products are presented in Table 19.3.5. It can be seen that the production branching fraction for the $D_0^{*+}$ is much smaller than for the $D_2^{*+}$.

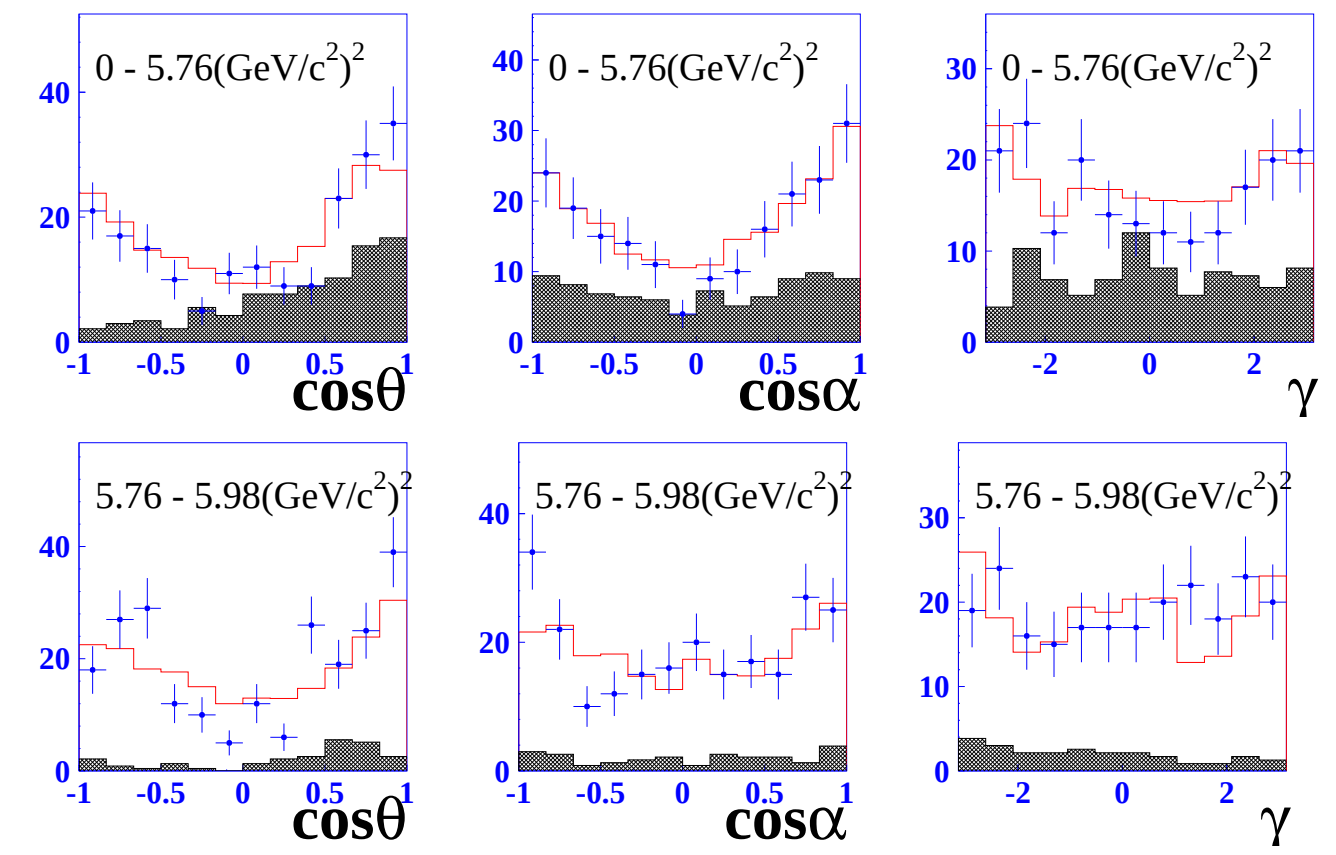

**Figure 19.3.7.** From (Abe, 2004f): Distributions of the $D^{**}$ helicity angle ($\cos\theta$), $\cos\alpha$ and $\gamma$ in regions of $M^2(D^*\pi)_{\min}$ with dominance of the $D_1'^0$ in $B^- \to D^{*+}\pi^-\pi^-$ decays: $M^2(D^*\pi)_{\min} < 5.76\,\mathrm{GeV}^2/c^4$ and the $D_1^0$: $5.76 < M^2(D^*\pi)_{\min} < 5.98\,\mathrm{GeV}^2/c^4$. Points with error bars correspond to data, the histogram is from MC simulation based on the fitted parameters and the hatched histogram shows the background contribution.

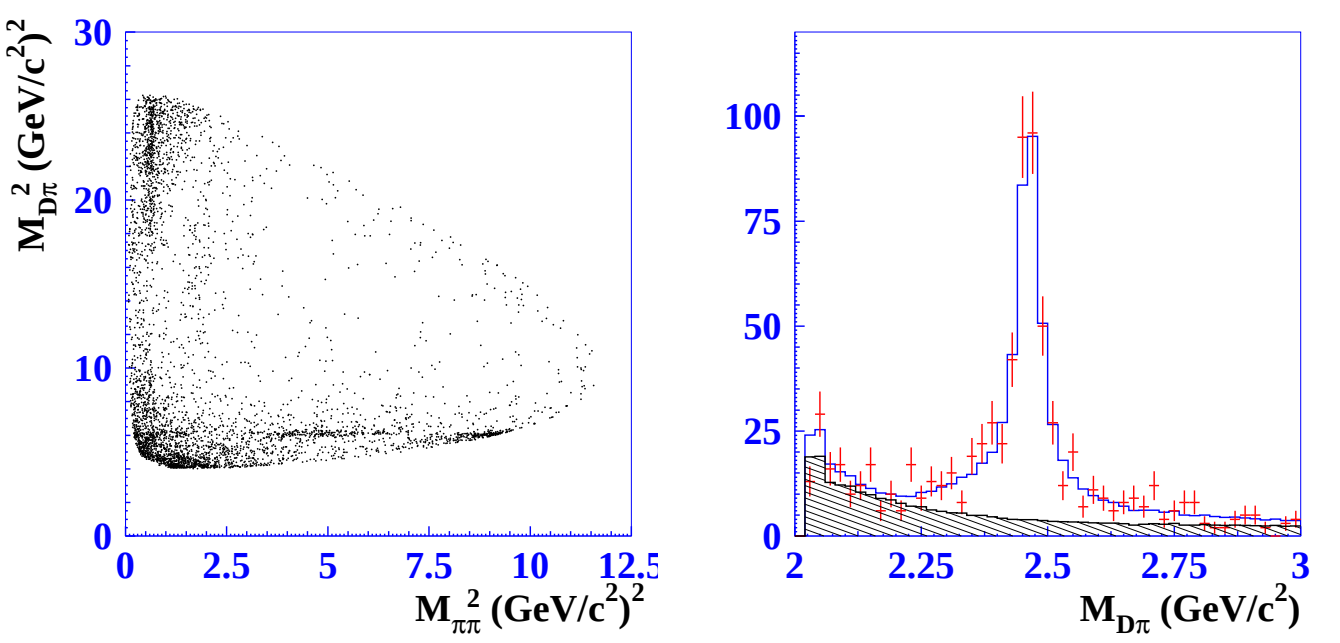

**Figure 19.3.8.** From (Kuzmin, 2007): Dalitz plot for the $\overline{B}^0 \to D^0\pi^+\pi^-$ signal candidates (left) and $M(D^0\pi^+)$ projection (right) of the events with $\cos\Theta_h > 0$. Points with error bars represent data, the hatched histogram shows background estimated from generic MC events normalized to the sideband data and the open histogram represents the fitted function.

### 19.3.3.5 The $D^{**}$ production rates

Measurements of the charmed meson production rates in $B$ decays provide tests of HQET and QCD sum rules. The measured branching ratio products (Table 19.3.5) show that the narrow mesons comprise $(36\pm 6)\%$ of the $B^- \to D^+\pi^-\pi^-$ and $(63\pm 6)\%$ of the $B^- \to D^{*+}\pi^-\pi^-$ decays, thus the production rates of the $j_q = \frac{1}{2}$ and $j_q = \frac{3}{2}$ $c\bar{u}$ mesons are similar. This is inconsistent with the QCD sum rule which predicts the dominance of the narrow, $j_q = \frac{3}{2}$ states. However, if the color-suppressed amplitude (left diagram in Fig. 19.3.2) contributes significantly, it would be enhanced for the $j_q = \frac{1}{2}$ states. This seems to be supported by the Belle measurements of the color-allowed $\overline{B}^0 \to D^{(*)0}\pi^+\pi^-$ decays, where production rates of the $j_q = \frac{3}{2}$ states are similar to the ones measured in charged $B$ decays, but are much lower for the broad $j_q = \frac{1}{2}$ states. The measured production rates (Table 19.3.5) give:

$$\frac{\mathcal{B}(B^-\to D_2^{*0}\pi^-)\mathcal{B}(D_2^{*0}\to D^+\pi^-)}{\mathcal{B}(B^-\to D_2^{*0}\pi^-)\mathcal{B}(D_2^{*0}\to D^{*+}\pi^-)} = \frac{\mathcal{B}(D_2^{*0}\to D^+\pi^-)}{\mathcal{B}(D_2^{*0}\to D^{*+}\pi^-)} = 1.9\pm 0.5, \quad (19.3.15)$$

which is consistent with a value predicted by theoretical models (Rosner, 1986; Godfrey and Kokoski, 1991; Falk and Peskin, 1994). Assuming that $D_2^*$ decay is saturated by the $D^{(*)}\pi$ transitions, whereas the $D_1$ decay is saturated by the $D^*\pi$ mode, one gets:

$$\frac{\mathcal{B}(B^-\to D_2^{*0}\pi^-)\mathcal{B}(D_2^{*0}\to D^+\pi^-, D^{*+}\pi^-)}{\mathcal{B}(B^-\to D_1^0\pi^-)\mathcal{B}(D_1^0\to D^{*+}\pi^-)} = 0.77\pm 0.15, \quad (19.3.16)$$

which is by a factor of two larger than the HQET prediction calculated in the factorization approximation (Leibovich, Ligeti, Stewart, and Wise, 1998; Neubert, 1998). From these measurements it is impossible to determine the size of the nonfactorized part for the tensor and axial mesons or whether higher order corrections to the leading factorized terms should be taken into account. More accurate measurements of the semileptonic $\overline{B} \to D^{**}l\nu$ decays (Section 17.1), which are free of nonfactorized contributions, may help to resolve this problem.

### 19.3.3.6 Other $D^{**}$ decays

Studies of subleading decay modes of the $D^{**}$ mesons are important for understanding heavy-light mesons and to further test theoretical models. Subleading decays could modify the ratio in Eq. (19.3.16). Belle observed the $D_1 \to D\pi^+\pi^-$ decays in $\overline{B} \to (D^{(*)}\pi^+\pi^-)\pi^-$, where $D = D^0$ or $D^+$, in a sample of $152\times 10^6$ $B\overline{B}$ pairs (Abe, 2005i). To suppress the large continuum background, the analysis uses a Fisher discriminant (see Section 9), that is based on the $B$ production angle, the thrust angle, as well as parameters characterizing the momentum flow in the event, originally developed by CLEO (Asner et al., 1996). The $M(D\pi^+\pi^-)$ spectra for the $B$ candidates in the $\Delta E$-$m_{\rm ES}$ signal region, shown in Fig. 19.3.9, demonstrate the prominent $D_1$ signals. No significant signals are observed for neither $D_1 \to D^*\pi^+\pi^-$ nor $D_2^* \to D^{(*)}\pi^+\pi^-$. Except for the $D_1$ peak, the signal-region data are consistent with the mass distributions for the $\Delta E$ sidebands. This suggests that there is no significant contribution from the broad $D^{**}$ mesons. The results of the fits to the $M(D^{(*)}\pi^+\pi^-)$ distributions and the measured $\mathcal{B}$ products are summarized in Table 19.3.5. The observed $D_1 \to D\pi^+\pi^-$ decays lower the ratio in Eq. (19.3.16) to $0.54 \pm 0.18$ which, thus, becomes consistent with the HQET predictions (see Section 19.3.3.5).

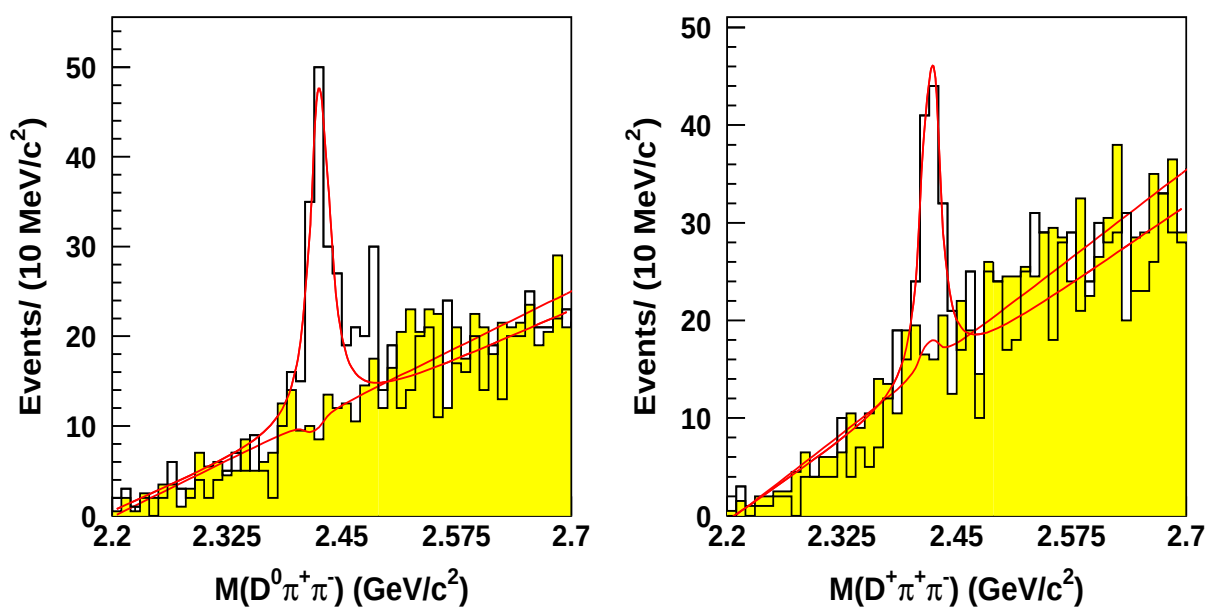

**Figure 19.3.9.** From (Abe, 2005i): $M(D\pi^+\pi^-)$ distributions for the $\overline{B} \to (D\pi^+\pi^-)\pi^-$ candidates in the signal region (open histogram) and $\Delta E$ sidebands (hatched-yellow).

The dynamics of the $D_1 \to D\pi^+\pi^-$ decays are examined in a simplified way by studying one-dimensional projections of mass and angular variables. The data, compared with MC simulations of various $D_1$ decays models, are found to be well described by the $D_1 \to D_0^*\pi^-$ decay.

### 19.3.3.7 New excited charmed mesons

To search for new excited charmed mesons, labeled as $D_J$, *BABAR* analyzed the inclusive production of the $D^+\pi^-$, $D^0\pi^+$, and $D^{*+}\pi^-$ final states in the reaction $e^+e^- \to c\bar{c} \to D^{(*)}\pi X$, where $X$ is any additional system (del Amo Sanchez, 2010i). They use a data sample consisting of approximately $590 \times 10^6$ $c\bar{c}$ events. In the $D\pi$ system, the $D^+ \to K^-\pi^+\pi^+$ and $D^0 \to K^-\pi^+$ decays are reconstructed. The $D^0$ candidates, when being combined with any additional pion in the event form a $D^*$, are rejected. To improve the signal purity for $D^0 \to K^-\pi^+$, it is required that $\cos\theta_K > -0.9$, where $\theta_K$ is the angle between the $K^-$ direction and the direction opposite to the $e^+e^-$ CM system in the $D^0$ rest frame. The $D^{*+}\pi^-$ system is reconstructed using the $D^{*+} \to D^0\pi^+$, $D^0 \to K^-\pi^+$ and $K^-\pi^+\pi^-\pi^+$ decay modes. Background from $e^+e^- \to B\overline{B}$ events and much of the combinatorial background, are removed by requiring the CM momentum of the $D^{(*)}\pi$ to be greater than 3.0 GeV/$c$.

The measured $D^+\pi^-$ and $D^0\pi^+$ mass spectra are presented in Fig. 19.3.10 and show similar features:

- Prominent $D_2^*$ peaks.
- The $M(D^+\pi^-)$ shows a peaking background at about 2.3 GeV/$c^2$ due to $D_1^0$ and $D_2^{*0}$ decays to $D^{*+}\pi^-$, with the $\pi^0$ from the $D^{*+} \to D^+\pi^0$ missing. Similarly, $M(D^0\pi^+)$ shows feeddown due to the $D_1^+$ and $D_2^{*+}$ decaying to $D^{*0}\pi^+$ where $D^{*0} \to D^0\pi^0$.
- Both $D^+\pi^-$ and $D^0\pi^+$ mass distributions show new structures around 2.60 and 2.76 GeV/$c^2$, labeled respectively as $D^*(2600)$ and $D^*(2760)$.

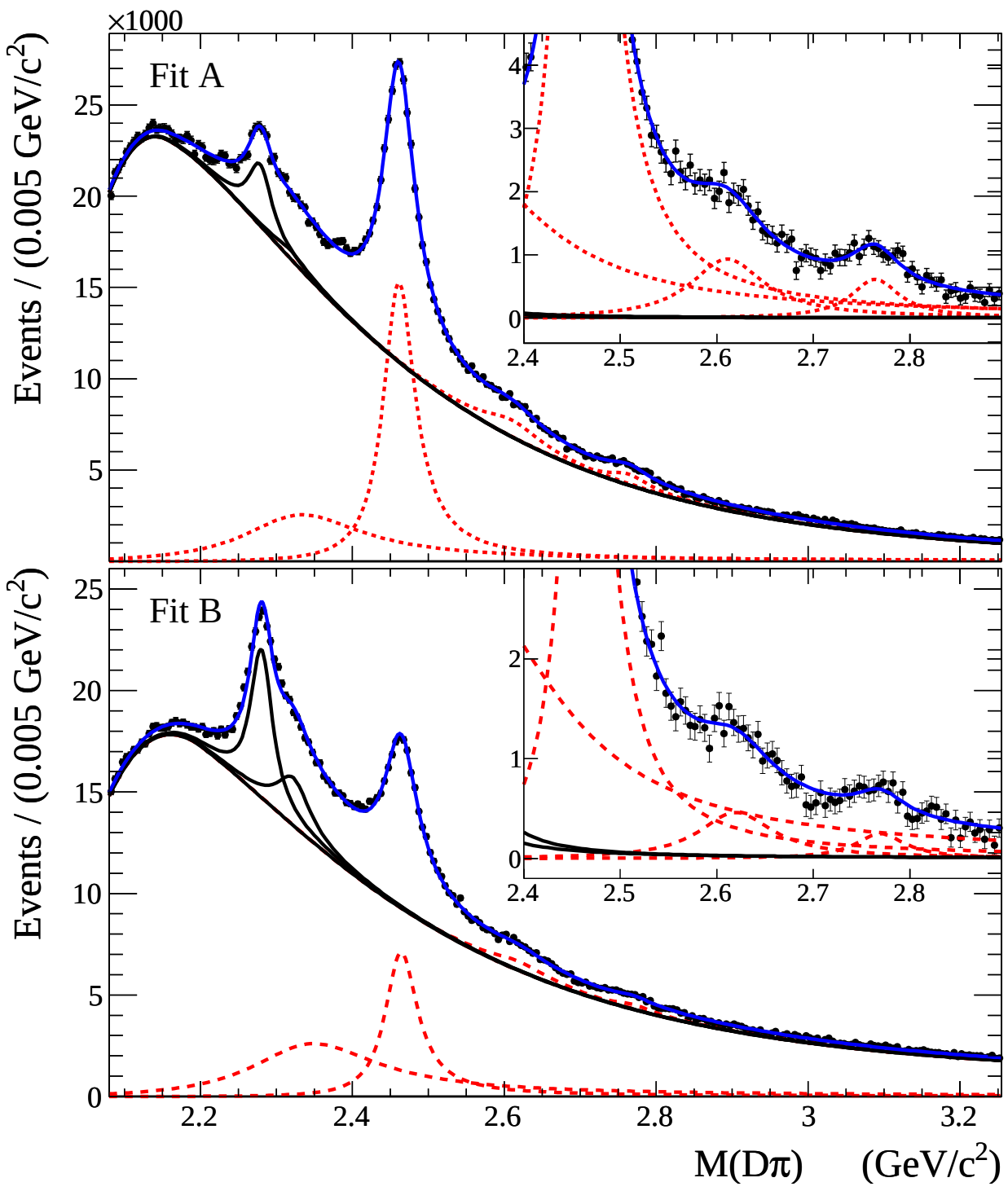

**Figure 19.3.10.** From (del Amo Sanchez, 2010i). Mass distribution for $D^+\pi^-$ (top) and $D^0\pi^+$ (bottom) candidates produced in the process $e^+e^- \to c\bar{c} \to D\pi X$, Points correspond to data, with the total fit overlaid as a solid curve. The dotted curves are the signal components. The lower solid curves correspond to the smooth combinatoric background and to the peaking backgrounds at 2.3 GeV/$c^2$. The inset plots show the distributions after subtraction of the combinatoric background.

The $M(D\pi)$ spectra are fitted with contributions from the $D_2^*$, $D^*(2600)$ and $D^*(2760)$ described with relativistic BW distributions. The smooth background is modeled using an exponential function multiplied by a two-body phase-space factor dropping toward the $D\pi$ mass threshold. The feeddown is described by convolving BW functions with a function describing the resolution and mass shift obtained from the MC simulation. The masses and widths of the $D_1$ and $D_2^*$ feeddowns are fixed to the values obtained respectively from the same $M(D\pi)$ distribution and from the $M(D^{*+}\pi^-)$ study described below. Finally, although not visible in the $M(D^+\pi^-)$ mass distribution, a BW function is included to account for the broad $D_0^*$.

The $D^{*+}\pi^-$ mass distribution is shown in Fig. 19.3.11 and exhibits the following features:

- Prominent $D_1^0$ and $D_2^{*0}$ peaks.
- Two enhancements at 2.60 GeV/$c^2$ and 2.75 GeV/$c^2$, which are denoted as $D^*(2600)^0$ and $D(2750)^0$.

The angular analysis of the $M(D^{*+}\pi^-) \approx 2.6\,\mathrm{GeV}/c^2$ region shows that it could not be described by a single resonance, instead two resonances with different helicity-angle distributions could be present. Thus, a new component, labeled as $D(2550)^0$, is included in the $M(D^{*+}\pi^-)$ fit. The $D(2550)^0$ parameters are obtained by requiring $|\cos\theta_H| > 0.75$ in order to suppress the other resonances (Fig. 19.3.11, top), where the helicity angle ($\theta_H$) is defined in the rest frame of the $D^*$ as the angle between the primary pion and the slow pion from the $D^*$ decay. In this fit, the parameters of the $D_2^{*0}$ and $D^*(2600)^0$ are fixed to those measured in the $D^+\pi^-$. This fit also determined the parameters of the $D_1^0$. A complementary fit with $|\cos\theta_H| < 0.5$, shown in Fig. 19.3.11 (middle), is performed to discriminate in favor of the $D^*(2600)^0$. To determine the final parameters of the $D(2750)^0$ signal the total $D^{*+}\pi^-$ sample is refitted (Fig. 19.3.11 (bottom)), while fixing the parameters of all other BW components to the values determined in the previous fits. The broad resonance $D_1'^0$ is known to decay to this final state, however, these fits were insensitive to its contribution due to its large width and because the background parameters are free. The fit results are summarized in Table 19.3.6.

**Table 19.3.6.** From (del Amo Sanchez, 2010i). Summary of the measurements of the old and newly discovered $D_J$ resonances. The first error is statistical and the second is systematic; "fixed" indicates parameters which were fixed to values obtained from other fits. The significance is defined as the yield divided by its total error.

| Resonance | Channel | Mass (MeV/$c^2$) | Width (MeV) | Significance |
|---|---|---|---|---|
| $D_1(2420)^0$ | $D^{*+}\pi^-$ | 2420.1±0.1±0.8 | 31.4±0.5±1.3 | |
| $D_2^*(2460)^0$ | $D^+\pi^-$ | 2462.2±0.1±0.8 | 50.5±0.6±0.7 | |
| $D(2550)^0$ | $D^{*+}\pi^-$ | 2539.4±4.5±6.8 | 130±12±13 | 3.0$\sigma$ |
| $D^*(2600)^0$ | $D^+\pi^-$ | 2608.7±2.4±2.5 | 93±6±13 | 3.9$\sigma$ |
| $D(2750)^0$ | $D^{*+}\pi^-$ | 2752.4±1.7±2.7 | 71±6±11 | 4.2$\sigma$ |
| $D^*(2760)^0$ | $D^+\pi^-$ | 2763.3±2.3±2.3 | 60.9±5.1±3.6 | 8.9$\sigma$ |
| $D_2^*(2460)^+$ | $D^0\pi^+$ | 2465.4±0.2±1.1 | 50.5 (fixed) | |
| $D^*(2600)^+$ | $D^0\pi^+$ | 2621.3±3.7±4.2 | 93 (fixed) | 2.8$\sigma$ |
| $D^*(2760)^+$ | $D^0\pi^+$ | 2769.7±3.8±1.5 | 60.9 (fixed) | 3.5$\sigma$ |

The $D^*(2760)^0$ signal observed in its decay to $D^+\pi^-$ is very close in mass to the $D(2750)^0$ signal observed in $D^{*+}\pi^-$.

To have information on the spin of the observed resonances, the data are divided into 10 sub-samples corresponding to $\cos\theta_H$ intervals of 0.2. Each sample is fitted with all shape parameters fixed to the values determined from the fits to the total samples. The yields extracted from these fits are plotted for each resonance in Fig. 19.3.12. The $\cos\theta_H$ distributions of the $D_2^*$ and $D^*(2600)$ are consistent with the expectations for natural parity, defined by $P = (-1)^J$, and leading to a $\sin^2\theta_H$-like distribution. This observation supports the assumption that the enhancement assigned to the $D^*(2600)$ observed in the $D^+\pi^-$ and $D^{*+}\pi^-$ mass spectra belong to the same state, as only states with natural parity can decay to both $D^+\pi^-$ and $D^{*+}\pi^-$. The $\cos\theta_H$ distribution for the $D(2550)^0$ is consistent with pure $\cos^2\theta_H$ as expected for a $J^P = 0^-$ state.

The branching fraction ratios, $\frac{\mathcal{B}(D_J\to D^+\pi^-)}{\mathcal{B}(D_J\to D^{*+}\pi^-)}$, may be used in the identification of the new states. Such ratios, computed using the yields obtained from the fits to the total samples and corrected for the reconstruction efficiency, are measured to be:

$$\frac{\mathcal{B}(D_2^*(2460)^0 \to D^+\pi^-)}{\mathcal{B}(D_2^*(2460)^0 \to D^{*+}\pi^-)} = 1.47 \pm 0.03 \pm 0.16,$$
$$\frac{\mathcal{B}(D^*(2600)^0 \to D^+\pi^-)}{\mathcal{B}(D^*(2600)^0 \to D^{*+}\pi^-)} = 0.32 \pm 0.02 \pm 0.09,$$
$$\frac{\mathcal{B}(D^*(2760)^0 \to D^+\pi^-)}{\mathcal{B}(D(2750)^0 \to D^{*+}\pi^-)} = 0.42 \pm 0.05 \pm 0.11. \quad (19.3.17)$$

The $D(2550)^0$ and the $D^*(2600)^0$ have mass values and $\cos\theta_H$ distributions that are consistent with the predicted radial excitations $D_0^1(2S)$ and $D_1^3(2S)$. The $D^*(2760)^0$ and the $D(2750)^0$ (assuming these are two different states) could be some of the four $L = 2$ states, predicted to lie in this mass region.

### 19.3.4 Charmed-strange mesons

#### 19.3.4.1 Introduction

The unexpected discovery of a narrow state, $D_{s0}^*(2317)^+ \to D_s^+\pi^0$, by the *BABAR* experiment (Aubert, 2003j), and a subsequent discovery of yet another narrow particle, $D_{s1}(2460)^+ \to D_s^*(2112)^+\pi^0$ by CLEO (Besson et al., 2003), Belle (Abe, 2004a) and *BABAR* (Aubert, 2004t) raised considerable interest in the spectroscopy of charmed mesons. These discoveries were a surprise because they contradicted the expectations of HQET which, till then, had been a very successful approach in describing the spectroscopy of $D_{(s)}$ and $B_{(s)}$ mesons. The expected spectrum for the $c\bar{s}$ states is sketched in Fig. 19.3.13. In this scheme, $L = 1$ $c\bar{s}$ excitations consist of a $J^P = (0^+, 1^+)$ doublet carrying the total light-quark angular momentum $j_q = \frac{1}{2}$ and a $(1^+, 2^+)$ doublet having $j_q = \frac{3}{2}$. Figure 19.3.13 shows also a comparison between HQET calculations and experimental measurements of $D_s$ meson masses. The two $j_q = \frac{3}{2}$ states are expected to be narrow and are identified as the $1^+$ $D_{s1}(2536)$ and the $2^+$ $D_{s2}^*(2573)$ seen in $D^*K$ and $DK$ decays, respectively. These states were first observed in the 1980-90s (Albrecht et al., 1989c; Alexander et al., 1993; Kubota et al., 1994).

Before the observations reported here became available, the two $j_q = \frac{1}{2}$ states were predicted to have masses above the $DK$ and $D^*K$ thresholds, respectively, and rather large widths of order of a few 100 MeV/$c^2$. The latter was blamed for making their detection hard.

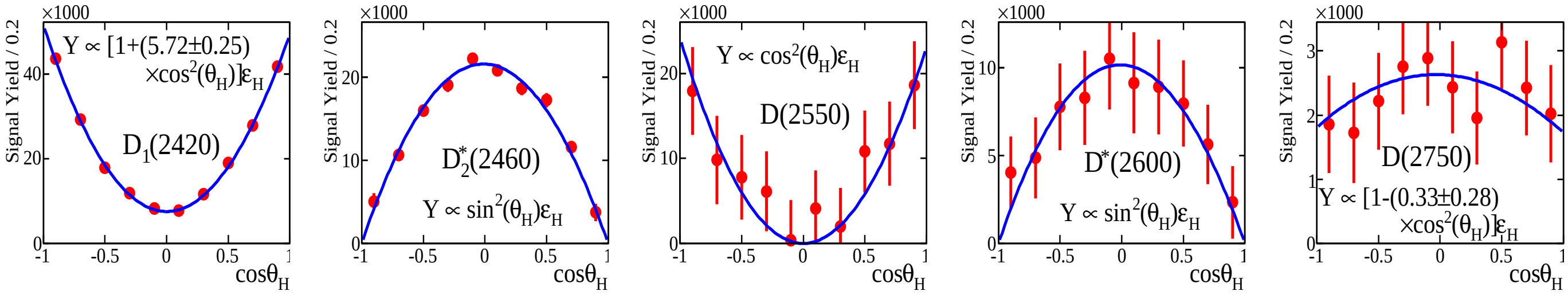

**Figure 19.3.12.** From (del Amo Sanchez, 2010i). Distribution in $\cos\theta_H$ for each signal observed in the $D^{*+}\pi^-$ final state. The error bars include statistical and correlated systematic uncertainties. The curve is a fit using the function Y shown in the plot; $\varepsilon_H$ is the efficiency as a function of $\cos\theta_H$.

As it can be seen from Fig. 19.3.13, the newly discovered states at 2.32 and 2.46 GeV/$c^2$ do not fit these predictions. Their masses are considerably lower and are below their relative $D^{(*)}K$ thresholds. The observed decay modes violates isospin conservation and is therefore likely to be of electromagnetic origin. This makes their widths extremely narrow (consistent with zero). Two possibilities then arise: the calculations are flawed, or these states have a parton composition other than the assumed conventional heavy-light $c\bar{s}$ quark-antiquark pair.

### 19.3.4.2 Discovery of $D^*_{s0}(2317)^+$ and $D_{s1}(2460)^+$

*BABAR* first investigated the $D_s^+\pi^0$ mass spectrum inclusively produced from $e^+e^-$ interactions, with 92 fb$^{-1}$ of data used (Aubert, 2003j). The $D_s^+$ is reconstructed in the $K^+K^-\pi^+$ decay mode, with $K^+K^-$ mass near the $\phi(1020)$ mass or $K^-\pi^+$ mass near the $\overline{K}^*(892)$ mass. The decay products of the $\phi(1020)$ and $\overline{K}^*(892)$ vector states exhibit a $\cos^2\theta_h$ helicity angle behavior and the signal-to-background ratio is improved by requiring $|\cos\theta_h| > 0.5$. Candidates for $\pi^0 \to \gamma\gamma$ are reconstructed using photons which do not belong to another acceptable $\pi^0$ candidate ("$\pi^0$ veto"). Only $D_s^+\pi^0$ candidates with their CM momenta $p^* > 3.5$ GeV/$c$ are retained to eliminate background from $B$ decays and reduce combinatorial background.

The resulting $D_s^+\pi^0$ mass spectrum, shown in Fig. 19.3.14a, exhibits a clear, narrow signal at a mass near 2.32 GeV/$c^2$, labeled as $D^*_{s0}(2317)^+$. No such signal is observed in the $M(D_s^+\pi^0)$ distribution obtained using candidates from either the $D_s^+$ or the $\pi^0$ mass sidebands. The same analysis procedure is applied to MC simulations of $e^+e^- \to c\bar{c}$ events which include all known charm states and decays, and yields no 2.32 GeV/$c^2$ peak. This proves that the $D^*_{s0}(2317)^+$ is not due to reflection from other charmed states.

The $M(D_s^+\pi^0)$ spectrum in Fig. 19.3.14a is fitted using a Gaussian function describing the $D^*_{s0}(2317)^+$ signal and a polynomial background, and yields about 1300 signal events with a mass of $(2316.8 \pm 0.4)$ MeV/$c^2$ and an experimental resolution of $(8.6 \pm 0.4)$ MeV/$c^2$. A clear $D^*_{s0}(2317)^+ \to D_s^+\pi^0$ signal is also observed for $D_s^+ \to K^+K^-\pi^+\pi^0$ (see Fig. 19.3.14b).

Using 13.5 fb$^{-1}$ of data, the $D^*_{s0}(2317)^+$ was readily confirmed by CLEO (Besson et al., 2003); in the same analysis they also claimed the discovery of the $D_{s1}(2460)^+ \to D_s^{*+}\pi^0$. In the original paper (Aubert, 2003j), *BABAR* also reports a narrow signal of 2.46 GeV/$c^2$ in the $M(D_s^+\pi^0\gamma)$ spectrum, with most of the peak events having $D_s^+\gamma$ masses consistent with the $D_s^{*+}$. However, MC simulations showed a complex kinematic superposition of signals and reflections between $D_{s1}(2460)^+$ and $D^*_{s0}(2317)^+$ (see below), and for this reason *BABAR* did not claim immediately the 2.46 GeV/$c^2$ structure being a new resonance.

A few months later, Belle confirmed both $D^*_{s0}(2317)^+$ and $D_{s1}(2460)^+$ using 70 fb$^{-1}$ of data sample (Abe, 2004a). The $D_{sJ}$ signals are studied in the $\Delta M(D_s^{(*)+}\pi^0) \equiv M(D_s^{(*)+}\pi^0) - M(D_s^{(*)+})$ mass difference spectra (shown in Fig. 19.3.15) as they have a better resolution than the invariant masses, while the secondary particles are reconstructed as $D_s^{*+} \to D_s^+\gamma$ and $D_s^+ \to \phi\pi^+$. Like in the *BABAR* study, MC simulation reveals that $D_{s1}(2460)^+ \to D_s^{*+}\pi^0$ decays produce a reflection in the $\Delta M(D_s^+\pi^0)$ distribution slightly below the $D^*_{s0}(2317)^+$ signal. On the other hand, the $D^*_{s0}(2317)^+ \to D_s^+\pi^0$ combined with a random photon passing the $D_s^{*+}$ selection, causes a peaking background to the $D_{s1}(2460)^+$. Another feed-down source is the $D_{s1}(2460)^+$ producing a wide structure at its nominal mass, which is due to a random $\gamma$ in the $D_s^{*+}$ reconstruction. Both these feed-downs are visible in the $\Delta M(D_s^{*+}\pi^0)$ distributions for the $D_s^{*+}$ sidebands in the right upper plot of Fig. 19.3.15.

In the $D^*_{s0}(2317)^+$ fit (left bottom plot in Fig. 19.3.15), both the signal and the feed-down are represented as Gaussian shapes, parameters of the latter are fixed according to MC expectation and normalized by the measured $D_{s1}(2460)^+$ signal. A fit to the $D_{s1}(2460)^+$ mass distribution is performed for the data with the $D_s^{*+}$ sideband subtracted bin-by-bin (right bottom plot in Fig. 19.3.15). Masses corresponding to the fitted peak positions and the upper limits set on the meson widths are summarized in Table 19.3.7.

*BABAR* developed a different method for solving an overlap between $D_{s1}(2460)^+$ and $D^*_{s0}(2317)^+$ (Aubert, 2004t). Considering the final state $D_s^+\pi^0\gamma$, the

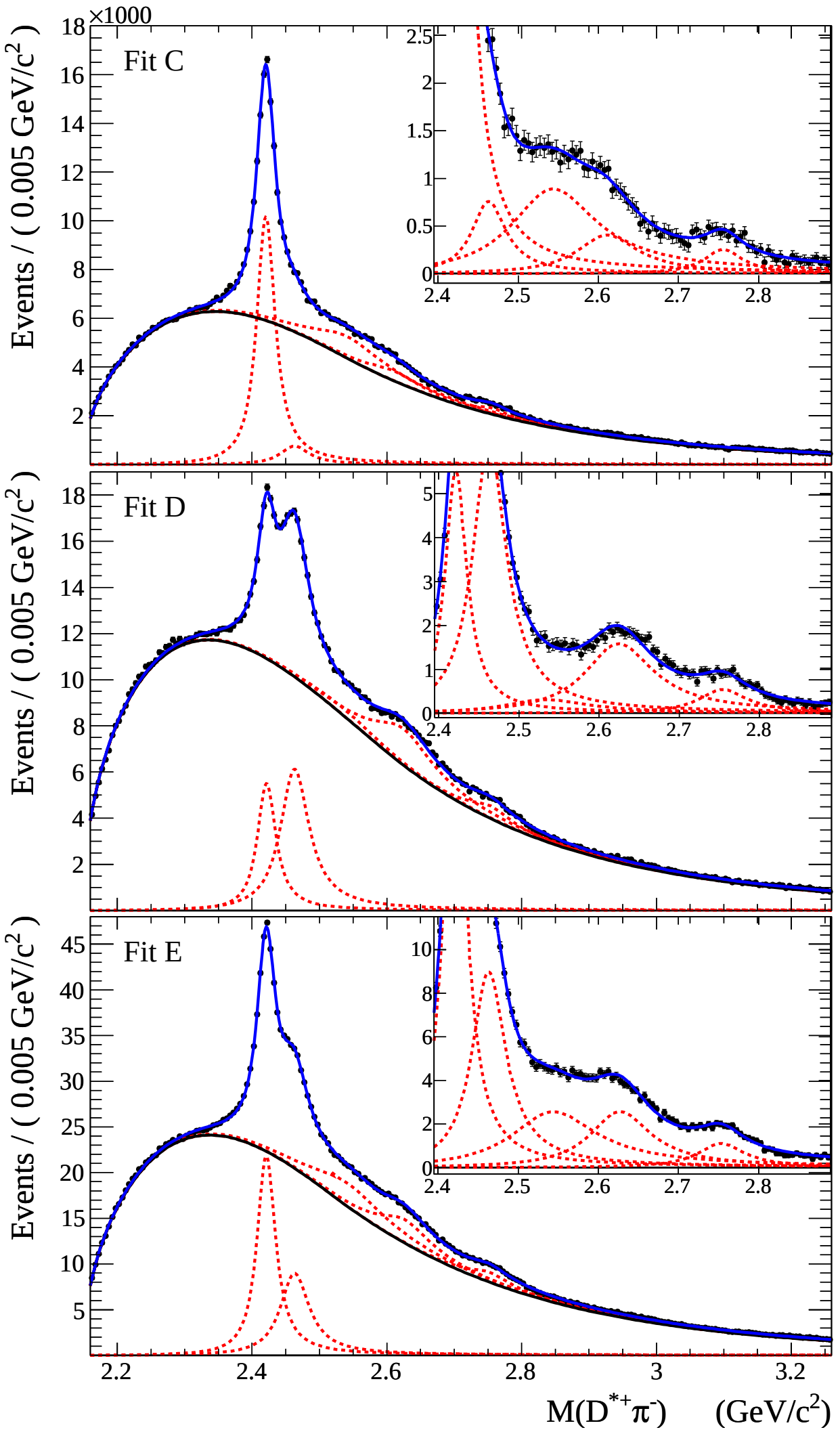

**Figure 19.3.11.** From (del Amo Sanchez, 2010i). Mass distributions for $D^{*+}\pi^-$ candidates produced in the process $e^+e^- \to c\bar{c} \to D^*\pi X$. Top: candidates with $|\cos\theta_H| > 0.75$. Middle: candidates with $|\cos\theta_H| < 0.5$. Bottom: all candidates. Points correspond to data, with the total fit overlaid as a solid curve. The lower solid curve is the combinatoric background, and the dotted curves are the signal components. The inset plots show the distributions after subtraction of the combinatoric background.

$D_{s1}(2460)^+$ may decay through either $D_s^{*+}\pi^0$ or $D_{s0}^*(2317)^+\gamma$. To disentangle these modes and reliably extract the parameters of the signal, *BABAR* applies an unbinned maximum likelihood fit simultaneously to the $M(D_s^+\pi^0\gamma)$, $M(D_s^+\pi^0)$ and $M(D_s^+\gamma)$ spectra of all the $D_s^+\pi^0\gamma$ combinations, using the channel likelihood method (Condon and Cowell, 1974). This fit describes the probability density function of the two $D_{s1}(2460)^+$ decay channels as the product of a Gaussian shape in the $M(D_s^+\pi^0\gamma)$ distribution and a Gaussian shape projected into the $M(D_s^+\pi^0)$ or $M(D_s^+\gamma)$ axes, as appropriate. Background sources included in the fit are: purely combinatorial background, $D_s^{*+} \to D_s^+\gamma$ decay combined with an unassociated $\pi^0$, $D_{s0}^*(2317)^+ \to D_s^+\pi^0$ decay combined with a random $\gamma$, and a contribution from $D_{s1}(2460)^+ \to D_s^{*+}\pi^0$ decay with a random $\gamma$ in $D_s^{*+}$ decay. The $D_{s1}(2460)^+$ signal for a particular decay mode is extracted by calculating for each $D_s^+\pi^0\gamma$ combination a weight proportional to the relative likelihood contributed by the decay mode of interest. Distributions of events so weighted, as well as the unweighted $M(D_s^+\pi^0\gamma)$ spectrum, are compared to the likelihood function in Fig. 19.3.16. The decay $D_{s1}(2460)^+ \to D_{s0}^*(2317)^+\gamma$ is found to be negligible, whereas the decay $D_{s1}(2460)^+ \to D_s^{*+}\pi^0$ saturated the $D_s^+\pi^0\gamma$ final state. The results of this study were later superseded by (Aubert, 2006e) described below.

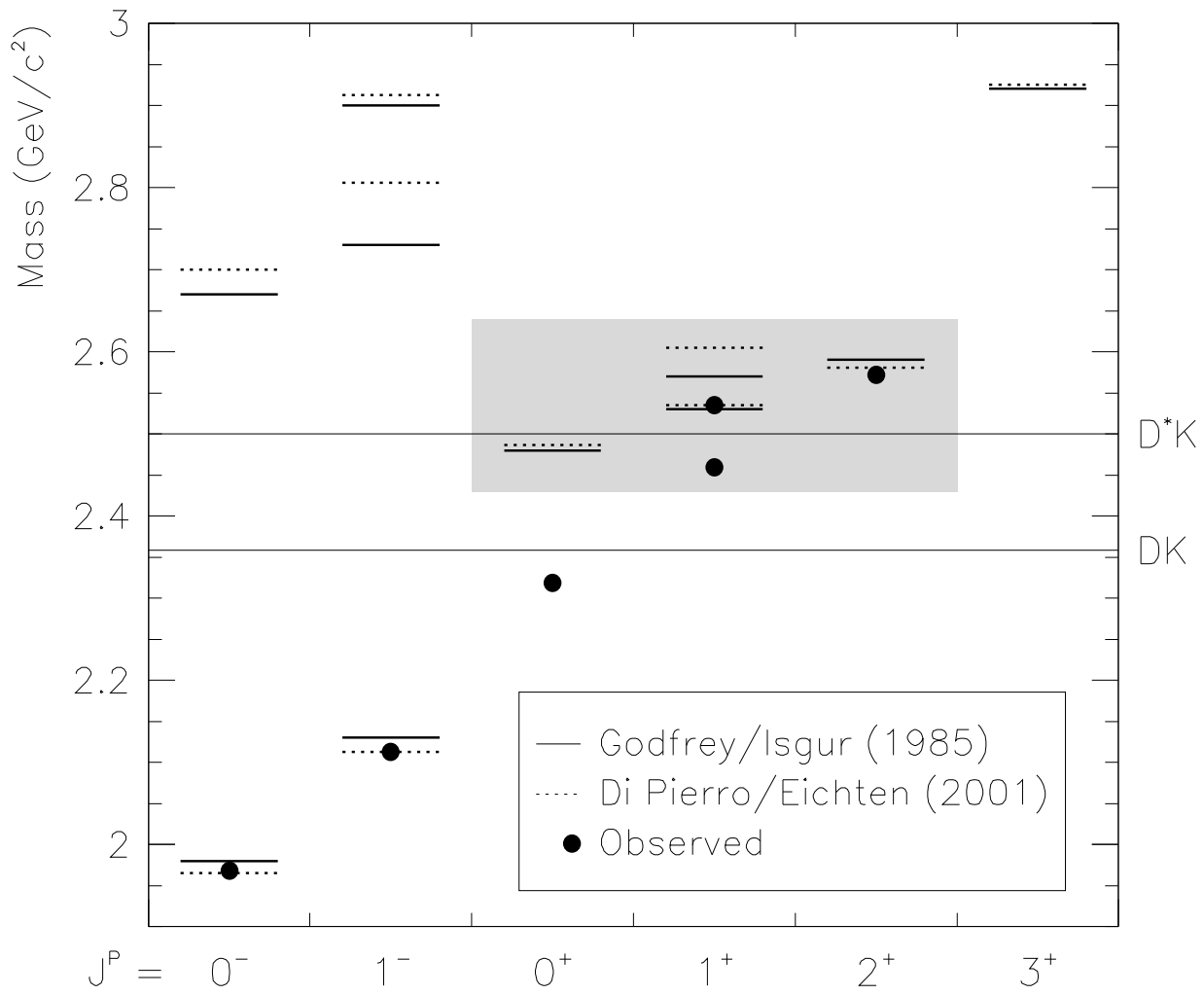

**Figure 19.3.13.** The $c\bar{s}$ spectrum according to the HQET scheme. The $P$-wave multiplet is shaded. Expectations on the masses according to HQET calculations (lines) are compared with experimental results (dots).

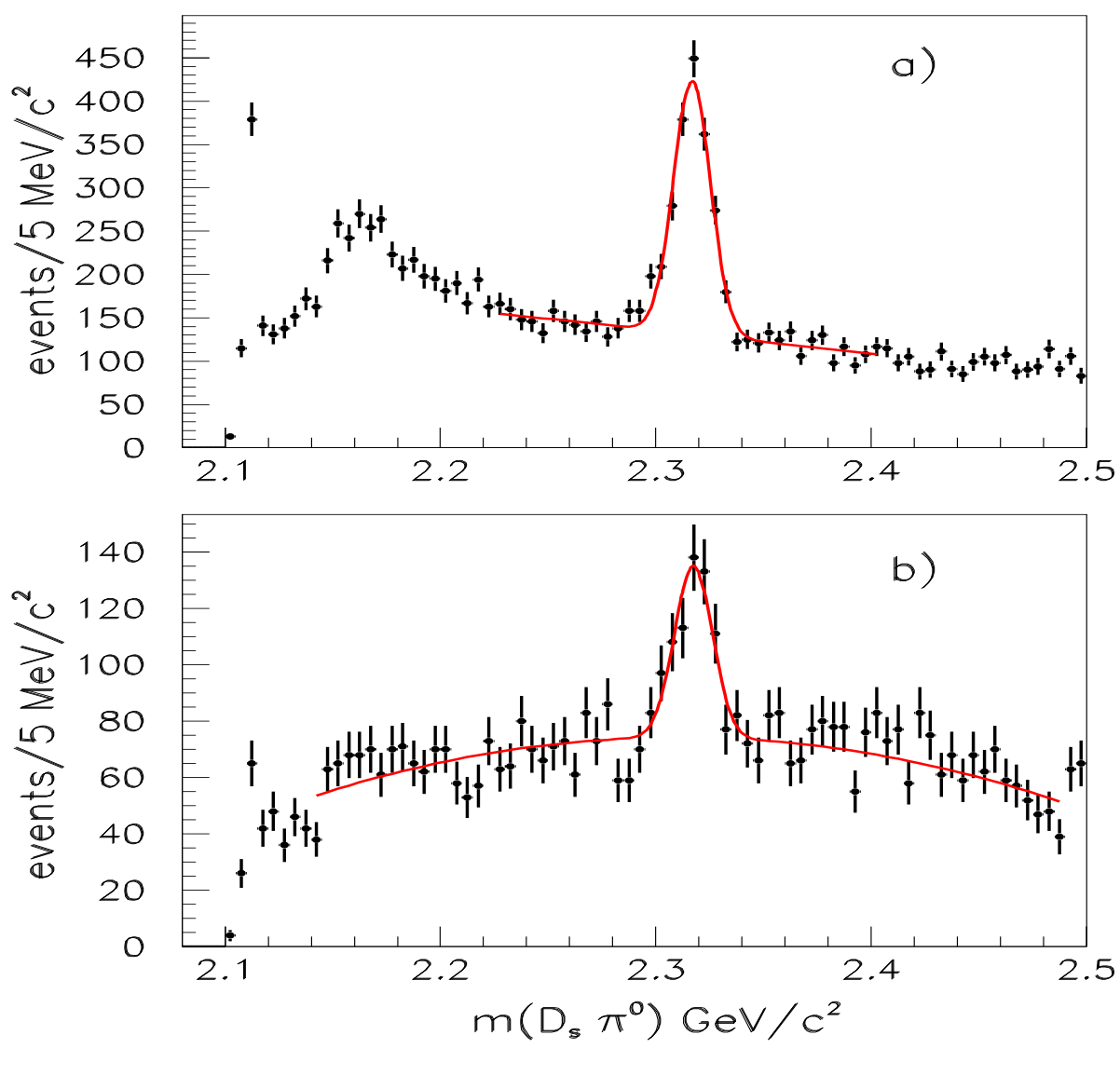

**Figure 19.3.14.** From (Aubert, 2003j). a) $D_s^+\pi^0$ mass spectrum for $D_s^+ \to K^+K^-\pi^+$ superimposed with the fit described in the text. b) $D_s^+\pi^0$ mass distribution for $D_s^+ \to K^+K^-\pi^+\pi^0$. The structure at 2.32 GeV is due to the $D_{s0}^*(2317)^+$ resonance. The narrow peak at threshold is due to the $D_s^{*+} \to D_s^+\pi^0$ decay. The broad structure is due to reflections from other states (see Section 19.3.4.3).

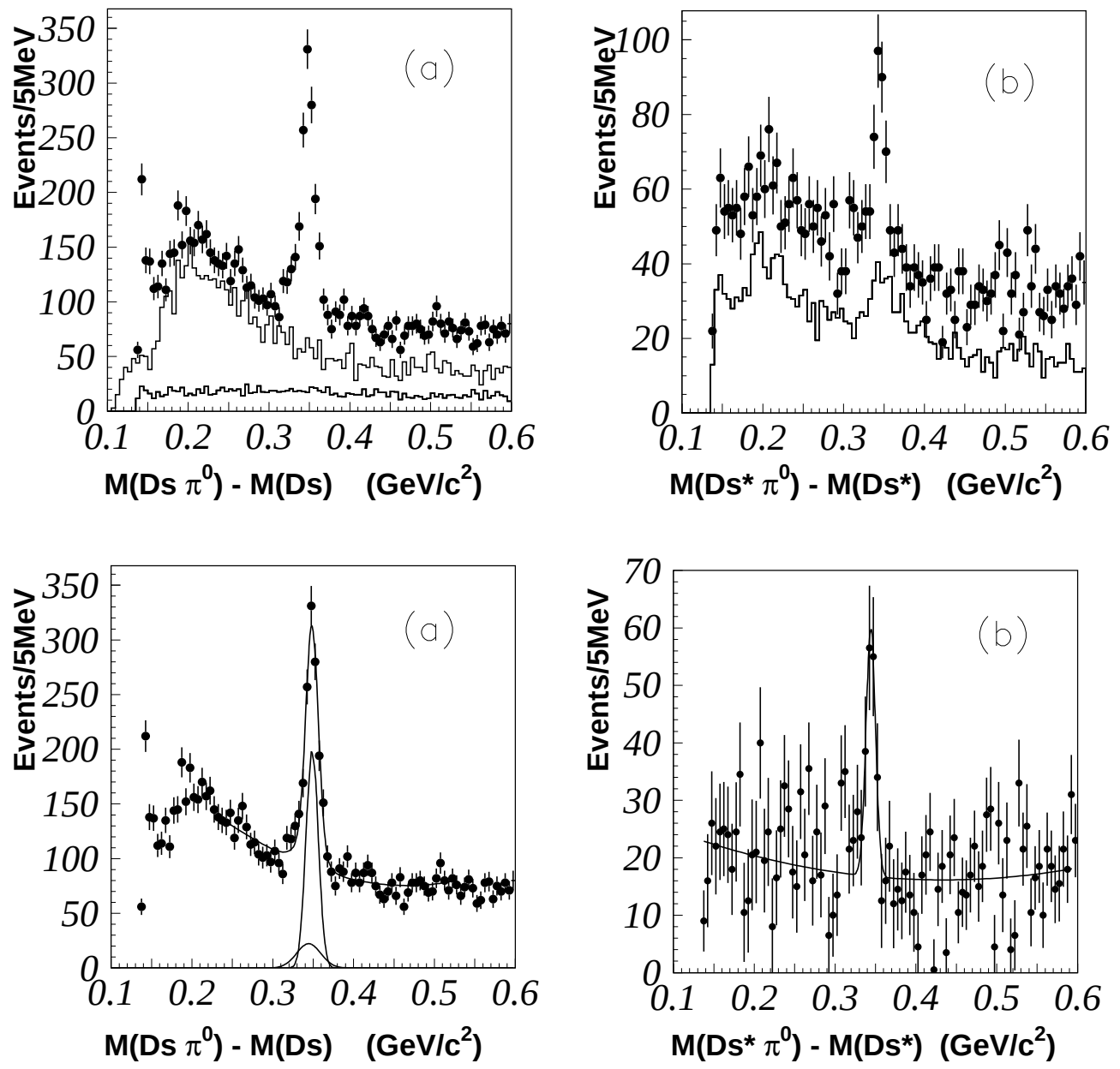

**Figure 19.3.15.** From (Abe, 2004a): Distributions of $\Delta M(D_s^+\pi^0)$ (left) and $\Delta M(D_s^{*+}\pi^0)$ (right). In the upper plots data from mass sidebands are also shown: the $D_s^+$ (left) and $D_s^{*+}$ (right) sideband regions (solid lines), $\pi^0$ sidebands (dashed). The bottom plots show fits to the $\Delta M(D_s^+\pi^0)$ (left) and $\Delta M(D_s^{*+}\pi^0)$ distributions after subtraction of the $D_s^{*+}$ sidebands.

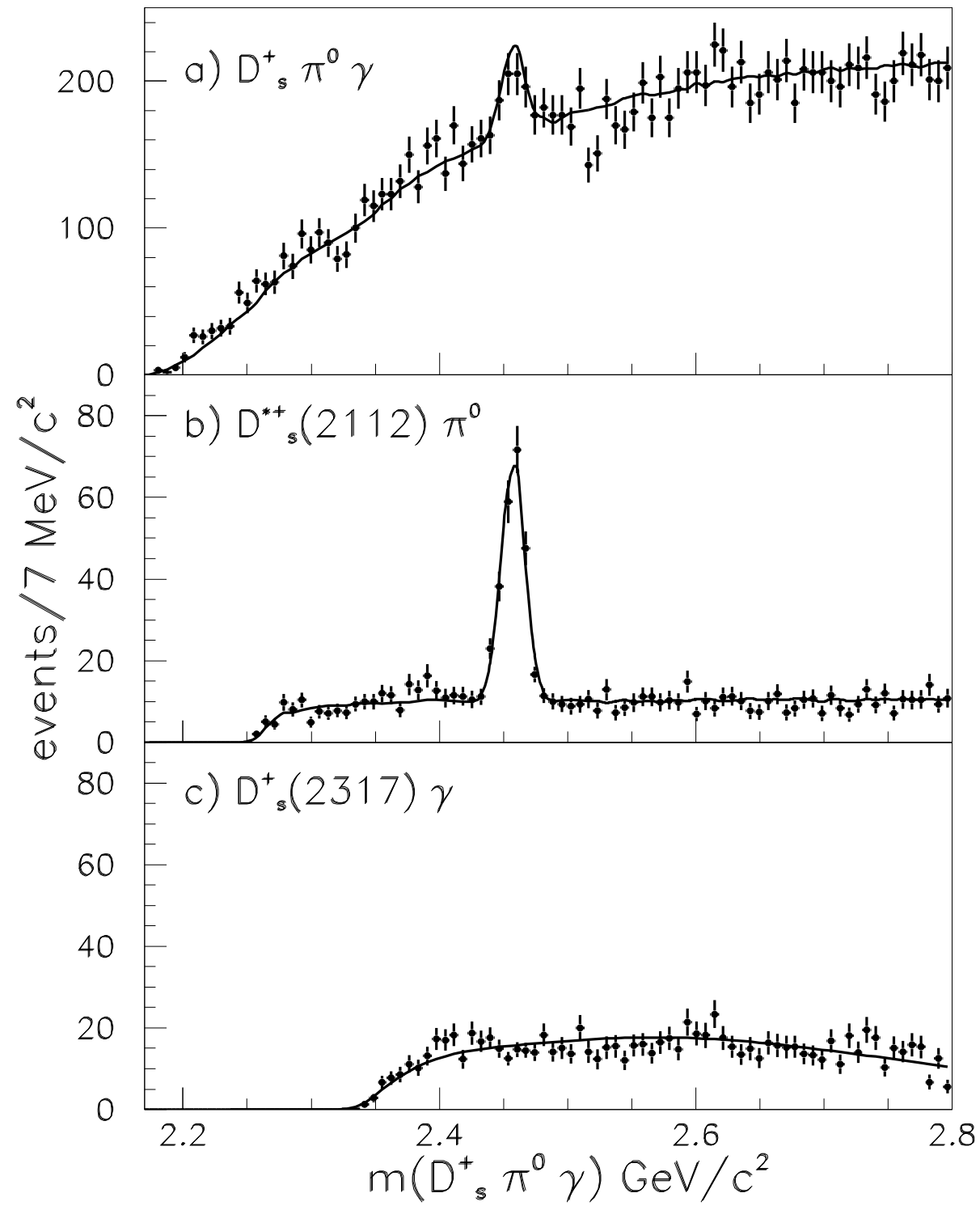

**Figure 19.3.16.** From (Aubert, 2004t). Maximum likelihood fit results overlaid on the $D_s^+\pi^0\gamma$ mass distribution with a) no weights and after applying weights corresponding to b) the $D_s^{*+}\pi^0$ and c) the $D_{s0}^*(2317)^+\gamma$ decays.

#### 19.3.4.3 High statistics study of $D_{s0}^*(2317)^+$ and $D_{s1}(2460)^+$

*BABAR* performed a complete analysis of the $D_s^{(*)+}\pi^0$ spectrum using 232 fb$^{-1}$ of data and with some of the selection criteria from the previous analyses reoptimized (Aubert, 2006e).

The invariant mass distribution of the $D_s^+\pi^0$ combinations is shown in Fig. 19.3.17. The $\pi^0$ momentum requirement to be greater than 350 MeV/$c$, removes the majority of the $D_s^{*+} \to D_s^+\pi^0$ decays, while keeping the entire $D_{s0}^*(2317)^+$ signal. The unbinned likelihood fit applied to the $M(D_s^+\pi^0)$ distribution includes the $D_s^{*+}$ and $D_{s0}^*(2317)^+$ signals and the following peaking components: a reflection at 2.17 GeV/$c^2$ from $D_s^{*+} \to D_s^+\gamma$ in which an unassociated $\gamma$ forms a false $\pi^0$ candidate, as well as a reflection appearing directly under the $D_{s0}^*(2317)^+$ signal. The latter originates from $D_{s1}(2460)^+ \to D_s^{*+}\pi^0$ with a missing photon from $D_s^{*+}$ decay. Parameters used to describe the former feed-down are determined directly from the data, while the shape of the latter one is based on the MC simulation of the $D_{s1}(2460)^+ \to D_s^+\pi^0\gamma$ Dalitz distribution. The $D_{s0}^*(2317)^+$ line shape used in the fit is derived from MC simulation configured with an intrinsic width (0.1 MeV/$c^2$) nearly indistinguishable from zero. Larger intrinsic widths assumed in the MC do not result in any significant improvement of the fit.

A study of the $D_s^+\pi^0\gamma$ mass (see Fig. 19.3.18) is performed with the $D_s^+\gamma$ mass required to be close to the $D_s^{*+}$ mass. Similarly to the Belle analysis, there were two reflections peaking near the $D_{s1}(2460)^+$: from $D_{s0}^*(2317)^+ \to D_s^+\pi^0$ decays combined with a random $\gamma$, and from $D_{s1}(2460)^+ \to D_s^+\pi^0\gamma$ with a wrong $\gamma$ chosen. In the fit to the $M(D_s^+\pi^0\gamma)$, these components are combined and described using MC simulations with the rates as measured in the previous analysis. A broad reflection at 2.34 MeV/$c^2$, coming from $D_s^{*+} \to D_s^+\gamma$ decays combined with an unassociated $\gamma$, is also considered in the fit. The $D_{s1}(2460)^+$ signal shape is obtained using MC, similarly to the $D_{s0}^*(2317)^+$ case.

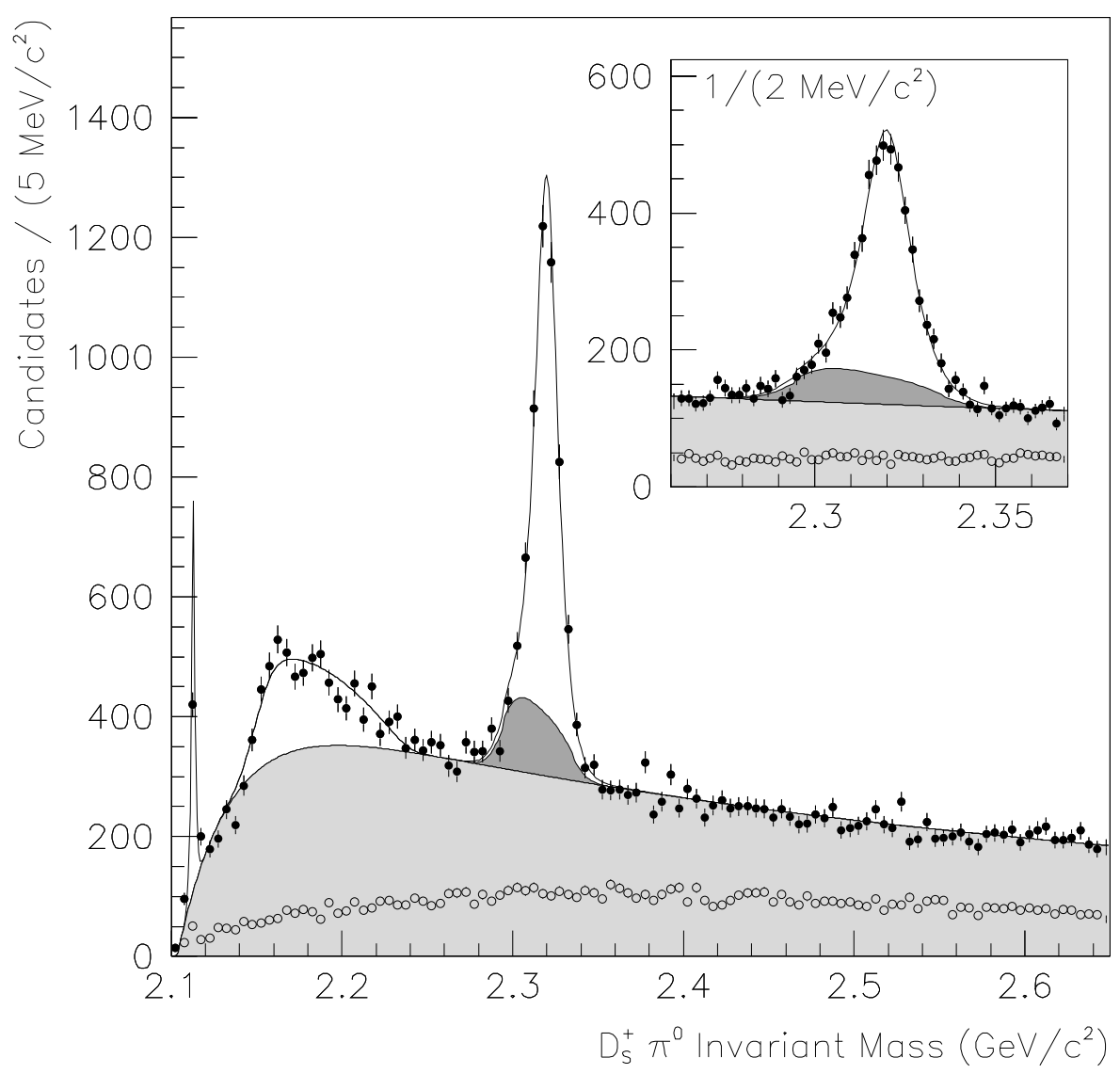

**Figure 19.3.17.** From (Aubert, 2006e). The invariant mass distribution for $D_s^+\pi^0$ candidates (solid points) and the equivalent using the $D_s^+$ sidebands (open points). The curve represents the likelihood fit described in the text and includes a contribution from combinatorial background (light shade) and the reflection from $D_{s1}(2460)^+ \to D_s^{*+}\pi^0$ decay (dark shade). The insert highlights the details near the $D_{s0}^*(2317)^+$ mass. The narrow peak at threshold is due to the $D_s^{*+} \to D_s^+\pi^0$ decay.

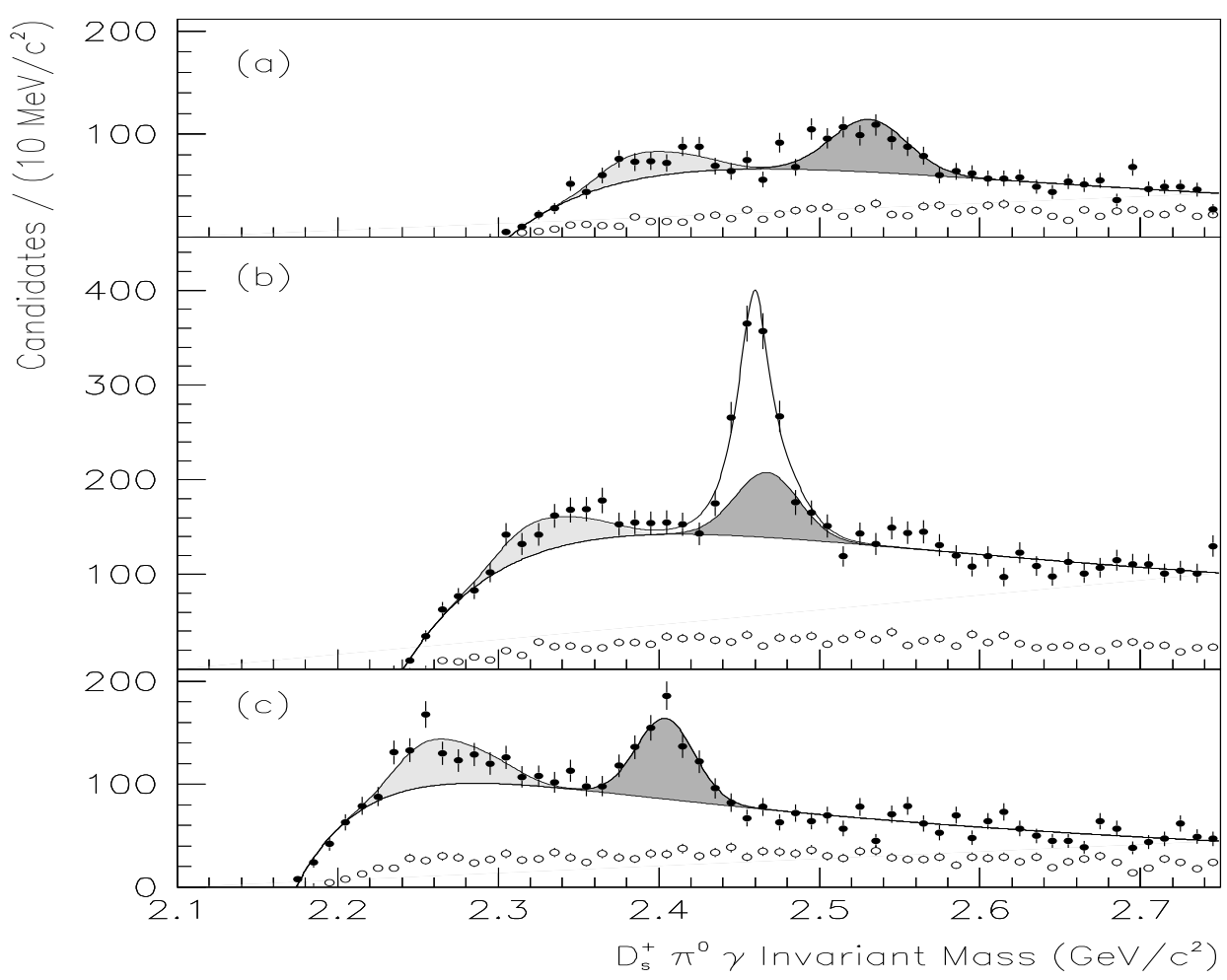

**Figure 19.3.18.** From (Aubert, 2006e). The $M(D_s^+\pi^0\gamma)$ invariant mass distribution of candidates in the (a) upper, (b) signal, and (c) lower $D_s^+\gamma$ mass selection windows for (solid points) the $D_s^+$ signal and (open points) $D_s^+$ sideband samples. The curves represent the fits described in the text. The gray regions correspond to the predicted reflections from $D_{s0}^*(2317)^+$ (dark) and $D_s^*(2112)^+$ (light).

**Table 19.3.7.** From (Aubert, 2006e) and (Abe, 2004a). The first section summarizes the combined mass and width results from *BABAR*. The $D_{s0}^*(2317)^+$ and $D_{s1}(2536)^+$ mesons are observed in only one decay mode covered by this analysis. The $D_{s1}(2460)^+$ mass is the average of that obtained from the $D_s^+\gamma$, $D_s^+\pi^0\gamma$, and $D_s^+\pi^+\pi^-$ final states, although the latter measurement dominates in the average due to superior systematic uncertainties. The second section gives Belle results based on the $D_s^{(*)+}\pi^0$ modes.

| Particle | Mass (MeV/$c^2$) | $\Gamma$ (MeV/$c^2$) |
|---|---|---|
| $D_{s0}^*(2317)^+$ | $2319.6 \pm 0.2 \pm 1.4$ | $< 3.8$ @ 95 % C.L. |
| $D_{s1}(2460)^+$ | $2460.1 \pm 0.2 \pm 0.8$ | $< 3.5$ @ 95 % C.L. |
| $D_{s1}(2536)^+$ | $2534.6 \pm 0.3 \pm 0.7$ | $< 2.5$ @ 95 % C.L. |
| $D_{s0}^*(2317)^+$ | $2317.2 \pm 0.5 \pm 0.9$ | $< 4.6$ @ 90 % C.L. |
| $D_{s1}(2460)^+$ | $2456.5 \pm 1.3 \pm 1.3$ | $< 5.5$ @ 90 % C.L. |

The resulting $D_{s0}^*(2317)^+$ and $D_{s1}(2460)^+$ masses and upper limits on their widths are given in Table 19.3.7. The $D_{s1}(2460)^+$ mass is the average obtained from the $D_s^+\pi^0\gamma$, $D_s^+\gamma$ and $D_s^+\pi^+\pi^-$ final states (see Section 19.3.4.4). The limit on the intrinsic $D_{s1}(2460)^+$ width is taken as the best limit obtained from these three decay modes.

#### 19.3.4.4 Other decay modes

The $D_{s1}(2460)^+$ decays to the $D_s^+\gamma$ and the $D_s^+\pi^+\pi^-$ final states were first observed by Belle (Abe, 2004a) and further confirmed by *BABAR* (Aubert, 2006e). Selected photons are required to pass the $\pi^0$ veto, while the $\pi^+\pi^-$ pairs are taken outside of the $K_S^0$ mass window. The left plot in Fig. 19.3.19 shows the $\Delta M(D_s^+\gamma) = M(D_s^+\gamma) - M(D_s^+)$ distribution measured by Belle. A peak at 490 MeV/$c^2$ corresponds to the $D_{s1}(2460)^+$, whereas no peak is present in the $D_{s0}^*(2317)^+$ region at 350 MeV/$c^2$. The observation of the radiative decay of the $D_{s1}(2460)^+$ rules out its spin-parity of $0^\pm$.

The invariant mass distribution of the $D_s^+\pi^+\pi^-$ candidates from *BABAR* presented in right plot in Fig. 19.3.19, shows clear signals from $D_{s1}(2460)^+$ and $D_{s1}(2536)^+$ and negligible contribution from $D_{s0}^*(2317)^+$. Searches for the decays of $D_{s0}^*(2317)^+$ and $D_{s1}(2460)^+$ to $D_s^+\pi^0\pi^0$ and $D_s^{*+}\gamma$ give no positive results. The branching fraction ratios for the studied channels, given in Table 19.3.8, are measured based on the fitted yields and detection efficiencies estimated with an assumption of the same fragmentation functions for the $D_{sJ}$ states.

The dominant decays observed, $D_{s0}^*(2317)^+ \to D_s^+\pi^0$ and $D_{s1}(2460)^+ \to D_s^{*+}\pi^0$, violate isospin conservation; that is often considered as a property of four-quark states, which have long been proposed (Jaffe, 1977b; Lipkin, 1977). The most unambiguous signature of a molecular interpretation of the $D_{s0}^*(2317)^+$ and the $D_{s1}(2460)^+$ would be an observation of their neutral and doubly-charged partners decaying to $D_s^{(*)+}\pi^\pm$. However a search for the $D_{s0}^*(2317)$ partners decaying to $D_s^+\pi^+$ and $D_s^+\pi^-$ resulted in no evidence (Aubert, 2006e).

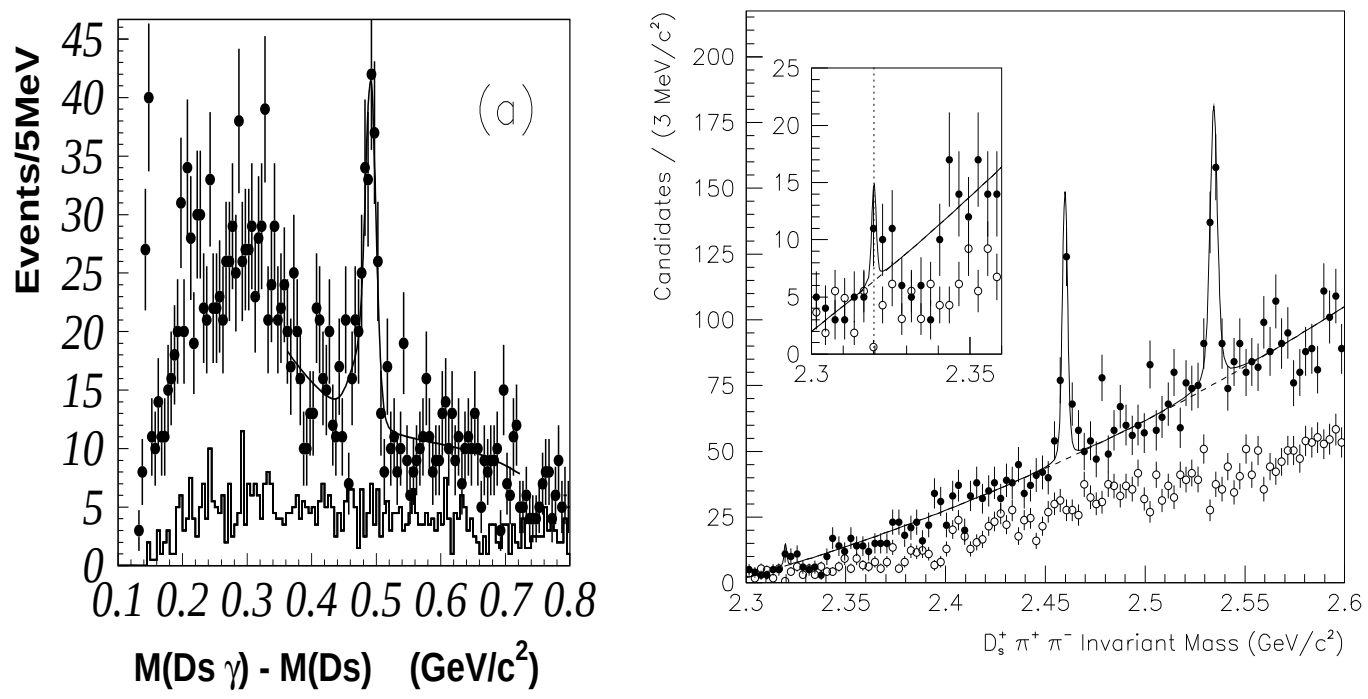

**Figure 19.3.19.** Left: The $\Delta M(D_s^+\gamma)$ distribution from (Abe, 2004a). Right: The $M(D_s^+\pi^+\pi^-)$ spectrum from (Aubert, 2006e). The insert focuses on the $D_{s0}^*(2317)^+$ region. The spectra for the $D_s^+$ sideband candidates are plotted with solid histogram in the left plot and open points in the right one.

**Table 19.3.8.** A summary of $D_s^{**}$ branching fraction ratios from Belle (first line) and *BABAR* (second line). For the $D_{s0}^*(2317)^+$ meson, only one decay mode has been observed; this is used as the denominator when calculating the $D_{s0}^*(2317)^+$ branching ratios. For the $D_{s1}(2460)^+$ meson, the $D_s^+\pi^0\gamma$ decay mode (consisting of possible decay through either $D_s^*(2112)^+\pi^0$ or $D_{s0}^*(2317)^+\gamma$) is chosen for this role.

| Ratio | Fraction or Limit |
|---|---|
| $\frac{\mathcal{B}(D_{s1}(2460)^+\to D_s^+\pi^0)}{\mathcal{B}(D_{s1}(2460)^+\to D_s^{*+}\pi^0)}$ | $< 0.21$ @ 90 % C.L.<br>$< 0.042$ @ 95 % C.L. |
| $\frac{\mathcal{B}(D_{s1}(2460)^+\to D_s^+\gamma)}{\mathcal{B}(D_{s1}(2460)^+\to D_s^{*+}\pi^0)}$ | $= 0.55 \pm 0.13 \pm 0.08$<br>$= 0.34 \pm 0.04 \pm 0.04$ |
| $\frac{\mathcal{B}(D_{s1}(2460)^+\to D_s^+\pi^+\pi^-)}{\mathcal{B}(D_{s1}(2460)^+\to D_s^{*+}\pi^0)}$ | $= 0.14 \pm 0.04 \pm 0.02$<br>$= 0.077 \pm 0.013 \pm 0.008$ |
| $\frac{\mathcal{B}(D_{s1}(2460)^+\to D_s^{*+}\gamma)}{\mathcal{B}(D_{s1}(2460)^+\to D_s^{*+}\pi^0)}$ | $< 0.31$ @ 90 % C.L.<br>$< 0.24$ @ 95 % C.L. |
| $\frac{\mathcal{B}(D_{s0}^*(2317)^+\to D_s^+\gamma)}{\mathcal{B}(D_{s0}^*(2317)^+\to D_s^+\pi^0)}$ | $< 0.05$ @ 90 % C.L.<br>$< 0.14$ @ 95 % C.L. |
| $\frac{\mathcal{B}(D_{s0}^*(2317)^+\to D_s^+\pi^+\pi^-)}{\mathcal{B}(D_{s0}^*(2317)^+\to D_s^+\pi^0)}$ | $< 0.004$ @ 90 % C.L.<br>$< 0.005$ @ 95 % C.L. |
| $\frac{\mathcal{B}(D_{s0}^*(2317)^+\to D_s^{*+}\gamma)}{\mathcal{B}(D_{s0}^*(2317)^+\to D_s^+\pi^0)}$ | $< 0.18$ @ 90 % C.L.<br>$< 0.16$ @ 95 % C.L. |

The observed decay pattern is consistent with the $J^P$ assignments for the $D_{s0}^*(2317)^+$ and $D_{s1}(2460)^+$ of respectively $0^+$ and $1^+$, as expected by the potential models for the $P$-wave $c\bar{s}$ mesons with $j_q = \frac{1}{2}$.

#### 19.3.4.5 $D_{sJ}$ production in $B$ decays

To clarify the nature of the $D_{s0}^*(2317)^+$ and $D_{s1}(2460)^+$ states, Belle and *BABAR* searched for their production in $B \to \overline{D}D_{sJ}$ decays. These reactions proceed via $\bar{b} \to \bar{c}W^+ \to \bar{c}c\bar{s}$ transition and are expected to be the dominant exclusive $c\bar{s}$ production mechanism. QCD sum rules predict that $P$-wave charmed mesons with $j_q = \frac{1}{2}$ should be more readily produced in $B$ decays than ones having $j_q = \frac{3}{2}$ (Le Yaouanc, Oliver, Pène, Raynal, and Morenas, 2001). Thus observation of the $B \to \overline{D}D_{s0}^*(2317)^+$ and $B \to \overline{D}D_{s1}(2460)^+$ would provide a confirmation of the $P$-wave nature of these $D_{sJ}$ states. Moreover, measurements of $B$ decays to the final states including the $j_q = \frac{3}{2}$ $D_{sJ}$ mesons, or higher orbital or radial $c\bar{s}$ excitations could be a further test of the theoretical predictions on the $c\bar{s}$ spectroscopy.

In a Belle study of the $B \to \overline{D}D_{sJ}$ decays, with $124 \times 10^6$ $B\overline{B}$ pairs used, the $\overline{D}$ is reconstructed as: $\overline{D}^0 \to K^+\pi^-$, $K^+\pi^-\pi^-\pi^+$, $K^+\pi^-\pi^0$, $D^- \to K^+\pi^-\pi^-$. The $D_{sJ}$ states are studied in the $D_s^{(*)+}\pi^0$, $D_s^{(*)+}\gamma$ and $D_s^{(*)+}\pi^+\pi^-$ final states, with $D_s^{*+} \to D_s^+\gamma$ and $D_s^+ \to \phi\pi^+, \overline{K}^{(*)0}K^+$ reconstructed (Krokovny, 2003b). To reduce background the $B$ candidates are required to have their $m_{\rm ES}$ consistent with the nominal $B$ mass (see Sections 9 and 7), while the signal events are identified in the $\Delta E$-$M(D_{sJ})$ space. The left plots in Fig. 19.3.20 show invariant mass distributions of the $D_{sJ}$ candidates within the $\Delta E$ signal region, for all the $D$ modes combined. The presented spectra are for the $D_{sJ}$ decay final states with significant signals found: $D_{s0}^*(2317)^+ \to D_s^+\pi^0$, $D_{s1}(2460)^+ \to D_s^{*+}\pi^0$ and $D_{s1}(2460)^+ \to D_s^+\gamma$. Unlike the continuum based $D_{sJ}$ studies, cross-feeds between the $D_{sJ}$ decay modes were not found, and thus the $M(D_{sJ})$ spectra are fitted with signal and background described respectively with a Gaussian and linear function. The measured masses of the $D_{s0}^*(2317)^+$ and $D_{s1}(2460)^+$ are respectively $2319.8 \pm 2.1 \pm 2.0\,{\rm MeV}/c^2$ and $2459.2 \pm 1.6 \pm 2.0\,{\rm MeV}/c^2$, while the fitted widths are consistent with the experimental resolution.

The $\Delta E$ distributions shown in Fig. 19.3.20, are projections for the $B$ candidates having the $M(D_{sJ})$ within the observed $D_{sJ}$ signals. Branching fractions given in Table 19.3.9, are extracted from the $\Delta E$ fits of the combined $B \to \overline{D}^0D_{sJ}^+$ and $B \to D^-D_{sJ}^+$ modes, with efficiencies of the individual modes taken into account and isospin invariance assumed. The determined ratio, $\frac{\mathcal{B}(D_{s1}(2460)^+\to D_s^+\gamma)}{\mathcal{B}(D_{s1}(2460)^+\to D_s^{*+}\pi^0)} = 0.38 \pm 0.11 \pm 0.04$, is consistent with the one obtained from $D_{s1}(2460)^+$ produced in the continuum (see Table 19.3.8).

A helicity angle ($\theta_{D_s\gamma}$) study is performed for the $D_{s1}(2460)^+ \to D_s^+\gamma$ decay, where $\theta_{D_s\gamma}$ is defined as the angle between the $D_s^+$ momentum and the opposite to the $B$ meson momentum in the $D_{s1}(2460)^+$ rest frame. Figure 19.3.21 shows the background-subtracted $\cos(\theta_{D_s\gamma})$ distribution, where the data points represent signal yields from the $\Delta E$ fits performed in respective $\cos(\theta_{D_s\gamma})$ bins.

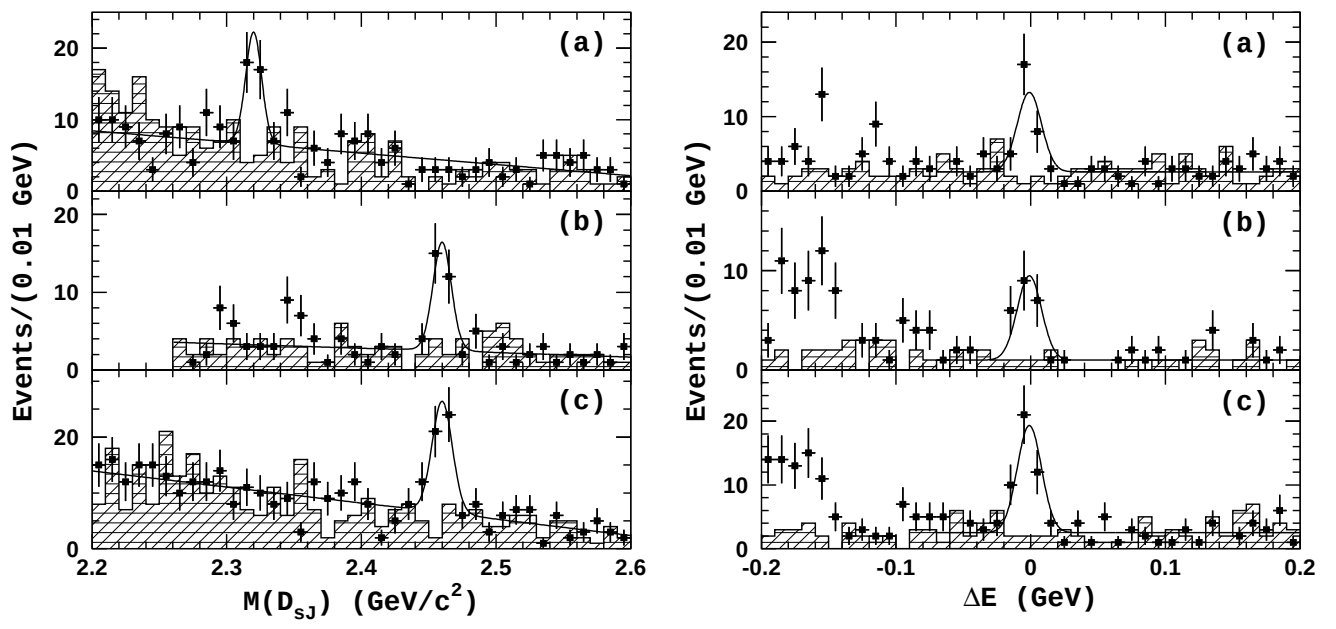

**Figure 19.3.20.** From (Krokovny, 2003b): The $M(D_{sJ})$ distribution for the $\Delta E$ signal region (left) and the $\Delta E$ distribution for the $M(D_{sJ})$ signal region (right) for the $B \to \overline{D} D_{sJ}$ candidates with (a) $D_{s0}^*(2317)^+ \to D_s^+\pi^0$, (b) $D_{s1}(2460)^+ \to D_s^{*+}\pi^0$ and (c) $D_{s1}(2460)^+ \to D_s^+\gamma$ decays observed. Hatched histograms in a given variable distribution show the sidebands of the other variable, lines represent the fit result.

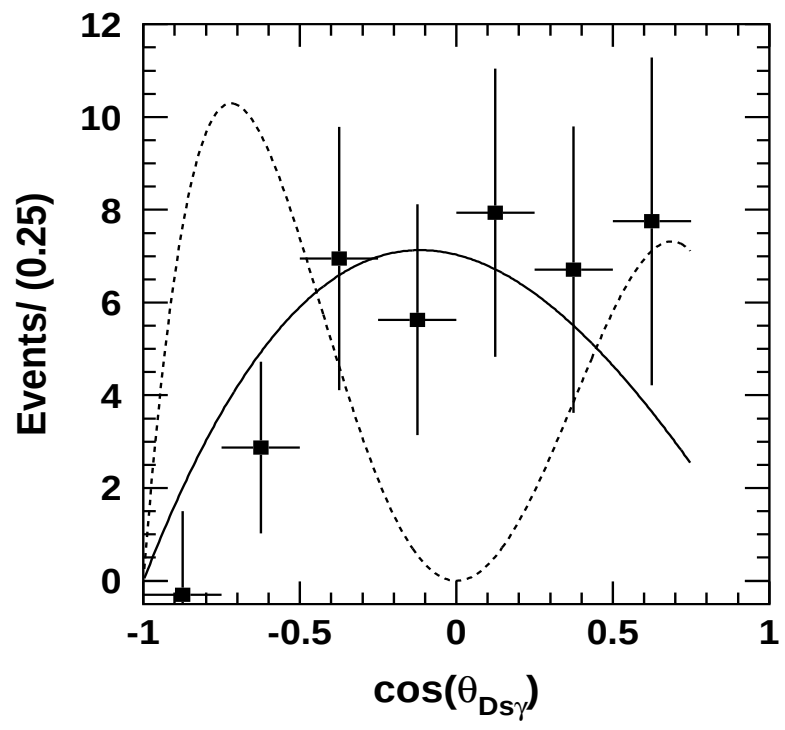

**Figure 19.3.21.** From (Krokovny, 2003b): Background-subtracted helicity distribution for the $D_{s1}(2460)^+ \to D_s^+\gamma$. Lines show MC expectations for spin hypotheses $J = 1$ (solid) and $J = 2$ (dashed); $J = 0$ is forbidden.

**Table 19.3.9.** From (Krokovny, 2003b). Measured branching fraction products $\mathcal{B}(B \to \overline{D} D_{sJ}) \times \mathcal{B}(D_{sJ} \to f)$ or 90 % C.L. limits. $f$ is the label of a given final state.

| Decay mode | Product $\mathcal{B}$ $[10^{-4}]$ |
|---|---|
| $B \to \overline{D} D_{s0}^*(2317)^+,\ D_{s0}^*(2317)^+ \to D_s^+\pi^0$ | $8.5^{+2.1}_{-1.9} \pm 2.6$ |
| $B \to \overline{D} D_{s0}^*(2317)^+,\ D_{s0}^*(2317)^+ \to D_s^{*+}\gamma$ | $2.5^{+2.0}_{-1.8} (< 7.5)$ |
| $B \to \overline{D} D_{s1}(2460)^+,\ D_{s1}(2460)^+ \to D_s^{*+}\pi^0$ | $17.8^{+4.5}_{-3.9} \pm 5.3$ |
| $B \to \overline{D} D_{s1}(2460)^+,\ D_{s1}(2460)^+ \to D_s^+\gamma$ | $6.7^{+1.3}_{-1.2} \pm 2.0$ |
| $B \to \overline{D} D_{s1}(2460)^+,\ D_{s1}(2460)^+ \to D_s^{*+}\gamma$ | $2.7^{+1.8}_{-1.5} (< 7.3)$ |
| $B \to \overline{D} D_{s1}(2460)^+,\ D_{s1}(2460)^+ \to D_s^+\pi^+\pi^-$ | $< 1.6$ |
| $B \to \overline{D} D_{s1}(2460)^+,\ D_{s1}(2460)^+ \to D_s^+\pi^0$ | $< 1.8$ |

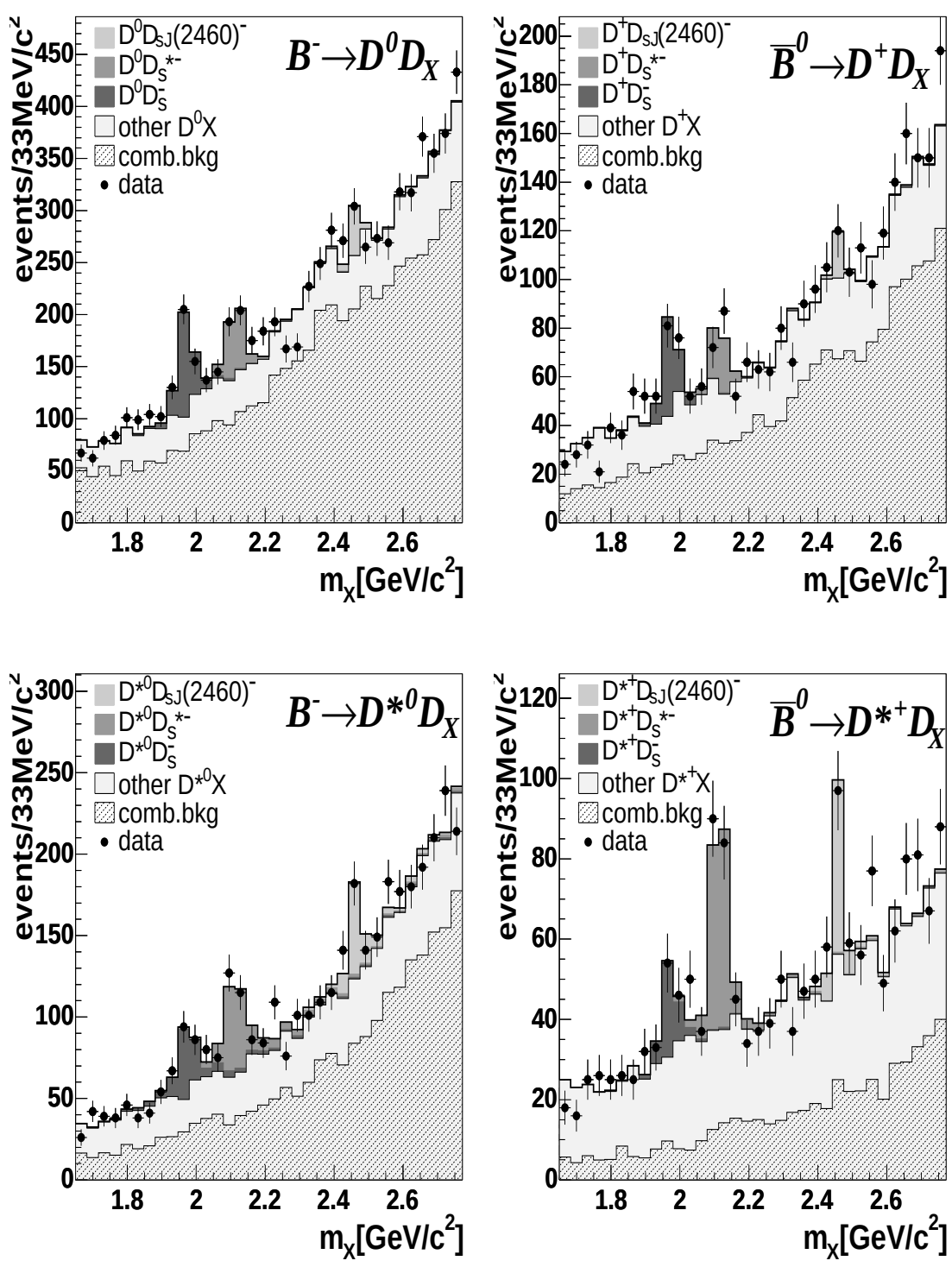

**Figure 19.3.22.** From (Aubert, 2006aw). Distributions of the $D_{sJ}$ invariant mass $m_X$(defined in the text). Fitted $\overline{B} \to D^{(*)+,0} D_s^{(*)-}$ and $\overline{B} \to D^{(*)+,0} D_{s1}(2460)^-$ signal contributions and background components are overlaid to the data points.

As it can be seen from the figure, the $J = 1$ hypothesis fits much better the data than the $J = 2$ hypothesis.

In the *BABAR* approach to the $B \to \overline{D}^{(*)} D_{sJ}^+$ study (Aubert, 2006aw), the $\overline{D}^{(*)} = \overline{D}^{(*)0}$, $D^{(*)-}$ mesons are fully reconstructed ($D_{\rm meas}$), while the invariant mass of the $D_{sJ}$ is inferred from the kinematics of the two-body $B$ decay, as well as the kinematics of the accompanying $\overline{B}$. The analysis used the $\Upsilon(4S) \to B\overline{B}$ events in which the $\overline{B}$ meson ($B_{\rm rec}$) decays into fully reconstructed hadronic final state $B_{\rm rec} \to D^{(*)}Y^-$, with the system $Y^-$ composed of combinations of kaons and pions (see Section 7). The $D_{sJ}$ invariant mass is derived from a missing four-momentum ($p_{\rm miss}$) as:

$$m_X \equiv \sqrt{p_{\rm miss}^2} = \sqrt{(p_{\Upsilon(4S)} - p_{B_{\rm rec}} - p_{D_{\rm meas}})^2}, \tag{19.3.18}$$

with all the momenta measured in the laboratory system. Such a method allows measurements of absolute $B \to \overline{D}^{(*)} D_{sJ}^+$ branching fractions, without any assumptions on the $D_{sJ}^+$ decays, at the cost of full reconstruction efficiency of about 0.3 % for $B^0\overline{B}^0$ and 0.2 % for $B^+B^-$, and the $m_X$ resolution being typically a few times worse than in a direct reconstruction. In the $m_X$ spectra, shown in Fig. 19.3.22, in addition to the $D_s^{(*)+}$, *BABAR* observes the $D_{s1}(2460)^+$ signal.

The measured $B \to \overline{D}^{(*)} D_{s1}(2460)^+$ branching fractions, calculated using PDG values of the $D^{(*)}$ decay rates,

**Table 19.3.10.** From (Aubert, 2006aw). Absolute branching fractions of the $B \to \overline{D}^{(*)} D_{s1}(2460)^+$ decays.

| Decay mode | $\mathcal{B}$ [%] |
|---|---|
| $B^0 \to D^- D_{s1}(2460)^+$ | $0.26 \pm 0.15 \pm 0.07$ |
| $B^0 \to D^{*-} D_{s1}(2460)^+$ | $0.88 \pm 0.20 \pm 0.14$ |
| $B^+ \to \overline{D}^0 D_{s1}(2460)^+$ | $0.43 \pm 0.16 \pm 0.13$ |
| $B^+ \to \overline{D}^{*0} D_{s1}(2460)^+$ | $1.12 \pm 0.26 \pm 0.20$ |

are summarized in Table 19.3.10. *BABAR* further combined these results with their measurements of the product $\mathcal{B}$ of the $B \to \overline{D}^{(*)} D_{s1}(2460)^+$ decays (Aubert, 2004ad), measured similarly to the described Belle analysis. This allowed a first measurement of the $D_{s1}(2460)^+$ decay rates:

$$\begin{aligned} \mathcal{B}(D_{s1}(2460)^+ \to D_s^{*+}\pi^0) &= (56 \pm 13 \pm 9)\%, \\ \mathcal{B}(D_{s1}(2460)^+ \to D_s^+\gamma) &= (16 \pm 4 \pm 3)\%. \end{aligned} \tag{19.3.19}$$

Belle studied the production of the $D_{s1}(2536)^+$ meson in $B \to \overline{D}^{(*)} D_{s1}(2536)^+$ decays, with $\overline{D}^{(*)}$ being either $\overline{D}^0$ or $D^{(*)-}$, using a data sample of $657 \times 10^6$ $B\overline{B}$ pairs (Aushev, 2011). The $D_{s1}(2536)^+$ is reconstructed in its dominant decay modes, $D^{*+}K_S^0$ and $D^{*0}K^+$. Figure 19.3.23 shows the $M(D_{s1}(2536))$ spectra for the $B$ candidates satisfying the $\Delta E$-$m_{\rm ES}$ selection, for each $\overline{D}^{(*)}$ flavor and the $D_{s1}(2536)^+$ decay mode separately. All these distributions are fitted simultaneously, with the signal $D_{s1}(2536)^+$ described as a BW function convolved with a double Gaussian function describing the mass resolution. The measured $D_{s1}(2536)^+$ mass and width were respectively $2534.1 \pm 0.6\,{\rm MeV}/c^2$ and $0.75 \pm 0.23\,{\rm MeV}/c^2$, consistent with their PDG values (Beringer et al., 2012). A fit with the $D_{s1}(2536)^+$ partial width ratio kept as a free parameter, yields:

$$\frac{\mathcal{B}(D_{s1}(2536)^+ \to D^{*0}K^+)}{\mathcal{B}(D_{s1}(2536)^+ \to D^{*+}K^0)} = 0.88 \pm 0.24 \pm 0.08, \tag{19.3.20}$$

in agreement with the *BABAR* study (Aubert, 2008bd). Table 19.3.11 summarizes ratios of branching fractions, calculated using the latest measurements of the $B \to \overline{D}^{(*)} D_{s(J)}^{(*)+}$ branching fractions, as well as the $D_{s1}(2536)^+$ measurements by Belle (Aushev, 2011) and *BABAR* (Aubert, 2008bd). In these calculations, 100% branching fractions are assumed for the $D_{s0}^*(2317)^+ \to D_s^+\pi^0$ and $D_{s1}(2536)^+ \to D^*K$ decay modes. Within the factorization model and in the heavy quark limit, these ratios should be of order unity for the $D_{s0}^*(2317)^+$ and $D_{s1}(2460)^+$, whereas for the $D_{s1}(2536)^+$ they are predicted to be very small (Datta and O'Donnell, 2003b; Le Yaouanc, Oliver, Pène, and Raynal, 1996). The decay pattern for the $D_{s1}(2536)^+$ follows these expectations, whereas for the $D_{s0}^*(2317)^+$ and $D_{s1}(2460)^+$ the ratios are rather different from unity and therefore such an approach does not really work. One possibility is that these states are not canonical $c\bar{s}$. However, the agreement with theory would be improved for the $D_{s0}^*(2317)^+$, if other prominent decay modes existed in addition to the $D_s^+\pi^0$.

**Table 19.3.11.** Ratios of $B$ decay branching ratios measured by Belle (Aushev, 2011) and *BABAR* (Aubert, 2008bd).

| Ratio | Fraction |
|---|---|
| $\mathcal{B}(B \to \overline{D} D_{s1}(2536)^+)/\mathcal{B}(B \to \overline{D} D_s^*)$ | $0.05 \pm 0.01$ |
| $\mathcal{B}(B \to \overline{D}^* D_{s1}(2536)^+)/\mathcal{B}(B \to \overline{D}^* D_s^*)$ | $0.04 \pm 0.01$ |
| $\mathcal{B}(B \to \overline{D} D_{s1}(2460)^+)/\mathcal{B}(B \to \overline{D} D_s^*)$ | $0.44 \pm 0.11$ |
| $\mathcal{B}(B \to \overline{D}^* D_{s1}(2460)^+)/\mathcal{B}(B \to \overline{D}^* D_s^*)$ | $0.58 \pm 0.12$ |
| $\mathcal{B}(B \to \overline{D} D_{s0}^*(2317)^+)/\mathcal{B}(B \to \overline{D} D_s)$ | $0.10 \pm 0.03$ |
| $\mathcal{B}(B \to \overline{D}^* D_{s0}^*(2317)^+)/\mathcal{B}(B \to \overline{D}^* D_s)$ | $0.15 \pm 0.06$ |

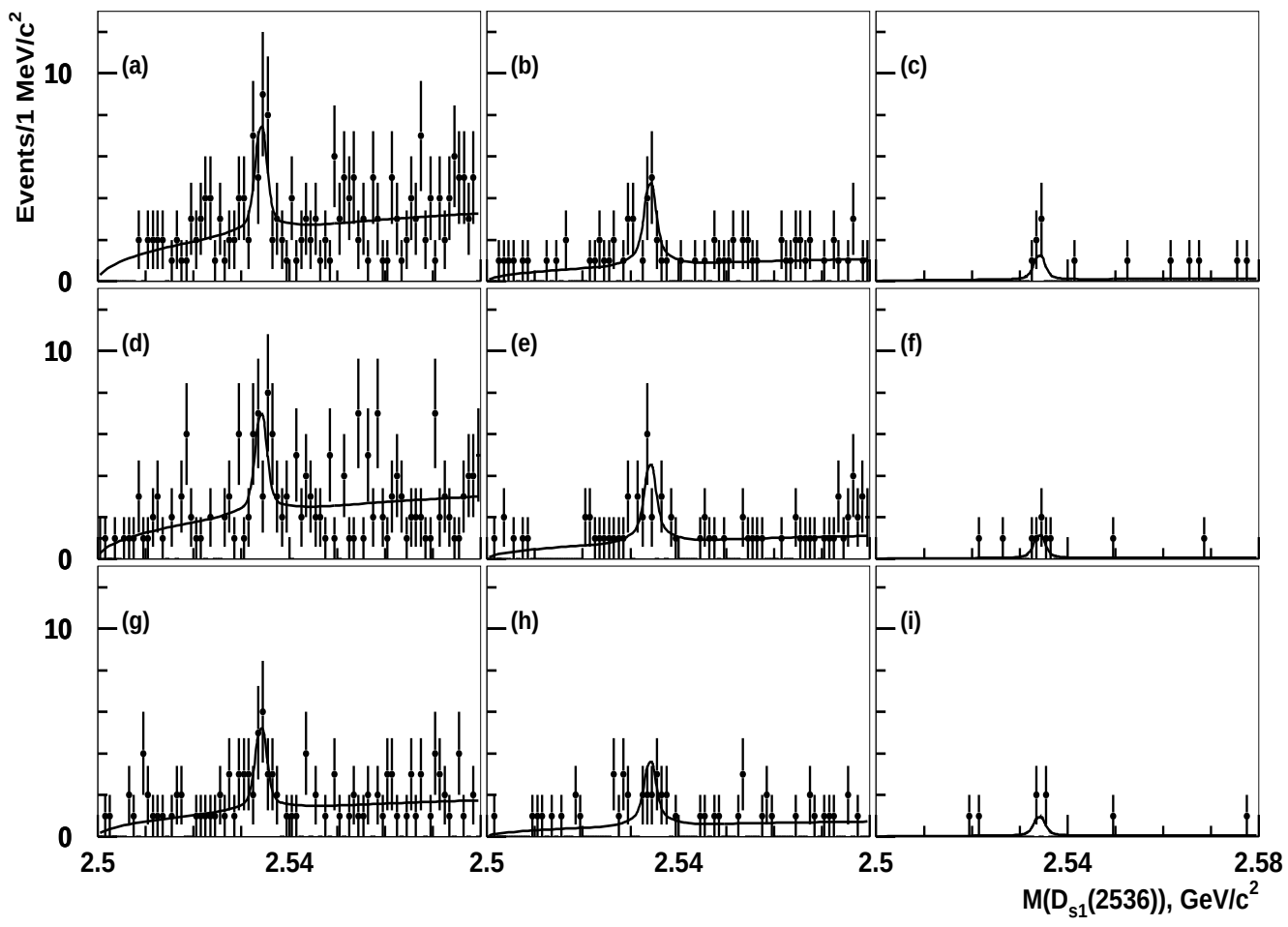

**Figure 19.3.23.** From (Aushev, 2011): The $D_{s1}(2536)^+$ mass distributions for: $B^+ \to \overline{D}^0 D_{s1}(2536)^+$ (a,b,c), $B^0 \to D^- D_{s1}(2536)^+$ (d,e,f), $B^0 \to D^{*-} D_{s1}(2536)^+$ (g,h,i) decays followed by the $D_{s1}(2536)^+$ decays to: $D^{*0}K^+$ with $D^{*0} \to D^0\gamma$ (a,d,g); $D^{*0}K^+$ with $D^{*0} \to D^0\pi^0$ (b,e,h) and $D^{*+}K_S^0$ with $D^{*+} \to D^0\pi^+$ (c,f,i). The curves show results of the simultaneous fit.

**Table 19.3.12.** From (Aushev, 2011). Fitted product $\mathcal{B}$: $\mathcal{B}(B \to \overline{D}^{(*)} D_{s1}(2536)^+) \times \mathcal{B}(D_{s1}(2536)^+ \to D^{*+}K_S^0 + D^{*0}K^+)$.

| Decay mode | Product $\mathcal{B}$ $[10^{-4}]$ |
|---|---|
| $B^+ \to \overline{D}^0 D_{s1}(2536)^+$ | $3.97 \pm 0.85 \pm 0.56$ |
| $B^0 \to D^- D_{s1}(2536)^+$ | $2.75 \pm 0.62 \pm 0.36$ |
| $B^0 \to D^{*-} D_{s1}(2536)^+$ | $5.01 \pm 1.21 \pm 0.70$ |

#### 19.3.4.6 Precision measurements of $D_{s1}(2536)^+$ properties

*BABAR* measured the mass of the $D_{s1}(2536)^+$ with a significant improvement compared to the world average, and, for the first time, measured directly its decay width, instead of reporting an upper limit only (Lees, 2011d).

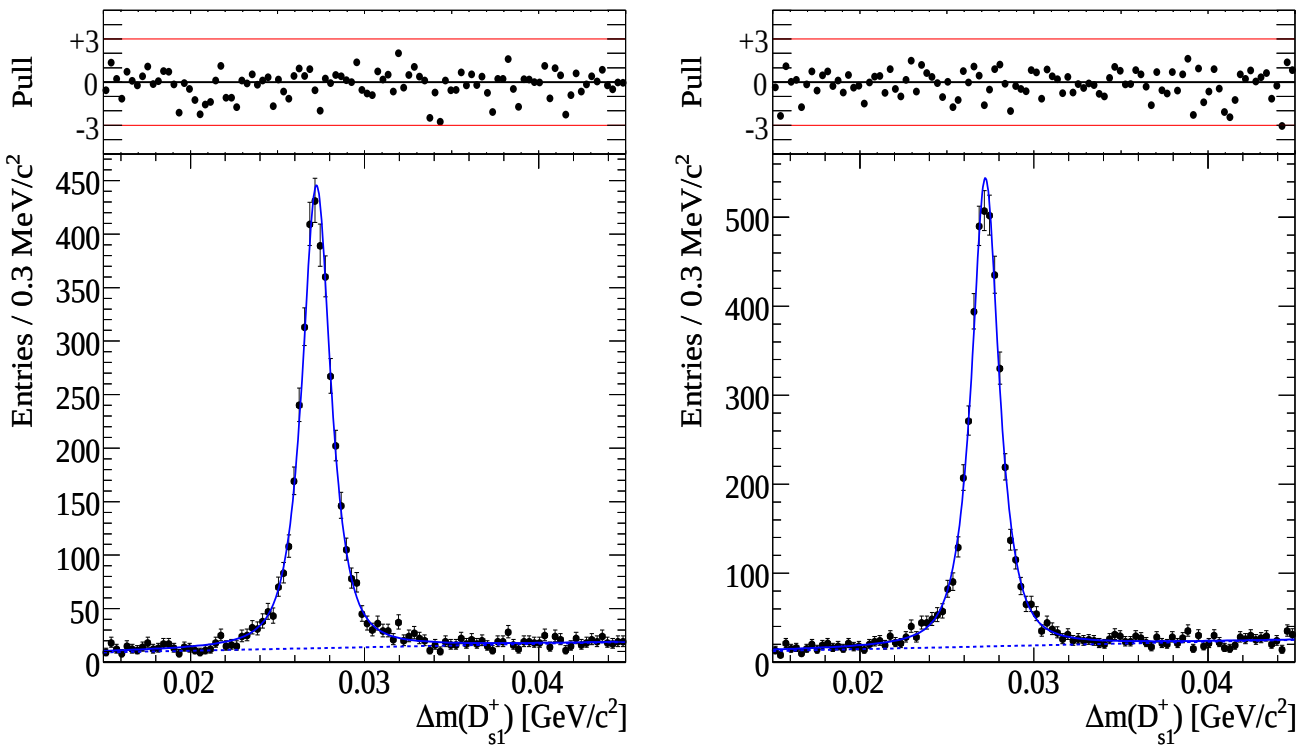

**Figure 19.3.24.** From (Lees, 2011d). $\Delta m(D_{s1}(2536)^+)$ distribution for (left) the $D^0 \to K^-\pi^+$ and (right) the $D^0 \to K^-\pi^+\pi^+\pi^-$ decay modes. The solid line is the fit function as described in the text, the dotted line indicates the background contribution. The normalized fit residuals are shown on top.

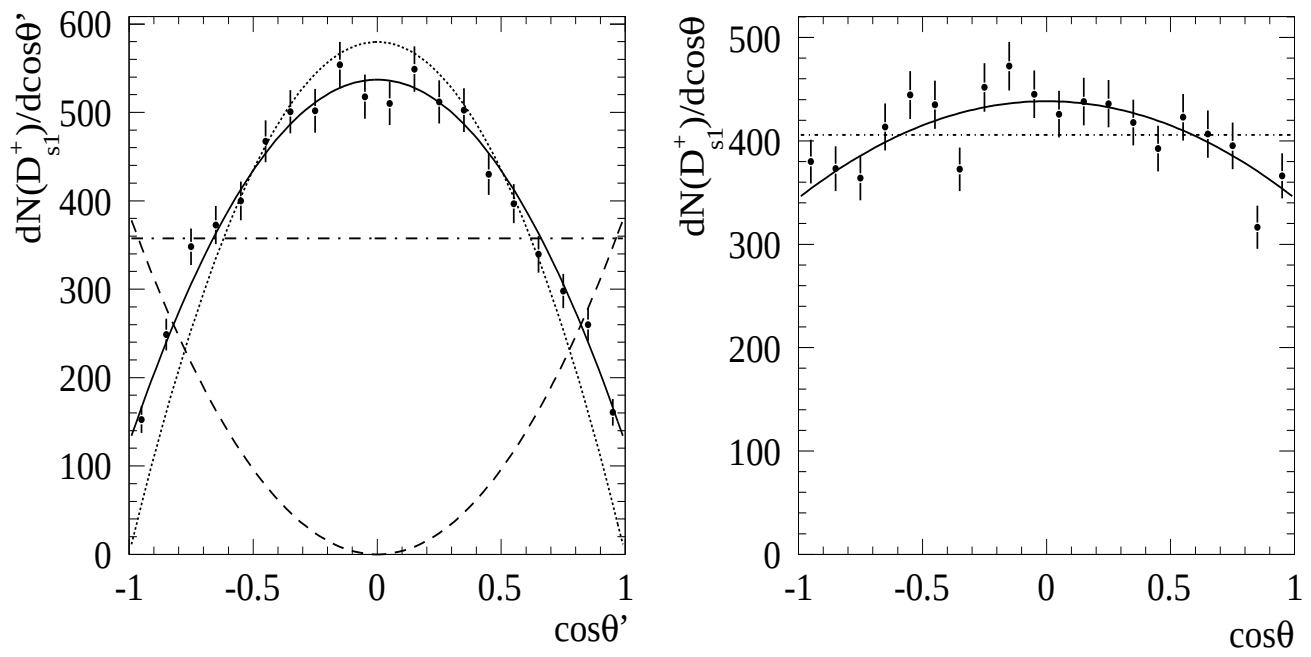

**Figure 19.3.25.** From (Lees, 2011d). Left: Efficiency-corrected $D_{s1}(2536)^+$ signal yield as function of $D^{*+}$ decay angle $\theta'$. Lines correspond to the following hypotheses fitted to the data: $J^P = 1^+, 2^-, 3^+, \ldots$ with $S$- and $D$-wave (solid); $1^+, 2^-, 3^+, \ldots$ with $S$-wave only (dash-dotted); $0^-$ (dashed); $1^-, 2^+, 3^-, \ldots$ (dotted). $0^+$ is forbidden. Right: Efficiency-corrected $D_{s1}(2536)^+$ signal yield as function of $D_{s1}(2536)^+$ decay angle $\theta$. Assuming $J^P = 1^+$, the data are fitted with hypotheses of pure $S$-wave (dotted line); $S$- and $D$-wave (solid line) $D_{s1}(2536)^+$ decays.

**Table 19.3.13.** From (Lees, 2011d). Combined results for mass and width of the $D_{s1}(2536)^+$.

| Parameter | Value [MeV/$c^2$] |
|---|---|
| $M(D_{s1}(2536)^+)$ | $2535.08 \pm 0.01 \pm 0.15$ |
| $M(D_{s1}(2536)^+) - M(D^{*+})$ | $524.83 \pm 0.01 \pm 0.04$ |
| $\Gamma(D_{s1}(2536)^+)$ | $0.92 \pm 0.03 \pm 0.04$ |

The $D_{s1}(2536)^+$ is reconstructed in $D^{*+}K^0_S$, with $D^{*+} \to D^0\pi^+$ and $D^0 \to K^-\pi^+$, $K^-\pi^+\pi^+\pi^-$. To improve the resolution, the mass difference $\Delta m(D_{s1}(2536)^+) = M(D_{s1}(2536)^+) - M(D^{*+}) - M(K^0_S)$ is examined. Combinatorial background and events from $B$ decays are suppressed by requiring a CM momentum $p^* > 2.7$ GeV/$c$.

The samples for the two $D^0$ decays are examined separately, the $\Delta m(D_{s1}(2536)^+)$ spectrum for the $K^-\pi^+$ and $D^0 \to K^-\pi^+\pi^+\pi^-$ modes are shown in Fig. 19.3.24. The signal is described with a convolution of a relativistic BW lineshape and a $p^*$ dependent, multi-Gaussian detector resolution parameterization. The dominant systematic uncertainties are related to track reconstruction, the signal lineshape modeling and detector resolution parameterization. Results from the fits to the two $\Delta m(D_{s1}(2536)^+)$ spectra are combined, and the final numbers are given in Table 19.3.13.

The large and clean sample of $D_{s1}(2536)^+$ signal candidates reconstructed by *BABAR*, enables a study of the spin-parity and decay properties of the $D_{s1}(2536)^+$. In this analysis the $D^{*+}K^0_S$ system is produced inclusively, and therefore the origin of the $D_{s1}(2536)^+$ is not known. Thus, for the $J^P$ study, the decay angle of the $D^{*+}$ ($\theta'$) is examined. The $\theta'$ is measured between the $D^0$ momentum in the $D^{*+}$ CM system and the $D^{*+}$ momentum in the $D_{s1}(2536)^+$ system; the resulting angular distribution is influenced by the spin of the $D_{s1}(2536)^+$. The $\cos\theta'$ distribution for both subsamples combined, is shown in Fig. 19.3.25 (left). Fits with several models used, show a clear preference for unnatural spin-parity values ($J^P = 1^+, 2^-, 3^+, \ldots$), described by a distribution proportional to $\cos^2\theta' + \beta\sin^2\theta'$. The parameter $\beta$ is the squared ratio of the $D^{*+}$ amplitudes with helicities 0 and $\pm 1$, and its measured value of $0.23 \pm 0.02$ clearly indicates a $D$-wave contribution to the decay $D_{s1}(2536)^+ \to D^{*+}K^0_S$ ($\beta = 1$ is the case of a pure $S$-wave decay.).

The $D_{s1}(2536)^+$ decay angle ($\theta$), defined as the angle between the $D^{*+}$ momentum in the $D_{s1}(2536)^+$ CM system and the $D_{s1}(2536)^+$ momentum in the $e^+e^-$ system, is shown in Fig. 19.3.25. It supports the hypothesis with both $S$- and $D$-wave contributing to the $D_{s1}(2536)^+$ decay amplitude. Assuming $J^P = 1^+$ for the $D_{s1}(2536)^+$, the data are expected to have a distribution proportional to $1 + t\cos^2\theta$. The fitted coefficient $t$ in combination with the measured $\beta$, yields $\rho_{00} = 0.48 \pm 0.03$, consistent with the Belle result (see below), with $\rho_{00}$ giving the probability that the $D_{s1}(2536)^+$ helicity is zero.

Belle studied the $D_{s1}(2536)^+$ state, inclusively produced in the $e^+e^- \to c\bar{c}$ continuum, using 462 fb$^{-1}$ of data (Balagura, 2008). A new decay mode, $D_{s1}(2536)^+ \to D^+\pi^-K^+$ is observed, with the $D^+\pi^-$ and $K^+\pi^-$ two-body systems consistent with phase-space distributions. Its branching fractions is measured with respect to the normalization mode $D_{s1}(2536)^+ \to D^{*+}K^0_S$ as:

$$\frac{\mathcal{B}(D_{s1}(2536)^+ \to D^+\pi^-K^+)}{\mathcal{B}(D_{s1}(2536)^+ \to D^{*+}K^0)} = (3.27 \pm 0.18 \pm 0.37)\%. \tag{19.3.21}$$

This ratio is obtained from the $D_{s1}(2536)^+$ signal yields measured from the $M(D^+\pi^-K^+)$ and $M(D^{*+}K^0_S)$ mass spectra, shown in Figure 19.3.26, with the secondary particles reconstructed as: $D^+ \to K^-\pi^+\pi^+$ and $K^0_S\pi^+$, $D^{*+} \to D^0\pi^+$, $D^0 \to K^-\pi^+$, $K^-\pi^+\pi^+\pi^-$ and $K^0_s\pi^+\pi^-$. Instead of the common $p^*$ cut, the analysts applies a requirement on the scaled momentum ($x_p$) of the $D_{s1}(2536)$ candidates to be larger than 0.8. The scaled momentum is defined as $x_p \equiv p^*/p^*_{\rm max}$, where $p^*$ is the $D_{s1}(2536)$ momentum in the $e^+e^-$ CM frame, while $p^*_{\rm max}$ indicates the maximum kinematically allowed momentum in this frame.

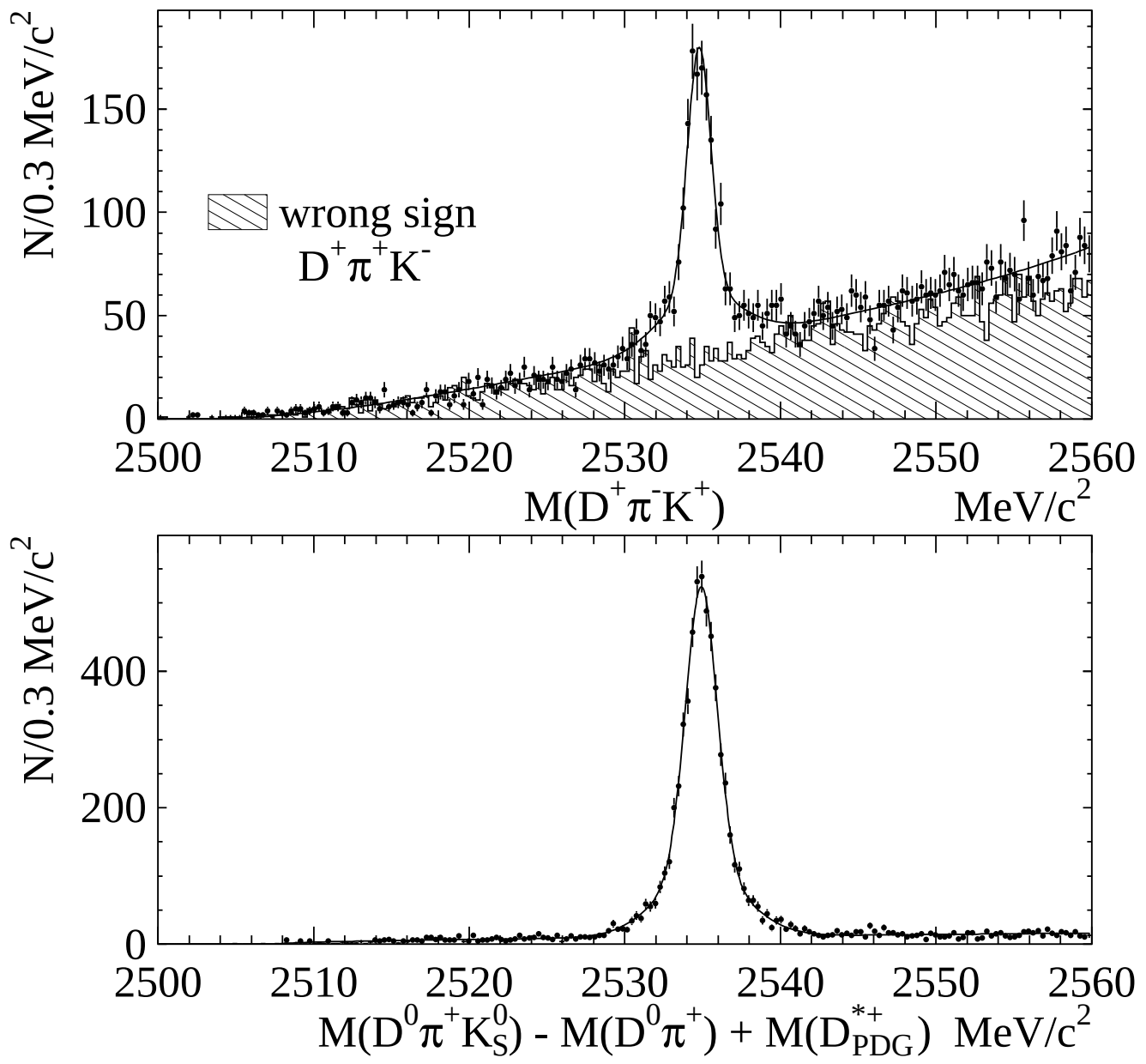

**Figure 19.3.26.** From (Balagura, 2008): Mass spectra of the $D^+\pi^-K^+$ (top) and $D^{*+}K^0_S$ (bottom) final states. The hatched histogram shows a spectrum of the *wrong-sign* $D^+\pi^+K^-$ combinations.

The mass distributions in Fig. 19.3.26 are fitted independently with a double Gaussian describing the $D_{s1}(2536)$ signal. The $D_{s1}(2536)$ mass, measured with respect to its PDG value of $2535.35 \pm 0.34 \pm 0.50\,{\rm MeV}/c^2$, is $-0.57 \pm 0.04\,{\rm MeV}/c^2$ for the $D_{s1}(2536) \to D^+\pi^-K^+$ and $-0.43 \pm 0.02\,{\rm MeV}/c^2$ for the $D_{s1}(2536) \to D^{*+}K^0_S$. The mass resolution is about 1.5 MeV/$c^2$, thus too large to measure the $D_{s1}(2536)$ width.

For the $D_{s1}(2536)^+ \to D^{*+}K^0_S$ sample as shown in Fig. 19.3.26, Belle performed a full three-dimensional angular analysis. Such an analysis allows to study the partial wave structure of the $D_{s1}(2536)$ decay. Assuming $J^P = 1^+$ for the $D_{s1}(2536)$, the kinematics of the $D_{s1}(2536) \to D^{*+}K^0_S$ decay can be described by three angles $\alpha$, $\beta$ and $\gamma$, defined as shown in Fig. 19.3.27. Then the angular distribution in the helicity formalism is expressed as:

$$N(\alpha,\beta,\gamma) = \tfrac{9}{4\pi(1+2R_\Lambda)} \times \Big( \cos^2\!\gamma \big[\rho_{00}\cos^2\!\alpha + \tfrac{1-\rho_{00}}{2}\sin^2\!\alpha\big] + R_\Lambda \sin^2\!\gamma \big[\tfrac{1-\rho_{00}}{2}\sin^2\!\beta + \cos^2\!\beta\,\big(\rho_{00}\sin^2\!\alpha + \tfrac{1-\rho_{00}}{2}\cos^2\!\alpha\big)\big] + \tfrac{\sqrt{R_\Lambda}(1-3\rho_{00})}{4} \sin2\alpha \sin2\gamma \cos\beta \cos\xi \Big), \tag{19.3.22}$$

where $\rho_{00}$ is the longitudinal $D_{s1}(2536)$ polarization, and $\sqrt{R_\Lambda}e^{i\xi} \equiv \frac{A_{1,0}}{A_{0,0}}$ denotes the ratio of the $D^{*+}$ amplitudes with helicities of respectively $\pm1$ and 0, which are related

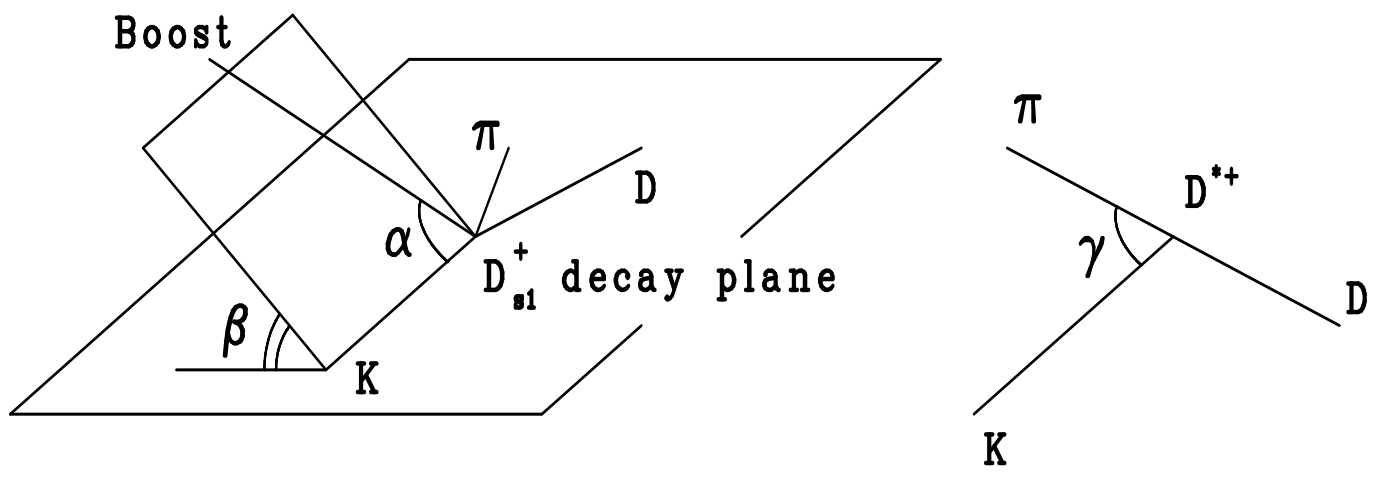

**Figure 19.3.27.** Definitions of the angles: $\alpha$ is the $D_{s1}(2536)$ helicity angle measured in the $D_{s1}(2536)$ CM frame as the angle between the boost direction in the $e^+e^-$ CM frame and the $K^0_S$ momentum; $\beta$ is the angle between the plane formed by these two vectors and the $D_{s1}(2536)$ decay plane measured also in the $D_{s1}(2536)$ rest frame; $\gamma$ is the $D^{*+}$ helicity angle between the $\pi^+$ and $K^0_S$ momenta in the $D^{*+}$ rest frame.

to $S$- and $D$-wave amplitudes in the $D_{s1}(2536)$ decay:

$$A_{1,0} = (S + D/\sqrt{2})/\sqrt{3} \quad A_{0,0} = (S - \sqrt{2}D)/\sqrt{3}. \tag{19.3.23}$$

To measure the phase $\xi$ and, thus, unambiguously determine the partial widths, the full three-dimensional angular analysis is necessary, as after integration over any of the angles, the $\cos\xi$ interference term in Eq. (19.3.22) vanishes.

The probability density function for the $D_{s1}(2536)$ signal is given by Eq. (19.3.22) which includes efficiency corrections obtained in the $(\cos\alpha, \beta, \cos\gamma)$ angular space determined from MC simulation; the background contribution is modeled in that space using the $M(D^{*+}K^0_S)$ sideband regions. The three-dimensional unbinned maximum likelihood fit to the $D_{s1}(2536) \to D^{*+}K^0_S$ signal region data yields:

$$\frac{A_{1,0}}{A_{0,0}} \equiv \sqrt{R_\Lambda}e^{i\xi} = a\ e^{\pm i\cdot b} \tag{19.3.24}$$

where $a = \sqrt{3.6 \pm 0.3 \pm 0.1}$, $b = 1.27 \pm 0.15 \pm 0.05$, and $\rho_{00} = 0.490 \pm 0.012 \pm 0.004$, showing that the $D_{s1}(2536)$ spin prefers to align transversely to the momentum in the $x_p > 0.8$ region. Figure 19.3.28 shows one-dimensional projections of the fitted data together with the fit result. The good agreement of the data with theoretical predictions for the angular distribution of the axial meson, identifies the spin-parity of the $D_{s1}(2536)$ to be $1^+$. The fit results given in Eq. (19.3.24) translate into the partial-wave amplitude ratio of:

$$\frac{D}{S} = ce^{\pm i\cdot d}, \tag{19.3.25}$$

where $c = 0.63 \pm 0.07 \pm 0.02$ and $d = 0.76 \pm 0.03 \pm 0.01$. This shows that the $S$-wave amplitude dominates and its contribution to the total width is $\frac{\Gamma_S}{\Gamma_S+\Gamma_D} = \frac{1}{1+|D/S|^2} = 0.72 \pm 0.05 \pm 0.01$, where $\Gamma_S$ and $\Gamma_D$ denote the $S$- and $D$-wave partial widths.

The result in Eq. (19.3.25) disagrees with HQET which predicts the pure $D$-wave decay of the $D_{s1}(2536)$. However, HQET breaking caused by finite $c$-quark mass, can lead to mixing between the axial $c\bar{s}$ states having $j_q = \frac{1}{2}$ and $j_q = \frac{3}{2}$. Then, similarly to Eq. (19.3.14), the physical states, $D_{s1}(2460)^+$ and $D_{s1}(2536)^+$, can be expressed as linear combinations of $S$- and $D$-wave amplitudes. As a result, the $D_{s1}(2536)$ can contain an admixture of the $J^P = 1^+$ state with $j_q = \frac{1}{2}$ and decaying in pure $S$ wave. Since the energy release in the $D_{s1}(2536)^+ \to D^{*+}K^0_S$ decay is small, the $D$-wave is strongly suppressed by the centrifugal barrier factor $(q/q_0)^5$ and the $S$-wave contribution, proportional to $q/q_0$, can be significantly enhanced even if the mixing itself is small. Here $q$ denotes the relative momentum of the $D_{s1}(2536)$ decay products in the $D_{s1}(2536)$ rest frame, while $q_0$ is a characteristic momentum scale of this reaction. Based on the measured $D/S$ and input on the parameter $q_0$, theoretical models can calculate the $c\bar{s}$ mixing angle (Godfrey, 2005b).

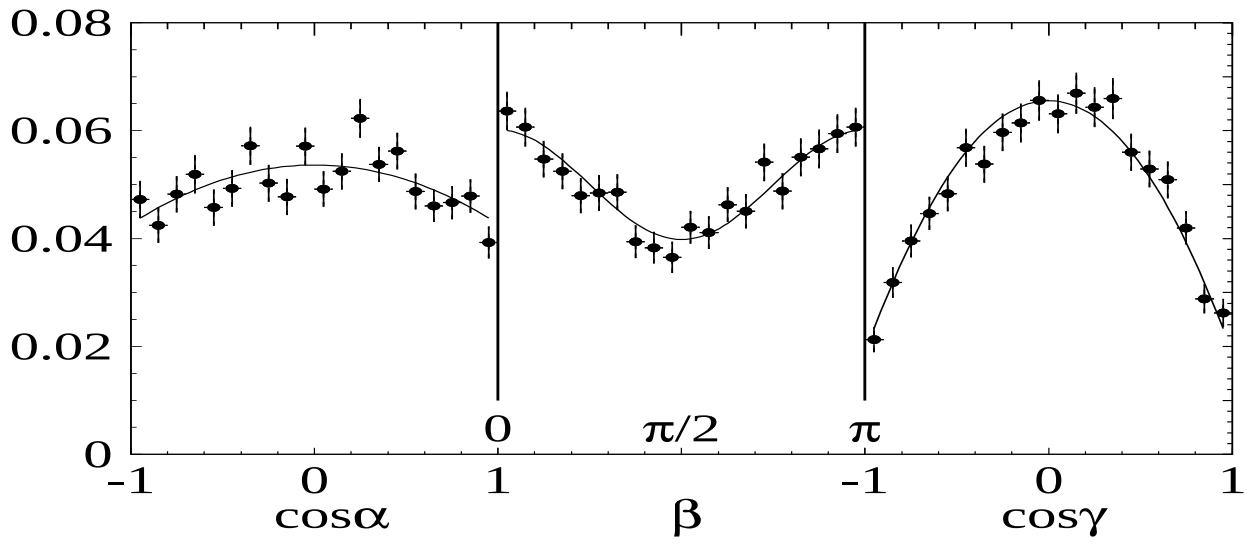

**Figure 19.3.28.** From (Balagura, 2008): Background-subtracted, efficiency-corrected and normalized one-dimensional projections of $(\cos\alpha, \beta, \cos\gamma)$ described in the text. The solid curves show projections of the three-dimensional fit.

Some information on the mixing could be also inferred from the ratio of branching fractions of the radiative decays $D_{s1}(2460)^+ \to D_s^+\gamma$ and $D_{s1}(2460)^+ \to D_s^{*+}\gamma$. However, only evidence exists for the $D_{s1}(2460)^+ \to D_s^{*+}\gamma$ (see Section 19.3.4.4), and an average of Belle results $\frac{\mathcal{B}(D_{s1}(2460)^+ \to D_s^{*+}\gamma)}{\mathcal{B}(D_{s1}(2460)^+ \to D_s^+\gamma)} = 0.31 \pm 0.14$ (Abe, 2004a; Krokovny, 2003b), gives a constraint of $\tan(\omega + \omega_0) = 0.8 \pm 0.4$, where $\omega$ is the mixing angle, while $\omega_0$ is a rotation angle between the $j_q$ and ${}^{(2S+1)}P_1$ bases, and $\tan\omega_0 = -\sqrt{2}$. The ${}^{(2S+1)}P_1$ basis is convenient here, as only the ${}^1P_1$ state in $D_{s1}(2460)^+$ undergoes an electric dipole transition to the $D_s^+$, while only the ${}^3P_1$ one to the $D_s^{*+}$ (Godfrey, 2005b; Yamada, Suzuki, Kazuyama, and Kimura, 2005).

### 19.3.4.7 New $D_{sJ}^+$ mesons decaying to $D^{(*)}K$.

The potential models predict higher orbitally-excited $c\bar{s}$ states, as well as a spectrum of states belonging to the next level ($n = 2$) of radial excitations. Within the $c\bar{s}$ spectrum above the $D^{(*)}K$ threshold, the radially-excited states $2\,{}^3S_1$ and $3\,{}^2S_1$ (possibly with admixtures of $D$-waves) are predicted to lie respectively at about $2.73\,\mathrm{GeV}/c^2$ and $3.1\,\mathrm{GeV}/c^2$ (Godfrey and Isgur, 1985). One expects a large total production rate of these $2S$ and $3S$ states in $B$ decays, respectively at a level of about 1% and 0.1% (Close and Swanson, 2005). However, since the potential model predictions for the $P$-wave $c\bar{s}$ states have failed, their evaluations must be revised to provide reliable masses and widths of the higher excitations.

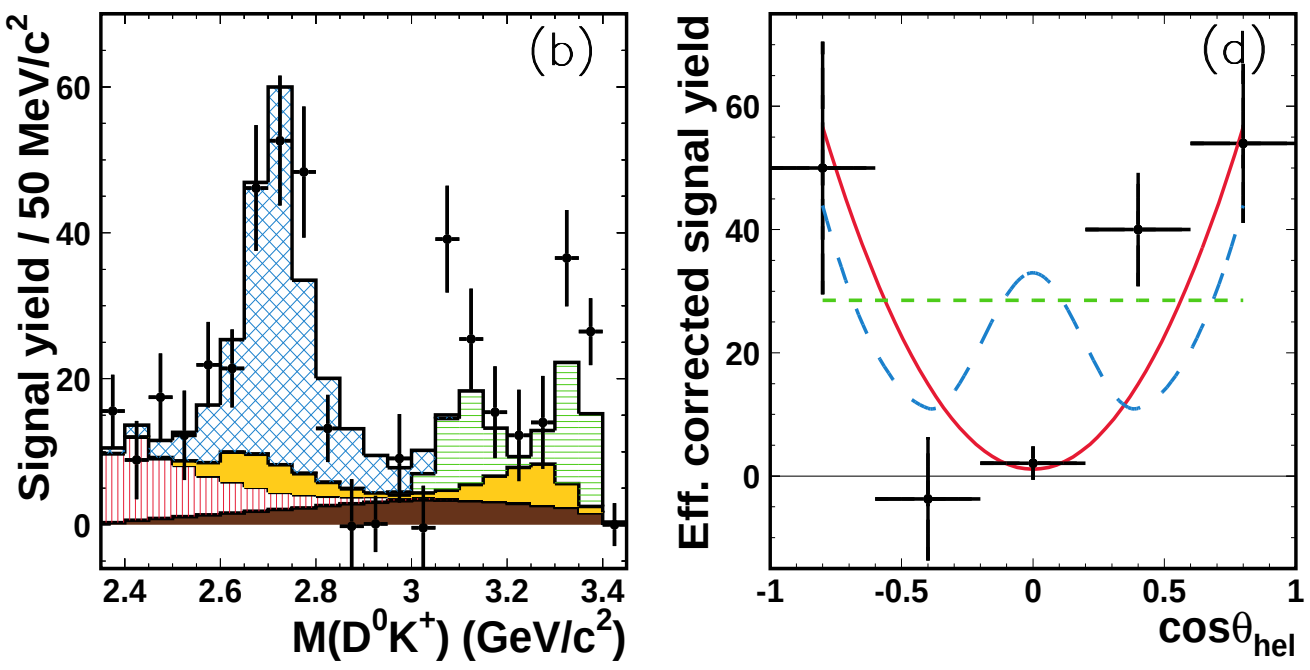

**Figure 19.3.29.** From (Brodzicka, 2008). Left: Background-subtracted $M(D^0K^+)$ distribution for $B^+ \to \overline{D}{}^0D^0K^+$ with contribution from $D_{s1}^*(2710)^+$ (blue , cross shaded histogram), reflections from $\psi(3770)$ (green, horizontal shaded) and $\psi(4160)$ (yellow, full light) and non-resonant contributions (brown (dark) and red (vertical shaded) obtained from MC simulations based on the fit results. Right: Background-subtracted and efficiency corrected $D_{s1}^*(2710)^+$ helicity-angle distribution compared to predictions for $J = 0$ (green, dashed), 1 (red, full line) and 2 (blue, large dash).

The mentioned excitations could appear as intermediate resonances in $B \to \overline{D}{}^{(*)}D^{(*)}K$ decays. Belle, in the study of the decay $B^+ \to \overline{D}{}^0D^0K^+$, observed a new $c\bar{s}$ resonance, the $D_{s1}^*(2710)^+$, decaying to $D^0K^+$ (Brodzicka, 2008). The $D_{s1}^*(2710)^+$ peak is clearly visible in Figure 19.3.29, showing the $M(D^0K^+)$ distribution of the $B$ signal candidates, which is obtained from fits to the $\Delta E$-$m_{\rm ES}$ distributions for a given $M(D^0K^+)$ bin. A limited reconstructed $B$ signal yield does not allow for a full Dalitz-plot analysis, instead the $D_{s1}^*(2710)^+$ parameters are measured from the fit to the background-subtracted $M(D^0K^+)$ spectrum, where reflections from charmonia decaying to $\overline{D}{}^0D^0$, $\psi(3770)$ and $\psi(4160)$, are also taken into account. The charmonium yields are estimated from fits to the $M(\overline{D}{}^0D^0)$ projection, while their shapes are based on $B^+ \to \psi K^+$ MC simulations. The $D_{s1}^*(2710)^+$ mass is measured to be $(2708 \pm 9^{+11}_{-10})\,\mathrm{MeV}/c^2$, its width $(108 \pm 23^{+36}_{-31})\,\mathrm{MeV}/c^2$. The systematic uncertainties include effects of possible interference between the $D_{s1}^*(2710)^+$ and the $\psi(4160)$. The spin-parity of the $D_{s1}^*(2710)^+$ is established to be $1^-$ from the helicity-angle distribution presented in Figure 19.3.29.

*BABAR* explored the $DK$ mass spectrum first using 240 $\mathrm{fb}^{-1}$ (Aubert, 2006ag) and then, with 470 $\mathrm{fb}^{-1}$, studied also the $D^*K$ system (Aubert, 2009au). In these analyses first observations of $D_{s1}^*(2710)^+$, $D_{sJ}^*(2860)^+$ and $D_{sJ}(3040)^+$ resonances were reported. The measured $D_{s1}^*(2710)^+$ parameters and properties are in agreement with those obtained for this state in $B$ decays, as reported above. The *BABAR* study is performed inclusively, with the $D^{(*)}K$ systems separated from the $B\overline{B}$ background by means of the CM momentum $p^* > 3.5$ GeV/$c$. To re-move combinations of the $K$ and $D^{(*)}$ mesons originating from opposite-side jets, $\cos\theta_K > -0.8$ is required, with $\theta_K$ defined as the angle between the $K$ direction and the direction opposite to the laboratory frame in the $D^{(*)}K$ rest frame.

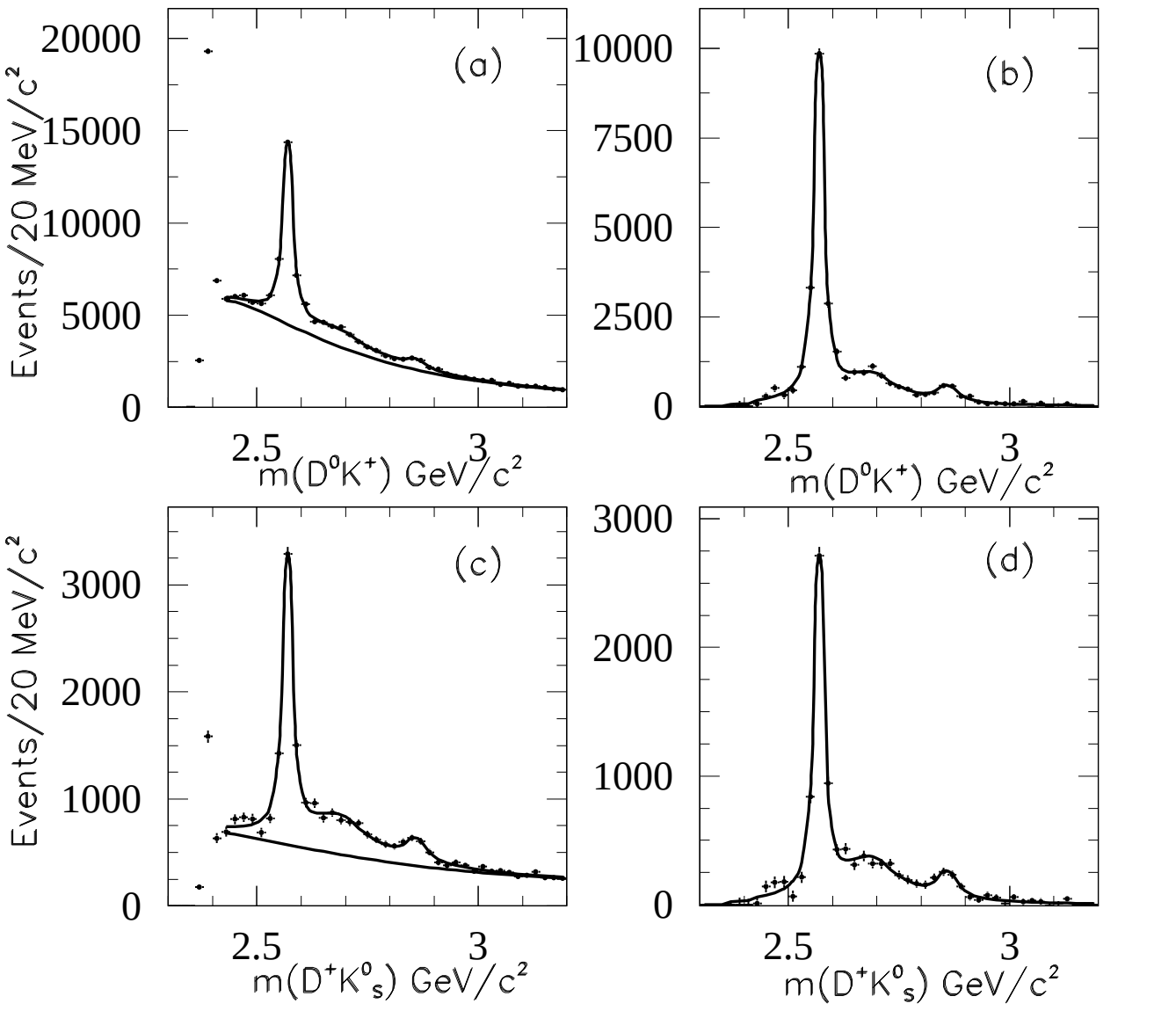

**Figure 19.3.30.** From (Aubert, 2009au). Sideband-subtracted $DK$ invariant mass distributions for (a) $D^0K^+$, (c) $D^+K^0_S$; (b) and (d) show the background-subtracted mass spectra, respectively.

**Table 19.3.14.** From (Aubert, 2009au). Resonance parameters obtained from the fits to the $DK$ and the $D^*K$ mass spectra. Masses and widths are given in units of MeV/$c^2$. Uncertainties are statistical only.

| System | $D^*_{s1}(2710)^+$ | $D^*_{sJ}(2860)^+$ | $D_{sJ}(3040)^+$ |
|---|---|---|---|
| D K | $m = 2710.0 \pm 3.3$<br>$\Gamma = 178 \pm 19$ | $m = 2860.0 \pm 2.3$<br>$\Gamma = 53 \pm 6$ | |
| $D^*K$ | $m = 2712 \pm 3$<br>$\Gamma = 103 \pm 8$ | $m = 2865.2 \pm 3.5$<br>$\Gamma = 44 \pm 8.3$ | $m = 3042 \pm 9$<br>$\Gamma = 214 \pm 34$ |

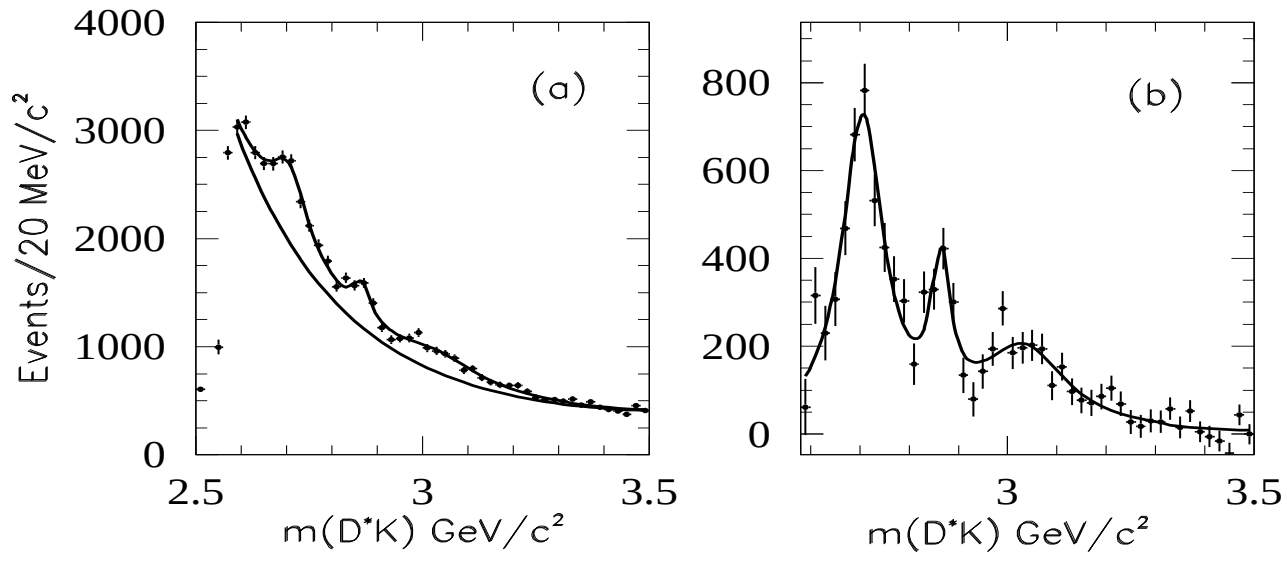

**Figure 19.3.31.** From (Aubert, 2009au). (a) Fit to the $D^*K$ invariant mass spectrum, (b) residuals after subtraction of the fitted background.

In a study of the $D^0K^+$ and $D^+K^0_S$ final states, $D^0 \to K^-\pi^+$ and $D^+ \to K^-\pi^+\pi^+$ are reconstructed. The $D^0K^+$ and $D^+K^0_S$ mass spectra, with $D$ sidebands subtracted, are shown in Fig. 19.3.30(a,c). A single bin peak at 2.4 GeV/$c^2$ is a reflection from decays of $D_{s1}(2536)^+$ to $D^*K$ in which the $\pi^0$ or $\gamma$ from the $D^*$ decay is missed; the $D_{s1}(2536)^+$ decay to $DK$ is forbidden. In addition to a prominent $D^*_{s2}(2573)^+$ signal, there are also broad structures associated with the $D^*_{s1}(2710)^+$ and $D^*_{sJ}(2860)^+$ mesons. A simultaneous binned $\chi^2$ fit was performed to the two mass spectra, with the background described by a threshold function and the $D^*_{s2}(2573)^+$, $D^*_{s1}(2710)^+$ and $D^*_{sJ}(2860)^+$ peaks parameterized with relativistic BW lineshapes where spin 2 was assumed for $D^*_{s2}(2573)^+$, 1 for $D^*_{s1}(2710)^+$ and spin 0 for $D^*_{sJ}(2860)^+$. Figures 19.3.30(b,d) show the $M(D^0K^+)$ and $M(D^+K^0_S)$ mass distributions with fitted background subtracted. The fit gives the parameters listed in Table 19.3.14.

In a study of the $D^*K$ system, $D^*$ resonances are reconstructed as $D^{*0} \to D^0\pi^0$, $D^{*+} \to D^+\pi^0$ and $D^{*+} \to D^0\pi^+$, with $D^0 \to K^-\pi^+$, $D^0 \to K^-\pi^+\pi^+\pi^-$ and $D^+ \to K^-\pi^+\pi^+$. The total $D^*K$ mass spectrum, $D^*$ sideband-subtracted and summed over all the channels, is shown in Fig. 19.3.31, where above the $D_{s1}(2536)^+$ signal (not shown because out of scale in the figure), there are structures present around 2.71, 2.86 and 3.04 GeV/$c^2$. A binned minimum $\chi^2$ fit is performed to the combined $D^*K$ mass spectrum in the region 2.58-3.48 GeV/$c^2$, with the background parameterized with an exponential function providing a good description of the MC in the same mass range. The $D^+_{sJ}$ peaks are described with relativistic BW lineshapes, with $J^P = 1^-$ and $J^P = 3^-$ assumed respectively for $D^*_{s1}(2710)^+$ and $D^*_{sJ}(2860)^+$, and an angular momentum $L = 1$, $L = 3$, and $L = 0$ for $D^*_{s1}(2710)^+$, $D^*_{sJ}(2860)^+$ and $D_{sJ}(3040)^+$, respectively. Since the width values for the resonances are much larger than the mass resolutions, effects of the latter are ignored in the fit. The resonance parameters resulting from the fit are given in Table 19.3.14 and the corresponding fitted curves are shown in Fig. 19.3.31. The width of the $D^*_{s1}(2710)^+$ differs somewhat between the $M(DK)$ and $M(D^*K)$ fits, while the parameter of the $D^*_{sJ}(2860)^+$ are consistent for both decay modes.

The observation of both $D^*_{s1}(2710)^+$ and $D^*_{sJ}(2860)^+$ decays to both $DK$ and $D^*K$, implies that they have natural parity $J^P = 1^-, 2^+, 3^-, \ldots$ ($J^P = 0^+$ is ruled out because of the $D^*K$ decay). A further test of the quantum numbers was performed through analysis of the helicity angle ($\theta_h$), computed as the angle between the $\pi$ from the $D^*$ decay and the kaon, in the $D^*$ rest frame. The efficiency-corrected $D^*_{s1}(2710)^+$ and $D^*_{sJ}(2860)^+$ yields are plotted in Fig. 19.3.32, together with the normalized expectations for the natural parity *i.e.* $1-\cos^2\theta_h$, which give $\chi^2$/NDF of 18.7/5 and 6.3/5, respectively. The large value for the $D^*_{s1}(2710)^+$ is related to the large uncertainties in the background parameterization.

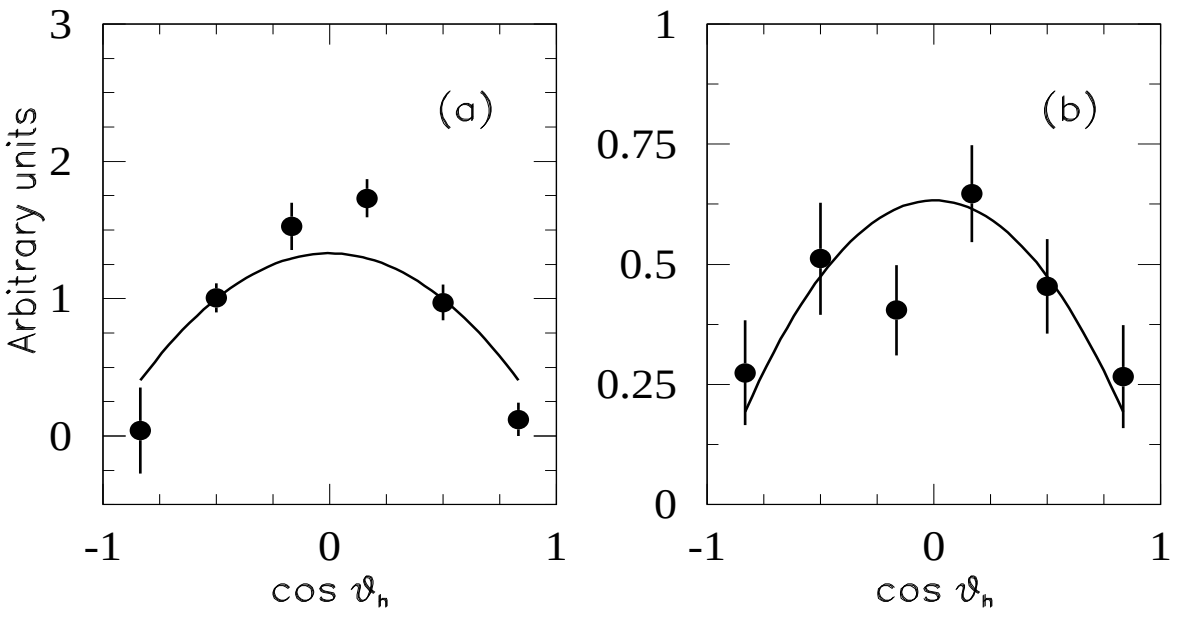

**Figure 19.3.32.** From (Aubert, 2009au). Distributions of $\cos\theta_h$ for (a) the $D^*_{s1}(2710)^+$ and (b) the $D^*_{sJ}(2860)^+$. The curves are expectations for the natural parity.

*BABAR* measures the following branching fraction ratios:

$$\frac{\mathcal{B}(D^*_{s1}(2710)^+ \to D^*K)}{\mathcal{B}(D^*_{s1}(2710)^+ \to DK)} = 0.91 \pm 0.13 \pm 0.12 \quad (19.3.26)$$

$$\frac{\mathcal{B}(D^*_{sJ}(2860)^+ \to D^*K)}{\mathcal{B}(D^*_{sJ}(2860)^+ \to DK)} = 1.10 \pm 0.15 \pm 0.19, \quad (19.3.27)$$

using the selected decay channels: $D^{*0}K^+$ with $D^{*0} \to D^0\pi^0$, $D^0 \to K^-\pi^+$, and $D^{*+}K^0_S$ with $D^{*+} \to D^+\pi^0$, $D^+ \to K^-\pi^+\pi^+$, to reduce systematic uncertainties.

The $D^*_{s1}(2710)^+$ can be either a radial excitation or an $L = 2$ orbital excitation, which are both predicted in this mass region. Observation of the $D^*_{s1}(2710)^+ \to D^*K$ decay with rate comparable to that for the $DK$, suggests that the $D^*_{s1}(2710)^+$ is a radial excitation of the $D^*_s$ state (Colangelo, De Fazio, Nicotri, and Rizzi, 2008). Interpretations of the $D^*_{sJ}(2860)^+$ and $D_{sJ}(3040)^+$ are still unknown.

### 19.3.5 Conclusions

Results from the $B$ Factories have given an important input to the experimental status of the charm meson spectroscopy. The observation of the broad states belonging to the $L = 1$ $c\bar{u}$ multiplets has validated outstanding predictions of the potential models. Properties of this multiplet need to be refined with higher statistics, while their production has to be studied in various processes. Finding new decay modes, especially radiative ones, could provide further tests of the theory. On the other hand, the $c\bar{s}$ spectroscopy has revealed some surprises. At the moment, it is still not completely clear if the $D_{s1}(2460)^+$ and $D^*_{s0}(2317)^+$ are fully understood in terms of $Q\bar{q}$ mesons, or if we need to include more complex quark configurations to understand their properties. In the simplest scenario, where the $D_{s1}(2460)^+$ and $D^*_{s0}(2317)^+$ are the conventional $L = 1$ $c\bar{s}$ mesons, the potential models may need some serious modifications.

The large data samples collected by Belle and *BABAR* also allowed for precision measurements of already known mesons, like the $D_{s1}(2536)^+$. The new excited $D_J$ and $D_{sJ}$ mesons observed open a spectrum of the higher orbital and radial excitations that can be studied. They however need confirmation and more studies to allow their final assignment. The LHCb experiment, dedicated charm factories like BES III, and super flavor factories could shed more light on this sector in the future.

## 19.4 Charmed baryon spectroscopy and decays

***Editors:***
*Matthew Charles (BABAR)*
*Ruslan Chistov (Belle)*

In this section we discuss the physics of charmed baryons at the $B$ Factories. We begin with the spectroscopy of these states, then turn to the weak decays of their lowest-lying ground states. Finally, we consider the use of these decays to study the properties of light baryons.

The data samples analysed include both baryons produced in the decays of $B$ mesons (see Section 17.12) and from the $e^+e^- \to c\bar{c}$ continuum (see Section 24.1). Both produce large samples of charmed baryons. In practice, most of the inclusive analyses discussed in this section use only the continuum sample, since a cut on the center-of-mass momentum of the charmed baryon, $p^*$, of around $E_{\rm CM}/4$ is very effective at suppressing combinatoric background, but also removes the entire $B$ sample in the process. However, there is an important exception: exclusive $B$ meson decays provide an initial state with known $J^P$, which is extremely helpful for measuring the $J^P$ quantum numbers of charmed baryons.

### 19.4.1 Spectroscopy

#### 19.4.1.1 Introduction

*Overview*

The spectroscopy of charmed baryons is beautiful and intricate. With three quarks there are numerous degrees of freedom, giving rise to many more states than in the charmed meson sector. At the same time, the large difference in mass between the charm quark and the light quarks provides a natural way to classify and understand these states by making use of the symmetries emerging in Heavy Quark Effective Theory (HQET). The spectrum of known singly-charmed states can be thought of in three broad regimes: the ground states, which are a vindication of the constituent quark model; the low-lying excited states, which are described well by heavy quark symmetries; and higher excited states, where the situation is murkier.

The naming convention for charmed baryons is to take a light baryon, replace one or more $s$ quarks with $c$ quarks, and add a $c$ subscript for every quark replaced. Isospin is unchanged. For example, $\Lambda$ denotes an *sud* baryon with isospin zero, and so $\Lambda_c^+$ denotes a *cud* baryon with isospin zero. Likewise, $\Xi_c^0$ denotes a *csd* baryon and $\Xi_{cc}^+$ denotes a *ccd* baryon. A summary of charmed baryon states is given in Table 19.4.1. Following this convention, the experimentally known $C = 1$ baryon states[147] are summarized in Fig. 19.4.1. Spin-parity assignments follow the PDG (Beringer et al. (2012)); note that in many cases these are assigned based on quark model expectations rather than measurements.

**Table 19.4.1.** Baryon flavor states, isospin, and quark content. The symbol $q$ denotes a $u$ or $d$ quark. Baryons with beauty are not included for brevity, but follow a similar pattern.

| Symbol | $I$ | Content |
|---|---|---|
| $N$ ($p$,$n$) | 1/2 | $udq$ |
| $\Delta$ | 3/2 | $qqq$ |
| $\Lambda$ | 0 | $sud$ |
| $\Sigma$ | 1 | $sqq$ |
| $\Xi$ | 1/2 | $ssq$ |
| $\Omega$ | 0 | $sss$ |
| $\Lambda_c$ | 0 | $cud$ |
| $\Sigma_c$ | 1 | $cqq$ |
| $\Xi_c$ | 1/2 | $csq$ |
| $\Omega_c$ | 0 | $css$ |
| $\Xi_{cc}$ | 1/2 | $ccq$ |
| $\Omega_{cc}$ | 0 | $ccs$ |
| $\Omega_{ccc}$ | 0 | $ccc$ |

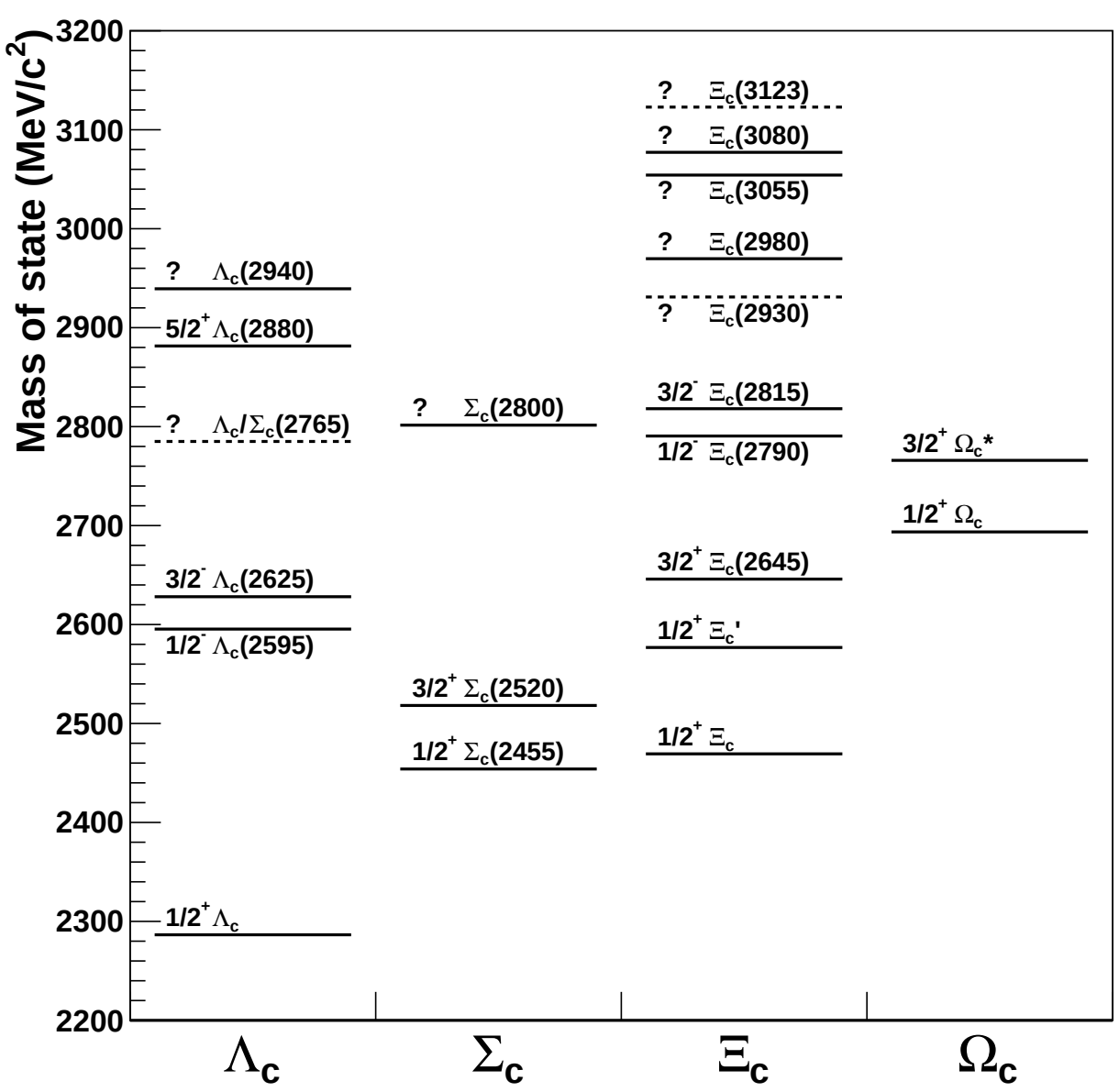

**Figure 19.4.1.** Summary of the known charmed baryon states. The spin-parity $J^P$ is given, or marked "?" if not known. States whose existence is unclear (one-star rating in PDG) are marked with a dashed line.

[147] Strongly-decaying states are distinguished by their mass following the PDG convention, *e.g.* $\Xi_c(2645)$. The $\Xi_c'$ and $\Omega_c^*$ do not decay strongly and are therefore not labelled by their mass, although the latter is sometimes referred to as the $\Omega_c(2770)$ in the literature.

*Quark model for ground states*

In the constituent quark model (Gell-Mann, 1964; Zweig, 1964a,b), baryons composed of $u, d, s, c$ quarks can be classified into $SU(4)$ multiplets according to the symmetry of their flavor, spin, and spatial wavefunctions. All states in a given $SU(4)$ multiplet have the same angular momentum $J$, and parity $P$, but can have different quark flavors. For excited states with multiple units of orbital angular momentum the number of possible multiplets becomes large, but for the ground states the picture is much simpler. This $SU(4)$ symmetry is badly broken due to the large charm mass. and thus different states with the same conserved quantum numbers will mix, and baryons are not pure three-quark objects—but it works remarkably well for the ground states.

Quarks are fermions, so the baryon wavefunction must be overall antisymmetric under quark interchange.[148] Baryons are color singlets, and so have an antisymmetric color wavefunction. In the ground state, the orbital angular momentum $L$ is zero ($S$-wave) and the spatial wavefunction is symmetric. Therefore, the product of the spin and flavor wavefunctions must also be symmetric for ground-state baryons. There are two ways this can be accomplished: both wavefunctions can be fully symmetric, or both can have mixed symmetry with the product being symmetric.

In concrete terms, we can consider a singly-charmed baryon to consist of a heavy $c$ quark and a light diquark with spin-parity $j^p$. Assuming isospin symmetry and letting $q$ denote a $u$ or $d$ quark, there are four possibilities for the flavor content of the diquark:

– $qq$ with isospin 0 (flavor antisymmetric);
– $qq$ with isospin 1 (flavor symmetric);
– $sq$ with isospin 1/2 (either);
– $ss$ with isospin 0 (flavor symmetric).

These correspond to the $\Lambda_c$, $\Sigma_c$, $\Xi_c$, and $\Omega_c$ states, respectively. The diquark wavefunction must be antisymmetric under quark interchange. Its color wavefunction is antisymmetric and in the ground state its spatial wavefunction is symmetric, so it may be either flavor-symmetric and spin-symmetric ($j^p = 1^+$) or flavor-antisymmetric and spin-antisymmetric ($j^p = 0^+$). Combining the diquark with the charm quark gives rise to the possible states set out in Table 19.4.2 and illustrated in Fig. 19.4.2, where the multiplets of the full $SU(3)$ symmetry (formed by the $u$, $d$, and $s$ quarks) are shown. Those with $J^P = 1/2^+$ are all members of the same multiplet as the proton, and those with $J^P = 3/2^+$ are all members of the same multiplet as the $\Delta$ and $\Omega$ (Fig. 19.4.3). Note that there is a second isospin doublet of $\Xi_c$ states with $J^P = 1/2^+$, denoted $\Xi_c'$.

The constituent quark model predicts relations between the masses of these states as well as their existence and quantum numbers. These were expressed for the light baryons as sum rules (see Gell-Mann (1962); Okubo (1962)):

$$(m_N + m_\Xi)/2 = (3m_\Lambda + m_\Sigma)/4, \qquad (19.4.1)$$

$$m_{\Sigma^*} - m_\Delta = m_{\Xi^*} - m_{\Sigma^*} = m_\Omega - m_{\Xi^*}, \qquad (19.4.2)$$

$$m_{\Sigma^*} - m_\Sigma = m_{\Xi^*} - m_\Xi, \qquad (19.4.3)$$

of which the first is the famous Gell-Mann-Okubo rule. These can be thought of as expressing the mass as the sum of the valence quark masses plus a hyperfine (spin-spin) coupling. This can be parameterized in various ways,[149] such as (De Rujula, Georgi, and Glashow (1975)):

$$M = A + B' \sum_i \Delta m_i + C' \sum_{i>j} \mathbf{s}_i \cdot \mathbf{s}_j \left(m_q - \Delta m_i - \Delta m_j\right), \qquad (19.4.4)$$

where $A$, $B'$, and $C'$ are constants, $m_q$ is the mass of a light quark, $\Delta m_i = m_i - m_q$ is the mass difference of the $i^{\text{th}}$ quark compared to $m_q$, and $\mathbf{s}_i$ is the spin of the $i^{\text{th}}$ quark. When evaluated for the ground state baryons, the sum rules given in Eq. 19.4.1–19.4.3 are recovered. This simple model can also be extended to baryons with heavy quarks. To illustrate its effectiveness, we note that the spectrum and decay pattern of singly-charmed baryon ground states was mapped out in an essentially correct way within about three months of the discovery of charm but it took three decades before all of the states were seen experimentally (Beringer et al., 2012). The last to be discovered was the $\Omega_c^*$. The equal-spacing mass rule still holds for the singly-charmed $J = 3/2^+$ multiplet:

$$m_{\Omega_c^*} - m_{\Xi_c^*} = m_{\Xi_c^*} - m_{\Sigma_c^*}, \tag{19.4.5}$$

but with additional flavors the hyperfine terms become more complicated so the analog of Eq. 19.4.3 is:

$$m_{\Omega_c^*} - m_{\Omega_c} = 2\left(m_{\Xi_c^*} - m_{\Xi_c'}\right) - \left(m_{\Sigma_c^*} - m_{\Sigma_c}\right). \tag{19.4.6}$$

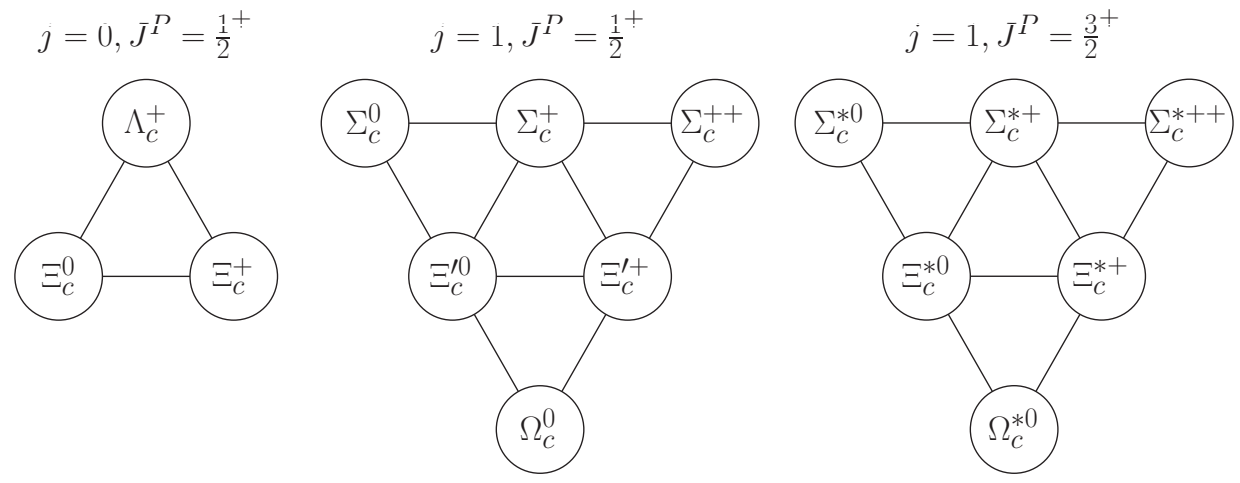

**Figure 19.4.2.** The $SU(3)$ multiplets containing the ground state baryons, grouped according to the spin $j$ of the light diquark and the spin-parity $J^P$ of the baryon.

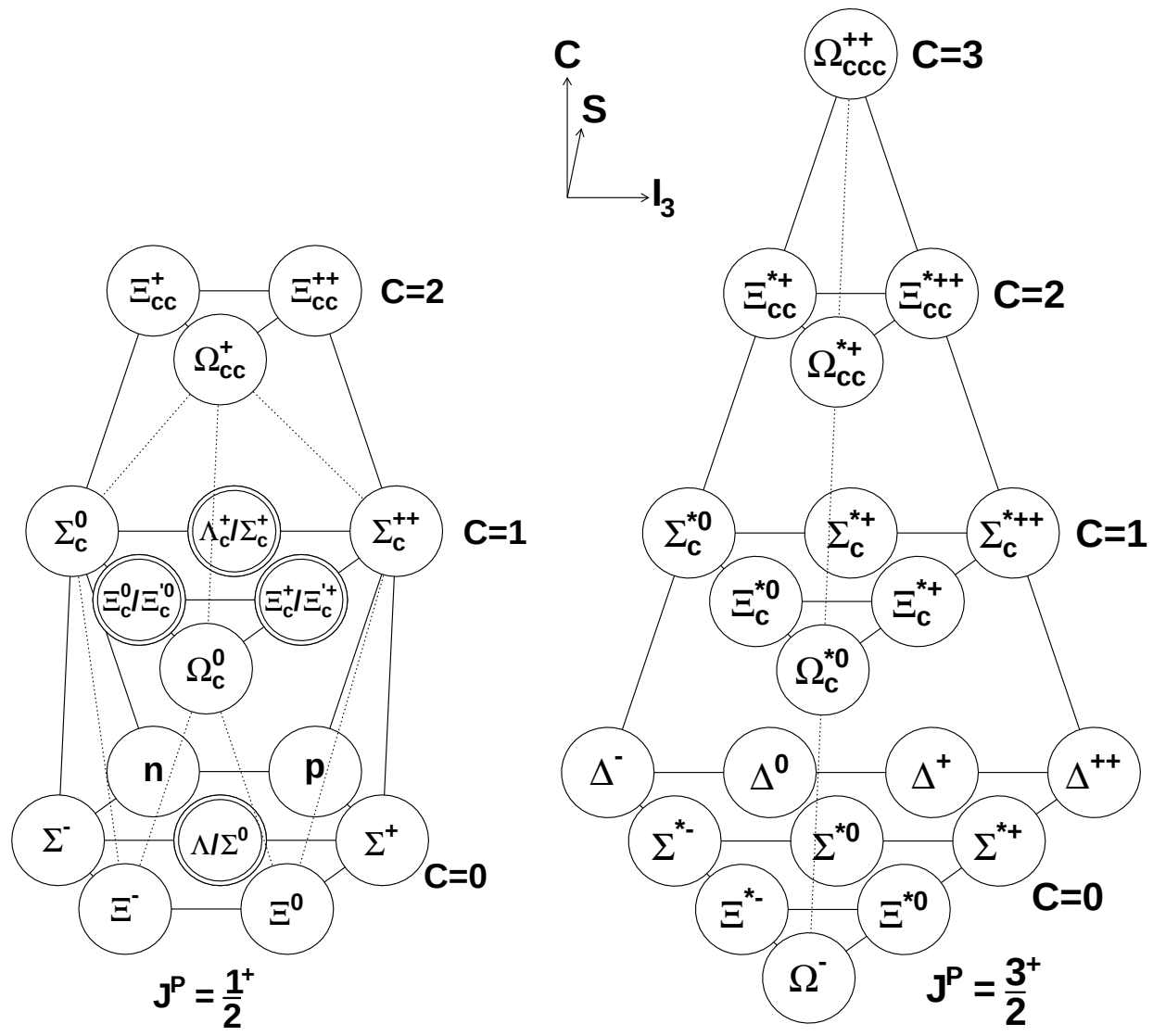

**Figure 19.4.3.** The $SU(4)$ multiplets containing the ground state baryons, arranged by spin-parity ($J^P$), isospin projection ($I_3$), strangeness ($S$), and charm ($C$). A double ring indicates that two states have the same $J^P$, $I_3$, $S$, and $C$ quantum numbers.

[148] Strictly, it only needs to be antisymmetric under interchange of equal-mass quarks, but in order to build the model we assume $SU(4)$ is a good symmetry.

[149] See also Gasiorowicz and Rosner (1981) for a nice review with a different hyperfine interaction term.

**Table 19.4.2.** Summary of the ground state singly-charmed baryons. $S$ denotes a wavefunction that is fully symmetric under interchange of any two quarks; $M_S$ and $M_A$ denote mixed overall symmetry with interchange of the two light quarks being symmetric or antisymmetric, respectively; and $A$ would denote a fully antisymmetric wavefunction.

| Baryon | Diquark | Diquark $I$ | Diquark $j^p$ | Baryon flavor symmetry | Baryon spin symmetry | Baryon $J^P$ |
|---|---|---|---|---|---|---|
| $\Lambda_c$ | $qq$ | 0 | $0^+$ | $M_A$ | $M_A$ | $1/2^+$ |
| $\Sigma_c$ | $qq$ | 1 | $1^+$ | $M_S$ | $M_S$ | $1/2^+$ |
| $\Sigma_c^*$ | $qq$ | 1 | $1^+$ | $S$ | $S$ | $3/2^+$ |
| $\Xi_c$ | $sq$ | 1/2 | $0^+$ | $M_A$ | $M_A$ | $1/2^+$ |
| $\Xi_c'$ | $sq$ | 1/2 | $1^+$ | $M_S$ | $M_S$ | $1/2^+$ |
| $\Xi_c^*$ | $sq$ | 1/2 | $1^+$ | $S$ | $S$ | $3/2^+$ |
| $\Omega_c$ | $ss$ | 0 | $1^+$ | $M_S$ | $M_S$ | $1/2^+$ |
| $\Omega_c^*$ | $ss$ | 0 | $1^+$ | $S$ | $S$ | $3/2^+$ |

Substituting in the current world-average experimental masses for states other than the $\Omega_c^*$ (Beringer et al., 2012), one would calculate $m_{\Omega_c^*}$ to be approximately 2774 MeV/$c^2$ from Eq. 19.4.5 or 2770 MeV/$c^2$ from Eq. 19.4.6. As we will show in Section 19.4.1.4, these simple estimates are in remarkably good agreement with the observed mass.

### *Higher states*

Baryons can be given orbital ($l$) or radial ($k$) excitations. Since in the simplest quark model they are three-body systems there are two degrees of freedom in each case (denoted $\rho$, $\lambda$). For baryons with one heavy quark (mass $M$) and two light quarks (mass $m$), a natural way to specify these is to divide the system into a light diquark and the heavy quark. Taking a simple potential model based on the harmonic oscillator, the energy levels are given by (Klempt and Richard (2010)):

$$E = \sqrt{\frac{K}{m}}\,(3 + 2l_\rho + 4k_\rho) + \sqrt{\frac{K}{\mu}}\,(3 + 2l_\lambda + 4k_\lambda)\,, \tag{19.4.7}$$

where $l_{\rho,\lambda} = 0, 1, 2, ...$ and $k_{\rho,\lambda} = 0, 1, 2, ...$, and $K$ is a constant describing the potential and $\mu = (2/3M + 1/3m)^{-1} \approx 3m$ in the heavy quark limit. Thus, the $\rho$ excitations (within the diquark) require roughly three times as much energy as the corresponding $\lambda$ excitations (between quark and diquark). Therefore the lowest-lying excitations are those with $l_\lambda = 1$ and the other quantum numbers zero, *i.e.* $L = 1$. (Within this band there will be further splitting, *e.g.* due to spin-spin and spin-orbit couplings.) The second band will consist of two groups of states that have comparable energy: those with $l_\lambda = 2$ ($L = 2$) and those with $k_\lambda = 1$ ($L = 0$), with the other quantum numbers being zero. Beyond the second band the degeneracy grows further, but we lack useful experimental data in this region in any case.

We can take this quark-diquark separation one step further for singly-heavy baryons by considering the heavy quark to be essentially a spectator and treating the diquark as a distinct object with its own conserved quantum numbers $j^p$ that is the main actor in decays (in HQET, $j^p$ is the total angular momentum of all light degrees of freedom, see Isgur and Wise (1991)). As a consequence, some transitions that would otherwise be allowed are now forbidden. For example, consider a heavier state with $(J^P, j^p) = (1/2^-, 1^-)$ and a lighter state with $(1/2^+, 0^+)$. If we considered only the overall $J^P$, an $S$-wave ($L = 0$) strong decay of the heavier state to the lighter state plus a pion, $(1/2^- \to 1/2^+\,0^-)$, would be allowed. This channel would dominate and, if well above threshold, would lead to a large width for the resonance. However, conservation of angular momentum forbids the corresponding $S$-wave diquark transition: $(1^- \to 0^+\,0^-)$. A $D$-wave ($L = 2$) transition is allowed, but would be kinematically suppressed. The HQET constraints thus have a concrete effect on the decay pattern of excited states, and imply that some will be narrow.

All this said, it is important to bear in mind that states which share all conserved, external quantum numbers ($J$, $P$, $I$, $C$, $S$) can mix. Therefore we should be careful when interpreting observed resonances as specific expected states, particularly for higher excitations.

### 19.4.1.2 $\Lambda_c$, $\Sigma_c$ families

The known $\Lambda_c$ and $\Sigma_c$ states are shown in Fig. 19.4.1. The lowest-lying is the $\Lambda_c^+$ ground state, which decays weakly (see next section). The most precise measurement of the $\Lambda_c^+$ mass was made by *BABAR* (Aubert, 2005a). Since $\Lambda_c^+$ are produced copiously at the $B$ Factories, the key challenge is control of the systematic uncertainties. These arise from effects which could change the momentum scale, principally uncertainties in the magnetic field and energy loss in material. By selecting $\Lambda_c^+$ decay modes with low energy release ($Q$ value), a large fraction of the reconstructed $\Lambda_c^+$ mass comes directly from the rest masses of the final-state daughters, which are known to high precision, and only a small fraction from the measured 3-momenta. Thus, the effect of the momentum uncertainty on the $\Lambda_c^+$ is reduced. Using the modes $\Lambda_c^+ \to \Lambda K_S^0 K^+$ and $\Lambda_c^+ \to \Sigma^0 K_S^0 K^+$, the mass is found to be $m(\Lambda_c^+) = (2286.46 \pm 0.14)\ \mathrm{MeV}/c^2$. At the time of writing, this is the most precise measurement of an open charm hadron mass and has significantly lower uncertainty than existing theoretical calculations based upon lattice QCD or advanced potential models.

In order of increasing mass, the next states are the ground states $\Sigma_c(2455)$ and $\Sigma_c(2520)$. In both cases these are isotriplets, and the only kinematically allowed strong decay is $\Sigma_c \to \Lambda_c^+\pi$. The $\Sigma_c(2455)$ is one of the few charmed baryons whose angular momentum has been measured. This was accomplished with a sample of fully reconstructed $B^- \to \Lambda_c^+\overline{p}\pi^-$ decays proceeding via an intermediate $\Sigma_c(2455)^0$ (Aubert, 2008aa). In this exclusive production environment where the initial state is known to have $J = 0$, the angular distributions for different spin hypotheses are fully determined (see Sections 12.1 and 19.4.3 for more on the helicity formalism). In this case the helicity angle, $\theta_h$, is defined as the angle between the direction of the $\Lambda_c^+$ in the $\Sigma_c$ rest frame and the direction of the $\Sigma_c$ in the $B^-$ rest frame. *BABAR* found that the $\Sigma_c(2455)^0$ is consistent with $J = 1/2$ and inconsistent with $J = 3/2$ as shown in Fig. 19.4.4, in line with the quark model predictions (Section 19.4.1.1).

The lowest-lying excited states are a pair of $\Lambda_c^+$ not far above the $\Lambda_c^+\pi^+\pi^-$ threshold, the $\Lambda_c(2595)$ and $\Lambda_c(2625)$. These are interpreted as a doublet with the light diquark in a spin-antisymmetric state and one unit of orbital angular momentum between the diquark and the heavy quark ($L = 1$), so that the total $j^p$ of the light degrees of freedom is $1^-$. Adding in the spin-1/2 heavy quark, the total $J^P$ of the baryon is $1/2^-$ for the $\Lambda_c(2595)$ and $3/2^-$ for the $\Lambda_c(2625)$ as shown in the first two rows of Table 19.4.3. The pattern of decays seen by ARGUS (Albrecht et al., 1993b, 1997), E687 (Frabetti et al., 1994a, 1996b), and CLEO (Edwards et al., 1995) leads to the following conclusions:

- The states decay to $\Lambda_c^+\pi^+\pi^-$ but not to $\Lambda_c^+\pi^0$, so they have isospin 0 ($\Lambda_c$) and not 1 ($\Sigma_c$).
- The $\Lambda_c(2595)$ decays predominantly to $\Sigma_c(2455)^0\pi^+$, which is a favored $S$-wave decay: $(1^-, 1/2^-) \to (1^+, 1/2^+)(0^-)$.

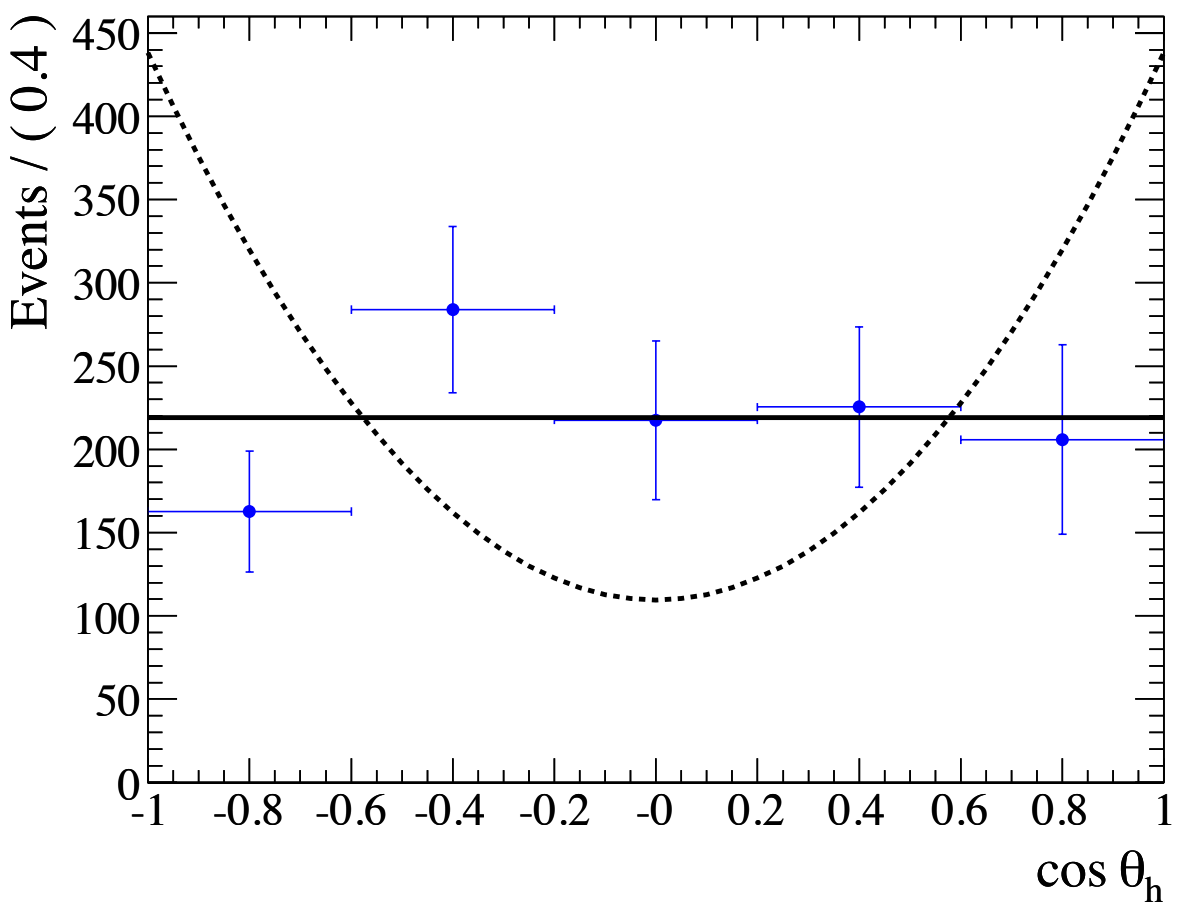

**Figure 19.4.4.** The helicity angle distribution for $\Sigma_c(2455)^0$ candidates in $B^- \to \Sigma_c(2455)^0\overline{p}$ at *BABAR*, corrected for efficiency (Aubert, 2008aa). The distributions expected for the spin-1/2 hypothesis (solid, horizontal line) and spin-3/2 hypothesis (dashed curve) are shown. The data are consistent with $J = 1/2$ and exclude $J = 3/2$ at the $4\sigma$ level.

- The $\Lambda_c(2625)$ does not decay to the kinematically-allowed $\Sigma_c(2455)^0\pi^+$ final state, since this would require a $D$-wave decay: $(1^-, 3/2^-) \to (1^+, 3/2^+)(0^-)$. Instead, it decays to the 3-body final state $\Lambda_c^+\pi^+\pi^-$ via a $P$-wave transition: $(1^-, 3/2^-) \to (0^+, 1/2^+)(0^-)(0^-)$.

**Table 19.4.3.** Possible low-lying excited states in HQET, classified according to the spin-alignment of the two light quarks, the spin-parity of the light degrees of freedom $j^p$, and the spin-parity of the baryon $J^P$.

| Diquark spin | $j^p$ | $J^P$ |
|---|---|---|
| 0 | $1^-$ | $1/2^-$ |
| 0 | $1^-$ | $3/2^-$ |
| 1 | $0^-$ | $1/2^-$ |
| 1 | $1^-$ | $1/2^-$ |
| 1 | $1^-$ | $3/2^-$ |
| 1 | $2^-$ | $3/2^-$ |
| 1 | $2^-$ | $5/2^-$ |

Following Table 19.4.3, we would then expect to see a set of five $\Lambda_c$ states in which the light diquark is in a spin-symmetric arrangement. In these configurations the unit of orbital angular momentum is between the two light quarks ($l_\rho$ in the notation of Eq. 19.4.7) and so the energy levels will be higher than those of the first two $L = 1$ $\Lambda_c$ states discussed above. These heavier states are often denoted $\Lambda_c'$ in the literature. We would also expect to see seven corresponding $\Sigma_c$ states, with an inverted hierarchy due to the different symmetry of the light flavor wavefunction (*i.e.* five lower $\Sigma_c$ isotriplets with $j = 1$ followed by two higher $\Sigma_c'$ isotriplets with $j = 0$). However, looking at Fig. 19.4.1 we see only a handful of known states at higher masses. By implication, the unobserved states either have too small a production cross-section or are too broad to be visible as distinct structures in the current data. Future flavor factories with larger data samples may be able to shed light on them.

Of the remaining known states, two are close in mass: the $\Lambda_c/\Sigma_c(2765)$, seen by CLEO and Belle in the $\Lambda_c^+\pi^+\pi^-$ final state (Artuso et al., 2001; Mizuk, 2007) and the $\Sigma_c(2800)$ isotriplet, seen in $\Lambda_c^+\pi^{+,0,-}$ by Belle (Mizuk, 2005). The isospin of the former state is not known experimentally: the $\Lambda_c^+\pi^+\pi^-$ final state is accessible to both $I = 0$ and $I = 1$, and no results are available for the related $\Lambda_c^+\pi^\pm\pi^0$ final states which would be populated for a $\Sigma_c$ state but not a $\Lambda_c$. The peak is also rather broad and could even be due to multiple overlapping resonances (see Fig. 19.4.5). A $\Sigma_c$ resonance in this region was also observed by *BABAR* in the analysis of $B^- \to \Lambda_c^+\overline{p}\pi^-$ decays mentioned previously. However, the fitted mass was $(2846\pm8\pm10)\,\mathrm{MeV}/c^2$, higher than the world-average mass of the $\Sigma_c(2800)^0$, $(2802^{+4}_{-7})\,\mathrm{MeV}/c^2$ by about $3\sigma$. If this difference is genuine and the states are distinct, the state seen by *BABAR* would be one of the missing $\Sigma_c$ states.

Finally, we turn to the last states: the $\Lambda_c(2880)$ and $\Lambda_c(2940)$, which were studied by both *BABAR* and Belle and are shown in Fig. 19.4.5. (Aubert, 2007ai; Mizuk, 2007). CLEO discovered the $\Lambda_c(2880)$ in the $\Lambda_c^+\pi^+\pi^-$ final state (Artuso et al., 2001) but did not establish its quantum numbers. *BABAR* observed both states in the $D^0p$ channel. This is notable as the first observation of the strong decay of a charmed baryon to a charmed meson and a light baryon. It also allowed the isospin to be determined straightforwardly: no corresponding resonances were seen in the isospin partner channel $D^+p$, thus excluding a $\Sigma_c$ interpretation.

Belle studied the same two resonances in the $\Lambda_c^+\pi^-\pi^+$ final state. The spin of the $\Lambda_c(2880)$, $J$, was measured by Belle in another application of the helicity formalism (Section 12.1)—but this time with an inclusive production environment. Because the initial state is not fixed, the elements of the density matrix were not known *a priori*. In the case that all diagonal elements were equal (unbiased production environment), one would get a flat distribution independent of $J$. However, if they were not equal, the result would be an even polynomial in $\cos\theta_h$ of order $\leq (2J-1)$, where $\theta_h$ is the angle between the direction of the $\Lambda_c^+$ in the $\Sigma_c$ rest frame and the direction of the $\Sigma_c$ in the $\Lambda_c(2880)^+$ rest frame. Thus, a flat distribution would give no discrimination between spins but a higher-order polynomial would exclude lower spins. Belle found that a polynomial of order at least 4 was required, excluding $J = 1/2$ and $3/2$ but consistent with $J = 5/2$ (or higher). They also showed that the relative branching fractions of $\Lambda_c(2880) \to \Sigma_c(2520)\pi$ and $\Lambda_c(2880) \to \Sigma_c(2455)\pi$ were more consistent with HQET predictions for $J^P = 5/2^+$ than $5/2^-$, favoring even parity. This would make the $\Lambda_c(2880)$ an $L = 2$ state from the second excitation band—though, as noted earlier, an admixture of states with the same external quantum numbers cannot be excluded (see Cheng and Chua (2007) for one example).

#### 19.4.1.3 $\Xi_c$ family

Since all three quark flavors are different for the $\Xi_c$, there are many allowed configurations. These may be divided into states for which the light diquark wavefunction is flavor-antisymmetric (analogous to $\Lambda_c$) or flavor-symmetric (analogous to $\Sigma_c$)—for the ground states, we saw this division between the $j = 0$ $\Xi_c$ and the $j = 1$ $\Xi_c'$ and $\Xi_c^*(2645)$ in Table 19.4.2

The masses of the weakly-decaying $\Xi_c^0$ and $\Xi_c^+$ were measured by Belle in several decay modes (Lesiak, 2005), and these results now dominate the current world-average. Several decay modes were considered, and as with the $\Lambda_c^+$ mass measurement those with smaller energy release (notably $\Xi_c^0 \to pK^-K^-\pi^+$) generally had smaller systematic uncertainties. Combining the decay modes, the mass values obtained were $m_{\Xi_c^+} = (2468.1 \pm 0.4^{+0.2}_{-1.4})\,\mathrm{MeV}/c^2$ and $m_{\Xi_c^0} = (2471.0 \pm 0.3^{+0.2}_{-1.4})\,\mathrm{MeV}/c^2$, where the first uncertainty is due to statistical, fitting, and selection effects and the second is due to mass scale uncertainty (evaluated with kinematically similar control modes and with Monte Carlo simulation). The mass splitting was measured to be $m_{\Xi_c^0} - m_{\Xi_c^+} = (2.9 \pm 0.5)\,\mathrm{MeV}/c^2$. Belle and *BABAR* also performed mass measurements of several of the higher states in a variety of decay modes, summarized in Table 19.4.4.

The $\Xi_c'$ and $\Xi_c(2645)$ ground states form a doublet analogous to the $\Sigma_c(2455)$ and $\Sigma_c(2520)$ with expected $(j^p, J^P)$ of $(1^+, 1/2^+)$ and $(1^+, 3/2^+)$, respectively. The former is too light to decay strongly, but the electromagnetic transition $\Xi_c' \to \Xi_c\gamma$ is allowed. *BABAR* performed an angular analysis of $\Xi_c'^0 \to \Xi_c^0(\Xi^-\pi^+)\gamma$ in the helicity formalism, similar to the $\Lambda_c(2880)$ discussed above, and found the data to be consistent with $J = 1/2$ (Aubert, 2006ba). However, due to the inclusive production environment higher spins could not be ruled out.

The low-lying excited states $\Xi_c(2790)$ and $\Xi_c(2815)$ are analogous to the $\Lambda_c(2595)$ and $\Lambda_c(2625)$, and their decays follow a corresponding pattern: $\Xi_c(2790) \to \Xi_c'\pi$, $\Xi_c(2815) \to \Xi_c(2645)\pi$. They were therefore identified as the $1/2^-, 3/2^-$ doublet with $j^p = 1^-$ and the diquark in a flavor-antisymmetric configuration (Alexander et al., 1999; Csorna et al., 2001). Following the $\Lambda_c/\Sigma_c$ analogy, we would then expect to see five low-lying $L = 1$ states with a flavor-symmetric diquark, followed by a small explosion of $L = 2$ states, radially excited states, and higher $L = 1$ states. We have a number of candidates for these states in Fig. 19.4.1, but less information to classify them this time since we cannot use isospin to distinguish them as we do for $\Lambda_c/\Sigma_c$.

The lightest of these, the $\Xi_c(2930)$, was seen in $B^- \to \Lambda_c^+\Lambda_c^- K^-$ decays (Aubert, 2008e). The Dalitz plot[150] is clearly not flat and the $\Lambda_c^+K^-$ projection is consistent with a single resonance with the parameters given in Table 19.4.4. However, given the small sample size and the inability to rule out other explanations (such as two overlapping $\Xi_c$ resonances or a complicated interference pattern between $\Xi_c$ and charmonium resonances) this is considered unconfirmed.

[150] See Section 13 for more on Dalitz plots.

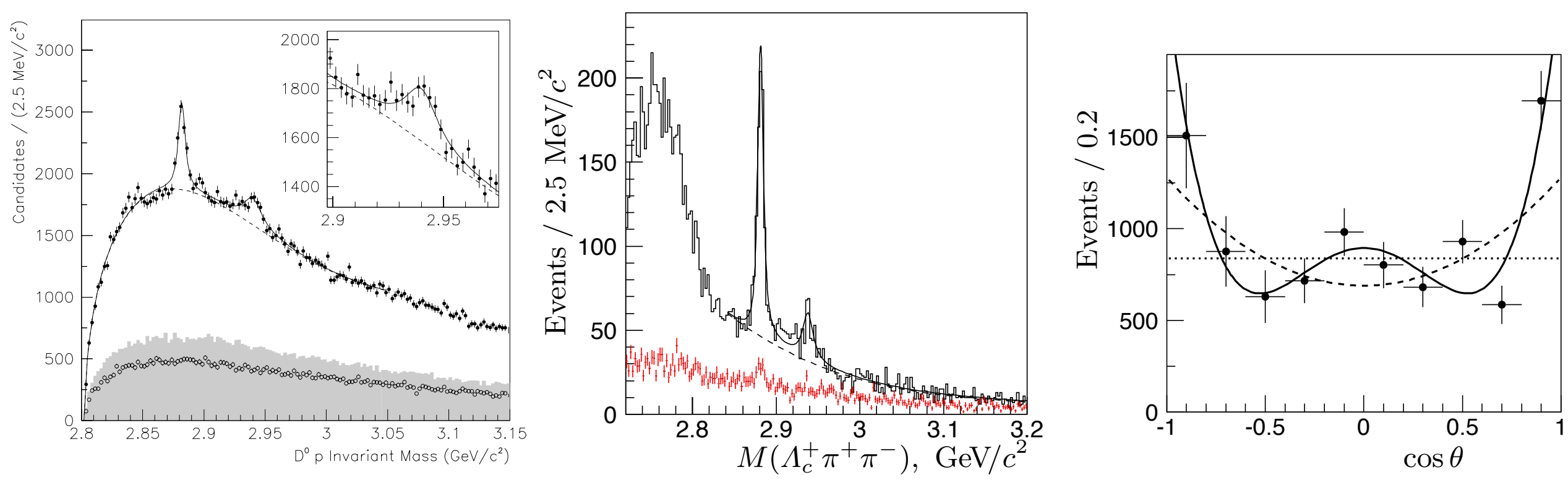

**Figure 19.4.5.** The $\Lambda_c(2880)^+$ and $\Lambda_c(2940)^+$. The two states are visible in the $D^0 p$ mass spectrum (left, *BABAR*) and the $\Lambda_c^+\pi^-\pi^+$ mass spectrum (center, Belle). The $\Lambda_c/\Sigma_c(2765)$ enhancement is also seen at the lower edge of the $m(\Lambda_c^+\pi^-\pi^+)$ spectrum. The helicity angle distribution for $\Lambda_c(2880) \to \Sigma_c(2455)\pi$ (right, Belle) is fitted assuming different spin hypotheses for the $\Lambda_c(2880)$: spin-1/2 (dotted line), spin-3/2 (dashed curve), and spin-5/2 (solid curve). The data are consistent with $J = 5/2$ and exclude $J = 1/2$ and $J = 3/2$ at the $5.5\sigma$ and $4.8\sigma$ levels, respectively. (Aubert, 2007ai; Mizuk, 2007)

**Table 19.4.4.** Mass and width measurements for strongly-decaying $\Xi_c$ states performed at the $B$ Factories. An asterisk indicates that the measurement has statistical significance below $5\sigma$, and a double-asterisk that the significance is below $3\sigma$.

| State | Decay mode | Mass (MeV/$c^2$) | Width (MeV) | Source |
|---|---|---|---|---|
| $\Xi_c(2645)^0$ | $\Xi_c^+\pi^-$ | $2465.7 \pm 0.2^{+0.6}_{-0.7}$ | neg. | Lesiak (2008) |
| $\Xi_c(2645)^+$ | $\Xi_c^0\pi^+$ | $2465.7 \pm 0.2^{+0.6}_{-0.7}$ | neg. | Lesiak (2008) |
| $\Xi_c(2815)^+$ | $\Xi_c(2645)^0\pi^+$ | $2817.0 \pm 1.2^{+0.7}_{-0.8}$ | neg. | Lesiak (2008) |
| $\Xi_c(2815)^0$ | $\Xi_c(2645)^+\pi^-$ | $2820.4 \pm 1.4^{+0.9}_{-1.0}$ | neg. | Lesiak (2008) |
| $\Xi_c(2930)^0$ | $\Lambda_c^+ K^-$ | $2931 \pm 3 \pm 5$ | $36 \pm 7 \pm 11$ | Aubert (2008e) |
| $\Xi_c(2980)^0$ | $\Lambda_c^+ K_S^0 \pi^-$ | $2977.1 \pm 8.8 \pm 3.5$ | 43.5 (fixed) | Chistov (2006b)** |
| $\Xi_c(2980)^0$ | $\Lambda_c^+ K_S^0 \pi^-$ | $2972.9 \pm 4.4 \pm 1.6$ | $31 \pm 7 \pm 8$ | Aubert (2008f)** |
| $\Xi_c(2980)^0$ | $\Xi_c(2645)^+\pi^-$ | $2965.7 \pm 2.4^{+1.1}_{-1.2}$ | $15 \pm 6 \pm 3$ | Lesiak (2008) |
| $\Xi_c(2980)^+$ | $\Lambda_c^+ K^-\pi^+$ | $2978.5 \pm 2.1 \pm 2.0$ | $43.5 \pm 7.5 \pm 7.0$ | Chistov (2006b) |
| $\Xi_c(2980)^+$ | $\Lambda_c^+ K^-\pi^+$ | $2969.3 \pm 2.2 \pm 1.7$ | $27 \pm 8 \pm 2$ | Aubert (2008f) |
| $\Xi_c(2980)^+$ | $\Xi_c(2645)^0\pi^+$ | $2967.7 \pm 2.3^{+1.1}_{-1.2}$ | $18 \pm 6 \pm 3$ | Lesiak (2008) |
| $\Xi_c(3055)^+$ | $\Lambda_c^+ K^-\pi^+$ | $3054.2 \pm 1.2 \pm 0.5$ | $17 \pm 6 \pm 11$ | Aubert (2008f) |
| $\Xi_c(3077)^0$ | $\Lambda_c^+ K_S^0 \pi^-$ | $3082.8 \pm 1.8 \pm 1.5$ | $5.2 \pm 3.1 \pm 1.8$ | Chistov (2006b)* |
| $\Xi_c(3077)^0$ | $\Lambda_c^+ K_S^0 \pi^-$ | $3079.3 \pm 1.1 \pm 0.2$ | $5.9 \pm 2.3 \pm 1.5$ | Aubert (2008f)* |
| $\Xi_c(3077)^+$ | $\Lambda_c^+ K^-\pi^+$ | $3076.7 \pm 0.9 \pm 0.5$ | $6.2 \pm 1.2 \pm 0.8$ | Chistov (2006b) |
| $\Xi_c(3077)^+$ | $\Lambda_c^+ K^-\pi^+$ | $3077.0 \pm 0.4 \pm 0.2$ | $5.5 \pm 1.3 \pm 0.6$ | Aubert (2008f) |
| $\Xi_c(3123)^+$ | $\Lambda_c^+ K^-\pi^+$ | $3122.9 \pm 1.3 \pm 0.3$ | $4.4 \pm 3.4 \pm 1.7$ | Aubert (2008f)* |

The remaining resonances were all seen in the $\Lambda_c^+\overline{K}\pi^+$ isodoublet of final states (and, in the case of the $\Xi_c(2980)$, in $\Xi_c(2645)\pi$). The $\Xi_c(2980)$ and $\Xi_c(3077)$ were discovered by Belle in $\Lambda_c^+\overline{K}\pi$ (see Fig. 19.4.6) and confirmed by *BABAR*. Since this is a three-body decay it could proceed via an intermediate $\Sigma_c$. *BABAR* tested this by fitting a two-dimensional *p.d.f.* in $m(\Lambda_c^+\pi), m(\Lambda_c^+\overline{K}\pi)$. It was found that approximately half of the $\Xi_c(2980)$ decays to this final state proceed through an intermediate $\Sigma_c(2455)$ with the rest non-resonant. By contrast, most if not all of the $\Xi_c(3077)$ decays to this final state proceed via $\Sigma_c(2455)$ or $\Sigma_c(2520)$ with approximately equal branching fractions to each. Because the $\Xi_c(2980)$ is close to threshold on the scale of its natural width, especially with an intermediate $\Sigma_c$, the available phase space changes significantly across the resonance. Different handling of this threshold behavior is the reason for the mild tension in the fitted $\Xi_c(2980)$ masses between (Chistov, 2006b) and (Aubert, 2008f). The masses measured in the $\Xi_c(2645)\pi^+$ final state (Lesiak, 2008), which is far from threshold, are consistent with the *BABAR* treatment, although the widths are smaller than either experiment saw in $\Lambda_c^+\overline{K}\pi$. Requiring an intermediate $\Sigma_c$ reduces the background levels, and by doing this *BABAR* was able to identify two further candidate states, the $\Xi_c(3055)$ and $\Xi_c(3123)$. The latter had a limited statistical significance ($3\sigma$), and needs confirmation.

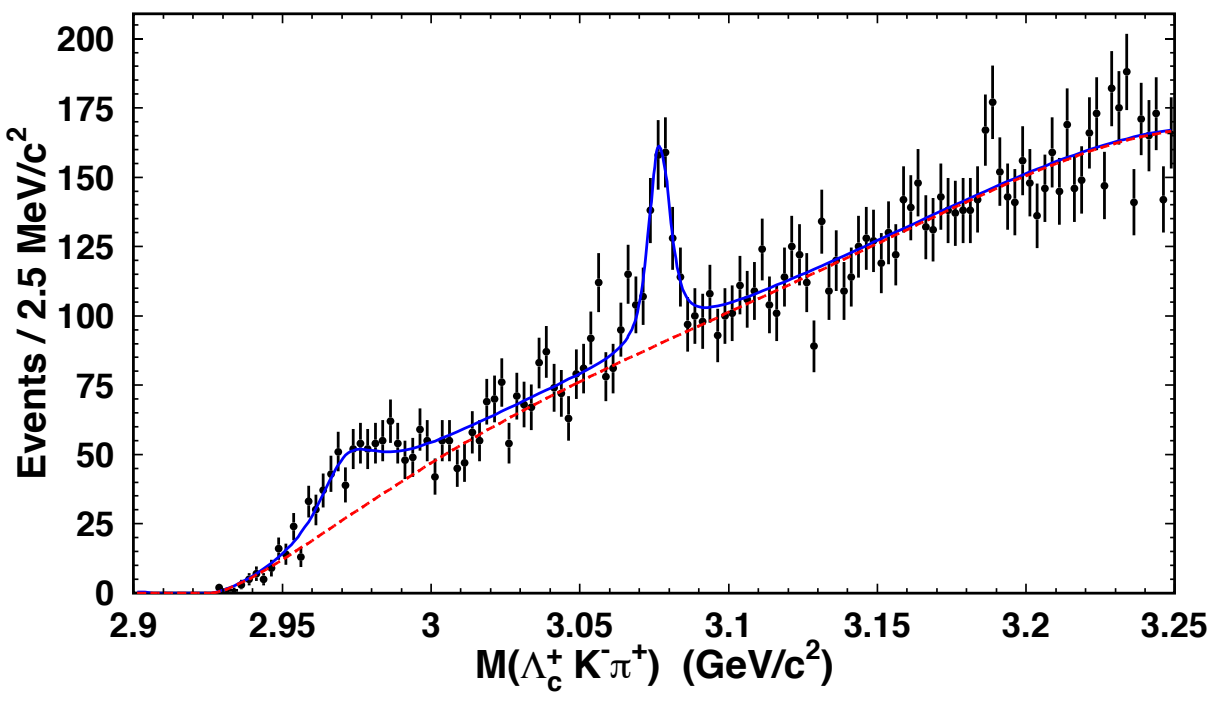

**Figure 19.4.6.** The $m(\Lambda_c^+K^-\pi^+)$ invariant mass spectrum at Belle. The $\Xi_c(2980)$ and $\Xi_c(3077)$ resonances are visible. (Chistov, 2006b)

No direct measurements of the $J^P$ of any of the excited $\Xi_c$ states are available. Mild constraints on the quantum numbers can be inferred from the decay pattern. For example, the observation of the $\Xi_c(3077)$ in $\Sigma_c(2455)\overline{K}$ and $\Sigma_c(2455)\overline{K}$ excludes states with diquark $j^p = 0^-$ ($0^- \not\to 1^+0^-$ for any $L$). However, many quantum numbers are still allowed for these states and there is a range of opinions on the best match to the data—see *e.g.* (Alexander et al., 1999; Cheng and Chua, 2007; Rosner, 2007).

### 19.4.1.4 $\Omega_c$ family

The available experimental data on the $\Omega_c$ ground states were limited before the $B$ Factories, in contrast to the $\Lambda_c$, $\Sigma_c$, and $\Xi_c$ families. The weakly-decaying $J = 1/2^+$ $\Omega_c^0$ had been seen in a number of different decay modes and production environments but with only limited statistics (typically samples of order 10 events in a given decay mode, and never more than 100) and the $J = 3/2^+$ $\Omega_c^{*0}$ had not been observed.

Belle carried out a precise measurement of the $\Omega_c^0$ mass using a sample of 725 decays to the $\Omega^-\pi^+$ final state (Solovieva, 2009), obtaining $(2693.6\pm0.3^{+1.8}_{-1.5})\,\mathrm{MeV}/c^2$. This decay channel was chosen because it is the most copious and cleanest; due to the limited production rate, it was not possible to employ low-rate modes close to threshold as was done for the $\Lambda_c^+$ in Section 19.4.1.2. As a result, the uncertainty is dominated by the mass scale uncertainty (evaluated based on the observed variation of the mass as a function of kinematic variables).

The $\Omega_c^{*0}$ is too light to undergo strong decay and so decays purely to $\Omega_c^0\gamma$ (see Fig. 19.4.7). It was discovered by *BABAR* and confirmed by Belle (Aubert, 2006ah; Solovieva, 2009). Both measured the mass difference $m(\Omega_c^{*0}) - m(\Omega_c^0)$, and the two results are in excellent agreement. The PDG average for the mass difference is $(70.7^{+0.8}_{-1.0})\,\mathrm{MeV}/c^2$. As we saw in equation 19.4.6, the naïve quark model prediction for this is:

$$m_{\Omega_c^*} - m_{\Omega_c} = 2\left(m_{\Xi_c^*} - m_{\Xi_c'}\right) - \left(m_{\Sigma_c^*} - m_{\Sigma_c}\right), \tag{19.4.8}$$

which we evaluate as $(74.0\pm4.3)\,\mathrm{MeV}/c^2$ based on current world-average measurements of $\Xi_c$ and $\Sigma_c$ mass differences (Beringer et al., 2012). This is in beautiful agreement with the experimental result.

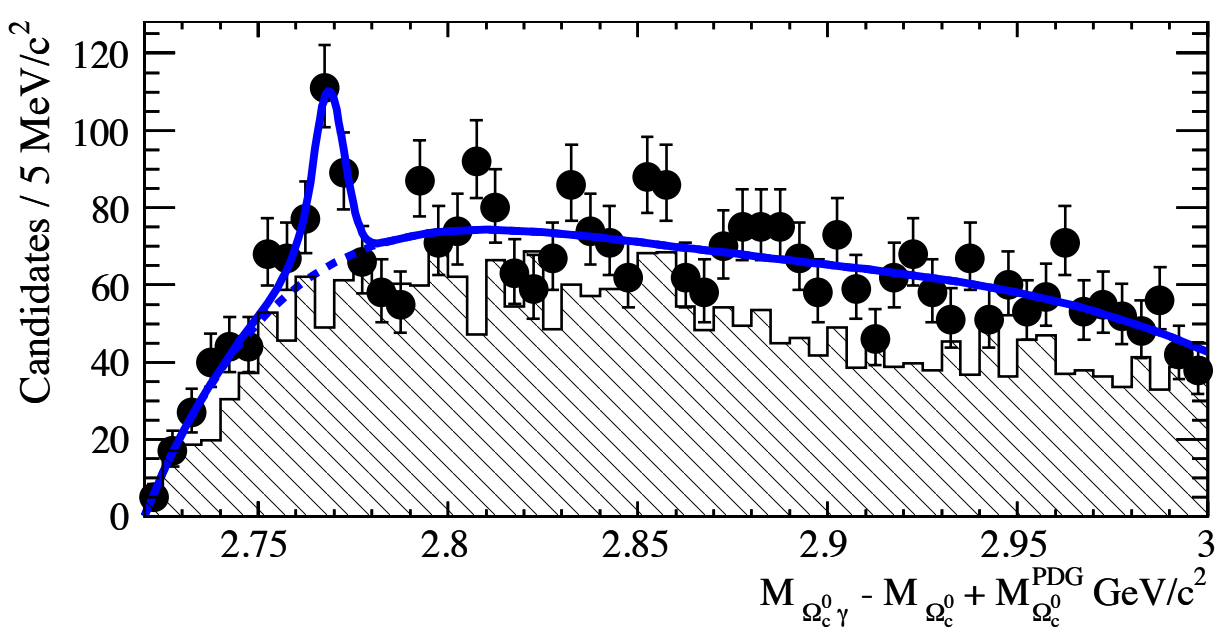

**Figure 19.4.7.** The $m(\Omega_c^0\gamma)$ invariant mass spectrum at *BABAR*, combining four decay modes of the $\Omega_c^0$: $\Omega^-\pi^+$, $\Omega^-\pi^+\pi^0$, $\Omega^-\pi^+\pi^-\pi^+$, $\Xi^-K^-\pi^+\pi^+$. The $\Omega_c^*$ resonance is visible. The shaded histogram represents the combinatoric background estimated from the $\Omega_c^0$ mass sidebands. (Aubert, 2006ah)

No radially or orbitally excited $\Omega_c$ have yet been discovered, but we would expect their masses to follow a similar pattern to the $\Sigma_c$ states discussed previously. However, there are fewer options for their decay: transitions of the form $\Omega_c \to \Omega_c\pi$ are isospin-suppressed (whereas $\Xi_c\overline{K}$ and $\Omega_c\pi\pi$ are allowed). This could result in $\Omega_c$ states that are narrow but whose $\Sigma_c$ analogs are too broad to resolve—this will be an interesting area for future flavor factories to search.

### 19.4.1.5 Searches for $\Xi_{cc}$

The quark model also predicts baryons with two charm quarks and one lighter quark, as shown in Fig. 19.4.2. Because two quark flavors are identical, fewer distinct configurations are possible than for the $C = 1$ baryons. We expect three weakly-decaying ground states with $J^P = \frac{1}{2}^+$ ($\Xi_{cc}^{+}$, $\Xi_{cc}^{++}$, $\Omega_{cc}^{+}$) and three further states with $J^P = \frac{3}{2}^+$ ($\Xi_{cc}^{*+}$, $\Xi_{cc}^{*++}$, $\Omega_{cc}^{*+}$). In the baryon mass parameterization of Eq. 19.4.4, the mass splittings between these states can be related to those of singly-charmed baryons:

$$m_{\Xi_{cc}^*} - m_{\Xi_{cc}} = m_{\Sigma_c^*} - m_{\Sigma_c} \approx 64\,\mathrm{MeV}/c^2 \quad (19.4.9)$$
$$m_{\Omega_{cc}^*} - m_{\Omega_{cc}} = m_{\Omega_c^*} - m_{\Omega_c} \approx 71\,\mathrm{MeV}/c^2 \quad (19.4.10)$$

This is well below threshold for strong decay, so the $\Xi_{cc}^*$ and $\Omega_{cc}^*$ states should decay electromagnetically like the $\Omega_c^*$.

More ambitiously, we can attempt to relate the masses of the weakly decaying ground states. Applying Eq. 19.4.4 to the $\Xi_{cc}$, $\Sigma_c$, $\Lambda_c$, and nucleon, we can write:

$$m_{\Xi_{cc}} = m_N + 2(m_{\Lambda_c} - m_N) + \frac{1}{2}(m_{\Sigma_c} - m_{\Lambda_c}), \quad (19.4.11)$$

where the terms represent the base nucleon mass, the mass offset for two charm quarks, and a hyperfine correction. Evaluating this we obtain $m_{\Xi_{cc}} \approx 3720\,\mathrm{MeV}/c^2$. However, this estimate should be treated with some skepticism: we have assumed that the coefficients $A$, $B'$, and $C'$ are the same for all baryons, but in practice these terms have some scale dependence (*e.g.* on the spatial extent of the wavefunction). There are numerous more rigorous theoretical predictions (see, *e.g.*, Roberts and Pervin (2008) and the references therein), including increasingly precise estimates from Lattice QCD (Liu, Lin, Orginos, and Walker-Loud, 2010). Most estimates lie between 3600 and 3700 MeV/$c^2$.

To date, sightings of $\Xi_{cc}$ states have been reported only at the SELEX experiment, a forward spectrometer in which a hyperon[151] beam (composed of $\Sigma^-$, $p$, and $\pi^-$) struck a fixed target of copper or diamond. SELEX claimed observation of $\Xi_{cc}^{+}$ at a mass of 3519 MeV/$c^2$ in the $\Lambda_c^+ K^- \pi^+$ and $pD^+K^-$ final states (Mattson et al., 2002; Ocherashvili et al., 2005). In each case the signature is a small, narrow signal on top of a smaller background: an excess of 15.9 events above an estimated background of $6.1 \pm 0.5$ for $\Lambda_c^+ K^- \pi^+$, and of 5.4 above $1.6 \pm 0.4$ for $pD^+K^-$. The observations were controversial (Kiselev and Likhoded, 2002), primarily because the lifetime and the production rate of $\Xi_{cc}$ at SELEX were far from expectations. The theory expectation for the $\Xi_{cc}^{+}$ lifetime is approximately 200–250 fs across a number of models (Chang, Li, Li, and Wang, 2008), compared to a reported upper limit of 33 fs. Even more surprising, by comparing the relative yields of $\Lambda_c^+$ and $\Xi_{cc}^{+}$ and correcting for acceptance and additional decay modes, SELEX estimated that 20% of its sample of 1,630 $\Lambda_c^+$ came from $\Xi_{cc}^{+}$ decays (presumably with a further contribution of similar order from $\Xi_{cc}^{++}$). This runs counter to expectations: it is much more difficult to produce a baryon with more than one unit of flavor because two heavy quark-antiquark pairs need to be created within a narrow enough kinematic window for them to coalesce into a baryon.[152] SELEX also reported preliminary observations of several other peaks in the $\Lambda_c^+ K^- \pi^+$ and $\Lambda_c^+ K^- \pi^+ \pi^+$ mass spectra, claiming a further $\Xi_{cc}^{+}$ state at 3443 MeV/$c^2$ and $\Xi_{cc}^{++}$ states at 3460, 3540, and 3780 MeV/$c^2$ (Russ, 2002, 2003), but did not publish these results.

*BABAR*, Belle, and the FOCUS photoproduction experiment carried out searches for $\Xi_{cc}$ in an attempt to reproduce the published SELEX observation (Aubert, 2006ao; Chistov, 2006b; Ratti, 2003). All three examined $\Lambda_c^+ K^- \pi^+$ along with a variety of other final states. None found any signal. Since the integrated luminosity and the production cross-section vary between experiments, upper limits were quoted in the form of a production rate relative to $\Lambda_c^+$. At the $B$ Factories this ratio is defined as

$$R_{\Xi_{cc}^+/\Lambda_c^+} \equiv \frac{\sigma(e^+e^- \to \Xi_{cc}^+ X)\,\mathcal{B}(\Xi_{cc}^+ \to \Lambda_c^+ K^- \pi^+)}{\sigma(e^+e^- \to \Lambda_c^+ X)}, \quad (19.4.12)$$

where $X$ represents the rest of the event and $\mathcal{B}$ the branching fraction, and the ratio is defined in an analogous way for FOCUS. The limits obtained are shown in Table 19.4.5. In each case, the samples of $\Lambda_c^+$ events used were much larger than that of SELEX: yields of 19k for FOCUS, 600k for *BABAR*, and 840k for Belle. However, because the production environments differ from that of SELEX it cannot be excluded that the double-charm baryon cross-section is dramatically higher with a hyperon in the initial state for reasons that are not understood theoretically.

**Table 19.4.5.** Upper limits on the production ratio $R_{\Xi_{cc}^+/\Lambda_c^+}$ defined in Eq. 19.4.12. SELEX reported a ratio of 9.6%.

| Experiment | Limit on $R_{\Xi_{cc}^+/\Lambda_c^+}$ | Kinematic cuts |
|---|---|---|
| *BABAR* | $6.9 \times 10^{-4}$ @ 95% C.L. | — |
| *BABAR* | $2.7 \times 10^{-4}$ @ 95% C.L. | $p^* > 2.3\,\mathrm{GeV}/c$ |
| Belle | $1.5 \times 10^{-4}$ @ 90% C.L. | $p^* > 2.5\,\mathrm{GeV}/c$ |
| FOCUS | $2.3 \times 10^{-3}$ @ 90% C.L. | — |

There is one further twist: several of the excited $\Xi_c$ states discussed in Section 19.4.1.3, notably the $\Xi_c(2980)$ and $\Xi_c(3077)$, were discovered in the $\Lambda_c^+ K^- \pi^+$ final state. None of these states were reported by SELEX (although they were not specifically excluded either). This poses a further challenge: if the production cross-section at SELEX were much larger for these excited $\Xi_c$ states than for $\Xi_{cc}$ they should have been seen clearly, so their non-observation raises questions about the $\Xi_{cc}$ signals; conversely, if the cross-section for the excited $\Xi_c$ is smaller than for $\Xi_{cc}$ the mechanism must be highly exotic since the $\Xi_c$ states are not only lighter than $\Xi_{cc}$ but also closer in flavor content to the initial-state $\Sigma^-$.

Ultimately, the nature of the $\Xi_{cc}$ states will become clear only when they are observed with high statistics at a flavor factory, either vindicating or excluding the SELEX results.

[151] A hyperon is a baryon containing at least one strange quark. The term predates the discovery of charm and is not normally used to refer to baryons with heavy flavor.

[152] For example, the production rates of $\Lambda$ and $\Lambda_c^+$ in $e^+e^-$ annihilation events are typically an order of magnitude larger than that of $\Xi^-$ (Beringer et al., 2012).

### 19.4.1.6 Conclusions

The spectroscopy of the ground state $C = 1$ baryons is now well established, and the lowest-lying $\Lambda_c$, $\Sigma_c$, and $\Xi_c$ excitations are reasonably well understood. There are still question marks about the nature of higher states which have been observed, and about the many states which are predicted but have not yet been seen. To extend our understanding, we need experimental information from three sources: from inclusive production with higher statistics (*e.g.* to be able to see the next $\Omega_c$ states, whose production cross-section at the $B$ Factories is bound to be small), from exclusive production in $b$-meson decays, and from exclusive production in $b$-baryon decays. The latter will open up angular analyses for some charmed baryon states that are difficult to produce in quasi-two-body decays of $B$ mesons (see Section 17.12), in much the same way that charmed baryon decays allow light baryons to be studied (see Section 19.4.3). We can look forward to results from these studies, as well as searches for doubly-charmed baryons, at LHCb and at future $e^+e^-$ flavor factories.

## 19.4.2 Weak decays

### 19.4.2.1 Introduction

The investigation of charmed baryon weak decays is more difficult then for charmed mesons due to their shorter lifetimes and smaller production rate. The reconstruction efficiency is often lower too, due to the frequent presence of hyperons in the final state which are long-lived and commonly have decay modes with a secondary neutron or $\pi^0$. The lightest charmed baryon $\Lambda_c^+$ was discovered long before the $B$ Factory era. Subsequently, many weak decay modes, mostly Cabibbo-favored tree diagrams, were observed for the $\Lambda_c^+$ and $\Xi_c^{0,+}$ baryons. By contrast, little was known about Cabibbo-suppressed and W-exchange decays[153] measurements with more than a dozen signal events had been made even in the early 2000's. Due to this lack of statistics, it was difficult to make definitive tests between theoretical models that predicted charmed baryon decay rates (Korner, Kramer, and Willrodt, 1979; Korner and Kramer, 1992; Uppal, Verma, and Khanna, 1994). Both current data and theoretical models point to non-factorizable amplitudes (*e.g.* $W$-exchange diagrams) having a significant impact on individual decay rates as well as the total widths and hierarchy of charmed baryon lifetimes. It is therefore important to gather as many measurements as possible on charmed baryon weak decays.

In this section we discuss the many new results on $\Lambda_c^+$, $\Xi_c^{0,+}$, and $\Omega_c^0$ weak decays obtained by Belle and *BABAR*. These measurements make use of the high statistics and the excellent pion-kaon-proton separation, secondary vertex reconstruction, and photon energy resolution available at the $B$ Factories.

[153] Fig. 17.4.1(E) illustrates the W-exchange decay of a meson; the baryon diagram is the same apart from a second spectator quark.

### 19.4.2.2 Results and discussion

#### $\Lambda_c^+$ *decays*

Belle and *BABAR* measured several ratios of branching fractions, in many cases Cabibbo-suppressed or $W$-exchange $\Lambda_c^+$ decay modes taken relative to topologically similar unsuppressed modes. The numerical results for the analyses discussed below are given in Table 19.4.6.

With the initial 32.6 fb$^{-1}$ of data, Belle observed two new Cabibbo-suppressed decays of $\Lambda_c^+$: $\Lambda_c^+ \to \Lambda K^+$ and $\Lambda_c^+ \to \Sigma^0 K^+$ (Abe, 2002h). The $\Lambda_c^+$ signals for these modes are shown in Fig. 19.4.8. *BABAR* subsequently confirmed these observarions with 125 fb$^{-1}$ of data, and set limits on two related four-body modes with the same sample: $\Lambda_c^+ \to \Lambda K^+\pi^+\pi^-$ and $\Lambda_c^+ \to \Sigma^0 K^+\pi^+\pi^-$ (Aubert, 2007ag). Both experiments measured the branching fractions relative to the Cabibbo-favored decay modes $\Lambda_c^+ \to \Lambda\pi^+$ and $\Lambda_c^+ \to \Sigma^0\pi^+$. The results were broadly consistent with predictions based on a flavor symmetry approach (Sharma and Verma, 1997):

$$\frac{\mathcal{B}(\Lambda_c^+ \to \Lambda K^+)}{\mathcal{B}(\Lambda_c^+ \to \Lambda\pi^+)} = [0.025 - 0.177], \quad (19.4.13)$$

$$\frac{\mathcal{B}(\Lambda_c^+ \to \Sigma^0 K^+)}{\mathcal{B}(\Lambda_c^+ \to \Sigma^0\pi^+)} = [0.069 - 0.78] \quad (19.4.14)$$

and on the constituent quark model approach (Uppal, Verma, and Khanna, 1994):

$$\frac{\mathcal{B}(\Lambda_c^+ \to \Lambda K^+)}{\mathcal{B}(\Lambda_c^+ \to \Lambda\pi^+)} = [0.039 - 0.056], \quad (19.4.15)$$

$$\frac{\mathcal{B}(\Lambda_c^+ \to \Sigma^0 K^+)}{\mathcal{B}(\Lambda_c^+ \to \Sigma^0\pi^+)} = [0.08 - 0.145] \quad (19.4.16)$$

although both models overshoot the second ratio.

Belle also made the first observation of the Cabibbo-suppressed mode $\Lambda_c^+ \to \Sigma^+K^+\pi^-$ (Abe, 2002h), measuring the branching fraction relative to $\Lambda_c^+ \to \Sigma^+\pi^+\pi^-$. The invariant mass distribution of the suppressed mode is shown in Fig. 19.4.9. Belle also investigated another Cabibbo-suppressed final state in the same paper: $\Lambda_c^+ \to pK^+K^-$. This was analysed inclusively and also separated into $p\phi$ and $p(K^+K^-)_{\text{non-}\phi}$. Finally, *BABAR* measured the Cabibbo-favored decays $\Lambda_c^+ \to \Sigma^0\pi^+$, $\Xi^-K^+\pi^+$, and $\Lambda\bar{K}^0K^+$ relative to $\Lambda_c^+ \to \Lambda\pi^+$ (Aubert, 2007ag).

Belle also investigated Cabibbo-favored, $W$-exchange $\Lambda_c^+$ decays to the final state $\Sigma^+K^+K^-$ (Abe, 2002h) which had previously been observed by CLEO (Avery et al., 1993). As well as measuring the overall branching fraction, they also extracted the rate of $\Lambda_c^+ \to \Sigma^+\phi$. Further, they observed a significant contribution of $\Lambda_c^+ \to \Xi(1690)^0K^+$ with $\Xi(1690)^0 \to \Sigma^+ K^-$, and confirmed this with the related decay $\Lambda_c^+ \to \Xi(1690)^0 K^+$, $\Xi(1690)^0 \to \Lambda K_S^0$. These demonstrate that $W$-exchange decays of charmed baryons occur at a non-negligible rate.

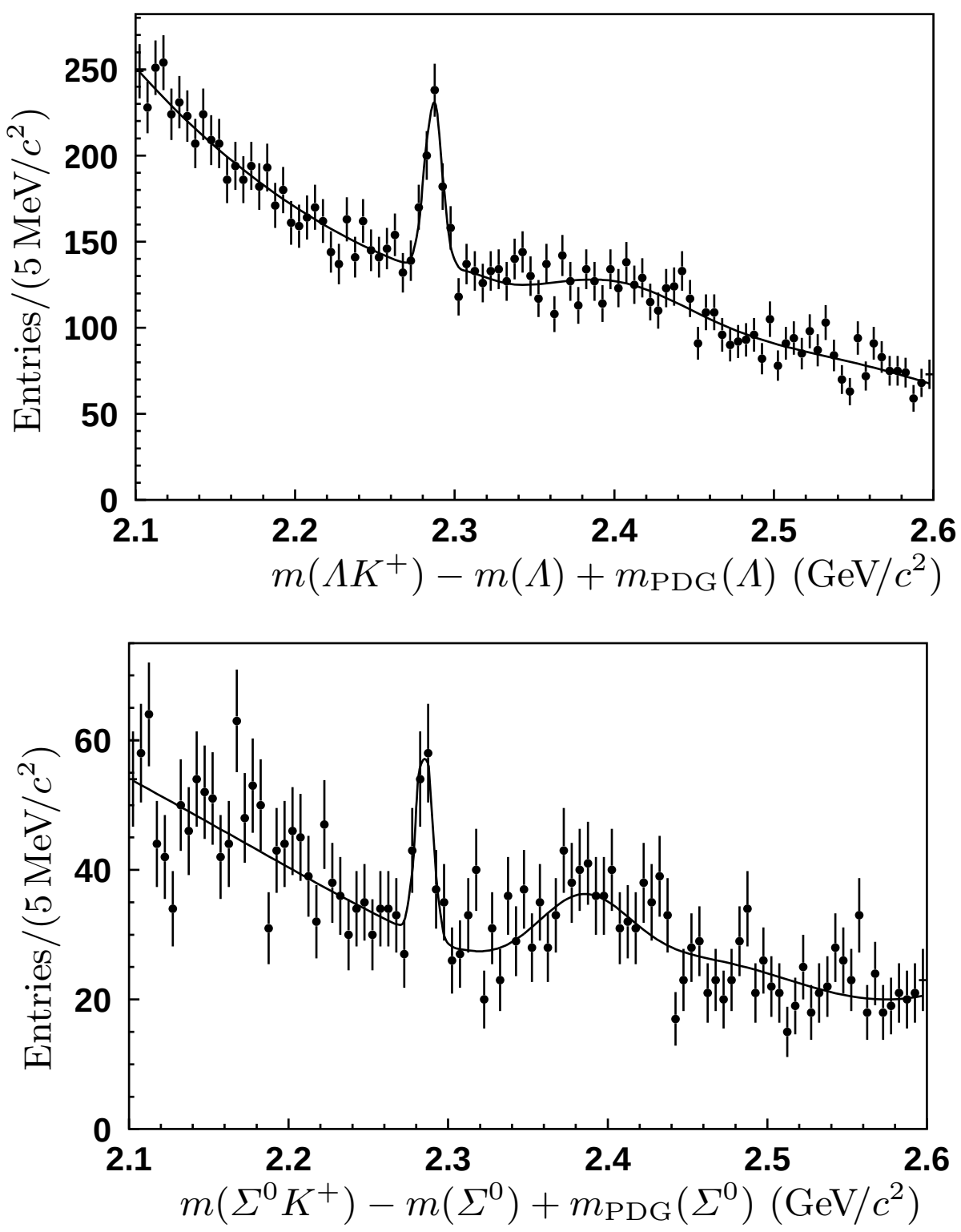

**Figure 19.4.8.** The invariant mass distribution of the selected $\Lambda K^+$ (upper) and $\Sigma^0 K^+$ (lower) combinations at Belle. The bumps to the right of the $\Lambda_c^+$ signal are due to the reflections from $\Lambda_c^+ \to \Lambda\pi^+$ and $\Lambda_c^+ \to \Sigma^0\pi^+$, where $\pi^+$ is missidentified as $K^+$. (Abe, 2002h)

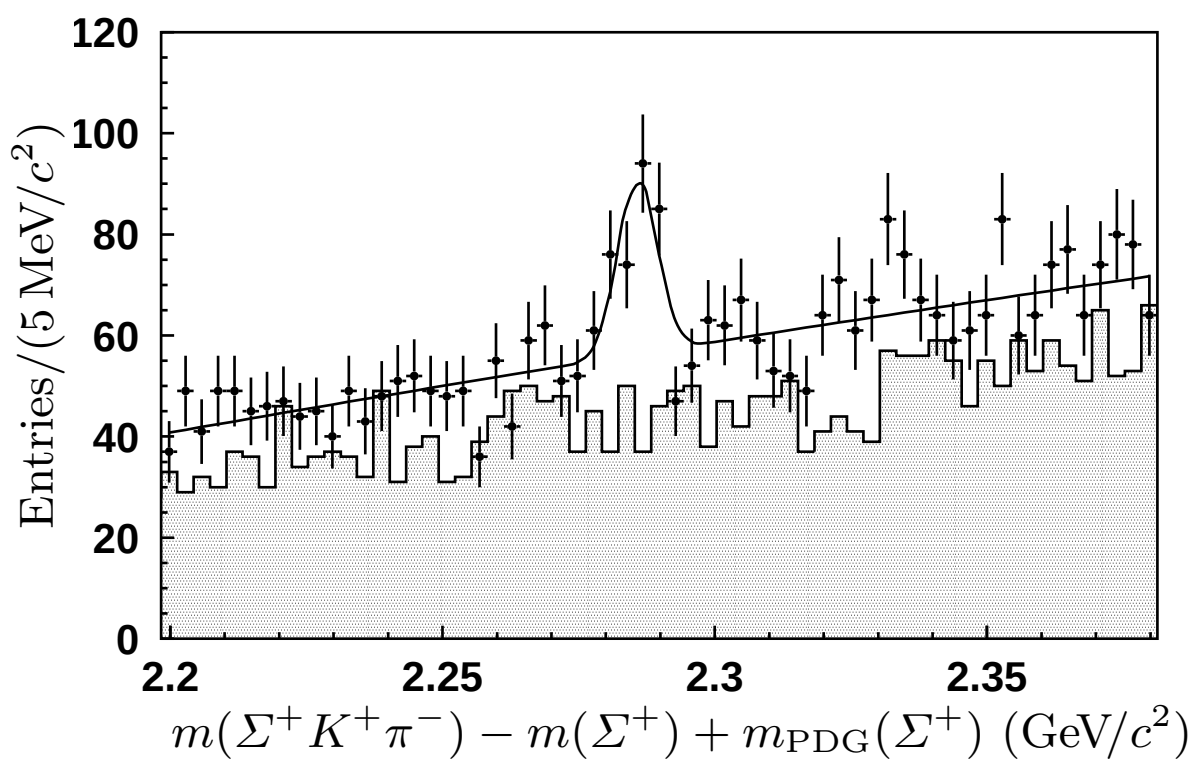

**Figure 19.4.9.** The invariant mass distribution of selected $\Sigma^+ K^+ \pi^-$ combinations at Belle, showing the $\Lambda_c^+$ signal. The shaded histogram presents the contribution from the $\Sigma^+$ sidebands. (Abe, 2002h)

*$\Xi_c^0$ and $\Xi_c^+$ decays*

In the charm strange baryon sector, Belle measured ratios of branching fractions for a suite of final states: $\Xi_c^+ \to \Xi^-\pi^+\pi^+$, $\Lambda K^-\pi^+\pi^+$, and $pK_S^0 K_S^0$; $\Xi_c^0 \to \Xi^-\pi^+$, $\Lambda K^-\pi^+$, $\Lambda K_S^0$, and $pK^-K^-\pi^+$ (Lesiak, 2005). The numerical results are given in Table 19.4.7. For the four-body decay $\Xi_c^0 \to pK^-K^-\pi^+$, Belle found that the 3-body resonant mode $\Xi_c^0 \to pK^-\bar{K}^*(892)^0$ was responsible for fully $0.51 \pm 0.03 \pm 0.01$ of the yield.

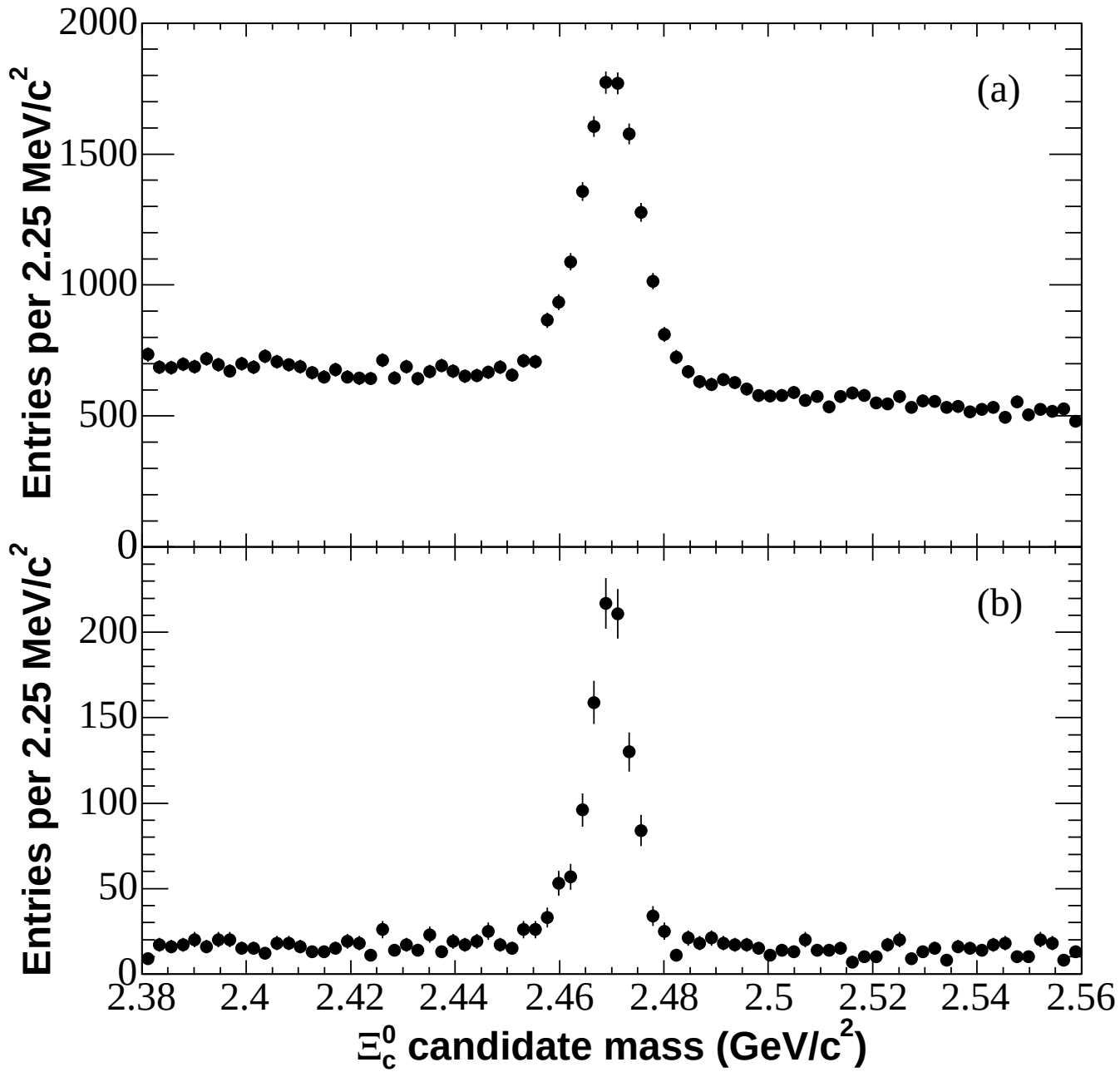

**Figure 19.4.10.** Invariant mass distribution for the $\Xi_c^0$ candidates reconstructed in the $\Xi^-\pi^+$ (upper) and $\Omega^- K^+$ (lower) decay modes at *BABAR*. (Aubert, 2005z)

*BABAR* studied the two decay modes $\Xi_c^0 \to \Omega^- K^+$ and $\Xi_c^0 \to \Xi^-\pi^+$. Both are Cabibbo-favored, but the former proceeds through a $W$-exchange diagram and the latter through a tree diagram. Fig. 19.4.10 shows the invariant mass distribution for $\Xi_c^0$ candidates in these two modes. The ratio $\mathcal{B}(\Xi_c^0 \to \Omega^- K^+)/\mathcal{B}(\Xi_c^0 \to \Xi^-\pi^+)$ was measured to be $0.294 \pm 0.018 \pm 0.016$ (Aubert, 2005z), and is consistent with a quark model prediction of 0.32 (Korner and Kramer, 1992). Note that this ratio is large, especially when considering that the difference in phase space favors $\Xi_c^0 \to \Xi^-\pi^+$ by a factor of 1.7, showing again that contributions from $W$-exchange processes cannot be neglected.

Using these modes *BABAR* measured the $\Xi_c^0$ production momentum spectrum in two data samples, one at the $\Upsilon(4S)$ resonance and one 40 MeV below. From these spectra the production rate of $\Xi_c^0$ baryon from $B$ decays was measured to be $\mathcal{B}(B \to \Xi_c^0 X) \times \mathcal{B}(\Xi_c^0 \to \Xi^-\pi^+) = (2.11 \pm 0.19 \pm 0.25) \times 10^{-4}$ (see Section 17.12) and the production cross-section from the continuum was measured to be $\sigma(e^+e^- \to \Xi_c^0 X) \times \mathcal{B}(\Xi_c^0 \to \Xi^-\pi^+) = (388 \pm 39 \pm 41)$ fb at $\sqrt{s} = 10.58$ GeV (see Section 24.1.2.2). One practical consequence is that the production rates of $\Xi_c$ from $B$ decays and from the $c\bar{c}$ continuuum are comparable, opening up the possibility of studying these states in $B$ decays.

### $\Omega_c^0$ decays

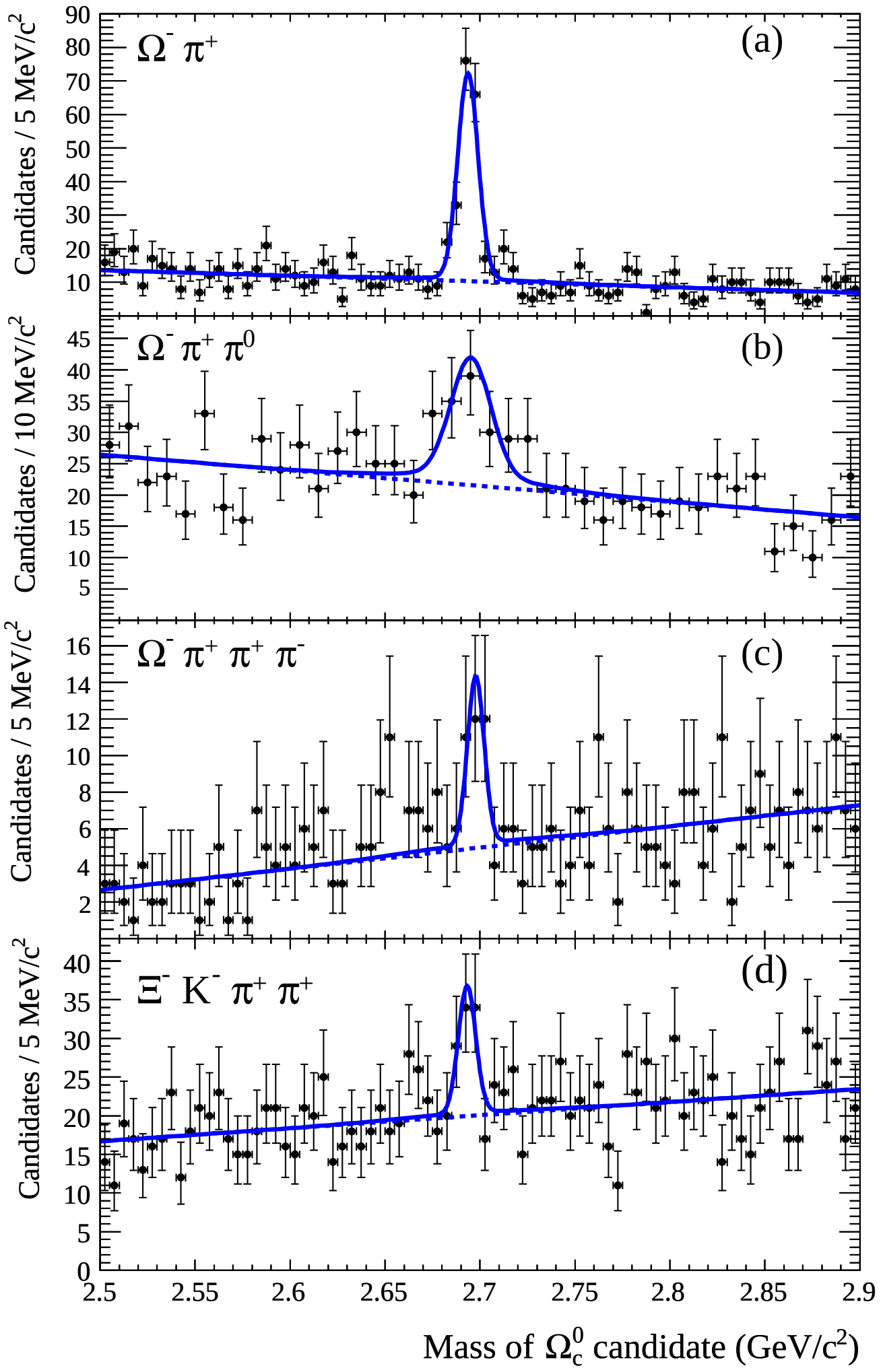

**Figure 19.4.11.** The invariant mass distributions of $\Omega_c^0$ candidates with 4 different decay modes at *BABAR*. (Aubert, 2007ao)

A thorough experimental study of the $\Omega_c$, the heaviest weakly-decaying $C = 1, B = 0$ hadron, was long overdue. Because of its heavy mass and triple flavor content, the production rate of $\Omega_c^0$ is low in comparison with other ground state charmed baryons.

*BABAR* studied four Cabibbo-favored decay modes with 230.5 fb$^{-1}$ of data: $\Omega_c^0 \to \Omega^-\pi^+$, $\Omega^-\pi^+\pi^0$, $\Omega^-\pi^+\pi^+\pi^-$, and $\Xi^-K^-\pi^+\pi^+$ (Aubert, 2007ao). Fig. 19.4.11 shows the $\Omega_c^0$ invariant mass distributions. The number of reconstructed events and the branching fractions relative to $\Omega^-\pi^+$ mode are presented in Table 19.4.7. These measurements represent a significant improvement upon the previous values. The $\Omega_c^0 \to \Omega^-\pi^+$ mode was also used by *BABAR* to study the $\Omega_c^0$ momentum spectrum. The pattern was similar to that seen for $\Xi_c^0$: comparable production rates of $\Omega_c^0$ baryons in the continuum and in the $B$ meson decays (see Fig. 19.4.12), although with overall yields a factor of $\sim 40$ smaller. At the time of writing, this remains the only observation of $\Omega_c^0$ in $B$ decays.

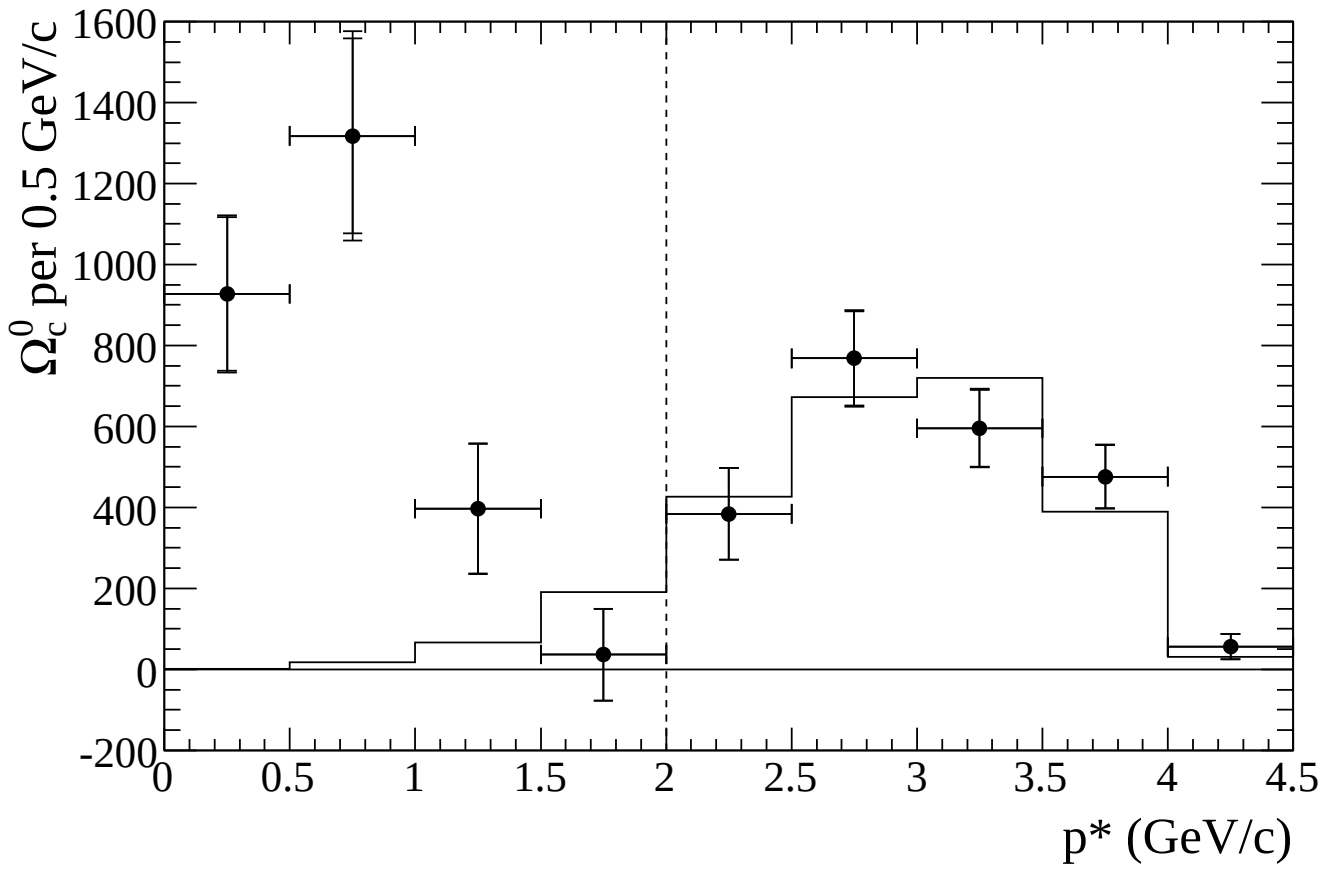

**Figure 19.4.12.** The background-subtracted and efficiency-corrected $\Omega_c^0$ yield in $p^*(\Omega_c^0)$ bins at *BABAR*. The histogram shows continuum contribution from Monte Carlo modelling. The double-peak structure in this spectrum signalizing about two production mechanisms. Lower $p^*$ peak is due to $\Omega_c^0$ production in $B$ decays. Higher-$p^*$ peak is due to $\Omega_c^0$ production in $c\bar{c}$ continuum. (Aubert, 2007ao)

### 19.4.2.3 Comments on absolute branching fractions

In the preceeding discussion, branching fractions of charmed baryon weak decays were quoted relative to reference modes. However, at present, there is limited experimental information on the absolute branching fractions of those reference modes for $\Lambda_c^+$ and none at all for $\Xi_c^0$, $\Xi_c^+$, and $\Omega_c^0$. Determining these branching fractions is highly important. At present, this limits many measurements involving charmed baryons, such as production cross-sections and branching fractions of $B$ decays.

The situation for $\Lambda_c^+$ is covered in the PDG review (Beringer et al., 2012), which obtains $\mathcal{B}(\Lambda_c^+ \to pK^-\pi^+) = (5.0 \pm 1.3)\%$. This was last updated in 2002 and there has been little progress since. The techniques outlined in the review are limited by systematic or theory uncertainties rather than statistical ones, as are many proposed methods. However, this is not universally true, and we discuss a few possibilities below.

The most direct method would be to operate an $e^+e^-$ collider at the $\Lambda_c^+\overline{\Lambda}_c^-$ threshold. In this environment, observing a $\Lambda_c^+$ necessarily implies the existence of a recoiling $\overline{\Lambda}_c^-$ with fully determined kinematics, so the branching fraction can simply be taken as the fraction of cases in which this decays to a particular final state.

**Table 19.4.6.** Summary of Belle and *BABAR* results on $\Lambda_c^+$ Cabibbo-favored (CF), Cabibbo-suppressed (CS) and $W$-exchange (WE) decays.

| $\Lambda_c^+$ mode | Experiment | Yield | $\Lambda_c^+$ reference mode | $\mathcal{B}_{signal}/\mathcal{B}_{ref.}$ |
|---|---|---|---|---|
| $\Lambda K^+$ (CS) | Belle | $265 \pm 35$ | $\Lambda\pi^+$ | $0.074 \pm 0.010 \pm 0.012$ |
| $\Sigma^0 K^+$ (CS) | Belle | $75 \pm 18$ | $\Sigma^0\pi^+$ | $0.056 \pm 0.014 \pm 0.008$ |
| $\Lambda K^+$ (CS) | *BABAR* | $1162 \pm 101$ | $\Lambda\pi^+$ | $0.044 \pm 0.004 \pm 0.003$ |
| $\Sigma^0 K^+$ (CS) | *BABAR* | $366 \pm 52$ | $\Sigma^0\pi^+$ | $0.038 \pm 0.005 \pm 0.003$ |
| $\Lambda K^+\pi^+\pi^-$ (CS) | *BABAR* | $160 \pm 62$ | $\Lambda\pi^+$ | $< 4.1 \times 10^{-2}$ @90% CL |
| $\Sigma^0 K^+\pi^+\pi^-$ (CS) | *BABAR* | $21 \pm 24$ | $\Sigma^0\pi^+$ | $< 2.0 \times 10^{-2}$ @90% CL |
| $\Sigma^+ K^+\pi^-$ (CS) | Belle | $105 \pm 24$ | $\Sigma^+\pi^+\pi^-$ | $0.047 \pm 0.011 \pm 0.008$ |
| $\Sigma^+ K^+ K^-$ (WE) | Belle | $246 \pm 20$ | $\Sigma^+\pi^+\pi^-$ | $0.076 \pm 0.007 \pm 0.009$ |
| $\Sigma^+\phi$ (WE) | Belle | $129 \pm 17$ | $\Sigma^+\pi^+\pi^-$ | $0.085 \pm 0.012 \pm 0.012$ |
| $\Xi(1690)^0 K^+$, $\Xi(1690)^0 \to \Sigma^+ K^-$ (WE) | Belle | $75 \pm 16$ | $\Sigma^+\pi^+\pi^-$ | $0.023 \pm 0.005 \pm 0.005$ |
| $\Xi(1690)^0 K^+$, $\Xi(1690)^0 \to \Lambda\bar{K}^0$ (WE) | Belle | $93 \pm 26$ | $\Lambda\bar{K}^0 K^+$ | $0.26 \pm 0.08 \pm 0.03$ |
| $\Sigma^+ K^+ K^-$ (non-res) (WE) | Belle | $11 \pm 16$ | $\Sigma^+\pi^+\pi^-$ | $< 0.018$ @90% CL |
| $pK^+K^-$ (CS) | Belle | $676 \pm 89$ | $pK^-\pi^+$ | $0.014 \pm 0.002 \pm 0.002$ |
| $p\phi$ (CS) | Belle | $345 \pm 43$ | $pK^-\pi^+$ | $0.015 \pm 0.002 \pm 0.002$ |
| $pK^+K^-$ (non-$\phi$) | Belle | $344 \pm 81$ | $pK^-\pi^+$ | $0.007 \pm 0.002 \pm 0.002$ |
| $\Sigma^0\pi^+$ (CF) | *BABAR* | $32693 \pm 324$ | $\Lambda\pi^+$ | $0.977 \pm 0.015 \pm 0.051$ |
| $\Xi^- K^+\pi^+$ (CF) | *BABAR* | $2665 \pm 84$ | $\Lambda\pi^+$ | $0.480 \pm 0.016 \pm 0.039$ |
| $\Lambda\bar{K}^0 K^+$ (CF) | *BABAR* | $460 \pm 30$ | $\Lambda\pi^+$ | $0.395 \pm 0.026 \pm 0.036$ |

**Table 19.4.7.** Summary of Belle and *BABAR* results on $\Xi_c^{+,0}$ and $\Omega_c^0$ decays.

| Decay mode | Experiment | Yield | Reference mode | $\mathcal{B}_{signal}/\mathcal{B}_{ref.}$ |
|---|---|---|---|---|
| $\Xi_c^+ \to \Lambda K^-\pi^+\pi^+$ | Belle | $1117 \pm 55$ | $\Xi_c^+ \to \Xi^-\pi^+\pi^+$ | $0.32 \pm 0.03 \pm 0.02$ |
| $\Xi_c^+ \to pK_S^0K_S^0$ | Belle | $168 \pm 27$ | $\Xi_c^+ \to \Xi^-\pi^+\pi^+$ | $0.087 \pm 0.016 \pm 0.014$ |
| $\Xi_c^0 \to pK^-K^-\pi^+$ | Belle | $1908 \pm 62$ | $\Xi_c^0 \to \Xi^-\pi^+$ | $0.33 \pm 0.03 \pm 0.03$ |
| $\Xi_c^0 \to \Lambda K_S^0$ | Belle | $465 \pm 37$ | $\Xi_c^0 \to \Xi^-\pi^+$ | $0.21 \pm 0.02 \pm 0.02$ |
| $\Xi_c^0 \to \Lambda K^-\pi^+$ | Belle | $3268 \pm 276$ | $\Xi_c^0 \to \Xi^-\pi^+$ | $1.07 \pm 0.12 \pm 0.07$ |
| $\Xi_c^0 \to \Omega^- K^+$ | *BABAR* | $\approx 650$ | $\Xi_c^0 \to \Xi^-\pi^+$ | $0.294 \pm 0.018 \pm 0.016$ |
| $\Omega_c^0 \to \Omega^-\pi^+\pi^0$ | *BABAR* | $64 \pm 15$ | $\Omega_c^0 \to \Omega^-\pi^+$ | $1.27 \pm 0.31 \pm 0.11$ |
| $\Omega_c^0 \to \Omega^-\pi^+\pi^+\pi^-$ | *BABAR* | $25 \pm 8$ | $\Omega_c^0 \to \Omega^-\pi^+$ | $0.28 \pm 0.09 \pm 0.01$ |
| $\Omega_c^0 \to \Xi^- K^-\pi^+\pi^-$ | *BABAR* | $45 \pm 12$ | $\Omega_c^0 \to \Omega^-\pi^+$ | $0.46 \pm 0.13 \pm 0.03$ |

A related method, applicable at $B$ Factories operating at the $\Upsilon(4S)$, is to identify events with baryon and charm content and then look for a recoiling $\overline{\Lambda}_c^-$. This approach was used by CLEO, requiring both a $D$ and a $p$ in the event and inferring that a $\overline{\Lambda}_c^-$ is present (Jaffe et al., 2000). However, this comes with two disadvantages: the kinematics of the $\overline{\Lambda}_c^-$ are not known, and there is a problematic background from other event types with the same signature (*e.g.* $e^+e^- \to Dp\overline{D}\overline{N}X$, $Dp\overline{\Xi}_c\overline{K}X$).[154] Alternatively, it is possible to use $e^+e^- \to \Lambda_c^+\overline{\Lambda}_c^- X$ events in which the $\Lambda_c^+$ and the $X$ system are fully reconstructed—the "popcorn" sample identified by *BABAR* in which $X$ consists of a small number of pions and has zero baryon and strangeness content is promising (Aubert, 2010b).

#### 19.4.2.4 Conclusions

The experimental and theoretical investigation of charmed baryon weak decays remains one step behind that of charmed mesons, where both experimental data and theoretical models are more abundant. Nonetheless, the results on charmed baryon weak decays obtained by *BABAR* and Belle could stimulate the development of theoretical models describing charmed baryon weak decays. They could also provide a roadmap for further studies of the charmed baryon sector at future super flavor factories.

[154] While we were finalising this book Belle submitted an absolute branching fraction analysis for publication, using a further development of this approach: reconstruction of $D^{(*)-}\overline{p}\pi^+$ events, identification of inclusive $\Lambda_c^+$ decays using the missing mass spectrum (the mass of the system recoiling against $D^{(*)-}\overline{p}\pi^+$), and reconstruction of the subset decaying to $pK^-\pi^+$ (Zupanc, 2013a). The result, $\mathcal{B}(\Lambda_c^+ \to pK^-\pi^+) = (6.84 \pm 0.24^{+0.21}_{-0.27})\%$, represents a significant advance on the precision of this quantity. At the time of writing, this analysis has not yet been published.

**Table 19.4.8.** Angular distributions for the decay chain $X_c \to RP$, $R \to HP$ for different spin hypotheses $J_R$. It is assumed that $X_c$ and $H$ have spin 1/2, that $P$ has spin $0^-$, and that there is no polarization in the initial state. (Aubert, 2006z)

| $J_R$ | $dN/\mathrm{d}\cos\theta_h \propto$ |
|---|---|
| 1/2 | $1+\beta\cos\theta_h$ |
| 3/2 | $1+3\cos^2\theta_h+\beta\cos\theta_h(5-9\cos^2\theta_h)$ |
| 5/2 | $1-2\cos^2\theta_h+5\cos^4\theta_h+\beta\cos\theta_h(5-26\cos^2\theta_h+25\cos^4\theta_h)$ |

## 19.4.3 Applications to light baryon spectroscopy

### 19.4.3.1 Introduction

A fully exclusive production environment is very helpful for determining the spin or parity of a resonance. We saw this used in Section 19.4.1.2 to measure the spin of the $\Sigma_c(2455)$ in quasi-two-body $B$ decays, for example. In the same way, charmed baryons may be used as a laboratory to study light baryons that are produced as intermediate resonances in their decays. The helicity formalism, discussed in Section 12.1, is used here; for more detail the reader is referred to Jacob and Wick (1959), Chung (1971), Richman (1984), and Ziegler (2007).

In the cases considered below, the decay chain is of the form $X_c \to RP$, $R \to HP$, where $X_c$ is a weakly decaying charmed baryon with $J^P = 1/2^+$, $R$ is the intermediate resonance to be studied, $H$ is a hyperon with $J^P = 1/2^+$, and $P$ are pseudoscalars.[155] The decay helicity angle $\theta_h$ is defined as the angle between the direction of $H$ in the rest frame of $R$ and the direction of $R$ in the rest frame of $X_c$, as illustrated in Fig. 19.4.13. The angular distributions expected under spin hypotheses $J_R = 1/2,\ 3/2,\ 5/2$ are given in Table 19.4.8. Parity violation is allowed in weak decays and introduces an asymmetry in the distributions, expressed by the parameter $\beta$:

$$\beta = \left[\frac{\rho_{\frac{1}{2},\frac{1}{2}} - \rho_{-\frac{1}{2},-\frac{1}{2}}}{\rho_{\frac{1}{2},\frac{1}{2}} + \rho_{-\frac{1}{2},-\frac{1}{2}}}\right]\left[\frac{|A^J_{\frac{1}{2}}|^2 - |A^J_{-\frac{1}{2}}|^2}{|A^J_{\frac{1}{2}}|^2 + |A^J_{-\frac{1}{2}}|^2}\right], \quad (19.4.17)$$

where the transition matrix element $A^J_{\lambda_f}$ represents the coupling of R to the final state with net helicity $\lambda_f$, and $\rho_{i,i}$ are the diagonal density matrix elements inherited from the charmed baryon. If $R \to HP$ is a strong decay then $|A^J_{\frac{1}{2}}| = |A^J_{-\frac{1}{2}}|$ and so $\beta$ vanishes.

In addition to the technique outlined above for measuring the spin of a resonance, charmed baryon decays can be used more generally to measure properties such as the mass and width of intermediate resonances in multi-body decays.

### 19.4.3.2 Spin of the $\Omega^-$

The method introduced in Section 19.4.3.1 was used by *BABAR* in an elegant way to measure the spin of the $\Omega^-$

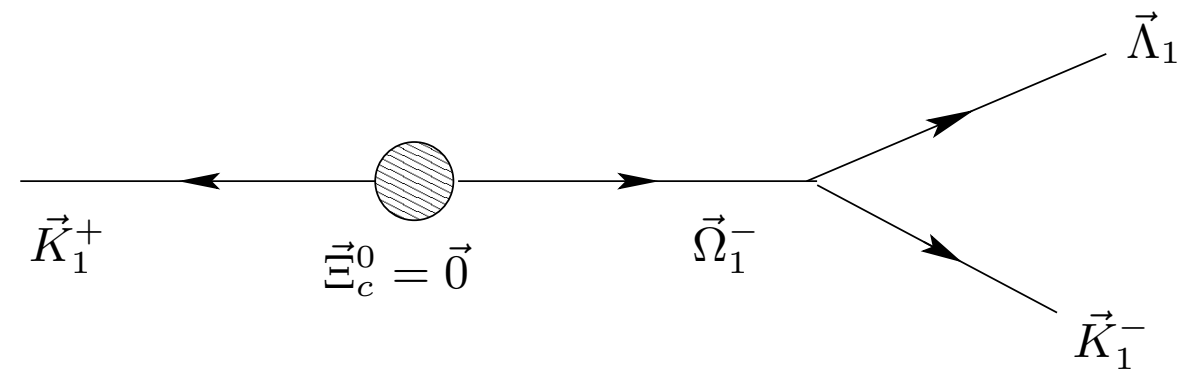

a) All decay products in the $\Xi_c^0$ rest-frame.

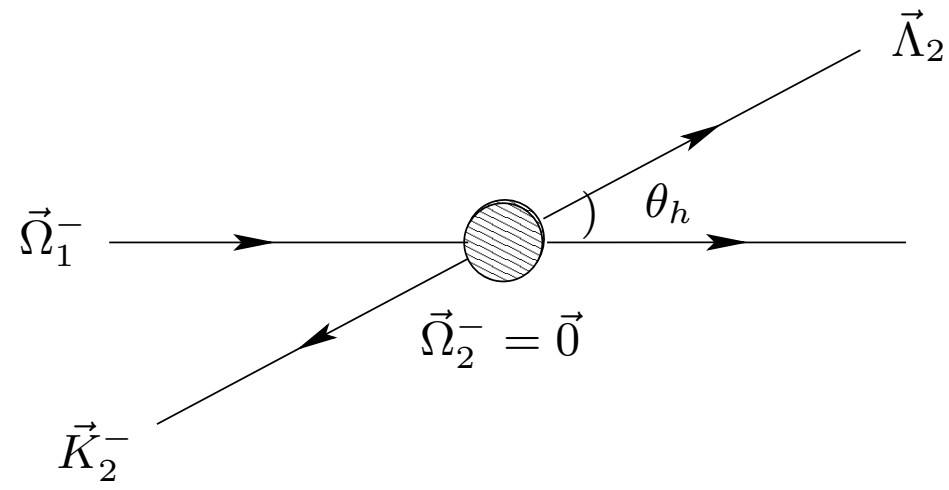

b) All decay products in the $\Omega^-$ rest-frame; in this frame, $\vec{\Omega}_1^- \to \vec{\Omega}_2^- = \vec{0}$, $\vec{\Lambda}_1 \to \vec{\Lambda}_2$, $\vec{K}_1^- \to \vec{K}_2^-$.

**Figure 19.4.13.** Illustration of the helicity angle $\theta_h$ for the case of $\Xi_c^0 \to \Omega^- K^+$, $\Omega^- \to \Lambda K^-$. $\theta_h$ is defined as the angle between the $\Lambda$ direction in the $\Omega^-$ rest frame and the $\Omega^-$ in the $\Xi_c^0$ rest frame. (Aubert, 2006z)

(Aubert, 2006z). $J_\Omega = 3/2$ is a key prediction of the quark model and was assumed to be correct, but had proved difficult to test definitively. A sample of approximately 770 decays of the form $\Xi_c^0 \to \Omega^- K^+$, $\Omega^- \to \Lambda K^-$ was selected. The mass distribution of these events is shown in Fig. 19.4.14(a), and the helicity angle distribution of these events after background subtraction and efficiency correction is shown in Fig. 19.4.14(b) and (c). The data are fully consistent with the $J_\Omega = 3/2$ hypothesis, and are highly inconsistent with the $J_\Omega = 1/2$ and 5/2 hypotheses. Higher spins would correspond to an even higher-order polynomial, which would not match the data. *BABAR* therefore concluded that $J_\Omega = 3/2$, as predicted.

As part of the analysis, *BABAR* tested for polarization of the $\Xi_c^0$ in the lab frame. Polarization would not directly affect the angular distribution, since the angle $\theta_h$ is Lorentz-invariant—but a large polarization could affect the weighted efficiency as a function of $\cos\theta_h$ (in particular because the *BABAR* detector is forward-backward asymmetric in the collision center-of-mass frame). In practice, no measureable polarization was found and this simplified

[155] Strictly speaking, the spin-parity of the charmed baryons themselves has not been measured—and indeed, some of the papers cited in this section allow for spins other than 1/2. But we do not consider that possibility here.

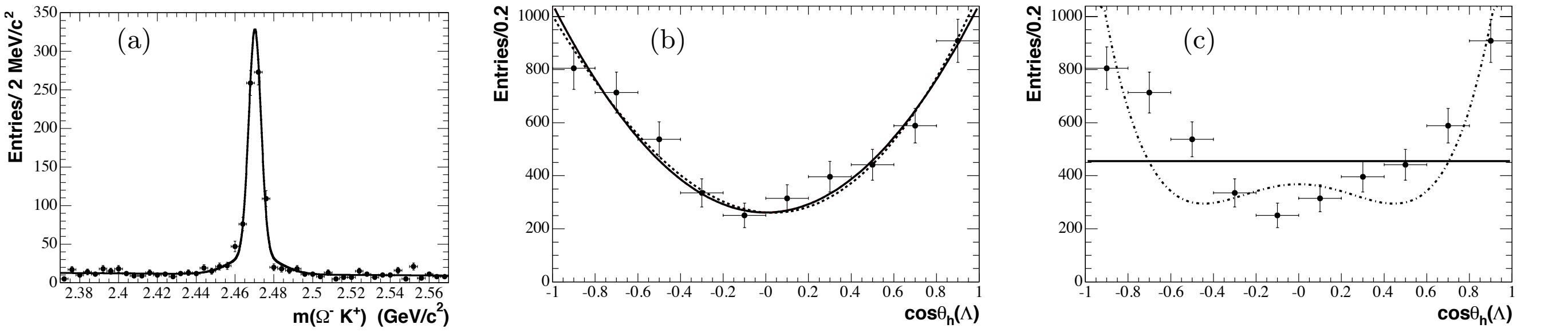

**Figure 19.4.14.** $\Xi_c^0 \to \Omega^- K^+$, $\Omega^- \to \Lambda K^-$ events at *BABAR*, showing the invariant mass distribution in (a) and the background-subtracted, efficiency-corrected angular distribution in (b) and (c). The curves in (b) show the distribution expected for $J_\Omega = 3/2$, with the asymmetry parameter $\beta$ fixed to zero (solid line, p-value 0.69) or floated (dashed line, p-value 0.64). The curves in (c) show other spin hypotheses: $J_\Omega = 1/2$ (solid line, p-value $1 \times 10^{-17}$), $J_\Omega = 5/2$ (dashed curve, p-value $3 \times 10^{-7}$). (Aubert, 2006z)

the analysis considerably. However, similar measurements at colliders such as the Tevatron and LHC would need to take polarization into account.

### 19.4.3.3 Properties of $\Xi(1530)$ and $\Xi(1690)$

*BABAR* also applied the method to study the $\Xi(1530)$ resonance, which is predicted by the quark model to have $J^P = 3/2^+$, in the decay $\Lambda_c^+ \to \Xi^- \pi^+ K^+$ (Aubert, 2008w). In the limit that the decay is pure quasi-two-body, the formalism described above applies—and superficially it is indeed quasi-two-body, as shown in Fig. 19.4.15(a) where the $\Xi(1530)^0$ is the only visible resonance. However, while the angular distribution is roughly quadratic and is clearly inconsistent with spin-1/2 or spin-5/2, it is not fully described by the spin-3/2 hypothesis either, as shown in Fig. 19.4.15(b). This implies interference with another resonance, which the paper speculates may be a high-mass $\Lambda$ or $\Sigma^0$ in the $\Xi^- K^+$ channel. The spin-3/2 hypothesis is corroborated with studies of the moments of the $m(\Xi^-\pi^+)$ distribution (weighting candidates by the $n^{\text{th}}$-order Legendre polynomial in $\cos\theta_h$).

A similar study of the $\Xi(1690)$ resonance was made in $\Lambda_c^+ \to \Lambda \overline{K}^0 K^+$ (Aubert, 2006y). As with the $\Xi(1530)$ analysis, interference from other resonances in the Dalitz plot was found to be significant and diluted the power of the angular analysis. Correcting for these effects, it was found that $J = 1/2$ was favored (p-value 0.30), but that higher spins could not be excluded (p-value 0.02 for 3/2, 0.01 for 5/2).

### 19.4.3.4 Conclusions

As illustrated in the previous sections, charmed baryons can be used as a clean, exclusive production environment to study the properties of light baryons. This allows angular analyses which would be nigh impossible in inclusive production. When applied to strongly decaying resonances in multi-body charmed baryon decays, however, interference effects cannot be neglected even for narrow states.

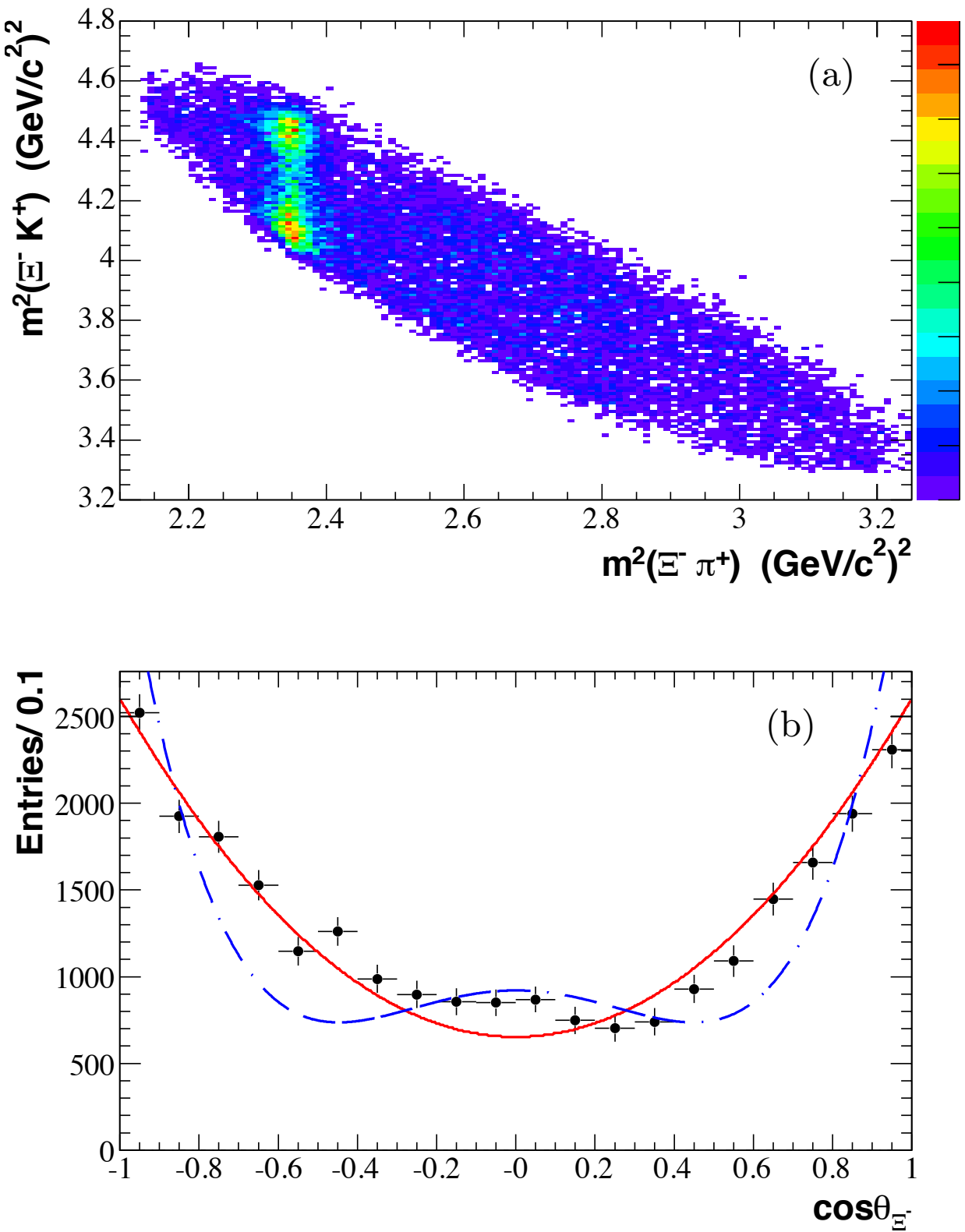

**Figure 19.4.15.** $\Lambda_c^+ \to \Xi^- \pi^+ K^+$ events at *BABAR*, showing (a) the Dalitz plot, and (b) the angular distribution in the $\Xi(1530)$ region. Curves are superimposed for the $\Xi(1530)$ spin hypotheses $J = 3/2$ (solid, p-value $3 \times 10^{-4}$) and $J = 5/2$ (dashed, p-value $6 \times 10^{-44}$). The hypothesis $J = 1/2$ would correspond to a flat distribution. Color online. (Aubert, 2008w)

# Chapter 20
# Tau physics

**Editors:**
*Mike Roney (BABAR)*
*Hisaki Hayashii (Belle)*
*Antonio Pich (theory)*

**Additional section writers:**
*Swagato Banerjee, Kiyoshi Hayasaka, George Lafferty, Alberto Lusiani, Boris Shwartz*

The $e^+e^-$ $B$ Factories, owing to the large cross section for producing tau lepton pairs, are also de facto "tau factories". The contributions to our understanding of a variety of sectors of the Standard Model and Beyond-the-Standard Model theories from studies of the tau lepton at Belle and *BABAR* are presented here. Following an introductory section that reviews the history of the tau lepton and some basics of its production and decay, this chapter proceeds with discussions of the status of tests of *CPT* and charged current lepton universality involving the tau in Section 20.2 and Section 20.3. We then discuss searches for new physics via lepton flavor violating processes in Section 20.4 and *CP* violation in tau production and decay in Section 20.5. Section 20.6 presents studies of hadronic decays of the tau. That section begins with a theoretical introduction before proceeding with presentations of measurements of branching fractions, measurements of hadronic mass spectra and spectral functions, and searches for second-class currents. These results are then interpreted in terms of tests of the Conserved Vector Current(CVC) hypothesis and used to extract the hadron vacuum polarization contribution to muonic $g-2$ in Section 20.7. Section 20.8 summarises values of $|V_{us}|$ extracted from strange decays of the tau. We close with a brief summary.[156]

## 20.1 Introduction

In this section, we give a brief history of the tau from its discovery to the status before $B$ Factory experiments. We also discuss the cross section of the tau-pair production in the $e^+e^-$ collisions, $e^+e^- \to \tau^+\tau^-$ as well as the commonly used techniques to select tau-pair events.

The tau lepton discovery was reported in 1974 by Martin Perl et al. (1975) using the SPEAR electron-positron collider at SLAC scanning over center-of-mass (CM) energies of 3 GeV to 7.8 GeV. The group initially observed 86 $e^+e^- \to e^\pm\mu^\mp +$ missing energy events above 3.8 GeV in CM with an expected background of 22 events from known non-tau sources using the "SLAC-LBL magnetic detector" (later called Mark I), which had a full $2\pi$ azimuthal angle and $50^\circ \le \theta \le 130^\circ$ polar angle acceptance. It consisted of barrels of "trigger counters" at two radii, cylindrical wire chambers inside a 0.4 T solenoidal magnetic field, lead-scintillator shower counters outside the 3 m×3 m long magnet coil and muon wire chambers surrounding the iron return-yoke of the magnet. The group collected more data and the following year published the cross section and lepton spectra from 105 signal events (above 34 background events) with which they demonstrated that the events were most economically described by $e^+e^- \to U^+U^-$ where $U^\pm$ is a heavy lepton with a mass between 1.6 GeV/$c^2$ and 1.8 GeV/$c^2$ decaying via $U^- \to \nu_U \ell^- \overline{\nu}_\ell$ (Perl et al., 1976). Under that hypothesis they also reported a value of the leptonic branching fraction, which was in excellent agreement with Paul Tsai's calculations for a third generation heavy lepton published a few years earlier in the classic and first comprehensive paper on tau physics (Tsai, 1971). Confirmation of the discovery came over the next two years. The Maryland-Princeton-Pavia magnetic detector group operating a single arm spectrometer at SPEAR reported an anomalous muon event rate (Cavalli-Sforza et al., 1976). (Snow, 1976) analysed the charged multiplicity of their "anomalous" events and concluded that a new heavy lepton was the simplest explanation. In 1977 the PLUTO (Burmester et al., 1977a,b) and DASP (Brandelik et al., 1977) groups reported confirmations of the discovery with their experiments operating at DESY's DORIS electron-positron collider in Hamburg. By early 1977 the new particle was being considered a sequential lepton and was first named the $\tau$: "Since there is now substantial evidence that it is a lepton, we wish to designate it by a lower case Greek letter. We use $\tau^\pm$ because it appears to be the third charged lepton to be found and τριτος means third in Greek." (Perl, 1977). Martin Perl was awarded the 1995 Nobel Prize in physics for this discovery.

Since its discovery, properties of the tau have primarily been determined with precision using the $e^+e^- \to \tau^+\tau^-$ process as a source of tau leptons. The cross section for this process at $\sqrt{s} = 10.58$ GeV is $0.919 \pm 0.003$ nb (Banerjee, Pietrzyk, Roney, and Wąs, 2008).

The text book (Stahl, 2000) is useful to learn more about the history of tau lepton physics.

## 20.2 Mass of the tau lepton

Masses of quarks and leptons are fundamental parameters of the Standard Model. They cannot be determined by the theory and must be measured. High precision measurements of the mass of the tau lepton are important for testing lepton universality and for calculating branching fractions that depend on the tau mass. Uncertainties in the tau mass have important consequences on the accuracy of the calculated leptonic-decay rate of the tau lepton, since it is proportional to $m_\tau^5$:

$$\Gamma(\tau^- \to \ell^- \nu_\tau \overline{\nu}_\ell) = \frac{G_\mu^2 m_\tau^5}{192\pi^3} f\left(\frac{m_\ell^2}{m_\tau^2}\right)\left(1 + \frac{3}{5}\frac{m_\tau^2}{m_W^2}\right) \left(1 + \frac{\alpha(m_\tau)}{2\pi}\left[\frac{25}{4} - \pi^2\right]\right), \quad (20.2.1)$$

[156] Through out this section, charge-conjugate $\tau$ decays are implied if it is not specified explicitly.

$$f(x) = 1 - 8x + 8x^3 - x^4 - 12x^2 \ln x, \quad (20.2.2)$$

where $\ell = e, \mu$ and $\alpha^{-1}(m_\tau) = 133.3$. $G_\mu$ is the Fermi coupling constant determined precisely from the muon lifetime (Marciano and Sirlin, 1988).

In addition to the fundamental importance of the tau lepton mass in the Standard Model, separate measurements of the masses of the $\tau^+$ and $\tau^-$ in $B$ Factory experiments allow us to test the *CPT* theorem. *CPT* invariance is a fundamental symmetry of any local field theory, including the Standard Model. Any evidence of *CPT* violation would be evidence of local Lorentz violation and a sign of physics beyond the Standard Model.

At present the precision of the tau mass is dominated by the KEDR (Shamov et al., 2009) and BES (Bai et al., 1996a) measurements where the mass value was derived from the energy dependence of the $e^+e^- \to \tau^+\tau^-$ cross section near production threshold. However, both $B$ Factories performed the tau mass determination using a different technique, the so called pseudomass method originally introduced by the ARGUS collaboration (Albrecht et al., 1992a).

In this technique, the pseudomass is defined in terms of the mass, energy and momenta of the tau decay products. For the hadronic decays of the $\tau^-$ ( $\tau^- \to h^-\nu_\tau$ and its charge conjugate), the tau mass, $m_\tau$, is given by

$$m_\tau = \sqrt{M_h^2 + 2(E_\tau^* - E_h^*)(E_h^* - P_h^* \cos\theta^*)}, (20.2.3)$$

where $M_h$, $E_h^*$, $P_h^*$ are the invariant mass, energy and the magnitude of the three-momentum of the hadronic system $h$ in the $e^+e^-$ CM frame, respectively. The energy of the tau lepton is given by $E_\tau^* = \sqrt{s}/2$, where $\sqrt{s} = 10.58$ GeV. $\theta^*$ is the angle between the hadronic system and the $\nu_\tau$ direction. Since the neutrino is undetected, one can not measure the angle $\theta^*$; thus one defines the pseudomass $M_{\rm min}$ by setting $\theta^* = 0$:

$$M_{\rm min} = \sqrt{M_h^2 + 2(E_{\rm beam} - E_h^*)(E_h^* - P_h^*)}, (20.2.4)$$

which is less than or equal to the tau lepton mass.

Figure 20.2.1 shows the typical pseudomass distribution of the combined $\tau^+$ and $\tau^-$ samples for $\tau^\pm \to \pi^\pm\pi^+\pi^-\nu_\tau$ candidates. A sharp kinematic cutoff is seen at $M_{\rm min} \sim m_\tau$. The smearing of the endpoint is caused by the initial and final state radiation and the detector resolution.

To determine the endpoint from the pseudomass distribution, a fit was performed to the data with an empirical function of the form

$$F(x) = (p_3 + p_4 x) \tan^{-1}\left(\frac{p_1 - x}{p_2}\right) + p_5 + p_6 x, \quad (20.2.5)$$

where $x$ is the pseudomass, and the $p_i$ are free parameters of the fit. Only the position of the endpoint, $p_1$, is important in determining the tau mass. The relation between the estimator $p_1$ and the true tau lepton mass is obtained by using several Monte Carlo (MC) samples with different values of tau mass. In the absence of initial and final state radiation (ISR/FSR) and with perfect detector resolution one expects the relation between the $p_1$ fit result and the generated tau mass to be linear with a slope of unity and zero offset. With the inclusion of ISR/FSR effects and detector resolution non-zero offset is expected.

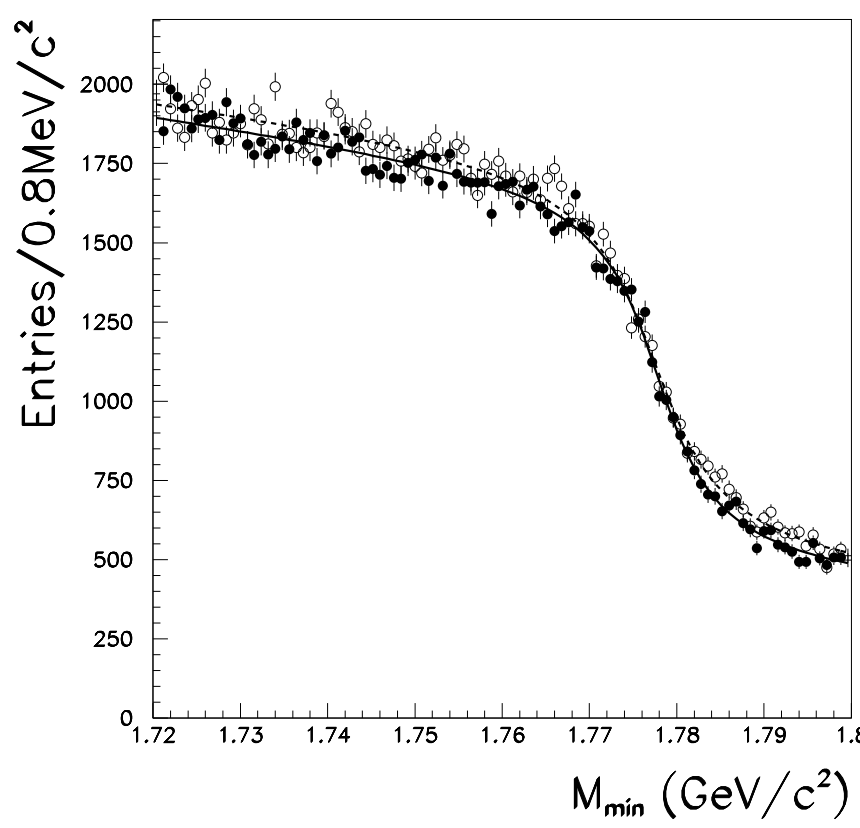

**Figure 20.2.1.** Pseudomass $M_{\rm min}$ distribution for $\tau^\pm \to \pi^\pm\pi^+\pi^-\nu_\tau$ candidates measured by the Belle, shown separately for positively and negatively charged tau decays. The solid points with error bars correspond to $\tau^+$ decays, while the open points with error bars are $\tau^-$ decays. The solid curve is the result of the fit to the $\tau^+$ pseudomass distribution (Abe, 2007c).

The current status of the tau mass measurements is summarized in Table 20.2.1 (see also Figure 20.2.2). In this table, we also include the results from the measurements by the BES and KEDR experiments, where the tau mass is measured from the $\tau^+\tau^-$ cross section around the production threshold. The systematics in the pseudomass technique and the threshold scan are quite different. Nevertheless, the results from the two methods are in good agreement with similar size of errors. The world average of the tau lepton mass is (Asner et al., 2010)

$$m_\tau = (1776.77 \pm 0.15)\,\text{MeV}. \quad (20.2.6)$$

**Table 20.2.1.** Summary of recent tau mass measurements.

| Experiment | $m_\tau$, MeV | Ref. |
|---|---|---|
| BES | $1776.96^{+0.18+0.25}_{-0.21-0.17}$ | Bai et al. (1996a) |
| KEDR | $1776.69^{+0.17}_{-0.19} \pm 0.15$ | Shamov et al. (2009) |
| Belle | $1776.61 \pm 0.13 \pm 0.35$ | Abe (2007c) |
| BaBar | $1776.68 \pm 0.12 \pm 0.41$ | Aubert (2009ac) |
| Average | $1776.77 \pm 0.15$ | Asner et al. (2010) |

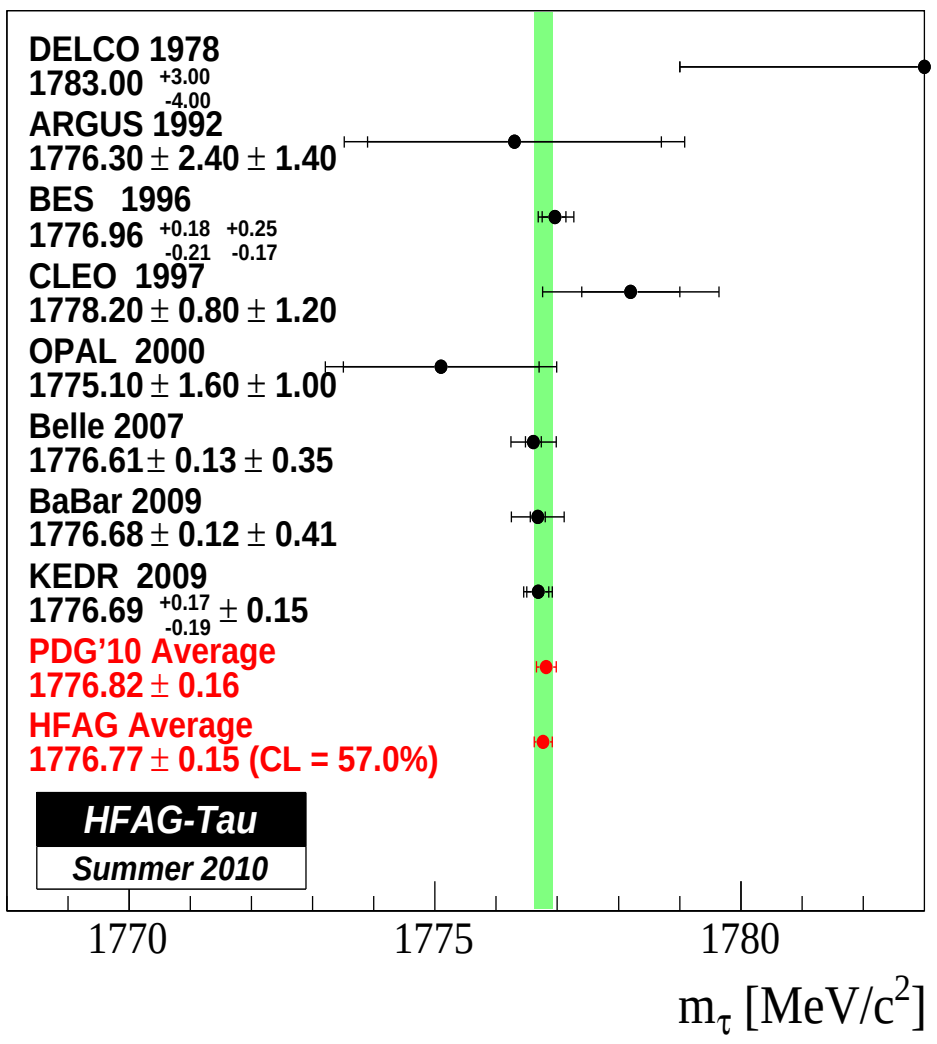

**Figure 20.2.2.** Measurements and average value of $m_\tau$ (Asner et al., 2010).

The mass difference between $\tau^+$ and $\tau^-$, $\triangle m = m_{\tau^+} - m_{\tau^-}$, can be measured precisely since many sources of systematic error are common for $\tau^+$ and $\tau^-$ and cancel out in $\triangle m$. The values of $\triangle m$ measured by Belle and *BABAR* collaborations are

$$\begin{aligned}\triangle m &= (\ \ 0.05 \pm 0.23(\text{stat}) \pm 0.14(\text{syst}))\ \text{MeV (Belle)},\\ \triangle m &= (-0.61 \pm 0.23(\text{stat}) \pm 0.06(\text{syst}))\ \text{MeV}\ (\textit{BABAR}).\end{aligned} \tag{20.2.7}$$

The values of $\triangle m$ obtained for both experiments are consistent with zero within the errors, where the precisions are dominated by statistical uncertainty. The systematic shift in the mass difference has been estimated from the mass differences for charged $D$ and $D_s$ mesons. The *BABAR* result shows some deviation, however this is interpreted as having a 1.2% chance of obtaining a result as different from zero as this under the condition of no *CPT* violation, which has been ascertained using MC simulation.

Combining both Belle and *BABAR* results, we obtain a mass difference of

$$\triangle m = (-0.24 \pm 0.18)\ \text{MeV}, \tag{20.2.8}$$

where the error is obtained by adding statistical and systematic errors from both experiments in quadrature. The mean value is the weighted average of the two experiments. From these results, we obtain an upper limit on the mass difference,

$$|m_{\tau+} - m_{\tau-}|/m^\tau_{\text{AVG}} < 3.0 \times 10^{-4}, \tag{20.2.9}$$

at 90% C.L. Where $m^\tau_{\text{AVG}}$ is the averaged mass of $m_{\tau+}$ and $m_{\tau-}$. The results improve upon the previous OPAL constraint (Abbiendi et al., 2000a) by one order of magnitude (see Table 20.2.2).

(In addition to tau mass, a precise measurement of the tau-lepton lifetime is reported recently by the Belle collaboration (Belous, 2014). For the measurement, they use an unique method that is only applicable in the asymmetric-energy $e^+e^-$ colliders.)

**Table 20.2.2.** Measured upper limit of the $\tau^+$ and $\tau^-$ mass difference at 90% C.L.

| Experiment | $\lvert m^+_\tau - m^-_\tau\rvert/m^\tau_{AVG}$ | Ref |
|---|---|---|
| OPAL | $< 3.0 \times 10^{-3}$ | Abbiendi et al. (2000a) |
| Belle | $< 2.8 \times 10^{-4}$ | Abe (2007c) |
| *BABAR* | $< 5.5 \times 10^{-4}$ | Aubert (2009ac) |

## 20.3 Tests of lepton universality

### 20.3.1 Charged current universality between $\mu$-$e$

Tests of $\mu - e$ universality can be expressed as

$$\left(\frac{g_\mu}{g_e}\right)^2 = \frac{\mathcal{B}(\tau^- \to \mu^-\overline{\nu}_\mu\nu_\tau)}{\mathcal{B}(\tau^- \to e^-\overline{\nu}_e\nu_\tau)}\frac{f(m_e^2/m_\tau^2)}{f(m_\mu^2/m_\tau^2)}, \tag{20.3.1}$$

where $f(x)$ is given by Eq. (20.2.2), assuming that the neutrino masses are negligible (Tsai, 1971). Also, in this equation, small corrections of the order $m^2_{e,\mu}/m^2_W$ and the difference between $\alpha(m_e)$ and $\alpha(m_\mu)$ are ignored, see Eq. (20.2.2). The relation between the weak coupling constant $g_l$ and the Fermi coupling constant $G_l$, for the lepton $l$, is given by

$$G_l = \frac{g_l^2}{4\sqrt{2}M_W^2}. \tag{20.3.2}$$

The HFAG group has performed a constrained fit (Amhis et al., 2012) using 157 branching fraction measurements and 47 constraint equations that fit 86 quantities. For example, there are measurements of the total branching fraction of all decays to three charged pions or kaons plus any number of neutrals. In addition, there are separate measurements of exclusive branching fractions to specific final states that have three identified charged mesons. One constraint is that the sum of exclusive 3-prong decays, the decays involving three charged particles in their final statess, must equal the inclusive 3-prong measurement. The fit is statistically consistent with the constraint that the sum of all base modes is equal to one, referred to as the "unitarity constraint", but the unitarity constraint is not explicitly applied. From that fit, which uses all available data including the recent *BABAR*'s results (Aubert, 2010f), we obtain $\mathcal{B}(\tau^- \to \mu^-\ \overline{\nu}_\mu\nu_\tau)/\mathcal{B}(\tau^- \to e^-\overline{\nu}_e\nu_\tau) = 0.9761 \pm 0.0028$, which includes a correlation coefficient of 23% between the branching fractions. This yields a value of $\left(\frac{g_\mu}{g_e}\right) = 1.0018 \pm 0.0014$, which is consistent with the SM value.

This prediction from tau decays is more precise than the other determinations:

– We average the measurements of $\mathcal{B}(\pi \to e\nu_e(\gamma))/\mathcal{B}(\pi \to \mu\nu_\mu(\gamma)) = (1.2265 \pm 0.0034\ (\text{stat}) \pm$

0.0044 (syst)) $\times 10^{-4}$ from TRIUMF (Britton et al., 1992) and = (1.2346 ± 0.0035 (stat) ± 0.0036 (syst)) $\times 10^{-4}$ from PSI (Czapek et al., 1993), to obtain a value of (1.2310 ± 0.0037) $\times 10^{-4}$. Comparing this with the prediction of (1.2352 ± 0.0001) $\times 10^{-4}$ from recent theoretical calculations (Cirigliano and Rosell, 2007), we obtain a value of $\left(\frac{g_\mu}{g_e}\right) = 1.0017 \pm 0.0015$.
- The ratio $\mathcal{B}(K \to e\nu_e(\gamma))/\mathcal{B}(K \to \mu\nu_\mu(\gamma))$ has recently been measured very precisely by the KLOE (Ambrosino et al., 2009b) and the NA62 (Goudzovski, 2011) collaborations. Using the new world average value of (2.487 ± 0.012) $\times 10^{-5}$ from Goudzovski (2010), and the predicted value of (2.477 ± 0.001) $\times 10^{-5}$ from Cirigliano and Rosell (2007), we obtain $\left(\frac{g_\mu}{g_e}\right) = 0.9980 \pm 0.0025$.
- From the report of the FlaviaNet Working Group on Kaon Decays (Antonelli et al., 2010b), we obtain $\left(\frac{g_\mu}{g_e}\right) = 1.0010 \pm 0.0025$ using measurements of $\mathcal{B}(K \to \pi\mu\overline{\nu})/\mathcal{B}(K \to \pi e\overline{\nu})$.
- From the report of the LEP Electroweak Working Group (Alcaraz et al., 2006), we obtain $\left(\frac{g_\mu}{g_e}\right) = 0.997 \pm 0.010$ using measurements of $\mathcal{B}(W \to \mu\overline{\nu}_\mu)/\mathcal{B}(W \to e\overline{\nu}_e)$.

### 20.3.2 Charged current universality between $\tau$-$\mu$

Tau-muon universality is tested with

$$\left(\frac{g_\tau}{g_\mu}\right)^2 = \frac{\mathcal{B}(\tau^- \to h^-\nu_\tau)}{\mathcal{B}(h^- \to \mu^-\overline{\nu}_\mu)} \frac{2m_h m_\mu^2 \tau_h}{(1+\delta_h) m_\tau^3 \tau_\tau} \left(\frac{1 - m_\mu^2/m_h^2}{1 - m_h^2/m_\tau^2}\right)^2 \tag{20.3.3}$$

where $h = \pi$ or $K$ and the radiative corrections are $\delta_\pi = (0.16 \pm 0.14)\%$ and $\delta_K = (0.90 \pm 0.22)\%$ (Decker and Finkemeier, 1994, 1995; Marciano and Sirlin, 1993).

Using the world average mass and lifetime values and meson decay rates (Nakamura et al., 2010) and our unitarity constrained fit including recent *BABAR* results (Aubert, 2010f), we determine $\left(\frac{g_\tau}{g_\mu}\right) = 0.9966 \pm 0.0030$ and $0.9860 \pm 0.0073$ from the pionic and kaonic branching fractions, respectively, where the correlation coefficient between these values is 13.10%. Combining these results, we obtain $\left(\frac{g_\tau}{g_\mu}\right) = 0.9954 \pm 0.0029$, which is 1.6 $\sigma$ below the SM expectation.

We also test lepton universality between $\tau$ and $\mu$ ($e$), by comparing the average electronic (muonic) branching fractions of the tau lepton with the predicted branching fractions from measurements of the $\tau$ and $\mu$ lifetimes and their respective masses (Nakamura et al., 2010), using known electroweak and radiative corrections (Marciano and Sirlin, 1993). This gives $\left(\frac{g_\tau}{g_\mu}\right) = 1.0011 \pm 0.0021$ and $\left(\frac{g_\tau}{g_e}\right) = 1.0030 \pm 0.0021$. The correlation coefficients between the determination of $\left(\frac{g_\tau}{g_\mu}\right)$ from the electronic branching fraction with the ones obtained from pionic and kaonic branching fractions are 48.16% and 21.82%, respectively. Averaging these three values, we obtain $\left(\frac{g_\tau}{g_\mu}\right) = 1.0001 \pm 0.0020$, which is consistent with the SM value. In Fig. 20.3.1, we compare these above determinations with each other and with the values obtained from W decays (Alcaraz et al., 2006).

## 20.4 Search for lepton flavor violation in tau decays

In order to progress beyond the Standard Model it is necessary to incorporate results from many different measurements and interpret them within a cohesive theoretical framework. This will include results from direct searches (and discoveries) of new particles at the energy frontier of the LHC, neutrino oscillation measurements, $g-2$ and electric dipole moment measurements, as well as searches (and discoveries) of LFV in the decays of leptons and mesons. Discoveries at the LHC alone will be insufficient to determine the underlying theoretical structures responsible for New Physics. Moreover, a discovery of $\mu^+ \to e^+\gamma$ alone will not provide sufficient information to nail down the underlying LFV mechanism or even to identify an underlying theory: it is critical to probe all LFV modes and searches for $\mu^+ \to e^+\gamma$ search need to be augmented by studies of $\tau^\pm \to \mu^\pm\gamma$ as well as $\tau^\pm \to e^\pm\gamma$. Even in the presence of the existing and projected $\mu^+ \to e^+\gamma$ bounds, $\tau^\pm \to \mu^\pm\gamma$ decays are predicted to occur at rates that are accessible at current experiments in many models (Aushev et al., 2010; Bona et al., 2007b). In fact, the full set of measurements of $\mu$ and $\tau$ LFV processes are required as in many models there are strong correlations between the expected rates of the different channels. In a supersymmetric seesaw model describing potential LFV (Babu and Kolda, 2002; Sher, 2002), for example, there is an expectation that the specific relative rates of $\mathcal{B}(\tau^\pm \to \mu^\pm\gamma) : \mathcal{B}(\tau^\pm \to \mu^\pm\mu^+\mu^-) : \mathcal{B}(\tau \to \mu\eta)$ are dependent on the model parameters. In the unconstrained minimal supersymmetric model (MSSM), which includes various correlations between the $\tau$ and $\mu$ LFV rates, $\tau$ LFV branching fractions can be as high as $10^{-7}$ (Brignole and Rossi, 2004; Goto, Okada, Shindou, and Tanaka, 2008) even with the strong experimental bounds on muon LFV.

### 20.4.1 Tau lepton data samples and search strategies

With 1,550 fb$^{-1}$ of data currently collected between the Belle and *BABAR* experiments and the $e^+e^- \to \tau^+\tau^-$ cross section of 0.919 nb (Banerjee, Pietrzyk, Roney, and Wąs, 2008), the world sample of $\tau$-leptons produced at the $e^+e^-$ colliders now exceeds $10^9$ which allows for experimental probing of LFV processes at the $\mathcal{O}(10^{-7})$ to $\mathcal{O}(10^{-8})$ levels.

The analyses typically select $\tau$-pair events with the appropriate charged-particle topology, removing non-$\tau$ events with an impact as minimal as possible on the signal efficiency. A candidate event is divided into hemispheres in the center-of-mass frame where each hemisphere contains either the $\tau^+$ or $\tau^-$ decay products. The $\tau$ decay associated with each hemisphere is then considered a possible candidate for the LFV decay under consideration, as can be seen in the *BABAR* detector display of a simulated $e^+e^- \to \tau^+\tau^-; \tau^+ \to e^+\bar{\nu}_\tau\nu_e; \tau^- \to \mu^-\gamma$ event depicted in Figure 20.4.1. Whereas Standard Model $\tau$-decays have at least one neutrino, the LFV decay products have a combined energy, $E_{\ell X}$, equal to the energy of the $\tau$ which is approximately equal to the beam energy in the center-of-mass, $\sqrt{s}/2$, and a mass ($m_{\ell X}$) equal to that of the $\tau$. Using a two dimensional signal region in the $m_{\ell X}$ *vs* $\Delta E$ plane, the signal is separated from the Standard Model $\tau$-decay backgrounds with minimal loss of efficiency, where $\Delta E = E_{\ell X} - \sqrt{s}/2$. The distributions for the $\tau^\pm \to e^\pm\gamma$ and $\tau^\pm \to \mu^\pm\gamma$ decays in that plane are shown in Figure 20.4.2 for the *BABAR* analysis, where the peaking at $\Delta E = 0$ and $m_{\ell X} = m_\tau = 1777\,\mathrm{MeV}/c^2$ is evident. For the lepton-photon invariant mass *BABAR* calculates $m_{\rm EC}$, which is obtained from a kinematic fit that requires the center-of-mass tau energy to be $\sqrt{s}/2$ after assigning the origin of the $\gamma$ candidate to the point of closest approach of the signal lepton track to the $e^+e^-$ collision axis. Use of a "signal box" in the $\Delta E$-$m_{\ell X}$ plane encompassing events within approximately two standard deviations of $\Delta E = 0$ and $m_{\ell X} = m_\tau = 1777\,\mathrm{MeV}/c^2$ serves as the most powerful requirement in the searches for LFV in $\tau$ decay. The signal peaks near zero in the distribution of $\Delta E = E_{\ell X} - \sqrt{s}/2$ and typically has a standard deviation of around 50 MeV. Using a beam-energy constrained mass and constraining photons to come from the same primary vertex as the charged particles in the event enables a resolution on $m_{\ell X}$ of 9 MeV to be achieved.

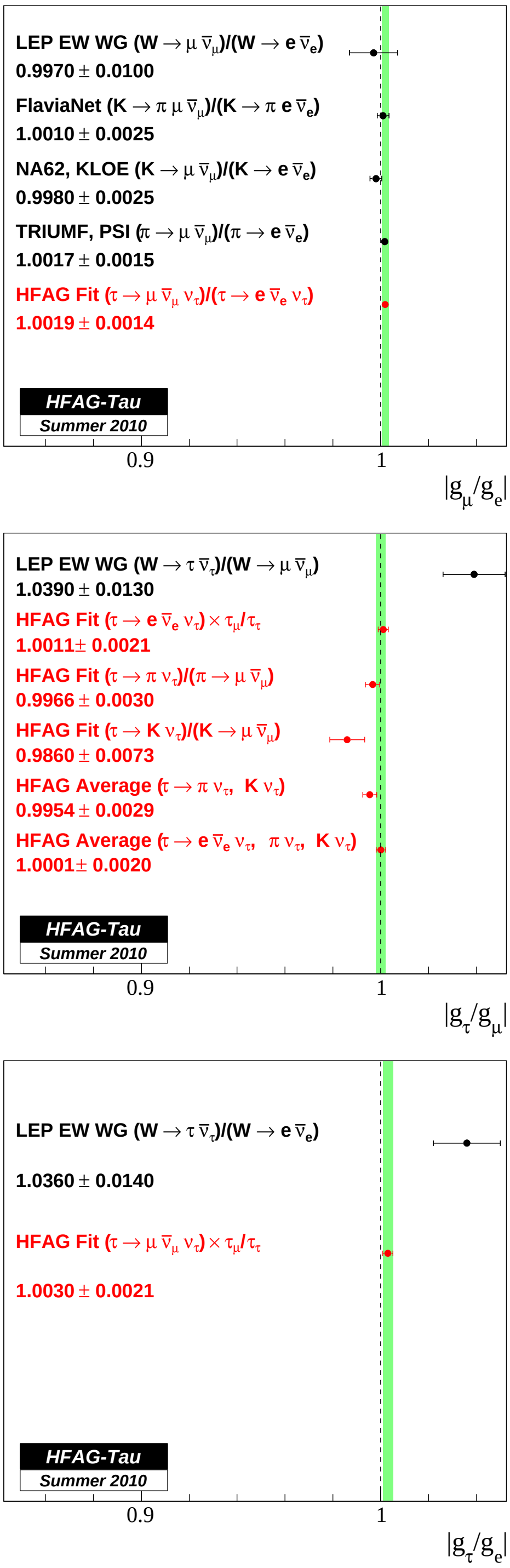

**Figure 20.3.1.** Measurements of lepton universality from W, kaon, pion, and tau decays.

The analyses are normally optimized to give the best "expected upper limit" using MC simulations of the signal and backgrounds. Signal efficiency ($\epsilon$) is initially estimated using simulated events and typically lies between 2% and 10%, depending on the channel under study. The components of a generic $\tau$ LFV decay selection efficiency are roughly: trigger (90%), acceptance/reconstruction (70%), charged-particle hemisphere topology (1-vs-1 or 1-vs-3: 70%), particle identification (50%), requirements apart from those on $\Delta E$ and $m_{\ell X}$ (50%), $\Delta E$ vs $m_{\ell X}$ signal box requirements (50%). Data-driven corrections are applied to the simulated signal efficiencies using the results of comparisons between data and simulated control samples.

Estimates of the expected number of background events ($N_{\rm bkd}$) are usually estimated using the distribution shapes from the Monte Carlo simulation of backgrounds with the normalization obtained from the data in the regions outside the signal box. These *BABAR* and Belle analyses are "blind" in the sense that the analysts have no knowledge of the data in the signal region when optimizing for a best "expected upper limit" and estimating systematic uncertainties. The data in the signal region is "unblinded" only after these steps are completed, and the analyst learns the number of events observed in the signal region ($N_{\rm obs}$), either making a discovery, or - as has been the case to date - setting an upper limit on the process (see Chapter 14 for a general discussion of blind analysis techniques).

$N_{\rm obs}$ and $N_{\rm bkd}$ together gives the number of signal events ($N_{\rm sig}$). When $N_{\rm obs}$-$N_{\rm bkd}$ is consistent with zero, an upper limit on $N_{\rm sig}$ ($N_{90}^{UL}$) is established. Conceptually, the 90%C.L. branching ratio upper limit is obtained from:

$$\mathcal{B}_{90}^{UL} = \frac{N_{90}^{UL}}{2N_{\tau\tau}\epsilon} = \frac{N_{90}^{UL}}{2\mathcal{L}\sigma_{\tau\tau}\epsilon}, \qquad (20.4.1)$$

where $N_{\tau\tau} = \mathcal{L}\sigma_{\tau\tau}$ is the number of $\tau$-pairs produced in $e^+e^-$ collisions obtained from the integrated luminosity, $\mathcal{L}$, and $\tau$-pair production cross section, $\sigma_{\tau\tau}$. In practice, when $N_{\rm bkd}$ is more than a few events, $N_{\rm sig}$ and $N_{\rm bkd}$ are determined from a fit.

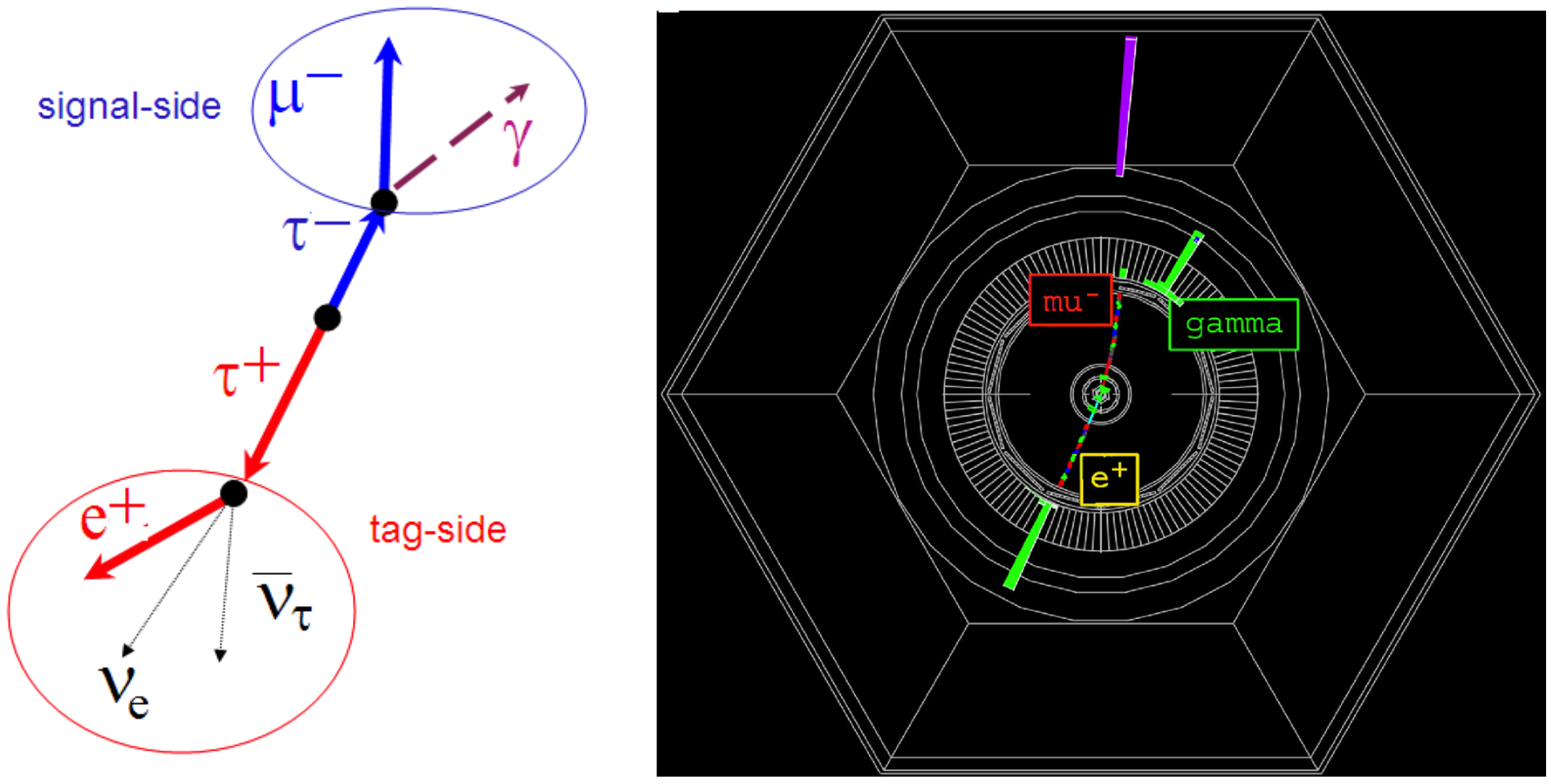

**Figure 20.4.1.** A *BABAR* event display with a simulated LFV $\tau^\pm \to \mu^\pm\gamma$ decay opposite a Standard Model $\tau^+ \to e^+\bar{\nu}_\tau\nu_e$ decay

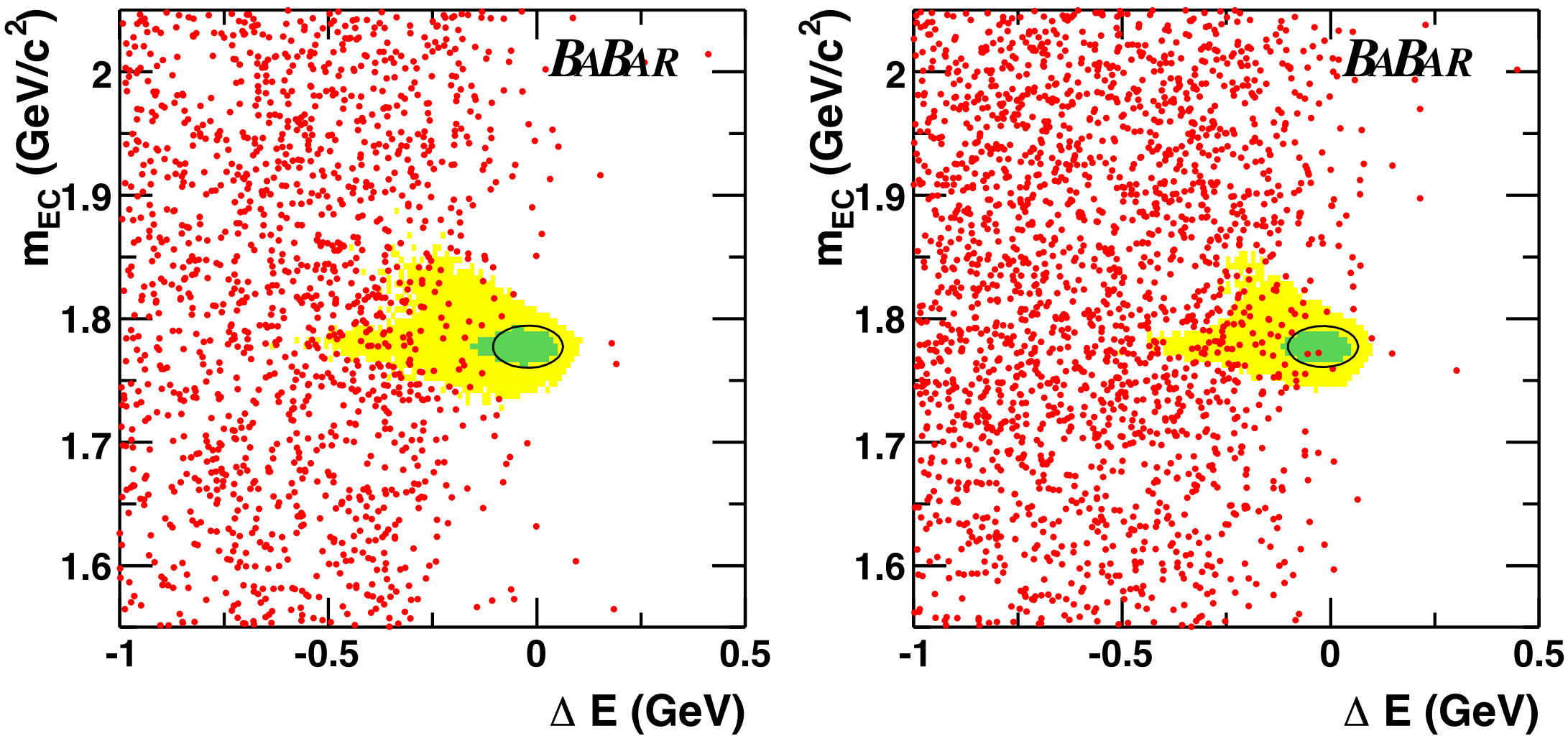

**Figure 20.4.2.** The $2\sigma$ elliptical signal-box for $\tau^\pm \to e^\pm\gamma$ (Left) and $\tau^\pm \to \mu^\pm\gamma$ (Right) decays in the $m_{\rm EC}$ vs. $\Delta E$ plane in the *BABAR* analysis. $m_{\rm EC}$ is the invariant mass of the lepton-photon pair as discussed in the text. Data are shown as red dots and contours containing 90% (50%) of signal MC events are shown as the yellow (green) shaded regions (Aubert, 2010i).

### 20.4.2 Results on LFV decays of the tau from Belle and BABAR

In performing searches, LFV decays can be conveniently classified as $\tau^\pm \to \ell^\pm\gamma$, $\tau^\pm \to \ell_1^\pm\ell_2^+\ell_3^-$ and $\tau^\pm \to \ell^\pm h^0$ where $\ell$ is either an electron or muon and $h^0$ represents a hadronic system. For the *BABAR* and Belle searches, the $h^0$ has been categorized in three ways: i) a pseudoscalar meson: *e.g.* $\pi^0$, $\eta$, $\eta'$, $K_S^0$; ii) a neutral vector meson: *e.g.* $\rho, \omega, K^*(892), \phi$; and iii) inclusive two charged meson decays, $h^0 = h_1^+h_2^-$ where $h_{1(2)}^\pm$ is either $\pi^\pm$ or $K^\pm$.

The most recent $\tau^\pm \to \mu^\pm\gamma$ and $\tau^\pm \to e^\pm\gamma$ results reported by Belle (Hayasaka, 2008) use a data sample having an integrated luminosity of 535 fb$^{-1}$ which corresponds to $492\times10^6$ $\tau$-pair events. Figure 20.4.3 shows the distribution in the $m_{\ell\gamma}$ vs $\Delta E$ plane for the selected sample in the Belle experiment. The main $\tau^\pm \to \mu^\pm\gamma$ backgrounds in these searches arise from $e^+e^- \to \mu^+\mu^-\gamma$ events and $e^+e^- \to \tau^+\tau^-\gamma$ events where one of the $\tau$'s decays via $\tau \to \mu\nu\bar{\nu}$. In both cases the photon, from initial state radiation in the latter and initial or final state radiation in the former, combines with a muon to accidentally fall within the signal box. The $e^+e^- \to \tau^+\tau^-\gamma; \tau^- \to \mu^-\nu_\tau\overline{\nu}_\mu$ events can be classified as "irreducible" because the events are genuine $\tau$-pair events and the $\mu$ and $\gamma$ are correctly identified and measured. A similar irreducible background source from $e^+e^- \to \tau^+\tau^-\gamma; \tau^- \to e^-\nu_\tau\overline{\nu}_e$ exists. Belle set a 90% C.L. upper limit on the number of signal events for $\tau \to \mu\gamma$ ($\tau \to e\gamma$) of 2.0 (3.34) events. These yield upper limits of $\mathcal{B}(\tau \to \mu\gamma) < 4.5 \times 10^{-8}$ and $\mathcal{B}(\tau \to e\gamma) < 1.2 \times 10^{-7}$. *BABAR*'s 2009 published 90% C.L. upper limits using a 534 fb$^{-1}$ data sample are $4.4 \times 10^{-8}$ and $3.3 \times 10^{-8}$ on $\mathcal{B}(\tau \to \mu\gamma)$ and $\mathcal{B}(\tau \to e\gamma)$, respectively (Aubert, 2010i). Both experiments report classical frequentist confidence intervals. These are reported in Table 20.4.1.

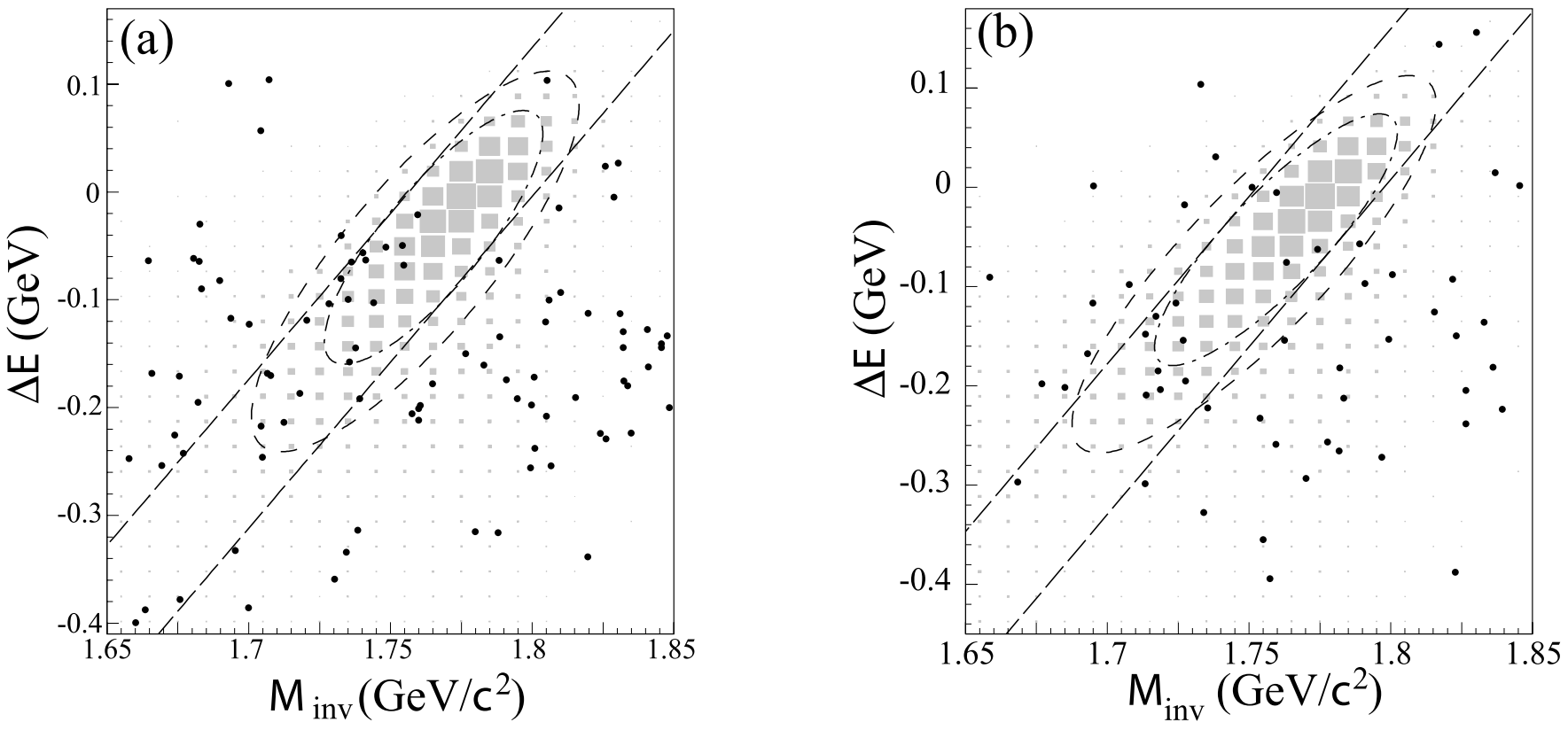

**Figure 20.4.3.** The distribution in the $m_{\ell\gamma}$ vs $\Delta E$ plane for (a) $\tau^{\pm} \to \mu^{\pm}\gamma$ and (b) $\tau^{\pm} \to e^{\pm}\gamma$ for the selected sample in the Belle analysis. The solid circles and the shaded boxes show the data and the signal MC, respectively. The outer (inner) ellipse shows the 3 (2)$\sigma$ signal region (Hayasaka, 2008).

**Table 20.4.1.** Summary of 90% C.L. upper limits on $\mathcal{B}(\tau^- \to \ell^-\gamma)$ and $\mathcal{B}(\tau^- \to \ell_1^-\ell_2^+\ell_3^-)$ LFV $\tau$ decays. $N_{\rm obs}$ and $N_{\rm bkg}$ are the number of events observed in the signal region and the estimated background, respectively. BF is the upper limit (90% C.L.) on the branching fraction.

| Channel | Belle (Hayasaka, 2008, 2010) | | *BABAR* (Aubert, 2010i; Lees, 2010a) | |
|---|---|---|---|---|
| | $N_{\rm obs}$ ($N_{\rm bkg}$) events | BF ($10^{-8}$) | $N_{\rm obs}$ ($N_{\rm bkg}$) events | BF ($10^{-8}$) |
| $\tau^- \to \mu^-\gamma$ | 10 ($13.9^{+6.0}_{-4.8}$) | 4.5 | 2 ($3.6 \pm 0.7$) | 4.4 |
| $\tau^- \to e^-\gamma$ | 5 ($5.14^{+3.86}_{-2.81}$) | 12 | 0 ($1.6 \pm 0.4$) | 3.3 |
| $\tau^- \to \mu^-e^+e^-$ | 0 (0.04±0.04) | 1.8 | 0 (0.64±0.19) | 2.2 |
| $\tau^- \to \mu^-\mu^+\mu^-$ | 0 (0.13±0.06) | 2.1 | 0 (0.44±0.17) | 3.3 |
| $\tau^- \to e^-\mu^+\mu^-$ | 0 (0.10±0.04) | 2.7 | 0 (0.54±0.14) | 3.2 |
| $\tau^- \to e^-e^+e^-$ | 0 (0.21±0.15) | 2.7 | 0 (0.12±0.02) | 2.9 |
| $\tau^- \to e^-\mu^+e^-$ | 0 (0.01±0.01) | 1.5 | 0 (0.34±0.12) | 1.8 |
| $\tau^- \to \mu^-e^+\mu^-$ | 0 (0.02±0.02) | 1.7 | 0 (0.03±0.02) | 2.6 |

Belle (Hayasaka, 2010) and *BABAR* (Lees, 2010a) also searched for $\tau \to \ell_1\ell_2\ell_3$. Fig. 20.4.4 shows the distributions in the $m_{\ell\ell\ell}$ vs $\Delta E$ plane for the $\tau \to \ell_1\ell_2\ell_3$ candidate events before the final selection. There is essentially no background in these samples, since the requirement for three leptons is tight and can reduce the background effectively. No evidence for a signal is seen by either experiment and the 90% C.L. upper limits on the branching fractions are presented in Table 20.4.1. Unlike the $\tau^{\pm} \to \mu^{\pm}\gamma$ and $\tau^{\pm} \to e^{\pm}\gamma$ searches, there is no irreducible background at the current luminosities. The Belle $\tau \to \ell_1\ell_2\ell_3$ analysis uses $719\times10^6$ $\tau$-pairs whereas BABAR reports on an analysis using $431\times10^6$ $\tau$-pairs. Note that, in addition to the reactions violating flavor, Table Table 20.4.1 also lists bounds of similar magnitude on $\tau^- \to \mathrm{e}^-\mu^+\mathrm{e}^-$ and $\tau^- \to \mu^-\mathrm{e}^+\mu^-$, which simultaneously violate the lepton-flavors, $L_e$, $L_\mu$ and $L_\tau$, but the total lepton number is conserved.

**Table 20.4.2.** Summary of 90% C.L. upper limit on $\mathcal{B}(\tau^- \to \ell^- h^0)$ in units of $(10^{-8})$ where $\ell = \mu$ or e and $h^0$ is either a pseudoscalar (upper half) or vector meson (lower half) from the Belle (Hayasaka, 2011; Miyazaki, 2006, 2010, 2011; Nishio, 2008) and BABAR (Aubert, 2007ar, 2008au, 2009o,ao) experiments.

| Channel | $\ell = \mathrm{e}$ Belle | $(10^{-8})$ BABAR | $\ell = \mu$ Belle | $(10^{-8})$ BABAR |
|---|---|---|---|---|
| $\tau^- \to \ell^- \pi^0$ | 2.2 | 13 | 2.7 | 11 |
| $\tau^- \to \ell^- \eta$ | 4.4 | 16 | 2.3 | 15 |
| $\tau^- \to \ell^- \eta'$ | 3.6 | 24 | 3.8 | 14 |
| $\tau^- \to \ell^- K_S^0$ | 2.6 | 3.3 | 2.3 | 4.0 |
| $\tau^- \to \ell^- \phi$ | 3.1 | 3.1 | 8.4 | 19.0 |
| $\tau^- \to \ell^- \rho^0$ | 1.8 | 4.6 | 1.2 | 2.6 |
| $\tau^- \to \ell^- \omega$ | 4.8 | 11 | 4.7 | 10 |
| $\tau^- \to \ell^- K^{*0}$ | 3.2 | 5.9 | 7.2 | 17.0 |
| $\tau^- \to \ell^- \overline{K}^{*0}$ | 3.4 | 4.6 | 7.0 | 7.3 |

Both Belle, using 901 fb$^{-1}$ (Hayasaka, 2011), and BABAR, using 339 fb$^{-1}$ (Aubert, 2007ar), have published bounds on LFV $\tau$ decays involving a lepton and a $\pi^0$, $\eta$ or $\eta'$ pseudoscalar. Belle has also published on searches for $\tau \to \ell K_S^0$ and $\tau \to \ell K_S^0 K_S^0$ (Miyazaki, 2010). Searches for LFV involving the $\omega$ vector meson, $\tau \to \ell\omega$, have been reported by both experiments with BABAR employing a data set of 384 fb$^{-1}$(Aubert, 2008au) whereas Belle used 854 fb$^{-1}$(Miyazaki, 2011). From the same data set Belle has also searched for $\tau \to \ell\rho$, $\tau \to \ell\phi$, $\tau \to \ell K^{*0}$ and $\tau \to \ell\overline{K}^{*0}$. The 90% C.L. upper limits on these processes are typically around $5\times10^{-8}$ and are listed in Table 20.4.2. BABAR, using a 221 fb$^{-1}$, sets limits on LFV inclusive decays with two charged mesons, $\tau^\pm \to \ell^\pm h_1^+ h_2^-$, where no assumptions are made on the resonance structure of the hadronic final state (Aubert, 2005ab). These bounds range from $1 \times 10^{-7}$ to $5 \times 10^{-7}$, depending on the final state. Belle's equivalent analysis used 854 fb$^{-1}$ and set bounds ranging from $2.0 \times 10^{-8}$ to $8.4 \times 10^{-8}$ (Miyazaki, 2013).

The status of searches for lepton flavor violation in $\tau$ decays is summarized in Figure 20.4.5. A table of these results and the corresponding references are provided by the HFAG report of Amhis et al. (2012).

### 20.4.3 Future Prospects

By the end of 2010 Belle and BABAR had collected a combined data sample of roughly 1.5 ab$^{-1}$, corresponding to the production of about $1.4\times10^9$ $\tau$ pairs, and those experiments can be expected to update their analyses with their complete data sets over the next year or two. However, SuperKEKB, a new significantly higher luminosity $e^+e^-$ collider designed to operate at the $\Upsilon$ resonances, but mainly on the $\Upsilon(4S)$, is on the horizon. It will provide exciting new opportunities for the discovery and potential study of LFV decays of the $\tau$ lepton. The physics potential of a super flavor factory operating with a luminosity of about $10^{36}\mathrm{cm}^{-2}\mathrm{s}^{-1}$ has been discussed extensively in (Aushev et al., 2010; Bona et al., 2007b) Such a facility is expected to probe LFV $\tau^\pm \to \ell^\pm\ell^+\ell^-$ and $\tau^\pm \to \ell^\pm h^0$ decays, which have no irreducible backgrounds, at the $\mathcal{O}(10^{-10})$ level. However, the initial state photon accidental background will prevent the $\tau^\pm \to \ell^\pm\gamma$ decays from being probed below the level of a few $10^{-9}$. However it should be noted that this irreducible background can be removed if one were to accumulate a large sample of tau leptons near production threshold (below about 4 GeV).

## 20.5 *CP* violation in the tau lepton system

Understanding the origin of *CP* violation is one of the most important outstanding questions in particle physics. To date *CP* violation has been observed only in the $K$ and $B$ meson system[157]. In the Standard Model, all observed *CP* violation effects can be accommodated by a single irreducible, complex phase in the CKM quark mixing matrix.

[157] The LHCb collaboration reported in 2011 the $3.5\sigma$ significance for the difference of the direct *CP* asymmetry $\Delta A_{CP} = A_{CP}(K^+K^-) - A_{CP}(\pi^+\pi^-)$, where $A_{CP}(K^+K^-)$ and $A_{CP}(\pi^+\pi^-)$ are the *CP* asymmetry for $D^0(\overline{D}^0) \to K^+K^-$ and $D^0(\overline{D}^0) \to \pi^+\pi^-$, respectively (Aaij et al., 2012c). However, their recent update does not confirm this observation (Aaij et al., 2013e). So more data are needed to establish *CP* violation in the $D$ meson system.

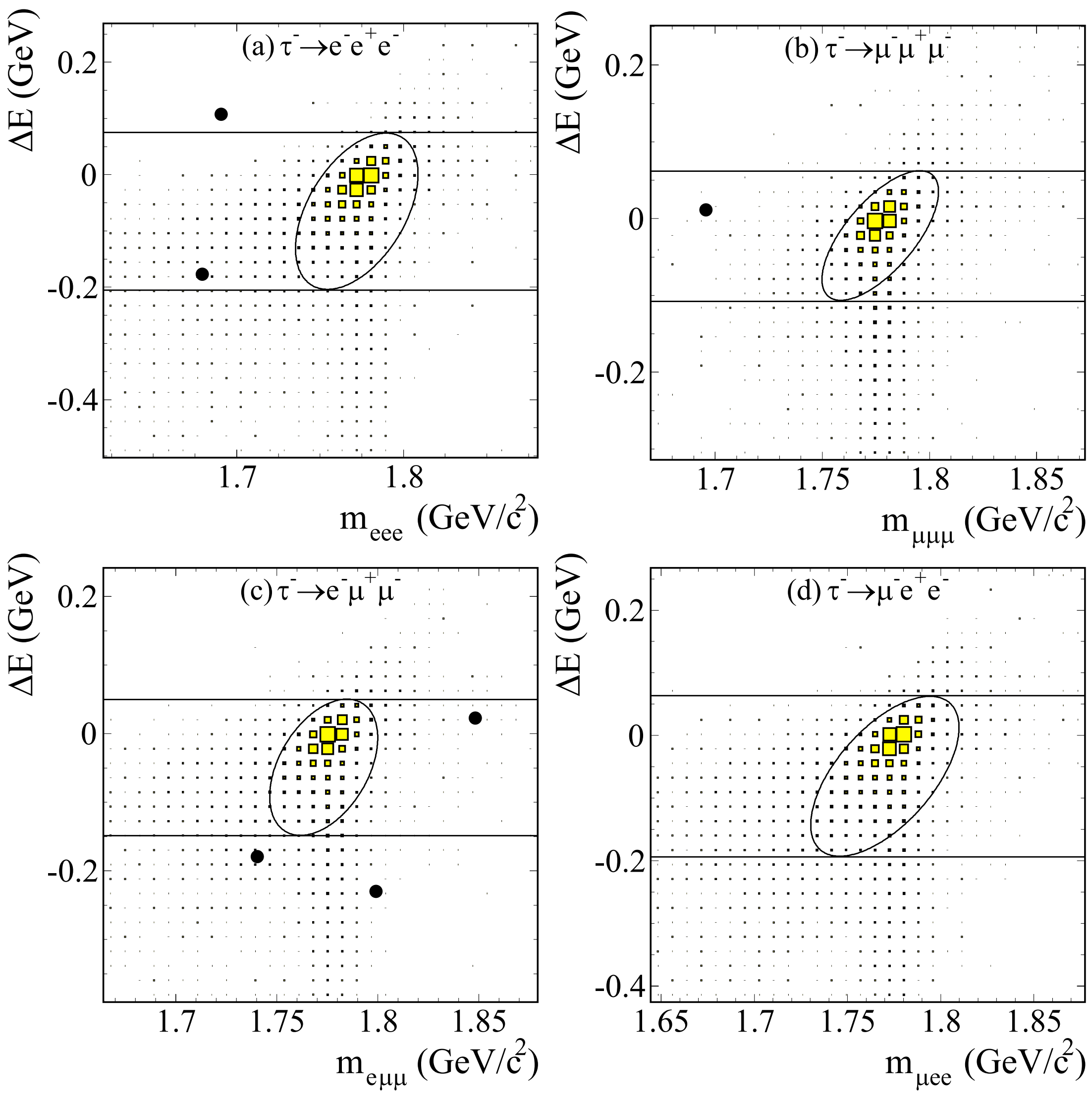

**Figure 20.4.4.** Distributions in the $M_{\ell\ell\ell}$ vs $\Delta E$ plane for the selected events for (a) $\tau^\pm \to e^\pm e^+ e^-$, (b) $\tau^\pm \to \mu^\pm \mu^+ \mu^-$, (c) $\tau^\pm \to e^\pm \mu^+ \mu^-$ and (d) $\tau^\pm \to \mu^\pm e^+ e^-$ modes in the Belle analysis (Hayasaka, 2010). The solid circles are data. The shaded boxes show the MC signal distribution with arbitrary normalization. The ellipse is the signal region used for evaluating the signal yield.

The CKM mechanism alone is however not sufficient enough to explain the observed matter-antimatter asymmetry in the universe, and thus new sources of *CP* violation are necessary. In this regard, one important area is the lepton sector. In the neutral lepton sector, the $3 \times 3$ neutrino mixing matrix can accommodate *CP* violation. A search for signs of *CP* violation in the neutrino sector is thus a primary task in future neutrino experiments. While in the charged lepton sector, there is no such mixing, so mixing-induced *CP* violation is not expected. However, physics beyond the Standard Model could produce *CP* violation in processes involving charged leptons; we would expect such effects to be enhanced for tau leptons because of their large mass.

In this section, we first discuss *CP* violation in tau-pair production, which is usually parameterized in terms of the electric dipole moment (EDM) of the tau lepton. We then describe various searches for the *CP* violation in tau decays.

### 20.5.1 Electric dipole moment of the tau lepton

If a particle has an EDM value of $d$, the Hamiltonian $H$ describing a non-relativistic particle of spin $S$ placed in a electric field $\boldsymbol{E}$ can be given by

$$H = -d\,\boldsymbol{E} \cdot \frac{\boldsymbol{S}}{S}. \qquad (20.5.1)$$

This interaction violates both parity and time-reversal invariance. Therefore, a non-zero $d$ can exist if and only if both parity and time-reversal invariance (or *CP* invariance under the *CPT* theorem) are broken.

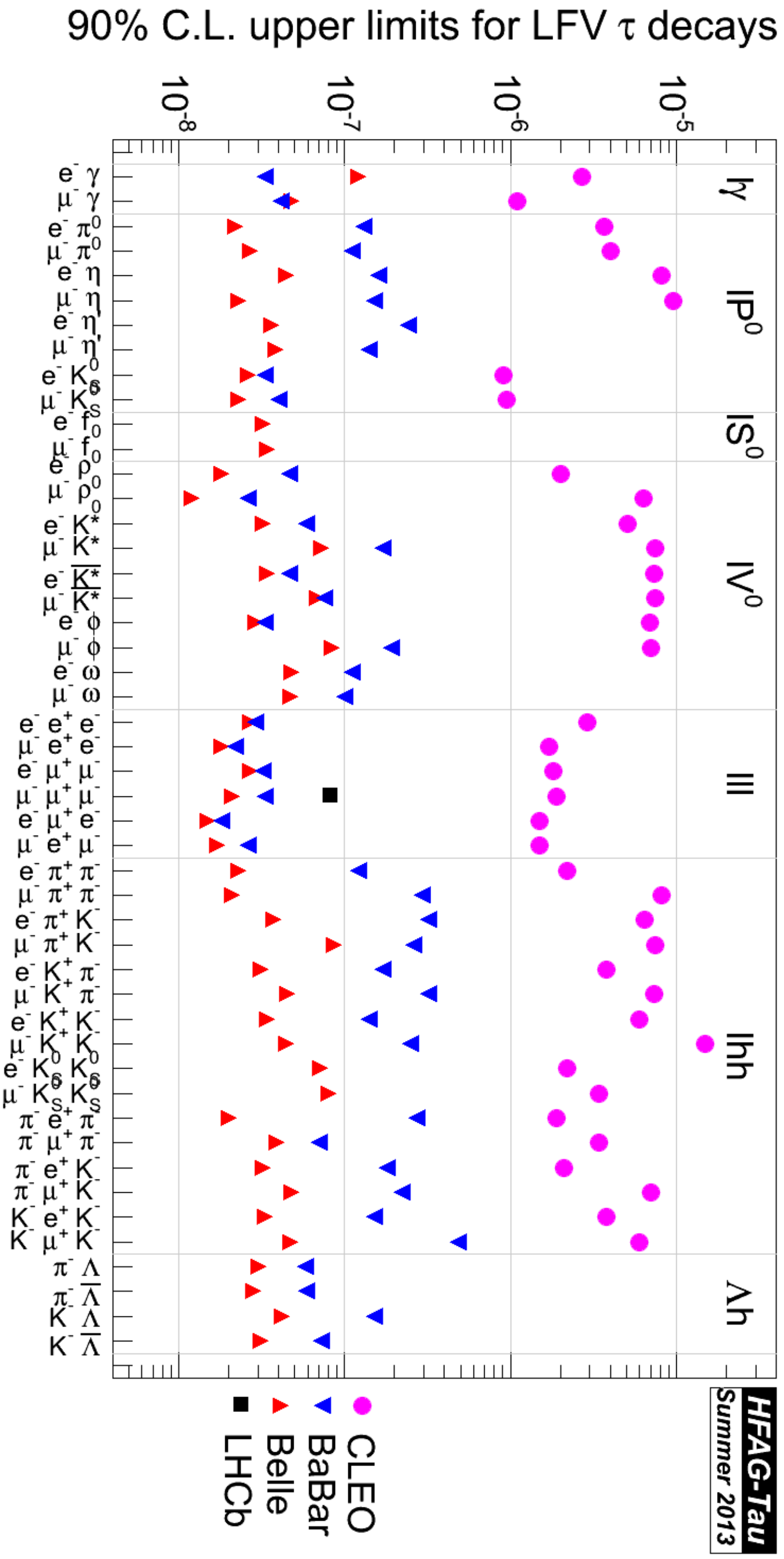

**Figure 20.4.5.** (color online) Limits on the branching fraction at 90 % C.L., obtained from searches for lepton flavor violation in $\tau$ decays. Results from CLEO (closed circle), *BaBar* (triangle-down), Belle (triangle-up) and LHCb (square) experiments are shown. The LHCb result is taken from Aaij et al. (2013f).

There are extensive studies of the EDM for electron, muon, neutron and various nuclei such as $^{199}$Hg,$^{205}$Tl, that provide stringent constraints on new *CP*-violating physics (Czarnecki and Marciano, 2010; Roberts and Mar-ciano, 2010). The current limit on the tau EDM, $d_\tau$, is many orders of magnitude less than that of the electron and nucleons, since it is difficult to measure its EDM due to the short lifetime. However, its size is interesting both theoretically and experimentally. In the Standard Model, the EDM of the tau can arise at the multi-loop level through CKM type quark mixing and is extremely small, $d_\tau < 10^{-34} e\,\mathrm{cm}$, (Hoogeveen, 1990; Pospelov and Khriplovich, 1991). A much larger contribution to the tau EDM is however predicted by various SM extensions such as supersymmetry (Pospelov and Ritz, 2005), unparticle physics (Moyotl, Rosado, and Tavares-Velasco, 2011), and mirror leptons (Ibrahim and Nath, 2010).

Experimentally, one can measure the tau EDM by using the momentum correlation of decay products in the tau-pair production $e^+e^- \to \tau^+\tau^-$ as explained below. The relativistic generalization of the interaction $d\boldsymbol{E}\cdot\boldsymbol{S}$ for a tau lepton (spin 1/2 particle) $\Psi$ can be expressed by an effective Lagrangian as

$$\mathcal{L}_{CP} = -d_\tau \frac{i}{2}\overline{\Psi}\sigma^{\mu\nu}\gamma_5\Psi F_{\mu\nu}, \quad (20.5.2)$$

where $d_\tau$ is the electric dipole moment of the tau lepton. Furthermore, $F^{\mu\nu}$ is the electric field tensor and $\sigma^{\mu\nu} = \frac{i}{2}(\gamma^\mu\gamma^\nu - \gamma^\nu\gamma^\mu)$. From this Lagrangian and the Standard Model one, the squared matrix element, $\mathcal{M} = \mathcal{M}_{\mathrm{SM}} + d_\tau\mathcal{M}_{\mathrm{EDM}}$, for the tau-pair production

$$e^+(\boldsymbol{p}) + e^-(-\boldsymbol{p}) \to \tau^+(\boldsymbol{k},\boldsymbol{S}_+) + \tau^-(-\boldsymbol{k},\boldsymbol{S}_-) \quad (20.5.3)$$

is given by the sum of the SM term $|\mathcal{M}|^2_{\mathrm{SM}}$, the EDM term $|d_\tau|^2|\mathcal{M}|^2_{\mathrm{EDM}}$ and the interference between them

$$\begin{aligned}|\mathcal{M}|^2 &= (\mathcal{M}^\dagger_{\mathrm{SM}} + d^\dagger_\tau\mathcal{M}^\dagger_{\mathrm{EDM}})(\mathcal{M}_{\mathrm{SM}} + d_\tau\mathcal{M}_{\mathrm{EDM}})\\ &= |\mathcal{M}|^2_{\mathrm{SM}} + Re(d_\tau)\mathcal{M}^2_{Re} + Im(d_\tau)\mathcal{M}^2_{Im}\\ &\quad +|d_\tau|^2|\mathcal{M}|^2_{\mathrm{EDM}}, \end{aligned} \quad (20.5.4)$$

where $Re(d_\tau)$ $[Im(d_\tau)]$ is the real [imaginary] part of the EDM. Since these terms vanish for the total integrated cross section, one needs to study *CP*-odd observables. The interference terms $\mathcal{M}_{Re/Im}$, being proportional to real and imaginary part of $d_\tau$, contain the following combination of spin-momentum correlations:[158]

$$\mathcal{M}^2_{Re} : (\boldsymbol{S}_+ \times \boldsymbol{S}_-)\cdot\widehat{\boldsymbol{k}} \text{ and } (\boldsymbol{S}_+ \times \boldsymbol{S}_-)\cdot\widehat{\boldsymbol{p}} \quad (20.5.5)$$

$$\mathcal{M}^2_{Im} : (\boldsymbol{S}_+ - \boldsymbol{S}_-)\cdot\widehat{\boldsymbol{k}} \text{ and } (\boldsymbol{S}_+ - \boldsymbol{S}_-)\cdot\widehat{\boldsymbol{p}}, \quad (20.5.6)$$

where $\widehat{\boldsymbol{p}}$ $(\widehat{\boldsymbol{k}})$ is the unit momentum vector of $e^+(\tau^+)$ in the CM frame and $\boldsymbol{S}_\pm$ are the spin vectors for $\tau^\pm$. To measure $Re(d_\tau)$ one needs a *CP*-odd and *T*-odd (*CPT*-even) operator, as shown in the first line. To measure $Im(d_\tau)$ one needs a *CP*-odd and *T*-even (*CPT*-odd) operator, as shown in the second equation (20.5.6).[159] These terms change their sign for the *CP* transformation (*i.e.* *CP*-odd terms). So if one of these terms is non-zero, the process violates *CP*.

In order to optimize the sensitivity to $d_\tau$, Belle employs a so-called optimal observable method, first proposed by Atwood and Soni (1992). In this method, the optimal observables can be defined as

$$\mathcal{O}_{Re} = \frac{\mathcal{M}^2_{Re}}{|\mathcal{M}_{\mathrm{SM}}|^2}, \quad \mathcal{O}_{Im} = \frac{\mathcal{M}^2_{Im}}{|\mathcal{M}_{\mathrm{SM}}|^2}. \quad (20.5.7)$$

The mean value of the observable $\mathcal{O}_{Re}$ is given by

$$\begin{aligned}< \mathcal{O}_{Re} > &\propto \int \mathcal{O}_{Re}|\mathcal{M}|^2 d\phi\\ &= \int \mathcal{M}^2_{Re} d\phi + Re(d_\tau)\int \frac{(\mathcal{M}^2_{Re})^2}{|\mathcal{M}_{\mathrm{SM}}|^2} d\phi, \end{aligned} \quad (20.5.8)$$

where the integration is over the phase space ($\phi$) spanned by the relevant kinematic variables. The expression for the imaginary part is similar. The first term containing the integral of $\mathcal{M}^2_{Re}$ and $\mathcal{M}^2_{Im}$ drops out because of their symmetry properties. The means of the observables $< \mathcal{O}_{Re} >$ and $< \mathcal{O}_{Im} >$ are therefore linear functions of $d_\tau$,

$$< \mathcal{O}_{Re} >= a_{Re}\cdot Re(d_\tau), \quad < \mathcal{O}_{Re} >= a_{Im}\cdot Im(d_\tau). \quad (20.5.9)$$

Eight different final states in the decays of $\tau$-pairs,

$$\begin{aligned}&(e^+\nu_e\overline{\nu}_\tau)(\mu^-\overline{\nu}_\mu\nu_\tau)\ ,\ (e^+\nu_e\overline{\nu}_\tau)(\pi^-\nu_\tau),\\ &(\mu^+\nu_\mu\overline{\nu}_\tau),(\pi^-\nu_\tau)\ ,\ (e^+\nu_e\overline{\nu}_\tau)(\rho^-\nu_\tau),\\ &(\mu^+\nu_\mu\overline{\nu}_\tau)(\rho^-\nu_\tau)\ ,\ (\pi^-\nu_\tau)(\rho^+\overline{\nu}_\tau),\\ &(\rho^-\nu_\tau)(\rho^+\overline{\nu}_\tau)\ ,\ (\pi^-\nu_\tau)(\pi^+\overline{\nu}_\tau) \end{aligned} \quad (20.5.10)$$

and their charge-conjugate modes are analyzed.

Because of the undetectable particles (neutrinos), one can not fully reconstruct the quantities $\boldsymbol{S}_\pm$ and $\widehat{\boldsymbol{k}}$. Therefore, for each event the mean values of $|\mathcal{M}_{\mathrm{SM}}|^2$, $\mathcal{M}^2_{Re}$ and $\mathcal{M}^2_{Im}$ are obtained by averaging over all possible kinematic configurations. In the case when both tau leptons decay hadronically, the tau flight direction can be determined with a two-hold ambiguity. In the case of leptonic tau decays, there is an additional ambiguity from the effective mass of the two daughter neutrinos. A Monte Carlo treatment is adopted to take into account the additional ambiguity in the effective mass of the $\nu\overline{\nu}$ system. Explicit formulae for reconstructing the tau's flight direction and the spin vectors are provided in Posthaus and Overmann (1998) and Ackerstaff et al. (1997b).

The obtained optimal observable distributions for the $\tau^+\tau^- \to (\pi\nu_\tau)(\rho\nu_\tau)$ mode in the Belle experiment (Inami, 2003) are shown in Fig. 20.5.1 (a) and (b) for real and imaginary part, respectively. Note that the width of the distribution is proportional to the sensitivity. The distributions do not show any apparent asymmetry.

The values of the EDM obtained from the mean values of the optimal observables are plotted in Figure 20.5.2 (a) and (b). All results are consistent with the EDM of zero within statistical errors. Taking the weighted average of eight different modes, Belle sets the 95% confidence level interval for the tau-lepton EDM (Inami, 2003) as

$$-2.2 \times 10^{-17} e\,\mathrm{cm} < Re(d_\tau) < 4.5 \times 10^{-17} e\,\mathrm{cm} \qquad (20.5.11)$$

and

$$-2.5 \times 10^{-17} e\,\mathrm{cm} < Im(d_\tau) < 0.8 \times 10^{-17} e\,\mathrm{cm}. \qquad (20.5.12)$$

[158] The complete form is given by Bernreuther, Nachtmann, and Overmann (1993).

[159] Even in the absence of *CPT* violation, a non-zero value of $Im(d_\tau)$ could be generated through absorptive contributions, *i.e.*, rescattering corrections from on-shell intermediate states.

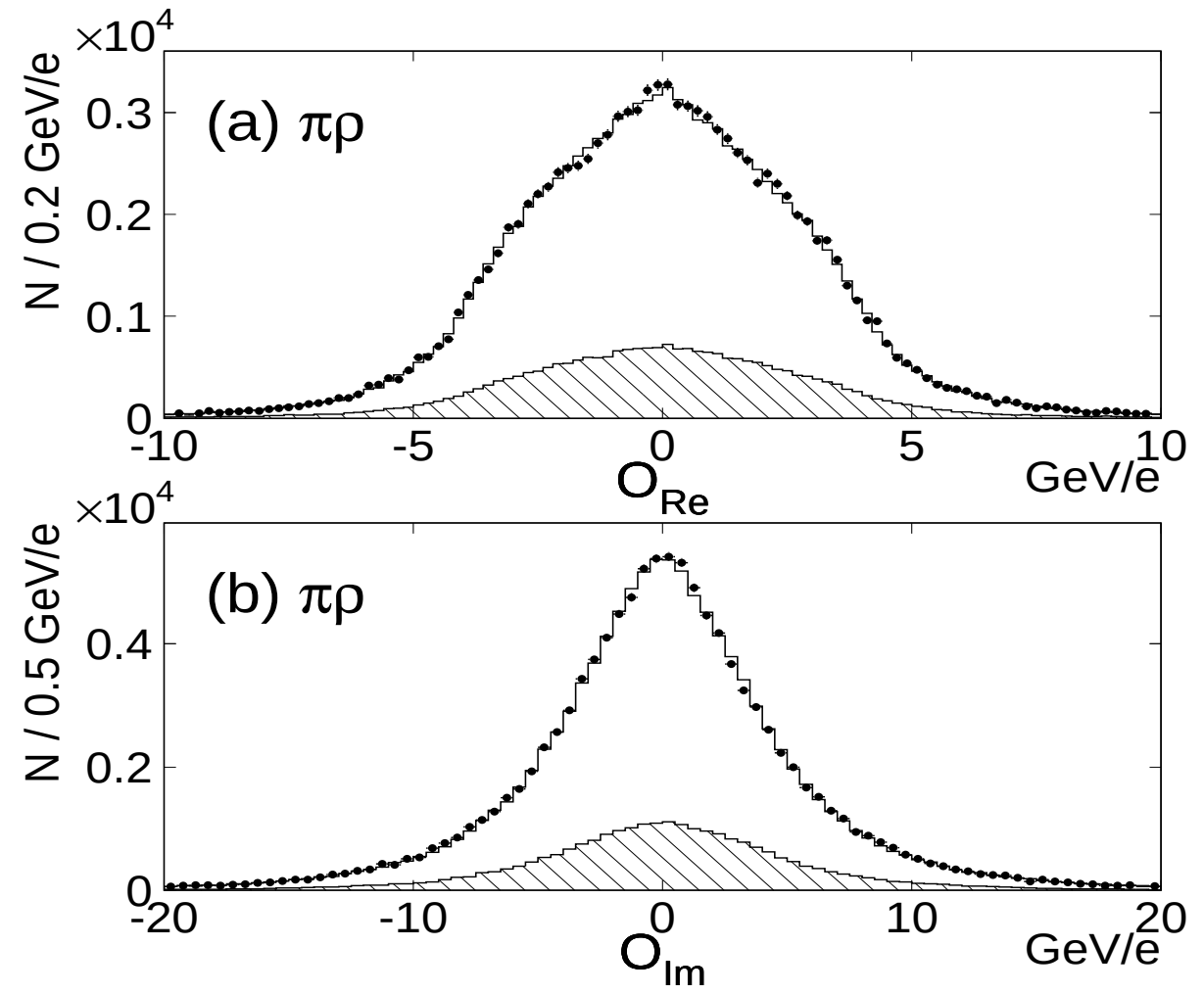

**Figure 20.5.1.** Distribution of optimal observables, (a) $\mathcal{O}_{Re}$ and (b) $\mathcal{O}_{Im}$, for $\pi\rho$ events $\tau^-\tau^+ \to (\pi^-\bar{\nu}_\tau)(\rho^+\nu_\tau)$ (Inami, 2003). Dots are Belle data and white and shaded histograms show the MC simulation for signal and background, respectively. If *CP* is violated, mean values $< \mathcal{O}_{Re} >$, $< \mathcal{O}_{Im} >$ are differ from zero.

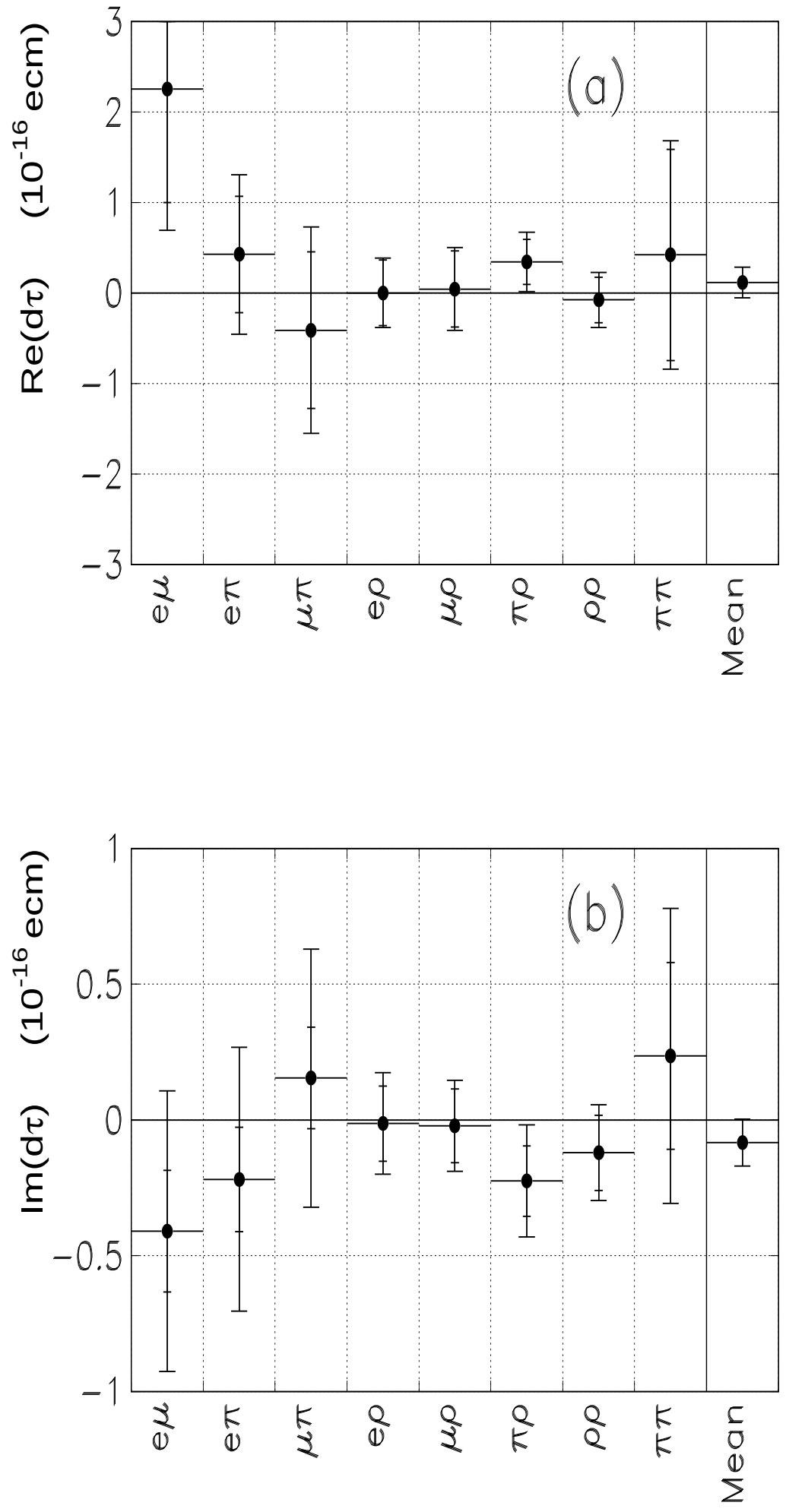

**Figure 20.5.2.** Results of the tau-lepton EDM for eight modes and the weighted mean for the (a) real and (b) imaginary parts. The error bars include both statistical and systematic errors. The small ticks on the error bars show the statistical errors (Inami, 2003).

These limits are ten times more stringent than the previous results given by L3 Acciarri et al. (1998), OPAL (Ackerstaff et al., 1998) and DELPHI (Abdallah et al., 2004) experiments:

$$|d_\tau| < 3.1 \times 10^{-16}\ e\,\mathrm{cm}\ \text{(L3)}, \qquad (20.5.13)$$

$$|d_\tau| < 3.7 \times 10^{-16}\ e\,\mathrm{cm}\ \text{(OPAL)}, \qquad (20.5.14)$$

$$|d_\tau| < 3.7 \times 10^{-16}\ e\,\mathrm{cm}\ \text{(DELPHI)}. \qquad (20.5.15)$$

### 20.5.2 *CP* violation in tau decay

In the Standard Model, no observable *CP* violation is expected in the hadronic decays of tau leptons except for the known *CP* violation in the neutral kaon system. Observing a signal would then be manifestation of some kind of new physics. For example, the *CP* violation could originate from the minimal supersymmetric Standard Model (MSSM) (Calderon, Delepine, and Castro, 2007; Ibrahim and Nath, 2008) or from multi-Higgs doublet models (MHDM) (Grossman, 1994; Kiers, Soni, and Wu, 2000; Weinberg, 1976). The charged Higgs bosons in these models play an important role in strangeness changing (Cabibbo-suppressed) processes with $\triangle S = 1$. In the $\tau^- \to K^0_S\pi^-\nu_\tau$ mode, a Standard Model *CP* asymmetry of 0.3%, due to the *CP* violation in $K_L \to \pi^+\pi^-$, is expected in the decay rates (Bigi and Sanda, 2005; Calderon, Delepine, and Castro, 2007). While in new physics models such as MSSM or MHDM, a non-zero *CP* asymmetry is not expected in the decay rates of $\tau^\pm$ even if the intermediate scalar bosons have *CP* violating couplings, but the bosons in the models introduce a *CP* asymmetry in the angular distribution of the tau decays.

#### 20.5.2.1 Decay rate asymmetry

A first search for *CP* violation in the decay rate of the tau lepton is carried out by *BABAR*. They use a dataset of 437 million tau lepton pairs (Lees, 2012q) and measure the decay-rate asymmetry

$$A_{CP} = \frac{\Gamma(\tau^+ \to K_S^0\pi^+\overline{\nu}_\tau) - \Gamma(\tau^- \to K_S^0\pi^-\nu_\tau)}{\Gamma(\tau^+ \to K_S^0\pi^+\overline{\nu}_\tau) + \Gamma(\tau^- \to K_S^0\pi^-\nu_\tau)} \tag{20.5.16}$$

for a $\tau^- \to K_S^0\pi^-(\geq 0\pi^0)\nu_\tau$ sample. The signal candidates are selected by dividing the event into two hemispheres using the event thrust axis in the $e^+e^-$ CM system. *BABAR* selects events with a charged track identified as a pion, a $K_S^0 \to \pi^+\pi^-$ and up to three $\pi^0$ candidates in the "signal" hemisphere. To reduce background from $q\bar{q}$ continuum, *BABAR* require that the momentum of the charged track in the "tag" hemisphere be less than 4 GeV/$c$ in the CM system and be identified as an electron or muon and the magnitude of the event thrust be between 0.92 and 0.99.

After all selection criteria are applied, a total of 199,064 (140,602) candidates are obtained in the $e$-tag ($\mu$-tag) sample, of which there are 99842 (70369) in the $\tau^-$ and 99222 (70233) in the $\tau^+$ sample. The background is estimated to be at the 1% level. After the subtraction of background composed of $q\bar{q}$ and non-$K_S^0$ tau decays, the decay-rate asymmetry is measured to be $(-0.32 \pm 0.23 \pm 0.13)\%$ for the $e$-tag sample and $(-0.05 \pm 0.27 \pm 0.10)\%$ for the $\mu$-tag sample, where the first errors are statistical and the second are systematic.

To these measured rate-asymmetries, two corrections are applied. One is the correction for the different nuclear-interaction cross section of the $K^0$ and $\overline{K}^0$ mesons with the material in the detector. A correction to the asymmetry is calculated using the momentum and polar angle of the $K_S^0$ candidate together with the nuclear-interaction cross section for the neutral kaons. The correction is found to be $(0.07 \pm 0.01)\%$ for both the $e$-tag and $\mu$-tag samples. The other is a correction for the dilution effect. The final sample includes other tau decay modes with one $K_S^0$. The decay-rate asymmetry for $\tau^- \to K^-K_S^0\nu_\tau$ is opposite to that of $\tau^- \to K_S^0\pi^-(\geq 0\pi^0)\nu_\tau$ in the Standard Model because the $K_S^0$ in the $\tau^- \to K_S^0\pi^-(\geq 0\pi^0)\nu_\tau$ is produced via a $\overline{K}^0$, whereas the $K_S^0$ in $\tau^- \to K^-K_S^0\nu_\tau$ is produced via a $K^0$. In addition, the decay asymmetry is zero in the Standard Model for the $\tau \to K^0\overline{K}^0\nu_\tau$ decay, because the asymmetries due to the $K^0$ and $\overline{K}^0$ will cancel each other. To obtain the genuine rate asymmetry for $\tau^- \to K_S^0\pi^-(\geq 0\pi^0)\nu_\tau$, the measured asymmetry is divided by $0.75 \pm 0.04$.

Finally, by applying these corrections and combining both the $e$-tag and $\mu$-tag samples, *BABAR* obtain the decay-rate asymmetry for the $\tau^- \to K_S^0\pi^-(\geq 0\pi^0)\nu_\tau$ decay to be

$$A_{CP} = (-0.36 \pm 0.23 \pm 0.11)\%. \tag{20.5.17}$$

As pointed out by Grossman and Nir (2012), the predicted decay-rate asymmetry is affected by the $K_S^0$ decay time dependence of the event selection efficiency. By taking into account the efficiency correction $(1.08 \pm 0.01)$ caused by the finite acceptance as a function of $K_S^0$ decay in the *BABAR* detector, the Standard Model decay-rate asymmetry is predicted to be $(0.36 \pm 0.01)\%$ (Bigi and Sanda, 2005). The sign of the asymmetry is different between experiment and the prediction. The measured value is 2.8 standard deviations from the Standard Model prediction.

#### 20.5.2.2 Asymmetry in angular distribution

A first search for the *CP* asymmetry in the angular distribution has been carried out by the CLEO collaboration for the $\tau^- \to K_S^0\pi^-\nu_\tau$ mode using 13 fb$^{-1}$ data (Bonvicini et al., 2002). Belle performs a similar search in the same mode using a data sample of 699 fb$^{-1}$ (Bischofberger, 2011).

In this section, we first introduce the generic formula for the *CP* asymmetry in the angular distribution of the tau decays and how the *CP* violating parameters are related to the observables. The relevant SM Hamiltonian for the Cabibbo-suppressed decays $\tau^\pm \to X_s^\pm\nu_\tau$, is given by

$$\mathcal{H}_{SM} = \frac{G_F}{\sqrt{2}} \sin\theta_c[\overline{\nu}_\tau\gamma_\mu(1-\gamma_5)\tau][\overline{s}\gamma^\mu(1-\gamma_5)u] \ + h.c., \tag{20.5.18}$$

where the form is determined by the vector-boson $W^\pm$ exchange. On the other hand, the Hamiltonian with scalar- or pseudoscalar-boson exchange has a form

$$\mathcal{H}_{NP} = \frac{G_F}{\sqrt{2}} \sin\theta_c[\overline{\nu}_\tau(1+\gamma_5)\tau][\overline{s}(\eta_S + \eta_P\gamma_5)u] \ + h.c. \tag{20.5.19}$$

where $\eta_P$ and $\eta_S$ are the complex parameters, in general, relevant for the pseudoscalar and scalar hadronic system, respectively. If either $\eta_P$ or $\eta_S$ has a non-zero imaginary part, *CP* is violated in this process (Kühn and Mirkes, 1997).

Among them, the scalar $\eta_S$ term can be measured in the modes decaying to two pseudoscalars such as $\tau^- \to K_S^0\pi^-\nu_\tau$. For $\tau^- \to K_S^0\pi^-\nu_\tau$, the full differential decay width, in the hadronic rest frame ($\boldsymbol{q}_1 + \boldsymbol{q}_2 = 0$), is given by

$$d\Gamma_{\tau^-} = \frac{G_F^2}{2m_\tau}\sin^2\theta_c \frac{1}{(4\pi)^3}\frac{(m_\tau^2 - Q^2)^2}{m_\tau^2}|\boldsymbol{q}_1| \times \frac{1}{2}\left(\sum_X \overline{L}_X W_X\right)\frac{dQ^2}{\sqrt{Q^2}}\frac{d\cos\theta}{2}\frac{d\cos\beta}{2}, \tag{20.5.20}$$

where $G_F$ is the Fermi coupling constant, $\theta_c$ is the Cabibbo angle, $m_\tau$ is the mass of the tau lepton, $\boldsymbol{q}_1$ and $\boldsymbol{q}_2$ denote the three-momenta of $K_S^0$ and $\pi^-$, respectively, and $Q^2 = (q_1 + q_2)^2$ is the invariant mass squared of the $K_S^0\pi^\pm$ system. $\beta$ is the helicity angle of $K_S^0$ in the $K_S^0\pi^\pm$ rest frame while $\theta$ is the helicity angle of the $K_S\pi^\pm$ system in the tau rest frame. The angular coefficients $L_X$ with $X = 1$ to 4 are known functions related to the leptonic current. The four hadronic functions $W_X$ are formed from the vector and scalar form factors $F(Q^2)$ and $F_S(Q^2)$ and are proportional to $|F|^2$, $|F_S|^2$, $Re(FF_S)$, and $Im(FF_S)$, respectively. Among them the last term $Im(FF_S)$ is most important for a $CP$ measurement, which involves the $CP$-odd term proportional to $Im(\eta_S)$ (Kühn and Mirkes, 1997). This term has an angular dependence of $\cos\beta\cos\psi$, where $\psi$ is the helicity angle of the tau lepton in the $K_S\pi^\pm$ rest-frame. Note that the angles $\theta$ and $\psi$ are correlated and their cosine can be determined from the energy of the $K_S\pi^\pm$ system without measuring the tau direction (Kühn and Mirkes, 1997).

In order to extract the $CP$ violating term proportional to $Im(FF_S)$, Belle measure the asymmetry, in bins of $Q^2$, defined as the difference of the differential $\tau^+$ and $\tau^-$ decay width weighted by $\cos\beta\cos\psi$:

$$A_i^{CP} = \frac{\int \cos\beta\cos\psi\left(\frac{d\Gamma_{\tau^-}}{d\omega} - \frac{d\Gamma_{\tau^+}}{d\omega}\right)d\omega}{\frac{1}{2}\int\left(\frac{d\Gamma_{\tau^-}}{dQ^2} + \frac{d\Gamma_{\tau^+}}{dQ^2}\right)dQ^2}$$
$$\simeq \langle\cos\beta\cos\psi\rangle_{\tau^-} - \langle\cos\beta\cos\psi\rangle_{\tau^+} \qquad (20.5.21)$$

with $d\omega = dQ^2 d\cos\theta d\cos\beta$. The asymmetry $A_i^{CP}$ is just the difference between the mean values $\langle\cos\beta\cos\psi\rangle$ for $\tau^+$ and $\tau^-$ events evaluated in bins of $Q^2$.

The measured $CP$ asymmetry $A_i^{CP}$ is related to the $CP$ parameter $Im(\eta_S)$ by

$$A_i^{CP} = \langle\cos\beta\cos\psi\rangle^i_{\tau^-} - \langle\cos\beta\cos\psi\rangle^i_{\tau^+}$$
$$= \frac{N_s}{n_i}\frac{1}{\epsilon_{\rm tot}\Gamma}\iiint_{Q^2_{1,i}}^{Q^2_{2,i}} \epsilon(Q^2,\cos\beta,\cos\theta)$$
$$\times\cos\beta\cos\psi\left[\frac{d\Gamma(\tau^-)}{d\omega} - \frac{d\Gamma(\tau^+)}{d\omega}\right]d\omega$$
$$\simeq Im(\eta_S)\frac{N_s}{n_i}\int_{Q^2_{1,i}}^{Q^2_{2,i}} C(Q^2)\frac{Im(FF_H^*)}{m_\tau}dQ^2, \qquad (20.5.22)$$

with $n_i = (n_i^- + n_i^+)/2$. Where $n_i^\pm$ denotes the observed number of $\tau^- \to K_S^0\pi^-\nu_\tau$ events in the $i$-th $Q^2$ bin, ($Q^2 \in [Q^2_{1,i}, Q^2_{2,i}]$), and $N_s = \sum_i n_i$. The function $C(Q^2)$ contains model-independent terms and detector efficiency effects, and is obtained after numerical integration over $\cos\beta$ and $\cos\theta$:

$$C(Q^2) = -\frac{1}{\Gamma}\frac{G_F^2}{2m_\tau}\sin^2\theta_c\frac{1}{(4\pi)^3}\frac{(m_\tau^2 - Q^2)^2}{Q^2}|\boldsymbol{q}_1|^2 \qquad (20.5.23)$$
$$\times\iint\frac{\epsilon(Q^2,\cos\beta,\cos\psi)}{\epsilon_{\rm tot}}\cos^2\beta\cos^2\psi\, d\cos\theta\, d\cos\beta,$$

here the coefficients $\epsilon_{\rm tot}$ and $\epsilon(Q^2,\cos\beta,\cos\psi)$ are the total and the three-dimensional detector efficiencies and $\Gamma$ is the total $\tau^- \to K_S^0\pi^-\nu_\tau$ decay width.

Belle selects events with a charged track identified as a pion, $K_S^0 \to \pi^+\pi^-$ and no additional photons with energy greater than 0.2 GeV in the signal hemisphere, and one charged track with the number of photons above 0.1 GeV less than five in the tag hemisphere. After all selection criteria are applied, a total of $162000 \pm 403$ $\tau^- \to K_S^0\pi^-\nu_\tau$ and $162200 \pm 403$ $\tau^+ \to K_S^0\pi^+\overline{\nu}_\tau$ candidates are selected from a 699 fb$^{-1}$ data sample. The background subtracted asymmetry as a function of $\sqrt{Q^2}$ is shown in Fig. 20.5.3 (a) and (b). The asymmetry is within two standard deviations ($\sigma$) from zero for all mass bins.

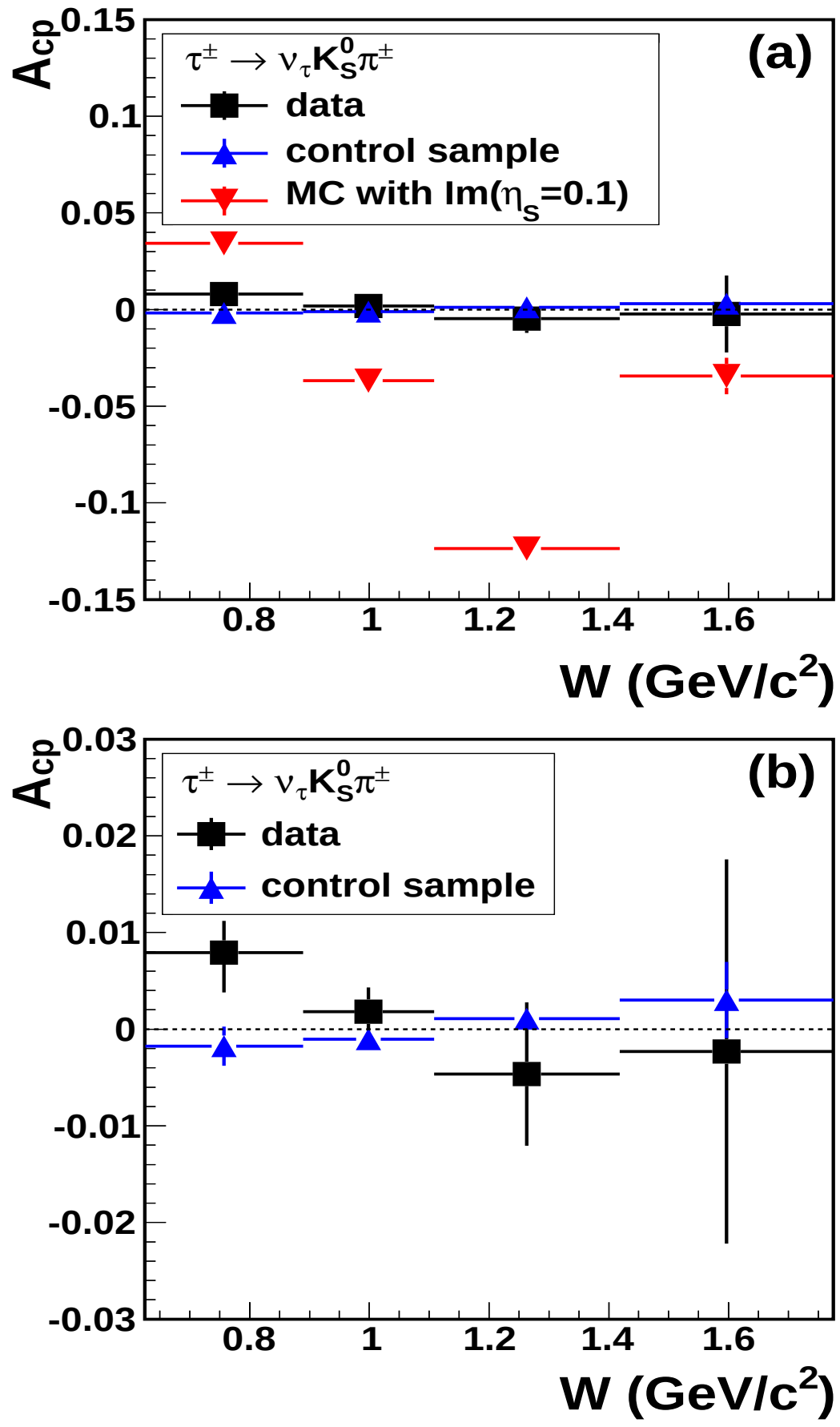

**Figure 20.5.3.** (a) Measured $CP$ asymmetry as a function of the mass of the hadron system $W = \sqrt{Q^2}$. The closed squares are data. The triangles indicate the expected asymmetry for $Im(\eta_S) = 0.1$ [$Re(\eta_S) = 0$]. (b) The same data with a zoomed vertical scale (×5) for the higher mass bins. The vertical lines indicate statistical and systematic errors added in quadrature (Bischofberger, 2011).

In the most precisely measured mass region, 0.9 GeV $< \sqrt{Q^2} <$ 1.1 GeV, the asymmetry is measured to be

$$A_{CP} = (1.8 \pm 2.1 \pm 1.4) \times 10^{-3}. \qquad (20.5.24)$$

From the measured values of $A_{CP}$, the parameter $Im(\eta_S)$ can be extracted from Eq. (20.5.22), where the results on the $K_S\pi^\pm$ mass spectra obtained by Epifanov (2007) are used for the values of the form factors $F$ and $F_S$. The resultant upper limit for the parameter $Im(\eta_S)$ is

$$|Im(\eta_S)| < (0.012 \text{ to } 0.026) \qquad (20.5.25)$$

at 90% C.L., where the range of the upper limit is due to the uncertainty of the parameterization used to describe the hadronic form factors and the unknown relative phase between the spin-one $|F|$ and the spin-zero $|F_S|$ form factors. The results improve upon the previous limit from the CLEO experiment (Bonvicini et al., 2002) by one order of magnitude.

Theoretical predictions for $Im(\eta_S)$ are available in the context of MHDM with three or more Higgs doublets (Choi, Hagiwara, and Tanabashi, 1995; Grossman, 1994). In such models $\eta_S$ is related to the model parameters as (Choi, Hagiwara, and Tanabashi, 1995),

$$\eta_S \simeq \frac{m_\tau m_s}{M^2_{H^\pm}} \cdot X^* Z \qquad (20.5.26)$$

if numerically small terms proportional to $m_u$ are ignored. Here, $M_{H^\pm}$ is the mass of the lightest charged Higgs boson and $X$ and $Z$ are the complex coupling constants shown in Fig. 20.5.4 (a). The limits for the $|Im(\eta_S)|$ result in the exclusion region shown in Fig. 20.5.4 (b). For the limit $|Im(\eta_S)| < 0.026$, this is equivalent to $|Im(XZ^*)| < 0.15 \times M^2_{H^\pm}$.

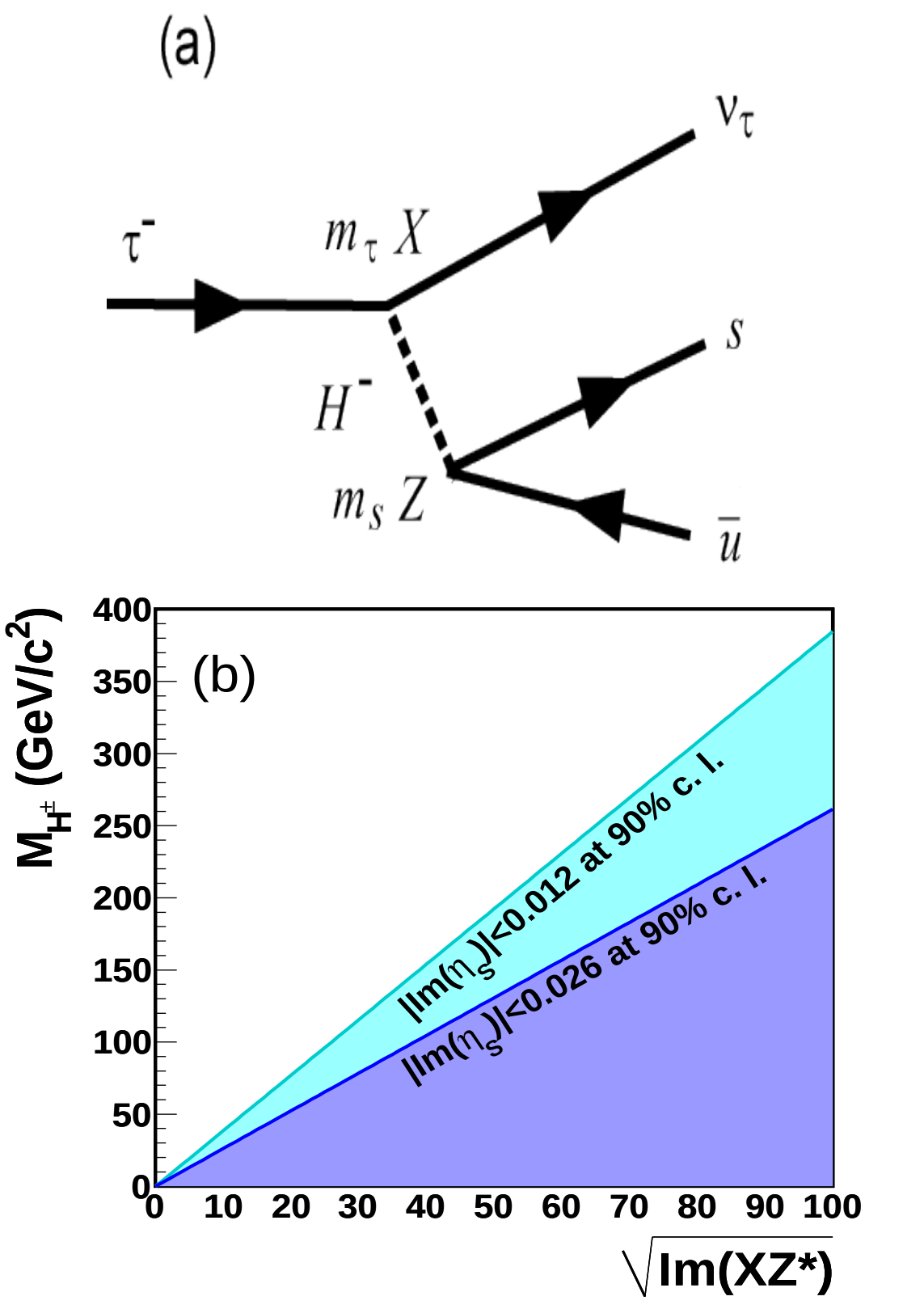

**Figure 20.5.4.** (a) Feynman diagram for a tau decay via exchange of a Higgs boson in Multi-Higgs-Doublet models. (b) Excluded region of the parameter space in Multi-Higgs-Doublet model from limits for $|Im(\eta_S)|$ (Bischofberger, 2011).

## 20.6 Hadronic tau decays

### 20.6.1 Theory

The hadronic tau decays turn out to be a beautiful laboratory for studying strong interaction effects at low energies (Pich, 1998). The tau is the only known lepton massive enough to decay into hadrons. Its semileptonic decays are then ideally suited to investigate the hadronic weak currents. The $\tau^- \to \nu_\tau h^-$ decay amplitude,

$$\mathcal{M}(\tau^- \to \nu_\tau h^-) \;=\; \frac{G_F}{\sqrt{2}}\,\mathcal{H}^\mu_h\;[\overline{\nu}_\tau \gamma_\mu (1-\gamma_5)\tau]\,, \qquad (20.6.1)$$

probes the matrix element of the left-handed charged current between the vacuum and the final hadronic state $h^-$,

$$\mathcal{H}^\mu_h \;\equiv\; \langle h^-|\left(V^*_{ud}\,\overline{d}\,\gamma^\mu(1-\gamma_5)u \,+\, V^*_{us}\,\overline{s}\,\gamma^\mu(1-\gamma_5)u\right)|0\rangle\,. \qquad (20.6.2)$$

For the decay modes with lowest multiplicity, $\tau^- \to \nu_\tau \pi^-$ and $\tau^- \to \nu_\tau K^-$, the relevant hadronic matrix elements are already known from the measured decays $\pi^- \to \mu^- \overline{\nu}_\mu$ and $K^- \to \mu^- \overline{\nu}_\mu$:

$$\langle \pi^-|\,\overline{d}\,\gamma^\mu u\,|0\rangle = -i\,f_\pi\,p^\mu_\pi\,,$$
$$\langle K^-|\,\overline{s}\,\gamma^\mu u\,|0\rangle = -i\,f_K\,p^\mu_K\,, \qquad (20.6.3)$$

where $f_\pi = (130.4\pm0.2)$ MeV and $f_K = (156.1\pm0.8)$ MeV are the so-called pion and kaon decay constants (Beringer et al., 2012). The corresponding tau decay widths can then be accurately predicted. The predictions are in good agreement with the measured values and provide a test of lepton universality. Assuming universality in the quark couplings, muonic decays of the pion and kaon determine the ratio (Cirigliano, Ecker, Neufeld, Pich, and Portoles, 2012)

$$\frac{|V_{us}|\,f_K}{|V_{cd}|\,f_\pi} \;=\; 0.2763 \pm 0.0005 \qquad (20.6.4)$$

to a higher precision than is currently obtained with tau decays. The determination of this ratio by tau decays is presently limited by the measurement of $\Gamma(\tau^- \to \nu_\tau K^-)$, but *BABAR* has significantly improved the precision of this mode, as discussed in Section 20.8.

For the two-pion final state, the hadronic matrix element is parameterized in terms of the so-called pion form factor $F_\pi(s)$, defined through $[s \equiv (p_{\pi^-} + p_{\pi^0})^2]$

$$\langle \pi^-\pi^0|\overline{d}\gamma^\mu u|0\rangle \equiv \sqrt{2}\,F_\pi(s)\;(p_{\pi^-} - p_{\pi^0})^\mu\,. \qquad (20.6.5)$$

Isospin symmetry relates this quantity to the analogous form factor measured in $e^+e^- \to \pi^+\pi^-$. Accurate measurements of $F_\pi(s)$ are a critical ingredient of the Standard Model prediction for the anomalous magnetic moment of the muon.

Owing to the different quarks involved, two form factors are needed to characterize the decays $\tau \to \nu_\tau K\pi$,

$$\langle \overline{K}^0\pi^-|\overline{s}\gamma^\mu u|0\rangle \equiv f^{K\pi}_+(s)\;(p_K - p_\pi)^\mu + f^{K\pi}_-(s)\;(p_K + p_\pi)^\mu \qquad (20.6.6)$$

with $s \equiv (p_\pi + p_K)^2$. The form factor $f_+^{K\pi}(s)$ corresponds to a hadronic final state with $J^P = 1^-$, while the scalar ($0^+$) combination $f_0^{K\pi}(s) = f_+^{K\pi}(s) + s\, f_-^{K\pi}(s)/(m_K^2 - m_\pi^2)$ vanishes in the $SU(3)$ limit because the vector current is conserved for equal quark masses. In $K \to \pi\ell\nu$ ($K_{\ell 3}$) decays one also measures the form factors in the timelike region. More precisely, in the kinematical region $m_\ell^2 < q^2 < (M_K - M_\pi)^2$. In tau decays this is extended up to the tau mass.

Higher-multiplicity modes involve a richer dynamical structure, providing a very valuable experimental window into the non-perturbative hadronization of the QCD currents. While $e^+e^-$ data only test the electromagnetic vector current, tau decays are sensitive to the vector and axial-vector currents, both in the Cabibbo-allowed and Cabibbo-suppressed channels.

A dynamical understanding of the hadronic matrix elements can be achieved using analyticity, unitarity and some general properties of QCD, such as chiral symmetry , the short-distance asymptotic behavior and the limit of a large number of QCD colors (Pich, 2007). The high-statistics $B$ Factory data samples provide very important information on the hadronic structure, allowing one to improve theoretical tools and get a better control of the strong interaction in the resonance region. These data have already triggered extensive theoretical activity (Boito, Escribano, and Jamin, 2009, 2010; Gómez Dumm, Pich, and Portolés, 2004; Gómez Dumm, Roig, Pich, and Portolés, 2010a,b; Guerrero and Pich, 1997; Guo and Roig, 2010; Jamin, Pich, and Portolés, 2006, 2008; Pich and Portolés, 2001). As a result, there has been considerable progress towards the development of a quantum field theory description of the resonance dynamics at the energy scales accessible through tau decays.

#### 20.6.1.1 Inclusive tau decay width

The inclusive character of the total tau hadronic width renders possible an accurate calculation of the ratio (Braaten, 1988, 1989; Braaten, Narison, and Pich, 1992; Le Diberder and Pich, 1992b; Narison and Pich, 1988)

$$R_\tau \equiv \frac{\Gamma(\tau^- \to \nu_\tau \,\text{hadrons})}{\Gamma(\tau^- \to \nu_\tau e^- \overline{\nu}_e)} = R_{\tau,V} + R_{\tau,A} + R_{\tau,S}\,. \tag{20.6.7}$$

$R_{\tau,V}$ ($R_{\tau,A}$) is the Cabibbo-allowed decay width into final states with $J^P = 1^-, 0^+$ ($1^+, 0^-$), while $R_{\tau,S}$ accounts for decays into states with strangeness $S = -1$.

The inclusive hadronic width,

$$\Gamma(\tau^- \to \nu_\tau \,\text{hadrons}) \propto L_{\mu\nu} \sum_h \int dQ_h \; \mathcal{H}_h^\mu \mathcal{H}_h^{\nu\dagger}\,, \tag{20.6.8}$$

involves a sum over all possible final hadronic states $h^-$ with the corresponding phase-space integration $dQ_h$. Unitarity and analyticity (optical theorem) relate this spectral distribution with the imaginary parts of the two-point correlation functions for the vector $V_{ij}^\mu = \overline{\psi}_j \gamma^\mu \psi_i$ and axial-vector $A_{ij}^\mu = \overline{\psi}_j \gamma^\mu \gamma_5 \psi_i$ color-singlet quark currents,

$$\Pi_{ij,\mathcal{J}}^{\mu\nu}(q) \equiv i \int d^4x\, e^{iqx} \langle 0 | T(\mathcal{J}_{ij}^\mu(x) \mathcal{J}_{ij}^\nu(0)^\dagger) | 0 \rangle\,, \tag{20.6.9}$$

which have the Lorentz decomposition ($\mathcal{J} = V, A$),

$$\Pi_{ij,\mathcal{J}}^{\mu\nu}(q) = \left(-g^{\mu\nu} q^2 + q^\mu q^\nu\right) \Pi_{ij,\mathcal{J}}^{(1)}(q^2) + q^\mu q^\nu\, \Pi_{ij,\mathcal{J}}^{(0)}(q^2)\,. \tag{20.6.10}$$

The superscript $J = 0, 1$ denotes the angular momentum in the hadronic rest frame and $i, j = u, d, s$.

The imaginary parts of $\Pi_{ij,\mathcal{J}}^{(J)}(q^2)$ are proportional to the spectral functions for hadrons with the corresponding quantum numbers. The hadronic decay rate of the tau can be written as an integral of these spectral functions over the invariant mass $s = q^2$ of the final-state hadrons:

$$R_\tau = 12\pi \int_0^{m_\tau^2} \frac{ds}{m_\tau^2} \left(1 - \frac{s}{m_\tau^2}\right)^2 \times \left[\left(1 + 2\frac{s}{m_\tau^2}\right) \mathrm{Im}\Pi^{(1)}(s) + \mathrm{Im}\Pi^{(0)}(s)\right], \tag{20.6.11}$$

where

$$\Pi^{(J)}(s) \equiv |V_{ud}|^2 \left(\Pi_{ud,V}^{(J)}(s) + \Pi_{ud,A}^{(J)}(s)\right) + |V_{us}|^2 \left(\Pi_{us,V}^{(J)}(s) + \Pi_{us,A}^{(J)}(s)\right). \tag{20.6.12}$$

The contributions coming from the first two terms correspond to $R_{\tau,V}$ and $R_{\tau,A}$, respectively, while $R_{\tau,S}$ contains the remaining Cabibbo-suppressed contributions.

The integrand in Eq. (20.6.11) cannot be calculated at present from QCD. Nevertheless the integral itself can be calculated systematically by exploiting the analytic properties of the correlators $\Pi^{(J)}(s)$. They are analytic functions of $s$ except along the positive real $s$-axis, where their imaginary parts have discontinuities. $R_\tau$ can then be written as a contour integral in the complex $s$-plane running counter-clockwise around the circle $|s| = m_\tau^2$ (Braaten, Narison, and Pich, 1992):

$$R_\tau = 6\pi i \oint_{|s|=m_\tau^2} \frac{ds}{m_\tau^2} \left(1 - \frac{s}{m_\tau^2}\right)^2 \times \left[\left(1 + 2\frac{s}{m_\tau^2}\right) \Pi^{(0+1)}(s) - 2\frac{s}{m_\tau^2} \Pi^{(0)}(s)\right]. \tag{20.6.13}$$

Cauchy's theorem guarantees that the integration along the closed contour shown in Fig. 20.6.1 gives zero. Up to a sign the integral along the circle is then equal to the sum of the integrals above and below the real axis, which reproduces Eq. (20.6.11) because $\Pi^{(J)}(s + i\epsilon) - \Pi^{(J)}(s - i\epsilon) = 2i\,\mathrm{Im}\Pi^{(J)}(s)$.

To compute the contour integral Eq. (20.6.13) we only need to know the correlators $\Pi^{(J)}(s)$ for complex values of $s$, with $|s| = m_\tau^2$ which is larger than the scale associated with non-perturbative effects. Therefore, the Operator Product Expansion (OPE) can be used to organize the perturbative and non-perturbative contributions into a systematic expansion in powers of $1/s$:

$$\Pi^{(J)}(s) \;=\; \sum_{D=0,2,4,\ldots} \frac{C_D^{(J)}}{(-s)^{D/2}}\,. \qquad (20.6.14)$$

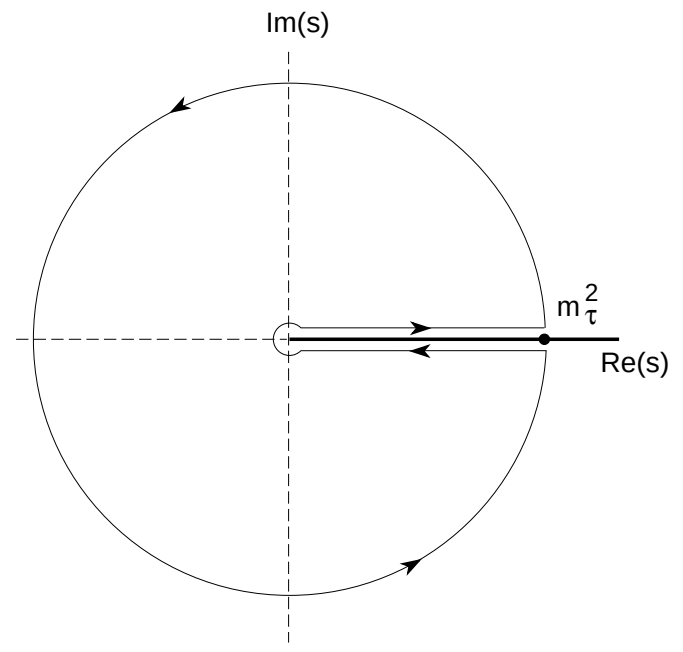

**Figure 20.6.1.** Integration contour used to derive Eq. (20.6.13).

$C_D^{(J)}$ parameterizes the contributions from operators with dimension $D$. Since parity is a symmetry of the strong interactions, only operators with even dimension contribute. The leading $D=0$ term is the perturbative contribution, while the corrections correspond to non-perturbative effects from operators with dimension $D \geq 4$. Since there are no gauge-invariant operators with D=2, the dominant non-perturbative effects appear at D=4 through the so-called gluon, $\langle \alpha_S G_{\mu\nu} G^{\mu\nu}\rangle$, and quark, $\langle m_q \bar{q} q\rangle$, vacuum condensates.

Inserting Eq. (20.6.14) into the contour integral, $R_\tau$ can be expressed as an expansion in powers of $1/m_\tau^2$. The uncertainties associated with the use of the OPE near the timelike axis are heavily suppressed by the presence of a double zero at $s = m_\tau^2$ in Eq. (20.6.13).

In the chiral limit ($m_{u,d,s} = 0$), the vector and axial-vector currents are conserved. This implies $s\,\Pi^{(0)}(s) = 0$. Therefore, only the correlator $\Pi^{(0+1)}(s)$ contributes to Eq. (20.6.13). Since $(1-x)^2(1+2x) = 1 - 3x^2 + 2x^3$ [$x \equiv s/m_\tau^2$], up to tiny logarithmic running corrections, the only non-perturbative contributions to the contour integration in Eq. (20.6.13) originate from operators of dimensions $D = 6$ and 8. The usually leading $D = 4$ operators can only contribute to $R_\tau$ with an additional suppression factor of $\mathcal{O}(\alpha_S^2)$, which makes their effect negligible (Braaten, Narison, and Pich, 1992).

### 20.6.1.2 Determination of $\alpha_S(m_\tau)$

The Cabibbo-allowed combination $R_{\tau,V+A}$ can be written as (Braaten, Narison, and Pich, 1992)

$$R_{\tau,V+A} \;=\; N_C\,|V_{ud}|^2\,S_{\rm EW}\,\{1 + \delta_{\rm P} + \delta_{\rm NP}\}\,, \qquad (20.6.15)$$

where $N_C = 3$ is the number of quark colors and $S_{\rm EW} = 1.0201 \pm 0.0003$ contains the electroweak radiative corrections (Braaten and Li, 1990; Erler, 2004; Marciano and Sirlin, 1988). The dominant correction ($\sim 20\%$) is the perturbative QCD contribution $\delta_{\rm P}$, which is already known to $O(\alpha_S^4)$ (Baikov, Chetyrkin, and Kühn, 2008; Braaten, Narison, and Pich, 1992). Quark mass effects are tiny for the Cabibbo-allowed current and amount to a negligible correction smaller than $10^{-4}$.

Non-perturbative contributions $\delta_{\rm NP}$, are suppressed by six powers of the tau mass and, therefore, are very small. Their numerical size has been determined from the invariant-mass distribution of the final hadrons in tau decay, through the study of weighted integrals (Le Diberder and Pich, 1992a),

$$R_\tau^{kl} \;\equiv\; \int_0^{m_\tau^2} ds\, \left(1 - \frac{s}{m_\tau^2}\right)^k \left(\frac{s}{m_\tau^2}\right)^l \frac{dR_\tau}{ds}\,, \qquad (20.6.16)$$

which can be calculated theoretically in the same way as $R_\tau$, but are more sensitive to OPE corrections. The predicted suppression of the non-perturbative contribution to $R_\tau$ has been confirmed by ALEPH (Barate et al., 1998; Buskulic et al., 1993a; Schael et al., 2005), CLEO (Coan et al., 1995) and OPAL (Ackerstaff et al., 1999). The most recent analysis gives (Davier, Hoecker, and Zhang, 2006)

$$\delta_{\rm NP} \;=\; -0.0059 \pm 0.0014\,, \qquad (20.6.17)$$

showing that non-perturbative corrections are below 1%.

The QCD prediction for $R_{\tau,V+A}$ is then completely dominated by $\delta_{\rm P}$; non-perturbative effects being smaller than the perturbative uncertainties from unknown higher-order corrections. Using $|V_{ud}| = 0.97425 \pm 0.00022$ (Hardy and Towner, 2009) and Eq. (20.6.17), the present experimental value $R_{\tau,V+A} = 3.4671 \pm 0.0084$ (Amhis et al., 2012), determines the purely perturbative contribution to $R_\tau$ to be

$$\delta_{\rm P} = 0.1995 \pm 0.0033\,. \qquad (20.6.18)$$

The predicted value of $\delta_{\rm P}$ turns out to be very sensitive to $\alpha_S(m_\tau^2)$, allowing for an accurate determination of the fundamental QCD coupling (Braaten, Narison, and Pich, 1992; Narison and Pich, 1988). The calculation of the $\mathcal{O}(\alpha_S^4)$ contribution (Baikov, Chetyrkin, and Kühn, 2008) has triggered a renewed theoretical interest on the $\alpha_S(m_\tau^2)$ determination, since it allows one to improve the accuracy to the four-loop level (Beneke and Jamin, 2008; Caprini and Fischer, 2009, 2011; Cvetic, Loewe, Martinez, and Valenzuela, 2010; Davier, Descotes-Genon, Höcker, Malaescu, and Zhang, 2008; Maltman and Yavin, 2008; Menke, 2009; Pich, 2011a). The value of $\delta_{\rm P}$ in Eq. (20.6.18) implies (Pich, 2011b)

$$\alpha_S(m_\tau^2) = 0.329 \pm 0.013\,, \qquad (20.6.19)$$

which is significantly larger than the values obtained at higher energies. After evolution up to the scale $M_Z$ (Rodrigo, Pich, and Santamaria, 1998), the strong coupling decreases to

$$\alpha_S(M_Z^2) \;=\; 0.1198 \pm 0.0015\,, \qquad (20.6.20)$$

in excellent agreement with the direct measurements at the $Z$ peak and with a better accuracy. The comparison of these two determinations of $\alpha_S$ in two very different energy regimes, $m_\tau$ and $M_Z$, provides a beautiful test of the predicted running of the QCD coupling; *i.e.*, a very significant experimental verification of asymptotic freedom.

#### 20.6.1.3 $|V_{us}|$ Determination

A separate measurement of the $|\Delta S| = 0$ and $|\Delta S| = 1$ tau decay widths provides a very clean determination of $V_{us}$ (Gamiz, Jamin, Pich, Prades, and Schwab, 2003, 2005, 2008). To a first approximation the Cabibbo mixing can be directly obtained from experimental measurements, without any theoretical input. Neglecting the small $SU(3)$-breaking corrections from the $m_s - m_d$ quark-mass difference, the measured ratio $R_{\tau,S} = 0.1612 \pm 0.0028$ (Amhis et al., 2012) implies

$$|V_{us}|^{SU(3)} = |V_{ud}| \left( \frac{R_{\tau,S}}{R_{\tau,V+A}} \right)^{1/2} = 0.210 \pm 0.002 \,. \tag{20.6.21}$$

The new branching ratios measured by *BABAR* and Belle are all smaller than the previous world averages, which translate into a smaller value of $R_{\tau,S}$ and $|V_{us}|$. For comparison, the previous value $R_{\tau,S} = 0.1686 \pm 0.0047$ (Davier, Hoecker, and Zhang, 2006) resulted in $|V_{us}|^{SU(3)} = 0.215 \pm 0.003$.

This rather remarkable determination is only slightly shifted by the small $SU(3)$-breaking contributions induced by the strange quark mass. These effects can be estimated through a QCD analysis of the differences (Baikov, Chetyrkin, and Kühn, 2005; Chen et al., 2001a; Chetyrkin, Kühn, and Pivovarov, 1998; Gamiz, Jamin, Pich, Prades, and Schwab, 2003, 2005, 2008; Kambor and Maltman, 2000; Korner, Krajewski, and Pivovarov, 2001; Maltman and Kambor, 2001; Maltman and Wolfe, 2006, 2007; Pich and Prades, 1998, 1999)

$$\delta R_\tau^{kl} \equiv \frac{R^{kl}_{\tau,V+A}}{|V_{ud}|^2} - \frac{R^{kl}_{\tau,S}}{|V_{us}|^2} \,. \tag{20.6.22}$$

The only non-zero contributions are proportional to the mass-squared difference $m_s^2 - m_d^2$ or to vacuum expectation values of $SU(3)$-breaking operators such as $\delta O_4 \equiv \langle 0|m_s\bar{s}s - m_d\bar{d}d|0\rangle \approx (-1.4 \pm 0.4) \cdot 10^{-3}$ GeV$^4$ (Gamiz, Jamin, Pich, Prades, and Schwab, 2003; Pich and Prades, 1998, 1999). The dimensions of these operators are compensated by corresponding powers of $m_\tau^2$, which implies a strong suppression of $\delta R_\tau^{kl}$ (Pich and Prades, 1998, 1999),

$$\delta R_\tau^{kl} \approx 24\, S_{\rm EW} \left\{ \frac{m_s^2(m_\tau^2)}{m_\tau^2} \left(1 - \epsilon_d^2\right) \Delta_{kl}(\alpha_S) - 2\pi^2 \frac{\delta O_4}{m_\tau^4} Q_{kl}(\alpha_S) \right\} , \tag{20.6.23}$$

where $\epsilon_d \equiv m_d/m_s = 0.053 \pm 0.002$ (Leutwyler, 1996). The perturbative corrections $\Delta_{kl}(\alpha_S)$ and $Q_{kl}(\alpha_S)$ are known to $O(\alpha_S^3)$ and $O(\alpha_S^2)$, respectively (Baikov, Chetyrkin, and Kühn, 2005; Pich and Prades, 1998, 1999).

The $J = 0$ contribution to $\Delta_{00}(\alpha_S)$ shows a rather pathological behavior, with clear signs of being a non-convergent perturbative series. Fortunately, the corresponding longitudinal contribution to $\delta R_\tau \equiv \delta R_\tau^{00}$ can be estimated phenomenologically with a much better accuracy, $\delta R_\tau|^L = 0.1544 \pm 0.0037$ (Gamiz, Jamin, Pich, Prades, and Schwab, 2003, 2005, 2008; Jamin, Oller, and Pich, 2006), because it is dominated by far by the well-known $\tau \to \nu_\tau \pi$ and $\tau \to \nu_\tau K$ contributions. To estimate the remaining transverse component, one needs an input value for the strange quark mass. Taking the conservative value $\delta R_{\tau,\rm th} = 0.239 \pm 0.030$ (Gamiz, 2013), one obtains

$$|V_{us}| = \left( \frac{R_{\tau,S}}{\frac{R_{\tau,V+A}}{|V_{ud}|^2} - \delta R_{\tau,\rm th}} \right)^{1/2} = 0.2173 \pm 0.0020_{\rm exp} \pm 0.0010_{\rm th} . \tag{20.6.24}$$

A larger central value, $|V_{us}| = 0.2217 \pm 0.0032$, is obtained with the old world average for $R_{\tau,S}$, *i.e.* $R_{\tau,S} = 0.1686 \pm 0.0047$ (Davier, Hoecker, and Zhang, 2006).

Sizeable changes on the experimental determination of $R_{\tau,S}$ could be expected from future analyses. In particular, the high-multiplicity decay modes are not well known at present. The recent decrease of several experimental tau branching ratios is also worrisome, since it could indicate some uncontrolled systematic effect. As pointed out by the PDG (Beringer et al., 2012), 18 of the 20 branching fractions measured at the $B$ Factories for which older non-$B$ Factory measurements exist are smaller than the previous non-$B$ Factory values. The average normalized difference between the two sets of measurements is $-1.30\,\sigma$. Thus, the result in Eq. (20.6.24) could easily fluctuate in the near future. In fact, combining the measured Cabibbo-suppressed tau distribution with hadronic $e^+e^-$ data, a slightly larger value of $|V_{us}|$ is obtained (Maltman, 2009; Maltman, Wolfe, Banerjee, Nugent, and Roney, 2009).

The final error of the $|V_{us}|$ determination from tau decay is dominated by the experimental uncertainties. This is in contrast with the standard determination from $K_{\ell 3}$ decays, where the achievable precision is limited by theoretical errors (Cirigliano, Ecker, Neufeld, Pich, and Portoles, 2012). If $R_{\tau,S}$ is measured with a 1% precision, the resulting $|V_{us}|$ uncertainty will get reduced to around 0.6%, *i.e.* $\pm 0.0013$, making tau decay the best source of information on $|V_{us}|$.

An accurate measurement of the invariant-mass distribution of the final hadrons could allow one to perform a simultaneous determination of $|V_{us}|$ and the strange quark mass, through a correlated analysis of several weighted differences $\delta R_\tau^{kl}$. However, the extraction of $m_s$ suffers from theoretical uncertainties related to the convergence of the perturbative series $\Delta_{kl}(\alpha_S)$. A better understanding of these corrections is needed.

#### 20.6.1.4 Spectral Functions

The invariant-mass distribution of the final state hadrons in tau decay contains very important information on the low-energy dynamics of QCD. While the separate measurement of each decay mode allows one to study the resonance structure of hadronic form factors with different $J^P$ and strangeness quantum numbers, the inclusive and semi-inclusive distributions associated with the different quark currents provide direct access to relevant perturbative and non-perturbative QCD parameters.

The precise determination of $\alpha_s(m_\tau^2)$ requires that one pins down the small non-perturbative contribution to $R_{\tau,V+A}$ in Eq. (20.6.17), which can only be done through an accurate measurement of the corresponding spectral distribution. Similarly, a precise experimental determination of the Cabibbo-suppressed distribution would help to improve the extraction of $|V_{us}|$, which at present is mainly obtained from the total decay width. The inclusive distribution of hadrons with $J^P = 1^-$ provides complementary information, which is needed to predict the hadronic contribution to the anomalous magnetic moment of the muon and the running of the electromagnetic coupling from low energies to the electroweak scale. While the present discrepancies between $e^+e^-$ and tau data are limiting the interpretation of the measured muon $g-2$, the uncertainties induced by this hadronic distribution on $\alpha(M_Z^2)$ are the largest source of error in the Higgs mass value extracted from precision electroweak tests, to be compared with its direct measurement at the LHC.

Moreover, the separate measurement of the vector and axial-vector spectral functions allows us to extract valuable information on the dynamical breaking of chiral symmetry. The chiral invariance of massless QCD guarantees that the two-point correlation function of a left-handed and a right-handed quark current vanishes identically to all orders in perturbation theory. The spontaneous breaking of chiral symmetry by the QCD vacuum generates a non-zero value of $\Pi_{LR}(s) = \Pi^{(0+1)}_{ud,V}(s) - \Pi^{(0+1)}_{ud,A}(s)$, which at large momenta manifests in its OPE through operators with dimension $D \geq 6$. At very low momenta, Chiral Perturbation Theory ($\chi$PT) dictates the low-energy expansion of $\Pi_{LR}(s)$ in terms of the pion decay constant and some $\chi$PT couplings (for a basic review of $\chi$PT, see Ecker (1995); Pich (1995)). Analyticity relates the short- and long-distance regimes through the dispersion relation

$$\frac{1}{2\pi i}\oint_{|s|=s_0} ds\, w(s)\, \Pi_{LR}(s) \;=\; -\int_{s_{th}}^{s_0} ds\, w(s)\, \rho(s) + 2 f_\pi^2\, w(m_\pi^2) + \mathrm{Res}[w(s)\Pi_{LR}(s), s=0] \;, \tag{20.6.25}$$

where $\rho(s) \equiv \frac{1}{\pi}\,\mathrm{Im}\Pi_{LR}(s)$ and $w(s)$ is an arbitrary weight function that is analytic in the whole complex plane except at the origin (where it can have poles). The last term in Eq. (20.6.25) accounts for the possible residue at the origin.

For $s_0 \leq m_\tau^2$, the integral along the real axis can be evaluated with the measured tau spectral functions. The residues at zero are determined by the $\chi$PT low-energy couplings and, taking $s_0$ large enough, the OPE can be applied in the entire circle $|s| = s_0$. Therefore, taking different weight functions one can determine the $\chi$PT couplings and the OPE coefficients from the tau spectral data (Davier, Girlanda, Höcker, and Stern, 1998; Gonzalez-Alonso, Pich, and Prades, 2008, 2010a,b). Moreover, one can explicitly check the predicted QCD asymptotic behavior (absence of $D < 6$ contributions in the OPE), which is reflected in the celebrated first and second Weinberg (1967) sum rules. The absence of perturbative contributions makes Eq. (20.6.25) an ideal tool to also investigate possible violations of quark-hadron duality, *i.e.* small departures from the OPE behavior at small values of $s_0$ (Cata, Golterman, and Peris, 2005; Shifman, 2000).

It is worth noting that the information extracted from $\Pi_{LR}(s)$ through Eq. (20.6.25) is needed to calculate the electromagnetic penguin contribution to the *CP*-violating ratio $\varepsilon'_K/\varepsilon_K$ in neutral kaon decays. Moreover, since the spontaneous breaking of the Standard Model electroweak symmetry has the same formal pattern as the QCD chiral symmetry breaking, $\Pi_{LR}(s)$ can also be used to investigate strongly-coupled scenarios of electroweak symmetry breaking (Peskin and Takeuchi, 1990, 1992).

Since we lack a $B$ Factory analysis of these spectral distributions, the ALEPH (Barate et al., 1999; Schael et al., 2005) and OPAL (Abbiendi et al., 2004; Ackerstaff et al., 1999) measurements are still being used at present. A change of this unfortunate situation is certainly possible and appears to be mandatory. Significant improvement of the LEP measurements should be possible with the much larger data samples collected at the $B$ Factories .

### 20.6.2 Tau lepton branching fractions

The $B$ Factories, thanks to the large and clean recorded tau pairs sample, provide improved measurements of hadronic branching fractions for many modes and also provide first measurements of several modes which have small branching fractions.

Table 20.6.1 shows the $B$ Factories results that have been included in the early 2012 HFAG report (Amhis et al., 2012), while Table 20.6.2 shows the results from two recent papers, which have not yet been included in the HFAG averages. The corresponding averages are also listed.

### 20.6.3 Hadronic spectral functions: Cabibbo-favored modes

A separate measurement of the mass distribution for each decay mode allows one to study the resonance structure of hadronic form factors with different $J^P$ and strangeness quantum numbers. In addition, as discussed in Sections 20.6.1.1 to 20.6.1.4, inclusive invariant-mass distributions of the final state hadrons in tau decay are very important for the determination of fundamental constants, such as $\alpha_S(m_\tau)$ and $|V_{us}|$. They also contain very important information about the low-energy dynamics of QCD. To measure the inclusive mass distribution experimentally, one needs to separate vector and axial vector final states based on the number of pions, while the Cabibbo-favored and Cabibbo-suppressed modes with kaons are distinguished by an odd or even number of kaons.

**Table 20.6.1.** Hadronic tau lepton branching fractions, $\mathcal{B}$, measured by the $B$ Factory experiments that were used in the early 2012 HFAG report (Amhis et al., 2012). The notation (ex.) $K^0$ means the branching fraction excludes the contribution $K_S^0 \to \pi^+\pi^-$ to a $\pi^+\pi^-$ contained in its final state.

| Decay mode | Source | $\mathcal{B}$ | Reference |
|---|---|---|---|
| $B(\tau \to \pi^-\nu_\tau)/B(\tau \to e^-\overline{\nu}_e\nu_\tau)$ | BaBar | $(59.45 \pm 0.57 \pm 0.25)\%$ | Aubert (2010f) |
| | HFAG | $(60.675 \pm 0.321)\%$ | Amhis et al. (2012) |
| $\tau \to h^-\pi^0\nu_\tau$ | Belle | $(25.67 \pm 0.01 \pm 0.39)\%$ | Fujikawa (2008) |
| | HFAG | $(25.93 \pm 0.09)\%$ | Amhis et al. (2012) |
| $\tau \to \pi^-\pi^0\nu_\tau$ | Belle | $(25.24 \pm 0.01 \pm 0.39)\%$ | Fujikawa (2008) |
| | HFAG | $(25.504 \pm 0.092) \cdot 10^{-2}$ | Amhis et al. (2012) |
| $\tau \to K^-\pi^0\nu_\tau$ | BaBar | $(0.416 \pm 0.003 \pm 0.18)\%$ | Aubert (2007ac) |
| | HFAG | $(0.432 \pm 0.015) \cdot 10^{-2}$ | Amhis et al. (2012) |
| $\tau \to \pi^-K^0\nu_\tau$ | Belle | $(0.808 \pm 0.004 \pm 0.026)\%$ | Epifanov (2007) |
| | BaBar | $(0.840 \pm 0.004 \pm 0.023)\%$ | Aubert (2009s) |
| | HFAG | $(0.8206 \pm 0.0182) \cdot 10^{-2}$ | Amhis et al. (2012) |
| $\tau \to \pi^-K^0\pi^0\nu_\tau$ | BaBar | $(0.342 \pm 0.006 \pm 0.015)\%$ | Paramesvaran (2009) |
| | Belle | $(0.384 \pm 0.004 \pm 0.016)\%$ | Ryu (2012) |
| | HFAG | $(0.365 \pm 0.011) \cdot 10^{-2}$ | Amhis et al. (2012) |
| $\tau \to K^-\pi^0K^0\nu_\tau$ | Belle | $(0.148 \pm 0.002 \pm 0.008)\%$ | Ryu (2012) |
| | HFAG | $(0.145 \pm 0.007) \cdot 10^{-2}$ | Amhis et al. (2012) |
| $\tau \to \pi^-\pi^-\pi^+\nu_\tau$(ex. $K^0$) | BaBar | $(8.833 \pm 0.007 \pm 0.127)\%$ | Aubert (2008k) |
| | Belle | $(8.420 \pm 0.003 \pm 0.259)\%$ | Lee (2010) |
| | HFAG | $(9.002 \pm 0.051) \cdot 10^{-2}$ | Amhis et al. (2012) |
| $\tau \to K^-\pi^-\pi^+\nu_\tau$(ex. $K^0$) | BaBar | $(0.272 \pm 0.0018 \pm 0.0092)\%$ | Aubert (2008k) |
| | Belle | $(0.330 \pm 0.0012 \pm 0.017)\%$ | Lee (2010) |
| | HFAG | $(0.293 \pm 0.007) \cdot 10^{-2}$ | Amhis et al. (2012) |
| $\tau \to \pi^-K^-K^+\nu_\tau$ | BaBar | $(1.346 \pm 0.010 \pm 0.036) \cdot 10^{-3}$ | Aubert (2008k) |
| | Belle | $(1.550 \pm 0.007 \pm 0.056) \cdot 10^{-3}$ | Lee (2010) |
| | HFAG | $(1.435 \pm 0.027) \cdot 10^{-3}$ | Amhis et al. (2012) |
| $\tau \to K^-K^-K^+\nu_\tau$ | BaBar | $(1.58 \pm 0.13 \pm 0.12) \cdot 10^{-5}$ | Aubert (2008k) |
| | Belle | $(3.29 \pm 0.17 \pm 0.20) \cdot 10^{-5}$ | Lee (2010) |
| | HFAG | $(2.18 \pm 0.80) \cdot 10^{-5}$ | Amhis et al. (2012) |
| $\tau \to \pi^-\phi\nu_\tau$ | BaBar | $(3.42 \pm 0.55 \pm 0.25) \cdot 10^{-5}$ | Aubert (2008k) |
| $\tau \to K^-\phi\nu_\tau$ | Belle | $(4.05 \pm 0.25 \pm 0.26) \cdot 10^{-5}$ | Inami (2006) |
| | BaBar | $(3.39 \pm 0.20 \pm 0.28) \cdot 10^{-5}$ | Aubert (2008k) |
| $\tau \to 3h^-2h^+\nu_\tau$(ex. $K^0$) | BaBar | $(8.56 \pm 0.05 \pm 0.42) \cdot 10^{-4}$ | Aubert (2005ag) |
| | HFAG | $(8.23 \pm 0.31) \cdot 10^{-4}$ | Amhis et al. (2012) |
| $\tau \to \pi^-\pi^0\eta\nu_\tau$ | Belle | $(1.35 \pm 0.03 \pm 0.07) \cdot 10^{-3}$ | Inami (2009) |
| | HFAG | $(1.39 \pm 0.07) \cdot 10^{-3}$ | Amhis et al. (2012) |
| $\tau \to K^-\eta\nu_\tau$ | BaBar | $(1.42 \pm 0.11 \pm 0.07) \cdot 10^{-4}$ | del Amo Sanchez (2011m) |
| | Belle | $(1.58 \pm 0.05 \pm 0.09) \cdot 10^{-4}$ | Inami (2009) |
| | HFAG | $(1.528 \pm 0.081) \cdot 10^{-4}$ | Amhis et al. (2012) |
| $\tau \to K^-\pi^0\eta\nu_\tau$ | Belle | $(4.6 \pm 1.1 \pm 0.4) \cdot 10^{-5}$ | Inami (2009) |
| | HFAG | $(4.8 \pm 1.2) \cdot 10^{-5}$ | Amhis et al. (2012) |
| $\tau \to \pi^-K^0\eta\nu_\tau$ | Belle | $(8.8 \pm 1.4 \pm 0.6) \cdot 10^{-5}$ | Inami (2009) |
| | HFAG | $(9.4 \pm 1.5) \cdot 10^{-5}$ | Amhis et al. (2012) |

**Table 20.6.2.** Recent hadronic tau lepton branching fractions, $\mathcal{B}$, measured by the BABAR and Belle experiments that are not yet used in the early 2012 HFAG report (Amhis et al., 2012). The reported HFAG averages do not use the measurements listed here. The notation (ex.)$K^0$ means the branching fraction excludes the contribution $K_S^0 \to \pi^+\pi^-$ to a $\pi^+\pi^-$ contained in its final state.

| Decay mode | Source | $\mathcal{B}$ | Reference |
|---|---|---|---|
| $\tau \to \pi^- K^0 \nu_\tau$ | Belle | $(0.832 \pm 0.002 \pm 0.016)\%$ | Ryu (2014) |
| $\tau \to K^- K^0 \nu_\tau$ | Belle | $(0.148 \pm 0.0014 \pm 0.0054)\%$ | Ryu (2014) |
| $\tau \to \pi^- K^0 \pi^0 \nu_\tau$ | Belle | $(0.386 \pm 0.004 \pm 0.014)\%$ | Ryu (2014) |
| $\tau \to K^- K^0 \pi^0 \nu_\tau$ | Belle | $(0.150 \pm 0.002 \pm 0.008)\%$ | Ryu (2014) |
| $\tau \to \pi^- K_S K_S \nu_\tau$ | BABAR | $(2.31 \pm 0.04 \pm 0.08) \cdot 10^{-4}$ | Lees (2012ae) |
| | Belle | $(2.33 \pm 0.03 \pm 0.09) \cdot 10^{-4}$ | Ryu (2014) |
| | HFAG | $(2.4 \pm 0.5) \cdot 10^{-4}$ | Amhis et al. (2012) |
| $\tau \to \pi^- K_S K_S \pi^0 \nu_\tau$ | BABAR | $(1.60 \pm 0.20 \pm 0.22) \cdot 10^{-5}$ | Lees (2012ae) |
| | Belle | $(2.00 \pm 0.22 \pm 0.20) \cdot 10^{-5}$ | Ryu (2014) |
| $\tau \to 3\pi^- 2\pi^+ \nu_\tau$ (ex. $K^0$) | BABAR | $(8.33 \pm 0.04 \pm 0.43) \cdot 10^{-4}$ | Lees (2012z) |
| $\tau \to 3\pi^- 2\pi^+ \pi^0 \nu_\tau$ (ex. $K^0$) | BABAR | $(1.65 \pm 0.05 \pm 0.09) \cdot 10^{-4}$ | Lees (2012z) |
| $\tau \to \pi^- \pi^- \pi^+ \eta \nu_\tau$ (ex. $K^0$) | BABAR | $(2.25 \pm 0.07 \pm 0.12) \cdot 10^{-4}$ | Lees (2012z) |
| | HFAG | $(1.492 \pm 0.097) \cdot 10^{-4}$ | Amhis et al. (2012) |
| $\tau \to \pi^- \pi^0 \pi^0 \eta \nu_\tau$ | BABAR | $(2.01 \pm 0.34 \pm 0.22) \cdot 10^{-4}$ | Lees (2012z) |
| $\tau \to \pi^- \pi^- \pi^+ \omega \nu_\tau$ (ex. $K^0$) | BABAR | $(8.4 \pm 0.4 \pm 0.6) \cdot 10^{-5}$ | Lees (2012z) |
| $\tau \to \pi^- \pi^0 \pi^0 \omega \nu_\tau$ | BABAR | $(7.3 \pm 1.2 \pm 1.2) \cdot 10^{-5}$ | Lees (2012z) |

#### 20.6.3.1 $\tau^- \to \pi^-\pi^0\nu_\tau$

Among the decay channels of the tau lepton, $\tau^- \to \pi^-\pi^0\nu_\tau$ has the largest branching fraction. From the CVC theorem, the $\pi^-\pi^0$ mass spectrum can be related to the cross section of the process $e^+e^- \to \pi^+\pi^-$ and thus used to improve a theoretical error on the anomalous magnetic moment of the muon $a_\mu = (g_\mu - 2)/2$.

Using a sample of 5,430,000 $\tau^- \to \pi^-\pi^0\nu_\tau$ decays, Belle measures the branching fraction and the $\pi\pi^0$ mass spectrum (Fujikawa, 2010). After unfolding performed using the singular-value-decomposition method (Höcker and Kartvelishvili, 1996), the distribution for the $\pi\pi^0$ mass spectrum shown in Fig. 20.6.2(a) is obtained. This precisely measured spectrum has a shape formed by $\rho(770)$, $\rho(1450)$ and $\rho(1700)$ resonances and their interference. Figure 20.6.2(b) shows the pion form factor in the $\rho(770)$ region obtained from the $\pi\pi^0$ mass spectrum. The measured branching fraction is

$$\mathcal{B}(\tau^- \to \pi^-\pi^0\nu_\tau) = (25.24 \pm 0.01(\mathrm{stat}) \pm 0.39(\mathrm{syst}))\%. \quad (20.6.26)$$

The discussion of the CVC relation and the evaluation of $a_\mu$ using tau-lepton data including this Belle measurement are provided in Section 20.7.

#### 20.6.3.2 $\tau^- \to (3\pi)^-\nu_\tau$

The three-pion final states form a dominant fraction of the axial-vector current. The unfolded $3\pi$ mass spectrum given by $(1/\Gamma)(\mathrm{d}\Gamma/\mathrm{d}M)$, measured in $\tau^- \to \pi^-\pi^+\pi^-\nu_\tau$ by Belle (Lee, 2010), is shown in Fig. 20.6.3(a). The solid histogram is the spectrum implemented in the current `TAUOLA` program (Gómez Dumm, Roig, Pich, and Portolés, 2010a). A systematic difference between data and MC is observed at the peak region of the $a_1(1260)$ resonance. Belle (red circle) and ALEPH (solid circles) show very good agreement in this peak region as can be seen in Fig. 20.6.3(b) and (c). This indicates that the model in the current `TAUOLA` should be updated. Figure 20.6.3(d) shows the ratio ALEPH/Belle−1. Both data show very good agreement for $M^2$ between 0.7 and 2.0 $\mathrm{GeV}^2/c^4$. Outside this range differences of up to 20% between the experiments have been observed. This can be attributed to the imperfect modeling of the background or the detector effects. Further studies are needed for the high mass region, where the mass spectrum plays an important role as the spectral function of the axial-vector current (see the discussion in Section 20.6.1.4, and 20.6.5 for the recent status).

#### 20.6.3.3 $\tau^- \to (K\overline{K}\pi)^-\nu_\tau$

The $K\overline{K}\pi$ mode has both vector and axial-vector components. The axial-vector component arises from the Wess-Zumino-Witten chiral-anomalous term (Wess and Zumino, 1971; Witten, 1983) and the non-anomalous odd-intrinsic-parity amplitude (Ruiz-Femenia, Pich, and Portoles, 2003).

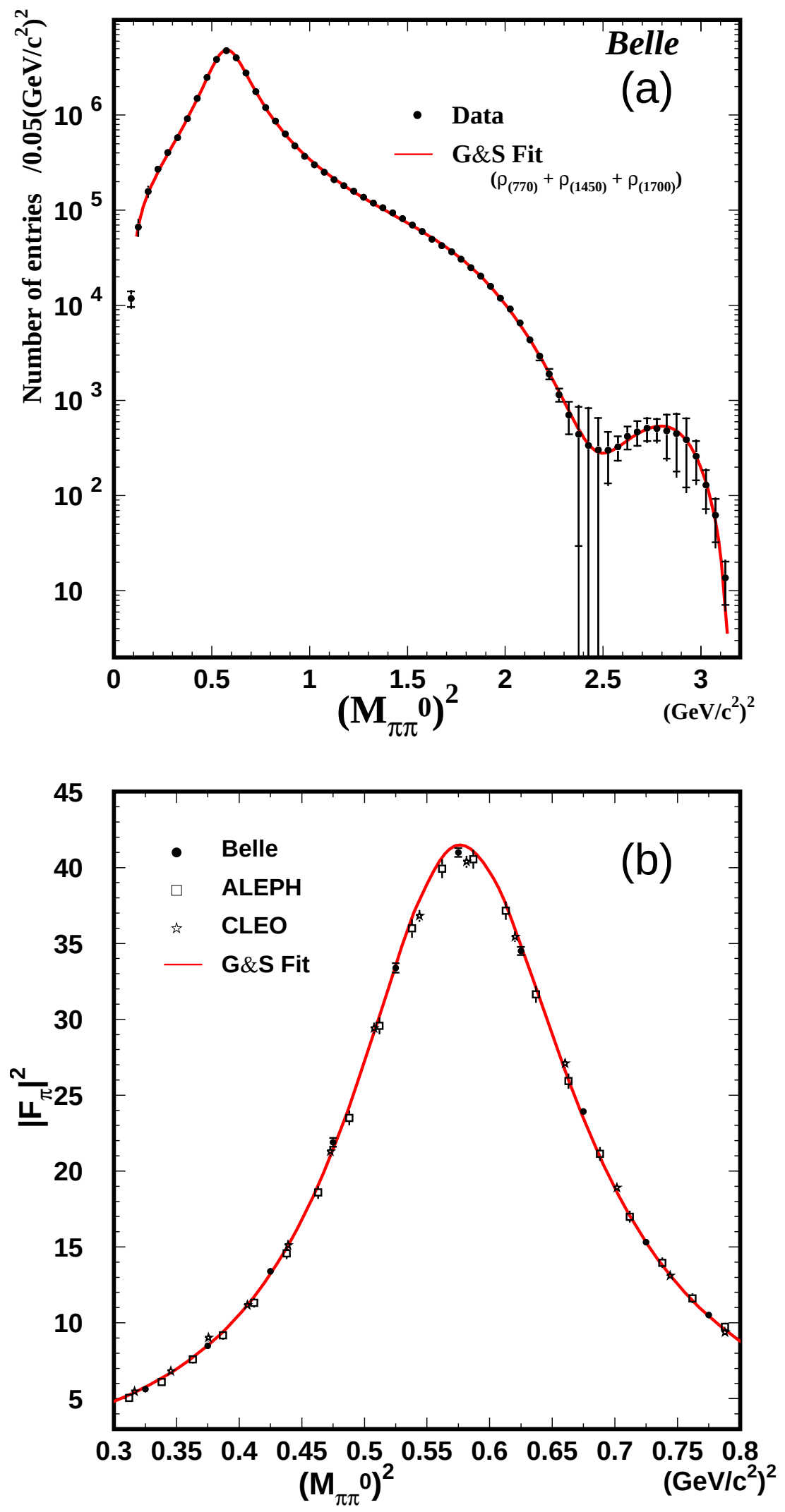

**Figure 20.6.2.** (a) Unfolded $\pi^{\pm}\pi^{0}$ mass spectrum for $\tau^{\pm} \to \pi^{\pm}\pi^{0}\nu_{\tau}$ obtained by the Belle experiment (Fujikawa, 2008). Solid circles are the data and the solid line is a fit based on the Gounaris-Sakurai (G&S) parameterization (Gounaris and Sakurai, 1968). The error bars include both statistical and systematic errors. (b) Pion form factor $|F_{\pi}(s)|^{2}$ in the $\rho(770)$ region derived from the $\pi^{\pm}\pi^{0}$ mass spectrum.

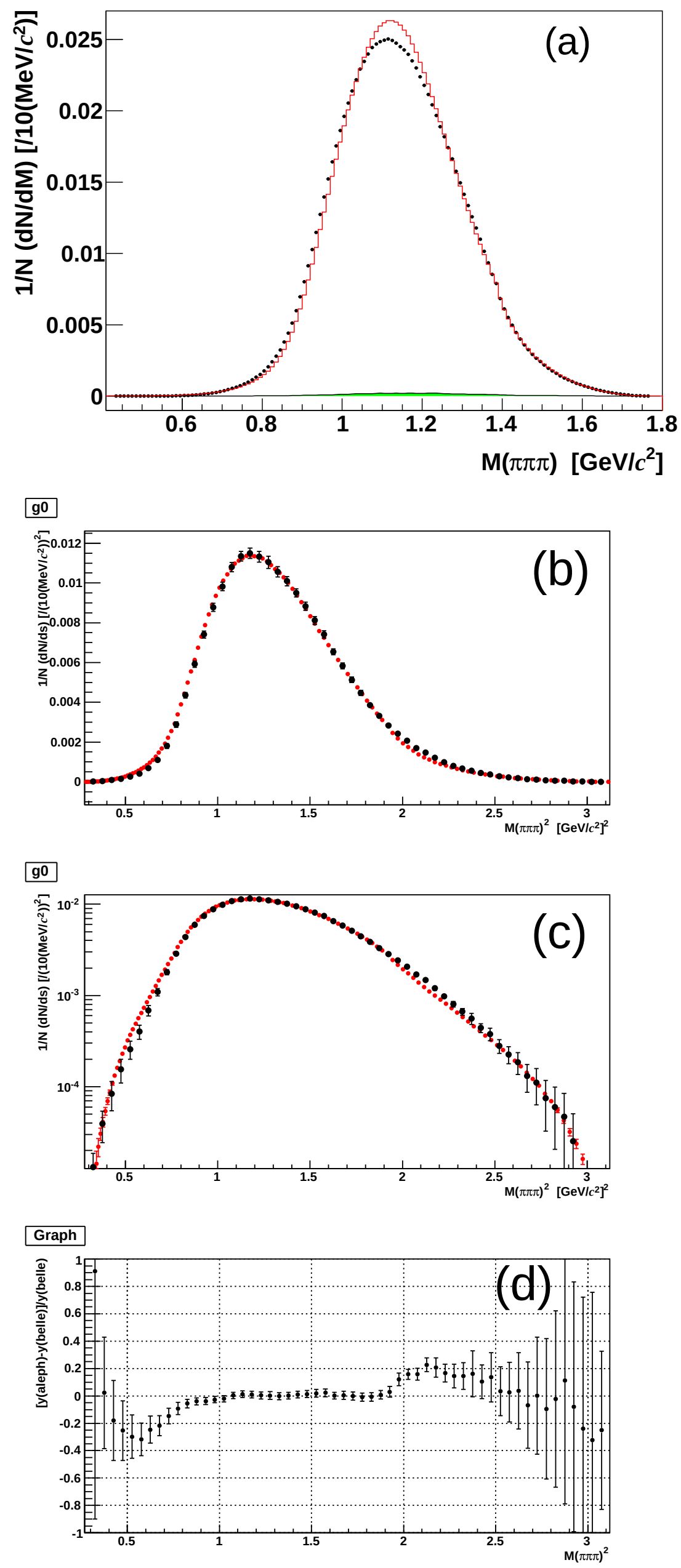

**Figure 20.6.3.** (color online) (a) Unfolded mass spectrum $(1/N)dN/dM$ for the $\pi^{-}\pi^{+}\pi^{-}$ system in $\tau^{-} \to \pi^{-}\pi^{+}\pi^{-}\nu_{\tau}$ measured by Belle (Lee, 2010). Solid circles with error bars are the data with statistical errors; solid-red histogram is the spectra implemented in the current `TAUOLA` program; the horizontal green band at the zero entry line shows the size of the systematic uncertainties for data. (b) Comparison of $\pi^{-}\pi^{-}\pi^{+}$ mass spectra as a function of mass-squared between Belle (red circle) and ALEPH (black circles). (c) in log-scale, (d) the ratio ALEPH/Belle−1.

See Shekhovtsova, Przedzinski, Roig, and Wąs (2012) for more details.

The unfolded $K\overline{K}\pi$ mass spectrum measured by Belle (Lee, 2010) for the $\tau^{-} \to K^{-}K^{+}\pi^{-}\nu_{\tau}$ is shown in Fig. 20.6.4 (a). In this process there are contributions from both vector and axial-vector currents. The current `TAUOLA` output (histogram) does not give a satisfactory description of the data. In Fig. 20.6.4 (b), the result is compared with a model prediction based on resonance chiral perturbation theory (Gómez Dumm, Roig, Pich, and Portolés, 2010a). The comparison indicates that a larger axial-vector component than the current prediction is needed to improve the agreement between the data and model.

#### 20.6.3.4 $\tau^{-} \to \pi^{-}\pi^{0}\eta\nu_{\tau}$

Using a data sample of 490 fb$^{-1}$, Belle has studied the $\tau^{-} \to \pi^{-}\pi^{0}\eta\nu_{\tau}$ decay where the $\eta$ meson is reconstructed through its decay into $\gamma\gamma$ or $\pi^+\pi^-\pi^0$ (Inami, 2009) and obtained the new world-average branching fraction of $(0.139 \pm 0.008)\%$. The branching fraction and the mass spectrum are compared with the prediction from the CVC theorem by Cherepanov and Eidelman (2011), using the data from the process $e^+e^- \to \eta\pi^+\pi^-$ including the recent results from *BABAR* (Aubert, 2007bb) and SND (Achasov et al., 2010) collaborations. Belle found that the expected branching fraction $(0.153\pm0.018)\%$ is compatible with the tau result and that the mass spectra of the $\eta\pi^+\pi^-$ system in $e^+e^-$ annihilation and tau decay are consistent with the CVC expectation.

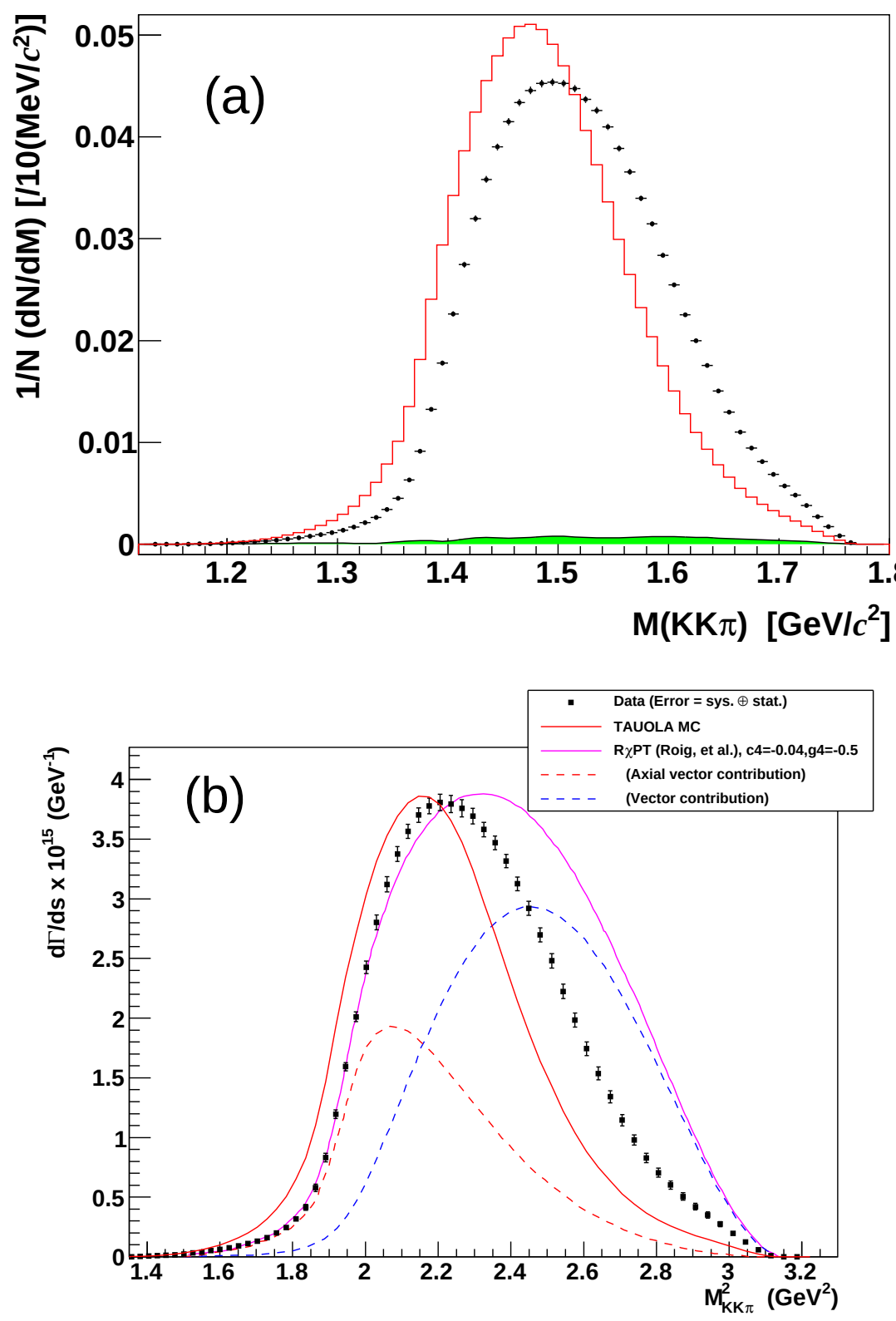

**Figure 20.6.4.** (color online) (a) Unfolded mass spectrum $(1/N)dN/dM$ for the $K^-K^+\pi^-$ system in $\tau^- \to K^-K^+\pi^-\nu_\tau$ measured by Belle (Lee, 2010). Solid circles with error bars are the data with statistical errors; solid-red histogram is the spectra implemented in the current `TAUOLA` program; the horizontal green band at the zero entry line shows the size of the systematic uncertainties for data. (b) Comparison of the result with the theoretical model based on resonance chiral-perturbation theory. Closed circles are Belle data, red-dashed and blue-dashed lines are the axial-vector and vector components from the chiral-perturbation theory, respectively. the pink-solid line is the sum of axial-vector and vector components. While the red-solid line is the model implemented in the `TAUOLA` progarm (Gómez Dumm, Roig, Pich, and Portolés, 2010a).

### 20.6.4 Hadronic spectral functions: Cabibbo-suppressed modes

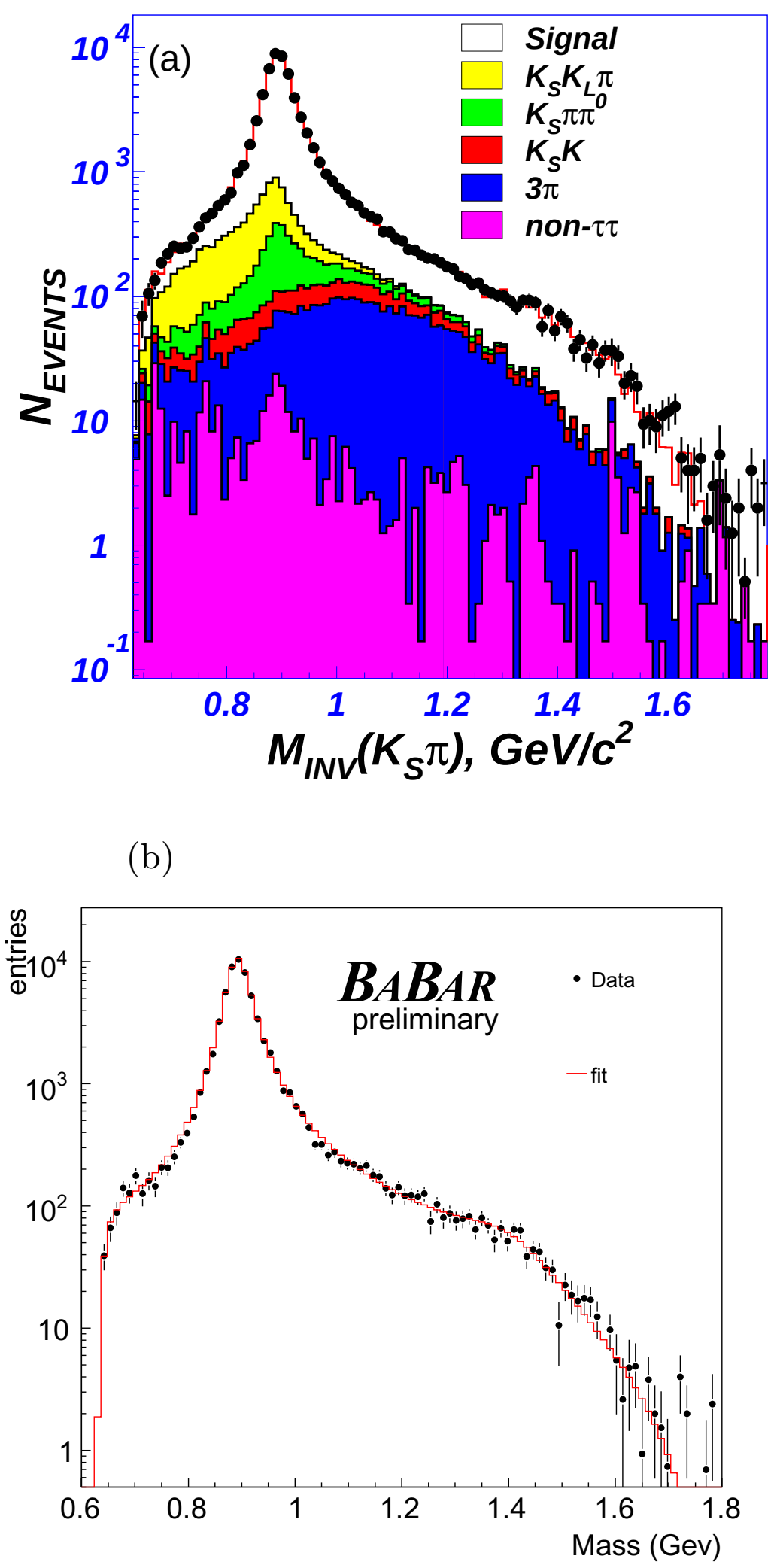

**Figure 20.6.5.** $K_S\pi$ mass distribution measured by (a) Belle (Epifanov, 2007) and by (b) *BABAR* (Adametz, 2011). Points with error bars in both figures are data, the histogram shows the fitted result for the spectrum expected in the model incorporating the $K^*(892) + K_0^*(800) + K^*(1410)$ model. In Belle data, different types of background are also shown, while in the *BABAR* plot, the background is already subtracted.

#### 20.6.4.1 $\tau^- \to (K\pi)^-\nu_\tau$

A data sample of 351 fb$^{-1}$ has been used by Belle to study the $K_S\pi^-\nu_\tau$ final state (Epifanov, 2007). As a result of the analysis, 53,110 lepton-tagged signal events have been selected. The measured branching fraction obtained is

$$\mathcal{B}(\tau^- \to K_S\pi^-\nu_\tau) = (0.404 \pm 0.002 \pm 0.013)\%$$

(20.6.27)

which is the most precise among all the published measurements and is somewhat lower than all of them although within errors consistent with the other results.

An analysis of the $K_S\pi^-$ invariant mass spectrum shown in Fig. 20.6.5 (top) reveals the dominant contribution from the $K^*(892)^-$ with additional contributions of higher-mass states at 1400 MeV. A satisfactory fit is obtained only if the existence of a broad scalar state, $K_0^*(800)$, is assumed. For the first time the $K^*(892)^-$ mass and width have been measured in tau decay:

$$\begin{aligned} M(K^*(892)^-) =&(895.47 \pm 0.20 \pm 0.44 \pm 0.59)\ \text{MeV},\\ \Gamma(K^*(892)^-) =&(46.2 \pm 0.6 \pm 1.0 \pm 0.7)\ \text{MeV}, \end{aligned} \tag{20.6.28}$$

where the first uncertainty is the statistical and the second systematic. The third uncertainty is model-based. The $K^*(892)^-$ mass is significantly higher than the world-average value based on various hadronic experiments and is much closer to the world average for the neutral $K^*(892)$ (Nakamura et al., 2010).

Recently the *BABAR* collaboration presented the $K_S\pi^-$ invariant mass spectrum in the $K_S\pi^-\nu_\tau$ decay (Adametz, 2011). Figure 20.6.5 (bottom) shows the $K_S\pi^-$-invariant mass distribution and the fit result. The mass and width of the $K^*(892)^-$ resonance determined using a $K^*(892) + K_0^*(800) + K^*(1410)$ model yield

$$\begin{aligned} M(K^*(892)^-) &= (894.57 \pm 0.19 \pm 0.19)\ \text{MeV},\\ \Gamma(K^*(892)^-) &= (45.56 \pm 0.43 \pm 0.57)\ \text{MeV}. \end{aligned} \tag{20.6.29}$$

The values of the $K^*(892)^-$ mass and width obtained by the $B$ Factories are in agreement with each other. However, it should be noted that *BABAR* did not present an uncertainty induced by the ambiguity in the model used for the mass spectrum fit.

#### 20.6.4.2 $\tau^- \to (K\pi\pi)^-\nu_\tau$

The unfolded $K^-\pi^+\pi^-$ mass spectrum measured by Belle (Lee, 2010), shown in Fig. 20.6.6 (a), has a peak at $1.25\,\text{GeV}/c^2$ and shoulder at $1.4\,\text{GeV}/c^2$, corresponding to $K_1(1270)$ and $K_1(1400)$, respectively. The model in the `TAUOLA` reproduces the qualitative features of the mass spectra but fails to reproduce their detail.

### 20.6.5 Inclusive non-strange spectral function

As was discussed in Section 20.6.1.2, the inclusive non-strange (Cabibbo-favored) spectral function plays an important role for the determination of $\alpha_S$ and various theoretical tests of QCD in the transition region between perturbative QCD and resonances. The results obtained by LEP experiments are shown in Fig. 20.6.7 for the vector current, and in Fig. 20.6.8, for the axial-vector current. The statistical errors in the high mass region above $2\,\text{GeV}^2/c^4$, where perturbative QCD plays an important role, are large. Mass spectra data can contribute significantly to the improvement of our knowledge of this region. See Boito et al. (2012) for a recent discussion of the importance of the new data in this region.

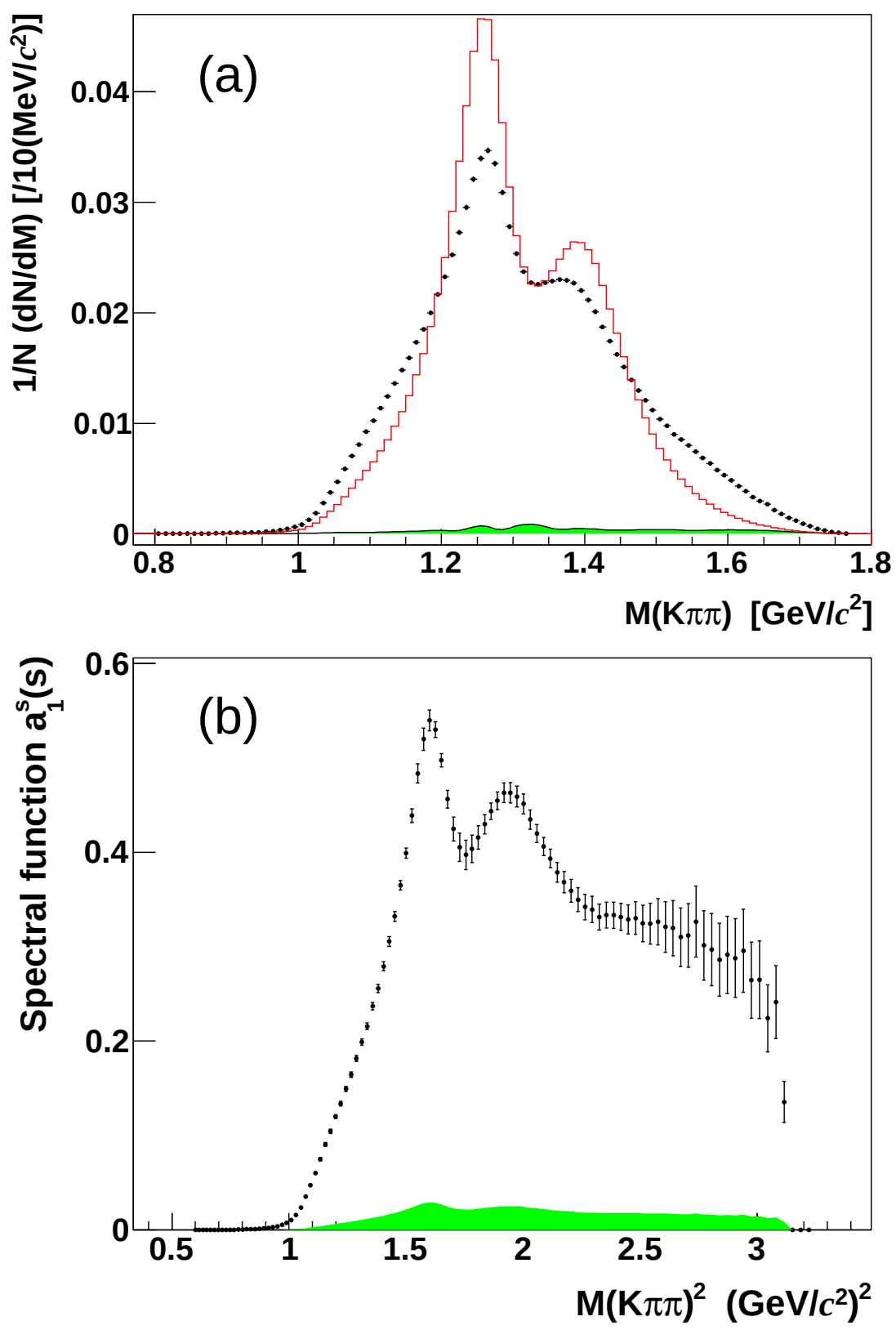

**Figure 20.6.6.** (color online) (a) Unfolded mass spectrum $(1/N)dN/dM$ for the $K^-\pi^+\pi^-$ system in $\tau^- \to K^-\pi^+\pi^-\nu_\tau$ measured by Belle (Lee, 2010). Solid circles with error bars are the data with statistical errors; solid-red histogram is the spectra implemented in the current `TAUOLA` program. (b) $K^-\pi^+\pi^-$ spectral function exstructed from the unfolded mass spectrum. The horizontal green band at the zero entry line shown in the both figures indicate the size of the systematic uncertainties for data.

### 20.6.6 Inclusive strange spectral functions

An accurate measurement of strange (Cabibbo-suppressed) spectral function plays an important role in the simultaneous determination of $|V_{us}|$ and the strange quark mass $m_s$ (see Section 20.6.1.3). The strange spectral function available now is obtained from the ALEPH and OPAL experiments at LEP (see Fig. 20.6.9) and has large errors with coarse binning. Exclusive spectral functions for $K^-K^+\pi^-$ (Fig. 20.6.4 (b)) and $K^-\pi^+\pi^-$ (Fig. 20.6.6 (b)) measured by the Belle, show a significant improvement over the previous LEP experiments. The measurements for all Cabibbo-suppressed modes, and their sum, to obtain the total inclusive spectral function are in progress.

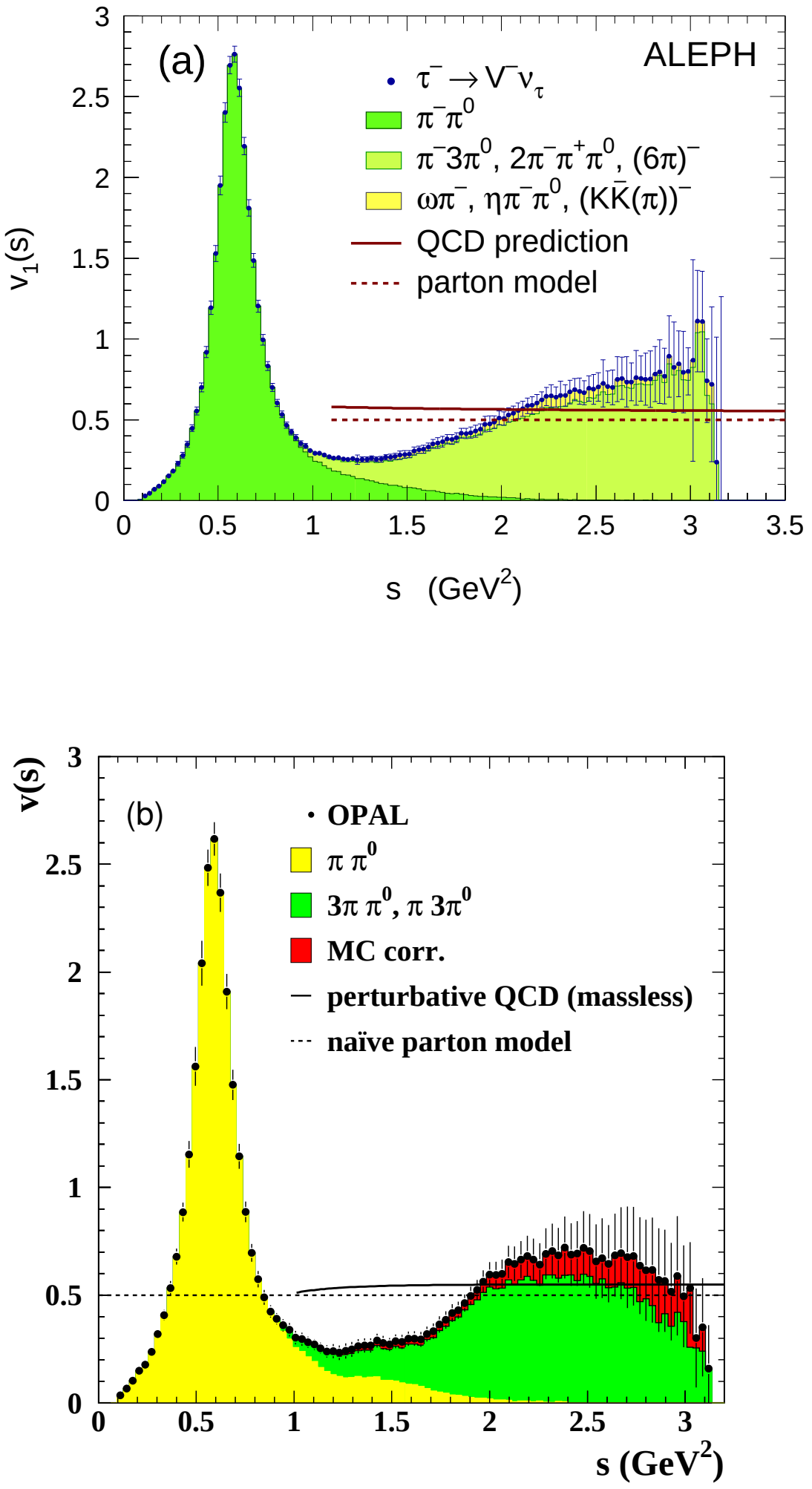

**Figure 20.6.7.** Vector spectral functions measured by (a) ALEPH (Davier, Hoecker, and Zhang, 2006; Schael et al., 2005) and (b) OPAL (Ackerstaff et al., 1999) experiments.

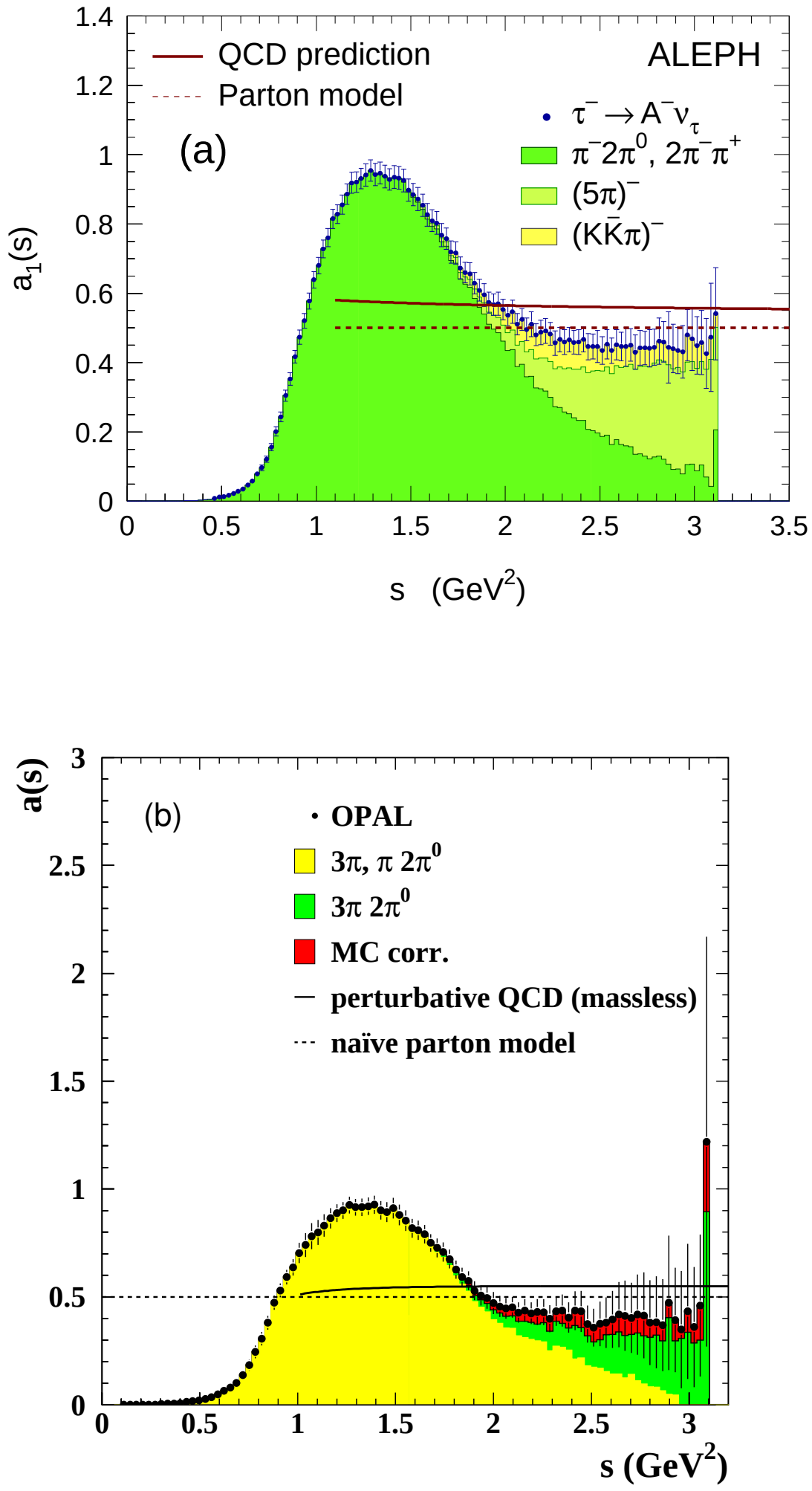

**Figure 20.6.8.** Axial Vector spectral functions measured by (a) ALEPH (Davier, Hoecker, and Zhang, 2006; Schael et al., 2005) and (b) OPAL (Ackerstaff et al., 1999) experiments.

### 20.6.7 Search for second-class currents

Standard Model processes that are mediated by the charged hadronic weak current, such as semihadronic tau decays, produce hadronic systems with a well-defined set of allowed quantum numbers for spin, parity and G-parity. In tau decays these so-called first-class currents (Weinberg, 1958) yield hadronic systems with $J^{PG} = 0^{++}, 0^{--}, 1^{+-}$ or $1^{-+}$.

The quantum numbers $J^{PG} = 0^{+-}, 0^{-+}, 1^{++}$ and $1^{--}$ would be associated with second-class currents. Small contributions to second-class currents, at the level of order $10^{-5}$ to $10^{-6}$, are expected from isospin violation (and hence G-parity violation) due to the mass difference between the up and down quarks (Berger and Lipkin, 1987; Nussinov and Soffer, 2008, 2009; Pich, 1987). However, second-class currents have not been seen to date in tau decays, nor indeed in any processes mediated by the hadronic weak current. Observation at a level significantly above that expected from isospin violation would be a signal for new physics contributions (Langacker, 1977).

Tau decay modes that have the quantum numbers associated with second-class currents include: $\tau^- \to \eta\pi^-\nu_\tau$ and $\tau^- \to \eta'(958)\pi^-\nu_\tau$, both of which correspond to $J^{PG} = 0^{+-}$ or $1^{--}$; and $\tau \to \omega(782)\pi^-\nu_\tau$ with the $\omega$ and $\pi^-$ in a relative $S$-wave or $D$-wave. However, the $\omega(782)\pi^-\nu_\tau$ mode is dominated by the $P$-wave $J^{PG} = 1^{-+}$ $\tau^- \to \rho(770)^-\nu_\tau$ and $\rho'\nu_\tau$ decays, and an angular analysis is required to search for any $S$-wave or $D$-wave (second-class) contribution.

Before the $B$ Factory experiments, the best upper limits on second-class currents in tau decays came from CLEO for the $\eta$ and $\eta'$ modes (Bartelt et al., 1996) and from ALEPH for the $\omega\pi^-$ mode (Buskulic et al., 1997). *BABAR* has published results for all of the above modes, providing significant improvements in upper limits for the branching fractions, but continues to see no evidence for the existence of second-class currents.

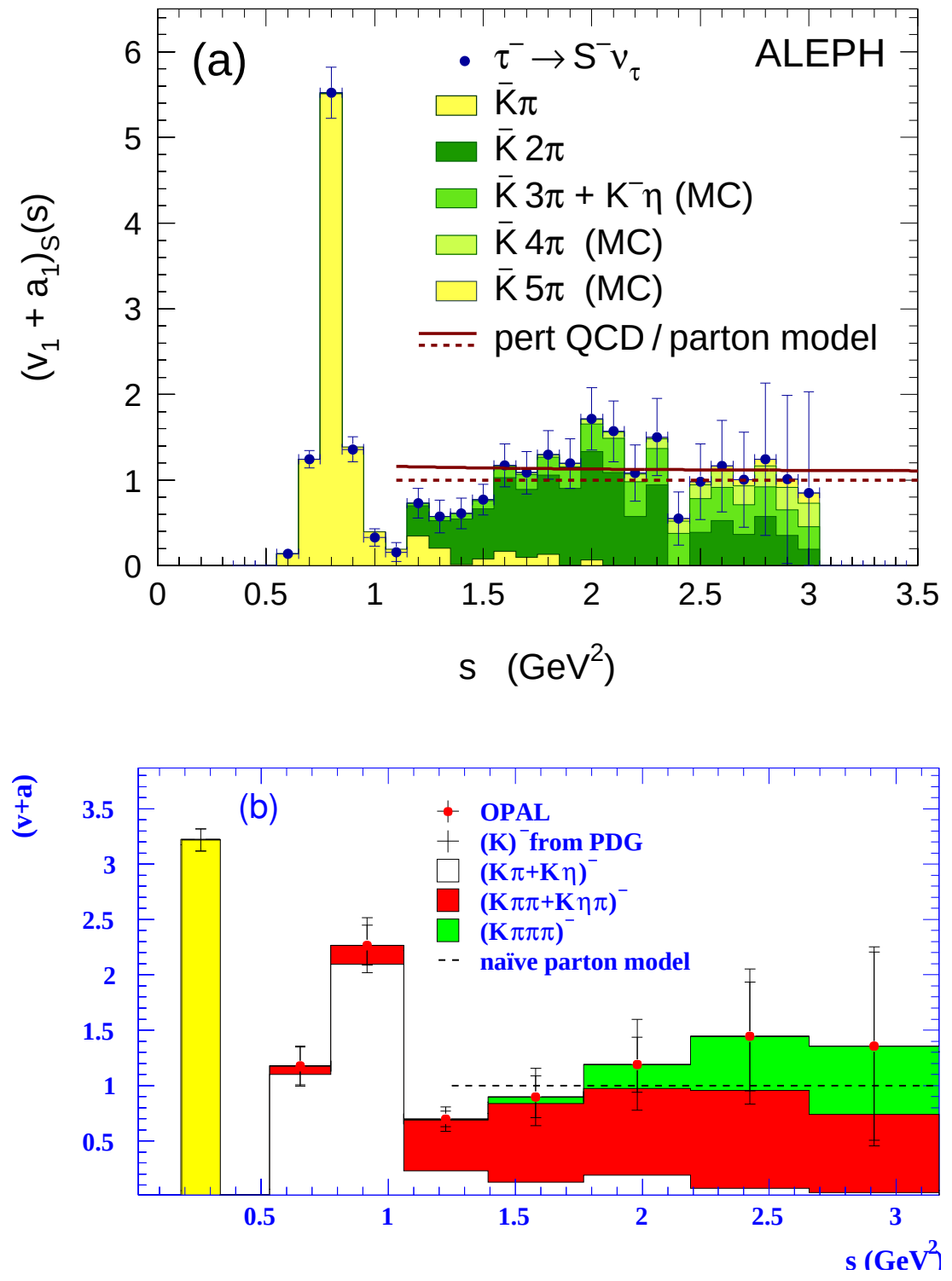

**Figure 20.6.9.** Strange spectral functions measured by (a) ALEPH (Davier, Hoecker, and Zhang, 2006) and (b) OPAL (Abbiendi et al., 2004) experiments.

The $\tau^- \to \eta\pi^-\nu_\tau$ channel has been studied by *BABAR*, using the $\eta \to \pi^+\pi^-\pi^0$ decay mode (del Amo Sanchez, 2011m). The basic method is to select a sample of candidate $\tau^- \to \pi^-\pi^-\pi^+\pi^0\nu_\tau$ decays and to fit the inclusive $\pi^+\pi^-\pi^0$ mass spectrum for the contribution from $\eta$. The limiting factors in this analysis are the relatively large combinatorial background in the $3\pi$ spectrum from tau decays to $\omega(782)\pi^-\nu_\tau$ and $\rho\pi\pi\nu_\tau$, and the background from $\tau^- \to \eta\pi^-\pi^0\nu_\tau$ where the additional $\pi^0$ is undetected. The former background cannot be removed without producing significant distortions of the phase space, and introducing a strong model dependence in any limit. For these reasons, the limit on the branching fraction obtained by *BABAR*, $< 0.9 \times 10^{-4}$ at 95% C.L., was dominated by systematic errors and represented only a small improvement over the previously existing limit from CLEO (Bartelt et al., 1996). It may be possible to improve limits further at both *BABAR* and Belle using the $\eta \to \gamma\gamma$ decay mode, but here again large backgrounds may be expected, particularly from $\tau^- \to \eta\pi^-\pi^0\nu_\tau$ and $\tau^- \to \pi^-\pi^0\nu_\tau$.

*BABAR* has also looked for the second-class current mode $\tau^- \to \eta'(958)\pi^-\nu_\tau$, with the decay mode $\eta' \to \eta\pi^+\pi^-$ and $\eta \to \pi^+\pi^-\pi^0$ (Lees, 2012z). A fit for a possible $\eta'$ contribution in the inclusive $\eta\pi^+\pi^-$ mass spectrum gave no significant signal above the expected background, allowing a limit to be set on the branching fraction at $< 4.0 \times 10^{-6}$ at a 90% C.L.. This is the lowest limit obtained for any possible second-class decay mode of the tau and is at the level where the effects of isospin violation may be expected to appear.

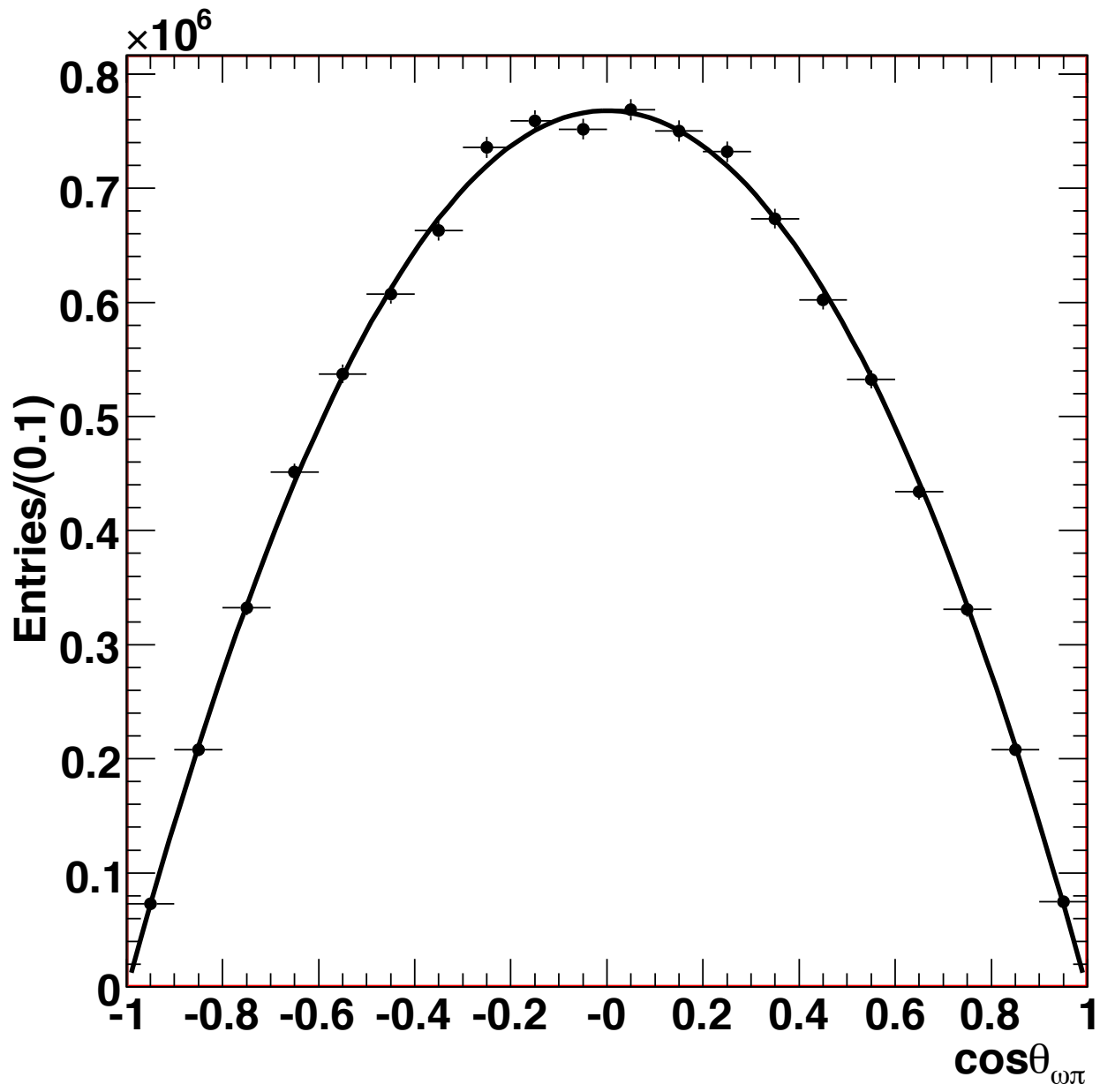

**Figure 20.6.10.** Distribution of $\cos\theta_{\omega\pi}$ from $\tau \to \omega(782)\pi^-\nu_\tau$ events in *BABAR* (Aubert, 2009ap). Here $\theta_{\omega\pi}$ is the angle, in the $\omega\pi^-$ rest frame, between the normal to the $\omega \to \pi^+\pi^-\pi^0$ decay plane and the direction of the pion. The curve shows a fit to determine the relative contributions from first-class and second-class currents.

Figure 20.6.10 is taken from the *BABAR* analysis of the channel $\tau \to \omega(782)\pi^-\nu_\tau$ (Aubert, 2009ap), and shows the distribution of $\cos\theta_{\omega\pi}$, where $\theta_{\omega\pi}$ is the angle, in the $\omega\pi^-$ rest frame, between the normal to the $\omega \to \pi^+\pi^-\pi^0$ decay plane and the direction of the pion. This distribution has been corrected for the combinatorial background, using distributions from the $\omega$ mass sidebands. The curve is a fit to a sum of $S$-wave and $P$-wave contributions. It is apparent that the data correspond to an almost pure $\sin^2\theta_{\omega\pi}$ distribution, as expected from a $P$-wave decay. The fit allows a limit to be put on an possible $S$-wave (*i.e.* second-class current) contribution to this mode at a level of $< 0.0069$ at a 95% C.L.. This corresponds to an absolute limit on the branching fraction for the decay $\tau \to \omega(782)\pi^-\nu_\tau$ via a second-class current of $< 1.4 \times 10^{-4}$ at 95% C.L.

## 20.7 Tests of CVC and vacuum hadronic polarization determination

The hypotheses of the CVC and isospin symmetry relate the isovector part of $e^+e^- \to$ hadrons and corresponding (vector current $J^P = 1^-$) hadronic decay of the tau lepton. This follows from the deep relation between weak and electromagnetic interactions. The weak vector current and the isovector part of the electromagnetic vector current are different components of the same vector current, so that the matrix element of these currents must be identical assuming $SU(2)$ symmetry. In this case the weak isovector current is assumed to be conserved in analogy with the electromagnetic current. This assumption is the CVC hypothesis and was first introduced by Feynman and Gell-Mann into their theory of weak interaction (Feynman and Gell-Mann, 1958).

As a consequence, hadronic currents describing the Cabibbo-allowed vector part of the tau hadronic decays, such as $2\pi$, $4\pi$, $\pi\pi\eta$ and $\pi\omega$ channels, and low-energy $e^+e^-$ annihilation are closely related to each other (Gilman and Rhie, 1985). In fact, this relation was used to predict the decays of a heavy lepton even before the discovery of the tau lepton (Thacker and Sakurai, 1971; Tsai, 1971).

### 20.7.1 CVC and vacuum hadronic polarization contribution in $(g-2)_\mu$

In addition, the CVC relations allow one to use an independent high-statistics data sample from tau decays for increasing accuracy of the prediction of the hadronic contributions to the muon anomalous magnetic moments $a_\mu = (g-2)/2$ (Alemany, Davier, and Hoecker, 1998). Here, we briefly review the basic formula required for this.

The leading-order hadronic contribution $a_\mu^{\text{had,LO}}$ can be obtained by using a combination of experimental data and perturbative QCD for the hadronic vacuum polarization (HVP) of the photon. At low energies, where QCD does not provide a reliable calculation, the HVP can be obtained as a sum over the production cross section of each $e^+e^- \to X^0$ channel (Jegerlehner and Nyffeler, 2009; Miller, de Rafael, and Roberts, 2007).

$$a_\mu^{\text{had,LO}}(e^+e^-) = \frac{\alpha^2}{3\pi^2}\int_{4m_\pi^2}^{\infty} ds \frac{K(s)}{s} R_{X^0}^{(0)}(s), \qquad (20.7.1)$$

where $s$ is the CM energy squared of the hadron system and $R_{X^0}^{(0)}$ is the ratio of hadronic $X^0$ to point-like $\mu^+\mu^-$ bare cross sections in $e^+e^-$ annihilation given by

$$R_{X^0}^{(0)}(s) = \frac{3s\sigma_{X^0}(s)}{4\pi\alpha^2} = 3v_0(s). \qquad (20.7.2)$$

The behavior of the QED kernel $K(s) \sim 1/s$ enhances the low-energy contributions to $a_\mu^{\text{had,LO}}$ (Jegerlehner and Nyffeler, 2009).

In the limit of isospin invariance ($v_0 = v_1$) and taking into account the isospin breaking effects, the spectral function of the vector current decay $\tau \to X^-\nu_\tau$ is related to the $e^+e^- \to X^0$ cross section of the corresponding isovector final sate $X^0$,

$$\sigma_{X^0}(s) = \frac{4\pi\alpha^2}{s} v_{1,X^-}(s), \qquad (20.7.3)$$

where $s$ is the invariant mass squared of the tau final state $X^-$ and $\alpha$ is the electromagnetic fine structure constant. The term $v_{1,X^-}(s)$ is the vector spectral function in the Cabibbo-allowed decays, which is given by

$$\begin{aligned} v_{1,X^-}(s) &= 3Im\left[\Pi_V^{(1)}(s)\right] \\ &= \frac{m_\tau^2}{6|V_{ud}|^2}\frac{\mathcal{B}_{X^-}}{\mathcal{B}_e}\frac{1}{N_X}\frac{dN_x}{ds} \\ &\times\left(1-\frac{s}{m_\tau^2}\right)^{-2}\left(1+\frac{2s}{m_\tau^2}\right)^{-1}\frac{R_{IB}(s)}{S_{EW}}, \end{aligned} \qquad (20.7.4)$$

where, $(1/N_X)dN_X/ds$ is the normalized invariant mass spectrum of the hadronic final state, $\mathcal{B}_{X^-}$ denotes the branching fraction of $\tau^- \to X^-\nu_\tau$. The values of other parameters, tau mass $m_\tau$, the CKM matrix element $|V_{ud}|$ and the electron branching fraction $\mathcal{B}_e$ are known precisely.

The last term $R_{IB}(s)/S_{EW}$ represents the correction for the isospin-breaking (IB) effects. Short-distance electroweak radiative effects lead to the correction $S_{EW} = 1.0235 \pm 0.0003$ (Davier, Eidelman, Hoecker, and Zhang, 2003b). All the $s$-dependent isospin-breaking corrections are included in $R_{IB}(s)$. In the dominant $\pi^+\pi^-$ decay channel, $R_{IB}(s)$ is given by

$$R_{IB}(s) = \frac{FSR(s)}{G_{EM}(s)}\frac{\beta_0^3(s)}{\beta_-^3(s)}\left|\frac{F_0(s)}{F_-(s)}\right|^2, \qquad (20.7.5)$$

where the subscripts $i = 0, -$ refer to the electric charge of the $2\pi$ system produced in $e^+e^-$ annihilation, and in $\tau^-$ lepton decay, respectively. $FSR(s)$ refers to the final state radiative corrections in the $e^+e^- \to \pi^-\pi^-$ channel, and $G_{EM}(s)$ denotes the long-distance radiative corrections to the inclusive $\pi^-\pi^0$ spectrum in $\tau \to \pi^-\pi^0\nu_\tau$. The second correction of the ratio of the pion velocities, $\beta_0^3(s)/\beta_-^3(s)$, arises from $\pi^\pm - \pi^0$ mass difference. The third IB correction term $|F_0/F_-|^2$ involves the ratio of electromagnetic ($F_0$) to weak ($F_-$) form factors and needs to be considered carefully. This ratio involves two sources of IB: (a) $\rho - \omega$ mixing effects and (b) the mass and width difference of neutral and charged $\rho$ mesons.

Taking these IB corrections into account, the shift in the lowest order hadronic contribution to the muon $g-2$ using tau data in the dominant $\pi\pi$ channel can be evaluated as

$$\begin{aligned} \triangle a_\mu^{\text{had,LO}}[\pi^+\pi^-,\tau] &= \frac{\alpha^2}{3\pi^2}\int_{4m_\pi^2}^{\infty} ds\frac{K(s)}{s}3v_-(s) \\ &\times\left[\frac{R_{IB}(s)}{S_{EW}} - 1\right]. \end{aligned} \qquad (20.7.6)$$

The most recent estimates for these effects are summarized in Table 20.7.1 (Castro, 2010; Davier et al., 2010). In the table, the last term, the radiative corrections for photon-inclusive $\rho \to \pi\pi$ is different from the previous estimate in (Davier, Eidelman, Hoecker, and Zhang, 2003b).

Using all available $\pi^+\pi^-$ data from tau lepton decays, ALEPH, CLEO, OPAL and Belle, and applying these IB

**Table 20.7.1.** Contributions to $\triangle a_\mu^{\rm had,LO}[\pi\pi,\tau](\times 10^{-10})$ from the isospin-breaking correction $R_{IB}(s)$. Corrections shown in two separate columns correspond to the results obtained using the Gounaris and Sakurai (1968)(GS), and Kuhn and Santamaria (1990)(KS) parameterization of pion form factors (Davier et al., 2010).

| Source | $\triangle a_\mu^{\rm had,LO}[\pi\pi,\tau](10^{-10})$ | |
|---|---|---|
| | GS model | KS model |
| $S_{EW}$ | $-12.21 \pm 0.15$ | |
| $G_{EM}$ | $-1.92 \pm 0.90$ | |
| FSR | $+4.67 \pm 0.47$ | |
| $\rho - \omega$ interference | $+2.80 \pm 0.19$ | $+2.80 \pm 0.15$ |
| $m_{\pi^\pm} - m_{\pi^0}$ effect on $\sigma$ | $-7.88$ | |
| $m_{\rho^\pm} - m_{\rho^0}$ | $0.20^{+0.27}_{-0.19}$ | $0.11^{+0.19}_{-0.11}$ |
| $m_{\pi^\pm} - m_{\pi^0}$ effect on $\Gamma_\rho$ | $+4.09$ | $+4.02$ |
| $\rho \to \pi\pi\gamma$ corr. | $-5.91 \pm 0.59$ | $-6.39 \pm 0.64$ |
| Total | $-16.07 \pm 1.22$ | $-16.70 \pm 1.23$ |
| | $-16.07 \pm 1.85$ | |

corrections, the lowest order hadronic contributions obtained are summarized in the $2^{nd}$ column of Table 20.7.2. The new tau-based estimate of $a_\mu^{\rm had,LO}$ is found to be 1.9 $\sigma$ lower than the results based on the $e^+e^-$ data. This updated estimate makes the difference between tau and $e^+e^-$ based estimates closer than the previous ones reported by Davier, Eidelman, Hoecker, and Zhang (2003b).

In addition to these IB corrections, the importance of the other effect that was not discussed is recently demonstrated by Jegerlehner and Szafron (2011). They argue that, in addition to $\rho - \omega$ mixing, $\rho - \gamma$ interference exists in the $e^+e^-$ reaction and they contribute to $a_\mu^{\rm had,LO}$, but that effect does not exist in tau-lepton decays. The size of the $\rho - \gamma$ interference effects is about 5% to 10% in the $\rho$-resonance region and changes the sign in the lower and upper side of the $\rho$ mass peak (see Fig.6 in Jegerlehner and Szafron (2011)). After taking this effect into account, the results of $a_\mu^{\rm had,LO}$ obtained from $e^+e^-$-based and tau-based data are summarized in the $3^{rd}$ column of Table 20.7.2. The table shows a good agreement between the $e^+e^-$ and tau-based results, if the $\rho - \gamma$ effects are taken into account. A similar estimation based on a Hidden Local Symmetry model is given by Benayoun, David, DelBuono, and Jegerlehner (2012). To confirm this interesting proposal, further investigation at the higher $2\pi$ mass region as well as precise experimental tests of the CVC relation for other modes such as $4\pi$, $\omega\pi$ and $\eta\pi\pi$ are important.

### 20.7.2 CVC and $\pi\pi$ branching fraction

The CVC relation allows one to predict the branching fraction of the decay $\tau \to \pi\pi^0\nu_\tau$ ($\mathcal{B}_{\pi\pi}$) in terms of the isovector part of the $e^+e^- \to \pi^+\pi^-$ cross section after taking into account the IB correction:

$$\mathcal{B}_{\pi\pi}^{CVC} = \frac{3}{2}\frac{\mathcal{B}_e|V_{ud}|^2}{\pi\alpha^2 m_\tau^2}\int_{s_{min}}^{m_\tau^2} ds\ s\sigma^0_{\pi^+\pi^-}(s)$$

**Table 20.7.2.** Lowest order hadronic (vacuum polarization) contribution $a_\mu^{\rm had,LO}[\pi\pi,\tau](\times 10^{-10}$ based on all $e^+e^-$ data including recent *BABAR* (Lees, 2012n) and KLOE (Ambrosino et al., 2009a; Babusci et al., 2013) results, and all tau data including recent Belle data (Fujikawa, 2010), obtained by Davier et al. (2010) and Jegerlehner and Szafron (2011).

| | Davier et al. | Jegerlehner and Szafron |
|---|---|---|
| $a_\mu^{\rm had,LO}[ee]$ | $690.9 \pm 5.2$ | $690.8 \pm 4.7$ |
| $a_\mu^{\rm had,LO}[\tau, ee]$ | $705.3 \pm 4.5$ | $691.0 \pm 4.7$ |

$$\times\left(1 - \frac{2}{m_\tau^2}\right)^2\left(1 + \frac{2s}{m_\tau^2}\right)\frac{S_{EW}}{R_{IB}}, \quad (20.7.7)$$

where $s_{min} = (m_{\pi^-} + m_{\pi^0})^2$. The meaning of the other parameters is the same as for Eq. (20.7.5). The result using all $e^+e^-$ data including recent *BABAR* (Lees, 2012n) and KLOE (Ambrosino et al., 2009a; Babusci et al., 2013) results is

$$\mathcal{B}_{\pi\pi}^{CVC} = (24.78 \pm 0.17 \pm 0.22)\%, \quad (20.7.8)$$

while the average of the measured $\mathcal{B}_{\pi\pi}$ is $(25.42 \pm 0.10)\%$ (Davier et al., 2010). The difference is $(0.64 \pm 0.10 \pm 0.28)\%$, which is still substantial, but less significant than the previous results (Davier, Eidelman, Hoecker, and Zhang, 2003b).

If the $\rho - \gamma$ mixing effect is included, the result is $\mathcal{B}_{\pi\pi}^{CVC} = (25.20 \pm 0.17 \pm 0.28)\%$ (Jegerlehner and Szafron, 2011), which is in good agreement with the measured branching fraction.

## 20.8 Measurement of $|V_{us}|$

Here we describe three ways to determine $|V_{us}|$ using tau decays : $\mathcal{B}(\tau^- \to K^-\nu_\tau)$, $\mathcal{B}(\tau^- \to K^-\nu_\tau)/\mathcal{B}(\tau^- \to \pi^-\nu_\tau)$, and the inclusive sum of tau branching fractions having net strangeness of unity in the final state:

1) We use the lattice QCD value of the kaon decay constant $f_K = 157 \pm 2$ MeV (Follana, Davies, Lepage, and Shigemitsu, 2008), and our value of

$$\mathcal{B}(\tau^- \to K^-\nu_\tau) = \frac{G_F^2 f_K^2 |V_{us}|^2 m_\tau^3 \tau_\tau}{16\pi\hbar} \times\left(1 - \frac{m_K^2}{m_\tau^2}\right)^2 S_{EW}, \quad (20.8.1)$$

where $S_{EW} = 1.0201 \pm 0.0003$ (Erler, 2004), to determine $|V_{us}| = 0.2204 \pm 0.0032$ from results of the unitarity constrained fit. This value is consistent with the estimate of $|V_{us}| = 0.2255 \pm 0.0010$ obtained using the unitarity constraint on the first row of the CKM matrix.

2) We use $f_K/f_\pi = 1.189 \pm 0.007$ from lattice QCD (Follana, Davies, Lepage, and Shigemitsu, 2008), $|V_{ud}| =$

$0.97425 \pm 0.00022$ (Hardy and Towner, 2009), and the long-distance correction $\delta_{LD} = (0.03 \pm 0.44)\%$, estimated (Banerjee, 2008) using corrections to $\tau \to h\nu_\tau$ and $h \to \mu\nu_\mu$ (Decker and Finkemeier, 1994, 1995; Marciano, 2004; Marciano and Sirlin, 1993), for the ratio

$$\frac{\mathcal{B}(\tau^- \to K^-\nu_\tau)}{\mathcal{B}(\tau^- \to \pi^-\nu_\tau)} = \frac{f_K^2 |V_{us}|^2}{f_\pi^2 |V_{ud}|^2} \frac{\left(1 - \frac{m_K^2}{m_\tau^2}\right)^2}{\left(1 - \frac{m_\pi^2}{m_\tau^2}\right)^2} (1 + \delta_{LD}), \tag{20.8.2}$$

where short-distance electroweak corrections cancel in this ratio.
From the unitarity constrained fit, we obtain $\mathcal{B}(\tau^- \to K^-\nu_\tau)/\mathcal{B}(\tau^- \to \pi^-\nu_\tau) = 0.0643 \pm 0.0009$, which includes a small correlation (coefficient of $-0.5\%$) between the branching fractions. This yields $|V_{us}| = 0.2229 \pm 0.0021$, which is also consistent with the value of $|V_{us}|$ from the CKM unitarity prediction.

3) The total hadronic width of the $\tau$ normalized to the electronic branching fraction, $R_{\rm had} = \mathcal{B}_{\rm had}/\mathcal{B}_{\rm e}$, can be written as $R_{\rm had} = R_{\rm non-strange} + R_{\rm strange}$. We can then measure

$$|V_{us}| = \sqrt{R_{\rm strange} / \left[\frac{R_{\rm non-strange}}{|V_{ud}|^2} - \delta R_{\rm theory}\right]}. \tag{20.8.3}$$

Here, we use $|V_{ud}| = 0.97425 \pm 0.00022$ (Hardy and Towner, 2009), and $\delta R_{\rm theory} = 0.240 \pm 0.032$ (Gamiz, Jamin, Pich, Prades, and Schwab, 2007) obtained with the updated average value of $m_s(2\,\text{GeV}) = 94 \pm 6$ MeV (Jamin, Oller, and Pich, 2006), which contributes to an error of 0.0010 on $|V_{us}|$. We note that this error is equivalent to half the difference between calculations of $|V_{us}|$ obtained using fixed order perturbation theory and contour improved perturbation theory calculations of $\delta R_{\rm theory}$ (Maltman, 2010), and twice as large as the theoretical error proposed in Gamiz, Jamin, Pich, Prades, and Schwab (2008).
As in Davier, Hoecker, and Zhang (2006), we improve upon the estimate of the electronic branching fraction by averaging its direct measurement with its estimates of $(17.899 \pm 0.040)\%$ and $(17.794 \pm 0.062)\%$ obtained from the averaged values of muonic branching fractions and the averaged value of the lifetime of the tau lepton $= (290.6 \pm 1.0) \times 10^{-15}$ s (Nakamura et al., 2010), assuming lepton universality and taking into account the correlation between the leptonic branching fractions. This gives a more precise estimate for the electronic branching fraction: $\mathcal{B}_{\rm e}^{\rm uni} = (17.852 \pm 0.027)\%$.
Assuming lepton universality, the total hadronic branching fraction can be written as: $\mathcal{B}_{\rm had} = 1 - 1.972558\,\mathcal{B}_{\rm e}^{\rm uni}$, which gives a value for the total $\tau$ hadronic width normalized to the electronic branching fraction as $R_{\rm had} = 3.6291 \pm 0.0086$.
The non-strange width is $R_{\rm non-strange} = R_{\rm had} - R_{\rm strange}$, where the estimate for the strange width $R_{\rm strange} = 0.1613 \pm 0.0028$ is obtained from the sum of the strange branching fractions with the unitarity constrained fit as listed in Table 199 in the HFAG report (Amhis et al., 2012). This gives a value of $|V_{us}| = 0.2174 \pm 0.0022$, which is 3.3 $\sigma$ lower than the CKM unitarity prediction.
A similar estimation using results from the unconstrained fit to the branching fractions gives $|V_{us}| = 0.2166 \pm 0.0023$, which is 3.6 $\sigma$ lower than the CKM unitarity prediction. Since the sum of base modes from our unconstrained fit is less than unity by 1.6 $\sigma$, instead of using $\mathcal{B}_{\rm non-strange} = 1 - \mathcal{B}_{\rm leptonic} - \mathcal{B}_{\rm strange}$, we also evaluate $|V_{us}|$ from the sum of the averaged non-strange branching fractions. This gives $|V_{us}| = 0.2169 \pm 0.0023$, which is 3.5 $\sigma$ lower than the CKM unitarity prediction.

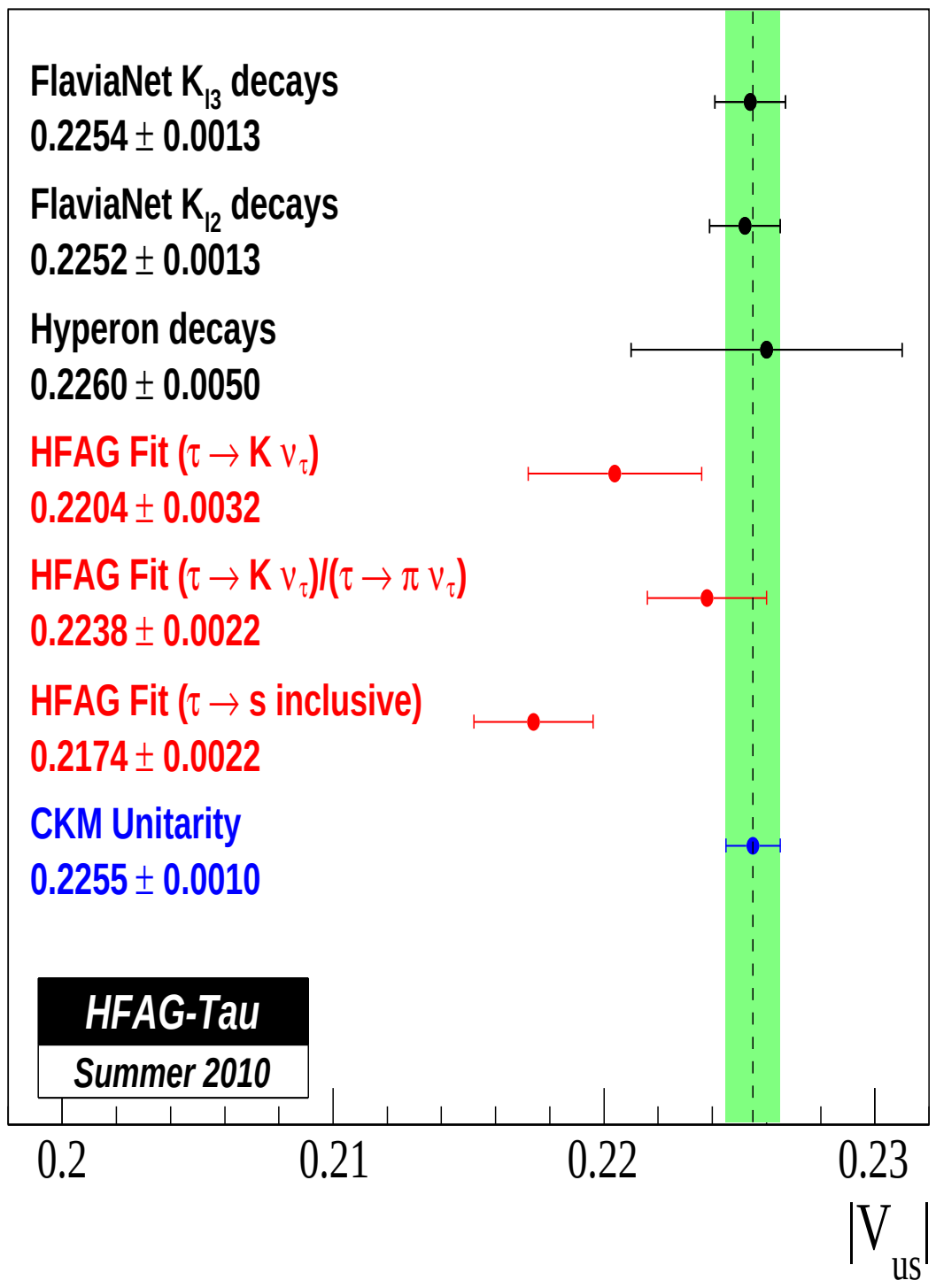

**Figure 20.8.1.** Measurements of $|V_{us}|$ from kaon, hyperon and tau decays (Asner et al., 2010).

A summary of these $|V_{us}|$ values is shown in Figure 20.8.1, where we also include values from kaon decays obtained from Antonelli et al. (2010b) and from hyperon decays obtained from Jamin (2007). The $|V_{us}|$ determination from hyperon decays was done in Mateu and Pich (2005).

## 20.9 Summary of the tau section

In summary, $B$ Factory experiments contribute to various aspects of physics related to tau leptons: The tau mass is measured with an accuracy of 0.15 MeV, and the mass difference of $\tau^+$ and $\tau^-$ is tested at the level of 0.03%. The charged current universality is tested at the level of 0.1 − 0.2 %. Two orders of magnitude more stringent limits are set for 50 lepton flavor violating tau-lepton decays. Measurements of the tau-EDM and tau mass-difference are improved significantly. For the *CP* violation of tau-lepton decays, the *BABAR* result shows a $2.8\sigma$ deviation from the Standard Model prediction for $\tau^- \to K_S^0\pi^-\nu_\tau$, while Belle find no indication of *CP* asymmetry in the decay angular distribution.

The hadronic tau decays provide a beautiful laboratory to measure the CKM matrix $V_{us}$ and to study the strong interactions. The values of $V_{us}$ have been estimated via various methods. The value derived from the inclusive strange branching fraction provides a result somewhat smaller than the one obtained from kaon decays. The branching fractions and the precise spectral functions have been measured for various decay modes. The second-class current has been searched for at the level of $< 4.0 \times 10^{-6}$ in the mode $\tau \to \eta'(958)\pi^-\nu_\tau$.

Yet, there are several on-going analyses, which include precise measurements of the tau-lepton lifetime, Michel parameters and Cabibbo-allowed and Cabibbo-suppressed inclusive spectral functions.

In addition, a set of form factors based on "Resonance Chiral Theory"($R\chi T$) is recently implemented in the `TAUOLA` MC program (Shekhovtsova, Przedzinski, Roig, and Wąs, 2012). The formulae of $R\chi T$ is designed to reproduce chiral perturbation theory at the low energy limit and has a smooth transition to the perturbative QCD results in the high energy region. They rely on the large-$N_C$ expansion of QCD. It is nice since these formulae are designed so that one can fit the data without violating the basic requirements of QCD. Tests of $R\chi T$ and the determination of the model parameters by the data may provide a new insight into the resonance region from QCD (see references in Shekhovtsova, Przedzinski, Roig, and Wąs (2012) for detailed discussion).

In the near future, an additional two orders of magnitude increase in available data sets is expected to be accumulated by future flavor factories.

# Chapter 21
# Initial state radiation studies

***Editors:***
*Fabio Anulli (BABAR)*
*Galina Pakhlova (Belle)*

**Additional section writers:**
*Michel Davier, Vladimir P. Druzhinin, Simon I. Eidelman, Bertrand Echenard, Mathew G. Graham, Simone Pacetti, Antimo Palano, Evgeni P. Solodov, Timofey Uglov, Shuwei Ye*

## 21.1 Introduction

Low energy $e^+e^-$ annihilation are among the most powerful ways to study the nature of hadrons, because of the very clean environment — with the perfectly known initial state and the low multiplicity of the produced final states — as has been shown since the first $e^+e^-$ accumulation ring, ADA, was built in Frascati (Bernardini, Corazza, Ghigo, and Touschek, 1960; Cabibbo and Gatto, 1961).

At low energies, the hadrons observed in the final state come from the hadronization of the original quark pair produced by the $e^+e^-$ annihilation via a single intermediate virtual photon. The process of hadronization is well described by Quantum Chromodynamics (QCD) for a relatively high center-of-mass (CM) energy of the $e^+e^-$ system. However, QCD fails to describe the low energy region, which is characterized by intense final state interactions and rich production of resonant states. Experimental data in this energy region are of fundamental importance both as an input and as validation for the various QCD-based theoretical models of hadronic interactions.

The total cross section of $e^+e^-$ annihilation into hadrons is also the experimental input to the calculation of the hadronic contribution to both the anomalous magnetic moment of the muon and the value of the running fine-structure constant at the $Z^0$ pole. Therefore, it provides high precision tests of the Standard Model and searches for New Physics effects.

Studies of the nature of known light and heavy vector mesons, and searches for new resonant states, can be performed measuring exclusive final states over a wide energy range.

A novel method of studying $e^+e^-$ annihilation using initial state radiation (ISR) has been developed in the last decade at various high-luminosity $e^+e^-$ colliders. Most of the results presented in this chapter rely on the ISR technique (exceptions include $D^{(*)+}D^{(*)-}$ production far from threshold, Section 21.4.1, and dark force searches, Section 21.6). The experimental method, with its theoretical foundations, is described in the next section. Section 21.3 reports on the measurement of a number of light meson final states, with implications for the value of $(g-2)_\mu$ discussed in Section 21.3.2. Measurements of time-like baryon form factors are presented in Section 21.3.7, while Sections 21.4 and 21.5 report on the studies of open-charm production via $e^+e^-$ annihilation and the searches for new, possibly exotic, vector states, respectively. A detailed discussion of the results of these two sections can be found in sections 18.2 and 18.3. Finally, Section 21.6 presents a search for $e^+e^-$ annihilation into multi-lepton final states at a CM energy of $\sim 10.6$ GeV, which could be a manifestation of dark boson decays.

## 21.2 The Initial State Radiation method

The $e^+e^-$ annihilation process is described at lowest order by the Feynman diagram shown in Fig. 21.2.1. The center-of-mass energy squared is given by $s = 4E_b^{*2}$, for a collider operating with beams of CM energy $E_b^*$. Exclusive and total hadronic cross sections as a function of $s$ have usually been obtained by scanning the accessible energy range, and collecting a certain amount of data at each value of the beam energies.

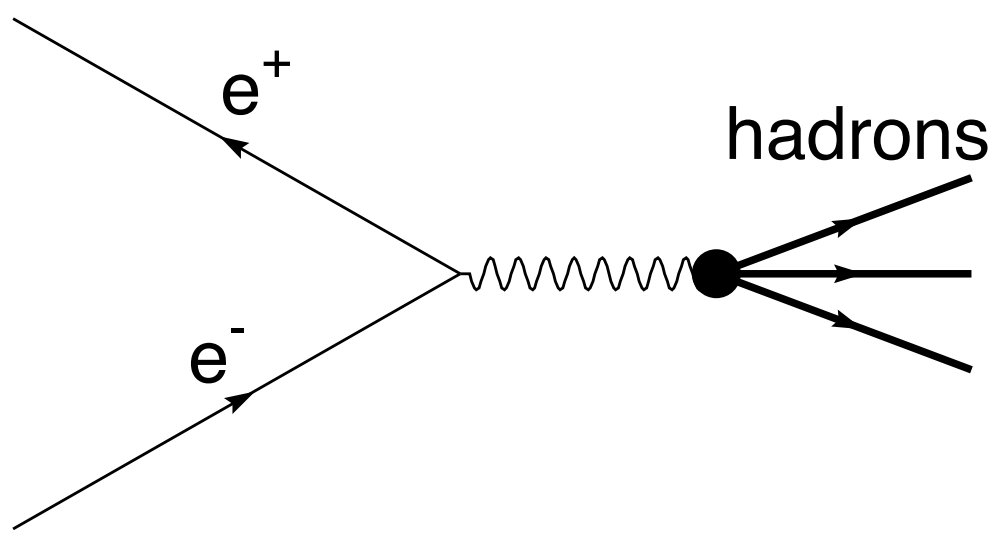

**Figure 21.2.1.** The lowest-order Feynman diagram describing the process of $e^+e^-$ annihilation into hadrons.

However, the colliding electrons can emit one or several photons from the initial state, so that the effective CM energy of the $e^+e^-$ collision can take any value from $m_{\rm th}$, the production threshold of the hadronic system, up to $\sqrt{s} = 2E_b^*$. The process effectively studied is thus $e^+e^- \to f + n\gamma$, $n = 0, 1, 2, ...$, where $f$ is a given final state; the cross section depends on the Born cross section at all energies below $\sqrt{s}$ (Kuraev and Fadin, 1985):

$$\sigma(s) = \int_0^{1-m_{\rm th}^2/s} W(s,x)\,\sigma_0(s(1-x))\,dx\,. \qquad (21.2.1)$$

Here, $x$ is the fraction of the beam energy carried by the photons emitted from the initial state, the radiator function $W(s,x)$ is the photon emission probability density function, which is fully calculable in QED (see *e.g.* Actis et al. (2010) and references therein), and $\sigma_0(s(1-x))$ is the Born cross section for the process $e^+e^- \to f$ at the reduced center-of-mass energy squared

$$s' = s(1-x). \qquad (21.2.2)$$

In energy-scan experiments, the contribution of ISR is normally suppressed by requiring energy and momentum balance between the final hadronic state and the initial $e^+e^-$ state. This limits the fraction of energy carried by radiated photons within the experimental resolution, and Eq.(21.2.1) can then be written as

$$\sigma(s) = \sigma_0(s)\,(1 + \delta(s)), \qquad (21.2.3)$$

where the factor $1 + \delta(s)$ summarizes the QED radiative corrections, which can be as large as 10% for slowly varying cross sections.

On the other hand, the emission of initial state radiation allows the study of $e^+e^-$ annihilation for a continuous spectrum of energies below the nominal beam energy, without changing the operating conditions of the collider, as outlined long ago (Baier and Khoze, 1965; Bonneau and Martin, 1971). This becomes clear when we write the differential form of the cross section

$$\frac{d\sigma(s,x)}{dx} = W(s,x)\,\sigma_0(s(1-x)), \qquad (21.2.4)$$

and note that the reduced CM energy after photon emission is just the invariant mass of the hadronic system: $m = \sqrt{s(1-x)}$. In terms of $m$, the differential cross section becomes

$$\frac{d\sigma(s,m)}{dm} = \frac{2m}{s}\,W(s,m)\,\sigma_0(m). \qquad (21.2.5)$$

It should also be noted that the dominant contribution from ISR processes comes from the diagram shown in Fig. 21.2.2, with a single photon emitted. From the experimental point of view, the Born differential cross section for the process $e^+e^- \to f$ as a function of $m$ is obtained from the measurement of the cross section for $e^+e^- \to \gamma_{\rm ISR} + f$.

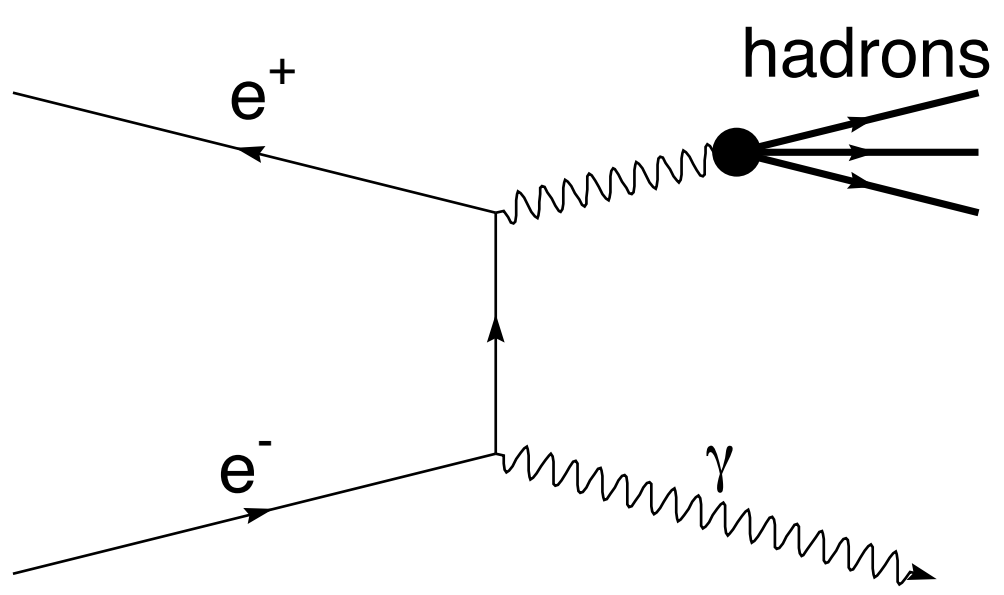

**Figure 21.2.2.** The lowest-order Feynman diagram describing the process of $e^+e^- \to \gamma_{\rm ISR}$ + hadrons.

The experimental method, and the potential for precise measurements of the hadronic cross sections and for low-energy spectroscopy at the forthcoming $\phi$- and $B$ Factories, were discussed in several papers at the end of the 90's: Arbuzov, Kuraev, Merenkov, and Trentadue (1998); Benayoun, Eidelman, Ivanchenko, and Silagadze (1999); Binner, Kühn, and Melnikov (1999); Konchatnij and Merenkov (1999). The high luminosities reached in these colliders provide substantial datasets despite the suppression due to the additional QED vertex in Fig. 21.2.2.

### 21.2.1 Radiator function and Monte Carlo generators

The dependence of the radiator function on the polar angle of the ISR photon with respect to the beam axis in the CM system is given at lowest order by (Bonneau and Martin, 1971)

$$W_0(s,x,\theta) = \frac{\alpha}{\pi x}\left[\frac{(2-2x+x^2)\sin^2\theta - \frac{x^2}{2}\sin^4\theta}{\left(\sin^2\theta + \frac{4m_e^2}{s}\cos^2\theta\right)^2} - \frac{4m_e^2}{s}\frac{(1-2x)\sin^2\theta - x^2\cos^4\theta}{\left(\sin^2\theta + \frac{4m_e^2}{s}\cos^2\theta\right)^2}\right], \qquad (21.2.6)$$

where $\alpha$ is the fine-structure constant, and $m_e$ is the electron mass. The ISR photons are emitted predominantly at small angles, however a significant fraction of them have large angles. In particular, at CM energies $\sqrt{s} \sim 10$ GeV more than 10% of the high-energy ISR photons are emitted within the fiducial volume of the detector. These features provide the basis for two different experimental approaches to the study of ISR processes: The *tagged* approach, with detection of the ISR photon, and the *untagged* one, where the detection of the ISR photon is not explicitly required. These two approaches will be discussed in detail in Section 21.2.4.

The radiator function at lowest order is obtained by integration of Eq.(21.2.6) over the polar angle in the CM frame, in the range appropriate to the experimental situation. In the tagged approach, $\theta_0 < \theta < \pi - \theta_0$, where $\theta_0 \gg m_e/\sqrt{s}$ is chosen to cover the fiducial volume of the electromagnetic calorimeter. Writing $C = \cos\theta_0$, the radiator function becomes

$$W_0(s,x,\theta_0) = \frac{\alpha}{\pi x}\left[(2-2x+x^2)\ln\frac{1+C}{1-C} - x^2 C\right], \qquad (21.2.7)$$

while for an untagged analysis ($0<\theta<\pi$)

$$W_0(s,x) = \frac{\alpha}{\pi x}(2-2x+x^2)\left[\ln\frac{s}{m_e^2} - 1\right]. \qquad (21.2.8)$$

Radiative corrections to $W_0$ are as large as 15% (Kuraev and Fadin, 1985). It is therefore necessary to include higher-order diagrams to reach the desired level of accuracy in the calculation of $W(s,x)$. The study of radiative corrections to ISR processes has been pursued in several theoretical works, some of which have been used as the basis for Monte Carlo (MC) generators to be used in analysis of experimental data.

Two different MC generators are used at the $B$ Factories: `AfkQed` and `PHOKHARA` . The `AfkQed` package is based on the `EVA` event generator (Binner, Kühn, and Melnikov, 1999; Czyz and Kühn, 2001). It was initially designed to simulate $2\pi$ and $4\pi$ production with the ISR photon emitted at large angles. Soft multi-photon emission in the initial state is generated with the structure function technique (Caffo, Czyz, and Remiddi, 1994), while final state radiation (FSR) is generated using the `PHOTOS` simulation package (Barberio, van Eijk, and Wąs, 1991). The AfkQed package provides the generation of a number of hadronic final states, including $\pi^+\pi^-$, $\pi^+\pi^-\pi^0$, $4\pi$, $5\pi$, $6\pi$, and modes with kaons and light baryons. Additional modes can be easily implemented. It also includes the process $e^+e^- \to \mu^+\mu^-\gamma$, for which both ISR and FSR diagrams and their interference are taken into account. This generator is used by *BABAR*, for all the analyses where a tagged ISR photon is required. In order to properly calculate the radiator function the ISR photon is generated in an angular range slightly larger than the acceptance of the electromagnetic calorimeter, typically with $20° < \theta < 160°$. The achieved accuracy, of the order of 1%, is sufficient for all the measured final states, with the exception of the $\pi^+\pi^-$ channel, for which sub-percent precision is required. This particular case will be discussed in detail in Section 21.3.3

The PHOKHARA event generator is based on theoretical work by Rodrigo, Czyz, Kühn, and Szopa (2002) and Czyz, Grzelinska, Kühn, and Rodrigo (2003). The calculations include one-loop corrections and next-to-leading order (NLO) ISR radiative corrections: that is, up to two hard ISR photons are generated. For the processes $e^+e^- \to \pi^+\pi^-\gamma$, $e^+e^- \to K^+K^-\gamma$, and $e^+e^- \to \mu^+\mu^-\gamma$, NLO FSR corrections are also implemented, with interference between ISR and FSR. The accuracy in the determination of the radiator function is estimated to be about 0.5%. The PHOKHARA generator is used for both tagged and untagged ISR studies, and it is particularly appropriate for measurement of the $\pi^+\pi^-$ and $K^+K^-$ final states. It is used in both $B$ Factory experiments, as well as in the KLOE experiment at the $\phi$ factory DAΦNE.

In the latest PHOKHARA version, several multi-hadron final states are implemented. However it should be noted that for these more complex channels the main theoretical uncertainty does not come from the treatment of the radiative corrections, but from the dependence of the matrix element on the hadronic model used to describe the process. Detection efficiencies estimated via MC simulation depend in fact on the angular and momentum distributions generated by the chosen model. In order to determine a systematic uncertainty due to the model dependence, the hadron distributions from MC simulations are reweighted with those from data, and the detection efficiencies resulting from different theoretical models are compared.

### 21.2.2 Cross section

Experimentally, the Born cross section as a function of $m$ for the process $e^+e^- \to f$, $\sigma_0(m)$, is obtained from the measured mass spectrum of the corresponding ISR process $e^+e^- \to \gamma_{\rm ISR} f$, taking into account the detection efficiency $\varepsilon(s,m)$, and the integrated luminosity $\mathcal{L}$:

$$\frac{dN(s,m)}{dm} = \varepsilon(s,m)\,\frac{d\sigma(s,m)}{dm}\,\mathcal{L}. \tag{21.2.9}$$

Replacing the differential cross section by the expression in Eq. (21.2.5), we obtain

$$\frac{dN(s,m)}{dm} = \varepsilon(s,m)\,(1+\delta_r(s,m))\,\sigma_0(m)\,\frac{d\mathcal{L}_{\rm ISR}(m)}{dm}, \tag{21.2.10}$$

where $1+\delta_r(s,m) = W(s,m)/W_0(s,m)$ is the radiative correction factor mentioned in the previous subsection, and we have introduced the so-called effective ISR differential luminosity:

$$\frac{d\mathcal{L}_{\rm ISR}(m)}{dm} \;=\; \frac{2m}{s}\,W_0(s,m)\,\mathcal{L}. \tag{21.2.11}$$

The Born radiator function is given by Eq.(21.2.7) or Eq.(21.2.8) depending on the experimental conditions.

The mass spectrum is then subdivided in small mass bins of width $\Delta m$, within which both $W$ and $\sigma_0$ vary little, and the cross section is extracted for each bin from the number of events $\Delta N$ falling in that bin:

$$\sigma_0(m_i) \;=\; \frac{\Delta N(m_i)}{\Delta m}\,\frac{1}{\varepsilon(s,m_i)\,(1+\delta_r)\,d\mathcal{L}_{\rm ISR}(m_i)/dm}. \tag{21.2.12}$$

The ISR luminosity is the quantity to be compared with the luminosity integrated by previous experiments via conventional energy scan. The solid line in Fig. 21.2.3 reports the mass dependence of the ISR differential luminosity calculated at an $e^+e^-$ CM energy of 10.58 GeV, for a tagged analysis, assuming a typical acceptance of 10% and the integrated luminosity of the *BABAR* data set of 470 fb$^{-1}$. The ISR luminosity, calculated considering an energy-bin width of 0.02 GeV, increases from about 0.2 pb$^{-1}$ at the $\pi^+\pi^-$ production threshold up to more than 3 pb$^{-1}$ at 3.5 GeV. Only part of this range has been covered by energy scans at previous experiments. In particular the region below $\sqrt{s} = 1.4$ GeV has been investigated with high precision by the SND and CMD-2 experiments at the VEPP-2M collider in Novosibirsk. They have collected a sample similar in size to that available at *BABAR*, in particular in the regions around the peaks of the vector meson resonances. There is much less data available at energies above 1.4 GeV, most of it collected by the experiments DM1 and DM2 at the DCI collider in Orsay. The histograms in the figure report the luminosity integrated by CMD-2 and DM2. Precise measurements of the $e^+e^- \to \pi^+\pi^-$ cross section have also been performed by the KLOE experiment by means of both ISR untagged and tagged analyses. Information about the data collected by energy scan experiments has been extracted from several published papers, whose main results are reported in comparison with $B$ Factories results in the following sections.

In the mass region of charm production the ISR luminosity ranges from tens to hundreds of pb$^{-1}$ per 100 MeV/$c^2$ wide mass bin, significantly exceeding the integrated luminosity collected by direct $e^+e^-$ experiments. By comparison the recent CLEO-c (Cronin-Hennessy et al., 2009) energy scan collected a total of 60 pb$^{-1}$ in twelve points between 3.97 and 4.26 GeV, and BESIII ac-cumulated 53 $\text{pb}^{-1}$ data at 3.900 GeV and 482 $\text{pb}^{-1}$ data at 4.009 GeV (Ablikim et al., 2013b).

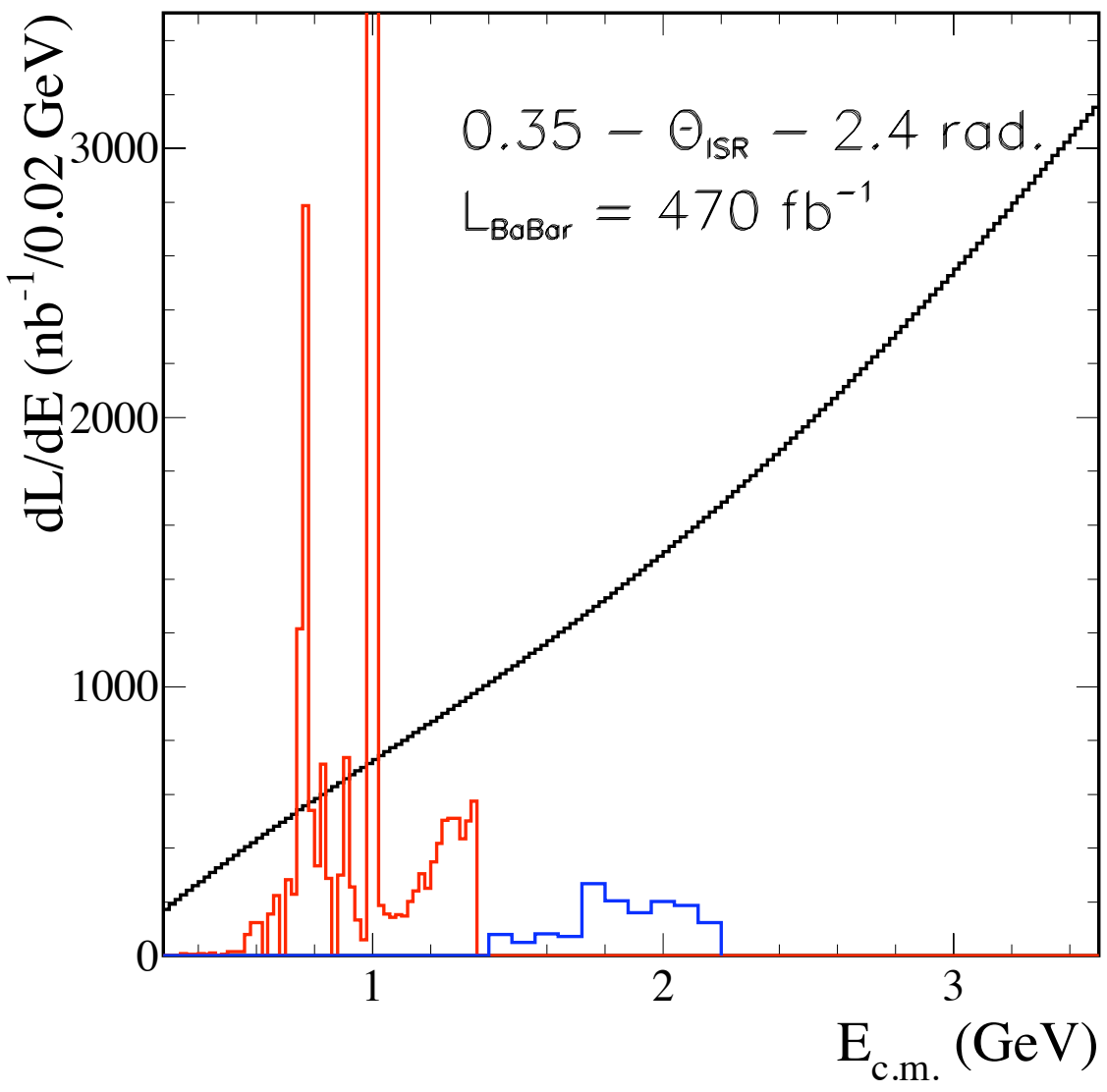

**Figure 21.2.3.** *BABAR* ISR luminosity versus equivalent CM energy calculated in energy bins 20 MeV wide, for an integrated machine luminosity of 470 $\text{fb}^{-1}$ and an angular acceptance for the ISR photon between 0.35 and 2.4 rad. For comparison, the histogram reports the luminosity integrated by the CMD-2 and DM2 experiment with a conventional energy scan, at energies below and above 1.4 GeV, respectively; these data are presented in many separate papers, as discussed in the text. The two high, narrow peaks correspond to the energy regions around the mass of the $\rho$ and $\phi$ resonances, where most of the data were collected.

In conclusion, the current data samples of ISR events available at the $B$ Factories are larger than those produced directly in $e^+e^-$ collisions for all masses with the exceptions of the regions around the narrow resonances ($\omega$, $\phi$, $J/\psi$, and $\psi(2S)$). In particular, the data in the mass regions above 1.4 $\text{GeV}/c^2$ for light quarks, and in the charm region, are unique both in terms of quantity and quality.

### 21.2.3 Mass resolution and energy scale

Mass resolution and absolute energy scale have to be kept under control in order to assess the accuracy of the cross sections measured with the ISR method.

The mass resolution is determined by the precision of the measurement of the parameters (angles and momenta) of the reconstructed tracks, and of the energy and direction of the photons from $\pi^0$ and $\eta$ decays. It is therefore expected that the mass resolution is best in the vicinity of the production threshold, due to the lower particle momenta, and degrades with increasing mass. The mass resolution is measured with simulated events, fitting the distribution of the difference between the reconstructed and generated invariant masses. It is then checked with experimental data by fitting the line shape of narrow resonances, such as the $\phi$ or the $J/\psi$.

For multi-hadron systems with only charged particles, typical values of the invariant mass resolution obtained at the $B$ Factories range from 4 up to 7 $\text{MeV}/c^2$, when the mass increases from 1.5 to 3 $\text{GeV}/c^2$. The presence of neutral pions worsens the resolution by a few $\text{MeV}/c^2$. The width of the mass bin chosen for the majority of the analyses in this energy region is 25 $\text{MeV}/c^2$, reducing in this way the effect of the mass resolution on the measured mass spectrum. A bin size of only 2 $\text{MeV}/c^2$ has been used by *BABAR* for the $\pi^+\pi^-$ final state close to the peak of the $\rho$ meson (Aubert, 2009ah), requiring a specific procedure for unfolding the resolution effects from the $\pi^+\pi^-$ mass-spectrum (Malaescu, 2009). A small bin size of 5 $\text{MeV}/c^2$ was also used for the process $e^+e^- \to p\overline{p}\gamma$ in the $p\overline{p}$ mass region close to the production threshold, where the resolution is less than 2 $\text{MeV}/c^2$, as shown in Fig. 21.2.4, allowing the study of the proton electromagnetic form factor with unprecedented accuracy (Aubert, 2005ah; see also the discussion in Section 21.3.7.3 below). In the charm mass region, the final-state hadrons have smaller momenta, and the mass resolution is of the order of 5 $\text{MeV}/c^2$, which is much smaller than the typical bin sizes used of 20–25 $\text{MeV}/c^2$.

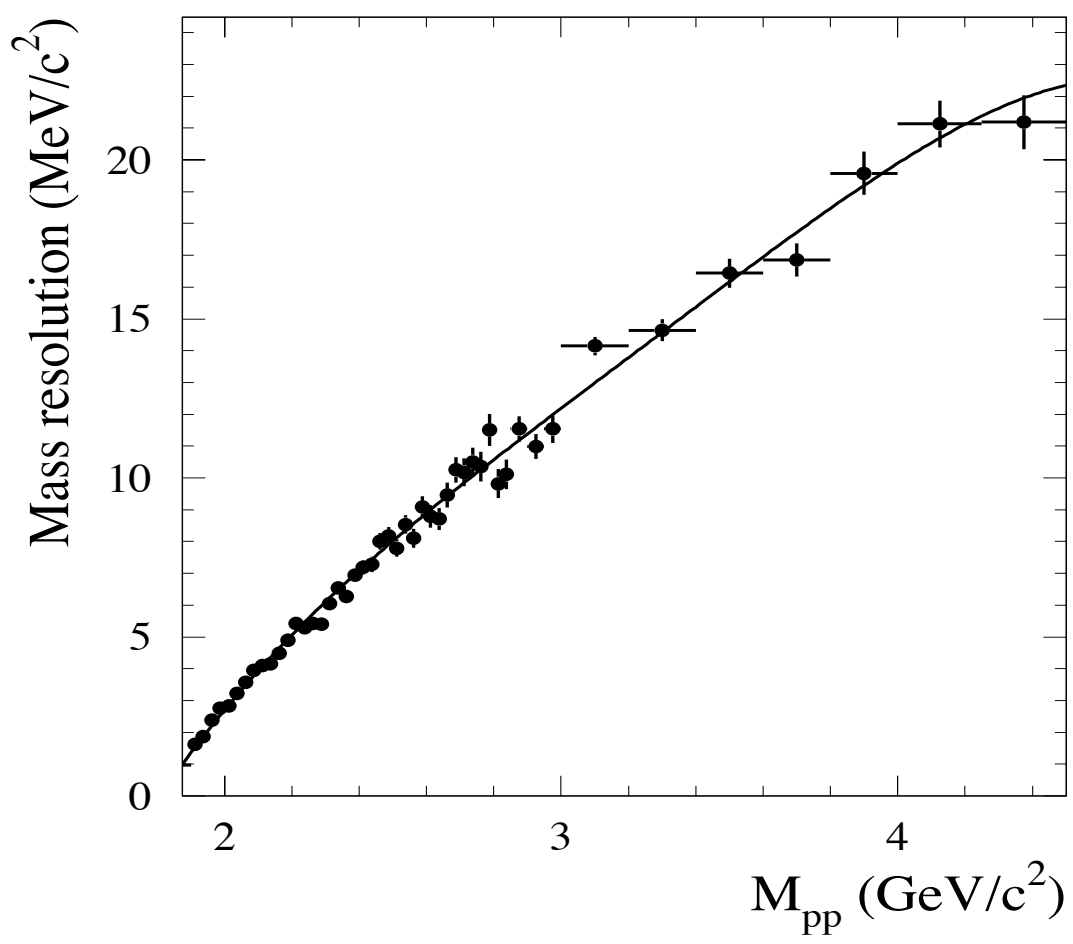

**Figure 21.2.4.** Mass resolution of the $p\overline{p}$ system as a function of the reconstructed $p\overline{p}$ mass obtained by *BABAR* for the process $e^+e^- \to p\overline{p}\gamma_{\text{ISR}}$. *BABAR* internal, prepared for (Aubert, 2006d) analysis.

The absolute mass scale is calibrated by comparison of the reconstructed mass values for known resonances with their nominal peak positions. In all cases a relative accuracy significantly better than $10^{-3}$ has been measured.

### 21.2.4 Comparison of tagged and untagged ISR measurements with direct $e^+e^-$ measurements

As already outlined, analyses of processes with hard photon emission in the initial state can be performed with or without detection of the ISR photon. In this section we discuss the general features of the two approaches, and compare them to direct $e^+e^-$ measurements.

One of the main issues with measurements of exclusive cross sections at $e^+e^-$ experiments is that each collider is able to scan only a limited range of center-of-mass energies. Significant normalization uncertainties are therefore present when data from different experiments, or even data from the same experiment at different energies, are combined. In the ISR technique, by contrast, exclusive cross sections are measured simultaneously over a continuous and very wide range of energies, with the same experimental conditions.

The requirement that the ISR photon is emitted within the detector angular acceptance results in an efficiency loss of about an order of magnitude relative to the full ISR production. In some cases, tagging of the ISR photon allows for analyses with partial reconstruction of the hadronic system, increasing the global detection efficiency while keeping the background at an affordable level.

In untagged analyses, the detection of the ISR photon is not required, while the hadronic system must be fully reconstructed. The detection efficiency is typically higher in this case with respect to the tagged analyses; however this is not true for all experimental conditions. In particular, for low invariant masses the hadronic system is subject to a strong boost, and the hadrons are produced in a narrow cone centered around the direction opposite to the ISR photon momentum. As a consequence, most of the events with the ISR photon emitted roughly collinear with the beam axis are rejected also in the case of an untagged analysis, because a fraction of the hadrons falls outside the detector acceptances. The overall detection efficiencies are therefore very similar for the tagged and untagged approach up to an invariant mass of about $3-3.5\,\mathrm{GeV}/c^2$, where the final state hadrons are emitted at large enough angles to be within the angular acceptance of the calorimeter and tracking system, and the small-angle ISR begins to contribute significantly.

The previous considerations about detection efficiencies constitute the main reason why, at the $B$ Factories, the untagged approach is used for measurements of exclusive cross sections of hadronic final states with an invariant mass above 3.5 $\mathrm{GeV}/c^2$, in particular to study the production of open charm and charmonium. By contrast, all the studies of $e^+e^-$ annihilation into light hadrons are performed requiring the detection of the ISR photon.

In ISR events, the final state hadrons have a measurable momentum even at production threshold, because of the boost of the hadronic system recoiling against the radiated high-energy photon. As a consequence, the detection efficiency differs from zero at threshold and, generally, varies smoothly over the whole measured range of invariant masses, unlike in direct $e^+e^-$ measurements where the detection efficiency drops to zero close to threshold. As an example, Fig. 21.2.5 shows the detection efficiency as a function of the $p\bar{p}$ mass, obtained at *BABAR* for the process $e^+e^- \to p\bar{p}\gamma_{\rm ISR}$ with the photon detected (Aubert, 2006d).

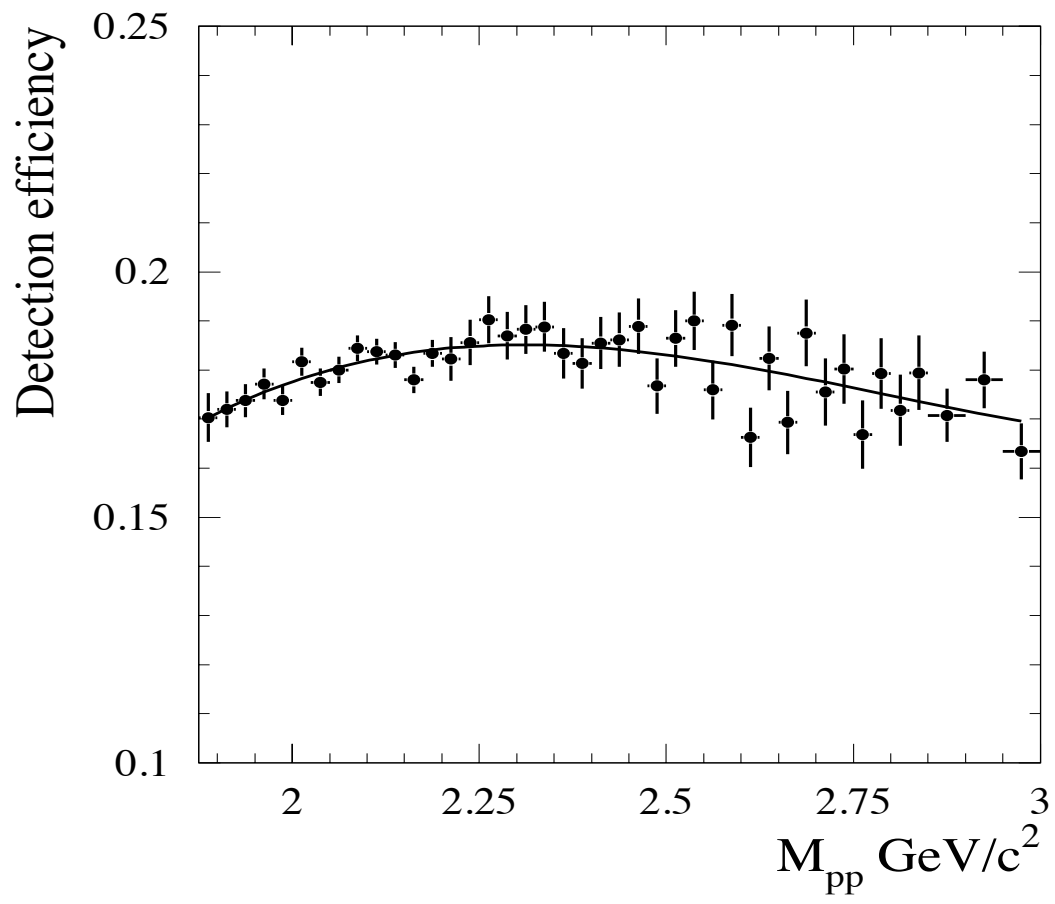

**Figure 21.2.5.** The detection efficiency, as measured by *BABAR* for the process $e^+e^- \to p\bar{p}\gamma_{\rm ISR}$ in a tagged analysis, varies slowly as a function of the reconstructed $p\bar{p}$ mass (Aubert, 2006d).

As we have already observed, for tagged ISR analyses the hadrons, produced in a cone around the direction opposite to the tagged photon, generally fall in an instrumented region of the detector. The detection efficiency is thus only weakly dependent on the final state hadrons' angular distribution in the reference frame where the hadronic system is at rest, and the uncertainties related to the theoretical models used for simulation are significantly reduced. This is in contrast with both untagged ISR analyses and direct $e^+e^-$ measurements, for which the region at small polar angles is largely inaccessible.

There are also prices to pay for using the ISR technique. As discussed in Section 21.2.3, a good mass resolution and absolute mass scale are obtained in ISR analyses, but in direct $e^+e^-$ measurements these quantities are given respectively by the beam energy spread and by the beam energy setting, which are determined far more precisely.

The sources of background events for ISR measurements are significantly larger than those for direct $e^+e^-$ measurements. In the latter case, the main backgrounds to a given final state come from other $e^+e^- \to hadrons$ reactions, due to undetected low momentum particles or to wrong particle identification, but such backgrounds are limited by the requirement of four-momentum conservation. For ISR events, a source of background of the same kind is due to mis-reconstructed events from other ISR processes where one or more particles escape detection. The constraint of four-momentum conservation is much less effective in this case because of the relatively poor resolution on the measurement of the energy of the ISR photon and the emission of secondary photons, which degrade the kinematic fits in both the tagged and untagged analyses.

A background source affecting mainly the tagged analyses comes from $e^+e^-$ annihilations at full energy (that is, without the emission of an ISR photon) but containing a high-energy $\pi^0$. If the $\pi^0$ is not correctly reconstructed because the two decay-photons are not resolved and therefore *merged* by the reconstruction algorithm, or one of them is undetected, the process $e^+e^- \to X\pi^0$ can mimic the ISR process $e^+e^- \to X\gamma_{\rm ISR}$. The contribution of this background to distributions needed in an analysis is estimated for each process from generic light quark continuum MC samples using for normalization a sample of $e^+e^- \to X\pi^0$ with a reconstructed $\pi^0$ selected from data. The background contribution is then subtracted. This is the dominant source of background at masses of the hadronic system higher than about 2 GeV/$c^2$, and limits the measurable mass range for light hadron final states to $m < 4.0$–$4.5$ GeV/$c^2$.

For untagged analyses, the background due to non-ISR events only partially reconstructed can be suppressed by requiring that the missing momentum of the event is collinear with the beam axis. Another significant source of background for untagged analysis is due to two-photon processes $e^+e^- \to e^+e^-\gamma^*\gamma^* \to e^+e^-X$, where the colliding electron and positron are scattered predominantly at small angles, and are therefore undetected (see Chapter 22 for a discussion of the two-photon reactions). In such events the missing momentum is roughly along the beam axis, but the missing mass is large, so effective suppression of this kind of background is obtained requiring a missing mass close to zero, as expected if the only missing particle is the hard ISR photon.

## 21.3 Exclusive hadronic cross-sections

The precision of the Standard Model calculation of the muon anomalous magnetic moment is limited by the uncertainty on the hadronic contribution; for many years this contribution was determined using only data from $e^+e^-$ scan experiments (Davier, Eidelman, Hoecker, and Zhang, 2003b). The discrepancy between $(g-2)_\mu$ calculations and the direct measurement by the E821 experiment (Bennett et al., 2006), on the order of three standard deviations, called for new and more precise measurements of the $e^+e^-$ hadronic cross section.

This has been the main physics motivation for the intensive *BABAR* program of measurements of exclusive $e^+e^-$ annihilation to light-quark hadrons, using ISR. In addition, the large data sample and good detector performance enable spectroscopic studies of unprecedented accuracy in the energy region below 3 GeV.

Many final states have already been studied at *BABAR*: from $\pi^+\pi^-$ (Aubert, 2009ah), the most important channel for $(g-2)_\mu$, to almost all of the possible channels with up to six hadrons in the final state. Exclusive production $e^+e^- \to \mathfrak{B}\overline{\mathfrak{B}}$ (where $\mathfrak{B} = p, \Lambda, \Sigma$) has also been measured in order to extract the time-like electromagnetic form factors of the corresponding baryons. Belle investigated the $\phi\pi^+\pi^-$ ($\phi \to K^+K^-$) final state (Shen, 2009). A few more states are under study at the time of writing of this book, which are essential to complete the main part of the program for the estimate of the hadronic contribution to the value of $(g-2)_\mu$, namely $\pi^+\pi^-\ \pi^0\ \pi^0$, $K^+K^-$, $K^0_SK^0_L+n\pi$, and $K^0_SK^\pm\pi^\mp + n\pi^0$, with $n = 0, 1, 2$.

This section is organized as follows. The general features common to most of these analyses are described in Section 21.3.1. The hadronic contribution to $(g-2)_\mu$, the measurement of the $\pi^+\pi^-$ cross-section, and the impact of this and other ISR results on $(g-2)_\mu$ and $\alpha(M_Z)$ are discussed in Sections 21.3.2–21.3.4 respectively. Light meson spectroscopy results from the study of multi-hadron final states are presented in Section 21.3.5; the search for the $f_J(2220)$ is discussed separately in Section 21.3.6. Finally, time-like baryon form factor measurements are described in Section 21.3.7.

### 21.3.1 Common analysis strategy

All the aforementioned analyses are performed with the ISR tagged approach. A loose pre-selection is applied to filter out ISR candidate events: The ISR event is tagged by the detection of a photon of CM energy $E^*_\gamma > 3$ GeV. A rough balance between the beam energies and the energy of the reconstructed event, and a well reconstructed primary vertex from the charged tracks are required. Additional photons are considered only if they have an energy above 0.03 GeV. In any case, the photon with highest energy is assumed to be the ISR photon. The above pre-selection works for most of the final states with a few exceptions, such as processes with long-lived particles or with only neutral particles in the final state, for which a dedicated selection has been implemented.

Each candidate event is then subject to a set of constrained kinematic fits, under different hypotheses for the final state. The fit results, along with information on charged-particle identification, are used to both select the final states of interest and measure backgrounds from other processes. The kinematic fits use the ISR photon direction and energy along with the four-momenta and covariance matrices of the colliding electrons and of the selected tracks and photons in the final state. Masses of narrow resonances, such as $\pi^0$, $\eta$ and $\phi$ mesons, are constrained in the fit to their nominal values.

In general, the main background at low invariant masses comes from other ISR processes; at higher masses the background is due to continuum $q\overline{q}$ production, as explained in Section 21.2.4 above.

### 21.3.2 Hadronic vacuum polarization

Here we briefly describe polarization of the vacuum due to fluctuations (Section 21.3.2.1), and its effects on the running of the electromagnetic coupling (Section 21.3.2.2) and on the muon magnetic anomaly (Section 21.3.2.3). The hadronic cross section in $e^+e^-$ annihilation is a crucial input to the calculation of both quantities. The measurement of the $e^+e^- \to \pi^+\pi^-$ cross section, and the effect of this and other ISR exclusive results on $(g-2)_\mu$ and $\alpha(M_Z)$, are then discussed in Sections 21.3.3 and 21.3.4 respectively.

#### 21.3.2.1 Quantum fluctuations

A virtual photon exchanged in an electromagnetic process can fluctuate into particle-antiparticle pairs leading to a polarization of the vacuum. While the effect of lepton pairs can be readily calculated with QED, hadronic effects can be treated with QCD only at large energies (quark-antiquark pairs). At low energies perturbative QCD cannot be employed any more, but fortunately hadronic vacuum polarization can still be evaluated with a dispersion integral (Bouchiat and Michel, 1961) involving experimental data on the cross section for $e^+e^- \to$ *hadrons*, usually expressed in terms of its ratio $R$ to the point like cross section. This technique applies to two situations of great importance in particle physics: the running of the electromagnetic coupling $\alpha(s)$, particularly its value at the $Z$ pole where precision tests of the electroweak physics are performed, and the calculation of the Standard Model prediction for the lepton magnetic anomaly, especially in the case of the muon because of its sensitivity to new physics.

The conservation of the vector current (CVC) allows one to use $\tau$ decay data to compute the dispersion integral (Alemany, Davier, and Hoecker, 1998), but in this case small corrections must be applied in order to take into account isospin symmetry breaking between the weak charged and electromagnetic hadronic currents. This subject is discussed in Section 20.7.

#### 21.3.2.2 The running of the electromagnetic coupling

The running of the electromagnetic fine structure constant $\alpha(s)$ is governed by the renormalized vacuum polarization function, $\Pi_\gamma(s)$. For the spin 1 photon, $\Pi_\gamma(s)$ is given by the Fourier transform of the time-ordered product of the electromagnetic currents $j^\mu_{\rm em}(s)$ in the vacuum:

$$(q^\mu q^\nu - q^2 g^{\mu\nu})\,\Pi_\gamma(q^2) = i\int d^4x\, e^{iqx}\langle 0|T(j^\mu_{\rm em}(x) j^\nu_{\rm em}(0))|0\rangle. \tag{21.3.1}$$

With $\Delta\alpha(s) = -4\pi\alpha\,{\rm Re}\,[\Pi_\gamma(s) - \Pi_\gamma(0)]$ and $\Delta\alpha(s) = \Delta\alpha_{\rm lep}(s) + \Delta\alpha_{\rm had}(s)$, which subdivides the running contributions into a leptonic and a hadronic part, one has

$$\alpha(s) = \frac{\alpha(0)}{1 - \Delta\alpha_{\rm lep}(s) - \Delta\alpha_{\rm had}(s)}\ , \tag{21.3.2}$$

where $4\pi\alpha(0)$ is the square of the electron charge in the long-wavelength Thomson limit.

The leptonic contribution at $s = M_Z^2$ is known precisely at three-loop order (Steinhauser, 1998): $\Delta\alpha_{\rm lep}(M_Z^2) = 314.98 \times 10^{-4}$. Using analyticity and unitarity, the dispersion integral for the contribution from the hadronic vacuum polarization reads

$$\Delta\alpha_{\rm had}(M_Z^2) = -\frac{\alpha(0)M_Z^2}{3\pi}\,{\rm Re}\int_{4m_\pi^2}^{\infty} ds\, \frac{R(s)}{s(s-M_Z^2) - i\epsilon}\ , \tag{21.3.3}$$

and, employing the identity $1/(x'-x-i\epsilon)_{\epsilon\to 0} = {\rm P}\{1/(x'-x)\} + i\pi\delta(x'-x)$, the above integral is evaluated using the principal value integration technique. Here, $R(s) \equiv R^{(0)}(s)$, denotes the ratio of the 'bare' cross section for $e^+e^-$ annihilation into hadrons to the point like muon-pair cross section. The 'bare' cross section, $\sigma^{(0)}(s) = \sigma(s)\,[\alpha(0)/\alpha(s)]$, is defined as the measured cross section with vacuum polarization effects from the photon propagator removed.

#### 21.3.2.3 The muon magnetic anomaly

A prediction from Dirac theory is that charged leptons have a magnetic moment equal to the Bohr magneton $(e\hbar/2m)$, corresponding to a gyromagnetic ratio $g = 2$. However the magnetic anomaly, defined as $a = (g-2)/2$, deviates from zero because of virtual corrections from higher orders in QED and from other interactions. It is convenient to separate the Standard Model prediction for the anomalous magnetic moment of the muon $a_\mu^{\rm SM}$ into its different contributions,

$$a_\mu^{\rm SM} = a_\mu^{\rm QED} + a_\mu^{\rm had} + a_\mu^{\rm weak}\ ; \tag{21.3.4}$$

the hadronic term is dominated by lowest-order (LO) vacuum polarization (VP), whose Feynman diagram is shown in Fig. 21.3.1. But $a_\mu^{\rm had}$ receives further contributions, smaller by a factor $\sim 30$, from higher-order (HO) vacuum polarization (several loops in photon propagators including at least one hadronic loop) and the so-called light-by-light (LBL) contribution, where the hadronic loop is connected to the QED part by four photon legs.

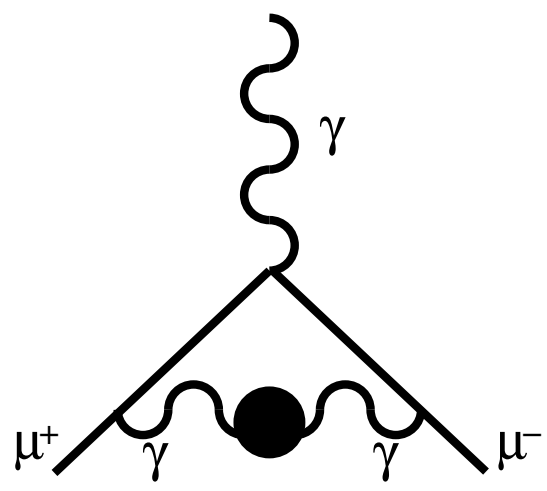

**Figure 21.3.1.** The Feynman diagram for the lowest-order contribution of hadronic vacuum polarization to the muon magnetic anomaly.

As in the case of $\alpha(M_Z)$, by virtue of the analyticity of the vacuum polarization correlator, the LO contribution of the hadronic vacuum polarization to $a_\mu$ can be calculated via a dispersion integral

$$a_\mu^{\rm had,LO} = \frac{\alpha^2(0)}{3\pi^2}\int_{4m_\pi^2}^{\infty} ds\, \frac{K(s)}{s} R(s)\ . \tag{21.3.5}$$

The QED kernel, $K(s)$ is given by (Brodsky and De Rafael, 1968)

$$K(s) = x^2\left(1-\frac{x^2}{2}\right) + (1+x)^2\left(1+\frac{1}{x^2}\right) \times\left(\ln(1+x) - x + \frac{x^2}{2}\right) + \frac{(1+x)}{(1-x)}x^2 \ln x \,, \tag{21.3.6}$$

with $x = (1-\beta_\mu)/(1+\beta_\mu)$ and $\beta_\mu = (1-4m_\mu^2/s)^{1/2}$. The function $K(s)$ decreases monotonically with increasing $s$. It gives a strong weight to the low energy part of the integral (21.3.5). About 92% of the total contribution to $a_\mu^{\rm had}$ is accumulated at CM energies $\sqrt{s}$ below 1.8 GeV, and 73% of $a_\mu^{\rm had}$ is covered by the two-pion final state which is dominated by the $\rho(770)$ resonance.

### 21.3.3 Measurement of $e^+e^- \to \pi^+\pi^-(\gamma)$

Precise results on the $e^+e^- \to \pi^+\pi^-(\gamma)$ cross section have been obtained by *BABAR* (Aubert, 2009ah), using 232 fb$^{-1}$ of recorded data. In this analysis, two-body ISR processes $e^+e^- \to \gamma_{\rm ISR}X$ with final states $X = \pi^+\pi^-(\gamma)$ and $X = \mu^+\mu^-(\gamma)$ are measured, where the charged particle pair can be accompanied by a final state radiation (FSR) photon. The $\pi\pi$ cross section is obtained from the ratio of pion to muon yields, thereby significantly reducing the systematic uncertainty. Furthermore the measured muon cross section can be compared to the QED prediction, providing a powerful cross check of the analysis.

In this approach the measurement of the cross section $\sigma_{\pi\pi(\gamma)}$ uses the effective ISR luminosity provided by the measured mass spectrum of $\mu\mu\gamma_{\rm ISR}(\gamma)$ events. For the muon QED test, the measurement of $\sigma_{\mu\mu(\gamma)}$ uses the ISR luminosity calculated from the $e^+e^-$ integrated luminosity and the radiator function obtained from `PHOKHARA` .

In addition to the pre-selection requirements, two-body ISR events are selected requiring exactly two tracks of opposite charge, each with a momentum $p > 1\,{\rm GeV}/c$ and within the polar angle range 0.40 to 2.45 rad in the laboratory frame. The charged-particle tracks are required to have at least 15 hits in the DCH, to originate within 5 mm of the collision axis, and to extrapolate to DIRC and IFR active areas, excluding low-efficiency regions.

MC simulation is used to compute acceptance and mass-dependent efficiencies for trigger, reconstruction, PID, and event selection. Corrections for differences between data and MC efficiencies amount to at most a few percent and are known to the few permill level or better.

The precision aimed for by this analysis requires dedicated studies of the detector performance, particularly regarding track reconstruction and particle identification efficiencies. These are determined taking advantage of the kinematic constraints of pair production. Two-prong ISR candidates are selected on the basis of the ISR photon and one detected track, and subjected to a kinematic fit to estimate the expected parameters of the second track. Comparison with the sample of reconstructed second-track candidates allows the measurement of the track reconstruction efficiency.

Pure samples of muon, pion, and kaon pairs are obtained from two-prong ISR events where one track is selected as a $\mu$-, $\pi$-, or $K$-candidate respectively, according to the output of cut-based and likelihood selectors (see Section 5.2). The other track is used to determine the efficiency and misidentification probabilities of the PID algorithm under test, as a function of momentum and position in the IFR or the DIRC. The efficiencies for $\mu$ are of the order of 90%, with 10% misidentified as $\pi$. The $\pi$ efficiency depends strongly on momentum, with the fraction of pions misidentified as $K$ increasing from 1% at $1\,{\rm GeV}/c$ to 20% at $6\,{\rm GeV}/c$; the fraction misidentified as $\mu$ is 5–6%, and that as $e$ around 2%.

The multi-hadronic background from $e^+e^- \to q\overline{q}$ is estimated as explained in Section 21.2.4. The background from other ISR processes is dominated by the $e^+e^- \to \pi^+\pi^-\pi^0\gamma_{\rm ISR}$ and $e^+e^- \to \pi^+\pi^-2\pi^0\gamma_{\rm ISR}$ reactions, which are estimated using MC simulation. Residual background sources from the $e^+e^- \to \gamma\gamma$ process with photon conversion and radiative Bhabha events are studied and taken into account.

The analysis allows for one additional ISR or FSR photon, and is thus effectively performed at NLO in $\alpha$. Each event is subjected to two kinematic fits to the $e^+e^- \to X\gamma_{\rm ISR}$ hypothesis, where $X$ allows for an additional photon from the initial or final state, detected or not. The first fit, called the 'ISR' fit, tests the consistency of the reconstructed event with the presence of an undetected ISR photon collinear with the collision axis. The second fit, called the 'FSR' fit, is performed only if an additional photon with $E_\gamma > 25\,{\rm MeV}$ is detected, and tests the hypothesis that this photon is radiated by the final state tracks. Misreconstructed events and residual background generally have large $\chi^2$ values for both fits, and can be separated from signal events. If the 'FSR' fit has the smaller $\chi^2$, the mass of the hadronic final state is calculated including the 4-momentum of the additional detected photon.

The computed detector acceptance and the selection efficiency of the kinematic fit procedure are sensitive to an imperfect description of radiative effects in the generator. The estimated rates of FSR for data and MC simulation are found to be consistent at better than the one permill level. The `AfkQed` generator simulates additional ISR production in the collinear approximation and applies an energy cut-off for very hard photons. The effects of these approximations are estimated by comparing the `AfkQed` generator at four-vector level with the `PHOKHARA` generator, which gives a full description of the process at next-to-leading order. Corrections to the acceptance of the order of a few percent are found for the individual $\pi^+\pi^-(\gamma)$ and $\mu^+\mu^-(\gamma)$ processes, but since photon emission from the initial state is common to the two channels, the $\pi^+\pi^-(\gamma)/\mu^+\mu^-(\gamma)$ ratio is affected only at the few permill level. Therefore the measurement of the pion cross section is to a large extent insensitive to the description of NLO effects in the generator.

A QED test is performed by comparing the $\mu^+\mu^-(\gamma)$ mass spectrum in data with that in MC-simulated events. In particular, the distribution of the data is background-subtracted, and the distribution of the AfkQed-based full simulation, normalized to the data luminosity, is corrected for all known data/MC detector and reconstruction differences and for the generator NLO limitations determined from the comparison between PHOKHARA and AfkQed. The ratio, shown in Fig. 21.3.2, is rather flat from threshold to $3\,\mathrm{GeV}/c^2$ and consistent with unity, as found by a fit to a constant value which returns

$$\frac{\sigma^{\rm data}_{\mu\mu\gamma(\gamma)}}{\sigma^{\rm NLO\ QED}_{\mu\mu\gamma(\gamma)}} - 1 = (40 \pm 20 \pm 55 \pm 94) \times 10^{-4}\ , \quad (21.3.7)$$

with $\chi^2/n_{\rm dof} = 55.4/54$; the errors are statistical, systematic from this analysis, and systematic from the integrated luminosity respectively. The QED test is thus satisfied within an overall accuracy of 1.1%.

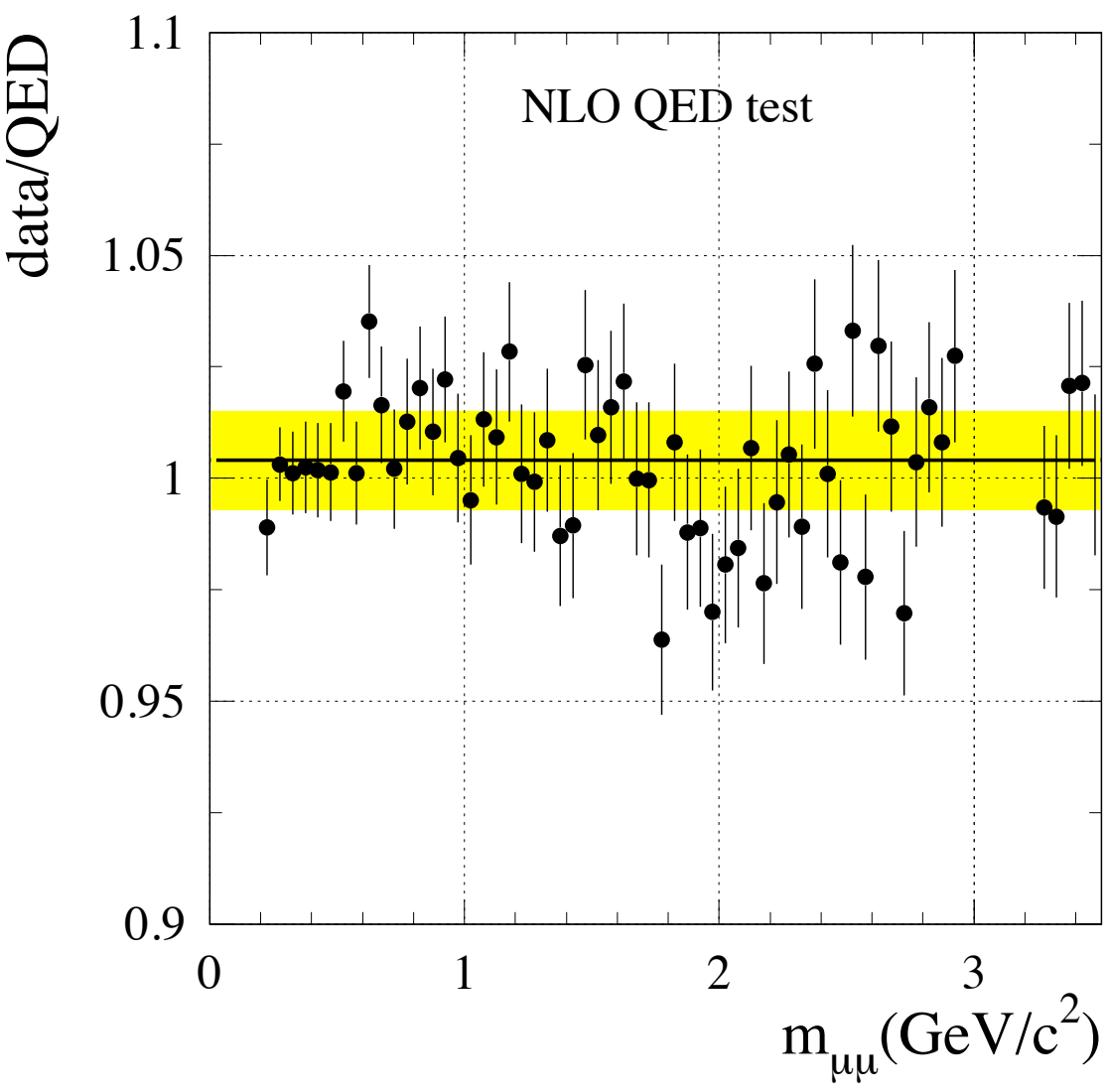

**Figure 21.3.2.** The ratio of the $e^+e^- \to \mu^+\mu^-\gamma(\gamma)$ cross section measured by *BABAR* to the NLO QED prediction (Aubert, 2009ah). The solid line and the shaded band represent the central value and errors given in Eq. (21.3.7).

Before extraction of the final cross section, an unfolding procedure, described in detail in Malaescu (2009), is applied to the background-subtracted and efficiency-corrected $m_{\pi\pi}$ spectrum. A mass-transfer matrix obtained using simulation provides the probability that an event generated in an interval $i$ of the reduced $e^+e^-$ CM energy $\sqrt{s'}$ is reconstructed in a $m_{\pi\pi}$ interval $j$. In the energy region around the $\rho$ peak, where the cross section is measured in energy intervals 2 MeV wide, the significant elements of the mass-transfer matrix lie near the diagonal over a typical range of 6 MeV, which corresponds to the energy resolution.

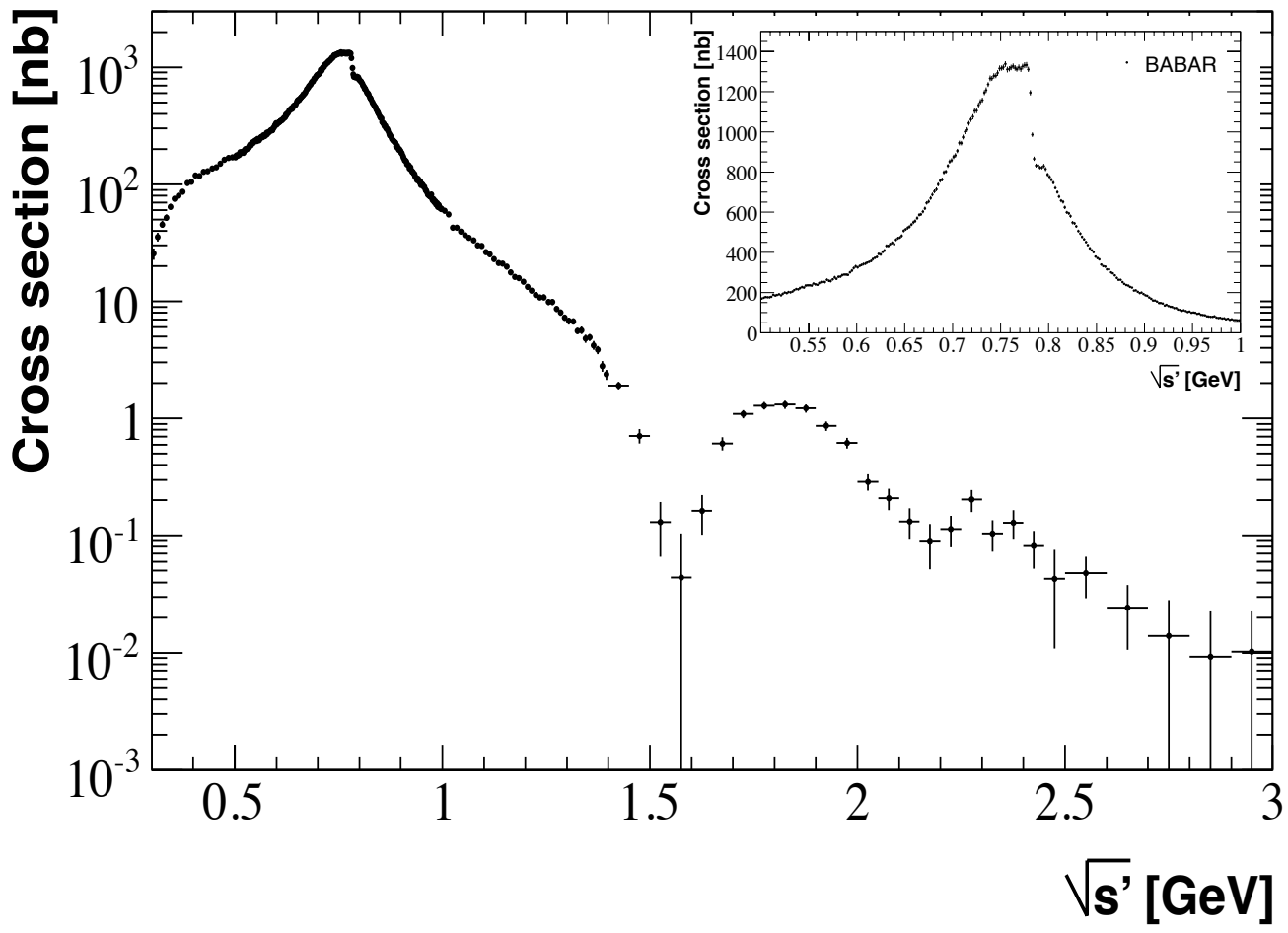

**Figure 21.3.3.** The bare cross section for $e^+e^- \to \pi^+\pi^-(\gamma)$ measured by *BABAR* (Aubert, 2009ah) in the full energy range. The inset shows an enlarged view of the energy region around the $\rho$ and $\omega$ masses. Total uncertainties are shown.

Figure 21.3.3 shows the $e^+e^- \to \pi^+\pi^-(\gamma)$ bare cross section including FSR measured by *BABAR* as a function of the CM energy. It is dominated by the $\rho$ resonance, and shows the effect of the $\rho-\omega$ interference at 0.78 GeV, a clear dip at 1.6 GeV resulting from interference with a heavier $\rho$ state, and additional structure above 2 GeV. The systematic uncertainty ranges from 0.5% in the energy region around the $\rho$ mass, up to 5% at the highest measured energies, and is smaller than the statistical error in the corresponding energy interval over the whole spectrum. The contributions of the various sources to the systematic uncertainties are shown in Table 21.3.1 for the central energy region $0.4 < \sqrt{s'} < 1.2$ GeV.

The square of the pion form factor is defined as usual by the ratio of the dressed cross section without FSR, and the lowest-order cross section for point-like spin 0 charged particles. Thus,

$$|F_\pi|^2(s') = \frac{3s'}{\pi\alpha^2(0)\beta_\pi^3}\,\sigma_{\pi\pi}(s')\ , \qquad (21.3.8)$$

with the pion velocity $\beta_\pi = \sqrt{1-4m_\pi^2/s'}$. A vector-meson-dominance (VMD) model is used to fit the *BABAR* pion form factor, correlating the observed structures to the effects from higher-mass isovector vector mesons. In addition to the $\rho$ and $\omega$ (isoscalar, but interfering with the $\rho$ through its isospin-violating $\pi^+\pi^-$ decay), three higher $\rho$ states at $(1493\pm15)$ MeV/$c^2$, $(1861\pm17)$ MeV/$c^2$, and $(2254\pm22)$ MeV/$c^2$ are required to fit the data. The fit is shown in Fig. 21.3.4.

**Table 21.3.1.** Relative systematic uncertainties (in $10^{-3}$) in the $e^+e^- \to \pi^+\pi^-(\gamma)$ cross section by $\sqrt{s'}$ intervals up to 1.2 GeV (Aubert, 2009ah). The statistical part of the efficiency uncertainties is included in the total statistical uncertainty in each interval, and, therefore, it is not reported in the table.

| source of uncertainty | $\sqrt{s'}$ (GeV) | | | |
|---|---|---|---|---|
| | 0.4–0.5 | 0.5–0.6 | 0.6–0.9 | 0.9–1.2 |
| trigger/ filter | 2.7 | 1.9 | 1.0 | 0.5 |
| tracking | 2.1 | 2.1 | 1.1 | 1.7 |
| $\pi$-ID | 2.5 | 6.2 | 2.4 | 4.2 |
| background | 4.3 | 5.2 | 1.0 | 3.0 |
| acceptance | 1.6 | 1.0 | 1.0 | 1.6 |
| kinematic fit ($\chi^2$) | 0.9 | 0.3 | 0.3 | 0.9 |
| correlated $\mu\mu$ ID loss | 2.0 | 3.0 | 1.3 | 2.0 |
| $\pi\pi/\mu\mu$ non-cancel. | 1.4 | 1.6 | 1.1 | 1.3 |
| unfolding | 2.7 | 2.7 | 1.0 | 1.3 |
| ISR luminosity ($\mu\mu$) | 3.4 | 3.4 | 3.4 | 3.4 |
| total uncertainty | 8.1 | 10.2 | 5.0 | 6.5 |

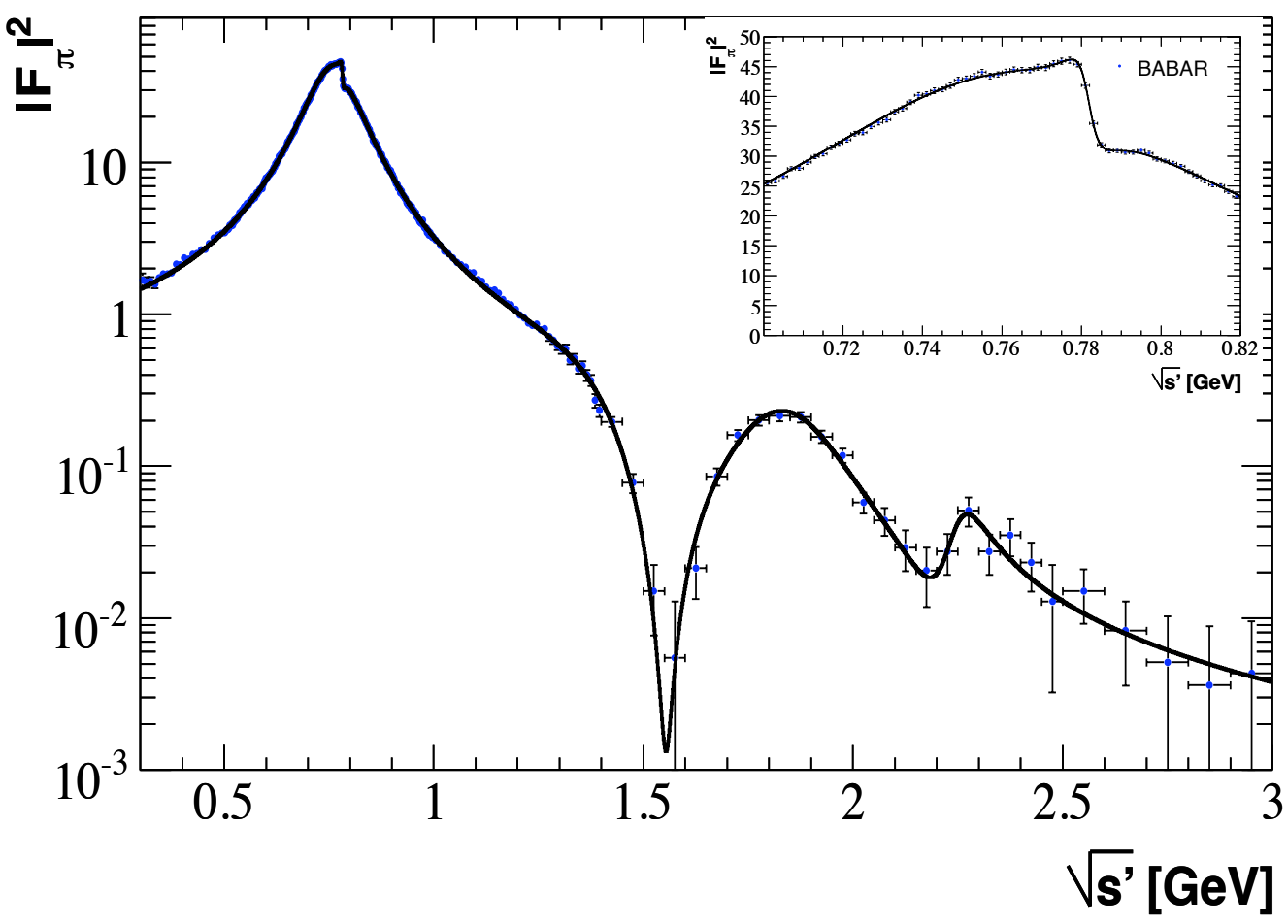

**Figure 21.3.4.** The pion form factor squared measured by *BABAR* (Aubert, 2009ah) as a function of $\sqrt{s'}$ in the full range, with details of the $\rho-\omega$ interference region shown in the inset. The line represents a VMD fit with the $\rho$, the $\omega$, and three higher $\rho$ states.

## 21.3.4 Impact of ISR results on $(g-2)_\mu$ and $\alpha(M_Z)$

### 21.3.4.1 The *BABAR* $\pi^+\pi^-$ contribution

The *BABAR* $2\pi$ results discussed above can be used in a straightforward way to compute the dispersion integral of Eq. 21.3.5. The errors are computed using the full statistical and systematic covariance matrices. The systematic uncertainties for each source are taken to be fully correlated over all mass regions. The upper range of integration (1.8 GeV) is chosen in accordance with previous evaluations (Davier, Eidelman, Hoecker, and Zhang, 2003a,b) in which the contribution of the higher energy region was computed using QCD. This procedure was justified by detailed studies using $\tau$ decay data (Barate et al., 1998). The contribution to $a_\mu$ in the 1.8–3 GeV range, obtained with the present *BABAR* data, is $(0.21\pm0.01)\times10^{-10}$, thus negligible with respect to the uncertainty in the main region. The contribution from threshold to 1.8 GeV is obtained for the first time from a single experiment:

$$a_\mu^{\pi\pi(\gamma),\mathrm{LO}} = (514.1\pm2.2\pm3.1)\times10^{-10}\ , \qquad (21.3.9)$$

where the errors are statistical and systematic.

### 21.3.4.2 Comparison to other determinations

Direct comparison with the results from other experiments is complicated by two facts: (i) $e^+e^-$ scan experiments provide cross section measurements at discrete and unequally spaced energy values, while the ISR method provides a continuous spectrum and, (ii) unlike *BABAR* no other experiment covers the complete mass spectrum from threshold up to energies where the contributions become negligible. Wherever gaps remain, they have been filled by using the weighted-average cross section values from the other experiments. This approach has been followed by Davier, Hoecker, Malaescu, Yuan, and Zhang (2010) from which the relevant integrals are extracted.

Correlations between systematic uncertainties have been taken into account, particularly for radiative corrections, when combining the results from all experiments. The combination is performed in small energy bins at the cross section level, taking into account possible disagreements leading to an increased uncertainty of the resulting average. The contribution of the $\pi^+\pi^-$ channel to $a_\mu$ obtained from the combination of all measurements of the $e^+e^-\to\pi^+\pi^-$ cross section is $(507.8\pm3.2)\times10^{-10}$. It is compared in Fig. 21.3.5 with the determinations of $a_\mu$ calculated using the data form the individual experiments. All determinations are indeed consistent within the uncertainties, *BABAR* and CMD-2 (Akhmetshin et al., 2006, 2007; Aulchenko et al., 2005) being almost a factor of two more precise than SND (Achasov et al., 2006) and KLOE (Ambrosino et al., 2009a, 2011).

The *BABAR* result is also consistent with determinations using $\tau$ decay with isospin-breaking corrections from Davier et al. (2010), which are also reported in Fig. 21.3.5. This reduces the previous tension between $e^+e^-$ and $\tau$ values (Davier, Eidelman, Hoecker, and Zhang, 2003a). Looking at the full picture it is important to note that the four inputs (CMD-2/SND, KLOE, *BABAR*, $\tau$) have completely independent systematic uncertainties.

### 21.3.4.3 Other exclusive channels

Remaining contributions from other exclusive channels up to 1.8 GeV amount to about 18% of $a_\mu^{\mathrm{had},LO}$. Previous results were obtained by CMD-2/SND from the $\omega$ and $\phi$ resonances, and from multihadrons up to 1.4 GeV. Data between 1.4 GeV and 2 GeV from the DM2 experiment were rather poor. However it was shown by the LEP experiments that perturbative QCD could be used at the $\tau$ mass scale with accuracies of about 1% (Ackerstaff et al., 1999;

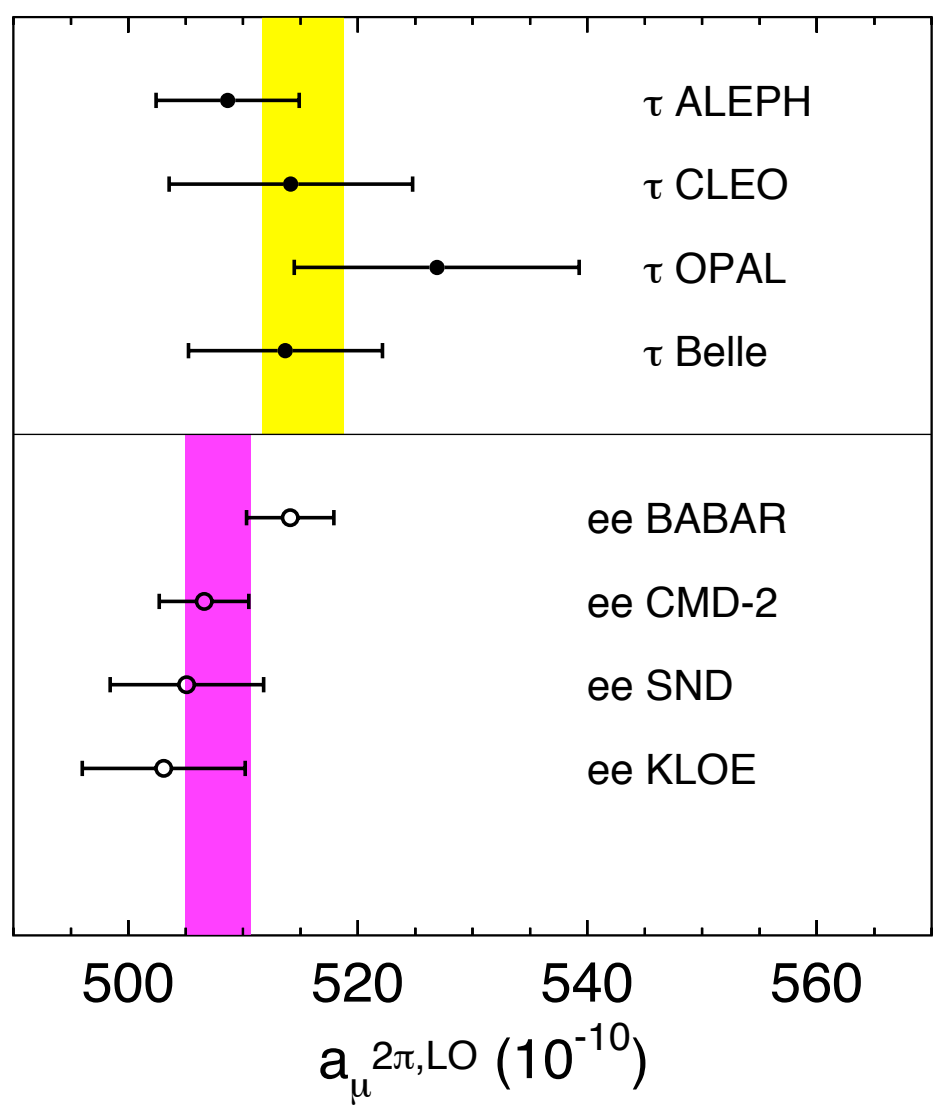

**Figure 21.3.5.** Evaluation of LO hadronic vacuum polarization $2\pi$ contributions to the muon magnetic anomaly in the energy range $[2m_\pi, 1.8\,\text{GeV}]$ from *BABAR*, other $e^+e^-$ experiments (Davier, Hoecker, Malaescu, Yuan, and Zhang, 2010), and $\tau$ experiments (Davier et al., 2010); see the text for details. The errors include both statistical and systematic sources. For the $\tau$ values, a common systematic error of 1.9 is included to account for uncertainties in the isospin-breaking corrections. The vertical bands represent the combined result, which amount to $(507.8 \pm 3.2) \times 10^{-10}$ for the $\pi\pi$ value, and $(515.2 \pm 3.0 \pm 1.9) \times 10^{-10}$ for the $\tau$ value. They are not obtained as the weighted average of the different values, but originate from a local combination of the respective spectral functions.

Barate et al., 1998), so it became advantageous (Davier and Hoecker, 1998) to use theory above 1.8 GeV.

The situation between 1 GeV and 1.8 GeV changed drastically with the advent of ISR *BABAR* data. In fact an almost complete set of precise measurements is available and a few remaining channels are being analyzed. These measurements benefit from the excellent particle identification, providing access to many previously unmeasured cross sections. They help to discriminate between older, less precise and sometimes contradictory results. Figure 21.3.6 gives a few examples of measured cross sections and demonstrates the impact of the *BABAR* results. The band shown on all these plots represents the combination by Davier, Hoecker, Malaescu, and Zhang (2011) of all existing data using the `HVPTools` package (Davier, Hoecker, Malaescu, Yuan, and Zhang, 2010): it is clearly dominated by the *BABAR* results.

The measurement using the full available data set of the production cross section for several final states are in progress at *BABAR*. Particularly relevant for the calculation of the muon anomaly are $\pi^+\pi^-\pi^0\pi^0$, $K^+K^-$, $K_SK_L$, and $K_SK_L\pi^+\pi^-$. Some channels involving $\pi^0$ multiplicities larger than 2 will probably remain unmeasured. The estimate of the missing channels, obtained using isospin relations or inequalities (Davier, Hoecker, Malaescu, and Zhang, 2011), is greatly facilitated by the studies of produced final states performed by *BABAR* on the related channels as discussed in the following sections.

#### 21.3.4.4 The complete muon anomaly prediction

Measurements from all experiments have been combined in the recent analysis of Davier, Hoecker, Malaescu, and Zhang (2011). The weight of the ISR *BABAR* data in the combination is the largest of all experiments: 41% for $2\pi$ and from 58 to 100% for the other measured exclusive channels below 1.8 GeV. More recently an independent analysis using the same data has been presented (Hagiwara, Liao, Martin, Nomura, and Teubner, 2011) with similar results. Adding all contributions (QED, electroweak, hadronic LO VP, hadronic HO VP, hadronic light-by-light) as given in Davier, Hoecker, Malaescu, and Zhang (2011), using for hadronic LO VP the combined result, one obtains the predicted value

$$a_\mu^{\rm SM} = (11\,659\,180.2 \pm 4.2 \pm 2.6 \pm 0.2) \times 10^{-10}\,, \quad (21.3.10)$$

where the three uncertainties come from hadronic VP ($e^+e^-$ data), the LBL calculations, and the sum of QED and Weak contributions, respectively, for a total uncertainty of $\pm 4.9 \times 10^{-10}$. The Standard Model prediction can be compared to the direct measurement (Bennett et al., 2006), slightly updated in (Nakamura et al., 2010):

$$a_\mu^{\rm exp} = (11\,659\,208.9 \pm 6.3) \times 10^{-10}\,. \quad (21.3.11)$$

The experimental value exceeds the theory prediction by $(28.7 \pm 8.0) \times 10^{-10}$, *i.e.* 3.6 standard deviations. Although the deviation is not significant enough to claim a departure from the Standard Model, it confirms the trend of earlier results using previous data (Davier, Eidelman, Hoecker, and Zhang, 2003b; Hagiwara, Martin, Nomura, and Teubner, 2007; Jegerlehner and Nyffeler, 2009).

Such a deviation could originate from new physics beyond the Standard Model. One possibility, much discussed in the literature, is the effect of contributions from supersymmetry (SUSY) involving scalar muons or neutrinos, and gauginos at a few hundred GeV mass scale. For the moment this explanation is not confirmed by the early LHC results as no evidence has yet been found for new particles in this mass range. While under tension this scenario is not yet ruled out as many SUSY versions can still be considered.

However it is clear that the muon $(g-2)$ discrepancy should be further explored. The progress will follow two lines. First, projects are considered at FNAL and J-PARC to extend the direct measurement to higher precision. Second, more precise measurements of $e^+e^- \to$ hadrons are possible with VEPP2000 and in the longer term with the new generation of $B$ Factories using the ISR method. Combining the two approaches could produce a very significant deviation which would unambiguously signal physics beyond the Standard Model.

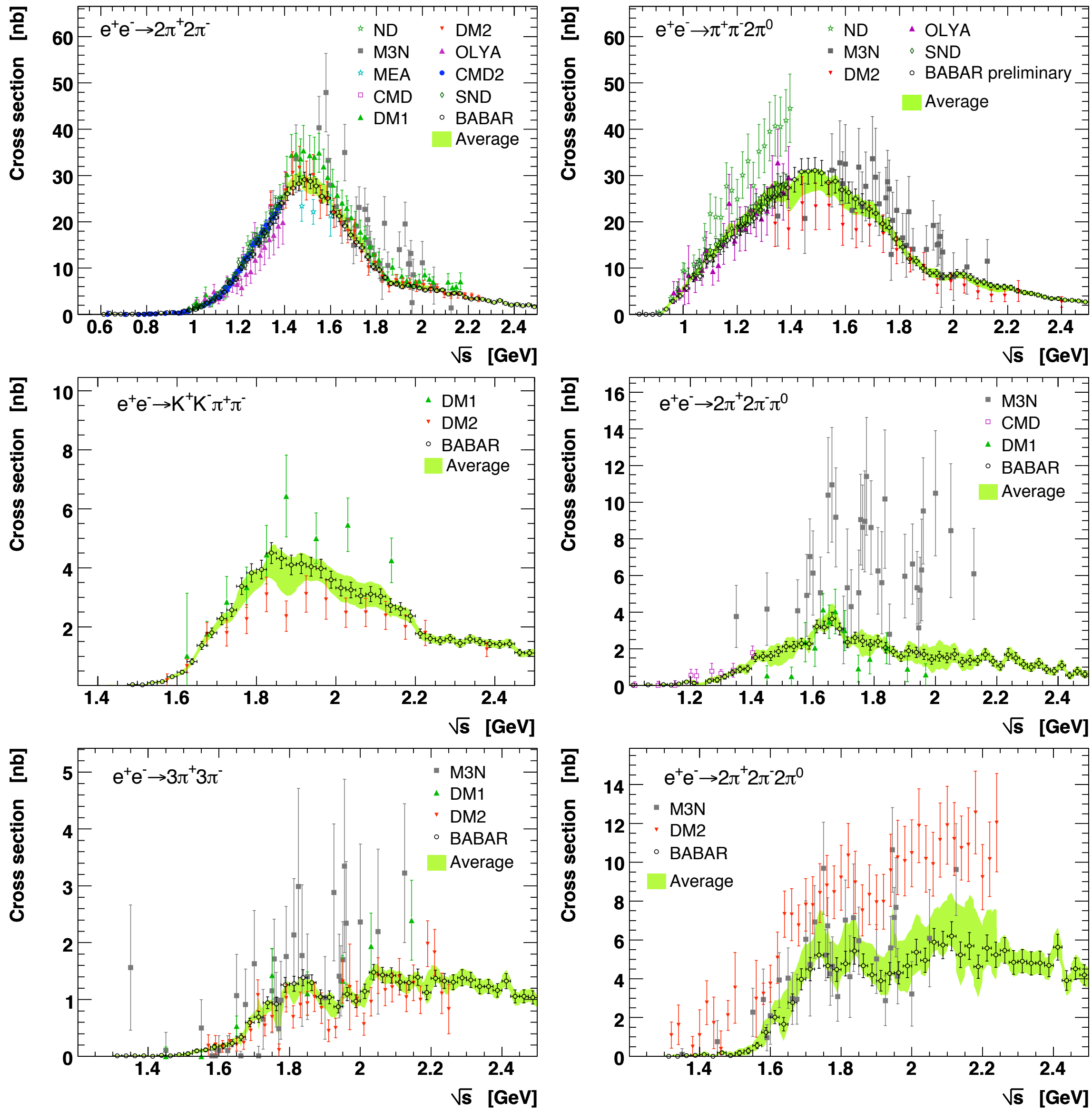

**Figure 21.3.6.** Cross sections versus center-of-mass energy for $e^+e^- \to \pi^+\pi^-\pi^+\pi^-$, $e^+e^- \to \pi^+\pi^-\pi^0\pi^0$, $e^+e^- \to K^+K^-\pi^+\pi^-$, $e^+e^- \to 2\pi^+2\pi^-\pi^0$, $e^+e^- \to 3\pi^+3\pi^-$, $e^+e^- \to 2\pi^+2\pi^-2\pi^0$. The open circles show data from *BABAR* which dominate in precision. The references for the earlier results displayed are given in Davier, Hoecker, Malaescu, and Zhang (2011). The error bars show the statistical and systematic uncertainties added in quadrature. The shaded (green online) band is the combined result $\pm 1\sigma$ taking all experiments into account using the `HVPTools` package (Davier, Hoecker, Malaescu, Yuan, and Zhang, 2010).

### 21.3.4.5 The prediction for $\alpha(M_Z^2)$

All hadronic contributions considered above are used as input to compute the dispersion relation in Eq. (21.3.3) with the result

$$\Delta\alpha_{\rm had}(M_Z^2) = (275.0 \pm 1.0) \times 10^{-4}\,, \qquad (21.3.12)$$

which, contrary to the evaluation of $a_\mu^{\rm had,LO}$, is not dominated by the uncertainty in the experimental low-energy data, but by contributions from all energy regions, where both experimental and theoretical errors have similar magnitude. Nevertheless the new ISR data provided by *BABAR* permits a significant improvement in precision. The result in Eq. (21.3.12) can be compared with the value obtained in Hagiwara, Liao, Martin, Nomura, and Teubner (2011), $(276.3 \pm 1.4) \times 10^{-4}$.

Adding the leptonic contribution $\Delta\alpha_{\rm lep}(M_Z^2)$, one finds

$$\alpha^{-1}(M_Z^2) = 128.952 \pm 0.014\,. \qquad (21.3.13)$$

The running electromagnetic coupling at $M_Z$ enters at various levels the global SM fit to electroweak precision data. It contributes to the radiator functions that modify the vector and axial-vector couplings in the partial $Z$ boson widths to fermions, and also to the SM prediction of the $W$ mass and the effective weak mixing angle. Overall, the fit exhibits a −39% correlation between the Higgs mass ($M_H$) and $\Delta\alpha_{\mathrm{had}}(M_Z^2)$ (Baak et al., 2012), so that the decrease in the value given in Eq.(21.3.12) and thus in the running electromagnetic coupling strength, with respect to earlier evaluations, leads to an increase in the most probable value of $M_H$ returned by the fit. Figure 21.3.7 shows the standard Gfitter result (green shaded band; Baak et al., 2012), using as the hadronic contribution the result obtained by using Eq. (21.3.12). The fitted Higgs mass shifts from $(84^{+30}_{-23})$ GeV/$c^2$ to $(91^{+30}_{-23})$ GeV/$c^2$. The stationary error of the latter value, in spite of the improved accuracy, is due to the logarithmic $M_H$ dependence of the fit observables. The new 95% upper limit on $M_H$ is 163 GeV/$c^2$. A new boson with properties compatible with the Higgs particle has been discovered by the ATLAS (Aad et al., 2012) and CMS (Chatrchyan et al., 2012b) experiments at the LHC. The fact that its mass of 126 GeV/$c^2$ is consistent with the range allowed above can be considered as a triumph for the Standard Model in the so-far hidden sector of gauge symmetry breaking.

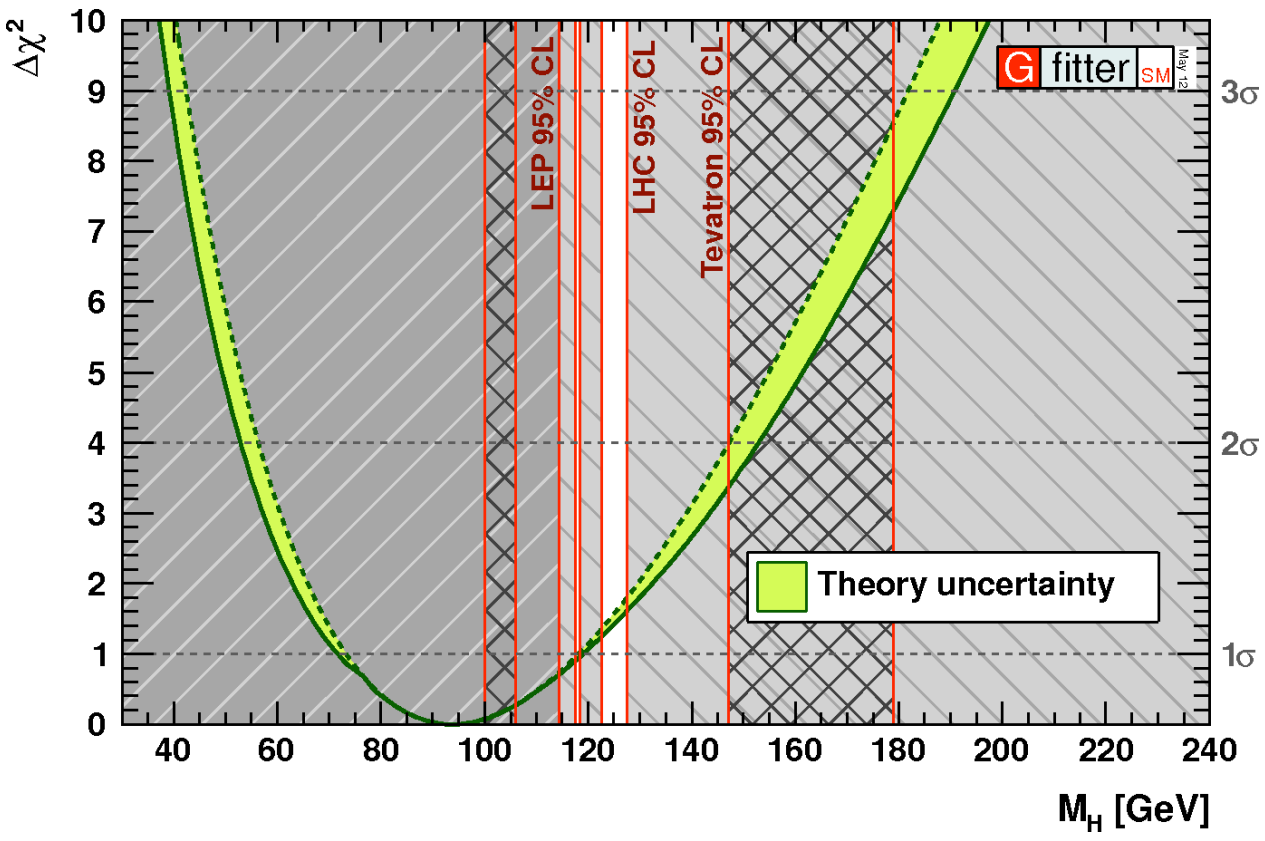

**Figure 21.3.7.** Overall $\chi^2$ for the Standard Gfitter electroweak fit (Baak et al., 2012; green shaded band) with the result obtained for the new evaluation of $\Delta\alpha_{\mathrm{had}}(M_Z^2)$. The shaded areas represent the excluded regions at 95% C.L. from the LEP and LHC experiments, leaving only a small window near 126 GeV where a very significant signal is observed by the ATLAS (Aad et al., 2012) and CMS (Chatrchyan et al., 2012b) experiments.

### 21.3.5 Light meson spectroscopy

Most of the multi-hadron final states feature a variety of internal sub-processes, with formation of several intermediate states, whose properties can be measured thanks to the large available statistics at the $B$ Factories.

In some cases, however, these studies are made difficult by the presence of broad interfering intermediate states, and have been performed only in a qualitative way. As an example, the study of the two- and three-pion invariant mass distributions of the process $e^+e^- \to \pi^+\pi^-\pi^0\pi^0$ shows important contributions from $\omega(780)\pi$, $a_1(1260)\pi$, and $\rho^+\rho^-$ intermediate states, which strongly interfere. A partial-wave analysis combining the data of $e^+e^- \to \pi^+\pi^-\pi^0\pi^0$ and $e^+e^- \to \pi^+\pi^-\pi^+\pi^-$ is required in order to separate the different sub-processes and to study the two excited $\rho$ states, $\rho(1450)$ and $\rho(1700)$, decaying into four pions.

In many other cases, more quantitative results have been obtained. A non-exhaustive summary of these results is presented below.

#### 21.3.5.1 Study of $\omega$-like resonances

The $e^+e^- \to \pi^+\pi^-\pi^0$ cross section is dominated by the production of the well-known vector states $\omega$, $\phi$, and $J/\psi$. Between 1 and 2 GeV the cross section is generally described as the sum of two $\omega$-like resonances: $\omega(1420)$ or $\omega'$, and $\omega(1650)$ or $\omega''$, whose parameters are not yet well established. The published *BABAR* results (Aubert, 2004af) are based on an integrated luminosity of only 89.3 fb$^{-1}$. Therefore, an update of the study using the full available dataset is desirable. The measured cross section in the 1.05–3.0 GeV/$c^2$ mass region is shown in Fig. 21.3.8. There is good agreement with previous results by the SND experiment (Achasov et al., 2002) below 1.4 GeV/$c^2$; significant disagreement with DM2 results (Antonelli et al., 1992) is observed at higher energies.

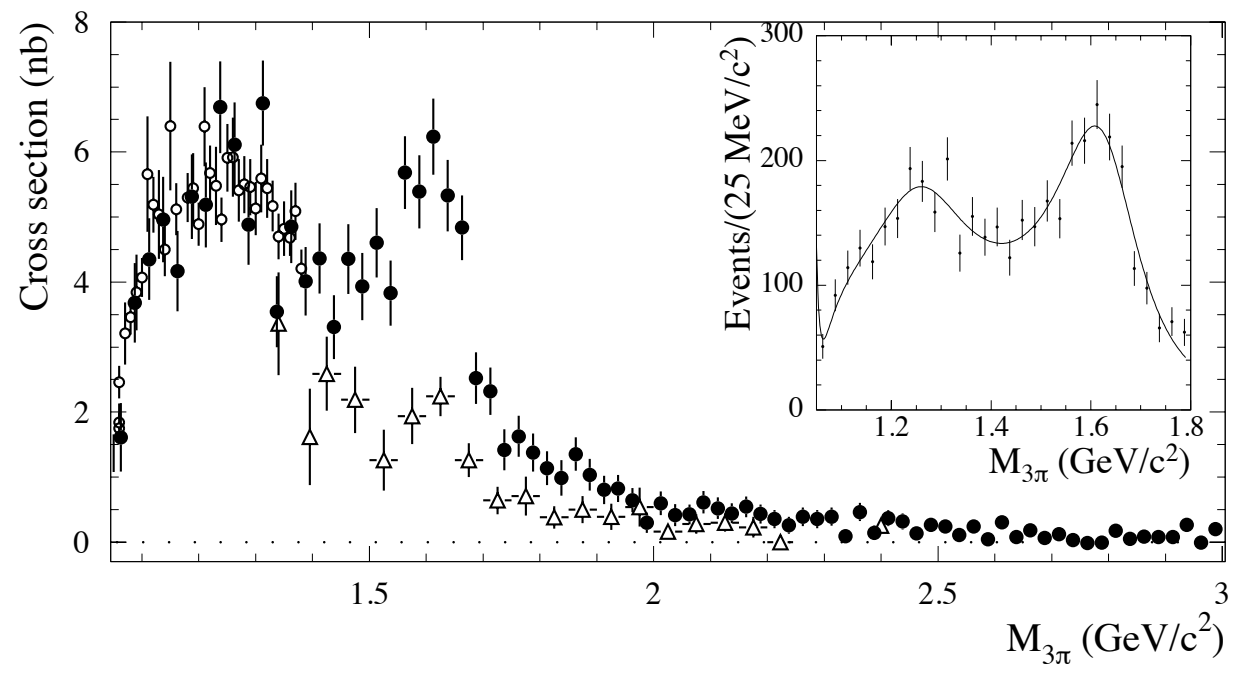

**Figure 21.3.8.** The $e^+e^- \to \pi^+\pi^-\pi^0$ cross section measured by the *BABAR* experiment (Aubert, 2004af; full circles) in the 1-3 GeV/$c^2$ range compared with the SND (Achasov et al., 2002; open circles) and DM2 (Antonelli et al., 1992; triangles) data. The inset shows the result of a fit to the mass distribution as explained in the text.

The three pion mass spectrum below 1.8 GeV/$c^2$, obtained by *BABAR*, is fitted as the sum of the four known vector resonances $\omega$, $\phi$, $\omega'$, and $\omega''$. The fit result in the $\omega'$ and $\omega''$ mass region is shown in the inset to Fig. 21.3.8, superimposed on the experimental data. The resonance parameters from the fit are reported in the first column of Table 21.3.2, together with the corresponding values from studies of ISR processes with five and six hadrons in the final state.

Clear $\eta \to 3\pi$ and $\omega \to 3\pi$ signals are observed in the $e^+e^- \to 2(\pi^+\pi^-)\pi^0$ process. The cross sections for production of the $\eta\pi^+\pi^-$ and $\omega\pi^+\pi^-$ final states, measured by *BABAR* (Aubert, 2007bb), are compared to previous, less precise data (Akhmetshin et al., 2000; Antonelli et al., 1988; Cordier et al., 1981; Druzhinin et al., 1986) in Fig. 21.3.9. Several new features are revealed by the *BABAR* data. In particular, the study of the $\pi^+\pi^-$ mass distribution shows a clear contribution of the intermediate state $\omega f_0(980)$ to the $\omega\pi^+\pi^-$ cross section. In addition, the $\omega\pi^+\pi^-$ cross section has been fitted, after removal of the $\omega f_0(980)$ contribution, with a sum of two BW functions, referring to the $\omega'$ and $\omega''$, as in the case of the $\pi^+\pi^-\pi^0$ final state. The fitted parameters are reported in the third column Table 21.3.2.

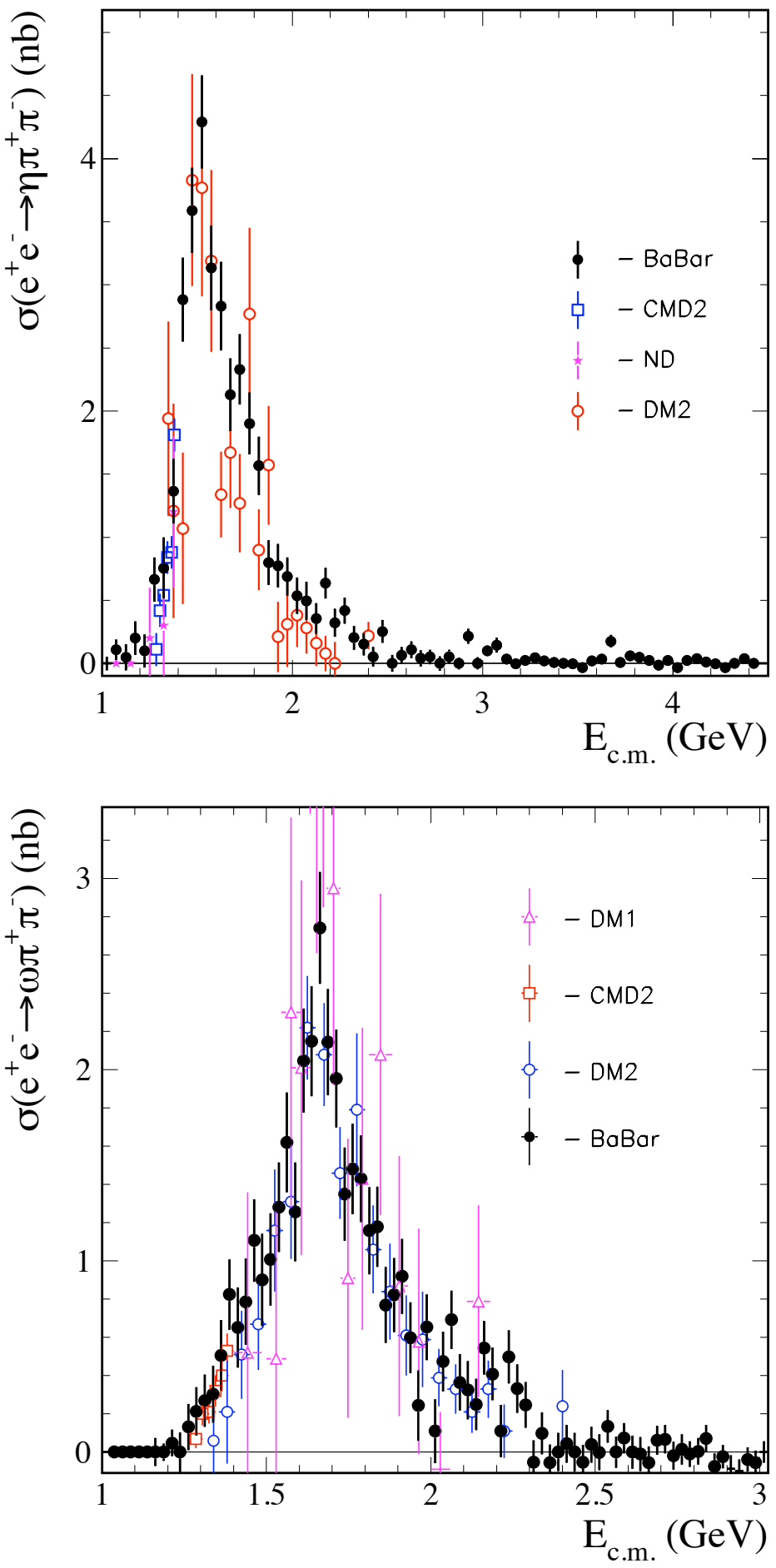

**Figure 21.3.9.** The $e^+e^- \to \eta\pi^+\pi^-$ (top) and $e^+e^- \to \omega\pi^+\pi^-$ (bottom) cross sections measured by *BABAR* (Aubert, 2007bb) in comparison with direct $e^+e^-$ measurements.

Finally, further structure compatible with $\omega$ excitations has been observed in $e^+e^- \to 2(\pi^+\pi^-\pi^0)$ (Aubert, 2006az). In fact, in addition to clear $\eta \to \pi^+\pi^-\pi^0$ and $\omega \to \pi^+\pi^-\pi^0$ signals, a small associated production of $\eta$ and $\omega$ is observed in this channel. The cross section for the $e^+e^- \to \omega\eta$ reaction, reported in Fig. 21.3.10, shows a peak in the $\omega(1650)$ energy region, which is fitted with a Breit-Wigner function.

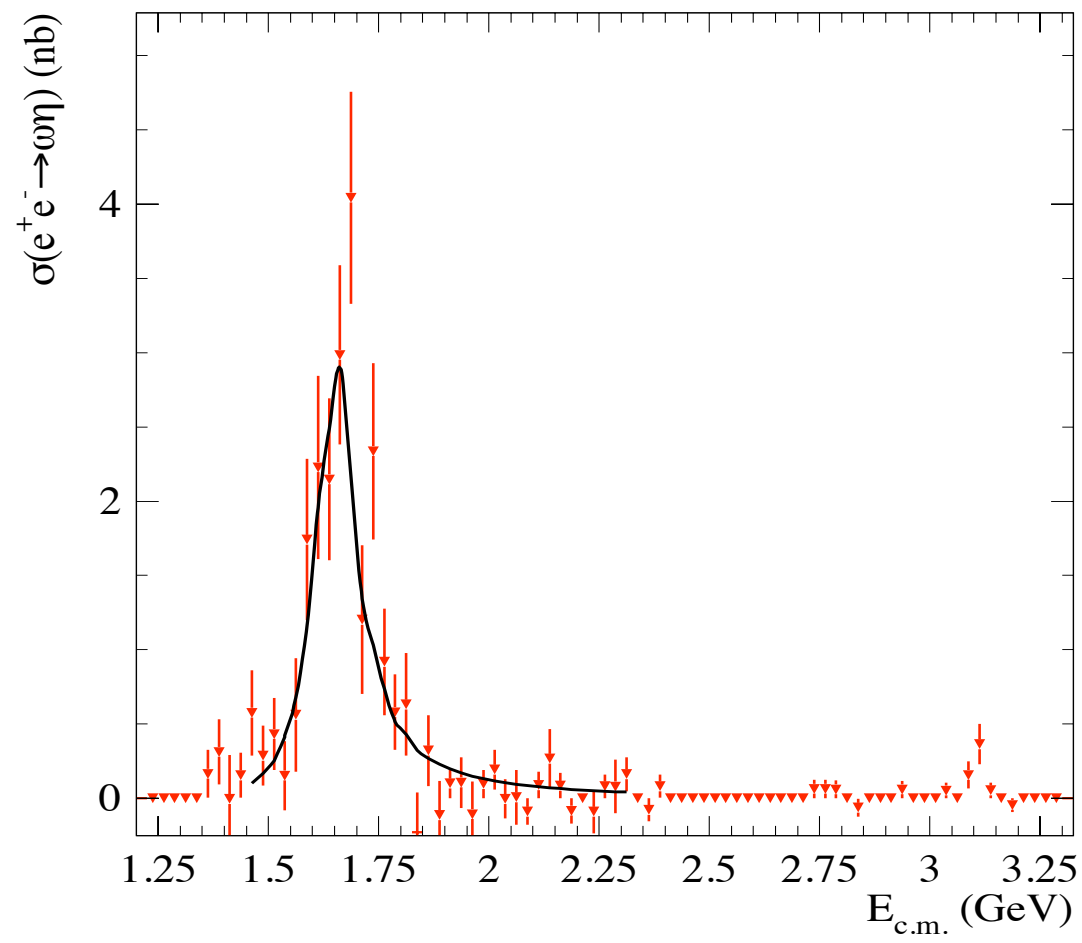

**Figure 21.3.10.** The $e^+e^- \to \omega\eta$ cross section extracted from the $2(\pi^+\pi^-\pi^0)$ final state measured by *BABAR* (Aubert, 2006az). The solid line is the result of the fit with a Breit-Wigner function.

As can be seen from Table 21.3.2, there is general consistency among the $\omega'$ and $\omega''$ parameters measured in the different channels. An update of the three pion final state measurement with the full available *BABAR* data set, and a combined fit to all channels, could give information on relative decay rates and significantly improve the knowledge of these states.

### 21.3.5.2 Study of excited $\rho$ and $\phi$ states in the $K\overline{K}\pi$ and $K\overline{K}\eta$ final states

By studying the Dalitz plots of the $e^+e^- \to K^0_S K^\pm\pi^\mp$ and $e^+e^- \to K^+K^-\pi^0$ final states, the DM1 and DM2 experiments have identified $e^+e^- \to \overline{K}K^*(892)$ and its charge conjugate as the dominant sub-process, and measured the contributions of the different isospin $(I = 0, 1)$ components. They also observed a resonant structure in the isoscalar component, which was interpreted as the first excitation of the $\phi$ resonance, thereafter called the $\phi(1680)$ or $\phi'$ (Bisello et al., 1991; Buon et al., 1982). *BABAR* performs a similar study of the $K^0_S K^\pm\pi^\mp$ and $K^+K^-\pi^0$ final states, using a $\simeq 220$ fb$^{-1}$ sample (Aubert, 2008ab). The large amount of data allows the measurement of the cross sections up to a CM energy of 4.5 GeV, and a much more accurate study of the Dalitz plots of the two processes, shown in Fig. 21.3.11. It can be seen that in both processes, the main contributions come from the $\overline{K}K^*(892)$ and $\overline{K}K^*_2(1430)$ intermediate states, and that the Dalitz plot population for the $K^0_S K^\pm\pi^\mp$ channel is strongly asymmetric. This is because both the neutral $\overline{K}^0K^{*0}$ and charged $K^\pm K^{*\mp}$ combinations are involved, and these are produced by, respectively, the sum and the difference of the iso-scalar and iso-vector amplitudes. By studying the Dalitz plots, the moduli and relative phase of the isospin components for both the $\overline{K}K^*(892)$ and $\overline{K}K_2^*(1430)$ cross sections have been obtained as a function of the CM energy. The isoscalar $\sigma_0$ and isovector $\sigma_1$ cross sections for $e^+e^- \to \overline{K}K^*(892)$ are shown in Fig. 21.3.12(a,b). The isoscalar component is dominant, and shows a clear resonant peak at a CM energy of $\sim$ 1.7 GeV, consistent with the $\phi(1680)$ meson. The isovector component is also incompatible with a pure phase space shape, and shows a resonant structure, as can be deduced by a study including the information on the relative phase between $\sigma_0$ and $\sigma_1$.

**Table 21.3.2.** Summary of the $\omega(1420)$ (or $\omega'$) and $\omega(1650)$ (or $\omega''$) resonance parameters obtained from the fits described in the text. $m_i$ and $\Gamma_i$ are the mass and the full width of state $i$, respectively, $\sigma_{0i}$ is the peak cross section, $\Gamma_{ee}B_{if}$ the dielectron width multiplied by the branching fraction for decays into the final state $f$, and $\phi_i$ is the phase w.r.t. the $\omega$ amplitude. The errors shown are combination of statistical and systematic uncertainties. The values without errors were fixed in the fits.

| Fit | $3\pi$ (Aubert, 2004af) | $\omega\eta$ (Aubert, 2006az) | $\omega\pi^+\pi^-$ (Aubert, 2007bb) | PDG (Amsler et al., 2008) |
|---|---|---|---|---|
| $\sigma_{0\omega'}$ (nb) | – | – | $1.01 \pm 0.29$ | – |
| $\Gamma_{ee}B_{\omega' f}$(eV) | 369 | – | $17.5 \pm 5.4$ | – |
| $m_{\omega'}$(GeV/$c^2$) | $1.350 \pm 0.03$ | – | $1.38 \pm 0.07$ | 1.40 −1.45 |
| $\Gamma_{\omega'}$(GeV) | $0.450 \pm 0.10$ | – | $0.13 \pm 0.05$ | 0.180 − 0.250 |
| $\phi_{\omega'}$ (rad) | $\pi$ | – | $\pi$ | – |
| $\sigma_{0\omega''}$ (nb) | – | $3.08 \pm 0.33$ | $2.47 \pm 0.18$ | – |
| $\Gamma_{ee}B_{\omega'' f}$(eV) | 286 | – | $103.5 \pm 8.3$ | – |
| $m_{\omega''}$(GeV/$c^2$) | $1.660 \pm 0.010$ | $1.645 \pm 0.008$ | $1.667 \pm 0.014$ | $1.670 \pm 0.030$ |
| $\Gamma_{\omega''}$(GeV) | $0.220 \pm 0.036$ | $0.114 \pm 0.014$ | $0.222 \pm 0.032$ | $0.315 \pm 0.035$ |
| $\phi_{\omega''}$ (rad) | 0 | 0 | 0 | – |

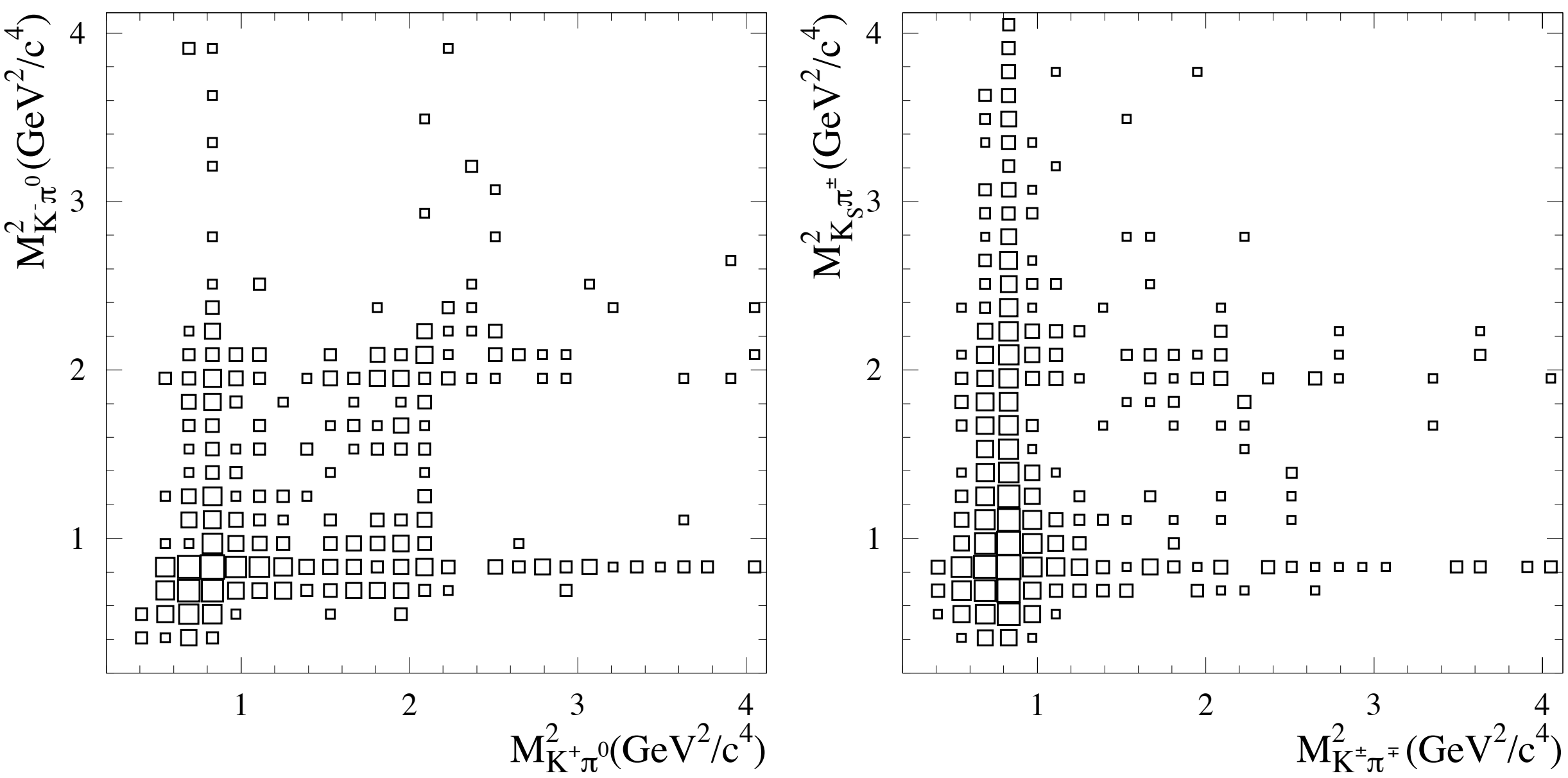

**Figure 21.3.11.** The Dalitz plot distribution for the $K^+K^-\pi^0$ (left) and $K^0_S K\pi$ final state (right) measured by BABAR (Aubert, 2008ab).

A global fit has been performed using six different sources of information: $\sigma_0$, $\sigma_1$ and their relative phase; the $K^+K^-\pi^0$ cross section shown in Fig. 21.3.12(c), and the $\phi\eta$ cross section measured reconstructing two different $\eta$ decay modes: $\eta \to \gamma\gamma$ (Aubert, 2008ab), and $\eta \to \pi^+\pi^-\pi^0$ (Aubert, 2007bb). The $e^+e^- \to \phi\eta$ reaction is well suited to study excited $\phi$ states, because the production of any $\omega$-like state, even if allowed by quantum-number conservation, is strongly suppressed by the OZI rule. The measured $\phi\eta$ cross section reported in Fig. 21.3.12(d) shows a broad peak at a CM energy of about 1.7 GeV, which can be identified as a new decay channel of the $\phi(1680)$ meson. The same dominant resonance $\phi'$ is therefore assumed to fit both the $\phi\eta$ and the isoscalar $\overline{K}K^*(892)$ cross sections. An additional resonance, $\phi''$ is included to account for a small peak seen at a CM energy of about 2.15 GeV in the $\phi\eta$ cross section. The fit results are superimposed on the cross section data in Fig. 21.3.12 and are listed in Table 21.3.3. The parameters obtained for the $\phi'$ and $\rho'$ are compatible with previous measurements (Amsler et al., 2008). Concerning the $\phi''$ resonance, which is seen only in the $\phi\eta$ channel with a significance of about $2.5\sigma$, the fitted parameters are close to those for the $Y(2175)$ state observed in the $\phi f_0(980)$ final state (Aubert, 2006c), discussed in Section 21.3.5.3.

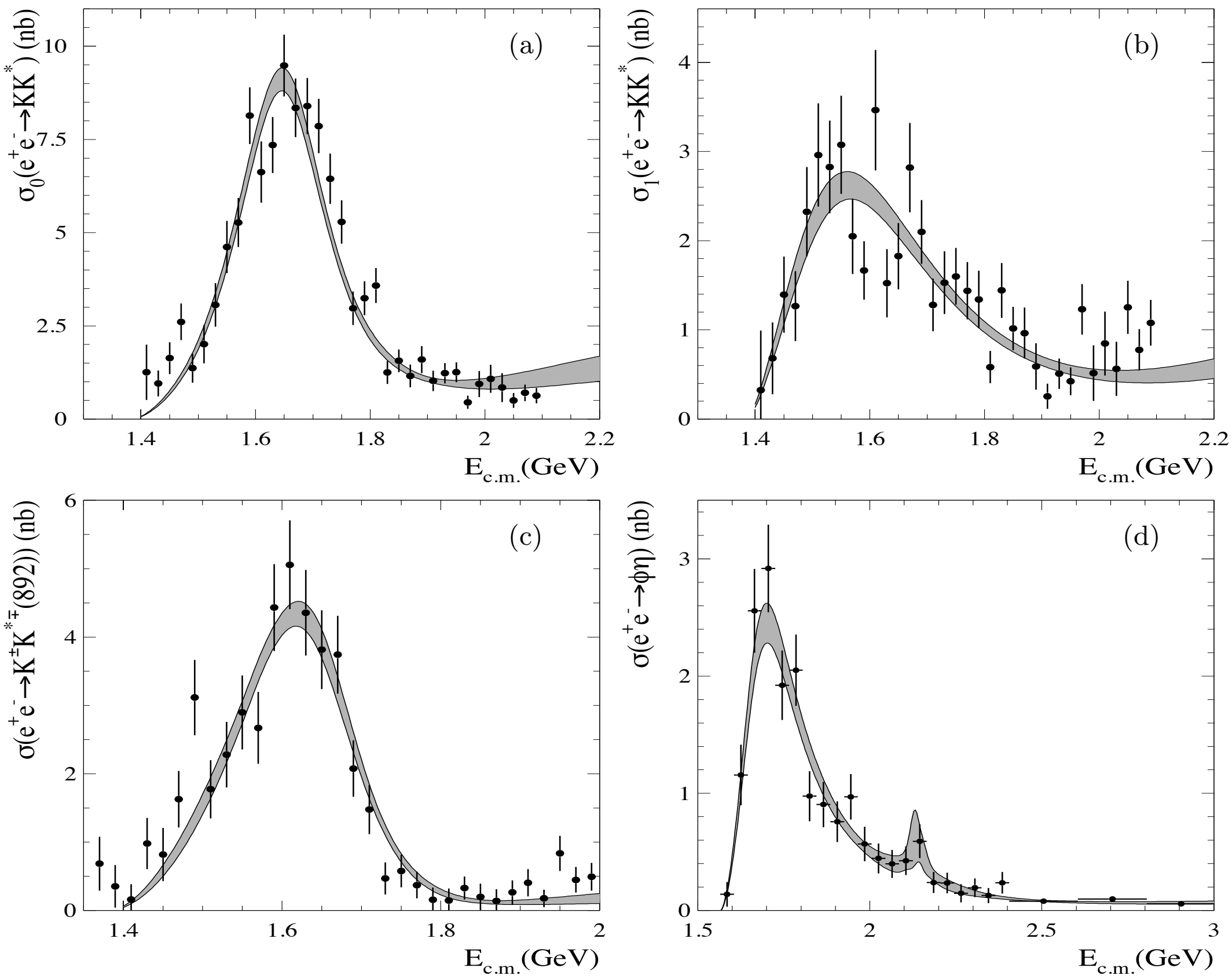

**Figure 21.3.12.** Isoscalar (a) and isovector (b) components of the $e^+e^- \to K_S^0 K\pi$ cross section; the $e^+e^- \to K^\pm K^*(892)^\mp$ cross section obtained by the *BABAR* experiment, using $e^+e^- \to K^+K^-\pi^0$ events (c), and the $e^+e^- \to \phi\eta$ cross section (d). The points with error bars are data and the gray bands represent the fit and its uncertainty (Aubert, 2008ab).

An interesting sub-process of the $K^+K^-\gamma\gamma$ final state measured for the first time by *BABAR* is $e^+e^- \to \phi\pi^0$ (Aubert, 2008ab). The decays of ordinary isovector resonances to $\phi\pi^0$ are suppressed by the OZI rule, so structure in this channel could serve as a signal for exotic resonant states. Two possible descriptions are considered to fit the $\phi\pi^0$ cross section, assuming respectively the presence of one or two radial excitations of the $\rho$ meson (despite being OZI suppressed).[160] In the first case, the parameters obtained for the unique isovector state are $1593 \pm 32$ MeV/$c^2$ for the mass and $203 \pm 97$ MeV for the width, which are compatible with those of the $\rho(1700)$ (Amsler et al., 2008). A slightly better fit quality is obtained if two resonances are assumed, as seen from the results shown in Table 21.3.3. The parameters obtained for the first resonance (indicated by $\rho''$ in the Table) are consistent with those of the $C(1480)$ state observed in $\pi^- p \to \phi\pi^0 n$ charge-exchange reaction (Bityukov et al., 1987). However, a firm conclusion cannot be drawn, and an OZI-violating decay of the $\rho(1700)$ is not excluded. The second structure, the $\rho(1900)$, is compatible with the "dip" already observed in other experiments, predominantly in multi-hadron final states (Antonelli et al., 1996; Frabetti et al., 2001), and by *BABAR* in the ISR production of six-pion final states (Aubert, 2006az). In the last-cited result, however, a significantly larger width has been measured, so the situation is still uncertain.

[160] It should be noted that in the region 1 GeV to 2 GeV several wide resonances, mixtures of $s\bar{s}$, $u\bar{u}$ and $d\bar{d}$ states, are present and hence the OZI rule may not be directly applicable.

#### 21.3.5.3 The discovery of the $Y(2175)$ in $K^+K^-\pi\pi$ final states.

The $e^+e^- \to K^+K^-\pi^+\pi^-$ and $e^+e^- \to K^+K^-\pi^0\pi^0$ reactions proceed through the production of numerous intermediate states. The invariant mass distributions of the two- and three-particle combinations indicate that the intermediate states $K^*(892)^0K^\pm\pi^\mp$ and $K^*(892)^\mp K^\pm\pi^0$ dominate in these reactions. A small $K_2^*(1430)K\pi$ contribution is also seen, while states with two $K^*$, namely $K^*(892)\overline{K}^*(892)$, $K^*(892)\overline{K}_2^*(1430)$, and $K_2^*(1430)\overline{K}_2^*(1430)$, account for less than 1% of the total reaction yield.

**Table 21.3.3.** Summary of parameters obtained for the $\rho$ and $\phi$ radial excitation from the study of the $K\overline{K}\pi$ and $K\overline{K}\eta$ final states (Aubert, 2008ab), including the data on the $\phi\eta$ cross section from Aubert (2007bb). The parameters for the $\rho''$ are taken from the fit to the $\phi\pi^0$ cross section with two resonances (see the text).

| Isospin | $R$ | $\Gamma^R_{ee}\mathcal{B}^R_{KK^*}$ (eV) | $\Gamma^R_{ee}\mathcal{B}^R_{\phi\eta}$ (eV) | $M_R$ (MeV) | $\Gamma_R$ (MeV) |
|---|---|---|---|---|---|
| 0 | $\phi'$ | $369 \pm 53 \pm 1$ | $138 \pm 33 \pm 28$ | $1709 \pm 20 \pm 43$ | $322 \pm 77 \pm 160$ |
| 0 | $\phi''$ | – | $1.7 \pm 0.7 \pm 1.3$ | $2125 \pm 22 \pm 10$ | $61 \pm 50 \pm 13$ |
| 1 | $\rho'$ | $127 \pm 15 \pm 6$ | – | $1505 \pm 19 \pm 7$ | $418 \pm 25 \pm 4$ |
| 1 | $\rho''$ | – | $3.5 \pm 0.9 \pm 0.3$ | $1570 \pm 36 \pm 62$ | $144 \pm 75 \pm 43$ |
| 1 | $\rho(1900)$ | – | $2.0 \pm 0.6 \pm 0.4$ | $1909 \pm 17 \pm 25$ | $48 \pm 17 \pm 2$ |

Among the most interesting studies of this final state performed by *BABAR* is the extraction of the relatively small contributions of the $\phi\pi^+\pi^-$ and $\phi\pi^0\pi^0$ ($\phi \to K^+K^-$) intermediate states (Aubert, 2006c, 2007bc). The original motivation was the search for decays of the then recently-discovered vector meson $Y(4260)$. As discussed in Section 18.3, the $Y(4260)$ was discovered by *BABAR* in the process $e^+e^- \to \gamma_{\rm ISR} Y(4260) \to \gamma_{\rm ISR} J/\psi\pi^+\pi^-$, but was not seen to decay to $D^{(*)}\overline{D}^{(*)}$, although this was expected for a wide conventional charmonium state with a mass well above the $D\overline{D}$ production threshold. A certain exotic-structure model for the $Y(4260)$ predicted a large branching fraction for the decay into $\phi\pi\pi$ (Zhu, 2005).

Since the $\phi$ resonance is relatively narrow, a clean sample of $\phi\pi\pi$ events can be easily separated. The scatter plot of the reconstructed masses, $m(\pi^+\pi^-)$ versus $m(K^+K^-)$, for selected events in a data sample corresponding to $232\,\text{fb}^{-1}$ is shown in Fig. 21.3.13(a) (Aubert, 2006c): a clear $\phi \to K^+K^-$ vertical band is visible, as well as an accumulation of events indicating correlated production of the $\phi$ and $f_0(980) \to \pi\pi$ resonances. A wide horizontal band corresponding to $\rho^0 \to \pi^+\pi^-$ production is also seen. The invariant mass distribution of the $\pi\pi$ system in $\phi\pi\pi$ events is obtained using the condition $|m(K^+K^-) - m_\phi| < 10$ MeV/$c^2$, where $m_\phi$ is the nominal $\phi$-mass. The background from true $K^+K^-\pi\pi$ events with non-resonant $K^+K^-$ pair is subtracted using the $\phi$ mass sidebands $10 < |m(K^+K^-) - m_\phi| < 20$ MeV/$c^2$; other backgrounds are subtracted based on MC simulation. The final mass spectrum for $\pi\pi$ pairs associated with $\phi$ production is shown in Fig. 21.3.13(b)(Aubert, 2007bc). Besides the clear $f_0(980)$ signal, and a concentration consistent with the $f_2(1270)$ resonance, a broad bump at lower mass values is observed, which can be interpreted as the controversial $f_0(600)$ scalar meson.

The $e^+e^- \to \phi\pi\pi$ mass spectrum is measured in 25 MeV/$c^2$ wide bins by extracting the number of reconstructed $\phi \to K^+K^-$ decays from a fit to the $K^+K^-$ mass spectrum. The corresponding cross section is then obtained applying Eq. (21.2.12) and taking into account the $\phi \to K^+K^-$ branching fraction. With an analogous procedure, but requiring in addition that $0.85 < m(\pi\pi) < 1.1$ GeV/$c^2$, a 90% pure sample of $\phi f_0(980)$ is selected, and, assuming a decay rate $\mathcal{B}(f_0(980) \to \pi^+\pi^-) = 2/3$, the cross section is measured. Similar distributions and results are obtained for the $\phi\pi^0\pi^0$ final state.

The cross section for the two $f_0(980)$ decay modes are consistent with each other; both are shown in Fig. 21.3.14. The data are successfully described by a relatively narrow resonance, called the $Y(2175)$, interfering with a non-resonant term. The result of the fit is shown as the solid line in the figure. By contrast, the attempt to fit the data with only the non-resonant term, accounting for the finite width of the $\phi$ and $f_0(980)$, and for their spin and phase space, is clearly unsatisfactory (see the dashed red line). The histogram is the result of a simulation of the non-resonant $e^+e^- \to \phi(1020)f_0(980)$ reaction, which also fails to reproduce the features seen in the data.

The $Y(2175)$ was confirmed by the BES Collaboration in the $\phi f_0(980)$ invariant mass spectrum from $J/\psi \to \eta\phi f_0(980)$ decays (Ablikim et al., 2008b), as well as in subsequent measurements making use of the full data sets now available to Belle and *BABAR*.

The new analyses at the two $B$ Factories select the $\phi\pi\pi$ and $\phi f_0(980)$ final states in a way similar to the previous *BABAR* analysis and measure rather consistent cross sections, and, thanks to the larger data samples (674 fb$^{-1}$ for Belle, and 475 fb$^{-1}$ for *BABAR*), a more precise study of the $e^+e^- \to \phi\pi\pi$ cross section and of the intermediate states involved is possible. In both cases, the $e^+e^- \to \phi(1020)\pi^+\pi^-$ cross section shows two clear peaks: the first, at about 1.7 GeV, can be attributed to the $\phi(1680)$, and the second, above 2 GeV, to the $Y(2175)$. Different models have been used by the two collaborations to fit the cross section distributions.

Belle (Shen, 2009) uses an incoherent sum of two Breit-Wigner functions, one for the $\phi(1680)$, which is assumed to decay into $\phi\pi^+\pi^-$, and the other for the $Y(2175)$ which decays predominantly into $\phi f_0(980)$. The result of the fit is shown as the solid line in Fig. 21.3.15, while the separate contributions of the two fitted resonances are given by the dotted lines.

In the *BABAR* analysis (Lees, 2012d), it is also noted that the second structure, associated to the $Y(2175)$, completely disappears if events with a dipion mass under the $f_0(980)$ peak are removed. Figure 21.3.16 shows the *BABAR* data and the result of a fit to a VMD-based model. This assumes that two vector mesons contribute to the cross sec-tion: the $\phi(1680)$ decaying both to $\phi f_0(600)$ and $\phi f_0(980)$, and the $Y(2175)$ decaying to $\phi f_0(980)$ only. Since the nominal $\phi(1680)$ mass lies below the $\phi f_0(980)$ threshold, the $\phi(1680) \to \phi f_0(980)$ decay will reveal itself as a smooth bump in the energy dependence of the $e^+e^- \to \phi f_0(980)$ cross section above 2 GeV, as shown by the dotted line in the plot. The solid line is the result of the total fit to the $\phi\pi\pi$ cross section, which clearly indicates the need for an additional resonance centered at about 2.2 GeV, on top of the dashed line showing the $\phi(1680)$ contribution.

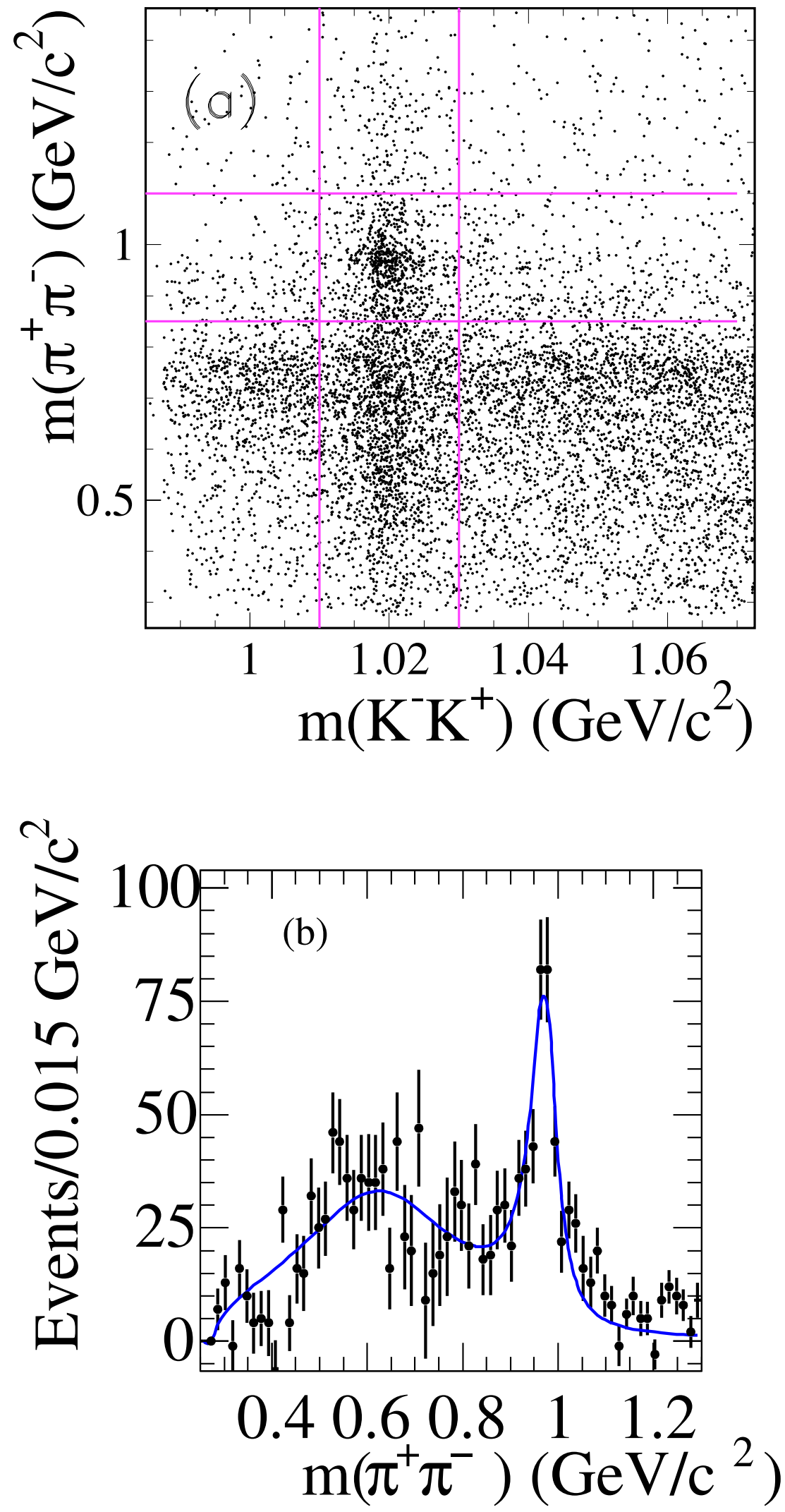

**Figure 21.3.13.** (a) The reconstructed $m(\pi^+\pi^-)$ *vs* $m(K^+K^-)$ distribution for the $e^+e^- \to K^+K^-\pi^+\pi^-$ reaction measured by *BABAR* (Aubert, 2006c). The vertical lines identify the selected region around the $\phi$ mass peak. (b) Background subtracted $m(\pi^+\pi^-)$ distribution of selected $e^+e^- \to \phi(1020)\pi^+\pi^-$ events (Aubert, 2007bc). The solid line is the results of the fit to the data with a coherent sum of two Breit-Wigner functions, referring to the $f_0(600)$ and $f_0(980)$.

Both collaborations also fitted the selected samples of $\phi f_0(980)$ events, finding results consistent with the fits to the whole $\phi\pi^+\pi^-$ sample. The final quoted values for the parameters of the $\phi(1680)$ and $Y(2175)$ resonances are in reasonable agreement between Belle and *BABAR* (considering slightly different modelling, for example (in)coherence of contributing amplitudes). In particular, for the mass and the width of the $Y(2175)$, Belle finds, respectively, $m_Y = (2.079 \pm 0.013^{+0.079}_{-0.028})$ GeV/$c^2$ and $\Gamma_Y = (192 \pm 23^{+25}_{-61})$ MeV, while *BABAR* finds a higher mass, $m_Y = (2.180 \pm 0.008 \pm 0.008)$ GeV/$c^2$, and a smaller width, $\Gamma_Y = (77 \pm 15 \pm 10)$ MeV.

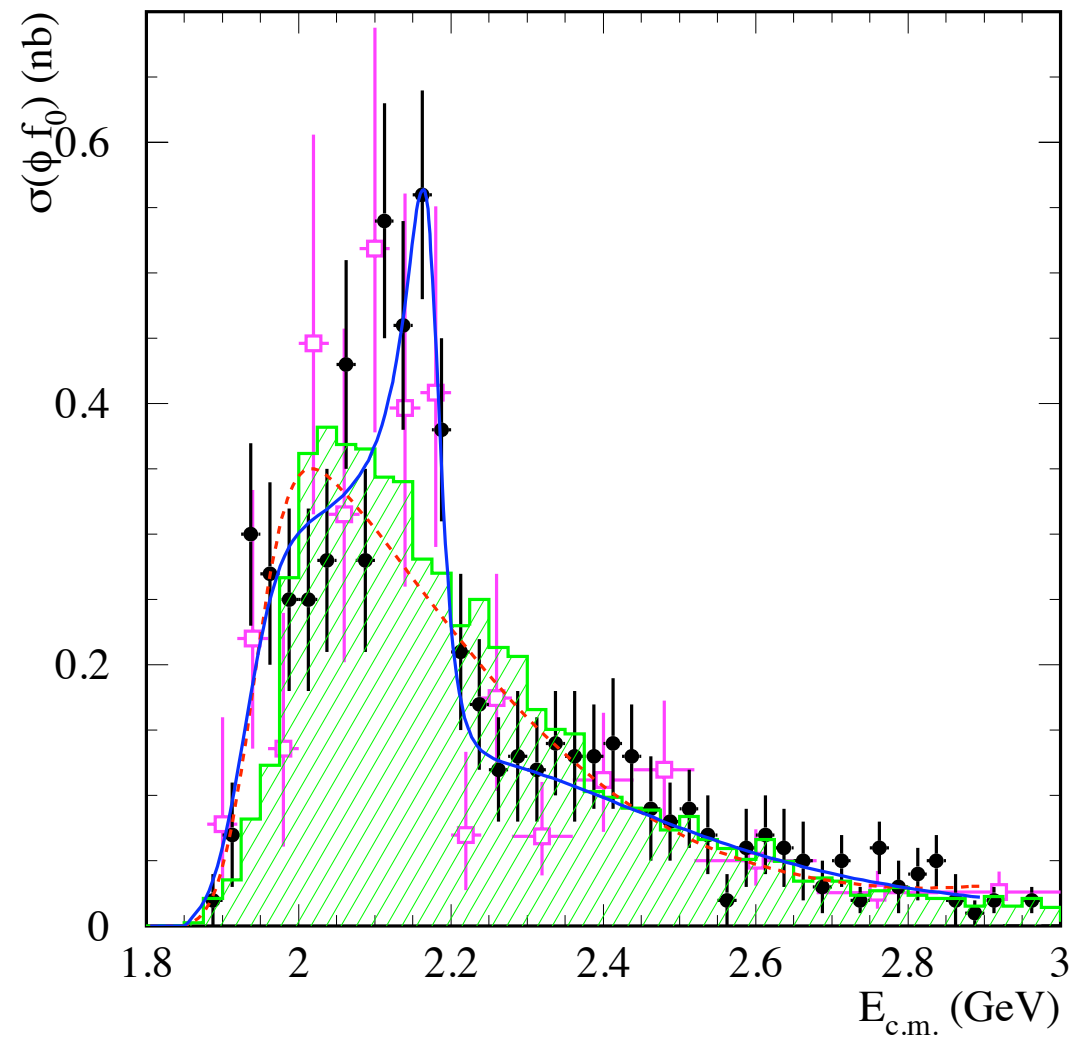

**Figure 21.3.14.** The $e^+e^- \to \phi f_0(980)$ cross sections measured in the $K^+K^-\pi^+\pi^-$ (circles) and $K^+K^-\pi^0\pi^0$ (squares) final states using an integrated luminosity of 232 fb$^{-1}$ by *BABAR* (Aubert, 2007bc). The hatched histogram shows the simulated cross section in the no-resonance hypothesis, which is consistent with a fit to the data with only a non-resonant component (dashed line). The solid line represents the result of the fit described in the text assuming the presence of the $Y(2175)$.

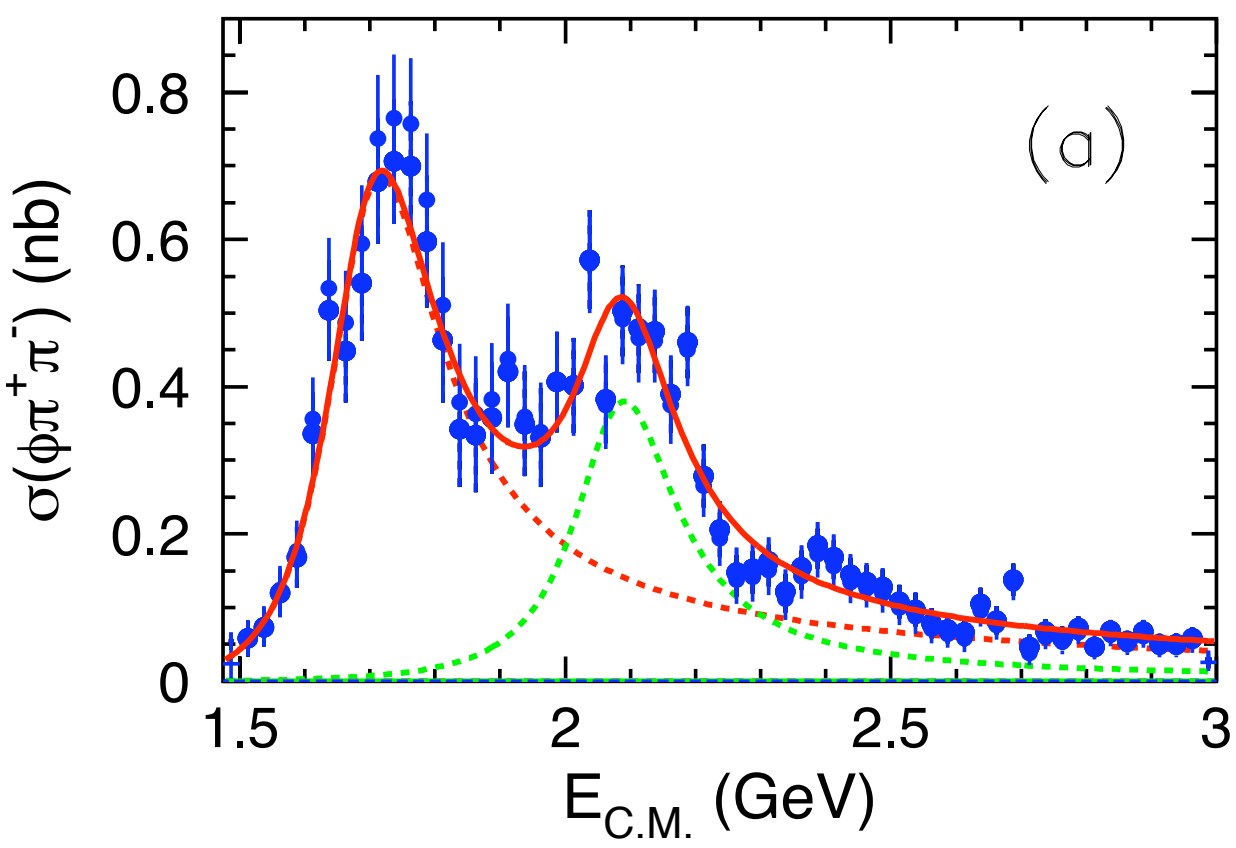

**Figure 21.3.15.** The $e^+e^- \to \phi\pi^+\pi^-$ cross section measured by Belle (Shen, 2009). The solid line is the result of a fit with two incoherent BW functions, one for the $\phi(1680)$ and the other for the $Y(2175)$. The dashed lines show the individual contributions of the two resonances.

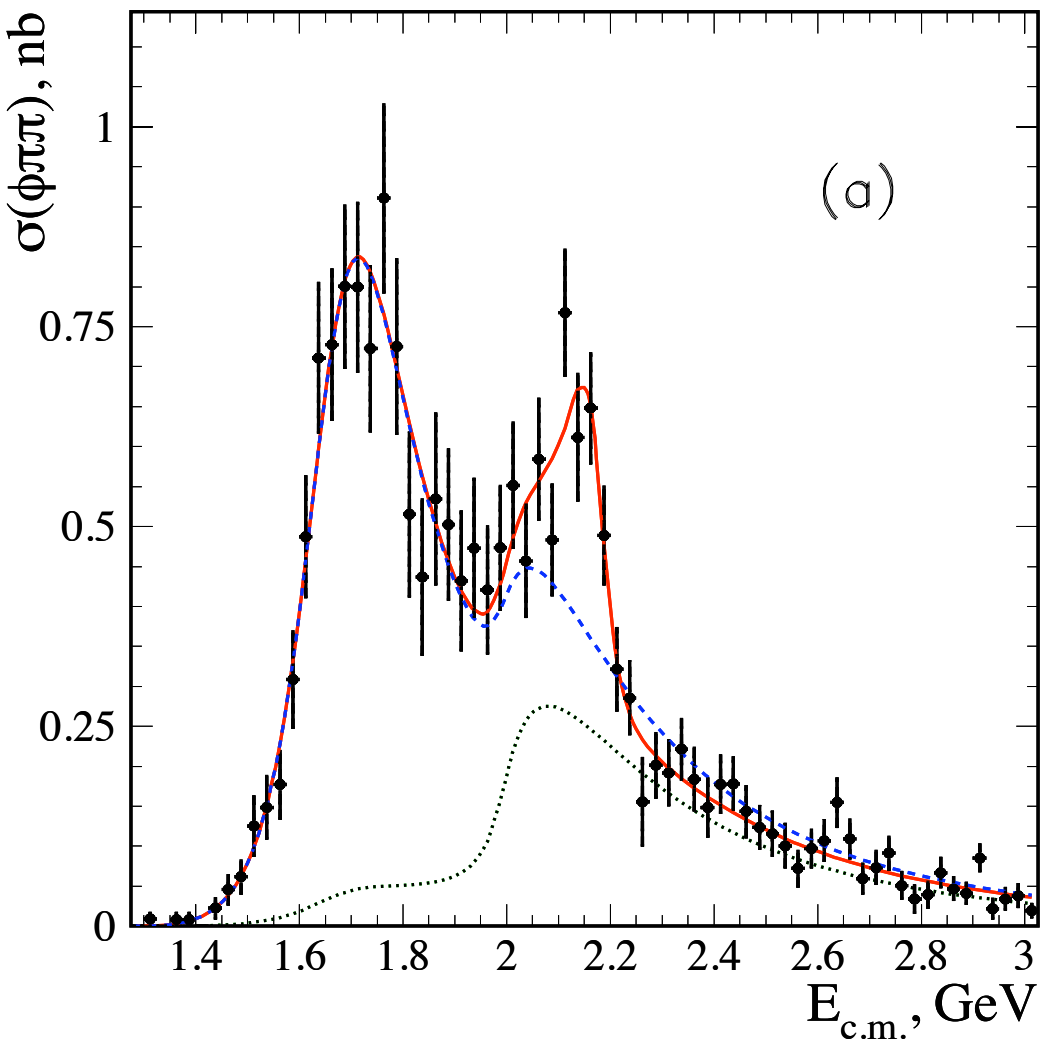

**Figure 21.3.16.** The fit to the $e^+e^- \to \phi\pi^+\pi^-$ cross section measured by *BABAR* (Lees, 2012d) in the two-resonance model described in the text (solid curve). The contribution of the first resonance ($\phi(1680)$) is shown by the dashed line. The dotted line shows the first resonance contribution in the $\phi f_0(980)$ decay mode only.

The nature of the $Y(2175)$ is still uncertain. The similar width ($\approx 200$ MeV) measured by Belle for the $\phi(1680)$ and the $Y(2175)$, even if with large uncertainties, may suggest that the $Y(2175)$ is a radially excited $s\bar{s}$ vector state. On the other hand, the significantly smaller width reported by *BABAR* and BES, and the different decay modes observed for the $\phi(1680)$ and the $Y(2175)$, do not favor this solution (Napsuciale, Oset, Sasaki, and Vaquera-Araujo, 2007). Several other interpretations have been proposed, such as a $s\bar{s}s\bar{s}$ four-quark state, or a gluon hybrid $s\bar{s}g$. For a review of this and other recently discovered hadrons see (Zhu, 2008) and references therein. The study of the $Y(2175)$ in other decay modes is needed to distinguish between the different possibilities.

#### 21.3.5.4 Summary of studies of $J/\psi$ and $\psi(2S)$ decays.

The clear $J/\psi$ and $\psi(2S)$ signals observed in the cross sections for $e^+e^-$ annihilation to almost all the final states presented in the previous sections allowed a systematic study of the decays of the two charmonium states to light hadrons with the *BABAR* detector.

The Born cross section for the production via ISR of a narrow resonance such as a $\psi$, and its subsequent decay to the final state $X$, is given by

$$\sigma_\psi = \frac{12\pi^2\Gamma(\psi \to e^+e^-)\mathcal{B}(\psi \to X)}{s m_\psi} W_0(s, x_\psi, \theta_0)\,, \tag{21.3.14}$$

where $m_\psi$, $\Gamma(\psi \to e^+e^-)$, and $\mathcal{B}(\psi \to X)$ are the mass of the $\psi$, its partial width to electrons, and its branching fraction to the final state $X$ respectively. The radiator function $W_0$ has been introduced in Section 21.2.1, and $x_\psi = 1 - m_\psi^2/s$ is the fraction of the CM energy carried by the photon in the case of radiative return to the $\psi$ mass. Therefore, for a given final state $X$, the product of the electronic width and the branching fraction $\Gamma(\psi \to e^+e^-)\mathcal{B}(\psi \to X)$ can be obtained by measuring the number of $\psi$ decays in the $e^+e^- \to X$ mass spectrum.

The samples of $J/\psi$ and $\psi(2S)$ available for these studies are significantly smaller than those collected at other facilities. The total cross section for $e^+e^- \to \gamma_{\rm ISR} J/\psi$ with a tagged ISR photon is about 3.4 pb, corresponding to 1.7 million $J/\psi$'s for an integrated luminosity of $\sim 500$ fb$^{-1}$; about 60 million $J/\psi$'s have been produced in the BES II experiment at the BEPC $e^+e^-$ collider. However, systematic uncertainties are significantly smaller at *BABAR* (typically 3–5%, compared to 10–15% at BES II), due in particular to superior particle identification.[161] As a result *BABAR* measurements are competitive and in many cases more accurate than previous data for all decays with branching fractions $O(10^{-3})$ and higher, and many decays to final states containing charged kaons have been studied for the first time. In summary, using the ISR method, the *BABAR* experiment has improved the precision on the measurement of a few tens of $J/\psi$ and $\psi(2S)$ branching fractions, and has observed about 20 new decay modes. A complete list of these results is found in the Review of Particle Physics (Beringer et al., 2012).

### 21.3.6 Search for $f_J(2220)$

Evidence for the $f_J(2220)$, a narrow resonance with a mass around 2.2 GeV/$c^2$ also known as $\xi(2230)$, was first presented by the Mark III Collaboration (Baltrusaitis et al., 1986). The $f_J(2220)$ was seen as a narrow signal above a broad enhancement in both $J/\psi \to \gamma f_J(2220)$, $f_J(2220) \to K^+K^-$ and $f_J(2220) \to K^0_S K^0_S$ decays with significance of 3.6 and 4.7 standard deviations, respectively. The BES Collaboration has also subsequently reported evidence in radiative $J/\psi$ decays at a comparable level of significance (Bai et al., 1996b), while searches for direct formation in $p\bar{p}$ collisions or two-photon processes were inconclusive (see for example Crede and Meyer, 2009, for an experimental review).

The unexpectedly narrow width of the $f_J(2220)$, approximately 20 MeV, triggered many conjectures about its nature (see for example Blundell and Godfrey, 1996). The possibility of a glueball (Ward, 1985), a bound state of gluons, is particularly attractive as several lattice QCD calculations predict a mass for the ground state $2^{++}$ glueball close to 2.2 GeV/$c^2$ (Chen and Su, 2004; Morningstar and Peardon, 1997). No glueball candidate has been unambiguously observed to date.

Based on a sample of 16 million $J/\psi$ mesons produced in ISR events, *BABAR* performed a search for $f_J(2220)$ production in radiative $J/\psi \to \gamma f_J(2220)$ decays (del Amo San-chez, 2010o). The $f_J(2220)$ is identified through its subsequent decay into a $K^+K^-$ or $K_S^0K_S^0$ pair. The ISR photon is not required to be detected.

[161] In 2009, the upgraded BESIII experiment at BEPCII has collected 225 million $J/\psi$ and 106 million $\psi(2S)$ events. The systematic error is significantly improved over BESII experiment (Ablikim et al., 2012b).

Requirements on particle identification, secondary vertex reconstruction, decay angles and global event information are used to improve the signal purity. The $J/\psi$ candidates are then fitted, constraining their mass and decay products to a common vertex. The fitted $K^+K^-$ and $K_S^0K_S^0$ mass spectra are shown in Fig. 21.3.17, together with the expected contributions of the inclusive $e^+e^- \to q\bar{q}(\gamma)$ $(q = u, d, s, c)$ background as well as $J/\psi \to \gamma f_2'(1525)$ and $\gamma f_0(1710)$ channels. A possible contribution from $J/\psi \to K^{*\pm}K^{\mp}$ decays was found to be negligible. Sideband data from the unconstrained $J/\psi$ mass distributions are used to model the non-resonant background. The sum of these components reproduces the data well at a global level. Remaining events are due mainly to generic $J/\psi$ decays producing additional undetected particles, mostly with pions misidentified as kaons in the charged mode.

The number of signal events is determined using an unbinned maximum likelihood fit in the range $1.9\,\mathrm{GeV}/c^2 < m_{KK} < 2.6\,\mathrm{GeV}/c^2$, fixing the mass and width of the $f_J(2220)$ to $2.231\,\mathrm{GeV}/c^2$ and $23\,\mathrm{MeV}$, respectively. No evidence of a $f_J(2220)$ signal is observed. Upper limits on the $J/\psi \to \gamma f_J(2220), f_J(2220) \to K^+K^-$ and $K_S^0K_S^0$ product branching fractions are derived at the 90% confidence level as a function of the spin and helicity assumed for the $f_J(2220)$. For all hypotheses of spin and helicity, these limits are below the central values reported by Mark III. Only one hypothesis (spin $J = 2$ and helicity $h = 0$) is compatible with the BES results for both final states, while all other possibilities are excluded.

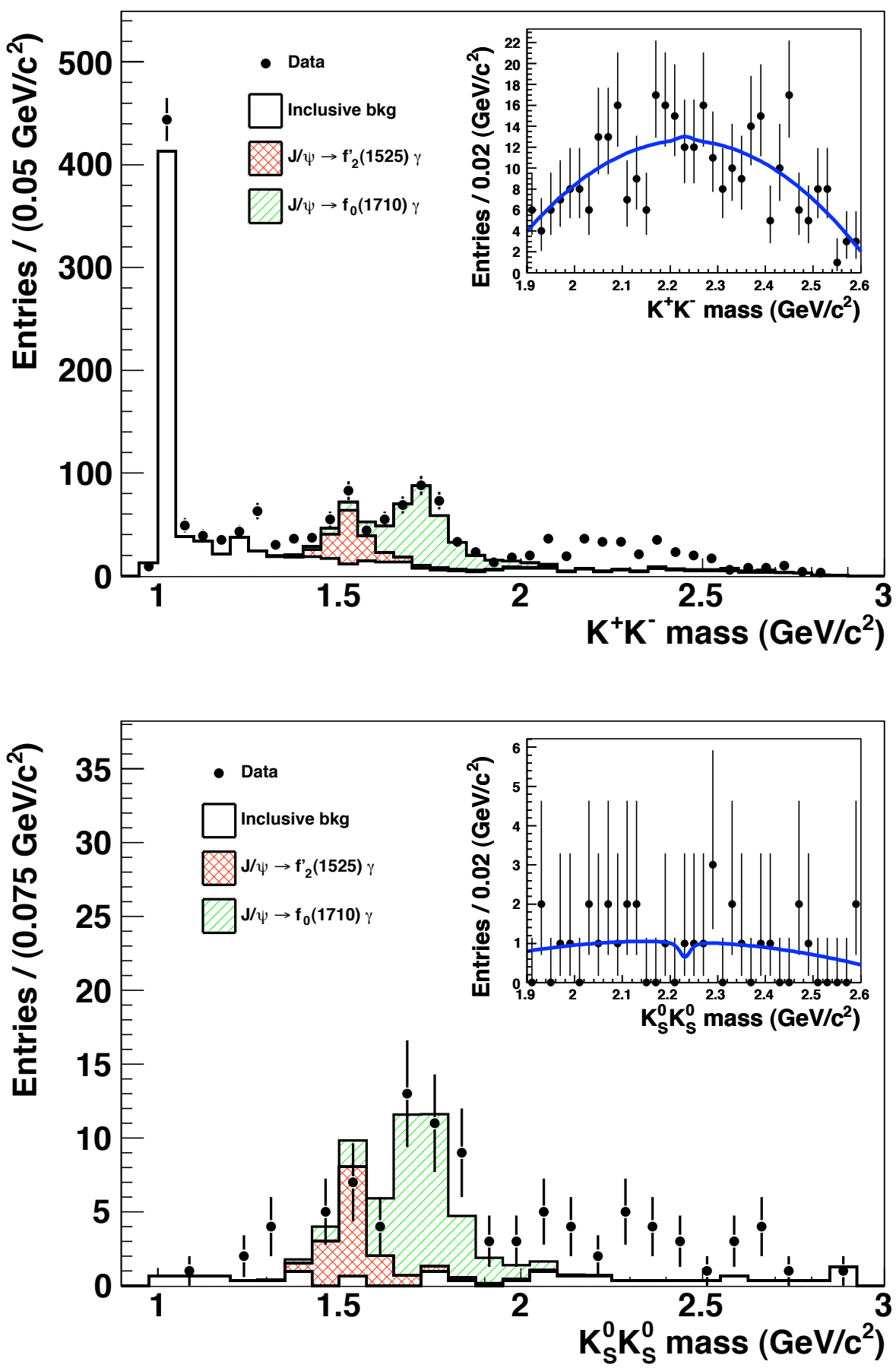

**Figure 21.3.17.** The $K^+K^-$ (top) and $K_S^0K_S^0$ (bottom) mass spectra obtained by BABAR in selected $J/\psi \to \gamma K^+K^-$ and $J/\psi \to \gamma K_S^0K_S^0$ decays, together with the expected contributions of the non-resonant background, $J/\psi \to \gamma f_2'(1525)$ and $J/\psi \to \gamma f_0(1710)$ reactions. The results of the fit to the data are shown in the inset (del Amo Sanchez, 2010o).

### 21.3.7 Measurement of time-like baryon form factors

Electromagnetic form factors (FFs) describe the modifications of pointlike photon-hadron vertices due to the structure of hadrons. The $e^+e^-$ annihilation in two hadrons or the electron scattering off a nucleon are described in QED by a product of an electronic and a hadronic electromagnetic currents. While the coupling of the photon with the electron is exactly calculable in QED, the coupling with the hadron is not. The dynamical content of the hadronic vertex can be however described with a set of form factors, and can be directly extracted from data.

#### 21.3.7.1 Nucleon Form Factors

The elastic scattering of an electron by a nucleon $e^-N \to e^-N$ is represented, in the Born approximation, by Fig. 21.3.18, with time flowing from the bottom to the top of the diagram. In this kinematic region the 4-momentum of the virtual photon is space-like and hence its squared value is negative: $q^2 = -2\omega_1\omega_2(1-\cos\theta_e) \le 0$, where $\omega_{1(2)}$ is the energy of the incoming (outgoing) electron and $\theta_e$ is the scattering angle. The same diagram, but with time flowing left to right (right to left), represents the annihilation $e^+e^- \to N\overline{N}$ $(N\overline{N} \to e^+e^-)$. For these processes the 4-momentum $q$ is time-like: $q^2 = (2\omega)^2 \ge 0$, where $\omega \equiv \omega_1 = \omega_2$ is the common value of the lepton energy in the $e^+e^-$ center-of-mass frame.

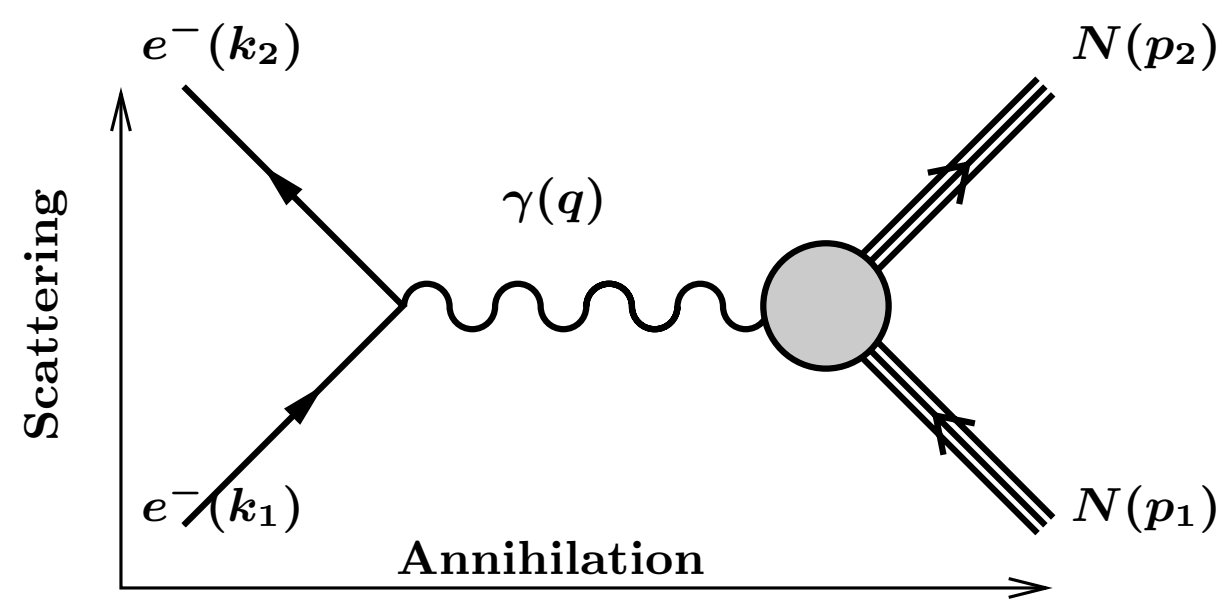

**Figure 21.3.18.** One-photon exchange Feynman diagram for scattering $e^-N \to e^-N$ and annihilation $e^+e^- \to N\overline{N}$.

The Feynman amplitude for the elastic scattering is

$$\mathcal{M} = \frac{1}{q^2}\left[e\,\overline{u}(k_2)\gamma^\mu u(k_1)\right]\left[e\,\overline{U}(p_2)\Gamma_\mu(p_1,p_2)U(p_1)\right], \tag{21.3.15}$$

where $k_i = (\omega_i,\, \boldsymbol{k}_i)$ and $p_i$ $(i = 1, 2)$ are the electron and nucleon four-vectors, $u$ and $U$ are the electron and nucleon spinors, and $\Gamma^\mu$ is a non-constant matrix which describes the nucleon vertex. Using gauge and Lorentz invariance the most general form of such a matrix is (Foldy, 1952)

$$\Gamma^\mu = \gamma^\mu F_1(q^2) + \frac{i\sigma^{\mu\nu}q_\nu}{2m}F_2(q^2), \tag{21.3.16}$$

where $m$ is the nucleon mass. $\Gamma^\mu$ depends on two Lorentz scalar functions of $q^2$, the Dirac ($F_1(q^2)$) and Pauli FFs ($F_2(q^2)$), that describe the helicity-conserving and the helicity-reversing parts of the hadronic current, respectively. Normalizations at $q^2 = 0$ follow from total charge and magnetic moment of the nucleon: $F_1(0) = Q_N$ and $F_2(0) = a_N$ respectively, where $Q_N$ is the electric charge (in units of $e$) and $a_N$ the anomalous magnetic moment of the nucleon $N$, in units of the nuclear magneton $e\hbar/2m$.[162]

Other pairs of FFs can be defined as combinations of $F_1$ and $F_2$: of particular interest are the so-called Sachs FFs $G_E$ and $G_M$ (Hand, 1963), defined as

$$\begin{aligned} G_E &= F_1 + \frac{q^2}{4m^2}F_2, \\ G_M &= F_1 + F_2\,. \end{aligned} \tag{21.3.17}$$

These expressions are obtained by considering the hadronic current in the Breit frame: the Fourier transformations of $G_E$ and $G_M$ give the spatial distributions of charge and magnetic moment of the nucleon; the normalizations are

$$G_E(0) = Q_N\,, \qquad G_M(0) = \mu_N\,, \tag{21.3.18}$$

where $\mu_N$ is the nucleon magnetic moment in units of the nuclear magneton.[163]

#### 21.3.7.2 Cross sections and data

The differential cross section of the annihilation process $e^+e^- \to N\overline{N}$, for unpolarized colliding beams, is given in the $e^+e^-$ CM by (Zichichi, Berman, Cabibbo, and Gatto, 1962)

$$\frac{d\sigma}{d\Omega} = \frac{\alpha^2\beta\,C}{4\,q^2}\left[(1+\cos^2\theta)\left|G_M(q^2)\right|^2 + \frac{1}{\tau}\sin^2\theta\left|G_E(q^2)\right|^2\right], \tag{21.3.19}$$

where $\tau = q^2/4m^2$, $\beta = \sqrt{1-1/\tau}$ is the velocity of the outgoing nucleon, and $C$ is the so-called Coulomb factor (Sakharov, 1948)

$$C = \begin{cases} \dfrac{\pi\alpha/\beta}{1-\exp(-\pi\alpha/\beta)} & \text{for } Q_N = 1 \\ 1 & \text{for } Q_N = 0\,, \end{cases} \tag{21.3.20}$$

which accounts for electromagnetic $N\overline{N}$ final state interactions; $C$ corresponds to the squared value of the Coulomb scattering wave function at the origin. Integration of Eq. (21.3.19) over the polar angle gives the total cross section:

$$\sigma(q^2) = \frac{4\pi\alpha^2\beta\,C}{3\,q^2}\left[\left|G_M(q^2)\right|^2 + \frac{1}{2\tau}\left|G_E(q^2)\right|^2\right]. \tag{21.3.21}$$

At production threshold $|G_E(4m^2)| = |G_M(4m^2)|$, while at high $q^2$ the contribution of $G_E$ to the cross section becomes negligible because of the suppression from the $1/2\tau$ term. Experiments usually quote measurements of $|G_M(q^2)|$ under the working hypothesis $|G_E(q^2)| = |G_M(q^2)|$, which is exactly true only at threshold. *BABAR* introduces an effective FF defined as

$$|F(q^2)|^2 = \frac{2\tau|G_M(q^2)|^2 + |G_E(q^2)|^2}{2\tau+1}, \tag{21.3.22}$$

which can be directly compared to the previous measurements of $|G_M(q^2)|$. The total cross section is thus written as

$$\sigma(q^2) = \frac{4\pi\alpha^2\beta C}{3q^2}\left(1+\frac{1}{2\tau}\right)|F(q^2)|^2\,. \tag{21.3.23}$$

Simultaneous extraction of $|G_E|$ and $|G_M|$ requires precise determination of the $|G_E/G_M|$ ratio, which is possible in principle by measuring the angular distributions of the outgoing particles, using Eq. (21.3.19). A $|G_E/G_M|$ measurement for the proton is discussed in Section 21.3.7.4 below.

#### 21.3.7.3 Measurement of $e^+e^- \to p\overline{p}$

The *BABAR* experiment performed an ISR tagged analysis of $e^+e^- \to \gamma_{\rm ISR}p\overline{p}$ based on a sample of $232\,{\rm fb}^{-1}$ of data, and selected more than 4000 events, measuring the $e^+e^- \to p\overline{p}$ cross section with highest accuracy in the range of $p\overline{p}$ invariant mass from threshold up to about $M^2_{p\overline{p}} = 4.5\ {\rm GeV}/c^2$ (Aubert, 2006d). As for the other ISR processes previously discussed, the reconstructed hadronic mass gives the CM energy of the hadronic process: $M^2_{p\overline{p}} = q^2$. Figure 21.3.19 shows a comparison of the measured cross section for $M_{p\overline{p}} < 2.9\ {\rm GeV}/c^2$ with previous measurements performed with lower precision by Castellano et al. (1973) and the Fenice experiment (Antonelli et al., 1998) at Adone, by DM1 (Delcourt et al., 1979) and DM2 (Bisello et al., 1990), and more recently by CLEO (Pedlar et al., 2005) and BES (Ablikim et al., 2005a).

[162] For proton $F_1(0) = 1$, $F_2(0) = 1.7928$, and for neutron $F_1(0) = 0$, $F_2(0) = -1.9130$.

[163] For proton $G_M(0) = 2.7928$ and for neutron $G_M(0) = -1.9130$.

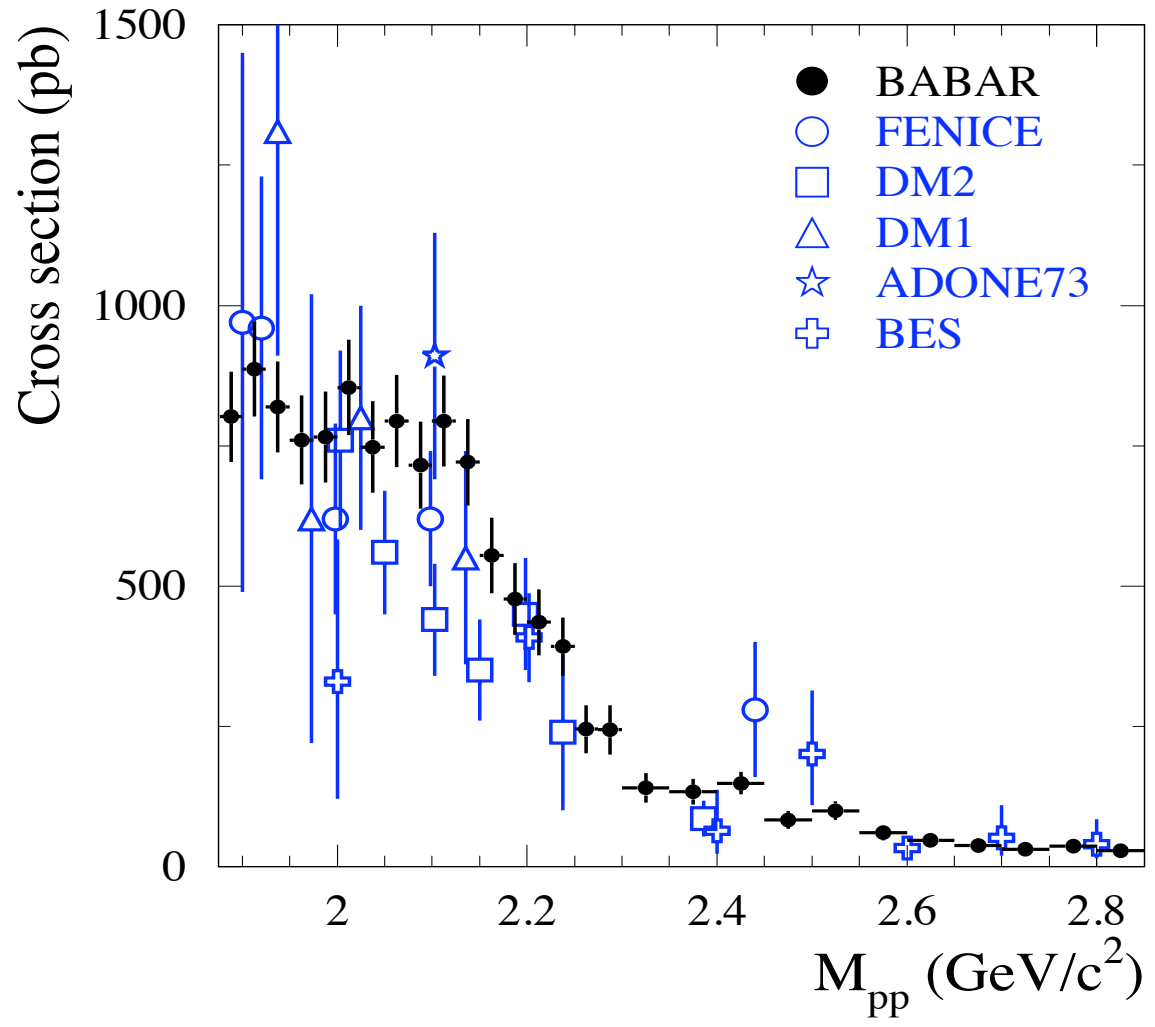

**Figure 21.3.19.** The $e^+e^- \to p\overline{p}$ cross section measured by *BaBar* (Aubert, 2006d) in the energy region from threshold to about 3 GeV. Data from previous experiments are shown for comparison.

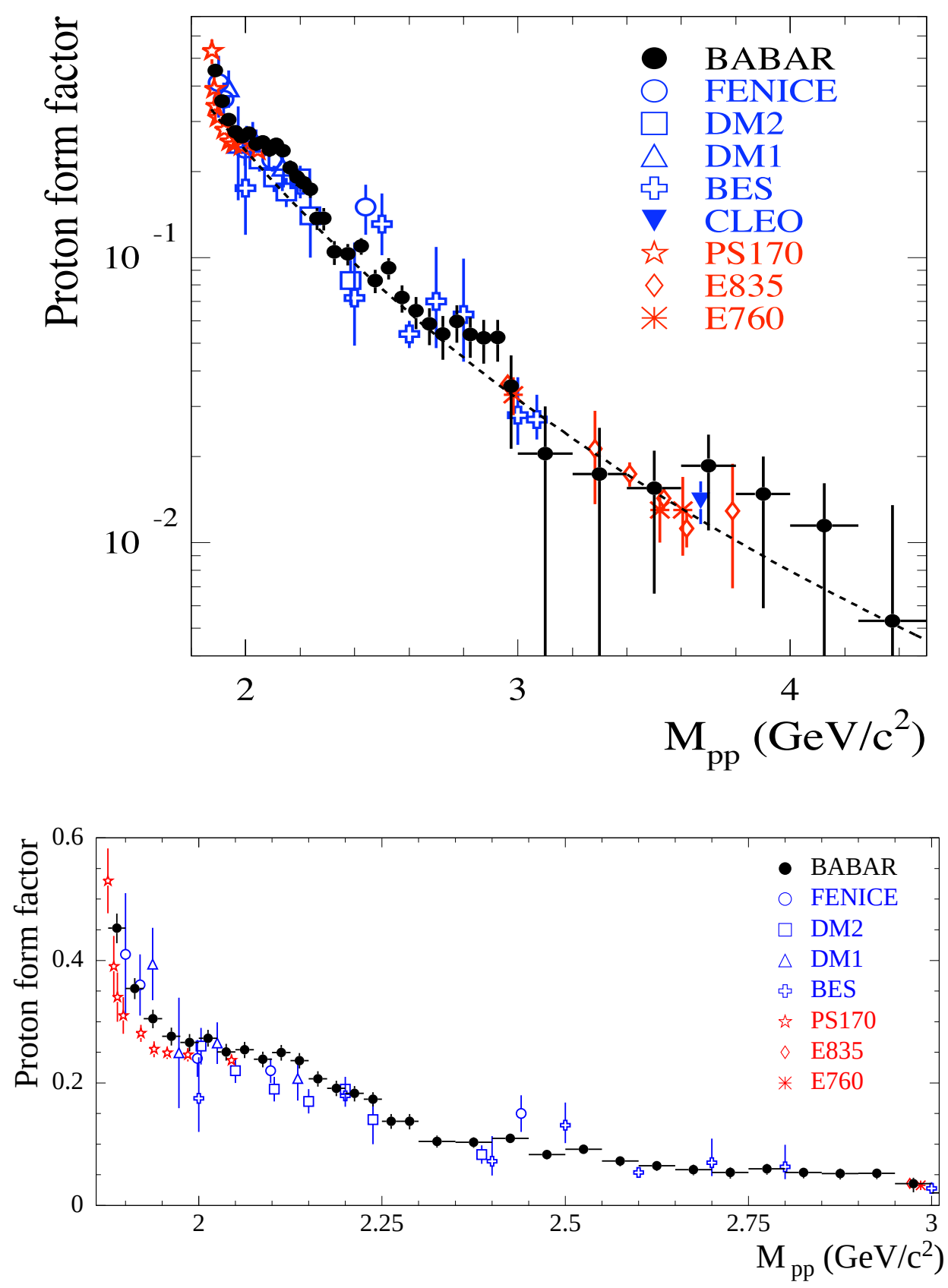

**Figure 21.3.20.** The proton form factor measured by *BaBar* (Aubert, 2006d) compared to data from previous $e^+e^- \to p\overline{p}$ (blue online) and $p\overline{p} \to e^+e^-$ (red online) experiments. The dashed line is the fit to all available data according to the perturbative QCD prediction. The bottom plot shows an expanded view of the energy region below 3 GeV.

One interesting feature, now apparent due to the precision of the data, is that the $p\overline{p}$ cross section is nearly constant at 0.8 nb for energies up to around 200 MeV above the threshold, before falling with increasing energy. The corresponding effective form factor results are presented in Fig. 21.3.20 as a function of $M_{p\overline{p}}$, together with results obtained from the $p\overline{p} \to e^+e^-$ experiments PS170 (Bardin et al., 1994), E760 (Armstrong et al., 1993), and E835 (Ambrogiani et al., 1999; Andreotti et al., 2003). General consistency among all these data is observed: in particular the precise data from *BaBar* and from the PS170 experiment ($p\overline{p}$ annihilation at LEAR) show a similar rise in the form factor when the energy approaches $p\overline{p}$ threshold. Several explanations of this intriguing feature have been proposed, such as that the sharp rise is due to final state interaction of the proton and antiproton (see Dmitriev and Milstein, 2007, and references therein); or that it is due to a contribution from a vector-meson resonance with a mass of about 1.9 GeV/$c^2$, just below the $p\overline{p}$ production threshold (such a state is observed as a dip in the cross section of the $e^+e^- \to 6\pi$ processes: Aubert, 2006az; Frabetti et al., 2001). The hypothesis of an incorrect evaluation of the Coulomb factor has also been advanced: see for example Ferroli, Pacetti, and Zallo (2012). The dashed line in Fig. 21.3.20 is the result of a fit of the form factor data according to the function $F_{\rm QCD} = A/(m^4 \log^2(m^2/\Lambda^2))$, which correspond to the perturbative QCD prediction for the asymptotic behavior of the baryon form factors (Chernyak and Zhitnitsky, 1977; Lepage and Brodsky, 1979b). Here, $\Lambda = 0.3$ GeV and $A$ is a free parameter of the fit. It is seen that the asymptotic result provides a reasonable description of the data even at these energies.

#### 21.3.7.4 Measurement of $|G_E^p/G_M^p|$

The distribution of the proton helicity angle in the $p\overline{p}$ rest frame for $e^+e^- \to \gamma_{\rm ISR} p\overline{p}$ events can be written as a function of the ratio of the electric and magnetic form factors:

$$\frac{dN}{d\cos\theta_p} = A\left(H_M(\cos\theta_p, q^2) + \left|\frac{G_E}{G_M}\right|^2 H_E(\cos\theta_p, q^2)\right). \tag{21.3.24}$$

The functions $H_E$ and $H_M$, which are determined using MC simulation, do not strongly differ from the terms $\sin^2\theta_p$ and $1+\cos^2\theta_p$ that appear in Eq. (21.3.19). The mass region from $p\overline{p}$ threshold up to 3 GeV/$c^2$ is divided into six intervals. The angular distribution in each interval is then fitted to Eq. (21.3.24), with $A$ and $|G_E/G_M|$ as free parameters. The functions $H_E$ and $H_M$ are modeled with the histograms obtained from MC simulation with the $p\overline{p}$ selection applied. The values of the ratio $|G_E/G_M|$ obtained by Aubert (2005ah) are significantly larger than unity, as shown in Fig. 21.3.21, in disagreement with previous results from PS170 at LEAR (Bardin et al., 1994).

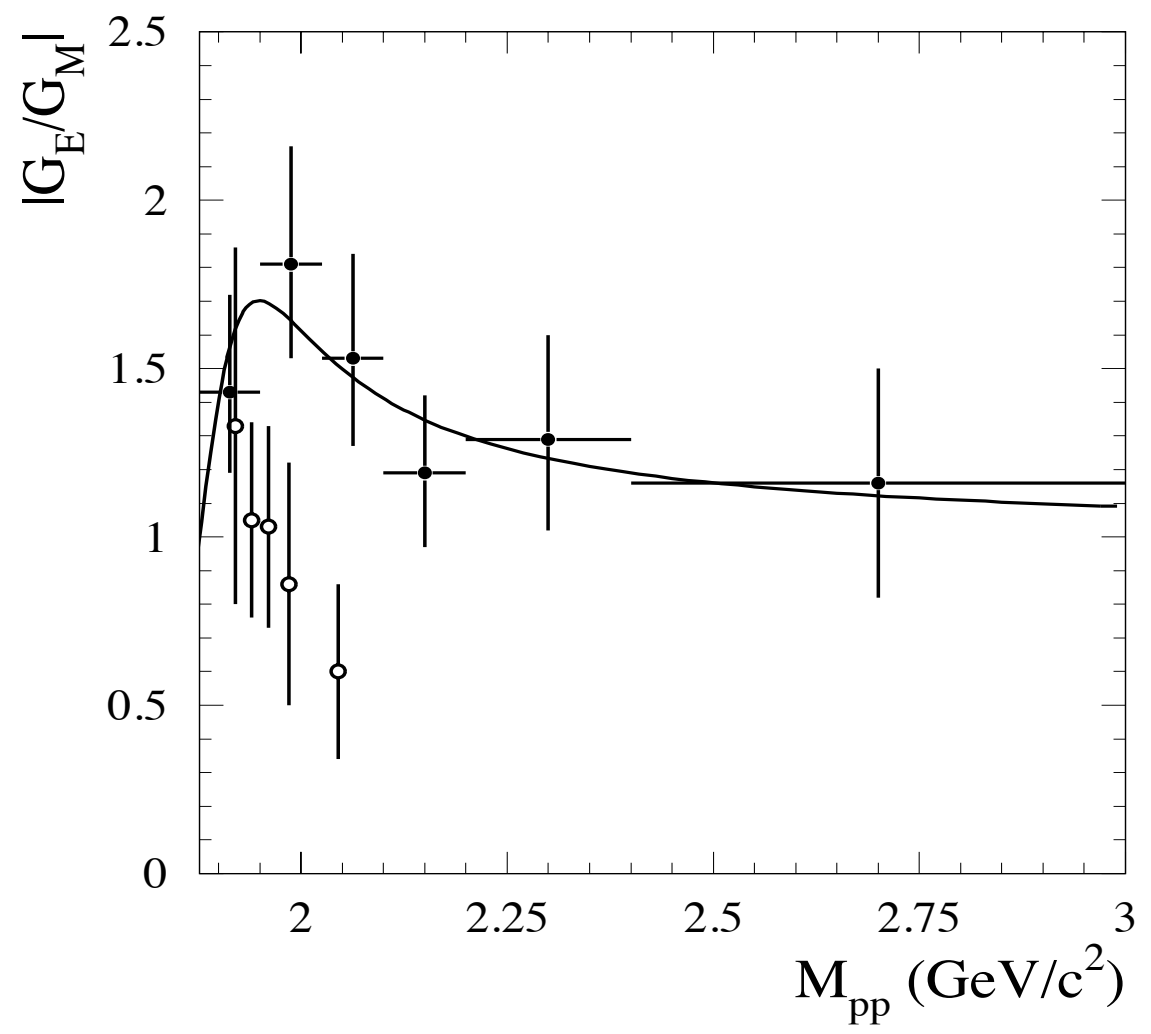

**Figure 21.3.21.** The proton $|G_E/G_M|$ ratio measured by *BABAR* (Aubert, 2006d; solid points) compared with PS170 data (Bardin et al., 1994; open circles). The curve is a result of a fit to *BABAR* data with the function $(1 + ax/(1 + bx^2))$.

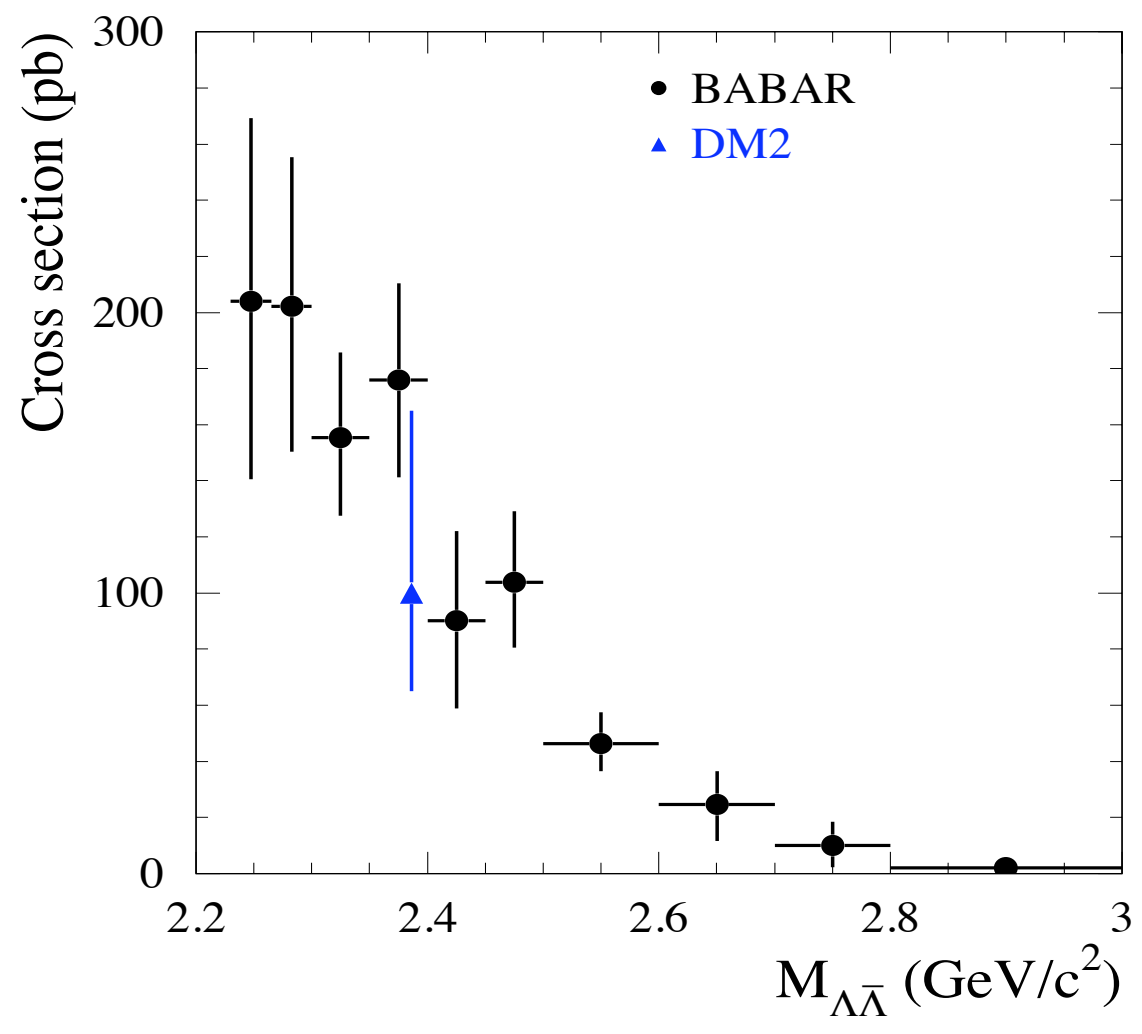

**Figure 21.3.22.** The $e^+e^- \to \Lambda\bar{\Lambda}$ cross section measured by *BABAR* (Aubert, 2007az) in comparison with the DM2 measurement (Bisello et al., 1990).

We should note that the PS170 data were limited by incomplete angular acceptance ($|\cos\theta_p| < 0.8$) and were affected by strong angular dependence of the detection efficiency: this limitation is not present when a tagged ISR analysis is performed.

#### 21.3.7.5 Measurement of strange baryon form factors.

The *BABAR* Collaboration has also studied the production via ISR of several final states made of strange baryon pairs, namely $e^+e^- \to \Lambda\bar{\Lambda}$, $\Sigma\bar{\Sigma}$, and $\Lambda\bar{\Sigma}^0$ ($\Sigma^0\bar{\Lambda}$) (Aubert, 2007az). Only a single measurement of $\Lambda\bar{\Lambda}$ production at 2.386 GeV, by the DM2 Collaboration, and upper limits on the other process, were previously available. The $\Lambda$ is reconstructed in the $p\pi^-$ decay, while for the $\Sigma$ the decay chain $\Sigma \to \Lambda\pi$, $\Lambda \to p\pi^-$ is used. The cross section measured for $e^+e^- \to \Lambda\bar{\Lambda}$, based on about 200 selected events, after background subtraction, is shown in Fig. 21.3.22. The measured cross section is consistent with a behavior similar to that in $p\bar{p}$ production: an almost flat distribution from threshold over a $\sim 200$ MeV range, and in particular a value different from zero at threshold. This behavior is contrary to expectations, as the Coulomb factor plays no role for neutral final-state particles. However, the large uncertainties due to the limited sample can not exclude a vanishing cross section at threshold.

The study of the angular distribution of produced $\Lambda$'s allows the measurement of $|G_E/G_M|$: the ratio is consistent with unity within the large statistical uncertainty.

Should the relative phase $\phi$ between $G_E$ and $G_M$ be different from zero, polarization of the outgoing baryons, perpendicular to the scattering plane of the process $e^+e^- \to \mathfrak{B}\overline{\mathfrak{B}}$, is expected (see Dubnickova, Dubnicka, and Rekalo, 1996, and Czyz, Grzelinska, and Kühn, 2007, for the specific case of ISR production). While the polarization of the outgoing protons in the $p\bar{p}$ production case cannot be measured in *BABAR*, the $\Lambda$ polarization $\zeta$ can be extracted from the proton angular distribution in the $\Lambda \to p\pi^-$ decay,

$$\frac{dN}{d\cos\theta_{p\zeta}} = A\,(1 + \alpha_\Lambda \zeta \cos\theta_{p\zeta})\;, \qquad (21.3.25)$$

where $\theta_{p\zeta}$ is the angle between the polarization axis and the proton momentum in the $\Lambda$ rest frame, and $\alpha_\Lambda = 0.642 \pm 0.013$ (Yao et al., 2006) the decays asymmetry parameter, with $\alpha_{\bar{\Lambda}} = -\alpha_\Lambda$. For unpolarized events in the MC simulation, the distribution is consistent with being isotropic (flat in $\cos\theta_{p\zeta}$), as expected.

The fit to the background subtracted data distribution returns a slope of $0.020 \pm 0.097$. Under the assumption of $|G_E| = |G_M|$ this measurement can be converted into a 90% C.L. interval for the relative phase of the two form factors: $-0.76 < \sin\phi < 0.98$. The obtained limits are very weak, but the method has been proven to work and could give interesting results when significantly larger samples become available.

Fig. 21.3.23 shows the strange-baryon effective form factors measured by *BABAR*. About 20 candidate events have been selected for both $e^+e^- \to \Sigma\bar{\Sigma}$ and $e^+e^- \to \Sigma^0\bar{\Lambda}$. It is seen that the $\Lambda$, $\Sigma^0$, and $\Sigma^0\Lambda$ form factors are of the same order. For comparison, the proton FF measured by *BABAR* is also shown: the energy-dependence of the $\Lambda$ and proton FFs differ. A fit to the $\Lambda$ FF with the power-law function $F(Q^2) \sim Q^{-n}$ returns $n \simeq 9$, showing that the asymptotic regime ($n = 4$) predicted by perturbative-QCD is not reached in the measured energy range below 3 GeV.

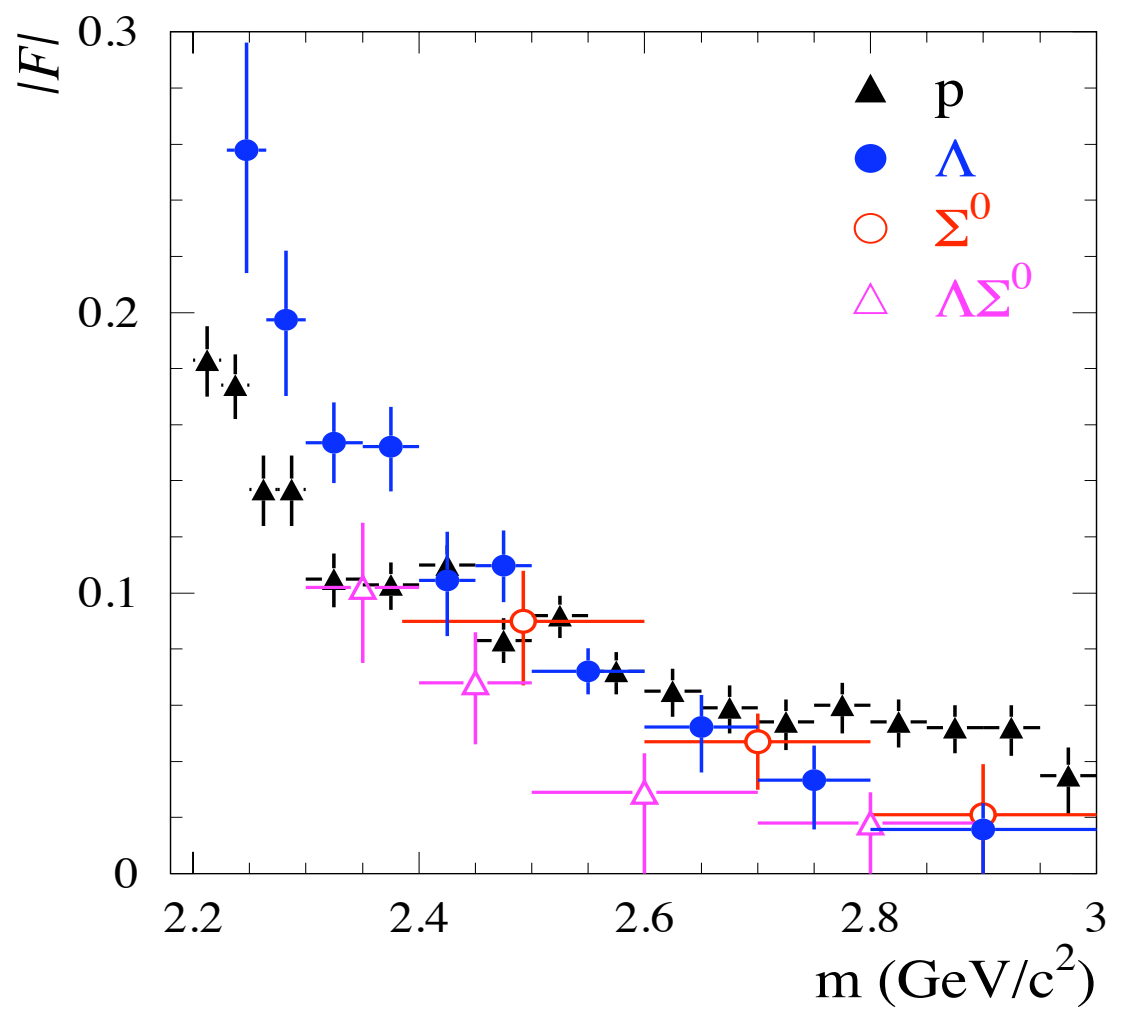

**Figure 21.3.23.** The baryon form factors measured by BABAR versus the dibaryon invariant mass. Data taken from Aubert (2005ah, 2007az).

## 21.4 Open charm production

Due to intriguing discoveries of exotic charmonium like states (see Secs. 21.5 and 18.3) with masses at which conventional charmonium states are expected to decay predominantly into pairs of open charm mesons it is of great interest to explore the ISR production of these final states.

Exclusive production of open charm has been studied in two different regimes at the $B$ Factories: far from threshold, in $e^+e^-$ annihilation at the CM energy of the collider (Section 21.4.1), and from threshold up to 5–6 GeV, depending on the final state, using the ISR technique (Sections 21.4.2–21.4.6). There is no general approach to these measurements: the method that yields maximum significance needs to be determined case-by-case. The measurement of $e^+e^- \to D^{(*)+}D^{(*)-}$ production at the CM energy (Section 21.4.1) introduced a partial reconstruction method that exploits the special properties of the $D^*$ decay. Here $D^{(*)}$ denotes a $D$ or $D^*$ meson. Measurement of $D\overline{D}$ production in ISR events, on the other hand, was performed by full reconstruction of both charmed mesons (Section 21.4.2). Analyses of $D^{(*)+}D^{*-}$ (Section 21.4.3), charmed-strange (Section 21.4.4), and three-body charmed meson cross sections (Section 21.4.5), and a study of charmed baryon production (Section 21.4.6), use variants of these techniques.

In ISR events of this type, the continuous spectrum of photons emitted from the initial state provides access to a range of energies above open charm threshold. This allows the measurement of cross sections without the additional systematic uncertainties due to variation of the detector and machine conditions from one energy point to another during the relatively long time of the data collection. However, the electromagnetic suppression of ISR processes and the reduced reconstruction efficiency due to the event topology present considerable challenges. Reconstruction of exclusive production at the CM energy does not face these problems, but in this regime the cross section itself is low. Together with the very high luminosities available at the $B$ Factories, careful choice of analysis methods has allowed many exclusive charmed hadron cross sections to be measured over a wide energy range, with good accuracy.

### 21.4.1 Measurement of exclusive $D^{(*)+}D^{(*)-}$ production far from threshold

Knowledge of exclusive charmed meson production in $e^+e^-$ annihilation is rather poor. The only exception is the near-threshold region, where the small phase space limits the number of particles in the final state. Heavy quark effective theory (HQET), based on heavy-quark spin symmetry, provides a description of these processes in terms of a universal form factor called the Isgur-Wise function. For large $q^2$, however, the leading-twist contribution, violating this symmetry, becomes dominant. In the intermediate-$q^2$ region the contribution of the symmetry-violating terms remains significant. A calculation that takes this effect into account (Grozin and Neubert, 1997) predicts cross sections of about $2.5\,\mathrm{pb}^{-1}$ for the $e^+e^- \to D_T^*\overline{D}_L^*$ and $e^+e^- \to D\overline{D}^*$ processes, where the subscripts T and L indicate transverse and longitudinal polarization of the $D^*$, respectively. The cross section of the $e^+e^- \to D\overline{D}$ process is estimated to be at least 1000 times smaller.

Exclusive meson production in $e^+e^-$ annihilation is difficult to study experimentally due to its extremely low cross-section and reconstruction efficiency. A partial reconstruction method is therefore used: one $D^{(*)}$ meson is fully reconstructed while the other remains unreconstructed. For definiteness, suppose the $D^{(*)+}$ is the fully reconstructed meson. The distribution of recoil mass $M_{\mathrm{recoil}}$, where

$$M^2_{\mathrm{recoil}}(D^{(*)+}) = (E_{\mathrm{CM}} - E(D^{(*)+}))^2 - p^2(D^{(*)+}), \tag{21.4.1}$$

can be used for identification of the process. Signal events are expected to cluster around the mass of the unreconstructed $D^{(*)-}$ meson. This method provides better reconstruction efficiency than the exclusive reconstruction of the event, but the background level is also much higher. If the unreconstructed meson is a $D^*$, one can reconstruct one of its decay products—usually the pion, $\pi^-_{slow}$, in $D^{*-} \to \overline{D}^0\pi^-_{slow}$—and construct the recoil mass difference

$$\Delta M_{\mathrm{recoil}} = M_{\mathrm{recoil}}(D^{(*)+}) - M_{\mathrm{recoil}}(D^{(*)+}\pi^-_{slow}). \tag{21.4.2}$$

Since most of the uncertainties cancel in the difference, the peak at the nominal mass difference $m_{D^*} - m_D$ in the $\Delta M_{\mathrm{recoil}}$ distribution remains narrow ($\sim 1\,\mathrm{MeV}/c^2$). The width is determined mostly by the slow pion reconstruction accuracy. The use of the recoil mass difference as a discriminating or a signal variable is a powerful tool for background suppression. It was used by the Belle collaboration in ISR analyses (see Section 21.4.3) and in the study of $D$-meson semileptonic decays (Widhalm, 2006; see Section 19.1.5).

Exclusive $e^+e^- \to D^{(*)+}D^{(*)-}$ processes have been studied by the Belle collaboration using $89\,\mathrm{fb}^{-1}$ of data (Uglov, 2004). Recoil mass spectra for events with recoil mass difference lying within $2\,\mathrm{MeV}/c^2$ of the nominal value are shown in Fig. 21.4.1(a) and (b) for $D^{*+}D^{*-}$ and $D^+D^{*-}$ final states. The recoil mass spectrum for the $e^+e^- \to D^+D^-$ process is shown in Fig. 21.4.1(c). The spectra are fitted with a sum of signal and background functions. ISR photon emission produces a tail in $M_{\rm recoil}$, which must be taken into account. The approach used is to divide events in the Monte Carlo into those with ($E_{\rm ISR} > 10\,\mathrm{MeV}$) and without ($E_{\rm ISR} < 10\,\mathrm{MeV}$) an ISR photon of significant energy, and to determine separate signal shapes for the two categories. The final measurement is of the Born cross-section, which does not depend on the ISR photon cutoff value, although experimental estimation of the ISR fraction does contribute to the systematic uncertainty.

The signal is parameterized as a sum of a signal Gaussian and a Monte Carlo-based shape for events with signficant ISR. The background is described by a threshold function $\alpha\cdot(x - m_{D^{*-}} - m_{\pi^0})^\beta$, where $\alpha$ and $\beta$ are fit parameters. Both signal and background contributions are convolved with the detector resolution. In the $D^{*+}D^{*-}$ and $D^+D^{*-}$ final states, the $D^*$ mesons can have either longitudinal or transverse polarization. To distinguish these cases the angular distributions of the $D^*$ decays are fitted with the sum of the Monte Carlo shapes for all possible polarizations. The fit results and extracted cross sections are summarized in Table 21.4.1.

Systematic uncertainties for $D^{*+}D^{*-}$ and $D^+D^{*-}$ final states, shown in Table 21.4.2, are dominated by the uncertainty in the tracking efficiency, and the estimation of the fraction of events with a hard ISR photon. The main uncertainty in the $e^+e^- \to D^+D^-$ cross-section is due to the non-resonant $e^+e^- \to D^+\overline{D}\pi$ process and cannot be reliably estimated.

The measured $e^+e^- \to D_T^{*+}D_L^{*-}$ and $D^+D_T^{*-}$ cross sections are 3–4 times smaller than predicted in Grozin and Neubert (1997). The upper limits set for the $e^+e^- \to D_L^{*+}D_L^{*-}$ and $D_T^{*+}D_T^{*-}$ processes are also lower than the HQET prediction; the limit on the $e^+e^- \to D^+D^-$ process does not contradict the prediction. Unlike the absolute values, the cross-section ratios (on which the theoretical uncertainties are much smaller) are in agreement with the measured values. Calculations in the pQCD framework (Liu, He, Zhang, and Chao, 2010) are in agreement with the measured $e^+e^- \to D^{*+}D^{*-}$ and $D^+D^{*-}$ cross-sections, although the predicted $D^*$ polarizations are far from the measured values. Another pQCD calculation taking into account the hard part of the meson wave function only (Berezhnoy and Likhoded, 2005) represents the data well, but being a rough estimation, requires a more accurate approach.

**Table 21.4.1.** Fit results and Born cross-sections for $e^+e^- \to D^{(*)+}D^{(*)-}$ from Uglov (2004). Upper limits are determined at the 90% confidence level. HQET predictions, which are approximate, are taken from Grozin and Neubert (1997); the $D_L^{*+}D^-$ final state is forbidden within this scheme. The pQCD prediction is from Liu, He, Zhang, and Chao (2010).

| final state | signal events | cross-section (pb) | HQET (pb) | pQCD (pb) |
|---|---|---|---|---|
| $D_T^{*+}D_T^{*-}$ | $5^{+15}_{-13}$ | $< 0.02$ | 0.05 | |
| $D_T^{*+}D_L^{*-}$ | $708 \pm 36$ | $0.55 \pm 0.03$ | 3.0 | 0.347 |
| $D_L^{*+}D_L^{*-}$ | $4^{+18}_{-17}$ | $< 0.02$ | 0.1 | |
| $D_T^{*+}D^-$ | $433 \pm 24$ | $0.62 \pm 0.03$ | 3.0 | 0.699 |
| $D_L^{*+}D^-$ | $-1.5 \pm 2.2$ | $< 0.006$ | – | |
| $D^+D^-$ | $-13 \pm 24$ | $< 0.04$ | 0.006 | 0.098 |

**Table 21.4.2.** Systematic uncertainty in the $D^{(*)+}D^{*-}$ cross-sections. From Uglov (2004).

| Source | $D^{*+}D^{*-}$ | $D^+D^{*-}$ |
|---|---|---|
| Tracking efficiency | 9% | 8% |
| Estimation of ISR-events fraction | 5% | 5% |
| $\mathcal{B}(D^{(*)})$ | 4% | 8% |
| $K/\pi$ misidentification | 2% | 2% |
| Background estimation | $^{+1}_{-0}\%$ | $^{+1}_{-0}\%$ |
| Form-factor energy dependence | 0.8% | 0.8% |
| Total | 11% | 13% |

## 21.4.2 Measurement of the $D\overline{D}$ cross section via full reconstruction

The simplest way to select signal events in the $e^+e^- \to D\overline{D}\gamma_{\rm ISR}$ process (where $D = D^0$ or $D^+$) is a full reconstruction of the final state, *i.e.* reconstruction of both $D$ and $\overline{D}$ mesons, and the ISR photon. Although this tagged ISR method provides almost complete background suppression due to the specific event topology (the final state contains an energetic photon of $(4-5)\,\mathrm{GeV}$ and a pair of charmed mesons), the efficiency is low as the photon escapes detection in $\sim 90\%$ of events. To increase the efficiency the presence of the ISR photon is inferred using energy-momentum conservation. The $\gamma_{\rm ISR}$ signature in this case is a peak at zero in the spectrum of the square of the mass recoiling against the reconstructed $D\overline{D}$ system:

$$M^2_{\rm recoil}(D\overline{D}) = (E_{\rm CM} - E_{D\overline{D}})^2 - p^2_{D\overline{D}}. \quad (21.4.3)$$

Here $E_{D\overline{D}}$ and $p_{D\overline{D}}$ are the CM energy and momentum of the $D\overline{D}$ combination, respectively. Good momentum resolution of the reconstructed charmed mesons provides a narrow peak in the recoil mass squared distribution, and a low background level. The remaining background contribution from $e^+e^- \to D\overline{D}(n)\pi\gamma_{\rm ISR}$ processes can be strongly suppressed by excluding events containing additional charged tracks not used in the $D$ or $\overline{D}$ reconstruction. To suppress the tail of the $e^+e^- \to D^{(*)}\overline{D}^{(*)}(n)\pi^0\gamma_{\rm ISR}$ spectrum, a tight requirement on $|M^2_{\rm recoil}(D\overline{D})|$ is applied.

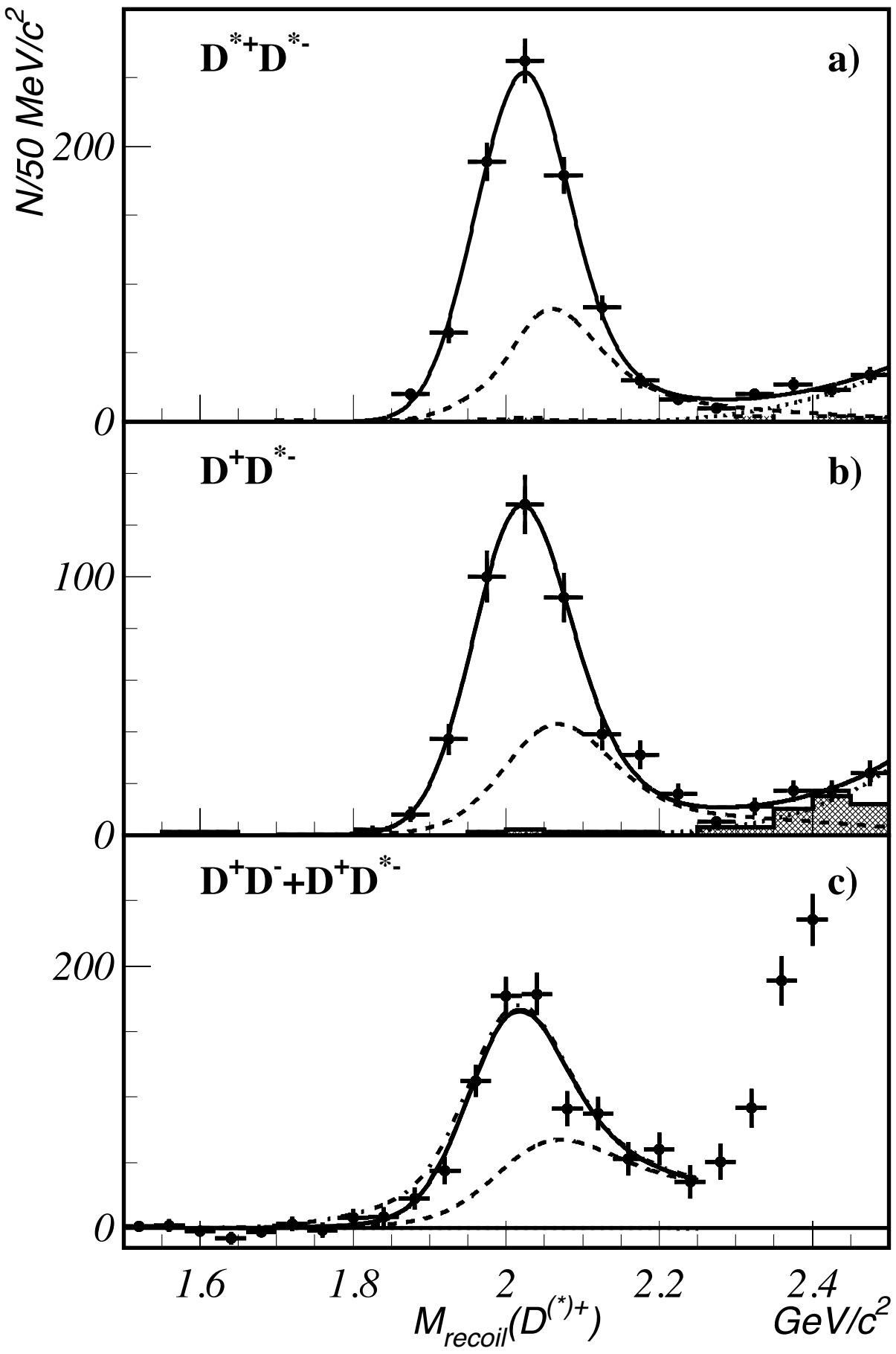

**Figure 21.4.1.** Recoil mass spectra for: a) $e^+e^- \to D^{*+}D^{*-}$ and b) $e^+e^- \to D^+D^{*-}$ processes with a $\Delta M_{\rm recoil}$ requirement, respectively; c) $e^+e^- \to D^+D^-$. Points with error bars represent the data. The solid curve corresponds to the fit described in the text; the hatched histogram shows the background contribution estimated from the sidebands; the dotted curve stands for the background fraction found by the fit. The dashed curve represents fraction of the events with significant ISR correction. From (Uglov, 2004).

Both *BABAR* and Belle collaborations use this method. The *BABAR* analysis is based on a $384\,{\rm fb}^{-1}$ data sample (Aubert, 2009n) in which $D\overline{D}$ candidates are reconstructed in seven combinations of $D^0$ and $D^+$ decay modes. Aside from $\pi^0$'s from $D^0$ decays, it is required that there be no more than one other $\pi^0$ candidate in the event. The tracks of each $D$ candidate are geometrically constrained to come from a common vertex. Subsequently, each $D\overline{D}$ pair is refitted to a common vertex with the constraint that they originate from the $e^+e^-$ interaction region.

The distribution of $M^2_{\rm recoil}(D\overline{D})$, summed over all $D\overline{D}$ channels, is shown in Fig. 21.4.2. The large bump to the right of the signal peak is due to $e^+e^- \to D\overline{D}\pi^0\gamma_{\rm ISR}$ events. The inset shows the distribution of the $D\overline{D}$ CM polar angle $\theta$ for events with $|M^2_{\rm recoil}(D\overline{D})| < 1\,{\rm GeV}^2/c^4$. The sharp peak at $\cos(\theta_{D\overline{D}}) = -1$ is typical for ISR production and agrees with Monte Carlo simulations.

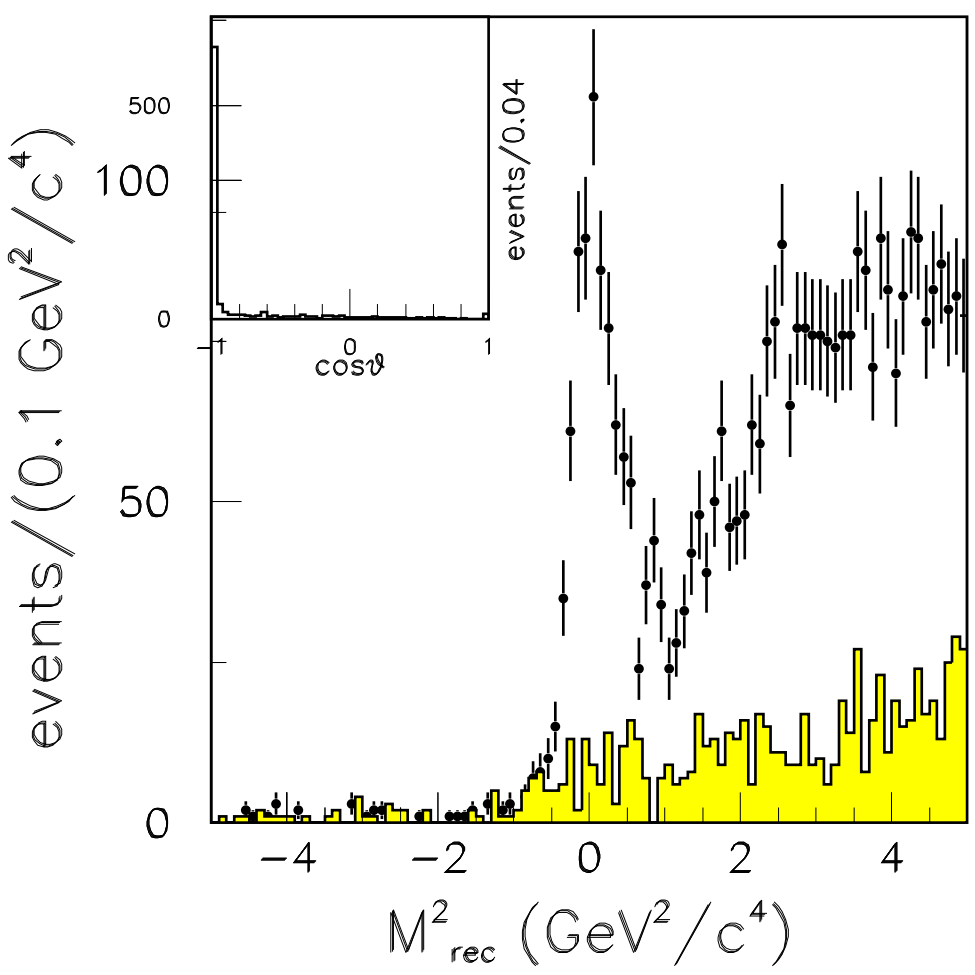

**Figure 21.4.2.** From *BABAR* (Aubert, 2009n). Recoil mass squared, summed over all $D\overline{D}$ channels for $e^+e^- \to D\overline{D}\gamma_{\rm ISR}$ event candidates. The shaded (yellow) histogram corresponds to combinatorial background estimated from the $D\overline{D}$ mass sidebands. The small inset shows the distribution of the center-of-mass polar angle of the $D\overline{D}$ system in the ISR region, $|M^2_{\rm recoil}(D\overline{D})| < 1\,{\rm GeV}^2/c^4$.

The $e^+e^- \to D\overline{D}$ cross sections (Fig. 21.4.3) are extracted from the $D\overline{D}$ mass distributions after background subtraction, using the method described in Section 21.2.2. The combinatorial background contribution is determined using $D\overline{D}$ sideband regions and amounts to 17.5% for $D^0\overline{D}^0$, and 7.1% for $D^+D^-$, of the signal candidates with $|M^2_{\rm recoil}(D\overline{D})| < 1\,{\rm GeV}^2/c^4$; this is the dominant source of background. Efficiencies and $D\overline{D}$ mass resolution are obtained from Monte Carlo simulation. The mass resolution determined from the difference between generated and reconstructed $D\overline{D}$ mass is found to be similar for all channels, and increases from 1.5 MeV/$c^2$ at threshold to 5 MeV/$c^2$ at $M_{D\overline{D}} = 6.0$ GeV/$c^2$.

The Belle analysis of the $e^+e^- \to D\overline{D}\gamma_{\rm ISR}$ process (Pakhlova, 2008a), based on a $673.8\,{\rm fb}^{-1}$ data sample, is similar. Belle defines a signal region by the requirement $|M^2_{\rm recoil}(D\overline{D})| < 0.7\,{\rm GeV}^2/c^4$. In cases when the ISR photon falls within the detector acceptance ($|\cos(\theta_{D\overline{D}})| < 0.9$), its detection is required and the difference between the CM energy and the invariant mass of the $D\overline{D}\gamma_{\rm ISR}$ combination must be smaller than $0.58\,{\rm GeV}/c^2$.

The resulting $e^+e^- \to D^0\overline{D}^0$ and $D^+D^-$ exclusive cross sections, averaged over each bin width, are shown in Fig. 21.4.3 with statistical uncertainties only. The total systematic uncertainties are 10% (*BABAR* obtained 10.9% for $D^0\overline{D}^0$ and 8.1% for $D^+D^-$) and comparable with the statistical errors in the cross section around the $\psi(3770)$ peak; elsewhere, statistical errors dominate.

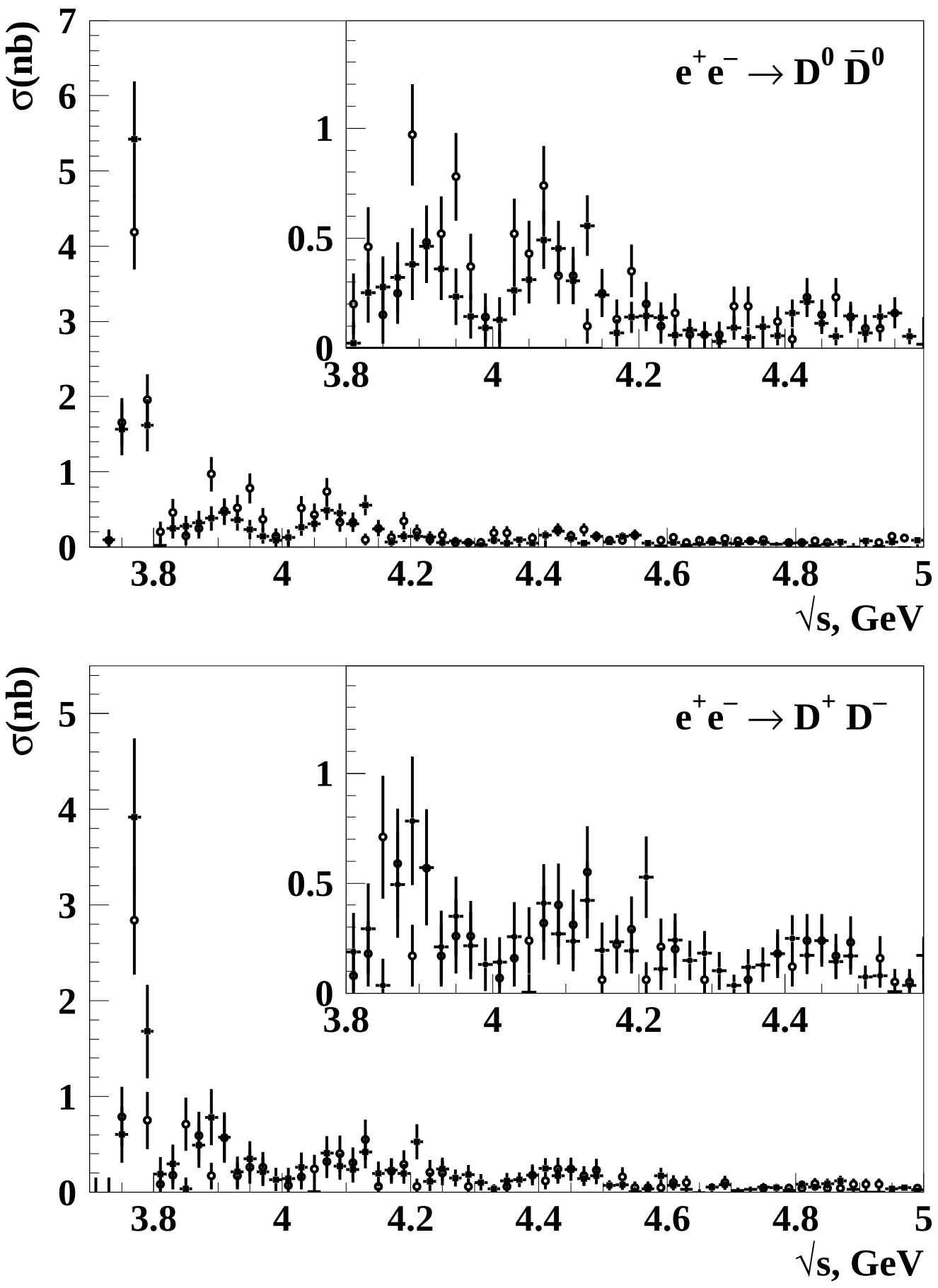

**Figure 21.4.3.** Exclusive cross sections vs. $\sqrt{s}$ for (upper plot) $e^+e^- \to D^0\overline{D}^0$ and (lower plot) $e^+e^- \to D^+D^-$, measured by Belle (solid squares) and *BABAR* (open circles). The region immediately above the $\psi(3770)$ is shown inset on an expanded scale. Prepared from the Pakhlova (2008a) and Aubert (2009n) data.

The $e^+e^- \to D\overline{D}$ exclusive cross sections observed by the two collaborations are in a good agreement. Both contain a clear $\psi(3770)$ signal and structures near the $\psi(4040)$ and $\psi(4415)$ masses. A significant peak at $3.9\,\mathrm{GeV}/c^2$, called $G(3900)$, is in qualitative agreement with predictions of the coupled-channel model (Eichten, Gottfried, Kinoshita, Lane, and Yan, 1980).

The cross section ratio $\sigma(e^+e^- \to D^+D^-)/\sigma(e^+e^- \to D^0\overline{D}^0)$ for $M_{D\overline{D}} \approx M_{\psi(3770)}$ is measured by Belle (in the bin $(3.76 - 3.78)$ $\mathrm{GeV}/c^2$) and by *BABAR* ($(3.74 - 3.80)\,\mathrm{GeV}/c^2$) to be $1.39 \pm 0.31 \pm 0.12$ and $1.78 \pm 0.33 \pm 0.24$ respectively. These values are in agreement with the world average value of $1.28 \pm 0.14$ (Beringer et al., 2012).

### 21.4.3 Partial reconstruction of $D^{(*)+}D^{*-}$ final states

In the case of $D^{(*)+}D^{*-}$ final states,[164] the full reconstruction method discussed in the previous section turns out to be too inefficient. The main reason for this is the very low reconstruction efficiency of the $D^{*\pm}$ when the ISR photon is emitted at small polar angles, due to the low reconstruction efficiency for slow pions from $D^{*\pm}$ decays. If the photon is emitted along the beam axis and the $D^{(*)+}D^{*-}$ system is close to threshold, the $D^{(*)+}$ and $D^{*-}$ meson transverse momenta are low. Because of the small energy release in $D^{*-} \to \overline{D}\pi^-_{\text{slow}}$ decay, the $\pi^-_{\text{slow}}$ transverse momentum is also very low, and such pions do not reach the instrumented parts of the detector. Therefore, reconstructable $D^{*\pm}$ mesons correspond to the $\gamma_{\text{ISR}}$ emitted at large angles, *i.e.* both the $\gamma_{\text{ISR}}$ and the $D^{(*)+}D^{*-}$ pair possess large transverse momenta. The reconstruction efficiency of such isolated energetic photons is high, and it is worth requiring the $\gamma_{\text{ISR}}$ to be detected.

The signal efficiency can be increased further by selecting events without explicitly reconstructing one of the charm mesons. In particular, full reconstruction of only the $D^{(*)+}$, together with the $\gamma_{\text{ISR}}$, allows the $D^{*-}$ meson to be identified using the peak around the $D^{*-}$ mass in the spectrum of masses recoiling against the $D^{(*)+}\gamma_{\text{ISR}}$ system:

$$M_{\text{recoil}}(D^{(*)+}\gamma_{\text{ISR}}) = \sqrt{(E_{\text{CM}} - E_{D^{(*)+}\gamma_{\text{ISR}}})^2 - p^2_{D^{(*)+}\gamma_{\text{ISR}}}}. \quad (21.4.4)$$

Here $E_{D^{(*)+}\gamma_{\text{ISR}}}$ and $p_{D^{(*)+}\gamma_{\text{ISR}}}$ are the CM energy and momentum, respectively, of the $D^{(*)+}\gamma_{\text{ISR}}$ combination. This peak is expected to be wide and asymmetric due to the $\gamma_{\text{ISR}}$ energy resolution and higher-order corrections to ISR cross section. The resolution of this peak (estimated to be $\sim 300\,\mathrm{MeV}/c^2$) is not sufficient to separate the $D\overline{D}^*$, $D^*\overline{D}^*$, and $D^{(*)}\overline{D}^*\pi$ final states. To disentangle these various contributions and to suppress combinatorial backgrounds, one can use the slow pion from the unreconstructed $D^{*-}$. The difference between the masses recoiling against $D^{(*)+}\gamma_{\text{ISR}}$ and $D^{(*)+}\pi^-_{\text{slow}}\gamma_{\text{ISR}}$ (recoil mass difference),

$$\Delta M_{\text{recoil}} = M_{\text{recoil}}(D^{(*)+}\gamma_{\text{ISR}}) - M_{\text{recoil}}(D^{(*)+}\pi^-_{\text{slow}}\gamma_{\text{ISR}}), \quad (21.4.5)$$

has a narrow distribution for signal events ($\sigma \sim 1.4\,\mathrm{MeV}/c^2$) around $m_{D^{*-}} - m_{\overline{D}^0}$, since the uncertainty in the $\gamma_{\text{ISR}}$ momentum partially cancels out.

The efficiency gain using the described partial reconstruction method over the full reconstruction method is $\sim 1/\epsilon_{D^0}$ and $\sim 2/\epsilon_{D^0}$, for $D^+D^{*-}$ and $D^{*+}D^{*-}$ final states respectively, where $\epsilon_{D^0}$ is the $D^0$ reconstruction efficiency.

In the case of full reconstruction, exclusive cross sections are obtained from $D^{(*)+}D^{*-}$ mass spectra. In the partial reconstruction case $D^{*-}$ is not reconstructed and the $D^{(*)+}D^{*-}$ mass can not be calculated directly. However, it is equivalent to $M_{\text{recoil}}(\gamma_{\text{ISR}})$, the mass recoiling against the ISR photon (ignoring higher-order QED processes). The problem of poor photon energy resolution (and, thus poor $M_{\rm recoil}(\gamma_{\rm ISR})$ resolution) is solved by applying a fit constraining $M_{\rm recoil}(D^{(*)+}\gamma_{\rm ISR})$ to the $D^{*-}$ mass. This refit procedure corrects the $\gamma_{\rm ISR}$ momentum and as a result, the $M_{\rm recoil}(\gamma_{\rm ISR}) = M(D^{(*)+}D^{*-})$ resolution is improved by a factor $\sim 10$: it varies from $\sim 6$ MeV at $D^{(*)+}D^{*-}$ threshold to $\sim 12$ MeV at $M(D^{(*)+}D^{*-}) = 5\,{\rm GeV}/c^2$. In addition the resolution of the recoil mass difference after the refit procedure ($\Delta M^{\rm fit}_{\rm recoil}$) is improved by a factor $\sim 2$.

[164] The notation represents the sum of $D^+D^{*-}$ and $D^{*+}D^{*-}$ final states.

This method, developed by Belle, is applied to the $e^+e^- \to D^{(*)+}D^{*-}$ cross section measurement using a data sample of $673.8\,{\rm fb}^{-1}$ (Abe, 2007d). The signal region is defined by the requirement that $M_{\rm recoil}(D^{*+}\gamma_{\rm ISR})$ lie within $\pm 0.2\,{\rm GeV}/c^2$ of the $D^{*-}$ mass and that $\Delta M^{\rm fit}_{\rm recoil}$ be within $\pm 2\,{\rm MeV}/c^2$ of $m_{D^{*-}} - m_{\overline{D}^0}$. The $e^+e^- \to D^{(*)+}D^{*-}$ cross sections are extracted from the $M_{\rm recoil}(\gamma_{\rm ISR})$ distributions after background subtraction. All background contributions are estimated from the data and the combinatorial background is found to be the dominant source.

The resulting exclusive $e^+e^- \to D^{(*)+}D^{*-}$ cross sections are shown in Figs 21.4.4 and 21.4.5 with statistical uncertainties only. The total systematic uncertainties are 11% for $D^+D^{*-}$ and 10% for $D^{*+}D^{*-}$, comparable to the statistical errors in the cross section.

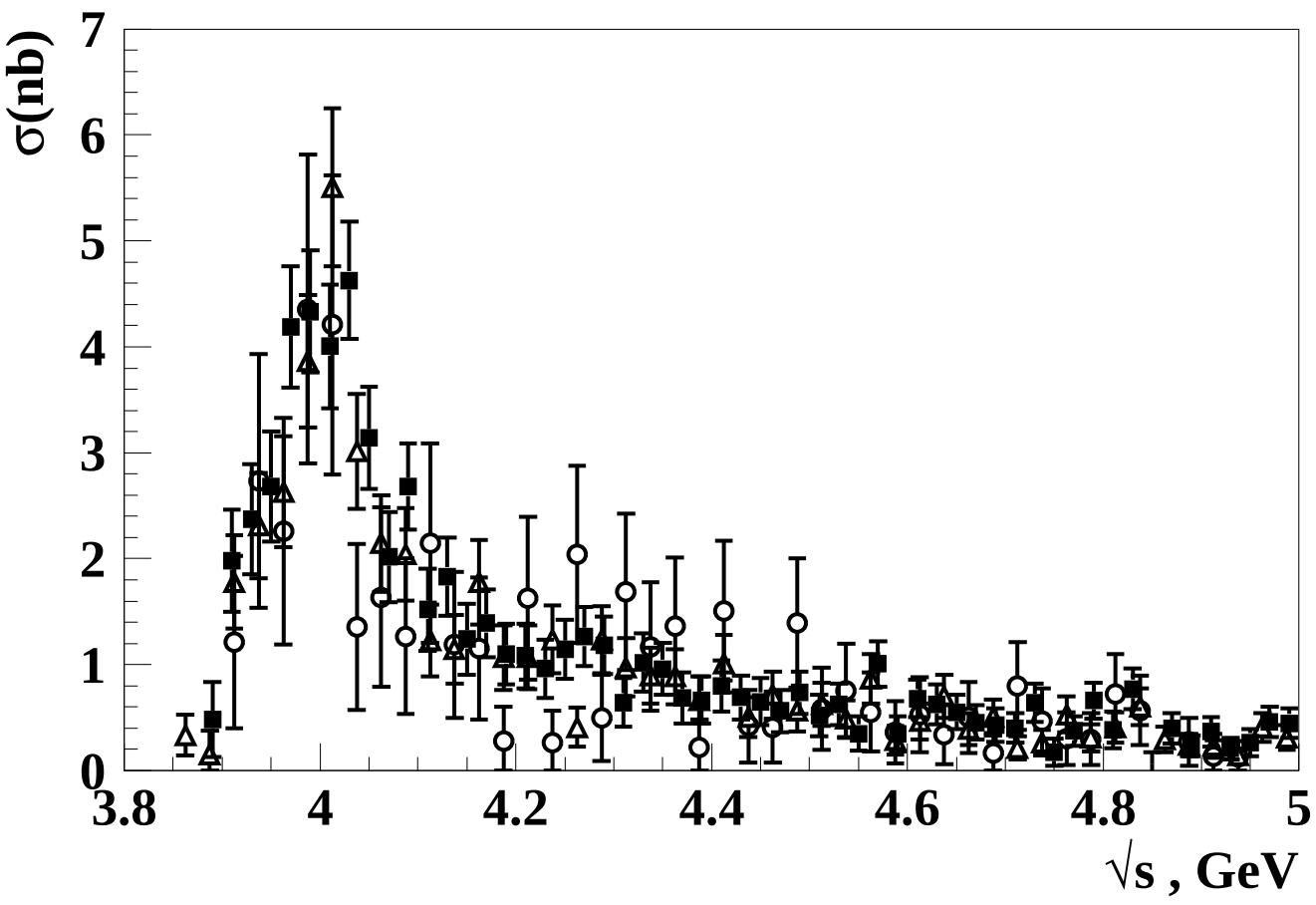

**Figure 21.4.4.** The exclusive cross sections for $e^+e^- \to D^+D^{*-}$ measured by Belle (Abe, 2007d; solid squares) and *BABAR* (Aubert, 2009n; open triangles); $e^+e^- \to D^0\overline{D}^{*0}$ measured by *BABAR* (open circles).

The results of the *BABAR* measurements (Aubert, 2009n) based on $384\,{\rm fb}^{-1}$ of data are shown in the same plots. Because the full reconstruction method was used, the statistical uncertainties in these measurements are significantly larger than those of Belle. The cross sections for $D^0\overline{D}^{*0}$, $D^+D^{*-}$, and $D^*\overline{D}^*$ (the latter being the sum of the neutral and charged modes, *i.e.* $D^{*0}\overline{D}^{*0}$ and $D^{*+}D^{*-}$) are presented in Fig. 21.4.4 and Fig. 21.4.5 respectively. The systematic uncertainties in the cross sections are estimated to be 10.9% for $D^0\overline{D}^{*0}$, 9.3% for $D^+D^{*-}$, and 12.4% for $D^*\overline{D}^*$.

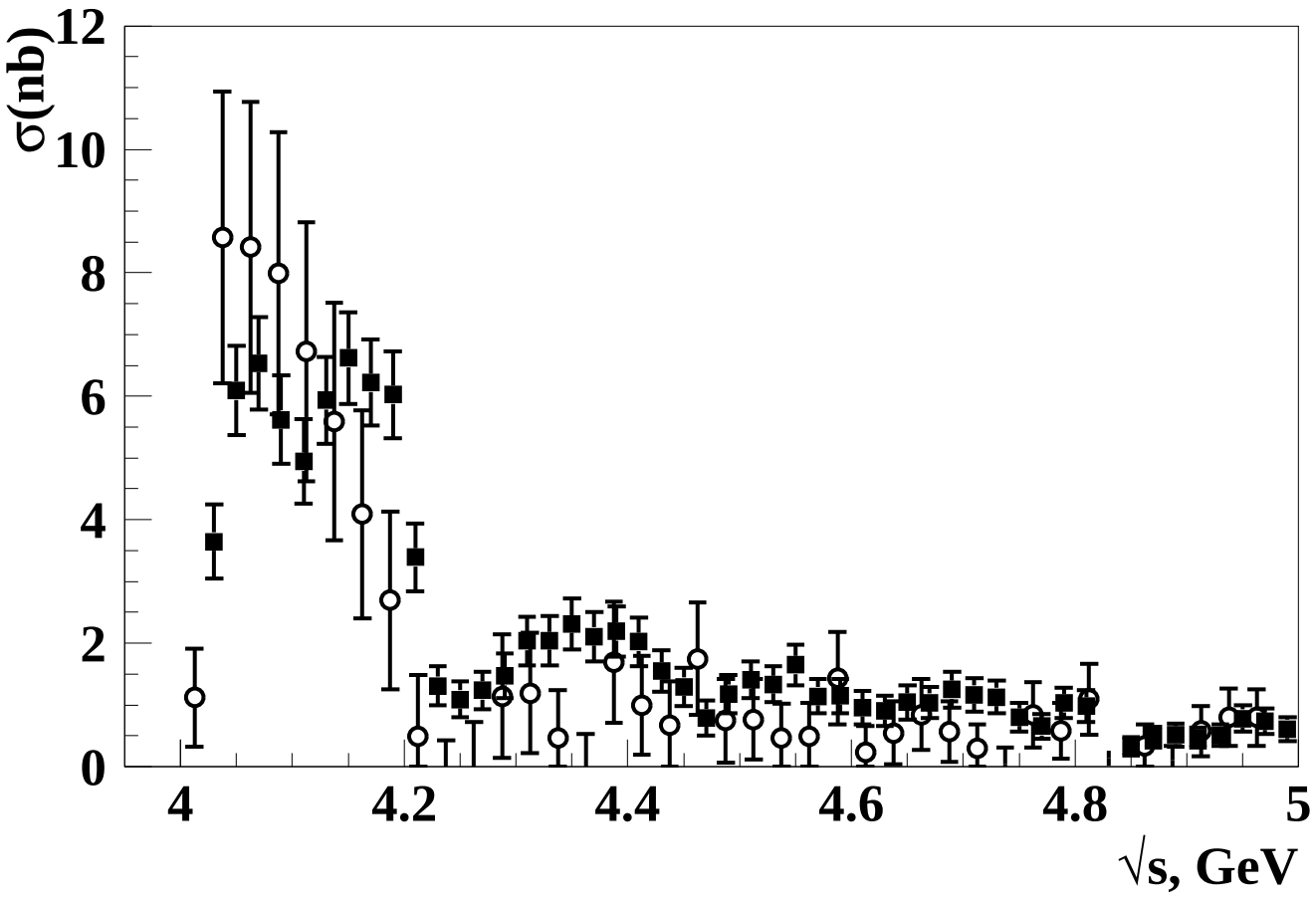

**Figure 21.4.5.** The exclusive cross sections for $e^+e^- \to D^{*+}D^{*-}$ measured by Belle (Abe, 2007d; solid squares) and $e^+e^- \to D^*\overline{D}^*$ measured by *BABAR* (Aubert, 2009n; open circles).

Belle and *BABAR* measurements are in good agreement, and compatible with the $D\overline{D}^*$ and $D^*\overline{D}^*$ exclusive cross sections measured by CLEO-c (Cronin-Hennessy et al., 2009); the CLEO-c measurements are more precise; they include, however, only the narrow energy range from 3.97 to 4.26 GeV. Aside from a prominent excess near the $\psi(4040)$ resonance, the $e^+e^- \to D^+D^{*-}$ cross section is relatively featureless. Integrating the cross sections from threshold to $6\,{\rm GeV}/c^2$, *BABAR* obtained

$$\frac{\sigma(D^+D^{*-})}{\sigma(D^0\overline{D}^{*0})} = 0.95 \pm 0.09 \pm 0.10, \qquad (21.4.6)$$

consistent with unity. The shape of the $e^+e^- \to D^{*+}D^{*-}$ cross section is complicated, with several local maxima and minima. Reliable interpretation will require more data.

### 21.4.4 $e^+e^- \to D_s^{(*)+}D_s^{(*)-}$

Both *BABAR* and Belle use the full reconstruction technique to measure $e^+e^- \to D_s^{(*)+}D_s^{(*)-}$ cross sections. Partial reconstruction of $D_s^{*+}$ decaying to $D_s^+\gamma$ is impractical because of a huge combinatorial background.

*BABAR* results based on a $525\,{\rm fb}^{-1}$ data sample (del Amo Sanchez, 2010d) are presented in Fig. 21.4.6. For each candidate event *BABAR* reconstructs a $D_s^+D_s^-$ pair. While one of the $D_s^+$ is required to decay to $K^+K^-\pi^+$, the second $D_s^-$ meson is reconstructed in three decay modes: $K^+K^-\pi^-$, $K^+K^-\pi^-\pi^0$, and $K_S^0K^-$. Belle measurements of the exclusive $e^+e^- \to D_s^{(*)+}D_s^{(*)-}$ cross sections (Pakhlova, 2011) based on a $967\,{\rm fb}^{-1}$ data sample are shown in Fig. 21.4.7. To increase the sample size Belle reconstructs both $D_s^+$ (+ c.c.) mesons in six decay modes: $K_S^0K^+$, $K^-K^+\pi^+$, $K^-K^+\pi^+\pi^0$, $K_S^0K^-\pi^+\pi^+$, $\eta\pi^+$ and $\eta'\pi^+$. The presented cross sections are averaged over the bin width; error bars show statistical uncertainties only. The systematic uncertainties are evaluated by *BABAR* (Belle) to be 23%(11%) for $D_s^+ D_s^-$, 13%(17%) for $D_s^+ D_s^{*-}$ and 13%(31%) for $D_s^{*+} D_s^{*-}$.

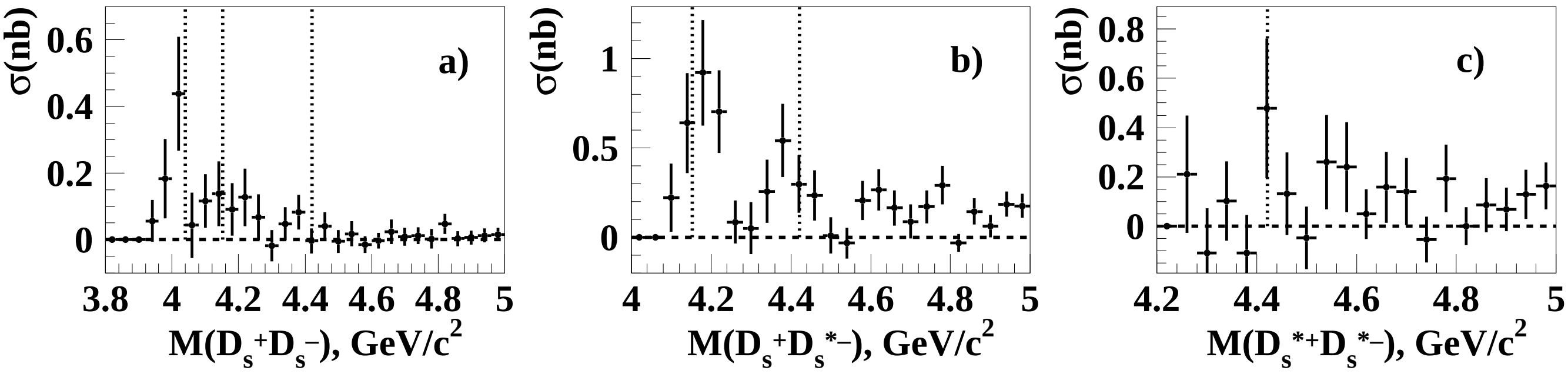

**Figure 21.4.7.** Exclusive cross sections for (a) $e^+e^- \to D_s^+ D_s^-$, (b) $e^+e^- \to D_s^+ D_s^{*-}$ (and charge conjugate), and (c) $e^+e^- \to D_s^{*+} D_s^{*-}$, from Belle data (Pakhlova, 2011). Error bars show statistical uncertainties only. The dotted lines show the masses of the $\psi(4040)$, $\psi(4160)$ and $\psi(4415)$ states.

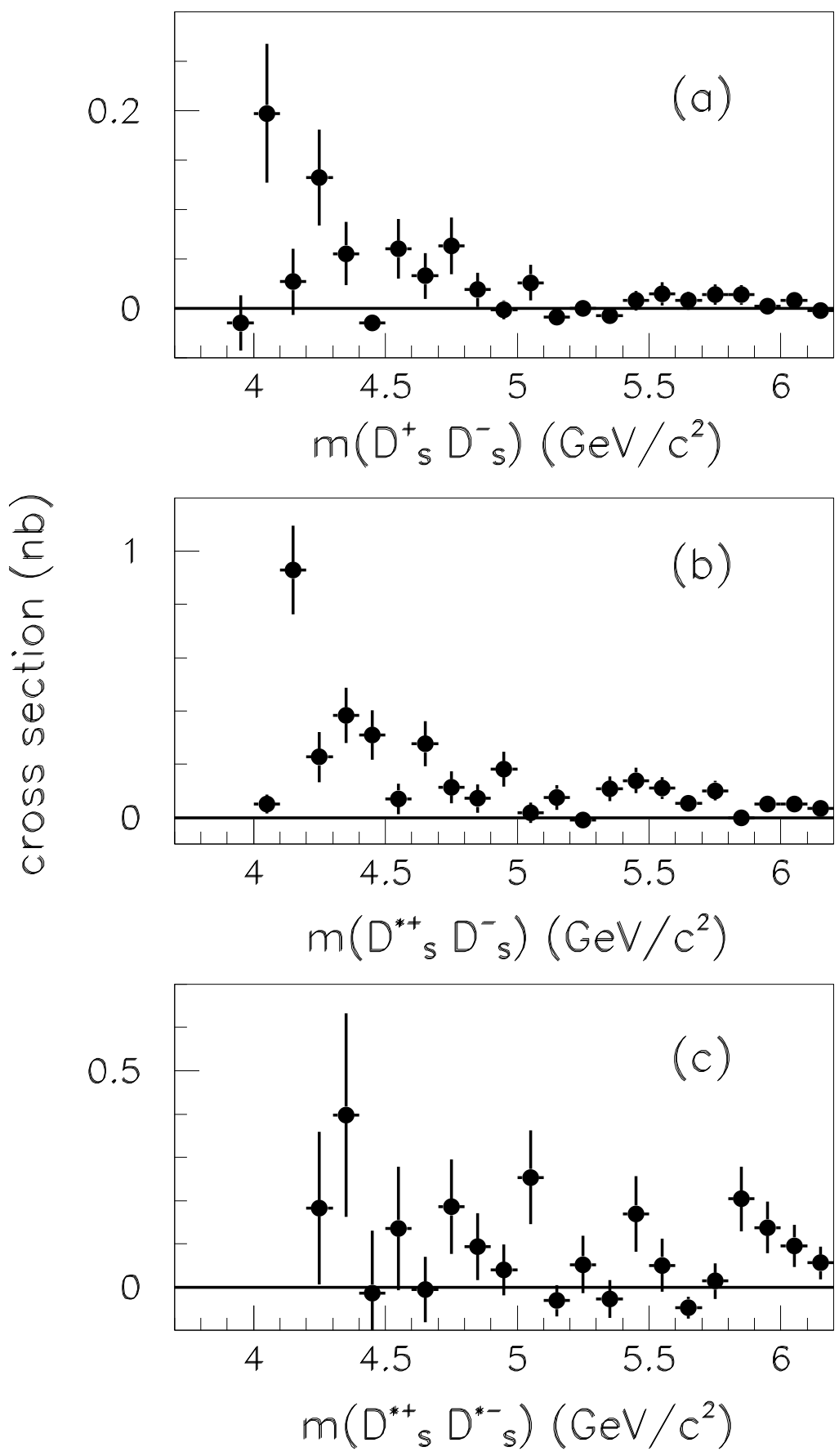

**Figure 21.4.6.** Exclusive cross sections for (a) $e^+e^- \to D_s^+ D_s^-$, (b) $e^+e^- \to D_s^+ D_s^{*-}$ (and charge conjugate), and (c) $e^+e^- \to D_s^{*+} D_s^{*-}$, from *BABAR* data (del Amo Sanchez, 2010d). The error bars correspond to statistical errors only.

The identification of relatively narrow $\psi$ states in *BABAR* spectra is complicated by the large bin size of 100 MeV. Nevertheless *BABAR* results are consistent with the more precise Belle measurements (in 40 MeV bins). A clear peak at threshold, around the $\psi(4040)$ mass, is seen in the $e^+e^- \to D_s^+ D_s^-$ cross section. In the $e^+e^- \to D_s^+ D_s^{*-}$ cross section two peaks are evident, around the $\psi(4160)$ and the $\psi(4415)$ masses. The $e^+e^- \to D_s^{*+} D_s^{*-}$ data sample is small, and there is no clear structure in the cross section. Both the $e^+e^- \to D_s^+ D_s^{*-}$ cross section and the sum of the exclusive $e^+e^- \to D_s^{(*)+} D_s^{(*)-}$ cross sections exhibit a dip near the $Y(4260)$ mass (see Section 18.3.5), similar to what is seen in $e^+e^- \to D^{*+} D^{*-}$ and in the total cross section for charm production.

### 21.4.5 Three-body charm final states

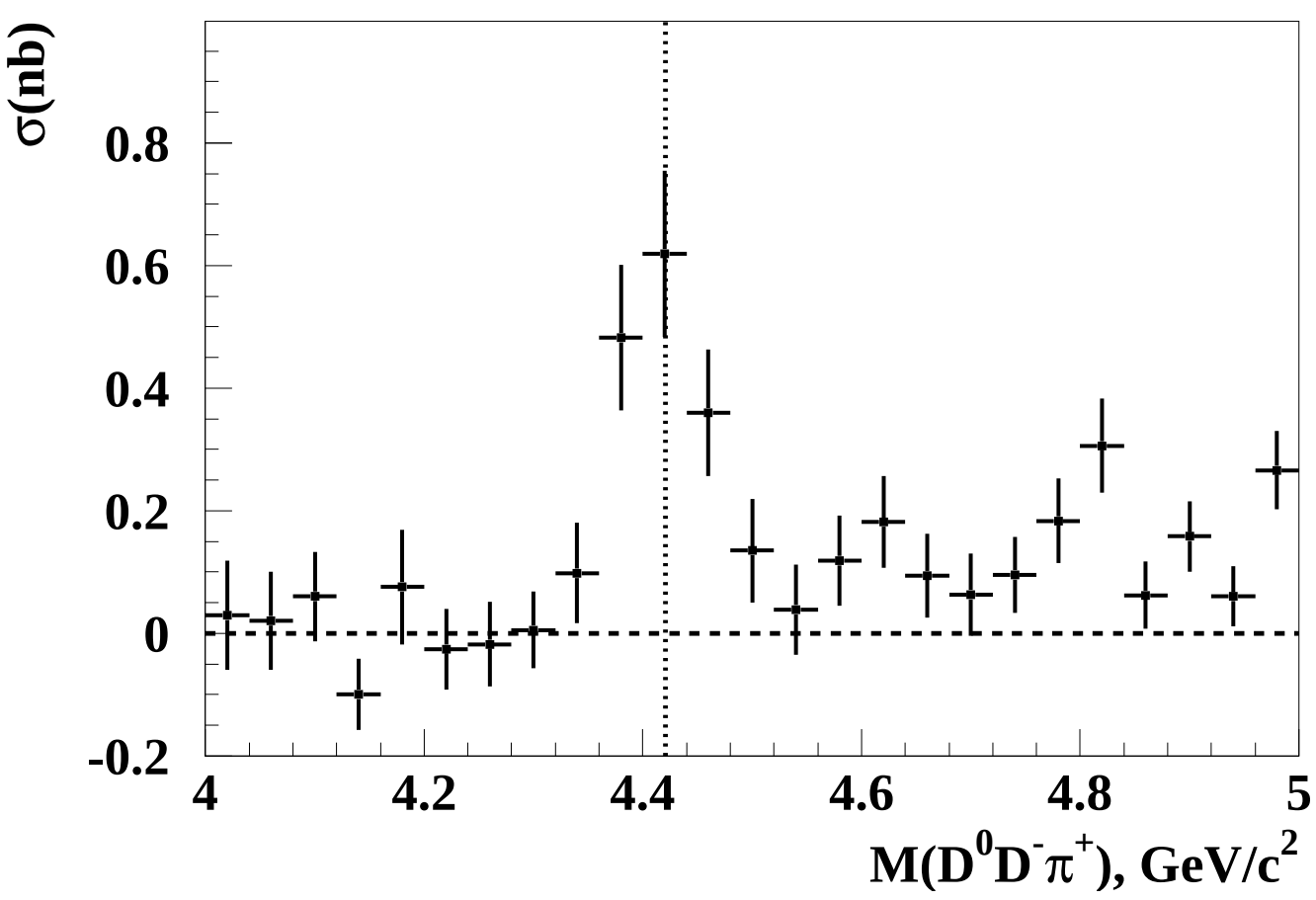

**Figure 21.4.8.** The exclusive cross section for $e^+e^- \to D^0 D^- \pi^+$ measured by Belle (Pakhlova, 2008c). The dotted line corresponds to the mass of the $\psi(4415)$.

The first measurements of three-body open charm final states in $e^+e^-$ annihilation have been performed using the full reconstruction method by Belle. The cross section for $e^+e^- \to D^0D^-\pi^+$ measured using a $673\,\mathrm{fb}^{-1}$ data sample (Pakhlova, 2008c) is shown in Fig. 21.4.8. A prominent $\psi(4415)$ peak is observed. From a study of the resonant structure in $\psi(4415)$ decays (discussed in Section 18.2.2.2) Belle concludes that the $\psi(4415) \to D^0D^-\pi^+$ process is dominated by $\psi(4415) \to D\bar{D}_2^*(2460)$. It was found that $\mathcal{B}(\psi(4415) \to D^0D^-\pi^+_{\mathrm{non-res}})/\mathcal{B}(\psi(4415 \to D\bar{D}_2^*(2460) \to D^0D^-\pi^+) < 0.22$ at the 90% C.L. The peak cross section for the $e^+e^- \to \psi(4415) \to D\bar{D}_2^*(2460)$ process at $E_{\mathrm{CM}} = m_{\psi(4415)}c^2$ is calculated to be $\sigma(e^+e^- \to \psi(4415)) \times \mathcal{B}(\psi(4415) \to D\bar{D}_2^*(2460)) \times \mathcal{B}(\bar{D}_2^*(2460) \to D\pi^+) = (0.74 \pm 0.17 \pm 0.08)\,\mathrm{nb}$.

The $e^+e^- \to D^0D^{*-}\pi^+$ exclusive cross section, based on a $695\,\mathrm{fb}^{-1}$ data sample (Pakhlova, 2009), is shown in Fig. 21.4.9. The main motivation of this study is the search for $Y(4260) \to D^0D^{*-}\pi^+$ decays discussed in Section 18.3.5. An estimate of the branching fraction for $\psi(4415) \to D^0D^{*-}\pi^+$ decay can be found in Section 18.2.2.2.

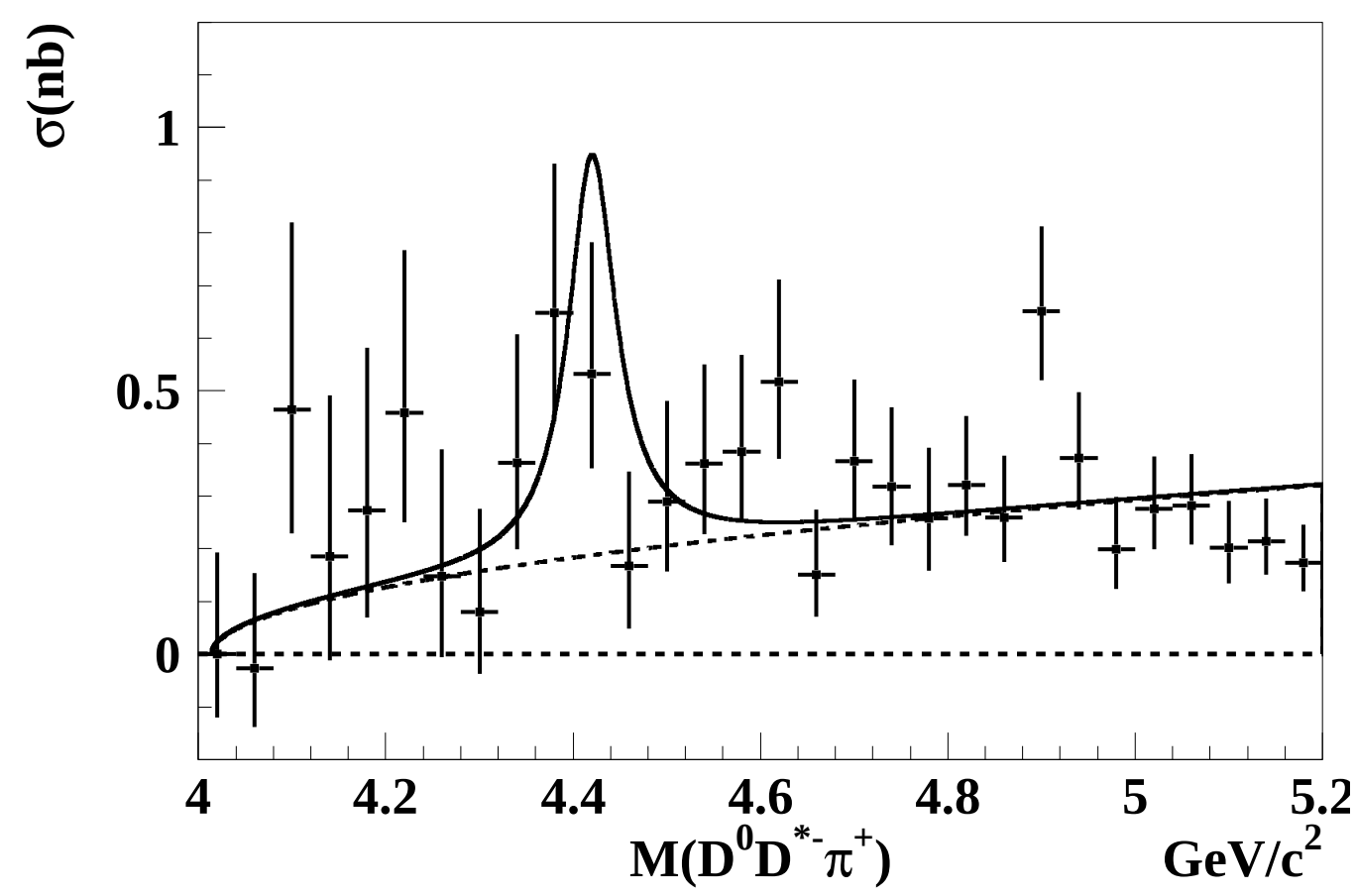

**Figure 21.4.9.** The exclusive cross section for $e^+e^- \to D^0D^{*-}\pi^+$ averaged over the bin width with statistical uncertainties only from Belle data (Pakhlova, 2009). The total systematic uncertainty is 10%. The fit function corresponds to the 90% C.L. upper limit on $\psi(4415)$ taking into account systematic uncertainties. The solid line represents the sum of the signal and threshold contributions. The threshold function is shown by the dashed line.

### 21.4.6 Charm baryon production in $e^+e^-$ annihilation

The first measurement of the $e^+e^- \to \Lambda_c^+\Lambda_c^-$ process near threshold has been performed by Belle in ISR events, using $695\,\mathrm{fb}^{-1}$ of data (Pakhlova, 2008b), with the partial reconstruction technique. Full reconstruction of both the $\Lambda_c^+$ and $\Lambda_c^-$ baryons suffers from the low $\Lambda_c$ reconstruction efficiency, and the small branching fractions for decays to accessible final states; Belle requires reconstruction of only one of the $\Lambda_c$ baryons (using $pK_S^0$, $pK\pi$ and $\Lambda\pi$ final states) and the ISR photon. The exclusive $e^+e^- \to \Lambda_c^+\Lambda_c^-$ cross section is determined from the recoil mass $M_{\mathrm{recoil}}(\gamma_{\mathrm{ISR}})$. A refit constraining $M_{\mathrm{recoil}}(\Lambda_c^+\gamma_{\mathrm{ISR}})$ to the nominal $\Lambda_c^-$ mass improves the $M_{\mathrm{recoil}}(\gamma_{\mathrm{ISR}})$ resolution: the final resolution varies from $\sim 3\,\mathrm{MeV}/c^2$ just above threshold to $\sim 8\,\mathrm{MeV}/c^2$ at $M_{\Lambda_c^+\Lambda_c^-} \sim 5.4\,\mathrm{GeV}/c^2$. Combinatorial background is suppressed by a factor $\sim 10$ by requiring the presence of at least one $\bar{p}$ in the event from the decay of the unreconstructed $\Lambda_c^-$; this requirement reduces the signal efficiency by $\sim 40\%$.

The resulting cross section is shown in Fig. 21.4.10. A significant enhancement, called $X(4630)$ by Belle, is seen at threshold, with a peak cross section $\sigma(e^+e^- \to X(4630)) \times \mathcal{B}(X(4630) \to \Lambda_c^+\Lambda_c^-) = (0.47^{+0.11+0.05}_{-0.10-0.08} \pm 0.19)\,\mathrm{nb}$. Belle obtains $\Gamma_{ee}/\Gamma_{\mathrm{tot}} \times \mathcal{B}(X(4630) \to \Lambda_c^+\Lambda_c^-) = (0.68^{+0.16+0.07}_{-0.15-0.11} \pm 0.28) \times 10^{-6}$. The last error is due to the uncertainties of $\Lambda_c$ branching fractions.

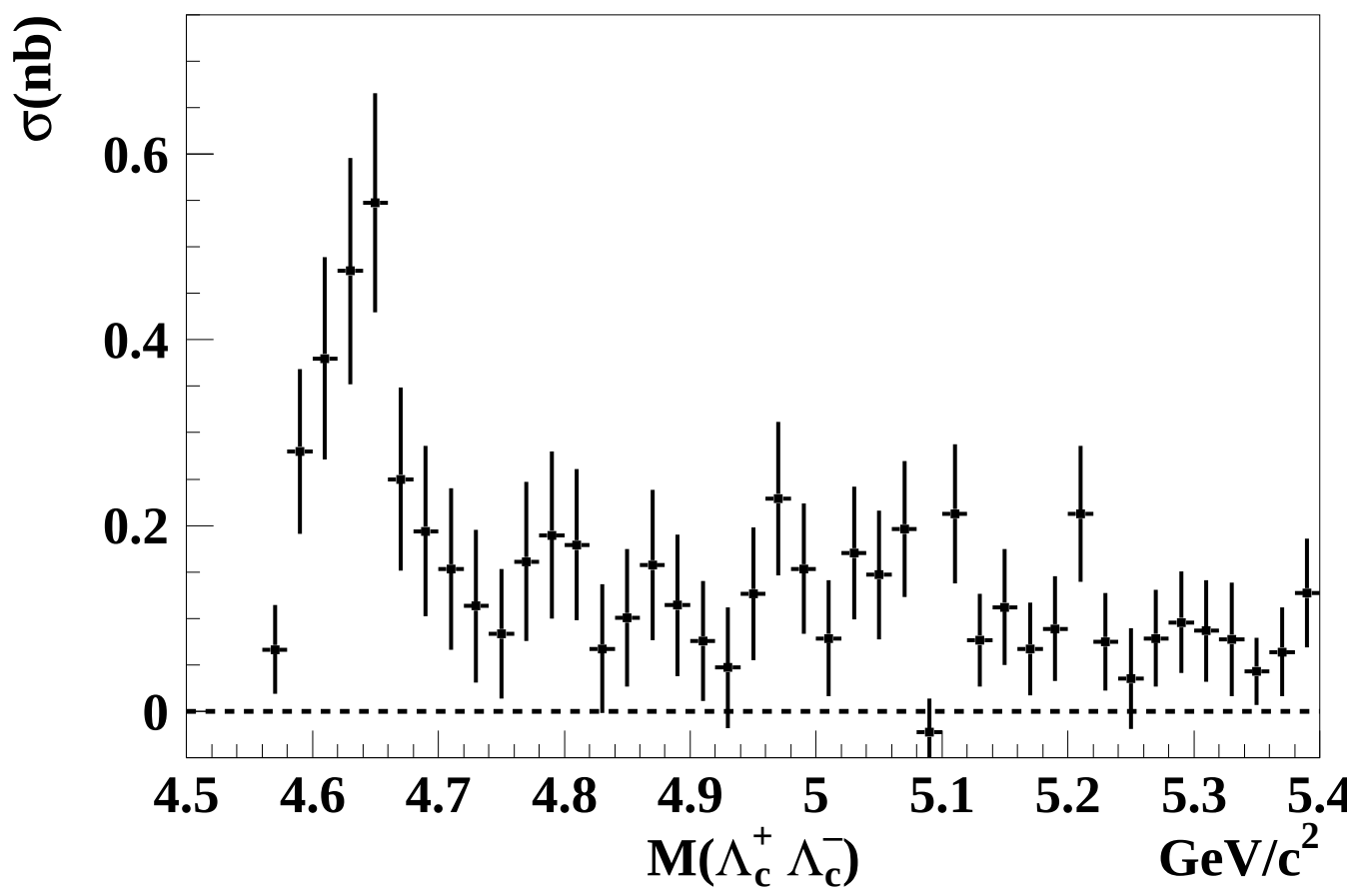

**Figure 21.4.10.** The exclusive cross section for $e^+e^- \to \Lambda_c^+\Lambda_c^-$ measured by Belle (Pakhlova, 2008b).

The nature of the observed enhancement remains unclear. More details on its parameters and corresponding discussion can be found in Section 18.3.5.

### 21.4.7 Sum of exclusive vs inclusive cross section

In conclusion, using the ISR method allows the measurement of nine cross sections for $e^+e^-$ annihilation to open charm final states over a wide energy range, beginning at threshold. For the first time the inclusive cross section to charm hadrons is decomposed into the sum of the exclusive components $e^+e^- \to D\bar{D}$, $D\bar{D}^*$, $D^*\bar{D}^*$, $D\bar{D}\pi$, $D\bar{D}^*\pi$, $D_s^+D_s^-$, $D_s^+D_s^{*-}$, $D_s^{*+}D_s^{*-}$, and $\Lambda_c^+\Lambda_c^-$. This sum, shown in Fig. 21.4.11, almost saturates the inclusive cross section; the $D\bar{D}^*$ and $D^*\bar{D}^*$ final states dominate.

## 21.5 Search for exotic charmonium

Initial state radiation events provide an ideal environment to study known vector states as well as search for additional such states. Searches for new vector mesons have commonly proceeded by looking for their decay to conventional charmonium states plus additional light hadrons. The study of the $\pi^+\pi^- J/\psi$ and $\pi^+\pi^-\psi(2S)$ final states is presented in Sections 21.5.1 and 21.5.2 respectively. Possible vector decays to open charm final states are treated in Section 21.4 above. A general discussion of the new, and possibly exotic vector states is presented elsewhere (Section 18.3.5).

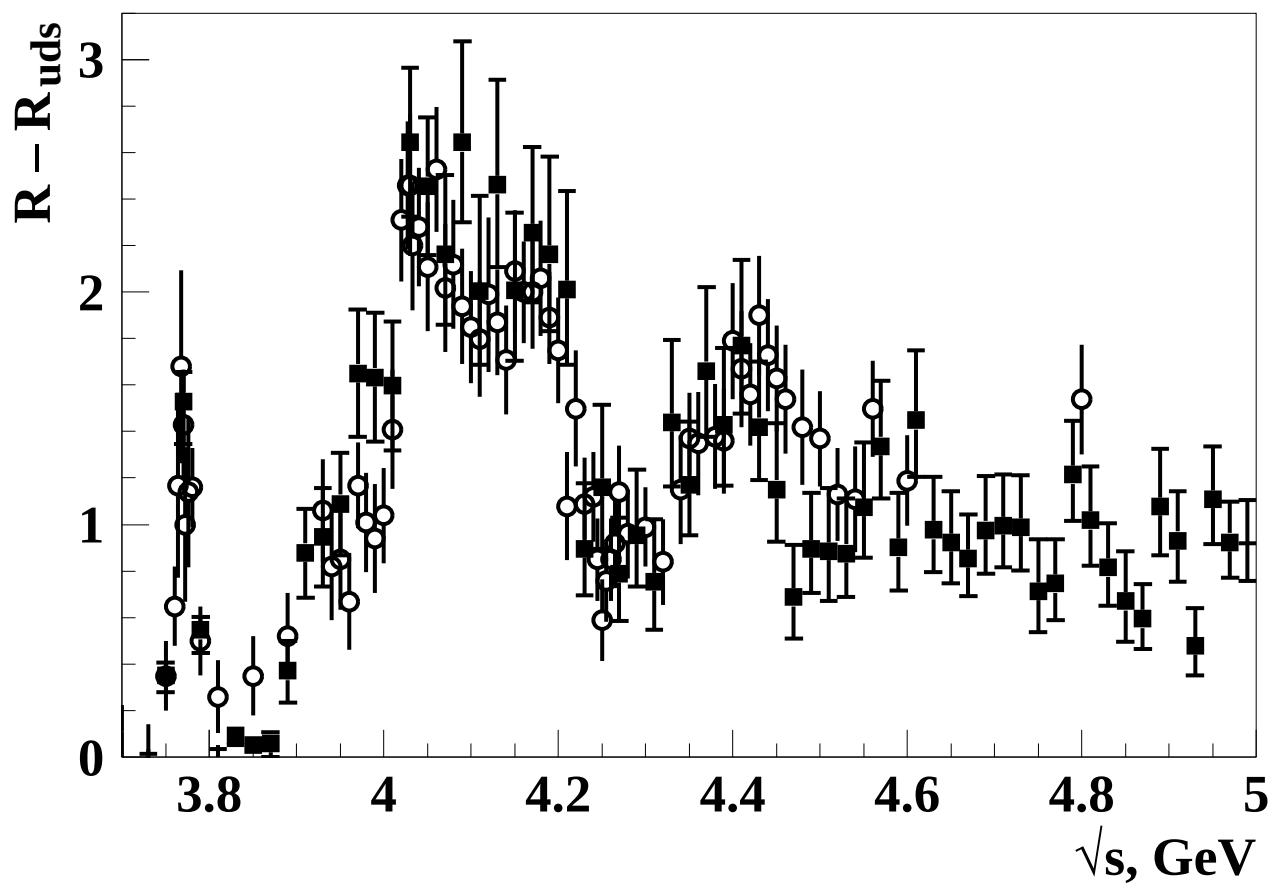

**Figure 21.4.11.** Inclusive measurements of $R - R_{uds}$, where $R = \sigma(e^+e^- \to \text{hadrons})/\sigma(e^+e^- \to \mu^+\mu^-)$ and $R_{uds} = 2.121 \pm 0.023 \pm 0.083$ by BES II (Ablikim et al., 2007; open circles), compared to the sum of exclusive cross sections (solid squares) measured by Belle.

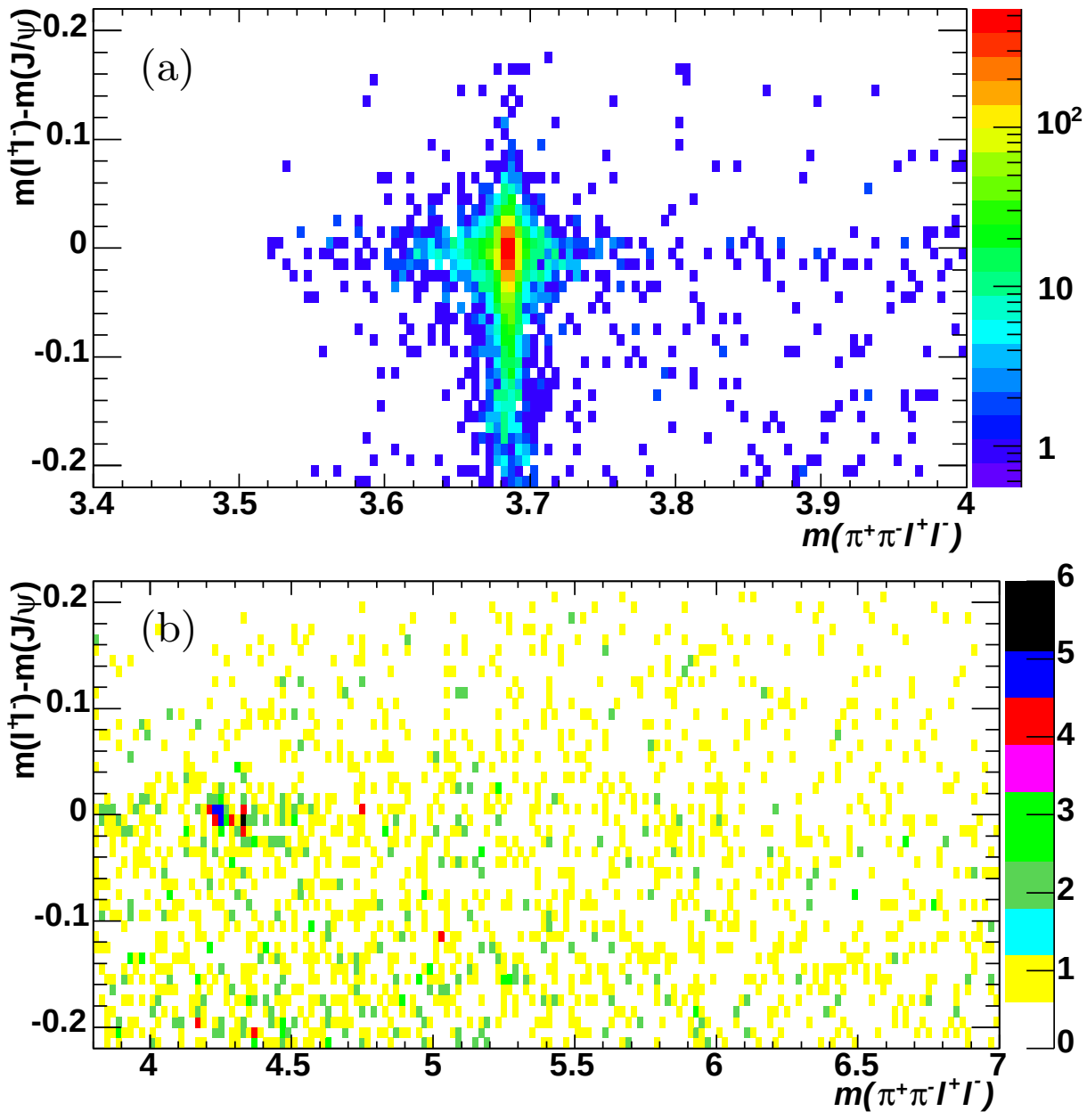

**Figure 21.5.1.** The scatter plot of $m(\ell^+\ell^-) - m(J/\psi)$ vs. $m(\pi^+\pi^- J/\psi)$ for ISR produced $\pi^+\pi^- J/\psi$ events collected by *BABAR* in 123 $\text{fb}^{-1}$. (a) low $m(\pi^+\pi^- J/\psi)$ range $[3.4, 4.0]\,\text{GeV}/c^2$ (log scale): A clean $\psi(2S)$ signal is observed while no evidence of a $X(3872)$ signal is seen. (b) high $m(\pi^+\pi^- J/\psi)$ range $[3.8, 7.0]\,\text{GeV}/c^2$ (linear scale): An enhancement around $4.3\,\text{GeV}/c^2$ is found. *BABAR* internal, from (Aubert, 2005y) analysis.

### 21.5.1 $Y$ family states in ISR $\pi^+\pi^- J/\psi$

An important exotic charmonium candidate, the $Y(4260)$, has been observed in the $\pi^+\pi^- J/\psi$ final state. The discovery and subsequent studies are reported in Sections 21.5.1.1 and 21.5.1.2 respectively. The claimed broad structure $Y(4008)$ is discussed in the latter section.

#### 21.5.1.1 The $Y(4260)$ discovery

The discovery by Belle of the surprisingly narrow $X(3872)$ resonance from the study of $B \to J/\psi\pi^+\pi^- K$ decays (Choi, 2003), discussed in Section 18.3.2, renewed experimental interest in charmonium spectroscopy. In order to understand the $X(3872)$ when its quantum numbers were hardly known in 2004, *BABAR* searched for $X(3872) \to \pi^+\pi^- J/\psi$ in the ISR process $e^+e^- \to \gamma_{\text{ISR}}\pi^+\pi^- J/\psi$, where $J/\psi$ decays to $\ell^+\ell^-$, using a data sample corresponding to 232 $\text{fb}^{-1}$ (Aubert, 2005y). The analysis was performed requiring exclusive reconstruction of the hadronic final state, but not explicit detection of the ISR photon. The large and clean ISR $\psi(2S) \to \pi^+\pi^- J/\psi$ sample provides a good control sample for validation and selection criteria optimization. In a subsample of 123 $\text{fb}^{-1}$ of data, as shown in Fig. 21.5.1, no evidence for the $X(3872)$ was found, but an enhancement was seen around 4.3 $\text{GeV}/c^2$.

In order to avoid any possible bias and to firmly establish this observation, the *BABAR* analysis blinds the enhancement region $[4.2, 4.4]$ $\text{GeV}/c^2$, and optimizes the selection criteria by maximizing the quantity $N/(3/2 + \sqrt{B})$ (Punzi, 2003b), where 3/2 corresponds to the search for a $3\sigma$ signal, $N$ is the total number of $\gamma_{\text{ISR}}\psi(2S)$, $\psi(2S) \to \pi^+\pi^- J/\psi$ candidates in the $20\,\text{MeV}/c^2$ $\pi^+\pi^- J/\psi$ mass range that brackets the $\psi(2S)$ nominal mass, and $B$ is the number of (background) events in the $\pi^+\pi^- J/\psi$ mass regions $[3.8, 4.2]$ $\text{GeV}/c^2$ and $[4.4, 4.8]$ $\text{GeV}/c^2$, scaled to the width of the excluded region. The selection criteria are optimized taking advantage of the features of ISR emission, that is, a recoil mass close to zero, and a small transverse component of the visible momentum in the $e^+e^-$ CM, including the ISR photon when it is reconstructed. Exactly four tracks consistent with production at the $e^+e^-$ interaction point are allowed: two oppositely charged tracks identified as pions, and a pair of identified leptons (either $e^+e^-$ or $\mu^+\mu^-$) whose reconstructed invariant mass is within an optimized interval around the $J/\psi$ peak. Additional cuts on kinematic variables of the hadronic system are applied to further reject background sources.

In order to improve the mass resolution, the four tracks are refitted with a constraint to a common vertex, and the lepton pair kinematically constrained to the $J/\psi$ mass. The resulting $\pi^+\pi^- J/\psi$ mass-resolution function is well-described by a Breit-Wigner distribution with a full width at half maximum increasing from $4.2\,\text{MeV}/c^2$ at the $\psi(2S)$ to 5.3 MeV/$c^2$ at 4.3 GeV/$c^2$. The $\pi^+\pi^- J/\psi$ invariant mass spectrum for candidates passing all criteria is shown in Fig. 21.5.2 as points with error bars. A signal of 11802 ± 110 $\psi(2S)$ events is observed, consistent with the expectation. An enhancement near 4.26 GeV/$c^2$, now known as the $Y(4260)$, is clearly observed; no other structures are evident at the masses of the known $J^{PC} = 1^{--}$ charmonium states (*i.e.* the $\psi(4040)$, $\psi(4160)$, and $\psi(4415)$), or at the $X(3872)$.

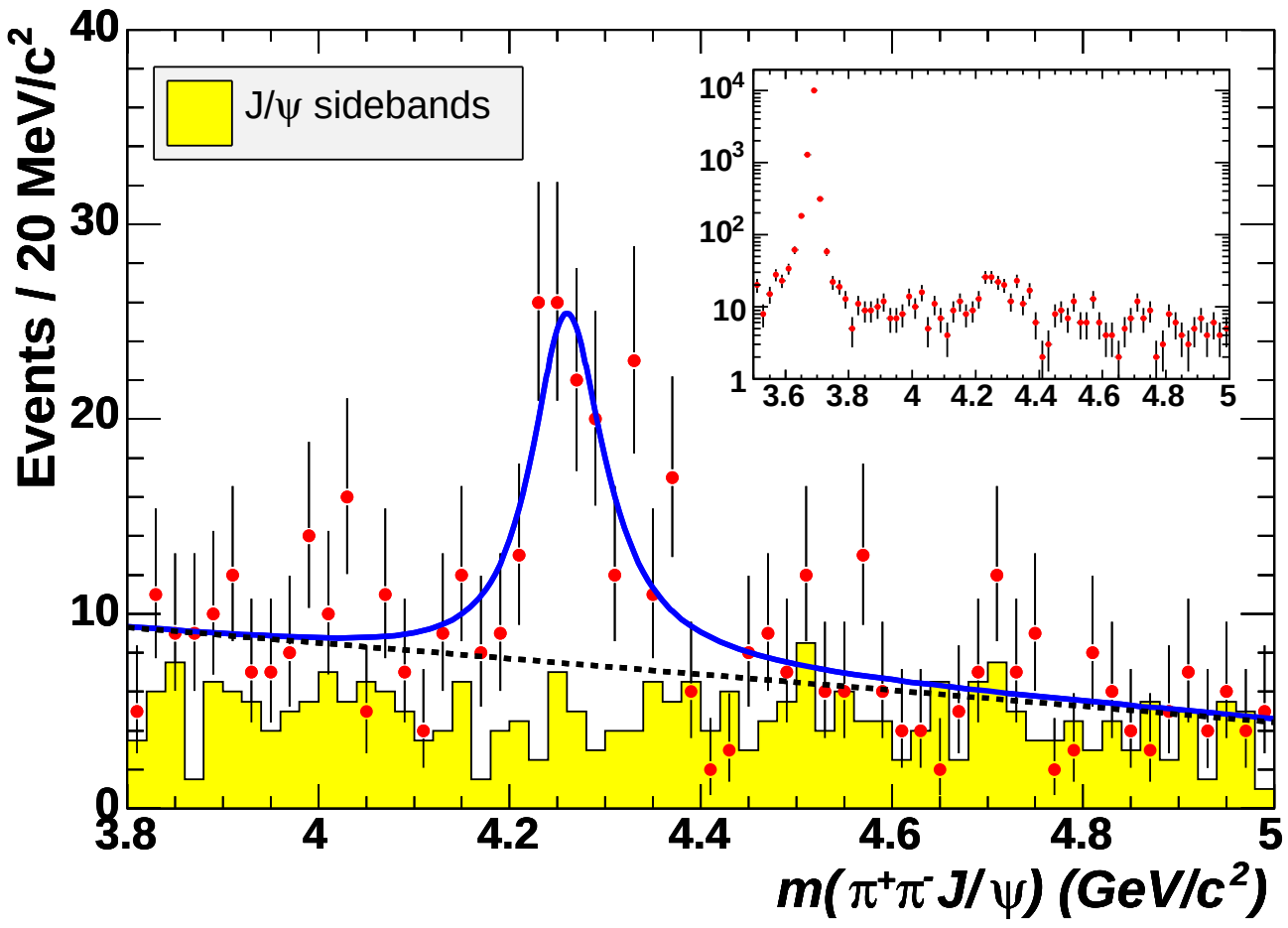

**Figure 21.5.2.** The $\pi^+\pi^- J/\psi$ invariant mass spectrum in the range 3.8–5.0 GeV/$c^2$ and (inset) over a wider range that includes the $\psi(2S)$, obtained from *BABAR* with a data sample of 232 fb$^{-1}$ (Aubert, 2005y). The points with error bars represent the selected data and the shaded histogram represents the scaled data from neighboring $J/\psi$ sidebands. A fit to the mass spectrum with a single Breit-Wigner and a polynomial function, shown as the solid line, clearly identifies the new $Y(4260)$ resonance. The dashed curve represents the background polynomial component.

An unbinned maximum likelihood fit to the $\pi^+\pi^- J/\psi$ mass spectrum in the range $[3.8, 5.0]$ GeV/$c^2$ is performed assuming only one broad resonance is present (Fig. 21.5.2). The fitting function consists of a relativistic Breit-Wigner function describing the peak, and a second-order polynomial background, convolved with a Cauchy resolution function. The fit finds (125±23) events in the peak, with a mass of (4259±8) MeV/$c^2$ and a width of (88±23) MeV/$c^2$. the significance of the $Y(4260)$ signal is above 8 $\sigma$.

The ISR photon is reconstructed in (24 ± 8)% of the $Y(4260)$ events, within the uncertainty the same as 25% observed for ISR $\psi(2S)$ events. Kinematic distributions for a sample of background subtracted data are compared to analogous quantities obtained for simulated ISR events, and found in good agreement, confirming that the ISR-production hypothesis – essential in order to assign the $J^{PC} = 1^{--}$ quantum numbers for the new resonant state – is correct.

#### 21.5.1.2 Subsequent $Y(4260)$ analyses

After the discovery of the $Y(4260)$, CLEO (Coan et al., 2006) performed an energy scan around $\sqrt{s} = 4.26$ GeV, and were able to confirm the process $Y(4260) \to \pi^+\pi^- J/\psi$ and observed also $Y(4260) \to \pi^0\pi^0 J/\psi$. The measured ratio, $\mathcal{B}(Y(4260) \to \pi^0\pi^0 J/\psi)/\ \mathcal{B}(Y(4260) \to \pi^+\pi^- J/\psi) \approx 0.5$, implies that the $Y(4260)$ has isospin zero. They also found the first evidence for $Y(4260) \to K^+K^- J/\psi$, and set upper limits on many other decay modes. These results are based on measurements at discrete energies, and do not resolve the $Y(4260)$ lineshape; it is assumed that the $Y(4260)$ saturates the $\pi^+\pi^- J/\psi$ and related cross-sections at the peak.

The observation of the $Y(4260) \to \pi^+\pi^- J/\psi$ decay in ISR-untagged studies has been confirmed by CLEO (He et al., 2006), by analysing a data sample of about 13.3 fb$^{-1}$, and with much more data by Belle (Yuan, 2007) and *BABAR* (Lees, 2012ab) . The selection criteria in the Belle and *BABAR* analyses are similar to those in the orig-inal *BABAR* measurement with some exceptions.[165] The resulting mass spectra shown in Fig. 21.5.3 present a clear peak at $\sim 4.26$ GeV/$c^2$, but have differences elsewhere in the analyzed energy region. The analysts of the two Collaborations have adopted different fit models to describe the data.

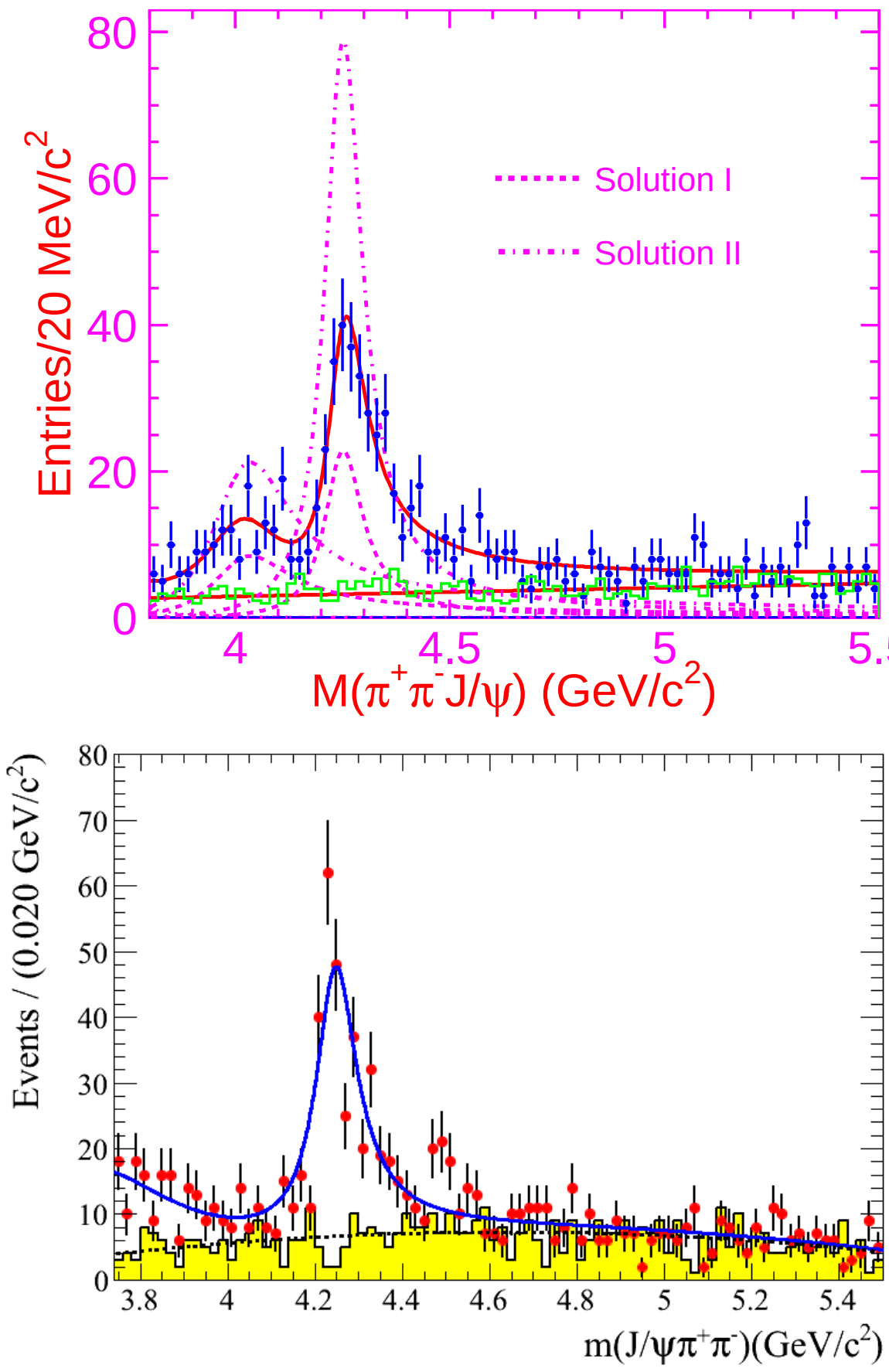

**Figure 21.5.3.** (upper) The $\pi^+\pi^- J/\psi$ mass spectrum measured by Belle (Yuan, 2007). Points with errors show selected events in the $J/\psi$ signal region, while the histogram shows the scaled $J/\psi$ sideband distribution. The curves show the best fit (solid line) and the contribution from each component for the two solutions, as described in the text. (lower) The $\pi^+\pi^- J/\psi$ mass spectrum measured by *BABAR* in a 454 fb$^{-1}$ data sample (Lees, 2012ab). The solid curve shows the result of a simultaneous fit to the data (points with errors) and to the background control sample obtained from the $J/\psi$ sidebands (shaded histogram).

**Table 21.5.1.** Results of the best fit to the $\pi^+\pi^- J/\psi$ mass spectrum obtained by the Belle experiment (Yuan, 2007). $M$, $\Gamma_{\rm tot}$, and $\mathcal{B}\times\Gamma_{e^+e^-}$ are respectively the mass, total width, and product of the branching fraction to $\pi^+\pi^- J/\psi$ and $e^+e^-$ partial width of the two interfering resonances $R_1$ and $R_2$, while $\phi$ is their relative phase. The results for both destructive and constructive interference solutions are reported.

| Parameters | | Solution I | Solution II |
|---|---|---|---|
| $R_1 : M$ | (MeV/$c^2$) | $4008 \pm 40^{+114}_{-28}$ | |
| $\Gamma_{\rm tot}$ | (MeV) | $226 \pm 44 \pm 87$ | |
| $\mathcal{B}\times\Gamma_{e^+e^-}$ | (eV) | $5.0 \pm 1.4^{+6.1}_{-0.9}$ | $12.4 \pm 2.4^{+14.8}_{-1.1}$ |
| $R_2 : M$ | (MeV/$c^2$) | $4247 \pm 12^{+17}_{-32}$ | |
| $\Gamma_{\rm tot}$ | (MeV) | $108 \pm 19 \pm 10$ | |
| $\mathcal{B}\times\Gamma_{e^+e^-}$ | (eV) | $6.0 \pm 1.2^{+4.7}_{-0.5}$ | $20.6 \pm 2.3^{+9.1}_{-1.7}$ |
| $\phi$ | ($^\circ$) | $+12 \pm 29^{+7}_{-98}$ | $-111 \pm 7^{+28}_{-31}$ |

In particular, an accumulation of events at a mass of about 4.01 GeV/$c^2$ is observed in about 548 fb$^{-1}$ of data analyzed by Belle. An unbinned maximum likelihood fit is performed to the mass spectrum corrected for the mass-dependent efficiency and normalized to the effective ISR luminosity, for masses above 3.8 GeV/$c^2$, as shown in Fig. 21.5.3 (top). The fit model consists of a coherent sum of two Breit-Wigner resonance functions (see Section 13.2.1), and assumes that there is no continuum production of $e^+e^- \to \pi^+\pi^- J/\psi$. The background is estimated from $J/\psi$ sidebands and fixed in the fit, while contributions from the tail of the $\psi(2S)$ and $\psi(3770)$ are estimated from the world average values of their parameters, added incoherently, and fixed. Two solutions with equally good fit quality are found, corresponding to constructive and destructive interference between the two resonances. They present equal masses and widths for the two resonances, but different partial widths and relative phase. The quality of the fit determined from the binned distribution is $\chi^2/n_{\rm dof} = 81/78$.

The results are summarized in Table 21.5.1. The significance of the resonance at lower mass is larger than $5\sigma$. Although its mass is close to that of the $\psi(4040)$, the fitted width is larger than the world average value ($80\pm10$ MeV) of the latter. The fit using two interfering resonances yields a much better description of the observed distribution than a fit with a single resonance. Using the same functional form of the fitted function as in Aubert (2005y) the parameters of the state at higher mass are consistent with those reported by *BABAR* (see Table 21.5.2).

[165] For example, Belle did not use the $J/\psi$ mass constraint and hence the sideband events can be used to estimate the background.

The *BABAR* update on the study of the ISR-produced $\pi^+\pi^- J/\psi$ final state is based on a data sample corresponding to 454 fb$^{-1}$ (Lees, 2012ab). An unbinned, extended-maximum likelihood fit is performed in the region $[3.74, 5.5]$ GeV/$c^2$ to the $\pi^+\pi^- J/\psi$ mass distribution from the $J/\psi$ signal region, shown in Fig. 21.5.3(bottom), and simultaneously to the background distribution from the $J/\psi$ sidebands. The shape of the background mass distribution is described by a third order polynomial function. The signal function consists of a coherent sum of a Breit-Wigner function for the $Y(4260)$, and an exponential function, which provides an empirical description of the $\psi(2S)$ tail and possible continuum production of the $\pi^+\pi^- J/\psi$ final state. Also in this case, the signal function accounts for the mass-dependency of the reconstruction efficiency and effective ISR luminosity, and is convolved with a Gaussian resolution function obtained from MC simulation. The parameters of the $Y(4260)$ resulting from the fit are in agreement with the previous measurements, and are reported in Table 21.5.2, which summarizes the ISR $Y(4260) \to \pi^+\pi^- J/\psi$ analyses from various experiments performed under the single-resonance hypothesis. The *BABAR* data do not support the Belle observation of a broad structure at 4.08 GeV/$c^2$.

The invariant mass distribution of the dipion system for events with a $\pi^+\pi^- J/\psi$ mass close to the $Y(4260)$ is found to deviate significantly from phase space in both *BABAR* and Belle data samples (see Fig. 18.3.10). It shows an accumulation around 0.95 GeV/$c^2$, followed by an abrupt fall to near zero at $\sim 1$ GeV/$c^2$, and then rises again. Such behavior is reproduced by *BABAR* (Lees, 2012ab) with a model assuming a coherent sum of a nonresonant $\pi^+\pi^-$ S-wave amplitude and a resonant amplitude describing the $f_0(980)$. The fit to the data sample assuming this simple model shows that the $f_0(980)$ contribution is produced in a fraction $0.17 \pm 0.13$ of $Y(4260) \to \pi^+\pi^- J/\psi$ decays, where only the statistical error is quoted.

In another study Belle observes also a significant accumulation of events in the $J/\psi\pi^\pm$ invariant mass of the $Y(4260) \to \pi^+\pi^- J/\psi$ decays (Liu, 2013). The state denoted as $Z(3900)^\pm$, which is clearly not a charmonium state (charge), is observed with a significance larger than 5 $\sigma$, and is in $Y(4260)$ decays produced with a branching ratio of $\mathcal{B}(Y(4260) \to Z(3900)^\pm\pi^\mp)\mathcal{B}(Z(3900)^\pm \to \pi^\pm J/\psi)/\mathcal{B}(Y(4260) \to \pi^+\pi^- J/\psi) = (29.0 \pm 8.9)\%$, where the uncertainty is statistical only. The properties of the state are consistent with the one observed by BESIII (Ablikim et al., 2013a).

Beside the $J/\psi\pi\pi\gamma_{\rm ISR}$ final state Belle studied also the ISR process with the production of $J/\psi KK$ (Yuan, 2008) and $J/\psi\eta$ (Wang, Han, Yuan, Shen, and Wang, 2013). The former measurement is the first observation of that final state in the ISR process. No evidence for the $Y(4260)$ is found and the upper limit on the product of the two-electron width and the branching fraction is determined to be $\Gamma(Y(4260) \to e^+e^-)\mathcal{B}(Y(4260) \to K^+K^- J/\psi) <$ 1.2 eV/ with 90% C.L. In the $J/\psi\eta$ final state clear signals of $\psi(4040)$ and $\psi(4160)$ are observed, and no signal of $Y(4260)$. For the latter the upper limit is $\Gamma(Y(4260) \to e^+e^-)\mathcal{B}(Y(4260) \to \eta J/\psi) < 14.2$ eV at 90% C.L.

**Table 21.5.2.** Summary of ISR $\pi^+\pi^- J/\psi$ measurements by various experiments under the single-resonance hypothesis, where $Y$ stands for $Y(4260)$. The first uncertainty is statistical and the second systematic (if only one is given it is the statistical uncertainty only).

| Experiments | $\mathcal{L}$ ($\text{fb}^{-1}$) | $N_Y$ | mass (MeV/$c^2$) | $\Gamma_{\text{tot}}$ (MeV) | $\mathcal{B}(Y \to \pi^+\pi^- J/\psi)\Gamma_{Y\to e^+e^-}$ (eV) |
|---|---|---|---|---|---|
| CLEO (He et al., 2006) | 13.3 | 37 | $4284^{+17}_{-16} \pm 4$ | $73^{+39}_{-25} \pm 5$ | $8.9^{+3.9}_{-3.1} \pm 1.8$ |
| Belle (Yuan, 2007) | 548 | $324 \pm 21$ | $4263 \pm 6$ | $126 \pm 18$ | $9.7 \pm 1.1$ |
| *BABAR* (Lees, 2012ab) | 454 | $344 \pm 39$ | $4244 \pm 5 \pm 4$ | $114^{+16}_{-15} \pm 7$ | $9.2 \pm 0.8 \pm 0.7$ |

### 21.5.2 $Y$ family states in ISR $\pi^+\pi^-\psi(2S)$

Since the $Y(4260)$ is above the mass threshold for $\psi(2S)$ plus $\pi^+\pi^-$, it is natural to ask if $Y(4260)$ also decays to $\pi^+\pi^-\psi(2S)$. In order to further clarify the nature of the $Y(4260)$, and search for similar states, *BABAR* (Aubert, 2007m) and Belle (Wang, 2007c) have studied the ISR process $e^+e^- \to \gamma_{\text{ISR}}\pi^+\pi^-\psi(2S)$. The $\psi(2S)$ is reconstructed in the $\pi^+\pi^- J/\psi$ decay mode, with $J/\psi \to \ell^+\ell^-$. The analysis procedure is similar to the case of the $\pi^+\pi^- J/\psi$ final state. An additional background source, produced by different combinations within the same $2(\pi^+\pi^-)J/\psi$ system where at least one of the primary pions is combined with the $J/\psi$ to form a $\pi^+\pi^- J/\psi$ candidate, must be subtracted from the selected sample. The clean sample of selected $\psi(2S)$ decays is used for signal estimation, while the sidebands of the $\psi(2S)$ mass distribution are used for background subtraction.

The *BABAR* analysis (Aubert, 2007m) is based on a sample of 298 $\text{fb}^{-1}$. The $\pi^+\pi^-\psi(2S)$ mass spectrum of 78 events selected within the $\psi(2S)$ mass window is represented by data points in Fig. 21.5.4. The dashed curve shows the result of the fit to the $Y(4260)$ using resonance parameters fixed to those of Aubert (2005y). The $\chi^2/n_{\text{dof}}$ of 21.3/8 quantifies the inconsistency of the data with the decay of $Y(4260)$ to this final state. However, a clear accumulation of events is seen at a mass of about 4.35 GeV/$c^2$, and a fit to a single resonance with free mass and width parameters returns $m = (4324 \pm 24)$ MeV/$c^2$ and $\Gamma = (172 \pm 33)$ MeV, with a much better fit quality: $\chi^2/n_{\text{dof}} = 7.3/7$ (the errors are statistical only). A fit to the known $\psi(4415)$ with fixed parameters is also of poor quality, supporting the hypothesis of a new state.

In a similar analysis, performed using twice the dataset, Belle (Wang, 2007c) confirms the state at about 4.35 GeV/$c^2$, and observes another structure at higher masses. A fit to the $\pi^+\pi^-\psi(2S)$ invariant-mass spectrum of selected events with two coherent vector resonances returns for the lower-mass resonance $m = (4361 \pm 9 \pm 9)$ MeV/$c^2$ and $\Gamma = (74 \pm 15 \pm 10)$ MeV, and for the state at higher mass $m = (4664 \pm 11 \pm 5)$ MeV/$c^2$ and $\Gamma = (48 \pm 15 \pm 3)$ MeV. A small enhancement around 4685 MeV/$c^2$ in a single 50 MeV/$c^2$ bin was visible also in the *BABAR* measurement, but not significant due to the lower integrated luminosity. A recent analysis performed by *BABAR* (Lees, 2012ac) using the whole data set confirms the state, now named $Y(4660)$, with parameters consistent with those measured by Belle.

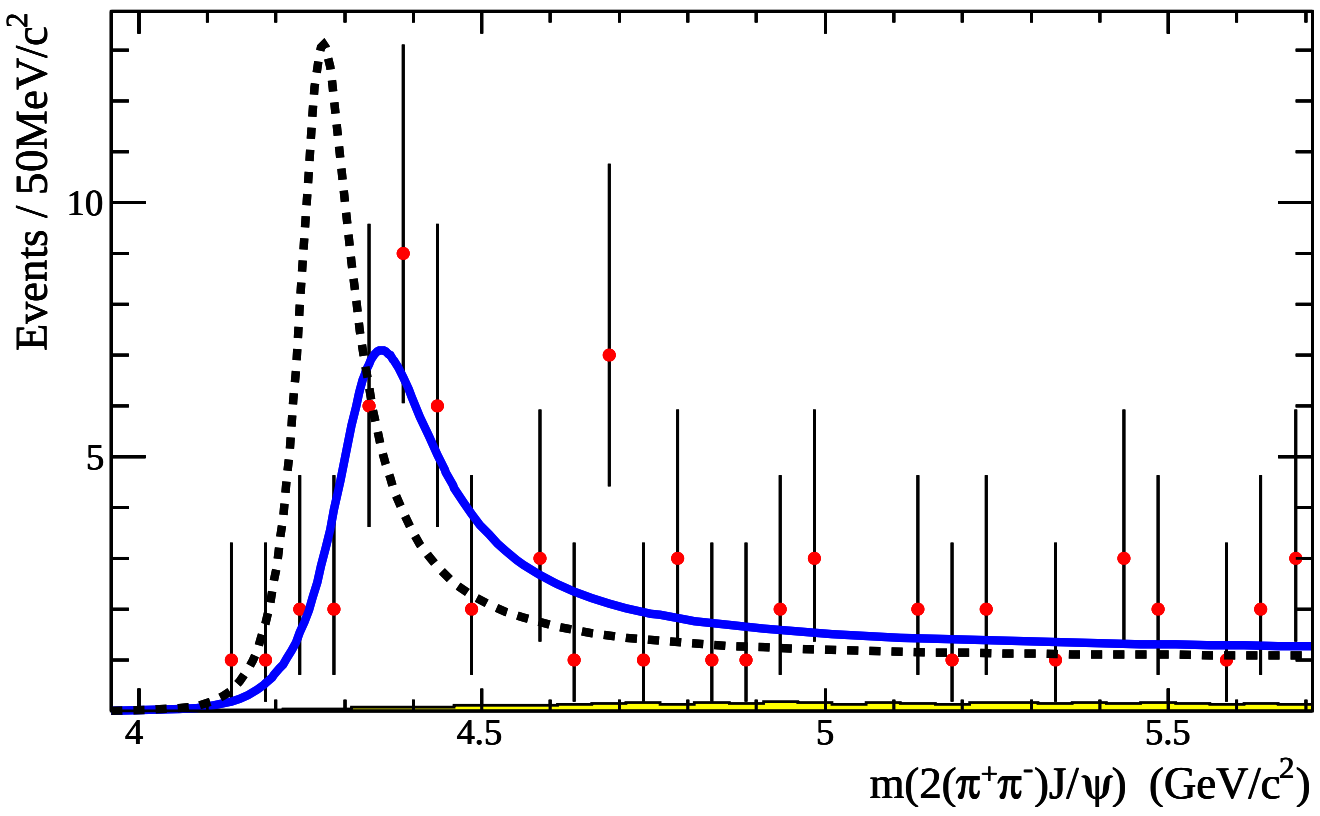

**Figure 21.5.4.** Invariant mass spectrum up to 5.7 GeV/$c^2$ for selected ISR-produced $2(\pi^+\pi^-)J/\psi$ candidates in 298 $\text{fb}^{-1}$ of *BABAR* data (Aubert, 2007m). The shaded histogram represents the background estimated from the sidebands of the $\psi(2S)$ mass spectrum, and the curves represent fits to the data (see the text).

## 21.6 Dark force searches

While the astrophysical evidence for dark matter is now overwhelming, its precise nature and origin remain elusive. Recent results from terrestrial and satellite experiments have motivated an interesting proposal in which WIMP-like dark matter particles carry charge of a new yet unknown force (Arkani-Hamed, Finkbeiner, Slatyer, and Weiner, 2009; Fayet, 2007; Pospelov, Ritz, and Voloshin, 2008). The corresponding gauge boson, the so-called dark photon $A'$, couples to the Standard Model photon through mixing between the photon and dark photon fields (kinetic mixing) with mixing strength $\epsilon$.

This opens the possibility of dark matter annihilation into a pair of dark photons which subsequently decay to SM particles. The mass of the dark photon is constrained to be at most a few GeV, to be compatible with the electron/positron excess observed by PAMELA (Adriani et al., 2010, 2009) and FERMI (Abdo et al., 2009; Ackermann et al., 2012), without a comparable anti-proton signal. Dark photons decay almost exclusively to lepton-pairs if their mass is below $\sim 500$ MeV, while the contribution of pion pairs is significant between $\sim 500$ MeV and $\sim 1$ GeV, and multi-hadron channels become dominant at higher masses. The dark boson masses are usually generated via the Higgs mechanism, adding one or more dark Higgs bosons ($h'$) to the theory.

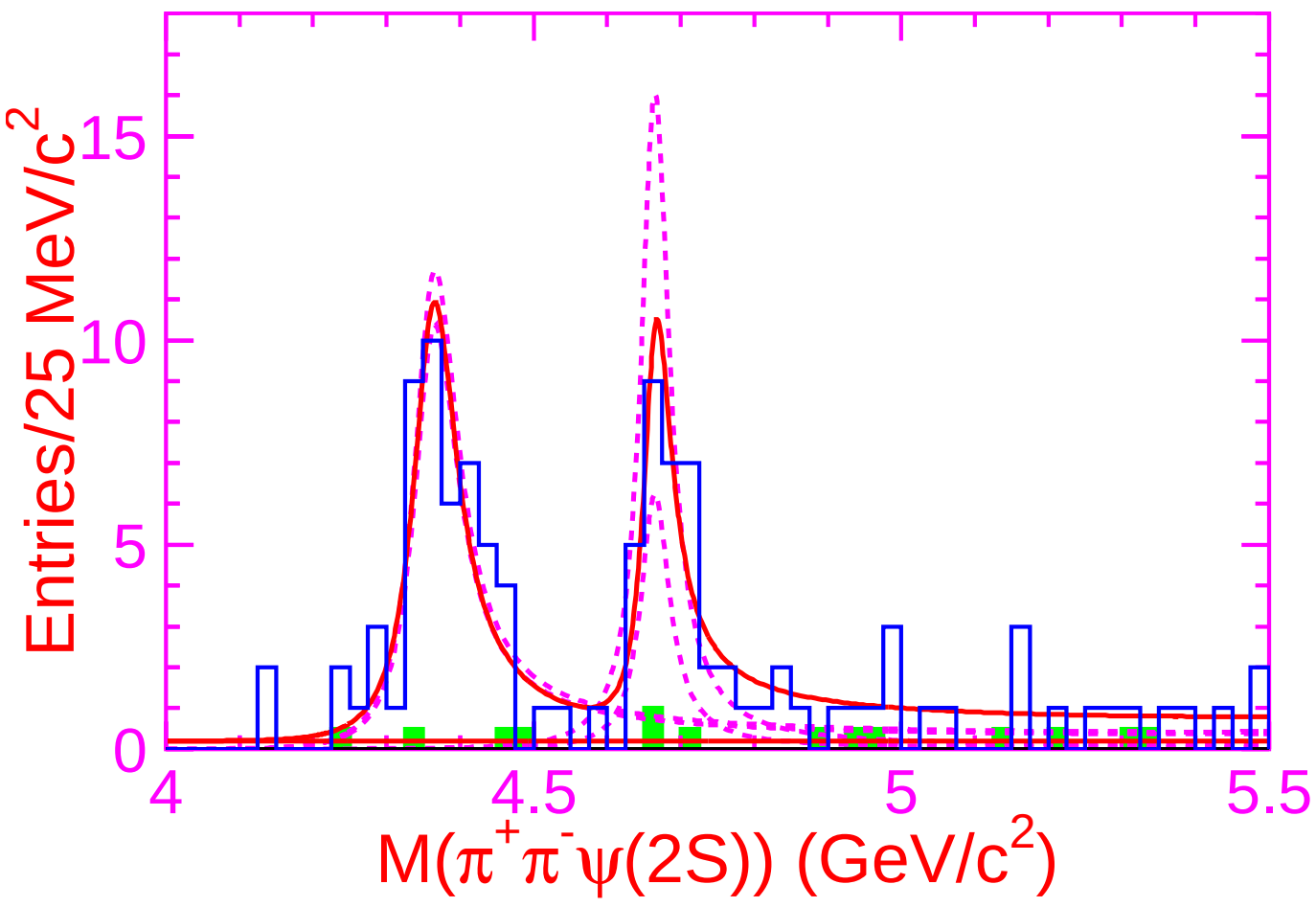

**Figure 21.5.5.** The $\pi^+\pi^-\psi(2S)$ invariant mass distribution for events that pass the $\psi(2S)$ selection in Belle 673 fb$^{-1}$ data (Wang, 2007c). The open histogram is the data while the shaded histogram is the normalized $\psi(2S)$ sidebands. The solid curve shows the result of the the best fit with two coherent $P$-wave resonances together with a constant incoherent background term. The two dashed curves at each peak show the two solutions for constructive and destructive interference.

### 21.6.1 Searches for a dark photon

A dark photon can be readily produced in the reaction $e^+e^- \to \gamma A'$, $A' \to \ell^+\ell^-$. The signature is similar to that of a light $CP$-odd Higgs boson, $A^0$, in $\Upsilon(2S,3S) \to \gamma A^0$, $A^0 \to \ell^+\ell^-$ (Aubert, 2009an). The $\Upsilon(2S,3S)$ candidates are reconstructed by combining two oppositely-charged tracks with a photon. At least one track must be identified as a muon and the energy of the photon in the CM frame is required to be larger than 0.5 GeV. No additional tracks or photons must be detected in the event. The signal yield is extracted as a function of the $A^0$ mass by a series of unbinned extended maximum likelihood fits to the distribution of the dimuon mass. No significant signal is observed, and upper limits on the branching fraction $\Upsilon(2S,3S) \to \gamma A^0$, $A^0 \to \ell^+\ell^-$ are derived. These results have been reinterpreted (Bjorken, Essig, Schuster, and Toro, 2009) as limits on the mixing strength $\epsilon$ at the level $10^{-3} - 10^{-2}$ (Fig. 21.6.1).

Additional searches have been performed which may be reinterpreted as limits on dark photon production, such as $e^+e^- \to \gamma$ invisible (Aubert, 2008at), $e^+e^- \to \gamma$ hadrons (Lees, 2011j), or $e^+e^- \to \gamma\tau^+\tau^-$ (see Section 18.4.7.1, and Aubert, 2009ai).

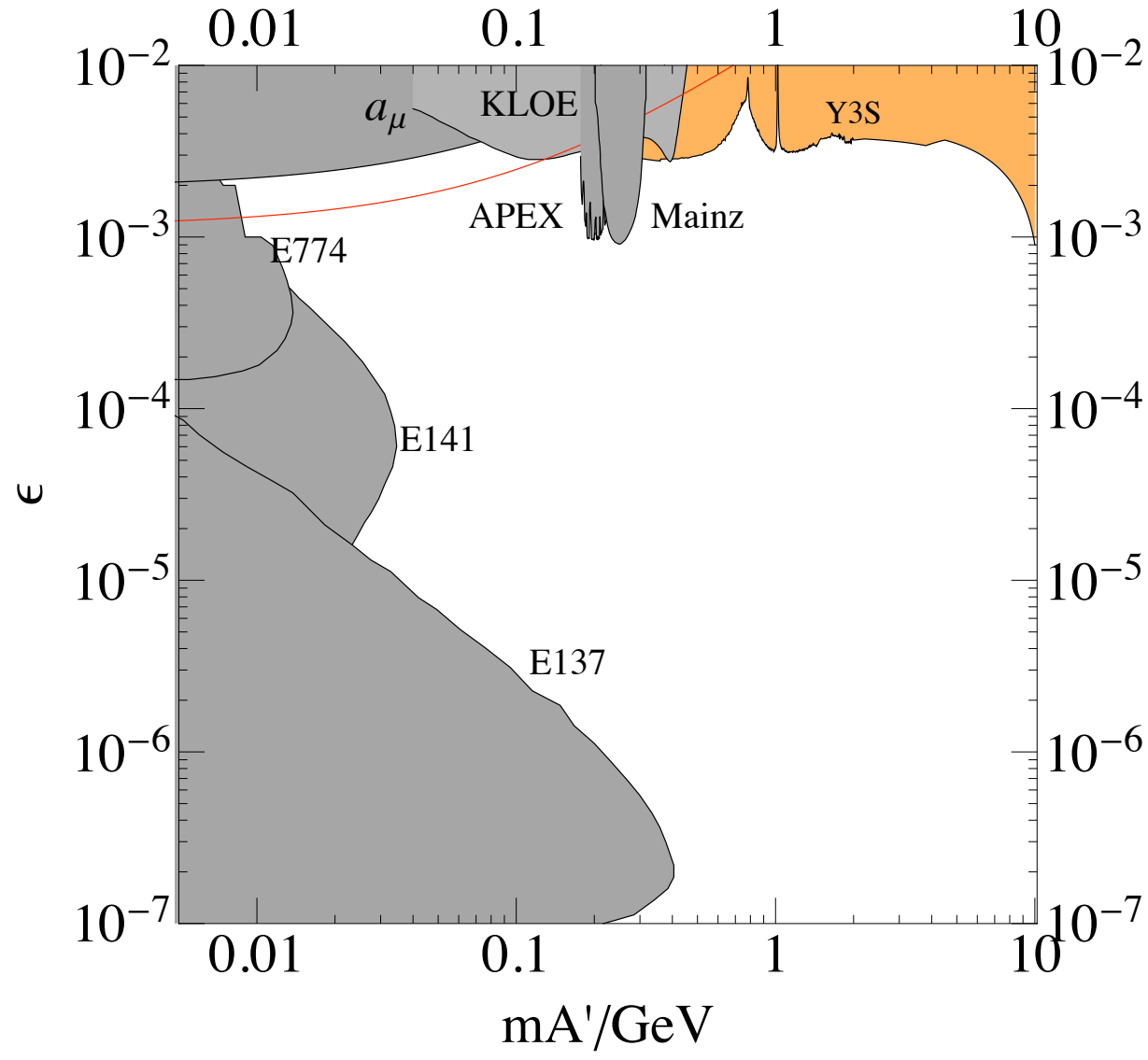

**Figure 21.6.1.** Adopted from (Bjorken, Essig, Schuster, and Toro, 2009). Constraints on the mixing strength, $\epsilon$, as a function of the dark photon mass. The red line shows the value of the coupling required to explain the discrepancy between the calculated and measured anomalous magnetic moment of the muon (Pospelov, 2009). The excluded regions obtained by reinterpreting the upper limits on the $Y(2S,3S) \to A^0\gamma$, $A^0 \to \ell^+\ell^-$ (Aubert, 2009an) branching fractions are shown as a yellow band.

### 21.6.2 A search for dark gauge bosons

Non-Abelian extensions of dark sectors introduce additional dark gauge bosons, generically denoted $W', W'', \ldots$. The detailed phenomenology depends on the structure of the model, but heavy dark bosons decay to lighter states if kinematically accessible, while the lightest bosons are metastable and decay to SM fermions via their mixing with the dark photon (Baumgart, Cheung, Ruderman, Wang, and Yavin, 2009; Essig, Schuster, and Toro, 2009).

*BABAR* has performed a search for di-boson production in the four lepton final state, $e^+e^- \to A'^* \to W'W''$, $W' \to \ell^+\ell^-$, $W'' \to \ell'^+\ell'^-$ with $\ell, \ell' = e, \mu$ (Aubert, 2009aj). The study, based on 513 fb$^{-1}$ of data collected mostly at the $\Upsilon(4S)$ resonance, has been performed in the context of inelastic dark matter models (Tucker-Smith and Weiner, 2001), searching for two bosons with similar masses. Events containing four leptons originating from the interaction point with a total invariant mass greater than 10 GeV are selected. Additional selection criteria on the boson decay angles and the angle between the decay planes of the two bosons are applied to further reject the background.

The signal is extracted as a function of the average dileptonic mass in the range $0.24 - 5.3$ GeV in 10 MeV steps. No significant signal is observed and 90% C.L. upper limits on the mixing strength at the $10^{-3}$ level have been set, assuming a dark sector coupling constant $\alpha_D =$ $g_D/4\pi = \mathcal{O}(10^{-2})$, and equal branching fractions of a dark gauge boson to $e^+e^-$ and $\mu^+\mu^-$.

### 21.6.3 A search for dark Higgs bosons

The Higgsstrahlung process, $e^+e^- \to A'h'$, $h' \to A'A'$, offers another gateway to dark sectors, as this process is one of the few suppressed by only a single power of the mixing strength, and the background is expected to be almost negligible (Batell, Pospelov, and Ritz, 2009). A search for dark Higgs boson production has been performed in the range $0.8 < m_{h'} < 10.0\,\text{GeV}$ and $0.25 < m_{A'} < 3.0\,\text{GeV}$ with the constraint $m_{h'} > 2m_{A'}$ (Lees, 2012s). The signal events are either fully reconstructed using $A' \to \ell^+\ell^-$ and $A' \to \pi^+\pi^-$ decays (exclusive mode), or partially reconstructed (inclusive mode). In the latter case, only two of the three dark photons are identified as dilepton resonances, with the four-momentum of the remaining dark photon identified with that of the recoiling system. The exclusive modes contain six tracks having an invariant mass close to $\sqrt{s}$, forming three dark photon candidates of similar mass. The six pion final state has a significantly larger background than the other channels and is excluded from the search. Inclusive modes are first identified by selecting two dileptonic resonances with similar mass, and requiring the mass of the recoiling system to be compatible with the dark photon hypothesis.

No significant signal is observed, and 90% C.L. upper limits on the product of the dark sector coupling constant and the mixing strength, $\alpha_D\epsilon^2$, are derived. The results are displayed in Fig. 21.6.2 as a function of the dark photon mass for selected values of the dark Higgs boson masses. Values as low as $10^{-10} - 10^{-8}$ are excluded for a large range of dark photon and dark Higgs masses. Assuming $\alpha_D = \alpha_{\text{EM}}$, these measurements translate into limits on the mixing strength in the range $10^{-4} - 10^{-3}$.

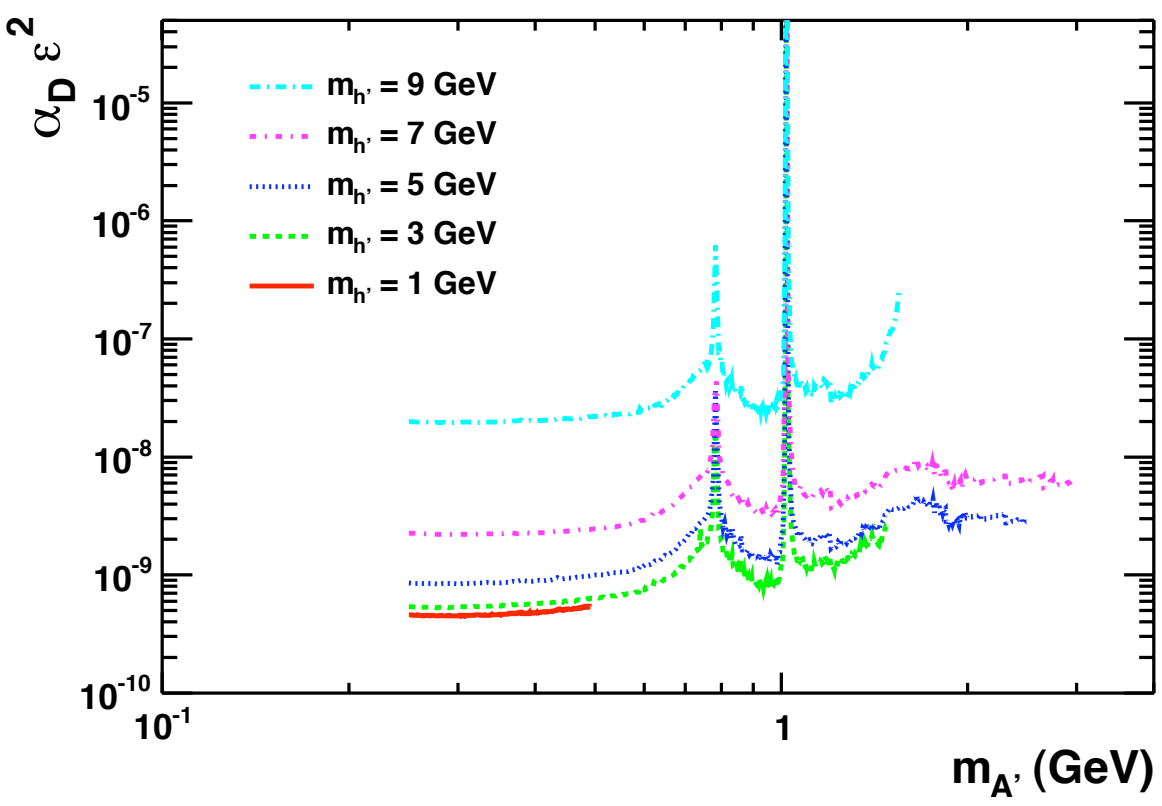

**Figure 21.6.2.** Upper limit at 90% C.L. set by *BABAR* (Lees, 2012s) on the product $\alpha_D\epsilon^2$ as a function of the dark photon mass for selected values of dark Higgs boson masses. The peaking structure arising near $m_{A'} \sim 0.8\,\text{GeV}$ and $m_{A'} \sim 1.0\,\text{GeV}$ reflects the presence of the $\omega$ and $\phi$ resonances. At these masses, dark photons decay predominantly to $3(\pi^+\pi^-\pi^0)$ and $3(K^+K^-)$ final states, not included in the search.

# Chapter 22
# Two-photon physics

***Editors:***
*Vladimir P. Druzhinin (BABAR)*
*Sadaharu Uehara (Belle)*

**Additional section writers:**
*Hideyuki Nakazawa, Cheng Ping Shen, Yasushi Watanabe, Chang Chun Zhang*

## 22.1 Descriptions of two-photon topics to be covered

### 22.1.1 Introduction for two-photon physics

An electron-positron collider is also a photon-photon collider. Since the photon couples directly to the electric charge of quarks, we can study hadron structures and QCD physics, effectively, in hadron production induced by two-photon collisions. The even $C$-parity of the two-photon system is complementary with the odd $C$-parity in $e^+e^-$ collisions. In an $e^+e^-$ collider, a virtual photon is emitted from each lepton; a collision of these photons produces final-state particles as shown in Figure 22.1.1. The two-photon center-of-mass (CM) energy, $W$, which is the same as the invariant mass of the final-state system, is continuously distributed between zero and just below the $e^+e^-$ CM energy. For practical purposes, the usable range is between a few 100 MeV and $\sim 4.5$ GeV. The lower side is limited by experimental trigger conditions and the upper by the luminosity and backgrounds from $e^+e^-$ annihilation events.

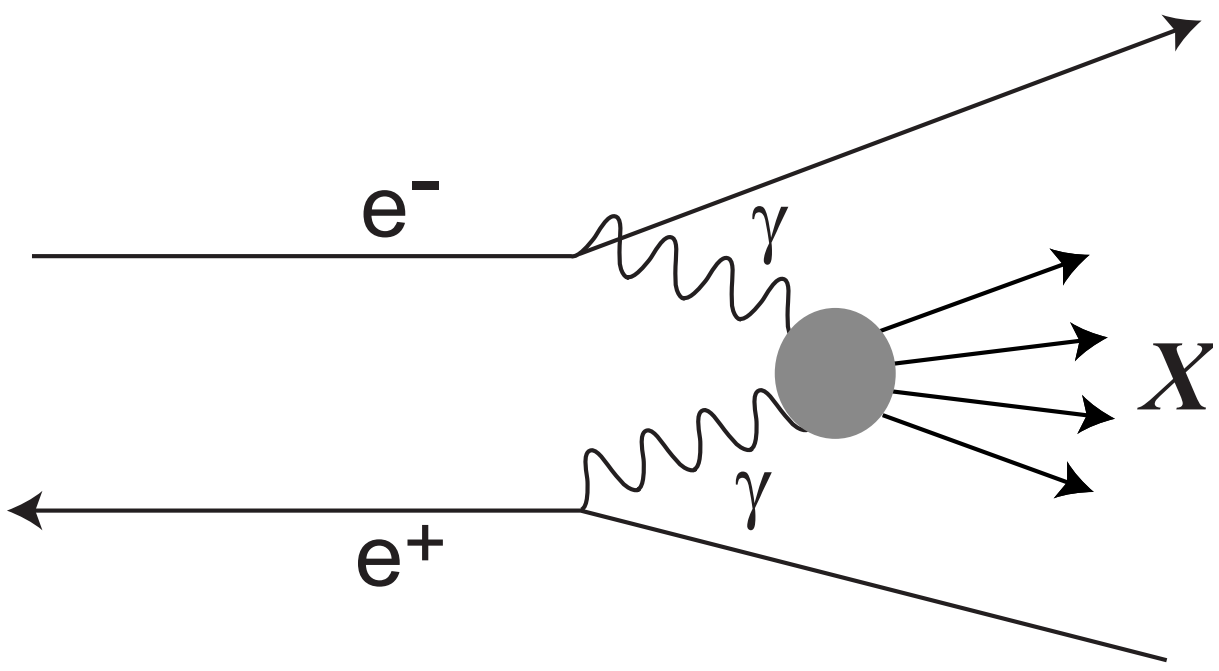

**Figure 22.1.1.** A Feynman diagram illustrating hadron production in two-photon collisions $e^+e^- \to e^+e^-X$.

The study of hadron production in two-photon processes at the $B$ Factories has contributed to better understanding of non-perturbative QCD and light-quark meson spectroscopy at low and intermediate energies. It also contributes to searches and measurements of charmonium(-like) states as well as a test of perturbative-QCD models for exclusive meson production at high energies (Brodsky and Lepage, 1981; Chernyak and Zhitnitsky, 1984).

### 22.1.2 Cross section for $\gamma\gamma$ collisions (zero-tag)

In hadron production via two-photon collisions,

$$e^+e^- \to e^+e^-X, \tag{22.1.1}$$

the cross sections of the processes are given as a function of $W$ and the $Q^2$ value of each incident photon, where $Q^2$ is the negative of the mass squared of the virtual photon and is the same as the absolute value of four-momentum transfer squared between the incident ($p_{e,\mathrm{ini}}$) and recoiled electron or positron ($p_{e,\mathrm{rec}}$), $Q^2 = -(p_{e,\mathrm{rec}} - p_{e.\mathrm{ini}})^2$. Hereafter we refer to both electrons and positrons as electrons for brevity.

Roughly speaking for the $B$ Factory energies, the size of the cross section is a comparable order of magnitude with that for $e^+e^-$ annihilation. The cross section is largest in the kinematic regions where the $Q^2$ value is very close to zero and $W$ is small when compared with the beam energy. In contrast, in regions where either $W$ or $Q^2$ is much larger than the typical QCD energy scale ($\sim 1$ GeV), the cross section decreases rapidly so that statistical uncertainty dominates the measurement errors at the $B$ Factories.

We discuss here the zero-tag method used to measure and derive the cross section corresponding to real two-photon collisions, where neither of the recoiling electrons are detected. The $Q^2$ of the emitted virtual photons has a continuous distribution and peaks very close to zero (*i.e.* smaller than the electron mass squared). Most events have $Q^2$ much smaller than the QCD scale or any hadron-mass squared, say $(100\ \mathrm{MeV})^2$, so the measured cross section approximates that of the collisions of real photons.

We approximate a real photon in $\gamma\gamma$ cross-section measurements with a virtual photon having a small $Q^2$ and extrapolate the cross section measured with finite $Q^2$ photons to the $Q^2 = 0$ limit, using an appropriate $Q^2$ dependence of the cross section that is interpreted as a photon-hadron form factor effect. In addition, when we adopt an equivalent-photon approximation (EPA), the effect from each photon is separated and factorized; also, the real two-photon cross section is decoupled from the photon-emission part (Berger and Wagner, 1987; Budnev, Ginzburg, Meledin, and Serbo, 1975). With these approximations, we can define and calculate a universal two-photon luminosity function $L_{\gamma\gamma}$ as a function of $W$:

$$L_{\gamma\gamma}(W) = \frac{d}{dW}\left(\int N(k_1, E_b) N(k_2, E_b) \frac{1}{Q_1^2}\frac{1}{Q_2^2} F(Q_1^2, W) F(Q_2^2, W) dk_1 dk_2 dQ_1^2 dQ_2^2\right), \tag{22.1.2}$$

where $N(k_i, E_b)$ is the probability density function (obtained from QED) for the virtual photon with index $i = 1, 2$ and energy $k_i$ emitted from the incident lepton with energy $E_b$, and $F(Q_i^2, W)$ is the factorized form factor effect for this photon (normalized to the real photon $F(0, W) = 1$). The two-photon luminosity function is used to translate the $e^+e^-$-based cross section $\sigma_{ee}$ at given $W$ to the corresponding $\gamma\gamma$ cross section $\sigma_{\gamma\gamma}(W)$ by the relation

$$\sigma_{\gamma\gamma}(W) = \frac{1}{L_{\gamma\gamma}(W)} \frac{d\sigma_{ee}}{dW}. \quad (22.1.3)$$

We also use several differential cross sections in the measurement of the angular and momentum distributions of the final-state particles, accounting for the efficiencies as a function of the measured variables. For example, the differential cross section for a process with a two-body final state is given by the following formula:

$$\frac{d\sigma_{\gamma\gamma}}{d|\cos\theta^*|} = \frac{\Delta N}{\Delta W \Delta|\cos\theta^*| \; \epsilon \; L_{\gamma\gamma}(W) \int \mathcal{L}dt}, \quad (22.1.4)$$

where $\theta^*$ is the scattering angle of a final-state particle in the two-photon CM frame, $\Delta N$ is the number of signal events in a two-dimensional bin with a bin size of $\Delta W \times \Delta|\cos\theta^*|$, $\epsilon$ is the efficiency for the bin and $\int \mathcal{L}dt$ is the integrated luminosity for $e^+e^-$ incident beams.

### 22.1.3 Resonance production

The single meson formation process $\gamma\gamma \to R$, in which only one meson $R$ is produced from the two-photon collision, is one of the most important processes in two-photon physics. The quantum numbers of the meson $R$ are limited: it must be electrically neutral and have even $C$-parity; the spin-parity of $J = 1^{\pm}$ or $J^P$ =(odd)$^-$ are prohibited in collisions of real photons. There are other restrictions for the helicity for the produced meson $R$.

This process allows us to measure the two-photon partial width of the final-state meson $R$. The cross section can be written as:

$$\sigma(W) = 8\pi(2J+1) \frac{\Gamma_{\gamma\gamma}\Gamma\mathcal{B}(R \to \text{final state})}{(W^2 - M_R^2)^2 + M_R^2\Gamma^2}, \quad (22.1.5)$$

where $M_R$, $\Gamma_{\gamma\gamma}$ and $\Gamma$ are the mass, two-photon decay width, and total width of the meson resonance $R$, respectively. $\mathcal{B}(R \to \text{ final state})$ is the branching fraction for the decay of the meson $R$. Here we assume that the resonance shape is represented by a conventional relativistic Breit-Wigner function (see Chapter 13). If we know the branching fraction of the meson's decay mode used in the measurement, then we can extract the two-photon decay width; otherwise, we measure the product of the two-photon decay width and the branching fraction. The two-photon partial decay width is a fundamental and direct observable to explore the $q\bar{q}$ or exotic nature of the neutral meson. Even for a non-exotic meson, the two-photon decay width is useful to study the quarks' quantum state inside the meson and to test QCD models (Munz, 1996).

In a zero-tag measurement, we can make the signal events almost free from other processes by applying a rather stringent transverse-momentum balance. This is a great advantage in searches for new resonances as well as new decay modes of known hadrons. The requirement for the transverse-momentum balance also restricts the $Q^2$ of the incident photon and ensures the processes originates from real photon collisions. Similar requirements are also useful in the single-tag cases described below to ensure a small $Q^2$ for the untagged photon.

### 22.1.4 Single-tag measurements

When one of the scattered electrons is detected, we refer to these two-photon processes as single-tag modes. In the single-tag process, we can probe the structure of the real photon or a hadron with a high-$Q^2$ photon; this process is very useful for studies of hadron and QCD physics such as meson transition form factors (Brodsky and Lepage, 1981). The $Q^2$ of the virtual photon is determined by measuring the scattering angle and energy of the recoil electron with the following Lorentz-invariant formula:

$$Q^2 = 4E_bE' \sin^2\frac{\theta}{2}, \quad (22.1.6)$$

where $E_b$ and $E'$ are the energies of the incident and recoiling (tagged) electrons and $\theta$ is the scattering angle of the tagged electron.

The formation of a resonance in the single-tag process is a fertile field of study at the $B$ Factories, especially in the high-$Q^2$ region where the production cross section is highly suppressed. We impose a kinematic requirement on the single-tag mode that all the final state particles be observed except for the non-tagged incident electron (which can be inferred from a missing-mass constraint). In the $B$ Factory experiments, the pseudoscalar transition form factors have been measured in single-tag modes, and are discussed in section 22.7. It is also possible to produce spin-1 (axial-vector) mesons in the single-tag process, although such measurements have not been reported from the $B$ Factories.

### 22.1.5 Monte-Carlo Techniques

No general-purpose Monte-Carlo (MC) generator for two-photon processes was available during the running of the $B$ Factories. For exclusive two-photon processes, we can explicitly specify the combination of the final-state particles exclusively as well as the $W$ distribution used in the event generation. In addition, angular distributions must be specified in the event generation.

For these purposes, several MC generators to simulate a resonance or an exclusive final-state system from two-photon collisions, such as `TREPS` (Uehara, 1996), `GGRESRC` (Druzhinin, Kardapoltsev, and Tayursky, 2010), `Gamgam` (Aubert, 2010g) etc., are prepared and are used in the analysis. In these generators, one can specify functional shapes of distributions for $W$, $Q^2$ and angles of final-state particles, or they can be generated by built-in default functions. Usually, an equivalent-photon approximation is adopted for zero-tag event generation while a $Q^2$ distribution not based on EPA is sometimes used for the single-tag cases. Even for the non-EPA cases, we can obtain the conversion factor to obtain the $\gamma\gamma^*$ physics variable, such as the square of a transition form factor, from the $e^+e^-$-based cross section through the MC calculations.

The $Q^2$ dependence or its maximum limit affects values or definitions of the luminosity function, the cross section on the $e^+e^-$-incident basis and the efficiency in the measurements of the zero-tag mode; hence consistency among these quantities must be considered in the analysis.

Background processes are modeled using generators which take into account the higher-order QED processes, parton-distribution functions and/or soft hadrons, such as `Pythia` (Sjöstrand, Mrenna, and Skands, 2006).

## 22.2 Pseudoscalar meson-pair production

The energy region of meson-pair production through two-photon processes can be naturally divided into a low energy region, where resonance production is dominant, and a high energy region, where cross sections tend to show asymptotic behavior and can be compared to QCD predictions.

In this section, we restrict ourselves to the pair production of pseudoscalar mesons (denoted as $P_1$ and $P_2$). Belle has made extensive studies of these processes for both charged-meson pairs,

(1) $\gamma\gamma \to \pi^+\pi^-$ (Mori, 2007a,b; Nakazawa, 2005),
(2) $\gamma\gamma \to K^+K^-$ (Abe, 2003d; Nakazawa, 2005),

and neutral-meson pairs,

(3) $\gamma\gamma \to K^0_S K^0_S$ (Chen, 2007c; Uehara, 2013),
(4) $\gamma\gamma \to \pi^0\pi^0$ (Uehara, 2008a, 2009b),
(5) $\gamma\gamma \to \eta\pi^0$ (Uehara, 2009a),
(6) $\gamma\gamma \to \eta\eta$ (Uehara, 2010a).

The polar angle ($\theta^*$) coverage is typically restricted to $|\cos\theta^*| < 0.6$; for neutral pairs decaying into photons this range is extended to $|\cos\theta^*| < 0.8$ or $|\cos\theta^*| < 1.0$, because small angle photons can be detected by the endcap calorimeters. The wider coverage provides better separation of the partial waves as discussed in Section 22.2.1.1.

As an example of the high statistics in raw data that have been recorded at the $B$ Factories, we show the $W$ dependence of the experimental yields for the $\gamma\gamma \to \pi^+\pi^-$ and $\gamma\gamma \to \eta\pi^0$ candidate events in Fig. 22.2.1 (a) and (b), respectively.

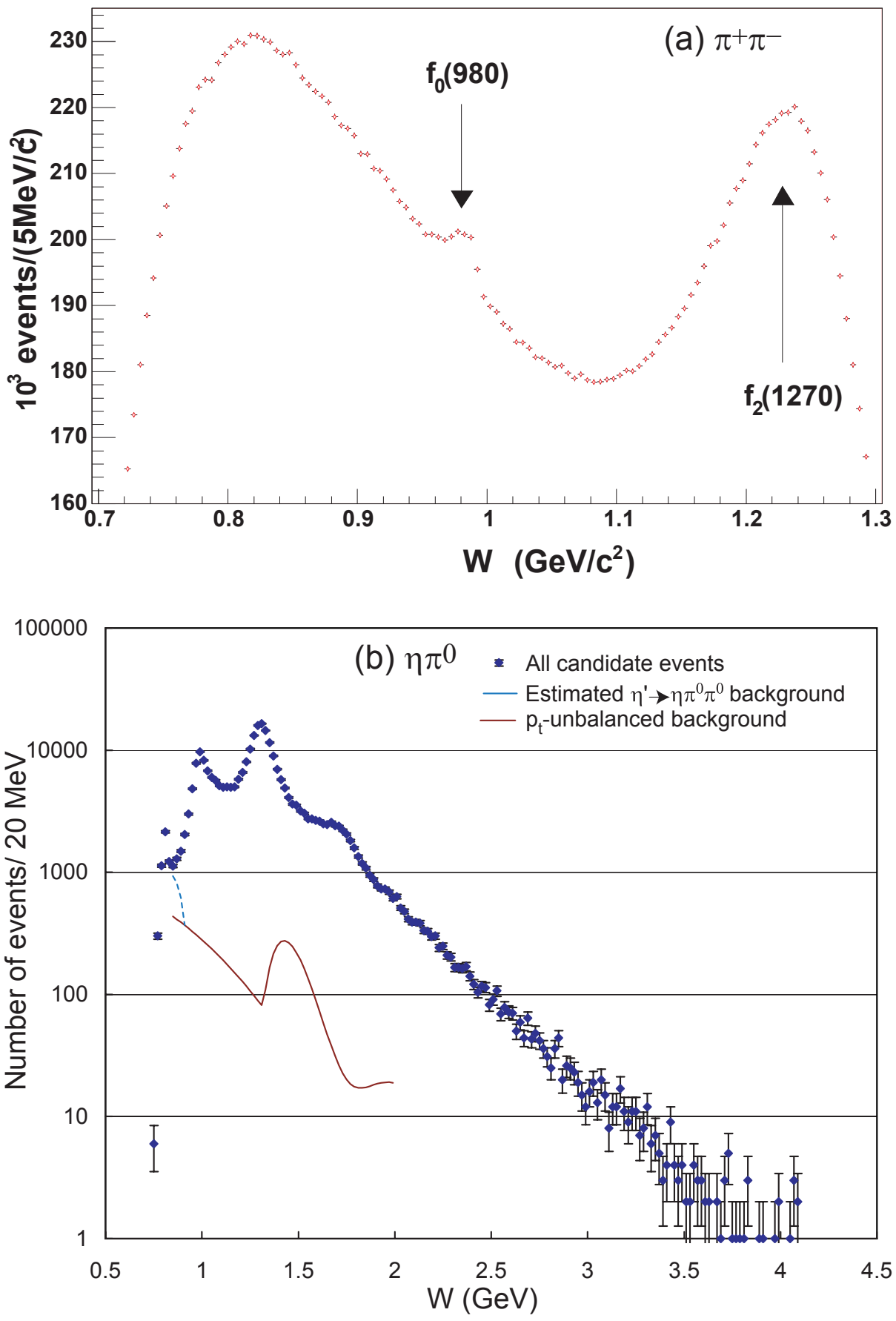

**Figure 22.2.1.** Invariant mass, $W$, distribution for (a) $\gamma\gamma \to \pi^+\pi^-$ and (b) $\gamma\gamma \to \eta\pi^0$ candidate events. The distribution in (a) includes the $\gamma\gamma \to \mu^+\mu^-$ background. In (b), the solid curve shows the background from other processes; that is determined experimentally from the transverse-momentum-balance distribution. Taken from (Mori, 2007a; Uehara, 2009a).

### 22.2.1 Light-quark meson resonances

A measurement of a meson resonance (denoted as $R$) decaying into two pseudoscalar mesons through two-photon production allows the determination of its parameters such as the mass, total width and, in particular, the product of the two-photon width and the branching fraction to the meson pair, $\Gamma_{\gamma\gamma}\mathcal{B}(R \to P_1P_2)$. If $\mathcal{B}(R \to P_1P_2)$ is known, a two-photon width is derived, which is otherwise difficult to obtain. The two-photon width of a meson is intimately related to its charge structure, giving in turn valuable information on its quark content and structure.

One of the longstanding puzzles of QCD is the low mass scalar nonet: $f_0(500)$, $K^*_0(800)$, $f_0(980)$ and $a_0(980)$, whose masses must be higher than 1.2 GeV/$c^2$ if they are ordinary $q\bar{q}$ bound states (Close and Tornqvist, 2002; 't Hooft, Isidori, Maiani, Polosa, and Riquer, 2008). One possible explanation exploits the attractive force between a quark pair (di-quark) in the color anti-triplet state. In this picture, a di-quark anti-di-quark pair forms a nonet with lower masses for the scalar states. In such a state, its two-photon width is expected to be an order of magnitude smaller compared to that of the $q\bar{q}$ state (Amsler and Tornqvist, 2004). [166]

[166] The typical size of the two-photon width for the usual $q\bar{q}$ state is given, for example, for $f_2(1270)$ meson; that is $\Gamma_{\gamma\gamma}(f_2(1270) = 3.03 \pm 0.35$ keV (Beringer et al., 2012).

#### 22.2.1.1 Differential cross sections and partial wave analysis

In this subsection, we present the formalism of the differential cross section for two-photon production of a meson pair in terms of partial waves and then discuss a possible method to extract the resonance parameters.

In the energy region $W \leq 3$ GeV, the partial waves with spin $J > 4$ may be neglected so that only $S$-, $D$- and $G$- waves need be considered. The differential cross section can be expressed as:

$$\frac{d\sigma}{d\Omega}(\gamma\gamma \to P_1 P_2) = \qquad (22.2.1)$$
$$\left|S\, Y_0^0 + D_0\, Y_2^0 + G_0\, Y_4^0\right|^2 + \left|D_2\, Y_2^2 + G_2\, Y_4^2\right|^2,$$

where $D_0$ and $G_0$ ($D_2$ and $G_2$) denote the helicity 0 (2) components of the $D$- and $G$- wave, respectively,[167] and $Y_J^\lambda$ are the spherical harmonics in which the helicity $\lambda$ is quantized for the $\gamma\gamma$ axis. The angular dependence of the cross section is governed by the spherical harmonics, while the energy dependence is determined by the partial waves. Since the absolute value of harmonics $|Y_J^\lambda|$ are not independent, the partial waves cannot be separated from the information on the differential cross sections alone.

We write Eq. (22.2.1) as

$$\frac{d\sigma}{4\pi d|\cos\theta^*|}\,(\,\gamma\gamma \to P_1 P_2) = \qquad (22.2.2)$$
$$\widehat{S}^2\, |Y_0^0|^2 + \widehat{D}_0^2\, |Y_2^0|^2 + \widehat{D}_2^2\, |Y_2^2|^2$$
$$+\widehat{G}_0^2\, |Y_4^0|^2 + \widehat{G}_2^2\, |Y_4^2|^2\,.$$

The amplitudes $\widehat{S}^2$, $\widehat{D}_0^2$, $\widehat{D}_2^2$, $\widehat{G}_0^2$ and $\widehat{G}_2^2$ can be expressed in terms of $S$, $D_0$, $D_2$, $G_0$ and $G_2$ (Uehara, 2008a). Since the squares of the spherical harmonics are orthogonal, we can fit differential cross sections to obtain $\widehat{S}^2$, $\widehat{D}_0^2$, $\widehat{D}_2^2$, $\widehat{G}_0^2$ and $\widehat{G}_2^2$ for each $W$ bin. Two types of fit are used: the "SD" fit and "SDG" fit; $G$-waves are neglected in the SD fit.

As an example of the analysis, we discuss the Belle results for $\gamma\gamma \to \eta\pi^0$ (Uehara, 2009a). The spectra of $\widehat{S}^2$, $\widehat{D}_0^2$ and $\widehat{D}_2^2$ obtained by the SD fit for this process are shown in Fig. 22.2.2. The spectrum for $\widehat{S}^2$ shows a clear peak of the spin zero $a_0(980)$ with a shoulder that may be due to the $a_0(1450)$ while the spectrum for $\widehat{D}_2^2$ is dominated by the spin two $a_2(1320)$ with a hint of the $a_2(1700)$. There is no clear structure for the $\widehat{D}_0^2$ component. The SDG fit reveals that $G$-waves are negligible in this energy region.

#### 22.2.1.2 Resonance parameter extraction by partial wave analysis

Here, we describe a method to derive information from resonances by parameterizing partial wave amplitudes and then fitting differential cross sections. Note that we do not fit the obtained $\widehat{S}^2$, $\widehat{D}_0^2$ and $\widehat{D}_2^2$ spectra; instead, we fit the differential cross sections directly, but the results are shown in the $\widehat{S}^2$, $\widehat{D}_0^2$, $\widehat{D}_2^2$ spectra. Two to three orders of magnitude higher statistics are available at $B$ Factories compared to pre-$B$ Factory experiments which permits such a luxury.[168]

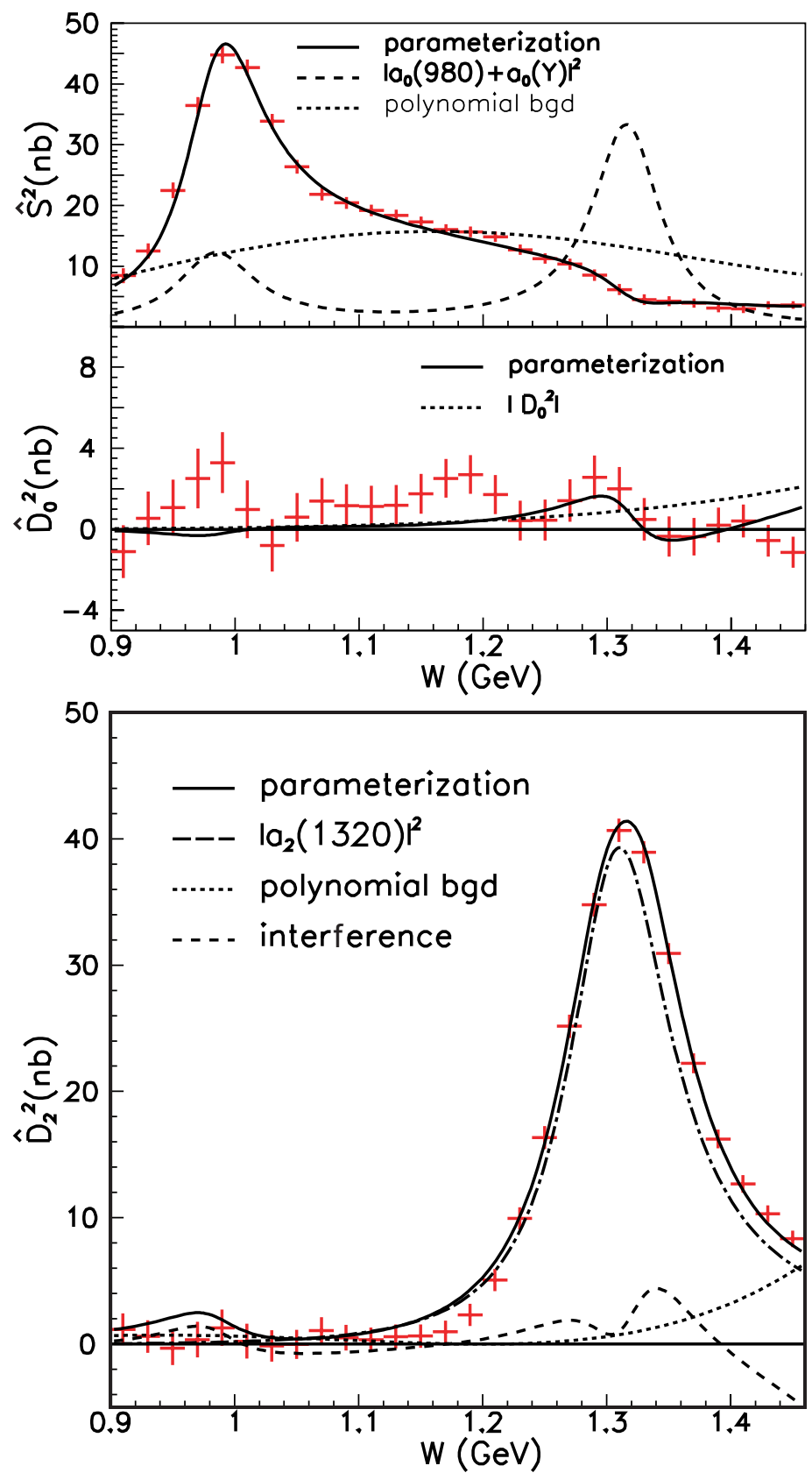

**Figure 22.2.2.** Spectra of (top) $\widehat{S}^2$, (middle) $\widehat{D}_0^2$ and (bottom) $\widehat{D}_2^2$ for $\gamma\gamma \to \eta\pi^0$. The solid lines are the results of the partial-wave analysis discussed in Section 22.2.1.2. The error bars show the diagonal components of error matrix for the fit parameters. From (Uehara, 2009a).

Once the functional forms of the amplitudes are prepared, we use Eq. (22.2.1) to fit the differential cross sections. The fundamental difficulty is the presence of interference among resonances and non-resonant amplitudes that are basically unknown. A fit with many free parameters often results in multiple solutions corresponding to constructive and destructive interference. Thus, one has to minimize the number of parameters to obtain a stable and unique solution. The effects of parameterization dependence are incorporated in the systematic uncertainties, which are typically obtained by changing the parameterization.

In the partial wave analysis for the low energy region $W < 1.5$ GeV where the $J > 2$ waves can be neglected safely, the $a_2(1320)$ contribution is assumed only in the

[167] In this chapter we denote individual partial waves by Roman style and parameterized waves by italics.

[168] See Whalley (2001) for a compilation for pre-$B$ Factory experiments.

$D_2$ wave, since $\widehat{D}_2^2$ is dominated by the $a_2(1320)$ resonance and the $\widehat{D}_0^2$ component is small. For the $S$-wave, the shoulder in the $\widehat{S}^2$ spectrum above the $a_0(980)$ peak may be understood due to the $a_0(1450)$. However, in the fit Belle introduces a new resonance $a_0(Y)$, since the resonance parameters are found to be quite different from those of the $a_0(1450)$.

As a result, Belle uses the following parameterization for the $S$-, $D_0$ and $D_2$ waves:

$$\begin{aligned} S &= A_{a_0(980)}e^{i\phi_{s0}} + A_{a_0(Y)}e^{i\phi_{s1}} + B_S \ , \\ D_0 &= B_{D0} \ , \\ D_2 &= A_{a_2(1320)}e^{i\phi_{d2}} + B_{D2} \ , \end{aligned} \quad (22.2.3)$$

where $A_{a_0(980)}$, $A_{a_0(Y)}$ and $A_{a_2(1320)}$ are the amplitudes of the $a_0(980)$, $a_0(Y)$ and $a_2(1320)$, respectively; $B_S$, $B_{D0}$ and $B_{D2}$ are non-resonant (hereafter called "background") amplitudes for $S$-, $D_0$ and $D_2$ waves; and $\phi_{s0}$, $\phi_{s1}$, and $\phi_{d2}$ are the phases of resonances relative to background amplitudes. The goal of the analysis is to obtain parameters of the $a_0(980)$ and $a_0(Y)$ and to check the consistency of the $a_2(1320)$ parameters that have been measured well in the past.

The background amplitudes are parameterized as second order polynomials in $W$ for both the real and imaginary parts of all waves. The arbitrary phases are fixed by choosing $\phi_{s0} = \phi_{d2} = 0.$ for $S$- and $D$- waves. We constrain all the background amplitudes to be zero at the threshold in accordance with the expectation that the cross section vanishes in the Thomson limit which was originally discussed in the context of low-energy Compton scattering. The relativistic Breit-Wigner resonance amplitude (see Chapter 13) is used for a resonance.

The data are fit using the minimizer `Minuit` (James and Roos, 1975). Many fits are done using different, randomly chosen starting parameters. In this way, Belle search for a global minimum and locate ambiguous solutions.

The resulting best fit obtained is displayed in Fig. 22.2.2. The measured spectra, $\widehat{S}^2$, $\widehat{D}_0^2$ and $\widehat{D}_2^2$, are reproduced fairly well by the fit.

Table 22.2.1 summarizes the fit results for $\gamma\gamma \to \eta\pi^0$ as well as the other processes measured by Belle: $\gamma\gamma \to \pi^+\pi^-$, $\gamma\gamma \to K^+K^-$, $\gamma\gamma \to K_S^0K_S^0$, $\gamma\gamma \to \pi^0\pi^0$ and $\gamma\gamma \to \eta\eta$.

The main results in this table are as follows. Belle measures the two-photon widths for the scalar resonances $f_0(980)$ and $a_0(980)$ for the first time with significant statistics; the $f_0(980)$ is observed as a clear peak both in the $\pi^+\pi^-$ and $\pi^0\pi^0$ modes; the $a_0(980)$ is measured clearly in the $\eta\pi^0$ mode. The measured two-photon widths are small compared to those of the $f_2(1270)$ and $a_2(1320)$. This supports the di-quark anti-di-quark hypothesis for these mesons. In addition, Belle finds several resonance states in the range 1.3–2.4 GeV with substantial coupling to two photons. Belle perform a generic partial wave analysis including possible interferences with non-resonant terms. Systematic errors for some resonance parameters are large, resulting from a preference for destructive interference between the resonance and other components. For example see $S$ and $D_2$ in Eq. 22.2.3, which tends to enhance variations of parameters. In addition, this generic analysis results in multiple solutions in some cases.

The Belle results provide good-quality unfolded data on the differential cross sections for six pseudoscalar modes, $\gamma\gamma \to \pi^+\pi^-$, $K^+K^-$, $\pi^0\pi^0$, $K_S^0K_S^0$, $\eta\pi^0$ and $\eta\eta$. These high-statistics data can be used to update partial wave analyses that employ low energy constraints and incorporate all available hadron data (Pennington, Mori, Uehara, and Watanabe, 2008). We expect that these high-statistics data will be used to derive more accurate resonance parameters.

### 22.2.2 Comparison with QCD predictions at high energy

Two-photon production of exclusive hadronic final states provides useful information about resonances, perturbative QCD, and non-perturbative QCD. From the theoretical point of view, a two-photon process is attractive because of the absence of strong interactions in the initial state.

Brodsky and Lepage (1981) (BL) numerically calculated the amplitude for the hard exclusive $\gamma\gamma \to M_1\overline{M}_2$ processes within the context of the perturbative-QCD (pQCD) for the first time. A similar formula is also discussed by Chernyak and Zhitnitsky (1984). In this pQCD framework, the amplitude for $\gamma\gamma \to M_1\overline{M}_2$ can be described in a factorized form:

$$\mathcal{M}_{\lambda_1\lambda_2}(s,\theta^*) = \int_0^1\int_0^1 dxdy\phi_M(x,Q_x)\phi_{\overline{M}}(y,Q_y)T_{\lambda_1\lambda_2}(x,y,\theta^*), \quad (22.2.4)$$

where $s$ is the squared invariant mass of the di-meson system, $\phi_M(x,Q_x)$ is a single-meson distribution amplitude for a meson $M$. The squared amplitude $|\phi_M(x,Q_x)|^2$ is proportional to a probability for finding a valence quark and antiquark in the meson, carrying a fraction $x$ and $1-x$, respectively, of the meson's momentum. $Q_x$ is the typical momentum scale in the process, $\sim \min(x, 1-x)\sqrt{s}\sin\theta^*$. The term $T_{\lambda_1\lambda_2}$ is a hard scattering amplitude for $\gamma_{\lambda_1}\gamma_{\lambda_2} \to q\bar{q}q\bar{q}$ with photon helicities $\lambda_1$ and $\lambda_2$. From the sum rule, the overall normalization is fixed as

$$\int_0^1 dx\phi_M(x,0) = f_M/2\sqrt{3}, \quad (22.2.5)$$

where $f_M$ is the decay constant for meson $M$.

For mesons with helicity zero the leading-term calculation gives the following dependence on $s$ and scattering angle $\theta^*$:

$$\frac{d\sigma}{d|\cos\theta^*|} = 16\pi\alpha^2\frac{|F_M(s)|^2}{s}\left\{\frac{[(e_1-e_2)^2]^2}{(1-\cos^2\theta^*)^2}\right. + \frac{2(e_1e_2)[(e_1-e_2)^2]}{1-\cos^2\theta^*}g(\theta^*)$$

**Table 22.2.1.** Summary of partial wave analyses in the energy region below 2.4 GeV. If two values are given for the two-photon partial decay width $\Gamma_{\gamma\gamma}$, the upper value is the one for the spin-helicity assignment $(J, \lambda) = (2, 2)$ and the lower one for $(0, 0)$. $\mathcal{B}$ is the branching fraction of the resonance to the corresponding decay mode otherwise explicitly noted. For the $\pi^0\pi^0$ mode, we provide the value $\mathcal{B}(f_2 \to \gamma\gamma)$ instead of $\Gamma_{\gamma\gamma}$ since the total width is known for $f_2(1270)$. The resonances $f_0(Y)$, $a_0(X)$ and $f_2(X)$ correspond to signals of new or unidentified resonances found. In the $K^+K^-$ mode, $f_J/f_0/a_2$ and $f_J/f_2$ mean that there are ambiguities in the signal assignment. Quoted upper limits are at 90% confidence level.

| Mode | Resonance | Mass (MeV/$c^2$) | Width (MeV) | $\Gamma_{\gamma\gamma}$ (eV), $(J,\lambda)=\begin{cases}(2,2)\\(0,0)\end{cases}$ | Reference |
|---|---|---|---|---|---|
| $\pi^+\pi^-$ | $f_0(980)$ | $985.6^{+1.2+1.1}_{-1.5-1.6}$ | $34.2^{+13.9+8.8}_{-11.8-2.5}$ | $205^{+95+147}_{-83-117}$ | Mori (2007b) |
| | $\eta'(958)$ | $\mathcal{B}(\pi^+\pi^-) < 2.9\times10^{-3}$ (with interference), $3.3\times10^{-4}$ (without) | | | |
| $K^+K^-$ | $f_2'(1525)$ | $1518\pm1\pm3$ | $82\pm2\pm3$ | $28.2\pm2.4\pm5.8/\mathcal{B}$ | Abe (2003d) |
| | $f_J/f_0/a_2$ | $1737\pm5\pm7$ | $151\pm22\pm24$ | $\begin{cases}10.3\pm2.1\pm2.3/\mathcal{B}\\76\pm15\pm17/\mathcal{B}\end{cases}$ | |
| | $f_2(2010)$ | $1980\pm2\pm14$ | $297\pm12\pm6$ | $61\pm2\pm3/\mathcal{B}$ | |
| | $f_J/f_2$ | $2327\pm9\pm6$ | $275\pm36\pm20$ | $\begin{cases}22\pm3\pm6/\mathcal{B}\\161\pm22\pm48/\mathcal{B}\end{cases}$ | |
| $K^0_SK^0_S$ | $f_2'(1525)$ | $1525.3^{+1.2+3.7}_{-1.4-2.1}$ | $82.9^{+2.1+3.1}_{-2.2-2.0}$ | $48^{+67+108}_{-8-12}/\mathcal{B}(K\overline{K})$ | Uehara (2013) |
| | $f_0(1710)$ | $1750^{+6+29}_{-7-18}$ | $139^{+11+96}_{-12-50}$ | $12^{+3+227}_{-2-8}/\mathcal{B}(K\overline{K})$ | |
| | $f_2(2200)$ | $2243^{+7+3}_{-6-29}$ | $145\pm12^{+27}_{-34}$ | $3.2^{+0.5+1.3}_{-0.4-2.2}/\mathcal{B}(K\overline{K})$ | |
| | $f_0(2500)$ | $2539\pm14^{+38}_{-14}$ | $274^{+77+126}_{-61-163}$ | $40^{+9+17}_{-7-40}/\mathcal{B}(K\overline{K})$ | |
| $\pi^0\pi^0$ | $f_0(980)$ | $982.2\pm1.0^{+8.1}_{-8.0}$ | | $286\pm17^{+211}_{-70}$ | Uehara (2008a) |
| | $f_2(1270)$ | fixed | fixed | | |
| | | $\mathcal{B}(f_2 \to \gamma\gamma) = (1.57 \pm 0.01^{+1.39}_{-0.14}) \times 10^{-5}$ | | | |
| | $f_0(Y)$ | $1470^{+6+72}_{-7-255}$ | $90^{+2+50}_{-1-22}$ | $11^{+4+603}_{-2-7}/\mathcal{B}$ | Uehara (2009b) |
| | $f_2(1950)$ | $2038^{+13}_{-11}$ | $441^{+27}_{-25}$ | $54^{+23}_{-14}/\mathcal{B}$ | |
| | $f_4(2050)$ | $1884^{+14+218}_{-13-25}$ | $453\pm20^{+31}_{-129}$ | $136^{+24+415}_{-22-91}$ | |
| $\eta\pi^0$ | $a_0(980)$ | $982.3^{+0.6+3.1}_{-0.7-4.7}$ | $75.6\pm1.6^{+17.4}_{-10.0}$ | $128^{+3+502}_{-2-43}/\mathcal{B}$ | Uehara (2009a) |
| | $a_0(Y)$ | $1316.8^{+0.7+24.7}_{-1.0-4.6}$ | $65.0^{+2.1+99.1}_{-5.4-32.6}$ | $432\pm6^{+1073}_{-256}/\mathcal{B}$ | |
| | $a_2(1320)$ | fixed | fixed | $145^{+97}_{-34}/\mathcal{B}$ | |
| $\eta\eta$ | $f_0(Y)$ | $1262^{+51+82}_{-78-103}$ | $484^{+246+246}_{-170-263}$ | $121^{+133+169}_{-53-106}/\mathcal{B}$ | Uehara (2010a) |
| | $f_2(1270)$ | fixed | fixed | $11.5^{+1.8+4.5}_{-2.0-3.7}/\mathcal{B}$ | |
| | $f_2(X)$ | $1737\pm9^{+198}_{-65}$ | $228^{+21+234}_{-20-153}$ | $5.2^{+0.9+37.3}_{-0.8-4.5}/\mathcal{B}$ | |

$$\left. +2(e_1e_2)^2 g^2(\theta^*) \right\}, \qquad (22.2.6)$$

where $e_1$ and $e_2$ are the quark charges and $F_M$ is the meson's electromagnetic form factor. Finally, $g(\theta^*)$ is a function of order one expressing the additional $\theta^*$ dependence.

[169] In principle, both $F_M$ and $g(\theta^*)$ can be obtained from the meson wave function $\phi_M(x, Q_x)$. However, at present they are unknown since the function $\phi_M(x, Q_x)$ is a non-perturbative quantity with an unknown $x$ dependence.

---

[169] See the original paper (Brodsky and Lepage, 1981) for the explicit formula.

Under the assumption that $\phi_K$ and $\phi_\pi$ are similar in shape, the differential cross section ratio for production of $K^+K^-$ and $\pi^+\pi^-$ depends only on the meson decay constants $f_K^4/f_\pi^4$. Benayoun and Chernyak (1990) (BC) employ different wave functions for $\phi_\pi(x)$ and $\phi_K(x)$, taking into account $SU(3)$ symmetry breaking effects. The next-to-leading order calculation is done by Duplancic and Nizic (2006).

As an alternative model for the meson-pair production in the two-photon processes, Diehl, Kroll, and Vogt (2002) (DKV) proposed a "handbag model". In the handbag model, the differential cross section for the process is given by

$$\frac{d\sigma}{d|\cos\theta^*|}(\gamma\gamma \to M\overline{M}) = \frac{8\pi\alpha^2}{s}\frac{1}{\sin^4\theta^*}|R_{M\overline{M}}(s)|^2, \tag{22.2.7}$$

where the transition amplitude is expressed as a hard scattering $\gamma\gamma \to q\bar{q}$ times a form factor $R_{M\overline{M}}(s)$ describing the soft transition $q\bar{q} \to M\overline{M}$. The main dynamical assumption of this model is that, at present energies $W \leq 4\,\mathrm{GeV}$, all amplitudes for $\gamma\gamma \to M\overline{M}$ are still dominated by "soft" components. In this model, the angular distribution is proportional to $1/\sin^4\theta^*$ for both charged and neutral meson pairs.

#### 22.2.2.1 Angular Dependence of Differential Cross Section

As can be seen from Eq. (22.2.6), there is distinct difference for charged and neutral meson pair production in pQCD. For charged meson pairs, the angular distribution is given by $1/\sin^4\theta^*$, since the first (leading) term dominates. This angular dependence is expected to be realized in a certain large $|t|$ region of the Mandelstam variable. Meanwhile, for neutral meson pairs, the first and the second terms are zero since the quark charges are the same ($e_1 = e_2$), so the angular dependence is given by the third (non-leading) term $\propto g^2(\theta^*)$. As a result, a complicated angular distribution depending on the meson wave amplitude is expected for the neutral meson pairs. On the other hand, the handbag model predicts a $1/\sin^4\theta^*$ dependence for large $W$ for both charged and neutral meson pairs.

The measured results from Belle are summarized in Table 22.2.2. For the charged meson pairs, $\gamma\gamma \to \pi^+\pi^-$, $K^+K^-$, the angular distributions are described by the $1/\sin^4\theta^*$ form quite well. On the other hand, for the neutral meson pairs, $\gamma\gamma \to \pi^0\pi^0$, $K_S^0K_S^0$, $\eta\pi^0$ and $\eta\eta$, the angular distributions show more complicated behavior. [170]

#### 22.2.2.2 Energy Dependence of Cross Section and ratio of Cross Sections

Other important predictions for the hard exclusive processes in QCD are the power-law dependence of the cross section, $\sigma_0 \sim W^{-n}$, and the ratio of the cross sections for the different processes. These results are summarized in Table 22.2.3.

For the $W$ dependence, pQCD predicts $n = 6$ for the charged meson pairs, and $n = 10$ for the neutral meson pairs in the energy region accessible in the $B$ Factory experiments (Benayoun and Chernyak, 1990). The data for $\gamma\gamma \to \pi^+\pi^-$, $K^+K^-$ show slightly higher $n$ than the predicted value of $n = 6$ but within the systematic errors as shown in Figs 22.2.3(a) and (b).

On the other hand, for the neutral meson pairs, the processes $\gamma\gamma \to K_S^0K_S^0$ shows a steeper $W$ dependence than that for the charged meson pairs as can be seen in Fig. 22.2.3(d). The measured value of the slope $n = 11.0 \pm 0.4 \pm 0.4$ (Uehara, 2013) shows a good agreement with the pQCD prediction by Chernyak (Chernyak, 2006, 2012).

For the ratio of the cross sections, pQCD predicts a large suppression for the neutral mesons compared to the charged mesons. In contrast, in the handbag model, the ratio is determined by soft dynamics such as the iso-spin, $SU(3)$ relation. As can be seen in Fig. 22.2.3(e), the data show a large suppression for the ratio

$$\frac{\sigma(\gamma\gamma \to K_S^0K_S^0)}{\sigma(\gamma\gamma \to K^+K^-)}. \tag{22.2.8}$$

This is consistent with the pQCD expectation, but does not agree with the handbag model. In addition, the ratio $\sigma(\gamma\gamma \to K^+K^-)$ over $\sigma(\gamma\gamma \to \pi^+\pi^-)$ is consistent with pQCD prediction by Benayoun and Chernyak (1990) [BC], where the difference of wave functions for pion and kaon is taken into account (see Fig. 22.2.3(c)).

One exception is the ratio

$$\frac{\sigma(\gamma\gamma \to \pi^0\pi^0)}{\sigma(\gamma\gamma \to \pi^+\pi^-)}. \tag{22.2.9}$$

This neutral to charged ratio is rather large as shown in Table 22.2.3. This result is inconsistent with the pQCD predictions where the neutral modes are expected to be suppressed. The result, however, can be explained by the handbag model (Diehl and Kroll, 2010) quite well.

Further study of the differential cross section data is needed to clarify these unresolved problems.

## 22.3 Vector meson-pair production

A clear signal for the production of a new state via the $\gamma\gamma$ process, $X(3915) \to \omega J/\psi$ (Uehara, 2010b),(Lees, 2012ad), and evidence for another state $X(4350) \to \phi J/\psi$ (Shen, 2010a) have been reported, thereby introducing new puzzles in charmonium(-like) spectroscopy (see also Sections 18.2 and 18.3). It is natural to extend the above theoretical picture to similar states coupling to $\omega\phi$, $\omega\omega$ or $\phi\phi$.

Measurements of the cross sections for $\gamma\gamma \to VV$ (Liu, 2012), where $V$ is a vector particle,in particular $VV = \omega\phi$, $\phi\phi$ and $\omega\omega$, are based on an analysis of the 870 $\mathrm{fb}^{-1}$ data sample taken at or near the $\Upsilon(nS)$ $(n = 1, ..., 5)$ resonances with the Belle detector.

[170] An update analysis for $\gamma\gamma \to K_S^0K_S^0$ using full Belle data shows clearly that the angular distribution for this mode can not be described by a $1/\sin^4\theta^*$ form. See Uehara (2013).

**Table 22.2.2.** Angular dependence of differential cross sections in comparison with $1/\sin^4\theta^*$ dependence.

| mode | $1/\sin^4\theta^*$ | energy range | $\lvert\cos\theta^*\rvert$ range | reference |
|---|---|---|---|---|
| $\pi^+\pi^-$ | Match well. | 3.0 - 4.1 | $< 0.6$ | Nakazawa (2005) |
| $K^+K^-$ | Match well. | 3.0 - 4.1 | $< 0.6$ | Nakazawa (2005) |
| $K_S^0 K_S^0$ | $\alpha$ varies from 4–8 for $1/\sin^\alpha\theta^*$ | 2.6 - 3.3 | $< 0.8$ | Uehara (2013) |
| $\pi^0\pi^0$ | $1/\sin^4\theta^* + b\cos\theta^*$ better.<br>Approaches $1/\sin^4\theta^*$ above 3.1 GeV. | 2.4 - 4.1$^\dagger$ | $< 0.8$ | Uehara (2008a) |
| $\eta\pi^0$ | Good agreement above 2.7 GeV. | 3.1 - 4.1 | $< 0.8$ | Uehara (2009a) |
| $\eta\eta$ | Poor agreement.<br>$1/\sin^6\theta^*$ better above 3.0 GeV. | 2.4 - 3.3 | $< 0.9$ | Uehara (2010a) |

† $\chi_{cJ}$ region, 3.3 - 3.6 GeV is excluded.

**Table 22.2.3.** The value of $n$ of $\sigma_0 \propto W^{-n}$ in various reactions fitted in the $W$ and $\lvert\cos\theta^*\rvert$ ranges indicated and the ratio of the cross sections in comparison with QCD predictions from Brodsky and Lepage (1981) [BL], Benayoun and Chernyak (1990) [BC], and Diehl, Kroll, and Vogt (2002) [DKV]. The first and second errors are statistical and systematic, respectively.

| Process | $n$ | $W$ (GeV) | $\lvert\cos\theta^*\rvert$ | BL | BC | DKV |
|---|---|---|---|---|---|---|
| $\pi^+\pi^-$ | $7.9 \pm 0.4 \pm 1.5$ | 3.0 - 4.1 | $< 0.6$ | 6 | 6 | |
| $K^+K^-$ | $7.3 \pm 0.3 \pm 1.5$ | 3.0 - 4.1 | $< 0.6$ | 6 | 6 | |
| $K_S^0 K_S^0$ # | $10.5 \pm 0.6 \pm 0.5$ | 2.4 - 4.0$^\dagger$ | $< 0.6$ | - | 10 | |
| $K_S^0 K_S^0$ ## | $11.0 \pm 0.4 \pm 0.4$ | 2.6 - 4.0$^\dagger$ | $< 0.8$ | - | 10 | |
| $\pi^0\pi^0$ | $8.0 \pm 0.5 \pm 0.4$ | 3.1 - 4.1$^\dagger$ | $< 0.8$ | - | 10 | |
| $\eta\pi^0$ | $10.5 \pm 1.2 \pm 0.5$ | 3.1 - 4.1 | $< 0.8$ | - | 10 | |
| $\eta\eta$ | $7.8 \pm 0.6 \pm 0.4$ | 2.4 – 3.3 | $< 0.8$ | - | 10 | |
| **Process** | **$\sigma_0$ ratio** | **$W$ (GeV)** | **$\lvert\cos\theta^*\rvert$** | **BL** | **BC** | **DKV** |
| $K^+K^-/\pi^+\pi^-$ | $0.89 \pm 0.04 \pm 0.15$ | 3.0 - 4.1 | $< 0.6$ | 2.3 | 1.06 | |
| $K_S^0 K_S^0/K^+K^-$ # | $\sim 0.13$ to $\sim 0.01$ | 2.4 - 4.0 | $< 0.6^\dagger$ | | 0.005 | 2/25 |
| $\pi^0\pi^0/\pi^+\pi^-$ | $0.32 \pm 0.03 \pm 0.06$ | 3.1 - 4.1 | $< 0.6^\dagger$ | | 0.04-0.07 | 0.5 |
| $\eta\pi^0/\pi^0\pi^0$ | $0.48 \pm 0.05 \pm 0.04$ | 3.1 - 4.0 | $< 0.8^\dagger$ | $0.24R_f(0.46R_f)^\ddagger$ | | |
| $\eta\eta/\pi^0\pi^0$ | $0.37 \pm 0.02 \pm 0.03$ | 2.4 - 3.3 | $< 0.8$ | $0.36R_f^2(0.62R_f^2)^\ddagger$ | | |

† $\chi_{cJ}$ region, 3.3 – 3.6 GeV is excluded.
‡ $\eta$ meson as a pure $SU(3)$ octet (mixture of octet and singlet with $\theta_p = -18°$), $R_f = f_\eta^2/f_{\pi^0}^2$.
# From Chen (2007c).
## From Uehara (2013).

After event selections, clear $\omega$ and $\phi$ signals are observed. We obtain the number of $VV$ events in each $VV$ invariant mass bin by fitting the $\lvert\sum \boldsymbol{P}_t^*\rvert$ distribution between zero and 0.9 GeV/$c$, where $\lvert\sum \boldsymbol{P}_t^*\rvert$ is the magnitude of the vector sum of the final transverse momenta of the particle in the $e^+e^-$ CM frame. The resulting $VV$ invariant mass distributions are shown in Fig. 22.3.1; some obvious structures are observed in the low $VV$ invariant mass region.

Two-dimensional (2D) angular distributions are investigated to obtain the $J^P$ quantum numbers of the structures. In the process $\gamma\gamma \to VV$, five angular variables are kinematically independent: $z$, $z^*$, $z^{**}$, $\phi^*$, and $\phi^{**}$. Using $\omega\phi$ as an example, $z$ is the cosine of the scattering polar angle of the $\phi$ meson in the $\gamma\gamma$ CM system; $z^*$ and $\phi^*$ are the cosine of the helicity angle of $K^+$ in the $\phi$ decay and the azimuthal angle defined in the $\phi$ rest frame with respect to the $\gamma\gamma \to \omega\phi$ scattering plane; $z^{**}$ and $\phi^{**}$ are the cosine of the helicity angle of normal direction to the decay plane of the $\omega \to \pi^+\pi^-\pi^0$ and the azimuthal angle defined in the $\omega$ rest frame. We use the transversity angle ($\phi_T$) and polar-angle product ($\Pi_\theta$) to analyze the angular distributions. They are defined as $\phi_T = \lvert\phi^* + \phi^{**}\rvert/2\pi$, $\Pi_\theta = [1-(z^*)^2][1-(z^{**})^2]$.

The number of signal events is obtained by fitting the $\lvert\sum \boldsymbol{P}_t^*\rvert$ distribution in each $\phi_T$ and $\Pi_\theta$ bin in the 2D space. The 2D space is divided into $4\times4$, $5\times5$, and $10\times10$ bins for $\omega\phi$, $\phi\phi$, and $\omega\omega$, respectively, in several wide $VV$ mass bins as shown in Fig. 22.3.2. The resulting 2D angular distributions are fitted with the signal shapes from MC-simulated samples with different $J^P$ assumptions ($0^+$, $0^-$, $2^+$, $2^-$). The following features are found: (1) for $\omega\phi$: $0^+$ ($S$-wave) or $2^+$ ($S$-wave) can describe the data with $\chi^2/ndf = 1.1$ or 1.2, while a mixture of $0^+$ ($S$-wave) and $2^+$ ($S$-wave) describes the data with $\chi^2/ndf = 0.9$ ($ndf$ being the number of degrees of freedom); (2) for $\phi\phi$: a mixture of $0^+$ ($S$-wave) and $2^-$ ($P$-wave) describes the data with $\chi^2/ndf = 1.3$; and (3) for $\omega\omega$: a mixture of $0^+$ ($S$-wave) and $2^+$ ($S$-wave) describes the data with $\chi^2/ndf = 1.3$.

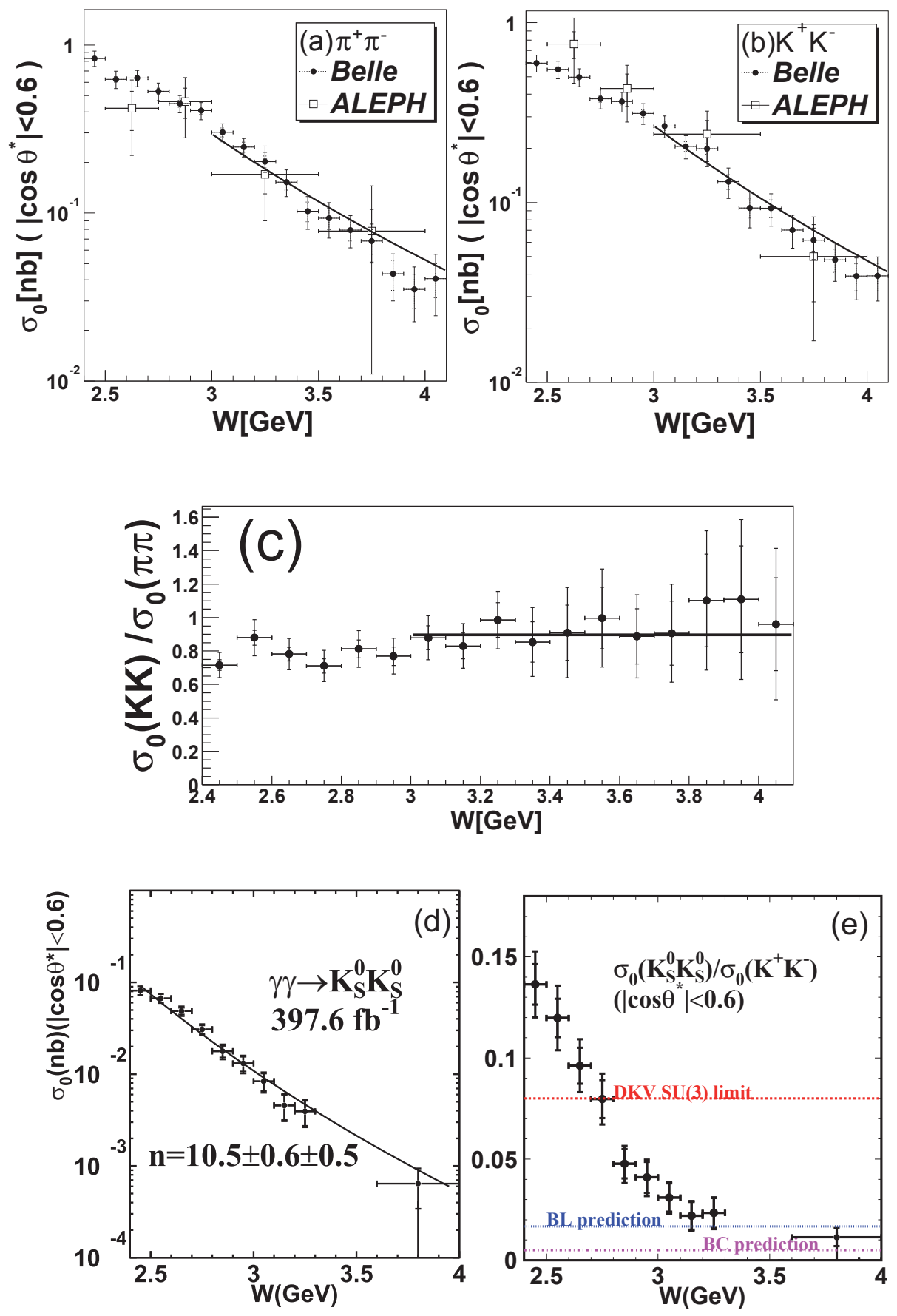

**Figure 22.2.3.** Cross section for the processes (a) $\gamma\gamma \to \pi^+\pi^-$, (b) $\gamma\gamma \to K^+K^-$ and (d) $\gamma\gamma \to K_S^0 K_S^0$, integrated over $|\cos(\theta^*| < 0.6$, as a function of $W$. The solid line in (a) and (b) is the prediction $n = 6$, while that in (d) is the best fit. The figures (c) and (e) show the ratio of the cross section among these processes. The solid line in (c) is a fit with a constant. The horizontal lines in (e) show various theoretical predictions. From (Chen, 2007c; Nakazawa, 2005).

The $\gamma\gamma \to VV$ cross sections are shown in Fig. 22.3.2. The cross sections for different $J^P$ values as a function of $M(VV)$ are also shown in this figure. While there are substantial spin-zero components in all three modes, there are also significant spin-two components, certainly in the $\phi\phi$ and $\omega\omega$ modes.

The cross sections for $\gamma\gamma \to \omega\phi$ are much lower than the prediction of the $q^2\bar{q}^2$ tetraquark model (Achasov and Shestakov, 1991) of 1 nb. The resonant structure in the $\gamma\gamma \to \phi\phi$ mode is found nearly at the predicted mass from the model. However, the $\phi\phi$ cross section is an order of magnitude lower than the expectation. On the other hand, the $t$-channel factorization model (Alexander, Levy, and Maor, 1986) predicts that the $\phi\phi$ cross sections vary between 0.001 nb and 0.05 nb in the mass region of 2.0 GeV/$c^2$ to 5.0 GeV/$c^2$, which are much lower than the experimental data. For $\gamma\gamma \to \omega\omega$, the $t$-channel factorization model predicts a broad structure between 1.8 GeV/$c^2$ and 3.0 GeV/$c^2$ with a peak cross section of 10-30 nb near 2.2 GeV/$c^2$, while the one-pion-exchange model (Achasov, Karnakov, and Shestakov, 1987) predicts an enhancement near threshold around 1.6 GeV/$c^2$ with a peak cross section of 13 nb using their preferred value of the slope parameter. Both the peak position and the peak height predictions from these models disagree with the Belle measurements. Therefore none of the models discussed here can explain the data (Chernyak, 2012).

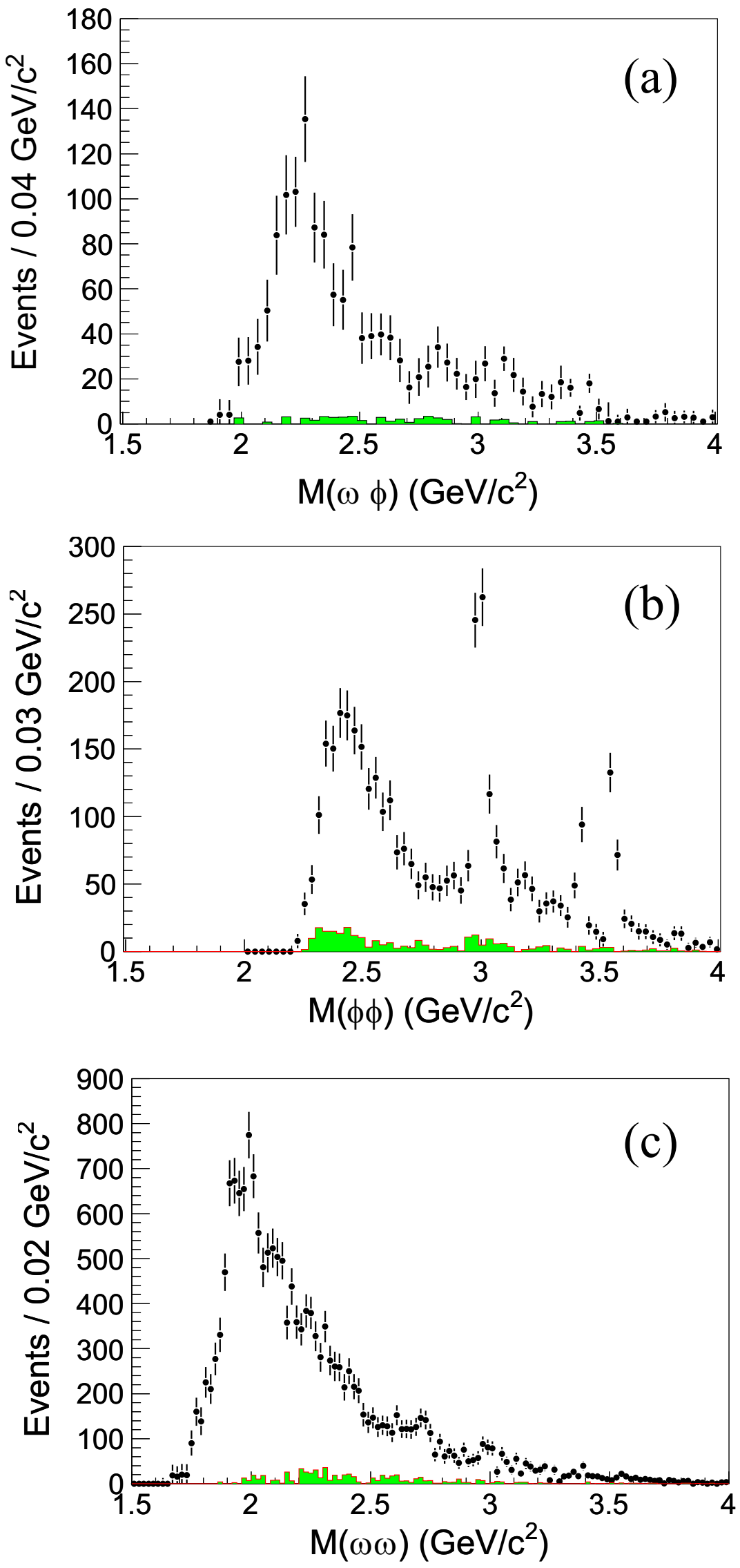

**Figure 22.3.1.** The (a) $\omega\phi$, (b) $\phi\phi$ and (c) $\omega\omega$ invariant mass distributions. The shaded histograms are from the corresponding normalized sidebands. From (Liu, 2012).

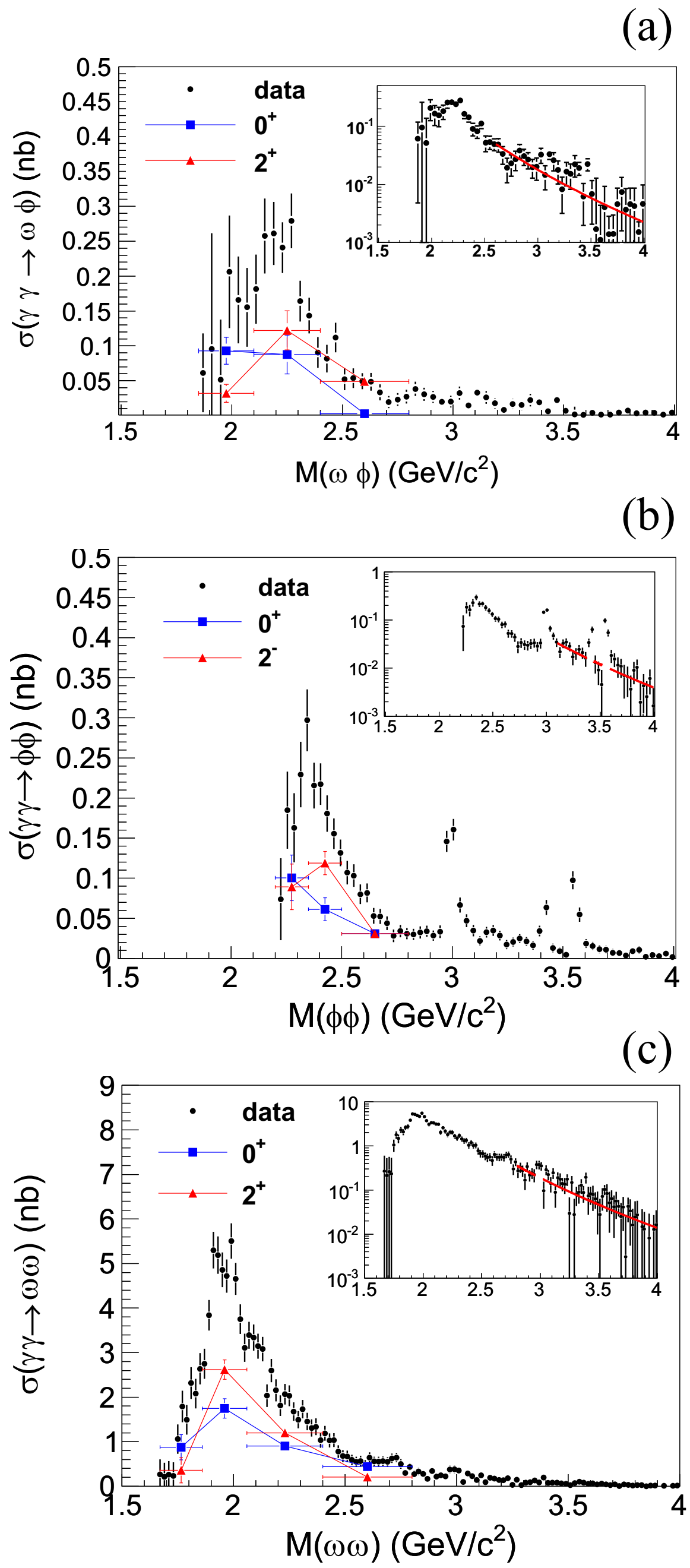

**Figure 22.3.2.** The cross sections of $\gamma\gamma \to \omega\phi$ (a), $\phi\phi$ (b), and $\omega\omega$ (c) are shown as points with error bars. The cross sections for different $J^P$ values as a function of $M(VV)$ are shown as the triangles and squares with error bars. The inset also shows the cross section on a semi-logarithmic scale. In the high energy region, the solid curve shows a fit to a $W_{\gamma\gamma}^{-n}$ dependence for the cross section after the significant charmonium contributions ($\eta_c$, $\chi_{c0}$ and $\chi_{c2}$) were excluded. From (Liu, 2012).

As shown Belle fits the $W$ dependence of the cross section with a form $W_{\gamma\gamma}^{-n}$. The results are shown, in the inset of Fig. 22.3.2 with solid curves. The fit gives $n = 7.2 \pm 0.6$, $8.4 \pm 1.1$, and $9.1 \pm 0.6$ for the $\omega\phi$, $\omega\omega$, and $\phi\phi$ modes, respectively. These results are consistent with the predictions from pQCD (Chernyak, 2010).

## 22.4 $\eta'\pi^+\pi^-$ production

The invariant mass spectrum of the $\eta'\pi^+\pi^-$ final state produced in two-photon collisions is obtained using a 673 fb$^{-1}$ data sample by the Belle experiment. In addition to the prominent $\eta_c$ signal, an enhanced shoulder is evident in the mass region below 2 GeV/$c^2$ in the background-subtracted distribution of Fig. 22.4.1 (Zhang, 2012).

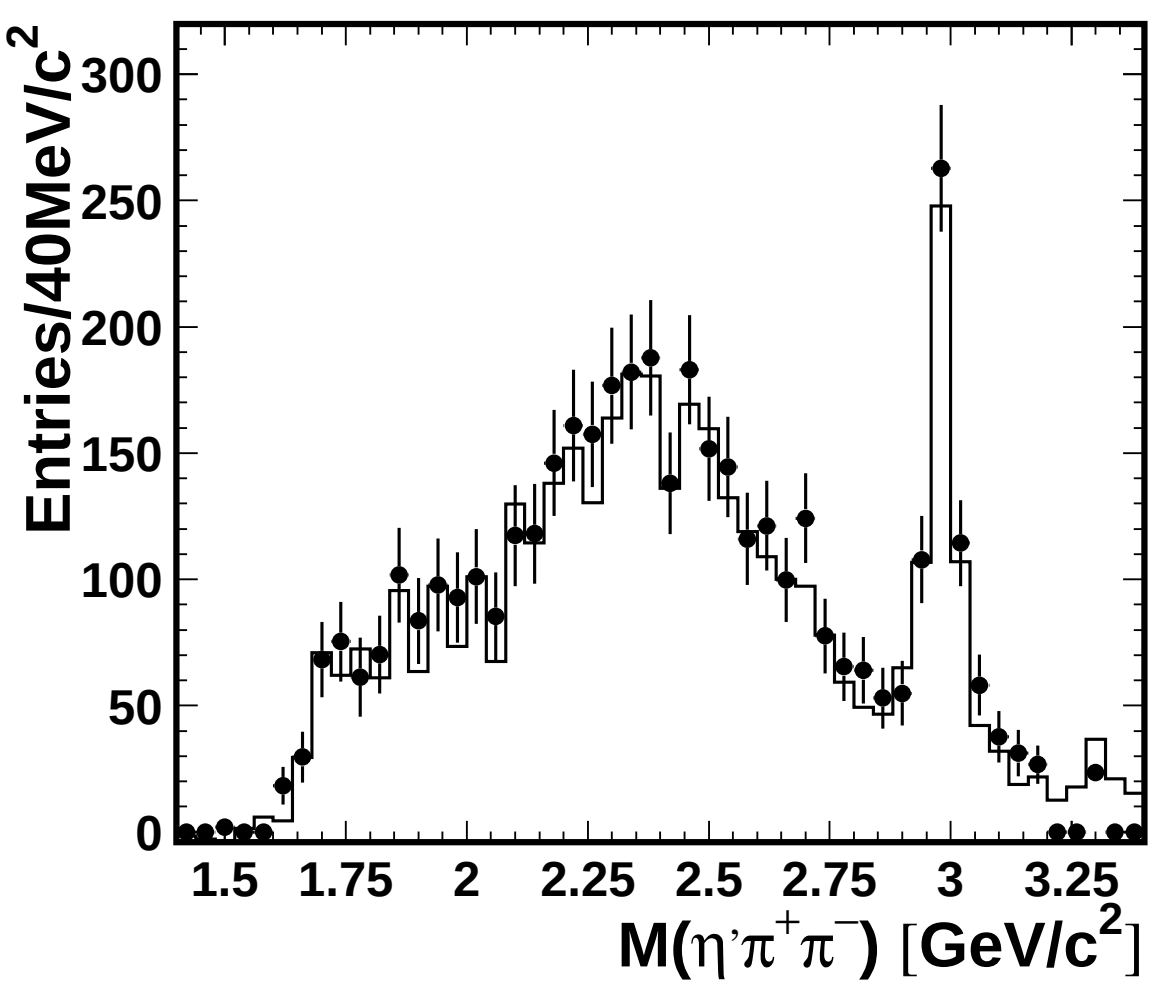

**Figure 22.4.1.** Background subtracted invariant mass distribution for the $\eta'\pi^+\pi^-$ candidates from (Zhang, 2012). The points with error bars are the $\eta'\pi^+\pi^-$ yields extracted from fitting the $|\sum \boldsymbol{p}_t^*|$ distribution in each sliced mass bin. Here, $|\sum \boldsymbol{p}_t^*|$ is determined by taking the absolute value of the vector sum of the transverse momenta of $\eta'$ and the $\pi^+\pi^-$ tracks in the $e^+e^-$ CM system.

The first evidence for decays of $\eta(1760)$ to $\eta'\pi^+\pi^-$ is reported, showing two solutions for its parameters, depending on the inclusion or not of the $X(1835)$ (Ablikim et al., 2005b, 2011), whose existence is marginal in our fits. The decay $\eta(1760) \to \eta'\pi^+\pi^-$ is found with a significance of $4.7\sigma$, with the assumption that the $X(1835)$ is not produced; the $\eta(1760)$ mass and width are determined to be $M = (1768^{+24}_{-25} \pm 10)$ MeV/$c^2$ and $\Gamma = (224^{+62}_{-56} \pm 25)$ MeV/$c^2$. The fitted $\eta(1760)$ mass is consistent with the existing measurements (Ablikim et al., 2006; Bai et al., 1999; Bisello et al., 1987, 1989). The product of the two-photon decay width and the branching fraction for the $\eta(1760)$ decay to $\eta'\pi^+\pi^-$ is determined to be $\Gamma_{\gamma\gamma}\mathcal{B}(\eta(1760) \to \eta'\pi^+\pi^-) = (28.2^{+7.9}_{-7.5} \pm 3.7)$ eV/$c^2$. When the mass spectrum is fitted with two coherent resonances, the $\eta(1760)$ and $X(1835)$, the $\eta(1760)$ mass and width are found to be $M = (1703^{+12}_{-11} \pm 1.8)$ MeV/$c^2$ and $\Gamma = (42^{+36}_{-22} \pm 15)$ MeV/$c^2$. The signal significances including the systematic error are estimated to be $4.1\sigma$ for the $\eta(1760)$ and $2.8\sigma$ for the $X(1835)$. Upper limits on the product $\Gamma_{\gamma\gamma}\mathcal{B}$ for the $X(1835)$ decay to $\eta'\pi^+\pi^-$ at the 90% confidence level for the two fit solutions are $\Gamma_{\gamma\gamma}\mathcal{B}(X(1835) \to \eta'\pi^+\pi^-) < 35.6$ eV/$c^2$ with $\phi = (287^{+42}_{-51})^\circ$ for constructive interference and $\Gamma_{\gamma\gamma}\mathcal{B}(X(1835) \to \eta'\pi^+\pi^-) < 83$ eV/$c^2$ with $\phi = (139^{+19}_{-9})^\circ$ for destructive interference. Here $\phi$ is the relative phase between $X(1835)$ and $\eta(1760)$. The fit results provide a marginal preference for the interpretation of the $X(1835)$ as a radial excitation of the $\eta'$ (Huang and Zhu, 2006; Klempt and Zaitsev, 2007).

## 22.5 Baryon-pair production

Two-photon collisions provide a clean environment for baryon pair production, which is a useful laboratory to study baryon production mechanisms and the perturbative QCD prediction. A measurement of the simplest process, $\gamma\gamma \to p\bar{p}$, has been reported by the Belle experiment (Kuo, 2005).

General theories of hard exclusive processes in QCD predict the dimensional counting rule (Sivers, Brodsky, and Blankenbecler, 1976) in the two-photon production processes of both meson and baryon pairs:

$$\frac{d\sigma}{dt} = s^{2-n_c} f(\theta^*) \tag{22.5.1}$$

at sufficiently high energy. Where $s$ is the invariant-mass square of the $\gamma\gamma$ system and $t$ is the Mandelstam variable between incident $\gamma$ and $p$ in the final state; $\theta^*$ is the meson scattering angle in the two-photon CM system as defined previously. The coefficient $n_c$ is the number of elementary constituents participating in the hard interaction. For the two-photonic baryon-pair production, $n_c$ is $n_c = 8$, which leads[171] to $\sigma \sim W^{-10}$ (for meson-pair production, $n_c = 6$ leads to $\sigma \sim W^{-6}$). The cross section is predicted to fall rapidly at high energies, and thus, its measurement is useful to test this behavior.

As another approach, the handbag model (Diehl, Kroll, and Vogt, 2003) — discussed earlier for meson-pair production — also provides a prediction for the baryon production process. Thus, we can apply the same models for both meson and baryon production phenomena from the vacuum. Quasi-elementary diquark ($qq$) models might have the potential to modify these predictions.

In Belle analysis, careful calibration of the trigger efficiency for both track and energy triggers and a base adjustment for time-of-flight measurement is needed in order to cope with the particular experimental conditions for an exclusive final state of a proton and an antiproton.

The cross section integrated over the CM angle in the range $|\cos\theta^*| < 0.6$ has been obtained. The fit, according to the power low ($\sigma \sim W^{-n}$), provides $n = 15.1^{+0.8}_{-1.1}$ and $n = 12.4^{+2.4}_{-2.3}$ in the range of $W = 2.5 - 2.9$ GeV and $3.2 - 4.0$ GeV, respectively, as shown in Fig. 22.5.1. These numbers are significantly larger than the meson-pair production case, and support the dimensional counting rule. The result for the higher energy region is consistent with the prediction with $n = 10$. While the measured cross section is more compatible with the calculation based on the diquark model with only helicity conserved amplitudes (Berger and Schweiger, 2003). Since precise data are available for baryon-pair production in two-photon processes at high energy, improved theoretical understanding is needed.

As for the angular dependence of the differential cross section, existing models can reproduce the general tendency, which shows a large-angle enhancement near the threshold (below 2.4 GeV) and a small-angle enhancement at the high energies (above 2.7 GeV). However, at the intermediate and the available highest energies, agreement between the model predictions and the results remains unsatisfactory.

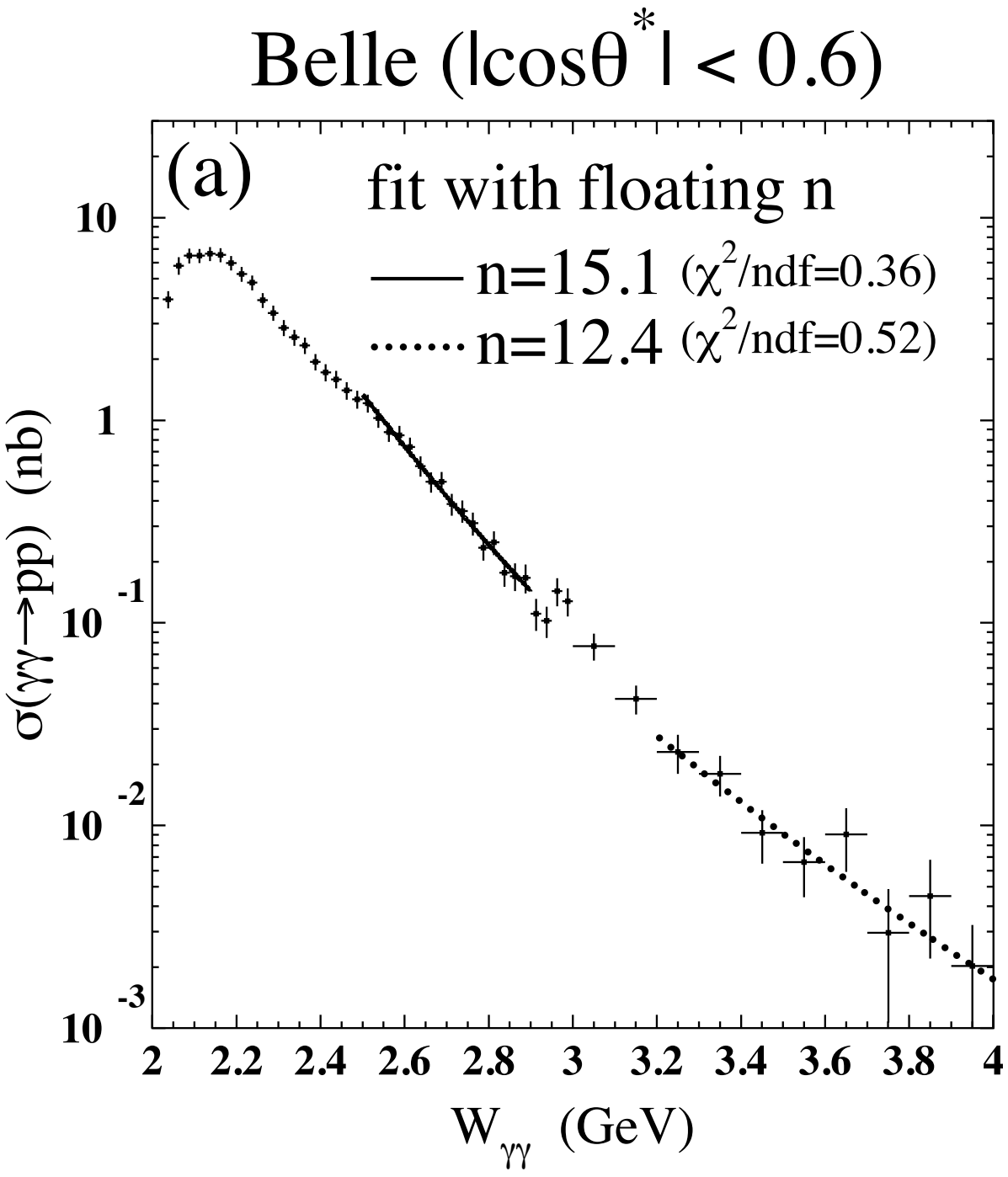

**Figure 22.5.1.** Cross section for $\gamma\gamma \to p\bar{p}$. The solid and dotted lines show the results of separate fits with $\sigma \propto W^{-n}$ to the data in the range of $W = 2.5 - 2.9$ GeV and $3.2 - 4.0$ GeV, respectively. The error bars include statistical and systematic errors. The $\chi^2/ndf$ values for each fit are indicated in the figure. From (Kuo, 2005).

## 22.6 Charmonium formation

In this section, we discuss only the two-photon decay width ($\Gamma_{\gamma\gamma}$) measurements of the charmonium(-like) states produced by photon-photon formation. Related studies of new Charmonia or exotic charmonium-like particles and the other properties of known or newly discovered states are discussed in other sections (18.2, 18.3).

[171] Note that integration over $t$ results in an additional factor of $W^2$.

The two-photon width of a charmonium(-like) particle probes the internal structure of the resonance because it is sensitive to electric charge of the constituents and the wave function near the central point. As a result, such measurements test QCD models that describe heavy quarkonia and help to identify possible exotic particles. Measurements of production from two-photon collisions give a direct way to measure this quantity in the presence of relatively small backgrounds.

The cross section for a single-resonance formation is proportional to the product of the two-photon decay width and the branching fraction to the final state(s) where the charmonium is identified, $\Gamma_{\gamma\gamma}(R)\mathcal{B}(R \to \text{final state})$ (see Eq. (22.1.5)). This is the direct observable from such measurements. Then, knowledge of either $\Gamma_{\gamma\gamma}(R)$ or $\mathcal{B}(R \to \text{final state})$ with a reasonable accuracy from independent measurements permits us to extract the other property.

We summarize the measurements from the zero-tag modes in Tables 22.6.1 and 22.6.2. The three well known Charmonia, $\eta_c$, $\chi_{c0}$ and $\chi_{c2}$, have been measured extensively in various decay modes. According to Beringer et al. (2012), the two-photon decay widths of these states, fitted to these measurements as well as branching fraction measurements in other processes, are more or less established and are dominated by the contributions from the $B$ Factory two-photon measurements. Inconsistencies among different kinds of measurement remain in some decay modes of the $\eta_c$. In addition, several new decay modes have been measured for the first time. For example, a new decay mode of $\eta_c(2S)$ to $K^+K^-\pi^+\pi^-\pi^0$ was reported by *BABAR* in the production process $\gamma\gamma \to \eta_c(2S)$ (Lees, 2010b).

The new states $\chi_{c2}(2P)$, also denoted by $Z(3930)$ (Uehara, 2006; Aubert, 2010g), $X(3915)$ (Uehara, 2010b),(Lees, 2012ad); and $X(4350)$ (Shen, 2010a) were found in the two-photon process and thus proved that this test bed is useful in searching for new states. Table 22.6.2 shows an upper limit for the $Y(4140)$ production (Shen, 2010a), which was observed in $B$ decays by CDF (Aaltonen et al., 2009a). It also shows the first-ever measurement reported by CLEO for the $\eta_c(2S)$ state (Asner et al., 2004a).

The measured size of $\Gamma_{\gamma\gamma}\mathcal{B}$ for $\chi_{c2}(2P) \to D\overline{D}$ is consistent with the theoretical prediction (Uehara, 2006), (Aubert, 2010g). The measured $\Gamma_{\gamma\gamma}\mathcal{B}$ values for the $\chi_{c2}(2P)$ and $\eta_c(2S)$ have been cited as evidence of their identification as radially excited states of the sequential Charmonia. In contrast, the large value of $\Gamma_{\gamma\gamma}\mathcal{B}$ for $X(3915) \to \omega J/\psi$ (Uehara, 2010b),(Lees, 2012ad) has attracted interest in exploring the true nature of this state.

## 22.7 Form factor measurements with single-tag processes

Two-photon production of the pseudoscalar mesons $\pi^0$ (Aubert, 2009y), $\eta$ (del Amo Sanchez, 2011f), $\eta'$ (del Amo Sanchez, 2011f), and $\eta_c$ (Lees, 2010b) in the single-tag mode has been studied by *BABAR*. Following these publications,

**Table 22.6.1.** Measurements of the product of the two-photon decay width and branching fraction ($\Gamma_{\gamma\gamma}\mathcal{B}$) for charmonium states. Upper limits correspond to a 90% confidence level.

| Decay | $\Gamma_{\gamma\gamma}\mathcal{B}$ (eV) | Reference |
|---|---|---|
| | $\eta_c$ | |
| $p\overline{p}$ | $7.20 \pm 1.53\ ^{+0.67}_{-0.75}$ | Kuo (2005) |
| $K\overline{K}\pi$ | $386 \pm 8 \pm 21$ | Lees (2010b) |
| $\eta'\pi^+\pi^-$ | $50.5^{+4.2}_{-4.1} \pm 5.6$ | Zhang (2012) |
| $\pi^+\pi^-\pi^+\pi^-$ | $40.7 \pm 3.7 \pm 5.3$ | Uehara (2008b) |
| $K^+K^-\pi^+\pi^-$ | $25.7 \pm 3.2 \pm 4.9$ | Uehara (2008b) |
| $K^+K^-K^+K^-$ | $5.6 \pm 1.1 \pm 1.6$ | Uehara (2008b) |
| $\rho\rho$ | $< 39$ | Uehara (2008b) |
| $f_2f_2$ | $69 \pm 17 \pm 12$ | Uehara (2008b) |
| $K^*\overline{K}^*$ | $32.4 \pm 4.2 \pm 5.8$ | Uehara (2008b) |
| $f_2f_2'$ | $49 \pm 9 \pm 13$ | Uehara (2008b) |
| $\phi\phi$ | $7.75 \pm 0.66 \pm 0.62$ | Liu (2012) |
| $\omega\omega$ | $8.67 \pm 2.86 \pm 0.96$ | Liu (2012) |
| $\omega\phi$ | $< 0.49$ | Liu (2012) |
| $K^+K^-\pi^+\pi^-\pi^0$ | $190 \pm 6 \pm 28$ | del Amo Sanchez† |
| | $\chi_{c0}$ | |
| $\pi^+\pi^-$ | $15.1 \pm 2.1 \pm 2.3$ | Nakazawa (2005) |
| $\pi^0\pi^0$ | $9.7 \pm 1.5 \pm 1.2$ | Uehara (2009b) |
| $K^+K^-$ | $14.3 \pm 1.6 \pm 2.3$ | Nakazawa (2005) |
| $K^0_SK^0_S$ | $5.57 \pm 0.56 \pm 0.39$ | Chen (2007c) |
| $\eta\eta$ | $9.4 \pm 2.3 \pm 1.2$ | Uehara (2010a) |
| $\pi^+\pi^-\pi^+\pi^-$ | $44.7 \pm 3.6 \pm 4.9$ | Uehara (2008b) |
| $K^+K^-\pi^+\pi^-$ | $38.8 \pm 3.7 \pm 4.7$ | Uehara (2008b) |
| $K^+K^-K^+K^-$ | $7.9 \pm 1.3 \pm 1.1$ | Uehara (2008b) |
| $K^{*0}K^-\pi^+$or c.c. | $16.7 \pm 6.1 \pm 3.0$ | Uehara (2008b) |
| $\rho\rho$ | $< 12$ | Uehara (2008b) |
| $K^*\overline{K}^*$ | $< 18$ | Uehara (2008b) |
| $\phi\phi$ | $1.72 \pm 0.33 \pm 0.14$ | Liu (2012) |
| $\omega\omega$ | $< 3.9$ | Liu (2012) |
| $\omega\phi$ | $< 0.34$ | Liu (2012) |
| $K^+K^-\pi^+\pi^-\pi^0$ | $26 \pm 4 \pm 4$ | del Amo Sanchez† |
| | $\chi_{c2}$ | |
| $\pi^+\pi^-$ | $0.76 \pm 0.14 \pm 0.11$ | Nakazawa (2005) |
| $\pi^0\pi^0$ | $0.18\ ^{+0.15}_{-0.14} \pm 0.08$ | Uehara (2009b) |
| $K^+K^-$ | $0.44 \pm 0.11 \pm 0.07$ | Nakazawa (2005) |
| $K^0_SK^0_S$ | $0.24 \pm 0.05 \pm 0.02$ | Chen (2007c) |
| $\eta\eta$ | $0.53 \pm 0.22 \pm 0.09$ | Uehara (2010a) |
| $\pi^+\pi^-\pi^+\pi^-$ | $5.01 \pm 0.44 \pm 0.55$ | Uehara (2008b) |
| $K^+K^-\pi^+\pi^-$ | $4.42 \pm 0.42 \pm 0.53$ | Uehara (2008b) |
| $K^+K^-K^+K^-$ | $1.10 \pm 0.21 \pm 0.15$ | Uehara (2008b) |
| $\rho^0\pi^+\pi^-$ | $3.2 \pm 1.9 \pm 0.5$ | Uehara (2008b) |
| $\rho\rho$ | $< 7.8$ | Uehara (2008b) |
| $K^*\overline{K}^*$ | $2.4 \pm 0.5 \pm 0.8$ | Uehara (2008b) |
| $\phi\phi$ | $0.62 \pm 0.07 \pm 0.05$ | Liu (2012) |
| $\omega\omega$ | $< 0.64$ | Liu (2012) |
| $\omega\phi$ | $< 0.04$ | Liu (2012) |
| $K\overline{K}\pi$ | $1.8 \pm 0.5 \pm 0.2$ | del Amo Sanchez† |
| $K^+K^-\pi^+\pi^-\pi^0$ | $6.5 \pm 0.9 \pm 1.5$ | del Amo Sanchez† |

† del Amo Sanchez (2011h)

**Table 22.6.2.** Measurements of the product of the two-photon decay width and branching fraction ($\Gamma_{\gamma\gamma}\mathcal{B}$) for charmonium-like states. Some results depend on the assumption of the spin-parity ($J^P$) assignments shown. Upper limits shown correspond to a 90% confidence level.

| Decay | $\Gamma_{\gamma\gamma}\mathcal{B}$ (eV) | Reference |
|---|---|---|
| $\eta_c(2S)$: | | |
| $K\overline{K}\pi$ | $73 \pm 20 \pm 8$ | CLEO* |
| | $41 \pm 4 \pm 6$ | del Amo Sanchez† |
| $\pi^+\pi^-\pi^+\pi^-$ | $< 6.5$ | Uehara (2008b) |
| $K^+K^-\pi^+\pi^-$ | $< 5.0$ | Uehara (2008b) |
| $K^+K^-K^+K^-$ | $< 2.9$ | Uehara (2008b) |
| $K^+K^-\pi^+\pi^-\pi^0$ | $30 \pm 6 \pm 5$ | del Amo Sanchez† |
| $\chi_{c2}(2P)(Z(3930))$: | | |
| $D\overline{D}$ | $180 \pm 50 \pm 30$ | Uehara (2006) |
| | $240 \pm 50 \pm 40$ | Aubert (2010g) |
| $K\overline{K}\pi$ | $< 2.1$ | del Amo Sanchez† |
| $K^+K^-\pi^+\pi^-\pi^0$ | $< 3.4$ | del Amo Sanchez† |
| others: | | |
| $X(3915) \to \omega J/\psi$ | $61 \pm 17 \pm 8$ $(0^+)$ | Uehara (2010b) |
| | $52 \pm 10 \pm 3$ $(0^+)$ | Lees (2012ad) |
| | $18 \pm 5 \pm 2$ $(2^+)$ | Uehara (2010b) |
| | $10.5\pm1.9\pm0.6$ $(2^+)$ | Lees (2012ad) |
| $Y(4140) \to \phi J/\psi$ | $< 36$ $(0^+)$ | Shen (2010a) |
| | $< 5.3$ $(2^+)$ | Shen (2010a) |
| $X(4350) \to \phi J/\psi$ | $6.7\,^{+3.2}_{-2.4}$ $(0^+)$ | Shen (2010a) |
| | $1.5\,^{+0.7}_{-0.6}$ $(2^+)$ | Shen (2010a) |

† del Amo Sanchez (2011h)
* derived from Asner et al. (2004a)

Belle reported their measurement of the $\pi^0$ production (Uehara, 2012). The amplitude of two-photon production of a pseudoscalar meson is written as follows

$$A = e^2 \varepsilon_{\mu\nu\alpha\beta} e_1^\mu e_2^\nu q_1^\alpha q_2^\beta F(q_1^2, q_2^2), \qquad (22.7.1)$$

where $\varepsilon_{\mu\nu\alpha\beta}$ is a Levi-Civita antisymmetric tensor and $e_i$ and $q_i$ are the polarization four-vectors and four-momenta of the photons. The effect of strong interactions in this process is described by the photon-meson transition form factor $F(q_1^2, q_2^2)$, which depends on the photon virtualities $q_1^2$ and $q_2^2$. In the single-tag mode, one of the photons is quasi-real, $q_2^2 \approx 0$. The form factor is measured as a function of the squared momentum transfer to the tagged (detected) electron $Q^2 = -q_1^2$.

In theory, the transition form factor $F(Q^2) \equiv F(-Q^2, 0)$ for the $\pi^0$ is calculated from first principles only in two extreme cases: at $Q^2 = 0$ from the axial anomaly in the chiral limit of QCD, $F(0) = \sqrt{2}/(4\pi^2 f_\pi)$ (Adler, 1969; Bell and Jackiw, 1969), and at $Q^2 \to \infty$ from perturbative QCD (pQCD), $Q^2F(Q^2) = \sqrt{2}f_\pi$ (Lepage and Brodsky, 1980), where $f_\pi \approx 0.131$ GeV is the pion decay constant. At large $Q^2$ ($Q^2 \gg \Lambda^2_{\rm QCD}$) in the framework of pQCD, the transition form factor can be represented as the convolution of a calculable amplitude for $\gamma\gamma^* \to q\overline{q}$ with a non-perturbative meson distribution amplitude (DA) $\phi_\pi(x, Q^2)$ (Lepage and Brodsky, 1980). The latter describes the transition of the meson with momentum $P$ into two quarks with momenta $Px$ and $P(1-x)$. In lowest order pQCD, the photon-pion transition form factor is given by

$$Q^2F(Q^2) = \frac{\sqrt{2}f_\pi}{3}\int_0^1 \frac{dx}{x}\phi_\pi(x, Q^2) + O(\alpha_s) + O\left(\frac{\Lambda^2_{\rm QCD}}{Q^2}\right), \qquad (22.7.2)$$

where $\Lambda_{\rm QCD}$ is the QCD scale parameter.

The meson DA plays an important role in theoretical descriptions of many QCD processes ($\gamma^* \to \pi^+\pi^-$, $\gamma\gamma \to \pi^+\pi^-$, and $B \to \pi l \nu_l$, for example). The shape of DA (*i.e.*, the $x$ dependence) is unknown, but its evolution with $Q^2$ is predicted by pQCD. The models for DA's can be tested using data on the photon meson transition form-factors.

It is instructive to estimate the sizes of the NLO pQCD and power corrections for Eq. (22.7.2). For this, we use results of the form-factor calculation with the asymptotic DA $\phi_{\rm ASY}(x) = 6x(1-x)$ (Lepage and Brodsky, 1979a) from Bakulev, Mikhailov, and Stefanis (2003, 2004). The NLO contribution (Braaten, 1983; del Aguila and Chase, 1981) varies from 15% to 10% in the $Q^2$ range of 4 to 50 $\rm GeV^2$ in the *BABAR* measurements. The power correction, estimated using the light-cone sum rule method (Khodjamirian, 1999; Schmedding and Yakovlev, 2000) (where the twist-4 contribution is also taken into account), is about 15% at 4 $\rm GeV^2$ and falls to about 2% at $Q^2 = 20$ $\rm GeV^2$. The calculation of the power correction is model-dependent. It is difficult to estimate the corresponding model uncertainty so it is important to perform form-factor measurements at the largest possible values of $Q^2$. At the center-of-mass energy $\sqrt{s} = 10.6$ GeV, the $e^+e^- \to e^+e^-\pi^0$ differential cross section $d\sigma/dQ^2$ is about 10 fb/$\rm GeV^2$ at $Q^2 = 10$ $\rm GeV^2$. It falls with increasing $Q^2$ as $Q^{-6}$. The most precise measurement of the form factors prior to *BABAR* was made by CLEO using data collected at $\Upsilon(4S)$ with an integrated luminosity of 3 $\rm fb^{-1}$. The $\gamma\gamma^*\pi^0$ form factor was measured for $Q^2$ up to 8 $\rm GeV^2$. The $B$ Factory measurements have extended the $Q^2$ region up to 40 $\rm GeV^2$. These measurements are described in the following section.

### 22.7.1 The $\gamma\gamma^*\pi^0$ transition form factor

The process $e^+e^- \to e^+e^-\pi^0$ in the single-tag mode has specific features that complicate its experimental study. The event contains only three detectable particles: two photons from $\pi^0$ decay and an electron. Such events, with only one charged track and low multiplicity, are rejected by the standard *BABAR* trigger and offline filters. Fortunately, a special trigger line was designed at *BABAR* to select so-called virtual Compton scattering (VCS) events for electromagnetic-calorimeter calibration. VCS is the $e^+e^- \to e^+e^-\gamma$ process in the specific kinematical configuration in which one of the final electrons moves along the collision axis while the other electron and photon are emitted at large angles. The VCS trigger selects events in which the detected electron plus photon system has a small transverse momentum and the recoil mass close to zero. For most of $e^+e^- \to e^+e^-\pi^0$ events, the close photons from $\pi^0$ decay cannot be separated by the trigger cluster algorithm and are identified as a single photon. Therefore, the VCS trigger has relatively large efficiency for events of the process under study (50–80%, depending on the $\pi^0$ energy).

The second feature is the large QED background. The main background source is VCS. There is also a sizable background from the $e^+e^- \to e^+e^-\gamma\gamma$ process in which one of the final electrons is soft and one of the photons is emitted along the beam axis. The photon from the QED process, together with a soft photon (from beam background, for example) may give the invariant mass close to the $\pi^0$ mass. Special selection criteria described in detail in Aubert (2009y) are applied to suppress QED background.

After QED background suppression, the signal events are selected by the requirements that there are electron and $\pi^0$ candidates in an event with energies above 2.0 and 1.5 GeV, respectively, and that the electron plus $\pi^0$ system has a small transverse momentum and a recoil mass close to zero. To avoid systematic uncertainty due to possible data-simulation differences in detector response near the detector edges, the $e^+e^- \to e^+e^-\pi^0$ cross section is measured in the region $Q^2 > 4$ GeV$^2$, where the detection efficiency for signal events is greater than 5%. The $Q^2$ region from 4 to 40 GeV$^2$ is divided into 17 intervals. For each interval, the number of signal events is determined from the fit to the two-photon invariant mass spectrum with a sum of a $\pi^0$ resolution function and a polynomial background distribution. For $Q^2 > 40$ GeV$^2$, no evidence of a signal over background is found in the two-photon invariant mass distribution. The total number of events with a $\pi^0$ in the $Q^2$ range 4–40 GeV$^2$ is about 14000.

Some events containing a $\pi^0$ may arise from background processes such as $e^+e^-$ annihilation, vector-meson bremsstrahlung $e^+e^- \to e^+e^-V$, and two-photon processes with higher multiplicity final states such as $e^+e^- \to e^+e^-\pi^0\pi^0$.

The $e^+e^-$ annihilation background is estimated using the difference in the distributions of the $e^\pm\pi^0$ momentum $z$-component for signal and background. In two-photon events with a tagged positron (electron), the momentum $z$-component is negative (positive), while annihilation events are produced symmetrically. The annihilation background is assumed to be equal to the number events with the wrong sign of the $e^\pm\pi^0$ momentum $z$-component and is found to be negligible.

The largest bremsstrahlung background is expected to arise from the process $e^+e^- \to e^+e^-\omega$ with $\omega$ decaying to $\pi^0\gamma$. The background is estimated from the number of data events with an extra photon, in which the invariant mass of the $\pi^0\gamma$ system is close to the $\omega$ mass. This background is also found to be negligible.

The main source of the peaking background is the process $e^+e^- \to e^+e^-\pi^0\pi^0$. To estimate this background, events with an extra $\pi^0$ are selected in data. The number of selected $e^+e^- \to e^+e^-\pi^0\pi^0$ data events is then scaled to the standard selection using a scale factor determined from MC simulation for the $e^+e^- \to e^+e^-\pi^0\pi^0$ process. The fraction of two-photon background events in the $e\pi^0$ data sample is found to be about 13% for $Q^2 < 10$ GeV$^2$ and decreases to 6–7% for $Q^2 > 10$ GeV$^2$.

The $Q^2$ dependence of the detection efficiency is determined from MC simulation. The MC efficiency is corrected for a possible data-MC simulation difference in electron identification, trigger inefficiency, $\pi^0$ detection efficiency, recoil-mass and transverse-momentum distributions. The identification and trigger corrections are determined from the control sample of VCS events. The $\pi^0$ efficiency is studied by using events of the process $e^+e^- \to \omega\gamma$, $\omega \to \pi^+\pi^-\pi^0$, which are selected and reconstructed using measured parameters of only the two charged tracks and the photon. A total efficiency correction is found to be about 7% and depends weakly on $Q^2$. The systematic uncertainty associated with the efficiency correction is about 2.5%.

The transition form factor $F(Q^2)$ is extracted by comparing the measured and calculated values of the differential cross section $d\sigma/dQ^2$. The cross-section measurement is performed at small (less than 0.18 GeV$^2$) but non-zero values of the momentum transfer to the untagged electron, $|q_2^2|$. This leads to a model uncertainty in the $F(Q^2)$ values extracted from the experiment due to the unknown dependence of the transition form factor on $q_2^2$. This model uncertainty is estimated from comparison of the $F(Q^2)$ values obtained with two different models for the $q_2^2$ dependence: the QCD-inspired model $F(q_1^2, q_2^2) \propto 1/(q_1^2 + q_2^2) \approx 1/q_1^2$ with the form factor practically independent of $q_2^2$, and the vector dominance model $F(q_2^2) \propto 1/(1 - q_2^2/m_\rho^2)$, where $m_\rho$ is the $\rho$ meson mass. The model uncertainty in the form factor is estimated to be 1.8%.

The $Q^2$ dependence of the scaled $\gamma\gamma^*\pi^0$ transition form factor $Q^2F(Q^2)$ obtained in the *BABAR* experiment, together with the CLEO (Gronberg et al., 1998) and CELLO (Behrend et al., 1991) results, is shown in Fig. 22.7.1.

The errors shown are statistical and $Q^2$-dependent systematic uncertainties combined in quadrature. The $Q^2$-dependent systematic uncertainty, which is of the same order of magnitude as the statistical one, is dominated by the uncertainties from the fitting procedure and background subtraction. The $Q^2$-independent systematic uncertainty, not shown in Fig. 22.7.1, is equal to 2.3% and includes uncertainties in the efficiency correction, radiative correction, integrated luminosity, and the model uncertainty discussed above. In the $Q^2$ region 4-9 GeV$^2$ the *BABAR* results are in reasonable agreement with the CLEO measurements (Gronberg et al., 1998), but have significantly better precision.

At $Q^2 > 10$ GeV$^2$, the measured form factor exceeds the asymptotic limit predicted by pQCD (Lepage and Brodsky, 1980), which is indicated in Fig. 22.7.1 by the horizontal line. Such behavior is specific for a “wide” pion

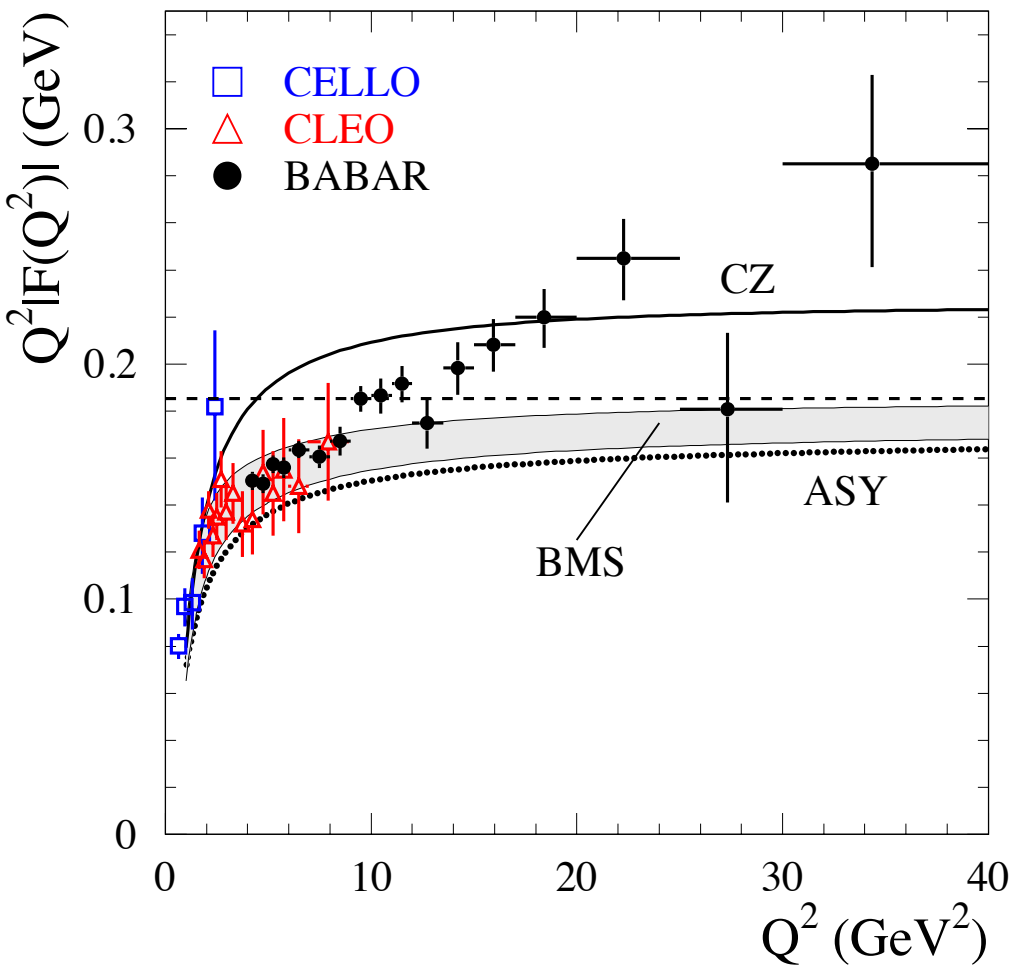

**Figure 22.7.1.** The scaled $\gamma\gamma^*\pi^0$ transition form factor. The dashed line indicates the asymptotic limit. The solid line, dotted line, and shaded band represent the predictions for the form factor (Bakulev, Mikhailov, and Stefanis, 2003, 2004) for the Chernyak and Zhitnitsky (1982) [CZ], asymptotic (Lepage and Brodsky, 1979a) [ASY], and Bakulev, Mikhailov, and Stefanis (2001) [BMS] models for the pion distribution amplitude, respectively.

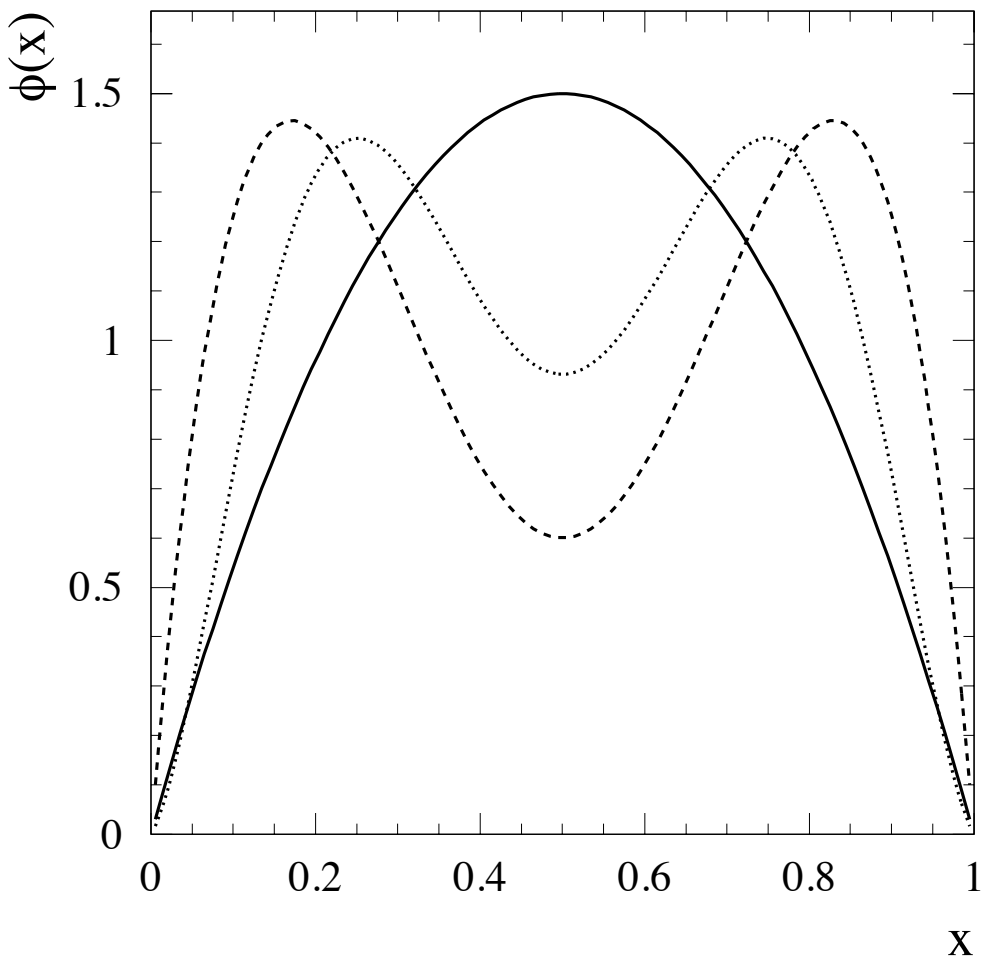

**Figure 22.7.2.** The pion distribution amplitudes at $Q^2 = 5.76\,\text{GeV}^2$ for three models: asymptotic (Lepage and Brodsky, 1979a) [solid line], Chernyak and Zhitnitsky (1982) [dashed line], and Bakulev, Mikhailov, and Stefanis (2001) [dotted line].

DA. The shapes of three selected DA's [asymptotic (Lepage and Brodsky, 1979a); Chernyak and Zhitnitsky (1982); and Bakulev, Mikhailov, and Stefanis (2001)] with different widths are shown in Fig. 22.7.2 for $Q^2 = 5.76$ GeV$^2$. The values of the Gegenbauer coefficients $a_2$ and $a_4$ used for calculation of the DA's are taken from Bakulev, Mikhailov, and Stefanis (2003, 2004). The form factors corresponding to these DA's (Bakulev, Mikhailov, and Stefanis, 2003, 2004) are presented in Fig. 22.7.1. None of these models describe the data satisfactorily in the full $Q^2$ range of the *BABAR* measurement. From the three models only one — with the widest Chernyak-Zhitnitsky DA — gives the form factor exceeding the asymptotic limit. The prediction of this model is not inconsistent with the *BABAR* data in the region $Q^2 > 15$ GeV$^2$, where power corrections are expected to be small.

Many theoretical papers devoted to the $\gamma\gamma^*\pi^0$ transition form factor appeared after the *BABAR* publication. There is no consensus on the theoretical description of the *BABAR* data at the time of writing. Some publications — for example, Bakulev, Mikhailov, Pimikov, and Stefanis (2011); Roberts, Roberts, Bashir, Gutierrez-Guerrero, and Tandy (2010)—argue that the form-factor data cannot be described by theory in the full $Q^2$ region covered by the CLEO and *BABAR* experiments, and that the *BABAR* data are likely to be incorrect for $Q^2 > 9$ GeV$^2$. Other authors obtain reasonable descriptions of the data using relatively wide DA's: see, for example, Agaev, Braun, Offen, and Porkert (2011); Kroll (2011). Note that Bakulev, Mikhailov, Pimikov, and Stefanis (2011) and Agaev, Braun, Offen, and Porkert (2011) use the same light-cone sum rule method to estimate power corrections but draw completely different conclusions. The third group of theoretical evaluations, Dorokhov (2010); Polyakov (2009); Radyushkin (2009), suggests the use of an unconventional pion DA, which is non-zero at the end points $x = 0, 1$. The transition form factor obtained with such DA increases logarithmically with increasing $Q^2$ and describes data well.

The *BABAR* measurement (Aubert, 2009y) covers the $Q^2$ region above 10 GeV$^2$ for the first time and, in this region, the form factor goes above the prediction of the asymptotic QCD value with a rather steep increase. This result has attracted the interest of many theoretical physicists as described above. An independent measurement of the pion form factor was much desired and was provided by Belle (Uehara, 2012) using data corresponding to an integrated luminosity of 759 fb$^{-1}$. The selection of the signal events, the background reduction, and the analysis were made using methods similar to those used in the *BABAR* analysis.

In this Belle analysis, event triggers to collect the signal events are provided by the electromagnetic calorimeter system. The main trigger, the total energy trigger with a high-energy threshold (high-energy trigger), is vetoed by the Bhabha trigger logic (the Bhabha trigger, to detect Bhabha events) to avoid polluting the high-energy trigger with Bhabha events. This mechanism brings a significant loss of the efficiency for the signal process $e^+e^- \to e(e)\pi^0$ and has the side effect of reducing a fraction of the acceptance. Here the symbol $(e)$ indicates the electron scattered in the forward direction and then is not detected. This condition in Belle is in contrast to the *BABAR* measurement, where a special salvaging mechanism for such events was used. Due to this situation in Belle, a complicated selection condition for the polar-angle combinations of the electron and the two-photon system were imposed to reduce the uncertainty of the trigger inefficiency caused by the Bhabha veto.

Radiative-Bhabha events with a VCS configuration were used for the calibration of the trigger system. Combining event samples collected by the high-energy trigger and by the Bhabha trigger (the latter taken with a prescale factor 50) and compensating for the effect from the Bhabha veto statistically, the trigger efficiency was determined as a function of the energy deposit. The MC events, generated by the Rabhat program (Tobimatsu and Shimizu, 1989) for the radiative-Bhabha process, are fed to the trigger simulator code. The thresholds of the Bhabha trigger have thus been tuned to reproduce the experimentally determined efficiencies. The cross section from the Rabhat program has also enabled a comparison of the absolute experimental yields with those of the MC sample after the tuning of the trigger simulation to verify the efficiency determination. This study validates the trigger efficiency at the 10% level.

In the Belle analysis, background contributions from both $e^+e^- \to e(e)\pi^0\pi^0$ and $e^+e^- \to e(e)\pi^0\gamma$ production processes are anticipated by studies. Belle actively collect these backgrounds and measures the yield of these processes by requiring the detection of an additional pion or photon. The observed yields are introduced into a MC generator for the background processes, and the contamination in the signal sample is estimated. The result is about 2% for the $e^+e^- \to e(e)\pi^0\pi^0$, and is 0.8% – 3% depending on $Q^2$ for $e^+e^- \to e(e)\pi^0\gamma$. These contributions are subtracted from the measured signal yield.

Among systematic uncertainties from different sources that are assigned to the cross section, the biggest contributions come from the extraction of the $\pi^0$ yield with the fit and the uncertainty of the trigger efficiency; they depend largely on the $Q^2$ regions. The total systematic uncertainty for the combined cross section is between 8% and 14% (and between 4% and 7% for the form factor), depending on the $Q^2$ region.

The Belle result for the transition form factor is shown in Fig. 22.7.3 together the results of the previous measurements. It is compared with the asymptotic QCD prediction shown by the dashed line. Belle has applied a fit to a parameterization with an asymptotic limit, $Q^2|F(Q^2)| = BQ^2/(Q^2 + C)$. The obtained result for the asymptotic value, $B = 0.209 \pm 0.016$ GeV, is slightly larger than the QCD prediction but still consistent with it. The fit curve is shown in the figure.

The values of $Q^2|F(Q^2)|$ measured by Belle agree with the previous measurements (Aubert (2009y), Behrend et al. (1991); Gronberg et al. (1998)), for $Q^2 \lesssim 9$ GeV$^2$. The apparent systematic shift between Belle and *BABAR* corresponds to a 2.3$\sigma$ difference in the $Q^2$ region between 9 GeV$^2$ and 20 GeV$^2$, taking into account both statistical and systematic uncertainties in the two measurements. The Belle result does not show a rapid growth of the form factor beyond the asymptotic prediction of QCD, in contrast to the *BABAR* result, and is closer to the theoretical predictions.

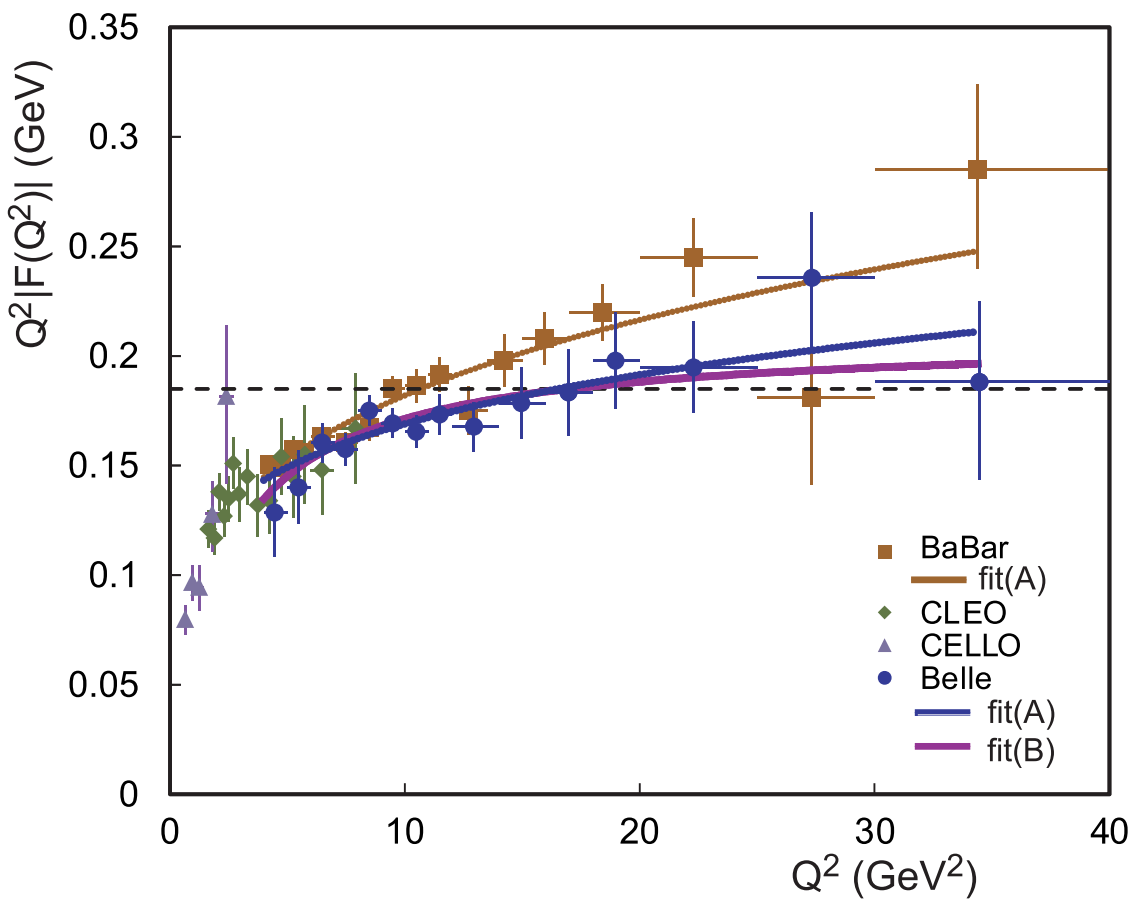

**Figure 22.7.3.** Comparison of the results for the product $Q^2|F(Q^2)|$ for the $\pi^0$ from different experiments. The curves are from the fits (A) to $\sim (Q^2/10\,\mathrm{GeV}^2)^\beta$ and (B) to $\sim Q^2/(Q^2+C)$. The dashed line shows the asymptotic prediction from pQCD ($\sim 0.185$ GeV).

### 22.7.2 The $\gamma\gamma^*\eta$ and $\gamma\gamma^*\eta'$ transition form factors

The meson-photon transition form factors for $\eta$ and $\eta'$ have been measured by *BABAR* in the $e^+e^- \to e^+e^-\eta^{(\prime)}$ (del Amo Sanchez, 2011f) and $e^+e^- \to \eta^{(\prime)}\gamma$ (Aubert, 2006w) reactions. The decay modes $\eta' \to \pi^+\pi^-\eta$, $\eta \to \gamma\gamma$ and $\eta \to \pi^+\pi^+\pi^0$ are used to reconstruct $\eta'$ and $\eta$ mesons, respectively. For single-tag $e^+e^- \to e^+e^-\eta$ events, $\eta \to \pi^+\pi^-\pi^0$ is the only decay mode available for analysis. The events with neutral $\eta$ decays, to $2\gamma$ and to $3\pi^0$, do not pass the *BABAR* trigger and background filters.

In contrast to the $e^+e^- \to e^+e^-\pi^0$ process, the QED background for the processes $e^+e^- \to e^+e^-\eta^{(\prime)}$ is almost fully rejected by the requirement that charged-pion candidates be identified as pions. The hadron background from $e^+e^-$ annihilation is suppressed by the requirement of electron identification. As a result, after applying the transverse-momentum and recoil-mass conditions $e^+e^- \to e^+e^-\eta^{(\prime)}$ events are selected with low non-peaking background. The numbers of events containing true $\eta$ and $\eta'$ are determined from the fit to the $\pi^+\pi^+\pi^0$ and $\pi^+\pi^-\eta$ mass distributions with a sum of an $\eta^{(\prime)}$ resolution function and a linear non-peaking background distribution. The fit is performed in 11 $Q^2$ intervals from the $Q^2$ range 4–40 GeV$^2$. Above 40 GeV$^2$ all observed $\eta^{(\prime)}$ candidates are expected to originate from background. The fitted number of $\eta$ and $\eta'$ events in the $Q^2$ range 4–40 GeV$^2$ is about 3000 and 5000, respectively.

For $\eta'$ events the only observed source of peaking background is $e^+e^-$ annihilation. The contribution of $e^+e^-$ annihilation is estimated using events with the wrong sign of the $e^\pm\eta'$ momentum $z$-component, and subtracted. This background is important only in the two highest-$Q^2$ intervals ($Q^2 > 25$ GeV$^2$), where it reaches about 10%.

For $\eta$ events, three sources of peaking background are studied and subtracted. These are the $e^+e^-$ annihilation and the two-photon processes $e^+e^- \to e^+e^-\eta\pi^0$ and $e^+e^- \to e^+e^-\eta'$ with $\eta'$ decaying to $\pi^0\pi^0\eta$. The contributions of the $e^+e^-$ annihilation and the $e^+e^- \to e^+e^-\eta\pi^0$ background are estimated using the difference between signal and background events in the distribution of the $e^\pm\eta$ recoil mass and are found to be about 15% of signal events.

The measured scaled $\gamma\gamma^*\eta$ and $\gamma\gamma^*\eta'$ transition form factors are shown in Fig. 22.7.4 as functions of $Q^2$. The quoted errors are statistical and $Q^2$-dependent systematic uncertainty combined in quadrature. The latter include the systematic uncertainty in the number of signal events due to the fitting procedure and background subtraction as well as the statistical errors on the efficiency correction and MC simulation, and do not exceed 50% of the statistical error. The $Q^2$-independent systematic error is about 3% and includes uncertainties in the efficiency correction, radiative correction, integrated luminosity, $\eta^{(\prime)}$ decay branching fractions, and the model uncertainty due to the unknown dependence of the transition form factor on the momentum transfer to the untagged electron. Figure 22.7.4 shows also results of the CLEO measurement (Gronberg et al., 1998). *BABAR* has improved significantly the precision and has extended the $Q^2$ region for form factor measurements relative to CLEO. For $\eta'$, the *BABAR* and CLEO data are in good agreement. For $\eta$, the agreement is worse. The CLEO point at 7 GeV$^2$ lies higher than the *BABAR* data by about $3\sigma$.

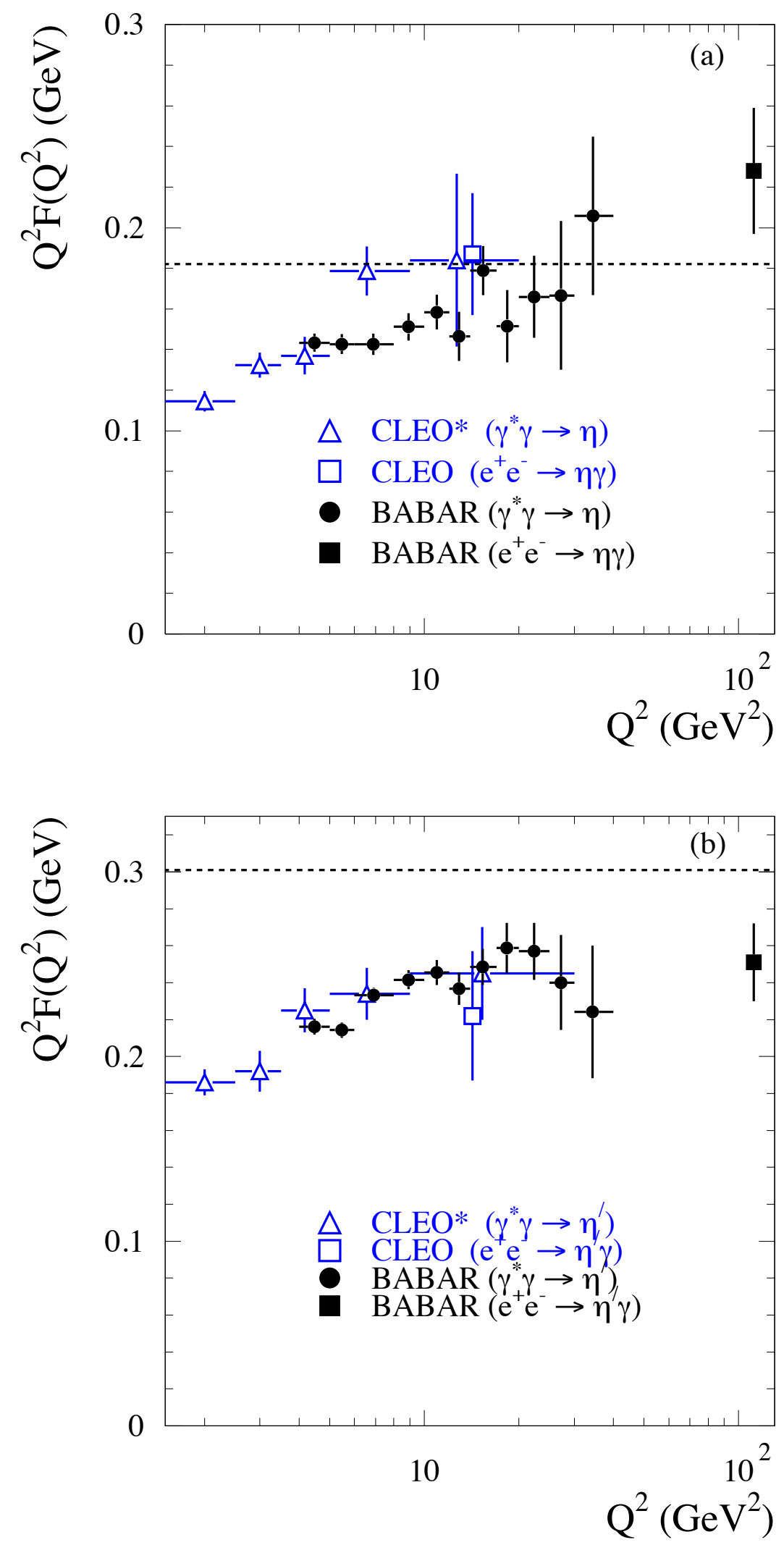

**Figure 22.7.4.** Scaled (a) $\gamma\gamma^*\eta$ and (b) $\gamma\gamma^*\eta'$ transition form factors. The dashed lines indicate the asymptotic limits for the form factors. From (del Amo Sanchez, 2011f) and (Aubert, 2006w).

The $e^+e^- \to \eta^{(\prime)}\gamma$ reactions can also be used to determine the transition form factors in the time-like region ($q^2 > 0$). The form factors at $Q^2 = 14.2$ GeV$^2$ are obtained from the values of the $e^+e^- \to \eta^{(\prime)}\gamma$ cross sections measured by CLEO (Pedlar et al., 2009) near the maximum of the $\psi(3770)$ resonance. The assumption is used that the contributions of the $\psi(3770) \to \eta^{(\prime)}\gamma$ decays to the cross sections are negligible. The CLEO form factors in both the time-like and space-like $q^2$ regions are compared in Fig. 22.7.4; they are expected to be close to each other at high $Q^2$. The CLEO measurements support this hypothesis. Therefore, the *BABAR* measurements of the $e^+e^- \to \eta^{(\prime)}\gamma$ cross sections (Aubert, 2006w) near the maximum of the $\Upsilon(4S)$ resonance can be used to extend the $Q^2$ region for the $\eta$ and $\eta'$ form-factor measurements up to 112 GeV$^2$. The time-like form-factor values at 112 GeV$^2$ are shown in Fig. 22.7.4.

Theoretical interpretation of the results on the $\eta$ and $\eta'$ form factors takes into account the $\eta - \eta'$ mixing and an admixture of the gluon component in the $SU(3)$-singlet state. The dashed lines in Fig. 22.7.4 indicate the asymptotic limits for the scaled $\eta$ and $\eta'$ form factors. They are calculated using mixing parameters from Kroll (2011). It is seen that $Q^2$ dependencies of the form factors for $\eta$ and $\eta'$ differ from those for $\pi^0$. On the other hand, there is an indication that the $\eta$ form factor exceeds the asymptotic limit at large $Q^2$. The theoretical analyses (see, for example, Agaev (2010); Brodsky, Cao, and de Teramond (2011); Kroll (2011); Noguera and Scopetta (2012)) show that the $\eta$ and $\eta'$ form-factor data are reasonably well reproduced by models with DA's not very different from the asymptotic one. The *BABAR* results on the meson-photon transition form factors for light pseudoscalars indicate that the pion DA is significantly wider than the DA's of $\eta$ and $\eta'$ mesons. However, Belle results do not exhibit such rapid growth in the higher $Q^2$ region seen in the *BABAR*. So further investigations are very important for understanding of the pion DA.

### 22.7.3 The $\gamma\gamma^*\eta_c$ transition form factor

The meson-photon transition form factor for $\eta_c$ has been measured by *BABAR* (Lees, 2010b) using single-tag events of the $e^+e^- \to e^+e^-\eta_c$ process. The $\eta_c$ is reconstructed via its decay to $K_SK^-\pi^+$. Since the branching fraction for $\eta_c \to K_SK^-\pi^+$ is known with low accuracy, it is impossible to perform an absolute measurement of the form factor. Therefore, the $Q^2$ distribution for selected $\eta_c$ events is divided by the number of no-tag two-photon $\eta_c \to K_S K^- \pi^+$ events, and the normalized form factor $F(Q^2)/F(0)$ is measured.

The number of single-tag events containing $\eta_c$ meson is determined from the fit to the $K_S K^- \pi^+$ invariant mass distributions with a sum of an $\eta_c$ resolution function, a $J/\psi$ resolution function, and a quadratic background distribution. The $J/\psi$'s are produced in the process $e^+e^- \to e^+e^- J/\psi$. The fitted number of $\eta_c$ events in the $Q^2$ region from 2 to 50 GeV$^2$ is about 500. To obtain the $Q^2$ distribution this region is divided into 11 intervals.

The process $e^+e^- \to e^+e^- J/\psi$ with $J/\psi$ decaying to $\eta_c\gamma$ is the dominant source of background peaking at $\eta_c$ mass. The background is estimated from the $Q^2$ distribution for $e^+e^- \to e^+e^- J/\psi$ events with $J/\psi \to K_S K^- \pi^+$. Its fraction changes from about 1.0% for $Q^2 < 10$ GeV$^2$ to about 5% at $Q^2 \sim 30$ GeV$^2$.

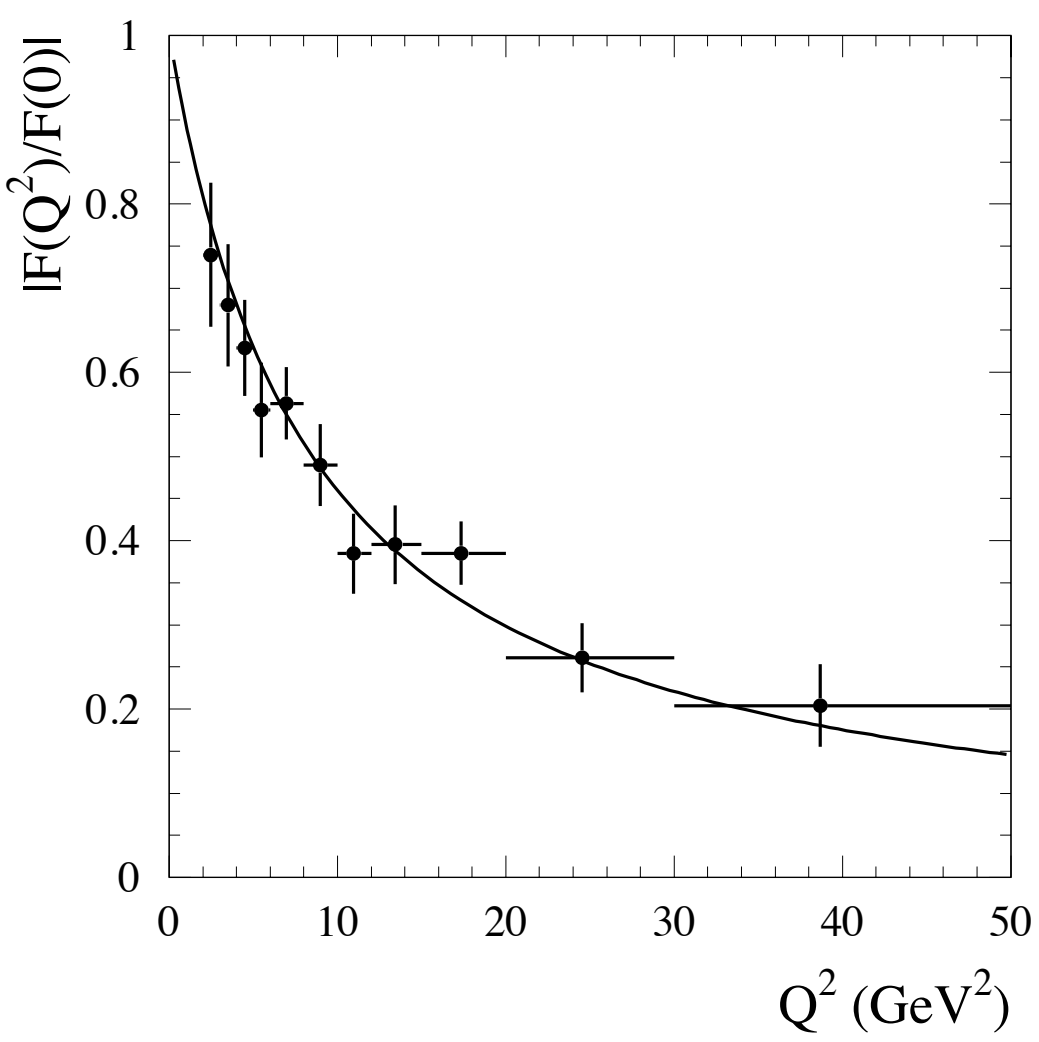

**Figure 22.7.5.** The normalized $\gamma\gamma^*\eta_c$ transition form factor measured by *BABAR*. The curve shows the fit with a monopole function. From (Lees, 2010b).

The normalized $\eta_c$ transition form factor is shown in Fig. 22.7.5. The errors shown are combined statistical and $Q^2$-dependent systematic uncertainty. There is also a $Q^2$-independent error equal to 4.3%. The main source of the systematic error is an uncertainty on the detection efficiency.

For the $\eta_c$ meson, the leading order formula for the light-meson transition form factor is modified to take into account the large mass of the $c$-quark. The term $1/x$ in Eq. (22.7.2) should be replaced with $Q^2/[xQ^2 + m_c^2(1 + 4x\overline{x})]$, where $\overline{x} = 1 - x$, and $m_c$ is the $c$-quark mass. As a consequence, the $\gamma\gamma^*\eta_c$ transition form factor can be predicted by pQCD starting from $Q^2 = 0$. However, the $Q^2$ dependence of the form factor becomes rather insensitive to the shape of the $\eta_c$ DA, and is described by a monopole function with a pole parameter $\Lambda \sim 10$ GeV$^2$ (Feldmann and Kroll, 1997). This value is close to the vector-meson dominance model (VDM) prediction $\Lambda = m^2_{J/\psi} = 9.6$ GeV$^2$.

The result of the fit to the *BABAR* data on the normalized $\eta_c$ form factor with a monopole function is shown in Fig. 22.7.5. The extracted pole parameter $\Lambda = 8.5 \pm 0.6 \pm 0.7$ GeV$^2$ is in agreement with both VDM and QCD (Feldmann and Kroll, 1997) predictions, and with the result of the lattice QCD calculation $\Lambda = 8.4 \pm 0.4$ GeV$^2$ (Dudek and Edwards, 2006).

### 22.7.4 Summary

Two-photon physics has entered a completely new level of maturity as a result of the $B$ Factory experiments. Prior to the $B$ Factories, two-photon couplings of resonances have been measured for a limited number of mesons only, and many of those were only known approximately. The $B$ Factory experiments have provided precise measurements and opened a path to the systematic study of spectroscopy including classification of mesons and exploration of exotic and new states.

In addition, the measurements of exclusive processes of hadron-pair production and of two-photon meson transition form factors have become practical tests of QCD using data from the the $B$ Factory experiments.

While many phenomena have been investigated, some subjects are still left to be clarified because of systematic uncertainties in experiments or yet-to-be developed theoretical frameworks. In some other subjects, insufficient statistics prevent us from obtaining a clear view. It is expected that future projects will be able to significantly advance our understanding in this area.

# Chapter 23
# $B_s^0$ physics at the $\Upsilon(5S)$

***Editors:***
*Alexey Drutskoy (Belle)*

**Additional section writers:**
*Sevda Esen, Brian Hamilton, Jin Li, Alan Schwartz*

## 23.1 Introduction

In this section we discuss the results of $B_s^0$ meson studies using *BABAR* and Belle data collected with a center-of-mass (CM) energy in the region of the $\Upsilon(5S)$ resonance.[172] The experimental exploration of this region was started in 1985 by the CUSB Collaboration (Lovelock et al., 1985) and since then much has been learned about $B_s^0$ decays using data from the $\Upsilon(5S)$. Future super flavor factories include $B_s^0$ studies in their physics proposals as an important part of their research program.

Experimental determination of the $B_s^0$ meson began circa 1990 at CUSB II (Lee-Franzini et al., 1990). Subsequently $B_s^0$ mesons were studied by the LEP experiments in $e^+e^-$ collisions at the CM energy of the $Z^0$ boson mass, however the statistical significance of those results is limited. Later, $B_s^0$ physics was explored with improved accuracy at the Tevatron experiments using $p\overline{p}$ collision data with a 1.8 TeV CM energy taken during the first run period, 1990–1995, and later with higher statistics during the run II period at 1.96 TeV. The Tevatron experiments CDF and DØ measure several $B_s^0$ decay branching fractions, determined for the first time the mixing parameter $\Delta m_s$, as opposed to placing lower bounds on this frequency, and obtained many other interesting results in this area.

Historically, $e^+e^-$ colliders near open beauty threshold were designed to run at the $\Upsilon(4S)$ resonance and focused on studies of $B_d^0$ and $B^+$ mesons. In particular, Belle and *BABAR* were designed to take data with asymmetric energy $e^+e^-$ beam collisions at a CM energy of 10.58 GeV. However, the new territory of $B_s^0$ physics could also be explored at the $B$ Factories. The $B_s^0\overline{B}_s^0$ production threshold lies only about 150 MeV higher than the $\Upsilon(4S)$ production energy, and therefore, is reachable by $B$ Factories without significant modification. Fig. 23.1.1 shows the ratio of hadronic to $e^+e^- \to \mu^+\mu^-$ cross-sections as a function of the CM energy in the region just above the $\Upsilon(4S)$, where two further resonances are clearly seen. These are usually called the $\Upsilon(5S)$ and $\Upsilon(6S)$. As the $\Upsilon(5S)$ has a $b\overline{b}$ quark content, $e^+e^-$ collisions at the $\Upsilon(5S)$ CM energy are expected to be an effective source of the $B_s^0$ meson production. However a very high luminosity must be delivered by the $B$ Factories to provide a sufficient number of $B_s^0$ mesons, for statistically significant measurements. The cross section for $B_s^0$ production at the $\Upsilon(5S)$ is measured to be $(0.340 \pm 0.016)$ nb (see Section 23.2.4).

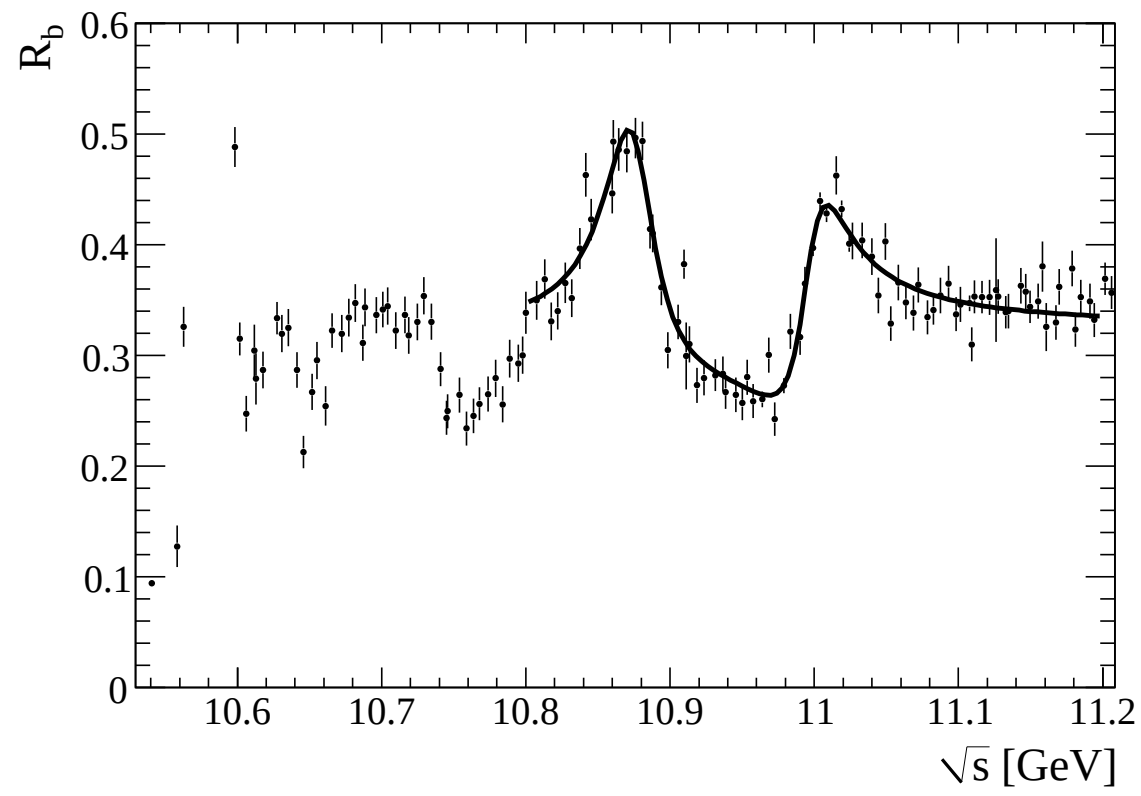

**Figure 23.1.1.** Measured ratio of hadronic to $e^+e^- \to \mu^+\mu^-$ cross-sections, $R_b$, as a function of the CM energy, from an energy scan performed by the *BABAR* collaboration (Aubert, 2009x). The result of a fit to a function including background and $\Upsilon(5S)$ and $\Upsilon(6S)$ resonances is shown by the curve. The error bars represent the statistical and the uncorrelated systematic uncertainties added in quadrature.

The opportunity to perform detailed studies of the $B$ and $B_s^0$ mesons using data collected at the $\Upsilon(5S)$ was discussed by theorists many years ago (Atwood and Soni, 2002; Falk and Petrov, 2000; Lee-Franzini, Ono, Sanda, and Tornqvist, 1985; Lellouch, Randall, and Sather, 1993). These ideas were explored experimentally from the outset, however the first significant tests were made in 2003 when the CLEO collaboration collected a small $\Upsilon(5S)$ data sample of 0.42 $\text{fb}^{-1}$. Using these data, CLEO found evidence of $B_s^0$ production at the $\Upsilon(5S)$ (Artuso et al., 2005a; Bonvicini et al., 2006; Huang et al., 2007). In 2005 the Belle collaboration collected 1.86 $\text{fb}^{-1}$ of data at the $\Upsilon(5S)$ and obtained statistically significant estimates of the $B_s^0$ production parameters (Drutskoy, 2007a,b) that established the feasibility of physics studies of $B_s^0$ mesons at Belle. Motivated by these results Belle started to accumulate data at the $\Upsilon(5S)$, collecting $\sim 22\,\text{fb}^{-1}$ in 2006, $\sim 27\,\text{fb}^{-1}$ in 2008 and $\sim 71\,\text{fb}^{-1}$ in 2009. In the last periods of running both Belle and *BABAR* performed energy scans in the region of the $\Upsilon(5S)$ and $\Upsilon(6S)$ resonances. Details of data taking for these experiments are found in Chapter 3, and the energy scan results are discussed in Section 18.4.3.

There are many reasons to develop a comprehensive physics program for the study of $B_s^0$ mesons at Belle. In 2005 the available information regarding $B_s^0$ decays was very limited, with only a few decay branching fractions measured. Studies of $B_s^0$ decays are important to build a more complete picture of $B$ physics: any significant difference between the behavior of $B_s^0$ decays and the corresponding decays of $B^0$ and $B^+$ mesons could indicate additional SM contributions such as annihilation penguins or large $W$-exchange topologies that play an important role. Formulae relating $B_s^0$, $B^0$, and $B^+$ production or decay parameters are tested and $SU(3)$ symmetry violating effects are estimated. $B_s^0$ decays are well suited for precise tests of the Standard Model, in particular there are unique features and processes that can be studied, such as the large lifetime difference between the short-lived and long-lived states and processes described by a penguin annihilation diagram.

[172] The $\Upsilon(5S)$ resonance is also referred to as the $\Upsilon(10860)$ in the literature.

In this chapter we discuss measurements of $B_s^0$ decays at the $\Upsilon(5S)$ that have been performed by the Belle collaboration since 2005. Currently the Belle sample comprises $121\,\text{fb}^{-1}$ of data, which is the world's largest dataset at the $\Upsilon(5S)$. At the time of writing most of the published Belle results were obtained using only $23.6\,\text{fb}^{-1}$ of data. However, it is worth noting that several measurements are in the process of being updated. The *BABAR* collaboration performed an energy scan in the range 10.54 GeV to 11.20 GeV, however *BABAR* did not take a large data sample of a fixed energy at the $\Upsilon(5S)$ peak. Section 23.3.1 discusses the *BABAR* measurement of the $B_s^0$ semileptonic branching fraction.

## 23.2 Basic $\Upsilon(5S)$ properties and beauty hadronization

In this section we give a classification of the basic processes which take place in $e^+e^-$ annihilation with CM energy close to the $\Upsilon(5S)$ mass peak (Section 23.2.1). The choice of the CM energy optimal for $B_s^0$ studies is then explained (Section 23.2.2). Then the procedures used to calculate the number of $B_s^0$ mesons in a data sample (Section 23.2.3) and to measure the $b\bar{b}$ cross section (Section 23.2.4) are discussed. We describe the method used to reconstruct $B$ and $B_s^0$ mesons exclusively in Section 23.2.6. The rates determined for specific $\Upsilon(5S)$ decay channels with the $B$ and $B_s^0$ mesons in the final states are summarized in Sections 23.2.5 and 23.2.7.

### 23.2.1 Event classification

The classification of hadronic events produced at the $\Upsilon(5S)$ is shown in Fig. 23.2.1: for simplicity, non-hadronic processes are not included. Final states with $B$ and $B_s^0$ mesons (designated as $b\bar{b}$ events) are formed through both resonant $\Upsilon(5S)$ production and $b\bar{b}$ continuum production. As these two event classes have the same final states, an individual event cannot be attributed to a specific class. The existence of two possible sources of $B$ and $B_s^0$ event production should always be taken into account in theoretical calculations. The $u\bar{u}$, $d\bar{d}$, $s\bar{s}$, and $c\bar{c}$ continuum is usually a significant source of background, when a specific $B$ or $B_s^0$ decay mode is reconstructed.

Thus $b\bar{b}$ events are divided into three categories: events containing $B_s^0$ mesons, $B$ mesons, and bottomonium states. A bottomonium state should have allowed quantum numbers and is accompanied by a light particle or a combination of light particles, such as $\pi^0$, $\eta$, $\gamma$, $\pi^+\pi^-$, $K^+K^-$, and so on. Taking into account the limited phase-space, only three final states with $B_s^0$ mesons are possible: $B_s^0\overline{B}_s^0$, $B_s^{*0}\overline{B}_s^0 + B_s^0\overline{B}_s^{*0}$, and $B_s^{*0}\overline{B}_s^{*0}$. Here the $B_s^{*0}\overline{B}_s^0$ and $B_s^0\overline{B}_s^{*0}$ states have the same mass combination and are almost indistinguishable experimentally, and are therefore counted as a single final state. For $B$ final states there is enough energy to produce three two-body final states, three three-body final states, and one four-body final state (see Fig. 23.2.1). Here the neutral $B^0$ and charged $B^+$ mesons are treated together as a $B$. Resonant states decay with $\sim 100\%$ probability via the modes $B_s^{*0} \to B_s^0\gamma$ and $B^{*0} \to B^0\gamma$. Additionally $e^+e^-$ collisions with a center-of-mass energy corresponding to the $\Upsilon(5S)$ resonance undergoing Initial State Radiation (ISR) can result in direct production of $B$ or $B_s^0$ mesons in association with an ISR photon.

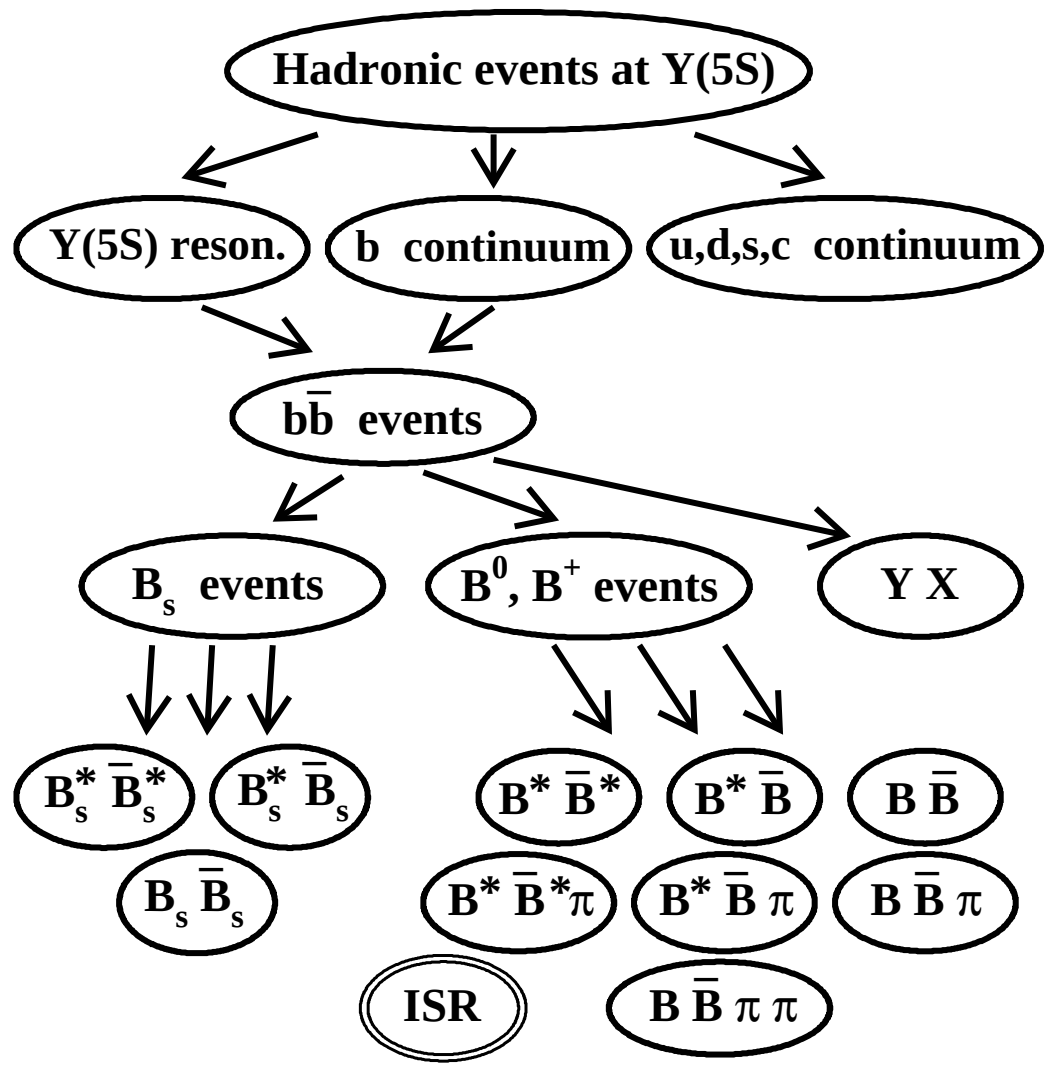

**Figure 23.2.1.** Classification of hadronic events produced in $e^+e^-$ collisions at a CM energy close to the $\Upsilon(5S)$ peak position. As noted in the text it is possible to have initial state radiation production of the lower mass states for CM energy collisions in the region of the $\Upsilon(5S)$ resonance, signified by the ISR term circled twice in this schematic.

Non-hadronic processes are not included in the above classification. Some of these processes have large cross-sections, in particular $e^+e^- \to e^+e^-$, $\mu^+\mu^-$, $\tau^+\tau^-$, $e^+e^+e^-e^-$, $\gamma\gamma$, and $\gamma\gamma X$. However these processes are strongly suppressed by the Belle trigger and HadronBJ (see Section 3.5.3) event selections, and therefore their residual contributions are negligible in most studies. The proposed classification reflects our best knowledge of possible mechanisms, and does not include some rare processes that have very low probabilities. For example, the strongly suppressed transition $e^+e^- \to s\bar{s} \to B_s^0\overline{B}_s^0$ is due to $s\bar{s}$ continuum production with subsequent $b\bar{b}$ pair creation, however here it is classified as $b\bar{b}$ continuum. The process of $b\bar{b}$ annihilation to lighter quarks also has a very low probability and is treated as non-$b\bar{b}$ continuum in the following.

### 23.2.2 Choice of CM energy for data taking at the $\Upsilon(5S)$

To maximize the $b\bar{b}$ and $B_s^0$ event production an optimal CM energy is chosen for taking data in the region of the $\Upsilon(5S)$. The suitable CM energy region was approximately known from the data collected by CLEO (Besson et al., 1985) and CUSB (Lovelock et al., 1985). However, to optimize the choice of CM energy, an energy scan in the region of the peak position of the $\Upsilon(5S)$ resonance was performed by Belle as a precursor to taking data at the $\Upsilon(5S)$ resonance. For technical reasons both KEKB beam energies were changed simultaneously, keeping the CM boost unchanged with respect to $\Upsilon(4S)$ running.

An integrated luminosity of $\sim 30\,\mathrm{pb}^{-1}$ was collected at five values with an $e^+e^-$ CM energy between 10825 MeV and 10905 MeV. The ratio of the number of hadronic events with second-to-zeroth Fox-Wolfram moment $R_2 < 0.2$ (see Chapter 9 for a definition of $R_2$) to the number of Bhabha events is measured as a function of the CM energy (Fig. 23.2.2). This ratio is expected to have a shape close to that of a Breit-Wigner function in the region of the $\Upsilon(5S)$ resonance, above a flat background. As a systematic check, the mean value of the mass obtained from the fit, $M = (10868 \pm 6 \pm 14)\,\mathrm{MeV}/c^2$, was found to be in good agreement with the PDG 2006 value $M_{\Upsilon(5S)} = (10865 \pm 8)\,\mathrm{MeV}/c^2$ (Yao et al., 2006). Finally, the energy of 10869 MeV was chosen for subsequent $\Upsilon(5S)$ runs.

The accuracy of the CM energy measurement based on the collider magnet currents is about $\pm 6$ MeV. Three methods are employed in order to measure the CM energy more precisely, all of these have an accuracy of about 1 MeV. The original method applied for data samples of $10\,\mathrm{fb}^{-1}$ or larger measures the energy by reconstructing $e^+e^- \to \Upsilon(1S)\pi^+\pi^-$ and $e^+e^- \to \Upsilon(2S)\pi^+\pi^-$ decays. As the masses of these $\Upsilon$ resonances are known to a precision better that $1\,\mathrm{MeV}/c^2$, these mass constraints are used to significantly improve the CM energy resolution relative to the collider measurement. Subsequent methods that are used include reconstructing specific $B$ and $B_s^0$ decay modes, and using $e^+e^- \to \mu^+\mu^-$. This last method requires that ISR corrections be determined accurately using Monte Carlo (MC) simulation.

The first two methods are applied post factum to obtain the CM energy; however, to keep the same energy from run to run, we need to know the energy accurately before taking data. Due to the proximity of the $B_s^0$ production threshold (in particular the $B_s^{*0}\bar{B}_s^{*0}$ channel opens up only $\sim 40\,\mathrm{MeV}$ below the chosen CM energy) even a few MeV CM energy shift can result in sizable variations of production rates of individual channels. Unfortunately, hysteresis can affect the CM energy setup process based on the magnet currents (taking into account the large uncertainty of the energy definition), even when the set-up procedure used to reach the same CM energy is similar for all data taking periods. To maximize luminosity the CM energy must be chosen at the beginning of a run. The $e^+e^- \to \mu^+\mu^-$ method is expected to provide the most robust and accurate value of the CM energy for future experiments.

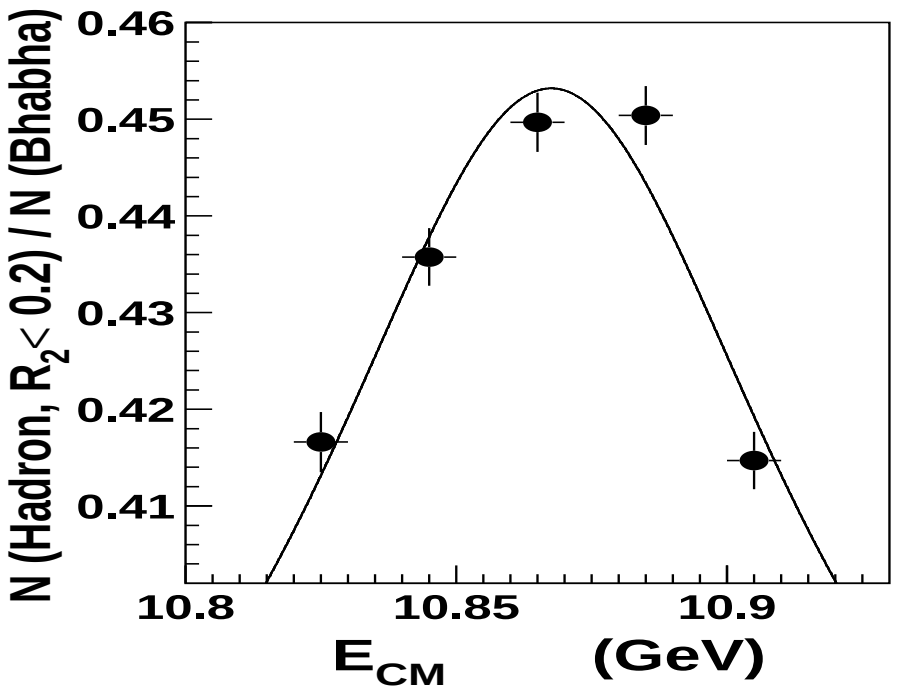

**Figure 23.2.2.** The ratio of the number of hadronic events ($R_2 < 0.2$) to the number of Bhabha events as a function of the $e^+e^-$ CM energy, from Drutskoy (2007a). Only statistical errors are shown. The curve is the result of the fit to a sum of a Breit-Wigner function and a constant.

### 23.2.3 Calculation of the number of $B_s^0$ mesons in a data sample

It is important to determine the number of $B_s^0$ mesons in a data sample with high precision because the corresponding uncertainty is usually the dominant systematic uncertainty in $B_s^0$ branching fraction measurements. Taking into account the hadronic event classification discussed above, the following parameters are required in order to obtain the number of $B_s^0$ mesons in a data sample taken at a given CM energy:

1. The integrated luminosity of the data sample $\mathcal{L}_{\mathrm{int}}$.
2. The $b\bar{b}$ production cross section $\sigma\,(e^+e^- \to b\bar{b})$, also denoted as $\sigma_{b\bar{b}}$.
3. The fraction of $b\bar{b}$ events containing a $B_s^{(*)0}\bar{B}_s^{(*)0}$ pair, usually referred to as $f_s$.
4. The fractions of events produced through specific production channels over all $B_s^{(*)0}\bar{B}_s^{(*)0}$ events: $f_{B_s^0\bar{B}_s^0}$, $f_{B_s^{*0}\bar{B}_s^0}$, and $f_{B_s^{*0}\bar{B}_s^{*0}}$.

These parameters are measured experimentally. The integrated luminosity of a data sample is precisely determined using the standard Belle procedure briefly discussed in Section 3.2.1. In the following two subsections we discuss the method used for the $\sigma_{b\bar{b}}$ cross section measurement and the experimental technique used to obtain $f_s$.

Using these parameters the number of $B_s^0$ mesons in a data sample is calculated:

$$N(B_s^0) = 2 \times \mathcal{L}_{\mathrm{int}} \times \sigma_{b\bar{b}} \times f_s. \qquad (23.2.1)$$

The number of $B_s^0$ mesons produced through a specific production channel is obtained by multiplying the number of $B_s^0$ mesons $N(B_s^0)$ by the corresponding channel fractions $f_{B_s^0\bar{B}_s^0}$, $f_{B_s^{*0}\bar{B}_s^0}$, or $f_{B_s^{*0}\bar{B}_s^{*0}}$.

Similar to the $B_s^0$ fraction $f_s$, the fractions of events with $B$ mesons ($f_B$) and a bottomonium ($f_{\rm bot}$) over all $b\bar{b}$ events is introduced. The sum of these parameters is fixed to unity: $f_s + f_B + f_{\rm bot} = 1$. Additionally the fractions of charged and neutral $B$ mesons produced per $b\bar{b}$ event, $f(B^+)$ and $f(B^0)$, is defined. As two $B$ mesons are produced per $\Upsilon(5S)$ decay and some channels include both charged and neutral $B$ mesons (such as $B^0B^-\pi^+$), a factor 2 is included in the definition, resulting in the equality

$$f_B = \frac{f(B^+) + f(B^0)}{2}. \qquad (23.2.2)$$

### 23.2.4 $b\bar{b}$ cross section at the $\Upsilon(5S)$

The $b\bar{b}$ production cross section at a fixed CM energy is obtained from the formula

$$\sigma_{b\bar{b}} = N_{\rm 5S}^{b\bar{b}} \ / \ \mathcal{L}_{\rm 5S}, \qquad (23.2.3)$$

where $N_{\rm 5S}^{b\bar{b}}$ is the number of $b\bar{b}$ events in the $\Upsilon(5S)$ data sample and $\mathcal{L}_{\rm 5S}$ is the integrated luminosity of the sample.

To obtain the number of $b\bar{b}$ events in a $\Upsilon(5S)$ data sample the $u\bar{u}$, $d\bar{d}$, $s\bar{s}$, and $c\bar{c}$ continuum are subtracted from the full number of hadronic events. The $u\bar{u}+d\bar{d}+s\bar{s}+c\bar{c}$ continuum contribution is estimated using the data collected at a CM energy 60 MeV below the $\Upsilon(4S)$ resonance (so-called "off-resonance" data). As the continuum cross section decreases with energy as $1/E_{\rm CM}^2$, the corresponding factor has to be applied to correct for the CM energy difference. Belle measured the $b\bar{b}$ cross section (Drutskoy, 2007a) using a data sample of $1.86\,{\rm fb}^{-1}$ taken at the $\Upsilon(5S)$ energy of $\sim 10869$ MeV and a data sample of $3.67\,{\rm fb}^{-1}$ collected at the off-resonance energy 10520 MeV.

The number of $b\bar{b}$ events is obtained from the formula

$$N_{\rm 5S}^{b\bar{b}} = \frac{1}{\epsilon_{\rm 5S}^{b\bar{b}}} \left( N_{\rm 5S}^{\rm had} - N_{\rm off}^{\rm had} \cdot \frac{\mathcal{L}_{\rm 5S}}{\mathcal{L}_{\rm off}} \cdot \frac{E_{\rm off}^2}{E_{\rm 5S}^2} \cdot \frac{\epsilon_{\rm 5S}^{\rm con}}{\epsilon_{\rm off}^{\rm con}} \right), \quad (23.2.4)$$

where $N_{\rm 5S}^{\rm had}$ and $N_{\rm off}^{\rm had}$ are the numbers of hadronic events in the $\Upsilon(5S)$ and off-resonance continuum data samples, respectively. Here $\epsilon_{\rm 5S}^{b\bar{b}}$ is the efficiency to select a $b\bar{b}$ event in the $\Upsilon(5S)$ data sample, which was estimated from MC simulation and found to be $\epsilon_{\rm 5S}^{b\bar{b}} = (99 \pm 1)\%$. The ratio of efficiencies to reconstruct continuum events in the $\Upsilon(5S)$ and off-resonance data samples (each one is around 79 %) was also obtained from MC: $\epsilon_{\rm 5S}^{\rm con}/\epsilon_{\rm off}^{\rm con} = 1.007 \pm 0.003$. The CM energy ratio $E_{\rm off}/E_{\rm 5S}$ is known with high accuracy, as discussed above. The integrated luminosity ratio $\mathcal{L}_{\rm 5S}/\mathcal{L}_{\rm off} = 0.5061 \pm 0.0020$ is calculated using the standard Belle luminosity measurement procedure with Bhabha events (Section 3.2.1).

The number of $b\bar{b}$ events obtained is very sensitive to these ratios of luminosities, energies, and efficiencies. The corresponding uncertainties are 0.4% for the luminosity ratio, 0.3% for the efficiency ratio, and less than 0.1% for the energy ratio. The combined uncertainty is about 0.5% and it is difficult to reduce this systematic uncertainty further. As the two terms in the subtraction are of the same order and about 10 times larger than the result (since continuum is about 10 times larger than $b\bar{b}$ production), the uncertainty of the $b\bar{b}$ event definition is dominated by the 0.5% uncertainty multiplied by a factor 10: $\sim 5\%$ in total. An improvement in the determination of the luminosity ratio is required if one is to reduce the total uncertainty of this method.

Potentially a precise luminosity ratio measurement is directly obtained from the ratio of normalized momentum distributions (for these two datasets) of high momentum charged pions, charged kaons, and $D^0$ mesons (Drutskoy, 2007a), however non-hadronic backgrounds must be accurately subtracted. Here the normalized momentum of a particle $h$ is defined as $x(h) = P(h)/P_{max}(h)$, where $P(h)$ is the measured momentum of the particle, and $P_{max}(h)$ is the expected value of its momentum if it were produced in the process $e^+e^- \to h\bar{h}$ at the same CM energy.

Using Eqs (23.2.3) and (23.2.4) Belle obtains the $b\bar{b}$ production cross section $\sigma_{b\bar{b}} = (0.302 \pm 0.015)\,{\rm nb}$ using $1.86\,{\rm fb}^{-1}$ $\Upsilon(5S)$ data sample (Drutskoy, 2007a). This number is used in the initial Belle $B_s^0$ studies based on the $23.6\,{\rm fb}^{-1}$ data sample. Later with the full data sample of $121.4\,{\rm fb}^{-1}$ a slightly higher $b\bar{b}$ production cross section was obtained, $\sigma_{b\bar{b}} = (0.340 \pm 0.016)\,{\rm nb}$, and is used in analyses of the full dataset.

### 23.2.5 Fraction of $b\bar{b}$ events with $B_s^0$ mesons

A method also used in inclusive particle spectra analyses is adopted to obtain the fraction of $b\bar{b}$ events resulting in final states with $B_s^0$ mesons in the $\Upsilon(5S)$ data. This method was developed by CLEO and was applied to the analysis of inclusive $D_{(s)}^+$ production (Artuso et al., 2005a), then later for the study of $\phi$ production (Huang et al., 2007). Belle measures the $D_{(s)}^+$ and $D^0$ inclusive spectra (Drutskoy, 2007a) and an estimate for $f_s$ is obtained from these studies. The inclusive spectra analysis method is based on the fact that the inclusive production rate of the $D_{(s)}^+$, $D^0$, and $\phi$ mesons is very different in $B_s^0$ and $B$ decays.

The analysis of the $D_{(s)}^+$ inclusive spectra will be discussed below to illustrate the method; the $D^0$ and $\phi$ spectra analyses are very similar. As Belle combines results of $D_{(s)}^+$ and $D^0$ production spectra to reduce the systematic uncertainty obtained for $f_s$, additional detail is given for the $D^0$ spectra analysis. The Belle measurement of $f_s$ was made using data samples of $1.86\,{\rm fb}^{-1}$ taken at the $\Upsilon(5S)$ and $3.67\,{\rm fb}^{-1}$ at the off-resonance energy of 10520 MeV. To avoid large backgrounds, only the clean decay mode $D_{(s)}^+ \to \phi\pi^+$, where $\phi \to K^+K^-$, is used in this analysis; similarly in order to select $D^0$, the decay $D^0 \to K^-\pi^+$ is used.

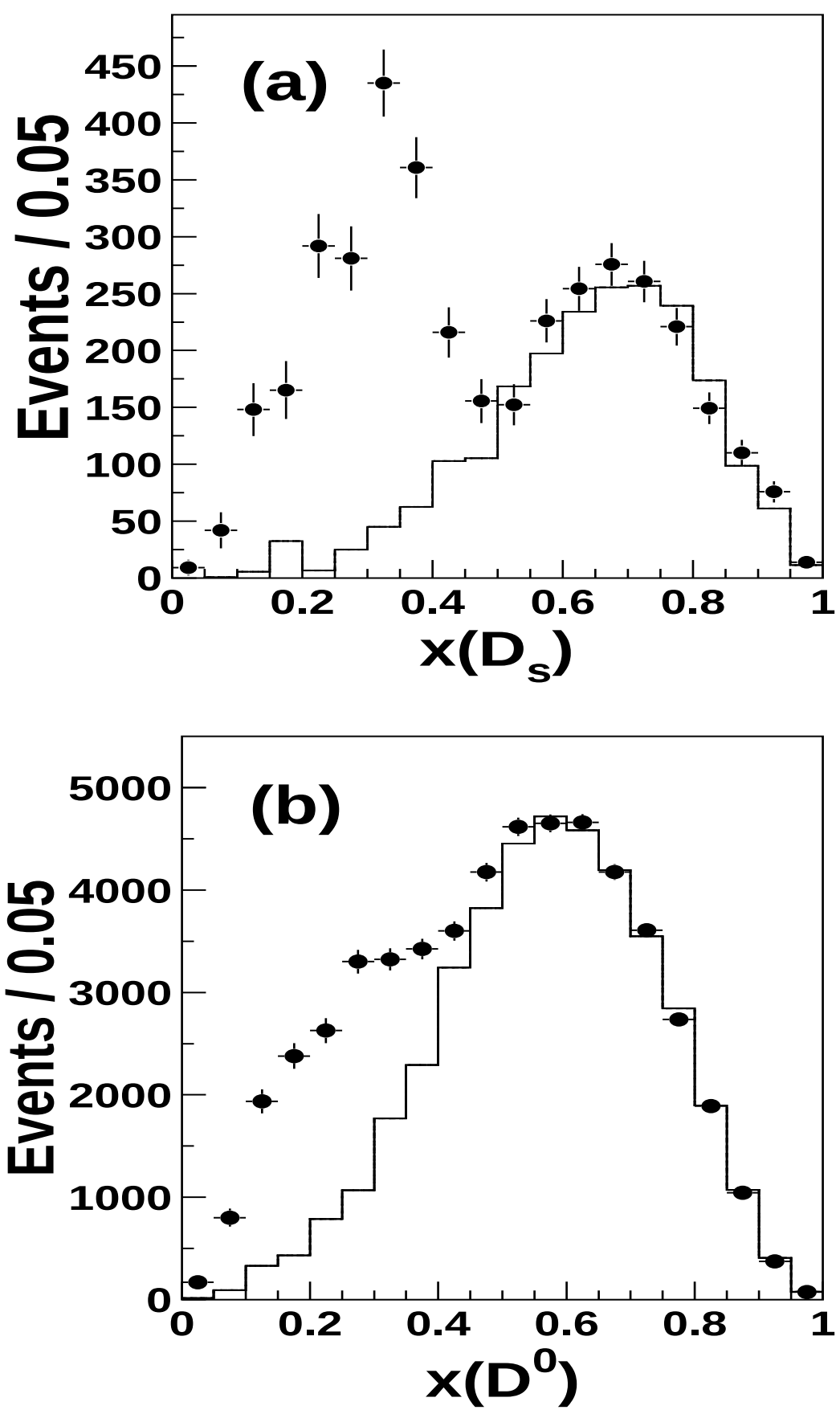

**Figure 23.2.3.** The $D_s^+$ normalized momentum $x(D_s^+)$ (a) and the $D^0$ normalized momentum $x(D^0)$ (b). The points with error bars are the $\Upsilon(5S)$ data, while the histograms show the normalized off-resonance data. Plots are from Drutskoy (2007a).

As the CM energy is different for the $\Upsilon(5S)$ and off-resonance data samples, a momentum scale normalization is applied when comparing the inclusive spectra. The normalized $D_{(s)}^+$ momentum distributions (as defined in the previous subsection) obtained for the $\Upsilon(5S)$ and off-resonance data samples are shown in Fig. 23.2.3(a). To obtain these distributions the $D_{(s)}^+$ signal yields were extracted from a fit in each bin of $x(D_s^+)$. The continuum distribution is normalized to the $\Upsilon(5S)$ distribution using the energy-corrected luminosity ratio. As we can see in Fig. 23.2.3(a) the $\Upsilon(5S)$ and off-resonance distributions agree well in the region $x(D_s^+) > 0.5$, where $b\bar{b}$ events cannot contribute. The excess of events in the region $x(D_s^+) < 0.5$ corresponds to inclusive $D_{(s)}^+$ production in $b\bar{b}$ events. Similar behavior is observed for the $D^0$ inclusive spectra as shown in Fig. 23.2.3(b).

After continuum subtraction and a bin-by-bin efficiency correction, the sum of events over all bins within the interval $x(D_s^+) < 0.5$ is divided by the $D_{(s)}^+$ and $\phi$ decay branching fractions and by the number of $b\bar{b}$ events in the $\Upsilon(5S)$ data sample to obtain the inclusive branching fraction:

$$\mathcal{B}(\Upsilon(5\mathrm{S}) \to D_s X) = \frac{\sum N_{bin}^{b\bar{b}}(D_s)/\epsilon_{bin}}{N_{5\mathrm{S}}^{b\bar{b}} \cdot \mathcal{B}(D_s^+ \to \phi\pi^+) \cdot \mathcal{B}(\phi \to K^+K^-)}. \quad (23.2.5)$$

From this formula Belle obtains the value $\mathcal{B}(\Upsilon(5\mathrm{S}) \to D_s^+ X)/2 = (23.6 \pm 1.2 \pm 3.6)\%$, which includes a factor of 1/2 to compare with $B_{(s)}$ branching fractions. This inclusive branching fraction gives the average number of $D_s^+$ mesons produced in $b\bar{b}$ events at the $\Upsilon(5S)$ energy.

The value of $\mathcal{B}(\Upsilon(5\mathrm{S}) \to D_s^+ X)/2$ is significantly larger than the branching fraction for $D_s^+$ production in $B$ decays, which was calculated in Drutskoy (2007a) as $\mathcal{B}(B \to D_s^+ X) = (8.7 \pm 1.2)\%$. The significant increase of $D_s^+$ production at the $\Upsilon(5S)$ compared to that at the $\Upsilon(4S)$ indicates a sizable $B_s^0$ production rate.

The fraction $f_s$ of $B_s^{(*)0}\overline{B}_s^{(*)0}$ events in all $b\bar{b}$ events produced at the $\Upsilon(5S)$ is extracted from the following relation:

$$\mathcal{B}(\Upsilon(5\mathrm{S}) \to D_s^+ X)/2 = f_s \cdot \mathcal{B}(B_s^0 \to D_s^+ X) + (1 - f_s) \cdot \mathcal{B}(B \to D_s^+ X), \quad (23.2.6)$$

where $\mathcal{B}(B_s^0 \to D_s^+ X)$ and $\mathcal{B}(B \to D_s^+ X)$ are the average fractions of $D_s^+$ mesons produced in $B_s^0$ and $B$ decays, respectively. Using the measurement of $\mathcal{B}(\Upsilon(5\mathrm{S}) \to D_s^+ X)$, the measured value of $\mathcal{B}(B \to D_s^+ X) = (8.7 \pm 1.2)\%$, and the model-dependent estimate $\mathcal{B}(B_s^0 \to D_s^+ X) = (92 \pm 11)\%$ (Artuso et al., 2005a), Belle determines $f_s = (17.9 \pm 1.4 \pm 4.1)\%$. The systematic uncertainty on $f_s$ is obtained by propagating the systematic uncertainties on the branching fractions included in Eq. (23.2.6), taking into account the correlation induced by $\mathcal{B}(D_{(s)}^+ \to \phi\pi^+)$. Bottomonium production in $\Upsilon(5S)$ decays, which occurs at the few percent level, is neglected in Eq. (23.2.6).

A similar procedure is applied to $D^0$ inclusive spectra and the inclusive branching fraction

$$\mathcal{B}(\Upsilon(5\mathrm{S}) \to D^0 X)/2 = (53.8 \pm 2.0 \pm 3.4)\% \quad (23.2.7)$$

is determined. Using the inclusive $D^0$ production branching fractions of the $\Upsilon(5S)$, $B$, and $B_s^0$ decays and replacing $D_s^+$ by $D^0$ in Eq. (23.2.6), the ratio $f_s = (18.1 \pm 3.6 \pm 7.5)\%$ of $B_s^{(*)0}\overline{B}_s^{(*)0}$ events to all $b\bar{b}$ events at the $\Upsilon(5S)$ is obtained. Combining these two $f_s$ measurements and taking into account the anti-correlated systematic uncertainty due to the number of $b\bar{b}$ events, an average value $f_s = (18.0 \pm 1.3 \pm 3.2)\%$ is obtained. This measurement is in good agreement with the CLEO measurement $f_s = (16.0 \pm 2.6 \pm 5.8)\%$ (Artuso et al., 2005a).[173] In the Belle measurements based on $23.6\,\mathrm{fb}^{-1}$ of data the value $f_s = (19.5^{+3.0}_{-2.3})\%$ is used, which was the PDG average of all $f_s$ measurements obtained at that time. The Belle analyses with the full data sample of 121.4 fb$^{-1}$ use run-dependent measurements of $f_s$, which are obtained from $D^+_{(s)}$ inclusive spectra studies.

[173] Subsequently CLEO used several methods to measure $f_s$ (Huang et al., 2007); however, all of these measurements have large uncertainties.

The uncertainty due to $f_s$ is currently the dominant systematic uncertainty on $B^0_s$ decay branching fraction measurements at the $\Upsilon(5S)$. There are several methods that can be used to potentially reduce this uncertainty to less than (4–5)%, however the choice of the most precise method requires further study. For the next generation of $B$ Factories this uncertainty is expected to be reduced to (2–3)%, which would result in a total systematic uncertainty of $\approx 5\%$ for $B^0_s$ branching fraction measurements.

### 23.2.6 Exclusive $B^0_s$ and $B$ decay reconstruction technique

The technique used for exclusive $B^0_s$ and $B$ meson reconstruction at the $\Upsilon(5S)$ is similar to the one used at the $\Upsilon(4S)$. However additional complexity appears at the $\Upsilon(5S)$ due to many new intermediate channels that open up at the increased CM energy. For a given $B^0_s$ or $B$ decay channel the energy and momenta for all final state particles are calculated in the CM system. Usually the low-momentum photons from $B^{*0}_s$ and $B^*$ decays are not reconstructed because of their low reconstruction efficiency. Two kinematic variables are used to reconstruct and identify exclusive $B$ (or $B^0_s$) signals at the $\Upsilon(5S)$ using a technique similar to that used at the $\Upsilon(4S)$ for isolating $B_{u,d}$ mesons (see Chapter 7). The first is the energy difference $\Delta E$ given by

$$\Delta E = E^{\star}_B - E^{\star}_{\rm beam}, \tag{23.2.8}$$

and the second is the beam-energy-substituted mass

$$m_{\rm ES} = \sqrt{E^{\star 2}_{\rm beam} - \boldsymbol{p}^{\star 2}_B}, \tag{23.2.9}$$

called "$M_{\rm bc}$" in Belle publications, where $E^{\star}_B$ and $\boldsymbol{p}^{\star}_B$ are the energy and momentum of the $B^0_s$ or $B$ candidate in the $e^+e^-$ CM system, and $E^{\star}_{\rm beam}$ is the CM beam energy.

Figure 23.2.4 shows the $B^0_s$ and $B$ signal distributions in the $m_{\rm ES}$ and $\Delta E$ plane for different intermediate $\Upsilon(5S)$ decay channels. The events shown are obtained from MC simulation of $B^0_s \to D^-_s\,\pi^+$ and $B^0 \to D^-\,\pi^+$ decays.

The three ellipsoidal regions on the right side of Figure 23.2.4 correspond to the intermediate $\Upsilon(5S)$ decay channels $B^{*0}_s\overline{B}{}^{*0}_s$, $B^{*0}_s\overline{B}{}^0_s + B^0_s\overline{B}{}^{*0}_s$, and $B^0_s\overline{B}{}^0_s$. These three $B^0_s$ signal regions are well separated, reflecting changes in kinematics corresponding to the cases where both, only one, or neither of the $B^0_s$ mesons originate from a $B^{*0}_s$ decay. A MC simulation indicates that the correlation between the reconstructed $m_{\rm ES}$ and $\Delta E$ variables within the ellipses is small and can usually be neglected. The central point of each ellipse corresponds to the decay parameters $M^{\rm cen}_{\rm bc} = M(B^0_s)$ and $\Delta E^{\rm cen} = 0$ for the channel $B^0_s\overline{B}{}^0_s$, and $M^{\rm cen}_{\rm bc} = M(B^{*0}_s)$ and $\Delta E^{\rm cen} = -E(\gamma)$ for the channel $B^{*0}_s\overline{B}{}^{*0}_s$, with the central value for the $B^{*0}_s\overline{B}{}^0_s + B^0_s\overline{B}{}^{*0}_s$ channel as the midpoint between the two individual contributions.

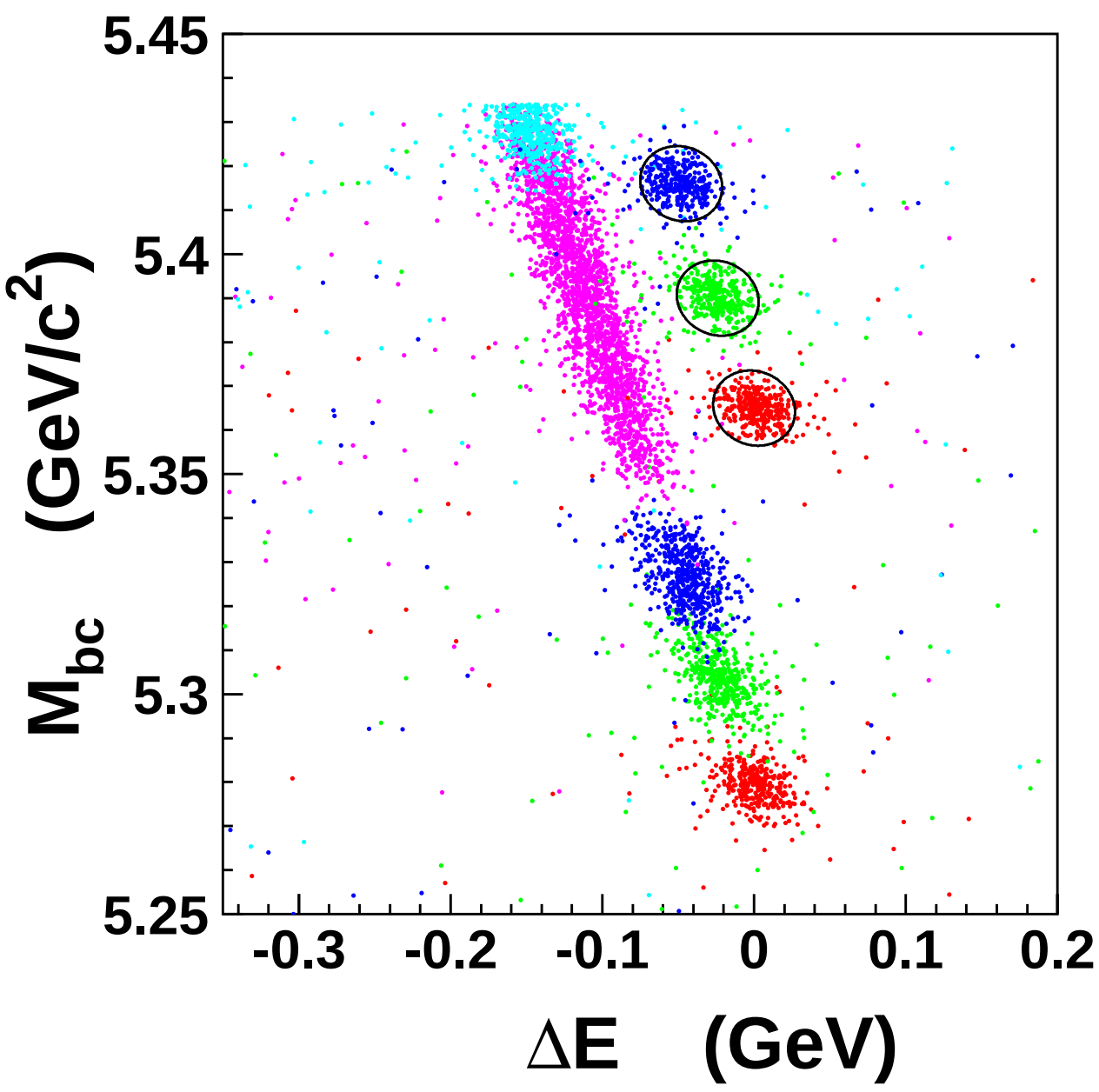

**Figure 23.2.4.** The $M_{\rm bc} = m_{\rm ES}$ and $\Delta E$ scatter plot obtained from MC simulations of different intermediate $\Upsilon(5S)$ decay channels with $B^0_s \to D^-_s\pi^+$ and $B^0 \to D^-\pi^+$ decays. The ellipses show the signal regions for the intermediate $B^{*0}_s\overline{B}{}^{*0}_s$ (top, blue), $B^{*0}_s\overline{B}{}^0_s$ and $B^0_s\overline{B}{}^{*0}_s$ (middle, green), and $B^0_s\overline{B}{}^0_s$ (bottom, red) channels. The band region includes (from bottom to top) the signals from $B\overline{B}$ (red), $B^*\overline{B}$ and $B\overline{B}{}^*$ (green), $B^*\overline{B}{}^*$ (blue), three-body $B^{(*)}\overline{B}{}^{(*)}\pi$ (violet), and four-body $B\overline{B}\pi\pi$ (turquoise) channels.

The band in the center of Fig. 23.2.4 corresponds to the intermediate $\Upsilon(5S)$ decay channels with non-strange $B$ mesons. The channels are located inside the band in the following order with increasing $m_{\rm ES}$: $B\overline{B}$, $B^*\overline{B} + B\overline{B}{}^*$, $B^*\overline{B}{}^*$, three-body $B^{(*)}\overline{B}{}^{(*)}\pi$, and four-body $B\overline{B}\pi\pi$. The two-body channels are well separated from each other in $m_{\rm ES}$ (for $B$ decays with only a few reconstructed photons). In contrast, the $m_{\rm ES}$ distributions of the three three-body channels overlap significantly with each other and partially overlap the distribution for the four-body channel.

The numbers of events inside and outside the elliptical regions is used to estimate the number of $B^0_s$ signal and background events. However, in general an unbinned extended maximum likelihood fit to $m_{\rm ES}$ and $\Delta E$ is applied to extract the number of signal events. The probability density functions used for this are adjusted using MC simulation and sideband data.

As for studies of $B$ mesons at the $\Upsilon(4S)$, the $B^0_s$ signal shapes at the $\Upsilon(5S)$ are usually modeled with a single Gaussian for $m_{\rm ES}$ and a double Gaussian with common mean for $\Delta E$. More complicated shapes have to be used to describe an electromagnetic tail, which appears if photons or electrons are present in the reconstructed $B^0_s$ final state of interest. Light quark continuum backgrounds are usually modeled with an ARGUS function (see Chapter 7) for $m_{\rm ES}$ and a polynomial function for $\Delta E$. Potentially it is possible to simultaneously fit all three $B_s^0$ signal regions (three intermediate channels, *i.e.* the three types of final state containing $B_s^0$ mesons introduced in Section 23.2.1). However, as shown below the rate of the channel $B_s^{*0}\overline{B}{}_s^{*0}$ is about 90% of all $B_s^0$ channels (if the CM energy is close to the $\Upsilon(5S)$ peak position), and in many cases it is reasonable to only include the region corresponding to this final state in the fit.

### 23.2.7 Fractions of events with $B$ mesons

Measurements of event fractions containing $B_s^0$ mesons in the final state produced at the CM energy of the $\Upsilon(5S)$ resonance (see Fig. 23.2.1) provide information about $b$-quark dynamics. Moreover, these fractions have to be precisely known to build a reliable MC model of $\Upsilon(5S)$ decays, which is required in order to make accurate background estimates in $B_{(s)}$ decay studies. Several measurements of the $B$ channel fractions at the $\Upsilon(5S)$ have been performed by Belle (Drutskoy, 2010). The experimental techniques used in these measurements are described below, and finally all results obtained are summarized in Table 23.2.1 and discussed together with the results of $B_s^0$ branching fraction measurements.

The measurements are based on a $23.6\,\mathrm{fb}^{-1}$ data sample. The five decay modes $B^+ \to J/\psi K^+$, $B^0 \to J/\psi K^{*0}$, $B^+ \to \overline{D}{}^0\pi^+$ (with $\overline{D}{}^0 \to K^+\pi^-$ and $K^+\pi^+\pi^-\pi^-$), and $B^0 \to D^-\pi^+$ ($D^- \to K^+\pi^-\pi^-$) were used to reconstruct $B$ mesons. These modes were chosen because they have large and precisely measured branching fractions and contain only charged particles in the final state; these characteristics result in small systematic uncertainties.

Belle measures the charged and neutral $B$ production rates normalized to the number of $b\bar{b}$ events. To obtain the sum of all possible channels, the $\Delta E + m_{\mathrm{ES}} - m_B$ projections of the two-dimensional scatter plots for all events within the allowed range $5.268\,\mathrm{GeV}/c^2 < m_{\mathrm{ES}} < 5.440\,\mathrm{GeV}/c^2$ are used. The $\Delta E + m_{\mathrm{ES}} - m_B$ projection works as a rotation in the $\Delta E$ and $m_{\mathrm{ES}}$ plane (see Fig. 23.2.4 for an illustration). Therefore all signal events from the inclined band contribute to the $\Delta E + m_{\mathrm{ES}} - m_B$ distribution as a single Gaussian peak. These inclined projections are fit to obtain the integrated $B$ decay event yields with a function including two terms: a Gaussian to describe the signal and a first-order polynomial to describe background.

Using the fit results, the charged and neutral $B$ production rates per $b\bar{b}$ event are obtained from the formula

$$f(B^{+,0}) = \frac{Y^{\mathrm{fit}}_{B\to X}}{(N^{b\bar{b}}_{5\mathrm{S}} \times \epsilon_{B\to X} \times \mathcal{B}_{B\to X})}, \qquad (23.2.10)$$

where $Y^{\mathrm{fit}}_{B\to X}$ is the event yield obtained from the fit for a specific mode $B \to X$, $N^{b\bar{b}}_{5\mathrm{S}}$ is the full number of $b\bar{b}$ events in the dataset, $\epsilon_{B\to X}$ is the reconstruction efficiency including all types of strange $B$ meson branching fractions, and $\mathcal{B}_{B\to X}$ is the corresponding $B$ decay branching fraction taken from the PDG (Beringer et al., 2012). The average production rates obtained for charged and neutral $B$ mesons are shown in Table 23.2.1. The $f(B^+)$ and $f(B^0)$ values are equal within uncertainties, which is consistent with isospin symmetry. The average of the charged and neutral $B$ modes is $(73.7 \pm 3.2 \pm 5.1)\%$.

**Table 23.2.1.** The $B$ and $B_s^0$ channel fractions at the CM energy of the $\Upsilon(5S)$.

| Channel | % / $b\bar{b}$ event | % / $B_s^0$ event |
|---|---|---|
| All $B_s^0$ events | $19.5^{+3.0}_{-2.3}$ | |
| $B_s^{*0}\overline{B}{}_s^{*0}$ | | $90.1^{+3.8}_{-4.0} \pm 0.2$ |
| $B_s^{*0}\overline{B}{}_s^0 + B_s^0\overline{B}{}_s^{*0}$ | | $7.3^{+3.3}_{-3.0} \pm 0.1$ |
| $B_s^0\overline{B}{}_s^0$ | | $2.6^{+2.6}_{-2.5}$ |
| All $B$ events | $73.7 \pm 3.2 \pm 5.1$ | |
| $B^+$ mesons | $72.1^{+3.9}_{-3.8} \pm 5.0$ | |
| $B^0$ mesons | $77.0^{+5.8}_{-5.6} \pm 6.1$ | |
| $B\overline{B}$ | $5.5^{+1.0}_{-0.9} \pm 0.4$ | |
| $B\overline{B}{}^*+B^*\overline{B}$ | $13.7 \pm 1.3 \pm 1.1$ | |
| $B^*\overline{B}{}^*$ | $37.5^{+2.1}_{-1.9} \pm 3.0$ | |
| $B\overline{B}\pi$ | $0.0 \pm 1.2 \pm 0.3$ | |
| $B\overline{B}{}^*\pi + B^*\overline{B}\pi$ | $7.3^{+2.3}_{-2.1} \pm 0.8$ | |
| $B^*\overline{B}{}^*\pi$ | $1.0^{+1.4}_{-1.3} \pm 0.4$ | |
| ISR to final $B$ | $9.2^{+3.0}_{-2.8} \pm 1.0$ | |

The two-body channel branching fractions are measured; values are averaged over charged and neutral $B$ mesons. The $m_{\mathrm{ES}}$ projections are obtained for events in the signal band (shown in Fig. 23.2.5(a)) and the projections are fit with a function including signal and background terms (Fig. 23.2.5(b)). The shapes of the signal components are taken from MC simulation, and those of combinatorial background are modeled using sideband data. The fit region is restricted to the interval $m_{\mathrm{ES}} \in [5.268, 5.348]\,\mathrm{GeV}/c^2$. The three peaks corresponding to the $B\overline{B}$, $B\overline{B}{}^*+B^*\overline{B}$ and $B^*\overline{B}{}^*$ channels (from left to right) are clearly seen in Fig. 23.2.5(b). The matrix elements responsible for three- and four-body decays are not known, and the rates of three- and four-body contributions cannot be obtained in a model-independent way from a fit to these $m_{\mathrm{ES}}$ distributions. The fit results are given in Table 23.2.1.

In order to reconstruct the three-body channels, an additional charged pion produced directly via $B^{(*)}\overline{B}{}^{(*)}\pi^+$ is combined with two $B^{(*)}$ mesons. For each charged pion not included in the reconstructed $B$ candidate, right-sign $B^+\pi^-$, $\overline{B}{}^0\pi^-$, $B^0\pi^+$, or $B^-\pi^+$ combinations are formed. The reconstructed $B$ candidates are selected from the three-body signal region given by $5.37\,\mathrm{GeV}/c^2 < m_{\mathrm{ES}} < 5.44\,\mathrm{GeV}/c^2$ and $|\Delta E + m_{\mathrm{ES}} - m_B| < 0.03\,\mathrm{GeV}$. Then a special variable $\Delta X = \Delta E^{\mathrm{mis}} + m^{\mathrm{mis}}_{\mathrm{ES}} - m_B$ was calculated for all selected $B\pi$ candidates, where the variables $m^{\mathrm{mis}}_{\mathrm{ES}}$ and $\Delta E^{\mathrm{mis}}$ are obtained for the missing $B$ meson using the energy and momentum of the reconstructed $B\pi$ combination in the CM frame. The $\Delta X$ variable reflects the missing mass in the reconstructed $B\pi$ system.

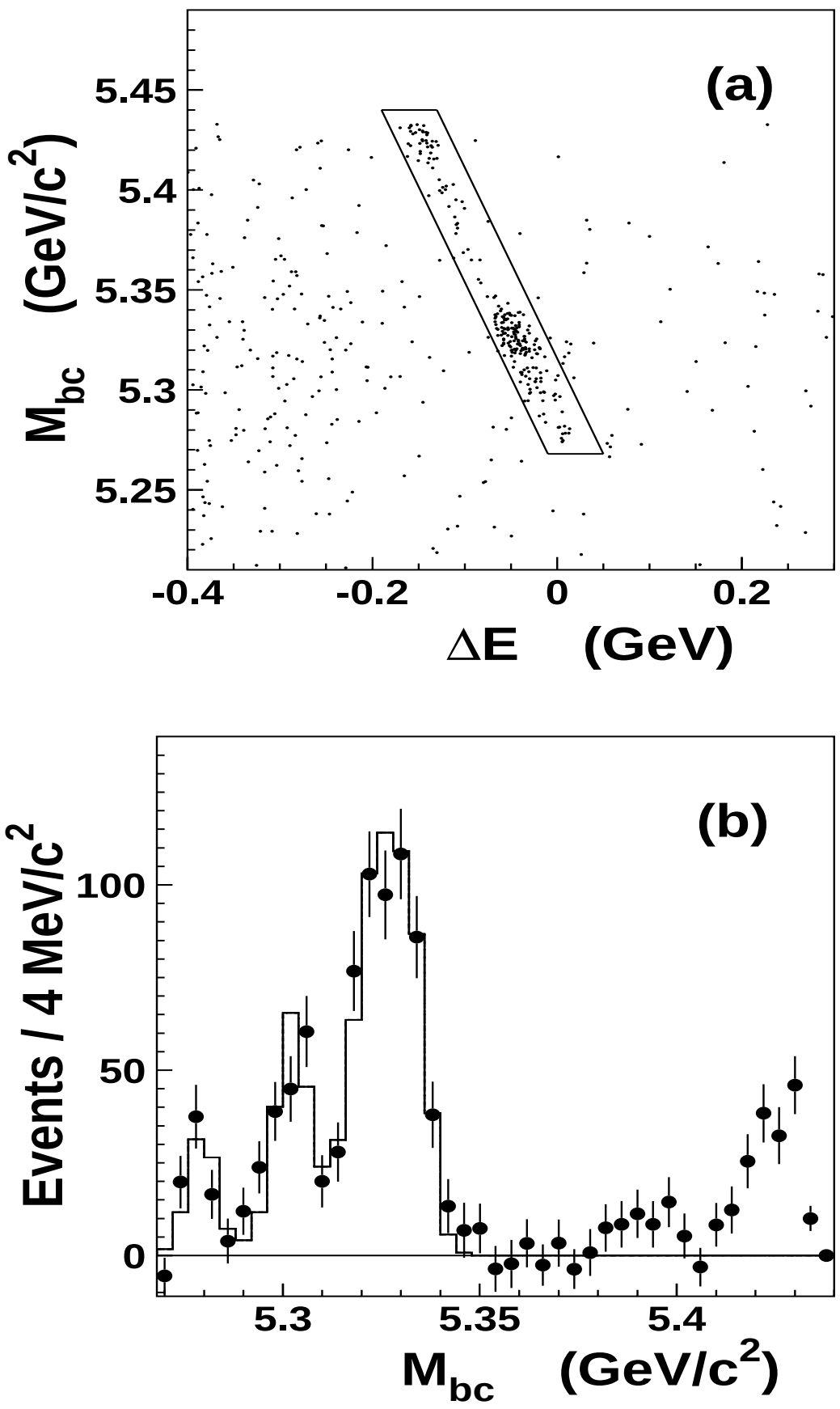

**Figure 23.2.5.** (a) The $M_{\rm bc} = m_{\rm ES}$ and $\Delta E$ scatter plot for the $B^+ \to J/\psi K^+$ mode (data) from Drutskoy (2010). The band indicates the signal region corresponding to the intervals $5.268\,{\rm GeV}/c^2 < m_{\rm ES} < 5.440\,{\rm GeV}/c^2$ and $|\Delta E + m_{\rm ES} - m_B| < 0.03\,{\rm GeV}$. (b) The $m_{\rm ES}$ distribution in data after background subtraction. The sum of the five significant $B$ decays (points with error bars, see Table 23.2.1) and results of the fit (histogram) used to extract the two-body channel fractions are shown.

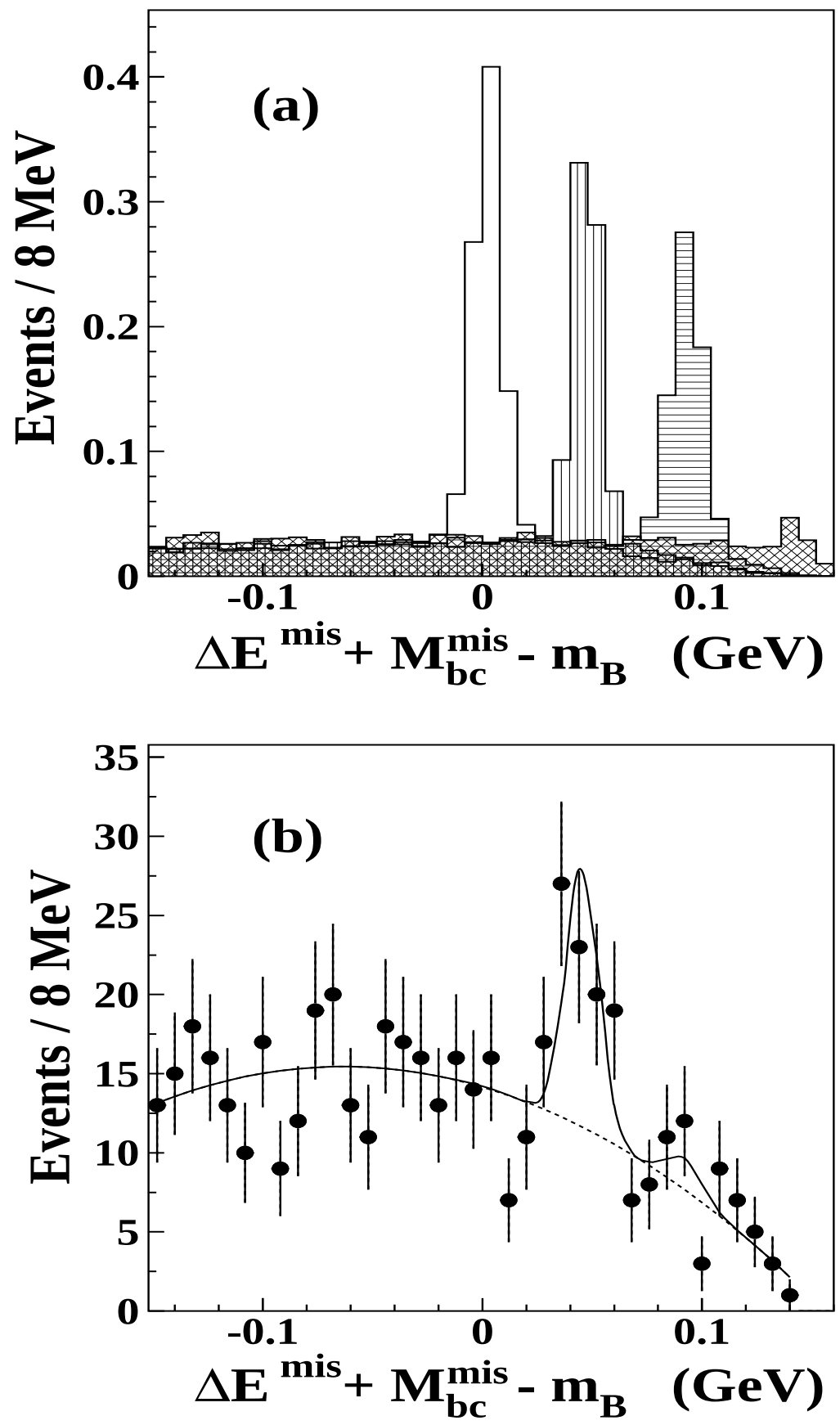

**Figure 23.2.6.** (a) The $\Delta X = \Delta E^{\rm mis} + m_{\rm ES}^{\rm mis} - m_B$ distribution normalized to unity for MC simulated $B^+ \to J/\psi K^+$ decays in the (peaks from left to right) $B\overline{B}\pi^+$, $B\overline{B}^*\pi^+ + B^*\overline{B}\pi^+$, $B^*\overline{B}^*\pi^+$, and $B\overline{B}\pi\pi$ channels. (b) The $\Delta X = \Delta E^{\rm mis} + m_{\rm ES}^{\rm mis} - m_B$ data distribution for right-sign $B^{-/0}\pi^+$ combinations for the sum of the five studied $B$ modes. Plots are from Drutskoy (2010).

Figure 23.2.6(a) shows the $\Delta X$ distributions obtained for MC simulated $B\overline{B}\pi^+$, $B\overline{B}^*\pi^+ + B^*\overline{B}\pi^+$, $B^*\overline{B}^*\pi^+$, and $B\overline{B}\pi\pi$ events where the $B^+ \to J/\psi K^+$ mode is generated. The background due to random charged tracks from the unobserved $B$ meson is also shown. The studied channel contributions are well separated as seen in Fig. 23.2.6(a). The reconstruction efficiency for the four-body channel (the small peak on the rightmost part of Fig. 23.2.6(a)) is small and model dependent.

Finally, the $\Delta X$ distribution obtained from the data is shown in Fig. 23.2.6(b). This distribution is fitted with a function including four terms: three Gaussian distributions with fixed shapes in order to describe the $B\overline{B}\pi^+$, $B\overline{B}^*\pi^+ + B^*\overline{B}\pi^+$, and $B^*\overline{B}^*\pi^+$ contributions, and a second-order polynomial to describe the background. The four-body channel contribution is negligible and is not included in the fit. The three-body channel fractions obtained from the fit are given in Table 23.2.1.

All results obtained on $B$ and $B_s^0$ channel fraction measurements are summarized in Table 23.2.1. The fractions for specific $B_s^0$ channels for all events with $B_s^0$ mesons are taken from Louvot (2009). The difference between the number of $B$ events and the sum of all two-body or three-body channels is assumed to be due to ISR producing a system of lower CM energy with subsequent $B\overline{B}$-pair production. The ISR rate obtained using this assumption agrees with theoretical expectations.

The measurement of fractions for specific $B_s^0$ channels of all events with $B_s^0$ mesons is discussed in the next section. An unexpected feature of $B_s^0$ production at the $\Upsilon(5S)$ is the strong dominance of the $B_s^{*0}\overline{B}_s^{*0}$ channel. The observed value is close to 90%, although initial theoretical estimates gives values of about 70%. Taking into account the $B$ event rate of $(73.7 \pm 3.2 \pm 5.1)\%$ and the $B_s^0$ event rate of $f_s = (19.5^{+3.0}_{-2.2})\%$ at the $\Upsilon(5S)$, there is still room for unobserved transitions to non-$B\overline{B}$ final states with a bottomonium meson. The large fraction for the three-body $B\overline{B}^*\pi + B^*\overline{B}\pi$ channel was not predicted theoretically and is not yet understood. The channel $B\overline{B}\pi$ was not observed and its production is probably suppressed due to the $0^-$ quantum numbers of all three final particles produced from the $1^-$ initial state, which results in two $P$-wave amplitudes.

## 23.3 Measurements of $B_s^0$ decays at $\Upsilon(5S)$

Branching fraction measurements discussed here are ordered in terms of decreasing value and cover semi-leptonic $B_s^0$ decays (Section 23.3.1); the Cabibbo favored decays $B_s^0 \to D_s^{(*)-}\pi^+(\rho^+)$ and $B_s^0 \to D_s^{(*)+}D_s^{(*)-}$ (Sections 23.3.2 and 23.3.3, respectively); color suppressed decays (Section 23.3.4); charmless two-body decays (Section 23.3.5); and finally loop (or penguin) decays (Section 23.3.6).

### 23.3.1 $B_s^0$ semileptonic branching fraction

Although *BABAR* performed no dedicated $\Upsilon(5S)$ running, a scan was performed in 2008 in 5 MeV steps covering the CM energy ($E_{\rm CM}$) range $10.54\,\text{GeV} \le E_{\rm CM} \le 11.2\,\text{GeV}$, integrating a total of $4.25\,\text{fb}^{-1}$ of data. These data are used to study $B_s^0$ production in this region and to obtain a measurement of the inclusive $B_s^0$ semileptonic branching fraction $\mathcal{B}(B_s^0 \to \ell\nu X)$. Semileptonic decays of $B_d^0$ mesons have a large branching fraction to final states including a charm meson. Similarly semileptonic decays of $B_s^0$ mesons have large transition rates to final states including a $D_s$ meson, which subsequently decays (15.7% of the time (Beringer et al., 2012)) into a $\phi$+anything final state. The *BABAR* analysis exploits the existence of the Cabibbo-favored decay chain $B_s \to D_s \to \phi$.

In 15 MeV wide bins of $E_{\rm CM}$, *BABAR* measures the yields of three inclusive processes in the scan data: the yield of multi-hadronic events containing at least three charged tracks and with Fox-Wolfram moment less than or equal to 0.2 (the event yield), the inclusive yield of events containing $\phi$ mesons (the $\phi$ yield), and the inclusive yield of events containing a $\phi$ in coincidence with a well-identified $e$ or $\mu$ having a CM momentum exceeding 900 MeV (the $\phi$-lepton yield). The $\phi$ candidates are reconstructed in the decay mode $\phi \to K^+K^-$ from pairs of tracks, with the highest probability to be kaons, that are fitted to a common vertex. Yields are determined by binned maximum likelihood fits to the $K^+K^-$ invariant mass distribution of the selected candidates in events with ($\phi$-lepton) and without ($\phi$) a track passing the lepton selection. The *p.d.f.* used to fit the distribution consists of a Voigt profile[174] signal peak on top of a background consisting of the product of a threshold function opening at $2m_{K^\pm}$ and a linear factor. Each yield is normalized to the number of events in the $E_{\rm CM}$ bin which pass criteria identifying them as $e^+e^- \to \mu^+\mu^-$ events.

[174] The convolution of a Breit-Wigner with a Gaussian resolution function.

In order to remove the contribution from continuum $e^+e^- \to q\overline{q}$ events, where $q = u, d, s, c$, *BABAR* performs the same measurements in a 7.89 $\text{fb}^{-1}$ sample of data taken 40 MeV below the $\Upsilon(4S)$ mass and subtracts the off-resonance yields from the yields obtained in each bin. The value to be subtracted is corrected as a function of $E_{\rm CM}$ for the variation of the selection efficiency with energy as deduced from simulation. With continuum contributions to the yields removed, the remaining yield is the sum of contributions from $B\overline{B}$ and $B_s^0\overline{B}_s^0$ events as follows:

$$C_h = R_B\left[f_s\epsilon_h^s + (1-f_s)\epsilon_h\right] \qquad (23.3.1)$$

$$C_\phi = R_B\left[f_s\epsilon_\phi^s P(B_s\overline{B}_s \to \phi X) + (1-f_s)\epsilon_\phi P(B\overline{B} \to \phi X)\right] \qquad (23.3.2)$$

$$C_{\phi\ell} = R_B\left[f_s\epsilon_{\phi\ell}^s P(B_s\overline{B}_s \to \phi\ell X) + (1-f_s)\epsilon_{\phi\ell} P(B\overline{B} \to \phi\ell X)\right], \qquad (23.3.3)$$

where

$$f_s \equiv \frac{N_{B_s}}{N_{B_u} + N_{B_d} + N_{B_s}} \qquad (23.3.4)$$

is the ratio of $B_s^0$ events to all $b$ hadron events, $R_B = \sigma(e^+e^- \to B_{(s)}\overline{B}_{(s)})/\sigma(e^+e^- \to \mu^+\mu^-)$, and the efficiencies $\epsilon_X^{(s)}$ and probabilities $P(B_{(s)}\overline{B}_{(s)} \to \phi(\ell)X)$ are presented schematically. The values of $C_h$, $C_\phi$ and $C_{\phi\ell}$ across the scanned region are shown in Fig. 23.3.1.

The ratio $f_s$ is obtained from the combination of Eqs (23.3.1) and (23.3.2). The probability that a $B\overline{B}$ pair produces a $\phi$ meson scaled by the corresponding efficiency, $\epsilon_\phi P(B\overline{B} \to \phi X)$, is obtained by direct measurement of the event and $\phi$ yields in an $18.55\,\text{fb}^{-1}$ sample of data taken at the $\Upsilon(4S)$ resonance, followed by the application of Eqs (23.3.1) and (23.3.2) with $f_s = 0$. The corresponding probability in $B_s^0$ events is estimated from previously measured branching fractions plus small corrections from estimated $B_s \to c\overline{c}\phi$ and $B_s \to DD_sX$ (followed by $D \to \phi$) rates. The result, presented in 45 MeV wide bins, is seen in Fig. 23.3.2. The ratio, above threshold, is observed to peak around the $\Upsilon(5S)$ mass and is small elsewhere.

In order to extract the semileptonic branching fraction, an estimate of $P(B_s^0\overline{B}_s^0 \to \phi\ell X)$ as a function of $\mathcal{B}(B_s \to \ell\nu X)$ is constructed using known branching fractions plus the estimated ones mentioned earlier. This estimate separately treats leptons from events with one $B_s^0$ semileptonic decay, events with two $B_s^0$ semileptonic decays, and events in which neither $B_s^0$ has decayed semileptonically but instead a lepton from a charmed meson has passed the lepton momentum selection. For the latter *BABAR* includes contributions from events with up to two leptons coming from $D^\pm$, $D^0$, or $D_s^\pm$ decays. Selection efficiencies for each case are determined separately from simulation. Again the $\Upsilon(4S)$ data mentioned earlier are used to obtain $\epsilon_{\phi\ell}P(B\overline{B} \to \phi\ell X)$. With these estimates, *BABAR* constructs a $\chi^2$ from the expected value of $\epsilon_{\phi\ell}P(B_s\overline{B}_s \to \phi\ell X)$, estimated as a function of $\mathcal{B}(B_s \to \ell\nu X)$, and the value obtained by the measured quantities and Eqs (23.3.1) and (23.3.3). Minimization of this $\chi^2$ with respect to the branching fraction yields $\mathcal{B}(B_s \to \ell\nu X) = (9.5^{+2.5}_{-2.0})\%$.

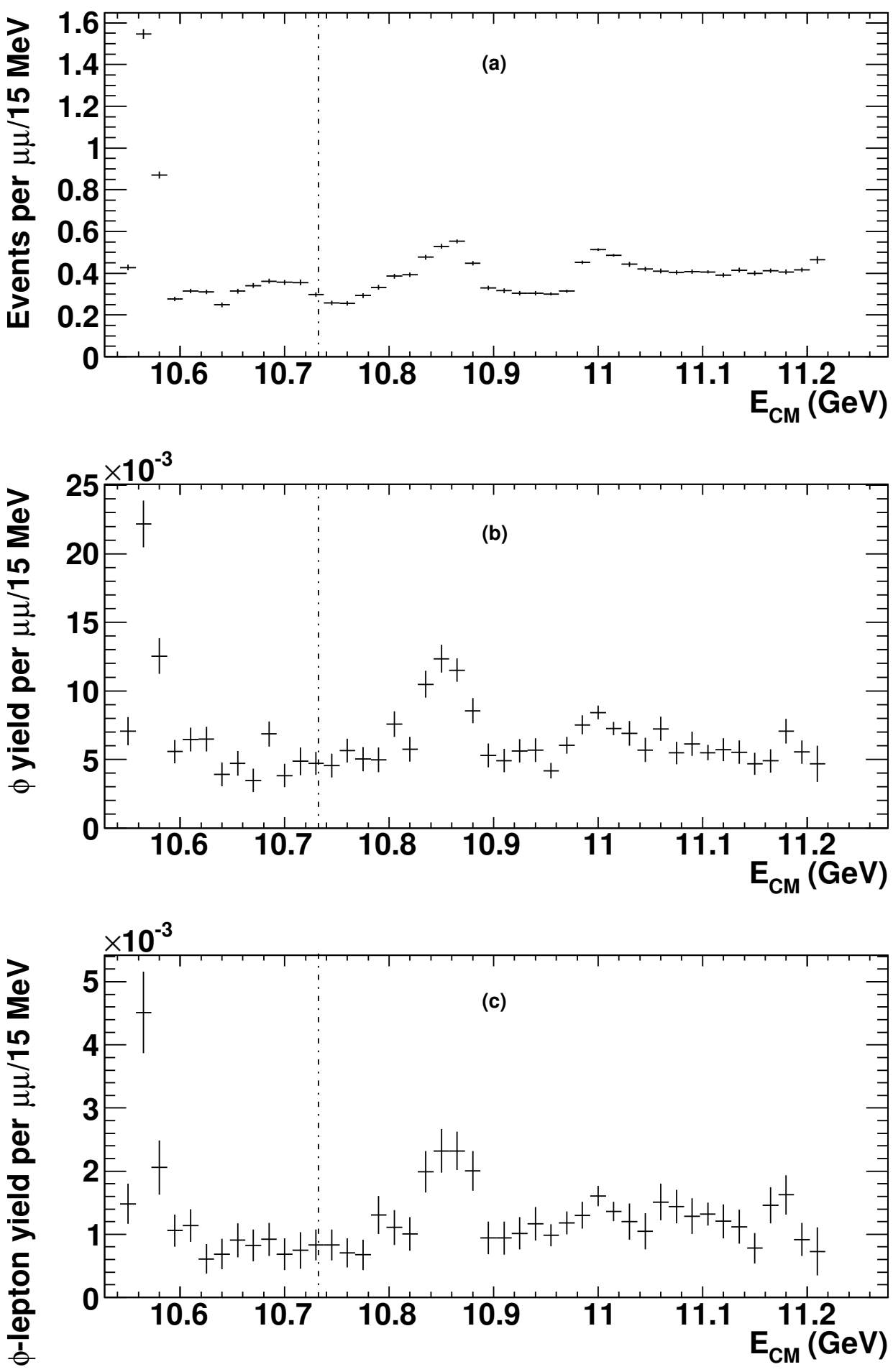

**Figure 23.3.1.** Relative values of the (a) event, (b) $\phi$, and (c) $\phi$-lepton yields after continuum subtraction is performed. Corrections for detector efficiency have not been applied. The $B_s^0$ production threshold is located at the dotted line. From Lees (2012a).

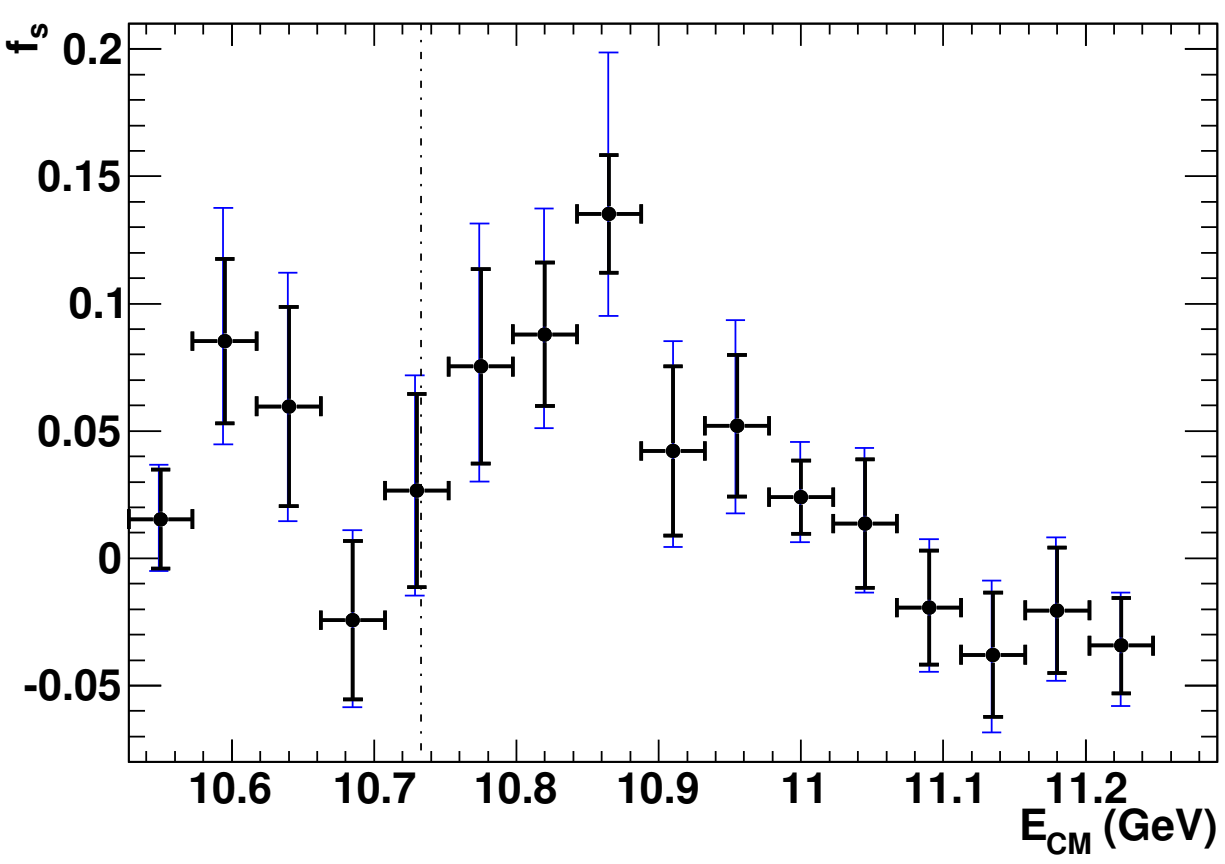

**Figure 23.3.2.** $f_s$ in 45 MeV wide bins of $E_{\rm CM}$. The larger blue error bars represent the sum in quadrature of statistical and systematic uncertainties, while the inner ones show the statistical uncertainty alone. The broken line denotes the $B_s^0$ threshold. From Lees (2012a).

The dominant contribution to the systematic uncertainty comes from the poor knowledge of the inclusive rate $\mathcal{B}(B_s \to D_s X)$ ($^{+8.72}_{-13.58}$ events). Other sources include uncertainties due to biases in the technique found using ensembles of simulated experiments ($^{+0.39}_{-10.00}$ events), see Section 11.5.2; the impact of neglecting ISR and two-photon contributions to the event subtraction ($^{+1.57}_{-7.14}$ events); estimated branching fractions used ($\pm 3.4$ events); uncertainties due to the use of particle identification ($\pm 3.21$ events); statistical uncertainties in the quantities determined from $\Upsilon(4S)$ data and daughter branching fractions ($\pm 3.1$ events); uncertainties in the selection efficiency obtained by variation of the $R_2$ and lepton momentum requirements ($^{+1.99}_{-2.85}$ events); sensitivity to the background parameterization and possible presence of scalar contributions in the threshold region ($\pm 0.93$ events); the uncertainty in other world-average branching fractions used ($^{+0.52}_{-0.54}$ events); and fixed parameters used in the fit to the $K^+K^-$ invariant mass distribution ($^{+0.49}_{-0.15}$ events).

With systematic uncertainties included, the final result is found to be $\mathcal{B}(B_s \to \ell\nu X) = (9.5^{+2.5+1.1}_{-2.0-1.9})\%$, in good agreement with expectations from spectator processes that would predict a similar branching fraction for the $B^0$ and the $B_s^0$. A more complete overview may be found in Lees (2012a).

### 23.3.2 Cabibbo favored decays $B_s^0 \to D_s^{(*)-}\pi^+(\rho^+)$

The $B_s^0 \to D_s^{(*)-}\pi^+$ and $B_s^0 \to D_s^{(*)-}\rho^+$ decays were the first processes studied by Belle at the $\Upsilon(5S)$. These decays are described by Cabibbo-favored tree diagrams, which have no suppression factors; therefore, relatively large branching fractions are expected for these decays. The decay $B_s^0 \to D_s^-\pi^+$ was previously observed and studied with high statistics at the Tevatron experiments. However to identify the other three modes the neutral particles ($\gamma$ from $D_s^{*-}$ and $\pi^0$ from $\rho$) have to be reconstructed, which was not possible at other experiments; LHCb was not in operation at that time. These decay modes were the primary goals for observation at Belle.

Although the theoretical models used to describe these decays are relatively simple, any experimental results on these modes are welcome because they test the applicability and accuracy of the basic approaches of $B$ meson theoretical calculations. In particular these decays are well suited to test heavy-quark theories that predict, based on $SU(3)$ symmetry, similarities between the decay parameters of $B_s^0$ mesons and their corresponding $B^0$ counterparts. These include the unitarized quark model (Tornqvist, 1984), heavy quark effective theory (HQET; Deandrea, Di Bartolomeo, Gatto, and Nardulli, 1993), and a more recent approach based on chiral symmetry (Bardeen, Eichten, and Hill, 2003).

Moreover, because of their large branching fractions, these decays can be used to measure parameters in $\Upsilon(5S)$ decays, such as the masses of the $B_s^0$ and $B_s^{*0}$ mesons and the relative fractions of different $\Upsilon(5S)$ decay channels. Precisely measured branching fractions of these decay modes can also be used as a primary normalization at hadron colliders, where the absolute branching fractions are difficult to measure.

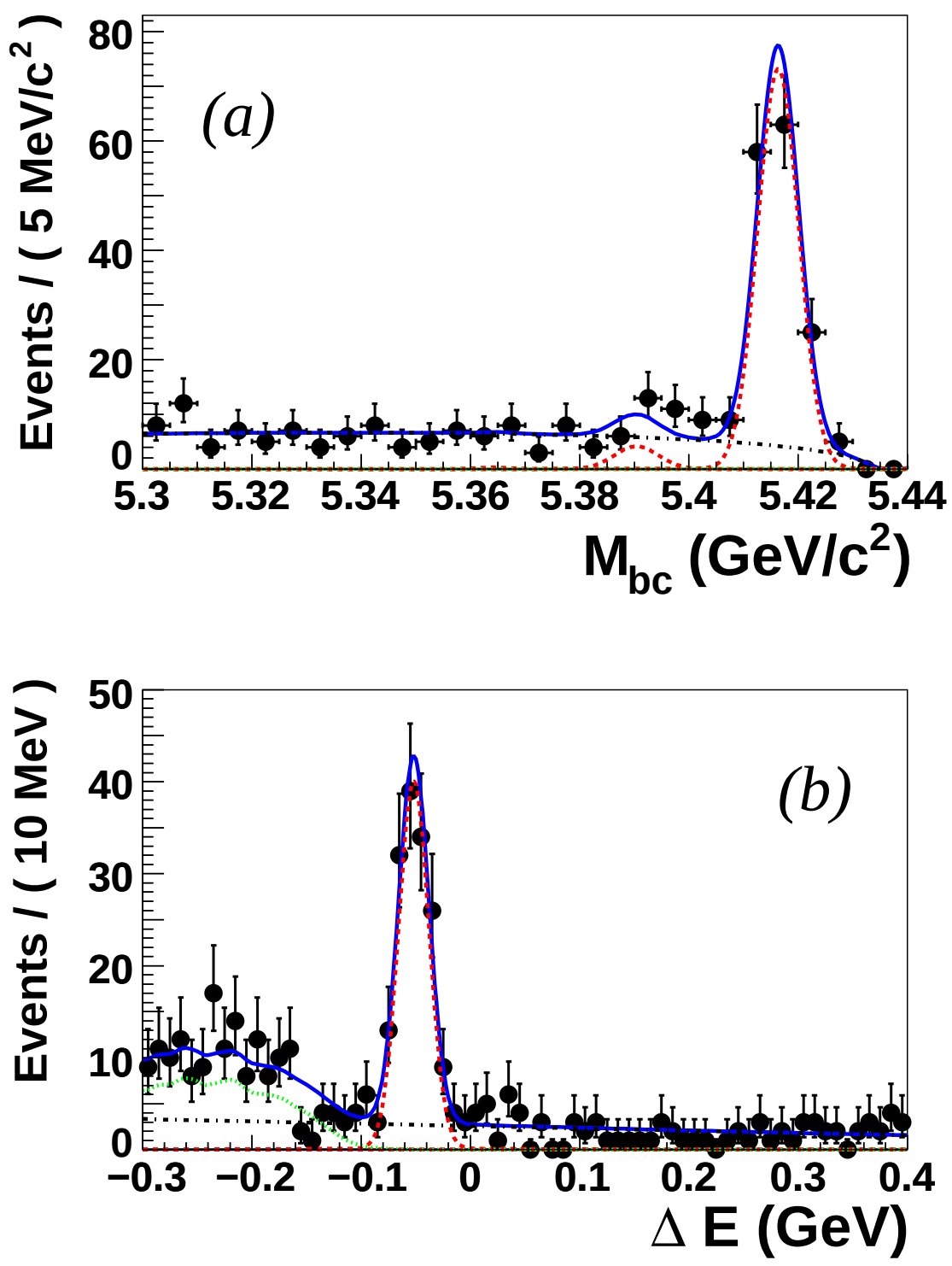

**Figure 23.3.3.** (a) $M_{\rm bc} = m_{\rm ES}$ distribution of the $B_s^0 \to D_s^- \pi^+$ candidates with $\Delta E$ in the $B_s^{*0}\overline{B}{}_s^{*0}$ signal region $[-80, -17]$ MeV. (b) $\Delta E$ distribution of the $B_s^0 \to D_s^- \pi^+$ candidates with $M_{\rm bc}$ in the $B_s^{*0}\overline{B}{}_s^{*0}$ signal region $[5.41, 5.43]$ GeV/$c^2$. The different fitted components are shown with dashed curves for the signal, dotted curves for the $B_s^0 \to D_s^{*-}\pi^+$ background, and dash-dotted curves for the continuum. The total fit is shown as a solid line. Plots are from Louvot (2009).

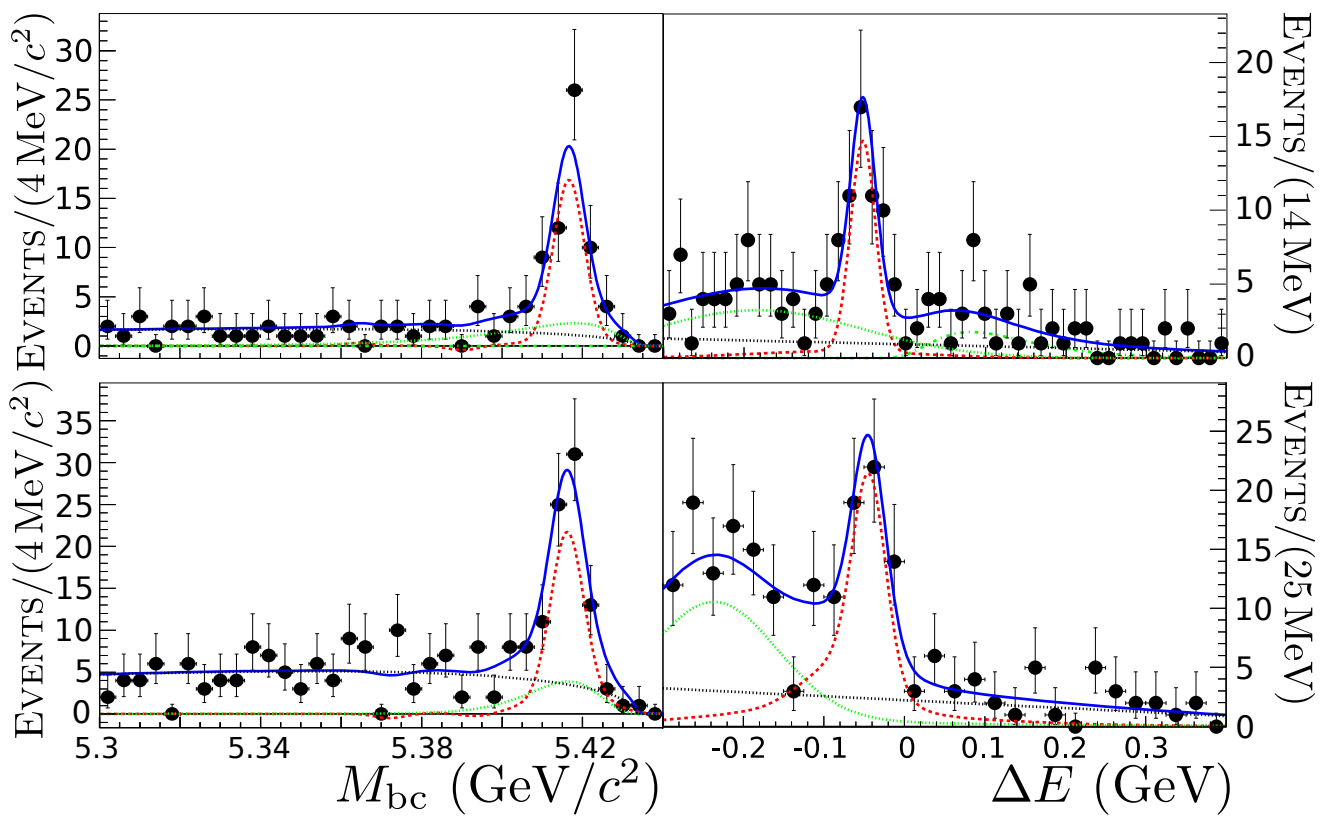

**Figure 23.3.4.** Left (right): $M_{\rm bc} = m_{\rm ES}$ ($\Delta E$) distributions for $B_s^0 \to D_s^{*-}\pi^+$ (top) and $B_s^0 \to D_s^- \rho^+$ (bottom) candidates with $\Delta E$ ($m_{\rm ES}$) restricted to the $\pm 2.5\sigma$ $B_s^{*0}\overline{B}{}_s^{*0}$ signal region. The blue solid curve is the total fitted *p.d.f.*, while the green (black) dotted curve is the peaking (continuum) background and the red dashed curve is the signal. Plots are from Louvot (2010).

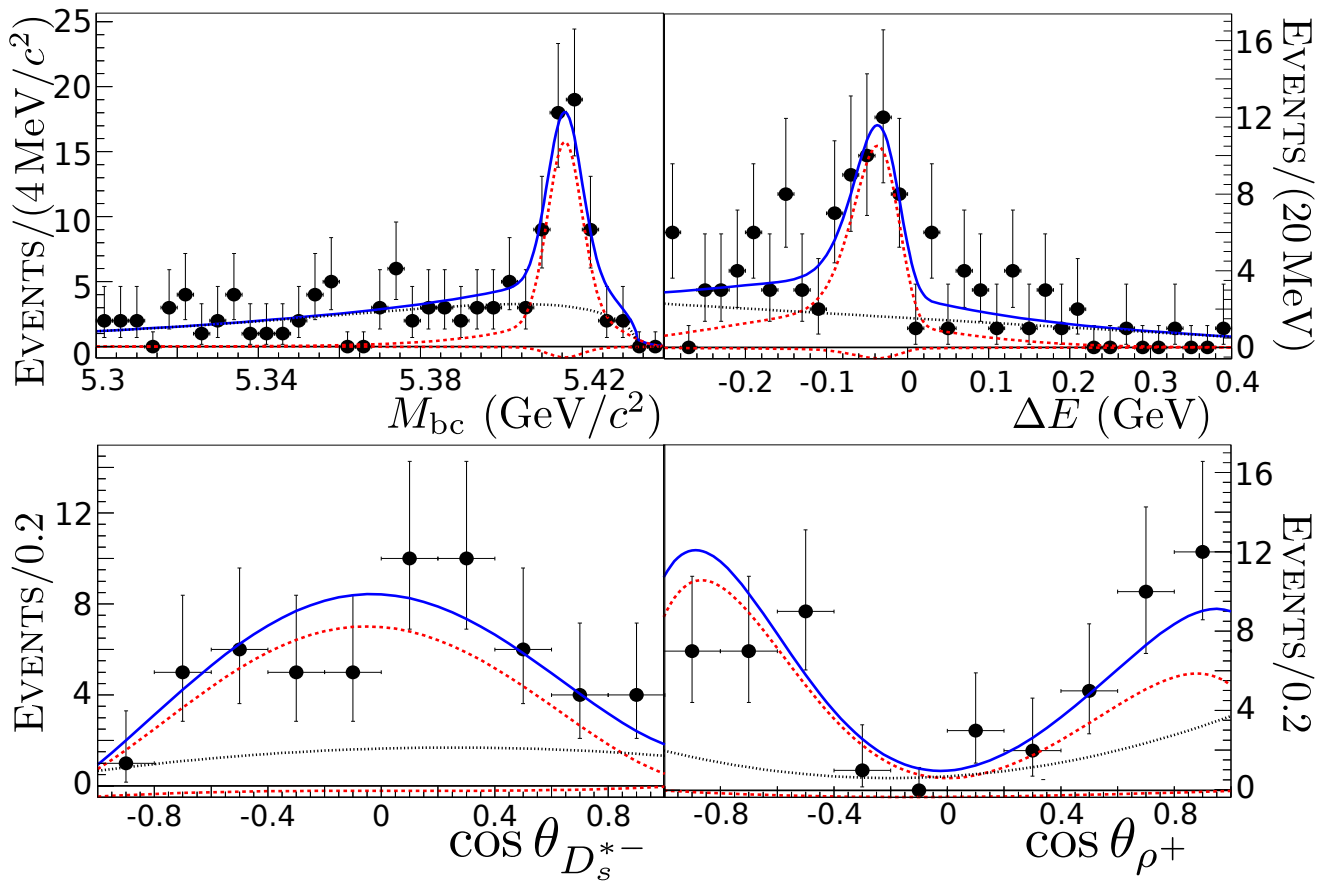

**Figure 23.3.5.** Distributions for the $B_s^0 \to D_s^{*-}\rho^+$ candidates. Top: $M_{\rm bc} = m_{\rm ES}$ and $\Delta E$ distributions, as in the previous figure. Bottom: distributions of the cosine of the helicity angles of the $D_s^{*-}$ (left) and $\rho^+$ (right) with $m_{\rm ES}$ and $\Delta E$ restricted to the $B_s^{*0}\overline{B}{}_s^{*0}$ kinematic region. The components of the total *p.d.f.* (blue solid line) are shown separately: the black-dotted curve is the background and the two red-dashed curves are the signal. The large (small) signal component corresponds to the longitudinal (transverse) signal. Plots are from Louvot (2010).

The $B_s^0 \to D_s^{(*)-}\pi^+$ and $B_s^0 \to D_s^{(*)-}\rho^+$ decays are observed and studied by Belle using 23.6 fb$^{-1}$ of data collected at the $\Upsilon(5S)$ resonance CM energy region (Louvot, 2009, 2010). Evidence for the $B_s^0 \to D_s^{(*)\mp}K^\pm$ decay is also found, at the level of $3.5\sigma$, in these studies. The technique used to identify $B_s^0$ signals is described in Section 23.2 and only final results are discussed here. Details of the analyses are found in the original papers.

Two-dimensional unbinned extended maximum likelihood fits are applied to obtain the $B_s^0 \to D_s^{(*)-}\pi^+$ and $B_s^0 \to D_s^{(*)-}\rho^+$ decay branching fractions. The $m_{\rm ES}$ and $\Delta E$ projections of the two-dimensional distribution for the decay $B_s^0 \to D_s^-\pi^+$ are shown in Fig. 23.3.3 together with the fit projections. Five additional parameters were measured using this decay mode: the fractions of the $B_s^0$ pair production modes at the $\Upsilon(5S)$ energy, $f_{B_s^{*0}\overline{B}{}_s^{*0}} = \left(90.1^{+3.8}_{-4.0} \pm 0.2\right)\%$, $f_{B_s^{*0}\overline{B}{}_s^0} = \left(7.3^{+3.3}_{-3.0} \pm 0.1\right)\%$, $f_{B_s^0\overline{B}{}_s^0} = \left(2.6^{+2.6}_{-2.5}\right)\%$, and the masses $m_{B_s^{*0}} = (5416.4 \pm 0.4 \pm 0.5)$ MeV/$c^2$ and $m_{B_s^0} = (5364.4 \pm 1.3 \pm 0.7)$ MeV/$c^2$. The $\Upsilon(5S) \to B_s^{*0}\overline{B}{}_s^{*0}$ channel fraction is found to be large and this channel strongly dominates over other channels.

A similar method is used to extract the branching fractions for the other three decay modes. The $m_{\rm ES}$ and $\Delta E$ projections of the two-dimensional distributions for the decays $B_s^0 \to D_s^{*-}\pi^+$ and $B_s^0 \to D_s^-\rho^+$ are shown in Fig. 23.3.4. The $m_{\rm ES}$ and $\Delta E$ projections and helicity dis-tributions for the $D_s^{*-}$ and $\rho$ are shown in Fig. 23.3.5 for the decay $B_s^0 \to D_s^{*-}\rho^+$. As the latter decay is that of a $B_s^0$ meson decaying into a final state with two vector particles, one can perform an angular analysis of the final state in terms of the helicity angles (and angle between decay planes) of the intermediate vector particles as described in Chapter 12. A simplified angular analysis (integrating over the angle between the two decay planes) of $B_s^0 \to D_s^{*-}\rho^+$ indicates strong dominance of the longitudinal polarization: the obtained fraction of this component is $f_L = 1.05^{+0.08}_{-0.10}(\text{stat})^{+0.03}_{-0.04}(\text{syst})$. This result is compatible with expectations from HQET and naïve factorization computed for some $B_{u,d}$ to two-vector particle final states (see Sections 17.3 and 17.4 for discussions of such states). All branching fractions measured and their ratios are summarized in Table 23.3.1. The results obtained are consistent with theoretical predictions based on HQET (Deandrea, Di Bartolomeo, Gatto, and Nardulli, 1993) and are similar to the corresponding $B^0$ decay branching fractions.

**Table 23.3.1.** Top: measured branching fractions with statistical, systematic (without $f_s$), and $f_s$ uncertainties, and HQET predictions from the factorization hypothesis (Deandrea, Di Bartolomeo, Gatto, and Nardulli, 1993). Bottom: branching fraction ratios where several systematic uncertainties cancel out.

| Mode | $\mathcal{B}$ $(10^{-3})$ | $\mathcal{B}_{\rm HQET}$ $(10^{-3})$ |
|---|---|---|
| $B_s^0 \to D_s^- \pi^+$ | $3.67^{+0.35+0.43}_{-0.33-0.42} \pm 0.49$ | 2.8 |
| $B_s^0 \to D_s^{*-} \pi^+$ | $2.4^{+0.5}_{-0.4} \pm 0.3 \pm 0.4$ | 2.8 |
| $B_s^0 \to D_s^- \rho^+$ | $8.5^{+1.3}_{-1.2} \pm 1.1 \pm 1.3$ | 7.5 |
| $B_s^0 \to D_s^{*-} \rho^+$ | $11.8^{+2.2}_{-2.0} \pm 1.7 \pm 1.8$ | 8.9 |
| $B_s^0 \to D_s^{\mp} K^{\pm}$ | $0.24^{+0.12}_{-0.10} \pm 0.03 \pm 0.03$ | |
| Ratios | | |
| $\mathcal{B}(B_s^0 \to D_s^{*-}\pi^+)/\mathcal{B}(B_s^0 \to D_s^-\pi^+) = 0.65^{+0.15}_{-0.13} \pm 0.07$ | | |
| $\mathcal{B}(B_s^0 \to D_s^-\rho^+)/\mathcal{B}(B_s^0 \to D_s^-\pi^+) = 2.3 \pm 0.4 \pm 0.2$ | | |
| $\mathcal{B}(B_s^0 \to D_s^{*-}\rho^+)/\mathcal{B}(B_s^0 \to D_s^-\pi^+) = 3.2 \pm 0.6 \pm 0.3$ | | |
| $\mathcal{B}(B_s^0 \to D_s^{*-}\rho^+)/\mathcal{B}(B_s^0 \to D_s^-\rho^+) = 1.4 \pm 0.3 \pm 0.1$ | | |

### 23.3.3 Cabibbo favored decays $B_s^0 \to D_s^{(*)+} D_s^{(*)-}$

Belle has also used $\Upsilon(5S)$ data to measure $B_s^0 \to D_s^{(*)+} D_s^{(*)-}$ decays. An initial study was done using $23.6\,\text{fb}^{-1}$ of data (Esen, 2010), while a subsequent study uses the full $121.4\,\text{fb}^{-1}$ data set (Esen, 2013). The final states reconstructed consist of $D_{(s)}^+ D_s^-$, $D_s^{*+} D_s^- + D_s^{*-} D_{(s)}^+$ ($\equiv D_s^{*\pm} D_s^{\mp}$), and $D_s^{*+} D_s^{*-}$. These are expected to be mostly *CP*-even, and their partial widths are expected to dominate the difference in widths between the two $B_s^0$ *CP* eigenstates, $\Delta\Gamma_s^{CP}$ (Aleksan, Le Yaouanc, Oliver, Pène, and Raynal, 1993). This parameter is equal to $\Delta\Gamma_s/\cos\phi_{12}$, where $\Delta\Gamma_s$ is the decay width difference between the mass eigenstates, and $\phi_{12}$ is the *CP*-violating phase in $B_s^0$-$\overline{B}{}_s^0$ mixing.[175] Thus the branching fraction gives a constraint in the $\Delta\Gamma_s$-$\phi_{12}$ parameter space. Both parameters can receive contributions from new physics (Buras, Carlucci, Gori, and Isidori, 2010; Lenz and Nierste, 2011; Ligeti, Papucci, Perez, and Zupan, 2010).

The decays $B_s^0 \to D_{(s)}^+ D_s^-$, $D_s^{*\pm} D_s^{\mp}$, and $D_s^{*+} D_s^{*-}$ are reconstructed via $D_{(s)}^+ \to \phi\pi^+$, $K_S^0 K^+$, $\overline{K}{}^{*0} K^+$, $\phi\rho^+$, $K_S^0 K^{*+}$, and $\overline{K}{}^{*0} K^{*+}$; charge-conjugate modes are implicitly included. The daughter mesons are reconstructed via $K_S^0 \to \pi^+\pi^-$, $K^{*0} \to K^+\pi^-$, $K^{*+} \to K_S^0\pi^+$, $\phi \to K^+K^-$, $\rho^+ \to \pi^+\pi^0$, and $\pi^0 \to \gamma\gamma$. For the three vector-pseudoscalar final states, it is required that $|\cos\theta_{\rm hel}| > 0.20$, where $\theta_{\rm hel}$ is the angle between the momentum of the charged daughter of the vector particle and the direction opposite the $D_s^+$ momentum, evaluated in the rest frame of the vector particle.

Belle combines $D_{(s)}^+$ candidates with photon candidates to reconstruct $D_s^{*+} \to D_{(s)}^+\gamma$ decays. The mass difference $M_{D_s^+\gamma} - M_{D_s^+}$ must be within 12.0 MeV/$c^2$ of the nominal value. Events are required to satisfy $5.25\,\text{GeV}/c^2 < m_{\rm ES} < 5.45\,\text{GeV}/c^2$ and $-0.15\,\text{GeV} < \Delta E < 0.10$ GeV. Only small contributions from $B_s^0\overline{B}{}_s^0$ and $B_s^0\overline{B}{}_s^{*0}$ events are expected, and these contributions are fixed relative to $B_s^{*0}\overline{B}{}_s^{*0}$ according to the Belle measurement of $B_s^0 \to D_s^-\pi^+$ decays (Louvot, 2009). Belle quotes fitted signal yields from $B_s^{*0}\overline{B}{}_s^{*0}$ only and uses these to determine the branching fractions. Approximately half of the selected events have multiple $B_s^0 \to D_s^{(*)+} D_s^{(*)-}$ candidates. These typically arise from photons produced via $\pi^0 \to \gamma\gamma$ that are wrongly assigned as $D_s^{*+}$ daughters. For these events the candidate that minimizes a $\chi^2$ constructed from the reconstructed $D_s^+$ and (if present) $D_s^{*+}$ masses is selected.

Background from $e^+e^- \to q\overline{q}$ ($q = u, d, s, c$) is rejected by using a Fisher discriminant based on a set of modified Fox-Wolfram moments (see Section 9.5). The remaining background consists of $\Upsilon(5S) \to B_s^{(*)}\overline{B}{}_s^{(*)} \to D_s^+ X$, $\Upsilon(5S) \to B\overline{B}X$ ($b\overline{b}$ hadronizes to $B^0$, $\overline{B}{}^0$, or $B^\pm$), and $B_s^0 \to D_{sJ}^{\pm}(2317) D_s^{(*)}$, $D_{sJ}^{\pm}(2460) D_s^{(*)}$, or $D_s^{\pm} D_s^{\mp}\pi^0$. The last three processes peak at negative $\Delta E$, and their yields are estimated to be small using analogous $B_d^0 \to D_{sJ}^{\pm} D^{(*)}$ branching fractions. They are considered only when evaluating the systematic uncertainty due to backgrounds.

The signal yields are determined via a two-dimensional unbinned maximum-likelihood fit to the $m_{\rm ES}$-$\Delta E$ distributions. The signal *p.d.f.*s have components for correctly reconstructed decays, "wrong combination" decays in which a non-signal track or $\gamma$ is included, and "cross-feed" decays in which a $D_s^{*\pm} D_s^{\mp}$ ($D_s^{*+} D_s^{*-}$) is reconstructed as a $D_s^+ D_s^-$ ($D_s^+ D_s^-$ or $D_s^{*\pm} D_s^{\mp}$), or a $D_s^+ D_s^-$ ($D_s^{*\pm} D_s^{\mp}$) is reconstructed as a $D_s^{*\pm} D_s^{\mp}$ or $D_s^{*+} D_s^{*-}$ ($D_s^{*+} D_s^{*-}$). All signal shape parameters are taken from the MC simulation and calibrated using $B_s^0 \to D_s^{(*)-}\pi^+$ and $B^0 \to D_s^{(*)+}D^-$ decays. The fractions of wrong combination signal and cross-feed down events are taken from MC (see Table 23.3.2); the fractions of cross-feed up events are floated as they are difficult to simulate accurately (many $B_s^0$ partial widths are unmeasured).[176] As the cross-feed down fractions are fixed, the separate $D_s^+D_s^-$, $D_s^{*\pm}D_s^{\mp}$, and $D_s^{*+}D_s^{*-}$ samples are fitted simultaneously.

[175] Specifically, $\phi_{12} = \arg(-M_{12}/\Gamma_{12})$, where $M_{12}$ and $\Gamma_{12}$ are the off-diagonal elements of the $B_s^0$-$\overline{B}{}_s^0$ mass and decay matrices. Also see Chapter 10.

**Table 23.3.2.** The fractions of $B_s^0$ signal events in % (from MC simulation) reconstructed as correctly reconstructed (CR) and wrong combination (WC) signal, and cross-feed events. Cross-feed up and down are denoted by † and ‡, respectively.

| Mode | CR | WC | Cross-feed |
|---|---|---|---|
| $D_s^+D_s^-$ | 76.1 | 6.0 | 17.1 $(D_s^{*\pm}D_s^{\mp})^‡$; 0.8 $(D_s^{*+}D_s^{*-})^‡$ |
| $D_s^{*\pm}D_s^{\mp}$ | 44.4 | 38.5 | 8.2 $(D_s^+D_s^-)^†$; 8.9 $(D_s^{*+}D_s^{*-})^‡$ |
| $D_s^{*+}D_s^{*-}$ | 31.8 | 37.6 | 2.0 $(D_s^+D_s^-)^†$; 28.6 $(D_s^{*\pm}D_s^{\mp})^†$ |

**Table 23.3.3.** $B_s^{*0}\overline{B}_s^{*0}$ correctly reconstructed signal yield ($Y$) and efficiency ($\varepsilon$), including intermediate branching fractions, and resulting branching fraction ($\mathcal{B}$). The first uncertainties listed are statistical; the others are systematic. The last error for the sum is due to external factors ($\Upsilon(5S) \to B_s^{*0}\overline{B}_s^{*0}$ and $D_s^+$ branching fractions).

| Mode | $Y$ (events) | $\varepsilon$ $(\times 10^{-4})$ | $\mathcal{B}$ (%) |
|---|---|---|---|
| $D_{(s)}^+D_s^-$ | $33.1^{+6.0}_{-5.4}$ | 4.72 | $0.58\,^{+0.11}_{-0.09} \pm 0.13$ |
| $D_s^{*\pm}D_s^{\mp}$ | $44.5^{+5.8}_{-5.5}$ | 2.08 | $1.76\,^{+0.23}_{-0.22} \pm 0.40$ |
| $D_s^{*+}D_s^{*-}$ | $24.4^{+4.1}_{-3.8}$ | 1.01 | $1.98\,^{+0.33}_{-0.31}\,^{+0.52}_{-0.50}$ |
| Sum | $102.0^{+9.3}_{-8.6}$ | | $4.32\,^{+0.42}_{-0.39}\,^{+0.56}_{-0.54} \pm 0.88$ |

The projections of the fit are shown in Fig. 23.3.6. The fitted correctly reconstructed signal yields are listed in Table 23.3.3 along with signal efficiencies (including intermediate branching fractions from Nakamura et al., 2010) and the resulting branching fractions. The significance is calculated as $\sqrt{2\ln(\mathcal{L}_{\rm max}/\mathcal{L}_0)}$, where $\mathcal{L}_{\rm max}$ and $\mathcal{L}_0$ are the likelihood values when the signal yield is floated and when it is set to zero, respectively. Systematic uncertainties on the yield are included in the significance by smearing the likelihood function by a Gaussian distribution with width corresponding to the total additive systematic uncertainty. The significance computed for $D_{(s)}^+D_s^-$ , $D_s^{*\pm}D_s^{\mp}$, and $D_s^{*+}D_s^{*-}$, is $11.5\sigma$, $10.1\sigma$ and $7.8\sigma$, respectively.

The systematic uncertainties are dominated by factors external to the analysis, *i.e.*, $f_s$ (18%) and the $D_{(s)}^+$ branching fractions (8.6%). The uncertainty due to the wrong combination and cross-feed fractions, which are taken from the MC simulation, is $\sim 4.6\%$ for $D_s^{*\pm}D_s^{\mp}$ and $\sim 10\%$ for $D_s^{*+}D_s^{*-}$. The fraction of longitudinal $D_s^{*+}D_s^{*-}$ polarization $f_L$ for this measurement is taken to be the value from the analogous decay $B_d^0 \to D_s^{*+}D^{*-}$: $0.52\pm 0.05$ (Beringer et al., 2012). The systematic error is taken to be the change in signal yield when $f_L$ is varied over a wide range: from $2\sigma$ higher than 0.52 down to the low central value measured by Belle (see below). This error is less than three events on each of the modes.

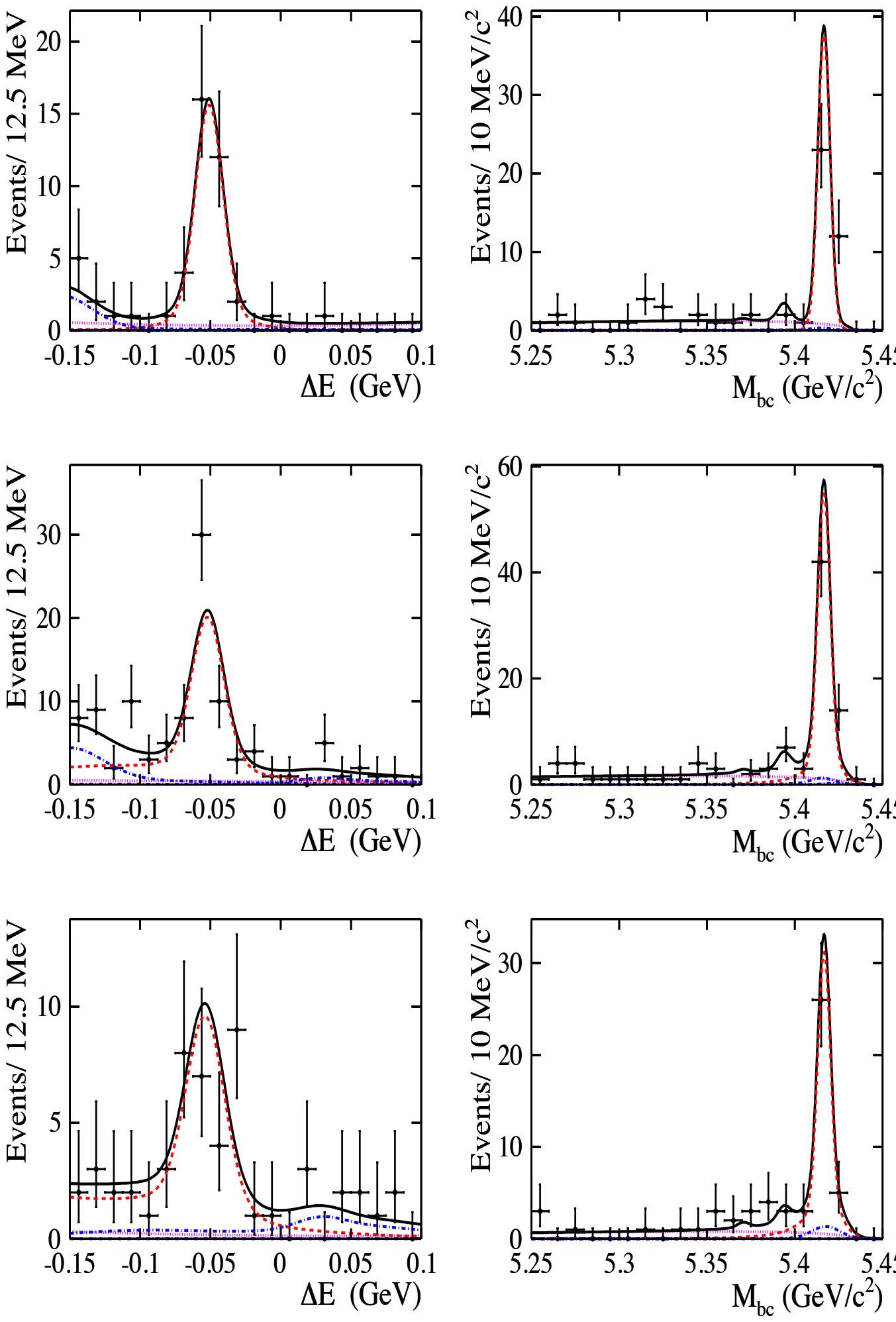

**Figure 23.3.6.** (left) $\Delta E$ fit projections for events satisfying $M_{\rm bc} = m_{\rm ES} \in [5.41, 5.43]\,{\rm GeV}/c^2$, and (right) $m_{\rm ES}$ fit projections for events satisfying $\Delta E \in [-0.08, -0.02]\,{\rm GeV}$. The top row shows $B_s^0 \to D_{(s)}^+D_s^-$; the middle row shows $B_s^0 \to D_s^{*\pm}D_s^{\mp}$; and the bottom row shows $B_s^0 \to D_s^{*+}D_s^{*-}$. The red dashed curves show correctly reconstructed and wrong combination signal; the blue dash-dotted curves show cross-feed; the magenta dotted curves show background; and the black solid curves show the total. Plots are from Esen (2013).

In the limits of $m_{b,c} \to \infty$ with $(m_b - 2m_c) \to 0$ and $N_c$ (number of colors)$\to \infty$, the $D_s^{*\pm}D_s^{\mp}$ and $D_s^{*+}D_s^{*-}$ modes are *CP*-even and (along with $D_s^+D_s^-$) saturate the width difference $\Delta\Gamma_s^{CP}$ (Aleksan, Le Yaouanc, Oliver, Pène, and Raynal, 1993). Assuming negligible *CP* violation ($\phi_s \approx 0$), the branching fraction is related to $\Delta\Gamma_s$ via $\Delta\Gamma_s/\Gamma_s =$

[176] Cross-feed up/down corresponds to a background event type that is mis-reconstructed so as to move upward/downward in $\Delta E$ in order to overlap with signal. For example a $D_s^+D_s^-$ final state would have to be mis-reconstructed, including an extra photon, in order to be reconstructed under the $D_s^{*+}D_s^-$ peak in $\Delta E$. This is an example of cross-feed up.

$2\mathcal{B}/(1-\mathcal{B})$. Inserting the total $\mathcal{B}$ from Table 23.3.3 gives $\Delta\Gamma_s/\Gamma_s = 0.090 \pm 0.009(\text{stat}) \pm 0.023(\text{syst})$, which is consistent with the HFAG average (Asner et al., 2011). This result has similar precision to that of recent measurements (Aaij et al., 2012h; Aaltonen et al., 2012b). The central value is consistent with, but lower than, the theoretical prediction (Lenz and Nierste, 2011); the difference may be due to the unknown *CP*-odd component in $B_s^0 \to D_s^{*+}D_s^{*-}$, and contributions from three-body final states. The former is estimated to be only 6% for analogous $B^0 \to D_s^{*+}D^{*-}$ decays (Rosner, 1990), but the latter are expected to be significant: Chua, Hou, and Shen (2011) calculate $\Delta\Gamma(B_s^0 \to D_s^{(*)}D^{(*)}K^{(*)})/\Gamma_s = 0.064 \pm 0.047$. This calculation predicts $\Delta\Gamma_s/\Gamma_s$ from $D_s^{(*)+}D_s^{(*)-}$ alone to be $0.102 \pm 0.030$, which agrees well with the Belle result.

In addition to measuring the branching fractions, Belle also measures the longitudinal polarization fraction ($f_L$) of $B_s^0 \to D_s^{*+}D_s^{*-}$. To measure $f_L$, Belle performs an unbinned maximum likelihood fit to the cosine of the helicity angles $\theta_1$ and $\theta_2$, where $\theta_{1,2}$ are the angles between the daughter $\gamma$ momentum and the direction opposite to the $B_s^0$ momentum in the $D_s^{*+}$ and $D_s^{*-}$ rest frames, respectively. The angular distribution is $\left(|A_+|^2 + |A_-|^2\right)\left(\cos^2\theta_1 + 1\right)\left(\cos^2\theta_2 + 1\right) + |A_0|^2 4\sin^2\theta_1 \sin^2\theta_2$, where $A_+$, $A_-$, and $A_0$ are the three polarization amplitudes in the helicity basis (see Chapter 12). The fraction $f_L$ equals $|A_0|^2/(|A_0|^2 + |A_+|^2 + |A_-|^2)$. To account for resolution and efficiency variation, the signal *p.d.f.*s are taken from MC. The result is

$$f_L = 0.06\,^{+0.18}_{-0.17} \pm 0.03\,, \qquad (23.3.5)$$

where the systematic errors are dominated by the fixed wrong combination signal fractions (+0.013, −0.015) and the fixed background level (±0.022). The helicity angle distributions and fit projections are shown in Fig. 23.3.7.

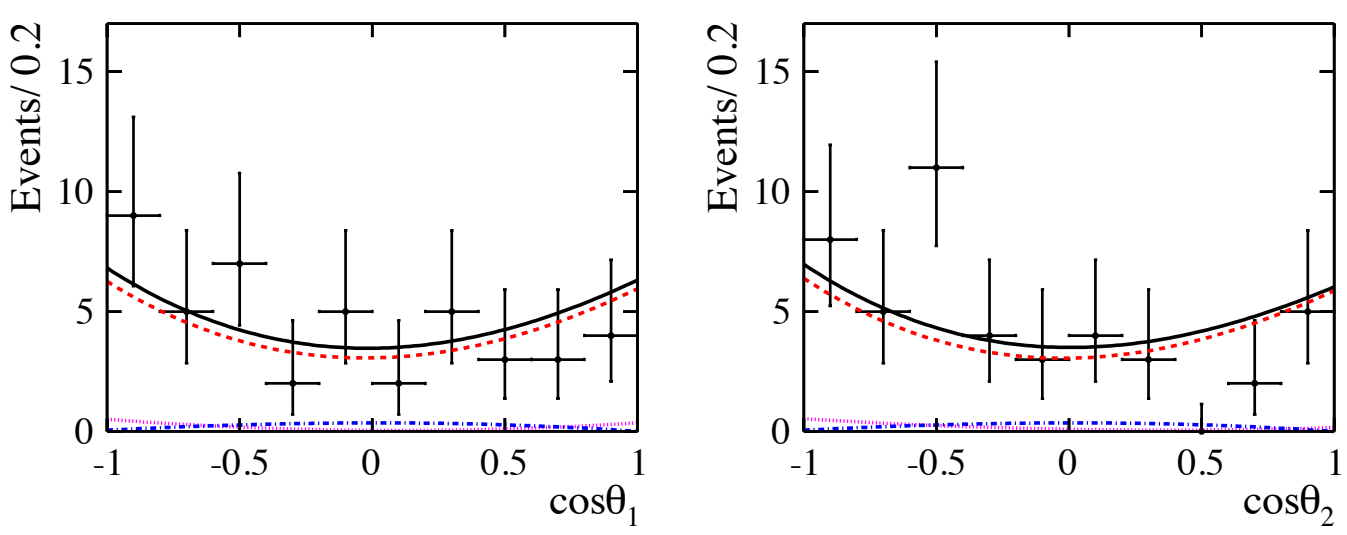

**Figure 23.3.7.** Helicity angle distributions and projections of the fit result for $B_s^0 \to D_s^{*+}D_s^{*-}$. The red dashed (blue dash-dotted) curves show the transverse (longitudinal) components; the magenta dotted curves show background; and the black solid curves show the total. Plots are from Esen (2013).

### 23.3.4 Color suppressed decays $B_s^0 \to J/\psi\eta^{(\prime)}$ and $B_s^0 \to J/\psi f_0(980)$

Belle uses $121.4\,\text{fb}^{-1}$ of data accumulated at the $\Upsilon(5S)$ resonance to search for, and observe, exclusive color-suppressed $B_s^0$ decays with an underlying $b \to c\bar{c}s$ quark transition. Among these, the pure *CP*-eigenstate final state decays $B_s^0 \to J/\psi f_0(980)$ and $B_s^0 \to J/\psi\eta^{(\prime)}$ are of special interest. These decays, which were not yet observed, could be used in the future to study time-dependent *CP* asymmetries. In addition the ratio of the $B_s^0 \to J/\psi\eta'$ and $J/\psi\eta$ branching fractions constitutes a test of $\eta - \eta'$ mixing.

The study of the $B_s^0 \to J/\psi f_0(980)$ decay is described in detail in Li (2011). The reconstruction of the decay mode $B_s^0 \to J/\psi f_0(980)$ includes $f_0(980)$ decaying into two charged pions. The main backgrounds are from the inclusive decays $B_s^0$, $B_d^0$, and $B^+ \to J/\psi X$, where the contributions from $B_s^0 \to J/\psi\eta'$ and $B^+ \to J/\psi(K^+,\pi^+)$ are modeled exclusively and set to their known branching fractions. Since the $f_0(980)$ resonance is wide, $\sim 60\,\text{MeV}/c^2$, we need to describe the $M_{\pi\pi}$ spectrum using the Flatté formula with a phase-space factor, and the $f_0(1370)$ lineshape is described using a relativistic Breit-Wigner. Both of these line shapes are described in Section 13.2.1. A two-dimensional fit with the variables $\Delta E$ and $M_{\pi\pi}$ is performed to extract the signal yield, where the $m_{\text{ES}}$ signal region corresponding to the $B_s^{*0}\overline{B}_s^{*0}$ channel was chosen. The data surprisingly also show an enhancement around $M_{\pi\pi} \sim 1400\,\text{MeV}/c^2$, where the $\Delta E$ distribution is also strongly peaked (see Fig. 23.3.8). Thus Belle includes the contribution from the $f_0(1370)$ resonance coherently with the $f_0(980)$ resonance in the $M_{\pi\pi}$ signal fit model.

Studies of the decay modes $B_s^0 \to J/\psi\eta$ and $J/\psi\eta'$ are described in (Li, 2012). Five $\eta$ and $\eta'$ sub-channels are reconstructed, $\eta \to \gamma\gamma$, $\eta \to \pi^+\pi^-\pi^0$, $\eta' \to \eta(\gamma\gamma)\pi^+\pi^-$, $\eta' \to \eta(\pi^+\pi^-\pi^0)\pi^+\pi^-$ and $\eta' \to \rho^0\gamma$. To use as much information as possible, a simultaneous fit to the two-dimensional $\Delta E$ - $m_{\text{ES}}$ distributions of all five sub-channels is performed in order to extract the branching fraction. The signal includes contributions from all three $B_s^0$ production channels $\Upsilon(5S) \to B_s^{(*)0}\overline{B}_s^{(*)0}$. The fit results are shown in Fig. 23.3.8.

Belle observes the three decay modes and measures their branching fractions:

$$\begin{aligned}
\mathcal{B}(B_s^0 \to J/\psi f_0(980); f_0(980) \to \pi^+\pi^-) \\
= (1.16^{+0.31}_{-0.19}\,^{+0.15}_{-0.17}\,^{+0.26}_{-0.18}) \times 10^{-4}, \\
\mathcal{B}(B_s^0 \to J/\psi\eta) = (5.10 \pm 0.50 \pm 0.25\,^{+1.14}_{-0.79}) \times 10^{-4}, \\
\mathcal{B}(B_s^0 \to J/\psi\eta') = (3.71 \pm 0.61 \pm 0.18\,^{+0.83}_{-0.57}) \times 10^{-4},
\end{aligned} \qquad (23.3.6)$$

where the three quoted uncertainties are statistical, systematic, and from $N_{B_s^{(*)0}\overline{B}_s^{(*)0}}$. Evidence for the decay with the $f_0(1370)$ is also found and the branching fraction

$$\begin{aligned}
\mathcal{B}(B_s^0 \to J/\psi f_0(1370); f_0(1370) \to \pi^+\pi^-) \\
= (0.34^{+0.11}_{-0.14}\,^{+0.03}_{-0.02}\,^{+0.08}_{-0.05}) \times 10^{-4}
\end{aligned} \qquad (23.3.7)$$

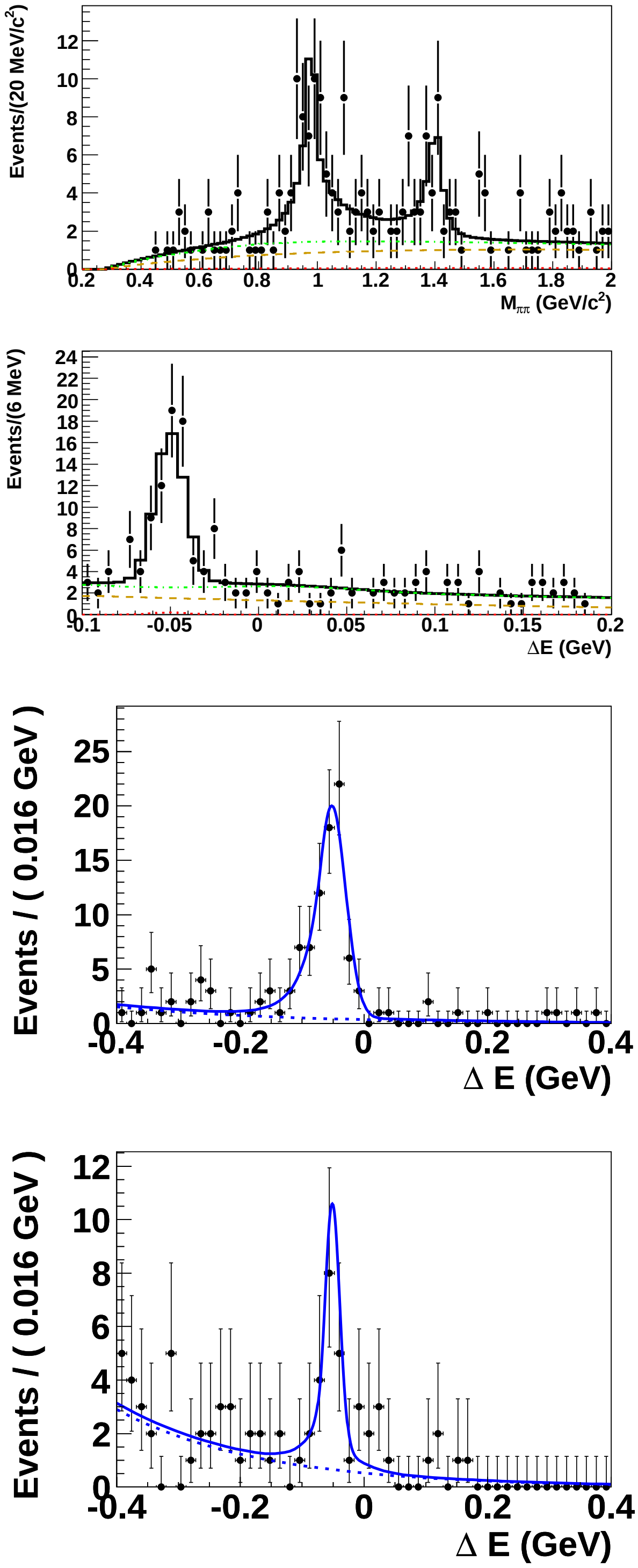

**Figure 23.3.8.** Data fit projections for $B_s^0 \to J/\psi\,\pi^+\pi^-$ (top two plots), $B_s^0 \to J/\psi\,\eta\,(\eta \to \gamma\gamma)$ and $B_s^0 \to J/\psi\,\eta'\,(\eta' \to \rho\gamma)$ (bottom two plots, respectively). The $J/\psi\,\pi^+\pi^-$ $M_{\pi\pi}$ ($\Delta E$) distributions are for events in the $\Delta E$ ($f_0(980)$) signal region. The $J/\psi\,\eta^{(\prime)}$ $\Delta E$ distributions are for events in the $m_{\rm ES}$ signal region. The dotted curves show the total background contribution. Plots are from Li (2011, 2012).

is measured, where the significance of the signal is $4.2\sigma$. The quoted uncertainties are statistical, systematic, and from $N_{B_s^{(*)0}\overline{B}_s^{(*)0}}$, respectively. The observed $J/\psi$ helicity distributions corresponding to $f_0(980)$ and $f_0(1370)$ signals are consistent with scalar $\pi\pi$ resonances.

The ratio obtained for the two branching fractions

$$\frac{\mathcal{B}(B_s^0 \to J/\psi\eta')}{\mathcal{B}(B_s^0 \to J/\psi\eta)} = 0.73 \pm 0.14(\text{stat}) \pm 0.02(\text{syst}) \qquad (23.3.8)$$

is smaller than the expected value $1.04 \pm 0.04$ calculated from other $\eta{-}\eta'$ mixing measurements at the level of $2.1\sigma$. The measured production channel fractions $f_{B_s^{*0}\overline{B}_s^{*0}}$ and $f_{B_s^{*0}\overline{B}_s^0}$ from the $B_s \to J/\psi\eta^{(\prime)}$ study are consistent with averages from other measurements.

### 23.3.5 Charmless decays $B_s^0 \to hh$, $h = \pi, K$

Belle uses $23.6\,\text{fb}^{-1}$ of data to search for the two-body charmless decays $B_s^0 \to K^+K^-$, $B_s^0 \to K^0\overline{K}^0$, $B_s^0 \to K^-\pi^+$, and $B_s^0 \to \pi^+\pi^-$. The branching fractions for these modes may exhibit direct *CP* asymmetries, as has been observed for $B_d^0 \to K^\pm\pi^\mp$ decays. In addition, measurement of the $K^+K^-$ and $\pi^+\pi^-$ time-dependent *CP* asymmetries yields information on the CKM phases $\phi_1$ and $\phi_3$ . While the all-charged final states have also been studied at hadron collider experiments (Aaij et al., 2012d; Aaltonen et al., 2009b; Abulencia et al., 2006c; Morello, 2007), the $K^0\overline{K}^0$ final state is difficult to reconstruct at a hadron collider but is well-suited to an $e^+e^-$ experiment.

Details of the analysis are found in Peng (2010). To select $B_s^0$ decays, events are required to satisfy $m_{\rm ES} \in [5.35, 5.45]\,\text{GeV}/c^2$ and $\Delta E \in [-0.20, 0.20]\,\text{GeV}$; this region is referred to as the *fitting* region. Within this region a smaller signal region is defined: $m_{\rm ES} \in [5.40, 5.43]\,\text{GeV}/c^2$ and $\Delta E \in [-0.10, 0.00]\,\text{GeV}$. The signal region corresponds to $e^+e^- \to B_s^{*0}\overline{B}_s^{*0}$ production.

To suppress the large backgrounds from $e^+e^- \to q\overline{q}$ continuum production, a Fisher discriminant based on a set of modified Fox-Wolfram moments is used. This discriminant is used to calculate the likelihood that an event is signal ($\mathcal{L}_s$) or background ($\mathcal{L}_{q\overline{q}}$). A requirement is then made on the ratio $\mathcal{L}_s/\mathcal{L}_{q\overline{q}}$. After this requirement, there are 300, 444, 188, and 345 candidates remaining in the fitting regions for $K^+K^-$, $K^-\pi^+$, $\pi^+\pi^-$, and $K^0\overline{K}^0$ modes, respectively.

The signal yields are obtained from an unbinned extended maximum likelihood fit to the $m_{\rm ES}$ and $\Delta E$ distributions. Projections of the fit are shown in Fig. 23.3.9, and the fit results along with corresponding branching fractions or 90% C.L. upper limits are listed in Table 23.3.4.

A significant signal is observed for the $K^+K^-$ final state; the significance is $5.8\sigma$. Systematic uncertainty is included in the significance by convolving the likelihood function with a Gaussian having a width equal to the total systematic uncertainty associated with the fitting procedure. The 90% C.L. upper limits ($\mathcal{B}_{90\%}$) in Table 23.3.4

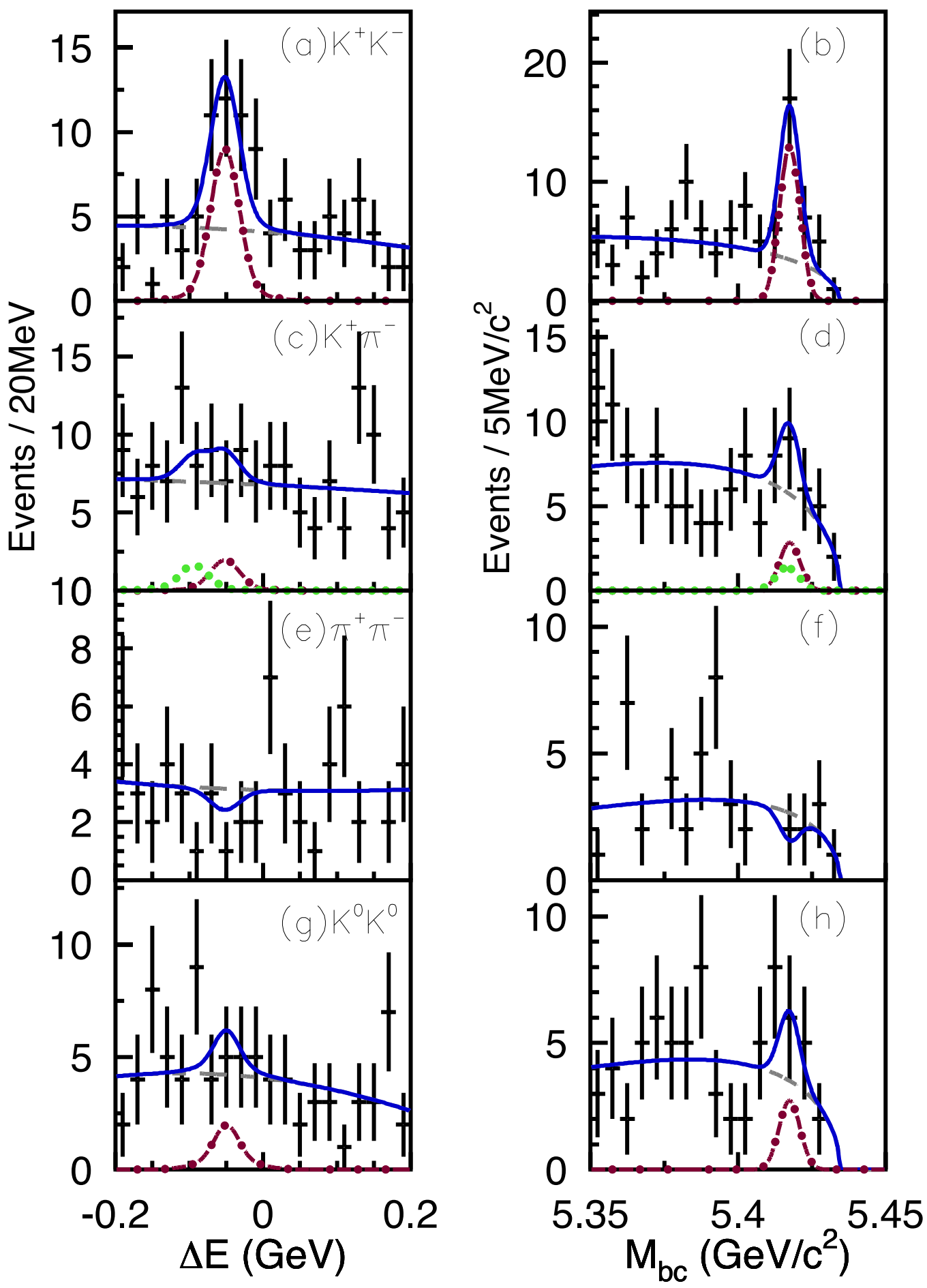

**Figure 23.3.9.** Distributions of $\Delta E$ (left) and $M_{\rm bc} = m_{\rm ES}$ (right) with fitted distributions superimposed for $K^+K^-$ (a,b), $K^+\pi^-$ (c,d), $\pi^+\pi^-$ (e,f), and $K^0\overline{K}^0$ (g,h) events. The $\Delta E$ ($m_{\rm ES}$) distributions are for events within the signal region for $m_{\rm ES}$ ($\Delta E$). The red dot-dashed curves show the signal component; the grey dashed curves show the $q\overline{q}$ background; the green dotted curves in the $K^-\pi^+$ plots show the $K^+K^-$ cross-feed; and the blue solid curves show the total. Plots are from Peng (2010).

**Table 23.3.4.** Signal yields, significances ($\Sigma$), reconstruction efficiencies ($\epsilon$), and either the branching fraction or 90% C.L. upper limit for charmless two-body $B_s^0$ decays. The first error listed is statistical, the second error is systematic, and the third error is due to the $B_s^{(*)0}\overline{B}_s^{(*)0}$ fraction $f_s$.

| Mode | Yield | $\Sigma$ | $\epsilon$(%) | $\mathcal{B}$ ($10^{-5}$) |
|---|---|---|---|---|
| $K^+K^-$ | $23.4^{+5.5}_{-6.3}$ | 5.8 | 24.5 | $3.8^{+1.0}_{-0.9} \pm 0.5 \pm 0.5$ |
| $K^-\pi^+$ | $5.4^{+5.1}_{-4.3}$ | 1.2 | 21.0 | $< 2.6$ |
| $\pi^+\pi^-$ | $-2.0^{+2.3}_{-1.5}$ | – | 14.4 | $< 1.2$ |
| $K^0\overline{K}^0$ | $5.2^{+5.0}_{-4.3}$ | 1.2 | 8.0 | $< 6.6$ |

are obtained by integrating the likelihood function:

$$\int_0^{\mathcal{B}_{90\%}} \mathcal{L}(\mathcal{B})\, d\mathcal{B} = 0.9 \times \int_0^1 \mathcal{L}(\mathcal{B})\, d\mathcal{B}\,. \qquad (23.3.9)$$

This method assumes a uniform prior distribution for $\mathcal{B}$. The $K^-\pi^+$ and $\pi^+\pi^-$ limits are consistent with, but have less sensitivity than, results from CDF (Aaltonen et al., 2009b; Morello, 2007). At the time of writing, there were no other limits on $B_s^0 \to K^0\overline{K}^0$.

### 23.3.6 Penguin decays $B_s^0 \to \phi\gamma$, $B_s^0 \to \gamma\gamma$

Belle uses $23.6\,\mathrm{fb}^{-1}$ of data to search for the radiative penguin decay $B_s^0 \to \phi\gamma$ and the penguin annihilation decay $B_s^0 \to \gamma\gamma$. The Standard Model predictions for these processes are $(3-6)\times 10^{-5}$ and $(5-10)\times 10^{-7}$, respectively (Ali, Pecjak, and Greub, 2008; Ball, Jones, and Zwicky, 2007; Bosch and Buchalla, 2002a; Chang, Lin, and Yao, 1997; Reina, Ricciardi, and Soni, 1997). Both decays proceed via internal loop diagrams and thus are sensitive to new physics at high energy scales occurring within the loops. For example, supersymmetric models with broken $R$ parity (Gemintern, Bar-Shalom, and Eilam, 2004) and two-Higgs doublet models (Aliev and Iltan, 1998) can increase the $B_s^0 \to \gamma\gamma$ branching fraction by an order of magnitude over the Standard Model prediction. As discussed in Section 17.2, a branching ratio measurement of $B_s^0 \to \gamma\gamma$ could be used in the determination of $|V_{td}/V_{ts}|$.

Details of these analyses are found in Wicht (2008). Neutral $\phi$ mesons are reconstructed via $\phi \to K^+K^-$, in which the $K^+K^-$ invariant mass is required to be within $12\,\mathrm{MeV}/c^2$ ($\sim 2.5\sigma$) of the nominal $\phi$ mass. For the $B_s^0 \to \gamma\gamma$ decay, only photons reconstructed within the barrel region of the ECL ($33^\circ < \theta < 128^\circ$) are used. To select $B_s^0$ decays, events are required to satisfy $m_{\rm ES} > 5.3\,\mathrm{GeV}/c^2$, $\Delta E < 0.40\,\mathrm{GeV}$, and either $\Delta E > -0.40\,\mathrm{GeV}$ for $B_s^0 \to \phi\gamma$ or $\Delta E > -0.70\,\mathrm{GeV}$ for $B_s^0 \to \gamma\gamma$. To reject large backgrounds from continuum production, especially $e^+e^- \to q\overline{q}\gamma$ in which a high energy $\gamma$ is produced, a Fisher discriminant based on a set of Fox-Wolfram moments is used.

The signal yields are obtained from an unbinned extended maximum likelihood fit to the $m_{\rm ES}$, $\Delta E$, and, for $B_s^0 \to \phi\gamma$, $\cos\theta_{\rm hel}$ distributions. The helicity angle $\theta_{\rm hel}$ is defined as the angle between the $B_s^0$ and the $K^+$ in the $\phi$ rest frame. Signal events should follow a $1-\cos^2\theta_{\rm hel}$ distribution, while $q\overline{q}$ continuum background events tend to be distributed flat in $\cos\theta_{\rm hel}$. The fit obtains separate signal yields for $e^+e^- \to B_s^0\overline{B}_s^0$, $B_s^*B_s^0$, and $B_s^*B_s^*$ production, but only the last of these is used to determine the branching fractions. The projections of the fit are shown in Figs 23.3.10 and 23.3.11, and the fit results are listed in Table 23.3.5.

A significant signal is observed for $B_s^0 \to \phi\gamma$. The significance is calculated as $\sqrt{2\ln(\mathcal{L}_{\rm max}/\mathcal{L}_0)}$, where $\mathcal{L}_{\rm max}$ and $\mathcal{L}_0$ are the likelihood values when the signal yield is floated and when it is set to zero, respectively; the result is $5.5\sigma$. Systematic uncertainties are evaluated by varying fixed parameters in the fit by $\pm1\sigma$, refitting, and taking the resulting change in the branching fraction as the systematic uncertainty associated with that parameter. The systematic uncertainty is included in the significance by calculating the significance using the lowest value of $\mathcal{L}_{\rm max}$ obtained when evaluating individual systematic uncertainties. The resulting $5.5\sigma$ significance constitutes the observation of a radiative decay of the $B_s^0$ meson. The measured branching fraction is in agreement with theoretical predictions.

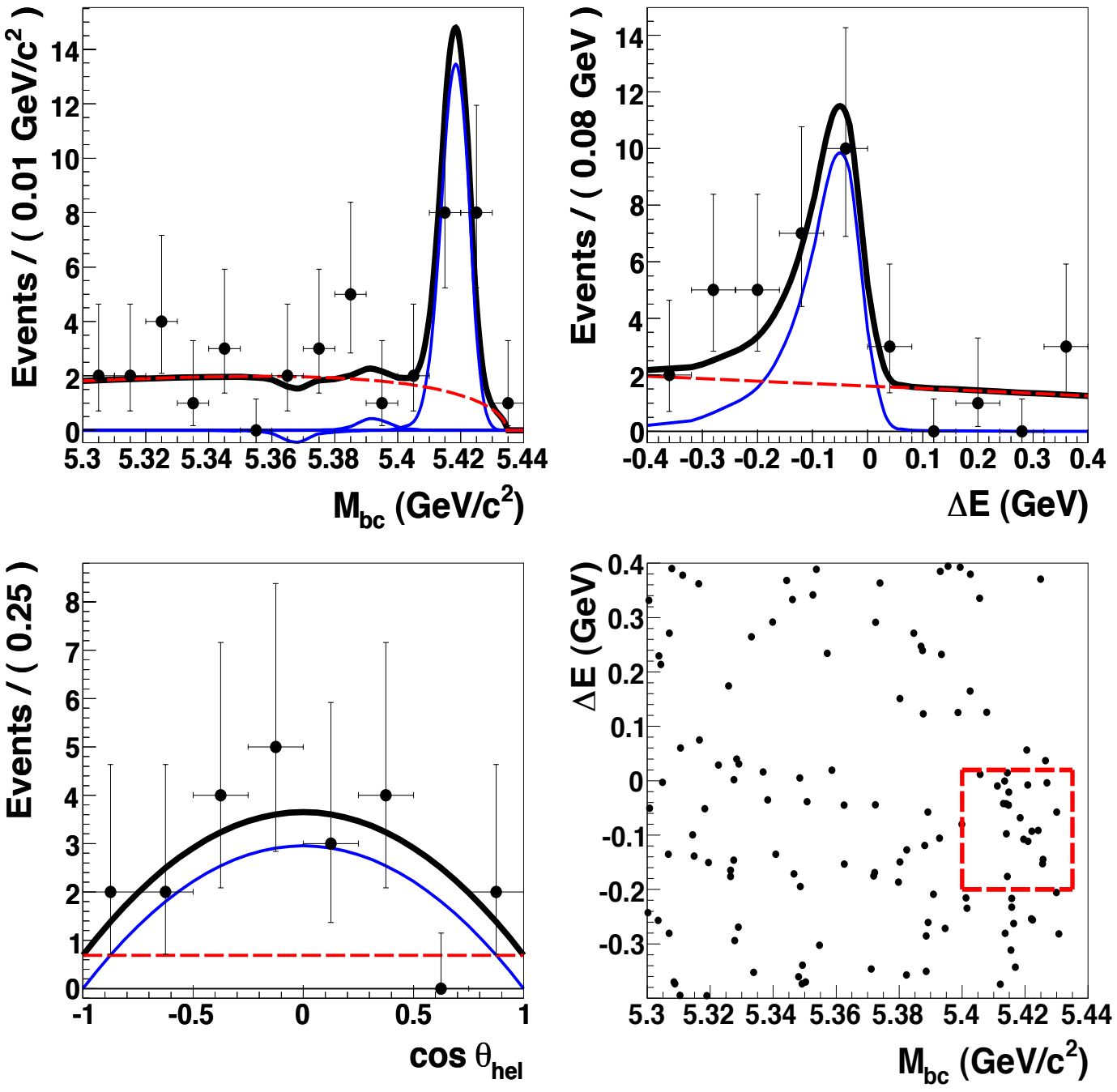

**Figure 23.3.10.** $M_{\rm bc} = m_{\rm ES}$, $\Delta E$, and $\cos\theta_{\rm hel}$ projections for $B_s^0 \to \phi\gamma$. The points with error bars show the data; the thin solid curves show the signal contribution; the dashed curves show the continuum contribution; and the thick solid curves show the total. The bottom right figure shows the ($m_{\rm ES}$, $\Delta E$) plane; the dashed lines denote the signal region used to calculate the branching fraction. Plots are from Wicht (2008).

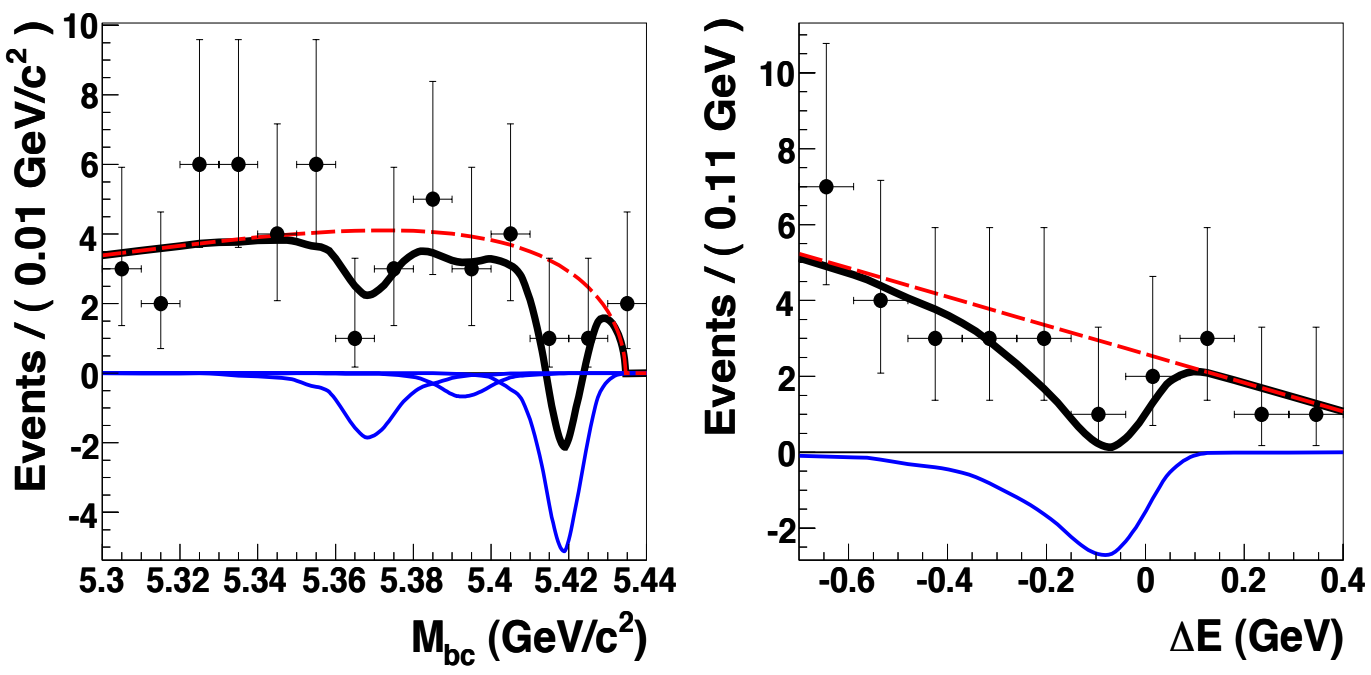

**Figure 23.3.11.** $M_{\rm bc} = m_{\rm ES}$ and $\Delta E$ projections for $B_s^0 \to \gamma\gamma$. The points with error bars show the data; the thin solid curves show the signal contribution; the dashed curves show the continuum contribution; and the thick solid curves show the total. In the $m_{\rm ES}$ figure, the signals (negative) from $e^+e^- \to B_s^0\overline{B}_s^0$, $B_s^{*0}\overline{B}_s^0$, and $B_s^{*0}\overline{B}_s^{*0}$ appear from left to right. Plots are from Wicht (2008).

**Table 23.3.5.** Fitted signal yields, and either the branching fraction or 90% C.L. upper limit. The first error listed is statistical, and the second is systematic.

| Mode | $S_{B_s^0\overline{B}_s^0}$ | $S_{B_s^{*0}\overline{B}_s^0}$ | $S_{B_s^{*0}\overline{B}_s^{*0}}$ | $\mathcal{B}$ ($10^{-6}$) |
|---|---|---|---|---|
| $\phi\gamma$ | $-0.7^{+2.5}_{-1.6}$ | $0.5^{+2.9}_{-1.9}$ | $18^{+6}_{-5}$ | $57^{+18+12}_{-15-11}$ |
| $\gamma\gamma$ | $-4.7^{+3.9}_{-2.8}$ | $-0.8^{+4.8}_{-3.8}$ | $-7.3^{+2.4}_{-2.0}$ | $< 8.7$ |

The 90% C.L. upper limit for the $B_s^0 \to \gamma\gamma$ branching fraction ($\mathcal{B}_{90\%}$) is obtained using the same method as for the two-body decays (see Eq. 23.3.9) by integrating the likelihood function assuming a uniform prior in $\mathcal{B}$. Systematic uncertainties are included by convolving the likelihood function with Gaussian distributions corresponding to each systematic error contribution. The resulting upper limit of $8.7 \times 10^{-6}$ is a significant improvement over previous limits and about one order of magnitude above the SM prediction.

## 23.4 Conclusion

$B_s^0$ mesons were studied using data collected in the $\Upsilon(5S)$ energy region with an $e^+e^-$ collider by the $B$ Factories. Many new $B_s^0$ decays are observed, and branching fractions for several $B_s^0$ decays are measured with a precision comparable with that of existing results.

$B_s^0$ measurements at the $\Upsilon(5S)$, when compared with the measurements at hadron-hadron colliders, have the following advantages:

1. Channels with neutral particles, such as photons, $\pi^0$'s, or $\eta$ mesons are precisely measured. The benefit is most prominent in the case of reconstruction of low energy photons or several neutral particles in an event.
2. Missing mass methods are used with the partial reconstruction technique, due to the low momenta of the $B_s^0$ mesons in the $\Upsilon(5S)$ rest frame.
3. The almost 100% trigger efficiency is an advantage of the $B$ Factory experiments. No special procedures have to be used to estimate the trigger efficiencies.
4. The full number of $B_s^0$ mesons in a data sample is calculated with a small systematic uncertainty. This allows one to measure absolute -rather than relative- branching fractions.

The most important disadvantages of $B_s^0$ meson studies at the $\Upsilon(5S)$ are the smaller initial number of $B_s^0$ mesons, the lack of ability to resolve $B_s^0$ oscillations (hence it is not possible to perform time-dependent $CP$ analyses), and the necessity to choose between data taking at the $\Upsilon(5S)$ and $\Upsilon(4S)$. In general $B_s^0$ meson studies with $e^+e^-$ colliders at the $\Upsilon(5S)$ and at hadron-hadron machines are complementary and together provide the best coverage of the whole spectrum of the most interesting decay channels.

**Table 23.4.1.** Measured $B_s^0$ branching fractions with statistical, systematic (without $f_s$), and $f_s$ uncertainties are shown. If no significant signal is observed the 90% C.L. upper limit is given. If the third uncertainty is absent, the second uncertainty includes both systematic uncertainties. The corresponding $B^0$ branching fractions from PDG Beringer et al. (2012) are also shown for comparison.

| $B_s^0$ mode | $\mathcal{B}$ | $B^0$ mode | $\mathcal{B}$ [PDG] |
|---|---|---|---|
| $B_s^0 \to D_s^- \pi^+$ | $(3.67^{+0.35+0.43}_{-0.33-0.42} \pm 0.49) \times 10^{-3}$ | $B^0 \to D^- \pi^+$ | $(2.68 \pm 0.13) \times 10^{-3}$ |
| $B_s^0 \to D_s^{*-} \pi^+$ | $(2.4^{+0.5}_{-0.4} \pm 0.3 \pm 0.4) \times 10^{-3}$ | $B^0 \to D^{*-} \pi^+$ | $(2.76 \pm 0.13) \times 10^{-3}$ |
| $B_s^0 \to D_s^- \rho^+$ | $(8.5^{+1.3}_{-1.2} \pm 1.1 \pm 1.3) \times 10^{-3}$ | $B^0 \to D^- \rho^+$ | $(7.8 \pm 1.3) \times 10^{-3}$ |
| $B_s^0 \to D_s^{*-} \rho^+$ | $(11.8^{+2.2}_{-2.0} \pm 1.7 \pm 1.8) \times 10^{-3}$ | $B^0 \to D^{*-} \rho^+$ | $(6.8 \pm 0.9) \times 10^{-3}$ |
| $B_s^0 \to D_s^{\mp} K^{\pm}$ | $(2.4^{+1.2}_{-1.0} \pm 0.3 \pm 0.3) \times 10^{-4}$ | $B^0 \to D^- K^+$ | $(1.97 \pm 0.21) \times 10^{-4}$ |
| $B_s^0 \to D_s^+ D_s^-$ | | $B^0 \to D^- D_s^+$ | $(7.2 \pm 0.8) \times 10^{-3}$ |
| $B_s^0 \to D_s^{*\pm} D_s^{\mp}$ | | $B^0 \to D^{*\pm} D_s^{\mp}$ | $(1.54 \pm 0.19) \times 10^{-2}$ |
| $B_s^0 \to D_s^{*+} D_s^{*-}$ | | $B^0 \to D^{*-} D_s^{*+}$ | $(1.77 \pm 0.14) \times 10^{-2}$ |
| $B_s^0 \to J/\psi \eta$ | $(5.10 \pm 0.50 \pm 0.25^{+1.14}_{-0.79}) \times 10^{-4}$ | $B^0 \to J/\psi K^0$ | $(8.74 \pm 0.32) \times 10^{-4}$ |
| $B_s^0 \to J/\psi \eta'$ | $(3.71 \pm 0.61 \pm 0.18^{+0.83}_{-0.57}) \times 10^{-4}$ | $B^0 \to J/\psi K^0$ | $(8.74 \pm 0.32) \times 10^{-4}$ |
| $B_s^0 \to J/\psi f_0(980)$, $f_0(980) \to \pi^+\pi^-$ | $(1.16^{+0.31+0.15+0.26}_{-0.19-0.17-0.18}) \times 10^{-4}$ | | |
| $B_s^0 \to K^+ K^-$ | $(3.8^{+1.0}_{-0.9} \pm 0.5 \pm 0.5) \times 10^{-5}$ | $B^0 \to K^+ \pi^-$ | $(1.94 \pm 0.06) \times 10^{-5}$ |
| $B_s^0 \to K^- \pi^+$ | $< 2.6 \times 10^{-5}$ | | |
| $B_s^0 \to \pi^+ \pi^-$ | $< 1.2 \times 10^{-5}$ | | |
| $B_s^0 \to K^0 \bar{K}^0$ | $< 6.6 \times 10^{-5}$ | | |
| $B_s^0 \to \phi \gamma$ | $(5.7^{+1.8+1.2}_{-1.5-1.1}) \times 10^{-5}$ | $B^0 \to K^{*0} \gamma$ | $(4.33 \pm 0.15) \times 10^{-5}$ |
| $B_s^0 \to \gamma\gamma$ | $< 8.7 \times 10^{-6}$ | | |

The $B_s^0$ decay branching fractions measured at the $\Upsilon(5S)$ are summarized in Table 23.4.1. The branching fractions of the corresponding $B^0$ decays are also shown for comparison.

The results obtained demonstrate that the $B_s^0$ and $B^0$ meson branching fractions are consistent within uncertainties. The $B_s^0$ branching fractions with $\eta$ and $\eta'$ mesons in the final state are expected to be about 1/3 of the corresponding $B^0$ branching fractions with a $K^0$, because the $\eta^{(\prime)}$ meson is assumed to be approximately one third $s\bar{s}$. The results indicate that the larger mass of the $s$-quark, compared with the mass of the $d$-quark, does not result in significant rearrangement of the intrinsic structure of the $B$ meson, consisting of both heavy and light quarks. It should also be noted that Belle did not find any significant deviations of the experimentally measured $B_s^0$ branching fractions from the corresponding theoretical predictions.

# Chapter 24
# QCD-related physics

Although the main objective of the $B$ Factories is the study of flavor physics and weak decays, some additional insights on QCD-related physics have been gained. From the data of both *BABAR* and Belle, the functions for the fragmentation of light quarks into light hadrons and for the fragmentation of the charm quark into charmed mesons and baryons have been extracted with improved precision. Not only the unpolarized, but also the spin-dependent fragmentation functions have been measured, adding a crucial input to studies of transverse quark polarization in the nucleon.

Another QCD-related issue is the search for exotic states. While charmonium-like exotic states are discussed in Section 18.3, searches for exotic states composed of light quarks are described in Section 24.2. Searches for such states have been performed by *BABAR* and Belle, showing no evidence for any of the previously claimed pentaquark states nor for any other member of the pentaquark family.

## 24.1 Fragmentation

***Editors:***
*Fabio Anulli (BABAR)*
*Ralf Seidl (Belle)*
*Shunzo Kumano (theory)*

**Additional section writers:**
*Dave Muller*

### 24.1.1 Introduction

A consequence of the confining property of the strong interaction is that energetic quarks and gluons produced in high-energy collisions appear as collimated "jets" of hadrons. The process by which this occurs, called fragmentation, is understood qualitatively, but there are few quantitative theoretical predictions. A better understanding of this process is desirable as a probe of the strong interaction, and an empirical understanding is essential to the interpretation of much current and future high-energy data, in which the observable products of interactions and decays of heavy particles, known and yet to be discovered, appear as hadronic jets. The intrinsic properties of jets are best studied in $e^+e^-$ annihilation, where hadrons originate from primordial quarks and antiquarks, which are produced back-to-back in the CM frame. This kind of process leads to final sates with two jets, in case hard gluons are emitted by the quarks three and more jet events emerge.

Jets can be characterized by their overall structure, *e.g.* shape, energy flow, etc., and by the number, types and momentum spectra of hadrons produced. These properties depend on the energy, mass, charge and spin of the quark or gluon that initiated the jet. The initial quark is contained in a leading hadron that carries its flavor, perhaps its spin, and a fraction of its energy that is higher, on average, for heavier quarks. Additional $q\bar{q}$ pairs are produced from the vacuum to form the other hadrons in the jet, with probabilities that depend strongly on the quark mass, with $s\bar{s}$ pairs being suppressed with respect to $u\bar{u}$ and $d\bar{d}$ pairs by a factor of roughly three, and heavier quarks by much larger factors. In this way, $c$ and $b$ jets generally contain a single $D$ or $B$ hadron with typically more than half the jet energy, along with a few softer hadrons. However, the leading heavy hadron decays into a number of softer particles that influence the jet structure. Lighter flavor jets contain more primary hadrons with a broader range of momentum.

Hadron-production processes in high-energy reactions are also important for investigating properties of quark-hadron matter in heavy-ion collisions and for finding the origin of the nucleon spin in polarized lepton-nucleon and nucleon-nucleon reactions. In describing the hadron-production cross sections in high-energy reactions, fragmentation functions (FFs) are essential quantities. A fragmentation function quantifies the probability of producing a particular hadron $h$ in a jet initiated by a particular parton. They are measured most directly by the hadron productions in electron-positron annihilation, $e^+e^- \to h\,X$, where $h$ is the hadron under investigation and $X$ is the rest of the hadronic final state.

In $e^+e^-$ annihilation the initial partonic state is rather simple and can be described by a quark-antiquark pair at leading order in the strong coupling $\alpha_s$. In single $\gamma^\star$ exchange (see Fig. 24.1.1), the relative production of quark flavors is given by the charge squared of the quarks. At energy scales below the open bottom threshold the production of $u\bar{u}$ and $c\bar{c}$ pairs amounts to 40% each, and that of $d\bar{d}$ and $s\bar{s}$ pairs to 10% each. The cross section for hadron production $e^+ + e^- \to h + X$ is described by a quark-antiquark pair creation by the reaction $e^+e^- \to q\bar{q}$ and higher-order corrections such as $e^+e^- \to q\bar{q}g$, and then by a fragmentation process to create a hadron $h$ from quark ($q$), antiquark ($\bar{q}$), or gluon ($g$). The hadron multiplicity[177] is defined by the hadron-production cross section and the total hadronic cross section (Ellis, Stirling, and Webber, 1996) $\sigma_{tot} = \sigma_{e^+e^-\to q\bar{q}}$:

$$F^h(z, Q^2) = \frac{1}{\sigma_{tot}} \frac{d\sigma(e^+e^- \to hX)}{dz}, \tag{24.1.1}$$

where the variable $Q^2$ is the virtual photon momentum squared in $e^+e^- \to \gamma^\star$, and it is given by $Q^2 = s$, with $\sqrt{s}$ being the CM energy. The variable $z$ is the hadron energy $E_h$ scaled to the beam energy $\sqrt{s}/2$:

$$z \equiv \frac{E_h}{\sqrt{s}/2} = \frac{2E_h}{Q}. \tag{24.1.2}$$

[177] The hadron multiplicity is often also called fragmentation function despite being a different object.

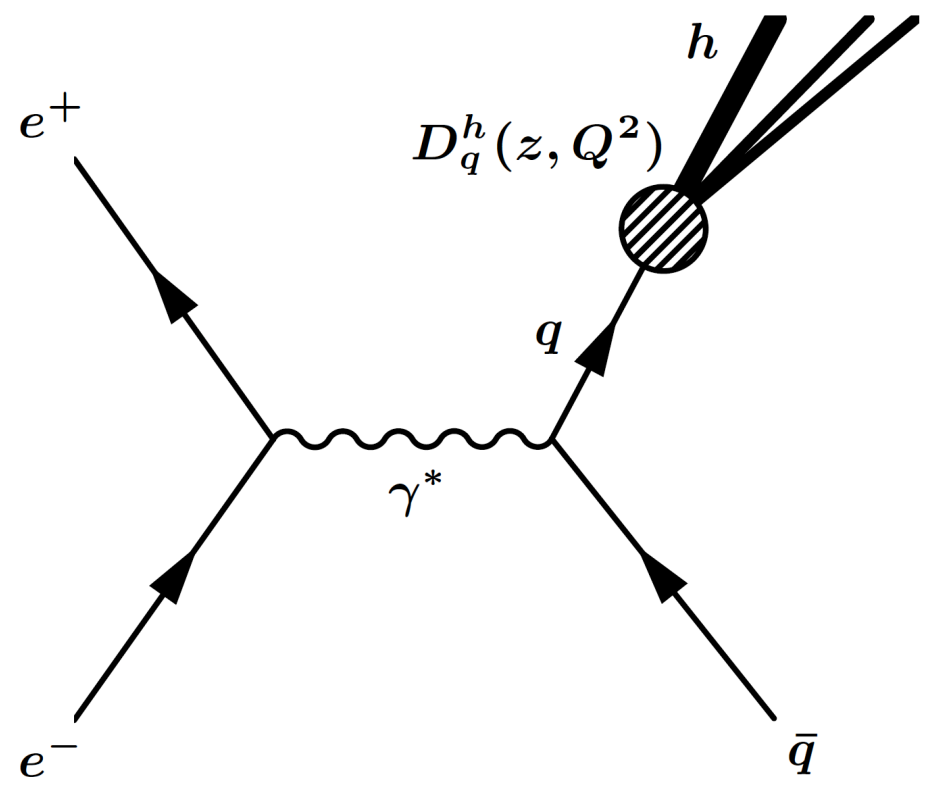

**Figure 24.1.1.** Typical hadron-production process in electron-positron annihilation ($e^+ + e^- \to h + X$).

The total cross section is described by the $q\bar{q}$-pair creation processes, $e^+e^- \to \gamma^\star \to q\bar{q}$ and higher-order corrections:

$$\sigma_{tot} = \frac{4\pi\alpha^2}{s} \sum_q e_q^2 \left[1 + \frac{\alpha_S(Q^2)}{\pi} + \cdots\right]. \quad (24.1.3)$$

At CM energies higher than those available at $B$ factories, $Z^0$ exchange must be taken into account, as it modifies the total cross section and the flavor composition, but not the FF for a given flavor.

The hadron multiplicities are related to the actual fragmentation functions which are defined as parton densities for inclusively detecting a hadron $h$ with fractional energy $z$ from an initial state parton $q$. Sometimes FFs are also obtained as a differential in the fractional energy and the hadron transverse momentum $P_{h,\perp}$ relative to the initial parton energy and can also depend on the parton and hadron spin orientations, $d^2\sigma/dzdP_{h,\perp}$. The fragmentation process is described by the sum of hadrons produced from primary quarks, antiquarks, and gluons (Ellis, Stirling, and Webber, 1996):

$$F^h(z, Q^2) = \sum_i C_i(z, \alpha_S) \otimes D_i^h(z, Q^2). \quad (24.1.4)$$

Here, $D_i^h(z, Q^2)$ is a fragmentation function of the hadron $h$ created by a parton $i$ ($= u,\ d,\ s,\ \cdots,\ g$), and it indicates the probability of producing the hadron $h$, from the parton $i$ with the energy fraction $z$ at the momentum squared scale $Q^2$. The convolution integral $\otimes$ is defined by

$$f(z) \otimes g(z) = \int_z^1 \frac{dy}{y} f(y) g\left(\frac{z}{y}\right). \quad (24.1.5)$$

The simplest FF is this unpolarized FF $D_i^h(z, Q^2)$. Polarized FFs also exist and are discussed in Section 24.1.3. Although the FF has the meaning of the production probability in the leading order (LO) of the running coupling constant $\alpha_S$, it is a scheme-dependent quantity if higher-order corrections are taken into account. The coefficient function $C_i(z, \alpha_S)$ is trivial at leading order. At next-to-leading order (NLO), quark-gluon splitting appears, the corresponding coefficient functions are needed, and the gluon FF appears. The NLO results are listed, for example, in Kretzer (2000) and Albino, Kniehl, and Kramer (2005, 2008) for the modified minimal subtraction ($\overline{\text{MS}}$) scheme. The coefficient functions are now known to the NNLO level for the unpolarized case (Mitov, Moch, and Vogt, 2006).

A sum rule exists for the FFs because of energy conservation. As the variable $z$ is the energy fraction for the produced hadron, its sum weighted by the fragmentation functions should be unity:

$$\sum_h \int_0^1 dz\, z\, D_i^h(z, Q^2) \equiv \sum_h M_i^h = 1, \quad (24.1.6)$$

where $M_i^h$ is the second moment of $D_i^h(z, Q^2)$. The meaning of this result is that the sum of all final state hadrons' fractional energies integrated over the energy fraction has to retain the initial parton's energy. Not all of the hadrons are observed experimentally, so that in practice it is not possible to confirm this sum rule precisely from measurements. However, it is a useful relation in determining the FFs when performing a global analysis by providing a constraint on their magnitude.

As this hadron formation takes place at small masses and low energies it can only be described non-perturbatively. In the parton model, the fragmentation function is defined by (Brock et al., 1995; Collins, 1993)

$$D_i^h(x) = \sum_X \int \frac{dy^-}{24\pi} e^{ik^+y^-} \,\mathrm{Tr}\left[\gamma^+ \langle\, 0 \,|\psi_i(0, y^-, 0_\perp)|\, h, X\rangle \times \langle h, X|\bar{\psi}_i(0)|\, 0 \,\rangle\right], \quad (24.1.7)$$

where $k$ is the parent quark momentum, the light-cone notation is defined by $a^\pm = (a^0 \pm a^3)/\sqrt{2}$, the variable $z$ is then given by $z = p_h^+/k^+$ with the hadron momentum $p_h$, and $\perp$ is the transverse direction to the third coordinate. To be precise, a gauge link needs to be introduced in Eq. (24.1.7) so as to satisfy color gauge invariance. In the parton model, fragmentation functions are generally defined in a similar way as the parton distribution functions, however they cannot be calculated by lattice simulations due to the hadron in the final state.

The $Q^2$ evolution for the fragmentation functions is calculated by perturbative QCD in the same way as the one for the parton distribution functions. It is given by the time-like DGLAP (Dokshitzer-Gribov-Lipatov-Altarelli-Parisi) evolution equations (Ellis, Stirling, and Webber,

1996; Hirai and Kumano, 2011):

$$\frac{\partial}{\partial \ln Q^2} D^h_{q_i^+}(z,Q^2) = \frac{\alpha_S(Q^2)}{2\pi}\left[\sum_j P_{q_j q_i}(z) \otimes D^h_{q_j^+}(z,Q^2) + 2P_{gq}(z) \otimes D^h_g(z,Q^2)\right],$$
$$\frac{\partial}{\partial \ln Q^2} D^h_g(z,Q^2) = \frac{\alpha_S(Q^2)}{2\pi}\left[P_{qg}(z) \otimes \sum_j D^h_{q_j^+}(z,Q^2) + P_{gg}(z) \otimes D^h_g(z,Q^2)\right], \quad (24.1.8)$$

where $D^h_{q^+}(x,Q^2)$ denotes the fragmentation-function combination $D^h_q(x,Q^2) + D^h_{\bar{q}}(x,Q^2)$. The functions $P_{q_jq_i}(z)$, $P_{gq}(z)$, $P_{qg}(z)$, and $P_{gg}(z)$ are splitting functions, where the off-diagonal elements $P_{gq}(z)$ and $P_{qg}(z)$ are interchanged in the splitting-function matrix from the parton distribution function case (Ellis, Stirling, and Webber, 1996). One should note that the time-like functions are slightly different from the space-like ones in the next-to-leading order (Ellis, Stirling, and Webber, 1996).

There are measurements of the FFs in electron-positron annihilation at various center-of-mass energies. However, they are not accurate enough to determine precise functional forms. Data were taken mainly in the $Z$-mass region at the SLAC Linear Collider (SLC) and Large Electron-Positron Collider (LEP), which means that the scaling violation has not been determined precisely for the multiplicities. The FFs still have large uncertainties, particularly for the so-called disfavored functions (such as $D^{\pi^-}_u(z,Q^2)$, *i.e.* the function describing a $\pi^-$ creation from an initial $u$ quark - as opposed to the favored FF $D^{\pi^+}_u(x,Q^2)$), even for the pion, and there are large discrepancies among the obtained FFs from different global analysis groups.

The Belle and *BABAR* data play an important role in the accurate determination of the FFs by extending the kinematical region of $z$ due to high-statistics measurements. The Belle and *BABAR* measurements are performed at CM energies around 10 GeV. Together with the accurate measurements at the Z mass the scaling behavior of FFs can be studied. The knowledge of the energy dependence will improve the physics results of high-energy experiments at LHC and RHIC, and other high-energy facilities.

Generally, fragmentation functions can be defined for any kind of final state hadron as long as it has been produced in strong processes only. Hadrons being produced in weak decays should in principle not be included, but given the difficulty of determining their relative fractions experimentally they are often included. So far, most of the fragmentation functions were obtained from $e^+e^-$ annihilation alone (Albino, Kniehl, and Kramer, 2008; Hirai, Kumano, Nagai, Oka, and Sudoh, 2007) due to their clean initial state. However, as the unpolarized parton distribution functions in the intermediate $x_{Bjorken}$ range, where $x_{Bjorken}$ is the momentum fraction a parton carries relative to the nucleon, are relatively well known, recent extractions of FFs from the world data also include some semi-inclusive deep inelastic scattering and proton-proton collision data, such as in de Florian, Sassot, and Stratmann (2008). Given its non-perturbative nature and the need for a DGLAP evolution to obtain all components, FFs are generally obtained from a global analysis of the world data where available. In such a global analysis the fragmentation functions are parameterized at an initial scale which is fit to all existing data. While the $e^+e^-$ data is usually quite precise, it is generally limited to the sum of quark and antiquark FFs as it is not known from which side a final state hadron emerged. Also, as mentioned above, the gluon FF only appears in NLO and therefore can only be obtained from $e^+e^-$ data by comparing very different scales via the evolution. Therefore the data from semi-inclusive DIS and proton-proton scattering gives some valuable additions to the global analysis, but also the large lever arm between data obtained close to the $Z^0$ resonance and the $B$ Factories' data is quite useful.

Belle and *BABAR* have studied the inclusive momentum spectra of a number of light and charmed hadrons in $e^+e^-$ annihilation. They have also studied spin-induced correlations between particles in opposite jets. These are described in the subsections below.

### 24.1.2 Unpolarized fragmentation functions

The production rate of a particular type of hadron in jets of a particular (set of) flavor(s) can be quantified by the multiplicities $F^h(z,Q^2)$ (see Eq. 24.1.1), which is the average number of hadrons of type $h$ produced per unit $z$ in a jet, and $z$ is a measure of the fraction of the quark's energy carried by the hadron. Apart from the normalized hadron energy $z$, a number of other definitions $x_h$ related to normalized momenta are in use, and the relevant one is defined below in each case.

FFs cannot be calculated perturbatively in QCD, so there are no firm theoretical predictions. The ansatz of local parton-hadron duality (LPHD) combined with calculations of gluon radiation in the modified leading logarithm approximation (MLLA) (Azimov, Dokshitzer, Khoze, and Troyan, 1985) predicts properties of the distributions of the dimensionless variable $\xi = \ln(\sqrt{s}/2p^*)$ for light hadrons, where $p^*$ is the magnitude of the hadron momentum in the CM system. The parameters depend on the hadron mass and the jet energy. There are several phenomenological models of fragmentation, involving three different hadron production methods. Here we consider representatives of each, the `HERWIG 5.8` (Marchesini et al., 1992), `Jetset 7.4` (Sjöstrand, 1994) and `UCLA 4.1` (Chun and Buchanan, 1998) event generators.

For sufficiently heavy quarks $q$, the high mass provides a convenient cut-off point in the perturbative regime and the multiplicity $F^q(x_q)$ of the heavy quark before hadronization can be calculated (Braaten, Cheung, Fleming, and Yuan, 1995; Colangelo and Nason, 1992; Collins and Spiller, 1986; Dokshitzer, Khoze, and Troian, 1996;

Mele and Nason, 1991). The observable heavy hadron multiplicity $F^H(x_H)$ is thought to be related by a simple convolution or hadronization model. Several phenomenological models of heavy-quark fragmentation have been proposed (Andersson, Gustafson, Ingelman, and Sjöstrand, 1983; Bowler, 1981; Kartvelishvili, Likhoded, and Petrov, 1978; Peterson, Schlatter, Schmitt, and Zerwas, 1983). Predictions depend on the quark mass, with $F^H(x_p)$ being much harder for $b$ hadrons than $c$ hadrons, and in some cases on the mass and quantum numbers of $H$. Hadrons containing the same heavy quark type are generally predicted to have similar $F^H(x_H)$, although differences between mesons and baryons have been suggested (Chun and Buchanan, 1998; Kartvelishvili and Likhoded, 1979).

The scaling properties, or $\sqrt{s}$ dependences, of hadron production are of particular interest. Since QCD is only weakly scale dependent, distributions of $x_h$ should be almost independent of $\sqrt{s}$, except for the effects of hadron masses/phase space and the running of $\alpha_S$. However, the quark flavor composition varies with $\sqrt{s}$ in $e^+e^-$ annihilation, and must be modeled for light hadrons. This provides a nice test of models and MLLA QCD.

So far, Belle and *BABAR* have measured multiplicities for several of the lightest and heaviest particles produced at around 10.5 GeV. These include the light, non-strange mesons $\pi^\pm$ and $\eta$, the lightest strange meson $K^\pm$, and the lightest baryon $p/\overline{p}$, which can be used to test MLLA QCD and hadronization models. The five charmed mesons $D^0$, $D^+$, $D_s^+$, $D^{*0}$ and $D^{*+}$, and three charmed baryons $\Lambda_c^+$, $\Xi_c^0$ and $\Omega_c^0$ can be used to test models and calculations for heavy quarks. In addition, the correlated production of $\Lambda_c^+$ and $\overline{\Lambda}_c^-$ has been studied, providing a stringent test of models in an extreme region. These are discussed in the following subsections.

#### 24.1.2.1 Light hadrons $\pi^\pm$, $K^\pm$, $p/\overline{p}$

*BABAR* and Belle have measured the inclusive production cross sections of charged pions and kaons in $e^+e^- \to q\overline{q}$ events at the off-resonance CM energies of 10.54 GeV and 10.52 GeV, respectively, while *BABAR* also measured protons and neutral $\eta$ meson at 10.54 GeV. The data sets used contain integrated luminosities of 0.91 $\text{fb}^{-1}$ for *BABAR* and 68.0 $\text{fb}^{-1}$ for Belle. Uncertainties of inclusive measurements of fragmentation are dominated by systematic uncertainties even in relatively small samples of data with good running conditions. Hence the data are selected from runs with very stable running conditions.

Inclusive measurements such as these require a clean sample of multihadron events with low bias against particles of any particular type, multiplicity or momentum. All results reported in this Section are displayed as a function of the normalized hadron energy $z$.

For the charged $\pi$/K/p analyses, *BABAR* (Muller, 2004) requires: three or more well reconstructed charged tracks that form a good vertex located within 5 mm of the beam axis and within 5 cm of the center of the collision region along the beam axis; a sum of charged plus neutral energy $E_{\text{tot}}$ in the range 5–14 GeV; $R_2$ (see Section 9.3) less than 0.9; the polar angle $\theta^*_{\text{thrust}}$ of the event thrust axis with respect to the electron beam direction in the CM frame to satisfy[178] $|\cos\theta^*_{\text{thrust}}| < 0.8$; the track with the highest momentum in the laboratory frame $p$, not to be identified as an electron in events with fewer than six good tracks, and neither of the two highest-$p$ tracks to be identified as an electron in events with only three tracks.

The requirements on $E_{tot}$ and $\cos\theta^*_{\text{thrust}}$ select events well contained within the sensitive volume of the detector with low bias on the momentum spectra. The efficiency of this selection is determined on a simulated sample of events, and corrected for differences between data and simulation. The result is 68% for $u\bar{u}$, $d\bar{d}$ and $s\bar{s}$ events, and 73% for $c\bar{c}$ events. Similarly, they estimate backgrounds of 5.1% and 0.1%, respectively, from $\tau$-pair and radiative Bhabha events, which contribute up to 20% and 8% of the charged tracks at the highest momenta. The background from two-photon processes is below 1%, and backgrounds from $\mu$-pairs, hard initial state radiation (*i.e.* photon energies of more than several 100 MeV), beam-gas and beam-wall interactions are negligible.

High quality charged tracks are selected and identified as pions, kaons or protons using the momentum and ionization energy loss measured in the DCH and the velocity measured via the Cherenkov angle in the DIRC (see Chapter 2). A global likelihood algorithm is used that considers the set of $\mathrm{d}E/\mathrm{d}x$ values for the reconstructed tracks in each event, along with the set of Cherenkov angles measured from photons detected in the DIRC (Chapter 5). The likelihood is optimized to keep the misidentification rates as low as reasonably possible, while maintaining high identification efficiencies that vary slowly with both momentum and polar angle. It identifies pions and kaons (protons) with efficiencies of over 99% for $p$ below 0.7 (1.0) GeV/$c$, over 90% for $p$ below 1.5 (4.5) GeV/$c$, and over 50% for $p$ below 4.0 (6.5) GeV/$c$. Misidentification rates are below 1%, 6% and 4% in these three regions.

Since the $e^+e^-$ system is boosted in the laboratory frame of reference, the analysis is performed separately for tracks in six different polar angle regions. The tracks in each region span different ranges of $p$, but each is transformed into the same range of momenta in the CM frame, $p^*$. This provides a set of powerful cross checks on the detector performance and material interactions, backgrounds, the true polar angle and $p^*$ distributions, and the boost value itself. In each region, the full matrix of hadron identification efficiencies $(\pi^\pm, K^\pm, p^\pm)$ is calibrated from the data as a function of $p$ using a set of control samples (Chapter 5). The corrected efficiency matrices are inverted and used to convert the numbers of identified pions, kaons and protons into differential production cross sections per hadronic event per unit momentum in the laboratory frame, $(1/N_{evt})dn_i/dp$, $i = \pi, K, p$. Each corrected cross section is then transformed into the $e^+e^-$ CM frame. Results from the six regions are compared as a cross check, and then combined to give the final measured cross sections, $(1/N_{evt})\ dn_i/dp^*$. There is a small correction for residual lepton contamination, and results are given in two ways, including or excluding the contributions from decays of $K_S^0$ and weakly decaying strange baryons; here we consider the latter.

[178] The event thrust axis is calculated using all particles in the event, or all charged tracks in the case of the *BABAR* light hadron analysis; it approximates the back-to-back direction of the two leading jets typically produced in events from continuum. We note that the thrust defined in Section 9.3, is instead built upon the decay products of the reconstructed $B$ meson, with $\theta_T$ the angle between the thrust of the $B$ decay products and the thrust of the rest of the event.

The total systematic uncertainty on the pion cross section is at the level of a few percent in the full momentum range. It is dominated at low momenta by tracking efficiencies, by background contamination between 0.75 and about 3 GeV/$c$, and by particle identification at the highest momenta. The uncertainties on the kaon and proton cross sections have similar patterns, but are significantly larger, in particular for momenta below 0.2 GeV/$c$ and above 4 GeV/$c$.

The uncertainties all have large point to point correlations. There is an overall normalization uncertainty of 0.9% that does not affect the shape of any cross section. Several uncertainties are fully correlated over the entire $p^*$ range, but vary slowly with $p^*$ and can have broad effects on the shape. The uncertainties from the calibration of the particle ID are correlated over ranges of a few bins, and can lead to apparent structures.

The Belle analysis (Leitgab, 2012, 2013) similarly requires events with at least 3 charged tracks, a visible energy above 7 GeV and either jet mass[179] above 1.8 GeV or the jet mass normalized by the visible energy to be above 0.25. Tracks are selected within the central detector with $-0.511 \leq \cos\theta < 0.842$, where the polar angle $\theta$ is calculated in the laboratory frame, with a minimum momentum of a track of 500 MeV and at least three hits in the vertex detector. Tracks are also required to originate from within distances of 1.3 cm radially and 4 cm longitudinally from the interaction point. The particles were identified as pions, kaons, protons, electrons or muons via likelihood ratios obtained from the information of the CDC, ACC, TOF, ECL, and KLM (see Chapter 2). The charge separated particle identification efficiencies and fake rates were evaluated using a data driven method by relying on known decays of $D^*, \Lambda$ and $J/\psi$'s (see Chapter 5). These matrices for $\pi, K, p, \mu$ and $e$ were obtained in a fine 17 $\times$ 9 $(p, \cos\theta)$ binning. Where not completely defined by data, an interpolation between adjacent bins or extrapolation based on MC simulation (`Pythia 6.2` for $u, d, s, c$ production, a dedicated $\tau^+\tau^-$ and electro-magnetic process generators) was used. The extracted matrices were inverted; the uncertainties arising from the limited size of the data control samples were assigned as systematic uncertainties. Another important uncertainty in the Belle analysis arises from momentum smearing which migrates the contents of a certain $z$-bin over several, mostly adjacent bins. Smearing was evaluated using MC and corrected for by inverting the smearing matrix (from generated $z$ bin a to reconstructed $z$ bin). The statistical uncertainties on the matrix elements were converted into systematic uncertainties of the multiplicities. Further corrections include in-flight decays, detector interactions as well as reconstruction efficiencies. Furthermore backgrounds from non-QCD processes were estimated using MC and subtracted. The effects of ISR events were removed by evaluating the fraction of events with center-of-mass energies less than 0.5% below the nominal energy in the MC and removing that fraction from the data sample. As the fraction might depend on how well the MC describes the data various MC parameter settings were considered and the spread obtained was assigned as systematic uncertainty. Acceptance effects were corrected by fitting both, data and MC $\cos\theta$ distributions within the measured range, to estimate the fraction of non-reconstructed tracks.

The *BABAR* analysis of the momentum spectrum of the $\eta$ meson begins with a similar hadronic event selection. The $\eta$ mesons are reconstructed in the $\gamma\gamma$ decay mode: high quality neutral clusters with energy above 0.15 GeV are selected, and all pairs of such photon candidates are considered. If any pair has an invariant mass in the range $0.11 < M_{\gamma\gamma} < 0.155$ GeV/$c^2$, consistent with a $\pi^0$ decay, then both photon candidates are rejected. A pair is also rejected if $|\cos\theta_\gamma| > 0.8$, where $\theta_\gamma$ is the angle between either photon momentum and the boost direction of the laboratory system in the $\eta$ rest frame.

Surviving pairs of photons are binned by the pair momentum in the CM frame $p^*$, and the invariant mass distribution in each bin is fitted with a sum of signal and background functions over the range $0.35 < M_{\gamma\gamma} < 0.75$ GeV/$c^2$. The signal function is the sum of a Gaussian distribution and Novosibirsk[180] distribution whose parameters depend on $p^*$ in such a way as to reproduce the line-shape induced by photon energy loss in front of the EMC. The efficiency, defined as the ratio of the yield fitted for the MC sample to the true number of $\eta \to \gamma\gamma$ decays produced, ranges between 24% and 34%. The relative systematic uncertainty includes a component due to normalization of 6.2%, arising from the single photon efficiency, the event selection and the $\eta \to \gamma\gamma$ branching fraction. Additional point-to-point systematic uncertainties arise from the fitting procedure and signal and background shapes. They are 27% at $p^* = 0$, where backgrounds are very high, but then drop rapidly to well below 6% at 1 GeV/$c$. Relative statistical uncertainty drops from 15% at low $p^*$ to 2% above 1.5 GeV/$c$.

### *Results*

Both *BABAR* (Lees, 2013f) and Belle (Leitgab, 2013) published results of the light hadron fragmentation recently,

[179] The jet mass squared is defined as the square of the sum of all particle four-momenta in one hemisphere: $M^2 = \left(\sum_i p_i\right)^2$ where the hemisphere is defined by the normal to the thrust axis.

[180] The Novosibirsk function is an empirical *p.d.f.* defined as $f(M_{\gamma\gamma}) = \frac{1}{\sqrt{2\pi}\sigma} \exp\left[-0.5\left(\ln^2(1 + \Lambda \cdot (M_{\gamma\gamma} - \mu)/\sigma\tau^2) + \tau^2\right)\right]$, where $\Lambda = \sinh(\tau\sqrt{\ln 4})/\sqrt{\ln 4}$, $\mu$ is the peak position, and $\tau$ is the tail parameter.

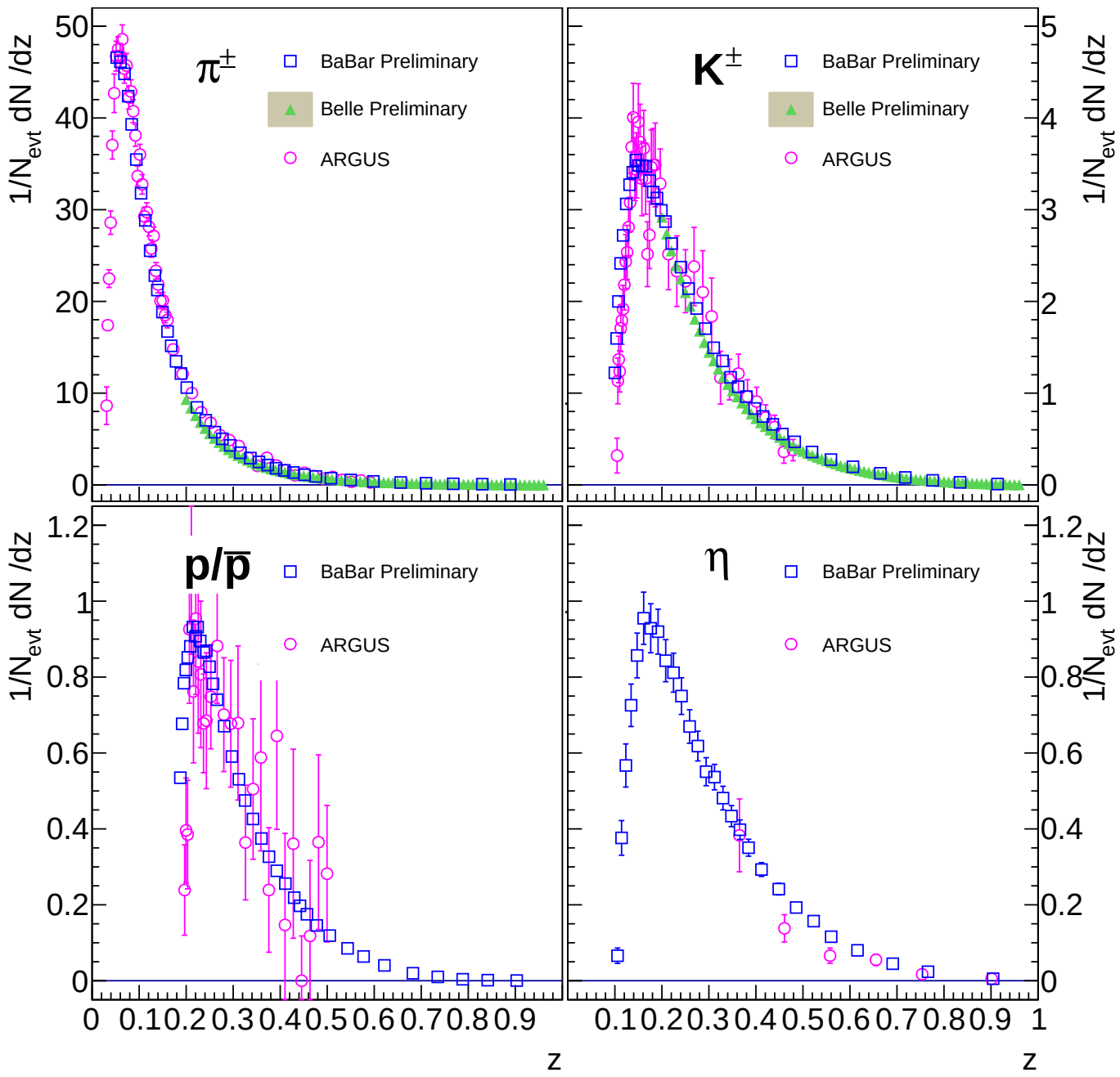

**Figure 24.1.2.** Comparison of the multiplicities for $\pi^{\pm}$, $K^{\pm}$, $p/\overline{p}$, and $\eta$, measured by *BABAR* at $\sqrt{s} = 10.54\,\mathrm{GeV}$ (compiled from Muller, 2004 and Lees, 2013f), Belle at 10.52 GeV (compiled from Leitgab, 2012 and Leitgab, 2013), and ARGUS at 9.98 GeV (Albrecht et al., 1989a, 1990b).

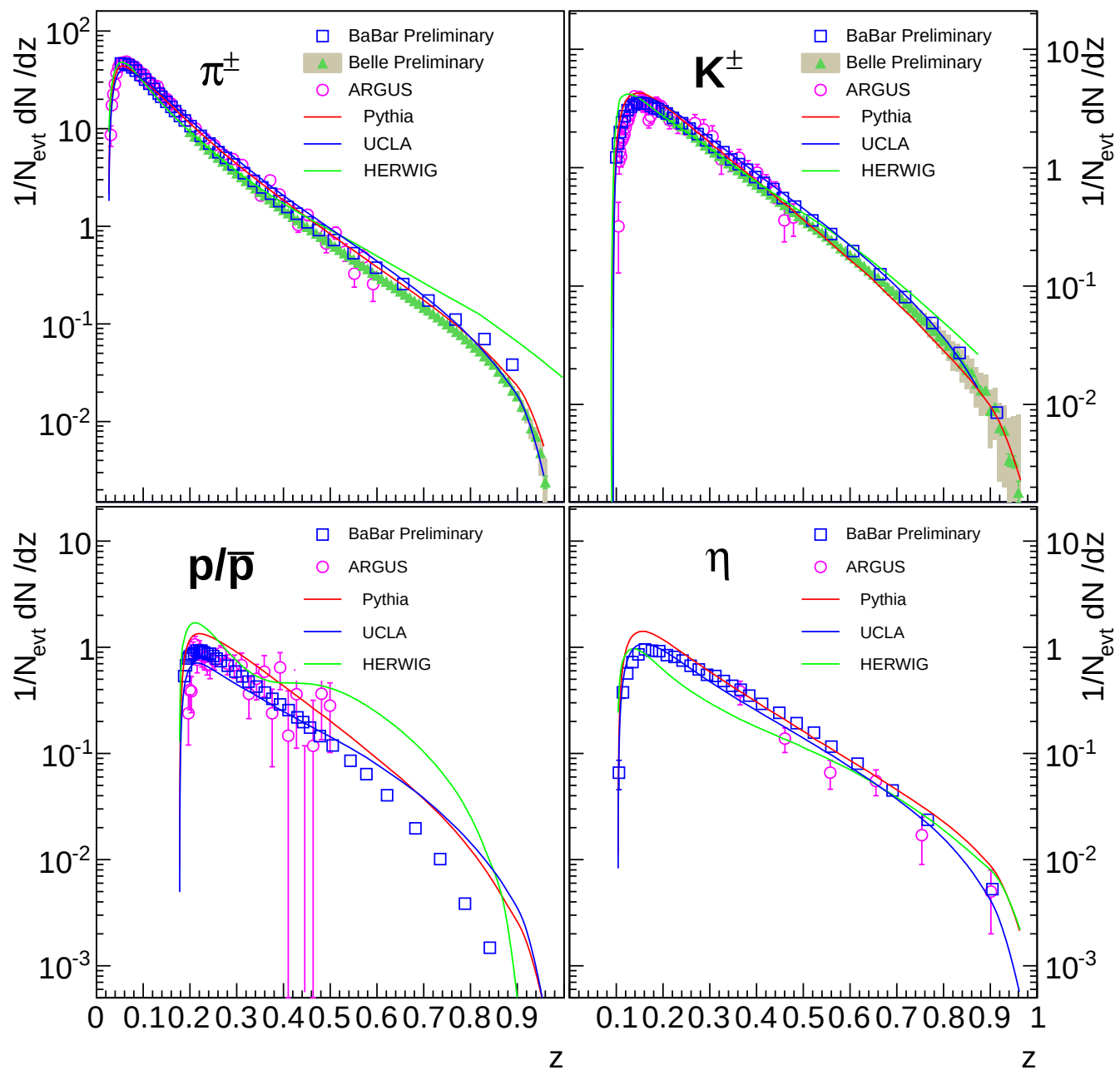

**Figure 24.1.3.** Comparison of the *BABAR* (Muller, 2004; Lees, 2013f), Belle (Leitgab, 2012, 2013), and ARGUS (Albrecht et al., 1989a, 1990b) $\pi^{\pm}$, $K^{\pm}$, $p/\overline{p}$, and $\eta$ multiplicities with the predictions of the `UCLA` (blue), `Jetset/Pythia` (red), and `HERWIG` (green) fragmentation models.

in the form of differential particles multiplicities ($(1/N_{\mathrm{evt}})\ dN/dz$) and in the form of differential cross sections ($(1/\sigma_{\mathrm{had}})\ d\sigma/dz$), respectively. The results presented in this section are compiled from earlier results leading to these publications. The measured particle spectra normalized to the number $N_{\mathrm{evt}}$ of hadronic events are shown as a function of $z$ in Figs 24.1.2 and 24.1.3 with linear and logarithmic vertical scales, respectively. The measured $\eta$ spectrum covers almost full kinematic range, and the others cover the range of $z$ from around 0.1 for pions and kaons and 0.15 for protons to the kinematic limit, which includes the bulk of the kaon and proton spectra, as well as the peak and high side of the pion spectrum. The only previous measurements at a nearby energy, from the ARGUS experiment at $\sqrt{s}=9.98$ GeV (Albrecht et al., 1989a, 1990b), are also shown in Fig. 24.1.2. Only statistical errors are shown for the *BABAR* data, as the correlated 3-6% systematic uncertainties are dominated by normalization. The results from the three experiments are in fair agreement within uncertainties, and the $B$ Factory results are far more precise in most ranges and extend the coverage significantly when compared with previous experiments. For pions, the total uncertainties are comparable and are correlated over significant $z$ ranges in the *BABAR* and ARGUS cases while they are significantly more precise in the Belle case. The ARGUS data extend to lower $z$ values, so that the majority of the spectrum is covered between the three experiments. At low $z$, *BABAR* and ARGUS data sets differ by up to $(7 \pm 4)\%$, which might indicate the expected small scaling violation.

Figure 24.1.3 compares the *BABAR*, Belle and ARGUS multiplicities with the predictions of the three fragmentation models discussed above. Default parameter values are used, which have been chosen based on previous data, mostly at higher energies. The shape of the bulk of the $\pi^{\pm}$ spectrum is described qualitatively by all three models, but no model describes the spectrum well in detail. `Jetset/Pythia` and `UCLA` also describe the $K^{\pm}$ cross section reasonably well, whereas `HERWIG` peaks at lower $z$ value. `Jetset` and `UCLA` also describe the $\eta$ cross section fairly well, though `UCLA`'s spectrum is slightly too soft and `Jetset`'s spectrum predicts higher multiplicity than the data both around the peak and at very high $z$ values. `HERWIG`'s spectrum shape does not agree with the data. The proton spectrum is quite problematic: `Jetset` describes the shape qualitatively, but is consistently above the data (*i.e.* predicts higher multiplicity); `UCLA` describes the shape in the peak region, but then falls much too slowly, rising above the data at high $z$; `HERWIG` also describes the shape in the peak region, but is far too high overall and exhibits an experimentally unobserved structure at high $z$ values.

Similar deficiencies in these models have been reported at higher energies (Abe et al., 1999; Abreu et al., 1998; Aihara et al., 1988; Akers et al., 1994a; Braunschweig et al., 1989; Buskulic et al., 1995; Itoh et al., 1995), although earlier MC versions were used and parameter values varied. One should note that the deviations of predictions from the data had the same sign at higher energies, suggesting that the scaling properties might be well simulated.

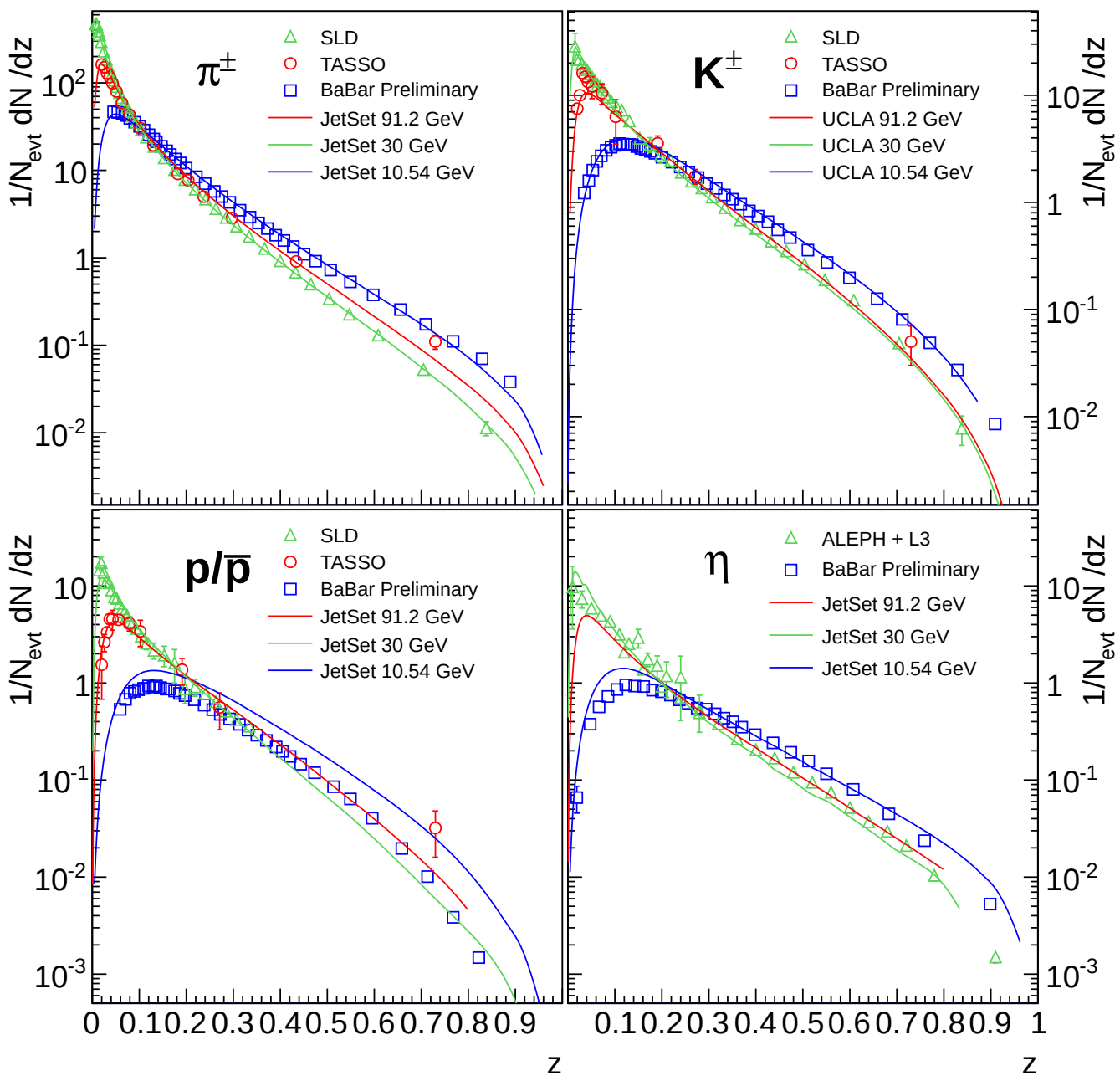

**Figure 24.1.4.** $\pi^{\pm}$, $K^{\pm}$, $p/\bar{p}$ and $\eta$ multiplicities measured at three different CM energies, compared with the predictions of the simulations described in the text. *BaBar* data are from Muller (2004).

To test the scaling properties of the models, each was run with its current default parameters at various energies and compared with the available data.

Figure 24.1.4 shows a scaling test of the `Jetset` model using $\pi^{\pm}$ cross sections from *BABAR*, TASSO (Braunschweig et al., 1989) and SLD (Abe et al., 1999). The latter two experiments provide the best precision and/or high-$z$ coverage at $\sqrt{s}$ near 30 GeV and at the $Z^0$, respectively. Data from other experiments are consistent and yield the same conclusions. Strong scaling violations are evident, both at low $z$ due to the pion mass and at high $z$ from the running of the strong coupling $\alpha_S$.`Jetset` provides an excellent description of all three data sets, with differences of only few percent at very low and very high $z$ values. `UCLA` and `HERWIG` also describe the scaling violation well, although they do not reproduce the spectrum as well at any energy.

Figure 24.1.4 also shows a similar test of the `UCLA` model for $K^{\pm}$ cross sections. `UCLA` describes the *BABAR* and Belle results best, and similar scaling is predicted by the other models. Here, the different flavor composition of the three samples modifies the expected scaling violation. Charged kaons from $b\bar{b}$ events, which are absent from the *BABAR* data, raise the TASSO cross section in the 0.1–0.3 region, but do not contribute at high $z$. At the $Z^0$, the relative production of up- and down-type quarks changes dramatically, and the larger $b\bar{b}$ and $s\bar{s}$ event fractions raise the simulated cross section to nearly the same level as at 35 GeV for $z$ above about 0.2. The flavor dependence has been shown (Abe et al., 1999; Abreu et al., 1998) to be modeled at the $Z^0$ at the level of about 10%. The change in the measured cross sections is about 15% less than predicted, but this could be due to issues with the flavor dependence.

Similar results are obtained for the $\eta$ meson from ALEPH and L3 data at 91 GeV (Adriani et al., 1992; Barate et al., 2000), and compared with the `Jetset` predictions. Again, other data and models give the same conclusions. The flavor dependence is smaller, and the discrepancy at the $Z^0$ is larger than for $K^{\pm}$, perhaps indicating a failure of the models.

For protons, also shown in Fig. 24.1.4, the `Jetset` model is tested with one parameter value changed, the probability for a given string break to produce a diquark-antidiquark, rather than quark-antiquark, pair, from 0.1 to 0.085, which provides a good description of the higher-energy data. Here, the simulated high-$x_p$ scaling violation between 10.54 and 34 GeV is about the same as for the pions, but that between 34 and 91 GeV is slightly larger since fast protons are expected to be produced predominantly in $u\bar{u}$ and $d\bar{d}$ events. The prediction for 10.54 GeV rises well above the *BABAR* data, exceeding it by as much as a factor of 4.5 at $z$ =0.9. Similar behavior is seen for `Jetset` with default parameters, `HERWIG`, and `UCLA` at high $x_p$. This indicates that we do not understand the scaling properties of protons, or perhaps of baryons or heavier hadrons in general.

These data can be used to test the predictions of MLLA QCD combined with the ansatz of LPHD (Azimov, Dokshitzer, Khoze, and Troyan, 1985), by transforming to the variable $\xi = \ln(\sqrt{s}/2p^*)$. This representation emphasizes the low momentum region (large $\xi$). It is predicted that: the $\xi$ distribution would be approximately Gaussian over a range of $\sim 1$ unit around it's peak position $\xi^*$; a distorted Gaussian should describe the distribution over a wider range; $\xi^*$ should decrease exponentially with hadron mass at a given CM energy $\sqrt{s}$ and $\xi^*$ should increase logarithmically with $\sqrt{s}$ for a given hadron. Conventionally, $\xi^*$ is found by fitting a Gaussian distribution to the data over sets of points within 0.5–1 units of the approximate peak position. Next, the widest roughly symmetric range about this position is found in which a Gaussian fit gives a good $\chi^2$, and this range is then extended as far as possible in one direction. Results of such fits and the ranges are listed in Table 24.1.1. Acceptable fits were found over ranges at least 1 unit wide, consistent with the prediction.

**Table 24.1.1.** Results of the Gaussian and distorted Gaussian (where a skewness term and a kurtosis term are added) fits to the $\xi$ distributions. The fit ranges and the peak positions $\xi^*$ are reported (Muller, 2004).

| Particle | Gaussian | | Distorted Gaussian | |
|---|---|---|---|---|
| | Fit range | $\xi^*$ | Fit range | $\xi^*$ |
| $\pi^{\pm}$ | 1.7 - 3.0 | 2.36±0.01 | 0.0 - 3.2 | 2.36±0.01 |
| $K^{\pm}$ | 1.0 - 2.2 | 1.64±0.01 | 0.0 - 3.2 | 1.64±0.01 |
| $\eta$ | 0.9 - 2.2 | 1.48±0.02 | | |
| $p/\bar{p}$ | 1.0 - 2.2 | 1.61±0.01 | 0.0 - 2.8 | 1.61±0.01 |

**Table 24.1.2.** The fraction of each particle's spectrum covered by the *BABAR* (Muller, 2004) measurement is given in the second column. The total multiplicity per $e^+e^- \to q\bar{q}$ event at 10.54 GeV measured by *BABAR* in the third column are compared with the predictions of fragmentation models and previous results from CLEO (at 10.49 GeV; Behrends et al., 1985) and ARGUS (at 9.98 GeV; Albrecht et al., 1989a). The first error on each *BABAR* result is experimental and the second is from the extrapolation procedure.

| Particle | Coverage | *BABAR* | `Jetset` | `UCLA` | `HERWIG` | CLEO | ARGUS |
|---|---|---|---|---|---|---|---|
| $\pi^\pm$ | 0.878±0.015 | 6.405±0.134±0.106 | 6.22 | 6.44 | 6.31 | 8.3±0.4 | 6.38±0.12 |
| $K^\pm$ | 0.985±0.006 | 0.910±0.017±0.006 | 0.934 | 1.010 | 1.010 | 1.3±0.2 | 0.888±0.030 |
| $\eta$ | 1.0 | 0.276±0.017±0.000 | 0.354 | 0.278 | 0.233 | – | 0.19±0.06 |
| $p/\bar{p}$ | 0.966±0.008 | 0.235±0.011±0.002 | 0.336 | 0.217 | 0.46 | 0.40±0.06 | 0.271±0.018 |

Table 24.1.1 reports also the results of the fits performed adding small skewness $s$ and kurtosis $\kappa$ terms to the Gaussian distribution.

$$G'(\xi) = \frac{N}{\sigma\sqrt{2\pi}} \exp\left(\frac{\kappa}{8} + \frac{s\delta}{2} - \frac{(2+\kappa)\delta^2}{4} + \frac{s\delta^3}{6} + \frac{\kappa\delta^4}{24}\right), \tag{24.1.9}$$

where $\delta = (\xi - \xi^*)/\sigma$, $\sigma$ is the square root of the variance. The fitted ranges are significantly larger, consistent with the MLLA QCD prediction. The values of $\xi^*$ measured by *BABAR* and previous experiments at higher energies for the different particles are shown in Fig. 24.1.5. The lines simply connect the precise points at the $Z^0$ with those from *BABAR*. The other data points are consistent with these lines, and hence with the expected logarithmic energy dependence, but more precise data at other energies are needed to test this prediction. There is a clear difference between the pions and kaons that increases slowly with energy; the values for $\eta$ (measured only at *BABAR* and at the $Z^0$), are slightly below those for $K^\pm$. However, the proton data are inconsistent with an overall decrease with hadron mass.

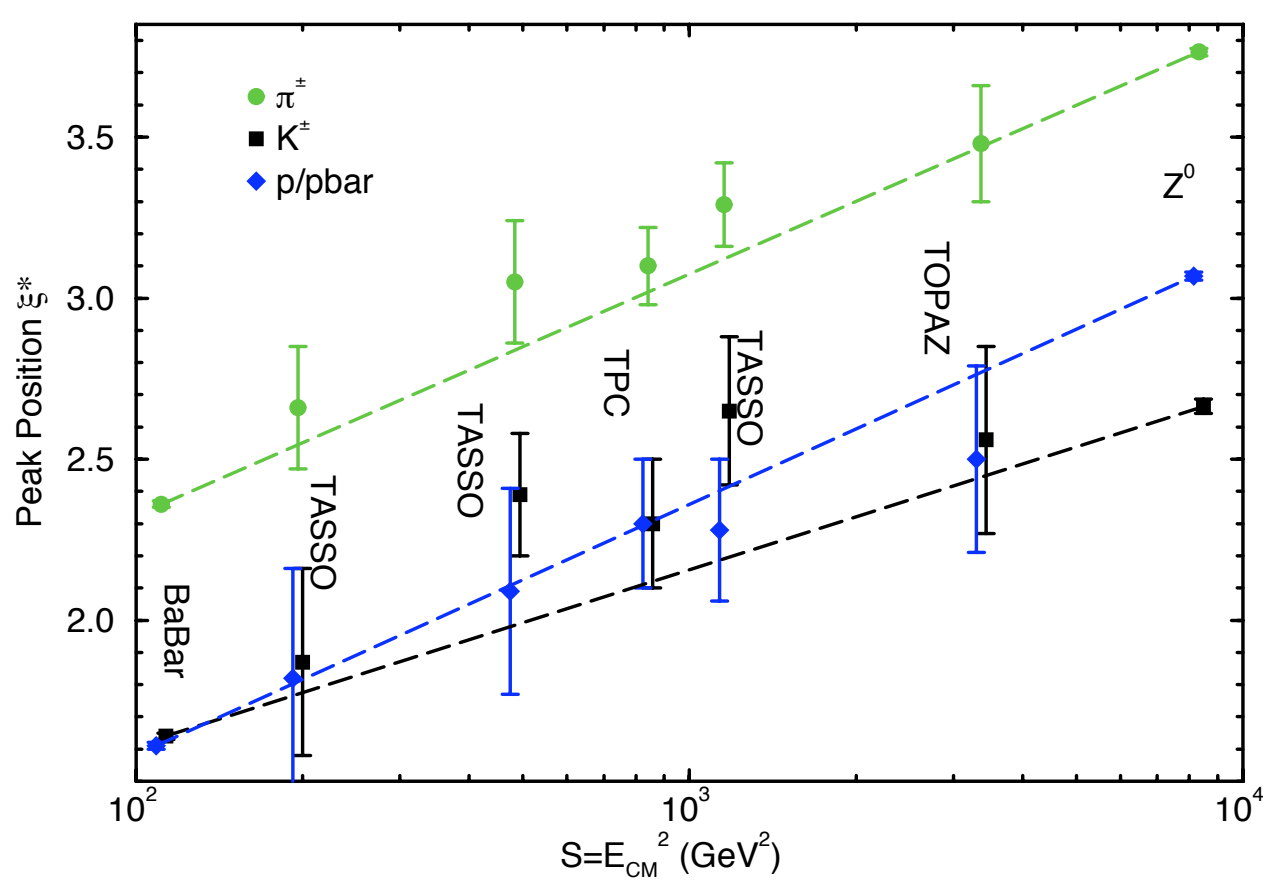

**Figure 24.1.5.** Peak position $\xi^*$ vs $e^+e^-$ CM energy for charged pions, kaons, and protons from various experiments. The lines connect the precise points at the $Z^0$ with those from *BABAR* (Muller, 2004).

Total multiplicities of each particle type per event are calculated by integrating the differential rates, taking all uncertainties and their correlations into account, and extrapolating into any unmeasured regions. The second step is model dependent: correction factors are evaluated using a combination of the three models and a number of fits to the $\xi$ distributions, and have large uncertainties when substantial fractions of the spectrum are not measured. The $\eta$ spectrum is measured over the full kinematic range, so this is not an issue. The good coverage for kaons and protons makes their corrections and their uncertainties fairly small. The coverage for pions at low momenta does not extend much below the peak (see Fig. 24.1.2), requiring large correction and giving the dominant uncertainty. The correction factors and the total rates are listed in Table 24.1.2, along with previous results from CLEO (Behrends et al., 1985) and ARGUS (Albrecht et al., 1989a) and the predictions of the three models. *BABAR* and ARGUS results are in good agreement, while CLEO measures significantly higher rates for all particle types.

Differential production ratios for pairs of particles are sensitive to specific features of the hadronization process, and many of the systematic uncertainties cancel at least partially. It is equivalent and conventional to report the fractions $f_\pi$, $f_K$ and $f_p$ of all charged hadrons that are pions, kaons and protons, respectively. Pions dominate the charged hadron production at low $z$, as is expected from their lower mass and the contributions from many decays of heavier hadrons. As $z$ increases, the pion fraction drops as the kaon and proton fractions rise toward values of about 35% and 8%, respectively. At higher $z$, the trend reverses due to kinematics, especially for protons which must be produced along with an anti-baryon. The three models describe the general trend of the data, but none describes either the shape or the magnitude at all momenta.

### *Global analysis for fragmentation functions*

Since there were measurements of the multiplicities given by Eq. (24.1.1) at various facilities, global analyses have been made using the available world data. From the analyses, the optimum FFs were determined, and even their uncertainties were estimated.

In the same way as global analyses for the parton distribution functions are determined, the FFs are expressed in terms of a number of parameters at the initial scale $Q^2$ ($\equiv Q_0^2$). Usually, a simple polynomial form is used:

$$D_i^h(z, Q_0^2) = N_i^h z^{\alpha_i^h} (1-z)^{\beta_i^h}, \qquad (24.1.10)$$

because the functions should vanish at $z = 1$. Here, $N_i^h$, $\alpha_i^h$, and $\beta_i^h$ are parameters to be determined by a $\chi^2$ minimization of $e^+ + e^- \to h + X$ data. The initial scale $Q_0^2$ is arbitrary. However, it is, for example, assumed that $Q_0^2 = 1\,\mathrm{GeV}^2$ for light quark and gluon functions, and above the mass thresholds $m_c^2$ and $m_b^2$ for charm and bottom functions, where $m_c$ and $m_b$ are charm- and bottom-quark masses, respectively. For light hadrons ($h$), one typically separates pions ($\pi^+ + \pi^-$), kaons ($K^+ + K^-$), and protons/anti-protons ($p+\bar{p}$). Because the second moments $M_i^h$ should satisfy the sum rule of Eq. (24.1.6), it is useful to take $M_i^h$ as one of the parameters instead of $N_i^h$. The parameters are related with each other by the relation

$$N_i^h = \frac{M_i^h}{B(\alpha_i^h + 2, \beta_i^h + 1)}, \qquad (24.1.11)$$

where $B(\alpha_i^h + 2, \beta_i^h + 1)$ is the beta function.

In analyzing the light hadrons, a common function is assumed for favored fragmentation functions (see Section 24.1.1) from up and down quarks while different parameters are allowed for a favored FF from a strange quark by considering the mass difference. Also different parameters are assigned for disfavored FFs. A flavor symmetric form is assumed for disfavored FFs from light quarks (up, down, and strange quarks) due to lack of experimental information, although the light antiquark distributions are not flavor symmetric in the unpolarized parton distribution functions (Kumano, 1998).

In Fig. 24.1.6, FFs for $(\pi^+ + \pi^-)/2$ determined by "HKNS" (Hirai, Kumano, Nagai, and Sudoh, 2007a,b) are shown, together with other parameterizations: "KKP" (Kniehl, Kramer, and Potter, 2000), "Kretzer" (Kretzer, 2000), "AKK" (Albino, Kniehl, and Kramer, 2005, 2008), and "DSS" (de Florian, Sassot, and Stratmann, 2007a,b; Epele, Llubaroff, Sassot, and Stratmann, 2012). Since neither *BABAR* and Belle results have been published at the time of this analysis, they were not yet taken into account. These functions were obtained in the NLO ($\overline{\mathrm{MS}}$) scheme and the uncertainty bands were obtained in the HKNS analysis by using the Hessian method (Pumplin, Stump, and Tung, 2001). The gluon and light-quark functions are shown at $Q^2 = 2\,\mathrm{GeV}^2$, where the uncertainties are generally large. The charm- and bottom-quark functions are shown at the scale of their mass thresholds $Q^2 = m_c^2$ or $m_b^2$. Disfavored-quark and gluon functions, for example $s$-quark functions of Kretzer and AKK, are completely different between the analysis groups; however, they agree within the uncertainties. One can notice that the disfavored-quark and gluon FFs have large uncertainties, which should be significantly improved by the recently published Belle and *BABAR* measurements once they are included. There are available codes from the different analysis groups for calculating the FFs at a given kinematical condition of $z$ and $Q^2$. A summary of the status of the various FFs is given by Albino et al. (2008); an online generator is provided by Arleo and Guillet (2008). Recent works on the analyses of the FFs can be found in Christova and Leader (2009) and Albino and Christova (2010).

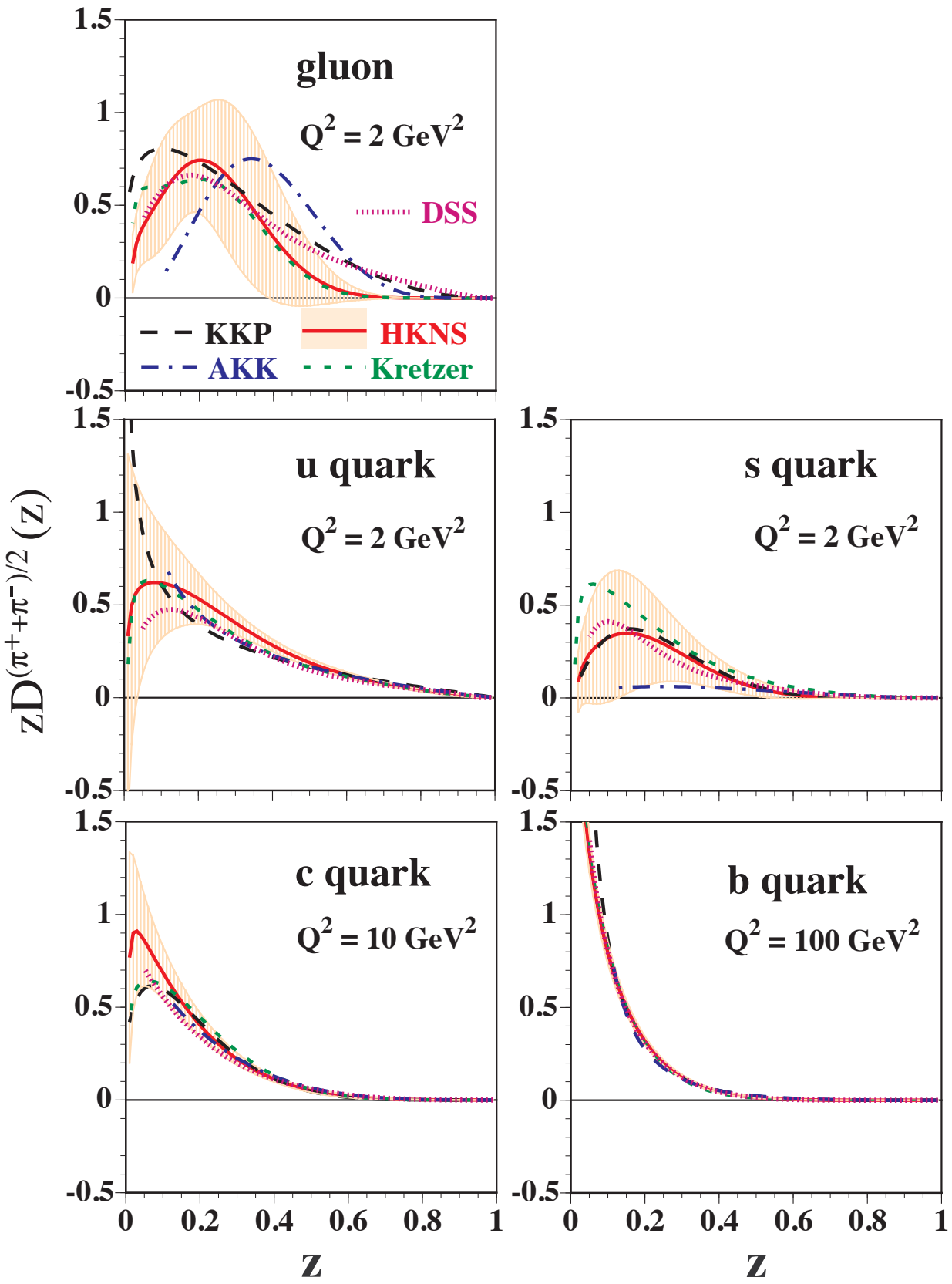

**Figure 24.1.6.** Determined fragmentation functions for the pion and their comparison with other parameterizations (Hirai, Kumano, Nagai, and Sudoh, 2007a,b). The shaded bands indicate estimated uncertainties of the HKNS.

The favored and disfavored FFs reflect internal flavor content of a hadron, which leads to an interesting suggestion that exotic hadrons could be found by investigating their FFs (Hirai, Kumano, Oka, and Sudoh, 2008). As an example, internal structure of the controversial $f_0(980)$ meson, which is possibly $q\bar{q}$ or tetra-quark ($qq\bar{q}\bar{q}$), could be determined if accurate data become available. For example, a future super flavor factory with forty times higher luminosity could provide the accurate information needed to distinguish the disfavored functions from the favored ones of the $f_0$ meson.

#### 24.1.2.2 Charmed hadrons

Belle and *BABAR* have measured fragmentation functions for several charmed hadrons, the heaviest particles avail-able for study below the $\Upsilon(4S)$. These are generally expressed in terms of $x_p = p^*/p_{\max}$, where $p_{\max} = \sqrt{s/4 - m_h^2}$ is the maximum momentum for production via the $e^+e^- \to q\overline{q}$ process. In this variable, all hadrons have the same kinematic range, $0 \le x_p \le 1$. In $B$ Factories where pairs of $B$ mesons decay nearly at rest in the CM system the $x_p$ for hadrons from $B$ decay cannot exceed 0.5 Charmed hadrons must be reconstructed in a particular decay mode. By convention, an analysis is described for a particular hadron and decay mode, but the inclusion of the charge conjugate state and decay mode is always implied.

Belle (Seuster, 2006) have studied several charmed mesons, the ground state $D^0$, $D^\pm$, $D_s^\pm$ and the excited $D^{*0}$ and $D^{*\pm}$, as well as the lowest-mass charmed baryon $\Lambda_c^\pm$. Events are selected by requiring at least three charged tracks and a calorimeter energy sum between 10% and 80% of the CM energy. There are also requirements on the average cluster energy, the invariant mass of the particles in each thrust hemisphere, and the position of the event vertex. This selection is 87% efficient for $c\overline{c}$ events. Charged tracks are required to be consistent with originating from the event vertex, and are identified as $\pi^\pm$, $K^\pm$ or $p\overline{p}$ using a combination of information from the drift chamber, time-of-flight and Cherenkov systems (see Chapter 5). A loose selection is used, in which identification efficiencies are above 95% (80% for protons) and misidentification rates are at most 26% (7%). Candidate $\pi^0$ mesons are formed from pairs of photon candidates with energy above 30 MeV and invariant mass near the $\pi^0$ mass.

Candidate $D^0 \to K^-\pi^+$ decays are formed by combining an identified $K^-$ with an identified $\pi^+$. Similarly, $D^+ \to K^-\pi^+\pi^+$ and $\Lambda_c^+ \to pK^-\pi^+$ candidates are formed by combining three identified tracks, and $D_s^+ \to K^+K^-\pi^+$ candidates are selected in which the $K^+K^-$ combination has an invariant mass within 7 MeV/$c^2$ of the nominal $\phi$ meson mass. Candidates of each type are binned in $x_p$, and the number of true charmed hadrons in each bin is estimated by fitting their invariant mass distribution in the region near the relevant hadron mass.

Candidate $D^{*+} \to D^0\pi^+$ decays are formed from those $D^0$ candidates with an invariant mass within 15 MeV/$c^2$ of the nominal $D^0$ mass, combined with each slow, positively charged track, assumed to be a $\pi^+$. Similarly, $D^{*+} \to D^+\pi^0$ and $D^{*0} \to D^0\pi^0$ candidates are formed from $D$ candidates with a mass within 15 MeV/$c^2$ of the relevant nominal mass, combined with a soft $\pi^0$ candidate. These $D^*$ candidates are binned in $x_p$, and the number of true $D^*$ mesons in each bin is extracted from a fit to the distribution of the mass difference $\Delta m$ between the $D^*$ and $D$ candidates. Since the true mass difference is close to threshold, $\Delta m$ has better resolution than any individual invariant mass. The yields are divided by the reconstruction efficiency and the relevant branching fraction(s) to give differential cross sections as functions of $x_p$.

*BABAR* have studied the charmed baryons $\Lambda_c^\pm$ (Aubert, 2007p), $\Xi_c^0$ (Aubert, 2005z), $\Omega_c^0$ (Aubert, 2007ao), containing zero, one and two strange valence quarks, respectively, in addition to the charm quark. Since charmed baryons can only be produced in $e^+e^- \to c\overline{c}$ events, there is no event selection other than the requirement of enough tracks in the event to reconstruct the particle in question in its target decay mode.

The $\Lambda_c^\pm$ study uses a sample of 9.5 fb$^{-1}$ of off-resonance data and reconstructs the 3-body decay mode $\Lambda_c^+ \to pK^-\pi^+$. High quality tracks are selected and identified as described in Section 24.1.2.1. Each set of an identified $p$, $K^-$ and $\pi^+$ is considered a $\Lambda_c^+$ candidate, and each track's momentum at its point of closest approach to the beam axis is corrected for energy loss using the correct mass. The invariant mass is calculated with a resolution that varies from 3.75 MeV/$c^2$ at low $x_p$ to 5.75 MeV/$c^2$ at high $x_p$.

The reconstruction efficiency varies rapidly near the edges of the detector acceptance, so a tight fiducial requirement is made that the polar angle of the $\Lambda_c^\pm$ candidate $\theta_\Lambda$ in the $e^+e^-$ CM frame satisfies $-0.7 < \cos\theta_\Lambda < 0.2$ This rejects all candidates in regions with efficiency below 5%, including those with momentum below 0.7 GeV/$c$ in the laboratory frame. A feature of the boosted CM system is that soft $\Lambda_c^\pm$ are boosted forward in the laboratory and often have all three decay tracks within the acceptance, giving access to the full range of $x_p$ with good efficiency and resolution. The efficiency for low-(high-)$x_p$ $\Lambda_c^\pm$ averages to 8% (17%).

In order to reduce model dependence, each candidate is given a weight equal to the inverse of its efficiency. The distribution of weighted invariant mass is then fitted in several bins of $x_p$ to extract a signal yield. Comparisons of results for $\Lambda_c^+$ and $\overline{\Lambda}_c^-$ and of results from different regions of $\cos\theta_\Lambda$ provide powerful consistency checks and constraints on systematic errors.

The $\Xi_c^0$ study uses a sample of 10.7 (105.4) fb$^{-1}$ of off-(on-)resonance data and reconstructs the two decay modes $\Xi_c^0 \to \Omega^-K^+$ and $\Xi^-\pi^+$, where $\Omega^- \to \Lambda^0K^-$, $\Xi^- \to \Lambda^0\pi^-$ and $\Lambda^0 \to p\pi^-$. Identified protons are combined with negatively charged tracks, assumed to be pions, to form $\Lambda^0$ candidates. Those with an invariant mass within $3\sigma$, where $\sigma$ is the fitted mass resolution, of the nominal $\Lambda^0$ mass, are accepted, and a kinematic fit is performed to each pair of tracks, in which their mass is constrained to the nominal value. Each of these reconstructed $\Lambda^0$ objects is then combined with all identified $K^-$ to form $\Omega^-$ ($\Xi^-$) candidates. Again, those with an invariant mass within $3\sigma$ of the nominal value are accepted, and subjected to a mass constrained fit (see Section 6.3). They are then combined with identified $K^+$ to form $\Xi_c^0$ candidates.

The position of the fitted $\Omega^-$ ($\Xi^-$) decay vertex must be at least 1.5 mm (2.5 mm) from the beamline, and the $\Omega^-$ and $\Lambda^0$ decay vertices must be separated by at least 3 mm. The $\Xi^-$ vertex must be farther from the beamline than the $\Lambda^0$ vertex, and the scalar product of the $\Lambda^0$ momentum vector and the displacement vector from its production vertex to its decay vertex must be positive. The surviving $\Xi_c^0$ candidates are binned in $x_p$, and signal yields are extracted by counting those with invariant mass in a signal window about the nominal $\Xi_c^0$ mass and subtracting backgrounds estimated from mass sidebands. The yields are divided by the simulated efficiency in each bin to give differential cross sections. Measurements are made using the on-resonance data for $x_p$ above 0.46, to remove the contribution of $B$ decays to $\Xi_c$ production. [181] Below the kinematic limit, the off-resonance data must be used, but are statistics limited and consistent with zero.

The $\Omega_c^0$ study uses a sample of 21.6 (208.9) $\text{fb}^{-1}$ of off-(on-)resonance data and reconstructs the four decay modes $\Omega_c^0 \to \Omega^-\pi^+$, $\Omega^-\pi^+\pi^0$, $\Omega^-\pi^+\pi^-\pi^+$, and $\Xi^-K^-\pi^+\pi^+$. Candidate $\Omega^-$ and $\Xi^-$ hyperons are reconstructed as above, with slightly different selection criteria. Photon clusters with energy above 80 MeV are paired, and those with total energy above 200 MeV and invariant mass in the range 120-150 $\text{MeV}/c^2$ are retained as $\pi^0$ candidates. The $\Omega^-$ and $\Xi^-$ candidates are then combined with identified $\pi^\pm$, $K^-$ and $\pi^0$ candidates to form $\Omega_c$ candidates in the four modes.

A likelihood is formed from the simulated signal and background distributions of: the flight distance of the $\Omega^-$ or $\Xi^-$ divided by its uncertainty; its momentum in the CM frame; the total momentum recoiling against it in the CM frame; and the $\pi^0$ momentum in the laboratory frame. This likelihood must exceed a threshold that depends on momentum. The surviving $\Omega_c^0$ candidates are binned in $x_p$, and signal yields are extracted and divided by the simulated efficiency. Measurements are made only for on-resonance data, assumed to be free from $B\overline{B}$ events above the kinematic limit of 0.44. The off-resonance data are statistics limited and consistent with zero below this momentum. There is a normalization uncertainty of 6.5%, and the statistical uncertainties are much larger than the remaining systematic uncertainties.

### *Results*

The differential cross sections measured for several charmed hadrons are shown as a function of $x_p$ in Fig. 24.1.7 and are not corrected for branching ratios into the displayed decay channels. Those for the $\Xi_c$ and $\Omega_c$ are for on-resonance data, and the low-$x_p$ regions left of the vertical lines are dominated by $B\overline{B}$ decays. $D_s^+$ mesons were also studied earlier in Aubert (2002d) but due to lack of numerical values for individual $x_p$ bins they were not included in this figure. These are first measurements for the $D_s^+$, $\Xi_c$, and $\Omega_c$; the others are consistent with previous measurements, but much more precise, except for similarly precise $D^{(*)0}$ and $D^{(*)+}$ results from CLEO. The $\Lambda_c^+$ spectra from Belle and *BABAR* are consistent.

In all cases, the cross section is very low at low $x_p$, rises steadily toward a broad peak near 0.6, then drops to zero as $x_p \to 1$. The contrast between this shape and those for the $\pi^\pm$, $K^\pm$ and $p/\overline{p}$ seen above is both striking and expected. The position of the peak at rather high $x_p$ is due to the large charm quark mass, and the low values at low $x_p$ is consistent with leading charmed hadrons from $e^+e^- \to c\overline{c}$ events being dominant.

A few $D^{(*)}$ mesons are observed with $x_p$ very near (and above) unity. These are due to the exclusive processes $e^+e^- \to D^{(*)}\overline{D}^{(*)}$, involving several combinations of ground state and excited $D$ mesons. Such events are interesting in their own right, but are a small minority, and not considered to be from the jet fragmentation process. There is no sign of the exclusive production of a charmed baryon and antibaryon.

The $D^{*0}$ and $D^{*+}$ spectra are consistent with each other, as expected due to the isospin symmetry and due to the fact that the mass difference amounts to only 3 $\text{MeV}/c^2$. They are similar in shape to the $D^0$ and $D^+$ spectra, but shifted to higher $x_p$ values by about 0.04 units, as expected since the $D^{*0}$ mesons are 40 $\text{MeV}/c^2$ heavier. The $D^0$ is produced twice as often as the $D^+$ and its spectrum peaks at a slightly lower $x_p$ than the $D^+$ spectrum. This is due to the fact that the $D^{*0}$ is too light to decay into $D^+\pi^-$, so its decays produce more $D^0$ mesons with a softer spectrum.

The $D_s^+$ accounts for 10% of the ground state mesons and their spectrum is both more strongly peaked and peaks at higher $x_p$ than any of the $D$ or $D^*$ spectra. This is not expected in current fragmentation models. The $\Xi_c$ and $\Lambda_c$ spectra have similar shapes, with the $\Xi_c$ spectrum shifted toward higher $x_p$ values, as shown in Fig. 24.1.7. The $\Omega_c$ spectrum is consistent with similar behavior, but statistics are low. From the distributions it is also clear that the baryon spectra have different shapes from the meson spectra: they are more sharply peaked, and fall to zero more rapidly at high $x_p$. Also, the meson and baryon spectra peak in the same region of $x_p$, even though the baryons are much heavier. Such differences are predicted by some fragmentation models.

The total cross sections for the continuum production $e^+e^- \to X_cY$ are summarized in Table 24.1.3.

Several fragmentation models exist with different assumptions on the functional form in the fractional energy $z$ and the transverse momentum relative to the initial parton.

Belle tested the models by Bowler (1981), Lund (Andersson, Gustafson, Ingelman, and Sjöstrand, 1983), "KLP" (Kartvelishvili, Likhoded, and Petrov, 1978), "CS" (Collins and Spiller, 1985), and Peterson (Peterson, Schlatter, Schmitt, and Zerwas, 1983). The model predictions were tested against both $\Lambda_c^+$ and meson spectra, and several model parameters were optimized in order to achieve a better agreement. No model gives a good $\chi^2$ for all of the spectra, but for most mesons, Bowler and Lund give reasonable qualitative descriptions, KLP and CS are somewhat worse, and Peterson is quite poor. Consequently, the Belle data ruled out the Peterson model which was widely used before while KLP and CS are rather disfavored. In both the `Jetset` (Lund) and `UCLA` models, the fitted parameter values differ substantially from the default values which give the best description of lighter hadron data.

[181] The maximal momentum of $\Xi_c^0$ in a $B$ meson decay conserving the baryon number would be achieved in a 2-body $B \to \Xi_c^0\overline{p}$ decay and would correspond to $x_p = 0.41$ for $\Xi_c^0$.

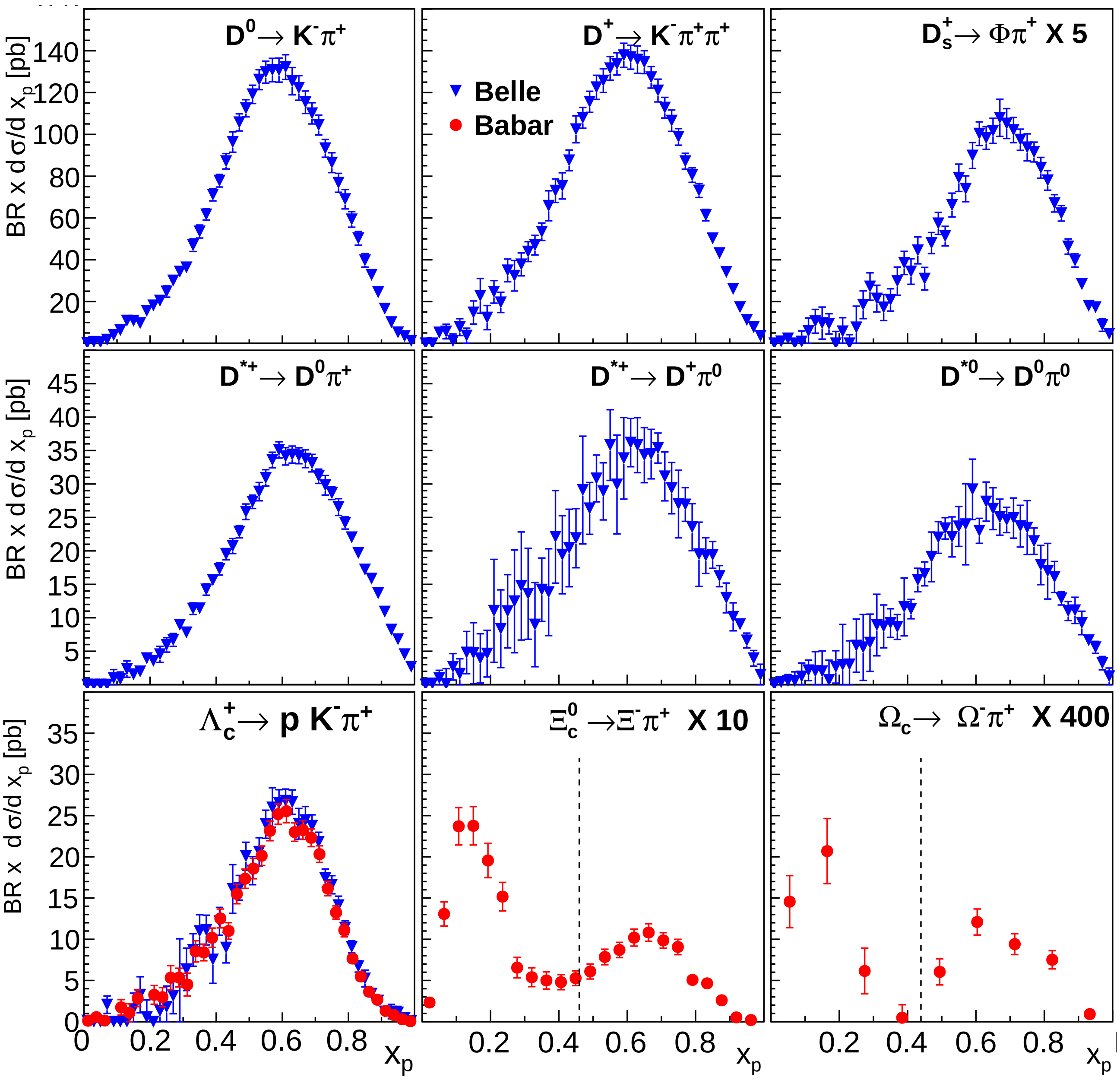

**Figure 24.1.7.** Fragmentation functions times branching ratios for charmed mesons and baryons as a function of the fractional momentum $x_p$. The Belle measurements (Seuster, 2006) are displayed in blue (triangles), while the *BABAR* measurements (Aubert, 2005z, 2007p,ao) are displayed in red (circles). Some measurements were performed on the $\Upsilon(4S)$ resonance and therefore contain B decays at $x_p$ values below the vertical lines displayed.

The models have also been tested directly against the data in *BABAR* for the $\Lambda_c^+$ spectrum (Aubert, 2007p). Eight models are tested within `Jetset` by re-weighting with varying parameter values so as to minimize the $\chi^2$ of a comparison with the data. The `UCLA` model (Chun and Buchanan, 1998) is fitted similarly, whereas `HERWIG` (Marchesini et al., 1992) has no relevant free parameters. In general, the fitted parameters are different between mesons and baryons, and between light and heavy hadrons.

These results indicate the need for different treatment of heavier hadrons and provide new, precise input for the development and tuning of such models. One has to keep in mind, however, that models contain a relatively large number of parameters which might affect the light and heavy hadron fragmentation descriptions.

#### $\Lambda_c^+ \overline{\Lambda}_c^- X$ *Events*

*BABAR* has studied events containing both a $\Lambda_c^+$ and a $\overline{\Lambda}_c^-$ (Aubert, 2010b), following a study by the CLEO collaboration (Bornheim et al., 2001). CLEO found ∼4 times more events than expected from the `Jetset` model, in which the $\Lambda_c^+$ and $\overline{\Lambda}_c^-$ are always accompanied by an antibaryon and baryon in their respective jets (4−baryon events). They interpreted this excess as evidence for a long-range baryon number correlation, but could not exclude either 4-baryon events or few-body processes, such as charmed pentaquark pair production, as a possible production mechanism.

**Table 24.1.3.** The total production cross-sections from Belle (Seuster, 2006) and *BABAR* (Aubert, 2002d, 2007p) for $e^+e^- \to X_cY$ (or $\Lambda_c^+ X$). The uncertainties are statistical, systematic and the uncertainty due to the knowledge of the branching ratios at the time of the measurement. The branching fractions used in Belle and *BABAR* measurements are listed in second rows of results. Older measurements by CLEO (Artuso et al., 2004; Bortoletto et al., 1988) are also given taking into account the world average of the respective product branching fractions at the time of the Belle publication.

| $X_c$ | $\sigma_{\rm PROD}$(Belle) [pb]<br>$\mathcal{B}$ used | $\sigma_{\rm PROD}$(*BABAR*) [pb] | $\sigma_{\rm PROD}$(CLEO) [pb] |
|---|---|---|---|
| $D^0 \to K^- \pi^+$ | $1449 \pm 2 \pm 64 \pm 38$ | - | $1521 \pm 16 \pm 62 \pm 36$ |
| | $\mathcal{B}(D^0 \to K^-\pi^+) = 0.0380 \pm 0.009$ | | |
| $D^+ \to K^- \pi^+ \pi^+$ | $654 \pm 1 \pm 36 \pm 46$ | - | $640 \pm 14 \pm 35 \pm 43$ |
| | $\mathcal{B}(D^+ \to K^-\pi^+\pi^+) = 0.092 \pm 0.006$ | | |
| $D_s^+ \to \Phi\pi^+$ | $231 \pm 2 \pm 92 \pm 77$ | $210 \pm 6 \pm 9 \pm 52$ | - |
| | $\mathcal{B}(D_s^+ \to \Phi\pi^+) = 0.036 \pm 0.009$ | | |
| | $\mathcal{B}(\Phi \to K^-K^+) = 0.491 \pm 0.006$ | | |
| $\Lambda_c^+ \to p\, K^- \pi^+$ | $189 \pm 1 \pm 66 \pm 66$ | $188 \pm 3 \pm 6 \pm 49$ | $270 \pm 90 \pm 70$ |
| | $\mathcal{B}(\Lambda_c^+ \to pK^-\pi^+) = 0.050 \pm 0.013$ | | |
| $D^{*0} \to D^0 \pi^0$ | $510 \pm 3 \pm 84 \pm 39$ | - | $559 \pm 24 \pm 35 \pm 39$ |
| | $\mathcal{B}(D^{*0} \to D^0\pi^0) = 0.619 \pm 0.029$ | | |
| $D^{*+} \to D^0 \pi^+$ | $598 \pm 2 \pm 77 \pm 20$ | - | $583 \pm 8 \pm 33 \pm 14$ |
| | $\mathcal{B}(D^{*+} \to D^0\pi^+) = 0.677 \pm 0.005$ | | |
| $D^{*+} \to D^+ \pi^0$ | $590 \pm 5 \pm 78 \pm 53$ | - | - |
| | $\mathcal{B}(D^{*+} \to D^+\pi^0) = 0.619 \pm 0.029$ | | |
| average $D^{*+}$ | $597 \pm 2 \pm 78 \pm 25$ | - | - |

Candidate $\Lambda_c^+$ are reconstructed in the $pK^-\pi^+$ (as in the *BABAR* study of inclusive $\Lambda_c^+$ production described above) and $pK_S^0$ decay modes. Additional decay modes were considered, but did not improve the result. Events with both a $\Lambda_c^+$ and $\overline{\Lambda}_c^-$ candidate with invariant mass within 190 MeV/$c^2$ of the nominal $\Lambda_c^+$ mass are selected if the opening angle between their momenta is over 90°.

The invariant mass distributions of the candidates and the two-dimensional distribution of the $\overline{\Lambda}_c^-$ vs. $\Lambda_c^+$ candidate masses are shown in Fig. 24.1.8. There are clear vertical and horizontal bands corresponding to single $\Lambda_c^+$ and $\overline{\Lambda}_c^-$ production, respectively, and an enhancement where they overlap. The signal region is considered by a circle of 12 MeV/$c^2$ radius around the two nominal masses, events well outside the two bands are used to characterize the combinatorial background, and events within the bands to characterize the background from events with either a real $\Lambda_c^+$ or a real $\overline{\Lambda}_c^-$. Of the 919 entries in the signal circle, $649 \pm 31$ are estimated to be $\Lambda_c^+\overline{\Lambda}_c^- X$ events.

If the charmed and anticharmed hadrons in a $c\overline{c}$ event are uncorrelated, then roughly 155 signal events would be expected, given the numbers of $c\overline{c}$ events in the sample and reconstructed $\Lambda_c^+/\overline{\Lambda}_c^-$. This estimate is independent of the branching fractions and average reconstruction efficiencies, but has ∼30% model dependence in the correlation between the two efficiencies in an event. Fragmentation models predict a 10–15% suppression of such events, due to the combined mass of the four baryons. The observed number of events is well above 155, by a factor of 4.2, consistent with the ratio reported by CLEO.

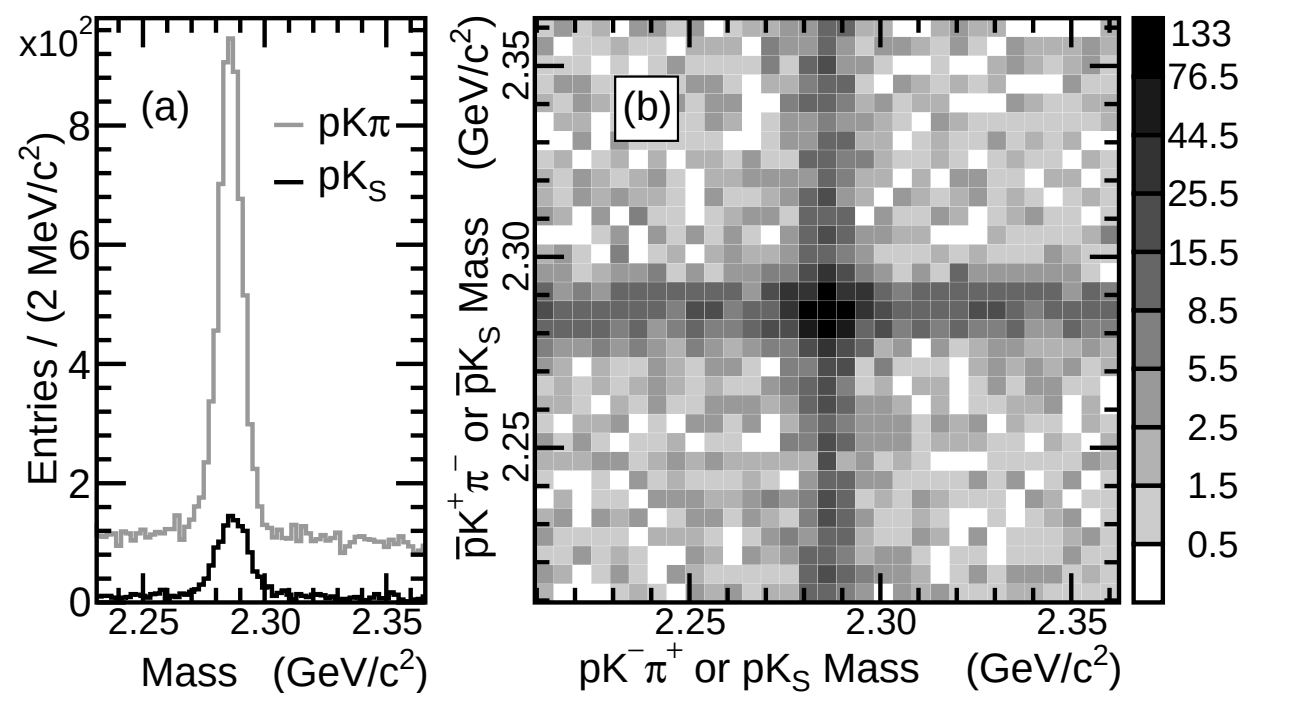

**Figure 24.1.8.** (a) Invariant mass distributions for $\Lambda_c^+ \to pK^-\pi^+$ (gray) and $\Lambda_c^+ \to pK_S^0$ (black) candidates in events with both a $\Lambda_c^+$ and a $\overline{\Lambda}_c^-$ candidate. (b) Invariant mass of the $\overline{\Lambda}_c^-$ candidate vs. that of the $\Lambda_c^+$ candidate, in 5 MeV/$c^2$ × 5 MeV/$c^2$ bins (Aubert, 2010b).

The additional tracks, not from the $\Lambda_c^+$ or $\overline{\Lambda}_c^-$ decay, are studied in order to characterize these events. The distribution of their multiplicity is shown in Fig. 24.1.9(a), along with those of identified $K^{\pm}$ and $p/\overline{p}$. The distribution of the missing mass per event is shown in Fig. 24.1.9(b), and can accommodate very few events with both a neutron and an antineutron. This, along with the small number of identified protons (anti-protons), leads to the conclusion that 4-baryon events are suppressed much more strongly than in the models, and that the selected sample is dominated by a new type of event containing a charmed baryon, a charmed antibaryon, several mesons, and no other (anti)baryons.

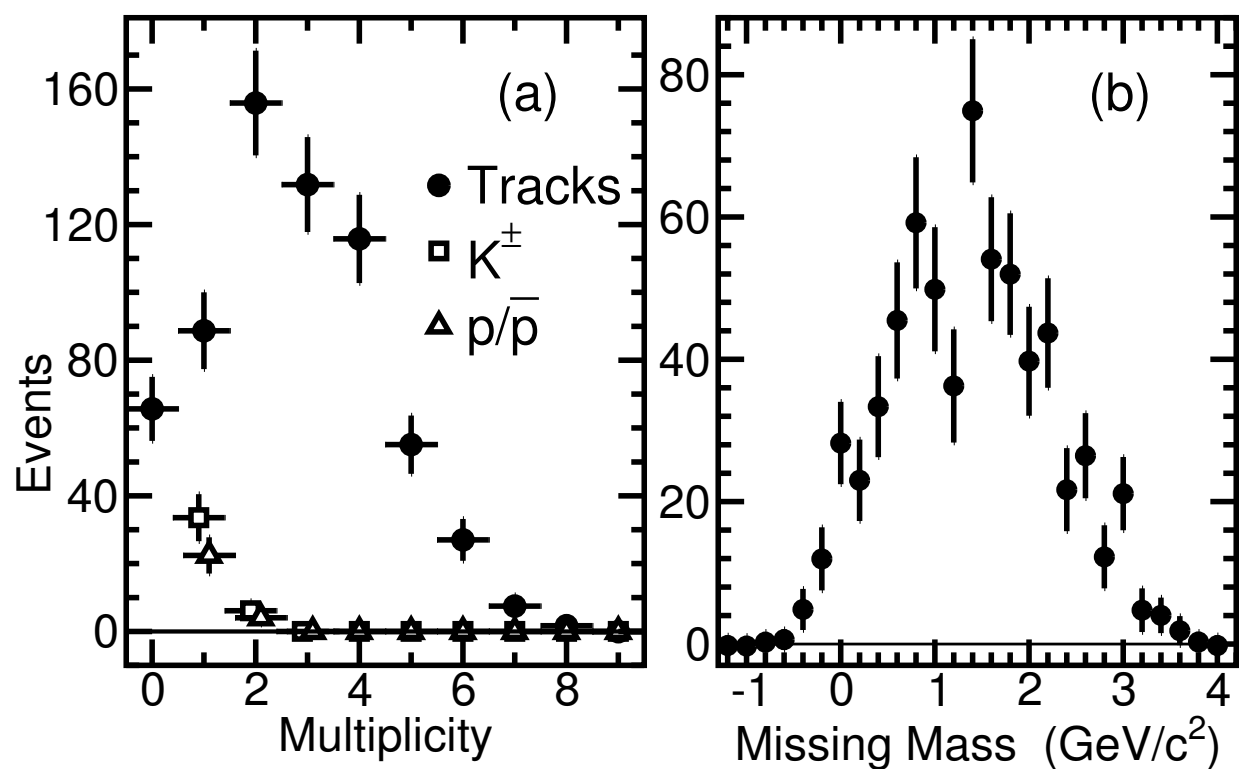

**Figure 24.1.9.** Background-subtracted distributions for $\Lambda_c^+\overline{\Lambda}_c^- X$ events of (a) the numbers of additional tracks in the event and those identified as $K^{\pm}$ or $p\overline{p}$, and (b) the missing mass in the event, with imaginary masses given negative real values. Most events have no identified $K^{\pm}$ or $p\overline{p}$, and the corresponding zero-multiplicity points are off the vertical scale in (a) (Aubert, 2010b).

The events containing an identified $p\overline{p}$ are used to estimate and subtract the contribution from 4-baryon events, and the features of the remaining $619 \pm 39$ events are studied. The broad multiplicity distribution and the distributions of energies, (transverse) momenta and rapidities of both the $\Lambda_c^+/\overline{\Lambda}_c^-$ and the additional tracks are similar to those in all hadronic events, the single-$\Lambda_c^+/\overline{\Lambda}_c^-$ sidebands, and models. Correcting for the decays of known heavier charmed baryons and tracking efficiency, an average of $2.63 \pm 0.21$ tracks per event is estimated. These tracks are combined, and only very small signals for $K_S^0$, $K^{*0}$ and $\rho$ mesons are observed. Along with the small number of $K^{\pm}$ observed, this indicates that fewer strange and vector mesons are produced than in normal fragmentation. Assuming that about half as many $\pi^0$ as $\pi^{\pm}$ are produced, which is consistent with the missing mass and total energy distributions, there is an average of about 4 additional mesons in these events.

It can be concluded that these events are indeed from a fragmenting $c\overline{c}$ system with a long-range baryon number correlation. Such events are not produced by `Jetset` or `HERWIG`. They are produced by `UCLA`, which also predicts an enhanced rate and suppressions of kaons and vector mesons. However, the predicted enhancements and suppressions are all of the wrong size, and the multiplicity distribution has the wrong shape and average value.

### 24.1.3 Polarized fragmentation functions

A second class of fragmentation functions depends on the spin of the fragmenting parton. In particular, in the case of transverse quark polarization this opens up a powerful tool to use such spin dependent fragmentation functions as an analyzer of transverse quark polarization in the nucleon, transversity. Transversity can be defined as the number density of transversely polarized partons inside a transversely polarized nucleon. It is a chiral odd distribution function, which means that there are mixed left-handed and right-handed quark fields and because of this spin-flip has been hard to access. The integrated transversity distribution provides the tensor charge of the nucleon, a fundamental charge just like the electric or the axial charge. In the non-relativistic case, the axial charge and the tensor charge are identical. Before the advent of spin dependent fragmentation functions the access to the transverse quark polarization of the nucleon was only possible via the Drell Yan process (Ralston and Soper, 1979). The reason is that the chiral-odd transversity parton distribution function needs to couple with another chiral-odd function to become observable as QCD conserves chirality. In the Drell Yan process that could be the antiquark transversity distribution from the second nucleon. In contrast, in the deep inelastic scattering process (DIS), which is the most abundant source for unpolarized and longitudinally polarized distribution functions of the nucleon transversity is not directly accessible. Using chiral-odd, transverse quark spin dependent fragmentation functions, it became possible to access quark transversity in transversely polarized semi-inclusive deep inelastic scattering (SIDIS) processes as well as in transversely polarized proton-proton collisions. In these cases one chiral-odd distribution function and one chiral-odd fragmentation function make the process chiral-even and thus observable. The $e^+e^-$ annihilation process allows one to cleanly study the fragmentation functions, including the transversely polarized fragmentation functions. However, due to the chiral-odd nature of this FF, one needs to couple it with another chiral-odd object, which in this case is a second chiral-odd fragmentation function. One therefore measures the product of two chiral-odd fragmentation functions, one related to the quark side and the other to the antiquark side. This combination of two chiral-odd fragmentation functions can be understood by the requirement that the initial quark-antiquark pair has its spins transverse to their momentum. Since this orientation is not directly observable, one can use the chiral-odd fragmentation function on one side to fix the spin direction via its spin dependence and analyze the other side with the second chiral-odd fragmentation function. The most prominent chiral-odd fragmentation functions are the Collins FF and the interference fragmentation function (IFF) which will be described below in Sections 24.1.3.1 and 24.1.3.2 respectively. Both were only proposed during the 90's and they remained unmeasured until recent transversely polarized DIS measurements (Airapetian et al., 2005, 2008; Alexakhin et al., 2005) and the $B$ Factories' (Abe, 2006a) measurements.

### 24.1.3.1 Collins fragmentation function

The most prominent transverse spin dependent fragmentation function is the Collins fragmentation function, $H^{\perp}_{1,q}(z, P_{h\perp})$, which Collins (1993) initially suggested as a possible way to explain the surprisingly large single spin asymmetries seen by the E704 experiment (Adams et al., 1991) in transversely polarized $p\,p$ collisions. The Collins fragmentation function translates the transverse spin of a quark $\mathbf{S}_q$ with momentum $\mathbf{k}$ into the azimuthal yield of final state hadrons with transverse momentum $\mathbf{P}_{h\perp}$ and fractional energy $z$. According to the Trento conventions (Bacchetta, D'Alesio, Diehl, and Miller, 2004) the overall number density for finding a hadron $h$ (with mass $M_h$) produced in the process $q^{\uparrow} \to hX$ can be defined as:

$$D^h_{q^{\uparrow}}(z, P^2_{h\perp}) = D^h_{1,q}(z, P^2_{h\perp}) + H^{\perp h}_{1,q}(z, P^2_{h\perp}) \frac{(\hat{\mathbf{k}} \times \mathbf{P}_{h\perp}) \cdot \mathbf{S}_q}{zM_h}, \tag{24.1.12}$$

where the first term is the unpolarized fragmentation function described in Section 24.1.2.1 before integration of the transverse momentum. The second term contains the Collins function $H^{\perp h}_{1,q}(z, P^2_{h\perp})$ and correlates the spin orientation of the quark $\mathbf{S}_q$, its momentum $\hat{k}$ and the transverse momentum of the hadron $\mathbf{P}_{h\perp}$. Upon flipping the quark spin this product changes sign and thus generates a single spin asymmetry proportional to a $\cos\phi$ modulation of the azimuthal angle spanned by the vectors $\hat{k}, \mathbf{P}_{h\perp}$ and $\mathbf{S}_q$. In an unpolarized case the second term vanishes and the conventional, unpolarized fragmentation function definition is recovered. The existence of the spin dependent term was found by the HERMES experiment when single spin asymmetries attributed to a convolution of quark transversity and the Collins function turned out to be nonzero (Airapetian et al., 2005). However, in order to explicitly measure it, $e^+e^-$ annihilation is used. A first study was performed on DELPHI data by Efremov, Smirnova, and Tkachev (1999), before the large amount of data accumulated by the $B$ Factories became available. The method which is used in both $B$ Factory experiments closely follows the prescription given by Boer (2009). The experimental selection requirements are also similar in the measurements Abe (2006a), Seidl (2008), and Garzia (2013). A product of Collins functions is accessed by selecting two charged pions in opposite hemispheres where the hemisphere is defined by the thrust axis (see Chapter 9). Experimentally this corresponds to the process $e^+e^- \to (\pi^{\pm})(\pi^{\pm,\mp})X$ where the brackets indicate the different hemispheres and X is the remainder of the final state. To select two-jet like events a minimum thrust value of 0.8 is required with the thrust axis being well within the barrel acceptance ($\cos\theta_T < 0.75$). Further selection criteria require the charged particles to be in the barrel part of the detector as well (for example, in Seidl, 2008, $-0.6 < \cos(\theta_{\rm lab}) < 0.9$ is used) with minimum fractional energies $z = 2E_h/\sqrt{s} > 0.2$. The azimuthal asymmetries are observed in a $\cos(\phi_1 + \phi_2)$ modulation in the normalized two-hadron yields, $R = N(\phi_1 + \phi_2)/\langle N_{12}\rangle$, where $\phi_{1,2}$ are the azimuthal angles defined in the CM by the two planes of hadrons relative to the plane containing both the $e^+e^-$ pair and the thrust axis as shown in Fig. 24.1.10. $N(\phi_1 + \phi_2)$ is the number of pion pairs with the sum of the azimuthal angles $\phi_1 + \phi_2$, and $\langle N_{12}\rangle$ is the average number of pion pairs over the whole $\phi_1 + \phi_2$ interval. Similar definitions can be obtained where the reference axis is defined by the plane containing the $e^+e^-$ pair and the second hadron in which case only one angle $\phi_0$ appears and the modulation becomes a $\cos(2\phi_0)$ modulation (and the corresponding normalized yields are denoted $R_0$).

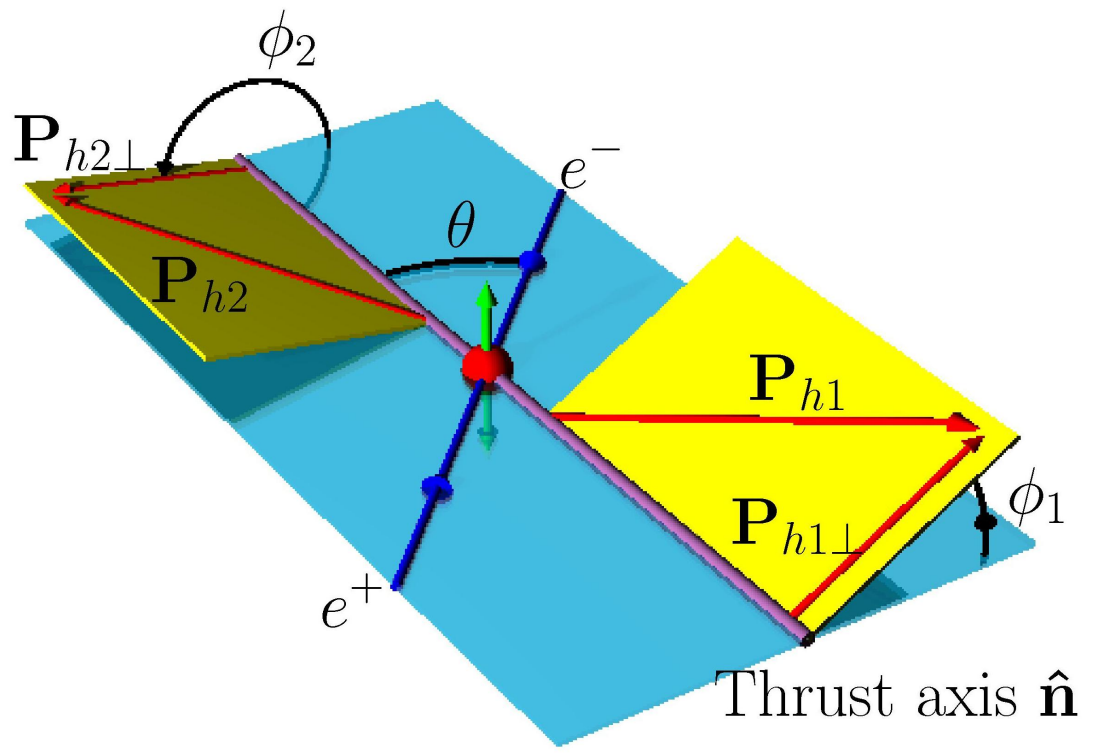

**Figure 24.1.10.** Azimuthal angles $\phi_1$ and $\phi_2$ defined for the two hadrons relative to the plane spanned by the lepton and thrust axis.

Acceptance effects and gluon radiation can also generate fake azimuthal modulations and were found to be substantial in MC (`Pythia 6.2` and `GEANT3` ). To isolate the spin dependent fragmentation effect from these background effects, the method of double ratios was applied using ratios of different pion charge combinations. The normalized yields for opposite-sign pion pairs as a function of the azimuthal angle $R^U(\cos(\phi_1 + \phi_2))$ (or $R^U_0(\cos(2\phi_0))$) were divided by the normalized yields of like-sign pairs, $R^U_{(0)}/R^L_{(0)}$ and fitted. Similar ratios between unlike sign pion pairs and any charged pion pairs ($R^U_{(0)}/R^C_{(0)}$) were also extracted. As both background effects are expected to be proportional to the unpolarized fragmentation functions, their contributions to the normalized yields are the same for both charge sign combinations and they cancel when building the double ratio. As the Collins functions are expected to be different for favored and disfavored fragmentation, a net asymmetry related to the Collins functions should remain. The double ratios as a function of the azimuthal angles were then fit with azimuthal modulations $R^U/R^L = A^{UL}_{12}\cos(\phi_1 + \phi_2) + B$ (or $R^U_0/R^L_0 = A^{UL}_0\cos(2\phi_0) + B_0$) of which the $\cos(\phi_1 + \phi_2)$ (or $\cos(2\phi_0)$) part is proportional to the Collins functions. In case of the any charge pion pairs the corresponding amplitudes are denoted $A^{UC}_{12(0)}$. An example of the normalized raw yield, double ratio and its modulation can be seen in Fig. 24.1.11 using the $\phi_0$ angle. The normalized raw yields are paremetrized similarly as the double ratios for like- and unlike-sign pairs. In order to distinguish the parameters the latter are labeled by $a, b$ instead of $A, B$.

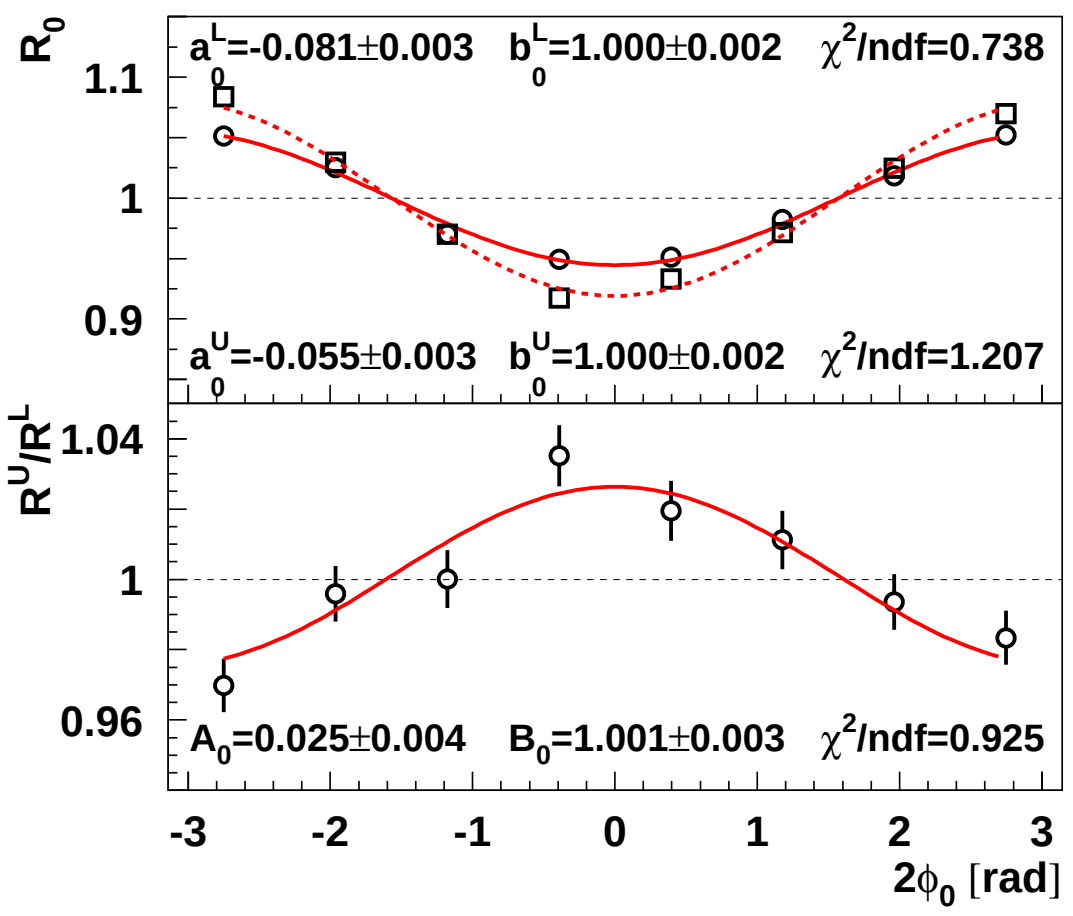

**Figure 24.1.11.** From Abe (2006a). Top: Example of uncorrected unlike-sign (open circles) and like-sign (open squares) di-pion normalized rate $R_0$ vs. $2\phi_0$ in the bin $z_1(z_2) \in [0.5, 0.7]$, $z_2(z_1) \in [0.3, 0.5]$. Bottom: The di-pion double ratio $R_0^U/R_0^L$ vs. $2\phi_0$ in the same $z_1, z_2$ bin. Resulting parameters of the fit described in the text (full and dashed lines) are also shown.

The double ratio method was tested using a sample of generic light quark MC production, where all effects except those by the Collins fragmentation were present and it was found, that the resulting raw asymmetries canceled as expected. Further tests include destroying the correlation between quark and anti-quark side by mixing pions from different events. Those were found to be consistent with zero as expected. Also artificial asymmetries were introduced in the MC to study the reconstruction efficiency. Some underestimation of the reconstructed asymmetries was found to be caused by the resolution smearing of the reconstructed thrust axis and the resulting azimuthal angles. The reason is that experimentally one cannot directly obtain the actual quark-antiquark axis, but has to rely on obtaining an approximate axis via the event shape variable thrust. As this is performed using the thrust algorithm summing over all reconstructed particles this experimental approximation reproduces the actual quark-antiquark axis with a finite accuracy. It is found that even on the generator level there is some discrepancy, which is further enhanced by detector resolutions. In the Belle experiment the average cosine of the angle formed by the reconstructed thrust axis and the generated quark-antiquark pair axis for light quarks is 0.990 with an RMS of 0.015 (Seidl, 2008), as obtained from a `Pythia` (Sjöstrand, 1995) simulated sample of events and `GEANT` (Brun, Bruyant, Maire, McPherson, and Zanarini, 1987) detector simulation. The resulting reduction of the extracted $\cos(\phi_1+\phi_2)$ asymmetries was corrected for by scaling them with a factor $1.66 \pm 0.04$ (Seidl, 2008) which was obtained from the weighted MC simulation asymmetry studies.

The contribution to the asymmetries by light quarks and charm quarks (the latter representing background for the light quark fragmentation measurements) were separated using, in addition to the main data sample, a charm enhanced data sample. In the latter candidate $\pi$ K pairs with the invariant mass in the range of $D$ meson and $D\ \pi_s$ combinations consistent with a $D^*$ meson were selected. The initial measurement (Abe, 2006a) was performed on 29.1 fb$^{-1}$ of data obtained 60 MeV below the $\Upsilon(4S)$ resonance, while the second publication (Seidl, 2008) utilized 551 fb$^{-1}$ of data including the resonance data. It was found that the thrust selection mentioned above removes most of the $B$ decay events (the remaining pollution of the sample with $B$ meson decays is around 2%). The results of the latter measurement are shown in Fig. 24.1.12.

It can be seen that the asymmetries are of the order of several percent and are rising with increasing fractional energy. The direct interpretation is not straightforward since the asymmetries are differences of products of favored and disfavored Collins and unpolarized fragmentation functions:

$$
\begin{aligned}
\frac{R_{12}^U}{R_{12}^C} &= 1 + \cos(\phi_1+\phi_2) A_{12}^{UC}, \\
A_{12}^{UC} &= \frac{\sin^2\theta}{1+\cos^2\theta} \\
&\times \left\{ \frac{f\left(H_1^{\perp,\mathrm{fav}} \bar{H}_2^{\perp,\mathrm{fav}} + H_1^{\perp,\mathrm{dis}} \bar{H}_2^{\perp,\mathrm{dis}}\right)}{\left(D_1^{\mathrm{fav}} \bar{D}_2^{\mathrm{fav}} + D_1^{\mathrm{dis}} \bar{D}_2^{\mathrm{dis}}\right)} \right. \\
&\left. - \frac{f\left((H_1^{\perp,\mathrm{fav}} + H_1^{\perp,\mathrm{dis}})(\bar{H}_2^{\perp,\mathrm{fav}} + \bar{H}_2^{\perp,\mathrm{dis}})\right)}{\left((D_1^{\mathrm{fav}} + D_1^{\mathrm{dis}})(\bar{D}_2^{\mathrm{fav}} + \bar{D}_2^{\mathrm{dis}})\right)} \right\},
\end{aligned} \tag{24.1.13}
$$

where a shorthand notation was used for the fragmentation functions of hemispheres 1 and 2, *i.e.* $H_2^\perp = H_{1,q}^{\perp h}(z_2)$ and similarly for the unpolarized fragmentation functions $D_2 = D_{1,q}^h(z_2)$; also the favored and disfavored FF's are denoted by superscripts fav and dis, respectively. A similar notation holds for the $A_{12}^{UL}$ amplitude. Anselmino et al. (2007) extracted the corresponding favored and disfavored Collins fragmentation functions from the Belle data and found that they are both sizeable and of opposite sign. Using this they were able to extract the quark transversity distribution from the HERMES (Airapetian et al., 2005) and COMPASS (Alexakhin et al., 2005) data for the first time.

In this extraction some improvements in the knowledge of the Collins function are still needed. For example, the intrinsic transverse momentum dependence is not known and was only estimated. Recent results have been shown by *BABAR* (Garzia, 2013), with an analysis similar to that performed by Belle and based on a data sample corresponding to an integrated luminosity of about 468 fb$^{-1}$ collected at the $\Upsilon(4S)$ and 40 MeV below. A general consistency between the *BABAR* and Belle asymmetries measured as a function of the fractional energies is observed. The $z$-range explored by *BABAR* extends from 0.15 to 0.9. In addition, *BABAR* performed a study of the azimuthal asymmetries as a function of the transverse momentum of the pions with respect to the thrust axis. As an example,

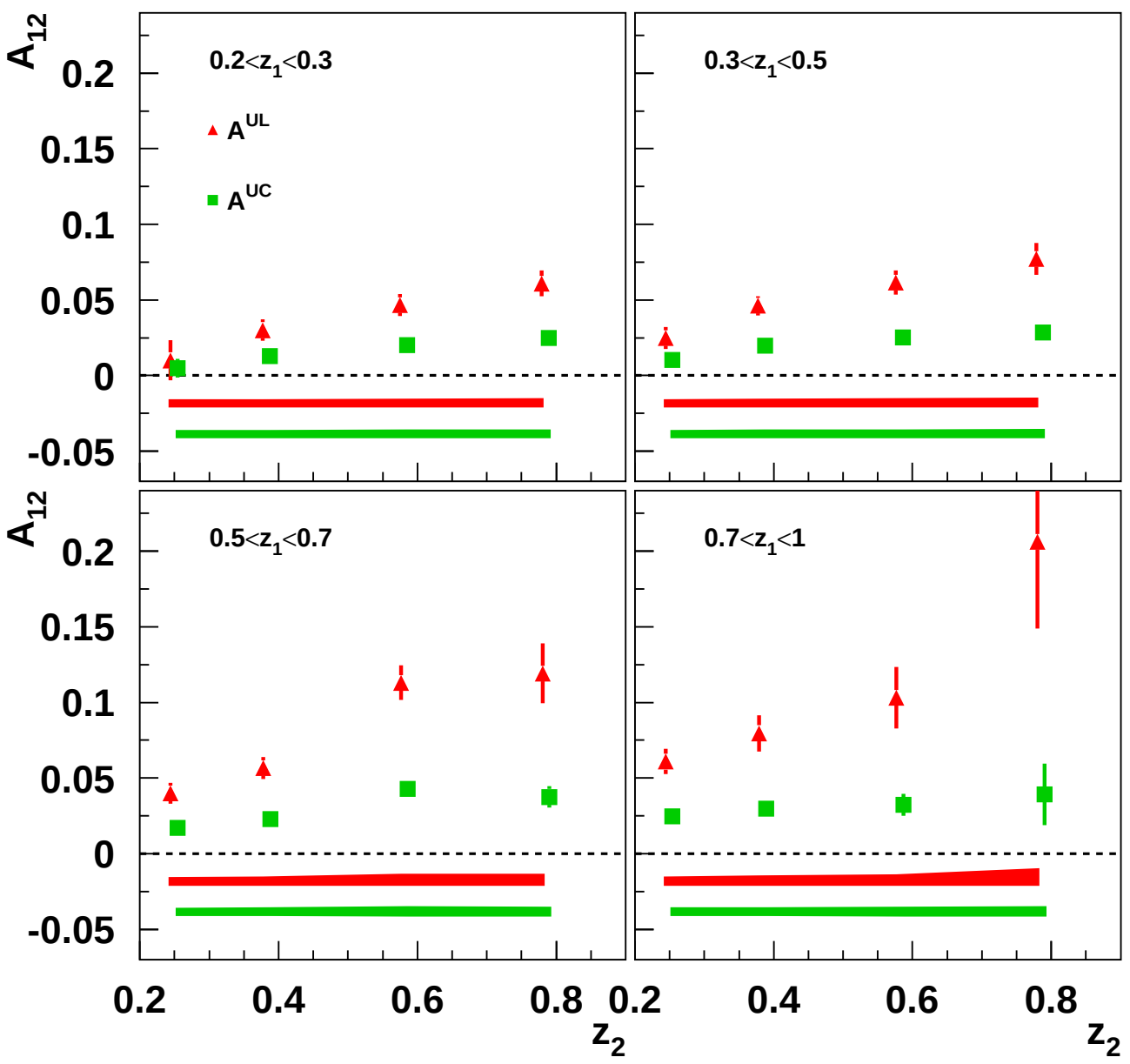

**Figure 24.1.12.** Azimuthal $\cos(\phi_1 + \phi_2)$ amplitude $A_{12}$ of the normalized yield ratios of unlike-sign pion pairs over like-sign pion pairs $A^{UL}$ (triangles), and of unlike-sign pion pairs over any charged pion pairs $A^{UC}$ (squares) as a function of the fractional energy $z_2$ for 4 different bins of $z_1$, from top left to bottom right, as measured by Belle (Seidl, 2008). The error bars represent the statistical errors, while systematic uncertainties are given as bands for each amplitude, the top band for $A^{UL}$, the bottom band for $A^{UC}$.

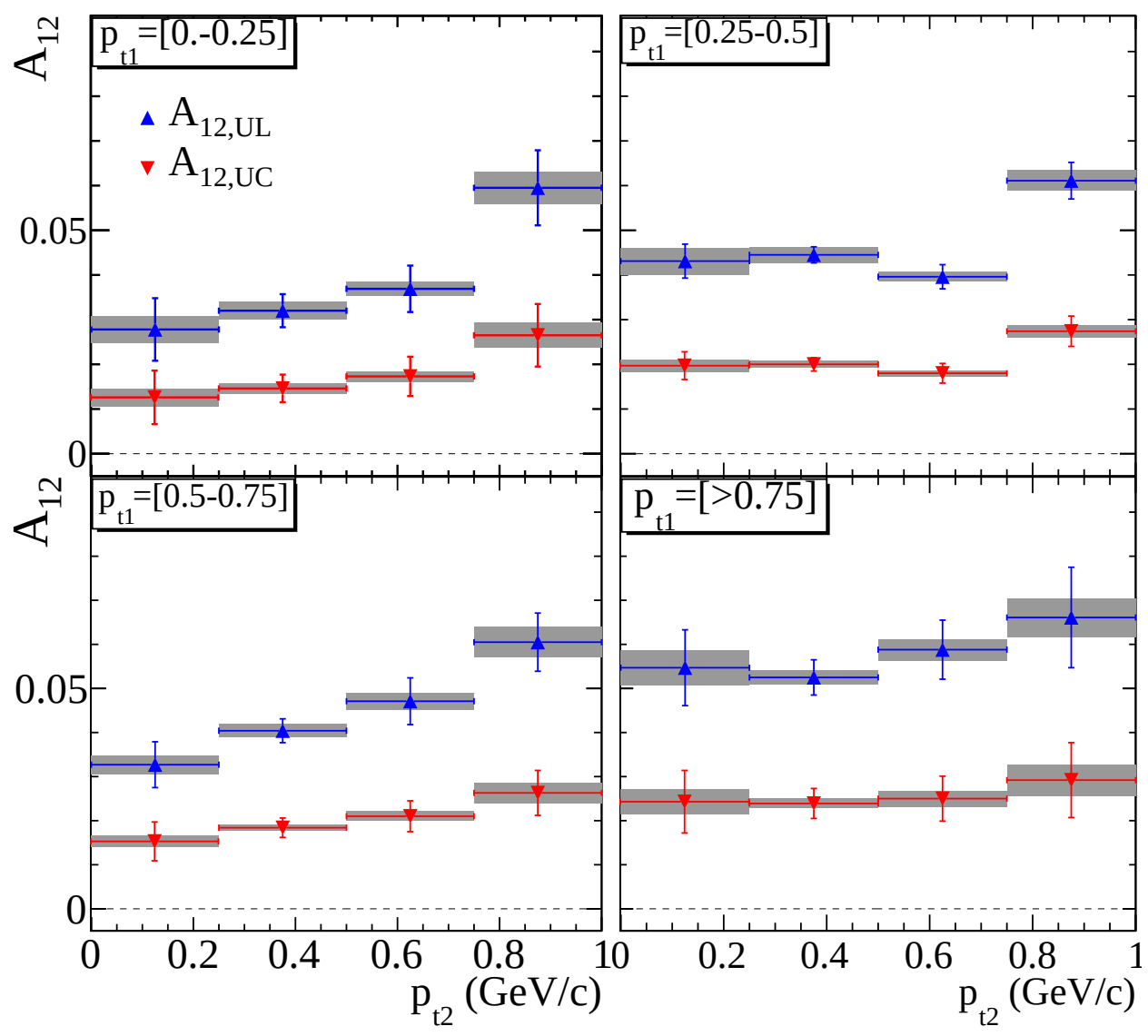

**Figure 24.1.13.** Preliminary *BABAR* results (Garzia, 2013) on azimuthal asymmetries measured from fits to the $R^U/R^L$ ($A_{12,UL}$), and $R^U/R^C$ ($A_{12,UC}$) ratios, as functions of the transverse momentum $p_{t2}$ for four bins of $p_{t1}$, from top left to bottom right. Statistical and systematic errors are shown as errors bars and shaded bands respectively.

Fig. 24.1.13 shows the azimuthal asymmetries measured in the thrust reference frame. The energy and transverse momentum dependence of the Collins asymmetries obtained by *BABAR* can be combined with the Belle data and the results from SIDIS experiments for an improved global analysis as done in Anselmino et al. (2007).

Collins functions for other final state hadrons still need to be extracted to improve the sensitivity for different quark flavors and to match the extracted asymmetries in SIDIS experiments. Also, an important test of the mechanism which creates this transverse spin effect (Collins effect) still needs to be performed. According to a model by Artru and Mekhfi (1990) which follows string fragmentation, the Collins effect for a transversely polarized vector meson should be of different sign as that for pseudoscalar mesons. Results from those studies are not yet available.

#### 24.1.3.2 Interference fragmentation function

The second chiral-odd fragmentation function is the interference fragmentation function (IFF), $H_1^{\sphericalangle}(z,m)$, which describes the fragmentation of a transversely polarized quark into a pair of hadrons of different charge with total fractional energy $z$ and an invariant mass $m$. A nonzero IFF can be created by the interference of two hadrons in a relative $S$- or $P$-wave state. For charged pion pairs this could therefore be either a simple $S$-wave (which might be related to the $\sigma$ resonance) interfering with the $P$-wave state related to the $\rho$ meson (which is observed as a dipion resonance). This potential interference also governs the invariant mass dependence according to theory predictions. Based on pion-pion phase shift analysis of data, Jaffe, Jin, and Tang (1998) suggest a sign change of the two-pion interference fragmentation function is required at the invariant mass of the $\rho$ meson. Radici, Jakob, and Bianconi (2002) suggest no sign change but the maximum of the IFF magnitude at the same mass. The difference between the interference fragmentation function and the Collins fragmentation function is that intrinsic transverse momenta of hadrons created in the fragmentation are integrated over in the former case, which enables the use of collinear factorization. This in turn leads to the QCD evolution of the interference fragmentation function being known and makes it easily applicable at various energies and in different processes such as in SIDIS or $p\,p$ collisions.

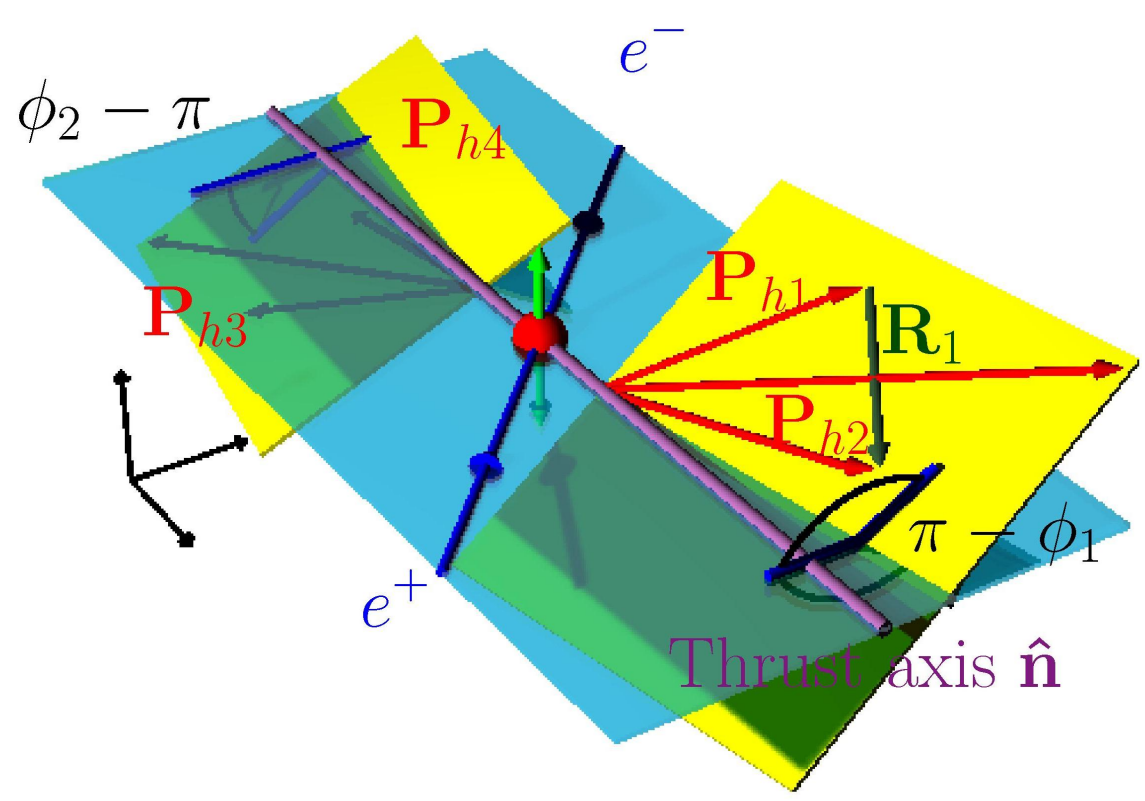

**Figure 24.1.14.** Azimuthal angles $\phi_1$ and $\phi_2$ defined for the two hadron pairs's planes (denoted by the yellow planes) relative to plane spanned by the lepton and thrust axis (blue).

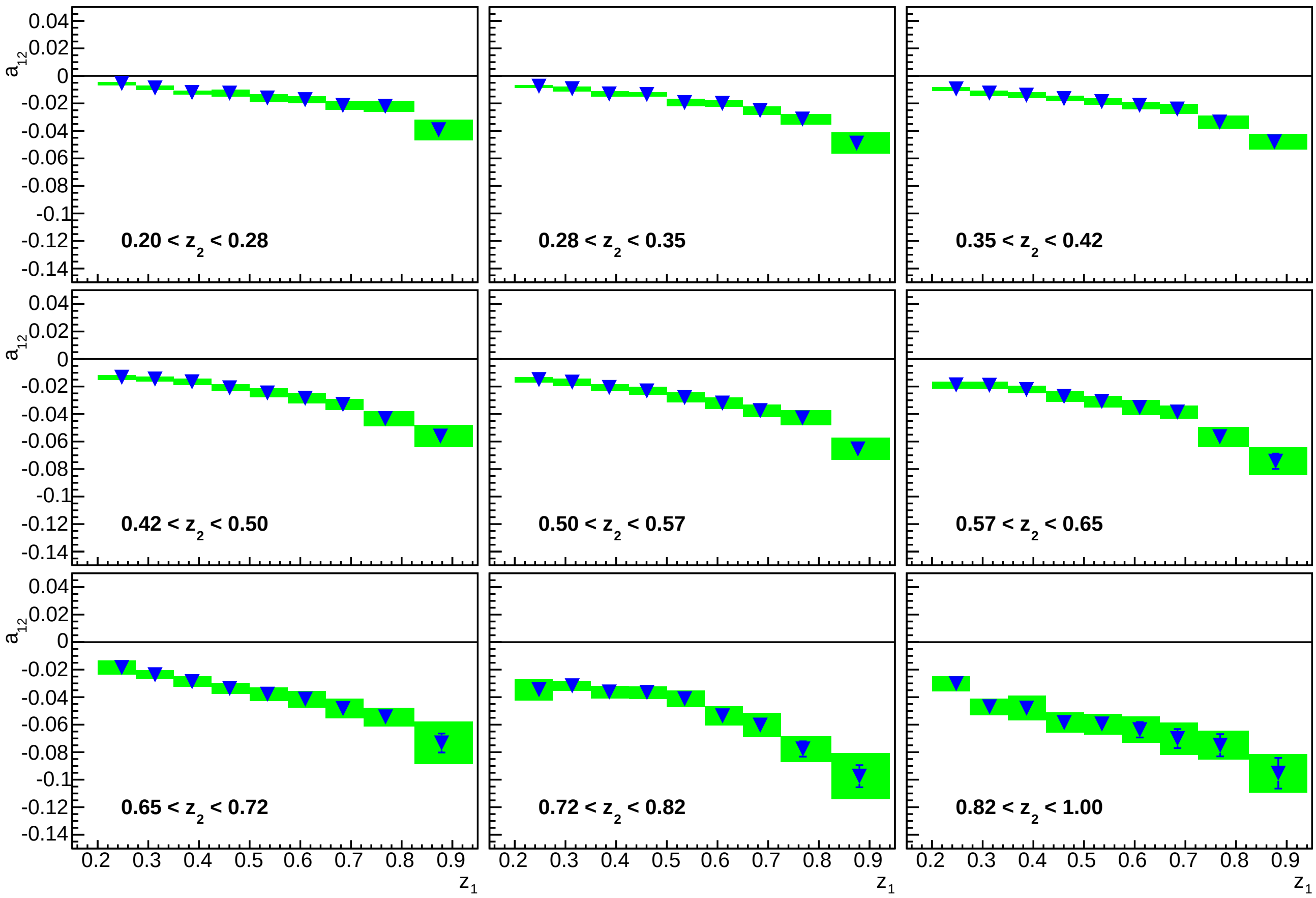

**Figure 24.1.15.** Azimuthal $\cos(\phi_1+\phi_2)$ modulations $a_{12}$ of normalized yield of charged pion pairs as a function of the fractional energy $z_2$ for 9 different bins of $z_1$ from top left to bottom right as measured by Belle (Vossen, 2011). The filled areas represent the systematic uncertainties.

IFF measurements were performed by HERMES (Airapetian et al., 2008), Compass (Adolph et al., 2012; Wollny, 2009), PHENIX (Yang, 2009) and STAR (Vossen, 2012). These measurements determine the product of two unknown quantities: quark transversity and the interference fragmentation functions. Therefore the measurement of the interference fragmentation function alone in $e^+e^-$ annihilation enable the access to quark transversity, independent of the methods using Collins FF. Similar to the Collins analysis, the interference fragmentation function can be reconstructed at $B$ Factories using the combination of two chiral-odd fragmentation functions in each hemisphere (Boer, Jakob, and Radici, 2003). Therefore, one measures inclusively two hadron pairs in opposite hemispheres in $e^+e^-$ annihilation. Again the two hemispheres are defined by the thrust axis and a thrust $> 0.8$ ensures two-jet like topology (Vossen, 2011). All four hadrons are required to be detected in the central part of the detector and to have a minimal fractional energy of 0.1. In addition, the invariant mass of each pair is required to be in the range of 0.25 to 2 GeV/$c^2$. To avoid acceptance effects at the edges of the detector, hadrons were only selected if they originated in a cone around the thrust axis of $\hat{\mathbf{n}} \cdot \mathbf{P}_h > 0.8$ where the thrust axis was again limited to the barrel parts of the detector, identical to the Collins analysis. Two azimuthal angles $\phi_{1,2}$ are calculated, defined by the plane of each hadron pair relative to the plane spanned by the $e^+e^-$ pair and the thrust axis, as displayed in Fig. 24.1.14.

Again the normalized yields as a function of the azimuthal angles are fitted. The $\cos(\phi_1+\phi_2)$ modulation is proportional to the product of the interference fragmentation functions for the quark and antiquark sides normalized by the corresponding unpolarized fragmentation functions. After the application of the opening angle selection around the thrust axis, nearly vanishing acceptance effects ($< 0.1\%$) were observed in MC simulations of light quark production. Therefore it is possible to directly obtain the IFFs without the need for double ratios. The azimuthal modulation is again fitted by $b_{12}+a_{12}\cos(\phi_1+\phi_2)$ of which the cosine modulation can be interpreted as

$$a_{12} \propto -\frac{\sin^2\theta}{1+\cos^2\theta}\frac{\sum_q e_q^2 H_1^{\sphericalangle,q}(z_1,m_1)H_1^{\sphericalangle,\bar{q}}(z_2,m_2)+c.c.}{\sum_q e_q^2 D_1^q(z_1,m_1)D_1^{\bar{q}}(z_2,m_2)+c.c.}, \tag{24.1.14}$$

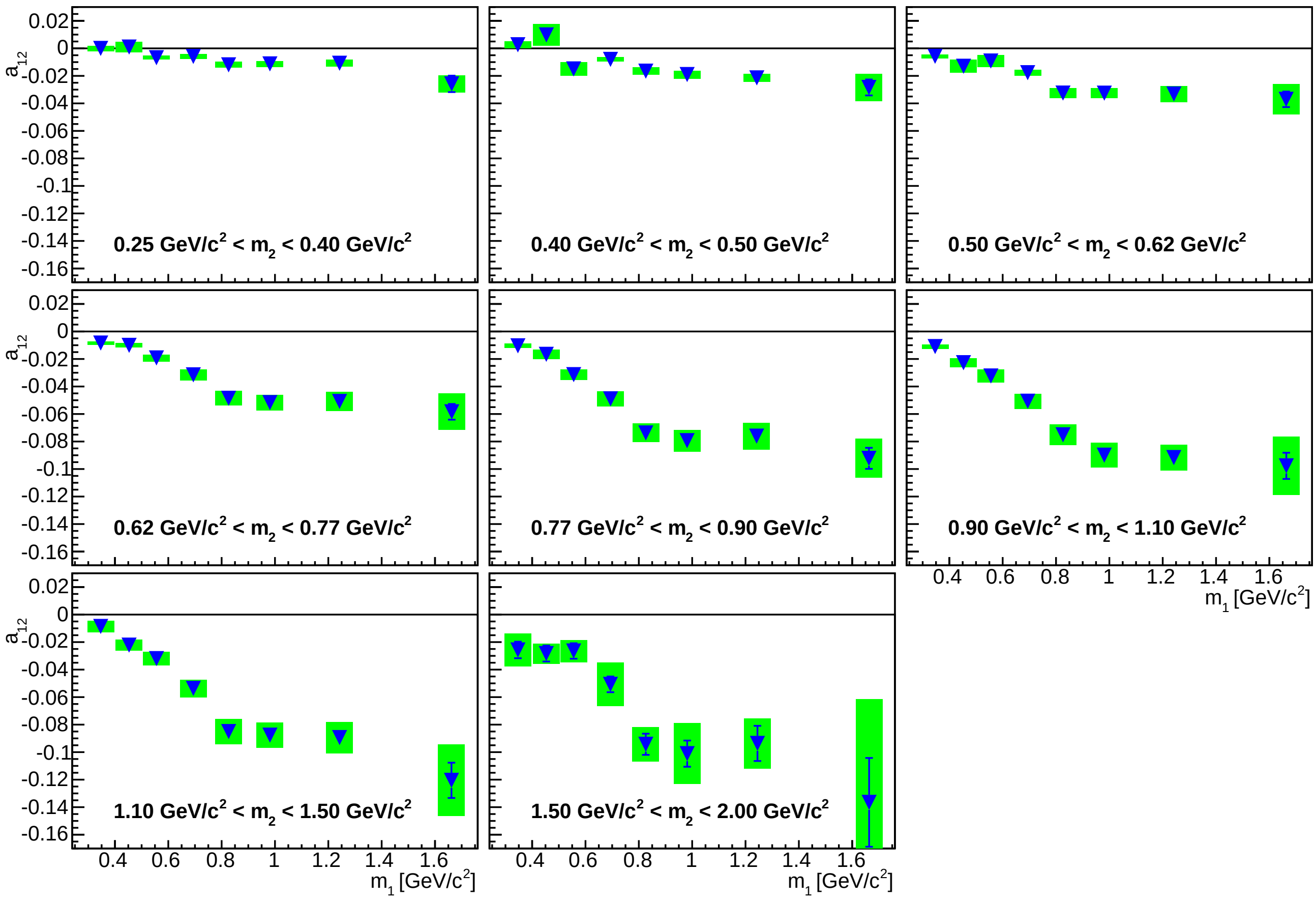

**Figure 24.1.16.** Azimuthal $\cos(\phi_1 + \phi_2)$ modulations $a_{12}$ of the normalized yield of charged pion pairs as a function of the invariant mass $m_2$ for 8 different bins of $m_1$ from top left to bottom right as measured by Belle (Vossen, 2011). The filled areas represent the systematic uncertainties.

where $\theta$ is the polar angle between the lepton and the thrust axis. Similar systematic studies as in the Collins analysis were performed, such as mixed-event tests where the two hadron pairs were selected from different events, the zero tests in MC simulations without any spin effects present as well as studies of the bias of the asymmetries for certain opening angle and thrust selections. The stability over various data-taking periods for on- and off-resonance data were also studied and found to be consistent. Again the thrust axis smearing reduces the magnitude of the asymmetries to 92% of the generated value as found in simulations (less than in the case of Collins FF measurement).

Results have been obtained so far by Belle, for charged pion pairs, using 672 $\text{fb}^{-1}$ (Vossen, 2011). The asymmetries are displayed, for example, as a function of the two fractional energies of the two pairs in Fig. 24.1.15. It can be seen that the asymmetries are increasing with fractional energy. This can again be explained if more of the quarks' spin information is contained at the highest fractional energies. The overall magnitude of the asymmetries is quite remarkable, reaching more than 10%. Given that the asymmetry parameter $a_{12}$ is proportional to the product of two IFFs (see Eq. 24.1.14) this means that the effect of a single IFF can be as large as 30%.

Figure 24.1.16 shows the asymmetries as a function of the invariant mass of the pion pairs. Here one sees an increase in the magnitude of asymmetries up to and slightly above the $\rho$ mass, where the asymmetries seem to level off. At the highest invariant masses, the fraction of charm events according to MC is largest, such that some effect might originate from charm events. However efforts to separate charm and *uds* events showed little difference between the extracted *uds* and all events. The asymmetries also show clearly that no sign change of the interference fragmentation function occurs at the invariant mass of the $\rho$ meson and therefore rules out Jaffe's prediction (Jaffe, Jin, and Tang, 1998). Theoretical efforts to obtain the quark transversity by combining the $e^+e^-$ IFF results with the corresponding SIDIS results performed by Bacchetta, Courtoy, and Radici (2011) and Courtoy, Bacchetta, Radici, and Bianconi (2012) show a quark transversity distribution similar to the one extracted using the Collins fragmentation functions, although uncertainties are currently still rather large. A recent update of that analysis by Courtoy, Bacchetta, and Radici (2012) shows excellent agreement when the recent COMPASS data (Adolph et al., 2012) are included. The similarity of the extracted transversity distributions suggests that the unknown QCD evolution of the Collins func-tion is not too different from the regular DGLAP evolution, otherwise the magnitudes at as different scales of $Q^2 = 110\,\mathrm{GeV}^2$ ($B$ Factories) and $2.4\,\mathrm{GeV}^2$ (HERMES) would also have been rather different.

Future measurements of IFF should include $\pi^0\pi^\pm$, $\pi^{\pm,0}K$ and $KK$, as well as combinations of different pairs in different hemispheres to gain additional information on the two hadron equivalents of favored and disfavored fragmentation.

### 24.1.4 Summary on fragmentation functions

Belle and *BABAR* have extracted high precision unpolarized fragmentation functions for light and charmed mesons and some of the charmed baryons. While light hadrons contain a relatively small energy fraction of the initial partons the same fraction for charmed hadrons amounts to about 60%. The measured light quark fragmentation functions will be used to significantly improve the global QCD fits on fragmentation and will in turn provide further access to the flavor structure of the nucleon in semi-inclusive DIS and $pp$ experiments. Spin dependent Collins and interference fragmentation functions were directly obtained for the first time and found to be sizeable. They give access to the transverse spin structure of the nucleon since they act as quark spin analyzers. They have also been used together with SIDIS world data to obtain the transversity distribution functions of the nucleon.

## 24.2 Pentaquark searches

***Editors:***
*Jonathon Coleman (BABAR)*
*Bruce Yabsley (Belle)*

**Additional section writers:**
*Roman Mizuk*

Since the early years of the quark model, there has been speculation concerning states with unusual valence quark content: something other than $qqq$ or $q\overline{q}$. Other color-singlet configurations are allowed by $SU(3)$, so the question becomes whether the dynamics of the strong interaction allows such states to form and to be at least metastable. So-called *pentaquark* states, with valence content $qqqq\overline{q}$ (or charge conjugate), are an important example. Theoretical work on pentaquark states is briefly summarized in Section 24.2.1.

The subject became important for the $B$ Factories due to positive claims in 2003, initially from experiments working at the boundary of particle and nuclear physics. The original $\Theta(1540)^+$ evidence was from a search in photoproduction by LEPS (Nakano et al., 2003), and an analysis of kaon interaction data from DIANA (Barmin et al., 2003). Many other positive claims for the $\Theta(1540)^+$ and other pentaquark states followed. These are summarized in Section 24.2.2.

A programme of searches then followed at both Belle and *BABAR*, which may be roughly divided by production mechanism: inclusive production (Section 24.2.3), searches in $B$ decays (Section 24.2.4), and interaction of primary particles in the material of the detector (Section 24.2.5). The latter provide the most direct challenge to the original pentaquark claims. The lessons from $B$ Factory data are summarized in Section 24.2.6.

### 24.2.1 Theoretical studies on pentaquarks

In the light-quark baryon sector, spin and flavor are combined to yield a flavor-spin $SU(6)$ representation of the spectroscopy. The baryon states are composed of three quarks, each of which is assumed to be a color triplet, with all baryons assumed to be color singlets. The allowed state vectors are thus anti-symmetric in color, and symmetric in space-spin-flavor, and in this way the quark configurations satisfy Fermi statistics. All of the known baryon states up to $\sim 2\,\mathrm{GeV}$, *i.e.* those with at least 3 stars in the Particle Data Group (PDG) evaluation, are accommodated in this scheme, and the non-relativistic three-quark potential model of Isgur and Karl can also explain the occurrence of "missing states" on the basis of their highly inelastic decay characteristics (Koniuk and Isgur, 1980). In fact, in the review article by Hey and Kelly (1983) it is pointed out that the successful description of the known baryon states in terms of confined triplets of spin-half quarks with a hidden color degree of freedom is perhaps the most significant outcome of all the attempts at describing the spectroscopy (and couplings) of the baryonic excitations. However, in the context of QCD, the apparent absence of baryons with composition $qqqq\overline{q}$ or $qqqg$ is not understood.

#### 24.2.1.1 Early partial wave analyses

The most obvious way to prove the existence of $qqqq\overline{q}$ states is to identify resonant structure in the $KN$ system, since an $S = +1$ baryon must at minimum contain five quarks. Since the early days of the quark model, such evidence has been sought in the Partial Wave Analysis (PWA) of $KN$ elastic, charge exchange, and inelastic scattering data. The results of these searches for $Z^*$ states, as they were called, are summarized briefly in Hey and Kelly (1983). Only the $P_{01}$ and $P_{13}$ amplitudes show hints of structure (in the mass region 1.8–1.9 $\mathrm{GeV}/c^2$), but it was then concluded that there is no convincing evidence of resonant behavior. In its 1986 review (Aguilar-Benitez et al., 1986) the PDG drew a line under these studies with the following comment:

> *... the [PWA] results permit no definite conclusion — the same story heard for 15 years. The standards of proof must simply be much more severe here than in a channel in which many resonances are already known to exist. The general prejudice against baryons not made of three quarks and the lack of any experimental activity in this area make it likely that it will be another 15 years before the issue is decided.*

The $Z^*$ listings appeared for the last time in that issue, and starting with the following review (Yost et al., 1988), only a reference to the 1986 edition was included. After that, the subject of exotic baryons did not receive much attention except from a few theorists motivated by the old chiral soliton ideas due to Skyrme (1962).

#### 24.2.1.2 The pentaquark revival; prediction of the $\Theta(1540)^+$

A decade later Diakonov, Petrov, and Polyakov (1997) made a remarkable prediction concerning the existence of a positive strangeness baryon state just above $KN$ threshold in mass ($\sim 1.53\,\mathrm{GeV}/c^2$) and of extremely narrow width ($\Gamma < 15\,\mathrm{MeV}$). Such a state would be manifestly exotic since it would have a minimal content of four quarks and an anti-quark ($\overline{s}$). The predictions were based on the generalization of a chiral soliton model (Skyrme, 1962), in which nucleons are viewed as solitons of the pion field. Quantization of the rotations of this field in ordinary and flavor $SU(3)$ space leads to a baryon ground state which is an octet with spin 1/2, and to a first excited state which is a spin 3/2 decuplet, just as happens to be the case in nature. In the case of three flavors, the next excitation corresponds to an anti-decuplet with spin 1/2. The structure of this anti-decuplet is shown in Fig. 24.2.1, with the exotic $S = +1$ state occupying the apex of the triangle. The states at the extreme edges of the base of the triangle are also manifestly exotic with minimal quark content as shown. As indicated, the mass splitting between the isospin multiplets of different strangeness is linear; in Diakonov, Petrov, and Polyakov (1997), it is estimated to be $\sim 180\,\mathrm{MeV}/c^2$. The overall mass scale was defined by identifying the nucleon member of the anti-decuplet with the $N(1710)$ resonance (Eidelman et al., 2004), and this led to an estimation of the $\Theta(1540)^+$ mass of $\sim 1.53\,\mathrm{GeV}/c^2$. A subsequent calculation of the width of this state led to the estimate that it should be $\sim 15\,\mathrm{MeV}$, which, if correct, should make it amenable to experimental detection provided that the cross section for production is large enough.

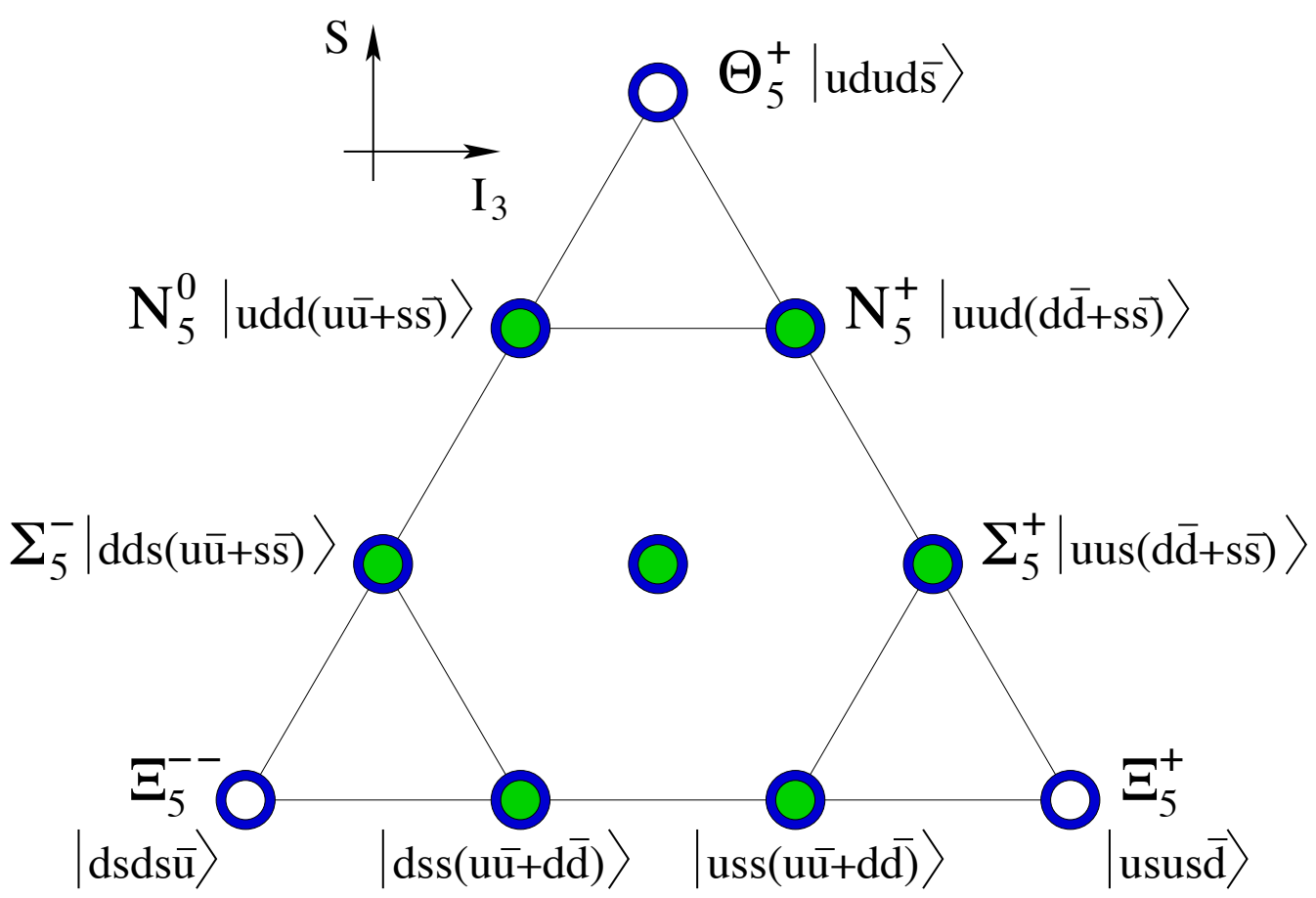

**Figure 24.2.1.** The anti-decuplet (annuli) and octet (filled circles) that are generally assumed for the lowest mass pentaquarks. The vertical axis is the strangeness and the horizontal axis is the isospin. The quark content of the anti-decuplet members is shown. (Reproduced from Aubert, 2004aa.)

On page 312 of Diakonov, Petrov, and Polyakov (1997), there is a comment that the data from the LASS experiment might be used to look for the $\Theta(1540)^+$. This was in fact done in 1997 using data selected for the reaction $K^+p \to \pi^+K^+n$ at an incident momentum $11\,\mathrm{GeV}/c$, but no signal was observed. The result shown at the $7^{th}$ International Symposium on Meson-Nucleon Physics and the Structure of the Nucleon in 1997 can be found in Napolitano, Cummings, and Witkowski (2004). Old, but high quality, bubble chamber data selected for the reaction $K^+p \to \pi^+K^0p$ in the momentum region around $1\,\mathrm{GeV}/c$ (Berthon et al., 1973) also fail to reveal a signal, and suggest cross section values less than $10\,\mu$b. Representative Dalitz plots in the $\Theta(1540)^+$ region reported in the 2004 PDG review article by George Trilling (Eidelman et al., 2004) are quite clear: there is no evidence of $\Theta(1540)^+$ production.

For more than five years after the publication of the Diakonov et al. paper there was no experimental evidence to support the prediction of the $\Theta(1540)^+$ but the situation changed dramatically in the fall of 2002 when the LEPS Collaboration claimed to have observed photoproduction of a $\Theta(1540)^+$ candidate (Nakano et al., 2003).

Then, during a pentaquark workshop at Jefferson Lab (JLab) in November 2003, K.Kadija, representing the NA49 Collaboration, presented evidence for the production of a $\Xi_5(1860)^{++}$ pentaquark candidate and a neutral partner in $p-p$ interactions at a CM energy of 17.2 GeV (later published in Alt et al., 2004). If this state is interpreted as belonging to the anti-decuplet of Diakonov et al., the mass ($\sim 1.862\,\mathrm{GeV}/c^2$) and width ($< 18\,\mathrm{MeV}$) values are much smaller than those predicted ($2.07\,\mathrm{GeV}/c^2$ and $> 140\,\mathrm{MeV}$, respectively). In a subsequent paper, Diakonov and Petrov (2004) no longer used the $N(1710)$ to set the absolute mass scale for their predictions, but used the mass values of the $\Theta(1540)^+$ and $\Xi_5(1860)$ to define a new anti-decuplet central mass and mass splitting ($\sim 108\,\mathrm{MeV}/c^2$). The reduction of the splitting from the $180\,\mathrm{MeV}/c^2$ value of Diakonov, Petrov, and Polyakov (1997) could be reproduced by increasing the value of the nucleon sigma term used in the calculation, and arguments were given to indicate that the NA49 width limit was reasonable if the true width of the $\Theta(1540)^+$ was $< 3\,\mathrm{MeV}$, and the two states were members of the same anti-decuplet.

#### 24.2.1.3 Subsequent studies

Stimulated by the flurry of experimental activity on the pentaquark front, other models of the "quark cluster" type soon appeared. The first of these, due to Karliner and Lipkin (2003), divided the pentaquark constituents into a di-quark and a tri-quark cluster with the quarks of identical flavor in different clusters. Each cluster has isospin zero and is a color non-singlet (separating the pairs of identical flavor); one unit of orbital angular momentum then yields $IJ^P = 0\frac{1}{2}^+$ as expected for the lowest anti-decuplet, and the centrifugal barrier keeps the clusters beyond the range of the repulsive color-magnetic force. The individual clusters bind together as a result of color-electric forces. The model yields a $\Theta(1540)^+$ mass estimate of $\sim 1.59\,\mathrm{GeV}/c^2$, and an anti-decuplet mass splitting which is only $\sim 50\,\mathrm{MeV}/c^2$. This is a quark-based model which led to a resonant $S = +1$ baryon state in the vicinity of $KN$ threshold.

A second model of this type is due to Jaffe and Wilczek (2003). The $\Theta(1540)^+$ is described in terms of two $ud$ diquarks and a bachelor $\bar{s}$ quark. The ground state diquark-diquark-antiquark configuration leads to a degenerate octet and anti-decuplet whose symmetry is broken as a result of the strange quark mass, leading to mixing of the two multiplets. Incorporating the $\Theta(1540)^+$ as the $Y = 2$ member of the anti-decuplet leads to a somewhat different spectroscopy than that of Diakonov, Petrov, and Polyakov (1997), and in particular yields a $J^P = 1/2^+$ nucleon state at a mass lower than the $\Theta(1540)^+$ which is associated with the broad Roper resonance.[182] However, the predicted mass of the $\Xi_5(1860)$ state is more than 100 MeV below the mass of the state claimed by the NA49 experiment.

[182] A Roper resonance is a broad baryon state with a mass circa 1440 MeV, denoted by $P_{11}(1440)$, see Alvarez-Ruso (2010) and references therein for more details.

Finally, it should be possible to employ Lattice Gauge techniques to investigate the hypothetical existence of an exotic pentaquark resonant state in the vicinity of $KN$ threshold. Holland and Juge (2006) discussed the status of such calculations, and concluded that there was as yet no evidence favoring the existence of any such state. However the paper cautions that "absence of evidence" should not be considered to be "evidence of absence" at the present early stage of these efforts.

Since the claim of evidence for $\Theta(1540)^+$ production, many models generating estimates of production cross section rates in photoproduction and hadroproduction reactions have been presented in the literature. Cross section estimates range from a fraction of a nanobarn to several hundred nb in photoproduction, and from a fraction of a microbarn to several millibarns in hadroproduction, depending on the reaction and details of the model. A sampling of such calculations can be found in Oh, Kim, and Lee (2004a,b), and in some of the references listed in these papers. Some of the calculations yield unacceptable results: *e.g.* for the reaction $K^+p \to \pi^+\Theta(1540)^+$ the predicted cross section is $\sim 1.5\,\text{mb}$ at low beam momentum, clearly inconsistent with the published data (Berthon et al., 1973).

### 24.2.2 Positive claims in 2003–2005

#### 24.2.2.1 $\Theta(1540)^+$

Table 24.2.1 is adapted from Dzierba, Meyer, and Szczepaniak (2005), and summarizes the various claims for the observation of the $\Theta(1540)^+$ pentaquark state. The corresponding collaborations and reactions studied are summarized; also listed are the mass and width estimates, and claimed significance.

The first three rows of the table result from photoproduction of the same exclusive final state, with the signal appearing in the $K^+n$ invariant mass distribution (Nakano et al., 2003; Stepanyan et al., 2003). The next two rows are from photoproduction of different exclusive final states on a proton target, with the signal again appearing in the $K^+n$ invariant mass distribution (Barth et al., 2003; Kubarovsky et al., 2004). The quoted mass and width values appear to be consistent.

The remaining measurements are for the $K^0_S p$ system. Only for the exclusive final states of Abdel-Bary et al. (2004) and Barmin et al. (2003) is it known with certainty that the $K^0_S$ is produced as a $K^0$. These eight measurements are obtained using a variety of incident particles ($K^+$, $p$, neutrino, and $e^\pm$) and targets, and six of them are from inclusive production processes. These measurements yield a discrepancy of at least $3\sigma$ between the mass values from the $K^+n$ and $K^0_S p$ systems.

The overall mean value of the mass from these measurements is $1535.3 \pm 2.6\,\text{MeV}/c^2$, where the error is estimated from the spread in the individual values, since this is larger than would be expected from the quoted errors on the measurements.

If the $\Theta(1540)^+$ exists, its effects should be seen in $K^+d$ scattering data for incident $K^+$ laboratory momenta around $440\,\text{MeV}/c$, unless its width is very small. Since the CM energy in this region is below pion production threshold, the cross section reaches the unitarity limit at resonance, and for spin 1/2 is $\mathcal{B}_i \times \mathcal{B}_f \times 68\,\text{mb}$, where $\mathcal{B}_i$ and $\mathcal{B}_f$ are the branching fractions to the initial and final states. These branching fractions are equal to 0.5 for the $\Theta(1540)^+$. Integrating over the resonance peak, the net effect results in a contribution $\Gamma \times (\mathcal{B}_i \times \mathcal{B}_f \times 107\,\text{mb})$, where $\Gamma$ is the width of the resonance. Several studies of the rather sparse scattering data in this region (Arndt, Strakovsky, and Workman, 2003; Cahn and Trilling, 2004; Haidenbauer and Krein, 2003; Nussinov, 2003, 2004) conclude that $\Gamma$ must be less than 5 MeV. In addition, in Cahn and Trilling (2004) the Xe bubble chamber data of Barmin et al. (2003), when interpreted in terms of $K^+n$ charge exchange scattering, yield the value $\Gamma = (0.9 \pm 0.3)\,\text{MeV}$, with no estimate of systematic uncertainty. The conclusion therefore is that if the $\Theta(1540)^+$ exists its width must be $< 5\,\text{MeV}$, and may even be as small as $\sim 1\,\text{MeV}$. The width estimates in Table 24.2.1 are consistent with such values.

Since the mass distributions of the claimed signals were obtained in many different contexts, it is difficult to believe that the signals might be spurious. However, the fact that the mass value estimates are spread over a range which seems too large for the uncertainties quoted, and the observation that in all of the distributions, the peak signal bin contains only 0–50 events above background, indicate the need to exercise caution. This is especially relevant in light of JLab results (DeVita, 2005) in what is essentially a much higher statistics repetition of the SAPHIR experiment. Cross section estimates for $\Theta(1540)^+$ production are either non-existent, unclear, or unreliable.

The COSY experiment (Abdel-Bary et al., 2004) quotes a cross section of $(0.4 \pm 0.1 \pm 0.1)\,\mu\text{b}$ but does not indicate clearly whether all branching fraction values and isospin Clebsch-Gordan coefficients have been taken into account. The SAPHIR experiment (Barth et al., 2003) initially quoted a cross section for the reaction $\gamma p \to \overline{K}^0\Theta(1540)^+$ of 300 nb, but this has since been reduced to 50 nb, and JLab measurements of the same reaction (DeVita, 2005) yield a 95% C.L. upper limit in the range 1–4 nb for the relevant region of photon laboratory energy. As of 2005, when the $B$ Factory studies were being performed, there was no other useful information.

#### 24.2.2.2 $\Xi_5(1860)$

These are the states contributing the base of the antidecuplet triangle in Fig. 24.2.1. Only one observation has been claimed to date, from the NA49 Collaboration studying the interactions of a 158 GeV/$c$ proton beam in a liquid hydrogen target (Alt et al., 2004). The combined invariant mass distributions for the systems $\Xi^-\pi^-$, $\Xi^-\pi^+$, and their anti-particle counterparts reveal a narrow signal of $\sim 68$ events over a background of $\sim 77$ events; the width is consistent with the detector resolution ($18\,\text{MeV}/c^2$). The fitted mass value is $(1.862 \pm 0.002)\,\mathrm{GeV}/c^2$. A $\Xi_5(1860)^{++}$ baryon is manifestly exotic; it occupies the lower-left vertex of the anti-decuplet, with the minimal quark composition shown. It is troubling that there have been no other observations of this state to date, but even more troubling that members of the same collaboration have publicly questioned the analysis (Fischer and Wenig, 2004).

**Table 24.2.1.** Results from experiments reporting the observation of the $\Theta(1540)^+$ in a $K^+n$ or $K^0_Sp$ invariant mass distribution, expanded from Table 3 of Dzierba, Meyer, and Szczepaniak (2005). Seven of these experiments involve real or virtual photoproduction (denoted by *), and four more involve hadroproduction on a hydrogen or nuclear target. $(\mathrm{Xe})'$ denotes a recoiling Xe. The second LEPS result (Nakano, 2004) was a conference presentation that did not include a mass and width.

| Experiment | Reaction | Mass (MeV) | Width (MeV) | Significance ($\sigma$) | Reference |
|---|---|---|---|---|---|
| LEPS(1)* | $\gamma^{12}\mathrm{C} \to K^+K^-X$ | $1540 \pm 10$ | $< 25$ | 4.6 | Nakano et al. (2003) |
| LEPS(2)* | $\gamma d \to K^+K^-X$ | – | −− | – | Nakano (2004) |
| CLAS($d$)* | $\gamma d \to K^+K^-(n)p$ | $1542 \pm 5$ | $< 21$ | 5.2 | Stepanyan et al. (2003) |
| CLAS($p$)* | $\gamma p \to K^+K^-(n)$ | $1555 \pm 10$ | $< 26$ | 7.8 | Kubarovsky et al. (2004) |
| SAPHIR* | $\gamma p \to \overline{K}^0K^+(n)$ | $1540 \pm 6$ | $< 25$ | 4.8 | Barth et al. (2003) |
| COSY | $pp \to \Sigma^+K^0p$ | $1530 \pm 5$ | $< 18$ | 4–6 | Abdel-Bary et al. (2004) |
| JINR | $p(\mathrm{C_3H_8}) \to K^0_SpX$ | $1530 \pm 5$ | $9.2 \pm 1.8$ | 5.5 | Aslanyan et al. (2005)† |
| SVD | $pA \to K^0_SpX$, ($A$ = C, Si, Pb) | $1540 \pm 8$ | $< 24$ | 5.6 | Aleev et al. (2005) |
| DIANA | $K^+\mathrm{Xe} \to K^0p(\mathrm{Xe})'$ | $1539 \pm 2$ | $< 9$ | 4.4 | Barmin et al. (2003) |
| $\nu$BC(ITEP) | $\nu\mathrm{Ne} \to K^0_SpX$ | $1533 \pm 5$ | $< 20$ | 6.7 | Asratyan et al. (2004)‡ |
| NOMAD | $\nu_\mu A \to K^0_SpX$, ($A$ = Fe, Al, Pb) | $1528.7 \pm 2.5$ | $< 21$ | 4.4 | Camilleri (2005) |
| HERMES* | $e^+d \to K^0_SpX$ | $1528 \pm 3$ | $13 \pm 9$ | $\sim 5$ | Airapetian et al. (2004) |
| ZEUS* | $e^-p \to e^-K^0_SpX$ | $1522 \pm 3$ | $8 \pm 4$ | $\sim 5$ | Chekanov et al. (2004a) |

† Aslanyan, Emelyanenko, and Rikhkvitzkaya (2005)
‡ Asratyan, Dolgolenko, and Kubantsev (2004)

#### 24.2.2.3 $\Theta_c(3100)$

Again there was only one experiment claiming evidence for this anti-charm baryon state (Aktas et al., 2004). The signal is observed by the H1 Collaboration at HERA in the $D^{*-}p$ and $D^{*+}\overline{p}$ invariant mass distributions; the fitted mass value is $(3.099 \pm 0.003 \pm 0.005)\,\mathrm{GeV}/c^2$ and the width is consistent with detector resolution ($\sim 12\,\mathrm{MeV}/c^2$). The minimal quark content is $uudd\overline{c}$, so that the state is a manifestly exotic pentaquark candidate. There has been no corroboration of this state to date; in particular, the ZEUS experiment operating under the same conditions at HERA has found no evidence of a signal (Chekanov et al., 2004b). The H1 result was later retracted.

#### 24.2.2.4 Negative searches

The experiments which have searched in vain for evidence of the three pentaquark states are summarized in Dzierba, Meyer, and Szczepaniak (2005). There is a fairly detailed discussion of the various non-observations in the reference, which we do not repeat here.

### 24.2.3 Inclusive production searches

#### 24.2.3.1 Strange pentaquark candidates

A dedicated search for inclusive production of the $\Theta(1540)^+$ was performed by *BABAR* (Aubert, 2005ac); searches for the doubly-strange states $\Xi_5(1860)^0$ and $\Xi_5(1860)^{++}$, reported by NA49 (Alt et al., 2004) and supposed to be partners of the $\Theta(1540)^+$ within a pentaquark $\overline{10}$ multiplet, were presented in the same paper.

In this analysis, the $pK^0_S$ system, with $K^0_S \to \pi^+\pi^-$, was investigated for evidence of $\Theta(1540)^+$ production in $e^+e^-$ collisions at a CM energy of 10.58 GeV, and 0.04 GeV below.

The $pK^0_S$ invariant mass distribution showed a large (~100,000 events) signal corresponding to the production of the $\Lambda(2285)^+$ charmed baryon, and this was used to verify that the dependence of mass resolution upon CM momentum $p^*$ was well reproduced in Monte Carlo simulation. The same simulation in the mass region of the $\Theta(1540)^+$ yields mass resolution values in the range 1.7–2.2 $\mathrm{MeV}/c^2$. The excellent agreement obtained for the $\Lambda(2285)^+$ data demonstrates that these resolution estimates are reliable. No $\Theta(1540)^+$ signal was observed, neither when the $pK^0_S$ mass distribution was taken as a whole, nor when it was examined as a function of $p^*$. This remained true for sub-samples of the data for which each event was required to contain an identified $K^-$ (in order to bias the $K^0_S$ sample toward $K^0$ rather than $\overline{K}^0$), and/or an anti-proton (in order to bias the sample toward conserved baryon number).

For a fixed $\Theta(1540)^+$ mass value of $1.54\,\mathrm{GeV}/c^2$, and fixed total width values of 1 MeV and 8 MeV, resolution-smeared $P$-wave Breit-Wigner lineshapes were used to obtain 95% C.L. cross section upper limits for each 0.5 GeV/$c$ momentum interval in the range 0–5 GeV/$c$.

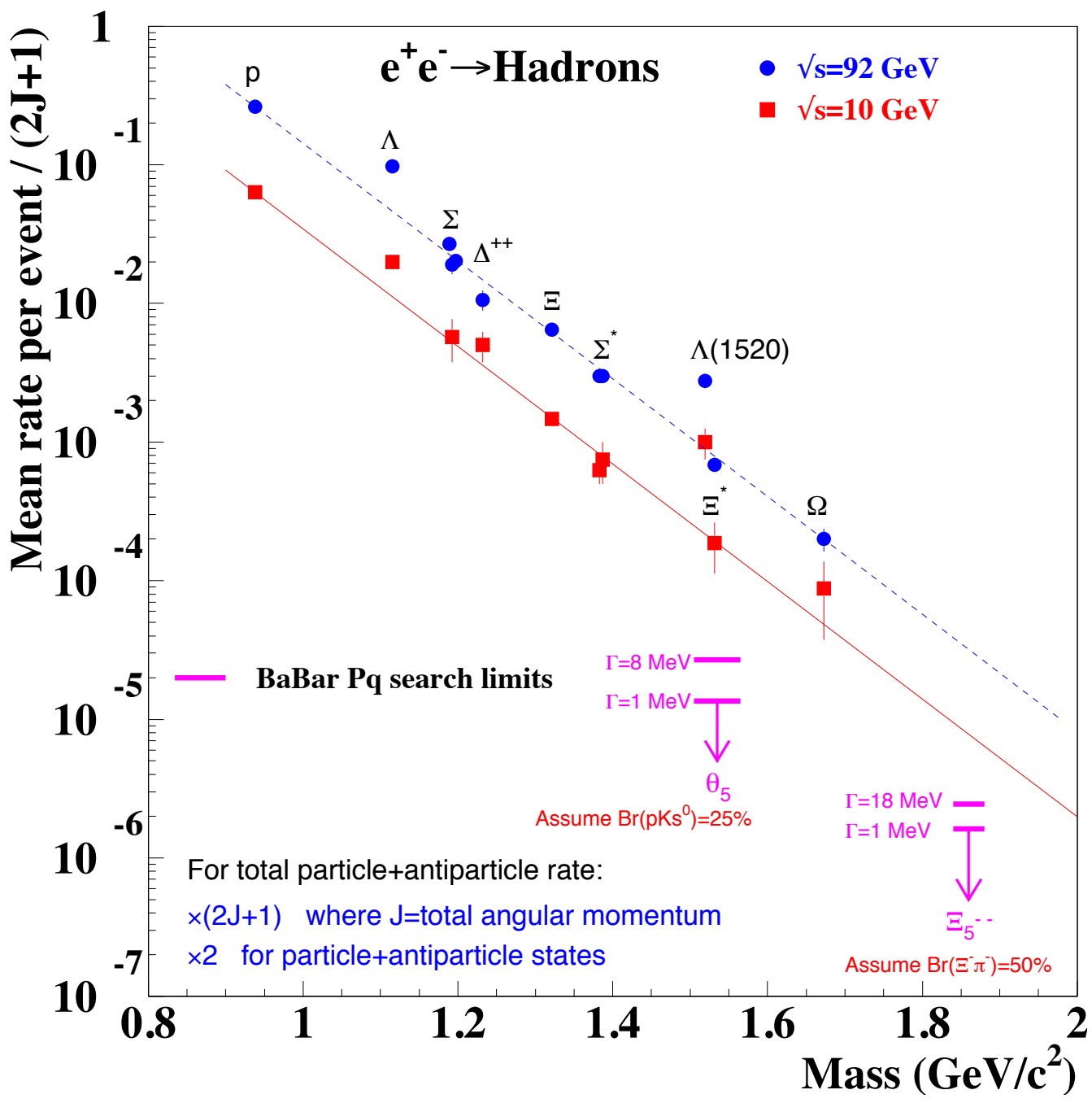

**Figure 24.2.2.** From Aubert (2004aa): Compilation of light (no $c$ or $b$ quarks) baryon production rates in $e^+e^-$ to hadrons from the PDG. Limits from *BABAR* on pentaquark production in $e^+e^- \to$ hadrons are shown, at least an order of magnitude below the expected trend for "normal" baryons.

Limits for the integrated cross section values of 80 fb and 60 fb were extracted for width values 1 MeV and 8 MeV, respectively. The values obtained for a mass choice of 1.53 GeV/$c^2$ were virtually identical to these. These upper limit values are significantly below the cross section values which would be expected for a particle of mass 1.54 GeV/$c^2$ on the basis of the observed production rates for "ordinary" hadrons (see Fig. 24.2.2). This suggests that, if the $\Theta(1540)^+$ pentaquark state does in fact exist, its production is highly suppressed in $e^+e^-$ interactions at 10.58 GeV with respect to that of well-established hadrons.

Two CLAS results testing for the possible existence of the $\Theta(1540)^+$ were found to be in disagreement with each other (McKinnon et al., 2006; Stepanyan et al., 2003). The 2003 result claimed observation of a narrow $K^+n$ resonance in the process $\gamma d \to K^+K^-pn$. The 2006 paper is the result of a dedicated high luminosity run by the CLAS collaboration designed specifically to test the validity of former publication in a data sample over 30 times larger than the original one. This subsequent paper found no evidence for a narrow resonance. In 2007 CLAS re-analysed their data using a consistent Bayesian methodology (Ireland et al., 2008) in order to understand if those previous results were compatible with each other or not, and to verify if there was any evidence for a $\Theta(1540)^+$ signal. They conclude that the results are indeed compatible, however that there is insufficient information in the 2003 data to support the original claim of the existence of a new state.

#### 24.2.3.2 The charmed pentaquark candidate $\Theta_c(3100)$

Searches for the charmed pentaquark candidate $\Theta_c(3100)^0$ were performed in inclusive production by *BABAR* (Aubert, 2006ar), and in $B$ decays by Belle (Abe, 2004e).

In *BABAR*, no evidence for the production of the $\Theta_c(3100)^0$ state in a sample of over 125,000 $pD^*$ combinations was found. Upper limits on the product of the inclusive $\Theta_c(3100)^0$ production cross section times branching fraction to this mode for two assumptions of its natural width, which are valid for any state in the vicinity of 3100 MeV/$c^2$, were set. It would be interesting to compare these limits with the rate expected for an ordinary charmed baryon of mass 3100 MeV/$c^2$. However, rates had been measured for only two charmed baryons, the $\Lambda_c^+(2285)$ (Seuster, 2006; see also Eidelman et al., 2004), and $\Sigma_c(2455)$ (Eidelman et al., 2004) at the time of publication, with a precision that does not allow a meaningful estimate of the mass dependence. The mass dependence observed (Eidelman et al., 2004) for non-charmed baryons in $e^+e^-$ annihilations predicts a rate for a 3100 MeV/$c^2$ baryon about 1,000 times smaller than that of the $\Lambda_c^+(2285)$. Belle limits for a narrow state in both $e^+e^- \to c\bar{c}$ and $\Upsilon(4S)$ events are roughly 1,000 and 500 times below the $\Lambda_c^+(2285)$ and $\Sigma_c(2455)$ rates, respectively.

As a result the existence of an ordinary charmed baryon with this mass and decay mode cannot be excluded.

### 24.2.4 Searches in $B$ decays

Parasitic searches were performed at the $B$ Factories for the $\Theta(1540)^+$ and a hypothetical partner $\Theta^{*++}$ (Wang, 2005; Aubert, 2005p), through analyses of $B \to p\bar{p}K$ decays (Section 17.12). These searches failed to provide any indication of either of these pentaquark candidates. Belle concludes from their data that a quark fragmentation interpretation is supported, while a resonant gluonic state origin is disfavored.

### 24.2.5 Searches using interactions in the detector material

If they exist, pentaquarks are thought to be produced at low energy, and as a result low energy searches are of paramount importance. To access the low energy domain, Belle performed a detailed investigation using secondary interactions of hadrons in the detector material. The protons and kaons that do not originate from the $e^+e^-$ interaction point were selected and $pK$ pairs that form high-quality vertices with the radial distance $R > 1$ cm were considered. The spatial distribution of the $pK_S^0$ pairs for the central part of the Belle detector is shown in Fig. 24.2.3 for the two running periods with different configurations of the inner detectors. The beam pipe, the layers of silicon vertex detector (three layers for SVD1 and four layers for SVD2) and inner support of the central drift chamber are clearly seen, which demonstrates that the secondary interactions are the dominant source of the selected $pK_S^0$ vertices. Similar "detector tomography" pictures were obtained for the $pK^-$ and $pK^+$ vertices.

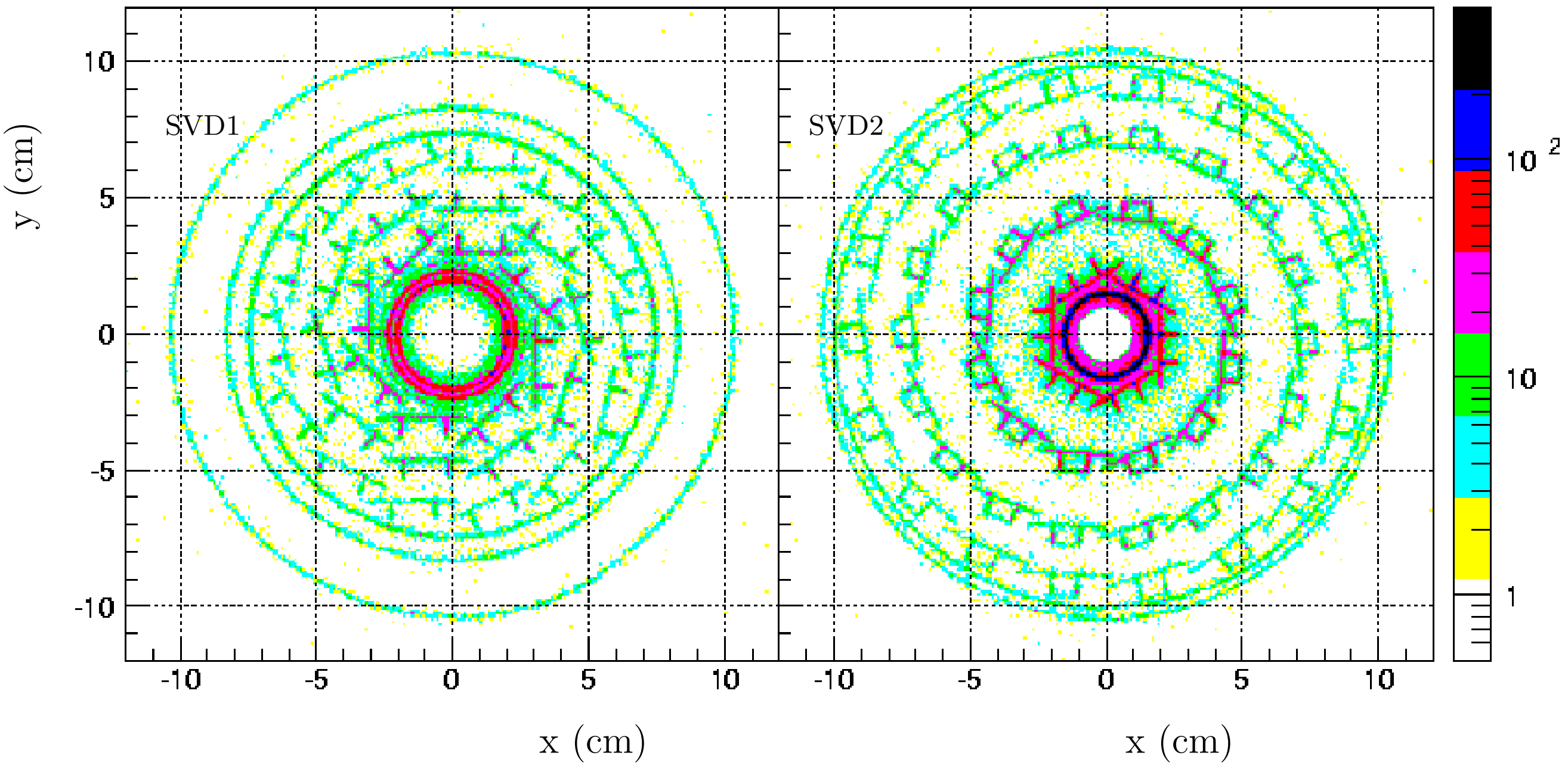

**Figure 24.2.3.** Distribution of reconstructed secondary $pK_S^0$ vertices in the Belle detector for both the SVD1 (left) and SVD2 (right) data samples, from Mizuk, Danilov (2006).

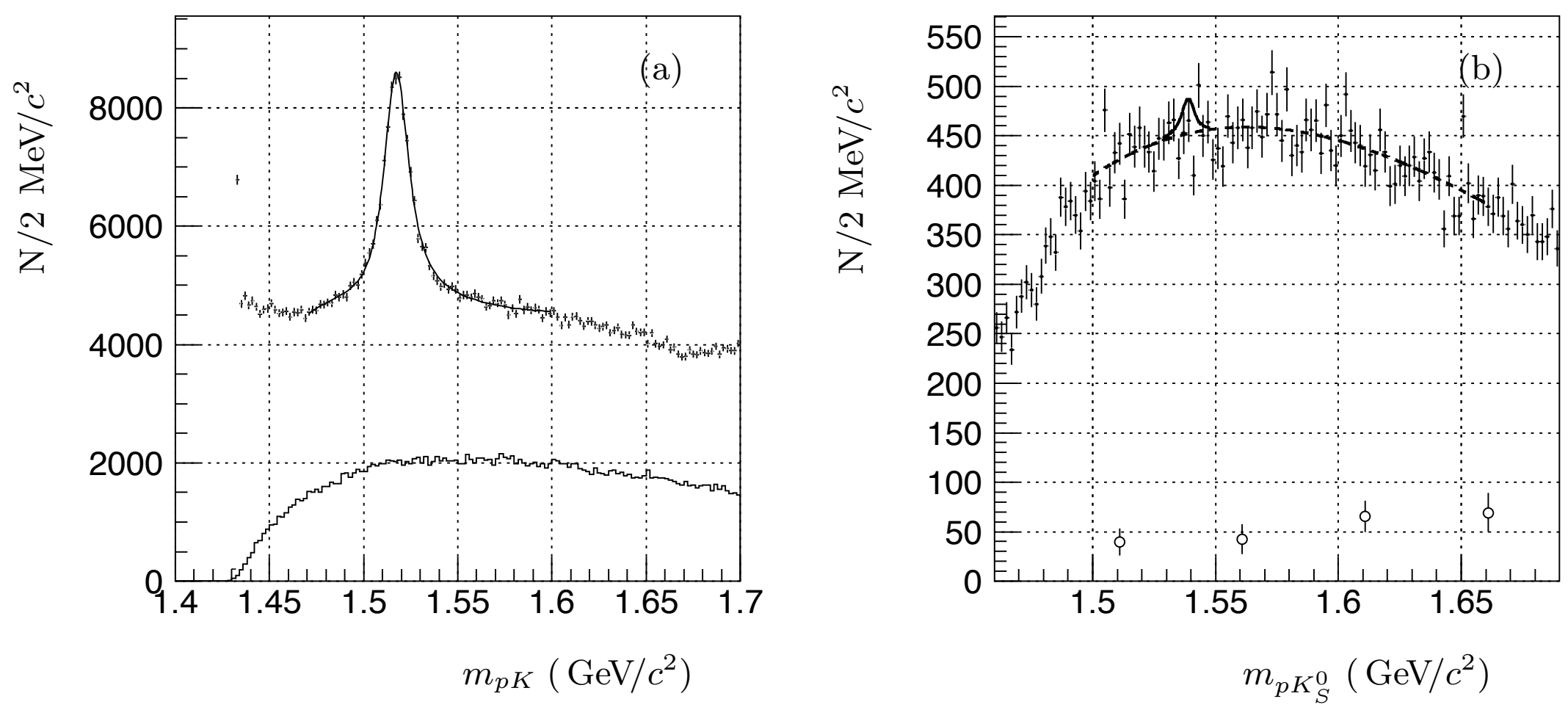

**Figure 24.2.4.** (a) Mass spectra of $pK^{\pm}$ (points with error bars) and $pK_S^0$ (histogram) of secondary pairs, and (b) secondary $pK_S^0$ pairs (small dots with error bars) and expected yield of the charge exchange reaction per 2 MeV/$c^2$ (open dots). The dashed line in (b) corresponds to the result of a fit to a third order polynomial and the $\Theta(1540)^+$ contribution expected from the DIANA result is shown (solid line). Figures are reproduced from Mizuk, Danilov (2006).

Belle performed the search both for inclusive production and for formation of the $\Theta(1540)^+$ (Mizuk, Danilov, 2006). The invariant mass distributions for the secondary $pK^-$ and $pK_S^0$ vertices are shown in Figure 24.2.4. There is a clear $\Lambda(1520)$ signal in the $pK^-$ distribution while there is no $\Theta(1540)^+$ signal in the $K_S^0$ distribution. Belle used the $\Lambda(1520)$ signal as a reference and placed a 90% C.L. upper limit on the ratio

$$\frac{\sigma(KN \to \Theta(1540)^+X)}{\sigma(KN \to \Lambda(1520)X} < 2.5\%. \qquad (24.2.1)$$

Belle finds that it is very rarely that there is an additional kaon track from the secondary $pK$ vertex. Therefore the interactions are induced by strange particles, primarily kaons, with the contribution of hyperons estimated to be negligible. Given that the typical projectile kaon momentum is only 1 GeV, this is a unique null result in the low energy domain. The upper limit is much smaller

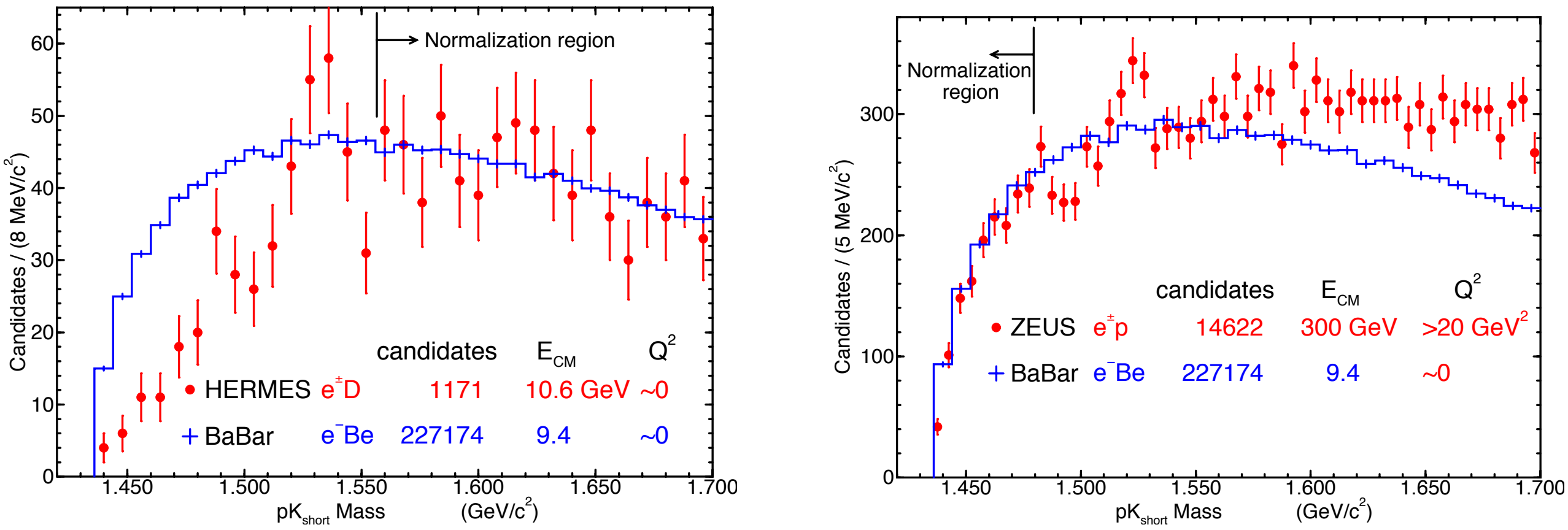

(a) Comparison with HERMES (Airapetian et al., 2004), using HERMES data above $1.58\,\mathrm{GeV}/c^2$ for normalization.

(b) Comparison with ZEUS (Chekanov et al., 2004a), using ZEUS data below $1.48\,\mathrm{GeV}/c^2$ for normalization.

**Figure 24.2.5.** From Coleman (2005): *BABAR* $K^0_S\,p$ mass distributions from electroproduction in Be, compared with results from previous experiments. In each case, the *BABAR* distribution is normalized to that of the other experiment in the region shown.

than the corresponding values from experiments with positive results, however, its interpretation remains model-dependent.

To obtain a model-independent constraint on the $\Theta(1540)^+$ parameters Belle searched for the formation of the $\Theta(1540)^+$ using the exclusive charge-exchange reaction $K^+p \to pK^0_S$. The yield of $\Theta(1540)^+$ formed in s-channel transitions is directly related to its width. In this search the projectile $K^+$ was not reconstructed and its momentum was determined from the energy-momentum of the secondary $pK^0_S$ pair and from the vertex constraints. The Fermi momentum of the struck neutron was also determined. The procedure was verified and calibrated using $D^{*+} \to \pi^+ D^0(\to \pi^- K^+)$ decays in which the $K^+$ had interacted in the detector material. To suppress the background from inelastic $K^+n \to pK^0_S X$ scattering, Belle applied veto on additional charged tracks from the secondary vertex and required that the Fermi momentum be in the range 50 to 300 MeV. These requirements suppress inelastic reactions by a factor of four. The charge exchange reaction accounts for about 10% of the resulting sample. No $\Theta(1540)^+$ signal is observed and an upper limit on the $\Theta(1540)^+$ width is set $\Gamma(K^+p \to \Theta(1540)^+ \to pK_s) < 0.64\,\mathrm{MeV}$ at 90% C.L. for a $\Theta(1540)^+$ mass of 1539 MeV. This upper limit is marginally consistent with the measurement by the DIANA experiment $(0.9 \pm 0.3)$ MeV (Barmin et al., 2003) and does not support the evidence reported by DIANA.

Likewise, the search for the $\Theta(1540)^+$ was extended to the interactions of secondary hadrons, background tracks of every type, and beam halo electrons and positrons in the material of the inner *BABAR* detector (Coleman, 2005). It was demonstrated that the candidate $(K^0_S,\, p)$ vertices reproduce the detector geometry very well, however the inclusive $K^0_S p$ mass distribution shows no pentaquark signal.

Sub-samples of the candidates with at least one associated charged track, and also those remaining after rejecting $(K^0_S,\, p)$ vertices with at least one associated baryon were examined. The study has been restricted to the regions which can be interpreted as corresponding to electroproduction in Be (mainly) and Ta, and has been repeated including in addition a small sample of vertices with an associated electron. Again, no $\Theta(1540)^+$ signal has been observed. Since there is no quantitative information on the flux of off-beam electrons and positrons, it is not possible to estimate upper limits for the production cross section as was done in Aubert (2004aa, 2005ac).

The *BABAR* electroproduction results have been compared to those of the HERMES (Airapetian et al., 2004) and ZEUS (Chekanov et al., 2004a) experiments, the results of which are shown in Fig. 24.2.5. These comparisons seem to indicate a significant loss of acceptance for the HERMES experiment in the $K^0_S p$ mass region below $\sim 1.52\,\mathrm{GeV}/c^2$. There is no evidence for the $\Lambda(1480)$ and $\Theta(1540)^+$ signals of the ZEUS analysis, and this creates serious reservations about the significance of the $\Theta(1540)^+$ observations claimed by HERMES and ZEUS. A subsequent search by H1 for the $\Theta(1540)^+$also found no evidence for a signal (Aktas et al., 2006).

Finally, the *BABAR* results on the electroproduction of the $K^0_S pK$ system have been compared to the SAPHIR (Barth et al., 2003) results on the photoproduction of the $K^+nK^0$ final state. A crude attempt at normalizing the production of $\Lambda(1520)$ observed in both analyses leads to the conclusion that a $\Theta(1540)^+$ signal as observed by SAPHIR would not be significant in the corresponding *BABAR* $K^0_S p$ mass distribution. The *BABAR* results are in complete accord with those presented by the CLAS Collaboration (DeVita, 2005).

### 24.2.6 Summary

In summary, the final result of the high statistics searches at the $B$ Factories summarized above is that no experiment has yielded any evidence for the production of the $\Theta(1540)^+$ or other members of the pentaquark family. Furthermore, a comparison of the *BABAR* results on electroproduction in Be to those from the HERMES ($e^+D$) and ZEUS ($e^+p$) experiments leads to the conclusion that prior claims for the observation of $\Theta(1540)^+$ in electroproduction are not convincing. The inclusive and exclusive $K^+p$ interaction searches from Belle also result in the conclusion that claims for the observation of $\Theta(1540)^+$ are unconvincing. In light of these results it would seem clear that the only way to clarify the issue brought about by the experiments which still claim a signal is for those experiments to collect and analyse significantly more data. However, many of the analyses have been carried out in experiments which are now decommissioned. Nevertheless, for those which can be repeated, there is a clear need for new high statistics data to be collected with well-calibrated, large-acceptance detectors. Proof of principle has been amply provided by the results from CLAS (DeVita, 2005), which so convincingly refute the earlier claim of $\Theta(1540)^+$ observation from SAPHIR (Barth et al., 2003).

The whole saga is succinctly summed up by the 2006 and 2008 PDG reports as given below. The 2006 Review of Particle Physics concluded (Yao et al., 2006):

> *... there has not been a high-statistics confirmation of any of the original experiments that claimed to see the $\Theta(1540)^+$; there have been two high-statistics repeats from Jefferson Lab that have clearly shown the original positive claims in those two cases to be wrong; there have been a number of other high-statistics experiments, none of which have found any evidence for the $\Theta(1540)^+$; and all attempts to confirm the two other claimed pentaquark states have led to negative results. The conclusion that pentaquarks in general, and the $\Theta(1540)^+$, in particular, do not exist, appears compelling.*

The 2008 Review of Particle Physics went even further (Amsler et al., 2008):

> *There are two or three recent experiments that find weak evidence for signals near the nominal masses, but there is simply no point in tabulating them in view of the overwhelming evidence that the claimed pentaquarks do not exist. The only advance in particle physics thought worthy of mention in the American Institute of Physics "Physics News in 2003" was a false alarm. The whole story — the discoveries themselves, the tidal wave of papers by theorists and phenomenologists that followed, and the eventual "undiscovery" — is a curious episode in the history of science.*

Despite these null results, LEPS results as of 2009 continue to claim the existence of a narrow state with a mass of $1524 \pm 4\,\mathrm{MeV}/c^2$, with a statistical significance of 5.1 $\sigma$ (Nakano et al., 2009).

# Chapter 25
# Global interpretation

The chapters in this book have described a wide range of measurements that have been made by the $B$ Factories. It is possible to relate a number of these to either expectations from Standard Model (SM) based calculations, or hypothesized scenarios of physics beyond the SM. This chapter describes in detail how key measurements from the $B$ Factories can be combined in order to constrain our understanding of the CKM matrix, and the role of the KM mechanism in the SM (Section 25.1); and how one may be able to go beyond the SM and constrain features and parts of the parameter space for postulated new physics models (Section 25.2).

Prior to the $B$ Factories there had already been a variety of results on flavor physics, which had produced a reference point to be improved upon by the $B$ Factories. The measurement of $\epsilon_K$ as well as measurements in the charm sector provided indirect constraints on $B$ decays. In particular, the charm physics performed at CLEO-c, running at the open charm threshold, produced some results that would not be surpassed by *BABAR* or Belle and gave some tight constraints. ARGUS, CLEO, and the LEP experiments investigated bottom hadrons directly and gave a first picture of what to expect in terms of the SM picture of the Unitarity Triangle. However, the question remained: would the SM expectation be borne out experimentally?

There are two requirements for an observable to be of interest for global fits. Firstly the observable must be something that is "theoretically clean", meaning that the theoretical uncertainties in predicting this quantity in the SM are negligible or at least small. If this is not satisfied, then one will have trouble incorporating theoretical uncertainties into the comparison of data with the model. The result may be apparent deviations from the SM that in reality have a more mundane origin, since they are really a manifestation of an approximate calculation in a SM framework. The second requirement is that one can make a significant measurement of the observable of interest. When looking for *CP* violation in order to test the KM mechanism this translates into finding evidence for a non-zero level of indirect *CP* violation. While direct *CP* violation is also of interest, hadronic uncertainties arising from strong phase differences, which are difficult to calculate, limit the constraints that can be placed on the SM using direct *CP* asymmetries.

In the context of searching for physics beyond the SM the requirement that an observable be theoretically clean is, again, of paramount importance. However, the requirement that one can perform a significant measurement may be relaxed in some circumstances. Broadly speaking there are two extremes that one can focus on. The first is to search for effects of forbidden or rare decays that would otherwise be absent or unobservable within the SM. If a large signal were to be found in such a mode, then that would unequivocally point to new physics. The second type of study requires the measurement of a process that is sensitive to new physics and in addition has a measurable SM contribution. Here one aims to compare a precisely measured observable with the SM expectation to see if there is agreement or not. A significant deviation from the SM would indicate new physics, and compatibility with the SM can be used to infer constraints on new physics models or even exclude them. Any deviation from the SM expectation can be used to constrain the ratio of the complex coupling divided by the square of the energy scale of the new physics appearing in the Lagrangian. This approach is complementary to direct searches at the LHC where one constrains the energy scale directly.

In recent years the focus on constraining benchmark new physics scenarios has matured with the realization that individual searches have so far failed to produce an unambiguous discovery. By combining constraints from a number of different modes one can learn about the allowed structure of beyond-SM scenarios — that is, which patterns of new physics predictions are compatible with the data. This aspect of the $B$ Factory program remains a focus for the next generation of experiments.

Experimental guidance is required in order to help theorists understand what possible behavior could be allowed by new physics, while still being compatible with the existing constraints on the SM. For this reason it is important to understand the SM description of *CP* violation, which is discussed in Section 25.1, and also to explore benchmark new physics models (Section 25.2), until such time as an experimental discovery is made that can be used to guide us toward an improved understanding of nature.

## 25.1 Global CKM fits

***Editors:***
*Gerald Eigen (BABAR)*
*Ryosuke Itoh (Belle)*
*Marcella Bona (theory)*

### 25.1.1 Introduction

The previous sections of this book present a plethora of analyses that provide measurements of various observables, which in turn can be related to fundamental theory parameters. In particular, some observables are connected with the elements of the CKM matrix, which for three quark families is specified by four independent parameters as discussed in detail in Section 16.4. One convenient parameterization of the CKM matrix is the small-angle approximation by Wolfenstein, Eq. (16.4.4). In this approximation, following the Wolfenstein-Buras redefinition, there are four parameters, $A$, $\lambda$, $\overline{\rho}$ and $\overline{\eta}$, that fully determine the CKM matrix, see Eqns (16.4.5), (16.4.6), and (16.4.8).

The parameter $A$ is of order one and is determined by $V_{cb}$, $\lambda$ is the expansion parameter and is related to the Cabibbo angle by $\lambda = \sin\theta_c$, and $\overline{\rho}$ and $\overline{\eta}$ represent the apex of the Unitarity Triangle (see Section 16.5). A non-zero value of $\overline{\eta}$ indicates *CP* violation in the Standard Model. These four parameters are simultaneously determined by combining various experimental results and theory parameters connecting the observables to the CKM formulation in a global CKM fit.

This section gives a brief overview of *CP* violation in the era of the $B$ Factories (Section 25.1.2) before describing two global fit strategies that were continually updated during that time (Section 25.1.3). Experimental and theoretical inputs required in order to perform a global fit are discussed in Sections 25.1.4 and 25.1.5, respectively. Results of SM-based global fits are given in Section 25.1.6, and concluding remarks can be found in Section 25.1.7.

### 25.1.2 *CP* violation in the era of the *B* Factories

One of the most important results of the $B$ Factory physics program is the observation that *CP* violation in quark flavor-changing processes is described by the Kobayashi-Maskawa (KM) mechanism. The KM mechanism has been tested to $\sim \mathcal{O}(10\%)$ by current measurements from the $B$ Factories. In the KM mechanism, only one fundamental weak phase (the Kobayashi-Maskawa phase $\delta$ in Eq. (16.4.3)) is present and $\delta$ is the single source of *CP* violation in the SM in the quark sector. The consistency of $\varepsilon_K$ with the observed *CP* violation in $B_d$ decays and measurements of sides and angles of the Unitarity Triangle all point to this common origin of *CP* violation. In addition, both the value of $\Delta m_s$ and the recent LHCb measurements of the weak phase describing mixing-induced *CP* violation in $B_s$ decays agree with the KM mechanism within errors. This agreement between $b \to d$, $b \to s$ and $s \to d$ transitions is nicely illustrated in Fig. 25.1.1. Another important confirmation of the KM mechanism is the observation that a global CKM fit with only *CP*-conserving observables and one with only *CP*-violating observables yield the same values of $\overline{\rho}$ and $\overline{\eta}$, as shown in Figs 25.1.2 and 25.1.3.

### 25.1.3 Methodology

Two main collaborations have performed global CKM fits, periodically releasing new results in concert with updated measurements from the $B$ Factories. Therefore the impact of new measurements from *BABAR* and Belle were constantly discussed in terms of their constraints on the SM, and of possible tensions that could point toward new physics. The methodologies employed by these two groups, CKMfitter and UTfit, are discussed below. Detailed references to the various global fit analyses performed are given in the sections concerned with the experimental and theoretical inputs later in this chapter. For the purpose of this book, the two groups used a unified set of inputs coming from the book averages of $B$ Factory data in the context of the Standard Model description of quark flavor-changing processes.

A third approach, called the scan method, has been described in a recent paper (Eigen, Dubois-Felsmann, Hitlin, and Porter, 2013). This method makes minimal assumptions as to the distribution of theory errors in fitting for the Unitarity Triangle parameters, instead scanning over a large range of theoretical parameters for fits found to have an acceptable $\chi^2$ value. It also extends the global fit, allowing the determination of correlations in the fit between underlying physical parameters that are related to more than one observable, *e.g.* $\phi_1$ and $\phi_2$. The statistical procedure used in the scan method follows the approach adopted in the *BABAR* physics book (Harrison and Quinn, 1998).

#### 25.1.3.1 CKMfitter

The CKMfitter group (Charles et al., 2005), a collaboration of experimental and theoretical physicists, performs phenomenological studies related to flavor physics and *CP* violation. To quantify the impact of measurements on the SM parameters as well as on some postulated extensions of the SM, they developed a global fit package based on a frequentist statistical approach. They use a specific model for theoretical uncertainties (most often related to the assessment of hadronic quantities) as biases bounded in a range, Rfit (Höcker, Lacker, Laplace, and Le Diberder, 2001).

The CKMfitter group adopted the following framework. They constrain a certain number of parameters in the model such as the Wolfenstein parameters describing the CKM matrix, quark masses, or hadronic quantities in order to compare a set of observations $\boldsymbol{x}_{\rm obs}$ with their theoretical predictions in a given model. The unknown parameters entering the theoretical predictions are split into two sets, parameters of interest such as $(\overline{\rho}, \overline{\eta})$ denoted by $\boldsymbol{\mu}$, and those of no interest called nuisance parameters, such as hadronic quantities, which are denoted by $\boldsymbol{\nu}$.

A general approach to constrain the fundamental parameters of interest is a hypothesis test, quantifying the compatibility of the data with the null hypothesis that the true value of a fundamental parameter $\mu_t$ is equal to a particular value $\mu$. In order to interpret the distribution of observables under the null hypothesis, a test statistic measuring whether the data are compatible or not is defined. In analogy with the case of simple hypotheses for which all the values of the fundamental parameters are defined, the CKMfitter framework uses the maximum likelihood ratio defined as

$$\Lambda(\boldsymbol{x}, \boldsymbol{\mu}) = \frac{\sup_{\boldsymbol{\nu}} L_{\boldsymbol{x}}(\boldsymbol{\mu}, \boldsymbol{\nu})}{\sup_{\boldsymbol{\mu},\boldsymbol{\nu}} L_{\boldsymbol{x}}(\boldsymbol{\mu}, \boldsymbol{\nu})}, \qquad \Delta\chi^2 = -2\ln(\Lambda), \tag{25.1.1}$$

where $L_{\boldsymbol{x}}(\boldsymbol{\mu}, \boldsymbol{\nu})$ denotes the likelihood built from experimental data and theoretical inputs $\boldsymbol{x}$, and sup refers to the supremum, otherwise known as least upper bound. Note that the $\Delta\chi^2$ of the test statistic does not necessarily follow a $\chi^2$ distribution, especially if the distributions of the observables exhibit significant non-Gaussian properties. In the asymptotic regime, however, Wilks' theorem states that the distribution of $\Delta\chi^2$ should converge to a $\chi^2$ distribution depending only on the number of parameters tested (Wilks, 1938).

From this test statistic, a $p$-value for the observations $\boldsymbol{x}_{\rm obs}$ is built under the null hypothesis $\mu_t = \mu$ corresponding to the probability that the test statistic is as large as, or larger than, that observed

$$p(\boldsymbol{x}_{\rm obs}; \boldsymbol{\mu}) = \mathcal{P}[\Delta\chi^2 \geq \Delta\chi^2(\boldsymbol{x}_{\rm obs}; \mu)]. \tag{25.1.2}$$

Small $p$-values provide evidence against the null hypothesis. This $p$-value can be computed using pseudo experiments with Monte Carlo methods, or directly in terms of incomplete $\Gamma$ functions if one assumes that the asymptotic regime has been reached. After having computed the $p$-value associated with each value of $\mu$, confidence intervals for a given confidence level can be defined by considering the region where $1 - \text{C.L.} \leq p(\boldsymbol{x}; \boldsymbol{\mu})$.

Practically the fit is performed by scanning $\boldsymbol{\mu}$, minimizing the likelihood $L_{\boldsymbol{x}}$ with respect to the other parameters, and identifying the "best-fit" value. Unless explicitly provided by the experiments, the likelihoods are built assuming the uncertainties in the experimental measurements are Gaussian, whereas the theoretical errors are treated in the Rfit scheme as constrained in a strict range. The $p$-values are extracted from the scan of the test statistic and are analyzed to provide confidence intervals or confidence regions for the fundamental parameters of interest.

#### 25.1.3.2 UTfit

The UTfit Collaboration is formed of experimental and theoretical physicists performing the Unitarity Triangle analysis following the method described in Ciuchini et al. (2001) and Bona et al. (2005). This section summarizes the basic ingredients of the UTfit analysis method that is developed in the framework of a Bayesian approach.

In the following sections, several equations are given that relate a constraint $c_j$ (where $c_j$ stands for one of $M$ constraints such as $|V_{ub}/V_{cb}|$, $\Delta m_d$, $\Delta m_s$, $|\varepsilon_K|$ and etc., for $j = 1, \ldots, M$) to the Unitarity Triangle parameters $\overline{\rho}$ and $\overline{\eta}$, via a set of $N$ ancillary parameters $\boldsymbol{x}$, where $\boldsymbol{x} = \{x_1, x_2, \ldots, x_N\}$ stand for all experimentally determined or theoretically calculated quantities on which the various $c_j$ depend:

$$c_j \equiv c_j(\overline{\rho}, \overline{\eta}; \boldsymbol{x}). \tag{25.1.3}$$

In the ideal case of exact knowledge of $c_j$ and $\boldsymbol{x}$, each of the constraints provides a curve in the $(\overline{\rho}, \overline{\eta})$ plane. In such a case, there would be no reason to favor any of the points on the curve, unless one has some further information or physical prejudice, which might exclude points outside a determined physical region, or, in general, assign different weights to different points. In reality one suffers from several uncertainties on the quantities $c_j$ and $\boldsymbol{x}$. However, there are values for $c_j$ and $\boldsymbol{x}$ which can be considered as ruled out. For example if one considers the measurement of $\phi_3 = (67 \pm 11)^\circ$, the value of this angle is currently constrained to lie within some *almost certain* range, $45^\circ < \phi_3 < 89^\circ$, and so a value of $\phi_3 \sim 20^\circ$ is excluded. Moreover, it is much more probable that the value of $\phi_3$ lies between $56^\circ$ and $78^\circ$ rather than in the rest of the interval, in spite of the fact that the two subintervals have the same widths. This means that, instead of a single curve in the $(\overline{\rho}, \overline{\eta})$ plane, one has a family of curves which depends on the distributions of $c_j$ and $\boldsymbol{x}$. As a result, different points in the $(\overline{\rho}, \overline{\eta})$ plane have different weights (even if they were taken to be equally probable *a priori*) and our confidence on the values of $\overline{\rho}$ and $\overline{\eta}$ clusters in a region of the plane.

The above considerations can be formalized using the Bayesian approach: the uncertainty is described in terms of a probability density function (*p.d.f.*) which quantifies our confidence on the values of a given quantity. The inference of $\overline{\rho}$ and $\overline{\eta}$ becomes a straightforward application of probability theory. In the following, all the *p.d.f.*s are called $f(\ldots)$ assuming a different functional form that depends on the specified arguments.

We can define the *p.d.f.* $f(\overline{\rho}, \overline{\eta})$ that takes into account the uncertainties on $c_j$ and $\boldsymbol{x}$:

$$f(\overline{\rho}, \overline{\eta}) \propto \int \frac{1}{\sqrt{2\pi}\,\sigma(c_j)} \exp\left[-\frac{(c_j(\overline{\rho}, \overline{\eta}, \boldsymbol{x}) - \widehat{c}_j)^2}{2\,\sigma^2(c_j)}\right] \cdot \\ \cdot f(x_1) \cdot f(x_2) \cdots f(x_N)\, \mathrm{d}\boldsymbol{x}\,, \tag{25.1.4}$$

where $\widehat{c}_j$ is the experimental best estimate of $c_j$, with uncertainty $\sigma(c_j)$. For simplicity, a Gaussian distribution can be assumed as an individual *p.d.f.* for each constraint $c_j$. In principle we should consider a joint $f(c_j, \boldsymbol{x})$, but we split it into the product of the individual *p.d.f.*s, assuming quantities are independent.

Although the above derivation of Eq. (25.1.4) is probably the most intuitive one, UTfit choose to define a global inference relating $\overline{\rho}$, $\overline{\eta}$, $c_j$ and $\boldsymbol{x}$, which is the usual way of performing Bayesian inference. This method is followed by a second step where marginalization is performed over those quantities which are not of interest. In this case Bayes theorem can be used to give

$$\begin{aligned} f(c_j, \overline{\rho}, \overline{\eta}, \boldsymbol{x} \,|\, \widehat{c}_j) &\propto f(\widehat{c}_j \,|\, c_j, \overline{\rho}, \overline{\eta}, \boldsymbol{x}) \cdot f(c_j, \overline{\rho}, \overline{\eta}, \boldsymbol{x}) \\ &\propto f(\widehat{c}_j \,|\, c_j) \cdot f(c_j \,|\, \overline{\rho}, \overline{\eta}, \boldsymbol{x}) \cdot f(\boldsymbol{x}, \overline{\rho}, \overline{\eta}) \\ &\propto f(\widehat{c}_j \,|\, c_j) \cdot \delta(c_j - c_j(\overline{\rho}, \overline{\eta}, \boldsymbol{x})) \cdot \\ &\quad \cdot f(\boldsymbol{x}) \cdot f_\circ(\overline{\rho}, \overline{\eta}) \,, \end{aligned} \tag{25.1.5}$$

where $f_\circ(\overline{\rho}, \overline{\eta})$ denotes the *prior* probability distribution. Equation (25.1.4) can be recovered by i) assuming a Gaussian error function for $\widehat{c}_j$ around $c_j$; ii) considering the various $x_i$ as independent; iii) taking a flat *a priori* distribution for $\overline{\rho}$ and $\overline{\eta}$; and iv) by integrating Eq. (25.1.5) over $c_j$ and $\boldsymbol{x}$.

At this point, the extension of the formalism to several constraints is straightforward. One can rewrite Eq. (25.1.5) as

$$\begin{aligned} f(\overline{\rho}, \overline{\eta}, \boldsymbol{x} \,|\, \widehat{c}_1, ..., \widehat{c}_M) \propto \prod_{j=1,M} f_j(\widehat{c}_j \,|\, \overline{\rho}, \overline{\eta}, \boldsymbol{x}) \times \prod_{i=1,N} f_i(x_i) \times \\ \times f_\circ(\overline{\rho}, \overline{\eta}) \ , \end{aligned} \tag{25.1.6}$$

where the conditioning on $f_j$ from the $c_j$ have been removed, since the $c_j$ act as intermediate variables which are integrated away to obtain a marginal probability distribution.

Integrating Eq. (25.1.6) over $\boldsymbol{x}$, one can rewrite the global inference in the following way:

$$f(\overline{\rho}, \overline{\eta} \,|\, \widehat{\mathbf{c}}, \mathbf{f}) \propto \mathcal{L}(\widehat{\mathbf{c}} \,|\, \overline{\rho}, \overline{\eta}, \mathbf{f}) \times f_\circ(\overline{\rho}, \overline{\eta}) \,, \tag{25.1.7}$$

where $\widehat{\mathbf{c}}$ stands for the set of measured constraints, $\mathbf{f}$ indicates the dependence from the set of *p.d.f.*s that can be explicitly written as:

$$\mathcal{L}(\widehat{\mathbf{c}} \,|\, \overline{\rho}, \overline{\eta}, \mathbf{f}) = \int \prod_{j=1,M} f_j(\widehat{c}_j \,|\, \overline{\rho}, \overline{\eta}, \boldsymbol{x}) \prod_{i=1,N} f_i(x_i) \, \mathrm{d}x_i \tag{25.1.8}$$

which is the effective overall likelihood taking into account all possible values of $x_j$, properly weighted. Hence the overall likelihood depends on the best knowledge of all $x_i$, described by $f(\boldsymbol{x})$.

In conclusion, while *a priori* all values for $\overline{\rho}$ and $\overline{\eta}$ are considered equally likely, *a posteriori* the probability clusters around the point which maximizes the likelihood. The final (unnormalized) *p.d.f.* obtained starting from a flat distribution of $\overline{\rho}$ and $\overline{\eta}$ is

$$f(\overline{\rho}, \overline{\eta}) \propto \int \prod_{j=1,M} f_j(\widehat{c}_j \,|\, \overline{\rho}, \overline{\eta}, \boldsymbol{x}) \prod_{i=1,N} f_i(x_i) \, \mathrm{d}x_i, \tag{25.1.9}$$

where the integration can be performed using Monte Carlo methods.

### 25.1.4 Experimental inputs

The following sections discuss the observables that are used in global fits in order to extract CKM parameters. The input values used for these observables are taken from the averages performed for this book and they are summarized in Tables 25.1.1 and 25.1.2.

#### 25.1.4.1 $|V_{ud}|$ and $|V_{us}|$

The magnitudes of the CKM elements $V_{ud}$ and $V_{us}$ required in the CKM global fits are extracted from semileptonic $u \to d$ and $u \to s$ transitions that are not measured at the $B$ Factories. These two CKM matrix elements are the most precisely measured parameters. The determination of $V_{ud}$ involves super-allowed $0^+ \to 0^+$ nuclear beta decays and $\pi^+ \to \pi^0 e^+ \nu$ decays. The determination of $V_{us}$ is based on semileptonic kaon decays. Results are provided by the Flavianet working group (Antonelli et al., 2010b). For the extraction of $V_{us}$, various structure constants are needed, which are calculated using lattice QCD analyses by averaging the latest $N_f = 2+1$ calculations by BMW (Durr et al., 2010), MILC'09 (Bazavov et al., 2009) and HPQCD/UKQCD (Follana, Davies, Lepage, and Shigemitsu, 2008). The latest results yield (Beringer et al., 2012)

$$|V_{ud}| = 0.97425 \pm 0.00022, \tag{25.1.10}$$

$$|V_{us}| = 0.2252 \pm 0.0009. \tag{25.1.11}$$

#### 25.1.4.2 $B$ Factory results

The experimental inputs from the $B$ Factories are:

Unitarity Triangle sides (see Section 16.5):
The sides of the Unitarity Triangle are:

$$R_u = \frac{|V_{ud}||V_{ub}|}{|V_{cd}||V_{cb}|} = \sqrt{\overline{\rho}^2 + \overline{\eta}^2} \tag{25.1.12}$$

and

$$R_t = \frac{|V_{td}||V_{tb}|}{|V_{cd}||V_{cb}|} = \sqrt{(1-\overline{\rho})^2 + \overline{\eta}^2}. \tag{25.1.13}$$

Since $|V_{tb}| \simeq 1$ and $|V_{cd}| \sim |V_{us}|$, we just need to focus on measurements of $|V_{cb}|$, $|V_{ub}|$ and $|V_{td}|$.

- $|V_{cb}|$: this CKM matrix element is measured in semileptonic $b \to c$ transitions as described in Section 17.1. The present average over inclusive and exclusive decays yields $|V_{cb}| = [41.67\ (1 \pm 0.009_{\rm exp} \pm 0.012_{\rm th})] \times 10^{-3}$.
- $|V_{ub}|$: this is the CKM matrix element measured in semileptonic $b \to u$ transitions as described in Section 17.1. The present average over inclusive and exclusive decays yields $|V_{ub}| = [3.95\ (1 \pm 0.096_{\rm exp} \pm 0.099_{\rm th})] \times 10^{-3}$.
- $|V_{td}|$: this CKM matrix element is extracted from the oscillation frequency $\Delta m_d$ in $B^0_d \overline{B}{}^0_d$ mixing as shown in Eq. (17.2.1). The definition of $\Delta m_d$ from first principles is given in Section 10.1, while its experimental extraction is described in Section 17.5.2. To relate $\Delta m_d$ to $|V_{td}|$, we need to know several other quantities: the Inami-Lim function $S_0(x_t)$ (Buras and Fleischer, 1998; Inami and Lim, 1981) with $x_t = m_t^2/M_W^2$, the top mass $m_t$ taken in the $\overline{MS}$ scheme ($\overline{m}_t$, see below), the perturbative QCD short-distance NLO correction $\eta_B$, and the non-perturbative QCD parameters $f_{B_d}$ and $B_{B_d}$. Table 25.1.3 lists the latest lattice QCD calculations for the latter two parameters (see Section 25.1.5), while $m_t$ is listed in Table 25.1.2.

Unitarity Triangle angles:

- $\phi_1$: this weak phase is defined in terms of CKM matrix elements in Eq. (16.5.3). The most precise measurements of this quantity are obtained via time-dependent *CP* analyses of $b \to c\bar{c}s$ processes. The analyses contributing to $\phi_1$ measurements are described in detail in Section 17.6 and yield $\sin 2\phi_1 = 0.677 \pm 0.020$. The ambiguities from the $\sin 2\phi_1$ measurements are resolved via measurements of final states that have an asymmetry dependence on $\cos 2\phi_1$ (see Section 17.6.8).
- $\phi_2$: this weak phase is defined in terms of CKM matrix elements in Eq. (16.5.4) and is measured in $b \to u\bar{u}d$ processes. The analyses contributing to $\phi_2$ measurement are described in detail in Section 17.7. Averaging results from charmless two-body $B$ decays into $\pi\pi$, $\rho\rho$, and $\rho\pi$ final states yields $\phi_2 = (88 \pm 5)^\circ$.
- $\phi_3$: this weak phase is defined in terms of CKM matrix elements in Eq. (16.5.5) and is extracted in $B \to D^{(*)}K$ and $B \to DK^*$ decays using various methods. The individual analyses are described in detail in Section 17.8. The present average from the $B$ Factories is $\phi_3 = (67 \pm 11)^\circ$.

Leptonic decays:

- $\mathcal{B}(B \to \tau\nu_\tau)$: this branching fraction is linked to the CKM matrix elements $|V_{ub}|$ and the $B$ decay constant $f_{B_d}$ via Eq. (17.10.4). Including the recent Belle results as discussed in Section 17.10.2.2, the present world average is $(1.15 \pm 0.23) \times 10^{-4}$.

#### 25.1.4.3 Other measurement inputs

A number of other experimental inputs are required in order to perform global fits. The values of these experimental inputs can be found in the review of particle physics compiled by the PDG (Beringer et al., 2012). The most relevant of these other observables are $\varepsilon_K$, $\Delta m_s$ and the quark masses. They are described in the following:

- $\varepsilon_K$: this parameter represents indirect *CP* violation in the mixing in the $K^0\overline{K}^0$ system. Defining the ratios of decay amplitudes of $K_S$ and $K_L$ into two pions as

$$\eta_{00} \equiv \frac{\mathcal{A}(K_L \to \pi^0\pi^0)}{\mathcal{A}(K_S \to \pi^0\pi^0)}, \quad \eta_{+-} \equiv \frac{\mathcal{A}(K_L \to \pi^+\pi^-)}{\mathcal{A}(K_S \to \pi^+\pi^-)}, \tag{25.1.14}$$

indirect (in the mixing) and direct (in the amplitudes) *CP* violation can be parameterized by

$$\varepsilon_K = \frac{\eta_{00} + 2\eta_{+-}}{3}, \quad \varepsilon'_K = \frac{-\eta_{00} + \eta_{+-}}{3}, \tag{25.1.15}$$

respectively.
Using the effective $\Delta S = 2$ Hamiltonian, $\varepsilon_K$ is related to CKM parameters by

$$\begin{aligned}|\varepsilon_K| = \frac{G_F^2 m_W^2 m_K f_K^2}{12\sqrt{2}\pi^2 \Delta m_K} \widehat{B}_K \big(&\eta_{cc} S(x_c, x_c) Im[(V_{cs}V_{cd}^*)^2] \\ &+\eta_{tt} S(x_t, x_t) Im[(V_{ts}V_{td}^*)^2] \\ &+2\eta_{tc} S(x_c, x_t) Im[V_{cs}V_{cd}^* V_{ts}V_{td}^*]\big),\end{aligned} \tag{25.1.16}$$

where $\Delta m_K$ is the $K^0\overline{K}^0$ oscillation frequency, $f_K$ is the kaon decay constant, $m_K$ is the kaon mass, $\widehat{B}_K$ parameterizes the value of the hadronic matrix element (bag parameter), $S(x_q, x_{q'})$ are Inami-Lim functions for top quark and charm quark contributions that introduce QCD correction factors $\eta_{tt}$, $\eta_{tc}$ and $\eta_{cc}$ and all other parameters are the same as those in Eq. (17.2.1). As for $\Delta m_d$, the quantity $x_q = \overline{m}_q(m_q)^2/m_W^2$ where $q = c$, $t$ and the quark masses are determined in the $\overline{MS}$ scheme discussed below.
While $\varepsilon'_K$ suffers from theory uncertainties that are too large to provide a useful constraint in the $(\overline{\rho}, \overline{\eta})$ plane, $\varepsilon_K$ provides a hyperbolic dependence between $\overline{\rho}$ and $\overline{\eta}$. This can be shown by rearranging the Wolfenstein parameters in Eq. (25.1.16):

$$|\varepsilon_K| \propto \overline{\eta}\big[(1 - \overline{\rho}) + P(A, \lambda)\big] \tag{25.1.17}$$

where $P$ is a function of $A$ and $\lambda$ and it does not depend on $\overline{\eta}$ or $\overline{\rho}$. A fit to $K \to \pi\pi$ data yields (Beringer et al., 2012)

$$|\varepsilon_K| = (2.228 \pm 0.011) \times 10^{-3}. \tag{25.1.18}$$

- $\Delta m_s$: this parameter is the oscillation frequency measured in $B_s^0\overline{B}_s^0$ mixing and it provides a determination of the CKM matrix element $|V_{ts}|$. It is calculated in a similar way to $\Delta m_d$, using the $\Delta B = 2$ effective Hamiltonian, yielding

$$\Delta m_s = \frac{G_F^2}{6\pi^2} \eta_B M_W^2 m_{B_s} f_{B_s}^2 \widehat{B}_{B_s} S_0(x_t) |V_{ts}V_{tb}^*|^2, \tag{25.1.19}$$

where $m_{B_s}$ is the $B_s^0$ mass, $f_{B_s}$ is the $B_s$-decay constant, $\widehat{B}_{B_s}$ is the bag parameter and all other parameters are the same as those for $\Delta m_d$ listed in Eq. (17.2.1). The numerical values of the theoretical input parameters are summarized in Table 25.1.3. The ratio $\Delta m_s/\Delta m_d$ provides a more accurate measurement of the Unitarity Triangle side $R_t$ than $\Delta m_d$ since the ratios of QCD parameters have smaller theory uncertainties than $f_{B_s}$ and $B_{Bs}$ themselves. The CDF experiment was the first to measure $\Delta m_s$, obtaining a value (Abulencia et al., 2006b) of:

$$17.77 \pm 0.10 \pm 0.07 \text{ ps}^{-1}. \tag{25.1.20}$$

This result has been subsequently confirmed by LHCb that obtains now an improved precision (Aaij et al., 2013d). Combining these results yields the present world average of $\Delta m_s = 17.719 \pm 0.043$ shown in Table 25.1.2.

**Table 25.1.1.** Input values for the global fit from $B$ Factory measurements.

| Input | Value | Reference |
|---|---|---|
| $\sin 2\phi_1$ | $0.677 \pm 0.020$ | Section 17.6. |
| $\phi_2$ [°] | $88 \pm 5$ | Section 17.7 |
| $\phi_3$ [°] | $67 \pm 11$ | Section 17.8 |
| $\Delta m_d$ [ps$^{-1}$] | $0.508 \pm 0.003 \pm 0.003$ | Section 17.5.2 |
| $\lvert V_{cb}\rvert$ [$10^{-3}$] | $41.67\ (1 \pm 0.009 \pm 0.012)$ | Section 17.1.6.1 |
| $\lvert V_{ub}\rvert$ [$10^{-3}$] | $3.95\ (1 \pm 0.096 \pm 0.099)$ | Section 17.1.6.2 |
| $\mathcal{B}(B \to \tau\nu_\tau)$ | $(1.15 \pm 0.23) \times 10^{-4}$ | Section 17.10.2.2 |

– top mass: the mass of the top quark $m_t$ has been measured by the Tevatron and the LHC experiments with a combined precision of about 0.6%, and its current average value is $m_t = (173.18 \pm 0.94)$ GeV/$c^2$ (Aaltonen et al., 2012a). Whatever the analysis method adopted, the quantity measured in data corresponds to the top quark mass scheme assumed in the Monte Carlo simulation used. As a consequence, there is no immediate connection between this measured value and any other mass scheme, such as the pole or $\overline{MS}$ mass scheme.
In QED the position of the pole in the propagator is the definition of the particle mass, while in QCD the quark propagator has no pole because the quarks are confined. So the problem of the definition of the quark mass can be addressed from two perspectives: the long-distance behavior which corresponds to the pole-mass scheme, and the short-distance behavior, which, for example, can be represented by the $\overline{MS}$ mass scheme.
The relation between the pole mass and any other mass scheme $m_t(R,\mu)$ is expressed as a perturbative series in $\alpha_S(m_t)$ and it can be written as $m_t^{pole} = m_t(R,\mu) + \delta m_t(R,\mu)$ (Hoang and Stewart, 2008) where

$$\delta m_t(R,\mu) = R \sum_{n=1}^{\infty} \sum_{k=0}^{n} a_{nk} \left[\frac{\alpha_S(\mu)}{4\pi}\right]^n \ln^k\left(\frac{\mu}{R}\right) \tag{25.1.21}$$

with $R$ being a dimension-one scale intrinsic to the scheme, $a_{nk}$ finite numerical coefficients and $\mu$ is the renormalization scale. An example of this conversion can be found in Beringer et al. (2012).
The experimentally measured $m_t$ is considered to be close to the pole mass and in some analyses it is assumed that $m_t$ measured is indeed equal to $m_t^{pole}$ (see for example ALEPH, CDF, D0, DELPHI, L3, OPAL, SLD and the LEP, Tevatron and SLD Electroweak and Heavy Flavour Working Groups, 2010). Conversely, the input value $\overline{m}_t$ used for the global fits is the top running mass calculated in the $\overline{MS}$ renormalization scheme and its value is obtained by pole-to-$\overline{MS}$ matching. At the 3-loop level with five quark flavors the relation is (Broadhurst, Gray, and Schilcher, 1991; Gray, Broadhurst, and Schilcher, 1990; Melnikov and van Ritbergen, 2000)

$$\overline{m}_t(m_t) = m_t\left(1 - \frac{4}{3}\left(\frac{\alpha_S(m_t)}{\pi}\right) - 9.12530\left(\frac{\alpha_S(m_t)}{\pi}\right)^2 - 80.4045\left(\frac{\alpha_S(m_t)}{\pi}\right)^3\right), \tag{25.1.22}$$

where $\alpha_S(m_t) = 0.1068 \pm 0.0018$. This yields an $\overline{MS}$ mass of $\overline{m}_t(m_t) = 163.3 \pm 0.9$ GeV/$c^2$. One can compare this result with the top mass obtained using the measured cross section, which has a similar central value, but slightly larger uncertainty $\overline{m}_t(m_t) = 163.3 \pm 2.7$ GeV/$c^2$, for example see (Moch, 2012).

– $\overline{m}_b(m_b)$, $\overline{m}_c(m_c)$, $\overline{m}_s(m_s)$: these are the running quark masses evaluated in the $\overline{MS}$ scheme and at the scale indicated. The values used are given in Table 25.1.2.

### 25.1.5 Theoretical inputs: derivation of hadronic observables

Most of the experimental inputs described above for the global fit rely upon knowledge of hadronic matrix elements that parameterize the non-perturbative QCD contributions to weak decays and mixing. These contributions represent the connection between the quark-level fundamental quantities and the hadronic-level experimental observables.

These hadronic matrix elements are obtained from numerical lattice QCD calculations. State-of-the-art lattice computations now regularly include the effects of the sea up, down and strange quarks. They also typically use simulations at pion masses below 300 MeV/$c^2$, or even below 200 MeV/$c^2$, in order to control the extrapolation to the physical pion mass. For many hadronic matrix elements of interest, there are now at least two or more reliable lattice calculation results. In order to use the lattice inputs in a global fit, it is necessary to average different results. However as there are potentially significant correlations between various lattice uncertainties, averaging is not straightforward, and correlations must be taken into account. Statistical and systematic errors may be correlated, and one needs to be sufficiently familiar with the lattice computational methods used in order to understand how to treat a given result in the averaging procedure.

A complete review of lattice techniques is beyond the scope of this book. We therefore rely on the averaging work of Laiho, Lunghi, and Van de Water (2010). They average the latest LQCD results for leptonic decay constants, meson mixing parameters, and semileptonic form factors. Only results from simulations with three dynamical quark flavors are included in the averages if there are associated proceedings or publications that include comprehensive error budgets. When computing averages, they assume that all errors are normally distributed and follow the prescription outlined by Schmelling (1995) to take the correlations into account. Moreover, they assume that any correlated source of error for two lattice calculations is 100% correlated. This assumption is conservative and will lead to an overestimate of the total error of the lattice averages; nevertheless, it is the most systematic treatment possible without knowledge of the correlation matrices, which do not exist, between the various calculations. Finally, they adopt the PDG prescription to combine several measurements whose spread is wider than that expected from the quoted errors: the error on the average is multiplied by the square root of the $\chi^2$ per degree of freedom.

**Table 25.1.2.** Input values for the global fit from other experimental measurements (non-$B$ Factory). The quark masses are the running $\overline{MS}$ quark masses as explained in the text.

| Input | Value | Reference |
|---|---|---|
| $\alpha_S(m_Z)$ | $0.1184 \pm 0.0007$ | (Beringer et al., 2012) |
| $m_t$ [GeV/$c^2$ ] | $163.3 \pm 0.9$ | Determined for this book |
| $m_b$ [GeV/$c^2$ ] | $4.18 \pm 0.03$ | (Beringer et al., 2012) |
| $m_c$ [GeV/$c^2$ ] | $1.275 \pm 0.025$ | (Beringer et al., 2012) |
| $m_s$ [GeV/$c^2$ ] | $0.0935 \pm 0.0025$ | (Beringer et al., 2012) |
| $\tau_{B_d}$ [ps] | $1.519 \pm 0.007$ | (Amhis et al., 2012) |
| $\tau_{B^+}$ [ps] | $1.642 \pm 0.008$ | (Amhis et al., 2012) |
| $\tau_{B_s}$ [ps] | $1.503 \pm 0.010$ | (Amhis et al., 2012) |
| $\varepsilon_K$ | $0.002228 \pm 0.000011$ | (Beringer et al., 2012) |
| $\Delta m_s$ | $17.719 \pm 0.043$ | (Amhis et al., 2012) |

The global fits use four inputs that depend on these hadronic observables: $\varepsilon_K$, $\Delta m_d$, $\Delta m_s$, and $B(B \to \tau\nu_\tau)$. These are related to two decay constants ($f_{B_d}$, $f_{B_s}$) and three hadronic matrix elements ($B_K$, $B_{B_d}$, $B_{B_s}$). The preferred choice of minimally correlated inputs is $B_K$, $f_{B_s}/f_{B_d}$, $B_{B_s}/B_{B_d}$, $B_{B_s}$ and $f_{B_s}$. In general the quantities related to $B^0_s\overline{B}^0_s$ mixing are preferred as current lattice QCD calculations can simulate directly at the physical $s$-quark mass, but must extrapolate to the $u$- and $d$-quark masses. Therefore the chiral extrapolation error, which is often the dominant systematic, is smaller for $B_{B_s}$ and $f_{B_s}$ than for the lighter $B$ meson parameters. The ratios are chosen in order to benefit from cancellations of uncertainties done on the lattice. A basic description of the hadronic quantities that must be included in the global fits follows:

– $B_K$ is related to the parameter $\varepsilon_K$ as defined in Section 25.1.4.3. When strong interactions are considered, $\Delta S = 2$ transitions can no longer be discussed at the quark level. Instead, an effective Hamiltonian must be considered between mesonic initial and final states. Since the strong coupling constant is large at typical hadronic scales, the resulting matrix element cannot be calculated in perturbation theory. However the OPE does factorize long- and short-distance effects. The dependence on the renormalization scheme and scale $\mu$ is canceled by that of the hadronic matrix element $\langle \overline{K}^0 | Q_R^{\Delta S=2}(\mu) | K^0 \rangle$. The latter corresponds to the long-distance effects of the effective Hamiltonian and must be computed non-perturbatively. For historical, as well as technical reasons, it is convenient to express it in terms of the parameter $B_K$, defined as:

$$B_K(\mu) = \frac{\langle \overline{K}^0 | Q_R^{\Delta S=2}(\mu) | K^0 \rangle}{\frac{8}{3} m_K^2 f_K^2}. \qquad (25.1.23)$$

The four-quark operator $Q^{\Delta S=2}(\mu)$ is renormalized at the scale $\mu$ in some regularization scheme, usually taken to be the naïve dimensional regularization. The renormalization group independent parameter $\widehat{B}_K$ is related to $B_K(\mu)$ by a function of the renormalized gauge coupling ($\overline{g}(\mu)$) and perturbative coefficients ($\beta_0$, $\beta_1$, $\gamma_0$, and $\gamma_1$):

$$\widehat{B}_K = \left(\frac{\overline{g}(\mu)^2}{4\pi}\right)^{-\frac{\gamma_0}{2\beta_0}} \left[1 + \frac{\overline{g}(\mu)^2}{(4\pi)^2}\left(\frac{\beta_1\gamma_0 - \beta_0\gamma_1}{2\beta_0}\right)\right] B_K(\mu). \qquad (25.1.24)$$

A more detailed discussion can be found in (Colangelo et al., 2011) and references therein.

– for $B_s$, the parameter of the renormalized operator is defined as

$$B_{B_s}(\mu) = \frac{\langle \overline{B}_s | Q_d^{\Delta B=2}(\mu) | B_s \rangle}{\frac{8}{3} m_{B_s}^2 f_{B_s}^2}, \qquad (25.1.25)$$

where $Q_s^{\Delta B=2} = (\bar{b}\gamma^\mu(1-\gamma_5)s)(\bar{b}\gamma_\mu(1-\gamma_5)s)$ and $\mu$ is the renormalization scale. This definition stems from the vacuum saturation approximation in which $B_{B_s} = 1$. One can define $B_{B_d}$ in a similar way.

The renormalization group invariant parameter $\widehat{B}_{B_s}$ of Eq. (25.1.19) is defined as

$$\widehat{B}_{B_s} = \alpha_S(\mu)^{-\frac{\gamma_0}{2\beta_0}} \left(1 + \frac{\alpha_S(\mu)}{4\pi} J\right) B_{B_s}(\mu). \qquad (25.1.26)$$

In all schemes $\gamma_0 = 4$, whereas $J$ depends on the scheme used for renormalizing $Q_s^{\Delta B=2}(\mu)$. In the theoretical expressions, the physical amplitudes are always defined in terms of $\widehat{B}_{B_s}$. The advantage is that this quantity is both renormalization scale and scheme independent.

However, the important quantities are not the $B$ parameters themselves but the combinations $f_x^2 B_x$ (where $x$ can be $K$ or $B_{d,s}$) which are simply related to the physical amplitudes by the factors $8m_x^2/3$. In the case of the kaon system, the decay constant is derived from experiments as one measures the product $|V_{us}|f_K$, while for the $B_{d,s}$ mesons, we rely on lattice QCD determinations.

The hadronic parameters from lattice QCD discussed above and used as inputs for the global fits can be found in Table 25.1.3.

### 25.1.6 Results from the global fits

Here we report the results for the global fits from the two main collaborations described in Section 25.1.3. Using all the inputs described in the previous sections and the statistical techniques specific to the two fitting groups, it is possible to extract the CKM matrix parameters from global fits to the set of relevant measurements. Particular attention is paid to the least precisely determined parameters $\overline{\rho}$ and $\overline{\eta}$, especially in light of their importance with regard to *CP* violation in the SM.

Table 25.1.4 shows the numerical results obtained from the two global fits performed using the inputs given in Tables 25.1.1, 25.1.2 and 25.1.3. Fig. 25.1.1 shows the ($\overline{\rho}$, $\overline{\eta}$) plane illustrating the constraints used along with the results from the two global fits. These results are consistent, indicating that there is sufficient experimental data to draw meaningful conclusions regarding the CKM matrix picture using either frequentist or Bayesian approaches. The combination of all the constraints gives a single preferred area, illustrating the exceptional agreement of measurements with SM predictions.

**Table 25.1.4.** Results from the global fits.

| Parameter | Output Value | |
|---|---|---|
| | CKMfitter | UTfit |
| $\overline{\rho}$ | $0.129^{+0.027}_{-0.022}$ | $0.130 \pm 0.020$ |
| $\overline{\eta}$ | $0.345 \pm 0.014$ | $0.348 \pm 0.013$ |
| $\sin 2\phi_1$ | $0.684 \pm 0.019$ | $0.689 \pm 0.018$ |
| $\phi_2$ [°] | $88.8^{+4.2}_{-3.6}$ | $88.4 \pm 2.8$ |
| $\phi_3$ [°] | $68.9^{+3.5}_{-4.2}$ | $69.5 \pm 3.0$ |

The compatibility of the constraints used can be evaluated in each of the two fit methods by excluding a single given constraint at a time and evaluating the value for that observable from the global fit performed using all the other constraints. The comparison between this inferred (or predicted) value and the value of the excluded constraint can be used to test the agreement of each individual constraint with respect to all the others, assuming that the SM is an adequate description of the underlying physics. These predictions are shown in Table 25.1.5.

A few small tensions are currently present in the global fits. None of these are statistically significant, however they should be kept in mind in light of future updates. The value of $\sin 2\phi_1$ is now obtained with such a small uncertainty that it is driving the other predictions, thus highlighting some inconsistencies. For example, $\varepsilon_K$ is showing some tension through the less precise determination of the $\widehat{B}_K$ parameter. Moreover, the two determinations of $|V_{ub}|$ and $|V_{cb}|$ still present marginal agreement between the inclusive and exclusive values and in particular the inclusive values show more significant discrepancies in the context of the global fit. Future improvements in lattice QCD determinations will be extremely interesting to assess these effects.

Historically there has been tension between the $B \to \tau\nu$ branching fraction and other constraints on the CKM matrix. For a number of years this tension was interpreted by the community as being a possible hint for new physics. At the time of writing, the experimental situation is compatible with the SM, and there is no significant tension evident between the constraints. The results shown in Table 25.1.4 include the branching fraction of $B \to \tau\nu$ as an input, however very similar results are obtained when this input is removed from the global fit. See Section 17.10 for more details regarding the experimental determination of the branching fraction of $B \to \tau\nu$, and the subsequent interpretation of this observable.

As stated at the start of this chapter, it is also possible to illustrate the success of the SM in describing the measurements of the $B$ Factories by comparing the values obtained for $\overline{\rho}$ and $\overline{\eta}$ for *CP*-conserving and *CP*-violating quantities. These constraints are shown in Figs 25.1.2 and 25.1.3. The agreement between these determinations of the apex of the Unitarity Triangle is good, and is taken as evidence that the CKM matrix and KM mechanism give the leading order description of quark mixing and *CP* violation in the SM. The dominant *CP*-violating constraints come from charmonium decay measurements of $\phi_1 = \beta$ (Section 17.6) and charmless $B$ decay measurements of $\phi_2 = \alpha$ (Section 17.7). Such a coherent picture is of course possible only after more than a decade of precise measurement by experimentalists and the corresponding improvement in theoretical control and understanding.

### 25.1.7 Conclusions

Data from the $B$ Factories have enabled the KM mechanism to be tested in a comprehensive way using direct measurements of *CP*-violating observables. The constraints from *CP*-violating observables alone is sufficient to verify that the KM mechanism is the dominant source of *CP* violation in the SM. The level of *CP* violation measured via Unitarity Triangle angles in $B$ decays is consistent with that obtained from other tests of the CKM matrix (sides: $V_{ub}$, $V_{cb}$, mixing parameters, and $\varepsilon_K$). The set of non-angle constraints related to the Unitarity Triangle in the

**Table 25.1.3.** Input values for the global fit from Lattice QCD.

| Input | Value | Reference |
|---|---|---|
| $\vert V_{ud}\vert$ | $0.97425 \pm 0.00022$ | (Colangelo et al., 2011) |
| $\vert V_{us}\vert$ | $0.2208 \pm 0.0039$ | (Colangelo et al., 2011) |
| $f_{B_s}$ [MeV ] | $227.6 \pm 2.2 \pm 4.5$ | (Laiho, Lunghi, and Van de Water, 2010) |
| $f_{B_s}/f_{B_d}$ | $1.201 \pm 0.012 \pm 0.012$ | (Laiho, Lunghi, and Van de Water, 2010) |
| $\widehat{B}_{B_s}$ | $1.33 \pm 0.06$ | (Laiho, Lunghi, and Van de Water, 2010) |
| $B_{B_s}/B_{B_d}$ | $1.05 \pm 0.07$ | (Laiho, Lunghi, and Van de Water, 2010) |
| $\widehat{B}_K$ | $0.7643 \pm 0.0034 \pm 0.0091$ | (Laiho, Lunghi, and Van de Water, 2010) |

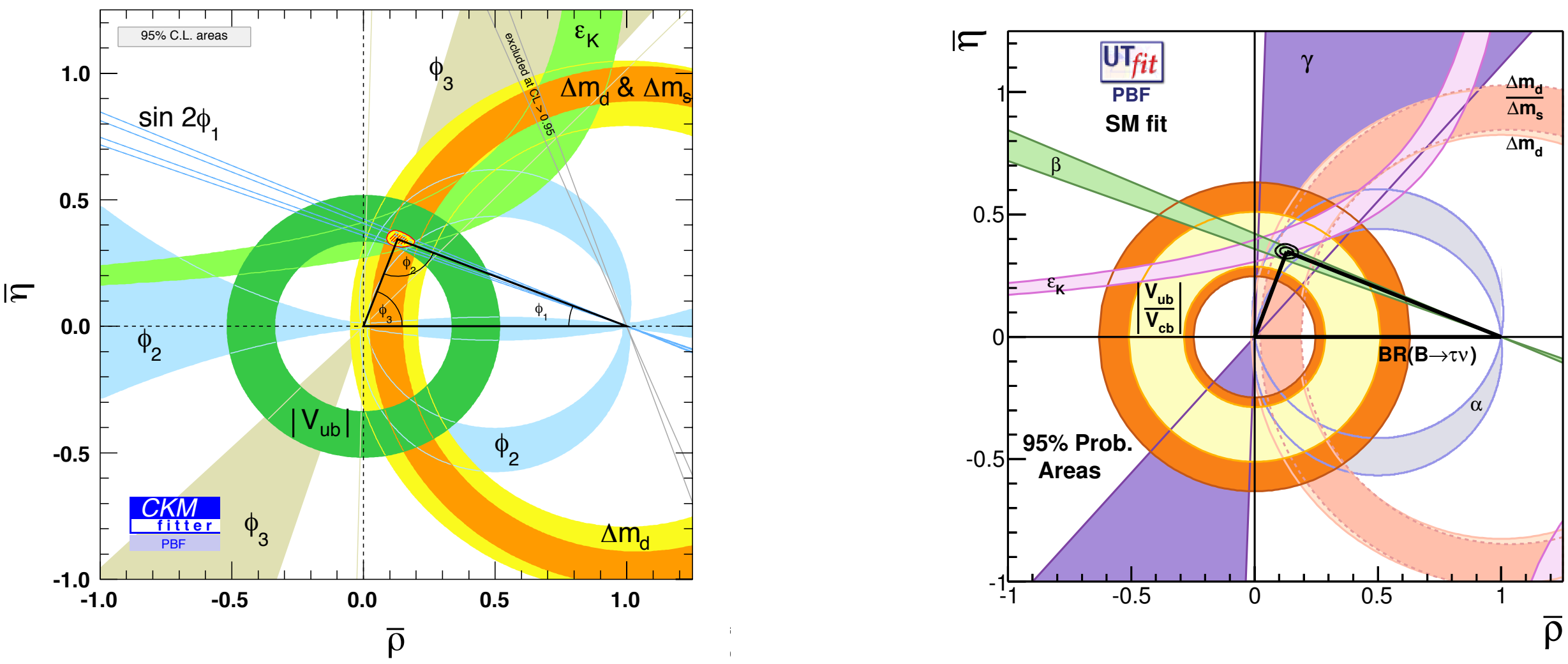

**Figure 25.1.1.** Results of global fits in the $(\bar{\rho}, \bar{\eta})$ plane, from CKMfitter and UTfit, showing the consistency of $b \to d$, $b \to s$ and $s \to d$ flavor-changing transitions with the Kobayashi-Maskawa mechanism for the common origin of the observed *CP* violation. The inputs of Tables 25.1.1 through 25.1.3 are used to obtain these plots. The second solution for the value of $\phi_1$ is suppressed using the measurements of final states that have an asymmetry dependence on $\cos 2\phi_1$. The corresponding numerical results from these fits can be found in Table 25.1.4.

**Table 25.1.5.** Compatibility of the individual inputs with their prediction from the global fit.

| Input | Input value | Predicted value UTfit [$\#\sigma$] |
|---|---|---|
| $\sin 2\phi_1$ | $0.677 \pm 0.020$ | $0.756 \pm 0.041$ [$1.7\sigma$] |
| $\phi_2$ [°] | $88 \pm 5$ | $88.7 \pm 3.3$ [$0.1\sigma$] |
| $\phi_3$ [°] | $67 \pm 11$ | $69.7 \pm 3.1$ [$0.2\sigma$] |
| $\Delta m_s$ [ps$^{-1}$] | $17.719 \pm 0.043$ | $17.35 \pm 1.05$ [$0.7\sigma$] |
| $\vert V_{cb}\vert$ [$10^{-3}$] | $41.67 \pm 0.63$ | $42.45 \pm 0.65$ [$0.8\sigma$] |
| $\vert V_{ub}\vert$ [$10^{-3}$] | $3.95 \pm 0.54$ | $3.61 \pm 0.11$ [$0.6\sigma$] |
| $\widehat{B}_K$ | $0.7643 \pm 0.0034 \pm 0.0091$ | $0.810 \pm 0.061$ [$0.3\sigma$] |
| $\mathcal{B}(B \to \tau\nu_\tau)$ $10^{-4}$ | $(1.15 \pm 0.23)$ | $0.818 \pm 0.062$ [$1.4\sigma$] |

SM provides a complementary test of the CKM mechanism, however those constraints require theoretical input in order to translate measurements into a constraint on the apex of the Unitarity Triangle. Hence the $B$ factories provided an experimentally and theoretically clean set of tests of the Standard Model in the measurements of the angles of the Unitarity Triangle. M. Kobayashi and T. Maskawa shared the 2008 Nobel Prize for their model of *CP* violation that inspired several generations of experimental exploration. During the lifetime of the $B$ Factories a number of constraints have been found to be in not-so-good agreement, and such tensions in the data have led to speculation of possible new physics scenarios, however the Standard Model persists.

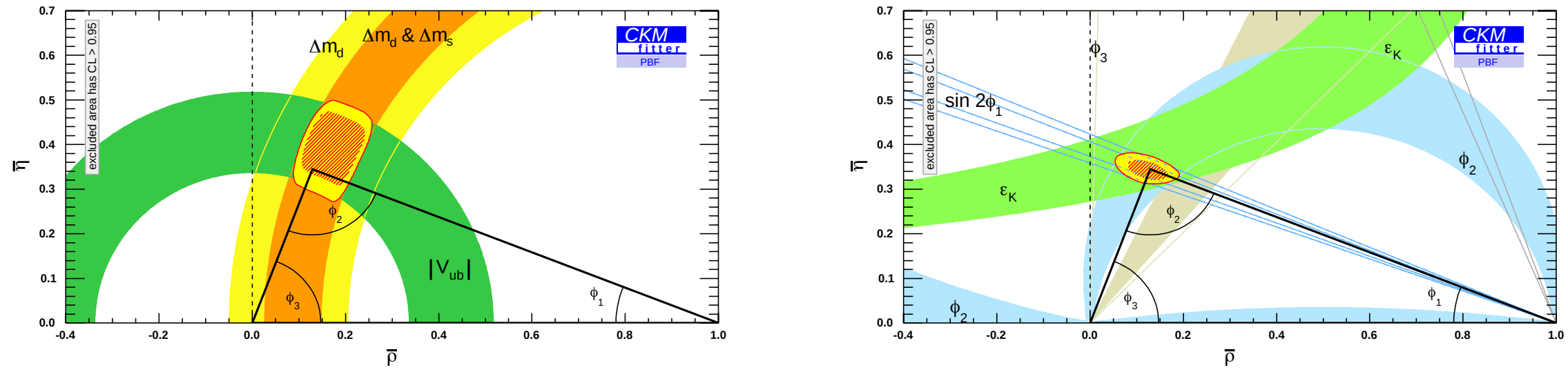

**Figure 25.1.2.** The consistency between the fit to *CP*-conserving observables (left) and *CP*-violating observables (right) from the CKMfitter group. The *CP*-conserving observables are the $B^0-\overline{B}^0$ and $B_s^0-\overline{B}_s^0$ mass differences, $\Delta m_d$ and $\Delta m_s$, respectively, and the measurement of $|V_{ub}|$ from semileptonic $b\to dl\nu_l$ decays. The *CP*-violating observables are from $K$ ($\varepsilon_K$) and $B$ ($\phi_1$, $\phi_2$, $\phi_3$) decays.

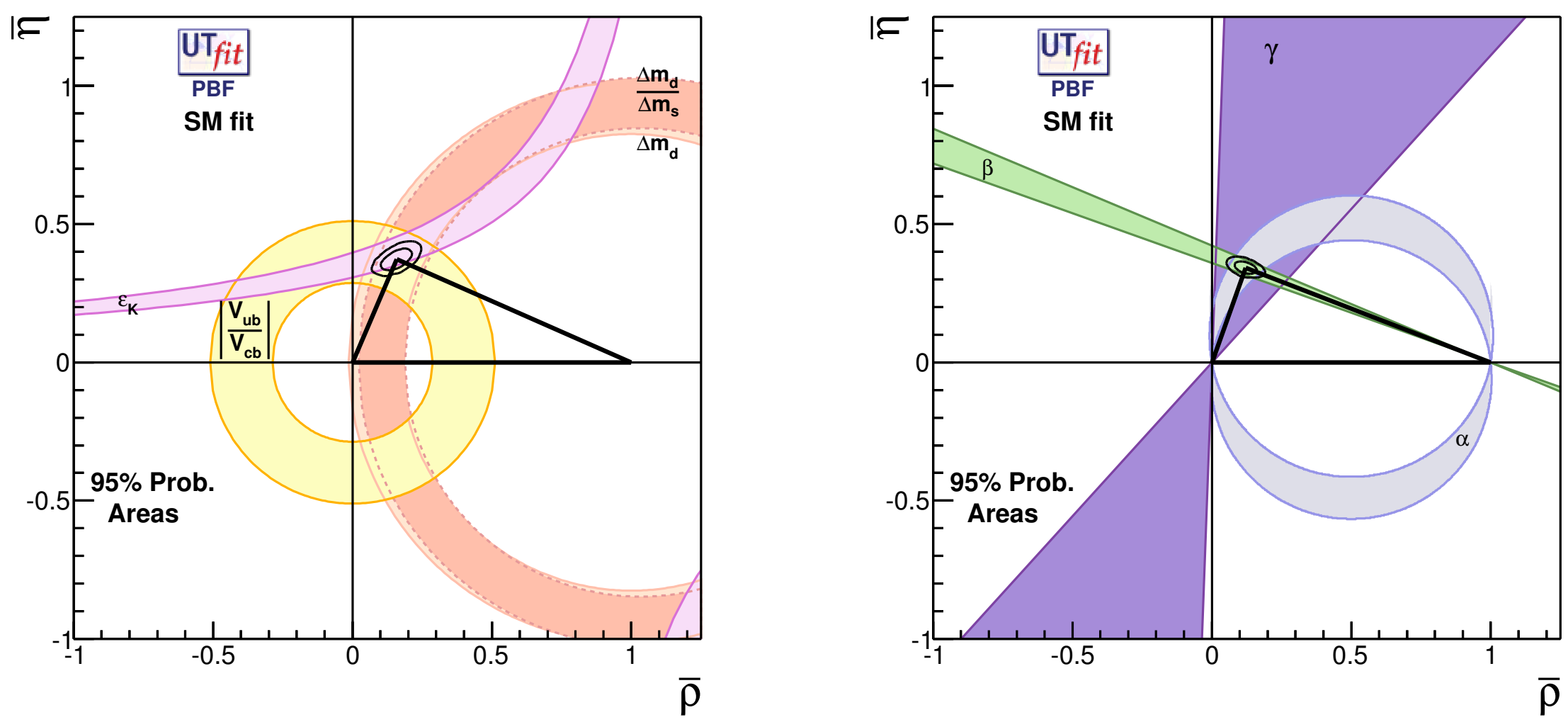

**Figure 25.1.3.** The consistency between the fit to *CP*-conserving observables and from $K$ measurements (on the left) and the angles ($B$ Factory-dominated on the right) from the UTfit group. The constraints used in the left plot are the $B^0-\overline{B}^0$ and $B_s^0-\overline{B}_s^0$ mass differences, $\Delta m_d$ and $\Delta m_s$, respectively, the measurements of $|V_{ub}|$ and $|V_{cb}|$ from semileptonic $B$ decays, and the *CP*-violation parameter $\varepsilon_K$. The constraints used in the right plot are the "angles" observables, *i.e.* measurements of $\phi_1$ ($=\beta$), $\phi_2$ ($=\alpha$), and $\phi_3$ ($=\gamma$).

The Tevatron and LHC experiments (CDF, DØ, ATLAS, CMS, and LHCb) have also produced results that can be used to test the *CP*-violating flavor parameters related to the Standard Model description of $B$ meson decays. While a detailed discussion is beyond the scope of this book, it should be noted that measurements from these experiments are compatible with results from the $B$ Factories. At the time of writing this book, there is no significant evidence for a departure from the KM picture of *CP* violation and the CKM matrix description of quark mixing.

It should be noted that it is still possible for sources of new physics to exist. Indeed given our understanding of the cosmological model of the Big Bang, it is thought that new sources of *CP* violation must exist. Any higher order contributions to *CP* violation in the quark sector that might be manifest in a hypothetical new physics scenario are constrained by the results discussed in this section: *i.e.* *CP* violation in the quark sector beyond the SM cannot be $\mathcal{O}(1)$. In the absence of experimental indications of a departure from the CKM picture the particle physics community chooses to explore the space of possible new physics models. Given the variety of such models, one has to focus on the predictions and behavior of specific benchmark models, a number of which are discussed in Section 25.2.

## 25.2 Benchmark new physics models

***Editors:***
*Emi Kou, Jure Zupan (theory)*

In this section we review the impact of the $B$ Factories on our understanding of the flavor structure in the Standard Model (SM) and on constraining new physics (NP) models. The most important overall result of the $B$ Factories physics program is the fact that the $CP$ violation observed in flavor changing processes with quarks is due to the Kobayashi-Maskawa (KM) mechanism (Kobayashi and Maskawa, 1973). For instance, prior to the $B$ Factories, the kaon sector was the only system where $CP$ violation was observed (see Section 16.1). The observed strength of the $CP$ violation in mixing, $\epsilon_K \simeq 2.3 \times 10^{-3}$, was consistent with the KM mechanism with an $\mathcal{O}(1)$ $CP$ phase in the CKM matrix. While encouraging, this by no means constituted a proof that the KM mechanism was really the origin of the observed $CP$ violation. The first test of the KM mechanism was then done by the measurement of $\sin 2\phi_1$ by the $B$ Factories.

By now the KM mechanism has been tested at the level of $\sim \mathcal{O}(10\%)$, while deviations from its predictions at levels smaller than this are still allowed. In the KM mechanism there is only one weak phase, providing a single source of $CP$ violation. The consistency of $\epsilon_K$ with the observed $CP$ violation in $B_d^0 - \overline{B}_d^0$ mixing and the measurements of the sides and angles of the standard CKM Unitarity Triangle all point to this common origin of $CP$ violation. In addition, the size of $\Delta m_s$ and the recent LHCb bound on the size of the weak phase in $B_s^0 - \overline{B}_s^0$ mixing both agree with the KM mechanism within errors. This agreement between $b \to d$, $b \to s$ and $s \to d$ transitions is nicely summarized in the CKM fit plot (see Fig. 25.1.1). Another important indicator that the $CP$ violation we are observing in flavor changing processes of quarks is due to the KM mechanism, comes from a comparison of a fit with only $CP$ conserving observables with a fit with only $CP$ violating observables. Both fits point to the same region in the $\bar{\rho}$ and $\bar{\eta}$ plane, which is a strong test of KM nature of $CP$ violation (see Fig. 25.1.2).

The measurements at the $B$ Factories also had a direct impact on new physics models. For instance, a measurement that $\sin 2\phi_1$ is $\mathcal{O}(1)$ immediately excluded approximate $CP$ models. In these models all the couplings which govern the low energy phenomena are real or almost real, with imaginary components always much smaller than the real ones. In the SM, the observed $\epsilon_K$ and $\epsilon'$, representing the $CP$ violation in the kaon sector, are small, which can be explained by the smallness of the CKM matrix elements entering into the description of these observables. In the approximate $CP$ models it would be small because $CP$ symmetry is only slightly broken and thus all $CP$ violating phases are small. A set of well motivated realizations in the SUSY framework was put forward (Abel and Frere, 1997; Babu and Barr, 1994; Babu, Dutta, and Mohapatra, 2000; Eyal, Masiero, Nir, and Silvestrini, 1999; Eyal and Nir, 1998). For instance approximate $CP$ conservation could naturally solve the "SUSY $CP$ problem" explaining why EDMs are so small. All these models were excluded once $\sin 2\phi_1$ was measured and its value turned out to be large, *i.e.* $\mathcal{O}(1)$.

Another set of models that was excluded by the fact that $\sin 2\phi_1$ was found to be bigger than 0.1, were the left-right symmetric models with spontaneous $CP$ breaking (Ball and Fleischer, 2000; Ball, Frere, and Matias, 2000; Bergmann and Perez, 2001). In these models Yukawa interactions are $CP$ conserving, while the $CP$ violation arises from the complex vacuum expectation values of the Higgs fields. Because of this the value of the CKM phase $\delta$ cannot take any value, but is found to be restricted to be below $|\delta \bmod \pi| < 0.25$ (Ball, Frere, and Matias, 2000). This was experimentally excluded as soon as $\sin 2\phi_1$ was measured with some precision.

After the measurement of $\sin 2\phi_1$ it was still possible that the observed large $\sin 2\phi_1$ value was due to a new superweak-like four fermion interaction and that direct $CP$ violation is small, just as it is small in kaon decays (Wolfenstein, 2002). This (somewhat artificial) possibility was excluded by the discovery of direct $CP$ violation in $B \to K^+\pi^-$ decays (Aubert, 2004u; Chao, 2005).

The constraints on general NP are best illustrated on the case of mixing. The NP contributions to $B_{d,s}^0 - \overline{B}_{d,s}^0$ mixing can be completely generally parameterized by

$$M_{12}^{d,s} = \left(M_{12}^{d,s}\right)^{\rm SM} \left(1 + h_{d,s}\, e^{2i\sigma_{d,s}}\right), \qquad (25.2.1)$$

where $\left(M_{12}^{d,s}\right)^{\rm SM}$ are the matrix elements of the SM effective weak Hamiltonian for $B_{d,s}^0 - \overline{B}_{d,s}^0$ mixing. The magnitudes of NP contributions relative to the SM are given by $h_d$ and $h_s$ for $B_d^0 - \overline{B}_d^0$ and $B_s^0 - \overline{B}_s^0$ mixing, respectively, while $\sigma_{d,s}$ are the corresponding NP weak phases. If NP models do not lead to diagrams which can induce the $B_d^0 - \overline{B}_d^0$ and $B_s^0 - \overline{B}_s^0$ mixing, then we have $h_s = h_d = 0$. Fig. 25.2.1 shows the present experimental constraints on the parameters $h_{d,s}$ and $\sigma_{d,s}$. One sees that contributions at the level of $\sim 10\% - 20\%$ are allowed for any weak phase (if one includes $B \to \tau\nu$ in the fit there is a slight preference for a nonzero phase), but corrections larger than 50% are also still possible.

Note that the SM contributions to the meson mixing are both loop and CKM suppressed. The CKM suppression for $B_d^0 - \overline{B}_d^0$ mixing is $|V_{td}^* V_{tb}|^2 \sim \mathcal{O}(\lambda^6)$. The expansion parameter, $\lambda \simeq 0.23$, in the Wolfenstein parameterization of the CKM matrix equals the sine of the Cabibbo angle. Similarly, there is a CKM suppression of $B_s^0 - \overline{B}_s^0$ mixing of $|V_{ts}^* V_{tb}|^2 \sim \mathcal{O}(\lambda^4)$ and a CKM suppression of $K^0 - \overline{K}^0$ mixing proportional to $|V_{td}^* V_{ts}|^2 \sim \mathcal{O}(\lambda^{10})$. This means that, if there is new physics at the TeV scale which contributes to the mixing amplitudes at tree level or at the loop level, it has to have a very non-generic flavor structure. The upper bounds on the coupling strengths for a 1 TeV suppression scale are given in Table 25.2.1. The same table also shows the alternative interpretation of the data – giving the lower bounds on NP scale, if the coupling strengths are assumed to be large, $\mathcal{O}(1)$.

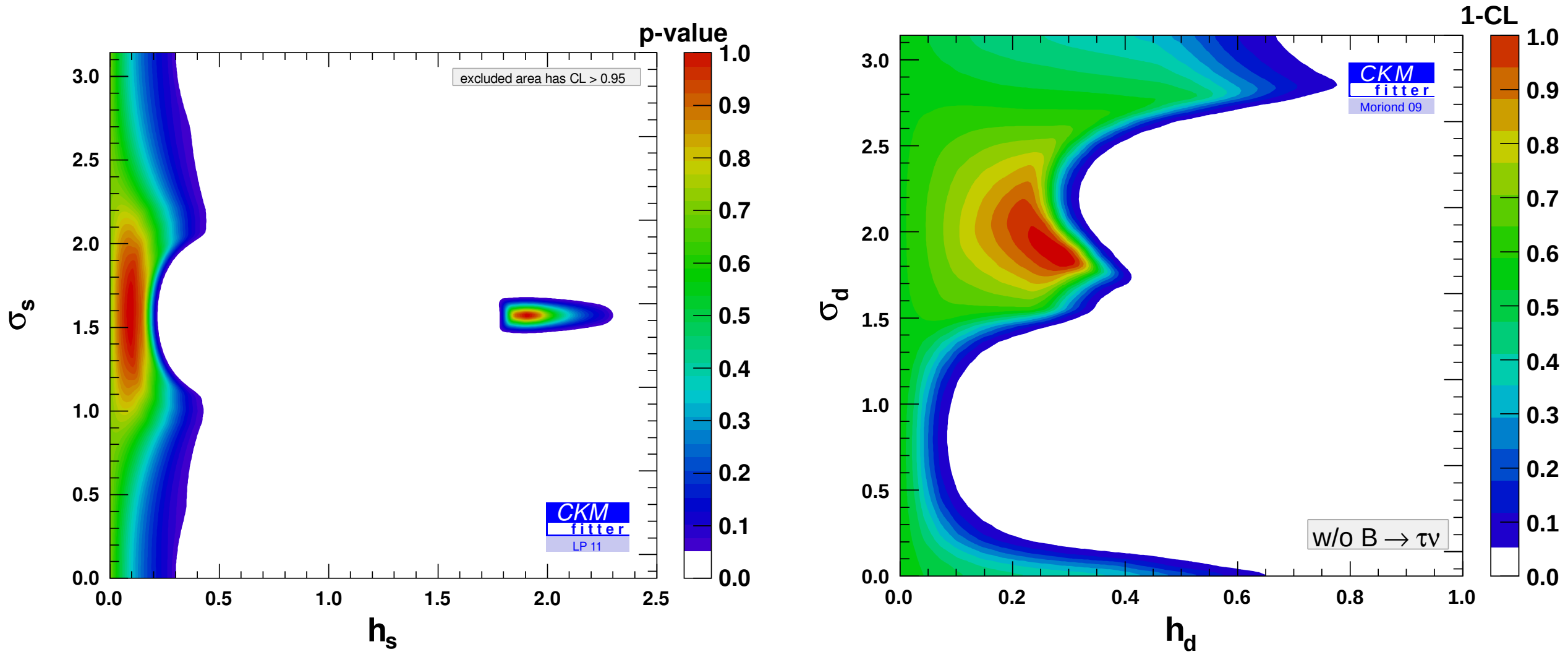

**Figure 25.2.1.** The present constraints on contributions from NP processes in $B_s^0$ mixing (left) and $B_d^0$ mixing (right), normalized to the SM contributions. From (Ligeti, 2011).

**Table 25.2.1.** Bounds on $\Delta F = 2$ operators of the form $(C/\Lambda^2)\,\mathcal{O}$, with $\mathcal{O}$ given in the first column. The bounds on $\Lambda$ assume $C = 1$, and the bounds on $C$ assume $\Lambda = 1\,\text{TeV}$. From (Hewett et al., 2012).

| Operator | Bounds on $\Lambda$ [TeV] $(C=1)$ | | Bounds on $C$ $(\Lambda = 1\,\text{TeV})$ | | Observables |
|---|---|---|---|---|---|
| | Re | Im | Re | Im | |
| $(\bar{s}_L\gamma^\mu d_L)^2$ | $9.8\times10^2$ | $1.6\times10^4$ | $9.0\times10^{-7}$ | $3.4\times10^{-9}$ | $\Delta m_K$; $\epsilon_K$ |
| $(\bar{s}_R\, d_L)(\bar{s}_L d_R)$ | $1.8\times10^4$ | $3.2\times10^5$ | $6.9\times10^{-9}$ | $2.6\times10^{-11}$ | $\Delta m_K$; $\epsilon_K$ |
| $(\bar{c}_L\gamma^\mu u_L)^2$ | $1.2\times10^3$ | $2.9\times10^3$ | $5.6\times10^{-7}$ | $1.0\times10^{-7}$ | $\Delta m_D$; $\lvert q/p\rvert, \phi_D$ |
| $(\bar{c}_R\, u_L)(\bar{c}_L u_R)$ | $6.2\times10^3$ | $1.5\times10^4$ | $5.7\times10^{-8}$ | $1.1\times10^{-8}$ | $\Delta m_D$; $\lvert q/p\rvert, \phi_D$ |
| $(\bar{b}_L\gamma^\mu d_L)^2$ | $5.1\times10^2$ | $9.3\times10^2$ | $3.3\times10^{-6}$ | $1.0\times10^{-6}$ | $\Delta m_d$; $S_{\psi K_S^0}$ |
| $(\bar{b}_R\, d_L)(\bar{b}_L d_R)$ | $1.9\times10^3$ | $3.6\times10^3$ | $5.6\times10^{-7}$ | $1.7\times10^{-7}$ | $\Delta m_d$; $S_{\psi K_S^0}$ |
| $(\bar{b}_L\gamma^\mu s_L)^2$ | $1.1\times10^2$ | $2.2\times10^2$ | $7.6\times10^{-5}$ | $1.7\times10^{-5}$ | $\Delta m_s$; $S_{\psi\phi}$ |
| $(\bar{b}_R\, s_L)(\bar{b}_L s_R)$ | $3.7\times10^2$ | $7.4\times10^2$ | $1.3\times10^{-5}$ | $3.0\times10^{-6}$ | $\Delta m_s$; $S_{\psi\phi}$ |

### 25.2.1 Short description of NP models

We first quickly describe the NP models we will consider. These are the models which could induce a significant deviation from the SM in the various observables measured at the $B$ Factories and thus, have been considered as benchmark NP models. The expected deviations in $B$ meson observables and the results from the $B$ Factories are described in the next section, where one can also find detailed references for further reading.

#### *Fourth generation*

Since the origin of the three generations in the SM is unclear, it is natural to consider a possibility that additional generations exist. An extension of the SM by a $4^{\text{th}}$ generation means that in addition to the particles in the SM there is an extra pair of heavy quarks, the $t'$ (up-type) and $b'$ (down-type), as well as a new heavy charged lepton and an additional heavy neutrino. We focus on the effects due to additional heavy quarks. The CKM matrix described in Chapter 16 is now a $4\times4$ unitary matrix. With three generations there are four physical parameters in the CKM matrix: three rotation angles and a single complex phase, where the phase is responsible for *CP* violation. In a four generation model there are 3 extra rotation angles and 2 new *CP* violating phases compared to the SM.[183] In this model, the CKM elements obey quadrilateral and not triangle unitarity relations as in the SM. The new *CP* violating phases can lead to additional *CP* violation beyond the SM predictions (*e.g.* in the penguin $b \to s$ transitions). The violation of Unitarity Triangle relations is checked by global CKM fits (see Section 25.1, the fits were also extended to include the $4^{\text{th}}$ generation).

[183] An $n\times n$ unitary matrix contains $2n^2-(n+(n^2-n)) = n^2$ real parameters while we can absorb $2n-1$ phases in the definition of the quark fields, which results in $(n-1)^2$ real parameters. Here we assign the $n(n-1)/2$ parameters as rotation angles and the rest as phases.

### Two Higgs Doublet Models

In the SM, the so-called Higgs field, an $SU(2)$ doublet scalar field, is introduced. The electroweak symmetry is broken spontaneously by its non-zero vacuum expectation value, which leads to the particle masses. This feature is retained also, if there is more than one Higgs doublet. The Two Higgs Doublet Model (2HDM) is the simplest extension of this kind, introducing one more Higgs doublet. In the particle spectrum we then have three neutral and one charged scalar. For $B$ physics the form of the Yukawa couplings is especially important. Several interesting limits are discussed in the literature. Here we mostly focus on the so-called Type II 2HDM where each of the two Higgs doublets only couple to down or up quarks. In this way phenomenologically unacceptable FCNCs due to tree level neutral Higgs exchanges do not arise. As a result, the main effects in flavor physics are due to the charged Higgs contributions. Even though the LHC searches for the charged Higgs directly, the constraints on the properties of this new particle, its mass and its couplings, come mainly indirectly from $B$ decays. Note that Type II 2HDM also describes the Higgs sector of the Minimal Supersymmetric Standard Model at tree level.

### Minimal Flavor Violation

Minimal Flavor Violation (MFV) is a general hypothesis that can apply to a large class of NP models. The central assumption is that the *only* source of flavor violation – also in the NP sector – are the SM Yukawa coupling matrices, $Y_{U,D}$. This is the minimal amount of flavor breaking in any NP model. The flavor breaking due to $Y_{U,D}$ will at least through loop corrections then also propagate to other sectors of the NP theory.

For MFV NP, generically the FCNCs are of the same order as in the SM (but can be smaller, if NP particles involved in the FCNC process are heavy). In MFV the form of flavor violation is fixed, implying strict correlations between different processes. This is especially true for the constrained MFV (cMFV) where *CP* violation is only due to the CKM phase and the effective weak Hamiltonian has exactly the same form as in the SM. For instance, in the cMFV the NP contributions to $B_d^0 - \overline{B}_d^0$ mixing and $B_s^0 - \overline{B}_s^0$ mixing, when normalized to the SM, are exactly the same. A sign of cMFV would be a deviation from the SM that can be described without new *CP* violating phases and without enlarging the SM operator basis. A discrepancy in $\phi_1$ determined from $B \to J/\psi K_S^0$ and the global Unitarity Triangle fit, on the other hand, would rule out the cMFV framework.

### Extensions of MFV

One can have viable TeV NP with not too large FCNCs even, if the flavor breaking is not just due to the SM Yukawa couplings. It suffices that all the flavor breaking has a structure similar to the SM one. The most important feature is that there is a hierarchy similar to the one in the quark flavor sector of the SM. The first two generations are much lighter than the third generation. Also, the mixing between the third and the first two generations is much smaller than the one between the first two generations (*i.e.* $V_{cb}, V_{ub} \ll V_{us}$). If this pattern is also present in NP with roughly the same hierarchies and directions of flavor breaking as in the SM, then FCNCs generated by NP will not be dangerously large. This insight can be formalized using symmetries and goes by the name of general MFV (GMFV). It is more general than cMFV, but coincides with the most general form of MFV, if Yukawa couplings are perturbative. Since GMFV is more general, there are also less correlations between observables. For instance, depending on which operators dominate, the new *CP* violating phases in $B_d^0 - \overline{B}_d^0$ and $B_s^0 - \overline{B}_s^0$ mixing are either exactly the same or there is a new *CP* violating phase only in $B_s^0 - \overline{B}_s^0$ mixing as we will see below.

### Supersymmetry

The supersymmetric (SUSY) extensions of the SM are some of the most popular NP models. SUSY relates fermions and bosons. For example, the gauge bosons have their fermion superpartners and fermions have their scalar superpartners. SUSY at the TeV scale is motivated by the fact that it solves the SM hierarchy problem. The quantum corrections to the Higgs mass are quadratically divergent and would drive the Higgs mass to the Planck scale $\sim 10^{19}$ GeV, unless the contributions are canceled. In SUSY models they are canceled by the virtual corrections from the superpartners. The minimal SUSY extension of the SM refers to the scenario where all the SM fields obtain superpartners but there are no other additional fields. This is the Minimal Supersymmetric Standard Model (MSSM). SUSY cannot be an exact symmetry since in that case superpartners would have the same masses as the SM particles, in clear conflict with observations. Different mechanisms of SUSY breaking have very different consequences for flavor observables. In complete generality the MSSM has more than a hundred parameters, most of them coming from the so-called soft SUSY breaking terms (the SUSY breaking terms with dimensionful couplings, such as *e.g.* masses, so that the divergence is at most logarithmic). If superpartners exist at the TeV scale the most general form with $\mathcal{O}(1)$ flavor breaking coefficients is excluded due to flavor constraints. This has been dubbed the SUSY flavor problem (or in general the NP flavor problem).

### MFV SUSY

A popular solution to the SUSY flavor problem is to assume that the SUSY breaking mechanism and the induced interactions are flavor "universal". The flavor universality is often imposed at a very high scale corresponding to the SUSY breaking mechanism. It could be at, for instance, the Planck scale ($\sim 10^{19}$ GeV), the GUT scale ($\sim 10^{16}$ GeV) or some intermediate scale such as the gauge mediation scale ($\sim 10^{6}$ GeV). The flavor breaking can then be transferred only from the SM Yukawa couplings to the other interactions through renormalization group running from the higher scale to the weak scale. As a result, the flavor breaking comes entirely from the SM Yukawa couplings (thus, an example of a concrete MFV NP scenario). Since the soft SUSY breaking terms are flavor-blind, the squark masses are degenerate at the high energy scale. The squark mass splitting occurs only due to quark Yukawa couplings, where only top Yukawa and potentially bottom Yukawa couplings are large. Thus the first two generation squarks remain degenerate to very good approximation, while the third generation squarks are split.

#### *nonMFV-SUSY*

While flavor blind SUSY breaking is well motivated, it is important to keep track of other possibilities – with more general flavor breaking patterns. A very useful approach is the mass insertion approximation (MIA). The approximation is easiest to explain in the basis, where the SM fermions have diagonal masses, while the sfermions have undergone exactly the same flavor transformations as their SM fermion partners (this is the so-called super-CKM basis). All neutral gauge interactions are still flavor diagonal, charged currents are proportional to the CKM elements, while the sfermion mass matrices have also off-diagonal flavor violating entries. Phenomenologically, the off-diagonal entries need to be small. They can thus be treated as perturbations, compared to the diagonal ones. The size of the resulting flavor violation is usually parametrized by dimensionless ratios of off-diagonal and (the average of) diagonal entries in mass matrices, $(\delta^q_{AB})_{ij}$, where $A, B$ are the chiralities $(L, R)$ and $q$ indicates the $(u, d)$ type. In principle $(\delta^q_{AB})_{ij}$ are only bounded by the experiments, but are otherwise completely general.

#### *SUSY alignment models*

If the squark mass basis is almost the same as the quark mass basis, then the FCNCs due to squarks and gluinos running in the loops are suppressed. Alignment of squark and quark mass matrices is easily achieved in flavor model building and could be a remnant of the underlying symmetry. The alignment models are very close in spirit to MFV models – that there is a relation between the flavor breaking in squark and quark sectors – but differ in details. For instance, in alignment models the squark matrices can carry arbitrary phases, the relation between squark and quark mass eigenstate bases is only approximate and most importantly, the first two generations squarks need not be degenerate but can have $\mathcal{O}(1)$ splitting in their mass spectra.

The bounds on FCNCs in up and down quark sectors put strong constraints on alignment models. For instance the measured values of $D^0 - \overline{D}^0$ and $K^0 - \overline{K}^0$ mixing require that the splitting between the first two generations of left-handed squarks is less than $\mathcal{O}(0.1)$ for TeV squarks masses.

#### *Randall-Sundrum models of flavor*

The Randall-Sundrum (RS) model introduces a "warped" extra dimension, *i.e.* a fifth dimension for which the metric of the five dimensional space time contains an exponential "warp" factor. The Higgs of the SM is confined to a four dimensional subspace, the "TeV brane", while the full five dimensional space is called the "bulk". This idea provides an interesting solution to the hierarchy problem by generating the weak scale from the Planck scale through this exponential warp factor. The RS model can also explain the hierarchy of fermion masses. The fermionic wave functions extend in the bulk with the heavier particles located closer to the TeV brane. Then, the fermion wave function profiles (locations of different fermions in the bulk) can also provide the exponential warp factor and an $\mathcal{O}(1)$ change in the parameters in the 5 dimensional theory can lead to the hierarchical flavor structure in the 4 dimensional theory. The existence of an extra dimension results in each SM particle to be accompanied by a whole tower of excited "Kaluza-Klein" (KK) states. The exchanges of KK excitations of gluons and the distortion of the $Z$ boson's 5D profile then generate FCNCs at tree level. These effects are naturally suppressed, however, by the same mechanism that explains the hierarchy of fermion masses. The resulting FCNCs are not too large for the light quarks, with the exception of $K^0 - \overline{K}^0$ mixing, where modest cancellations between different contributions are required if the KK mass scale is 2-3 TeV. In this case large effects in several $B$ physics observables are also expected. The 5D masses are complex in general, so that there are many new sources of $CP$ violation. $\mathcal{O}(1)$ NP weak phases in $B^0_d - \overline{B}^0_d$ mixing and $B^0_s - \overline{B}^0_s$ mixing would thus be expected.

#### *Little Higgs Models*

Little Higgs models are an alternative way to solve the hierarchy problem. The mass of a scalar particle is stabilized, if it is a Goldstone boson of a spontaneously broken global symmetry. The Goldstone boson is massless, if the original global symmetry is exact, and has a small nonzero mass, if there is already a small explicit breaking of the global symmetry. Little Higgs models implement this idea by constructing an extension of the SM such that the SM Higgs particle is the Goldstone mode of an enlarged symmetry. This symmetry is also explicitly broken in order to achieve a small mass for the Goldstone modes. Unlike SUSY, the solution to the hierarchy problem in Little Higgs models is only partial – just for 1-loop corrections to the Higgs mass. The models therefore require UV completion at a scale of a few TeV, *i.e.* the models are not valid to arbitrary high energies but need to be supplemented with extra fields and/or interactions. One of the more interesting realizations is the Littlest Higgs model with $T$ parity. In it the SM fields are supplemented by a new heavy top quark ($T_+$), a triplet of heavy scalars ($\Phi$) as well as a new set of heavy gauge bosons $W_H^\pm, Z_H^0, A_H$. It has an interesting non-MFV flavor structure with only 10 new parameters in the quark sector. As a result there are still correlations between FCNC processes in the down and up-quark sectors. The constraints from $B \to X_s\gamma$ are easily satisfied, while significant effects would be expected in $B_s$ mixing and in $K \to \pi\nu\bar{\nu}$ and $K_L \to \pi^0\ell^+\ell^-$ decays.

### 25.2.2 Detailed description of NP models

We now give a more detailed description of the models, focusing especially on the impact the $B$ Factory observables had on constraining the models. We highlight the following observables in particular, $\sin 2\phi_1$, $B \to X_s\gamma$, $B \to \tau\nu$, $D^0-\overline{D}^0$ mixing, $B \to \phi K_S^0$, all of which are listed in Table 25.2.2. We also include the anomalous moment of the muon, $(g-2)_\mu$, as another important observable. Even though it was not measured at the $B$ Factories, the ISR results obtained at the $B$ Factories provide crucial inputs to the SM predictions (see Section 21.3.4 for detailed discussions on the impact of the $B$ Factories ISR result on the muon $g-2$). For ease of comparison a star system is used in Table 25.2.2, where more stars mean that the generic predictions of the model agree better with the observations (from 1 to 3 stars).

#### 25.2.2.1 Fourth generation

There is no compelling theoretical reason for having only three generations of fermions. It is thus important to search for additional heavier quarks. The simplest possibility is a sequential $4^{\rm th}$ generation, where new heavy quarks have the same quantum numbers as in the SM – the left-handed $t'$ and $b'$ form an $SU(2)_L$ doublet, while the right-handed $t'$ and $b'$ are singlets (for a review see (Frampton, Hung, and Sher, 2000)). In this model the $3\times 3$ CKM matrix is no longer unitary since it is only a part of the full $4\times 4$ matrix

$$\left(\begin{array}{ccc|c} V_{ud} & V_{us} & V_{ub} & V_{ub'} \\ V_{cd} & V_{cs} & V_{cb} & V_{cb'} \\ V_{td} & V_{ts} & V_{tb} & V_{tb'} \\ \hline V_{t'd} & V_{t's} & V_{t'b} & V_{t'b'} \end{array}\right). \qquad (25.2.2)$$

This means that the global fits of the CKM unitarity must be supplemented with $3\times 3$ unitarity relaxed, see, *e.g.*, (Bobrowski, Lenz, Riedl, and Rohrwild, 2009).

The heavy quarks can contribute to any loop type diagrams. In particular, since the loop function of the box and the penguin diagrams grow with the mass of the heavy quark in the loop, these contributions can be large. Therefore, the precise measurements obtained by the $B$ Factories lead to very strong constraints on the fourth row and column of the enlarged $4\times 4$ quark mixing matrix.

We first focus on the impact of the $\sin 2\phi_1$ measurement by the $B$ Factories. The $B_d^0 - \overline{B}_d^0$ box diagram now also has a heavy top quark, $t'$, running in the loop. After imposing the $4\times 4$ unitarity the mixing matrix element is given by

$$M_{12} = \frac{G_F^2 m_W^2}{6\pi^2}\hat{\eta}_B m_B \hat{B}_{B_d} f_{B_d}^2 \big[\lambda_t^2 S_0(x_t) + 2\lambda_t\lambda_{t'} S_0(x_t, x_{t'}) + \lambda_{t'}^2 S_0(x_{t'})\big], \qquad (25.2.3)$$

where the Inami-Lim functions, $S_0$, describe the $t$ and $t'$ quark contributions in the loop in terms of their masses $x_i \equiv m_i^2/m_W^2$ (Inami and Lim, 1981). The CKM matrix elements are included in $\lambda_i \equiv V_{ib}^* V_{id}$ and the NLO QCD correction $\hat{\eta}_B$ is taken as $\hat{\eta}_B = 0.55$ (Buchalla, Buras, and Lautenbacher, 1996). The non-perturbative QCD effects are absorbed in the bag parameter $\hat{B}_{B_d}$ and the decay constant $f_{B_d}$, for which we use $(\hat{B}_{B_d})^{1/2} f_{B_d} = (216\pm 16)$ MeV (Nakamura et al., 2010). Introduction of the fourth generation quarks has two effects. First, the value of the CKM element $\lambda_t$ multiplying the top contribution can differ from the one obtained from the CKM fits where $3\times 3$ unitarity is assumed. Second, there is an additional contribution from the $t'$ in the loop.

The two effects are related through the $4\times 4$ unitarity condition, $\lambda_u + \lambda_c + \lambda_t + \lambda_{t'} = 0$ with $\lambda_u = (3.8\pm 0.5)\times 10^{-3} e^{i(-68\pm 10)^\circ}$ and $\lambda_c = (-9.4\pm 0.5)\times 10^{-3}$ extracted from the processes involving only tree level diagrams (Nakamura et al., 2010). The predictions

$$\Delta m_d = 2|M_{12}^{\rm 4SM}|, \quad S_{b\to c\bar{c}s} = -{\rm Im}\sqrt{\frac{M_{12}^{*\rm 4SM}}{M_{12}^{\rm 4SM}}}, \qquad (25.2.4)$$

then depend only one complex parameter, $\lambda_{t'}$ (or equivalently $\lambda_t$ by using the unitarity relation), and the mass of $t'$. The experimental world averages, $\Delta m_d = (0.510\pm 0.004)$ ps$^{-1}$ and $S_{b\to c\bar{c}s} = 0.677\pm 0.020$ (see Sections 17.5.2 and 17.6 for the experimental extraction of these values), are consistent with the SM predictions within errors, $\Delta m_d^{\rm SM} = (0.51\pm 0.12)$ ps$^{-1}$ and $S_{b\to c\bar{c}s} = 0.74\pm 0.09$. Fixing the mass of $t'$ to 600 GeV the constraints on $\lambda_{t'}$ are shown in Fig. 25.2.2. The dashed line is the constraint from the mass difference $\Delta m_d$, which was known before the $B$ Factories but with much larger uncertainty. The $\sin 2\phi_1$ measurement (solid line) gives a strong constraint $|\lambda_{t'}| < 0.005$. Combined constraints exclude the colored regions.

The next example is the $B \to X_s\gamma$ branching ratio, which also receives a contribution from a $t'$ quark running in the loop. The experimental measurements agree with the SM prediction from up, charm and top quarks running in the loop. However, as in the case of $B^0-\overline{B}^0$ mixing, the top contribution in the $4^{\rm th}$ generation scenario can differ from the SM due to a different $V_{tb}^* V_{ts}$ obtained from $4\times 4$ CKM fits. This can then leave some more space for the non-zero $t'$ contribution. The obtained constraint on $V_{t'b}^* V_{t's}$ is not so strong since the branching ratio is dominated by the tree level $\mathcal{O}_2$ contributions which are proportional to $V_{cb}^* V_{cs}$. The $B \to X_s\gamma$ branching ratio allows $V^*_{t'b}V_{t's} \sim \mathcal{O}(\lambda)$ which is a much weaker bound than the constraints obtained from other measurements such as $B_s^0 - \overline{B}_s^0$ oscillation.

**Table 25.2.2.** The agreement of NP models and the SM with the experimental results including the measurements from the $B$ Factories (more stars means better agreement). See text for further explanations. The − sign means there is no clear expectation.

| Observable | $4^{\text{th}}$ gen. | 2HDM | MFV | eMFV | MFV-SUSY | genSUSY | aligSUSY | RS | Little H | SM |
|---|---|---|---|---|---|---|---|---|---|---|
| $\sin 2\phi_1$ | ★★★ | ★★★ | ★★★ | ★★ | ★★★ | ★ | ★★★ | ★★ | ★★ | ★★★ |
| $B \to X_s\gamma$ | ★★★ | ★ | ★★ | ★★ | ★★ | ★ | ★★★ | ★★ | ★★★ | ★★★ |
| $B \to \tau\nu$ | ★★ | ★★ | ★★ | ★★ | ★★ | ★★ | ★★ | ★★ | ★★ | ★★ |
| $D^0 - \overline{D}^0$ mixing | ★★ | ★★★ | ★★★ | ★★★ | ★★★ | ★ | ★ | ★★ | ★★ | ★★ |
| $B \to \phi K_S^0$ | ★★★ | ★★ | ★★ | ★★★ | ★★★ | ★★ | ★★ | ★★★ | ★ | ★★ |
| $(g-2)_\mu$ | ★ | ★ | − | − | ★★ | ★★ | − | ★★ | ★★ | ★ |

Another example is $D^0 - \overline{D}^0$ mixing where the $b'$ quark could contribute in the loop diagram in addition to the SM particles. Saturating the experimental value of $x_D$ gives

$$|V^*_{cb'}V_{ub'}|^2 \lesssim 10^{-2}, \tag{25.2.5}$$

for a 600 GeV $b'$.

The other observables receive some contributions from the $4^{\text{th}}$ generation, though so far, the constraints are not as strong as from $B^0 - \overline{B}^0$ mixing and $D^0 - \overline{D}^0$ mixing. Namely, the *CP* asymmetry of $B \to \phi K_S^0$, to which $t'$ can contribute both in $B^0 - \overline{B}^0$ oscillations (box diagram) and the $b \to s$ decay (gluon penguin diagram). The oscillation part is constrained very strongly by the $B \to J/\psi K_S^0$ channel. The decay part could constrain the $4^{\text{th}}$ generation parameters although it turns out that the obtained constraint is not very strong. Another observable, the decay $B \to \tau\nu$ is tree dominated. Thus, the $4^{\text{th}}$ generation contribution appears only through the modification of the CKM matrix element, namely, $V_{ub}$, due to the broken unitarity. This can be detected by observing different values of $V_{ub}$ in the measurements through tree and loop processes. So far, the observed difference is not statistically significant. The muon $g-2$ can receive a contribution from the $4^{\text{th}}$ generation neutrino (Lynch, 2001). However, the contribution is not large enough to explain the observed deviation from the SM.

The main message of this section is that the CKM matrix elements for the heavy fermions can be very strongly constrained from the $B$ Factory observables. A sizable mixing of SM quarks with $4^{\text{th}}$ generation quarks is excluded. If, on the other hand, one naïvely extrapolates the Wolfenstein structure of the CKM to 4 generations, one may expect that $\lambda_{t'} \sim \lambda^4$ so that an order of magnitude improvement in precision of flavor observables would be most welcome. Also, the direct searches at the LHC now exclude heavy fermions with masses below $\sim 600$ GeV. This limit is high enough that the unitarity mass limit has been reached – the fermion-fermion scattering amplitude becomes larger than one and the fermions become strongly coupled. This means that the loop corrections we discussed above must be taken with a grain of salt as they were obtained using perturbative calculations. Finally, the discovery of the Higgs-like particle with a mass of about 125 GeV also excludes the perturbative $4^{\text{th}}$ generation, since in that case the production and decays of the Higgs would be modified significantly (Eberhardt et al., 2012; Kuflik, Nir, and Volansky, 2012).

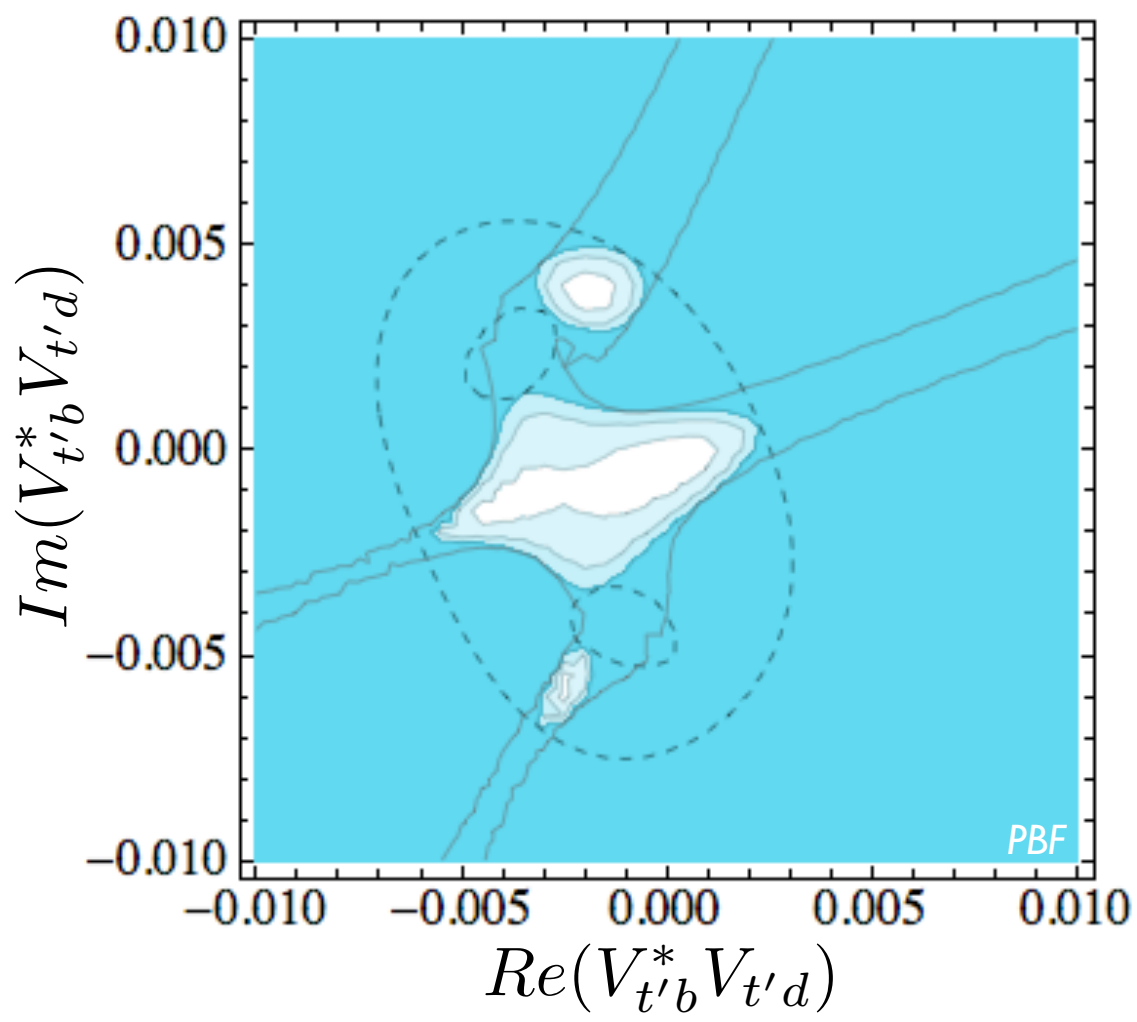

**Figure 25.2.2.** Constraints on the real and imaginary parts of $V^*_{t'b}V_{t'd}$ from $\Delta m_d$ (dashed) and $\sin 2\phi_1$ (solid) are shown for $m_{t'} = 600$ GeV. The blue regions are excluded at $1\sigma$, $2\sigma$ and $3\sigma$ (from lighter to darker) assuming Gaussian errors.

### 25.2.2.2 Two Higgs Doublet Models

A simple extension of the Standard Model is to add an extra Higgs doublet to the field content. Despite being a very simple modification, it can lead to drastic changes in low energy flavor phenomenology. If all Yukawa interactions between the two Higgs doublets and quarks are allowed, then this leads to FCNCs from neutral Higgs exchanges that are orders of magnitude above the experimentally allowed values. No FCNCs arise if only one Higgs doublet couples to quarks, while the other is inert (this is the Type I 2HDM). The other option is that one of the Higgs doublets, $H_1$, couples to the right-handed down quarks, while the other Higgs doublet, $H_2$ couples to the right-handed up quarks only. This is the Type II 2HDM, and corresponds to the Higgs sector of the MSSM when loop corrections from sparticles are neglected. A more general way to avoid tree-level FCNCs is to require alignment between the Yukawa couplings of the two Higgs doublets. Alignment is parameterized by three complex parameters and covers the above two types of the 2HDM models as special cases (Pich and Tuzon, 2009).

From now on we focus on the Type II 2HDM for which the expectations regarding flavor observables are collected in Table 25.2.2. In the 2HDM there are 3 real scalars, two *CP*-even and one *CP*-odd neutral Higgs bosons, and one charged Higgs boson $H^{\pm}$. In Type II 2HDM the interactions of the neutral Higgses are flavor diagonal, so that the flavor violation arises only at loop level like in the SM. The exchange of a charged Higgs, $H^{\pm}$, can lead to significant contributions to the $B$ physics observables. The interactions with quarks are given by

$$\begin{aligned}\mathcal{L} =& (2\sqrt{2}G_F)^{1/2} \sum_{i,j=1}^{3} \overline{u}_i \big(A_u m_{u_i} V_{ij} P_L \\ &- A_d V_{ij} m_{d_i} P_R\big) d_j H^+ + h.c.,\end{aligned} \tag{25.2.6}$$

where $P_{L,R} \equiv (1 \mp \gamma_5)/2$. $u_i$ and $d_j$ are the mass eigenstate fields, $i, j$ the generation indices, $m_{u,d}$ the diagonal quark mass matrices and $V_{ij}$ the CKM matrix elements. In the Type II 2HDM the coefficients $A_{u,d}$ depend only on the ratio of the $H_2$ and $H_1$ Higgs vacuum expectation values, $\tan\beta = v_2/v_1$,

$$A_u = \cot\beta, \quad A_d = -\tan\beta. \tag{25.2.7}$$

We expect a significant contribution from the loop induced diagrams with a charged Higgs and a top quark in the loop such as $B_d - \overline{B}_d$ oscillation and the decay $B \to X_s\gamma$. A large enhancement of the charged Higgs contribution is possible because of the first term in Eq. (25.2.6) with $u_i = t$ due to the large top quark mass. While the experimental measurements of the $B_d - \overline{B}_d$ oscillation frequency can constrain part of the parameter space, in particular for small $\tan\beta$, the constraint from $B \to X_s\gamma$ is generically more important. At LO the Wilson coefficient relevant for $B \to X_s\gamma$ including the NP contributions is (Ciuchini, Degrassi, Gambino, and Giudice, 1998b; Grinstein, Springer, and Wise, 1990; Hou and Willey, 1988)

$$\begin{aligned}C_{7,8}(M_W) =& C_{7,8}^{\rm SM}\left(\frac{m_t^2}{m_{W^{\pm}}^2}\right) + \frac{A_u^2}{3} G_{7,8}\left(\frac{m_t^2}{m_{H^{\pm}}^2}\right) \\ &- A_u A_d F_{7,8}\left(\frac{m_t^2}{m_{H^{\pm}}^2}\right).\end{aligned} \tag{25.2.8}$$

The second and the third terms give charged Higgs contributions and are enhanced by $m_t^2$, as expected. The signs of $A_u$ and $A_d$ are opposite in Type II 2HDM (see Eq. 25.2.7) so that both terms interfere constructively with the SM term. The term proportional to $A_u^2$ is relevant only for small $\tan\beta$, while the term proportional to $A_u A_d$ is important almost independently of $\tan\beta$.

The SM prediction for the branching ratio of $B \to X_s\gamma$ has been drastically improved in the past 10 years. The NNLO predictions are available both for the SM and for the charged Higgs contributions (Misiak et al., 2007). The most recent analysis shows that the SM prediction is roughly 1 sigma below the experimental value (see Section 17.9 for more details) and the following lower limit on the charged Higgs mass is obtained for any value of $\tan\beta$

$$m_{H^{\pm}} > 295 \text{ GeV}. \tag{25.2.9}$$

Another very important constraint on the Type II 2HDM has been obtained from the $B$ Factory measurements of $\mathcal{B}(B \to \tau\nu)$ (see Section 17.10 for more discussions on this process). This tree level process occurs in the SM with a diagram in which the $B$ meson annihilates into a $W$ boson followed by its decay into $\tau\nu$. In the 2HDM, a similar process is possible but the $W$ is replaced with the charged Higgs. The resulting branching ratio can be expressed as

$$\mathcal{B}(B \to \tau\nu) = \mathcal{B}(B \to \tau\nu)_{\rm SM}\left(1 - \tan^2\beta \frac{m_B^2}{m_{H^{\pm}}^2}\right)^2, \tag{25.2.10}$$

where

$$\mathcal{B}(B \to \tau\nu)_{\rm SM} = \frac{G_F^2 m_B m_\tau^2}{8\pi}\left(1 - \frac{m_\tau^2}{m_B^2}\right)^2 f_B^2 |V_{ub}|^2 \tau_B. \tag{25.2.11}$$

The second term in the parenthesis of Eq. (25.2.10) is due to the charged Higgs.

Replacing the left hand side of Eq. (25.2.10) with the experimental bound obtained by the $B$ Factories, a combination of the parameters $m_{H^{\pm}}/\tan\beta$ can be constrained. The two solutions

$$\frac{m_{H^{\pm}}}{m_B \tan\beta} = \left(1 \pm \sqrt{r_{\rm exp}^{B\to\tau\nu}}\right)^{-1/2}, \tag{25.2.12}$$

with

$$r_{\rm exp}^{B\to\tau\nu} = \frac{\mathcal{B}(B \to \tau\nu)_{\rm exp}}{\mathcal{B}(B \to \tau\nu)_{\rm SM}} \tag{25.2.13}$$

are plotted, respectively, as grey and black lines in Fig. 25.2.3. Here we use the SM prediction for the branching ratio, $\mathcal{B}(B \to \tau\nu)_{\rm SM} = (1.01 \pm 0.29) \times 10^{-4}$, as quoted in Section 17.10. The vertical lines indicate the current world average (see also Section 17.10)

$$\mathcal{B}(B \to \tau\nu)_{\rm exp} = (1.15 \pm 0.23) \times 10^{-6}, \tag{25.2.14}$$

with $1\sigma$ (dotted), $2\sigma$ (dashed) and $3\sigma$ (solid) errors. The two vertical yellow regions are excluded at 95% C.L. Note that the theoretical error to the SM prediction denoted above is indicated in the right top corner of Fig. 25.2.3. It results in additional uncertainties on top of the experimental ones. The horizontal yellow region is excluded by the $\mathcal{B}(B \to X_s\gamma)$ measurement.

The first solution (gray lines) are already excluded by the constraints from $B \to X_s\gamma$ as well as from $B \to D\tau\nu$. The second solutions (black lines) can give a constraint on $m_H^{\pm}$ stronger than $\mathcal{B}(B \to X_s\gamma)$. By taking into account the $3\sigma$ experimental error, for instance, one obtains a lower bound $m_{H^\pm} \geq 370$ (556) GeV for $\tan\beta = 40$ (60). Note that this bound is very sensitive to the theoretical input and could be lowered by over 100 (150) GeV by taking into account the theoretical uncertainty discussed above. It is clear that further reduction of both theoretical and experimental errors will shed light on the charged Higgs searches in this model in the future.

We discuss next briefly the other observables, for details see (Barger, Hewett, and Phillips, 1990; Grossman, 1994; Krawczyk and Pokorski, 1991; Wahab El Kaffas, Osland, and Ogreid, 2007). For the $\Delta F = 2$ ($F$ is strangeness or bottomness) processes, the impact on the $B^0 - \bar{B}^0$ mixing frequency and phase is small while the $D^0 - \bar{D}^0$ mixing could receive a large contribution, in particular for large $\tan\beta$ (Golowich, Hewett, Pakvasa, and Petrov, 2007). However, since the SM value of the $D^0 - \bar{D}^0$ mixing parameter is not well determined it is difficult to obtain a strict constraint, even though order of magnitude estimates are already interesting. Since this model does not include a large *CP* violating phase in the $b \to s$ transition part, the *CP* violation in $B \to \phi K^0_S$ does not receive a significant NP contribution. The recent *BABAR* measurement of the ratio $\mathcal{R}(D^{(*)}) \equiv \mathcal{B}(B \to D^{(*)}\tau\nu)/\mathcal{B}(B \to D^{(*)}l\nu)$ shows some deviation from the SM (Lees, 2012e). The charged Higgs can contribute to this process, however, the deviation in $\mathcal{R}(D)$ and $\mathcal{R}(D^*)$ cannot be explained simultaneously within Type II 2HDM (Lees, 2012e) (see Section 17.10 for more details). It is, however, possible to explain the observed pattern using 2HDM with more general flavor structure (Fajfer, Kamenik, Nisandzic, and Zupan, 2012). The muon magnetic moment, $(g-2)_\mu$, is found not to provide a significant contribution after taking into account the constraints discussed in this Section (Jegerlehner and Nyffeler, 2009; Krawczyk, 2002; Wahab El Kaffas, Osland, and Ogreid, 2007).

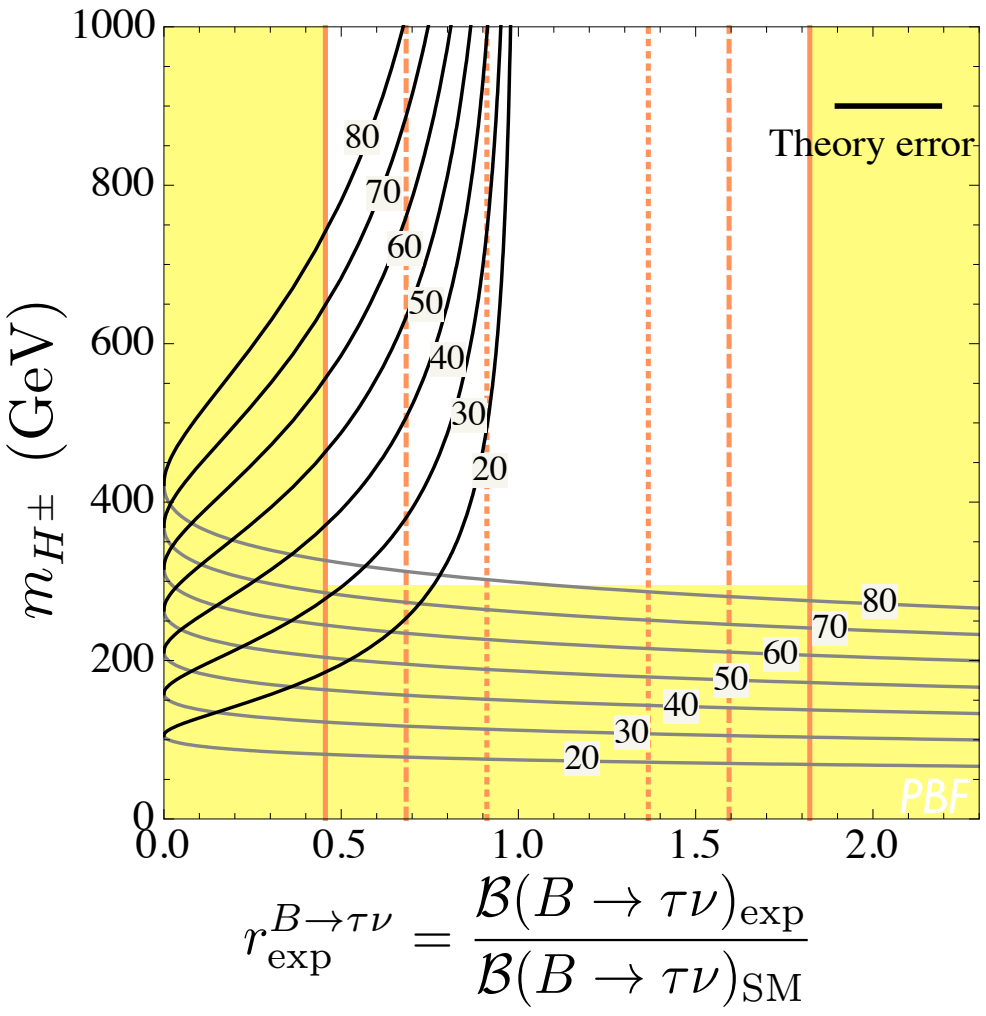

**Figure 25.2.3.** Constraints on the charged Higgs mass in the Type II 2HDM from the $B$ Factory measurements of $\mathcal{B}(B \to \tau\nu)$ and $\mathcal{B}(B \to X_s\gamma)$. The $x$-axis represents the present experimental value of $\mathcal{B}(B \to \tau\nu)$ normalized to the SM prediction (see text for details). The vertical yellow regions are excluded by the current world average experimental value, $\mathcal{B}(B \to \tau\nu)_{\rm exp} = (1.15 \pm 0.23) \times 10^{-6}$, at 95% C.L., while the $1\sigma$, $2\sigma$, $3\sigma$ errors on the same experimental value are denoted by the dotted, dashed, and solid lines, respectively. The horizontal yellow region is excluded by the $\mathcal{B}(B \to X_s\gamma)$ measurement (see Eq. 25.2.9). The grey and the black lines correspond to the predictions of the Type II 2HDM given in Eq. (25.2.12), respectively, with labels denoting the different values of $\tan\beta$.

#### 25.2.2.3 Minimal Flavor Violation

The MFV hypothesis states that flavor violation in NP comes from the same source as in the SM, the SM Yukawa couplings (Buras, 2003; Buras, Gambino, Gorbahn, Jager, and Silvestrini, 2001; Chivukula and Georgi, 1987; Ciuchini, Degrassi, Gambino, and Giudice, 1998a; D'Ambrosio, Giudice, Isidori, and Strumia, 2002; Hall and Randall, 1990). The reasoning behind this hypothesis on one hand is that this is the minimal amount of flavor breaking that needs to be present, since it is already seen in the SM. On the other hand, it also leads to relatively small deviations from the SM in flavor observables and is not excluded experimentally. In principle many different NP models can be of MFV type, but the best known example is low energy SUSY with gauge mediated SUSY breaking (that we discuss separately below). Another example are universal extra dimensions with universal boundary conditions. In both cases the only source of flavor violation are the SM Yukawa couplings.

An important benefit of the MFV hypothesis is that the effects of NP on flavor observables can be worked out without committing to a concrete model. This is done using a so-called spurion analysis (D'Ambrosio, Giudice, Isidori, and Strumia, 2002), which we review quickly. The SM Yukawa interactions for quarks are

$$\mathcal{L}_Y = \bar{Q}_L Y_D d_R H + \bar{Q}_L Y_U u_R H^c + h.c., \quad (25.2.15)$$

where the generation indices $i$ on the left-handed quarks $Q_i = (u_L, d_L)_i$, and on right-handed quarks $(u_R)_i, (d_R)_i$ were suppressed, while $H^c \equiv i\tau_2 H^*$, where $\tau_2$ is the $SU(2)$ generator, was used. If the Yukawa coupling matrices $Y_D, Y_U$ were zero, it would not be possible to distinguish the three generations of quarks. Thus, the theory would have a global symmetry, $G_F = SU(3)_Q \times SU(3)_{U_R} \times SU(3)_{D_R}$, since any of the quark fields $Q_L, u_R, d_R$ can be rotated independently. In other words, the global symmetry $G_F$ is explicitly broken by the fact that Yukawa couplings are not zero – quarks have nonzero masses in the SM. One can then use a formal trick and pretend that Yukawa coupling matrices do transform under $G_F$ as $Y_U \sim (3, \bar{3}, 1), Y_D \sim (3, 1, \bar{3})$ (The jargon used is that $Y_{U,D}$ are promoted to be spurions, where the name comes from the fact that these are now fictitious or spurious auxiliary fields.). All the interactions are then formally invariant under $G_F$. Under the MFV hypothesis also NP is assumed to be invariant under $G_F$.

Integrating out NP particles we obtain corrections to the effective weak Hamiltonian. However, since NP is formally invariant under $G_F$ we know its flavor structure. We just need to construct an effective weak Hamiltonian that is $G_F$ invariant, and all the breaking comes from $Y_U$ and $Y_D$. This then also fixes the allowed flavor breaking. The Yukawa coupling matrices $Y_U$ and $Y_D$ are not aligned and are diagonal in different bases for $Q_L$. The misalignment leads to flavor changing charged currents, $J_C^\mu = \bar{u}_L \gamma^\mu V d_L$, with $V$ being the same as the CKM matrix.

In this section we focus on a particular realization of MFV – the so-called constrained minimal flavor violation (cMFV) (Blanke, Buras, Guadagnoli, and Tarantino, 2006; Buras, Gambino, Gorbahn, Jager, and Silvestrini, 2001). The assumptions that underlie cMFV are (i) the SM fields are the only light degrees of freedom in the theory, (ii) there is only one light Higgs and (iii) the SM Yukawa couplings are the only sources of flavor violation. The NP effective Hamiltonian for $q_j \to q_i$ processes following from these assumptions thus has exactly the same CKM suppression and form of the effective operators as in the SM. This is sometimes taken to be the definition of cMFV (Blanke, Buras, Guadagnoli, and Tarantino, 2006; Buras, 2003; Buras, Gambino, Gorbahn, Jager, and Silvestrini, 2001). For instance, the effective Hamiltonian for the $\Delta F = 2$ transitions is

$$\mathcal{H}_{\rm eff}^{\rm NP} = \frac{C^{\rm NP}}{\Lambda_{\rm NP}^2} \left(V_{ti}^* V_{tj}\right)^2 Q_{ij}, \qquad (25.2.16)$$

where the Wilson coefficient $C^{\rm NP}$ is real, while $Q_{ij}$ are exactly the same operators as in the SM effective weak Hamiltonian. For $B_d^0 - \overline{B}_d^0$ this is $Q_{bd} = (\bar{b}_L \gamma^\mu d_L)^2$. Note that $C^{\rm NP}$ is universal – it is the same for $K^0 - \overline{K}^0$, $B_d^0 - \overline{B}_d^0$ and $B_s^0 - \overline{B}_s^0$ mixing. The relative sizes of the NP contributions are given entirely by the CKM matrix elements which are the same as in the SM. As a consequence Eq. (25.2.16) provides a very strong constraint on the scale of NP masses. Note that two-Higgs doublet models or MFV MSSM even with small $\tan\beta$ does not fit in the cMFV and sizable contributions from operators with non-SM chiral structures in addition to Eq. (25.2.16) are possible.

Because cMFV is a very constrained modification of the weak Hamiltonian, Eq. (25.2.16), one can experimentally distinguish it from the other BSM scenarios by looking at the correlations between observables in $K$ and $B$ decays. A sign of cMFV would be a deviation from the SM predictions that can be described without new *CP* violating phases and without enlarging the SM operator basis. For instance, in $B_d^0 - \overline{B}_d^0$ and $B_s^0 - \overline{B}_s^0$ mixing observables the discrepancy from the SM is possible only in the value of $\Delta m_d$ and $\Delta m_s$ and not in the mixing phases. Furthermore, the corrections normalized to the SM are universal, so that $h_s = h_d$ in Eq. (25.2.1), while the additional weak phases are zero, *i.e.* $\sigma_s = \sigma_d = 0$ or $\sigma_s = \sigma_d = \pi/2$. A discrepancy in $\phi_1$ determined from $B \to J/\psi K_S^0$ and the global Unitarity Triangle fit would rule out the cMFV framework. Similarly, a sizable $B_s^0 - \overline{B}_s^0$ mixing phase would rule out cMFV NP.

Since the CKM-like suppression is automatically encoded in the NP contributions to flavor transitions, no large flavor violations are expected from TeV scale NP. The results from the $B$ Factories are precise enough, however, that the energy scale probed is in the multi-TeV regime already. The most stringent constraints are coming from $b \to s\gamma$ and $b \to s l^+ l^-$ decays. Setting the NP Wilson coefficients $C_i^{\rm NP} = 1$ in Eq. (25.2.16), one has $\Lambda_{\rm NP} > 6.1$ TeV (Hurth, Isidori, Kamenik, and Mescia, 2009 and update by Hurth and Mahmoudi, 2012). If the NP states are exchanged at tree level, this would mean that they need to be heavier than about 6 TeV. If they only contribute through loops, they need to be heavier than about $\sim 6$ TeV$/4\pi = 0.5$ TeV for $\mathcal{O}(1)$ couplings. The precise value depends on the spin and charge of the exchanged particle.

Similarly, one has a bound $\Lambda_{\rm NP} > 5.9$ TeV for contributions to meson mixing (Bona et al., 2006, 2008). The NP modifications to $K^0 - \overline{K}^0$, $D^0 - \overline{D}^0$, $B_d^0 - \overline{B}_d^0$ and $B_s^0 - \overline{B}_s^0$ are rigidly related in cMFV. There is only one operator, $(\bar{q}_{Lj} \gamma^\mu q_{Lj})^2$, and the flavors of quarks fix the CKM suppressions. The bound on $\Lambda_{\rm NP}$ is predominantly due to the $\epsilon_K$ constraint. This means that no effects in $D^0 - \overline{D}^0$, $B_d^0 - \overline{B}_d^0$ and $B_s^0 - \overline{B}_s^0$ mixing are expected in cMFV with the present experimental precision. Similarly, since there are no new *CP* violating phases beyond the CKM phase, the $\phi_1$ phase determined from $b \to s$ penguin transitions $B \to \phi K_S^0$ is expected to be the same as obtained from $B \to J/\psi K_S^0$.

The MFV also relates the $b \to u\tau\nu$ and $b \to c\tau\nu$ charged current transitions. In the SM the amplitudes for these two transitions are proportional to the $V_{ub}$ and $V_{cb}$ matrix elements. The same is true for NP contributions. Normalized to the SM the deviations in both of these two transitions thus need to be the same, if NP is of MFV type. There are indications of deviations from the SM in $B^- \to \tau^- \nu$ and in $B \to D^{(*)} \tau\nu$ (Lees, 2012e), which proceed through $b \to u\tau\nu$ and $b \to c\tau\nu$ quark level transitions, respectively, (see Sections 17.10 and 25.1). The relative sizes of the two discrepancies, however, differ from the universal behavior predicted by the MFV. Therefore MFV is not preferred as an explanation of the anomalies (Fajfer, Kamenik, Nisandzic, and Zupan, 2012).

Finally, the minimal incarnation of MFV – cMFV – is a hypothesis about the flavor violation in the quark sector. Therefore there is no clear prediction about the size of the muon anomalous magnetic moment, $(g-2)_\mu$.

#### 25.2.2.4 Extensions of MFV

The phenomenologically most important extensions of cMFV hypothesis are: i) relaxing the assumption that the *CP* violation is only due to the CKM phase, allowing also for nonzero flavor diagonal weak phases, and ii) to allow for larger higher order terms in the spurion expansion. This General MFV (GMFV) hypothesis was formalized by Kagan, Perez, Volansky, and Zupan (2009), who identified the new small spurions: the off-diagonal CKM matrix elements and the masses of the first two generation quarks. For an earlier discussion of flavor diagonal *CP* phases within the MSSM see (Colangelo, Nikolidakis, and Smith, 2009), for a nonlinear realization of MFV see (Albrecht, Feldmann, and Mannel, 2010; Feldmann, Jung, and Mannel, 2009; Feldmann and Mannel, 2008)

The important difference between cMFV and GMFV is that in cMFV there are relations between $s \leftrightarrow d$, $b \leftrightarrow d$ and $b \leftrightarrow s$ transitions, in GMFV only $b \leftrightarrow d$ and $b \leftrightarrow s$ transitions are directly related. There are two classes of NP contributions. Class-1 operators *do not* contain light right-handed quarks, and class-2 operators do. An example of a class-1 operator is, for instance, $(\bar{q}_L\gamma_\mu b_L)^2$, and an example of a class-2 operator is $(\bar{q}_L b_R)(\bar{q}_R b_L)$. The distinction is phenomenologically important, since the Wilson coefficients of the class-2 operators are proportional to light-quark masses. In $B^0_{d,s} - \overline{B}^0_{d,s}$ mixing then the class-2 operators only contribute to $B^0_s - \overline{B}^0_s$ mixing (up to $m_d/m_s$ corrections) and would give $h_d \ll h_s$. Thus one would expect a deviation from the SM in the *CP* violating phase of the $B_s$ oscillation measurements but not in measurements of $\sin 2\phi_1$ from $B \to J/\psi K^0_S$. In contrast, class-1 operators contribute universally to both (relative to the SM) and would give $h_d = h_s$ and $\sigma_d = \sigma_s$ in Eq. (25.2.1). In many realistic models we would expect both class-1 and class-2 NP contributions, so the predictions would be somewhere in between: smaller effects in $B^0_d - \overline{B}^0_d$ mixing than in $B^0_s - \overline{B}^0_s$ mixing, yet still nonzero.

An example of such GMFV NP is MSSM with $U(2)^3$ flavor symmetry (Barbieri, Isidori, Jones-Perez, Lodone, and Straub, 2011), which is broken by the light-quark masses and the off-diagonal CKM elements. Gluino mediated amplitudes are the dominant non-standard effect in $\Delta F = 2$ observables and are of class-1. All class-2 contributions are suppressed. As a result the size of the correction is proportional to the CKM combination of the corresponding SM amplitude, a signature of class-1 MFV contributions. The proportionality coefficients are the same for the $B_d$ and $B_s$ systems, while it may be different in the kaon system – a signature of GMFV. Another GMFV characteristic is that new *CP* violating phases can only appear in the $B_d$ and $B_s$ systems. Since in the $U(2)^3$ symmetric MSSM they would come from class-1 contributions, the phase shifts would be universal. From the still allowed deviations in $S_{B_d\to\psi K^0_S}$ from $\sin 2\phi_1$ Barbieri, Isidori, Jones-Perez, Lodone, and Straub (2011) deduce that $0.05 \lesssim S_{B_s\to\psi\phi} \lesssim 0.2$.

The expectation for the $B \to X_s\gamma$ branching ratio are the same as in cMFV, discussed in Section 25.2.2.3. For weak scale NP particles with masses of a few 100 GeV we would thus expect a deviation in $B \to X_s\gamma$ already in the present measurements, even though the particles only enter in loops. On top of this, in GMFV there are additional *CP* violating phases, which can lead to an enhanced direct *CP* asymmetry in $B \to X_{d,s}\gamma$. The *CP* violating effects are also expected in $D^0 - \overline{D}^0$ mixing, with $\arg(M_{12}/\Gamma_{12}) \sim \mathcal{O}(5\%)$ for $\Lambda_{\rm GMFV} = 1$ TeV (Kagan, Perez, Volansky, and Zupan, 2009). At present, the experimental error is roughly twice as large as this expectation. There is also no clear prediction for the flavor diagonal observable $(g-2)_\mu$.

There are also other extensions of the MFV hypothesis, beside GMFV. At the practical level the GMFV is equivalent to the Next-to-Minimal Flavor Violation (NMFV) hypothesis, even though the original motivations were different. NMFV was put forward in (Agashe, Papucci, Perez, and Pirjol, 2005) by demanding that NP contributions only roughly obey the CKM hierarchy, and in particular can have $\mathcal{O}(1)$ new weak phases. The consequences of spurions that transform differently under $G_F$ than the SM Yukawa coupling matrices have been worked out by Feldmann and Mannel (2007). The MFV hypothesis has also been extended to the leptonic sector (MLFV) in (Cirigliano and Grinstein, 2006; Cirigliano, Grinstein, Isidori, and Wise, 2005). In MLFV the most sensitive FCNC probe in the leptonic sector is $\mu \to e\gamma$, while $\tau \to \mu\gamma$ could be suppressed below the sensitivity of future super flavor factories.

#### 25.2.2.5 MFV SUSY

Low energy supersymmetry (SUSY), where the superpartners have ∼TeV scale masses is one of the most popular solutions to the hierarchy problem. Since this model is perturbative one can make reliable predictions. This aids the popularity of SUSY among theorists. Already its minimal incarnation – the Minimal Supersymmetric Standard Model – has the salient features of gauge coupling unification and contains a viable dark matter candidate. The “Minimal” in the MSSM refers to the field content. Each SM particle obtains only one superpartner, and also the extension of the Higgs sector is minimal. However, the flavor structure need not be minimal. The parameters that describe the supersymmetry breaking, *e.g.*, the squark masses and trilinear couplings can in principle carry very different flavor structures from the one seen in the quark sector of the SM. In total there are 124 parameters in the MSSM, much more than the 19 parameters of the SM (Berger and Grossman, 2009; Dimopoulos and Sutter, 1995; Haber, 2001). Of these parameters, 110 are in the flavor sector: 30 masses, 39 real mixing angles and 41 phases. If all of the mixing angles and phases were $\mathcal{O}(1)$ this would lead to FCNCs that are orders of magnitude larger than the experimental bounds. The SUSY breaking does have to be non-generic and further assumptions about its structure are required in order to have an acceptable phenomenology. An attractive hypothesis is MFV, which we discussed in general terms in the previous subsections. The flavor breaking is assumed to arise only from the Yukawa interactions (in this case from the superpotential), while the SUSY breaking is flavor blind. This means that the squark masses can be written as

$$\begin{aligned}\tilde{m}^2_{q_L} =& \tilde{m}^2(a_1 1 + b_1 Y_U Y_U^\dagger + b_2 Y_D Y_D^\dagger \\ & b_3 Y_D Y_D^\dagger Y_U Y_U^\dagger + b_4 Y_U Y_U^\dagger Y_D Y_D^\dagger + \cdots),\end{aligned} \qquad (25.2.17)$$

and similarly for right-handed squarks. The coefficients $a_1, b_{1,2}$ are real from the hermiticity of the Hamiltonian,

while $b_3$ and $b_4$ can in general be complex and be sources of additional *CP* violating weak phases. For small $\tan\beta$ these terms are negligible since $Y_D$ is much smaller than $Y_U$. The values of the coefficients are fixed by the model of the SUSY breaking. They can be zero at some high scale $M$, but are then generated due to the renormalization group running effect from this high scale to the low scale $b_i \sim (1/4\pi^2)\log(M^2/\tilde{m}^2)$. In gauge-mediated SUSY breaking the scale $M$ would be given by the masses of the messengers particles between the SUSY breaking sector and low energy sector.

Since MFV SUSY is an example of the MFV theory our general discussion in the previous two subsections applies. The superpartners can be integrated out and matched onto the effective weak Hamiltonian. Because MFV MSSM is a concrete model the predictions can in fact be more precise (for a review see, *e.g.*, Isidori and Straub, 2012). For $\tan\beta \gg 1$ and/or if the $\mu$ parameter, the coupling of bilinear term of the MSSM Higgs sector, is large enough, the resulting low energy operator basis does not contain only the SM operators. This means that in this limit it is not a cMFV model. In the large $\tan\beta$ limit the most sensitive observables are the branching fractions for $B_s \to \mu^+\mu^-$ and $B \to \tau\nu$. For most of the parameter space $\mathcal{B}(B \to \tau\nu)$ is reduced by the charged Higgs correction compared to the SM.

An interesting prediction of MFV MSSM is that the contributions to $\Delta m_{d,s}$ are always positive and increase the oscillation frequency above the SM (Altmannshofer, Buras, and Guadagnoli, 2007). However, the NP effects in $S_{J/\psi K^0_S}$ and $\Delta m_d/\Delta m_s$ are very small and thus do not disrupt the Unitarity Triangle of the SM. There are also only small effects in $B_s \to J/\psi\phi$ expected. Similarly, the contributions to $D^0 - \overline{D}^0$ mixing are small (Altmannshofer, Buras, Gori, Paradisi, and Straub, 2010). If the flavor blind phases are nonzero, then EDMs and $\mathcal{B}(b \to s\gamma)$ are the strongest constraints (Altmannshofer, Buras, and Paradisi, 2008). The dominant NP source in $S_{\phi K^0_S}$ arises from the chromomagnetic operator $C_{8g}^{\rm NP}$, and the effects in $S_{\phi K^0_S}$ are expected to be significantly larger than in $S_{\eta' K^0_S}$ (both in the same direction), with even $\mathcal{O}(1)$ corrections not too hard to achieve. The effect is also strongly correlated with the size of the direct *CP* asymmetry $A_{CP}(b \to s\gamma)$, with an effect of up to 0.05 typical. There is also a natural explanation of the small deviation from the SM observed in the muon-magnetic moment $(g-2)_\mu$ as long as sleptons are not much heavier than squarks (see *e.g.* Jegerlehner and Nyffeler, 2009).

#### 25.2.2.6 non-MFV SUSY

In investigations of SUSY models, strong assumptions (such as the MFV ansatz discussed in the previous section) are often imposed in order to reduce the large number of parameters introduced by the unknown SUSY breaking mechanism. A common motivation behind the assumptions is to avoid an unwanted excess of *CP* violation and FCNC. However, by working within those assumptions, one could potentially also miss a signal of SUSY particles. In this section we follow a more model independent approach – the mass insertion approximation (MIA) (Hall, Kostelecký, and Raby, 1986). In this approach the flavor off-diagonal part of the squark mass matrix in the Super-CKM basis is parameterized by the mass insertion parameter, $(\delta^q_{AB})_{ij}$, where $A, B$ denote the chirality $(L, R)$ and $q$ indicates the $(u, d)$ type. Assuming that the off-diagonal elements are smaller than the diagonal ones, the sfermion propagator can be expanded as

$$\begin{aligned}\langle \tilde{q}_{Ai}\tilde{q}^*_{Bj}\rangle &= i(k^2\mathbf{1} - \tilde{m}^2\mathbf{1} - \tilde{m}^2\delta^q_{AB})^{-1}_{ij} \\ &\simeq \frac{i\delta_{ij}}{k^2 - \tilde{m}^2} + \frac{i\tilde{m}^2(\delta^q_{AB})_{ij}}{(k^2-\tilde{m}^2)^2} + \cdots ,\end{aligned} \quad (25.2.18)$$

where $\mathbf{1}$ is the unit matrix and $\tilde{m}$ is the averaged squark mass, used also to normalize the off-diagonal mass matrix elements, so that $(\delta^q_{AB})_{ij}$ are dimensionless. In a general analysis all the mass insertion parameters $(\delta^q_{AB})_{ij}$ should be taken into account and are then only constrained from various flavor experiments.

The $B$ Factory observables are particularly sensitive to the down type mass insertion elements with $ij = 13$ ($b \to d$ transitions) and $ij = 23$ ($b \to s$ transitions). We also note that the source of flavor violation in this framework comes from the loop diagrams with gluinos and neutralinos whereas the former is dominant due to the large strong coupling constant.

For a general flavor structure significant excesses in FCNC and *CP* violation are possible for various $B$ Factory observables. The non-observation of large deviations from the SM therefore stringently constrains the mass insertion parameters. As an example let us take the $(\delta^d_{AB})_{13}$ mass insertions. These are constrained by the $\Delta m_d$ and $\sin 2\phi_1$ measurements. The $m_{\tilde{g}} \simeq m_{\tilde{q}} = 500$ GeV gluino contribution to $\Delta m_d$ normalized to the SM is (Gabbiani, Gabrielli, Masiero, and Silvestrini, 1996; Gabrielli and Khalil, 2003)

$$\begin{aligned}\frac{M_{12}^{\rm SUSY}}{M_{12}^{\rm SM}} \simeq \frac{1}{(V_{tb}V_{td}^*)^2}\Big\{ &4.0\times 10^{-3}\left[(\delta^d_{23})^2_{LL} + (\delta^d_{23})^2_{RR}\right] \\ &+ 8.1\times 10^{-2}\left[(\delta^d_{23})^2_{LR} + (\delta^d_{23})^2_{RL}\right] \\ &- 1.3\times 10^{-1}\left[(\delta^d_{23})_{LR}(\delta^d_{23})_{RL}\right] \\ &- 5.0\times 10^{-1}\left[(\delta^d_{23})_{LL}(\delta^d_{23})_{RR}\right]\Big\}. \end{aligned} \quad (25.2.19)$$

Note that in terms of observables one has

$$\Delta m_d = 2|M_{12}^{\rm SM} + M_{12}^{\rm SUSY}|, \quad (25.2.20)$$

$$S_{b\to c\bar{c}s} = -{\rm Im}\left(\frac{M_{12}^{*\rm SM} + M_{12}^{*\rm SUSY}}{M_{12}^{\rm SM} + M_{12}^{\rm SUSY}}\right)^{1/2}. \quad (25.2.21)$$

The resulting constraints following from the experimental world averages $\Delta m_d = (0.510 \pm 0.004)\ {\rm ps}^{-1}$ and $S_{b\to c\bar{c}s} = 0.677 \pm 0.020$ (see Section 17.5.2 and Section 17.6 for the experimental extraction of these values) are shown in Fig. 25.2.4 (top panels). We use $V_{tb}V_{td}^* = (8.7\pm 0.8)\times 10^{-3} e^{i(0.41\pm 0.06)}$, which is obtained in a similar way as in Section 25.2.2.1 by using the unitarity condition with $V_{cb}V_{cd}^*$ and $V_{ub}V_{ud}^*$ extracted from the tree level processes. We assume that only one of four mass insertions is nonzero, and show the constraints on $LL$ and $LR$ mass insertions as representative examples. The $RR$ and $RL$ mass insertion have similar constraints, respectively. The dashed lines show the constraints from $\Delta m_d$ measurements, while the solid lines show the constraints from the latest $\sin 2\phi_1$ measurement from the $B$ Factories. One finds that typically the chirality preserving mass insertions, $(\delta_{13}^d)_{LL/RR} \lesssim 10^{-1}$, are less constrained than the chirality flipping ones, $(\delta_{13}^d)_{LR/RL} \lesssim 10^{-2}$, (for both their real and imaginary parts).

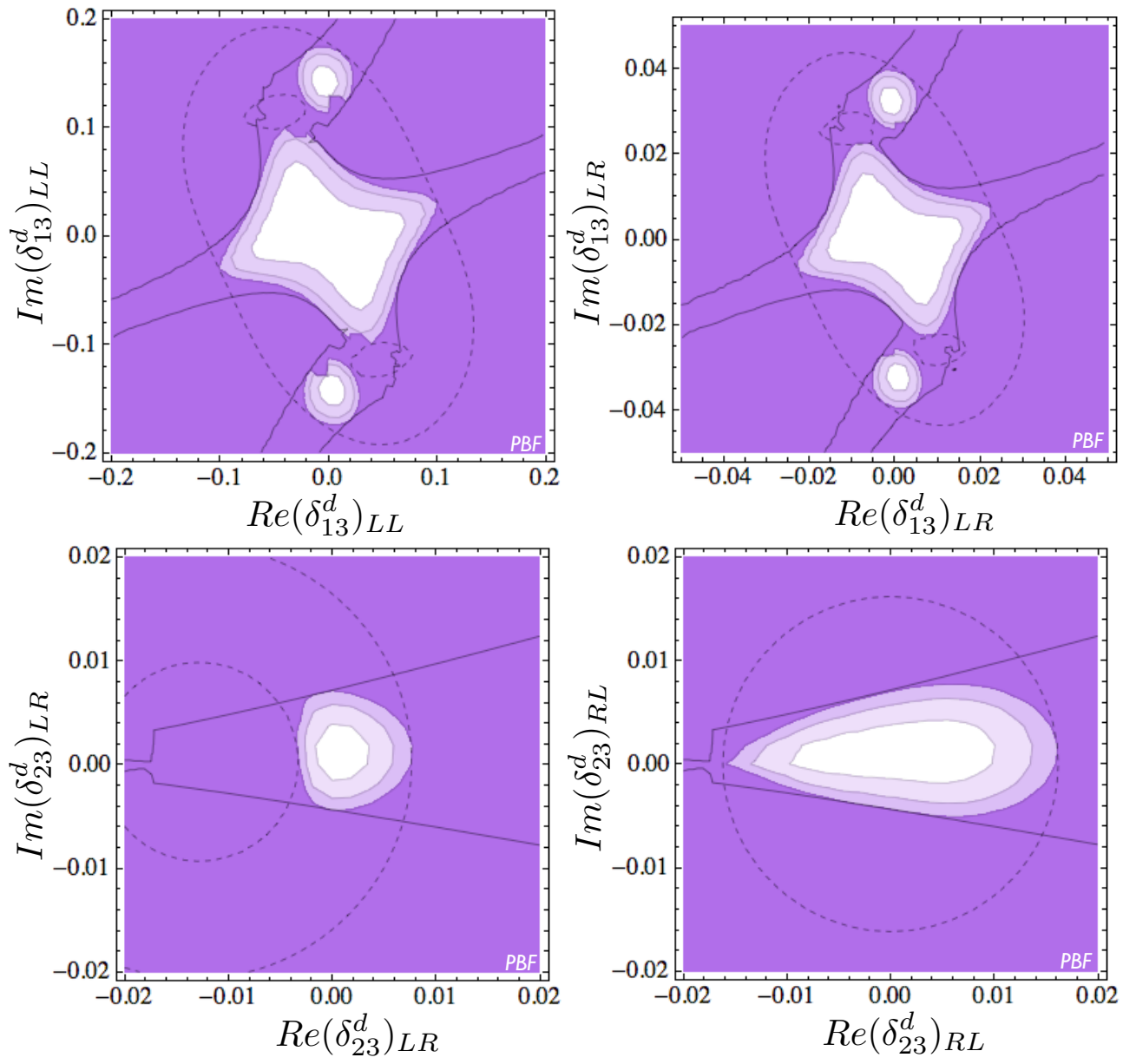

**Figure 25.2.4.** Top: constraints on the down type $ij = 13$ mass insertions from $B_d^0 - \overline{B}{}_d^0$ oscillation measurements at the $B$ Factories. The dashed lines represent $3\sigma$ bounds from $\Delta m_d$ and the solid lines from $\sin 2\phi_1$. Bottom: constraints on the down type $ij = 23$ mass insertions from the $\mathcal{B}(B \to X_s\gamma)$ measurement (dashed lines) and the time-dependent $CP$ asymmetry in the penguin-dominated $B \to \phi K_S^0$ transition (solid line). The colored regions are excluded at 1, 2 and 3 $\sigma$ combining the corresponding two measurements.

In the lower panels of Fig. 25.2.4 we show the impact that the $B$ Factory measurements of the $B \to X_s\gamma$ branching ratio have on the chirality flipping ($AB = LR/RL$) mass insertions for $b \to s$ transitions. Again, the constraints are at the $\mathcal{O}(0.01)$ level. The $(\delta_{AB}^d)_{23}$ mass insertions are also constrained by the $B_s^0 - \overline{B}{}_s^0$ oscillation measurement. However, as we show below, for the case of $AB = LR/RL$ the chiral enhancement makes the $B \to X_s\gamma$ constraints much stronger than the ones from $B_s^0 - \overline{B}{}_s^0$ oscillations.

The gluino contribution to the $b \to s$ ($\Delta B = 1$) transition is described by the following effective Hamiltonian (Khalil and Kou, 2003)

$$\mathcal{H}_{\rm eff}^{\Delta B=1} = -\frac{G_F}{\sqrt{2}} V_{tb}V_{ts}^* \left[ \sum_{i=3}^{6} C_i O_i \right. \tag{25.2.22}$$

$$\left. + C_\gamma O_\gamma + C_g O_g + \sum_{i=3}^{6} \tilde{C}_i \tilde{O}_i + \tilde{C}_\gamma \tilde{O}_\gamma + \tilde{C}_g \tilde{O}_g \right],$$

with

$$O_{3/4} = \left(\bar{s}_{\alpha/\alpha} b_{\alpha/\beta}\right)_{V-A} \left(\bar{s}_{\beta/\beta} s_{\beta/\alpha}\right)_{V-A}, \tag{25.2.23}$$

$$O_{5/6} = \left(\bar{s}_{\alpha/\alpha} b_{\alpha/\beta}\right)_{V-A} \left(\bar{s}_{\beta/\beta} s_{\beta/\alpha}\right)_{V+A}, \tag{25.2.24}$$

$$O_\gamma = \frac{-1}{3}\frac{e}{4\pi^2} m_b \left(\bar{s}_\alpha \sigma^{\mu\nu} P_R b_\alpha\right) F_{\mu\nu}, \tag{25.2.25}$$

$$O_g = \frac{g_s}{4\pi^2} m_b \left(\bar{s}_\alpha \sigma^{\mu\nu} P_R T_{\alpha\beta}^A b_\beta\right) G_{\mu\nu}^A, \tag{25.2.26}$$

where the Dirac structure of four-fermion operators is $\left(\bar{s}b\right)_{V-A}\left(\bar{s}s\right)_{V\mp A} = 4\left(\bar{s}\gamma^\mu P_L b\right)\left(\bar{s}\gamma_\mu P_R s\right)$, and the summation over the color indices $\alpha, \beta$ is understood. The terms with a tilde are obtained from $C_{i,g}$ and $O_{i,g}$ through a $L \leftrightarrow R$ replacement. The $B \to X_s\gamma$ branching fraction receives SUSY contributions mainly from $O_\gamma$, $\tilde{O}_\gamma$, $\tilde{O}_g$ and $O_g$. These dimension 5 operators are chirality flipping – the external $b$ and $s$ quark fields have different chiralities. In the SM, the $W$ boson couples only to the left-handed fermions ($V - A$ current) so that the chirality flip comes from the quark mass insertion on the external quark fields. The $O_\gamma$ and $O_g$ operators are thus suppressed in the SM by one power of $m_b/m_W$. In non-MFV SUSY, on the other hand, there are additional interactions that can induce the chirality flip inside the loop. The $m_b$ factor is then replaced by the internal heavy particle masses and the chirality flipping coupling (in MIA these will be $(\delta_{LR}^q)_{ij}$). The resulting Wilson coefficients from the gluino-squark loop are (Gabbiani, Gabrielli, Masiero, and Silvestrini, 1996):

$$C_\gamma^{\tilde{g}}(M_S) = -\frac{\sqrt{2}\alpha_s\pi}{2G_F V_{tb}V_{ts}^* m_{\tilde{q}}^2} \tag{25.2.27}$$
$$\times \left\{ (\delta_{LR}^d)_{23} \frac{m_{\tilde{g}}}{m_b} \frac{8}{3} M_1(x) + (\delta_{LL}^d)_{23} \frac{8}{3} M_3(x) \right\}.$$

$$C_g^{\tilde{g}}(M_S) = -\frac{\sqrt{2}\alpha_s\pi}{2G_F V_{tb}V_{ts}^* m_{\tilde{q}}^2} \tag{25.2.28}$$
$$\times \left\{ (\delta_{LR}^d)_{23} \frac{m_{\tilde{g}}}{m_b} \left[ \frac{1}{3} M_1(x) + 3M_2(x) \right] \right.$$
$$\left. + (\delta_{LL}^d)_{23} \left[ \frac{1}{3} M_3(x) + 3M_4(x) \right] \right\}.$$

The $\tilde{C}_\gamma$ and $\tilde{C}_g$ coefficients are obtained by making the $L \leftrightarrow R$ replacement. One can see that indeed the terms with the chirality-flipping $LR/RL$ mass insertions are enhanced by the $m_{\tilde{g}}/m_b$ factor. This term could potentially induce large contributions to $b \to s$ penguin transition processes. On the bottom panels of Fig 25.2.4, we present the constraints on the LR and RL mass insertions (dashed line) from the $\mathcal{B}(B \to X_s\gamma)$ measurement. The difference between the two cases comes from the fact that the LR contribution adds coherently to the SM at the amplitude level while the RL contribution does not interfere with the SM and adds in the amplitudes squared.

The corrections to the chromo-magnetic operators $O_g$ and $\tilde{O}_g$ can also have significant effects on the hadronic $B$ decays. An important additional constraint on $(\delta^d_{LR/RL})_{23}$ is thus obtained from the penguin-dominated $B \to \phi K^0_S$ decay. In the SM, the time dependent $CP$ asymmetry for this channel is the same as the one obtained from the decay $B \to c\bar{c}K^0_S$ (where $c\bar{c}$ represents any charmonium state such as $J/\psi$). If $S_{\phi K^0_S} \neq S_{c\bar{c}K^0_S}$ is found, this would be an indication of new physics. The deviation needs to be larger than the theoretical errors on the difference. A list of estimates for the theory errors on $S_{\phi K^0_S} - S_{c\bar{c}K^0_S}$ and for other $b \to s$ penguin transition dominated channels, *e.g.*, $S_{\eta' K^0_S}$, etc., can be found in (Zupan, 2007). Historically there was an indication for a deviation from the SM, however by now, the deviation has diminished and the current world averages (see Section 17.6.6 for details of experimental extraction of these values),

$$S_{\phi K^0_S} - S_{c\bar{c}K^0_S} = \quad 0.06 \pm 0.12, \qquad (25.2.29)$$

$$S_{\eta' K^0_S} - S_{c\bar{c}K^0_S} = -0.09 \pm 0.07, \qquad (25.2.30)$$

are consistent with the SM, though the experimental errors are still relatively large. The constraints on the LR and RL mass insertions are nontrivial and are comparable to the ones following from $B \to X_s\gamma$, see lower panels in Fig. 25.2.4. The regions excluded by combining the measurements of the $B \to X_s\gamma$ branching fraction and $S_{\phi K^0_S}$ are also shown in Fig. 25.2.4 as colored regions.

Further constraints on the corresponding mass insertions can be obtained from the other loop-induced observables. For instance, $D^0-\overline{D}^0$ mixing constrains the up-type mass insertions $(\delta^u_{AB})_{12}$, while $(g-2)_\mu$ constrains the slepton mass insertions $(\delta^l_{AB})_{22}$, see, *e.g.*, (Chang, Chang, Keung, Sinha, and Sinha, 2002; Chankowski, Lebedev, and Pokorski, 2005; Gabbiani, Gabrielli, Masiero, and Silvestrini, 1996; Hisano and Tobe, 2001).

### 25.2.2.7 SUSY Alignment models

The measurement of $D^0 - \overline{D}^0$ mixing at Belle and *BABAR* had important implications for the flavor structure of the MSSM (Ciuchini et al., 2007; Nir, 2007a,b). The squark contributions to $\Delta F = 2$ processes involving the first two generations, *i.e.* to $K^0-\overline{K}^0$ and to $D^0-\overline{D}^0$ mixing, have to be sufficiently suppressed in order not to generate contributions to the mixing larger than what is experimentally observed. The contributions arising from the box diagram with a gluino and the first two generation squark doublets $\tilde{Q}_{L1,2}$ are given by, see *e.g.* (Raz, 2002),

$$M^D_{12} \propto \frac{1}{m^2_{\tilde{u}}}\frac{(\Delta m^2_{\tilde{u}})^2}{m^4_{\tilde{u}}}(K^u_{21}K^{u*}_{11})^2, \qquad (25.2.31)$$

$$M^K_{12} \propto \frac{1}{m^2_{\tilde{d}}}\frac{(\Delta m^2_{\tilde{d}})^2}{m^4_{\tilde{d}}}(K^d_{21*}K^d_{11})^2. \qquad (25.2.32)$$

Here $m_{\tilde{u},\tilde{d}}$ are the averaged squark masses of the first two up and down generation squarks, $\Delta m^2_{\tilde{u},\tilde{d}}$ are the corresponding mass squared differences, while $K^{u(d)}$ is the mixing matrix for the gluino coupling to left-handed up (down) quarks and the squark partners. The proportionality coefficients depend on the $D$ and $K$ decay constants, the bag parameters and a function of $m_{\tilde{q}}/m_{\tilde{u},\tilde{d}}$.

There are three generic ways how these contributions can be suppressed. The first possibility is that the first two generation squarks are *heavy*, $m_{\tilde{q}} \gg 1$ TeV. Since they contribute to dimension 6 operators, these contributions scale as $\propto 1/m^2_{\tilde{q}}$ and become irrelevant when squarks are much heavier than the weak scale. The second possibility is that the squarks are *degenerate*, *i.e.* that the mass splitting between the first two generations is small, $\Delta m^2_{\tilde{q}} \ll m^2_{\tilde{q}}$. If they were exactly degenerate, one would be free to choose the flavor basis for squarks anyway one wants – in particular to coincide with the mass basis of first two generation quarks. The flavor breaking effects thus need to be proportional to the splitting between the squarks. Finally, the squarks could be *aligned* with the quarks, so that the mass eigenstate basis for squarks almost coincides with the mass eigenstate basis of quarks and thus $K^{d,u}_{21} \ll 1$.

Alignment naturally arises in Froggatt-Nielsen type flavor models of squark masses (Leurer, Nir, and Seiberg, 1994; Nir and Seiberg, 1993). For left-handed squarks there is also a relation between the matrices that diagonalize up and down squarks. Up to corrections of $m^2_c/m^2_{\tilde{q}} \sim \mathcal{O}(10^{-5})$ one has

$$K^u K^{d\dagger} = V_{\rm CKM}. \qquad (25.2.33)$$

For the mixing between the first two generations this means that

$$K^u_{21} - K^d_{21} \simeq \sin\theta_C = 0.23. \qquad (25.2.34)$$

Therefore the 21 entries in the quark-squark-gluino coupling matrices cannot be smaller than the mixing between the first two generations in the SM. If squarks have masses of around 1 TeV and non-degenerate this is at odds with either $K^0 - \overline{K}^0$ mixing and $D^0 - \overline{D}^0$ mixing. Barring cancellations there are two possibilities. The first one is that squarks are quasi-degenerate. The level of degeneracy required is $\Delta m_{\tilde{q}}/m_{\tilde{q}} \lesssim 0.12$ (Gedalia, Grossman, Nir, and Perez, 2009; Nir, 2007b). The other option is that squarks are heavy. If one sets $K^u_{21} = 0.23$ and $K^d_{12} \simeq 0$ as in the original alignment model by Nir and Seiberg (1993), then $m_{\tilde{q}} \gtrsim 2$ TeV, and much heavier, if $K^d_{12}$ is nonzero. We thus, can state model-independently that barring cancellations, if the squarks are light enough to be observed at the LHC, then they must be quasi-degenerate. Note that, in order to reach this conclusion the experimental information on $D^0 - \overline{D}^0$ mixing parameters provided by the $B$ Factories (in particular that they are small, $x, y \sim \mathcal{O}(1\%)$), was essential. In the original alignment model the gluino-squark loop induced FCNCs are absent in the down quark sector. Therefore among the observables in Table 25.2.2 the only place we would expect deviations is $D^0 - \overline{D}^0$ mixing.

#### 25.2.2.8 Randall-Sundrum models of flavor

The Randall-Sundrum (RS) models of flavor have an ambitious goal. They strive to simultaneously solve the hierarchy problem, the flavor problem of new physics and explain the flavor structure in the SM. The hierarchy problem refers to the Planck scale $\sim 10^{19}$ GeV being so much bigger than the electroweak scale $\sim$ 1 TeV. The flavor problem of new physics is that new physics at 1 TeV (which solves the hierarchy problem) must not enhance FCNCs above the SM level. At the same time the RS models of flavor also provide an explanation for the flavor structure in the Standard Model – the origin of the hierarchy of masses and mixing – through localization of quark fields in the 5$^{\text{th}}$ dimension (Gherghetta and Pomarol, 2000; Grossman and Neubert, 2000).

The RS models are extra dimensional models where a slice of 5-dimensional (5D) space-time (bulk) is truncated by flat 4-dimensional boundaries – the two 4-dimensional (4D) branes. The Planck brane is on the UV side of the bulk, while the TeV brane is on the IR side of the bulk. This setup results in a warped metric for the bulk (Randall and Sundrum, 1999)

$$ds^2 = e^{-2kr_c|\phi|}\eta_{\mu\nu}dx^\mu dx^\nu - r_c^2 d\phi^2, \qquad (25.2.35)$$

where $k$ is the 5D curvature scale, $r_c$ the compactification radius, and $\phi \in [-\pi, \pi]$ the coordinate along the 5$^{\text{th}}$ dimension. The hierarchy problem is solved by the presence of the warp factor $e^{-2kr_c|\phi|}$, which suppresses the fundamental Planck scale $\sim 10^{19}$ GeV down to the weak scale $\sim$ 1 TeV for $kr_c \approx 12$.

In the initial RS models all the SM fields were assumed to be localized on the IR brane and only gravity propagated in the bulk (Davoudiasl, Hewett, and Rizzo, 2000). This immediately lead to phenomenological problems, because the cut-off of the effective 4D theory is also warped down to TeV scale. A viable model is obtained, if the fermions and gauge bosons are allowed to propagate in the bulk, while only the Higgs boson is localized on the IR brane. The 5D profiles of the SM fermions (the zero modes) have an exponential form

$$f_i^{(0)} \sim e^{(1/2-c_i)kr_c\phi}. \qquad (25.2.36)$$

The light fermions have bulk mass parameters $c_i > 1/2$, and are thus localized near the Planck brane. This has two beneficial consequences. On one hand it suppresses FCNCs due to virtual exchanges of Kaluza-Klein (KK) states – the excitations of the SM fields that arise because of the compact extra dimension. The suppression is due to different localizations of zero mode fermions, that peak near the UV brane, and KK modes that peak near the IR brane. Therefore there is only a small overlap between the two (Gherghetta and Pomarol, 2000). Since the Higgs boson is localized on the IR brane this also explains the smallness of the masses of UV localized light fermions – it is due to the fact that they have only a small overlap with the Higgs wave function. The 4D Yukawa coupling matrices are given by

$$(Y_{u,d}^{4D})_{ij} = (Y_{u,d}^{5D})_{ij} f_{Q_i} f_{u_j,d_j}, \qquad (25.2.37)$$

with $f_{Q_i}$ ($f_{u_j,d_j}$) the values of wave functions for the left-handed (right-handed) fermions at the IR brane, where the Higgs is situated, cf. Eq. (25.2.36). The hierarchy of the SM quark masses is naturally obtained for $\mathcal{O}(1)$ values of 5D Yukawa parameters $(Y_{u,d}^{5D})_{ij}$ and values of $c_i$, where an order unity change in $c_i$ results in an exponential change in the value of quark masses (Gherghetta and Pomarol, 2000; Grossman and Neubert, 2000; Huber, 2003).

The zero mode (*i.e.* SM) gluons and photons have flat wave functions in the $\phi$ direction due to unbroken $SU(3)_C \times U(1)_{\text{EM}}$ gauge invariance. The $Z$ and $W^\pm$ wave functions, on the other hand, are distorted near the IR brane, since electroweak symmetry is spontaneously broken. This generates tree level FCNCs mediated by the $Z$ and the KK gauge bosons. This is in contrast to the SM, where due to the GIM mechanism the FCNCs arise only at the one loop level. Still, the FCNCs in the RS models are suppressed despite the fact that they arise at tree level. The reason is that the light quarks are localized at the UV brane, while both the KK modes and the distortion of the $Z$ shape function are all localized near the IR brane. The FCNCs are then suppressed by the small wave function overlaps. This so-called "RS-GIM mechanism" suffices to avoid disastrously large FCNCs, but with some tension in the kaon sector (Agashe, Perez, and Soni, 2004).

The above setup has several phenomenological implications. Because the SM gauge bosons and fermions mix with the corresponding KK modes, the CKM matrix is no longer unitary. The corrections to CKM unitarity are of order $\mathcal{O}(v^2/m_{\text{KK}}^2)$. For KK masses in the few TeV range these corrections are thus very small, at percent level or smaller.

Easier to observe are effects due to additional flavor and $CP$ violating sources in the model. The 5D mass matrices $C^Q$, $C^u$, $C^d$ have 18 new mixing angles and 9 complex phases beyond the SM Yukawa couplings (Agashe, Perez, and Soni, 2005). The new $CP$ violating weak phases can affect low energy $CP$ violating observables. For a low KK mass scale, $m_{\text{KK}} \lesssim 3$ TeV, the tree level KK gluon exchanges lead to contributions to $B_d^0 - \overline{B}_d^0$ and $B_s^0 - \overline{B}_s^0$ mixing that are roughly of the same size as the SM contributions, but with arbitrary weak phases (Agashe, Perez, and Soni, 2005). The possibility of such large contributions to the $B_d^0 - \overline{B}_d^0$ mixing was excluded by the $B$ Factories, while large phases in $B_s^0 - \overline{B}_s^0$ mixing are severely constrained by the LHCb results. For $K^0 - \overline{K}^0$ mixing by far the most important is the generation of operators with the left-right (LR) chiral structure from KK gluon exchanges. These are enhanced by the renormalization group running and by chirally enhanced matrix elements. They give a contribution that is a factor 140 larger than the SM LL operator, if the KK scale is around 3 TeV (Bauer, Casagrande, Haisch, and Neubert, 2010; Blanke, Buras, Duling, Gori, and Weiler, 2009; Casagrande, Goertz, Haisch, Neubert, and Pfoh, 2008; Csaki, Falkowski, and Weiler, 2008).

The KK masses below 3 TeV would be allowed by the RS models with additional custodial symmetry which can satisfy the electroweak precision data constraints. (Agashe, Contino, Da Rold, and Pomarol, 2006; Agashe, Delgado,

May, and Sundrum, 2003; Csaki, Grojean, Pilo, and Terning, 2004). In contrast the data on $\epsilon_K$ imply a generic lower bound on $M_{\rm KK}$ of roughly 20 TeV for anarchic 5D masses $c_i$ with $\mathcal{O}(1)$ coefficients (Csaki, Falkowski, and Weiler, 2008). With modest fine tuning KK mass scales of 2 to 3 TeV are still allowed (Blanke, Buras, Duling, Gori, and Weiler, 2009), or if additional flavor symmetries are introduced in the 5D Yukawa coupling matrices (Cacciapaglia et al., 2008; Csaki, Falkowski, and Weiler, 2009; Csaki, Perez, Surujon, and Weiler, 2010; Fitzpatrick, Perez, and Randall, 2007). Constraints from $K^0-\overline{K}^0$ mixing are also reduced in "soft wall" RS models, where the IR brane is removed and the Higgs is free to propagate in the bulk (Archer, Huber, and Jager, 2011).

In RS models there are also potential effects in $\Delta F = 1$ modes (Bauer, Casagrande, Haisch, and Neubert, 2010; Blanke, Buras, Duling, Gori, and Weiler, 2009). Rare $B_{s,d} \to \mu^+\mu^-$ and $B_{s,d} \to X_{s,d}\nu\bar{\nu}$ are not affected much and remain SM like, with the corrections to the branching fractions below 15%. Larger effects in $\mathcal{B}(B \to X_s\gamma)$ are in principle possible from contributions to dipole operators for a somewhat tuned parameter set, which however, are excluded by the data from $B$ Factories. The contributions to EDMs are also generically large, about a factor 20 above the present experimental bounds, leading to additional constraints on RS parameter space. The branching fractions $\mathcal{B}(K \to \pi^0\nu\bar{\nu})$ and $\mathcal{B}(K^+ \to \pi^+\nu\bar{\nu})$ can be enhanced by a factor $\sim 2$ compared to the SM, however simultaneous enhancements of $S_{B_s\to J/\psi\,\phi}$ and $K \to \pi\nu\bar{\nu}$ branching ratios are not likely. The corrections to the CKM matrix are dominated by the effects due to the mixing with the KK gauge bosons. The deviations from the SM are small: even for the largest effects for the CKM elements involving third generation quarks, they can be up to $1-2\%$. Modest effects at the order of $\mathcal{O}(5\%)$ are expected on $S_{B\to\phi K^0_S}$ and other $b \to s$ penguin transitions (Agashe, Perez, and Soni, 2005; Bauer, Casagrande, Haisch, and Neubert, 2010). The constraints from $Z \to b\bar{b}$, $\epsilon_K$ and $B^0-\overline{B}^0$ mixing also suffice to predict the mass difference in $D^0-\overline{D}^0$ mixing to be not much larger then what is observed, however a significant spread in the $CP$ violating mixing phase is possible, with the bulk of the predictions on the phase $|\arg(M^D_{12}/\Gamma^D_{12})| \lesssim 90°$. The $B$ Factories constraints from $D^0-\overline{D}^0$ mixing are thus nontrivial and exclude a significant part of the parameter space (Bauer, Casagrande, Haisch, and Neubert, 2010). The corrections to $B \to \tau\nu$ can be at most 1% (Bauer, Casagrande, Haisch, and Neubert, 2010). The first complete calculation of one loop contributions to $(g-2)_\mu$ was completed recently by Beneke, Dey, and Rohrwild (2012), and the result is about an order of magnitude below the present experimental error.

#### 25.2.2.9 Little Higgs models

The Little Higgs models present another direction for a potential solution to the hierarchy problem (for a pedagogical review see Schmaltz and Tucker-Smith, 2005). We focus on a particular model – the Littlest Higgs Model with $T$-parity (Cheng and Low, 2003, 2004; Low, 2004) – whose flavor structure was studied in detail by the Munich group (Bigi, Blanke, Buras, and Recksiegel, 2009; Blanke, Buras, Duling, Poschenrieder, and Tarantino, 2007; Blanke, Buras, Duling, Recksiegel, and Tarantino, 2010; Blanke et al., 2007, 2006; Blanke, Buras, Recksiegel, and Tarantino, 2008; Blanke, Buras, Recksiegel, Tarantino, and Uhlig, 2007a,b) and supplemented by Goto, Okada, and Yamamoto (2009) and by del Aguila, Illana, and Jenkins (2009).

The Littlest Higgs Model with $T$-parity (LHT) has a relatively small number of new parameters that describe the flavor sector – 10 in the quark sector (Blanke, Buras, Duling, Recksiegel, and Tarantino, 2010). The relevant operators in the effective weak Hamiltonian that are generated by integrating out NP contributions are the same as in the SM. The fact that the model has $T$-parity means that the NP scale $f$ can be quite low, $f = 500$ GeV. This is quite different from for instance the RS models that were described in the previous section (where even in the models with custodial protection the KK scale is in the range of $2-3$ TeV). Another interesting difference with the RS models is that the constraints from $B \to X_s\gamma$ and neutron electric dipole moment are not very strong and are easily satisfied.

In the LHT there are new flavor interactions beyond the CKM matrix $V_{\rm CKM}$. These new interactions involve the heavy gauge bosons $W^\pm_H, Z_H, A_H$ which get emitted from the SM quarks when they convert to a mirror quark. These interactions of mirror and SM quarks are described by the two $3\times 3$ unitary mixing matrices $V_{H_d}$ and $V_{H_u}$ that are related by $V^\dagger_{H_u}V_{H_d} = V_{\rm CKM}$. This means that the FCNCs in down-quark and up-quark sectors are related.

The main phenomenological features of LHT contributions in the flavor observables are as follows (Buras, 2009). The rare $B$ decays are SM-like to a good extend. For instance $B_{s,d} \to \mu^+\mu^-$ can be enhanced by $\mathcal{O}(30\%)$ compared to the SM, where the largest corrections come from the $T$-even sector. Typical deviations in $S_{J/\psi\phi}$ are at the order of $\mathcal{O}(5-10\%)$, with the details depending on the spectrum of the mirror fermions. They are thus smaller then in RS. Similar effects would be expected in $\sin 2\phi_1$ determination from $B \to J/\psi K^0_S$ with the corrections to a large extend uncorrelated with the ones in $B^0_s-\overline{B}^0_s$ mixing (Blanke, Buras, Recksiegel, and Tarantino, 2008). The contributions to $\mathcal{B}(B \to X_s\gamma)$ are relatively small, at the order of up to about 3% of the SM value, which is smaller than the theoretical uncertainty on the SM prediction. The corrections to $B \to \tau\nu$ are very small since there are no tree level contributions. The effects in $S_{\phi K^0_S}$ are also expected to be small, since both $b \to sg$ and electroweak penguin corrections are not sizable.

The $\mathcal{B}(\mu \to e\gamma)$ can reach $2\times 10^{-11}$ so that some fine tuning of the parameters is required to satisfy MEG bounds (Adam et al., 2013). Also, the contributions in the lepton flavor violating decays, $\mu-e$ conversion, $\mu^- \to e^-e^+e^-$, $\tau \to \mu\gamma$, $\tau \to 3\mu$ clearly distinguish LHT from SUSY. The contributions to $(g-2)_\mu$ are negligible (see (Jegerlehner and Nyffeler, 2009) and references therein). There are $CP$ violating contributions to the $D^0 - \overline{D}^0$ mixing amplitudes. The weak phase in the mixing can still be large (and would be even much larger if the $\epsilon_K$ constraint is omitted). It can lead to effects in $D^0 - \overline{D}^0$ mixing that are of several percent, *e.g.* $-0.02 \lesssim S_{D\to K^0_S\phi} \lesssim +0.01$ (Bigi, Blanke, Buras, and Recksiegel, 2009).

### 25.2.3 Summary

Flavor physics has a significant potential to discover new physics by its sensitivity to high energy scales through virtual effects. At present, there is no solid experimental hint of an effect beyond the SM. Lacking any preferred theoretical foundation for the observed flavor structure, an analysis of the new physics effects in low energy precision observables must thus make use of well defined and commonly agreed benchmark models. While clearly it is not possible to cover all the possibilities, a large enough set of representative benchmark models gives a picture of what kind of effects are possible. Many of the currently discussed scenarios are already highly constrained or can be strongly constrained at the currently planned experiments.

One of the main motivations to extend the Standard Model with new particles with TeV masses is to solve the hierarchy problem – to stabilize the electroweak scale against radiative corrections. The flavor structure in most cases is not fixed by the rationale behind the model and hence remains mostly arbitrary from the theoretical considerations. Experimentally, on the other hand, the flavor structure in the new physics sector is tightly constrained. The legacy of the $B$ Factories program is that in low energy flavor violating processes the dominant contributions are from the SM. The low energy effects of a viable new physics model have to be minimally flavor violating, or at least have to be close to this limit. In fact, mainly due to the data of the two $B$ Factories, the corners of phase space for non-MFV effects in low energy processes have become very sparse.

Both *BABAR* and Belle have performed a test of the flavor structure, in many cases at a precision level. In this respect the two experiments have performed a similar task in the flavor sector as LEP did for the gauge couplings; still we do not have any substantial hint for a crack in the structure of the SM, neither in the gauge nor in the flavor sector.

Future experiments at both the energy as well as the intensity frontier will have an extended reach and a larger sensitivity. In particular, super flavor factories will refine many of the measurements performed at *BABAR* and Belle and thus improve the reach for new physics. Complementary to this, there will be measurements of leptonic processes at dedicated experiments, focusing especially on lepton-number and lepton-flavor violating processes. These efforts will be augmented further by experiments at the energy frontier, which will be mainly the LHC experiments ATLAS and CMS for the next decade. A direct discovery of new degrees of freedom at the energy frontier – beyond the discovery of a single Higgs particle – will clearly have a significant impact on our understanding of flavor physics in the future.

# Appendix A
# Glossary of terms

This part of the book summarizes commonly used terms, abbreviations, quantities, and acronyms found elsewhere in the book as a quick reference. References to places where terms are first introduced have been made. Where appropriate the second cross reference is given to sections in which a term is described in more details.

**ACC** : The Belle aerogel Cherenkov counter (1.4.4; 2.2.3).

**Acoplanarity** : The acoplanarity of a two-particle final state is defined as $\phi_2 - \phi_1 - \pi$. The azimuthal angles of the final state particles are $\phi_i$, where $i = 1, 2$ (17.4.3).

**AWG** : Analysis Working Group; a physics sub-group within the *BABAR* Collaboration (2.1).

`basf` : Belle Analysis and Simulation Framework (3.1).

**BDT** : Boosted (or bagged) decision tree. This is an MVA classification algorithm used widely in the latter years of data analysis at the $B$ Factories (5; 4).

**BSM** : Beyond the Standard Model (18.1; 25.2).

**CDC** : The Belle central drift chamber (1.4.4; 2.2.2).

**CLEO Fisher** : The CLEO Fisher discriminant formed of nine energy flow cones. This has been widely used as a variable to discriminate between $B$ meson signal-like events and light quark continuum background. Also see Fisher discriminant (9.3).

**CKM matrix** : Cabibbo-Kobayashi Maskawa quark mixing matrix (1; 16).

**CM** : Centre of mass (1.2.2).

**Continuum background** : This is the term given to backgrounds from $e^+e^-$ transitions to light fermion anti-fermion pairs in collisions. Typically continuum background refers to light-quark pairs $q\overline{q}$, where $q = u$, $d$, $s$, and $c$ (2.2.5).

$\cos\boldsymbol{\theta}_{\mathbf{B}}$ : Cosine of the angle between beam axis and $B$ momentum in the $\Upsilon(4S)$ rest frame (9.3).

$\cos\boldsymbol{\theta}_{\mathbf{S}}$ : Cosine of the angle between sphericity axes of ROE and $B$ candidate (9.3).

$\cos\boldsymbol{\theta}_{\mathbf{T}}$ : Cosine of the angle between beam and $B$ thrust axes (9.3).

**DAQ** : Data acquisition (1.4.3.1; 2.2.7).

**DCH** : The *BABAR* drift chamber (2.1; 2.2.2).

**DIRC** : The *BABAR* detector of internally reflected Cherenkov light, used for charged particle identification (in particular the $\pi$/K separation) in the barrel region (1.4; 2.2.3).

**ECOC** : Error-correcting output codes (5.2).

**ECL** : The Belle electromagnetic calorimeter (1.4.4.2; 2.2.4).

**EMC** : The *BABAR* electromagnetic calorimeter (2.1; 2.2.4).

**EML** : Extended maximum likelihood (11).

**Experiment** : An Experiment, with an upper case "E", is the name given in Belle to the different data taking periods. *BABAR* analogy of an Experiment is a Run. Only odd numbers were used for Experiments, there are 31 Belle Experiments. See also Run (3.2).

$\boldsymbol{\mathcal{F}}$ : Generic Fisher discriminant: a linear combination of variables (4).

**Fisher discriminant** : A linear combination of variables which is often used to compute a variable to discriminate between signal-like $B$ meson events and light quark continuum background; see also CLEO Fisher (4).

**FPGA** : Field-Programmable Gate Array, a configurable integrated circuit (2.2.2).

**FOM** : Figure of merit, a test statistic often used during optimization (4.3).

**FSR** : Final state radiation (17.4.4).

$\boldsymbol{H_i}$ : The $i^{th}$ Fox-Wolfram moments (9).

$\boldsymbol{h_l^k}$ : Normalized Fox-Wolfram moments given by Eq. (9.5.2).

**HER** : High energy ring (1.4.3.1).

**HQET** : Heavy Quark Effective Theory (17.1; 17.9.1.2).

**IFR** : The instrumented flux return of *BABAR*, used for $K_L^0$ and muon detection (1.4.3.1; 2.2.5).

**IP** : The interaction point (2.1).

**IR** : The interaction region (1.3).

**ISR** : Initial state radiation (15.1.1).

**KLM** : The instrumented flux return of Belle, used for $K_L^0$ and muon detection (1.4.4; 2.2.5).

**KSFW** : Fisher discriminant: a linear combination of modified Fox-Wolfram moments given by Eq. (9.5.3).

$\boldsymbol{L_i}$ : The $i^{th}$ 'monomial' corresponding to an angle-weighted energy flow variable given by Eq. (9.4.1).

**L1, L3** : The first (hardware-based) and the second (software-based) trigger level, respectively (2.1; 2.2.6).

**LCSR** : Light cone sum rule, a formulation of QCD sum rules specifically suited for the calculation of heavy-to-light form factors. (17.1).

**LER** : Low energy ring (1.4.3.1).

**Local operator** : In quantum field theory local operators are the product of field operators evaluated at the same space-time point (17.1.3).

**LQCD** : Lattice QCD (17.1).

**LST** : Limited Streamer Mode, the technology selected by *BABAR* to replace the whole IFR barrel because of a dramatic decrease in performance of the original RPCs (1.4; 2.2.5).

**LTDA** : *BABAR* long-term data analysis system (3.7).

**MC** : Monte Carlo (3.1).

**MFV** : Minimal flavor violation model (25.2).

**ML** : Maximum likelihood (11).

**MLP** : A multi-layer perceptron is a common type of artificial neural network used in particle physics. Neural networks have been used widely at the $B$ Factories, most notably in terms of flavor tagging (4; 8).

**MSSM** : Minimal super-symmetric standard model (25.2).

**MVA** : A multi-variate analysis is the study of a multi-dimensional problem space in the context of discriminating between different types of event. Practical demonstrations of the use of MVA techniques can be found throughout this book, in particular in the context of PID (4; 5; 9).

**NN** : Neural Network : multi-layered combination of variables. See also MLP (4; 9).

**NP** : "New physics", which is any physics not described by the Standard Model (2.2.6; 17.2).

**NRQCD** : Non-relativistic QCD (17.1.1).

**OPE** : Operator Product Expansion (17.1.1).

***p.d.f.*** : Probability density function (7.4.4; 11.1).

**Penguin** : Loop contribution mediating a flavour changing neutral current (7.4).

**Penguin pollution** : Penguin contributions which carry a weak phase different form the tree contribution, thereby introducing a "pollution" (i,e, hadronic uncertainties) into the extraction of CKM phases. (17.4.4; 17.7).

**PID** : Charged particle identification (2.1; 2.2.3).

**PV** : Primary Vertex (6.4).

**Planarity** : The planarity (or aplanarity) of the event is a measure of the transverse component of momentum of of the event plane. This is related ot the smallest eigenvalue of the sphericity tensor $\lambda_3$, where the aplanarity $A$ is $3\lambda_3/2$. For a planar event $A = 0$. For an isotropic event $A = 1/2$ (17.4.3).

**QCDF** : QCD Factorization (17.4; 17.9.1).

**Quasi-Two-Body** : For a decay to a final state with on resonance and a long-lived particle *e.g.* $B^0 \to \rho^+\pi^-$, the quasi-two-body approximation is sometimes invoked. This approximation is the assumption that the resonance can be treated as a particle with definite mass. In practice this means that any interference between the reconstructed resonance of interest and other amplitudes contributing to a same body final state is not explicitly accounted for in a fit to data using a Dalitz plot, but is treated as a systematic effect, or if deemed appropriate neglected. This approximation is commonly used in the study of charmless $B$ decays (17.4).

$\boldsymbol{\mathcal{R}}$ : Signal-to-background likelihood ratio used by Belle and given by Eq. (9.5.11).

$\boldsymbol{R_i}$ : Normalized Fox-Wolfram moments (*BABAR* notation; 9).

$\boldsymbol{R_l^{s0}}$**,** $\boldsymbol{R_l^{00}}$ : Modified Fox-Wolfram moments (9).

**ROE** : Rest of the event: Particles found in the detector that are not associated with the reconstructed signal candidate (6.5).

**RPC** : Resistive plate chamber, the technology selected by *BABAR* and Belle to instrument their muon detectors (1.4; 2.2.5).

**Run** : A Run, with an upper case "R", is the name given in *BABAR* to the different data taking periods. An analogy of Run at Belle is an Experiment. There are seven *BABAR* Runs, each several month-long. See also Experiment (3.2).

**run** : A run, with a lower case "r", is the basic unit of *BABAR* and Belle data collection. The full *BABAR* physics dataset contains more than 38,000 such runs (3.2).

**SCET** : Soft Collinear Effective Theory (17.4; 17.9.1.4).

**SFW** : Fisher discriminant : linear combination of Fox-

Wolfram moments given by Eq. (9.5.1).

**SM** : Standard Model of Particle Physics (1).

**SNNS** : Stuttgart Neural Network Simulator, an implementation of a neural network algorithm (4).

**Spectator** : A quark, which does not change its flavor in a weak decay process (17.4.4; 17.7)

**SOB** : The so-called *stand off box* of the *BABAR* DIRC (2.1).

**Sphericity** : Tensorial representation of energy flow, given by Eq. 9.3.2,

**$_s\mathcal{P}lot$** : The $_s\mathcal{P}lot$ technique is an event re-weighting technique that is used in order to project out fit components, such as signal or background. The technique was developed at the $B$ Factories. This technique has often been used when presenting results from *BABAR* (11; 11.2.3).

**SPR** : StatPatternRecognition, a ROOT-based package with a number of implemented multivariate methods (4).

**Strong phase** : A phase that is invariant under the operator *CP* (13.2.4).

**SVT** : The *BABAR* silicon vertex tracker (2.1; 2.2.1).

**SVTRAD** : The *BABAR* silicon vertex radiation monitoring system (2.2.1).

**SVD** : The Belle silicon vertex detector (2.1; 2.2.1).

**Three-Body** : Decay to a final state with three long-lived particles *e.g.* $B^+ \to \pi^+\pi^+\pi^-$ (9.4.2).

**Thrust** : Vectorial representation of energy flow, given by Eq. 9.3.1.

**Tree** : Contribution to a decay which is mediated by Feynman diagrams without loops. (17.4.4; 17.7)

**Two-Body** : Decay to a final state with two long-lived particles *e.g.* $B^0 \to \pi^+\pi^-$ (7.1.1).

**TMVA** : Toolkit for Multivariate Analysis, a ROOT-based package with a number of implemented multivariate methods (4.4.4; 4.5).

**TOF** : The Belle time-of-flight detector (1.4.4.1; 2.2.3).

**Twist** : In quantum field theory the twist of an operator is defined as the difference between its dimension and its spin (17.1.4.1).

**2HDM** : Two-Higgs doublet model (17.10.2.1; 25.2).

**VM** : Virtual Machine (3.7).

**Weak phase** : A phase that changes sign under the operator *CP* (13.2.4).

## Appendix B: The BABAR Collaboration author list

B. Aubert[a], R. Barate[a], D. Boutigny[a], F. Couderc[a], P. del Amo Sanchez[a], J.-M. Gaillard[a], A. Hicheur[a], Y. Karyotakis[a], J. P. Lees[a], V. Poireau[a], X. Prudent[a], P. Robbe[a], V. Tisserand[a], A. Zghiche[a], E. Grauges[b], J. Garra Tico[b], L. Lopez[c,d], M. Martinelli[c,d], A. Palano[c,d], M. Pappagallo[c,d], A. Pompili[c,d], G. P. Chen[e], J. C. Chen[e], N. D. Qi[e], G. Rong[e], P. Wang[e], Y. S. Zhu[e], G. Eigen[f], B. Stugu[f], L. Sun[f], G. S. Abrams[g], M. Battaglia[g], J. Beringer[g,v], A. W. Borgland[g], A. B. Breon[g], D. N. Brown[g], J. Button-Shafer[g], R. N. Cahn[g], E. Charles[g], M. V. Chistiakova[g], A. R. Clark[g], C. T. Day[g], M. Furman[g], M. S. Gill[g], Y. Groysman[g], B. Hooberman[g], R. G. Jacobsen[g], F. Jensen[g], R. W. Kadel[g], J. A. Kadyk[g], L. T. Kerth[g], Yu. G. Kolomensky[g], J. F. Kral[g], G. Kukartsev[g], C. LeClerc[g], M. J. Lee[g], M. E. Levi[g], G. Lynch[g], A. M. Merchant[g], L. M. Mir[g], P. J. Oddone[g], T. J. Orimoto[g], I. L. Osipenkov[g], E. Petigura[g], M. Pripstein[g], N. A. Roe[g], A. Romosan[g], M. T. Ronan[†g], V. G. Shelkov[g], A. Suzuki[g], K. Tackmann[g], T. Tanabe[g], D. Troost[g], W. A. Wenzel[g], M. Zisman[g], P. G. Bright-Thomas[h], K. E. Ford[h], T. J. Harrison[h], A. J. Hart[h], C. M. Hawkes[h], A. Kirk[h], D. J. Knowles[h], S. E. Morgan[h], S. W. O'Neale[†h], R. C. Penny[h], D. Smith[h], N. Soni[h], A. T. Watson[h], N. K. Watson[h], K. Goetzen[i], T. Held[i], H. Koch[i], M. Kunze[i], B. Lewandowski[†i], M. Pelizaeus[i], K. Peters[i], H. Schmuecker[i], T. Schroeder[i], M. Steinke[i], A. Fella[j], E. Antonioli[j], J. C. Andress[k], J. T. Boyd[k], N. Chevalier[k], W. N. Cottingham[k], N. Dyce[k], B. Foster[k], C. Mackay[k], A. Mass[k], J. D. McFall[k], D. Walker[k], D. Wallom[k], K. Abe[l], D. J. Asgeirsson[l], T. Cuhadar-Donszelmann[l], C. Hearty[l], N. S. Knecht[l], T. S. Mattison[l], J. A. McKenna[l], R. Y. So[l], D. Thiessen[l], M. Barrett[m], B. Camanzi[m], S. Jolly[m], A. Khan[m], P. Kyberd[m], A. K. McKemey[m], M. Saleem[m], D. J. Sherwood[m], L. Teodorescu[m], V. E. Blinov[n,o,p], A. A. Botov[n], A. D. Bukin[†n,p], A. R. Buzykaev[n], V. P. Druzhinin[n,p], V. B. Golubev[n,p], V. N. Ivanchenko[n], A. A. Korol[n,p], E. A. Kravchenko[n,p], A. P. Onuchin[n,o,p], S. I. Serednyakov[n,p], Yu. I. Skovpen[n,p], E. P. Solodov[n,p], V. I. Telnov[n,p], K. Yu. Todyshev[n,p], A. N. Yushkov[n], D. S. Best[q], M. Bondioli[q], J. Booth[q], M. Bruinsma[q], M. Chao[q], S. Curry[q], I. Eschrich[q], D. Kirkby[q], A. J. Lankford[q], M. Mandelkern[q], E. C. Martin[q], R. K. Mommsen[q], J. Schultz[q], D. P. Stoker[q], G. Zioulas[q], S. Abachi[r], K. Arisaka[r], C. Buchanan[r], S. Chun[r], B. L. Hartfiel[r], H. Atmacan[s], B. Dey[s], S. D. Foulkes[s], J. W. Gary[s], J. Layter[s], F. Liu[s], O. Long[s], E. Mullin[s], B. C. Shen[†s], G. M. Vitug[s], K. Wang[s], Z. Yasin[s], L. Zhang[s], H. K. Hadavand[t], E. J. Hill[t], H. P. Paar[t], S. Rahatlou[t], U. Schwanke[t], V. Sharma[t], J. W. Berryhill[u], C. Campagnari[u], A. Cunha[u], B. Dahmes[u], J. M. Flanigan[u], M. Franco Sevilla[u], T. M. Hong[u], D. Kovalskyi[u], N. Kuznetsova[u], S. L. Levy[u], A. Lu[u], M. A. Mazur[u], J. D. Richman[u], Y. Rozen[u], W. Verkerke[u], C. A. West[u], T. W. Beck[v], A. M. Eisner[v], C. J. Flacco[v], A. A. Grillo[v], M. Grothe[v], C. A. Heusch[v], J. Kroseberg[v], W. S. Lockman[v], A. J. Martinez[v], G. Nesom[v], T. Schalk[v], R. E. Schmitz[v], B. A. Schumm[v], A. Seiden[v], E. Spencer[v], P. Spradlin[v], M. Turri[v], W. Walkowiak[v], L. Wang[v], M. Wilder[v], D. C. Williams[v], M. G. Wilson[v], L. O. Winstrom[v], D. S. Chao[w], E. Chen[w], C. H. Cheng[w], D. A. Doll[w], M. P. Dorsten[w], A. Dvoretskii[w], B. Echenard[w], R. J. Erwin[w], F. Fang[w], K. T. Flood[w], J. E. Hanson[w], D. G. Hitlin[w], S. Metzler[w], J. S. Minamora[w], I. Narsky[w], P. Ongmongkolkul[w], J. Oyang[w], T. Piatenko[w], F. C. Porter[w], A. Y. Rakitin[w], A. Ryd[w], A. Samuel[w], S. Yang[w], R. Y. Zhu[w], R. Andreassen[x], S. Devmal[x], M. S. Dubrovin[x], C. Fabby[x], T. L. Geld[x], Z. Huard[x], S. Jayatilleke[x], G. Mancinelli[x], B. T. Meadows[x], K. Mishra[x], M. D. Sokoloff[x], L. Sun[x], T. Abe[y], E. A. Antillon[y], T. Barillari[y], J. Becker[y], F. Blanc[y], P. C. Bloom[y], B. Broomer[y], S. Chen[y], Z. C. Clifton[y], I. M. Derrington[y], J. Destree[y], M. O. Dima[y], E. Erdos[y], S. Fahey[y], W. T. Ford[y], F. Gaede[y], A. Gaz[y], J. D. Gilman[y], J. Hachtel[y], J. F. Hirschauer[y], D. R. Johnson[y], A. Kreisel[y], A. K. Michael[y], M. Nagel[y], U. Nauenberg[y], A. Olivas[y], H. Park[y], A. Penzkofer[y], P. Rankin[y], D. M. Rodriguez[y], J. Roy[y], W. O. Ruddick[y], S. Sen[y], J. G. Smith[y], E. W. Thomas[y], E. W. Tomassini[y], K. A. Ulmer[y], W. C. van Hoek[y], D. L. Wagner[y], S. R. Wagner[y], C. G. West[y], J. Zhang[y], R. Ayad[z], J. Blouw[z], A. Chen[z], E. A. Eckhart[z], J. L. Harton[z], T. Hu[z], W. H. Toki[z], R. J. Wilson[z], F. Winklmeier[z], Q. L. Zeng[z], D. Altenburg[aa], A. Hauke[aa], H. Jasper[aa], T.M. Karbach[aa], J. Merkel[aa], A. Petzold[aa], B. Spaan[aa], K. Wacker[aa], T. Brandt[ab], J. Brose[ab], T. Colberg[ab], G. Dahlinger[ab], M. Dickopp[ab], P. Eckstein[ab], H. Futterschneider[ab], S. Kaiser[ab], M. J. Kobel[ab], R. Krause[ab], W. F. Mader[ab], E. Maly[ab], R. Müller-Pfefferkorn[ab], R. Nogowski[ab], S. Otto[ab], J. Schubert[ab], K. R. Schubert[ab], R. Schwierz[ab], J. E. Sundermann[ab], A. Volk[ab], L. Wilden[ab], L. Behr[ac], D. Bernard[ac], F. Brochard[ac], J. Cohen-Tanugi[ac], F. Dohou[ac], S. Ferrag[ac], G. Fouque[ac], F. Gastaldi[ac], E. Latour[ac], A. Mathieu[ac], P. Matricon[ac], P. Mora de Freitas[ac], C. Renard[ac], E. Roussot[ac], S. Schrenk[ac], S. T'Jampens[ac], Ch. Thiebaux[ac], G. Vasileiadis[ac], M. Verderi[ac], A. Anjomshoaa[ad], R. Bernet[ad], P. J. Clark[ad], D. R. Lavin[ad], F. Muheim[ad], S. Playfer[ad], A. I. Robertson[ad], J. E. Swain[ad], J. E. Watson[ad], Y. Xie[ad], M. Falbo[ae], D. Andreotti[af], M. Andreotti[af,ag], D. Bettoni[af], C. Bozzi[af], R. Calabrese[af,ag], V. Carassiti[af], A. Cecchi[af], G. Cibinetto[af], A. Cotta Ramusino[af], F. Evangelisti[af], E. Fioravanti[af], P. Franchini[af], I. Garzia[af], L. Landi[af,ag], E. Luppi[af,ag], R. Malaguti[af], M. Munerato[af,ag], M. Negrini[af], C. Padoan[af,ag], A. Petrella[af], L. Piemontese[af], V. Santoro[af], A. Sarti[af,ag], E. Treadwell[ah], F. Anulli[ai,ce], R. Baldini-Ferroli[ai], M. E. Biagini[ai], A. Calcaterra[ai], G. Finocchiaro[ai], S. Martellotti[ai], P. Patteri[ai], I. M. Peruzzi[ai,by], M. Piccolo[ai], M. Rama[ai], R. de Sangro[ai], Y. Xie[ai], A. Zallo[ai], S. Bagnasco[aj,ak], A. Buzzo[aj], R. Capra[aj,ak], R. Contri[aj,ak], G. Crosetti[aj,ak], E. Guido[aj,ak], M. Lo Vetere[aj,ak], M. M. Macri[aj], S. Minutoli[aj], M. R. Monge[aj,ak], P. Musico[aj], S. Passaggio[aj], F. C. Pastore[aj,ak], C. Patrignani[aj,ak],

M. G. Pia[aj], E. Robutti[aj], A. Santroni[aj,ak], S. Tosi[aj,ak], B. Bhuyan[al], V. Prasad[al], S. Bailey[am], G. Brandenburg[†am], K. S. Chaisanguanthum[am], C. L. Lee[am], M. Morii[am], E. Won[am], J. Wu[am], A. J. Edwards[an], A. Adametz[ao], R. S. Dubitzky[ao], U. Langenegger[ao], J. Marks[ao], S. Schenk[ao], U. Uwer[ao], V. Klose[ap], H. M. Lacker[ap], M. L. Aspinwall[aq], W. Bhimji[aq], D. A. Bowerman[aq], P. D. Dauncey[aq], U. Egede[aq], R. L. Flack[aq], J. R. Gaillard[aq], N. J. W. Gunawardane[aq], G. W. Morton[aq], J .A. Nash[aq], M. B. Nikolich[aq], W. Panduro Vazquez[aq], P. Sanders[aq], D. Smith[aq], G. P. Taylor[aq], M. Tibbetts[aq], P. K. Behera[ar], X. Chai[ar], M. J. Charles[ar], G. J. Grenier[ar], R. Hamilton[ar], S.-J. Lee[ar], U. Mallik[ar], N. T. Meyer[ar], C. Chen[as], J. Cochran[as], H. B. Crawley[as], L. Dong[as], V. Eyges[as], P.-A. Fischer[as], J. Lamsa[as], W. T. Meyer[as], S. Prell[as], E. I. Rosenberg[as], A. E. Rubin[as], Y. Y. Gao[at], A. V. Gritsan[at], Z. J. Guo[at], C. K. Lae[at], G. Schott[au], J. N. Albert[av], N. Arnaud[av], C. Beigbeder[av], M. Benkebil[av], D. Breton[av], R. Cizeron[av], M. Davier[av], D. Derkach[av], S. Dû[av], J. Firmino da Costa[av], G. Grosdidier[av], A. Höcker[av], S. Laplace[av], F. Le Diberder[av], V. Lepeltier[†av], A. M. Lutz[av], B. Malaescu[av], J. Y. Nief[av], T. C. Petersen[av], S. Plaszczynski[av], S. Pruvot[av], S. Rodier[av], P. Roudeau[av], M. H. Schune[av], J. Serrano[av], V. Sordini[av,ce,cf], A. Stocchi[av], V. Tocut[av], S. Trincaz-Duvoid[av], A. Valassi[av], L. L. Wang[av], G. Wormser[av], R. M. Bionta[aw], V. Brigljević[aw], D. J. Lange[aw], M. Mugge[aw], M. C. Simani[aw], K. van Bibber[aw], D. M. Wright[aw], I. Bingham[ax], J. P. Burke[ax], M. Carroll[ax], C. A. Chavez[ax], J. P. Coleman[ax], P. Cooke[ax], I. J. Forster[ax], J. R. Fry[ax], E. Gabathuler[ax], R. Gamet[ax], M. George[ax], D. E. Hutchcroft[ax], M. Kay[ax], S. McMahon[ax], A. Muir[ax], R. J. Parry[ax], D. J. Payne[ax], K. C. Schofield[ax], R. J. Sloane[ax], P. Sutcliffe[ax], C. Touramanis[ax], D. E. Azzopardi[ay], G. Bellodi[ay], A. J. Bevan[ay], C. K. Clarke[ay], C. M. Cormack[ay], F. Di Lodovico[ay], P. Dixon[ay], K. A. George[ay], W. Menges[ay], D. Newman-Coburn[†ay], R. J. L. Potter[ay], R. Sacco[ay], H. W. Shorthouse[ay], M. Sigamani[ay], P. Strother[ay], P. B. Vidal[ay], M. I. Williams[ay], C. L. Brown[az], G. Cowan[az], H. U. Flaecher[az], S. George[az], M. G. Green[az], D. A. Hopkins[az], P. S. Jackson[az], A. Kurup[az], C. E. Marker[az], P. McGrath[az], T. R. McMahon[az], S. Paramesvaran[az], F. Salvatore[az], G. Vaitsas[az], M. A. Winter[az], A. C. Wren[az], J. Bougher[ba], D. N. Brown[ba], C. L. Davis[ba], Y. Li[ba], J. Pavlovich[ba], A. G. Denig[au,bb], M. Fritsch[bb], W. Gradl[bb], K. Griessinger[bb], A. Hafner[bb], E. Prencipe[bb], J. Allison[bc], K. E. Alwyn[bc], D. S. Bailey[bc], N. R. Barlow[bc], R. J. Barlow[bc], Y. M. Chia[bc], C. L. Edgar[bc], A. C. Forti[bc], J. Fullwood[bc], P. A. Hart[bc], M. C. Hodgkinson[bc], F. Jackson[bc], G. Jackson[bc], M. P. Kelly[bc], S. D. Kolya[bc], G. D. Lafferty[bc], A. J. Lyon[bc], M. T. Naisbit[bc], N. Savvas[bc], J. H. Weatherall[bc], T. J. West[bc], J. C. Williams[bc], J. I. Yi[bc], J. Anderson[bd], E. Behn[bd], A. Farbin[bd], B. Hamilton[bd], W. D. Hulsbergen[bd], A. Jawahery[bd], V. Lillard[bd], D. A. Roberts[bd], J. R. Schieck[bd], J. M. Tuggle[bd], G. Blaylock[be], C. Dallapiccola[be], S. S. Hertzbach[be], R. Kofler[be], V. B. Koptchev[be], X. Li[be], C. S. Lin[be], T. B. Moore[be], E. Salvati[be], S. Saremi[be], H. Staengle[be], S. Y. Willocq[be], J. Wittlin[be], R. Cowan[bf], D. Dujmic[bf], P. H. Fisher[bf], S. W. Henderson[bf], K. Koeneke[bf], M. I. Lang[bf], G. Sciolla[bf], M. Spitznagel[bf], F. Taylor[bf], R. K. Yamamoto[†bf], M. Yi[bf], M. Zhao[bf], Y. Zheng[bf], D. I. Britton[bg], R. Cheaib[bg], M. Klemetti[bg], D. J. J. Mangeol[bg], S. E. Mclachlin[†bg], M. Milek[bg], P. M. Patel[†bg], S. H. Robertson[bg], M. Schram[bg], P. Biassoni[bh,bi], G. Cerizza[bh,bi], P. Gandini[bh,bi], F. Lanni[bh,bi], A. Lazzaro[bh,bi], V. Lombardo[bh,bi], N. Neri[bh], F. Palombo[bh,bi], R. Pellegrini[bh,bi], S. Stracka[bh,bi], J. M. Bauer[bj], M. Booke[†bj], L. Cremaldi[bj], V. Eschenburg[bj], R. Kroeger[bj], M. Reep[bj], J. Reidy[bj], D. A. Sanders[bj], P. Sonnek[bj], D. J. Summers[bj], H. W. Zhao[bj], R. Godang[bk], J. F. Arguin[bl], M. Beaulieu[bl], S. Brunet[bl], D. Cote[bl], J. P. Martin[bl], X. Nguyen[bl], S. Sabik[bl], R. Seitz[bl], E. Sicard[bl], M. Simard[bl], P. Taras[bl], B. Viaud[bl], A. Woch[bl], V. Zacek[bl], H. Nicholson[bm], N. Cavallo[bn], G. De Nardo[bn,bo], F. Fabozzi[bn], C. Gatto[bn], L. Lista[bn], D. Monorchio[bn,bo], G. Onorato[bn,bo], P. Paolucci[bn], D. Piccolo[bn,bo], C. Sciacca[bn,bo], M. A. Baak[bp], H. Bulten[bp], G. Raven[bp], H. L. Snoek[bp], N. M. Cason[bq], C. P. Jessop[bq], K. J. Knoepfel[bq], J. M. LoSecco[bq], J. R. G. Alsmiller[br], T. A. Gabriel[br], T. Allmendinger[bs], G. Benelli[bs], B. Brau[bs], L. A. Corwin[bs], K. K. Gan[bs], K. Honscheid[bs], D. Hufnagel[bs], H. Kagan[bs], R. Kass[bs], J. P. Morris[bs], A. M. Rahimi[bs], J. J. Regensburger[bs], D. S. Smith[bs], R. Ter-Antonyan[bs], Q. K. Wong[bs], N. L. Blount[bt], J. Brau[bt], R. Frey[bt], O. Igonkina[bt], M. Iwasaki[bt], J. A. Kolb[bt], M. Lu[bt], C. T. Potter[bt], R. Rahmat[bt], N. B. Sinev[bt], D. Strom[bt], J. Strube[bt], E. Torrence[bt], E. Borsato[bu,bv], G. Castelli[bu], F. Colecchia[bu,bv], A. Crescente[bu], F. Dal Corso[bu], A. Dorigo[bu], C. Fanin[bu], E. Feltresi[bu,bv], F. Furano[bu], N. Gagliardi[bu,bv], F. Galeazzi[bu,bv], M. Margoni[bu,bv], M. Marzolla[bu], G. Michelon[bu,bv], M. Morandin[bu], M. Posocco[bu], M. Rotondo[bu], G. Simi[bu], F. Simonetto[bu,bv], P. Solagna[bu], E. Stevanato[bu], R. Stroili[bu,bv], G. Tiozzo[bu], E. Torassa[bu], C. Voci[bu,bv], S. Akar[bw], P. Bailly[bw], E. Ben-Haim[bw], M. Benayoun[bw], M. Bomben[bw], G.R. Bonneaud[bw], H. Briand[bw], L. Del Buono[bw], G. Calderini[bw], J. Chauveau[bw], P. David[bw], J.-F. Genat[bw], O. Hamon[bw], M. J. J. John[bw], H. Lebbolo[bw], Ph. Leruste[bw], J. Lory[bw], J. Malclès[bw], G. Marchiori[bw], L. Martin[bw], J. Ocariz[bw], M. Pivk[bw], J. Prendki[bw], L. Roos[bw], S. Sitt[bw], J. Stark[bw], G. Thérin[bw], C. De la Vaissière[bw], A. Vallereau[bw], S. Versillé[bw], B. Zhang[bw], M. Biasini[bx,by], R. Covarelli[bx,by], E. Manoni[bx], S. Pacetti[bx,by], S. Pennazzi[bx,by], M. Pioppi[bx,by], A. Rossi[bx], C. Angelini[bz,ca], G. Batignani[bz,ca], S. Bettarini[bz,ca], F. Bosi[bz], F. Bucci[bz,ca], E. Campagna[bz,ca], M. Carpinelli[bz,ca], G. Casarosa[bz,ca], R. Cenci[bz,ca], A. Cervelli[bz,ca], V. Del Gamba[bz,ca], F. Forti[bz,ca], M. A. Giorgi[bz,ca], A. Lusiani[bz,cb], M. Morganti[bz,ca], F. Morsani[bz], B. Oberhof[bz,ca], E. Paoloni[bz,ca], A. Perez[bz], F. Raffaelli[bz], G. Rizzo[bz,ca], F. Sandrelli[bz,ca], G. Triggiani[bz,ca], J. J. Walsh[bz,ca], M. Haire[cc], D. Judd[cc], K. Paick[cc], L. Turnbull[cc], D. E. Wagoner[cc], J. Biesiada[cd], N. Danielson[cd], P. Elmer[cd], R. E. Fernholz[cd], Y. P. Lau[cd], C. Lu[cd], V. Miftakov[cd], J. Olsen[cd],

D. Lopes Pegna[cd], W. R. Sands[cd], S. F. Schaffner[cd], A. J. S. Smith[cd], A. V. Telnov[cd], A. Tumanov[cd], E. W. Varnes[cd], E. Baracchini[ce,cf], F. Bellini[ce,cf], G. Cavoto[ce], A. D'Orazio[ce,cf], E. Di Marco[ce,cf], R. Faccini[ce,cf], F. Ferrarotto[ce], F. Ferroni[ce,cf], K. Fratini[ce], M. Gaspero[ce,cf], P. D. Jackson[ce,cf], E. Lamanna[ce,cf], E. Leonardi[ce], L. Li Gioi[ce,cf], M. A. Mazzoni[ce], S. Morganti[ce], G. Piredda[ce], F. Polci[ce,cf,av], D. del Re[ce,cf], F. Renga[ce,cf], F. Safai Tehrani[ce], M. Serra[ce], C. Voena[ce], C. Bünger[cg], S. Christ[cg], S. Dittrich[cg], O. Grünberg[cg], T. Hartmann[cg], M. Heß[cg], T. Leddig[cg], H. Schröder[†cg], C. Voß[cg], G. Wagner[cg], R. Waldi[cg], T. Adye[ch], M. Bly[ch], C. Brew[ch], B. Claxton[ch], C. Condurache[ch], N. De Groot[ch], J. Dowdell[ch], B. Franek[ch], S. Galagedera[ch], N. I. Geddes[ch], G. P. Gopal[ch], J. Kay[ch], J. Lidbury[ch], S. Madani[ch], G. Markey[ch], E. O. Olaiya[ch], P. Olley[ch], S. Ricciardi[ch], W. Roethel[ch], M. Watt[ch], F. F. Wilson[ch], S. M. Xella[ch], R. Aleksan[ci], P. Besson[†ci], P. Bourgeois[ci], P. Convert[ci], G. De Domenico[ci], S. Emery[ci], M. Escalier[ci], L. Esteve[ci], A. Gaidot[ci], S. F. Ganzhur[ci], Z. Georgette[ci], P.-F. Giraud[ci], L. Gosset[ci], P. Graffin[ci], G. Graziani[ci], G. Hamel de Monchenault[ci], S. Hervé[ci], M. Karolak[ci], W. Kozanecki[ci], M. Langer[ci], M. Legendre[ci], A. de Lesquen[ci], G. W. London[ci], V. Marques[ci], B. Mayer[ci], P. Micout[ci], J. P. Mols[ci], J. P. Mouly[ci], Y. Penichot[ci], J. Rolquin[ci], B. Serfass[ci], J. C. Toussaint[ci], M. Usseglio[ci], G. Vasseur[ci], Ch. Yèche[ci], M. Zito[ci], I. Adam[cj], I. J. R. Aitchison[cj], M. T. Allen[cj], R. Akre[†cj], P. L. Anthony[cj], D. Aston[cj], T. Azemoon[cj], D. J. Bard[cj], J. Bartelt[cj], R. Bartoldus[cj], P. Bechtle[cj], J. Becla[cj], R. Bell[†cj], J. F. Benitez[cj], N. Berger[cj], K. Bertsche[cj], E. Bloom[cj], C. T. Boeheim[cj], K. Bouldin[cj], A. M. Boyarski[cj], R. F. Boyce[cj], M. Browne[cj], O. L. Buchmueller[cj], W. Burgess[cj], Y. Cai[cj], C. Cartaro[cj], A. Ceseracciu[cj], R. Claus[cj], M. R. Convery[cj], D. P. Coupal[cj], W. W. Craddock[cj], G. Crane[cj], M. Cristinziani[cj], S. DeBarger[cj], F. J. Decker[cj], H. DeStaebler[†cj], J. C. Dingfelder[cj], M. Donald[cj], J. Dorfan[cj], G. P. Dubois-Felsmann[cj], W. Dunwoodie[cj], M. Ebert[cj], S. Ecklund[cj], R. Erickson[cj], S. Fan[cj], R. C. Field[cj], A. Fisher[cj], J. Fox[cj], B. G. Fulsom[cj], A. M. Gabareen[cj], I. Gaponenko[cj], T. Glanzman[cj], S. J. Gowdy[cj], M. T. Graham[cj], P. Grenier[cj], T. Hadig[cj], V. Halyo[cj], G. Haller[cj], J. Hamilton[cj], A. Hanushevsky[cj], A. Hasan[cj], T. Haas[cj], C. Hast[cj], C. Hee[cj], T. Himel[cj], T. Hryn'ova[cj], M. E. Huffer[cj], T. Hung[cj], W. R. Innes[cj], R. Iverson[cj], J. Kaminski[cj], M. H. Kelsey[cj], H. Kim[cj], P. Kim[cj], D. Kharakh[cj], M. L. Kocian[cj], A. Krasnykh[cj], J. Krebs[cj], W. Kroeger[cj], A. Kulikov[cj], N. Kurita[cj], D. W. G. S. Leith[cj], P. Lewis[cj], S. Li[cj], J. Libby[cj], D. Lindemann[cj], B. Lindquist[cj], V. Lüth[cj], S. Luitz[cj], H. L. Lynch[cj], D. B. MacFarlane[cj], H. Marsiske[cj], M. McCulloch[cj], J. McDonald[cj], R. Melen[cj], S. Menke[cj], R. Messner[†cj], S. Metcalfe[cj], L. J. Moss[cj], R. Mount[cj], D. R. Muller[cj], H. Neal[cj], D. Nelson[cj], S. Nelson[cj], M. Nordby[cj], Y. Nosochkov[cj], A. Novokhatski[cj], C. P. O'Grady[cj], F. G. O'Neill[cj], I. Ofte[cj], V. E. Ozcan[cj], T. Pavel[cj], A. Perazzo[cj], M. Perl[cj], S. Petrak[cj], M. Piemontese[cj], S. Pierson[cj], T. Pulliam[cj], H. Quinn[cj], B. N. Ratcliff[cj], S. Ratkovsky[cj], R. Reif[cj], C. Rivetta[cj], R. Rodriguez[cj], A. Roodman[cj], A. A. Salnikov[cj], O. H. Saxton[cj], T. Schietinger[cj], R. H. Schindler[cj], H. Schwarz[cj], J. Schwiening[cj], J. Seeman[cj], V. V. Serbo[cj], D. Smith[cj], A. Snyder[cj], E. Soderstrom[cj], A. Soha[cj], M. Stanek[cj], J. Stelzer[cj], D. Su[cj], M. K. Sullivan[cj], S. Sun[cj], K. Suzuki[cj], S. K. Swain[cj], H. A. Tanaka[cj], D. Teytelman[cj], J. M. Thompson[cj], J. S. Tinslay[cj], A. Trunov[cj], J. Turner[cj], N. van Bakel[cj], D. van Winkle[cj], J. Va'vra[cj], A. P. Wagner[cj], W. F. Wang[cj], M. Weaver[cj], T. Weber[cj], A. J. R. Weinstein[cj], U. Wienands[cj], W. J. Wisniewski[cj], M. Wittgen[cj], W. Wittmer[cj], D. H. Wright[cj], H. W. Wulsin[cj], Y. Yan[cj], A. K. Yarritu[cj], K. Yi[cj], G. Yocky[cj], C. C. Young[cj], V. Ziegler[cj], X. R. Chen[ck], N. Copty[ck], H. Liu[ck], W. Park[ck], M. V. Purohit[ck], H. Singh[ck], A. W. Weidemann[ck], R. M. White[ck], J. R. Wilson[ck], F. X. Yumiceva[ck], A. Randle-Conde[cl], S. J. Sekula[cl], M. Bellis[cm], P. R. Burchat[cm], S. A. Majewski[cm], T. I. Meyer[cm], T. S. Miyashita[cm], B. A. Petersen[cm], E. M. T. Puccio[cm], C. Roat[cm], M. Ahmed[cn], S. Ahmed[cn], M. S. Alam[cn], R. Bula[cn], J. A. Ernst[cn], V. Jain[cn], J. Liu[cn], B. Pan[cn], M. A. Saeed[cn], F. R. Wappler[cn], S. B. Zain[cn], R. Gorodeisky[co], N. Guttman[co], D. R. Peimer[co], A. Soffer[co], R. Henderson[cp], A. De Silva[cp], W. Bugg[cq], H. Cohn[cq], P. Lund[cq], M. Krishnamurthy[cq], G. Ragghianti[cq], S. M. Spanier[cq], B. J. Wogsland[cq], R. Eckmann[cr], J. L. Ritchie[cr], A. M. Ruland[cr], A. Satpathy[cr], C. J. Schilling[cr], R. F. Schwitters[cr], B. C. Wray[cr], B. W. Drummond[cs], J. M. Izen[cs], I. Kitayama[cs], X. C. Lou[cs], G. Williams[cs], S. Ye[cs], F. Bianchi[ct,cu], M. Bona[ct,cu], F. De Mori[ct,cu], A. Filippi[ct,cu], F. Gallo[ct,cu], D. Gamba[ct,cu], M. Pelliccioni[ct,cu], S. Zambito[ct,cu], F. Daudo[ct], B. Di Girolamo[ct], P. Grosso[ct], A. Smol[ct], P. P. Trapani[ct], D. Zanin[ct], C. Borean[cv,cw], L. Bosisio[cv,cw], F. Cossutti[cv], G. Della Ricca[cv,cw], S. Dittongo[cv,cw], S. Grancagnolo[cv,cw], L. Lanceri[cv,cw], P. Poropat[†cv,cw], M. Prest[cv], I. Rashevskaya[cv], E. Vallazza[cv], L. Vitale[cv,cw], G. Vuagnin[cv,cw], P. F. Manfredi[cx], V. Re[cx], V. Speziali[cx], E. D. Frank[cy], L. Gladney[cy], Q. H. Guo[cy], J. Panetta[cy], R. S. Panvini[†cz], V. Azzolini[da], J. Bernabeu[da], N. Lopez-March[da], F. Martinez-Vidal[da], D. A. Milanes[da], A. Oyanguren[da], P. Villanueva-Perez[da], A. Agarwal[db], H. Ahmed[db], J. Albert[db], Sw. Banerjee[db], F. U. Bernlochner[db], C. M. Brown[db], H. H. F. Choi[db], D. Fortin[db], K. B. Fransham[db], K. Hamano[db], G. J. King[db], R. Kowalewski[db], M. J. Lewczuk[db], C. Lindsay[db], C. B. Locke[db], T. Lueck[db], I. M. Nugent[db], J. M. Roney[db], R. J. Sobie[db], N. Tasneem[db], J. J. Back[dc], T. J. Gershon[dc], P. F. Harrison[dc], J. Ilic[dc], T. E. Latham[dc], G. B. Mohanty[dc], M. R. Pennington[dc], H. R. Band[dd], X. Chen[dd], B. Cheng[dd], S. Dasu[dd], M. Datta[dd], A. M. Eichenbaum[dd], J. J. Hollar[dd], H. Hu[dd], J. R. Johnson[dd], P. E. Kutter[dd], H. Li[dd], R. Liu[dd], B. Mellado[dd], A. Mihalyi[dd], A. K. Mohapatra[dd], Y. Pan[dd], M. Pierini[dd], R. Prepost[dd], I. J. Scott[dd], P. Tan[dd], C. O. Vuosalo[dd], J. H. von Wimmersperg-Toeller[dd], S. L. Wu[dd], Z. Yu[dd], M. G. Greene[de], T. M. B. Kordich[de], Y. Rozen[df]

[a]Laboratoire d'Annecy-le-Vieux de Physique des Particules (LAPP), Université de Savoie, CNRS/IN2P3, F-74941 Annecy-le-Vieux, France
[b]Universitat de Barcelona, Facultat de Fisica, Departament ECM, E-08028 Barcelona, Spain
[c]INFN Sezione di Bari, I-70126 Bari, Italy
[d]Dipartmento di Fisica, Università di Bari, I-70126 Bari, Italy
[e]Institute of High Energy Physics, Beijing 100039, China
[f]University of Bergen, Institute of Physics, N-5007 Bergen, Norway
[g]Lawrence Berkeley National Laboratory and University of California, Berkeley, California 94720, USA
[h]University of Birmingham, Birmingham, B15 2TT, United Kingdom
[i]Ruhr Universität Bochum, Institut für Experimentalphysik 1, D-44780 Bochum, Germany
[j]INFN CNAF I-40127 Bologna, Italy
[k]University of Bristol, Bristol BS8 1TL, United Kingdom
[l]University of British Columbia, Vancouver, British Columbia, Canada V6T 1Z1
[m]Brunel University, Uxbridge, Middlesex UB8 3PH, United Kingdom
[n]Budker Institute of Nuclear Physics SB RAS, Novosibirsk 630090, Russia
[o]Novosibirsk State Technical University, Novosibirsk 630092, Russia
[p]Novosibirsk State University, Novosibirsk 630090, Russia
[q]University of California at Irvine, Irvine, California 92697, USA
[r]University of California at Los Angeles, Los Angeles, California 90024, USA
[s]University of California at Riverside, Riverside, California 92521, USA
[t]University of California at San Diego, La Jolla, California 92093, USA
[u]University of California at Santa Barbara, Santa Barbara, California 93106, USA
[v]University of California at Santa Cruz, Institute for Particle Physics, Santa Cruz, California 95064, USA
[w]California Institute of Technology, Pasadena, California 91125, USA
[x]University of Cincinnati, Cincinnati, Ohio 45221, USA
[y]University of Colorado, Boulder, Colorado 80309, USA
[z]Colorado State University, Fort Collins, Colorado 80523, USA
[aa]Technische Universität Dortmund, Fakultät Physik, D-44221 Dortmund, Germany
[ab]Technische Universität Dresden, Institut für Kern- und Teilchenphysik, D-01062 Dresden, Germany
[ac]Laboratoire Leprince-Ringuet, CNRS/IN2P3, Ecole Polytechnique, F-91128 Palaiseau, France
[ad]University of Edinburgh, Edinburgh EH9 3JZ, United Kingdom
[ae]Elon University, Elon University, North Carolina 27244-2010, USA
[af]INFN Sezione di Ferrara, I-44100 Ferrara, Italy
[ag]Dipartimento di Fisica e Scienze della Terra, Università di Ferrara, I-44100 Ferrara, Italy
[ah]Florida A&M University, Tallahassee, Florida 32307, USA
[ai]INFN Laboratori Nazionali di Frascati, I-00044 Frascati, Italy
[aj]INFN Sezione di Genova, I-16146 Genova, Italy
[ak]Dipartimento di Fisica, Università di Genova, I-16146 Genova, Italy
[al]Indian Institute of Technology Guwahati, Guwahati, Assam, 781 039, India
[am]Harvard University, Cambridge, Massachusetts 02138, USA
[an]Harvey Mudd College, Claremont, California 91711, USA
[ao]Universität Heidelberg, Physikalisches Institut, D-69120 Heidelberg, Germany
[ap]Humboldt-Universität zu Berlin, Institut für Physik, D-12489 Berlin, Germany
[aq]Imperial College London, London, SW7 2AZ, United Kingdom
[ar]University of Iowa, Iowa City, Iowa 52242, USA
[as]Iowa State University, Ames, Iowa 50011-3160, USA
[at]Johns Hopkins University, Baltimore, Maryland 21218, USA
[au]Universität Karlsruhe, Institut für Experimentelle Kernphysik, D-76021 Karlsruhe, Germany
[av]Laboratoire de l'Accélérateur Linéaire, IN2P3/CNRS et Université Paris-Sud 11, Centre Scientifique d'Orsay, F-91898 Orsay Cedex, France
[aw]Lawrence Livermore National Laboratory, Livermore, California 94550, USA
[ax]University of Liverpool, Liverpool L69 7ZE, United Kingdom
[ay]Queen Mary, University of London, London, E1 4NS, United Kingdom
[az]University of London, Royal Holloway and Bedford New College, Egham, Surrey TW20 0EX, United Kingdom
[ba]University of Louisville, Louisville, Kentucky 40292, USA
[bb]Johannes Gutenberg-Universität Mainz, Institut für Kernphysik, D-55099 Mainz, Germany
[bc]University of Manchester, Manchester M13 9PL, United Kingdom
[bd]University of Maryland, College Park, Maryland 20742, USA
[be]University of Massachusetts, Amherst, Massachusetts 01003, USA
[bf]Massachusetts Institute of Technology, Laboratory for Nuclear Science, Cambridge, Massachusetts 02139, USA
[bg]McGill University, Montréal, Québec, Canada H3A 2T8
[bh]INFN Sezione di Milano, I-20133 Milano, Italy
[bi]Dipartimento di Fisica, Università di Milano, I-20133 Milano, Italy
[bj]University of Mississippi, University, Mississippi 38677, USA
[bk]University of South Alabama, Mobile, Alabama 36688, USA
[bl]Université de Montréal, Physique des Particules, Montréal, Québec, Canada H3C 3J7
[bm]Mount Holyoke College, South Hadley, Massachusetts 01075, USA
[bn]INFN Sezione di Napoli, I-80126 Napoli, Italy
[bo]Dipartimento di Scienze Fisiche, Università di Napoli Federico II, I-80126 Napoli, Italy
[bp]NIKHEF, National Institute for Nuclear Physics and High Energy Physics, NL-1009 DB Amsterdam, The Netherlands
[bq]University of Notre Dame, Notre Dame, Indiana 46556, USA
[br]Oak Ridge National Laboratory, Oak Ridge, Tennessee 37831, USA

[bs]Ohio State University, Columbus, Ohio 43210, USA
[bt]University of Oregon, Eugene, Oregon 97403, USA
[bu]INFN Sezione di Padova, I-35131 Padova, Italy
[bv]Dipartimento di Fisica, Università di Padova, I-35131 Padova, Italy
[bw]Laboratoire de Physique Nucléaire et de Hautes Energies, IN2P3/CNRS, Université Pierre et Marie Curie-Paris6, Université Denis Diderot-Paris7, F-75252 Paris, France
[bx]INFN Sezione di Perugia I-06123 Perugia, Italy
[by]Dipartimento di Fisica, Università di Perugia, I-06123 Perugia, Italy
[bz]INFN Sezione di Pisa, I-56127 Pisa, Italy
[ca]Dipartimento di Fisica, Università di Pisa, I-56127 Pisa, Italy
[cb]Scuola Normale Superiore di Pisa, I-56127 Pisa, Italy
[cc]Prairie View A&M University, Prairie View, Texas 77446, USA
[cd]Princeton University, Princeton, New Jersey 08544, USA
[ce]INFN Sezione di Roma, I-00185 Roma, Italy
[cf]Dipartimento di Fisica, Università di Roma La Sapienza, I-00185 Roma, Italy
[cg]Universität Rostock, D-18051 Rostock, Germany
[ch]Rutherford Appleton Laboratory, Chilton, Didcot, Oxon, OX11 0QX, United Kingdom
[ci]CEA, Irfu, SPP, Centre de Saclay, F-91191 Gif-sur-Yvette, France
[cj]SLAC National Accelerator Laboratory, Stanford University, Menlo Park, California 94025, USA
[ck]University of South Carolina, Columbia, South Carolina 29208, USA
[cl]Southern Methodist University, Dallas, Texas 75275, USA
[cm]Stanford University, Stanford, California 94305-4060, USA
[cn]State University of New York, Albany, New York 12222, USA
[co]Tel Aviv University, Tel Aviv, 69978, Israel
[cp]TRIUMF, Vancouver, BC, Canada V6T 2A3
[cq]University of Tennessee, Knoxville, Tennessee 37996, USA
[cr]University of Texas at Austin, Austin, Texas 78712, USA
[cs]University of Texas at Dallas, Richardson, Texas 75083, USA
[ct]INFN Sezione di Torino, I-10125 Torino, Italy
[cu]Dipartimento di Fisica Sperimentale, Università di Torino, I-10125 Torino, Italy
[cv]INFN Sezione di Trieste, I-34127 Trieste, Italy
[cw]Dipartimento di Fisica, Università di Trieste, I-34127 Trieste, Italy
[cx]Università di Pavia, Dipartimento di Elettronica and INFN, I-27100 Pavia, Italy
[cy]University of Pennsylvania, Philadelphia, Pennsylvania 19104, USA
[cz]Vanderbilt University, Nashville, Tennessee 37235, USA
[da]IFIC, Universitat de Valencia-CSIC, E-46071 Valencia, Spain
[db]University of Victoria, Victoria, British Columbia, Canada V8W 3P6
[dc]Department of Physics, University of Warwick, Coventry CV4 7AL, United Kingdom
[dd]University of Wisconsin, Madison, Wisconsin 53706, USA
[de]Yale University, New Haven, Connecticut 06511, USA
[df]Technion, Haifa, Israel

[†]Deceased

## Appendix C: The Belle Collaboration author list

A. Abashian[†g], K. Abe[u], K. Abe[cg], N. Abe[ci], R. Abe[bh], T. Abe[ch,u], I. Adachi[u], K. Adamczyk[s], B. S. Ahn[ar], H. S. Ahn[by], H. Aihara[j], K. Akai[u], M. Akatsu[q], M. Akemoto[u], R. Akhmetshin[i,b], J. P. Alexander[u], G. Alimonti[cq], Q. An[cv], D. Anipko[b], K. Aoki[u], K. Arinstein[b,bj], K. Asai[bc], M. Asai[v], Y. Asano[cx], D. M. Asner[bn], T. Aso[cm], V. Aulchenko[b,bj], T. Aushev[l,ah], T. Aziz[cd], S. Bahinipati[co,aa], A. M. Bakich[bx], A. Bala[bo], V. Balagura[ah], Y. Ban[bp], E. Banas[s], S. Banerjee[cd], E. Barberio[bw], M. Barbero[cq], M. Barrett[cq], W. Bartel[k,u], A. Bay[l], I. Bedny[b,bj], S. Behari[u], P. K. Behera[cy], D. Beiline[b], K. Belous[af], V. Bhardwaj[bo,bc], B. Bhuyan[y], M. Bischofberger[bc], U. Bitenc[al], I. Bizjak[al], S. Blyth[bd,bg,i], A. Bondar[b,bj], G. Bonvicini[cz], A. Bozek[s], M. Bračko[cs,al,u], J. Brodzicka[s,u], T. E. Browder[cq], B. C. K. Casey[cq], M. C. Chang[h,i,ch], P. Chang[i], Y. H. Chang[bd], Y. W. Chang[i], Y. Chao[i], V. Chekelian[aw], A. Chen[bd], H. F. Chen[cv], K. F. Chen[i], J. -H. Chen[i], P. Chen[i], W. T. Chen[bd], Y. Q. Chen[i], B. G. Cheon[e,t,cb], C. C. Chiang[i], S. Chidzik[bq], K. Chilikin[ah], R. Chistov[ah], I. S. Cho[dc], K. Cho[aq], V. Chobanova[aw], K. S. Choi[dc], S. K. Choi[r], Y. K. Choi[cb], Y. Choi[cb], P. H. Chu[i], A. Chuvikov[bq], D. Cinabro[cz], S. Cole[bx], J. Crnkovic[cr], J. Dalseno[bw,u,aw,m], M. Danilov[ah,ay], A. Das[cd], M. Dash[g], J. Dingfelder[cn], L. Dneprovsky[†b], Y. Doi[u], Z. Doležal[n], L. Y. Dong[ai], R. Dowd[bw], Z. Drásal[n], J. Dragic[bw,u], A. Drutskoy[co,ah,ay], Y. C. Duh[h], Y. T. Duh[i], W. Dungel[aj], D. Dutta[y], S. Eidelman[b,bj], V. Eyges[ah], Y. Enari[q], R. Enomoto[u], D. Epifanov[b,bj,j], S. Esen[co], C. W. Everton[bw], F. Fang[cq], H. Farhat[cz], J. E. Fast[bn], M. Feindt[ae], T. Ferber[k], R. E. Fernholz[bq], J. Flanagan[u], S. Fratina[al], A. Frey[w], M. Friedl[aj], H. Fujii[u], M. Fujikawa[bc], Y. Fujita[u], Y. Fujiyama[ci], C. Fukunaga[cj], M. Fukushima[u], Y. Funahashi[u], Y. Funakoshi[u], K. Furukawa[u], N. Gabyshev[b,bj,u], S. Ganguly[cz], A. Garmash[b,bj,ce,bq], V. Gaur[cd], T. J. Gershon[u], R. Gillard[cz], F. Giordano[cr], R. Glattauer[aj], A. Go[bd], Y. M. Goh[t], G. Gokhroo[cd], P. Goldenzweig[co], B. Golob[o,al], A. Gordon[bw], A. Gorišek[al], V. I. Goriletsky[ag], K. Gotow[g], B.V. Grinyov[ag], H. Guler[cq,ax], R. S. Guo[be], H. C. Ha[ar], H. Ha[ar], J. Haba[u], C. Hagner[g], F. Haitani[cg], T. Haji[j], H. Hamasaki[u], B. Y. Han[ar], Y. L. Han[ai], H. Hanada[ch], K. Hanagaki[bq], F. Handa[ch], K. Hara[u,q,bm], T. Hara[u,bm], Y. Harada[bh], B. Harrop[bq], T. Haruyama[u], Y. Hasegawa[bz], N. C. Hastings[bw,u,j], K. Hasuko[bu], K. Hayasaka[ap,q], K. Hayashi[u], H. Hayashii[bc], M. Hazumi[u,bm], E. M. Heenan[bw], D. Heffernan[bm], Y. Higashi[u], Y. Higasino[q], I. Higuchi[ch], T. Higuchi[ao,u], S. Hikita[ck], L. Hinz[l], T. Hirai[ci], H. Hirano[ck], N. Hitomi[u], C. T. Hoi[i], T. Hojo[bm], T. Hokuue[q], Y. Horii[ch,ap], Y. Hoshi[cg], K. Hoshina[ck], S. Hou[i,bd], W. S. Hou[i], Y. B. Hsiung[i], C. L. Hsu[i], S. C. Hsu[i], H. C. Huang[i], T. J. Huang[i], Y. C. Huang[be], H. J. Hyun[at], S. Ichizawa[ci], T. Igaki[q], A. Igarashi[cx], S. Igarashi[u], Y. Igarashi[u], T. Iijima[ap,q], K. Ikado[q], H. Ikeda[u], H. Ikeda[u,bc], K. Ikeda[bc], K. Inami[q], Y. Inoue[bl], A. Ishikawa[u], A. Ishikawa[q,bv,ch,j], H. Ishino[ci], K. Itagaki[ch], S. Itami[q], K. Itoh[j], R. Itoh[u], M. Iwabuchi[ce,dc,c], G. Iwai[bh], M. Iwai[u], S. Iwaida[cx], M. Iwamoto[c], H. Iwasaki[u], M. Iwasaki[j], Y. Iwasaki[u], T. Iwashita[bc], D. J. Jackson[bm], C. Jacoby[l], I. Jaegle[cq], P. Jalocha[s], H. K. Jang[by], C. M. Jen[i], X. B. Ji[ai], M. Jones[cq], K. K. Joo[u], N. J. Joshi[cd], N. Joshi[cd], T. Julius[bw], R. Kagan[ah], D. H. Kah[at], H. Kaji[q], S. Kajiwara[bm], H. Kakuno[j,ci,cj], T. Kameshima[cx], T. Kamitani[u], J. Kaneko[ci], J. H. Kang[dc], J. S. Kang[ar], T. Kani[q], P. Kapusta[s], K. Kasami[u], G. Katano[u], S. U. Kataoka[bc], N. Katayama[u], E. Kato[ch], Y. Kato[q], H. Kawai[c], H. Kawai[j], M. Kawai[u], N. Kawamura[a], T. Kawasaki[bm,bh], N. Kent[cq], H. R. Khan[bv,ci], A. Kibayashi[u,ci], H. Kichimi[u], C. Kiesling[aw], M. Kikuchi[u], E. Kikutani[u], B. H. Kim[by], C. H. Kim[by], D. W. Kim[cb], H. J. Kim[at], H. J. Kim[dc], H. O. Kim[at,cb], H. W. Kim[ar], J. B. Kim[ar], J. H. Kim[cb,aq], K. T. Kim[ar], M. J. Kim[at], S. K. Kim[by], S. M. Kim[cb], T. H. Kim[dc], Y. I. Kim[at], Y. J. Kim[ce,aq], K. Kinoshita[co], J. Klucar[al], B. R. Ko[ar], N. Kobayashi[bt,ci], S. Kobayashi[bv], T. Kobayashi[u], S. Koblitz[aw], P. Kodyš[n], S. Koike[u], S. Koishi[ci], H. Koiso[u], Y. Kondo[u], H. Konishi[ck], P. Koppenburg[u], K. Korotushenko[bq], S. Korpar[cs,al], R. T. Kouzes[bn], Y. Kozakai[q], M. Kreps[ae], P. Križan[o,al], P. Krokovny[b,bj,u], B. Kronenbitter[ae], T. Kubo[u], T. Kuhr[ae], R. Kulasiri[co], R. Kumar[bo,br], S. Kumar[bo], T. Kumita[cj], T. Kuniya[bv], C. C. Kuo[bd], T. -L. Kuo[i], H. Kurashiro[ci], E. Kurihara[c,u], Y. Kuroki[bm], A. Kusaka[j], A. Kuzmin[b,bj], P. Kvasnička[n], Y. J. Kwon[dc], S. H. Kyeong[dc], J. S. Lange[cp,bu,am], G. Leder[aj], J. S. Lee[cb], J. Lee[by], M. C. Lee[i], M. H. Lee[u], M. J. Lee[by], S. E. Lee[by], S. H. Lee[by,ar], Y. J. Lee[i], M. Leitgab[cr,bu], R. Leitner[n], C. Leonidopoulos[bq], T. Lesiak[s,cc], H. B. Li[ai], J. Li[cv,cq,by], X. Li[by], Y. Li[g], J. Libby[z], C. L. Lim[dc], A. Limosani[bw,u], J. Y. Lin[h], S. W. Lin[i], Y. S. Lin[i], C. Liu[cv], H. M. Liu[ai], T. Liu[bq], Y. Liu[ce,q,co,i], Z. Q. Liu[ai], D. Liventsev[ah,u], R. Louvot[l], R. S. Lu[i], P. Lukin[b,bj], O. Lutz[ae], V. R. Lyubinsky[ag], J. MacNaughton[u,aj], G. Majumder[cd], Y. Makida[u], H. Mamada[ck], A. Manabe[u], F. Mandl[aj], Z. P. Mao[ai], D. Marlow[bq], M. Masuzawa[u], T. Matsubara[j], T. Matsuda[ba], Takeshi Matsuda[u], H. Matsumoto[bh], S. Matsumoto[f], T. Matsumoto[cj,q], H. Matsuo[†as], D. Matvienko[b,bj], A. Matyja[s], S. McOnie[bx], T. Medvedeva[ah], S. Michizono[u], Y. Mikami[ch], T. Mimashi[u], C. Mindas[bq], W. Mitaroff[aj], K. Miyabayashi[bc], H. Miyake[bm], H. Miyata[bh], Y. Miyazaki[q], R. Mizuk[ah,ay,az], L. C. Moffitt[bw], G. B. Mohanty[cd], A. Mohapatra[cy], D. Mohapatra[bn,g], A. Moll[aw,m], G. R. Moloney[bw], G. F. Moorhead[bw], N. Morgan[g], S. Mori[cx], T. Mori[f,q,ci], J. Mueller[u,cu], A. Murakami[bv], T. Murakami[u], N. Muramatsu[bm,bt,bs], R. Mussa[ac,ad], I. Nagai[q], T. Nagamine[ch], Y. Nagasaka[v], Y. Nagashima[bm], S. Nagayama[u], T. Nakadaira[j], Y. Nakahama[j], M. Nakajima[ch], T. Nakajima[ch], I. Nakamura[u], T. T. Nakamura[u], T. Nakamura[ci], E. Nakano[bl], M. Nakao[u], H. Nakayama[u], H. Nakazawa[f,ce,bd], J. W. Nam[cb], S. Narita[ch], Z. Natkaniec[s], M. Nayak[z], E. Nedelkovska[aw], K. Neichi[cg], N. K. Nisar[cd], S. Neubauer[ae], C. Ng[j], C. Niebuhr[k], M. Niiyama[as], S. Nishida[u,as], K. Nishimura[cq], Y. Nishio[q], O. Nitoh[ck],

S. Noguchi[bc], T. Nomura[as], S. Nozaki[ch], T. Nozaki[u], A. Ogawa[bu], K. Ogawa[u], S. Ogawa[cf], Y. Ogawa[u], R. Ohkubo[u], K. Ohmi[u], Y. Ohnishi[u], F. Ohno[ci], T. Ohshima[q], Y. Ohshima[ci], N. Ohuchi[u], K. Oide[u], N. Oishi[q], T. Okabe[q], N. Okazaki[ck], T. Okazaki[bc], S. Okuno[an], S. L. Olsen[by,cq,ai], S. Ono[ci], Y. Onuki[bh,bu,ch,j], T. Ooba[c], T. Oshima[q], W. Ostrowicz[s], C. Oswald[cn], H. Ozaki[u], P. Pakhlov[ah,ay,az], G. Pakhlova[ah], H. Palka[†s], A. I. Panova[ag], E. Panzenböck[bc,w], C. S. Park[by], C. W. Park[ar,cb], H. K. Park[at], H. Park[at], K. S. Park[cb], N. Parslow[bx], L. S. Peak[bx], T. K. Pedlar[av], C. C. Peng[i], J. C. Peng[i], K. C. Peng[i], T. Peng[cv], M. Grosse Perdekamp[cr,bu], M. Pernicka[†aj], J.-P. Perroud[l], R. Pestotnik[al], M. Peters[cq], M. Petrič[al], L. E. Piilonen[g], A. Poluektov[b,bj], E. Prebys[bq], M. Prim[ae], K. Prothmann[aw,m], M. Röhrken[ae], J. Raaf[co], R. Rabberman[bq], B. Reisert[aw], M. Ritter[aw], J. L. Rodriguez[cq], L. Romanov[b], F. J. Ronga[l,u], N. Root[b], M. Rosen[cq], A. Rostomyan[k], M. Rozanska[s], K. Rybicki[s], S. Ryu[by], J. Ryuko[bm], H. Sagawa[u], H. Sahoo[cq], S. Sahu[i], M. Saigo[ch], T. Saito[ch], S. Saitoh[c,ce], K. Sakai[u,bh], Y. Sakai[u], H. Sakamoto[as], H. Sakaue[bl], S. Sandilya[cd], W. Sands[bq], M. Sanpei[cg], D. Santel[co], L. Santelj[al], T. Sanuki[ch], T. R. Sarangi[cy,ce], T. Sasaki[u], N. Sasao[as], M. Satapathy[cy], Noriaki Sato[q], Nobuhiko Sato[u], Y. Sato[ch], N. Satoyama[bz], A. Satpathy[u,co], V. Savinov[cu], K. Sayeed[co], P. Schönmeier[ch], J. Schümann[u,bg,i], T. Schietinger[l], S. Schmid[aj], O. Schneider[l], G. Schnell[cw,x], S. Schrenk[g,co], C. Schwanda[u,aj], A. J. Schwartz[co], R. Seidl[cr,bu], T. Seki[cj], A.I. Sekiya[bc], S. Semenov[ah], D. Semmler[am], K. Senyo[da,q], O. Seon[q], Y. Settai[f], R. Seuster[cq], M. E. Sevior[bw], K. V. Shakhova[ag], L. Shang[ai], M. Shapkin[af], V. Shebalin[b,bj], C. P. Shen[cq,ai,q], D. Z. Shen[d], Y. T. Shen[i], T. A. Shibata[bt,ci], T. Shibata[bh], H. Shibuya[cf], T. Shidara[u], K. Shimada[bh], M. Shimoyama[bc], S. Shinomiya[bm], J. G. Shiu[i], L. I. Shpilinskaya[ag], B. Shwartz[b,bj], A. Sibidanov[bx], A. Sidorov[b], V. Sidorov[†b], V. Siegle[bu], F. Simon[aw,m], J. B. Singh[bo], R. Sinha[ak], P. Smerkol[al], Y. S. Sohn[dc], A. Sokolov[af], E. Solovieva[ah], A. Somov[co], N. Soni[bo], R. Stamen[u], S. Stanič[ct,u,cx], M. Starič[al], M. Steder[k], H. Steininger[aj], R. Stock[cp], H. Stoeck[bx], J. Stypula[s], R. Suda[cj], R. Sugahara[u], A. Sugi[q], T. Sugimura[u], A. Sugiyama[q,bv], S. Suitoh[q], M. Sumihama[bt,p], K. Sumisawa[bm,u], T. Sumiyoshi[u,cj], H. F. Sung[i], Y. Susaki[q], J. I. Suzuki[u], J. Suzuki[u], K. Suzuki[c,u,q], S. Y. Suzuki[u], S. Suzuki[db,bv,q], S. K. Swain[cy,cq], M. Tabata[c], H. Tajima[j], O. Tajima[ch,u], K. Takahashi[ci], S. Takahashi[bh], T. Takahashi[bl], F. Takasaki[u], T. Takayama[ch], M. Takita[bm], K. Tamai[u], U. Tamponi[ac,ad], N. Tamura[bk,bh], N. Tan[cl], K. Tanabe[j], J. Tanaka[j], M. Tanaka[u], S. Tanaka[u], Y. Tanaka[bb], K. Tanida[by], N. Taniguchi[u,as], G. Tatishvili[bn], T. Tatomi[u], M. Tawada[u], G. N. Taylor[bw], Y. Teramoto[bl], F. Thorne[aj], X. C. Tian[bp], I. Tikhomirov[ah], M. Tomoto[u,q], T. Tomura[j], S. N. Tovey[bw], K. Trabelsi[u,cq,bm], W. Trischuk[bq], K. L. Tsai[i], Y. T. Tsai[i], T. Tsuboyama[u], Y. Tsujita[cx], K. Tsukada[u], T. Tsukamoto[u], Y. W. Tung[i], K. Uchida[cq], M. Uchida[bt,ci], Y. Uchida[ce], S. Uehara[u], M. Ueki[ch], K. Ueno[u], K. Ueno[i], T. Uglov[ah,ay], N. Ujiie[u], Y. Unno[e,c,t,u], S. Uno[u], P. Urquijo[bw,cn], Y. Ushiroda[u,as], Y. Usov[b,bj], Y. Usuki[q], S. E. Vahsen[bq,cq], P. Vanhoefer[aw], C. Van Hulse[cw], G. Varner[cq], K. E. Varvell[bx], Y. S. Velikzhanin[i], K. Vervink[l], S. Villa[l], E. L. Vinograd[ag], A. Vinokurova[b,bj], V. Vorobyev[b,bj], A. Vossen[ab,cr], M. N. Wagner[am], C. C. Wang[i,ai], C. H. Wang[bf,bg], J. G. Wang[g], J. Wang[bp], M. Z. Wang[i], P. Wang[ai], T. J. Wang[ai], X. L. Wang[ai,g], Y. F. Wang[cv], M. Watanabe[bh], Y. Watanabe[an,ci], R. Wedd[bw], J. T. Wei[i], E. White[co], J. Wicht[l,u], L. Widhalm[†aj], J. Wiechczynski[s], K. M. Williams[g], R. Wixted[bq], E. Won[ar,by], C. H. Wu[i], Q. L. Xie[ai], Z. Z. Xu[cv], B. D. Yabsley[bx,g], Y. Yamada[u], M. Yamaga[ch], A. Yamaguchi[ch], H. Yamaguchi[u], T. Yamaki[ca], H. Yamamoto[cq,ch], N. Yamamoto[u], S. Yamamoto[cj], T. Yamanaka[bm], H. Yamaoka[u], J. Yamaoka[cq], Y. Yamaoka[u], Y. Yamashita[bi,j], M. Yamauchi[u], D. S. Yan[d], H. Yanai[bh], S. Yanaka[ci], H. Yang[by], R. Yang[bq], S. Yashchenko[k], J. Yashima[u], Y. Yasu[u], S. W. Ye[cv], P. Yeh[i], Z. W. Yin[d], J. Ying[bp], K. Yokoyama[u], M. Yokoyama[j], T. Yokoyama[ck], K. Yoshida[q], M. Yoshida[u], Y. Yoshimura[u], C. X. Yu[ai], C. Z. Yuan[ai], Y. Yuan[ai], Y. Yusa[bh,ch,g], H. Yuta[a], D. Zürcher[l], D. Zander[ae], S. L. Zang[ai], B. G. Zaslavsky[ag], C. C. Zhang[ai], J. Zhang[u,cx], L. M. Zhang[cv], S. Q. Zhang[ai], Z. P. Zhang[cv], H. W. Zhao[ai,u], Z. G. Zhao[cv], Y. H. Zheng[cq], Z. P. Zheng[ai], V. Zhilich[b,bj], P. Zhou[cz], Z. M. Zhu[bp], V. Zhulanov[b,bj], T. Ziegler[bq], A. Zupanc[ae,al], N. Zwahlen[l], O. Zyukova[b,bj], T. Živko[al], D. Žontar[o,al,u,cx]

[a] *Aomori University, Aomori 030-0943, Japan*
[b] *Budker Institute of Nuclear Physics SB RAS, Novosibirsk 630090, Russian Federation*
[c] *Chiba University, Chiba 263-8522, Japan*
[d] *Chinese Academy of Science, Beijing 100864, PR China*
[e] *Chonnam National University, Gwangju 500-757, South Korea*
[f] *Chuo University, Tokyo 192-0393, Japan*
[g] *Virginia Polytechnic Institute and State University, Blacksburg, VA 24061, USA*
[h] *Department of Physics, Fu Jen Catholic University, Taipei 24205, Taiwan*
[i] *Department of Physics, National Taiwan University, Taipei 10617, Taiwan*
[j] *Department of Physics, University of Tokyo, Tokyo 113-0033, Japan*
[k] *Deutsches Elektronen-Synchrotron, 22607 Hamburg, Germany*
[l] *École Polytechnique Fédérale de Lausanne (EPFL), 1015 Lausanne, Switzerland*
[m] *Excellence Cluster Universe, Technische Universität München, 85748 Garching, Germany*
[n] *Faculty of Mathematics and Physics, Charles University, 121 16 Prague, The Czech Republic*
[o] *Faculty of Mathematics and Physics, University of Ljubljana, 1000 Ljubljana, Slovenia*
[p] *Gifu University, Gifu 501-1193, Japan*
[q] *Graduate School of Science, Nagoya University, Nagoya 464-8602, Japan*
[r] *Gyeongsang National University, Chinju 660-701, South Korea*

[s]H. Niewodniczanski Institute of Nuclear Physics, Krakow 31-342, Poland
[t]Hanyang University, Seoul 133-791, South Korea
[u]High Energy Accelerator Research Organization (KEK), Tsukuba 305-0801, Japan
[v]Hiroshima Institute of Technology, Hiroshima 731-5193, Japan
[w]II. Physikalisches Institut, Georg-August-Universität Göttingen, 37073 Göttingen, Germany
[x]Ikerbasque, 48011 Bilbao, Spain
[y]Indian Institute of Technology Guwahati, Assam 781039, India
[z]Indian Institute of Technology Madras, Chennai 600036, India
[aa]Indian Institute of Technology Bhubaneswar, SatyaNagar, 751007, India
[ab]Indiana University, Bloomington, IN 47408, USA
[ac]INFN - Sezione di Torino, 10125 Torino, Italy
[ad]Dipartimento di Fisica, Università di Torino, I-10125 Torino, Italy
[ae]Institut für Experimentelle Kernphysik, Karlsruher Institut für Technologie, 76131 Karlsruhe, Germany
[af]Institute for High Energy Physics, Protvino 142281, Russian Federation
[ag]Institute for Single Crystals, National Academy of Sciences of Ukraine, Kharkov 61001, Ukraine
[ah]Institute for Theoretical and Experimental Physics, Moscow 117218, Russian Federation
[ai]Institute of High Energy Physics, Chinese Academy of Sciences, Beijing 100049, PR China
[aj]Institute of High Energy Physics, 1050 Vienna, Austria
[ak]Institute of Mathematical Sciences, Chennai 600113, India
[al]J. Stefan Institute, 1000 Ljubljana, Slovenia
[am]Justus-Liebig-Universität Gießen, 35392 Gießen, Germany
[an]Kanagawa University, Yokohama 221-8686, Japan
[ao]Kavli Institute for the Physics and Mathematics of the Universe (WPI), University of Tokyo, Kashiwa 277-8583, Japan
[ap]Kobayashi-Maskawa Institute, Nagoya University, Nagoya 464-8602, Japan
[aq]Korea Institute of Science and Technology Information, Daejeon 305-806, South Korea
[ar]Korea University, Seoul 136-713, South Korea
[as]Kyoto University, Kyoto 606-8502, Japan
[at]Kyungpook National University, Daegu 702-701, South Korea
[au]Lawrence Berkeley National Laboratory, Berkeley, CA 94720, USA
[av]Luther College, Decorah, IA 52101, USA
[aw]Max-Planck-Institut für Physik, 80805 München, Germany
[ax]McGill University, Montréal H3A 0G4, Canada
[ay]Moscow Institute of Physics and Technology, Moscow Region 141700, Russian Federation
[az]Moscow Physical Engineering Institute, Moscow 115409, Russian Federation
[ba]University of Miyazaki, Miyazaki 889-2192, Japan
[bb]Nagasaki Institute of Applied Science, Nagasaki 851-0123, Japan
[bc]Nara Women's University, Nara 630-8506, Japan
[bd]National Central University, Chung-li 32054, Taiwan
[be]National Kaohsiung Normal University, Kaohsiung 80201, Taiwan
[bf]National Lien-Ho Institute of Technology, Miao Li 360, Taiwan
[bg]National United University, Miao Li 36003, Taiwan
[bh]Niigata University, Niigata 950-2181, Japan
[bi]Nippon Dental University, Niigata 951-8580, Japan
[bj]Novosibirsk State University, Novosibirsk 630090, Russian Federation
[bk] Okayama University, Okayama 700-8530, Japan
[bl]Osaka City University, Osaka 558-8585, Japan
[bm]Osaka University, Osaka 565-0871, Japan
[bn]Pacific Northwest National Laboratory, Richland, WA 99352, USA
[bo]Panjab University, Chandigarh 160014, India
[bp]Peking University, Beijing 100871, PR China
[bq]Princeton University, Princeton, NJ 08542, USA
[br]Punjab Agricultural University, Ludhiana 141004, India
[bs]Research Center for Electron Photon Science, Tohoku University, Sendai 980-8578, Japan
[bt]Research Center for Nuclear Physics, Osaka University, Osaka 567-0047, Japan
[bu]RIKEN BNL Research Center, Brookhaven, NY 11973, USA
[bv]Saga University, Saga 840-8502, Japan
[bw]School of Physics, University of Melbourne, Victoria 3010, Australia
[bx]School of Physics, University of Sydney, NSW 2006, Australia
[by]Seoul National University, Seoul 151-742, South Korea
[bz]Shinshu University, Nagano 390-8621, Japan
[ca]Sugiyama Jogakuen University, Aichi 470-0131, Japan
[cb]Sungkyunkwan University, Suwon 440-746, South Korea
[cc]T. Kościuszko Cracow University of Technology, Krakow 31-342, Poland
[cd]Tata Institute of Fundamental Research, Mumbai 400005, India
[ce]The Graduate University for Advanced Studies, Hayama 240-0193, Japan
[cf]Toho University, Funabashi 274-8510, Japan
[cg]Tohoku Gakuin University, Tagajo 985-8537, Japan
[ch]Tohoku University, Sendai 980-8578, Japan
[ci]Tokyo Institute of Technology, Tokyo 152-8550, Japan
[cj]Tokyo Metropolitan University, Tokyo 192-0397, Japan
[ck]Tokyo University of Agriculture and Technology, Tokyo 184-8588, Japan

[cl] Tokyo University of Science, Chiba 278-8510, Japan
[cm] Toyama National College of Maritime Technology, Toyama 933-0293, Japan
[cn] University of Bonn, 53115 Bonn, Germany
[co] University of Cincinnati, Cincinnati, OH 45221, USA
[cp] University of Frankfurt, 60318 Frankfurt am Main, Germany
[cq] University of Hawaii, Honolulu, HI 96822, USA
[cr] University of Illinois at Urbana-Champaign, Urbana, IL 61801, USA
[cs] University of Maribor, 2000 Maribor, Slovenia
[ct] University of Nova Gorica, 5000 Nova Gorica, Slovenia
[cu] University of Pittsburgh, Pittsburgh, PA 15260, USA
[cv] University of Science and Technology of China, Hefei 230026, PR China
[cw] University of the Basque Country UPV/EHU, 48080 Bilbao, Spain
[cx] University of Tsukuba, Tsukuba 305-0801, Japan
[cy] Utkal University, Bhubaneswar, India
[cz] Wayne State University, Detroit, MI 48202, USA
[da] Yamagata University, Yamagata 990-8560, Japan
[db] Yokkaichi University, Yokkaichi 512-8045, Japan
[dc] Yonsei University, Seoul 120-749, South Korea

[†] Deceased

# Appendix D
# Acknowledgments

The preparation of this book has been directly supported by the US Department of Energy, MEXT (Japan), the Natural Sciences and Engineering Research Council (Canada), and Commissariat à l'Energie Atomique and Institut National de Physique Nucléaire et de Physique des Particules (France). Individuals have been supported by the Royal Society (UK).

The authors of this book wish to thank the KEK and SLAC laboratories for their support with regard to preparation of this manuscript, and in particular providing meeting and computing facilities to aid preparation of this manuscript. In addition we wish to thank the following institutes who have hosted meetings of the book collaboration and general editors: Johannes Gutenberg Universität, Mainz; Iowa State University; LAAP; University of Siegen; University of Ljubljana; Queen Mary, University of London, and the University of Melbourne. We would also like to thank the reprographics department at Queen Mary, University of London for their contribution in preparation of the artwork for the cover page of this book. Adrian Bevan and Soeren Prell would specifically like to thank Patricia Burchat, Francois Le Diberder, and A.J. Stewart Smith for their counsel as members of the *BABAR* advisory committee on preparations for this book and J. Michael Roney and David B. MacFarlane for their support as *BABAR* spokesperson and SLAC director of particle physics and astrophysics during the writing of the book. Boštjan Golob and Bruce Yabsley would specifically like to thank Tom Browder, Hisaki Hayashii, Toru Iijima, Leo Piilonen, Yoshihide Sakai and Masanori Yamauchi for their support of the project in the role of Belle spokespersons. We would also like to thank Nittsia Harrison, Donna Hernandez, Chihiro Imai, Homer Neal, and Shinobu Oishi, for their local support for meetings, Homer Neal and Charlotte Hee for computing support at SLAC and in particular Sarodia Vydelingum for her many efforts with regard to organizing meetings, websites and travel arrangements during this project.

The *BABAR* and Belle authors are grateful for the tremendous support they have received from their home institutions during the operation of the $B$ Factories.

The *BABAR* Collaboration is grateful for the extraordinary contributions of their PEP-II colleagues in achieving the excellent luminosity and machine conditions that have made this work possible. The success of this project also relies critically on the expertise and dedication of the computing organizations that support *BABAR*. The collaborating institutions wish to thank SLAC for its support and the kind hospitality extended to them. This work is supported by the US Department of Energy and National Science Foundation, the Natural Sciences and Engineering Research Council (Canada), the Commissariat à l'Energie Atomique and Institut National de Physique Nucléaire et de Physique des Particules (France), the Bundesministerium für Bildung und Forschung and Deutsche Forschungsgemeinschaft (Germany), the Istituto Nazionale di Fisica Nucleare (Italy), the Foundation for Fundamental Research on Matter (The Netherlands), the Research Council of Norway, the Ministry of Education and Science of the Russian Federation, Ministerio de Economía y Competitividad (Spain), the Science and Technology Facilities Council (United Kingdom), and the Binational Science Foundation (U.S.-Israel). Individuals have received support from the Marie-Curie IEF program (European Union) and the A. P. Sloan Foundation (USA).

The Belle Collaboration wishes to thank the KEKB group for the excellent operation of the accelerator; the KEK cryogenics group for the efficient operation of the solenoid; and the KEK computer group, the National Institute of Informatics, and the PNNL/EMSL computing group for valuable computing and SINET4 network support. We acknowledge support from the Ministry of Education, Culture, Sports, Science, and Technology (MEXT) of Japan, the Japan Society for the Promotion of Science (JSPS), and the Tau-Lepton Physics Research Center of Nagoya University; the Australian Research Council and the Australian Department of Industry, Innovation, Science and Research; Austrian Science Fund under Grant No. P 22742-N16; the National Natural Science Foundation of China under Contracts No. 10575109, No. 10775142, No. 10825524, No. 10875115, No. 10935008 and No. 11175187; the Ministry of Education, Youth and Sports of the Czech Republic under Contract No. LG14034; the Carl Zeiss Foundation, the Deutsche Forschungsgemeinschaft and the Volkswagen-Stiftung; the Department of Science and Technology of India; the Istituto Nazionale di Fisica Nucleare of Italy; the WCU program of the Ministry of Education, Science and Technology, National Research Foundation of Korea Grants No. 2011-0029457, No. 2012-0008143, No. 2012R1A1A2008330, No. 2013R1A1A3007772; the BRL program under NRF Grant No. KRF-2011-0020333, No. KRF-2011-0021196, Center for Korean J-PARC Users, No. NRF-2013K1A3A7A06056592; the BK21 Plus program and the GSDC of the Korea Institute of Science and Technology Information; the Polish Ministry of Science and Higher Education and the National Science Center; the Ministry of Education and Science of the Russian Federation and the Russian Federal Agency for Atomic Energy; the Slovenian Research Agency; the Basque Foundation for Science (IKERBASQUE) and the UPV/EHU under program UFI 11/55; the Swiss National Science Foundation; the National Science Council and the Ministry of Education of Taiwan; and the U.S. Department of Energy and the National Science Foundation. This work is supported by a Grant-in-Aid from MEXT for Science Research in a Priority Area ("New Development of Flavor Physics") and from JSPS for Creative Scientific Research ("Evolution of Tau-lepton Physics").

## BaBar publications

## Belle publications

# Index